\documentclass[12pt,openany]{book}
\usepackage[russian]{babel}
\usepackage[cp1251]{inputenc}
\usepackage{amscd,amsmath,amsxtra,amssymb,amstext,latexsym,amsthm}
\usepackage{dsfont,diagrams,exscale}
\usepackage{array,graphicx,bbm}
\usepackage{makeidx,multicol}
\makeindex\textwidth160mm
\newcommand {\al}   {\alpha}       \newcommand {\bt}  {\beta}
\newcommand {\g }   {\gamma}       
\newcommand {\dl}   {\delta}       \newcommand {\e }  {\epsilon}
\newcommand {\z }   {\zeta}        \newcommand {\et}  {\eta}
\newcommand {\ve}   {\varepsilon}  
\newcommand {\lm}   {\lambda}      
          \newcommand {\x }  {\xi}
\newcommand {\s }   {\sigma}       \newcommand {\vr } {\varrho}
\newcommand {\ta}   {\tau}         
\newcommand {\vf }  {\varphi}      
         \newcommand {\om}  {\omega}
\newcommand {\Lm}   {\Lambda}      \newcommand {\Om}  {\Omega}
       
\newcommand {\pl}   {\partial}     \newcommand {\nb}  {\nabla}
\renewcommand {\sin}{{\sf\,sin\,}}       \renewcommand {\cos}{{\sf\,cos\,}}
\renewcommand {\tg}{{\sf\,tg\,}}         \renewcommand {\ctg}{{\sf\,ctg\,}}
\renewcommand {\arcsin}{{\sf\,arcsin\,}}
\renewcommand {\arctg}{{\sf\,arctg\,}}
\renewcommand {\sh}{{\sf\,sh\,}}         \renewcommand {\ch}{{\sf\,ch\,}}
\renewcommand {\tanh}{{\sf\,th\,}}       \renewcommand {\cth}{{\sf\,cth\,}}
   
\newcommand   {\arcth}{{\sf\,arcth\,}}  \newcommand   {\arccth}{{\sf\,arccth\,}}
\newcommand   {\class}{{\sf\,class\,}}
\renewcommand {\min}{{\sf\,min\,}}       \renewcommand {\max}{{\sf\,max\,}}
\renewcommand {\inf}{{\sf\,inf\,}}       \renewcommand {\sup}{{\sf\,sup\,}}
\renewcommand {\ln}{{\sf\,ln}}           
\renewcommand {\det}{{\sf\,det\,}}       \renewcommand {\exp}{{\sf\,exp\,}}
\renewcommand {\dim}{{\sf\,dim\,}}       \renewcommand {\deg}{{\sf\,deg\,}}
\renewcommand {\ker}{{\sf\,ker\,}}       \renewcommand {\mod}{{\sf\,mod\,}}
\renewcommand {\div}{{\sf\,div\,}}       \renewcommand {\lim}{{\sf\,lim\,}}
\renewcommand {\arg}{{\sf\,arg\,}}   
\newcommand   {\ex}{{\sf\,e}}            \newcommand   {\pr}{{\sf\,pr}}

\newcommand   {\re}{{\sf\,re\,}}         \newcommand   {\im}{{\sf\,im\,}}
\newcommand   {\sdet}{{\sf\,sdet\,}}     \newcommand   {\str}{{\sf\,str\,}}
\newcommand   {\aut}{{\sf\,aut\,}}       \newcommand   {\End}{{\sf\,end\,}}
\newcommand   {\sign}{{\sf\,sign\,}}     \newcommand   {\sgn}{{\sf\,sgn\,}}
\newcommand   {\grad}{{\sf\,grad\,}}     \newcommand   {\rot}{{\sf\,rot\,}}
         \newcommand   {\Lie}{{\sf\,L}}
\newcommand   {\Int}{{\sf\,int\,}}       \newcommand   {\supp}{{\sf\,supp\,}}
\newcommand   {\const}{{\sf\,const}}     \newcommand   {\diag}{{\sf\,diag\,}}
\newcommand   {\id}{{\sf\,id\,}}         \newcommand   {\tr}{{\sf\,tr\,}}
\newcommand   {\diff}{{\sf\,diff\,}}     \newcommand   {\rank}{{\sf\,rank\,}}
\newcommand   {\ad}{{\sf\,ad\,}}         
\newcommand   {\inm}{{\sf\,i}}           
\newcommand   {\rel}{{\sf\,rel}}         \newcommand   {\gr}{{\sf\,gr\,}}
\newcommand   {\hor}{{\sf h}}           \newcommand   {\ver}{{\sf v}}
             \renewcommand   {\P}{{\sf P}}
\newcommand   {\loc}{{\sf loc}}         \newcommand   {\nf}{{\sf inf}}
\newcommand   {\osmall}{{\sf o}}        \newcommand   {\obig}{{\sf O}}

\newcommand {\MA}  {{\mathbb A}}   \newcommand {\MB}  {{\mathbb B}}
\newcommand {\MC}  {{\mathbb C}}   \newcommand {\MD}  {{\mathbb D}}
\newcommand {\ME}  {{\mathbb E}}   \newcommand {\MF}  {{\mathbb F}}
\newcommand {\MG}  {{\mathbb G}}   \newcommand {\MH}  {{\mathbb H}}
\newcommand {\MI}  {{\mathbb I}}   \newcommand {\MJ}  {{\mathbb J}}
\newcommand {\MK}  {{\mathbb K}}   \newcommand {\ML}  {{\mathbb L}}
\newcommand {\MM}  {{\mathbb M}}   \newcommand {\MN}  {{\mathbb N}}
\newcommand {\MO}  {{\mathbb O}}   \newcommand {\MP}  {{\mathbb P}}
\newcommand {\MQ}  {{\mathbb Q}}   \newcommand {\MR}  {{\mathbb R}}
\newcommand {\MS}  {{\mathbb S}}   \newcommand {\MT}  {{\mathbb T}}
\newcommand {\MU}  {{\mathbb U}}   \newcommand {\MV}  {{\mathbb V}}
\newcommand {\MW}  {{\mathbb W}}   \newcommand {\MX}  {{\mathbb X}}
   \newcommand {\MZ}  {{\mathbb Z}}

\newcommand {\GA}  {\mathfrak{A}}   \newcommand {\GB}  {\mathfrak{B}}
\newcommand {\GC}  {\mathfrak{C}}   \newcommand {\GD}  {\mathfrak{D}}
\newcommand {\GE}  {\mathfrak{E}}   \newcommand {\GF}  {\mathfrak{F}}
\newcommand {\GG}  {\mathfrak{G}}   \newcommand {\GH}  {\mathfrak{H}}
\newcommand {\GI}  {\mathfrak{I}}   \newcommand {\GJ}  {\mathfrak{J}}
\newcommand {\GK}  {\mathfrak{K}}   \newcommand {\GL}  {\mathfrak{L}}
\newcommand {\GM}  {\mathfrak{M}}   \newcommand {\GN}  {\mathfrak{N}}
\newcommand {\GO}  {\mathfrak{O}}   \newcommand {\GP}  {\mathfrak{P}}
\newcommand {\GQ}  {\mathfrak{Q}}   \newcommand {\GR}  {\mathfrak{R}}
\newcommand {\GS}  {\mathfrak{S}}   \newcommand {\GT}  {\mathfrak{T}}
\newcommand {\GU}  {\mathfrak{U}}   \newcommand {\GV}  {\mathfrak{V}}
\newcommand {\GW}  {\mathfrak{W}}   \newcommand {\GX}  {\mathfrak{X}}
\newcommand {\GY}  {\mathfrak{Y}}   \newcommand {\GZ}  {\mathfrak{Z}}

\newcommand {\Ga}  {\mathfrak{a}}   \newcommand {\Gb}  {\mathfrak{b}}
\newcommand {\Gc}  {\mathfrak{c}}   \newcommand {\Gd}  {\mathfrak{d}}
\newcommand {\Ge}  {\mathfrak{e}}   \newcommand {\Gf}  {\mathfrak{f}}
\newcommand {\Gg}  {\mathfrak{g}}   \newcommand {\Gh}  {\mathfrak{h}}
\newcommand {\Gi}  {\mathfrak{i}}   \newcommand {\Gj}  {\mathfrak{j}}
\newcommand {\Gk}  {\mathfrak{k}}   \newcommand {\Gl}  {\mathfrak{l}}
\newcommand {\Gm}  {\mathfrak{m}}   \newcommand {\Gn}  {\mathfrak{n}}
\newcommand {\Go}  {\mathfrak{o}}   \newcommand {\Gp}  {\mathfrak{p}}
\newcommand {\Gq}  {\mathfrak{q}}   \newcommand {\Gr}  {\mathfrak{r}}
\newcommand {\Gs}  {\mathfrak{s}}   \newcommand {\Gt}  {\mathfrak{t}}
\newcommand {\Gu}  {\mathfrak{u}}   \newcommand {\Gv}  {\mathfrak{v}}
\newcommand {\Gw}  {\mathfrak{w}}   \newcommand {\Gx}  {\mathfrak{x}}
\newcommand {\Gy}  {\mathfrak{y}}   \newcommand {\Gz}  {\mathfrak{z}}
\newcommand {\BA}  {\boldsymbol{A}}   
   
\newcommand {\BE}  {\boldsymbol{E}}   \newcommand {\BF}  {\boldsymbol{F}}
   \newcommand {\BH}  {\boldsymbol{H}}

   \newcommand {\BT}  {\boldsymbol{T}}
   \newcommand {\BV}  {\boldsymbol{V}}

\newcommand {\Ba}  {\boldsymbol{a}}   \newcommand {\Bb}  {\boldsymbol{b}}
   
\newcommand {\Be}  {\boldsymbol{e}}   \newcommand {\Bf}  {\boldsymbol{f}}
\newcommand {\Bg}  {\boldsymbol{g}}   \newcommand {\Bh}  {\boldsymbol{h}}
   
\newcommand {\Bk}  {\boldsymbol{k}}   
   \newcommand {\Bn}  {\boldsymbol{n}}
   \newcommand {\Bp}  {\boldsymbol{p}}
\newcommand {\Bq}  {\boldsymbol{q}}   \newcommand {\Br}  {\boldsymbol{r}}
   
\newcommand {\Bu}  {\boldsymbol{u}}   \newcommand {\Bv}  {\boldsymbol{v}}
\newcommand {\Bw}  {\boldsymbol{w}}   \newcommand {\Bx}  {\boldsymbol{x}}
\newcommand {\By}  {\boldsymbol{y}}   \newcommand {\Bz}  {\boldsymbol{z}}
\newcommand {\CA }  {{\cal A}}      \newcommand {\CB}  {{\cal B}}
\newcommand {\CC }  {{\cal C}}      \renewcommand {\CD}  {{\cal D}}
\newcommand {\CE }  {{\cal E}}      \newcommand {\CF}  {{\cal F}}
\newcommand {\CG }  {{\cal G}}      \newcommand {\CH}  {{\cal H}}
\newcommand {\CI }  {{\cal I}}      \newcommand {\CJ}  {{\cal J}}
      \newcommand {\CL}  {{\cal L}}
\newcommand {\CM }  {{\cal M}}      \newcommand {\CN}  {{\cal N}}
      \newcommand {\CP}  {{\cal P}}
      
      \newcommand {\CT}  {{\cal T}}
      \newcommand {\CV}  {{\cal V}}
      \newcommand {\CX}  {{\cal X}}
      
\newcommand {\Sa}  {{\textsc{a}}}   \newcommand {\Sb}  {{\textsc{b}}}
\newcommand {\Sc}  {{\textsc{c}}}   \newcommand {\Sd}  {{\textsc{d}}}
\newcommand {\Se}  {{\textsc{e}}}   \newcommand {\Sf}  {{\textsc{f}}}
\newcommand {\Sg}  {{\textsc{g}}}   \newcommand {\Sh}  {{\textsc{h}}}
\newcommand {\Si}  {{\textsc{i}}}   \newcommand {\Sj}  {{\textsc{j}}}
\newcommand {\Sk}  {{\textsc{k}}}   \newcommand {\Sl}  {{\textsc{l}}}
\newcommand {\Sm}  {{\textsc{m}}}   \newcommand {\Sn}  {{\textsc{n}}}
   \newcommand {\Sp}  {{\textsc{p}}}
\newcommand {\Sq}  {{\textsc{q}}}   \newcommand {\Sr}  {{\textsc{r}}}
\newcommand {\Ss}  {{\textsc{s}}}   \newcommand {\St}  {{\textsc{t}}}
   
\newcommand {\Sw}  {{\textsc{w}}}   
\newcommand {\Sy}  {{\textsc{y}}}   \newcommand {\Sz}  {{\textsc{z}}}
\def\No{\hbox{\cyr\char'031\hskip2pt}}   
\newcommand {\opm}  {\lefteqn{\,\,\pm}\bigcirc}
\newcommand {\omp}  {\lefteqn{\,\,\mp}\bigcirc}
\newcommand {\one}  {\mathbbm{1}}
\newcommand {\zero} {{\it 0}}
\newcommand {\vol}  {\sqrt{|g|}}
\newtheorem{lemma}{Лемма}[section]
\newtheorem{prop}{Предложение}[section]
\newtheorem{theorem}{Теорема}[section]

\theoremstyle{definition}
\newtheorem*{cor}{Следствие}
\newtheorem*{com}{Замечание}

\newtheorem*{defn}{Определение}
\newtheorem{exa}{Пример}[section]
\begin{document}
\title     {\huge\bf Геометрические методы \\[3mm] в математической физике.
            \\[3mm] \today}
\author    {\Large Катанаев Михаил Орионович\thanks{Любые замечания, указания на
            ошибки, неточности и опечатки прошу отправлять на e-mail:
            katanaev@mi.ras.ru}\\[5mm]
    Математический институт имени В.\ А.\ Стеклова\\ Российской Академии Наук}
\maketitle

\frontmatter
\tableofcontents
\chapter{Основные обозначения и соглашения                       \label{snotat}}
Дифференцирование выполняется раньше алгебраических операций.\newline
\vskip2mm
\noindent\begin{tabular}{ll}
$\MN=\lbrace 1,2,\dotsc\rbrace$ &~-- множество натуральных чисел,\\
$\MZ=\lbrace \dotsc-1,0,1,\dotsc\rbrace$ &~-- группа целых чисел по сложению,\\
$\MQ$, $\MR$, $\MC$ &~-- поле рациональных, вещественных, комплексных чисел,\\
$\MR_+$ &~-- множество положительных вещественных чисел,\\
$\dagger$  &~-- комплексное или эрмитово сопряжение,\\
$ :=$ &~-- равно по-определению,\\
$x~\text{и}~(x^1,\dotsc,x^n)$ &~-- точка многообразия и ее координаты,\\
$\lbrace x^\al\rbrace=\lbrace x^0,x^\mu\rbrace=\lbrace x^0,\Bx\rbrace $
&~-- декартовы координаты в пространстве Минковского,\\
  &\qquad или координаты на псевдоримановом многообразии,\\
$\pl_\al:={\frac\pl{\pl x^\al}}$ &~-- частная производная,\\ \smallskip
$\pl^2_{\al\bt}:={\frac{\pl^2}{\pl x^\al\pl x^\bt}}$ &~--
частная производная второго порядка,\\  \smallskip
$\nb_\al$ &~-- ковариантная производная,\\ \smallskip
$\triangle$&~-- оператор Лапласа или Лапласа--Бельтрами,\\
$\square$&~-- оператор Даламбера, конец доказательства, примера \\
&\qquad или определения,\\  \medskip
$(\pl f)^2:=g^{\al\bt}\pl_\al f\pl_\bt f$ &~--
квадрат градиента функции $f$,\\
$g_{\al\bt}$ &~-- компоненты метрики, \\
$g:=\det(g_{\al\bt})$ &~-- определитель метрики,\\
$e_\al{}^a$ &~-- компоненты репера,\\
$\sqrt{|g|}=\det(e_\al{}^a)$ &~-- элемент объема (определитель репера),\\
$\upsilon:=dx^1\wedge\dotsc\wedge dx^n\sqrt{|g|}$ &~-- форма объема на
(псевдо-)римановом многообразии $\MM$,\\
$e$ &~-- единица группы,\\
$\ex$ &~-- основание натурального логарифма,\\
$\Gamma_{\al\bt}{}^\g$ &~-- компоненты аффинной связности,\\
$\om_{\al a}{}^b$ &~-- компоненты линейной или лоренцевой связности,\\
$\sgn\s(\al_1,\dotsc,\al_n)$ &~-- знак перестановки $\s$ индексов
$\al_1,\dotsc,\al_n$.\\
\end{tabular}

Некоторые многообразия и классы объектов имеют специальные обозначения:\newline
\vskip2mm
\begin{tabular}{ll}
$\MR^n$ &~-- $n$-мерное евклидово пространство,\\
$\MR^n_+$ &~-- подпространство в $\MR^n$, определяемое условием $x^n>0$,\\
$\MR^{1,n-1}$ &~-- $n$-мерное пространство Минковского,\\
$\MS^n_r$ &~-- $n$-мерная сфера радиуса $r$,\\
$\MB^n_r$ &~-- $n$-мерный шар радиуса $r$,\\
$\CC^k(\MM)$ &~-- класс функций на многообразии $\MM$, непрерывных вместе с
         производными\\
        &\quad вплоть до $k$-го порядка,\\
$\CX(\MM)$ &~-- множество гладких векторных полей на многообразии $\MM$,\\
$\CT(\MM)$ &~-- множество гладких тензорных полей на многообразии $\MM$,\\
$\MM\approx\MN$ &~-- многообразие $\MM$ диффеоморфно (гомеоморфно) многообразию
                  $\MN$,\\
$\MG\simeq\MH$ &~-- группа (алгебра, векторное пространство, \dots) $\MG$
                 изоморфна \\
               &\quad группе (алгебре, векторному пространству, \dots)
                  $\MH$,\\
$\MM\rightarrow\MN$ &~-- отображение множеств,\\
$\MM\ni a\mapsto b\in\MN$ &~-- отображение элементов множеств,\\
\end{tabular}

Антисимметризация по индексам обозначается квадратными скобками:
\begin{equation*}                                                 \label{eantis}
\begin{split}
  A^{[ab]}&:=\frac12(A^{ab}-A^{ba}),
\\
  A^{[abc]}&:=\frac16(A^{abc}+A^{bca}+A^{cab}-A^{bac}-A^{acb}-A^{cba}).
\end{split}
\end{equation*}
В общем случае, когда имеется $n$ индексов, сумма берется по всем $n!$
перестановкам и делится на $n!$. При этом четные перестановки индексов входят со
знаком плюс, а нечетные -- со знаком минус.

Симметризация индексов обозначается круглыми скобками:
\begin{equation*}
\begin{split}
  A^{(ab)}&:=\frac12(A^{ab}+A^{ba}),
\\
  A^{(abc)}&:=\frac16(A^{abc}+A^{bca}+A^{cab}+A^{bac}+A^{acb}+A^{cba}).
\end{split}
\end{equation*}

Символ Кронекера $\dl_a^b$ является тождественным оператором, действующим в
векторном пространстве, и также равен единичной матрице. Например, в $\MR^4$
\begin{equation}                                                  \label{ekrosy}
  \dl_a^b:=\diag(++++):=\begin{pmatrix}
  1&0&0&0\\0&1&0&0\\0&0&1&0\\0&0&0&1 \end{pmatrix}.
\end{equation}
Для краткости, произведение символов Кронекера обозначается одним символом
\begin{equation}                                                  \label{ekrmul}
  \dl_{ef\dots h}^{ab\dots d}:=\dl_e^a\dl_f^b\dots\dl_h^d.
\end{equation}
Иногда используется обобщенный символ Кронекера, помеченный шляпкой,
\begin{equation}                                                  \label{ekrmug}
  \hat\dl_{ef\dots h}^{ab\dots d}:=n!\dl_{~e}^{[a}\dl_f^b\dots\dl_h^{d]}
  =n!\dl_{[e}^{~a}\dl_f^b\dots\dl_{h]}^d,
\end{equation}
который получается из произведения (\ref{ekrmul}) антисимметризацией по
верхним или нижним индексам (что эквивалентно).

Евклидова метрика $\dl_{ab}$ имеет два нижних индекса и равна единичной
матрице. Например, в четырехмерном евклидовом пространстве $\MR^4$
\begin{equation}                                                  \label{eumetr}
  \dl_{ab}:=\diag(++++):=\begin{pmatrix}
  1&0&0&0\\0&1&0&0\\0&0&1&0\\0&0&0&1\end{pmatrix}.
\end{equation}

Метрика Минковского в четырехмерном пространстве-времени $\MR^{1,3}$ имеет вид
\begin{equation}                                                  \label{emimme}
  \et_{ab}:=\diag(+---):=\begin{pmatrix}
  1&0&0&0\\0&-1&0&0\\0&0&-1&0\\0&0&0&-1\end{pmatrix}.
\end{equation}

Каноническая симплектическая форма в евклидовом пространстве $\MR^{2n}$
имеет вид
\begin{equation}                                                  \label{ecasyf}
  \varpi:=\begin{pmatrix}0&-\one\\ \one&0\end{pmatrix},
\end{equation}
где $\one$ -- единичная $n\times n$ матрица.

\centerline{\bf Готический шрифт}
\begin{align*}
  &\GA~\Ga&&\GB~\Gb&&\GC~\Gc&&\GD~\Gd&&\GE~\Ge&&\GF~\Gf&&\GG~\Gg&&\GH~\Gh&&\GI
  ~\Gi
\\
  &\GJ~\Gj&&\GK~\Gk&&\GL~\Gl&&\GM~\Gm&&\GN~\Gn&&\GO~\Go&&\GP~\Gp&&\GQ~\Gq&&\GR
  ~\Gr
\\
  &\GS~\Gs&&\GT~\Gt&&\GU~\Gu&&\GV~\Gv&&\GW~\Gw&&\GX~\Gx&&\GY~\Gy&&\GZ~\Gz&&
\end{align*}
\centerline{\bf Используемый греческий шрифт}
\begin{align*}
  &\quad \al&&\quad \bt&&\Gamma~\g&&\Delta~\dl&&\quad\e&&\quad\ve&&\quad\zeta&
  &\quad\eta
\\
  &\Theta~\theta&&\quad \vartheta&&\quad \iota&&\quad \kappa&&\Lm~\lm&&\quad\mu&
  &\quad \nu&&\Xi~\xi
\\
  &\quad o&&\Pi~\pi&&\quad \varpi&&\quad \rho&&\quad \varrho&&\Sigma~\sigma&
  &\quad ~\varsigma&&\quad \tau
\\
  &\Upsilon~\upsilon&&\Phi~\phi&&\quad \varphi&&\quad \chi&&\Psi~\psi&&\Omega~\omega
\end{align*}
\mainmatter
\chapter{Введение                                                \label{sintro}}
После напоминания основных понятий теории множеств и введения обозначений, будет
достаточно подробно рассмотрено евклидово пространство с различных точек зрения.
Это сделано по двум причинам. Во-первых, чтобы подчеркнуть, что в евклидовом
пространстве можно задавать различные структуры, которые в дальнейшем будут
обобщаться в аффинной геометрии. Во-вторых, чтобы при изложении аффинной
геометрии не напоминать относительно сложные понятия геометрии Евклида. Далее
рассматриваются евклидовы пространства и пространства Минковского низших
размерностей, которые играют важную роль в приложениях. В заключительной главе
кратко изложены основы специальной теории относительности. При написании
Введения, которое содержит хорошо известный материал, использованы, в основном,
монографии [1--17].
\nocite{Arnold75R,ArKoNe02R,Vladim88R,Gelfan98R,GeMiSh58R,DroZav06R,DuNoFo98R,
Zharin08R,Kelley57R,Kirill78R,Kostri00R,KosMan86R,NovTai05R,Postni79R,Rashev67R,
RohFuk77R,Gantma88R}
\section{Множества                                               \label{ssetsd}}
В математике некоторые исходные понятия не имеют определения. Эти понятия
основаны на интуиции и служат для определения других, более сложных,
конструкций. Такими интуитивными понятиями являются {\em множество} и
{\em элемент множества}. Под множеством понимают произвольную совокупность
объектов, которые называются элементами множества. Говорят, что множество
состоит из своих элементов. Один из способов задания множества состоит просто
в перечислении его элементов.
\index{Множество (set)}\index{Элемент множества (element of a set)}%
\begin{exa}
Натуральные числа являются множеством, которое будем обозначать следующим
образом:
\begin{equation*}
  \MN:=\left\lbrace 1,2,\dots\right\rbrace.
\end{equation*}
Множество натуральных чисел не определяется и лежит в основе всей математики.
\qed\end{exa}
\begin{exa}
Множество целых чисел, включающее нуль и все отрицательные числа, будем
обозначать
\begin{equation*}
  \MZ:=\left\lbrace\dotsc-2,-1,0,1,2,\dotsc\right\rbrace
\end{equation*}
Как правило, мы рассматриваем множество целых чисел $\MZ$ как абелеву группу по
отношению к сложению.
\qed\end{exa}
\begin{exa}
Пусть $m,n$ -- произвольные целые числа, причем $n\ne0$. Тогда множество чисел
вида $m/n$ называется множеством рациональных чисел и обозначается $\MQ$.
\qed\end{exa}
\begin{exa}
Вещественные и комплексные числа также являются множествами. Для них приняты
обозначения $\MR$ и $\MC$.
\qed\end{exa}
Множества натуральных, целых и рациональных чисел являются бесконечными
{\em счетными множествами}, т.е.\ их элементы можно пронумеровать натуральными
числами. Множества вещественных $\MR$ и комплексных чисел $\MC$ представляют
собой примеры {\em несчетных множеств}, поскольку их элементы невозможно
пронумеровать натуральными числами. Множество называется {\em конечным}, если
оно состоит из конечного числа элементов.
\index{Счетное множество (countable set)}%
\index{Множество счетное (countable set)}%
\index{Множество несчетное (uncountable set)}%
\index{Несчетное множество (uncountable set)}%
\index{Множество конечное (finite set)}\index{Конечное множество (finite set)}%
\begin{exa}
Множество поворотов евклидовой плоскости $\MR^2$ вокруг начала координат на угол
$\pi/2$ является конечным и состоит из четырех элементов. Это -- циклическая
группа четвертого порядка.
\qed\end{exa}

В большинстве случаев для обозначения множеств, таких как многообразия, группы,
и др.\ мы будем употреблять латинские буквы, напечатанные ажурным шрифтом,
$\MA,\MB,\MC,\dotsc$. Иногда будут использоваться также заглавные буквы,
напечатанные курсивом $\CA,\CB,\CC,\dotsc$.

Синонимами понятий множества и элемента множества являются {\em пространство} и
{\em точка пространства}. Разница в употреблении терминов множество и
пространство сложилась исторически. О множестве векторов говорят, как о
векторном пространстве. Множество, снабженное топологией, называют
топологическим пространством. Часто, говоря о пространстве, подразумевают, что
оно, в отличие от множества, снабжено какой-либо дополнительной структурой, будь
то топология или структура векторного пространства. Множество элементов
называется также {\em семейством}. Как правило, термин семейство употребляется
тогда, когда его элементами являются некоторые множества.
\index{Пространство (space)}\index{Точка пространства (space point)}%
\index{Семейство (family)}%

Если элемент $x$ принадлежит множеству $\MU$, то мы пишем $x\in\MU$. В противном
случае применяется обозначение $x\notin\MU$.
\begin{defn}
Если каждый элемент множества $\MU$ принадлежит также множеству $\MV$, то $\MU$
является {\em подмножеством} $\MV$. Это обозначается $\MU\subset\MV$ или
$\MV\supset\MU$. При этом совпадение или равенство множеств $\MU=\MV$ значит,
что $\MU\subset\MV$ и $\MV\subset\MU$. Это записывается в виде
\begin{equation*}
  \MU=\MV\quad \Leftrightarrow\quad\MU\subset\MV\quad\text{и}\quad\MV\subset\MU,
\end{equation*}
где стрелка $\Leftrightarrow$ обозначает утверждение ``тогда и только тогда''.
Множество, не содержащее ни одного элемента, называется {\em пустым} и
обозначается символом $\emptyset$. Оно, по-определению, является подмножеством
любого множества, $\emptyset\subset\MU$, $\forall\MU$. Подмножество
$\MU\subset\MV$ называется {\em собственным}, если оно не пусто и не совпадает
со всем $\MV$: $\MU\ne\emptyset$ и $\MU\ne\MV$. Мы также пишем
$\MU\subseteq\MV$, если либо $\MU\subset\MV$, либо $\MU=\MV$.
\qed\end{defn}
\index{Подмножество (subset)}%
\index{Подмножество собственное (proper subset)}%
\index{Собственное подмножество (proper subset)}%
\index{Пустое множество (empty set)}\index{Пустое множество (empty set)}%
\begin{exa}
Множество целых чисел $\MZ$ имеет много подмножеств. Среди них есть
подмножества чисел, которые делятся без остатка на натуральное число $n\in\MN$,
отличное от нуля. Это подмножество обозначается $n\MZ\subset\MZ$ и получается
умножением каждого целого числа на $n$.
\qed\end{exa}

\begin{figure}[h,b,t]
\hfill\includegraphics[width=.7\textwidth]{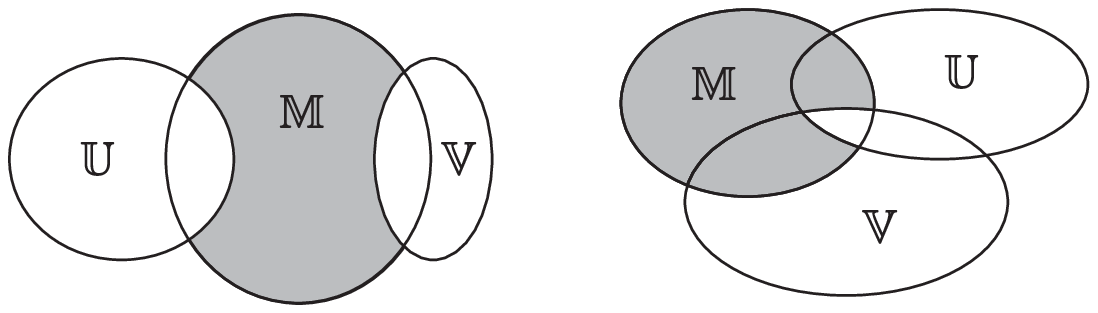}
\hfill {}
\centering \caption{Объединение, пересечение и дополнение множеств. Слева и
справа затемнены те области, которые соответствуют равенствам (\ref{emorga}).
\label{funins}}
\end{figure}

Подмножество элементов $x\in\MU$, удовлетворяющих некоторому свойству $P$
обозначается следующим образом
$$
  \lbrace x\in\MU:~P\rbrace ,
$$
где свойство $P$, как правило, задается некоторым уравнением или неравенством
для элементов $x$.
\begin{exa}
Множество положительных и отрицательных вещественных чисел определяется
равенствами:
\begin{align*}
  \MR_+&:=\lbrace x\in\MR:~x>0\rbrace,
\\                                                                   \tag*{\qed}
  \MR_-&:=\lbrace x\in\MR:~x<0\rbrace.
\end{align*}
\end{exa}
\begin{defn}
{\em Объединение, пересечение} и {\em разность} двух множеств $\MU$ и $\MV$
обозначается соответственно через
\begin{align*}
  \MU\cup\MV&:=\lbrace x\in\MU\quad \text{или}\quad x\in\MV\rbrace,
\\
  \MU\cap\MV&:=\lbrace x\in\MU\quad \text{и}\quad x\in\MV\rbrace,
\\
  \MU\setminus\MV&:=\lbrace x\in\MU:~x\notin\MV\rbrace.
\end{align*}
Если $\MU\cap\MV=\emptyset$, т.е.\ множества $\MU$ и $\MV$ не содержат общих
элементов, то они называются {\em непересекающимися}. Если множество $\MV$
является подмножеством в $\MU$, $\MV\subset\MU$, то разность $\MU\setminus\MV$
называется {\em дополнением} к $\MV$ в $\MU$.
\qed\end{defn}
\index{Объединение множеств (union of sets)}%
\index{Пересечение множеств (intersection of sets)}%
\index{Разность множеств (difference of sets)}%
\index{Непересекающиеся множества (nonoverlapping sets}%
\index{Множества непересекающиеся (nonoverlapping sets}%
\index{Дополнение (complement of a set)}%
Из определения, очевидно, следует
\begin{prop}
Пусть $\MV\subset\MU$ и $\MW=\MU\setminus\MV$, тогда $\MW\cup\MV=\MU$ и
$\MW\cap\MV=\emptyset$.
\end{prop}
\begin{com}
Операции объединения, пересечения и дополнения соответствуют логическим связкам
``или'', ``и'', ``нет''.
\qed\end{com}
\begin{prop}
Операции объединения, пересечения и дополнения множеств удовлетворяют
тождествам:
\begin{equation}                                                  \label{emorga}
\begin{split}
  \MM\setminus(\MU\cup\MV)&=(\MM\setminus\MU)\cap(\MM\setminus\MV),
\\
  \MM\setminus(\MU\cap\MV)&=(\MM\setminus\MU)\cup(\MM\setminus\MV)
\end{split}
\end{equation}
и
\begin{align*}
  \MM\cap(\MU\cup\MV)&=(\MM\cap\MU)\cup(\MM\cap\MV),
\\                                                                   \tag*{\qed}
  \MM\cup(\MU\cap\MV)&=(\MM\cup\MU)\cap(\MM\cup\MV).
\end{align*}
\end{prop}
\begin{proof}
Выполняется простой проверкой. Рис.\ref{funins} иллюстрирует формулы
(\ref{emorga}).
\end{proof}
\begin{defn}
Пусть $\MM_1$ и $\MM_2$ -- два непустых множества, тогда их {\em прямым} или
{\em декартовым} произведением $\MM_1\times\MM_2$ называется множество
упорядоченных пар $(x_1,x_2)$, где $x_1\in\MM_1$ и $x_2\in\MM_2$:
\begin{equation*}
  \MM_1\times\MM_2:=\lbrace(x_1,x_2):\quad x_1\in\MM_1,\quad x_2\in\MM_2\rbrace.
\end{equation*}
Если одно из множеств $\MM_1$ или $\MM_2$ пусто, то их прямое произведение
пусто.
\qed\end{defn}
\index{Прямое произведение (direct product)}%
\index{Произведение прямое (direct product)}%
\index{Декартово произведение (Cartesian product)}%
\index{Произведение декартово (Cartesian product)}%
Для того, чтобы понятие множества было более содержательным, на нем вводятся
различные структуры. Важным примером является упорядочение.
\begin{defn}
Множество $\MM$ называется {\em частично упорядоченным}, если на нем задано
бинарное отношение $\le$, удовлетворяющее следующим условиям:
\begin{align*}
  &x\le x, &&\text{-- рефлексивность},
\\
  &x\le y\quad \text{и}\quad y\le z\quad \Rightarrow\quad x\le z, &&
  \text{-- транзитивность},
\\
  &x\le y\quad \text{и}\quad y\le x\quad \Rightarrow\quad x=y, &&
  \text{-- антисимметричность}.
\end{align*}
В общем случае может оказаться, что для некоторой пары элементов соотношение
$\le$ не определено. Если для любой пары элементов $x,y\in\MM$ либо $x\le y$,
либо $y\le x$, то множество $\MM$ называется {\em линейно} упорядоченным или
{\em цепью}.
\qed\end{defn}
\index{Частично упорядоченное множество (partially ordered set, poset)}%
\index{Множество частично упорядоченное (partially ordered set, poset)}%
\index{Линейно упорядоченное множество (linearly ordered set)}%
\index{Множество линейно упорядоченное (linearly ordered set)}%
\index{Цепь (chain)}%
\begin{exa}
Множество всех подмножеств $\MP(\MM)$ множества $\MM$ является частично
упорядоченным по отношению включения $\subseteq$. Если множество состоит из двух
элементов $\MM:=\lbrace x,y\rbrace$, то его подмножествами являются $\emptyset$,
$\lbrace x\rbrace$, $\lbrace y\rbrace$ и $\lbrace x,y\rbrace$. Оно не является
линейно упорядоченным, т.к.\ для подмножеств $\lbrace x\rbrace$ и
$\lbrace y\rbrace$ отношение $\subseteq$ не определено.
\qed\end{exa}

\begin{exa}
Множество вещественных чисел $\MR$, где символ $\le$ означает ``меньше или
равно'', является линейно упорядоченным.
\qed\end{exa}
\section{Поле вещественных чисел $\MR$ и прямая                  \label{srealn}}
В геометрии к неопределяемым понятиям относится {\em прямая линия} или, короче,
прямая, которую будем обозначать буквой $\MR$. При этом ее представляют как
отрезок, начерченный по линейке и мысленно продолженный до бесконечности в обе
стороны. Прямая линия находится во взаимно однозначном соответствии с полем
вещественных (действительных) чисел. Другими словами, прямая является наглядным
изображением поля вещественных чисел. На ней выбирается произвольная точка,
{\em начало отсчета}, которой ставится в соответствие число нуль. Затем каждому
положительному числу $x\in\MR_+$ ставится в соответствие точка, лежащая
справа от нуля на расстоянии, равном этому числу. Каждому отрицательному числу
$x\in\MR_-$ ставится в соответствие точка, лежащая слева от начала отсчета на
расстоянии, равном модулю этого числа. При этом расстояние измеряется с помощью
линейки, а число $x$ называется координатой точки. Таким образом между точками
прямой и вещественными числами устанавливается взаимно однозначное соответствие.
Суммируя, можно сказать, что прямая это не более, чем наглядный образ
вещественных чисел.
\index{Линия прямая (straight line)}\index{Прямая линия (straight line)}%
\index{Начало отсчета (origin)}%

Расстояние $l$ между двумя точками $a\in\MR$ и $b\in\MR$, по-определению, равно
модулю разности двух вещественных чисел $l(a,b)=|b-a|$. Его можно записать в
виде интеграла
\begin{equation}                                                  \label{emetca}
  l(a,b):=\left|\int\limits_a^bdxg\right|=|b-a|,\qquad g=1.
\end{equation}
Здесь введена функция $g$, равная единице, которая называется {\em метрикой}.
Метрика в (\ref{emetca}) всюду равна единице и называется {\em евклидовой}. В
дифференциальной геометрии понятие расстояния обобщается за счет расширения
класса рассматриваемых метрик. Понятие же евклидовой метрики чрезвычайно важно,
т.к.\ служит той точкой отсчета, с которой сравниваются все остальные метрики.
\index{Метрика (metric)} \index{Метрика евклидова (Euclidean metric)}%
\index{Евклидова метрика (Euclidean metric)}%

Координаты точек складываются и умножаются так же, как и обычные вещественные
числа. Для вещественных чисел определены две операции: сложение и умножение.
По отношению к сложению вещественные числа образуют абелеву группу. Напомним
общие определения и основные сведения из теории групп.
\begin{defn}
Непустое множество $\MG$ называется {\em группой}, если выполнены четыре
условия:

1) \parbox[t]{.92\linewidth}{
  {\em Закон композиции.} Каждой паре элементов $a,b\in \MG$ сопоставляется
  третий элемент этого же множества, называемый {\em произведением} элементов и
  обозначаемый $a\cdot b$ или $ab$. Закон композиции называется также
  {\em бинарной операцией}.}

2) \parbox[t]{.92\linewidth}{
  {\em Закон ассоциативности.} Для любых трех элементов
  $a,b,c\in \MG$ имеет место равенство
\begin{equation*}
  (ab)c=a(bc).
\end{equation*}}

3) \parbox[t]{.92\linewidth}{
  В $\MG$ существует левая {\em единица} $e$:
\begin{equation*}
  ea=a,\qquad \forall a\in \MG.
\end{equation*}}

4) \parbox[t]{.92\linewidth}{
  Для каждого элемента $a\in \MG$ существует по крайней мере один
  левый {\em обратный элемент} $a^{-1}\in \MG$:
\begin{equation*}
  a^{-1}a=e.
\end{equation*}}\newline
Множество элементов с одной бинарной операцией, которая удовлетворяет только
условию ассоциативности, называется {\em полугруппой}. Полугруппа с единичным
элементом называется {\em моноидом}. Если для любых двух элементов $ab=ba$, то
группа (или полугруппа) называется {\em коммутативной} или {\em абелевой}.
В противном случае группа (или полугруппа) называется {\em неабелевой}.
\qed\end{defn}
\index{Группа (group)}\index{Полугруппа (semigroup)}\index{Моноид (monoid)}%
\index{Закон композиции (composition rule)}\index{Композиция (composition)}%
\index{Бинарная операция (binary operation)}%
\index{Операция бинарная (binary operation)}%
\index{Произведение (product)}\index{Единица (identity)}%
\index{Закон ассоциативности (associativity law)}%
\index{Ассоциативности закон (associativity law)}%
\index{Обратный элемент (inverse element)}%
\index{Элемент обратный (inverse element)}%
\index{Коммутативная группа (commutative group)}%
\index{Группа коммутативная (commutative group)}%
\index{Абелева группа (Abelian group)}\index{Группа абелева (Abelian group)}%
\index{Неабелева группа (non-Abelian group)}%
\index{Группа неабелева (non-Abelian group)}%
\begin{exa}
Пусть $n\MZ$ -- множество целых чисел, делящихся на $n$, где $n$ -- произвольное
натуральное число. Это множество содержит число $0$ при всех $n$, и в нем
определены операции сложения $(+)$ и умножения $(\cdot)$. Пара $(n\MZ,+)$
является коммутативной группой, где роль единицы выполняет число $0$. Обозначим
через $n\MZ_+$ все неотрицательные числа из $n\MZ$. Тогда пара $(n\MZ_+,+)$
будет коммутативным моноидом. Если $n\ge2$, то пара $(n\MZ,\cdot)$ является
коммутативной полугруппой без единицы.
\qed\end{exa}
\begin{exa}
Рассмотрим трехмерное евклидово пространство $\MR^3$ с декартовыми координатами
$x,y,z$. Пусть $\MG$ -- множество всех отображений $\MR^3$ в себя. Под
композицией двух отображений мы понимаем их последовательное выполнение. В
предложении \ref{passco} будет доказано, что композиция отображений
произвольного множества является ассоциативной операцией. Множество отображений
$\MG$ содержит единицу, которой является тождественное отображение, оставляющее
точки евклидова пространства неподвижными. Ясно, что множество всех отображений
представляет собой некоммутативный моноид. В общем случае множество отображений
$\MG$ группу не образует, т.к.\ содержит, например, проекцию на плоскость $x,y$,
которая не имеет обратного отображения.
\qed\end{exa}

В дифференциальной геометрии изучаются чаще всего такие преобразования
многообразий, которые образуют группу. Поэтому опишем некоторые свойства групп
более подробно.
\begin{prop}
Каждая левая единица $e\in\MG$ является одновременно и правой единицей, $ae=a$
для всех $a\in\MG$. Единица в группе единственна. Каждый левый обратный элемент
$a^{-1}\in\MG$ одновременно является и правым обратным элементом, $aa^{-1}=e$.
Обратный элемент $a^{-1}$ для всех $a\in\MG$ единственен. Справедливо правило
$(ab)^{-1}=b^{-1}a^{-1}$.
\end{prop}
\begin{proof}
Приведено в большинстве учебников по теории групп.
\end{proof}
\begin{defn}
Группа, состоящая из конечного числа элементов, называется {\em конечной}. При
этом количество элементов в группе называется {\em порядком} группы. Группа,
состоящая из степеней одного элемента $a$ называется {\em циклической}. Если
циклическая группа имеет порядок $n$, то она обозначается $\MC_n$, и $a^n=e$.
Циклические группы обязательно абелевы. {\em Кручением} конечно порожденной
абелевой группы $\MG$ (т.е.\ когда любой элемент группы $\MG$ представим в виде
конечного произведения некоторых элементов и их обратных), называется ее
подгруппа, состоящая их всех элементов
конечного порядка. Говорят, что конечно порожденная абелева группа не имеет
кручения, если в ней нет элементов конечного порядка, т.е.\ не существует
элемента конечная положительная степень которого равна единице.
\qed\end{defn}
\index{Группа конечная (finite group)}\index{Конечная группа (finite group)}%
\index{Порядок группы (order of a group)}%
\index{Циклическая группа (cyclic group)}%
\index{Группа циклическая (cyclic group)}%
\index{Кручение группы (group torsion)}%
\begin{com}
В дифференциальной геометрии термин кручение имеет совершенно другой смысл (см.\
раздел \ref{saffco}).
\qed\end{com}
\begin{exa}
Множество невырожденных (с отличным от нуля определителем) квадратных
$n\times n$ матриц над полем вещественных или комплексных чисел представляют
собой группы по отношению к умножению матриц, которые обозначаются
$\MG\ML(n,\MR)$ и $\MG\ML(n,\MC)$. Эти группы при $n>1$ некоммутативны
(неабелевы).
\qed\end{exa}
\begin{exa}
Группа вращений трехмерного евклидова пространства $\MO(3)$ и группа Лоренца
$\MO(1,3)$ являются неабелевыми.
\qed\end{exa}
\begin{exa}                                                       \label{xzetat}
Два целых числа $1$ и $-1$ с обычным умножением образуют циклическую группу
$\MZ_2$. Ее также отождествляют с группой одномерных вращений, $\MZ_2=\MO(1)$.
\qed\end{exa}
\begin{exa}
Множество всех целых чисел $\MZ$ образует бесконечномерную циклическую группу по
отношению к сложению. Она порождена одним элементом -- числом 1 -- и не имеет
кручения.
\qed\end{exa}
\begin{exa}
Дробно-линейные преобразования расширенной комплексной плоскости
\begin{equation}                                                  \label{emebgr}
  \overline\MC\ni\qquad z~\mapsto~\frac{az+b}{cz+d}\qquad\in\overline\MC,\qquad
  a,b,c,d\in\MC,\qquad ad-bc\ne0
\end{equation}
образуют группу Ли, которая называется {\em группой Мёбиуса}. Эта группа шести
параметрическая, т.к.\ числитель и знаменатель можно разделить на произвольное
отличное от нуля комплексное число, и изоморфна группе невырожденных
комплексных матриц $\MS\ML(2,\MC)$.
\qed\end{exa}
\index{Группа Мёбиуса (M\"obius group)}\index{Мёбиуса группа (M\"obius group)}%
\begin{defn}
Пусть $\MH\subset\MG$ -- {\em подгруппа} группы $\MG$, т.е.\ элементы $\MH$ сами
по себе образуют группу. При этом единица подгруппы $\MH$ совпадает с единицей
группы $\MG$. Обозначим через $a\MH$ и $\MH a$ множество элементов вида $ah$ и
$ha$, где $a$ -- некоторый фиксированный элемент группы $\MG$, а элемент
$h\in\MH$ пробегает всю подгруппу $\MH$. Множества элементов $a\MH$ и $\MH a$,
где $a\in\MG$, называются {\em левым} и {\em правым смежными классами} группы
$\MG$ по подгруппе $\MH$. Если $a\in\MH$, то левый и правый смежный класс
совпадает с $\MH$.
\qed\end{defn}
\index{Подгруппа (subgroup)}\index{Смежный класс (coset)}%
\index{Левый смежный класс (left coset)}%
\index{Правый смежный класс (right coset)}%

Нетрудно проверить, что два левых смежных класса по $\MH$ либо совпадают, либо
не имеют ни одного общего элемента. Поэтому любой элемент группы принадлежит
одному и только одному левому смежному классу, и его можно рассматривать как
представитель этого класса. Под произведением двух левых смежных классов $a\MH$
и $b\MH$ понимается множество всех элементов вида $cd$, где $c\in a\MH$ и
$d\in b\MH$.

Сказанное выше верно и для правых смежных классов.
\begin{defn}
Два элемента $a'$ и $a$ группы $\MG$ называются {\em сопряженными}, если они
связаны преобразованием {\em подобия}
\index{Сопряженный элемент (conjugate element)}%
\index{Элемент сопряженный (conjugate element)}%
\index{Преобразование подобия (similarity transformation)}%
\index{Подобия преобразование (similarity transformation)}%
\begin{equation*}
  a'=bab^{-1},
\end{equation*}
где $b$ -- некоторый элемент из $\MG$. Сопряженность элементов является
отношением эквивалентности (см.\ раздел \ref{smappi}) и определяет разбиение
группы $\MG$ на классы сопряженных элементов. Подгруппа $\MH'\subset\MG$
называется {\em сопряженной} подгруппе $\MH$, если
$$
  \MH'=b\MH b^{-1},
$$
для некоторого $b\in\MG$. Очевидно, что единичные элементы сопряженных
подгрупп $\MH$ и $\MH'$ совпадают. Подгруппа $\MH$ отображается на себя для всех
$b\in\MG$ тогда и только тогда, когда подгруппа $\MH$ содержит все элементы,
сопряженные с ее элементами, что можно записать в виде $b\MH=\MH\,b$. Такая
подгруппа называется {\em нормальной} подгруппой или {\em нормальным делителем}.
Ее также называют {\em инвариантной подгруппой}. Левые и правые смежные классы
по нормальному делителю совпадают и образуют группу по отношению к операции
умножения смежных классов. Действительно, пусть $\MH$ -- нормальный делитель
группы $\MG$. Тогда для любых смежных классов $a\MH$ и $b\MH$ определено
умножение
\begin{equation*}
  a\MH b\MH=ab\MH\MH=ab\MH.
\end{equation*}
То есть множество смежных классов образует группу само по себе, при этом роль
единицы выполняет нормальная подгруппа $\MH$. Эта группа называется
{\em факторгруппой} и обозначается $\MG/\MH$.
\qed\end{defn}
\index{Подгруппа сопряженная (conjugate subgroup)}%
\index{Сопряженная подгруппа (conjugate subgroup)}%
\index{Нормальная подгруппа (normal subgroup)}%
\index{Подгруппа нормальная (normal subgroup)}%
\index{Нормальный делитель (normal subgroup)}%
\index{Делитель нормальный (normal subgroup)}%
\index{Инвариантная подгруппа (invariant subgroup)}%
\index{Подгруппа инвариантная (invariant subgroup)}%
\index{Факторгруппа (factor group, quotient group))}%
\begin{exa}
Если в качестве подгруппы выбрать саму группу, $\MH=\MG$, то факторгруппа
состоит из одного элемента -- единицы, $\MG/\MG=e$.
\qed\end{exa}
\begin{defn}
Множество всех элементов группы $\MG$, перестановочных с любым элементом этой
группы, является нормальным делителем и называется {\em центром} группы. Центр
всегда является абелевой подгруппой группы $\MG$.

Для абелевых групп композицию двух элементов часто называют {\em сложением} и
пишут $a+b$. Тогда группу $\MG$ называют {\em аддитивной группой} или
{\em модулем}. Вместо единичного элемента здесь фигурирует {\em нулевой
элемент}:
\begin{equation*}
  0+a=a,\qquad \forall a\in \MG,
\end{equation*}
а обратный элемент обозначают $-a$:
\begin{equation*}                                                    \tag*{\qed}
  -a+a=0.
\end{equation*}
\end{defn}
\index{Центр группы (group center)}\index{Группы центр (group center)}%
\index{Сложение (addition)}\index{Модуль (module)}%
\index{Аддитивная группа (additive group)}%
\index{Группа аддитивная (additive group)}
\index{Нулевой элемент (zero element)}\index{Элемент нулевой (zero element)}%
Очевидно, что любая подгруппа абелевой группы является нормальным делителем.

\begin{exa}
Множество натуральных чисел $\MN$ группы по отношению к сложению не образует,
т.к.\ не содержит ни нулевого, ни обратных элементов.
\qed\end{exa}
\begin{exa}
Множества целых $\MZ$ и вещественных чисел $\MR$ образуют аддитивные группы по
отношению к сложению. По сути дела, отсюда и пошло название ``сложение'' для
групповой композиции в абелевых группах.
\qed\end{exa}
\begin{exa}
По отношению к умножению вещественные числа группу не образуют, т.к.\ у нуля
обратного элемента не существует. В то же время множество вещественных чисел без
нуля образует абелеву группу по отношению к умножению. Эта группа является
одномерной группой Ли и состоит из двух несвязных компонент: отрицательных и
положительных чисел чисел $\MR_-\cup\MR_+$. При этом $e=1\in\MR_+$.
\qed\end{exa}
\begin{exa}
Множество целых чисел с операцией сложения и отношением эквивалентности
$n+p\sim n$, где натуральное число $p\in\MN$ фиксировано, и $n\in\MZ$ любое,
представляет собой конечную циклическую группу $\MZ_p$, состоящую из $p$
элементов. В качестве элементов группы можно выбрать числа $0,1,\dotsc,p-1$. В
этом случае мы пишем $n+p=n$ $\mod p$.

Группу $\MZ_p$ можно описать также следующим образом. Обозначим через $p\MZ$
подгруппу целых чисел, состоящую из тех чисел, которые делятся на $p$ без
остатка. Множество $p\MZ$ является нормальной подгруппой в $\MZ$. Тогда
множества левых и правых смежных классов совпадают и образуют фактор группу
\begin{equation*}
  \MZ_p=\MZ/p\MZ.
\end{equation*}
При $p=2$ эта группа изоморфна группе, рассмотренной в примере \ref{xzetat}.
\qed\end{exa}
\begin{defn}
{\em Прямым произведением} двух групп $\MG_1$ и $\MG_2$ называется группа
$\MG=\MG_1\times\MG_2$, образованная всеми упорядоченными парами $(a_1,a_2)$,
где $a_1\in\MG_1$ и $a_2\in\MG_2$, с умножением, определяемым формулой
\begin{equation*}
  (a_1,a_2)(b_1,b_2)=(a_1b_1,a_2b_2),
\end{equation*}
для всех $a_1,b_1\in\MG_1$ и $a_2,b_2\in\MG_2$.
\qed\end{defn}
\index{Прямое произведение групп (direct product of groups)}%
\index{Произведение групп прямое (direct product of groups)}%
Если группы $\MG_1$ и $\MG_2$ конечны, то порядок группы $\MG=\MG_1\times\MG_2$
равен произведению порядков групп $\MG_1$ и $\MG_2$. Единицей группы $\MG$
является пара $(e_1,e_2)$, где $e_1$ и $e_2$ -- единицы соответственно в группах
$\MG_1$ и $\MG_2$. При этом элементы вида $(a_1,e_2)$, где $a_1$ пробегает всю
группу $\MG_1$, образуют подгруппу в прямом произведении $\MG_1\times\MG_2$,
которая изоморфна $\MG_1$. Аналогично, элементы вида $(e_1,a_2)$ образуют
подгруппу в $\MG_1\times\MG_2$, изоморфную $\MG_2$.

Если у групп $\MG_1$ и $\MG_2$ нет общих элементов, то элемент их прямого
произведения $(a_1,a_2)$ можно записывать просто $a_1a_2$, причем $a_1e_2=a_1$ и
$e_1a_2=a_2$.
\begin{exa}
Любая размерная величина в физике является прямым произведением числа и единицы
измерения. Выражения вида ``$\text{сила}=\text{масса}\times\text{ускорение}$''
также являются прямыми произведениями.
\qed\end{exa}
Вещественные числа $\MR$ представляют собой множество с двумя бинарными
операциями. А именно, для двух произвольных чисел $a,b\in\MR$ однозначно
определена их сумма $a+b$ и произведение $ab$. Вещественные числа представляют
собой частный случай множества, которое называется в алгебре кольцом.
\begin{defn}
  Непустое множество с двумя бинарными операциями называется {\em кольцом},
  если выполняются следующие условия.

Законы сложения.

1) Закон ассоциативности: $a+(b+c)=(a+b)+c$.

2) Закон коммутативности: $a+b=b+a$.

3) Разрешимость уравнения $a+x=b$ для всех $a$ и $b$.

Закон умножения.

1) Закон ассоциативности: $a(bc)=(ab)c$.

Законы дистрибутивности.

1) $a(b+c)=ab+ac$.

2) $(b+c)a=ba+ca$.\newline
Если умножение в кольце коммутативно, то говорят о {\em коммутативном кольце}.
\qed\end{defn}
\index{Кольцо (ring)}\index{Законы дистрибутивности (distributivity rules)}%
\index{Кольцо коммутативное (commutative ring)}%
\index{Коммутативное кольцо (commutative ring)}%
Три закона сложения означают в совокупности, что элементы кольца образуют
абелеву группу (модуль) по отношению к сложению. Если кольцо обладает правым и
левым единичным элементом одновременно, $ea=ae=a$, то он называется
{\em единицей}, и говорят о {\em кольце с единицей}.
\index{Единица кольца (identity of a ring)}%
\index{Кольцо с единицей (ring with identity)}%
В общем случае левые и правые единицы у кольца могут различаться, а единиц может
быть несколько.
\begin{exa}
Минимальное конечное кольцо состоит из двух элементов $0$ и $1$ со следующей
таблицей сложения и умножения
\begin{align*}
  0+0&=0, & 0+1&=1, & 1+1&=0,
\\
  0\cdot0&=0, & 0\cdot1&=0, & 1\cdot1&=1.
\end{align*}
Это кольцо коммутативно, т.к.\
\begin{equation*}                                                    \tag*{\qed}
  1\cdot0=1\cdot(1+1)=1\cdot1+1\cdot1=1+1=0.
\end{equation*}
\end{exa}
\begin{exa}
Множество целых чисел с операцией сложения и умножения образуют коммутативное
кольцо с единицей.
\qed\end{exa}
\begin{defn}
Кольцо, для каждого элемента которого $a^2=a$, называется {\em булевым}.
\qed\end{defn}
\index{Булево кольцо (Boolean ring)}\index{Кольцо булево (Boolean ring)}%
\begin{defn}
  Кольцо называется {\em телом}, если:

1) в нем есть по крайней мере один элемент, отличный от нуля;

2) уравнения $ax=b$ и $xa=b$ при $a\ne0$ разрешимы.\newline
Коммутативное тело называется {\em полем} или {\em рациональным кольцом}.
\qed\end{defn}
\index{Тело (division ring)}\index{Поле (field)}%
\index{Рациональное кольцо (rational ring)}%
\index{Кольцо рациональное (rational ring)}%
\begin{com}
Тело, как и кольцо, является аддитивной группой (модулем) по отношению к
сложению, однако не является группой по отношению к умножению, т.к.\ у нуля в
общем случае нет обратного элемента. Операции сложения и умножения в поле
связаны дистрибутивными законами.
\qed\end{com}
\begin{exa}                                                       \label{ecofil}
Множества рациональных $\MQ$ и действительных $\MR$ чисел с естественными
операциями сложения и умножения являются полями. При этом $\MQ\subset\MR$. Целые
числа образуют подкольцо поля рациональных чисел $\MZ\subset\MQ$.
\qed\end{exa}
Следующий пример показывает, что из одного поля можно построить другие поля.
\begin{exa}                                                       \label{ecofie}
В дальнейшем мы будем рассматривать не только вещественные многообразия, но и
комплексные. Поэтому напомним основные определения для комплексных чисел.
Рассмотрим упорядоченную пару вещественных чисел $(a,b)$, где $a,b\in\MR$. Будем
считать две пары $(a,b)$ и $(c,d)$ равными, если $a=c$ и $b=c$. Введем на
множестве пар операции сложения и умножения следующим образом:
\begin{equation*}
\begin{split}
  (a,b)+(c,d)&:=(a+b,c+d),
\\
  (a,b)(c,d)&:=(ac-bd,ad+bc).
\end{split}
\end{equation*}
Нетрудно проверить, что все аксиомы поля для множества пар выполнены.
\begin{defn}
Множество пар $(a,b)$ с введенными выше операциями сложения и умножения
называется {\em полем комплексных чисел} и обозначается $z:=(a,b)\in\MC$.
\qed\end{defn}
\index{Поле комплексных чисел (complex number field)}%

Множество комплексных чисел вида $(a,0)\in\MC$ можно отождествить с множеством
вещественных чисел $a\in\MR$. Тогда поле вещественных чисел является подполем
поля комплексных чисел, $\MR\subset\MC$.

Определим умножение пар $(a,b)\in\MC$ на вещественные числа:
\begin{equation*}
  c(a,b):=(a,b)c:=(ca,cb),\qquad c\in\MR.
\end{equation*}
Множество комплексных чисел с операцией сложения и умножения на вещественные
числа образует двумерное векторное пространство. В качестве базиса этого
векторного пространства выберем пары:
\begin{equation*}
  e_1:=(1,0),\qquad e_2:=(0,1).
\end{equation*}
Теперь произвольное комплексное число можно представить в виде
\begin{equation*}
  (a,b)=ae_1+be_2.
\end{equation*}
Базисный вектор $e_1$ при умножении ведет себя как единица: $ze_1=e_1z=z$.
Поэтому его можно отождествить с единицей. Для второго элемента базиса принято
обозначение $i:=(0,1)$. Его называют {\em мнимой единицей}.
\index{Мнимая единица (imaginary unit)}%
Нетрудно проверить, что $i^2=-1$. Обычно комплексные числа записывают в виде
\begin{equation*}
  z=a+ib,\qquad a,b\in\MR,\quad z\in\MC.
\end{equation*}
\begin{defn}
Множество комплексных чисел $z=(a,b)$ отождествляется с евклидовой плоскостью
$\MC\approx\MR^2$, где вещественные числа $a$ и $b$ рассматриваются в качестве
декартовых координат. В этом случае множество комплексных чисел называется
{\em комплексной плоскостью}. При этом ось абсцисс называется {\em
действительной осью}, а ось ординат -- {\em мнимой осью}.
\qed\end{defn}
\index{Комплексная плоскость (complex plane)}%
\index{Плоскость комплексная (complex plane)}%
\index{Действительная ось (real axis}%
\index{Ось действительная (real axis}%
\index{Мнимая ось (imaginary axis}%
\index{Ось мнимая (imaginary axis}%
Линейная структура евклидовой плоскости $\MR^2$, рассматриваемой как векторное
пространство (см.\ раздел \ref{seucve}), совпадает с линейной структурой
комплексных чисел $\MC$. Однако скалярное умножение векторов в $\MR^2$ не имеет
никакого отношения к умножению комплексных чисел.

Пусть $z=a+ib\in\MC$. Тогда $|z|:=\sqrt{a^2+b^2}$ -- это евклидово расстояние от
начала координат до точки $z$. Пусть $\vf$ -- угол между положительным
направлением оси абсцисс и радиус-вектором точки $z$ определяется уравнениями:
\begin{equation*}
    \cos\vf=\frac a{\sqrt{a^2+b^2}},\qquad \sin\vf=\frac b{\sqrt{a^2+b^2}},
    \qquad |z|\ne0,\qquad \vf\in[0,2\pi).
\end{equation*}
Тогда приняты следующие обозначения и названия:
\begin{equation}                                                  \label{enacox}
\begin{split}
  a&:=\re z\quad \text{-- действительная часть } z,
\\
  b&:=\im z\quad \text{-- мнимая часть } z,
\\
  \bar z&:=a-ib\quad \text{-- число, комплексно сопряженное к } z,
\\
  |z|&:=\sqrt{a^2+b^2}\quad \text{-- модуль } z,
\\
  \arg z&:=\vf+2\pi k,~k\in\MZ,\quad \text{-- аргумент $z$}.
\end{split}
\end{equation}
Функция $\arg z$ определена с точностью до прибавления целого кратного $2\pi$.

Для комплексных чисел часто используют тригонометрическую запись:
\begin{equation*}
  z=r\ex^{i\vf}=r\cos\vf+ir\sin\vf,\qquad r:=|z|.
\end{equation*}
При этом
\begin{equation*}
  \bar z=r\ex^{-i\vf}=r\cos\vf-ir\sin\vf.
\end{equation*}
\begin{defn}
В теории функций комплексного переменного часто используют {\em расширенную
комплексную плоскость} $\overline\MC$, которая получается из комплексной
плоскости $\MC$ добавлением бесконечно удаленной точки $z=\infty$. На этом
множестве вводится следующая топология. Если подмножество
$\MU\subset\overline\MC$ не содержит бесконечно удаленной точки $\infty$, то оно
считается открытым, если оно открыто в $\MC$. Если подмножество $\MU$ содержит
точку $\infty$, то оно считается открытым в $\overline\MC$, если его дополнение
является компактом в $\MC$. Как правило, открытыми окрестностями точки $\infty$
мы будем считать дополнения $\overline\MC\setminus\overline\MB_R$ замкнутых
дисков $\overline\MB_R$ радиуса $R$ до всей расширенной комплексной плоскости
$\overline\MC$. Расширенную комплексную плоскость называют также
{\em комплексной сферой} или {\em сферой Римана}.
\qed\end{defn}
\index{Расширенная комплексная плоскость (extended complex plane)}%
\index{Комплексная плоскость расширенная (extended complex plane)}%
\index{Комплексная сфера (complex sphere)}%
\index{Сфера комплексная (complex sphere)}%
\index{Сфера Римана (Riemann sphere)}%
\index{Римана сфера (Riemann sphere)}%
Комплексная плоскость $\MC$ с естественной топологией евклидова пространства
является некомпактным многообразием. При добавлении бесконечной точки мы меняем
топологию комплексной плоскости (теперь, например, точки $(-\infty,0)$ и
$(\infty,0)$ близки). В результате получаем компактное топологическое
пространство -- сферу Римана $\overline\MC$ (см.\ раздел \ref{seucto}).
Комплексная сфера является компактификацией комплексной плоскости (см.\ раздел
\ref{smacos}), которая осуществляется с помощью стереографической проекции.

На сфере Римана $\overline\MC$ мы определили топологию, которая отличается от
топологии комплексной плоскости $\MC$. Эта топология является метрической.
Соответствующую метрику можно определить следующим образом. С помощью
стереографической проекции (см.\ раздел \ref{sphere}) устанавливаем взаимно
однозначное соответствие между точками расширенной комплексной плоскости и
обычной сферой, вложенной в трехмерное евклидово пространство,
$\MS^2\hookrightarrow\MR^3$. Затем определяем расстояние между точками
расширенной комплексной плоскости как обычное евклидово расстояние между точками
сферы $\MS^2$. Это расстояние определяет на расширенной комплексной плоскости
метрику, которая называется {\em хордовой} и отличается от исходной евклидовой
метрики. Явные выражения для хордовой метрики нам не понадобятся, и мы не
будем их приводить (см., например, \cite{Evgraf91R}). Заметим, что при работе с
расширенной комплексной плоскостью в ТФКП используют, как правило, не хордову
метрику, которая определяет топологию, а обычную евклидову метрику как более
наглядную.

Очевидно, что сфера Римана не несет структуры поля или векторного пространства.
Эти структуры можно определить только на подмножестве $\MC\subset\overline\MC$.
Тем не менее для каждой точки $p\in\MC$ отображение $z\mapsto\frac1{z-p}$,
определенное обычным образом при $z\in\MC\setminus\lbrace p\rbrace$, можно
продолжить до биекции множества $\overline\MC$ на себя, если положить
$\frac10:=\infty$ и $\frac1\infty:=0$.
\qed\end{exa}
\index{Хордова метрика (chord metric)}\index{Метрика хордова (chord metric)}%
\begin{defn}
Говорят, что поле $\MF$ имеет {\em характеристику} $p$, если существует такое
простое число $p$, что выполнено равенство
$$
  \underbrace{a+\dotsc+a}_p=0,\qquad \forall a\in\MF.
$$
В этом случае поле $\MF$ содержит подполе $\MZ_p$. Если такого числа $p$ не
существует, то говорят, что поле $\MF$ имеет характеристику нуль. В последнем
случае оно содержит подполе, состоящее из рациональных чисел.
\qed\end{defn}
\index{Характеристика поля (characteristic of a field)}%
\begin{exa}
Поля рациональных $\MQ$, действительных $\MR$ и комплексных $\MC$ чисел имеют
характеристику нуль.
\qed\end{exa}
\section{Евклидово пространство $\MR^{n}$                        \label{seucls}}
Основным понятием дифференциально геометрии является дифференцируемое
многообразие, которое является обобщением евклидова пространства. Евклидово
пространство наделено различными структурами: евклидовой метрикой, топологией,
структурой векторного и аффинного пространства. Поэтому в настоящем разделе мы
изучим это пространство с различных точек зрения, чтобы в дальнейшем было ясно,
что именно и как обобщается в дифференциальной геометрии.
\subsection{$\MR^n$ как метрическое пространство                 \label{seucme}}
Пусть $\MR$ -- поле вещественных чисел. В геометрии поле вещественных чисел
наглядно изображается в виде прямой линии на рисунках. При этом каждая точка
прямой $x\in\MR$ находится во взаимно однозначном соответствии с вещественным
числом, которое обозначается той же буквой $x$ и называется координатой точки.
На прямой выбирается произвольная точка -- {\em начало отсчета},
\index{Начало отсчета (origin)}%
которой ставится в соответствие число нуль $0\in\MR$. Обычно предполагают, что
на рисунках координаты точек упорядочены и возрастают слева направо, что
отмечают стрелкой.

Евклидово расстояние $l$ между двумя точками $x,y\in\MR$, по-определению, равно
модулю разности двух вещественных чисел $l(x,y)=|x-y|$.
\begin{defn}
Обозначим через $\MR^n$ прямое произведение $n$ прямых:
\begin{equation}                                                  \label{eucdef}
  \MR^n:=\underbrace{\MR\times \MR\times\dots\times \MR}_n,
\end{equation}
где $n\in\MN$ -- произвольное натуральное число, которое называется
{\em размерностью} пространства $\MR^n$. Точкой $x\in \MR^n$ является
упорядоченный набор $n$ вещественных чисел $x^\al\in\MR$, $\al=1,\dots,n$,
которые называются {\em декартовыми координатами} данной точки. Мы записываем
координаты точки в виде строки,
$$
  x=\lbrace x^\al\rbrace =(x^1,\dots,x^n)\in \MR^n.
$$
Каждый из сомножителей, входящих в определение пространства $\MR^n$
(\ref{eucdef}), называется {\em координатной прямой}, а точка с нулевыми
координатами $(0,\dots,0)\in\MR^n$ -- {\em началом координат}.

Под $0$-мерным пространством $\MR^0$ понимают одну точку -- число нуль.
Одномерное пространство представляет собой прямую, $\MR^1=\MR$. Двумерное
пространство $\MR^2$ называется {\em плоскостью}.
\qed\end{defn}
\index{Размерность евклидова пространства
(dimensionality of the Euclidean space)}%
\index{Координаты декартовы (Cartesian coordinates)}%
\index{Декартовы координаты (Cartesian coordinates)}%
\index{Координатная прямая (coordinate line)}%
\index{Прямая координатная (coordinate line)}%
\index{Начало координат (origin of coordinates)}%
\index{Плоскость (plane)}%
\begin{com}
Номер координаты обозначается с помощью верхнего индекса так же, как и
показатель степени. Как правило, различие в значении индексов ясно из контекста.
\qed\end{com}
В определении пространства $\MR^n$ точку и ее координаты можно отождествить.
Однако, определив таким образом $\MR^n$, мы можем затем перейти в другую
систему координат (см.\ раздел \ref{scooch}). Тогда той же точке пространства
$\MR^n$ будет соответствовать другой набор вещественных чисел. Поэтому
следует различать точку пространства $\MR^n$ и ее координаты, которые зависят
от выбора системы координат.

Расстояние $l$, т.е.\ отображение $l:~\MR^n\times\MR^n\rightarrow\MR$, между
двумя произвольными точками $p,q\in \MR^n$ с декартовыми координатами $p^\al$ и
$q^\al$ определяется следующей формулой
\begin{equation}                                                  \label{ecldis}
  l(p,q):=|q-p|=\sqrt{(q^1-p^1)^2+\dotsc+(q^n-p^n)^2}.
\end{equation}
Между двумя бесконечно близкими точками $x^\al$ и $x^\al+dx^\al$ расстояние
задается {\em интервалом}, который представляет симметричная квадратичная форма,
\index{Интервал (interval)}%
\begin{equation}                                                  \label{einter}
  ds^2:=l^2(x,x+dx)=dx^\al dx^\bt g_{\al\bt},
\end{equation}
где компоненты метрики $g_{\al\bt}$ в декартовой системе координат не зависят от
точки пространства $\MR^n$ и представляют собой единичную матрицу, которую будем
обозначать следующим образом
\begin{equation}                                                  \label{eclmet}
  g_{\al\bt}:=\dl_{\al\bt}=\diag(\underbrace{1,\dots,1}_{\mbox{$n$}}).
\end{equation}

В формуле (\ref{einter}) и в дальнейшем по повторяющимся индексам, один из
которых пишется сверху, а другой -- снизу, производится суммирование, если не
оговорено противное. Это правило называется {\em правилом суммирования
Эйнштейна}.
\index{Правило суммирования Эйнштейна (Einstein's summation rule)}%
\index{Эйнштейна правило суммирования (Einstein's summation rule)}%
Матрица (\ref{eclmet}) называется {\em евклидовой метрикой} и имеет
специальное обозначение $\dl_{\al\bt}$.
\index{Метрика евклидова (Euclidean metric)}%
\index{Евклидова метрика (Euclidean metric)}%

В дифференциальной геометрии роль индексов чрезвычайно важна. Поэтому отметим
ряд общих правил, которые всюду используются в дальнейшем. Эти правила связаны
с группами преобразований, которые действуют на геометрические объекты.

1) \parbox[t]{.92\linewidth}{Каждое слагаемое может содержать некоторый индекс
один или два раза. В первом случае он называется {\em свободным}, а во втором
случае -- {\em немым}.}

2) \parbox[t]{.92\linewidth}{Если некоторое выражение состоит из суммы
нескольких слагаемых, то каждое слагаемое должно содержать один и тот же набор
свободных индексов. При этом значения этих индексов во всех слагаемых должно
фиксироваться одновременно.}

3) \parbox[t]{.92\linewidth}{Немой индекс обязательно встречается один раз
сверху и один раз снизу в каждом слагаемом. Значение этого индекса в каждом
слагаемом не может быть зафиксировано, т.к.\ по нему проводится суммирование.
В разных слагаемых немые индексы можно обозначать различными буквами, а число
их пар может различаться.}
\index{Индекс свободный (free index)}\index{Свободный индекс (free index)}%
\index{Индекс немой (damn index)}\index{Немой индекс (damn index)}%

Иногда мы все же будем писать повторяющиеся индексы как нижние или верхние.
Например, запись $g_{\al\al}$ обозначает диагональный элемент метрики,
стоящий на $\al$-том месте. При этом суммирование не проводится.

Выше мы ввели евклидову метрику в декартовой системе координат, с помощью
которой было определено пространство $\MR^n$. С помощью расстояния, т.е.\
отображения $\MR^n\times\MR^n~\rightarrow~\MR$, понятие декартовой системы
координат можно обобщить следующим образом. Существуют преобразования
координат пространства $\MR^n$, которые сохраняют вид расстояния (\ref{ecldis})
и, следовательно, интервала (\ref{einter}). Эти преобразования образуют
неоднородную группу вращений $\MI\MO(n)$, состоящую из вращений пространства
$\MR^n$,
\begin{equation*}
  x^{\prime\al}=x^\bt S_\bt{}^\al,\qquad S_\bt{}^\al\in\MO(n),
\end{equation*}
и сдвигов,
\begin{equation*}
  x^{\prime\al}=x^\al+a^\al,\qquad a^\al=\const.
\end{equation*}
Поэтому в дальнейшем под декартовой системой координат мы будем понимать любую
систему координат, в которой расстояние между точками пространства $\MR^n$ имеет
вид (\ref{eclmet}).
\begin{defn}
Система координат $x^\al$, $\al=1,\dotsc,n$ пространства $\MR^n$, в которой
расстояние между двумя произвольными точками $a,b\in\MR^n$ имеет вид
(\ref{ecldis}), называется {\em декартовой системой координат}.
\qed\end{defn}
\index{Декартова система координат (Cartesian coordinate system)}%
\index{Система координат декартова (Cartesian coordinate system)}%
\index{Координаты декартовы (Cartesian coordinates)}%
\index{Декартовы координаты (Cartesian coordinates)}%

После того, как пространство $\MR^n$ определено, в нем можно строить
произвольные криволинейные, например, сферические или цилиндрические системы
координат в зависимости от специфики той или иной задачи. В таких системах
координат метрика $g_{\al\bt}(x)$ в (\ref{einter}) будет зависеть от точки
пространства $\MR^n$.
\begin{defn}
{\em Кривой} $\g=x(t)=\lbrace x^\al(t)\rbrace$ в пространстве $\MR^n$ называется
отображение замкнутого единичного отрезка $[0,1]$ в пространство $\MR^n$,
\begin{equation}                                                  \label{ecurve}
  \g:\quad [0,1]\ni\quad t\mapsto x(t)=\lbrace x^\al(t)\rbrace\quad\in\MR^n,
\end{equation}
где $t$ -- вещественный параметр вдоль кривой. Все функции $x^\al(t)$
предполагаются достаточно гладкими. Говорят, что кривая соединяет две
точки $p$ и $q$, где $x(0)=p$, $x(1)=q$. Если граничные точки кривой совпадают,
$p=q$, то кривая называется {\em замкнутой}. Совокупность функций $\dot x^\al$,
где точка обозначает дифференцирование по параметру $t$, определяет касательный
вектор к кривой,
\begin{equation*}
  X_{(\g)}=\lbrace X_{(\g)}^\al:=\dot x^\al\rbrace,
\end{equation*}
который называется {\em вектором скорости} кривой. Кривая $\g$ называется также
{\em путем}, при этом точка $p$ является началом, а $q$ -- концом пути.
\qed\end{defn}
\index{Кривая (curve)}%
\index{Замкнутая кривая (closed curve)}%
\index{Кривая замкнутая (closed curve)}%
\index{Вектор скорости (velocity vector)}%
\index{Путь (path)}%
Не следует смешивать понятие кривой с множеством точек, через которые она
проходит. Согласно определению понятие кривой включает также порядок прохождения
точек данного множества.

В общем случае кривая может иметь точки самопересечения.
\begin{defn}
Кривая, не имеющая точек самопересечения, называется {\em простой кривой}.
Совпадение начала и конца замкнутой кривой мы не будем рассматривать как точки
самопересечения. Поэтому можно говорить о {\em простой замкнутой кривой}.
\qed\end{defn}
\index{Простая кривая (tame curve, simple curve)}%
\index{Кривая простая (tame curve, simple curve)}%

\begin{defn}
Кривая $\g$ называется {\em гладкой (дифференцируемой)}, если все координатные
функции $x^\al(t)$ являются гладкими (дифференцируемыми). Мы предполагаем, что
вектор скорости дифференцируемой кривой отличен от нуля, т.е.\ отлична от нуля
по крайней мере одна из компонент $\dot x^\al$, что соответствует погружению
отрезка $[0,1]$ в $\MR^n$. Кривая $\g$ называется {\em кусочно гладкой
(дифференцируемой)}, если ее можно представить в виде объединения конечного
числа простых замкнутых гладких (дифференцируемых) кривых.
\qed\end{defn}
\index{Гладкая кривая (smooth curve)}%
\index{Кривая гладкая (smooth curve)}%
\index{Дифференцируемая кривая (differentiable curve)}%
\index{Кривая дифференцируемая (differentiable curve)}%
\index{Кусочно гладкая кривая (piecewise smooth curve)}%
\index{Кривая кусочно гладкая (piecewise smooth curve)}%
\index{Кусочно дифференцируемая кривая (piecewise differentiable curve)}%
\index{Кривая кусочно дифференцируемая (piecewise differentiable curve)}%
Согласно данным определениям не всякая гладкая кривая является кусочно гладкой,
т.к.\ может иметь бесконечное число точек самопересечения.

Единичный отрезок в определении кривой выбран для удобства определения
произведения путей и фундаментальной группы многообразия (см.\ раздел
\ref{spaths}). Выбор другого замкнутого интервала соответствует
{\em перепараметризации кривой}. Под этим понимается замена параметра
$t\mapsto t'=t'(t)$, где $t'(t)$ -- произвольная достаточно гладкая
монотонная функция такая, что $dt'/dt\ne0$. При этом вектор скорости кривой
преобразуется по-правилу дифференцирования сложных функций:
$$
  X^\al_{(\g)}=\frac{dx^\al}{dt}=\frac{dx^\al}{dt'}\frac{dt'}{dt}.
$$
\index{Перепараметризация кривой (curve reparameterization)}%
\index{Кривой перепараметризация (curve reparameterization)}%

\begin{defn}
{\em Длиной} дифференцируемой кривой $\g$ называется интеграл
\begin{equation}                                                  \label{ediseu}
  l(\g):=\int_p^q \!\!\!ds
  =\int_0^1\!\!\! dt\sqrt{\dot x^\al\dot x^\bt g_{\al\bt}}.
\end{equation}
Если кривая является кусочно дифференцируемой, то ее длиной называется сумма
длин каждой дифференцируемой части. В евклидовом пространстве $\MR^n$ длину
кривой $s(t)$ от начала $p$ до текущей точки $x(t)\in\g$ всегда можно выбрать в
качестве параметра вдоль кривой. В этом случае $s$ называют {\em каноническим
параметром}. Он однозначно определяется уравнением
$$
  ds=dt\sqrt{\dot x^\al\dot x^\bt g_{\al\bt}}
  =dt\sqrt{X^\al_{(\g)}X^\bt_{(\g)} g_{\al\bt}},
$$
с начальным условием $s(0)=0$.
\qed\end{defn}
\index{Длина кривой (length of a curve)}%
\index{Канонический параметр (canonical parameter)}%
\index{Параметр канонический (canonical parameter)}%
Из формулы (\ref{ediseu}) следует, что определение длины кривой не зависит от
выбора ее параметризации.

Если кривая параметризована каноническим параметром $s$, то вектор скорости
$X^\al_{(\g)}=dx^\al/ds$ имеет единичную длину:
\begin{equation*}
  X^\al_{(\g)}X^\bt_{(\g)}g_{\al\bt}=1.
\end{equation*}

Можно доказать, что расстояние (\ref{ecldis}) является точной нижней гранью
интегралов (\ref{ediseu}) по всем возможным путям, соединяющим точки $p$ и $q$,
\begin{equation}                                                  \label{eucldi}
  l(p,q)=\underset{\g}{\inf}\int\limits_p^q ds.
\end{equation}
Кривая, вдоль которой интеграл (\ref{eucldi}) принимает наименьшее значение
называется отрезком {\em прямой линии}, соединяющим точки $p$ и $q$.
\index{Прямая линия (straight line)}%
\index{Линия прямая (straight line)}%

Точная нижняя грань (\ref{eucldi}) по всем кривым, соединяющим две точки
$p,q\in\MR^n$, или явная формула (\ref{ecldis}) задают расстояние или метрику
пространства $\MR^n$ в топологическом смысле. Напомним определение метрики
для произвольного множества.
\begin{defn}
{\em Метрикой} на множестве $\MM$ называется функция $l(p,q)$, определенная для
любых двух точек $p,q\in \MM$, которая удовлетворяет следующим условиям:
  \begin{align*}
  1)\quad l(p,p) &=0;\qquad l(p,q)>0,\quad (p\ne q) &
  &\text{-- положительная определенность}\\
  2)\quad l(p,q) &=l(q,p) & &\text{-- симметричность},\\
  3)\quad l(p,q) &\le l(p,r)+l(r,q),\quad \forall r\in \MM &
  &\text{-- неравенство треугольника}.
  \end{align*}
Значение $l(p,q)$ называется {\em расстоянием} между точками $p$ и $q$. Пара
$(\MM,l)$, т.е.\ множество $\MM$ с заданной метрикой $l$, называется
{\em метрическим пространством}.
\qed\end{defn}
\index{Метрика топологического пространства (metric of a topological space)}%
\index{Расстояние (distance)}%
\index{Неравенство треугольника (triangle inequality)}%
\index{Треугольника неравенство (triangle inequality)}%
\index{Топологическая метрика (topological metric)}              \label{ptopme}%
\index{Метрическое пространство (metric space)}%
\index{Пространство метрическое (metric space)}%
\begin{com}
Эту метрику мы будем называть {\em топологической метрикой}, чтобы отличать ее
от дифференциально-геометрической метрики (\ref{eclmet}) и ее обобщения, которое
будет сделано в разделе \ref{smetri}.
\qed\end{com}
\index{Метрика топологическая (topological metric)}%
\index{Топологическая метрикой (topological metric)}%

Понятие топологической метрики очень важно и может быть использовано для
определения сходимости последовательностей. Пусть $(\MM,l)$ -- метрическое
пространство, и $\lbrace x_i\rbrace$, $i\in\MN$ -- последовательность точек в
$\MM$.
\begin{defn}
Назовем $\lbrace x_i\rbrace$ {\em фундаментальной последовательностью} или
{\em последовательностью Коши}, если для любого $\e>0$ существует такое
натуральное число $N_\e\in\MN$, что $l(x_i,x_k)<\e$ при всех $i>N_\e$ и
$k>N_\e$. Фундаментальная последовательность, по-определению, называется
сходящейся к некоторой точке, которая называется {\em пределом}
последовательности. Пространство $\MM$ называется {\em метрически полным},
если любая фундаментальная последовательность в $(\MM,l)$ сходится к некоторой
точке из $\MM$.
\qed\end{defn}
\index{Фундаментальная последовательность (fundamental sequence)}%
\index{Последовательность фундаментальная (fundamental sequence)}%
\index{Последовательность Коши (Cauchy sequence)}%
\index{Коши последовательность (Cauchy sequence)}%
\index{Предел последовательности (limit of a sequence)}%
\index{Метрическая полнота (metric completeness)}%
\index{Полнота метрическая (metric completeness)}%
Другими словами полнота означает, что предел любой последовательности из $\MM$,
если он существует, тоже принадлежит $\MM$. Нетрудно показать, что если предел
существует, то он единственен.

На одном и том же множестве $\MM$ можно задавать различные топологические
метрики. Метрика $l$ на множестве $\MM$ называется {\em полной}, если
метрическое пространство $(\MM,l)$ полное.
\index{Полная метрика (complete metric)}%
\index{Метрика полная (complete metric)}%
\begin{exa}
Все пространство $\MR^n$, а также любое замкнутое подмножество в нем с метрикой
(\ref{eucldi}) является метрически полным. Открытые подмножества $\MR^n$,
отличные от всего пространства, метрически неполны, т.к.\ фундаментальные
последовательности могут сходиться к граничным точкам, которые не принадлежат
данным подмножествам.
\qed\end{exa}

В пространстве $\MR^n$, используя понятие расстояния (топологической метрики)
(\ref{ecldis}), можно задавать различные подмножества, которые широко
используются в дальнейшем и играют большую роль в приложениях. Эти подмножества
мы определим в декартовой системе координат.
\begin{exa}[\bf Шар]
Подмножество $\MB_r^n(p)\subset\MR^n$, определяемое неравенством:
\begin{equation}                                                  \label{eballs}
  \MB_r^n(p):=\lbrace x\in \MR^n:~|x-p|<r\rbrace,\qquad r=\const>0,
\end{equation}
называется $n$-мерным открытым {\em шаром} радиуса $r$ с центром в точке $p$.
В одномерном случае, $n=1$, шар называется {\em интервалом}\footnote{Этот
термин употребляется в дифференциальной геометрии также для обозначения
квадрата расстояния между двумя бесконечно близкими точками (\ref{einter})}.
При $n=2$ шар $\MB_r^2$ называется {\em диском} или {\em кругом}.
\index{Шар (ball)}\index{Интервал (interval)}%
\index{Круг (circle)}\index{Диск (disc)}%

Если пара $(\MM,l)$ -- произвольное метрическое пространство, то в нем
также можно определить шар радиуса $r$ с центром в точке $p$:
\begin{equation}                                                  \label{eballm}
  \MB_r(p):=\lbrace x\in \MM:~l(x,p)<r\rbrace.
\end{equation}
Последнее определение не зависит от того имеет ли множество точек $\MM$
размерность или нет.
\qed\end{exa}
\begin{exa}[\bf Сфера]
Подмножество $\MS_r^{n-1}(p)\subset\MR^n$, определяемое равенством:
\index{Сфера (sphere)}%
\begin{equation}                                                  \label{esphde}
  \MS_r^{n-1}(p):=\lbrace x\in \MR^n:~|x-p|=r\rbrace,\qquad r=\const>0,
\end{equation}
называется $(n-1)$-мерной {\em сферой} радиуса $r$ с центром в
точке $p$. При $n=1$ сфера вырождается в две точки, являющиеся концами
интервала $(p-r,p+r)\subset\MR$. В дальнейшем для сферы единичного радиуса
нижний индекс мы будем опускать, $\MS^n:=\MS^n_1$.
\qed\end{exa}
\begin{exa}[\bf Куб]
\index{Куб (cube)}%
Подмножество $\MU^n_a(p)\subset\MR^n$, определяемое неравенством:
\begin{equation}                                                  \label{edecub}
  \MU^n_a(p):=\lbrace x\in\MR^n:~|x^\al-p^\al|<a/2,~\al=1,\dots,n\rbrace,\qquad
  a=\const>0,
\end{equation}
называется $n$-мерным открытым {\em кубом} со стороной $a$ и
центром в точке $p$. Куб представляет собой прямое произведение $n$ интервалов
$(-a/2,a/2)\subset\MR$.
\qed\end{exa}
\begin{com}
Условия (\ref{eballs})--(\ref{edecub}) определяют только множества точек шара,
сферы и куба, ничего не говоря о том, как устроена топология на этих множествах.
Мы всегда предполагаем, что топология шара, сферы, куба, а также других
подмножеств евклидова пространства индуцирована их вложением в $\MR^n$
(см.\ следующий раздел).
\qed\end{com}
\begin{exa}
Обозначим через $\MR^n_+\subset\MR^n$ подпространство в $\MR^n$, определяемое
условием $x^n>0$. Его замыкание $\overline\MR^n_+$ имеет край, который является
гиперплоскостью $\MR^{n-1}$ и определяется уравнением $x^n=0$. Мы считаем, что
топологии на $\MR^n_+$ и $\overline\MR^n_+$ индуцированы вложениями
$\MR^n_+,\overline\MR^n_+\hookrightarrow\MR^n$. Заметим, что открытые
подмножества $\overline\MR^n_+$ могут содержать точки края.
\qed\end{exa}
\begin{defn}
Подмножество $\MU\subset\MR^n$ называется {\em ограниченным}, если существует
шар $\MB^n_r$ конечного радиуса $r$, целиком содержащий $\MU$.
\qed\end{defn}
\index{Ограниченное подмножество (bounded subset)}%
\index{Подмножество ограниченное (bounded subset)}%
\begin{exa}
Все $\MR^n$, а также $\MR^n_+,\overline\MR^n_+$ является неограниченными
подмножествами в $\MR^n$.
\qed\end{exa}
\subsection{$\MR^n$ как топологическое пространство              \label{seucto}}
В предыдущем разделе мы использовали понятия открытого и замкнутого множеств,
не дав им определений. Ниже мы восполним этот пробел и приведем необходимые
сведения из общей топологии, которые необходимы для определения основного
понятия дифференциальной геометрии -- многообразия.

Начнем с общего определения топологического пространства и топологии.
\begin{defn}
{\em Топологическим пространством} называется пара $(\MM,\CI)$, состоящая из
множества точек $x\in \MM$ и некоторого семейства
$\lbrace\MU_i\rbrace_{i\in I}=\CI$ своих подмножеств $\MU_i\subset\MM$, которые
удовлетворяют следующим условиям:

1) пустое множество $\emptyset$ и все множество $\MM$ принадлежат $\CI$;

2) пересечение {\em любой пары} подмножеств из $\CI$ принадлежит $\CI$;

3) объединение {\em любого семейства} подмножеств из $\CI$ принадлежит $\CI$.
\newline
Элементы семейства $\CI$ называются {\em открытыми множествами}
пространства $\MM$, а семейство открытых множеств $\CI$ -- {\em топологией}
пространства $\MM$. Некоторое семейство $\lbrace\MB_j\rbrace_{j\in J}=\CJ$
открытых множеств называется {\em базой} топологии пространства $\MM$, если
каждое множество из $\CI$ есть объединение каких-либо множеств из $\CJ$.
\qed\end{defn}
\index{Пространство топологическое (topological space)}%
\index{Топологическое пространство (topological space)}%
\index{Открытое множество (open set)}\index{Множество открытое (open set)}%
\index{Топология (topology)}%
\index{База топологии (base of topology)}%
\index{Топологии база (base of topology)}%

Ясно, что $\MM=\cup_{i\in I}\MU_i$. Множество индексов $I$ может быть
произвольным, в том числе несчетным.

Из условия 2) следует, что пересечение любого, но конечного числа открытых
множеств также является открытым. В примере \ref{eopser} показано,
что пересечение бесконечного числа открытых множеств может не быть открытым.
В условии 3) допускается объединение бесконечного числа открытых множеств.

Одна и та же топология $\CI$ на множестве $\MM$ может иметь много баз.

На любом множестве можно задать топологию, и не одну. Исключение составляют
пустое множество $\emptyset$ (топология состоит из одного открытого множества
-- самого $\emptyset$) и множество, состоящее из одного элемента. В
последнем случае топология единственна и состоит из двух открытых множеств:
пустого множества и самого элемента.
\begin{exa}
Пусть $\MM$ -- множество. Будем считать, что каждая точка $x\in\MM$ является
открытым множеством и их совокупность образует базу топологии $\MM$. Тогда
любое подмножество $\MU\subset\MM$ будет открытым. Такую топологию называют
{\em дискретной}.
\qed\end{exa}
\index{Дискретная топология (discrete topology)}%
\index{Топология дискретная (discrete topology)}%
\begin{exa}
Пусть $\MM$ -- множество. Будем считать, что вся топология $\MM$ состоит из
двух множеств: пустого множества и всего $\MM$. Это -- пример другой крайности,
и поэтому такую топологию называют {\em антидискретной}. Антидискретную
топологию называют также {\em тривиальной}, потому что она слишком бедна.
\qed\end{exa}
\index{Антидискретная топология (indiscrete topology)}%
\index{Топология антидискретная (indiscrete topology)}%
\index{Тривиальная топология (trivial topology)}%
\index{Топология тривиальная (trivial topology)}%
\begin{defn}
Если на множестве $\MM$ задано две топологии $\CI_1$ и $\CI_2$, причем
$\CI_1\subset\CI_2$, то говорят, что топология $\CI_1$ {\em слабее (грубее)}
топологии $\CI_2$, или что топология $\CI_2$ {\em сильнее (тоньше)} топологии
$\CI_1$.
\qed\end{defn}
\index{Топология более слабая (coarser topology)}%
\index{Топология более грубая (coarser topology)}%
\index{Топология более сильная (stronger topology)}%
\index{Топология более тонкая (stronger topology)}%
Очевидно, что дискретная топология является наиболее тонкой, а антидискретная
топология -- наиболее грубой.
\begin{com}
Явное описание всех открытых множеств, т.е.\ топологии на множестве точек $\MM$
в общем случае является сложной задачей. Поэтому на практике, чтобы определить
топологию в пространстве $\MM$, сначала задают базу топологии, а затем объявляют
все возможные объединения элементов базы открытыми множествами. Поэтому понятие
базы топологии помогает конструктивно подойти к заданию топологии на множестве
точек $\MM$.
\qed\end{com}
\begin{exa}
Пусть $\MR$ -- вещественная прямая. Семейство всех открытых интервалов
$(a,b)\subset\MR$ является базой некоторой топологии на $\MR$, которая
занимает промежуточное положение между дискретной и антидискретной
топологией. Такая топология на вещественной прямой называется
{\em естественной}. Она лежит в основе математического анализа.
\qed\end{exa}
\index{Естественная топология (natural topology)}%
\index{Топология естественная (natural topology)}%
\begin{defn}
Пусть $(\MM,\CI)$ и $(\MN,\CJ)$ -- два топологических пространства. На прямом
произведении $\MM\times\MN$, образованном всеми упорядоченными парами $(x,y)$,
где $x\in\MM$ и $y\in\MN$, можно ввести топологию. А именно, множество всех
пар подмножеств $(\MU,\MV)$, где $\MU$ и $\MV$ открыты соответственно в $\MM$
и $\MN$, образует базу некоторой топологии в прямом произведении $\MM\times\MV$.
Такое произведение называется {\em топологическим}.
\qed\end{defn}
\index{Топологическое произведение (topological product)}%
\index{Произведение топологическое (topological product)}%
\begin{defn}
{\em Евклидовым} пространством $\MR^n$ размерности $n$ называется топологическое
произведение $n$ вещественных прямых $\MR$:
\begin{equation}                                                  \label{euclde}
  \MR^n:=\underbrace{\MR\times \MR\times\dots\times \MR}_n,
\end{equation}
каждая из которых снабжена естественной топологией. Топологию $\MR^n$ также
будем называть {\em естественной}. Будем писать $\dim\MR^n=n$.
\qed\end{defn}
\index{Размерность евклидова пространства %
(dimensionality of the Euclidean space}%
Для топологического произведения мы будем использовать тот же символ, что и для
прямого, подразумевая, что на прямом произведении задается топология, которая
определяется топологией сомножителей. Поэтому для евклидова пространства мы
сохранили общепринятое прежнее обозначение $\MR^n$.

Базой естественной топологии в евклидовом пространстве $\MR^n$ является
множество всех кубов (\ref{edecub}) с произвольными сторонами и центрами.

Аналогично определяется комплексное пространство
\begin{equation}                                                  \label{eucldc}
  \MC^n:=\underbrace{\MC\times\MC\times\dots\times\MC}_n,
\end{equation}
где $\MC=\MR\times\MR$ -- комплексная плоскость с естественной топологией,
совпадающей с топологией евклидовой плоскости $\MR^2$. Очевидно, что, как
топологическое пространство, комплексное пространство $\MC^n$ диффеоморфно
евклидову пространству вдвое большего числа измерений, $\MC^n\approx\MR^{2n}$.
\begin{defn}
Рассмотрим некоторое подмножество $\MD\subset\MM$ топологического пространства
$(\MM,\CI)$. Будем говорить, что на $\MD$ задана {\em индуцированная
топология} $(\MD,\CJ)$, если подмножество $\MV\in\CJ$ открыто тогда и
только тогда, когда $\MV=\MU\cap\MD$, где $\MU\in\CI$ -- некоторое открытое
подмножество в $\MM$.
\qed\end{defn}
\index{Индуцированная топология (induced topology)}%
\index{Топология индуцированная (induced topology)}%
\begin{exa}
Пусть $\MD\subset\MR^n$ -- конечное множество точек евклидова пространства
$\MR^n$. Тогда топология на $\MD$, индуцированная естественной топологией в
$\MR^n$, является дискретной.
\qed\end{exa}
\begin{com}
При рассмотрении подмножеств евклидова пространства $\MR^n$ как топологических
пространств мы будем предполагать, что они снабжены индуцированной топологией,
если не оговорено противное. В частности, говоря о шарах, сферах и других
подмножествах в $\MR^n$, мы предполагаем, что на них задана естественная
топология, индуцированная вложением в евклидово пространство.
\qed\end{com}

Обсудим связь между метрическими пространствами, рассмотренными в предыдущем
разделе, и топологическими пространствами.
В произвольном метрическом пространстве $(\MM,l)$ можно выбрать в качестве
базы топологии семейство шаров, определенных формулой (\ref{eballs}). Тем
самым любое метрическое пространство превращается в топологическое.
Обратное утверждение неверно. Не на всяком топологическом пространстве
можно ввести метрику такую, чтобы она определяла заданную топологию.
\begin{defn}
Топологическое пространство $(\MM,\CI)$ называется {\em метризуемым}, если
на множестве его точек $\MM$ можно ввести такую метрику $l(p,q)$, что
множество всех шаров (\ref{eballm}) является базой топологии $\CI$.
\qed\end{defn}
\index{Метризуемое топологическое пространство (metrizable topological space)}%
\index{Топологическое пространство метризуемое (metrizable topological space)}%
\begin{exa}
Топологическое пространство с дискретной топологией не является метризуемым.
\qed\end{exa}

Если на множестве $\MM$ заданы две различные топологические метрики (расстояния)
$l_1$ и $l_2$, то они могут индуцировать одно и то же топологическое
пространство $(\MM,\CI)$.

Наличие евклидова расстояния (\ref{ecldis}) в евклидовом пространстве $\MR^n$
позволяет определить шар (\ref{eballs}). Совокупность всех шаров можно выбрать
в качестве базы некоторой топологии евклидова пространства. Любой куб можно
представить в виде объединения бесконечного числа шаров. Поэтому множество всех
шаров также является базой естественной топологии евклидова пространства
$\MR^n$. Таким образом, евклидово пространство является метризуемым. На нем
определена евклидова метрика (\ref{ecldis}). В дальнейшем, мы, как правило,
предполагаем, что на евклидовом пространстве задана также евклидова метрика.

Можно показать, что минимальная база естественной топологии евклидова
пространства $\MR^n$ состоит из шаров с рациональными радиусами. Эта база
топологии является счетной.

В общем случае, если топологическое пространство $(\MM,\CI)$ имеет счетную базу,
то говорят, что $\MM$ удовлетворяет {\em второй аксиоме счетности}.
\index{Вторая аксиома счетности (second axiom of countability)}%

В топологическом пространстве $\MM$ {\em окрестностью} точки $x\in\MM$
называют любое открытое множество в $\MM$, содержащее $x$.
\index{Окрестность (neighborhood)}%
Обычно в качестве окрестности точки евклидова пространства $x\in\MR^n$
выбирается некоторый шар с центром в этой точке. Нетрудно показать, что
подмножество $\MU\subset\MM$ открыто тогда и только тогда, когда каждая точка
$x\in\MU$ имеет окрестность $\MU_x$, целиком содержащуюся в $\MU$.
Точка $x$ есть {\em предельная точка}
\index{Точка предельная (limit point)}%
\index{Предельная точка (limit point)}
подмножества $\MU\subset \MM$, если каждая окрестность точки $x$
содержит точки множества $\MU$, отличные от $x$.

Дадим два эквивалентных определения замкнутого множества.
\begin{defn}
Подмножество $\MU\subset\MM$ топологического пространства $(\MM,\CI)$
{\em замкнуто}, если\\
\indent 1) $\MU$ есть дополнение в $\MM$ некоторого открытого множества,\\
или\\
\indent 2) $\MU$ содержит все свои предельные точки.
\qed\end{defn}
\index{Множество замкнутое (closed set)}%
\index{Замкнутое множество (closed set)}%
Нетрудно доказать, что объединение двух и, следовательно, произвольного
конечного числа замкнутых множеств является замкнутым множеством. Кроме того,
пересечение любого семейства, в том числе бесконечного, замкнутых множеств есть
замкнутое множество.

Пустое множество включается в определение топологического пространства и,
по-определению, является открытым. Отсюда вытекает, что все топологическое
пространство, как дополнение пустого множества, является одновременно открытым
и замкнутым. Аналогично, само пустое множество также является открытым и
замкнутым одновременно. Отдельная точка евклидова пространства $\MR^n$ является
замкнутым подмножеством как дополнение открытого подмножества.
\begin{com}
Топологию пространства можно определить в терминах замкнутых подмножеств.
Определение аналогично приведенному выше, только открытые множества надо
заменить на замкнутые, а операции пересечения и объединения поменять местами.
\qed\end{com}
\begin{com}
Отрезок $[0,1]\subset\MR$ является замкнутым подмножеством в $\MR$. Однако, если
рассматривать отрезок $[0,1]$ как самостоятельное топологическое пространство с
топологией, индуцированной из $\MR$, то он будет одновременно и замкнутым, и
открытым.
\qed\end{com}
\begin{defn}
{\em Дискретным подмножеством} топологического пространства $\MM$ называется
такое его подпространство $\MU\subset\MM$, каждое подмножество которого замкнуто
в $\MM$.
\qed\end{defn}
\index{Дискретное подмножество (discreet subset)}%
\index{Подмножество дискретное (discreet subset)}%
\begin{exa}
Набор изолированных точек в евклидовом пространстве $\MR^n$ является дискретным
подмножеством. Он всегда конечен или счетен.
\qed\end{exa}

Рассмотрим произвольное подмножество $\MU\subset\MR^n$, которое не совпадает со
всем евклидовым пространством. В общем случае оно может не быть ни открытым, ни
замкнутым. {\em Замыканием} множества называется объединение самого множества
$\MU$ и всех его предельных точек или, что эквивалентно, пересечение всех
замкнутых подмножеств, содержащих $\MU$. Оно обозначается через
$\overline{\MU}$. Из определения следует, что всегда $\MU\subset\overline{\MU}$.
\index{Замыкание множества (closure of a set)}%
\begin{prop}                                                     \label{lpcous}
Точка $x\in\overline{\MU}$ тогда и только тогда, когда любая окрестность
$\MV_x$ точки $x$ имеет непустое пересечение с $\MU$,
$\MV_x\bigcap\MU\ne\emptyset$.
\end{prop}
\begin{proof}
См., например, \cite{Kelley57R}.
\end{proof}
\begin{defn}
Если некоторое множество совпадает со своим замыканием, $\MU=\overline\MU$, то
оно является замкнутым. {\em Внутренностью} множества $\MU$ называется
наибольшее открытое множество $\Int \MU$, целиком содержащееся в $\MU$, или, что
эквивалентно, объединение всех открытых множеств, содержащихся в $\MU$. Точка
$x\in\MU$ называется внутренней, если $x\in\Int\MU$. {\em Границей} множества
$\MU$ называется разность
\begin{equation*}                                                    \tag*{\qed}
  \pl \MU:=\overline{\MU}\setminus\Int \MU.
\end{equation*}
\end{defn}
\index{Внутренность множества (set interior)}%
\index{Граница множества (boundary of a set)}%
\begin{exa}
Шар, определенный равенством (\ref{eballs}), является {\em открытым} шаром.
Замыкание шара $\overline{\MB}_r^n$ включает в себя точки $(n-1)$-мерной сферы
(\ref{esphde}). При этом сфера $\MS_r^{n-1}$ является границей как открытого,
так и замкнутого шара.
\qed\end{exa}
\begin{exa}                                                       \label{eopser}
Покажем, что пересечение бесконечного числа открытых множеств может быть
замкнутым. Рассмотрим последовательность открытых шаров переменного радиуса
$\MB_{r_i}^n(0)$, $i\in\MN$, в
евклидовом пространстве $\MR^n$, $n\ge2$, с центром в начале координат. Пусть
все радиусы больше единицы, $r_i>1$, и их последовательность сходится к единице.
Тогда пересечением этой последовательности шаров является замкнутый шар
единичного радиуса $\overline{\MB}_1^n(0)$, включающий граничную окружность
$\MS^{n-1}_1(0)$.
\qed\end{exa}
\begin{exa}
Объединение бесконечного числа замкнутых подмножеств может быть открыто.
Действительно, рассмотрим последовательность замкнутых шаров переменного радиуса
$\overline{\MB}_{r_i}^n(0)$, $i\in\MN$, в евклидовом пространстве $\MR^n$,
$n\ge2$, с центром в начале координат. Пусть все радиусы ограничены, $0<r_i<1$,
и их последовательность сходится к единице. Тогда объединением этой
последовательности шаров является открытый шар единичного радиуса $\MB_1^n(0)$.
\qed\end{exa}
\begin{prop}
Граница $\pl\MU$ любого подмножества $\MU\subset\MM$ является замкнутым
подмножеством.
\end{prop}
\begin{proof}
Поскольку $\MM\setminus\overline\MU$ и $\Int\MU$ открыты в $\MM$, то граница
$\pl\MU$ всегда замкнута, как дополнение открытых множеств.
\end{proof}
Дадим эквивалентное определение внутренних и граничных точек.
\begin{defn}
Пусть $\MU\subset\MM$ -- собственное подмножество. Точка $x\in\MU$ называется
{\em внутренней}, если подмножество $\MU$ содержит некоторую окрестность
$\MU_x$, содержащую точку $x$. Точка $y\in\MM$ называется {\em внешней} по
отношению к подмножеству $\MU\subset\MM$, если существует окрестность
$\MU_y\ni y$, которая не имеет общих точек с $\MU$. Точка $x\in\MM$ называется
{\em граничной} точкой подмножества $\MU\subset\MM$, если любая окрестность
$\MU_x\ni x$ содержит как внутренние, так и внешние точки подмножества
$\MU\subset\MM$.
\qed\end{defn}\index{Внутренняя точка (internal point)}%
\index{Точка внутренняя (internal point)}%
\index{Внешняя точка (external point)}%
\index{Точка внешняя (external point)}%
\index{Граничная точка (boundary point)}%
\index{Точка граничная (boundary point)}%
Дадим еще несколько менее наглядных определений.
\begin{defn}
Точка $x\in\MM$ называется {\em изолированной}, если у нее есть окрестность, не
содержащая других точек $\MM$. Ясно, что эта окрестность состоит только из одной
точки $x$. Подмножество $\MU\subset\MM$ называется
{\em всюду плотным} в $\MM$, если его замыкание $\overline{\MU}$ совпадает со
всем $\MM$. Эквивалентно, множество $\MU$ {\em всюду плотно} в $\MM$, если в
каждом открытом множестве пространства $\MM$ содержится хотя бы одна точка из
$\MU$.
\qed\end{defn}
\index{Изолированная точка (isolated point)}%
\index{Точка изолированная (isolated point)}%
\index{Всюду плотное подмножество (everywhere dense subset)}%
\index{Подмножество всюду плотное (everywhere dense subset)}%
\begin{exa}
Множество рациональных чисел $\MQ$ всюду плотно в $\MR$.
\qed\end{exa}
\begin{defn}
Семейство $\lbrace\MU_i\rbrace_{i\in I}$ подмножеств пространства $\MM$
называется {\em покрытием} множества $\MM$, если
\begin{equation*}
  \MM=\bigcup_{i\in I}\MU_i.
\end{equation*}
При этом множество значений индекса $i\in I$ не обязано быть счетным. Если $\MM$
-- топологическое пространство, то покрытие называется {\em открытым}, если все
множества $\MU_i$ открыты. Мы говорим, что некоторое покрытие имеет
{\em подпокрытие} $\lbrace\MU_i\rbrace_{i\in J}$, где $J\subset I$, если оно
является покрытием само по себе. Если множество значений индекса $J$ является
конечным, то подпокрытие будет конечным. Покрытие
$\lbrace\MV_i\rbrace_{i\in I}$ называется {\em измельчением} покрытия
$\lbrace\MU_i\rbrace_{i\in I}$, если для всех значений индекса
$\MV_i\subset\MU_i$. При этом некоторые из множеств $\MV_i$ могут быть пустыми.
\qed\end{defn}
\index{Покрытие (covering)}%
\index{Открытое покрытие (open covering)}%
\index{Покрытие открытое (open covering)}%
\index{Подпокрытие (subcovering)}%
\index{Измельчение (refinement)}%
\begin{exa}
База топологии как и сама топология на $\MM$ являются покрытиями.
\qed\end{exa}
\begin{exa}
Совокупность интервалов $\MU_1=(0,1)$, $\MU_2=(0,3/4)$, $\MU_3=(1/4,1)$
является покрытием единичного интервала $(0,1)$. Это покрытие имеет много
измельчений, в том числе $\MV_1=\emptyset$, $\MV_2=(0,2/3)$, $\MV_3=(1/3,1)$.
\qed\end{exa}
\begin{com}
В дифференциальной геометрии важную роль играют счетные измельчения покрытий.
\qed\end{com}

В предыдущем разделе мы определили сходимость последовательностей в метрических
пространствах $(\MM,l)$. Дадим определение сходимости последовательностей в
топологических пространствах $(\MM,\CI)$, которое не опирается на метрическую
сходимость.
\begin{defn}
Последовательность точек $\lbrace x_i\rbrace$, $i=1,2,\dotsc$, топологического
пространства $\MM$ сходится к точке $x\in\MM$, если для каждой окрестности
$\MU_x$ точки $x$ существует такое натуральное число $N_\MU$, что
$x_i\in\MU_x$ для всех $i>N_\MU$.
\qed\end{defn}
\begin{theorem}
Пусть топологическое пространство $(\MM,\CI)$ метризуемо, т.е.\ на $\MM$
существует метрика $l$, которая индуцирует данную топологию. Тогда
последовательность точек $\lbrace x_i\rbrace$ сходится к точке $x\in\MM$ в
метрике $l$ тогда и только тогда, когда она сходится в топологии $\CI$.
\end{theorem}
\begin{proof}
См., например, \cite{Engelk85R}.
\end{proof}
Из этой теоремы следует, что для метрических пространств определения сходимости
по метрике и индуцированной топологии эквивалентны. В то же время
определение сходимости последовательностей в топологических пространствах
является более общим, т.к.\ не всякое топологическое пространство метризуемо.
С другой стороны, в метрических пространствах можно ввести понятие
фундаментальной последовательности и полноты.

Продолжим общее рассмотрение.
\begin{defn}
Топологическое пространство $\MM$ называется {\em компактным}, если каждое его
открытое покрытие содержит конечное подпокрытие. Хаусдорфово компактное
топологическое пространство называется {\em компактом}. Подмножество $\MU$
топологического пространства $\MM$ называется {\em компактным}, если оно
компактно в индуцированной топологии. Подмножество $\MU$ топологического
пространства $\MM$ называется {\em относительно компактным}, если его замыкание
$\overline\MU$ компактно.
\end{defn}
\index{Пространство компактное (compact space)}%
\index{Компактное пространство (compact space}%
\index{Компакт (compact set)}%
\index{Компактное подмножество (compact subset)}%
\index{Подмножество компактное (compact subset)}%
\index{Относительно компактное подмножество (relatively compact subset)}%
\index{Подмножество относительно компактное (relatively compact subset)}%
Компактные топологические пространства обладают рядом замечательных свойств.
Доказательства следующих пяти теорем можно найти, например, в
\cite{Kelley57R}.
\begin{theorem}                                                   \label{tcomam}
Топологическое пространство компактно тогда и только тогда, когда всякая
непрерывная функция на нем ограничена, т.е.\ достигает на нем
своего минимального и максимального значения.
\end{theorem}
\begin{theorem}
Топологическое пространство компактно тогда и только тогда, когда каждое
бесконечное его подмножество имеет предельную точку.
\end{theorem}
\begin{theorem}
Если $\MU$ -- компактное подмножество хаусдорфова пространства $\MM$ и
$x\in\MM\setminus\MU$, то у точки $x$ и подмножества $\MU$ существуют
непересекающиеся окрестности.
\end{theorem}
\begin{cor}
Каждое компактное подмножество $\MU$ хаусдорфова пространства $\MM$ замкнуто:
$\MU=\overline\MU$.
\end{cor}
\begin{theorem}
У любых двух непересекающихся компактных подмножеств $\MU$ и $\MV$ хаусдорфова
пространства $\MM$ существуют непересекающиеся окрестности.
\end{theorem}
\begin{theorem}
Пусть $\lbrace\MU_i\rbrace$ -- открытое покрытие компактного подмножества $\MV$
метрического пространства $(\MM,l)$. Тогда существует такое положительное число
$r$, что открытый шар радиуса $r$ с центром в произвольной точке подмножества
$\MU$ содержится в некотором элементе покрытия $\MU_i$.
\end{theorem}
\begin{cor}[\bf Лемма Лебега]
Для любого открытого покрытия $\lbrace\MU_i\rbrace$ замкнутого интервала
вещественных чисел существует такое положительное число $r$, что если $|x-y|<r$,
то в покрытии найдется такой элемент $\MU_i$, который содержит обе точки $x$ и
$y$.
\end{cor}

\begin{exa}
Все евклидово пространство $\MR^n$ является некомпактным, т.к.\ можно выбрать
последовательность точек, которая не содержит ни одной предельной точки.
Например, множество натуральных чисел, лежащих на любой из координатных
прямых не содержит предельной точки.
\qed\end{exa}
\begin{exa}
Открытый шар $\MB^n_r\subset\MR^n$ является некомпактным множеством, т.к.\
последовательность точек, сходящихся к некоторой граничной точке, не имеет в нем
предельной точки. В то же время замкнутый шар $\overline{\MB}^n_r$ будет уже
компактным множеством.
\qed\end{exa}
\begin{exa}
Множество с дискретной топологией компактно тогда и только тогда, когда оно
конечно.
\qed\end{exa}
\begin{theorem}[\bf Гейне--Борель--Лебег]
Любой компакт является компактным множеством.
\end{theorem}
\begin{proof}
См., например, \cite{Kelley57R}.
\end{proof}
\index{Теорема Гейне--Бореля--Лебега (Heine--Borel--Lebesgue theorem}%
\begin{theorem}[\bf Тихонов]
Топологическое произведение произвольного множества непустых топологических
пространств является компактным тогда и только тогда, когда каждый сомножитель
является компактным пространством.
\end{theorem}
\index{Теорема Тихонова (Tychonoff theorem)}%
\begin{proof}
См.\ \cite{Tychon35}.
\end{proof}
\begin{com}
Количество сомножителей в сформулированной теореме может быть несчетным.
\qed\end{com}
Теорема Тихонова позволяет дать критерий компактности подмножеств евклидова
пространства с индуцированной топологией.
\begin{theorem}
Подпространство $\MM$ $n$-мерного евклидова пространства $\MR^n$ является
компактом в том и только в том случае, если множество $\MM$ замкнуто и
ограничено.
\end{theorem}
\begin{proof}
См., например, \cite{Engelk85R}.
\end{proof}

\begin{defn}
Топологическое пространство $\MM$ называется  {\em локально компактным},
если каждая точка $x\in\MM$ имеет окрестность, замыкание которой компактно.
\qed\end{defn}
\index{Топологическое пространство локально компактное %
(locally compact topological space)}%
\index{Локально компактное топологическое пространство %
(locally compact topological space)}%
Конечно, каждое компактное топологическое пространство является локально
компактным. Обратное утверждение неверно.
\begin{exa}
Евклидово $\MR^n$ и комплексное $\MC^n$ пространства некомпактны, но
локально компактны.
\qed\end{exa}
\begin{exa}
Бесконечномерное гильбертово пространство не является локально компактным.
\qed\end{exa}

В дальнейшем при рассмотрении многообразий необходимо будет воспользоваться
существованием разбиения единицы. Для этой цели нам понадобится понятие
паракомпактности, которое является слабейшим требованием, достаточным для
существования разбиения единицы.
\begin{defn}
Топологическое пространство $\MM$ называется {\em паракомпактным}, если любое
его покрытие открытыми множествами $\lbrace\MU_i\rbrace_{i\in I}$ имеет локально
конечное измельчение $\lbrace\MV_i\rbrace_{i\in I}$. {\em Локальная конечность}
означает, что для каждой точки $x\in\MM$ существует окрестность
$\MW_x\subset\MM$ такая, что $\MV_i\cap\MW_x\ne\emptyset$ только для конечного
числа индексов. Или, любое компактное подмножество $\MM$ пересекается с конечным
числом открытых множеств $\MV_i$.
\qed\end{defn}
\index{Паракомпактность (paracompactness}%
\index{Локальная конечность (local finiteness)}%
\index{Конечность локальная (local finiteness)}%
\begin{exa}
Любое метризуемое топологическое пространство паракомпактно. В частности,
евклидово пространство $\MR^n$ является паракомпактным.
\qed\end{exa}
Перечислим некоторые топологические свойства $\MR^n$. Часть этих свойств,
например, сепарабельность и хаусдорфовость, наследуется всеми многообразиями.
Другие же свойства, такие как связность и односвязность, различны для
различных многообразий.
\begin{defn}
Топологическое пространство называется {\em сепарабельным}, если оно содержит
счетное всюду плотное подмножество.
\qed\end{defn}
\index{Сепарабельное пространство (separable space)}%
\index{Пространство сепарабельное (separable space)}%
\index{Множество всюду плотное (everywhere dense set)}%
\index{Всюду плотное множество (everywhere dense set)}%
\begin{exa}
Евклидово пространство $\MR^n$ является сепарабельным, т.к.\ в качестве счетного
всюду плотного подмножества можно выбрать, например, множество точек с
рациональными координатами, которое, как известно, счетно при конечном $n$ и
всюду плотно в $\MR^n$.
\qed\end{exa}
\begin{theorem}
Пространство, топология которого обладает счетной базой, сепарабельно.
\end{theorem}
\begin{proof}
См., например, \cite{Kelley57R}.
\end{proof}
\begin{defn}
Топологическое пространство $\MM$ называется {\em хаусдорфовым}, если для любой
пары различных точек $x,y\in\MM$ существуют открытые подмножества $\MU_x$ и
$\MU_y$, содержащие соответственно точки $x$ и $y$, такие, что
$\MU_x\cap \MU_y=\emptyset$.
\qed\end{defn}
\index{Пространство хаусдорфово (Hausdorff space)}%
\index{Хаусдорфово пространство (Hausdorff space)}%
\begin{exa}
Евклидово пространство является хаусдорфовым пространством.
\qed\end{exa}
Хаусдорфовость топологического пространства важна при определении предела
последовательностей. Если мы хотим, чтобы у любой последовательности точек мог
существовать не более, чем один предел, необходимо потребовать, чтобы
пространство было хаусдорфовым.
\begin{theorem}                                                   \label{thausd}
Топологическое пространство является хаусдорфовым тогда и только
тогда, когда никакая последовательность в этом пространстве не
сходится к двум различным точкам.
\end{theorem}
\begin{proof}
См., например, \cite{Kelley57R}.
\end{proof}
\begin{exa}                                                       \label{enhaus}
Дадим пример нехаусдорфова топологического пространства. Рассмотрим множество
точек, состоящее из объединения вещественной прямой $y=0$ и точки
$(0,1)\in\MR^2$ на евклидовой плоскости, изображенных на
рис.\ref{fnhaus},{\it a}.
\label{pnonha}
\begin{figure}[h,b,t]
\hfill\includegraphics[width=.8\textwidth]{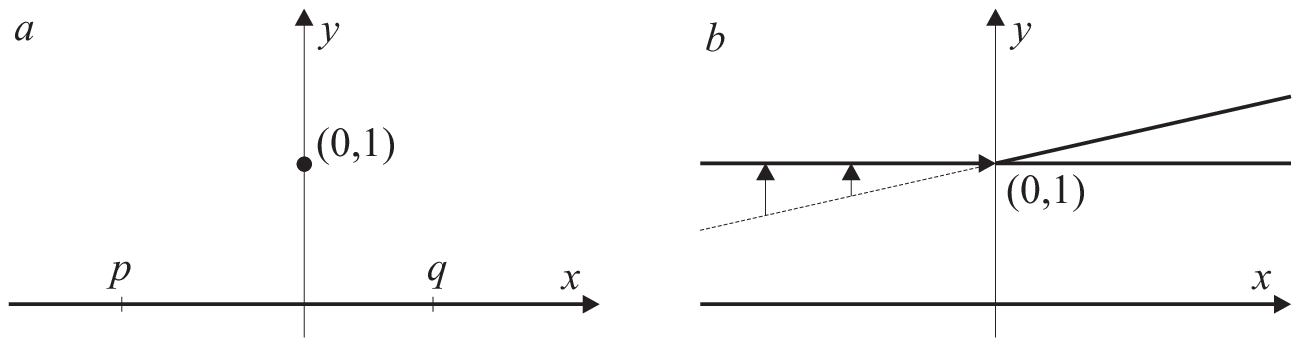}
\hfill {}
\\
\centering \caption{Примеры \ref{enhaus} и \ref{enhaub} нехаусдорфовых
                    топологических пространств. \label{fnhaus}}
\end{figure}
Определим топологию на рассматриваемом множестве следующим образом. Совокупность
открытых интервалов $(p,q)$ на оси $x$, а также множества, состоящие из
объединения точки $(0,1)$ с интервалами $(p,q)$, $p<0$, $q>0$ оси $x$ с
выколотым началом координат, будем считать базой топологии. По-построению, любые
окрестности начала координат $(0,0)\in\MR^2$ и точки $(0,1)\in\MR^2$ имеют
непустое пересечение, и, следовательно, построенное топологическое пространство
не является хаусдорфовым. Любая последовательность точек, сходящаяся к началу
координат $(0,0)$, сходится также и к точке $(0,1)$, и наоборот. Другими
словами, с точки зрения сходящихся последовательностей две точки $(0,0)$ и
$(0,1)$ неразличимы.
\qed\end{exa}
\begin{exa}                                                       \label{enhaub}
Рассмотрим две прямые на плоскости с естественной топологией, проходящие через
точку $(0,1)\in\MR^2$, и отождествим их точки c отрицательными абсциссами,
$x<0$, как показано на рис.\ref{fnhaus},{\it b}. В результате получим некоторое
топологическое пространство $\MM$. Тогда одна точка плоскости $(0,1)\in\MR^2$
соответствует двум различным точкам в $\MM$, которые лежат на разных прямых и
неотождествлены. Эти точки различны, и в тоже время не имеют непересекающихся
окрестностей.
\qed\end{exa}
\begin{com}
В приведенных примерах нехаусдорфовых пространств их топология не является
топологией, индуцированной из $\MR^2$.
\qed\end{com}
\begin{theorem}
Любое метризуемое топологическое пространство $\MM$ является хаусдорфовым.
\end{theorem}
\begin{proof}
Выберем две произвольные различные точки $x\ne y$ в $\MM$ и построим два
непересекающихся шара с центрами в $x$ и $y$.
\end{proof}

Дадим два эквивалентных определения связности топологических пространств.
\begin{defn}
  Топологическое пространство $\MM$ называется {\em связным},

1) \parbox[t]{.92\linewidth}{если его нельзя представить в виде объединения
двух непересекающихся множеств, каждое из которых не содержит предельных
  точек другого, или}

2) \parbox[t]{.92\linewidth}{если единственными подмножествами $\MM$,
открытыми и замкнутыми одновременно, являются $\emptyset$ и $\MM$.}\newline
{\em Компонентой} топологического пространства называется максимальное связное
подпространство.
\qed\end{defn}
\index{Топологическое пространство связное (connected topological space)}%
\index{Связное топологическое пространство (connected topological space)}%
\index{Компонента топологического пространства %
(component of a topological space)}%
Если пространство связно, то оно состоит из одной компоненты.
\begin{prop}
Если топологическое пространство является объединением двух непустых
непересекающихся открытых подмножеств, $\MM=\MU_1\cup\MU_2$, где
$\MU_1\cap\MU_2=\emptyset$, то оно не является связным.
\end{prop}
\begin{proof}
От противного. Пусть $x\in\MU_1$ -- предельная точка для $\MU_2$. Выберем
окрестность $\MU_x\subset\MU_1$, содержащую точку $x$. Тогда она содержит точки
множества $\MU_2$, и, следовательно, подмножества $\MU_1$ и $\MU_2$
пересекаются.
\end{proof}
\begin{defn}
Произвольное открытое связное и односвязное (см.\ раздел \ref{sfundg})
подмножество $\MU\subset\MR^n$ называется {\em областью}.
\qed\end{defn}
\index{Несвязное топологическое пространство (disconnected topological space)}%
\index{Топологическое пространство несвязное (disconnected topological space)}%
\index{Область (domain)}%
\begin{exa}
Рассмотрим сферу, вложенную в евклидово пространство,
$\MS^{n-1}\hookrightarrow\MR^n$. Будем считать, что топология на сфере
индуцирована вложением. Тогда сфера -- связное топологическое пространство.
\qed\end{exa}
{\em Несвязные} топологические пространства $\MM$ представляют собой объединение
конечного или бесконечного числа компонент. В последнем случае число компонент
может быть счетным или несчетным. Каждая точка принадлежит только одной
компоненте, которая называется компонентой данной точки. Всякая компонента
является замкнутым подмножеством в $\MM$.
\begin{com}
В общем случае компонента топологического пространства может не быть открытой.
Экзотический пример описан в \cite{Engelk85R}.
\qed\end{com}

В дальнейшем будут использованы следующие два утверждения, доказательство
которых приведено, например, в \cite{Engelk85R}.
\begin{theorem}                                                   \label{tconun}
Если произвольное семейство $\lbrace\MU_i\rbrace$, $i\in I$, связных
подпространств топологического пространства $\MM$ имеет непустое пересечение, то
его объединение $\cup_{i\in I}\MU_i$ связно.
\end{theorem}
\begin{theorem}
Топологическое произведение произвольного множества непустых топологических
пространств связно в том и только в том случае, если все сомножители связны.
\end{theorem}
\begin{defn}
Топологическое пространство $\MM$ называется {\em локально связным}, если для
каждой точки $x\in\MM$ и любой ее окрестности $\MU_x$ существует связное
множество $\MV\subset\MU_x$, такое, что $x\in\Int\MV$.
\qed\end{defn}
\index{Локально связное пространство (locally connected space)}%
\index{Пространство локально связное (locally connected space)}%
\begin{exa}
Евклидово пространство $\MR^n$ связно и локально связно.
\qed\end{exa}
\begin{exa}
Рассмотрим подмножество точек плоскости $(x,y)\in\MR^2$, состоящее из прямых,
проходящих через начало координат и заданных уравнениями $mx+ny=0$ с целыми
коэффициентами $m$ и $n$. Будем считать, что топология этого подмножества
индуцирована плоскостью. Это множество связно, как объединение связных множеств,
имеющих общую точку. В то же время оно не является локально связным, т.к.\ любое
связное подпространство обязано содержать начало координат.
\qed\end{exa}
\begin{theorem}
Топологическое пространство $\MM$ локально связно в том, и только в том случае,
если компоненты открытых множеств открыты. В частности, любая окрестность любой
точки локально связного пространства содержит связную окрестность этой точки.
\end{theorem}
\begin{proof}
См., например, \cite{RohFuk77R}.
\end{proof}
\begin{theorem}                                                   \label{tlocco}
В локально связном топологическом пространстве $\MM$ каждая компонента является
одновременно открытой и замкнутой.
\end{theorem}
\begin{proof}
См., например, \cite{Engelk85R}.
\end{proof}
\subsection{$\MR^n$ как векторное пространство                   \label{seucve}}
Если на евклидовом пространстве $\MR^n$ задана евклидова метрика и
индуцированная топология, то этого достаточно для определения дифференцируемого
многообразия. Более того, метрику можно и не задавать, понимая под евклидовым
пространством просто топологическое произведение $n$ прямых. Вместе с этим на
евклидовом пространстве помимо метрики и топологии можно ввести дополнительную
алгебраическую структуру -- структуру векторного или линейного пространства.
Она понадобится для определения касательного расслоения к многообразию. Кроме
этого, структура векторного пространства играет большую роль в приложениях,
т.к.\ пространство-время в нерелятивистских моделях и специальной теории
относительности имеет структуру векторного пространства.

Пусть $\MR^n$ -- евклидово пространство с декартовыми координатами $x^\al$,
$\al=1,\dotsc,n$.
\begin{defn}
Назовем {\em вектором} $\overrightarrow{\Sp\Sq}$ евклидова пространства $\MR^n$
с началом в точке $\Sp=\lbrace x^\al_\Sp\rbrace $ и концом в точке
$\Sq=\lbrace x^\al_\Sq\rbrace$ упорядоченный набор чисел
$\lbrace x^\al_\Sq-x^\al_\Sp\rbrace $, которые называются компонентами вектора.
\qed\end{defn}
\index{Вектор (vector)}%
В евклидовой геометрии вектор представляется в виде направленного отрезка прямой
линии, соединяющего точки $\Sp$ и $\Sq$.

Рассмотрим множество векторов $\MV$ с началом в начале системы координат. Для их
обозначения будем использовать жирный наклонный шрифт
$$
  \Bx=\lbrace x^\al\rbrace \in \MV.
$$
Пространство векторов $\MV$ находится во взаимно однозначном соответствии с
точками евклидова пространства $\MR^n$, при этом вектор $\Bx\in\MV$ называется
{\em радиусом-вектором} точки $\lbrace x^\al\rbrace \in\MR^n$.
Назовем {\em нулевым} вектором ${\bf 0}$
\index{Радиус-вектор (radius vector)}%
\index{Нулевой вектор (zero vector)}\index{Вектор нулевой (zero vector)}%
вектор с нулевыми компонентами, ${\bf 0}=(0,\dotsc,0)\in\MV$. Вектор $-\Bx$ с
компонентами $\lbrace -x^\al\rbrace \in \MV$ назовем обратным вектором к вектору
$\Bx$. Для любых двух векторов $\Bx,\By\in\MV$ определим их сумму $\Bx+\By$ как
вектор $\Bz$ с компонентами
\begin{equation}                                                  \label{evecsu}
   \Bz=\Bx+\By=\lbrace z^\al:=x^\al+y^\al\rbrace\in\MV.
\end{equation}
Из свойств сложения вещественных чисел следует, что операция сложения векторов
коммутативна: $\Bx+\By=\By+\Bx$. Пространство векторов $\MV$ с введенной
операцией сложения образует абелеву группу или модуль, поскольку сложение
векторов коммутативно и выполнены необходимые групповые аксиомы:
\begin{equation}                                                  \label{edevec}
\begin{aligned}
&1)\quad \Bx+\By\in\MV & &\text{-- замкнутость по отношению к сложению};\\
&2)\quad (\Bx+\By)+\Bz=\Bx+(\By+\Bz) & &\text{-- ассоциативность};\\
&3)\quad \Bx+{\bf 0}=\Bx & &\text{-- существование единичного элемента};\\
&4)\quad \Bx+(-\Bx)={\bf 0} & &\text{-- существование обратного элемента}.
\end{aligned}
\end{equation}
Напомним, что для абелевых групп групповую операцию принято называть сложением,
а единичный элемент -- нулем.
\begin{com}
Понятие вектора в дифференциальной геометрии является инвариантным и не зависит
от выбора системы координат. Приведенные выше определения были сделаны в
декартовой системе координат сознательно, т.к.\ линейная структура наиболее
просто задается в декартовых координатах. Векторное пространство $\MV$ можно
было бы описать и в произвольных координатах. Однако в криволинейной системе
координат операция сложения векторов выглядит довольно громоздко.
\qed\end{com}

Определим новую операцию на множестве векторов $\MV$. А именно, введем операцию
умножения векторов на действительные числа $a,b,\dots$$\in \MR$, которые в
данном случае принято называть {\em скалярами},
\index{Скаляр (scalar)}%
\begin{equation}                                                  \label{evecnu}
  \MR\times\MV\ni\quad a,\Bx\mapsto a\Bx:=\lbrace ax^\al\rbrace\quad\in\MV.
\end{equation}
Здесь каждая компонента вектора умножается на одно и то же число $a$. Эта
операция обладает всеми свойствами, которые перечислены в следующем общем
определении векторного пространства.
\begin{defn}
Абелева группа $\MV=\lbrace\Bx,\By,\Bz,\dotsc\rbrace$ с операциями сложения
(\ref{edevec}) и умножения на скаляры, которые удовлетворяют
свойствам:
\begin{equation}                                        \label{edevam}
\begin{aligned}
&1)\quad a\Bx\in\MV & &\text{-- замкнутость по отношению к умножению на скаляры};\\
&2)\quad a(\Bx+\By)=a\Bx+a\By & &\text{-- дистрибутивность по
отношению к сложению векторов};\\
&3)\quad (a+b)\Bx=a\Bx+b\Bx & &\text{-- дистрибутивность по
отношению к сложению скаляров};\\
&4)\quad (ab)\Bx=a(b\Bx) & &\text{-- ассоциативность};\\
&5)\quad 1\cdot\Bx=\Bx & &\text{-- умножение на единицу.}
\end{aligned}
\end{equation}
называется {\em линейным пространством}, {\em линейным векторным пространством}
или просто {\em векторным пространством}. Элементы пространства $\MV$ называются
{\em векторами}. Мы также говорим, что векторное пространство -- это модуль над
полем вещественных чисел.
\qed\end{defn}
\index{Линейное пространство (linear space)}%
\index{Пространство линейное (linear space)}%
\index{Линейное векторное пространство (linear vector space)}%
\index{Векторное пространство (vector space)}%
\index{Пространство векторное (vector space)}%
\index{Вектор (vector)}%
\begin{com}
В определении сложения векторов евклидова пространства (\ref{evecsu}) и их
умножения на скаляры (\ref{evecnu}) мы складываем и умножаем компоненты как
вещественные числа. В абстрактном подходе рассматривают множество векторов $\MV$
как произвольную абелеву группу, в которой задано умножение на скаляры со
свойствами (\ref{edevam}). В этом случае можно забыть про исходное евклидово
пространство и говорить про декартовы или криволинейные координаты не имеет
смысла. Если в определении векторного пространства заменить вещественные числа
на комплексные, то получим векторное пространство над полем комплексных чисел.
\qed\end{com}

В произвольном векторном пространстве можно ввести важное понятие {\em базиса},
как такой набор линейно независимых векторов $\lbrace \Be_\al\rbrace$, что
произвольный вектор можно представить в виде линейной комбинации базисных
векторов:
\begin{equation}                                                  \label{evedeb}
  \Bx=x^\al\Be_\al.
\end{equation}
В этом разложении числа $x^\al\in\MR$ называются компонентами вектора $\Bx$ по
отношению к базису $\Be_\al$. Мы будем говорить, что векторное пространство
$\MV$ натянуто на векторы $\Be_\al$, т.е.\ состоит из всех линейных комбинаций
вида (\ref{evedeb}). Базис векторного пространства определен неоднозначно, но
число базисных векторов от выбора базиса не зависит. Это число называется
{\em размерностью} векторного пространства, и мы пишем
\begin{equation*}
  \dim\MV=n,\qquad n\in\MN.
\end{equation*}
\index{Базис векторного пространства (basis of vector space)}%
\index{Размерность векторного пространства (dimensionality of vector space)}%
Другими словами, размерностью линейного пространства называется максимальное
число линейно независимых векторов. Эти векторы можно выбрать в качестве базиса
векторного пространства. В общем случае размерность $n$ может быть бесконечна,
однако в настоящей монографии мы рассматриваем, как правило, конечномерные
векторные пространства. По построению, базисные векторы имеют следующие
компоненты:
\begin{equation}                                                  \label{ebasis}
  \Be_\al=(\underbrace{0,\dotsc,0,}_{\al-1}1,0,\dotsc,0),
\end{equation}
где на $\al$-том месте стоит единица, а остальные компоненты равны нулю.

Пусть задано произвольное векторное пространство $\MV$, $\dim\MV=n$, с
фиксированным базисом. Тогда каждый вектор $\Bx\in\MV$ взаимно однозначно
задается своими компонентами в этом базисе,
$\Bx\leftrightarrow\lbrace x^\al\rbrace$, $\al=1,\dotsc,n$. Сложение векторов и
умножение на скаляры в компонентах задается при этом точно так же, как и для
векторов евклидова пространства $\MR^n$. Поэтому произвольное векторное
пространство $\MV$ с фиксированным базисом изоморфно евклидову пространству
$\MR^n$ той же размерности. Тем самым на любом векторном пространстве можно
ввести структуру многообразия, при этом оно становится тривиальным
многообразием, диффеоморфном $\MR^n$.
\begin{exa}
Произвольное поле $\MF$ можно рассматривать как одномерное векторное
пространство с базисом, состоящем из одного вектора $\Be$. Базисным вектором в
этом случае может являться любой элемент $\Be\in\MF$, отличный от нуля.
\qed\end{exa}
\begin{exa}
Естественным базисом евклидова пространства $\MR^n$, которое рассматривается,
как векторное пространство, является набор единичных векторов, направленных
вдоль декартовых координатных осей.
\qed\end{exa}
\begin{exa}
Множество абсолютно интегрируемых функций на отрезке $[-\pi,\pi]$ с поточечным
сложением и умножением на действительные числа является примером
бесконечномерного векторного пространства. При этом разложение в ряд Фурье
представляет собой разложение по счетному базису, состоящему из
тригонометрических функций.
\qed\end{exa}
\begin{defn}
Отображение двух векторных пространств
\begin{equation*}
  f:\quad \MV\ni\quad\Bx\mapsto\By\quad\in\MW
\end{equation*}
называется {\em линейным}, если
\begin{equation*}
  f(a\Bx_1+b\Bx_2)=af(\Bx_1)+bf(\Bx_2),\qquad \forall\Bx_1,\Bx_2\in\MV,\quad
  a,b\in\MR.
\end{equation*}
Отображение $f$ называется также {\em линейным оператором}.
\qed\end{defn}
\index{Линейный оператор (linear operator)}%
\index{Оператор линейный (linear operator)}%
Если в пространствах $\MV$ и $\MW$ заданы базисы $\Be_\al$,
$\al=1,\dotsc,\dim\MV$, и $\Be_i$, $i=1,\dotsc,\dim\MW$, то произвольный
линейный оператор задается матрицей $A_\al{}^i$:
\begin{equation*}
  f:\quad \MV\ni\quad\Bx=x^\al\Be_\al\mapsto\By=y^i\Be_i\quad\in\MW,\qquad
  y^i=x^\al A_\al{}^i,
\end{equation*}
где $f(\Be_\al):=A_\al{}^i\Be_i$.

Если компоненты всех векторов (\ref{evedeb}) преобразовать с помощью
невырожденной матрицы
\begin{equation}                                                  \label{elintv}
  x^\al\mapsto x^\bt A_\bt{}^\al,\qquad \det A\ne0,
\end{equation}
то получим взаимно однозначное отображение векторного пространства на себя, при
котором все линейные свойства пространства сохраняются. То есть каждая
невырожденная матрица задает некоторый автоморфизм векторного пространства.
Верно и обратное утверждение. Любой автоморфизм векторного пространства в
компонентах задается некоторой невырожденной матрицей. Совокупность матриц $A$
образует группу, которая называется группой линейных однородных преобразований
векторного пространства и обозначается $\MG\ML(n,\MR)$. При преобразовании
координат (\ref{elintv}) с произвольной матрицей $A\in\MG\ML(n,\MR)$ нулевой
вектор не меняется. Поэтому группа общих линейных преобразований действует
эффективно, но не свободно (см.\ раздел \ref{stragr}).

В приложениях часто используются также комплексные векторные пространства. Если
задано вещественное векторное пространство, то оно определяет также комплексное
векторное пространство.
\begin{defn}
{\em Комплексификацией} вещественного векторного пространства $\MV$,
$\dim\MV=n$, называется комплексное векторное пространство ${}^\MC\MV$,
комплексной размерности $\dim_\MC{}(^\MC\MV)=n$ и вещественной размерности
$\dim_\MR({}^\MC\MV=2n$, состоящее из пар $(\Bx,\By)$, которые
обозначаются $\Bz:=\Bx+i\By$, где $\Bx,\By\in\MV$, с обычными операциями
сложения и умножения на комплексные числа. Комплексификацией линейного оператора
$A:~\MV\rightarrow\MV$ называется линейный оператор
\begin{equation*}                                                    \tag*{\qed}
  {}^\MC\!A:\quad {}^\MC\MV\ni\quad\Bx+i\By\mapsto
  {}^\MC\!A(\Bx+i\By):=A\Bx+iA\By\quad\in{}^\MC\MV.
\end{equation*}
\end{defn}
\index{Комплексификация (complexification)}%

Если некоторое подмножество векторного пространства $\MV_1\subset\MV$ само
является векторным пространством, то оно называется {\em линейным
подпространством}. Пусть $\MV_1$ и $\MV_2$ -- два линейных подпространства
векторного пространства $\MV$. Нетрудно проверить, что их пересечение
$\MV_1\cap\MV_2$ также является векторным пространством. В то же время их
объединение в общем случае векторным пространством не будет. Тем не менее мы
будем писать $\MV_1\cup\MV_2$, когда понадобиться необходимость сказать, что
некоторый элемент принадлежит либо пространству $\MV_1$, либо пространству
$\MV_2$.
\begin{exa}
Пусть $\MV=\MR^3$. Рассмотрим две плоскости $\MV_1$ и $\MV_2$ натянутые
на базисные векторы $(\Be_1,\Be_2)$ и $(\Be_2,\Be_3)$ соответственно. Тогда их
пересечение $\MV_1\cap\MV_2$ является прямой линией с направляющим вектором
$\Be_2$. Тем самым пересечение является одномерным векторным пространством.
Объединение $\MV_1\cup\MV_2$ представляет собой объединение двух
перпендикулярных плоскостей, что, конечно, не совпадает с трехмерным евклидовым
пространством $\MR^3$ и не является векторным пространством.
\qed\end{exa}
Очевидно, что
\begin{equation*}
  \dim(\MV_1\cap\MV_2)\le\min\lbrace\dim\MV_1,\dim\MV_2\rbrace.
\end{equation*}

\begin{defn}
Два векторных подпространства $\MV_1$ и $\MV_2$ конечномерного векторного
пространства $\MV$ называются {\em трансверсальными} друг к другу, если они
порождают все $\MV$, т.е.\
\begin{equation*}                                                    \tag*{\qed}
  \dim(\MV_1\cap\MV_2)+\dim\MV=\dim\MV_1+\dim\MV_2.
\end{equation*}
\end{defn}
\index{Трансверсальные векторные пространства (transversal vector spaces)}%
\index{Векторные пространства трансверсальные (transversal vector spaces)}%
В частном случае, если подпространства не пересекаются,
$\MV_1\cap\MV_2=\emptyset$, то $\MV=\MV_1\oplus\MV_2$.

Векторное пространство $\MV$, $\dim\MV=n$, над полем вещественных чисел образует
абелеву группу по отношению к сложению или модуль. Рассмотрим его
подпространство $\MH\subset\MV$, состоящее из всех линейных комбинаций
\begin{equation*}
  \sum_{i=1}^p a^i\Bx_i\in\MH,\qquad a^i\in\MR,\quad \Bx_i\in\MV,
\end{equation*}
где $p<n$. Не ограничивая общности, можно считать, что все векторы $\Bx_i$
линейно независимы. Тем самым мы предполагаем, что $\dim\MH=p$. Подпространство
$\MH$ образует подгруппу векторного пространства $\MV$ или подмодуль. Эта
подгруппа является нормальной, поскольку группа абелева. Отсюда следует, что
существует факторгруппа или факторпространство $\MV/\MH$. Элементами
факторпространства являются все линейные комбинации вида
\begin{equation*}
  \Bx+\sum_{i=1}^p a^i\Bx_i\in\Bx+\MH,\qquad \text{где}~\Bx\in\MV,
\end{equation*}
со следующим отношением эквивалентности. Два элемента факторпространства
$\Bx+\MH$ и $\By+\MH$ совпадают, если их разность лежит в $\MH$:
\begin{equation*}
  \By-\Bx\in\MH.
\end{equation*}
В факторпространстве $\MV/\MH$ естественным образом вводится умножение на
вещественные числа, что превращает его само в векторное пространство. Очевидно,
что $\dim\MV/\MH=\dim\MV-\dim\MH=n-p$.
\begin{exa}
Пусть подпространство $\MH\subset\MV$ натянуто на $p<n$ первых базисных
векторов векторного пространства $\MV$
\begin{equation*}
  \MH=\lbrace\Bx\in\MV:\quad \Bx=\sum_{i=1}^p a^i\Be_i\rbrace.
\end{equation*}
Тогда факторпространство $\MV/\MH$ изоморфно векторному пространству,
натянутому на оставшиеся базисные векторы:
\begin{equation*}
  \frac\MV\MH\simeq\lbrace\Bx\in\MV:\quad
  \Bx=\sum_{k=p+1}^na^k\Be_k\rbrace. \tag*{\qed}
\end{equation*}
\end{exa}
\begin{defn}
{\em Прямой линией}, проходящей через две различные точки векторного
пространства $\Bx,\By\in\MV$ называется множество векторов
\begin{equation}                                        \label{estrli}
  \lbrace \Bz\in\MV:\quad \Bz=t\Bx+(1-t)\By,\quad \forall t\in\MR\rbrace.
\end{equation}
Подмножество векторного пространства $\MU\subset\MV$ называется {\em плоским},
если вместе с любыми двумя точками, оно содержит прямую, проходящую через эти
точки. Если параметр пробегает единичный отрезок, $t\in[0,1]$, то мы имеем
отрезок, соединяющий точки $\Bx$ и $\By$. Подмножество точек $\MU$ векторного
пространства называется  {\em выпуклым}, если вместе с двумя произвольными
точками $\Bx,\By\in\MU$ оно содержит и отрезок, соединяющий эти точки.
\qed\end{defn}
\index{Прямая линия (straight line)}\index{Линия прямая (straight line)}%
\index{Плоское подмножество (planar subset)}%
\index{Подмножество плоское (planar subset)}%
\index{Выпуклое множество (convex set)}\index{Множество выпуклое (convex set)}%
\subsubsection{Прямая сумма}
Если задано два векторных пространств $\MV_1$ и $\MV_2$, то с их помощью можно
построить новые векторные пространства: прямую сумму и тензорное произведение.
\begin{defn}
{\em Прямой суммой} двух векторных пространств называется векторное пространство
$\MV=\MV_1\oplus\MV_2$ образованное всеми парами $\Bx_1\oplus\Bx_2$, где
$\Bx_1\in\MV_1$ и $\Bx_2\in\MV_2$, с векторным сложением и умножением на
скаляры, определяемыми формулами:
\begin{align*}
  \Bx_1\oplus\Bx_2+\By_1\oplus\By_2&=(\Bx_1+\By_1)\oplus(\Bx_2+\By_2),
\\
  a(\Bx_1\oplus\Bx_2)&=(a\Bx_1)\oplus(a\Bx_2),
\end{align*}
где $\By_1\oplus\By_2$ также принадлежит $\MV_1\oplus\MV_2$.
\qed\end{defn}
\index{Прямая сумма векторных пространств (direct sum of vector spaces)}%
\index{Сумма прямая векторных пространств (direct sum of vector spaces)}%
Нетрудно проверить, что все аксиомы векторного пространства для прямой суммы
выполнены.

Размерность прямой суммы векторных пространств равна сумме размерностей
пространств $\MV_1$ и $\MV_2$:
\begin{equation*}
  \dim(\MV_1\oplus\MV_2)=\dim\MV_1+\dim\MV_2.
\end{equation*}

Пусть в векторных пространствах заданы базисы $\Be_{\al_1}$,
$\al_1=1,\dotsc,\dim\MV_1$, и $\Be_{\al_2}$, $\al_2=1,\dotsc,\dim\MV_2$. Тогда
прямые суммы $\Be_{\al_1}\oplus\Be_{\al_2}$ образуют базис в $\MV_1\oplus\MV_2$.
Компонентами прямой суммы двух векторов в этом базисе является упорядоченный
набор $\dim\MV_1+\dim\MV_2$ чисел, состоящий из компонент первого и второго
вектора:
$$
  \Bx_1\oplus\Bx_2=\lbrace x_1^{\al_1},x_2^{\al_2}\rbrace.
$$
\begin{exa}
Вещественную прямую $\MR$ можно естественным образом рассматривать, как
одномерное векторное пространство. Тогда векторное пространство
$\MR^n$ есть прямая сумма $n$ одномерных векторных пространств.
\begin{equation*}
  \MR^n=\underbrace{\MR\oplus\dotsc\oplus\MR}_n. \tag*{\qed}
\end{equation*}
\end{exa}

Векторные пространства $\MV_1$ и $\MV_2$ можно рассматривать, как
подпространства $\MV_1\oplus\bf0$ и ${\bf0}\oplus\MV_2$ в прямой сумме
$\MV_1\oplus\MV_2$. Поэтому они изоморфны следующим факторпространствам
\begin{equation*}
  \MV_1\simeq\frac{\MV_1\oplus\MV_2}{\MV_2}\qquad \text{и}\qquad
  \MV_2\simeq\frac{\MV_1\oplus\MV_2}{\MV_1}
\end{equation*}

Обобщением разложения вектора по базису (\ref{evedeb}) служит градуировка
векторного пространства.
\begin{defn}
{\em Градуированным векторным (линейным) пространством} называется векторное
пространство $\MV$ вместе с его разложением в прямую сумму подпространств
\begin{equation}                                                  \label{egravs}
  \MV=\bigoplus\limits_{i=1}^\infty\MV_i,
\end{equation}
где некоторые подпространства могут быть пустыми. Суммарная размерность
подпространств в сумме должна быть равна размерности $\MV$. В общем случае
бесконечномерных векторных пространств эта сумма бесконечна, но каждый отдельный
вектор $\Bx\in\MV$ однозначно представляется в виде конечной суммы
$$
  \Bx=\sum_{i=1}^\infty\Bx_i,\qquad \Bx_i\in\MV_i.
$$
где все $\Bx_i$ кроме конечного числа равны нулю. Вектор $\Bx_i$ называется
{\em однородной компонентой} вектора $\Bx$ степени $i$. Если $x\in\MV_i$, то
вектор $\Bx$ называется {\em однородным элементом} степени $i$.
\qed\end{defn}
\index{Градуированное векторное пространство (graded vector space)}%
\index{Векторное пространство градуированное (graded vector space)}%
\index{Градуированное линейное пространство (graded linear space)}%
\index{Линейное пространство градуированное (graded linear space)}%
\index{Однородная компонента вектора (homogeneous vector component)}%
\index{Однородный элемент (homogeneous element)}%
\index{Элемент однородный (homogeneous element)}%
\begin{exa}
Евклидово векторное пространство $\MR^n$ представляет собой прямую сумму $n$
прямых, что можно рассматривать как градуировку. При этом количество
подпространств в разложении (\ref{egravs}) является конечным. На одном и том же
векторном пространстве можно задать несколько различных градуировок. Например,
евклидово пространство можно представить в виде суммы двух подпространств:
\begin{equation*}                                                    \tag*{\qed}
  \MR^n=\MR^m\oplus\MR^{n-m},\qquad 1\le m\le n-1.
\end{equation*}
\renewcommand{\qed}{}\end{exa}
\begin{defn}
Одним из способов изучения множеств $\MM$ с заданными алгебраическими
структурами состоит в выделении в них последовательности подмножеств
$\MM_0\subset\MM_1\subset\MM_2\subset\dotsc$ или
$\MM_0\supset\MM_1\supset\MM_2\supset\dotsc$ таким образом, что переход от
одного подмножества к другому устроен к каком то смысле просто. Общее название
таких последовательностей -- {\em фильтрации (возрастающая и убывающая}
соответственно). В теории линейных пространств строго возрастающая
последовательность подпространств $\MV_0\subset\MV_1\subset\dotsc\subset\MV_n$
пространства $\MV$ называется {\em флагом}. Число $n$ называется длиной флага.
Флаг $\MV_0\subset\MV_1\subset\dotsc\subset\MV_i\subset\dotsc$ называется
{\em максимальным}, если $\MV_0=\lbrace0\rbrace$, $\cup_i\MV_i=\MV$ и между
$\MV_i$ и $\MV_{i+1}$ нельзя вставить подпространство, т.е.\ если
$\MV_i\subset\MU\subset\MV_{i+1}$, то либо $\MU=\MV_i$, либо $\MU=\MV_{i+1}$.
\qed\end{defn}
\index{Фильтрация (filtration)}%
\index{Флаг (flag)}%
\begin{com}
Мотивировка названия: флаг $\lbrace\text{точка 0}\rbrace\subset
\lbrace\text{прямая}\rbrace\subset\lbrace\text{плоскость}\rbrace$ -- это
``гвоздь'', ``древко'' и ``полотнище''.
\qed\end{com}
По всякому базису $\Be_\al$, $\al=1,\dotsc,n$, векторного пространства $\MV$,
$\dim\MV=n$, можно построить флаг следующим образом. Положим
$\MV_0:=\lbrace0\rbrace$ и пусть $\MV_i$ -- линейная оболочка первых $i$
базисных векторов $\lbrace\Be_1,\dotsc,\Be_i\rbrace$ при $1\le i\le n$. Нетрудно
проверить, что построенный флаг максимален.
\begin{prop}
Размерность векторного пространства $\MV$ равна длине любого его максимального
флага.
\end{prop}
\begin{proof}
См., например, \cite{KosMan86R}.
\end{proof}
В конечномерном линейном пространстве $\MV$ любой флаг можно дополнить до
максимального, и поэтому его длина всегда меньше или равна $\dim\MV$.
\subsubsection{Тензорная алгебра}
Перейдем к описанию более сложного понятия тензорного произведения векторных
пространств. Обозначим через $\MV_1\times\MV_2$ прямое произведение векторных
пространств. Оно не снабжено структурой линейного пространства и просто
обозначает множество упорядоченных пар элементов $(\Bx_1,\Bx_2)$, где
$x_1\in\MV_1$ и $x_2\in\MV_2$.
\begin{defn}
Пусть $\MF(\MV_1,\MV_2)$ -- векторное пространство над полем вещественных чисел,
свободно порожденное элементами вида $(\Bx_1,\Bx_2)\in\MV_1\times\MV_2$. То
есть $\MF(\MV_1,\MV_2)$ состоит из всех конечных линейных комбинаций пар
$(\Bx_1,\Bx_2)$. А именно, мы рассматриваем все конечные линейные комбинации
\begin{equation*}
  \sum_{ij}a_{ij}(\Bx_{1i},\Bx_{2j}),\qquad \Bx_{1i}\in\MV_1,
  \quad \Bx_{2j}\in\MV_2,\quad a_{ij}\in\MR.
\end{equation*}
Построенное векторное пространство $\MF(\MV_1,\MV_2)$ является бесконечномерным,
т.к.\ базисом этого пространства являются все элементы вида $(\Bx_1,\Bx_2)$,
которых бесконечно много.

Пусть $\MH(\MV_1,\MV_2)$ -- линейное подпространство в $\MF(\MV_1,\MV_2)$,
свободно порожденное всеми элементами вида
\begin{equation}                                                  \label{eprtep}
\begin{split}
  (\Bx_1,\Bx_2+\By_2)-(\Bx_1,\Bx_2)-(\Bx_1,\By_2),
\\
  (\Bx_1+\By_1,\Bx_2)-(\Bx_1,\Bx_2)-(\By_1,\Bx_2),
\\
  (a\Bx_1,\Bx_2)-a(\Bx_1,\Bx_2),
\\
  (\Bx_1,a\Bx_2)-a(\Bx_1,\Bx_2),
\end{split}
\end{equation}
где $a\in\MR$. Это пространство также бесконечномерно. Тогда факторпространство
$$
  \MV_1\otimes \MV_2:=\frac{\MF(\MV_1,\MV_2)}{\MH(\MV_1,\MV_2)}
$$
называется {\em тензорным произведением} векторных пространств. При этом каждой
паре элементов $\Bx_1\in\MV_1$ и $\Bx_2\in\MV_2$ ставится в соответствие их
{\em тензорное произведение} $\Bx_1\otimes\Bx_2\in\MV_1\otimes\MV_2$.
\qed\end{defn}
\index{Тензорное произведение (tensor product)}%
\index{Произведение тензорное (tensor product)}%
Тензорное произведение некоммутативно, т.к.\ порядок сомножителей важен и
фиксирован.

Из определения следует, что тензорное произведение векторов {\em билинейно},
т.е.\ линейно по каждому из сомножителей при фиксированном другом:
\begin{equation}                                                  \label{eteprp}
\begin{split}
  \Bx_1\otimes(\Bx_2+\By_2)&=\Bx_1\otimes\Bx_2+\Bx_1\otimes\By_2,
\\
  (\Bx_1+\By_1)\otimes\Bx_2&=\Bx_1\otimes\Bx_2+\By_1\otimes\Bx_2,
\\
  (a\Bx_1)\otimes\Bx_2&=a(\Bx_1\otimes\Bx_2),
\\
  \Bx_1\otimes(a\Bx_2)&=a(\Bx_1\otimes\Bx_2),
\end{split}
\end{equation}
так как элементы $\MF(\MV_1,\MV_2)$ вида (\ref{eprtep}) отождествлены.
\index{Билинейный (bilinear)}%

Пусть $\MV_1$ и $\MV_2$ -- конечномерные векторные пространства с заданными
базисами $\Be_\al$, $\al=1,\dotsc,\dim\MV_1$ и $\Be_i$, $i=1,\dotsc,\dim\MV_2$.
Тогда их элементы имеют вид $\Bx_1=x_1^\al\Be_\al$ и $\Bx_2=x_2^i\Be_i$. Из
билинейности (\ref{eteprp}) вытекает, что тензорное произведение произвольных
векторов имеет вид
\begin{equation*}
  \Bx_1\otimes\Bx_2=x^{\al i}\Be_\al\otimes\Be_i,
\end{equation*}
где $x^{\al i}=x_1^\al x_2^i$. Это означает, что векторы
$\lbrace\Be_\al\otimes\Be_i\rbrace$ образуют базис тензорного пространства
$\MV_1\otimes\MV_2$, и компонентами тензорного произведения векторов
$\Bx_1\otimes\Bx_2$ относительно этого базиса являются простые произведения
компонент сомножителей $\lbrace x_1^\al x_2^i\rbrace$. Отсюда следует, что
размерность тензорного произведения конечна и равна произведению размерностей
векторных пространств:
\begin{equation*}
  \dim(\MV_1\otimes\MV_2)=\dim\MV_1\times\dim\MV_2.
\end{equation*}

Заметим, что тензорное произведение различных пар векторов может совпадать.
\begin{exa}
\begin{equation*}                                                    \tag*{\qed}
  \Bx_1\otimes\Bx_2=(a\Bx_1)\otimes\left(\frac{\Bx_2}a\right),\qquad
  \forall a\ne0.
\end{equation*}
\end{exa}

\begin{defn}
Элементы тензорного произведения векторных пространств $\MV_1\otimes\MV_2$
называются {\em тензорами}, а элементы вида $\Bx_1\otimes\Bx_2$ --
{\em разложимыми тензорами}.
\qed\end{defn}
\index{Тензор (tensor)}%
\index{Тензор разложимый (reducible tensor)}%
\index{Разложимый тензор (reducible tensor)}%
Поскольку элементы вида $(\Bx_1,\Bx_2)$ составляют базис $\MF(\MV_1,\MV_2)$, то
разложимые тензоры $\Bx_1\otimes\Bx_2$ порождают все тензорное произведение
$\MV_1\otimes\MV_2$. Однако они не являются базисом, т.к.\ между ними существует
много линейных зависимостей.
\begin{defn}
Каждой паре элементов $(\Bx_1,\Bx_2)\in\MV_1\times\MV_2\subset\MF(\MV_1,\MV_2)$
мы поставили в соответствие их тензорное произведение $\Bx_1\otimes\Bx_2$,
которое определяется естественной проекцией
$$
  \MF(\MV_1,\MV_2)\rightarrow\MV_1\otimes\MV_2.
$$
Это отображение называется {\em каноническим билинейным отображением} векторных
пространств и обозначается следующим образом
\begin{equation*}                                                    \tag*{\qed}
  \MV_1\times\MV_2\ni\qquad(\Bx_1,\Bx_2)\mapsto\vf(\Bx_1,\Bx_2)
  :=\Bx_1\otimes\Bx_2\qquad\in\MV_1\otimes\MV_2.
\end{equation*}
\end{defn}
\index{Каноническое билинейное отображение (canonical bilinear map)}%

Тензорное произведение обладает следующим свойством универсальности. Пусть $\MW$
-- некоторое векторное пространство и $\psi$ -- билинейное (\ref{eteprp})
отображение векторных пространств $\MV_1$ и $\MV_2$:
$$
  \psi:\quad \MV_1\times\MV_2\rightarrow\MW.
$$
Тогда для любых $\MW$ и $\psi$ существует единственное линейное отображение
\begin{equation*}
  \phi:\quad \MV_1\otimes\MV_2\rightarrow\MW
\end{equation*}
такое, что $\psi=\phi\circ\vf$, где $\vf$ -- каноническое билинейное
отображение. Это означает коммутативность диаграммы
\begin{equation*}
\begin{diagram}
  \MV_1\times\MV_2 & \rTo^\vf & \MV_1\otimes\MV_2 \\
   & \rdTo_\psi & \dTo_\phi \\
   &  & \MW
\end{diagram}
\end{equation*}

Последовательно умножая векторные пространства можно получить тензорное
произведение произвольного конечного числа сомножителей. При этом выполняется
закон ассоциативности
$(\MV_1\otimes\MV_2)\otimes\MV_3=\MV_1\otimes(\MV_2\otimes\MV_3)$ и можно просто
писать $\MV_1\otimes\MV_2\otimes\MV_3$.

Два вектора из одного векторного пространства $\Bx,\By\in\MV$ можно
рассматривать как элементы двух одинаковых векторных пространств,
$\MV_1=\MV_2=\MV$, и построить их тензорное произведение $\Bx\otimes\By$.
Оно, как было отмечено выше, некоммутативно,
$$
  \Bx\otimes\By\ne\By\otimes\Bx,
$$
поскольку в тензорном произведении важен порядок сомножителей.

\begin{defn}
Рассмотрим векторное пространство $\MV$ и его тензорное произведение на себя
$$
  \MV^r:=\underbrace{\MV\otimes\dotsc\otimes\MV}_r.
$$
Элемент $\Bx\in\MV^r$ называется {\em тензором} ранга $r$. Прямая сумма всех
тензорных пространств
\begin{equation}                                                  \label{etenal}
  \otimes\MV:=\bigoplus_{r=0}^\infty\MV^r,
\end{equation}
где $\MV^0$ -- поле вещественных чисел, с операциями сложения, умножения на
числа и тензорным умножением называется {\em тензорной алгеброй}.
\qed\end{defn}
\index{Тензор (tensor)}%
\index{Тензорная алгебра (tensor algebra)}%
\index{Алгебра тензорная (tensor algebra)}%
В тензорной алгебре определены три операции: сложение, умножение на числа
(элементы из $\MV^0$), а также тензорное произведение. При этом сложение двух
тензоров одного ранга понимается, как сложение в соответствующем векторном
пространстве $\MV^r$. В то же время для тензоров разного ранга используется
прямая сумма.

Тензорная алгебра некоммутативна и ассоциативна. Если $\Bx\in\MV^r$, то мы пишем
$\deg\Bx=r$. Тензорная алгебра (\ref{etenal}) бесконечномерна, т.к.\ ранг
тензоров неограничен. По построению она обладает естественной
$\MZ$-градуировкой.
Гомоморфизм $\MZ\rightarrow\MZ_2$, делящий все целые числа на четные и
нечетные, индуцирует $\MZ_2$-градуировку тензорной алгебры.

Напомним общее определение алгебры.
\begin{defn}
{\em Ассоциативной алгеброй} над полем вещественных чисел называется кольцо
$\MA$, которое является векторным пространством над полем вещественных чисел
$\MR$ и удовлетворяет условию
\begin{equation*}
  a(fg)=(af)g=f(ag),\qquad \forall f,g\in\MA,\quad \forall a\in\MR.
\end{equation*}
Если в определении кольца исключить условие ассоциативности, то получим общее
определение алгебры. Алгебра называется {\em коммутативной}, если $fg=gf$.
{\em Размерностью} алгебры называется ее размерность как векторного
пространства.
\qed\end{defn}
\index{Ассоциативная алгебра (associative algebra)}%
\index{Алгебра ассоциативная (associative algebra)}%
\index{Алгебра (algebra)}%
\index{Коммутативная алгебра (commutative algebra)}%
\index{Алгебра коммутативная (commutative algebra)}%
\index{Размерность алгебры (dimensionality of an algebra)}%
\index{Алгебра размерность (dimensionality of an algebra)}%
Таким образом в алгебре определено три операции: сложение, умножение и
умножение на числа (скаляры).

Отметим, что коммутативная алгебра может не быть ассоциативной.
\begin{exa}
Рассмотрим множество квадратных $n\times n$-матриц, элементами которых являются
вещественные $\MR$ или комплексные $\MC$ числа. В этих множествах определены
обычные операции сложения и умножения матриц. Вместе с умножением матриц на
скаляры, в роли которых выступают соответственно вещественные или комплексные
числа, мы получаем ассоциативные некоммутативные (при $n\ge2$) алгебры над
полем $\MR$ или $\MC$. Эти алгебры обозначаются соответственно
$\Gm\Ga\Gt(n,\MR)$ и $\Gm\Ga\Gt(n,\MC)$. Отметим, что умножение матриц не
является групповой операцией в алгебре, т.к.\ среди матриц есть вырожденные (с
нулевым определителем), которые не имеют обратных. Размерность алгебры
$\Gm\Ga\Gt(n,\MR)$ равна числу независимых
вещественных параметров, однозначно определяющих соответствующие матрицы,
\begin{equation*}
  \dim\Gm\Ga\Gt(n,\MR)=n^2.
\end{equation*}
Комплексная размерность алгебры $\Gm\Ga\Gt(n,\MC)$ также равна $n^2$. Алгебру
комплексных матриц $\Gm\Ga\Gt(n,\MC)$ можно также рассматривать над полем
вещественных чисел. В этом случае ее вещественная размерность равна $2n^2$.
Подмножества обратимых матриц в алгебрах $\Gm\Ga\Gt(n,\MR)$ и $\Gm\Ga\Gt(n,\MC)$
представляют собой общие линейные группы $\MG\ML(n,\MR)$, $\MG\ML(n,\MC)$.
В них сохраняется только одна бинарная операция -- умножение матриц.
\qed\end{exa}

Для любых двух элементов $f,g$ ассоциативной алгебры $\MA$ можно ввести их
{\em коммутатор}
\index{Коммутатор (commutator)}%
\begin{equation*}
  [f,g]:=fg-gf.
\end{equation*}
Ассоциативность необходима для того, чтобы выполнялись {\em тождества Якоби}
\index{Тождества Якоби (Jacobi identity)}%
\index{Якоби тождества (Jacobi identity)}%
\begin{equation*}
  \big[f,[g,h]\big]+\big[g,[h,f]\big]+\big[h,[f,g]\big]=0,\qquad f,g,h\in\MA,
\end{equation*}
которые проверяются прямой проверкой. Это означает, что на любой ассоциативной
алгебре можно ввести структуру алгебры Ли (см.\ раздел \ref{salgvf}). Обратное
утверждение неверно. Не всякую алгебру Ли можно рассматривать, как ассоциативную
алгебру.
\begin{exa}
На алгебре Ли векторных полей нельзя определить структуру ассоциативной алгебры.
\qed\end{exa}

Напомним определение идеала.
\begin{defn}
Подмножество $\MI$ кольца $\MR$ называется {\em идеалом}, если выполнены два
условия:

1) $\MI$ есть подгруппа $\MR$ по отношению к сложению;

2) \parbox[t]{.92\linewidth}{
$\MI$ содержит все произведения $ab$ ({\em левый идеал}), или все произведения
$ba$ ({\em правый идеал}), или все произведения $ab$ и $ba$ ({\em двусторонний
идеал}), где $a$ -- любой элемент кольца $\MR$ и $b$ -- любой элемент идеала
$\MI$.\qed}
\end{defn}
\index{Идеал (ideal)}%
\index{Левый идеал (left ideal)}\index{Идеал левый (left ideal)}%
\index{Идеал правый (right ideal)}\index{Правый идеал (right ideal)}%
\index{Двусторонний идеал (two-sided ideal)}%
\index{Идеал двусторонний (two-sided ideal)}%

Роль двусторонних идеалов в алгебрах аналогична роли нормальных подгрупп в
теории групп. В частности, можно проверить, что фактор пространство $\MA/\MI$
является алгеброй.

\begin{defn}
Алгебра $\MA$ называется {\em фильтрованной}, если в ней выделены
подпространства $\MA_k$, индексированные элементами линейно упорядоченной группы
$\MG$ таким образом, что $\MA_k\subseteq\MA_l$ при $k<l$ и
$\MA_k\MA_l\subseteq\MA_{k+l}$.

С каждой фильтрацией данной алгебры ассоциируется {\em градуированная алгебра}
\begin{equation*}
  \gr\MA=\oplus_k\bar\MA_k,
\end{equation*}
где
\begin{equation*}
  \bar\MA_k:=\frac{\MA_k}{\sum_{l<k}\MA_l},
\end{equation*}
а произведение элементов $\bar f\in\bar\MA_k$ и $\bar g\in\bar\MA_l$
определяется по формуле $\bar f\bar g:=\overline{fg}$, где $f$ и $g$ --
представители смежных классов $\bar f$ и $\bar g$, а
$\overline{fg}\in\bar\MA_{k+l}$ -- смежный класс, порожденный элементом
$fg\in\MA_{k+l}$.
\qed\end{defn}
\index{Фильтрация алгебры (filtered algebra)}%
\index{Градуированная алгебра (graded algebra)}%
\index{Алгебра градуированная (graded algebra)}%
\begin{com}
Чаще всего в качестве линейно упорядоченной группы выступает группа целых чисел
по сложению $\MZ$. В этом случае $\bar\MA_k=\MA_k/\MA_{k-1}$.
\qed\end{com}
Если в алгебре $\MA$ выполняется какое либо полилинейное тождество (например,
коммутативность, ассоциативность или тождество Якоби), то в градуированной
алгебре $\gr\MA$ также выполняется это тождество.
\begin{exa}
Рассмотрим алгебру Клиффорда $\MC\ML(\MV,Q)$ с образующими $e^a$,
$a=1,\dotsc,n$. Обозначим через $\MA_k$, $k\in\MZ$,
множество элементов алгебры Клиффорда, представимых в виде многочленов степени
$\le k$ от образующих. Множества $\MA_k$ при $k<0$ являются пустыми. Тогда
подмножества $\MA_k\subset\MC\ML(\MV,Q)$ задают фильтрацию алгебры Клиффорда.
Ассоциированной градуированной алгеброй будет внешняя алгебра $\Lm_n(\MV)$.
\qed\end{exa}
\subsubsection{Нормированные векторные пространства}
Векторное евклидово пространство $\MR^n$ можно снабдить естественной
топологической метрикой. Определим расстояние между двумя векторами
$\Bx,\By\in\MR^n$ с началом в начале системы координат по следующей формуле
\begin{equation}                                                  \label{edisve}
  l(\Bx,\By):=\sqrt{(x^1-y^1)^2+\dotsc+(x^n-y^n)^2},
\end{equation}
что совпадает с евклидовым расстоянием (\ref{ecldis}) между точками в $\MR^n$,
соответствующими концам векторов. При этом все свойства метрики, очевидно,
выполнены. Таким образом векторное пространство $\MR^n$ становится метрическим
пространством, и, следовательно, на нем можно определить естественную топологию.
Можно доказать, что любая топология в евклидовом пространстве, в которой
операции сложения и умножения на числа непрерывны, является естественной
\cite{DroZav06R}.

Метрика (\ref{edisve}) обладает двумя важными свойствами:
\begin{equation}                                                  \label{qkodas}
\begin{aligned}
&1)\quad l(\Bx,\By)=l(\Bx+\Bz,\By+\Bz) & &\text{-- \parbox[t]{.5\linewidth}
{инвариантность относительно сдвига на вектор $\Bz\in\MR^n$;}}\\
&2)\quad l(a\Bx,a\By)=|a|l(\Bx,\By), & &\text{-- \parbox[t]{.5\linewidth}
{умножение на скаляр увеличивает расстояние в $|a|$ раз.}}
\end{aligned}
\end{equation}
Заметим, что не каждая метрика обладает такими свойствами.

Если в векторном пространстве определена метрика, то определено отображение
\begin{equation}                                                  \label{qzvarf}
  \MV\ni\quad\Bx\mapsto \|\Bx\|:=l(\Bx,{\bf 0})\quad\in\MR.
\end{equation}
Если при этом метрика обладает свойствами (\ref{qkodas}), то построенное
отображение имеет следующие свойства:
\begin{equation}                                                  \label{qadsqw}
\begin{aligned}
&1)\quad \|{\bf 0}\|=0,\quad \|\Bx\|>0,\qquad
   \forall\Bx\ne{\bf 0} & &\text{-- положительная определенность}; \\
&2)\quad \|\Bx\|=\|-\Bx\| & &\text{-- четность;}\\
&3)\quad \|\Bx+\By\|\le\|\Bx\|
   +\|\By\| & &\text{-- неравенство треугольника}; \\
&4)\quad \| a\Bx\|=|a|\|\Bx\| &
     &\text{-- умножение на скаляры.}
\end{aligned}
\end{equation}
Свойство 2) является следствием 4) при $a=-1$. Оно выделено, потому что является
также следствием инвариантности метрики относительно сдвига и симметрии
относительно перестановки аргументов.

\begin{defn}
Векторное пространство $\MV$ называется {\em нормированным}, если оно снабжено
нормой, т.е.\ отображением
\begin{equation}                                                  \label{enorvs}
  \MV\ni\quad\Bx\mapsto \|\Bx\|\quad\in\MR,
\end{equation}
удовлетворяющим условиям (\ref{qadsqw}). При этом число $\|\Bx\|$ называется
{\em нормой} или {\em длиной} вектора $\Bx\in\MV$. Полное (см.\ раздел
\ref{seucme}) нормированное векторное пространство называется {\em банаховым}.
\qed\end{defn}
\index{Нормированное векторное пространство (normed vector space)}%
\index{Векторное пространство нормированное (normed vector space)}%
\index{Норма вектора (norm of vector)}\index{Вектора норма (norm of vector)}%
\index{Длина вектора (length of vector)}%
\index{Вектора длина (length of vector)}%
\index{Банахово пространство (Banach space)}%
\index{Пространство банахово (Banach space)}%

Если в векторном пространстве $\MV$ задана норма, то она порождает следующую
метрику $l(\Bx,\By):=\|\By-\Bx\|$ и, следовательно, некоторую топологию.
Нетрудно проверить, что все аксиомы метрики при этом выполнены. Конечно, не
каждая метрика порождается некоторой нормой.

Наличие нормы в векторном пространстве позволяет определить сходимость
последовательностей, используя соответствующую метрику. Если на некотором
линейном пространстве задано две нормы, то они называются  {\em эквивалентными},
если сходимость последовательности по любой из них влечет за собой сходимость по
другой норме.
\index{Эквивалентная норма (equivalent norm)}
\index{Норма эквивалентная (equivalent norm)}
\begin{theorem}
Все нормы в конечномерных вещественных или комплексных векторных пространствах
эквивалентны.
\end{theorem}
\begin{proof}
См., например, \cite{HorJoh86R}.
\end{proof}

\begin{exa}
В евклидовом пространстве наиболее часто используются следующие нормы:
\begin{equation}                                                  \label{enorms}
\begin{aligned}
  \|\Bx\|_1&:=|x^1|+|x^2|+\dotsc+|x^n|, &&\qquad L_1-\text{норма}, \\
  \|\Bx\|_2&:=\sqrt{|x^1|^2+|x^2|^2+\dotsc+|x^n|^2}, &&\qquad L_2-\text{норма},\\
  \|\Bx\|_p&:=\left(\sum_{\al=1}^n |x^\al|^p\right)^{1/p},\quad p\ge1 &&
  \qquad L_p-\text{норма}, \\
  \|\Bx\|_\infty&:=\max\lbrace|x^1|,|x^2|,\dotsc,|x^n|\rbrace, &&
  \qquad L_\infty-\text{норма}.
\end{aligned}
\end{equation}
\index{Норма $L_1$ ($L_1$ norm)}\index{$L_1$-норма ($L_1$ norm)}%
\index{Норма $L_2$ ($L_2$ norm)}\index{$L_2$-норма ($L_2$ norm)}%
\index{Норма $L_p$ ($L_p$ norm)}\index{$L_p$-норма ($L_p$ norm)}%
\index{Норма $L_\infty$ ($L_\infty$ norm)}%
\index{$L_\infty$-норма ($L_\infty$ norm)}%
Все выражения, как нетрудно проверить, удовлетворяют свойствам 1) -- 4) норм.
$L_1$-норма не порождается никаким скалярным произведением. Обычная длина
вектора в евклидовом пространстве есть ни что иное, как $L_2$-норма, которая
порождается евклидовым скалярным произведением. $L_p$-норма называется также
нормой {\em Гёльдера} с показателем $p\ge1$. Она порождается скалярным
произведением
\index{Норма Гёльдера (H\"older norm)}%
\index{Гёльдера норма (H\"older norm)}%
\begin{equation*}
  l(\Bx,\By):=\left(\sum_{\al=1}^n|x^\al-y^\al|^p\right)^{1/p}.
\end{equation*}
При $0<p<1$ выражение $\|\Bx\|_p$ удовлетворяет всем свойствам нормы кроме
неравенства треугольника. $L_\infty$-норма является пределом $L_p$-нормы.
А именно,
\begin{equation*}
  \|\Bx\|_\infty=\underset{p\to\infty}{\lim}\|\Bx\|_p,\qquad\forall\Bx\in\MR^n.
\end{equation*}
\index{Норма Гёльдера}\index{Гёльдера норма}%
Все приведенные выше нормы прямо обобщаются на комплексное пространство $\MC^n$
Пространства $\MR^n$ и $\MC^n$ с любой из приведенных выше норм являются
банаховыми.
\qed\end{exa}
Рассмотренный пример показывает, что в одном и том же векторном пространстве
можно задавать различные нормы и скалярные произведения. Выбор той или иной
нормы зависит от конкретной задачи.
\begin{exa}
Понятие нормы можно ввести и на бесконечномерном векторном пространстве.
На множестве $\CC[a,b]$ всех непрерывных функций $f(x)$ на отрезке $[a,b]$,
которое является бесконечномерным векторным пространством, обычно рассматривают
следующие нормы:
\begin{equation}                                        \label{enormf}
\begin{aligned}
  \|f\|_1&=\int_a^b dx|f(x)|, &&\qquad L_1-\text{норма}, \\
  \|f\|_2&=\left[\int_a^b dx|f(x)|^2\right]^{1/2}, &&\qquad L_2-\text{норма}, \\
  \|f\|_p&=\left[\int_a^b dx|f(x)|^p\right]^{1/p},\quad p\ge1 &&
  \qquad L_p-\text{норма}, \\
  \|f\|_\infty&=\max\lbrace|f(x)|:~x\in[a,b]\rbrace, &&
  \qquad L_\infty-\text{норма}.
\end{aligned}
\end{equation}
При этом, как и в предыдущем примере, выполнено равенство
\begin{equation*}
  \|f\|_\infty=\underset{p\to\infty}{\lim}\|f\|_p,\qquad\forall f\in\CC[a,b].
\tag*{\qed}
\end{equation*}
\renewcommand{\qed}{}\end{exa}
\begin{defn}
Метрику (\ref{edisve}) (а также любую другую метрику) можно использовать для
определения базы топологии векторного пространства. В результате будет построено
{\em топологическое векторное пространство}. Топологическое векторное
пространство $\MV$ называется {\em метризуемым}, если его топология может быть
индуцирована трансляционно инвариантной метрикой
$l(\Bx,\By)$$=$$l(\Bx+\Bz,\By+\Bz)$, $\Bx,\By,\Bz\in\MV$. В общем случае
топологическое векторное пространство, которое является метризуемым и полным,
называется {\em пространством Фреше}.
\qed\end{defn}
\index{Топологическое векторное пространство (topological vector space)}%
\index{Векторное пространство топологическое (topological vector space)}%
\index{Пространство векторное топологическое (topological vector space)}%
\index{Метризуемое топологическое векторное пространство %
(metrizable topological vector space)}%
\index{Топологическое векторное пространство метризуемое %
(metrizable topological vector space)}%
\index{Пространство Фреше (Frech\'et space)}%
\index{Фреше пространство (Frech\'et space)}%
\begin{exa}
Евклидово пространство $\MR^n$, рассматриваемое как векторное пространство с
метрикой (\ref{edisve}), является пространством Фреше.
\qed\end{exa}

\begin{defn}
Пусть $f:~\MV\rightarrow\MW$ -- отображение, возможно, нелинейное, двух
нормированных векторных пространств, и пусть $\Bx\in\MV$ -- некоторая точка.
Тогда, если отображение $f$ вблизи точки $\Bx$ можно представить в виде
\begin{equation*}
  f(\Bx+\Bh)=f(\Bx)+f^\prime_{\Bx}(\Bh)+\osmall(\Bh),
\end{equation*}
где $f'_{\Bx}$ -- линейное отображение $\MV\rightarrow\MW$ и
$\osmall(\Bh)\in\MW$ -- некоторый вектор, такой, что
\begin{equation}                                         \label{eostfr}
  \underset{\|\Bh\|\to0}\lim\frac{\|\osmall(\Bh)\|}{\|\Bh\|}=0,
\end{equation}
то отображение $f'_{\Bx}:\quad \MV\rightarrow\MW$ называется {\em производной по
Фреше} отображения $f$ в точке $\Bx$.
\qed\end{defn}
\index{Линейное отображение (linear map)}%
\index{Отображение линейное (linear map)}%
\index{Производная по Фреше (Frech\'et derivative)}%
\index{Фреше производная (Frech\'et derivative)}%
В частности, если $\MW=\MR$ -- вещественная прямая, то мы имеем производную по
Фреше от функционала на нормированном векторном пространстве.

Нетрудно показать, что если производная по Фреше существует, то она единственна.
\begin{exa}
Пусть в векторных пространствах $\MV$ и $\MW$ заданы базисы $\Be_\al$,
$\al=1,\dotsc,n$, и $\Be_i$, $i=1,\dotsc,m$, соответственно. Тогда отображение
$f$ задается $m$ функциями $f^i(\Bx)$ от $n$ переменных, которые в общем случае
могут быть нелинейными. Если функции $f^i$ дифференцируемы по всем аргументам,
то $n\times m$-матрица Якоби, составленная из частных производных
$\lbrace\pl_\al f^i|_{\Bx}\rbrace$ является производной по Фреше отображения
$f$ в точке $\Bx\in\MV$.
\qed\end{exa}
Для производной по Фреше справедлива теорема о дифференцировании сложной
функции.

Если отображение $f$ двух банаховых пространств непрерывно дифференцируемо по
Фреше в некоторой окрестности точки $\Bx$ и в этой точке производная Фреше
$f'_{\Bx}$ является гомеоморфизмом, то в окрестности данной точки
существует обратное отображение, которое также является гомеоморфизмом.

Наряду с производной Фреше отображений двух линейных пространств используется
также производная Гато, для которой необходимо только наличие топологии.
\begin{defn}
Пусть $f:~\MV\rightarrow\MW$ -- отображение двух топологических векторных
пространств, и пусть $\Bx\in\MV$ некоторая точка. Если отображение $f$ двух
линейных пространств вблизи точки $\Bx$ можно представить в виде
\begin{equation*}
  f(\Bx+\Bh)=f(\Bx)+f^\prime_{G,\Bx}(\Bh)+\osmall(\Bh),
\end{equation*}
где $f^\prime_{G,\Bx}(\Bh)$ -- линейное отображение $\MV\rightarrow\MW$ и
\begin{equation}                                                  \label{eostga}
  \underset{t\to0}\lim\frac{\osmall(t\Bh)}t={\bf 0}\in\MW,\qquad t\in\MR,
\end{equation}
где сходимость определяется топологией пространства $\MW$, то линейное
отображение $f'_{G,\Bx}:\quad \MV\rightarrow\MW$ называется {\em производной
по Гато} отображения $f$ в точке $\Bx$.
\qed\end{defn}
\index{Производная по Гато (G\^ateaux derivative)}%
\index{Гато производная (G\^ateaux derivative)}%
В частности, если $\MW=\MR$ -- вещественная прямая, то мы имеем производную по
Гато от функционала на топологическом векторном пространстве.

Отличие производной Гато от производной Фреше сводится только к определению
сходимости остаточных членов (\ref{eostfr}) и (\ref{eostga}).

Поскольку каждое нормированное векторное пространство является топологическим
с естественной топологией, то производная по Фреше является одновременно и
производной по Гато. Обратное утверждение в общем случае неверно, т.к.\ наличие
топологии в векторном пространстве совсем не означает наличие нормы. По этой
причине производную Фреше называют также {\em сильной производной}, а
производную Гато -- {\em слабой}.
\index{Слабая производная (weak derivative)}%
\index{Производная слабая (weak derivative)}%
\index{Сильная производная (strong derivative)}%
\index{Производная сильная (strong derivative)}%

Для производной Гато теорема о дифференцируемости сложной функции, вообще
говоря, не верна.
\subsubsection{Скалярное произведение в векторном пространстве}
\begin{defn}
Скалярным произведением $(\Bx,\By)$ двух элементов векторного пространства
$\Bx,\By\in\MV$ называется билинейное отображение
\begin{equation*}
  \MV\times\MV\ni\quad\Bx,\By\mapsto(\Bx,\By)\quad\in\MR,
\end{equation*}
которое обладает следующими свойствами:
\begin{align*}
&1)\quad (\Bx,\By)=(\By,\Bx)&& \text{-- симметричность,}
\\
&2)\quad (\Bx,\Bx)>0,~\forall\Bx\ne{\bf 0};\qquad ({\bf 0},{\bf 0})=0&&
\text{-- положительная определенность.}
\end{align*}
Если скалярное произведение двух векторов равно нулю $(\Bx,\By)=0$, то они
называются {\em ортогональными}.
\qed\end{defn}
\index{Скалярное произведение (scalar product)}%
\index{Произведение скалярное (scalar product)}%
\index{Ортогональные векторы (orthogonal vectors)}%
\index{Векторы ортогональные (orthogonal vectors)}%
В компонентах скалярное произведение задается положительно определенной
квадратичной формой
\begin{equation}                                                  \label{escpve}
  (\Bx,\By)=x^\al y^\bt g_{\al\bt},
\end{equation}
где $g_{\al\bt}:=(\Be_\al,\Be_\bt)$ -- произвольная симметричная положительно
определенная матрица.
\begin{exa}
В евклидовом пространстве $\MR^n$ естественное скалярное произведение векторов
задается в декартовых координатах с помощью евклидовой метрики (\ref{eclmet})
Это скалярное произведение симметрично и положительно определено. При этом норму
(длину) вектора можно записать в виде $\|\Bx\|=\sqrt{(\Bx,\Bx)}$.
\qed\end{exa}

Отметим, что скалярное произведение различных векторов $(\Bx,\By)$ может быть
отрицательно. Нулевой вектор ортогонален самому себе и всем другим векторам.
\begin{prop}
Пусть $\MV$ -- произвольное векторное пространство с положительно определенным
скалярным произведением (\ref{escpve}). Тогда на $\MV$ можно задать норму или
{\em длину вектора}
\index{Длина вектора (length of a vector)}%
\begin{equation}                                                  \label{enosca}
  \|\Bx\|:=\sqrt{(\Bx,\Bx)}.
\end{equation}
\end{prop}
\begin{proof}
Из линейности и положительной определенности скалярного произведения немедленно
следуют свойства 1), 2) и 4) в определении нормы. Для доказательства неравенства
треугольника воспользуемся неравенством Коши--Буняковского (\ref{enschw}),
которое будет доказано чуть позже:
\begin{align*}
  \|\Bx+\By\|^2=(\Bx+\By,\Bx+\By)&\le\|\Bx\|^2+\|\By\|^2+2|(\Bx,\By)|
\\
  &\le\|\Bx\|^2+\|\By\|^2+2\|\Bx\|\|\By\|
\\                                                                   \tag*{\qed}
  &=\big(\|\Bx\|+\|\By\|\big)^2.
\end{align*}
\renewcommand{\qed}{}\end{proof}
Обсудим некоторые свойства векторных пространств с положительно определенным
скалярным произведением. Следующие два утверждения проверяются прямой проверкой.
\begin{prop}[\bf Правило параллелограмма]
Пусть $\MV$ -- векторное пространство с положительно определенным скалярным
произведением. Тогда справедлива формула
\begin{equation*}
  \|\Bx+\By\|^2+\|\Bx-\By\|^2=2\|\Bx\|^2+2\|\By\|^2,
\end{equation*}
где $\Bx,\By\in\MV$ -- произвольные векторы.
\end{prop}
\index{Правило параллелограмма (the parallelogram law)}%
\index{Параллелограмма правило (the parallelogram law)}%
\begin{prop}[\bf Теорема Пифагора]
Пусть $\MV$ -- векторное пространство с положительно определенным скалярным
произведением. Тогда справедлива формула
\begin{equation*}
  \|\Bx+\By\|^2=\|\Bx\|^2+\|\By\|^2,
\end{equation*}
где $\Bx,\By\in\MV$ -- произвольные ортогональные векторы, $(\Bx,\By)=0$.
\end{prop}
\index{Теорема Пифагора (the Pythagorean theorem}%
\index{Пифагора теорема (the Pythagorean theorem}%
Допустим, что в векторном пространстве $\MV$, $\dim\MV=n$, задан произвольный
набор взаимно ортогональных векторов $\Bx_i$, $i=1,\dotsc,\Sn\le n$:
\begin{equation*}
  (\Bx_i,\Bx_j)=0,\qquad i\ne j.
\end{equation*}
Тогда из теоремы Пифагора по индукции следует равенство
\begin{equation*}
  \left\|\sum_{i=1}^\Sn\Bx_i\right\|^2=\sum_{i=1}^\Sn\|\Bx_i\|^2.
\end{equation*}

В произвольном векторном пространстве $\MV$, $\dim\MV=n$, со скалярным
произведением можно выбрать {\em ортонормальный базис} $\Be_\al$,
$\al=1,\dotsc,n$, состоящий из взаимно ортогональных векторов единичной длины:
\begin{equation*}
  (\Be_\al,\Be_\bt)=\dl_{\al\bt},
\end{equation*}
где $\dl_{\al\bt}$ -- евклидова метрика (\ref{eclmet}).
\index{Ортонормальный базис (orthonormal basis}%
\index{Базис ортонормальный (orthonormal basis}%
Произвольный вектор $\Bx\in\MV$ можно разложить по ортонормальному базису
\begin{equation*}
  \Bx=x^\al\Be_\al,
\end{equation*}
где $x^\al:=\dl^{\al\bt}(\Bx,\Be_\bt)$ -- компоненты данного вектора в
ортонормальном базисе.
\begin{prop}[\bf Неравенство Бесселя]
Пусть $\Bx_i\in\MV$, $i=1,\dotsc,\Sn\le n$, -- произвольный набор ортонормальных
векторов. Тогда справедливо неравенство
\begin{equation*}
  \sum_{i=1}^\Sn\left|(\Bx,\Bx_i)\right|^2\le\|\Bx\|^2,
\end{equation*}
где $\Bx\in\MV$ -- произвольный вектор.
\end{prop}
\begin{proof}
Из неравенства
\begin{equation*}
  0\le\left\|\Bx-\sum_{i=1}^\Sn(\Bx,\Bx_\al)\Bx_\al\right\|^2
  =\|\Bx\|^2-2\sum_{i=1}^\Sn(\Bx,\Bx_i)^2
  +\sum_{i=1}^\Sn\sum_{j=1}^\Sn(\Bx,\Bx_i)(\Bx,\Bx_j)(\Bx_i,\Bx_j).
\end{equation*}
вытекает неравенство Бесселя, т.к.\ векторы $\Bx_i$ ортонормальны,
$(\Bx_i,\Bx_j)=\dl_{ij}$.
\end{proof}
\index{Неравенство Бесселя (Bessel's inequality}%
\index{Бесселя неравенство (Bessel's inequality}%
Знак равенства в неравенстве Бесселя достигается тогда и только тогда, когда
набор векторов $\Bx_i$ образует ортонормальный базис в $\MV$, т.е.\ $\Sn=n$.
\begin{prop}
Пусть $\Be_\al\in\MV$, $\al=1,\dotsc,n$, -- ортонормальный базис векторного
пространства. Тогда справедливо равенство
\begin{equation*}
  \sum_{\al=1}^n\left|(\Bx,\Be_\al)\right|^2=\|\Bx\|^2,
\end{equation*}
где $\Bx\in\MV$ -- произвольный вектор. То есть квадрат длины произвольного
вектора равен сумме квадратов компонент относительно ортонормального базиса.
\end{prop}
\begin{proof}
В доказательстве неравенства Бесселя надо заменить знак неравенства на
равенство.
\end{proof}
\begin{prop}[\bf Неравенство Коши--Буняковского]
Пусть $\Bx,\By\in\MV$ -- произвольные векторы. Тогда справедливо неравенство
\begin{equation}                                                  \label{enschw}
  |(\Bx,\By)|\le\|\Bx\|\|\By\|.
\end{equation}
\end{prop}
\index{Неравенство Коши--Буняковского (Cauchy--Bunyakovskii inequality)}%
\index{Коши--Буняковского неравенство  (Cauchy--Bunyakovskii inequality)}%
Неравенство Коши--Буняковского называют также {\em неравенством Шварца}.
\index{Неравенство Шварца (Schwartz inequality)}%
\index{Шварца неравенство (Schwartz inequality)}%
\begin{proof}
Если $\By={\bf 0}$, то неравенство Коши--Буняковского (\ref{enschw}) выполнено.
Допустим, что $\By\ne{\bf 0}$. В этом случае можно построить единичный вектор
$\By_0:=\By/\|\By\|$, $\|\By_0\|=1$. Тогда из неравенства Бесселя следует
неравенство
\begin{equation*}
  |(\Bx,\By_0)|\le\|\Bx\|.
\end{equation*}
Поскольку
\begin{equation*}
  |(\Bx,\By_0)|=\frac{(\Bx,\By)}{\|\By\|},
\end{equation*}
то отсюда вытекает неравенство Коши--Буняковского.
\end{proof}
\begin{com}
В моделях математической физики, например, при рассмотрении лоренцевых
многообразий, под скалярным произведением понимают произвольное симметричное
билинейное отображение, отбрасывая требование положительной определенности.
В этом случае два вектора также называются ортогональными, если их скалярное
произведение равно нулю. При этом отличный от нуля вектор $\Bx\ne{\bf 0}$ может
оказаться ортогональным самому себе, если $(\Bx,\Bx)=0$. Такой вектор называется
{\em нулевым}. Если метрика имеет лоренцеву сигнатуру
$\sign g_{\al\bt}=(+-\dotsc-)$, то нулевой вектор называется также
{\em светоподобным}.
\end{com}
\index{Нулевой вектор (null vector)}\index{Вектор нулевой (null vector)}%
\index{Светоподобный вектор (lightlight vector)}%
\index{Вектор светоподобный (lightlight vector)}%

Из симметрии матрицы $g_{\al\bt}:=(\Be_\al,\Be_\bt)$ следует, что все ее
собственные числа вещественны. Для того, чтобы матрица была положительно
определена необходимо и достаточно, чтобы все ее собственные числа были
вещественны и положительны.

Если в векторном пространстве $\MV$ задано скалярное произведение, то в нем
всегда можно выбрать базис $\Be_\al$, состоящий из взаимно ортогональных
векторов. Эти векторы будут собственными векторами матрицы $g_{\al\bt}$:
\begin{equation*}
  \sum_{\bt=1}^ng_{\al\bt}\Be_\bt=\lm_\al\Be_\al,
\end{equation*}
где проводится суммирование по индексу $\bt$, но отсутствует суммирование по
$\al$, и $\lm_\al>0$ -- собственные значения матрицы $g_{\al\bt}$. В общем
случае часть собственных значений может совпадать. Для метрики евклидова
пространства $\MR^n$ в декартовых координатах все собственные значения совпадают
и равны единице.

Рассмотрим два векторных пространства $\MV_1$ и $\MV_2$ с элементами
$\Bx=x^\al\Be_\al\in\MV_1$, $\al=1,\dotsc,m$ и $\By=y^i\Be_i\in\MV_2$,
$i=1,\dotsc,n$. Выберем в каждом пространстве базис, состоящий из ортогональных
векторов. Пусть в них заданы скалярные произведения матрицами $g_{\al\bt}$
и $g_{ij}$ соответственно. Тогда в тензорном произведении $\MV_1\otimes\MV_2$
можно определить скалярное произведение. Действительно, пусть задано два вектора
\begin{equation*}
  X=X^{\al i}\Be_\al\otimes\Be_i,\qquad Y=Y^{\al a}\Be_\al\otimes\Be_i.
\end{equation*}
Тогда билинейное отображение
\begin{equation*}
  (X,Y):=X^{\al i}Y^{\bt j}g_{\al\bt}g_{ij}
\end{equation*}
симметрично и положительно определено. Последнее утверждение следует из того,
что векторы базиса $\Be_\al\otimes\Be_i$ являются собственными векторами
матрицы $g_{(\al i)(\bt j)}:=g_{\al\bt}g_{ij}$, определяющей скалярное
произведение,
\begin{equation*}
  \sum_{\bt,j}g_{\al\bt}g_{ij}\Be_\bt\otimes\Be_j
  =\lm_\al\lm_i\Be_\al\otimes\Be_i,
\end{equation*}
с положительными собственными числами $\lm_\al\lm_i>0$.

Как уже отмечалось, векторные пространства можно рассматривать над полем
комплексных чисел. В этом случае говорят о комплексном векторном пространстве.
\begin{defn}
{\em Скалярным произведением} в комплексном векторном пространстве $\MH$
называется отображение
\begin{equation*}
  \MH\times\MH\ni\quad\Bx,\By\mapsto(\Bx,\By)\quad\in\MC,
\end{equation*}
которое линейно по второму аргументу,
\begin{equation*}
  (\Bx,a\By+b\Bz)=a(\Bx,\By)+b(\Bx,\Bz),\qquad a,b\in\MC,
\end{equation*}
и удовлетворяет равенству
\begin{equation*}
  (\Bx,\By)^\dagger=(\By,\Bx),
\end{equation*}
где символ $\dagger$ обозначает комплексное сопряжение. Если скалярное
произведение положительно определено, т.е.\ $(\Bx,\Bx)\ge0$ для всех
$\Bx\in\MH$, то оно задает норму в $\MH$:
$$
  \|\Bx\|:=\sqrt{(\Bx,\Bx)}.
$$
Эта норма называется строго положительной, если из условия $\|\Bx\|=0$ следует
$\Bx=0$. Полное комплексное векторное пространство со строго положительной
нормой называется {\em гильбертовым}.
\qed\end{defn}
\index{Скалярное произведение (scalar product)}%
\index{Произведение скалярное (scalar product)}%
\index{Гильбертово пространство (Hilbert space)}%
\index{Пространство гильбертово (Hilbert space)}%
Гильбертовы пространства лежат в основе квантовой механики, где каждое состояние
физической системы отождествляется с вектором соответствующего гильбертова
пространства. Гильбертовы пространства в квантовой механике обычно
бесконечномерны.
\begin{exa}
В нерелятивистской квантовой механике в качестве гильбертова пространства, как
правило, рассматривается пространство $\MH=\CL_2(\MR^n)$ достаточно гладких
квадратично интегрируемых функций в евклидовом пространстве
\begin{equation*}
  \psi:\quad \MR^n\rightarrow\MC.
\end{equation*}
Скалярное произведение в этом пространстве определяется интегралом
\begin{equation*}
  \MH\times\MH\ni\quad\psi_1,\psi_2\mapsto(\psi_1,\psi_2):=
  \int_{\MR^n}\!\!\!dx\,\psi_1^\dagger\psi_2\quad\in\MC.
\end{equation*}
Можно доказать, что это пространство является полным.
\qed\end{exa}
\subsubsection{Сопряженные пространства}
\begin{defn}
{\em Линейным функционалом} или {\em линейной формой} $\Bf$ на вещественном
векторном пространстве $\MV$ называется линейное отображение
$$
  \MV\ni\quad\Bx\mapsto\Bf(\Bx)\quad\in\MR,
$$
обладающее свойствами:
$$
  \Bf(\Bx+\By)=\Bf(\Bx)+\Bf(\By),\qquad \Bf(a\Bx)=a\Bf(\Bx),\qquad a\in\MR.
$$
Множество $\MV^*$ всех линейных функционалов на $\MV$ снабжено естественной
структурой векторного пространства над полем вещественных чисел относительно
сложения и умножения на числа:
\begin{align*}
  (\Bf_1+\Bf_2)(\Bx)&:=\Bf_1(\Bx)+\Bf_2(\Bx),
\\
  (a\Bf)(\Bx)&:=a\Bf(\Bx).
\end{align*}
Векторное пространство $\MV^*$ называется {\em сопряженным}, {\em дуальным} или
{\em двойственным} пространством к $\MV$.
\qed\end{defn}
\index{Линейный функционал (linear functional)}%
\index{Функционал линейный (linear functional)}%
\index{Линейная форма (linear form)}\index{Форма линейная (linear form)}%
\index{Сопряженное пространство (conjugate space)}%
\index{Пространство сопряженное (conjugate space)}%
\index{Дуальное пространство (dual space)}%
\index{Пространство дуальное (dual space)}%
\index{Двойственное пространство (dual space)}%
\index{Пространство двойственное (dual space)}%

Можно показать, что сопряженное пространство имеет ту же размерность, что и
векторное пространство $\MV$. Для конечномерных векторных пространств взятие
сопряжения дважды приводит к векторному пространству, которое изоморфно
исходному, $\MV^{**}\simeq\MV$.
\begin{defn}
Пусть $\MU\subset\MV$, тогда {\em ортогональным дополнением} или
{\em аннулятором} множества $\MU$ (не обязательно подпространства) называется
подмножество
\begin{equation}                                                  \label{eannul}
  \MU^\bot:=\lbrace \Bf\in\MV^*:\quad \Bf(\Bx)=0,\qquad \forall\Bx\in\MU\rbrace. \qed
\end{equation}
\end{defn}
\index{Ортогональное дополнение (orthogonal complement)}%
\index{Дополнение ортогональное (orthogonal complement)}%
\index{Аннулятор (annihilator)}%
\begin{prop}
Если $\MU$ -- линейное подпространство в $\MV$ размерности $m$, $1\le m\le n$,
то его ортогональное дополнение $\MU^\perp$ является линейным подпространством
в $\MV^*$ и имеет размерность  $n-m$.
\end{prop}
\begin{proof}
Выберем в подпространстве $\MU$ $m$ линейно независимых векторов $\Bx_i$,
$i=1,\dotsc,m$. Тогда система линейных уравнений $\Bf(\Bx_i)=0$ будет иметь
$n-m$ линейно независимых решений, которые можно выбрать в качестве базиса в
$\MU^\perp$.
\end{proof}

Аналогично определяется ортогональное дополнение некоторого подмножества
$\MU^*\subset\MV^*$:
\begin{equation}                                                  \label{eannvs}
  \MU^{*\bot}:=\lbrace \Bx\in\MV:\quad \Bf(\Bx)=0,\qquad \forall\Bf\in\MU^*\rbrace.
\end{equation}
Это означает, что операцию ортогонального дополнения можно применить дважды.
Тогда для конечномерных векторных пространств справедливо равенство
\begin{equation*}
  (\MU^\bot)^\bot=\MU.
\end{equation*}
Если $\MV_1,\MV_2\subset\MV$ -- два подпространства векторного пространства,
то верна формула
\begin{equation}                                                  \label{ebotin}
\begin{split}
  (\MV_1\cap\MV_2)^\bot&=\MV_1^\bot\oplus
  \left[\MV_2^\bot\setminus(\MV_1^\bot\cap\MV_2^\bot)\right]
\\
  &=\left[\MV_1^\bot\setminus(\MV_1^\bot\cap\MV_2^\bot)\right]\oplus
  \MV_2^\bot.
\end{split}
\end{equation}
Здесь из одного из слагаемых вычтено пересечение ортогональных дополнений,
чтобы не учитывать его дважды.
\begin{defn}
{\em Гиперплоскостью} $\MS$ в векторном пространстве $\MV$, ортогональной
некоторому фиксированному элементу из сопряженного пространства $\Bf\in\MV^*$,
называется множество точек
\begin{equation}                                                  \label{ehypep}
  \MS:=\lbrace \Bx\in\MV:\quad \Bf(\Bx)=a,\qquad \Bf\ne0,~\Bf\in\MV^*,
  \quad a\in\MR\rbrace.\qed
\end{equation}
\end{defn}
\index{Гиперплоскость (hyperplane)}%
Гиперплоскости являются подпространствами размерности $n-1$ и определены для
всех ненулевых линейных функционалов.
\begin{exa}
В евклидовом пространстве $\MR^n$ скалярное произведение векторов при одном
фиксированном векторе задает линейный функционал. Верно и обратное утверждение:
любой линейный функционал можно задать, как скалярное произведение с некоторым
вектором. Это означает, что дуальное пространство $\MR^{n*}$ изоморфно самому
евклидову пространству,
\begin{equation*}
  \MR^{n*}\simeq\MR^n.
\end{equation*}
Поэтому понятия ортогонального дополнения и гиперплоскости, ортогональной
заданному вектору, в евклидовом пространстве имеют наглядный геометрический
смысл.
\qed\end{exa}

Если $\Be_\al$ -- базис векторного пространства $\MV$, то множество функционалов
$\Be^\al$, определенное соотношением
\begin{equation}                                                  \label{edualb}
  \Be^\al(\Be_\bt)=\dl_\bt^\al,
\end{equation}
определяет единственный {\em сопряженный (дуальный) базис} сопряженного
пространства.
\index{Дуальный базис (dual basis)}\index{Базис дуальный (dual basis)}%
При этом любой функционал взаимно однозначно представим в виде
\begin{equation}                                                  \label{efunde}
  \Bf=\Be^\al f_\al,
\end{equation}
с некоторыми компонентами $f_\al\in\MR$, а значение функционала на векторе
равно сумме компонент
\begin{equation}                                                  \label{efuvec}
  \Bf(\Bx)=x^\al f_\al.
\end{equation}

Метрика в евклидовом пространстве $\MR^n$ задает двойственное пространство
$\MR^{n*}$, элементы которого определяются компонентами
$x_\al$$:=$$x^\bt g_{\bt\al}$. Наличие метрики позволяет установить изоморфизм
$\MR^n$ и его дуального пространства $\MR^{n*}$, рассматриваемых как векторные
пространства.

Используя понятие линейного функционала, можно дать новое эквивалентное
определение тензорного произведения векторных пространств.
\begin{defn}
Пусть $\MV^*_1$ и $\MV^*_2$ -- векторные пространства, дуальные к $\MV_1$ и
$\MV_2$. Тогда  {\em тензорным произведением} векторных пространств
$\MV_1\otimes\MV_2$ называется множество всех билинейных отображений
упорядоченной пары дуальных пространств в поле вещественных чисел
\begin{equation}                                                  \label{etenvt}
  \MV_1\otimes\MV_2:\quad (\MV^*_1,\MV^*_2)\rightarrow\MR. \qed
\end{equation}
\end{defn}
\index{Тензорное произведение (tensor product)}%
\index{Произведение тензорное (tensor product)}%

В компонентах отображение (\ref{etenvt}) записывается следующим образом. Если
$\Bf_1=f_{1\al}\Be^\al\in\MV_1$ и $\Bf_2=f_{2i}\Be^i\in\MV_2$, то элемент
тензорного произведения $x^{\al i}\Be_\al\otimes\Be_i\in\MV_1\otimes\MV_2$
задает отображение $x^{\al i}f_{1\al}f_{2i}\in\MR$.

Пусть задано векторное пространство $\MV$ и сопряженное к нему пространство
$\MV^*$. Тензором типа $(r,s)$ мы будем называть элемент тензорного
произведения
\begin{equation*}
  \underbrace{\MV\otimes\dotsc\otimes\MV}_r\otimes
  \underbrace{\MV^*\otimes\dotsc\otimes\MV^*}_s,
\end{equation*}
где пространства $\MV$ и $\MV^*$ встречаются соответственно $r$ и $s$ раз.
Для определенности векторное пространство $\MV$ выбрано в качестве первых
$r$ сомножителей, хотя возможен и любой другой порядок множителей.
Если в пространствах $\MV$ и $\MV^*$ заданы базисы $\Be_\al$ и $\Be^\al$, то
тензор типа $(r,s)$ будет иметь $r$ верхних и $s$ нижних индексов
\begin{equation*}
  X=X^{\al_1\dotsc\al_r}{}_{\bt_1\dotsc\bt_s}\Be_{\al_1}\otimes\dotsc\otimes
  \Be_{\al_r}\otimes \Be^{\bt_1}\otimes\dotsc\otimes \Be^{\bt_s},
\end{equation*}
которые называются {\em контравариантными} и {\em ковариантными},
соответственно.
\index{Контравариантный индекс}\index{Индекс контравариантный}%
\index{Ковариантный индекс}\index{Индекс ковариантный}%
Поскольку тензорное произведение некоммутативно, то порядок индексов
является существенным. Так $X^{\al\bt}\ne X^{\bt\al}$ и
$X^\al{}_\bt\ne X_\bt{}^\al$.
\begin{defn}
Пусть задано линейное отображение двух векторных пространств
\begin{equation*}
  \vf:\quad \MV~\rightarrow~\MW,
\end{equation*}
которые могут иметь разные размерности: $\dim\MV=n$, $\dim\MW=m$. В компонентах
это отображение задается $n\times m$-матрицей $A=(A_\al{}^i)$
\begin{equation*}
  \vf:\quad \MV\quad\ni\lbrace x^\al\rbrace\mapsto
  \lbrace y^i:=x^\al A_\al{}^i\rbrace\quad\in\MW.
\end{equation*}
Пусть $\Bf$ и $\Bg$ -- линейные формы на $\MV$ и $\MW$, соответственно. Тогда
отображение $\vf$ индуцирует {\em возврат отображения} $\vf^*$, действующий в
обратную сторону
\begin{equation*}
  \vf^*:\quad \MW^*~\rightarrow~\MV^*
\end{equation*}
по следующему правилу
\begin{equation*}
  (\Bf=\vf^*\Bg)(\Bx):=\Bg\big(\vf(\Bx)\big).
\end{equation*}
В компонентах возврат отображения имеет вид
\begin{equation*}                                                    \tag*{\qed}
  \vf^*:\quad \MW^*\ni\quad\lbrace g_i\rbrace\mapsto
  \lbrace f_\al:=A_\al{}^i g_i\rbrace\quad\in\MV^*.
\end{equation*}
\end{defn}
\index{Возврат отображения (pullback)}\index{Отображения возврат (pullback)}%
\begin{com}
Матрица $A_\al{}^i$ в общем случае является прямоугольной, и говорить об
обратной матрице не имеет смысла.
\qed\end{com}

Понятие тензоров с ковариантными и контравариантными индексами будет
использовано в дальнейшем при построении тензорных полей на многообразиях.
\subsection{$\MR^n$ как аффинное пространство                    \label{seucaf}}
Прямая в евклидовом пространстве обладает двумя важными и независимыми
свойствами. Во-первых, отрезок, соединяющий две точки евклидова пространства,
имеет наименьшую длину среди всех кривых, соединяющих эти точки (метрическое
свойство.) Во-вторых, касательный вектор к прямой остается касательным при
параллельном переносе вдоль прямой. В настоящем разделе мы придадим смысл
последнему утверждению. Это важно, т.к.\ параллельный перенос векторов в
дифференциальной геометрии является нетривиальным обобщением параллельного
переноса векторов в аффинном пространстве.

В предыдущем разделе было установлено, что точки евклидова пространства $\MR^n$
находятся во взаимно однозначном соответствии с множеством векторов $\MV$
евклидова пространства, имеющих начало в начале системы координат. Компоненты
этих векторов совпадают с координатами точек -- концов векторов -- и поэтому их
можно складывать с координатами произвольных точек из $\MR^n$, которые сами уже
не рассматриваются, как элементы векторного пространства. Введение этой операции
приводит к понятию аффинного пространства. Чтобы подчеркнуть разницу между
точками и векторами дадим сначала общее определение.
\begin{defn}
{\em Аффинным пространством} над полем вещественных чисел называется тройка
$(\MA,\MV,+)$, состоящая из множества $\MA$, элементы которого называются
точками, ассоциированного векторного пространства $\MV$ и отображения
\begin{equation}                                                  \label{etrvex}
  \MA\times\MV\ni\quad\Sp,\Bx\mapsto\Sp+\Bx\quad\in\MA,
\end{equation}
удовлетворяющего следующим аксиомам:

1) \parbox[t]{.92\linewidth}{$(\Sp+\Bx)+\By=\Sp+(\Bx+\By)\qquad
   \forall\Sp\in\MA,\qquad \forall\Bx,\By\in\MV,$}

2) \parbox[t]{.92\linewidth}{$\Sp+{\bf 0}=\Sp, \qquad \forall\Sp\in\MA$,}

3) \parbox[t]{.92\linewidth}{для двух произвольных точек $\Sp,\Sq\in\MA$
уравнение $\Sq=\Sp+\Bx$ на вектор $\Bx$ имеет единственное решение.\qed}
\end{defn}
\index{Аффинное пространство (affine space)}%
\index{Пространство аффинное (affine space)}%
В рассматриваемом случае множество точек представляет собой множество точек
евклидова пространства $\MA=\MR^n$. Векторное пространство $\MV$ также совпадает
с евклидовым пространством $\MV=\MR^n$, которое рассматривается как векторное
пространство. При этом отображение (\ref{etrvex}) называется {\em сдвигом}
аффинного пространства на вектор $\Bx$. Сдвиг на противоположный вектор $-\Bx$
обозначается знаком минус: $\Sp\mapsto\Sp-\Bx$.
\index{Сдвиг (translation)}%
\begin{com}
Отметим, что в определении аффинного пространства не фигурирует явно структура
умножения на скаляры. Вообще говоря, для точек аффинного пространства не
определено понятие суммы $\Sp+\Sq$ и умножения на числа $a\Sp$.
\qed\end{com}

Тот единственный вектор $\Bx\in\MV$, для которого $\Sq=\Sp+\Bx$, удобно
обозначать $\Sq-\Sp$. Эта операция вычитания
$$
  \MA\times\MA\ni\quad\Sq,\Sp\mapsto\Sq-\Sp\quad\in\MV
$$
обладает следующими свойствами:

1) $(\Sr-\Sq)+(\Sq-\Sp)=\Sr-\Sp$;

2) $\Sp-\Sp={\bf 0}$;

3) $(\Sp+\Bx)-(\Sq+\By)=(\Sp-\Sq)+(\Bx-\By)$.

Свойство 1), справедливое для трех произвольных точек аффинного пространства,
называется {\em соотношением Шаля} и его можно записать в эквивалентной форме
\index{Соотношение Шаля (Chasles relation)}%
\index{Шаля соотношение (Chasles relation)}%
$$
  (\Sr-\Sq)+(\Sq-\Sp)+(\Sp-\Sr)={\bf 0}.
$$

Интуитивно аффинное пространство $(\MA,\MV,+)$ можно представлять себе как
векторное пространство $\MV$ с забытым началом координат ${\bf 0}$, в котором
введена операция сдвига на вектор.

В аффинном пространстве можно ввести аффинные координаты, состоящие из
фиксированной точки $\Sp_0\in\MA$ (начала координат) и базиса $\Be_\al$
соответствующего векторного пространства. Это является следствием третьего
свойства в определении аффинного пространства. Произвольная точка $\Sp\in\MA$
имеет единственное представление
\begin{equation}                                                  \label{epoasp}
  \Sp=\Sp_0+x_\Sp^\al\Be_\al,
\end{equation}
где числа $x_\Sp^\al$ называются координатами точки $\Sp$ в данной системе
аффинных координат. Если предположить, что точка $\Sp_0$ совпадает с началом
координат в евклидовом векторном пространстве $\MR^n$, то произвольная точка
$\Sp$ определяется только координатами соответствующего вектора
$\Sp=\lbrace x_\Sp^\al\rbrace $. Таким образом, мы имеем взаимно однозначное
соответствие между точками аффинного пространства и векторами из $\MV$. При этом
сдвиг точки $\Sp$ (\ref{etrvex}) на вектор $\Bx=\lbrace x^\al\rbrace $
записывается в виде
$$
  x_\Sp^\al\mapsto x_\Sp^\al+x^\al.
$$
После отождествления точек аффинного и векторного пространства можно считать,
что в общем случае точка $\Sp$ имеет координаты
$$
  \Sp=\lbrace x_0^\al+x_\Sp^\al\rbrace ,
$$
где $x_0^\al$ -- координаты точки $\Sp_0$ в разложении (\ref{epoasp}). Поскольку
количество аффинных координат точек аффинного пространства равно $n$, то
{\em размерностью} аффинного пространства называется размерность
ассоциированного векторного пространства, $\dim\MA:=\dim\MV$.
\index{Размерность аффинного пространства (dimensionality of an affine space)}%

Пусть аффинное пространство имеет конечную размерность $n$. Если координаты всех
точек аффинного пространства преобразовать по правилу
\begin{equation}                                                  \label{elinas}
  x^\al\mapsto x^{\prime\,\al}=x^\bt A_\bt{}^\al+a^\al,
\end{equation}
где $A\in\MG\ML(n,\MR)$ -- произвольная невырожденная матрица и
$\Ba=a^\al\Be_\al\in\MV$ -- произвольный фиксированный вектор, то получим
взаимно однозначное отображение аффинного пространства на себя. Последовательное
выполнение двух преобразований (\ref{elinas}) с параметрами $(A,\Ba)$
и $(B,\Bb)$ имеет вид
$$
  x^{\prime\prime\,\al}=x^\g A_\g{}^\bt B_\bt{}^\al+a^\bt B_\bt{}^\al+b^\al.
$$
\begin{defn}
Совокупность преобразований вида (\ref{elinas}) с законом композиции
$$
  (B,\Bb)\circ(A,\Ba)=(AB,\Ba B+\Bb)
$$
образует группу Ли {\em аффинных преобразований} $\MA(n)$ аффинного пространства
$\MA$. Преобразования (\ref{elinas}) при $A_\bt{}^\al=\dl_\bt^\al$ называются
{\em трансляциями} на вектор $\Ba\in\MV$.
\qed\end{defn}
\index{Группа аффинных преобразований (affine group)}%
\index{Аффинная группа (affine group)}%
\index{Трансляция (translation)}%
Множество трансляций образует инвариантную (нормальную) подгруппу аффинной
группы, а сама аффинная группа представляет собой полупрямое произведение,
которое будем обозначать символом $\ltimes$, группы невырожденных матриц
$\MG\ML(n,\MR)$ на подгруппу трансляций, которая естественным образом
отождествляется с векторным пространством $\MV$. Таким образом группа аффинных
преобразований имеет вид
$$
  \MA(n)=\MG\ML(n,\MR)\ltimes\MV.
$$
Подгруппа $\MA(n)$, сохраняющая начало координат, совпадает с $\MG\ML(n,\MR)$.

Аффинная группа является неабелевой и имеет размерность $\dim\MA(n)=n^2+n$. Она
действует в аффинном пространстве транзитивно и эффективно (см.\ раздел
\ref{stragr}). Аффинные преобразования называются также {\em движениями
аффинного пространства}. Если $\det A>0$ или $\det A<0$, то движения аффинного
пространства (\ref{elinas}) называются соответственно {\em собственными} или
{\em несобственными}.
\index{Собственное движение (proper motion)}%
\index{Движение собственное (proper motion)}%
\begin{com}
В векторном пространстве действует только группа $\MG\ML(n,\MR)$, т.к.\ в нем
отсутствует понятие сдвига (трансляции).
\qed\end{com}
\begin{defn}
Подгруппа сдвигов действует на $\MR^n$ свободно. Рассмотрим подгруппу
$\MG\subset\MV=\MR^n$, где $\MV$ рассматривается, как группа трансляций,
состоящую из всех сдвигов
\begin{equation}                                                  \label{etrnac}
  \sum_{i=1}^m n^i\Ba_i,\qquad n^i\in\MZ,
\end{equation}
на линейно независимые векторы $\Ba_i$, $i=1,\dots m\le n$ с целыми
коэффициентами. Эта подгруппа действует собственно разрывно
(см.\ раздел \ref{stragr}). Поэтому, отождествляя точки евклидова пространства,
связанные трансляциями (\ref{etrnac}), получим важный класс многообразий,
называемых {\em цилиндрами}. Они являются фактор пространствами $\MR^n/\MG$.
При $m=n$ цилиндр компактен и называется {\em тором} $\MT^n$.
\qed\end{defn}
\index{Цилиндр (cylinder)}\index{Тор (torus)}%
\label{ptorde}
\begin{exa}
В частном случае, сдвиги евклидовой плоскости $\MR^2$ с декартовыми координатами
$x,y$ на постоянный вектор $\Ba=\lbrace a,0\rbrace $, $a\ne0$, вдоль оси
абсцисс:
$$
  x\mapsto x+a,\qquad y\mapsto y,
$$
приводит к простейшему двумерному цилиндру, хорошо известному из курса
элементарной геометрии.
\qed\end{exa}
В одномерном случае, $n=1$, тор совпадает с окружностью $\MT^1=\MS^1$. В общем
случае при $1\le m<n$ векторное пространство разлагается в прямую сумму
$\MV=\MU\oplus\MW$, где $\MW$ -- подпространство, натянутое на векторы $\Ba_i$.
Отсюда следует
$$
  \frac{\MR^n}\MG=\frac{\MR^{n-m}\times\MR^m}\MG=\MR^{n-m}\times\frac{\MR^m}\MG.
$$
То есть цилиндры представляют собой прямое произведение $(n-m)$-мерного
аффинного пространства и $m$-мерного тора.

Вернемся к аффинным преобразованиям (\ref{elinas}).
\begin{theorem}                                                   \label{tmovas}
Всякое аффинное преобразование $n$-мерного евклидова пространства $\MA=\MR^n$
может быть представлено в виде композиции трех отображений:\\
1) \parbox[t]{.92\linewidth}{$n$ растяжений с положительными коэффициентами
вдоль $n$ попарно ортогональных осей, проходящих через некоторую точку
$\Sp\in\MA$;}\\
2) \parbox[t]{.92\linewidth}{движения, оставляющего неподвижной точку
 $\Sp$;}\\
3) \parbox[t]{.92\linewidth}{сдвига.}
\end{theorem}
\begin{proof}
Первые два преобразования вытекают из полярного разложения матриц, теорема
\ref{tpomad}. Остаются еще сдвиги. Детали доказательства приведены, например, в
\cite{KosMan86R}.
\end{proof}
\begin{defn}
В аффинном пространстве {\em прямой линией}, проходящей через точку $\Sp\in\MA$
в направлении вектора $\Bx\in\MV$, называется множество точек вида
\begin{equation}                                                  \label{estlia}
  \Sp+a\Bx,\qquad a\in\MR.
\end{equation}
Две прямые линии называются {\em параллельными}, если они задаются уравнениями
\begin{equation*}
  \Sp+a\Bx\qquad\text{и}\qquad\Sq+a\Bx,\qquad a\in\MR,\quad \Sp\ne\Sq.
\end{equation*}
Подмножество $\MB\subset\MA$ называется {\em аффинным подпространством}, если
множество векторов $(\Sq-\Sp)$, $\forall\Sq,\Sp\in\MB$ образует подпространство
$\MU\subset\MV$ в ассоциированном векторном пространстве.
\qed\end{defn}
\index{Прямая линия (straight line)}\index{Линия прямая (straight line)}%
\index{Параллельная прямая (parallel straight line)}%
\index{Прямая параллельная (parallel straight line)}%
\index{Аффинное подпространство (affine subspace)}%
\index{Подпространство аффинное (affine subspace)}%

Легко проверить, что две прямые линии параллельны тогда и только тогда, когда
они не имеют общих точек.

Ясно, что каждое аффинное подпространство $\MB\subset\MA$ состоит из точек вида
$$
  \MB:=\lbrace\Sp+\Bx:\quad \Bx\in\MU\rbrace,
$$
где $\Sp$ -- произвольная точка из $\MB$.
\begin{exa}
Прямая линия является одномерным аффинным подпространством в $\MA$.
\qed\end{exa}
При аффинном преобразовании прямые линии переходят в прямые, причем
пересекающиеся линии переходят в пересекающиеся, а параллельные -- в
параллельные.
\begin{defn}
Пусть $\Sp_1,\dots\Sp_k$ -- произвольные точки аффинного пространства, и числа
$a_1,\dots a_k$ удовлетворяют условию $\sum_{i=1}^k a_i=1$. Определим сумму
\begin{equation}                                                  \label{ebarpo}
  \Sp=\sum_{i=1}^k a_i\Sp_i:=\Sp_0+\sum_{i=1}^k a_i(\Sp_i-\Sp_0),
\end{equation}
где $\Sp_0$ -- произвольная точка из $\MA$. Поскольку $(\Sp-\Sp_0)\in\MV$ и
выражение (\ref{ebarpo}) не зависит от выбора $\Sp_0$, то оно корректно
определяет точку аффинного пространства $\Sp\in\MA$. Эта точка называется
{\em барицентрической комбинацией} точек $\Sp_1\dots\Sp_k$ с коэффициентами
$a_1\dots a_k$.
\qed\end{defn}
\index{Барицентрическая комбинация точек (barycentric combination of points)}%
\begin{exa}
В евклидовом аффинном пространстве барицентрическая сумма
$$
  \sum_{i=1}^k\frac{m_i}M\Sp_i,\qquad M:=\sum_{i=1}^k m_i
$$
представляет собой положение центра масс системы масс $m_i$, расположенных в
точках $\Sp_i$.
\qed\end{exa}

\begin{figure}[h,b,t]
\hfill\includegraphics[width=.6\textwidth]{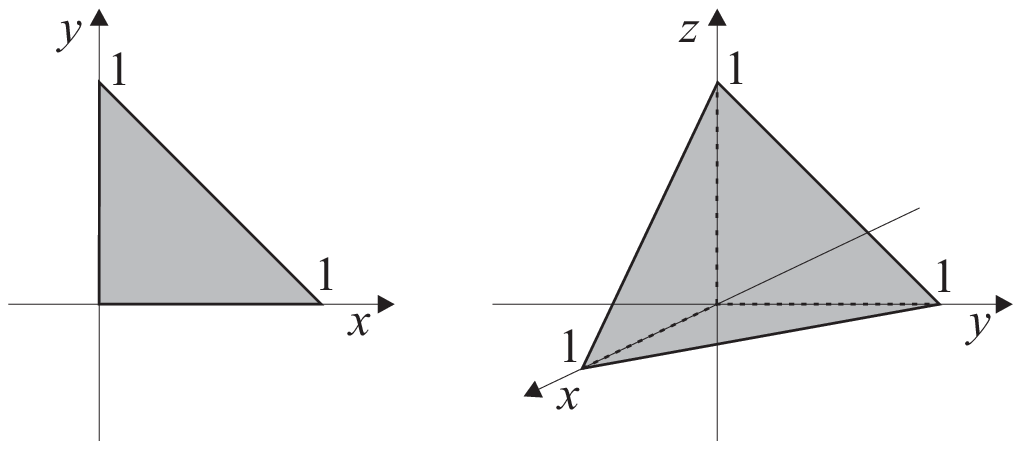}
\hfill {}
\centering \caption{Симплексы в двух и трех измерениях. \label{fntant}}
\end{figure}

Из определения (\ref{ebarpo}) следует, что система
$$
  \lbrace\Sp_0,~(\Sp_1-\Sp_0),\dots,(\Sp_n-\Sp_0)\rbrace ,
$$
состоящая из точки $\Sp_0$ и множества векторов $(\Sp_i-\Sp_0)$, образует
систему аффинных координат тогда и только тогда, когда любая точка $\Sp\in\MA$
представима в виде барицентрической суммы
$$
  \Sp=\sum_{i=1}^n x_i\Sp_i,\qquad \sum_{i=1}^n x_i=1.
$$
Если это условие выполнено, то множество точек $\lbrace\Sp_0,\dots,\Sp_n\rbrace$
называется {\em барицентрической} системой координат в аффинном пространстве, а
числа $x_i\in\MR$ -- барицентрическими координатами точки $\Sp\in\MA$.
\index{Барицентрическая система координат (barycentric coordinate system)}%
\index{Координаты барицентрические (barycentric coordinates)}%
\begin{defn}
Пусть точки $\Sp_i$, $i=1,\dots n$ имеют аффинные координаты
$$
  \Sp_i=\lbrace \underbrace{0\dots0}_{i-1}10\dots0\rbrace .
$$
Эти точки вместе с началом системы координат образуют барицентрическую систему
координат в аффинном пространстве. Рассмотрим пересечение множества точек с
барицентрическими координатами, не превосходящими единицы, $0\le x_i\le1$, с
положительным $2^n$-тантом. Это множество называется {\em стандартным
$(n-1)$-мерным симплексом}. В общем случае множество точек
\begin{equation}                                                  \label{esimpl}
  \left\lbrace \sum_{i=1}^n x_i \Sp_i:\quad \sum_{i=1}^nx_i=1,\quad
  0\le x_i\le1\right\rbrace
\end{equation}
называется {\em замкнутым симплексом} с вершинами в точках $\Sp_i$ в аффинном
пространстве $\MA$. Симплекс называется {\em вырожденным}, если векторы
$(\Sp_2-\Sp_1),\dots,(\Sp_n-\Sp_1)$ линейно зависимы.
\qed\end{defn}
\index{Симплекс (simplex)}%
\index{Замкнутый симплекс (closed simplex)}%
\index{Симплекс замкнутый (closed simplex)}%
\index{Симплекс вырожденный (degenerate simplex)}%
\index{Вырожденный симплекс (degenerate simplex)}%
\begin{exa}
Одномерный симплекс -- это отрезок прямой, двумерный -- треугольник, трехмерный
-- тетраэдр (см.~рис.\ref{fntant}).
\qed\end{exa}
Определим теперь параллельный перенос векторов в аффинном пространстве.
\begin{defn}
Рассмотрим произвольную кривую $\g=x(t)$, соединяющую две произвольно выбранные
точки $\Sp$ и $\Sq$ в аффинном пространстве. С каждой точкой аффинного
пространства и, следовательно, с каждой точкой кривой $\g$ ассоциировано
векторное пространство $\MV$. Будем говорить, что вектор $\Bx\in\MV$
{\em параллельно переносится} вдоль кривой $\g$, если его компоненты
относительно некоторого фиксированного базиса в $\MV$ не меняются при
переходе от точки к точке.
\qed\end{defn}
\index{Параллельный перенос (parallel transport)}%
\index{Перенос параллельный (parallel transport)}%
\begin{com}
В данном определении важно различать точки аффинного и векторного пространств.
\qed\end{com}
Очевидно, что параллельный перенос вектора из точки $\Sp$ в точку $\Sq$ не
зависит от кривой, соединяющей эти точки.

В настоящем и предыдущем разделах векторы обозначались жирным шрифтом. В
дальнейшем мы упростим обозначения. Если у символа индекс присутствует, то это
значит, что рассматривается соответствующая компонента вектора. Например,
$x^\al$ обозначает компоненту вектора с индексом $\al$. В то же время, если
индекс опущен или используются фигурные скобки, то подразумевается весь вектор
$x=\lbrace x^\al\rbrace=(x^1,\dotsc,x^n)$, и мы, как правило, не будем
использовать жирный шрифт.
\section{Отображения                                             \label{smappi}}
\begin{defn}
Рассмотрим два множества $\MM_1$ и $\MM_2$ и их {\em отображение}
\begin{equation}                                                  \label{esemap}
  f:\quad \MM_1\rightarrow\MM_2\qquad \text{или}\qquad
  \MM_1\stackrel{f}{\rightarrow}\MM_2,
\end{equation}
которое ставит каждому элементу $x\in\MM_1$ единственный элемент
$f(x)\in\MM_2$. Отображение $f$ называется также {\em функцией}, а переменная
$x\in\MM_1$ --{\em аргументом}. Вообще говоря, функция может отображать
различные элементы из $\MM_1$ в один и тот же элемент множества $\MM_2$.
Ситуация, когда один элемент из $\MM_1$ отображаются сразу в несколько элементов
множества $\MM_2$, не допускается. Пространство $\MM_2$ в (\ref{esemap})
называется {\em пространством-мишенью}. {\em Тождественное} отображение
$$
  \id(\MM):\quad \MM\rightarrow\MM
$$
определяется равенством $\id(x)=x$ для всех $x\in\MM$. Отображение $f$
называется {\em взаимно однозначным} или {\em инъективным}, если различные точки
$x_1\ne x_2$ из $\MM_1$ отображаются в различные точки $f(x_1)\ne f(x_2)$ из
$\MM_2$. Отображение $f$ является отображением $\MM_1$ {\em на\/} $\MM_2$ или
{\em сюрьективным}, если $f(\MM_1)=\MM_2$. Если отображение $f$ инъективно и
сюрьективно одновременно, то оно называется {\em биективным отображением} или
{\em биекцией}. Каждое отображение порождает некоторое подмножество в прямом
произведении $\MM_1\times\MM_2$,
\begin{equation*}
  {\sf Gr}_f:=\lbrace x,f(x)\in\MM_1\times\MM_2\rbrace,
\end{equation*}
которое называется {\em графиком} отображения $f$.
\qed\end{defn}
\index{Отображение множеств (set mapping)}\index{Функция (function)}%
\index{Аргумент (argument)}%
\index{Пространство-мишень (target space)}%
\index{Отображение тождественное (identity mapping)}%
\index{Тождественное отображение (identity mapping)}%
\index{Отображение взаимно однозначное (one-to-one mapping)}%
\index{Взаимно однозначное отображение (one-to-one mapping)}%
\index{Отображение инъективное (injective (фр.) mapping)}%
\index{Инъективное отображение (injective (фр.) mapping)}%
\index{Отображение сюрьективное (surjective (фр.) mapping)}%
\index{Сюрьективное отображение (surjective (фр.) mapping)}%
\index{Биекция (bijective (фр.) mapping, bijection)}%
\index{График отображения (graph of a map)}%
\begin{exa}
Рассмотрим дифференцируемую функцию одной переменной $f(x)\in\CC^1(\MR)$. Если
эта функция монотонно возрастает, $f'>0$, или монотонно убывает, $f'<0$, на
интервале $x\in(p,q)$, то она задает биективное отображение интервалов
$f:~(p,q)\rightarrow\big(f(p),f(q)\big)$.
\qed\end{exa}
\begin{exa}
Преобразование координат с невырожденным якобианом является биекцией двух
областей евклидова пространства (теорема \ref{tcoord}).
\qed\end{exa}
Сопоставление аргументу $x\in\MM_1$ в (\ref{esemap}) значения функции
$f(x)\in\MM_2$ будем обозначать ограниченной стрелкой:
\begin{equation*}
  f:\quad \MM_1\ni\quad x\mapsto f(x)\quad\in\MM_2.
\end{equation*}
\begin{defn}
Множество $f(\MM_1)=\MU\subset\MM_2$, которое может и не совпадать со всем
$\MM_2$, называется {\em образом} множества $\MM_1$ или {\em областью значений}
функции $f$. Множество $\MM_1$ при этом называется {\em прообразом} $\MU$ или
{\em областью определения} $f$ и обозначается $\MM_1$$=f^{-1}(\MU)$. Вообще,
прообразом произвольного подмножества $\MV\subset\MU\subset\MM_2$ называется
множество тех точек $\MW\subset\MM_1$, для которых $f(\MW)=\MV$. При этом символ
$f^{-1}$ является отображением
\begin{equation*}
  f^{-1}:\quad \MU\ni\quad f(x)\mapsto x\quad\in\MM_1
\end{equation*}
тогда и только тогда, когда отображение $f$ является взаимно однозначным. В этом
случае оно единственно и называется {\em обратным отображением}.
\qed\end{defn}
\index{Образ множества (set image)}%
\index{Область значений функции (range of function values)}%
\index{Прообраз множества (set inverse image)}%
\index{Область определения функции (domain of function definition)}%
\index{Отображение обратное (inverse mapping)}%
\index{Обратное отображение (inverse mapping)}%

Для биективного отображения $f:~\MM_1\rightarrow\MM_2$ всегда существует
обратное отображение
\begin{equation*}
  f^{-1}:\quad \MM_2\rightarrow\MM_1.
\end{equation*}
При этом отображение $f^{-1}$ также биективно и выполнено равенство
\begin{equation*}
  (f^{-1})^{-1}=f.
\end{equation*}
Если отображение $f$ не является взаимно однозначным, то прообразом одной точки
$f^{-1}(x)$, где $x\in\MU$, может быть несколько элементов из $\MM_1$ или их
бесконечное число. В этом случае символ $f^{-1}$ не является отображением.
\begin{exa}
Рассмотрим три отображения:
\begin{equation*}
  f:~\MR\rightarrow\MR,\qquad g:~\MR\rightarrow\MR_+\cup\lbrace0\rbrace,\qquad
  h:~\MR_+\rightarrow\MR_+,
\end{equation*}
которые определены одним и тем же правилом $x\mapsto x^2$. Эти отображения
различны: $f$ не сюрьективно, не инъективно; $g$ сюрьективно, но не инъективно;
$h$ биективно. Этот пример показывает, что задание области определения и области
значений является существенной частью определения отображения (функции).
\qed\end{exa}
\begin{defn}
Пусть $f$ -- функция (\ref{esemap}) и $\MU_1$ -- некоторое подмножество $\MM_1$,
тогда отображение
$$
  f|_{\MU_1}:\quad \MU_1\rightarrow\MM_2,
$$
называется {\em сужением} функции (или отображения) на $\MU_1\subset\MM_1$.
Функция $f$ является {\em продолжением} функции
$$
  g:\quad \MU_1\rightarrow\MM_2,
$$
заданной на некотором подмножестве $\MU_1\subset\MM_1$, если ее сужение
на $\MU_1$ совпадает с $g$, $f|_{\MU_1}=g$.
\qed\end{defn}
\index{Сужение функции (restriction of function)}%
\index{Сужение отображения (restriction of mapping)}%
\index{Продолжение функции (continuation of function)}%
\index{Продолжение отображения (continuation of mapping)}%

Ясно, что сужение функции единственно, а продолжение - нет.

\begin{defn}
{\em Композицией}, ({\em произведением} или {\em суперпозицией}) двух
отображений $f:~\MM_1$$\rightarrow\MM_2$ и $g:~\MM_2\rightarrow\MM_3$ называется
отображение
$$
  g\circ f:\quad \MM_1\rightarrow\MM_3,
$$
которое состоит в последовательном применении этих отображений:
\begin{equation*}                                                    \tag*{\qed}
  (g\circ f)(x)=g\big(f(x)\big).
\end{equation*}
\renewcommand{\qed}{}\end{defn}
\index{Композиция отображений (map composition)}%
\index{Произведение отображений (composition of transformations)}%
\index{Суперпозиция отображений (composition of transformations)}%
\begin{com}
Произведение определено не для всех отображений, а только для тех, у которых
множество $\MM_2$ в предыдущих обозначениях общее. Если рассматриваются
отображения некоторого множества в себя, то для них композиция всегда
определена.
\qed\end{com}
Композиция отображений есть не что иное, как сложная функция.

Если задано отображение $f:~\MM_1\rightarrow\MM_2$, то
\begin{equation*}
  f\circ\id(\MM_1)=f\qquad \text{и}\qquad \id(\MM_2)f=f.
\end{equation*}
Для биективного отображения выполнены равенства:
\begin{equation*}
  f^{-1}\circ f=\id(\MM_1)\qquad \text{и}\qquad f\circ f^{-1}=\id(\MM_2).
\end{equation*}
Если заданы два биективных отображения $f$ и $g$ и определена их композиция,
т.е.\ область значений $f$ совпадает с областью определения $g$, то обратное
отображение для композиции существует и задается формулой
\begin{equation*}
  (g\circ f)^{-1}=f^{-1}\circ g^{-1}.
\end{equation*}

В дальнейшем вместо $g\circ f$ мы часто будем писать просто $gf$.

Произведение отображений играет огромную роль в математике, и его изображают в
виде диаграммы
\begin{equation*}
\begin{diagram}
  \MM_1 & &\rTo^{g\circ f} & & \MM_3 \\
  & \rdTo_f & &\ruTo_g & \\ &&\MM_2 &&
\end{diagram},
\end{equation*}
которая называется {\em коммутативной}.
\index{Коммутативная диаграмма (commutative diagram)}%
\index{Диаграмма коммутативная (commutative diagram)}%
Это значит, что результат перехода от
$\MM_1$ к $\MM_3$ не зависит от того, по какому пути мы движемся: то ли вдоль
стрелки $g\circ f$, то ли последовательно вдоль стрелок $f$ и $g$. Изображение
произведений отображений в виде коммутативных диаграмм наглядно и полезно. Это
будет продемонстрировано в следующем утверждении.
\begin{prop}                                                      \label{passco}
Пусть определено произведение трех отображений:
\begin{equation*}
  f:~\MM_1\rightarrow\MM_2,\qquad g:~\MM_2\rightarrow\MM_3,\qquad
  h:~\MM_3\rightarrow\MM_4.
\end{equation*}
Тогда произведение этих отображений ассоциативно:
\begin{equation*}
  h(gf)=(hg)f
\end{equation*}
и его обозначают $hgf:~\MM_1\rightarrow\MM_4$.
\end{prop}
\begin{proof}
Все необходимые рассуждения содержатся в следующей диаграмме:
\begin{equation*}
\begin{diagram}
  \MM_1 && \rTo^{hgf} && \MM_4 \\ & \rdLine^{gf} & & \ruTo^{hg} & \\
  \dTo^f && \rdLine && \uTo_h \\
  & \ruLine && \rdTo & \\ \MM_2 && \rTo^g && \MM_3
\end{diagram}
\end{equation*}
Для доказательства необходимо проследить куда отобразится произвольный элемент
$x\in\MM_1$ под действием композиций $h(gf)$ и $(hg)f$. Формально доказательство
выглядит следующим образом:
\begin{equation*}                                                    \tag*{\qed}
  \big(h(gf)\big)x=h\big((gf)x\big)=h\big(g(fx)\big)=(hg)(hx)=\big((hg)f\big)x.
\end{equation*}
\renewcommand{\qed}{}\end{proof}

Отображения конечных множеств обладают простыми свойствами.
\begin{prop}
Если отображение $f:~\MM\rightarrow\MM$ конечного множества $\MM$ в себя
инъективно или сюрьективно, то оно биективно.
\end{prop}
\begin{proof}
Пусть отображение $f$ инъективно. Покажем, что оно является также сюрьективным.
Для произвольного элемента множества $x\in\MM$ определим последовательность
\begin{equation*}
  f^0(x):=x,\qquad \text{и}\qquad
  f^k(x):=\underbrace{f\big(\dotsc f}_k(x)\dotsc\big)=f\big(f^{k-1}(x)\big),
  \qquad k=1,2,\dotsc.
\end{equation*}
Из конечности множества $\MM$ вытекает, что в этой последовательности есть
повторения. Допустим, скажем, что $f^m(x)=f^n(x)$ при некоторых $m>n$. Если
$n>0$, то из равенства $f\big(f^{m-1}(x)\big)=f\big(f^{n-1}(x)\big)$ и
инъективности $f$ следует $f^{m-1}(x)=f^{n-1}(x)$. Сократив отображение $f$ $n$
раз, приходим к равенству
\begin{equation*}
  f^{m-n}(x)=x.
\end{equation*}
В этом случае для любого $x\in\MM$ найдется такое $x'=f^{m-n-1}(x)$, что
$f(x')=x$, т.е.\ отображение $f$ сюрьективно.

Второе утверждение предложения докажем от противного. Пусть отображение $f$
сюрьективно, но не инъективно. Тогда существует такой элемент $x\in\MM$, что
его прообраз состоит не менее, чем из двух элементов. Пусть множество $\MM$
содержит $N$ элементов. Тогда $f$ должно отобразить оставшиеся элементы, которых
не более, чем $N-2$, на $N-1$ элемент. Для сюрьективного отображения это
невозможно.
\end{proof}
В приведенном доказательстве существенна конечность множества $\MM$.
\begin{exa}
{\em Отображение следования}, определенное для натуральных чисел сдвигом на
единицу,
\begin{equation*}
  \s:\quad \MN\ni\quad n\mapsto n+1\quad\in\MN,
\end{equation*}
является инъективным, но не сюрьективным, поскольку единица не принадлежит
образу $\s(\MN)$.
\qed\end{exa}
\index{Отображение следования (succession map)}%
\begin{defn}
Мы говорим, что множества $\MM_1$ и $\MM_2$ имеют одинаковую {\em мощность},
если существует биективное отображение $f:~\MM_1\rightarrow\MM_2$. Множества,
имеющие ту же мощность, что и множество натуральных чисел $\MN$ называется
{\em счетным}.
\qed\end{defn}
\index{Счетное множество (countable set)}%
\index{Множество счетное (countable set)}%
\index{Мощность (cardinality)}%
\subsection{Отображения топологических пространств               \label{smacos}}
При отображении топологических пространств важную роль играют отображения,
согласованные с их топологией.
\begin{defn}
Отображение (\ref{esemap}) двух топологических пространств называется
{\em непрерывным} в точке $x\in\MM_1$, если для каждой окрестности $\MU_2$
точки $f(x)\in\MM_2$ существует такая окрестность $\MU_1$ точки $x$, что ее
образ $f(\MU_1)$ содержится в $\MU_2$. Отображение называется непрерывным в
области, если оно непрерывно в каждой точке этой области.
\qed\end{defn}
\index{Непрерывное отображение (continuous mapping)}%
\index{Отображение непрерывное (continuous mapping)}%
\begin{com}
В общем случае окрестность $\MU_2\subset\MM$ может содержать точки, не лежащие
в образе $f(\MM_1)$ и образ $f(\MU_1)$ совсем не обязан совпадать с $\MU_2$.
\qed\end{com}

\begin{figure}[h,b,t]
\hfill\includegraphics[width=.3\textwidth]{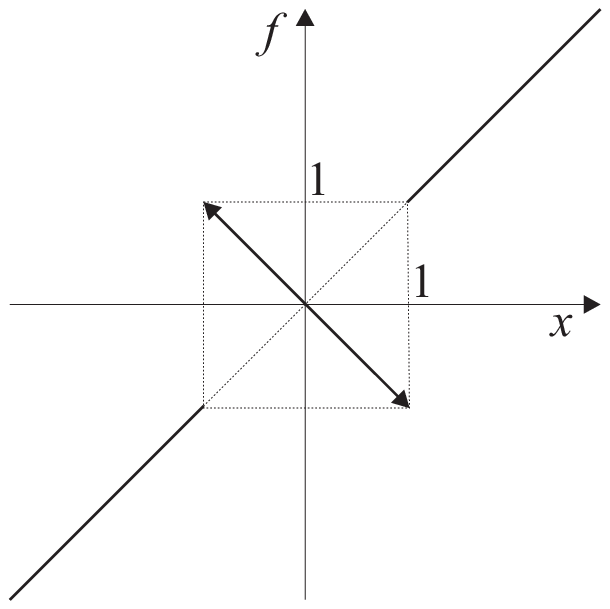}
\hfill {}
\centering\caption{Биективное отображение прямых, которое не является
гомеоморфизмом.}
\label{fbieknongo}
\end{figure}

\begin{defn}
Говорят, что два топологических пространства $\MM_1$ и $\MM_2$
{\em гомеоморфны}, если существуют взаимно обратные непрерывные отображения
$f:\quad \MM_1\rightarrow\MM_2$ и $g:\quad \MM_2\rightarrow\MM_1$ такие, что
$g\circ f=\id(\MM_1)$ и $f\circ g=\id(\MM_2)$. Тогда отображения $f$ и $g$
называются {\em гомеоморфизмами}.
\qed\end{defn}
\index{Гомеоморфизм (homeomorphism)}%
Ясно, что любой гомеоморфизм является биекцией. Обратное утверждение неверно.
\begin{exa}
Рассмотрим функцию
\begin{equation*}
  f(x)=\begin{cases} \quad x, & x\le-1~\text{или}~x\ge1, \\ -x, & x\in(-1,1),
       \end{cases}
\end{equation*}
показанную на рис.\ref{fbieknongo}. Эта функция задает отображение
$\MR\rightarrow\MR$, которое биективно, но не задает гомеоморфизм.
\qed\end{exa}

Справедливы две фундаментальные теоремы, которые приведем без доказательства.
\begin{theorem}[\bf Инвариантность размерности]                   \label{tinvdi}
Открытое подмножество в $\overline\MR^m_+$ не может быть гомеоморфно никакому
открытому подмножеству в $\overline\MR^n_+$, если $m\ne n$.
\end{theorem}
\index{Инвариантность размерности (invariance of dimensionality)}%
В этой теореме замкнутое полупространство $\overline\MR^n_+$ можно заменить на
все евклидово пространство $\MR^n$.
\begin{theorem}[\bf Инвариантность края]                          \label{tinvbo}
Пусть $\MU$ и $\MV$ -- два открытых подмножества $\overline\MR^n_+$ и $f$ --
гомеоморфизм $\MU$ на $\MV$, тогда
$f(\MU\bigcap\MR^{n-1})=\MV\bigcap\MR^{n-1}$, где $\MR^{n-1}$ -- гиперплоскость
(край), определяемая условием $x^n=0$. Другими словами, при гомеоморфизмах точки
края отображаются в край.
\end{theorem}
\index{Инвариантность края (invariance of boundary}%

Приведем пять критериев (необходимых и достаточных условий) непрерывности
отображения, которые будут использоваться в дальнейшем.
\begin{theorem}                                                   \label{tconma}
Если отображение $f:~\MM_1\rightarrow\MM_2$ сюрьективно, то оно непрерывно тогда
и только тогда, когда выполнено одно из следующих условий:

1) \parbox[t]{.92\linewidth}{
прообраз любого замкнутого подмножества $\MM_2$ замкнут в $\MM_1$;}

2) \parbox[t]{.92\linewidth}{
прообраз любого открытого подмножества $\MM_2$ открыт в $\MM_1$;}

3) \parbox[t]{.92\linewidth}{
$\forall\MU_1\subset\MM_1$\quad справедливо включение \quad
$f(\overline{\MU}_1)\subset\overline{f(\MU_1)}$;}

4) \parbox[t]{.92\linewidth}{
$\forall\MU_2\subset\MM_2$\quad справедливо включение \quad
$\overline{f^{-1}(\MU_2)}\subset f^{-1}(\overline{\MU}_2)$;}

5) \parbox[t]{.92\linewidth}{
$\forall\MU_2\subset\MM_2$\quad справедливо включение \quad
$f^{-1}(\Int\MU_2)\subset \Int f^{-1}(\MU_2)$.}
\end{theorem}
\begin{proof}
См., например, \cite{Kelley57R}.
\end{proof}

\begin{figure}[h,b,t]
\hfill\includegraphics[width=.3\textwidth]{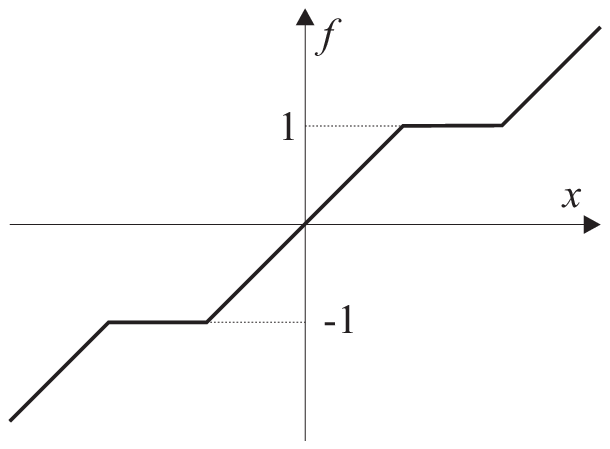}
\hfill {}
\centering\caption{Пример замкнутого, но не открытого непрерывного сюрьективного
  отображения.}
\label{fopclm}
\end{figure}

Определение и критерии непрерывности отображений говорят о топологических
свойствах прообраза, исходя из свойств образа. При этом нельзя сделать
однозначного утверждения относительно свойств образа при заданных свойствах
прообраза. Действительно, из критерия 3) следует, что образ замкнутого
множества, вообще говоря, не является замкнутым. Критерий 5), записанный в виде
$$
  \Int\MU_2\subset f\big(\Int f^{-1}(\MU_2)\big),
$$
говорит о том, что образ открытого множества может не быть открытым.
\begin{defn}
Непрерывное отображение (\ref{esemap}) называется {\em открытым}
({\em замкнутым}) если образ любого открытого (замкнутого) подмножества $\MM_1$
открыт (замкнут) в $\MM_2$.
\qed\end{defn}
\index{Отображение открытое (open mapping)}%
\index{Открытое отображение (open mapping)}%
\index{Отображение замкнутое (closed mapping)}%
\index{Замкнутое отображение (closed mapping)}%
\begin{exa}
Гомеоморфизм двух топологических пространств является открытым и замкнутым
одновременно, т.е.\ открытое множество отображается в открытое, а замкнутое --
в замкнутое.
\qed\end{exa}
\begin{exa}
Приведем пример замкнутого, но не открытого непрерывного сюрьективного
отображения $f:~\MR\rightarrow\MR$ (см.~рис.\ref{fopclm}), которое задано
следующим образом
\begin{equation*}
  f(x)=\begin{cases}
  x+1, & \qquad \qquad x\le-2, \\ -1, & -2\le x\le-1, \\ \quad x, & -1\le x\le1,
  \\\quad 1, & \quad 1\le x\le 2, \\ x-1, & \quad  2\le x.
  \end{cases}
\end{equation*}
При таком отображении открытый интервал $(a,b)$, где $a\in(-2,-1)$ и $b\in(1,2)$
отображается в замкнутый отрезок $[-1,1]$.
\qed\end{exa}
\begin{defn}
Рассмотрим отображение $f:~\MM_1\rightarrow\MM_2$ топологического пространства
$\MM_1$ в некоторое множество $\MM_2$. Если отображение $f$ является
сюрьективным, т.е.\ отображением на, то на множестве $\MM_2$ можно определить
{\em топологию идентификации} или {\em фактортопологию}. А именно, назовем
подмножество $\MU\subset\MM_2$ открытым, если его прообраз
$f^{-1}(\MU)\subset\MM_1$ открыт в $\MM_1$. Это самая тонкая топология на
$\MM_2$, в которой отображение $f$ является непрерывным.
\qed\end{defn}
\index{Фактортопология (quotient topology)}%
\index{Топология идентификации (identification topology)}%
\index{Идентификации топология (identification topology)}%
\begin{exa}
Пусть $\MM$ -- топологическое пространство, на котором задано некоторое
отношение эквивалентности $x_1\sim x_2$. Пусть $\MM/\sim$ -- множество
классов эквивалентности, и $f:~\MM\rightarrow\MM/\sim$ -- каноническая проекция
элемента $x\in\MM$ на его класс эквивалентности. Тогда отображение $f$
сюрьективно и на фактор пространстве $\MM/\sim$ можно ввести топологию
идентификации.
\qed\end{exa}
На самом деле, рассмотренный пример является общим. Если задано сюрьективное
отображение $f:~\MM_1\rightarrow\MM_2$, то на множестве $\MM_1$ всегда можно
определить отношение эквивалентности: будем считать точки $x_1,x_2\in\MM_1$
эквивалентными, если $f(x_1)=f(x_2)$. Другими словами, в множестве $\MM_1$ мы
идентифицируем те точки, которые отображаются в одну точку пространства-мишени
$\MM_2$. Этим объясняется название ``топология идентификации''.
\begin{defn}
Если некоторое свойство $P$ топологического пространства сохраняется при
гомеоморфизме, то оно называется {\em топологическим инвариантом} или
{\em свойством}. Другими словами, топологическое пространство $\MM$ обладает
свойством $P$ в том и только в том случае, если этим свойством обладает любое
другое гомеоморфное ему топологическое пространство.
\qed\end{defn}
\index{Топологический инвариант (topological invariant)}%
\index{Инвариант топологический (topological invariant)}%
\index{Топологическое свойство (topological property)}%
\index{Свойство топологическое (topological property)}%
Так как гомеоморфизм устанавливает взаимно однозначное соответствие не только
между точками, но и между открытыми множествами, то любое свойство, определенное
в терминах открытых множеств, является топологическим инвариантом.

\begin{exa}
Число связных компонент топологического пространства является топологическим
инвариантом.
\qed\end{exa}
Если отображение (\ref{esemap}) непрерывно, но не является гомеоморфизмом,
то число связных компонент может уменьшаться, но не увеличиваться.
\begin{exa}                                                       \label{eontwm}
Если пространство $\MM_1=\MI_1\cup\MI_2$ является объединением двух
непересекающихся открытых интервалов $\MI_1$ и $\MI_2$, то его можно
непрерывно отобразить на произвольный открытый интервал $\MI_3$. При
этом $\MI_1\rightarrow\MI_3$ и $\MI_2\rightarrow\MI_3$, т.е.\ каждый из
интервалов $\MI_1$ и $\MI_2$ отображается на весь интервал $\MI_3$. Обратное
утверждение неверно. Один интервал невозможно непрерывно отобразить
на два. Действительно, если существует непрерывное отображение
$$
  g:\quad \MI_3\rightarrow\MI_1\cup\MI_2,
$$
то в силу непрерывности $g$ прообразы $\MI_1$ и $\MI_2$ являются открытыми
и замкнутыми непустыми подмножествами в $\MI_3$ одновременно. Поскольку
интервал $\MI_3$ связен, то $\MI_3$$=g^{-1}(\MI_1)$$=g^{-1}(\MI_2)$. Это
означает, что каждой точке интервала $\MI_3$ соответствует две точки в
объединении $\MI_1\bigcup\MI_2$, что противоречит определению отображения
(\ref{esemap}).
\qed\end{exa}
\begin{defn}
{\em Компактификацией} топологического пространства $\MM$ называется
пара $(\MU,f)$, где $\MU$ -- компактное топологическое пространство, а $f$ --
гомеоморфизм $\MM$ на всюду плотное подмножество в $\MU$.
\qed\end{defn}
\index{Компактификация (compactification)}%
\begin{exa}
Отображение евклидовой плоскости на сферу, $f:$ $\MR^2$ $\rightarrow$ $\MS^2$,
которое задается стереографической проекцией (см.\ раздел \ref{sphere})
является компактификацией. При этом все бесконечно удаленные точки
евклидовой плоскости отображаются на одну точку сферы -- северный полюс.
\qed\end{exa}
\begin{exa}
Добавим к евклидову пространству $\MR^n$ бесконечно удаленную точку $\infty$
и объявим ее окрестностями множества вида $(\MR^n\setminus \MB)\cup\infty$
для всех ограниченных замкнутых подмножеств $\MB\subset\MR^n$. Тогда объединение
$\MR^n\cup\infty$ будет компактным пространством, гомеоморфным сфере $\MS^n$.
В этом примере пара $\big((\MR^n\cup\infty),f\big)$, где отображение
$f:~\MR^n\rightarrow\MR^n\cup\infty$ имеет вид
\begin{equation*}
  f:\quad \MR^n\ni\quad x\mapsto(x,\infty)\quad\in\MR^n\cup\infty,
\end{equation*}
является компактификацией евклидова пространства.
\qed\end{exa}

{\bf Отображения хаусдорфовых топологических пространств.}
Поскольку многообразия, по определению, являются хаусдорфовыми топологическими
пространствами, то отображения хаусдорфовых пространств будут играть в
дальнейшем важную роль.

Доказательство следующих двух теорем и следствий содержится, например, в
\cite{Engelk85R}.
\begin{theorem}
Если компактное пространство $\MM_1$ непрерывно отображается на пространство
$\MM_2$, то $\MM_2$ компактно.
\end{theorem}
\begin{cor}
Если $f:~\MM_1\rightarrow\MM_2$ -- непрерывное отображение компактного
пространства $\MM_1$ в хаусдорфово пространство $\MM_2$, то
$\overline{f(\MU)}=f(\overline\MU)$ для каждого подмножества $\MU\subset\MM_1$.
\end{cor}
\begin{cor}
Каждое непрерывное отображение компактного пространства в хаусдорфово
пространство замкнуто.
\end{cor}
\begin{theorem}
Каждое непрерывное взаимно однозначное отображение компактного пространства
на хаусдорфово пространство является гомеоморфизмом.
\end{theorem}

Сформулируем три теоремы, доказательство которых можно найти, например, в
\cite{Kosnio80R}.
\begin{theorem}
Компактное подмножество $\MU$ хаусдорфова топологического пространства $\MM$
хаусдорфово и замкнуто.
\end{theorem}
\begin{theorem}
Пусть $\MM_1$ и $\MM_2$ -- топологические пространства. Тогда $\MM_1$ и $\MM_2$
хаусдорфовы в том и только в том случае, если произведение $\MM_1\times\MM_2$
хаусдорфово.
\end{theorem}
Хотя подпространства и произведения хаусдорфовых пространств хаусдорфовы,
факторпространство хаусдорфова пространства, вообще говоря, нехаусдорфово.
\begin{exa}                                                       \label{enonha}
Рассмотрим вещественную прямую $\MR$ и открытый интервал $(0,1)$. Введем
отношение эквивалентности $\sim$ на $\MR$, при котором $x\sim x'$ тогда и
только тогда, когда $x=x'$ или $x,x'\in(0,1)$. Грубо говоря, фактор пространство
$X/\sim$ -- это вещественная прямая $\MR$ со стянутым в точку интервалом
$(0,1)$. Если снабдить $\MR/\sim$ фактортопологией относительно естественной
проекции $\pi:~\MR\ni x\mapsto\tilde x\in\MR/\sim$, то прообраз точки
$\tilde x_0$, где $x_0\in(0,1)$ есть интервал $(0,1)$, который открыт в $\MR$.
Следовательно, точка $\tilde x_0$ открыта в $\MR/\sim$. Таким образом фактор
пространство $\MM/\sim$ представляет собой объединение двух лучей $(-\infty,0]$,
$[1,\infty)$ и точки $\tilde x_0$, которая открыта. Любые окрестности точек
$0$ и $1$ содержат точку $\tilde x_0$ и, потому, пересекаются. Следовательно,
факторпространство $\MR/\sim$ нехаусдорфово. Подчеркнем, что в рассматриваемом
примере фактор пространство $\MR/\sim$ снабжено фактор топологией (топологией
идентификации), которая отличается от топологии, индуцируемой вложением в
$\MR^2$.
\qed\end{exa}
В рассмотренном примере факторпространство $\MR/\sim$ не является многообразием,
хотя проекция $\pi$ непрерывна. Для того, чтобы обеспечить хаусдорфовость
факторпространства $\MN$ хаусдорфова пространства $\MM$ по некоторому отношению
эквивалентности необходимо наложить дополнительные условия. Достаточное условие
дает следующая
\begin{theorem}                                                   \label{thaufa}
Пусть $\MN$ -- факторпространство топологического пространства $\MM$,
определенное при помощи сюрьективного отображения $f:~\MM\rightarrow\MN$. Если
пространство $\MM$ компактно и хаусдорфово, а отображение $f$ замкнуто, то $\MN$
компактно и хаусдорфово.
\end{theorem}
\begin{cor}
Если $\MM$ -- компактное хаусдорфово пространство, на котором действует конечная
группа преобразований $\MG$, то $\MM/\MG$ -- компактное хаусдорфово
пространство.
\qed\end{cor}
\begin{exa}
Проективное пространство $\MR\MP^n$ получается из сферы $\MS^n$ отождествлением
диаметрально противоположных точек, $\MR\MP^n=\MS^n/\MZ_2$. Это есть компактное
хаусдорфово пространство.
\qed\end{exa}
Чтобы сформулировать другое следствие теоремы \ref{thaufa}, рассмотрим
пространство $\MM$ вместе с его подмножеством $\MU\subset\MM$. Обозначим через
$\MM/\MU$ фактор пространство $\MM/\sim$, где $\sim$ -- отношение
эквивалентности на $\MM$, при котором $x\sim x'$ тогда и только тогда, когда
$x=x'$ или $x,x'\in\MU$.
\begin{cor}
Если $\MM$ -- компактное хаусдорфово пространство и $\MU$ -- его замкнутое
подмножество, то $\MM/\MU$ -- компактное хаусдорфово пространство.
\qed\end{cor}
\begin{com}
В примере \ref{enonha} подмножество $(0,1)\subset\MR$ открыто, и фактор
пространство $\MR/\sim$ не является хаусдорфовым.
\qed\end{com}
\section{Преобразования координат                                \label{scooch}}
Важным примером отображений являются преобразования координат, которые
рассмотрены в настоящем разделе. Евклидово пространство $\MR^n$ было определено
в разделе \ref{seucto} как топологическое произведение $n$ прямых. При
этом каждая точка $x\in\MR^n$ задается набором $n$ вещественных чисел:
\begin{equation*}
  x=(x^1,\dotsc,x^n)=\lbrace x^\al\rbrace,\qquad \al=1,\dotsc,n,
\end{equation*}
-- декартовых координат, которые покрывают все евклидово пространство.
Рассмотрим $n$ дифференцируемых функций (преобразование координат)
\begin{equation}                                                  \label{ecootr}
  x^{\al'}=x^{\al'}(x)\quad \in\CC^1(\MU),\qquad \al'=1,\dotsc,n,
\end{equation}
от декартовых координат $x^\al$, заданных в некоторой области $\MU\subset\MR^n$.
В общем случае эти функции отображают область $\MU\subset\MR^n$ на некоторое
множество точек $\MU'\subset\MR^n$ -- образ отображения. В наших обозначениях
запись $x^{\al'}$ эквивалентна записи $x^{\prime\,\al}$. При этом штрихованные
индексы пробегают те же значения, что и нештрихованные $\al,\al'=1,\dotsc,n$.
Такая запись удобнее, т.к.\ позволяет во многих случаях опускать букву $x$.
В сокращенной записи мы опускаем индексы:
\begin{equation}                                                  \label{edifdx}
  \MU\ni\quad x\mapsto x'(x)\quad\in\MU'.
\end{equation}
\begin{defn}
Матрица, составленная из частных производных,
\begin{equation}                                                  \label{ejacma}
  J_\al{}^{\al'}=\frac{\pl x^{\al'}}{\pl x^\al}=\pl_\al x^{\al'}
\end{equation}
называется {\em матрицей Якоби}. Здесь индекс $\al$ считается первым, а индекс
$\al'$ -- вторым. Определитель матрицы Якоби называется {\em якобианом},
\begin{equation}                                                  \label{ejacob}
  J=\frac{\pl(x^{\al'})}{\pl(x^\al)}
  :=\det\left(\frac{\pl x^{\al'}}{\pl x^\al}\right). \qed
\end{equation}
\end{defn}
\index{Матрица Якоби (Jacobi matrix)}\index{Якоби матрица (Jacobi matrix)}%
\index{Якобиан (Jacobian)}%

Из курса математического анализа известно следующее утверждение.
\begin{theorem}                                               \label{tcoord}
Пусть функции (\ref{ecootr}) задают дифференцируемое отображение
класса $\CC^k$ открытого множества $\MU\subset\MR^n$ в $\MR^n$. Если якобиан
этого отображения не обращается в нуль в точке $x_0\in\MU$, то существуют такие
окрестности $\MU_0\subset\MU$ и $\MU_0^\prime\subset\MR^n$ точек $x_0$ и
$x_0^\prime=x_0^\prime(x_0)$, что функции $x^{\al^\prime}(x)$ задают взаимно
однозначное отображение окрестности $\MU_0$ на $\MU_0^\prime$ (биекцию), и
обратное ему отображение дифференцируемо и того же класса гладкости $\CC^k$ в
области $\MU_0^\prime$. В частности, возможно $k=\infty$.
\end{theorem}
\begin{cor}
Пусть $x^\prime(x)$ -- дифференцируемое отображение открытого множества
$\MU\subset\MR^n$ в $\MR^n$. Если якобиан этого отображения отличен от нуля на
всем $\MU$, то образ этого множества $\MU'$ также является открытым множеством.
В частности, окрестность точки $x$ отображается в окрестность точки $x'$.
\qed\end{cor}

Пусть задана последовательность биекций:
\begin{equation*}
  \MU\xrightarrow{x'(x)}\MU'\xrightarrow{x''(x')}\MU''\rightarrow\dotsc
\end{equation*}
Мы говорим, что точка $x\in\MU\subset\MR^n$ евклидова пространства
задается $n$ вещественными числами
\begin{equation*}
 x=\lbrace x^\al\rbrace=\lbrace x^{\al'}\rbrace=\lbrace x^{\al''}\rbrace=\dotsc,
\end{equation*}
которые являются координатами одной и той же точки в различных системах
координат. Теперь мы можем расширить понятие евклидова пространства,
допустив произвольные замены координат. Тогда мы считаем, что евклидово
пространство $\MR^n$ покрывается некоторыми областями (картами)
\begin{equation*}
  \MR^n=\bigcup_i\MU_i,
\end{equation*}
на каждой из которых выбрана своя система координат. Рассмотрим две
пересекающиеся карты $\MU_i$ и $\MU_j$ в $\MR^n$. В частном случае одна или обе
области могут совпадать со всем евклидовым пространством. Обозначим координаты
точек, принадлежащих $\MU_i$ и $\MU_j$, соответственно через $x^\al$ и
$x^{\al'}$. Тогда в области пересечения карт $\MU_i\cap\MU_j$ координаты точек
будут связаны некоторым преобразованием вида (\ref{ecootr}). По этой причине
функции (\ref{ecootr}) называются {\em функциями перехода} к новой системе
координат или {\em функциями склейки}.
\index{Перехода функции (transition function)}%
\index{Функции перехода (transition function)}%
\index{Функции склейки (sewing function)}%
\index{Склейки функции (sewing function)}%
Заметим, что их одновременно можно рассматривать и как функции, осуществляющие
склейку двух областей и как замену координат на пересечении $\MU_i\cap \MU_j$.
\begin{com}
В дифференциальной геометрии такие объекты как векторные поля, метрика,
связность и другие геометрические структуры вводятся инвариантным образом,
независимо от выбранной системы координат. В приложениях выбор той или иной
системы координат диктуется симметрией задачи, и эти координаты, как правило, не
покрывают все евклидово пространство. Например, сферические и цилиндрические
координаты определены всюду в $\MR^3$ за исключением оси $z$.
\qed\end{com}
Функции (\ref{ecootr}) осуществляют замену или {\em преобразование координат}.
\index{Преобразование координат (coordinate transformation)}%
\index{Координат преобразование (coordinate transformation)}%
Это преобразование в зависимости от обстоятельств можно рассматривать как
{\em пассивное} или {\em активное}.
\index{Преобразование координат активное (active coordinate transformation}%
\index{Преобразование координат пассивное (passive coordinate transformation)}%
При пассивном рассмотрении считается, что одной и той же точке области
соответствуют различные координаты в различных системах отсчета. Именно
так мы и рассматривали преобразование координат до сих пор. Во втором случае
считается, что заданная точка евклидова пространства с координатами
$x^\al$ в результате некоторого преобразования или деформирования области
занимает новое положение с координатами $x^{\al'}$ относительно старой
координатной системы. Такая точка зрения принята в теории групп преобразований.
\begin{exa}
Вращение евклидова пространства относительно начала в данной декартовой системе
координат обычно рассматривают как активное. С математической точки зрения оба
подхода равноправны, т.к.\ определяются одним и тем же набором функций
(\ref{ecootr}), а деление преобразований на активные и пассивные зависит от
физической интерпретации и традиций.
\qed\end{exa}

Обычно от функций перехода (\ref{ecootr}) требуется, чтобы они и их обратные
$x^\al=x^\al(x')$ были достаточно гладкими. В частности, непрерывность функций
перехода означает гомеоморфность отображения (\ref{ecootr}). В дальнейшем, если
не оговорено противное, мы рассматриваем преобразования координат класса
$\CC^\infty$. В этом случае гомеоморфизм (\ref{edifdx}) называется
{\em диффеоморфизмом}.
\index{Диффеоморфизм (diffeomorphism)}%

Отличие от нуля якобиана (\ref{ejacob}) означает, что функции (\ref{ecootr}),
определяющие преобразования координат, функционально независимы.
\begin{defn}
Набор функций $u^i(x^\al)$, $i=1,\dotsc,\Sn$, в некоторой области $\MU$
называется {\em функционально зависимым}, если на любом компакте из $\MU$
существует функция от $\Sn$ переменных $F(u^i)$, определенная в области значений
$u^i$, которая непрерывна вместе со всеми частными производными первого порядка,
не равна тождественно нулю ни в какой подобласти и для которой выполнено
соотношение $F\big(u^i(x^\al)\big)=0$.
\qed\end{defn}
\index{Функциональная зависимость (functional dependence)}%
\index{Зависимость функциональная (functional dependence)}%
Дифференцируя уравнение $F\big(u^i(x^\al)\big)=0$ по $x^\al$, получим равенство
\begin{equation*}
  \frac{\pl F}{\pl u^i}\frac{\pl u^i}{\pl x^\al}=0.
\end{equation*}
Поскольку среди функций $\pl F/\pl u^i$ по крайней мере одна отлична от нуля,
то для функциональной зависимости необходимо, чтобы ранг матрицы
$\pl u^i/\pl x^\al$ был меньше $\Sn$. Это условие является и достаточным.
В частности, при $\Sn>n$ функции всегда будут функционально зависимы. Тем
самым можно дать эквивалентное
\begin{defn}
$\Sn$ функций, для которых ранг матрицы $\pl u^i/\pl x^\al$ равен $\Sn$ в каждой
точке области $\MU\subset\MR^n$ называются {\em функционально независимыми} в
этой области.
\qed\end{defn}
\index{Независимость функциональная (functional independence)}%
Таким образом, $n$ достаточно гладких функционально независимых функций
(\ref{ecootr}) задают преобразование координат в некоторой области
$\MU\subset\MR^n$.

Преобразование координат (\ref{ecootr}) называется {\em невырожденным}
\index{Невырожденное преобразование координат%
(nondegenerate coordinate transformation)}%
\index{Преобразование координат невырожденное%
(nondegenerate coordinate transformation)}%
в некоторой точке или области, если якобиан преобразования отличен от нуля в
этой точке или области, соответственно. Если преобразование координат
невырождено в некоторой области $\MU$, то согласно теореме \ref{tcoord} в
достаточно малой окрестности произвольной точки $x\in\MU$ это преобразование
является взаимно однозначным. Для невырожденных преобразований координат
локально существует обратное преобразование
$$
  x^{\al'}\mapsto x^\al=x^\al(x'),
$$
при этом якобиан обратного преобразования равен
$$
  J^{-1}=\det\left(\frac{\pl x^\al}{\pl x^{\al'}}\right).
$$
Если якобиан преобразования обращается в нуль в некоторой точке, то в этой
точке преобразование координат является вырожденным, и обратного
преобразования не существует.

Совершим два преобразования координат $x\mapsto x'\mapsto x''$. Тогда координаты
точки $x$ можно выразить, как в системе координат $x'$, так и в $x''$. При этом,
по-определению, справедливы равенства $x^\al(x'')=x^\al\big(x'(x'')\big)$.
Отсюда следует, что каждая координатная функция $x^\al$ при преобразовании
координат $x'\rightarrow x''$ ведет себя, как скалярное поле (см.\ раздел
\ref{scfden}). Верно также и обратное утверждение. Если в некоторой области
евклидова пространства $\MU\subset\MR^n$ задано $n$ функций (скалярных полей)
$f^a(x)$, $a=1,\dotsc,n$, таких, что $\det(\pl f^a/\pl x^\al)\ne0$, то их всегда
можно выбрать в качестве новой системы координат на $\MU$.

Обозначим множество невырожденных преобразований координат в $\MR^n$ через
$\MG(\MR^n)$. При этом будем рассматривать преобразования координат как
активные, т.е.\ отображающие области в одном фиксированном евклидовом
пространстве. Для заданного преобразования координат $f=x'(x)$ будем писать
$f\in\MG(\MR^n)$. Эти преобразования согласованы с топологией в $\MR^n$.
\begin{defn}
Множество преобразований $\MG(\MR^n)$ топологического пространства $\MR^n$
называется {\em псевдогруппой преобразований}, если выполнены следующие условия:

1) \parbox[t]{.92\linewidth}{
каждое преобразование $f\in\MG(\MR^n)$ является гомеоморфизмом открытого
множества $\MU_i$ (область определения $f$) на другое открытое множество
$f(\MU_i)=\MU_j$ (область значений $f$);}

2) \parbox[t]{.92\linewidth}{
если $f\in\MG(\MR^n)$, то сужение $f$ на произвольное открытое подмножество
области определения $f$ также принадлежит $\MG(\MR^n)$;}

3) \parbox[t]{.92\linewidth}{
пусть множество $\MU$ есть объединение произвольного числа открытых
множеств, $\MU=\bigcup_i\MU_i$, тогда гомеоморфизм $f$ множества $\MU$
принадлежит $\MG(\MR^n)$, если сужение $f$ на $\MU_i$ принадлежит $\MG(\MR^n)$
для каждого $i$;}

4) \parbox[t]{.92\linewidth}{
тождественное преобразование каждого открытого множества
$\MU\subset\MR^n$ принадлежит $\MG(\MR^n)$;}

5) \parbox[t]{.92\linewidth}{
для каждого $f\in\MG(\MR^n)$ существует обратное преобразование
$f^{-1}\in\MG(\MR^n)$;}

6) \parbox[t]{.92\linewidth}{
если $f_1$ есть гомеоморфизм $\MU_1\rightarrow\MV_1$, а $f_2$ --
гомеоморфизм $\MU_2\rightarrow\MV_2$, и пересечение $\MV_1\cap\MU_2$
не пусто, то гомеоморфизм $f_2f_1$ множества $f_1^{-1}(\MV_1\cap\MU_2)$
на множество $f_2(\MV_1\cap\MU_2)$ также принадлежит $\MG(\MR^n)$.\qed}
\end{defn}
\index{Псевдогруппа преобразований (pseudogroup of transformations)}%
\begin{com}
Первые три свойства являются топологическими, а остальные -- групповыми, если
под композицией преобразований понимать их последовательное выполнение в тех
точках, где они определены. В общем случае преобразования координат группы не
образуют, т.к.\ при композиции двух преобразований необходимо следить за их
областью определений и значений. Например, если область значений преобразования
$f_1$ не имеет общих точек с областью определения $f_2$, то композиция
$f_2\circ f_1$ не определена. Следовательно, преобразования координат группы не
образуют, так как закон композиции в группе должен быть определен для всех
элементов группы. Тем не менее часто употребляется термин группа преобразований
координат. Для того, чтобы преобразования координат действительно образовали
группу необходимо предположить, что области определений и значений
$\MU\subset\MR^n$ для всех $f\in\MG(\MR^n)$ совпадают и отказаться от условия
2). Эта группа бесконечномерна, т.к.\ параметризуется $n$ функциями от $n$
переменных.
\qed\end{com}
Отображение $f$ областей евклидова пространства называется
{\em диффеоморфизмом}.
\index{Диффеоморфизм (diffeomorphism)}%
Если все диффеоморфизмы $f\in\MG(\MR^n)$ одного класса гладкости $\CC^k$, то
будем писать $f\in\diff^k(\MR^n)$. Для гладкой псевдогруппы преобразований
координат индекс обычно опускают $\diff(\MR^n):=\diff^\infty(\MR^n)$. Ее часто
называют просто {\em группой диффеоморфизмов} евклидова пространства $\MR^n$.
\index{Группа диффеоморфизмов (group of diffeomorphisms)}%
\index{Диффеоморфизмов группа (group of diffeomorphisms)}%
Очевидно, что $\diff^l(\MR^n)$ -- подпсевдогруппа $\diff^k(\MR^n)$ при $l>k$.

Множество преобразований координат с положительным определителем
$\diff^k_0(\MR^n)$ также является подпсевдогруппой:
$\diff^k_0(\MR^n)\subset\diff^k(\MR^n)$. Все невырожденные преобразования
координат можно разделить на два класса: с положительным, $\diff^k_0(\MR^n)$, и
отрицательным, $\diff^k(\MR^n)\setminus\diff^k_0(\MR^n)$, якобианом. Говорят,
что координатный базис касательного пространства (см.\ раздел \ref{svecfi})
имеет одинаковую {\em ориентацию} в двух системах координат, если якобиан
соответствующего преобразования положителен. Если якобиан преобразования
отрицателен, то при переходе между системами координат базис касательного
пространства, по-определению, меняет ориентацию.
\index{Ориентация (orientation)}%
\section{Группа двумерных вращений $\MO(2)$                      \label{stwodr}}
Рассмотрим евклидову плоскость $\MR^2$ с декартовыми координатами
$\lbrace x,y\rbrace=\lbrace x^\al\rbrace$, $\al=1,2$, и евклидовой метрикой
$g_{\al\bt}=\dl_{\al\bt}$.
\begin{defn}
Невырожденные линейные однородные преобразования координат,
$$
  x^{\al}\mapsto x^{\prime\al}=x^\bt S_\bt{}^\al,\qquad \det S_\bt{}^\al\ne0,
$$
относительно которых евклидова метрика остается инвариантной,
\begin{equation}                                                  \label{edeotw}
  \dl_{\al\bt}=S_\al{}^\g S_\bt{}^\e\dl_{\g\e},
\end{equation}
образуют группу, которая называется {\em группой двумерных вращений} и
обозначается $\MO(2)$. Квадратные матрицы, удовлетворяющие условию
(\ref{edeotw}), которое можно переписать в матричном виде
$$
  SS^\St=S^\St S=1,
$$
где $S^\St$ обозначает транспонированную матрицу, называются
{\em ортогональными}.
\qed\end{defn}
\index{Матрица ортогональная (orthogonal matrix)}%
\index{Ортогональная матрица (orthogonal matrix)}%

Группа $\MO(2)$ является группой ортогональных $2\times2$ матриц с обычным
правилом умножения матриц. Вычисляя определитель левой и правой части уравнения
(\ref{edeotw}) получим, что определитель матриц $S$ по модулю равен единице,
$$
  \det S=\pm1.
$$
\begin{prop}
Любое решение уравнения (\ref{edeotw}) принадлежит одному из двух классов:
\begin{align}                                                     \label{eormtr}
  S_+(\om)&:=\begin{pmatrix}  \quad \cos\om & \sin\om \\ -\sin\om & \cos\om
  \end{pmatrix}, & \det S_{+}&=1,
\\                                                                \label{eormtm}
  S_-(\om)&:=\begin{pmatrix} \quad \cos\om & -\sin\om \\ -\sin\om & -\cos\om
  \end{pmatrix}, & \det S_{-}&=-1,
\end{align}
при некотором {\em угле вращения} $\om\in[0,2\pi]$. При этом точки $\om=0$ и
$\om=2\pi$ отождествляются. Следовательно, параметр $\om$ лежит на окружности,
$\om\in\MS^1$.
\end{prop}
\index{Угол вращения (rotational angle)}%
\index{Вращения угол (rotational angle)}%
\begin{proof}
Уравнение (\ref{edeotw}) -- это система трех квадратных уравнений:
\begin{equation*}
\begin{split}
  (S_\al{}^1)^2+(S_\al{}^2)^2&=1,\qquad \al=1,2,
\\
  S_1{}^1S_2{}^1+S_1{}^2S_2{}^2&=0.
\end{split}
\end{equation*}
Перебор всех возможностей показывает, что с точностью до переопределения угла
поворота формулы (\ref{eormtr}), (\ref{eormtm}) дают общее решение.
\end{proof}

Для краткости обозначим множество матриц вида (\ref{eormtr}), (\ref{eormtm})
теми же, но ажурными буквами:
\begin{equation*}
\begin{split}
  \MS_+:&=\lbrace S_+(\om):\quad \om\in\MS^1\rbrace,
\\
  \MS_-:&=\lbrace S_-(\om):\quad \om\in\MS^1\rbrace.
  \end{split}
\end{equation*}
Тогда $\MO(2)=\MS_+\cup\MS_-$, и справедлива следующая таблица умножения:
\begin{equation*}
  \MS_+\MS_+=\MS_+,\qquad \MS_+\MS_-=\MS_-,\qquad \MS_-\MS_+=\MS_-,
  \qquad \MS_-\MS_-=\MS_+.
\end{equation*}
Обратные матрицы для $\MS_+$ имеют вид
\begin{equation}                                                  \label{einvro}
  S_+^{-1}(\om)=S_+(-\om)
\end{equation}
и соответствует вращению на тот же угол $\om$, но в противоположном направлении.
При этом единичная матрица $\one=S_+(0)\in\MS_+$.

Таким образом, группа двумерных вращений является однопараметрической компактной
группой Ли, состоящей из двух компонент. Матрицы $\MS_+$ с единичным
определителем образуют нормальную подгруппу группы вращений, которая называется
группой {\em собственных} вращений и обозначается $\MS\MO(2)=\MS_+$. Эти матрицы
образуют связную компоненту единицы. В этом нетрудно убедиться, заменив параметр
$\om$ на $t\om$ в (\ref{eormtr}). Тогда мы получим непрерывное семейство
ортогональных матриц, связывающее заданную матрицу при $t=1$ с единичной
матрицей при $t=0$.

Отметим, что матрица $-\one$ также принадлежат связной компоненте единицы,
$-\one=S_+(\pi/2)$.

Матрицы несобственных вращений $S_-$ группы не образуют, т.к.\ не содержат,
например, единицы. Их можно представить в виде призведения двух матриц
\begin{equation}                                                  \label{erebcc}
  S_-=S_+P,
\end{equation}
где
\begin{equation}                                                  \label{eparop}
  P=\begin{pmatrix}  1 & \quad 0 \\ 0 & -1 \end{pmatrix}=S_-(0),\qquad
  \det P=-1,
\end{equation}
-- {\em оператор четности}, который отражает ось ординат,
\index{Оператор четности (parity operator)}%
\index{Четности оператор (parity operator)}%
$$
  P:\quad (x,y)~\mapsto~(x,-y).
$$
При этом меняется ориентация декартовых координат.

Таким образом, матрицы с положительным и отрицательным определителем образуют
две связные компоненты группы вращений $\MO(2)$. Связная компонента единицы
образует группу собственных вращений $\MS\MO(2)$, а оператор четности определяет
диффеоморфизм между компонентами.

Легко проверить, что группа собственных вращений $\MS\MO(2)$ является абелевой.
В то же время полная группа вращений $\MO(2)$ неабелева. В этом просто
убедиться, вычислив коммутатор
\begin{equation*}
  \big[S_-(\om_1),S_-(\om_2)\big]=2\begin{pmatrix}
  0 & \sin(\om_1-\om_2) \\ -\sin(\om_1-\om_2) & 0 \end{pmatrix},
\end{equation*}
который не лежит ни в $\MS_+$, ни в $\MS_-$.

Элементы группы $\MS\MO(2)$ можно представить в виде
\begin{equation}                                                  \label{eirota}
  S_{+\al}{}^\bt=(\ex^{\om\ve})_\al{}^\bt
  =\dl_\al^\bt\cos\om+\ve_\al{}^\bt\sin\om,
\end{equation}
где экспонента определена с помощью ряда Тейлора, и $\ve_{\al\bt}$
-- полностью антисимметричный тензор второго ранга (см.\ приложение
\ref{scomat}).

Подгруппа $\MS\MO(2)\subset\MO(2)$ является нормальной подгруппой.
Следовательно, фактор пространство $\MO(2)/\MS\MO(2)$ является группой,
состоящей из двух элементов
\begin{equation*}
  \MZ_2\simeq\frac{\MO(2)}{\MS\MO(2)},
\end{equation*}
где $\MZ_2:=\lbrace1,-1\rbrace$ -- группа из двух элементов по умножению.

Матрицы (\ref{eormtr}), (\ref{eormtm}) можно рассматривать как представление
группы вращений $\MO(2)$ в двумерном векторном пространстве. Оно неприводимо и
называется {\em фундаментальным} или {\em векторным} представлением.
\index{Фундаментальное представление $\MO(2)$%
 (fundamental representation of $\MO(2)$}%
\index{Векторное представление группы $\MO(2)$%
 (vector representation of $\MO(2)$}%

Бесконечно малое вращение на угол $\dl\om\ll1$ в линейном приближении имеет вид
\begin{equation}                                                  \label{einfro}
  S_{+\al}{}^\bt\approx\dl_\al^\bt+\dl\om L_\al{}^\bt,
\end{equation}
где $L_\al{}^\bt=\ve_\al{}^\bt$ -- генератор (вектор базиса алгебры Ли) группы
$\MS\MO(2)$ в фундаментальном представлении.

\begin{figure}[h,b,t]
\hfill\includegraphics[width=.8\textwidth]{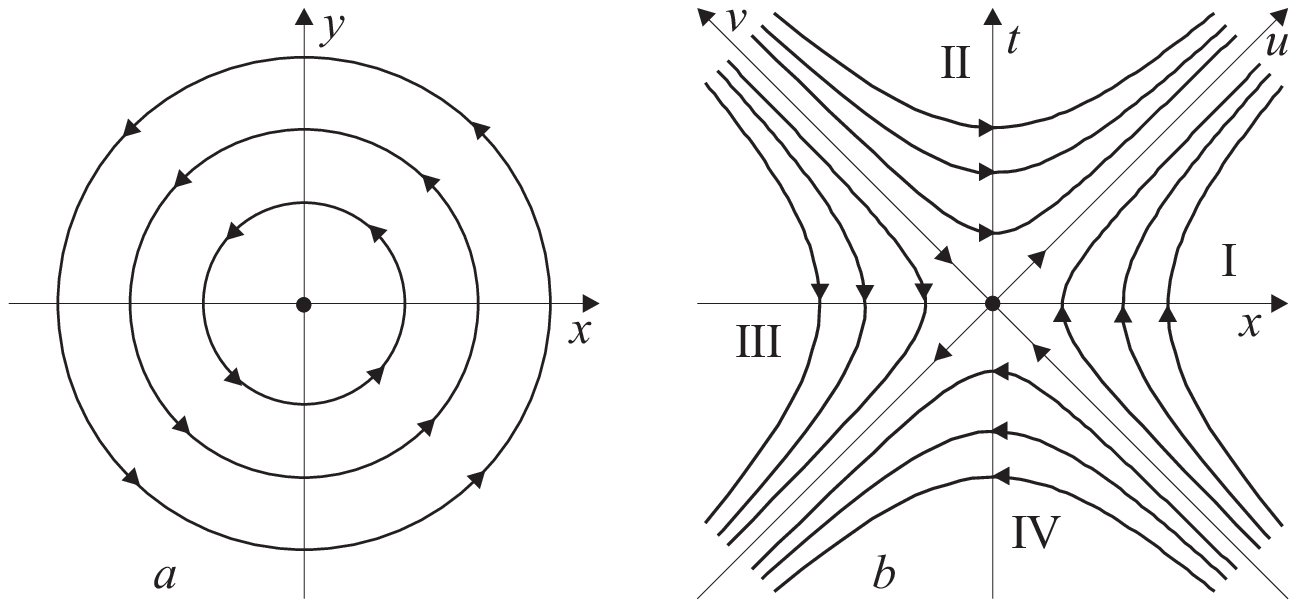}
\hfill {}
\centering \caption{\label{fplrot} Траектории Киллинга $=$ орбиты действия
 групп $\MS\MO(2)$ и $\MS\MO_\uparrow(1,1)$ на евклидовой плоскости
 {\it (a)} и плоскости Минковского {\it (b)}.}
\end{figure}

Вращения евклидовой плоскости показаны на рис.~\ref{fplrot},{\it a}.
Орбитами действия и связной компоненты единицы группы $\MS\MO(2)$, и полной
группы $\MO(2)$ являются концентрические окружности,
\begin{equation}                                                  \label{ecircl}
  x^2+y^2=r^2,\qquad r\in\MR_+,
\end{equation}
с центром в начале координат. Само начало координат является неподвижной точкой
относительно действия группы вращений.

Определим {\em полярные координаты} на евклидовой плоскости
\index{Полярные координаты (polar coordinates)}%
\index{Координаты полярные (polar coordinates)}%
\begin{equation}                                                  \label{epolco}
  x=r\cos\vf, \qquad y=r\sin\vf,
\end{equation}
где радиус точки $r$ и полярный угол $\vf$ определены в интервалах:
$r\in(0,\infty)$, $\vf\in[0,2\pi)$. Обратное преобразование координат имеет вид
\begin{equation}                                                  \label{epcint}
  r=\sqrt{x^2+y^2},\qquad \vf=\arctg\left(\frac yx\right).
\end{equation}
\begin{com}
Эти координаты особенно удобны в задачах, обладающих вращательной симметрией,
поскольку при вращении меняется только полярный угол, а радиус точки остается
прежним. Например, для собственных вращений
\begin{equation}                                                  \label{eropan}
  S_+(\om):\quad (r,\vf)~\mapsto~(r,\vf+\om). \qed
\end{equation}
\end{com}
Евклидов интервал в полярных координатах имеет нетривиальный вид
\begin{equation}                                                  \label{eumepc}
  ds^2=dr^2+r^2d\vf^2.
\end{equation}
Недостатком полярных координат является то, что они вырождены в начале
координат, поскольку для этой точки полярный угол (\ref{epcint}) не определен.
Соответственно, якобиан преобразования (\ref{epolco}),
$$
  J=\frac{\pl(x,y)}{\pl(r,\vf)}=r,
$$
равен нулю в начале координат. Во всех остальных точках преобразование координат
является взаимно однозначным и сохраняет ориентацию, поскольку якобиан
положителен.

Если исключить (выколоть) сингулярную точку, которой является начало координат,
то евклидова плоскость в полярных координатах представляет собой топологическое
произведение окружности $\MS^1$ на полубесконечный интервал $\MR_+=(0,\infty)$,
$$
  \MR^2\setminus\lbrace0\rbrace=\MS^1\times\MR_+.
$$

Рассмотрим два ненулевых вектора в начале координат, $X_1=(x_1,y_1)=(r_1,\vf_1)$
и $X_2=(x_2,y_2)=(r_2,\vf_2)$. Их длины равны $\|X_1\|=r_1$ и $\|X_2\|=r_2$.
Тогда {\em угол между векторами} $\vf_{12}$, по-определению, равен
\begin{equation}                                                  \label{eangvp}
  \cos\vf_{12}:=\frac{(X_1,X_2)}{\|X_1\|\|X_2\|},
\end{equation}
где круглые скобки обозначают скалярное произведение векторов
$(X_1,X_2):=x_1x_2+y_1y_2$. Это соотношение определяет угол между векторами с
точностью до знака, т.к.\ $\cos$ является четной функцией. Чтобы устранить
возникшую неоднозначность, будем считать угол $\vf_{12}$ между векторами $X_1$
и $X_2$ положительным, если вращение от $X_1$ к $X_2$ происходит против часовой
стрелки. По-определению, вращение от оси абсцисс к оси ординат по меньшему углу
называется {\em вращением против часовой стрелки}.
\index{Вращение против часовой стрелки}%
При вращении по часовой стрелке угол считается отрицательным. Тогда
\begin{equation}                                                  \label{eadang}
  \vf_{12}:=\vf_2-\vf_1.
\end{equation}
Отсюда следует, что углы между векторами складываются. То есть, если углы между
векторами $X_1$, $X_2$ и $X_2$, $X_3$ равны, соответственно, $\vf_{12}$ и
$\vf_{23}$, то угол между векторами $X_1$ и $X_3$ равен сумме углов
\begin{equation}                                                  \label{eangad}
  \vf_{13}:=\vf_{12}+\vf_{23}.
\end{equation}
Это правило сложения углов эквивалентно абелевости группы собственных вращений
$\MS\MO(2)$.

Рассмотрим бесконечно малые вращения евклидовой плоскости на угол
$|\dl\om|$$\ll1$. Разлагая матрицу (\ref{eormtr}) в ряд, в линейном приближении
получим следующие приращения координат:
\begin{equation}                                                  \label{ecroin}
\begin{split}
  \dl x &= -y\dl\om,
\\
  \dl y &=  x\dl\om.
\end{split}
\end{equation}
Забегая вперед, скажем, что, если евклидова плоскость рассматривается, как
многообразие, то генератором вращений является векторное поле Киллинга
(см.\ главу \ref{skilve})
$$
  K=-y\pl_x+x\pl_y.
$$
Изменение формы функции $f(x,y)$ (см.\ раздел \ref{sinfct}) в линейном
приближении можно выразить через вектор Киллинга:
$$
  \dl f(x,y):=-\dl x^\al\pl_\al f=-\dl\om Kf|_{(x,y)},
$$
где векторное поле действует на функцию как дифференцирование. В полярных
координатах, генератор вращений (векторное поле Киллинга) есть дифференцирование
по полярному углу
$$
  K=\pl_\vf.
$$
При этом интегральные кривые для генератора вращений (траектории Киллинга)
совпадают с орбитами действия группы вращений и являются концентрическими
окружностями (\ref{ecircl}) с центром в начале координат, как показано на
рис.~\ref{fplrot},{\it a}.

В комплексных координатах на евклидовой плоскости,
\begin{equation}                                                  \label{epofcn}
  z:=x+iy=r\ex^{i\vf},\qquad \bar z:=x-iy=r\ex^{-i\vf},
\end{equation}
где черта обозначает комплексное сопряжение, собственные и несобственные
вращения на угол $\om$ выглядят соответственно
\begin{equation}                                                  \label{eccprr}
  S_+(\om):\quad
  \begin{aligned}[t] z&\mapsto z'=z\ex^{i\om}, \\
  \bar z&\mapsto z'=\bar z\ex^{-i\om},
  \end{aligned} \hspace{10mm}
  S_-(\om):\quad
  \begin{aligned}[t] z&\mapsto z'=\bar z\ex^{-i\om}, \\
  \bar z&\mapsto z'=z\ex^{i\om}.
  \end{aligned}
\end{equation}
Отсюда следует, что связная компонента единицы $\MS\MO(2)$ изоморфна группе
$\MU(1)$ комплексных чисел, равных по модулю единице, с обычным правилом
умножения.

Приведем для справки ряд формул. Интервал на евклидовой плоскости в комплексных
координатах (\ref{epofcn}) имеет вид
$$
  ds^2=dzd\bar z.
$$
Операторы дифференцирования по декартовым и комплексным координатам связаны
между собой соотношениями:
\begin{equation}                                                  \label{epldcc}
\begin{aligned}
  \pl_z &=\frac12(\pl_x-i\pl_y), &   \qquad\pl_{\bar z} &=\frac12(\pl_x+i\pl_y),
\\
  \pl_x &=\pl_z+\pl_{\bar z},  & \pl_y &=i(\pl_z-\pl_{\bar z}).
\end{aligned}
\end{equation}
Отсюда следует выражение для лапласиана
\begin{equation}                                                  \label{elapop}
  \triangle=\pl^2_{xx}+\pl^2_{yy}=4\pl^2_{z\bar z}.
\end{equation}
Из формул преобразования комплексных координат (\ref{epldcc}) сразу следует,
что оператор Лапласа инвариантен относительно полной группы вращений $\MO(2)$.
\section{Группа двумерных преобразований Лоренца $\MO(1,1)$      \label{stwodl}}
\begin{defn}
{\em Плоскостью Минковского} $\MR^{1,1}$ называется евклидова плоскость с
декартовыми координатами $\lbrace x^\al\rbrace=\lbrace t,x\rbrace$, $\al=0,1$, в
которых задана двумерная метрика Лоренца
\begin{equation}                                                  \label{etdmim}
  \eta=\lbrace\eta_{\al\bt}\rbrace:=\diag(+-):=
  \begin{pmatrix} 1 & \quad 0 \\ 0 &-1\end{pmatrix}.
\end{equation}
В пространстве Минковского декартова система координат, в которой задана метрика
(\ref{etdmim}), называется  {\em инерциальной} системой отсчета. Метрике Лоренца
соответствует интервал
\begin{equation}                                                  \label{einmpl}
  ds^2=dx^\al dx^\bt \eta_{\al\bt}=dt^2-dx^2=dudv,
\end{equation}
где введены {\em координаты светового конуса} (см.\ рис.~\ref{fplrot},{\it b}):
\begin{equation}                                                  \label{econco}
  u=t+x,\qquad v=t-x,
\end{equation}
которые являются аналогом комплексных координат (\ref{epofcn}) на евклидовой
плоскости. Координатные оси $t$ и $x$ называют соответственно {\em временем} и
{\em пространством}.
\qed\end{defn}
\index{Плоскость Минковского (Minkowskian plane)}%
\index{Минковского плоскость (Minkowskian plane)}%
\index{Инерциальная система отсчета (inertial coordinate system)}%
\index{Система отсчета инерциальная (inertial coordinate system)}%
\index{Координаты светового конуса}\index{Светового конуса координаты}%
Якобиан преобразования координат (\ref{econco}) равен
$$
  J=\frac{\pl(u,v)}{\pl(x,t)}=2.
$$
Это преобразование сохраняет ориентацию, если координаты упорядочены следующим
образом: $x,t\mapsto u,v$. В координатах светового конуса линии, параллельные
осям координат, $u=\const$ и $v=\const$, имеют нулевую длину, т.е.\
светоподобны.
\begin{com}
На плоскости Минковского $\MR^{1,1}$ время и пространство играют равноправную
роль, поскольку пространство одномерно, и переобозначение $t\leftrightarrow x$
приводит только к изменению общего знака метрики.
\qed\end{com}

\begin{figure}[htb]
\hfill\includegraphics[width=.3\textwidth]{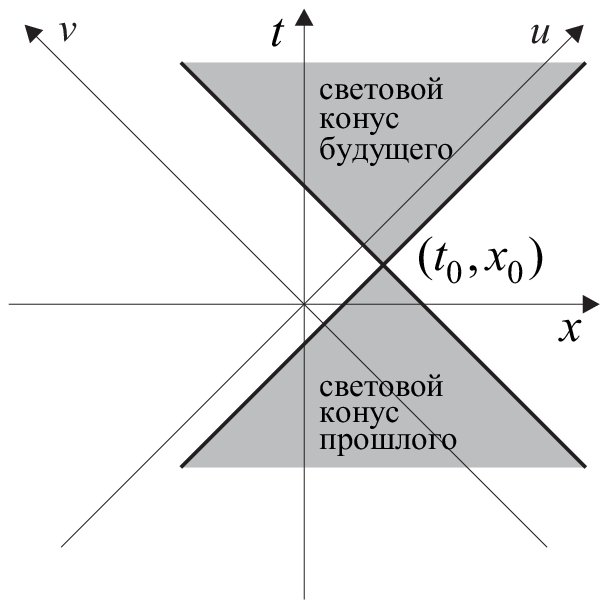}
\hfill {}
\\
\centering \caption{Световой конус на плоскости Минковского. \label{flicon}}
\end{figure}

Подчеркнем, что на плоскости Минковского $\MR^{1,1}$ задано две метрики:
евклидова метрика и метрика Минковского (\ref{etdmim}). Первая необходима
для задания топологии и определения дифференцируемых в обычном смысле функций на
$\MR^{1,1}$.
\begin{prop}
На плоскости Минковского любая изотропная кривая $\lbrace t(s),x(s)\rbrace$,
т.е.\ линия, касательный вектор к которой имеет нулевую длину:
\begin{equation}                                                  \label{enulmp}
  \left(\frac{dt}{ds}\right)^2-\left(\frac{dx}{ds}\right)^2=0,
\end{equation}
является прямой линией вида $t=\pm x+\const$.
\end{prop}
\begin{proof}
Из уравнения (\ref{enulmp}) следует, что $dt=\pm dx$. Общим решением данного
уравнения являются прямые линии $t=\pm x+\const$.
\end{proof}
\begin{defn}
Уравнение
$$
  (t-t_0)^2-(x-x_0)^2=0
$$
определяет на плоскости Минковского $\MR^{1,1}$ две перпендикулярные прямые,
пересекающиеся в точке $(t_0,x_0)$. Они называются {\em световым конусом} в
данной точке, рис.~\ref{flicon}. Световой конус состоит из двух связных
компонент: светового конуса {\em прошлого}, $t\le t_0$, и {\em будущего},
$t\ge t_0$. Световые конусы прошлого и будущего имеют общую вершину в точке
$(t_0,x_0)$.
\qed\end{defn}
\index{Световой конус (light cone)}\index{Конус световой (light cone)}%
Любая времениподобная кривая $\lbrace x^\al(s)\rbrace$,
$$
  \dot x^\al\dot x^\bt\eta_{\al\bt}>0,\qquad \dot x^\al:=\frac{dx^\al}{ds},
$$
проходящая через точку $(t_0,x_0)$, будет целиком лежать внутри светового конуса
(затемненная область на рисунке).
\begin{defn}
Невырожденные линейные однородные преобразования координат
\begin{equation}                                                  \label{elotrp}
  x^{\al}\mapsto x^{\prime\al}=x^\bt S_\bt{}^\al,\qquad \det S_\bt{}^\al\ne0,
\end{equation}
относительно которых метрика Минковского (\ref{etdmim}) остается инвариантной,
\begin{equation}                                                  \label{edeotp}
  \et_{\al\bt}=S_\al{}^\g S_\bt{}^\e\et_{\g\e},
\end{equation}
образуют группу, которая называется {\em группой двумерных преобразований
Лоренца} и обозначается $\MO(1,1)$.
\qed\end{defn}
Из уравнения (\ref{edeotp}) следует, что определитель матриц $S$, как и в случае
вращений евклидовой плоскости, по модулю равен единице, $\det S=\pm1$.
\begin{prop}
Любое решение уравнения (\ref{edeotp}) принадлежит одному из четырех классов:
\begin{align}                                                     \label{elorpu}
  S_{+\uparrow}(\om)&:=\begin{pmatrix} \ch\om & \sh\om \\ \sh\om & \ch\om
  \end{pmatrix}, & &\det S_{+\uparrow}=1,
\\                                                                \label{elormu}
  S_{-\uparrow}(\om)&:=\begin{pmatrix} \ch\om & -\sh\om \\ \sh\om & -\ch\om
  \end{pmatrix}, & &\det S_{-\uparrow}=-1,
\\                                                                \label{elorpd}
  S_{+\downarrow}(\om)&:=\begin{pmatrix} -\ch\om & -\sh\om
  \\ -\sh\om & -\ch\om \end{pmatrix}, & &\det S_{+\downarrow}=1,
\\                                                                \label{elormd}
  S_{-\downarrow}(\om)&:=\begin{pmatrix} -\ch\om & \sh\om \\ -\sh\om & \ch\om
  \end{pmatrix}, & &\det S_{-\downarrow}=-1,
\end{align}
при некотором $\om\in\MR$. Параметр $\om$ называется {\em гиперболическим} углом
вращения.
\end{prop}
\index{Угол вращения гиперболический (hyperbolic rotational angle)}%
\index{Гиперболический угол вращения (hyperbolic rotational angle)}%
\begin{proof}
Уравнение (\ref{etdmim}) -- это система трех квадратных уравнений:
\begin{equation*}
\begin{split}
  (S_1{}^1)^2-(S_1{}^2)^2&=1,
\\
  (S_2{}^1)^2-(S_2{}^2)^2&=-1,
\\
  S_1{}^1S_2{}^1-S_1{}^2S_2{}^2&=0.
\end{split}
\end{equation*}
Перебор всех возможных решений показывает, что с точностью до переопределения
угла поворота формулы (\ref{elorpu})--(\ref{elormd}) дают общее решение.
\end{proof}
Индекс $\pm$ у матрицы преобразований Лоренца соответствует знаку определителя,
а смысл стрелки будет ясен из дальнейшего рассмотрения.

Из доказанного утверждения следует, что двумерная группа Лоренца $\MO(1,1)$
является однопараметрической некомпактной группой Ли и состоит из четырех
несвязных компонент.

Заметим, что
\begin{equation*}
  S_{-\uparrow}^\St(\om)=S_{-\uparrow}(-\om)\qquad \text{и}\qquad
  S_{-\downarrow}^\St(\om)=S_{-\downarrow}(-\om).
\end{equation*}

Как и в случае группы вращений евклидовой плоскости $\MO(2)$, обозначим
множества матриц (\ref{elorpu})--(\ref{elormd}) теми же, но ажурными буквами.
Например,
\begin{equation*}
  \MS_{+\uparrow}:=\lbrace S_{+\uparrow}(\om):\quad \om\in\MR\rbrace.
\end{equation*}
Тогда выполнена следующая таблица умножения
\begin{equation}                                                  \label{eklegs}
    \begin{tabular}{|c|c|c|c|c|}                                          \hline
&$\MS_{+\uparrow}$&$\MS_{-\uparrow}$&$\MS_{-\downarrow}$&$\MS_{+\downarrow}$
 \\ \hline
$\MS_{+\uparrow}$&$\MS_{+\uparrow}$&$\MS_{-\uparrow}$&$\MS_{-\downarrow}$
&$\MS_{+\downarrow}$ \\ \hline
$\MS_{-\uparrow}$&$\MS_{-\uparrow}$&$\MS_{+\uparrow}$&$\MS_{+\downarrow}$
&$\MS_{-\downarrow}$ \\ \hline
$\MS_{-\downarrow}$&$\MS_{-\downarrow}$&$\MS_{+\downarrow}$&$\MS_{+\uparrow}$
&$\MS_{-\uparrow}$ \\ \hline
$\MS_{+\downarrow}$&$\MS_{+\downarrow}$&$\MS_{-\downarrow}$&$\MS_{-\uparrow}$
&$\MS_{+\uparrow}$ \\ \hline
    \end{tabular}
\end{equation}
Обратные матрицы из $\MS_{+\uparrow}$ имеют вид
\begin{equation*}
  S^{-1}_{+\uparrow}(\om)=S_{+\uparrow}(-\om),
\end{equation*}
т.е.\ получаются заменой $\om\mapsto-\om$.

Единичная матрица $\one=S_{+\uparrow}(0)$ и матрица $-\one=S_{+\downarrow}(0)$
принадлежат разным компонентам группы.

\begin{defn}
Введем оператор {\em обращения времени} и  {\em пространственного отражения
(четности)}
\begin{equation}                                                  \label{etirop}
\begin{split}
  T&:=\begin{pmatrix}-1&0\\0&1\end{pmatrix}=S_{-\downarrow}(0),\qquad \det T=-1,
\\
  P&:=\begin{pmatrix}1&0\\0&-1\end{pmatrix}=S_{-\uparrow}(0),\qquad \det P=-1.
\end{split}
\end{equation}
Действие этих операторов мы записываем в виде $x^\al\mapsto x^\bt T_\bt{}^\al$ и
$x^\al\mapsto x^\bt P_\bt{}^\al$. Они отражают соответственно временн\'ую и
пространственную координаты:
\begin{align*}
  T:\quad (t,x)~&\mapsto~(-t,x),
\\
  P:\quad (t,x)~&\mapsto~(t,-x).
\end{align*}
Введем также обозначение для их композиции
\begin{equation}                                                  \label{einvfo}
 R:=PT=TP=\begin{pmatrix}-1&0\\0&-1\end{pmatrix}=S_{+\uparrow}(0),\qquad
 \det R=1,
\end{equation}
которая соответствует полному отражению координат $R:~(t,x)\mapsto(-t,-x)$.
\qed\end{defn}
\index{Оператор четности (parity operator)}%
\index{Оператор обращения времени (time reversal operator)}%
\index{Обращение времени (time reversal)}%
Тогда матрицы преобразований Лоренца (\ref{elormu})--(\ref{elormd}) можно
представить в виде
\begin{equation}                                                 \label{eepldow}
\begin{split}
  S_{-\uparrow}&=S_{+\uparrow}P,
\\
  S_{-\downarrow}&=S_{+\uparrow}T,
\\
  S_{+\downarrow}&=S_{+\uparrow}R.
\end{split}
\end{equation}
Преобразования (\ref{elorpu}) и (\ref{elormu}) не меняют направления времени и
называются {\em ортохронными}, что отмечено направленной вверх стрелкой.
\index{Ортохронные преобразования (orthochronous transformations)}%
\index{Преобразования ортохронные (orthochronous transformations)}%
Эти преобразования образуют компоненты $S_{+\uparrow}$ и $S_{-\uparrow}$, для
которых $S_0{}^0>0$. Матрицы из компонент $S_{+\downarrow}$ и $S_{-\downarrow}$
не являются ортохронными, и для них $S_0{}^0<0$.

Из таблицы умножения (\ref{eklegs}) следует, что только одна из четырех связных
компонент $\MS_{+\uparrow}$ (связная компонента единицы) образует группу. Эта
группа обозначается
\begin{equation}                                                  \label{eprorl}
  \MS\MO_\uparrow(1,1)\approx\MS_{+\uparrow},\qquad \det S=1,\quad S_0{}^0>0,
\end{equation}
и называется {\em собственной ортохронной} группой Лоренца.
\index{Собственная ортохронная группа Лоренца%
 (proper orthochronous Lorentz group)}%
Кроме этого группа Лоренца $\MO(1,1)$ имеет еще три нетривиальные подгруппы,
состоящие из двух компонент, которые включают отражение времени $T$,
преобразование четности $P$ и полного отражения $R$. Введем для этих подгрупп
следующие обозначения:
\begin{align}                                                          \nonumber
  \MO_\downarrow(1,1)&\approx \MS_{+\uparrow}\cup \MS_{-\downarrow},
  && \det S=\pm1,
\\                                                                \label{esubso}
  \MO_\uparrow(1,1)&\approx \MS_{+\uparrow}\cup \MS_{-\uparrow},
  && \det S=\pm1,\quad S_0{}^0>0,
\\                                                                     \nonumber
  \MS\MO(1,1)&\approx \MS_{+\uparrow}\cup \MS_{+\downarrow},
  && \det S=\quad 1.
\end{align}
Матрицы подгрупп $\MS\MO(1,1)$ и $\MS\MO_\uparrow(1,1)$ имеют положительный
определитель. Легко проверить, что они являются абелевыми. Подгруппы
$\MO_\uparrow(1,1)$ и $\MO_\downarrow(1,1)$, а также полная группа Лоренца
$\MO(1,1)$ являются неабелевыми.
\begin{com}
Знание подгрупп полной группы Лоренца $\MO(1,1)$ важно для описания
представлений, в том числе спинорных.
\qed\end{com}
\begin{defn}
Если при пространственных отражениях скаляр, вектор и вообще произвольный
тензор меняет знак, то их принято называть  {\em псевдоскаляр},
{\em псевдовектор} и {\em псевдотензор}.
\qed\end{defn}
\index{Псевдоскаляр (pseudoscalar)}%
\index{Псевдовектор (pseudovector)}%
\index{Псевдотензор (pseudotensor)}%

Все подгруппы в полной группе Лоренца $\MO(1,1)$ являются нормальными. Поэтому
определена факторгруппа
\begin{equation*}
  \MK_4:=\frac{\MO(1,1)}{\MS\MO_\uparrow(1,1)},
\end{equation*}
состоящая из четырех элементов. Из выражений (\ref{eepldow}) следует, что в
качестве представителей смежных классов можно выбрать четыре матрицы
$\one,P,T,R$. Эти матрицы, как легко проверить, имеют следующую таблицу
умножения
\begin{equation}                                                  \label{eklegu}
    \begin{tabular}{|c|c|c|c|c|}                                          \hline
      & $\one$ & $P$ & $T$ & $R$ \\ \hline
    $\one$ & $\one$ &P& $T$ & $R$ \\ \hline
    $P$ & $P$ & $\one$ & $R$ & $T$ \\ \hline
    $T$ & $T$ & $R$ & $\one$ & $P$ \\ \hline
    $R$ & $R$ & $T$ & $P$ & $\one$ \\ \hline
    \end{tabular}~,
\end{equation}
которая с точностью до обозначений совпадает с (\ref{eklegs}).
\begin{defn}
Группа состоящая из четырех элементов, удовлетворяющих таблице умножения
(\ref{eklegu}), называется {\em 4-группой Клейна}.
\qed\end{defn}
\begin{com}
Группа Клейна не является циклической.
\qed\end{com}
\index{4-группа Клейна (Klein 4-group)}\index{Клейна 4-группа (Klein 4-group)}%

Элементы группы $\MS\MO_\uparrow(1,1)$ можно представить в виде
\begin{equation}                                                  \label{eirotl}
  S_{+\uparrow\al}{}^\bt=(\ex^{-\om\ve})_\al{}^\bt
  =\dl_\al^\bt\ch\om-\ve_\al{}^\bt\sh\om,
\end{equation}
где $\ve_{\al\bt}=\ve_\al{}^\g\et_{\g\bt}$, $\e_{01}=1$, -- полностью
антисимметричный тензор второго ранга в пространстве Минковского
(см.\ приложение \ref{scomat}).

В координатах светового конуса (\ref{econco}) преобразования из группы Лоренца
имеют вид:
\begin{equation}                                                  \label{elorof}
\begin{aligned}
  S_{+\uparrow}:\quad u\mapsto u'&=u\ex^{\om},  \\
             v\mapsto v'&=v\ex^{-\om}, \\
  S_{-\uparrow}:\quad u\mapsto u'&=v\ex^{-\om}, \\
             v\mapsto v'&=u\ex^{\om},
\end{aligned}\qquad
\begin{aligned}
  S_{+\downarrow}:\quad u\mapsto u'&=-u\ex^{\om}, \\
             v\mapsto v'&=-v\ex^{-\om}, \\
  S_{-\downarrow}:\quad u\mapsto u'&=-v\ex^{-\om}, \\
             v\mapsto v'&=-u\ex^{\om}.
\end{aligned}
\end{equation}

Матрицы (\ref{elorpu})--(\ref{elormd}) можно рассматривать как представление
группы Лоренца $\MO(1,1)$ в двумерном векторном пространстве. Оно неприводимо и
называется {\em фундаментальным} или {\em векторным} представлением.
\index{Фундаментальное представление $\MO(1,1)$%
 (fundamental representation of $\MO(1,1)$)}%
\index{Векторное представление $\MO(1,1)$%
 (vector representation of $\MO(1,1)$)}%
Относительно группы собственных ортохронных преобразований
$\MS\MO_\uparrow(1,1)$ и группы собственных преобразований $\MS\MO(1,1)$ это
представление является приводимым и распадается на два неприводимых одномерных
представления (\ref{elorof}).
\begin{com}
Все неприводимые представления собственных групп Лоренца $\MS\MO_\uparrow(1,1)$
и $\MS\MO(1,1)$ являются одномерными, поскольку группы абелевы. Это является
причиной того, что выбор конусных координат часто приводит к существенному
упрощению вычислений.
\qed\end{com}

Бесконечно малые собственные ортохронные лоренцевы вращения на угол
$\dl\om\ll1$ имеют вид
\begin{equation}                                                  \label{einlor}
  S_{+\uparrow\al}{}^\bt=\dl_\al{}^\bt+\dl\om L_\al{}^\bt,
\end{equation}
где $L_\al{}^\bt=-\ve_\al{}^\bt$ -- генераторы группы Лоренца в фундаментальном
представлении.

Гиперболические вращения плоскости Минковского при действии собственной
ортохронной группы Лоренца $\MS\MO_\uparrow(1,1)$ показаны на
рис.~\ref{fplrot},{\it b}. При этом плоскость разделяется на четыре квадранта
I--IV. Собственные ортохронные преобразования Лоренца преобразуют точки внутри
каждого квадранта. Орбитами этих точек являются гиперболы
\begin{equation}                                                  \label{ehyper}
  t^2-x^2=\pm r^2,\qquad r\in\MR_+.
\end{equation}
Здесь знак плюс соответствует II и IV, а минус -- I и III квадрантам. Кроме
того, орбитами группы $\MS\MO_\uparrow(1,1)$ являются четыре луча $t=\pm x$,
выходящих из начала координат $(0,0)$, которое является неподвижной точкой.
В каждом из квадрантов можно ввести гиперболическую полярную систему координат:
\begin{equation}                                                  \label{epcmip}
\begin{aligned}
    \text{I}:\quad t &=r\sh\vf,  \\
               x &=r\ch\vf,  \\
   \text{II}:\quad t &=r\ch\vf,  \\
               x &=r\sh\vf,
\end{aligned}\qquad
\begin{aligned}
    \text{III}:\quad t &=-r\sh\vf, \\
                 x &=-r\ch\vf, \\
   \text{IV}:\quad   t &=-r\ch\vf, \\
                 x &=-r\sh\vf,
\end{aligned}
\end{equation}
где $r\in(0,\infty)$, $\vf\in(-\infty,\infty)$ в каждом квадранте.
При этом ``началом'' координат, $r=0$, являются две перпендикулярные прямые
$$
  r=0~\leftrightarrow~t=\pm x,
$$
т.е.\ световой конус в начале координат. На рис.~\ref{fplrot},{\it b} стрелками
показано направление возрастания угла $\vf$ внутри квадрантов и возрастание
гиперболического угла вращения $\om$ на световом конусе.

Обратные преобразования координат, например, во II квадранте имеют вид
$$
  r=\sqrt{t^2-x^2},\qquad \vf=\arcth\left(\frac xt\right).
$$
Знаки для преобразования координат в различных квадрантах (\ref{epcmip})
подобраны таким образом, чтобы при лоренцевом преобразовании (\ref{elorpu})
полярный угол менялся по закону
\begin{equation}                                                  \label{eropal}
  \vf\quad \mapsto\quad \vf'=\vf+\om,
\end{equation}
при этом радиус точки остается прежним. Лоренцев интервал в гиперболических
полярных координатах отличается знаком в разных квадрантах:
\begin{equation}                                                  \label{eloinp}
\begin{split}
  \text{I,III}:\quad ds^2 &=-dr^2+r^2d\vf^2, \\
  \text{II,IV}:\quad ds^2 &= dr^2-r^2d\vf^2.
\end{split}
\end{equation}
Отсюда следует, что радиус точки играет роль пространственной координаты в I и
III квадранте и времениподобной координаты во II и IV квадранте. Угол, наоборот,
времениподобен в I и III квадранте и пространственноподобен во II и IV
квадранте. Якобиан преобразования (\ref{epcmip}) имеет различные знаки в разных
квадрантах:
$$
  \text{I,III}:\quad J=\frac{\pl(x,t)}{\pl(r,\vf)}=r,\qquad
  \text{II,IV}:\quad J=\frac{\pl(x,t)}{\pl(r,\vf)}=-r
$$
и вырожден на линиях $r=0$. Переход к гиперболическим полярным координатам
сохраняет ориентацию базиса в {\rm I} и {\rm III} квадрантах и меняет ориентацию
во {\rm II} и {\rm IV} квадрантах.

С топологической точки зрения внутренность каждого квадранта является прямым
произведением полубесконечного интервала и прямой
$$
  \Int(\text{I})=\MR_+\times\MR.
$$

Нетрудно выразить координаты светового конуса через полярные координаты.
Например, во втором квадранте
\begin{equation*}
  u=r\ex^\vf,\qquad v=r\ex^{-\vf},
\end{equation*}
что является аналогом полярной записи комплексных чисел (\ref{epofcn}).

При пространственном отражении, обращении времени и полном отражении квадранты
отображаются друг на друга следующим образом:
\begin{align*}
  P:\quad && {\rm I}&\leftrightarrow{\rm III}, & {\rm II}&\rightarrow{\rm II},
      & {\rm IV}&\rightarrow{\rm IV},
\\
  T:\quad && {\rm I}&\rightarrow{\rm I}, & {\rm II}&\leftrightarrow{\rm IV},
      & {\rm III}&\rightarrow{\rm III},
\\
  R:\quad && {\rm I}&\leftrightarrow{\rm III}, & {\rm II}&\leftrightarrow{\rm IV},
       &&
\end{align*}

Определим угол $\vf_{12}$ между векторами $X_1$ и $X_2$ из одного квадранта
следующим образом:
\begin{equation}                                                  \label{eangmp}
  \ch\vf_{12}:=\frac{|(X_1,X_2)|}{|X_1||X_2|}.
\end{equation}
Этот угол определен с точностью до знака. Будем считать угол $\vf_{12}$ между
векторами $X_1=\lbrace r_1,\vf_1\rbrace$ и $X_2=\lbrace r_2,\vf_2\rbrace$
положительным, если поворот от $X_1$ к $X_2$ происходит в сторону увеличения
полярного угла. Тогда
\begin{equation}                                                  \label{eadanl}
  \vf_{12}=\vf_2-\vf_1.
\end{equation}
Отсюда следует, что углы между векторами складываются (\ref{eangad}), что
соответствует абелевости собственной ортохронной группы Лоренца
$\MS\MO_\uparrow(1,1)$. Подчеркнем, что понятие угла вводится только для
векторов из одного квадранта.

Бесконечно малые гиперболические вращения плоскости Минковского $\MR^{1,1}$ на
угол $\dl\om\ll1$ можно получить, разлагая матрицу (\ref{elorpu}) в ряд Тейлора.
В линейном приближении получим следующие приращения координат:
\begin{equation}                                                  \label{ecroil}
  \dl t =x\dl\om,\qquad \dl x =t\dl\om.
\end{equation}
Отсюда следует, что генератором лоренцевых преобразований является векторное
поле Киллинга
$$
  K=x\pl_t+t\pl_x.
$$
В полярных координатах, т.е.\ внутри каждого квадранта, генератор вращений есть
дифференцирование по полярному углу:
$$
  K=\pl_\vf.
$$
При этом интегральные кривые для генератора вращений (траектории Киллинга)
совпадают с орбитами действия группы и представляют собой гиперболы
(\ref{ehyper}), как показано на рис.~\ref{fplrot},{\it b}.

В физике собственные ортохронные преобразования Лоренца (\ref{eirotl}) обычно
записывают для размерных координат $x^0=ct$, $x^1=x$, где $c$ -- скорость света.
Они имеют вид
\begin{equation}                                                  \label{eloust}
  t'=\frac{t-Vx/c^2}{\sqrt{1-V^2/c^2}},\qquad
  x'=\frac{x-Vt}{\sqrt{1-V^2/c^2}},
\end{equation}
где $V$ -- скорость движения штрихованной системы координат вдоль оси $x$. При
этом угол гиперболического поворота $\om$ выражается через скорость
относительного движения $V$ с помощью соотношения
\begin{equation}                                                  \label{dvdefi}
  \tanh\om=-\frac Vc.
\end{equation}
Поскольку $-1<\tanh\om<1$, то преобразования Лоренца определены для скоростей,
меньших скорости света $V<c$.
Поэтому из постулата ковариантности законов Природы относительно преобразований
Лоренца вытекает, что скорость света является предельной скоростью для частиц и
волн. Это предположение является постулатом специальной теории относительности и
находится в прекрасном согласии с известными экспериментальными данными.

\begin{figure}[htb]
\hfill\includegraphics[width=.25\textwidth]{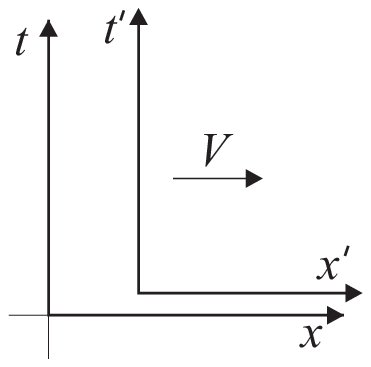}
\hfill {}
\centering \caption{Физическая интерпретация преобразований Лоренца.
 Штрихованная инерциальная система отсчета движется равномерно и прямолинейно
 относительно нештрихованной со скоростью $V$. \label{flotrf}}
\end{figure}

Преобразования Лоренца (\ref{eloust}) имеют следующую физическую интерпретацию.
Дополним двумерное пространство Минковского до четырехмерного, введя
дополнительные декартовы координаты $y,z$. Предположим, что дополнительные
координаты не меняются
\begin{equation*}
  y'=y,\qquad z'=z.
\end{equation*}
Тогда координаты $t,x,y,z$ и $t',x',y',z'$ являются координатами одного и того
же события в двух инерциальных системах отсчета c параллельными осями координат
(см.~рис.\ref{flotrf}). При этом штрихованная система координат движется
равномерно и прямолинейно со скоростью $V$ относительно нештрихованной системы
вдоль оси $x$. Системы координат выбраны таким образом, что их начала совпадают:
\begin{equation*}
  t=0,~x=0\qquad \Leftrightarrow\qquad t'=0,~x'=0.
\end{equation*}

Преобразования Лоренца (\ref{eloust}) при $c\rightarrow\infty$ переходят в
преобразования Галилея
\begin{equation}                                                  \label{egalto}
  x'=x-vt,\qquad t'=t.
\end{equation}
Эти преобразования хорошо знакомы из курса классической механики, где время
является абсолютным и не зависит от выбора системы координат.

Вернемся к двумерному пространству Минковского. Поскольку в приложениях часто
используются конусные координаты (\ref{econco}), в заключение настоящего раздела
приведем для справок некоторые формулы:
\begin{equation}                                                  \label{einvcc}
  t=\frac12(u+v),\qquad x=\frac12(u-v),
\end{equation}
\begin{equation}                                                  \label{eccdif}
\begin{aligned}
  \pl_u&=\frac12(\pl_t+\pl_x),\\
  \pl_t&=\pl_u+\pl_v,
\end{aligned}\qquad
\begin{aligned}
  \pl_v&=\frac12(\pl_t-\pl_x),\\
  \pl_x&=\pl_u-\pl_v.
\end{aligned}
\end{equation}
Оператор Даламбера на плоскости Минковского имеет вид
\begin{equation}                                                  \label{edalot}
  \square:=\pl^2_{tt}-\pl^2_{xx}=4\pl^2_{uv}.
\end{equation}
Он инвариантен относительно действия полной группы Лоренца $\MO(1,1)$.
\section{Трехмерное евклидово пространство                       \label{sthecs}}
Рассмотрим трехмерное евклидово пространство $\MR^3$ с декартовыми координатами
$\lbrace x^i\rbrace$$=\lbrace x,y,z\rbrace$, $i=1,2,3$.
\begin{com}
В настоящем разделе для обозначения индексов используются буквы из середины
латинского алфавита $i,j,k,\dotsc$. Такое изменение обозначений связано с тем,
что в дальнейшем координаты в пространстве Минковского $\MR^{1,3}$ часто будут
обозначаться через $\lbrace x^a\rbrace=\lbrace x^0,x^i\rbrace$, где явно
выделено время $x^0=t$, $a=0,1,2,3$ и $i=1,2,3$. В обозначениях мы используем
следующее правило. Множество чисел $\lbrace 1,2,3\rbrace$ является подмножеством
чисел $\lbrace 0,1,2,3\rbrace$, так же как и буквы $\lbrace i,j,\dotsc\rbrace$
являются подмножеством всего алфавита $a,b,\dotsc$.
\qed\end{com}

Дифференциально геометрическая евклидова метрика в декартовой системе координат
задается единичной матрицей
\begin{equation}                                                  \label{eclmth}
  ds^2=\dl_{ij}dx^i dx^j=dx^2+dy^2+dz^2.
\end{equation}
\begin{defn}
Группа линейных неоднородных преобразований декартовых координат, оставляющая
метрику (\ref{eclmth}) инвариантной, называется {\em неоднородной группой
вращений} и обозначается $\MI\MO(3)$.
В общем случае преобразование из неоднородной группы вращений имеет вид
$$
  x^i\mapsto x^{\prime i}=x^j S_j{}^i+a^i,
$$
где $S_j{}^i\in\MO(3)$ -- вещественная ортогональная матрица, определяемая
условием
\begin{equation}                                                  \label{eormat}
  S^\St S=SS^\St=1,
\end{equation}
и $a^i\in\MR^3$ -- произвольный вектор. Матрицы $\MO(3)$ образуют группу,
которая называется {\em группой трехмерных вращений}. Абелева подгруппа в
$\MI\MO(3)$, параметризуемая вектором $a$, называется {\em группой трансляций}.
\qed\end{defn}
\index{Неоднородная группа вращений (inhomogeneous rotation group)}%
\index{Группа вращений неоднородная (inhomogeneous rotation group)}%
\index{Группа трансляций (translation group)}%
\index{Трансляций группа (translation group)}%
Поскольку группа вращений сохраняет метрику (\ref{eclmth}), то при поворотах и
трансляциях евклидова пространства сохраняются длины и углы между векторами.
Конечно, трансляции можно было бы ввести и ранее в двумерном случае.

Можно показать, что любое преобразование декартовых координат, оставляющее
метрику (\ref{eclmth}) инвариантной, является линейным. Поэтому группа
$\MI\MO(3)$ является максимальной группой преобразований симметрии евклидова
пространства $\MR^3$ с заданной метрикой. Группа неоднородных вращений, как
будет показано в дальнейшем, является группой Ли.

Группа $\MI\MO(3)$ представляет собой полупрямое произведение группы вращений
$\MO(3)$ вокруг начала координат на абелеву группу трансляций. Группа
трансляций действует в $\MR^3$ свободно и транзитивно. Как многообразие она
диффеоморфна евклидову пространству $\MR^3$. Группа вращений $\MO(3)$ действует
в $\MR^3$ эффективно. При этом начало координат является неподвижной точкой.
Многообразие группы вращений будет описано немного позже.

Из условия ортогональности (\ref{eormat}) следует, что определитель
ортогональной матрицы равен по модулю единице
$$
  \det S=\pm1.
$$
Группа трехмерных вращений $\MO(3)$ является неабелевой и состоит из двух
компонент. Связная компонента единицы обозначается $\MS\MO(3)$ и называется
группой {\em собственных} вращений. К ней относятся ортогональные матрицы с
единичным определителем. Вторая связная компонента состоит из ортогональных
матриц с отрицательным определителем и сама по себе группы не образует.
\index{Группа собственных вращений (proper rotation group)}%
\index{Собственных вращений группа (proper rotation group)}%
\begin{defn}
{\em Оператором пространственных отражений} называется матрица
\begin{equation}                                                  \label{espinb}
  P:=\begin{pmatrix}-1 &\quad 0 &\quad 0 \\\quad 0 & -1 &\quad 0 \\\quad 0
  &\quad 0 & -1\end{pmatrix}
  \qquad \text{или}\qquad P_j{}^i=-\dl_j^i,
\end{equation}
при действии которой все координаты меняют знак $x^i\mapsto x^jP_j{}^i=-x^i$.
\qed\end{defn}
\index{Оператором пространственных отражений (space reflection operator)}%
Произвольную ортогональную матрицу с отрицательным определителем можно
однозначно представить в виде произведения собственно ортогональной матрицы и
оператора пространственных отражений. Это значит, что множество ортогональных
матриц с отрицательным определителем представляет собой смежный класс матрицы
отражений по подгруппе $\MS\MO(3)$.

Связная компонента единицы $\MS\MO(3)$ является нормальной подгруппой в группе
вращений $\MO(3)$. Поэтому определена фактор группа
\begin{equation*}
  \MZ_2=\frac{\MO(3)}{\MS\MO(3)},
\end{equation*}
состоящая из двух элементов $\MZ_2=\lbrace \one,P\rbrace$. Отображение
\begin{equation*}
  \MO(3)\ni\quad S\mapsto\det S\quad\in\MZ_2
\end{equation*}
есть гомоморфизм группы вращений на мультипликативную группу $\MZ_2$. Как
многообразие полная группа вращений гомеоморфна прямому произведению:
$\MO(3)\approx\MZ_2\times\MS\MO(3)$.

Группа $\MS\MO(3)$ проста (т.е.\ не содержит инвариантных абелевых подгрупп),
и, значит, нетривиальные гомоморфизмы на другие группы Ли невозможны (см.\
раздел \ref{slieho}).

\begin{figure}[h,b,t]
\hfill\includegraphics[width=.4\textwidth]{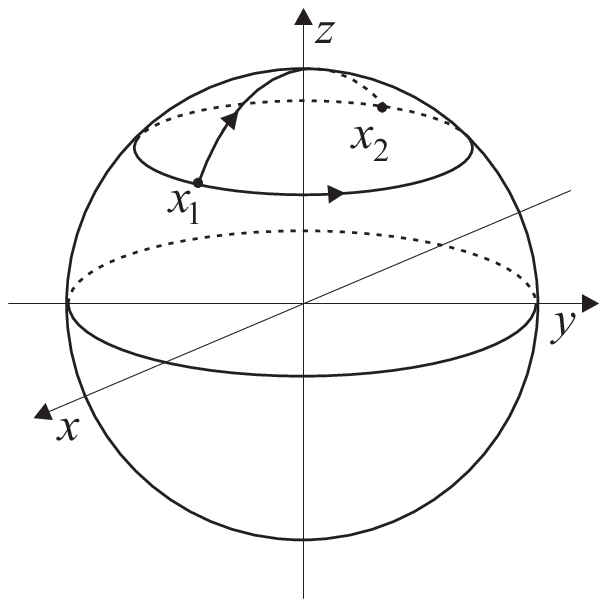}
\hfill {}
\centering\caption{Отображение двух точек на сфере $x_1\mapsto x_2$ не
определяет трехмерное вращение однозначно.}
\label{fxonxt}
\end{figure}

При вращениях произвольная точка $x\in\MR^3$, не совпадающая с началом
координат, пробегает все точки сферы $\MS^2$ с центром в начале координат и
содержащей точку $x\in\MS^2$. При этом сфера является орбитой точки
$x\in\MR^3$ при действии групп $\MS\MO(3)$ и $\MO(3)$. Отметим, что две
точки сферы $x_1,x_2\in\MS^2$ не определяют вращение однозначно (см.\
рис.~\ref{fxonxt}). Действительно, пусть точки $x_1$ и $x_2$ лежат в
плоскости $x,z$ и симметричны относительно оси $z$. Тогда точку $x_2$
можно получить из точки $x_1$ вращением вокруг оси $z$ на угол $\pm\pi$.
Этого же можно добиться вращением вокруг оси $y$ на меньший угол, причем
траекторией точки $x_1$ будет часть большой окружности, соединяющей точки
$x_1$ и $x_2$. В общем случае осью вращения может служить любой единичный
вектор, лежащий в плоскости, перпендикулярной отрезку, соединяющему точки
$x_1$ и $x_2$ и проходящей через начало координат. В рассматриваемом случае
ось вращения должна лежать в плоскости $y,z$. Если на сфере задать перемещение
не двух, а всех точек одновременно, тогда вращение будет определено однозначно.

Генераторы группы вращений $\MS\MO(3)$ могут быть представлены в евклидовом
пространстве в виде дифференциальных операторов (векторных полей Киллинга)
\begin{equation}                                                  \label{esoroe}
  L_{ij}=x_i\pl_j-x_j\pl_i.
\end{equation}
Прямое вычисление коммутатора приводит к следующей алгебре Ли
\begin{equation}                                                  \label{esoall}
  [L_{ij},L_{kl}]=-\dl_{ik}L_{jl}+\dl_{il}L_{jk}+\dl_{jk}L_{il}-\dl_{jl}L_{ik}.
\end{equation}
Отсюда следует, что группа трехмерных вращений $\MS\MO(3)$ является неабелевой
группой Ли. Эта группа трехмерна, т.к.\, ввиду антисимметрии по индексам,
существуют только три независимых вектора: $L_{12}$, $L_{13}$ и $L_{23}$.
Алгебру Ли (\ref{esoall}) можно переписать в более компактном виде для
эквивалентного набора генераторов
\begin{equation}                                                  \label{ealsor}
  J_i:=\frac12\ve_{ijk}L^{jk},\qquad L_{ij}=\ve_{ijk}J^k,
\end{equation}
где $\ve_{ijk}$ -- полностью антисимметричный тензор третьего ранга,
$\ve_{123}=1$, (см.\ Приложение \ref{scomat}), а подъем и опускание индексов
производится с помощью евклидовой метрики. Тогда алгебра Ли группы $\MS\MO(3)$
принимает вид
\begin{equation}                                                  \label{ealatr}
  [J_i,J_j]=-\ve_{ijk}J^k.
\end{equation}
Алгебры Ли групп Ли $\MS\MO(3)$ и $\MO(3)$, естественно, совпадают.

При проведении расчетов с группой $\MO(3)$ удобно использовать явную
параметризацию элементов группы элементами ее алгебры. Элемент алгебры
$\Gs\Go(3)$ может быть представлен в виде произвольной антисимметричной
$3\times3$-матрицы
\begin{equation}                                                  \label{epeans}
  (\om\ve)_i{}^j=(\om^k \ve_k)_i{}^j=\om^k\ve_{ki}{}^j\quad \in\Gs\Go(3),
\end{equation}
где $\boldsymbol{\om}=\lbrace\om^k\rbrace\in\MR^3$ -- {\em вектор вращения}.
Здесь первый индекс $k$ нумерует базис алгебры, а индексы $i,j$ рассматриваются,
как матричные. Для сравнения с (\ref{ealatr}) заметим, что полностью
антисимметричный тензор третьего ранга можно рассматривать в качестве базиса
алгебры $\Gs\Go(3)$:
\index{Вектор вращения (rotational vector)}%
\begin{equation*}
  (J_k)_i{}^j:=\ve_{ki}{}^j.
\end{equation*}
Элемент алгебры параметризуется трехмерным вектором $\boldsymbol{\om}\in\MR^3$,
и, значит, группа $\MO(3)$ является трехмерной группой Ли.
\begin{prop}
Каждому элементу алгебры Ли (вектору вращения)
$\boldsymbol{\om}=\lbrace\om^i\rbrace\in\Gs\Go(3)$ соответствует элемент группы
Ли $S=\lbrace S_i{}^j\rbrace\in\MS\MO(3)$ (связной компоненты единицы):
\begin{equation}                                                  \label{elsogr}
  S_i{}^j(\boldsymbol{\om})=(\ex^{(\om\ve)})_i{}^j=\dl_i^j\cos\om
  +\frac{(\om\ve)_i{}^j}\om\sin\om
  +\frac{\om_i\om^j}{\om^2}(1-\cos\om)~          \in\MS\MO(3),
\end{equation}
где $\om:=\sqrt{\om^i\om_i}$ -- модуль вектора вращения $\boldsymbol{\om}$.
\end{prop}
\begin{proof}
Последнее равенство в (\ref{elsogr}) доказывается путем разложения обеих частей
равенства в ряды, которые равномерно сходятся для всех $\om<\infty$. С помощью
прямых вычислений нетрудно убедиться, что $S_i{}^j$ является действительно
ортогональной матрицей.
\end{proof}

Можно также доказать обратное утверждение, что любая ортогональная матрица с
единичным определителем имеет вид (\ref{elsogr}) для некоторого вектора
$\boldsymbol{\om}$.
\begin{com}
В отличие от элемента алгебры, элемент группы имеет как симметричную, так и
антисимметричную части.
\qed\end{com}
Элемент группы $\MS\MO(3)$ инвариантен относительно сдвига вектора вращения
$\om^i\mapsto\om^i+2\pi\om^i/\om$:
\begin{equation*}
  S_i{}^j\left(\om^k+2\pi\frac{\om^k}\om\right)=S_i{}^j(\om^k),
\end{equation*}
и это единственная инвариантность. При сдвиге вектора $\boldsymbol{\om}$
меняется только его длина $\om\mapsto\om+2\pi$, а направление остается
неизменным. В инвариантности матрицы вращений нетрудно убедиться, если заметить,
что отношение $\om^i/\om$, определяющее направление вектора $\boldsymbol{\om}$,
не меняется при произвольном сдвиге:
\begin{equation*}
  \frac{\om^i}\om\mapsto\frac{\om^i+c\om^i/\om}{\om+c}=\frac{\om^i}\om,
  \qquad c\in\MR.
\end{equation*}
Тем самым элемент группы вращений параметризуется точками евклидова пространства
$\boldsymbol{\om}\in\MR^3$ с единственным отношением эквивалентности
\begin{equation}                                                  \label{eqrsog}
  \om^i\sim\om^i+2\pi\frac{\om^i}\om.
\end{equation}
При этом неопределенность в нуле раскрывается по радиальным направлениям
$\om^i=\e k^i$, $\e\to0$. То есть начало координат отождествляется со всеми
сферами радиуса $2\pi m$, $m=1,2,\dotsc$

Вектор вращения $\boldsymbol{\om}$ параметризует группу $\MS\MO(3)$ следующим
образом. Направление вектора $\boldsymbol{\om}$ определяет ось вращения, а
модуль вектора $\om$ равен углу поворота. Таким образом, каждый элемент группы
$\MS\MO(3)$ отождествляется с точкой трехмерного шара $\MB^3_\pi\subset\MR^3$
радиуса $\pi$ с центром в начале координат. При этом различным внутренним точкам
шара соответствуют различные вращения, а диаметрально противоположные точки
граничной сферы $\MS^2_\pi$ необходимо отождествить, т.к.\ повороты вокруг
фиксированной оси на углы $\pi$ и $-\pi$ приводят к одному и тому же результату.

Таким образом, мы построили экспоненциальное отображение (\ref{elsogr}) алгебры
Ли в группу Ли:
\begin{equation*}
  \Gs\Go(3)\ni\qquad\boldsymbol{\om}\mapsto S\qquad\in\MS\MO(3).
\end{equation*}
Это отображение, конечно, не является взаимно однозначным. Оно также не является
накрытием. Действительно, прообразом каждого нетривиального вращения вокруг
некоторой оси является счетное число точек, лежащих на этой оси на расстоянии
$2\pi$ друг от друга. В то же время прообразом единицы группы $e\in\MS\MO(3)$
является начало координат в $\MR^3\approx\Gs\Go(3)$ и все сферы $\MS^2_{2\pi m}$
радиусов $2\pi m$, $m=1,2,\dotsc$.

Полная группа трехмерных вращений $\MO(3)$ состоит из двух связных компонент:
ортогональных матриц $S_+$ с положительным и $S_-$ отрицательным определителем.
Элементы полной группы вращений $\MO(3)$ параметризуются элементом алгебры
(\ref{epeans}) следующим образом:
\begin{equation}                                                  \label{elsogt}
  S_{\pm i}{}^j=\pm S_i{}^j~\in\MO(3).
\end{equation}
То есть матрицы $S_-$ получаются из матриц связной компоненты единицы умножением
на оператор пространственного отражения $P:=-\one$. При этом каждому элементу
алгебры $\boldsymbol{\om}\in\Gs\Go(3)$ соответствуют два элемента группы
$S_{\pm}(\om)\in\MO(3)$ -- по одному из каждой компоненты.

Обратные матрицы собственных вращений имеют вид
\begin{equation}                                                  \label{elsogi}
  S^{-1}_{\quad i}{}^j(\boldsymbol{\om})=S_i{}^j(-\boldsymbol{\om})=\dl_i^j\cos\om
  -\frac{(\om\ve)_i{}^j}\om\sin\om
  +\frac{\om_i\om^j}{\om^2}(1-\cos\om)~\in\MS\MO(3),
\end{equation}
т.е.\ соответствуют противоположному вектору вращения
$-\boldsymbol{\om}\in\MR^3$. Другими словами, обратный элемент группы
$\MS\MO(3)$ представляет собой поворот евклидова пространства вокруг той же оси,
но на противоположный угол.

Свернем матрицы вращений $S_\pm$ с вектором $\lbrace\om_j\rbrace$
\begin{equation*}
  S_{\pm i}{}^j\om_j=\pm\om_i.
\end{equation*}
Это значит, что вектор $\boldsymbol{\om}$ является собственным вектором
матриц вращений с собственными значениями $\pm1$. Другими словами, вращения
$S_+$ и $S_-$ оставляют, соответственно, ось вращения без изменения и меняют ее
направление на противоположное.

Допустим, что элемент алгебры $\Gs\Go(3)$ (вектор поворота) дифференцируемо
зависит от точки некоторого многообразия $\MM$, т.е.\
\begin{equation*}
  \MM\ni\quad x\mapsto \boldsymbol{\om}(x)\quad\in\Gs\Go(3).
\end{equation*}
Каждому вектору поворота по формуле (\ref{elsogr}) ставится в соответствие
матрица вращений $S\in\MS\MO(3)$. Введем обозначение
\begin{equation}                                                  \label{ealroe}
  l_{\al i}{}^j=(\pl_\al S_\pm^{-1}S_\pm)_i{}^j
  =(\pl_\al S^{-1}S)_i{}^j,
\end{equation}
где $\pl_\al:=\pl/\pl x^\al$ -- частная производная в локальной системе
координат. Последнее равенство следует из (\ref{elsogt}). С помощью прямых
вычислений можно убедиться в справедливости следующей формулы
\begin{multline}                                                  \label{ealgrs}
  l_{\al i}{}^j=-\frac{(\pl_\al\om\ve)_i{}^j}\om\sin\om
  -\frac{\pl_\al\om(\om\ve)_i{}^j}\om\left(1-\frac{\sin\om}\om\right)+
\\
  +\frac{\pl_\al\om_i\om^j-\om_i\pl_\al\om^j}{\om^2}(1-\cos\om)
  \quad \in\Gs\Go(3).
\end{multline}
Эта матрица антисимметрична по своим индексам, и, значит, является элементом
алгебры $\Gs\Go(3)$. Она представляет собой тривиальную $\MS\MO(3)$-связность,
для которой тензор кривизны тождественно равен нулю (чистая калибровка
(\ref{elofla})). Если в качестве многообразия $\MM$ выбрать многообразие самой
группы вращений $\MS\MO(3)$, то 1-форма $dx^\al l_{\al i}{}^j$ будет являться
канонической 1-формой на $\MS\MO(3)$ (см.\ раздел \ref{sleacg}).

Построим гомоморфизм группы двумерных унитарных матриц $\MS\MU(2)$ на
группу собственных трехмерных вращений $\MS\MO(3)$.  Алгебра Ли $\Gs\Gu(2)$
состоит из антиэрмитовых матриц $A=-A^\dagger$ с нулевым следом $\tr A=0$.
Элемент алгебры $\Gs\Gu(2)$ имеет вид
\begin{equation}                                                  \label{oantim}
  r^k\frac i2(\s_k)_\Sa{}^\Sb~\in\Gs\Gu(2),\qquad
  \boldsymbol{r}=\lbrace r^k\rbrace\in\MR^3,
\end{equation}
где $\s_k$ -- матрицы Паули (см.\ приложение \ref{spamat}), и $\Sa,\Sb=1,2$ --
матричные индексы. Множитель $i/2$ выбран с тем, чтобы коммутаторы векторов
базиса $\frac i2\s_k$ имели вид (\ref{ealatr}) как и для алгебры $\Gs\Go(3)$.
Это доказывает, что алгебры $\Gs\Gu(2)$ и $\Gs\Go(3)$ изоморфны. Задать этот
изоморфизм можно в явном виде. Для этого заметим, что произвольную антиэрмитову
матрицу можно взаимно однозначно представить виде (\ref{oantim}):
\begin{equation}                                                  \label{ematap}
  A=\frac i2\begin{pmatrix}r^3 & r^1-ir^2 \\ r^1+ir^2 & -r^3 \end{pmatrix}.
\end{equation}
\begin{prop}
Отображение
\begin{equation}                                                  \label{ososol}
  \vf:\quad \Gs\Gu(2)\ni\qquad A\mapsto\vf(A):=(\om\e)\qquad\in\Gs\Go(3),
\end{equation}
где $\boldsymbol{\om}=\boldsymbol{r}$ из представления (\ref{ematap}) и элемент
алгебры $(\om\e)$ определен формулой (\ref{epeans}), является изоморфизмом
алгебр $\Gs\Gu(2)\simeq\Gs\Go(3)$.
\end{prop}
\begin{proof}
Взаимная однозначность и линейность отображения $\vf$ очевидны. Сохранение
коммутатора при отображении,
\begin{equation*}
  \vf\big([A,B]\big)=\big[\vf(A),\vf(B)\big],\qquad A,B\in\Gs\Gu(2),
\end{equation*}
проверяется прямой проверкой. Это также следует из того, что коммутаторы
базисных векторов в алгебрах совпадают.
\end{proof}
Очевидно, что отображение $\vf$ является гладким. Заметим также, что
$$
  \det A=\frac14\big((\om^1)^2+(\om^2)^2+(\om^3)^2\big).
$$
Пусть $U\in\MS\MU(2)$ -- произвольная унитарная матрица,
$UU^\dagger=U^\dagger U=1$, с единичным определителем, $\det U=1$. Построим
новую матрицу
$$
  A'=UAU^\dagger.
$$
Эта матрица также антиэрмитова, $A'=-A^{\prime\dagger}$, и ее след равен нулю,
$\tr A'=0$. Поэтому матрицу $A'$ также можно представить в виде
$$
  A'=\begin{pmatrix}r^{\prime3} & r^{\prime1}-ir^{\prime2}
  \\ r^{\prime1}+ir^{\prime2} & -r^{\prime3} \end{pmatrix}.
$$
Поскольку $\det A'=\det A$, то
$$
  (\om^1)^2+(\om^2)^2+(\om^3)^2=(\om^{\prime1})^2+(\om^{\prime2})^2
  +(\om^{\prime3})^2
$$
Следовательно, вектор вращения $\boldsymbol{\om}'$ получаются из вектора
$\boldsymbol{\om}$ с помощью
некоторой ортогональной матрицы. Поскольку группа $\MS\MU(2)$ связна, то из
непрерывности следует, что каждой унитарной матрице $U\in\MS\MU(2)$ однозначно
ставится в соответствие некоторая ортогональная матрица $S\in\MS\MO(3)$ из
связной компоненты единицы. Полученное отображение гомоморфно, т.к.\ групповые
операции, как нетрудно проверить, согласованы. Ядро гомоморфизма состоит из двух
элементов $\lbrace\one,-\one\rbrace$, где $\one$ -- двумерная единичная матрица.
Это значит, что каждой ортогональной матрице $S\in\MS\MO(3)$ соответствуют две
унитарные матрицы $U$ и $-U$ группы $\MS\MU(2)$. Тем самым отображение
$\MS\MU(2)\rightarrow\MS\MO(3)$ является двулистным накрытием, и это накрытие
универсально, т.к.\ многообразие группы $\MS\MU(2)$ связно и односвязно (см.\
раздел \ref{scover}). Таким образом, справедлива
\begin{theorem}                                                   \label{tsusot}
Существует гомоморфизм $\MS\MU(2)\rightarrow\MS\MO(3)$, при котором группа
$\MS\MU(2)$ является двулистным универсальным накрытием группы $\MS\MO(3)$.
\end{theorem}

Произвольную унитарную матрицу с единичным определителем $U\in\MS\MU(2)$ можно
параметризовать двумя комплексными числами $a,b\in\MC$:
\begin{equation}                                                  \label{emattw}
  U=\begin{pmatrix} \quad a & b \\-\bar b & \bar a \end{pmatrix},
\end{equation}
которые удовлетворяют условию $|a|^2+|b|^2=1$. Это значит, что элементы группы
$\MS\MU(2)$ находятся во взаимно однозначном соответствии с точками трехмерной
сферы $\MS^3$, вложенной в четырехмерное евклидово пространство $\MR^4$, т.е.\
$$
  \MS\MU(2)\approx\MS^3.
$$
Другими словами, трехмерную сферу можно оснастить групповой структурой. При
гомоморфизме $\MS\MU(2)\rightarrow\MS\MO(3)$ два элемента $U$ и $-U$
отображаются в один элемент группы вращений, и этим элементам соответствуют две
противоположные точки на сфере $\MS^3$. Это значит, что с топологической точки
зрения группа $\MS\MO(3)$ представляет собой сферу $\MS^3$ с отождествленными
диаметрально противоположными точками, т.е.\ проективную плоскость
$$
  \MS\MO(3)\approx\MR\MP^3=\frac{\MS^3}{\MZ_2}.
$$

Важно отметить, что из каждой пары элементов $U$ и $-U$ невозможно выбрать по
одному элементу группы $\MS\MU(2)$ так, чтобы эти элементы образовали бы группу
сами по себе. Это связано с тем, что при вращении на угол $2\pi t$, где
$t\in[0,1]$, при $t=1$ мы имеем тождественное преобразование в группе
$\MS\MO(3)$, а матрицы $U\in\MS\MU(2)$ меняют свой знак $U\rightarrow-U$.
Другими словами, когда параметр $t$ пробегает единичный отрезок мы имеем
замкнутый путь в $\MS\MO(3)$ и отрезок в группе $\MS\MU(2)$. Это доказывает, в
частности, что группа $\MS\MO(3)$ не является подгруппой в $\MS\MU(2)$.

Гомоморфизм $\MS\MU(2)\rightarrow\MS\MO(3)$ имеет следующую геометрическую
интерпретацию. Каждое вращение трехмерного евклидова пространства $\MR^3$
порождает вращение сферы $\MS^2$ с центром в начале координат. Верно и
обратное утверждение: каждое вращение сферы однозначно определяет вращение
евклидова пространства. При стереографической проекции все точки сферы за
исключением северного полюса проектируются на евклидову плоскость $\MR^2$,
на которой можно ввести комплексные координаты $z\in\MC\approx\MR^2$ (детали
конструкции содержатся в \cite{Nimark58R}).
\begin{theorem}
Каждому собственному вращению $S\in\MS\MO(3)$ соответствуют два дробно линейных
преобразования комплексных координат
\begin{equation*}
  z'=\pm\frac{\quad az+b}{-\bar b\bar z+\bar a},\qquad |a|^2+|b|^2=1.
\end{equation*}
Обратно. Каждой унитарной матрице (\ref{emattw}) соответствует некоторое
вращение.
\end{theorem}
\begin{proof}
См., например, \cite{Nimark58R}.
\end{proof}
\begin{com}
Матрицу (\ref{emattw}) можно рассматривать, как представление кватернионов
(\ref{equare}) с единичной нормой.
\qed\end{com}

Отображение $U\mapsto S$ унитарной матрицы (\ref{emattw}) в ортогональную
матрицу $S\in\MS\MO(3)$ можно записать в явном виде \cite{GeMiSh58R}
\begin{equation}                                                  \label{eormau}
  S=\begin{pmatrix}
    \frac12(a^2-b^2+\bar a^2-\bar b^2) & \frac i2(-a^2-b^2+\bar a^2+\bar b^2)
    & -ab-\bar a\bar b \\[2mm]
    \frac i2(a^2-b^2-\bar a^2+\bar b^2) & \frac12(a^2+b^2+\bar a^2+\bar b^2)
    & -i(ab-\bar a\bar b) \\[2mm]
    a\bar b+\bar a\bar b & i(-a\bar b+\bar a b) & a\bar a-b\bar b
\end{pmatrix}.
\end{equation}

Выше было доказано, что группа $\MS\MU(2)$ дважды накрывает группу вращений
$\MS\MO(3)$. Покажем, как это выглядит при использовании явной параметризации
унитарной группы $\MS\MU(2)$.
\begin{prop}
Пусть элемент алгебры $\Gs\Gu(2)$ имеет вид (\ref{oantim}) с
$\boldsymbol{r}=\boldsymbol{\om}$.
Тогда ему соответствует элемент группы $\MS\MU(2)$
\begin{equation}                                                  \label{esutgr}
  U_\Sa{}^\Sb(\boldsymbol{\om})=(\ex^{(i\om^k\s_k/2)})_\Sa{}^\Sb
  =\dl_\Sa^\Sb\cos\frac\om2+i\frac{\om^k\s_{k\Sa}{}^\Sb}\om\sin\frac\om2
  \quad \in\MS\MU(2),
\end{equation}
где $\om:=\sqrt{\om^i\om_i}$ -- модуль вектора вращения $\boldsymbol{\om}$.
\end{prop}
\begin{proof}
Прямая проверка путем разложения в ряды.
\end{proof}

Верно также и обратное утверждение: любая унитарная матрица с единичным
определителем имеет вид (\ref{esutgr}) для некоторого вектора
$\boldsymbol{\om}\in\MR^3$. Из вида элемента группы следует равенство
\begin{equation*}
  U_\Sa{}^\Sb\left(\om^i+4\pi\frac{\om^i}\om\right)=U_\Sa{}^\Sb(\om^i).
\end{equation*}
То есть для группы $\MS\MU(2)$ в пространстве параметров
$\boldsymbol{\om}\in\MR^3$ (алгебре Ли) имеется отношение эквивалентности
\begin{equation}                                                  \label{eqrsut}
  \om^i\sim\om^i+4\pi\frac{\om^i}\om.
\end{equation}
При этом начало координат отождествляется со всеми сферами радиуса $4\pi m$,
$m=1,2,\dotsc$. По сравнению с отношением эквивалентности (\ref{eqrsog}) для
группы $\MS\MO(3)$ сдвиг происходит на вектор вдвое большей длины. Тем самым
групповое многообразие параметризуется внутренними точками шара
$\MB^3_{2\pi}\subset\MR^3$ радиуса $2\pi$. При этом все точки граничной сферы
$\MS^3_{2\pi}$ необходимо отождествить, т.к.\ второе слагаемое в (\ref{esutgr})
пропадает. Это приводит к второму отношению эквивалентности в пространстве
параметров
\begin{equation}                                                  \label{equspe}
  \om_1^i|_{\om_1=2\pi}\sim\om_2^i|_{\om_2=2\pi}.
\end{equation}
Дополнительное соотношение эквивалентности связано с отсутствием третьего
слагаемого в (\ref{esutgr}) по сравнению с (\ref{elsogr}). В этом также есть
отличие от группы вращений, для которой отождествляются только диаметрально
противоположные точки граничной сферы, что входит в отношение эквивалентности
(\ref{eqrsog}). Можно показать, что в пространстве параметров для группы
$\MS\MU(2)$ других отношений эквивалентности кроме (\ref{eqrsut}) и
(\ref{equspe}) не существует.

Выше было показано, что групповое многообразие унитарной группы $\MS\MU(2)$
представляет собой трехмерную сферу $\MS^3_{2\pi}$. Если начало координат в
приведенной параметризации соответствует южному полюсу сферы $\MS^3_{2\pi}$, то
граничная сфера $\om=2\pi$ -- северному. Это является следствием отношения
эквивалентности (\ref{equspe}).

Из гомоморфизма $\MS\MU(2)\rightarrow\MS\MO(3)$ следует гомоморфизм
представлений этих групп. При этом каждой матрице представления группы вращений
$\MS\MO(3)$ соответствуют две матрицы представления группы $\MS\MU(2)$, которые
отличаются знаком, если представление точное. В физической литературе принято
говорить, что точное представление группы $\MS\MU(2)$ является двузначным
представлением группы вращений $\MS\MO(3)$. Оно называется {\em спинорным
представлением} группы $\MS\MO(3)$. (См.\ также пример \ref{esuspi})
\index{Спинорное представление $\MS\MO(3)$%
 (spinor representation) of $\MS\MO(3)$}%
\index{Представление спинорное $\MS\MO(3)$%
 (spinor representation) of $\MS\MO(3)$}%
Это представление двузначно, т.к.\ одному вращению $S\in\MS\MO(3)$ соответствует
два элемента унитарной группы $\pm U\in\MS\MU(2)$. И эту неоднозначность нельзя
устранить. Действительно, если вращению $S$ поставить в соответствие только одну
унитарную матрицу $U$, то после тождественного преобразования (поворота на угол
$\om=2\pi$), матрица $U$ изменит знак. Это следует из явного вида унитарного
преобразования (\ref{esutgr}), т.к.\ аргумент тригонометрических функций
равен половине угла поворота.

Поскольку группа вращений компактна, то все ее неприводимые представления
унитарны и конечномерны. При этом любое другое ее представление разлагается в
прямую сумму неприводимых представлений. Подробное описание всех представлений
группы вращений содержится в \cite{GeMiSh58R,Nimark58R}.

В заключение настоящего раздела скажем несколько слов о группе вращений
$\MO(n)$ $n$-мерного евклидова пространства $\MR^n$. Под группой
$\MO(1)\approx\MZ_2$ удобно понимать группу, состоящую из двух элементов
$\lbrace 1,-1\rbrace$, которые соответствуют тождественному преобразованию и
отражению вещественной прямой относительно начала координат. При $n\ge2$ группа
$\MO(n)$ сохраняет евклидову метрику и состоит из двух несвязных компонент при
любом $n\ge2$: матриц с положительным и отрицательным определителем. Матрица
пространственных отражений $-\one$ при четных $n$ имеет положительный
определитель, принадлежит связной компоненте единицы $\MS\MO(n)$ и не меняет
ориентации декартовых осей. Поэтому преобразование четности в этом случае
целесообразно определить, как отражение только одной, например, первой
координаты
\begin{equation*}
  P=\begin{pmatrix} -1 & 0 \\ 0 & \one \end{pmatrix},
\end{equation*}
где $\one$ -- единичная матрица размера $(n-1)\times(n-1)$.
\section{Пространство Минковского                                \label{sminsp}}
В настоящем разделе мы рассмотрим пространство-время Минковского произвольного
числа измерений и его общие свойства. Затем остановимся на специфических
свойствах пространств Минковского трех и четырех измерений, которые наиболее
часто встречаются в физических приложениях.

\begin{figure}[h,t]
 \begin{center} 
\includegraphics[width=.8\textwidth]{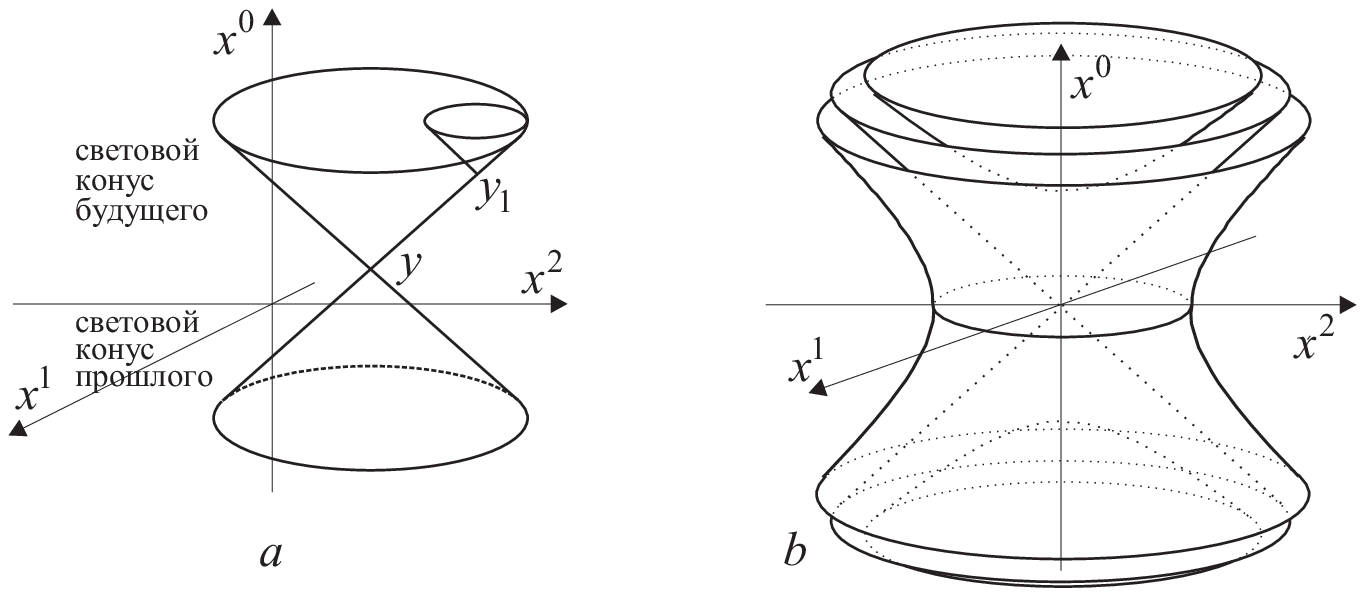}
 \end{center}
 \caption{Световой конус будущего и прошлого в трехмерном пространстве
 Минковского {\it(a)}. Орбитами точек пространства Минковского, лежащих вне
 светового конуса, относительно действия собственной ортохронной группы Лоренца
 $\MS\MO_\uparrow(1,n-1)$ являются однополостные гиперболоиды {\it(b)}.
 \label{flight}}
\end{figure}

\begin{defn}
{\em Пространством} или {\em пространством-временем Минковского} $\MR^{1,n-1}$
размерности $n$ называется евклидово пространство $\MR^n$, на котором в
декартовых координатах задана {\em лоренцева метрика}
\begin{equation}                                                  \label{emimet}
  \eta=\lbrace\et_{ab}\rbrace=\diag(1\underbrace{-1,\dots,-1}_{\mbox{$n-1$}})
\end{equation}
Декартовы системы координат в пространстве Минковского называются
{\em инерциальными}.
\qed\end{defn}
\index{Инерциальная система координат (inertial coordinate system)}%
\index{Система координат инерциальная (inertial coordinate system)}%
\index{Метрика Лоренца}\index{Лоренцева метрика}%
\index{Пространство Минковского (Minkowskian space)}%
\index{Минковского пространство (Minkowskian space)}%
\begin{com}
Выбор общего знака в метрике (\ref{emimet}) является условным, и часто
используется метрика противоположного знака, когда время входит в интервал
со знаком минус, а пространственные координаты -- со знаком плюс.
\qed\end{com}
Таким образом в пространстве Минковского задано две метрики: евклидова и
лоренцева. Евклидова метрика задается естественным образом на прямом
произведении прямых и определяет топологию $\MR^{1,n-1}$. Лоренцева метрика
(\ref{emimet}) не является положительно определенной, и ее нельзя использовать
для определения расстояния в топологическом смысле (см.\ раздел \ref{seucme}).
Поэтому не следует рассматривать пространство Минковского просто как прямое
произведение прямых, на котором задана только метрика (\ref{emimet}), поскольку
евклидова метрика необходима для задания топологии. На самом деле это всегда
подразумевается, т.к.\ непрерывность функций в пространстве Минковского
понимается относительно естественной топологии евклидова пространства.
Инерциальные координаты, в которых метрика имеет вид (\ref{emimet}), обозначим
$\lbrace x^0,x^1,\dots x^{n-1}\rbrace=\lbrace x^a\rbrace$, $a=0,1,\dotsc,n-1$.
Здесь первая координата $t:=x^0$ называется временем. Остальные координаты
$x^i$, $i=1,\dotsc,x^{n-1}$, называются пространственноподобными и параметризуют
пространственные сечения $x^0=\const$. Очевидно, что пространство Минковского
есть прямое произведение
$$
  \MR^{1,n-1}=\MR\times \MR^{n-1},
$$
где первый сомножитель соответствует времени, а второй -- $(n-1)$-мерному
евклидову пространству с отрицательно определенной метрикой.
\begin{com}
Здесь и в дальнейшем пространственные индексы будут нумероваться латинскими
буквами из середины алфавита $\lbrace i,j,k,\dotsc\rbrace$. Мнемоническое
правило следующее. Эти индексы образуют подмножество всего алфавита
$\lbrace a,b,c,\dotsc\rbrace$, так же как множество чисел
$\lbrace 1,2,\dotsc,n-1\rbrace$ является подмножеством чисел
$\lbrace 0,1,2,\dotsc,n-1\rbrace$.
\qed\end{com}
\begin{com}
Между евклидовым пространством и пространством Минковского существует связь.
Если время $x^0$ умножить на мнимую единицу $x^0\mapsto ix^0$ и изменить общий
знак метрики, то получим евклидову метрику. К тому же результату можно прийти,
если на мнимую единицу умножить все пространственные координаты
$x^i\mapsto ix^i$. Оба преобразования называются {\em комплексным поворотом}.
При комплексном повороте оператор Даламбера переходит в оператор Лапласа, что
приводит к качественному отличию решений уравнений, содержащих эти операторы.
В приложениях важную роль играет связь между этими решениями.
\qed\end{com}
\index{Комплексный поворот (complex rotation)}%
\index{Поворот комплексный (complex rotation)}%
\begin{defn}
В каждой точке пространства Минковского $y\in\MR^{1,n-1}$ уравнение
$$
  \et_{ab}(x^a-y^a)(x^b-y^b)=0,
$$
задает {\em световой конус} с вершиной в точке $y$ (см.\
рис.\ref{flight},{\it a}). Конус представляет собой объединение двух связных
компонент: светового конуса {\em прошлого} $x^0\le y^0$ и {\em будущего}
$x^0\ge y^0$ с общей вершиной в точке $y$.
\qed\end{defn}
\index{Световой конус (light cone)}\index{Конус световой (light cone)}%
\index{Световой конус прошлого (past light cone)}%
\index{Световой конус будущего (future light cone)}%
Световые конусы прошлого и будущего являются $(n-1)$-мерными подмногообразиями в
пространстве Минковского $\MR^{1,n-1}$. В целом световой конус подмногообразием
не является из-за общей вершины. Очевидно, что образующие световых конусов имеют
нулевую длину.

\begin{defn}
Дифференцируемая кривая $\lbrace x^a(s)\rbrace$ в пространстве Минковского
называется  {\em времениподобной}, {\em пространственноподобной} или
{\em светоподобной (нулевой)}, если касательный вектор, соответственно,
удовлетворяет условиям:
\begin{equation}                                                  \label{edefcu}
  \dot x^a\dot x^b\eta_{ab}>0,\qquad \dot x^a\dot x^b\eta_{ab}<0,\qquad
  \dot x^a\dot x^b\eta_{ab}=0,
\end{equation}
для всех значений параметра $s\in\MR$. Если в некоторой точке $y$, через которую
проходит кривая, выполнены два условия $\dot x^a\dot x^b\eta_{ab}>0$ и
$\dot x^0>0$, то мы говорим, что касательный вектор к времениподобной кривой в
точке $y$ времениподобен и направлен в будущее.
\qed\end{defn}
\index{Времениподобная кривая (timelike curve)}%
\index{Кривая времениподобная (timelike curve)}%
\index{Пространственноподобная кривая (spacelike curve)}%
\index{Кривая пространственноподобная (spacelike curve)}%
\index{Кривая светоподобная (lightlike curve)}%
\index{Светоподобная кривая (lightlike curve)}%
\index{Нулевая кривая (null curve)}\index{Кривая нулевая (null curve)}%
\begin{prop}
Любая дифференцируемая времениподобная кривая в пространстве Минковского
$\MR^{1,n-1}$, проходящая через точку $y$, целиком лежит внутри светового конуса
с вершиной в этой точке.
\end{prop}
\begin{proof}
Если кривая дифференцируема, времениподобна и проходит через точку $y$, то из
дифференцируемости следует, что она будет лежать внутри светового конуса в
некоторой окрестности точки $y$. Кроме того она не может касаться или пересекать
световой конус, т.к.\ в такой точке она не была бы времениподобна.
\end{proof}

Для полноты картины обобщим пространство Минковского на метрику произвольной
сигнатуры.
\begin{defn}
Псевдоевклидовым пространством $\MR^{p,q}$, $p+q=n$, называется евклидово
пространство $\MR^n$ с метрикой
\begin{equation}                                                  \label{epsecm}
  \eta=\lbrace\eta_{ab}\rbrace
  =\diag(\underbrace{+\dotsc+}_p\underbrace{-\dotsc-}_q),\qquad a,b=1,\dots,n,
\end{equation}
заданной в декартовой системе координат.
\qed\end{defn}
Многие свойства пространства Минковского естественным образом переносятся на
псевдоевклидово пространство. При этом группа Лоренца $\MO(1,n-1)$ заменяется на
группу псевдовращений $\MO(p,q)$, оставляющую метрику (\ref{epsecm})
инвариантной. Группы псевдовращений $\MO(p,q)$ при $1<p<n-1$ устроены сложнее
группы Лоренца и в настоящей монографии рассматриваться не будут.
\subsection{Группа Пуанкаре                                      \label{spungr}}
\begin{defn}
Линейные неоднородные преобразования декартовых координат пространства
Минковского $\MR^{1,n-1}$, оставляющие инвариантной квадратичную форму
\begin{equation}                                                  \label{equfms}
  ds^2=\et_{ab}dx^a dx^b
\end{equation}
образуют группу Ли, которая называется {\em группой Пуанкаре} и обозначается
$\MI\MO(1,n-1)$. Эта группа состоит из лоренцевых вращений с матрицами
$S=\lbrace S_b{}^a\rbrace$ и сдвигов на вектор $a=\lbrace a^a\rbrace$:
\begin{equation}                                                  \label{epotrm}
  x^a\mapsto x^{\prime a}=x^b S_b{}^a+a^a,
\end{equation}
где постоянная матрица $S$ является решением уравнения
\begin{equation}                                                  \label{eloreq}
  \et_{ab}=S_a{}^c S_b{}^d\et_{cd},
\end{equation}
и $a\in\MR^{1,n-1}$ -- постоянный вектор. Множество матриц $S$ образует группу
Ли, которая называется {\em группой Лоренца} и обозначается $\MO(1,n-1)$.
Абелева подгруппа, параметризуемая вектором $a$, называется {\em группой
трансляций}.
\qed\end{defn}
\index{Группа Пуанкаре $\MI\MO(1,n-1)$ (Poincar\'e group)}%
\index{Пуанкаре группа $\MI\MO(1,n-1)$ (Poincar\'e group)}%
\index{Группа Лоренца $\MO(1,n-1)$ (Lorentz group)}%
\index{Лоренца группа $\MO(1,n-1)$  (Lorentz group)}%
\index{Группа трансляций (translational group)}%
\begin{com}
В отличие от группы аффинных преобразований (\ref{elinas}) матрицы $S$ являются
элементами группы Лоренца $\MO(1,n-1)$, а не общей линейной группы
$\MG\ML(n,\MR)$.
\qed\end{com}
\begin{prop}
Любое (возможно, нелинейное) преобразование декартовых координат пространства
Минковского, оставляющее квадратичную форму (\ref{equfms}) инвариантной,
является аффинным, т.е.\ имеет вид (\ref{epotrm}) с некоторой невырожденной
матрицей $S$.
\end{prop}
\begin{proof}
См., например, \cite{NovTai05R}.
\end{proof}
Это утверждение доказывает, что преобразования из группы Пуанкаре исчерпывают
все возможные преобразования, сохраняющие метрику Лоренца (\ref{emimet}).
Другими словами, преобразования (\ref{eloreq}) являются движениями пространства
Минковского общего вида.

Два последовательных преобразования $S_1,a_1$ и $S_2,a_2$, которые отображают
$x\mapsto x'\mapsto x''$ имеют вид
\begin{equation}                                                  \label{etwpot}
  x^{\prime\prime a}=x^{\prime b}S_{2b}{}^a+a_2^a
  =x^c(S_{1c}{}^b S_{2b}{}^a)+(a_1^b S_{2b}{}^a+a_2^a).
\end{equation}
Группа Лоренца является подгруппой группы Пуанкаре и имеет размерность
$$
  \dim\MO(1,n-1)=\frac{n(n-1)}2.
$$
Она неабелева и некомпактна. Сдвиги образуют нормальную абелеву подгруппу группы
Пуанкаре размерности $n$. При этом параметр сдвига $a$ преобразуется как вектор
относительно преобразований Лоренца. В целом группа Пуанкаре представляет собой
полупрямое произведение группы Лоренца на подгруппу сдвигов и имеет размерность
$$
  \dim\MI\MO(1,n-1)=\frac{n(n+1)}2.
$$

Из определяющего уравнения (\ref{eloreq}) следует, что $\det S=\pm1$. Кроме
этого, $00$ компоненту данного уравнения можно переписать в виде
\begin{equation*}
  (S_0{}^0)^2=1+\sum_{i=1}^{n-1}(S_0{}^i)^2.
\end{equation*}
Заметим также, что из условия инвариантности обратной метрики Лоренца:
\begin{equation*}
  \eta^{ab}=\eta^{cd}S_c{}^aS_d{}^b,
\end{equation*}
которое эквивалентно (\ref{eloreq}), вытекает равенство
\begin{equation}                                                  \label{emazzs}
  (S_0{}^0)^2=1+\sum_{i=1}^{n-1}(S_i{}^0)^2,
\end{equation}
где суммирование проходит по нижнему индексу.
Отсюда вытекает, что для любой матрицы лоренцевых вращений либо $S_0{}^0\ge1$,
либо $S_0{}^0\le-1$. Ясно, что никакую матрицу преобразований Лоренца с
определителем $\det S=1$ нельзя непрерывно деформировать в матрицу с
определителем $\det S=-1$, т.к.\ множество из двух элементов
$\lbrace1,-1\rbrace$ не является связным. Аналогично, матрицу с $S_0{}^0\ge1$
нельзя непрерывно деформировать в матрицу с $S_0{}^0\le-1$. Отсюда вытекает, что
группа Лоренца состоит не менее, чем из четырех связных компонент.
\begin{theorem}
Группа Лоренца $\MO(1,n-1)$ при любом $n\ge2$ состоит из четырех связных
компонент.
\end{theorem}
\begin{proof}
При $n=2$ теорема была доказана в разделе \ref{stwodl}. Доказательство для
произвольного $n$ содержится, например, в \cite{NovTai05R}.
\end{proof}

\begin{defn}
Преобразования Лоренца с положительным определителем называются
{\em собственными}. Преобразования Лоренца, для которых $S_0{}^0\ge1$ называются
{\em ортохронными}.
\qed\end{defn}
\index{Собственное преобразование Лоренца (proper Lorentz transformation)}%
\index{Преобразование Лоренца собственное (proper Lorentz transformation)}%
\index{Ортохронное преобразование Лоренца %
(orthochronous Lorentz transformation)}%
\index{Преобразование Лоренца ортохронное %
(orthochronous Lorentz transformation)}%
\begin{prop}
Преобразование Лоренца является ортохронным тогда и только тогда, когда оно
всякий времениподобный вектор $X=\lbrace X^0,X^1,\dotsc,X^{n-1}\rbrace$ в начале
координат, направленный в будущее, $X^0>0$, переводит во времениподобный вектор,
который также направлен в будущее.
\end{prop}
\begin{proof}
Пусть $S:~X^a\mapsto X^{\prime a}=X^bS_b{}^a$ -- преобразование Лоренца. При
этом времениподобный вектор переходит во времениподобный, т.к.\ квадрат вектора
сохраняется при любом преобразовании Лоренца. Для произвольного времениподобного
вектора выполнено неравенство
\begin{equation*}
  (X^0)^2-(X^1)^2-\dotsc-(X^{n-1})^2>0.
\end{equation*}
Применим неравенство Коши--Буняковского:
\begin{multline*}
  (X^1S_1{}^0+\dotsc+X^{n-1}S_{n-1}{}^0)^2
  \le\big((S_1{}^0)^2+\dotsc+(S_{n-1}{}^0)^2\big)
  \big((X^1)^2+\dotsc+(X^{n-1})^2\big)\le
\\
  \le\big((S_0{}^0)^2-1\big)(X^0)^2<(S_0{}^0)^2(X^0)^2,
\end{multline*}
где мы воспользовались равенством (\ref{emazzs}). Отсюда вытекает, что нулевая
компонента преобразованного вектора
\begin{equation*}
  X^{\prime0}:=X^0S_0{}^0+X^1S_1{}^0+\dotsc+X^{n-1}S_{n-1}{}^0
\end{equation*}
имеет тот же знак, что и $S_0{}^0$, если $X^0>0$. Отсюда же следует, что если
$X^0>0$ и $X^{\prime0}$, то $S_0{}^0>0$.
\end{proof}

\begin{defn}
Введем оператор {\em обращения времени}  $T$ и {\em пространственного
отражения (четности)} $P$:
\begin{equation}                                                  \label{etptra}
\begin{aligned}
  &T:\quad (x^0,x^i)\mapsto(-x^0,x^i), &&\qquad\det T=-1,
\\
  &P:\quad (x^0,x^1,x^2,\dotsc,x^{n-1})\mapsto
  (x^0,-x^1,x^2,\dotsc,x^{n-1}), &&\qquad \det P=-1.
\end{aligned}
\end{equation}
Введем также обозначение для их композиции
\begin{equation}                                                  \label{einvpt}
  R:=PT=TP:\quad (x^0,x^1,x^2,\dotsc,x^{n-1})\mapsto
  (-x^0,-x^1,x^2,\dotsc,x^{n-1}), \qquad \det R=1,
\end{equation}
которая отражает только первые две координаты.
\qed\end{defn}
\begin{com}
Пространственные отражения при четных $n$ можно было бы определить, как
отражение всех пространственных координат, т.к.\ в этом случае $\det P=-1$, и
ориентация осей координат меняется. Тогда оператор $R$ соответствует полному
отражению всех координат. При нечетных $n$ отражение всех пространственных
координат имеет положительных определитель и не подходит, т.к.\ принадлежит
связной компоненте единицы группы. Отражение одной (любой) из координатных осей
меняет ориентацию пространства Минковского и может быть использовано в качестве
оператора четности. Такое определение подходит для пространства Минковского
произвольного числа измерений.
\qed\end{com}
Так же, как и в двумерном случае, полная группа Лоренца $\MO(1,n-1)$ при $n\ge3$
состоит из четырех связных компонент, которые получаются из связной компоненты
единицы $\MS_{+\uparrow}(1,n-1)$ обращением времени, пространственным отражением
и их композицией:
\begin{equation}                                                  \label{ecolop}
\begin{split}
  \MS_{-\downarrow}=\MS_{+\uparrow}(1,n-1)T,
\\
  \MS_{-\uparrow}=\MS_{+\uparrow}(1,n-1)P,
\\
  \MS_{+\downarrow}=\MS_{+\uparrow}(1,n-1)R.
\end{split}
\end{equation}

Единица полной группы Лоренца $\MO(1,n-1)$ содержится в компоненте
$\MS_{+\uparrow}$. Эта компонента является связной группой Ли, называется
{\em собственной ортохронной группой Лоренца} и обозначается
$\MS\MO_\uparrow(1,n-1)$.
\index{Собственная ортохронная группа Лоренца %
(proper orthochronous Lorentz group}%
\index{Группа Лоренца собственная ортохронная %
(proper orthochronous Lorentz group}%
Кроме того, полная группа Лоренца $\MO(1,n-1)$ содержит еще три подгруппы,
состоящие из двух компонент каждая:
\begin{align}                                                          \nonumber
  \MO_\downarrow(1,n-1)&\approx \MS_{+\uparrow}\cup \MS_{-\downarrow},
  && \det S=\pm1,
\\                                                                \label{esubss}
  \MO_\uparrow(1,n-1)&\approx \MS_{+\uparrow}\cup \MS_{-\uparrow},
  && \det S=\pm1,\quad S_0{}^0>0,
\\                                                                     \nonumber
  \MS\MO(1,n-1)&\approx \MS_{+\uparrow}\cup \MS_{+\downarrow},
  && \det S=\quad 1.
\end{align}
Все четыре подгруппы являются нормальными подгруппами в $\MO(1,n-1)$.

Очевидно, что отображения (\ref{ecolop}) взаимно однозначны и гладки. Поэтому
как многообразия все четыре связные компоненты групп $\MO(1,n-1)$ диффеоморфны
между собой:
\begin{equation*}
  \MS_{+\uparrow}\approx\MS_{-\downarrow}\approx\MS_{-\uparrow}
  \approx\MS_{+\downarrow}.
\end{equation*}
Компоненты (\ref{ecolop}) являются смежными классами в группе $\MO(1,n-1)$ по
нормальной подгруппе $\MS\MO_\uparrow(1,n-1)$. В качестве их представителей в
фактор группе $\MO(1,n-1)/\MS\MO_\uparrow(1,n-1)$ можно выбрать матрицы $T,P,R$.
Фактор группа $\MO(1,n-1)/\MS\MO_\uparrow(1,n-1)$ -- это 4-группа Клейна
$\MK_4=\lbrace\one,T,P,R\rbrace$, рассмотренная в разделе \ref{stwodl}.

Рассмотрим автоморфизмы связной компоненты единицы группы Лоренца произвольной
размерности. Пусть $S\in\MS\MO_\uparrow(1,n-1)$, тогда отображение
\begin{equation*}
  S\mapsto S_0SS_0^{-1},
\end{equation*}
где $S_0\in\MS\MO_\uparrow(1,n-1)$ -- произвольный фиксированный элемент из
связной компоненты единицы, задает внутренний автоморфизм группы собственных
преобразований Лоренца. Отображения
\begin{equation}                                                  \label{emapts}
  S\mapsto TST^{-1}\qquad \text{и}\qquad S\mapsto PSP^{-1},
\end{equation}
определяемые обращением времени и пространственным отражением, также задают
автоморфизм собственных преобразований Лоренца. Однако этот автоморфизм будет
внешним, т.к.\ преобразования $T$ и $P$ не принадлежат $\MS\MO_\uparrow(1,n-1)$.
Можно доказать, что внешние автоморфизмы (\ref{emapts}) совпадают между собой с
точностью до внутреннего автоморфизма\footnote{Если в пространстве Минковского
четной размерности оператор пространственного отражения определен, как отражение
всех пространственных координат, то внешние автоморфизмы (\ref{emapts}) просто
совпадают.}.
Чтобы найти явный вид внешнего автоморфизма, перепишем определение лоренцевых
вращений (\ref{eloreq}), не делая различия между верхними и нижними индексами,
\begin{equation*}
  \eta=S\eta S^\St\qquad \text{или}\qquad
  (S^\St)^{-1}=\eta^{-1}S\eta.
\end{equation*}
Поскольку $\eta^{-1}=-T$, то отсюда следует формула для внешнего автоморфизма
\begin{equation}                                                  \label{extalo}
  TST^{-1}=(S^\St)^{-1}.
\end{equation}
Конечно, внешние автоморфизмы связной компоненты единицы
$\MS\MO_\uparrow(1,n-1)$ являются внутренними для полной группы Лоренца
$\MO(1,n-1)$.
\begin{com}
Автоморфизм (\ref{extalo}) играет большую роль при рассмотрении спинорных
представлений полной группы Лоренца $\MO(1,n-1)$, когда требуется определить
оператор обращения времени и четности.
\qed\end{com}

Орбитами точек пространства Минковского (см.\ рис.\ref{flight},{\it b}),
лежащих вне светового конуса, относительно действия собственной ортохронной
группы Лоренца $\MS\MO_\uparrow(1,n-1)$ являются однополостные гиперболоиды,
определяемые уравнением
\begin{equation*}
  (x^0)^2-(x^1)^2-\dotsc-(x^{n-1})^2=-r^2,\qquad r>0.
\end{equation*}
Для точек, лежащих внутри конусов будущего и прошлого, орбитами являются,
соответственно, верхняя и нижняя полы двуполостного гиперболоида
\begin{equation*}
  (x^0)^2-(x^1)^2-\dotsc-(x^{n-1})^2=r^2,\qquad r>0.
\end{equation*}
Сами конусы будущего и прошлого представляют собой орбиты лежащих на них точек.
Начало координат является неподвижной точкой относительно лоренцевых вращений.
На каждой из орбит собственная группа Лоренца $\MS\MO_\uparrow$ действует
транзитивно, т.е.\ для любых двух точек, лежащих на одной орбите, найдется по
крайней мере одно собственное ортохронное преобразование Лоренца, переводящее
одну точку в другую. В целом действие группы $\MS\MO_\uparrow(1,n-1)$ в
пространстве Минковского $\MR^{1,n-1}$ эффективно, но не свободно, т.к.\ начало
координат является неподвижной точкой.
\begin{com}
Инвариантность законов Природы относительно действия группы Пуанкаре является
фундаментальным требованием к современным моделям математической физики и
составляет основное содержание специальной теории относительности. Эта группа не
является полупростой, т.к.\ содержит нормальную абелеву подгруппу (сдвиги).
Следовательно ее форма Киллинга--Картана вырождена, и это создает существенные
трудности при построении физических моделей, т.к.\ на многообразии параметров
группы Пуанкаре не существует двусторонне инвариантной невырожденной метрики.
\qed\end{com}

Бесконечно малые преобразования группы Пуанкаре в пространстве Минковского
$\MR^{1,n-1}$ в линейном приближении можно записать с помощью дифференциальных
операторов $L_{ab}=-L_{ba}$ и $P_a$:
\begin{equation}                                                  \label{einpot}
  \dl x^a=\left(\frac12\dl\om^{bc}L_{bc}+\dl a^b P_b\right)x^a,
\end{equation}
где $\dl\om^{ab}=-\dl\om^{ba}\ll1$ и $\dl a^a\ll1$ -- параметры преобразований
Лоренца и сдвигов, соответственно. Множитель $\frac12$ перед $\dl\om^{bc}L_{bc}$
связан с тем, что независимые параметры входят в эту сумму дважды. Например,
\begin{equation*}
  \dl\om^{01}L_{01}+\dl\om^{10}L_{10}=2\dl\om^{01}L_{01}.
\end{equation*}
Генераторы лоренцевых вращений и сдвигов представляются дифференциальными
операторами (векторными полями Киллинга) на $\MR^{1,n-1}$
\begin{align}                                                     \label{elorot}
  L_{ab}&=x_a\pl_b-x_b\pl_a,
\\                                                                \label{etrand}
  P_a&=\pl_a,
\end{align}
где подъем и опускание индексов осуществляется с помощью метрики Лоренца. При
этом формула (\ref{einpot}) переходит в равенство
\begin{equation*}
  \dl x^a=x^b\dl\om_b{}^a+\dl a^a.
\end{equation*}
Дифференциальные операторы (\ref{elorot}) и (\ref{etrand}) действуют в алгебре
функций $\CC^1(\MR^{1,n-1})$, заданных в пространстве Минковского $\MR^{1,n-1}$.
Например, изменение формы функции (см.\ раздел \ref{sinfct}) при бесконечно
малых преобразованиях в линейном приближении имеет вид
\begin{equation*}
  \dl f(x):=-\dl x^a\pl_a f(x)=-(x^b\dl\om_b{}^a+\dl a^a)\pl_a f(x).
\end{equation*}

Матрица бесконечно малых вращений $S\in\MS\MO_\uparrow(1,n-1)$ в линейном
приближении имеет вид
\begin{equation}                                                  \label{einlog}
  S_b{}^a\approx\dl_b^a+\om_b{}^a.
\end{equation}
Отметим, что антисимметрия параметров лоренцевых вращений следует из уравнения
(\ref{eloreq}) в линейном приближении. Каждый элемент антисимметричной матрицы
$\om^{ab}$ представляет собой угол бесконечно малого поворота в плоскости
$x^a,x^b$. Лоренцевы вращения с параметрами $\om^{0i}$, $i=1,\dotsc,n-1$
называют {\em бустом}, чтобы отличать их от чисто пространственных вращений,
соответствующих параметрам $\om^{ij}$.
\index{Буст (boost)}%

Генераторы вращений и сдвигов удовлетворяют {\em алгебре Пуанкаре}
\index{Алгебра Пуанкаре (Poincar\'e algebra)}%
\index{Алгебра Пуанкаре (Poincar\'e algebra)}%
\begin{align}                                                     \label{epoalo}
  [L_{ab},L_{cd}]&=-\et_{ac}L_{bd}+\et_{ad}L_{bc}+\et_{bc}L_{ad}-\et_{bd}L_{ac},
\\                                                                \label{epoals}
  [L_{ab},P_c]&=-\et_{ac}P_b+\et_{bc}P_a,
\\                                                                \label{epoalt}
  [P_a,P_b]&=0,
\end{align}
что проверяется прямой проверкой.

При действии вращений на векторы в пространстве Минковского $\MR^{1,n-1}$
генераторы лоренцевых вращений, но не сдвиги, можно представить также в
матричном виде
$$
  S_{b}{}^a\approx\dl_b^a+\frac12\dl\om^{cd}L_{cd\,b}{}^a,
$$
где
\begin{equation}                                                  \label{elorom}
  L_{cd\,b}{}^a=\et_{d b}\dl_c^a-\et_{cb}\dl_d^a.
\end{equation}
Это -- фундаментальное или векторное представление образующих алгебры Ли группы
Лоренца. Нетрудно проверить, что матричное представление генераторов вращений
также удовлетворяет алгебре Ли (\ref{epoalo}), которую можно записать в виде
\begin{equation}                                                  \label{eloama}
  [L_{ab},L_{cd}]=f_{ab\,cd}{}^{ef}L_{ef},
\end{equation}
где $f_{ab\,cd}{}^{ef}$ -- структурные константы группы Лоренца для
коммутационных соотношений (\ref{epoalo}). Трансляции не имеют матричного
представления в пространстве Минковского. Группа Лоренца при $n>2$ является
неабелевой и простой. Ее форма Киллинга--Картана невырождена
\begin{equation}                                                  \label{ekikal}
  \et_{ab\,cd}
  :=f_{ab\,ef}{}^{gh}f_{cd\,gh}{}^{ef}=4(-\et_{ac}\et_{bd}+\et_{ad}\et_{bc}),
\end{equation}
и поэтому может быть использована в качестве инвариантной метрики для построения
инвариантов. Для матричного представления генераторов (\ref{elorom}) справедливо
равенство
\begin{equation}                                                  \label{etrlrm}
  \tr(L_{ab}L_{cd})=\frac12\et_{ab\,cd}.
\end{equation}
\begin{com}
При $n=2$ собственная ортохронная группа Лоренца $\MS\MO_\uparrow(1,1)$ является
абелевой, ее структурные константы равны нулю и форма Киллинга--Картана
вырождена. Заметим, что правая часть (\ref{ekikal}) отлична от нуля и в этом
случае. Формула для следа (\ref{etrlrm}) справедлива также при $n=2$, если под
$\et_{ab\,cd}$ понимать правую часть (\ref{ekikal}), а не форму
Киллинга--Картана.
\qed\end{com}

Группа Пуанкаре в $n$-мерном пространстве Минковского связана с группой Лоренца
в $(n+1)$-мерном пространстве следующим образом. Пусть индекс
$\Sa=\lbrace 0,1,\dotsc,n-1,n\rbrace $ нумерует координаты $(n+1)$-мерного
пространства Минковского $\MR^{1,n}$ с метрикой
\begin{equation}                                                  \label{elomnp}
  \lbrace\et_{\Sa\Sb}\rbrace=\diag(1,\underbrace{-1,\dotsc,-1}_{n-1},-k^2),
  \qquad k>0.
\end{equation}
Обозначим генераторы лоренцевых вращений, затрагивающих $n$-тую координату,
через $\widetilde P_a:=L_{an}$. Тогда алгебра Лоренца в $(n+1)$-мерном
пространстве примет вид
\begin{align}                                                          \nonumber
  [L_{ab},L_{cd}]&=-\et_{ac}L_{bd}+\et_{ad}L_{bc}+\et_{bc}L_{ad}-\et_{bd}L_{ac},
\\                                                                \label{eloals}
  [L_{ab},\widetilde P_c]&=-\et_{ac}\widetilde P_b+\et_{bc}\widetilde P_a,
\\                                                                     \nonumber
  [\widetilde P_a,\widetilde P_b]&=-k^2L_{ab}.
\end{align}
Эта алгебра отличается от алгебры Пуанкаре тем, что ``сдвиги'' $\widetilde P_a$
уже не коммутируют. Алгебра Пуанкаре получается после формального предела
$k\rightarrow0$, который называют {\em контракцией}. В этом пределе
$\widetilde P\rightarrow P$ и метрика Лоренца (\ref{elomnp}) в $n+1$ мерном
пространстве Минковского $\MR^{1,n}$ вырождается.
\index{Контракция (contraction)}%
\subsection{Группа Галилея}
Модели математической физики, которые инвариантны относительно действия группы
Пуанкаре называются {\em релятивистскими}. Постулат о том, что физические модели
должны быть инвариантны относительно группы Пуанкаре, лежит в основе
{\em специальной теории относительности}, которая была предложена А.~Эйнштейном
в 1905 году \cite{Einste05R}. До создания специальной теории относительности
ньютонова механика точечных частиц рассматривалась как основная фундаментальная
модель. В механике Ньютона важнейшую роль играют преобразования Галилея, которые
также образуют группу симметрии пространства-времени.
\index{Релятивистская модель (relativistic model)}%
\index{Модель релятивистская (relativistic model)}%

Рассмотрим пространство-время Галилея $\MR^1\times\MR^{n-1}=\MR^n$, где первый
сомножитель соответствует времени, $t=x^0\in\MR$, а второй -- пространству.
Пространственные декартовы координаты занумеруем буквами из середины латинского
алфавита $x^i$, $i=1,\dotsc,n-1$.
\begin{defn}
Зададим преобразования в $\MR^1\times\MR^{n-1}$ следующими формулами:
\begin{equation}                                                  \label{egaltr}
\begin{split}
  t&\mapsto t'=t+a,
\\
  x&\mapsto x^{\prime i}=x^j S_j{}^i+v^i t+b^i,
\end{split}
\end{equation}
где $S_j{}^i\in\MO(n-1)$ -- матрица ортогональных вращений, действующая в
пространстве. Преобразования (\ref{egaltr}) содержат дополнительные параметры:
{\em скорость} $\lbrace v^i\rbrace\in\MR^{n-1}$, сдвиги по времени $a\in\MR$ и
сдвиги пространства $\lbrace b^i\rbrace\in\MR^{n-1}$. Эти преобразования
образуют {\em группу Галилея} $\MG(1,n-1)$. Декартовы системы координат,
связанные преобразованием (\ref{egaltr}), как и в пространстве Минковского,
называются {\em инерциальными}.

Подгруппа группы Галилея, для которой начало координат является неподвижной
точкой, называется {\em однородной группой Галилея} и обозначается
$\MH\MG(1,n-1)$. Она состоит из пространственных вращений с матрицей $S_j{}^i$ и
преобразований, которые параметризуются вектором скорости $v^i$. Последние
преобразования называются {\em галилеевыми бустами}.
\qed\end{defn}
\index{Группа Галилея (Galelian group)}\index{Галилея группа (Galelian group)}%
\index{Группа Галилея однородная (Homogeneous Galelian group)}%
\index{Буст галилеев (Galilean boost)}\index{Галилеев буст (Galilean boost)}%
\index{Инерциальные координаты (inertial coordinates)}%
\index{Координаты инерциальные (inertial coordinates)}%
Посчитаем размерность групп. Пространственные вращения параметризуются
$(n-1)(n-2)/2$ параметрами. Галилеевы бусты -- $(n-1)$ параметром. Кроме того,
сдвиги времени и пространства задаются $n$ параметрами. Таким образом,
преобразования из полной группы Галилея задаются $n(n+1)/2$ параметрами, также
как и преобразования из группы Пуанкаре. Размерность однородной группы Галилея
равна размерности группы Лоренца,
\begin{equation*}
  \dim\MH\MG(1,n-1)=\frac{n(n-1)}2.
\end{equation*}

Пусть в галилеевом пространстве-времени произошло два события в точках
$(t_1,x_1)$ и $(t_2,x_2)$. Эти события разделены во времени
$\triangle t:=t_2-t_1$ и пространстве $\triangle x^i:=x^i_2-x^i_1$.
Тогда преобразования Галилея -- это такие линейные неоднородные преобразования
пространства-времени, которые оставляют инвариантными временной интервал
$\triangle t$ и расстояние между двумя одновременными событиями
$\dl_{ij}\triangle x^i\triangle x^j$, при $t=\const$.

Генераторы полной группы Галилея можно представить в виде дифференциальных
операторов (векторных полей):
\begin{equation*}
\begin{split}
  L_{ij}&=x_i\pl_j-x_j\pl_i,
\\
  P_i&=\pl_i,
\\
  B_i&=t\pl_i,
\\
  P_0&=\pl_0.
\end{split}
\end{equation*}
Эти генераторы удовлетворяют алгебре Галилея:
\begin{equation}                                                  \label{egalal}
\begin{split}
  [L_{ij},L_{kl}]&=-\dl_{ik}L_{jl}+\dl_{il}L_{jk}+\dl_{jk}L_{il}-\dl_{jl}L_{ik},
\\
  [L_{ij},P_k]&=-\dl_{ik}P_j+\dl_{jk}P_i,
\\
  [L_{ij},B_k]&=-\dl_{ik}B_j+\dl_{jk}B_i,
\\
  [B_i,P_0]&=-P_i,
\end{split}
\end{equation}
где выписаны только отличные от нуля коммутаторы. Отсюда следует, что при
пространственных вращениях генераторы галилеевых бустов $B_i$ преобразуются,
как ковекторы $\lbrace 0,x_i\rbrace\in\MR^{1,n-1}$.

В этом представлении группа Галилея действует в алгебре функций
$\CC^1(\MR^{1,n-1})$. При этом временн\'ая координата может быть сдвинута
только на постоянный вектор. Механика Ньютона инварианта относительно
преобразований Галилея. Поэтому говорят, что в механике Ньютона время имеет
абсолютное значение.

Однородная группа Галилея представляет собой полупрямое произведение группы
пространственных вращений $\MO(n-1)$ на группу галилеевых бустов. Группа
галилеевых бустов образует абелеву нормальную подгруппу однородной группы
Галилея, которая, следовательно, не является полупростой. Отсюда следует, что в
галилеевом пространстве $\MR\times\MR^{n-1}$ не существует метрики, инвариантной
относительно преобразований Галилея.
\subsection{Группа конформных преобразований                     \label{scotrm}}
Рассмотрим пространство Минковского $\MR^{1,n-1}$ с декартовыми координатами
$x^a$, $a=0,1,\dotsc,n-1$. В настоящем разделе подъем и опускание
индексов производится с помощью метрики Лоренца (\ref{emimet}). Рассмотрим
преобразования координат $x\mapsto x'(x)$ и ослабим требование инвариантности
лоренцевой метрики (\ref{eloreq}), заменив его следующим условием
\begin{equation}                                                  \label{ecotme}
  \et_{ab}dx^{\prime a}dx^{\prime b}=\Om^2(x)\et_{ab}dx^a dx^b,
\end{equation}
где $\Om(x)$ -- произвольная отличная от нуля функция (конформный множитель).
Для определенности, будем считать, что $\Om>0$.
\begin{defn}
Преобразования координат пространства Минковского, при которых метрика Лоренца
умножается на некоторый отличный от нуля множитель (\ref{ecotme}) образуют
{\em конформную группу}.
\qed\end{defn}
\index{Конформная группа (conformal group)}%
\index{Группа конформная (conformal group)}%

При конформных преобразованиях длины векторов меняются, а углы между ними
сохраняются.

Решения уравнения (\ref{ecotme}) для функций преобразования координат $x'(x)$
зависят от размерности пространства-времени.

В двумерном пространстве-времени, $n=2$, уравнение (\ref{ecotme}) удобно
переписать в светоподобных координатах (\ref{econco})
\begin{equation*}
  du'dv'=\Om^2(u,v)dudv.
\end{equation*}
Ясно, что конформные преобразования
\begin{equation}                                                  \label{edcmol}
   u\mapsto u'(u),\qquad v\mapsto v'(v)
\end{equation}
где $u'(u)$ и $v'(v)$ -- произвольные монотонные функции с отличными от нуля
производными, удовлетворяют поставленному условию. Действительно,
\begin{equation*}
  du'dv'=\frac{du'}{du}\frac{dv'}{dv}dudv.
\end{equation*}
То есть для конформных преобразований
\begin{equation*}
  \Om(u,v)=\frac{du'}{du}\frac{dv'}{dv}.
\end{equation*}

Кроме того, двумерные преобразования координат из группы Пуанкаре также
удовлетворяют равенству (\ref{ecotme}). Для этих преобразований $\Om=1$.
Можно доказать и обратное утверждение. Преобразования из группы Пуанкаре и
конформные преобразования (\ref{edcmol}) исчерпывают все возможные
преобразования координат, для которых выполнено равенство (\ref{ecotme}).

Группа конформных преобразований (\ref{edcmol}) является бесконечномерной
группой Ли.

Теперь рассмотрим случай $n\ge3$.

Нетрудно доказать, что преобразования пространства Минковского (возможно,
нелинейные), удовлетворяющие условию (\ref{ecotme}), образуют группу Ли.

\begin{prop}
Конформная группа при $n\ge3$ состоит из подгруппы преобразований из группы
Пуанкаре (\ref{epotrm}), для которых $\Om=1$, {\em специальных конформных
преобразований}, которые параметризуются постоянным вектором $b=\lbrace
b^a\rbrace\in\MR^{1,n-1}$,
\begin{equation}                                                  \label{espctr}
  x^{\prime a}=\frac{x^a+b^a x^2}{1+2bx+b^2x^2},\qquad b\ne0,
  \quad \Om=\frac1{1+2bx+b^2x^2},
\end{equation}
где $bx:=b^a x_a$, $b^2:=b^ab_a$, $x^2:=x^ax_a$, и {\em дилатаций}
\begin{equation}                                                  \label{ediltr}
  x^{\prime a}=\Om x^a,\qquad \Om=\const\ne1. \qed
\end{equation}
\renewcommand{\qed}{}\end{prop}
\begin{proof}
См., например, \cite{}.
\end{proof}
\index{Специальное конформное преобразование %
(special conformal transformation)}%
\index{Конформное преобразование специальное %
(special conformal transformation)}%
\index{Дилатация (dilatation)}%
Дилатации называются также {\em гомотетией}.
\index{Гомотетия (homothety)}%

Специальные конформные преобразования (\ref{espctr}) определены при
\begin{equation*}
  1+2bx+b^2x^2\ne0.
\end{equation*}

Нетрудно проверить, что из формулы преобразования координат (\ref{espctr})
следует равенство
\begin{equation*}
  x^{\prime2}=x^2\frac 1{1+2bx+b^2x^2}.
\end{equation*}
Если $x^2\ne0$, то отсюда вытекает правило
\begin{equation}                                                  \label{ebvcoi}
  \frac{x^{\prime a}}{x^{\prime 2}}=\frac{x^a}{x^2}+b^a,\qquad x^2\ne0.
\end{equation}
То есть специальные конформные преобразования -- это сдвиг ``обращенных''
координат
\begin{equation*}
  x^a\mapsto \frac{x^a}{x^2},
\end{equation*}
для которых бесконечно удаленная точка отображается в начало координат.

Генераторы бесконечно малых специальных конформных преобразований $K_a$ и
дилатаций $D$ можно представить в виде дифференциальных операторов (векторных
полей) на $\MR^{1,n-1}$:
\begin{align}                                                     \label{espcog}
  K_{a}&=x^2\pl_a-2x_a x^b\pl_b,
\\                                                                \label{ediltg}
  D&=x^a\pl_a.
\end{align}
Эти векторные поля не являются полями Киллинга, т.к.\ при соответствующих
преобразованиях метрика меняется. Прямые вычисления с учетом представления
(\ref{elorot}), (\ref{etrand}) для генераторов группы Пуанкаре приводят к
следующим коммутационным соотношениям
\begin{align}                                                          \nonumber
  [L_{ab},K_c]&=\et_{ac}K_b-\et_{bc}K_a,
\\                                                                     \nonumber
  [P_a,K_b]&=-2\et_{ab}D-2L_{ab},
\\                                                                \label{ecoath}
  [P_a,D]&=P_a,
\\                                                                     \nonumber
  [K_a,D]&=-K_a,
\\                                                                     \nonumber
  [L_{ab},D]&=[K_a,K_b]=[D,D]=0.
\end{align}
Эти соотношения вместе с коммутационными соотношениями
(\ref{epoalo})--(\ref{epoalt}) для группы Пуанкаре показывают, что генераторы
$L_{ab}$, $P_a$, $K_a$ и $D$ образуют алгебру Ли, соответствующую конформной
группе преобразований. Она имеет размерность
$$
  \frac12(n+1)(n+2).
$$

В отличие от группы Пуанкаре группа конформных преобразований проста и ее
связная компонента единицы изоморфна связной компоненте единицы группы
псевдовращений $\MS\MO(2,n)$, действующей в пространстве $\MR^{2,n}$,
размерность которого на два превышает размерность исходного пространства
Минковского $\MR^{1,n-1}$. Изоморфизм алгебр Ли можно задать следующим
образом
\begin{equation}                                                  \label{eisclg}
  L_{\Sa\Sb}=\begin{pmatrix}
  \frac12L_{ab} & \frac14(P_a+K_a) & \frac14(P_a-K_a)
\\[2mm]
  -\frac14(P_b+K_b) & 0 & \frac12D
\\[2mm]
  -\frac14(P_b-K_b) & -\frac12D & 0   \end{pmatrix},
\end{equation}
где индексы $\Sa,\Sb$ пробегают значения $\Sa=\lbrace a,n,n+1\rbrace=\lbrace
0,1,\dotsc,n+1\rbrace$, и метрика в пространстве $\MR^{2,n}$ имеет вид
\begin{equation*}
  \eta_{\Sa\Sb}=\begin{pmatrix} \eta_{ab} &  0 & 0 \\
                  0 & -1 & 0 \\ 0 &  0 & 1 \end{pmatrix}
\end{equation*}
Поскольку конформная группа проста, то на групповом многообразии существует
двусторонне инвариантная метрика. Это -- существенное отличие от группы
Пуанкаре.

Оператор Даламбера в $n$-мерном пространстве Минковского
\begin{equation*}
  \square=\pl_0^2-\sum_{i=1}^{n-1}\pl_i^2
\end{equation*}
инвариантен относительно группы Пуанкаре $\MI\MO(1,n-1)$, причем сдвиги
действуют на оператор дифференцирования тривиально $\pl_a\mapsto\pl_a$. При
специальных конформных преобразованиях и дилатациях оператор Даламбера
умножается на конформный множитель:
\begin{equation*}
  \square=\Om^2\square',
\end{equation*}
где $\square':=\eta^{ab}\pl'_a\pl'_b$. Это означает, что, если некоторая
функция удовлетворяет уравнению Даламбера $\square f=0$, то она будет
удовлетворять уравнению Даламбера после произвольного конформного
преобразования координат.
\begin{com}
Конформные преобразования были введены в физику в 1909 году Куннигамом и
Бейтменом \cite{Cunnin09,Batema09}, которые показали, что уравнения Максвелла
ковариантны относительны конформных преобразований. Позже Дирак показал
конформную инвариантность уравнения для нейтрино (безмассового уравнения
Дирака) \cite{Dirac36R}. В настоящее время принято считать, что конформная
инвариантность играет большую роль в асимптотических режимах, где отсутствуют
размерные параметры. Подробное изложение конформной квантовой теории поля
можно найти в монографии \cite{FraPal96}. \qed\end{com}
\begin{com}
Конформные преобразования координат можно рассматривать также в евклидовом
пространстве. Для евклидовой плоскости конформные преобразования (\ref{edcmol})
заменяются на конформные преобразования комплексной плоскости $z\mapsto w(z)$.
При $n\ge3$ связная компонента единицы группы конформных преобразований
изоморфна группе Лоренца $\MS\MO_\uparrow(1,n+1)$.
\end{com}
\subsection{Трехмерное пространство Минковского                  \label{sthmio}}
Рассмотрим трехмерное пространство Минковского $\MR^{1,2}$ с декартовыми
координатами $x^a$, $a=0,1,2$, и метрикой Лоренца
$\lbrace\eta_{ab}\rbrace=\diag(+--)$. В трехмерном случае группа Лоренца
$\MO(1,2)$ имеет всего три независимых генератора: $L_{01},L_{02}$ и $L_{12}$.
Первые два генератора соответствуют лоренцевым бустам, а третий --
пространственным вращениям в плоскости $x^1,x^2$. Генераторы группы Лоренца
удовлетворяют алгебре (\ref{epoalo}), которая в рассматриваемом случае
существенно упрощается:
\begin{equation*}
\begin{split}
  [L_{01},L_{02}]&=L_{12},
\\
  [L_{01},L_{12}]&=L_{02},
\\
  [L_{02},L_{12}]&=-L_{01},
\end{split}
\end{equation*}
а все остальные коммутаторы равны нулю.

Используя полностью антисимметричный тензор третьего ранга $\ve^{abc}$,
$\ve^{012}=1$, (см. приложение \ref{scomat}) удобно перейти к дуальному
базису алгебры Ли $\Gs\Go(1,2)$
\begin{equation*}
  M^a:=\frac12\ve^{abc}L_{bc},\qquad L_{bc}=\ve_{bca}M^a,
\end{equation*}
где подъем и опускание индексов осуществляется с помощью метрики Лоренца. При
этом генератор $M^0=L_{12}$ соответствует вращениям в пространственной
плоскости, а генераторы $M^1=-L_{02}$ и $M^2=L_{01}$ -- лоренцевым бустам
соответственно в плоскостях $x^0,x^2$ и $x^0,x^1$. Нетрудно проверить, что
новый базис удовлетворяет коммутационным соотношениям
\begin{equation*}
  [M^a,M^b]=\ve^{abc}M_c.
\end{equation*}
В отличие от алгебры Ли группы трехмерных вращений (\ref{ealatr}), подъем и
опускание индексов в рассматриваемом случае осуществляется не с помощью
евклидовой метрики, а с помощью метрики Лоренца: $M_a:=\eta_{ab}M^b$ ($M_0=M^0$,
$M_{1,2}=-M^{1,2}$).

Группа Лоренца $\MO(1,2)$ неабелева и некомпактна.

Собственная ортохронная группа Лоренца $\MS\MO_\uparrow(1,2)$ имеет двумерные
неабелевы некомпактные подгруппы. В качестве такой подгруппы можно выбрать
группу, генерируемую, например, базисными векторами $M_1$ и $K=M_0+M_2$, которые
соответствуют бусту и вращению в перпендикулярной нулевой плоскости. Эти векторы
образуют базис двумерной неабелевой алгебры Ли $\Gg\subset\Gs\Go(1,2)$ и
удовлетворяют следующим коммутационным соотношениям
\begin{equation*}
\begin{split}
  [M_1,M_1]&=0,\qquad [K,K]=0,
\\
  [M_1,K]&=K.
\end{split}
\end{equation*}
Это алгебра генерирует простейшую двумерную неабелеву группу Ли $\MG$ (группу
аффинных преобразований прямой). Ее
групповое многообразие некомпактно и подробно изучено в разделе \ref{stwnog}.
С топологической точки зрения эта подгруппа диффеоморфна евклидовой плоскости
$\MR^2$. Любое преобразование из группы $\MS\MO_\uparrow(1,2)$ можно представить
в виде композиции некоторого преобразования из данной подгруппы и некоторого
пространственного вращения из подгруппы $\MS\MO(2)\subset\MS\MO_\uparrow(1,2)$.
Поэтому в соответствии с общей теоремой \ref{tnolig} как многообразие трехмерная
группа Лоренца диффеоморфна прямому произведению
\begin{equation*}
  \MS\MO_\uparrow(1,2)\approx\MS^1\times\MR^2,
\end{equation*}
так как $\MS\MO(2)\approx\MS^1$. Этот диффеоморфизм описан явно в примере
\ref{esltre}.

Поскольку трехмерная группа Лоренца $\MS\MO_\uparrow(1,2)$ является подгруппой
всех групп Лоренца более высоких размерностей $\MS\MO_\uparrow(1,n-1)$, $n>3$,
то соответствующая двумерная подгруппа является подгруппой всех групп Лоренца
$\MG\subset\MS\MO_\uparrow(1,n-1)$, $n\ge3$.

Построим гомоморфизм группы вещественных $2\times2$ матриц с единичным
определителем $\MS\ML(2,\MR)$ на связную компоненту единицы трехмерной группы
Лоренца $\MS\MO_\uparrow(1,2)$. Для этого рассмотрим произвольную симметричную
$2\times2$ матрицу $A$ (поскольку матрица вещественна, то она также эрмитова),
которую параметризуем следующим образом
\begin{equation}                                                  \label{esympa}
  A=\begin{pmatrix} t+x & y \\ y & t-x \end{pmatrix},\qquad t,x,y\in\MR.
\end{equation}
Рассмотрим числа $t,x,y$ как декартовы координаты в пространстве Минковского
$\MR^{1,2}$. Поскольку
\begin{equation*}
  \det A=t^2-x^2-y^2,
\end{equation*}
то определитель $\det A$ задает квадратичную форму Лоренца в $\MR^{1,2}$.

Пусть $M\in\MS\ML(2,\MR)$ -- произвольная вещественная $2\times2$ матрица с
единичным определителем. Тогда каждому преобразованию
\begin{equation}                                                  \label{elyreq}
  A\mapsto A'=MAM^\St
\end{equation}
можно сопоставить некоторое преобразование Лоренца в пространстве Минковского.
Действительно, при таком преобразовании матрица $A'$ остается симметричной, и ее
можно представить в виде
\begin{equation*}
  A'=\begin{pmatrix} t'+x' & y' \\ y' & t'-x' \end{pmatrix}
\end{equation*}
с некоторыми новыми координатами $t',x',y'$. Поскольку при преобразовании
(\ref{elyreq}) определитель не меняется,
\begin{equation*}
  \det A'=\det A,
\end{equation*}
то каждому элементу $M\in\MS\ML(2,\MR)$ однозначно ставится в соответствие
элемент из группы Лоренца $\MS\MO_\uparrow(1,2)$. При этом групповые операции,
как нетрудно видеть, согласованы. Обратно, по заданному преобразованию Лоренца
уравнение (\ref{elyreq}) определяет матрицу $M$, имеющую единичный определитель,
с точностью до знака. Поскольку группа $\MS\ML(2,\MR)$ связна, то
это устанавливает изоморфизм групп
\begin{equation}                                                  \label{eistrl}
  \MS\MO_\uparrow(1,2)\simeq\frac{\MS\ML(2,\MR)}{\MZ_2}.
\end{equation}

Отображение $\MS\ML(2,\MR)\rightarrow\MS\MO_\uparrow(1,2)$ является двулистным
накрытием. Группа $\MS\ML(2,\MR)$ как многообразие диффеоморфна прямому
произведению $\MS^1\times\MR^2$ (см.\ пример \ref{esltre}). Поэтому она является
связной, но не односвязной. Следовательно, накрытие (\ref{eistrl}) не является
универсальным. Универсальная накрывающая для групп $\MS\MO_\uparrow(1,2)$ и
$\MS\ML(2,\MR)$ построена в примере \ref{esltre}.

Построим явную параметризацию элементов собственной ортохронной группы Лоренца
$\MS\MO_\uparrow(1,2)$ элементами ее алгебры (экспоненциальное отображение). Для
этого рассмотрим произвольный элемент алгебры $X\in\Gs\Gu(2)$ (\ref{oantim})
\begin{equation*}
  X=\om^k\frac i2\s_k=\frac i2 \begin{pmatrix}
    \om^3 & \om^1-i\om^2 \\ \om^1+i\om & -\om^3 \end{pmatrix}
\end{equation*}
где $\s_k$, $k=1,2,3$, -- матрицы Паули. Положим
\begin{equation}                                                  \label{echomx}
  \om^1=-2ix,\qquad\om^2=-2t,\qquad \om^3=-2iy,
\end{equation}
где $t,x,y$ -- некоторые вещественные числа, тогда матрица $X$ примет вид
\begin{equation*}
  X=\begin{pmatrix} y & x-t \\ x+t & -y \end{pmatrix},\qquad t,x,y\in\MR.
\end{equation*}
Здесь мы используем те же буквы, что и в параметризации (\ref{esympa}), но они
имеют другой смысл. Полученная матрица вещественна, симметрична и имеет нулевой
след. То есть представляет собой элемент алгебры $\Gs\Gl(2,\MR)$ общего вида,
т.к.\ параметризуется тремя параметрами.

Поскольку экспонента от произвольной комплексной $n\times n$ матрицы равномерно
сходится в любой ограниченной области комплексного пространства $\MC^{n^2}$, то
экспоненциальной отображение (\ref{esutgr}) можно переписать для группы
$\MS\ML(2,\MR)$, просто произведя замену координат (\ref{echomx}). Поскольку
\begin{equation*}
  \sin(ia)=i\sh a,\qquad \cos(ia)=\ch a,\qquad a\in\MR,
\end{equation*}
то преобразованные матрицы выглядят по-разному в различных точках пространства
Минковского $(t,x,y)\in\MR^{1,2}$. Внутри светового конуса $t^2>x^2+y^2$
экспоненциальное отображение имеет вид
\begin{equation}                                                  \label{einlcm}
  M=\ex^X=\begin{pmatrix}
    \cos s+\frac ys\sin s & \frac{x-t}s\sin s \\[2mm]
    \frac{x+t}s\sin s & \cos s-\frac ys\sin s
    \end{pmatrix}\quad \in\MS\ML(2,\MR),
\end{equation}
где $s:=\sqrt{t^2-x^2-y^2}$. На световом конусе $t^2=x^2+y^2$
\begin{equation}                                                  \label{elicmr}
  M=\ex^X=\begin{pmatrix} 1+y & x-t \\ x+y & 1-y \end{pmatrix}\quad \in\MS\ML(2,\MR).
\end{equation}
Вне светового конуса
\begin{equation}                                                  \label{einlcs}
  M=\ex^X=\begin{pmatrix}
    \ch r+\frac yr\sh r & \frac{x-t}r\sh r \\[2mm]
    \frac{x+t}r\sh r & \ch r-\frac yr\sh r
    \end{pmatrix}\quad \in\MS\ML(2,\MR),
\end{equation}
где $r:=\sqrt{x^2+y^2-t^2}$. Матрицы (\ref{einlcm}) и (\ref{einlcs}) имеют
одинаковый предел при $s\rightarrow0$ и $r\rightarrow0$, который совпадает с
матрицей (\ref{elicmr}).

Экспоненциальное отображение
\begin{equation*}
  \Gs\Gl(2,\MR)\ni\quad X\mapsto M\quad\in\MS\ML(2,\MR)
\end{equation*}
не является однозначным, и многим элементам алгебры соответствует один и тот же
элемент группы. То есть групповое многообразие $\MS\MO_\uparrow(1,2)$ получается
из $\MR^3$ после отождествления некоторых точек. Например, единице группы
соответствует начало координат и все двуполостные гиперболоиды
\begin{equation*}
  s=2\pi m,\qquad m=1,2,\dotsc.
\end{equation*}
Из вида матрицы (\ref{einlcm}) следует, что внутри светового конуса существует
отношение эквивалентности
\begin{equation*}
  x^a\sim x^a+2\pi\frac{x^a}s.
\end{equation*}
Как и в случае группы $\MS\MU(2)$ существует второе отношение эквивалентности.
А именно, необходимо отождествить также точки, лежащие на гиперболоиде $s=\pi$,
т.к.\ всем этим точкам соответствует одна и та же матрица $-\one$.
\subsection{Четырехмерное пространство Минковского               \label{sfomin}}
Обсудим ряд специфических свойств группы Лоренца в четырехмерном пространстве
Минковского $\MR^{1,3}$, которые часто используются в физических приложениях.

\begin{prop}
В четырехмерном пространстве Минковского $\MR^{1,3}$ любая дифференцируемая
изотропная кривая $\lbrace x^a(s)\rbrace$, определяемая уравнением
\begin{equation}                                                  \label{equlil}
  \et_{ab}\dot x^a\dot x^b=0,
\end{equation}
представима в виде
\begin{equation}                                                  \label{enulfm}
\begin{aligned}
  x^0 &=\int^s\!\!\!du\,R,
\\
  x^1 &=\int^s\!\!\!du\,R\sin\theta\cos\vf,
\end{aligned}\qquad
\begin{aligned}
  x^2 &=\int^s\!\!\!du\, R\sin\theta\sin\vf,
\\
  x^3 &=\int^s\!\!\!du\, R\cos\theta,
\end{aligned}
\end{equation}
где каждая из координат определена с точностью до постоянной, а
$R(u)$, $\theta(u)$ и $\vf(u)$ -- произвольные непрерывные функции от $u$. Эти
линии являются прямыми тогда и только тогда, когда функции $R,\theta,\vf$
постоянны.
\end{prop}
\begin{proof}
Явное решение алгебраического уравнения (\ref{equlil}) относительно производных
и последующее интегрирование.
\end{proof}
Отсюда следует, что, в отличии от плоскости Минковского, класс изотропных кривых
состоит не только из прямых линий. Поскольку все экстремали (см.\ главу
\ref{sgextr}) в пространстве Минковского и только они являются прямыми, то это
доказывает, что не всякая изотропная кривая является экстремалью. Исключение
составляет только двумерное пространство Минковского, где все изотропные кривые
являются прямыми и, следовательно, экстремалями.
\begin{prop}
Любая изотропная кривая в пространстве Минковского, проходящая через точку $y$,
не может выйти за пределы светового конуса с вершиной в этой точке.
\end{prop}
\begin{proof}
Допустим, что в некоторой точке $y_1$, лежащей на световом конусе будущего и не
совпадающей с $y$, изотропная кривая покидает световой конус (см.\
рис.\ref{flight}). Тогда в некоторой окрестности данной точки условие
(\ref{equlil}) будет нарушено. Это ясно из того, что световой конус будущего с
вершиной в точке $y_1$ касается светового конуса будущего в точке $y$ и целиком
лежит внутри него.
\end{proof}

Четырехмерное пространство Минковского, $\MR^{1,3}=\MR^{1,1}\times \MR^2$, можно
представить как прямое произведение двумерной плоскости Минковского $\MR^{1,1}$
и евклидовой плоскости $\MR^2$ с координатами $(x^0,~x^1)$ и $(x^2,~x^3)$
соответственно. При таком разбиении можно ввести координаты светового конуса
(\ref{econco}) на $\MR^{1,1}$ и комплексные координаты (\ref{epofcn}) на
$\MR^2$. Тогда лоренцев интервал примет вид
\begin{equation}                                                  \label{eintmc}
  ds^2=dudv-dzd\bar z,
\end{equation}
где $u:=x^0+x^1$, $v:=x^0-x^1$, $z:=x^2+ix^3$ и $\bar z:=x^2-ix^3$.
Соответствующая метрика в координатах
$\lbrace x^a\rbrace=\lbrace u,v,z,\bar z\rbrace$ имеет вид
$$
  \eta=\frac12\begin{pmatrix}
  0&1&0&0\\1&0&0&0\\0&0&0&-1\\0&0&-1&0 \end{pmatrix}.
$$
В моделях гравитации используется комплексная изотропная тетрада, которая
соответствует введенным выше координатам в касательном пространстве к
пространству-времени.

Коммутационные соотношения, определяющие алгебру Пуанкаре
(\ref{epoalo})--(\ref{epoalt}), в четырехмерном пространстве-времени можно
переписать в другом, эквивалентном, виде, выделив явно пространственные и
временн\'ые компоненты. Среди координат пространства Минковского, как и ранее,
явно выделяем время $\lbrace x^a\rbrace=\lbrace x^0,x^i\rbrace$, $i=1,2,3$.
Группа Пуанкаре содержит подгруппу трехмерных вращений, алгебра которой
определяется генераторами $L_{ij}$. В связи с этим вместо генераторов $L_{ab}$
можно рассматривать 6 генераторов
\begin{equation*}
  J_i:=\frac12\ve_i{}^{jk}L_{jk},\qquad N_i:=L_{0i},
\end{equation*}
где подъем и опускание латинских индексов $i,j=1,2,3$ осуществляется с помощью
евклидовой метрики $\dl_{ij}$. Образующие алгебры Ли группы Лоренца $J_i$ и
$N_i$ представляют собой генераторы пространственных вращений и бустов. Нетрудно
проверить, что они удовлетворяют следующим коммутационным соотношениям
\begin{equation}                                                  \label{eloraz}
\begin{split}
  [J_i,J_j]&=-\ve_{ij}{}^kJ_k,
\\
  [N_i,N_j]&=\quad \ve_{ij}{}^kJ_k,
\\
  [J_j,N_j]&=-\ve_{ij}{}^kN_k.
\end{split}
\end{equation}
Последнее коммутационное соотношение показывает, что базисные векторы $N_i$
преобразуются как компоненты ковектора при пространственных вращениях.
Остальные коммутационные соотношения (\ref{epoalo}), (\ref{epoals}) примут вид
\begin{equation}                                                  \label{epoalw}
\begin{split}
  [J_i,P_0]&=0,
\\
  [J_i,P_j]&=-\ve_{ijk}P^k,
\\
  [N_i,P_0]&=P_i,
\\
  [N_i,P_j]&=\dl_{ij}P_0.
\end{split}
\end{equation}

Построим гомоморфизм группы комплексных $2\times2$-матриц с единичным
определителем $\MS\ML(2,\MC)$ на связную компоненту единицы группы Лоренца
$\MS\MO_\uparrow(1,3)$. Это построение аналогично доказательствам существования
накрытий $\MS\MU(2)\rightarrow\MS\MO(3)$ и
$\MS\ML(2,\MR)\rightarrow\MS\MO_\uparrow(1,2)$, проведенным в разделах
\ref{sthecs} и \ref{sthmio}. Для построения гомоморфизма заметим, что
произвольную эрмитову матрицу $A=A^\dagger$ можно представить в виде
\begin{equation}                                                  \label{ermapa}
  A=\begin{pmatrix} t+z & x-iy \\ x+iy & t-z \end{pmatrix}=t+x\s_1+y\s_2+z\s_3,
  \qquad t,x,y,z\in\MR,
\end{equation}
где $\s_1$, $\s_2$ и $\s_3$ -- матрицы Паули.
Рассмотрим числа $t,x,y,z$ как координаты в пространстве-времени Минковского
$\MR^{1,3}$. Очевидно, что
$$
  \det A=t^2-x^2-y^2-z^2.
$$
Пусть $M\in\MS\ML(2,\MC)$ -- произвольная комплексная матрица с единичным
определителем. Тогда каждому преобразованию
\begin{equation}                                                  \label{ermtrm}
  A\mapsto A'=MAM^\dagger,
\end{equation}
можно однозначно поставить в соответствие некоторое собственное ортохронное
преобразование Лоренца. Действительно, при таком преобразовании матрица $A'$
остается эрмитовой, и ее можно записать в виде
$$
  A'=\begin{pmatrix} t'+z' & x'-iy' \\ x'+iy' & t'-z' \end{pmatrix}
$$
с некоторыми новыми координатами $t',x',y',z'$ в $\MR^{1,3}$. Поскольку при этом
определитель матрицы не меняется,
$$
  \det A'=\det A,
$$
то каждому элементу $M\in\MS\ML(2,\MC)$ однозначно ставится в соответствие
элемент из группы Лоренца $\MS\MO_\uparrow(1,3)$. Прямая проверка показывает,
что отображение
\begin{equation}                                                  \label{ecovos}
  \MS\ML(2,\MC)\rightarrow\MS\MO_\uparrow(1,3)
\end{equation}
является гомоморфизмом групп. Обратно, по заданному преобразованию Лоренца
уравнение (\ref{ermtrm}) определяет матрицу $M$ с точностью до знака, причем
отображение (\ref{ecovos}) сюрьективно \cite{Naimar58R}. Поскольку
группа $\MS\ML(2,\MC)$ связна, то это устанавливает изоморфизм групп
\begin{equation}                                                  \label{esoslr}
  \MS\MO_\uparrow(1,3)\simeq\frac{\MS\ML(2,\MC)}{\MZ_2}.
\end{equation}
При этом тождественному преобразованию Лоренца соответствуют две матрицы $\one$
и $-\one$ из группы $\MS\ML(2,\MC)$. Можно доказать (см., например,
\cite{BarRac77R}), что группа $\MS\ML(2,\MC)$ является односвязной. Это значит,
что построенное отображение представляет собой двулистное универсальное накрытие
(см.\ раздел \ref{scover}). Таким образом справедлива
\begin{theorem}
Существует гомоморфизм групп (\ref{ecovos}), который представляет собой
двулистное универсальное накрытие.
\end{theorem}
\begin{defn}
Точное представление универсальной накрывающей $\MS\ML(2,\MC)$ называется
{\em спинорным} представлением собственной ортохронной группы Лоренца
$\MS\MO_\uparrow(1,3)$. При этом каждому вращению $S\in\MS\MO_\uparrow(1,3)$
ставится в соответствие неупорядоченная пара элементов $\pm M\in\MS\ML(2,\MC)$.
\qed\end{defn}
\index{Спинорное представление (spinor representation)}%
\index{Представление спинорное (spinor representation)}%

Избавиться от неоднозначности спинорного представления, т.е.\ упорядочить пару
элементов $\pm M$, нельзя, т.к.\ при тождественном преобразовании
(пространственное вращение на угол $2\pi$) матрица $M$ изменит знак. Поэтому
говорят, что оно является двузначным представлением собственной ортохронной
группы Лоренца. Таким образом, спинорное представление связной компоненты
единицы группы Лоренца $\MS\MO_\uparrow(1,3)$ является комплексным двумерным.
Спинорные представления полной группы Лоренца $\MO(1,3)$ являются комплексными
четырехмерными.

Собственные вращения $\MS\MO(3)$ в трехмерном евклидовом пространстве,
определяемом уравнением $t=0$ в пространстве Минковского, образуют подгруппу
группы Лоренца $\MS\MO_\uparrow(1,3)$. Пусть $\tilde S\in\MS\MO(3)$. Тогда
каждой матрице вращений $\tilde S$ соответствуют две матрицы $\pm U$ из группы
специальных унитарных матриц $\MS\MU(2)$, которая является подгруппой в
$\MS\ML(2,\MC)$. Это следует из параметризации (\ref{ermapa}). Заметим, что для
преобразований (\ref{ermtrm}) с унитарной матрицей
$U\in\MS\MU(2)\subset\MS\ML(2,\MC)$ сохраняется не только определитель, но и
след, $\tr A'=\tr A$. Так как $\tr A=2t$, то каждому преобразованию из подгруппы
унитарных унимодулярных матриц ставится в соответствие такое преобразование
Лоренца, которое не затрагивает время, т.е.\ некоторое вращение
$\tilde S\in\MS\MO(3)$.

Аналогичное рассуждение применимо к подгруппе вещественных матриц
$\MS\ML(2,\MR)\subset\MS\ML(2,\MC)$. Если $M\in\MS\ML(2,\MR)$, то
$M^\St=M^\dagger$, и преобразование (\ref{ermtrm}) переходит в $A'=MAM^\St$,
которое отображает вещественные симметричные матрицы на себя. Это соответствует
таким преобразованиям пространства Минковского, которые не затрагивает
координату $y$ в параметризации (\ref{ermapa}). Следовательно, каждой
вещественной унимодулярной матрице $M$ из $\MS\ML(2,\MR)$ ставится в
соответствие некоторое собственное ортохронное преобразование Лоренца из
$\MS\MO_\uparrow(1,2)$, которое действует только на координаты $t,x,z$ в
пространстве Минковского.

Посмотрим, что представляют из себя группа Лоренца $\MS\MO_\uparrow(1,3)$ и ее
универсальная накрывающая $\MS\ML(2,\MC)$ как многообразия.
\begin{prop}
Группа специальных комплексных $2\times2$ матриц $\MS\ML(2,\MC)$ и собственная
ортохронная группа Лоренца $\MS\MO_\uparrow(1,3)$ как многообразия представляют
собой прямые произведения:
\begin{align}                                                     \label{edipsl}
  \MS\ML(2,\MC)&\approx\MS^3\times\MR^3.
\\                                                                \label{edilof}
  \MS\MO_\uparrow(1,3)&\approx\MR\MP^3\times\MR^3.
\end{align}
\end{prop}
\begin{proof}
Согласно полярному разложению матриц (теорема \ref{tpomac}) произвольная матрица
$M\in\MS\ML(2,\MC)$ однозначно представима в виде
\begin{equation}                                                  \label{epomal}
  M=RU,
\end{equation}
где $R=\sqrt{MM^\dagger}$ -- положительно определенная эрмитова матрица и $U$ --
унитарная матрица. Поскольку $\det M=1$, то выполнены равенства $\det R=1$ и
$\det U=1$. Следовательно, $U\in\MS\MU(2)$. Множество эрмитовых $2\times2$
матриц с единичным определителем имеет вид (\ref{ermapa}) с единственным
ограничением
\begin{equation*}
  \det R=t^2-x^2-y^2-z^2=1.
\end{equation*}
Поскольку матрица $R$ положительно определена, то нам необходимо выбрать верхнюю
полу однополостного гиперболоида, соответствующую значениям $t>0$. Она, как
многообразие, диффеоморфна трехмерному евклидову пространству $\MR^3$. Ранее
было установлено, что многообразие группы $\MS\MU(2)$ представляет собой сферу
$\MS^3$. Таким образом построен диффеоморфизм (\ref{edipsl}).

Собственная ортохронная группа Лоренца $\MS\MO_\uparrow(1,3)$ возникает после
отождествления $M\sim-M$. В полярном разложении (\ref{epomal}) это означает
отождествление унитарных матриц $U\sim-U$, т.к.\ матрица $R$ положительно
определена. Это отождествление соответствует накрытию трехмерного проективного
пространства $\MR\MP^3$ трехмерной сферой $\MS^3$ (см.\ раздел \ref{sthecs}).
Следовательно, существует диффеоморфизм (\ref{edilof}).
\end{proof}

Существование диффеоморфизмов (\ref{edipsl}) и (\ref{edilof}) согласуется с
общей теоремой \ref{tnolig}. Тем самым мы показали, что подгруппы $\MS\MU(2)$ и
$\MS\MO(3)$ являются максимальными компактными подгруппами в $\MS\ML(2,\MC)$ и
$\MS\MO_\uparrow(1,3)$ соответственно.

Определим оператор пространственного отражения (четности) в пространстве
Минковского как отражение всех пространственных координат
\begin{equation*}
  P:\quad (x^0,x^1,x^2,x^3)\mapsto (x^0,-x^1,-x^2,-x^3).
\end{equation*}
Тогда внешние автоморфизмы (\ref{emapts}), вызванные обращением времени и
преобразованием четности, совпадают, т.к.\ $P=-T$. При инверсии пространственных
координат матрица преобразований Лоренца меняется по правилу (\ref{extalo}):
\begin{equation*}
  P:\quad S\mapsto PSP^{-1}=(S^\St)^{-1}.
\end{equation*}
Пусть пространственному вращению $\tilde S\in\MS\MO(3)$ соответствуют две
комплексные матрицы $\pm\tilde A\in\MS\MU(2)$. Тогда внешнему автоморфизму
(\ref{extalo}) соответствует автоморфизм в группе $\MS\ML(2,\MC)$
\begin{equation*}
  P:\quad \pm \tilde A~\mapsto~\pm(\tilde A^\St)^{-1}.
\end{equation*}
Последний будет внутренним, т.к.\ группа $\MS\ML(2,\MC)$ связна. Это значит, что
представление группы $\MS\ML(2,\MC)$ является спинорным представлением не только
для связной компоненты единицы $\MS\MO_\uparrow(1,3)$, но и группы
$\MO_\uparrow(1,3)$, включающей преобразование четности.

Группа Лоренца является некомпактной, ее представления могут быть неунитарны, а
среди ее неприводимых представлений есть бесконечномерные. Известно, что все
конечномерные неприводимые представления, за исключением тривиального
единичного, группы Лоренца неунитарны \cite{Nimark58R}. Это существенно отличает
ее от группы вращений в евклидовом пространстве и представляет существенные
трудности при построении физических моделей, где требуется положительная
определенность энергии и скалярного произведения в гильбертовом пространстве
соответствующей квантовой теории. Из изоморфизма (\ref{esoslr}) следует, что
точное представление группы $\MS\ML(2,\MC)$ является двузначным или спинорным
представлением группы $\MS\MO_\uparrow(1,3)$. При этом представление собственной
ортохронной группы Лоренца $\MS\MO_\uparrow(1,3)$ однозначно или двузначно
одновременно с порожденным им представлением группы вращений $\MS\MO(3)$.

Построим параметризацию группы $\MS\ML(2,\MC)$ элементами ее алгебры. Алгебра Ли
$\Gs\Gl(2,\MC)$ состоит из комплексных $2\times2$-матриц с нулевым следом (6
вещественных параметров). Алгебру Ли $\Gs\Gl(2,\MC)$ можно рассматривать как
трехмерное векторное пространство над полем комплексных чисел. В качестве базиса
выберем матрицы Паули $\s_i$ (см.\ приложение \ref{spamat}). Тогда произвольный
элемент алгебры Ли $\Gs\Gl(2,\MC)$ параметризуется вектором $\Bz$ в трехмерном
комплексном пространстве
\begin{equation*}
  \Bz=z^k\s_k,\qquad \Bz=\lbrace z^k\rbrace\in\MC^3,\quad k=1,2,3.
\end{equation*}
\begin{prop}
Каждому элементу алгебры Ли $\Bz=\lbrace z^k\rbrace\in\Gs\Gl(2,\MC)\approx\MC^3$
соответствует элемент группы Ли $S=\lbrace S_\Sa{}^\Sb\rbrace\in\MS\ML(2,\MC)$
\begin{equation}                                                  \label{esltce}
\begin{split}
  S_\Sa{}^\Sb(\Bz)=\left(\ex^{iz^k\s_k/2}\right)_\Sa{}^\Sb
  &=\dl_\Sa^\Sb\cos\frac{\sqrt{\Bz^2}}2
  +i\frac{z^k\s_{k\Sa}{}^\Sb}{\sqrt{\Bz^2}}\sin\frac{\sqrt{\Bz^2}}2=
\\
  &=\begin{pmatrix}
  \cos\frac{\sqrt{\Bz^2}}2+i\frac{z^3}{\sqrt{\Bz^2}}\sin\frac{\sqrt{\Bz^2}}2 &
  i\frac{z^1-iz^2}{\sqrt{\Bz^2}}\sin\frac{\sqrt{\Bz^2}}2 \\[2mm]
  i\frac{z^1+iz^2}{\sqrt{\Bz^2}}\sin\frac{\sqrt{\Bz^2}}2 &
  \cos\frac{\sqrt{\Bz^2}}2-i\frac{z^3}{\sqrt{\Bz^2}}\sin\frac{\sqrt{\Bz^2}}2
\end{pmatrix},
\end{split}
\end{equation}
где $\Bz^2:=(z^1)^2+(z^2)^2+(z^3)^2$ -- комплексной число.
\end{prop}
\begin{proof}
Формальное переписывание (\ref{esutgr}), т.к.\ все разложения равномерно
сходятся в произвольном ограниченном шаре в $\MC^3$.
\end{proof}
Многообразие группы $\MS\ML(2,\MC)$ возникает после отождествления точек в
$\MC^3$, т.к.\ некоторым элементам алгебры соответствует один и тот же элемент
группы. В рассматриваемом случае это отождествление выглядит громоздко, и мы не
будем его проводить.
\section{Специальная теория относительности}
\index{Специальная теория относительности (special theory of relativity)}%
\index{Теория относительности специальная (special theory of relativity)}%
Специальная теория относительности -- это учение о пространстве-времени, в
котором строятся современные модели всех взаимодействий за исключением
гравитационных. В двух словах суть специальной теории относительности состоит в
утверждении, что пространство-время представляет собой пространство Минковского
$\MR^{1,3}$, в котором определена метрика Лоренца $\eta_{ab}=\diag(+---)$, и все
законы Природы записываются в виде некоторой системы уравнений, которые
ковариантны относительно преобразований Лоренца.

Мы начнем с описания основных свойств нерелятивистских моделей, а затем обсудим
некоторые свойства пространства Минковского, которые затем будут обобщены в
теории тяготения Эйнштейна. Кроме того, рассмотренные примеры из специальной
теории относительности помогают развить определенную интуицию, которая полезна
при исследовании многообразий общего вида, на которых задана метрика лоренцевой
сигнатуры.
\subsection{Нерелятивистские модели}
До создания специальной теории относительности модели математической физики
были нерелятивистскими и строились в пространстве-времени, в котором время
играет выделенную роль.
\begin{defn}
{\em Галилеевым} пространством-временем называется топологически тривиальное
четырехмерное многообразие $\MM=\MR\times\MR^3$, где первый сомножитель $\MR$
соответствует {\em времени} $t$, а второй $\MR^3$ -- {\em пространству}, со
следующей структурой. Мы
предполагаем,  что в пространстве $\MR^3$ задана евклидова метрика. Пусть $x^i$,
$i=1,2,3$, -- декартовы координаты в пространстве $\MR^3$, в которых метрика
имеет диагональный вид $\dl_{ij}=\diag(+++)$. Каждая точка пространства-времени
$x=\lbrace t,x^i\rbrace\in\MM$ называется {\em событием} и говорит о том, что
нечто произошло в момент времени $t$ в точке пространства $\lbrace x^i\rbrace$.
События, соответствующие фиксированному значению $t$, называются
{\em одновременными}. Пусть задано два произвольных события  $x_1,x_2\in\MM$ с
координатами $x_1=\lbrace t_1,x^i_1\rbrace$ и $x_2=\lbrace t_2,x^i_2\rbrace$.
Тогда временн\'ой интервал между этими событиями определяется разностью
\begin{equation}                                                  \label{etintd}
  \triangle t:=t_2-t_1.
\end{equation}
Если $t_2>t_1$, то мы говорим, что событие $x_2$ произошло позже события $x_1$.
В противном случае, $t_2<t_1$, событие $x_2$ произошло раньше события $x_1$.
Если $t_2=t_1$, то мы говорим, что события произошли одновременно.
Пространственное расстояние между двумя одновременными событиями
$x_1=\lbrace t,x^i_1\rbrace$ и $x_2=\lbrace t,x^i_2\rbrace$ равно евклидову
расстоянию в $\MR^3$:
\begin{equation}                                                  \label{espdis}
  \triangle x:=\sqrt{\dl_{ij}(x^i_2-x^i_1)(x^j_2-x^j_1)}.
\end{equation}
Система координат $t,x^i$ называется {\em инерциальной} и обозначается $O$.
Кроме того, мы предполагаем, что галилеево пространство-время снабжено
естественной линейной и аффинной структурой, такой же как и евклидово
пространство $\MR^4$ в разделах \ref{seucve} и \ref{seucaf}.
\qed\end{defn}
\index{Галилеево пространство-время (Galilean space-time)}%
\index{Пространство-время галилеево (Galilean space-time)}%
\index{Время (time)}\index{Пространство (space)}\index{Событием (event)}%
\index{События одновременные (simultaneous events)}%
\index{Одновременные события (simultaneous events)}%
\index{Инерциальная система координат (inertial coordinate system)}%
\index{Система координат инерциальная (inertial coordinate system)}%

Одновременные события составляют пространственные сечения пространства-времени
$\MM$, соответствующие фиксированному значению времени $t$. Из определения
следует, что понятия ``раньше'' и ``позже'' не зависят от выбора событий в
каждом сечении, т.е.\ временн\'ой интервал определен для двух произвольных
сечений $t_1$ и $t_2$. Другими словами, на оси времени $t\in R$ задана обычная
топологическая евклидова метрика $l(t_1,t_2)=|t_2-t_1|$, которая определяет
расстояние между двумя пространственными сечениями. Единственное отличие -- это
наличие знака у временн\'ого интервала, который соответствует понятиям
``раньше'' и ``позже''.

Подчеркнем, что пространственное расстояние можно определить только для
одновременных событий. Если две системы координат движутся друг относительно
друга с ненулевой скоростью и события не являются одновременными, то, как легко
видеть, пространственное расстояние (\ref{espdis}) между этими событиями,
измеренное в этих системах координат, будет различным.

Данное выше определение галилеева пространства-времени содержит одну
инерциальную систему отсчета $O$. Расширим это понятие. Временн\'ой
интервал между двумя произвольными событиями и пространственное расстояние между
двумя одновременными событиями инвариантны относительно преобразований Галилея
(\ref{egaltr}). Обратное утверждение в общем случае неверно.

Обсудим это. Сохранение временн\'ого интервала оставляет возможность только для
сдвигов вдоль оси времени. Эти сдвиги не могут зависеть от точки пространства,
т.к.\ мы требуем сохранение интервала (\ref{etintd}) не для отдельных событий,
а для сечений.

Пространственная метрика (\ref{espdis}) инвариантна только относительно
преобразований из неоднородной группы вращений $\MI\MO(3)$. В общем случае
преобразований пространства-времени $\MR\times\MR^3$ и матрица вращений, и
вектор сдвига пространства могут зависеть от времени, причем произвольно.
Однако, если ограничиться линейными преобразованиями, то остаются только
преобразования из группы Галилея. Снабдим галилеево пространство-время обычной
структурой линейного пространства из $\MR^4$. Тогда два произвольных различных
события определяют единственную прямую (\ref{estrli}), проходящую через эти
события. При преобразованиях Галилея линейная структура сохраняется и прямые
переходят в прямые. Верно также обратное утверждение: любое преобразование
пространства-времени, которое сохраняет временн\'ой интервал (\ref{etintd}),
пространственную метрику (\ref{espdis}) и линейную структуру
пространства-времени, называется преобразованием Галилея.

\begin{defn}
Любая система координат в галилеевом пространстве-времени $\MM=\MR\times\MR^3$,
связанная с системой координат $O$ преобразованием из группы Галилея,
называется {\em инерциальной}.
\qed\end{defn}

Точечные частицы движутся в пространстве по траекториям
$\lbrace x^i(t)\rbrace\in\MR^3$. Эти три функции задают мировую линию частицы в
пространстве-времени $\lbrace t,x^i(t)\rbrace\in\MM$. Мировые линии частиц
вводятся точно таким же образом и в специальной, и в общей теории
относительности. Различие заключается лишь в том, что время в нерелятивистской
механике имеет абсолютный характер, и его естественно выбрать в качестве
параметра вдоль мировых линий всех частиц. Такой выбор общепринят и нагляден с
физической точки зрения.

\begin{defn}
Если в некоторой инерциальной системе координат мировая линия частицы
представляет собой прямую линию вида $\lbrace t,x^i=v^it+x^i_0\rbrace\in\MM$,
где $v^i,x^i_0$ -- некоторые постоянные, то такая частица называется {\em
свободной}. Точка $\lbrace x^i_0\rbrace\in\MR^3$ является точкой пространства,
в которой частица расположена в начальный момент времени $t=0$, а векторное поле
$\Bv=\lbrace v^i\rbrace\in\MT(\MR^3)$, определенное вдоль траектории
частицы, называется {\em наблюдаемой скоростью}.
\qed\end{defn}
\index{Свободная частица (free particle)}%
\index{Частица свободная (free particle)}%
\index{Наблюдаемая скорость (observed velocity)}%
\index{Скорость наблюдаемая (observed velocity)}%

Из определения инерциальной системы координат следует, что, если мировая линия
частицы является прямой в какой либо инерциальной системе отсчета, то она будет
прямой и в любой другой инерциальной системе.
\begin{com}
В физической литературе понятие инерциальной системы координат часто определяют
следующим образом. Систему координат называют инерциальной, если свободная
частица движется в ней равномерно и прямолинейно. В таком определении не
сказано, что значит свободная частица. Поэтому мы пошли другим путем. Сначала
было построено пространство-время и инерциальные системы отсчета. Это позволило
дать определение свободной частицы.
\qed\end{com}
\begin{com}
В пространстве-времени $\MM$ не существует метрики, инвариантной относительно
преобразований Галилея, потому что эта группа не является полупростой. Поэтому
говорить про экстремали не имеет смысла. В галилеевом пространстве-времени
заданы естественные линейная и аффинная структуры, и параллельный перенос мы
отождествляем с трансляциями. Это определяет связность на $\MM$,
которая в инерциальной системе отсчета имеет нулевые компоненты. Эта связность
плоская, а геодезические являются прямыми линиями.
\qed\end{com}

Линейность функций $x^i(t)$ означает, что свободная частица движется в
пространстве равномерно и прямолинейно. Это утверждение известно как {\em первый
закон Ньютона}.
\index{Первый закон Ньютона (First Newton's law)}%
\index{Закон Ньютона первый (First Newton's law)}%
Мировая линия свободной частицы в пространстве-времени в любой инерциальной
системе координат представляет собой прямую линию. Мы говорим, что частица не
является свободной, если ее мировая линия отличается от прямой. В этом случае
траектория частицы имеет ненулевое ускорение $\Ba:=d\Bv/dt$. Мы говорим, что на
частицу действует сила $\BF$, которая, по-определению, пропорциональна
ускорению:
\begin{equation}                                                  \label{esenel}
  \BF=m\Ba,
\end{equation}
где коэффициент пропорциональности $m$ называется {\em инертной массой}. Это
соотношение называется {\em вторым законом Ньютона}.
\index{Второй закон Ньютона (second Newton's law)}%
\index{Закон Ньютона второй (second Newton's law)}%
\index{Инертная масса (inertial mass)}%
\index{Масса инертная (inertial mass)}%

Если имеется две взаимодействующих частицы, то между ними действует сила. Пусть
$\BF$ -- сила, действующая на первую частицу со стороны второй. Тогда сила,
действующая на вторую частицу со стороны первой, равна $-\BF$. Это утверждение
составляет {\em третий закон Ньютона}. Коротко говорят: ``Сила действия равна
силе противодействия''.
\index{Третий закон Ньютона (third Newton's law)}%
\index{Закон Ньютона третий (third Newton's law)}%

Для механики Галилея справедливо следующее правило сложения скоростей. Если
частица движется относительно инерциальной системы координат с постоянной
скоростью $\Bv=\lbrace v^i\rbrace$, то ее скорость движения относительно другой
инерциальной системы координат, движущейся со скоростью $\BV=\lbrace V^i\rbrace$
относительно первой, равна разности $\Bv-\BV=\lbrace v^i-V^i\rbrace$.

Модели математической физики, которые инвариантны или ковариантны относительно
преобразований Галилея называются {\em нерелятивистскими}.
\begin{exa}
В теории тяготения Ньютона две точечные частицы с массами $m_1$ и $m_2$,
находящиеся соответственно в точках $\Bx_1$ и $\Bx_2$ (в инерциальной системе
отсчета), испытывают гравитационное притяжение. При этом сила, действующая со
стороны второй частицы на первую, равна
\begin{equation}                                                  \label{enewgr}
  \BF_{12}=G\frac{m_1m_2}{\Br^2}\frac\Br{|\Br|},
\end{equation}
где $G$ -- {\em гравитационная постоянная}, $\Br:=\Bx_2-\Bx_1$ и $|\Br|$ --
расстояние между частицами. Выражение (\ref{enewgr}) называется законом
{\em всемирного тяготения} или {\em законом тяготения Ньютона}.
\index{Закон всемирного тяготения (Newton's gravitational law)}%
\index{Закон гравитации Ньютона (Newton's gravitational law)}%
\index{Ньютона закон гравитации (Newton's gravitational law)}%
\index{Гравитационная постоянная (gravitational constant)}%
\index{Постоянная гравитационная (gravitational constant)}%
Теория тяготения Ньютона является нерелятивистской моделью, в которой время
имеет абсолютный характер. Она инвариантна относительно преобразований Галилея,
поскольку сила гравитационного взаимодействия двух тел в каждый момент времени
зависит только от их масс и расстояния между ними. Массы частиц $m_1$ и $m_2$
называются {\em гравитационными массами}. В механике Ньютона инертная и
гравитационная массы частиц считаются равными. В этом случае из закона
всемирного тяготения и второго закона Ньютона (\ref{esenel}) следует, что
ускорение, которая испытывает частица массы $m_1$ в поле тяжести частицы $m_2$,
не зависит от ее массы. Это следствие проверено в поле тяжести Земли и Солнца с
высокой степенью точности.
\qed\end{exa}
\index{Гравитационная масса (gravitational mass)}%
\index{Масса гравитационная (gravitational mass)}%

В классической механике взаимодействие частиц описывается энергией
потенциального взаимодействия, которая является функцией от координат частиц.
Тем самым изменение положения одной из взаимодействующих частиц отразится на
других частицах в тот же момент времени. Это соответствует предположению о том,
что взаимодействие распространяется с бесконечной скоростью. Опыт, однако,
показывает, что мгновенных взаимодействий в природе не существует. Если сдвинуть
одну частицу, то это отразится на других частицах только через некоторое время,
которое определяется скоростью распространения взаимодействий.

Электромагнитные взаимодействия распространяются со скоростью света
\begin{equation*}
  c=2,998\cdot 10^{10}~\text{см/с}\approx 300 000~\text{км/с}.
\end{equation*}
В 1887 году Майкельсон и Морли \cite{MicMor87} экспериментально установили с
точностью до 5 км/с, что скорость света постоянна и не зависит от того, движется
ли он параллельно траектории Земли или в перпендикулярном направлении. Точность
этого результата сейчас доведена до 1 км/с \cite{JaJaMuTo64}.

Постоянство скорости света находится в явном противоречии с правилом сложения
скоростей в механике Галилея и с предположением о мгновенности распространения
взаимодействий. Это и привело к созданию специальной теории относительности.
\subsection{Релятивистские модели}
Специальная теория относительности изменила наше представление о структуре
пространства-времени. В релятивистских моделях отсутствует понятие одновременных
событий и пространственного расстояния между двумя одновременными событиями. Эти
обстоятельства являются наиболее трудными в процессе понимания релятивистских
моделей, т.к.\ наша интуиция основана на механике Ньютона.
\begin{defn}
{\em Пространством Минковского} $\MR^{1,3}$ называется четырехмерное
топологически тривиальное многообразие с заданной лоренцевой метрикой в
инерциальной системе отсчета и естественными линейной и аффинной структурами
евклидова пространства.
\index{Пространство Минковского (Minkowskian space)}%
\index{Минковского пространство (Minkowskian space)}%
\qed\end{defn}
В специальной теории относительности мы предполагаем, что все законы природы
формулируются в пространстве Минковского. Как линейное пространство
пространство-время $\MR^{1,3}$ такое же, как и в нерелятивистских моделях. С
топологической точки зрения это просто прямое произведение,
$\MR^{1,3}=\MR\times\MR^3$. Точка пространства-времени, как и раньше,
называется событием, и каждая точечная частица движется вдоль своей мировой
линии в $\MR^{1,3}$. Отличие заключается в том, что вместо понятий
одновременности и пространственного расстояния между одновременными событиями,
мы постулируем существование инвариантной метрики в пространстве Минковского
$\MR^{1,3}$. В декартовой системе координат $x^a$, $a=0,1,2,3$, лоренцева
метрика, по-определению, имеет диагональный вид $\eta_{ab}:=\diag(+,-,-,-)$.

Как и в нерелятивистских моделях, в специальной теории относительности вводится
понятие инерциальной системы координат. Однако ее определение другое.
\begin{defn}
Система координат в пространстве Минковского $\MR^{1,3}$, в которой метрика
имеет вид $\eta_{ab}=\diag(+---)$, называется {\em инерциальной}. Если
мировая линия частицы в инерциальной системе координат представляет собой прямую
линию, то она называется {\em свободной}.
\qed\end{defn}
\index{Инерциальная система координат (inertial coordinate system)}%
\index{Система координат инерциальная (inertial coordinate system)}%
\index{Свободная частица (free particle)}%
\index{Частица свободная (free particle)}%
\begin{com}
Напомним, что прямые линии в пространстве Минковского и только они являются
экстремалями. Поэтому в определении свободной частицы ``прямую линию'' можно
заменить на ``экстремаль''.
\qed\end{com}
\begin{com}
Данное определение инерциальной системы координат по своей сути совпадает с
определением декартовой системы координат, которое было дано в разделе
\ref{seucme} в случае евклидовой метрики.
\qed\end{com}

Инерциальные системы координат определены неоднозначно. Любые две инерциальные
системы координат связаны между собой преобразованием из полной группы Пуанкаре,
которая включает в себя преобразования Лоренца (собственные и несобственные),
сдвиги, преобразование четности и обращение времени (см.\ раздел \ref{sminsp}).
Обратно, если некоторая система координат связана с инерциальной преобразованием
из группы Пуанкаре, то она сама является инерциальной. Это следует из
определений группы Пуанкаре и инерциальной системы координат.

Вторым постулатом специальной теории относительности является предположение о
том, что все физические явления описываются некоторым набором полей на
пространстве Минковского $\MR^{1,3}$, которые преобразуются по какому то,
возможно, приводимому представлению группы Пуанкаре.

В физической литературе принято формулировать\newline
{\bf Принцип относительности.} Все физические законы природы (уравнения
движения, равновесия и т.д.) инвариантны или ковариантны относительно
преобразований из группы Пуанкаре.
\qed
\index{Принцип относительности (relativity principle)}%
\index{Относительности принцип (relativity principle)}%
\begin{exa}
Уравнение Клейна--Гордона--Фока для скалярного поля (\ref{eqmrsc}) в
пространстве Минковского инвариантно относительно преобразований из группы
Пуанкаре. Уравнение Дирака для спинорного поля и уравнения Максвелла для
электромагнитного поля в пространстве Минковского ковариантны относительно
преобразований из группы Пуанкаре.
\qed\end{exa}
\begin{com}
Часто принцип относительности формулируют следующим образом: законы природы
выглядят одинаково во всех инерциальных системах отсчета. В такой формулировке
за кадром остается определение инерциальной системы отсчета и смысл термина
``одинаково''.
\qed\end{com}

Приведенные рассуждения приводят к следующему итогу. В специальной теории
относительности имеется три основных постулата:
\begin{enumerate}
\item Пространство-время, в котором происходят все окружающие нас явления,
представляет собой четырехмерное пространство Минковского $\MR^{1,3}$.
\item Модели математической физики строятся из соответствующего набора полей на
пространстве Минковского $\MR^{1,3}$, которые преобразуются по некоторому
представлению группы Пуанкаре.
\item Все законы природы описываются некоторыми уравнениями, которые либо
инвариантны, либо ковариантны относительно преобразований из группы Пуанкаре.
\end{enumerate}

Мы всегда предполагаем, что точечные частицы, на которые не действуют никакие
силы, кроме гравитационных, движутся в пространстве-времени вдоль экстремалей. В
специальной теории относительности гравитационные эффекты не учитываются.
Поэтому свободные частицы, на которые не действуют никакие силы, движутся в
пространстве Минковского вдоль экстремалей. В пространстве Минковского
экстремали совпадают с геодезическими и являются прямыми линиями. Поэтому
свободные точечные частицы в специальной теории относительности в инерциальной
системе координат, как и в нерелятивистских моделях, движутся равномерно и
прямолинейно.

Мы говорим, что пространство-время специальной теории относительности однородно
и изотропно. Математически это значит, что метрика инвариантна относительно
сдвигов из группы Пуанкаре и вращений пространства Минковского, т.е.\
преобразований Лоренца. Другими словами, физические свойства
пространства-времени не зависят от выбора начала отсчета декартовой системы
координат и направления.

Понятие инерциальной системы координат является математической абстракцией.
Поскольку во вселенной существуют материальные тела, то гравитационное поле
не устранимо. Поэтому системы координат, которые используются в экспериментах,
можно считать инерциальными только с определенной степенью точности.
\begin{exa}
Система координат, покоящаяся относительно Земли, с хорошей степенью точности
является инерциальной при изучении электромагнитных явлений.
\qed\end{exa}
Несмотря на то, что инерциальных систем отсчета, строго говоря, в природе не
существует, это понятие чрезвычайно важно в современной теоретической физике.
Модели математической физики (например, электродинамика), сформулированные в
пространстве Минковского, приводят к следствиям, которые находятся в прекрасном
согласии с экспериментами. В настоящее время известно только об одном случае
несогласия специальной теории относительности с опытом: нарушение четности в
слабых взаимодействиях. Поэтому считается, что специальная теория
относительности находится в прекрасном согласии с экспериментом, и все
взаимодействия, кроме гравитационных, инвариантны относительно трансляций и
преобразований Лоренца.

Наиболее успешные модели математической физики: модели электромагнитных, слабых,
сильных (квантовая хромодинамика) взаимодействий, а также их объединения, не
претендующие на описание гравитационных взаимодействий, построены в пространстве
Минковского. С этой целью выбирается некоторое представление группы группы
Пуанкаре и строится инвариантное действие в пространстве Минковского, которое
приводит к ковариантным уравнениям движения. В последние десятилетия такой
подход оказался самым распространенным и успешным.

Обсудим понятие одновременности и пространственного расстояния в пространстве
Минковского. Зафиксируем два события $x_1=\lbrace ct_1,x^i_1\rbrace$ и
$x_2=\lbrace ct_2,x^i_2\rbrace$ в $\MR^{1,3}$, где мы ввели скорость света $c$,
чтобы измерять время в секундах, а не в метрах. Если сказать, что
эти события разделены временн\'ым интервалом $\triangle t:=t_2-t_1$, то такое
определение будет некорректным, т.к.\ $\triangle t$ зависит от выбора
инерциальной системы отсчета. Это следует прямо из выражения для преобразований
координат (\ref{eloust}) (лоренцевых бустов в плоскости $x^0,x^1$). Поэтому
понятие одновременности в специальной теории относительности отсутствует.
Как следствие, нельзя также корректно ввести понятие пространственного
расстояния между одновременными событиями, т.к.\ одновременные события не
определены. Вместо этого вводится понятие интервала между двумя событиями
\begin{equation}                                                  \label{eintsr}
  \triangle s^2
  =(x^0_2-x^0_1)^2-(x^1_2-x^1_1)^2-(x^2_2-x^2_1)^2-(x^3_2-x^3_1)^2.
\end{equation}
Кроме этого, введение скорости света полезно для определения нерелятивистского
предела. Данное определение интервала корректно, т.к.\ не зависит от выбора
инерциальной системы отсчета. Интервал для двух событий может быть положителен,
равен нулю или отрицателен. Мы говорим, что два события причинно связаны, если
интервал между ними положителен, $\triangle s^2>0$. В этом случае существует
такая инерциальная система координат, в которой оба события происходят в одной
точке. Если свет испущен в точке $x_1$ и получен в точке $x_2$, то для этих
событий интервал равен нулю, $\triangle s^2=0$. Если интервал отрицателен,
$\triangle s^2<0$, то события являются причинно не связанными. Для этих событий
существует такая инерциальная система координат, в которой оба события
происходят одновременно.

Рассмотрим две инерциальные системы координат $O$ и $O'$ с декартовыми
координатами $t,x,y,z$ и $t',x',y',z'$ соответственно. Пусть начала систем
координат совпадают и система координат $O'$ движется со скоростью $V$
относительно $O$ вдоль оси $x$. Тогда события в этих системах координат связаны
между собой преобразованием Лоренца (\ref{eloust}):
\begin{equation}                                                  \label{elotra}
  t'=\frac{t-Vx/c^2}{\sqrt{1-V^2/c^2}},\qquad
  x'=\frac{x-Vt}{\sqrt{1-V^2/c^2}},\qquad y'=y,\quad z'=z.
\end{equation}
Обратные преобразования координат от системы координат $O'$ к $O$ получаются
простой заменой $V\rightarrow-V$:
\begin{equation}                                                  \label{elotrb}
  t=\frac{t'+Vx'/c^2}{\sqrt{1-V^2/c^2}},\qquad
  x=\frac{x'+Vt'}{\sqrt{1-V^2/c^2}},\qquad y=y',\quad z=z'.
\end{equation}
Напомним, что эти преобразования координат называются бустами.

Аналогичные формулы преобразования координат справедливы, если система координат
$O'$ движется относительно $O$ либо вдоль оси $y$, либо оси $z$.

В приложениях иногда необходимо знать выражение для преобразований Лоренца в
общем случае, когда система координат $O'$ движется относительно $O$ равномерно
и прямолинейно, но в произвольном направлении. Эти формулы легко получить из
(\ref{elotra}) в виде, который инвариантен относительно пространственных
вращений. Пусть $\BV=\lbrace V^i\rbrace\ne0$ -- скорость движения штрихованной
системы координат. Введем векторное обозначение
$\Bx=\lbrace x^i\rbrace=\lbrace x,y,z\rbrace$ для пространственных координат
событий и аналогичные обозначения в штрихованной системе координат. Введем также
обычное евклидово скалярное произведение пространственных векторов:
\begin{equation*}
  \BV^2:=-V^iV_i,\qquad (\Bx,\BV):=-\Bx^iV_i.
\end{equation*}
Знак минус в этих выражениях следует из того, что при опускании пространственных
индексов знак компонент меняется: $V_i=V^j\eta_{ji}=-V^i$. Тогда радиус-вектор
можно разложить на составляющие, $\Bx=\Bx_\parallel+\Bx_\perp$, которые
параллельны и перпендикулярны вектору скорости:
\begin{equation*}
  \Bx_\parallel:=\frac{(\Bx,\BV)\BV}{\BV^2},\qquad
  \Bx_\perp:=\Bx-\frac{(\Bx,\BV)\BV}{\BV^2}.
\end{equation*}
Аналогично
\begin{equation*}
  \Bx'_\parallel:=\frac{(\Bx',\BV)\BV}{\BV^2},\qquad
  \Bx'_\perp:=\Bx'-\frac{(\Bx',\BV)\BV}{\BV^2}.
\end{equation*}
Из преобразований Лоренца (\ref{elotra}) следует, что перпендикулярная
составляющая радиус-вектора не меняется, а параллельная меняется так же, как
координата $x$. Отсюда вытекает общее правило преобразования координат:
\begin{equation}                                                  \label{elobog}
  t'=\frac{t-(\Bx,\BV)/c^2}{\sqrt{1-\BV^2/c^2}},\qquad
  \Bx'_\parallel=\frac{\Bx_\parallel-\BV t}{\sqrt{1-\BV^2/c^2}},\qquad
  \Bx'_\perp=\Bx_\perp.
\end{equation}
Для полного радиуса-вектора получаем следующую формулу преобразования
\begin{equation*}
  \Bx'=\Bx+\frac{(\Bx,\BV)\BV}{\BV^2}\left(\frac1{\sqrt{1-\BV^2/c^2}}-1\right)
  -\frac{\BV t}{\sqrt{1-\BV^2/c^2}}
\end{equation*}
или, в компонентах,
\begin{equation}                                                  \label{elobor}
  x^{\prime i}=x^i-\frac{x^jV_jV^i}{\BV^2}\left(\frac1{\sqrt{1-\BV^2/c^2}}
  -1\right)-\frac{V^i t}{\sqrt{1-\BV^2/c^2}}.
\end{equation}
В общем случае координаты события преобразуются по правилу
\begin{equation*}
  t'=tS_0{}^0+x^iS_i{}^0,\qquad x^{\prime i}=tS_0{}^i+x^jS_j{}^i,
\end{equation*}
где $\lbrace S_a{}^b\rbrace=\lbrace S_0{}^0,S_i{}^0,S_0{}^i,S_i{}^j\rbrace$ --
матрица преобразований Лоренца. Сравнение этого выражения с формулами
(\ref{elobog}) и (\ref{elobor}) дает довольно симметричные выражения:
\begin{equation}                                                  \label{elobug}
\begin{split}
  S_0{}^0&=\frac1{\sqrt{1-\BV^2/c^2}},
\\
  S_i{}^0&=\frac{V_i}{c^2\sqrt{1-\BV^2/c^2}},
\\
  S_0{}^i&=-\frac{V^i}{\sqrt{1-\BV^2/c^2}},
\\
 S_i{}^j&=\dl_i^j-\frac{V_iV^j}{\BV^2}\left(\frac1{\sqrt{1-\BV^2/c^2}}-1\right).
\end{split}
\end{equation}
Нетрудно проверить инвариантность лоренцевой метрики $\eta_{ab}$ относительно
этих преобразований:
\begin{equation*}
  \eta_{ab}=S_a{}^cS_b{}^d\eta_{bd}.
\end{equation*}
Эти формулы для лоренцевых бустов получены в \cite{Herglo11}.

Полученные общие формулы для лоренцевых бустов (\ref{elobug}) будут использованы
в разделе \ref{sadmpa} при АДМ параметризации репера.

В релятивистской механике скорость света $c$ является универсальной постоянной,
и не зависит от выбора инерциальной системы отсчета. Если отношение
$\BV^2/c^2$ стремится к нулю, что соответствует стремлению скорости света к
бесконечности, то лоренцевы бусты (\ref{elotra}) стремятся к преобразованиям
Галилея.
\subsection{Замедление времени и лоренцево сокращение}
Сначала выведем формулу для замедления времени в движущейся системе отсчета.
Пусть часы покоятся в системе координат $O'$ и, следовательно, движутся
равномерно и прямолинейно в системе $O$. Время $t'$, которое показывают часы в
той системе координат, где они покоятся, называется собственным временем.
Введем для него специальное обозначение $t_0=t'$. Это время совпадает с длиной
мировой линии часов, деленной на $c$, которая является прямой, параллельной оси
времени. Поскольку для часов $x'=\const$, то из первой формулы преобразований
Лоренца (\ref{elotrb}) следует, что для произвольных промежутков времени
справедливо равенство
\begin{equation*}
  \triangle t_0=\triangle t\sqrt{1-V^2/c^2}\quad \Leftrightarrow\quad
  \triangle t=\frac{\triangle t_0}{\sqrt{1-V^2/c^2}}.
\end{equation*}
Это соотношение показывает, что в движущейся системе координат время течет
медленнее. Это свойство называется эффектом {\em замедления времени}.
\index{Замедление времени (clock retardation)}%

\begin{figure}[h,b,t]
 \begin{center}
\includegraphics[width=.25\textwidth]{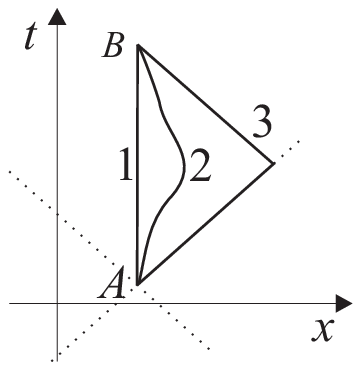}
 \end{center}
 \caption{Мировые линии близнецов 1 и 2. Мировая линия первого близнеца
 ``длиннее'' мировой линии второго.}
 \label{ftwins}
\end{figure}

Эффект замедления времени в движущейся системе координат породил множество
``парадоксов''.\newline
{\bf Парадокс близнецов.}
Допустим, что в инерциальной системе координат в некоторый момент времени
родились два близнеца, событие $A$ на рис.\ref{ftwins}. Первый близнец остался
на месте, а второго посадили в ракету и отправили в космическое путешествие с
большой скоростью. Через некоторое время ракета вернулась, и близнецы
встретились снова, событие $B$. На рисунке мировые линии близнецов помечены
цифрами 1 и 2 соответственно. С точки зрения первого близнеца ракета быстро
двигалась и из-за замедления времени второй близнец должен оказаться более
молодым, чем он сам. С точки зрения второго близнеца ситуация прямо
противоположная: он покоился относительно ракеты, а двигался первый близнец.
Поэтому именно первый близнец должен оказаться моложе.

Чтобы разъяснить парадокс, часто говорят, что рассуждения второго близнеца
неверны, т.к.\ его система координат не может быть инерциальной (ракета должна
замедлить свой полет и повернуть обратно). Однако это замечание ничего не
объясняет. Действительно, если оба близнеца отправятся в космические путешествие
в разные стороны, а потом вернутся и встретятся. Кто из них будет моложе ? Они
оба находились в неинерциальных системах отсчета, и с этой точки зрения
совершенно равноправны. Для того, чтобы разрешить парадокс, необходимо дать
\begin{defn}
Возрастом человека называется промежуток собственного времени, который прошел с
момента рождения. Или, эквивалентно, длина мировой линии человека от момента
рождения, деленная на скорость света $c$.
\qed\end{defn}
Это определение корректно, так как инвариантно относительно произвольных
преобразований координат. Поэтому возраст можно вычислить в любой, в том числе
неинерциальной системе отсчета.

Теперь все становится на свои места. Поскольку на плоскости $x,t$ задана метрика
Лоренца, то длина любого наклонного отрезка между двумя моментами времени,
меньше длины вертикального отрезка.  Из рисунка ясно, что длина (собственное
время) вертикального отрезка между точками $A$ и $B$, который является мировой
линией первого близнеца, максимальна. Поэтому длина любой (времениподобной)
мировой линии, соединяющей точки $A$ и $B$, меньше длины вертикального отрезка.
Заметим, что если бы на плоскости $x,t$ была задана евклидова метрика, то длина
вертикального отрезка была бы наименьшей среди всех кривых, соединяющих точки
$A$ и $B$.

Разница в возрасте может оказаться большой, если ракета движется
достаточно быстро. В предельном случае, когда ракета удаляется от первого
близнеца со скоростью света, затем отражается и возвращается обратно также со
скоростью света, второй близнец за время своего путешествия вообще не
состарится (траектория 3 на рисунке).
\qed

Теперь предположим, что в движущейся системе координат $O'$ на оси $x'$ покоится
линейка длины $l_0:=x'_2-x'_1$. Собственная длина линейки $l_0$ измеряется при
постоянном времени $t'=\const$ в сопутствующей системе координат $O'$. В системе
координат $O$ линейка равномерно движется и ее длина измеряется при постоянном
времени $t$. Из второй формулы в (\ref{elotra}) вытекает соотношение между
длинами:
\begin{equation*}
  l_0=\frac l{\sqrt{1-V^2/c^2}}\quad \Leftrightarrow\quad l=l_0\sqrt{1-V^2/c^2}.
\end{equation*}
Отсюда следует, что движущаяся линейка в покоящейся системе координат имеет
длину $l<l_0$ и выглядит короче. Это явление называется {\em лоренцевым
сокращением}.

Из формул для лоренцевых преобразований (\ref{elotra}) следует, что поперечные
наблюдаемые размеры тел вдоль осей $y$ и $z$ не испытывают никаких сокращений.
Отсюда следует, что объем движущегося тела $W$ уменьшается так же, как и его
длина:
\begin{equation*}
  W:=\triangle x\triangle y\triangle z=W_0\sqrt{1-V^2/c^2},
\end{equation*}
где $W_0$ -- собственный объем в системе координат, связанной с телом.

Сокращение длин при движении, так же, как и замедление времени, породило много
``парадоксов''.\newline
\index{Лоренцево сокращение}%
{\bf Парадокс машины и гаража.}
Предположим, что машина и гараж имеют одинаковую собственную длину $L$. Водитель
на большой скорости $V$ въезжает в гараж, а дежурный по гаражу закрывает ворота
в тот момент, когда задний бампер машины войдет в гараж. С точки зрения
дежурного никаких проблем не возникает: машина испытывает лоренцево сокращение
и, следовательно, без труда поместится в гараже. Однако с точки зрения водителя
ситуация противоположная: лоренцево сокращение испытывает гараж и поэтому машина
никак в гараже поместится не может.

\begin{figure}[h,t]
\hfill\includegraphics[width=.6\textwidth]{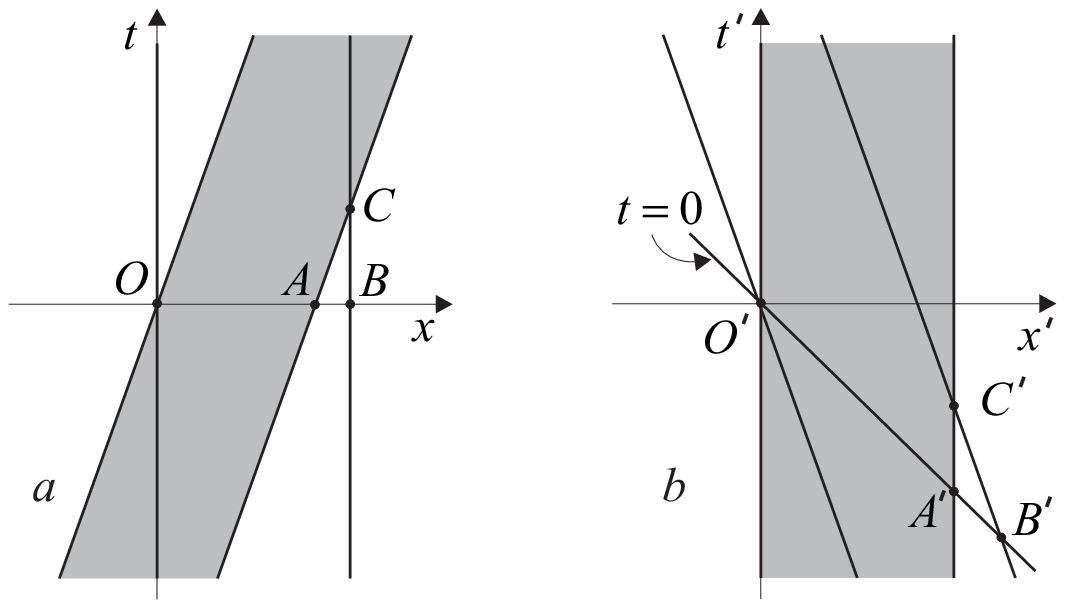}
\hfill {}
\centering \caption{\label{fcarga} Пространственно-временн\'ые диаграммы машины
и гаража в различных системах координат. Общее начало систем координат, точка
$O$, соответствует событию закрытия ворот. Затемненная полоса соответствует
движению машины. Машина и гараж с точки зрения дежурного (\textit{a}).
Машина и гараж с точки зрения водителя (\textit{b}).}
\end{figure}

Разрешение ``парадокса'' заключается в отсутствии понятия одновременности в
специальной теории относительности. Для простоты мы предполагаем, что машина
свободно пробивает заднюю стену гаража и продолжает движение. На рисунках
\ref{fcarga}{\it a,b} показаны пространственно-временн\'ые диаграммы событий с
точки зрения дежурного по гаражу и водителя. Для удобства, мы выбрали общее
начало координат (точка $O$) таким образом, что оно соответствует моменту
закрытия ворот. С точки зрения дежурного гараж покоится и ему соответствует
вертикальная полоса, границы которой пересекают ось $x$ в точках $O$ и $B$.
Машине соответствует затемненная наклонная полоса, пересекающая ось $x$ в точках
$O$ и $B$. В системе координат дежурного машина испытывает лоренцево сокращение
и поэтому событие $A$ расположено левее $B$. Это значит, что в момент закрытия
ворот машина будет расположена в гараже. Через некоторое время передним бампер
машины достигнет стены гаража и пробьет его, событие $C$.

Нетрудно проверить, что события $A,B$ и $C$ в нештрихованной системе координат,
связанной с гаражом, на рис.\ref{fcarga}{\it a} имеют следующие координаты:
\begin{equation*}
\begin{split}
  A&=\left(0,L\sqrt{1-V^2/c^2}\right),
\\
  B&=\left(0,L\right)
\\
  C&=\left(\frac LV\left(1-\sqrt{1-V^2/c^2}\right),L\right).
\end{split}
\end{equation*}

С точки зрения водителя гараж равномерно приближается к машине со скоростью $V$.
Используя формулы для лоренцевых преобразований координат (\ref{elotra}),
вычислим координаты событий $A,B$ и $C$ в штрихованной системе координат,
привязанной к водителю:
\begin{equation*}
\begin{split}
  A'&=\left(-\frac{VL}{c^2},L\right),
\\
  B'&=\left(-\frac{VL}{c^2\sqrt{1-V^2/c^2}},\frac L{\sqrt{1-V^2/c^2}}\right)
\\
  C'&=\left(-\frac LV\left(1-\sqrt{1-V^2/c^2}\right),L\right).
\end{split}
\end{equation*}
На рис.\ref{fcarga}{\it b} показана схема событий с точки зрения водителя.
Машине и гаражу соответствуют вертикальная и наклонная полосы. С точки зрения
водителя передний бампер машины сначала пробивает стену гаража, событие $C'$, и
лишь после этого закрываются ворота, событие $O'$.

Таким образом, последовательность событий с точки зрения дежурного и водителя
прямо противоположная. С точки зрения дежурного сначала закрываются ворота и
лишь потом машина пробивает стену. Водитель же видит, что сначала машина
пробивает стену, и только потом закрываются ворота.

Заметим, что, если скорость света равна бесконечности, то все три события $A,B$
и $C$ имеют одинаковые координаты $(0,L)$ в обеих системах отсчета.
\qed

Рассмотренный пример наглядно показывает, что понятие одновременности событий в
специальной теории относительности отсутствует. К этому трудно привыкнуть, т.к.\
наша интуиция основана на повседневной жизни, где скорости движения частиц малы
по сравнению со скоростью света.
\subsection{Сложение скоростей и эффект Доплера}
Пусть, как и ранее, штрихованная система координат $O'$ движется с постоянной
скоростью $V$ вдоль оси $x$ в покоящейся системе координат $O$. Следовательно,
координаты событий в этих системах отсчета связаны преобразованиями Лоренца
(\ref{elotra}). Предположим, что в штрихованной системе координат точечная
частица движется равномерно и прямолинейно с постоянной скоростью
$v'=(v'_x,v'_y,v'_z)$:
\begin{equation*}
  x'=v'_xt',\qquad y'=v'_yt',\qquad z'=v'_zt'.
\end{equation*}
Тогда в покоящейся системе координат мировая линия частицы будет прямой линией,
заданной уравнениями:
\begin{equation*}
\begin{aligned}
  t&=\frac{1+v'_xV/c^2}{\sqrt{1-V^2/c^2}}t',&  \qquad y&=v'_y t',
\\
  x&=\frac{v'_x+V}{\sqrt{1-V^2/c^2}}t', & z&=v'_z t',
\end{aligned}
\end{equation*}
где время $t'$ рассматривается, как параметр вдоль мировой линии частицы. Отсюда
вытекает, что скорость частицы, наблюдаемая в покоящейся системе координат,
имеет следующие компоненты:
\begin{equation}                                                  \label{eveadd}
\begin{split}
  v_x&:=\frac xt=\frac{v'_x+V}{1+v'_xV/c^2},
\\
  v_y&:=\frac yt=\frac{v'_y}{1+v'_xV/c^2}\sqrt{1-V^2/c^2},
\\
  v_z&:=\frac yt=\frac{v'_z}{1+v'_xV/c^2}\sqrt{1-V^2/c^2}.
\end{split}
\end{equation}
Это и есть правило сложения скоростей в специальной теории относительности.
Из полученных формул сразу следуют правила сложения скоростей в частных случаях,
когда скорости параллельны или перпендикулярны друг другу.

Обозначим углы, которые составляют скорости $v'$ и $v$ с осями $x'$ и $x$,
соответственно через $\al'$ и $\al$, т.е.
\begin{equation*}
  v'_x=v'\cos\al',\qquad v_x=v\cos\al.
\end{equation*}
Здесь мы предполагаем, что углы меняются в интервале: $0\le\al',\al\le\pi$.
Для модуля скорости $v:=\sqrt{v_x^2+v_y^2+v_z^2}$, наблюдаемой в покоящейся
системе отсчета, из (\ref{eveadd}) следует равенство
\begin{equation}                                                  \label{eangre}
  v=\frac{\sqrt{v^{\prime2}+V^2+2v'V\cos\al'
  -\left(v'V\sin\al'/c\right)^2}}{1+v'V\cos\al'/c^2}.
\end{equation}
Далее, простые вычисления дают
\begin{equation}                                                  \label{evelmo}
  \sqrt{1-v^2/c^2}=
  \frac{\sqrt{1-v^{\prime2}/c^2}\sqrt{1-V^2/c^2}}{1+v'V\cos\al'/c^2}.
\end{equation}
\begin{prop}                                                      \label{prelve}
Если скорость частицы в какой либо инерциальной системе отсчета меньше скорости
света, $v'<c$, то она будет меньше скорости света и в любой другой инерциальной
системе координат, $v<c$. Если в движущейся системе координат распространяется
свет, $v'=c$, то в покоящейся системе координат его скорость такая же, $v=c$.
\end{prop}
\begin{proof}
Если $v'<c$ и $V<c$, то знаменатель в (\ref{evelmo}) положителен, т.к.\
$\cos\al'$ ограничен сверху единицей. Следовательно, правая часть (\ref{evelmo})
положительна. Оценим ее сверху. Максимальное значение принимается,
когда $\cos\al'=-1$, т.е.\ частица движется в сторону, противоположную движению
штрихованной системы координат. Сверху правая часть (\ref{evelmo}) ограничена
единицей, поскольку справедливо неравенство
\begin{equation*}
  \left(1-\frac{v^{\prime2}}{c^2}\right)\left(1-\frac{V^2}{c^2}\right)
  \le\left(1-\frac{v'V}{c^2}\right),
\end{equation*}
которое просто проверяется. Таким образом, правая часть  лежит в пределах от $0$
до $1$. Поэтому $0\le v<c$. При этом $v=0$ тогда и только тогда, когда $v'=V$ и
$\cos\al'=-1$.

Если $v'=c$, то из уравнения (\ref{eangre}) вытекает, что $v=c$.
\end{proof}
Из формулы (\ref{evelmo}) следует, что если частица движется в штрихованной
системе координат со скоростью $V$ в противоположную сторону, $\cos\al'=-1$, то
ее скорость в неподвижной системе координат равна нулю. Это означает, что для
произвольной частицы, движущейся со скоростью, меньшей скорости света, всегда
найдется сопутствующая система координат, в которой она покоится. Ясно также,
что для света сопутствующей системы координат не существует.

В разделе \ref{stwodl} была введена скорость относительного движения двух
инерциальных систем отсчета (\ref{dvdefi}). Из этой формулы вытекает, что
скорость относительного движения двух инерциальных систем отсчета не превосходит
скорости света. Вместе с предложением \ref{prelve} это означает, что специальная
теория относительности описывает движение частиц со скоростью, меньшей скорости
света, самосогласованным образом. Если не постулировать невозможность движения
частиц со скоростью, превышающей скорость света, то специальная теория
относительности такое движение допускает. Ясно, что для таких частиц не может
существовать сопутствующей системы координат. Частицы, движущиеся со скоростью,
превышающей скорость света называются {\em тахионами}. Такие частицы, если они
существуют, представляют большие теоретические трудности. Например, возникает
проблема с положительной определенностью канонического гамильтониана. Тем не
менее они возникают в различных моделях математической физики. В настоящее время
тахионы экспериментально не обнаружены, и принято считать, что в природе они
отсутствуют.
\index{Тахион (tachyon)}%

Из правила сложения скоростей (\ref{eveadd}) следует также соотношение между
углами:
\begin{equation*}
  \tg\al=\frac{\sqrt{v_y^2+v_z^2}}{v_x}=
  \frac{v'\sin\al'}{v'\cos\al'+V}\sqrt{1-V^2/c^2}.
\end{equation*}

Теперь кратко обсудим два эффекта, которые подтверждаются экспериментально.

Если в системе отсчета $O'$ распространяется луч света, т.е.\ $v'=c$, то для
угла, наблюдаемого в неподвижной системе координат, получаем выражение
\begin{equation}                                                  \label{etgrel}
  \tg\al=\frac{\sin\al'}{\cos\al'+V/c}\sqrt{1-V^2/c^2}.
\end{equation}
Эту формулу можно переписать в более симметричном виде:
\begin{equation*}
  \sin\al=\sin\al'\frac{\sqrt{1-V^2/c^2}}{1+\frac Vc\cos\al'}
\end{equation*}
или
\begin{equation*}
  \tg\frac\al2=\tg\frac{\al'}2\sqrt{\frac{1-V/c}{1+V/c}}.
\end{equation*}
Из последней формулы, в частности, следует, что $\al'>\al$ при $V>0$.
Изменение угла, под которым наблюдается луч света, в движущейся и покоящейся
системах координат, называется {\em аберрацией света}.
\begin{com}
Если источник света покоится в системе координат $O$, а наблюдатель движется
вместе с системой отсчета $O'$, то бесконечно удаленный источник света будет
казаться ему смещенным на угол $\al'-\al$. Для наблюдательной астрономии это
означает, что звезды на небе будут описывать эллипсы в соответствии с годовым
вращением Земли вокруг Солнца. Этот эффект действительно наблюдается.
\qed\end{com}
\index{Аберрация света (light aberration)}%

В качестве еще одного приложения преобразований Лоренца рассмотрим {\em эффект
Доплера}.
\index{Эффект Доплера (Doppler effect)}\index{Доплера эффект(Doppler effect)}%
Допустим, что луч света (плоская электромагнитная волна) распространяется в
покоящейся системе координат $O$ в плоскости $x,y$ под углом $\al$ к оси $x$.
В такой волне напряженность электрического поля $E(t,x)$ меняется по правилу
\begin{equation*}
  E(t,x)=E_0\ex^{i\vf},
\end{equation*}
где $E_0=\const$ -- амплитуда электрического поля, и
\begin{equation*}
  \vf(t,x,y)=\om\left(t-\frac{x\cos\al+y\sin\al}c\right)
\end{equation*}
-- фаза электромагнитной волны. В приведенной формуле $\om$ -- это частота
волны, которая наблюдается в покоящейся системе координат. Электромагнитное
поле, которое описывает плоскую волну удовлетворяет уравнениям Максвелла.
Поскольку уравнения Максвелла ковариантны относительно преобразований Лоренца и
напряженность $E$ подчиняется тензорному закону преобразования, то фаза волны
$\vf$ является скалярным полем на пространстве Минковского. Следовательно, фазы
волны в покоящейся и движущейся системах координат совпадают. Для штрихованной
системы координат, которая равномерно движется вдоль оси $x$ и была описана
ранее, равенство фаз имеет вид
\begin{equation*}
  \om\left(t-\frac{x\cos\al+y\sin\al}c\right)
  =\om'\left(t'-\frac{x'\cos\al'+y'\sin\al'}c\right),
\end{equation*}
где $\om'$ -- частота электромагнитной волны в движущейся системе координат.
Полученное уравнение должно выполняться при всех значениях $t,x,y$ и $t',x'y'$.
Подставляя выражения для $t,x,y$ из преобразований Лоренца (\ref{elotrb}) и
приравнивая коэффициенты при $t',x'$ и $y'$, получаем формулы преобразования для
частоты
\begin{equation}                                                  \label{edoeff}
  \om'=\om\frac{1-\frac Vc\cos\al}{\sqrt{1-V^2/c^2}}
\end{equation}
и углов
\begin{equation*}
  \cos\al'=\frac{\cos\al-V/c}{1-\frac Vc\cos\al},\qquad
  \sin\al'=\sin\al\frac{\sqrt{1-V^2/c^2}}{1-\frac Vc\cos\al}.
\end{equation*}
Из последних двух равенств следует соотношение между углами:
\begin{equation*}
  \tg\al'=\frac{\sin\al}{\cos\al-V/c}\sqrt{1-V^2/c^2}.
\end{equation*}
Отличие в знаках в знаменателе полученной формулы и равенства (\ref{etgrel})
обусловлено тем, что углы $\al$ и $\al'$ поменялись местами, что соответствует
изменению знака скорости, $V\to-V$.

Изменение частоты света (\ref{edoeff}), связанное с движением системы отсчета,
называется {\em эффектом Доплера}. Пусть источник света находится в покоящейся
системе отсчета $O$, и свет распространяется вдоль оси $x$. Допустим, что
наблюдатель находится в движущейся системе отсчета $O'$ и удаляется от источника
света, т.е.\ $V>0$ и $\cos\al=1$. Тогда наблюдаемая им частота света уменьшится.
Этот эффект называется {\em красным смещением}. Если же, наоборот, наблюдатель
приближается к источнику света, $V>0$ и $\cos\al=-1$, то частота света
увеличится. Такое поведение частоты называется {\em голубым смещением}.
\index{Красное смещение (red shift)}\index{Смещение красное (red shift)}%
\index{Голубое смещение (blue shift)}\index{Смещение голубое (blue shift)}%
\begin{com}
Изучение спектра далеких галактик показывает, что он смещен в красную сторону.
Это приводит к выводу о том, что галактики удаляются от Земли и вселенная
расширяется.
\qed\end{com}
\subsection{Равноускоренное движение                             \label{srausk}}
В настоящем разделе мы определим 4-скорость и наблюдаемую скорость точечной
частицы, а также ее ускорение. Затем рассмотрим равноускоренное движение в
специальной теории относительности и сравним его с равноускоренным движением в
механике Ньютона.

Обозначим инерциальные координаты в пространстве Минковского $\MR^{1,3}$ через
$x^\al$, $\al=0,1,2,3$. Пусть задана мировая линия частицы
\begin{equation*}
  \MR\ni\quad\s\mapsto\lbrace q^\al(\s)\rbrace\quad\in\MR^{1,3},
\end{equation*}
где $\s$ -- некоторый параметр.
По-определению, мировая линия частицы времениподобна, а ее длина $s$ при
$\s\in[0,\tau]$ равна интегралу
\begin{equation*}
  s:=\int_0^\tau\!\!\!
  d\s\sqrt{\eta_{\al\bt}\frac{dq^\al}{d\s}\frac{dq^\bt}{d\s}}.
\end{equation*}
Выберем длину траектории $s$, которая пропорциональна собственному времени
частицы, в качестве параметра вдоль мировой линии. Тогда мировая линия задается
четырьмя функциями $q^\al(s)$.

Для времениподобных кривых в качестве параметра вдоль кривой можно выбрать также
время $x^0$. Действительно, поскольку для произвольной траектории частицы
$dq^0/ds>0$, то параметр $s$ можно рассматривать как некоторую функцию
координаты $s=s(x^0)$. Тогда мировая линии частицы задается функциями
\begin{equation*}
  q^0=x^0,\qquad q^\mu=q^\mu(q^0),\qquad \mu=1,2,3.
\end{equation*}
\begin{com}
Физическая интерпретация приведенных параметризаций мировых линий следующая.
Время для наблюдателя, который движется вместе с частицей, совпадает с
собственным временем $s/c$. Поэтому проведенные им измерения соответствуют
сопутствующей системе координат. Во втором случае $x^0$ -- это время
наблюдателя, покоящегося в выбранной исходной инерциальной системе отсчета.
Поэтому его измерения соответствуют внешнему наблюдателю.
\qed\end{com}
\begin{defn}
4-вектор $u$ с компонентами
\begin{equation}                                                  \label{efovel}
  u^\al:=c\frac{dq^\al}{ds}
\end{equation}
называется {\em 4-скоростью} или просто {\em скоростью} частицы. 4-вектор $w$ с
компонентами
\begin{equation}                                                  \label{efoacs}
  w^\al:=c\frac{du^\al}{ds}=c^2\frac{d^2 q^\al}{ds^2}
\end{equation}
называется {\em 4-ускорением} или просто {\em ускорением} частицы.
{\em Наблюдаемой скоростью} частицы $\Bv$ в системе координат $x^\al$ называется
3-вектор с компонентами
\begin{equation}                                                  \label{ethvec}
  v^\mu:=\frac{dq^\mu}{dq^0}=\frac{u^\mu}{u^0},\qquad \mu=1,2,3.
\end{equation}
3-вектор $\Ba$ с компонентами
\begin{equation}                                                  \label{ethvea}
  a^\mu:=\frac{dv^\mu}{dq^0}=\frac{d^2q^\mu}{(dq^0)^2}.
\end{equation}
называется {\em наблюдаемым ускорением} частицы.
\qed\end{defn}
\index{4-скорость (4-velocity)}\index{Скорость (velocity)}%
\index{4-ускорение (4-acceleration)}\index{Ускорение (acceleration)}%
\index{Наблюдаемая скорость (observed velocity)}%
\index{Скорость наблюдаемая (observed velocity)}%
\index{Наблюдаемое ускорение (observed acceleration)}%
\index{Ускорение наблюдаемое (observed acceleration)}%
Скорость (\ref{efovel}) и ускорение (\ref{efoacs}) точечной частицы являются
4-векторами относительно преобразований из группы Лоренца $\MO(1,3)$,
действующей в пространстве Минковского $\MR^{1,3}$. Наблюдаемая скорость
(\ref{ethvec}) и наблюдаемое ускорение (\ref{ethvea}) преобразуются как векторы
только относительно подгруппы вращений $\MO(3)$, действующей на пространственных
сечениях $x^0=\const$. Они нековариантны относительно лоренцевых преобразований
и не являются компонентами каких либо 4-векторов.

Мы предполагаем, что траектория частицы времениподобна и направлена в будущее.
Поэтому временн\'ая компонента скорости положительна, $u^0>0$.

В настоящем разделе полезно следить за размерностями различных величин:
\begin{align*}
  [x^0]&=\text{сек}, & [x^\mu]&=[s]=\text{см},
\\
  [u^0]&=1,          & [u^\mu]&=[v^\mu]=\frac{\text{см}}{\text{сек}},
\\
  [w^0]&=\frac1{\text{сек}}, & [w^\mu]&=[a^\mu]=\frac{\text{см}}{\text{сек}^2}.
\end{align*}

Из определения скорости (\ref{efovel}) сразу следует  равенство
\begin{equation}                                                  \label{esquve}
  u^\al u_\al=c^2(u^0)^2-\Bu^2=c^2,
\end{equation}
где $\Bu^2=-u^\mu u_\mu\ge0$ -- обычный квадрат трехмерного вектора в евклидовом
пространстве. Дифференцирование этого равенства по $s$ приводит к соотношению
\begin{equation*}
  u^\al w_\al=0,
\end{equation*}
т.е.\ 4-ускорение частицы всегда ортогонально ее 4-скорости.

Умножим равенство (\ref{esquve}) на $m^2c^2$, где $m>0$ -- масса частицы,
$[m]=$г, и перепишем его в виде
\begin{equation*}
  E=mc^2\sqrt{1+\frac{\Bu^2}{c^2}},
\end{equation*}
где введено обозначение $E:=mc^2u^0>0$. В нерелятивистском пределе, когда
пространственные компоненты скорости малы по сравнению со скоростью света,
$\Bu^2/c^2\ll1$, квадратный корень можно разложить в ряд Тейлора. В первом
приближении получим равенство
\begin{equation*}
  E=mc^2+\frac{m\Bu^2}2.
\end{equation*}
Поскольку второе слагаемое совпадает с выражением для кинетической энергии
точечной частицы в механике Ньютона, то отсюда следует, что величина $E$
представляет собой энергию частицы в специальной теории относительности. В
сопутствующей системе координат, где $\Bu=0$, получаем знаменитую формулу
Эйнштейна
\begin{equation*}
  E=mc^2,
\end{equation*}
которая дает выражение для энергии покоя точечной частицы. Формулу
(\ref{esquve}) можно переписать также в виде
\begin{equation*}
  \frac{E^2}{c^2}-\Bp^2=m^2c^2,
\end{equation*}
где введен пространственный импульс частицы $\Bp:=m\Bu$. Если ввести
{\em 4-импульс} частицы
\index{4-импульс (4-momentum)}\index{Импульс (momentum)}%
\begin{equation*}
  p^\al:=mu^\al=(mu^0,mu^\mu)=\left(\frac E{c^2},p^\mu\right),
\end{equation*}
то равенство (\ref{esquve}) примет вид
\begin{equation*}
  p^2=p^\al p_\al=m^2c^2.
\end{equation*}

Согласно определению (\ref{ethvec}) для наблюдаемой скорости выполнено равенство
$u^\mu=u^0v^\mu$. Поэтому из уравнения (\ref{esquve}) вытекает соотношение
\begin{equation}                                                  \label{ezecvo}
  u^0=\frac1{\sqrt{1-\Bv^2/c^2}}.
\end{equation}

Заметим, что в нерелятивистском пределе, который мы определили как
$\Bu^2/c^2\to0$, из равенства (\ref{esquve}) вытекает, что при этом $u^0\to1$.
Поэтому из соотношения $u^\mu=u^0v^\mu$ следует, что пределы $\Bu^2\to0$ и
$\Bv^2\to0$ эквивалентны.

Теперь найдем формулы, связывающие 4-ускорение с наблюдаемым ускорением частицы.
Для этого в определении 4-ускорения надо заменить производную
\begin{equation*}
  c\frac d{ds}=u^0\frac d{dq^0}
\end{equation*}
и воспользоваться формулой (\ref{ezecvo}). В результате получим равенства
\begin{equation}                                                  \label{etracc}
\begin{split}
  w^0&=\frac{(\Bv,\Ba)}{c^2(1-\Bv^2/c^2)^2}.
\\
  \Bw&=\Ba\frac1{1-\Bv^2/c^2}
  +\Bv\frac{(\Bv,\Ba)}{c^2}\frac1{(1-\Bv^2/c^2)^2}
\end{split}
\end{equation}
В нерелятивистском пределе, очевидно, $\Bw=\Ba$ и $w^0=0$.

Для определения равноускоренного движения нам понадобятся формулы преобразования
компонент наблюдаемого ускорения при переходе от покоящейся системы координат к
движущейся.
Как было отмечено, компоненты наблюдаемого ускорения не являются компонентами
какого либо 4-вектора, и поэтому при лоренцевых бустах преобразуются не по
тензорным правилам. Пусть система координат $O'$ связана с частицей, т.е.\ в
момент времени $ct=x^0$, ее скорость совпадает с мгновенной скоростью частицы,
$\BV=\Bv$. Тогда в штрихованной системе координат скорость $\Bv'=0$, и формулы
(\ref{etracc}) принимают вид
\begin{equation}                                                  \label{erelwv}
  \Bw'=\Ba',\qquad w^{\prime 0}=0.
\end{equation}
Для определенности, предположим, что скорость $\Bv$ направлена вдоль оси $x$,
т.е.\ $\Bv=(v^1,0,0)$ и $\Bv^2=(v^1)^2$. Тогда из формул (\ref{etracc}) следуют
соотношения:
\begin{equation}                                                  \label{ewdovr}
  w^1=\frac{a^1}{1-\Bv^2/c^2}+v^1w^0,\qquad
  w^2=\frac{a^2}{1-\Bv^2/c^2},\qquad w^3=\frac{a^3}{1-\Bv^2/c^2}.
\end{equation}
Теперь вспомним, что 4-ускорение $w$ -- лоренцев вектор. Его компоненты
преобразуются, как компоненты вектора $(x^0,x^1,x^2,x^3)$. Поэтому компоненты
4-ускорения в штрихованной и нештрихованной системах координат связаны
преобразованием Лоренца (\ref{elotrb}):
\begin{equation*}
  w^0=\frac{v^1w^{\prime1}}{c\sqrt{1-\Bv^2/c^2}},\qquad
  w^1=\frac{w^{\prime1}}{\sqrt{1-\Bv^2/c^2}},\qquad w^2=w^{\prime2},\qquad
  w^3=w^{\prime3},
\end{equation*}
где мы учли, что $w^{\prime0}=0$. Подставляя в эти формулы соотношения между
4-ускорениями и наблюдаемыми ускорениями (\ref{erelwv}) и (\ref{ewdovr})
окончательно получаем формулы преобразования наблюдаемых ускорений:
\begin{equation}                                                  \label{eobacr}
  a^1=a^{\prime1}(1-\Bv^2/c^2)^{3/2},\qquad
  a^2=a^{\prime2}(1-\Bv^2/c^2),\qquad
  a^3=a^{\prime3}(1-\Bv^2/c^2).
\end{equation}
Эти формулы преобразования, как и следовало ожидать, отличаются от
преобразований Лоренца.

Рассмотрим равноускоренное движение.
\begin{defn}
Движение точечной частицы называется {\em равноускоренным}, если наблюдаемое
ускорение $\Ba'$ в системе координат $O'$, движущейся вместе с данной частицей,
постоянно, $\Ba'=\const$.
\qed\end{defn}
\index{Равноускоренное движение (uniformly accelerated motion)}%
\index{Движение равноускоренное (uniformly accelerated motion)}%

Теперь найдем мировую линию частицы, которая движется равноускоренно в
покоящейся системе координат. Пусть система отсчета $O$ фиксирована. Поскольку
частица движется равноускоренно, то штрихованная система координат $O'$,
связанная с частицей, меняется с течением времени.
\begin{figure}[h,b]
\hfill\includegraphics[width=.2\textwidth]{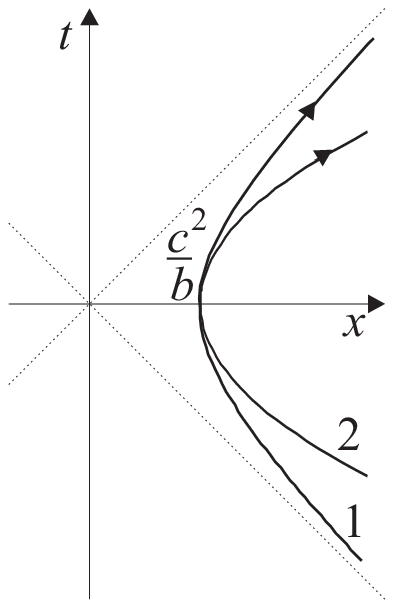}
\hfill {}
\centering \caption{Мировые линии частиц при прямолинейном равноускоренном
движении в специальной теории относительности (1) и в механике Ньютона (2).
\label{funacli}}
\end{figure}

Рассмотрим случай прямолинейного движения, когда ускорение $\Ba'$ постоянно и
параллельно скорости $\Bv$. Для этого достаточно, чтобы ускорение было
постоянным и частица в начальный момент времени покоилась. Для определенности,
предположим, что ускорение и скорость параллельны оси $x=x^1$, т.е.\ мировая
линия частицы лежит в плоскости $t,x$. Положим $\Ba'=(b,0,0)$, где $b=\const$, и
$\Bv=\big(v(t),0,0\big)$. Тогда из (\ref{eobacr}) следует одно дифференциальное
уравнение
\begin{equation*}
  \frac{dv}{dt}=b\left(1-v^2/c^2\right)^{3/2}.
\end{equation*}
Оно легко интегрируется:
\begin{equation*}
  b(t-t_0)=\frac v{\sqrt{1-v^2/c^2}}.
\end{equation*}
где $t_0$ -- постоянная интегрирования. Подставляя равенство $v=dx/dt$, получаем
дифференциальное уравнение на функцию $x(t)$. Интегрирование полученного
уравнения дает уравнение на мировую линию частицы:
\begin{equation*}
  (x-x_0)^2-c^2(t-t_0)^2=\frac{c^4}{b^2},
\end{equation*}
где $x_0$ -- вторая постоянная интегрирования. Таким образом, мировая линия
частицы при прямолинейном равноускоренном движении представляет собой гиперболу
в плоскости движения $t,x$. При $t_0=0$ и $x_0=0$ мировая линия показана на
рис.\ref{funacli}. Для сравнения показана также мировая линия частицы в механике
Ньютона, которая задана хорошо известным уравнением
\begin{equation*}
  x=\frac{c^2}b+\frac12bt^2
\end{equation*}
и представляет собой параболу.

При равноускоренном движении наблюдаемая скорость частицы при $t\to\infty$
в специальной теории относительности стремится к скорости света, а в механике
Ньютона -- к бесконечности.

\chapter{Многообразия и тензорные поля                           \label{sdifma}}
В настоящей главе начинается изложение основных понятий дифференциальной
геометрии. Изложение часто носит координатный характер, т.е.\ все соотношения
записываются в локальной системе координат и приводятся правила преобразований
геометрических объектов при переходе от одной системы координат к другой.
Параллельно даются также глобальные определения, не зависящие от выбора системы
координат, и показывается связь координатного и инвариантного описания.
Глобальный бескоординатный подход компактен в обозначениях, прозрачен и удобен
для определения геометрических объектов. Координатный подход является более
громоздким, однако это не является признаком меньшей строгости. Он незаменим в
моделях математической физики, основанных на дифференциальной геометрии, где
требуется проведение расчетов.
\section{Многообразие                                            \label{sdefma}}
Основным понятием дифференциальной геометрии является дифференцируемое
многообразие $\MM$, которое обобщает понятие евклидова пространства $\MR^n$.
Топологически нетривиальные многообразия не покрываются одной системой
координат, однако локально устроены так же, как и евклидовы пространства. Это
позволяет использовать математический анализ для построения и анализа многих
важных моделей современной математической физики.
\begin{defn}
Топологическое хаусдорфово пространство $\MM$ со счетной базой, каждая точка
которого имеет окрестность, гомеоморфную открытому $n$-мерному шару единичного
радиуса $\MB_1^n$ в евклидовом пространстве $\MR^n$, называется $n$-мерным
{\em многообразием}. Число $n$ называется {\em размерностью} многообразия. Мы
пишем $\dim\MM=n$.
\qed\end{defn}
\index{Многообразие (manifold)}%
\index{Размерность многообразия (dimensionality of a manifold)}%
\begin{com}
В определении многообразия требование хаусдорфовости существенно и исключает
случаи подобные примерам на с.~\pageref{pnonha}. Согласно теореме \ref{thausd},
любая сходящаяся последовательность в хаусдорфовом пространстве имеет не более
одного предела.

Предположение о счетности базы также существенно, потому что это требование
обеспечивает паракомпактность многообразия, которая в рассматриваемом случае
достаточна для существования разбиения единицы. В свою очередь, существование
разбиения единицы важно, т.к.\ позволяет определять геометрические объекты на
всем многообразии, исходя из их задания в локальных координатах. Оно часто
используется для доказательства теорем существования.

Выбор шаров единичного радиуса $\MB^n_1$ сделан для определенности и не является
существенным, т.к.\ шар единичного радиуса гомеоморфен всему евклидову
пространству $\MR^n$, а также любой (связной и односвязной) области в $\MR^n$.
\qed\end{com}
\begin{com}
Поскольку многообразие, по-определению, является хаусдорфовым топологическим
пространством, то для многообразий понятие компактного пространства и компакта
совпадают.
\qed\end{com}

Окрестность произвольной точки многообразия $\MM$ устроена так же, как и
окрестность точки в евклидовом пространстве $\MR^n$. Однако в отличие от
евклидова пространства, которое, по-определению, может быть покрыто одной
картой, многообразие в общем случае одной картой не покрывается. Поэтому на
многообразии общего вида нельзя ввести структуру векторного или аффинного
пространства.
\begin{prop}
Многообразие представляет собой объединение конечного или счетного числа
областей, $\MM=\bigcup_i\MU_i$, каждая из которых гомеоморфна $n$-мерному шару,
и, следовательно, любой области в $\MR^n$.
\end{prop}
\begin{proof}
Поскольку каждая точка имеет окрестность, гомеоморфную $\MR^n$, то все
многообразие можно покрыть, возможно, несчетным числом карт $\MU_i$. Выберем
счетную базу на $\MM$, которая также является покрытием. Каждая координатная
окрестность $\MU_i$ является объединением счетного числа элементов базы и,
поэтому, на каждом элементе базы задан гомеоморфизм в $\MR^n$. Теперь можно
выбрать базу топологии в качестве координатного покрытия, а она -- счетна.
\end{proof}
В приложениях рассматриваются, как правило, многообразия, которые покрываются
конечным числом карт.
\begin{com}
Поскольку многообразие представляет собой объединение не более, чем счетного
числа областей, гомеоморфных $n$-мерному шару, то любое многообразие является
сепарабельным топологическим пространством.
\qed\end{com}

Согласно данному определению, многообразие, как объединение открытых множеств,
не имеет границы, которую в случае многообразий принято называть краем
(см.\ раздел \ref{smabou}).

Поскольку каждая окрестность многообразия гомеоморфна шару в евклидовом
пространстве, то всякое многообразие является локально связным. Согласно теореме
\ref{tlocco} всякое многообразие представляет собой объединение связных
компонент, каждая из которых является одновременно открытой и замкнутой. Число
этих компонент может быть не более, чем счетным, т.к.\ мы предполагаем счетность
базы многообразия. В дальнейшем, однако, мы иногда будем рассматривать
многообразия, состоящие из несчетного числа компонент.
\begin{defn}
Под $0$-мерным многообразием мы будем понимать счетное множество точек с
дискретной топологией. Одномерное и двумерное многообразия называются
соответственно {\em линией} и {\em поверхностью}.
\qed\end{defn}
\index{Линия (line)}\index{Поверхность (surface)}%

Размерность многообразия является топологическим инвариантом: два гомеоморфных
многообразия имеют одинаковую размерность. Это утверждение является прямым
следствием теоремы \ref{tinvdi} об инвариантности размерности.

Данное выше определение задает {\em топологическое многообразие}, т.к.\ в нем
говорится только о непрерывности. Теперь перейдем к описанию дифференцируемых
многообразий.
\index{Топологическое многообразие (topological manifold)}%
\begin{defn}
Из определения многообразия следует, что существует гомеоморфизм (непрерывная
биекция)
\begin{equation*}
  \vf_i:\quad\MM\supset\quad\MU_i\rightarrow\vf_i(\MU_i)\quad\subset\MR^n,
\end{equation*}
области $\MU_i$ на ее образ $\vf_i(\MU_i)$ (суммирования нет) в евклидовом
пространстве $\MR^n$. Поскольку в евклидовом пространстве есть система
координат, например, декартова, то данный гомеоморфизм можно записать в виде
\begin{equation*}
  \vf_i:\quad\MM\supset\MU_i\ni\quad x\mapsto\vf_i(x)
  =(x^1,\dotsc,x^n)\quad\in\vf_i(\MU_i)\subset\MR^n,
\end{equation*}
где точку многообразия $x$ и ее координаты $(x^1,\dotsc,x^n)$ мы обозначили
одной и той же буквой. Области $\MU_i$, покрывающие многообразие, называются
{\em координатными окрестностями}, а набор чисел $(x^1,\dotsc,x^n)$ --
{\em локальными координатами}. Пара $(\MU_i,\vf_i)$ называется {\em картой}.

Если две карты пересекаются, $\MU_i\bigcap\MU_j\ne\emptyset$, то произвольная
точка из пересечения $x\in\MU_i\bigcap\MU_j$ имеет свой набор координат в каждой
карте. Отображение областей евклидова пространства,
\begin{equation*}
  \vf_j\circ\vf^{-1}_i:\quad\MR^n\supset\quad\vf_i(\MU_i\cap\MU_j)\rightarrow
  \vf_j(\MU_i\cap\MU_j)\quad\subset\MR^n,
\end{equation*}
задается набором $n$ функций $x^{\al'}(x)$ от $n$ переменных (\ref{ecootr}), где
$x^{\al}$ и $x^{\al'}$ -- координаты соответственно на $\MU_i$ и $\MU_j$. Они
называются {\em функциями склейки}, поскольку склеивают между собой различные
карты. Совокупность всех карт, покрывающих многообразие, $\MM=\bigcup_i\MU_i$,
называется {\em координатным покрытием} или {\em атласом}
$\lbrace\MU_i,\vf_i\rbrace$, $i\in I$ многообразия $\MM$. Атлас, который не
содержится ни в каком другом атласе, называется {\em полным}.
\qed\end{defn}
\index{Координаты локальные (local coordinates)}%
\index{Локальные координаты (local coordinates)}%
\index{Окрестность координатная (coordinate neighborhood)}%
\index{Координатная окрестность (coordinate neighborhood)}%
\index{Карта (chart)}\index{Атлас (atlas)}%
\index{Функция склейки (sewing function)}%
\index{Склейки функции (sewing function)}%
\index{Координатное покрытие (coordinate covering)}%
\index{Покрытие координатное (coordinate covering)}%
\index{Атлас полный (complete atlas)}\index{Полный атлас (complete atlas)}%

При проведении вычислений в одной карте точку многообразия $x\in\MM$ и ее
координаты $\vf(x)\in\MR^n$ можно отождествлять. Но всегда следует помнить, что
точка одна, а координат много.
\begin{com}
В силу счетности топологической базы многообразия, координатное покрытие всегда
можно выбрать счетным. Однако, даже евклидово пространство можно при желании
покрыть несчетным числом карт. Это значит, что полный атлас любого многообразия
всегда несчетен.
\qed\end{com}

Мы требуем, чтобы каждая функция склейки $\vf_j\circ\vf^{-1}_i$ заданного атласа
определяла некоторый диффеоморфизм, т.е.\ являлась элементом псевдогруппы
преобразований координат $\diff^k(\MR^n)$ для некоторого $k$, определенной в
разделе \ref{scooch}. Поэтому мы говорим, что атлас $\lbrace\MU_i,\vf_i\rbrace$
совместим с псевдогруппой преобразований координат $\diff^k(\MR^n)$.
\begin{defn}
Многообразие $\MM$ вместе с полным атласом $\lbrace\MU_i,\vf_i\rbrace$
называется {\em дифференцируемым многообразием} класса $\CC^k$, $k\in\MN$, если
функции склейки (\ref{ecootr}) для всех пересекающихся карт непрерывны вместе
со своими частными производными $k$-того порядка:
$\vf_j\circ\vf^{-1}_i\in\diff^k(\MR^n)$ . Полный атлас
$\lbrace\MU_i,\vf_i\rbrace$ называется {\em дифференцируемой структурой}
многообразия $\MM$.
\qed\end{defn}
\index{Многообразие дифференцируемое (differentiable manifold)}%
\index{Дифференцируемое многообразие (differentiable manifold)}%
\index{Структура дифференцируемая (differentiable structure)}%
\index{Дифференцируемая структура (differentiable structure)}%
Аналогично определяются гладкие $\CC^\infty$ и вещественно аналитические
многообразия $\CC^\om$. На $\CC^\om$-многообразиях функции склейки задаются
сходящимися степенными рядами. Напомним, что функции склейки имеют ненулевой
якобиан и поэтому осуществляют взаимно однозначное отображение областей в
$\MR^n$ (теорема \ref{tcoord}). Это означает, что обратное преобразование
существует, и дифференцируемость обратных функций такая же, как и самих функций
склейки. В дальнейшем, если не оговорено противное, под многообразием мы будем
понимать $\CC^\infty$ дифференцируемые многообразия. Многие из рассмотренных
ниже утверждений справедливы и при более слабых ограничениях на
дифференцируемость функций склейки. Для проверки достаточно следить лишь за
числом производных при вычислениях.
\begin{com}
Отображения $\vf_i$ являются гомеоморфизмами. Это означает, что дифференцируемые
структуры, согласованы с естественной топологией евклидова пространства $\MR^n$,
т.е.\ можно считать, что топология многообразия индуцирована отображениями
областей $\MU_i$ в евклидово пространство $\MR^n$ с естественной топологией.
\qed\end{com}
\begin{com}
В определение дифференцируемого многообразия входит понятие евклидова
пространства $\MR^n$. В евклидовом пространстве можно определить много
различных структур. Для определения топологического многообразия нам достаточно
наличия топологии в $\MR^n$. Для определения дифференцируемого многообразия нам
достаточно дополнительно предположить наличие структуры поля на вещественной
прямой $\MR$ (для определения производной необходимо вычитать и делить числа).
Конечно, в евклидовом пространстве всегда можно ввести евклидову метрику и
структуру векторного пространства. Однако для определения дифференцируемого
многообразия это не является необходимым условием.
\qed\end{com}

Отметим некоторые свойства функций склейки.
\begin{prop}
Функции склейки удовлетворяют тождествам:
\begin{align}                                                     \label{esefui}
  f_{ij}&=f_{ji}^{-1}, && \forall\quad x\in\MU_i\cap\MU_j,
\\                                                                \label{esefus}
  f_{ij}f_{jk}f_{ki}&=\id, && \forall\quad x\in\MU_i\cap\MU_j\cap\MU_k,
\end{align}
где $\id$ -- тождественное отображение.
\end{prop}
\begin{proof}
Прямая проверка.
\end{proof}
\begin{cor}
Справедливо тождество
\begin{equation*}
  f_{ii}=\id.
\end{equation*}
\end{cor}
\begin{proof}
Из равенства (\ref{esefus}) с учетом (\ref{esefui}) получаем тождество
\begin{equation*}                                                    \tag*{\qed}
  f_{ii}=f_{ii}f^{-1}_{ki}f_{ki}=\id.
\end{equation*}
\renewcommand{\qed}{}\end{proof}
\begin{prop}
Любой атлас класса $\CC^k$ можно дополнить до полного атласа того же класса
гладкости, и соответствующий полный атлас единственный. Если на многообразии
задано два диффеоморфных атласа $\lbrace\MU_i,\vf_i\rbrace$ и
$\lbrace\MV_\Sa,\psi_\Sa\rbrace$, т.е.\ все отображения
\begin{equation*}
  \psi_\Sa\circ\vf^{-1}_i:\qquad
  \vf_i(\MU_i\cap\MV_\Sa)\rightarrow\psi_\Sa(\MU_i\cap\MV_\Sa)
\end{equation*}
являются диффеоморфизмами областей евклидова пространства $\MR^n$, как только
пересечения $\MU_i\cap\MV_\Sa$ не пусты, то они содержатся в одном полном
атласе. Любые два атласа, содержащиеся в одном полном атласе связаны между собой
диффеоморфизмом.
\end{prop}
\begin{proof}
Пусть $\lbrace\MW,\chi\rbrace$ -- семейство всех пар таких, что $\chi$ есть
гомеоморфизм открытого подмножества $\MW\subset\MM$ на открытое подмножество в
евклидовом пространстве $\MR^n$ и что отображение
\begin{equation*}
  \vf_i\circ\chi^{-1}:\qquad \chi(\MW\cap\MU_i)\rightarrow\vf_i(\MW\cap\MU_i)
\end{equation*}
есть элемент $\diff^k(\MR^n)$, как только $\MW\cap\MU_i$ не пусто. Тогда
$\lbrace\MW,\chi\rbrace$ есть полный атлас, содержащий
$\lbrace\MU_i,\vf_i\rbrace$. Этот атлас единственный по-построению.

Если заданы два диффеоморфных атласа $\lbrace\MU_i,\vf_i\rbrace$ и
$\lbrace\MV_\Sa,\psi_\Sa\rbrace$, то их объединение также является атласом и
того же класса гладкости. Поэтому они содержатся в одном полном атласе.

Все атласы, содержащиеся в одном полном атласе, получаются путем отбрасывания
некоторого количества карт. Они диффеоморфны между собой, поскольку все функции
склейки полного атласа принадлежат $\diff^k(\MR^n)$.
\end{proof}
Данное предложение означает, что для задания дифференцируемой структуры на
многообразии достаточно задать один атлас, не обязательно полный. Если на
многообразии $\MM$ можно задать два атласа одного класса гладкости, которые не
диффеоморфны между собой, то это означает, что на $\MM$ существуют различные
дифференцируемые структуры.
\begin{defn}
Пусть на многообразии $\MM$ введено две дифференцируемые структуры разных
классов гладкости $\CC^k$ и $\CC^l$, где $l>k$, причем полный атлас класса
$\CC^l$ является одновременно атласом класса $\CC^k$. Если $l=\infty$, то
ограничение класса гладкости $\CC^k$-многообразия до $\CC^\infty$-многообразия
называется {\em сглаживанием}. При этом сглаживания, которые связаны $\CC^k$
автоморфизмами, рассматриваются, как эквивалентные.
\qed\end{defn}
\index{Сглаживание многообразия (smoothing of a manifold)}%
\begin{theorem}[\bf Уитни]
Каждое $\CC^k$-многообразие имеет ровно одно (с точностью до
$\CC^k$-диффеоморфизмов) сглаживание.
\end{theorem}
\begin{proof}
См.\ \cite{Whitne36}.
\end{proof}
В силу этой теоремы во многих случаях можно игнорировать различие между $\CC^k$
дифференцируемыми и $\CC^\infty$ гладкими структурами.

В низших размерностях дифференцируемая структура на многообразиях единственна.
Доказано, что двух- и трехмерные многообразия допускают ровно одну
дифференцируемую структуру \cite{Munkre60}. Для многообразий более высоких
размерностей это не так. Данное утверждение иллюстрирует пример Милнора
\cite{Milnor56R}, который показал, что на семимерной сфере $\MS^7$,
рассматриваемой, как топологическое многообразие, существует 28 недиффеоморфных
дифференцируемых структур. Построение этих структур является сложным, и
мы его не приводим. Было также доказано, что четырехмерное евклидово
пространство $\MR^4$ допускает бесконечное множество недиффеоморфных
дифференцируемых структур \cite{DonKro91}. Существует также теорема о том,
что каждое гладкое многообразие имеет единственную вещественно аналитическую
структуру \cite{Whitne36}.

Если в определении многообразия отбросить требование счетности базы, то получим
{\em локально евклидово топологическое пространство}.
\index{Локально евклидово топологическое пространство%
(Locally Euclidean topological space)}%
\index{Топологическое пространство локально евклидово%
(Locally Euclidean topological space)}%
Существуют такие топологические локально евклидовы пространства, которые
вообще не допускают дифференцируемой структуры \cite{Kervai60}.

Любое открытое подмножество $\MD$ дифференцируемого многообразия $\MM$ само
является многообразием. Дифференцируемая структура на $\MD$ состоит из
карт $(\MU_i\cap\MD,\vf_i|_\MD)$, возникающих после сужения гомеоморфизмов
$\vf_i$ на $\MD$.

Поскольку многообразия являются топологическими пространствами, то их можно
умножать как топологические пространства. При этом произведение двух
многообразий $\MM\times\MN$ также является многообразием размерности $m+n$, где
$\dim\MM=m$ и $\dim\MN=n$. Дифференцируемая структура на прямом произведении
многообразий $\MM\times\MN$ строится, как прямое произведение дифференцируемых
структур на $\MM$ и $\MN$. А именно, пусть дифференцируемая структура на
многообразии $\MM$ определяется атласом $\lbrace\MU_i,\vf_i\rbrace$, а на $\MN$
-- атласом $\lbrace\MV_\Sa,\phi_\Sa\rbrace$. Тогда естественная дифференцируемая
структура на топологическом произведении $\MM\times\MN$ определяется атласом
$\lbrace\MU_i\times\MV_\Sa,\vf_i\times\phi_\Sa\rbrace$, где отображение
\begin{equation*}
  \vf_i\times\phi_\Sa:
  \quad \MU_i\times\MV_\Sa~\rightarrow~\MR^{m+n}=\MR^m\times\MR^n
\end{equation*}
определяется естественным образом. Отметим, что этот атлас не будет полным даже
если атласы $\lbrace\MU_i,\vf_i\rbrace$ и $\lbrace\MV_\Sa,\phi_\Sa\rbrace$
полны.
\begin{defn}
Если связное многообразие покрыто совокупностью карт $\lbrace\MU_i,\vf_i\rbrace$
с координатами $\vf_i(x)=\lbrace x^\al_i\rbrace$, причем якобианы функций
перехода (\ref{ejacob}) для всех пересекающихся карт $\MU_i$ и $\MU_j$
положительны,
$$
  \det\left(\frac{\pl x^\al_i}{\pl x^\bt_j}\right)>0,\qquad \forall i,j
$$
то многообразие называется {\em ориентированным}. Многообразие называется
{\em неориентируемым}, если атласа со всеми положительными якобианами функций
склейки не существует. При неудачно выбранном атласе на ориентируемом
многообразии якобианы могут быть разных знаков, однако атлас с положительными
якобианами существует. Такие многообразия называются {\em ориентируемыми}.
Несвязное многообразие $\MM$ называется ориентируемым, если ориентируема каждая
его компонента.
\qed\end{defn}
\index{Многообразие ориентированное (oriented manifold)}%
\index{Ориентированное многообразие (oriented manifold)}%
\index{Многообразие неориентируемое (nonorientable manifold)}%
\index{Неориентируемое многообразие (nonorientable manifold)}%
\index{Многообразие ориентируемое (orientable manifold)}%
\index{Ориентируемое многообразие (orientable manifold)}%
Связное ориентируемое многообразие допускает в точности две ориентации. Чтобы
поменять ориентацию ориентированного многообразия достаточно заменить каждую
карту $(\MU_i,\vf_i)$ ориентированного атласа на карту $(\MU_i,\psi_i)$, где
гомеоморфизм $\psi_i$ является композицией $\vf_i$ и отражения первой (или любой
другой) координаты: $(x^1,x^2,\dotsc,x^n)\mapsto$ $(-x^1,x^2,\dotsc,x^n)$.
(Более подробно ориентация многообразий будет обсуждаться в разделе
\ref{sfunor}.) Ориентация несвязного многообразия -- это выбор ориентации на
каждой компоненте.

Рассмотрим простейшие примеры многообразий.
\begin{exa}
Все евклидово пространство $\MR^n$ является простейшим $n$-мерным многообразием,
которое можно покрыть одной картой (а можно и несколькими). Дифференцируемая
структура -- это полный атлас, содержащий естественную карту
$\big(\MU=\MR^n,\vf=\id(\MR^n)\big)$. При этом класс гладкости многообразия
определяется классом гладкости допустимых преобразований координат. Любое
многообразие $\MM$, $\dim\MM=n$, которое покрывается одной картой, диффеоморфно
(см.\ раздел \ref{smapma}) $\MR^n$ и называется {\em тривиальным}.
\qed\end{exa}
\index{Тривиальное многообразие (trivial manifold)}%
\index{Многообразие тривиальное (trivial manifold)}%
\begin{exa}
Рассмотрим произвольное вещественное векторное пространство $\MV$, $\dim\MV=n$.
Если в векторном пространстве выбран базис $\lbrace e_a\rbrace$, $a=1,\dotsc,n$,
то каждая точка векторного пространства задается упорядоченным набором
вещественных чисел $(x^1,\dotsc,x^n)$. Тогда его можно отождествить с
евклидовым пространством $\MR^n$ и рассматривать, как гладкое многообразие.
Дифференцируемая структура в $\MV$ не зависит от выбора базиса, т.к.\ замена
базиса задается невырожденной матрицей, и соответствует преобразованию
координат класса $\CC^\infty$.
В дальнейшем мы всегда будем считать, что все векторные пространства
снабжены естественной дифференцируемой структурой евклидова пространства.
\qed\end{exa}

\begin{figure}[h,b,t]
\hfill\includegraphics[width=.7\textwidth]{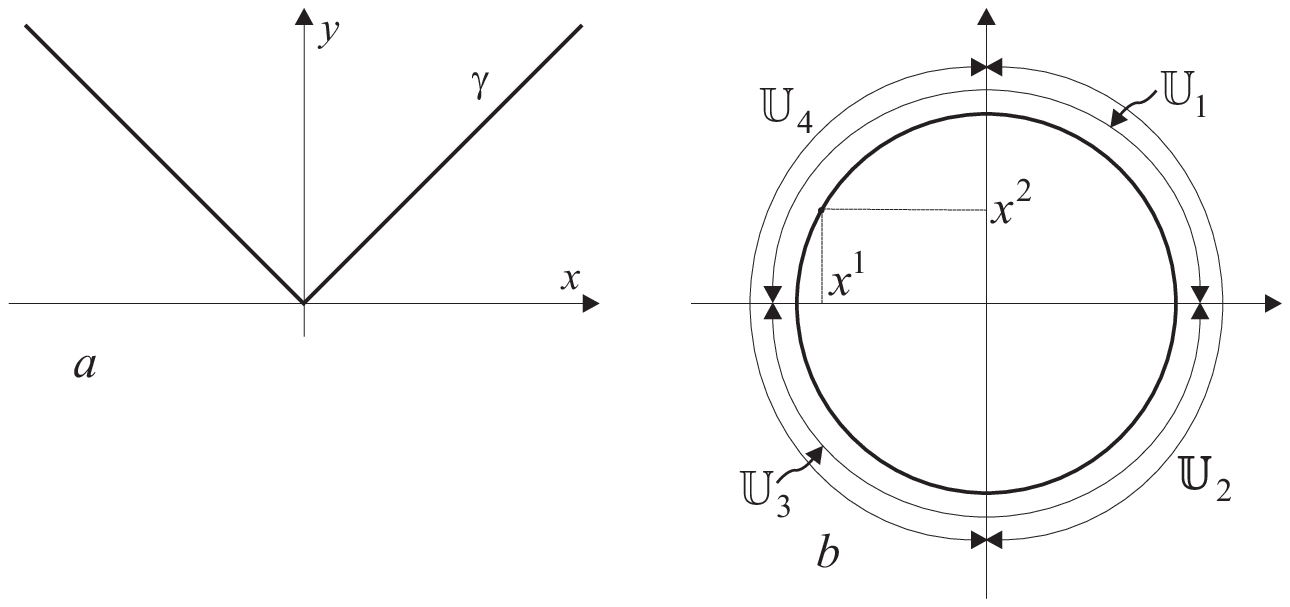}
\hfill {}
\centering \caption{Ломаная прямая является $\CC^\infty$ многообразием {\it(a)}.
Покрытие окружности $\MS^1$ четырьмя картами {\it(b)}. Показана точка,
принадлежащая пересечению $\MU_1\cap\MU_4$.}
\label{fzigcu}
\end{figure}

\begin{exa}
Рассмотрим ломанную линию $\g$ на евклидовой плоскости $\MR^2$, заданную
уравнением $y=|x|$ (см.\ рис.~\ref{fzigcu},{\it a}). Топология, индуцированная
из $\MR^2$, превращает ее в связное хаусдорфово топологическое пространство.
Ломаную $\g$ можно покрыть одной картой $(\MU,\vf)$, спроектировав ломаную линию
на ось $x$,
\begin{equation*}
  \MU=\lbrace(x,y)\in\MR^2:\quad y=|x|\rbrace, \qquad \vf(x,y)=x.
\end{equation*}
Таким образом $\g$ становится одномерным тривиальным многообразием класса
$\CC^\infty$. Однако вложение $\g\hookrightarrow\MR^2$ (см.\ раздел
\ref{smapma}) является только непрерывным, а не дифференцируемым, и это является
причиной излома.
\qed\end{exa}

\begin{exa}
$n$-мерная сфера $\MS^n_r\hookrightarrow\MR^{n+1}$ радиуса $r$ с центром в
начале координат, вложенная в $(n+1)$-мерное евклидово пространство, задается
уравнением
\begin{equation}                                                  \label{esphmd}
  \MS^n_r:=\lbrace x\in\MR^{n+1}:\quad
  (x^1)^2+(x^2)^2+\dotsc+(x^{n+1})^2=r^2\rbrace,\qquad r=\const>0.
\end{equation}
Само по себе это уравнение задает только множество точек в $\MR^{n+1}$, а никак
не многообразие. Зададим на нем топологию, сказав, что топология индуцирована
вложением $\MS^n_r\hookrightarrow\MR^{n+1}$. Пусть $N=(0,\dotsc,0,r)$ и
$S=(0,\dotsc,0,-r)$ -- северный и южный полюс сферы. Гладкая дифференцируемая
структура на $\MS^n_r$ -- это полный атлас, содержащий две карты:
$(\MS^n_r\setminus N,\vf_N)$ и $(\MS^n_r\setminus S,\vf_S)$, где $\vf_N$ и
$\vf_S$ -- стереографические проекции из северного и южного полюса. Тогда сфера
становится $n$-мерным компактным ориентируемым многообразием.Это многообразие
нетривиально и покрывается не менее, чем двумя картами.
\qed\end{exa}
\begin{exa}
В примере \ref{ecofil}, была построена сфера Римана $\overline\MC$. Она
представляет собой вещественное двумерное многообразие, диффеоморфное обычной
сфере, $\overline\MC\approx\MS^2$. С точки зрения комплексной геометрии она
является одномерным голоморфным комплексным многообразием комплексной
размерности $\dim_\MC\overline\MC=1$. Действительно, покроем сферу Римана
двумя картами: $\MU_0:=\MC$ -- окрестность нуля и
$\MU_\infty:=\overline\MC\setminus\lbrace0\rbrace$ -- окрестность бесконечности.
Они имеют непустое пересечение
$\MU_0\cap\MU_\infty=\MC\setminus\lbrace0\rbrace$.
В качестве координат на $\MU_0$ и $\MU_\infty$ выберем, соответственно, $z$ и
$1/z$. В области пересечения функция склейки $f_{0\infty}(z)=1/z$ голоморфны.
Тем самым мы построили голоморфный атлас на $\overline\MC$.

Как комплексное многообразие сфера Римана диффеоморфна одномерному комплексному
проективному пространству (проективной прямой), $\overline\MC\approx\MC\MP^1$.
Напомним, что проективная прямая -- это множество комплексных прямых
$az^1+bz^2=0$, где $(z^1,z^2)\in\MC\times\MC$ и $a,b\in\MC$, в двумерном
комплексном многообразии $\MC^2:=\MC\times\MC$. Если $b\ne0$, то из уравнения
прямой следует равенство $z^2=-az^1/b$, т.е.\ каждая прямая параметризуется
комплексным числом $-a/b=z\in\MC$. При $a\ne0$, аналогично, каждая прямая
параметризуется числом $-b/a=1/z\in\MC$. Тем самым мы покрыли проективное
пространство $\MC\MP^1$ двумя картами. В области пересечения $ab\ne0$ функция
склейки имеет вид $f(z)=1/z$. Таким образом, сфера Римана и одномерное
комплексное проективное пространство имеют одинаковое координатное покрытие и,
следовательно, диффеоморфны.
\qed\end{exa}

\begin{exa}                                                       \label{xcircl}
Рассмотрим задание окружности $\MS^1\hookrightarrow\MR^2$ единичного радиуса с
помощью четырех карт:
\begin{equation*}
\begin{split}
  \MU_1&=\lbrace x\in\MS^1:\quad x^2>0\rbrace,\quad \vf_1(x)=x^1,
\\
  \MU_2&=\lbrace x\in\MS^1:\quad x^1>0\rbrace,\quad \vf_2(x)=x^2,
\\
  \MU_3&=\lbrace x\in\MS^1:\quad x^2<0\rbrace,\quad \vf_3(x)=x^1,
\\
  \MU_4&=\lbrace x\in\MS^1:\quad x^1<0\rbrace,\quad \vf_4(x)=x^2.
\end{split}
\end{equation*}
Очевидно, что совокупность областей $\lbrace\MU_1,\MU_2,\MU_3,\MU_4\rbrace$
является конечным открытым покрытием окружности $\MS^1$. Четыре карты
$\lbrace(\MU_1,\vf_1),(\MU_2,\vf_2),(\MU_3,\vf_3),(\MU_4,\vf_4)\rbrace$
представляют собой атлас. В области пересечения двух карт $\MU_1\cap\MU_4$
функции преобразования координат имеют вид (см.\ рис.~\ref{fzigcu},{\it b})
\begin{equation*}
\begin{split}
  x^{2'}&=\quad \sqrt{1-(x^1)^2}>0,
\\
  x^{1'}&=-\sqrt{1-(x^2)^2}<0,
\end{split}
\end{equation*}
где нештрихованные и штрихованные координаты относятся соответственно к
областям $\MU_1$ и $\MU_4$.
Аналогично выписываются функции склейки для всех других пересечений карт.
Все функции склейки являются бесконечно дифференцируемыми. Это значит, что
построенный атлас принадлежит классу $\CC^\infty$.
\qed\end{exa}

\begin{figure}[h,b,t]
\hfill\includegraphics[width=.7\textwidth]{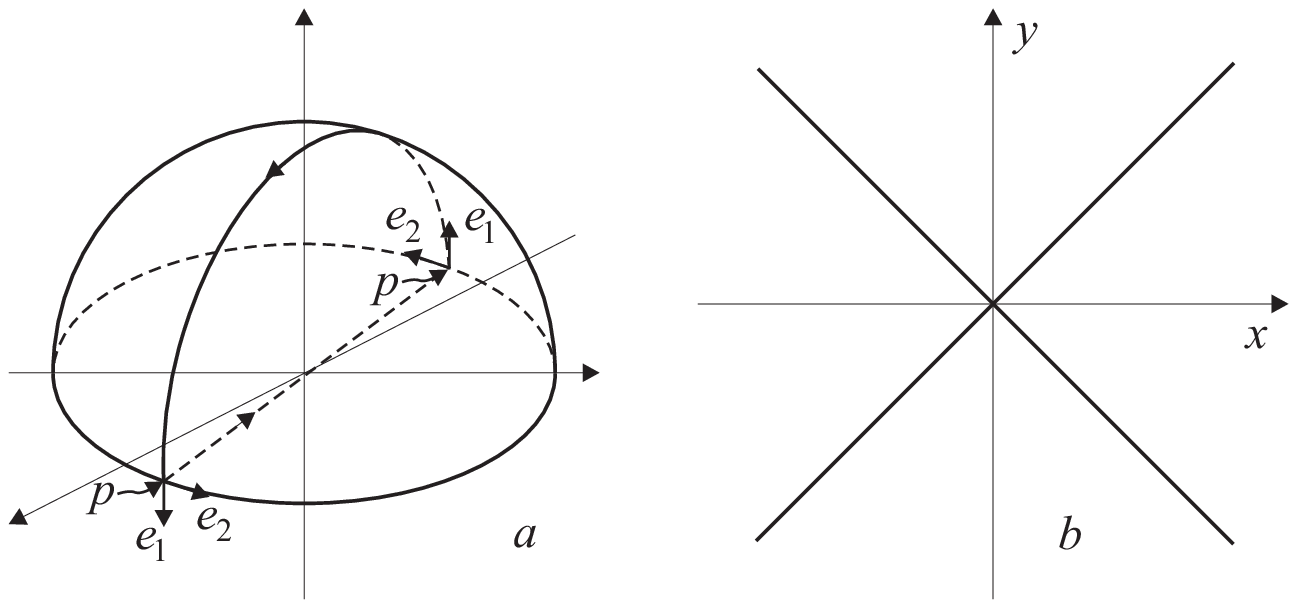}
\hfill {}
\centering \caption{Проективная плоскость $\MR\MP^2$ как полусфера в $\MR^3$.
Показан замкнутый путь, проходящий через граничную точку $p$, и перенос базиса
вдоль пути {\it(a)}. Две пересекающиеся прямые не являются многообразием
{\it(b)}.\label{fpropl}}
\end{figure}

\begin{exa}
Тор $\MT^n$ представляет собой прямое произведение $n$ окружностей,
\begin{equation*}
  \MT^n=\underbrace{\MS\times\MS\times\cdots\times\MS}_n,
\end{equation*}
и является $n$-мерным компактным ориентируемым многообразием. Дифференцируемая
структура на $\MT^n$ задается как на прямом произведении многообразий.
\qed\end{exa}
\begin{exa}                                                       \label{eprspa}
{\em Проективным пространством} $\MR\MP^n$ над полем
\index{Проективное пространство (projective space) $\MR\MP^n$}%
\index{Пространство проективное (projective space) $\MR\MP^n$}%
вещественных чисел называется множество прямых евклидова пространства
$\MR^{n+1}$, проходящих через начало координат. Проективное пространство
$\MR\MP^n$ представляет собой многообразие размерности $n$. Его можно
представлять себе как сферу $\MS^n$ с отождествленными диаметрально
противоположными точками. Действительно, любая прямая, проходящая
через начало координат, пересекает единичную сферу с центром в начале
координат ровно в двух диаметрально противоположных точках. Обратно,
любая из этих двух точек однозначно определяет прямую, проходящую через
начало координат. Таким образом
$$
  \MR\MP^n\approx\frac{\MS^n}{\MZ_2},
$$
где циклическая группа $\MZ_2$ состоит из двух элементов $\lbrace 1,-1\rbrace$.
Проективное пространство $\MR\MP^n$ можно представить также в виде
полусферы $x\in\MS^n$, $x^n\ge0$, у которой отождествлены диаметрально
противоположные ``граничные'' точки, т.е.\ точки $(n-1)$-мерной сферы
$$
  \MS^{n-1}=\lbrace x\in\MS^n:~x^n=0\rbrace.
$$

Рассмотрим замкнутую кривую в проективном пространстве, проходящую через одну из
``граничных'' точек. На рис.~\ref{fpropl},{\it a}, для наглядности изображена
проективная плоскость $\MR\MP^2$ в трехмерном евклидовом пространстве и
возможная кривая. Выберем ортонормированный базис вдоль кривой, включающий
единичный касательный вектор $e_1$. При прохождении через ``граничную'' точку
$p$ касательный вектор $e_1$ не меняет ориентации относительно кривой, в то
время как все остальные базисные векторы $e_2,e_3,\dotsc$ меняют направление.
Это значит, что ориентация ортонормированного базиса при прохождении вдоль этой
замкнутой кривой изменится при четных $n$ и сохранится при нечетных $n$. Если
замкнутая кривая целиком лежит в верхней полусфере, то ориентация базиса вдоль
кривой сохраняется. Тем самым мы показали, что проективные пространства четного
числа измерений неориентируемы, а нечетного числа измерений -- ориентируемы.
\qed\end{exa}

\begin{exa}
Продемонстрируем отличие топологического пространства от многообразия. Пусть
множество точек на евклидовой плоскости состоит из двух пересекающихся прямых
$y^2-x^2=0$, изображенных на рис.~\ref{fpropl},{\it b}. Пусть топология на этих
прямых индуцирована вложением. Тогда это множество представляет собой связное
хаусдорфово топологическое пространство. В то же время оно не является
многообразием, потому что окрестность точки пересечения прямых нельзя
взаимно однозначно отобразить на интервал вещественной прямой $\MR$.
\qed\end{exa}

\begin{figure}[h,b,t]
\hfill\includegraphics[width=.7\textwidth]{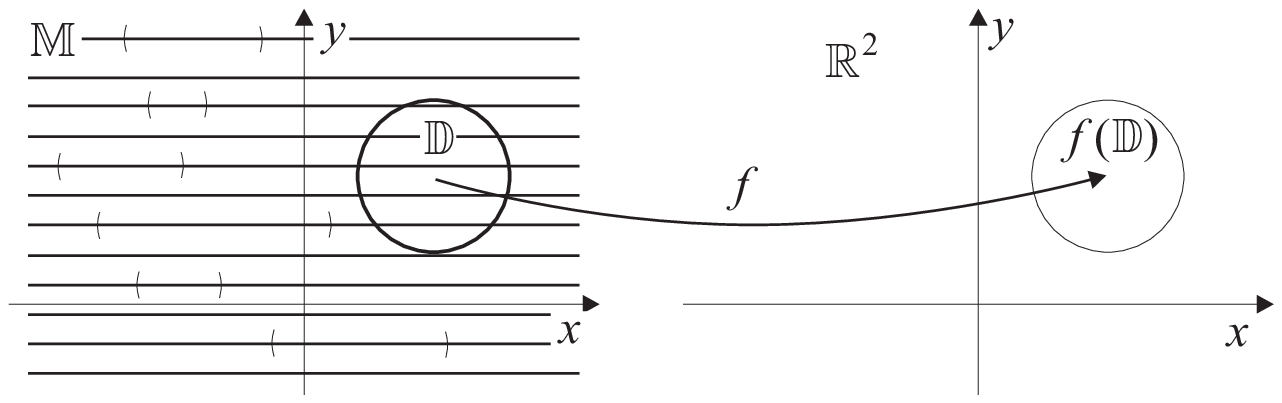}
\hfill {}
\centering \caption{Отображение $f$ объединения $\MM$ несчетного числа прямых,
параллельных оси $x$, на евклидову плоскость $\MR^2$.\label{fsloen}}
\end{figure}

\begin{exa}
Рассмотрим прямое произведение двух прямых $\MM=\MR\times\MR_\Sd$. Будем
считать, что на первом сомножителе задана естественная топология а на втором
-- дискретная, что отмечено индексом $\Sd$. Тогда, как множество,
многообразие $\MM$ совпадает с двумерным евклидовым пространством $\MR^2$,
а как многообразие -- нет. База топологии $\MM$ состоит из всех интервалов
на всех прямых, параллельных оси $x$ и проходящих через все точки оси $y$
(см.\ рис.~\ref{fsloen}). На каждой прямой база топологии счетна. Рассмотрим
открытый диск $\MD\subset\MM$. Так же как и в евклидовом пространстве
$\MR^2$, он является открытым множеством, но на этот раз как объединение
несчетного числа интервалов. Рассмотрим отображение $f$ многообразия $\MM$ на
евклидову плоскость $\MR^2$, которое задается простым отождествлением
координат: $\MM\ni(x,y)\overset{f}{\mapsto}(x,y)\in\MR^2$. Это отображение
биективно и непрерывно. Однако обратное отображение $f^{-1}$ не является
непрерывным, т.к.\ на евклидовой плоскости не существует открытого множества
$\MU\subset\MR^2$, образ которого при обратном отображении $f^{-1}(\MU)$
лежал бы в интервале. Поэтому отображение $f$ не является гомеоморфизмом
и множество $\MM=\MR\times\MR_\Sd$ не является двумерным многообразием. По
построению, множество $\MM$ является одномерным несвязным многообразием,
состоящим из несчетного числа одномерных многообразий -- прямых $\MR$.

Рассмотренный пример показывает, что на одном и том же множестве точек можно
задавать различные структуры. Мы говорим, что на евклидовой плоскости $\MR^2$,
рассматриваемой, как двумерное многообразие, задана структура слоения, т.е.\ мы
представляем плоскость в виде объединения несчетного числа одномерных
подмногообразий -- прямых $\MR$. Каждая прямая является листом слоения, которые
параметризуются точками другой прямой $\MR_\Sd$. Более подробно слоения
рассмотрены в разделе \ref{sfolia}.
\qed\end{exa}

В дальнейшем мы будем изучать различные свойства многообразий. Условимся о
терминологии. Будем говорить, что данное свойство выполняется на многообразии
$\MM$ {\em глобально}, если оно выполнено во всех точках $x\in\MM$. Гораздо чаще
встречаются свойства, которые выполнены только {\em локально}. А именно, для
каждой точки $x\in\MM$ существует координатная окрестность $\MU_x$ такая, что
данное свойство выполнено на $\MU_x$. В этом случае можно говорить, что данное
свойство выполнено в фиксированной системе координат. Конечно, любое свойство,
выполненное глобально, справедливо и локально, но не наоборот.
\index{Глобальность (globality)}\index{Локальность (locality)}%
\section{Разбиение единицы}
Один из способов задания геометрических структур, например, метрики, на
многообразии $\MM$ заключается в следующем. Сначала выбирается некоторый атлас,
покрывающий $\MM$. Затем в каждой карте данного атласа в координатах задается
некоторая геометрическая структура. Чтобы задать данную геометрическую структуру
на всем многообразии $\MM$, ее необходимо склеить в областях пересечения карт.
Для этого используется разбиение единицы.

Как уже отмечалось, для существования разбиения единицы нам достаточно
рассматривать паракомпактные многообразия. Счетность базы многообразия является
достаточным условием паракомпактности. Напомним некоторые определения и
утверждения из общей топологии, которые понадобятся для формулировки теоремы о
существовании разбиения единицы.
\begin{defn}
Топологическое пространство $\MM$ является {\em счетным в бесконечности}, если
существует счетное семейство компактных множеств $\MU_i$, $i\in\MN$, таких, что
\begin{equation*}
  \MU_1\subset\Int\MU_2\subset\MU_2\subset\dotsc\subset\MU_i\subset
  \Int\MU_{i+1}\subset\MU_{i+1}\dotsc
\end{equation*}
и
\begin{equation*}                                                    \tag*{\qed}
  \MM=\bigcup_{i=1}^\infty\MU_i.
\end{equation*}
\renewcommand{\qed}{}\end{defn}
\index{Топологическое пространство счетное в бесконечности%
(locally compact $\s$-compact topological space}%
\begin{theorem}
Паракомпактное топологическое пространство является объединением семейства
связных паракомпактных топологических пространств, которые являются счетными в
бесконечности.
\end{theorem}
\begin{proof}
См., например, \cite{Aubin01}.
\end{proof}
\begin{theorem}                                                   \label{tlocon}
Любое покрытие паракомпактного многообразия имеет счетное локально конечное
измельчение.
\end{theorem}
\begin{proof}
Существование локально конечного измельчения, которое также является
покрытием, входит в определение паракомпактности и его доказывать не
надо. Нетривиальность утверждения теоремы в том, что локально конечное
покрытие можно выбрать счетным. Доказательство приведено в \cite{Warner83R}.
\end{proof}
Перед тем как ввести разбиение единицы напомним
\begin{defn}
{\em Носителем} функции $f$, заданной на многообразии $\MM$, называется
замыкание множества тех точек, в которых она отлична от нуля. Носитель функции
обозначается $\supp f\subset\MM$. Если носитель функции компактен, то функция
называется {\em финитной}.
\qed\end{defn}
\index{Носитель функции (support of a function)}%
\index{Функции носитель (support of a function)}%
\index{Финитная функция (compactly supported function)}%
\index{Функция финитная (compactly supported function)}%
Из определения следует, что носитель произвольной функции $f$ всегда замкнут в
$\MM$.
\begin{defn}
{\em Разбиением единицы}, подчиненным заданному покрытию
$\lbrace\MU_i\rbrace_{i\in I}$ многообразия $\MM$, называется семейство функций
$\lbrace f_i\rbrace$, удовлетворяющих следующим условиям:\newline
\indent 1) $\supp f_i\subset\MU_i$ для всех $i$;\newline
\indent 2) \parbox[t]{.92\linewidth}{каждая точка имеет окрестность
$\MW$ такую, что $\MW\cap\supp f_i=\emptyset$ за исключением конечного
числа функций $f_i$;}\newline
\indent 3) $0\le f_i(x)\le1, \quad \sum_i f_i(x)=1,\qquad \forall x\in\MM$.
\qed\end{defn}
\index{Разбиение единицы (partition of unity}%
\begin{com}
В данном определении покрытие произвольно и не обязательно счетное и локально
конечное. Сумма в условии 3) определена, т.к.\ содержит только конечное число
слагаемых в силу условия 2).
\qed\end{com}
\begin{theorem}[\bf Разбиение единицы]                            \label{trazed}
На любом паракомпактном многообразии класса $\CC^k$ существует $\CC^k$
разбиение единицы, подчиненное заданному счетному локально конечному покрытию.
\end{theorem}
\begin{proof}
См., например, \cite{ChChLa00,Warner83R}.
\end{proof}
\begin{com}
Разбиение единицы существует как для ориентируемых, так и для неориентируемых
многообразий.
\qed\end{com}
Для заданного покрытия $\lbrace\MU_i\rbrace$ существует много разбиений единицы.
Пусть $\lbrace f_i\rbrace$ и $\lbrace g_i\rbrace$ -- два разбиения единицы,
подчиненные одному покрытию $\lbrace\MU_i\rbrace$. Тогда очевидна формула
\begin{equation*}
  \sum_i f_i\sum_j g_j=\sum_{ij}f_i g_j=\sum_j g_j\sum_i f_i,
\end{equation*}
поскольку в каждой точке $x\in\MM$ суммы содержат только конечное число
слагаемых.

Существование разбиения единицы на многообразии является эффективным средством
доказательства существования геометрических структур на многообразии путем
склеивания этих структур, заданных в отдельных картах. Например, в теореме
\ref{trimex} доказано существование римановой метрики на произвольном
многообразии.
\begin{com}
Условие паракомпактности, являющееся достаточным условием существования
разбиения единицы можно заменить на вторую аксиому счетности, утверждающую
существование счетной базы топологии. Если потребовать, чтобы многообразие $\MM$
являлось хаусдорфовым топологическим пространством, удовлетворяющим второй
аксиоме счетности, то отсюда будет следовать паракомпактность многообразий и,
кроме того, нормальность и метризуемость. (Доказательство метризуемости
топологических пространств, удовлетворяющих второй аксиоме счетности, приведено,
например, в \cite{Kelley57R}.)
\qed\end{com}
\section{Многообразия с краем                                    \label{smabou}}
\begin{defn}
Пусть $f(x)\in C^\infty(M)$ -- гладкая функция на многообразии $\MM$. Тогда
замкнутое множество $\MA\subset\MM$, выделяемое в многообразии $\MM$
неравенством $f(x)\le0$ (или $f(x)\ge0$) называется {\em многообразием с краем}.
Подмногообразие $\pl\MA\subset\MM$, задаваемое уравнением $f(x)=0$,
называется {\em краем} $\MA$. При этом мы предполагаем, что градиент функции $f$
на крае $\pl\MA$ отличен от нуля.
\qed\end{defn}
\index{Многообразие с краем (manifold with boundary)}%
\index{Край многообразия (boundary of a manifold)}%

\begin{figure}[h,b,t]
\hfill\includegraphics[width=.7\textwidth]{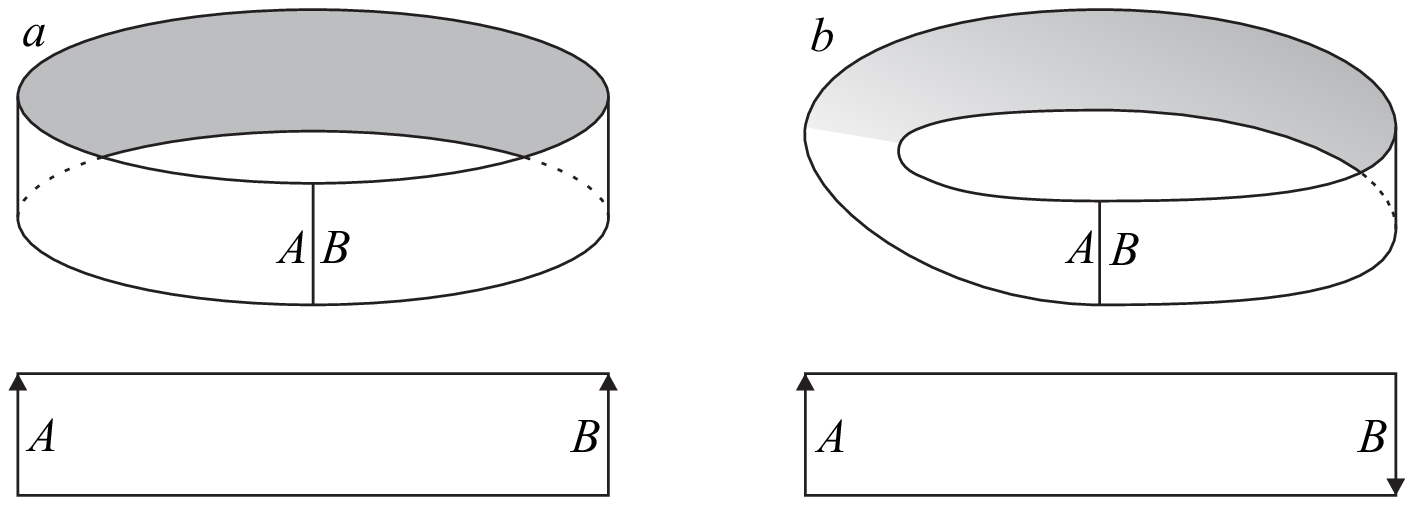}
\hfill {}
\centering\caption{Цилиндр {\it(a)} и лента Мебиуса {\it(b)} получаются при
склеивании двух сторон прямоугольника с сохранением и изменением направления
сторон при склейке, как показано стрелками.}
\label{fcymst}
\end{figure}

Если функция положительна, $f>0$, на $\MM$, то край пустой, $\pl\MA=\emptyset$,
и $\MA=\MM$. Нетривиальное многообразие с краем получается, если область
значений функции $f$ включает нуль.
\begin{exa}
Замыкание любой ограниченной открытой области $\MU\subset\MR^n$ в евклидовом
пространстве $\MR^n$ является многообразием с краем $\overline\MU$. При этом
краем является граница области: $\pl\overline\MU=\overline\MU\setminus\MU$.
\qed\end{exa}
\begin{com}
Условие отличия градиента функции от нуля на крае является достаточным условием
того, что край является $(n-1)$-мерным подмногообразием в $\MM$.
По сути дела это определение является инвариантным обобщением понятия области
и ее границы в евклидовом пространстве $\MR^n$. Если многообразие с краем можно
покрыть одной картой, то оба понятия в точности совпадают.
\qed\end{com}
\begin{exa}
Ориентируемым многообразием с краем является цилиндр конечной высоты, который
получается при склейке двух краев прямоугольника, показанного на
рис.~\ref{fcymst},{\it a}. При этом направление склеиваемых сторон, которое
показано стрелками, сохраняется. Цилиндр можно покрыть двумя картами, которые
пересекаются по двум областям, и в обеих областях якобиан перехода положителен.
Краем цилиндра является объединение двух окружностей.
\qed\end{exa}
\begin{exa}
Если перед склейкой прямоугольника изменить направление одной из сторон, как
показано стрелками на рис.~\ref{fcymst},{\it b}, то получится неориентируемая
поверхность с краем, которая называется {\em листом Мебиуса}. Лист Мебиуса также
можно покрыть двумя картами, пересекающимися по двум областям, но в этом случае
якобиан перехода в этих областях будет иметь разный знак. У листа Мебиуса край
диффеоморфен одной окружности и является связным. Окружность, возникающая при
склеивании середин отрезков $A$ и $B$, называется центральной.
\qed\end{exa}
\index{Лист Мебиуса (M\"obius band)}\index{Мебиуса лист (M\"obius band)}%
Если $\MA$ -- многообразие с краем, то разность $\MA\setminus\pl\MA$ также
является многообразием, но уже без края.

Дадим эквивалентное определение многообразия с краем. Обозначим полупространство
евклидова пространства $\MR^n$, определяемое уравнением $x^n>0$, через
$\MR^n_+$. Будем считать, что на замыкании $\overline\MR^n_+\subset\MR^n$ задана
индуцированная топология. При этом открытые множества в $\overline\MR^n_+$ могут
как содержать, так и не содержать точки края, определяемого уравнением $x^n=0$.
\begin{defn}
Топологическое хаусдорфово пространство $\MM$ со счетной базой, каждая точка
которого имеет окрестность, гомеоморфную открытому множеству в
$\overline{\MR}^n_+$, называется $n$-мерным {\em многообразием с краем}. Точки,
которые имеют окрестность, гомеоморфную $\MR^n$, называются {\em внутренними}.
Остальные точки называются {\em краевыми}.
\qed\end{defn}
\index{Внутренняя точка многообразия (interior point of a manifold)}%
\index{Точка многообразия внутренняя (interior point of a manifold)}%
\index{Краевая точка многообразия (boundary point of a manifold)}%
\index{Точка многообразия краевая (boundary point of a manifold)}%
Дифференцируемая структура на многообразии с краем вводится так же как и на
многообразии без края. Если не оговорено противное, то мы рассматриваем гладкие
дифференцируемые структуры класса $\CC^\infty$.
\begin{theorem}
Пусть $\MM$, $\dim\MM=n$, -- многообразие с краем. Если край $\pl\MM$
не является пустым, то $\pl\MM$ представляет собой многообразие размерности
$n-1$ и без края $\pl(\pl\MM)=\emptyset$.
\end{theorem}
\begin{proof}
См., например, \cite{MilSta74R}.
\end{proof}
\begin{defn}
Многообразие с краем $\MM$ называется {\em ориентируемым}, если ориентируемо
соответствующее ему многообразие без края $\MM\setminus\pl\MM$.
\qed\end{defn}
\index{Ориентируемое многообразие с краем (orientable manifold with boundary)}%
\index{Многообразие с краем ориентируемое (orientable manifold with boundary)}%
\begin{theorem}                                                   \label{toribo}
Если многообразие с краем $\MM$ ориентируемо, то его край $\pl\MM$ также
является ориентируемым многообразием. Ориентация на $\MM$ индуцирует
каноническую ориентацию края $\pl\MM$.
\end{theorem}
\begin{proof}
Пусть на $\MM\setminus\pl\MM$ задана какая либо ориентация, которую назовем
положительной. Рассмотрим естественное вложение края $h:~\pl\MM\rightarrow\MM$.
Тогда дифференциал отображения действует на касательные векторы
$h_*:~\MT_x(\pl\MM)\rightarrow\MT_x(\MM)$. Для каждой точки края $x\in\pl\MM$
выберем первый базисный вектор $e_1\in\MT_x(\MM)$, $e_1\notin\MT_x(\pl\MM)$
таким образом, чтобы он был ориентирован наружу. Это значит, что для любой
дифференцируемой функции, удовлетворяющей условиям $f(\pl\MM)=0$ и
$f(\MM\setminus\pl\MM)\le0$, производная вдоль $e_1$ неотрицательна $e_1f\ge0$.
Дополним этот вектор базисными векторами края $\lbrace e_2,\dotsc,e_n\rbrace$
таким образом, чтобы базис $\lbrace e_1,\dotsc,e_n\rbrace$ имел положительную
ориентацию в $\MM$. Тогда ориентация $\lbrace e_2,\dotsc,e_n\rbrace$ задает
{\em каноническую ориентацию} края.
\index{Каноническая ориентация края (canonical orientation of the boundary)}%
\index{Ориентация края каноническая (canonical orientation of the boundary)}%
\end{proof}
На языке дифференциальных форм задание согласованной ориентации многообразия
$\MM$ и его края $\pl\MM$ означает следующее. Пусть $\lbrace e^a\rbrace$,
$a=1,\dotsc,n$, -- набор 1-форм, дуальных к базису $e_a$, построенному в
доказательстве теоремы \ref{toribo}: $e^a(e_b)=\dl^a_b$. Поскольку 1-формы
линейно независимы, то $n$-форма $e^1\wedge e^2\wedge\dotsc\wedge e^n$ отлична
от нуля и задает ориентацию на $\MM$. Формы $e^2,\dotsc,e^n$, по-построению,
линейно независимы на крае $\pl\MM$ и задают каноническую ориентацию края.
В дальнейшем мы всегда предполагаем, что ориентация края $\pl\MM$ индуцирована
ориентацией самого многообразия $\MM$.

Из определения края следует выражение для края прямого произведения двух
многообразий (правило Лейбница)
\begin{equation*}
  \pl\big(\MM_1\times\MM_2\big)
  =\big(\pl\MM_1\times\MM_2\big)\cup\big(\MM_1\times\pl\MM_2\big).
\end{equation*}
\begin{defn}
В моделях гравитации принято называть вселенную {\em замкнутой}, если она
представляет собой компактное многообразие без края. {\em Открытая вселенная}
является  некомпактным многообразием без края. В общем случае будем называть
компактное многообразие без края {\em замкнутым}. Такие многообразия являются,
конечно, замкнутыми множествами в топологическом смысле. Однако термин замкнутый
в данном определении и в определении замкнутого множества в топологии имеют
разный смысл.
\qed\end{defn}
\index{Замкнутая вселенная (closed universe)}%
\index{Вселенная замкнутая (closed universe)}%
\index{Замкнутое многообразие (closed manifold)}%
\index{Многообразие замкнутое (closed manifold)}%
\index{Открытая вселенная (open universe)}%
\index{Вселенная открытая (open universe)}%
\begin{exa}
Прямая представляет собой замкнутое (и одновременно открытое) множество в
естественной топологии. В то же время она не является замкнутым многообразием,
поскольку некомпактна.
\qed\end{exa}
\section{Расслоения                                              \label{sfibun}}
В разделе \ref{sdefma} было определено прямое произведение $\MM\times\MF$ двух
дифференцируемых многообразий $\MM$ и $\MF$, которое также является
дифференцируемым многообразием. При этом дифференцируемая структура на прямом
произведении многообразий определяется дифференцируемыми структурами
сомножителей. В настоящем разделе мы обобщим понятие прямого произведения
многообразий.
\begin{defn}
{\em Расслоением} называется четверка $\ME(\MM,\pi,\MF)$, где $\ME,\MM,\MF$ --
многообразия, а $\pi:~\ME\rightarrow\MM$ -- дифференцируемое отображение,
которые удовлетворяют следующим условиям:\newline
\indent 1) \parbox[t]{.93\linewidth}{каждая точка $x\in\MM$ имеет окрестность
$\MU_x$ такую, что существует диффеоморфизм
$\chi:~\pi^{-1}(\MU_x)\rightarrow\MU_x\times\MF$
(локальная тривиальность);}\newline
\indent 2) \parbox[t]{.93\linewidth}{композиция отображений
$\pi\circ\chi^{-1}:~\MU_x\times\MF\rightarrow\MM$ есть проекция на первый
сомножитель: $(y,v)\mapsto y$ для всех $y\in\MU_x$ и $v\in\MF$.} \newline
Многообразие $\ME$ называется {\em пространством расслоения}, $\MM$ --
{\em базой} расслоения, $\MF$ -- {\em типичным слоем} и $\pi$ --
{\em проекцией}.
\qed\end{defn}
\index{Расслоение (fiber bundle)}%
\index{Пространство расслоения (fiber bundle space)}%
\index{Типичный слой (typical fiber)}\index{Слой типичный (typical fiber)}%
\index{База расслоения (base of a fiber bundle}%
\index{Проекция (projection)}%

Поскольку отображение $\chi$ в условии 1) является диффеоморфизмом, то
\begin{equation*}
  \dim\ME=\dim\MM+\dim\MF.
\end{equation*}

Мы предполагаем, что все многообразия являются достаточно гладкими. Кроме того
база предполагается связным многообразием. В противном случае расслоения можно
рассматривать над каждой связной компонентой. Расслоение также называется
{\em расслоенным пространством} и обозначается $\ME\xrightarrow{\pi}\MM$. Мы
считаем, что проекция $\pi$ является дифференцируемым сюрьективным отображением
на базу $\MM$. В противном случае можно рассмотреть расслоение с
базой $\pi(\ME)\subset\MM$. Поскольку отображение $\pi$ непрерывно, то согласно
теореме \ref{tconma} прообраз $\pi^{-1}(x)$ является замкнутым подмногообразием
в $\ME$, которое называется {\em слоем} над точкой базы $x\in\MM$.
\index{Слой (fiber)}%

\begin{figure}[h,b,t]
\hfill\includegraphics[width=.5\textwidth]{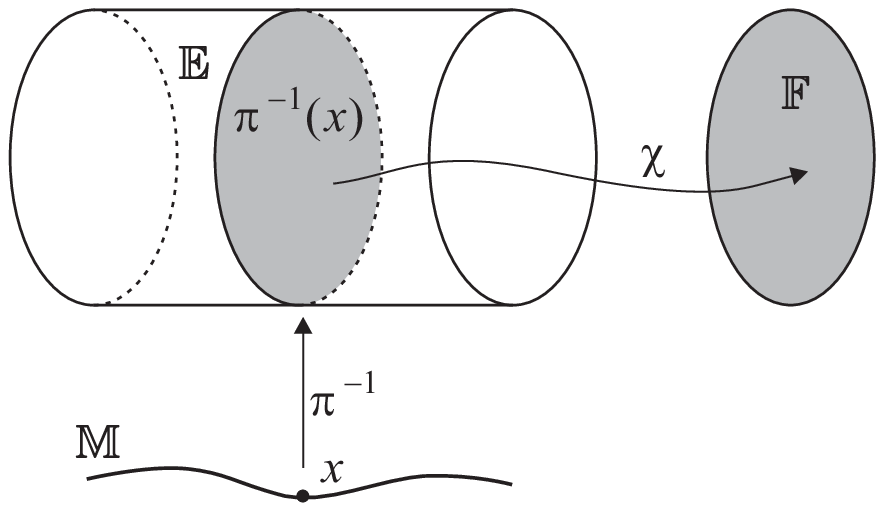}
\hfill {}
\centering\caption{Схематичное изображение структуры расслоения. $\ME$ --
пространство расслоения, $\MM$ -- база, $\MF$ -- типичный слой,
$\pi:~\ME\rightarrow\MM$ -- проекция, $\chi:~\pi^{-1}(x)\rightarrow\MF$ --
диффеоморфизм.}
\label{ffibun}
\end{figure}

В дальнейшем, для краткости, мы иногда будем обозначать расслоение одной буквой
$\ME$. На рис.\ \ref{ffibun} схематично показана структура расслоения.
\index{Расслоенное пространство (fiber bundle)}%
\index{Пространство расслоенное (fiber bundle)}%
\begin{com}
Атласы, заданные на базе и типичном слое, определяют атлас на пространстве
расслоения в силу локальной тривиальности расслоения. Тем самым первое условие в
определении расслоения является достаточным для того, чтобы дифференцируемые
структуры на трех многообразиях $\ME,\MM$ и $\MF$ были согласованы между собой.
Второе требование является условием коммутативности диаграммы
\begin{equation*}
\begin{diagram}
  \ME\supset\pi^{-1}(\MU_x) & \rTo^{\chi} & \MU_x\times\MF \\
  & \rdTo^\pi & \dTo_{\pr=\pi\circ\chi^{-1}} \\ & & \MU_x
\end{diagram}
\end{equation*}
где $\pr$ -- естественная проекция прямого произведения
$\MU_x\times\MF\rTo^\pr \MU_x$ на первый сомножитель. В общем случае это могло
бы быть не так. В приведенной диаграмме окрестность $\MU_x$ нельзя заменить на
все многообразие $\MM$, так как отображение $\chi$ определено локально.
\qed\end{com}
\begin{exa}
Прямое произведение двух многообразий $\ME=\MM\times\MF$ с проекцией на первый
сомножитель $\MM\times\MF\xrightarrow\pi\MM$ является расслоением с базой $\MM$
и типичным слоем $\MF$, которое называется {\em тривиальным}.
\index{Тривиальное расслоение (trivial fiber bundle)}%
\index{Расслоение тривиальное (trivial fiber bundle)}%
С равным успехом четверка $\MM\times\MF\xrightarrow{\pi'}\MF$ с проекцией на
второй сомножитель является тривиальным расслоением с базой $\MF$ и типичным
слоем $\MM$.
\qed\end{exa}

Многообразия $\MM,\MF,\ME$ могут быть как с краем, так и без края. Мы допускаем
также в качестве базы или типичного слоя 0-мерные многообразия, т.е.\ конечные
или счетные наборы точек с дискретной топологией.

По-построению, пространство расслоения $\ME$ представляет собой объединение
несчетного числа слоев, $\ME=\bigcup_{x\in\MM}\pi^{-1}(x)$, каждый из которых
диффеоморфен типичному слою $\MF$ и ``нумеруется'' точкой базы. Это --
частный случай слоений, рассмотренных в разделе \ref{sfolia}.

\begin{figure}[h,b,t]
\hfill\includegraphics[width=.6\textwidth]{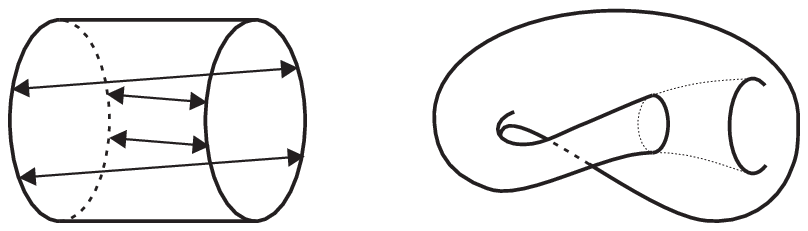}
\hfill {}
\centering\caption{Бутылка Клейна}
\label{fkleinbottle}
\end{figure}

Как было отмечено, дифференцируемые структуры на базе и слое согласованы с
дифференцируемой структурой на пространстве расслоения $\ME$. Опишем это более
подробно. Пусть $\ME(\MM,\pi,\MF)$ -- расслоение, $\dim\MM=n$, $\dim\MF=k$.
По-определению, отображение $\chi$ действует на каждую точку $p\in\ME$:
$\chi(p)=\big(x=\pi(p),v\big)\in\MM\times\MF$. Поскольку база и типичный слой --
многообразия, то существуют карты:
\begin{align*}
  (\MU_x,\vf)&:\quad \MM\supset\quad\MU_x\xrightarrow{\vf}\vf(\MU_x)
  \quad\subset\MR^n,
\\
  (\MV_v,\phi)&:\quad \MF\supset\quad\MV_v\xrightarrow{\phi}\phi(\MV_v)
  \quad\subset\MR^k.
\end{align*}
Локально определено отображение из пространства расслоения $\ME$ в евклидово
пространство $\MR^{n+k}$:
\begin{equation*}
  \rho:\quad\chi^{-1}(\MU_x\times\MV_v)~\rightarrow~\vf(\MU_x)\times\phi(\MV_v)
  \in\MR^{n+k}.
\end{equation*}
Таким образом, для каждой точки расслоения $p\in\ME$ определена карта
$\big(\chi^{-1}(\MU_x\times\MV_v),\rho\big)$. Определено также отображение,
действующее из $\MR^n\times\MR^k$ в $\MR^n$:
\begin{equation*}
  \tilde\pi=\vf\circ\pi\circ\rho^{-1}:\quad
  \vf(\MU_x)\times\phi(\MV_v)\ni(y,w)~\mapsto~y\in\vf(\MU_x),
\end{equation*}
где $y\in\vf(\MU_x)\subset\MR^n$ и $w\in\phi(\MV_f)\subset\MR^k$.
\begin{defn}
Дифференцируемое отображение $\MM\xrightarrow{\s}\ME$ называется {\em сечением}
или {\em глобальным сечением} расслоения $\ME\xrightarrow{\pi}\MM$, если
$\pi\circ\s=\id(\MM)$. Аналогичным образом, дифференцируемое отображение
$\MU\xrightarrow{\s}\ME$, где $\MU$ -- открытое подмножество базы $\MM$,
называется {\em локальным сечением} расслоения $\ME\xrightarrow{\pi}\MM$, если
$\pi\circ\s=\id(\MU)$.
\qed\end{defn}
\index{Сечение (cross-section)}%
\index{Глобальное сечение (global cross-section)}%
\index{Сечение глобальное (global cross-section)}%
\index{Локальное сечение (local cross-section)}%
\index{Сечение локальное (local cross-section)}%
Локальные сечения существуют у любого расслоения -- это функции со значениями
в пространстве расслоения $\ME$, которые определены в областях $\MU\subset\MM$.
Глобальные сечения расслоений, как мы увидим в дальнейшем, существуют далеко не
всегда.
\begin{exa}
Пусть $\ME=\MM\times\MR$ -- тривиальное расслоение, типичным слоем которого
является поле вещественных чисел. Тогда множество всех гладких сечений
\begin{equation*}
  f:\quad \MM\ni\quad x\mapsto\big(x,f(x)\big)\quad\in\ME=\MM\times\MR
\end{equation*}
совпадает с множеством графиков всех вещественнозначных бесконечно
дифференцируемых функций $\CC^\infty(\MM)$
\qed\end{exa}
\begin{exa}[\bf Цилиндр]
Цилиндр единичной высоты -- это прямое произведение окружности на единичный
отрезок $\MS^1\times[0,1]$. Его можно рассматривать как расслоение с базой
$\MM=\MS^1$ и типичным слоем $[0,1]$. У этого расслоения существуют глобальные
сечения -- гладкие функции на окружности со значениями в единичном отрезке
$[0,1]$. Например, $f(\vf)=\sin\vf$, где $\vf\in[0,2\pi]$ -- координата на
окружности.
\qed\end{exa}
\begin{exa}[\bf Лист Мёбиуса]
Лист Мебиуса, изображенный на рис.\ref{fcymst},{\it b}, является расслоением с
базой $\MS^1$, типичным слоем которого, как и у цилиндра, является единичный
отрезок $[0,1]$. Это расслоение нетривиально, т.к.\ не имеет вида прямого
произведения.
\qed\end{exa}
\begin{exa}[\bf Бутылка Клейна]
Построение бутылки Клейна изображено на рис.\ref{fkleinbottle}. Мы берем цилиндр
конечной высоты и отождествляем точки граничных окружностей, предварительно
отобразив точки окружности с одной стороны цилиндра относительно произвольного
диаметра. Эту поверхность нельзя вложить в трехмерное евклидово пространство
$\MR^3$ и поэтому трудно представить. Бутылку Клейна можно рассматривать, как
расслоение с базой $\MM=\MS^1$ и типичным слоем $\MF=\MS^1$. Это расслоение
нетривиально.
\qed\end{exa}
\index{Бутылка Клейна (Klein bottle)}\index{Клейна бутылка (Klein bottle)}%

\begin{exa}
Пусть задана группа Ли $\MG$ и ее нормальная подгруппа $\MH\subset\MG$.
Рассмотрим отображение группового многообразия $\MG$ на пространство правых
(или левых) смежных классов $\pi:\MG\rightarrow\MG/\MH$, определенное правилом
$\pi(g)=\MH g$, где $g\in\MG$. Тогда $\MG\xrightarrow\pi\MG/\MH$ -- расслоение.
Пространством расслоения является группа Ли $\MG$, базой -- факторгруппа
$\MG/\MH$ и типичным слоем -- нормальная подгруппа $\MH$. Дифференцируемая
структура на фактор пространстве $\MG/\MH$ будет определена позже в теореме
\ref{tfacma}. Если размерность группы равна размерности нормальной подгруппы,
$\dim\MG=\dim\MH$, то базой является конечный или счетный набор точек, т.е.\
0-мерное многообразие.
\qed\end{exa}
\begin{com}
В определении расслоения общего вида мы не предполагаем наличия каких либо
структур в типичном слое, кроме структуры дифференцируемого многообразия. В
дальнейшем мы рассмотрим частные случаи расслоенных пространств, когда типичным
слоем является векторное пространство (векторное расслоение) или группа Ли
(главное расслоение).
\qed\end{com}
\section{Скалярные поля и плотности                              \label{scfden}}
В моделях математической физики, как правило, постулируется, что пространство и
пространство-время, в котором мы живем, являются многообразиями. Само по себе
это очень глубокое предположение. Однако для построения физических моделей его
недостаточно. Для описания движения и взаимодействия различных физических
объектов в пространстве-времени необходимо задание дополнительных структур на
многообразии. Такими структурами являются скалярные и векторные поля,
$r$-формы, тензорные поля, метрика и связность, которые, в частности,
характеризуются различными трансформационными свойствами при преобразовании
координат. Начнем с простейшего объекта -- скалярного поля (функции).

Рассмотрим вещественнозначную функцию $f$ на многообразии $\MM$, $\dim\MM=n$,
т.е.\ отображение
\begin{equation}                                                  \label{escfde}
  f:\quad \MM\rightarrow\MR.
\end{equation}
Это отображение часто, особенно в физических приложениях, называют
{\em скалярным полем} на $\MM$. По-определению отображения (\ref{esemap})
скалярное поле должно быть однозначно. Функция называется {\em дифференцируемой}
класса $\CC^k$, если отображение (\ref{escfde}), заданное в координатах,
\index{Скалярное поле (scalar field)}\index{Поле скалярное (scalar field)}%
\index{Дифференцируемая функция (differentiable function)}%
\index{Функция дифференцируемая (differentiable function)}%
\begin{equation*}
  f\circ\vf^{-1}_i:\quad \MR^n\supset\quad\vf_i(\MU_i)\ni x\mapsto f(x)
  \quad\in\MR
\end{equation*}
$k$ раз непрерывно дифференцируемо в каждой карте атласа
$\lbrace\MU_i,\vf_i\rbrace$. Конечно, не имеет смысла говорить о степени
гладкости функции, которая превышает степень гладкости дифференцируемой
структуры многообразия. Поэтому мы предполагаем, что степень гладкости функции
меньше или равна степени гладкости многообразия.

В двух областях $\MU_i$ и $\MU_j$ скалярное поле задается, соответственно, двумя
функциями $f(x)$ и $f'(x')$ от $n$ переменных $x=\lbrace x^\al\rbrace$ и
$x'=\lbrace x^{\al'}\rbrace$, $\al,\al'=1,\dotsc,n$. Если области пересекаются,
то в области пересечения согласно, (\ref{ecootr}), справедливо равенство
\begin{equation}                                                  \label{escfun}
  f'\big(x'(x)\big)=f(x),
\end{equation}
поскольку в каждой точке функция имеет только одно значение. Формулу
(\ref{escfun}) можно интерпретировать, как правило преобразования функции при
замене координат $x^\al\mapsto x^{\al'}(x)$. Другими словами, значение функции
после преобразования в точке $x'$ равно ее прежнему значению в точке $x$.
\begin{com}
В дальнейшем мы будем позволять себе некоторую вольность в обозначениях.
Запись $f(x)$ в зависимости от контекста будет пониматься двояко. Во-первых,
$f(x)$ обозначает значение функции $f$ в произвольной точке многообразия
$x\in\MM$ безотносительно какой либо карты. Это не есть функция $n$ вещественных
переменных. Во-вторых, запись $f(x)$ обозначает также значение функции
$f\circ\vf^{-1}$ в точке $\lbrace x^\al\rbrace\in\MR^n$. Это -- обычная функция
от $n$ вещественных переменных (координат). Строго говоря, частная производная
от функции на многообразии $\pl_\al f$ не определена, т.к.\ мы не знаем, что
такое разность двух точек $x_1-x_2$ на многообразии. Тем не менее мы будем
употреблять запись $\pl_\al f$, принимая
\begin{equation*}
  \pl_\al f:=\pl_\al(f\circ\vf^{-1}).
\end{equation*}
Фактически это означает, что в каждой отдельно взятой карте мы отождествляем
точки многообразия с точками евклидова пространства:
$x\sim\vf(x)=\lbrace x^\al\rbrace\in\MR^n$, и функции: $f\sim f\circ\vf^{-1}$.
При проведении вычислений в одной карте это не приводит к какой либо путанице.
\qed\end{com}

Множество всех функций класса $\CC^k$ на многообразии $\MM$ обозначим
$\CC^k(\MM)$. Множество гладких (бесконечно дифференцируемых) функций
обозначим $\CC^\infty(\MM)$. Скалярное поле называется {\em тривиальным}, если
оно равно нулю на $\MM$.
\index{Тривиальное скалярное поле (trivial scalar field)}%
\index{Скалярное поле тривиальное (trivial scalar field)}%

На множестве функций $\CC^k(\MM)$ определим две поточечные операции: сложение и
умножение:
$$
  (f+g)(x):=f(x)+g(x),\qquad(fg)(x):=f(x)g(x),\qquad\forall x\in\MM.
$$
То есть значения суммы и произведения двух функций в данной точке равно
соответственно сумме и произведению значений этих функций в той же точке.
Очевидно, что сумма и произведение двух функций снова дает функцию. По отношению
к этим операциям функции образуют коммутативное кольцо. Кроме этого функции
можно умножать на действительные числа. Умножение на числа вместе с операцией
сложения превращает множество функций в векторное пространство над полем
вещественных чисел. Если на множестве функций рассматривать все три операции
(умножение на числа, сложение и умножение функций), то оно образует
коммутативную ассоциативную алгебру с единицей над полем вещественных чисел,
которую так же будем обозначать $\CC^k(\MM)$. Эта алгебра является
бесконечномерной.

В разделе \ref{scooch} было показано, что координаты точки евклидова
пространства сами можно рассматривать, как набор функций. Для каждой карты
$(\MU,\vf)$ многообразия $\MM$ определен набор функций $x^\al(x)$,
$\al=1,\dotsc,n$ от точки многообразия $x\in\MU$,
\begin{equation}                                                  \label{ecoorf}
  \vf(x)=\lbrace x^\al(x)\rbrace\quad\in\MR^n,
\end{equation}
которые называются {\em координатными функциями}. Эти функции свои для каждой
карты.
\index{Координатные функции (coordinate functions)}%
\index{Функции координатные (coordinate functions)}%
Во многих случаях координатные функции, определенные на $\MU\subset\MM$,
упрощают запись, позволяя опускать символ отображения $\vf(x)$.

Определим новый геометрический объект -- скалярную плотность $h$. С этой целью
рассмотрим две пересекающиеся карты $\MU_i\cap\MU_j\ne\emptyset$. При
преобразовании координат можно умножить функцию $h_i$, заданную на карте
$\MU_i$, на якобиан преобразования (\ref{ejacob}) в степени $p$:
\begin{equation}                                                  \label{escfud}
  h_j\big(x'(x)\big):=J^p_{ji}(x)h_i(x),
\end{equation}

Пусть $\lbrace \MU_i,\vf_i\rbrace$ -- некоторый атлас на многообразии $\MM$. В
каждой карте $(\MU_i,\vf_i)$ зададим отображение
\begin{equation*}
  h_i:\quad \MR^n\supset\vf_i(\MU_i)\ni\quad x=\lbrace x^\al\rbrace\mapsto
  h_i(x)\quad\in\MR
\end{equation*}
таким образом, что в области пересечения карт $\MU_i\cap\MU_j$ выполнен закон
преобразования (\ref{escfud}) для всех $x\in\MU_i\cap\MU_j$. Это определение
корректно, т.к.\ в области пересечения трех карт $\MU_i\cap\MU_j\cap\MU_k$ с
координатами $x=\lbrace x^\al\rbrace$, $x'=\lbrace x^{\al'}\rbrace$ и
$x''=\lbrace x^{\al''}\rbrace$ выполнено равенство
\begin{equation*}
  h_k\big(x''\big(x'(x)\big)\big)=J^p_{kj}h_j\big(x'(x)\big)
  =J^p_{kj}J^p_{ji}h(x)=J^p_{ki}h(x),
\end{equation*}
где мы воспользовались равенством $J_{kj}J_{ji}=J_{ki}$ для якобианов
преобразования координат $x\mapsto x'(x)$, $x'\mapsto x''(x')$ и
$x\mapsto x''(x)$.

Преобразования (\ref{escfud}) образуют группу. Действительно, якобианы
преобразований координат, по-определению, отличны от нуля, якобиан двух
последовательных преобразований равен произведению якобианов, а якобиан
обратного преобразования равен $J^{-1}$.
\begin{defn}
Геометрический объект $h=\lbrace h_i\rbrace$, заданный в некотором атласе
$\lbrace\MU_i,\vf_i\rbrace$ многообразия $\MM$ с правилом преобразования
(\ref{escfud}) в области пересечения любых двух карт называется скалярной
{\em плотностью степени $p$}, и мы будем писать $\deg h=p$.
\qed\end{defn}
\index{Плотность степени (density of weight) $p$}%

Строго говоря, скалярная плотность не является функцией в смысле определения
(\ref{escfde}), и мы не можем писать $h(x)$, где $x$ -- точка многообразия.
Имеет смысл лишь запись $h(x^\al)$ для функции $h\circ\vf^{-1}$, которая задана
в координатном евклидовом пространстве. Несмотря на это, использование скалярных
плотностей, например, при интегрировании, бывает удобным. Кроме того,
производить вычисления с плотностями часто бывает проще, чем с тензорами,
как, например, в общей теории относительности.

Забегая вперед, отметим, что поскольку определитель репера
$\det e_\al{}^a=\vol\ne0$
является скалярной плотностью веса $-1$, то произвольную скалярную плотность
степени $p$ можно представить в виде
$$
  h=\vol^{-p}f,
$$
где $f=\vol^p h$ -- скалярное поле (функция).
\begin{com}
Множество скалярных плотностей фиксированной степени алгебры не образует, т.к.\
произведение двух плотностей степеней $p_1$ и $p_2$ дает скалярную плотность
степени $p_1+p_2$.
\qed\end{com}
\section{Векторные поля и 1-формы                                \label{svecfi}}
\subsection{Локальное определение}
Начнем с локального определения векторных полей и 1-форм, которое является
более наглядным. Рассмотрим многообразие $\MM$, $\dim\MM=n$. Ограничим наше
рассмотрение двумя пересекающимися картами $(\MU_i,\vf_i)$ и $(\MU_j,\vf_j)$ с
координатами $x^\al$ и $x^{\al'}$ соответственно. В области пересечения этих
карт (или при преобразовании координат $x^\al\rightarrow x^{\al'}(x)$)
дифференциалы умножаются на матрицу Якоби, а частные производные -- на ее
обратную:
\begin{align}                                                     \label{etrdif}
  dx^{\al'}&=dx^\al\frac{\pl x^{\al'}}{\pl x^\al},
\\                                                                \label{etrpar}
  \pl_{\al'}&=\frac{\pl x^\al}{\pl x^{\al'}}\pl_\al.
\end{align}
Матрицы преобразования дифференциалов и частных производных являются
взаимно обратными по правилу дифференцирования сложных функций:
\begin{equation}                                                  \label{edifcf}
  \frac{\pl x^\al}{\pl x^\bt}=\frac{\pl x^\al}{\pl x^{\al'}}
  \frac{\pl x^{\al'}}{\pl x^\bt}=\dl^\al_\bt.
\end{equation}

Векторные поля и $1$-формы на многообразии определяются, исходя из правила
преобразования дифференциалов и частных производных. А именно, $n$ достаточно
гладких функций $X^\al(x)$, заданных на карте $(\MU_i,\vf_i)$ и преобразующихся
по правилу (\ref{etrdif}),
\begin{equation}                                                  \label{evectr}
  X^{\al'}:=X^\al\frac{\pl x^{\al'}}{\pl x^\al},
\end{equation}
при преобразовании координат, называются {\em компонентами векторного} или
{\em контравариантного векторного} поля.
\index{Векторное поле (vector field)}\index{Поле векторное (vector field)}%
\index{Контравариантное векторное поле (contravariant vector field)}%
\index{Поле векторное контравариантное (contravariant vector field)}%
Аналогично, $n$ достаточно гладких функций $A_\al(x)$, преобразующихся по
правилу (\ref{etrpar}),
\begin{equation}                                                  \label{eonftr}
  A_{\al'}:=\frac{\pl x^\al}{\pl x^{\al'}}A_\al.
\end{equation}
называются {\em компонентами ковекторного}, или {\em ковариантного векторного}
поля, или {\em $1$-формы}.
\index{$1$-форма ($1$-form)}%
\index{Ковекторное поле (covector field)}%
\index{Поле ковекторное (covector field)}%
\index{Ковариантное векторное поле (covariant vector field)}%
\index{Поле векторное ковариантное (covariant vector field)}%
1-формы называются также {\em формами Пфаффа}.
\index{Форма Пфаффа (Pfaffian form)}\index{Пфаффа форма (Pfaffian form)}%
В общем случае (ко-)векторное поле имеет $n$ независимых компонент.
Каждая из функций $X^\al$ или $A_\al$ является компонентой векторного
или ковекторного поля относительно координатных базисов $e_\al=\pl_\al$ и
$e^\al=dx^\al$. Смысл обозначения координатных базисов частными производными
и дифференциалами не случаен и будет ясен из дальнейшего.

Для того, чтобы задать компоненты (ко-)векторного поля на всем многообразии, их
необходимо задать в каждой карте некоторого атласа $\lbrace\MU_i,\vf_i\rbrace$
таким образом, чтобы во всех областях пересечения карт $\MU_i\cap\MU_j$ они были
связаны преобразованием (\ref{evectr}). Это определение непротиворечиво.
Действительно, если точка $x$ лежит в пересечении трех карт $(\MU_i,\vf_i)$,
$(\MU_j,\vf_j)$ и $(\MU_k,\vf_k)$ с координатами $x^\al$, $x^{\al'}$ и
$x^{\al''}$ соответственно, то компоненты векторов преобразуются по правилу:
\begin{equation*}
  X^{\al''}=X^{\al'}\frac{\pl x^{\al''}}{\pl x^{\al'}}
  =X^\al\frac{\pl x^{\al'}}{\pl x^\al}\frac{\pl x^{\al''}}{\pl x^{\al'}}
  =X^\al\frac{\pl x^{\al''}}{\pl x^\al}.
\end{equation*}
Данное равенство является следствием правила дифференцирования сложных функций и
означает коммутативность следующей диаграммы
\begin{equation*}
\begin{diagram}
  \MU_i & \rTo^{\vf^{-1}_j\circ\vf_i} & \MU_j \\
   & \rdTo_{\vf^{-1}_k\circ\vf_i} & \dTo_{\vf^{-1}_k\circ\vf_j} \\
   & & \MU_k
\end{diagram}
\end{equation*}
для всех точек $x\in\MU_i\cap\MU_j\cap\MU_k$.

Само векторное поле $X$ на многообразии $\MM$ является инвариантным
геометрическим объектом и не зависит от выбора системы координат. Выше мы
определили {\em компоненты} векторного поля на многообразии $\MM$ путем их
задания в некотором атласе. Далее мы должны показать, что таким образом
определенный геометрический объект -- векторное поле -- не зависит от выбора
атласа. Это также обеспечено правилом преобразования (\ref{evectr}). Таким
образом, для задания векторного поля $X$ на многообразии $\MM$ достаточно
задать его компоненты в некотором атласе и указать правило преобразования
(\ref{evectr}) или (\ref{eonftr}). Аналогично дается глобальное определение
ковекторного поля.
\begin{exa}
Если векторное поле имеет нулевые компоненты в одной системе координат, то они
равны нулю и во всех других системах. Нулевое векторное поле, компоненты
которого равны нулю во всех картах, называется {\em тривиальным}. Это
единственное векторное поле, компоненты которого инвариантны относительно
преобразований координат. Аналогично определяется нулевая 1-форма.
\qed\end{exa}
\index{Тривиальное векторное поле (trivial vector field)}%
\index{Векторное поле тривиальное (trivial vector field)}%
\begin{exa}
Дифференциалы $dx^\al$, рассматриваемые как функции от точки $x\in\MU$,
являются компонентами гладкого векторного поля. Это векторное поле определено в
произвольной карте, а в областях пересечения карт справедливо равенство
(\ref{etrdif}). Сами координатные функции $x^\al(x)$, хотя и имеют векторный
индекс, векторного поля не образуют. Это -- набор скалярных полей.
\qed\end{exa}
\begin{exa}
Частные производные от произвольной функции $f\circ\vf$ на образе
$\vf(\MU)\subset\MR^n$ являются компонентами ковариантного векторного поля
$\lbrace\pl_\al f\rbrace$, которое называется градиентом функции. Это
ковекторное поле определено в произвольной карте и имеет правильный закон
преобразования (\ref{eonftr}).
\qed\end{exa}
\begin{exa}
Примером векторного поля на кривой $\g=\lbrace x^\al(t)\rbrace\in\MM$ является
вектор скорости (\ref{ecurve}). Действительно, при преобразовании координат
компоненты вектора скорости $\dot x^\al$ преобразуются, как дифференциалы. При
этом вектор скорости рассматривается в точке кривой $\g\in\MM$. В то же время
сами координатные функции $x^\al(t)$ определены на отрезке $t\in[0,1]$, а не на
многообразии и векторного поля не образуют. Векторное поле скорости называется
также касательным векторным полем к кривой $\g$.
\qed\end{exa}
Преобразования векторных полей (\ref{evectr}) и 1-форм (\ref{eonftr})
различны, поэтому контравариантные и ковариантные индексы необходимо
различать и они всегда будут писаться, соответственно, сверху и снизу.

Преобразования векторов (\ref{evectr}) и 1-форм (\ref{eonftr}) линейны и
однородны, причем матрица $\pl x^\al/\pl x^{\al'}$ в каждой точке многообразия
невырождена и поэтому принадлежит группе невырожденных матриц $\MG\ML(n,\MR)$.
То есть каждому преобразованию координат соответствует $\MG\ML(n,\MR)$
преобразование компонент векторного поля в касательном пространстве. Поскольку
элементы группы $\MG\ML(n,\MR)$ зависят от точки многообразия, то такие
преобразования называются локальными. Обратное утверждение, вообще говоря,
неверно. Не каждому локальному $\MG\ML(n,\MR)$ преобразованию компонент
векторного поля можно сопоставить некоторое преобразование координат. Это видно
из подсчета параметров: локальное $\MG\ML(n,\MR)$ преобразование параметризуется
$n^2$ функциями по числу элементов $n\times n$ матрицы, в то время как
преобразования координат параметризуются $n$ функциями.

Из закона преобразования частных производных (\ref{etrpar}) и дифференциалов
(\ref{etrdif}) следует, что суммы
\begin{align}                                                     \label{evecfi}
  X&=X^\al\pl_\al=X^\al e_\al,
\\                                                                \label{eonefo}
  A&=dx^\al A_\al=e^\al A_\al
\end{align}
инвариантны относительно преобразований координат. Эти формулы представляют
собой разложения векторов и 1-форм по координатному базису, определенному далее
в разделе \ref{svecdi}.
\begin{com}
Скажем несколько слов об обозначениях. Там, где это возможно, мы будем
записывать индексы суммирования по правилу ``с десяти до четырех'' (имеется
ввиду циферблат часов), т.е.\
сначала будем писать верхний индекс, а затем -- нижний. Это правило является
следствием записи {\em оператора внешнего дифференцирования} в виде
\begin{equation}                                                  \label{extdif}
  d=dx^\al\pl_\al.
\end{equation}
В обратном порядке запись $d=\pl_\al dx^\al$ выглядит чрезвычайно неуклюже.
В дифференциальной геометрии, когда все координаты являются вещественными
числами, это правило не играет существенной роли, т.к.\ компоненты векторов и
1-форм можно менять местами. Однако, если часть координат антикоммутирует, то
порядок индексов существенен, и необходимо придерживаться какого-либо
фиксированного правила. Правило ``с десяти до четырех'' не является
единственным, однако оно широко используется при построении моделей
супергравитации в суперпространстве, где часть координат является
антикоммутирующей.
\qed\end{com}
\index{Оператор внешнего дифференцирования %
(operator of exterior differentiation)}%
\index{Внешнего дифференцирования оператор %
(operator of exterior differentiation)}%
Дифференцируемое векторное поле, заданное в какой нибудь одной карте
$\MU\subset\MM$, не всегда может быть продолжено до дифференцируемого векторного
поля на всем многообразии.
\begin{exa}                                                       \label{evenoo}
Рассмотрим сферу $\MS^2\hookrightarrow\MR^3$ единичного радиуса, вложенную в
евклидово пространство (рис.\ref{fspheresing}):
\begin{equation*}
  x^2+y^2+z^2=1,\qquad (x,y,z)\in\MR^3.
\end{equation*}
Рассмотрим открытую верхнюю полусферу:
\begin{equation*}
  \MU=\lbrace (x,y,z)\in\MS^2:~z>0\rbrace.
\end{equation*}
\begin{figure}[h,b,t]
\hfill\includegraphics[width=.4\textwidth]{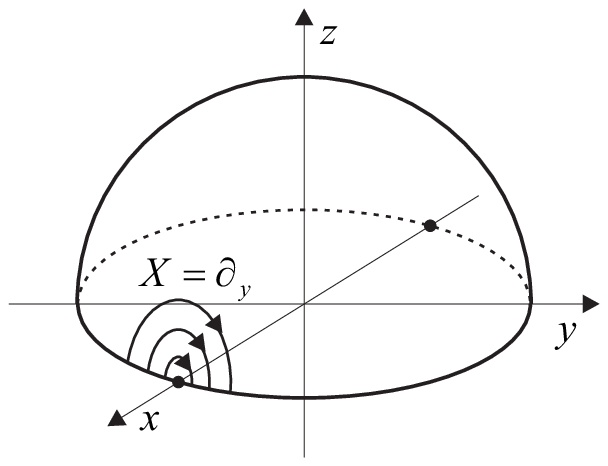}
\hfill {}
\centering\caption{Полусфера, вложенная в трехмерное евклидово пространство.
Точки $(1,0,0)$ и $(-1,0,0)$ являются особыми для векторного поля $X=\pl_y$.}
\label{fspheresing}
\end{figure}
Зададим на $\MU$ систему координат, спроектировав точки полусферы на плоскость
$x,y$: $(x,y,z)\mapsto(x,y)$. Таким образом мы построили карту $(\MU,\vf)$.
Зададим в этой карте векторное поле $X=\pl_y$. В координатном базисе оно имеет
компоненты $(0,1)$ и поэтому является гладким векторным полем на $\MU$. Точки
сферы $(1,0,0)$ и $(-1,0,0)$ являются предельными для $\MU$ и существенно
особыми точками для векторного поля $X$, поскольку предел зависит от пути, по
которому мы стремимся к данным точкам. Пока речь идет об открытом подмножестве
$\MU$ проблем не возникает, т.к.\ указанные точки не принадлежат $\MU$. Однако
любое продолжение векторного поля на окрестность, содержащую любую из точек
$(1,0,0)$ и $(-1,0,0)$ приведет к векторному полю с особенностью. В данном
случае особенность означает не обращение компонент векторного поля в
бесконечность, а то, что в указанных точках векторное поле не будет однозначно
определено.
\qed\end{exa}
\subsection{Глобальное определение векторных и ковекторных полей \label{sglove}}
Дадим глобальное определение векторных полей и 1-форм на многообразии $\MM$, как
это обычно делается в современных курсах дифференциальной геометрии. Рассмотрим
дифференцируемую кривую $\g_1:[-1,1]\ni t\rightarrow\MM$, проходящую через
некоторую точку $p\in\MM$. Пусть $(\MU,\vf)$ -- карта, содержащая точку
$p\in\MU\subset\MM$. Тогда в координатах кривая задается набором функций:
\begin{equation*}
  \vf\circ\g_1=\lbrace x^\al_1(t)\rbrace,\quad p=\g_1(0).
\end{equation*}
Для определенности мы выбрали такую параметризацию кривой, что точке $p$
соответствует значение $t=0$. Рассмотрим другую дифференцируемую кривую $\g_2$,
также проходящую через точку $p$,
\begin{equation*}
  \vf\circ\g_2=\lbrace x^\al_2(t)\rbrace,\quad p=\g_2(0).
\end{equation*}
Мы говорим, что две кривые касаются друг друга в точке $p$, если векторы
скорости кривых в этой точке совпадают:
\begin{equation*}
  \dot x^\al_1|_{t=0}=\dot x^\al_2|_{t=0},\qquad \al=1,\dotsc,n.
\end{equation*}
Поскольку векторы скорости при преобразованиях координат преобразуются
одинаково, то данное определение не зависит от карты, покрывающей точку
$p\in\MM$. Касание кривых в точке $p$ является отношением эквивалентности в
классе всех кривых, проходящих через эту точку. Обозначим класс эквивалентности
кривых, проходящих через точку $p$, который соответствует некоторому
представителю $\g_p$, квадратными скобками $[\g_p]$. Каждый класс
эквивалентности в координатах взаимно однозначно характеризуется набором чисел
$\lbrace\dot x^\al(0)\rbrace$.
\begin{defn}
{\em Касательным вектором} $X_{(\g)}(p)$ к многообразию $\MM$ в точке $p\in\MM$
называется
\index{Касательный вектор (tangent vector)}%
\index{Вектор касательный (tangent vector)}%
класс эквивалентности кривых $[\g_p]$, проходящих через эту точку. Множество
всех касательных векторов в точке $p$ называется {\em касательным пространством}
\index{Касательное пространство (tangent space)}%
\index{Пространство касательное (tangent space)}%
к многообразию в точке $p$ и обозначается $\MT_p(\MM)$. Объединение всех
касательных пространств
\begin{equation}                                                  \label{edetma}
  \MT(\MM)=\bigcup_{p\in\MM}\MT_p(\MM).
\end{equation}
называется {\em касательным расслоением} с базой $\MM$ и естественной проекцией
\index{Касательное расслоение (tangent bundle)}%
\index{Расслоение касательное (tangent bundle)}%
$\pi:\MT(\MM)\rightarrow\MM$, которая задана отображением
$(p,[\g_p])\mapsto p$. Слоем касательного расслоения в точке $p$ является
касательное пространство $\pi^{-1}(p)=\MT_p(\MM)$. {\em Векторным полем} на
многообразии $\MM$ называется сечение касательного расслоения $\MT(\MM)$.
\index{Векторное поле (vector field)}\index{Поле векторное (vector field)}%
\qed\end{defn}
То, что каждый слой является многообразием и диффеоморфен $n$-мерному векторному
пространству, мы покажем чуть ниже.

Пусть $f$ -- функция на многообразии $\MM$. Тогда каждому вектору $X_{(\g)}$ в
точке $p$ мы ставим в соответствие производную функции вдоль вектора
\begin{equation}                                                  \label{ederax}
  X_{(\g)}f|_p=\left.\frac{df\big(\g(t)\big)}{dt}\right|_{t=0}
  =\left.X^\al(p)\pl_\al(f\circ\vf^{-1})\right|_{t=0},
\end{equation}
где $X^\al(p):=\dot x^\al(0)$ -- компоненты вектора $X(p)$ в точке $p$ в
некоторой карте. Поскольку векторы скорости для всех кривых из одного класса
эквивалентности совпадают, то вектор $X(p)$ в точке $p$ взаимно однозначно
характеризуется своими компонентами $\lbrace X^\al(p)\rbrace$. Отсюда следует,
что векторное поле в произвольной карте $(\MU,\vf)$ взаимно однозначно задается
своими компонентами $X(x)=\lbrace X^\al(x)\rbrace$, $x\in\MU$. При этом
компоненты векторного поля при переходе от одной системы координат к другой
преобразуются по-правилу (\ref{evectr}), что является следствием правила
дифференцирования сложных функций. Таким образом, локальное определение
векторного поля, данное ранее, вытекает из глобального определения
настоящего раздела.

В каждой точке многообразия $p\in\MM$ множество векторов $\MT_p(\MM)$ обладает
естественной структурой векторного пространства.
\begin{theorem}
Касательное пространство $\MT_p(\MM)$ имеет естественную структуру
вещественного векторного пространства той же размерности, что и само
многообразие, $\dim\MT_p(\MM)=\dim\MM=n$.
\end{theorem}
\begin{proof}
Рассмотрим два вектора $X_1=[\g_1]\in\MT_p(\MM)$ и $X_2=[\g_2]\in\MT_p(\MM)$
в произвольной точке $p\in\MM$. Пусть $(\MU,\vf)$ -- координатная окрестность
точки $p$. Пусть $\g_1\in[\g_1]$ и $\g_2\in[\g_2]$ -- две произвольные кривые из
классов эквивалентности $[\g_1]$ и $[\g_2]$. Тогда в евклидовом пространстве
определены две кривые $\vf\circ\g_1$ и $\vf\circ\g_2$, которые задаются набором
функций $\vf\circ\g_1=\lbrace x^\al_1(t)\rbrace$ и
$\vf\circ\g_2=\lbrace x^\al_2(t)\rbrace$. Поскольку координаты являются
вещественными числами, то их можно складывать и умножать. Определим сумму двух
векторов и умножение на число $a\in\MR$ как следующие классы эквивалентности
\begin{equation*}
\begin{split}
  X_1+X_2&:=[\vf^{-1}\circ(\vf\circ\g_1+\vf\circ\g_2)],
\\
  aX&:=[\vf^{-1}\circ(a\vf\circ\g)].
\end{split}
\end{equation*}
Касательные векторы к кривым $\vf\circ\g_1+\vf\circ\g_2$ и $a\vf\circ\g$ в точке
$p$ имеют, соответственно, компоненты:
$\lbrace\dot x^\al_1(0)+\dot x^\al_2(0)\rbrace$ и
$\lbrace a\dot x^\al(0)\rbrace$. Это определение суммы векторов и умножения на
числа не зависит от выбора карты $(\MU,\vf)$ и представителя $\g\in[\g]$.
Тем самым касательное пространство $\MT_p(\MM)$ снабжается структурой
векторного пространства. Поскольку каждый вектор $X(p)$ в карте взаимно
однозначно задается набором $n$ чисел $\lbrace X^\al(p)\rbrace$, то размерность
касательного пространства совпадает с размерностью самого многообразия,
$\dim\MT_p(\MM)=\dim\MM=n$.
\end{proof}
Таким образом, мы установили, что касательное пространство $\MT_p(\MM)$ в каждой
точке многообразия имеет естественную структуру векторного пространства $\MR^n$.
Определим в этом векторном пространстве евклидову топологию, которая
является единственной топологией, согласованной с линейной структурой.
Тем самым касательное пространство $\MT_p(\MM)$ так же является многообразием.
В дальнейшем мы всегда предполагаем, что касательное пространство $\MT_p(\MM)$
снабжено естественной структурой векторного пространства $\MR^n$ и
евклидовой топологией.

Типичным слоем касательного расслоения $\MT(\MM)\xrightarrow\pi\MM$ является
евклидово пространство $\MR^n$, на котором введена структура векторного
пространства. Это частный случай векторных расслоений, рассмотренных далее в
разделе \ref{svecbu}.

Касательное расслоение является многообразием размерности
$2n$. При этом, если на базе $\MM$ задана дифференцируемая структура класса
$\CC^k$, $k\ge1$, то касательное расслоение $\MT(\MM)$ является дифференцируемым
многообразием класса $\CC^{k-1}$. Понижение класса дифференцируемости связано с
тем, что, если функции преобразования координат принадлежат классу $\CC^k$, то
матрица Якоби преобразования координат, которая действует в касательном
пространстве, принадлежит классу $\CC^{k-1}$.

Пусть $\MM$ -- гладкое многообразие. Векторное поле $X(x)$ называется гладким,
если в любой карте атласа $\lbrace\MU_i,\vf_i\rbrace$ компоненты $X^\al(x)$,
задающие векторное поле, являются гладкими функциями. Множество всех гладких
векторных полей на многообразии $\MM$ обозначим $\CX(\MM)$. Это множество, так
же как и множество всех векторов в фиксированной точке $\MT_p(\MM)$, обладает
структурой вещественного линейного пространства с поточечным определением
сложения и умножения на числа. Более того, вместо умножения на числа можно
рассматривать умножение на гладкие функции $f\in\CC^\infty(\MM)$. Легко
проверить, что, если $X(x)$ -- векторное поле, то $fX(x)$ так же является
векторным полем. Таким образом, множество векторных полей $\CX(\MM)$ является
модулем над алгеброй гладких функций $\CC^\infty(\MM)$. Как линейное
пространство множество векторных полей является бесконечномерным.

Теперь нетрудно дать глобальное определение ковариантных векторных полей.
\begin{defn}
Множество линейных функционалов на касательном пространстве $\MT_p(\MM)$ в точке
$p\in\MM$ называется {\em кокасательным векторным пространством} и обозначается
\index{Кокасательное векторное пространство (cotangent vector space)}%
\index{Векторное пространство кокасательное (cotangent vector space)}%
$\MT^*_p(\MM)$. Объединение всех кокасательных пространств
\begin{equation}                                                  \label{edetmc}
  \MT^*(\MM)=\bigcup_{p\in\MM}\MT^*_p(\MM).
\end{equation}
называется {\em кокасательным расслоением} с базой $\MM$ и естественной
проекцией $\pi:\MT^*(\MM)\rightarrow\MM$. Слоем кокасательного расслоения в
точке $p$ является кокасательное пространство $\pi^{-1}(p)=\MT^*_p(\MM)$.
\index{Кокасательное расслоение (cotangent bundle)}%
\index{Расслоение кокасательное (cotangent bundle)}%
{\em Кокасательным векторным полем} или {\em $1$-формой} на многообразии $\MM$
\index{$1$-форма ($1$-form)}%
\index{Кокасательное векторное поле (cotangent vector field)}%
\index{Векторное поле кокасательное (cotangent vector field)}%
называется сечение кокасательного расслоения $\MT^*(\MM)$ или линейное
отображение множества векторных полей
\begin{equation*}                                                    \tag*{\qed}
  A:~\CX(\MM)\ni\quad X\mapsto A(X)\quad\in\CC^\infty(\MM).
\end{equation*}
\renewcommand{\qed}{}\end{defn}
В координатах ковекторное поле задается набором $n$ функций $A_\al(x)$, которые
при преобразовании координат преобразуются по правилу (\ref{eonftr}). Тогда
линейное отображение задается простым суммированием компонент:
\begin{equation*}
  A(X)=X^\al A_\al.
\end{equation*}
Кокасательное векторное поле $A(x)$ называется гладким, если его компоненты
$\lbrace A_\al(x)\rbrace$ являются гладкими функциями во всех картах.
Кокасательное пространство $\MT^*_p(\MM)$ в точке $p\in\MM$ снабжается
естественной структурой векторного пространства $\MR^n$ и евклидовой топологией.
Множество всех кокасательных векторных полей, которое обозначим $\Lm_1(\MM)$,
так же как и множество векторных полей $\CX(\MM)$, образует модуль над алгеброй
гладких функций $\CC^\infty(\MM)$.

В дальнейшем нам понадобится
\begin{prop}
Касательное пространство к прямому произведению двух многообразий $\MM\times\MN$
в точке $(p,q)\in\MM\times\MN$ естественно изоморфно прямой сумме касательных
пространств:
\begin{align*}
  \MT_{(p,q)}(\MM\times\MN)\ni\quad X
  \simeq\big(\pi_{1*}(X),\pi_{2*}(X)\big)\quad\in\MT_p(\MM)\oplus\MT_q(\MN),
\end{align*}
где введены проекции на первый и второй сомножитель
\begin{align*}
  \pi_1:~\MM\times\MN&\ni\quad(p,q)\mapsto p\quad\in\MM,
\\
  \pi_2:~\MN\times\MN&\ni\quad(p,q)\mapsto q\quad\in\MN,
\end{align*}
и $\pi_{1*}$, $\pi_{2*}$ -- дифференциалы соответствующих отображений
(см. следующий раздел).
\end{prop}
\begin{proof}
Достаточно спроектировать кривую в прямом произведении $\MM\times\MN$ на каждый
из сомножителей.
\end{proof}
\subsection{Кокасательные векторные поля и ростки}
Дадим также независимое определение кокасательного пространства 1-форм
$\MT^*_p(\MM)$ в точке $p\in\MM$, без обращения к понятию дуального
пространства как это было сделано в предыдущем разделе. Рассмотрим алгебру
гладких функций $\CC^\infty(\MM)$. Зафиксируем точку $p\in\MM$. Будем считать
две функции $f,g\in\CC^\infty(\MM)$ эквивалентными, $f\sim g$, если существует
окрестность $\MU\ni x$ такая, что $f|_\MU=g|_\MU$. Очевидно, что отношение
$\sim$ является отношением эквивалентности в алгебре функций $\CC^\infty(\MM)$.
Класс эквивалентности функции $f$ обозначается $[f_p]$ и называется $\CC^\infty$
{\em ростком} в точке $p\in\MM$.
\index{Росток (germ)}%
Обозначим множество всех ростков в точке $p$ через
\begin{equation}                                                  \label{egermd}
  \CC_p=\CC^\infty(\MM)/\sim~=\lbrace [f_p]:\quad f\in\CC^\infty(\MM)\rbrace.
\end{equation}
Это множество естественным образом снабжается структурой векторного
пространства, которая переносится из алгебры функций,
\begin{equation*}
  [f_p]+[g_p]:=[f_p+g_p],\qquad a[f_p]:=[af_p],\qquad a\in\MR.
\end{equation*}
Таким образом, множество $\CC^\infty$ ростков $\CC_p$ в точке $p\in\MM$
превращается в бесконечномерное вещественное векторное пространство.

Обозначим множество гладких кривых $\g$, которые задаются набором функций
$\lbrace x^\al(t)\rbrace$, проходящих через точку $\g(0)=p$, символом
$\Gamma_p$.
Тогда производная функции вдоль кривой в точке $p$ равна (\ref{ederax}).
Выражение $\pl_\al(f\circ\vf^{-1})$ в правой части равенства зависит только от
ростка $[f]\in\CC_p$, но не от представителя $f\in[f_p]$. Поэтому будем писать
\begin{equation*}
  X_{(\g)}[f_p]:\quad \CC_p\rightarrow\MR.
\end{equation*}
Это отображение линейно по росткам:
\begin{equation*}
  X_{(\g)}([f_p]+[g_p])=X_{(\g)}[f_p]+X_{(\g)}[g_p],\qquad
  X_{(\g)}(a[f_p])=aX_{(\g)}[f_p],\qquad a\in\MR.
\end{equation*}
Введем обозначение для тех ростков, производные которых вдоль всех кривых
$\g\in\Gamma_p$, проходящих через точку $p$, равны нулю
\begin{equation*}
  \CH_p=\lbrace [f_p]\in\CC_p:\quad X_{(\g)}[f_p]=0\rbrace.
\end{equation*}
Множество $\CH_p$ является линейным подпространством в $\CC_p$. В координатах
принадлежность $[f_p]\in\CH_p$ задается равенством $\pl_\al(f\circ\vf^{-1}=0)$.
То есть подпространство $\CH_p$ состоит в точности из тех ростков, у
которых все частные производные равны нулю.
\begin{defn}
{\em Кокасательным пространством} в точке $p\in\MM$ называется фактор
пространство $\MT^*_p(\MM)=\CC_p/\CH_p$.
\qed\end{defn}
\index{Кокасательное пространство (cotangent space)}%
\index{Пространство кокасательное (cotangent space)}%
Из данного определения следует, что кокасательное пространство состоит из тех
ростков $[f_p]$, для которых хотя бы одна частная производная была отлична от
нуля. Поскольку две функции принадлежат одному ростку $f\sim g$ тогда и только
тогда, когда совпадают все их частные производные, то каждый росток взаимно
однозначно определяется градиентом функции. Это доказывает эквивалентность
независимого определения кокасательного пространства определению, данному ранее,
т.к.\ градиент функции есть 1-форма в смысле прежнего определения.
В основе данного определения, так же, как и в определении вектора, лежит понятие
кривой на многообразии и вектора скорости.

В дальнейшем, в целях упрощения обозначений, мы почти всегда будем писать
\begin{equation*}
  \pl_\al f:=\frac{\pl f}{\pl x^\al}=\frac{\pl(f\circ\vf^{-1})}{\pl x^\al},
\end{equation*}
несмотря на то, что функция $f(x)$ определена на точках многообразия $x\in\MM$,
а не в евклидовом пространстве. Это является общепринятой записью и не приводит
к путанице. В тех местах, где нужно подчеркнуть различие точки многообразия и
точки евклидова пространства, мы будем использовать полную запись.
\subsection{Векторные поля и дифференцирования                   \label{svecdi}}
Дадим второе, теперь уже алгебраическое, глобальное определение векторных полей
на многообразии $\MM$. Зафиксируем произвольную точку $p\in\MM$ и рассмотрим
некоторую координатную окрестность этой точки $(\MU,\vf)$. С каждым векторным
полем $X\in\CX(\MU)$ естественным образом связывается оператор дифференцирования
(\ref{evecfi}) в алгебре гладких функций $\CC^\infty(\MU)$. Его действие на
функцию в произвольной карте
\begin{equation}                                                  \label{ediavf}
  \CC^\infty(\MU)\ni\quad f\mapsto Xf:=X^\al\pl_\al f\quad\in\CC^\infty(\MU)
\end{equation}
представляет собой {\em дифференцирование вдоль векторного поля}.
\index{Дифференцирование вдоль векторного поля %
(differentiation along a vector field)}%
Это дифференцирование не зависит от выбора карты, т.к.\ запись (\ref{ediavf})
инвариантна относительно преобразований координат, и удовлетворяет свойствам:
\begin{align*}
  1)\quad & X(af+bg)=aXf+bXg  &&\text{-- линейность},
\\
  2)\quad & X(fg)=(Xf)g+f(Xg) &&\text{-- правило Лейбница},
\end{align*}
для всех $a,b\in\MR$ и $f,g\in\CC^\infty(\MU)$.
\begin{defn}
Непрерывное линейное отображение алгебры функций $\CC^\infty(\MU)$ в $\MR$:
\begin{equation*}
  X_p:\qquad\CC^\infty(\MU)\ni\quad f\mapsto X_pf\quad\in\MR,
\end{equation*}
удовлетворяющее правилу Лейбница $2)$, называется {\em дифференцированием} в
точке $p\in\MM$. Множество всех дифференцирований в данной точке обозначим
$\MD_p(\MM)$.
\qed\end{defn}
\index{Дифференцирование (differentiation)}%
\begin{com}
Каждое векторное поле на $\MU\subset\MM$ отображает алгебру функций
$\CC^\infty(\MU)$ в себя. Обратим внимание, что в определении дифференцирования
в точке $p$ стоит отображение не в алгебру функций $\CC^\infty(\MU)$, а в
вещественную прямую $\MR$. Ясно, что если два векторных поля касаются друг друга
в точке $p$, то они порождают одно и то же дифференцирование.
\qed\end{com}

Множество дифференцирований $\MD_p(\MM)$ снабжается естественной структурой
вещественного векторного пространства:
\begin{equation}                                                  \label{elindi}
\begin{split}
  (X_1+X_2)f&:=X_1 f+X_2 f,
\\
  (aX)f&:=a(Xf),
\end{split}
\end{equation}
где $X_1,X_2,X\in\CX(\MM)$ и $a\in\MR$.
\begin{theorem}
Касательное пространство $\MT_p(\MM)$ и пространство дифференцирований
$\MD_p(\MM)$ изоморфны как векторные пространства.
\end{theorem}
\begin{proof}
То, что каждому вектору $X\in\MT_p(\MM)$ однозначно ставится в соответствие
дифференцирование было показано выше. Нетрудно проверить, что это отображение
сохраняет линейную структуру.

Докажем обратное утверждение. С этой целью рассмотрим важный пример отображений
$\CC^\infty(\MU)\rightarrow\MR$:
\begin{equation}                                                  \label{ecjjot}
  \left(\frac\pl{\pl x^\al}\right)_p f(p)
  :=\pl_\al(f\circ\vf^{-1})|_{\vf(p)},\qquad \al=1,\dotsc,n,
\end{equation}
Эти отображения линейны, удовлетворяют правилу Лейбница и, следовательно,
являются дифференцированиями. Подчеркнем, что символ
$\left(\frac\pl{\pl x^\al}\right)_p$ не является частной производной, т.к.\
определен в точке многообразия $p\in\MM$, а не евклидова пространства $\MR^n$.
В правой же части равенства (\ref{ecjjot}) стоит частная производная от функции,
определенной в евклидовом пространстве. Теперь докажем два утверждения.
\begin{lemma}                                                     \label{ldecon}
Пусть $X$ -- дифференцирование в $\CC^\infty(\MU)$ и $f_c(x)=c=\const$ --
постоянная функция на $\MM$. Тогда $Xc=0$.
\end{lemma}
\begin{proof}
Представим постоянную функцию на $\MM$ в виде $f_c=cf_1=cf_1\cdot f_1$, где
$f_1(x)=1$ -- функция, равная единице на всем многообразии $\MM$. Используем
линейность дифференцирования и правило Лейбница,
\begin{equation*}
  Xc=cX(1\cdot 1)=cX(1)+cX(1)=2cX(1)=2Xc,
\end{equation*}
что возможно только при $Xc=0$. Это -- нетривиальное использование, казалось бы,
тривиального тождества: $1\cdot 1=1$.
\end{proof}
\begin{lemma}                                                     \label{lrecfu}
Для любой гладкой функции $f\in\MC^\infty(\MU)$ существует такой набор функций
$f_\al\in\CC^\infty(\MU)$, $\al=1,\dotsc,n$, что для любой точки $x$ в некоторой
окрестности точки $p\in\MU$ выполнены равенства:
\begin{align}                                                     \label{elefon}
  f_\al(p)&=\left(\frac\pl{\pl x^\al}\right)_p f(p),
\\                                                                \label{elefot}
  f(x)&=f(p)+x^\al(x)f_\al(x),
\end{align}
где $\lbrace x^\al(x)\rbrace:~\MU\rightarrow\MR^n$ -- координатные функции.
\end{lemma}
\begin{proof}
Для определенности будем считать, что образ точки $p\in\MU$ совпадает с началом
координат евклидова пространства, $\vf(p)=(0,\dotsc,0)\in\MR^n$.
Пусть $F(x^1,\dotsc,x^n)=f\circ\vf^{-1}$ -- координатное представление функции
$f$ в некоторой окрестности точки $p$. Тогда справедливо тождество
\begin{align*}
  F(x^1,\dotsc,x^n)&=F(x^1,\dotsc,x^n)-F(x^1,\dotsc,x^{n-1},0)+
\\
  &+F(x^1,\dotsc,x^{n-1},0)-F(x^1,\dotsc,x^{n-2},0,0)+
\\
  &+ \dotsc
\\
  &+ F(x^1,0,\dotsc,0)-F(0,\dotsc,0)+
\\
  &+F(0,\dotsc,0).
\end{align*}
Это тождество перепишем в виде
\begin{align}                                                          \nonumber
  F(x^1,\dotsc,x^n)=&F(0,\dotsc,0)
  +\sum_{\al=1}^n F(x^1,\dotsc,tx^\al,0,\dotsc,0)\big|_{t=0}^{t=1}=
\\                                                                     \nonumber
  =&F(0,\dotsc,0)+\sum_{\al=1}^n\int_0^1 dt\frac{\pl F}{\pl(tx^\al)}
  (x^1,\dotsc,x^{\al-1},tx^\al,0,\dotsc,0)x^\al=
\\                                                                \label{qsawdc}
  =&F(0,\dotsc,0)+x^\al F_\al(x^1,\dotsc,x^n),
\end{align}
где
\begin{equation*}
  F_\al(x^1,\dotsc,x^n)
  :=\int_0^1 dt\frac{\pl F}{\pl(tx^\al)}(x^1,\dotsc,x^{\al-1},tx^\al,0,\dotsc,0)
\end{equation*}
-- набор гладких функций в некоторой окрестности начала координат евклидова
пространства. Теперь вернемся на многообразие и определим набор функций
$f_\al:=F_\al\circ\vf$. Тогда из последнего равенства (\ref{qsawdc}) следует
равенство (\ref{elefot}).

Теперь надо определить вид функций $f_\al(x)$. С этой целью применим
дифференцирование $\left(\frac\pl{\pl x^\al}\right)_x$ к равенству
(\ref{elefot})
\begin{equation*}
  \left(\frac\pl{\pl x^\al}\right)_x f(x)
  =\left(\frac\pl{\pl x^\al}\right)_x f(p)
  +\left(\frac\pl{\pl x^\al}\right)_x x^\bt(x) f_\bt(x)
  +x^\bt(x)\left(\frac\pl{\pl x^\al}\right)_x f_\bt(x).
\end{equation*}
Первое слагаемое равно нулю, как следствие леммы \ref{ldecon}. Поскольку
$\left(\frac\pl{\pl x^\al}\right)_x x^\bt(x)=\dl_\al^\bt$, то в точке $p$ имеем
равенство (\ref{elefon}) т.к.\ $x^\al(p)=0$.
\end{proof}
\begin{cor}
Если $X_p\in\MD_p(\MU)$ -- дифференцирование в точке $p\in\MM$, то
\begin{equation}                                                  \label{ederde}
  X_p=X_p\, x^\al(p)\left(\frac\pl{\pl x^\al}\right)_p
  =X^\al_p\left(\frac\pl{\pl x^\al}\right)_p.
\end{equation}
\renewcommand{\qed}{}\end{cor}
\begin{proof}
Пусть $(\MU,\vf)$ -- произвольная карта в окрестности точки $p\in\MM$. Тогда,
возможно, $x^\al(p)\ne0$. В этом случае сдвинем начало координат в евклидовом
пространстве: $y^\al:=x^\al-x^\al(p)$. Тогда из леммы \ref{lrecfu} следует
представление
\begin{equation*}
  f(x)=f(p)+\big(x^\al(x)-x^\al(p)\big)f_\al(x).
\end{equation*}
Применяя дифференцирование $X$ к этому равенству и переходя в точку $p$, получим
(\ref{ederde}).
\end{proof}
Таким образом, множество всех дифференцирований $\MD_p(\MU)$ в произвольной
точке $p\in\MM$ представляет собой конечномерное векторное пространство,
$\dim\MD_p(\MU)=n$, с базисом $\left(\frac\pl{\pl x^\al}\right)_p$. Это
пространство изоморфно касательному пространству $\MT_p(\MM)$, и теорема
доказана.
\end{proof}
\begin{defn}
{\em Координатным базисом} векторных полей $\CX(\MU)$ на карте $(\MU,\vf)$
многообразия $\MM$ называется набор гладких векторных полей
\begin{equation*}
  \left\lbrace e_\al(x)
  :=\left(\frac\pl{\pl x^\al}\right)_x\right\rbrace\in\CX(\MU),\qquad
  \al=1,\dotsc,n,
\end{equation*}
определенных формулой (\ref{ecjjot}). Дуальный базис
$\left\lbrace e^\al(x)\right\rbrace\in\Lm_1(\MU)$, $e^\al(e_\bt)=\dl^\al_\bt$,
называется {\em координатным базисом} ковекторных полей (1-форм). Координатный
базис для ковекторных полей обозначается $dx^\al_p$ или просто $dx^\al$.
Координатные базисы называют также {\em голономными}.
\qed\end{defn}
\index{Координатный базис (coordinate basis)}%
\index{Базис координатный (coordinate basis)}%
\index{Голономный базис (holonomic basis)}%
\index{Базис голономный (holonomic basis)}%
\begin{com}
Подчеркнем, что координатный базис -- это не набор частных производных, а
векторные поля на многообразии. Их действие, как дифференцирований, определено
только для достаточно гладких функций $\CC^k(\MU)$. Действие векторных полей
$e_\al(x)$ на тензоры более высокого ранга не определено.
\qed\end{com}
\begin{com}
Обозначение координатных базисов векторных и ковекторных полей через $\pl_\al$ и
$dx^\al$ оправдано простой формулой из математического анализа
\begin{equation*}
  \frac{\pl x^\al}{\pl x^\bt}=\dl^\al_\bt.
\end{equation*}
Для тривиальных многообразий $\MM\approx\MR^n$, покрытых одной картой, векторы
$e_\al$ можно отождествить с операторами частных производных $\pl_\al$, которые
действуют на дифференцируемые функции. Тогда дуальный базис $e^\al$ ковекторных
полей естественным образом отождествляется с дифференциалами координатных
функций $dx^\al$.
\qed\end{com}
Из разложения по базису (\ref{ederde}) следует, что компонента векторного поля
$X$ в точке $p\in\MM$ -- это результат действия векторного поля на координатную
функцию $X^\al(p)=Xx^\al(x)|_{x=p}$.

В дальнейшем мы будем писать сокращенно $X=X^\al(x)\pl_\al$, имея в виду, что
на функции $f\circ\vf$ в евклидовом пространстве координатный базис
действительно действует, как частная производная. Для 1-форм в координатном
базисе мы часто будем использовать общепринятую запись $A=dx^\al A_\al$,
$e^\al:=dx^\al$.

Пусть $D_x(\MU)\in\MD_x(\MM)$ -- некоторое дифференцирование в точке $x\in\MM$.
Рассмотрим объединение $D(\MM):=\bigcup_{x\in\MM}D_x(\MU)$ по всем точкам
многообразия, которое соответствует некоторому векторному полю $X\in\CX(\MM)$.
Оно задает отображение
\begin{equation}
  D(\MM):~ \CC^\infty(\MM)\rightarrow\CC^\infty(\MM),
\end{equation}
которое называется дифференцированием в алгебре функций $\CC^\infty(\MM)$,
т.е.\ непрерывный линейный эндоморфизм в $\CC^\infty(\MM)$, удовлетворяющий
правилу Лейбница. Поэтому каждому векторному полю $X\in\CX(\MM)$ ставится в
соответствие некоторое дифференцирование $D(\MM)\in\CD(\MM)$, где $\CD(\MM)$ --
множество всех дифференцирований в алгебре функций. Верно также и обратное
утверждение: любому дифференцированию в алгебре функций $\CC^\infty(\MM)$
соответствует единственное векторное поле. Мы доказали аналогичное утверждение
в фиксированной точке многообразия, где пространство $\MD_x(\MU)$ является
конечномерным. Доказательство в рассматриваемом бесконечномерном случае сложнее
и приведено в \cite{Zharin08R}. Линейная структура на $\CD(\MM)$ вводится также,
как и в точке (\ref{elindi}). Таким образом множество векторных полей $\CX(\MM)$
биективно отображается на множество дифференцирований $\CD(\MM)$, при этом как
векторные пространства эти множества изоморфны. Эта биекция позволяет дать
эквивалентное определение векторного поля.
\begin{defn}
{\em Векторным полем} $X\in\CX(\MM)$ на многообразии $\MM$ называется
дифференцирование в алгебре функций $\CC^\infty(\MM)$.
\qed\end{defn}
\index{Векторное поле (vector field)}%
\subsection{Векторные поля и интегральные кривые                 \label{svechs}}
Начнем с локального описания. Пусть $x^\al$, $\al=1,\dotsc,n$, -- локальные
координаты в окрестности точки $x\in\MU_x\subset\MM$. Рассмотрим векторное поле
$X=X^\al\pl_\al$, все компоненты которого отличны от нуля в окрестности $\MU_x$.
Бесконечно малые перемещения $dx^\al$ точки $x$ вдоль этого векторного поля
должны быть пропорциональны компонентам $X^\al$ и поэтому удовлетворять системе
уравнений
\begin{equation}                                                  \label{evesye}
  \frac{dx^1}{X^1}=\frac{dx^2}{X^2}=\dots=\frac{dx^n}{X^n}.
\end{equation}
Эту систему уравнений можно переписать в виде равенства нулю 1-форм:
\begin{equation*}
  A^\Sm:=\frac{dx^\Sm}{X^\Sm}-\frac{dx^1}{X^1}=0,\qquad \Sm=2,\dotsc,n.
\end{equation*}
В каждой точке $x\in\MM$ векторное поле $X$ задает одномерное подпространство в
касательном пространстве $\MT_x(\MM)$, а совокупность 1-форм
$\lbrace A^\Sm\rbrace$ -- $(n-1)$-мерное ортогональное дополнение в сопряженном
пространстве $\MT^*_x(\MM)$, поскольку $A^\Sm(X)=0$.

Согласно теории дифференциальных уравнений, уравнения (\ref{evesye}) допускают
$n-1$ функционально независимых решений
\begin{equation}                                                  \label{eindso}
  \vf^\Sm(x^\al)=c^\Sm=\const,\qquad \Sm=2,\dots,n,
\end{equation}
для которых $A^\Sm=d\vf^\Sm=0$. При этом прямоугольная $n\times(n-1)$-матрица,
составленная из производных $\pl_\al\vf^\Sm$, имеет ранг $n-1$, и каждая из
функций $\vf^\Sm$ является решением уравнения в частных производных
\begin{equation}                                                  \label{eqpdvf}
  X^\al\pl_\al\vf^\Sm=0.
\end{equation}

Совершим преобразование координат $x^\al\mapsto y^\al(x)$, выбрав в качестве
последних $n-1$ координат функции (\ref{eindso})
$(y^2,\dotsc,y^n):=(\vf^2,\dotsc,\vf^n)$, а координату $x^1$ оставим без
изменения. Якобиан этого преобразования отличен от нуля в силу функциональной
независимости функций $\vf^\Sm$. Тогда из закона преобразования векторного поля
(\ref{evectr}) и уравнения (\ref{eqpdvf}) следует, что в новой системе
координат все компоненты векторного поля, кроме первой, равны нулю:
$$
  X^1\ne0,\quad X^2=\dotsc=X^n=0.
$$
Заменим теперь координату $x^1$ на функцию $y^1(x)$, которая удовлетворяет
дифференциальному уравнению
\begin{equation*}
  \frac{dy^1}{dx^1}X^1=1.
\end{equation*}
Это уравнение локально разрешимо, и, значит, в новой системе координат $X^1=1$.

Если у векторного поля $X^\al$ часть компонент равнялась нулю до преобразования
координат, то все, сказанное выше, можно повторить для ненулевых компонент.
Отсюда следует частный случай теоремы Фробениуса, которая будет сформулирована
в разделе \ref{sfrote}.
\begin{theorem}                                                   \label{tvecfi}
Для произвольного отличного от нуля векторного поля $X\in\CX(\MM)$ в некоторой
окрестности $\MU_x$ произвольной точки $x\in\MM$ существует такая система
координат, в которой все компоненты $X$, кроме одной (например, первой)
обращаются в нуль. Координатную функцию $x^1(x)$, $x\in\MU_x$ можно подобрать
таким образом, чтобы первая компонента векторного поля $X^1$ была равна единице
в этой окрестности, т.е.\ $X=\pl_1$.
\end{theorem}
\begin{com}
Эта теорема показывает, что векторное поле в окрестности точки, в которой
оно отлично от нуля, устроено довольно просто. Если в некоторой точке векторное
поле обращается в нуль, то в окрестности этой точки оно может быть устроено
очень сложно \cite{Chern67A}. Нули гладкого касательного поля к многообразию
связаны с топологическими свойствами. Например, на четномерной сфере не
существует векторного поля, нигде не обращающегося в нуль (см.\ теорему
\ref{tvecsp}). В то же время такое поле всегда можно задать на торе.
\qed\end{com}

Если в уравнения (\ref{eindso}) подставить координаты некоторой фиксированной
точки $p=\lbrace{x^\al_p}\rbrace\in\MM$, то определятся значения постоянных
$c^\Sm$. При этих значениях постоянных система $n-1$ трансцендентных уравнений
относительно $x^\al$ (\ref{eindso}) определяет кривую, проходящую через точку
$p$. Если кривая $\g$ параметризуется параметром $t$,
$\g=\lbrace x^\al(t)\rbrace$, то уравнения (\ref{evesye}) эквивалентны системе
обыкновенных дифференциальных уравнений
\begin{equation}                                                  \label{eintcv}
  \frac{dx^\al}{dt}=X^\al(x),
\end{equation}
с начальными условиями
\begin{equation}                                                  \label{eincov}
  x^\al|_{t=t_p}=x^\al_p,
\end{equation}
где $t_p$ -- значение параметра вдоль кривой, при котором она проходит через
точку $p$.

Из теории дифференциальных уравнений хорошо известна
\begin{theorem}                                                   \label{tdifve}
Если векторное поле $X$ на многообразии $\MM$ дифференцируемо, то через каждую
точку $p\in\MM$ проходит одна и только одна интегральная кривая этого векторного
поля.
\end{theorem}
На языке теории дифференциальных уравнений это утверждение означает, что решение
задачи Коши (\ref{eintcv}), (\ref{eincov}) существует и единственно.

Общее решение системы уравнений (\ref{eintcv}) зависит от $n$ постоянных
интегрирования. Одна из постоянных соответствует сдвигу параметра вдоль кривой
$t\mapsto t+\const$, а оставшиеся $n-1$ постоянных определяются положением точки
$p$ на гиперповерхности (\ref{eindso}), проходящей через точку $p$.

Если решение системы уравнений (\ref{eintcv}) представимо в виде ряда, то
вблизи точки $p$ оно выглядит очень просто
\begin{equation*}
  x^\al(t)=x^\al_p+X^\al_p (t-t_p)+\dotsc,\qquad (t-t_p)\ll1.
\end{equation*}
То есть компоненты векторного поля определяют главную линейную часть
интегральной кривой.
\begin{com}
Интегральные кривые векторного поля являются ни чем иным, как
{\em характеристиками} (линиями уровня) для решений дифференциального
уравнения в частных производных первого порядка (\ref{eqpdvf}).
\qed\end{com}
\index{Характеристика (characteristic)}%

Условие дифференцируемости векторного поля в теореме \ref{tdifve} можно
ослабить.
\begin{defn}
Функция (отображение)
\begin{equation*}
  f:\quad \MR^n\supset\MU\ni\quad x \mapsto f(x)\quad\in\MR^m
\end{equation*}
удовлетворяет {\em условию Липшица}, если существует такая положительная
постоянная $C$, что выполнено неравенство
\begin{equation*}
  |f(x_2)-f(x_1)|\le C|x_2-x_1|,
\end{equation*}
для всех $x_{1,2}\in\MU$.
\qed\end{defn}
\index{Условие Липшица (Lipschitz condition)}%
\index{Липшица условие (Lipschitz condition)}%
В условии Липшица знак модуля обозначает обычный модуль вектора в евклидовом
пространстве.
\begin{theorem}
Пусть правая часть системы уравнений (\ref{eintcv}) непрерывна и удовлетворяет
условию Липшица в области $\MU$. Тогда через каждую точку $\MU$ проходит одна и
только одна интегральная кривая системы уравнений (\ref{eintcv}).
\end{theorem}
Если правая часть системы уравнений (\ref{eintcv}) только непрерывна, то и тогда
через каждую точку $p\in\MU$ проходит хотя бы одна интегральная кривая. Однако
единственность может быть нарушена.
\begin{exa}
Рассмотрим обыкновенное дифференциальное уравнение
\begin{equation*}
  \frac23\dot x=x^{1/3}.
\end{equation*}
Его решения имеют вид
\begin{equation*}
  x=0\qquad\text{и}\qquad x=\pm(x+c)^{3/2}.
\end{equation*}
Отсюда следует, что через каждую точку $x\ne0$ проходят две интегральные кривые.
Через точку $x=0$ проходит даже три интегральные кривые: $x=0$, $x=\pm t^{3/2}$.
Если положить в условии Липшица $x_1=0$, то оно примет вид
\begin{equation*}
  |x_2|^{1/3}\le C|x_2|.
\end{equation*}
Ясно, что такой постоянной $C$ не существует, т.к.\ правая часть неравенства
при $x_2\to0$ стремится к нулю быстрее.
\qed\end{exa}
В дальнейшем мы всегда предполагаем, что векторное поле по крайней мере
дифференцируемо.
\begin{prop}                                                      \label{polvec}
Если дифференцируемое векторное поле $X$ обращается в нуль в некоторой точке
многообразия $p\in\MM$, то эта точка является неподвижной относительно потока
векторного поля, т.е.\ $x(t)=p$ для всех $t\in\MR$. Обратно. Если точка
$p\in\MM$ является неподвижной на интегральной кривой некоторого векторного
поля, то в этой точке векторное поле обращается в нуль, $X(p)=0$.
\end{prop}
\begin{proof}
Постоянные функции $x^\al(t)=x^\al_p$ удовлетворяют системе уравнений
(\ref{eintcv}), если $X^\al_p=0$. Обратное утверждение очевидно.
\end{proof}

Если векторное поле умножить на произвольную достаточно гладкую отличную от нуля
функцию: $X\mapsto fX$, то уравнение для интегральной кривой $x(\tau)$ примет
вид
\begin{equation*}
  \frac{dx^\al}{d\tau}=fX^\al.
\end{equation*}
Введем новый параметр $t(\tau)$ вдоль кривой, который является решением
уравнения
\begin{equation*}
  \frac{dt}{d\tau}=f.
\end{equation*}
Решение этого уравнения существует и является монотонным, т.к.\ $f\ne0$. В новой
параметризации уравнение для интегральной кривой принимает прежний вид
(\ref{eintcv}). Таким образом, два векторных поля $X$ и $fX$, отличающиеся
умножением на отличную от нуля функцию, имеют интегральные кривые, которые
совпадают, как подмножества в $\MM$. Отличие сводится только к различным
параметризациям кривых.

Перейдем к глобальному описанию.
\begin{defn}
{\em Интегральной кривой} векторного поля $X\in\CX(\MM)$, проходящей через
точку $p\in\MM$ называется кривая $\g:~(a,b)\ni t\mapsto x(t)\in\MM$ такая,
что $\g(t_p)=p$ и
\begin{equation}                                                  \label{einglo}
  \g_*\pl_t=X\big(x(t)\big),
\end{equation}
где $\g_*$ -- дифференциал отображения некоторого открытого интервала
$(a,b)\subset\MR$, содержащего точку $t_p$, и $\pl_t$ -- касательный вектор к
интервалу в точке $t\in(a,b)$. Вектор $X\big(x(t)\big)$ называется
{\em касательным вектором} к кривой $\g$ в точке $x(t)$ (вектором скорости).
\qed\end{defn}
\index{Интегральная кривая (integral curve)}%
\index{Кривая интегральная (integral curve)}%
\index{Касательный вектор (tangent vector)}%
\index{Вектор касательный (tangent vector)}%
Нетрудно проверить, что в каждой карте уравнение (\ref{einglo}) записывается в
виде системы обыкновенных дифференциальных уравнений (\ref{eintcv}).

Вообще говоря, интегральные кривые существуют только локально, даже для
гладких векторных полей, заданных на всем многообразии. Другими словами,
параметр $t$ в общем случае определен лишь на некотором конечном или
полубесконечном интервале $(a,b)\subset\MR$.
\begin{defn}
Векторное поле $X\in\CX(\MM)$ называется {\em полным}, если все интегральные
кривые этого поля определены при всех значениях $t\in\MR$.
\qed\end{defn}
\index{Полное векторное поле (complete vector field)}%
\index{Векторное поле полное (complete vector field)}%
\begin{theorem}                                                   \label{tcovcr}
На компактном многообразии $\MM$ любое гладкое векторное поле $X\in\CX(\MM)$, не
обращающееся в нуль, является полным.
\end{theorem}
\begin{proof}
См., например, \cite{Arnold75R}.
\end{proof}
\begin{com}
Для многообразий понятие компакта и компактного пространства совпадают, т.к.\
многообразие, по-определению, является хаусдорфовым пространством.
\end{com}

На некомпактном многообразии векторное поле может быть либо полным, либо
неполным.
\begin{exa}
Векторному полю $\pl_x$ на вещественной прямой $\MR$ соответствуют интегральные
кривые $x=t+\const$. Они полны на всей прямой $\MR$. Однако они неполны
на полупрямой $\MR_+$ или любом конечном открытом интервале $(a,b)\subset\MR$.
\qed\end{exa}
\begin{exa}
Рассмотрим гладкое векторное поле $X=x^2\pl_x$ на вещественной прямой $\MR$.
Общее решение уравнения интегральной кривой $\dot x=x^2$ имеет вид
\begin{equation*}
  x=-\frac1{t-c},\qquad c=\const.
\end{equation*}
Таким образом, для каждого значения постоянной $c$, имеются две никак не
связанные между собой интегральные кривые:
\begin{align*}
  \g_1:\quad (-\infty,c)~\rightarrow~(0,\infty),
\\
  \g_2:\quad (c,\infty)~\rightarrow~(-\infty,0).
\end{align*}
При этом точка $x=0$ соответствует бесконечному значению параметра вдоль
интегральной кривой $t=\pm\infty$, а бесконечно удаленные точки $x=\pm\infty$
-- конечному значению параметра $t=c$. Поэтому векторное поле $X=x^2\pl_x$
неполно.
\qed\end{exa}
\begin{defn}
Точка $p\in\MM$, в которой векторное поле обращается в нуль, $X(p)=0$,
называется {\em особой точкой векторного поля}. Если компоненты векторного поля
разлагаются в ряд Тейлора, то система уравнений (\ref{eintcv})
в координатной окрестности особой точки в линейном приближении имеет вид
\begin{equation}                                                  \label{eliosy}
  \frac{dx^\al}{dt}=x^\bt B_\bt{}^\al,\qquad
  B_\bt{}^\al:=\pl_\bt X^\al\big|_{x=p}.
\end{equation}
Если $\det B\ne0$, то особая точка называется {\em невырожденной}.
\qed\end{defn}
\index{Особая точка векторного поля (singular point of a vector field)}%
\index{Невырожденная особая точка (nondegenerate singular point)}%
При преобразовании координат $x^\al\mapsto x^{\al'}(x)$ уравнение (\ref{eliosy})
сохраняет свой вид. При этом компоненты матрицы $B$ преобразуются по-правилу
\begin{equation*}
  B_\bt{}^\al\mapsto B_{\bt'}{}^{\al'}
  =\frac{\pl x^\bt}{\pl x^{\bt'}} B_\bt{}^\al\frac{\pl x^{\al'}}{\pl x^\al},
\end{equation*}
т.е.\ подвергаются преобразованию подобия.

Напомним, что любое дифференциальное уравнение $n$-того порядка можно записать в
виде эквивалентной ей системы уравнений первого порядка, состоящей из $n$
уравнений. Следовательно, теория обыкновенных
дифференциальных уравнений произвольного порядка, разрешенных относительно
старшей производной, сводится к нахождению и исследованию свойств интегральных
кривых векторных полей. При этом поведение интегральных кривых в окрестностях
особых точек представляет исключительный интерес, т.к.\ позволяет понять
качественное поведение решений.

В общем случае поведение интегральных кривых в окрестности особой точки довольно
сложно. Особые точки можно классифицировать, приведя матрицу $B$ к какому либо
каноническому виду с помощью преобразования подобия, что означает переход в
новую систему координат.
\begin{exa}
Рассмотрим двумерное многообразие (поверхность) $\MM$, на котором задано
дифференцируемое векторное поле $X\in\CX(\MM)$. Пусть $p\in\MM$ -- невырожденная
особая точка векторного поля $X$. Выберем систему координат в окрестности особой
точки так, чтобы она находилась в начале координат. Обозначим собственные числа
матрицы $B$ через $\lm_1$ и $\lm_2$. В общем случае они комплексны. Из
невырожденности следует, что $\lm_1\lm_2\ne0$. Невырожденные особые точки в
рассматриваемом случае делятся на шесть классов.

\textit{a).} {\em Седло.} $\lm_1\lm_2<0$.
\index{Седло (saddle point)}%
Пусть собственные числа вещественны и разных знаков. Тогда существует система
координат в которой матрица $B$ диагональна и уравнения для интегральных кривых
(\ref{eliosy}) примут вид
\begin{equation}                                                  \label{esadpo}
  \dot x^1=x^1\lm_1,\qquad \dot x^2=x^2\lm_2.
\end{equation}
Они легко интегрируются:
\begin{equation}                                                  \label{esosad}
  x^1=C_1\ex^{\lm_1t},\qquad x^2=C_2\ex^{\lm_2t},\qquad t\in\MR,
\end{equation}
где $C_{1,2}$ -- постоянные интегрирования. Касательный вектор к интегральной
кривой имеет угол наклона
\begin{equation*}
  \frac{dx^2}{dx^1}=\frac{C_2\lm_2}{C_1\lm_1}\ex^{(\lm_2-\lm_1)t}.
\end{equation*}
Соответствующие интегральные кривые для $\lm_1<0<\lm_2$ показаны на
рис.\ref{fsingpoint},\textit{a}. Стрелки указывают направление возрастания
параметра $t$. Ни одна из интегральных кривых не проходит через начало
координат, которое является неподвижной точкой.
\begin{figure}[h,b,t]
\hfill\includegraphics[width=.95\textwidth]{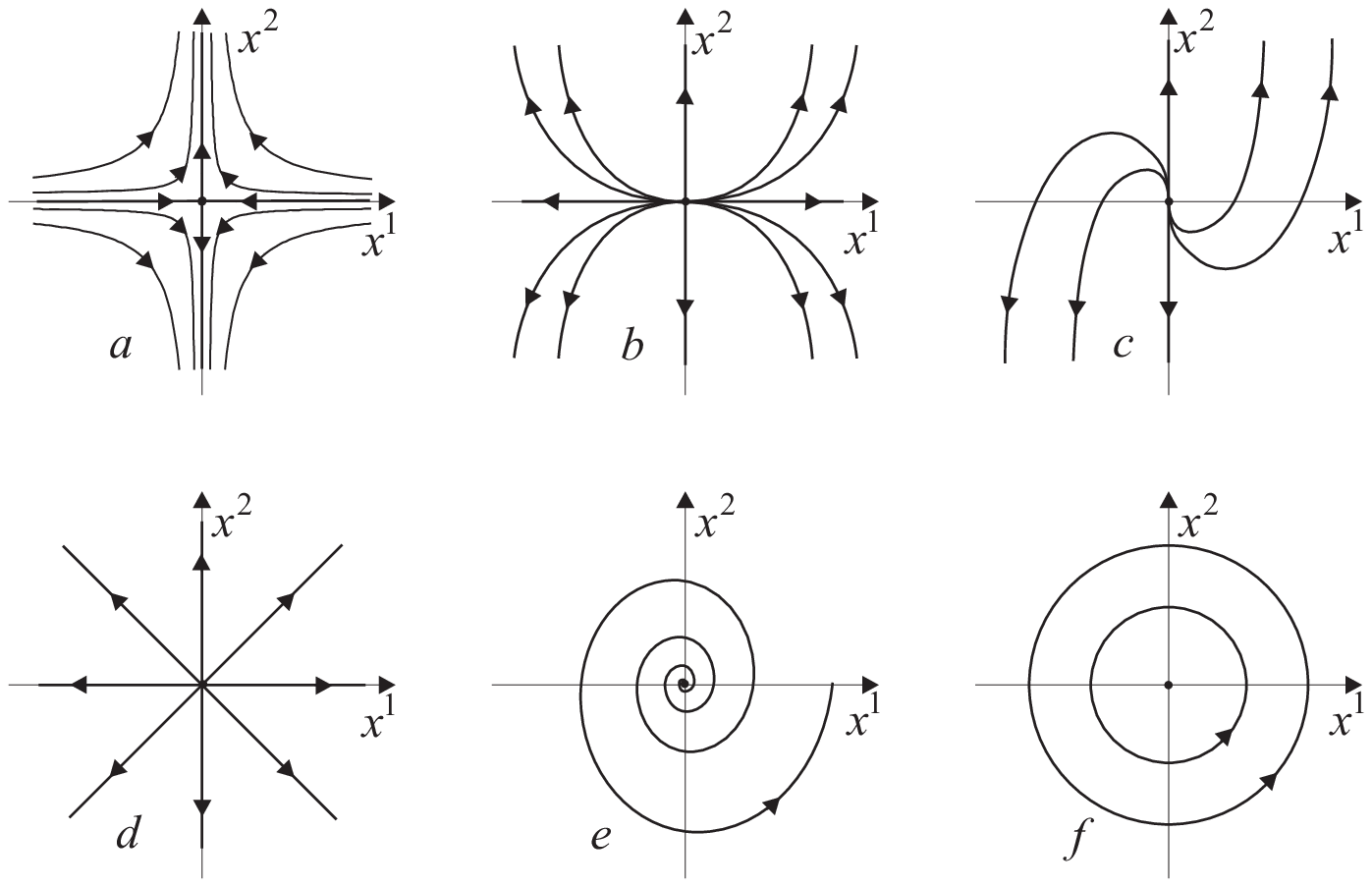}
\hfill {}
\\
\centering \caption{Седло (\textit{a}), узел (\textit{b}), жорданов узел
 (\textit{c}), дикритический узел (\textit{d}), фокус (\textit{e}),
 центр (\textit{f}).\label{fsingpoint}}
\end{figure}

\textit{b).} {\em Узел.} $\lm_1\lm_2>0$, $\lm_1\ne\lm_2$.
\index{Узел (knot)}%
Если собственные числа матрицы $B$ вещественны одного знака и различны, то ее
также можно диагонализировать. При этом уравнения для интегральных кривых и
решений имеют прежний вид (\ref{esadpo}), (\ref{esosad}). Меняется только знак
одного из собственных чисел. На рис.\ref{fsingpoint},\textit{b} показаны
интегральные кривые для $0<\lm_1<\lm_2$.

\textit{c).} {\em Жорданов узел.} $\lm_1=\lm_2$, матрица $B$ недиагонализируема.
\index{Жорданов узел (Jordan knot)}%
Пусть собственные числа матрицы $B$ вещественны и равны, и матрицу $B$ нельзя
диагонализировать преобразованием подобия. Тогда матрицу $B$ можно привести к
жордановой клетке (см.\ дополнение \ref{smatri})
\begin{equation*}
  B=\begin{pmatrix} \lm_1 & 1 \\ 0 & \lm_1 \end{pmatrix}.
\end{equation*}
Соответствующие уравнения интегральных кривых
\begin{equation*}
  \dot x^1=x^1\lm_1,\qquad \dot x^2=x^1+x^2\lm_1,
\end{equation*}
легко интегрируются:
\begin{equation*}
  x^1=C_1\ex^{\lm_1t},\qquad x^2=(C_2+C_1t)\ex^{\lm_1t},\qquad t\in\MR,
\end{equation*}
Касательная к интегральной кривой имеет угол наклона
\begin{equation*}
  \frac{dx^2}{dx^1}=\frac{C_2}{C_1}+\frac1{\lm_1}+t.
\end{equation*}
На рис.\ref{fsingpoint},\textit{c} показаны интегральные кривые для $0<\lm_1$.

\textit{d).} {\em Дикритический узел.} $\lm_1=\lm_2$, матрица $B$
кратна единичной.
\index{Дикритический узел (critical knot)}%
Пусть собственные числа матрицы $B$ вещественны и равны, и матрица $B$ кратна
единичной,
\begin{equation*}
  B=\begin{pmatrix} \lm_1 & 0 \\ 0 & \lm_1 \end{pmatrix}.
\end{equation*}
При преобразовании координат матрица $B$ не меняется, т.к.\ пропорциональна
единичной матрице. В этом случае уравнения для интегральных кривых,
\begin{equation*}
  \dot x^1=x^1\lm_1,\qquad \dot x^2=x^2\lm_1,
\end{equation*}
легко интегрируются:
\begin{equation*}
  x^1=C_1\ex^{\lm_1t},\qquad x^2=C_2\ex^{\lm_1t},\qquad t\in\MR,
\end{equation*}
Касательная к интегральной кривой имеет постоянный угол наклона
\begin{equation*}
  \frac{dx^2}{dx^1}=\frac{C_2}{C_1}.
\end{equation*}
На рис.\ref{fsingpoint},\textit{d} показаны интегральные кривые для $0<\lm_1$.

\textit{e).} {\em Фокус.} $\lm_{1,2}$ комплексны, $\re \lm_{1,2}\ne0$.
\index{Фокус (focus)}%
Поскольку комплексные собственные значения могут встречаться только комплексно
сопряженными парами, то $\lm_1=\mu+i\tilde\om$ и $\lm_2=\mu-i\tilde\om$, где
$\mu,\tilde\om\in\MR$ и $\mu\ne0$. В этом случае матрицу $B$ можно привести к
виду
\begin{equation*}
  B=\begin{pmatrix} 0 & -\om^2 \\ 1 & 2\mu \end{pmatrix},
\end{equation*}
где $\om^2:=\tilde\om^2+\mu^2$. Соответствующие уравнения для интегральных
кривых принимают вид
\begin{align*}
  \dot x^1&=x^2,
\\
  \dot x^2&=-\om^2x^1+2\mu x^2.
\end{align*}
Дифференцирование первого уравнения приводит к уравнению для осциллятора с
трением
\begin{equation*}
  \ddot x^1-2\mu\dot x^1+\om^2x^1=0.
\end{equation*}
Отсюда следует, что интегральные кривые имеют вид
\begin{align*}
  x^1&=C_1\ex^{\mu t}\cos(\tilde\om t+C_2),
\\
  x^2&=C_1\ex^{\mu t}\left[\mu\cos(\tilde\om t+C_2)
  -\tilde\om\sin(\tilde\om t+C_2)\right].
\end{align*}
На рис.\ref{fsingpoint},\textit{д} показаны интегральные кривые для
$\tilde\om>0$, $\mu>0$.

\textit{f).} {\em Центр.} $\lm_{1,2}$ комплексны, $\re \lm_{1,2}=0$.
\index{Фокус (focus)}%
Пусть собственные числа чисто мнимые: $\lm_1=-\lm_2=i\om$, $\om\in\MR$. В этом
случае матрицу $B$ можно привести к виду
\begin{equation*}
  B=\begin{pmatrix} 0 & -\om \\ \om & 0 \end{pmatrix}.
\end{equation*}
Уравнения для интегральных кривых,
\begin{equation*}
  \dot x^1=\om x^2,\qquad \dot x^2=-\om x^1,
\end{equation*}
соответствуют гармоническому осциллятору
\begin{equation*}
  \ddot x^1+\om^2x^1=0
\end{equation*}
и легко интегрируются:
\begin{equation*}
  x^1=C_1\cos(\om t+C_2),\qquad x^2=-C_1\sin(\om t+C_2),\qquad t\in\MR.
\end{equation*}
Интегральные кривые являются окружностями,
\begin{equation*}
  (x^1)^2+(x^2)^2=C_1^2.
\end{equation*}
Они изображены на рис.\ref{fsingpoint},\textit{f} при $\om>0$.

Во всех случаях интегральные кривые определены при всех значениях параметра $t$.
Это означает, что векторные поля полны. Для узлов и фокуса интегральные кривые
стремятся к началу координат при $t\to-\infty$. Начало координат в соответствии
с предложением \ref{polvec} является неподвижной точкой.
\qed\end{exa}

Множество всех полных векторных полей на некомпактном многообразии $\MM$
является подмножеством в $\CX(\MM)$. В отличие от всего множества $\CX(\MM)$
это подмножество не образует абелеву группу (модуль) по отношению к сложению.
\begin{exa}                                                       \label{ecomlv}
Рассмотрим два векторных поля $X=y^2\pl_x$ и $Y=x^2\pl_y$ на евклидовой
плоскости $\MR^2$. Оба поля являются полными, однако их сумма $X+Y$ неполна.
Действительно, векторное поле $X$ определяет интегральные кривые:
\begin{align*}
  \dot x=y^2\quad  \Rightarrow\quad   x&=y^2t+C_0, &&C_0,C_1=\const,
\\
  \dot y=0\quad ~\Rightarrow\quad     y&=C_1. &&
\end{align*}
Эти интегральные кривые параллельны оси $x$ и проходят через все точки $y$,
кроме $y=0$, см.\ рис.\ref{fintegcurve},$a$ (стрелки показывают возрастание
параметра $t$). Поскольку все интегральные прямые определены при всех
$t\in(-\infty,\infty)$, то векторное поле $X$ полно. Все точки оси абсцисс
$(x,0)\in\MR^2$ для векторного поля $X$ являются вырожденными особыми точками и
неподвижны. Через них не проходит ни одна интегральная кривая.
\begin{figure}[h,b,t]
\hfill\includegraphics[width=.8\textwidth]{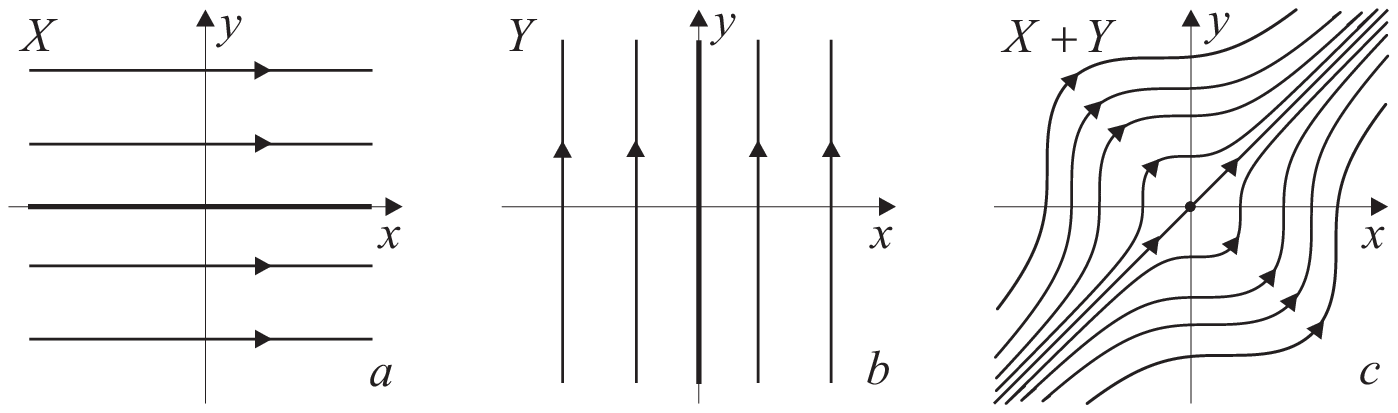}
\hfill {}
\centering\caption{Интегральные кривые для векторного поля $X$ $(a)$, $Y$ $(b)$
и $X+Y$ $(c)$.}
\label{fintegcurve}
\end{figure}

Аналогично, векторное поле $Y$ определяет интегральные кривые:
\begin{align*}
  \dot x=0\quad \Rightarrow\quad   x&=C_2, &&C_0,C_2=\const,
\\
  \dot y=x^2\quad \Rightarrow\quad     y&=x^2t+C_0. &&
\end{align*}
Интегральные кривые параллельны оси $y$ и проходят через все точки $x$, кроме
$x=0$, см.\ рис.\ref{fintegcurve},$b$. Соответствующее векторное поле $Y$ полно.
Все точки оси ординат $(0,y)\in\MR^2$ являются вырожденными особыми точками и
неподвижны.

Уравнения для интегральных кривых, определяемых суммой векторных полей, имеют
вид:
\begin{align}                                                     \label{eseinc}
  \dot x&=y^2,
\\                                                                \label{esecin}
  \dot y&=x^2.
\end{align}
При $x\ne0$ и $y\ne0$ отсюда следует дифференциальное уравнение для формы
интегральной кривой:
\begin{equation*}
  x^2dx=y^2dy\quad \Rightarrow\quad y^3=x^3+c,\qquad c=\const.
\end{equation*}
Подставляя это решение в уравнение (\ref{eseinc}), получаем равенство
\begin{equation}                                                  \label{eqiney}
  \dot x=(x^3+c)^{2/3}\quad \Leftrightarrow\quad t=\int\frac{dx}{(x^3+c)^{2/3}}.
\end{equation}
При $c=0$ это уравнение легко интегрируется
\begin{equation}                                                  \label{easppy}
  x=-\frac1{t-c_0},\qquad c_0=\const.
\end{equation}
После подстановки в (\ref{esecin}) получаем уравнение
\begin{equation}                                                  \label{easyku}
  y=-\frac1{t-c_0}+c_1,\qquad c_1=\const.
\end{equation}
Таким образом мы получили интегральные кривые для векторного поля $X+Y$, которые
неполны, т.к.\ уходят в бесконечность $x=\pm\infty$, $y=\pm\infty$ при конечном
значении параметра $t=c_0$. Остальные интегральные кривые можно не исследовать,
поскольку, по-определению, векторное поле полно, если {\em все} интегральные
кривые определены при всех $t\in\MR$.

Интегральные кривые для суммы векторных полей показаны на
рис.\ref{fintegcurve},$c$. Все интегральные кривые неполны в бесконечности. Это
легко видеть, т.к.\ при $x\to\pm\infty$ постоянной $c$ в интеграле
(\ref{eqiney}) можно пренебречь, и, следовательно, справедлива асимптотика
(\ref{easppy}). Матрица $B$ (\ref{eliosy}), определяющая линейное приближение,
для суммы векторных полей $X+Y$ имеет вид
\begin{equation*}
  B=\begin{pmatrix} 0 & 2x \\ 2y & 0 \end{pmatrix}.
\end{equation*}
Она вырождена только в начале координат. Поэтому точка $(0,0)$ является
вырожденной особой точкой. К ней подходят интегральные кривые (\ref{easppy}),
(\ref{easyku}) при бесконечном значении параметра $t$.
\qed\end{exa}

Ранее мы показали, что всюду отличное от нуля дифференцируемое векторное поле,
заданное на многообразии $\MM$, определяет семейство интегральных кривых,
проходящих через каждую точку $\MM$, причем через каждую точку проходит
единственная кривая.

\begin{defn}
Пусть $s(t,p)$ -- интегральная кривая векторного поля $X\in\CX(\MM)$, проходящая
через точку $p$, $s(0,p)=p$. Будем считать, что векторное поле полно. Тогда
отображение
\begin{equation}                                                  \label{eflode}
  s:\quad \MR\times\MM\ni\quad t,p\mapsto s(t,p)\quad\in\MM,
\end{equation}
генерируемое векторным полем $X$, называется {\em потоком векторного поля}.
\qed\end{defn}
\index{Поток векторного поля (vector field flow)}%
\index{Векторного поля поток (vector field flow)}%
\begin{prop}                                                      \label{planxu}
Отображение (\ref{eflode}) удовлетворяет тождеству
\begin{equation}                                                  \label{eflide}
  s\big(t_1,s(t_2,p)\big)=s(t_1+t_2,p),
\end{equation}
для всех значений $t_1,t_2\in\MR$, для которых формула (\ref{eflide}) имеет
смысл.
\end{prop}
\begin{proof}
Предложение следует из единственности решения системы дифференциальных
уравнений. Действительно, в произвольной карте выполнено равенство
\begin{align*}
  \frac d{dt_1}s^\al\big(t_1,s(t_2,p)\big)&=X^\al\big(s(t_2,p)\big)
\\
  s\big(0,s(t_2,p)\big)&=s(t_2,p).
\end{align*}
С другой стороны
\begin{align*}
  \frac d{dt_1}s^\al(t_1+t_2,p)&=\frac d{d(t_1+t_2)}s^\al(t_1+t_2,p)
  =X^\al\big(s(t_1+t_2,p)\big),
\\
  s(0+t_2,p)&=s(t_2,p).
\end{align*}
Тем самым и правая, и левая часть равенства (\ref{eflide}) удовлетворяют
одной и той же системы уравнений с одинаковыми начальными условиями.
\end{proof}
\begin{com}
Поток векторного поля можно представлять себе, как стационарный поток жидкости.
В этом случае параметр $t$ является временем, а $X$ -- векторным полем скорости
частиц жидкости.
\qed\end{com}
\begin{exa}
Рассмотрим гладкое векторное поле $K=-y\pl_x+x\pl_y$ на евклидовой плоскости
$\MR^2$. Нетрудно проверить, что поток этого векторного поля имеет вид
\begin{equation*}
  s:\quad \MR\times\MR^2\ni\quad t\times(x,y)
  \mapsto(x\cos t-y\sin t,x\sin t+y\cos t)\quad\in\MR^2.
\end{equation*}
Интегральная кривая, проходящая через точку $(x,y)$, представляет собой
окружность с центром в начале координат. В начале координат векторное поле
обращается в нуль, и интегральная кривая вырождается в точку. Матрица
(\ref{eliosy}) имеет вид
\begin{equation*}
  B=\begin{pmatrix} 0 & 1 \\ -1 & 0 \end{pmatrix}.
\end{equation*}
Ее собственные числа являются мнимыми $\lm_{1,2}=\pm i$. Поэтому начало
координат является для векторного поля $K$ невырожденной особой точкой --
центром.

Векторное поле $K$ является ничем иным, как векторным полем Киллинга двумерных
вращений евклидовой плоскости и принимает особо простой вид в полярных
координатах, $K=\pl_\vf$. Интегральные кривые поля $K$ являются в данном случае
траекториями Киллинга. Поток векторного поля на $\MR^2$ определяется независимо
от наличия метрики. Однако интерпретация векторного поля $K$, как поля Киллинга
уже связана с наличием на плоскости евклидовой метрики.
\qed\end{exa}
\begin{exa}
Рассмотрим гладкое векторное поле $K=y\pl_x+x\pl_y$ на плоскости $\MR^2$.
Поток этого векторного поля имеет вид
\begin{equation*}
  s:\quad \MR\times\MR^2\ni\quad t\times(x,y)
  \mapsto(x\ch t+y\sh t,x\sh t+y\ch t)\quad\in\MR^2.
\end{equation*}
Интегральная кривая, проходящая через точку $(x,y)$, является ветвью
гиперболы $x^2-y^2=\const$ с центром в начале координат. При $y=\pm x$ гиперболы
вырождаются в прямые линии, проходящие через начало координат под углом
$\pm\pi/4$. Матрица (\ref{eliosy}) имеет вид
\begin{equation*}
  B=\begin{pmatrix} 0 & 1 \\ 1 & 0 \end{pmatrix}.
\end{equation*}
Она невырождена и имеет различные вещественные собственные значения
$\lm_{1,2}=\pm1$. Поэтому начало координат является невырожденной особой точкой
-- седлом.

Векторное поле $K$ является векторным полем Киллинга для метрики Лоренца,
заданной на плоскости $\MR^2$, а интегральные кривые -- траекториями Киллинга.
\qed\end{exa}
\begin{defn}
При фиксированном значении параметра $t$ поток $s(t,x)$ представляет
собой диффеоморфизм, обозначаемый также
\begin{equation*}
  s_t:\quad \MM\rightarrow\MM.
\end{equation*}
Из предложения \ref{planxu} следует, что он представляет собой абелеву группу:

1) $s_{t_1}s_{t_2}=s_{t_1+t_2}$;

2) $s_0$ -- единичный элемент;

3) $s^{-1}_t=s_{-t}$ -- обратный элемент.\newline
Эта группа называется  {\em однопараметрической группой преобразований},
генерируемой векторным полем $X$. Действительно, из системы уравнений
(\ref{eintcv}) следует, что при малых значениях параметра поток имеет вид
\begin{equation*}
  s_\e:\quad x^\al\rightarrow x^\al+\e X^\al.
\end{equation*}
То есть векторное поле $X$ является генератором бесконечно малых преобразований
многообразия $\MM$.
\qed\end{defn}
\index{Однопараметрическая группа преобразований %
(one-parameter transformation group)}%
\index{Группа преобразований однопараметрическая%
(one-parameter transformation group)}%
\begin{com}
Псевдогруппа гладких преобразований координат $\diff\MM$ на многообразии $\MM$,
бесконечномерна. С соответствующими оговорками множество гладких векторных
полей $\CX(\MM)$ можно рассматривать, как бесконечномерную алгебру Ли для
$\diff\MM$.
\qed\end{com}

Сам поток часто обозначают
\begin{equation*}
  s^\al(t,x)=\exp(tX)x^\al
\end{equation*}
и называют  {\em экспоненциальным отображением}.
\index{Экспоненциальное отображение (exponential map)}%
\index{Отображение экспоненциальное (exponential map)}%
Это обозначение оправдано следующим образом. Разложим функцию $s^\al(t,x)$ в ряд
Тейлора по $t$:
\begin{equation*}
\begin{split}
  s^\al(t,x)
  &=x^\al+t\left.\frac{\pl s^\al(\tau,x)}{\pl\tau}\right|_{\tau=0}
  +\frac{t^2}{2!}\left.\frac{\pl^2s^\al(\tau,x)}{\pl\tau^2}
  \right|_{\tau=0}+\dotsc
\\
  &=\left.\left[1+t\frac\pl{\pl\tau}+\frac{t^2}{2!}\frac{\pl^2}{\pl\tau^2}
    +\dotsc\right]s^\al(\tau,x)\right|_{\tau=0}
\\
  &=\left.\exp\left(t\frac\pl{\pl\tau}\right)s^\al(\tau,x)\right|_{\tau=0}
  =\exp(tX)x^\al.
\end{split}
\end{equation*}
При этом выполнены формальные свойства экспоненты:
\begin{align*}
  \exp(0X)x&=x,
\\
  \exp(t_1X)\exp(t_2X)x&=\exp[(t_1+t_2)X]x,
\\
  \frac d{dt}\exp(tX)x&=X\exp(tX)x.
\end{align*}

Выше было показано, что любое полное векторное поле генерирует единственную
однопараметрическую группу преобразований. Верно и обратное утверждение:
любая однопараметрическая группа $s^\al(t,x)$ определяет векторное поле.
Для этого достаточно положить
\begin{equation*}
  X^\al:=\left.\frac{ds^\al}{dt}\right|_{t=0}.
\end{equation*}

Если векторное поле $X(\MM)$ является неполным, то понятие потока и
однопараметрической группы преобразований можно ввести только локально (см.,
например, \cite{Isham99}.)
\subsection{1-формы и гиперповерхности                           \label{sofhys}}
Рассмотрим отличную от нуля в каждой точке 1-форму $dx^\al A_\al(x)$ в некоторой
карте на многообразии $\MM$, $\dim\MM=n$. Тогда линейное алгебраическое
уравнение
\begin{equation}                                                  \label{eqonfo}
  dx^\al A_\al=0
\end{equation}
относительно дифференциалов $dx^\al$ имеет $n-1$ линейно независимых решений в
каждой точке $x\in\MM$. При этом любое решение уравнения (\ref{eqonfo}) является
линейной комбинацией данных решений. Отсюда следует, что совокупность векторов
$dx^\al$, удовлетворяющих соотношению (\ref{eqonfo}), задает $(n-1)$-мерное
подпространство в касательном пространстве $\MT_x(\MM)$. Таким образом,
отличная от нуля 1-форма задает распределение $(n-1)$-мерных подпространств в
касательном расслоении $\MT(\MM)$.

Казалось бы, что существуют такие $(n-1)$-мерные подмногообразия в $\MM$, что
касательные векторы к ним образуют $(n-1)$-мерное распределение, задаваемое
1-формой. Однако в общем случае это не так. Критерий существования таких
подмногообразий дает теорема Фробениуса \ref{tfroth}, которая будет рассмотрена
несколько позже. Сейчас мы остановимся на простейшем случае.

Пусть 1-форма (\ref{eqonfo}) является точной, т.е.\ имеет вид,
\begin{equation}                                                  \label{exactf}
  dx^\al A_\al=dx^\al \pl_\al f,
\end{equation}
для некоторой функции $f(x)\in\CC^1(\MM)$. Тогда уравнение (\ref{eqonfo})
можно рассмотреть, как дифференциальное уравнение на $f$, любое решение
которого имеет вид
\begin{equation}                                                  \label{esolef}
  f(x)=\const.
\end{equation}
Уравнение (\ref{esolef}) при разных значениях константы определяет семейство
$(n-1)$-мерных подмногообразий в $\MM$, которые называются
{\em гиперповерхностями}. При $n=2$ эти подмногообразия называются {\em линиями
уровня}. В этом случае касательные векторы к гиперповерхностям определяют те же
$(n-1)$-мерные подпространства в касательном пространстве, что и 1-форма.
Говорят также, что гиперповерхность имеет коразмерность один.
\index{Гиперповерхность (hypersurface)}%
\index{Линия уровня (level line, level curve)}%

Не ограничивая общности, для непостоянной функции можно считать, что только
$\pl_1 f\ne0$ в некоторой области. Поэтому функцию $f$ можно выбрать в качестве
первой координаты $f:=x^1$. Тогда семейство гиперповерхностей будет задано
уравнением $x^1=\const$, а направления вдоль остальных координатных осей будут
определять касательные к гиперповерхности направления.
\subsection{Алгебра Ли векторных полей                           \label{salgvf}}
Ранее мы дали два эквивалентных определения векторных полей $\CX(\MM)$ как
сечений касательного расслоения $\MT(\MM)$ и как дифференцирований в алгебре
функций $\CC^\infty(\MM)$. Это -- разные определения, одно из которых может
иметь определенные преимущества в той или иной ситуации. При рассмотрении
алгебраических вопросов, как правило, удобнее использовать алгебраическое
определение векторных полей через дифференцирования. Используя это определение,
мы введем на множестве векторных полей $\CX(\MM)$ структуру алгебры Ли.
\begin{defn}
Последовательное применение двух двух дифференцирований (векторных полей) $X$ и
$Y$ к некоторой функции $f$ снова дает функцию из $\CC^\infty(\MM)$. Определим
композицию двух дифференцирований $X\circ Y$ формулой $(X\circ Y)f:=X(Yf)$.
Отображение $(X\circ Y)$ является линейным, однако Правило Лейбница для него
не выполнено:
\begin{equation*}
  (X\circ Y)fg=XfYg+fX(Yg)+XgYf+gX(Yf)\ne f(X\circ Y)g+g(X\circ Y)f.
\end{equation*}
Это означает, что композиция векторных полей $X\circ Y$ не является векторным
полем. По другому, отображение $X\circ Y$ в координатах содержит не только
первые, но и вторые производные.
Рассмотрим композицию этих дифференцирований в другом порядке,
\begin{equation*}
  (Y\circ X)fg=YfXg+fY(Xg)+YgXf+gY(Xf).
\end{equation*}
Нетрудно проверить, что разность $Y\circ X-Y\circ X$ удовлетворяет правилу
Лейбница
\begin{equation*}
  (X\circ Y-Y\circ X)fg=f(X\circ Y-Y\circ X)g+g(X\circ Y-Y\circ X)f,
\end{equation*}
т.е.\ является векторным полем. Эта разность называется {\em коммутатором
векторных полей} или {\em скобкой Ли} и обозначатся
\begin{equation*}                                                    \tag*{\qed}
  [X,Y]:=X\circ Y-Y\circ X.
\end{equation*}
\renewcommand{\qed}{}\end{defn}
\index{Коммутатор векторных полей (commutator of vector fields)}%
\index{Скобка Ли векторных полей (Lie bracket of vector fields)}%
Из определения следует, что коммутатор двух векторных полей антисимметричен,
\begin{equation}                                                  \label{eantco}
  \left[X,Y\right]=-\left[Y,X\right],
\end{equation}
и коммутаторы трех произвольных векторных полей удовлетворяют
{\em тождеству Якоби}
\index{Якоби тождество (Jacobi identity)}%
\index{Тождество Якоби (Jacobi identity)}%
\begin{equation}                                                  \label{ejacid}
  \big[[X,Y],Z\big]+\big[[Y,Z],X\big]+\big[[Z,X],Y\big]=0,
\end{equation}
где слагаемые отличаются циклической перестановкой.
\begin{com}
Антисимметрия коммутатора (\ref{eantco}) эквивалентна условию $[X,X]=0$.
Действительно, то, что это условие вытекает из (\ref{eantco}) очевидно. Для
доказательства обратного утверждения достаточно рассмотреть уравнение
$[X+Y,X+Y]=0$.
\qed\end{com}

Алгебра Ли является неассоциативной алгеброй, при этом условие ассоциативности
заменяется на тождества Якоби.

Рассмотрим векторные поля в произвольной карте
$X=X^\al\pl_\al$, $Y=Y^\al\pl_\al$. Тогда коммутатор дает новое векторное поле
\begin{equation}                                                  \label{ecomvf}
  Z:=\left[X,Y\right]=(X^\bt\pl_\bt Y^\al-Y^\bt\pl_\bt X^\al)\pl_\al.
\end{equation}
Используя закон преобразования компонент векторных полей (\ref{evectr}),
нетрудно проверить, что выражение в правой части инвариантно относительно
преобразования координат.
\begin{defn}
Множество векторных полей с операцией сложения и коммутирования, которое
удовлетворяет условиям (\ref{eantco}) и (\ref{ejacid}) образует {\em кольцо Ли}.
Умножение векторного поля на числа снова дает векторной поле, при этом
коммутатор (\ref{ecomvf}) билинеен
\begin{align*}
  [aX+bY,Z]&=a[X,Z]+b[Y,Z].
\\
  [X,aY+bZ]&=a[X,Y]+b[X,Z],
\end{align*}
где $a,b\in\MR$. Множество векторных полей с операциями умножения на
вещественные числа, сложения и коммутирования образует {\em алгебру Ли} над
полем вещественных чисел. Эта алгебра бесконечномерна и также обозначается
$\CX(\MM)$.
\qed\end{defn}
\index{Кольцо Ли (Lie ring)}\index{Ли кольцо (Lie ring)}%
\index{Алгебра Ли (Lie algebra)}\index{Ли алгебра (Lie algebra)}%

Алгебры Ли образуют не только векторные поля на многообразиях. Структуру алгебры
Ли можно также ввести на абстрактном векторном пространстве.
\begin{exa}
Двумерное векторное пространство с базисом $e_1$ и $e_2$ становится алгеброй
Ли, если положить
\begin{equation*}
  [e_1,e_1]=[e_2,e_2]=0,\qquad [e_1,e_2]=e_2
\end{equation*}
и продолжить эту операцию по линейности.
\qed\end{exa}
\begin{exa}
Трехмерное векторное пространство $\MR^3$ с ортонормальным базисом $e_i$,
$i=1,2,3$, является алгеброй Ли, если в качестве коммутатора двух векторов
$X,Y\in\MR^3$ выбрать их векторное произведение
\begin{equation*}
  [X,Y]^i:=-\ve^{ijk}X_jY_k,
\end{equation*}
где $\ve^{ijk}$ -- полностью антисимметричный тензор третьего ранга и опускание
индексов производится с помощью евклидовой метрики $\dl_{ij}$. Эта алгебра Ли
совпадает с алгеброй Ли группы трехмерных вращений $\MS\MO(3)$ (\ref{ealatr}).
\qed\end{exa}
\begin{exa}
Векторное пространство $\Gg\Gl(n,\MR)$ всех вещественных
$n\times n$-матриц образует конечномерную алгебру Ли, если положить
\begin{equation*}
  [A,B]:=AB-BA,\qquad A,B\in\Gg\Gl(n,\MR),
\end{equation*}
где $AB$ -- обычное произведение матриц. При этом $\dim\Gg\Gl(n,\MR)=n^2$.
\qed\end{exa}
Забегая вперед (см.\ раздел \ref{stornm}), заметим, что коммутатор векторных
полей (\ref{ecomvf}) можно записать в эквивалентном виде, используя ковариантную
производную и тензор кручения
\begin{equation*}
  [X,Y]
  =(X^\bt\nb_\bt Y^\al-Y^\bt\nb_\bt X^\al-X^\bt Y^\g T_{\bt\g}{}^\al)\pl_\al.
\end{equation*}
В таком виде правая часть этого равенства явно ковариантна.
\section{Тензорные поля                                          \label{stenfi}}
Рассмотрим многообразие $\MM$, $\dim\MM=n$. В каждой точке $x\in\MM$ у нас есть
два $n$-мерных векторных пространства: касательное $\MT_x(\MM)$ и кокасательное
$\MT^*_x(\MM)$ пространства. Рассмотрим их тензорное произведение
\begin{equation}                                                  \label{etavht}
 \MT^r_{s,x}(\MM):=\underbrace{\MT_x(\MM)\otimes\dotsc\otimes\MT_x(\MM)}_r
 \otimes\underbrace{\MT^*_x(\MM)\otimes\dotsc\otimes\MT^*_x(\MM)}_s,
\end{equation}
где мы взяли $r$ экземпляров касательного и $s$ экземпляров кокасательного
пространства. Для определенности мы фиксировали порядок сомножителей. Таким
образом в каждой точке многообразия мы построили векторное пространство
размерности $\dim\MT^r_{s,x}(\MM)=n^{r+s}$.
\begin{defn}
Объединение
\begin{equation*}
  \MT^r_s(\MM):=\bigcup_{x\in\MM}\MT^r_{s,x}(\MM),
\end{equation*}
взятое по всем точкам многообразия, называется {\em расслоением тензоров типа}
$(r,s)$ на многообразии $\MM$. Сечение этого расслоения $T^r_s(x)$ называется
{\em тензорным полем типа} $(r,s)$ или $r$ раз {\em контравариантным}
и $s$ раз {\em ковариантным} тензорным полем на многообразии $\MM$. Число $r+s$
называется {\em рангом} тензорного поля.
\qed\end{defn}
\index{Расслоение тензоров типа (tensor bundle of type) $(r,s)$}%
\index{Тензорное поле типа (tensor field of type) $(r,s)$}%
\index{Ранг тензорного поля (rank of a tensor field)}%
\index{Тензора ранг (rank of a tensor)}%
\index{Контравариантное векторное поле (contravariant vector field)}%
\index{Векторное поле контравариантное (contravariant vector field)}%
\index{Ковариантное векторное поле (covariant vector field)}%
\index{Векторное поле ковариантное (covariant vector field)}%

Базой этого расслоения является многообразие $\MM$, типичным слоем -- векторное
пространство
\begin{equation*}
  \underbrace{\MR^n\otimes\dotsc\otimes\MR^n}_r\otimes
  \underbrace{\MR^n\otimes\dotsc\otimes\MR^n}_s,
\end{equation*}
где $\MR^n$ -- типичный слой касательного расслоения. Слоем над $x\in\MM$
является векторное пространство (\ref{etavht}) (тем самым мы определили
проекцию). Дифференцируемая структура на расслоении тензоров задается
дифференцируемыми структурами на базе и в типичном слое аналогично тому, как она
была построена для касательного расслоения.

Координатные базисы в касательном и кокасательном пространствах, $e_\al=\pl_\al$
и $e^\al=dx^\al$, индуцируют координатный базис в тензорном произведении,
который мы обозначим
\begin{equation}                                                  \label{ebaste}
  e_{\al_1}\otimes\dotsc\otimes e_{\al_r}\otimes
  e^{\bt_1}\otimes\dots\otimes e^{\bt_s}.
\end{equation}
Напомним, что тензорное произведение векторов не является коммутативным,
$$
  e_\al\otimes e^\bt\ne e^\bt\otimes e_\al,
$$
поэтому порядок следования базисных векторов в произведении (\ref{ebaste})
фиксирован: сначала мы пишем базисные векторы касательного, а затем
кокасательного пространств.

Рассмотрим произвольную карту $(\MU,\vf)$ на многообразии. Тогда тензорное
поле типа $(r,s)$ в координатах имеет вид
\begin{equation}                                                  \label{etenfi}
  T^r_s(x)=T_{\bt_1\dotsc\bt_s}{}^{\al_1\dotsc\al_r}(x)\,
  e_{\al_1}\otimes\dotsc\otimes e_{\al_r}\otimes
  e^{\bt_1}\otimes\dotsc\otimes e^{\bt_s}.
\end{equation}
Нижние и верхние индексы называют соответственно {\em ковариантными} и
{\em контравариантными}.
\index{Ковариантный индекс (covariant index)}%
\index{Индекс ковариантный (covariant index)}%
\index{Контравариантный индекс (contravariant index)}%
\index{Индекс контравариантный (contravariant index)}%
Общее число индексов $r+s$ равно рангу тензорного поля.
\begin{com}
Для определенности, у компонент $T_{\bt_1\dotsc\bt_s}{}^{\al_1\dotsc\al_r}(x)$
мы сначала выписали все ковариантные индексы, а затем -- все контравариантные.
Порядок индексов зафиксирован порядком сомножителей в правой части
(\ref{etavht}) и принятом нами соглашением для записи тензорного поля типа
$(1,1)$:
\begin{equation*}
  dx^\bt T_\bt{}^\al\pl_\al.
\end{equation*}
Контравариантные индексы, так же, как и ковариантные, упорядочены между собой.
В разделе \ref{smetri} будет введена операция опускания и подъема индексов с
помощью метрики. Она будет неоднозначной, если порядок контравариантных и
ковариантных индексов не фиксирован.

Ниже мы построим тензорную алгебру для тензорных полей вида (\ref{etenfi}).
Аналогично можно построить тензорную алгебру для произвольного расположения
сомножителей в правой части (\ref{etavht}), когда касательные и кокасательные
пространства чередуются в произвольном порядке. Мы будем предполагать, что все
индексы упорядочены определенным образом. В этом случае обозначение векторного
поля $T^r_s(x)$ является грубым, т.к.\ учитывает только общее число ковариантных
и контравариантных индексов, а не их последовательность.
\qed\end{com}

Набор функций $X_{\bt_1\dots\bt_s}{}^{\al_1\dots\al_r}(x)$ с $r$
верхними и $s$ нижними индексами называется {\em компонентами} тензорного поля
типа $(r,s)$ в карте $(\MU,\vf)$.
\index{Компоненты тензорного поля (components of a vector field)}%
Тензорное поле называется гладким, если все компоненты -- гладкие функции.
При преобразованиях координат каждый контравариантный индекс умножается на
матрицу Якоби (\ref{ejacma}), так же, как и компоненты вектора, а каждый
ковариантный индекс -- на обратную матрицу Якоби, так же, как и 1-форма.
\begin{exa}
Компоненты тензорного поля типа (1,1) при преобразовании координат
$x^\al\mapsto x^{\al'}(x)$ преобразуются по правилу:
\begin{equation}                                                  \label{etentv}
  X_{\bt'}{}^{\al'}=\frac{\pl x^\bt}{\pl x^{\bt'}}
  X_\bt{}^\al\frac{\pl x^{\al'}}{\pl x^\al}.
\end{equation}
Аналогично преобразуются компоненты тензорных полей произвольного типа.
\qed\end{exa}
В дальнейшем, для краткости, тензорные поля мы часто будем называть просто
тензорами.

Очевидно, что, если все компоненты тензорного поля равны нулю в какой-то одной
системе координат, то они равны нулю во всех остальных системах отсчета. У
нетривиальных тензоров хотя бы одна компонента должна быть отлична от нуля. В
общем случае у тензора типа $(r,s)$ на многообразии размерности $n$ имеется
$n^{r+s}$ независимых компонент в каждой точке.

Обозначим множество гладких тензорных полей типа $(r,s)$ символом
$\CT^r_s(\MM)$. При этом $\CT^0_0(\MM)=\CC^\infty(\MM)$, $\CT^1_0=\CX(\MM)$ и
$\CT^0_1=\Lm_1(\MM)$. В дальнейшем индекс $0$ у множеств тензорных полей,
имеющих только контравариантные или ковариантные индексы, писаться не будет:
$\CT^r_0(\MM)=\CT^r(\MM)$ и $\CT^0_s(\MM)=\CT_s(\MM)$.

Тензорные поля фиксированного типа в каждой точке можно складывать и умножать на
числа, т.е.\ они образуют (бесконечномерное) векторное пространство над полем
вещественных чисел. Кроме того, тензорное поле произвольного типа можно
поточечно умножать на произвольные функции, при этом получится новое тензорное
поле того же типа. Таким образом они образуют модуль над алгеброй гладких
функций $\CC^\infty(\MM)$.

Введем обозначение для прямой суммы тензорных полей
\begin{equation*}
  \CT(\MM):=\bigoplus_{r,s=0}^\infty \CT^r_s(\MM).
\end{equation*}
На множестве $\CT(\MM)$ можно ввести поточечное тензорное умножение,
которое двум тензорам типа $(r,s)$ и $(p,q)$ ставит в соответствие
тензорное поле типа $r+p,s+q$. А именно, зафиксируем точку $x\in\MM$.
Из универсального факторизационного свойства тензорного произведения следует,
что существует единственное билинейное отображение из
$\MT^r_{s,x}\times\MT^p_{q,x}$ в $\MT^{r+p}_{s+q,x}$, которое
отображает пару тензоров
\begin{equation*}
\begin{split}
(X_1\otimes\cdots\otimes X_r\otimes A_1\otimes\cdots\otimes A_s)&\in\MT^r_{s,x},
\\
 (Y_1\otimes\cdots\otimes Y_p\otimes B_1\otimes\cdots\otimes B_q)&\in\MT^p_{q,x}
\end{split}
\end{equation*}
в тензор типа $r+p,s+q$:
\begin{equation}                                                  \label{etenpr}
  X_1\otimes\cdots\otimes X_r\otimes Y_1\otimes\cdots\otimes Y_p\otimes
  A_1\otimes\cdots A_s\otimes B_1\otimes\cdots\otimes B_q\quad
   \in\MT^{r+p}_{s+q,x}.
\end{equation}
Это отображение называется тензорным произведением тензоров в данной точке.
\begin{com}
Мы зафиксировали порядок сомножителей в произведении (\ref{etenpr}), который
соответствует тензорному произведению (\ref{etavht}).
\end{com}
\begin{defn}
{\em Тензорным произведением} тензорных полей $\CT^r_s(\MM)$ и $\CT^p_q(\MM)$
называется тензорное поле типа $\CT^{r+p}_{s+q}(\MM)$, полученное поточечным
тензорным произведением (\ref{etenpr}).
\qed\end{defn}
\index{Тензорное произведение тензорных полей%
 (tensor product of tensor fields)}%
Чтобы получить выражение для компонент тензорного произведения в определенной
карте, достаточно в качестве векторных $X,Y$ и ковекторных $A,B$ полей в
определении тензорного произведения выбрать координатный базис. Пусть в
некоторой карте задано два тензорных поля:
\begin{align*}
  K&=K_{\bt_1\cdots\bt_s}{}^{\al_1\cdots\al_r}\,e_{\al_1}\otimes e_{\al_r}
  \otimes e^{\bt_1}\cdots\otimes e^{\bt_s}\quad \in\CT^r_s(\MM),
\\
  L&=L_{\dl_1\cdots\dl_q}{}^{\g_1\cdots\g_p}\,e_{\g_1}\otimes e_{\g_p}
  \otimes e^{\dl_1}\cdots\otimes e^{\dl_q}\qquad \in\CT^p_q(\MM).
\end{align*}
Тогда компоненты их тензорного произведения
\begin{equation*}
  (K\otimes L)_{\bt_1\cdots\bt_s\dl_1\cdots\dl_q}
  {}^{\al_1\cdots\al_r\g_1\cdots\g_p}=K_{\bt_1\cdots\bt_s}{}^{\al_1\cdots\al_r}
  L_{\dl_1\cdots\dl_q}{}^{\g_1\cdots\g_p}
\end{equation*}
просто равны произведению компонент каждого сомножителя, как чисел.
\begin{exa}
Произведение двух векторных полей $X\otimes Y$ дает контравариантный тензор
второго ранга с компонентами
$$
  T(x)=T^{\al\bt}(x)e_\al\otimes e_\bt=X(x)\otimes Y(x)
  =X^\al(x)Y^\bt(x)e_\al\otimes e_\bt,
$$
Эта операция является некоммутативной, поскольку первый индекс поля $T(x)$
относится к векторному полю $X$, а не $Y$.
\qed\end{exa}
Вместе с тензорным умножением, множество тензорных полей $\CT(\MM)$ образует
некоммутативную ассоциативную {\em тензорную алгебру} над полем
\index{Тензорная алгебра (tensor algebra)}%
\index{Алгебра тензорная (tensor algebra)}%
вещественных чисел. Эта алгебра бесконечномерна, поскольку векторное
пространство тензоров фиксированного типа бесконечномерно само по себе и,
вдобавок, ранг тензоров неограничен. Алгебра тензоров имеет естественную
градуировку, как прямая сумма тензоров фиксированного типа. Образующими
тензорной алгебры являются векторные поля и 1-формы.
\begin{defn}
Каждой паре индексов $(i,j)$ таких, что $1\le i\le r$ и $1\le j\le s$, мы ставим
в соответствие линейное отображение $C^{ij}:$
$\CT^r_s(\MM)\rightarrow\CT^{r-1}_{s-1}(\MM)$ с помощью следующей формулы
\begin{multline*}
  X_1\otimes\dotsc\otimes X_r\otimes A_1\otimes\dotsc\otimes A_s\quad \mapsto
\\
  \mapsto (A_j,X_i) X_1\otimes\dotsc X_{i-1}\otimes X_{i+1}
  \otimes\dotsc\otimes X_r\otimes
  A_1\otimes\dotsc\otimes A_{j-1}\otimes\dotsc\otimes A_{j+1}\otimes A_s,
\end{multline*}
где $(A_j,X_i)=A_j(X_i)$ -- значение 1-формы $A_j$ на векторе $X_i$.
Это отображение называется {\em сверткой} и обозначается $C^{ij}$.
\qed\end{defn}
\index{Свертка тензоров (contraction of tensors)}%
Компоненты свернутого тензора $K\in\CT^r_s(\MM)$ имеют вид
\begin{equation*}
  K_{\bt_1\cdots\bt_{j-1}\g\bt_{j+1}\cdots\bt_s}{}
  ^{\al_1\cdots\al_{i-1}\g\al_{i+1}\cdots\al_r}
  = \dl^{\bt_j}_{\al_i}K_{\bt_1\cdots\bt_s}{}^{\al_1\cdots\al_r},
\end{equation*}
где произведена свертка по одному верхнему и одному нижнему индексу.
\begin{exa}
Тензору типа $(1,1)$ ставится в соответствие скалярное поле $\tr K=K_\al{}^\al$,
которое называется {\em следом} тензора $K=K_\bt{}^\al e_\al\otimes e^\bt$.
\qed\end{exa}
\index{След тензора (trace of a tensor)}%
\begin{exa}
Значением 1-формы $A=dx^\al A_\al$ на векторном поле $X=X^\al\pl_\al$
является свертка тензорного произведения $A\otimes X$:
$(A,X):=A(X)=X^\al A_\al$.
\qed\end{exa}
\begin{defn}
Тензорное поле $K\in\CT^r_s(\MM)$ называется {\em разложимым}, если его можно
представить в виде
\begin{equation*}
  K=X_1\otimes\cdots\otimes X_r\otimes A_1\otimes\cdots\otimes A_s,
\end{equation*}
для некоторых векторов: $X_i\in\CT^1(\MM)=\CX(\MM)$, $i=1,\cdots,r$ и 1-форм:
$A_j\in\CT_1(\MM)=\Lm_1(\MM)$, $j=1,\cdots,s$.
\qed\end{defn}
\index{Разложимое тензорное поле (decomposable vector field)}%
\index{Тензорное поле разложимое (decomposable vector field)}%
\begin{exa}
Сумма двух разложимых контравариантных тензоров второго ранга
\begin{equation*}
  X_1\otimes Y_1+X_2\otimes Y_2
  =(X^\al_1 Y^\bt_1+X^\al_2Y^\bt_2)e_\al\otimes e_\bt
\end{equation*}
может не быть разложимым тензором.
\qed\end{exa}

Если тензор имеет два или более индексов одного типа, то с помощью симметризации
или антисимметризации по верхним или нижним индексам можно строить новые
тензорные поля. Поскольку преобразование координат действует одинаково на каждый
ковариантный и контравариантный индекс, то симметризация и антисимметризация
индексов является инвариантной операцией и свойство симметрии по индексам
сохраняется при преобразовании координат.
\begin{defn}
Тензорное поле называется {\em неприводимым}, если нельзя найти такие линейные
комбинации его компонент с постоянными коэффициентами, которые сами образовывали
бы тензор.
\qed\end{defn}
\index{Неприводимое тензорное поле (irreducible tensor field)}%
\index{Тензорное поле неприводимое (irreducible tensor field)}%
\begin{exa}
Скалярные, векторные поля и 1-формы являются неприводимыми тензорными полями.
\qed\end{exa}
\begin{exa}
Ковариантные или контравариантные тензорные поля второго ранга приводимы, т.к.\
их компоненты можно разложить на симметричную и антисимметричную неприводимую
части:
\begin{equation}                                                  \label{edecst}
  X_{\al\bt}=X_{(\al\bt)}+X_{[\al\bt]},
\end{equation}
где
\begin{equation*}                                                    \tag*{\qed}
  X_{(\al\bt)}:=\frac12(X_{\al\bt}+X_{\bt\al}),\qquad
  X_{[\al\bt]}:=\frac12(X_{\al\bt}-X_{\bt\al}).
\end{equation*}
\end{exa}
\begin{com}
Выделение следа у тензоров со всеми ковариантными или контравариантными
индексами невозможно без наличия метрики.
\qed\end{com}
\begin{exa}
Тензорное поле $X_\al{}^\bt$ типа (1,1) также приводимо, поскольку у него можно
выделить след $\tr X$ и бесследовую часть $Y_\al{}^\bt$ $(Y_\al{}^\al=0)$:
\begin{equation}                                                  \label{etrsrt}
  X_\al{}^\bt=Y_\al{}^\bt+\frac1n\dl_\al^\bt\tr X,
\end{equation}
где
\begin{equation*}                                                    \tag*{\qed}
  \tr X:=X_\al{}^\al,\qquad Y_\al{}^\bt:=X_\al{}^\bt-\frac1n\dl_\al^\bt\tr X.
\end{equation*}
\end{exa}
Тензорные поля ранга три и выше, в общем случае, приводимы. Если на
многообразии не задано никаких других объектов, кроме тензорного поля, то
разложение на неприводимые компоненты может осуществляться только с помощью
взятия следа, симметризации или антисимметризации по индексам.
\begin{com}
При проведении вычислений с тензорными полями важно иметь ввиду следующее
обстоятельство. Если тензорное поле приводимо, то его разложение на неприводимые
компоненты в большинстве случаев упрощает вычисления и проясняет математическую
структуру модели.
\qed\end{com}
\begin{exa}
{\em Символ Кронекера}, компоненты которого в каждой карте многообразия $\MM$
составляют $n$-мерную единичную матрицу,
\begin{equation}                                                  \label{ekrode}
  \dl_\bt^\al=\left\lbrace
  \begin{array}{rl}
  1,\quad  & \al=\bt,\\
  0,\quad  & \al\ne\bt, \end{array} \right.
\end{equation}
и имеют один верхний и один нижний индекс, определяет тензорное поле типа
$(1,1)$. Он инвариантен относительно преобразований координат
$$
  \dl_{\bt'}^{\al'}=\frac{\pl x^\bt}{\pl x^{\bt'}}\dl_\bt^\al
  \frac{\pl x^{\al'}}{\pl x^\al}
  =\frac{\pl x^\bt}{\pl x^{\bt'}}\frac{\pl x^{\al'}}{\pl x^\bt}
  =\begin{cases}1,\quad\al'=\bt',\\ 0,\quad\al'\ne\bt',
  \end{cases}
$$
т.к.\ верхний и нижний индексы преобразуется с помощью взаимно обратных матриц.
Символ Кронекера представляет собой исключение в двух отношениях. Во-первых, он
инвариантен относительно преобразований координат и, во-вторых, его индексы
можно писать один под другим, поскольку подъем и опускание индексов с помощью
метрики приводит к симметричным тензорам.
\qed\end{exa}
\index{Символ Кронекера (Kronecker symbol)}%
\index{Кронекера символ (Kronecker symbol)}%
Тензорное поле $T_\al{}^\bt\in\CT^1_1(\MM)$ типа $(1,1)$ можно рассматривать,
как линейный оператор (эндоморфизм), действующий в пространстве векторов
$\CT^1(\MM)=\CX(\MM)$ и 1-форм $\CT_1(\MM)=\Lm_1(\MM)$. В компонентах действие
оператора задается правилами:
$$
  A^\prime_\al=T_\al{}^\bt A_\bt,\qquad X^{\prime\al}=X^\bt T_\bt{}^\al.
$$
Действие оператора $T_\al{}^\bt$ естественным образом распространяется на
тензоры произвольного типа.
\begin{exa} Символ Кронекера представляет собой тождественный оператор.
\qed\end{exa}
\begin{exa}
Проекционные операторы (\ref{epropv}) являются
тензорными полями типа $(1,1)$.
\qed\end{exa}
На многообразии $\MM$ можно также определить {\em тензорные плотности}
степени $p\in\MZ$ и ранга $(r,s)$, если
\index{Тензорная плотность (tensor density)}%
\index{Плотность тензорная (tensor density)}%
при преобразовании координат все их компоненты умножить на якобиан
преобразования в степени $p$, как и в случае скалярных полей (\ref{escfud}).
Например, тензорная плотность типа $(1,1)$ и степени $p$ преобразуется
по-правилу
\begin{equation}                                                  \label{etedtr}
  X_{\al'}{}^{\bt'}=J^p\frac{\pl x^\al}{\pl x^{\al'}}
  X_\al{}^\bt\frac{\pl x^{\bt'}}{\pl x^\bt}.
\end{equation}
В каждой точке многообразия тензорные плотности фиксированного типа и степени
образуют векторное пространство над полем вещественных чисел и модуль над
алгеброй гладких функций. По аналогии с тензорным произведением тензоров можно
ввести тензорное произведение плотностей, которое двум плотностям типа
$(r_1,s_1)$, $(r_2,s_2)$ и степеней $p_1$ и $p_2$ ставит в соответствие
тензорную плотность типа $(r_1+r_2,s_1+s_2)$ и степени $p_1+p_2$. Множество
тензорных плотностей и всех их линейных комбинаций в
фиксированной точке образует некоммутативную ассоциативную алгебру над полем
вещественных чисел. Эта алгебра имеет естественную градуировку, как прямая сумма
тензорных плотностей фиксированного типа и степени.

Так же, как и в случае скалярных плотностей тензорные плотности
$X_{\bt_1\dotsc\bt_s}{}^{\al_1\dotsc\al_r}$ степени $p$ можно представить в виде
$$
  X_{\bt_1\dotsc\bt_s}{}^{\al_1\dotsc\al_r}
  =\vol^{-p}Y_{\bt_1\dotsc\bt_s}{}^{\al_1\dotsc\al_r},
$$
где $\vol=\det e_\al{}^a$ -- определитель репера, а
$Y_{\bt_1\dotsc\bt_s}{}^{\al_1\dotsc\al_r}$ --
тензорное поле того же типа, что и исходная тензорная плотность.

Множество тензорных полей можно интерпретировать, как множество полилинейных
отображений.
\begin{theorem}
Множество ковариантных тензорных полей $\CT_s(\MM)$ можно рассматривать, как $s$
линейное отображение \big($\CC(\MM)$-модуль\big) из
$\underbrace{\CX\times\dotsc\times\CX}_s$ в алгебру непрерывных функций
$\CC(\MM)$ такое, что
\begin{equation*}
  K(f_1X_1,\dotsc,f_sX_s)=f_1\dotsc f_sK(X_1,\dotsc,X_s),\qquad K(x)\in\CT_s(\MM),
\end{equation*}
для всех $f_i\in\CC(\MM)$ и $X_i\in\CX(\MM)$. Обратно, каждое такое
отображение можно рассматривать, как тензорное поле типа $(0,s)$.
\end{theorem}
\begin{proof}
См., например, \cite{KobNom6369R}.
\end{proof}

Аналогично можно интерпретировать тензоры произвольного типа $(r,s)$, как
множество всех $r+s$ линейных отображений:
\begin{equation}                                                  \label{epoltm}
  \CT^r_s(\MM):\qquad \underbrace{\CX\times\dotsc\times\CX}_s\times
  \underbrace{\Lm_1\times\dotsc\times\Lm_1}_r~\rightarrow~\CC(\MM).
\end{equation}
Поскольку между тензорными полями и полилинейными отображениями существует
взаимно однозначное соответствие, то некоторые авторы принимают эти отображения
в качестве глобального определения тензорных полей.
\section{Полностью антисимметричные тензоры                      \label{santst}}
В настоящем разделе мы рассмотрим полностью антисимметричные тензоры, которые
играют очень важную роль в различных приложениях дифференциальной геометрии.
\begin{defn}
Рассмотрим тензорные поля типа $(r,0)$ или $(0,r)$ при $r\le n$ на многообразии
$\MM$, $\dim\MM=n$, компоненты которых антисимметричны относительно перестановки
любой пары индексов. В инвариантном виде условие антисимметричности для
ковариантных тензоров записывается в виде:
\begin{equation*}
  T(X_1,\dotsc,X_i,\dotsc,X_j,\dotsc,X_r)
  =-T(X_1,\dotsc,X_j,\dotsc,X_i,\dotsc,X_r),\qquad 1\le i<j\le r,
\end{equation*}
где $T\in\CT_r(\MM)$, $X_1,\dotsc,X_r\in\CX(\MM)$. Эти тензоры неприводимы и
называются {\em полностью антисимметричными} ковариантными тензорами ранга $r$.
\qed\end{defn}
\index{Полностью антисимметричный тензор (totally antisymmetric tensor)}%
\index{Тензор полностью антисимметричный (totally antisymmetric tensor)}%
Компонента полностью антисимметричного тензорного поля может быть отлична от
нуля только в том случае, если все индексы различны, поскольку при совпадении
двух или более индексов соответствующая компонента равна нулю. На многообразии
размерности $n$ не существует полностью антисимметричного тензора ранга
большего, чем размерность многообразия, т.к.\ в этом случае по крайней мере два
индекса будут совпадать.

Очевидно, что число независимых компонент полностью антисимметричного тензора
ранга $r$ равно числу выборок $r$ различных индексов из $n$:
$C_n^r$$=n!/r!(n-r)!$. В частности, полностью антисимметричный тензор
$X_{\al_1\dots\al_n}$ максимального ранга $n$ имеет только одну независимую
компоненту. Нетрудно проверить, что при преобразовании координат
$x\mapsto y(x)$ полностью антисимметричный тензор типа $(0,n)$ преобразуется
по закону
\begin{equation}                                                  \label{etotat}
  X_{\al_1\dotsc\al_n}^\prime
  =\frac{\pl x^{[\bt_1}}{\pl y^{\al_1}}\dots
  \frac{\pl x^{\bt_n]}}{\pl y^{\al_n}}X_{\bt_1\dotsc\bt_n}
  =X_{\al_1\dotsc\al_n}J^{-1},
\end{equation}
где $J:=\det (\pl y^\al/\pl x^\bt)$ -- якобиан преобразования координат. То есть
каждая компонента полностью антисимметричного ковариантного тензора
максимального ранга умножается на якобиан преобразования в минус первой степени,
и ее фиксированную компоненту можно рассматривать как скалярную плотность веса
$-1$.

В каждой карте $(\MU,\vf)$ можно построить полностью антисимметричный объект,
компоненты которого равны по модулю единице
\begin{equation}                                                  \label{etoatd}
  \hat\ve_{\al_1\dotsc\al_n}:=\sgn\s(\al_1\dotsc\al_n),\qquad
  \hat\ve_{1\dotsc n}=1,
\end{equation}
где $\sgn\s$ -- {\em знак перестановки} $\s$,
\index{Знак перестановки (sign of permutation)}%
\index{Перестановки знак (sign of permutation)}%
который равен $+1$ или $-1$, если для получения последовательности индексов
$\al_1,\dotsc,\al_n$ из последовательности натуральных чисел $1,\dots,n$
необходимо переставить соответственно четное и нечетное число пар индексов.
Объект $\hat\ve_{\al_1\dotsc\al_n}$ не может быть тензором, т.к.\ в общем
случае якобиан преобразования $J$ отличен от единицы. Из закона преобразования
тензорных плотностей (\ref{escfud}) следует, что каждую фиксированную компоненту
$\hat\ve_{\al_1\dots\al_n}$ можно рассматривать, как скалярную тензорную
плотность степени $1$. Поскольку компоненты антисимметричной тензорной плотности
постоянны, то в произвольной системе координат справедливо равенство
\begin{equation}                                                  \label{ecoatd}
  \pl_\al\hat\ve_{\al_1\dotsc\al_n}=0.
\end{equation}

В (псевдо-)римановом пространстве при наличии метрики $g_{\al\bt}$ (см.\ раздел
\ref{smetde}) можно построить полностью антисимметричный тензор
\begin{equation}                                                  \label{etoate}
  \ve_{\al_1\dots\al_n}=\sqrt{|g|}\hat\ve_{\al_1\dots\al_n},
\end{equation}
где введено сокращенное обозначение для определителя метрики, которое будет
часто использоваться в дальнейшем
\begin{equation}                                                  \label{emodem}
  g:=\det g_{\al\bt}.
\end{equation}
Полностью антисимметричный тензор (\ref{etoate}) преобразуется по стандартному
закону (\ref{etotat}).
\begin{com}
Здесь и в дальнейшем мы примем следующее обозначение: шляпка над символом
означает, что рассматривается тензорная плотность, а не тензор.
\qed\end{com}
Полностью антисимметричные тензоры типа $(k,0)$ со всеми контравариантными
индексами называются {\em поливекторами}.
\index{Поливектор (multivector)}%
Поливектор максимального ранга $n$ имеет одну независимую компоненту,
которая при преобразовании координат преобразуется по правилу
\begin{equation}                                                  \label{etotal}
  X^{\prime\al_1\dotsc\al_n}=X^{\al_1\dotsc\al_n}J,
\end{equation}
В пространстве контравариантных тензоров можно ввести полностью
антисимметричную тензорную плотность степени $-1$ с компонентами, равными
по модулю единице, аналогично тому, как это было сделано для
ковариантных тензоров. Мы положим
\begin{equation}                                                  \label{eantdc}
  \hat\ve^{\al_1\dots\al_n}:=\hat\ve_{\al_1\dots\al_n}\sgn,
\end{equation}
где множитель $\sgn$ зависит от того задана ли на многообразии метрика или нет
\begin{equation}                                                  \label{esgnmn}
  \sgn:=\begin{cases}
  1,                     & \text{если метрика не задана},\\
  \sgn(\det g_{\al\bt}), & \text{если метрика задана}.
  \end{cases}
\end{equation}
Компоненты этой плотности также равны по модулю единице и постоянны:
$\pl_\al\hat\ve^{\al_1\dotsc\al_n}=0$. При наличии метрики можно
ввести полностью антисимметричный контравариантный тензор
\begin{equation}                                                  \label{etoanc}
  \ve^{\al_1\dotsc\al_n}=\frac1\vol\hat\ve^{\al_1\dotsc\al_n}.
\end{equation}
\begin{com}
Отметим, что равенство (\ref{eantdc}) имеет смысл, несмотря на то, что индексы
слева контравариантны, а справа ковариантны, т.к.\ компоненты тензорных
плотностей $\hat\ve$ не зависят от выбора системы координат. Появление
множителя (\ref{esgnmn}) в определении (\ref{eantdc}) объясняется тем, что на
(псевдо-)римановом многообразии мы требуем, чтобы тензор с контравариантными
индексами можно было бы получить из тензора с ковариантными индексами простым
подъемом индексов:
\begin{equation*}
  \ve^{\al_1\dotsc\al_n}=g^{\al_1\bt_1}\dotsc g^{\al_n\bt_n}
  \ve_{\bt_1\dotsc\bt_n}.
\end{equation*}
Отсюда следует равенство (\ref{eantdc}).
\qed\end{com}
Наличие тензора $\ve_{\al_1\dots\al_n}$ позволяет представить компоненты
произвольного ковариантного антисимметричного тензора максимального ранга в виде
\begin{equation}                                                  \label{etaant}
  X_{\al_1\dots\al_n}=X^*(x)\ve_{\al_1\dots\al_n},\qquad
  X^*:=\ve^{\al_1\dotsc\al_n}X_{\al_1\dotsc\al_n}\sgn,
\end{equation}
где $X^*(x)$ -- некоторое (псевдо-)скалярное поле.

Аналогичное представление имеет место для произвольного контравариантного
тензора максимального ранга. Это означает, что полностью антисимметричные
тензоры максимального ранга имеют ровно одну независимую компоненту.
\section{Отображения многообразий                                \label{smapma}}
Рассмотрим отображение $h$ многообразия $\MM$, $\dim\MM=m$, в многообразие
$\MN$, $\dim\MN=n$,
\begin{equation}                                                  \label{emamap}
  h:\quad \MM\ni\quad x\mapsto y\quad\in\MN.
\end{equation}
Многообразие $\MN$ мы будем называть {\em пространством-мишенью}.
\index{Пространство-мишень (target space)}%
Пусть при этом отображении карта $(\MU,\vf)$ многообразия $\MM$ отображается в
некоторую карту $(\MV,\phi)$ многообразия $\MN$, $h(\MU)\subset\MV$. Обозначим
координаты на $\MU$ и $\MV$ через $x^\al$, $\al=1,\dotsc,m$, и $y^\mu$,
$\mu=1,\dotsc,n$. Тогда отображение
\begin{equation}                                                  \label{esmofu}
  \phi\circ h\circ\vf^{-1}:\quad \MR^m\supset\quad\vf(\MU)\rightarrow
  \phi(\MV)\quad\subset\MR^n
\end{equation}
двух областей евклидова пространства $\vf(\MU)\subset\MR^m$ и
$\phi(\MV)\subset\MR^n$ задается $n$ функциями от $m$ переменных $y^\mu(x)$. При
этом размерность многообразия $\MM$ может быть меньше, равна или больше
размерности $\MN$.
\begin{defn}
Отображение гладких многообразий называется {\em гладким (дифференцируемым)},
если задается гладкими (дифференцируемыми) функциями (\ref{esmofu}) в полных
атласах на $\MM$ и $\MN$.
\qed\end{defn}
В дальнейшем, если не оговорено противное, мы будем рассматривать гладкие
отображения.
\index{Гладкое отображение (smooth map)}%
\index{Отображение гладкое (smooth map)}%
\index{Дифференцируемое отображение (differentiable map)}%
\index{Отображение дифференцируемое (differentiable map)}%
\begin{defn}
Дифференцируемое отображение (\ref{emamap}) индуцирует линейное отображение
касательных пространств
\begin{equation}                                                  \label{etamap}
  h_*:\quad \MT_x(\MM)\ni\quad X=X^\al\pl_\al\mapsto
  Y=h_*X=Y^\mu\pl_\mu\quad\in\MT_{h(x)}(\MN)
\end{equation}
следующим образом. Рассмотрим кривую $\g\subset\MM$, проходящую через
произвольную точку $p\in\MM$ в направлении произвольного вектора $X_{(\g)}(p)$.
Эта кривая отобразится в некоторую кривую $h(\g)$ на $\MN$. По определению,
вектор $X_{(\g)}(p)\in\MT_p(\MM)$ отображается в тот вектор
$Y_{(h\circ\g)}\in\MT_{h(p)}(\MN)$, который касается кривой $h(\g)$ в точке
$h(p)$. Поскольку вектор в точке $p$ -- это класс эквивалентности кривых, то это
условие записывается в виде
\begin{equation*}
  h_*\big(X_{(\g)}(p)\big)
  =[h\circ\g]=Y_{(h\circ\g)}\big(h(p)\big),\qquad X_{(\g)}(p)=[\g].
\end{equation*}
Теперь мы упростим обозначения, опустив индекс кривой у вектора и обозначение
точки $p$. Поскольку по правилу дифференцирования сложной функции
\begin{equation*}
  \dot y^\mu\big(\phi\circ h\circ\g(t)\big)
  =\frac{\pl y^\mu}{\pl x^\al}\dot x^\al\big(\g(t)\big),
\end{equation*}
то вектор $Y$ единственен и не зависит от представителя класса эквивалентности
кривых, определяющих вектор $X$ в точке $x\in\MM$. Это отображение касательных
пространств (\ref{etamap}) называется {\em дифференциалом отображения}.
\qed\end{defn}
\index{Дифференциал отображения (map differential)}%
\index{Отображения дифференциал (map differential)}%

В компонентах дифференциал отображения задается матрицей Якоби:
\begin{equation}                                                  \label{edidma}
  Y^\mu=X^\al\frac{\pl y^\mu}{\pl x^\al}.
\end{equation}

Дифференциал отображения $h_*$ является линейным отображением:
\begin{align*}
  h_*(X+Y)&=h_*(X)+h_*(Y),\qquad X,Y\in\CX(\MM),
\\
  h_*(aX)&=ah_*(X),\qquad a\in\MR.
\end{align*}
множества всех векторных полей $\CX(\MM)$ на $\MM$ в множество векторных полей
$\CX\big(h(\MM)\big)$ на образе $h(\MM)\subset\MN$, который может не совпадать
со всем $\MN$.
\begin{prop}
Дифференциал отображения согласован со структурой алгебры Ли в пространствах
векторных полей, т.е.\
\begin{equation}                                                  \label{ecomhy}
  h_*[X,Y]=[h_*X,h_*Y].
\end{equation}
\end{prop}
\begin{proof}
Простая проверка.
\end{proof}
Рассмотрим два отображения $\MM\xrightarrow h\MN\xrightarrow g\MP$. Если
обозначить координаты на многообразиях $\MM$, $\MN$ и $\MP$ соответственно через
$x$, $y$ и $z$, то по правилу дифференцирования сложной функции в
соответствующих областях определения справедливо равенство
\begin{equation*}
  \frac{\pl z}{\pl x}=\frac{\pl z}{\pl y}\frac{\pl y}{\pl x}.
\end{equation*}
Здесь, для краткости, мы опустили индексы. Отсюда следует, что дифференциал
произведения равен произведению дифференциалов каждого отображения
\begin{equation}                                                  \label{emumap}
  (g\circ h)_*=g_*\circ h_*.
\end{equation}
В координатах мы имеем обычное произведение матриц Якоби. Это уравнение говорит
о том, что прямое отображение $g\circ h:~\MM\rightarrow \MP$ не зависит от
выбора промежуточного многообразия $\MN$: $\MM\rightarrow\MN\rightarrow\MP$,
т.е.\ диаграмма
\begin{equation*}
\begin{diagram}
  \MT_x(\MM) & \rTo^{h_*} & \MT_{h(x)}(\MN) \\
   & \rdTo_{g_*\circ h_*} & \dTo_{g_*} \\
   &  & \MT_{g\big(h(x)\big)}(\MP)
\end{diagram}
\end{equation*}
коммутативна для всех точек $x\in\MM$.
\begin{com}
Поскольку сами многообразия в общем случае не являются векторными
пространствами, то дифференциал отображения не имеет смысла производной по Фреше
отображения $h$.
\qed\end{com}

Дифференциал отображения естественным образом обобщается на случай произвольных
тензорных полей типа $(r,0)$, имеющих только контравариантные индексы, и
обозначается $(h_*)^r$. При этом каждый контравариантный индекс суммируется с
матрицей Якоби отображения.
\begin{exa}
Для компонент контравариантных тензоров второго ранга имеем следующий закон
преобразования
\begin{equation*}
  Y^{\mu\nu}=X^{\al\bt}\frac{\pl y^\mu}{\pl x^\al}\frac{\pl y^\nu}{\pl x^\bt},
\end{equation*}
где $X=X^{\al\bt}(x)\pl_\al\otimes\pl_\bt\in\CT^2(\MM)$ и
$Y=Y^{\mu\nu}(y)\pl_\mu\otimes\pl_\nu\in\CT^2(\MN)$
\qed\end{exa}
\begin{com}
Если векторное поле рассматривается, как дифференцирование в алгебре функций,
то дифференциал отображения определяется следующей формулой
\begin{equation}                                                  \label{edifde}
  (h_*X)f=X(f\circ h),\qquad \forall~ f\in\CC^1(\MN),~X\in\CX(\MM).
\end{equation}
При таком определении дифференциала отображения, его линейность очевидна.
Это еще раз говорит о том, что алгебраические определения удобнее в тех случаях,
когда исследуются алгебраические свойства.
\qed\end{com}
\begin{defn}
Инъективное отображение многообразий $h:~\MM\rightarrow\MN$ индуцирует
{\em возврат отображения} в кокасательных пространствах, который мы обозначим
тем же символом, но со звездочкой вверху:
\begin{equation}                                                  \label{epulba}
  h^*:\quad \MT^*_{h(x)}(\MN)\ni\quad B\mapsto A=h^*B\quad\in\MT^*_x(\MM).
\end{equation}
Возврат отображения $h^*$ дуален к дифференциалу отображения $h_*$ и
определяется следующим равенством
\begin{equation}                                                  \label{epubde}
  A(X)=B(Y)\qquad \text{или}\qquad h^*B(X)=B(h_*X),
\end{equation}
где $Y=h_*X$.
\qed\end{defn}
\index{Возврат отображения (pull back map)}%
\index{Отображения возврат (pull back map)}%
\begin{com}
Возврат отображения действует в сторону, обратную самому отображению $h$.
\qed\end{com}

При определении возврата отображения $h^*$ мы требуем, чтобы исходное
отображение $h$ было инъективным. В противном случае прообраз
$h^{-1}\big(h(x)\big)$ для некоторого $x\in\MM$ состоит не из одного элемента, и
возврат отображения не определен.
Возврат отображения определен не на всех формах из $\CT_1(\MN)$, а только
на формах из образа инъективного отображения $\CT_1\big(h(\MM)\big)$. Возврат
отображения $h^*$ будет определен на множестве всех форм $\CT_1(\MN)$ тогда и
только тогда, когда отображение $h$ биективно.

В компонентах возврат отображения записывается в виде
\begin{equation}                                                  \label{epubco}
  A_\al=\frac{\pl y^\mu}{\pl x^\al}B_\mu,
\end{equation}
т.е.\, так же, как и дифференциал отображения, определяется матрицей Якоби.

Возврат отображения естественным образом обобщается на тензорные поля из
$\CT_s(\MN)$ типа $(0,s)$, имеющие только ковариантные индексы, и обозначается
$(h^*)^s$.
\begin{exa}
Для компонент ковариантных тензоров второго ранга имеем следующий закон
преобразования
\begin{equation*}
  X_{\al\bt}=\frac{\pl y^\mu}{\pl x^\al}\frac{\pl y^\nu}{\pl x^\bt}Y_{\mu\nu},
\end{equation*}
где $X=X_{\al\bt}(x)dx^\al\otimes dx^\bt\in\CT_2(\MM)$ и
$Y=Y_{\mu\nu}(y)dy^\mu\otimes dy^\nu\in\CT_2(\MN)$
\qed\end{exa}

Из правила дифференцирования сложных функций следует, что возврат произведения
отображений равен произведению возвратов отображений, но взятых в обратном
порядке
\begin{equation}                                                  \label{qpuiiy}
  (g\circ f)^*=f^*\circ g^*.
\end{equation}

Для тензоров смешанного типа в общем случае не существует индуцированного
отображения, поскольку дифференциал и возврат отображения действуют в разные
стороны.
\begin{defn}
Назовем {\em рангом отображения} $h:~\MM\rightarrow\MN$ ранг соответствующей
матрицы Якоби
\begin{equation*}
  \rank h:=\rank\frac{\pl y^\mu}{\pl x^\al}.
\end{equation*}
Если $\rank h=m$ во всех точках многообразия $\MM$, то отображение $h$
называется {\em невырожденным}. Для этого необходимо, чтобы $m\le n$.

Рассмотрим отображение (\ref{emamap}) двух многообразий одинаковой размерности
$\dim\MM$$=\dim\MN$ и одного класса дифференцируемости $\CC^k$. Если отображение
$h$ биективно, и оба отображения $h$ и $h^{-1}$ в координатах задаются функциями
класса $\CC^k$, то отображение $h$ называется {\em диффеоморфизмом}.
\qed\end{defn}
\index{Ранг отображения (rank of map)}%
\index{Невырожденное отображение (nondegenerate map)}%
\index{Отображение невырожденное (nondegenerate map)}%
\index{Диффеоморфизм (diffeomorphism)}%
\begin{exa}
Пусть $(\MU,\vf)$ карта на дифференцируемом многообразии $\MM$. Тогда
отображение $\vf:~\MU\rightarrow\vf(\MU)\subset\MR^n$ является диффеоморфизмом.
\qed\end{exa}
\begin{exa}
Преобразование координат $\MU_0\rightarrow\MU_0'$ в теореме \ref{tcoord}
является диффеоморфизмом.
\qed\end{exa}
Композиция двух диффеоморфизмов снова будет диффеоморфизмом. Таким образом,
диффеоморфизм является отношением эквивалентности в категории дифференцируемых
многообразий. С точки зрения дифференциальной геометрии два диффеоморфных между
собой многообразия можно рассматривать, как одно многообразие, заданное в
различных координатах, поэтому мы пишем $\MM\approx\MN$\footnote{Мы не
используем обычный знак равенства, т.к.\ множества точек $\MM$ и $\MN$ могут
отличаться по другим признакам, например, по наличию групповой структуры.}.
В этом случае индуцированные отображения касательных пространств представляют
собой не что иное, как правила преобразования тензорных полей при преобразовании
координат.
\begin{com}
Не имеет смысла рассматривать степень гладкости отображения, которая превышает
гладкость самих многообразий.
\qed\end{com}
\begin{com}
Любой диффеоморфизм представляет собой гомеоморфизм многообразий,
рассматриваемых, как топологические пространства, т.к.\ функции $h$ и $h^{-1}$
непрерывны. Обратное утверждение неверно. Как отмечено в разделе \ref{sdefma},
на семимерной сфере можно задать несколько различных дифференцируемых структур.
То есть гомеоморфные многообразия могут быть недиффеоморфны.
\qed\end{com}

\begin{figure}[h,b,t]
\hfill\includegraphics[width=.5\textwidth]{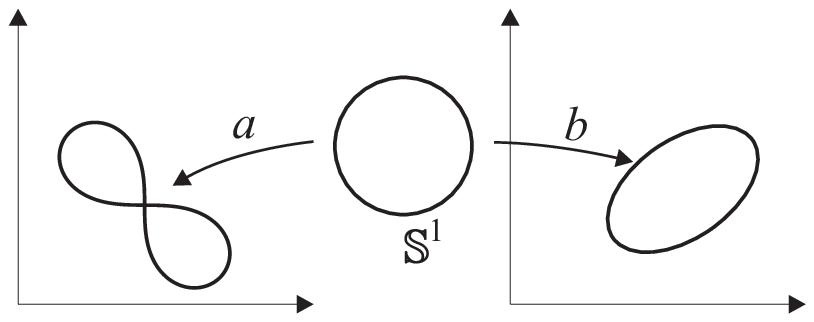}
\hfill {}
\centering \caption{Отображение окружности на плоскость в виде восьмерки
является погружением, но не вложением, {\it a}. Вложение окружности в плоскость
в виде гладкой замкнутой кривой без самопересечений,{\it b}.}
\label{fcirem}
\end{figure}

Множество невырожденных гладких отображений $h:~\MM\rightarrow\MM$ многообразия
$\MM$ в себя образует группу преобразований многообразия, которая обозначается
$\diff(\MM)$. Если многообразие тривиально (т.е.\ покрывается одной картой), то
это просто группа преобразований координат $\diff(\MR^n)$.
\begin{defn}
Пусть для отображения (\ref{emamap}) $m<n$. Если отображение касательного
пространства $\MT_x(\MM)$ для всех точек $x\in\MM$ на его образ в касательном
пространстве $\MT_{h(x)}(\MN)$ является взаимно однозначным, т.е.\
$\rank h=\dim\MM$, то отображение $h$ называется  {\em погружением}.
По-определению, любое невырожденное отображение задает погружение. При
погружении само отображение $h$ может не быть взаимно однозначным. Размерность
многообразия $\MM$ не может превосходить размерности многообразия $\MN$, т.к.\
в этом случае не может быть взаимной однозначности дифференциала отображения.
Если само отображение $h$ на его образ $h(\MM)$ и его дифференциал являются
взаимно однозначными, то отображение $h$ называется {\em вложением}. В
дальнейшем вложение многообразий мы будем обозначать специальным символом:
$\MM\hookrightarrow\MN$.
\qed\end{defn}
\index{Погружение многообразия (immersion of a manifold)}%
\index{Многообразие погружение (immersion of a manifold)}%
\index{Вложение многообразий (imbedding of a manifold)}%
\index{Многообразие вложение (imbedding of a manifold)}%
Конечно, каждое вложение одновременно является и погружением.
\begin{exa}
Отображение окружности $\MS^1$ на плоскость $\MR^2$ в виде восьмерки является
погружением, но не вложением, см.\ рис.~\ref{fcirem},{\it a}. В то же время
отображение окружности в гладкую замкнутую кривую без самопересечений
представляет собой вложение, рис.~\ref{fcirem},{\it b}.
\qed\end{exa}
Если $m=n$, отображение $h$ биективно и дифференцируемо, а образ $h(\MM)$
совпадает со всем $\MN$, то точки многообразий $\MM$ и $\MN$ можно отождествить
и рассматривать вложение $\MM\overset{h}{\hookrightarrow}\MN\approx\MM$ как
диффеоморфизм. Нетривиальная ситуация может возникнуть, если многообразие $\MM$
отображается только на часть $\MN$. Тогда мы имеем диффеоморфизм между
многообразием $\MM$ и его образом $h(\MM)$.
\section{Подмногообразия                                         \label{subman}}
Важным классом отображений многообразий являются подмногообразия.
\begin{defn}
Пусть $f:~\MM\hookrightarrow\MN$ -- вложение многообразия $\MM$ в $\MN$,
размерностей $m$ и $n$, при этом $m\le n$, тогда пара $(f,\MM)$ называется {\em
подмногообразием} в $\MN$.
\end{defn}
\index{Подмногообразие (submanifold)}%
В дальнейшем, если не оговорено противное, под вложением мы понимаем гладкое
вложение, когда отображение $f$ определяется гладкими функциями.
\begin{exa}[\bf Лемниската Бернулли]
Кривая четвертого порядка, заданная в декартовых координатах $x,y$ на евклидовой
плоскости $\MR^2$ уравнением
\begin{equation}                                                  \label{eledef}
  (x^2+y^2)^2-2a^2(x^2-y^2)=0,\qquad a\in\MR,
\end{equation}
называется {\em лемнискатой Бернулли}, рис.\ref{flemniscata}. Эта кривая
обладает следующим свойством. Произведение расстояний от произвольной точки
лемнискаты до двух заданных точек (фокусов) постоянно и равно квадрату половины
расстояния между фокусами. Верно и обратное утверждение. На плоскости можно
выбрать такую систему декартовых координат, что произвольная кривая с данным
свойством будет задана уравнением (\ref{eledef}).
\index{Лемниската Бернулли (Bernoulli lemniscate)}%
\index{Бернулли лемниската (Bernoulli lemniscate)}%

В полярных координатах $r,\vf$ уравнение лемнискаты имеет вид:
\begin{equation*}
  r^2=2a^2\cos(2\vf).
\end{equation*}
Лемниската Бернулли не является одномерным многообразием, т.к.\ имеет точку
самопересечения -- начало координат. Точка самопересечения имеет касательные
$y=\pm  x$ и является точкой перегиба. Площадь каждой петли равна $a^2$.
\begin{figure}[h,b,t]
\hfill\includegraphics[width=.3\textwidth]{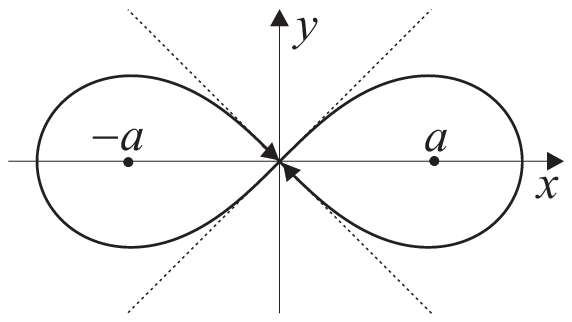}
\hfill {}
\centering\caption{Лемниската Бернулли.}
\label{flemniscata}
\end{figure}

Рассмотрим вложение прямой в евклидову плоскость
\begin{equation*}
  \g:~\MR\ni\quad t\mapsto(x,y)\quad\in\MR^2,
\end{equation*}
гладко отобразив прямую в точки лемнискаты, как показано на
рис.\ref{flemniscata}. При этом
мы считаем, что точка $t=0$ отображается в начало координат $(0,0)\in\MR^2$.
Стрелки показывают, что концы прямой $t=\pm\infty$ при вложении стремятся к
началу координат $(0,0)\in\MR^2$. Это -- гладкое вложение. Однако топология
прямой не совпадает с топологией, которая индуцирована вложением. Действительно,
любая последовательность $\lbrace t_k\rbrace$ такая, что
$\lim_{k\to\infty}t_k=\pm\infty$ в естественной топологии на прямой, будет
сходится к точке $t=0$ в индуцированной топологии. Конечно, вместо лемнискаты
Бернулли можно было бы выбрать произвольную гладкую кривую в виде восьмерки.
\qed\end{exa}
\begin{exa}[\bf Всюду плотная обмотка тора]                       \label{edetor}
Двумерный тор $\MT^2=\MS^1\times\MS^1$ можно рассматривать, как двумерное
многообразие, полученное путем отождествления противоположных сторон
единичного квадрата на евклидовой плоскости. Тем самым точка тора задается
упорядоченной парой чисел $(x,y)$, каждое из которых определено по модулю один,
$x\sim x+1$ и $y\sim y+1$. Зафиксируем пару чисел $a$ и $b\ne0$ и рассмотрим
отображение вещественной прямой в тор
\begin{equation*}
  f:\quad \MR\ni\quad t\mapsto (at\mod 1,bt\mod1)\quad\in\MT^2.
\end{equation*}
Если отношение $a/b$ иррационально, то отображение $f$ является вложением,
причем образ прямой $f(\MR)$ всюду плотен в $\MT^2$ ({\em всюду плотная
обмотка тора}).
\index{Всюду плотная обмотка тора (everywhere dense winding of torus)}%
\index{Обмотка тора всюду плотная (everywhere dense winding of torus)}%
Если отношение $a/b$ рационально, то $f$ представляет собой погружение,
и его образ диффеоморфен окружности. Согласно данному выше определению всюду
плотная обмотка тора является одномерным подмногообразием на торе. Для всюду
плотной обмотки тора индуцированная топология на образе прямой не совпадает с
естественной топологией на $\MR$.
\qed\end{exa}
При вложении дифференцируемая структура и топология, заданные на $\MM$,
естественным образом переносятся на образ $f(\MM)$, поскольку отображение
$f$ инъективно и дифференцируемо. Тем самым многообразие $\MM$ диффеоморфно
своему образу $f(\MM)$. С другой стороны, поскольку образ $f(\MM)$ является
некоторым подмножеством в пространстве-мишени $\MN$, то на нем определяется
индуцированная из $\MN$ топология. Рассмотренные выше примеры вложения прямой в
плоскость и всюду плотной обмотки тора показывает, что топология на $f(\MM)$,
наследуемая из $\MM$ при вложении, совсем не обязательно совпадает с топологией,
индуцированной из $\MN$. В общем случае топология из $\MM$ является более
тонкой, чем топология, индуцированная из $\MN$. Это наблюдение мотивирует
следующее
\begin{defn}
Пусть вложение $f:~\MM\hookrightarrow\MN$ является также и гомеоморфизмом. Тогда
пара $(f,\MM)$ называется {\em регулярным подмногообразием}, а $f$ --
{\em регулярным вложением} $\MM$ в $\MN$.
\qed\end{defn}
\index{Регулярное подмногообразие (regular submanifold)}%
\index{Подмногообразие регулярное (regular submanifold)}%
\index{Регулярное вложение (regular embedding)}%
\index{Вложение регулярное (regular embedding)}%
Доказательство следующих трех теорем приведено в \cite{ChChLa00}.
\begin{theorem}                                                  \label{tresub}
Пусть $(f,\MM)$ является $m$-мерным подмногообразием в $n$-мерном многообразии
$\MN$, причем $m<n$. Для того, чтобы пара $(f,\MM)$ была регулярным
подмногообразием в $\MN$, необходимо и достаточно, чтобы оно было замкнутым
подмногообразием некоторого открытого подмногообразия в $\MN$.
\end{theorem}
\begin{theorem}                                                   \label{tregvl}
Пара $(f,\MM)$ является регулярным подмногообразием в $\MN$ тогда и только
тогда, когда для каждой точки $x\in\MM$ существует система координат $y^\al$
на $\MU_{f(x)}\subset\MN$, $y^\al\big(f(x)\big)=0$, такая, что пересечение
$f(\MM)\cap\MU_{f(x)}$ определяется системой, состоящей из $n-m$ уравнений:
\begin{equation*}
  y^{m+1}=y^{m+2}=\dotsc=y^n=0.
\end{equation*}
\end{theorem}
\begin{theorem}
Пусть пара $(f,\MM)$ является подмногообразием в $\MN$. Если многообразие
$\MM$ компактно, тогда вложение $f:~\MM\rightarrow\MN$ регулярно.
\end{theorem}

В дальнейшем под вложением мы будем понимать регулярное вложение, для
краткости не оговаривая этого явно. В этом случае мы будем отождествлять
подмногообразие $\MM$ и его образ $f(\MM)$ и будем писать $\MM\subset\MN$.
\begin{defn}
Пусть $\MU$ является открытым подмножеством $\MN$. Если ограничить гладкую
структуру $\MN$ на $\MU$, то $\MU$ становится многообразием той же размерности
$\dim\MU=\dim\MN$. Тогда вложение $f:~\MU\hookrightarrow\MN$ осуществляется
тождественным отображением, а подмногообразие $\MU\subset\MN$ называется
{\em открытым подмногообразием} в $\MN$.

Пусть $f(\MM)$ является подмногообразием в $\MN$ таким, что выполнены два
условия:\newline
\indent 1) $f(M)$ -- замкнутое подмножество в $\MN$;\newline
\indent 2) \parbox[t]{.93\linewidth}{для каждой точки $y\in f(\MM)$ существует
такая система координат $y^\al$ на $\MU_y\subset\MN$, что пересечение
$f(\MM)\cap\MU_y$ задается уравнениями $y^{m+1}=y^{m+2}=\dotsc=y^n=0$,
где $m=\dim\MM$.}\newline
Тогда подмногообразие $f(\MM)$ называется {\em замкнутым подмногообразием}
в $\MN$. Размерность замкнутого подмногообразия $f(\MM)$ всегда меньше
размерности $\MN$.
\qed\end{defn}
\index{Открытое подмногообразие (open submanifold)}%
\index{Подмногообразие открытое (open submanifold)}%
\index{Замкнутое подмногообразие (closed submanifold)}%
\index{Подмногообразие замкнутое (closed submanifold)}%
\begin{com}
Если подмногообразие (без края) $\MM\subset\MN$ имеет ту же размерность, что и
$\MN$, то оно будет обязательно открытым подмногообразием, т.к.\ в противном
случае оно имело бы край. Теорема \ref{tregvl} утверждает, что любое
подмногообразие $\MM\subset\MN$ меньшей размерности $1<m<n$ является замкнутым
подмногообразием.
\qed\end{com}

Мы видим, что подмногообразия меньшей размерности $1<m<n$ в евклидовом
пространстве $\MM\subset\MR^n$ можно задавать с помощью системы алгебраических
уравнений на координаты, покрывающие все $\MR^n$. Вернее, алгебраические
уравнения задают только множества точек. Чтобы превратить это множество в
многообразие мы предполагаем, что топология на $\MM$ задается вложением
$f:~\MM\hookrightarrow\MR^n$. Тогда подмногообразие определяется системой
уравнений по крайней мере локально. Обратное утверждение неверно. Не каждая
система алгебраических уравнений на координаты в евклидовом пространстве $\MR^n$
определяет некоторое подмногообразие.
\begin{exa}
Рассмотрим уравнение $x^2=y^2$ на евклидовой плоскости $\MR^2$. Оно определяет
две пересекающиеся прямые, которые не являются многообразием.
\qed\end{exa}

Приведем критерий, который часто используется в приложениях.
\begin{theorem}                                                   \label{tsuman}
Для того, чтобы подмножество $\MM\subset\MR^n$, $\dim\MM=m\le n$,
было подмногообразием класса $\CC^k$, необходимо и достаточно, чтобы для
каждой точки $x\in\MM$ существовала окрестность $\MU$ этой точки и система
$n-m$ функций $f^\Sm(x)\in\CC^k(\MU)$, $\Sm=m+1,\dots,n$,
определенных на $\MU$, и обладающих следующими свойствами:

1) функции $f^\Sm$ -- функционально независимы;

2) пересечение $\MU\bigcap\MM$ в точности определяется системой уравнений
\begin{equation}                                                  \label{esubma}
  f^\Sm=0,\qquad \Sm=m+1,\dotsc,n.
\end{equation}
\end{theorem}
\begin{proof}
См., например, \cite{Schwar67R}.
\end{proof}
Заметим, что если $m=n$, то уравнений не будет, а $\MM$ будет некоторым открытым
подмножеством в $\MM$. При $m=0$ число уравнений равно размерности многообразия.
Тогда, если в точке $x\in\MR^n$ система уравнений (\ref{esubma}) удовлетворена,
то в некоторой окрестности точки $x$ других решений кроме самой точки не будет.
В этом случае подмногообразие $\MM$ имеет размерность $0$ и представляет собой
множество корней системы уравнений $f^\Sm=0$.
\begin{defn}
Подмногообразие $(f,\MM)$ размерности $n-1$ в $\MN$ называется
{\em гиперповерхностью} в $\MN$.
\qed\end{defn}
\index{Гиперповерхность (hypersurface)}%

Напомним, что функции $f^\Sm$ функционально независимы в области $\MU$ тогда и
только тогда, когда $n\times m$ матрица, составленная из частных производных
\begin{equation}                                                  \label{efumat}
  (\pl_\al f^\Sm)=\left(\begin{array}{cccc}
  \pl_1f^{m+1} & \pl_1f^{m+2} & \dotsc & \pl_1f^n \\
  \pl_2f^{m+1} & \pl_2f^{m+2} & \dotsc & \pl_2f^n \\
  \vdots   & \vdots   & \vdots & \vdots   \\
  \pl_nf^{m+1} & \pl_nf^{m+2} & \dotsc & \pl_nf^n
  \end{array}\right),
\end{equation}
имеет постоянный ранг $m$ во всех точках $\MU$.
\begin{com}
Забегая вперед, переформулируем условие 1) в теореме \ref{tsuman} в терминах
1-форм. По-определению, 1-формы $df^\Sm=dx^\al\pl_\al f^\Sm$, заданные на
области $\MU$, называются {\em линейно независимыми}  в области $\MU$,
\index{Линейно независимые $1$-формы (linearly independent $1$-forms)}%
\index{$1$-формы линейно независимые (linearly independent $1$-forms)}%
если в каждой точке этой области ранг матрицы (\ref{efumat}) равен $m$.
Поэтому условие функциональной независимости функций можно
заменить условием линейной независимости 1-форм $df^\Sm$.
\qed\end{com}
Если подмногообразие $\MM\subset\MR^n$ задано с помощью $n-m$ алгебраических
уравнений (\ref{esubma}), то множество функций $f^\Sm$, в окрестности
произвольной точки $x\in\MM$ всегда можно дополнить $m$ функциями
$g^1(x),\dotsc$, $g^m(x)$ такими, что весь набор функций
$g^1,\dotsc,g^m,f^{m+1},\dotsc,f^n$ будет функционально независим.
В этом случае можно перейти в новую систему координат
\begin{equation}                                                  \label{enemco}
\begin{split}
  \lbrace y^\al\rbrace&=\lbrace y^\Sa,y^\Sm\rbrace
  =\lbrace g^\Sa,f^\Sm\rbrace=(g^1,\dotsc,g^m,f^{m+1},\dotsc,f^n),
\\
  &\Sa=1,\dotsc,m,\quad \Sm=m+1,\dotsc,n,
\end{split}
\end{equation}
в которой подмногообразие $\MM$ будет задаваться особенно просто
$y^\Sm:=f^\Sm=0$. При этом функции $y^\Sa=g^\Sa$ образуют систему координат на
подмногообразии $\MM\subset\MR^n$. Конечно, выбор координат $g^\Sa$ на
$\MM$ является неоднозначным. Отметим, что $\lbrace y^\al\rbrace$ -- это те
координаты, которые фигурируют в теореме \ref{tregvl}.
\begin{exa}
Двумерная сфера $\MS^2_r$ радиуса $r$ с центром в начале координат является
замкнутым подмногообразием трехмерного евклидова пространства $\MR^3$, которое
задается одним уравнением $x^2+y^2+z^2=r^2$. Она является гиперповерхностью в
$\MR^3$. В сферической системе координат уравнение сферы задается уравнением
$r=\const$, а угловые координаты образуют систему координат на сфере.
\qed\end{exa}

Функции $f^\Sm(x)$ определяют $n-m$ точных 1-форм. В силу их линейной
независимости, в каждой точке $x\in\MM$ они определяют $(n-m)$-мерное
подпространство в кокасательном пространстве $\MT^*_x(\MR^n)$. Ортогональное
дополнение этого подпространства является касательным к подмногообразию $\MM$ и
образовано всеми векторами $X^\al\pl_\al\in\MT_x(\MR^n)$, для которых выполнено
условие $X^\al\pl_\al f^\Sm=0$. Это подпространство касательного пространства
имеет размерность $m$, такую же, как и само подмногообразие. Нетрудно проверить,
что, если два векторных поля $X_1$ и $X_2$ касательны к подмногообразию $\MM$,
то их коммутатор $[X_1,X_2]$ также касателен.

Ответ на обратный вопрос, в каком случае распределение векторных полей на
многообразии определяет касательные подпространства к некоторому
подмногообразию, дает теорема Фробениуса (см.\ следующий раздел).

В координатах $\lbrace g^\Sa,f^\Sm\rbrace$ дополнение касательного пространства
к подмногообразию $\MM$ до полного касательного пространства к $\MR^n$ имеет
простой геометрический смысл. Пусть в $\MR^n$ задано векторное поле
$X=X^\al\pl_\al$. Тогда в координатах (\ref{enemco}), связанных с
подмногообразием, оно будет иметь компоненты
\begin{equation*}
  X=(X^\al\pl_\al g^\Sa)\pl_\Sa+(X^\al\pl_\al f^\Sm)\pl_\Sm.
\end{equation*}
При этом первое слагаемое лежит в касательном пространстве $\MT(\MM)$.

В дальнейшем нам понадобится следующее
\begin{defn}
Пусть $f\in\CC^\infty(\MN)$ и $\MM\subset\MN$ -- подмногообразие, тогда
отображение
\begin{equation*}
  f|_\MM:\quad \MM\rightarrow\MR.
\end{equation*}
называется {\em сужением} функции на подмногообразие. Функция $f$ является
{\em продолжением} функции $g$ с некоторого подмногообразия $\MM$ на все $\MN$,
если ее сужение на $\MM$ совпадает с $g$, $f|_\MM=g$. Аналогично, сужением
произвольного тензорного поля $T\in\CT(\MN)$ на подмногообразие называется
тензорное поле $T|_\MM$, рассматриваемое только в точках $x\in\MM$. Обратно,
тензорное поле $T\in\CT(\MN)$ является продолжением тензорного поля
$K\in\CT(\MN)$, заданного на подмногообразии $\MM$, если его сужение совпадает
с $K$, $T|_\MM=K$.
\qed\end{defn}
\index{Сужение функции (restriction of a function)}%
\index{Продолжение функции (continuation of a function)}%
\begin{defn}
Два подмногообразия $\MM_1$ и $\MM_2$ многообразия $\MN$ называются
{\em трансверсальными} в точке $x\in\MM_1\cap\MM_2$, если касательные
подпространства $\MT_x(\MM_1)$ и $\MT_x(\MM_2)$ порождают все касательное
пространство $\MT_x(\MN)$, т.е.\ касательные подпространства $\MT_x(\MM_1)$ и
$\MT_x(\MM_2)$ трансверсальны.
\qed\end{defn}
\index{Трансверсальные подмногообразия (transversal submanifolds)}%
\index{Подмногообразия трансверсальные (transversal submanifolds)}%
Трансверсальность подмногообразий означает, что в некоторой окрестности $\MU$
точки $x\in\MM_1\cap\MM_2$ существует такая система координат
$x^\al$, $\al=1,\dotsc,n$, что подмногообразия задаются условиями:
\begin{align*}
  \MM_1\cap\MU&=\lbrace x\in\MU:\quad x^{\dim\MM_1+1}=0,\dotsc,x^n=0\rbrace,
\\
  \MM_2\cap\MU&=\lbrace x\in\MU:\quad x^1=0,\dotsc,x^{n-\dim\MM_2}=0\rbrace.
\end{align*}
\begin{exa}
Пусть $\MN:=\MM_1\times\MM_2$. Тогда подмногообразия $(\MM_1,x_2)\subset\MN$,
где $x_2\in\MM_2$, и $(x_1,\MM_2)\subset\MN$, где $x_1\in\MM_1$, трансверсальны
в точке $x=(x_1,x_2)$. Если в окрестностях $\MU_1\subset\MM_1$ и
$\MU_2\subset\MM_2$, где $x_1\in\MU_1$ и $x_2\in\MU_2$, заданы системы
координат, $x_1=\lbrace x_1^\mu\rbrace$ и $x_2=\lbrace x_2^\Sa\rbrace$, то
подмногообразия задаются уравнениями
\begin{align*}
  (\MU_1,x_2)&=\lbrace x\in\MU_1\times\MU_2:\quad x_2^\Sa=\const,~\forall\Sa\rbrace,
\\
  (x_1,\MU_2)&=\lbrace x\in\MU_1\times\MU_2:\quad x^\mu_1=\const,~\forall\mu\rbrace.
\end{align*}
Конечно, системы координат можно выбрать таким образом, что точки $x_1$ и $x_2$
будут находиться в началах систем отсчета.
\qed\end{exa}
\begin{defn}
Отображение $f:~\MM\rightarrow\MN$ {\em трансверсально} к подмногообразию
$\ML\subset\MN$ в точке $x\in f^{-1}\big(\ML\cap f(\MM)\big)$, если
образ$f_*\big(\MT_x(\MM)\big)$ трансверсален к $\MT_{f(x)}(\ML)$. Два
отображения $f_1:~\MM_1\rightarrow\MN$ и $f_2:~\MM_2\rightarrow\MN$
{\em трансверсальны друг к другу} в точке $(x_1,x_2)\in\MM_1\times\MM_2$, где
$f_1(x_1)=f_2(x_2)$, если образы $f_{1*}\big(\MT_{x_1}(\MM_1)\big)$ и
$f_{2*}\big(\MT_{x_2}(\MM_2)\big)$ порождают все касательное пространство
$\MT_{f_1(x_1)}(\MN)$.
\qed\end{defn}
\index{Трансверсальное отображение (transversal map)}%
\index{Отображение трансверсальное (transversal map)}%
Понятия трансверсальных подмногообразий и трансверсальных отображений
естественным образом сводятся друг к другу. Если
$f_{1,2}:~\MM_{1,2}\hookrightarrow\MN$ -- вложения, то трансверсальность
вложений эквивалентна трансверсальности подмногообразий $f_1(\MM_1)\subset\MN$ и
$f_2(\MM_2)\subset\MN$.
\section{Теорема Фробениуса                                      \label{sfrote}}
В разделе \ref{svechs} было показано, что у любого дифференцируемого векторного
поля $X$ существуют интегральные кривые. Если векторное поле нигде не обращается
в нуль на многообразии $\MM$, $\dim\MM=n$, то интегральная кривая, проходящая
через некоторую точку $x\in\MM$, представляет собой одномерное подмногообразие,
и в окрестности точки $x$ существует такая система координат, что $X=\pl_1$.
Поставим более общую задачу. Пусть в некоторой окрестности $\MU\subset\MM$
задано $k$ векторных полей $(X_1,\dotsc,X_k)=\lbrace X_\Sa\rbrace$,
$\Sa=1,\dotsc,k<n$, которые линейно независимы в каждой точке $x\in\MU$.
Возникает вопрос о том, существуют ли такие подмногообразия в $\MM$, касательные
пространства к которым в каждой точке $x\in\MU$ совпадают с подпространством,
натянутым на векторы $\lbrace X_\Sa\rbrace$ ? Ответ на этот вопрос дает теорема
Фробениуса. Чтобы ее сформулировать введем несколько новых понятий.
\begin{defn}
{\em Распределением} $\CL_k(\MM)$ размерности $k$ на многообразии $\MM$
называется сопоставление каждой точке $x\in\MM$ $k$-мерного подпространства в
касательном пространстве $\ML_x(\MM)\subset\MT_x(\MM)$. Распределение называется
{\em дифференцируемым}, если каждая точка $x\in\MM$ имеет окрестность $\MU$ и
$k$ дифференцируемых векторных полей $(X_1,\dotsc,X_k)$ на $\MU$, которые
образуют базис в $\ML_x(\MM)$ для всех $x\in\MU$. Множество векторных полей
$(X_1,\dotsc,X_k)$ называется {\em локальным базисом} распределения $\CL_k(\MM)$
на $\MU$. Векторное поле $X$ принадлежит распределению, если $X(x)\in\ML_x(\MM)$
для всех $x\in\MM$. Распределение называется {\em инволютивным} или {\em вполне
интегрируемым}, если для любых векторных полей из распределения
$X,Y\in\CL_k(\MM)$ их коммутатор также принадлежит распределению:
$[X,Y]\in\CL_k(\MM)$.
\qed\end{defn}
\index{Распределение (distribution)}%
\index{Базис распределения (basis of distribution)}%
\index{Инволютивное распределение (involutory distribution)}%
\index{Распределение инволютивное (involutory distribution)}%
\index{Вполне интегрируемое распределение (totally integrable distribution)}%
\index{Распределение вполне интегрируемое (totally integrable distribution)}%
\begin{com}
Ни одно из векторных полей $X_\Sa$ не может обращаться в нуль, т.к.\ в
соответствующей точке векторные поля были бы линейно зависимы.
\qed\end{com}
\begin{com}
Инволютивность векторных полей, принадлежащих распределению, означает, что
эти векторные поля образуют подалгебру Ли алгебры Ли векторных полей $\CX(\MM)$.
\qed\end{com}
\begin{defn}
Связное подмногообразие $f:~\MN\hookrightarrow\MM$ называется {\em интегральным
многообразием} распределения $\CL_k(\MM)$, если
$f_*\big(\MT_x(\MN)\big)=\ML_{f(x)}(\MM)$, для всех $x\in\MN$. Если не
существует других интегральных многообразий, содержащих $\MN$, то $\MN$
называется {\em максимальным интегральным многообразием} для распределения
$\CL_k(\MM)$.
\qed\end{defn}
\index{Максимальное интегральное многообразие (maximal integral manifold)}%
\index{Интегральное многообразие максимальное (maximal integral manifold)}%
\index{Интегральное многообразие (integral manifold)}%
\begin{com}
Поскольку интегральное многообразие -- это подмногообразие и, следовательно
задается парой $(f,\MN)$, где $f$ -- вложение $\MN$ в $\MM$, то определение
максимального интегрального многообразия требует уточнения. А именно, мы
говорим, что интегральное многообразие $(f_1,\MN_1)$ содержит интегральное
многообразие $(f_2,\MN_2)$, если $f_2(\MN_2)$ есть подмножество в $f_1(\MN_1)$:
$f_2(\MN_2)\subset f_1(\MN_1)$. Из определения максимального интегрального
многообразия сразу следует его единственность.
\qed\end{com}
\begin{prop}                                                      \label{pfrobe}
Пусть $\CL_k(\MM)$ -- гладкое распределение на многообразии $\MM$ такое, что
через каждую точку $x\in\MM$ проходит интегральное многообразие, тогда
распределение $\CL_k(\MM)$ инволютивно.
\end{prop}
\begin{proof}
Пусть $X,Y$ -- два произвольных векторных поля, принадлежащих распределению
$\CL_k(\MM)$. Пусть $(f,\MN)$ -- интегральное многообразие, проходящее через
точку $f(x)\in\MM$, $x\in\MN$. Поскольку отображение $f$ в каждой точке
интегрального многообразия $x\in\MN$ задает изоморфизм
$\MT_x(\MN)\simeq\ML_{f(x)}(\MM)$, то на подмногообразии $\MN$ существуют
векторные поля $\tilde X(x)=f_*^{-1}X\big(f(x)\big)$ и
$\tilde Y=f_*^{-1}Y\big(f(x)\big)$. Поскольку коммутатор касательных векторов к
$\MN$ также касателен к $\MN$, то
\begin{equation*}
  [X,Y]\big(f(x)\big)=f_*[\tilde X,\tilde Y](x)\in\ML_x(\MM).  \tag*{\qed}
\end{equation*}
\renewcommand{\qed}{}\end{proof}
\begin{theorem}[\bf Фробениус]                                    \label{tfrofi}
Пусть $\CL_k(\MM)$ -- $k$-мерное гладкое инволютивное распределение на
многообразии $\MM$. Тогда для каждой точки $x\in\MM$ существует интегральное
многообразие распределения $\CL_k(\MM)$, проходящее через точку $x$. Кроме того,
существует такая система координат в некоторой окрестности точки $x$, что базис
распределения имеет вид $(\pl_1,\dotsc,\pl_k)$.
\end{theorem}
\index{Теорема Фробениуса (Frobenius theorem)}%
\index{Фробениуса теорема (Frobenius theorem)}%
\begin{proof}
См., например, \cite{Warner83R}.
\end{proof}
\begin{com}
Предложение \ref{pfrobe} дает необходимое условие для существования интегральных
многообразий, а теорема \ref{tfrofi} -- достаточное. Обе теоремы локальны, т.к.\
в них говорится об интегральных многообразиях, проходящих через точку
многообразия.
\qed\end{com}

Если распределение $\CL_k(\MU)$ задано набором гладких линейно независимых
векторных полей $(X_1,\dotsc,X_k)=\lbrace X_\Sa\rbrace$,
$\Sa=1,\dotsc,k<n$, в каждой точке $x\in\MU\subset\MM$, то будем писать
$\CL_k(\MU)=\lbrace X_\Sa\rbrace$. В этом случае будем говорить, что
распределение задано {\em распределением векторных полей}.
\index{Распределение векторных полей (distribution of vector fields)}%
Тогда теорему Фробениуса вместе с предложением \ref{pfrobe} можно
переформулировать.

\begin{theorem}[\bf Фробениус]
Пусть $\CL_k(\MU)=\lbrace X_\Sa\rbrace$, $\Sa=1,\dotsc,k$, является $k$-мерным
гладким распределением векторных полей в области $\MU\subset\MM$. Тогда для
существования такой системы координат в некоторой подобласти $\MV\subset\MU$,
что распределение имеет вид
\begin{equation}                                                  \label{efrthv}
  \CL_k(\MV)=(\pl_1,\dotsc,\pl_k)
\end{equation}
необходимо и достаточно, чтобы коммутатор векторных полей также принадлежал
распределению,
\begin{equation}                                                  \label{ecovef}
  [X_\Sa,X_\Sb]=f_{\Sa\Sb}{}^\Sc X_\Sc\in\ML_x(\MM),\qquad \forall x\in\MU,
\end{equation}
где $f_{\Sa\Sb}{}^\Sc(x)$ -- некоторые гладкие функции на $\MU$.
\end{theorem}
\begin{proof}
См., например, \cite{ChChLa00}.
\end{proof}
\begin{com}
Конечно, функции $f_{\Sa\Sb}{}^\Sc$ в теореме Фробениуса не могут быть
произвольными. Из определения (\ref{ecovef}) следует, что они антисимметричны
по нижним индексам и удовлетворяют тождествам, вытекающим из тождеств Якоби для
коммутатора векторных полей.
\qed\end{com}

Допустим, что на многообразии $\MM$ существует гладкое распределение
$\CL_k(\MM)$. Тогда существование системы координат (\ref{efrthv}) в окрестности
каждой точки $x\in\MM$ означает следующее. Уравнения
\begin{equation*}
  x^\al=\const,\qquad \al=k+1,\dotsc,n
\end{equation*}
определяют $k$-мерные подмногообразия, которые являются интегральными
многообразиями распределения $\CL_k(\MM)$, причем каждое векторное поле $X_\Sa$
касается одного из подмногообразий. Интегральные многообразия не пересекаются,
т.к.\ в точке пересечения у нас не было бы подмногообразия. Кроме того, через
каждую точку $\MM$ проходит одно подмногообразие. Тем самым $k$-мерное
распределение $\CL_k(\MM)$ расслаивает область $\MV\subset\MM$, представляя ее в
виде объединения несчетного числа $k$-мерных подмногообразий. Очевидно, что
интегральная кривая для любого векторного поля из заданного инволютивного
распределения, проходящая через точку $x\in\MM$ целиком лежит в интегральном
подмногообразии, содержащем эту точку.
\begin{exa}
Произвольное дифференцируемое векторное поле $X\in\CX(\MM)$, не равное нулю ни в
одной точке, задает одномерное распределение на $\MM$. Это распределение
инволютивно. Интегральными многообразиями одномерного распределения являются
интегральные кривые векторного поля $X$.
\qed\end{exa}
\begin{exa}
Рассмотрим трехмерное евклидово пространство с выколотым началом
$\MR^3\setminus\lbrace0\rbrace$ в сферической системе координат $r,\theta,\vf$.
Отличные от нуля векторные поля $\pl_\theta$ и $\pl_\vf$ коммутируют и, значит,
находятся в инволюции. Интегральными подмногообразиями этого распределения
являются сферы $r=\const$ с центром в начале координат. Угловые координаты
$\theta,\vf$ и есть те координаты, которые фигурируют в теореме Фробениуса.

Векторные поля
\begin{equation}                                                  \label{egenro}
  J_1=\frac12(y\pl_z-z\pl_y),\qquad J_2=\frac12(z\pl_x-x\pl_z),\qquad
  J_3=\frac12(x\pl_y-y\pl_x),
\end{equation}
заданные в декартовых координатах, являются генераторами алгебры $\Gs\Go(3)$
и находятся в инволюции. Интегральными подмногообразиями для них также
являются сферы. Однако поля (\ref{egenro}) не задают базис распределения, т.к.\
линейно зависимы:
\begin{equation*} \tag*{\qed}
  xJ_1+yJ_2+zJ_3=0.
\end{equation*}
\renewcommand{\qed}{}\end{exa}

Сформулируем глобальный вариант теоремы теоремы Фробениуса.
\begin{theorem}[\bf Фробениус]
Пусть $\CL_k(\MM)$ -- гладкое инволютивное $k$-мерное распределение. Тогда
через каждую точку $x\in\MM$ проходит единственное максимальное $k$-мерное
интегральное многообразие $f:~\MN\hookrightarrow \MM$ распределения
$\CL_k(\MM)$. Любое другое связное интегральное многообразие этого
распределения, проходящее через точку $x$, содержится в $(f,\MN)$.
\end{theorem}
\begin{proof}
См., например, \cite{Cheval46R}).
\end{proof}
\begin{com}
В общем случае интегральное многообразие распределения $\CL_k(\MM)$ может не
быть регулярным подмногообразием размерности $k$ в $\MM$. Примером является
иррациональная обмотка тора в примере \ref{edetor}.
\qed\end{com}

Эта теорема означает, что, если на всем многообразии $\MM$ существует гладкое
инволютивное $k$-мерное распределение, то оно представляет собой слоение, т.е.\
объединение несчетного числа $k$-мерных максимальных интегральных многообразий
(листов) данного распределения (см.\ следующий раздел).

Переформулируем теорему Фробениуса на языке дифференциальных форм. Дополним поля
$\lbrace X_\Sa\rbrace=(X_1,\dotsc,X_k)$ в окрестности $\MU\subset\MM$ $n-k$
векторными полями $\lbrace X_\Sm\rbrace=(X_{k+1},\dotsc,X_n)$ так, чтобы вся
совокупность векторных полей $\lbrace X_\Sa,X_\Sm\rbrace=(X_1,\dotsc,X_n)$ была
линейно независима в каждой точке $x\in\MU$. Пусть
$\lbrace A^\Sa,A^\Sm\rbrace=(A^1,\dotsc,A^n)$ -- соответствующие дуальные
1-формы. Тогда в каждой точке $x\in\MU$ 1-формы $A^\Sm$ определяют ортогональное
дополнение $(\CL_k)^\bot$. Это значит, что локально задание распределения
векторных полей эквивалентно нахождению решения системы $n-k$ уравнений на
дифференциалы $dx^\al$,
\begin{equation}                                                  \label{epfafe}
  A^\Sm=dx^\al A_\al{}^\Sm=0,\qquad \Sm=k+1,\dotsc,n.
\end{equation}
Эта система называется {\em пфаффовой системой уравнений}.
\index{Пфаффова система уравнений (Pfaffian system of equations)}%
\index{Система уравнений пфаффова (Pfaffian system of equations)}%

Из тождества (\ref{edacov}) следуют равенства
\begin{equation}                                                  \label{epfasy}
  2dA^\Sm(X_\Sa,X_\Sb)=X_\Sa\big(A^\Sm(X_\Sb)\big)-X_\Sb\big(A^\Sm(X_\Sa)\big)
  -A^\Sm\big([X_\Sa,X_\Sb]\big)=0,
\end{equation}
т.к.\ распределение векторных полей удовлетворяет условию теоремы
Фробениуса (\ref{ecovef}). Это значит, что распределение векторных полей
$\CL_k$ удовлетворяет условию теоремы Фробениуса тогда и только тогда, когда
\begin{equation*}
  dA^\Sm(X_\Sa,X_\Sb)=0,\qquad\forall\Sa,\Sb=1,\dotsc,k,\quad\forall\Sm=k+1,
  \dotsc,n.
\end{equation*}

Для любой 2-формы, в том числе и для $dA^\Sm$, справедливо представление
\begin{equation*}
  dA^\Sm=B^\Sm{}_\Sn\wedge A^\Sn+g^\Sm{}_{\Sb\Sa}A^\Sa\wedge A^\Sb,
\end{equation*}
где $\lbrace B^\Sm{}_\Sn\rbrace$ и $\lbrace g^\Sm{}_{\Sb\Sa}\rbrace$
-- некоторые наборы, соответственно, 1-форм и функций. Поскольку
$A^\Sa(X_\Sb)=\dl^\Sa_\Sb$, то из равенства (\ref{epfasy}) следует, что
$g^\Sm{}_{\Sb\Sa}=0$. Таким образом, справедливо разложение
\begin{equation*}
  dA^\Sm=B^\Sm{}_\Sn\wedge A^\Sn,
\end{equation*}
которое эквивалентно условию коммутативности векторных полей (\ref{ecovef}).
\begin{defn}
Система уравнений Пфаффа называется {\em вполне интегрируемой}, если существует
такая система координат $y^\al(x)$, что для подмногообразий, определяемых
системой уравнений
\begin{equation}                                                  \label{efrosu}
  y^\Sm=\const\quad \Leftrightarrow\quad dy^\Sm=dx^\al\pl_\al y^\Sm=0,
  \qquad \Sm=k+1,\dotsc,n,
\end{equation}
выполнялась система уравнений Пфаффа (\ref{epfafe}).
\qed\end{defn}
\index{Вполне интегрируемая пфаффова система уравнений %
(totally integrable Pfaffian system of equations)}%
\index{Пфаффова система уравнений вполне интегрируемая %
(totally integrable Pfaffian system of equations)}%
Для вполне интегрируемой системы уравнений Пфаффа распределение векторных полей
$\CL_k$ задается в точности первыми $k$ координатными полями
$\pl_1,\dotsc,\pl_k$. Обратное утверждение также справедливо. Таким образом
теорему Фробениуса можно переформулировать полностью в терминах 1-форм.
\begin{theorem}[\bf Фробениус]                                    \label{tfroth}
Для того, чтобы система уравнений Пфаффа
\begin{equation}                                                  \label{epfequ}
  A^\Sm=0,\qquad \Sm=k+1,\dotsc,n,
\end{equation}
была вполне интегрируемой, необходимо и достаточно, чтобы существовал
такой набор 1-форм $B^\Sm{}_\Sn$, что
\begin{equation}                                                  \label{efrnes}
  dA^\Sm=B^\Sm{}_\Sn\wedge A^\Sn.
\end{equation}
\end{theorem}
\begin{cor}
Если все 1-формы $A^\Sm$ замкнуты, $dA^\Sm=0$, то система уравнений Пфаффа
(\ref{epfequ}) вполне интегрируема.
\qed\end{cor}
\begin{exa}
Рассмотрим уравнение Пфаффа на 1-форму заданную на евклидовой плоскости $\MR^2$
в декартовой системе координат
\begin{equation}                                                  \label{eonpfp}
  dxP+dyQ=0,
\end{equation}
где $P(x,y),Q(x,y)\in\CC^1(\MR^2)$. Для определенности будем считать, что
$P\ne0$. Тогда 1-формы
\begin{equation*}
  A^1:=dxP+dyQ\qquad\text{и}\qquad A^2:=dy
\end{equation*}
можно выбрать в качестве базиса 1-форм на плоскости. Соответствующий ему
дуальный базис векторных полей имеет вид
\begin{equation*}
  X_1=\left(\frac1P,0\right),\qquad X_2=\left(-\frac QP,1\right).
\end{equation*}
Действительно, нетрудно проверить равенство $A^i(X_j)=\dl^i_j$, где $i,j=1,2$.
Векторное поле $X_2$ определяет ортогональное дополнение к $A^1$ и задает
уравнения, определяющие интегральные кривые в параметрическом виде
\begin{equation*}
  \dot x=-\frac QP,\qquad \dot y=1.
\end{equation*}
Отсюда следует уравнение на форму интегральной кривой
\begin{equation*}
  \frac{dx}{dy}=-\frac QP,
\end{equation*}
которое эквивалентно уравнению Пфаффа (\ref{eonpfp}). Это уравнение для
достаточно гладких функций $P$ и $Q$ всегда интегрируемо. Нетрудно
проверить, что в двумерном случае 1-формы $B$ в условии теоремы Фробениуса
(\ref{efrnes}) всегда существуют.

Отметим, что интегрируемость уравнения Пфаффа (\ref{eonpfp}) совсем не означает,
что 1-форма $A^1$ является точной, т.е.\ $A^1=df$ для некоторой функции
$f\in\CC^\infty(\MR^2)$. Критерием точности 1-формы на плоскости являются
нетривиальные условия интегрируемости $\pl_xQ=\pl_yP$, которые не имеют
отношения к интегрируемости уравнений Пфаффа. Полная интегрируемость уравнения
Пфаффа $A^1=0$ означает существование такой функции $f\in\CC^\infty(\MR^2)$, что
$df=0$ для любой траектории, определяемой уравнением Пфаффа (\ref{eonpfp}), а
не то, что $A=df$.
\qed\end{exa}

\begin{exa}
Рассмотрим уравнение Пфаффа на 1-форму в трехмерном евклидовом пространстве
$\MR^3$ в декартовой системе координат
\begin{equation}                                                  \label{epfeqe}
  A=dxP+dyQ+dzR=0,
\end{equation}
где $P,Q,R\in\CC^1(\MR^3)$. Полная интегрируемость этого уравнения означает
существование функции $f\in\CC^2(\MR^3)$ такой, что равенство $f=\const$
является первым интегралом уравнения Пфаффа (\ref{epfeqe}). Из теоремы
Фробениуса следует критерий интегрируемости, который можно (только для
трехмерного многообразия) записать в виде
\begin{equation}                                                  \label{eqfrte}
  dA=B\wedge A\quad \Leftrightarrow\quad dA\wedge A=0.
\end{equation}
Поскольку
\begin{equation*}
  dA=dx\wedge dy(\pl_xQ-\pl_yP)+dy\wedge dz(\pl_yR-\pl_zQ)
  +dz\wedge dx(\pl_zP-\pl_xR),
\end{equation*}
то необходимым и достаточным условием интегрируемости уравнения (\ref{epfeqe})
является равенство
\begin{equation*}                                                    \tag*{\qed}
  dA\wedge A=dx\wedge dy\wedge dz\big[P(\pl_yR-\pl_zQ)
  +Q(\pl_zP-\pl_xR)+R(\pl_xQ-\pl_yP)\big]=0.
\end{equation*}
\end{exa}
\begin{exa}
Покажем, что в общем случае, когда распределение не является инволютивным,
интегральные многообразия отсутствуют. Пусть в трехмерном евклидовом
пространстве $\MR^3$ задана 1-форма
\begin{equation*}
  A=dx+ydz.
\end{equation*}
Ее внешняя производная равна
\begin{equation*}
  dA=dy\wedge dz.
\end{equation*}
Поскольку
\begin{equation*}
  dA\wedge A=dx\wedge dy\wedge dz\ne0,
\end{equation*}
то условие теоремы Фробениуса не выполняются, которое в трехмерном случае
имеет вид (\ref{eqfrte}). Это значит, что уравнение Пфаффа $A=0$ в данном
случае неинтегрируемо.

Посмотрим на эту задачу с точки зрения векторных полей. Ортогональное
дополнение к форме $A$ натянуто, например, на векторные поля
\begin{equation*}
  X=-y\pl_x+\pl_z\qquad \text{и}\qquad Y=\pl_y,
\end{equation*}
поскольку они линейно независимы и
\begin{equation*}
  A(X)=0,\qquad A(Y)=0.
\end{equation*}
Эти векторные поля не находятся в инволюции, т.к.\ их коммутатор нельзя
представить в виде линейной комбинации
\begin{equation*}
  [X,Y]=\pl_x\ne fX+gY,\qquad f,g\in\CC^\infty(\MR^3).
\end{equation*}
Это значит, что условие теоремы Фробениуса нарушено.

Докажем от противного, что у векторных полей $X,Y$ интегральных поверхностей
не существует. Допустим, что через начало координат проходит интегральная
поверхность. Это значит, что, двигая начало координат вдоль интегральных
кривых, задаваемых векторными полями, мы никогда не покинем эту поверхность.
Векторные поля $X,Y$ задают потоки $s_X(t),s_Y(t)$, проходящие через точку
$(x_0,y_0,z_0)$ (см.\ раздел \ref{svechs}),
\begin{align*}
  s_X:\qquad &t\times(x_0,y_0,z_0)\mapsto(-y_0t+x_0,y_0,t+z_0),
\\
  s_Y:\qquad &t\times(x_0,y_0,z_0)\mapsto(x_0,t+y_0,z_0).
\end{align*}
Коммутатору векторных полей $[Y,X]$ соответствует отображение
\begin{equation*}
  s^{-1}_Y(t_2)s^{-1}_X(t_1)s_Y(t_2)s_X(t_1)
\end{equation*}
При этом начало координат перейдет в точку
\begin{equation*}
  (0,0,0)\stackrel{s_X(t_1)}{\longmapsto}(0,0,t_1)
  \stackrel{s_Y(t_2)}{\longmapsto}(0,t_2,t_1)
  \stackrel{s^{-1}_X(t_1)}{\longmapsto}(t_1t_2,t_2,0)
  \stackrel{s^{-1}_Y(t_2)}{\longmapsto}(t_1t_2,0,0).
\end{equation*}
Ясно, что не существует такой линейной комбинации векторных полей $fX+gY$,
чей поток переводил бы начало координат в точку $(t_1t_2,0,0)$.
Тем самым мы пришли к противоречию, что и доказывает отсутствие интегральной
поверхности, проходящей через начало координат. Аналогичное построение
можно провести для произвольной точки $\MR^3$. По сути дела доказательство
теоремы Фробениуса сводится к такому же построению в общем случае.
\qed\end{exa}
\section{Слоения}                                                 \label{sfolia}
Слоения обобщают понятие расслоения, введенное в разделе \ref{sfibun}. Они
часто возникают и играют большую роль в теории динамических систем. Мы начнем с
простейшего примера слоения, который поможет понять данное ниже общее
определение.
\begin{exa}[\bf Тривиальное слоение]
Представим евклидово пространство $\MR^n$ с естественной топологией и декартовой
системой координат в виде прямого произведения:
\begin{equation*}
  \MR^n\ni\quad x=(y,z)\quad\in\MR^k\times\MR^{n-k},\qquad 1\le k<n.
\end{equation*}
Построим многообразие $\MM^k$ меньшей размерности $k$, которое как множество
совпадает со всем $\MR^n$, т.е.\ $\MM^k=\MR^n$, и задает на нем структуру
слоения. Базу топологии $\MM^k$ зададим открытыми множествами вида
\begin{equation*}
  \MB_z:=\lbrace (y,z)\in\MR^n:~y\in\MO,\rbrace,
\end{equation*}
где $\MO$ -- открытое подмножество в $\MR^k$, для каждой фиксированной точки
$z\in\MR^{n-k}$. Эта топология тоньше исходной топологии в $\MR^n$.
Назовем листом тривиального слоения сечение евклидова пространства $\MR^n$
гиперплоскостью $z=\const$:
\begin{equation*}
  \MV_z:=\lbrace (y,z)\in\MR^n:~y\in\MR^k,~z=\const\rbrace.
\end{equation*}
Тогда многообразие $\MM^k$ представляет собой несвязное объединение всех листов
\begin{equation*}
  \MM^k=\bigcup_{z\in\MR^{n-k}}\MV_z
\end{equation*}
и покрывается картами
\begin{equation*}
  \psi_z:\quad \MV_z\ni\quad(y,z)\mapsto y\quad\in\MR^k.
\end{equation*}

Рассмотренную конструкцию можно представить по-другому. Рассмотрим два евклидова
пространства: $\MR^k$ с обычной топологией и $\MR^{n-k}_\Sd$ с дискретной
топологией, что отмечено индексом $\Sd$. Тогда многообразие $\MM^k$ является
топологическим произведением $\MM^k=\MR^k\times\MR^{n-k}_\Sd$ и как множество
совпадает с $\MR^n$.
\qed\end{exa}
В рассмотренном примере многообразие $\MM^k$ несвязно и состоит из несчетного
множества связных компонент (листов) $\MV_z$, которые параметризуются точками
$z\in\MR^{n-k}$. Эти свойства лежат в основе общего определения.
\begin{defn}
Пусть на множестве $\MM$ заданы две структуры многообразия разных размерностей.
То есть у нас есть два многообразия: $\MM^n$, $\dim\MM^n=n$, и $\MM^k$,
$\dim\MM^k=k$, где $1\le k<n$, которые совпадают как множества:
$\MM^n=\MM^k=\MM$. Многообразие $\MM^n$ называется {\em слоением}, если на нем
дополнительно определена другая более тонкая топология и вторая соответствующая
ей структура многообразия меньшей размерности $\MM^k$. Поскольку размерность
многообразия $\MM^k$ меньше размерности многообразия $\MM^n$, то многообразие
$\MM^k$ не может быть связным, и $\MM^n$ представляет собой несчетное
объединение связных компонент $\MM^k$. Связная компонента многообразия $\MM^k$,
содержащая точку $p\in\MM^n$, называется {\em листом слоения}, проходящим через
эту точку. Мы говорим, что многообразие $\MM^k$ {\em слоит} многообразие
$\MM^n$. Числа $k$ и $n-k$ называются соответственно {\em размерностью} и
{\em коразмерностью} слоения. Слоение мы будем обозначать парой $(\MM^n,\MM^k)$.
\qed\end{defn}
\index{Слоение (foliation)}\index{Слоить (foliate)}%
\index{Локальный лист (local leaf)}\index{Лист локальный (local leaf)}%
\index{Лист слоения (foliation leaf)}%
\index{Размерность слоения (dimensionality of a foliation)}%
\index{Коразмерность слоения (codimensionality of a foliation)}%
По-определению, через каждую точку $p\in\MM^n$ проходит один и только один лист.
В общем случае листы устроены довольно сложно. Как покажет дальнейший пример
слоения тора, листы могут не быть даже регулярными подмногообразиями в $\MM^n$.
Многообразие $\MM^n$ является объединением своих листов, число которых
бесконечно и не может быть счетным из-за разных размерностей. Многообразие
$\MM^k$ также является объединением листов и не может быть связным, т.к.\
каждый лист является открытым подмножеством и представляет собой компоненту
связности. В этом смысле многообразие $\MM^n$ слоится на листы.

\begin{figure}[h,t]
\includegraphics{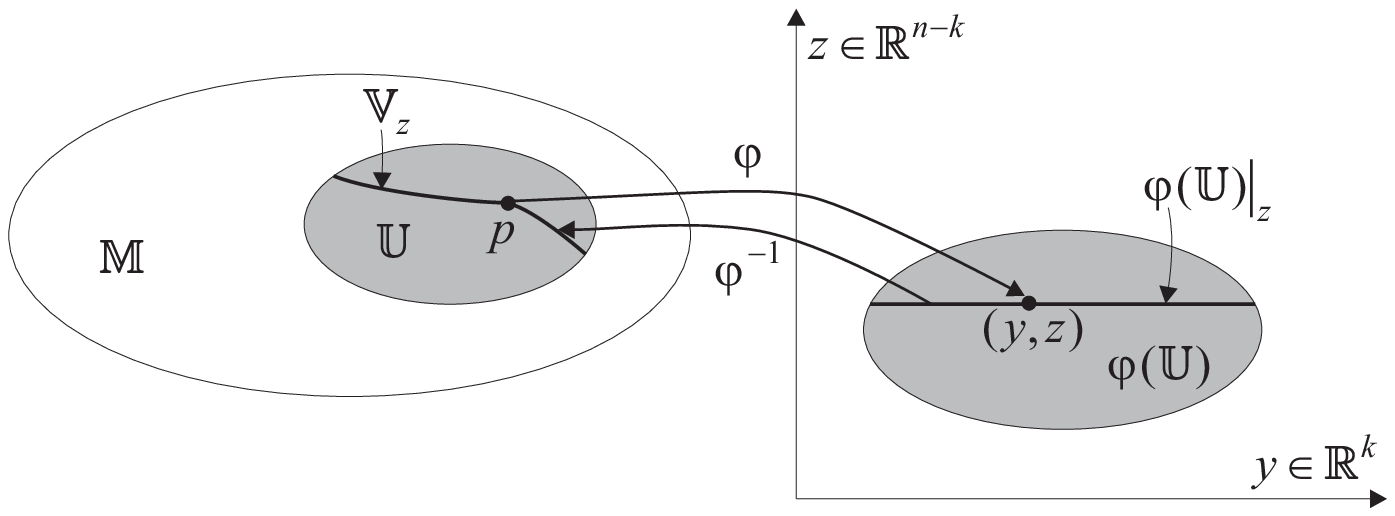}
\centering \caption{\label{fleaffoliation} Карта и локальный лист слоения}
\end{figure}

Можно доказать, что если $(\MM^n,\MM^k)$ -- слоение, то на многообразии $\MM^n$
существует атлас $\lbrace\MU_i,\vf_i\rbrace$, $i\in I$, со следующими
свойствами. Пусть
\begin{equation*}
  \vf:\quad \MM^n\supset\quad\MU\rightarrow\vf(\MU)\quad\subset\MR^n
\end{equation*}
-- карта исходной дифференцируемой структуры многообразия $\MM^n$. Представим
евклидово пространство $\MR^n$ в виде прямого произведения
$\MR^n=\MR^k\times\MR^{n-k}$, тогда для всех $p\in\MU$
\begin{equation*}
  \vf(p)=(y,z)\in\MR^k\times\MR^{n-k}.
\end{equation*}
Для каждой фиксированной точки $z\in\MR^{n-k}$ обозначим
\begin{equation*}
  \vf(\MU)|_z=\lbrace (y,z)\in\vf(\MU):~z=\const\rbrace
\end{equation*}
сечение образа $\vf(\MU)$ гиперплоскостью $z=\const$. Соответствующий прообраз
\begin{equation*}
  \MV_z:=\vf^{-1}\big(\vf(\MU)|_z\big)
\end{equation*}
называется {\em локальным листом}. Локальный лист является частью листа,
проходящего через точку $\vf^{-1}(y,z)$. Каждый локальный лист -- это открытое
подмногообразие (возможно пустое) в $\MM^k$, а пара $(\MV_z,\psi_z)$ для
непустого $\MV_z$ является картой на многообразии $\MM^k$:
\begin{equation*}
  \psi_z:\quad \MM^k\supset\quad\MV_z\rightarrow\psi_z(\MV_z)\quad\subset\MR^k,
\end{equation*}
где отображение $\psi_z:=\pi_k\circ\vf|_{\MV_z}$ определено естественной
проекцией
\begin{equation*}
  \pi_k:\quad \MR^k\times\MR^{n-k}\ni\quad(y,z)\mapsto y\quad\in\MR^k
\end{equation*}
и сужением $\vf|_{\MV_z}$ отображения $\vf$ на $\MV_z$. Описанная конструкция
изображена на рис.\ref{fleaffoliation}.

Функции склейки данного атласа для любых пересекающихся карт имеют вид
\begin{equation*}
  \vf_j\circ\vf^{-1}_i:\quad \MR^k\times\MR^{n-k}\supset\vf_i(\MU_i\cap\MU_j)
  \quad\ni y,z\mapsto y',z'
  \quad\in\vf_j(\MU_i\cap\MU_j)\subset\MR^k\times\MR^{n-k},
\end{equation*}
где $y'=y'(y,z)$, но $z'=z'(z)$. В этом случае локальные листы для этих карт
принадлежат одному листу слоения тогда и только тогда, когда они содержат хотя
бы одну точку из пересечения $p\in\MU_i\cup\MU_j$. Условие того, что функции
перехода $z'=z'(z)$ зависят только от $z$ необходимо и достаточно, для того,
чтобы склеивание локальных листов не зависело от выбора точки $p$ из
пересечения. Действительно, если $q,p\in\MU_i\cup\MU_j$ -- две различные точки
из пересечения координатных окрестностей, то в этих картах они имеют координаты
$(y_p,z_p)$, $(y'_p,z'_p)$ и $(y_q,z_q)$, $(y'_q,z'_q)$. Если $z_p=z_q$, то
$z'_p=z'_q$ тогда и только тогда, когда $z'=z'(z)$.

Теперь рассмотрим несколько типичных примеров слоений.
\begin{exa}
Пусть $dy/dx=f(x,y)$ -- обыкновенное дифференциальное уравнение, где правая
часть $f(x,y)$ -- функция класса $\CC^r$ в некоторой области $\MU\subset\MR^2$.
Из теории обыкновенных дифференциальных уравнений известно, что через каждую
точку области $\MU$ проходит одна и только одна интегральная кривая (решение)
данного уравнения класса $\CC^{r+1}$. Поскольку решение является функцией класса
$\CC^r$ от начальных данных, то объединение всех интегральных кривых образует
одномерное многообразие $\MV$, которое слоит $\MU$.

Аналогичный результат справедлив и для систем обыкновенных дифференциальных
уравнений. Для уравнений в частных производных ситуация существенно отличается.
\qed\end{exa}
\begin{exa}[\bf Распределение]
Пусть на многообразии $\MM^n$ задано $k$-мерное распределение векторных полей
$\CL_k(\MM^n)$, где $1\le k<n$, которое находится в инволюции (см.\ раздел
\ref{sfrote}). Напомним, что $k$-мерное распределение векторных полей ставит
каждой точке многообразия некоторое подпространство размерности $k$ в
касательном пространстве к данной точке:
\begin{equation*}
  \MM^n\ni\quad p\mapsto\ML_p(\MM^n)\quad\subset\MT_p(\MM^n)
\end{equation*}
Тогда согласно теореме Фробениуса
через каждую точку $p\in\MM^n$ проходит интегральное подмногообразие данного
распределения размерности $k$. Эти интегральные подмногообразия можно
рассматривать как листы слоения $(\MM^n,\MM^k)$.
\qed\end{exa}
Рассмотрим слоение $\MM^n$, которое слоится многообразием $\MM^k$. Тогда
определено касательное пространство к листу $\MT_p(\MM^k)$ в произвольной точке
$p\in\MM^n$. Размерность этого пространства равна $k$, и оно является
$k$-мерным подпространством касательного пространства $\MT_p(\MM^n)$.
Следовательно, определено $k$-мерное распределение векторных полей
$\CL_k(\MM^n)$ на многообразии $\MM^n$. Вместе с предыдущим примером это
доказывает следующее утверждение.
\begin{theorem}
Любое гладкое слоение $(\MM^n,\MM^k)$, $1\le k<n$, находится во взаимно
однозначном соответствии с $k$-мерным гладким инволютивным распределением
векторных полей $\CL_k(\MM^n)$, которое касается $\MM^k$, т.е.\
$\ML_p(\MM^n)=\MT_p(\MM^k)$. При этом лист слоения, проходящий через точку
$p\in\MM^n$, совпадает с интегральным подмногообразием данного распределения,
проходящим через ту же точку.
\end{theorem}
\begin{com}
В силу данной теоремы слоение часто определяют с помощью распределений и их
интегральных подмногообразий.
\qed\end{com}

\begin{figure}[h,b,t]
\hfill\includegraphics[width=.35\textwidth]{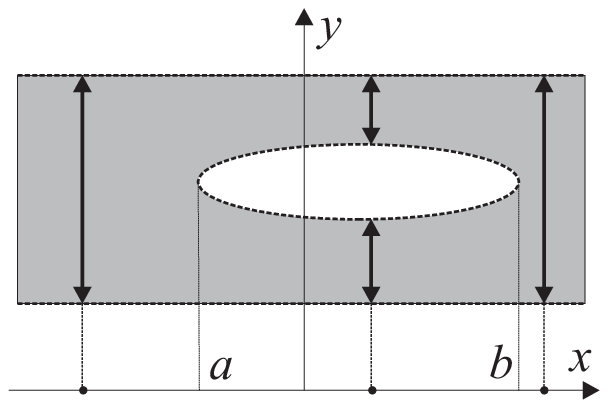}
\hfill {}
\centering\caption{Субмерсия ленты с вырезанной дыркой на ось $x$.}
\label{fsubme}
\end{figure}

Заметим, что не каждое многообразие допускает структуру слоения.
\begin{exa}
Теорема \ref{tvecsp} утверждает, что на четномерной сфере $\MS^{2n}$,
$n=1,2,\dotsc$, не существует непрерывного векторного поля, которое всюду
отлично от нуля. Это означает, что на $\MS^{2n}$ невозможно задать одномерное
распределение векторных полей $\CL_1(\MS^{2n})$, и, следовательно одномерных
слоений на четномерной сфере не существует. В частности, на двумерной сфере
$\MS^2$ нельзя задать структуру слоения.
\qed\end{exa}
Далее мы рассмотрим более подробно связь слоений и расслоений.
\begin{exa}[\bf Расслоение]
Рассмотрим дифференцируемое расслоение $\ME(\MM,\pi,\MF)$ (см., раздел
\ref{sfibun}) с $n$-мерным пространством расслоения $\ME$, $k$-мерным типичным
слоем $\MF$ и $(n-k)$-мерной базой $\MM$. Через каждую точку расслоения
$p\in\ME$ проходит слой $\MV_{\pi(p)}:=\pi^{-1}\big(\pi(p)\big)$, диффеоморфный
типичному слою $\MF$, и, значит, имеющий структуру дифференцируемого $k$-мерного
многообразия. Известно, что слои расслоения либо совпадают:
$\MV_{\pi(p)}=\MV_{\pi(q)}$, если $\pi(p)=\pi(q)$, либо не пересекаются:
$\MV_{\pi(p)}\cap\MV_{\pi(q)}=\emptyset$, если $\pi(p)\ne\pi(q)$. Рассмотрим
множество, которое состоит из объединения всех слоев
\begin{equation*}
  \ME^k=\bigcup_{x\in\MM}\MV_x.
\end{equation*}
Как множество оно совпадает с пространством расслоения $\ME$, но мы введем на
нем другую топологию и дифференцируемую структуру. А именно, будем считать, что
в этой топологии и дифференцируемой структуре каждый слой является открытым
подмножеством и открытым подмногообразием. Для того, чтобы проверить, что
многообразие $\ME^k$ слоит расслоение $\ME$ следует воспользоваться картами на
$\ME$, которые тривиализируют расслоение. Как и в предыдущем примере
многообразие $\ME^k$ несвязно и состоит из несчетного числа листов, которые
совпадают со слоями расслоения $\MV_x$ и параметризуются точками базы $x\in\MM$.
В этом слоении все листы $\MV_x$ диффеоморфны между собой и диффеоморфны
типичному слою, $\MV_x\approx\MF$.
\qed\end{exa}
Рассмотренный пример показывает, что на каждом расслоении определена
естественная структура слоения. Обратное утверждение неверно: не каждое слоение
является расслоением. Чтобы построить соответствующий пример, нам понадобится
\begin{defn}
Дифференцируемое отображение двух многообразий $f:\quad \MM\rightarrow\MN$
называется {\em субмерсией}, если дифференциал отображения
$f_*:\quad \MT_p(\MM)\rightarrow\MT_{f(p)}(\MN)$ является сюрьективным
отображением для всех точек $p\in\MM$.
\qed\end{defn}
\index{Субмерсия (submersion)}%
\begin{exa}[\bf Субмерсия]
Рассмотрим субмерсию двух многообразий  $f:\quad \MM\rightarrow\MN$. Из
определения следует, что $\dim\MM\ge\dim\MN$. Мы положим $\dim\MM=n$ и
$\dim\MN=n-k$, где $1\le k<n$. Каждая точка $p\in\MM$ порождает лист или
несвязное объединение листов $\MV_p:=f^{-1}\big(f(p)\big)$. При этом прообразы
либо совпадают: $\MV_p=\MV_q$, если $f(p)=f(q)$, либо не пересекаются:
$\MV_p\cap\MV_q=\emptyset$, если $f(p)\ne f(q)$. Предположим, что отображение
$f$ сюрьективно. Тогда слои можно параметризовать точками $x\in\MN$. Из условия
сюрьективности $f_*\big(\MT_p(\MM)\big)=\MT_{f(p)}(\MN)$ для всех $p\in\MM$
следует, что для каждой точки $p\in\MM$ найдутся карты:
\begin{align*}
  \vf:&\quad \MM\supset\MU\ni\quad p\mapsto \vf(p)\quad\in\MR^n,
\\
  \chi:&\quad \MN\supset\MW\ni\quad x=f(p)\mapsto \chi(x)\quad\in\MR^{n-k}
\end{align*}
со следующими свойствами:
\begin{enumerate}
\itemindent0mm\itemsep-1mm\parsep0mm
\item $f(\MU)\subset\MW$,
\item $\vf(q)=(y,z)\in\MR^k\times\MR^{n-k},\quad \forall q\in\MU$,
\item $\chi(x)\in\MR^{n-k},\quad \forall x\in\MW$,
\item $(\chi\circ f\circ\vf^{-1})(y,z)
        =z\in\chi(\MW),\quad \forall (y,z)\in\vf(\MU)$,
\item $\vf(\MV_p)=\lbrace(y,z)\in\vf(\MU):~z=\chi\big(f(p)\rbrace,\quad
       \forall f(p)\in\MW$.
\end{enumerate}
Отсюда легко выводится, что на множестве $\MM$ определена структура $k$-мерного
слоения $\MM^k$, в котором листы, содержащиеся в $f^{-1}(x)$, являются открытыми
подмногообразиями.
\qed\end{exa}
Ранее было показано, что любое расслоение является слоением. Приведем пример
слоения, которое не является расслоением.
\begin{exa}
Рассмотрим субмерсию, изображенную на рис.\ref{fsubme}, где лента в $\MR^2$ с
вырезанной дыркой проектируется на ось $x$. Ленту с вырезанной дыркой мы
рассматриваем, как открытое подмногообразие $\MM$ в $\MR^2$ с декартовыми
координатами $x,y$. Субмерсия задается проекцией $f:~(x,y)\mapsto x$ и
определяет одномерное слоение $\MM$. Соответствующее распределение векторных
полей имеет вид $\pl_y$. Все интегральные кривые этого распределения, которые
являются листами слоения, параллельны оси $y$. Если $x<a$ или $x>b$, то
листами слоения $f^{-1}(x)$ являются интервалы. В то же время каждый прообраз
$f^{-1}(x)$ для $a\le x\le b$ представляет несвязное объединение двух листов.
Ясно, что не все прообразы $f^{-1}(x)$ данного слоения диффеоморфны между собой,
и поэтому рассматриваемое слоение не может быть расслоением.
\qed\end{exa}
Следующий пример показывает, что листы слоения даже в простых случаях
могут быть устроены достаточно сложно.
\begin{exa}[\bf Слоения тора]
Рассмотрим двумерный тор, как квадрат на плоскости:
$\MT^2=[0,1]\times[0,1]\subset\MR^2$
с отождествленными противоположными сторонами. Пусть на нем задано векторное
поле $X=X^x\pl_x+X^y\pl_y$ с постоянными компонентами $X^x=a\ne0$ и $X^y=b$. Тем
самым определено одномерное распределение векторного поля. Оно, конечно,
находится в инволюции. Обозначим интегральные кривые этого векторного поля через
$\g_{ab}$. Они же являются листами одномерного слоения тора. Если отношение
$b/a$
рационально, то каждый лист диффеоморфен окружности $\MS^1$ и является
регулярным замкнутым подмногообразием в $\MT^2$. Если отношение $b/a$
иррационально (всюду плотная обмотка тора в примере \ref{edetor}), то каждый
лист диффеоморфен прямой $\MR$. При этом отображение
$\MR\rightarrow\g_{ab}\in\MT^2$ не является регулярным. В этом случае лист не
является одномерным регулярным подмногообразием и замкнутым подмножеством в
$\MT^2$.
\qed\end{exa}
\section{Бесконечно малые преобразования координат               \label{sinfct}}
В настоящем разделе рассматриваются свойства различных полей на многообразии при
бесконечно малых преобразованиях координат. Полученные ниже формулы важны в
приложениях, в частности, они будут использованы в следующем разделе для
вычисления производных Ли.

Поскольку рассмотрение носит локальных характер, то для простоты и наглядности
мы рассмотрим евклидово пространство $\MR^n$ с фиксированной декартовой системой
координат. Будем рассматривать преобразования координат (\ref{ecootr}) с
активной точки зрения. То есть будем считать, что в евклидовом пространстве
точка с координатами $x^\al$ перемещается в новую точку с координатами
$x^{\prime\al}(x)$ в той же системе координат. Бесконечно малые
(инфинитезимальные) преобразования координат
\begin{equation}                                                  \label{einfct}
  x^{\prime\al}=x^\al+ u^\al(x)
\end{equation}
определяются векторным полем
$$
   u=u^\al\pl_\al.
$$
Обозначение $u^\al(x)$ заимствовано из теории упругости, где этот вектор
называется {\em вектором смещения}. Будем считать, что малы сами смещения,
$|u^\al|\ll1$\footnote{В физике координаты являются размерными величинами, и
условие малости смещений записывается в виде $|u^\al|\ll l$, где $l$ --
выбранная единица длины.}, и относительные смещения $|\pl_\al u^\bt|\ll1$ для
всех значений индексов $\al,\bt$.
\index{Вектор смещения (displacement vector)}%
\index{Смещения вектор (displacement vector)}%

Рассмотрим, как меняются тензорные поля при инфинитезимальных преобразованиях
координат (\ref{einfct}). Начнем с простейшего случая скалярного поля (функции)
$f\in\CC^\infty(\MR^n)$. Из закона преобразования скалярного поля (\ref{escfun})
следует равенство
\begin{equation}                                                  \label{esctra}
  f'(x+u)=f(x).
\end{equation}
Разлагая левую часть этого равенства в ряд Тейлора\footnote{Здесь и далее мы
предполагаем, что разложение всех функций и компонент тензорных полей в ряд
Тейлора имеет смысл в некоторой окрестности точки $x$.}, в первом порядке по
$u$ получим следующее выражение
\begin{equation}                                                  \label{einftr}
  f'=f-u^\al\pl_\al f,
\end{equation}
где все поля рассматриваются в точке $x$. Полученное соотношение
интерпретируется следующим образом. При инфинитезимальном преобразовании
координат (\ref{einfct}) значение функции $f(x)$ в точке $x$ меняется
на бесконечно малую величину
\begin{equation}                                                  \label{einfts}
  \dl f(x):=f'(x)-f(x)=-u^\al\pl_\al f=-u f.
\end{equation}
которая называется {\em приращением} или {\em вариацией формы} функции $f$ в
\index{Вариация формы (form variation)}%
\index{Приращение функции (increment of a function)}%
точке $x$. Правая часть этого равенства с обратным знаком представляет собой
производную от функции $f$ вдоль векторного поля $u$, определяющего
преобразование координат.

Из закона преобразования векторных полей (\ref{evectr}), который для
инфинитезимальных преобразований можно переписать в виде
$$
  X^{\prime\al}(x+u)=X^\bt(\dl^\al_\bt+\pl_\bt u^\al),
$$
следует, что вариация формы компонент векторного поля имеет вид
\begin{equation}                                                  \label{eintvf}
  \dl X^\al=X^\bt\pl_\bt u^\al-u^\bt\pl_\bt X^\al
  =[X,u]^\al.
\end{equation}
То есть приращение векторного поля в точке $x$ равно коммутатору
этого векторного поля с инфинитезимальным векторным полем смещения $u$.

Закон преобразования компонент 1-формы (\ref{eonftr}) приводит к следующему
приращению компонент
\begin{align}                                                     \label{eintof}
  \dl\om_\al&=-\pl_\al u^\bt\om_\bt-u^\bt\pl_\bt\om_\al
\\ \intertext{или в ковариантном виде}                                 \nonumber
  \dl\om_\al&=-\nb_\al u^\bt\om_\bt-u^\bt\nb_\bt\om_\al
  +u^\bt T_{\al\bt}{}^\g\om_\g
\\                                                                \label{eintct}
  &=-\widetilde\nb_\al u^\bt\om_\bt-u^\bt\widetilde\nb_\bt\om_\al.
\end{align}
При этом использовано явное выражение для аффинной связности (\ref{elicon}).
Напомним, что $\nb$ и $\widetilde\nb$ обозначают ковариантные производные
производные с аффинной связностью и символами Кристоффеля, соответственно.
\begin{com}
Приращения компонент тензорных полей, а также связности при инфинитезимальных
преобразованиях координат являются тензорными полями. Поскольку бесконечно малые
преобразования координат параметризуются векторным полем $u$, то вариации полей
всегда можно записать в явно ковариантном виде.
\qed\end{com}

Аналогично находится явный вид бесконечно малых преобразований координат
для тензоров произвольного ранга. Рассмотрим бесконечно малые преобразования
основных геометрических объектов: метрики и аффинной связности, а также
репера и линейной связности. В частности, для метрики справедливо равенство
\begin{align}                                                     \label{eintrm}
  \dl g_{\al\bt}&=-\pl_\al u^\g g_{\g\bt}-\pl_\bt u^\g g_{\al\g}
  -u^\g\pl_\g g_{\al\bt}
\\ \intertext{или в ковариантном виде}                                 \nonumber
  \dl g_{\al\bt}&=-\nb_\al u_\bt-\nb_\bt u_\al
  +u^\g(T_{\al\g\bt}+T_{\bt\g\al}
  -Q_{\al\g\bt}-Q_{\bt\g\al}+Q_{\g\al\bt})
\\                                                                \label{eitcms}
  &=-\widetilde\nb_\al u_\bt-\widetilde\nb_\bt u_\al.
\end{align}
Отсюда следует, что, если метрика инвариантна относительно бесконечно малых
преобразований координат, $\dl g_{\al\bt}=0$, то соответствующее векторное поле
должно удовлетворять уравнению Киллинга (\ref{ekileq}).

Из уравнения (\ref{eintrm}) следует правило преобразования определителя метрики
\begin{equation}                                                  \label{edemet}
  \dl g=-u^\al\pl_\al g-2\pl_\al u^\al g.
\end{equation}

Рассмотрим, как меняются компоненты аффинной связности при инфинитезимальных
преобразованиях координат. Из закона преобразования (\ref{econtr}) следует
выражение для вариации формы компонент аффинной связности:
\begin{equation}                                                  \label{eiitac}
  \dl\Gamma_{\al\bt}{}^\g=-\pl_\al u^\dl\Gamma_{\dl\bt}{}^\g
  -\pl_\bt u^\dl\Gamma_{\al\dl}{}^\g
  +\Gamma_{\al\bt}{}^\dl\pl_\dl u^\g
  -\pl^2_{\al\bt}u^\g-u^\dl\pl_\dl\Gamma_{\al\bt}{}^\g.
\end{equation}
В ковариантном виде это изменение имеет вид
\begin{equation}                                                  \label{ecoitc}
  \dl\Gamma_{\al\bt}{}^\g=-\nb_\al\nb_\bt u^\g
  +\nb_\al u^\dl T_{\bt\dl}{}^\g
  +u^\dl(R_{\al\dl\bt}{}^\g+\nb_\al T_{\bt\dl}{}^\g).
\end{equation}
В частном случае для символов Кристоффеля справедливо равенство
\begin{equation}                                                  \label{eecost}
  \dl\widetilde\Gamma_{\al\bt}{}^\g=
  -\widetilde\nb_\al\widetilde\nb_\bt u^\g
  +u^\dl\widetilde R_{\al\dl\bt}{}^\g.
\end{equation}
Эту вариацию можно переписать в виде
\begin{equation}                                                  \label{evasch}
  \dl\widetilde\Gamma_{\al\bt\g}=\frac12(\widetilde\nb_\al\dl g_{\bt\g}
  +\widetilde\nb_\bt\dl g_{\al\g}-\widetilde\nb_\g\dl g_{\al\bt}).
\end{equation}
В римановой геометрии, где связность однозначно определяется метрикой,
инвариантность метрики является достаточным условием инвариантности
символов Кристоффеля. Необходимое и достаточное условие инвариантности символов
Кристоффеля относительно бесконечно малых преобразований координат имеет вид
$\dl\widetilde\Gamma_{\al\bt}{}^\g=0$, где вариация определяется формулой
(\ref{eecost}).

Антисимметризация уравнения (\ref{ecoitc}) по индексам $\al,\bt$ дает приращение
компонент тензора кручения:
\begin{align}                                                     \nonumber
  \dl T_{\al\bt}{}^\g&=\nb_\al u^\dl T_{\bt\dl}{}^\g
  -\nb_\bt u^\dl T_{\al\dl}{}^\g+T_{\al\bt}{}^\dl\nb_\dl u^\g
\\                                                                \label{etoint}
  &+u^\dl(-R_{\al\bt\dl}{}^\g-R_{\dl\al\bt}{}^\g
  -R_{\bt\dl\al}{}^\g
  +\nb_\al T_{\bt\dl}{}^\g-\nb_\bt T_{\al\dl}{}^\g)
\end{align}
или, с учетом (\ref{ecuath}),
\begin{align}                                                     \nonumber
  \dl T_{\al\bt}{}^\g&=\nb_\al u^\dl T_{\bt\dl}{}^\g
  -\nb_\bt u^\dl T_{\al\dl}{}^\g+T_{\al\bt}{}^\dl\nb_\dl u^\g
\\                                                      \label{etoist}
  &-u^\dl(\nb_\dl T_{\al\bt}{}^\g+T_{\al\bt}{}^\e T_{\e\dl}{}^\g
  +T_{\bt\dl}{}^\e T_{\e\al}{}^\g+T_{\dl\al}{}^\e T_{\e\bt}{}^\g)
\end{align}
Нетрудно получить также вариацию формы компонент тензора неметричности при
инфинитезимальных преобразованиях координат:
\begin{align}                                                     \nonumber
  \dl Q_{\al\bt\g}&=-\nb_\al u^\dl Q_{\dl\bt\g}
  -\nb_\bt u^\dl Q_{\al\dl\g}-\nb_\g u^\dl Q_{\al\bt\dl}
\\                                                                \label{enmint}
  &+u^\dl(T_{\al\dl}{}^\e Q_{\e\bt\g}+T_{\bt\dl}{}^\e Q_{\al\e\g}
  +T_{\g\dl}{}^\e Q_{\al\bt\e}).
\end{align}
В аффинной геометрии при заданной связности можно поставить задачу
нахождения таких преобразований координат, которые не меняют связность,
кручение или неметричность. Для этого необходимо решить, соответственно,
уравнения (\ref{ecoitc}), (\ref{etoint}) или (\ref{enmint}).

Репер и лоренцева связность являются ковекторами относительно преобразований
координат. Поэтому их приращения такие же, как и приращения 1-форм
(\ref{eintof}). Приведем их явный вид, т.к.\ они часто встречаются в приложениях
\begin{align}                                                     \label{eintre}
  \dl e_\al{}^a&=-\pl_\al u^\bt e_\bt{}^a-u^\bt\pl_\bt e_\al{}^a
  =-\widetilde\nb_\al u^a+u^\bt\widetilde\om_{\bt b}{}^ae_\al{}^b,
\\                                                                \label{eintrl}
  \dl \om_\al{}^{ab}&=-\pl_\al u^\bt\om_\bt{}^{ab}
  -u^\bt\pl_\bt\om_\al{}^{ab}
  =-\widetilde\nb_\al u^\bt\om_\bt{}^{ab}
  -u^\bt(\pl_\bt\om_\al{}^{ab}-\widetilde\Gamma_{\bt\al}{}^\g\om_\g{}^{ab}).
\end{align}
Из формулы (\ref{eintre}) следует выражение для вариации формы элемента объема
\begin{equation}                                                  \label{eintdt}
  \dl\vol=-u^\al\pl_\al\vol-\vol\pl_\al u^\al
  =-\vol\widetilde\nb_\al u^\al,
\end{equation}

Бесконечно малые приращения различных полей, рассмотренные в настоящем
разделе, называются вариацией формы, т.к.\ показывают как меняется значение
полей в фиксированной точке многообразия. Следующее свойство проверяется прямой
проверкой и используется при доказательстве теоремы Нетер.
\begin{prop}
Вариация формы производной скалярного поля равно производной от вариации
\begin{equation*}                                                  \tag*{\qed}
  \dl(\pl_\al f)=\pl_\al(\dl f).
\end{equation*}
\renewcommand{\qed}{}\end{prop}

Рассмотрим коммутатор двух бесконечно малых преобразований координат. Пусть эти
преобразования задаются векторными полями $u=u^\al\pl_\al$ и $v=v^\al\pl_\al$.
Для простоты ограничимся их последовательным действием на функцию $f(x)$.
Поскольку преобразования координат евклидова пространства $\MR^n$ образуют
псевдогруппу, то коммутатор двух преобразований также является преобразованием
координат с параметром, который квадратичен по компонентам $u^\al$ и $v^\al$.
Поэтому, в отличие от предыдущего рассмотрения, для вычисления коммутатора
необходимо удерживать квадратичные слагаемые.

Сначала совершим преобразование координат, определяемое векторным полем $u$.
Из закона преобразования функции (\ref{esctra}) с учетом квадратичных
слагаемых получаем равенство
\begin{equation*}
  f'+u^\al\pl_\al f'+\frac12u^\al u^\bt\pl^2_{\al\bt}f'=f.
\end{equation*}
Во втором слагаемом в левой части равенства $f'$ можно выразить через $f$,
воспользовавшись линейным приближением (\ref{einftr}), а в третьем слагаемом
заменим $f'$ на $f$, потому что коэффициент перед ним уже квадратичен по $u$.
В результате с точностью до квадратичных слагаемых получаем
\begin{equation}                                                  \label{equstr}
  f':=(1+T_u)f=f-u^\al\pl_\al f+u^\al\pl_\al u^\bt\pl_\bt f
  +\frac12u^\al u^\bt\pl^2_{\al\bt}f+\dotsc,
\end{equation}
где мы ввели генератор общих преобразований координат $T_u$ и выписали его
действие на функцию в квадратичном приближении.

Совершим теперь второе преобразование координат с параметром $v$
\begin{equation*}
\begin{split}
  (1+T_v)f'&=(1+T_v)(1+T_u)f=
\\
  &=f'-v^\al\pl_\al f'+v^\al\pl_\al v^\bt\pl_\bt f'
  +\frac12v^\al v^\bt\pl^2_{\al\bt}f'+\dotsc=
\\
  &=f-(u^\al+v^\al-u^\bt\pl_\bt u^\al-v^\bt\pl_\bt v^\al
  -v^\bt\pl_\bt u^\al)\pl_\al f
\\
  &\qquad +\frac12(u^\al u^\bt+2u^\al v^\bt+v^\al v^\bt)\pl^2_{\al\bt}f+\dotsc.
\end{split}
\end{equation*}
Вычитая из этого выражения результат тех же преобразований в обратном порядке,
получим явное выражение для коммутатора двух преобразований координат
\begin{equation}                                                  \label{ecotrc}
  [T_v,T_u]f=T_{[v,u]}f=(v^\bt\pl_\bt u^\al-u^\bt\pl_\bt v^\al)\pl_\al f.
\end{equation}
Таким образом, коммутатор двух преобразований координат в евклидовом
пространстве $\MR^n$, определяемых инфинитезимальными векторными полями $u$ и
$v$, является преобразованием координат, которое задается коммутатором векторных
полей $[v,u]$.

Из групповых соображений следует, что коммутатор двух преобразований координат
не зависит от представления. Другими словами, это же выражение для коммутатора
имеет место не только для функций, но и для тензорных полей или плотностей более
высокого ранга. В последнем случае вычисления являются более громоздкими.

Рассмотрим, как меняются при бесконечно малых преобразованиях тензорные
плотности, заданные на многообразии $\MM$. Мы уже получили правила
преобразования определителей метрики (\ref{edemet}) и репера (\ref{eintdt}),
которые являются тензорными плотностями соответственно степеней $-2$ и $-1$. В
общем случае, если на $\MM$ задана скалярная плотность $f$ степени $\deg f=p$,
то из правила преобразования тензорных плотностей (\ref{escfud}) следует правило
\begin{equation*}
  \dl f=-u^\al\pl_\al f+p\pl_\al u^\al f,
\end{equation*}
т.к.\ якобиан бесконечно малых преобразований координат (\ref{einfct})
в первом порядке по вектору смещения имеет вид
\begin{equation}                                                  \label{ejaint}
  J=\det(\dl_\al^\bt+\pl_\al u^\bt)\approx 1+\pl_\al u^\al.
\end{equation}
Если на многообразии $\MM$ задана тензорная плотность $T$ произвольного
ранга и степени $p$, то ее вариация формы равна
\begin{equation*}
  \dl T=\dotsc+p\pl_\al u^\al T,
\end{equation*}
где точки обозначают набор обычных тензорных слагаемых для всех ковариантных
и контравариантных индексов.
\section{Производная Ли                                          \label{sliede}}
Понятие потока $\MR\times\MM\xrightarrow s\MM$ или экспоненциального отображения
$s^\al(t,x)=\exp(tX)x^\al$, генерируемого отличным от нуля дифференцируемым
векторным полем $X\in\CX(\MM)$ (раздел \ref{svechs}), позволяет определить
производную Ли $\Lie_X$ от произвольного тензорного поля $T$ вдоль векторного
поля $X$. Для этого определения достаточно, чтобы поток существовал локально.

Рассмотрим произвольную точку $x\in\MM$. Тогда у нее существует окрестность
$\MU_x$ такая, что отображение $s:~\MU_x\rightarrow s(\MU_x)$ является
диффеоморфизмом. Поэтому для него определены дифференциал отображения $s_*$ и
обратное отображение $s^{*-1}$ к возврату отображения $s^*$.
Пусть в некоторой окрестности точки $x\in\MM$ задано тензорное поле
$T\in\CT^r_s(\MM)$ типа $(r,s)$. В результате экспоненциального отображения
$s(\e,x)$ с малым параметром $\e$ тензор $T(x)$ в точке $x$ отобразится в тензор
$\tilde sT(x)$ в точке $s(\e,x)$, где $\tilde sT$ обозначает продолжение
отображения $(s_*)^r(s^{*-1})^sT$, заданного в касательном и
кокасательном пространствах, на всю тензорную алгебру (в компонентах:
на каждый контравариантный индекс действует дифференциал отображения $s_*$, а на
каждый ковариантный индекс -- обратное отображение $s^{*-1}$). Это значит, что в
точку $x$ отобразится тензор из точки $s(-\e,x)$:
\begin{equation*}
  T\big(s(-\e,x)\big)~\mapsto~\tilde s\big(\e,s(-\e,x)\big)T\big(s(-\e,x)\big).
\end{equation*}
\begin{defn}
{\em Производной Ли} от тензорного поля $T$ вдоль векторного поля $X$ в точке
$x$ называется предел
\begin{equation}                                                  \label{elided}
  \Lie_XT:=\underset{\e\to0}\lim
  \frac{T(x)-\tilde s\big(\e,s(-\e,x)\big)T\big(s(-\e,x)\big)}\e.
\end{equation}
В упрощенной записи мы пишем
\begin{equation}                                                  \label{eliesi}
  \Lie_XT=\underset{\e\to0}\lim\frac{T(x)-\tilde s(\e)T}\e,
\end{equation}
где мы опустили аргумент $s(-\e,x)\in\MM$.
\qed\end{defn}
\index{Производная Ли (Lie derivative)}\index{Ли производная (Lie derivative)}%
Для дифференцируемых векторных и тензорных полей этот предел существует.
\begin{com}
При малых $\e$ экспоненциальное отображение имеет вид
$x^\al\mapsto s^\al(\e,x)=x^\al+\e X^\al+\dotsc$, т.е.\ соответствует
бесконечно малым преобразованиям координат, рассмотренным в предыдущем разделе,
с вектором смещения $u^\al=\e X^\al$. При этом выражение, стоящее в числителе
производной Ли (\ref{elided}), совпадает с вариацией формы тензорного поля $T$,
взятой с обратным знаком. Это следует непосредственно из определения вариации
формы тензорного поля. Поэтому определение (\ref{elided}) для компонент
тензорного поля типа $(r,s)$ принимает вид
\begin{equation}                                                  \label{elidte}
  \Lie_X T_{\bt_1\dotsc\bt_s}{}^{\al_1\dotsc\al_r}(x)=-\underset{\e\to0}\lim
  \frac{\dl T_{\bt_1\dotsc\bt_s}{}^{\al_1\dotsc\al_r}(x)}\e,
\end{equation}
где $\dl T_{\bt_1\dotsc\bt_s}{}^{\al_1\dotsc\al_r}(x)$ -- вариация формы
компонент тензорного поля типа $(r,s)$, рассмотренная в разделе \ref{sinfct}.
Несложные вычисления приводят к следующему выражению для производной Ли
компонент тензорного поля
\begin{equation}                                                  \label{elidet}
\begin{split}
  \Lie_X T_{\bt_1\dotsc\bt_s}{}^{\al_1\dotsc\al_r}
  &=X^\g\pl_\g T_{\bt_1\dotsc\bt_s}{}^{\al_1\dotsc\al_r}
\\
  &+\pl_{\bt_1}X^\g T_{\g\bt_2\dotsc\bt_s}{}^{\al_1\dotsc\al_r}+\dotsc
  +\pl_{\bt_s}X^\g T_{\bt_1\dotsc\bt_{s-1}\g}{}^{\al_1\dotsc\al_r}
\\
  &-T_{\bt_1\dotsc\bt_s}{}^{\g\al_2\dotsc\al_r}\pl_\g X^{\al_1}-\dotsc
  -T_{\bt_1\dotsc\bt_s}{}^{\al_1\dotsc\al_{r-1}\g}\pl_\g X^{\al_1}.
\end{split}
\end{equation}
Первое слагаемое в правой части (\ref{elidet}) соответствует смещению самой
точки $x$, слагаемые во второй и третьей строках (\ref{elidet}) возникают при
действии отображений  $s^{*-1}$ и $s_*$ на каждый ковариантный и
контравариантный индекс, соответственно.
\qed\end{com}

Формула (\ref{elidet}) конструктивна и позволяет получить явные выражения для
производных Ли различных тензорных полей в координатах.
\begin{exa}
В простейшем случае скалярного поля $f(x)$ производная Ли совпадает с
производной функции вдоль векторного поля. Покажем это. Поскольку на
скалярное поле ни дифференциал отображения, ни его возврат не действуют, то из
определения (\ref{elided}) следует
\begin{equation*}
  \Lie_Xf=\underset{\e\to0}\lim\frac{f(x)-f(x-\e X)}\e=X^\al\pl_\al f.
\end{equation*}
Таким образом, производная Ли от функции -- это просто производная
вдоль векторного поля.
\qed\end{exa}
\begin{exa}
Из уравнения (\ref{elidet}) следует выражение для производной Ли от векторного
поля $Y$ в координатах:
$$
  \Lie_X Y^\al=X^\bt\pl_\bt Y^\al-Y^\bt\pl_\bt X^\al,
$$
что совпадает с коммутатором векторных полей (\ref{ecomvf})
\begin{equation}                                                  \label{elicom}
  \Lie_X Y=[X,Y].
\end{equation}
Определение производной Ли (\ref{elided}) для векторного поля принимает вид
\begin{equation}                                                  \label{elieve}
  \Lie_X Y=\underset{\e\to0}\lim \frac{Y-s_*(\e)Y}\e. \qed
\end{equation}
\end{exa}

Производная Ли от векторного поля обладает следующими свойствами:
\begin{equation*}
\begin{split}
  \Lie_{fX}Y&=f[X,Y]-(Yf)X,
\\
  \Lie_X(fY)&=f[X,Y]+(Xf)Y,
\end{split}
\end{equation*}
где $f(x)\in\CC^1(\MM)$ -- произвольное скалярное поле.

Отметим, что тождества Якоби (\ref{ejacid}) в алгебре Ли векторных полей можно
переписать в эквивалентном виде, используя производную Ли,
\begin{equation*}
  \Lie_X[Y,Z]=[\Lie_XY,Z]+[Y,\Lie_XZ].
\end{equation*}
В таком виде тождества Якоби аналогичны правилу Лейбница.

Можно также проверить следующее свойство производной Ли:
\begin{equation}                                                  \label{eliecl}
  \Lie_{[X,Y]}=\Lie_X\circ\Lie_Y-\Lie_Y\circ\Lie_X.
\end{equation}
Действие этого равенства на векторное поле сводится к тождеству Якоби.

Производная Ли отличается от ковариантной производной вдоль векторного поля
(\ref{evefde}).
\begin{prop}
Для компонент векторного поля справедливо равенство
\begin{equation}                                                  \label{ediflv}
  \Lie_X Y^\al-\nb_X Y^\al=-Y^\bt(\nb_\bt X^\al+X^\g T_{\bt\g}{}^\al),
\end{equation}
где $\nb_\al$ -- ковариантная производная и $T_{\bt\g}{}^\al$ -- тензор
кручения.
\end{prop}
\begin{proof}
Простая проверка.
\end{proof}
\begin{exa}
Из формулы (\ref{eintof}) следует, что производная Ли от $1$-формы
$A=dx^\al A_\al$ имеет вид
$$
  \Lie_X A_\al=X^\bt\pl_\bt A_\al+\pl_\al X^\bt A_\bt.
$$
При этом разность с ковариантной производной вдоль векторного поля имеет
другой знак по сравнению с (\ref{ediflv})
\begin{equation*}                                                    \tag*{\qed}
  \Lie_X A_\al-\nb_X A_\al=(\nb_\al X^\bt+X^\g T_{\al\g}{}^\bt)A_\bt.
\end{equation*}
\end{exa}
\begin{defn}
{\em Дифференцированием} $D$ тензорной алгебры $\CT(\MM)$ называется линейный
эндоморфизм $\CT(\MM)$, удовлетворяющий следующим свойствам:

1) $D$ сохраняет тип тензорных полей: $D\CT^r_s(\MM)\subset\CT^r_s(\MM)$.

2) \parbox[t]{.92\linewidth}{Дифференцирование удовлетворяет правилу Лейбница.
Если $Y$ и $Z$ -- два произвольных тензорных поля на $\MM$, то
\begin{equation*}
  D(Y\otimes Z)=D Y\otimes Z+Y\otimes DZ.
\end{equation*}}

3) $D$ перестановочен с каждым свертыванием: $DC=CD$.
\qed\end{defn}
\index{Дифференцирование тензорной алгебры (differentiation in tensor algebra)}%
\begin{prop}
Производная Ли $\Lie_X$ является дифференцированием в тензорной алгебре
$\CT(\MM)$.
\end{prop}
\begin{proof}
Прямая проверка.
\end{proof}

Рассмотрим тензорное поле $S=S_\bt{}^\al(x)e_\al\otimes e^\bt$ типа $(1,1)$. В
каждой точке $x\in\MM$ оно задает линейный эндоморфизм касательного $\MT_x(\MM)$
и кокасательного $\MT^*_x(\MM)$ пространств, который естественным образом
продолжается до линейного эндоморфизма $D_S$ тензорной алгебры в этой точке.
Если $T\in\CT^r_s(\MM)$ -- тензорное поле типа $(r,s)$, то в компонентах
\begin{equation*}
\begin{split}
  (D_ST)_{\bt_1\dotsc\bt_s}{}^{\al_1\dotsc\al_r}
  &=S_{\bt_1}{}^\g T_{\g\bt_2\dotsc\bt_s}{}^{\al_1\dotsc\al_r}+\dotsc
  +S_{\bt_s}{}^\g T_{\bt_1\dotsc\bt_{s-1}\g}{}^{\al_1\dotsc\al_r}-
\\
  &-T_{\bt_1\dotsc\bt_s}{}^{\g\al_2\dotsc\al_r}S_\g{}^{\al_1}-\dotsc
  -T_{\bt_1\dotsc\bt_s}{}^{\al_1\dotsc\al_{r-1}\g}S_\g{}^{\al_r}.
\end{split}
\end{equation*}
Тогда $D_S$ представляет собой дифференцирование тензорной алгебры $\CT(\MM)$,
индуцированное тензорным полем $S$ типа $(1,1)$.
\begin{theorem}                                                   \label{tditea}
Каждое дифференцирование $D$ тензорной алгебры $\CT(\MM)$ допускает единственное
разложение
\begin{equation*}
  D=\Lie_X+D_S,
\end{equation*}
где $X$ есть векторное поле, а $S$ -- тензорное поле типа $(1,1)$.
\end{theorem}
\begin{proof}
См., например, \cite{KobNom6369R}.
\end{proof}
\begin{com}
Поскольку любое дифференцирование $D$ линейно отображает алгебру функций
$\CC^\infty(\MM)$ в себя, и выполняется правило Лейбница, то векторное поле
$X$ существует и определяется единственным образом из условия $Df=Xf$ для всех
$f\in\CC^\infty(\MM)$.
\qed\end{com}
\begin{cor}
Разность двух произвольных дифференцирований есть дифференцирование,
индуцированное некоторым тензорным полем типа $(1,1)$.
\qed\end{cor}

Таким образом, у нас есть два дифференцирования в тензорной алгебре $\CT(\MM)$:
производная Ли вдоль векторного поля $\Lie_X$ и дифференцирование $D_S$,
индуцированное тензорным полем $S$ типа $(1,1)$. В разделе \ref{saffco} будет
введено еще одно дифференцирование в тензорной алгебре -- ковариантное
дифференцирование $\nb$. Согласно теореме \ref{tditea}, эти три вида
дифференцирований связаны между собой. Нетрудно убедиться, что в общем случае
производная Ли $\Lie_X$ и ковариантная производная вдоль векторного поля $\nb_X$
связаны следующим соотношением
\begin{equation*}
\begin{split}
  &(\Lie_X-\nb_X)Y_{\bt_1\dotsc\bt_s}{}^{\al_1\dotsc\al_r}=
\\
  &=(\nb_{\bt_1}X^\g+T_{\bt_1\dl}{}^\g X^\dl)
  Y_{\g\bt_2\dotsc\bt_s}{}^{\al_1\dotsc\al_r}{}+\dotsc
  +(\nb_{\bt_s}X^\g+T_{\bt_s\dl}{}^\g X^\dl)
  Y_{\bt_1\dotsc\bt_{s-1}\g}{}^{\al_1\dotsc\al_r}
\\
  &-Y_{\bt_1\dotsc\bt_s}{}^{\g\al_2\dotsc\al_r}
  (\nb_\g X^{\al_1}+T_{\g\dl}{}^{\al_1}X^\dl)-\dotsc
  -Y_{\bt_1\dotsc\bt_s}{}^{\al_1\dotsc\al_{r-1}\g}
  (\nb_\g X^{\al_r}+T_{\g\dl}{}^{\al_r}X^\dl).
\end{split}
\end{equation*}
То есть разность ковариантного дифференцирования и производной Ли является
дифференцированием, индуцированном тензорным полем с компонентами
$S_\bt{}^\al:=\nb_\bt X^\al+T_{\bt\g}{}^\al X^\g$.

Как следствие этой формулы или непосредственной проверкой можно убедиться в том,
что в правой части (\ref{elidet}) все частные производные можно выразить через
ковариантные. Это естественно, т.к.\ производная Ли является инвариантным
оператором. В этом случае появляются дополнительные слагаемые, содержащие тензор
кручения для каждого тензорного индекса. Если кручение аффинной связности равно
нулю, то все частные производные в производной Ли (\ref{elidet}) можно просто
заменить на ковариантные.
\begin{exa}
Для иллюстрации приведем формулу производной Ли от компонент тензора
$Y_\al{}^\bt$ типа (1,1)
\begin{equation*}
\begin{split}
  \Lie_X Y_\al{}^\bt=&X^\g\pl_\g Y_\al{}^\bt+\pl_\al X^\g Y_\g{}^\bt
  -Y_\al{}^\g\pl_\g X^\bt
\\
  =&X^\g\nb_\g Y_\al{}^\bt+\nb_\al X^\g Y_\g{}^\bt
  -Y_\al{}^\g\nb_\g X^\bt
\\
  &+X^\g Y_\al{}^\dl T_{\dl\g}{}^\bt-X^\g Y_\dl{}^\bt T_{\al\g}{}^\dl=
\\
  =&X^\g\widetilde\nb_\g Y_\al{}^\bt+\widetilde\nb_\al X^\g Y_\g{}^\bt
  -Y_\al{}^\g\widetilde\nb_\g X^\bt.
\end{split}
\end{equation*}
Аналогичные формулы справедливы для произвольных тензорных полей.
\qed\end{exa}
\begin{com}
Производная Ли не зависит ни от метрики, ни от аффинной связности, которые
могут быть заданы на многообразии совершенно независимо.
\qed\end{com}
\begin{exa}
Приведем также явное выражение для производной Ли от метрики
\begin{equation*}
  \Lie_X g_{\al\bt}=X^\g\pl_\g g_{\al\bt}
  +\pl_\al X^\g g_{\g\bt}+\pl_\bt X^\g g_{\al\g}.
\end{equation*}
Это выражение можно переписать в виде
\begin{equation}                                                  \label{elidme}
  \Lie_X g_{\al\bt}=\widetilde\nb_\al X_\bt+\widetilde\nb_\bt X_\al,
\end{equation}
где $\widetilde\nb_\al$ -- ковариантная производная со связностью
Леви--Чивиты (раздел \ref{stornm}).
\qed\end{exa}
\begin{defn}
Пусть $T\in\CT^r_s(\MM)$ и  $X\in\CX(\MM)$ -- произвольное тензорное поле типа
$(r,s)$ и полное векторное поле на многообразии $\MM$. Полное векторное поле
порождает однопараметрическую группу преобразований $s(t,x)$. Если значение
тензорного поля $T(s)$ в точке $s(t,x)$ равно $\tilde s(t,x)T(x)$, то мы
говорим, что тензорное поле $T(x)$ {\em инвариантно} при действии
однопараметрической группы преобразований.
\qed\end{defn}
\index{Инвариантное тензорное поле (invariant tensor field)}%
\index{Тензорное поле инвариантное (invariant tensor field)}%

Из определения производной Ли следует
\begin{prop}
Тензорное поле $T\in\CT^r_s(\MM)$ инвариантно относительно однопараметрической
группы преобразований $s(t,x)$, порожденной векторным полем $X$, для всех $t$
тогда и только тогда, когда производная Ли равна нулю, $\Lie_X T=0$.
\end{prop}
\chapter{Дифференциальные формы и интегрирование                 \label{sdifin}}
В дифференциальной геометрии и приложениях важную роль играют дифференциальные
формы. Они используются для определения связностей на расслоениях,
интегрирования по многообразию, характеристических классов, когомологий де Рама,
в гамильтоновом формализме и других областях. В настоящей главе будут
даны определения и рассмотрены основные свойства дифференциальных форм.
Будут также сформулированы две фундаментальные теоремы: теорема Дарбу
и формула Стокса, играющие исключительно важную роль в приложениях.
\section{Внешняя алгебра}
Рассмотрим векторное и сопряженное к нему пространства $\MV$ и $\MV^*$. В
разделе \ref{seucve} над произвольным векторным пространством, например, над
$\MV^*$, была построена тензорная алгебра $\otimes\MV^*$. В этой алгебре
существуют двусторонние идеалы, и соответствующие фактор пространства также
представляют собой алгебры. Наиболее важными из этих фактор алгебр являются
внешняя алгебра и алгебры Клиффорда. В настоящем разделе мы рассмотрим внешнюю
алгебру $\Lm(\MV)$, которая лежит в основе теории дифференциальных форм.

Рассмотрим $n$-мерное векторное пространство $\MV$ над полем вещественных чисел
с базисом $e_a$, $a=1,\dotsc,n$. Дуальный базис сопряженного пространства
$\MV^*$ обозначим через $e^a$, $e^a(e_b)=\dl^a_b$. В следующем разделе мы
отождествим пространства $\MV$ и $\MV^*$ с касательным $\MT_x(\MM)$ и
кокасательным $\MT_x^*(\MM)$ пространствами в фиксированной точке $x\in\MM$
многообразия $\MM$. Пока же под $\MV$ будем понимать абстрактное векторное
пространство.

Начнем с длинного, но конструктивного определения внешней алгебры над
векторным пространством $\MV$. Сначала введем вспомогательные понятия.
\begin{defn}
Множество полностью антисимметричных ковариантных тензоров типа $(0,r)$,
$0\le r\le n$, называется множеством {\em форм степени} $r$ над векторным
пространством $\MV$ и обозначается $\Lm_r(\MV)$.
\end{defn}
\index{Форма степени $r$ (form of degree $r$, $r$-form)}%
В данном определении антисимметрия формы означает следующее. Пусть
$A\in\Lm_r(\MV)$. Тогда ее значение на произвольном наборе векторов
$X_1,\dotsc,X_r\in\MV$ антисимметрично относительно перестановки любой пары
векторов:
\begin{equation*}
  A(X_1,\dotsc,X_i,\dotsc,X_j,\dotsc,X_r)
  =-A(X_1,\dotsc,X_j,\dotsc,X_i,\dotsc,X_r),\qquad 1\le i<j\le r.
\end{equation*}
В компонентах это условие означает антисимметрию относительно перестановки
любой пары индексов.

Множество форм $\Lm_r(\MV)\subset\otimes\MV^*$ степени $r$ является векторным
пространством над полем вещественных чисел размерности $C^r_n$.

По определению, вещественные числа (скаляры) образуют множество $0$-форм,
$\Lm_0(\MV)=\MR$. Отметим, что 1-формы $A\in\Lm_1(\MV)=\MV^*$ (ковекторы
или линейные формы) являются элементами сопряженного пространства и
представляются в виде $A=e^a A_a$. Максимальная степень формы совпадает
с размерностью векторного пространства $n$, поскольку для более высоких
степеней по крайней мере два индекса будут совпадать, а это невозможно
для антисимметричных тензоров.

Из определения тензора следует, что $r$-форма $A\in\Lm_r(\MV)$ представляет
собой полилинейное отображение $r$ экземпляров векторного пространства
$\MV$ в вещественную прямую
\begin{equation*}
  A:\qquad \underbrace{\MV\times\dotsc\times\MV}_r\quad \rightarrow\quad \MR.
\end{equation*}
{\em Полилинейность} означает, что значение $r$-формы на $r$ векторах
\index{Полилинейность (multilinearity)}%
$X_1,\dotsc,X_r\in\MV$ линейно по каждому аргументу при фиксированных
остальных.

Ввиду антисимметрии $r$-форм относительно перестановки индексов, они имеют
меньше независимых компонент, чем ковариантные тензоры типа $(0,r)$. Введем
обозначение
\begin{equation}                                                  \label{ebakfo}
  e^{a_1}\wedge\dotsc\wedge e^{a_r}
  :=\sum_\s e^{\s(a_1}\otimes\dotsc\otimes e^{a_r)}\sgn\s,
\end{equation}
где символ $\otimes$ обозначает тензорное произведение, сумма берется по всем
перестановкам $\s(a_1\dotsc a_r)$ индексов, а $\sgn\s=\pm1$ обозначает знак
перестановки. Здесь мы ввели знак внешнего умножения $\wedge$, которое будет
определено ниже. Пока он рассматривается просто, как некоторый символ.

По-построению, выражение (\ref{ebakfo}) является ковектором типа $(0,r)$ и
антисимметрично относительно перестановки любой пары индексов:
\begin{equation*}
  e^{\s(a_1}\wedge\dotsc\wedge e^{a_r)}=e^{a_1}\wedge\dotsc\wedge e^{a_r}\sgn\s.
\end{equation*}
Поэтому множество ковекторов с упорядоченным набором индексов
\begin{equation}                                                  \label{ebarfo}
  e^{a_1}\wedge\dotsc\wedge e^{a_r},\qquad
  1\le a_1<\dotsc<a_r\le n,
\end{equation}
образует базис в пространстве $r$-форм.
Тогда произвольная форма степени $r$ имеет вид
\begin{align}                                                     \label{erform}
  A&=\sum_{a_1<\dotsc<a_r}
  e^{a_1}\wedge\dotsc\wedge e^{a_r}A_{a_1\dots a_r},
\\                                                                \label{erfoba}
  &=\frac1{r!}e^{a_1}\wedge\dots\wedge e^{a_r}A_{a_1\dotsc a_r}.
\end{align}
В разложении (\ref{erform}) при фиксированной выборке индексов суммирование
по индексам отсутствует, а сумма подразумевается только по различным выборкам.
В последнем выражении (\ref{erfoba}) суммирование проводится по всем
значениям индексов, и поэтому введен компенсирующий множитель $1/r!$.
Это часто бывает удобнее, поскольку запись $r$-формы $A$ в виде
(\ref{erfoba}) имеет инвариантный вид в отличие от (\ref{erform}).
\begin{exa}
Рассмотрим 2-форму $A=e^a\otimes e^b A_{ab}$, где $A_{ab}=-A_{ba}$. Тогда
\begin{equation*}
  A=\frac12e^a\otimes e^b(A_{ab}-A_{ba})=\frac12 e^a\wedge e^b A_{ab},
\end{equation*}
где
\begin{equation}                                                  \label{eweprd}
  e^a\wedge e^b=-e^b\wedge e^a
  =e^a\otimes e^b-e^b\otimes e^a.\qed
\end{equation}
\end{exa}
В компонентах значение $r$-формы $A$ на векторных полях
$\lbrace X_1,\dotsc,X_r\rbrace$ определяется сверткой компонент:
\begin{equation*}
  A(X_1,\dotsc,X_r):=X^{a_1}_1\dotsc X^{a_r}_r A_{a_1\dotsc a_r}.
\end{equation*}

Построим $2^n$-мерное векторное пространство, равное прямой сумме форм
всех степеней
\begin{equation}                                                  \label{edeffs}
  \Lm(\MV):=\bigoplus_{r=0}^n\Lm_r(\MV).
\end{equation}
Базис этого пространства имеет вид
\begin{equation}                                                  \label{ebaclf}
  1,\quad e^a,\quad e^{a_1}\wedge e^{a_2},\quad
  \dotsc,\quad e^{a_1}\wedge\dotsc\wedge e^{a_r},
  \dotsc,\quad e^1\wedge\dotsc\wedge e^n,
\end{equation}
где последовательность $a_1\dots a_r$ является упорядоченным набором
различных индексов, расположенных в порядке возрастания
$1\le a_1<\dots<a_r\le n$. Первый базисный вектор является единицей,
поскольку поле вещественных чисел $\MR$ является подпространством
рассматриваемого векторного пространства. Последний базисный вектор
единственен, т.к.\ содержит ровно $n$ индексов.
Элементы из $\Lm(\MV)$ представимы в виде
\begin{equation}                                                  \label{evesfo}
  A=\sum_{r=0}^n \frac1{r!}
  e^{a_1}\wedge\dots\wedge e^{a_r}
  A_{a_1\dots a_r},
\end{equation}
где подразумевается суммирование по всем значениям индексов.

Введем на множестве форм {\em внешнее умножение}, отображающее
$\Lm(\MV)\rightarrow\Lm(\MV)$. Пусть $A$ и $B$ -- формы фиксированных
\index{Внешнее умножение (exterior product, wedge product)}%
\index{Умножение внешнее (exterior product, wedge product)}%
степеней $r$ и $s$, соответственно. Их тензорное произведение $A\otimes B$
является ковариантным тензором типа $(0,r+s)$, однако не будет формой, т.к.\
не будет антисимметрично относительно всех перестановок индексов. Чтобы
исправить ситуацию после тензорного умножения форм необходимо произвести
полную антисимметризацию. Тогда внешнее умножение станет отображением
$\Lm_r\times\Lm_s\rightarrow\Lm_{r+s}$.
\begin{defn}
Внешним произведением двух форм $A\in\Lm_r(\MV)$ и $B\in\Lm_s(\MV)$ называется
форма $C\in\Lm_{r+s}(\MV)$, построенная по правилу:
\begin{align}                                                     \label{extpro}
  A\wedge B&=\frac1{(r+s)!}e^{a_1}\wedge\dots\wedge e^{a_{r+s}}
  C_{a_1\dotsc a_{r+s}},
\\ \intertext{где компоненты внешнего произведения равны}              \nonumber
  C_{a_1\dotsc a_{r+s}}&:=\frac1{r!s!}\sum_\s
  A_{\s(a_1\dotsc a_r}B_{a_{r+1}\dotsc a_{r+s})}\sgn\s,
\\                                                                \label{ecoepr}
  &=C^r_{r+s}A_{[a_1\dotsc a_r}B_{a_{r+1}\dotsc a_{r+s}]}.
\end{align}
Здесь $\s(a_1\dotsc a_r a_{r+1}\dotsc a_{r+s})$ обозначает перестановку
$\s$ всех индексов $(a_1\dotsc a_{r+s})$, $\sgn\s$ -- знак перестановки,
суммирование ведется по всем перестановкам, а квадратные скобки обозначают
антисимметризацию всех индексов.
\qed\end{defn}
Формулу внешнего произведения (\ref{extpro}) можно переписать в эквивалентном
виде
\begin{equation}                                                  \label{ewpprd}
  A\wedge B=\frac1{r!s!}e^{a_1}\wedge\dots\wedge e^{a_{r+s}}
  A_{a_1\dotsc a_r}B_{a_{r+1}\dotsc a_{r+s}}.
\end{equation}
\begin{com}
Множитель $1/(r+s)!$ в (\ref{extpro}) связан с тем, что суммирование
проводится по всем значениям индексов. Необходимость введения множителя
$1/r!s!$ в (\ref{ecoepr}) будет ясна из дальнейшего.
\qed\end{com}
\begin{exa}
Внешнее умножение произвольной $r$-формы $A\in\Lm_r(\MV)$ на $0$-форму (число)
$f\in\MR=\Lm_0(\MV)$ сводится к обычному умножению компонент формы $A$ на число,
\begin{equation*}
  f\wedge A=\frac1{r!}e^{a_1}\wedge\dots\wedge e^{a_r}
  (fA_{a_1\dots a_r}). \tag*{\qed}
\end{equation*}
\end{exa}
\begin{exa}
Внешнее произведение двух 1-форм $A=e^a A_a$ и $B=e^a B_a$ равно
\begin{equation}                                                  \label{ewepro}
  A\wedge B=\frac12e^a\wedge e^b (A_a B_b-A_b B_a)
  =e^a\wedge e^b A_a B_b. \qed
\end{equation}
\end{exa}
\begin{exa}
Внешнее произведение 1-формы $A=e^a A_a$ на 2-форму
$B=\frac12e^a\wedge e^b B_{ab}$ равно
\begin{align}                                                     \nonumber
  A\wedge B&=\frac16e^a\wedge e^b \wedge e^c
  (A_a B_{bc}+A_b B_{ca}+A_c B_{ab})
\\                                                                \label{eweprt}
  &=\frac12e^a\wedge e^b\wedge e^c A_a B_{bc}. \qed
\end{align}
\end{exa}
Если суммарная степень форм превосходит размерность векторного
пространства $r+s>n$, то внешнее произведение этих форм дает нуль.

Можно проверить, что внешнее умножение обладает следующими свойствами:
\begin{align}
  A\wedge B &\quad \text{линейно по}~A~\text{и}~B,
\\                                                                \label{exprtw}
  A\wedge B &=(-1)^{rs}B\wedge A,
\\
  (A\wedge B)\wedge C &=A\wedge(B\wedge C).
\end{align}
Первые два свойства очевидны. Из второго свойства (\ref{exprtw}) следует, что,
если $A$ -- форма нечетной степени, то $A\wedge A=0$.

Третье свойство доказывается прямым вычислением.
Для трех форм, степеней $r,s$ и $t$, соответственно, справедливо равенство
\begin{align*}
  &(A\wedge B)\wedge C=A\wedge(B\wedge C)=A\wedge B\wedge C
\\
  &=\frac1{(r+s+t)!}e^{a_1}\wedge\dotsc\wedge e^{a_{r+s+t}}
  \frac1{r!s!t!}\sum_\s\sgn(\s)
  A_{\s(a_1\dotsc a_r}B_{a_{r+1}\dotsc a_{r+s}}
  C_{a_{r+s+1}\dotsc a_{r+s+t})}
\\
  &=\frac1{r!s!t!}e^{a_1}\wedge\dots\wedge e^{a_{r+s+t}}
  A_{a_1\dots a_r}B_{a_{r+1}\dotsc a_{r+s}}
  C_{a_{r+s+1}\dotsc a_{r+s+t}}.
\end{align*}
Последнее выражение симметрично относительно перестановок $r,s$ и $t$, что
соответствует ассоциативности внешнего умножения и определяют выбор
коэффициента в равенстве (\ref{ecoepr}): они подобраны таким образом, чтобы
внешнее умножение было ассоциативным.
\begin{defn}
Форма фиксированной степени называется {\em однородной}. Прямая сумма форм
разных степеней называется {\em неоднородной}.
\qed\end{defn}
\index{Однородная форма (homogeneous form)}%
\index{Форма однородная (homogeneous form)}%
\index{Неоднородная форма (inhomogeneous form)}%
\index{Форма неоднородная (inhomogeneous form)}%
Внешнее умножение (\ref{extpro}), определенное для однородных форм, продолжается
на неоднородные формы общего вида (\ref{evesfo}) по линейности. Тем самым
множество форм $\Lm(\MV)$ над произвольным векторным пространством $\MV$,
$\dim\MV=n$, с операцией внешнего умножения представляет собой ассоциативную
алгебру над полем вещественных чисел.
\begin{exa}
Пусть $A,B,C,D$ -- четыре однородные формы различных степеней, тогда
\begin{equation*}                                                 \tag*{\qed}
  (A\oplus B)\wedge(C\oplus D):=(A\wedge C)\oplus(B\wedge C)
  \oplus(A\wedge D)\oplus(B\wedge D).
\end{equation*}
\renewcommand{\qed}{}\end{exa}

\begin{defn}
Множество форм вида (\ref{evesfo}) с внешним умножением (\ref{extpro}),
продолженным на формы общего вида по линейности, называется
{\em внешней алгеброй} $\Lm(\MV)$ над векторным пространством $\MV$,
$\dim\MV=n$.
\qed\end{defn}
\index{Внешняя алгебра (exterior algebra)}%
\index{Алгебра внешняя (exterior algebra)}%
Эта алгебра ассоциативна, градуирована, антикоммутативна и содержит единичный
элемент $1\in\Lm_0(\MV)$. Внешняя алгебра является алгеброй Грассмана с
образующими $e^a$ (частным случаем алгебр Клиффорда).

В обозначении базисных векторов внешней алгебры (\ref{ebarfo})
был использован знак внешнего умножения, которое было определено позже.
Покажем, что это обозначение обосновано. Рассмотрим две базисные
1-формы $A^a=e^a$ и $B^b=e^b$. Тогда их внешнее произведение равно
\begin{equation}                                                  \label{expbav}
  A^a\wedge B^b=e^a\wedge e^b.
\end{equation}
Более общо, для двух базисных форм
$A^{a_1\dotsc a_r}=e^{a_1}\wedge\dotsc\wedge e^{a_r}$ и
$B^{b_1\dotsc b_s}=e^{b_1}\wedge\dotsc\wedge e^{b_s}$ справедливо равенство
$$
  A^{a_1\dotsc a_r}\wedge B^{b_1\dotsc b_s}
  =e^{a_1}\wedge\dotsc\wedge e^{a_r}
  \wedge e^{b_1}\wedge\dotsc\wedge e^{b_s}.
$$
Это оправдывает обозначение для базиса (\ref{ebakfo}) и выбор
коэффициентов для внешнего умножения (\ref{ecoepr}).
\begin{com}
Для определения ассоциативного умножения в пространстве форм $\Lm(\MV)$
выбор коэффициента в определении внешнего умножения не является единственным.
При определении внешнего умножения (\ref{extpro}), (\ref{ecoepr})
можно было бы выбрать другой коэффициент в (\ref{ecoepr}):
\begin{equation}                                                  \label{ecoepd}
  C_{a_1\dotsc a_{r+s}}:=\frac1{(r+s)!}\sum_\s\sgn\s
  A_{\s(a_1\dotsc a_r}B_{a_{r+1}\dotsc a_{r+s})}.
\end{equation}
В этом случае произведение трех форм
\begin{align*}
  &(A\wedge B)\wedge C=A\wedge(B\wedge C)=A\wedge B\wedge C
\\
  &=\frac1{(r+s+t)!}e^{a_1}\wedge\dots\wedge e^{a_{r+s+t}}
  A_{a_1\dots a_r}B_{a_{r+1}\dotsc a_{r+s}}C_{a_{r+s+1}\dotsc a_{r+s+t}}
\end{align*}
также симметрично относительно перестановок $r,s,t$ и, значит,
ассоциативно. Однако в произведении базисных векторов (\ref{expbav})
появился бы дополнительный множитель.
\qed\end{com}
К определению внешней алгебры можно подойти с другой, более абстрактной,
точки зрения. Рассмотрим две произвольные 1-формы $A,B\in\Lm_1(\MV)=\MV^*$.
Тогда для их тензорного произведения справедливо тождество
\begin{equation}                                                  \label{eteexp}
  A\otimes B=A\wedge B+\frac12\big[(A+B)\otimes(A+B)-A\otimes A-B\otimes B\big].
\end{equation}
Выражение в квадратных скобках является элементом тензорной алгебры,
и каждое слагаемое представляет собой тензорное произведение одинаковых
1-форм. Рассмотрим элементы более общего вида
$X\otimes A\otimes A\otimes Y$, где $A\in\MV^*$ -- линейная форма, а
$X,Y\in\otimes\MV^*$ -- произвольные элементы тензорной алгебры. Нетрудно
проверить, что элементы такого вида образуют двусторонний идеал $\MI$ в
тензорной алгебре $\otimes\MV^*$, которая рассматривается, как кольцо по
отношению к сложению и тензорному умножению. Тогда из представления
(\ref{eteexp}) следует, что внешняя алгебра является фактор-пространством
$$
  \Lm(\MV)=\otimes\MV^*/\MI.
$$
То есть два элемента $A,B\in\otimes\MV^*$ являются эквивалентными
$A\sim B$, если $A=B+C$, где $C\in\MI$. Выражение в квадратных скобках
в (\ref{eteexp}) принадлежит идеалу, и, значит,
$$
  A\otimes B\sim A\wedge B.
$$
Таким образом каждый элемент внешней алгебры взаимно однозначно
определяет класс эквивалентности в тензорной алгебре.

Сформулируем четыре утверждения, доказательства которых сводятся к
алгебраическим выкладкам.
\begin{prop}
Для того, чтобы набор 1-форм $\lbrace A_i\rbrace$, $i=1,\dotsc,\Sn$,
был линейно зависим, необходимо и достаточно, чтобы выполнялось равенство
\begin{equation}                                                  \label{elidef}
  A_1\wedge\dotsc\wedge A_\Sn=0.
\end{equation}
\end{prop}
\begin{prop}
Пусть $\lbrace A_i\rbrace$ и $\lbrace B_i\rbrace$, $i=1,\dotsc,\Sn$, --
два набора 1-форм таких, что выполнено равенство
\begin{equation*}
  \sum_{i=1}^\Sn A_i\wedge B_i=0.
\end{equation*}
Если 1-формы $\lbrace A_i\rbrace$ линейно независимы, то 1-формы
$\lbrace B_i\rbrace$ можно выразить в виде линейной комбинации 1-форм
$\lbrace A_i\rbrace$:
\begin{equation*}
  B_i=\sum_{j=1}^\Sn C_{ij}A_j,
\end{equation*}
причем $C_{ij}=C_{ji}$.
\end{prop}
\begin{prop}
Пусть $\lbrace A_i\rbrace$, $i=1,\dotsc,\Sn$, -- набор $\Sn$ линейно независимых
1-форм и $B\in\Lm_s$. Для того, чтобы $s$-форму $B$ можно было представить в
виде линейной комбинации
\begin{equation*}
  B=A_1\wedge C_1+\dotsc+A_\Sn\wedge C_\Sn,
\end{equation*}
где $\lbrace C_i\rbrace$ -- некоторый набор $(s-1)$-форм, необходимо и
достаточно, чтобы
\begin{equation*}
  A_1\wedge\dotsc\wedge A_\Sn\wedge B=0.
\end{equation*}
\end{prop}
\begin{com}
При $\Sn+s>n$ эта теорема тривиально выполняется.
\qed\end{com}
\begin{prop}
Пусть $\lbrace A_i,B_i\rbrace$ и $\lbrace A^\prime_i,B^\prime_i\rbrace$,
$i=1,\dotsc,\Sn$, -- два набора по $2\Sn$ 1-форм. Если 1-формы
$\lbrace A_i,B_i\rbrace$ линейно независимы и выполнено равенство
\begin{equation*}
  \sum_{i=1}^\Sn A_i\wedge B_i=\sum_{i=1}^\Sn A^\prime_i\wedge B^\prime_i,
\end{equation*}
то набор 1-форм $\lbrace A^\prime_i,B^\prime_i\rbrace$ также линейно независим
и представляется в виде линейных комбинаций 1-форм $\lbrace A_i,B_i\rbrace$.
\end{prop}

Пусть некоторое семейство 1-форм $\lbrace A_i\rbrace$, $i=1,\dotsc,\Sn$,
выражается в виде линейной комбинации 1-форм $\lbrace B_i\rbrace$:
\begin{equation*}
  A_i=M_i{}^j B_j,
\end{equation*}
где $M_i{}^j$ -- некоторая квадратная $\Sn\times\Sn$ матрица. Тогда нетрудно
проверить следующую формулу
\begin{equation*}
  A_1\wedge\dotsc\wedge A_\Sn=\det(M_i{}^j)B_1\wedge\dotsc\wedge B_\Sn.
\end{equation*}

\begin{defn}
Пусть $A\in\Lm_r(\MV)$ -- внешняя $r$-форма, тогда внешнее произведение
\begin{equation*}
  A^s:=\underbrace{A\wedge\dotsc\wedge A}_s
\end{equation*}
называется {\em внешней степенью} $r$-формы $A$.

На множестве 2-форм $\Lm_2(\MV)$ можно определить {\em внешнюю экспоненту}
$\widehat\exp$, которая каждой 2-форме $A\in\Lm_2(\MV)$ ставит в соответствие
прямую сумму форм четной степени по правилу
\begin{equation}                                                  \label{expfor}
  \widehat\exp A:=1\oplus A\oplus\frac12A\wedge A\oplus
  \frac1{3!}A\wedge A\wedge A\oplus\dots  \qed
\end{equation}
\renewcommand{\qed}{}\end{defn}
\index{Внешняя степень (external power)}%
\index{Степень внешняя (external power)}%
\index{Внешняя экспонента (exterior exponent)}%
\index{Экспонента внешняя (exterior exponent)}%

Ряд (\ref{expfor}) содержит конечное число слагаемых, т.к.\ максимальное число
сомножителей в последнем отличном от нуля слагаемом не может превосходить $n/2$.
Внешняя экспонента обладает следующими свойствами:
\begin{align*}
  \widehat\exp A\wedge\widehat\exp B
  &=\widehat\exp(A+B),\qquad \forall A,B\in\Lm_2(\MV).
\\
  \frac d{dt}\widehat\exp(tA)&=A\wedge\widehat\exp(tA)
  =\widehat\exp(tA)\wedge A.
\end{align*}
При доказательстве этих формул использовано равенство $A\wedge B=B\wedge A$,
справедливое для форм четной степени.

\begin{defn}
В пространстве векторов и $r$-форм введем {\em внутреннее умножение} как
билинейное отображение
$$
  \inm:\quad \MV\times\Lm_r(\MV)\ni\quad(X,A)\mapsto\inm_X A
  \quad\in\Lm^{r-1}(\MV).
$$
Оно определяется следующим образом. Пусть $X\in\MV$ и $A\in\Lm_r(\MV)$.
Положим
\begin{equation}                                                  \label{einder}
  \inm_XA(X_1,\dotsc,X_{r-1}):=A(X,X_1,\dotsc,X_{r-1}).
\end{equation}
Другими словами, компоненты вектора $X$ нужно просто свернуть с первым
индексом $r$-формы. Правая часть выражения (\ref{einder})
$(r-1)$-линейна и антисимметрична, как функция $X_1,\dotsc,X_{r-1}$,
а также билинейна по $X$ и $A$. Удобно считать, что $\inm_XA=0$, если
$A\in\Lm_0$ -- форма нулевой степени. Внутреннее умножение обозначают
также $\inm_XA=X\rfloor A$. Определение (\ref{einder}) можно переписать
в эквивалентном виде
\begin{equation}                                                  \label{einpro}
  \inm_XA=X\rfloor A=\frac1{(r-1)!}e^{a_1}\wedge\dotsc\wedge e^{a_{r-1}}
  X^b A_{ba_1\dotsc a_{r-1}}. \qed
\end{equation}
\end{defn}
\index{Внутреннее умножение (internal multiplication)}%
\index{Умножение внутреннее (internal multiplication)}%

Внутреннее умножение линейно
\begin{align*}
  \inm_{X+Y}&=\inm_X+\inm_Y,
\\
  \inm_{aX}&=a\inm_X,\qquad a\in\MR,
\end{align*}
и антисимметрично
\begin{equation*}
  \inm_X\inm_Y=-\inm_Y\inm_X\quad \Leftrightarrow\quad \inm_X\inm_X=0,
\end{equation*}
для всех $X,Y\in\MV$.

Нетрудно проверить, что внутреннее умножение удовлетворяет правилу Лейбница
$$
  \inm_X(A\wedge B)=\inm_XA\wedge B+(-1)^rA\wedge \inm_XB,\qquad A\in\Lm_r(\MV).
$$
Поэтому внутреннее умножение называют также {\em внутренним
дифференцированием}.
\index{Внутреннее дифференцирование (internal differentiation)}%
\index{Дифференцирование внутреннее (internal differentiation)}%

Внутреннее умножение удобно использовать для характеристики внешней алгебры,
построенной на фактор пространстве $\MV/\MU$, где $\MU\subset\MV$ -- некоторое
линейное подпространство в $\MV$. А именно, внешняя алгебра $\Lm(\MV/\MU)$
состоит из тех форм $A\in\Lm(\MV)$, для которых выполнено равенство
$\inm_X A=0$ для всех $X\in\MU$. Верно и обратное утверждение. Пусть задано
подпространство $\MU\subset\MV$. Тогда множество векторов $X$, удовлетворяющих
условию $\inm_X A=0$ для всех $A\in\Lm(\MV/\MU)$ образует подпространство,
которое совпадает с $\MU$.

В заключение обсудим свойства внешнего умножения при отображении векторных
пространств. Пусть $f$ -- линейное отображение векторных пространств
\begin{equation*}
  f:\quad\MV~\rightarrow~\MW,
\end{equation*}
и пусть на векторном пространстве-мишени $\MW$ заданы две формы
$A,B\in\Lm(\MW)$, тогда для возврата отображения справедливо равенство
\begin{equation*}
  f^*(A\wedge B)=(f^*A)\wedge(f^*B)~\in\Lm(\MV),
\end{equation*}
которое просто проверяется.
\section{Дифференциальные формы                                  \label{sdfint}}
Рассмотрим многообразие $\MM$ размерности $\dim\MM=n$. В каждой точке
многообразия мы имеем касательное и кокасательное векторные пространства.
Построим над касательным векторным пространством $\MV=\MT_x(\MM)$ с
координатным базисом $\pl_\al$ в каждой точке $x\in\MM$ векторное пространство
форм степени $r$: $\Lm_{r,x}(\MM)$, как это было сделано в предыдущем разделе.
Объединение форм степени $r$ по всем точкам многообразия
$\bigcup_{x\in\MM}\Lm_{r,x}(\MM)$ приведет к векторному расслоению, сечениями
которого являются полностью антисимметричные ковариантные тензорные поля ранга
$r$ (дифференциальные формы). Очевидно, что множество дифференциальных форм
фиксированной степени с поточечным сложением и умножением на числа образует
бесконечномерное линейное пространство. Прямая сумма по $r$ множества всех
дифференциальных форм даст внешнюю алгебру дифференциальных форм $\Lm(\MM)$ на
многообразии $\MM$, в которой все операции (умножение на числа, сложение и
внешнее умножение) определены поточечно. Ясно, что все свойства внешнего
умножения без изменений переносятся на внешнюю алгебру дифференциальных форм, и
мы не будем их переписывать. Отметим изменение обозначений и терминологии.
\begin{defn}
Множество полностью антисимметричных ковариантных тензорных полей типа $(0,r)$,
$0\le r\le n$ называется множеством {\em дифференциальных форм степени} $r$
на многообразии $\MM$ и обозначается $\Lm_r(\MM)\subset\CT_r(\MM)$.
\qed\end{defn}
\index{Дифференциальная форма степени (differential form of degree) $r$}%
В дальнейшем дифференциальные формы на многообразии $\MM$ степени $r$
мы, для краткости, будем называть просто $r$-формами.

По определению, гладкие функции на многообразии образуют множество гладких
$0$-форм, $\Lm_0(\MM)=\CC^\infty(\MM)$. 1-формы $A\in\Lm_1(\MM)=\CT_1(\MM)$
(ковариантные векторные поля $=$ сечения кокасательных расслоений)
представляются в виде
$$
  A(x)=dx^\al A_\al(x),
$$
где дифференциалы $dx^\al$ представляют собой координатный базис
кокасательного пространства, который дуален к $\pl_\al$.
Дифференциальные 1-формы называются также {\em формами Пфаффа}.
\index{Форма Пфаффа (Pfaffian form)}\index{Пфаффова форма (Pfaffian form)}%
Максимальная степень нетривиальной формы совпадает с размерностью многообразия
$n$, поскольку для более высоких степеней по крайней мере два индекса
будут совпадать, а это невозможно для антисимметричных тензоров.
\begin{com}
Напомним, что $\pl_\al$ это сокращенное обозначение набора векторных полей
$(\pl_\al)_p\in\CX(\MU)$, $p\in\MU$, на карте $(\MU,\vf)$ многообразия $\MM$
(см.\ раздел \ref{svecdi}), а $dx^\al=(dx^\al)_p$ -- дуальный базис 1-форм,
$dx^\al(\pl_\bt)=\dl^\al_\bt$.
\qed\end{com}

Как и для векторных пространств, алгебра дифференциальных форм равна
прямой сумме форм всех степеней
\begin{equation}                                                  \label{edeffm}
  \Lm(\MM):=\bigoplus_{r=0}^n\Lm_r(\MM).
\end{equation}
Координатный базис этого пространства имеет вид
\begin{equation}                                                  \label{ebaclm}
  1,\quad dx^\al,\quad dx^{\al_1}\wedge dx^{\al_2},\quad
  \dots,\quad dx^{\al_1}\wedge\dots\wedge dx^{\al_r},
  \dots,\quad dx^1\wedge\dots\wedge dx^n,
\end{equation}
где индексы упорядочены по возрастанию $1\le\al_1<\dotsc\al_r\le r$. Элементы
внешней алгебры $\Lm(\MM)$, в соответствии с разложением (\ref{edeffm}),
представимы в виде
\begin{equation}                                                  \label{evesfm}
  A=\sum_{r=0}^n \frac1{r!}
  dx^{\al_1}\wedge\dots\wedge dx^{\al_r}A_{\al_1\dots\al_r},
\end{equation}
где подразумевается суммирование по всем значениям индексов. Форма $A$
называется {\em дифференцируемой}, если все антисимметричные тензорные поля
$A_{\al_1\dots\al_r}(x)$ (компоненты формы) дифференцируемы. Если не оговорено
противное, мы будем предполагать, что все компоненты форм гладко зависят от
точки многообразия.
\begin{com}
Запись дифференциальной формы в виде (\ref{evesfm}) имеет смысл только в
определенной карте многообразия, где определен координатный базис кокасательного
пространства $dx^\al$. Однако она универсальна, т.к.\ инвариантна относительно
преобразований координат, поскольку в точках пересечения двух карт с
координатами $x^\al$ и $x^{\al'}$ справедливо равенство
\begin{equation*}
  dx^{\al_1}\wedge\dots\wedge dx^{\al_r}A_{\al_1\dots\al_r}
  =dx^{\al'_1}\wedge\dots\wedge dx^{\al'_r}A_{\al'_1\dots\al'_r}.
\end{equation*}
Запись дифференциальных форм в виде (\ref{evesfm}) является общепринятой.
\qed\end{com}

Введем несколько новых понятий, которые являются полезными с точки зрения
дифференциальной геометрии. Конечно, часть из них можно было бы ввести и
для дифференциальных форм над векторным пространством еще в предыдущем разделе.

Можно рассматривать дифференциальные формы со значениями в произвольном
векторном пространстве $\MW$. Пусть $e_i$ -- базис векторного
пространства $\MW$. Тогда $r$-форма $A$ со значениями в $\MW$ имеет вид
$A=A^ie_i$. При каждом значении индекса $i$ коэффициенты этого разложения
представляют собой $r$-формы $A^i\in\Lm_r(\MM)$. По индексу $i$ эти
формы преобразуются, как компоненты вектора из пространства $\MW$.
\begin{exa}
Произвольную $r$-форму на многообразии $\MM$
\begin{equation*}
  A=\frac1{r!}dx^{\al_1}\wedge\dots\wedge dx^{\al_r}A_{\al_1\dots\al_r}
  \in\Lm_r(\MM),
\end{equation*}
можно поточечно рассматривать, как внешнее произведение базисной $r$-формы
$dx^{\al_1}\wedge\dotsc\wedge dx^{\al_r}$ на 0-форму $A_{\al_1\dotsc\al_r}$
со значениями в векторном пространстве
$\MW=\MT^*_x(\MM)\wedge\dotsc\wedge\MT^*_x(\MM)$.
\qed\end{exa}
Пусть на многообразии $\MM$ помимо форм заданы векторные поля. Тогда
значение произвольной $r$-формы $A$ на $r$ векторных полях
$X_1,\dotsc,X_r\in\CX(\MM)$ равно свертке компонент,
$$
  A(X_1,\dots,X_r)=X_1^{\al_1}\dotsc X_r^{\al_r}A_{\al_1\dotsc\al_r}
  \in\CC^\infty(\MM).
$$
(Данное определение не зависит от выбора карты и потому корректно.) Отсюда
следует, что произвольная $r$-форма в каждой точке $x\in\MM$ определяет
полилинейное отображение ($\CC^\infty$-модуль):
$\big(\MT_x(\MM)\big)^r\rightarrow\MR$, которое антисимметрично относительно
перестановки аргументов.
\begin{exa}
Значение 1-формы $A=dx^\al A_\al$ на векторном поле $X=X^\al\pl_\al$ определяет
функцию, которая равна сумме компонент: $A(X)=X^\al A_\al$.
\qed\end{exa}

Пусть $A_i$ и $X_i$, $i=1,\dotsc,r\le n$ -- произвольный набор 1-форм и
векторных полей на $\MM$. Тогда из определения внешнего умножения следует, что
значение внешнего произведения 1-форм на векторных полях равно определителю
матрицы с элементами $A_i(X_j)$
$$
  (A_1\wedge\dots\wedge A_r)(X_1\dotsc X_r)=\det\big(A_i(X_j)\big).
$$

Пусть две 1-формы $A$ и $B$ линейно зависимы, т.е.\ $A=fB$, где
$f\in\Lm_0(\MM)$, тогда их внешнее произведение, очевидно, равно нулю. Верно
также и обратное утверждение, из которого следует, что, если $A\wedge B\ne0$ в
некоторой области $\MU\subset\MM$, то 1-формы линейно независимы в каждой точке
$\MU$. Более общо, справедливо следующее утверждение.
\begin{prop}
Для того, чтобы линейные формы $e^i=dx^\al e_\al{}^i$, $i=1,\dotsc,r\le n$ были
линейно независимы в области $\MU\subset\MM$, необходимо и достаточно, чтобы в
каждой точке этой области $e^1\wedge\dotsc\wedge e^r\ne0$.
\end{prop}

Дифференциальные формы часто бывает удобно рассматривать в неголономном
базисе. Выберем в качестве базиса кокасательного пространства 1-формы
$e^a:=dx^\al e_\al{}^a$, $a=1,\dotsc,n$, где $e_\al{}^a(x)$ -- поле репера,
тогда соответствующий базис $r$-форм имеет вид
\begin{equation}                                        \label{enhoba}
  e^{a_1}\wedge\dots\wedge e^{a_r},\qquad 1\le a_1<\dots<a_r\le n.
\end{equation}

Мы используем для неголономного базиса те же обозначения, что и для базиса
форм над векторным пространством (\ref{ebarfo}). Разница заключается в том,
что теперь базис $e^a(x)$ зависит от точки многообразия, и это всегда ясно
из контекста.

Для того, чтобы 1-формы $e^a$ действительно образовывали базис, необходимо
и достаточно, чтобы
\begin{equation*}
  \det e_\al{}^a\ne0\quad \Leftrightarrow\quad e^1\wedge\dotsc\wedge e^n\ne0
\end{equation*}
во всех точках многообразия $\MM$, что совпадает с определением репера, который
будет введен позже в разделе \ref{scorep}. Компоненты разложения $r$-формы
$$
  A=\frac1{r!}e^{a_1}\wedge\dots\wedge e^{a_r}A_{a_1\dotsc a_r}
$$
по этому базису связаны с компонентами в голономном базисе простым соотношением
$$
  A_{a_1\dots a_r}
  :=e^{\al_1}{}_{a_1}\dotsc e^{\al_r}{}_{a_r}A_{\al_1\dotsc\al_r},
$$
где $e^\al{}_a$ -- матрица обратного репера: $e^\al{}_ae_\al{}^b=\dl_a^b$.
\section{Внешнее дифференцирование                               \label{sextpr}}
Во внешней алгебре дифференциальных форм $\Lm(\MM)$ на многообразии $\MM$
определим {\em внешнее дифференцирование} как линейное отображение
\index{Внешнее дифференцирование (exterior differentiation)}%
\index{Дифференцирование внешнее (exterior differentiation)}%
\begin{equation}                                                   \label{exdir}
  d:\quad \Lm_r(\MM)~\rightarrow~\Lm_{r+1}(\MM)
\end{equation}
множества $r$-форм в множество $(r+1)$-форм. Начнем с локального определения.
Пусть в некоторой карте $(\MU,\vf)$ задана $r$-форма
\begin{equation*}
  A=\frac1{r!}dx^{\al_1}\wedge\dots
  \wedge dx^{\al_r}A_{\al_1\dots\al_r}\in\Lm_r(\MU).
\end{equation*}
Положим
\begin{align}                                                      \nonumber
  dA&:=\frac1{r!}dA_{\al_1\dots\al_r}\wedge dx^{\al_1}
  \wedge\dots\wedge dx^{\al_r}
\\                                                                \label{extder}
  &=\frac1{r!}dx^{\al_1}\wedge\dots\wedge dx^{\al_{r+1}}
  \pl_{\al_1}A_{\al_2\dots\al_{r+1}}.
\end{align}
Здесь под $dA_{\al_1\dots\al_r}=dx^\al\pl_\al A_{\al_1\dots\al_r}$ понимается
обычный дифференциал. Внешнее дифференцирование обладает следующими свойствами:
\begin{align}                                                     \label{eptdfi}
  1)&\qquad \text{если}\quad f~-~0\text{-форма, то}\quad df=dx^\al\pl_\al f,
\\                                                                \label{eptdse}
  2)&\qquad \text{если}\quad A~-~r\text{-форма, то}\quad
  d(A\wedge B)=dA\wedge B+(-1)^rA\wedge dB,
\\                                                                \label{eptdth}
  3)&\qquad d(dA)=0\quad \text{для любой формы}~A.
\end{align}
Первые два свойства легко проверяются, исходя из определения
(\ref{extder}). Доказательство третьего свойства сводится к тому, что
вторая частная производная $\pl^2_{\al\bt}$ симметрична по индексам и
дает нуль при свертке с антисимметричным базисным вектором.

Формула для внешнего дифференцирования (\ref{extder}) является прямым
следствием свойств 1)--3). Действительно, представив $r$-форму как
внешнее произведение 0-формы $A_{\al_1\dotsc\al_n}$ со значениями в
векторном пространстве на базисную $r$-форму
$dx^{\al_1}\wedge\dotsc\wedge dx^{\al_n}$ сразу приходим к (\ref{extder}).
Поэтому она задает единственное дифференцирование во внешней алгебре,
удовлетворяющее условиям 1)--3).

Глобальное определение внешнего дифференцирования сформулируем в виде теоремы.
\begin{theorem}[\bf Внешнее дифференцирование]
Существует единственное линейное отображение (\ref{exdir}) такое, что выполнены
условия (\ref{eptdfi})--(\ref{eptdth}). Оно называется внешним
дифференцированием.
\end{theorem}
\begin{proof}
Выше мы доказали существование и единственность внешнего дифференцирования в
определенной карте. Осталось доказать, что существование и единственность не
зависят от выбора карты. Действительно, компоненты в разложении внешней
производной от $r$-формы (\ref{extder})
$$
  \pl_{[\al_1}A_{\al_2\dots\al_{r+1}]},
$$
где квадратные скобки обозначают антисимметризацию всех индексов, являются
компонентами полностью антисимметричного тензора  типа $(0,r+1)$. Это
утверждение нетривиально, потому что обычная частная производная от компонент
тензора не дает тензора. Для 1-формы имеем следующий закон преобразования
частной производной в области пересечения двух карт
$$
  \pl_{\al'}A_{\bt'}=\frac{\pl x^\al}{\pl x^{\al'}}
  \frac{\pl x^\bt}{\pl x^{\bt'}}\pl_\al A_\bt
  +\frac{\pl^2 x^\al}{\pl x^{\al'}\pl x^{\bt'}}A_\al.
$$
Первое слагаемое соответствует тензорному закону преобразования, а второе --
нет. Однако второе слагаемое исчезает при антисимметризации и мы получаем
тензорный закон преобразования для внешней производной. Аналогично сокращаются
дополнительные слагаемые для произвольной $r$-формы. Подробности доказательства
приведены в \cite{Warner83R}.
\end{proof}
Для практических вычислений внешней производной, которые проводятся в
координатах, достаточно формулы (\ref{extder}).
\begin{exa}
Внешний дифференциал от 1-формы $A=dx^\al A_\al$ равен
\begin{equation}                                                  \label{exdeof}
  dA=dA_\al\wedge dx^\al=\frac12dx^\al\wedge dx^\bt
  (\pl_\al A_\bt-\pl_\bt A_\al)=dx^\al\wedge dx^\bt\pl_\al A_\bt.\qed
\end{equation}
\end{exa}
\begin{exa}
Внешний дифференциал от 2-формы $B=\frac12dx^\al\wedge dx^\bt B_{\al\bt}$
равен
\begin{equation}                                                  \label{exdeos}
\begin{split}
  dB&=\frac12dB_{\al\bt}\wedge dx^\al\wedge dx^\bt
  =\frac12dx^\al\wedge dx^\bt\wedge dx^\g\pl_\al B_{\bt\g}=
\\
  =&\frac1{12}dx^\al\wedge dx^\bt\wedge dx^\g
  (\pl_\al B_{\bt\g}+\pl_\bt B_{\g\al}+\pl_\g B_{\al\bt}
  -\pl_\al B_{\g\bt}-\pl_\g B_{\bt\al}-\pl_\bt B_{\al\g}).\qed
\end{split}
\end{equation}
\end{exa}

Формулу внешнего дифференцирования легко запомнить с помощью следующего
мнемонического правила. Сравнивая последнее выражение в (\ref{extder}) c
формулой для внешнего умножения (\ref{ewpprd}), можно записать
\begin{equation*}
  dA=\pl\wedge A.
\end{equation*}

Из определения \ref{extder} следует, что внешний дифференциал можно записать
с помощью ковариантной производной, определяемой символами Кристоффеля
или другой произвольной {\em симметричной} аффинной связностью
\begin{equation}                                                  \label{eexdco}
  dA=\frac1{r!}dx^{\al_1}\wedge\dots\wedge dx^{\al_{r+1}}
  \widetilde\nb_{\al_1}A_{\al_2\dots\al_{r+1}}.
\end{equation}
Для несимметричной связности возникнут дополнительные слагаемые с тензором
кручения.

Приведем некоторые свойства внешнего  дифференцирования. С помощью прямых
вычислений доказывается формула
\begin{equation}                                                  \label{edacov}
  2dA(X,Y)=X\big(A(Y)\big)-Y\big(A(X)\big)-A\big([X,Y]\big),
\end{equation}
где $A$ -- произвольная 1-форма, а $X,Y$ -- любые векторные поля на $\MM$.
В общем случае справедлива
\begin{theorem}                                                   \label{tedvec}
Пусть $A$ -- гладкая дифференциальная форма степени $r$ и $X_1\dotsc X_{r+1}$
-- гладкие векторные поля. Тогда значение $r+1$-формы $dA$ на этих векторных
полях может быть найдено по формуле
\begin{align}                                                     \nonumber
  dA(X_1,\dotsc,X_{r+1})&
  =\frac1{r+1}\sum_i(-1)^{i-1}X_i^\al\pl_\al A(X_1,\dotsc,\breve X_i,
  \dotsc,X_{r-1})+
\\                                                                \label{evadfv}
  &+\frac1{r+1}\sum_{i<j}(-1)^{i+j}A([X_i,X_j],X_1,\dotsc,\breve X_i,
  \dotsc,\breve X_j,\dotsc,X_{r+1}),
\end{align}
где символ $\breve X_i$ означает, что соответствующее векторное поле
отсутствует.
\end{theorem}
\begin{proof}
Прямая проверка.
\end{proof}
\begin{defn}
Форма $A$ степени $r\ge1$ называется {\em точной}, если существует такая
$(r-1)$-форма $B$, что
\begin{equation}                                                  \label{exafor}
  A=dB.
\end{equation}
Форма $A$ степени $r\ge0$ называется {\em замкнутой}, если выполнено условие
\begin{equation}                                                  \label{eclfor}
  dA=0.
\end{equation}
По-определению, точных 0-форм не существует.
\qed\end{defn}
\index{Точная форма (exact form)}\index{Форма точная (exact form)}%
\index{Замкнутая форма (closed form)}\index{Форма замкнутая (closed form)}%
\begin{prop}
Множество всех замкнутых $0$-форм (функций) на связном многообразии состоит из
функций, которые постоянны на всем многообразии.
\end{prop}
\begin{proof}
Пусть $f\in\CC^\infty(\MM)$. Тогда замкнутость $0$-формы в произвольной системе
координат означает
\begin{equation*}
  df=0\quad\Leftrightarrow\quad \pl_\al f=0.
\end{equation*}
Откуда следует $f=\const$ и в данной карте и на всем многообразии.
\end{proof}

Любая точная форма является замкнутой в силу свойства внешнего дифференцирования
(\ref{eptdth}). Обратное утверждение в общем случае неверно. Для замкнутой формы
представление (\ref{exafor}) справедливо только локально, т.е.\ в некоторой
окрестности произвольной точки (лемма Пуанкаре), которая будет сформулирована
ниже. При этом глобальное
представление может не иметь места. В общем случае множество замкнутых форм
больше множества точных форм. В дальнейшем будет показано, что отличие множества
замкнутых и точных форм связано с топологическими свойствами многообразий.
\begin{com}
Если форма $A$ точна, $A=dB$, и принадлежит классу $\CC^k$, то это не значит,
что форма $B$ принадлежит классу $\CC^{k+1}$, потому что в определении
внешней производной (\ref{extder}) содержится только часть частных производных.
Например, если $B\in\Lm_n(\MM)$ -- форма максимальной степени и класса $\CC^1$,
то $A=dB=0$, и, следовательно, принадлежит классу $\CC^\infty$, а это ничего не
говорит о степени гладкости формы $B$. В лемме Пуанкаре нам встретится ситуация,
когда обе формы $A$ и $B$ одного класса гладкости.
\qed\end{com}
\begin{exa}
Пусть на компактном многообразии $\MM$ задана гладкая функция
$f\in\CC^\infty(\MM)$. Тогда 1-форма $A=dx^\al\pl_\al f$ точна, замкнута и имеет
по крайней мере два нуля. Наличие нулей следует из теоремы \ref{tcomam}
о том, что на компактном многообразии всякая непрерывная функция принимает
свое минимальное и максимальное значения, где все частные производные
обращаются в нуль. Этот пример показывает, что нули замкнутых форм
связаны с топологией многообразий.
\qed\end{exa}
\begin{prop}
Внешнее произведение замкнутых форм замкнуто, а если одна из форм, кроме того,
точна, то это произведение является точной формой.
\end{prop}
\begin{proof}
Прямая проверка.
\end{proof}
Приведем важную теорему, имеющую многочисленные применения в дифференциальной
геометрии.
\begin{theorem}[\bf Лемма Пуанкаре]                               \label{tlempo}
Пусть $A\in\Lm_r(\MR^n)$, $r\ge1$, -- замкнутая $r$-форма, $dA=0$, класса
$\CC^1$, задана в евклидовом пространстве $\MR^n$. Тогда существуют
$(r-1)$-формы $B\in\Lm_{r-1}(\MR^n)$, класса $\CC^1$, такие, что $A=dB$. При
этом все формы $B$ отличаются друг от друга на точные формы $B_1=B_2+dC$,
где $C\in\Lm_{r-2}(\MR^n)$, класса $\CC^1$ при $r\ge2$. При $r=1$,
0-формы $B$ определены с точностью до константы.
\end{theorem}
\index{Лемма Пуанкаре (Poincar\'e lemma)}%
\index{Пуанкаре лемма (Poincar\'e lemma)}%
\begin{proof}
Доказательство проводится по индукции по размерности евклидова пространства
путем явного построения формы $B$ с помощью интеграла вдоль кривой
\cite{Schwar67R}.
\end{proof}

Эта теорема локальна, т.к.\ все евклидово пространство диффеоморфно конечной
области в $\MR^n$. В случае, когда многообразие $\MM$ нетривиально
и не покрывается одной картой, лемма Пуанкаре справедлива для каждой карты, но
не для всего $\MM$. Отличие классов замкнутых и точных форм служит для
глобальной характеристики многообразий и лежит в основе теории когомологий
де Рама -- одного из подходов к теории гомологий.
\begin{com}
Согласно лемме Пуанкаре, если форма $A$ определена на области $\MU\subset\MM$,
гомеоморфной евклидову пространству $\MR^n$, то она всегда точна. Поэтому
различие между точными и замкнутыми формами проявляется только для форм,
заданных на всем многообразии $\MM$.
\qed\end{com}
\begin{exa}
Поясним лемму Пуанкаре на примере 1-формы $A$ и 0-формы $f$. В этом случае
условие точности формы (\ref{exafor}) сводится к уравнению
\begin{equation}                                                  \label{exfone}
  A_\al=\pl_\al f.
\end{equation}
Из теории дифференциальных уравнений известно, что это уравнение локально
разрешимо тогда и только тогда, когда
\begin{equation}                                                  \label{eclofo}
  \pl_\al A_\bt-\pl_\bt A_\al=0\quad \text{или}\quad dA=0.
\end{equation}
Таким образом, замкнутость некоторой формы обеспечивает ее точность
только локально, но не глобально.

Дифференциальная 1-форма $dx^\al A_\al$ на $n$-мерном многообразии задается $n$
независимыми компонентами, а скалярное поле $f$ -- только одной. С другой
стороны, если форма $A$ точна, то она параметризуется только одной функцией,
т.к.\ выполнено равенство (\ref{exfone}). На первый взгляд, число уравнений на
компоненты $A_\al$ в (\ref{eclofo}) равно $C^2_n$, что превышает разность в
числе независимых компонент, которая равна $n-1$, при $n\ge3$. Кажущееся
противоречие заключается в наивном подсчете числа уравнений. Подсчитаем число
независимых уравнений в (\ref{eclofo}). Общее число уравнений равно $C^2_n$,
однако независимых уравнений меньше, поскольку имеется дифференциальное
тождество $d^2 A=0$. Число этих тождеств равно $C^3_n$, однако среди них также
есть зависимые в силу тождества $d^3A=0$. Продолжая этот процесс до порядка
$n-1$ (дальнейшая антисимметризация по индексам не имеет смысла) получим, что
число независимых уравнений среди (\ref{eclofo}) равно
$$
  C_n^2-C_n^3+C_n^4-\dotsc(-1)^n C_n^n=C_n^1-1=n-1.
$$
Это равенство следует из (\ref{ebicpr}).
То есть среди $n$ компонент замкнутой 1-формы $A$ (ковекторного поля)
только одна является независимой, в качестве которой можно выбрать $f$.
\qed\end{exa}
\begin{exa}
Продемонстрируем локальность леммы Пуанкаре и важность нетривиальных замкнутых
форм для описания глобальных свойств многообразий.
Рассмотрим евклидову плоскость $\MR^2$ в декартовых координатах $x,y$. Тогда
дифференциал полярного угла $\vf$ имеет вид (\ref{epcint})
\begin{equation*}
  A=\frac{xdy-xdy}{x^2+y^2}.
\end{equation*}
1-форма $A$ определена и замкнута, $dA=0$, всюду, за исключением начала
координат, и, значит, не удовлетворяет условиям леммы Пуанкаре. Отсюда следует,
что не существует дифференцируемой функции $\vf(x,y)$, определенной на всей
плоскости $\MR^2$, такой, что $A=d\vf$. Действительно, при обходе начала
координат по замкнутому контуру один раз, функция $\vf$ получит приращение
$2\pi$. В то же время на любой односвязной области, не содержащей начала
координат, такая функция существует.
\qed\end{exa}
\begin{prop}
Если на многообразии задано векторное поле $X$, то производная Ли (см.\
раздел \ref{sliede}) от произвольной $r$-формы $A\in\Lm_r(\MM)$ представляет
собой симметричную комбинацию внешней и внутренней производной
\begin{equation}                                                  \label{eldefo}
  \Lie_X=d\,\inm_X+\inm_Xd.
\end{equation}
Это соотношение называется {\em основной формулой гомотопии}.
\end{prop}
\index{Основная формула гомотопии}\index{Формула гомотопии основная}%
\begin{proof}
Достаточно показать, что правая часть этого выражения является
дифференцированием, коммутирующим с внешней производной $d$, и что оно совпадает
с производной Ли на функциях. Действительно, пусть $A\in\Lm_r(\MM)$ и
$B\in\Lm_s(\MM)$, тогда нетрудно проверить равенство
\begin{equation}                                                  \label{elieex}
  (d\inm_X+\inm_Xd)(A\wedge B)=(d\inm_X+\inm_Xd)A\wedge B
  +A\wedge(d\inm_X+\inm_Xd)B.
\end{equation}
На функциях $f\in\Lm_0(\MM)$
$$
  (d\inm_X+\inm_Xd)f=\inm_Xdf=Xf=\Lie_Xf.
$$
Коммутативность $(d\inm_X+\inm_Xd)$ с внешней производной легко проверяется.
\end{proof}

Формулу (\ref{elieex}) можно переписать в виде
\begin{equation*}
  \Lie_X(A\wedge B)=\Lie_XA\wedge B+A\wedge\Lie_XB.
\end{equation*}
\begin{prop}
Пусть задан диффеоморфизм двух многообразий $h:~\MM\rightarrow\MN$. Тогда
возврат отображения
\begin{equation*}
  h^*:\quad \MT^*(\MN)\rightarrow\MT^*(\MM),
\end{equation*}
действует на формы в обратную сторону и имеет следующие свойства.

1) \parbox[t]{.92\linewidth}{Для произвольных форм $A,B\in\Lm(\MN)$, заданных на
многообразии $\MN$, справедливо равенство
\begin{equation*}
  h^*(A\wedge B)=h^*A\wedge h^*B.
\end{equation*}  }

2) \parbox[t]{.92\linewidth}{Возврат отображения коммутирует с внешней
производной
\begin{equation}                                                  \label{epufor}
  h^*dA=d(h^*A),
\end{equation}
где $A$ -- произвольная $r$-форма на многообразии $\MN$.}

3) \parbox[t]{.92\linewidth}{Пусть задан набор векторных полей $X_1,\dotsc,X_r$
на $\MM$. Тогда
\begin{equation*}
  h^*A(X_1,\dotsc,X_r)=A(h_*X_1,\dotsc,h_*X_r),
\end{equation*}
где $A$ -- произвольная $r$-форма на $\MN$.}
\end{prop}
\begin{proof}
Прямая проверка.
\end{proof}
Поскольку производная Ли от форм определяется возвратом отображения, то из
свойства 1) вытекает, что производная Ли от произвольной $r$-формы коммутирует
с внешним дифференцированием:
$$
  \Lie_Xd=d\Lie_X.
$$
Это свойство было использовано в доказательстве равенства (\ref{elieex}).
\section{Теорема Дарбу}
Пусть на многообразии $\MM$, $\dim\MM=n$, задана произвольная $r$-форма
\begin{equation*}
  A=\frac1{n!}dx^{\al_1}\wedge\dotsc\wedge dx^{\al_r}A_{\al_1\dotsc\al_r}
  \quad \in\Lm_r(\MM).
\end{equation*}
Зададим вопрос о том, каково минимальное число функций
$f^1(x),\dotsc,f^q(x)$ таких, что $r$-форма $A$ представима в виде
\begin{equation}                                        \label{efospf}
  A=\sum_{1\le\al_1<\dotsc<\al_r\le q}\!\!\!
  df^{\al_1}\wedge\dotsc\wedge df^{\al_r}A_{\al_1\dotsc\al_r}(f^1\dotsc f^q).
\end{equation}
Ясно, что при $q=n$ любая форма уже имеет такой вид. С другой стороны, число
функций не может быть меньше степени формы. Поэтому представляет интерес случай
$r\le q<n$. Важность этого вопроса заключается в том, что функции
$f^1,\dotsc,f^q$ можно выбрать в качестве части координат некоторой новой
координатной системы и тем самым привести форму к каноническому виду, в котором
она будет зависеть не от всех координат, а только от их части. Ответ на
поставленный вопрос дает теорема \ref{efaqfa}, сформулированная в настоящем
разделе. Для 1- и 2-форм ответ дают теоремы Дарбу.

Для формулировки теорем мы введем новые понятия: ранг и класс дифференциальных
форм. Начнем с форм над произвольным векторным пространством $\MV$ с базисом
$e_\al$, $\al=1,\dotsc,n$. Дадим несколько определений.
\begin{defn}
Назовем $r$-форму $A\in\Lm_r(\MV)$ {\em разложимой}, если существует
$r$ линейно независимых 1-форм $a^1,\dotsc,a^r\in\Lm_1(\MV)$ таких, что
\begin{equation*}                                                    \tag*{\qed}
  A=a^1\wedge\dotsc\wedge a^r.
\end{equation*}
\renewcommand{\qed}{}\end{defn}
\index{Разложимая $r$-форма}\index{Форма разложимая}%

Ясно, что всякую $r$-форму можно представить в виде суммы разложимых
$r$-форм, например, разложив по базису. Однако, не всякая $r$-форма
разложима.
\begin{exa}
Рассмотрим четырехмерное векторное пространство $\MV$ с базисом
$e_\al$, $\al=1,2,3,4$. Базис сопряженного пространства $\MV^*$ обозначим через
$e^\al$. Рассмотрим 2-форму
\begin{equation*}
  A=e^1\wedge e^2+e^3\wedge e^4~\in\Lm_2(\MV).
\end{equation*}
Если форма $A$ разложима, то существуют две 1-формы $a=e^\al a_\al$ и
$b=e^\al b_\al$ такие, что выполнено условие
\begin{equation*}
  A=a\wedge b=e^\al\wedge e^\bt a_\al b_\bt=e^1\wedge e^2+e^3\wedge e^4.
\end{equation*}
Отсюда вытекает следующая система уравнений на компоненты
\begin{align*}
  a_1b_2&=1, & a_2b_3&=0,
\\
  a_1b_3&=0, & a_2b_4&=0,
\\
  a_1b_4&=0, & a_3b_4&=1.
\end{align*}
Не ограничивая общности, можно положить $a_1=1$. Тогда из трех уравнений в
первой колонке следует $b_2=1$, $b_3=b_4=0$. Подставив эти решения в остальные
уравнения, приходим к противоречию. Следовательно, 2-форма $A$ является
неразложимой.

Рассмотрим другую 2-форму над тем же векторным пространством $\MV$
\begin{equation*}
  A=e^1\wedge e^2+e^1\wedge e^3~\in\Lm_2(\MV).
\end{equation*}
Ее можно представить в виде $A=e^1\wedge(e^2+e^3)$. Отсюда следует, что эта
форма разложима: $A=e^1\wedge b$, где $b=e^2+e^3\in\Lm_1(\MV)$.
\qed\end{exa}

\begin{defn}
Назовем подпространство $\MA(A)\subset\MV$ {\em ассоциированным} с формой $A$,
если это наибольшее из тех подпространств $\MH\subset\MV$, для которых
$A\in\Lm_r(\MV/\MH)$. Другими словами, пространство $\MA(A)$ образовано теми
векторами $X\in\MV$, чье внутреннее произведение с формой равно нулю,
$\inm_XA=0$. Это значит, что подпространство, ассоциированное с формой $A$, есть
подпространство в $\MV$, которое определяется системой уравнений $\inm_XA=0$.

Каждому ассоциированному пространству $\MA(A)$ можно поставить в соответствие
{\em ассоциированную с формой систему} $\MA^*(A)$, как ортогональное дополнение
сопряженного пространства
\begin{equation*}
  \MA^*(A):=\MA^\bot(A)\in\MV^*.
\end{equation*}
Размерность ассоциированной системы называется {\em рангом} формы $A$
\begin{equation*}                                                    \tag*{\qed}
  \rank A:=\dim\MA^*(A).
\end{equation*}
\renewcommand{\qed}{}\end{defn}
\index{Подпространство, ассоциированное с $r$-формой}%
\index{Ассоциированное с формой $A$ подпространство}%
\index{Система, ассоциированная с $r$-формой}%
\index{Ранг $r$-формы (rank of $r$-form)}%

Ясно, что ранг $r$-формы $A$ равен минимальному числу линейно независимых
1-форм таких, что форма $A$ равна линейной комбинации их внешних произведений.
\begin{exa}
Для 2-форм $A=\frac12e^\al\wedge e^\bt A_{\al\bt}$ понятие ранга особенно
наглядно: это просто ранг матрицы компонент,
\begin{equation*}                                                    \tag*{\qed}
  \dim\MA^*(A)=\rank A_{\al\bt}.
\end{equation*}
\end{exa}
\begin{exa}
Рассмотрим 0-формы $A\in\Lm_0(\MV)$, т.е.\ $A\in\MR$. Тогда ассоциированное
подпространство $\MA(A)=\MV$, т.к.\ факторпространство $\MV/\MV$ состоит из
одной точки -- начала координат, и $\Lm_0(\MV/\MV)=\MR$. Соответствующая система
$\MA^*(A)$, ассоциированная с формой $A$, также состоит из одной точки -- начала
координат сопряженного пространства $\MA^*(A)=\lbrace0\rbrace$. Таким образом,
ранг произвольной 0-формы равен нулю.
\qed\end{exa}
\begin{exa}
Пусть $\MV=\MR^n$ с базисом $\lbrace e_1,\dotsc,e_n\rbrace$ и
$A=e^1+e^2\in\Lm_1(\MR^n)$, тогда ассоциированное подпространство
$\MA(A)=\MR^{n-1}$ порождено базисными векторами
$\lbrace e_1-e_2,e_3,\dotsc,e_n\rbrace$. Соответствующая ассоциированная
система $\MA^*(A)$ представляет собой одномерное подпространство сопряженного
пространства $\MV^*=\MR^n$, порожденное формой $A$, т.е.\ натянутое на базисный
вектор $e^1+e^2$. Ранг 1-формы $e^1+e^2$ равен единице. Ясно, что ранг
произвольной 1-формы равен единице.
\qed\end{exa}
\begin{exa}
Пусть $\MV=\MR^n$ и $A=e^1\wedge e^2+e^3\wedge e^4\in\Lm_2(\MR^n)$.
Эта 2-форма неразложима, поэтому ассоциированное с ней подпространство
имеет размерность $n-4$. Следовательно, ассоциированное подпространство имеет
вид $\MA(A)=\MR^{n-4}$ и порождено векторами $e_5,\dotsc,e_n$. Соответствующая
ассоциированная система четырехмерна и порождена векторами $e^1,e^2,e^3,e^4$.
Таким образом, ранг 2-формы $e^1\wedge e^2+e^3\wedge e^4$ равен четырем.
\qed\end{exa}
В общем случае справедлива следующая теорема, дающая ограничения на
возможный ранг формы степени $r$.
\begin{theorem}
Если $A\in\Lm_r(\MV)$ -- ненулевая $r$-форма на $\MV$, то ее ранг не меньше $r$
и не больше $n$
\begin{equation*}
  r\le\rank A\le n.
\end{equation*}
Ранг $r$-формы $A$ равен $r$ тогда и только тогда, когда форма $A$ разложима.
\end{theorem}
\begin{proof}
См., например, \cite{Godbil69R}.
\end{proof}

Как следствие получаем, что, поскольку любая $n$-форма разложима, то ее ранг
равен $n$. Можно доказать, что любая $(n-1)$-форма разложима и ее ранг,
следовательно, равен $n-1$. А также, что ранг $(n-2)$-формы равен либо $n-2$,
либо $n$.

Доказательства следующих трех предложений для внешних форм второй степени
$A\in\Lm_2(\MV)$, $\dim(\MV)=n$, приведены в \cite{Godbil69R}.
\begin{prop}
Пусть $A\in\Lm_2(\MV)$ -- внешняя 2-форма. Тогда существует целое четное число
$2s$ такое, что $2\le 2s\le n$, и $2s$ независимых линейных форм
$a^1,\dotsc,a^{2s}$ таких, что
\begin{equation*}
  A=a^1\wedge a^2+\dotsc+a^{2s-1}\wedge a^{2s}.
\end{equation*}
При этом 1-форму $a^1$ можно выбрать произвольно в ассоциированной
системе $\MA^*(A)$.
\end{prop}
\begin{prop}
Внешняя 2-форма имеет четный ранг.
\end{prop}
\begin{prop}                                                      \label{pseome}
Пусть $A\in\Lm_2(\MV)$ -- внешняя 2-форма. Для того, чтобы 2-форма $A$
имела ранг $2s$, необходимо и достаточно, чтобы
\begin{equation*}
  A^s\ne0\quad \text{и}\quad A^{s+1}=0,
\end{equation*}
где в левой части соотношений стоят внешние степени формы $A$.
\end{prop}

Приведенные выше определения и утверждения даны для внешней алгебры $\Lm(\MV)$
над произвольным векторным пространством $\MV$ и применимы в каждой точке
многообразия $\MM$ для внешней алгебры $\Lm(\MM)$. При рассмотрении форм на
многообразии эти понятия, в том числе ранг формы, могут меняться от точки к
точке.

Введем несколько новых понятий, для которых уже важна зависимость компонент
формы от точки многообразия.
\begin{defn}
{\em Характеристическим подпространством} $r$-формы $A\in\Lm_r(\MM)$ в точке
$x\in\MM$ называется подпространство касательного пространства
$\MC_x(A)\subset\MT_x(\MM)$, которое является пересечением ассоциированных
подпространств для формы $A$ и ее внешней производной $dA$,
\begin{equation*}
  \MC_x(A):=\MA_x(A)\cap\MA_x(dA),
\end{equation*}
в точке $x$.
{\em Характеристической системой} формы $A\in\Lm_r(\MM)$ в точке $x\in\MM$
называется подпространство $\MC^*_x(A)\subset\MT^*_x(\MM)$, которое ортогонально
характеристическому подпространству этой формы $\MC_x(A)$.
\qed\end{defn}
\index{Характеристическое подпространство $r$-формы %
(characteristic subspace for $r$-form)}%
\index{Характеристическая система $r$-формы %
(characteristic system for $r$-form)}%

Из определения и формулы (\ref{ebotin}) вытекает, что характеристическая система
$\MC^*_x(A)$ строится по ассоциированным системам $\MA^*_x(A)$ и $\MA^*_x(dA)$
следующим образом
\begin{equation*}
\begin{split}
  \MC^*_x(A)&=\MA^*_x(A)\oplus\left[\MA^*_x(dA)\setminus
  \left(\vphantom{A^1_1}\MA^*_x(A)\cap\MA^*_x(dA)\right)\right]
\\
  &=\left[\MA^*_x(A)\setminus\left(\vphantom{A^1_1}
  \MA^*_x(A)\cap\MA^*_x(dA)\right)\right]\oplus\MA^*_x(dA).
\end{split}
\end{equation*}
\begin{defn}
{\em Классом} $r$-формы $A\in\Lm_r(\MM)$ в точке $x\in\MM$ называется
размерность характеристической системы: $\class(A):=\dim\MC^*_x(A)$.
\qed\end{defn}
\index{Класс $r$-формы (rank of $r$-form)}%
Из определения следует, что класс ненулевой $r$-формы $A$ не меньше ее ранга и,
следовательно, не меньше степени $r$
\begin{equation*}
  r\le\rank(A)\le\class(A).
\end{equation*}
В общем случае класс формы может зависеть от точки многообразия.
\begin{exa}
Рассмотрим 1-форму $A=dy(x^2+y^2)$ на евклидовой плоскости $\MR^2$ с декартовыми
координатами $x,y$. Ее внешний дифференциал равен $dA=dx\wedge dy\, x$.
Существуют три случая.

1) $x\ne0$. В этом случае ассоциированное подпространство $\MA(A)$ порождено
вектором $\pl_x$, а ассоциированное подпространство внешнего дифференциала
нульмерно, $\MA(dA)=\lbrace0\rbrace$. Поэтому характеристическое подпространство
состоит из одной точки $\MC_x(A)=\lbrace0\rbrace$, а характеристическая система
совпадает с кокасательным пространством, $\MC_x(A)=\MT^*_x(\MR^2)$.
Следовательно, класс формы $A$ равен двум, $\class(A)=2$.

2) $x=0$, $y\ne0$. Ассоциированное подпространство $\MA(A)$ по-прежнему
порождено вектором $\pl_x$. Внешний дифференциал формы $A$ равен нулю, $dA=0$, и
ассоциированное подпространство внешнего дифференциала совпадает с касательным
пространством, $\MA(dA)=\MT_x(\MR^2)$. Следовательно, характеристическое
подпространство $\MC_x(A)$ порождено вектором $\pl_x$, а характеристическая
система -- дифференциалом $dy$. Поэтому класс формы $A$ равен единице,
$\class(A)=1$.

3) $x=y=0$. В этом случае и форма, и ее внешний дифференциал обращаются в нуль:
$A=0$, $dA=0$. Поэтому ассоциированное подпространство совпадает с касательным
пространством, $\MC_x(A)=\MT_x(\MR^2)$, а ассоциированная система состоит из
одной точки $\MC^*_x=\lbrace0\rbrace$. Следовательно, класс формы $A$ равен
нулю, $\class(A)=0$.

Таким образом, класс рассматриваемой формы зависит от точки плоскости и равен:
\newline
\indent\indent \class(A)=2, если $x\ne0$;\newline
\indent\indent \class(A)=1, если $x=0$ и $y\ne0$;\newline
\indent\indent \class(A)=0, если $x=y=0$.
\qed\end{exa}

Продолжим общее рассмотрение. Нетрудно проверить, что если класс $r$-формы $A$
равен ее степени $r$ в некоторой точке $x\in\MM$, то эта форма замкнута,
$dA(x)=0$. В обратную сторону справедливо другое утверждение. Если форма
замкнута, $dA(x)=0$, в точке $x$, то в этой точке ее класс равен рангу, а не
степени. Отсюда, в частности, следует, что замкнутая форма второй степени имеет
четный класс.

Теперь сформулируем основное утверждение настоящего раздела.
\begin{theorem}                                                   \label{ethelo}
Пусть $A\in\Lm_r(\MM)$ -- дифференциальная $r$-форма на многообразии $\MM$
постоянного класса, $\class(A)=q$, $r\le q\le n$. Тогда в окрестности любой
точки $x\in\MM$ существует такая система координат $y^1,\dotsc,y^n$, что форма
$A$ представима в виде
\begin{equation}                                                  \label{efaqfa}
  A=\sum_{1\le\al_1<\dotsc<\al_r\le q}\!\!\!
  dy^{\al_1}\wedge\dotsc\wedge dy^{\al_r}
  A_{\al_1\dotsc\al_r}(y^1\dotsc y^q).
\end{equation}
Для форм непостоянного класса представления (\ref{efaqfa}) не существует.
\end{theorem}
\begin{proof}
См., например, \cite{Godbil69R}
\end{proof}
Нетривиальность этой теоремы заключается в том, что форма записана в виде,
содержащим только первые $q$ координат. Эта теорема дает ответ на вопрос,
поставленный в начале раздела. Другими словами, класс формы равен минимальному
числу независимых функций, необходимых для явного выражения $r$-формы
постоянного класса.
\begin{cor}
Пусть $A\in\Lm_r(\MM)$ -- дифференциальная $r$-форма на многообразии $\MM$
постоянного класса, который равен степени, $\class(A)=r$. Тогда в окрестности
любой точки $x\in\MM$ существует такая система координат $y^1,\dotsc,y^n$,
что форма $A$ имеет вид
\begin{equation}                                                  \label{efaqfr}
  A=dy^{\al_1}\wedge\dotsc\wedge dy^{\al_r}
\end{equation}
\end{cor}
\begin{proof}
См., например, \cite{Godbil69R}
\end{proof}

Для замкнутых 1- и 2-форм постоянного класса теорему \ref{ethelo} можно
уточнить.
\begin{theorem}[\bf Дарбу]                                        \label{tdarbo}
Пусть $A$ -- произвольная 1-форма без нулей на многообразии $\MM$ постоянного
класса $2s+1$ (или $2s$). Тогда для любой точки $x\in\MM$ существуют $2s+1$ (или
$2s$) функций $y^1,\dotsc,y^{2s+1}$ (или $y^1,\dotsc,y^{2s}$), заданных в
некоторой окрестности $\MU\subset\MM$, содержащей точку $x$, такие, что
\begin{align*}
  &y^1(x)=\dotsc=y^{2s+1}(x)=0,
\\
  \text{или}\qquad &y^1(x)=\dotsc=y^{2s}(x)=0,
\\ \intertext{и}
  A|_\MU&=dy^1+y^2dy^3+\dotsc+y^{2s}dy^{2s+1},
\\
  \text{или}\qquad A|_\MU&=(1+y^1)dy^2+y^3dy^4+\dotsc+y^{2s-1}dy^{2s}.
\end{align*}
\end{theorem}
\begin{proof}
См., например, \cite{Godbil69R}.
\end{proof}
\index{Теорема Дарбу (Darboux theorem)}\index{Дарбу теорема (Darboux theorem)}%
\begin{com}
Если 1-форма имеет постоянный нечетный класс, то она не имеет нулей на
многообразии. Напротив, если 1-форма имеет постоянный четный класс, то она может
иметь нули. В этом случае невозможно привести общую локальную модель 1-формы.
\qed\end{com}
\begin{theorem}[\bf Дарбу]                                        \label{tdartw}
Пусть $\om$ -- замкнутая дифференциальная 2-форма на многообразии $\MM$
постоянного класса $\class\om=2s$. Тогда для любой точки $x\in\MM$ существуют
$2s$ дифференцируемых функций $y^1,\dotsc,y^{2s}$, заданных в некоторой
окрестности $\MU\subset\MM$, содержащей точку $x$, такие, что
\begin{align*}
  &y^1(x)=\dotsc=y^{2s}(x)=0,
\\
  &\om|_\MU=dy^1\wedge dy^2+\dotsc+dy^{2s-1}\wedge dy^{2s}.
\end{align*}
\end{theorem}
\begin{proof}
Из замкнутости 2-формы $\om$ по лемме Пуанкаре следует, что она локально
представима в виде $\om=dA$, где 1-форма $A$ в точке $x$ имеет класс $2s$ или
$2s+1$. Тогда утверждение теоремы следует из теоремы \ref{tdarbo}.
\end{proof}
\begin{exa}
Рассмотрим симплектическое многообразие $(\MM,\om)$, $\dim\MM=2n$.
По-определению, симплектическая форма $\om$ на многообразии $\MM$ невырождена,
и, следовательно, ее ранг и класс постоянны и равны размерности многообразия
\begin{equation*}
  \rank\om=\class\om=\dim\MM=2n.
\end{equation*}
Симплектическая форма $\om$ также замкнута. Поэтому из теоремы Дарбу следует,
что для любой точки $x\in\MM$ существуют $2n$  дифференцируемых функций
$y^1,\dotsc,y^{2n}$, заданных в некоторой окрестности  $\MU\subset\MM$,
содержащей точку $x$, таких, что
\begin{align}                                                          \nonumber
  &y^1(x)=\dotsc=y^{2n}(x)=0,
\\
  &\om|_\MU=dy^1\wedge dy^2+\dotsc+dy^{2n-1}\wedge dy^{2n}. \label{ecasif}
\end{align}
Другими словами, для любой симплектической формы в окрестности произвольной
точки существует такая система координат, в которой она имеет канонический
вид. Именно это следствие теоремы Дарбу, которое часто тоже называют теоремой
Дарбу, обычно используют в гамильтоновой динамике.
\qed\end{exa}
\begin{defn}
Система координат, в которой симплектическая форма имеет канонический вид
(\ref{ecasif}), называется {\em координатами Дарбу}.
\qed\end{defn}
\index{Координаты Дарбу (Darboux coordinates)}%
\index{Дарбу координаты (Darboux coordinates)}%
\begin{com}
В римановой геометрии $(\MM,g)$ с метрикой $g$ общего вида всегда можно выбрать
такие координаты, чтобы метрика совпала с евклидовой метрикой (единичной
матрицей) в любой заданной точке многообразия. В общем случае этого можно
добиться вдоль экстремали, но не в окрестности. На симплектическом
многообразии симплектическую форму можно привести к каноническому виду в
окрестности произвольной точки, что является значительно более сильным
утверждением.
\qed\end{com}
\section{Оператор Лапласа--Бельтрами}                              \label{slabe}
\begin{defn}
Пространства форм $\Lm_r(\MM)$ и $\Lm_{n-r}(\MM)$ имеют одинаковую размерность
$C_n^r=C_n^{n-r}$, и при наличии метрики $g_{\al\bt}$ между их элементами можно
установить взаимно однозначное соответствие (изоморфизм
$\CC^\infty(\MM)$-модулей), которое задается оператором
\begin{equation*}
   \ast:\quad \Lm_r(\MM)~\leftrightarrow~\Lm_{n-r}(\MM).
\end{equation*}
Рассмотрим дифференциальную $r$-форму
$A=\frac1{r!}dx^{\al_1}\wedge\dots\wedge dx^{\al_r}A_{\al_1\dotsc\al_r}$
в (псевдо-)римановом пространстве, т.е.\ при наличии метрики. Поставим
ей в соответствие $(n-r)$-форму по следующему правилу
\begin{equation}                                                  \label{estard}
  \ast A:=\frac1{(n-r)!}dx^{\al_1}\wedge\dots\wedge dx^{\al_{n-r}}
  (\ast A)_{\al_1\dotsc\al_{n-r}},
\end{equation}
где компоненты $(n-r)$-формы определены выражением
\begin{equation}                                                  \label{estaco}
  (\ast A)_{\al_{r+1}\dotsc\al_n}:=\frac1{r!}\ve_{\al_1\dots\al_n}
  A^{\al_1\dotsc\al_r},\qquad
  A^{\al_1\dotsc\al_r}=
  g^{\al_1\bt_1}\dotsc g^{\al_r\bt_r}A_{\bt_1\dots\bt_r}.
\end{equation}
Действие оператора $\ast$ можно записать также в виде
\begin{equation}                                                  \label{eastop}
  *A=\frac1{(n-r)!}dx^{\al_{r+1}}\wedge\dots\wedge dx^{\al_n}
  \frac1{r!}\ve_{\al_1\dotsc\al_n}A^{\al_1\dots\al_r}.
\end{equation}
Эти формулы определяют {\em оператор $\ast$}, который также называется {\em
оператором Ходжа}. Форма $\ast A$ называется {\em дуальной} к форме $A$.
\qed\end{defn}
\index{Оператор $\ast$ ($\ast$-operator)}%
\index{Оператор Ходжа (Hodge operator)}\index{Ходжа оператор (Hodge operator)}%
\index{Дуальная форма (dual form)}\index{Форма дуальная (dual form)}%
\begin{com}
Метрика риманова пространства входит в определение оператора $\ast$ дважды.
Во-первых, с ее помощью определяется полностью антисимметричный тензор
$\ve_{\al_1\dots\al_n}=\vol\hat\ve_{\al_1\dots\al_n}$. Во-вторых, метрика
используется для подъема индексов в уравнении (\ref{estaco}). Отметим, что для
определения оператора $\ast$ необходимо не только наличие метрики на
многообразии, но и его ориентируемость. В противном случае плотность $\vol$
глобально не определена.
\qed\end{com}
Оператор Ходжа $\ast$ линеен:
\begin{equation*}
  \ast(fA+gB)=f(\ast A)+g(\ast B),\qquad \forall~f,g\in\CC^\infty(\MM),
  \quad A,B\in\Lm_r(\MM).
\end{equation*}

Оператор $\ast$, действуя на $r$-форму, обладает следующим свойством
\begin{equation}                                                  \label{estapr}
  \ast\ast A=(-1)^{r(n-r)}A\sgn,
\end{equation}
где множитель $\sgn$ равен знаку определителя метрики (\ref{esgnmn}). Вид
этой формулы сохраняется при замене $r\mapsto n-r$. Отсюда следует, что
обратный оператор, действующий на $(n-r)$-форму в порядке $\ast^{-1}\ast$ или
на $r$-форму в обратном порядке $\ast\ast^{-1}$, имеет одинаковый вид
$$
  \ast^{-1}=(-1)^{r(n-r)}\ast\sgn.
$$
\begin{com}
Отметим, что при проведении расчетов необходимо учитывать, что множитель
$(-1)^{r(n-r)}$ зависит от степени формы, на которую действует оператор
$\ast^{-1}$. Для форм четной степени он всегда равен единице.
\qed\end{com}
\begin{com}
Оператор $\ast$ можно определить в произвольной внешней алгебре, снабженной
метрикой, поскольку он не содержит дифференцирования.
\qed\end{com}
Поскольку на $\MM$ задана метрика, то определена свертка двух $r$-форм $A$ и
$B$:
$$
  (A,B):=g^{\al_1\bt_1}\dotsc g^{\al_r\bt_r}
  A_{\al_1\dotsc\al_r}B_{\bt_1\dotsc\bt_r},
$$
которая дает скалярное поле (функцию).
\begin{prop}
Для двух $r$-форм $A,B\in\Lm_r(\MM)$ справедлива формула
\begin{equation}                                                  \label{escptf}
  A\wedge\ast B=dx^1\wedge\dots\wedge dx^n \vol\frac{(A,B)}{r!}.
\end{equation}
\end{prop}
\begin{proof}
Прямая проверка.
\end{proof}
\begin{defn}
Введем {\em скалярное произведение}
\begin{equation*}
  \langle\cdot,\cdot\rangle:\quad \Lm_r(\MM)\times\Lm_r(\MM)\rightarrow\MR
\end{equation*}
в пространстве $r$-форм $A,B\in\Lm_r(\MM)$ с помощью интеграла
\begin{equation}                                                  \label{escpfo}
  \langle A,B\rangle:=\int_\MM A\wedge\ast B=\int_\MM B\wedge\ast A
  =\frac1{r!}\int_\MM dx\vol(A,B)\sgn.
\end{equation}
Мы предполагаем, что этот интеграл сходится. В противном случае скалярное
произведение неопределено.
\qed\end{defn}
\index{Скалярное произведение форм (scalar product of forms)}%
\index{Произведение форм скалярное (scalar product of forms)}%
Скалярное произведение симметрично, $\langle A,B\rangle=\langle B,A\rangle$.
Если метрика риманова, т.е.\ положительно определена, то скалярное произведение
$\langle A,B\rangle$ также положительно определено.
\begin{prop}
Оператор $\ast$ ортогонален:
$$
  \langle\ast A,\ast B\rangle=\langle A, B\rangle\sgn.
$$
\end{prop}
\begin{proof}
Прямая проверка.
\end{proof}
\begin{defn}
Введем дифференциальный оператор {\em кограницы $\dl$}:
$\Lm_r(\MM)\rightarrow\Lm_{r-1}(\MM)$
\begin{equation}                                                  \label{edefdl}
  \dl:=(-1)^{n(r+1)+1}\sgn\ast d\,\ast,
\end{equation}
понижающий степень формы на единицу. Его можно записать также в виде
$$
  \dl=(-1)^r\ast^{-1}d\,\ast,
$$
учитывая, что обратный оператор $\ast^{-1}$ действует на $(n-r+1)$-форму.
По-определению, действие оператора кограницы на функцию дает нуль $\dl f=0$.
\qed\end{defn}
\index{Оператор $\dl$ (coboundary operator)}%
\index{Оператор кограницы (coboundary operator)}%
В компонентах действие оператора $\dl$ на $r$-форму $A$ имеет вид
\begin{align}                                                          \nonumber
  \dl A&=-\frac1{(r-1)!}dx^{\al_2}\wedge\dots\wedge dx^{\al_r}
  \frac1\vol(g_{\al_2\bt_2}\dotsc g_{\al_r\bt_r})
  \pl_{\al_1}(\vol A^{\al_1\bt_2\dotsc\bt_r})
\\                                                                \label{eacdlo}
  &=-\frac1{(r-1)!}dx^{\al_2}\wedge\dots\wedge dx^{\al_r}
  \widetilde\nb_{\al_1}A^{\al_1}{}_{\al_2\dotsc\al_r}.
\end{align}
Отсюда следует, что оператор $\dl$ можно интерпретировать, как ковариантную
дивергенцию антисимметричного тензора с верхними индексами.
\begin{prop}
Для замкнутых (компактных и без края) многообразий оператор $\dl$ сопряжен с
оператором внешнего дифференцирования относительно скалярного произведения
(\ref{escpfo})
\begin{equation}                                                  \label{ecopst}
  \langle dA,B\rangle=\langle A,\dl B\rangle,
\end{equation}
где $A\in\Lm_{r-1}(\MM)$ и $B\in\Lm_r(\MM)$.
\end{prop}
\begin{proof}
Прямая проверка. Компактность многообразия достаточна для существования
интеграла (\ref{escpfo}), а отсутствие края позволяет отбросить граничные
слагаемые, возникающие при интегрировании по частям.
\end{proof}
\begin{prop}
Квадрат оператора кограницы $\dl$ равен нулю:
\begin{equation}                                                  \label{esqcob}
  \dl\dl=0,
\end{equation}
\end{prop}
\begin{proof}
Прямое следствием определения (\ref{edefdl}) и нильпотентности внешнего
дифференцирования (\ref{eptdth}).
\end{proof}
\begin{defn}
Оператором {\em Лапласа--Бельтрами} в римановом пространстве называется оператор
\begin{equation}                                                  \label{elabeo}
  \triangle_\Sf=-(d+\dl)^2=-d\dl-\dl d,
\end{equation}
который действует в пространстве форм,
$\triangle_\Sf:~\Lm_r(\MM)\rightarrow\Lm_r(\MM)$, $0\le r\le n$,
\qed\end{defn}
\index{Оператор Лапласа--Бельтрами (Laplace-Beltrami operator)}%
\index{Лапласа--Бельтрами оператор (Laplace-Beltrami operator)}%
\begin{prop}
Для замкнутых многообразий оператор Лапласа--Бельтрами самосопряжен:
$$
  \langle\triangle_\Sf A,B\rangle=\langle A,\triangle_\Sf B\rangle,
$$
\end{prop}
\begin{proof}
Из симметричности скалярного произведения и формулы (\ref{ecopst}) следуют
равенства
\begin{equation*}
  \langle d\dl A,B\rangle=\langle\dl A,\dl B\rangle=\langle A,d\dl B\rangle.
\end{equation*}
Аналогичные равенства справедливы для оператора $\dl d$.
Отсюда вытекает самосопряженность оператора Лапласа--Бельтрами.
\end{proof}
\begin{com}
Оператор Лапласа--Бельтрами является дифференциальным оператором второго порядка
эллиптического типа в римановом пространстве с положительно определенной
метрикой. Если на многообразии задана метрика лоренцевой сигнатуры, то
инвариантный оператор (\ref{elabeo}) также определен. В этом случае он будет
гиперболического типа. Если для метрики лоренцевой сигнатуры скалярное
произведение определено, то оператор (\ref{elabeo}) будет также самосопряжен,
поскольку в доказательстве сигнатура метрики не используется.
\qed\end{com}
\begin{prop}
Оператор Лапласа--Бельтрами удовлетворяет следующим перестановочным
соотношениям:
\begin{equation}                                                  \label{elabec}
  \triangle_\Sf d=d\triangle_\Sf,\qquad \triangle_\Sf\dl=\dl\triangle_\Sf,\qquad
  \triangle_\Sf\,\ast=\ast\,\triangle_\Sf,
\end{equation}
\end{prop}
\begin{proof}
Прямое следствие определения (\ref{elabeo}).
\end{proof}
\begin{prop}
Формулы действия оператора Лапласа--Бельтрами на 0-форму $f$, 1-форму $A$ и
произвольную $r$-форму $B$ в компонентах можно выразить через ковариантные
производные и тензор кривизны следующим образом:
\begin{align}                                                     \nonumber
  \triangle_\Sf f&=\tilde\triangle f,
\\                                                                \label{elabef}
  \triangle_\Sf A&=dx^\al(\tilde\triangle A_\al
  +\widetilde R_\al{}^\bt A_\bt),
\\                                                                \nonumber
  \triangle_\Sf B&=\frac1{r!}dx^{\al_1}\wedge\dotsc\wedge dx^{\al_r}
  \left(\tilde\triangle A_{\al_1\dotsc\al_r}
  +r\widetilde R_{\al_1}{}^\bt A_{\bt\al_2\dotsc\al_r}
  -\frac{r(r-1)}2\widetilde R_{\al_1\al_2}{}^{\bt\g}
  A_{\bt\g\al_3\dotsc\al_r}\right),
\end{align}
где $\tilde\triangle:=g^{\al\bt}\widetilde\nb_\al\widetilde\nb_\bt$
-- оператор Лапласа--Бельтрами, построенный с помощью символов Кристоффеля.
\end{prop}
\begin{proof}
Прямые вычисления.
\end{proof}
\begin{com}
В аффинной геометрии мы имеем три инвариантных дифференциальных оператора
второго порядка $\triangle_\Sf$, $\tilde\triangle$ и $\triangle$, действующих из
$\Lm_r(\MM)\rightarrow\Lm_r(\MM)$. Первый оператор определен только на формах и
поэтому помечен индексом $\Sf$. Два других определены для произвольных тензорных
полей. В (псевдо-)римановой геометрии при нулевом кручении и неметричности
$\tilde\triangle=\triangle$. Отметим также, что действие операторов
$\triangle_\Sf$ и $\tilde\triangle$ на функции совпадает. Как правило, под
оператором Лапласа--Бельтрами мы понимаем оператор $\tilde\triangle$,
действующий на произвольные тензорные поля и зависящий только от метрики.
\qed\end{com}
\begin{com}
В геодезических (римановых) координатах (см.\ раздел \ref{srinog}) символы
Кристоффеля и все симметризованные частные производные от них обращаются в
нуль в некоторой точке $p\in\MM$, поэтому в этой точке оператор
Лапласа--Бельтрами принимает вид
$\tilde\triangle|_{x=p}=g^{\al\bt}\pl_\al\pl_\bt$
\qed\end{com}
\begin{exa}                                                       \label{etheuc}
Рассмотрим трехмерное евклидово пространство $\MR^3$ с декартовыми
координатами $x^\al=\lbrace x,y,z\rbrace $. Приведем явные формулы
действия операторов, введенных ранее, на формы. В трехмерном пространстве
могут быть заданы только 0-, 1-, 2- и 3-формы:
\begin{align*}
  f&                                &-&~\text{0-форма},
\\
  A&=dx^\al A_\al=dxA_x+dyA_y+dzA_z &-&~\text{1-форма},
\\
  B&=\frac12dx^{\al_1}\wedge dx^{\al_2}B_{\al_1\al_2}
    =dx\wedge dyB_{xy}+dy\wedge dzB_{yz}+dz\wedge dxB_{zx} &-&~\text{2-форма},
\\
  C&=\frac1{3!}dx^{\al_1}\wedge dx^{\al_2}\wedge dx^{\al_3}C_{\al_1\al_2\al_3}
    =dx\wedge dy\wedge dzC_{xyz} &-&~\text{3-форма}.
\end{align*}
Формы более высокого порядка тождественно равны нулю в силу трехмерности
пространства. Пространства 0- и 3-форм являются одномерными, а пространства
1- и 2-форм -- трехмерны.

Оператор внешнего дифференцирования (\ref{extder}) повышает на единицу степень
формы и действует следующим образом:
\begin{align*}
  df&=dx\pl_xf+dy\pl_yf+dz\pl_zf,
\\
  dA&=dx\wedge dy(\pl_xA_y-\pl_yA_x)+dy\wedge dz(\pl_yA_z-\pl_zA_y)
     +dz\wedge dx(\pl_zA_x-\pl_xA_z),
\\
  dB&=dx\wedge dy\wedge dz(\pl_xB_{yz}+\pl_yB_{zx}+\pl_zB_{xy}),
\\
  dC&=0.
\end{align*}
Нетрудно проверить, что оператор $\ast$, определенный соотношением
(\ref{estard}), действует на формы по-правилам:
\begin{align*}
  \ast f&=dx\wedge dy\wedge dzf,
\\
  \ast A&=dx\wedge dyA_z+dy\wedge dzA_x+dz\wedge dxA_y,
\\
  \ast B&=dxB_{yz}+dyB_{zx}+dzB_{xy},
\\
  \ast C&=C_{xyz}.
\end{align*}
Этот оператор устанавливает взаимно однозначное соответствие между
пространствами $\Lm_0(\MR^3)$ и $\Lm_3(\MR^3)$, а также между $\Lm_1(\MR^3)$ и
$\Lm_2(\MR^3)$. Заметим, что для любой формы в трехмерном евклидовом
пространстве квадрат оператора $\ast$ является тождественным оператором:
$$
  \ast\ast =1.
$$

Оператор кограницы $\dl$, определенный соотношением (\ref{edefdl}), понижает
степень форм на единицу:
\begin{align*}
  \dl f&=0,
\\
  \dl A&=-(\pl_xA_x+\pl_yA_y+\pl_zA_z),
\\
  \dl B&=dx(\pl_yB_{xy}-\pl_zB_{zx})+dy(\pl_zB_{yz}-\pl_xB_{xy})
        +dz(\pl_xB_{zx}-\pl_yB_{yz}),
\\
  \dl C&=-(dx\wedge dy\pl_zC_{xyz}+dy\wedge dz\pl_xC_{xyz}
        +dz\wedge dx\pl_yC_{xyz}).
\end{align*}
Теперь нетрудно найти явное действие оператора Лапласа--Бельтрами в пространстве
форм:
\begin{align*}
  \triangle_\Sf f&=\triangle f,
\\
  \triangle_\Sf A&=dx\triangle A_x+dy\triangle A_y+dz\triangle A_z,
\\
  \triangle_\Sf B&=dx\wedge dy\triangle B_{xy}+dy\wedge dz\triangle B_{yz}
               +dz\wedge dx\triangle B_{zx},
\\
  \triangle_\Sf C&=dx\wedge dy\wedge dz\triangle C_{xyz},
\end{align*}
где
$$
  \triangle=\pl^2_x+\pl^2_y+\pl^2_z
$$
-- обычный оператор Лапласа в трехмерном евклидовом пространстве.
\qed\end{exa}
\begin{defn}
Действие оператора внешнего дифференцирования на 0-формы совпадает с
определением градиента функции, поэтому назовем {\em градиентом} произвольной
$r$-формы ее внешнюю производную
\begin{equation}                                                  \label{egradd}
  \grad:=d.
\end{equation}

Действие оператора $\dl$ на 1-формы с точностью до знака совпадает с
дивергенцией 1-формы. Поэтому примем в качестве инвариантного определения
{\em дивергенции} от произвольной $r$-формы соотношение
\begin{equation}                                                  \label{edivde}
  \div:=-\dl. \qed
\end{equation}
\end{defn}
\index{Градиент (gradient)}\index{Дивергенция (divergence)}%

В заключение рассмотрим еще один инвариантный оператор, действующий в
пространстве форм произвольной степени,
\begin{equation}                                                  \label{erotde}
  \rot:=\ast d:\quad\Lm_r\rightarrow\Lm_{n-r-1}.
\end{equation}
В частном случае трехмерного евклидова пространства (пример \ref{etheuc}) он
отображает 1-формы в 1-формы:
$$
  \ast dA=dx(\pl_yA_z-\pl_zA_y)+dy(\pl_zA_x-\pl_xA_z)+dz(\pl_xA_y-\pl_yA_x),
$$
что совпадает с определением ротора от 1-формы. Поэтому примем выражение
(\ref{erotde}) в качестве инвариантного определения {\em ротора}. Тогда
определение оператора Лапласа--Бельтрами (\ref{elabeo}) сводится к хорошо
известному тождеству из векторного анализа
\begin{equation}                                                  \label{evecid}
  \triangle_\Sf=\grad\div-\rot\rot.
\end{equation}
\index{Ротор (curl)}%
\section{Разложение Ходжа}
Если на многообразии $\MM$, $\dim\MM=n$, задана положительно определенная
риманова метрика, то можно доказать ряд важных результатов для дифференциальных
форм, которые не имеют места в общем случае. Для формулировки этих результатов
введем несколько понятий.
\begin{defn}
Форма $A\in\Lm_r(\MM)$ является {\em коточной}, если ее можно представить в виде
\begin{equation*}
  A=\dl B,\qquad \text{где}\quad B\in\Lm_{r+1}(\MM).
\end{equation*}
Форма $A\in\Lm_r(\MM)$ называется {\em козамкнутой}, если
\begin{equation*}
  \dl A=0.
\end{equation*}
Форма $A\in\Lm_r(\MM)$ называется {\em гармонической}, если
\begin{equation*}
  \triangle_\Sf A=0. \qed
\end{equation*}
\end{defn}
\index{Коточная форма (coexact form)}\index{Форма коточная (coexact form)}%
\index{Козамкнутая форма (coclosed form)}%
\index{Форма козамкнутая (coclosed form)}%
\index{Гармоническая форма (harmonic form)}%
\index{Форма гармоническая (harmonic form)}%
Любая функция $f\in\CC^1(\MM)$, очевидно, является козамкнутой, $\dl f=0$. Любая
коточная форма является козамкнутой в силу свойства (\ref{esqcob}).
\begin{theorem}
Любая гармоническая форма $A\in\Lm_r(\MM)$ на компактном ориентируемом
римановом многообразии является замкнутой и козамкнутой.
\end{theorem}
\begin{proof}
Из определения оператора Лапласа--Бельтрами (\ref{elabeo}) для
скалярного произведения (\ref{escpfo}) следует равенство
\begin{equation}                                                  \label{ehaccl}
  \langle\triangle_\Sf A,A\rangle=-\langle\dl A,\dl A\rangle
                                -\langle dA,dA\rangle.
\end{equation}
Отсюда следует, что условие гармоничности $\triangle_\Sf A=0$ влечет за собой
равенства $\dl A=0$ и $dA=0$. При этом ориентируемость и компактность
многообразия достаточны для существования интеграла, входящего в скалярное
произведение. Положительная определенность метрики необходима для того, чтобы
оба слагаемых в правой части равенства (\ref{ehaccl}) обращались в нуль по
отдельности, и из равенства $\langle\dl A,\dl A\rangle=0$ вытекало
$\dl A=0$ (аналогично для $dA$).
\end{proof}
\begin{cor}
Любая гармоническая функция на компактном ориентируемом римановом многообразии
является замкнутой $df=0$ и, следовательно, равна константе.
\qed\end{cor}
\begin{com}
Сформулированная теорема справедлива также для компактных неориентируемых
многообразий. Действительно, она верна для двулистного ориентируемого накрытия
$\widetilde\MM\rightarrow\MM$, которое всегда существует и тоже компактно, и при
проекции замкнутость и козамкнутость форм сохраняется.
\qed\end{com}
\begin{theorem}[\bf Разложение Ходжа]
Пусть $\MM$ -- компактное ориентируемое риманово многообразие. Тогда любую
$r$-форму $A\in\Lm_r(\MM)$ можно представить в виде суммы точной, коточной и
гармонической формы
\begin{equation}                                                  \label{ehodde}
  A=dB+\dl C+H,\qquad B\in\Lm_{r-1}(\MM),\quad C\in\Lm_{r+1}(\MM),\quad H\in\Lm_r(\MM).
\end{equation}
Это разложение единственно.
\end{theorem}
\begin{proof}
См., например, \cite{deRham46}.
\end{proof}
\index{Разложение Ходжа (Hodge decomposition)}%
\index{Ходжа разложение (Hodge decomposition)}%

Единственность разложения вытекает из ортогональности трех подпространств
относительно скалярного произведения форм:
\begin{equation*}
  \langle A,A\rangle=\langle dB,dB\rangle+\langle\dl C,\dl C\rangle
                    +\langle H,H\rangle.
\end{equation*}
\begin{cor}
Уравнение
\begin{equation*}
  \triangle_\Sf A=D,\qquad A,D\in\Lm_r(\MM),
\end{equation*}
на компактном ориентируемом многообразии имеет решение тогда и только тогда,
когда $r$-форма $D$ ортогональна пространству гармонических $r$-форм.
\qed\end{cor}
\begin{theorem}
На компактном ориентируемом римановом многообразии существует только конечное
число линейно независимых гармонических форм.
\end{theorem}
\begin{proof}
См., например, \cite{deRham55R}.
\end{proof}
\begin{defn}
Размерность пространства гармонических $r$-форм
\begin{equation*}
  \CH_r(\MM)=\lbrace H\in\Lm_r(\MM):\quad \triangle_\Sf H=0\rbrace
\end{equation*}
называется $r$-тым {\em числом Бетти}
\begin{equation}                                                  \label{ebetnu}
  b_r(\MM):=\dim\CH_r(\MM).  \qed
\end{equation}
\end{defn}
\index{Число Бетти (Betti number)}\index{Бетти число (Betti number)}%
Отметим несколько свойств чисел Бетти. Поскольку оператор Лапласа--Бельтрами
коммутирует с оператором Ходжа (\ref{elabec}), а оператор Ходжа устанавливает
изоморфизм $\CH_r(\MM)$ и $\CH_{n-r}(\MM)$, то $b_r(\MM)=b_{n-r}(\MM)$. Так как
все гармонические $0$-формы на компактном ориентируемом римановом многообразии
являются константами, то $b_0(\MM)=b_n(\MM)=1$. Отметим, что пространство
$\CH_n(\MM)$ состоит из $n$-форм, пропорциональных форме объема $\upsilon$
(\ref{evolel}). Если многообразие $\MM$ связно, то $b_n(\MM)=1$ или
$b_n(\MM)=0$ соответственно для ориентируемых и неориентируемых многообразий.

\begin{defn}
Число
\begin{equation}                                                  \label{eulnum}
  \chi(\MM)=\sum_{r=0}^n(-1)^r b_r(\MM)
\end{equation}
называется {\em эйлеровой характеристикой} компактного ориентируемого риманова
многообразия $\MM$.
\qed\end{defn}
\index{Эйлерова характеристика (Euler characteristic)}%
\index{Характеристика эйлерова (Euler characteristic)}%
\begin{theorem}
Пусть $\MM$ -- компактное ориентируемое риманово многообразие и $A$ --
дифференцируемая 1-форма. Тогда
\begin{equation*}
  \int\upsilon\dl A=0.
\end{equation*}
В частности, для произвольной функции $f\in\CC^2(\MM)$
\begin{equation*}
  \int\upsilon\triangle_\Sf f=0.
\end{equation*}
\end{theorem}
\begin{proof}
Справедливо равенство
\begin{equation*}                                                    \tag*{\qed}
  \int_\MM\upsilon\dl A=\langle\dl A,1\rangle=\langle A,d1\rangle=0.
\end{equation*}
\renewcommand{\qed}{}\end{proof}
\begin{theorem}
Пусть $\MM$ -- компактное риманово ориентируемое многообразие с положительной
гауссовой кривизной $K>0$. Тогда первое число Бетти равно нулю $b_1(\MM)=0$.
\end{theorem}
\begin{proof}
Напомним, что $K=-\frac12\widetilde R$, где $\widetilde R$ -- скалярная
кривизна, определенная в разделе \ref{scurpr}. Доказательство приведено,
например, в \cite{Aubin01}.
\end{proof}
\begin{prop}
Любая ковариантно постоянная форма является замкнутой, козамкнутой и
гармонической.
\end{prop}
\begin{proof}
Это следует из выражений для внешней производной (\ref{eexdco}), оператора
кограницы (\ref{eacdlo}) и определения оператора Лапласа--Бельтрами
(\ref{elabeo}).
\end{proof}
\begin{exa}
Форма объема $\upsilon$ (\ref{evolel}) замкнута, козамкнута и гармонична,
поскольку $\widetilde\nb_\al\vol=0$.
\qed\end{exa}
\section{Интегрирование дифференциальных форм                    \label{sintdf}}
Начнем с обсуждения интегрирования в евклидовом пространстве $\MR^n$. Для наших
целей достаточно знакомства с определением интеграла по Риману, т.к.\ мы будем
интегрировать непрерывные функции по достаточно хорошим областям.

Напомним правило замены переменных интегрирования. Пусть $h$ -- диффеоморфизм
ограниченного открытого подмножества $\MU\subset\MR^n$ на ограниченное
открытое подмножество $h(\MU)\subset\MR^n$. В координатах диффеоморфизм $h$
задается $n$ гладкими функциями $x^\al\mapsto y^\al(x)$, $\al=1,\dotsc,n$, где
$x^\al$ -- координаты точки $x\in\MU$. Пусть $f$ -- ограниченная непрерывная
функция на $\overline\MU$, включая границу. Тогда справедлива формула замены
переменных интегрирования, которая хорошо известна из курса математического
анализа,
\begin{equation}                                                  \label{qasdqb}
  \int_\MU\! dx^1\dots dx^n f(x)=
  \int_{h(\MU)}\!\!\!dy^1\dots dy^n |J|^{-1}f\big(x(y)\big).
\end{equation}
где $J:=\det(\pl y^\bt/\pl x^\al)$ -- якобиан преобразования координат.
Поскольку мы предполагаем, что отображение $h$ является диффеоморфизмом, то
якобиан преобразования всюду отличен от нуля. В частном случае, если $f=1$, то
равенство (\ref{qasdqb}) принимает вид
\begin{equation*}
  \int_\MU\! dx^1\dots dx^n=
  \int_{h(\MU)}\!\!\!dy^1\dots dy^n |J|^{-1}.
\end{equation*}
Если предположить, что $x^\al$ -- это декартовы координаты в $\MR^n$, то
последняя формула представляет собой объем области $\MU\subset\MR^n$. Его можно
вычислить как в декартовой системе координат $x^\al$, так и в произвольной
криволинейной системе координат $y^\al$. В последнем случае появляется
нетривиальный множитель, который равен модулю якобиана преобразования координат.

В дальнейшем мы ограничимся преобразованиями координат с положительным
якобианом, $J>0$, т.е.\ теми преобразованиями координат, которые сохраняют
ориентацию. Кроме этого будем использовать сокращенное выражение для элемента
интегрирования
\begin{equation*}
  dx:=dx^1\dots dx^n.
\end{equation*}
\begin{defn}
Рассмотрим произвольную $n$-форму на многообразии $\MM$, $\dim\MM=n$.
На карте $(\MU,\vf)$ она имеет вид
\begin{equation*}
  A=dx^1\wedge\dotsc\wedge dx^n A_{1\dotsc n}
  =\frac1{n!}dx^{\al_1}\wedge\dotsc\wedge dx^{\al_n} A_{\al_1\dotsc\al_n},
\end{equation*}
где в последнем выражении производится суммирование по всем значениям индексов.
Положим
\begin{equation}                                                  \label{eintan}
  \int_\MU\! A
  =\int_{\vf(\MU)}\!\!\! dx^1\wedge\dotsc\wedge dx^n A_{1\dotsc n}
  :=\int_{\vf(\MU)} \!\!\!dx^1\dotsc dx^n A_{1\dotsc n}(x)
  =\int_{\vf(\MU)} \!\!\!dx\, A_{1\dotsc n}(x).
\end{equation}
Правая часть этого равенства определена, т.к.\ $\vf(\MU)\subset\MR^n$, а запись
$A_{1\dotsc n}(x)$ обозначает значение единственной независимой компоненты
$n$-формы $A_{1\dotsc n}\circ\vf^{-1}$ в точке
$\lbrace x^\al\rbrace \in\vf(\MU)$. Мы, конечно, предполагаем, что интеграл в
правой части приведенной формулы сходится. Формула (\ref{eintan}) корректна в
том смысле, что не зависит от выбора системы координат, т.к.\ при преобразовании
координат $x^\al\mapsto y^\al(x)$ (замене переменных интегрирования)
единственная нетривиальная компонента $n$-формы $A_{1\dotsc n}$ преобразуется,
как тензорная плотность степени $-1$:
\begin{equation*}
  A_{1\dotsc n}(x)\mapsto J^{-1}A_{1\dotsc n}(y),
\end{equation*}
что совпадает с правилом замены переменных интегрирования (\ref{qasdqb}) при
положительном якобиане. Равенство (\ref{eintan}) примем за определение
{\em интеграла от $n$-формы} по области $\MU\subset\MM$.
\qed\end{defn}
\index{Интеграл (integral)}%
\begin{com}
Определение интеграла от $n$-формы по области $\MU\subset\MM$ не зависит от
того задана на многообразии метрика или нет.
\qed\end{com}
Из равенства
\begin{equation}                                                  \label{evoele}
  dx^{\al_1}\wedge\dotsc\wedge dx^{\al_n}=dx^1\wedge\dotsc\wedge dx^n
  \hat\ve^{\al_1\dotsc\al_n},
\end{equation}
где $\hat\ve^{\al_1\dotsc\al_n}$ -- полностью
антисимметричная тензорная плотность степени $-1$, следует, что интеграл
(\ref{eintan}) можно записать в виде
\begin{equation*}
  \int_\MU\! A=\frac1{n!}\int_{\vf(\MU)}\!\!\!dx\,\hat\ve^{\al_1\dotsc\al_n}
  A_{\al_1\dotsc\al_n}.
\end{equation*}
Поскольку компоненты тензорной плотности $\hat\ve^{\al_1\dotsc\al_n}$ во
всех системах координат по модулю равны единице, то этот интеграл не
зависит от метрики на $\MM$.

Правило замены переменных интегрирования можно переписать иначе на языке
возврата отображения. Пусть $h:~\MU\rightarrow h(\MU)$ диффеоморфизм двух
областей многообразия $\MM$, и пусть $n$-форма $A$ задана на области $h(\MU)$.
Тогда справедлива формула
\begin{equation}                                                  \label{qhwaqn}
  \int_{h(\MU)}\!\!\!A=\pm\int_\MU h^*A,
\end{equation}
где $h^*$ -- возврат отображения $h$. Если диффеоморфизм $h$ сохраняет
ориентацию, то в правой части равенства необходимо выбрать знак $``+''$.
В противном случае -- знак $``-''$.
\begin{com}
Запись интеграла $\int dx\,A_{1\dotsc n}$ в виде
$\int dx^1\wedge\dotsc\wedge dx^n A_{1\dotsc n}$ удобнее, т.к.\ якобиан
преобразования координат возникает автоматически при замене дифференциалов:
\begin{equation}                                                  \label{editra}
  dy^1\wedge\dotsc\wedge dy^n
  =\left(\frac{\pl y^1}{\pl x^{\al_1}}dx^{\al_1}\right)\wedge\dotsc
  \wedge\left(\frac{\pl y^n}{\pl x^{\al_n}}dx^{\al_n}\right)
  =Jdx^1\wedge\dotsc\wedge dx^n.\qed
\end{equation}
\end{com}

Теперь определим интеграл по произвольному ориентируемому многообразию с
помощью разбиения единицы.
\begin{defn}
Пусть на многообразии $\MM$, $\dim\MM=n$, заданы: ориентация, некоторый атлас
$\lbrace\MU_i\rbrace$, согласованный с ориентацией, и разбиение единицы
$\lbrace f_i\rbrace$, подчиненное выбранному покрытию $\lbrace\MU_i\rbrace$.
Рассмотрим произвольную непрерывную $n$-форму $A\in\Lm_n(\MM)$ с компактным
носителем. Тогда
\begin{equation*}
  A=\left(\sum_i f_i\right)A=\sum_i(f_i A).
\end{equation*}
Поскольку $\supp(f_i A)\subset\supp f_i$, то справедливо равенство
\begin{equation*}
  \int_\MM\! f_i A:=\int_{\MU_i}\!f_i A,
\end{equation*}
где интеграл в правой части был определен ранее (\ref{eintan}). Этот интеграл
определен корректно, т.к.\ не зависит от выбора системы координат на $\MU_i$.
Определим {\em интеграл по ориентируемому многообразию} от $n$-формы с
компактным носителем следующей формулой
\begin{equation}                                                  \label{eintdx}
  \int_\MM\! A=\sum_i\int_\MM\! f_i A.
\end{equation}
Для любого заданного разбиения единицы правая часть равенства однозначно
определена.
\qed\end{defn}
\index{Интеграл по многообразию (integral over a manifold)}%
Нетрудно проверить, что данное определение не зависит от разбиения единицы.
Требование компактности носителя формы $A$ является достаточным для
существования интеграла.
\begin{com}
При изменении ориентации системы координат в области $\MU$ компонента $n$-формы
$A_{1\dotsc n}$ изменит знак. Поэтому правая часть формулы (\ref{eintan}) также
изменит знак, что недопустимо. Поскольку на неориентируемом многообразии
не существует возможности однозначно задать ориентацию, то правая часть
равенства (\ref{eintdx}) будет зависеть от выбора ориентации областей и,
следовательно, определена неоднозначно. Поэтому требование ориентируемости
$\MM$ является существенным.
\qed\end{com}
Из свойств интеграла следует, что для любых двух $n$-форм $A$ и $B$ c компактным
носителем справедливы равенства
\begin{align*}
  \int_\MM\!(A+B)&=\int_\MM\! A+\int_\MM \!B,
\\
  \int_\MM\! aA&=a\int_\MM\! A,\qquad\qquad a\in\MR.
\end{align*}
Поэтому интеграл является линейным функционалом на множестве $n$-форм с
компактным носителем.

Если $\MM$ -- ориентируемое многообразие с краем $\pl\MM$, то положим
\begin{equation*}
  \int_\MM \!A:=\int_{\MM\setminus\pl\MM}\!\!\!A.
\end{equation*}
Правая часть этого равенства определена выше, т.к.\ $\MM\setminus\pl\MM$ --
ориентируемое многообразие без края.

Если ориентируемое многообразие $\MM$, $\dim\MM=n$, можно представить в виде
объединения
\begin{equation*}
  \MM=\big(\cup_i\MU_i\big)\cup\big(\cup_\Sa\MV_\Sa\big)
\end{equation*}
непересекающихся открытых подмногообразий $\MU_i$ (и, следовательно,
многообразий той же размерности $\dim\MU_i=n$, $\forall i$) и произвольного
набора подмногообразий меньшей размерности $\MV_\Sa$, $\dim\MV_\Sa<n$,
$\forall\Sa$, тогда
\begin{equation*}
  \int_\MM\!A=\sum_i\int_{\MU_i}\!\!\!A.
\end{equation*}
Это свойство можно использовать для вычисления интегралов, используя
``глобальные'' координаты, заданные на всем многообразии, за исключением
подмногообразий меньшей размерности, которые не дают вклада в интеграл.
\begin{exa}
В задачах со сферической симметрией в $\MR^3$ интегралы удобнее вычислять
в сферических координатах, которые покрывают все многообразие, за исключением
оси $z$. В этом случае евклидово пространство $\MR^3$ представляется в виде
\begin{equation*}
  \MR^3=\MU\cup\MV,
\end{equation*}
где $\MV=\MR$ -- ось $z$, и $\MU=\MR^3\setminus\MV$ -- неодносвязное
открытое подмногообразие в $\MR^3$, $\dim\MU=3$.
\qed\end{exa}
\subsection{Форма объема}
Пусть на ориентируемом многообразии $\MM$, $\dim\MM=n$, задана метрика
$dx^\al\otimes dx^\bt g_{\al\bt}$\footnote{Согласно
теореме \ref{trimex} на любом гладком многообразии можно задать риманову
(положительно определенную) метрику.} произвольной сигнатуры. Рассмотрим карту
$(\MU,\vf)$. В этой карте метрика имеет компоненты $g_{\al\bt}$, и ее
определитель при преобразовании координат $x\mapsto y(x)$ (в областях
пересечения двух карт с координатами $x^\al$ и $y^\al$) преобразуется по правилу
\begin{equation*}
  \det g_{\al\bt}(y)=\det g_{\al\bt}(x) J^{-2},
\end{equation*}
где  $J:=\det(\pl y^\bt/\pl x^\al)$ -- якобиан преобразования координат. Поэтому
определитель метрики является тензорной плотностью степени $-2$. Обозначим
\begin{equation}                                                  \label{emoege}
  g:=\det g_{\al\bt}.
\end{equation}
Тогда $\vol$ при преобразованиях координат умножается на модуль якобиана,
\begin{equation}                                                  \label{etraxy}
  \vol(y)=\vol(x)|J|^{-1}.
\end{equation}
Мы ограничимся преобразованиями координат с положительными якобианами, что
позволяет рассматривать $\vol$ как тензорную плотность степени $-1$. Если задано
координатное покрытие многообразия $\MM=\cup_i\MU_i$, то в каждой карте
определена плотность $\vol$. Для того, чтобы плотность $\vol$ была определена
глобально, необходимо существование такого координатного покрытия, в которым все
якобианы преобразований координат были положительны, а это влечет за собой
ориентируемость многообразия. Таким образом, плотность $\vol$ определена только
на ориентируемых многообразиях. Она всюду отлична от нуля, что следует из
невырожденности метрики.
\begin{defn}
$n$-форма на ориентируемом многообразии $\MM$ с заданной метрикой $g_{\al\bt}$
\begin{equation}                                                  \label{evolel}
  \upsilon:=dx^1\wedge\dots\wedge dx^n \vol
  =\frac1{n!}dx^{\al_1}\wedge\dots\wedge dx^{\al_n}
  \ve_{\al_1\dotsc\al_n},
\end{equation}
называется {\em формой объема} $\upsilon$, задаваемой метрикой $g_{\al\bt}$.
\qed\end{defn}
\index{Форма объема (volume form)}\index{Объема форма (volume form)}%
\begin{com}
Форма объема на ориентируемом многообразии $\MM$ задается произвольной метрикой,
независимо от ее сигнатуры. Ясно, что различные метрики могут определять одну и
ту же форму объема из-за наличия знака модуля. Например, евклидова метрика и
метрика Лоренца в евклидовом пространстве $\MR^n$ задают одинаковую форму
объема.
\qed\end{com}
\begin{com}
Для обозначения формы объема мы используем греческую букву эпсилон $\upsilon$,
которая по написанию очень похожа на латинскую букву $v$.
\qed\end{com}

Форма объема определена глобально на ориентируемых многообразиях и всюду отлична
от нуля, поэтому она задает ориентацию многообразия и выбор знака в форме
(\ref{evolel}) соответствует выбору ориентации. При замене координат с
отрицательным якобианом ориентация формы объема меняется на противоположную,
а интеграл меняет знак.

Метрика евклидова пространства $\MR^n$ в декартовых координатах равна единичной
матрице (\ref{eclmet}). Поэтому форма объема евклидова пространства $\MR^n$ в
декартовых координатах имеет канонический вид
\begin{equation}                                                  \label{ecaore}
  \upsilon_0=dx^1\wedge\dotsc\wedge dx^n.
\end{equation}
Будем говорить, что данная форма объема задает {\em каноническую ориентацию}.
\index{Каноническая ориентация (canonical orientation)}%
\index{Ориентация каноническая (canonical orientation)}%
\begin{com}
Покажем, что форма объема $\upsilon$ на многообразии $\MM$ соответствует
нашему обычному представлению об объеме. Для этого зафиксируем произвольную
точку $x$ риманова многообразия $\MM$ с положительно определенной метрикой и
диагонализируем в ней метрику, что всегда возможно в произвольной фиксированной
точке. Тогда все базисные векторы в точке $x$ будут ортогональны, а длина
базисного вектора $\pl_\al$ будет равна $\sqrt{g_{\al\al}}$. В этом случае
объем бесконечно малого параллелепипеда со сторонами $\pl_\al$ равен
$\prod_\al\sqrt{g_{\al\al}}=\sqrt{g}$.
\qed\end{com}
Интеграл (\ref{eintan}) от произвольной $n$-формы по многообразию можно
переписать, используя форму объема,
\begin{equation}                                                  \label{eintin}
  \int_\MU\!A=
  \frac1{n!}\int_{\vf(\MU)}\!\!\! dx^{\al_1}\wedge\dots\wedge dx^{\al_n}
  A_{\al_1\dotsc\al_n}
  =\frac1{n!}\int_{\vf(\MU)}\!\!\! d x\,\vol\ve^{\al_1\dotsc\al_n}
  A_{\al_1\dotsc\al_n}\sgn,
\end{equation}
где множитель $\sgn$ равен знаку определителя метрики (\ref{esgnmn}).

Приведем два свойства формы объема. Нетрудно проверить, что форма объема
дуальна к единице
$$
  \upsilon=\frac1{n!}dx^{\al_1}\wedge\dots\wedge dx^{\al_n}
  \ve_{\al_1\dotsc\al_n}=*1.
$$
Кроме того она является гармонической формой $\triangle_\Sf\upsilon=0$, т.к.\
форма объема точна $d\upsilon=0$, как форма максимальной размерности, и коточна
$\dl\upsilon=0$, поскольку $(n+1)$-форм на $\MR^n$ не существует и
$\widetilde\nb_\al \vol=0$.

Если на многообразии задан репер, $g_{\al\bt}=e_\al{}^a e_\bt{}^b\eta_{ab}$, где
$\eta_{ab}$ -- евклидова или лоренцева метрика, то $\vol=|\det e_\al{}^a|$.
Тогда репер $e_\al{}^a$ однозначно определяет форму объема $\upsilon$. В
обратную сторону утверждение неверно, т.к.\ форма объема определяет только
определитель репера. Даже если метрика и форма объема на ориентируемом
многообразии $\MM$ существуют, то репер может не существовать. Локально репер
всегда существует, однако глобально это не так.
\begin{exa}
В теореме \ref{tvecsp} о невозможности причесать ежа доказано, что на
четномерной сфере не существует непрерывных векторных векторных полей, нигде не
обращающихся в нуль. Поскольку репер представляет собой набор из $n$ линейно
независимых в каждой точке векторных полей на многообразии $\MM$, $\dim\MM=n$,
то на четномерной сфере не существует глобально определенного репера. В то же
время метрика определена глобально.
\qed\end{exa}

Наличие формы объема на (псевдо-)римановом ориентируемом многообразии позволяет
определить инвариантным образом интеграл от произвольной функции
$f(x)\in\CC(\MM)$ с компактным носителем:
$$
  \int_\MM\!\upsilon f=\int_\MM \!dx^1\wedge\dotsc\wedge dx^n \vol f(x)
  =\int_\MM\! dx\,\vol f(x).
$$
Используя определение скалярного произведения (\ref{escpfo}), этот интеграл
можно переписать в виде
\begin{equation*}
  \int_\MM\!\upsilon f=\langle f,1\rangle.
\end{equation*}
\begin{com}
Именно это свойство чаще всего используется для построения моделей
математической физики. А именно, мы фиксируем некоторый набор полей
$\lbrace\vf^a(x)\rbrace$, $a=1,2,\dotsc$, на многообразии $\MM$, среди
которых содержится метрика пространства-времени, как в общей теории
относительности. Затем строится лагранжева плотность
$L(\vf,\pl\vf,\pl^2\vf,\dotsc)$, которая является скалярным полем (функцией),
зависящим от данного набора полей и их частных производных.
Определяем инвариантное действие на многообразии
\begin{equation*}
  S:=\int_\MM\! \upsilon L.
\end{equation*}
Уравнения Эйлера--Лагранжа для данного действия принимаются в качестве уравнений
движения (уравнений равновесия и т.д.) для рассматриваемой модели.
По-построению, действие инвариантно относительно общих преобразований координат
и, возможно, других преобразований симметрии. Соответствующие уравнения
Эйлера--Лагранжа ковариантны, т.е.\ преобразуются, как тензорные поля, и из
второй теоремы Нетер следует зависимость уравнений движения. В релятивистских
моделях теории поля, не описывающих гравитационные взаимодействия,
предполагается, что многообразие $\MM$ является пространством-временем
Минковского $\MR^{1,3}$ с заданной лоренцевой метрикой, по которой варьирование
не проводится. Соответствующие модели инвариантны относительно преобразований из
группы Пуанкаре и, возможно, других групп симметрии, а инвариантность
относительно общих преобразований координат в них отсутствует. В этом случае из
первой теоремы Нетер следуют законы сохранения.
\qed\end{com}

Приведем без доказательства два свойства формы объема $\upsilon$ на
ориентированном римановом многообразии $\MM$, $\dim\MM=n$. Пусть
$X_1,\dotsc,X_n$ и $Y_1,\dotsc,Y_n$ -- два набора векторных полей на
$\MM$. Тогда
\begin{align*}
  \upsilon(X_1,\dotsc,X_n)\cdot\upsilon(Y_1,\dotsc,Y_n)&
  =\det\lbrace(X_i,Y_j)\rbrace,\qquad i,j=1,\dotsc,n,
\\
  \upsilon(X_1,\dotsc,X_n)\cdot\upsilon&=\tilde X_1\wedge\dotsc\wedge\tilde X_n,
\end{align*}
где $(X_i,Y_j):=X^\al_i Y^\bt_j g_{\al\bt}$ и $\tilde X_i:=dx^\al X_{i\al}$
обозначает 1-форму, сопряженную в смысле римановой метрики к векторному полю
$X_i=X^\al_i\pl_\al$: $X_{i\al}:=X^\bt_i g_{\bt\al}$.
\begin{defn}
Интеграл от формы объема (\ref{evolel}) по всему многообразию, если он
существует, называется {\em объемом} ориентируемого многообразия $\MM$
\begin{equation}                                                  \label{evoldu}
  V_\MM:=\int_\MM\!\upsilon=\int_\MM\!dx\,\vol. \qed
\end{equation}
\end{defn}
\index{Объем (volume)}%
Объем многообразия определяется метрикой, входящей в определение формы объема,
и всегда положителен, т.к.\ ориентация многообразия фиксирована.
\subsection{Формула Стокса                                       \label{stokes}}
Пусть $\MN$, $\dim\MN=n$ -- ориентируемое многообразие, на котором задана
произвольная $r$-форма $A\in\Lm_r(\MN)$. Рассмотрим $r$-мерное ориентируемое
подмногообразие $(f,\MM)$ в $\MN$. Поскольку $r$-форма задана также на
подмногообразии $\MM\subset\MN$, то ее можно по нему проинтегрировать.
Пусть подмногообразие $\MM\subset\MN$ задано параметрически\footnote{Чтобы не
усложнять ситуацию склейкой карт, которую можно провести, будем считать, что
каждое многообразие покрывается одной картой.}
$$
  x^\al=x^\al(u^i),\qquad \al=1,\dotsc,n,\quad i=1,\dotsc,r\le n.
$$
Тогда интеграл от произвольной $r$-формы $A$ степени $r$ по этому
подмногообразию равен
\begin{equation}                                                  \label{edefir}
\begin{split}
  \int_{f(\MM)}\!\!\!A&=
  \frac1{r!}\int_{f(\MM)} \!\!\!dx^{\al_1}\wedge\dotsc\wedge dx^{\al_r}
  A_{\al_1\dotsc\al_r}=
\\
  &=\frac1{r!}\int_\MM \!du^{i_1}\wedge\dotsc\wedge du^{i_r}
  \frac{dx^{\al_1}}{du^{i_1}}\dotsc\frac{dx^{\al_r}}{du^{i_r}}
  A_{\al_1\dotsc\al_r}.
\end{split}
\end{equation}
Этот интеграл имеет явно ковариантный вид и не зависит ни от выбора
системы координат $x^\al$, ни от параметризации подмногообразия $u^i$.
\begin{exa}
Рассмотрим интеграл от 1-формы $A=dx^\al A_\al$ вдоль кривой $x^\al(t)$,
$t\in[a,b]$, соединяющей точки $x_1$ и $x_2$ в евклидовом пространстве $\MR^n$.
Тогда интеграл (\ref{edefir}) принимает стандартный вид
$$
  \int_{x_1}^{x_2}\!\!\!dx^\al A_\al=\int_a^b \!\!\!dt\frac{dx^\al}{dt}A_\al.
$$
Этот интеграл инвариантен относительно перепараметризации кривой и общего
преобразования координат на многообразии.
\qed\end{exa}
\begin{exa}                                                       \label{xinttw}
Интеграл от 2-формы $A=\frac12dx^\al\wedge dx^\bt A_{\al\bt}$ по поверхности в
евклидовом пространстве $\MR^n$, заданной параметрически
$x^\al=x^\al(u,v)$, $(u,v)\in\MU$, равен
\begin{equation}                                                  \label{exsute}
  \frac12\int_{f(\MU)} \!\!\!dx^\al\wedge dx^\bt A_{\al\bt}
  =\frac12\int_\MU \!du\wedge dv\left(\frac{dx^\al}{du}\frac{dx^\bt}{dv}
  -\frac{dx^\al}{dv}\frac{dx^\bt}{du}\right)A_{\al\bt}.
\end{equation}
В трехмерном евклидовом пространстве $\MR^3$ с декартовыми координатами $x^\al$
этот интеграл можно преобразовать к виду, хорошо знакомому из курса
математического анализа. Касательные векторы к поверхности имеют вид
$$
  \x=\frac{dx^\al}{du}\Be_\al,\qquad \eta=\frac{dx^\al}{dv}\Be_\al,
$$
где $\Be_\al$ -- ортонормированные базисные векторы декартовой системы
координат. Их векторное произведение
\begin{equation*}
  n^\al:=[\x,\eta]^\al=\ve^{\al\bt\g}\x_\bt\eta_\g
\end{equation*}
нормально к поверхности. Длина нормального вектора равна
$\sqrt{n^\al n_\al}=\sqrt{\x^2\eta^2-(\x\eta)^2}$, что совпадает с элементом
площади $\sqrt{g}$, где $g=\det g_{ij}$ -- определитель индуцированной метрики
на поверхности
$$
  g_{ij}=\frac{dx^\al}{dw^i}\frac{dx^\bt}{dw^j}\dl_{\al\bt},\qquad
  \lbrace w^i\rbrace=\lbrace u,v\rbrace,\quad i,j=1,2.
$$
Так как в трехмерном евклидовом пространстве 2-форма $A$ взаимно однозначно
определяется векторным полем $X=X^\al\pl_\al$, где
$A_{\al\bt}=\ve_{\al\bt\g}X^\g$, то интеграл (\ref{exsute}) можно переписать в
виде
\begin{equation}                                                  \label{einsur}
  \frac12\int_{f(\MU)} \!\!\!dx^\al\wedge dx^\bt A_{\al\bt}
  =\int_\MU \!du\wedge dv\, \vol\,(X^\al N_\al),
\end{equation}
где
$$
  N^\al:=\frac{[\x,\eta]^\al}\vol
$$
-- единичный вектор нормали к поверхности. Таким образом интеграл от 2-формы по
поверхности в трехмерном евклидовом пространстве $\MR^3$ сводится к суммарному
потоку дуального векторного поля через эту поверхность.
\qed\end{exa}
Сформулируем общую теорему, связывающую интеграл от $(n-1)$-формы $A$ по границе
некоторой области $\pl\MU$ с интегралом от $n$-формы $dA$ по самой области
$\MU$.
\begin{theorem}[\bf Формула Стокса]                               \label{tstote}
Пусть $\MU\subset\MM$ -- открытое подмногообразие в $\MM$ такое, что его
замыкание $\overline\MU$ компактно. Пусть $\pl\MU=\overline\MU\setminus\MU$ --
его край, который предполагается кусочно гладким. Тогда справедлива формула
Стокса
\begin{equation}                                                  \label{eforst}
  \int_{\pl\MU}\!\!\!A=\int_\MU \!dA.
\end{equation}
где $A\in\Lm_{n-1}(\MM)$ -- произвольная гладкая дифференциальная $(n-1)$-форма
на $\MM$. При этом предполагается, что ориентация границы задана каноническим
образом.
\end{theorem}
\index{Формула Стокса (Stokes formula)}\index{Стокса формула (Stokes formula)}%
\begin{proof}
Доказательство формулы Стокса довольно громоздко и приведено, например,
в \cite{ChChLa00}. Каноническая ориентация края определена в доказательстве
теоремы \ref{toribo}.
\end{proof}

Формулу Стокса можно переписать в другом виде. Пусть на многообразии задано
векторное поле $X=X^\al\pl_\al$. Тогда ему можно взаимно однозначно поставить в
соответствие $(n-1)$-форму:
$$
  A:=\ast X=\frac1{(n-1)!}dx^{\al_2}\wedge\dots\wedge dx^{\al_n}
  \ve_{\bt\al_2\dotsc\al_n}X^\bt.
$$
Внешний дифференциал этой формы равен
\begin{align*}                                                         \nonumber
  dA&=\frac1{(n-1)!}dx^{\al_1}\wedge\dots\wedge dx^{\al_n}
  \pl_{\al_1}(\ve_{\bt\al_2\dotsc\al_n}X^\bt)=
\\
  &=dx^1\wedge\dots\wedge dx^n \vol\widetilde\nb_\al X^\al.
\end{align*}
Отсюда следует, что формулу Стокса можно записать в виде
\begin{equation}                                                  \label{estsec}
  \int_{\pl\MU}\!\!\!X^\al ds_\al=
  \int_\MU \!dx^1\wedge\dots\wedge dx^n \vol\widetilde\nb_\al X^\al,
\end{equation}
где
\begin{equation}                                                  \label{ehypsu}
  ds_\al:=\frac1{(n-1)!}dx^{\al_2}\wedge\dots\wedge dx^{\al_n}
  \ve_{\al\al_2\dotsc \al_n}
\end{equation}
-- ориентированный {\em элемент площади} гиперповерхности.
\index{Элемент площади гиперповерхности (area element of hypersurface)}%

Рассмотрим частные случаи формулы Стокса.
\begin{exa}
Пусть многообразие $\MM$ одномерно, $n=1$, а форма $A=f(x)$ имеет нулевую
степень. Тогда граница области $\MU=(a,b)$ состоит из двух точек
$\pl\MU=\lbrace a,b\rbrace$, и формула Стокса принимает вид
$$
  \int_{\pl\MU}\!\!\!f=f(b)-f(a)=\int_a^b\!df=\int_a^b \!dx\pl_xf.
$$
Здесь под интегралом от функции (0-формы) по подмногообразию
нулевой размерности, состоящему из отдельных точек, понимается
алгебраическая сумма значений функции в этих точках.
Таким образом мы получили хорошо известную {\em формулу Ньютона--Лейбница}.
\index{Формула Ньютона--Лейбница (Newton--Leibnitz formula)}%
\index{Ньютона--Лейбница формула (Newton--Leibnitz formula)}%
\qed\end{exa}
\begin{exa}
Пусть задана 1-форма $dx^\al A_\al$ на евклидовой плоскости с декартовыми
координатами $\lbrace x^\al\rbrace=\lbrace x,y\rbrace$. Рассмотрим
область $\MU$, ограниченную замкнутой кусочно гладкой кривой $x^\al(t)$. Тогда
\begin{equation}                                                  \label{egreef}
  \int_{\pl\MU}\!\!\!dt\frac{dx^\al}{dt}A_\al
  =\int_{\pl\MU}\!(dxA_x+dyA_y)=\int_\MU dx\wedge dy
  \left(\frac{\pl A_y}{\pl x}-\frac{\pl A_x}{\pl y}\right).
\end{equation}
Таким образом мы получили {\em формулу Грина}.
\index{Формула Грина (Green's formula)}\index{Грина формула (Green's formula)}%
\qed\end{exa}
\begin{exa}
Пусть задана 1-форма $A=dx^\al A_\al$ в трехмерном евклидовом пространстве c
декартовыми координатами $x^\al$. Рассмотрим поверхность $\MU\subset\MR^3$, с
кусочно гладкой границей $\pl\MU$. Тогда общая формула Стокса (\ref{eforst})
принимает вид
\begin{multline*}
  \oint_{\pl\MU}\!\!\!dx^\al A_\al=
\\
  =\int_\MU\!\big[dx^1\wedge dx^2(\pl_1A_2-\pl_2A_1)
  +dx^1\wedge dx^3(\pl_1A_3-\pl_3A_1)+dx^2\wedge dx^3(\pl_2A_3-\pl_3A_2)\big].
\end{multline*}
Если поверхность задана параметрически $x^\al=x^\al(u,v)$, то последний
интеграл можно преобразовать к виду
\begin{equation}                                                  \label{estofo}
  \oint_{\pl\MU}\!\!\!dx^\al A_\al
  =\int_\MU\! du\wedge dv\vol\,(\rot A,N),
\end{equation}
где введен ротор (\ref{erotde}) 1-формы:
$[\rot A]^\al:=\ve^{\al\bt\g}\pl_\bt A_\g$ и $N=N^\al\pl_\al$ -- единичный
вектор нормали. Таким образом получена {\em формула Стокса}.
\qed\end{exa}
\index{Формула Стокса (Stokes formulae)}%
\index{Стокса формула (Stokes formulae)}%
\begin{exa}
Пусть в трехмерном евклидовом пространстве $\MR^3$ с декартовыми координатами
$x^\al$ задана 2-форма $A=\frac12dx^\al\wedge dx^\bt A_{\al\bt}$. Рассмотрим
двумерную компактную поверхность $\MU\subset\MR^3$ с кусочно дифференцируемой
границей $\pl\MU$. Будем считать, что поверхность задана параметрически
$x^\al=x^\al(u,v)$. Тогда общая формула Стокса (\ref{eforst}) принимает вид
\begin{equation}                                                  \label{exastf}
  \frac12\int_{\pl\MU}\!\!\!dx^\al\wedge dx^\bt A_{\al\bt}
  =\int_\MU \!dx^1\wedge dx^2\wedge dx^3
  (\pl_1A_{23}+\pl_2A_{31}+\pl_3A_{12}).
\end{equation}
Как уже отмечалось 2-форма в трехмерном пространстве взаимно однозначно
определяется векторным полем $A_{\al\bt}=\ve_{\al\bt\g}X^\g$.
Интеграл в левой части равенства сводится к интегралу по поверхности
от скалярного произведения вектора $X^\al$ на единичную нормаль $N=N^\al\pl_\al$
к поверхности (\ref{einsur}). Подынтегральное выражение в правой части
(\ref{exastf}) представляет собой дивергенцию от векторного поля
$$
  \pl_1A_{23}+\pl_2A_{31}+\pl_3A_{12}=\pl_\al X^\al.
$$
Таким образом окончательно получаем равенство
\begin{equation}                                                  \label{egaosf}
  \int_{\pl\MU}\!\!\!du\wedge dv\,\vol\, X^\al N_\al
  =\int_\MU \!dx^1\wedge dx^2\wedge dx^3\, \pl_\al X^\al,
\end{equation}
где $\vol$ -- квадратный корень из индуцированной метрики (см.\ пример
\ref{xinttw}). Это есть {\em формула Гаусса--Остроградского} в
трехмерном евклидовом пространстве.
\index{Формула Гаусса--Остроградского (Gauss--Ostrogradski\u\i formula)}%
\index{Гаусса--Остроградского формула (Gauss--Ostrogradski\u\i formula)}%
\qed\end{exa}
\chapter{Метрика                                                 \label{smetri}}
Одним из важнейших понятий в геометрии и физике является дифференциально
геометрическая метрика или просто метрика. Трудно переоценить ту роль, которую
метрика играет в физических приложениях. С ее помощью строятся инварианты,
определяется форма объема. Симметрии метрики являются симметриями
пространства-времени, которые определяют сохраняющиеся токи. В частности,
фундаментальные законы сохранения энергии-импульса и момента количества движения
связаны с наличием у метрики определенных симметрий (см.\ раздел \ref{sfinet}).
\section{Определение и свойства                                  \label{smetde}}
Рассмотрим многообразие $\MM$, $\dim\MM=n$.
\begin{defn}
{\em Метрикой} $g\in\CT_2(\MM)$ на многообразии $\MM$ называется достаточно
гладкое ковариантное тензорное поле типа $(0,2)$, которое является
симметричным и невырожденным в каждой точке многообразия:

1) \parbox[t]{.93\linewidth}{$g_{\al\bt}=g_{\bt\al}$ -- симметричность,}

2) \parbox[t]{.93\linewidth}{$\det g_{\al\bt}\ne0$ -- невырожденность,}
где $g=dx^\al\otimes dx^\bt g_{\al\bt}$ -- выражение для метрики в координатном
базисе. Метрика называется {\em римановой}, если матрица $g_{\al\bt}(x)$
положительно определена во всех точках $x\in\MM$. В противном случае метрика
называется {\em псевдоримановой}.
\qed\end{defn}
\index{Метрика (metric)}%
\index{Риманова метрика (Riemannian metric)}%
\index{Метрика риманова (Riemannian metric)}%
\index{Псевдориманова метрика (pseudo-Riemannian metric)}%
\index{Метрика псевдориманова (pseudo-Riemannian metric)}%
\begin{com}
Метрика в данном определении не совпадает с тем же термином (топологическая
метрика), определенном в разделе \ref{seucme}. Слово метрика широко используется
в научной литературе в обоих значениях. Как правило, смысл термина ясен из
контекста. Там, где мы хотим подчеркнуть различие терминов, мы будем говорить
дифференциально геометрическая метрика для метрики, которая определена в
настоящем разделе, и топологическая метрика, определенная в разделе
\ref{seucme}. Далее в подавляющем большинстве случаев термин метрика будет
употребляться в дифференциально геометрическом смысле.
\qed\end{com}

Метрика определяет билинейное невырожденное и симметричное отображение
\begin{equation*}
  g:\quad\CX(\MM)\times\CX(\MM)\ni\quad X,Y\mapsto (X,Y)\quad\in\CC^\infty(\MM),
\end{equation*}
которое называется {\em скалярным произведением} векторных полей.
\index{Скалярное произведение векторных полей (scalar product of vector fields)}%
В компонентах  скалярное произведение векторных полей
$X=X^\al\pl_\al$ и $Y=Y^\al\pl_\al$ задается сверткой индексов:
\begin{equation}                                                  \label{escvec}
  (X,Y):=X^\al Y^\bt g_{\al\bt},
\end{equation}
для которого мы также будем иногда употреблять обозначение $(X,Y)=g(X,Y)$, чтобы
отметить метрику, с помощью которой проводится свертка. Скалярное произведение,
очевидно, симметрично $(X,Y)=(Y,X)$. Невырожденность метрики формулируется так:
не существует отличного от нуля векторного поля $X\in\CX(\MM)$ такого, что
скалярное произведение $(X,Y)=0$ для всех $Y\in\CX(\MM)$ и во всех точках
$x\in\MM$. Для римановой метрики скалярное произведение является положительно
определенным в каждой точке, т.е.\ $(X,X)\ge0$, причем равенство $(X,X)=0$ имеет
место только для тривиального векторного поля $X=0$. Скалярное произведение
базисных векторных полей $\pl_\al$ равно компонентам метрики:
\begin{equation}                                                  \label{escbap}
  (\pl_\al,\pl_\bt)=g_{\al\bt}.
\end{equation}

Поскольку метрика невырождена, то существует {\em обратная метрика},
\index{Обратная метрика (inverse metric)}%
\index{Метрика обратная (inverse metric)}%
т.е.\ симметричное невырожденное контравариантное тензорное поле
$g^{\al\bt}\pl_\al\otimes\pl_\bt\in\CT^2(\MM)$ типа $(2,0)$,
компоненты которого в каждой точке многообразия удовлетворяет условию
\begin{equation}                                                  \label{einvme}
  g^{\al\bt}g_{\bt\g}=g_{\g\bt}g^{\bt\al}=\dl^\al_\g.
\end{equation}
При преобразовании координат $x^\al\rightarrow x^{\al'}(x)$ компоненты метрики
и ее обратной преобразуются по обычным правилам для тензорных полей:
\begin{align}                                                     \label{emetra}
  g_{\al'\bt'}&=\frac{\pl x^\al}{\pl x^{\al'}}
  \frac{\pl x^\bt}{\pl x^{\bt'}}g_{\al\bt},
\\                                                                \label{eimetr}
  g^{\al'\bt'}&=g^{\al\bt}\frac{\pl x^{\al'}}{\pl x^\al}
  \frac{\pl x^{\bt'}}{\pl x^\bt}.
\end{align}
Отсюда следует, что определитель метрики
\begin{equation}                                                  \label{edetme}
  g(x):=\det g_{\al\bt}(x)
\end{equation}
преобразуется по правилу $g'=J^{-2}g$, где $J$ -- якобиан преобразования
координат (\ref{ejacob}). То есть определитель метрики является скалярной
плотностью степени $-2$.

В общем случае, ввиду симметрии по индексам, метрика $g_{\al\bt}$ задается
$$
  [g_{\al\bt}]=\frac{n(n+1)}2
$$
произвольными компонентами с единственным условием невырожденности. Здесь и
в дальнейшем число компонент тензора мы будем обозначать квадратными скобками.

С помощью метрики и ее обратной можно изменить тип тензора путем опускания или
подъема всех или части индексов произвольного тензорного поля.
\begin{exa}
Произвольному векторному полю $X\in\CX(\MM)$ можно взаимно однозначно поставить
в соответствие 1-форму и наоборот. В компонентах отображение задается
следующими формулами:
\begin{equation}                                                  \label{epopin}
  X_\al:=X^\bt g_{\bt\al},\qquad X^\al=g^{\al\bt}X_\bt.
\end{equation}
Это отображение $\CX(\MM)\leftrightarrow\Lm_1(\MM)$, очевидно, линейно, не
зависит от выбора системы координат и, значит, является изоморфизмом
$\CC^\infty(\MM)$-модулей. Операция подъема и опускания индексов естественным
образом продолжается на тензоры произвольного типа.
\qed\end{exa}
\begin{com}
Если на многообразии задана метрика, то между контравариантными и ковариантными
тензорными полями одного ранга $r$ существует взаимно однозначное соответствие,
точнее, изоморфизм $\CC^\infty(\MM)$-модулей:
\begin{equation*}
  \CT^r(\MM)\simeq\CT_r(\MM)\simeq\CT^p_q(\MM),\qquad\forall p,q:~p+q=r.
\end{equation*}
По этой причине будем считать, что тензоры одного ранга, но с различным типом
индексов, описывают один и тот же геометрический объект и будем обозначать их,
как правило, одной буквой. Например,
\begin{equation*}
  T_{\al\bt}=T_\al{}^\g g_{\g\bt}=T^\g{}_\bt g_{\g\al}
  =T^{\g\dl}g_{\g\al}g_{\dl\bt},
\end{equation*}
где $T_{\al\bt}$ и $T^{\al\bt}$ -- ковариантные и контравариантные компоненты
тензора второго ранга. Компоненты $T_\al{}^\bt$ и $T^\al{}_\bt$ называются
смешанными. Поскольку при наличии метрики появилась возможность опускать и
поднимать индексы, то необходимо следить за порядком ковариантных и
контравариантных индексов. В общем случае $T_\al{}^\bt\ne T^\bt{}_\al$. Поэтому
не стоит писать контра- и ковариантные индексы один под другим.
\qed\end{com}
\begin{defn}
Скалярное произведение вектора в каждой точке $x\in\MM$ с самим собой
называется {\em квадратом} вектора и обозначается
\begin{equation}                                                  \label{esqucf}
  X^2:=(X,X).
\end{equation}
{\em Длиной вектора} $X$ назовем выражение
\begin{equation}                                                  \label{elenve}
  |X|:=\sqrt{|X^2|}.
\end{equation}
Знак модуля в этом определении необходим в том случае, если метрика не является
положительно определенной. Пусть $X,Y$ -- два векторных поля на римановом
многообразии $\MM$, т.е.\ с положительно определенной метрикой. Если в точке
$x\in\MM$ они отличны от нуля, то определим {\em угол} $\vf$ между ними в данной
точке с помощью следующего соотношения
\begin{equation}                                                  \label{evecan}
  \cos\vf:=\frac{(X,Y)}{\sqrt{X^2 Y^2}}.   \qed
\end{equation}
\end{defn}
\index{Квадрат вектора (square of a vector)}
\index{Длина вектора (length of a vector)}%
\index{Вектора длина (length of a vector)}%
\index{Угол между векторами (angle between vectors)}%

Определение угла (\ref{evecan}) корректно. Действительно, для положительно
определенной метрики правая часть соотношения (\ref{evecan}) не превышает
единицу, поскольку справедливо неравенство треугольника
$$
  (X,Y)\le\sqrt{X^2 Y^2}.
$$
Определение угла инвариантно относительно выбора координат, т.к.\ правая часть
(\ref{evecan}) содержит только инвариантные комбинации векторных полей.
\begin{defn}
Если $(X,Y)=0$, то два вектора называются {\em ортогональными} или
{\em перпендикулярными}, независимо от положительной определенности метрики.
\qed\end{defn}
\index{Ортогональные векторы (orthogonal vectors)}%
\index{Векторы ортогональные (orthogonal vectors)}%
\index{Перпендикулярные векторы (perpendicular vectors)}%
\index{Векторы перпендикулярные (perpendicular vectors)}%
\begin{defn}
Две метрики $g'=dx^\al\otimes dx^\bt g'_{\al\bt}$ и
$g=dx^\al\otimes dx^\bt g_{\al\bt}$ на многообразии $\MM$ называются
{\em конформно эквивалентными} или связанными преобразованием Вейля, если они
пропорциональны
\begin{equation}                                                  \label{ecoeqm}
  g'=\ex^{2\phi}g,
\end{equation}
где $\phi(x)$ -- некоторая функция на $\MM$. Преобразование (\ref{ecoeqm})
называют {\em конформным преобразованием} метрики или
{\em преобразованием Вейля}.
\qed\end{defn}
\index{Конформно эквивалентные метрики (conformally equivalent metrics)}%
\index{Метрики конформно эквивалентные (conformally equivalent metrics)}%
\index{Преобразование Вейля (Weyl transformation)}%
\index{Вейля преобразование (Weyl transformation)}%

Это определение не зависит от выбора системы координат и поэтому задает
вейлевскую (конформную) эквивалентность глобально.

Угол между векторами определяется только метрикой и не изменится, если метрику
умножить на произвольную функцию, отличную от нуля (преобразование Вейля).
\begin{com}
Термин конформная эквивалентность метрик широко используется, однако он не имеет
отношения к конформным преобразованиям в комплексном анализе. Конформные
преобразования комплексных переменных -- это подпсевдогруппа общих
преобразований координат, в то время как при преобразовании (\ref{ecoeqm})
координаты не меняются. Поэтому мы будем употреблять термин вейлевская
инвариантность, так как в двумерных моделях математической физики
термин конформная инвариантность употребляется в своем первоначальном значении,
как в комплексном анализе.
\qed\end{com}

Поскольку метрика невырождена, то у нее существует обратная метрика
(\ref{einvme}), которая естественным образом определяет скалярное произведение
в пространстве 1-форм. Скалярное произведение двух 1-форм $A=dx^\al A_\al$ и
$B=dx^\bt B_\bt$ по определению равно
\begin{equation}                                                  \label{esupfo}
  (A,B)=g(A,B):=g^{\al\bt}A_\al B_\bt.
\end{equation}
Это скалярное произведение согласовано с операцией подъема и опускания
индексов (\ref{epopin}):
\begin{equation*}
  g^{\al\bt}A_\al B_\bt=g_{\al\bt}A^\al B^\bt,\qquad\text{где}\quad
  A^\al:=g^{\al\bt}A_\bt,\quad B^\al:=g^{\al\bt}B_\bt.
\end{equation*}
Скалярные произведения векторов (\ref{escvec}) и 1-форм (\ref{esupfo})
естественным образом продолжаются на тензорные поля произвольного типа $(r,s)$.

Метрика на многообразии определяет инвариантную квадратичную форму
дифференциалов, которая называется {\em интервалом}
\index{Интервал (interval)}%
\begin{equation}                                                  \label{einteg}
  ds^2:=g_{\al\bt}dx^\al dx^\bt.
\end{equation}
Интервал задает расстояние между двумя бесконечно близкими точками с
координатами $x^\al$ и $x^\al+dx^\al$. Это расстояние зависит от точки
многообразия, но не от выбора системы координат. В случае псевдоримановой
метрики выражение (\ref{einteg}) не задает метрику (расстояние) в топологическом
смысле, определенную в разделе \ref{seucme}, т.к.\ квадратичная форма
(\ref{einteg}) не является положительно определенной. С помощью положительно
определенной римановой метрики топологическое расстояние между двумя точками
$p,q\in\MM$ можно определить как точную нижнюю грань интеграла
\begin{equation}                                                  \label{edista}
  l(p,q):=\inf\int\limits_p^q ds,\qquad
  ds:=\sqrt{ds^2}=dt\sqrt{g_{\al\bt}\frac{dx^\al}{dt}\frac{dx^\bt}{dt}},
\end{equation}
где интегрирование ведется по всем кусочно дифференцируемым кривым $x(t)$,
соединяющим точки $p$ и $q$. Доказательство свойств расстояния при этом
существенно опирается на положительную определенность римановой метрики.
Топология, определяемая топологической метрикой (\ref{edista}), совпадает с
топологией многообразия $\MM$

В заключение раздела обсудим вопрос о существовании римановых метрик.
\begin{theorem}                                                   \label{trimex}
На любом $n$-мерном дифференцируемом многообразии $\MM$ существует риманова
метрика.
\end{theorem}
\begin{proof}
Выберем локально конечный атлас $\lbrace\MU_i\rbrace$ на $\MM$. Такой атлас
всегда существует в силу теоремы \ref{tlocon}. Обозначим координаты на каждой
карте через $x^\al_i$, $\al=1,\dotsc,n$. Пусть $\lbrace f_i\rbrace$ --
разбиение единицы, подчиненное данному атласу, такое, что
$\supp f_i\subset\MU_i$. Выберем евклидову метрику на каждой карте и склеим
карты с помощью разбиения единицы
\begin{align}                                                     \label{epauem}
  ds^2_i&:=\dl_{\al\bt}dx^\al_idx^\bt_i,
\\                                                                \label{eglmee}
  ds^2&:=\sum_i f_i ds^2_i,
\end{align}
где выражение $f_i ds^2_i$ определено
\begin{equation*}
  (f_i ds^2_i)(x)=\left\lbrace
\begin{aligned}
  &f_i(x)ds^2_i,& &x\in\MU_i,
\\
  &0, & & x\notin\MU_i.
\end{aligned}\right.
\end{equation*}
Уравнения (\ref{epauem}), (\ref{eglmee}) определяют гладкое симметричное
ковариантное тензорное поле второго ранга на $\MM$. Поскольку правая часть
(\ref{eglmee}) содержит конечное число слагаемых в каждой точке $x\in\MM$, то
сумма корректно определена. Выберем координатную окрестность $\MU$ с
координатами $x^\al$ такую, что замыкание области $\overline{\MU}$ является
компактным. Тогда область $\MU$ пересекается с конечным числом карт
$\MU_{i_1},\dotsc,\MU_{i_r}$, т.к.\ атлас является локально конечным покрытием.
Поэтому ограничение (\ref{eglmee}) на $\MU$ можно записать в виде
\begin{equation*}
  ds^2=\sum_{\lm=1}^r f_{i_\lm}ds^2_{i_\lm}=g_{\al\bt}dx^\al dx^\bt,
\end{equation*}
где
\begin{equation*}
  g_{\al\bt}:=\sum_{\lm=1}^r f_{i_\lm}\dl_{\g\dl}
  \frac{\pl x^\g_{i_\lm}}{\pl x^\al}\frac{\pl x^\dl_{i_\lm}}{\pl x^\bt}.
\end{equation*}
Поскольку $0\le f_i\le1$ и $\sum_i f_i=1$, то существует такой
индекс $j$, что $f_j(x)>0$. Поэтому
\begin{equation*}
  ds^2(x)\ge f_jds^2_j.
\end{equation*}
Таким образом метрика $ds^2$ является положительно определенной на $\MM$.
\end{proof}
\begin{com}
Доказательство этой теоремы свелось к переносу евклидовой метрики $\dl_{ij}$ из
евклидова пространства $\MR^n$ на многообразие $\MM$ с помощью возврата
отображения, которое фигурирует в определении многообразия, как гомеоморфизм.
При этом положительная определенность евклидовой метрики существенна.
Доказательство не проходит, если таким же образом попытаться перенести
лоренцеву метрику из $\MR^{1,n-1}$ на $\MM$.
\qed\end{com}
\section{Метрика на лоренцевых многообразиях                     \label{signme}}
За счет выбора системы координат метрику всегда можно привести к диагональному
виду в любой наперед заданной точке $x\in\MM$. Действительно, при замене
координат компоненты метрики преобразуются по-правилу (\ref{emetra}). При этом
матрица Якоби преобразования координат $\pl x^\al/\pl x^{\al'}$ в фиксированной
точке многообразия может быть выбрана произвольным образом.
\begin{exa}
Однородное линейное преобразование координат
$x^\al=x^{\al'}M_{\al'}{}^\al$ с постоянной невырожденной матрицей
$M_{\al'}{}^\al=\const$ дает
\begin{equation*}                                                    \tag*{\qed}
  g_{\al'\bt'}=M_{\al'}{}^\al M_{\bt'}{}^\bt g_{\al\bt}.
\end{equation*}
\renewcommand{\qed}{}\end{exa}
\begin{com}
В общем случае метрику можно привести к диагональному виду в фиксированной
точке, но не в окрестности. Это связано с тем, что $n$ функций преобразований
координат, которыми можно воспользоваться, недостаточно для фиксирования
$n(n-1)/2$ функций, параметризующих недиагональные элементы метрики. Исключение
составляют многообразия двух и трех измерений. В двумерном случае метрика имеет
только одну недиагональную компоненту и ее можно привести к диагональному виду в
окрестности произвольной точки. Более того, ее можно преобразовать к конформно
плоскому виду. На трехмерном многообразии метрика имеет три недиагональные
компоненты, что равно числу произвольных функций, параметризующих преобразования
координат. Можно показать, что ее также можно привести к диагональному виду не
только в заданной точке, но и в некоторой окрестности этой точки.
\qed\end{com}
Привести метрику к диагональному виду в точке можно многими способами. При этом
сигнатура не зависит от выбора системы координат в которой метрика диагональна.
Действительно, если в точке $p\in\MM$ метрика диагональна в двух системах
координат и при этом имеет различную сигнатуру, то эти системы координат не
могут быть связаны никаким преобразованием координат. В различных системах
координат диагональные компоненты метрики могут иметь различные (ненулевые)
значения, может меняться последовательность положительных и отрицательных
компонент, однако число положительных и отрицательных компонент остается
неизменным.

\begin{defn}
{\em Сигнатурой метрики}, заданной на многообразии $\MM$, $\dim\MM=n$,
называется пара натуральных чисел $(p,q)$ таких, что $p+q=n$, где
$p$ и $q$ -- количество, соответственно, положительных и отрицательных чисел,
стоящих на диагонали метрики после ее диагонализации в какой либо точке
многообразия $x\in\MM$.
\qed\end{defn}
\index{Сигнатура метрики (signature of a metric)}%
\index{Метрики сигнатура (signature of a metric)}%
Если $g_{\al\bt}$ -- матрица, составленная из компонент метрики в некоторой
системе координат, то числа $p$ и $q$ равны, соответственно, числу положительных
и отрицательных собственных значений. При общих преобразованиях координат
собственные числа могут менять свою величину, но не знак. Нулевых собственных
значений быть не может, т.к.\ в этом случае метрика была бы вырожденной.
\begin{prop}
Сигнатура метрики не зависит от точки линейно связного многообразия.
\end{prop}
\begin{proof}
Допустим, что в некоторых точках $x_1,x_2\in\MM$ метрика имеет разную сигнатуру,
и соединим эти точки произвольной кривой. Тогда из непрерывности метрики
следует, что ее определитель обратился бы в нуль в некоторой точке кривой, что
недопустимо.
\end{proof}

\begin{com}
В зависимости от сигнатуры метрики скалярное произведение векторов может быть
положительно определено или нет. Риманова метрика на многообразии размерности
$n$ имеет сигнатуру $(n,0)$. Отрицательно определенная метрика имеет сигнатуру
$(0,n)$. В римановой геометрии $(\MM,g)$ положительно и отрицательно
определенные метрики, по существу, можно не различать, т.к.\ все геометрические
характеристики многообразий связаны простым преобразованием -- изменением знака
метрики. Поэтому многообразия с отрицательно определенной метрикой мы также
будем называть римановыми. В то же время многие модели математической физики
неинвариантны относительно изменения знака метрики
\begin{equation}                                                  \label{emetin}
  g_{\al\bt}\mapsto-g_{\al\bt},
\end{equation}
поскольку действие, которым описывается модель, помимо метрики содержит также
другие поля. Поэтому решения уравнений Эйлера--Лагранжа с положительно и
отрицательно определенной метрикой, вообще говоря, неэквивалентны.
\qed\end{com}
Сигнатура метрики инвариантна, не зависит от выбора системы координат и точки
линейно связного многообразия. Будем считать, что после диагонализации номера
координат выбраны таким образом, что сначала идут все положительные, а затем --
отрицательные собственные значения. Будем писать
\begin{equation}                                                  \label{emetsi}
  \sign g_{\al\bt}=(\underbrace{+\dots+}_p\underbrace{-\dots-}_q).
\end{equation}
\begin{defn}
Пара $(\MM,g)$ называется {\em римановым многообразием}, если метрика является
знакоопределенной $p=n$ или $q=n$, и {\em псевдоримановым многообразием}, если
метрика не является знакоопределенной, $p\ne0$ и $q\ne0$.
Если положительный элемент на диагонали один,
$$
  \sign g_{\al\bt}=(+-\dots-),
$$
то говорят, что метрика имеет {\em лоренцеву сигнатуру}. Если на многообразии
задана метрика лоренцевой сигнатуры, то будем говорить, что многообразие
{\em лоренцево}.
\qed\end{defn}
\index{Риманово многообразие (Riemannian manifold)}%
\index{Многообразие риманово (Riemannian manifold)}%
\index{Псевдориманово многообразие (pseudo-Riemannian manifold)}%
\index{Многообразие псевдориманово (pseudo-Riemannian manifold)}%
\index{Лоренцева сигнатура (Lorentzian signature)}%
\index{Сигнатура лоренцева (Lorentzian signature)}%
\index{Лоренцево многообразие (Lorentzian manifold)}%
\index{Многообразие лоренцево (Lorentzian manifold)}%
\begin{com}
Подчеркнем, что задание метрики на многообразии может быть произвольным. В
частности, на одном многообразии можно задать несколько метрик одновременно,
причем разной сигнатуры, если такие существуют.
\qed\end{com}

В настоящей монографии псевдоримановы многообразия размерности $n\ge4$ с $p\ge2$
и $q\ge2$ не рассматриваются, так как они недостаточно хорошо изучены и не имеют
широкого применения в математической физике.

Для лоренцевой метрики скалярное произведение двух векторов может быть
положительно, отрицательно или равно нулю, а из условия $X^2=0$ не следует,
что $X=0$.
\begin{defn}
Назовем {\em пространством-временем} псевдориманово многообразие $\MM$ с
заданной метрикой $g=dx^\al\otimes dx^\bt g_{\al\bt}$ лоренцевой сигнатуры
(лоренцево многообразие). В пространстве-времени векторное поле $X$ в точке
$x\in\MM$ называется:
\begin{equation}                                                  \label{evecor}
\begin{array}{ll}
\text{\it времениподобным,}                      & \text{если}\qquad (X,X)>0, \\
\text{\it светоподобным~(изотропным, нулевым),}  & \text{если}\qquad (X,X)=0, \\
\text{\it пространственноподобным,}              &
\text{если}\qquad (X,X)<0.
\end{array}
\end{equation}
Это определение распространятся на область $\MU\subset\MM$, если во всех
точках $x\in\MU$ выполнены соответствующие соотношения.
\qed\end{defn}
\index{Пространство-время (space-time)}%
\index{Времениподобное векторное поле (time-like vector field)}%
\index{Векторное поле времениподобное (time-like vector field)}%
\index{Светоподобное векторное поле (light-like vector field)}%
\index{Векторное поле светоподобное (light-like vector field)}%
\index{Изотропное векторное поле (isotropic vector field)}%
\index{Векторное поле изотропное (isotropic vector field)}%
\index{Пространственноподобное векторное поле (space-like vector field)}%
\index{Векторное поле пространственноподобное (space-like vector field)}%
\index{Нулевое векторное поле (null vector field)}%
\index{Векторное поле нулевое (null vector field)}%
\begin{com}
Определение (\ref{evecor}) инвариантно относительно замены координат. В общем
случае гладкое векторное поле может иметь различный тип в различных областях
связного многообразия.
\qed\end{com}
\begin{defn}
Так же, как и в римановом пространстве, два вектора в пространстве-времени
называются {\em ортогональными (перпендикулярными)}, если их скалярное
произведение равно нулю. В частности, любой изотропный вектор перпендикулярен
самому себе.
\index{Ортогональные векторы (orthogonal vectors)}%
\index{Векторы ортогональные (orthogonal vectors)}%
\index{Перпендикулярные векторы (perpendicular vectors)}%
\index{Векторы перпендикулярные (perpendicular vectors)}%

Определим тип координатного векторного поля $\pl_0$. Из (\ref{escbap}) следует,
что $(\pl_0,\pl_0)=g_{00}$. Если в данной системе координат $g_{00}>0$, то
векторное поле $\pl_0$ времениподобно. В этом случае назовем координату $x^0=t$
{\em времениподобной} или {\em временем}. Другими словами, временем называется
любой параметр вдоль интегральной кривой произвольного времениподобного
векторного поля. Противоположно направленное векторное поле $-\pl_0$ также
времениподобно и связано с $\pl_0$ преобразованием координат
$(x^0,x^1,\dotsc,x^{n-1})\mapsto(-x^0,x^1,\dotsc,x^{n-1})$, которое назовем
{\em обращением времени}. Выберем (произвольно) ориентацию координаты $x^0$ и
будем говорить, что векторное поле $\pl_0$ направлено в будущее, а $-\pl_0$ -- в
прошлое. Тогда на времениподобных векторных полях можно ввести ориентацию.
Произвольное времениподобное поле $X$ направлено в будущее, если
$(X,\pl_0)$$=$$X^0>0$. В противном случае, когда $X^0<0$, будем говорить, что
времениподобное векторное поле направлено в прошлое.
\qed\end{defn}
\index{Время (time)}%
\index{Обращение времени (time inversion)}%
\index{Времени обращение (time inversion)}%
\begin{com}
В общем случае метрика, индуцированная на сечениях $t=x^0=\const$ лоренцева
многообразия, может быть отрицательно определена или быть знаконеопределенной.
То есть сечения пространства-времени, соответствующие постоянному времени,
совсем не обязательно пространственноподобны (= все касательные векторы
пространственноподобны). Это зависит от выбора остальных координат. В конце
настоящего раздела мы рассмотрим простой пример.
\qed\end{com}

Можно доказать, что каждая точка многообразия $\MM$ лоренцевой сигнатуры
$(1,n-1)$ имеет такую координатную окрестность с координатами
$(t,u^1,\dots,u^{n-1})$, что в этих координатах метрика имеет блочно
диагональный вид
\begin{equation}                                                  \label{etigam}
  ds^2=dt^2+g_{\mu\nu}(t,u)du^\mu du^\nu,\qquad \mu,\nu=1\dotsc n-1.
\end{equation}
При этом метрика $g_{\mu\nu}$, индуцированная на сечениях $t=\const$, является
отрицательно определенной. Другими словами, все сечения $t=\const$ для метрики
вида (\ref{etigam}) являются пространственноподобными подмногообразиями, т.е.\
все касательные векторы к сечениями пространственноподобны. Эти сечения являются
вложенными римановыми многообразиями с локальной системой координат
$\lbrace u^\mu\rbrace$ и отрицательно определенной метрикой $g_{\mu\nu}$.
Отметим также, что все векторы, касательные к этому подмногообразию ортогональны
времениподобному векторному полю $\pl_0$.
\begin{com}
Здесь и в дальнейшем мы будем использовать следующие обозначения координат
на лоренцевом многообразии. Греческие буквы из начала алфавита $\al,\bt,\dotsc$
будут использоваться для нумерации всех координат, а буквы из середины алфавита
$\mu,\nu,\dotsc$ -- для нумерации остальных координат,
$\lbrace x^\al\rbrace=\lbrace x^0,x^\mu\rbrace$. Это правило легко
запомнить по следующим включениям:
\begin{equation*}
  \lbrace 1,\dotsc,n-1\rbrace\subset\lbrace 0,1,\dotsc,n-1\rbrace,\qquad
  \lbrace \mu,\nu,\dotsc\rbrace\subset\lbrace\al,\bt,\dotsc\rbrace.
\end{equation*}
Как правило, мы будем считать, что координата $x^0$ является временем, и все
остальные координатные линии $x^\mu$ -- пространственноподобны. Поэтому
координаты $x^\mu$ будем называть пространственноподобными.
\qed\end{com}
В общем случае метрика на лоренцевом многообразии
\begin{equation}                                                  \label{elomeg}
  g_{\al\bt}=\begin{pmatrix}
  g_{00} & g_{0\nu} \\ g_{\mu0} & g_{\mu\nu} \end{pmatrix}
\end{equation}
не имеет блочно диагонального вида (\ref{etigam}). Приведем критерий того, что
метрика (\ref{elomeg}), заданная на некотором многообразии $\MM$, имеет
лоренцеву сигнатуру.
\begin{theorem}                                                   \label{tlosim}
Пусть в некоторой окрестности $\MU\subset\MM$ задана метрика (\ref{elomeg})
такая, что $g_{00}>0$. Эта метрика имеет лоренцеву сигнатуру тогда и только
тогда, когда матрица
\begin{equation}                                                  \label{elomec}
  g_{\mu\nu}-\frac{g_{0\mu}g_{0\nu}}{g_{00}}
\end{equation}
отрицательно определена в каждой точке $\MU$.
\end{theorem}
\begin{proof}
Достаточно рассмотреть произвольную точку из $\MU$. Пусть на $\MU$ задана
метрика (\ref{elomeg}) для которой $g_{00}>0$. Интервал в окрестности
$\MU\subset\MM$ имеет вид
\begin{equation*}
  ds^2=g_{00}dx^0dx^0+2g_{0\mu}dx^0dx^\mu+g_{\mu\nu}dx^\mu dx^\nu,
  \qquad g_{00}>0.
\end{equation*}
Зафиксируем произвольную точку $x\in\MU$. Введем вместо $x^0$ новую координату
$\tilde x^0$, для которой в точке $x\in\MU$ выполнено соотношение
\begin{equation*}
  dx^0=d\tilde x^0-\frac{dx^\mu g_{0\mu}}{g_{00}}.
\end{equation*}
В фиксированной точке этого всегда можно добиться линейным преобразованием
координат. Тогда интервал примет вид
\begin{equation}                                                  \label{evidsp}
  ds^2=g_{00}d\tilde x^0d\tilde x^0
  +\left(g_{\mu\nu}-\frac{g_{0\mu}g_{0\nu}}{g_{00}}\right)dx^\mu dx^\nu.
\end{equation}

Если метрика (\ref{elomeg}) имеет лоренцеву сигнатуру, то существует такая
система координат, что в точке $x\in\MM$ метрика диагональна, причем $g_{00}>0$,
а все остальные диагональные компоненты $g_{\mu\mu}$ отрицательны. Поскольку
метрика (\ref{evidsp}) связана с диагональной метрикой также невырожденным
преобразованием координат, то матрица (\ref{elomec}) отрицательно определена.

Обратно. Если матрица (\ref{elomec}) отрицательно определена, то дальнейшим
линейным преобразованием координат $x^\mu$ ее всегда можно преобразовать к
диагональному виду в фиксированной точке, причем на диагонали будут стоять
отрицательные числа. Следовательно, метрика имеет лоренцеву сигнатуру.
\end{proof}
\begin{com}
В разделе \ref{sadmpa} мы докажем, что отрицательная определенность матрицы
(\ref{elomec}) эквивалентна отрицательной определенности
``пространственного блока'' $g^{\mu\nu}$ обратной метрики
\begin{equation*}
  g^{\al\bt}=\begin{pmatrix} g^{00} & g^{0\nu} \\ g^{\mu 0} & g^{\mu\nu}
\end{pmatrix}.    \qed
\end{equation*}
\end{com}

Матрица (\ref{elomec}) на лоренцевом многообразии симметрична и невырождена,
т.к.\ отрицательно определена. Она имеет следующий геометрический смысл. Вдоль
координатных линий времени $t=x^0$ всегда можно определить единичное векторное
поле
\begin{equation*}
  n:=\frac1{\sqrt{g_{00}}}\pl_0,\qquad(n,n)=1.
\end{equation*}
Рассмотрим произвольное векторное поле $X=X^0\pl_0+X^\mu\pl_\mu$. У него есть
составляющая, перпендикулярная времениподобному векторному полю $n$,
\begin{equation*}
  X_\perp=X-(X,n)n=X^\mu\pl_\mu-X^\mu g_{0\mu}\pl_0.
\end{equation*}
Если задано два произвольных векторных поля $X$ и $Y$, то скалярное произведение
их перпендикулярных составляющих равно
\begin{equation*}
  (X_\perp,Y_\perp)
  =\left(g_{\mu\nu}-\frac{g_{0\mu}g_{0\nu}}{g_{00}}\right)X^\mu Y^\nu.
\end{equation*}
Таким образом, матрица (\ref{elomec}) играет роль метрики для перпендикулярных
составляющих векторных полей. Из теоремы \ref{tlosim} следует, что, если $\MM$
-- лоренцево многообразие и $x^0$ -- время, то составляющие векторных полей,
перпендикулярные $n$, всегда пространственноподобны.
\begin{com}
Не следует думать, что матрицу (\ref{elomec}) можно рассматривать, как метрику
на некотором $(n-1)$-мерном подмногообразии, касательные векторы к которому
всюду перпендикулярны времениподобному векторному полю $n$. Дело в том, что
такие подмногообразия могут не существовать. Можно проверить, что коммутатор
двух перпендикулярных векторных полей $[X_\perp,Y_\perp]$ в общем случае не
будет ортогонален $n$. Для этого достаточно, не ограничивая общности, положить
$g_{00}=1$ и проверить, что коммутатор $[X_\perp,Y_\perp]$ имеет нетривиальную
составляющую вдоль $n$. Тем самым множество всех векторных полей $X_\perp$ не
находится в инволюции, и, согласно теореме Фробениуса, подмногообразия,
касательные векторы к которому всюду перпендикулярны $n$, не существует.
\qed\end{com}

Приведем еще один критерий того, что метрика на многообразии является
лоренцевой.
\begin{theorem}
Для того, чтобы метрика (\ref{elomeg}) с $g_{00}>0$ в точке $x\in\MM$ имела
лоренцеву сигнатуру необходимо и достаточно, чтобы выполнялись неравенства:
\begin{equation}                                        \label{ecrlsi}
  \det\begin{pmatrix} g_{00} & g_{01} \\ g_{10} & g_{11}\end{pmatrix}<0,~
  \det\begin{pmatrix} g_{00} & g_{01} & g_{02}\\ g_{10} & g_{11} & g_{12} \\
  g_{20} & g_{21} & g_{22}\end{pmatrix}>0,~\dotsc,~\det(g_{\al\bt})
  \begin{cases}
    >0, & n-\text{нечетно},\\
    <0, & n-\text{четно}.
\end{cases}
\end{equation}
\end{theorem}
\begin{proof}
См., например, \cite{}.
\end{proof}
\begin{exa}
Рассмотрим плоскость Минковского $\MR^{1,1}$ с декартовой системой
координат $\lbrace x^0,x^1\rbrace$ и метрикой
\begin{equation*}
  \eta_{\al\bt}:=\begin{pmatrix}1&0\\ 0&-1\end{pmatrix}.
\end{equation*}
Введем новую систему координат $\lbrace\tilde x^0,\tilde
x^1\rbrace=\lbrace x^0-\sqrt{2}x^1,x^1\rbrace$. Тогда координатные
базисные векторы преобразуются по-правилу (см.\ рис.~\ref{fmiplg})
\begin{equation*}
  \lbrace \pl_0,\pl_1\rbrace\mapsto\lbrace\tilde\pl_0,\tilde\pl_1\rbrace
  =\lbrace\pl_0,\sqrt2\pl_0+\pl_1\rbrace.
\end{equation*}
\begin{figure}[h,b,t]
\hfill\includegraphics[width=.25\textwidth]{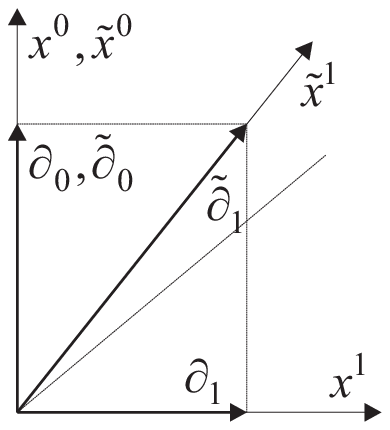}
\hfill {}
\centering\caption{Пример времениподобного сечения $\tilde x^0=\const$.}
\label{fmiplg}
\end{figure}
В новой системе координат сечения $\tilde x^0=\const$ являются
времениподобными прямыми, а метрика имеет вид
\begin{equation*}
  g_{\al\bt}=\begin{pmatrix} 1 & \sqrt2 \\ \sqrt2 & 1 \end{pmatrix}.
\end{equation*}
Матрица, соответствующая выражению (\ref{elomec}), состоит из одного
элемента $g_{11}-g_{01}^2/g_{00}=-1$.

Обратная метрика имеет вид
\begin{equation*}
  g^{\al\bt}=\begin{pmatrix} -1 & \sqrt2 \\ \sqrt2 & -1 \end{pmatrix}.
\end{equation*}
Отметим, что в этих координатах обе компоненты обратной метрики на
диагонали $g^{00}$ и $g^{11}$ отрицательны.
\qed\end{exa}

Рассмотренный пример показывает, что ``пространственные'' компоненты метрики
$g_{\mu\nu}$ в общем случае не образуют отрицательно определенной матрицы. Роль
пространственной части метрики играет выражение (\ref{elomec}), и она должна
быть отрицательно определена на лоренцевом многообразии. Кроме того, из примера
следует, что, даже если координата $x^0$ на лоренцевом многообразии является
временем, тем не менее временн\'ая компонента обратной метрики $g^{00}$ может
быть отрицательна.

В дальнейшем мы всегда предполагаем, что координаты на лоренцевом многообразии
выбраны таким образом, что координата $x^0$ является временем, и все сечения
$x^0=\const$ пространственноподобны. Такой выбор координат удобен, например,
при постановке задачи Коши в различных моделях математической физики. На языке
компонент метрики эти условия означают, что $g_{00}>0$, и матрица $g_{\mu\nu}$
отрицательно определена. В этом случае нетрудно доказать, что временн\'ая
компонента обратной метрики также положительна $g^{00}>0$. В обратную сторону
утверждение неверно: условий положительности $g_{00}>0$ и $g^{00}>0$
недостаточно для отрицательной определенности матрицы $g_{\mu\nu}$.
\section{Векторные поля и вложения                               \label{sumhsi}}
В разделе \ref{sofhys} было установлено, что отличная от нуля 1-форма
$A=dx^\al A_\al$ определяет в касательном расслоении $\MT(\MM)$ распределение
$(n-1)$-подпространств, натянутое на линейно независимые векторы с компонентами
$Y^\al$, удовлетворяющими
уравнению $Y^\al A_\al=0$. Наличие метрики позволяет каждой 1-форме поставить в
соответствие векторное поле $X^\al=g^{\al\bt}A_\bt$. По-построению, это
векторное поле перпендикулярно любому вектору из $(n-1)$-мерного
подпространства в касательном пространстве, т.к.\ $X^\al Y^\bt g_{\al\bt}=0$.
Таким образом, при наличии метрики, задание ненулевого векторного поля
эквивалентно заданию $(n-1)$-мерного распределения на многообразии. В общем
случае это распределение не является инволютивным, и поэтому у него не
существует интегральных подмногообразий. То есть векторное поле не определяет
ортогональные интегральные подмногообразия. Обратное утверждение верно. Каждое
$(n-1)$-мерное подмногообразие (гиперповерхность) определяет нормальное
векторное поле, заданное на подмногообразии.

\begin{defn}
Пусть $\MM$ -- риманово многообразие. Тогда {\em углом} между двумя
пересекающимися гиперповерхностями в точках их пересечения называется угол между
соответствующими нормальными векторами в данных точках.
\qed\end{defn}
\begin{defn}
Если на многообразии задана метрика и векторное поле, то определим проекционные
операторы на заданное векторное поле и перпендикулярные направления следующим
образом. Пусть $X^\al$ -- компоненты произвольного векторного поля, такого, что
$X^2\ne0$. Тогда операторы
\begin{equation}                                                  \label{epropv}
  \Pi_{\al}^\Sl{}^\bt:=\frac{X_\al X^\bt}{X^2},\qquad
  \Pi_{\al}^\St{}^\bt:=\dl_\al^\bt-\frac{X_\al X^\bt}{X^2},
\end{equation}
являются {\em проекционными операторами} на направление векторного поля и
перпендикулярные направления, соответственно. При этом выполнены равенства,
определяющие набор проекционных операторов:
$$
  (\Pi^\Sl)^2=\Pi^\Sl,\qquad(\Pi^\St)^2=\Pi^\St,
  \qquad\Pi^\Sl\Pi^\St=0,\qquad\Pi^\Sl+\Pi^\St=1. \qed
$$
\end{defn}
\index{Проекционный оператор (projection operator)}%
\index{Оператор проекционный (projection operator)}%

\begin{exa}
Векторное поле $Y=Y^\al\pl_\al$ имеет следующее разложение
\begin{equation*}
  Y=Y^\Sl+Y^\St,
\end{equation*}
где его компоненты определены следующими выражениями:
\begin{align*}
  Y^{\Sl\al}&:=Y^\bt\Pi_\bt^{\Sl\al}=\frac{(Y,X)X^\al}{X^2},
\\
  Y^{\St\al}&:=Y^\bt\Pi_\bt^{\St\al}=Y^\al-\frac{(Y,X)X^\al}{X^2}.
\end{align*}

Аналогично раскладывается $1$-форма $A=dx^\al A_\al$:
\begin{equation*}
  A=A^\Sl+A^\St,
\end{equation*}
где
\begin{align*}
  A_\al^\Sl&:=\Pi_\al^{\Sl\bt}A_\bt=\frac{A(X)X_\al}{X^2},
\\
  A_\al^\St&:=\Pi_\al^{\St\bt}A_\bt=A_\al-\frac{A(X)X_\al}{X^2}.
\end{align*}

Для того, чтобы разложить тензоры более высокого ранга, необходимо произвести
разложение по каждому индексу. Например, контравариантный тензор второго ранга
$T$ имеет следующее разложение
\begin{equation*}
  T=T^{\Sl\Sl}+T^{\Sl\St}+T^{\St\Sl}+T^{\St\St},
\end{equation*}
где
\begin{align*}
  T^{\Sl\Sl\al\bt}&:=T^{\g\dl}\Pi_\g^{\Sl\al}\Pi_\dl^{\Sl\bt}
  =\frac{T^{\g\dl}X_\g X_\dl X^\al X^\bt}{X^4},
\\
  T^{\Sl\St\al\bt}&:=T^{\g\dl}\Pi_\g^{\Sl\al}\Pi_\dl^{\St\bt}
  =\frac{T^{\g\bt}X_\g X^\al}{X^2}-\frac{T^{\g\dl}X_\g X_\dl X^\al X^\bt}{X^4},
\\
  T^{\St\Sl\al\bt}&:=T^{\g\dl}\Pi_\g^{\St\al}\Pi_\dl^{\Sl\bt}
  =\frac{T^{\al\g}X_\g X^\bt}{X^2}-\frac{T^{\g\dl}X_\g X_\dl X^\al X^\bt}{X^4},
\\
  T^{\St\St\al\bt}&:=T^{\g\dl}\Pi_\g^{\St\al}\Pi_\dl^{\St\bt}
  =T^{\al\bt}-\frac{T^{\al\g}X_\g X^\bt}{X^2}-\frac{T^{\g\bt}X_\g X^\al}{X^2}
  +\frac{T^{\g\dl}X_\g X_\dl X^\al X^\bt}{X^4}.
\end{align*}
Аналогично раскладываются тензоры с произвольным числом контра- и ковариантных
индексов.
\qed\end{exa}
Поскольку нетривиальные ковекторы с компонентами
\begin{equation*}
  V_\al-X_\al\frac{(V,X)}{X^2}\qquad\text{и}\qquad X_\al
\end{equation*}
являются собственными векторами проекционных операторов $\Pi^\Sl$ и $\Pi^\St$
с нулевыми собственными значениями, то проекционные операторы вырождены:
\begin{equation*}
  \det \Pi^\Sl=\det\Pi^\St=0.
\end{equation*}

Рассмотрим, как меняется метрика при отображениях. Пусть многообразие $\MM$
вложено\footnote{На самом деле достаточно погружения.} в $\MN$,
$\dim\MM\le\dim\MN$, и на пространстве мишени $\MN$ задана метрика $g$. Тогда
возврат отображения $f:~\MM\hookrightarrow\MN$ задает метрику на $\MM$.
Бескоординатное определение индуцированной метрики $f^*g$ имеет вид
\begin{equation*}
  (f^*g)(X,Y)=g(f_*X,f_*Y),\qquad\forall X,Y\in\CX(\MM),
\end{equation*}
где $f_*$ -- дифференциал отображения, отображающий векторные поля $X,Y$ на
$\MM$ в векторные поля $f_*X,f_*Y$ на $\MN$. Скалярные произведения на $\MM$ и
$\MN$ заданы соответственно метриками $f^*g$ и $g$. Это значит, что при вложении
длины векторов и углы между ними сохраняются. Обозначив координаты на $\MM$ и
$\MN$ через $x^\al$ и $y^\Sa$, получим явное выражение для компонент
индуцированной метрики $f^*g$:
\begin{equation}                                                  \label{eindmm}
  (f^*g)_{\al\bt}(x)=\frac{\pl y^\Sa}{\pl x^\al}\frac{\pl y^\Sb}{\pl x^\bt}
  g_{\Sa\Sb}(y).
\end{equation}
\begin{defn}
Пусть на многообразии $\MM$ задана некоторая метрика $h$. Если при вложении
$\MM$ в многообразие $\MN$ с метрикой $g$ исходная метрика совпадает с
индуцированной, $h=f^*g$, то такое вложение называется {\em изометрическим}.
\qed\end{defn}
\index{Изометрическое вложение (isometric embedding)}%
\index{Вложение изометрическое (isometric embedding)}%
Представляет большой интерес задача о нахождении изометрических вложений
заданного многообразия с определенной на нем метрикой в евклидово пространство.
\begin{exa}
Пусть двумерная сфера $\MS^2_r$ радиуса $r$ вложена в трехмерное
евклидово пространство $\MR^3$. Вложение можно записать в параметрическом виде
\begin{align*}
  x&=r\sin\theta\cos\vf,
\\
  y&=r\sin\theta\sin\vf,
\\
  z&=r\cos\vf,
\end{align*}
где $0\le\theta\le\pi$ и $0\le\vf<2\pi$ -- координаты на сфере и $r=\const>0$.
Тогда метрика, индуцированная на сфере, примет вид
\begin{equation}                                                  \label{emeins}
  ds^2=dx^2+dy^2+dz^2=r^2d\theta^2+r^2\sin^2\theta d\vf^2.
\end{equation}
То есть в выражение для евклидова интервала вместо дифференциалов $dx$, $dy$ и
$dz$ необходимо просто подставить их выражения через дифференциалы
полярного и азимутального углов. \qed\end{exa}
\begin{exa}
Пусть двумерный тор $\MT^2$ радиуса $r$ с направляющей окружностью
радиуса $R>r$ вложен в трехмерное евклидово пространство $\MR^3$
(см.\ рис.~\ref{ftorin}):
\begin{align*}
  x&=(R+r\cos\theta)\cos\vf,
\\
  y&=(R+r\cos\theta)\sin\vf,
\\
  z&=r\sin\theta,
\end{align*}
\begin{figure}[h,b,t]
\hfill\includegraphics[width=.4\textwidth]{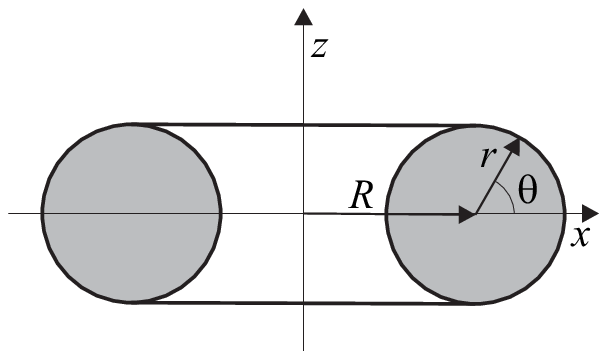}
\hfill {}
\centering\caption{Сечение двумерного тора, вложенного в трехмерное евклидово
пространство.}
\label{ftorin}
\end{figure}
где $0\le\theta<2\pi$ и $0\le\vf<2\pi$ -- координаты на торе.
Тогда индуцированная метрика имеет вид
\begin{equation}                                                  \label{etorim}
  ds^2=dx^2+dy^2+dz^2=r^2d\theta^2+(R+r\cos\theta)^2d\vf^2.\qed
\end{equation}
\end{exa}
\section{Выбор системы координат                                 \label{scoeum}}
Если на многообразии задана метрика, которую преобразованием координат нельзя
привести к евклидовой или лоренцевой форме, тогда понятие декартовой системы
координат отсутствует, и это создает определенные трудности для наглядного
представления многообразия. С практической точки зрения удобнее сначала
совершить преобразование координат, после которого метрика примет какой-либо
относительно простой вид, связанный, например, с симметрией задачи, и уже
потом дать физическую интерпретацию пространству или пространству-времени.
\begin{defn}
Риманова метрика называется {\em конформно евклидовой} или {\em вейлевски
евклидовой} в некоторой области $\MU\subset\MM$, если существует такая система
координат на $\MU$, в которой метрика имеет вид
\begin{equation}                                                  \label{ecoeum}
  g_{\al\bt}=\ex^{2\phi}\dl_{\al\bt}.
\end{equation}
Здесь $\phi(x)$ -- некоторая достаточно гладкая функция от $x^\al$. Функция
$\ex^{2\phi}$ называется {\em конформным множителем}. В таком виде метрика
невырождена при $\phi\ne\pm\infty$. Если у каждой точки $x\in\MM$
существует окрестность, на которой метрика является конформно евклидовой, то
метрика называется {\em локально конформно евклидовой} или локально вейлевски
евклидовой.
\qed\end{defn}
\index{Метрика конформно евклидова (conformally Euclidean metric)}%
\index{Конформно евклидова метрика (conformally Euclidean metric)}%
\index{Метрика вейлевски евклидова (Weyl Euclidean metric)}%
\index{Вейлевски евклидова метрика (Weyl Euclidean metric)}%
\index{Конформный множитель (conformal factor)}%
\index{Множитель конформный (conformal factor)}%
Метрика, обратная к метрике (\ref{ecoeum}) имеет вид
\begin{equation*}
  g^{\al\bt}=\ex^{-2\phi}\dl^{\al\bt}.
\end{equation*}
\begin{com}
Подчеркнем, что $\phi(x)$ не является скалярным полем, хотя и не имеет ни
одного индекса, т.к.\ в другой системе координат конформно евклидова метрика
может не иметь вида (\ref{ecoeum}), и, значит, функция $\phi$ не определена
вовсе.
\qed\end{com}
Если метрика записана в конформно евклидовом виде, то в этой системе координат
угол (\ref{evecan}) между двумя произвольными векторами тот же, что и в
евклидовом пространстве. Это обстоятельство позволяет более наглядно представить
себе свойства многообразия.

Конформно евклидова метрика является частным случаем метрики, и далеко не
каждую метрику можно привести к такому виду путем преобразования системы
координат. Это видно уже из того, что в общем случае конформно евклидова метрика
зависит от $n+1$ независимой функции (здесь $n$ функций соответствует выбору
системы координат и одна функция -- конформному множителю), в то время как
метрика общего вида зависит от $n(n+1)/2$ функций. Исключение представляет
только случай двух измерений.
\begin{theorem}                                                   \label{tlococ}
Если риманова метрика принадлежит классу $\CC^3(\MM)$ на двумерном многообразии
$\MM$ (поверхности), то для любой точки $x\in\MM$ существует окрестность
$\MU_x$, в которой можно выбрать конформно евклидовы координаты.
\end{theorem}
\begin{proof}
См., например, \cite{Wolf72R}.
\end{proof}
Локально конформно евклидовы координаты $x^1,x^2$ в теореме \ref{tlococ}
определены на некоторой окрестности $\MU_x$ с точностью до конформных
преобразований комплексных координат, которые вводятся на $\MU_x$ следующим
образом: $z:=x^1+ix^2$, что оправдывает название ``конформно евклидова'' для
метрики.
\begin{defn}
Если в пространстве-времени метрика имеет вид (\ref{ecoeum}), где символ
Кронекера $\dl_{\al\bt}$ заменен на лоренцеву метрику $\eta_{\al\bt}$, то
пространство-время называется {\em конформно минковским}.
\qed\end{defn}
\index{Конформно минковское пространство-время %
(conformally Minkowskian space-time}%

В общем случае, для того, чтобы задать метрику в некоторой окрестности,
необходимо задать $n(n+1)/2$ функций от $n$ переменных (компоненты метрики). Во
многих случаях,
например, в общей теории относительности, метрика ищется как решение некоторой
системы уравнений, которая ковариантна относительно выбора системы координат.
То есть система уравнений движения выглядит по-разному в различных системах
отсчета, однако пространства решений находятся во взаимно однозначном
соответствии. При этом соответствие устанавливается преобразованием координат.
При решении ковариантной системы уравнений удобно выбрать ту или иную систему
отсчета с тем, чтобы уменьшить число неизвестных функций. Поскольку переход
между системами координат характеризуется $n$ функциями от $n$ переменных, то
максимальное число функций, которое можно зафиксировать в метрике общего вида,
равно $n$. В результате метрика будет определяться $n(n-1)/2$ функциями. В этом
случае говорят, что выбрана система координат или {\em зафиксирована калибровка}
для метрики\footnote{Это название пришло из электродинамики, инвариантной
относительно локальных преобразований, которые называются калибровочными
(см.\ раздел \ref{selema}).}. Как правило, координаты выбираются таким образом,
чтобы система уравнений имела наиболее простой вид.
\index{Калибровка (gauge)}\index{Фиксирование калибровки (gauge fixing)}%

Выбор системы координат не означает, что $n$ функций, входящих в метрику, можно
зафиксировать произвольным образом. Для того, чтобы определить, существует ли
система отсчета, где метрика имеет заданный вид, необходимо проанализировать
систему уравнений для функций перехода к такой системе координат. Если эта
система уравнений имеет решение, то такая калибровка называется
{\em допустимой}. Часто система уравнений на функции перехода, имеет только
локальные решения. Это означает, что соответствующая система координат может
быть выбрана только локально.
\index{Калибровка допустимая (admissible gauge)}%
\index{Допустимая калибровка (admissible gauge)}%

Различные удобные выборы систем координат в общей теории относительности
рассмотрены в разделе \ref{sdeawh}.
\chapter{Связность на векторном расслоении и расслоении реперов}
Большинство моделей математической физики сводится к решению некоторых
дифференциальных уравнений, которыми являются уравнения движения или уравнения
равновесия. Мы предполагаем, что дифференциальные уравнения являются
ковариантными объектами, потому что при преобразовании координат преобразуются
по тензорным правилам. В этом случае физические следствия не будут зависеть от
выбора системы отсчета. Чтобы строить такие модели используется понятие
связности и соответствующей ковариантной производной, т.к.\ что обычные частные
производные от тензорных полей на многообразии не приводят к тензорным объектам.
Это относится не только к уравнениям общей теории относительности, но и к
другим моделям математической физики. По сути дела ковариантные производные
появляются и в плоском пространстве при переходе к криволинейным системам
координат.
\section{Векторные расслоения                                    \label{svecbu}}
В разделе (\ref{sfibun}) были определены расслоения общего вида, когда типичным
слоем является произвольное многообразие. Ниже мы рассмотрим частный случай
расслоений, типичными слоями которых являются векторные пространства, и введем
понятие связности. Рассмотренные ранее касательное $\MT(\MM)$, кокасательное
$\MT^*(\MM)$ и тензорные $\MT^r_s(\MM)$ расслоения с базой $\MM$ представляет
собой частные примеры векторных расслоений, а понятие связности приводит к
оператору ковариантного дифференцирования произвольных тензорных полей --
исключительно важному инструменту для построения моделей математической физики.
\begin{defn}
Дифференцируемое многообразие $\ME(\MM,\pi,\MV)$ называется {\em векторным
расслоением} с {\em базой} $\MM$, {\em проекцией} $\pi$ и {\em типичным слоем}
$\MV$, где $\MM$ -- дифференцируемое многообразие и $\MV\simeq\MR^n$ --
векторное пространство. Проекция $\pi$ является дифференцируемым сюрьективным
отображением $\ME\xrightarrow{\pi}\MM$. При этом требуется, чтобы для любого
атласа $\MM=\bigcup_i\MU_i$ существовали отображения $\lbrace\chi_i\rbrace$,
удовлетворяющие следующим условиям:

1) \parbox[t]{.92\linewidth}{Отображение $\chi_i$ есть диффеоморфизм
$\chi_i:~\pi^{-1}(\MU_i)\rightarrow\MU_i\times\MV$ такой, что
$\pi\circ\chi^{-1}_i(x,v)=x$ для всех точек $x\in\MU_i$ и $v\in\MV$.}

2) \parbox[t]{.92\linewidth}{Сужение отображения $\chi_i$ на каждый слой
$\chi_{i,x}:\pi^{-1}(x)\rightarrow\MV$ есть гомоморфизм векторных
пространств для всех $x\in\MU_i$. При этом, если $x\in\MU_i\cap\MU_j$, то
отображение
\begin{equation*}
  f_{ij}(x):=\chi_{j,x}\chi^{-1}_{i,x}:~\MV\rightarrow\MV,
\end{equation*}
является автоморфизмом, т.е.\ $f_{ij}(x)\in\MG\ML(n,\MR)$.}

3) \parbox[t]{.92\linewidth}{Если $\MU_i\cap\MU_j\ne\emptyset$,
то отображение $f_{ij}(x):~\MU_i\cap\MU_j\rightarrow\MG\ML(n,\MR)$
является достаточно гладким.}\newline
Отображения $\lbrace f_{ij}\rbrace$ называется {\em функциями перехода}.
\qed\end{defn}
\index{Векторное расслоение (vector bundle)}%
\index{Расслоение векторное (vector bundle)}%
\index{Функция перехода (transition function)}%
\index{Перехода функция (transition function)}%
Пусть $p\in\pi^{-1}(x)$ -- некоторая точка векторного расслоения $\ME$ из слоя
над $x\in\MU_i\cap\MU_j$. Тогда $\chi_i(p)=(x,v_i)$ и $\chi_j(p)=(x,v_j)$.
Следовательно,
\begin{equation}                                                  \label{enecon}
  p=\chi^{-1}_i(x,v_i)=\chi^{-1}_j(x,v_j).
\end{equation}
Зафиксируем базис $\hat e_a$, $a=1,\dotsc,n$ в векторном пространстве $\MV$ и
разложим по нему векторы: $v_i=v_i^a\hat e_a$ и $v_j=v_j^a\hat e_a$. Тогда из
условия 2) следует равенство
\begin{equation*}
  v_j^a=v_i^bf_{ij\,b}{}^a,
\end{equation*}
где введена невырожденная матрица
$f_{ij}(x)=\lbrace f_{ij\,b}{}^a(x)\rbrace\in\MG\ML(n,\MR)$. Эта матрица
достаточно гладко зависит от точки $x$ в силу условия 3 из определения
расслоения.
\begin{com}
В данном определении оба свойства из определения расслоения общего вида в
разделе \ref{sfibun} объединены в одно свойство 1). Свойство 2) предполагает,
что слой векторного расслоения $\pi^{-1}(x)$ в каждой точке $x\in\MM$, сам
является векторным пространством, {\em гомоморфным} типичному слою, а в областях
пересечения карт допускаются автоморфизмы, то есть линейные отображения $\MV$ на
себя. С учетом свойства 1) получаем, что каждый слой изоморфен векторному
пространству, $\pi^{-1}(x)\simeq\MV$. В условии 2) гомоморфизм можно заменить на
гомеоморфизм и не требовать изначально наличия структуры векторного пространства
в слое $\pi^{-1}(x)$. Гомеоморфизма достаточно для переноса структуры векторного
пространства из $\MV$ на слой $\pi^{-1}(x)$, при этом каждый слой будет,
конечно, изоморфен $\MV$. Свойство 3) говорит о том, что автоморфизмы задаются
матрицей $\MG\ML(n,\MR)$, элементы которой достаточно гладко
зависят от координат. Поэтому в определении векторного расслоения был выбран
атлас вместо окрестности точки $x\in\MM$ в определении расслоения общего вида.
\qed\end{com}
Если в векторном пространстве $\MV$, $\dim\MV=n$, зафиксирован базис
$\lbrace\hat e_a\rbrace$, $a=1,\dotsc,n$, то его можно отождествить с
евклидовым пространством $\MR^n$. После этого топология и дифференцируемая
структура евклидова пространства с помощью отображений $\lbrace\chi_i\rbrace$
естественным образом переносятся на пространство расслоения $\ME$, превращая
его в многообразие. Фактически, требование дифференцируемости отображений
$\chi_i$ означает, что дифференцируемая структура на пространстве расслоения
$\ME$ согласована с дифференцируемой структурой, индуцированной отображениями
$\chi_i$.

В общем случае размерности базы $\MM$ и векторного пространства $\MV$ (типичного
слоя) могут не совпадать. Если $\dim\MM=m$ и $\dim\MV=n$, то $\dim\ME=m+n$.

В силу условия 1) в определении расслоения многообразие $\pi^{-1}(\MU_i)$
является тривиальным расслоением над каждой областью $\MU_i$.
\begin{exa}
Построим двумерные расслоения $\ME(\MS^1,\pi,\MR)$, базой которых является
окружность $\MS^1$, а типичным слоем -- вещественная прямая $\MR$, которая
рассматривается, как векторное пространство. Отождествим каждую точку окружности
с полярным углом $\MS^1\supset\al\in[0,2\pi)$. Покроем окружность двумя картами
(см.\ рис.\ref{fprinfibucircle}):
$\MS^1\supset\MU_1=(2\pi-\dl,2\pi)\cup[0,\pi+\dl)$ и
$\MS^1\supset\MU_2=(\pi-\dl,2\pi)\cup[0,\dl)$,
где $0<\dl<\pi/2$, которые пересекаются по двум областям
$\MU_1\cap\MU_2=\MU_{12}\cup\MU_{21}$, где $\MU_{12}=(\pi-\dl,\pi+\dl)$ и
$\MU_{21}=(2\pi-\dl,2\pi)\cup[0,\dl)$, при этом координатные функции определяют
диффеоморфизмы:
\begin{equation*}
\begin{split}
  \vf_1(\MU_1)&=(-\dl,\pi+\dl)\in\MR,
\\
  \vf_2(\MU_2)&=(\pi-\dl,2\pi+\dl)\in\MR.
\end{split}
\end{equation*}
\begin{figure}[h,b,t]
\hfill\includegraphics[width=.4\textwidth]{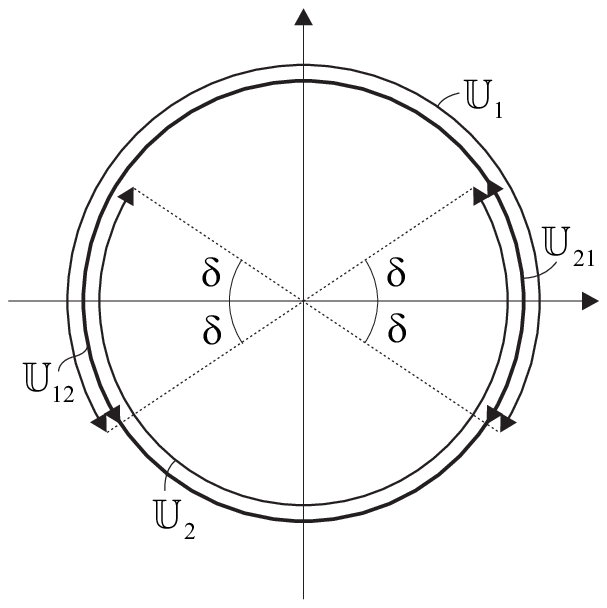}
\hfill {}
\centering\caption{Покрытие окружности двумя картами $\MS^1=\MU_1\cup\MU_2$.
$\MU_{12}$ и $\MU_{21}$ -- области пересечения карт.}
\label{fprinfibucircle}
\end{figure}
Определим векторное расслоение $\ME$ его локальным вложением в $\MR^3$:
\begin{equation*}
\begin{split}
  \pi^{-1}(\MU_1)&=\lbrace p=(x,v^1,v^2)\in\MR^3:\quad x\in\vf_1(\MU_1),~
  v^2\cos(kx/2)=v^1\sin(kx/2)\rbrace,
\\
  \pi^{-1}(\MU_2)&=\lbrace p=(x,v^1,v^2)\in\MR^3:\quad x\in\vf_2(\MU_2),~
  v^2\cos(kx/2)=v^1\sin(kx/2)\rbrace,
\end{split}
\end{equation*}
где $k=0,1,2,\dotsc$ и $v^1,v^2$ -- декартовы координаты на плоскости $\MR^2$.
Для большей ясности мы обозначили точки окружности $\al\in\MS^1$ и их координаты
$x=\vf(\al)\in\MR$ разными буквами. Словами, при движении точки по окружности
$\MS^1$ вещественная прямая $\MR\subset\MR^2$ поворачивается в плоскости
$\MR^2$ на угол $kx/2$. Если $v\in\MR$ -- точка на прямой, то ей соответствует
точка $\Bv=(v^1,v^2)=\big(v\cos(kx/2),v\sin(kx/2)\big)\in\MR^2$ на плоскости.
Направляющим вектором прямой является
$\Bn=\big(\cos(kx/2),\sin(kx/2)\big)\in\MR^2$. Теперь определим отображения:
\begin{equation*}
\begin{split}
  \chi_1:\quad \pi^{-1}(\MU_1)\ni\quad p\mapsto(\al,v)\quad\in\MS^1\times\MR,
\\
  \chi_2:\quad \pi^{-1}(\MU_2)\ni\quad p\mapsto (\al,v)\quad\in\MS^1\times\MR,
\end{split}
\end{equation*}
где $v:=(\Bn,\Bv)=v^1\cos(kx/2)+v^2\sin(kx/2)$. Для определения гомоморфизмов
$f_{ij}$ необходимо учесть, что точки $x$ и $x+2\pi$ в пересечении
$\MU_{21}$ соответствуют одной и той же точке окружности, и
\begin{equation*}
  \cos\big(k(2\pi+x)/2\big)=(-1)^k\cos(kx/2),\qquad
  \sin\big(k(2\pi+x)/2\big)=(-1)^k\sin(kx/2).
\end{equation*}
Поэтому $f_{12}=\id(\MR)$ и $f_{21}=(-1)^k\id(\MR)$. Отсюда следует, что все
расслоения для четных $k$ диффеоморфны между собой и диффеоморфны тривиальному
расслоению $\ME=\MS^1\times\MR$, которое получается при $k=0$. Расслоения при
нечетных $k$ также диффеоморфны между собой и диффеоморфны листу Мебиуса,
соответствующего $k=1$.
\qed\end{exa}

При задании расслоения функции перехода должны удовлетворять условиям
совместности:

1) для всех $x\in\MU_i$, $f_{ii}=\id$;

2) для всех $x\in\MU_i\cap\MU_j\cap\MU_k\ne\emptyset$,
$\quad f_{ij}f_{jk}f_{ki}=\id$.\newline
Эти условия совместности являются, очевидно, необходимыми. Второе свойство
означает коммутативность диаграммы
\begin{equation*}
\begin{diagram}
  \MU_i & \rTo^{f_{ij}} & \MU_j \\
  & \rdTo_{f_{ik}} & \dTo_{f_{jk}} \\
  &  & \MU_k
\end{diagram}
\end{equation*}
где $f_{ik}=f^{-1}_{ki}$. При задании расслоения с помощью функций перехода
необходимо требовать выполнения данных свойств.
\begin{theorem}
Пусть $\lbrace\MU_i\rbrace$ -- открытое координатное покрытие (атлас)
дифференцируемого многообразия $\MM$, и $\MV$ -- векторное пространство. Если
для всех пар пересекающихся координатных областей $\MU_i\cap\MU_j\ne\emptyset$
заданы достаточно гладкие отображения 
$f_{ij}:~\MU_i\cap\MU_j\rightarrow\MG\ML(\MV)$,
которые удовлетворяет условиям совместности 1) и 2), тогда существует
векторное расслоение $\ME(\MM,\pi,\MV)$ с функциями перехода $f_{ij}$.
\end{theorem}
\begin{proof}
См., например, \cite{Steenr51}.
\end{proof}

Поскольку слоями векторных расслоений являются векторные пространства, то
из заданных векторных расслоений с одной и той же базой можно строить новые
векторные расслоения, производя со слоями операции, которые допустимы для
векторных пространств, поточечно.
\begin{exa}
[{\bf Прямая сумма $\ME_1\oplus\ME_2$ и тензорное произведение
$\ME_1\otimes\ME_2$ расслоений.}]
Пусть $\ME_1(\MM,\pi_1,\MV_1)$ и $\ME_2(\MM,\pi_2,\MV_2)$ -- два векторных
расслоения с одной и той же базой $\MM$. Определим прямую сумму векторных
расслоений
\begin{equation}                                                  \label{ediveb}
  \ME_1\oplus\ME_2(\MM,\pi,\MV_1\oplus\MV_2)
  :=\bigcup_{x\in\MM}\pi^{-1}_1(x)\oplus\pi^{-1}_2(x).
\end{equation}
Если точка $p\in\ME_1\oplus\ME_2$, то она имеет вид $p=p_1\oplus p_2$, где
$p_1\in\pi^{-1}_1(x)$, $p_2\in\pi^{-1}_2(x)$ для некоторого $x\in\MM$. Тем
самым определена проекция $\pi(p)=x$. Типичным слоем прямой суммы векторных
расслоений является векторное пространство $\MV_1\oplus\MV_2$. Пусть
$\lbrace\MU_i\rbrace$ -- координатное покрытие $\MM$ и
$\lbrace f_{ij}^{(1)}\rbrace$ и $\lbrace f^{(2)}_{ij}\rbrace$ -- функции
перехода для расслоений $\ME_1$ и $\ME_2$, соответственно. Функции перехода
для прямой суммы векторных расслоений имеют вид
\begin{equation*}
  h_{ij}:=\begin{pmatrix}f_{ij}^{(1)}&0\\0&f^{(2)}_{ij}\end{pmatrix},
\end{equation*}
которые задают автоморфизм в $\MV_1\oplus\MV_2$.
\index{Прямая сумма расслоений (direct sum of fiber bundles)}%

Аналогично определяется тензорное произведение векторных расслоений:
\begin{equation}                                                  \label{etiveb}
  \ME_1\otimes\ME_2(\MM,\pi,\MV_1\oplus\MV_2)
  :=\bigcup_{x\in\MM}\pi^{-1}_1(x)\otimes\pi^{-1}_2(x).
\end{equation}
Функции перехода для тензорного произведения имеют вид
\begin{equation*}
  h_{ij}:=f^{(1)}_{ij}\otimes f_{ij}^{(2)},
\end{equation*}
которые действуют в тензорном произведении $\MV_1\otimes\MV_2$.
\index{Тензорное произведение расслоений (tensor product of fiber bundles)}%
\qed\end{exa}

Размерности прямой суммы и тензорного произведения расслоений равны:
\begin{equation*}
\begin{split}
  \dim(\ME_1\oplus\ME_2)&=\dim\MM+\dim\MV_1+\dim\MV_2,
\\
  \dim(\ME_1\otimes\ME_2)&=\dim\MM+\dim\MV_1\times\dim\MV_2.
\end{split}
\end{equation*}

В заключение данного раздела обсудим касательное расслоение, построенное в
разделе \ref{sglove}.
\begin{theorem}
Касательное расслоение $\MT(\MM)$ является векторным расслоением
$\ME(\MM,\pi,\MR^n)$ с базой $\MM$, $\dim\MM=n$, типичным слоем $\MR^n$ и
проекцией
$\pi:~(x,X)\rightarrow x$, где $X\in\MT_x(\MM)$.
\end{theorem}
\begin{proof}
Проверка свойств 1)--3) в определении векторного расслоения.
\end{proof}
\begin{com}
Для касательного расслоения размерности базы и типичного слоя совпадают:
\begin{equation*}                                              \tag*{\qed}
  \dim\MM=\dim\MR^n=n.
\end{equation*}
\renewcommand{\qed}{}\qed\end{com}

Касательное расслоение имеет свою специфику. Вообще говоря, в определении
векторного расслоения ничего не говорится о том, какое именно линейное
преобразование векторного пространства происходит в областях пересечения карт.
Говорится лишь о том, что оно возможно. Для касательного расслоения реализован
естественный способ гладко сопоставить каждому преобразованию координат
автоморфизм касательного пространства. А именно, каждому преобразованию
координат, соответствует тождественное преобразование касательного пространства,
$f_{ij}=\id\big(\MT_x(\MM)\big)$ для всех $x\in\MU_i\cap\MU_j$. Если в
касательном пространстве в точке $x\in\MM$ (слое над $x$) выбран координатный
(голономный) базис $\lbrace\pl_\al\rbrace$, то при преобразовании координат в
базе компоненты вектора умножаются на матрицу Якоби, а голономный базис
касательного пространства -- на обратную матрицу Якоби (\ref{etrpar}). Поэтому
сам вектор остается без изменения. Если вектор касательного пространства
рассматривается, как набор компонент $\lbrace X^\al\rbrace$, считая, что базис
векторного пространства фиксирован, то при преобразовании координат компоненты
вектора умножаются на матрицу Якоби $\pl_\al x^{\al'}$ справа. Тогда функциями
перехода являются матрицы Якоби. Обе точки зрения на функции перехода допустимы.
Это зависит от того, как рассматривать базис векторного пространства $\MV$. В
общем случае базис касательного пространства $e_a$, $a=1,\dotsc,n$, можно
зафиксировать произвольным образом (неголономный базис). Тогда мы считаем, что
при преобразовании координат базис и координаты векторов относительно этого
базиса: $X=X^a e_a$ не меняются. Такой подход часто бывает удобнее и
соответствует использованию репера $e_a=e^\al{}_a\pl_\al$, который образует
базис касательного пространства.

Дифференцируемое сечение векторного расслоения $\ME(\MM,\pi,\MV)$, которое всюду
отлично от нуля, существует не всегда. Существование таких сечений отражает
определенные топологические свойства базы $\MM$. Множество сечений векторного
расслоения $\ME$ будем обозначать $\CV_\ME(\MM)$.

Понятие (псевдо-)риманова многообразия естественным образом обобщается на
произвольные векторные расслоения.
\begin{defn}
Если в каждой точке $x\in\MM$ задана невырожденная симметричная билинейная форма
$g$ на слое $\pi^{-1}(x)$, и значение $g(V_1,V_2)$ является достаточно гладкой
функцией на $\MM$ для произвольных достаточно гладких сечений
$V_1,V_2\in\CV_\ME(\MM)$, то $\ME$ называется  {\em (псевдо-)римановым векторным
расслоением}.
\qed\end{defn}
\index{Риманово векторное расслоение (Riemannian vector bundle)}%
\index{Расслоение векторное риманово (Riemannian vector bundle)}%
\index{Псевдо риманово векторное расслоение (pseudo-Riemannian vector bundle)}%
\index{Расслоение векторное псевдо риманово (pseudo-Riemannian vector bundle)}%
\begin{exa}
Тензорное расслоение на римановом многообразии является римановым
векторным расслоением. Например, для тензорных полей
$X=\lbrace X_\al{}^\bt\rbrace$ и $\lbrace Y_\al{}^\bt\rbrace$
типа $(1,1)$ квадратичная форма задается римановой метрикой
\begin{equation*}
  g(X,Y)=g^{\al\bt}X_\al{}^\g Y_\bt{}^\dl g_{\g\dl}.\tag*{\qed}
\end{equation*}
\end{exa}
\section{Связность на векторном расслоении                       \label{scovec}}
В настоящем разделе мы определим связность на векторном расслоении в
инвариантной форме. Рассмотрим многообразие $\MM$, $\dim\MM=m$, и векторное
пространство $\MV$, $\dim\MV=n$. Пусть $\ME(\MM,\pi,\MV)$ -- векторное
расслоение. Сечениями этого расслоения являются векторные поля
$V(x)\in\CV_\ME(\MM)$. На множестве сечений векторного расслоения поточечно
вводится сложение и умножение на гладкие функции. Таким образом, множество
сечений $\CV_\ME(\MM)$ является $\CC^\infty(\MM)$-модулем.

Все дальнейшие конструкции мы будем иллюстрировать записью соответствующих
выражений в компонентах, поскольку это, во-первых, наглядно и, во-вторых,
необходимо при проведении вычислений. Выберем в каждом слое $\pi^{-1}(x)$,
$x\in\MU\subset\MM$, базис $e_a(x)\in\CV_\ME(\MU)$, $a=1,\dotsc,n$. Этот базис
называется {\em репером}. Для каждой точки $x$ существует окрестность
$\MU\subset\MM$, в которой репер задается достаточно гладкими функциями от $x$.
Вообще говоря, такой репер глобально существует не для всех многообразий, и это
зависит от их топологических свойств. В компонентах сечение векторного
расслоения имеет вид $V=V^a(x)e_a$.
\index{Репер (frame)}%

Кокасательное расслоение $\MT^*(\MM)$ имеет ту же базу, что и
$\ME(\MM,\pi,\MV)$, поэтому можно построить тензорное произведение расслоений
$\MT^*(\MM)\otimes\ME$. Пусть $\CV_{\MT^*\otimes\ME}(\MM)$ -- множество сечений
тензорного произведения расслоений. В компонентах сечение этого расслоения
задается векторным полем с дополнительным ковариантным индексом
$dx^\al V_\al{}^a(x)e_a$, где $dx^\al$, $\al=1,\dotsc,m$, -- координатный базис
1-форм.
\begin{defn}
{\em Связностью $\nb$ на векторном расслоении} $\ME(\MM,\pi,\MV)$
называется отображение
\begin{equation}                                                  \label{edacoc}
  \nb:\quad \CV_\ME(\MM)\rightarrow\CV_{\MT^*\otimes\ME}(\MM),
\end{equation}
которое удовлетворяет двум условиям:\newline
\indent 1) для любых двух сечений $V_1,V_2\in\CV_\ME(\MM)$
\begin{equation*}
  \nb(V_1+V_2)=\nb V_1+\nb V_2;
\end{equation*}
\indent 2) для произвольного сечения $V\in\CV_\ME(\MM)$ и произвольной функции
$f\in\CC^\infty(\MM)$
\begin{equation}                                                  \label{edecse}
  \nb(fV)=df\otimes V+f\nb V. \qed
\end{equation}
\end{defn}
\index{Связность на векторном расслоении (connection on a vector bundle)}%
Если в определении связности положить $f=\const$, то из условий 1) и 2) следует
линейность отображения $\nb$.
\begin{defn}
Сечение $\nb V\in\Lm_1(\MM)\otimes\MV$ представляет собой
1-форму на $\MM$ со значениями в векторном пространстве $\MV$. Рассмотрим
касательное векторное поле к базе $X\in\CX(\MM)$. Тогда определено значение
значение 1-формы $\nb V$ на векторном поле $X$
\begin{equation}                                                  \label{ecovva}
  \nb_XV:=\nb V(X),
\end{equation}
которое называется {\em ковариантной производной} векторного поля
$V\in\CV_\ME(\MM)$ вдоль касательного векторного поля $X\in\CX(\MM)$.
\qed\end{defn}
\index{Ковариантная производная вдоль векторного поля %
(covariant derivative along a vector field)}%

Ковариантная производная вдоль векторного поля $\nb_X$ является отображением
$\CV_\ME(\MM)\rightarrow\CV_\ME(\MM)$. Пусть $X,Y\in\CX(\MM)$ -- два
произвольных векторных поля на $\MM$, и $f\in\CC^\infty(\MM)$ -- гладкая
функция. Тогда ковариантная производная имеет следующие свойства:
\begin{align*}
  \nb_{X+Y}V&=\nb_XV+\nb_YV,
\\
  \nb_{fX}V&=f\nb_XV,
\\
  \nb_X(V_1+V_2)&=\nb_XV_1+\nb_XV_2,
\\
  \nb_X(fV)&=(Xf)V+f\nb_XV.
\end{align*}

Построим выражение для ковариантной производной в компонентах. Запишем
отображение (\ref{edacoc}) для векторов репера:
\begin{equation}                                                  \label{edeomv}
  \nb e_a=dx^\al\otimes\om_{\al a}{}^b e_b,
\end{equation}
где $\om_{\al a}{}^b(x)$ -- некоторые функции на координатной окрестности $\MU$.
Введем 1-формы
\begin{equation}                                                  \label{ecovof}
  \om_a{}^b:=dx^\al\om_{\al a}{}^b.
\end{equation}
Тогда соотношение (\ref{edeomv}) можно переписать в виде
\begin{equation}                                                  \label{ecofnj}
  \nb e_a=\om_a{}^b\otimes e_b.
\end{equation}

Рассмотрим, как меняются компоненты $\om_{\al a}{}^b$ при локальном вращении
репера
\begin{equation}                                                  \label{elobac}
  e^\prime_a=S_a{}^b e_b,
\end{equation}
где $S_a{}^b(x)\in\MG\ML(n,\MR)$ -- некоторая матрица преобразования,
элементы которой могут зависеть от точки $x\in\MU$. В новом базисе справедливо
равенство
\begin{equation*}
  \nb e^\prime_a=\om^\prime_a{}^b\otimes e^\prime_b.
\end{equation*}
С другой стороны, из определения связности (\ref{edecse}) следует
\begin{equation*}
  \nb(S_a{}^be_b)=dS_a{}^b\otimes e_b+S_a{}^b\nb e_b.
\end{equation*}
Сравнивая эти выражения, получаем правило преобразования:
\begin{align}                                                     \label{etrcoc}
  \om^\prime_a{}^b&=S_a{}^c\om_c{}^dS^{-1}_{\quad d}{}^b+dS_a{}^cS^{-1}_{\quad c}{}^b,
\\ \intertext{или, в комопонентах,}                               \label{etrcor}
  \om^\prime_{\al a}{}^b&=S_a{}^c\om_{\al c}{}^dS^{-1}_{\quad d}{}^b
  +\pl_\al S_a{}^cS^{-1}_{\quad c}{}^b.
\end{align}
\begin{com}
Это -- одна из самых важных формул дифференциальной геометрии, которая широко
используется в калибровочных моделях математической физики. Подчеркнем, что
преобразование 1-форм $\om_a{}^b$ содержит неоднородное слагаемое и поэтому не
является тензорным. Следовательно, компоненты $\om_{\al a}{}^b$ не определяют
никакого тензора. Они определяют связность.
\qed\end{com}
\begin{theorem}
На любом векторном расслоении $\ME(\MM,\pi,\MV)$ существует связность.
\end{theorem}
\begin{proof}
Доказательство проводится путем явного построения связности с использованием
разложения единицы \cite{ChChLa00}.
\end{proof}

Пусть на многообразии $\MM$ задан атлас $\lbrace \MU_i\rbrace$. Тогда
связность в каждой карте задается компонентами $\om_{\al a}{}^b$, которые в
областях пересечения карт связаны преобразованием (\ref{etrcor}). Верно также и
обратное утверждение. Набор компонент $\om_{\al a}{}^b$, связанных
преобразованием (\ref{etrcor}) в областях пересечения карт, однозначно
определяет связность на векторном расслоении.
\begin{defn}
Компоненты $dx^\al\om_{\al a}{}^b$ называются {\em локальной формой связности}.

Определив действие связности на репер, можно выписать явное выражение
для действия связности на произвольное векторное поле $V^ae_a\in\CV_\ME(\MM)$:
\begin{equation}                                                  \label{ecovec}
  \nb V=(dV^a+V^b\om_b{}^a)\otimes e_a
  =dx^\al(\pl_\al V^a+V^b\om_{\al b}{}^a)\otimes e_a,
\end{equation}
где мы воспользовались определением (\ref{edecse}). Это выражение называется
{\em ковариантной производной} векторного поля. Выражение в скобках,
\begin{equation}                                                  \label{ecodvd}
  \nb_\al V^a=\pl_\al V^a+V^b\om_{\al b}{}^a,
\end{equation}
называется {\em ковариантной производной компонент} векторного поля.
\qed\end{defn}
\index{Локальная форма связности (local connection form)}%
\index{Форма связности локальная (local connection form)}%
\index{Ковариантная производная (covariant derivative)}%
\index{Производная ковариантная (covariant derivative)}%
Замечательным свойством ковариантной производной является то, что при локальном
изменении базиса (\ref{elobac}) ковариантная производная меняется по тензорному
закону. Действительно, при изменении базиса (репера) (\ref{elobac}) компоненты
вектора $V=V^ae_a=V^{\prime a}e^\prime_a$ преобразуются по векторному
представлению группы $\MG\ML(n,\MR^n)$:
\begin{equation}                                                  \label{egatrx}
  V^{\prime a}=V^b S^{-1}_{\quad b}{}^a,\qquad S_b{}^a(x)\in\MG\ML(n,\MR^n).
\end{equation}
Используя преобразование связности (\ref{etrcor}), нетрудно проверить
справедливость следующей формулы преобразования ковариантной производной
\begin{equation*}
  \nb'_\al V^{\prime a}
  =\pl_\al V^{\prime a}+V^{\prime b}\om^{\prime}_{\al b}{}^a
  =(\nb_\al V^b)S^{-1}_{\quad b}{}^a,
\end{equation*}
где ковариантная производная со штрихом $\nb_\al^\prime$ берется с локальной
формой связности $\om'_{\al b}{}^a$.
Тем самым ковариантная производная компонент векторного поля преобразуется по
векторному представлению группы $\MG\ML(n,\MR^n)$.

Ковариантная производная от компонент векторного поля $V$ вдоль касательного
векторного поля $X$ (\ref{ecovva}) в компонентах имеет вид
\begin{equation*}
  \nb_X V^a=X^\al\nb_\al V^a.
\end{equation*}

Рассмотрим некоторые свойства связности. Легко доказывается
\begin{prop}
Пусть на векторном расслоении $\ME(\MM,\pi,\MV)$ задана связность $\nb$ и
зафиксирована точка базы $x\in\MM$. Тогда в окрестности точки $x$ существует
такой репер, что компоненты локальной формы связности в этой точке обращаются
в нуль $\om_{\al a}{}^b(x)=0$.
\end{prop}
\begin{com}
Подчеркнем, что в данном предложении речь идет о фиксированной точке
многообразия, а не об окрестности.
\qed\end{com}

Продолжим изучение локальной формы связности. Перепишем соотношение
(\ref{etrcoc}) в виде
\begin{equation}                                                  \label{eanftr}
  \om^\prime S=S\om+dS,
\end{equation}
где, для краткости, мы опустили векторные индексы, предполагая всюду
суммирование ``с десяти до четырех'' (имеется ввиду расположение чисел на
циферблате часов). Внешняя производная от этого равенства имеет вид
\begin{equation*}
  d\om^\prime S-\om^\prime\wedge dS
  =dS\wedge\om+Sd\om.
\end{equation*}
Подстановка выражения для $dS$ из (\ref{eanftr}) приводит к равенству
\begin{equation}                                                  \label{etrcul}
  d\om^\prime-\om^\prime\wedge\om^\prime
  =S(d\om-\om\wedge\om)S^{-1},
\end{equation}
которое приводит к следующему важному понятию.
\begin{defn}
2-форма на $\MM$
\begin{equation}                                                  \label{edecuf}
  R_a{}^b:=d\om_a{}^b-\om_a{}^c\wedge\om_c{}^b
\end{equation}
называется {\em локальной формой кривизны} связности $\nb$. В компонентах:
\begin{equation*}
  R_a{}^b=\frac12dx^\al\wedge dx^\bt R_{\al\bt a}{}^b,
\end{equation*}
где
\begin{equation}                                                  \label{ecucot}
  R_{\al\bt a}{}^b:=\pl_\al\om_{\bt a}{}^b-\pl_\bt\om_{\al a}{}^b
  -\om_{\al a}{}^c\om_{\bt a}{}^c+\om_{\bt a}{}^c\om_{\al c}{}^b. \qed
\end{equation}
\end{defn}
\index{Форма кривизны локальная (local curvature form)}%
\index{Локальная форма кривизны (local curvature form)}%
Локальная форма кривизны -- это 2-форма на многообразии $\MM$ со значениями в
тензорном произведении $\MV^*\otimes\MV$, где $\MV^*$ -- векторное пространство,
сопряженное к $\MV$.

Формула (\ref{etrcul}) показывает, что при локальном преобразовании репера форма
кривизны преобразуется по тензорному закону, т.е.\ однородно, несмотря на то,
что преобразование формы связности (\ref{etrcoc}) содержит неоднородное
слагаемое.

Пусть на многообразии $\MM$ задано два векторных поля $X,Y\in\CX(\MM)$. Тогда
значение формы кривизны на этих полях $R(X,Y)_a{}^b$ определяет линейное
преобразование из $\CV_\ME(\MM)$ в $\CV_\ME(\MM)$ (эндоморфизм). Если
$V=V^ae_a\in\CV_\ME(\ME)$, то
\begin{equation*}
  R(X,Y):\quad V^a\rightarrow V^bR(X,Y)_b{}^a.
\end{equation*}
Это отображение обладает очевидными свойствами:\newline
\indent\indent 1) $R(X,Y)=-R(Y,X)$;\newline
\indent\indent 2) $R(fX,Y)=fR(X,Y)$;\newline
\indent\indent 3) $R(X,Y)(fV)=fR(X,Y)V$;\newline
где $f\in\CC^\infty(\MM)$.
\begin{prop}                                                      \label{tcuexp}
Пусть на векторном расслоении $\ME(\MM,\pi,\MV)$ задана связность $\nb$ с
локальной формой кривизны $R$. Тогда для любых векторных полей $X,Y\in\CX(\MM)$
справедлива формула
\begin{equation}                                                  \label{ecucod}
  R(X,Y)=\nb_X\nb_Y-\nb_Y\nb_X-\nb_{[X,Y]}.
\end{equation}
\end{prop}
\begin{proof}
Прямая проверка в компонентах.
\end{proof}
\begin{theorem}[\bf Тождества Бианки]
Форма кривизны $R$ удовлетворяет тождествам Бианки:
\begin{equation}                                                  \label{ebidth}
  dR_a{}^b+R_a{}^c\wedge\om_c{}^b-\om_a{}^c\wedge R_c{}^b=0.
\end{equation}
\end{theorem}
\index{Тождества Бианки (Bianchi identities)}%
\index{Бианки тождества (Bianchi identities)}%
\begin{proof}
Внешняя производная от определения кривизны (\ref{edecuf}) равна
\begin{align*}
  dR&=-d\om\wedge\om+\om\wedge d\om=
\\
  &=-(R+\om\wedge\om)\wedge\om+\om\wedge(R+\om\wedge\om)=
\\
  &=-R\wedge\om+\om\wedge R. \tag*{\qed}
\end{align*}
\renewcommand{\qed}{}\end{proof}

В компонентах тождества Бианки имеют вид:
\begin{equation*}
  dx^\al\wedge dx^\bt\wedge dx^\g\pl_\al R_{\bt\g a}{}^b=
  dx^\al\wedge dx^\bt\wedge dx^\g(\om_{\al a}{}^c R_{\bt\g c}{}^b
  -R_{\al\bt a}{}^c\om_{\g c}{}^b)
\end{equation*}
или
\begin{equation*}
  \nb_\al R_{\bt\g a}{}^b+\nb_\bt R_{\g\al a}{}^b+\nb_\g R_{\al\bt a}{}^b=0,
\end{equation*}
где
\begin{equation*}
  \nb_\al R_{\bt\g a}{}^b:=\pl_\al R_{\bt\g a}{}^b
  +R_{\bt\g a}{}^c\om_{\al c}{}^b-\om_{\al a}{}^c R_{\bt\g c}{}^b.
\end{equation*}
-- ковариантная производная от тензора (2-формы) кривизны.
\begin{com}
Построенная конструкция имеет важное приложение в калибровочных моделях теории
поля. Если матрицы $S_a{}^b(x)\in\MG\ML(n,\MR)$ в (\ref{egatrx}) образуют
представление $\rho$ некоторой полупростой группы Ли $\MG$, то эта группа
называется калибровочной. В этом случае локальная форма связности
$\om_{\al a}{}^b$ называется {\em полем Янга-Миллса}, а локальная форма кривизны
$R_{\al\bt a}{}^b$ -- {\em напряженностью} поля Янга-Миллса.
\index{Поле Янга-Миллса (Yang--Mills field)}%
\index{Янга-Миллса поле (Yang--Mills field)}%
\index{Напряженность поля Янга-Миллса  (Yang--Mills field strength)}%
Векторные поля $V=V^a e_a$ называются {\em полями материи},
\index{Поля материи (matter fields)}%
преобразующимися по представлению $\rho$ калибровочной группы. Эти поля
называются скалярными полями, так как представляют собой набор функций на
многообразии, не меняющихся при преобразовании координат,
$V^a\in\CC^\infty(\MM)$ для всех значений индекса $a$.
\qed\end{com}
\begin{defn}
Рассмотрим сечение векторного расслоения $\ME$, т.е.\ векторное поле
$V\in\CV_\ME(\MM)$. Сечение $V$ называется {\em параллельным или горизонтальным}
относительно связности $\nb$, если ковариантная производная от него равна нулю
\begin{equation}                                                  \label{eparse}
  \nb V=0. \qed
\end{equation}
\end{defn}
\index{Параллельное сечение (parallel section)}%
\index{Сечение параллельное (parallel section)}%
\index{Горизонтальное сечение (horizontal section)}%
\index{Сечение горизонтальное (horizontal section)}%
Нулевое сечение, которое каждой точке базы сопоставляет нулевой вектор,
очевидно, всегда является параллельным. В то же время отличное от нуля
параллельное сечение в общем случае может отсутствовать. Рассмотрим этот вопрос
подробно. Разложим векторное поле по локальному базису: $V=V^a e_a$. Тогда
уравнение (\ref{eparse}) эквивалентно пфаффовой системе уравнений на компоненты:
\begin{equation}                                                  \label{epfvef}
  dV^a+V^b\om_b{}^a=0.
\end{equation}
Для анализа этой системы уравнений введем обозначение для левой части
\begin{equation*}
  A^a:=dV^a+V^b\om_b{}^a.
\end{equation*}
Внешняя производная от этой 1-формы равна
\begin{equation*}
  dA^a=A^b\wedge\om_b{}^a+V^b R_b{}^a.
\end{equation*}
Из этого выражения и теоремы Фробениуса (\ref{tfroth}), записанной в терминах
дифференциальных форм, следует, что, если кривизна связности равна нулю, то
пфаффова система уравнений (\ref{eparse}) является вполне интегрируемой. В этом
случае существует $n$ линейно независимых параллельных сечений векторного
расслоения. Мы видим, что для существования отличных от нуля локальных решений
системы уравнений (\ref{epfvef}) локальная форма кривизны данной связности
должна обращаться в нуль. Другими словами, равенство нулю формы кривизны
является необходимым и достаточным условием разрешимости системы уравнений в
частных производных (\ref{epfvef}).
\begin{defn}
Пусть на многообразии $\MM$ задана дифференцируемая кривая
$\g=\lbrace x^\al(t)\rbrace$, $\al=1,\dotsc,m$. Обозначим касательный вектор
к кривой (вектор скорости) через $u$. Тогда сечение векторного расслоения
$V\in\CV_\ME(\MM)$ называется {\em параллельным вдоль кривой} $\g$, если
ковариантная производная от него вдоль $u$ равна нулю
\begin{equation}                                                  \label{eparge}
  \nb_uV=0.  \qed
\end{equation}
\end{defn}
\index{Сечение, параллельное вдоль кривой (parallel along a curve section)}%
\index{Параллельное вдоль кривой сечение (parallel along a curve section)}%
Касательный вектор к кривой (вектор скорости) имеет вид $u:=\dot x^\al\pl_\al$.
В компонентах условие параллельности сечения вдоль кривой $\g$ сводится
к системе уравнений на компоненты сечения
\begin{equation}                                                  \label{epascr}
  \dot V^a+\dot x^\al V^b\om_{\al b}{}^a=0.
\end{equation}
Поскольку это -- система обыкновенных дифференциальных уравнений, то локально у
нее существует единственное решение при заданных начальных данных. Таким
образом, если задан вектор $V_p$ в некоторой точке $p$ на кривой $\g$, то он
однозначно определяет векторное поле вдоль кривой $\g$, которое называется
{\em параллельным переносом} вектора $V_p$ вдоль кривой $\g$. Очевидно, что
параллельный перенос задает изоморфизм слоев векторного расслоения $\ME$ в
различных точках кривой $\g$.
\index{Параллельный перенос (parallel transfer)}%
\index{Перенос параллельный (parallel transfer)}%

Связность $\nb$ на векторном расслоении $\ME(\MM,\pi,\MV)$ индуцирует
связность на сопряженном векторном расслоении $\ME^*(\MM,\pi,\MV^*)$,
которую мы также обозначим $\nb$. Это делается следующим образом.
Пусть заданы сечения $V\in\CV_\ME(\MM)$ и $V^*\in\CV_{\ME^*}(\MM)$. Тогда
каждое сечение $V^*(x)$ задает поточечное линейное отображение
\begin{equation*}
  V^*(x):\quad \CV_\ME(\MM)\ni\quad V(x)\mapsto(V^*,V)\quad\in\CC^\infty(\MM).
\end{equation*}
Определим индуцированную связность на сопряженном расслоении $\ME^*$ следующей
формулой
\begin{equation}                                                  \label{eindco}
  d(V^*,V)=(V^*,\nb V)+(\nb V^*,V),\qquad \forall V\in\CV_\ME(\MM).
\end{equation}

Найдем выражение для индуцированной связности в компонентах. Пусть $e^a$
-- базис сопряженного векторного пространства $\MV^*$: $(e^a,e_b)=\dl^a_b$.
Этот базис называется {\em корепером}.
\index{Корепер (coframe)}%
Тогда подстановка корепера $e^a$ и репера $e_b$ в определение (\ref{eindco})
приводит к равенству
\begin{equation}                                                  \label{eincos}
  \nb e^a=-e^b\otimes\om_b{}^a.
\end{equation}
Отсюда следует выражение для ковариантной производной сечения $V^*=e^aV_a$
сопряженного расслоения
\begin{equation}                                                  \label{ecoded}
  \nb V^*=e^a\otimes(dV_a-\om_a{}^bV_b).
\end{equation}
Эта ковариантная производная отличается знаком у слагаемого со связностью
от соответствующего выражения для сечения векторного расслоения (\ref{ecovec}).
\begin{defn}
Пусть на векторных расслоениях $\ME_1(\MM,\pi_1,\MV_1)$ и
$\ME_2(\MM,\pi_2,\MV_2)$ с одинаковой базой $\MM$ независимо заданы связности
$\nb_1$ и $\nb_2$. Пусть заданы сечения $V_1\in\CV_{\ME_1}(\MM)$ и
$V_2\in\CV_{\ME_2}(\MM)$. Тогда на прямой сумме $\ME_1\oplus\ME_2$ и тензорном
произведении $\ME_1\otimes\ME_2$ расслоений определена связность
\begin{equation}                                                  \label{etecon}
\begin{split}
  \nb(V_1\oplus V_2)&=\nb_1V_1\oplus\nb_2V_2,
\\
  \nb(V_1\otimes V_2)&=\nb_1V_1\otimes V_2+V_1\otimes\nb_2V_2,
\end{split}
\end{equation}
которая называется {\em индуцированной}.
\qed\end{defn}
\index{Индуцированная связность (induced connection)}%
\index{Связность индуцированная (induced connection)}%
\section{Аффинная связность                                      \label{saffco}}
Касательное расслоение $\MT(\MM)$, $\dim\MM=n$, является векторным расслоением,
типичным слоем которого является векторное пространство $\MR^n$. Поэтому
определение связности на векторном расслоении без изменений переносится на
касательные расслоения. Связность на касательном расслоении называется
{\em аффинной связностью}. Если в касательном пространстве и сопряженном к нему
\index{Аффинная связность (affine connection)}%
\index{Связность аффинная (affine connection)}%
пространстве 1-форм выбрать координатный базис $\pl_\al$ и $dx^\al$, то формулы
для компонент связности (\ref{edeomv}) и (\ref{eincos}) определяют аффинную
связность
\begin{equation}                                                  \label{ecofac}
\begin{split}
  \nb(\pl_\al)&=\quad dx^\bt\otimes\Gamma_{\bt\al}{}^\g\pl_\g,
\\
  \nb(dx^\al)&=-dx^\g\otimes dx^\bt\Gamma_{\bt\g}{}^\al.
\end{split}
\end{equation}
У компонент аффинной связности $\Gamma_{\al\bt}{}^\g$ все индексы являются
координатными, и для них принято обозначение $\Gamma$ вместо $\om$. Формула
преобразования компонент аффинной связности при
преобразовании координат получается из формулы преобразования компонент
связности на векторном расслоении (\ref{etrcor}) после замены матрицы
$S_b{}^a(x)$ на матрицу Якоби $\pl_\al x^{\al'}$. При этом необходимо учесть,
что при преобразовании координат преобразуется также базис 1-форм, что приводит
к дополнительному общему множителю:
\begin{equation}                                                  \label{econts}
  \Gamma_{\bt'\g'}{}^{\al'}
  =(\pl_{\bt'} x^\bt \pl_{\g'} x^\g \Gamma_{\bt\g}{}^\al
  +\pl^2_{\bt'\g'}x^\al)\pl_\al x^{\al'}.
\end{equation}
\begin{defn}
Функции $\Gamma_{\al\bt}{}^\g$, преобразующихся по правилу (\ref{econts}) при
преобразовании координат, называются {\em компонентами локальной формы аффинной
связности} или, короче, аффинной связностью на $\MM$.
\qed\end{defn}
\begin{com}
Аффинная связность, так же как и связность на произвольном векторном расслоении,
определенная в предыдущем разделе, не имеет ничего общего с метрикой. Ее можно
задать независимо от метрики, которая может вообще отсутствовать на
многообразии.
\qed\end{com}

Аффинная связность $\nb$ осуществляет отображение
\begin{equation}                                                  \label{ecomap}
  \nb:\quad \CX(\MM)\times\CX(\MM)\ni\quad X,Y~\mapsto\nb_XY\quad\in\CX(\MM),
\end{equation}
где $\nb_X Y$ -- ковариантная производная от векторного поля $Y$ вдоль
векторного поля $X$. Как и в разделе \ref{scovec}, можно убедиться, что это
отображение обладает следующими свойствами:
\begin{equation}                                                  \label{eprcod}
\begin{split}
  \nb_{X+Y}Z&=\nb_XZ+\nb_YZ,
\\
  \nb_{fX}Y&=f\nb_XY,
\\
  \nb_X(Y+Z)&=\nb_XY+\nb_XZ,
\\
  \nb_X(fY)&=(Xf)Y+f\nb_XY,
\end{split}
\end{equation}
где $f\in\CC^\infty(\MM)$ и $X,Y,Z\in\CX(\MM)$. Верно также обратное
утверждение: отображение $\nb$ со свойствами (\ref{eprcod}) определяет аффинную
связность на касательном расслоении. Иногда его принимают в качестве определения
аффинной связности.

Связность, определенная на касательном и кокасательном расслоении индуцирует
связность на произвольных тензорных расслоениях с помощью правил (\ref{etecon}).
В компонентах ковариантная производная от тензорных полей будет выписана в
следующей главе. Индуцированная ковариантная производная линейна,
$$
  \nb(aY\oplus bZ)=a\nb Y\oplus b\nb Z,\qquad a,b\in\MR,
$$
где $Y,Z\in\CT(\MM)$ -- произвольные тензорные поля. Справедливо правило
Лейбница
\begin{equation}                                                  \label{eleiru}
  \nb(Y\otimes Z)
  =(\nb Y)\otimes Z+Y\otimes(\nb Z).
\end{equation}
Нетрудно проверить, что ковариантная производная перестановочна с каждым
свертыванием: $\nb C=C\nb$, где $C$ -- оператор свертки. Кроме того,
ковариантная производная вдоль векторного поля $\nb_X$ сохраняет тип тензорных
полей $\nb_X\CT^r_s(\MM)\subset\CT^r_s(\MM)$. Таким образом, ковариантная
производная вдоль векторного поля $\nb_X$ является дифференцированием в
тензорной алгебре, которое было определено в разделе \ref{sliede}. В силу
теоремы \ref{tditea} ковариантная производная вдоль векторного поля $\nb_X$
отличается от производной Ли $\Lie_X$ на дифференцирование, порожденное
тензорным полем типа $(1,1)$.

Тензор кривизны (\ref{ecucot}) для аффинной связности (\ref{ecofac}) в
компонентах принимает вид
\begin{equation}                                                  \label{ecuaft}
  R_{\al\bt\g}{}^\dl=\pl_\al\Gamma_{\bt\g}{}^\dl-\pl_\bt\Gamma_{\al\g}{}^\dl
  -\Gamma_{\al\g}{}^\e\Gamma_{\bt\e}{}^\dl+\Gamma_{\bt\g}{}^\e\Gamma_{\al\e}{}^\dl.
\end{equation}

Рассмотрим тензор кривизны (\ref{ecuaft}), как отображение касательных
пространств
\begin{equation*}
  \CX(\MM)\times\CX(\MM)\times\CX(\MM)\rightarrow\CX(\MM)
\end{equation*}
и обозначим
\begin{equation*}
  R(X,Y,Z):=X^\al Y^\bt Z^\g R_{\al\bt\g}{}^\dl\pl_\dl.
\end{equation*}
Тогда нетрудно проверить справедливость следующего тождества:
\begin{equation*}
  R(X,Y,Z)=\nb_X\nb_Y Z-\nb_Y\nb_X Z-\nb_{[X,Y]}Z,
\end{equation*}
которое является частным случаем предложения \ref{tcuexp}. Это соотношение
иногда принимают в качестве определения кривизны $R$ аффинной связности $\nb$.
\begin{defn}
2-форма со значениями в касательном пространстве,
\begin{equation*}
  T=\frac12dx^\al\wedge dx^\bt T_{\al\bt}{}^\g\pl_\g,
\end{equation*}
определенная равенством:
\begin{equation}                                                  \label{etorde}
  T(X,Y):=\nb_X Y-\nb_Y X-[X,Y],\qquad \forall X,Y\in\CX(\MM).
\end{equation}
называется {\em тензором кручения}.
\qed\end{defn}
\index{Тензор кручения (torsion tensor)}%
\index{Кручения тензор (torsion tensor)}%
\index{Кручение (torsion)}%
Компоненты тензора кручения равны антисимметричной части компонент аффинной
связности:
\begin{equation}                                                  \label{etoind}
  T_{\al\bt}{}^\g=\Gamma_{\al\bt}{}^\g - \Gamma_{\bt\al}{}^\g.
\end{equation}
Поскольку при преобразовании координат неоднородное слагаемое в (\ref{econts})
симметрично по индексам $\bt',\g'$, то компоненты кручения действительно
преобразуются по тензорному закону. Таким образом, аффинная связность определяет
два тензора: тензор кривизны и тензор кручения.
\section{Связность на расслоении реперов                         \label{scorep}}
Существует тесная взаимосвязь между касательным расслоением и расслоением
реперов. Начнем с конструктивного построения расслоения реперов. Предположим,
что задано дифференцируемое многообразие $\MM$, $\dim\MM=n$.
\begin{defn}
{\em Репером} в точке $x\in\MM$ называется упорядоченный набор объектов
$p=(x,e_1,\dotsc,e_n)=\lbrace x,e_a\rbrace$, $a=1,\dotsc,n$, где
$\lbrace e_a\rbrace$ -- набор линейно независимых касательных векторов в точке
$x$. Множество всех реперов на $\MM$ обозначим через $\ML(\MM)$. Ниже мы
построим дифференцируемую структуру на $\ML(\MM)$, тем самым превратив множество
всех реперов в дифференцируемое многообразие размерности $\dim\ML(\MM)=n+n^2$
такое, что естественная проекция
\begin{equation}                                                  \label{eprrep}
  \pi(x,e_1,\dotsc,e_n)=x,
\end{equation}
является гладким отображением $\ML(\MM)\rightarrow\MM$. Тогда $\ML(\MM)$
называется {\em расслоением реперов} на $\MM$.
\qed\end{defn}
\index{Репер (frame)}%
\index{Расслоение реперов (frame bundle)}%
\index{Реперов расслоение (frame bundle)}%
\begin{com}
Забегая вперед, скажем, что расслоение реперов представляет собой главное
расслоение $\ML(\MM)=\MP\big(\MM,\pi,\MG\ML(n,\MR)\big)$ с базой $\MM$,
проекцией $\pi$ и структурной группой $\MG\ML(n,\MR)$, с которым ассоциировано
касательное расслоением $\MT(\MM)$.
\qed\end{com}

Процедура построения дифференцируемой структуры на $\ML(\MM)$ аналогична
построению дифференцируемой структуры на тензорных расслоениях. Пусть
$\MU\subset\MM$ -- некоторая карта с координатами $x^\al$, $\al=1,\dotsc,n$.
Координатный базис касательного расслоения $\pl_\al$ образует координатный репер
в каждой точке. Тогда произвольный репер можно разложить по координатному
базису:
\begin{equation}                                                  \label{ederec}
  e_a=e^\al{}_a\pl_\al.
\end{equation}
По-определению, совокупность чисел $e^\al{}_a$ образует невырожденную
$n\times n$ матрицу, т.е.\ представляет собой элемент группы $\MG\ML(n,\MR)$ в
каждой точке $x\in\MM$. Поэтому можно определить отображение
$\chi:~\pi^{-1}(\MU)\rightarrow\MU\times\MG\ML(n,\MR)$ такое, что для каждой
точки $(x,e_1,\dotsc,e_n)\in\ML(\MM)$ и $x\in\MU$, $e^\al{}_a\in\MG\ML(n,\MR)$
имеем равенство
\begin{equation*}
  \chi(x,e_1,\dotsc,e_n)=(x,e^\al{}_a),
\end{equation*}
где репер $e_a$ задан равенством (\ref{ederec}). Очевидно, что отображение
$\chi$ является взаимно однозначным.

Теперь выберем атлас $\lbrace\MU_i\rbrace$ на $\MM$ с построенными выше
отображениями $\lbrace\chi_i\rbrace$. Прообразы всех открытых подмножеств
топологических произведений $\MU_i\times\MG\ML(n,\MR)$ для отображений $\chi_i$
образует базу топологии на $\ML(\MM)$. При этом все отображения $\chi_i$ будут
гомеоморфизмами.

Благодаря отображению $\chi_i$, прообраз проекции $\pi^{-1}(\MU_i)$ становится
координатной окрестностью расслоения реперов $\ML(\MM)$. Пусть в пересекающихся
областях $\MU_i\cap\MU_j\ne\emptyset$ заданы координаты $x^\al$ и $x^{\al'}$
соответственно. Тогда координатами репера $(x,e_1,\dotsc,e_n)$ в пересечении
$\MU_i\cap\MU_j$ будут наборы из $n+n^2$ чисел $\lbrace x^\al,e^\bt{}_a\rbrace$
и $\lbrace x^{\al'},e^{\bt'}{}_a\rbrace$, которые связаны преобразованием
\begin{equation}                                                  \label{etrrec}
\begin{split}
  x^{\al'}&=x^{\al'}(x),
\\
  e^{\al'}{}_a&=e^\al{}_a\pl_\al x^{\al'}.
\end{split}
\end{equation}
Если преобразование координат на $\MM$ класса $\CC^\infty$, то мы получаем
бесконечно дифференцируемую структуру на $\ML(\MM)$. Таким образом расслоение
реперов становится дифференцируемым $(n+n^2)$-мерным многообразием класса
$\CC^\infty$. При этом естественная проекция $\pi:~\ML(\MM)\rightarrow\MM$
является гладким сюрьективным отображением.

Сечение расслоения реперов $\ML(\MM)$ мы будем обозначать $e_a(x)$.
Его также можно разложить по координатному базису $e_a(x)=e^\al{}_a(x)\pl_\al$.
Мы предполагаем, что все компоненты этого разложения являются гладкими
функциями: $e^\al{}_a(x)\in\CC^\infty(\MU)$. В разделе \ref{scovec} под репером
понималось именно сечение расслоения реперов.
\begin{com}
В формуле (\ref{ederec}) компоненты разложения $e^\al{}_a$ рассматриваются,
как координаты (числа) на пространстве главного расслоения $\ML(\MM)$, а не как
функции на базе $e^\al{}_a(x)$. Формулы (\ref{etrrec}) -- это также
преобразование координат на $\ML(\MM)$. В физической литературе сечение
расслоения реперов $e_a(x)$ принято называть просто репером. Для краткости мы
также будем употреблять термин репер для обозначения сечения расслоения реперов
там, где это не вызывает двусмысленности.
\qed\end{com}

Отображение $\chi$ задает локально структуру прямого произведения на расслоении
реперов $\ML(\MM)$. То есть прообраз $\pi^{-1}(\MU)$ диффеоморфен прямому
произведению $\MU\times\MG\ML(n,\MR)$.

Сужение отображения $\chi$ на каждый слой
$\chi_x:~\pi^{-1}(x)\rightarrow\MG\ML(n,\MR)$ является гомоморфизмом групп.
Если $x\in\MU_i\cap\MU_j$ тогда отображения
$f_{ij}(x)=\chi_{j,x}\circ\chi^{-1}_{i,x}$ (функции перехода) представляют
собой автоморфизмы группы $\MG\ML(n,\MR)$. Из (\ref{etrrec}) следует, что
преобразование $\chi_{j,x}\circ\chi^{-1}_{i,x}$ является умножением справа в
группе $\MG\ML(n,\MR)$ на матрицу Якоби
$J_{ij\,\al}{}^{\al'}:=\pl x^{\al'}_{(j)}/\pl x^\al_{(i)}$. Это значит, что
набор матриц Якоби $J_{ij}$ представляет собой совокупность функций склейки на
расслоении реперов.
\begin{com}
Типичным слоем расслоения реперов является группа $\MG\ML(n,\MR)$, на которой
нет структуры векторного пространства. Поэтому расслоение реперов не является
векторным расслоением.

Поскольку каждый слой $\pi^{-1}(x)$ диффеоморфен и гомоморфен $\MG\ML(n,\MR)$,
то он изоморфен группе Ли: $\pi^{-1}(x)\simeq\MG\ML(n,\MR)$.
\qed\end{com}
Структурная группа $\MG\ML(n,\MR)$ естественным образом действует на расслоении
реперов и представляет собой группу преобразований $\ML(\MM)$. Пусть
$S_a{}^b\in\MG\ML(n,\MR)$. Тогда правое действие $r_S$ структурной
группы на $\ML(\MM)$ определим следующим образом
\begin{equation*}
  r_S(x,e_a)=(x,e^\prime_a),\quad \text{где}\quad e^\prime_a=S_a{}^be_b.
\end{equation*}
\begin{com}
Мы записываем правое действие структурной группы в виде матричного умножения
слева. В этом нет неоднозначности, т.к.\ суммирование всегда проводится по
одному нижнему и одному верхнему индексу, $S_a{}^b e_b=e_bS_a{}^b$. Наши
обозначения вызваны принятым правилом суммирования ``с десяти до четырех''. Это
правило является следствием общепринятой записи $dx^\al\pl_\al$. Другая запись
$\pl_\al dx^\al$ может вызвать недоразумения. Кроме того, векторы репера принято
нумеровать нижним индексом, поскольку координатные векторы $\pl_\al$ образуют
координатный репер.
\qed\end{com}
Очевидно, что каждое преобразование $r_S$ задает диффеоморфизм $\ML(\MM)$ и
сохраняет слои: $\pi\circ r_S=\pi$. Действие $r_S$ называется {\em правым
действием структурной группы}. Если $S,T\in\MG\ML(n,\MR)$, то
\begin{equation*}
  r_{ST}=r_S\circ r_T.
\end{equation*}
Действие группы преобразований $\MG\ML(n,\MR)$ свободно, т.е. любой элемент
группы, отличный от единичного, перемещает все точки многообразия $\ML(\MM)$.

Обозначим матрицу, обратную к реперу $e^\al{}_a$, через $e_\bt{}^b$:
\begin{equation*}
  e^\al{}_ae_\al{}^b=\dl_a^b,\qquad e_\al{}^ae^\bt{}_a=\dl_\al^\bt.
\end{equation*}
Тогда при преобразовании координат (\ref{etrrec}) на пространстве расслоения
$\ML(\MM)$ она преобразуется, как ковекторное поле
\begin{equation*}
  e_{\al'}{}^a=\pl_{\al'}x^\al e_\al{}^a.
\end{equation*}
Отсюда следует, что 1-формы (корепер)
\begin{equation}                                                  \label{edecja}
  e^a=dx^\al e_\al{}^a
\end{equation}
не зависят от выбора координат на $\MM$. Это значит, что $e^a$ являются
дифференциальными 1-формами, заданными в каждой координатной окрестности
согласованным образом и, следовательно, определяющими некоторую 1-форму на
всем пространстве расслоения $\ML(\MM)$. Напомним, что числа $e^\al{}_a$ в
(\ref{edecja}) рассматриваются в качестве координат на $\ML(\MM)$. Поэтому
$e_\al{}^a$ представляют собой набор хорошо определенных функций на $\ML(\MM)$,
поскольку $\det e^\al{}_a\ne0$.

Пфаффова система уравнений
\begin{equation}                                                  \label{epfves}
  e^a=0
\end{equation}
определяет $n^2$-мерное касательное подпространство $\MV_p(\ML)$ в
касательном пространстве $\MT_p(\ML)$ к каждой точке расслоения
$p=(x,e)\in\ML(\MM)$, которое называется {\em вертикальным пространством}.
\index{Вертикальное пространство}\index{Пространство вертикальное}%
Система уравнений (\ref{epfves}) в каждой координатной окрестности
эквивалентна системе уравнений
\begin{equation*}
  dx^\al=0.
\end{equation*}
Эта система уравнений является вполне интегрируемой, и ее максимальное
интегральное многообразие задается уравнениями
\begin{equation*}
  x^\al=\const,
\end{equation*}
т.е.\ является слоем $\pi^{-1}(x)$ в точке $x\in\MM$. Это значит, что
вертикальные подпространства являются касательными пространствами к слоям.

Векторное поле на пространстве расслоения $X\in\CX(\ML)$ в компонентах имеет вид
\begin{equation*}
  X=X^\al\frac\pl{\pl x^\al}+X^\al{}_a\frac\pl{\pl e^\al{}_a}.
\end{equation*}
Поэтому векторное поле $Y$ принадлежит вертикальному подпространству
$Y\in\MV_p(\ML)$ в каждой точке $p=(x,e)\in\ML(\MM)$ тогда и только тогда, когда
оно имеет компоненты только вдоль координат слоя:
\begin{equation*}
  Y=Y^\al{}_a\frac\pl{\pl e^\al{}_a}
\end{equation*}

Дифференциал $\pi_*$ проекции (\ref{eprrep}) отображает касательные пространства
$\MT_p(\ML)\rightarrow\MT_x(\MM)$. Очевидно, что все векторы из
вертикального подпространства $\MV_p(\ML)$ отображаются в нулевой вектор
$\MT_x(\MM)$. Верно и обратное утверждение. Если какой либо вектор из
$\MT_p(\ML)$ отображается в нулевой вектор, то он принадлежит вертикальному
подпространству. Это значит, что вертикальные подпространства являются ядром
дифференциала проекции $\pi_*$.

Пусть на базе $\MM$ задана аффинная связность $\Gamma$ или, что эквивалентно,
1-формы $\Gamma_\al{}^\bt=dx^\g\Gamma_{\g\al}{}^\bt$. Тогда ковариантный
дифференциал векторного поля $e_a(x)$ равен
\begin{equation*}
  De_a=(de^\al{}_a+e^\bt{}_a\Gamma_\bt{}^\al)\otimes\pl_\al.
\end{equation*}
Если рассматривать $e^\al{}_a$, как координаты на $\ML(\MM)$ (что мы и делаем),
то выражение
\begin{equation*}
  De^\al{}_a=de^\al{}_a+e^\bt{}_a\Gamma_\bt{}^\al
\end{equation*}
является 1-формой на координатной окрестности $\pi^{-1}(\MU)$ расслоения
реперов $\ML(\MM)$. Поскольку 1-форма $De^\al{}_a$ при преобразовании координат
является вектором по индексу $\al$, то 1-форма
\begin{equation}                                                  \label{eonfab}
  \om_a{}^b=e_\al{}^bDe^\al{}_a
\end{equation}
не зависит от выбора системы координат и, значит, задает дифференциальную
1-форму на всем расслоении реперов $\ML(\MM)$. 1-форма $\om_a{}^b$ называется
{\em формой связности} на расслоении реперов.
\index{Форма связности (connection form)}%
\index{Связности форма (connection form)}%
Так как $(x^\al,e^\bt{}_a)$ являются координатами на $\ML(\MU)$, то
$(dx^\al,de^\bt{}_a)$ представляют собой координатный базис кокасательного
пространства к $\ML(\MU)$. Поэтому $(x^\al,e^\bt{}_a,dx^\al,de^\bt{}_a)$ задают
систему координат кокасательного расслоения $\MT^*(\ML)$ к $\ML(\MU)$. Теперь
$e^a$ и $\om_a{}^b$ задают $n+n^2$ дифференциальных 1-форм на расслоении реперов
$\ML(\MU)$. В каждой координатной окрестности $\pi^{-1}(\MU)$ их можно выразить
в виде линейной комбинации 1-форм $dx^\al$ и $de^\al{}_a$, и наоборот. Поскольку
1-формы $e^a$ и $\om_a{}^b$ линейно независимы, то они задают поле корепера на
всем пространстве $\ML(\MM)$. Дуальный к нему базис определяет поле репера на
пространстве расслоения $\ML(\MM)$ глобально.
\begin{com}
В общем случае сечение расслоения реперов $e_a(x)$ может не существовать
глобально. Например, на двумерной сфере $\MS^2$ не существует непрерывного
векторного поля, которое не обращается в нуль ни в одной точке (теорема
\ref{tvecsp}). Следовательно, глобального сечения $\ML(\MS^2)$ не существует.
В то же время, поскольку на произвольном многообразии $\MM$ существует аффинная
связность, то на расслоении реперов $\ML(\MM)$ всегда существует корепер $e^a$
глобально. В этом смысле многообразие расслоения реперов $\ML(\MM)$ устроено
проще, чем база $\MM$.
\qed\end{com}
В локальной системе координат из определения 1-форм $e^a$ и $\om_a{}^b$ следуют
равенства
\begin{equation*}
\begin{split}
  dx^\al&=e^a e^\al{}_a,
\\
  de^\al{}_a&=-e^\bt{}_a\Gamma_\bt{}^\al +\om_a{}^be^\al{}_b.
\end{split}
\end{equation*}
Здесь, по-прежнему, все формы рассматриваются на пространстве главного
расслоения $\ML(\MM)$.
Взятие внешней производной от равенств (\ref{edecja}) и (\ref{eonfab}) приводит
к {\em структурным уравнениям} для аффинной связности:
\begin{align}                                                     \label{etoste}
  de^a-e^b\wedge\om_b{}^a&
  =\frac12e^b\wedge e^c\,T_{bc}{}^a,
\\                                                                \label{ecusre}
  d\om_a{}^b-\om_a{}^c\wedge\om_c{}^b&
  =\frac12e^c\wedge e^d\,R_{cda}{}^b,
\end{align}
\index{Структурные уравнения (structure equations)}%
\index{Уравнения структурные (structure equations)}%
где $T_{bc}{}^a:=e^\bt{}_b e^\g{}_c T_{\bt\g}{}^\al e_\al{}^a$ и
$R_{cda}{}^b:=e^\g{}_c e^\dl{}_d e^\al{}_aR_{\g\dl\al}{}^\bt e_\bt{}^b$
-- компоненты тензоров кручения и кривизны. Компоненты тензоров кручения
$T_{\bt\g}{}^\al$ и кривизны $R_{\g\dl\al}{}^\bt$ в координатном базисе были
определены ранее (\ref{etoind}) и (\ref{ecuaft}).
Введем обозначения для 2-форм кручения и кривизны:
\begin{equation}                                                  \label{etocuf}
\begin{split}
  T^a&:=\frac12 e^b\wedge e^c T_{bc}{}^a,
\\
  R_a{}^b&:=\frac12 e^c\wedge e^d R_{cda}{}^b.
\end{split}
\end{equation}
Взятие внешней производной от уравнений структуры (\ref{etoste}), (\ref{ecusre})
приводит к {\em  тождествам Бианки} для аффинной связности:
\index{Тождества Бианки (Bianchi identities}%
\begin{equation}                                                  \label{ebiacn}
\begin{split}
  dT^a+T^b\wedge\om_b{}^a&=e^b\wedge R_b{}^a,
\\
  dR_a{}^b+R_a{}^c\wedge\om_c{}^b-\om_a{}^c\wedge R_c{}^b&=0.
\end{split}
\end{equation}
Второе из этих тождеств совпадает с полученной ранее формулой (\ref{ebidth}). В
компонентах тождества Бианки будут подробно рассмотрены в разделе \ref{sbianc}.
\begin{com}
Структурные уравнения (\ref{etoste}) и (\ref{ecusre}) часто принимают за
определение тензоров кручения и кривизны.
\qed\end{com}

В координатном базисе $e^a\mapsto dx^\al$,
$\om_a{}^b\mapsto\Gamma_\al{}^\bt:=dx^\g\Gamma_{\g\al}{}^\bt$, и структурные уравнения
принимают вид
\begin{equation}                                                  \label{estrec}
\begin{split}
  -dx^\bt\wedge\Gamma_\bt{}^\al&
  =\frac12dx^\bt\wedge dx^\g\, T_{\bt\g}{}^\al,
\\
  d\Gamma_\al{}^\bt-\Gamma_\al{}^\g\wedge\Gamma_\g{}^\bt&
  =\frac12dx^\g\wedge dx^\dl\, R_{\g\dl\al}{}^\bt,
\end{split}
\end{equation}
где в правых частях стоят тензоры кручения $T_{\bt\g}{}^\al$ и кривизны
$R_{\g\dl\al}{}^\bt$ и учтено равенство $ddx^\al=0$.
\begin{theorem}
Пусть на расслоении реперов $\ML(\MM)$ заданы $n^2$ дифференциальных 1-форм
$\om_a{}^b$, которые вместе с корепером $e^a$ удовлетворяют структурным
уравнениям (\ref{etoste}), (\ref{ecusre}), где $T_{bc}{}^a$ и $R_{cda}{}^b$ --
некоторые функции, определенные на $\ML(\MM)$. Тогда существует аффинная
связность $\nb$ такая, что выполнено равенство
(\ref{eonfab}).
\end{theorem}
\begin{proof}
См., например, \cite{ChChLa00}.
\end{proof}
\begin{com}
От компонент, определяющих кручение и кривизну, нет надобности требовать
выполнения тождеств Бианки, т.к.\ они являются следствием структурных
уравнений, выполнение которых является условием теоремы.
\qed\end{com}

Ранее было отмечено, что пфаффова система уравнений $e^a=0$ определяет
распределение вертикальных подпространств $\MV_p(\ML)\subset\MT_p(\ML)$
на расслоении реперов $\ML(\MM)$.
В каждой точке $(x,e)\in\ML(\MM)$ система пфаффовых уравнений
\begin{equation}                                                  \label{ehopae}
  \om_a{}^b=0
\end{equation}
определяет $n$-мерное касательное подпространство
$\MH_p(\ML)\subset\MT_p(\ML)$, которое называется {\em горизонтальным}.
\index{Горизонтальное пространство (horizontal space)}%
\index{Пространство горизонтальное (horizontal space)}%
Таким образом, мы имеем распределение горизонтальных подпространств $\MH(\ML)$,
однозначно определенное аффинной связностью $\nb$ на расслоении реперов
$\ML(\MM)$. Оно имеет следующие свойства.
\begin{theorem} Пусть задано расслоение реперов $\ML(\MM)$ и на базе $\MM$
определена аффинная связность $\nb$. Тогда определены распределения
вертикальных $\MV_p(\ML)$ и горизонтальных $\MH_p(\ML)$ подпространств,
которые удовлетворяют свойствам:

1) \parbox[t]{.92\linewidth}{В каждой точке $(x,e)\in\ML(\MM)$ касательное
пространство $\MT_p(\ML)$ разлагается в прямую сумму
\begin{equation}                                                  \label{edishv}
  \MT_p(\ML)=\MV_p(\ML)\oplus\MH_p(\ML).
\end{equation}
При этом образ горизонтального пространства $\MH_p(\ML)$ при проекции $\pi$
изоморфен касательному пространству к базе $\MT_x(\MM)$.}

2) \parbox[t]{.92\linewidth}{Горизонтальные пространства инвариантны
относительно правого действия группы преобразований $r_S$, $S\in\MG\ML(n,\MR)$.
То есть
\begin{equation*}
  (r_S)_*\MH_p(\ML)=\MH_{r_S(x,p)}(\ML).
\end{equation*}}
\end{theorem}
\begin{proof}
Докажем свойство 1). Поскольку сумма размерностей горизонтального
$\MH_p(\ML)$ и вертикального $\MV_p(\ML)$ подпространств равна
размерности касательного пространства $\MT_p(\ML)$, то достаточно доказать,
что $\MH_p(\ML)\cap\MV_p(\ML)=\emptyset$. От противного. Пусть
$X\in\MH_p(\ML)\cap\MV_p(\ML)$. Тогда, по-определению, выполнены
уравнения
\begin{equation*}
  e^a(X)=0,\qquad \om_a{}^b(X)=0.
\end{equation*}
Поскольку 1-формы $\lbrace e^a,\om_b{}^c\rbrace$ образуют базис кореперов
на $\ML(\MM)$, то $X=0$. Это доказывает разложение (\ref{edishv}). Поскольку
проекция $\pi$ является гладким и сюрьективным отображением, то ее
дифференциал $\pi_*$ является сюрьективным гомоморфизмом. Поскольку
$\pi_*\big(\MV_p(\ML)\big)=0$, то отображение
$\pi_*:~\MH_p(\ML)\rightarrow\MT_x(\MM)$ является изоморфизмом.

Для доказательства свойства 2) необходимо проверить, что правое действие
группы преобразований на 1-формы $\om_a{}^b$ имеет вид
\begin{equation*}
  (r_S)^*\om_a{}^b=S_a{}^c\om_c{}^dS^{-1}_d{}^b.
\end{equation*}
Действительно, это следует из определения (\ref{eonfab}) и равенства
\begin{equation*}
  de^{\prime\al}{}_a=\frac{\pl e^{\prime\al}{}_a}{\pl e^\bt{}_b} de^\bt{}_b
  =S_a{}^bde^\al{}_b.
\end{equation*}
Поскольку распределение горизонтальных подпространств $\MH(\ML)$ является
аннигилятором подпространства $\om_a{}^b$, то из правила действия левых
преобразований следует свойство 2).
\end{proof}
Верно также обратное утверждение. Если в касательном расслоении $\MT(\ML)$
задано дифференцируемое $n$-мерное распределение горизонтальных подпространств
$\MH$, удовлетворяющих свойствам 1) и 2), то на $\MM$ существует аффинная
связность $\nb$ такая, что $\MH$ является распределением горизонтальных
подпространств расслоения реперов $\ML(\MM)$ для связности $\nb$. Поэтому, с
точки зрения расслоения реперов, задание аффинной связности эквивалентно заданию
распределения горизонтальных подпространств со свойствами 1), 2). Далее эти
свойства лягут в основу определения связности на главных расслоениях общего
вида.
\begin{com}
Поскольку аффинная связность $\nb$ зависит дифференцируемо от $p\in\ML(\MM)$, то
распределение горизонтальных подпространств $\MH_p(\ML)$ также дифференцируемо
зависит от точки $p$.
\qed\end{com}

В формулах (\ref{etoste})--(\ref{ebiacn}) все формы рассматривались на
пространстве расслоения $\ML(\MM)$. В частности компоненты тензоров кручения
$T_{bc}{}^a$ и кривизны $R_{cda}{}^b$ в структурных уравнениях (\ref{etoste}) и
(\ref{ecusre}) явно зависят от координат слоя $e^\al{}_a$. Эти формы можно
спустить на базу с помощью возврата отображения.
\begin{defn}
Рассмотрим сечение $\s:\MM\rightarrow\ML(\MM)$, которому в локальных координатах
соответствует репер $e^\al{}_a(x)$. Тогда дифференциал сечения $\s_*$ отображает
касательные векторы к базе в касательные векторы к сечению:
\begin{equation*}
  \CX(\MM)\ni\quad X=X^\al\pl_\al\mapsto\s_*X
  =X^\al\pl_\al+X^\bt\pl_\bt e^\al{}_a\frac{\pl}{\pl e^\al{}_a}\quad
  \in\CX\big(\s(\MM)\big).
\end{equation*}
Следовательно, возврат отображения $\s^*$ спускает форму связности для каждого
сечения на базу:
\begin{equation}                                                  \label{edecol}
  \s^*\om_a{}^b:=dx^\al\om_{\al a}{}^b
  =dx^\al e_\g{}^b(\pl_\al e^\g{}_a+e^\bt{}_a\Gamma_{\al\bt}{}^\g).
\end{equation}
Форма $dx^\al\om_{\al a}{}^b$ называется {\em локальной формой связности} на
расслоении реперов. Она получается из формы связности (\ref{eonfab}) простой
\index{Локальная форма связности (local connection form)}%
\index{Форма связности локальная (local connection form)}%
заменой: $e^\al{}_a\mapsto e^\al{}_a(x)$ и
$de^\al{}_a\mapsto dx^\bt\pl_\bt e^\al{}_a$. Аналогично спускаются
на базу 2-формы кручения и кривизны (\ref{etocuf}):
\begin{align}                                                     \label{etoext}
  T^a&=\frac12dx^\al\wedge dx^\bt\, T_{\al\bt}{}^a
  =de^a-e^b\wedge\om_b{}^a,
\\                                                                \label{ecuext}
  R_a{}^b&=\frac12dx^\al\wedge dx^\bt\, R_{\al\bt a}{}^b
  =d\om_a{}^b-\om_a{}^c\wedge\om_c{}^b,
\end{align}
где введены 1-формы локальной связности и корепера:
\begin{equation}                                                  \label{eomome}
  \om_a{}^b=dx^\al\om_{\al a}{}^b,\qquad e^a=dx^\al e_\al{}^a,
\end{equation}
и
\begin{align}                                                     \label{ecurcv}
  T_{\al\bt}{}^a&=\pl_\al e_\bt{}^a-e_\al{}^b\om_{\bt b}{}^a
                    -(\al\leftrightarrow\bt),
\\                                                                \label{ecucav}
  R_{\al\bt a}{}^b&=\pl_\al \om_{\bt a}{}^b-\om_{\al a}{}^c
                       \om_{\bt c}{}^b-(\al\leftrightarrow\bt).
\end{align}
2-формы (\ref{etoext}) и (\ref{ecuext}) называются {\em локальными формами}
кручения и кривизны.
\qed\end{defn}
\index{Локальная форма кручения (local torsion form)}%
\index{Форма кручения локальная (local torsion form)}%
\index{Локальная форма кривизны (local curvature form)}%
\index{Форма кривизны локальная (local curvature form)}%

В приложениях выражения (\ref{ecurcv}) и (\ref{ecucav}) часто называют тензорами
кручения и кривизны.

\begin{com}
Для локальных форм мы сохранили прежние обозначения, чтобы не вводить новых.
Фактически, в приложениях при проведении вычислений используются локальные
формы, заданные на базе $\MM$, а не на расслоении реперов $\ML(\MM)$.
\qed\end{com}
Тождества Бианки (\ref{ebiacn}) после спуска на базу имеют точно такой же вид,
только вместо компонент корепера $e_\al{}^a$ надо рассматривать сечение
$e_\al{}^a(x)$, вместо формы связности $\om_a{}^b$, определенную в
(\ref{eonfab}), -- локальную форму связности $dx^\al\om_{\al a}{}^b(x)$ и вместо
форм кручения и кривизны -- соответствующие им локальные формы.
\begin{defn}
Поля $e_\al{}^a(x)$ и $\om_{\al a}{}^b(x)$, заданные на $\MM$, в аффинной
геометрии называются {\em переменными Картана}.
\qed\end{defn}
\index{Переменные Картана (Cartan variables)}%
\index{Картана переменные (Cartan variables)}%
Переменные Картана часто используются в приложениях, т.к.\ позволяют упростить
вычисления. Формулы (\ref{ecurcv}) и (\ref{ecuext}) дают выражение для тензоров
кручения и кривизны в переменных Картана.
\section{Критерий локальной тривиальности                        \label{slojnf}}
Тензоры кривизны и кручения играют исключительно важную роль в аффинной
геометрии. Ниже мы покажем, что их обращение в нуль является критерием локальной
тривиальности связности и существования такой системы координат, в которой репер
совпадает с координатным базисом касательного пространства.

Рассмотрим расслоение реперов $\ML(\MM)$ с заданной связностью $\nabla$ и
координатную окрестность $\MU$ произвольной точки базы $x\in\MM$. Пусть
задано некоторое локальное сечение $\s:~\MM\supset\MU\rightarrow\ML$. Тогда в
области $\MU$ определен репер $e_\al{}^a(x)$ и компоненты локальной формы
связности $\om_{\al a}{}^b(x)$. Кроме того, определены тензоры кривизны
$R_{\al\bt a}{}^b$ и кручения $T_{\al\bt}{}^a$, которые заданы формулами
(\ref{ecurcv}) и (\ref{ecucav}).

При локальном вращении репера (\ref{elobac}) с матрицей $S(x)\in\MG\ML(n,\MR)$
компоненты линейной связности преобразуются по правилу (\ref{etrcoc}). Поставим
следующий вопрос: ``Существует ли такая матрица вращений $S$, что после
преобразования компоненты локальной формы связности $\om'_{\al a}{}^b$ обратятся
в нуль в некоторой окрестности точки $x$?''. Ответ на этот вопрос дает следующая
\begin{theorem}
Пусть на $\MM$ задана линейная связность. Тогда равенство нулю тензора кривизны
$R_{\al\bt a}{}^b=0$ в некоторой (односвязной) окрестности $\MU$ произвольной
точки $x\in\MM$ является необходимым и достаточным условием существования такой
матрицы $S(x)$, что после локального вращения репера (\ref{elobac}) компоненты
локальной формы связности обратятся в нуль в, возможно, меньшей окрестности
точки $x$.
\end{theorem}
\begin{proof}
Если после вращения $\om'_{\al a}{}^b=0$, то равенство (\ref{etrcoc}) после
умножения справа на $S$ приводит к уравнению на компоненты матрицы вращений:
\begin{equation}                                                  \label{qeahcb}
  \pl_\al S_a{}^b=-S_a{}^c\om_{\al c}{}^b.
\end{equation}
Для того, чтобы получить критерий локальной разрешимости этой системы уравнений,
ее нужно продифференцировать по $x^\bt$ и антисимметризовать по индексам $\al$ и
$\bt$. После исключения первых производных от матрицы $S$ в правой части с
помощью исходного уравнения, получим равенство
\begin{equation*}
  (\pl_\bt\pl_\al-\pl_\al\pl_\bt)S=S(\pl_\al\om_\bt-\pl_\bt\om_\al
  -\om_\al\om_\bt+\om_\bt\om_\al),
\end{equation*}
где мы, для краткости, опустили матричные индексы. Сравнивая это равенство с
определением компонент локальной формы кривизны (\ref{ecurcv}), находим искомый
критерий.
\end{proof}

На этом этапе можно забыть о существовании репера. Достаточно считать, что на
$\MU$ заданы только компоненты локальной формы связности $\om_{\al a}{}^b(x)$ с
соответствующим правилом преобразования. Про метрику можно вообще не вспоминать:
она просто отсутствует на многообразии $\MM$.

При нулевой кривизне зафиксируем матрицу вращений таким образом, что
$\om_{\al a}{}^b=0$. Тогда после локального поворота компоненты локальной формы
связности станут нетривиальными в соответствии с правилом преобразования
(\ref{etrcoc}). Это доказывает следующее утверждение.
\begin{cor}
Пусть тензор кривизны для некоторой линейной связности обращается в нуль в
некоторой (односвязной) области $\MU\subset\MM$. Тогда, возможно, в меньшей
окрестности существует такая матрица вращений $S(x)\in\MG\ML(n,\MR)$, что
компоненты связности представимы в следующем виде
\begin{equation}                                                  \label{elphfs}
  \om_{\al a}{}^b=\pl_\al S_a{}^c S^{-1}_{\quad c}{}^b. \qed
\end{equation}
\end{cor}
Такая связность называется чистой калибровкой и является плоской, т.к. ее тензор
кривизны тождественно равен нулю.

Теперь вернемся к вопросу, поставленному в начале раздела. Рассмотрим
односвязную область $\MU\subset\MM$ с координатами $x^\al$. Если в этой области
существует такая система координат $y^a(x)$, что соответствующий координатный
базис касательного пространства совпадает с заданным репером, то должно
выполняться равенсто
\begin{equation}                                                  \label{egnnmj}
  \pl_\al y^a=e_\al{}^a.
\end{equation}
Это -- система уравнений на функции перехода к новой системе координат $y^a(x)$.
Критерием ее разрешимости являются условия
\begin{equation}                                                  \label{qgqwes}
  \pl_\al e_\bt{}^a-\pl_\bt e_\al{}^a=0.
\end{equation}
Левая часть этого равенства совпадает с тензором кручения (\ref{ecucav}) при
нулевой связности $\om_{\al a}{}^b=0$. Поэтому справедлива
\begin{theorem}                                                   \label{tghbyr}
Пусть на $\MM$ заданы репер и линейная связность. Тогда одновременное обращение
в нуль тензоров кривизны и кручения в некоторой односвязной области
$\MU\subset\MM$ является необходимым и достаточным условием существования в,
возможно, меньшей области такой матрицы $S(x)\in\MG\ML(n,\MR)$ и системы
координат $y^a(x)$, что компоненты связности обратятся в нуль,
$\om_{\al a}{}^b=0$, и репер совпадет с координатным базисом касательного
пространства, $e_\al{}^a=\pl_\al y^a$.
\end{theorem}
\begin{com}
Подчеркнем, что на многообразии $\MM$ не предполагается наличие какой либо
метрики.
\end{com}
В дальнейшем мы покажем, что последняя теорема лежит в основе геометрической
теории дефектов в упругой среде.

Теперь предположим дополнительно, что на $\MM$ задана метрика с компонентами
$g_{\al\bt}$. Для определенности, предположим, что метрика положительно
определена, т.е. риманова. В такой ситуации в приложениях рассматривают, как
правило, не всю совокупность реперов, а их подмножество, которое определяется
уравнением
\begin{equation}                                                  \label{egdasx}
  g_{\al\bt}=e_\al{}^a e_\bt{}^b\dl_{ab},
\end{equation}
где $\dl_{ab}=\diag(+\dotsc+)$ -- евклидова метрика. Решения данного квадратного
уравнения относительно репера существуют и определены с точностью до локальных
$\MO(n)$ вращений. То есть, если репер $e_\al{}^a$ удовлетворяет уравнению
(\ref{egdasx}), то повернутый репер $e'_\al{}^a=e_\al{}^b S_b{}^a$, где
$S\in\MO(n)$ -- произвольная матрица ортогональных вращений из полной группы
$\MO(n)$, также удовлетворяет данному уравнению. Поскольку произвол в выборе
репера теперь сужен до подгруппы вращений $\MO(n)\subset\MG\ML(n,\MR)$, то
естественно рассматривать не полную линейную связность, а $\MS\MO(n)$-связность,
которая удовлетворяет условию
\begin{equation}                                                  \label{qplkib}
  \nb_\al\dl_{ab}=-\om_{\al a}{}^c\dl_{cb}-\om_{\al b}{}^c\dl_{ac}=0.
\end{equation}
То есть компоненты связности $\om_{\al a}{}^b$ принимают значения в алгебре
вращений $\Gs\Go(n)$. Эта связность является метрической, поскольку
ковариантная производная от метрики (\ref{qplkib}) равна нулю, а соответствующая
геометрия -- геометрией Римана--Картана.

Напомним, что алгебры Ли собственных вращений $\Gs\Go(n)$ и полной группы
вращений $\Go(n)$ изоморфны. Поэтому $\MS\MO(n)$- и $\MO(n)$-связности -- это
одно и то же.

Согласно теореме \ref{tghbyr} при нулевой кривизне и кручении существует такая
матрица вращений и система координат $y^a$, что будет выполнено равенство
\begin{equation*}
  g_{\al\bt}=\pl_\al y^a\pl_\bt y^b\dl_{ab}.
\end{equation*}
Это означает, что существует такая система координат в которой метрика является
локально евклидовой. Тем самым доказана
\begin{theorem}                                                   \label{tiolcv}
Пусть на многообразии $\MM$ задана геометрия Римана--Картана, т.е. в
произвольной области определены компоненты репера $e_\al{}^a$, метрики
$g_{\al\bt}=e_\al{}^a e_\bt{}^b\dl_{ab}$ и $\MS\MO(n)$-связности
$\om_{\al a}{}^b$. Если тензоры кривизны и кручения в некоторой (односвязной)
области $\MU\subset\MM$ обращаются в нуль, то, возможно, в меньшей области
существует такая система координат $y^a(x)$, в которой метрика является
евклидовой, $g_{ab}=\dl_{ab}$. При этом локальное лоренцево вращение можно
подобрать таким образом, что $e_\al{}^a=\dl_\al^a$ и $\om_{\al a}{}^b=0$.
\end{theorem}

Если выбрать произвольный репер для заданной метрики $g_{\al\bt}$, то условия
интегрируемости (\ref{qgqwes}) в общем случае выполнены не будут. Однако
равенство нулю тензора кривизны гарантирует существование такого вращения, что
после поворота репера условия интегрируемости окажутся выполнены. Этот
повернутый репер и определит функции перехода $y^a(x)$ к нужной системе
координат.

В настоящем разделе мы подчеркивали, что рассматриваются только односвязные
области. В дальнейшем мы увидим, что для неодносвязных областей приведенные
выше утверждения в общем случае не имеют места.

В римановой геометрии кручение равно нулю с самого начала. И, конечно, теорема
о приведении метрики к локально евклидову виду верна и в этом случае. Достаточно
потребовать равенства нулю только тензора кривизны.

В доказательстве теоремы \ref{tiolcv} сигнатура метрики никак не использовалась.
Поэтому сформулированное утверждение верно для метрик произвольной сигнатуры.
Например, в равенстве (\ref{egdasx}) евклидову метрику можно заменить на метрику
Лоренца $\eta_{ab}$. Тогда вместо $\MS\MO(n)$-связности следует рассматривать
лоренцеву $\MS\MO(1,n-1)$-связность, и все предыдущее построение остается в
силе.
\chapter{Аффинная геометрия. Локальное рассмотрение}
В современной дифференциальной геометрии принят бескоординатный язык. Это
оправдано тем, что все геометрические объекты, например, тензорные поля,
являются понятиями, которые не зависят от выбора системы координат. Этот язык
удобен для определений и формулировки утверждений. Однако, он абстрактен, и
требуется немалое время для его усвоения. Кроме того, при проведении вычислений,
особенно в моделях математический физики, используется координатный подход, в
котором геометрические объекты отождествляются с набором своих компонент. Этот
язык действительно необходим для проведения вычислений и является более
наглядным. В настоящем разделе мы дадим определения и опишем основные свойства
объектов аффинной геометрии в компонентах. По сути дела это означает, что мы
рассматриваем топологически тривиальное многообразие $\MM$, которое диффеоморфно
евклидову пространству $\MR^n$ и покрывается одной картой. Это не значит, что
координатный подход является менее строгим. Как только установлено, что
некоторый объект или утверждение не зависят от выбора системы координат в
$\MR^n$, то все сказанное сразу переносится на нетривиальные многообразия.
\section{Локальное определение аффинной связности}
Ввиду важности понятия аффинной связности покажем, как она вводится локально
без обращения к общей теории связностей в векторных расслоениях. Пусть на
многообразии $\MM$, $\dim\MM=n$ в некоторой карте заданы: функция
$f\in\CC^1(\MM)$, дифференцируемое векторное поле $X=X^\al\pl_\al\in\CX(\MM)$ и
дифференцируемая  1-форма $A=dx^\al A_\al\in\Lm_1(\MM)$. Пусть на этом
многообразии задана также аффинная связность $\nb$. Тогда в компонентах
ковариантные производные имеют вид:
\begin{align}                                                     \label{ecodes}
  \nb_\al f&=\pl_\al f
\\                                                                \label{ecodev}
  \nabla_\al X^\bt&=\pl_\al X^\bt+X^\g\Gamma_{\al\g}{}^\bt,
\\                                                                \label{ecodec}
  \nabla_\al A_\bt&=\pl_\al A_\bt-\Gamma_{\al\bt}{}^\g A_\g,
\end{align}
где $\Gamma_{\al\bt}{}^\g(x)$ -- компоненты локальной формы аффинной связности,
которые, мы предполагаем, являются достаточно гладкими функциями от координат
$x^\al$. То есть ковариантная производная для скалярного поля совпадает с
обычной частной производной, а для векторного поля и 1-формы появляются
дополнительные слагаемые, линейные по компонентам. Отметим, что дополнительные
слагаемые в (\ref{ecodev}) и (\ref{ecodec}) имеют разные знаки.

Покажем, что ковариантная производная от компонент векторного поля
преобразуется, как тензор. Рассмотрим преобразование координат (\ref{ecootr}).
Тогда компоненты векторного поля преобразуется по правилу (\ref{evectr}).
Дифференцируя соотношение (\ref{evectr}) по $x^{\bt'}$, получаем два слагаемых
\begin{equation}                                                  \label{eplvef}
  \pl_{\bt'}X^{\al'}=\pl_{\bt'}x^\bt\pl_\bt X^\al\pl_\al x^{\al'}+
  \pl_{\bt'}x^\bt X^\al\pl^2_{\bt\al} x^{\al'}.
\end{equation}
Первое слагаемое в правой части соответствует тензорному закону преобразования
для производной от векторного поля, в то время как второе слагаемое этот закон
нарушает.

Чтобы получить тензорный закон
преобразования в общем случае вводится понятие ковариантной производной, которая
содержит дополнительное слагаемое в (\ref{ecodev}). Если потребовать, чтобы
ковариантная производная после преобразования координат имела тот же вид с
некоторыми новыми компонентами $\Gamma_{\bt'\g'}{}^{\al'}$,
$$
  \pl_{\bt'}X^{\al'}+\Gamma_{\bt'\g'}{}^{\al'}X^{\g'}
  =\pl_{\bt'}x^\bt(\pl_\bt X^\al+\Gamma_{\bt\g}{}^\al X^\g)\pl_\al x^{\al'},
$$
то с учетом уравнения (\ref{eplvef}) получим следующий закон преобразования
компонент аффинной связности:
\begin{equation}                                                  \label{econtr}
  \Gamma_{\bt'\g'}{}^{\al'}
  =\pl_{\bt'} x^\bt \pl_{\g'} x^\g \Gamma_{\bt\g}{}^\al\pl_\al x^{\al'}
  -\pl_{\bt'} x^\bt\pl_{\g'} x^\g\pl^2_{\bt\g}x^{\al'},
\end{equation}
С учетом тождества
\begin{equation*}
  \pl_\al(\pl_\bt x^{\bt'}\pl_{\bt'} x^\g)=\pl_\al\dl^\bt_\g=0
  \quad \Leftrightarrow\quad
  \pl^2_{\al\bt}x^{\bt'}=-\pl_\al x^{\al'}\pl_\bt x^{\bt'}\pl^2_{\al'\bt'}x^\g
  \pl_\g x^{\bt'}
\end{equation*}
закон преобразования (\ref{econtr}) можно переписать в эквивалентной форме
(\ref{econts}), отличающейся вторым слагаемым.

Закон преобразования компонент аффинной связности отличается от тензорного
закона наличием неоднородных слагаемых в (\ref{econts}) и (\ref{econtr}),
которые содержат вторые производные от функций перехода. Если ограничить класс
допустимых преобразований координат аффинными (линейными неоднородными), то
аффинная связность будет преобразовываться как тензор.

Сами по себе компоненты аффинной связности не являются компонентами тензора,
однако они позволяют строить новые тензорные поля из заданных с помощью
ковариантного дифференцирования. Собственно, название ``ковариантное'' и
отражает это обстоятельство. Нетрудно проверить, что ковариантная производная
от 1-формы (\ref{ecodec}) является тензорным полем второго ранга типа $(0,2)$.

Ковариантное дифференцирование было продолжено на тензоры произвольного ранга
с помощью формул (\ref{etecon}). В компонентах это выглядит следующим образом.
Помимо обычной частной производной в ковариантную производную входят слагаемые
с компонентами аффинной связности со знаком плюс для каждого контравариантного
и минус для каждого ковариантного индекса:
\begin{equation*}
\begin{split}
  \nb_\al K_{\bt_1\dotsc\bt_s}{}^{\g_1\dotsc\g_r}
  =\pl_\al K_{\bt_1\dotsc\bt_s}{}^{\g_1\dotsc\g_r}
  &-\Gamma_{\al\bt_1}{}^\dl K_{\dl\bt_2\dotsc\bt_s}{}^{\g_1\dotsc\g_r}-\dotsc
  -\Gamma_{\al\bt_s}{}^\dl K_{\bt_1\dotsc\bt_{s-1}\dl}{}^{\g_1\dotsc\g_r}
\\
  &+K_{\bt_1\dotsc\bt_s}{}^{\dl\g_2\dotsc\g_r}\Gamma_{\al\dl}{}^{\g_1}+\dotsc
  +K_{\bt_1\dotsc\bt_s}{}^{\g_1\dotsc\g_{r-1}\dl}\Gamma_{\al\dl}{}^{\g_r}.
\end{split}
\end{equation*}
\begin{exa} Для тензора второго ранга типа $(1,1)$ ковариантная производная
имеет вид
\begin{equation}                                                  \label{ecodmt}
  \nb_\al K_\bt{}^\g=\pl_\al K_\bt{}^\g
  -\Gamma_{\al\bt}{}^\dl K_\dl{}^\g
  +\Gamma_{\al\dl}{}^\g K_\bt{}^\dl.
\end{equation}
В частном случае, ковариантная производная от символа Кронекера (\ref{ekrode})
тождественно равна нулю
$$
  \nb_\al\dl_\bt^\g=0,
$$
поскольку равна нулю частная производная $\pl_\al\dl_\bt^\g=0$, а
слагаемые со связностью сокращаются. То есть символ Кронекера ковариантно
постоянен (параллелен) относительно произвольной аффинной связности.
\qed\end{exa}
Можно показать с помощью прямых вычислений (если доказательство в общем виде,
которое было дано ранее, кого то не убеждает), что определенная таким образом
ковариантная производная от тензора произвольного типа $(r,s)$ действительно
дает тензор типа $(r,s+1)$.

Поскольку произвольный тензор типа $(r,s)$ является полилинейным отображением
(\ref{epoltm}), то вид ковариантной производной от тензорного поля в компонентах
однозначно определяется четырьмя условиями: видом ковариантной производной от
функции, векторного и ковекторного полей (\ref{ecodes})--(\ref{ecodec})
и правилом Лейбница (\ref{eleiru}).
\begin{exa}
Найдем вид ковариантной производной от тензорного поля второго ранга
$K_\al{}^\bt$. Ковариантная производная от функции
$K_\g{}^\bt X^\g A_\bt\in\CC^\infty(\MM)$, где $X^\g\in\CX(\MM)$ и
$A_\bt\in\Lm_1(\MM)$ -- произвольные векторное и ковекторное поля, совпадает с
обычной:
$$
  \nb_\al(K_\g{}^\bt X^\g A_\bt)=\pl_\al(K_\g{}^\bt X^\g A_\bt).
$$
Воспользовавшись правилом Лейбница и видом ковариантных производных
(\ref{ecodev}), (\ref{ecodec}), получим формулу для ковариантной
производной от тензора второго ранга (\ref{ecodmt}).
\qed\end{exa}

Ковариантная производная коммутирует с произвольным свертыванием:
\begin{equation*}
  \nb_\al(\dl^{\bt_j}_{\al_i}K_{\bt_1\cdots\bt_s}{}^{\al_1\cdots\al_r})=
  \dl^{\bt_j}_{\al_i}\nb_\al K_{\bt_1\cdots\bt_s}{}^{\al_1\cdots\al_r},
  \qquad 1\le i\le r,~1\le j\le s.
\end{equation*}
что следует из ковариантного постоянства символа Кронекера. В частном случае,
\begin{equation*}
  \pl_\al(X^\bt A_\bt)=\nb_\al(X^\bt A_\bt)
  =(\nb_\al X^\bt)A_\bt+X^\bt(\nb_\al A_\bt).
\end{equation*}
\begin{defn}
Произвольному векторному полю можно поставить в соответствие бесконечно малую
величину
\begin{equation}                                                  \label{ecodif}
  DX^\al=dx^\bt\nb_\bt X^\al
  =dx^\bt(\pl_\bt X^\al+\Gamma_{\bt\g}{}^\al X^\g),
\end{equation}
которая называется {\em ковариантным дифференциалом}. Аналогично определяется
ковариантный дифференциал для произвольного тензорного поля.
\qed\end{defn}
\index{Ковариантный дифференциал (covariant differential)}%
\index{Дифференциал ковариантный (covariant differential)}%

Для функции ковариантный дифференциал совпадает с обычным.
\begin{defn}
Рассмотрим произвольную дифференцируемую кривую $\g=\lbrace x^\al(t)\rbrace$
в римановом пространстве с положительно определенной метрикой. Тогда длина дуги
кривой отлична от нуля
\begin{equation}                                                  \label{eculei}
  ds=dt\sqrt{\dot x^2}\ne0,\qquad
  \text{где}\quad \dot x^2:=\dot x^\al\dot x^\bt g_{\al\bt}.
\end{equation}
Выражение
\begin{equation}                                                  \label{ecodcu}
  \frac{D X^\al}{ds}:=u^\bt\nb_\bt X^\al,
\end{equation}
где
\begin{equation}                                                  \label{euntav}
  u^\al:=\frac{\dot x^\al}{\sqrt{\dot x^2}},\qquad u^2=1,
\end{equation}
-- единичный касательный вектор к кривой, называется {\em ковариантной
производной вектора $X$ вдоль кривой $\g$.}
\qed\end{defn}
\index{Ковариантная производная вдоль кривой %
(covariant derivative along a curve)}%

Если метрика имеет лоренцеву сигнатуру, то ковариантную производную можно
определить либо вдоль времениподобных, либо вдоль пространственноподобных
кривых. В последнем случае $\dot x^2$ заменяется на $-\dot x^2$ под знаком
корня.

Аналогично определяется ковариантная производная вдоль кривой от произвольного
тензорного поля.

Пусть на многообразии задано векторное поле $X\in\CX(\MM)$. Тогда
{\em ковариантная производная вдоль векторного поля} $X$ от тензорного поля
$K_{\al_1\dotsc\al_r}{}^{\bt_1\dots\bt_s}$ типа $(r,s)$ дает тензорное поле
того же типа с компонентами
\index{Производная вдоль векторного поля (derivative along a vector field)}%
\begin{equation}                                                  \label{evefde}
  (\nb_X K)_{\al_1\dotsc\al_r}{}^{\bt_1\dots\bt_s}
  =X^\g\nb_\g K_{\al_1\dotsc\al_r}{}^{\bt_1\dots\bt_s}.
\end{equation}
\begin{exa}
В частности, для двух векторных полей справедливо равенство
\begin{equation}                                                  \label{edeavx}
  (\nb_XY)^\al=X^\bt\nb_\bt Y^\al. \qed
\end{equation}
\end{exa}
Рассмотренная выше ковариантная производная от тензора вдоль кривой является
частным случаем ковариантной производной вдоль векторного поля, которое касается
кривой. Ковариантная производная вдоль кривой определена только в тех точках
многообразия $\MM$, через которые проходит кривая.

Наличие неоднородного слагаемого в законе преобразования компонент аффинной
связности (\ref{econtr}) или (\ref{econts}) не позволяет их складывать как
тензорные поля и умножать даже на числа. Например, если на многообразии задано
две аффинные связности $\Gamma_{1\al\bt}{}^\g$ и $\Gamma_{2\al\bt}{}^\g$, то их сумма в
общем случае связностью не является. В то же время нетрудно проверить, что сумма
\begin{equation}                                                  \label{esucon}
  \Gamma_{\al\bt}{}^\g=f\Gamma_{1\al\bt}{}^\g+(1-f)\Gamma_{2\al\bt}{}^\g,
\end{equation}
где $f\in\CC^1(\MM)$, задает компоненты некоторой аффинной связности.

Разность двух аффинных связностей является тензорным полем типа $(1,2)$,
т.к.\ неоднородные слагаемые сокращаются.
\begin{defn}
{\em Вариацией} аффинной связности $\dl\Gamma_{\al\bt}{}^\g$ называется разность
двух связностей, заданных на многообразии $\MM$.
\qed\end{defn}
\index{Вариация аффинной связности (variation of an affine connection)}%
Вариация связности является тензорным полем типа $(1,2)$.
\section{Кручение и неметричность                                \label{stornm}}
Продолжим локальное изучение аффинной связности. В общем случае компоненты
связности $\Gamma_{\al\bt}{}^\g$ никакой симметрии по индексам не имеет и никак
не связаны с метрикой $g_{\al\bt}$, поскольку эти понятия определяют на
многообразии $\MM$ разные геометрические операции. А именно, метрика
многообразия определяет в каждой точке скалярное произведение векторов из
касательного пространства, а аффинная связность определяет ковариантное
дифференцирование и параллельный перенос тензоров. Геометрия на многообразии
$\MM$ определяется метрикой и аффинной связностью. Будем говорить, что на $\MM$
задана  {\em аффинная геометрия}, если заданы достаточно гладкие метрика и
аффинная связность, т.е.\ задано три объекта $(\MM,g,\Gamma)$.
\index{Аффинная геометрия (affine geometry)}%
\index{Геометрия аффинная (affine geometry)}%
\begin{com}
В общем случае метрика и аффинная связность задаются произвольным образом и
являются совершенно независимыми геометрическими объектами. Поэтому при
построении физических моделей их можно рассматривать как независимые поля,
имеющие разную физическую интерпретацию. В настоящее время принято считать, что
метрика описывает гравитационное взаимодействие. Физический смысл аффинной
связности пока неясен. Это связано с тем, что физическая интерпретация связности
зависит от конкретной модели. Соответствующие модели сложны с математической
точки зрения и в настоящее время изучены недостаточно хорошо.
\qed\end{com}
По-определению, кручение многообразия в локальной системе координат равно
антисимметричной части аффинной связности (\ref{etoind})
\begin{equation}                                                  \label{qfabcz}
  T_{\al\bt}{}^\g:=\Gamma_{\al\bt}{}^\g-\Gamma_{\bt\g}{}^\g.
\end{equation}
Из закона преобразования связности (\ref{econts}) следует, что кручение является
тензорным полем типа $(1,2)$.
\begin{defn}
Если на многообразии задана аффинная геометрия, то можно построить {\em тензор
неметричности} $Q_{\al\bt\g}$. Он равен ковариантной производной от метрики
\begin{equation}                                                  \label{enonme}
  -Q_{\al\bt\g}:=\nb_\al g_{\bt\g}=\pl_\al g_{\bt\g}
  -\Gamma_{\al\bt}{}^\dl g_{\dl\g}-\Gamma_{\al\g}{}^\dl g_{\bt\dl}. \qed
\end{equation}
\end{defn}
\index{Тензор неметричности (nonmetricity tensor)}%
\index{Неметричности тензор (nonmetricity tensor)}%

Тензор неметричности, по-построению, симметричен относительно перестановки двух
последних индексов
$$
  Q_{\al\bt\g}=Q_{\al\g\bt}.
$$
Заметим, что для определения неметричности необходимы оба объекта:
и метрика, и связность.

Таким образом по заданной метрике и аффинной связности построено два
тензорных поля: кручение и тензор неметричности. Докажем, что по
заданной метрике, кручению и тензору неметричности можно однозначно
восстановить соответствующую аффинную связность. Уравнение (\ref{enonme})
всегда можно решить относительно связности $\Gamma$. Действительно,
линейная комбинация
$$
  \nb_\al g_{\bt\g}+\nb_\bt g_{\g\al}-\nb_\g g_{\al\bt}
$$
приводит к следующему общему решению для аффинной связности
с опущенным верхним индексом
\begin{align}                                                      \nonumber
  \Gamma_{\al\bt\g}:=\Gamma_{\al\bt}{}^\dl g_{\dl\g}&=
  \frac12(\pl_\al g_{\bt\g}+\pl_\bt g_{\g\al}-\pl_\g g_{\al\bt})
  +\frac12(T_{\al\bt\g} - T_{\bt\g\al} + T_{\g\al\bt} )
\\                                                                \label{elicon}
  &+\frac12(Q_{\al\bt\g} + Q_{\bt\g\al} - Q_{\g\al\bt} ).
\end{align}
Правая часть этого равенства симметрична по индексам $\al$ и $\bt$
за исключением одного слагаемого, $\frac12T_{\al\bt\g}$, что согласуется
с определением тензора кручения (\ref{qfabcz}). Заметим, что слагаемые
с метрикой и тензором неметричности в (\ref{elicon}) имеют одинаковый
порядок индексов и знаков, а слагаемые с кручением при том же порядке
индексов отличаются знаками. Таким образом, для того, чтобы на
многообразии $\MM$ задать аффинную геометрию, необходимо и достаточно
задать три тензорных поля: метрику, кручение и неметричность. Подчеркнем
еще раз, что все три объекта можно задать совершенно независимым
образом, и в моделях математической физики их можно рассматривать как
независимые динамические переменные.
\begin{defn}
Второе слагаемое в (\ref{elicon})
\begin{equation}                                                  \label{ecotor}
  K_{\al\bt\g}:=\frac12(T_{\al\bt\g} - T_{\bt\g\al} + T_{\g\al\bt})
\end{equation}
называется тензором {\em кокручения}.
\qed\end{defn}
\index{Тензор кокручения (contorsion tensor)}%
\index{Кокручения тензор (contorsion tensor)}%

Заметим, что уравнение (\ref{enonme}) симметрично по индексам $\bt,\g$,
т.е.\ число алгебраических уравнений в точности равно числу компонент
симметричной части связности $\Gamma_{\bt\g}{}^\al$. Поэтому весь произвол
в решении этой системы уравнений определяется тензором кручения.

Вычислив ковариантную производную от тождества
$g^{\al\bt}g_{\bt\g}=\dl^\al_\g$, находим, что ковариантная производная
от обратной метрики,
\begin{equation}                                                  \label{ecodim}
  \nb_\al g^{\bt\g}=Q_\al{}^{\bt\g},\qquad
  Q_\al{}^{\bt\g}:=g^{\bt\dl}g^{\g\e}Q_{\al\dl\e},
\end{equation}
отличается знаком от (\ref{enonme}). Если тензор неметричности отличен
от нуля, то подъем и опускание индексов не коммутирует с ковариантным
дифференцированием.
\begin{exa} \qquad $\nb_\al X_\bt=(\nb_\al X^\g)g_{\g\bt}-Q_{\al\g\bt}X^\g$.
\qed\end{exa}

Для задания аффинной геометрии на многообразии необходимо задать метрику
и аффинную связность. Рассмотрим частные случаи аффинной геометрии.
\begin{defn}
При попытке объединить гравитацию с электромагнетизмом Г.~Вейль
рассмотрел тензор неметричности специального вида \cite{Weyl18AR}
\begin{equation}                                                  \label{ewevec}
  Q_{\al\bt\g}=W_\al g_{\bt\g},
\end{equation}
где $W_\al$ -- {\em форма Вейля},
\index{Форма Вейля (Weyl form)}\index{Вейля форма (Weyl form)}%
отождествленная с электромагнитным потенциалом, (при этом
предполагалось, что кручение тождественно равно нулю). Будем говорить,
что на многообразии задана {\em геометрия Римана--Картана--Вейля},
\index{Геометрия Римана--Картана--Вейля (Riemann--Cartan--Weyl geometry)}%
\index{Римана--Картана--Вейля геометрия (Riemann--Cartan--Weyl geometry)}%
если на нем задана метрика, кручение и неметричность специального
вида (\ref{ewevec}).

Если тензор неметричности тождественно равен нулю, а метрика и кручение
нетривиальны, то будем говорить, что на многообразии задана {\em геометрия
Римана--Картана}.
\index{Геометрия Римана--Картана (Riemann--Cartan geometry)}%
\index{Римана--Картана геометрия (Riemann--Cartan geometry)}%
В этом случае из уравнения (\ref{elicon}) следует, что аффинная
связность однозначно определяется метрикой и кручением:
\begin{equation}                                                  \label{emecon}
  \Gamma_{+\al\bt\g}=
  \frac12(\pl_\al g_{\bt\g}+\pl_\bt g_{\al\g}-\pl_\g g_{\al\bt})
  +\frac12(T_{\al\bt\g} - T_{\bt\g\al} + T_{\g\al\bt} ).
\end{equation}
Такую связность называют {\em метрической},
\index{Метрическая связность (metrical connection)}%
\index{Связность метрическая (metrical connection)}%
поскольку ковариантная производная от метрики тождественно равна нулю,
\begin{equation}                                                  \label{emetco}
  \nb_\al g_{\bt\g}=\pl_\al g_{\bt\g}-\Gamma_{\al\bt}{}^\dl g_{\dl\g}-
                       \Gamma_{\al\g}{}^\dl g_{\bt\dl} = 0.
\end{equation}
Это равенство называют {\em условием метричности},
\index{Условие метричности (metricity condition)}%
\index{Метричность (metricity)}%
и оно обеспечивает коммутативность ковариантного дифференцирования с
опусканием и подъемом индексов. Условие метричности (\ref{emetco})
эквивалентно условию $\nb_\al g^{\bt\g}=0$ в силу уравнения (\ref{ecodim}).

Если тензор кручения равен нулю, $T_{\al\bt}{}^\g=0$, а неметричность имеет вид
(\ref{ewevec}), то будем говорить, что задана геометрия {\em Римана--Вейля}. В
геометрии Римана--Вейля выражение для аффинной связности можно записать в
следующем виде:
\index{Геометрия Римана--Вейля (Riemann--Weyl geometry)}%
\index{Римана--Вейля геометрия (Riemann--Weyl geometry)}%
\begin{equation}                                                  \label{elicrw}
  \Gamma_{\al\bt\g}=\frac12\left[(\pl_\al+W_\al)g_{\bt\g}
  +(\pl_\bt+W_\bt)g_{\al\g}-(\pl_\g+W_\g)g_{\al\bt}\right].
\end{equation}
При этом она симметрична по двум первым индексам.

Если и тензор неметричности, и кручение тождественно равны нулю, а метрика
нетривиальна и положительно определена, то будем говорить, что на многообразии
задана {\em геометрия Римана}. В этом случае метрическая связность также
симметрична по двум первым индексам и однозначно определяется метрикой:
\index{Геометрия Римана (Riemannian geometry)}%
\index{Риманова геометрия (Riemannian geometry)}%
\begin{equation}                                                  \label{echris}
  \widetilde\Gamma_{\al\bt}{}^\g=
  \frac12g^{\g\dl}(\pl_\al g_{\bt\dl}+\pl_\bt g_{\al\dl}-\pl_\dl g_{\al\bt})
\end{equation}
Эта связность называется {\em связностью Леви--Чивиты}
\index{Связность Леви--Чивита (Levi--Civita connection)}%
\index{Леви--Чивита связность (Levi--Civita connection)}%
или {\em символами Кристоффеля}.
\index{Символы Кристоффеля (Christoffel's symbols)}%
\index{Кристоффеля символы (Christoffel's symbols)}%
Заметим, что в определении метрики мы требовали невырожденность матрицы
$g_{\al\bt}$.  Из выражения для символов Кристоффеля видно, что это условие
необходимо, в частности, для того, чтобы метрика определяла связность
Леви--Чивиты. Если метрика не является положительно определенной, то геометрия
называется {\em псевдоримановой}. Для связности при этом сохраняются прежние
названия.
\qed\end{defn}
\index{Псевдориманова геометрия (pseudo-Riemannian geometry)}%
\index{Геометрия псевдориманова (pseudo-Riemannian geometry)}%
\begin{com}
Если на многообразии задана аффинная геометрия общего вида $(\MM,g,\Gamma)$, то
на нем определены две связности: аффинная связность $\Gamma$ и связность
Леви--Чивиты $\widetilde\Gamma$, поскольку задана метрика. В такой ситуации над
связностью Леви--Чивиты и построенных с ее помощью геометрических объектах мы
будем писать знак тильды.
\qed\end{com}
\begin{exa}
Связность Леви--Чивита имеет наглядный геометрический смысл для двумерных
поверхностей, изометрически вложенных в трехмерное евклидово пространство.
Рассмотрим два вектора $X_x$ и $Y_y$, касательных к поверхности в близких точках
$x$ и $y$. Леви-Чивита \cite{LeviCi17A} предложил считать вектор $Y_y$
параллельным вектору $X_x$, если его проекция в евклидовом пространстве на
касательную плоскость в точке $x$ параллельна вектору $X_x$. Можно проверить,
что такая связность согласована с метрикой, индуцированной вложением, и имеет
нулевое кручение.
\qed\end{exa}

В римановой геометрии ковариантная производная определяется компонентами
связности с двумя нижними и одним верхним индексом,
$\widetilde\Gamma_{\al\bt}{}^\g:=\widetilde\Gamma_{\al\bt\dl}g^{\g\dl}$ и
поэтому одинакова для двух метрик, отличающихся постоянным множителем.

Символы Кристоффеля (\ref{echris}) симметричны по первым двум индексам. Это
свойство выполняется в голономном базисе. В неголономном базисе эта симметрия
нарушается (см.\ раздел \ref{sunhba}).

Из выражения для символов Кристоффеля (\ref{echris}) или условия
метричности (\ref{emetco}) следует равенство
\begin{equation}                                                  \label{ederga}
  \pl_\al g_{\bt\g}
  =\widetilde\Gamma_{\al\bt\g}+\widetilde\Gamma_{\al\g\bt}.
\end{equation}
Это доказывает следующее
\begin{prop}
Для того, чтобы символы Кристоффеля в некоторой системе координат были равны
нулю, необходимо и достаточно, чтобы в этой системе координат компоненты метрики
были постоянны.
\end{prop}
Поскольку символы Кристоффеля не являются компонентами тензора, то
в другой системе координат они могут быть нетривиальны.
\begin{exa}
Символы Кристоффеля для евклидова пространства в декартовой системе
координат равны нулю, но, скажем, в сферической или цилиндрической
системе координат они отличны от нуля (см.\ раздел \ref{scnhyu}).
\qed\end{exa}
\begin{defn}
В случае, когда тензор неметричности и кручение тождественно равны нулю, и при
этом в окрестности любой точки существует система координат, в которой метрика
является единичной диагональной матрицей, и, следовательно, символы Кристоффеля
равны нулю, геометрия называется {\em локально евклидовой} (или {\em
псевдоевклидовой}, если часть единиц входят в метрику с отрицательным знаком).
Соответствующая система координат называется {\em декартовой}.
\qed\end{defn}
\index{Локально евклидова геометрия (locally Euclidean metric)}%
\index{Геометрия локально евклидова (locally Euclidean metric)}%
\index{Псевдоевклидова геометрия (pseudo-Euclidean geometry)}%
\index{Декартовы координаты (Cartesian coordinates)}%
\index{Координаты декартовы (Cartesian coordinates)}%
\begin{exa}
Рассмотрим двумерный тор, вложенный в трехмерное евклидово пространство. Если
предположить, что кручение и неметричность на торе равны нулю, а метрика
индуцирована вложением, то тензор кривизны тора равен нулю. В этом случае
тор представляет собой нетривиальное локально евклидово многообразие.
\qed\end{exa}

Нетрудно посчитать число независимых компонент у связности,
тензора кручения и неметричности:
$$
  [\Gamma_{\al\bt}{}^\g]=n^3,\qquad
  [T_{\al\bt}{}^\g]=\frac{n^2(n-1)}2,\qquad
  [Q_{\al\bt\g}]=\frac{n^2(n+1)}2.
$$
Отсюда следует, что суммарное число независимых компонент кручения и
неметричности равно числу компонент аффинной связности.
\section{Ковариантная производная тензорных плотностей           \label{scoded}}
Ковариантная производная тензорного поля естественным образом
обобщается на тензорные плотности произвольной степени $p$. Для
определенности рассмотрим тензорную плотность $X_\al{}^\bt$ степени $p$
и типа $(1,1)$. По-определению, при преобразовании координат она
преобразуется по-правилу
$$
  X_{\al'}{}^{\bt'}=J^p \frac{dx^\al}{dx^{\al'}}X_\al{}^\bt
  \frac{dx^{\bt'}}{dx^\bt},
$$
где $J$ -- якобиан преобразования (\ref{ejacob}). Потребуем, чтобы
ковариантная производная от тензорной плотности была тензорной
плотностью той же степени,
\begin{equation}                                                  \label{etdcdt}
  \nb_{\al'}X_{\bt'}{}^{\g'}=J^p\frac{dx^\al}{dx^{\al'}}
  \frac{dx^\bt}{dx^{\bt'}}\nb_\al X_\bt{}^\g\frac{dx^{\g'}}{dx^\g}.
\end{equation}
Отсюда следует, что ковариантная производная от тензорной плотности
имеет вид
\begin{equation}                                                  \label{ecodtd}
  \nb_\al X_\bt{}^\g=\pl_\al X_\bt{}^\g
  -\Gamma_{\al\bt}{}^\dl X_\dl{}^\g
  +X_\bt{}^\dl\Gamma_{\al\dl}{}^\g+p\Gamma_\al X_\bt{}^\g,
\end{equation}
где $\Gamma_\al:=\Gamma_{\al\bt}{}^\bt$ -- след аффинной связности. Сравнивая эту
ковариантную производную с ковариантной производной (\ref{ecodmt}) от
тензорного поля типа $(1,1)$, находим, что различие состоит в появлении
дополнительного слагаемого, пропорционального степени тензорной плотности и
следу аффинной связности. Чтобы проверить выполнение закона преобразования
(\ref{etdcdt}) заметим, что в законе преобразования следа аффинной связности
(\ref{etrltc}) содержится неоднородное слагаемое, которое компенсирует
производную от якобиана
$$
  \pl_\al J^p=pJ^p\frac{dx^\bt}{dx^{\bt'}}
  \pl_\al\left(\frac{dx^{\bt'}}{dx^\bt}\right).
$$
Отсюда следует, что ковариантная производная от тензорной плотности
произвольного типа отличается от ковариантной производной соответствующего
тензора одним дополнительным слагаемым, пропорциональным степени тензорной
плотности и следу аффинной связности.
\begin{defn}
Параллельный перенос для тензорных плотностей определяется так же, как и для
тензоров. Мы говорим, что тензорная плотность {\em параллельно переносится}
вдоль заданной кривой, если ее ковариантная производная вдоль этой кривой
равна нулю.
\qed\end{defn}
\index{Параллельный перенос (parallel transport)}%
\index{Перенос параллельный (parallel transport)}%

Ковариантная производная от линейной комбинации тензорных плотностей $X$ и $Y$
одинакового ранга и степени равна линейной комбинации ковариантных производных
\begin{equation*}
  \nb_\al(aX+bY)=a\nb_\al X+b\nb_\al Y,\qquad a,b\in\MR.
\end{equation*}

Естественным образом определяется тензорное произведение $X\otimes Y$ тензорных
плотностей произвольных рангов и степеней. При этом степени тензорных плотностей
складываются. Нетрудно проверить, что для ковариантного дифференцирования
справедливо правило Лейбница
\begin{equation*}
  \nb_\al(X\otimes Y)=(\nb_\al X)\otimes Y+X\otimes(\nb_\al Y).
\end{equation*}
Так же, как и для тензоров, свертка по индексам тензорных плотностей
перестановочна с операцией ковариантного дифференцирования.
\begin{exa}
Рассмотрим частные случаи. Определители метрики $g$ и элемент объема 
$\vol=\det e_\al{}^a$ являются тензорными плотностями степеней $-2$ и $-1$,
соответственно. Используя выражение для следа аффинной связности (\ref{etrlio}),
получаем следующие равенства:
\begin{align}                                                     \label{ecoddm}
  \nb_\al g&=\pl_\al g-2\Gamma_\al g=-Q_\al\vol,
\\                                                                \label{ecoddr}
  \nb_\al \vol&=\pl_\al\vol-\Gamma_\al\vol=-\frac12Q_\al\vol,
\end{align}
где $Q_\al:=Q_{\al\bt}{}^\bt$ -- след тензора неметричности.
Отсюда следует, что для метрической связности определители метрики и репера
ковариантно постоянны. То есть в геометрии Римана--Картана и, в частности, в
римановой геометрии ковариантные производные от определителя метрики и элемента
объема равны нулю:
\begin{equation*}
  \widetilde\nb_\al g=0,\qquad \widetilde\nb_\al\vol=0.
\end{equation*}
При этом условие метричности для связности является достаточным, но не
необходимым. Из выражения (\ref{ecoddr}) следует, что для сохранения
элемента объема при параллельном переносе вдоль произвольной кривой,
необходимо и достаточно, чтобы след тензора неметричности был равен нулю.
В геометрии Римана--Картана--Вейля справедлива формула
$$
  \nb_\al\vol=-\frac n2W_\al\vol,
$$
где $n$ -- размерность многообразия. То есть форма Вейля в этом случае
определяет изменение элемента объема при параллельном переносе.
\qed\end{exa}
\section{Параллельный перенос                                    \label{spartx}}
В разделе \ref{scovec} был определен параллельный перенос векторов. С помощью
аффинной связности на многообразии $\MM$ можно определить {\em параллельный
перенос} касательных векторов, а также тензоров произвольного ранга вдоль
кривой.
\index{Параллельный перенос (parallel transport)}%
\index{Перенос параллельный (parallel transport)}%
\begin{defn}
Пусть дифференцируемая кривая $\g=x(t)=\lbrace x^\al(t)\rbrace$, $t\in[0,1]$
соединяет две точки многообразия $p,q\in\MM$: $x(0)=p$, $x(1)=q$. Касательный
вектор к кривой (вектор скорости) имеет компоненты $u^\al:=\dot x^\al$ и
предполагается отличным от нуля. Инвариантное условие параллельного переноса
(\ref{eparge}),
\begin{equation}                                                  \label{epartc}
  \nb_u X=0,
\end{equation}
в компонентах (\ref{epascr}) принимает вид
\begin{equation}                                                  \label{ediept}
  \dot X^\al=-\dot x^\bt X^\g\Gamma_{\bt\g}{}^\al.
\end{equation}
Это -- система линейных дифференциальных уравнений на компоненты векторного поля
$X^\al(t)$, которая решается с некоторым начальным условием $X^\al(0)=X^\al_0$.
Решение этой задачи мы называем {\em параллельным переносом} вектора $X_0$ в
точку $x(t)$ вдоль кривой $\g$. Результат параллельного переноса вектора вдоль
кривой однозначен в силу единственности решения системы обыкновенных
дифференциальных уравнений с (\ref{ediept}) с заданными начальными условиями.
\qed\end{defn}

Условие параллельного переноса (\ref{ediept}) можно переписать в интегральной
форме
\begin{equation}                                                  \label{eparin}
  X^\al(t)=X^\al_0-\int_0^t\!\!\! ds\,\dot x^\bt X^\g\Gamma_{\bt\g}{}^\al,
\end{equation}
где все функции в подынтегральном выражении рассматриваются, как функции от
параметра $s$ вдоль кривой и точка обозначает дифференцирование по $s$. При
$t=1$ получим компоненты вектора в точке $q$.

Результат параллельного переноса вектора из точки $p$ в точку $q$ не зависит ни
от параметризации кривой, что очевидно, ни от выбора системы координат.
Действительно, в новой штрихованной системе координат справедливо равенство
\begin{equation*}
  X^{\al'}(t)
  =X^{\al'}_0-\int_0^t\!\!\! ds\,\dot x^{\bt'} X^{\g'}\Gamma_{\bt'\g'}{}^{\al'}.
\end{equation*}
Это интегральное уравнение связано с уравнением (\ref{eparin}) преобразованием
координат. Чтобы доказать независимость параллельного переноса, представим
неоднородное слагаемое в преобразовании компонент аффинной связности
(\ref{econtr}), умноженное на $\dot x^{\bt'}X^{\g'}$, в виде
\begin{equation*}
  \dot x^\bt X^\g\pl^2_{\bt\g}x^{\al'}=X^\g\pl_t(\pl_\g x^{\al'})
\end{equation*}
и проинтегрируем по частям.

В то же время результат параллельного переноса вектора из точки $p$ в точку $q$
в общем случае зависит от кривой $\g$, соединяющей эти точки.

Параллельный перенос вектора вдоль кривой можно естественным образом обобщить
на кусочно гладкие кривые, как последовательный параллельный перенос от одного
излома к другому.

Аналогично определяется параллельный перенос 1-форм и тензоров произвольного
ранга.

При параллельном переносе вектора из точки $x$ в близкую точку $x+dx$, где
инфинитезимальное приращение $dx^\al$ мы рассматриваем, как касательный вектор
к кривой в точке $x$, компоненты вектора получат приращение
\begin{equation}                                                  \label{epatrv}
  \dl X^\al=-dx^\bt X^\g\Gamma_{\bt\g}{}^\al.
\end{equation}
Эти приращения линейны по дифференциалам $dx^\al$ и самому векторному полю.
Формула (\ref{epatrv}) позволяет дать следующую интерпретацию ковариантной
производной (\ref{ecodev}). Пусть на многообразии задано произвольное
векторное поле $X(x)$. Чтобы получить ковариантную производную от него
в точке $x$, необходимо взять значение векторного поля в точке $x+dx$,
вычесть из него результат параллельного переноса вектора из точки $x$ в
точку $x+dx$ и разделить эту разность на $dx$.

\begin{exa}
При параллельном переносе скаляра $f$ получаем постоянное скалярное поле
вдоль кривой $\dl f=0$. Результат параллельного переноса скаляра для
любой аффинной связности не зависит от пути переноса
\qed\end{exa}
\begin{exa}
При параллельном переносе 1-форма получает приращение
\begin{equation}                                                  \label{eonfpa}
  \dl A_\al=dx^\bt\Gamma_{\bt\al}{}^\g A_\g,
\end{equation}
которое отличается знаком от приращения вектора (\ref{epatrv}).
\qed\end{exa}
\begin{exa}
При параллельном переносе тензора типа $(1,1)$ он получает приращение
\begin{equation*}
  \dl T^\al{}_\bt=dx^\g(-\Gamma_{\g\dl}{}^\al T^\dl{}_\bt
  +\Gamma_{\g\bt}{}^\dl T^\al{}_\dl).
\end{equation*}
Аналогично для тензора произвольного ранга в правой части будем иметь
по одному слагаемому со знаком минус и плюс для каждого контра- и
ковариантного индекса соответственно.
\qed\end{exa}
\begin{exa}
В геометрии Римана--Картана, по-определению, ковариантная производная от метрики
равна нулю. Поэтому можно считать, что метрика получается в результате
параллельного переноса симметричной невырожденной матрицы из произвольной точки
многообразия. При этом результат параллельного переноса заданной матрицы не
зависит от пути для любой метрической связности. Это будет ясно из дальнейшего
рассмотрения.
\qed\end{exa}

Пусть на многообразии $\MM$ помимо аффинной связности задана метрика.
Рассмотрим, как меняется скалярное произведение $(X,Y)$ двух векторов при
параллельном переносе вдоль произвольной кривой $\g$.
\begin{prop}                                                      \label{pscver}
Зависимость скалярного произведения $(X,Y)$ двух векторов, которые параллельно
переносятся вдоль $\g$, от точки кривой определяется только тензором
неметричности:
\begin{equation}                                                  \label{eparsx}
  \pl_u(X,Y)=\nb_u(X^\al Y^\bt g_{\al\bt})=-u^\g X^\al Y^\bt Q_{\g\al\bt}.
\end{equation}
Отсюда следует, что в геометрии Римана--Картана ($Q=0$) скалярное произведение
двух векторов при параллельном переносе вдоль произвольной кривой сохраняется.
В частности, квадрат вектора скорости $u^2$ кривой $\g$ в геометрии
Римана--Картана постоянен вдоль нее.
\end{prop}
\begin{proof}
Следует из определения тензора неметричности (\ref{enonme}).
\end{proof}
\begin{cor}
В римановой геометрии и геометрии Римана--Картана длины векторов и углы между
ними сохраняются при параллельном переносе вдоль произвольной кривой $\g$.
\end{cor}

Задание аффинной связности позволяет сравнивать компоненты тензоров в бесконечно
близких точках, причем делать это ковариантным образом. Трудность сравнения
тензоров в различных точках связана с тем, что при преобразовании координат
тензоры в разных точках преобразуются по-разному, и их сравнение (сложение
компонент) теряет всякий смысл.

Ниже мы дадим геометрическую интерпретацию тензора кривизны, и с этой целью
выполним следующее построение.

\begin{figure}[h,b,t]
\hfill\includegraphics[width=.3\textwidth]{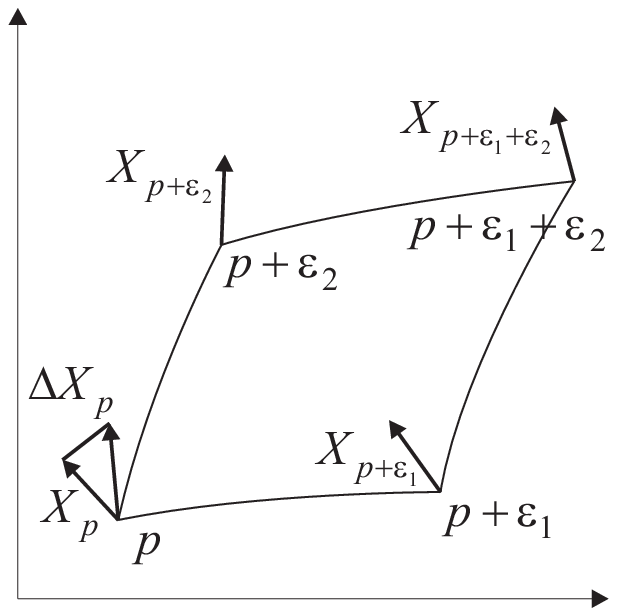}
\hfill {}
\centering\caption{Геометрический смысл кривизны: при параллельном переносе
вектора $X_p$ вдоль замкнутого контура он получает приращение $\triangle X_p$.}
\label{fgeomecur}
\end{figure}

Выберем точку многообразия $p=\lbrace p^\al\rbrace\in\MM$ и замкнутую кривую
$\g_p$ с началом и концом в точке $p$. Возьмем произвольный вектор $X_p$ в
точке $p$ и перенесем его параллельно вдоль кривой $\g_p$. В результате получим
новый вектор $X_p+\triangle X_p$, который может отличаться от исходного.
Рассмотрим класс замкнутых кривых $\g_p=\lbrace x^\al(t)\rbrace$ малой
``длины'', для которых выполнено неравенство
\begin{equation}                                                  \label{edecur}
  \oint_{\g_p}\!\!\!dt\sqrt{(\dot x^1)^2+\dotsc+(\dot x^n)^2}<\e\ll1.
\end{equation}
Это выражение нековариантно, т.е.\ мы работаем в фиксированной системе координат
некоторой координатной окрестности. Тогда все рассматриваемые кривые заведомо
лежат в малой окрестности точки $p\in\MU_{p,\e}\subset\MM$, координаты точек
которой удовлетворяют неравенствам:
\begin{equation*}
  \MU_{p,\e}:=\lbrace x\in\MM:\quad |x^\al-p^\al|<\e,~\al=1,\dotsc,n\rbrace.
\end{equation*}
Эта окрестность $\MU_{p,\e}$ диффеоморфна шару и поэтому односвязна. Рассмотрим
двумерную поверхность $\MS_\g\subset\MM$ в многообразии $\MM$, с границей $\g_p$
и целиком лежащую в $\MU_{p,\e}$, которая задана параметрически:
$\lbrace x^\al(u,v)\rbrace$, где $u,v$ -- координаты на поверхности. Тогда
справедлива
\begin{theorem}[\bf Геометрический смысл кривизны]
Пусть на многообразии $\MM$ в некоторой координатной окрестности задана аффинная
связность $\Gamma_{\al\bt}{}^\g$ с тензором кривизны $R_{\al\bt\g}{}^\dl$. Перенесем
вектор $X_p$ вдоль замкнутой кривой $\g_p$, для которой выполнено неравенство
(\ref{edecur}). Тогда компоненты вектора получат приращение порядка не выше
$\e^2$:
\begin{equation}                                                  \label{eparal}
  \triangle X^\dl_p=-X^\g_p R_{\al\bt\g}{}^\dl(p)\int_{\MS_\g}\!\!\!
  du\wedge dv\frac{\pl x^\al}{\pl u}\frac{\pl x^\bt}{\pl v}+\osmall(\e^2),
\end{equation}
где $\MS_\g\subset\MU_{p,\e}$ -- поверхность, ограниченная кривой $\g_p$.
\end{theorem}
\index{Геометрический смысл кривизны (geometrical meaning of curvature)}%
\begin{proof}
При параллельном переносе вектора вдоль замкнутого пути приращение его
компонент задается интегралом (\ref{eparin})
\begin{equation}                                                  \label{eittrv}
  \triangle X^\al_p=-\oint_{\g_p}\!\!\!dt\,\dot x^\bt X^\g_\parallel
  \Gamma_{\bt\g}{}^\al
  =-\oint_{\g_p}\!\!\!dx^\bt X^\g_\parallel\Gamma_{\bt\g}{}^\al,
\end{equation}
где $X^\g_\parallel(t)$ -- результат параллельного переноса вектора $X_p$ вдоль
кривой $\g$ до точки $x(t)$. Разложим подынтегральную функцию в ряд Тейлора
\begin{equation*}
  X^\g_\parallel\Gamma_{\bt\g}{}^\al=(X^\g_\parallel\Gamma_{\bt\g}{}^\al)_p
  +\big[\pl_\dl(X^\g_\parallel\Gamma_{\bt\g}{}^\al)\big]_p(x^\dl-p^\dl)+\dotsc
\end{equation*}
Интеграл от первого слагаемого равен нулю, т.к.\ кривая замкнута, и, поэтому
порядок приращения вектора при параллельном переносе не может превышать $\e^2$.
Это значит, что при вычислении интеграла в (\ref{eittrv}) с точностью $\e^2$ в
подынтегральном выражении достаточно учитывать члены порядка $\e$.

Разнесем вектор $X_p$ из точки $p$ на всю окрестность $\MU_{p,\e}$ с помощью
следующего соотношения
\begin{equation}                                                  \label{enevef}
  X^\al(x):=X^\al_p-(x^\bt-p^\bt)X^\g_p\Gamma_{\bt\g}{}^\al(p).
\end{equation}
Это -- параллельный перенос вектора $X_p$ на окрестность $\MU_{p,\e}$ с
точностью $\e$. С указанной точностью параллельный перенос вектора $X_p$ не
зависит от пути, так как приращение компонент вектора по замкнутому пути
порядка $\e^2$. В результате мы получаем гладкое векторное поле $X(x)$ на
$\MU_{p,\e}$. Поэтому в подынтегральном выражении (\ref{eittrv}) можно заменить
векторное поле вдоль кривой $X^\al_\parallel$, которое в общем случае не
является непрерывным, на гладкое векторное поле $X$. Тогда можно воспользоваться
формулой Грина
(\ref{egreef}):
\begin{equation*}
  \oint_{\g_p}\!\!\!dt\,\dot x^\bt X^\g\Gamma_{\bt\g}{}^\al
  =\int_{\MS_\g}\!\!\!du\wedge dv\left[
  \frac\pl{\pl u}\left(\frac{\pl x^\bt}{\pl v}X^\g\Gamma_{\bt\g}{}^\al\right)
-\frac\pl{\pl v}\left(\frac{\pl x^\bt}{\pl u}X^\g\Gamma_{\bt\g}{}^\al\right)\right].
\end{equation*}
Для векторного поля (\ref{enevef}) выполнено равенство
\begin{equation*}
  \frac{\pl X^\g}{\pl u}=-\frac{\pl x^\dl}{\pl u}X^\bt_p\Gamma_{\dl\bt}{}^\g(p)
\end{equation*}
и аналогичную формулу для частной производной по $v$. В результате с точностью
$\e^2$ получаем формулу (\ref{eparal}).
\end{proof}
\begin{com}
С точностью $\e^2$ приращение компонент вектора (\ref{eparal}) ковариантно.
По построению, это приращение не зависит от выбора поверхности
$\MS_\g\subset\MU_{p,\e}$, натянутой на контур $\g_p$.
\qed\end{com}
\begin{exa}                                                       \label{exapar}
Рассмотрим ``параллелограмм'' $\g_p$ со сторонами $\e^\al_1,\e^\al_2<\e$ и
вершинами в точках $\lbrace p^\al\rbrace$, $\lbrace p^\al+\e^\al_1\rbrace$,
$\lbrace p^\al+\e^\al_1+\e^\al_2\rbrace$ и $\lbrace p^\al+\e^\al_2\rbrace$.
Пусть на него натянута поверхность
$\MS_\g=\lbrace p^\al+\e^\al_1u+\e^\al_2v\rbrace$; $u,v\in[0,1]$. Тогда при
обходе по ``параллелограмму'' вектор получит приращение
\begin{equation*}
  \triangle X^\dl_p=-\e^\al_1\e^\bt_2X^\g_p R_{\al\bt\g}{}^\dl(p),
\end{equation*}
что следует из (\ref{eparal}). См.\ рис.\ref{fgeomecur}.
\qed\end{exa}

\begin{com}
Теорема о геометрическом смысле кривизны не зависит от того задана на
многообразии $\MM$ метрика или нет, т.к.\ параллельный перенос определяется
только связностью.
\qed\end{com}
Пусть на многообразии $\MM$ задана также метрика $g_{\al\bt}$. Тогда на
поверхности $\MS_\g$ индуцируется метрика
$g_{ij}=\pl_i x^\al\pl_j x^\bt$ ($i,j=1,2$, где $\lbrace u^i\rbrace=(u,v)$).
Следовательно, определена форма объема (площади поверхности) (\ref{evolel})
\begin{equation*}
  \upsilon=\frac12du^i\wedge du^j\ve_{ij}.
\end{equation*}
Тогда интеграл в (\ref{eparal}) можно переписать в виде
\begin{equation*}
  \frac12\int_{\MS_\g}\!\!\!du^i\wedge du^j\,\frac{\pl x^\al}{\pl u^i}
  \frac{\pl x^\bt}{\pl u^j}=\frac12\int_{\MS_\g}\!\!\!dudv\,\vol\ve^{ij}
  \frac{\pl x^\al}{\pl u^i}\frac{\pl x^\bt}{\pl u^j}\sgn,
\end{equation*}
где $\sgn:=\sign(\det g_{ij})$ и $\vol=\sqrt{|\det g_{ij}|}$. С точностью $\e^2$
подынтегральное выражение можно вынести из под знака интегрирования и
формула для приращения компонент вектора при параллельном переносе вдоль
малого замкнутого пути (\ref{eparal}) принимает вид
\begin{equation}                                                  \label{eajhpu}
  \triangle X^\dl_p=-\frac1{\sqrt{|g_p|}}
  \left(\frac{\pl x^\al}{\pl u}\frac{\pl x^\bt}{\pl v}\right)_p
  X^\g_p R_{\al\bt\g}{}^\dl(p)S,
\end{equation}
где
\begin{equation*}
  S:=\int_{\MS_\g}\!\!\!dudv\,\vol
\end{equation*}
-- площадь поверхности $\MS_\g$, которая имеет порядок $\e^2$. Полученная
формула ковариантна относительно преобразований координат на многообразии
$x^\al$ и инвариантна относительно выбора координат $u,v$ на поверхности.
\begin{com}
Геометрический смысл тензора кривизны лег в основу его физической интерпретации,
как поверхностной плотности вектора Франка, характеризующего распределение
дисклинаций в упругой среде со спиновой структурой (см.\ \cite{Katana05R}).
\qed\end{com}
Формулы переноса вектора вдоль малого замкнутого контура без труда обобщаются на
произвольные тензорные поля.
\begin{exa}
Рассмотрим такой же ``параллелограмм'' со сторонами $\e_1$ и $\e_2$ как и в
примере \ref{exapar}. При параллельном переносе вдоль него компоненты тензора
$T$ типа $(1,1)$ получает приращение
\begin{equation*}
  \triangle T_p{}^\al{}_\bt=-\e_1^\g\e_2^\dl\big(R_{\g\dl\e}{}^\al(p)
  T_p{}^\e{}_\bt-R_{\g\dl\bt}{}^\e(p) T_p{}^\al{}_\e\big).
\end{equation*}
Для тензора произвольного ранга в правой части равенства будем иметь по одному
слагаемому со знаками плюс и минус соответственно для каждого контравариантного
и ковариантного индекса.
\qed\end{exa}
\section{Геометрический смысл кручения                           \label{sgeotc}}
\begin{figure}[h,b,t]
\hfill\includegraphics[width=.3\textwidth]{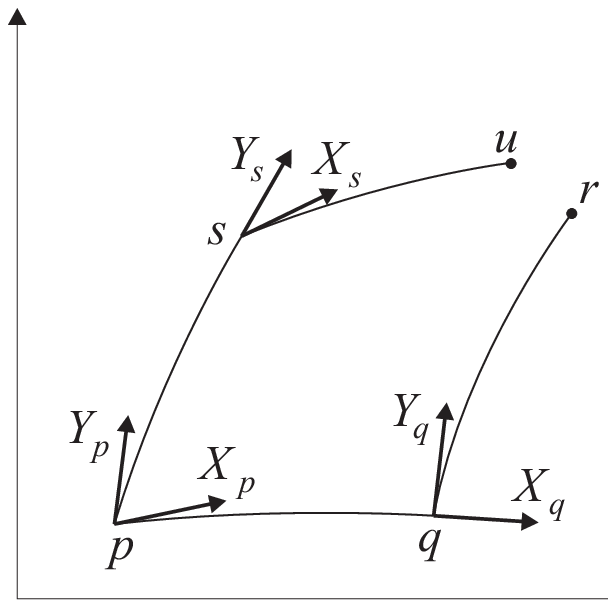}
\hfill {}
\centering\caption{Геометрический смысл кручения: ``параллелограмм'' с
  направляющими $X_p\e$ и $Y_p\e$ разомкнут.}
\label{fgemto}
\end{figure}

Рассмотрим многообразие $\MM$ с заданной аффинной связностью. Пусть
$\MU\subset\MM$ -- координатная окрестность и $\Gamma_{\al\bt}{}^\g$ -- компоненты
связности. Пусть в произвольной точке $p\in\MU$ заданы два касательных вектора
$X_p=\lbrace X^\al_p\rbrace$ и $Y_p=\lbrace Y^\al_p\rbrace$,
рис.~\ref{fgemto}.
Чтобы дать геометрическую интерпретацию тензору кручения, произведем следующее
построение. Перенесем вектор $Y_p$ вдоль геодезической линии, которая касается
вектора $X_p$, в близкую точку $q$, соответствующую параметру $t_q=t_p+\e$,
$\e\ll0$. Затем проведем геодезическую в точке $q$, которая касается вектора
$Y_q$, и отметим на ней точку $r$, соответствующую параметру $t_q+\e$.
Затем проведем то же построение, но в обратном порядке: перенесем вектор
$X_p$ вдоль геодезической для $Y_p$ в точку $s$, соответствующую параметру
$t_p+\e$, выпустим из нее геодезическую вдоль $X_s$ и отметим точку $u$,
соответствующую параметру $t_s+\e$. Если кручение аффинной связности равно нулю,
то точки $r$ и $u$ совпадут. При ненулевом кручении разность координат точек $r$
и $u$ имеет порядок $\e^2$ и, как будет показано, определяется тензором
кручения.

Начнем построение. Рассмотрим геодезическую линию $\lbrace x^\al(t)\rbrace$ в
$\MM$, которая проходит через $p$ и касается вектора $X_p$. Перенесем вектор
$X_p$ вдоль геодезической. В результата получим векторное поле $X(t)$ на
геодезической, компоненты которого, по-построению, удовлетворяет уравнению
\begin{equation}                                                  \label{egevee}
  \nb_{\dot x}X^\al=\frac{DX^\al}{dt}=\dot x^\bt(\pl_\bt X^\al
  +X^\g\Gamma_{\bt\g}{}^\al)=\dot X^\al+\dot x^\bt X^\g\Gamma_{\bt\g}{}^\al=0,
\end{equation}
где $DX$ -- ковариантный дифференциал векторного поля (\ref{ecodif}) и $t$ --
произвольный параметр вдоль геодезической. Тогда геодезическая будет
интегральной кривой полученного векторного поля, проходящей через точку $p$.
\begin{com}
Параметр $t$ может не совпадать с каноническим параметром вдоль геодезической,
а векторное поле $X(t)$ -- с полем скорости, соответствующим каноническому
параметру.
\qed\end{com}
Геодезическая удовлетворят уравнению
\begin{equation*}
  \ddot x^\al=-\Gamma_{\bt\g}{}^\al\dot x^\bt\dot x^\g
\end{equation*}
 с начальными условиями:
\begin{equation}                                                  \label{eperal}
  \dot x^\al|_{t=t_p}=X^\al,\qquad x^\al|_{t=t_p}=x^\al_p.
\end{equation}
Для наших целей вычисления достаточно провести во втором порядке по $\e$. Ниже
мы будем писать знак равенства, понимая его с точностью $\osmall(\e^2)$. В
этом порядке координаты точки $q$ равны
\begin{equation*}
  x^\al_q=x^\al_p+X^\al_p\e+\frac12\dot X^\al_p\e^2
  =x^\al_p+X^\al_p\e-\frac12X^\bt_p X^\g_p\Gamma_{p\,\bt\g}{}^\al\e^2,
\end{equation*}
где мы учли уравнения (\ref{egevee}) и (\ref{eperal}). Теперь параллельно
перенесем вектор $Y_p$ в точку $q$:
\begin{equation*}
  Y^\al_q=Y^\al_p-(x^\bt_q-x^\Sb_p)Y^\g_p\Gamma_{p\bt\g}{}^\al
  =Y^\al_p-X^\bt_p Y^\g_p\Gamma_{p\,\bt\g}{}^\al\e,
\end{equation*}
где достаточно учесть только линейный член разложения. Теперь найдем координаты
точки $r$ вдоль геодезической, выпущенной из точки $q$:
\begin{equation}                                                  \label{egeoya}
\begin{split}
  x^\al_r&=x^\al_q+Y^\al_q\e-\frac12Y^\bt_q Y^\g_q\Gamma_{q\,\bt\g}{}^\al
  \e^2=
\\
  &=x^\al_p+(X^\al_p+Y^\bt_p)\e^2-\left(\frac12 X^\bt_p X^\g_p
  +X^\bt_p Y^\g_p+\frac12Y^\bt_p Y^\g_p\right)\Gamma_{p\,\bt\g}{}^\al\e^2.
\end{split}
\end{equation}
Аналогично, координаты точки $u$ во втором порядке по $\e$ равны
\begin{equation*}
  x^\al_u=x^\al_p+(X^\al_p+Y^\bt_p)\e^2-\left(\frac12 X^\bt_p X^\g_p
  +Y^\bt_p X^\g_p+\frac12Y^\bt_p Y^\g_p\right)\Gamma_{p\,\bt\g}{}^\al\e^2.
\end{equation*}
Таким образом получаем равенство
\begin{equation}                                                  \label{egepto}
  x^\al_u-x^\al_r=X^\bt_p Y^\g_p T_{p\,\bt\g}{}^\al\e^2,
\end{equation}
где $T_{p\,\bt\g}{}^\al$ -- компоненты тензора кручения в точке $p$.

Формула (\ref{egepto}) позволяет дать геометрическую интерпретацию тензору
кручения: главная часть разомкнутости бесконечно малого ``параллелограмма'' с
направляющими $X_p\e$ и $Y_p\e$ квадратична по $\e$ и определяется тензором
кручения.
\begin{com}
Геометрический смысл кручения лег в основу физической интерпретации
кручения, как поверхностной плотности вектора Бюргерса, характеризующего
дислокации в упругой среде (см.\ \cite{Katana05R}).
\qed\end{com}
\section{Свойства аффинной связности                             \label{susefg}}
Во многих важных моделях математической физики пространство-время
рассматривается, как многообразие $\MM$, $\dim\MM=n$, на котором задана
аффинная геометрия, т.е.\ метрика и аффинная связность. В настоящем разделе
приведены тождества, включающие аффинную связность, которые полезны для
приложений при проведении вычислений. Кроме того, определены инвариантные
дифференциальные операторы второго порядка и приведена формула интегрирования
по частям. Все формулы настоящего раздела доказываются прямыми вычислениями.

Определим {\em след аффинной связности},
\index{След аффинной связности (trace of an affine connection)}%
\index{Аффинной связности след (trace of an affine connection)}%
который получается после свертки последней пары индексов:
\begin{equation}                                                  \label{eafctr}
  \Gamma_\al:=\Gamma_{\al\bt}{}^\bt.
\end{equation}
Он не является ковекторным полем, поскольку закон преобразования
содержит неоднородное слагаемое:
\begin{equation}                                                  \label{etrltc}
  \Gamma_{\al'}=\pl_{\al'}x^\al\Gamma_\al
  -\pl_{\al'}x^\al\pl_{\bt'}x^\bt\pl^2_{\al\bt}x^{\bt'}
  =\pl_{\al'}x^\al\Gamma_\al+\pl^2_{\al'\bt'}x^\al\pl_\al x^{\bt'}.
\end{equation}

Из выражения (\ref{elicon}) для компонент аффинной связности следует, что след
аффинной связности (\ref{eafctr}) равен
\begin{equation}                                                  \label{etrlio}
  \Gamma_\al=\widetilde\Gamma_\al+\frac12 Q_\al,
\end{equation}
где след символов Кристоффеля,
$\widetilde\Gamma_\al:=\widetilde\Gamma_{\al\bt}{}^\bt
=\widetilde\Gamma_{\bt\al}{}^\bt$,
имеет вид
\begin{equation}                                                  \label{etrchs}
  \widetilde\Gamma_\al=\frac12g^{\bt\g}\pl_\al g_{\bt\g}
  =-\frac12g_{\bt\g}\pl_\al g^{\bt\g}=\frac12\frac{\pl_\al g}g
  =\frac{\pl_\al\vol}\vol=e^\bt{}_b\pl_\al e_\bt{}^b,
\end{equation}
и введен след тензора неметричности по последним индексам:
\begin{equation}                                                  \label{enomtr}
  Q_\al:=Q_{\al\bt}{}^\bt.
\end{equation}
В тождествах (\ref{etrchs})
$$
  g:=\det g_{\al\bt},\qquad \sqrt{|g|}=\det e_\al{}^a.
$$
\begin{com}
След аффинной связности не зависит от кручения.
\qed\end{com}

Приведем также несколько полезных формул, справедливых в (псевдо-)римановой
геометрии:
\begin{align}                                                     \label{epldet}
  \pl_\al g&=gg^{\bt\g}\pl_\al g_{\bt\g},
\\                                                                \label{eplrde}
  \pl_\al\vol&=\frac12\vol g^{\bt\g}\pl_\al g_{\bt\g}
  = \vol\widetilde \Gamma_{\bt\al}{}^\bt  = \vol e^\bt{}_a\pl_\al e_\bt{}^a,
\\                                                                \label{eplinm}
  \pl_\al g^{\bt\g}&=-g^{\bt\dl}g^{\g\e}\pl_\al g_{\dl\e}
  =-g^{\bt\dl}e^\g{}_a\pl_\al e_\dl{}^a-g^{\g\dl}e^\bt{}_a\pl_\al e_\dl{}^a,
\\                                                                \label{epleim}
  \pl_\al(\vol g^{\bt\g})&=\vol(g^{\bt\g}\widetilde\Gamma_\al
  -g^{\bt\dl}\widetilde\Gamma_{\al\dl}{}^\g-g^{\g\dl}
  \widetilde\Gamma_{\al\dl}{}^\bt),
\\                                                      \label{eplimc}
  \pl_\bt(\vol g^{\bt\al})&=-\vol g^{\bt\g}\widetilde\Gamma_{\bt\g}{}^\al,
\\                                                      \label{ederit}
  e_\al{}^a\pl_\bt(\vol e^\al{}_a)&=(n-1)\vol e^\al{}_a\pl_\bt e_\al{}^a.
\end{align}

В пространстве Римана--Картана--Вейля след неметричности пропорционален форме
Вейля $W_\al$ из (\ref{ewevec}):
$$
  Q_\al=nW_\al.
$$

Свертка аффинной связности (\ref{elicon}) по первому и третьему индексам
приводит к равенству
\begin{equation}                                                  \label{etracx}
  \Gamma_{\bt\al}{}^\bt=\widetilde\Gamma_\al+T_\al+\frac12 Q_{\al}
  =\Gamma_\al+T_\al,
\end{equation}
где введен след тензора кручения
\begin{equation}                                                  \label{etotra}
  T_\al:=T_{\bt\al}{}^\bt.
\end{equation}

\begin{prop}
Справедливо равенство
\begin{equation}                                                  \label{ecodet}
  \nb_\al\ve_{\al_1\dotsc\al_n}=-\frac12Q_\al\ve_{\al_1\dotsc\al_n}.
\end{equation}
\end{prop}
\begin{proof}
Из определения ковариантной производной и связи между полностью антисимметричным
тензором и тензорной плотностью (\ref{etoate}) следует равенство
\begin{equation*}
  \nb_\al\ve_{\al_1\dotsc\al_n}=\hat\ve_{\al_1\dotsc\al_n}\pl_\al\vol
  -\Gamma_{\al\al_1}{}^\bt\ve_{\bt\al_2\dotsc\al_n}-\dotsc
  -\Gamma_{\al\al_n}{}^\bt\ve_{\al_1\dotsc\al_{n-1}\bt}.
\end{equation*}
Это выражение антисимметрично по индексам $\al_1,\dotsc,\al_n$ и, следовательно,
пропорционально полностью антисимметричному тензору:
\begin{equation*}
  \nb_\al\ve_{\al_1\dotsc\al_n}=\ve_{\al_1\dotsc\al_n}X_\al,
\end{equation*}
где $X_\al$ -- компоненты некоторого ковектора. Свертка полученного равенства с
контравариантным тензором $\ve^{\al_1\dotsc\al_n}$ с учетом равенств
(\ref{etrlio}), (\ref{etrchs}) и (\ref{etoprz}) определяет
\begin{equation*}                                                    \tag*{\qed}
  X_\al=-\frac12Q_\al.
\end{equation*}
\renewcommand{\qed}{}\end{proof}
\begin{cor}
Для метрической связности полностью антисимметричный тензор является ковариантно
постоянным.
\qed\end{cor}

В дифференциальной геометрии естественным образом вводится понятие градиента и
дивергенции.
\begin{defn}
Назовем {\em градиентом} скалярного поля (функции) $f$ ковекторное поле
(1-форму):
\begin{equation}                                                  \label{egradv}
  \grad f:=dx^\al\nb_\al f=df=dx^\al\pl_\al f.
\end{equation}
Назовем {\em дивергенцией} векторного поля $X$ скалярное поле
\begin{equation}                                                  \label{edivvf}
  \div X:=\nb_\al X^\al=\pl_\al X^\al+\Gamma_{\bt\al}{}^\bt X^\al,
\end{equation}
построенное с помощью ковариантной производной.
\qed\end{defn}
\index{Градиент скалярного поля (gradient of a scalar field)}
\index{Дивергенция векторного поля (divergence of a vector field)}%
С учетом выражения аффинной связности через кручение и неметричность
(\ref{elicon}) дивергенция векторного поля (\ref{edivvf}) принимает вид
\begin{equation}                                                  \label{edivef}
  \nb_\al X^\al=\frac1\vol\pl_\al\left(\vol X^\al\right)
  +T_\al X^\al+\frac12Q_\al X^\al.
\end{equation}
В частности, в (псевдо-)римановой геометрии
\begin{equation}                                                  \label{edivrf}
  \widetilde\nb_\al X^\al=\frac1\vol\pl_\al\left(\vol X^\al\right).
\end{equation}
Эта полезная формула позволяет переписать инвариантный интеграл от дивергенции
произвольного векторного поля в (псевдо-)римановой геометрии в виде
\begin{equation*}
  \int\! dx\vol \widetilde\nb_\al X^\al=\int\! dx\pl_\al(\vol X^\al)
\end{equation*}
и воспользоваться формулой Стокса (см.\ раздел \ref{stokes}). В пространстве
аффинной связности после применения формулы Стокса возникают дополнительные
объемные интегралы с кручением и неметричностью (\ref{edivef}).
\begin{com}
В разделе \ref{slabe} мы определили дивергенцию (\ref{edivde}) для внешних форм.
Пусть на многообразии $\MM$ задана метрика $g_{\al\bt}$ и 1-форма
$A=dx^\al A_\al$, соответствующая вектору $X$, где $A_\al:=X^\bt g_{\al\bt}$.
Тогда из сравнения формул (\ref{eacdlo}) и (\ref{edivvf}) следует, что
$\div A=\div X$ тогда и только тогда, когда
\begin{equation*}
  T_\al X^\al+\frac12 Q_\al X^\al=0.
\end{equation*}
В частности, равенство $\div A=\div X$ имеет место в (псевдо-)римановой
геометрии.
\qed\end{com}

Дивергенция от антисимметричного тензорного поля, $F^{\al\bt}=-F^{\bt\al}$,
равна
\begin{equation}                                                  \label{edivat}
  \nb_\al F^{\al\bt}=\frac1\vol\pl_\al\left(\vol F^{\al\bt}\right)
  +F^{\al\bt}\left(T_\al+\frac12Q_\al\right)+\frac12F^{\al\g}T_{\al\g}{}^\bt.
\end{equation}
В римановой геометрии это тождество упрощается:
\begin{equation}                                                  \label{edivrt}
  \widetilde\nb_\al F^{\al\bt}=\frac1\vol\pl_\al\left(\vol F^{\al\bt}\right).
\end{equation}
Приведем также формулу для антисимметричной ковариантной производной от 1-формы:
\begin{equation}                                                  \label{eroifo}
  \nb_\al X_\bt-\nb_\bt X_\al=\pl_\al X_\bt-\pl_\bt X_\al
  -T_{\al\bt}{}^\g X_\g.
\end{equation}
Из этого равенства следует, что антисимметризация обычной производной от 1-формы
дает тензорное поле типа $(0,2)$. Это процедура положена в основу определения
внешней производной от произвольной формы (см.\ раздел \ref{sextpr}).

Наличие ковариантной производной $\nb_\al$ и метрики $g_{\al\bt}$ позволяет
строить ковариантные дифференциальные операторы, действующие на произвольные
тензорные поля на многообразии. Для их построения достаточно взять произвольный
дифференциальный оператор в (псевдо-)евклидовом пространстве, заменить частные
производные $\pl_\al$ на ковариантные $\nb_\al$ и (псевдо-)евклидову метрику
$\eta_{\al\bt}$ на метрику многообразия $g_{\al\bt}$.

Инвариантный дифференциальный оператор второго порядка, действующий на
произвольные тензорные поля на многообразии имеет вид
\index{Оператор Лапласа--Бельтрами (Laplace--Beltrami operator)}%
\index{Лапласа--Бельтрами оператор (Laplace--Beltrami operator)}%
\begin{equation}                                                  \label{elabep}
  \triangle:=g^{\al\bt}\nb_\al\nb_\bt
  =\nb_\al g^{\al\bt}\nb_\bt-Q_\al{}^{\al\bt}\nb_\bt.
\end{equation}
Он часто называется {\em оператором Лапласа--Бельтрами} независимо от сигнатуры
метрики. В римановой геометрии оператор Лапласа--Бельтрами устроен проще:
\begin{equation}                                                  \label{elabri}
  \tilde\triangle:=g^{\al\bt}\widetilde\nb_\al\widetilde\nb_\bt
  =\widetilde\nb_\al g^{\al\bt}\widetilde\nb_\bt.
\end{equation}
\begin{com}
Если метрика имеет лоренцеву сигнатуру, то мы будем обозначать оператор
Лапласа--Бельтрами квадратом $\square$, а не треугольником $\triangle$, по
аналогии с оператором Даламбера. В этом случае уравнение
$\square K_{\bt_1\dotsc\bt_s}{}^{\al_1\dotsc\al_r}=0$ для каждой
компоненты тензорного поля $K\in\CT^r_s(\MM)$ будет гиперболического типа.
\qed\end{com}
Действие оператора Лапласа--Бельтрами на скалярное поле можно записать в виде
\begin{equation}
  \triangle f=\frac1\vol\pl_\al\left(\vol g^{\al\bt}\pl_\bt f\right)
  +\left(T^\bt+\frac12Q^\bt-Q_\al{}^{\al\bt}\right)\pl_\bt f.
\end{equation}
В римановой геометрии правая часть этого равенства записывается через обычную
производную от тензорной плотности:
\begin{equation}                                                  \label{elassr}
  \tilde\triangle f=\frac1\vol\pl_\al\left(\vol g^{\al\bt}\pl_\bt f\right).
\end{equation}

Приведем также формулу интегрирования по частям. Используя (\ref{etrchs}) и
(\ref{etracx}), проверяется, что
$$
  \pl_\al\left(\vol X^{\al\Sa}Y_\Sa\right)=\vol\nb_\al X^{\al\Sa}Y_\Sa
  -\vol\left(T_\al+\frac12Q_\al\right)X^{\al\Sa}Y_\Sa
  +\vol X^{\al\Sa}\nb_\al Y_\Sa.
$$
Здесь индекс $\Sa$ обозначает произвольную совокупность ковариантных и
контравариантных индексов, по которым подразумевается суммирование.
Интегрируя это соотношение по многообразию и пренебрегая граничными
членами, получим равенство
\begin{equation}                                                  \label{eintpa}
  \int\! dx\,\vol X^{\al\Sa}\nb_\al Y_\Sa
  =-\int\! dx\,\vol\left[\nb_\al X^{\al\Sa}Y_\Sa
  -\left(T_\al+\frac12Q_\al\right)X^{\al\Sa}Y_\Sa\right].
\end{equation}
В римановой геометрии эта формула имеет тот же вид, что и обычное интегрирование
по частям:
\begin{equation}                                                  \label{eintpr}
  \int\! dx\,\vol X^{\al\Sa}\widetilde\nb_\al Y_\Sa
  =-\int\! dx\,\vol\left[\widetilde\nb_\al X^{\al\Sa}Y_\Sa\right].
\end{equation}
Отсюда следует, что с точностью до граничных членов имеет место равенство
\begin{equation}                                                  \label{elabrp}
  \int\! dx\,\vol X^\Sa\tilde\triangle Y_\Sa
  =\int\!dx\,\vol \tilde\triangle X^\Sa Y_\Sa.
\end{equation}
\section{Локальное определение тензора кривизны                  \label{sloccu}}
Помимо кручения аффинная связность $\Gamma_{\al\bt}{}^\g$ на многообразии $\MM$
задает еще один важный геометрический объект -- {\em тензор кривизны} аффинной
связности (\ref{ecuaft}) или {\em тензор Римана--Кристоффеля}.
\index{Тензор кривизны (curvature tensor)}%
\index{Кривизны тензор (curvature tensor)}\index{Кривизна (curvature)}%
\index{Тензор Римана--Кристоффеля (Riemann--Christoffel tensor)}%
\index{Римана--Кристоффеля тензор (Riemann--Christoffel tensor)}%
В локальной системе координат он имеет следующие компоненты:
\begin{equation}                                                  \label{ecurva}
  R_{\al\bt\g}{}^\dl:=\pl_\al\Gamma_{\bt\g}{}^\dl-\Gamma_{\al\g}{}^\e
                     \Gamma_{\bt\e}{}^\dl-(\al\leftrightarrow\bt),
\end{equation}
где скобки $(\al\leftrightarrow\bt)$ обозначают предыдущие слагаемые с
переставленными индексами $\al$ и $\bt$.
\begin{com}
Тензор кривизны никакого отношения к метрике не имеет и определяется только
связностью.
\qed\end{com}

Тензор кривизны играет очень важную роль в дифференциальной геометрии и
возникает в различных контекстах. Покажем, что тензор кривизны позволяет
сформулировать критерий локальной тривиальности аффинной связности.

Пусть на многообразии $\MM$ задана аффинная связность с нулевым кручением,
$T_{\al\bt}{}^\g=0$. При этом неметричность, если задана также метрика, может
быть отлична от нуля. Рассмотрим соотношения (\ref{econts}) как уравнения на
функции перехода $x^{\al}(x')$. Потребуем, чтобы в новой системе координат
компоненты связности обращались в нуль в некоторой односвязной области. Тогда
функции перехода должны удовлетворять уравнению
\begin{equation}                                                  \label{eqtrfc}
  \frac{\pl^2x^\al}{\pl x^{\bt'}\pl x^{\g'}}
  =-\frac{\pl x^\bt}{\pl x^{\bt'}}\frac{\pl x^\g}{\pl x^{\g'}}
  \Gamma_{\bt\g}{}^\al.
\end{equation}
Дифференцируя это соотношение по $x^{\dl'}$ и исключая вторые производные
от функций перехода с помощью исходного уравнения (\ref{eqtrfc}), получим
равенство
$$
  \frac{\pl^3x^\al}{\pl x^{\bt'}\pl x^{\g'}\pl x^{\dl'}}
  =\frac{\pl x^\bt}{\pl x^{\bt'}}\frac{\pl x^\g}{\pl x^{\g'}}
  \frac{\pl x^\dl}{\pl x^{\dl'}}\big(-\pl_\dl\Gamma_{\bt\g}{}^\al
  +\Gamma_{\bt\dl}{}^\e\Gamma_{\e\g}{}^\al+\Gamma_{\dl\g}{}^\e\Gamma_{\bt\e}{}^\al\big).
$$
Условия интегрируемости уравнений (\ref{eqtrfc}) получаются из этих уравнений
антисимметризацией выражения в круглых скобках по индексам $\bt',\dl'$ или
$\g',\dl'$ и приравниванием результата нулю. Обе антисимметризации приводят к
единственному условию: равенству нулю тензора кривизны (\ref{ecurva}),
$R_{\al\bt\g}{}^\dl=0$.

Равенство нулю тензора кривизны является необходимым и достаточным условием
локальной разрешимости системы уравнений (\ref{eqtrfc}) относительно матриц
Якоби преобразования координат. После ее решения возникнет система равенств
\begin{equation}                                                  \label{qknzfd}
  \frac{\pl x^\al}{\pl x^{\al'}}=f_{\al'}{}^\al,
\end{equation}
где в правой части стоят некоторые функции от $x'$. Эта система уравнений на
функции перехода, в свою очередь, имеет свои условия интегрируемости:
\begin{equation*}
  \pl_{\al'}f_{\bt'}{}^\al-\pl_{\bt'}f_{\al'}{}^\al=0.
\end{equation*}
Эти условия интегрируемости в общем случае выполнены не будут. Однако, можно
доказать, что правая часть уравнений (\ref{qknzfd}), как решений исходных
уравнений (\ref{eqtrfc}) определена неоднозначно. Кроме того, этот произвол в
выборе решений всегда можно зафиксировать таким образом, чтобы условия
интегрируемости для функций перехода были выполнены. Соответствующее
доказательство в переменных Картана было дано в разделе \ref{slojnf}. Таким
образом, справедлива
\begin{theorem}
Пусть на многообразии $\MM$ задана геометрия Римана--Картана. Если тензоры
кривизны и кручения равны нулю в некоторой односвязной области $\MU\subset\MM$,
то, возможно, в меньшей окрестности существует такая система координат, в
которой компоненты связности обратятся в нуль.
\end{theorem}
\begin{defn}
Назовем аффинную связность на многообразии $\MM$ {\em локально тривиальной},
если для любой точки $x\in\MM$ найдется окрестность $\MU_x\ni x$ и такая система
координат, что компоненты связности $\Gamma_{\al\bt}{}^\g$ на $\MU_x$ равны нулю.
Такая аффинная связность называется также {\em интегрируемой}.
\qed\end{defn}
\index{Локально тривиальная связность (locally trivial connection)}%
\index{Связность локально тривиальная (locally trivial connection)}%
\index{Интегрируемая связность (integrable connection)}%
\index{Связность интегрируемая (integrable connection)}%

Таким образом, найден критерий локальной тривиальности аффинной связности.
\index{Локальная тривиальность аффинной связности %
(local triviality of an affine connection)}%
\begin{theorem}
Для локальной тривиальности аффинной связности на многообразии $\MM$ необходимо
и достаточно, чтобы ее кручение и тензор кривизны равнялись нулю на $\MM$.
\end{theorem}
\begin{com}
Тензор кручения является тензором, и его компоненты нельзя обратить в нуль
никаким преобразованием координат. Поэтому равенство нулю тензора кривизны
необходимо для локальной тривиальности аффинной связности.
\qed\end{com}

Пусть на многообразии $\MM$ помимо связности $\Gamma_{\al\bt}{}^\g$ задана также
метрика $g_{\al\bt}$. Тогда определен тензор неметричности (\ref{enonme}). При
ненулевом тензоре неметричности локальная тривиальность аффинной связности не
означает, что многообразие является локально (псевдо-)евклидовым. Действительно,
при $\Gamma_{\al\bt}{}^\g=0$ и $T_{\al\bt}{}^\g=0$ уравнение (\ref{enonme}) дает
соотношение между метрикой и неметричностью
\begin{equation}                                                  \label{emenom}
  \pl_\al g_{\bt\g}=-Q_{\al\bt\g}.
\end{equation}
В частности, метрику можно задать произвольно, и она будет определять
неметричность. В римановой геометрии $T_{\al\bt}{}^\g=0$ и $Q_{\al\bt\g}=0$, и
равенство нулю тензора кривизны, $\widetilde R_{\al\bt\g}{}^\dl=0$, означает,
что найдется такая система координат, в которой все частные производные метрики
обратятся в нуль, $\pl_\al g_{\bt\g}=0$, т.е.\ метрика имеет постоянные
компоненты в некоторой области $\MU\subset\MM$. С помощью последующего линейного
преобразования координат ее всегда можно привести к диагональному виду, когда на
диагонали будут стоять $\pm1$ в соответствии с исходной сигнатурой метрики. Это
дает критерий локальной (псевдо-)евклидовости (псевдо-)риманова пространства.
Сформулируем сразу глобальное утверждение.
\begin{theorem}
Пусть на многообразии $\MM$ задана аффинная геометрия. Тогда, если тензоры
кручения, неметричности и кривизны равны нулю на всем многообразии, то
$\MM$ изометрично либо (псевдо-)евклидову пространству $\MR^n$, либо
факторпространству $\MR^n/\MG$, где $\MG$ -- группа преобразований, действующая
на $\MR^n$ свободно и собственно разрывно.
\end{theorem}
\begin{proof}
См., например, \cite{Wolf72R}.
\end{proof}
\section{Свойства тензора кривизны                               \label{scurpr}}
Помимо тождеств Бианки тензоры кручения и кривизны удовлетворяют ряду других
дифференциальных соотношений. Выведем ряд полезных тождеств, исходя из
определения тензора кривизны (\ref{ecurva}). Прямые вычисления с учетом
выражения аффинной связности через метрику, кручение и неметричность показывают,
что антисимметризация тензора кривизны по первым трем индексам определяется
только тензором кручения и его ковариантными производными:
\begin{align}                                                     \nonumber
  R_{\al\bt\g}{}^\dl+R_{\bt\g\al}{}^\dl+R_{\g\al\bt}{}^\dl &=
  \nb_\al T_{\bt\g}{}^\dl+\nb_\bt T_{\g\al}{}^\dl
  +\nb_\g T_{\al\bt}{}^\dl
\\                                                                \label{ecuath}
  &+ T_{\al\bt}{}^\e T_{\e\g}{}^\dl+T_{\bt\g}{}^\e T_{\e\al}{}^\dl
  +T_{\g\al}{}^\e T_{\e\bt}{}^\dl.
\end{align}
Это означает, что в римановой геометрии, а также в аффинной геометрии
без кручения тензор кривизны удовлетворяет тождеству
\begin{equation}                                                  \label{eidcrg}
  R_{\al\bt\g}{}^\dl+R_{\bt\g\al}{}^\dl+R_{\g\al\bt}{}^\dl = 0.
\end{equation}
Обратное утверждение в общем случае неверно. То есть из равенства
(\ref{eidcrg}) не следует, что кручение равно нулю.

Пусть на многообразии $\MM$ задана аффинная геометрия, т.е.\ метрика и
связность. Приведем явное выражение для тензора кривизны со всеми опущенными
индексами. С учетом разложения связности (\ref{elicon}) и тождества
(\ref{ederga}) получаем равенство
\begin{equation}                                                  \label{ecurlo}
  R_{\al\bt\g\dl}:=R_{\al\bt\g}{}^\e g_{\e\dl}
  =\pl_\al\Gamma_{\bt\g\dl}+\Gamma_{\al\g}{}^\e\Gamma_{\bt\dl\e}
  -\Gamma_{\al\g}{}^\e(T_{\bt\dl\e}+Q_{\bt\dl\e})-(\al\leftrightarrow\bt).
\end{equation}
Если у тензора кривизны со всеми опущенными индексами произвести
симметризацию по последней паре индексов, то получим тождество
\begin{equation}                                                  \label{ecantl}
  R_{\al\bt\g\dl}+R_{\al\bt\dl\g}
  =\nb_\al Q_{\bt\g\dl}-\nb_\bt Q_{\al\g\dl}+T_{\al\bt}{}^\e Q_{\e\g\dl}.
\end{equation}
Свертывая это тождество по последней паре индексов, получаем равенство
\begin{align}                                                     \nonumber
  2R_{\al\bt\g}{}^\g=2(\pl_\al\Gamma_\bt-\pl_\bt\Gamma_\al) &
  =\nb_\al Q_\bt-\nb_\bt Q_\al+T_{\al\bt}{}^\g Q_\g
\\                                                                \label{ewfist}
  &=\pl_\al Q_\bt-\pl_\bt Q_\al,
\end{align}
где $Q_\al:=Q_{\al\bt}{}^\bt$. Отсюда следует, что в геометрии
Римана--Картана--Вейля свертка тензора кривизны по последним двум
индексам дает напряженность для формы Вейля.

Рассмотрим симметрии тензора кривизны относительно перестановок индексов. В
общем случае аффинной геометрии единственная симметрия тензора кривизны со
всеми опущенными индексами -- это антисимметрия по первой паре индексов:
\begin{equation}                                                  \label{esymco}
  R_{\al\bt\g\dl}=-R_{\bt\al\g\dl},
\end{equation}
что сразу вытекает из определения (\ref{ecurva}). Отсюда следует, что в
аффинной геометрии число линейно независимых компонент тензора кривизны  равно
\begin{equation*}
  [R_{\al\bt\g\dl}]=\frac{n^3(n-1)}2.
\end{equation*}

В геометрии Римана--Картана тензор кривизны антисимметричен также и по
второй паре индексов, что является следствием уравнения (\ref{ecantl}),
\begin{equation}                                                  \label{ecusym}
  R_{\al\bt\g\dl}=-R_{\al\bt\dl\g}.
\end{equation}
Поэтому число его линейно независимых компонент меньше:
\begin{equation*}
  [R_{\al\bt\g\dl}]=\frac{n^2(n-1)^2}4.
\end{equation*}

В (псевдо-)римановой геометрии тензор кривизны обладает дополнительной
симметрией: его антисимметризация по первым трем индексам тождественно
обращается в нуль
\begin{equation}                                                  \label{eanfti}
  \widetilde R_{[\al\bt\g]\dl}=0.
\end{equation}
что следует из уравнения (\ref{ecuath}). Следовательно, число его линейно
независимых компонент равно
\begin{equation*}
  [\widetilde R_{\al\bt\g\dl}]=\frac{n^2(n^2-1)}{12}.
\end{equation*}
Из свойств (\ref{esymco}), (\ref{ecusym}) и (\ref{eanfti}) следует, что в
(псевдо-)римановой геометрии тензор кривизны симметричен также относительно
перестановки первой пары индексов со второй,
\begin{equation}                                                  \label{esymct}
  \widetilde R_{\al\bt\g\dl}=\widetilde R_{\g\dl\al\bt}.
\end{equation}
\begin{com}
Обратные утверждения, связывающие геометрию с симметрией тензора кривизны
неверны. Например, тензор кривизны может быть антисимметричен по второй паре
индексов и в то же время тензор неметричности может быть нетривиальным.
\qed\end{com}
В (псевдо-)римановой геометрии тензор кривизны со всеми опущенными индексами
следующим образом выражается через метрику:
\begin{equation}                                                  \label{ecutrl}
  \widetilde R_{\al\bt\g\dl}=\frac12
  (\pl_{\al\g}^2g_{\bt\dl}-\pl_{\al\dl}^2g_{\bt\g}
  -\pl_{\bt\g}^2g_{\al\dl}+\pl_{\bt\dl}^2g_{\al\g})
  +\widetilde\Gamma_{\al\g}{}^\e\widetilde\Gamma_{\bt\dl\e}
  -\widetilde\Gamma_{\al\dl}{}^\e\widetilde\Gamma_{\bt\g\e}.
\end{equation}
Как видим, он линеен по вторым производным от метрики и квадратичен по первым
производным.

\begin{defn}
По заданному тензору кривизны путем свертки пары индексов можно построить
{\em тензор Риччи}
\begin{equation}                                                  \label{ericte}
  R_{\al\g}:=R_{\al\bt\g}{}^\bt. \qed
\end{equation}
\end{defn}
\index{Тензор Риччи (Ricci tensor)}\index{Риччи тензор (Ricci tensor)}%
\begin{com}
Тензор кривизны и тензор Риччи строятся только по аффинной связности, без
использования метрики.
\qed\end{com}
В общем случае тензор Риччи не обладает никакой симметрией по своим индексам.
В (псевдо-)римановой геометрии он симметричен относительно перестановки
индексов:
$$
  \widetilde R_{\al\g}=\widetilde R_{\g\al}.
$$
\begin{defn}
Свертывая тензор Риччи с обратной метрикой, получаем {\em скалярную кривизну}
многообразия
\begin{equation}                                                  \label{escurv}
  R:=R_{\al\bt}g^{\al\bt}.
\end{equation}
Скалярная кривизна зависит и от связности, и от метрики. Функция
\begin{equation}                                                  \label{egaucu}
  K:=-\frac12 R,
\end{equation}
где $R$ --скалярная кривизна, называется {\em гауссовой кривизной} многообразия.
\qed\end{defn}
\index{Скалярная кривизна (scalar curvature)}%
\index{Кривизна скалярная (scalar curvature)}%
\index{Гауссова кривизна (Gaussian curvature)}%
\index{Кривизна гауссова (Gaussian curvature)}%
\begin{exa}
Пусть двумерная поверхность задана в трехмерном евклидовом пространстве $\MR^3$
уравнением $z=f(x,y)$ с дважды непрерывно дифференцируемой правой частью, причем
\begin{equation*}
  f(0,0)=0,\qquad f_x(0,0)=0,\qquad f_y(0,0)=0,
\end{equation*}
где $f_x:=\pl_x f$ и $f_y:=\pl_y f$.
Тогда скалярная кривизна в начале координат, соответствующая индуцированной
метрике,
\begin{equation*}
\begin{split}
  ds^2&=dx^2+dy^2+dz^2=
\\
  &=(1+f_x^2)dx^2+2f_xf_ydxdy+(1+f_y^2)dy^2,
\end{split}
\end{equation*}
пропорциональна определителю гессиана
\begin{equation*}
  R=-2\det \begin{pmatrix}f_{xx} & f_{xy} \\ f_{xy} & f_{yy} \end{pmatrix}=-2K,
\end{equation*}
где $K$ -- гауссова кривизна поверхности. Поворотом евклидова пространства в
плоскости $x,y$ гессиан всегда можно привести к диагональному виду
\begin{equation*}
  \begin{pmatrix}f_{xx} & 0 \\ 0 & f_{yy} \end{pmatrix},
\end{equation*}
где, скажем, $f_{xx}(0,0)\ge f_{yy}(0,0)$. Если $f_{xx}>f_{yy}$, то оси $x,y$
называются {\em главными направлениями} поверхности в данной точке.
\index{Главные направления поверхности (principal directions of a surface)}%
\index{Направления поверхности главные (principal directions of a surface)}%
При $f_{xx}=f_{yy}$, главные направления не определены. Если гауссова кривизна
положительна, $K>0$, то поверхность в начале координат имеет локальный
экстремум и лежит по одну сторону плоскости $z=0$. Если кривизна меньше нуля,
$K<0$, то поверхность в начале координат имеет седловую точку. При $K=0$ по
крайней мере одна из вторых частных производных обращается в нуль, и мы имеем
касание более высокого порядка.
\qed\end{exa}
\begin{com}
Многие авторы определяют тензор кривизны многообразия таким образом, чтобы
скалярная кривизна совпала с гауссовой кривизной. Мы используем определения
тензора кривизны, тензора Риччи и скалярной кривизны, которые широко
используется в дифференциальной геометрии и приложениях. Это не приводит к
недоразумениям. Однако надо помнить, что в наших обозначениях скалярная и
гауссова кривизна двумерной сферы $\MS^2_r$ радиуса $r$ равны: $R=-2K=-2/r^2$ и
$K=1/r^2$ (см.\ вычисления далее в разделе \ref{sphere}).
\qed\end{com}
В аффинной геометрии компоненты связности $\Gamma_{\al\bt}{}^\g$
инвариантны относительно преобразования
\begin{equation}                                                  \label{emeinv}
  g_{\al\bt}\mapsto-g_{\al\bt},\qquad
  T_{\al\bt}{}^\g\mapsto T_{\al\bt}{}^\g,\qquad
  Q_{\al\bt}{}^\g\mapsto Q_{\al\bt}{}^\g.
\end{equation}
Поэтому тензор кривизны и тензор Риччи также инвариантны. При этом
скалярная кривизна (\ref{escurv}) меняет знак.

Рассмотрим на многообразии $\MM$ произвольное векторное поле
$X=X^\al\pl_\al\in\CX(\MM)$ и 1-форму $A=dx^\al A_\al\in\Lm_1(\MM)$.
Ковариантные производные от их компонент имеют вид (\ref{ecodev}),
(\ref{ecodec}). Прямые вычисления показывают, что коммутатор двух ковариантных
производных определяется тензором кривизны и кручения:
\begin{align}                                                     \label{emeacu}
  [\nb_\al,\nb_\bt]X^\g&=R_{\al\bt\dl}{}^\g X^\dl
  -T_{\al\bt}{}^\dl\nb_\dl X^\g,
\\                                                                \label{eonfco}
  [\nb_\al,\nb_\bt]A_\g&=-R_{\al\bt\g}{}^\dl A_\dl
  -T_{\al\bt}{}^\dl\nb_\dl A_\g.
\end{align}
и отличается только знаком перед слагаемым с кривизной. Коммутатор ковариантных
производных от скалярного поля $\vf(x)$ проще:
\begin{equation}                                                  \label{ecocds}
  [\nb_\al,\nb_\bt]\vf=-T_{\al\bt}{}^\g\nb_\g\vf
\end{equation}
и определяется только тензором кручения.

Формулы (\ref{emeacu}), (\ref{eonfco}) обобщаются на тензоры произвольного
ранга.
\begin{exa}
Коммутатор ковариантных производных от компонент тензора $K\in\CT^1_1(\MM)$
типа $(1,1)$ имеет вид
\begin{equation*}
  [\nb_\al,\nb_\bt]K^\g{}_\dl=R_{\al\bt\e}{}^\g K^\e{}_\dl
  -R_{\al\bt\dl}{}^\e K^\g{}_\e-T_{\al\bt}{}^\e\nb_\e K^\g{}_\dl.
\end{equation*}
Для тензоров более высокого ранга будем иметь по одному слагаемому с кривизной
со знаком плюс и минус соответственно для каждого контравариантного и
ковариантного индекса и одно общее слагаемое с тензором кручения.
\qed\end{exa}

Формулы для коммутатора ковариантных производных можно использовать для
доказательства некоторых тождеств. Докажем несколько формул, справедливых в
(псевдо-)римановой геометрии. Пусть $F^{\al\bt}$ -- произвольный
антисимметричный тензор второго ранга. Тогда в (псевдо-)римановой геометрии
справедливо равенство
\begin{equation*}
  \widetilde\nb_\al\widetilde\nb_\bt F^{\al\bt}=0.
\end{equation*}
Действительно,
\begin{equation*}
  [\widetilde\nb_\al,\widetilde\nb_\bt]F^{\al\bt}
  =\widetilde R_{\al\bt\g}{}^\al F^{\g\bt}
  +\widetilde R_{\al\bt\g}{}^\bt F^{\al\g}=2\widetilde R_{\al\bt}F^{\al\bt}=0,
\end{equation*}
поскольку тензор Риччи в этом случае симметричен.
\begin{prop}
В (псевдо-)римановой геометрии имеет место следующая формула
\begin{equation*}
  \widetilde\nb_\al\widetilde\nb_\bt\widetilde R^{\al\bt\g\dl}=0.
\end{equation*}
\end{prop}
\begin{proof}
Справедливы следующие равенства:
\begin{equation*}
\begin{split}
  2&\widetilde\nb_\al\widetilde\nb_\bt\widetilde R^{\al\bt\g\dl}
  =[\widetilde\nb_\al,\widetilde\nb_\bt]R^{\al\bt\g\dl}=
\\
  &=\widetilde R_{\al\bt\e}{}^\al\widetilde R^{\e\bt\g\dl}
  +\widetilde R_{\al\bt\e}{}^\bt\widetilde R^{\al\e\g\dl}
  +\widetilde R_{\al\bt\e}{}^\g\widetilde R^{\al\bt\e\dl}
  +\widetilde R_{\al\bt\e}{}^\dl\widetilde R^{\al\bt\g\e}.
\end{split}
\end{equation*}
Первые два слагаемых обращаются в нуль в силу симметрии тензора Риччи.
Последние два слагаемых сокращаются:
\begin{equation*}
  \widetilde R_{\al\bt\e}{}^\g\widetilde R^{\al\bt\e\dl}
  -\widetilde R_{\al\bt\e}{}^\dl\widetilde R^{\al\bt\e\g}=0
\end{equation*}
в силу симметрии слагаемых по индексам $\g,\dl$.
\end{proof}
В заключение настоящего раздела приведем важную формулу, связывающую скалярную
кривизну $R(g,\Gamma)$ в геометрии Римана--Картана со скалярной кривизной
$\widetilde R(g)$ в соответствующей (псевдо-)римановой геометрии, вычисленной
только по метрике при нулевом кручении:
\begin{equation}                                                  \label{eident}
  R+\frac14T_{\al\bt\g}T^{\al\bt\g}-\frac12T_{\al\bt\g}T^{\g\al\bt}-T_\al T^\al
    -\frac2\vol\pl_\al(\vol T^\al)=\widetilde R,
\end{equation}
где $T_\al:=T_{\bt\al}{}^\bt$ -- след тензора кручения. Это тождество
проверяется путем прямых вычислений и справедливо в пространстве произвольной
размерности $n\ge2$ и для метрики произвольной сигнатуры. Приведенное тождество
можно переписать в явно ковариантном виде
\begin{equation*}
  R+\frac14T_{\al\bt\g}T^{\al\bt\g}-\frac12T_{\al\bt\g}T^{\g\al\bt}-T_\al T^\al
    -2\tilde\nb_\al T^\al=\widetilde R,
\end{equation*}
где мы воспользовались тождеством (\ref{edivrf}). Важность этой формулы
заключается в том, что, поскольку обе скалярные кривизны приводят к уравнениям
движения не выше второго порядка (в переменных Картана), то при выборе
лагранжиана можно ограничиться одной скалярной кривизной и квадратами тензоров
кручения в моделях гравитации с динамическим кручением.
\section{Неголономный базис                                      \label{sunhba}}
Аффинная геометрия на многообразии $\MM$ задается метрикой и аффинной связностью
или, что эквивалентно, метрикой, кручением и неметричностью. При таком описании
каждое преобразование координат сопровождаются соответствующим преобразованием
компонент тензорных полей относительно координатного базиса. Существует также
другой способ описания геометрии, когда компоненты тензорных полей
рассматриваются относительно репера, который не меняется при преобразовании
координат. В этом случае на компоненты тензорных полей действует группа
локальных преобразований $\MG\ML(n,\MR)$, что соответствует вращению репера.
В результате геометрия на многообразии $\MM$ будет задана репером $e_\al{}^a$ и
линейной или $\MG\ML(n,\MR)$-связностью $\om_{\al a}{}^b$
(см.\ раздел \ref{scorep}).
\begin{defn}
Переменные репер $e_\al{}^a(x)$ и $\MG\ML(n,\MR)$-связность
$\om_{\al a}{}^b(x)$, задающие на многообразии $\MM$ аффинную геометрию,
называются {\em переменными Картана}. В четырехмерном пространстве-времени репер
называется {\em тетрадой}. В двумерном и трехмерном пространстве репер
называется соответственно {\em диадой} и {\em триадой}.
\qed\end{defn}
\index{Переменные Картана (Cartan variables)}%
\index{Картановы переменные (Cartan variables)}%
\index{Тетрада (tetrad)}\index{Диада (diad)}\index{Триада (triad)}%
\begin{com}
В моделях математической физики переменные Картана, как правило, упрощают
вычисления и необходимы при рассмотрении спинорных полей на многообразии $\MM$.
\qed\end{com}

Напомним, что координатный базис касательных пространств $\MT_x(\MM)$ во всех
точках многообразия $x\in\MM$ мы обозначаем ${\pl_\al}$, и он называется
{\em голономным}. Важным свойством координатных базисных векторов $\pl_\al$
является их коммутативность:
\index{Голономный базис (holonomic basis)}%
\index{Базис голономный (holonomic basis)}%
$$
  [\pl_\al,\pl_\bt]=0.
$$
Предположим, что в каждой точке многообразия $x\in\MM$ задан произвольный базис
касательного пространства $e_a(x)$ (репер) и дуальный к нему базис $1$-форм
$e^a(x)$ (корепер), $a=1,\dots,n$. Дуальность означает, что значение 1-форм
$e^a$ на векторных полях $e_b$ равно символу Кронекера: $e^a(e_b)=\dl^a_b$.
\index{Репер (frame)}\index{Корепер (coframe)}%
\begin{com}
Как уже отмечалось, репер может существовать глобально не для всех многообразий.
Например, его не существует на неориентируемых многообразиях. В таких случаях
все, сказанное ниже, имеет локальный характер. Тем не менее полученные формулы
важны для вычислений, которые, как правило, проводятся в какой либо системе
координат.
\qed\end{com}
Репер и корепер можно разложить по координатному базису:
$$
  e_a=e^\al{}_a\pl_\al,\qquad e^a=dx^\al e_\al{}^a,
$$
где $e^\al{}_a$ и $e_\al{}^a$ -- взаимно обратные невырожденные матрицы, что
является следствием дуальности:
$$
  e^\al{}_a e_\al{}^b=\dl_a^b,\qquad e^\al{}_a e_\bt{}^a=\dl_\bt^\al.
$$
По-предположению, матрицы $e^\al{}_a$ и $e_\al{}^a$ невырождены и достаточно
гладко зависят от точки многообразия.

В общем случае репер представляет собой {\em неголономный базис} касательного
пространства, т.е.\ не существует такой системы координат $x^a=x^a(x)$, что
\index{Неголономный базис (nonholonomic basis)}%
\index{Базис неголономный (nonholonomic basis)}%
\begin{equation}                                                  \label{eholba}
  e^\al{}_a=\pl_ax^\al.
\end{equation}
Репер определен с точностью до локальных $\MG\ML(n,\MR)$ преобразований,
действующих на латинские индексы. Его важнейшей характеристикой являются
{\em компоненты неголономности} $c_{ab}{}^c$,
\index{Компоненты неголономности (nonholonomicity components)}%
\index{Неголономности компоненты (nonholonomicity components)}%
которые определяются коммутатором базисных векторных полей:
\begin{equation}                                                  \label{eunhoc}
  [e_a,e_b]:=c_{ab}{}^c e_c
\end{equation}
и антисимметричны по нижним индексам,
$$
  c_{ab}{}^c(x)=-c_{ba}{}^c(x).
$$
Из тождеств Якоби для алгебры Ли векторных полей,
$$
  \big[e_a[e_b,e_c]\big]+\big[e_b[e_c,e_a]\big]+\big[e_c[e_a,e_b]\big]=0,
$$
следуют тождества для компонент неголономности:
\begin{equation}                                                  \label{ecounc}
  \pl_a c_{bc}{}^d+\pl_b c_{ca}{}^d+\pl_c c_{ab}{}^d
  +c_{ab}{}^e c_{ce}{}^d+c_{bc}{}^e c_{ae}{}^d+c_{ca}{}^e c_{be}{}^d=0,
\end{equation}
где $\pl_a:=e^\al{}_a\pl_\al$.

Из определения (\ref{eunhoc}) следует явное выражение для компонент
неголономности через компоненты репера и их производные
\begin{equation}                                                  \label{eunhoe}
  c_{ab}{}^c:=\big(e^\al{}_a\pl_\al e^\bt{}_b-e^\al{}_b\pl_\al e^\bt{}_a\big)
  e_\bt{}^c.
\end{equation}
Умножив это соотношение на обратные матрицы $e_\al{}^a$, получим эквивалентную
формулу
\begin{equation}                                                  \label{equnhc}
  c_{\al\bt}{}^c=e_\al{}^a e_\bt{}^b c_{ab}{}^c
  =-\pl_\al e_\bt{}^c+\pl_\bt e_\al{}^c,
\end{equation}
которую можно переписать в виде
\begin{equation}                                                  \label{eexrep}
  de^c=\frac12e^a\wedge e^b\, c_{ba}{}^c,
\end{equation}
где использовано определение внешнего умножения и дифференцирования форм
(см.\ раздел \ref{sdifin}).

Многие формулы содержат след компонент неголономности, который определяется
следующим образом:
\begin{equation}                                                  \label{etranc}
  c_a:=c_{ba}{}^b=\pl_\al e^\al{}_a+\frac{\pl_a\vol}\vol.
\end{equation}

Компоненты неголономности ковариантны относительно преобразования координат
$x^\al$$\rightarrow$$x^{\al'}(x)$, но не являются компонентами тензора
относительно локальных $\MG\ML(n,\MR)$ преобразований.

Нетрудно проверить, что
равенство нулю компонент неголономности является необходимым и достаточным
условием локальной разрешимости системы уравнений (\ref{eholba}). Это означает,
что, если компоненты неголономности равны нулю в некоторой области, то для
любой точки из этой области существует окрестность, в которой можно выбрать
такую систему координат, что базис станет голономным $e_a=\pl_a$.

Использование неголономного базиса вместо координатного бывает значительно
удобнее и часто используется в приложениях. Поэтому получим основные формулы
дифференциальной геометрии в неголономном базисе.

Произвольное векторное поле можно разложить как по координатному,
так и по некоординатному (неголономному) базису
$$
  X=X^\al\pl_\al=X^ae_a,
$$
где $X^\al=e^\al{}_a X^a$ и $X^a=X^\al e_\al{}^a$. Предположим, что переход от
греческих индексов к латинским и наоборот у компонент тензорных полей
произвольного ранга всегда осуществляется с помощью компонент репера и корепера.
При этом все симметрии относительно перестановок индексов, конечно, сохраняются.
Компоненты метрики в неголономном базисе имеют вид
\begin{equation}                                                  \label{emeunb}
  g_{ab}=e^\al{}_a e^\bt{}_b g_{\al\bt}.
\end{equation}
В общем случае компоненты метрики $g_{ab}(x)$ зависят от точки многообразия.
Метрика $g_{ab}$ всегда имеет ту же сигнатуру, что и метрика $g_{\al\bt}$,
т.к.\ матрица $e^\al{}_a$ невырождена. Подъем и опускание греческих и латинских
индексов осуществляется с помощью метрик $g_{\al\bt}$ и $g_{ab}$,
соответственно.

Как правило, репер используют в тех случаях, когда матрица $g_{ab}$ является
диагональной и постоянной, а на диагонали расположены плюс и минус единицы:
\begin{equation*}
  g_{ab}=\eta_{ab}=\diag(\underbrace{+\dots+}_p\underbrace{-\dots-}_q).
\end{equation*}
Локально такой репер существует, поскольку уравнение (\ref{emeunb}) при
одинаковых сигнатурах метрик $g_{ab}$ и $g_{\al\bt}$ всегда разрешимо
относительно репера. Такой репер называется {\em ортонормальным} и определен с
точностью до $\MO(p,q)$ вращений.
\index{Ортонормальный репер (orthonormal frame)}%
\index{Репер ортонормальный (orthonormal frame)}%
Ортонормальный базис часто бывает более удобным, т.к.\ метрика в этом базисе
постоянна.

Для римановой метрики множество реперов делится на два класса: с положительным и
отрицательным определителем. Для многообразий с метрикой лоренцевой сигнатуры
множество реперов можно разбить на четыре класса, по числу несвязных компонент
группы Лоренца (см.\ раздел \ref{spungr}).

Компоненты тензоров второго и более высокого рангов могут содержать одновременно
и греческие и латинские индексы. По построению, ковариантная производная от
компонент такого тензора содержит по одному слагаемому с аффинной связностью
для каждого греческого индекса и одному слагаемому с линейной связностью для
каждого латинского индекса. Из определения (локальной формы) линейной связности
(\ref{edecol}) следует взаимно однозначная связь между линейной и аффинной
связностью:
\begin{equation}                                                  \label{eliafc}
  \om_{\al a}{}^b=\Gamma_{\al\bt}{}^\g e^\bt{}_a e_\g{}^b
  -\pl_\al e_\bt{}^b e^\bt{}_a.
\end{equation}
Эту формулу можно переписать в виде равенства нулю ковариантной производной
от компонент корепера:
\begin{equation}                                                  \label{erelal}
  \nb_\al e_\bt{}^a=\pl_\al e_\bt{}^a-\Gamma_{\al\bt}{}^\g e_\g{}^a
  +e_\bt{}^b\om_{\al b}{}^a=0.
\end{equation}
Отсюда следует, что ковариантная производная от компонент репера также
обращается в нуль
\begin{equation*}
  \nb_\al e^\bt{}_a=\pl_\al e^\bt{}_a+\Gamma_{\al\g}{}^\bt e^\g{}_a
  -\om_{\al a}{}^b e^\bt{}_b=0.
\end{equation*}
Тогда, используя правило Лейбница, можно свободно переходить от греческих
индексов к латинским и наоборот под знаком ковариантного дифференцирования:
\begin{equation*}
\begin{split}
  \nb_\al X^a&=\nb_\al(X^\bt e_\bt{}^a)=(\nb_\al X^\bt) e_\bt{}^a,
\\
  \nb_\al X_a&=\nb_\al(X_\bt e^\bt{}_a)=(\nb_\al X_\bt) e^\bt{}_a,
\end{split}
\end{equation*}
где
\begin{equation}                                                  \label{ecovan}
\begin{split}
  \nb_\al X^a&=\pl_\al X^a+\om_{\al b}{}^a X^b,
\\
  \nb_\al X_a&=\pl_\al X_a-\om_{\al a}{}^b X_b
\end{split}
\end{equation}
-- ковариантные производные от компонент векторного поля относительно
неголономного базиса.

Если аффинная связность не является метрической, то операция подъема и опускания
индексов с помощью метрик $g_{\al\bt}$ и $g_{ab}$ не коммутирует с ковариантной
производной.
\begin{com}
Формулу (\ref{eliafc}) можно рассматривать, как калибровочное преобразование
(вращение) $\Gamma\mapsto\om$ в касательном пространстве, которое совпадает
с преобразованием калибровочных полей Янга--Миллса (\ref{qymtrf}). При этом
репер $e_\al{}^a\in\MG\ML(n,\MR)$ играет роль матрицы преобразования
(локального вращения), а координаты многообразия не затрагиваются.
\qed\end{com}
В координатном базисе преобразование координат сопровождается преобразование
компонент тензорных полей. Введение репера позволяет отделить преобразование
координат от преобразований в касательном пространстве. Это достигается путем
введения $n^2$ новых полей $e_\al{}^a(x)$. В результате появляется
дополнительная возможность совершать локальные $\MG\ML(n,\MR)$ преобразования,
зависящие также от $n^2$ функций, в касательном пространстве, не затрагивая
координат многообразия. Очевидно, что всегда можно совершить такое
преобразование, что в результате репер совпадет с координатным базисом
$e_\al{}^a=\dl_\al^a$. В этом случае линейная связность совпадет с аффинной
$\om_{\al a}{}^b=\Gamma_{\al a}{}^b$, а выражения для кривизны (\ref{euncur}) и
кручения (\ref{euntot}) перейдут в уже знакомые формулы аффинной геометрии,
т.к.\ компоненты неголономности обратятся в нуль: $c_{ab}{}^c=0$.

Формулы для кривизны (\ref{ecucav}) и кручения (\ref{ecurcv}) содержат два
греческих индекса. Эти индексы также можно преобразовать в неголономные.
Простые вычисления приводят к следующим компонентам тензора кривизны и
кручения в неголономном базисе:
\begin{align}                                                     \label{euncur}
  R_{abc}{}^d&=\pl_a\om_{bc}{}^d-\pl_b\om_{ac}{}^d
  -\om_{ac}{}^e\om_{be}{}^d+\om_{bc}{}^e\om_{ae}{}^d
  -c_{ab}{}^e\om_{ec}{}^d,
\\                                                                \label{euntot}
  T_{ab}{}^c&=\om_{ab}{}^c-\om_{ba}{}^c-c_{ab}{}^c,
\end{align}
где $\om_{ab}{}^c:=e^\al{}_a\om_{\al b}{}^c$ и $\pl_a:=e^\al{}_a\pl_\al$. Эти
формулы также часто используются в приложениях, особенно тогда, когда компоненты
линейной связности $\om_{abc}$ являются постоянными относительно некоторого
неголономного базиса. В разделе \ref{sliegr} мы используем их для вычисления
тензора кручения и кривизны групп Ли.

Если на многообразии задана метрика, то ковариантные компоненты тензора
кривизны в неголономном базисе равны
\begin{equation}                                                  \label{ecurcn}
  R_{abcd}=\pl_a\om_{bcd}-\pl_b\om_{acd}-\om_{bc}{}^e\om_{ade}
  +\om_{ac}{}^e\om_{bde}-c_{ab}{}^e\om_{ecd}.
\end{equation}
В этом случае линейную связность, аналогично аффинной, можно выразить через
репер, кручение и неметричность. Из уравнения (\ref{elicon}) и определения
(\ref{eliafc}) следует выражение для линейной связности со всеми неголономными
индексами:
\begin{align}                                                     \nonumber
  \om_{abc}&=\frac12(\pl_a g_{bc}+\pl_b g_{ca}-\pl_c g_{ab})
  +\frac12(c_{abc}-c_{bca}+c_{cab})
\\                                                                \label{elicou}
  &+\frac12(T_{abc}-T_{bca}+T_{cab})+\frac12(Q_{abc}+Q_{bca}-Q_{cab}),
\end{align}
где $c_{abc}:=c_{ab}{}^d g_{dc}$.

Если (псевдо-) риманово многообразие допускает векторное поле Киллинга
$K=K^a e_a$, то уравнение Киллинга в неголономном базисе принимает вид
\begin{align}                                                     \label{eunbak}
  \widetilde\nb_a K_b&+\widetilde\nb_b K_a=0,
\\ \intertext{где}
  \widetilde\nb_a K_b&=e^\al{}_a\pl_\al K_b-\widetilde\om_{ab}{}^c K_c,
  \qquad K_b:=K^a g_{ab}.
\end{align}
Как и ранее знак тильды означает, что связность построена по метрике при нулевом
кручении и неметричности. Эта формула следует из того, что переход между
индексами можно проводить под знаком ковариантного дифференцирования.

В (псевдо-)римановой геометрии кручение и неметричность равны нулю. Рассмотрим
ортонормальные реперы, для которых $g_{ab}=\dl_{ab}$ или $\eta_{ab}$, если
сигнатура отличается от евклидовой. Такие реперы определены с точностью до
локальных $\MO(n)$ вращений (или $\MO(p,q)$ вращений, $p+q=n$, для неевклидовой
сигнатуры). Тогда из формулы (\ref{elicou}) следует выражение для
соответствующей $\MS\MO(n)$- или $\MS\MO(p,q)$-связности через компоненты
 неголономности:
\begin{equation}                                                  \label{elocac}
  \widetilde\om_{abc}=\frac12(c_{abc}-c_{bca}+c_{cab}).
\end{equation}
\section{Тождества Бианки                                        \label{sbianc}}
Тождества Бианки играют большую роль в приложениях и поэтому в настоящем разделе
мы рассмотрим их в компонентах. Громоздкие, но прямые вычисления позволяют
записать тождества (\ref{etoste}) и (\ref{ecusre}) после спуска на базу в
переменных Картана:
\begin{align}                                                     \label{eBianc}
  \nb_\al R_{\bt\g a}{}^b+\nb_\bt R_{\g\al a}{}^b
  +\nb_\g R_{\al\bt a}{}^b&=T_{\al\bt}{}^\dl R_{\g\dl a}{}^b
  +T_{\bt\g}{}^\dl R_{\al\dl a}{}^b+T_{\g\al}{}^\dl R_{\bt\dl a}{}^b,
\\                                                                \nonumber
  \nb_\al T_{\bt\g}{}^a+\nb_\bt T_{\g\al}{}^a+\nb_\g T_{\al\bt}{}^a
  &=T_{\al\bt}{}^\dl T_{\g\dl}{}^a+T_{\bt\g}{}^\dl T_{\al\dl}{}^a
  +T_{\g\al}{}^\dl T_{\bt\dl}{}^a
\\                                                                \label{ebiant}
  &+R_{\al\bt\g}{}^a+R_{\bt\g\al}{}^a+R_{\g\al\bt}{}^a.
\end{align}
\index{Тождества Бианки (Bianchi identities)}%
\index{Бианки тождества (Bianchi identities)}%
В аффинной геометрии их можно свернуть, соответственно, с репером $e^\g{}_b$
и $e^\g{}_a$:
\begin{equation*}
\begin{split}
  \nb_\al R_{\bt a}-\nb_\bt R_{\al a}
  +\nb_\g R_{\al\bt a}{}^\g&=-T_{\al\bt}{}^\dl R_{\dl a}
  -T_{\al\g}{}^\dl R_{\al\dl a}{}^\g+T_{\bt\g}{}^\dl R_{\al\dl a}{}^\g,
\\                                                                \label{eBiant}
  \nb_\al T_\bt-\nb_\bt T_\al-\nb_\g T_{\al\bt}{}^\g&=
  T_{\al\g}{}^\dl T_{\bt\dl}{}^\g-T_{\bt\g}{}^\dl T_{\al\dl}{}^\g
  -T_{\al\bt}{}^\g T_\g+R_{\al\bt}-R_{\bt\al}-R_{\al\bt\g}{}^\g.
\end{split}
\end{equation*}
В геометрии Римана--Картана, когда выполнено условие метричности, последнее
слагаемое в последнем тождестве обращается в нуль. Тогда свернутые тождества
Бианки можно переписать в виде
\begin{align}                                                     \label{ecobic}
  \nb_\al R_{\bt\g}-\nb_\bt R_{\al\g}+\nb_\dl R_{\al\bt\g}{}^\dl
  &=-T_{\al\bt}{}^\dl R_{\dl\g}-T_{\al\dl}{}^\e R_{\bt\e\g}{}^\dl
  +T_{\bt\dl}{}^\e R_{\al\e\g}{}^\dl,
\\                                                                \label{ecodit}
  \nb_\al T_\bt-\nb_\bt T_\al-\nb_\g T_{\al\bt}{}^\g&=
  T_{\al\g}{}^\dl T_{\bt\dl}{}^\g-T_{\bt\g}{}^\dl T_{\al\dl}{}^\g
  -T_{\al\bt}{}^\g T_\g+R_{\al\bt}-R_{\bt\al},
\end{align}
где мы перешли ко всем греческим индексам.

Из последнего тождества (\ref{ecodit}) следует, что антисимметричная часть
тензора Риччи в (псевдо-)римановой геометрии (при нулевом кручении) всегда равна
нулю: $\widetilde R_{\al\bt}=\widetilde R_{\bt\al}$.

Далее, свертка (\ref{ecobic}) с $g^{\al\g}$ (при нулевом тензоре неметричности)
приводит к равенству
\begin{equation}                                                  \label{ecobir}
  2\nb_\bt R_\al{}^\bt-\nb_\al R=2T_{\al\bt\g}R^{\g\bt}
  -R_{\al\bt\g\dl}T^{\g\dl\bt}.
\end{equation}
В (псевдо-)римановой геометрии это тождество упрощается:
\begin{equation}                                                  \label{ebieit}
  \widetilde\nb_\bt G^{\bt\al}=0,
\end{equation}
где
\begin{equation*}
  G_{\al\bt}:=\widetilde R_{\al\bt}-\frac12 g_{\al\bt}\widetilde R
\end{equation*}
-- тензор Эйнштейна, построенный из тензора Риччи и скалярной кривизны при
нулевом кручении и тензоре неметричности.

Тождество (\ref{ecodit}) антисимметрично по индексам $\al\bt$ и
поэтому свертка с $g^{\al\bt}$ приводит к тождеству $0=0$.
\chapter{Криволинейные координаты в $\MR^3$                      \label{scnhyu}}
Формализм дифференциальной геометрии, развитый в предыдущих разделах, становится
полезным и естественным даже в евклидовом пространстве, если вычисления
проводятся в криволинейных системах координат. В настоящем главе мы
продемонстрируем это на примере сферической и цилиндрической систем координат в
трехмерном евклидовом пространстве $\MR^3$. В частности, понятие метрики и
символов Кристоффеля чрезвычайно полезно при нахождении явного вида ковариантных
дифференциальных операторов, которые часто используются в приложениях.
\section{Сферические координаты                                  \label{spheco}}
\begin{figure}[h,b,t]
\hfill\includegraphics[width=.4\textwidth]{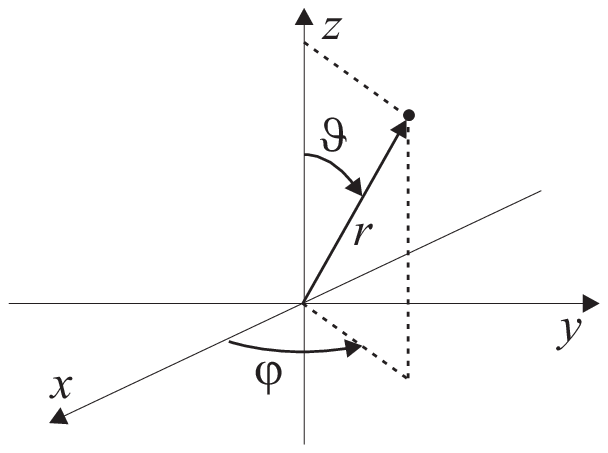}
\hfill {}
\centering\caption{Сферические координаты $r,\theta,\vf$ в трехмерном евклидовом
пространстве $\MR^3$.}
\label{fsphco}
\end{figure}

Рассмотрим {\em сферические координаты} $r,\theta,\vf$ в трехмерном евклидовом
\index{Сферические координаты (spherical coordinates)}%
\index{Координаты сферические (spherical coordinates)}%
пространстве $\MR^3$ с декартовыми координатами $x,y,z$, рис.~\ref{fsphco}.
Функции перехода от сферических координат к декартовым имеют хорошо известный
вид
\begin{equation}                                                  \label{esphct}
\begin{split}
  x &= r\sin\theta\cos\vf,\\
  y &= r\sin\theta\sin\vf,\\
  z &= r\cos\theta,
\end{split}
\end{equation}
и определены при всех значениях $r,~\theta,~\vf$. Углы $\theta$ и $\vf$
называются соответственно {\em азимутальным} и {\em полярным}.
\index{Азимутальный угол (azimuth angle)}%
\index{Угол азимутальный (azimuth angle)}%
\index{Полярный угол (polar angle)}\index{Угол полярный (polar angle)}%
Якобиан этого преобразования легко вычислить
\begin{equation}                                                  \label{ejspco}
  J=r^2\sin\theta.
\end{equation}
Отсюда следует, что преобразование координат (\ref{esphct}) вырождено при $r=0$
или $\theta=0$, $\pi$, т.е.\ на оси $z$. Действительно, обратные преобразования,
\begin{equation}                                                  \label{eispct}
\begin{split}
  r      &= \sqrt{x^2+y^2+z^2},\\
  \theta &= \arctg\displaystyle\frac{\sqrt{x^2+y^2}}z,\\
  \vf     &= \arctg\displaystyle\frac yx,
\end{split}
\end{equation}
однозначно определены всюду, кроме оси $z$. При этом все евклидово пространство,
из которого удалена полуплоскость $y=0$, $x\ge0$, включающая ось $z$, взаимно
однозначно отображается на открытую область
\begin{equation}                                                  \label{edefdo}
  0<r<\infty,\qquad 0<\theta<\pi,\qquad 0<\vf<2\pi,
\end{equation}
Если полуплоскость не удалять, то точки с координатами $\vf$ и $\vf+2\pi$
необходимо отождествить.

Переход к сферическим координатам сохраняет ориентацию, поскольку якобиан
перехода (\ref{ejspco}) положителен в области (\ref{edefdo}).

Из формул (\ref{esphct}) следуют правила преобразования дифференциалов:
\begin{equation}                                                  \label{espdic}
\begin{split}
  dx &=\sin\theta\cos\vf dr+r\cos\theta\cos\vf d\theta
  -r\sin\theta\sin\vf d\vf,
\\
  dy &=\sin\theta\sin\vf dr+r\cos\theta\sin\vf d\theta
  +r\sin\theta\cos\vf d\vf,
\\
  dz &=\cos\theta dr-r\sin\theta d\theta.
\end{split}
\end{equation}
Подставляя эти выражения в евклидову метрику
\begin{equation}                                                  \label{euctin}
  ds^2=dx^2+dy^2+dz^2,
\end{equation}
получим следующее выражение для метрики в сферической системе координат
\begin{equation}                                                  \label{euspin}
  ds^2=dr^2+r^2d\Om,
\end{equation}
где
\begin{equation}                                                  \label{eangdi}
  d\Om:=d\theta^2+\sin^2\theta d\vf^2
\end{equation}
-- дифференциал телесного угла. Отсюда следует, что компоненты евклидовой
метрики в сферической системе координат
\begin{equation}                                                  \label{eucmsp}
  g_{\al\bt}=\begin{pmatrix}
  1 & 0 & 0 \\ 0 & r^2 & 0 \\ 0 & 0 & r^2\sin^2\theta
  \end{pmatrix}
\end{equation}
являются функциями, а не константами. Отметим, что определитель метрики равен
квадрату якобиана преобразования координат, $\det g_{\al\bt}=J^2$.

Помимо инвариантной квадратичной формы декартовых дифференциалов (\ref{euctin})
существует еще одна линейно с ней независимая квадратичная форма
$$
  (xdx+ydy+zdz)^2,
$$
инвариантная относительно $\MS\MO(3)$ вращений. В сферических координатах также
имеются две независимые инвариантные квадратичные формы, которые даются,
например, выражениями $dr^2$ и $d\Om$. Между этими формами существует связь:
\begin{equation}                                                  \label{erelqw}
\begin{split}
  dr^2&=\frac1{r^2}(xdx+ydy+zdz)^2,
\\
  d\Om&=\frac1{r^2}(dx^2+dy^2+dz^2)-\frac1{r^4}(xdx+ydy+zdz)^2.
\end{split}
\end{equation}

Обсудим, какие структуры существуют в евклидовом пространстве $\MR^3$,
инвариантные относительно $\MS\MO(3)$ вращений. Общая теория инвариантных
структур будет рассмотрена позже в разделе \ref{sinvst}.

Из декартовых координат можно составить только одну функционально независимую
сферически симметричную комбинацию -- это радиус $r:=\sqrt{x^2+y^2+z^2}$.
Нетрудно показать, что любая сферически симметричная функция или плотность $h$
может зависеть только от радиуса, $h=h(r)$.

Сферически симметричное векторное поле в сферической системе координат имеет
только одну нетривиальную компоненту, направленную по радиусу,
$$
  X=\lbrace X^r(r):=f(r),0,0\rbrace.
$$
В декартовой системе координат оно имеет вид
$$
  X=\lbrace xf,yf,xf\rbrace
$$
или
$$
  X^i=x^if(r),
$$
где $\lbrace x^i\rbrace=\lbrace x,y,z\rbrace$, $i=1,2,3$, а $f(r)$ -- некоторая
функция только от радиуса.

Инвариантный симметричный тензор второго ранга удобно записывать в декартовой
системе координат, выделив явно след и бесследовую часть (неприводимые
компоненты):
$$
  X^{ij}=\dl^{ij}f_1(r)+\left(\dl^{ij}-3\frac{x^ix^j}{r^2}\right)f_2(r),
$$
где $\dl^{ij}$ -- евклидова метрика и $f_{1,2}$ -- некоторые функции. Первое
слагаемое в этом разложении пропорционально следу тензора $X^{ij}\dl_{ij}=3f_1$,
а второе слагаемое имеет нулевой след. То есть сферически инвариантный тензор
второго ранга взаимно однозначно определяется двумя функциями только от радиуса:
$f_1(r)$ и $f_2(r)$.

Антисимметричный тензор второго ранга $Y^{ij}=-Y^{ji}$, инвариантный
относительно полной группы вращений $\MO(3)$, определяется одной псевдоскалярной
функцией $f^*(r)$ от радиуса:
$$
  Y^{ij}=\ve^{ijk}x_k f^*(r),
$$
где $\ve^{ijk}$ -- полностью антисимметричный псевдотензор третьего ранга. При
отражении одной из декартовых осей координат $\ve^{ijk}$ и $f^*$ меняют знаки,
оставляя антисимметричный тензор без изменения.

Дифференциал телесного угла (\ref{eangdi}) задает метрику на сфере единичного
радиуса с центром в начале координат. С топологической точки зрения евклидово
пространство $\MR^3$ с выколотым началом координат является прямым произведением
сферы $\MS^2$ и положительной полуоси $\MR_+$:
$$
  \MR^3\backslash\lbrace0\rbrace=\MS^2\times\MR_+.
$$

Приведем явные формулы преобразования частных производных (координатного базиса
векторных полей), которые часто используются в приложениях:
\begin{align*}
  \pl_r&=\sin\theta\cos\vf\pl_x+\sin\theta\sin\vf\pl_y+\cos\theta\pl_z,
\\
  \pl_\theta&=r\cos\theta\cos\vf\pl_x+r\cos\theta\sin\vf\pl_y
  -r\sin\theta\pl_z,
\\
  \pl_\vf&=-r\sin\theta\sin\vf\pl_x+r\sin\theta\cos\vf\pl_y,
\\ \intertext{и их обратные:}
  \pl_x&=\sin\theta\cos\vf\pl_r+\frac{\cos\theta\cos\vf}r\pl_\theta
  -\frac{\sin\vf}{r\sin\theta}\pl_\vf,
\\
  \pl_y&=\sin\theta\sin\vf\pl_r+\frac{\cos\theta\sin\vf}r\pl_\theta
  +\frac{\cos\vf}{r\sin\theta}\pl_\vf,
\\
  \pl_z&=\cos\theta\pl_r-\frac{\sin\theta}r\pl_\theta.
\end{align*}

В приложениях тензорные поля часто рассматриваются в ортонормированном базисе
векторных полей и 1-форм:
\begin{equation}                                                  \label{eortba}
\begin{aligned}
  e_r&:=\pl_r, &\qquad e_\theta&:=\frac1r\pl_\theta, &\qquad
  e_\vf&:=\frac1{r\sin\theta}\pl_\vf,
\\
  e^r&:=dr, & e^\theta&:=rd\theta, & e^\vf&:=r\sin\theta d\vf,
\end{aligned}
\end{equation}
что следует непосредственно из вида метрики (\ref{euspin}). Этому базису
соответствуют следующие компоненты ортонормированного репера и его обратного:
\begin{equation}                                                  \label{eorvei}
\begin{aligned}
  e_r{}^{\hat r}&=1, &\qquad e_\theta{}^{\hat\theta}&=r, &\qquad
  e_\vf{}^{\hat\vf}&=r\sin\theta,
\\
  e^r{}_{\hat r}&=1, & e^\theta{}_{\hat\theta}&=\frac1r, &
  e^\vf{}_{\hat\vf}&=\frac1{r\sin\theta}.
\end{aligned}
\end{equation}
При этом все недиагональные компоненты репера равны нулю. Здесь и далее шляпками
обозначены индексы относительно этого репера. Векторное поле можно разложить как
по координатному, так и по ортонормированному базису:
$$
  X=X^r\pl_r+X^\theta\pl_\theta+X^\vf\pl_\vf
  =\hat X^re_r+\hat X^\theta e_\theta+\hat X^\vf e_\vf,
$$
причем переход от компонент $X^\al$ к $\hat X^\al$ осуществляется с помощью
репера. Отметим, что, поскольку репер ортонормирован, то компоненты векторных
полей и соответствующих 1-форм совпадают:
\begin{equation*}
  \hat X_r=\hat X^r,\qquad \hat X_\theta=\hat X^\theta,\qquad \hat X_\vf
  =\hat X^\vf.
\end{equation*}

Получим явные формулы для градиента, ротора, дивергенции и лапласиана в
сферической системе координат, которые используются в задачах со сферической
симметрией. Чтобы получить явное выражение для инвариантных дифференциальных
операторов в сферической системе координат нам понадобятся символы Кристоффеля
(\ref{echris}). Они имеют девять ненулевых компонент:
\begin{align}                                                          \nonumber
  \Gamma_{r\theta}{}^\theta&=\Gamma_{\theta r}{}^\theta=\frac1r, &
  \Gamma_{\theta\vf}{}^\vf&=\Gamma_{\vf\theta}{}^\vf=\frac{\cos\theta}{\sin\theta},
\\                                                                \label{espchs}
  \Gamma_{r\vf}{}^\vf&=\Gamma_{\vf r}{}^\vf=\frac1r, &
  \Gamma_{\vf\vf}{}^r&=-r\sin^2\theta,
\\                                                                     \nonumber
  \Gamma_{\theta\theta}{}^r&=-r, &
  \Gamma_{\vf\vf}{}^\theta&=-\sin\theta\cos\theta.
\end{align}

Из условия равенства нулю ковариантной производной репера (\ref{eorvei})
\begin{equation}                                                  \label{ecovve}
  \nb_\al e_\bt{}^a=\pl_\al e_\bt{}^a-\Gamma_{\al\bt}{}^\g e_\g{}^a
  +\om_{\al b}{}^a e_\bt{}^b=0,
\end{equation}
где $\om_{\al b}{}^a$ -- $\MS\MO(3)$-связность (без кручения), находим, что у
$\MS\MO(3)$-связности только 6 компонент отличны от нуля:
\begin{equation}                                                  \label{esotco}
\begin{split}
  \om_{\theta\hat r}{}^{\hat\theta}&=-\om_{\theta\hat\theta}{}^{\hat r}=1,
\\
  \om_{\vf\hat r}{}^{\hat\vf}&=-\om_{\vf\hat\vf}{}^{\hat r}=\sin\theta,
\\
  \om_{\vf\hat\theta}{}^{\hat\vf}&=-\om_{\vf\hat\vf}{}^{\hat\theta}=\cos\theta.
\end{split}
\end{equation}

Градиент скалярного поля $f$ дает 1-форму:
\begin{equation}                                                  \label{egrsca}
  d f=e^r\pl_rf+e^\theta\frac1r\pl_\theta f+e^\vf\frac1{r\sin\theta}\pl_\vf f.
\end{equation}
Для получения этой формулы в дифференциале
$df=dr\pl_r f+d\theta\pl_\theta f+d\vf\pl_\vf f$ дифференциалы $dr$, $d\theta$ и
$d\vf$ необходимо заменить на ортонормированный базис $e^r$, $e^\theta$ и
$e^\vf$ с помощью соотношений (\ref{eortba}).

Дивергенция вектора (\ref{edivrf}) дает функцию
\begin{equation}                                                  \label{edivsp}
  \nb_\al X^\al=\frac1{r^2}\pl_r(r^2\hat X^r)
  +\frac1{r\sin\theta}\pl_\theta(\sin\theta\hat X^\theta)
  +\frac1{r\sin\theta}\pl_\vf\hat X^\vf.
\end{equation}
Эта и последующие формулы получаются подстановкой символов Кристоффеля
(\ref{espchs}) в ковариантные производные и выражением компонент вектора в
координатном базисе через его компоненты в ортонормированном базисе.

В инвариантном виде ротор 1-формы $A=dx^\al A_\al$ определяется оператором
Ходжа (\ref{estard}) и внешним дифференцированием (\ref{exdeof}):
\begin{equation}                                                  \label{erocac}
  \rot A=\ast dA=e^r\hat B_r+e^\theta\hat B_\theta+e^\vf\hat B_\vf,
\end{equation}
где компоненты разложения задаются формулами:
\begin{equation}
\begin{split}
  \hat B_r&=\hat B^r=\frac1{r\sin\theta}\pl_\theta(\sin\theta\hat A_\vf)
  -\frac1{r\sin\theta}\pl_\vf\hat A_\theta,
\\
  \hat B_\theta&=\hat B^\theta=\frac1{r\sin\theta}\pl_\vf\hat A_r
  -\frac1r\pl_r(r\hat A_\vf),
\\
  \hat B_\vf&=\hat B^\vf=\frac1r\pl_r(r\hat A_\theta)
  -\frac1r\pl_\theta\hat A_r.
\end{split}
\end{equation}

Лапласиан функции (\ref{elassr}) дает также функцию:
\begin{equation}                                                  \label{elacac}
  \triangle f=\frac1{r^2}\pl_r(r^2\pl_r f)
  +\frac1{r^2\sin\theta}\pl_\theta(\sin\theta\pl_\theta f)
  +\frac1{r^2\sin^2\theta}\pl^2_\vf f.
\end{equation}
Лапласиан 1-формы снова дает 1-форму:
$\triangle\hat A_\al=g^{\bt\g}\widetilde\nb_\bt\widetilde\nb_\g\hat A_\al$.
После громоздких вычислений можно получить явные выражения для компонент:
\begin{equation}                                                  \label{elaofs}
\begin{split}
  \triangle\hat A_r&=\frac1{r^2}\pl_r(r^2\pl_r\hat A_r)
  +\frac1{r^2\sin\theta}\pl_\theta(\sin\theta\pl_\theta\hat A_r)
  +\frac1{r^2\sin^2\theta}\pl^2_\vf\hat A_r-\frac2{r^2}\hat A_r
\\
  &-\frac2{r^2\sin\theta}\pl_\theta(\sin\theta\hat A_\theta)
  -\frac2{r^2\sin\theta}\pl_\vf\hat A_\vf,
\\
  \triangle\hat A_\theta&=\frac1r\pl^2_r(r\hat A_\theta)
  +\frac1{r^2\sin\theta}\pl_\theta(\sin\theta\pl_\theta\hat A_\theta)
  +\frac1{r^2\sin^2\theta}\pl^2_\vf\hat A_\theta
  -\frac1{r^2\sin^2\theta}\hat A_\theta
\\
  &+\frac2{r^2}\pl_\theta\hat A_r
  -\frac{2\cos\theta}{r^2\sin^2\theta}\pl_\vf\hat A_\vf,
\\
  \triangle\hat A_\vf&=\frac1r\pl^2_r(r\hat A_\vf)
  +\frac1{r^2\sin\theta}\pl_\theta(\sin\theta\pl_\theta\hat A_\vf)
  +\frac1{r^2\sin^2}\pl^2_\vf\hat A_\vf
  -\frac1{r^2\sin^2\theta}\hat A_\vf
\\
  &+\frac{2\cos\theta}{r^2\sin^2\theta}\pl_\vf\hat A_\theta
  +\frac2{r^2\sin\theta}\pl_\vf\hat A_r.
\end{split}
\end{equation}
Отметим появление дополнительных слагаемых по сравнению с лапласианом функции
(\ref{elacac}).

Приведенные выше формулы для дифференциальных операторов можно также получить
путем элементарных, но громоздких геометрических построений. Такой подход
эквивалентен введенному ранее понятию ковариантной производной. Это показывает,
что дифференциальная геометрия предоставляет естественный и конструктивный
подход к получению явного вида дифференциальных операторов в различных
криволинейных системах координат. Разумеется, геометрия пространства при этом
остается евклидовой.
\section{Цилиндрические координаты}
\begin{figure}[h,b,t]
\hfill\includegraphics[width=.4\textwidth]{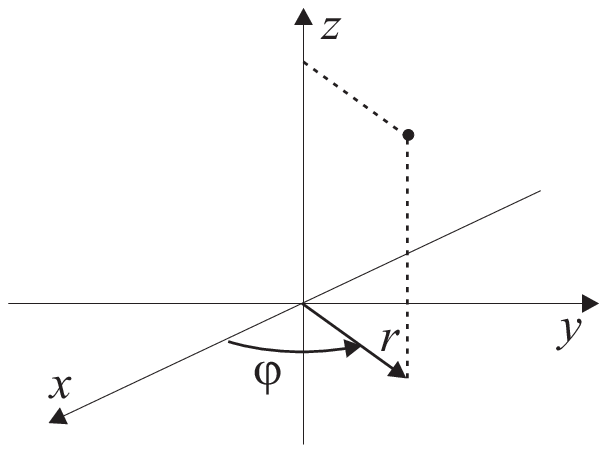}
\hfill {}
\centering\caption{Цилиндрические координаты $r,\vf,z$ в трехмерном евклидовом
пространстве $\MR^3$.}
\label{fcylco}
\end{figure}

Рассмотрим {\em цилиндрические координаты} $r,\vf,z$ в трехмерном евклидовом
пространстве $\MR^3$, рис.~\ref{fcylco}. При переходе к цилиндрическим
координатам мы вводим полярные координаты на плоскости $x,y$, оставляя
координату $z$ без изменений. Функции перехода и их обратные имеют вид (за
третьей координатой $z$ можно не следить):
\begin{align*}
  x&=r\cos\vf, & r&=\sqrt{x^2+y^2},
\\
  y&=r\sin\vf, & \vf&=\arctg(y/x).
\end{align*}
Угол $\vf$, как и в случае сферических координат, называется {\em полярным}.
Якобиан преобразования координат равен $J=r$ и является вырожденным в начале
полярных координат, что соответствует оси $z$. Область определения
цилиндрических координат задается неравенствами:
$$
  0<r<\infty,\qquad 0<\vf<2\pi,\qquad -\infty<z<\infty,
$$
что соответствует евклидову пространству с удаленной полуплоскостью $y=0$,
$x\ge0$.
\index{Цилиндрические координаты (cylindrical coordinates)}%
\index{Координаты цилиндрические (cylindrical coordinates)}%
\index{Полярный угол (polar angle)}\index{Угол полярный (polar angle)}%

Дифференциалы декартовых и полярных координат (координатные базисы 1-форм)
связаны преобразованием:
\begin{equation}                                                  \label{ecyldi}
\begin{aligned}
  dx&=\cos\vf dr-r\sin\vf d\vf, &\qquad
  dr&=\hphantom{-}\cos\vf dx+\sin\vf dy,
\\
  dy&=\sin\vf dr+r\cos\vf d\vf, &
  d\vf&=-\frac{\sin\vf}r dx+\frac{\cos\vf}r dy.
\end{aligned}
\end{equation}
Частные производные в декартовых и полярных координатах (координатные базисы
векторных полей) связаны простыми соотношениями:
\begin{equation}                                                  \label{ecyldd}
\begin{aligned}
  \pl_x&=\cos\vf\pl_r-\frac{\sin\vf}r\pl_\vf, &\qquad
  \pl_r&=\quad \cos\vf\pl_x+\sin\vf\pl_y,
\\
  \pl_y&=\sin\vf\pl_r+\frac{\cos\vf}r\pl_\vf, &
  \pl_\vf&=-r\sin\vf\pl_x+r\cos\vf\pl_y.
\end{aligned}
\end{equation}
Прямая проверка показывает, что справедлива формула
\begin{equation*}
  (xdx+ydy)^2=r^2dr^2.
\end{equation*}

Соотношения (\ref{ecyldi}) приводят к следующему выражению для евклидовой
метрики (\ref{euctin}) в цилиндрических координатах:
$$
  ds^2=dr^2+r^2d\vf^2+dz^2.
$$
В матричных обозначениях метрика и ее обратная имеют вид
$$
  g_{\al\bt}=\begin{pmatrix} 1&0&0\\0&r^2&0\\0&0&1 \end{pmatrix},\qquad
  g^{\al\bt}=\begin{pmatrix} 1&0&0\\0&1/r^2&0\\0&0&1 \end{pmatrix}.
$$
Из символов Кристоффеля только три отличны от нуля:
$$
  \Gamma_{r\vf}{}^\vf=\Gamma_{\vf r}{}^\vf=\frac1r,\qquad \Gamma_{\vf\vf}{}^r=-r.
$$

Ортонормированный базис векторных полей и 1-форм имеет вид (для полярных
координат)
\begin{equation*}
\begin{aligned}
  e_r&=\pl_r,&\qquad e_\vf&=\frac1r\pl_\vf,
\\
  e^r&=dr, & e^\vf&=rd\vf.
\end{aligned}
\end{equation*}
Этому базису соответствует ортонормированный репер
\begin{equation*}
\begin{aligned}
  e_r{}^{\hat r}&=1,&\qquad e_\vf{}^{\hat\vf}&=r,
\\
  e^r{}_{\hat r}&=1, & e^\vf{}_{\hat\vf}&=\frac1r.
\end{aligned}
\end{equation*}
При этом все недиагональные компоненты равны нулю.

Из условия (\ref{ecovve}) находим компоненты $\MS\MO(2)$-связности. Несложные
вычисления показывают, что только две компоненты отличны от нуля:
\begin{equation*}
  \om_{\vf\hat r}{}^{\hat\vf}=-\om_{\vf\hat\vf}{}^{\hat r}=1.
\end{equation*}

Вычисления, аналогичные случаю сферических координат, приводят к следующим
выражениям для простейших дифференциальных операторов в цилиндрических
координатах:
\begin{align}                                                     \label{egracy}
  df&=e^r\pl_rf+e^\vf\frac1r\pl_\vf f+e^z\pl_z f,
\\                                                                \label{edivcy}
  \nb_\al X^\al&=\frac1r\pl_r(r\hat X^r)
  +\frac1r\pl_\vf\hat X^\vf+\pl_z\hat X^z,
\\                                                                \label{erotcy}
  \rot A&=e^r\left[\frac1r\pl_\vf\hat A_z-\pl_z\hat A_\vf\right]
  +e^\vf[\pl_z\hat A_r-\pl_r\hat A_z]
  +e^z\frac1r[\pl_r(r\hat A_\vf)-\pl_\vf\hat A_r],
\\                                                                \label{elapcy}
  \triangle f&=\frac1r\pl_r(r\pl_r f)+\frac1{r^2}\pl^2_\vf f+\pl^2_z f,
\\                                                                \label{elarcy}
  \triangle\hat A_r&=\frac1r\pl_r(r\pl_r\hat A_r)
  +\frac1{r^2}\pl^2_\vf\hat A_r+\pl^2_z\hat A_r
  -\frac1{r^2}\hat A_r-\frac2{r^2}\pl_\vf\hat A_\vf,
\\                                                                \label{elefcy}
  \triangle\hat A_\vf&=\frac1r\pl_r(r\pl_r\hat A_\vf)
  +\frac1{r^2}\pl^2_\vf\hat A_\vf+\pl^2_z\hat A_\vf
  -\frac1{r^2}\hat A_\vf+\frac2{r^2}\pl_\vf\hat A_r,
\\                                                                \label{elazcy}
  \triangle\hat A_z&=\frac1r\pl_r(r\pl_r\hat A_z)
  +\frac1{r^2}\pl^2_\vf\hat A_z+\pl^2_z\hat A_z.
\end{align}

Напомним, что шляпки над символами означают, что компоненты (ко-)векторов
рассматриваются относительно ортонормированного базиса.

\chapter{Группы Ли                                              \label{sliegr}}
Группы Ли образуют один из наиболее важных классов многообразий и имеют широкие
приложения в математической физике. В настоящем разделе мы дадим формальные
определения, опишем основные свойства, а также рассмотрим группы Ли с
дифференциально геометрической точки зрения. В частности, многообразия
полупростых групп Ли будут рассмотрены, как римановы пространства и пространства
абсолютного параллелизма, когда групповая операция отождествляется с
параллельным переносом. В конце главы без доказательств приведена классификация
простых групп Ли.
\section{Группы Ли и локальные группы Ли                         \label{slolir}}
\begin{defn}
Множество элементов $a,b,c,\dotsc\in\MG$ называется {\em группой Ли}, если оно
является одновременно и группой, и гладким многообразием, при этом требуется,
чтобы была определена гладкая групповая операция
\begin{equation}                                                  \label{qskfuy}
  \MG\times\MG\ni\qquad(a,b)\mapsto ab\qquad\in\MG,\qquad \forall a,b\in\MG,
\end{equation}
со следующими свойствами:

1) \parbox[t]{.92\linewidth}{Ассоциативность:
$(ab)c=a(bc),\quad \forall a,b,c\in\MG$.}

2) \parbox[t]{.92\linewidth}{Существование единицы $e\in\MG$:
$ea=ae=a,\quad \forall a\in\MG$.}

3) \parbox[t]{.92\linewidth}{Для каждого элемента $a\in\MG$ существует обратный
элемент $a^{-1}\in\MG$ такой, что $a^{-1}a=aa^{-1}=e$, причем отображение
$\MG\ni a\mapsto a^{-1}\in\MG$ гладкое.\qed}
\end{defn}
\index{Группа Ли (Lie group)}\index{Ли группа (Lie group)}%
\begin{com}
В определении оба условия гладкости отображений:
\begin{align}                                                     \label{eficom}
  \MG\times\MG\ni\qquad(a,b)&\mapsto ab\quad\in\MG,
\\                                                                \label{esecom}
  \MG\ni\quad a&\mapsto a^{-1}\quad\in\MG,
\end{align}
можно объединить в эквивалентное условие гладкости одного отображения:
\begin{equation}                                                  \label{ethcom}
  \MG\times\MG\ni\qquad(a,b)\mapsto ab^{-1}\quad\in\MG.
\end{equation}
То, что (\ref{ethcom}) следует из (\ref{eficom}) и (\ref{esecom}), очевидно.
Обратно. Из (\ref{ethcom}) при $a=e$ следует гладкость отображения
(\ref{esecom}). Гладкость отображения (\ref{eficom}) следует из (\ref{ethcom})
после гладкого отображения  $b^{-1}\mapsto b$.
\qed\end{com}
Каждому элементу группы $b\in\MG$ поставим в соответствие отображение
\begin{equation}                                                  \label{elimap}
  \MG\ni\quad a\mapsto ab\quad\in\MG,\qquad \forall b\in\MG.
\end{equation}
Тогда из определения группы Ли следует, что это отображение взаимно однозначно и
гладко, т.е.\ является диффеоморфизмом $\MG\xrightarrow{b}\MG$. Следовательно,
отображение (\ref{qskfuy}) можно рассматривать как группу преобразований
многообразия $\MG$.

Перефразируем определение: группой Ли является группа, снабженная гладкой
структурой многообразия. При этом не всякая группа допускает существование
такой структуры.

\begin{com}
Если на произвольной группе задать дискретную топологию, то она будет группой
Ли, рассматриваемой, как $0$-мерное многообразие (если отбросить требование
счетности базы топологии). В дальнейшем, чтобы исключить подобные ситуации, при
рассмотрении групп Ли мы не будем рассматривать $0$-мерные многообразия. Поэтому
множество элементов группы Ли не может быть счетным, т.к.\ множество точек
области в $\MR^n$ несчетно.
\qed\end{com}
\begin{exa}
Группа перестановок конечного числа элементов не является группой Ли, т.к.\
содержит конечное число элементов.
\qed\end{exa}

Можно сказать также, что группой Ли является многообразие, на котором
задана групповая операция. При этом не на всяком многообразии можно
определить гладкую бинарную операцию, удовлетворяющую групповым аксиомам.
\begin{exa}
Двумерная сфера $\MS^2$ не может быть наделена групповой структурой.
Действительно, зафиксируем произвольный отличный от нуля вектор в касательном
пространстве к единице группы $\MT_e(\MG)$. С помощью дифференциала отображения
(\ref{elimap}) этот вектор можно разнести по всему групповому многообразию. В
результате получим гладкое векторное поле на $\MG$, которое всюду отлично от
нуля, т.к.\ отображение (\ref{elimap}) является диффеоморфизмом. Это
противоречит теореме \ref{tvecsp}, утверждающей, что на двумерной сфере не
существует непрерывного отличного от нуля векторного поля.
\qed\end{exa}
Поскольку группы Ли являются многообразиями, то на них без изменения
переносится большинство характеристик многообразий.
\begin{defn}
{\em Размерностью} группы Ли называется ее размерность как многообразия.
Группа Ли называется {\em компактной}, если групповое многообразие компактно.
\qed\end{defn}
\index{Размерность группы Ли (dimensionality of a Lie group)}%
\index{Компактная группа Ли (compact Lie group)}%
\index{Группа Ли компактная (compact Lie group)}%

Из непрерывности групповой операции следует, что, если подмножество
$\MU\subset\MG$ -- открыто или замкнуто в $\MG$, то таковым же будет
подмножество $\MU^{-1}$, состоящее из обратных элементов, а также подмножества
$a\MU$ и $\MU a$ для всех $a\in\MG$. Отсюда вытекает, что для открытого
подмножества $\MU$ подмножества $\MV\MU$ и $\MU\MV$ также открыты в $\MG$
каково бы ни было подмножество $\MV\in\MG$.
\begin{com}
В определении группы Ли мы потребовали гладкость т.е.\ бесконечную
дифференцируемость групповой операции и, соответственно, гладкость группы Ли,
как многообразия. На самом деле ситуация следующая. Согласно теореме
Глисона--Монтгомери--Циппина \cite{Gleaso52,MonZip52} на всякой непрерывной
группе (класса $\CC^0$) можно ввести структуру аналитического многообразия
(класса $\CC^\om$), совместимую с групповой структурой. Эта теорема дает
положительное решение пятой проблемы Гильберта: следует ли из существования
в группе $\MG$ каких-нибудь координат существование в ней дифференцируемых
координат. Поэтому, не ограничивая общности, можно считать, что группы Ли
являются вещественно аналитическими многообразиями, а групповая операция --
вещественно аналитическое отображение.
\qed\end{com}
Таким образом, группы Ли представляют собой специальный класс групп, для которых
групповая операция может быть записана в координатной форме. Это важное свойство
позволяет применить методы математического анализа для исследования свойств
групп Ли.

Приведем несколько примеров групп Ли.
\begin{exa}
Прямая $\MR$, на которой определены сдвиги (сложение) $a+b=c$, где
$a,b,c\in\MR$, является одномерной связной односвязной некомпактной абелевой
группой Ли.
\qed\end{exa}
\begin{exa}
Проколатая прямая $\MR_0:=\MR\setminus\lbrace0\rbrace$ является группой Ли по
отношению к умножению, $xy=z$, где $x,y,z\in\MR_0$. Эта группа одномерна абелева
несвязна и состоит из двух компонент. Каждая компонента связности некомпактна.
Связная компонента единицы $\MR_+$ (положительные числа) изоморфна группе
вещественных чисел по сложению, $\MR_+\simeq\MR$. Изоморфизм задается
показательной функцией с произвольным положительным основанием, отличным от
единицы. Например, $x=\ex^a$.
\qed\end{exa}
\begin{exa}
Евклидово пространство $\MR^n$ (как и произвольное векторное пространство)
является $n$-мерной связной односвязной некомпактной абелевой группой Ли
относительно сложения векторов.
\qed\end{exa}
\begin{exa}
Комплексная плоскость с выколотым началом координат
$\MC_0:=\MC\setminus\lbrace 0\rbrace$ является комплексной группой Ли по
отношению к обычному умножению комплексных чисел. Ее комплексная размерность
равна единице. Эта группа некомпактна связна, но неодносвязна. Ее
фундаментальная группа равна $\MZ$ (число обходов вокруг начала координат).
\qed\end{exa}
\begin{exa} Единичная окружность $\MS^1$, точками которой являются комплексные
числа $z=e^{i\vf}$ является абелевой компактной группой Ли по отношению к
умножению. Она обозначается через $\MU(1)$ (унитарные матрицы размера
$1\times1$) и изоморфна группе собственных двумерных вращений $\MS\MO(2)$. Эта
группа связна, но не односвязна. Ее фундаментальная группа изоморфна группе
целых чисел $\MZ$.
\qed\end{exa}
\begin{exa}Трехмерная сфера $\MS^3$ в пространстве кватернионов, точками которой
являются кватернионы $q$, для которых $|q|=1$, может быть снабжена групповой
структурой (см.\ раздел \ref{sthecs}). Это -- трехмерная связная односвязная
неабелева компактная группа Ли, которая изоморфна группе унитарных матриц
размерности $2\times2$ с единичным определителем $\MS\MU(2)$.
\qed\end{exa}
Можно показать, что сферы только двух размерностей $\MS^1$ и $\MS^3$ исчерпывают
все сферы на которых можно задать структуру группы Ли.

В рассмотренных примерах есть две одномерные абелевы группы Ли: $\MR$ и $\MS^1$.
Любая связная одномерная группа Ли является абелевой, и изоморфна либо
вещественной прямой $\MR$, либо окружности $\MS^1$. Этот результат имеет широкие
применения, т.к.\ каждая группа Ли имеет одномерные подгруппы. Если одномерная
группа Ли не является связной, то она может быть неабелевой. Примеры дают группа
вращений $\MO(2)$ и группа Лоренца $\MO(1,1)$ (см.\ разделы \ref{stwodr} и
\ref{stwodl}).

В случае связных одномерных групп Ли можно описать все возможные гомоморфизмы
между ними:
\begin{equation*}
\begin{aligned}
  \MR\ni\quad x&\mapsto ax\quad\in\MR,&&a\in\MR;
\\
  \MR\ni\quad x&\mapsto\ex^{iax}\quad\in\MS^1,&&a\in\MR;
\\
  \MS^1\ni\quad z&\mapsto 0\quad\in\MR;&&
\\
  \MS^1\ni\quad z&\mapsto z^n\quad\in\MS^1,&&n\in\MZ;
\end{aligned}
\end{equation*}
где $x$ -- точка вещественной прямой, $z$ -- комплексное число, равное по
модулю единице, а числа $a$ и $n$ параметризуют гомоморфизмы.

Прямое произведение групп Ли $\MG_1\times\MG_2$ является группой Ли с
дифференцируемой структурой прямого произведения многообразий и
групповой структурой прямого произведения групп.
\begin{exa} Тор $\MT^n:=\underbrace{\MS^1\times\dotsc\times\MS^1}_n
\approx\underbrace{\MU(1)\times\dotsc\times\MU(1)}_n$ для всех $n$ является
$n$-мерной абелевой компактной группой Ли как прямое произведение окружностей.
Эта группа связна, но не односвязна.
\qed\end{exa}
\begin{prop}
Всякая компактная связная абелева группа Ли $\MG$ размерности $n$ изоморфна
$n$-мерному тору $\MT^n$.
\qed\end{prop}
\begin{proof}
См., например, \cite{Warner83R}.
\end{proof}

С топологической точки зрения группы Ли $\MG$ могут состоять из нескольких
компонент связности\footnote{Поскольку мы предполагаем, что многообразие
обладает счетной базой, то число компонент связности у групп Ли может быть
не более, чем счетно.}.
При этом каждая компонента имеет одинаковую размерность, равную размерности
группы. Связная компонента единицы $\MG_0$ является открытой подгруппой в $\MG$
и представляет собой нормальный делитель. Остальные компоненты представляют
собой смежные классы по этой подгруппе, и фактор группа $\MG/\MG_0$ состоит не
более, чем из счетного числа элементов. Если группу Ли рассматривать, как группу
преобразований, то каждая связная компонента представляет собой орбиту
произвольного элемента из этой компоненты относительно действия элементов группы
из связной компоненты единицы $\MG_0$. Ясно, что все компоненты связности группы
Ли диффеоморфны между собой.

Теперь перейдем к координатному описанию.
Поскольку группа Ли является многообразием, то групповое умножение можно
записать в координатах. Рассмотрим связную компоненту единицы $\MG_0$ группы Ли
размерности $\Sn$. Пусть в окрестности единицы задана некоторая система координат
$\lbrace a^{\Sa}\rbrace$, $\Sa=1,\dotsc,\Sn$. Не ограничивая общности, будем
считать, что начало координат $a^\Sa=0$ совпадает с единицей группы. Тогда
групповое умножение
\begin{equation}                                                  \label{ecofut}
  c=ab=f(a,b),\qquad \quad a,b,c\in\MG_0,
\end{equation}
и групповые аксиомы (см.\ раздел \ref{srealn}) можно записать в координатном
виде (в достаточно малой окрестности единицы):
\begin{equation}                                                  \label{ecoruc}
\begin{aligned}
  1)\quad &c^\Sa=f^\Sa(a,b)                  &&\text{-- закон композиции},
\\
  2)\quad &f^\Sa\big(a,f(b,c)\big)=f^\Sa\big(f(a,b),c\big)
                                         &&\text{-- ассоциативность},
\\
  3)\quad &a^\Sa=f^\Sa(a,0)=f^\Sa(0,a)        &&\text{-- существование единицы},
\\
  4)\quad &f^\Sa(a,a^{-1})=f^\Sa(a^{-1},a)=0
  &&\text{-- существование обратного элемента}.
\end{aligned}
\end{equation}
Набор $\Sn$ функций $f^\Sa$ от $2\Sn$ переменных, удовлетворяющих условиям
(\ref{ecoruc}), называется {\em функцией композиции} для группы Ли $\MG$. Она
определена в некоторой окрестности единицы группы.
\index{Функция композиции (composition function)}%
\index{Композиции функция (composition function)}%

Связная компонента группы Ли $\MG_0$ может оказаться нетривиальным многообразием
и не покрываться одной картой. Поэтому введем новое понятие.
\begin{defn}
{\em Локальной группой Ли} называется пара $(\MU,f)$, где $\MU\subset\MR^\Sn$ --
область евклидова пространства, содержащая начало координат, и $f$ -- гладкое
отображение $\MU\times\MU\rightarrow\MR^\Sn$ такое, что выполнены условия
(\ref{ecoruc}). При этом мы предполагаем, что все равенства выполнены в той
области, где определены все выражения, входящие в данное равенство.
\qed\end{defn}
\index{Локальная группа Ли (local Lie group)}%
\index{Группа Ли локальная (local Lie group)}%
\begin{com}
В определении локальной группы Ли мы не предполагаем замкнутости области $\MU$
относительно группового умножения, т.е.\ возможно существование таких точек в
$\MU$, что их произведение не лежит в $\MU$. Конечно, для таких точек про
ассоциативность в определении ничего не говорится.
\qed\end{com}
Каждая группа Ли $\MG$ порождает бесконечное множество локальных групп Ли
следующим образом. Пусть $(\MV,\vf)$ -- карта на $\MG$, содержащая единицу,
которая отображается в начало координат евклидова пространства $\MR^\Sn$.
Возьмем такую окрестность $\MU\subset\MV$, что выполнены включения
$\MU\cdot\MU\subset\MV$ и $\MU^{-1}\subset\MV$, где под произведением
$\MU\cdot\MU$ мы понимаем множество всех произведений $ab$, где $a,b\in\MU$.
Существование такой окрестности следует из непрерывности отображений
$(a,b)\mapsto ab$ и $a\mapsto a^{-1}$. Тогда каждая пара $(\MU,f)$, где мы
отождествили область $\MU\subset\MG$ с ее образом $\vf(\MU)\in\MR^\Sn$, а $f$ --
функция композиции на $\MG$, записанная в координатах, является локальной
группой Ли.
\begin{defn}
Две локальные группы Ли $(\MU_1,f_1)$ и $(\MU_2,f_2)$ называются
{\em изоморфными}, если существуют такие окрестности нуля $\MU_1'\subset\MU_1$
и $\MU_2'\subset\MU_2$ и такой диффеоморфизм $h:~\MU_1'\rightarrow\MU_2'$, что
\index{Изоморфизм локальных групп Ли (isomorphism of local Lie groups)}%
диаграмма
\begin{equation*}
\begin{diagram}
  \MU_1'\times\MU_1' & \rTo^{f_1} \MR^\Sn\supset & ~\MU_1' \\
  \dTo^{h\times h} & & \dTo_h \\
  \MU_2'\times\MU_2' & \rTo^{f_2} \MR^\Sn\supset & ~\MU_2'
\end{diagram}
\end{equation*}
коммутативна, т.е.\ $h\big(f_1(a,b)\big)=f_2\big(h(a),h(b)\big)$ при
$a,b\in\MU_1'$, там, где обе части равенства определены.
\qed\end{defn}
Ясно, что все локальные группы Ли, полученные описанным выше способом из одной
группы Ли, изоморфны между собой. Поэтому введем
\begin{defn}
Две группы Ли $\MG_1$ и $\MG_2$ называются {\em локально изоморфными},
если порождаемые ими локальные группы Ли изоморфны.
\qed\end{defn}
\index{Локальный изоморфизм групп Ли (local isomorphism of Lie groups)}%
Конечно, две изоморфные группы Ли являются также локально изоморфными.
Нетривиальность данного выше определения заключается в том, что существуют
локально изоморфные группы Ли, которые в то же время не изоморфны между собой.
\begin{exa}
Пусть $\MG_1=\MR$ и $\MG_2=\MS^1$. В качестве $\MU_1'\subset\MR$ возьмем
интервал $(-\pi,\pi)$, а в качестве $\MU_2'\subset\MS^1$ -- дополнение к точке
$-1\in\MS^1$. Отображение
\begin{equation*}
  h:\quad \MR\supset\MU_1'\ni\quad x~\mapsto~\ex^{ix}\quad\in\MU_2'\subset\MS^1
\end{equation*}
устанавливает диффеоморфизм между $\MU_1'$ и $\MU_2'$, перестановочный с
умножением. Таким образом, группы Ли $\MR$ и $\MS^1$ локально изоморфны. В то же
время они не изоморфны между собой.
\qed\end{exa}
\begin{prop}
Всякая $\Sn$-мерная абелева группа Ли локально изоморфна группе трансляций
$\Sn$-мерного векторного пространства, т.е.\ евклидову пространству $\MR^\Sn$,
рассматриваемому как группа сдвигов.
\end{prop}
\begin{proof}
См., например, \cite{Pontry84R}.
\end{proof}

Структура группы Ли накладывает жесткие условия на функцию композиции
$f^\Sa$. Как уже было отмечено, функция композиции $f^\Sa(a,b)$ вещественно
аналитична (т.е.\ разлагается в сходящиеся степенные ряды) во всех точках
группового многообразия по всем $2\Sn$ переменным.
\begin{exa}                                                       \label{eglnex}
Рассмотрим группу невырожденных вещественных $n\times n$ матриц $\MG\ML(n,\MR)$
с обычным правилом умножения. Произвольную матрицу $M\in\MG\ML(n,\MR)$
можно представить в виде
\begin{equation*}
  M=\one+\big(X_a{}^b\big),\qquad a,b=1,\dotsc,n,
\end{equation*}
где $\one$ -- единичная матрица и $X_a{}^b$ -- набор $n^2$ чисел (элементы
$n\times n$ матрицы), которые мы примем за координаты матрицы $M$. Полученное
таким образом отображение всей группы $\MG\ML(n,\MR)$ на область евклидова
пространства $\MR^{n^2}$ переводит единицу группы в начало координат. Групповое
многообразие при этом представляет собой открытую область в $\MR^{n^2}$, которая
является дополнением замкнутого множества точек, определяемого уравнением
$\det M=0$. Функция композиции для группы $\MG\ML(n,\MR)$ принимает вид
\begin{equation*}
  f_a{}^b(M,N)=X_a{}^b+Y_a{}^b+X_a{}^c Y_c{}^b,
\end{equation*}
где $N=\one+Y$. Таким образом, функция композиции является аналитической
функцией. Аналогично, множество невырожденных $n\times n$ комплексных матриц
$\MG\ML(n,\MC)$ образует аналитическую группу Ли вещественной размерности
$2n^2$.
\qed\end{exa}

Проанализируем функцию композиции.
Из условия существования единицы (\ref{ecoruc}) следуют равенства:
\begin{equation*}
  \left.\frac{\pl f^\Sa}{\pl a^\Sb}\right|_{b=0}=\dl^\Sa_\Sb,\qquad
  \left.\frac{\pl f^\Sa}{\pl b^\Sb}\right|_{a=0}=\dl^\Sa_\Sb.
\end{equation*}
Из этих уравнений и дифференцируемости функции композиции вытекает, что вблизи
единицы группы функция композиции разрешима относительно координат $a^\Sa$ и
$b^\Sa$ соответствующих элементов группы $cb^{-1}$ и $a^{-1}c$, где $c=f(a,b)$.
При этом класс дифференцируемости функций $a^\Sa(c,b)$ и $b^\Sa(a,c)$ такой же,
как и у функции композиции (бесконечный).

Разложим функцию композиции в ряд Тейлора в окрестности единицы группы с
точностью до членов третьего порядка
\begin{equation}                                                  \label{ecofgl}
  f^\Sa(a,b)=a^\Sa+b^\Sa+a^\Sc b^\Sb\hat f_{\Sb\Sc}{}^\Sa
  +a^\Sd a^\Sc b^\Sb g_{\Sb\Sc\Sd}{}^\Sa
  +a^\Sd b^\Sc b^\Sb h_{\Sb\Sc\Sd}{}^\Sa+\dotsc
\end{equation}
В этом разложении $\hat f_{\Sb\Sc}{}^\Sa$,
$g_{\Sb\Sc\Sd}{}^\Sa=g_{\Sb\Sd\Sc}{}^\Sa$,
$h_{\Sb\Sc\Sd}{}^\Sa=h_{\Sc\Sb\Sd}{}^\Sa$ -- некоторые постоянные. Из
существования единичного элемента следует отсутствие нулевого члена разложения и
вид линейных слагаемых. Слагаемые более высоких порядков обязательно должны
содержать смешанное произведение $a^\Sa b^\Sb$, т.к.\ в противном случае условие
3) в определении функции композиции не может быть выполнено. Подстановка
разложения (\ref{ecofgl}) в условие ассоциативности накладывает условие на
коэффициенты разложения:
\begin{equation}                                                  \label{ethiro}
  \hat f_{\Sb\Sc}{}^\Se\hat f_{\Se\Sd}{}^\Sa+2h_{\Sb\Sc\Sd}{}^\Sa
  =\hat f_{\Sc\Sd}{}^\Se\hat f_{\Sb\Se}{}^\Sa+2g_{\Sb\Sc\Sd}{}^\Sa.
\end{equation}
При антисимметризации этого выражения по индексам $\Sb,\Sc,\Sd$ слагаемые
$h_{\Sb\Sc\Sd}{}^\Sa$ и $g_{\Sb\Sc\Sd}{}^\Sa$ выпадают из-за симметрии
по паре индексов. В результате получаем ограничение на {\em структурные
константы} группы Ли,
\index{Структурные константы (structure constants)}%
\index{Константы структурные (structure constants)}%
\begin{equation}                                                  \label{estrco}
  f_{\Sa\Sb}{}^\Sc:=-\hat f_{\Sa\Sb}{}^\Sc+\hat f_{\Sb\Sa}{}^\Sc,\qquad
  f_{\Sa\Sb}{}^\Sc=-f_{\Sb\Sa}{}^\Sc,
\end{equation}
известные как {\em тождества Якоби}
\index{Тождества Якоби (Jacobi identities)}%
\index{Якоби тождества (Jacobi identities)}%
\begin{equation}                                                  \label{ejatoj}
  f_{\Sa\Sb}{}^\Sd f_{\Sc\Sd}{}^\Se+f_{\Sb\Sc}{}^\Sd f_{\Sa\Sd}{}^\Se
  +f_{\Sc\Sa}{}^\Sd f_{\Sb\Sd}{}^\Se=0,
\end{equation}
где слагаемые отличаются циклической перестановкой индексов $\Sa,\Sb,\Sc$.
При свертке тождеств Якоби по индексам $\Sc,\Se$ последние два слагаемых
сокращаются, и мы получаем тождество
$$
  f_{\Sa\Sb}{}^\Sd f_{\Sd\Sc}{}^\Sc=0.
$$

Для абелевых групп функция композиции симметрична $f^\Sa(a,b)=f^\Sa(b,a)$,
и структурные константы тождественно обращаются в нуль.

Координаты обратного элемента $a^{-1\Sa}(a)$ зависят от элемента $a$. Разложив
это выражение по $a^\Sa$ и подставив в определение обратного элемента, получим
следующее разложение для координат обратного элемента с точностью до членов
третьего порядка:
\begin{equation}                                                  \label{einvel}
  a^{-1\Sa}=-a^\Sa+a^\Sc a^\Sb\hat f_{\Sb\Sc}{}^\Sa
  -\frac12a^\Sd a^\Sc a^\Sb\hat f_{\Sb\Sc}{}^\Se
  (\hat f_{\Sd\Se}{}^\Sa+\hat f_{\Se\Sd}{}^\Sa)+\dotsc.
\end{equation}
При получении этого выражения было использовано соотношение (\ref{ethiro}).
\section{Действие группы слева                                   \label{sleacg}}
Рассмотрим группу Ли как группу преобразований группового многообразия,
действующую слева. При этом каждому элементу $a\in\MG$ ставится в
соответствие отображение
$$
  a:\qquad\MG\ni\quad b~\mapsto~l_a(b):=ab\quad\in\MG.
$$
Это отображение не является автоморфизмом группы Ли $\MG$, т.к.\ нетривиально
действует на единичный элемент. Из единственности единичного элемента в группе
следует, что группа левых преобразований действует свободно и транзитивно (см.\
раздел \ref{stragf}).

Функция $F(a)\in\CC^\infty(\MG)$, заданная на групповом многообразии, будет
инвариантна относительно действия группы, если выполнено условие $F(ab)=F(b)$,
$\forall a\in\MG$, и, следовательно, также для всех $b\in\MG$. Очевидно, что
инвариантные функции на группе тождественно равны константе. Другими словами,
значение инвариантной функции, например, в единице $F(e)$ разносится по всему
групповому многообразию с помощью действия группы слева. При этом группа Ли
$\MG$ может состоять из нескольких компонент.

Перейдем к рассмотрению левоинвариантных векторных полей на групповом
многообразии. Поскольку на группе задано отображение $l_a$, которое является
диффеоморфизмом, то в касательном пространстве определен соответствующий
дифференциал отображения $l_{a*}$:
\begin{equation}                                                 \label{elevei}
  X\big(l_a(b)\big)=l_{a*}X(b),
\end{equation}
который переводит вектор $X(b)$ в точке $b$ в некоторый вектор
$X\big(l_a(b)\big)$ в точке $ab$.
\begin{defn}
Векторное поле $X\in\CX(\MG)$ на группе Ли $\MG$ называется
{\em левоинвариантным}, если выполнено условие (\ref{elevei}) для всех $a\in\MG$
и, следовательно, для всех $b\in\MG$.
\qed\end{defn}
\index{Левоинвариантное векторное поле (left invariant vector field)}%
\index{Векторное поле левоинвариантное (left invariant vector field)}%
Левоинвариантные векторные поля определены глобально, т.е.\ на всем групповом
многообразии.
\begin{prop}
Любое левоинвариантное векторное поле является гладким.
\end{prop}
\begin{proof}
Следствие гладкости групповой операции. См., например, \cite{Warner83R}.
\end{proof}
Проиллюстрируем действие дифференциала отображения для соответствующей локальной
группы Ли, на которой задана функция композиции в явном виде. В компонентах
дифференциал отображения записывается в виде
$$
  X^\Sa(b)~\mapsto~X^\Sa(c)=X^\Sb(b)U_\Sb{}^\Sa(a,b),\qquad c=ab,
$$
где $X^\Sa(b)$ -- компоненты касательного вектора в точке $b$ относительно
координатного базиса $\pl_\Sa$, и дифференциал отображения задается матрицей
$$
  U_\Sb{}^\Sa(a,b):=\frac{\pl f^\Sa(a,b)}{\pl b^\Sb},
$$
зависящей от координат двух точек $a,b\in\MG$. Поскольку групповая операция,
$(a,b)\mapsto f(a,b)$, является диффеоморфизмом группового многообразия на себя
при фиксированном $a$ или $b$, то матрица $U_\Sa{}^\Sb$ невырождена для всех
$a,b\in\MG$, $\det U_\Sa{}^\Sb\ne0$ (там, где определена функция композиции).

Пусть $X_0=\lbrace X_0^\Sa\rbrace$ -- произвольный вектор из касательного
пространства к единице группы. Тогда этот вектор можно разнести по
всему групповому многообразию с помощью дифференциала левого действия элементов
группы. По построению, это будет левоинвариантное векторное поле
$X(a):=l_{a*}X_0$. Оно определено глобально.

Компоненты левоинвариантного векторного поля вблизи единицы группы имеют вид
\begin{align*}
  X^\Sa(a)&=X_0^\Sb L_\Sb{}^\Sa(a),
\\ \intertext{где}
  L_\Sb{}^\Sa(a)&:=U_\Sb{}^\Sa(a,0)
  =\left.\frac{\pl f^\Sa(a,b)}{\pl b^\Sb}\right|_{b=0}.
\end{align*}

Тем самым, каждому вектору из касательного пространства к единице однозначно
ставится в соответствие левоинвариантное векторное поле, и, наоборот, каждое
левоинвариантное векторное поле однозначно определяет некоторый вектор в
касательном пространстве к единице группы.

Матрица $U_\Sb{}^\Sa(a,b)$ удовлетворяет простому тождеству. Дифференцируя
равенство $f^\Sa(0,b)=b^\Sa$ по $b^\Sb$, получаем тождество
\begin{equation}                                                  \label{ematru}
  U_\Sb{}^\Sa(0,b)=\dl_\Sb^\Sa,
\end{equation}
которое нам понадобится в дальнейшем.

Множество векторных полей на группе Ли $\MG$, как и на любом другом
многообразии, образует бесконечномерную алгебру Ли $\CX(\MG)$ (см.\ раздел
\ref{salgvf}) и $\CC^\infty(\MG)$-модуль. Поскольку дифференциал отображения
сохраняет скобку Ли (коммутатор) (\ref{ecomhy}), то коммутатор двух
левоинвариантных векторных полей будет левоинвариантным векторным полем.
Следовательно, множество всех левоинвариантных векторных полей образует
подалгебру Ли в $\CX(\MG)$. Поскольку, как векторное пространство, множество
левоинвариантных векторных полей изоморфно касательному пространству в единице
группы $\MT_e(\MG)$, то множество левоинвариантных векторных полей представляет
собой конечномерную алгебру Ли такой же размерности $\Sn$, что и сама группа.
\begin{com}
Алгебра Ли левоинвариантных векторных полей не выдерживает умножения на
гладкие функции $f(a)\in\CC^\infty(\MG)$, и поэтому представляет собой линейное
пространство, а не $\CC^\infty(\MG)$-модуль.
\qed\end{com}
Структура алгебры Ли (коммутатор) левоинвариантных векторных полей естественным
образом переносится на касательное пространство к единице группы:
\begin{equation*}
  [X_0,Y_0]:=[l_{a*}X_0,l_{a*}Y_0]_e.
\end{equation*}
Словами. Берем два произвольных вектора $X_0$ и $Y_0$ из касательного
пространства к единице группы, разносим их с помощью действия группы слева по
всему многообразию, вычисляем коммутатор получившихся левоинвариантных векторных
полей и выбираем в касательном пространстве к единице группы тот вектор,
который соответствует коммутатору.

Множество левоинвариантных векторных полей образует $\Sn$-мерное векторное
пространство над полем вещественных чисел, которое изоморфно касательному
пространству в начале координат (или в любой другой точке группового
многообразия). В качестве базиса левоинвариантных векторных полей удобно выбрать
$\Sn$ левоинвариантных векторных полей ({\em образующих алгебры Ли})
\index{Образующие алгебры Ли (generators of Lie algebra)}%
\begin{equation}                                                  \label{elefba}
  L_\Sa(a):=l_{a*}\big(\pl_\Sa|_e\big)=L_\Sa{}^\Sb\pl_\Sb|_a,
\end{equation}
где мы разложили левоинвариантный базис по координатному базису $\pl_\Sb|_a$ в
точке $a\in\MG$. В построенном базисе произвольное левоинвариантное векторное
поле $X$ имеет постоянные компоненты, равные его компонентам в начале координат
относительно координатного базиса:
\begin{equation*}
  X=X_0^\Sa L_\Sa.
\end{equation*}

Поскольку левоинвариантные векторные поля образуют алгебру Ли, то коммутатор
двух базисных левоинвариантных векторных полей $L_\Sa$ будет левоинвариантным
векторным полем и, следовательно, его можно разложить по базису $L_\Sa$ с
некоторыми постоянными коэффициентами:
\begin{equation}                                                  \label{eliall}
  [L_\Sa,L_\Sb]=f_{\Sa\Sb}{}^\Sc L_\Sc.
\end{equation}
То, что в правой части этого равенства стоят структурные константы
$f_{\Sa\Sb}{}^\Sc$, которые были определены ранее через функцию композиции
(\ref{estrco}) не случайно.

Докажем это, рассмотрев окрестность единицы группы и соответствующую локальную
группу Ли. Вблизи единицы группы левоинвариантный базис в компонентах имеет вид
$$
  L_\Sa(a)=L_\Sa{}^\Sb(a)\pl_\Sb,
$$
где $\pl_\Sb$ -- координатный базис касательных пространств, что оправдывает
выбранные обозначения в (\ref{elefba}). Из разложения (\ref{ecofgl}) следует
разложение для компонент левоинвариантного базиса:
\begin{equation}                                                  \label{expliv}
  L_\Sa{}^\Sb=\dl_\Sa^\Sb+a^\Sc\hat f_{\Sa\Sc}{}^\Sb
  +a^\Sd a^\Sc g_{\Sa\Sc\Sd}{}^\Sb+\dotsc.
\end{equation}
Подстановка этого разложения в (\ref{eliall}) доказывает, что в правой части
стоят структурные константы (\ref{estrco}), потому что коэффициенты разложения
$f_{\Sa\Sb}{}^\Sc$ постоянны. При этом возникают также ограничения на
коэффициенты разложения при высших степенях $a^\Sa$, на которых мы
останавливаться не будем.
\begin{com}
Для всех $a\in\MG$ выполнено равенство $\det L_\Sa{}^\Sb(a)\ne0$, потому что
групповая операция задает диффеоморфизм группы $\MG$ на себя. При малых $a^\Sa$
это следует также из (\ref{expliv}).
\qed\end{com}
Продолжим рассмотрение локальной группы Ли и дадим независимый вывод формулы
(\ref{eliall}), исходя только из свойств функции композиции.
\begin{prop}
Базис алгебры Ли (\ref{elefba}) удовлетворяет коммутационным соотношениям
(\ref{eliall}) с некоторым набором структурных констант, удовлетворяющих
тождеству Якоби.
\end{prop}
\begin{proof}
Рассмотрев тождество $b=a^{-1}ab$ в касательном пространстве, получим равенство
$$
  L^{-1}_{~~\Sa}{}^\Sb(a)=U_\Sa{}^\Sb(a^{-1},a).
$$
Далее, распишем условие ассоциативности $(ab)c=a(bc)$ в касательном
пространстве:
$$
  U_\Sa{}^\Sb\big(f(a,b),c\big)=U_\Sa{}^\Sc(b,c)U_\Sc{}^\Sb\big(a,f(b,c)\big).
$$
Полагая $c=0$, получим равенства
\begin{equation}                                                  \label{elieqs}
  \frac{\pl f^\Sc(a,b)}{\pl b^\Sa}=U_\Sa{}^\Sc(a,b)
  =L^{-1}_{~~\Sa}{}^\Sd(b)L_\Sd{}^\Sc(f),
\end{equation}
где $f=f(a,b)$. Рассмотрим эту систему уравнений, как уравнения на функцию
композиции. Система уравнений (\ref{elieqs}) разрешима тогда и только тогда,
когда выполнены условия совместности. Чтобы получить эти условия,
продифференцируем уравнение (\ref{elieqs}) по $b^\Sb$ и антисимметризуем по
индексам $\Sa$ и $\Sb$. После несложных алгебраических преобразований получим
уравнение
$$
  [L_\Sa(b),L_\Sb(b)]^\Sd L^{-1}_{~~\Sd}{}^\Sc(b)
  =[L_\Sa(f),L_\Sb(f)]^\Sd L^{-1}_{~~\Sd}{}^\Sc(f),
$$
где квадратные скобки обозначают коммутатор векторных полей. Поскольку левая и
правая часть этого равенства рассматриваются в различных точках группового
многообразия, и точки могут выбираться произвольно, то они должны быть равны
константам. Эти константы проще всего определить, рассмотрев линейное
приближение для левоинвариантных векторных полей (\ref{expliv}). В результате
получим уравнение для образующих алгебры Ли (\ref{eliall}).

Тождества Якоби для структурных констант следуют из тождеств Якоби для векторных
полей.
\end{proof}

Таким образом, мы доказали, что множество левоинвариантных векторных полей
образует $\Sn$-мерную алгебру Ли над полем вещественных чисел, которая
является подалгеброй бесконечномерной алгебры всех векторных полей $\CX(\MG)$.
\begin{defn}
Алгебра левоинвариантных векторных полей называется {\em алгеброй Ли} $\Gg$
группы Ли $\MG$. Базисные векторы (образующие) $L_\Sa$ алгебры Ли $\Gg$
называются {\em генераторами группы Ли}.
\qed\end{defn}
\index{Алгебра Ли (Lie algebra)}\index{Ли алгебра (Lie algebra)}%
\index{Генератор группы Ли (generator of Lie group)}%
\begin{exa}
Вещественная прямая $x\in\MR$ является абелевой группой по сложению. Ее алгебра
Ли состоит из постоянных векторных полей $X=a\pl_x\in\CX(\MR)$, $a\in\MR$.
\qed\end{exa}

Алгебра Ли определяется с точностью до изоморфизма (выбора базиса
левоинвариантных векторных полей) только окрестностью единицы группы. Поэтому,
если две группы Ли локально изоморфны, то их алгебры Ли также изоморфны. В то
же время сами группы Ли могут не быть изоморфными.
\begin{exa}
Группы $\MS\MU(2)$, $\MS\MO(3)$ и $\MO(3)$ имеют изоморфные алгебры Ли, однако
сами не изоморфны. При
этом группа $\MS\MU(2)$ связна и односвязна и является универсальной накрывающей
группы $\MS\MO(3)$. Группа вращений $\MS\MO(3)$ связна, но не односвязна. Группа
$\MO(3)$ состоит из двух компонент связности.
\qed\end{exa}

Левоинвариантные векторные поля $L_\Sa$ (генераторы) образуют базис алгебры Ли.
Этот базис определен с точностью до выбора базиса в касательном пространстве к
единице группы, т.е.\ с точностью до действия группы $\MG\ML(\Sn,\MR)$. При этом
вид структурных констант зависит от выбора базиса. Легко проверить, что при
преобразовании базиса структурные константы преобразуются как компоненты тензора
третьего ранга с двумя ковариантными индексами и одним контравариантным.

Выше мы показали, как групповая операция определяет алгебру Ли левоинвариантных
векторных полей. Проведем обратное построение и покажем, что структурные
константы (\ref{estrco}) со свойством (\ref{ejatoj}) позволяют восстановить
функцию композиции $f^\Sa(a,b)$ в окрестности единицы. Это построение дает

{\bf Доказательство теоремы \ref{tlocis}}.
Перепишем соотношение (\ref{eliall}) в компонентах:
\begin{equation}                                                  \label{ecocvl}
  L_\Sa{}^\Sd\pl_\Sd L_\Sb{}^\Sc-L_\Sb{}^\Sd\pl_\Sd L_\Sa{}^\Sc
  =f_{\Sa\Sb}{}^\Sd L_\Sd{}^\Sc.
\end{equation}
Свертка этого выражения с тремя обратными матрицами $L^{-1}$ по индексам
$\Sa,\Sb,\Sc$ приводит к соотношению
\begin{equation}                                                  \label{ecjinl}
  \pl_\Sa L^{-1}_\Sb{}^\Sc-\pl_\Sb L^{-1}_\Sa{}^\Sc
  =-L^{-1}_\Sa{}^\Se L^{-1}_\Sb{}^\Sd f_{\Sd\Se}{}^\Sc.
\end{equation}
Это равенство является дифференциальным уравнением на матрицу $L^{-1}(a)$.
Левая часть уравнения антисимметрична по индексам $\Sa$ и $\Sb$ и,
следовательно, представляет собой 2-форму на $\MG_0$. Для того, чтобы найти
необходимые и достаточные условия интегрируемости этой системы уравнений, ее
необходимо продифференцировать, скажем, по $a^\Sd$, и антисимметризировать по
индексам $\Sa,\Sb$ и $\Sd$. После несложных вычислений, получим, что необходимым
и достаточным условием интегрируемости уравнений (\ref{ecjinl}) являются
тождества Якоби для структурных констант. Таким образом, матрицы $L^{-1}(a)$ и,
следовательно, $L(a)$ определяются структурными константами с точностью до
некоторой константы, которая фиксируется условием $L_\Sa{}^\Sb(0)=\dl_\Sa^\Sb$.
После этого функция композиции однозначно находится из уравнения (\ref{elieqs})
с фиксированными условиями в нуле. Условия интегрируемости этих уравнений, как
было показано выше, опять-таки сводятся к тождествам Якоби. Таким образом, если
задан произвольный набор структурных констант
$f_{\Sa\Sb}{}^\Sc=-f_{\Sb\Sa}{}^\Sc$, удовлетворяющих тождествам Якоби
(\ref{ejatoj}), то функция композиции однозначно определяется в окрестности
единицы группы и, следовательно, групповая структура задается в некоторой
окрестности единицы группы $\MG$. Таким образом, доказано, что произвольная
алгебра Ли определяет локальную группу Ли с точностью до изоморфизма.

В обратную сторону: каждая локальная группа Ли определяет алгебру Ли с точностью
до изоморфизма. Это сразу следует из определения алгебры Ли, как множества
левоинвариантных векторных полей.
\qed
\begin{com}
Явный вид функции композиции известен только в самых простейших случаях,
например, для абелевой группы сложения векторов евклидова пространства
$f^\Sa=a^\Sa+b^\Sa$. В примере \ref{eglnex} функция композиции была построена
для группы $\MG\ML(n,\MR)$. В разделе \ref{stwnog} будет явно построена
функция композиции в простейшем случае двумерной неабелевой группы Ли (группа
аффинных преобразований прямой). В более сложных случаях решить явно уравнения
(\ref{ecjinl}) и (\ref{elieqs}) не удается. Но это обычно и не нужно. В
приложениях, как правило, работают с матричным представлением алгебры Ли и
структурными константами.
\qed\end{com}
\begin{defn}
Пусть $X=X_0^\Sa L_\Sa{}^\Sb(a)\pl_\Sb$ -- произвольное левоинвариантное
векторное поле. Форма $\om=dx^\Sa\om_\Sa(a)$ называется {\em левоинвариантной,}
если ее значение на левоинвариантном векторном поле равно константе,
\begin{equation*}                                                    \tag*{\qed}
  \om(X)=X_0^\Sa L_\Sa{}^\Sb\om_\Sb=\const.
\end{equation*}
\end{defn}
\index{Левоинвариантная $1$-форма (left invariant $1$-form)}%
\index{$1$-форма левоинвариантная (left invariant $1$-form)}%
\begin{prop}
Любая левоинвариантная $1$-форма является гладкой.
\end{prop}
\begin{proof}
Следствие гладкости групповой операции. См., например, \cite{Warner83R}.
\end{proof}

Из определения следует, что компоненты левоинвариантной $1$-формы в координатном
базисе имеют вид
\begin{equation}                                                  \label{eleinf}
  \om_\Sa(a)=L^{-1}_{~~\Sa}{}^\Sb(a)\om_{0\Sb},
\end{equation}
где $\om_{0\Sb}$ -- компоненты левоинвариантной $1$-формы в начале координат.
При работе с левоинвариантными 1-формами удобно ввести левоинвариантный базис,
который дуален к левоинвариантным векторным полям: $\om^\Sa(L_\Sb)=\dl^\Sa_\Sb$.
Он имеет следующий вид в координатном базисе:
\begin{equation*}
  \om^\Sa(a)=dx^\Sb L^{-1}_{~~\Sb}{}^\Sa(a),
\end{equation*}
Тогда произвольная левоинвариантная 1-форма будет иметь в левоинвариантном
базисе постоянные компоненты $\om:=\om^\Sa\om_{0\Sa}$.
\begin{prop}[\bf Формула Маурера--Картана]
Пусть $\om$ -- произвольная левоинвариантная 1-форма и $X$, $Y$ -- два
произвольных левоинвариантных векторных поля. Тогда значение внешней производной
$d\om$ на векторных полях $X,Y$ пропорционально значению $1$-формы $\om$ на
коммутаторе этих полей:
\begin{equation}                                                  \label{emacaf}
  d\om(X,Y)=-\frac12\om\big([X,Y]\big).
\end{equation}
Это тождество известно, как {\em формула Маурера--Картана}. Ее можно переписать
для левоинвариантного базиса 1-форм:
\begin{equation}                                                  \label{emacar}
  d\om^\Sa=-\frac12\om^\Sb\wedge\om^\Sc f_{\Sb\Sc}{}^\Sa.
\end{equation}
\end{prop}
\begin{proof}
Прямая проверка с использованием тождества (\ref{ecjinl}).
\end{proof}
\index{Формула Маурера--Картана (Maurer--Cartan formulae)}%
\index{Маурера--Картана формула (Maurer--Cartan formulae)}%

При рассмотрении главных расслоений нам понадобится рассматривать формы,
заданные на некотором многообразии $\MM$, со значениями в алгебре Ли $\Gg$.
Введем для таких форм понятие коммутатора. Пусть $L_\Sa$, $\Sa=1,\dotsc,\Sn$, --
базис алгебры Ли. Пусть $A$ и $B$ -- соответственно $r$ и $s$ формы на $\MM$ со
значениями в алгебре Ли $\Gg$. Тогда они имеют вид
\begin{equation*}
  A=A^\Sa L_\Sa,\quad B=B^\Sa L_\Sa,
\end{equation*}
где $A^\Sa\in\Lm_r(\MM)$ и $B^\Sa\in\Lm_s(\MM)$ для всех значений индекса
$\Sa=1,\dotsc,\Sn$. Определим коммутатор заданных форм следующим
равенством
\begin{equation}                                                  \label{erolfo}
  [A,B]:=A^\Sa\wedge B^\Sb[L_\Sa,L_\Sb]
  =A^\Sa\wedge B^\Sb f_{\Sa\Sb}{}^\Sc L_\Sc.
\end{equation}
Выражение в правой части равенства представляет собой $r+s$ форму на $\MM$ со
значениями в алгебре Ли $\Gg$.

Если на многообразии $\MM$ заданы три формы $A$, $B$ и $C$ степеней $r$, $s$ и
$t$, соответственно, со значениями в алгебре Ли $\Gg$, то из тождеств Якоби
вытекает равенство
\begin{equation}                                                  \label{ejacfo}
  (-1)^{rt}\big[[A,B],C\big]+(-1)^{rs}\big[[B,C],A\big]
  +(-1)^{st}\big[[C,A],B\big].
\end{equation}
\begin{exa}
Пусть $A$ и $B$ -- 1-формы на $\MM$ со значениями в алгебре Ли $\Gg$ и
$X,Y\in\CX(\MM)$ -- два векторных поля на $\MM$. Тогда значение коммутатора форм
на векторных полях равно
\begin{multline*}
  [A,B](X,Y)=A^\Sa\wedge B^\Sa(X,Y)[L_\Sa,L_\Sb]=
\\
  =\frac12\big(A^\Sa(X)B^\Sb(Y)-A^\Sa(Y)B^\Sb(X)\big)
  =\frac12[A(X),B(Y)]-\frac12[A(Y),B(X)].
\end{multline*}
В частности,
\begin{equation*}                                                    \tag*{\qed}
  [A,A](X,Y)=[A(X),A(Y)].
\end{equation*}
\end{exa}
\begin{prop}
Если $A$ и $B$ -- соответственно $r$ и $s$ формы со значениями в алгебре Ли
$\Gg$, то справедливы следующие формулы:
\begin{align*}
  [A,B]&=(-1)^{rs+1}[B,A],
\\
  d[A,B]&=[dA,B]+(-1)^r[A,dB].
\end{align*}
\end{prop}
\begin{proof}
Прямая проверка.
\end{proof}

В дальнейшем особую роль будут играть формы, заданные на группе Ли
$\hat\MG$, со значениями в ее алгебре Ли $\Gg$. То есть в качестве
многообразия $\MM$ мы рассматриваем саму группу $\hat\MG$, которую для
отличия мы отметили шляпкой. Другими словами, рассмотрим два экземпляра
одной и той же группы Ли: $\hat\MG=\MG$. Пусть $\hat\om^\Sa$
-- левоинвариантный базис кокасательного расслоения к $\hat\MG$. Реализуем
алгебру Ли $\Gg$, как алгебру левоинвариантных векторных полей на $\MG$ с
базисом $L_\Sa$.
\begin{defn}
Левоинвариантная 1-форма на группе Ли $\hat\MG$ со значениями в алгебре
Ли $\Gg$,
\begin{equation}                                                  \label{exanfo}
  \theta=\hat\om^\Sa L_\Sa,
\end{equation}
называется {\em канонической}.
\qed\end{defn}
\index{Каноническая левоинвариантная форма (left invariant canonical form)}%
\index{Левоинвариантная каноническая форма (left invariant canonical form)}%
\index{Форма каноническая левоинвариантная (left invariant canonical form)}%
Если $X(\hat a)\in\CX(\hat\MG)$ -- векторное поле на $\hat\MG$, то
его можно разложить по левоинвариантному базису: $X=X^\Sa(\hat a)\hat L_\Sa$.
Тогда значение канонической формы на этом поле равно
\begin{equation*}
  \theta(X)(\hat a)=X^\Sa(\hat a) L_\Sa.
\end{equation*}
То есть каждой точке $\hat a\in\hat\MG$ ставится в соответствие
левоинвариантное векторное поле на $\MG$. Эта форма будет использована при
изучении связностей на главных расслоениях в разделе \ref{scofib}. Формулу
Маурера--Картана (\ref{emacar}) можно переписать для канонической формы
\begin{equation}                                                  \label{emucac}
  d\theta=-\frac12[\theta,\theta],
\end{equation}
где справа стоит коммутатор в алгебре Ли $\Gg$ и внешние произведения 1-форм
$\hat\om^\Sa$.

Наличие левоинвариантных форм на группе Ли позволяет доказать следующее
\begin{prop}
Любая группа Ли $\MG$ является ориентируемым многообразием.
\end{prop}
\begin{proof}
Левоинвариантные 1-формы $\om^\Sa$ определены на всем групповом многообразии
и линейно независимы. Поэтому $\Sn$-форма $\om^1\wedge\dotsc\wedge\om^\Sn$
нигде не обращается в нуль. Следовательно, согласно теореме \ref{torivo}, любая
группа Ли ориентируема.
\end{proof}
\section{Действие группы справа                                  \label{sriact}}
Все, сказанное относительно действия группы на себя слева, естественным образом
переносится на действие группы справа. Ниже мы рассмотрим отличия, которые при
этом возникают. Для простоты, рассмотрим локальную группу Ли, т.е.\
окрестность единицы группы.

Рассмотрим группу Ли $\MG$, как группу преобразований, действующую справа:
\begin{equation*}
  \MG\ni a:\quad \MG\ni\quad b~\mapsto~r_a(b):=ba\quad\in\MG.
\end{equation*}
Пусть
$$
  V_\Sb{}^\Sa(a,b):=\frac{\pl f^\Sa(a,b)}{\pl a^\Sb}.
$$
Дифференцируя тождество $f^\Sa(a,0)=a^\Sa$ по $a^\Sb$, получаем равенство
\begin{equation*}
  V_\Sb{}^\Sa(a,0)=\dl_\Sb^\Sa,
\end{equation*}
которое нам понадобится в дальнейшем.

Аналогично случаю левого действия группы, построим правоинвариантный базис
векторных полей
\begin{equation}                                                  \label{erinve}
  R_\Sa(a)=R_\Sa{}^\Sb(a)\pl_\Sb,
\end{equation}
где
\begin{equation}                                                  \label{ergrom}
  R_\Sa{}^\Sb(a):=V_\Sa{}^\Sb(0,a)
  =\left.\frac{\pl f^\Sb(b,a)}{\pl b^\Sa}\right|_{b=0}.
\end{equation}
В этом базисе произвольное правоинвариантное векторное поле имеет постоянные
компоненты. Как и для левоинвариантных векторных полей,
$\det R_\Sa{}^\Sb(a)\ne0$ для всех $a\in\MG$.

Как и в случае левого действия группы, можно доказать, что правоинвариантные
векторные поля образуют $\Sn$-мерную алгебру Ли над полем вещественных чисел.
При этом коммутатор векторов правоинвариантного базиса,
\begin{equation}                                                  \label{erinba}
  [R_\Sa,R_\Sb]=-f_{\Sa\Sb}{}^\Sc R_\Sc,
\end{equation}
отличается знаком структурных констант от коммутатора левоинвариантного базиса.
Алгебры Ли лево- и правоинвариантных векторных полей изоморфны, поскольку
переходят друг в друга после отображения $R_\Sa\mapsto-R_\Sa$.

Формула для коммутатора правоинвариантных векторных полей (\ref{erinba})
справедлива, как легко видеть, глобально на всей группе Ли.

Из разложения функции композиции (\ref{ecofgl}) следует разложение для
компонент правоинвариантных векторных полей относительно координатного базиса
\begin{equation}                                                  \label{exprin}
  R_\Sb{}^\Sa(a)=\dl_\Sb^\Sa+a^\Sc\hat f_{\Sc\Sb}{}^\Sa
  +a^\Sd a^\Sc h_{\Sc\Sd\Sb}{}^\Sa+\dotsc.
\end{equation}
Из тождества $b=baa^{-1}$, записанного в касательном пространстве, следует
равенство
\begin{equation}                                                  \label{elainr}
  R^{-1}_\Sb{}^\Sa(a)=V_\Sb{}^\Sa(a,a^{-1}).
\end{equation}

Рассмотрим функцию $F(a)\in\CC^\infty(\MG)$ на группе Ли $\MG$. Подействуем на
точки многообразия бесконечно малым элементом $\e\in\MG$ справа. Тогда точка $a$
перейдет в точку $a'=a+da$, где
$$
  da^\Sa:=f^\Sa(a,\e)-a^\Sa=\e^\Sb L_\Sb{}^\Sa(a).
$$
Поскольку для функции $F'(a')=F(a)$, то изменение формы функции равно:
$$
  \dl F(a):=F'(a)-F(a)=-\e^\Sa L_\Sa{}^\Sb\pl_\Sb F(a)=-\e^\Sa L_\Sa F(a).
$$
Таким образом, левоинвариантные векторные поля, с точностью до знака, являются
генераторами действия группы Ли справа. Аналогично доказывается, что
генераторами групповых преобразований слева являются правоинвариантные векторные
поля.

Поскольку преобразования, вызванные действием группы слева и справа коммутируют,
то отсюда следует, что лево- и правоинвариантные векторные поля коммутируют
между собой,
\begin{equation}                                                  \label{elrcom}
  [L_\Sa,R_\Sb]=0.
\end{equation}
Это равенство справедливо, конечно, глобально.

Из формулы для производной Ли от векторного поля (\ref{elicom}) и равенства
(\ref{elrcom}) следует, что производная Ли от правоинвариантного векторного
поля вдоль левоинвариантного векторного поля равна нулю и наоборот.

В дальнейшем под алгеброй Ли $\Gg$ группы Ли $\MG$ всегда понимается множество
левоинвариантных векторных полей, которые генерируют действие группы справа.
\section{Присоединенное представление                            \label{sadpre}}
Рассмотрим автоморфизм группы Ли,
\begin{equation}                                                  \label{eadrep}
  \ad(a):\quad \MG\ni\quad b\mapsto \ad(a) b:=aba^{-1}\quad\in\MG,
\end{equation}
который сопоставляется каждому элементу группы $a\in\MG$. Каждый автоморфизм
группы Ли порождает автоморфизм ее алгебры Ли, потому что левоинвариантные
векторные поля получены с помощью группового действия. Поэтому автоморфизм
$\ad a$ порождает автоморфизм алгебры Ли $\Gg$, который также обозначается
$\ad a\in\MG\ML(\Sn,\MR)$.
\begin{defn}
Представление $a\mapsto\ad a$, где
\begin{equation}                                                  \label{eadjoi}
  \ad a:\quad \Gg\ni\quad X
  =X^\Sa_0 L_\Sa\mapsto X'=X^\Sa_0 S_\Sa{}^\Sb(a)L_\Sb\quad\in\Gg,
\end{equation}
индуцированное групповым действием (\ref{eadrep}), называется
{\em присоединенным представлением} группы Ли $\MG$ в ее алгебре Ли $\Gg$.
\qed\end{defn}
\index{Присоединенное представление (adjoint representation)}%
\index{Представление присоединенное (adjoint representation)}%
Присоединенное представление каждому элементу группы $a\in\MG$ ставит в
соответствие $\Sn\times\Sn$ матрицу $S_\Sa{}^\Sb(a)$. Для краткости мы будем
писать
\begin{equation*}
  \ad(a)=S_\Sa{}^\Sb(a).
\end{equation*}
Ясно, что единичному элементу группы соответствует единичная матрица
\begin{equation*}
  \ad(e)=S_\Sa{}^\Sb(e)=\dl_\Sa^\Sb,
\end{equation*}
а обратному элементу группы -- обратная матрица
\begin{equation*}
  \ad(a^{-1})=S_\Sa{}^\Sb(a^{-1})=S^{-1}_{~~\Sa}{}^\Sb(a).
\end{equation*}
В наших обозначениях матрица для произведения двух элементов группы равна
произведению матриц для каждого элемента,
\begin{equation*}
  S_\Sa{}^\Sb(ab)=S_\Sa{}^\Sc(b)S_\Sc{}^\Sb(a),
\end{equation*}
взятых в обратном порядке.

Для абелевых групп присоединенное представление всегда тривиально.

Ядром отображения (\ref{eadrep}) является центр группы, т.е.\ множество всех
тех элементов группы, которые коммутируют со всеми элементами группы $\MG$.
Каждому элементу центра группы соответствует единичная матрица присоединенного
представления.

Для локальных неабелевых групп Ли можно построить явный вид матриц
присоединенного представления в терминах функции композиции. С этой целью
рассмотрим действие преобразования (\ref{eadrep}) в касательном пространстве в
последовательности $b\mapsto(ab)a^{-1}$:
$$
  X^\Sa(b)\mapsto X^\Sc(b)U_\Sc{}^\Sb(a,b)V_\Sb{}^\Sa\big(f(a,b),a^{-1}\big).
$$
Полагая $b=0$ и учитывая (\ref{elainr}), получаем отображение
$$
  X_0^\Sa\mapsto X_0^\Sb L_\Sb{}^\Sc(a)R^{-1}_\Sc{}^\Sa(a)
  :=X_0^\Sb S_\Sb{}^\Sa(a),
$$
где введена матрица присоединенного представления
$S_\Sb{}^\Sa(a)\in\MG\ML(n,\MR)$, сопоставляемая каждому элементу $a\in\MG$.
Таким образом получаем явные выражение для матрицы присоединенного
представления через производные от функции композиции:
\begin{equation}                                                  \label{eadsma}
  S_\Sb{}^\Sa(a)=L_\Sb{}^\Sc(a)R^{-1}_{~~\Sc}{}^\Sa(a)
  =R_\Sb{}^\Sc(a^{-1})L^{-1}_{~~\Sc}{}^\Sa(a^{-1}),
\end{equation}
где последнее выражение получено в результате рассмотрения действия
$\ad a$ в последовательности $a(ba^{-1})$. Из этого представления
следует, что
$$
  S_\Sb{}^\Sa(a^{-1})=S^{-1}_{~~\Sb}{}^\Sa(a).
$$

Разложения (\ref{expliv}) и (\ref{exprin}) приводят к следующему разложению
матрицы присоединенного представления вблизи начала координат
\begin{equation}                                                  \label{etelsf}
  S_\Sb{}^\Sa(a)=\dl_\Sb^\Sa+a^\Sc f_{\Sc\Sb}{}^\Sa+\dotsc.
\end{equation}
Отсюда следует, что структурные константы, взятые с обратным знаком, можно
рассматривать, как генераторы присоединенного представления:
$L_{(\Sa)\Sb}{}^\Sc=-f_{\Sa\Sb}{}^\Sc$. При этом первый индекс $\Sa$ нумерует
генераторы и, для отличия, взят в скобки. Второй и третий индексы структурных
констант $\Sb,\Sc$ рассматриваются, как матричные. Тождества Якоби
(\ref{ejatoj}) переписываются в виде
\begin{equation}                                                  \label{ejjcma}
  [L_{(\Sa)},L_{(\Sb)}]_\Sc{}^\Sd=f_{\Sa\Sb}{}^\Se L_{(\Se)\Sc}{}^\Sd,
\end{equation}
где квадратные скобки обозначают коммутатор матриц.

Рассмотрим два последовательных автоморфизма группы Ли $\MG$, соответствующих
элементам $b$ и $a$:
$$
  c~\mapsto~bcb^{-1}~\mapsto~abcb^{-1}a^{-1}=(ab)c(ab)^{-1}.
$$
Соответствующее преобразование в алгебре Ли $\Gg$ задается матрицей
присоединенного представления для произведения $f(a,b)$
\begin{equation}                                                  \label{epradr}
  S_\Sa{}^\Sb\big(f(a,b)\big)=S_\Sa{}^\Sc(b)S_\Sc{}^\Sb(a),
\end{equation}
которая равна произведению матриц для каждого из элементов, но взятых в обратном
порядке.

Пусть $b\ll1$. Разложив равенство (\ref{epradr}) в ряд по $b$ с использованием
формулы (\ref{etelsf}) и разложения для функции композиции,
$$
  f^\Sa(a,b)=a^\Sa+b^\Sb L_\Sb{}^\Sa(a)+\dotsc,
$$
в первом порядке по $b$ получим правило дифференцирования матрицы
присоединенного представления
\begin{equation}                                                  \label{eaddil}
  L_\Sc S_\Sa{}^\Sb=L_\Sc{}^\Sd\pl_\Sd S_\Sa{}^\Sb=f_{\Sc\Sa}{}^\Sd S_\Sd{}^\Sb.
\end{equation}
Из этого равенства с учетом тождества
$L_\Sc(S_\Sa{}^\Sd S^{-1}_{~~\Sd}{}^\Sb)=0$ получаем правило действия
левоинвариантных векторных полей на обратные матрицы присоединенного
представления:
\begin{equation}                                                  \label{ediadm}
  L_\Sc S^{-1}_{\quad \Sa}{}^\Sb=-S^{-1}_{~~\Sa}{}^\Sd f_{\Sc\Sd}{}^\Sb.
\end{equation}

Аналогично, раскладывая соотношение (\ref{epradr}) в ряд по $a\ll1$, получим
\begin{equation}                                                  \label{eaddir}
  R_\Sc S_\Sa{}^\Sb=R_\Sc{}^\Sd\pl_\Sd S_\Sa{}^\Sb=S_\Sa{}^\Sd f_{\Sc\Sd}{}^\Sb.
\end{equation}
С учетом формулы $R_\Sc(S_\Sa{}^\Sd S^{-1}_{~~\Sd}{}^\Sb)=0$ это дает правило
дифференцирования обратной матрицы
\begin{equation*}
  R_\Sc S^{-1}_{~~\Sa}{}^\Sb=-f_{\Sc\Sa}{}^\Sd S^{-1}_{~~\Sd}{}^\Sb.
\end{equation*}
\begin{com}
Полученные формулы дифференцирования матриц присоединенного представления
(\ref{eaddil}) и (\ref{eaddir}) нековариантны, поскольку частная производная
от тензора не является тензором. Однако при рассмотрении алгебр Ли
левоинвариантные векторные поля имеют постоянные компоненты относительно
левоинвариантного базиса и поэтому достаточно ограничиться аффинными
преобразованиями координат с постоянными коэффициентами. В этом случае частные
производные преобразуются по тензорным правилам.
\qed\end{com}
\begin{prop}
Структурные константы группы Ли $\MG$ инвариантны относительно присоединенного
действия группы:
\begin{equation}                                                  \label{eadiac}
  f_{\Sa\Sb}{}^\Sc
  =S_\Sa{}^\Sd S_\Sb{}^\Se f_{\Sd\Se}{}^\Sf S^{-1}_{~~\Sf}{}^\Sc.
\end{equation}
\end{prop}
\begin{proof}
Подействуем присоединенным представлением (\ref{eadrep}) на групповое
многообразие $\MG$. Тогда базис левоинвариантных векторных полей преобразуется
в соответствии с присоединенным представлением
\begin{equation*}
  L_\Sa\mapsto S_\Sa{}^\Sb L_\Sb.
\end{equation*}
При этом коммутационные соотношения (\ref{eliall}) сохранят свой вид, поскольку
действие дифференциала отображения сохраняет коммутатор векторных полей.
Следовательно, для структурных констант справедливо соотношение
\begin{equation*}
  S_\Sa{}^\Sd S_\Sb{}^\Se f_{\Sd\Se}{}^\Sc=f_{\Sa\Sb}{}^\Sd S_\Sd{}^\Sc.
\end{equation*}
Умножив это равенство справа на $S^{-1}$ получим (\ref{eadiac}).
\end{proof}
\section{Группы Ли как (псевдо-)римановы пространства            \label{sligri}}
На групповом многообразии $\MG$ можно определить метрику, превратив тем самым
группу Ли в (псевдо-)риманово пространство. Среди всех возможных метрик важную
роль играют метрики, инвариантные относительно групповых преобразований. Эти
метрики описываются следующим образом. Зададим в начале координат произвольную
симметричную невырожденную матрицу $\theta_{\Sa\Sb}$ и разнесем ее по групповому
многообразию с помощью действия группы слева или справа. В результате получим
соответственно лево- и правоинвариантные метрики. Компоненты этих метрик в лево-
и правоинвариантном базисе имеют тот же вид $\theta_{\Sa\Sb}$, что и начале
координат. В координатном базисе $g=dx^\Sa\otimes dx^\Sb g_{\Sa\Sb}$ и
компоненты метрик нетривиально зависят от точки $a\in\MG$:
\begin{equation}                                                  \label{elerim}
\begin{split}
  g^L_{\Sa\Sb}(a)&=L^{-1}{}_\Sa{}^\Sc(a)L^{-1}{}_\Sb{}^\Sd(a)\theta_{\Sc\Sd},
\\
  g^R_{\Sa\Sb}(a)&=R^{-1}{}_\Sa{}^\Sc(a)R^{-1}{}_\Sb{}^\Sd(a)\theta_{\Sc\Sd},
\end{split}
\end{equation}
где индексами $L$ и $R$ отмечены соответственно лево- и правоинвариантная
метрики. В общем случае лево- и правоинвариантные метрики различны и их
компоненты в координатном базисе нетривиально зависят от точки группового
многообразия.

Для абелевых групп лево- и правоинвариантные метрики равны, т.к.\ совпадают
действия группы слева с справа.

Скалярное произведение левоинвариантных векторных полей: $X=X^\Sa_0L_\Sa$ и
$Y=Y^\Sa_0L_\Sa$, определенное левоинвариантной метрикой, равно
$$
  g^L(X,Y)=X^\Sa_0Y^\Sb_0g^L(L_\Sa,L_\Sb)=X^\Sa_0Y^\Sb_0\theta_{\Sa\Sb}
$$
и не зависит от точки группы Ли $\MG$. Аналогично, скалярное произведение
правоинвариантных векторных полей, определенное правоинвариантной метрикой равно
константе.

Условия левой и правой инвариантности метрик в инвариантном виде имеют
следующий вид. Пусть $X,Y\in\CX(\MG)$ -- два произвольных векторных поля на
группе Ли $\MG$. Тогда лево- и правоинвариантные метрики определены
соотношениями
\begin{equation*}
\begin{split}
  l^*_a g^L(X,Y)&=g^L(l_{a*}X,l_{a*}Y),
\\
  r^*_a g^R(X,Y)&=g^R(r_{a*}X,r_{a*}Y),
\end{split}
\end{equation*}
где $l^*_a$ и $r^*_a$ -- возвраты левого и правого действия группы.

Особую роль играют двусторонне инвариантные метрики $g^L=g^R$ на неабелевых
группах Ли. Для таких метрик скалярное произведение левоинвариантных векторных
полей инвариантно относительно действия группы справа. Двусторонне инвариантные
метрики существуют далеко не на всех группах Ли. Условие двусторонней
инвариантности немедленно следует, например, из (\ref{elerim}),
$$
  \theta_{\Sa\Sb}=S_\Sa{}^\Sc(a)S_\Sb{}^\Sd(a)\theta_{\Sc\Sd},\qquad \quad
  \forall a\in\MG,
$$
где $S_\Sa{}^\Sb$ -- матрица присоединенного представления. Для малых $a$, т.е.\
вблизи единицы группы это уравнение сводится к условию
\begin{equation}                                                  \label{ekcfoi}
  f_{\Sa\Sb}{}^\Sd\theta_{\Sd\Sc}+f_{\Sa\Sc}{}^\Sd\theta_{\Sd\Sb}=0.
\end{equation}
То есть, если двусторонне инвариантная метрика на группе Ли существует, то
структурные константы со всеми опущенными индексами должны быть антисимметричны
по всем трем индексам.
\begin{com}
Двусторонне инвариантные метрики на группе Ли играют ту же роль, что и метрика
Лоренца в пространстве Минковского $\MR^{1,n-1}$ или евклидова метрика в
евклидовом пространстве $\MR^n$.
\qed\end{com}

Построим двусторонне инвариантную метрику на неабелевой группе Ли в том случае,
когда это возможно.
\begin{defn}
Свертка структурных констант алгебры Ли
\begin{equation}                                                  \label{ekilcf}
  \eta_{\Sa\Sb}:=-f_{\Sa\Sc}{}^\Sd f_{\Sb\Sd}{}^\Sc=-\tr(f_\Sa f_\Sb)
\end{equation}
называется {\em формой Киллинга--Картана}.
\qed\end{defn}
\index{Форма Киллинга--Картана (Killing--Cartan form)}%
\index{Киллинга--Картана форма (Killing--Cartan form)}%
Здесь мы рассматриваем структурные константы $f_{\Sa\Sc}{}^\Sd$ в виде набора
матриц. При этом второй и третий индексы рассматриваются, как матричные, а
первый -- нумерует матрицы. Форма (\ref{ekilcf}) двусторонне инвариантна, что
следует из инвариантности структурных констант (\ref{eadiac}).
\begin{prop}                                                      \label{panstr}
Для произвольной группы Ли $\MG$ структурные константы со всеми опущенными
индексами,
\begin{equation*}
  f_{\Sa\Sb\Sc}:=f_{\Sa\Sb}{}^\Sd\eta_{\Sd\Sc}
\end{equation*}
антисимметричны по всем трем индексам,
\begin{equation*}
  f_{\left[abc\right]}=0.
\end{equation*}
\end{prop}
\begin{proof}
Прямое следствие тождеств Якоби (\ref{ejatoj}), которое сводится к простой
проверке.
\end{proof}

Таким образом, если форма Киллинга--Картана является невырожденной, то она
задает двусторонне инвариантную метрику на группе Ли. Форма Киллинга--Картана
является невырожденной не для всякой группы Ли. Напомним, что связная группа Ли
называется полупростой, если она не имеет нетривиальных инвариантных связных
абелевых подгрупп.
\begin{theorem}[{\bf Э.\ Картан}]                                 \label{tcarki}
Форма Киллинга--Картана для связной группы Ли является невырожденной тогда и
только тогда, когда группа Ли является полупростой.
\end{theorem}
\begin{proof}
См., например, \cite{BarRac77R}.
\end{proof}

Положительная определенность формы Киллинга--Картана связана с компактностью
группового многообразия. А именно, справедлива
\begin{theorem}                                                   \label{tcomnh}
Связная полупростая группа Ли компактна тогда и только тогда, когда ее форма
Киллинга--Картана положительно определена.
\end{theorem}
\begin{proof}
См., например, \cite{GotGro78R}.
\end{proof}
Отсюда следует, что форма Киллинга--Картана превращает компактную полупростую
группу Ли в риманово многообразие с инвариантной метрикой. Для некомпактных
полупростых групп Ли форма Киллинга--Картана не является знакоопределенной.
Следовательно, в этом случае групповое пространство становится псевдоримановым
многообразием.

Полупростые группы не исчерпывают весь класс групп Ли, на которых можно задать
двусторонне инвариантную метрику. Например, на абелевых группах любая
невырожденная матрица в единице, разнесенная по групповому многообразию, задает
двусторонне инвариантную метрику.
\begin{exa}
Двумерная связная абелева группа Ли изоморфна либо евклидовой плоскости $\MR^2$
со сложением в качестве групповой операции, либо цилиндру, либо тору. Пусть
$x,y$ -- декартовы координаты на плоскости. Тогда лево- и правоинвариантные
векторные поля совпадают и равны
\begin{equation*}
  L_1=R_1=\pl_x,\qquad L_2=R_2=\pl_y.
\end{equation*}
Двусторонне инвариантная метрика на $\MR^2$ получается путем разнесения
произвольной симметричной невырожденной $2\times2$ матрицы $\theta_{\Sa\Sb}$,
$\Sa,\Sb=1,2$, заданной в начале координат, с помощью действия группы.
В результате получаем двусторонне инвариантную метрику, которая в декартовых
координатах имеет вид $\theta_{\Sa\Sb}$ для всех точек $\MR^2$. Евклидова
плоскость с такой метрикой является в общем случае однородным, но не изотропным
пространством. При $\theta_{\Sa\Sb}=\dl_{\Sa\Sb}$ евклидова плоскость становится
также изотропной, т.е.\ инвариантной также относительно действия группы вращений
$\MO(2)$. Цилиндр и тор в этом случае неинвариантны относительно вращений.
\qed\end{exa}

Форма Киллинга--Картана (\ref{ekilcf}) задает метрику на групповом многообразии
полупростой группы Ли $\MG$ либо в лево-, либо в правоинвариантном базисе. Для
определенности будем считать, что метрика задана в левоинвариантном базисе.
Тогда в координатном базисе она будет нетривиально зависеть от точки группового
многообразия:
$$
  g_{\Sa\Sb}(a)=L^{-1}_{~~\Sa}{}^\Sc(a)L^{-1}_{~~\Sb}{}^\Sd(a)\eta_{\Sc\Sd}.
$$
Двусторонняя инвариантность формы Киллинга--Картана приводит к существованию, по
крайней мере, $2\Sn$ векторов Киллинга $L_\Sa$ и $R_\Sa$. Уравнения Киллинга в
неголономном базисе (\ref{eunbak}) выполнены ввиду антисимметрии символов
Кристоффеля (\ref{elicgl}) по первым двум индексам.

Вычислим геометрические характеристики группового многообразия неабелевой
полупростой группы Ли, рассматривая его, как (псевдо-)риманово пространство с
нулевым кручением и неметричностью. Вычисления можно проводить в любом базисе,
но в данном случае удобнее рассматривать все характеристики многообразия в
неголономном базисе, определяемом левоинвариантными векторными полями $L_\Sa$,
поскольку компоненты метрики в этом базисе постоянны. Соответствующие формулы
для вычислений приведены в разделе \ref{sunhba}.

Компоненты неголономности базиса определяются коммутатором левоинвариантных
векторных полей (\ref{eliall}) и совпадают со структурными константами:
$$
  c_{\Sa\Sb}{}^\Sc=f_{\Sa\Sb}{}^\Sc.
$$
Символы Кристоффеля в неголономном базисе, вычисленные по формуле
(\ref{elicou}), равны
\begin{equation}                                                  \label{elicgl}
  \widetilde\om_{\Sa\Sb\Sc}=\frac12f_{\Sa\Sb\Sc},
\end{equation}
где знак тильды означает, что кручение и неметричность положены равными нулю.
В рассматриваемом случае символы Кристоффеля антисимметричны по первым
двум индексам ввиду неголономности используемого базиса.
Соответствующий тензор кривизны также задается структурными константами
$$
  \widetilde R_{\Sa\Sb\Sc}{}^\Sd=\frac14f_{\Sa\Sb}{}^\Se f_{\Sc\Se}{}^\Sd.
$$
Отсюда вытекают выражения для тензора Риччи и скалярной кривизны:
\begin{equation*}
  \widetilde R_{\Sa\Sb}=\frac14\eta_{\Sa\Sb},\qquad
  \widetilde R=\frac\Sn4,
\end{equation*}
где $\Sn$ -- размерность полупростой группы Ли.

Покажем, что полупростая группа Ли с формой Киллинга--Картана в качестве
метрики, представляет собой пространство постоянной кривизны. В этом случае
структурные константы $f_{\Sa\Sb\Sc}$ задают полностью антисимметричный
ковариантный тензор третьего ранга в левоинвариантном базисе. Тогда ковариантная
производная этого тензора со связностью (\ref{elicgl}),
$$
  \widetilde\nb_\Sa f_{\Sb\Sc\Sd}=L_\Sa f_{\Sb\Sc\Sd}
  -\frac12f_{\Sa\Sb}{}^\Se f_{\Se\Sc\Sd}-\frac12f_{\Sa\Sc}{}^\Se f_{\Sb\Se\Sd}
  -\frac12f_{\Sa\Sd}{}^\Se f_{\Sb\Sc\Se}=0,
$$
равна нулю в силу тождеств Бианки. Поскольку связность является метрической, а
подъем и опускание индексов можно переставлять с оператором ковариантного
дифференцирования, то ковариантная производная от тензора кривизны равна нулю,
$$
  \widetilde\nb_\Sa \widetilde R_{\Sb\Sc\Sd}{}^\Se=0.
$$
Это и означает, что группа Ли как (псевдо-)риманово пространство является
пространством постоянной кривизны.

Тензор кривизны со всеми опущенными индексами в неголономном базисе имеет вид
\begin{equation}                                                  \label{eculig}
  \widetilde R_{\Sa\Sb\Sc\Sd}=-\frac14 f_{\Sa\Sb}{}^\Se f_{\Sc\Sd\Se},
\end{equation}
в котором явно прослеживаются все свойства симметрии относительно перестановок
индексов.
\begin{com}
Широкий класс (псевдо-)римановых пространств $(\MM,g)$ постоянной кривизны
описывается метрикой, удовлетворяющей следующему соотношению
\begin{equation}                                                  \label{qdfrss}
  \widetilde R_{\al\bt\g\dl}=C(g_{\al\g}g_{\bt\dl}-g_{\al\dl}g_{\bt\g}),
\end{equation}
где $C$ -- некоторая постоянная. Группы Ли с формой Киллинга--Картана в качестве
метрики дают пример пространств постоянной кривизны другого типа, для которых
равенство (\ref{qdfrss}) не выполнено. Действительно, равенство
$$
  f_{\Sa\Sb}{}^\Se f_{\Sc\Sd\Se}
  =C(\eta_{\Sa\Sc}\eta_{\Sb\Sd}-\eta_{\Sa\Sd}\eta_{\Sb\Sc}),\qquad C=\const,
$$
не может быть выполнено ни при каких значениях константы $C$. Для
доказательства достаточно свернуть это равенство сначала с $\eta^{\Sb\Sd}$,
а затем с $f^{\Sc\Sd}{}_\Sf$, что приведет к противоречию.
\qed\end{com}
\section{Группы Ли как пространства Римана--Картана              \label{sligrc}}
В предыдущем разделе связность на групповом многообразии определялась по метрике
при нулевом кручении и неметричности. Ниже мы рассмотрим другой способ
определения связности, который во многих отношениях является более естественным
для групп Ли. А именно, отождествим групповое действие справа с параллельным
переносом. Это означает, что результат параллельного переноса вектора из точки
$a$ в точку $c$ совпадает с дифференциалом группового преобразования $c=f(a,b)$
под действием элемента $b:=a^{-1}c$.

Формализуем данное определение связности. В рассматриваемом случае
правоинвариантные векторные поля можно рассматривать, как результат
параллельного переноса векторов из начала координат по всему групповому
многообразию. Это означает, что в правоинвариантном базисе компоненты векторов
постоянны, и, следовательно, компоненты связности равны нулю. Отсюда сразу
вытекает равенство нулю тензора кривизны. То есть группа Ли с аффинной
связностью, заданной групповой операцией, является пространством абсолютного
параллелизма. Так и должно быть, потому что результат параллельного переноса
вектора из точки $a$ в точку $c=ab$ однозначно определяется начальной и конечной
точкой и не зависит от пути, вдоль которого осуществляется параллельный перенос.

Кручение для рассматриваемой связности отлично от нуля. В правоинвариантном
базисе, который удовлетворяет коммутационным соотношениям (\ref{erinba}),
компоненты тензора кручения, согласно формулам (\ref{euntot}) имеют вид
\begin{equation*}
  T_{\Sa\Sb}{}^\Sc=f_{\Sa\Sb}{}^\Sc.
\end{equation*}
\begin{com}
В данном определении связности наличие или отсутствие метрики на групповом
многообразии не играет никакой роли.
\qed\end{com}

Посмотрим, как выглядят основные геометрические объекты в левоинвариантном
базисе. В левоинвариантном базисе правоинвариантное векторное поле имеет
непостоянные компоненты:
$$
  X=X_0^\Sa R_\Sa{}^\Sb(a)\pl_\Sb=X_0^\Sa S_\Sa{}^\Sb(a^{-1})L_\Sb,
$$
где мы воспользовались представлением (\ref{eadsma}) для матрицы присоединенного
представления. Несложные вычисления показывают, что разница компонент векторов в
соседних точках равна
$$
  \dl X^\Sa(a):=X^\Sa(a+da)-X^\Sa(a)=-X^\Sc da^\Sb f_{\Sb\Sc}{}^\Sa.
$$
С другой стороны, при параллельном переносе компоненты вектора получают
приращение, определяемое аффинной связностью:
$$
  \dl X^\Sa=-X^\Sc da^\Sb\om_{\Sb\Sc}{}^\Sa.
$$
Отсюда следует явное выражение для компонент аффинной связности
в левоинвариантном базисе
\begin{equation}                                                  \label{elicli}
  \om_{\Sa\Sb}{}^\Sc=f_{\Sa\Sb}{}^\Sc.
\end{equation}

Связность (\ref{elicli}) отличается от связности Леви-Чивиты (\ref{elicgl})
множителем $1/2$. Соответствующие тензоры кривизны и кручения, вычисленные по
формулам для неголономного базиса (\ref{euncur}) и (\ref{euntot}), равны
\begin{align}                                                     \label{ecupal}
  R_{\Sa\Sb\Sc}{}^\Sd&=0,
\\                                                                \label{etocpa}
  T_{\Sa\Sb}{}^\Sc&=f_{\Sa\Sb}{}^\Sc.
\end{align}
Равенство компонент тензора кривизны нулю соответствует тому, что групповое
пространство является пространством абсолютного параллелизма. Компоненты тензора
кручения в левоинвариантном базисе имеют тот же вид, что и в правоинвариантном.
Это соответствует тому обстоятельству, что структурные константы инвариантны
относительно действия присоединенного представления (\ref{eadiac}).

Компоненты связности (\ref{elicli}) в левоинвариантном базисе определены
независимо от метрики и для произвольных, не обязательно полупростых, групп.
На полупростых группах Ли эта связность является метрической:
$$
  \nb_\Sa\eta_{\Sb\Sc}=-\om_{\Sa\Sb}{}^\Sd\eta_{\Sd\Sc}
  -\om_{\Sa\Sc}{}^\Sd\eta_{\Sb\Sd},
$$
где $\eta_{\Sb\Sc}$ -- двусторонне инвариантная форма Киллинга--Картана.
Это значит, что групповое многообразие полупростых групп Ли со связностью,
определенной компонентами (\ref{elicli}), представляет собой пространство
Римана--Картана.

Группа Ли, рассматриваемая как пространство Римана--Картана является
пространством постоянной (нулевой) кривизны и постоянного кручения.
Действительно, ковариантная производная от тензора кривизны, очевидно, равна
нулю. Ковариантная производная от тензора кручения со связностью (\ref{elicli}),
$$
  \nb_\Sa T_{\Sb\Sc}{}^\Sd=L_\Sa T_{\Sb\Sc}{}^\Sd
  -f_{\Sa\Sb}{}^\Se T_{\Se\Sc}{}^\Sd-f_{\Sa\Sc}{}^\Se T_{\Sb\Se}{}^\Sd
  +f_{\Sa\Se}{}^\Sd T_{\Sb\Sc}{}^\Se=0,
$$
также равна нулю в силу тождеств Якоби.

Данное определение связности не зависит от задания метрики на групповом
многообразии, т.к.\ связность была определена исключительно групповой операцией.
По-построению, интегральные кривые правоинвариантных векторных полей на группе
Ли являются геодезическими линиями для этой связности. Действительно,
правоинвариантное векторное поле является касательным к интегральной кривой и в
то же время является результатом параллельного переноса со связностью
(\ref{elicli}).
\begin{com}
Для задания на группе Ли инвариантной аффинной связности общего вида можно
задать в некоторой точке метрику, кручение и тензор неметричности, а затем все
эти объекты разнести по групповому многообразию.
\qed\end{com}
\section{Группа аффинных преобразований прямой                   \label{stwnog}}
Проиллюстрируем общие свойства групп Ли на примере простейшей связной неабелевой
группы Ли $\MG$. Размерность этой группы минимальна и равна двум. Этот
пример интересен, поскольку нетривиален и в то же время достаточно прост
для явного построения всех геометрических конструкций. Это -- один из немногих
случаев, когда функцию композиции можно записать в явном виде на всем групповом
многообразии.

Простейшая двумерная алгебра Ли $\Gg$ с образующими
$\lbrace L_\Sa\rbrace=(L_x,L_y)$, $\Sa=x,y$, задается коммутационными
соотношениями:
\begin{equation}                                                  \label{elitwd}
\begin{split}
  [L_x,L_x]&=[L_y,L_y]=0,
\\
  [L_x,L_y]&=\al L_y,\qquad \qquad \al\in\MR.
\end{split}
\end{equation}
При $\al=0$ эта группа является абелевой. При $\al\ne0$ растяжкой координат,
которая соответствует преобразованию $L_x\rightarrow\al L_x$, можно обратить
структурную константу в единицу, $\al=1$. Мы этого делать не будем, чтобы
следить за пределом $\al\to 0$. Группа, соответствующая этой
алгебре является подгруппой группы Лоренца $\MS\MO(1,2)$.

Алгебра Ли (\ref{elitwd}) содержит инвариантную абелеву подгруппу, которая
порождается образующей $L_y$. Поэтому она не является полупростой.
\begin{prop}
Алгебра Ли (\ref{elitwd}) является единственной двумерной неабелевой алгеброй
Ли с точностью до изоморфизма.
\end{prop}
\begin{proof}
В общем случае нетривиальное коммутационное соотношение для неабелевой алгебры
Ли определяется двумя отличными от нуля постоянными $\al$ и $\bt$:
\begin{equation*}
  [L_x,L_y]=\bt L_x+\al L_y,
\end{equation*}
где хотя бы одна из постоянных отлична от нуля. Пусть $\bt\ne 0$. Тогда сдвигом
\begin{equation*}
  L_y\mapsto L_x+\frac\al\bt L_y
\end{equation*}
это коммутационное соотношение приводится к виду (\ref{elitwd}).
\end{proof}

Покажем, что алгебра Ли (\ref{elitwd}) является алгеброй Ли группы аффинных
преобразований прямой. Аффинные преобразования прямой $x\in\MR$ параметризуются
двумя числами $a,b\in\MR$:
\begin{equation*}
  \MR\ni\quad x\mapsto ax+b\quad\in\MR,\qquad a\ne0.
\end{equation*}
Параметр $a$ параметризует дилатации, а $b$ -- сдвиги. Генераторы дилатаций и
сдвигов имеют, соответственно, следующий вид
\begin{equation*}
  L_1:=x\pl_x,\qquad L_2:=\pl_x.
\end{equation*}
Нетрудно вычислить соответсвующую алгебру Ли:
\begin{equation*}
  [L_1,L_2]=-L_2,\qquad [L_1,L_1]=[L_2,L_2]=0,
\end{equation*}
которая совпадает с алгеброй Ли (\ref{elitwd}) при $\al=-1$. Таким образом,
алгебра Ли (\ref{elitwd}) изоморфна алгебре Ли аффинных преобразований прямой.

Группа аффинных преобразований прямой двумерна и неабелева. Она состоит из двух
компонент связности: $a>0$ и $a<0$. Каждая компонента связности диффеоморфна
$\MR^2$.

Опишем явно связную компоненту единицы аффинных преобразований прямой.
Зададим алгебру Ли (\ref{elitwd}) двумя левоинвариантными векторными полями
(генераторами действия группы справа) на евклидовой плоскости $\MR^2$ в
декартовых координатах $x,y$:
\begin{equation}                                                  \label{exavel}
  L_x:=\pl_x,\qquad L_y:=\ex^{\al x}\pl_y.
\end{equation}
Нетрудно проверить, что эти векторные поля действительно удовлетворяют алгебре
(\ref{elitwd}). При $\al=0$ группа Ли является абелевой и совпадает с группой
трансляций плоскости. Поэтому ограничимся неабелевым случаем $\al>0$.
Структурные константы группы Ли равны
\begin{equation*}
\begin{split}
  f_{xy}{}^x&=-f_{yx}{}^x=0,
\\
  f_{xy}{}^y&=-f_{yx}{}^y=\al.
\end{split}
\end{equation*}
Простые вычисления дают форму Киллинга--Картана (\ref{ekilcf})
\begin{equation*}
\eta_{\Sa\Sb}=
\begin{pmatrix}
  \al^2 & 0 \\ 0 & 0
\end{pmatrix}.
\end{equation*}
Эта форма вырождена, поскольку группа Ли $\MG$ не является полупростой.

Постоим функцию композиции $f^\Sa(a,b)$ в явном виде. Обозначим координаты
элементов группы $a,b\in\MG$ через $a=(x_a,y_a)$ и $b=(x_b,y_b)$. Тогда система
дифференциальных уравнений на функцию композиции (\ref{elieqs}) примет вид
\begin{equation*}
\begin{split}
  \frac{\pl f^x(a,b)}{\pl x_b}&=1,
\\
  \frac{\pl f^x(a,b)}{\pl y_b}&=0,
\end{split}
\qquad
\begin{split}
  \frac{\pl f^y(a,b)}{\pl x_b}&=0,
\\
  \frac{\pl f^y(a,b)}{\pl y_b}&=\ex^{\al(f^x-x_b)}.
\end{split}
\end{equation*}
Два уравнения на $f^x$ с учетом начального условия $f^x(a,0)=x_a$ имеют
единственное решение $f^x(a,b)=x_a+x_b$. Аналогично, оставшиеся уравнения
определяют вторую компоненту функции композиции $f^y(a,b)=y_a+\ex^{\al x_a} y_b$.
Таким образом, дифференциальные уравнения (\ref{elieqs}) с начальными условиями
$a^\Sa=f^\Sa(a,0)$ однозначно определили функцию композиции по левоинвариантным
векторным полям $L_\Sa$:
\begin{equation}                                                  \label{econtc}
  \lbrace f^\Sa(a,b)\rbrace=(x_a+x_b,y_a+\ex^{\al x_a}y_b).
\end{equation}
Поскольку функция композиции определена для всех $a,b\in\MR^2$ и  отображает
$\MR^2\times\MR^2$ на всю плоскость $\MR^2$, то группа Ли $\MG$, как
многообразие, диффеоморфна евклидовой плоскости $\MR^2$ и, следовательно,
некомпактна.

Несложные вычисления показывают, что построенная функция композиции
(\ref{econtc}) действительно удовлетворяет всем групповым аксиомам
(\ref{ecoruc}) с обратным элементом:
\begin{equation*}
  a^{-1}=(-x_a,-\ex^{-\al x_a}y_a).
\end{equation*}

Таким образом, мы построили неабелеву групповую структуру на евклидовой
плоскости $\MR^2$. Вспомним также, что плоскость $\MR^2$ можно рассматривать,
как абелеву группу Ли двумерных трансляций. Тем самым мы показали, что на
одном и том же многообразии возможно задание различных групповых структур.

Знание функции композиции позволяет вычислить все объекты на групповом
многообразии, которые задаются групповым действием. В частности,
правоинвариантные векторные поля (\ref{erinve}) имеют вид:
\begin{equation}                                                  \label{erinvf}
  R_x=\pl_x+\al y\pl_y,\qquad R_y=\pl_y,
\end{equation}
и удовлетворяют коммутационным соотношениям:
\begin{equation*}
\begin{split}
  [R_x,R_x]&=[R_y,R_y]=0
\\
  [R_x,R_y]&=-\al R_y,
\end{split}
\end{equation*}
которые отличаются знаком от коммутационных соотношений алгебры Ли
левоинвариантных векторных полей (\ref{elitwd}). Нетрудно также проверить,
что лево- и правоинвариантные векторные поля коммутируют, $[L_\Sa,R_\Sb]=0$.

\begin{figure}[h,b,t]
\hfill\includegraphics[width=.3\textwidth]{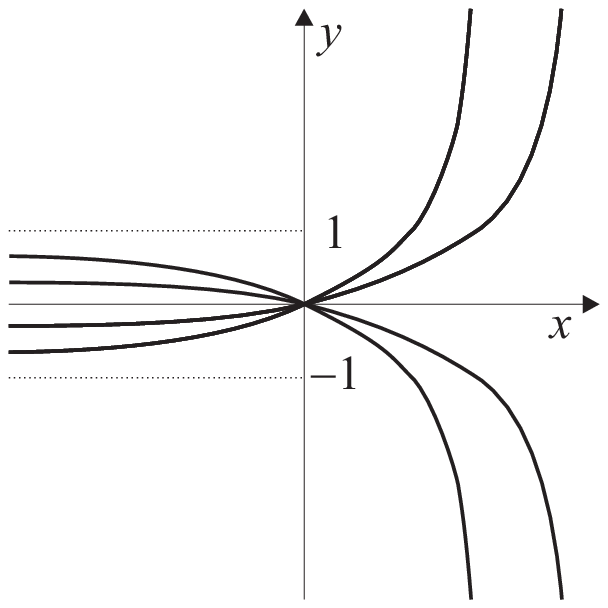}
\hfill {}
\centering\caption{Интегральные кривые для левоинвариантных векторных полей,
 проходящие через начало координат, $x_0=0$, $y_0=0$, под углом
 $\arctg(Y_0/X_0)$.}
\label{fintct}
\end{figure}

Построим интегральные кривые левоинвариантных векторных полей. Произвольное
левоинвариантное векторное поле задается двумя постоянными $X_0,Y_0\in\MR$
\begin{equation*}
  L=X_0\pl_x+Y_0\ex^{\al x}\pl_y.
\end{equation*}
Соответствующие интегральные кривые $\big(x(t),y(t)\big)$ определяются системой
уравнений
\begin{equation*}
  \dot x=X_0,\qquad \dot y=Y_0\ex^{\al x}.
\end{equation*}
При $X_0\ne0$, эти уравнения имеют общее решение, задающее интегральную
кривую в параметрическом виде
\begin{align*}
  x&=X_0t+x_0,
\\
  y&=\frac{Y_0}{\al X_0}\ex^{\al x_0}(\ex^{\al X_0t}-1)+y_0,
\end{align*}
где постоянные интегрирования $x_0,y_0$ являются координатами точки, через
которую проходит интегральная кривая при $t=0$. Отсюда следует соотношение,
определяющее форму интегральной кривой
\begin{equation*}
  y=\frac{Y_0}{\al X_0}(\ex^{\al x}-\ex^{\al x_0})+y_0.
\end{equation*}
При $X_0=0$ интегральные кривые параллельны оси $y$. Вид интегральных
кривых, проходящих через начало координат под разными углами, которые
определяются постоянными $X_0,Y_0$, показан на рис.~\ref{fintct}
Интегральные кривые определены для всех значений параметра $t\in\MR$ и
поэтому левоинвариантные векторные поля полны.

Зададим теперь геометрию, т.е.\ метрику и аффинную связность, на группе Ли
аффинных преобразований прямой $\MG\approx\MR^2$. Конечно, интерес представляют
те геометрические объекты, которые согласованы с групповой операцией. Начнем с
метрики. Поскольку группа Ли $\MG$ неабелева и не является полупростой, то на
ней не существует двусторонне инвариантной метрики. Лучшее, что можно сделать,
это построить право- или левоинвариантную метрику. С этой целью зададим матрицу
$\theta_{\Sa\Sb}$ в начале координат и разнесем ее, например, с помощью действия
группы справа. Правоинвариантная метрика в координатном базисе (\ref{elerim})
имеет вид
\begin{equation}                                                  \label{erinml}
  g_{\Sa\Sb}:=g^R_{\Sa\Sb}(a)
  =R^{-1}_{~~\Sa}{}^\Sc(a)R^{-1}_{~~\Sb}{}^\Sd(a)\theta_{\Sc\Sd},
\end{equation}
где матрица $\theta_{\Sa\Sb}$ симметрична и невырождена, а в остальном
совершенно произвольна. В частности, она может иметь лоренцеву сигнатуру,
т.е.\ на группе Ли $\MG$ можно построить как риманову, так и псевдориманову
геометрию.

Рассмотрим случай, когда в начале координат задана единичная матрица,
$\theta_{\Sa\Sb}=\dl_{\Sa\Sb}$. Тогда формула (\ref{erinml}), в которой
матрицы $R_\Sa{}^\Sb$ определены равенствами (\ref{erinvf}), задает
следующую правоинвариантную метрику в декартовых координатах на плоскости
$\MR^2$
\begin{equation}                                                  \label{emegri}
  g_{\Sa\Sb}=\begin{pmatrix} 1+\al^2y^2 & -\al y \\ -\al y & 1 \end{pmatrix}.
\end{equation}
Эта метрика имеет единичный определитель, $\det g_{\Sa\Sb}=1$, и ее
обратная имеет вид
\begin{equation*}
  g^{\Sa\Sb}=
\begin{pmatrix}
    1 & \al y \\ \al y & 1+\al^2y^2
\end{pmatrix}.
\end{equation*}
Прямые вычисления дают следующие выражения для символов Кристоффеля:
\begin{equation*}
\begin{aligned}
  \widetilde\Gamma_{xx}{}^x&=-\al^3 y^2, &\qquad \widetilde\Gamma_{xx}{}^y
  &=-\al^2y(1+\al^2y^2),
\\
  \widetilde\Gamma_{xy}{}^x&=\widetilde\Gamma_{yx}{}^x=\al^2y,
  & \widetilde\Gamma_{xy}{}^y&=\widetilde\Gamma_{yx}{}^y=\al^3y^2,
\\
  \widetilde\Gamma_{yy}{}^x&=-\al, & \widetilde\Gamma_{yy}{}^y&=-\al^2y.
\end{aligned}
\end{equation*}
У соответствующего тензора кривизны со всеми опущенными индексами только
одна компонента отлична от нуля:
\begin{equation*}
  \widetilde R_{xyxy}=\al^2,
\end{equation*}
что можно переписать в ковариантном виде
\begin{equation*}
  \widetilde R_{\Sa\Sb\Sc\Sd}=\al^2(g_{\Sa\Sc}g_{\Sb\Sd}-g_{\Sa\Sd}g_{\Sb\Sc}).
\end{equation*}
Тем самым групповое многообразие $\MG\approx\MR^2$ с метрикой (\ref{emegri})
становится пространством постоянной кривизны.
Соответствующий тензор Риччи и скалярная кривизна нетривиальны:
\begin{equation*}
\begin{split}
  \widetilde R_{xx}&=\al^2(1+\al^2y^2),
\\
  \widetilde R_{xy}&=\widetilde R_{yx}=-\al^3y,
\\
  \widetilde R_{yy}&=\al^2,
\\
  \widetilde R&=-2K=2\al^2,
\end{split}
\end{equation*}
где $K=-\al^2$ -- гауссова кривизна двумерной группы Ли $\MG$. Как многообразие,
это -- поверхность постоянной отрицательной кривизны (одна компонента
двуполостного гиперболоида), которая рассмотрена в разделе \ref{sutwse}.
\section{Гомоморфизмы групп Ли                                   \label{slieho}}
\begin{defn}
Отображение $f:~\MG\rightarrow\MH$ называется {\em гомоморфизмом групп Ли}, если
$f$ гладко и в то же время является гомоморфизмом групп. Если, кроме того, $f$
является диффеоморфизмом, то отображение $f$ называется {\em изоморфизмом групп
Ли}. Изоморфизм группы Ли на себя называется ее {\em автоморфизмом}. {\em Ядром}
гомоморфного отображения называется то множество элементов группы $\MG$, которые
отображаются в единичный элемент группы группы $\MH$.
\qed\end{defn}
\index{Гомоморфизм групп Ли (homomorphism of Lie groups)}%
\index{Изоморфизм групп Ли (isomorphism of Lie groups)}%
\index{Автоморфизм группы Ли (automorphism of Lie groups)}%
\index{Ядро отображения (homomorphism kernel)}%
\index{Отображения ядро (homomorphism kernel)}%
Из определения следует, что при гомоморфизме групп единичный элемент группы
$\MG$ всегда отображается в единичный элемент группы $\MH$.

Можно доказать, что из непрерывности гомоморфизма $f:~\MG\rightarrow\MH$
следует его гладкость \cite{Warner83R}. Поэтому в определении гомоморфизма
достаточно потребовать его непрерывность.

\begin{theorem}
Ядром гомоморфного отображения $\MG_1\rightarrow\MG_2$ всегда является
инвариантная подгруппа $\MH_1$ группы $\MG_1$. При этом имеет место изоморфизм
$$
  \frac{\MG_1}{\MH_1}\simeq\MG_2.
$$
\end{theorem}
\begin{proof}
См., например, \cite{Pontry84R}.
\end{proof}
\begin{defn}
Гомоморфизм $\rho:~\MG\rightarrow\MH$ называется {\em представлением} группы
Ли $\MG$, если $\MH=\aut(\MV)$ для некоторого векторного пространства
$\MV$, или $\MH=\MG\ML(n,\MC)$, или $\MH=\MG\ML(n,\MR)$. Если гомоморфизм
$\rho:~\MG\rightarrow f(\MG)\subset\MH$ является изоморфизмом, то представление
называется {\em точным}. Представление $\rho$ группы Ли $\MG$ в векторном
пространстве $\MV$ мы обозначаем парой $(\rho,\MV)$.
\qed\end{defn}
\index{Представление группы Ли (representation of a Lie group)}%
\index{Представление группы Ли точное (exact representation of a Lie group)}%
\index{Точное представление группы Ли (exact representation of a Lie group)}%

В определении представления во всех трех случаях каждому элементу группы $\MG$
ставится в соответствие матрица. Если элементами матриц являются вещественные
или комплексные числа, то говорят, соответственно, о {\em вещественном} или
{\em комплексном} представлении. При этом все матрицы имеют отличный от нуля
определитель, т.к.\ у каждого элемента группы должен быть обратный элемент.
Следовательно, каждой матрице представления соответствует некоторый автоморфизм
векторного пространства. Для точного представления группы образ
$f(\MG)\subset\MH$ совсем не обязательно совпадает со всем $\MH$.

Пусть $f:~\MG\rightarrow\MH$ -- гомоморфизм групп Ли. Так как под действием
отображения $f$ единица группы $\MG$ переходит в единицу группы $\MH$, то
дифференциал отображения отображает касательные пространства к единицам групп:
\begin{equation*}
  f_*:\quad \MT_e(\MG)~\rightarrow~\MT_e(\MH).
\end{equation*}
\begin{com}
Единицы разных групп мы обозначаем одним и тем же символом $e$, т.к.\ это не
приводит к путанице.
\qed\end{com}
Поскольку касательные пространства к единицам изоморфны алгебрам Ли,
то дифференциал отображения индуцирует отображение алгебр, которое мы
также будем обозначать $f_*$. Таким образом,
\begin{equation}                                                  \label{ealgma}
  f_*:\quad \Gg~\rightarrow~\Gh,
\end{equation}
где образом $f_*(X)$ для $X\in\Gg$ является единственное левоинвариантное
векторное поле на $\MH$, такое, что в единице группы $e\in\MH$ оно определено
соотношением
\begin{equation*}
  f_*(X)(e):=f_*\big(X(e)\big).
\end{equation*}
\begin{prop}
Пусть $\MG$ и $\MH$ -- группы Ли с алгебрами Ли $\Gg$ и $\Gh$ соответственно,
и пусть $f:~\MG\rightarrow\MH$ -- гомоморфизм групп Ли. Тогда отображение
(\ref{ealgma}) является гомоморфизмом алгебр Ли.
\end{prop}
\begin{proof}
Дифференциал отображения линеен и сохраняет скобку Ли (\ref{ecomhy}).
\end{proof}
Пусть $f:~\MG\rightarrow\MH$ -- гомоморфизм групп Ли. Тогда возврат отображения
$f^*$ отображает левоинвариантные 1-формы на $\MH$ в левоинвариантные 1-формы на
$\MG$, как отображение сопряженного пространства к $\Gh$ в сопряженное
пространство к $\Gg$.
\begin{exa}[\bf Обозначения]
Запишем вектор из векторного пространства $\MV$ в виде $X=X^i\Be_i$, где
$\Be_i$, $i=1,\dotsc,\dim\MV$, -- некоторый фиксированный базис. Если задано
представление $(\rho,\MV)$ группы Ли $\MG$, то мы пишем
\begin{equation*}
  \MG\ni a:\quad \MV\ni\quad X=X^i\Be_i\mapsto X'=X^iS_i{}^j\Be_j\quad\in\MV.
\end{equation*}
То есть каждому элементу группы $a\in\MG$ соответствует некоторая матрица
$S_i{}^j(a)$. По определению, все элементы матрицы гладко зависят от $a$. Мы
пишем
\begin{equation*}
  \rho:\quad \MG\ni\quad a\mapsto S_i{}^j(a)\quad\in\aut\MV.
\end{equation*}
Ясно, что единичному элементу группы соответствует единичная матрица,
\begin{equation*}
  S_i{}^j(e)=\dl_i^j,
\end{equation*}
а обратному элементу группы -- обратная матрица,
\begin{equation*}
  S_i{}^j(a^{-1})=S^{-1}_{~~j}{}^j(a).
\end{equation*}
Произведению двух элементов группы $ab$ соответствует матрица
\begin{equation*}
  S_i{}^j(ab)=S_i{}^k(b)S_k{}^j(a),
\end{equation*}
которая равна произведению матриц для каждого элемента, взятых в обратном
порядке. Порядок матриц является следствием принятого нами соглашения в теории
групп преобразований, где элемент группы, по-определению, действует на точку
многообразия справа (см.\ раздел \ref{stragf}).

Дифференциал отображения $\rho_*$ отображает алгебры Ли. В частности, генераторы
алгебры Ли отображаются в матрицы,
\begin{equation*}
  \rho_*:\quad \Gg\ni\quad L_\Sa\mapsto L_{\Sa i}{}^j\quad\in\End\MV.
\end{equation*}
Матрицы представления $L_{\Sa i}{}^j$ генераторов $L_\Sa$ удовлетворяют тем же
коммутационным соотношениям, что и левоинвариантный базис векторных полей на
группе Ли:
\begin{equation}                                                  \label{ecomar}
  [L_\Sa,L_\Sb]_i{}^j=f_{\Sa\Sb}{}^\Sc L_{\Sc i}{}^j,
\end{equation}
где $f_{\Sa\Sb}{}^\Sc$ -- структурные константы группы Ли $\MG$, определенные в
(\ref{eliall}).

Можно доказать, что матрицы представления группы Ли $\MG$ вблизи единицы группы
в линейном приближении имеют вид
\begin{equation}                                                  \label{esmrem}
  S_i{}^j(da)=\dl_i^j-da^\Sa L_{\Sa i}{}^j+\dotsc,
\end{equation}
где $da\in\MG$ -- элемент группы, близкий к единице.

Найдем правила дифференцирования матриц представления. С одной стороны,
используя разложение (\ref{esmrem}), имеем равенство
\begin{equation}                                                  \label{eqmatr}
  S_{i}{}^j\big(f(a,da)\big)=S_i{}^k(da)S_k{}^j(a)
  =\left(\dl_i^k-da^\Sa L_{\Sa i}{}^k\right)S_k{}^j(a).
\end{equation}
С другой стороны, матрицы представления можно разложить в ряд Тейлора:
\begin{equation}                                                  \label{edigen}
  S_i{}^j\big(f(a,da)\big)=S_i{}^j(a)+da^\Sa L_\Sa S_i{}^j,
\end{equation}
где левоинвариантные векторные поля $L_\Sa$ действуют на матрицы представления,
как дифференцирования, и мы воспользовались равенством
\begin{equation*}
  f^\Sa(a,da)= a^\Sa+da^\Sb L_\Sb{}^\Sa+\dotsc.
\end{equation*}
Сравнивая формулы (\ref{eqmatr}) и (\ref{edigen}), получаем правило
дифференцирования матриц представления
\begin{equation}                                                  \label{eredim}
  L_\Sa S_i{}^j=-L_{\Sa i}{}^k S_k{}^j.
\end{equation}
Дифференцируя равенство $S^{-1}S=\one$, получаем правило дифференцирования
обратных матриц представления
\begin{equation}                                                  \label{ediinm}
  L_\Sa S^{-1}_{~~i}{}^j=S^{-1}_{~~i}{}^k L_{\Sa k}{}^j.
\end{equation}

Каждому автоморфизму $\rho(a)$ векторного пространства $\MV$ соответствует
гомоморфизм алгебр $\rho_*(a)$. Поскольку гомоморфизм алгебр сохраняет скобку
Ли, то матрицы представления генераторов группы инвариантны:
\begin{equation}                                                  \label{einvge}
  L_{\Sa i}{}^j=S^{-1}_{~~\Sa}{}^\Sb S^{-1}_{~~i}{}^k L_{\Sb k}{}^l S_l{}^j,
\end{equation}
где $S_\Sa{}^\Sb$ -- матрица присоединенного представления.
\qed\end{exa}

Продолжим общее рассмотрение.
\begin{defn}
Пара $(f,\MH)$ называется {\em подгруппой Ли} группы Ли $\MG$, если
выполнены следующие условия:

1) $\MH$ -- группа Ли;

2) $(f,\MH)$ -- подмногообразие в $\MG$;

3) $f:~\MH\rightarrow\MG$ -- групповой гомоморфизм.\newline
Пара $(f,\MH)$ называется {\em замкнутой подгруппой Ли} в $\MG$, если
к тому же $f(\MH)$ -- замкнутое подмногообразие в $\MG$.
\qed\end{defn}
\index{Подгруппа Ли (Lie subgroup)}\index{Ли подгруппа (Lie subgroup)}%
\index{Замкнутая подгруппа Ли (closed Lie subgroup)}%
\index{Подгруппа Ли замкнутая (closed Lie subgroup)}%
Определение замкнутого подмногообразия было дано в разделе \ref{subman}.
Если группа Ли $\MG$ связна и $\MH$ не изоморфна $\MG$, то размерность замкнутой
подгруппы Ли всегда меньше размерности самой группы.

Пусть $(f,\MH)$ -- подгруппа Ли в $\MG$, и $\Gh$ и $\Gg$ -- соответствующие
алгебры Ли. Тогда дифференциал отображения $f_*$ задает изоморфизм $\Gh$ с
подалгеброй $f_*(\Gh)$ в алгебре $\Gg$.

Гомоморфизмы групп можно описать следующим образом. Пусть $f:~\MG\rightarrow\MH$
-- сюрьективное отображение. Если это отображение является взаимно однозначным,
то мы имеем изоморфизм групп, $\MG\simeq\MH$. Если отображение $f$ не является
взаимно однозначным, то ядро этого отображения $\ker f$ представляет собой
нормальную подгруппу в $\MG$. Тогда фактор группа $\MG/\ker f$ изоморфна $\MH$.
Если отображение $f$ не является сюрьективным, то группа $\MG$ сюрьективно
отображается на какую то подгруппу $\MH_1\subset\MH$. Если отображение $\MG$ на
$\MH_1$ взаимно однозначно, то отображение $f$ является мономорфизмом. В
противном случае, $\MG/\ker f\simeq\MH_1$.

Напомним, что подалгеброй Ли $\tilde\Gh$ в $\Gg$ называется линейное
подпространство в $\Gg$, для которого выполнено включение
$[\tilde\Gh,\tilde\Gh]\subset\tilde\Gh$.
\begin{theorem}
Пусть $\MG$ -- группа Ли с алгеброй Ли $\Gg$, и пусть $\tilde\Gh$ -- подалгебра
Ли в $\Gg$. Тогда существует единственная связная подгруппа Ли $(f,\MH)$ группы
$\MG$ с алгеброй $\Gh$, такая, что $f_*(\Gh)=\tilde\Gh$.
\end{theorem}
\begin{proof}
Основано на теореме Фробениуса \cite{Warner83R}. Левоинвариантные векторные поля
из подалгебры $\Gh$, согласно определению, находятся в инволюции. Из теоремы
Фробениуса следует, что в каждой точке группы Ли они задают интегральные
подмногообразия. Максимальное интегральное подмногообразие, проходящее через
единицу группы, и есть подгруппа Ли $(f,\MH)$. Если $\dim\tilde\Gh<\dim\Gg$, то
подалгебра Ли $\tilde\Gh$ обязательно является замкнутым подмножеством в $\MG$.
\end{proof}
\begin{cor}
Существует взаимно однозначное соответствие между связными подгруппами Ли группы
Ли и подалгебрами ее алгебры Ли.
\qed\end{cor}
\begin{theorem}
Пусть $(f,\MH)$ -- подгруппа Ли группы Ли $\MG$. Тогда отображение $f$ является
регулярным вложением тогда и только тогда, когда $(f,\MH)$ -- замкнутая
подгруппа в $\MG$.
\end{theorem}
\begin{proof}
См., например, \cite{Warner83R}.
\end{proof}
\section{Экспоненциальное отображение для групп Ли               \label{sexpli}}
\begin{defn}
Гомоморфизм $f:~\MR\rightarrow\MG$ вещественной прямой, рассматриваемой
как группа Ли по сложению, в группу Ли называется {\em однопараметрической
подгруппой} группы Ли $\MG$.
\qed\end{defn}
\index{Однопараметрическая подгруппа (one parameter subgroup)}%
\index{Подгруппа однопараметрическая (one parameter subgroup)}%
Подчеркнем, что однопараметрическая подгруппа Ли является не подмножеством в
$\MG$, а отображением.

Рассмотрим отображение соответствующих алгебр Ли. Элемент алгебры Ли $\Gr$
вещественной прямой $t\in\MR$ имеет вид $s\pl_t\in\Gr$, где $s\in\MR$ и $\pl_t$
-- базисный вектор касательного пространства к прямой $\MR$ в нуле. Пусть $\Gg$
-- алгебра Ли группы Ли $\MG$, которую в данном случае удобнее отождествить с
касательным пространством к единице группы. Рассмотрим произвольный элемент
алгебры Ли $X\in\Gg$. Тогда отображение
\begin{equation*}
  \Gr\ni\quad s\pl_t\mapsto sX\quad\in\Gg
\end{equation*}
является гомоморфизмом алгебры Ли $\Gr$ в алгебру Ли $\Gg$. Так как
вещественная прямая односвязна, то по утверждению теоремы~\ref{talgrl}
существует единственная однопараметрическая подгруппа, для которой мы введем
специальное обозначение
\begin{equation}                                                  \label{expsug}
  \exp_X:\quad \MR\rightarrow\MG,
\end{equation}
такая, что дифференциал соответствующего отображения отображает элементы алгебры
Ли по следующему правилу
\begin{equation}                                                  \label{expalk}
  (\exp_X)_*(s\pl_t)=sX.
\end{equation}
\begin{defn}
{\em Экспоненциальным отображением} алгебры Ли $\Gg$ в группу Ли $\MG$
называется отображение
\begin{equation}                                                  \label{expalg}
  \exp:\quad \Gg\ni\quad X\mapsto\exp(X):=\exp_X(1)\quad\in\MG. \qed
\end{equation}
\end{defn}
\index{Экспоненциальное отображение (exponential map)}%
\index{Отображение экспоненциальное (exponential map)}%
Происхождение такой терминологии будет ясно из дальнейшего, когда будет
показано, что для линейной группы $\MG\ML(n,\MR)$ экспоненциальное
отображение задается показательной функцией от матрицы.

Рассмотрим экспоненциальное отображение (\ref{expalg}) с точки зрения
интегральных кривых для векторных полей, введенных в разделе \ref{svechs}.
Обозначим координаты в окрестности единицы группы через
$a^\Sa$, $\Sa=1,\dotsc,\Sn$, такие, что единице группы $e\in\MG$ соответствует
начало координат. Тогда отображение (\ref{expsug}) в координатах задается
набором функций $a^\Sa(t)$, которые определяются системой обыкновенных
дифференциальных уравнений
\begin{equation*}
  \dot a^\Sa=X^\Sa,
\end{equation*}
где $X^\Sa$ -- компоненты левоинвариантного векторного поля $X$, с начальными
условиями $a^\Sa(0)=0$. Другими словами, отображение $t\mapsto\exp_X(t)$
является единственной однопараметрической подгруппой в $\MG$ с касательным
вектором $X(0)$ в нуле.

Между интегральными кривыми левоинвариантных векторных полей и операцией
коммутирования в алгебре Ли существует связь.
\begin{prop}
Пусть задано два левоинвариантных векторных поля $X,Y\in\Gg$. Обозначим через
$a(t)$ и $b(t)$ интегральные кривые этих векторных полей, которые проходят
через единицу группы. Тогда кривая
\begin{equation*}
  c(t):=a(\sqrt t)b(\sqrt t)a^{-1}(\sqrt t)b^{-t}(\sqrt t)
\end{equation*}
касается векторного поля $[X,Y]\in\Gg$ в единице группы.
\end{prop}
\begin{proof}
Прямая проверка.
\end{proof}

Свойства экспоненциального отображения дает следующая
\begin{theorem}
Рассмотрим произвольный элемент алгебры Ли $X\in\Gg$ на группе Ли $\MG$.
Тогда \newline
\indent 1) $\exp(tX)=\exp_X(t)$, \quad$\forall t\in\MR$;\newline
\indent 2) $\exp[(t_1+t_2)X]=\exp(t_1X)\exp(t_2X)$,\quad
$\forall t_1,t_2\in\MR$;\newline
\indent 3) $\exp(-tX)=[\exp(tX)]^{-1}$, \quad$\forall t\in\MR$;\newline
\indent 4) \parbox[t]{.92\linewidth}{отображение $\exp:~\Gg\rightarrow\MG$
гладкое, и дифференциал отображения $\exp_*:~\MT_0(\Gg)\rightarrow\MT_e(\MG)$
-- тождественное отображение (при обычных отождествлениях), так что $\exp$
задает диффеоморфизм окрестности нуля в алгебре $\Gg$ на окрестность
единицы в группе $\MG$;}\newline
\indent 5) \parbox[t]{.92\linewidth}{одномерное подмногообразие, полученное
с помощью действия левых сдвигов $l_a\circ\exp_X\in\MG$, -- это единственная
интегральная кривая левоинвариантного векторного поля $X$, проходящая при $t=0$
через точку $a\in\MG$;}\newline
\indent 6) \parbox[t]{.92\linewidth}{однопараметрическая группа
диффеоморфизмов $X_t$, порожденная левоинвариантным векторным полем $X$,
задается правыми сдвигами
\begin{equation*}
  X_t=r_{\exp_X(t)}.
\end{equation*}}
\end{theorem}
\begin{proof}
См., например, \cite{Warner83R}.
\end{proof}
Из свойства 5), в частности, следует, что все левоинвариантные
векторные поля полны.
\begin{com}
Согласно теореме \ref{tcovcr}, векторное поле на компактном многообразии, не
обращающееся в нуль, является полным. В то же время существует много примеров
неполных векторных полей на некомпактных многообразиях. В случае групп Ли каждое
право- или левоинвариантное векторное поле полно, даже ясли группа Ли
некомпактна.
\qed\end{com}

Для двух коммутирующих элементов алгебры Ли,
\begin{equation*}
  [X,Y]=0,\qquad X,Y\in\Gg,
\end{equation*}
справедлива формула
\begin{equation*}
  \exp(X+Y)=\exp X\exp Y.
\end{equation*}

Экспоненциальное отображение согласовано с гомоморфизмом групп.
\begin{theorem}
Пусть задан гомоморфизм групп Ли $f:~\MH\rightarrow\MG$, тогда
следующая диаграмма коммутативна
\begin{equation*}
\begin{diagram}
  \MH & \rTo^f & \MG \\
  \uTo^\exp & & \uTo_\exp \\
  \Gh & \rTo^{f_*} & \Gg \\
\end{diagram}
\end{equation*}
\end{theorem}
\begin{proof}
См., например, \cite{Warner83R}.
\end{proof}
\begin{cor}
Любая однопараметрическая подгруппа в $\MG$ имеет вид экспоненциального
отображения $t\mapsto \exp_X(t)$ для некоторого $X\in\Gg$.
\qed\end{cor}

Если группа Ли $\MG_0$ компактна и связна, то экспоненциальное отображение
является сюрьективным, т.е.\ отображает алгебру Ли на все многообразие группы.
Как показывают явные формулы для групп $\MS\MO(3)$ и $\MS\MU(2)$ (см.\ раздел
\ref{sthecs}) экспоненциальное отображение не является взаимно однозначным:
одному элементу группы Ли $\MG_0$ в общем случае соответствует много различных
элементов соответствующей алгебры Ли $\Gg$. Поскольку $n$-мерная алгебра Ли
как векторное пространство представляет собой вещественное $n$-мерное
пространство $\MR^n$, то при экспоненциальном отображении возникает
параметризация связной компоненты единицы компактной группы Ли $\MG_0$ точками
евклидова пространства $\MR^n$ с некоторым отношением эквивалентности. А именно,
мы отождествляем те точки евклидова пространства, которые отображаются в один и
тот же элемент группы. Экспоненциальное отображение является взаимно однозначным
только в некоторой окрестности единицы группы $e\in\MG_0$.

Если связная группа Ли некомпактна, то экспоненциальное отображение в общем
случае не является сюрьективным. Пример приведен в \cite{Helgas62}.
\section{Интегрирование на группах Ли}
Рассмотрим группу Ли $\MG$ размерности $\Sn$. В разделе \ref{sleacg} было
показано, что любая группа Ли является ориентируемым многообразием. Зафиксируем
некоторую ориентацию на $\MG$ и выберем левоинвариантную дифференциальную
$\Sn$-форму $\om$ на $\MG$, согласованную с ориентацией, которая нигде не
обращается в нуль. Для этого достаточно выбрать ненулевую $\Sn$-форму правильной
ориентации, например, в единице группы, и разнести ее по групповому многообразию
с помощью действия группы слева. Напомним, что действие группы и слева, и справа
сохраняет ориентацию выбранного базиса.

Запишем условие левой инвариантности формы объема $\om$ с помощью возврата
отображения. Пусть задано левое действие группы
\begin{equation*}
  l_a:\quad\MG\ni\quad b\mapsto l_a b:=ab\quad\in\MG.
\end{equation*}
Тогда условие левой инвариантности принимает вид
\begin{equation*}
  \om(b)=l_a^*\om(ab),
\end{equation*}
где $l_a^*$ -- возврат отображения $l_a$.

Пусть на группе Ли $\MG$ задана функция $f$ с компактным носителем. Тогда
определен интеграл
\begin{equation}                                                  \label{eingro}
  \int_\MG \!\om(b) f(b),
\end{equation}
где аргумент $b$ соответствует переменной интегрирования. Этот интеграл зависит
от формы $\om$, которую можно умножить на постоянный положительный множитель. В
случае компактных групп Ли форму $\om$ можно однозначно фиксировать с помощью
дополнительного условия
\begin{equation*}
  \int_\MG\!\om(b)=1.
\end{equation*}

Заметим, что если $\om$ является формой объема, то $\om f$ -- это $n$-форма, и
на нее также действует возврат отображения..

Поскольку левое действие группы является диффеоморфизмом, то справедлива
формула замены переменных интегрирования (\ref{qhwaqn}). В рассматриваемом
случае она принимает вид
\begin{equation*}
  \int_{a\MG}\!\om(ab)f(ab)=\int_\MG\!l_a^*\big(\om(ab)f(ab)\big),
\end{equation*}
где в первом интеграле интегрирование ведется по $ab$, а во втором -- по $b$.
Поскольку $l_a^*\big(\om(ab)f(ab)\big)=\big(l_a^*\om(ab)\big)f(ab)$, то из
равенства $a\MG=\MG$ и левой инвариантности формы объема вытекает равенство
интегралов
\begin{equation*}
  \int_\MG\!\om(b)f(b)=\int_\MG\!\om(b)f(ab).
\end{equation*}
Это означает что интеграл (\ref{eingro}) от функции $f$ является
левоинвариантным.

Поставим вопрос о том, когда интеграл (\ref{eingro}) является одновременно
и правоинвариантным, т.е.\ выполнено равенство
\begin{equation*}
  \int_\MG\!\om(b)f(b)=\int_\MG\!\om(b)f(ba).
\end{equation*}
Форма $r_a^*\om$ по-прежнему левоинвариантна, потому что левые и правые сдвиги
коммутируют между собой. Следовательно, форма $r_a^*\om$ отличается от $\om$
только постоянным положительным множителем, который может зависеть от $a$. Это
значит, что на $\MG$ существует положительная вещественная функция $\lm(a)>0$
такая, что
\begin{equation*}
  r_a^*\om(ba)=\lm(a)\om(b).
\end{equation*}
\begin{prop}
Функция $\lm(a)$ задает гомоморфизм группы Ли $\MG$ в мультипликативную группу
положительных чисел:
\begin{equation}                                                  \label{qnhjjh}
  \lm(ab)=\lm(a)\lm(b).
\end{equation}
\end{prop}
\begin{proof}
Вычислим действие двух возвратов отображений
\begin{equation*}
  r_a^*r_b^*\om(cab)=r_a^*\lm(b)\om(ca)=\lm(a)\lm(b)\om(c).
\end{equation*}
С другой стороны, для возвратов отображений выполнено равенство
\begin{equation*}
  r_a^*r_b^*=(r_br_a)^*=r_{ab}^*,
\end{equation*}
где мы воспользовались формулой (\ref{qpuiiy}). Поэтому
\begin{equation*}
  r_a^*r_b^*\om(cab)=r_{ab}^*\om(cab)=\lm(ab)\om(c).
\end{equation*}
Следовательно, равенство (\ref{qnhjjh}) выполнено.
\end{proof}
\begin{defn}
Функция $\lm(a)$ называется {\em модулярной функцией}.
\qed\end{defn}
\index{Модулярная функция (modular function)}%
\index{Функция модулярная (modular function)}%
\begin{prop}
Модулярная функция связана с присоединенным представлением следующим
соотношением
\begin{equation*}
  \lm(a)=|\det\ad (a)|,\qquad a\in\MG,
\end{equation*}
где $\ad(a)$ -- матрица присоединенного представления элемента группы $a$.
\end{prop}

Поскольку групповое действие является диффеоморфизмом, то справедливы равенства:
\begin{equation*}
  \int_{\MG a}\!\om(ba)f(ba)=\int_\MG\!r_a^*\big(\om(ba)f(ba)\big)
  =\int_\MG\!\om(b)\lm(a)f(ba).
\end{equation*}
Отсюда следует, что левоинвариантный интеграл (\ref{eingro}) является также
правоинвариантным тогда и только тогда, когда $\lm(a)=1$ для всех $a\in\MG$.
Группа Ли $\MG$, для которой $\lm(a)=1$ называется {\em унимодулярной}.
\index{Группа Ли унимодулярная (unimodular Lie group)}%
\index{Унимодулярная группа Ли (unimodular Lie group)}%
\begin{prop}
Каждая компактная группа Ли является унимодулярной.
\end{prop}
\begin{proof}
Поскольку левоинвариантная форма объема на компактной группе Ли нормирована, то
справедливы равенства:
\begin{equation*}
  1=\int_{\MG a}\!\om(ba)=\int_\MG\!r_a^*\om(ba)=\lm(a)\int_\MG\om(b)=\lm(a).
\end{equation*}
Отсюда следует утверждение предложения.
\end{proof}
Следовательно, интеграл от функции (\ref{eingro}) по компактной группе Ли
является одновременно и право-, и левоинвариантным.

Компактные группы Ли не исчерпывают весь класс унимодулярных групп.
\begin{prop}
Любая полупростая группа Ли является унимодулярной.
\end{prop}
\begin{proof}
См., например, \cite{Helgas62R}.
\end{proof}
Существуют также другие унимодулярные группы.

\begin{defn}
Мера интегрирования $\om$ на группе Ли называется {\em левоинвариантной мерой
Хаара}. В случае унимодулярной группы Ли мера $\om$ называется
{\em двусторонне инвариантной мерой Хаара}
\qed\end{defn}
\index{Мера Хаара (Haar measure)}\index{Хаара мера (Haar measure)}%
Существование лево- или правоинвариантной меры на группе Ли, {\em меры
Хаара}, возможно только для локально компактных групп Ли.

Поскольку левые и правые сдвиги на группе Ли коммутируют, то всякая абелева
локально компактная группа Ли унимодулярна.
\begin{exa}
Рассмотрим вещественную прямую $x\in\MR$ как группу сдвигов. Эта группа абелева,
и ее двусторонне инвариантная мера Хаара имеет стандартный вид $\om=dx$. Условие
левой и правой инвариантности интеграла от непрерывной функции с компактным
носителем хорошо известно:
\begin{equation*}
  \int_{\MR}\!dx\, f(x)=\int_{\MR}\!dx\, f(x+h),
\end{equation*}
где $h\in\MR$.
\qed\end{exa}
\begin{exa}[\bf Группа Вейля]
\index{Группа Вейля (Weyl group)}\index{Вейля группа (Weyl group)}%
Группа Вейля состоит из верхне треугольных матриц вида
\begin{equation*}
  g(a,b,c):=\begin{pmatrix} 1 & a & c \\ 0 & 1 & b \\ 0 & 0 & 1 \end{pmatrix},
\end{equation*}
где $a,b,c\in\MR$, с обычным правилом умножения. Эта группа трехмерна и
диффеоморфна евклидову пространству $\MR^3$. Функция композиции для группы
Вейля записывается в явном виде:
\begin{equation*}
  (a,b,c)(a',b',c')=(a+a',b+b',c+c'+ab').
\end{equation*}
Отсюда видно, что группа Вейля является неабелевой.

Нетрудно проверить, что якобиан отображений $g\mapsto g'g$ и $g\mapsto gg'$
равен единице. Поэтому евклидова мера в $\MR^3$, заданная формулой
\begin{equation*}
  dg(a,b,c):=da\,db\,dc,
\end{equation*}
является двусторонне инвариантной мерой Хаара. Следовательно, группа Вейля
унимодулярна.
\qed\end{exa}
\begin{prop}
Для унимодулярной группы справедливо равенство
\begin{equation*}
  \int_\MG\!\om(b)f(b)=\int_\MG\!\om(b)f(b^{-1}).
\end{equation*}
\begin{proof}
См., например, \cite{BarRac77R}.
\end{proof}
\end{prop}
Типичное применение интеграла по компактной группе Ли состоит в следующем.
\begin{defn}
Рассмотрим представление $\rho:~\MG\rightarrow\aut(\MV)$ группы Ли $\MG$ в
группе автоморфизмов вещественного или комплексного векторного пространства
$\MV$, наделенного скалярным произведением. Представление $\rho$ называется
{\em унитарным} (когда $\MV$ комплексное) или {\em ортогональным} (когда $\MV$
вещественное), если для всех $u,v\in\MV$ и всех $a\in\MG$ выполнено равенство
\begin{equation}                                                  \label{eunrep}
  \big(\rho(a)u,\rho(a)v\big)=(u,v). \qed
\end{equation}
\end{defn}
\index{Унитарное представление (unitary representation)}%
\index{Представление унитарное (unitary representation)}%
\index{Ортогональное представление (orthogonal representation)}%
\index{Представление ортогональное (orthogonal representation)}%
\begin{theorem}
Пусть $\MG$ -- компактная группа Ли, и $\MV$ -- комплексное (вещественное)
векторное пространство, в котором задано представление $\rho$ группы $\MG$.
Тогда на $\MV$ существует скалярное произведение, относительно которого
представление Ядро $\rho$ унитарно (ортогонально).
\end{theorem}
\begin{proof}
Рассмотрим произвольное скалярной произведение $\langle\cdot,\cdot\rangle$ на
$\MV$. С его помощью определим новое скалярное произведение следующей формулой
\begin{equation*}
  (u,v):=\int_\MG\!\om(a)\,\langle\rho(a)u,\rho(a)v\rangle,
\end{equation*}
где интегрирование ведется по $a\in\MG$ с двусторонне инвариантной мерой Хаара
$\om$. Используя правую инвариантность интеграла, нетрудно убедиться в
справедливости равенства (\ref{eunrep}).
\end{proof}
\section{Некоторые общие свойства групп Ли}
Существует определенная связь между группами Ли, локальными группами Ли и
алгебрами Ли, которые будут определены в следующем разделе \ref{sleacg}. Сейчас
мы приведем несколько утверждений об общем устройстве групп Ли и о связи между
группами Ли и их алгебрами Ли.
\begin{theorem}                                                   \label{tloglo}
Пусть $\MG$ -- связная группа Ли, и $\MU$ -- некоторая окрестность единицы.
Тогда
\begin{equation}                                                  \label{elocli}
  \MG=\bigcup_{k=1}^\infty\MU^k,
\end{equation}
где множество элементов $\MU^k$ состоит из всех возможных произведений
$k$ элементов из $\MU$.
\end{theorem}
\begin{proof}
См., например, \cite{Warner83R}.
\end{proof}

Основываясь на этой теореме, иногда говорят, что произвольная
окрестность единицы группы Ли (соответствующая локальная группа Ли) порождает
всю связную компоненту единицы. Это не так. Знание локальной группы Ли
недостаточно для построения всей связной компоненты единицы группы. Например,
группы Ли $\MS\MO(3)$ и $\MS\MU(2)$ имеют одинаковые локальные группы Ли, однако
не изоморфны. Дело в том, что в формуле (\ref{elocli}) есть произведения
большого числа элементов локальной группы Ли, которые лежат вне $\MU$. То есть
для построения связной компоненты единицы по формуле (\ref{elocli}) необходимо
знание правила умножения не только элементов локальной группы Ли, но и всех
других элементов группы.

В то же время локальная группа Ли, как мы увидим ниже, однозначно с точностью до
изоморфизма определяет связную и односвязную группу Ли (универсальную
накрывающую).

\begin{theorem}[\bf Э.\ Картан]
Каждая алгебра Ли $\Gg$ является алгеброй Ли некоторой группы Ли $\MG$.
\end{theorem}
\index{Теорема Картана (Cartan theorem)}%
\index{Картана теорема (Cartan theorem)}%
\begin{proof}
См., например, \cite{Pontry84R}.
\end{proof}
Важность понятия локальной группы Ли заключается в следующем утверждении,
которое является следствием теоремы Картана.
\begin{theorem}                                                   \label{tlocis}
Каждая алгебра Ли является алгеброй Ли некоторой локальной группы Ли. Локальные
группы Ли изоморфны тогда и только тогда, когда изоморфны их алгебры Ли.
\end{theorem}
\begin{proof}
Приведено в следующем разделе.
\end{proof}

Другими словами, каждая алгебра Ли определяет локальную группу Ли с точностью
до изоморфизма, но ни в коем случае группу Ли в целом.

В общем случае справедлива следующая
\begin{theorem}                                                   \label{qbhgtd}
Каждой алгебре Ли $\Gg$ соответствует единственная, с точностью до изоморфизма,
связная односвязная группа Ли $\widetilde\MG$ (универсальная накрывающая), для
которой $\Gg$ является алгеброй Ли. Все связные группы Ли, для которых $\Gg$
является алгеброй Ли имеют вид $\widetilde\MG/\MD$, где $\MD$ -- дискретный
нормальный делитель, лежащий в центре группы Ли $\MG$.
\end{theorem}
\begin{proof}
См., например, \cite{Kirill78R}
\end{proof}
Напомним, что центром группы $\MG$ называется множество всех элементов группы,
которые перестановочны со всеми элементами из $\MG$. Слово дискретный в
сформулированной теореме означает, что подмножество $\MD\subset\MG$ состоит из
отдельных точек.

Таким образом, существует взаимно однозначное соответствие между классами
изоморфных алгебр Ли и классами изоморфных связных и односвязных групп Ли.

При описании связи между группами Ли, локальными группами Ли и алгебрами Ли
используется
\begin{theorem}[\bf о монодромии]
Пусть $\MG$ -- связная односвязная группа Ли, и $\MH$ -- произвольная группа Ли.
Тогда всякий локальный гомоморфизм $\MG$ в $\MH$ (т.е.\ гомоморфизм
соответствующих локальных групп), однозначно продолжается до глобального
гомоморфизма $\MG$ в $\MH$.
\end{theorem}
\begin{proof}
См., например, \cite{Kirill78R}.
\end{proof}

В теореме \ref{qbhgtd} мы упомянули универсальную накрывающую группы Ли.
Общая теория накрытий для многообразий рассмотрена далее в главе \ref{scover}.
Ниже мы сформулируем несколько утверждений, касающихся накрытий групп Ли.
\begin{defn}
Непрерывное отображение топологических пространств
$p:~\widetilde\MM\rightarrow\MM$ называется {\em накрытием}, если выполнены
следующие условия:

1) \parbox[t]{.92\linewidth}{$p$ сюрьективно;}

2) \parbox[t]{.92\linewidth}{для любого $x\in\MM$ найдется открытая окрестность
$\MU_x$ точки $x$ такая, что $p^{-1}(\MU_x)=\bigcup_{j\in\CJ}\MU_j$
для некоторого семейства открытых подмножеств $\MU_j\subset\widetilde\MM$,
удовлетворяющих условиям $\MU_j\bigcap\MU_k=\emptyset$ при $j\ne k$ и
сужение отображения $p|_{\MU_j}:~\MU_j\rightarrow\MU_x$ -- гомеоморфизм для
всех $j\in\CJ$.}\newline
Топологическое пространство $\MM$ называется {\em базой} накрытия, а
$\widetilde\MM$ -- {\em накрывающим пространством}. Если топологическое
пространство $\widetilde\MM$ является односвязным, то накрытие называется
{\em универсальным}.
\qed\end{defn}
\index{Накрывающая группа (covering group)}%
\index{Группа накрывающая (covering group)}%
В рассматриваемом случае базой накрытия является группа Ли $\MM=\MG$. Если
накрывающее пространство односвязно, то на нем можно индуцировать групповую
структуру, которая превращает накрывающее пространство
$\widetilde\MM=\widetilde\MG$ в группу Ли, а накрывающее отображение $p$ в
гомоморфизм групп. Доказательство следующих трех теорем приведено в
\cite{Warner83R}.
\begin{theorem}
Каждая связная группа Ли $\MG$ обладает универсальным накрывающим пространством
$\widetilde\MG$, которое само является группой Ли, причем накрывающее
отображение -- гомоморфизм групп Ли.
\end{theorem}
\begin{theorem}
Пусть $\MG$ и $\MH$ -- связные группы Ли, и $f:~\MG\rightarrow\MH$ --
гомоморфизм. Отображение $f$ является накрытием тогда и только тогда,
когда дифференциал отображения $f_*:~\MT_e(\MG)\rightarrow\MT_e(\MH)$
является изоморфизмом касательных пространств к единицам групп.
\end{theorem}
\begin{theorem}                                                   \label{talgrl}
Пусть $\MG$ и $\MH$ -- связные группы Ли с алгебрами Ли $\Gg$ и $\Gh$, и пусть
группа Ли $\MG$ является односвязной. Пусть $\vf:~\Gg\rightarrow\Gh$ --
гомоморфизм алгебр Ли. Тогда существует единственный гомоморфизм групп
Ли $f:~\MG\rightarrow\MH$, такой, что $f_*=\vf$.
\end{theorem}
\begin{cor}
Если алгебры Ли связных и односвязных групп Ли $\MG$ и $\MH$ изоморфны, то
сами группы $\MG$ и $\MH$ изоморфны.
\qed\end{cor}
В сформулированных утверждениях важно, чтобы группы Ли были связны, т.к.\ в
противном случае теоремы не верны.

Пусть $e\in\MG$ -- единица группы Ли. Согласно определению накрытия, множество
прообразов $p^{-1}(e)\in\widetilde\MG$ является конечным или счетным. В случае
гомоморфизма групп $\widetilde\MG\rightarrow\MG$ справедливо равенство
\begin{equation*}
  p(\tilde a\tilde b)=p(\tilde a)p(\tilde b),\qquad\forall\tilde a,\tilde b\in
  \widetilde\MG.
\end{equation*}
Из этого равенства немедленно следует, что элементы прообраза $p^{-1}(e)$
образуют абелеву подгруппу в $\widetilde\MG$. Более того, эта абелева подгруппа
лежит в центре группы Ли $\widetilde\MG$. Введем для нее обозначение
$\MD:
=p^{-1}(e)$. Поскольку $\MD$ -- нормальная подгруппа, то универсальное
накрытие групп Ли всегда можно представить в виде фактор группы
\begin{equation*}
  \MG=\frac{\widetilde\MG}{\MD},
\end{equation*}
где $\widetilde\MG$ -- универсальная накрывающая группы Ли $\MG$, а $\MD$ --
дискретный нормальный делитель, лежащий в центре группы Ли $\MG$.

Забегая вперед, отметим, что из теоремы \ref{tfrdkk} немедленно вытекает
\begin{theorem}
Для любой связной группы Ли $\MG$ ее фундаментальная группа $\pi(\MG)$ изоморфна
группе $\MD$ и, следовательно, абелева и лежит в центре универсальной
накрывающей группы $\widetilde\MG$.
\end{theorem}

Следующая теорема говорит о том как устроены некоторые некомпактные группы Ли с
топологической точки зрения.
\begin{theorem}                                                   \label{tnolig}
Пусть $\MG$ -- связная некомпактная полупростая группа Ли $\MG$ и
$\MK\subset\MG$ -- ее некоторая максимальная компактная подгруппа. Тогда
существует такое подмногообразие $\MU$ в $\MG$, диффеоморфное евклидову
пространству, что отображение
\begin{equation*}
  \MK\times\MU\ni\quad(k,u)\mapsto ku\quad\in\MG
\end{equation*}
является диффеоморфизмом прямого произведения $\MK\times\MU$ на $\MG$.
\end{theorem}
\begin{proof}
См., например, \cite{Helgas01R}.
\end{proof}
Данная теорема означает, что как многообразие любая связная некомпактная
полупростая группа Ли представляет собой прямое произведение
\begin{equation}                                                  \label{esegrd}
  \MG\approx\MK\times\MR^{\dim\MG-\dim\MK}
\end{equation}
некоторого компактного многообразия и евклидова пространства.
\begin{exa}
Рассмотрим произвольную $n\times n$ матрицу $M\in\MS\ML(n,\MR)$ с единичным
определителем. Согласно теореме о полярном разложении матриц \ref{tpomad}, она
единственным образом представляется в виде произведения $M=SA$, где
$S\in\MS\MO(n)$ -- некоторая специальная ортогональная матрица и $A$ --
симметричная положительно определенная матрица с единичным определителем,
$\det A=1$. Группа $\MS\ML(n,\MR)$ связна, некомпактна и проста. Группа
$\MS\MO(n)$ связна, компактна и является подгруппой в $\MS\ML(n,\MR)$. Ее
размерность равна $n(n-1)/2$. Множество симметричных положительно определенных
матриц с единичным определителем диффеоморфно евклидову пространству
$\MR^{\frac{n(n+1)}2-1}$. Таким образом, как многообразие группа $\MS\ML(n,\MR)$
диффеоморфна прямому произведению
\begin{equation*}                                                    \tag*{\qed}
  \MS\ML(n,\MR)\approx\MS\MO(n)\times\MR^{\frac{n(n+1)}2-1}.
\end{equation*}
\end{exa}
\section{Полупрямое произведение групп}
Введем важное понятие полупрямого произведения двух групп, которое обобщает
прямое произведение групп, определенное в разделе \ref{srealn}. Для этого нам
понадобится несколько предварительных определений и утверждений. Напомним, что
сюрьективное гомоморфное отображение группы (в том числе группы Ли) на себя
называется автоморфизмом. При этом единица группы отображается в себя.
\begin{exa}
Рассмотрим группу $\MG$ и произвольный элемент $b\in\MG$. Тогда преобразование
подобия
$$
  a\mapsto bab^{-1},
$$
как легко проверить, задает автоморфизм. Такие автоморфизмы называются
{\em внутренними.}
\index{Внутренний автоморфизм (inner automorphism)}%
\index{Автоморфизм внутренний (inner automorphism)}%
Все другие виды автоморфизмов называются {\em внешними.}
\index{Внешний автоморфизм (outer automorphism)}%
\index{Автоморфизм внешний (outer automorphism)}%
\qed\end{exa}
\begin{theorem}
Множество автоморфизмов группы $\MG$ само образуют группу, которая
называется группой автоморфизмов группы $\MG$ и обозначается $\aut\MG$.
\end{theorem}
\begin{proof}
Проводится простой проверкой групповых аксиом, причем единицей
группы $\aut\MG$ служит тождественное отображение $\id:~\MG\rightarrow\MG$.
\end{proof}
Поскольку внутренние автоморфизмы сами образуют группу, то они образуют
подгруппу группы $\aut\MG$. Если группа $\MG$ допускает гомоморфное отображение
на свою подгруппу $\MH\subset\MG$, то говорят, что она допускает
{\em эндоморфизм.}
\index{Эндоморфизм (endomorphism)}%

Прямое произведение групп (см.\ раздел \ref{srealn}), было определено для
произвольных групп. Полупрямое произведение двух групп $\MG$ и $\MH$ определено
только в том случае, когда $\MG$ является группой автоморфизмов $\MH$.
\begin{defn}
Рассмотрим две группы $\MH$ и $\MG\subset\aut\MH$. Обозначим их элементы через
$a,b,\dotsc$ и $\al,\bt,\dotsc$ соответственно. Тогда {\em полупрямым
произведением} $\MG\ltimes\MH$ групп $\MG$ и $\MH$ называются все возможные
 упорядоченные пары $(\al,a)$ со следующим законом композиции
\begin{equation*}
  (\MG\ltimes\MH)\times(\MG\ltimes\MH)\ni\quad(\al,a)\ltimes(\bt,b)
  :=\big(\al\bt,\al(b)a\big)\quad\in\MG\ltimes\MH,
\end{equation*}
где $\al(b)$ -- образ элемента $b$ при автоморфизме, соответствующем элементу
$\al$.
\qed\end{defn}
\index{Полупрямое произведение групп (semidirect product of groups)}%
\index{Произведение групп полупрямое (semidirect product of groups)}%
Нетрудно проверить групповые аксиомы. Прямые вычисления показывают, что
полупрямое произведение трех элементов ассоциативно:
\begin{equation*}
  (\al,a)\ltimes(\bt,b)\ltimes(\g,c)=\big(\al\bt\g,\al(c)\bt(c)\al(b)a\big).
\end{equation*}
Единица полупрямого произведения определяется единицами в группах $\MG$ и $\MH$:
\begin{equation*}
  \MG\ltimes\MH\ni e=(e_\MG,e_\MH),\qquad e_\MG\in\MG,~e_\MH\in\MH.
\end{equation*}
Обратный элемент имеет вид
\begin{equation*}
  (\al,a)^{-1}=\big(\al^{-1},\al^{-1}(a^{-1})\big).
\end{equation*}

Перечислим некоторые свойства полупрямого произведения.

1). Группы $\MG$ и $\MH$ естественно вложены в $\MG\ltimes\MH$. Это вложение
задается отображениями $\al\mapsto(\al,e_\MH)$ и $a\mapsto(e_\MG,a)$. При этом
группа $\MH$ является нормальной подгруппой в $\MG\ltimes\MH$, и существует
изоморфизм
\begin{equation*}
  \MG\simeq\frac{\MG\ltimes\MH}\MH.
\end{equation*}

2). Каждый элемент полупрямого произведения групп $g\in\MG\ltimes\MH$ взаимно
однозначно представим в виде $g=(\al,a)$, где $\al\in\MG$ и $a\in\MH$. Это
свойство оправдывает название ``полупрямое произведение''.

3). Заданное действие группы $\MG$ в группе $\MH$ с помощью автоморфизмов
$\aut\MH$ совпадает с действием $\MG$ в $\MH$, которое определяется сужением
внутренних автоморфизмов в полупрямом произведении $\MG\ltimes\MH$ на группу
$\MH$.

Можно доказать, что любая группа, обладающая свойствами 1)--3), изоморфна
полупрямому произведению некоторых подгрупп (свойство универсальности
полупрямого произведения).

Введенное выше полупрямое произведение относится к произвольным группам, в том
числе и к группам Ли.

Заметим, что прямое произведение является частным случаем полупрямого
произведения групп. Оно возникает, если гомоморфизм $\MG\rightarrow\aut\MH$
тривиален, т.е.\ каждому элементу группы $\MG$ соответствует тождественное
преобразование группы $\MH$.
\begin{exa}
Группа Ли аффинных преобразований прямой, рассмотренная в разделе \ref{stwnog},
является полупрямым произведением группы дилатаций (группа $\MG$) на группу
сдвигов (группа $\MH$). Действительно, совершим два преобразования с параметрами
$(\bt,b)$ и $(\al,a)$:
\begin{equation*}
  x~\mapsto~\bt x+b~\mapsto~\al(\bt x+b)+a=\al\bt x+\al b+a,
\end{equation*}
где греческими буквами обозначены параметры дилатаций. Тогда закон умножения в
группе запишется в виде
\begin{equation*}
  (\al,a)\ltimes(\bt,b)=(\al\bt,\al b+a),
\end{equation*}
что соответствует полупрямому произведению.
\qed\end{exa}
\begin{exa}
В физике важную роль играет группа Пуанкаре $\MI\MO(1,n-1)$, $n\ge2$,
рассмотренная в разделе \ref{spungr} и равная полупрямому произведению группы
Лоренца на группу сдвигов. При этом каждому элементу группы Лоренца
соответствует вращение (автоморфизм) пространства Минковского, которое
рассматривается как группа сдвигов.

Ограничимся собственной ортохронной группой Лоренца $\MS\MO_\uparrow(1,n-1)$
(связной компонентой единицы). Тогда группа Пуанкаре является связной как
полупрямое произведение связных подгрупп. Она не является односвязной, поскольку
собственная ортохронная группа Лоренца не является односвязной.

Алгебра Ли группы Пуанкаре $\Gl+\Gt$ состоит из преобразований Лоренца $\Gl$ и
сдвигов $\Gt$ со следующими коммутационными соотношениями:
\begin{equation*}
  [\Gl,\Gl]=\Gl,\qquad[\Gl,\Gt]=\Gt,\qquad[\Gt,\Gt]=0.
\end{equation*}
Мы видим, что алгебра Ли группы Пуанкаре имеет две подалгебры $\Gl$ и $\Gt$. При
этом подгруппа сдвигов является инвариантной, и ей соответствует идеал $\Gt$.
Факторалгебра Ли $(\Gl+\Gt)/\Gt$ изоморфна алгебре Лоренца $\Gs\Go(1,n-1)$.
Линейное факторпространство $(\Gl+\Gt)/\Gl$ изоморфно пространству Минковского,
но не является факторалгеброй Ли.
\qed\end{exa}
\section{Алгебры Ли                                                            }
Алгебры Ли уже давно стали важным самостоятельным разделом математики. В
настоящем разделе мы приведем определения и кратко рассмотрим основные свойства
алгебр Ли.
\begin{defn}
Пусть $\Gg$ -- конечномерное векторное (линейное) пространство над полем
вещественных $\MR$ или комплексных $\MC$ чисел. Если в $\Gg$ задана билинейная
операция (коммутатор), которую мы обозначим квадратными скобками,
\begin{equation}                                                  \label{qmupjz}
  \Gg\times\Gg\ni\quad X,Y\mapsto Z:=[X,Y]\quad\in\Gg,
\end{equation}
такая, что выполнены два условия:

1) $[X,Y]=-[Y,X]$  -- антисимметрия,

2) $\big[[X,Y],Z\big]+\big[[Y,Z],X\big]+\big[[Z,X],Y\big]=0$ -- тождества Якоби,
\newline то множество векторов $\Gg$ называется вещественной или комплексной
{\em алгеброй Ли}. {\em Размерностью} алгебры Ли называется ее размерность как
векторного пространства. Алгебра Ли называется {\em абелевой} или
{\em коммутативной}, если $[X,Y]=0$ для всех $X,Y\in\Gg$.
\qed\end{defn}
\index{Алгебра Ли (Lie algebra)}%
\index{Ли алгебра (Lie algebra)}%
\index{Абелева алгебра Ли (Abelian Lie algebra)}%
\index{Алгебра абелева Ли (Abelian Lie algebra)}%

Мы уже достаточно подробно рассмотрели алгебру Ли векторных полей. Ниже мы
рассматриваем алгебры Ли с абстрактной точки зрения.

В приложениях, как правило, используются вещественные алгебры и группы Ли.
Поскольку алгебры Ли являются векторными пространствами, то они допускают
естественную комплексификацию, которая будет описана ниже. Оказывается, что
простые комплексные алгебры Ли допускают классификацию, что приводит к
классификации простых вещественных алгебр Ли. Поэтому в настоящем
разделе мы рассматриваем алгебры Ли над полем как вещественных, так и
комплексных чисел. При этом комплексные алгебры Ли следует рассматривать, в
первую очередь, как средство для изучения вещественных алгебр Ли.

Как многообразие алгебра Ли диффеоморфна евклидову пространству $\MR^\Sn$ и
поэтому некомпактна. С другой стороны, каждая алгебра Ли является алгеброй Ли
некоторой группы Ли. Поэтому введем
\begin{defn}
Алгебра Ли, изоморфная алгебре Ли некоторой компактной группы Ли, называется
{\em компактной}.
\qed\end{defn}
\index{Компактная алгебра Ли (compact Lie algebra)}%
\index{Алгебра Ли компактная (compact Lie algebra)}%
\begin{exa}
Рассмотрим абелеву группу Ли $\MR^\Sn$ по сложению. Эта группа некомпактна.
С другой стороны, $\Sn$-мерный тор $\MT^\Sn=\MU(1)\times\dotsc\times\MU(1)$
также является абелевой группой, но уже компактной. Их алгебры Ли изоморфны и,
согласно определению, компактны.
\qed\end{exa}
Таким образом, компактная алгебра Ли может быть также алгеброй Ли некомпактной
группы. Это связано с наличием абелевых инвариантных подгрупп.
\begin{theorem}
Всякая связная группа Ли с компактной полупростой алгеброй Ли компактна.
\end{theorem}
\begin{proof}
См., например, \cite{Pontry84R}. Weyl
\end{proof}

Пусть $e_\Sa$, $\Sa=1,\dotsc,\Sn$, -- базис в алгебре Ли
$\Gg$, $\dim_\MC\Gg=\Sn$ или $\dim_\MR\Gg=\Sn$, рассматриваемой как векторное
пространство. Число $\Sn$ является (комплексной) размерностью алгебры Ли. Любой
элемент алгебры Ли представим виде $X=X^\Sa e_\Sa\in\Gg$, где $X^\Sa$
вещественные или комплексные компоненты вектора $X$. Коммутатор двух базисных
векторов всегда можно разложить по базису:
\begin{equation}                                                  \label{qmdwrs}
  [e_\Sa,e_\Sb]=f_{\Sa\Sb}{}^\Sc e_\Sc,
\end{equation}
где $f_{\Sa\Sb}{}^\Sc$ -- структурные константы алгебры Ли. Из антисимметрии
коммутатора и тождеств Якоби следует, что структурные константы удовлетворяют
следующим тождествам:
\begin{align}                                                          \nonumber
  f_{\Sa\Sb}{}^\Sc&=-f_{\Sb\Sa}{}^\Sa,
\\                                                                \label{qdmetg}
  f_{\Sa\Sb}{}^\Sd f_{\Sc\Sd}{}^\Se+f_{\Sb\Sc}{}^\Sd f_{\Sa\Sd}{}^\Se+
  f_{\Sc\Sa}{}^\Sd f_{\Sb\Sd}{}^\Se&=0.
\end{align}
Равенство $Z=[X,Y]$ в компонентах принимает вид
\begin{equation*}
  Z^\Sa=[X,Y]^\Sa=X^\Sb Y^\Sc f_{\Sb\Sc}{}^\Sa.
\end{equation*}

Несмотря на свое название структурные константы не являются постоянными. При
изменении базиса в алгебре Ли $\Gg$ они преобразуются как компоненты тензора
третьего ранга с одним контравариантным и двумя ковариантными индексами.

Алгебра Ли $\Gg$ является кольцом (со сложением и умножением) и одновременно
векторным пространством над полем вещественных или комплексных чисел. Нулевой
элемент алгебры Ли $0\in\Gg$ является единичным элементом по отношению к
сложению векторов в $\Gg$. Однако он не является единичным элементом по
отношению к умножению (коммутатору), поскольку $[X,0]=0\ne X$ для всех отличных
от нуля элементов $X\in\Gg$. Следовательно, единица в алгебре Ли $\Gg$, которую
мы рассматриваем в данном случае как кольцо, отсутствует.

В алгебре Ли $\Gg$ умножение задается коммутатором, и поэтому в общем случае она
является некоммутативной алгеброй. Ясно также, что умножение в алгебре Ли,
которое задается коммутатором, неассоциативно. Условие ассоциативности при этом
заменяется на тождества Якоби.

Если $A$ -- произвольная конечномерная ассоциативная алгебра с законом умножения
\begin{equation*}
  A\times A\ni\quad X,Y\mapsto X\cdot Y\quad\in A,
\end{equation*}
то ее можно снабдить также структурой алгебры Ли, если положить
\begin{equation*}
  [X,Y]:=X\cdot Y-Y\cdot X.
\end{equation*}
Обратное утверждение, вообще говоря, неверно. Не каждую алгебру Ли можно
снабдить структурой ассоциативной алгебры. Примером является алгебра Ли
векторных полей.

Пусть $\Gh\subset\Gg$ и $\Gf\subset\Gg$ -- два линейных подпространства в $\Gg$.
Обозначим через $\Gh+\Gf$ и $[\Gh,\Gf]$ множество всех векторов вида $X+Y$ и
$[X,Y]$, где $X\in\Gh$ и $Y\in\Gf$.

Если алгебра Ли $\Gg$ абелева, то $[\Gg,\Gg]=0$.

Каждая алгебра Ли содержит нетривиальную абелеву подалгебру $\Gh\subset\Gg$.
Например, все элементы вида $X=aX_0$, где $a\in\MR$ или $a\in\MC$, которые
пропорциональны некоторому фиксированному элементу $X_0\in\Gg$, образуют абелеву
подалгебру $X\in\Gh$, поскольку $[\Gh,\Gh]=0$.

Пусть $\Gh$ и $\Gf$ -- линейные подпространства в алгебре Ли $\Gg$, тогда
нетрудно проверить следующие соотношения:
\begin{align}                                                     \label{qgaiok}
  [\Gh_1+\Gh_2,\Gf]&\subset[\Gh_1,\Gf]+[\Gh_2,\Gf],
\\                                                                \label{qnabsw}
  [\Gh,\Gf]&\subset[\Gf,\Gh],
\\                                                                \label{qpswdn}
  \big[[\Gh,\Gf],\Gg\big]&\subset\big[[\Gf,\Gg],\Gh\big]
  +\big[[\Gg,\Gh],\Gf\big].
\end{align}
Второе включение является следствием того обстоятельства, что если $X\in\Gh$, то
также $-X\in\Gh$. Последнее включение является следствием тождеств Якоби.
\begin{defn}
Линейное подпространство $\Gh\subset\Gg$ называется {\em подалгеброй Ли}, если
выполнено включение $[\Gh,\Gh]\subset\Gh$. То есть оно само является алгеброй
Ли. Линейное подпространство $\Ga\subset\Gg$ называется {\em идеалом}, если
выполнено включение $[\Gg,\Ga]\subset\Ga$.
\qed\end{defn}
\index{Подалгебра Ли (Lie subalgebra)}%
\index{Ли подалгебра (Lie subalgebra)}%
\index{Идеал (ideal)}%
\begin{com}
Пусть $\Ga\ni X$ является левым идеалом в алгебре Ли $\Gg$, т.е.
$[\Gg,\Ga]\subset\Ga$. Тогда все элементы вида $-X$ также лежат в $\Ga$, т.к.\
элементы идеала должны быть подгруппой по отношению к сложению. Поскольку для
левого идеала $[\Gg,\Ga]=[\Ga,\Gg]\subset\Ga$, то каждый левый идеал совпадает
с правым и наоборот. Поэтому в алгебре Ли все идеалы являются двусторонними.
\qed\end{com}

Каждая алгебра Ли $\Gg$ содержит по крайней мере два идеала, которыми являются
нулевой элемент $0$ и вся алгебра Ли $\Gg$.

Пусть $\Ga\subset\Gg$ и $\Gb\subset\Gg$ -- два идеала в $\Gg$. Тогда из
включения (\ref{qgaiok}) следует, что их сумма $\Ga+\Gb$ также является идеалом
в $\Gg$. Аналогично, из формулы (\ref{qpswdn}) вытекает, что коммутатор двух
идеалов $[\Ga,\Gb]$ также является идеалом.

Пусть $\Gh$ -- подалгебра в некоторой алгебре Ли $\Gg$. Введем в пространстве
$\Gg$ отношение эквивалентности
\begin{equation*}
  X\sim Y~(\mod\Gh),
\end{equation*}
если $X-Y\in\Gh$.
Тогда вся алгебра $\Gg$ разлагается в непересекающиеся классы эквивалентности
$[X]:=X+\Gh$. В общем случае множество классов эквивалентности $[X]$ не
образует алгебры Ли. Действительно, если
\begin{align*}
  X_1\sim Y_1~(\mod\Gh),\quad\text{т.е.}\quad X_1=Y_1+H_1,
\\
  X_2\sim Y_2~(\mod\Gh),\quad\text{т.е.}\quad X_2=Y_2+H_2,
\end{align*}
где $H_{1,2}\in\Gh$, то
\begin{equation}                                                  \label{qkpqsw}
  [X_1,X_2]=[Y_1,Y_2]+[Y_1,H_2]+[H_1,Y_2]+[H_1,H_2].
\end{equation}
Отсюда следует, что отношение эквивалентности для коммутаторов
\begin{equation}                                                  \label{qlgykd}
  [X_1,X_2]\sim[Y_1,Y_2]~(\mod\Gh)
\end{equation}
в общем случае не выполняется. Однако, если подалгебра является идеалом,
$\Gh=\Ga$, то последние три слагаемых содержатся в $\Ga$ и условие
(\ref{qlgykd}) выполнено. Тогда множество классов эквивалентности представляет
собой алгебру Ли. В этом случае множество классов эквивалентности $[X]$
называется {\em факторалгеброй Ли} и обозначается $\Gg/\Ga$.
\index{Факторалгебра Ли (quotient algebra}%
\index{Ли факторалгебра (quotient algebra}%
\subsection{Операции над алгебрами Ли                            \label{shsxwd}}
\begin{defn}
Пусть $\Gg$ и $\Gh$ -- алгебры Ли. Тогда отображение $\vf:~\Gg\rightarrow\Gh$
называется {\em гомоморфизмом алгебр Ли}, если оно линейно и сохраняет скобку
Ли, т.е.\ $\vf([X,Y])=[\vf(X),\vf(Y)]\in\Gh$ для всех $X,Y\in\Gg$. Если, кроме
того,  $\vf$ является взаимно однозначным отображением ``на'' (биекцией), то
$\vf$ называется {\em изоморфизмом алгебр Ли}. Изоморфизм алгебры Ли на себя
называется ее {\em автоморфизмом}. Автоморфизм $\vf$ алгебры Ли $\Gg$ называется
{\em инволютивным}, если $\vf^2=\id$.

Множество векторов
\begin{equation*}
  \Gn:=\lbrace X\in\Gg:\quad \vf(X)=0\rbrace
\end{equation*}
называется {\em ядром} гомоморфизма $\vf$.
\qed\end{defn}
\index{Гомоморфизмом алгебр Ли (homomorphism of Lie algebras)}%
\index{Изоморфизм алгебр Ли (isomorphism of Lie algebras)}%
\index{Автоморфизм алгебры Ли (automorphism of Lie algebras)}%
\index{Ядро гомоморфизма (kernel oh homomorphism}%
\index{Инволютивный автоморфизм (involute automorphism}%
\index{Автоморфизм инволютивный (involute automorphism}%
В определении гомоморфизма алгебр Ли мы не требуем гладкости отображения $\vf$,
т.к.\ алгебры Ли рассматриваются, как векторные пространства, а не многообразия.
С другой стороны, на любом векторном пространстве можно ввести гладкую структуру
многообразия, тогда из линейности отображения $\vf$ будет следовать его
гладкость.
\begin{prop}                                                      \label{pkfsve}
Ядро $\Gn$  всякого гомоморфизма алгебр Ли $\vf:~\Gg\rightarrow\Gh$ является
идеалом в $\Gg$. При этом факторалгебра $\Gg/\Gn$ изоморфна $\vf(\Gg)$.
\end{prop}
\begin{proof}
Пусть $X\in\Gg$ и $Y\in\Gh$. Тогда справедливы равенства
\begin{equation*}
  \vf([X,Y]_\Gg)=[\vf(X),0]_\Gh=0.
\end{equation*}
Следовательно, $[X,Y]\in\Gn$. Изоморфизм алгебр просто проверяется.
\end{proof}
\begin{defn}
Гомоморфизм $\phi:~\Gg\rightarrow\Gh$ называется {\em представлением алгебры Ли}
$\Gg$, если $\Gh=\End(\MV)$ для некоторого векторного пространства $\MV$, или
$\Gh=\Gg\Gl(n,\MC)$, или $\Gh=\Gg\Gl(n,\MR)$. Если гомоморфизм
$\phi:~\Gg\rightarrow \vf(\Gg)\subset\Gh$ является изоморфизмом, то
представление называется {\em точным}.
\qed\end{defn}
\index{Представление алгебры Ли (representation of a Lie algebra)}%
\index{Точное представление алгебры Ли (exact representation of a Lie algebra)}%
\index{Представление алгебры Ли точное (exact representation of a Lie algebra)}%

Как и в случае представления группы каждый элемент представления алгебры
является матрицей. Однако эти матрицы могут быть вырождены и, соответственно,
задавать только эндоморфизм векторного пространства.

\begin{defn}
Отображение $\s:~\Gg\rightarrow\Gg$ комплексной алгебры Ли $\Gg$ на себя,
удовлетворяющее условиям:
\begin{align*}
  \s(aX+bY)&:=\bar a\s(X)+\bar b\s(Y),
\\
  \s([X,Y])&:=[\s(X),\s(Y)],
\end{align*}
где $a,b\in\MC$ и черта обозначает комплексное сопряжение, такое, что
$\s^2=\id$, называется {\em сопряжением} в алгебре Ли.
\qed\end{defn}
\index{Сопряжение в алгебре Ли (conjugation in Lie algebra)}%
Заметим, что отображение $\s$ не является автоморфизмом, т.к.\ оно антилинейно.

\begin{exa}
Пусть ${}^\MC\Gg$ -- комплексификация вещественной алгебры Ли $\Gg$. Тогда
отображение
\begin{equation*}
  {}^\MC\Gg\ni\quad X+iY\mapsto \s(X+iY):=X-iY\quad\in{}^\MC\Gg,
\end{equation*}
где $X,Y\in\Gg$, является сопряжением в алгебре Ли ${}^\MC\Gg$.
\qed\end{exa}
\begin{defn}
{\em Дифференцированием} $D$ в алгебре Ли $\Gg$ называется линейное отображение
алгебры Ли в себя, удовлетворяющее условию (правило Лейбница)
\begin{equation*}
  D[X,Y]=[DX,Y]+[X,DY],
\end{equation*}
для всех $X,Y\in\Gg$.
\qed\end{defn}
\index{Дифференцирование в алгебре Ли (differentiation in Lie algebra)}%
Пусть в алгебре Ли $\Gg$ задано два дифференцирования $D_1$ и $D_2$. Тогда их
линейная комбинация $aD_1+bD_2$, как легко проверить, также будет
дифференцированием в $\Gg$. Кроме этого справедливо равенство
\begin{multline}
  D_1D_2[X,Y]=D_1([D_2X,Y]+[X,D_2Y]=
\\
  =[D_1D_2X,Y]+[D_2X,D_1Y]+[D_1X,D_2Y]+[X,D_1D_2Y].
\end{multline}
Если из этого равенства вычесть такое же равенство с заменой индексов
$1\leftrightarrow2$, то получим соотношение
\begin{equation*}
  [D_1,D_2][X,Y]=\big[[D_1,D_2]X,Y\big]+\big[X,[D_1,D_2]Y\big].
\end{equation*}
То есть коммутатор двух дифференцирований снова является дифференцированием в
алгебре Ли $\Gg$. Таким образом, множество всех дифференцирований алгебры Ли
$\Gg$ само образует алгебру Ли, которую мы обозначим $\Gg_A$.
\begin{prop}
Алгебра дифференцирований $\Gg_A$ является алгеброй Ли группы Ли $\MG_A$ всех
автоморфизмов исходной алгебры Ли $\Gg$.
\end{prop}
\begin{proof}
Пусть $\vf_t:=\exp(At)$, -- однопараметрическая группа автоморфизмов $\Gg$.
Тогда выполнено равенство
\begin{equation}                                                  \label{qkjuug}
  \vf_t([X,Y])=[\vf_t(X),\vf_t(Y)].
\end{equation}
Дифференцирование этого соотношения по $t$ при $t=0$ приводит к равенству
\begin{equation*}
  A[X,Y]=[AX,Y]+[X,AY].
\end{equation*}
Это означает, что генератор $A$ однопараметрической группы диффеоморфизмов
является дифференцированием в алгебре Ли $\Gg$, т.е.\ $A\in\Gg_A$.

Обратно. Пусть $A\in\Gg_A$ -- дифференцирование. Тогда можно доказать, что
соответствующая однопараметрическая группа преобразований удовлетворяет
равенству (\ref{qkjuug}).
\end{proof}

\begin{defn}
Пусть $\Gg$ -- алгебра Ли. Каждому элементу алгебры Ли $X\in\Gg$ поставим в
соответствие отображение по следующему правилу
\begin{equation}                                                  \label{qhujgg}
  \Gg\ni\quad Y\mapsto\ad X(Y):=[X,Y]\quad\in\Gg.
\end{equation}
Следовательно, определено отображение
\begin{equation}                                                  \label{qjjhyd}
  \ad:\quad \Gg\ni X~\mapsto~\ad_X\in\End\Gg
\end{equation}
Это отображение называется {\em присоединенным} представлением алгебры Ли $\Gg$.
\qed\end{defn}
\index{Присоединенное представление алгебры Ли%
(adjoint representation of Lie algebra}%
\index{Представление присоединенное алгебры Ли%
(adjoint representation of Lie algebra}%
Проверим, что данное отображение действительно является представлением. Из
тождеств Якоби следует равенство
\begin{equation*}
  \ad[X,Y](Z)=\ad X\big(\ad Y(Z)\big)-\ad Y\big(\ad X(Z)\big).
\end{equation*}
Это означает, что
\begin{equation*}
  \ad[X,Y]=\ad X\big(\ad Y\big)-\ad Y\big(\ad X\big),
\end{equation*}
и отображение $X\mapsto\ad X$ -- действительно представление.

Если в алгебре Ли $\Gg$ выбран некоторый базис $e_\Sa$, $\Sa=1,\dotsc,\Sn$, то
каждое отображение $\ad X$ будет задаваться некоторой $\Sn\times\Sn$ матрицей.

С другой стороны, тождества Якоби можно переписать в виде равенства
\begin{equation*}
  \ad X([Y,Z])=[\ad X(Y),Z]+[Y,\ad X(Z)].
\end{equation*}
Это означает, что присоединенное представление алгебры Ли является
дифференцированием в $\Gg$. Таким
образом, множество элементов присоединенного представления образует некоторую
подалгебру в алгебре всех дифференцирований $\Gg_A$.
\begin{defn}
Множество элементов присоединенного представления
\begin{equation*}
  \Gg_\ad:=\lbrace \ad X,\quad X\in\Gg\rbrace
\end{equation*}
называется {\em присоединенной} алгеброй Ли.
\qed\end{defn}
\index{Присоединенная алгебра Ли (adjoint Lie algebra)}%
\index{Алгебра Ли присоединенная (adjoint Lie algebra)}%

Отображение
\begin{equation*}
  \psi:\quad \Gg\ni\quad X\mapsto\ad X\quad\in\Gg_\ad
\end{equation*}
представляет собой гомоморфизм алгебр Ли. Ядро этого гомоморфизма является
центром алгебры Ли $\Gg$.

Если $\vf:~\Gg\rightarrow\Gg$ -- произвольный автоморфизм алгебры Ли, то
справедлива следующая цепочка равенств:
\begin{equation*}
  \ad\vf(X)(Y)=[\vf(X),Y]=\vf[X,\vf^{-1}(Y)]
  =\vf\Big(\ad X\big(\vf^{-1}(Y)\big)\Big).
\end{equation*}
Это означает, что
\begin{equation}                                                  \label{qgplsw}
  \ad\vf(X)=\vf\ad X\vf^{-1},
\end{equation}
как и положено присоединенному представлению.

Если у алгебры Ли $\Gg$ нулевой центр (т.е.\ единственный элемент, коммутирующий
со всеми элементами алгебры -- это нуль), то гомоморфизм (\ref{qjjhyd})
инъективен, и присоединенное представление является точным представлением
алгебры Ли $\Gg$ в $\End\Gg$.
\begin{exa}
Рассмотрим трехмерную алгебру Ли $\Gs\Gu(2)$. Выберем в ней базис
\begin{equation*}
  e_i:=\frac i2\s_i\in\Gs\Gu(2),
\end{equation*}
где $\s_i$, $i=1,2,3$, -- матрицы Паули. Этот базис удовлетворяет коммутационным
соотношениям
\begin{equation*}
  [e_i,e_j]=-\ve_{ijk}e^k,
\end{equation*}
где $\ve_{ijk}$ -- полностью антисимметричный тензор третьего ранга и подъем
индексов осуществляется с помощью евклидовой метрики, $e^k:=\dl^{ki}e_i$.
Согласно определению, $\ad e_1(e_1)=0$, $\ad e_1(e_2)=-e_3$ и
$\ad e_1(e_3)=e_2$. Аналогичные соотношения можно выписать для $\ad e_2$ и
$\ad e_3$. Следовательно, присоединенная алгебра $\Gs\Gu_\ad(2)$ трехмерна,
и присоединенное представление базисных векторов имеет вид
\begin{equation*}
  \ad e_1=\begin{pmatrix}0 & 0 & 0 \\ 0 & 0 & 1 \\ 0 &-1 & 0\end{pmatrix},\qquad
  \ad e_2=\begin{pmatrix}0 & 0 &-1 \\ 0 & 0 & 0 \\ 1 & 0 & 0\end{pmatrix},\qquad
  \ad e_3=\begin{pmatrix}0 & 1 & 0 \\-1 & 0 & 0 \\ 0 & 0 & 0\end{pmatrix}.
\end{equation*}
Можно проверить, что эти матрицы удовлетворяют тем же коммутационным
соотношениям
\begin{equation*}
  [\ad e_i,\ad e_j]=-\ve_{ijk}\ad e^k.
\end{equation*}
Заметим, что присоединенное представление алгебры $\Gs\Gu(2)$ совпадает с
фундаментальным представлением алгебры Ли группы вращений $\Gs\Go(3)$, состоящим
из всех антисимметричных $3\times3$ матриц.
\qed\end{exa}
\begin{exa}
Рассмотрим вещественную алгебру Ли $\Gg$ с базисом $e_\Sa$, $\Sa=1,\dotsc,\Sn$.
Базис алгебры Ли удовлетворяет коммутационным соотношениям (\ref{qmdwrs}) с
некоторыми вещественными структурными константами $f_{\Sa\Sb}{}^\Sc$.
Рассмотрим отображение
\begin{equation*}
  \Gg\ni\quad e_\Sa\mapsto E_\Sa\quad\in\Gg\Gl(\Sn,\MR),
\end{equation*}
где матрицы $E_\Sa$ определяются структурными константами:
$E_{\Sa\,\Sb}{}^\Sc:=-f_{\Sa\Sb}{}^\Sc$. При этом индексы $\Sb$, $\Sc$
рассматриваются как матричные. Тогда из тождеств Якоби для структурных
констант (\ref{qdmetg}) следует следующее правило коммутации
\begin{equation*}
  [E_\Sa,E_\Sb]=f_{\Sa\Sb}{}^\Sc E_\Sc.
\end{equation*}
Это означает, что построенное отображение является представлением алгебры Ли:
\begin{equation*}
  \ad:\quad\Gg\ni\quad X=X^\Sa e_\Sa\mapsto\ad X
  =X^\Sa E_\Sa\quad\in\Gg\Gl(\Sn,\MR).
\end{equation*}
Нетрудно проверить, что это действительно присоединенное представление.
\qed\end{exa}

Теперь введем новое понятие для двух алгебр Ли.
\begin{defn}
Пусть $\Gg$ и $\Gh$ -- две алгебры Ли. Рассмотрим их прямую сумму как векторных
пространств $\Gg\oplus\Gh$. Введем в этом пространстве операцию коммутирования с
помощью коммутирований, определенных в $\Gg$ и $\Gh$:
\begin{equation*}
  [X_1\oplus Y_1,X_2\oplus Y_2]:=[X_1,X_2]\oplus[Y_1,Y_2],
\end{equation*}
для всех $X_{1,2}\in\Gg$ и $Y_{1,2}\in\Gh$. Тогда алгебра Ли $\Gg\oplus\Gh$
называется прямой суммой алгебр Ли $\Gg$ и $\Gh$.
\qed\end{defn}
Корректность определения коммутатора в прямой сумме, т.е.\ билинейность,
антисимметрия и тождества Якоби, легко проверяется.
\begin{defn}
Если алгебру Ли $\Gg$ как векторное пространство можно представить в виде прямой
суммы векторных подпространств
\begin{equation*}
  \Gg=\bigoplus_{i=1}^k\Gg_i
\end{equation*}
и, кроме того, справедливы включения
\begin{equation*}
  [\Gg_i,\Gg_i]\subset\Gg_i,\qquad [\Gg_i,\Gg_j]=0,\quad i\ne j,
\end{equation*}
то мы говорим, что алгебра Ли $\Gg$ разлагается в прямую сумму подалгебр
$\Gg_i$.
\qed\end{defn}

Ясно, что все подалгебры Ли $\Gg_i$ являются идеалами в $\Gg$, потому что
\begin{equation*}
  [\Gg,\Gg_i]=[\Gg_i,\Gg_i]\subset\Gg_i.
\end{equation*}
Кроме того, если $\Ga$ -- идеал в одной из подалгебр $\Gg_i$, то он также
является идеалом во всей алгебре Ли $\Gg$.

Рассмотрим более сложную конструкцию.
\begin{defn}
Пусть $\Gg$ и $\Gt$ -- две алгебры Ли, и пусть $D:~\Gg\rightarrow\Gt_A$ --
гомоморфизм алгебры Ли $\Gg$ в алгебру дифференцирований $\Gt_A$, т.е.\
$D(X)\in\Gt_A$ для всех $X\in\Gg$. Возьмем прямую сумму векторных пространств
$\Gg\dotplus\Gt$ и снабдим ее структурой алгебры Ли с помощью следующего
коммутатора
\begin{equation}                                                  \label{qbtygf}
  [X_1\dotplus T_1,X_2\dotplus T_2]:=[X_1,X_2]\dotplus \Big([T_1,T_2]
  +D(X_1)T_2-D(X_2)T_1\Big),
\end{equation}
для всех $X_{1,2}\in\Gg$ и $T_{1,2}\in\Gt$. Тогда $\Gg\dotplus\Gt$ называется
{\em полупрямой суммой} алгебр Ли $\Gg$ и $\Gt$.
\qed\end{defn}
\index{Полупрямая сумма (semidirect sum)}%
\index{Сумма полупрямая (semidirect sum)}%

Если дифференцирование тривиально, т.е.\ $D(X)=0$ для всех $X\in\Gg$, то
полупрямая сумма алгебр Ли сводится к прямой сумме.

Проверим корректность данного определения. Билинейность и антисимметрия
коммутатора очевидны. Необходимо проверить только тождества Якоби. Если все три
вектора имеют вид $X\dotplus 0$ или $0\dotplus T$, где $X\in\Gg$ и $T\in\Gt$,
то тождества Якоби выполняются, т.к.\ они выполнены в каждом слагаемом $\Gg$ и
$\Gt$. Следовательно, ввиду линейности коммутатора, достаточно проверить
тождества Якоби для двух троек векторов:
$\lbrace X\dotplus0,Y\dotplus0,0\dotplus T\rbrace$ и
$\lbrace X\dotplus0,0\dotplus T,0\dotplus P\rbrace$. Двойной коммутатор для
каждой тройки будет иметь вид $0\dotplus *$, где $*$ -- некоторый элемент из
$\Gt$. Поэтому для упрощения записи в следующих формулах мы опустим два первых
символа $0$ и $\dotplus$. Для первой тройки векторов $\lbrace X,Y,T\rbrace$
тождества Якоби принимают вид
\begin{multline*}
  \big[[X,Y],T\big]+\big[[Y,T],X\big]+\big[[T,X],Y\big]=
\\
  =D\big([X,Y])T-D(X)D(Y)T+D(Y)D(X)T=0.
\end{multline*}
Равенство нулю следует из того, что $D(X)$ -- это дифференцирование в $\Gt$.
Аналогично, для второй тройки векторов $\lbrace X,T,P\rbrace$ тождества Якоби
принимают вид
\begin{multline*}
  \big[[X,T],P\big]+\big[[T,P],X\big]+\big[[P,X],T\big]=
\\
  =[D(X)T,P]-D(X)[T,P]+[T,D(X)P]=0.
\end{multline*}
Равенство нулю здесь также является следствием свойств дифференцирования.

Полупрямая сумма двух алгебр $\Gg\dotplus\Gt$ содержит по крайней мере две
подалгебры: $\Gg$ и $\Gt$. Более того, подалгебра $\Gt$ является идеалом в
$\Gg\dotplus\Gt$, что является следствием определения коммутатора
(\ref{qbtygf}).
\subsection{Простые и полупростые алгебры Ли}
Простые алгебры Ли наиболее часто встречаются в математической физике и поэтому
представляют особенный интерес. В настоящее время все они классифицированы.
Ниже мы дадим необходимые определения и приведем некоторые свойства простых и
полупростых алгебр Ли.

Начнем с необходимой конструкции. Напомним, что алгебра Ли $\Gg$ является
идеалом в себе самой и коммутатор двух идеалов также является идеалом. Поэтому
формулы
\begin{equation}                                                  \label{qhfrde}
  \Gg^{(1)}:=\Gg,\quad \Gg^{(2)}:=[\Gg^{(1)},\Gg^{(1)}],\quad \dotsc,\quad
  \Gg^{(k)}:=[\Gg^{(k-1)},\Gg^{(k-1)}],\quad \dotsc
\end{equation}
определяют в алгебре Ли $\Gg$ невозрастающую последовательность идеалов:
\begin{equation*}
  \Gg=\Gg^{(1)}\supset\Gg^{(2)}\supset\dotsc\supset\Gg^{(k)}\supset\dotsc
\end{equation*}
\begin{defn}
Алгебра Ли $\Gg$ называется {\em разрешимой}, если существует такое число
$k\ge1$, что $\Gg^{(k)}=0$.
\qed\end{defn}
\index{Разрешимая алгебра Ли (solvable Lie algebra)}%
\index{Алгебра Ли разрешимая (solvable Lie algebra)}%

\begin{exa}                                                       \label{ewqsag}
Рассмотрим алгебру Ли $\Gi\Go(2)$ неоднородной группы вращений евклидовой
плоскости $\MR^2$. Она состоит из вращений $L=x\pl_y-y\pl_x$ и сдвигов
$P_x=\pl_x$, $P_y=\pl_y$. Нетривиальные коммутационные соотношения имеют вид
\begin{equation*}
  [L,P_x]=-P_y,\qquad [L,P_y]=P_x.
\end{equation*}
Все остальные коммутаторы равны нулю. Поэтому последовательность идеалов
(\ref{qhfrde}) обрывается после второго шага:
\begin{equation*}
  \Gg^{(1)}=\Gi\Go(2),\qquad\Gg^{(2)}=\Gp,\qquad\Gg^{(3)}=0,
\end{equation*}
где $\Gp$ -- двумерная абелева алгебра Ли с образующими $P_x$ и $P_y$
(подалгебра сдвигов). Следовательно, алгебра неоднородных вращений плоскости
$\Gi\Go(2)$ разрешима.
\qed\end{exa}
\begin{exa}
Рассмотрим алгебру Ли трехмерных вращений $\Gs\Go(3)$ с образующими $e_i$,
$i=1,2,3$. Поскольку коммутационные соотношения имеют вид
\begin{equation*}
  [e_i,e_j]=-\ve_{ijk}e^k,
\end{equation*}
где $\ve_{ijk}$ -- полностью антисимметричный тензор третьего ранга, то
последовательность идеалов (\ref{qhfrde}) никогда не оборвется:
\begin{equation*}
  \Gg^{(1)}=\Gg^{(2)}=\Gg^{(3)}=\dotsc=\Gs\Go(3).
\end{equation*}
Следовательно, алгебра трехмерных вращений $\Gs\Go(3)$ неразрешима.
\qed\end{exa}

Теперь введем другую последовательность идеалов, которую пронумеруем нижними
индексами:
\begin{equation}                                                  \label{qhfrds}
  \Gg_{(1)}:=\Gg,\quad \Gg_{(2)}:=[\Gg_{(1)},\Gg],\quad \dotsc,\quad
  \Gg_{(k)}:=[\Gg_{(k-1)},\Gg],\quad \dotsc.
\end{equation}
Она также является невозрастающей:
\begin{equation*}
  \Gg=\Gg_{(1)}\supset\Gg_{(2)}\supset\dotsc\supset\Gg_{(k)}\supset\dotsc
\end{equation*}
\begin{defn}
Алгебра Ли $\Gg$ называется {\em нильпотентной}, если существует такое число
$k\ge1$, что $\Gg_{(k)}=0$.
\qed\end{defn}
\index{Нильпотентная алгебра Ли (nilpotent Lie algebra)}%
\index{Алгебра Ли нильпотентная (nilpotent Lie algebra)}%

Ясно, что любая абелева алгебра Ли является и разрешимой и нильпотентной.

\begin{prop}
$\Gg^{(n)}\subset\Gg_{(n)}$.
\end{prop}
\begin{proof}
Проверяем по индукции. В самом деле $\Gg^{(1)}=\Gg_{(1)}$. Если
$\Gg^{(n)}\subset\Gg_{(n)}$, то
\begin{equation*}                                                    \tag*{\qed}
  \Gg^{(n+1)}=[\Gg^{(n)},\Gg^{(n)}]\subset[\Gg_{(n)},\Gg]\subset\Gg_{(n+1)}.
\end{equation*}
\renewcommand{\qed}{}\end{proof}
\begin{cor}
Всякая нильпотентная алгебра разрешима.
\qed\end{cor}
Обратное утверждение неверно.
\begin{exa}
Рассмотрим двумерную неабелеву алгебру Ли $\Gg$ из раздела \ref{stwnog},
соответствующую аффинным преобразованиям прямой. Ее алгебра Ли имеет две
образующие $L_x$, $L_y$ с коммутационными соотношениями
\begin{equation*}
  [L_x,L_y]=L_y,\qquad[L_x,L_x]=[L_y,L_y]=0.
\end{equation*}
Легко вычислить последовательности идеалов
\begin{equation*}
\begin{split}
  &\Gg^{(1)}=\Gg,\qquad\Gg^{(2)}=\lbrace L_y\rbrace,\qquad\Gg^{(3)}=0,
\\
  &\Gg^{(1)}=\Gg,\qquad\Gg^{(2)}=\Gg^{(3)}=\dotsc=\lbrace L_y\rbrace,
\end{split}
\end{equation*}
где $\lbrace L_y\rbrace$ -- одномерная абелева алгебра Ли с образующей $L_y$.
Таким образом, алгебра Ли $\Gg$ разрешима, но не нильпотентна.
\qed\end{exa}
\begin{exa}
Вычислим последовательность идеалов (\ref{qhfrds}) для алгебры Ли $\Gi\Go(2)$ из
примера \ref{ewqsag}:
\begin{equation*}
  \Gg^{(1)}=\Gi\Go(2),\qquad\Gg^{(2)}=\Gg^{(3)}=\dotsc=\Gp.
\end{equation*}
Поэтому алгебра Ли $\Gi\Go(2)$ разрешима, но не нильпотентна.
\qed\end{exa}
\begin{theorem}[\bf Энгель]
Алгебра Ли нильпотентна тогда и только тогда, когда ее форма Киллинга--Картана
тождественно равна нулю.
\end{theorem}
\begin{proof}
См., например, \cite{Postni82R}.
\end{proof}

Напомним, что для любых двух идеалов $\Ga\subset\Gg$ и $\Gb\subset\Gg$ их сумма
$\Ga+\Gb$ также является идеалом в $\Gg$.
\begin{prop}
Пусть $\Ga$ и $\Gb$ -- два идеала в алгебре Ли $\Gg$. Тогда факторалгебра
$(\Ga+\Gb)/\Gb$ изоморфна фактор алгебре $\Ga/(\Ga\cap\Gb)$ и поэтому разрешима,
если идеал $\Ga$ разрешим.
\end{prop}
\begin{proof}
Пусть $\vf$ -- естественный гомоморфизм алгебры $\Ga+\Gb$ на факторалгебру
$(\Ga+\Gb)/\Gb$. Тогда $\vf(\Ga)=(\Ga+\Gb)/\Gb$. Ядром гомоморфизма $\vf$
алгебры $\Ga$ является пересечение $\Ga\cap\Gb$. Поэтому из предложения
\ref{pkfsve} следует изоморфизм
\begin{equation*}                                                    \tag*{\qed}
  (\Ga+\Gb)/\Gb\simeq \Ga/(\Ga\cap\Gb).
\end{equation*}
\renewcommand{\qed}{}\end{proof}
\begin{cor}
Если идеал $\Gb$ также разрешим, то разрешим и идеал $\Ga+\Gb$.
\qed\end{cor}
Таким образом, сумма $\Ga+\Gb$ двух разрешимых идеалов $\Ga$ и $\Gb$ является
разрешимым идеалом. Поэтому в любой конечномерной алгебре Ли $\Gg$ существует
наибольший разрешимый идеал $\Gr$, который называется {\em радикалом},
содержащий все разрешимые идеалы: им является сумма всех разрешимых идеалов.
\index{Радикал алгебры Ли (radical of a Lie algebra)}%
\begin{defn}
Алгебра Ли $\Gg$ называется полупростой, если ее радикал тривиален $\Gr=0$.
\qed\end{defn}
Другими словами, полупростая алгебра Ли не содержит разрешимых идеалов, кроме
тривиального $\Gr=0$.
\begin{theorem}
Пусть $\Gr$ -- радикал алгебры Ли $\Gg$, тогда фактор алгебра Ли $\Gg/\Gr$
полупроста.
\end{theorem}
\begin{theorem}[\bf Картан]                                       \label{tsemis}
Для того, чтобы алгебра Ли $\Gg$ была полупростой, необходимо
и достаточно, чтобы она не содержала абелевых идеалов, отличных от $0$.
\end{theorem}
\begin{proof}
См., например, \cite{Pontry84R}.
\end{proof}
Конечно, каждая полупростая алгебра Ли является алгеброй Ли некоторой
полупростой группы Ли, и любая полупростая группа Ли имеет полупростую алгебру
Ли.
\begin{defn}
{\em Центром} алгебры Ли $\Gg$ называется ее аннулятор, т.е.\ наибольшее
\index{Центр алгебры Ли (center of a Lie algebra)}%
подпространство $\Gz\subset\Gg$, для которого $[\Gz,\Gg]=0$.
\qed\end{defn}
У каждой алгебры ли $\Gg$ есть нулевой элемент $0\in\Gg$, для которого
$[0,\Gg]=0$. Поэтому $0\in\Gz$. Очевидно, что центр алгебры Ли $\Gg$ является
абелевым идеалом, и алгебра Ли $\Gg$ абелева тогда и только тогда, когда ее
центр совпадает со всей алгеброй $\Gz=\Gg$. Из теоремы \ref{tsemis} следует, что
алгебра Ли является полупростой тогда и только тогда, когда ее центр равен нулю.

Введем понятие простой алгебры и группы Ли.
\begin{defn}
Алгебра Ли $\Gg$ называется {\em простой}, если выполнены следующие два условия:
алгебра $\Gg$ неабелева, и единственные ее идеалы есть $0$ и $\Gg$.

Группа Ли $\MG$ называется {\em простой}, если она неабелева и не имеет
инвариантных подгрупп, отличных от единицы и всей группы.
\qed\end{defn}
\index{Простая алгебра Ли (simple Lie algebra)}%
\index{Алгебра Ли простая (simple Lie algebra)}%
\index{Простая группа Ли (simple Lie group)}%
\index{Группа Ли простая (simple Lie group)}%
Условие неабелевости алгебры Ли $\Gg$ эквивалентно условию $\Gg^{(2)}\ne0$. Это
условие исключает алгебры Ли размерности 1. Алгебры ли размерности 1 без
условия неабелевости являлись бы в этом случае простыми, но не полупростыми.

Всякие простые алгебры и группы Ли являются также полупростыми. Обратное
утверждение неверно, но справедливы следующие теоремы.
\begin{theorem}
Полупростая алгебра Ли $\Gg$ представима одним и только одним способом
в виде прямой суммы конечного числа простых идеалов
$$
  \Gg=\bigoplus_{i=1}^k\Gs_i
$$
для некоторого $k$. При этом каждый идеал алгебры Ли $\Gg$ является прямой
суммой некоторых идеалов $\Gs_i$.
\end{theorem}
\begin{theorem}                                                   \label{tsemsi}
Полупростая группа Ли $\MG$ представима одним и только одним способом в
виде прямого произведения конечного числа простых групп Ли
$$
  \MG=\bigotimes_{i=1}^k\MG_i.
$$
При этом каждая инвариантная подгруппа группы Ли $\MG$ является прямым
произведением некоторых подгрупп $\MG_i$.
\end{theorem}
\subsection{Квадратичные формы}
Поскольку алгебра Ли является векторным пространством, то на ней можно задать
билинейную симметричную квадратичную форму
\begin{equation*}
  \Gg\times\Gg\ni\quad X,Y\mapsto (X,Y)\quad\in\MR,\MC.
\end{equation*}
Если $e_\Sa$, $\Sa=1,\dotsc\Sn$ -- базис алгебры Ли, то в компонентах эта форма
имеет вид
\begin{equation*}
  (X,Y)=X^\Sa Y^\Sb\eta_{\Sa\Sb},
\end{equation*}
где $\eta_{\Sa\Sb}$ -- некоторая матрица.

В разделе \ref{shsxwd} было введено присоединенное представление $Z\mapsto\ad Z$
для любого элемента алгебры Ли $Z\in\Gg$. Это позволяет дать следующее
\begin{defn}
Квадратичная форма $(X,Y)$ называется {\em инвариантной}, если выполнено
равенство
\begin{equation*}
  \big(\ad Z(X),Y\big)+\big(X,\ad Z(Y)\big)=0
\end{equation*}
для всех $X,Y,Z\in\Gg$.
\qed\end{defn}
\index{Инвариантная форма (invariant form)}%
\index{Форма инвариантная (invariant form)}%

Поскольку алгебра Ли изоморфна касательному пространству в произвольной точке
группового многообразия, то задание невырожденной симметричной квадратичной
формы в алгебре Ли однозначно определяет метрику на соответствующей группе Ли.
Эта метрика в левоинвариантном базисе (см.\ раздел \ref{sleacg}) имеет
постоянные компоненты $\eta_{\Sa\Sb}$ и, по-построению, является
левоинвариантной. В общем случае метрика на групповом многообразии в
левоинвариантном базисе имеет компоненты $g_{\Sa\Sb}(a)$ и зависит от точки
$a\in\MG$. То есть каждой точке $a\in\MG$ соответствует своя метрика в алгебре
Ли. Задание инвариантной квадратичной формы в алгебре Ли соответствует заданию
двусторонне инвариантной метрики на групповом многообразии.

Присоединенное представление алгебры Ли позволяет определить следующую
симметричную квадратичную форму.
\begin{defn}
Симметричная билинейная квадратичная форма в алгебре Ли
\begin{equation}                                                  \label{qlojjy}
  (X,Y):=-\tr(\ad X\ad Y)
\end{equation}
называется {\em формой Киллинга--Картана} алгебры Ли $\Gg$.
\qed\end{defn}
\index{Форма Киллинга--Картана (Killing--Cartan form}%
\index{Киллинга--Картана форма (Killing--Cartan form}%
В компонентах форма Киллинга--Картана имеет вид
\begin{equation*}
  \eta_{\Sa\Sb}=-f_{\Sa\Sc}{}^\Sd f_{\Sb\Sd}{}^\Sc.
\end{equation*}
Форма Киллинга--Картана играет фундаментальную роль в теории алгебр Ли и их
представлений.
\begin{com}
Знак минус в определении формы Киллинга--Картана необходим для того, чтобы
метрика на соответствующих полупростых вещественных компактных группах Ли была
положительно, а не отрицательно определена. Например, для группы $\MS\MO(3)$,
структурными константами которой является полностью антисимметричный тензор
третьего ранга, $f_{\Sa\Sb}{}^\Sc\mapsto \ve_{ij}{}^k$, форма Киллинга--Картана
имеет вид
\begin{equation*}
  \eta_{ij}=-\ve_{ik}{}^l\ve_{jl}{}^k=2\dl_{ij},
\end{equation*}
где мы воспользовались формулами Приложения \ref{ecotof}.
\qed\end{com}
В общем случае форма Киллинга--Картана может быть вырождена или невырождена. Это
зависит от алгебры Ли. Для абелевых алгебр Ли форма Киллинга--Картана равна
нулю.
\begin{prop}
Форма Киллинга--Картана (\ref{qlojjy}) инвариантна.
\end{prop}
\begin{proof}
Следствие симметрии следа матриц
\begin{equation*}                                                    \tag*{\qed}
  \tr(\ad X\ad Y\ad Z)=\tr(\ad Z\ad X\ad Y).
\end{equation*}
\renewcommand{\qed}{}\end{proof}

Форма Киллинга--Картана инвариантна относительно произвольных автоморфизмов
алгебры Ли $\vf:~\Gg\rightarrow\Gg$. Действительно, поскольку при автоморфизме
матрица присоединенного представления преобразуется по-правилу (\ref{qgplsw}),
то форма Киллинга--Картана инвариантна:
\begin{equation*}
  \big(\vf(X),\vf(Y)\big)=(X,Y).
\end{equation*}

Форма Киллинга--Картана позволяет сформулировать важный критерий.
\begin{theorem}[\bf Картан]
Алгебра Ли $\Gg$ полупроста тогда и только тогда, когда ее форма
Киллинга--Картана невырождена.
\end{theorem}
\begin{proof}
См., например, \cite{BarRac77R}.
\end{proof}
В заключение приведем критерий компактности алгебр Ли.
\begin{theorem}
Алгебра Ли $\Gg$ компактна тогда и только тогда, когда в $\Gg$ существует
положительно определенная инвариантная квадратичная форма.
\end{theorem}
\begin{proof}
См., например, \cite{BarRac77R}.
\end{proof}
\section{Группа Ли $\MG\ML(n,\MC)$}
В настоящем разделе мы приведем некоторые сведения из теории линейных групп Ли
$\MG\ML(n,\MC)$ преобразований (автоморфизмов) комплексного $n$-мерного
пространства $\MC^n$ и их алгебр Ли $\Gg\Gl(n,\MC)$, которые важны с точки
зрения дифференциальной геометрии и физических приложений. Напомним, что при
общих преобразованиях координат в касательных пространствах к точкам
многообразия на компоненты тензоров действует матрица Якоби (\ref{ejacma}),
которая является элементом группы $\MG\ML(n,\MR)\subset\MG\ML(n,\MC)$. Кроме
того, удобным выбором независимых переменных в аффинной геометрии являются
переменные Картана: репер и линейная или $\MG\ML(n,\MR)$-связность (см.\ раздел
\ref{scorep}).

Сначала мы рассмотрим алгебры Ли, как более простые объекты, а затем перейдем к
изучению соответствующих групп Ли.
\subsection{Алгебра Ли $\Gg\Gl(n,\MC)$}
\index{Алгебра Ли $\Gg\Gl(n,\MC)$}%

Обозначим множество всех квадратных $n\times n$ матриц с комплексными элементами
через $\Gg\Gl(n,\MC)$. Это обозначение обосновано в дальнейшем тем, что
множество комплексных квадратных матриц можно рассматривать, как алгебру Ли
группы Ли $\MG\ML(n,\MC)$. Множество матриц $\Gg\Gl(n,\MC)$ с обычным умножением
матриц на числа и сложением можно рассматривать как векторное пространство
размерности $n^2$ над полем комплексных чисел (комплексная размерность) или как
вещественное векторное пространство размерности $2n^2$ (вещественная
размерность). Чтобы превратить это множество в алгебру необходимо ввести
дополнительную бинарную операцию -- умножение. Если в качестве умножения
рассматривать обычное умножение матриц, то мы получим ассоциативную алгебру
матриц над полем комплексных чисел. Однако это не единственная возможность. В
качестве алгебраической операции мы будем рассматривать коммутатор матриц, т.е.\
любым двум матрицам $X,Y\in\Gg\Gl(n,\MC)$ мы ставим в соответствие их
коммутатор:
\begin{equation*}
  \Gg\Gl(n,\MC)\times\Gg\Gl(n,\MC)\ni\quad X,Y\mapsto [X,Y]:=XY-YX\quad
  \in\Gg\Gl(n,\MC).
\end{equation*}
Эта операция антисимметрична, $[X,Y]=-[Y,X]$, и для нее выполняется тождество
Якоби
\begin{equation*}
  \big[[X,Y],Z\big]+\big[[Y,Z],X\big]+\big[[Z,X],Y\big]=0.
\end{equation*}
Тем самым множество всех матриц, включая вырожденные, становится алгеброй Ли
$\Gg\Gl(n,\MC)$. Эта алгебра неассоциативна, что следует из тождеств Якоби.

В дальнейшем мы, как правило, будем рассматривать множество комплексных
$n\times n$ матриц, как алгебру Ли $\Gg\Gl(n,\MC)$.

Множество матриц можно рассматривать, как множество операторов, действующих
в комплексном векторном пространстве $\MV$. Обозначим элементы
векторного пространства через $x=x^ae_a\in\MV$, где $e_a$, $a=1,\dotsc,n$
-- некоторый фиксированный базис, и $x^a\in\MC$ -- комплексные компоненты
вектора. Комплексная размерность этого векторного пространства равна $n$, а
вещественная -- $2n$. При фиксированном базисе векторное пространство $\MV$
естественным образом отождествляется с комплексным евклидовым пространством
$\MC^n$ и вещественным евклидовым пространством удвоенной размерности
$\MR^{2n}$. Мы предполагаем, что топология $\MV$ индуцируется взаимно
однозначным отображением $\MV\leftrightarrow\MR^{2n}$.

Преобразование элементов векторного пространства $x=x^ae_a\in\MC^n$ мы
записываем в виде
\begin{equation*}
  x^a\mapsto x^b X_b{}^a,\qquad X=(X_b{}^a)\in\Gg\Gl(n,\MC),
\end{equation*}
т.е.\ вектор-строка $(x^a)$ умножается справа на матрицу преобразований
$X=(X_b{}^a)$. Такая запись вызвана принятыми ранее правилами: компоненты
вектора мы нумеруем верхним индексом и придерживаемся правила записи индексов
суммирования ``с десяти до четырех''.

Базис алгебры Ли $\Gg\Gl(n,\MC)$ можно выбрать из квадратных матриц $\Be_a{}^b$,
имеющих один ненулевой элемент, который равен единице и находится на $a$-той
строке и в $b$-том столбце. Очевидно, что любую матрицу можно записать в виде
\begin{equation*}
  X=X_a{}^b\Be_b{}^a\quad\in\Gg\Gl(n,\MC),\qquad X_a{}^b\in\MC.
\end{equation*}
Выбранный базис удовлетворяет соотношениям коммутации
\begin{equation}                                                  \label{ecogln}
  [\Be_a{}^b,\Be_c{}^d]=\dl_c^b\Be_a{}^d-\dl_a^d\Be_c{}^b,
\end{equation}
что проверяется прямой проверкой.

Рассмотренная параметризация естественным образом отождествляет алгебру Ли
$\Gg\Gl(n,\MC)$ с комплексным евклидовым пространством $\MC^{n^2}$ и
вещественным евклидовым пространством $\MR^{2n^2}$. Мы предполагаем, что
топология алгебры Ли $\Gg\Gl(n,\MC)$ индуцирована вложением
$\Gg\Gl(n,\MC)\hookrightarrow\MR^{2n^2}$.

Базис $\Be_a{}^b$ алгебры Ли $\Gg\Gl(n,\MC)$ можно также представить в виде
дифференциальных операторов
\begin{equation*}
  \Be_a{}^b=-x^b\pl_a,
\end{equation*}
действующих на компоненты векторов из $\MV$. Это представление определяет
действие генераторов алгебры Ли $\Gg\Gl(n,\MC)$ на любую дифференцируемую
функцию $f(x)\in\CC^1(\MC^n)$ от декартовых координат точки $\MC^n$.
\begin{defn}
Матрица $X$ называется {\em невырожденной} или {\em регулярной}, если для нее
существует обратная матрица $X^{-1}$, т.е.\ выполнено равенство
$XX^{-1}=X^{-1}X=\one$.
\qed\end{defn}
\index{Невырожденная матрица (nondegenerate matrix)}%
\index{Матрица невырожденная (nondegenerate matrix)}%
\index{Регулярная матрица (regular matrix)}%
\index{Матрица регулярная (regular matrix)}%
Для того, чтобы матрица $X$ была регулярной необходимо и достаточно, чтобы
ее определитель был отличен от нуля, $\det X\ne0$. Если эндоморфизм $X$
векторного пространства $\MC^n$ отображает $\MC^n$ {\em на} себя (сюрьективен),
а не на какое-нибудь подпространство низшей размерности, то соответствующая
матрица $X$ регулярна, и существует обратный эндоморфизм $X^{-1}$. В этом
случае эндоморфизм $X$ является автоморфизмом.

Пусть $X=(X_a{}^b)$ -- квадратная $n\times n$ матрица, тогда для определителя
справедливо разложение, в частности, по первой строке
\begin{equation*}
  \det X=\sum_\s X_1{}^{\s(1}\dotsc X_n{}^{n)}\sgn\s,
\end{equation*}
где сумма берется по всем перестановкам $\s(1,\dotsc,n)$.

Перепишем соотношения коммутации (\ref{ecogln}) в виде
\begin{equation*}
  [\Be_\Sa,\Be_\Sb]=f_{\Sa\Sb}{}^\Sc\Be_\Sc,
\end{equation*}
где пару индексов мы для краткости обозначили одной буквой $\Be_\Sa:=\Be_a{}^b$,
$\Sa=1,\dotsc,n^2$. Тогда
\begin{equation*}
  f_{\Sa\Sb}{}^\Sc=f_a{}^b{}_c{}^d{}_e{}^f=\dl_a^f{}_c^b{}_e^d
  -\dl_a^d{}_c^f{}_e^b.
\end{equation*}
Простые вычисления приводят к следующей форме Киллинга--Картана
\begin{equation*}
  \eta_{\Sa\Sb}:=-f_{\Sa\Sc}{}^\Sd f_{\Sb\Sd}{}^\Sc=\eta_a{}^b{}_c{}^d
  =-2\left(n\dl_{ac}^{db}+\dl_{ac}^{bd}\right).
\end{equation*}
Форма Киллинга--Картана задает инвариантное скалярное произведение в
алгебре Ли $\Gg\Gl(n,\MC)$:
\begin{equation}                                                  \label{escpgl}
  (X,Y)=-2n\tr(XY)+2\tr X\tr Y.
\end{equation}
Инвариантность в данном случае означает независимость результата скалярного
произведения от преобразования подобия:
\begin{equation*}
  (X',Y')=(X,Y),
\end{equation*}
где
\begin{equation*}
  X'=SXS^{-1},\quad Y'=SYS^{-1},\qquad \forall S\in\MG\ML(n,\MC).
\end{equation*}

Скалярное произведение (\ref{escpgl}) вырождено. Действительно, поскольку
$\tr\one=n$, то скалярное произведение всех матриц, кратных единичной матрице
$E=a\one$, $a\in\MC$, равно нулю со всеми матрицами из алгебры Ли
$\Gg\Gl(n,\MC)$,
\begin{equation*}
  (E,X)=a(\one,X)=-2an\tr X+2an\tr X=0,\qquad \forall X\in\Gg\Gl(n,\MC).
\end{equation*}
Отсюда следует, что форма Киллинга--Картана вырождена, $\det\eta_{\Sa\Sb}=0$,
и поэтому алгебра Ли $\Gg\Gl(n,\MC)$ не является полупростой. Это значит, что на
групповом многообразии группы Ли $\MG\ML(n,\MC)$ не существует двусторонне
инвариантной метрики и, следовательно, возникают серьезные проблемы с
построением инвариантов.

Поскольку единичная матрица коммутирует со всеми матрицами, то множество
матриц, кратных единице, $\Ga=\lbrace E\rbrace$ является центром и образует
идеал в алгебре Ли $\Gg\Gl(n,\MC)$. Его вещественная размерность равна двум,
$\dim\Ga=2$.

Алгебра Ли $\Gg\Gl(n,\MC)$ содержит много подалгебр. Рассмотрим некоторые
подалгебры, которые наиболее часто встречаются в приложениях. Нетрудно видеть,
что множество всех квадратных вещественных $n\times n$ матриц также является
алгеброй Ли, которую мы обозначим через $\Gg\Gl(n,\MR)$. Эта алгебра является
подалгеброй в алгебре комплексных матриц: $\Gg\Gl(n,\MR)\subset\Gg\Gl(n,\MC)$.
Базис $\Be_a{}^b$ в $\Gg\Gl(n,\MR)$ можно выбрать таким же, как и в случае
$\Gg\Gl(n,\MC)$, только коэффициенты разложения теперь будут не комплексные, а
вещественные числа. Таким образом, алгебра Ли $\Gg\Gl(n,\MR)$ естественным
образом отождествляется с евклидовым пространством $\MR^{n^2}$. Формы
Киллинга--Картана для $\Gg\Gl(n,\MC)$ и $\Gg\Gl(n,\MR)$ совпадают и,
следовательно, алгебра Ли $\Gg\Gl(n,\MR)$ также не является полупростой.

В разделе \ref{skiiut} будет показано, что вещественная алгебра Ли
$\Gg\Gl(n,\MR)={}^r\Gg\Gl(n,\MC)$ является вещественной формой комплексной
алгебры Ли $\Gg\Gl(n,\MC)$ и, наоборот, комплексная алгебра Ли
$\Gg\Gl(n,\MC)={}^\MC\Gg\Gl(n,\MR)$ является комплексификацией вещественной
алгебры Ли $\Gg\Gl(n,\MR)$.

Максимальные полупростые подалгебры Ли в $\Gg\Gl(n,\MC)$ и $\Gg\Gl(n,\MR)$ --
это алгебры комплексных и вещественных матриц с нулевым следом:
\begin{align}                                                     \label{esplco}
  \Gs\Gl(n,\MC)&=\lbrace X\in\Gg\Gl(n,\MC):~\tr X=0\rbrace,
\\                                                                \label{esprea}
  \Gs\Gl(n,\MR)&=\lbrace X\in\Gg\Gl(n,\MR):~\tr X=0\rbrace.
\end{align}
\index{Алгебра (algebra) $\Gs\Gl(n,\MC)$}%
\index{Алгебра (algebra) $\Gs\Gl(n,\MR)$}%
Они имеют следующие вещественные размерности:
\begin{equation*}
  \dim\Gs\Gl(n,\MC)=2n^2-2,\qquad \dim\Gs\Gl(n,\MR)=n^2-1.
\end{equation*}

Каждую матрицу можно однозначно представить в виде
\begin{equation*}
  X=\widetilde X+\frac1n\tr X\one,\quad \tr\widetilde X=0,
\end{equation*}
выделив из нее след. Будем считать две матрицы $X$ и $Y$ эквивалентными, если
равны их бесследовые части, $\widetilde X=\widetilde Y$. Такие матрицы связаны
соотношением
\begin{equation*}
  X=Y+\frac1n(\tr X-\tr Y)\one.
\end{equation*}
Поэтому
\begin{equation*}
  \Gs\Gl(n,\MC)\simeq \frac{\Gg\Gl(n,\MC)}\Ga,
\end{equation*}
где $\Ga$ -- идеал в алгебре Ли $\Gg\Gl(n,\MC)$, состоящий из матриц, кратных
единичной.

Чтобы доказать полупростоту этих подалгебр, вычислим форму Киллинга--Картана.
Выберем базис в алгебрах Ли $\Gs\Gl(n,\MC)$ и $\Gs\Gl(n,\MR)$
\begin{equation*}
  \tilde\Be_a{}^b:=\Be_a{}^b-\dl_a^b\Be_n{}^n,\quad a,b=1,\dotsc,n,\qquad
  \text{(суммирования по $n$ нет)},
\end{equation*}
который удовлетворяет условию $\tr\tilde\Be_a{}^b=0$. Это соотношение определяет
только $n^2-1$ базисных векторов, поскольку $\Be_n{}^n=0$. Коммутационные
соотношения принимают вид
\begin{equation*}
  [\tilde\Be_a{}^b,\tilde\Be_c{}^d]
  =\dl_c^b\tilde\Be_a{}^d-\dl_a^d\tilde\Be_c{}^b
  -\dl_{ac}^{bn}\tilde\Be_n{}^d+\dl_{an}^{bd}\tilde\Be_c{}^n
  -\dl_{cn}^{db}\tilde\Be_a{}^n+\dl_{ca}^{dn}\tilde\Be_n{}^b.
\end{equation*}
Теперь нетрудно вычислить форму Киллинга--Картана
\begin{equation*}
  \eta_{\Sa\Sb}=\eta_a{}^b{}_c{}^d=-2n\left(\dl_{ac}^{db}+\dl_{ac}^{bd}
  -\dl_{acn}^{bnd}-\dl_{can}^{dnb}\right).
\end{equation*}
Отсюда следует, что, если хотя бы одна из пар индексов имеет вид $(a,b)=(n,n)$
или $(c,d)=(n,n)$, то
\begin{equation*}
  \eta_a{}^b{}_n{}^n=\eta_n{}^n{}_c{}^d=0.
\end{equation*}
Форма Киллинга--Картана задает инвариантное скалярное произведение в алгебрах
Ли:
\begin{equation*}
  (X,Y)=-2n \tr(XY),\qquad X,Y\in\Gs\Gl(n,\MC)\quad \text{или}\quad
  X,Y\in\Gs\Gl(n,\MR).
\end{equation*}
Эта форма Киллинга--Картана невырождена, и, значит, алгебры матриц с
нулевым следом полупросты.

Более того, алгебра Ли $\Gs\Gl(n,\MC)$ проста.

В настоящее время все простые комплексные алгебры Ли классифицированы. Согласно
теореме Адо классификация алгебр Ли сводится к классификации матричных алгебр
Ли. И эта задача решена. Существует четыре классические бесконечные серии:
\begin{equation}                                                  \label{qjhpsw}
  \Ga_n~(n\ge1),\qquad\Gb_n~(n\ge2),\qquad\Gc_n~(n\ge3),\qquad
  \Gd_n~(n\ge 4)
\end{equation}
и пять исключительных алгебр
\begin{equation}                                                  \label{qwgars}
  \Gg_2,\qquad\Gf_4,\qquad\Ge_6,\qquad\Ge_7,\qquad\Ge_8.
\end{equation}
\begin{theorem}
Любая простая комплексная алгебра Ли изоморфна одной из алгебр (\ref{qjhpsw}),
(\ref{qwgars}). Алгебры  (\ref{qjhpsw}) и (\ref{qwgars}) между собой попарно
не изоморфны.
\end{theorem}
\begin{proof}
См., например, \cite{Pontry84R}.
\end{proof}

Нижний индекс в используемых обозначениях для простых алгебр Ли имеет особый
смысл: он равен комплексной размерности максимальной коммутативной подалгебры.

Алгебры Ли комплексных матриц с нулевым следом представляют собой первую из
классических серий (\ref{qjhpsw}),
\begin{equation*}
  \Ga_n:=\Gs\Gl(n+1,\MC),
\end{equation*}
и, следовательно, просты.

Три оставшиеся серии классических комплексных простых алгебр Ли строятся с
помощью квадратичных форм.

Пусть в комплексном векторном пространстве $\MV$ задана симметричная билинейная
форма
\begin{equation}                                                  \label{qoyyuh}
  \MV\times\MV\ni\quad x,y\mapsto (x,y)\quad\in\MV.
\end{equation}
Рассмотрим линейные преобразования $X$ пространства $\MV$, которые сохраняют
заданную квадратичную форму в следующем смысле:
\begin{equation}                                                  \label{qkhiut}
  (X x,y)+(x,Xy)=0,\qquad\forall x,y\in\MV.
\end{equation}
Множество таких преобразований образуют алгебру Ли с коммутатором
$[X,Y]:=XY-YX$. Действительно, если для двух преобразований $X$ и $Y$ выполнено
равенство (\ref{qkhiut}), то оно будет выполнено и для их коммутатора:
\begin{equation*}
  ([X,Y]x,y)=(XYx,y)-(YXx,y)=-(x,[X,Y]y).
\end{equation*}
Алгебры Ли преобразований векторного пространства $\MV$, сохраняющих билинейную
форму, определяются выбором этой формы.

Пусть комплексная размерность векторного пространства равна $\dim_\MC\MV=m$.
Зададим в $\MV$ невырожденную билинейную симметричную положительно определенную
квадратичную форму
\begin{equation*}
  (x,y):=(y,x).
\end{equation*}
Тогда в пространстве $\MV$ существует базис $e_a$, $a=1,\dotsc,m$, такой,
что квадратичная форма имеет вид
\begin{equation*}
  (x,y)=x^a y^b\dl_{ab},
\end{equation*}
где $x=x^ae_a$, $y=y^ae_a$ и $\dl_{ab}:=\diag(+\dotsc+)$ -- обычная
евклидова метрика. Алгебры Ли, сохраняющие эту форму называются
{\em ортогональными} и обозначаются $\Go(m,\MC)$.

Вторая и четвертая из классических серий (\ref{qjhpsw}) определяются следующими
равенствами:
\begin{align*}
  \Gb_n&:=\Go(2n+1,\MC),
\\
  \Gd_n&:=\Go(2n,\MC).
\end{align*}

Пусть в векторном пространстве $\MV$ задана невырожденная антисимметричная
билинейная квадратичная форма, $(x,y)=-(y,x)$. В этом случае размерность
векторного пространства должна быть четной $\dim\MV=2n$ для невырожденности.
Тогда в пространстве $\MV$ существует такой базис $e_a$, где $a=1,\dotsc,2n$,
что квадратичная форма имеет вид
\begin{equation*}
  (x,y)=x^a y^b\varpi_{ab},
\end{equation*}
где
\begin{equation*}
  \varpi=(\varpi_{ab}):=\begin{pmatrix}
    0 & -\one \\ \one & 0 \end{pmatrix}
\end{equation*}
-- каноническая симплектическая форма. Комплексные алгебры Ли, сохраняющие эту
квадратичную форму, называются {\em симплектическими} и обозначаются
$\Gs\Gp(n,\MC)$. Они дают третью классическую серию
\begin{equation*}
  \Gc_n:=\Gs\Gp(n,\MC).
\end{equation*}

Описание комплексных исключительных алгебр Ли (\ref{qwgars}) довольно сложно.
Интересующийся читатель может найти его, например, в \cite{Pontry84R}.

Теперь обсудим более элементарные свойства матриц.
\begin{defn}
Пусть $A\in\Gg\Gl(n,\MC)$ и $\lm\in\MC$. Алгебраическое уравнение
\begin{equation}                                                  \label{eveceq}
  \det(A-\lm\one)=0
\end{equation}
$n$-того порядка относительно $\lm$ называется {\em характеристическим} или
{\em вековым} уравнением. Решения этого уравнения называются {\em собственными
числами} матрицы $A$.
\qed\end{defn}
\index{Характеристическое уравнение (secular equation)}%
\index{Уравнение характеристическое (secular equation)}%
\index{Вековое уравнение (secular equation)}%
\index{Уравнение вековое (secular equation)}%
\index{Собственное число (eigenvalue)}%
\index{Число собственное (eigenvalue)}%
Согласно основной теореме алгебры каждая матрица $A$ имеет в точности $n$
собственных чисел $\lm_1,\lm_2,\dotsc,\lm_n$ с учетом их кратности. В общем
случае собственные числа комплексны даже для вещественной матрицы
$A\in\Gg\Gl(n,\MR)$.

Матрицы $A$ и $A'$ называются {\em подобными} или {\em сопряженными}, если
\index{Подобная матрица (conjugate matrix)}%
\index{Матрица подобная (conjugate matrix)}%
\index{Сопряженная матрица (conjugate matrix)}%
\index{Матрица сопряженная (conjugate matrix)}%
существует невырожденная матрица $B$ такая, что $A'=BAB^{-1}$.
Из уравнения (\ref{eveceq}) следует, что собственные числа подобных матриц
совпадают.

Опишем экспоненциальное отображение матриц.
\begin{defn}
Пусть $A\in\Gg\Gl(n,\MC)$ -- произвольная матрица с ограниченными элементами.
Тогда ряд\begin{equation*}
  \ex^A=\exp A:=\one+A+\frac{A^2}{2!}+\frac{A^3}{3!}+\dotsc
  =\sum_{k=0}^\infty \frac{A^k}{k!}
\end{equation*}
равномерно сходится, если $A$ остается в ограниченной области пространства
$\Gg\Gl(n,\MC)\approx\MR^{2n^2}$, т.е.\ каждый элемент матрицы ограничен. Этот
ряд называется {\em экспоненциалом} матрицы $A$.
\qed\end{defn}
Функция $\exp A$ определена и непрерывна на $\Gg\Gl(n,\MC)$ и отображает
$\Gg\Gl(n,\MC)$ в себя.
\index{Экспоненциал матрицы (matrix exponent)}%

Сформулируем некоторые свойства экспоненциала матрицы \cite{Cheval46R}.
\begin{prop}
Пусть $B$ -- невырожденная $n\times n$ матрица. Тогда
\begin{equation*}
  \ex^{BAB^{-1}}=B\ex^A B^{-1}.
\end{equation*}
\end{prop}
\begin{prop}
Для произвольной матрицы $A\in\Gg\Gl(n,\MC)$ существует такая невырожденная
матрица $B$, что матрица $BAB^{-1}$ является верхнетреугольной.
Тогда матрица $\ex^{BAB^{-1}}$ также является верхнетреугольной. При этом
на диагонали матрицы $BAB^{-1}$ стоят собственные числа $\lm_1,\dotsc,\lm_n$
матрицы $A$, а на диагонали матрицы $\ex^{BAB^{-1}}$ стоят собственные числа
$\ex^{\lm_1},\dotsc,\ex^{\lm_n}$ матрицы $\ex^A$.
\end{prop}
\begin{cor}
Для любой матрицы $A\in\Gg\Gl(n,\MC)$ справедливо равенство
\begin{equation}                                                  \label{elndet}
  \det(\exp A)=\exp(\tr A)\quad \Leftrightarrow\quad \det B=\exp(\tr\ln B),
\end{equation}
где $B:=\ex^A$.
\qed\end{cor}
Отсюда следует, в частности, что $\det(\exp A)\ne0$. То есть экспоненциал
произвольной матрицы является невырожденной (регулярной) матрицей и,
следовательно, принадлежит группе общих преобразований $\exp A\in\MG\ML(n,\MC)$
(см.\ следующий раздел).
\begin{prop}
Если матрицы $A,B$ коммутируют, то
\begin{equation*}
  \ex^{A+B}=\ex^A\ex^B.
\end{equation*}
\end{prop}
\begin{cor}
Экспоненциальное отображение $t\mapsto\ex^{tA}$, где $t\in\MR$ и
$A\in\Gg\Gl(n,\MC)$ -- фиксированная матрица, есть гладкое гомоморфное
отображение аддитивной группы вещественных чисел в группу $\MG\ML(n,\MC)$.
\qed\end{cor}
Касательным вектором к отображению $t\mapsto\ex^{tA}$ в точке $t=0$ является
матрица $A$ (достаточно почленно продифференцировать степенной ряд).
Экспоненциальное отображение $\MR\rightarrow\MG\ML(n,\MC)$ является единственной
однопараметрической подгруппой в $\MG\ML(n,\MC)$ с касательным вектором $A$
в нуле.

Приведем несколько очевидных формул для экспоненциального отображения матриц
\begin{align*}
  \exp A^\St&=(\exp A)^\St,
\\
  \exp A^\dagger&=(\exp A)^\dagger,
\\
  \exp(-A)&=(\exp A)^{-1},
\end{align*}
где индексы $\St$ и $\dagger$ обозначают транспонирование и эрмитово сопряжение
матриц.

Экспоненциальное отображение матриц аналогично экспоненциальному отображению,
которое генерируется векторными полями (см.\ раздел \ref{sexpli}). Оно
задает отображение алгебры Ли $\Gg\Gl(n,\MC)$ в группу Ли $\MG\ML(n,\MC)$.
\begin{prop}
В алгебре Ли $\Gg\Gl(n,\MC)$ существует окрестность $\MU$ нулевой матрицы
$0\in\MU\subset\Gg\Gl(n,\MC)$, которая удовлетворяет следующим условиям:

1) при экспоненциальном отображении $A\mapsto\exp A$, где $A\in\MU$, окрестность
$\MU$ непрерывно отображается на некоторую окрестность единичной матрицы
$\one\in\MG\ML(n,\MC)$;

2) $|\tr A|<2\pi$;

3) при $A\in\MU$ справедливы включения $-A,A^\St,A^\dagger\in\MU$.
\end{prop}

При построении экспоненциального отображение поле комплексных чисел можно
заменить на поле вещественных чисел. При этом конструкция не изменится,
если собственные числа матрицы с вещественными элементами также вещественны.
(В общем случае это, конечно, не так.)

Экспоненциальное отображение задает отображение множества эндоморфизмов
векторного пространства в множество его автоморфизмов
\begin{equation*}
  \exp:\quad \End(\MV)\rightarrow\aut(\MV),
\end{equation*}
где $\MV$ -- произвольное векторное пространство над полем вещественных
или комплексных чисел.

Можно доказать \cite{Warner83R}, что экспоненциальное отображение для группы
$\MG\ML(n,\MC)$ является сюрьективным. В то же время для группы общих линейных
преобразований над полем вещественных чисел $\MG\ML(n,\MR)$ это не так.
\begin{exa}
Матрицу
\begin{equation*}
\begin{pmatrix}-2 & \quad 0\\\quad 0&-1\end{pmatrix}
\end{equation*}
нельзя представить в виде $\ex^A$ ни для какой матрицы $A\in\Gg\Gl(2,\MR)$.
\qed\end{exa}

В алгебре Ли операцией умножения является коммутатор. Посмотрим, что ему
соответствует в группе Ли.
\begin{prop}
В окрестности единицы группы справедливо равенство
\begin{equation*}
  \ex^{tA}\ex^{tB}\ex^{-tA}\ex^{-tB}=[A,B]t^2+\obig(t^3),\qquad t\to 0.
\end{equation*}
\end{prop}
\begin{proof}
Прямая проверка.
\end{proof}
Аналогичное утверждение справедливо не только для матричных, но и произвольных
групп Ли, которое можно также проверить прямыми вычислениями.
\subsection{Группа Ли $\MG\ML(n,\MC)$}
\begin{defn}
Группой {\em общих линейных преобразований} $\MG\ML(n,\MC)$ $n$-мерного
векторного пространства над полем комплексных чисел называется множество всех
невырожденных квадратных $n\times n$ матриц с комплексными элементами и обычным
правилом умножения. Множество всех невырожденных $n\times n$ матриц с
вещественными элементами образует группу общих линейных преобразований
$n$-мерного векторного вещественного пространства $\MG\ML(n,\MR)$.
\qed\end{defn}
\index{Группа общих линейных преобразований %
(group of general linear transformations)}%
\index{Группа Ли (Lie group) $\MG\ML(n,\MC)$}%
\index{Группа Ли (Lie group) $\MG\ML(n,\MR)$}%
Единственное условие, которое накладывается на матрицы -- это отличие от нуля их
определителей, которое необходимо и достаточно для существования обратных
матриц.

Умножение компонент векторов, которые мы записываем в виде строки, на матрицы
сохраняет линейную структуру векторного пространства $\MV$ и поэтому является
автоморфизмом $\MV$, т.к.\ нулевой вектор остается неподвижным.

Множество всех, в том числе вырожденных, комплексных $n\times n$ матриц, как
многообразие, представляет собой евклидово пространство $\MR^{2n^2}$, т.к.\
параметризуется $n^2$ комплексными числами, на которые не наложено никаких
ограничений. Поскольку определитель матрицы является непрерывной функцией от
матрицы, то множество матриц с нулевым определителем представляет собой
замкнутое подмножество в $\MR^{2n^2}$, поскольку $0$ является замкнутым
подмножеством в $\MR$. Поэтому множество невырожденных матриц $\MG\ML(n,\MC)$
как дополнение замкнутого подмножества образует в евклидовом пространстве
$\MR^{2n^2}$ открытое подмногообразие и, следовательно, имеет ту же размерность
$2n^2$, что и евклидово пространство.

Нетрудно проверить, что групповая операция является гладкой, и, следовательно,
группа $\MG\ML(n,\MC)$ является группой Ли. Эта группа связна некомпактна, но
локально компактна.

Аналогично, группа всех невырожденных вещественных $n\times n$ матриц
$\MG\ML(n,\MR)$, как многообразие, представляет собой открытое подмногообразие в
евклидовом пространстве $\MR^{n^2}$ с гладкой групповой операцией. В отличие от
группы комплексных матриц $\MG\ML(n,\MC)$, группа вещественных матриц не
является связной. Она состоит из двух связных компонент: матриц с положительным
и отрицательным определителем. Множество матриц с положительным определителем
$\MG\ML_+(n,\MR)$ само является группой Ли и представляет собой связную
компоненту единицы в $\MG\ML(n,\MR)$. Вторая компонента представляет собой ее
смежный класс и состоит из матриц с отрицательным определителем.

Это различие легко понять. В комплексном случае матрицу $A$ с вещественным
положительным определителем, $\det A>0$, можно непрерывно трансформировать в
матрицу $B$ с вещественным отрицательным определителем, $\det B<0$, вдоль пути,
целиком лежащим в $\MG\ML(n,\MC)$ и состоящим из матриц с комплексным
определителем. Здесь прослеживается связь с вещественной прямой и комплексной
плоскостью. Если из вещественной прямой удалить точку $0$, то она распадается на
два несвязных многообразия. В то же время, удаление начала координат из
комплексной плоскости оставляет ее связной.

Очевидно что группа вещественных матриц $\MG\ML(n,\MR)$ является подгруппой в
$\MG\ML(n,\MC)$. Поскольку размерность $\dim\MG\ML(n,\MR)<\dim\MG\ML(n,\MC)$, то
согласно теоремам \ref{tresub} и \ref{tregvl} эта подгруппа замкнута.

Теперь опишем несколько других подгрупп в $\MG\ML(n,\MC)$.

Произвольную невырожденную матрицу $A=(A_a{}^b)$, $a,b,\dotsc=1,2,\dotsc,n$,
можно представить в виде
\begin{equation*}
  A=a^{1/n}B,\qquad\text{где}~a:=\det A.
\end{equation*}
Тогда
\begin{equation*}
  \det B=1.
\end{equation*}
\begin{defn}
Совокупность всех квадратных $n\times n$ матриц с комплексными элементами,
определитель которых равен единице, является группой, которая называется
группой {\em специальных линейных преобразований} $n$-мерного векторного
пространства над полем комплексных чисел и обозначается $\MS\ML(n,\MC)$. Если
элементы матриц вещественны, то соответствующая группа обозначается
$\MS\ML(n,\MR)$. Она является подгруппой в  $\MS\ML(n,\MC)$.
\qed\end{defn}
\index{Группа специальных линейных преобразований %
(group of special linear transformations)}%
\index{Группа (group) $\MS\ML(n,\MC)$}%
\index{Группа (group) $\MS\ML(n,\MR)$}%
Очевидно, что $\MS\ML(n,\MC)\subset\MG\ML(n,\MC)$. Это максимальная полупростая
подгруппа группы $\MG\ML(n,\MC)$, и ее размерность равна $2n^2-1$. Можно
доказать, что она является простой. Соответствующая алгебра Ли $\Gs\Gl(n,\MC)$
изоморфна алгебре комплексных матриц с нулевым следом и была рассмотрена в
предыдущем разделе.

Группа $\MS\ML(n,\MR)$ является подгруппой группы общих линейных преобразований
$\MG\ML(n,\MR)$ и является связной некомпактной, но локально компактной группой
Ли размерности $n^2-1$. Эта группа является максимальной полупростой подгруппой
группы $\MG\ML(n,\MR)$ (и простой). Алгебра Ли $\Gs\Gl(n,\MR)$ изоморфна алгебре
вещественных матриц с нулевым следом.

В некоторой окрестности единичной матрицы $\one\in\MS\ML(n,\MK)$, где $\MK=\MR$
или $\MK=\MC$, элемент группы $S\in\MS\ML(n,\MK)$ взаимно однозначно
представляется экспоненциальным отображением элемента алгебры
$A\in\Gs\Gl(n,\MK)$:
\begin{equation*}
  S=\ex^A,\qquad \tr A=0.
\end{equation*}

У группы общих линейных преобразований $\MG\ML(n,\MC)$ есть много других
подгрупп.

\begin{exa}
\index{Группа (group) $\MO(n)$}%
Группа вещественных ортогональных $n\times n$ матриц $\MO(n)$ состоит из матриц
$S\in\MO(n)\subset\MG\ML(n,\MR)$, удовлетворяющих условию ортогональности
$$
  SS^\St=S^\St S=1,
$$
где $S^\St$ обозначает транспонированную матрицу. След от этого равенства
приводит к ограничению на матричные элементы $S_i{}^jS^i{}_j=n$. Это значит,
что групповое многообразие является замкнутым подмножеством сферы
$\MS^{n^2-1}\subset\MR^{n^2}$. Отсюда, по теореме Гейне--Бореля--Лебега, следует
компактность группы вращений. Группа $\MO(n)$ является компактной группой
Ли размерности $n(n-1)/2$ и состоит из двух компонент связности. Матрицы с
единичным определителем образуют связную компоненту единицы $\MS\MO(n)$
(собственные вращения). Вторую компоненту связности образует ее смежный класс,
состоящий из несобственных ортогональных матриц с определителем, равным $-1$.

Группы вращений $\MS\MO(n)$ при $n\ge2$ являются группами симметрий сфер
$\MS^{n-1}$, вложенных в евклидово пространство $\MR^n$. Можно доказать, что
фундаментальные группы групп вращений $\MS\MO(n)$ при всех $n\ge3$ равны
$\MZ_2$. Следовательно, они не являются односвязными.

Алгебры Ли групп $\MO(n)$ и $\MS\MO(n)$ совпадают и изоморфны алгебре
антисимметричных $n\times n$ матриц $\Gs\Go(n)$. Вблизи единицы
$\one\in\MS\MO(n)$ матрица вращений $S\in\MS\MO(n)$ взаимно однозначно
представляется в виде экспоненциала отображения некоторого элемента алгебры
$A\in\Gs\Go(n)$:
\begin{equation*}                                                    \tag*{\qed}
  S=\ex^A,\qquad\text{где}~ A=-A^\St.
\end{equation*}
\renewcommand{\qed}{}\end{exa}
\begin{exa}
Группа двумерных вращений $\MS\MO(2)$ как многообразие представляет собой
окружность $\MS^1$. Ее фундаментальная группа изоморфна группе целых
чисел $\MZ$. Полная группа вращений $\MO(2)$ состоит из двух компонент
связности. Как многообразие она представляет
собой два экземпляра окружности $\MO(2)\approx\MS^1\times\MZ_2$. Фундаментальная
группа окружности совпадает с группой целых чисел по сложению $\MZ$.
\qed\end{exa}
\begin{exa}
Группа трехмерных вращений $\MS\MO(3)$ является группой симметрии сферы
$\MS^2$, вложенной в трехмерное евклидово пространство $\MR^3$. Каждое вращение
можно параметризовать вектором $\vec n$, направленным вдоль оси вращения,
длина которого равна углу поворота против часовой стрелки $|\vec n|\le\pi$.
Множество векторов поворота заполняет шар $\MB_{\pi}^3\subset\MR^3$. При этом
противоположные точки граничной сферы $\MS^2_\pi$ радиуса $\pi$ необходимо
отождествить, т.к.\ поворот на угол $\pi$ совпадает с поворотом на угол $-\pi$.
Таким образом, как многообразие группа $\MS\MO(3)$ представляет собой трехмерный
шар с отождествленными противоположными точками граничной сферы $\MS^2_\pi$.
Фундаментальная группа группы $\MS\MO(3)$ является группа $\MZ_2$. Полная группа
трехмерных вращений $\MO(3)$ как многообразие состоит из двух одинаковых
компонент связности. Фундаментальная группа каждой компоненты связности равна
$\MZ_2$.
\qed\end{exa}

\begin{exa}
Пусть в декартовых координатах евклидова пространства $\MR^n$ задана
метрика, представляющая собой диагональную матрицу, у которой на
диагонали стоят $p$ положительных и $q$ отрицательных единиц
\begin{equation}                                                  \label{epseme}
  \eta_{ab}=\diag(\underbrace{+\dotsc+}_p\underbrace{-\dotsc-}_q),
  \qquad p+q=n.
\end{equation}
Для определенности мы считаем, что все положительные единицы идут вначале.
Сигнатура этой метрики равна $p-q$. Матрицы преобразований евклидова
пространства $S_a{}^b$,  сохраняющие метрику (\ref{epseme}),
\begin{equation}                                                  \label{epsrog}
  \eta_{ab}=S_a{}^cS_b{}^d\eta_{cd},
\end{equation}
образуют группу, которая называется группой вращений $\MO(p,q)$.
\index{Группа (group) $\MO(p,q)$}%
В частности, при $q=0$ группа $\MO(n,0)=\MO(n)$, а при $p=1$ она становится
группой Лоренца $\MO(1,n-1)$. Из определения (\ref{epsrog}) следует, что
$\det S=\pm1$. То есть группа вращений $\MO(p,q)$ состоит по крайней мере из
двух компонент связности: матриц с положительным и отрицательным определителем.
Можно показать, что группа Лоренца, состоит из четырех компонент связности.
При $q\ne0,n$ группа вращений $\MO(p,q)$ является некомпактной.

Группа вращений $\MO(p,q)$ является группой симметрии пространств
постоянной кривизны: гиперболоидов $\MH$ размерности $\dim\MH=n-1$,
вложенных в псевдоевклидово пространство $\MR^{p,q}$ уравнением
\begin{equation*}
  (x^1)^2+\dotsc+(x^p)^2-(x^{p+1})^2-\dotsc-(x^n)^2=\pm r^2,\qquad r=\const.
\end{equation*}
Это утверждение является следствием определения (\ref{epsrog}).
В зависимости от знака правой части мы получаем различные гиперболоиды,
которые в общем случае состоят из нескольких компонент связности.
\qed\end{exa}
\begin{exa}
Рассмотрим комплексное векторное пространство $\MC^n$ с базисом $e_a$,
$a=1,\dotsc,n$. Пусть $X=X^ae_a$ и $Y=Y^ae_a$ -- два произвольных вектора из
$\MC^n$. Определим {\em эрмитово} скалярное произведение в $\MC^n$ следующим
соотношением
\index{Эрмитово скалярное произведение (Hermitian scalar product)}%
\index{Скалярное произведение эрмитово (Hermitian scalar product)}%
\begin{equation}                                                  \label{ermsca}
  (X,Y):=X^{a\dagger} Y^b\dl_{ab},
\end{equation}
где символ $\dagger$ обозначает комплексное сопряжение. Отсюда следует, что
$(X,X)\ge0$ и $(X,X)=0$ тогда и только тогда, когда $X=0$. Вещественное число
\begin{equation*}
  \|X\|=\sqrt{(X,X)}
\end{equation*}
называется {\em длиной} вектора $X\in\MC^n$.
\index{Длина вектора (length of a vector)}%

Группа унитарных матриц $\MU(n)$ состоит из комплексных $n\times n$
матриц $U\in\MU(n)$, которые сохраняют эрмитово скалярное произведение
(\ref{ermsca}). В матричных обозначениях это условие записывается в виде
\index{Группа $\MU(n)$}%
{\em условия унитарности}
\index{Унитарность (unitarity)}
\index{Условие унитарности (unitarity condition)}%
$$
  UU^\dagger=U^\dagger U=1,
$$
где символ $\dagger$ обозначает эрмитово сопряжение (транспонирование и
комплексное сопряжение всех элементов). Из условия унитарности следует,
что определитель унитарной матрицы равен по модулю единице. Группа $\MU(n)$
является связной группой Ли размерности $\dim\MU(n)=n^2$. Так же как и
группа вещественных ортогональных матриц эта группа компактна. Группа
$\MU(n)$ не является односвязной, и ее фундаментальная группа изоморфна
группе целых чисел $\MZ$ \cite{Postni82R}. Как многообразие группа $\MU(n)$
диффеоморфна прямому произведению окружности на группу специальных унитарных
матриц с единичным определителем $\MU(n)\approx\MS^1\times\MS\MU(n)$, $n\ge2$.

При $n=1$ имеем абелеву группу $\MU(1)$, которая состоит из комплексных чисел,
равных по модулю единице, с обычным правилом умножения комплексных чисел. Как
многообразие эта группа представляет собой окружность, $\MU(1)\approx\MS^1$.
Легко построить изоморфизм групп, $\MU(1)\simeq\MS\MO(2)$:
\begin{equation*}
  \MU(1)\ni\quad\ex^{i\al}\leftrightarrow \begin{pmatrix}
    \quad \cos\al & \sin\al \\-\sin\al & \cos\al \end{pmatrix}\quad\in\MS\MO(2),
    \qquad \al\in[0,2\pi).
\end{equation*}

Алгебра Ли $\Gu(n)$ изоморфна алгебре антиэрмитовых $n\times n$ матриц. Вблизи
единицы $\one\in\MU(n)$ произвольная унитарная матрица $U\in\MU(n)$ взаимно
однозначно представляется в виде экспоненциала от некоторой антиэрмитовой
матрицы $A\in\Gu(n)$
\begin{equation*}                                                    \tag*{\qed}
  U=\ex^A,\qquad A=-A^\dagger.
\end{equation*}
\renewcommand{\qed}{}\end{exa}
\begin{exa}                                                       \label{esungr}
Группа специальных унитарных матриц $\MS\MU(n)\subset\MU(n)$, состоит из
унитарных $n\times n$ матриц $U\in\MS\MU(n)$ с единичным определителем
\index{Группа (group) $\MS\MU(n)$}%
$$
  UU^\dagger=U^\dagger U=1,\qquad \det U=1.
$$
Эта группа является связной компактной группой Ли размерности
$\dim\MS\MU(n)=n^2-1$.
Она образует замкнутое подмножество в $\MU(n)$. Группа $\MS\MU(n)$ является
также односвязной. Доказательство этого утверждения сложно, и интересующийся
читатель может найти его в \cite{Postni82R}.

Группа $\MS\MU(1)$ тривиальна и состоит из единственного элемента -- единицы.

Алгебра Ли $\Gs\Gu(n)$ изоморфна алгебре антиэрмитовых $n\times n$ матриц
с нулевым следом. Вблизи единицы $\one\in\MS\MU(n)$ каждая унитарная матрица
с единичным определителем $U\in\MS\MU(n)$ взаимно однозначно представляется
экспоненциальным отображением элемента алгебры $A\in\Gs\Gu(n)$:
\begin{equation*}                                                    \tag*{\qed}
  U=\exp A,\qquad \text{где}~A=-A^\dagger,\quad \tr A=0.
\end{equation*}
\renewcommand{\qed}{}\end{exa}

Множество эрмитовых матриц играет важную роль и в математике и в физике.
Опишем их свойства подробнее. Напомним
\begin{defn}
Матрица $A\in\Gg\Gl(n,\MC)$ называется {\em эрмитовой}, если
выполнено условие
\begin{equation}                                                  \label{ermatr}
  A=A^\dagger. \qed
\end{equation}
\end{defn}
\index{Эрмитова матрица (Hermitian matrix)}%
\index{Матрица эрмитова (Hermitian matrix)}%

В вещественном случае множество эрмитовых матриц соответствует симметричным
матрицам, а множество антиэрмитовых -- антисимметричным.

Множество эрмитовых матриц $A,B,\dotsc$ в отличие от множества антиэрмитовых
матриц не образует подалгебры Ли в $\Gg\Gl(n,\MC)$, т.к.\
\begin{equation*}
  [A,B]^\dagger=[B,A]=-[A,B].
\end{equation*}
Заметим также, что отображение $A\rightarrow A^\dagger$ не является
автоморфизмом группы $\MG\ML(n,\MC)$, т.к.\ при эрмитовом сопряжении меняется
порядок матриц, и что эрмитовы матрицы не образуют в $\MG\ML(n,\MC)$ подгруппы.

Сформулируем некоторые свойства эрмитовых матриц \cite{Cheval46R}.
\begin{prop}
Если $A$ -- эрмитова матрица, то $UAU^{-1}$ -- также эрмитова для любой
унитарной матрицы $U\in\MU(n)$. Далее, существует такая унитарная матрица
$U_0$ (не единственная), что $U_0AU_0^{-1}$ является диагональной матрицей.
Если матрица $A$ вещественна, то матрицу $U_0$ можно выбрать ортогональной.
\end{prop}
\begin{prop}
Все собственные числа эрмитовой матрицы вещественны.
\end{prop}
\begin{prop}
Если $A$ -- эрмитова матрица, то в пространстве $\MC^n$ существует
ортонормальный базис, составленный из ее собственных векторов.
\end{prop}

Эрмитова матрица называется {\em положительно определенной}, если все ее
собственные числа положительны.
\index{Положительно определенная эрмитова матрица %
(positive definite Hermitian matrix)}%
\index{Эрмитова матрица положительно определенная %
(positive definite Hermitian matrix)}%
\begin{prop}[\bf Полярное разложение матриц]
Любая невырожденная матрица $A\in\MG\ML(n,\MC)$ может быть записана, и притом
лишь единственным способом, в виде произведения $A=UP$ или $A=PU$ унитарной
матрицы $U\in\MU(n)$ и положительно определенной матрицы $P$. Сомножители
$U$ и $P$ являются непрерывными функциями $A$.
\end{prop}

Поскольку множество всех положительно определенных матриц гомеоморфно
$\MR^{n^2}$, то группа $\MG\ML(n,\MC)$, как многообразие, гомеоморфна прямому
произведению $\MG\ML(n,\MC)\approx\MU(n)\times\MR^{n^2}$.
\section{Универсальная накрывающая $\widetilde{\MS\ML}(2,\MR)$   \label{esltre}}
Построим универсальную накрывающую группу $\widetilde{\MS\ML}(2,\MR)$ для
группы $\MS\ML(2,\MR)$. Это построение будет конструктивным, т.е.\ мы определим
групповую операцию на связном и односвязном многообразии
$\widetilde{\MS\ML}(2,\MR)$, а затем построим гомоморфизм групп
$\widetilde{\MS\ML}(2,\MR)\rightarrow\MS\ML(2,\MR)$ (см., например,
\cite{Postni82R}).

Рассмотрим трехмерное многообразие $\widetilde{\MS\ML}(2,\MR)=\MR\times\MD$,
которое представляет собой прямое произведение вещественной прямой $\MR$ на
единичный диск $\MD$ плоскости комплексного переменного, состоящего из таких
комплексных чисел $z$, что $|z|<1$. Поскольку каждый из сомножителей является
связным и односвязным многообразием, то и их произведение также связно и
односвязно.

Функция
\begin{equation*}
  w=\frac{1+z}{1+\bar z}
\end{equation*}
осуществляет непрерывное отображение диска $\MD$ на окружность $|w|=1$ с
выколотой точкой $w=-1$. Поэтому для любого числа $z\in\MD$ существует
единственное число $t$, $-\pi<t<\pi$, для которого
\begin{equation*}
  \ex^{it}=\frac{1+z}{1+\bar z}.
\end{equation*}
Обозначим это число символом
\begin{equation*}
  t=\frac1i\ln\frac{1+z}{1+\bar z}.
\end{equation*}

Определим на многообразии $\widetilde{\MS\ML}(2,\MR)$ умножение следующей
формулой
\begin{equation}                                                  \label{emulsl}
  (x,u)(y,v)=\left(x+y+t,\frac{u+\ex^{2iy}v}{\ex^{2iy}+u\bar v}\right),\qquad
  x,y\in\MR,\quad u,v\in\MD,
\end{equation}
где
\begin{equation*}
  t=\frac1{2i}\ln\frac{1+\ex^{-2iy}u\bar v}{1+\ex^{2iy}\bar u v}.
\end{equation*}
Так как
\begin{equation*}
  |\ex^{2iy}+u\bar v|^2-|u+\ex^{2iy}v|^2
  =\left(1-|u|^2\right)\left(1-|v|^2\right)>0,
\end{equation*}
то
\begin{equation*}
  \left|\frac{u+\ex^{2iy}v}{\ex^{2iy}+u\bar v}\right|<1.
\end{equation*}
Поэтому формула (\ref{emulsl}) корректно определяет в
$\widetilde{\MS\ML}(2,\MR)=\MR\times\MD$ умножение.

Это умножение обладает единицей $(0,0)$, и для любого элемента $(x,u)$
существует обратный элемент
\begin{equation*}
  (x,u)^{-1}=(-x,-\ex^{2ix}u).
\end{equation*}
Кроме того, прямые вычисления показывают, что умножение (\ref{emulsl})
ассоциативно. Это умножение гладко, и тем самым доказано, что многообразие
$\widetilde{\MS\ML}(2,\MR)=\MR\times\MD$ является группой Ли относительно
умножения (\ref{emulsl}).

Центр этой группы состоит из таких элементов $(x,u)$, для которых, в частности,
равенство
\begin{equation}                                                  \label{equuvy}
  \frac{u+\ex^{2iy}v}{\ex^{2iy}+u\bar v}=\frac{v+\ex^{2ix}u}{\ex^{2ix}+v\bar u}
\end{equation}
выполнено для всех $y\in\MR$ и $v\in\MD$. В частном случае при $v=0$ должно
выполняться равенство $u=\ex^{2iy}u$ для всех $y$. Это возможно только при
$u=0$. Тогда из (\ref{equuvy}) следует равенство $v=\ex^{2ix}v$, которое должно
выполняться для всех $v$, что возможно только при $x=\pi n$, $n\in\MZ$. Нетрудно
проверить, что все элементы вида $(\pi n,0)$ принадлежат центру группы
$\widetilde{\MS\ML}(2,\MR)$. Следовательно, центр группы
$\widetilde{\MS\ML}(2,\MR)$ бесконечен.

Теперь построим гомоморфизм $\widetilde{\MS\ML}(2,\MR)\rightarrow\MS\ML(2,\MR)$.
Легко проверить, что матрица
\begin{equation}                                                  \label{ematsl}
  A=\frac1{\sqrt{1-|u|^2}}\begin{pmatrix}
    \cos x+|u|\cos(x+\vf) & |u|\sin(x+\vf)-\sin x \\[2mm]
    |u|\sin(x+\vf)+\sin x &\cos x-|u|\cos(x+\vf)
    \end{pmatrix},
\end{equation}
где $\vf=\arg u$, имеет единичный определитель. Прямые вычисления позволяют
проверить, что отображение $\widetilde{\MS\ML}(2,\MR)\rightarrow\MS\ML(2,\MR)$
определено корректно и является гомоморфизмом групп. Его ядро состоит из
элементов вида $(2\pi n,0)$ и, значит, дискретно. Поскольку размерности групп
совпадают, то гомоморфизм $\widetilde{\MS\ML}(2,\MR)\rightarrow\MS\ML(2,\MR)$
является групповым накрытием. Таким образом мы построили универсальную
накрывающую для группы $\MS\ML(2,\MR)$:
\begin{equation*}
   \MS\ML(2,\MR)=\frac{\widetilde{\MS\ML}(2,\MR)}{\MZ},
\end{equation*}
где группа преобразований
\begin{equation*}
  \MZ:\qquad x\rightarrow x+2\pi n
\end{equation*}
действует на прямой $\MR$. С топологической точки зрения группа $\MS\ML(2,\MR)$
является произведением окружности на диск
\begin{equation*}
   \MS\ML(2,\MR)\approx\MS^1\times\MD.
\end{equation*}

Группа специальных матриц $\MS\ML(2,\MR)$ дважды накрывает собственную
ортохронную группу Лоренца $\MS\MO_\uparrow(1,2)$ (\ref{eistrl}). При этом
одному лоренцеву вращению соответствует пара матриц (\ref{ematsl}), отличающихся
знаком. Изменению знака матрицы $A$ соответствует сдвиг координаты $x$ на $\pi$.
Поэтому
\begin{equation*}
  \MS\MO_\uparrow(1,2)=\frac{\MS\ML(2,\MR)}{\MZ_2}
  =\frac{\widetilde{\MS\ML}(2,\MR)}{\MZ},
\end{equation*}
где группа преобразований действует по следующему правилу:
\begin{equation*}
  \MZ:\qquad x\rightarrow x+\pi n.
\end{equation*}
Группа $\MZ_2$ действует на окружности $\MS^1$ путем отождествления
противоположных точек и превращает окружность в проективное пространство
$\MR\MP^1$. Поскольку одномерное проективное пространство с топологической
точки зрения является окружностью, то
\begin{equation*}
  \MS\MO_\uparrow(1,2)\approx\MS^1\times\MD.
\end{equation*}
Поскольку диск диффеоморфен всей плоскости $\MD\approx\MR^2$, то
\begin{equation*}
  \MS\MO_\uparrow(1,2)\approx\MS^1\times\MR^2.
\end{equation*}

Предложенная конструкция универсальной накрывающей имеет следующее объяснение.
Согласно теореме о полярном разложении матриц любая унимодулярная матрица
$A\in\MS\ML(2,\MR)$ единственным образом раскладывается в произведение
$A=UP$ некоторой матрицы вращения $U$ и некоторой положительно определенной
унимодулярной матрицы $P$. Матрица $U$ задается углом поворота $x$, а матрица
$P$, имеющая вид
\begin{equation*}
\begin{pmatrix} a & b \\ b & c \end{pmatrix},
\end{equation*}
где $a>0$ и $ac-b^2=1$, задается двумя числами $a>0$ и $b\in\MR$, т.е.\
комплексным числом $z=a+ib$, принадлежащим правой полуплоскости $a>0$.
Переходя с помощью дробно-линейного преобразования от правой полуплоскости
к единичному кругу, мы получаем, следовательно, что каждая матрица
$A\in\MS\ML(2,\MR)$ однозначно характеризуется парой $(x,u)$, где $x\in\MR$
и $u\in\MD$. При этом умножению матриц соответствует умножение пар, которое
задается формулой (\ref{emulsl}). Чтобы получить теперь универсальную
накрывающую $\widetilde{\MS\ML}(2,\MR)$ достаточно считать координату $x$
не углом, а точкой вещественной прямой $\MR$.

Теорема Адо, которая будет сформулирована в следующем разделе, утверждает, что
любая алгебра Ли имеет точное матричное представление. То есть классификация
матричных алгебр Ли влечет за собой классификацию всех алгебр Ли. С группами
дело обстоит иначе.
\begin{theorem}
Группа $\widetilde{\MS\ML}(2,\MR)$ не вкладывается ни в какую группу
$\MG\ML(n,\MR)$ и поэтому не является матричной группой Ли.
\end{theorem}
\begin{proof}
См., например, \cite{NovTai05R}.
\end{proof}
\section{Классификация простых алгебр и групп Ли                 \label{skiiut}}
Классификация простых алгебр Ли сводится к классификации матричных алгебр Ли,
благодаря следующему утверждению.
\begin{theorem}[\bf Адо]
Каждая алгебра Ли $\Gg$ над полем комплексных чисел $\MC$ имеет точное
(инъективное) представление в алгебре Ли $\Gg\Gl(n,\MR)$ для некоторого $n$.
\end{theorem}
\begin{proof}
См.\ \cite{Ado47R}.
\end{proof}
\index{Теорема Адо (Ado theorem)}\index{Адо теорема (Ado theorem)}%
\begin{cor}
Для каждой алгебры Ли $\Gg$ существует группа Ли $\MG$ (и, в частности,
односвязная группа Ли) с данной алгеброй Ли.
\qed\end{cor}
Теорема Адо важна, поскольку позволяет свести изучение абстрактных алгебр Ли к
матричным.

Напомним, что комплексификацией вещественного векторного пространства $\MV$
называется комплексное векторное пространство ${}^\MC\MV$, состоящее из векторов
вида $Z=X+iY$, где $X,Y\in\MV$. При этом сложение векторов $Z_1=X_1+iY_1$ и
$Z_2=X_2+iY_2$ задается формулой
\begin{equation*}
  Z_1+Z_2:=X_1+X_2+i(Y_1+Y_2).
\end{equation*}
Умножение вектора $Z$ на комплексное число $c=a+ib$ определяется равенством
\begin{equation*}
  cZ:=aX-bY+i(aY+bX).
\end{equation*}
\begin{defn}
Комплексификацией ${}^\MC\Gg$ вещественной алгебры Ли $\Gg$ называется
комплексная алгебра Ли, удовлетворяющая двум условиям:

1) ${}^\MC\Gg$ является комплексификацией вещественного векторного пространства
$\Gg$;

2) умножение в ${}^\MC\Gg$ задается формулой
\begin{equation}                                                  \label{qamftd}
  [Z_1,Z_2]=[X_1+iY_1,X_2+iY_2]:=[X_1,X_2]-[Y_1,Y_2]+i[X_1,Y_2]+i[Y_1,X_2].\qed
\end{equation}
\end{defn}

Если $\dim\Gg=\Sn$, то вещественная размерность ее комплексификации вдвое
больше: $\dim{}^\MC\Gg=2\Sn$. При этом комплексная размерность
комплексифицированной алгебры Ли остается прежней
$\dim_\MC({}^\MC\Gg)=\dim\Gg=\Sn$.

С другой стороны, комплексную алгебру Ли $\Gh$ комплексной размерности $\Sn$ с
базисом $e_\Sa$, $\Sa=1,\dotsc\Sn$, можно рассматривать как вещественную алгебру
Ли вещественной размерности $2\Sn$ с базисом $e_1$, $ie_1$, $\dotsc$,
$e_\Sn$, $ie_\Sn$. Эту овеществление комплексной алгебры Ли обозначим
${}^\MR\Gh$. Можно поступить по-другому. Назовем вещественной формой
комплексной алгебры Ли $\Gh$ такую вещественную алгебру Ли ${}^r\Gh$, которая
после комплексификации дает $\Gh$. В этом случае размерность вещественной формы
комплексной алгебры Ли в два раза меньше вещественной размерности исходной
алгебры Ли.
\begin{exa}
Рассмотрим вещественную алгебру Ли $\Gg\Gl(n,\MR)$, состоящую из вещественных
$n\times n$ матриц с обычным правилом коммутирования. Тогда ее комплексификацией
будет алгебра Ли ${}^\MC\Gg\Gl(n,\MR)=\Gg\Gl(n,\MC)$, состоящая из комплексных
$n\times n$ матриц. Теперь рассмотрим комплексную алгебру Ли $\Gg\Gl(n,\MC)$. Ее
овеществление будет изоморфно прямой сумме алгебр Ли:
${}^\MR\Gg\Gl(n,\MC)\simeq\Gg\Gl(n,\MR)\oplus\Gg\Gl(n,\MR)$ той же размерности.
В то же время вещественной формой алгебры Ли $\Gg\Gl(n,\MC)$ будет вещественная
алгебра Ли ${}^r\Gg\Gl(n,\MC)=\Gg\Gl(n,\MR)$ вдвое меньшей размерности.
\qed\end{exa}

В предыдущем разделе была приведена классификация всех простых комплексных
алгебр Ли. Поэтому возникает вопрос следует ли отсюда классификация вещественных
алгебр Ли. Ответ на этот вопрос положителен в силу следующего утверждения.
\begin{theorem}
Все вещественные простые алгебры Ли получаются из простых комплексных алгебр Ли
либо овеществлением, либо они являются вещественными формами простых комплексных
алгебр Ли.
\end{theorem}
\begin{proof}
См., например, \cite{BarRac77R}.
\end{proof}
Таким образом, для классификации простых вещественных алгебр Ли необходимо найти
все вещественные неизоморфные между собой формы комплексных алгебр Ли.
Этот вопрос сложен, но решен. Доказывается, что каждая комплексная простая
алгебра имеет только конечное число вещественных форм, и все они найдены
\cite{Cartan29}. Среди этих вещественных форм только одна является компактной.
Таким образом, существует взаимно однозначное соответствие между простыми
комплексными алгебрами Ли и простыми вещественными компактными алгебрами Ли.
Поэтому классификация простых вещественных компактных алгебр Ли сводится к
классификации простых комплексных алгебр Ли.

Приведем классификацию всех простых вещественных алгебр Ли (доказательство
можно найти, например, в \cite{BarRac77R}).

\subsubsection{Вещественные формы алгебры Ли $\Ga_n\simeq\Gs\Gl(n+1,\MC)$,
$n\ge1$}
\begin{enumerate}
\item $\Gs\Gu(n+1)$ -- компактная алгебра Ли всех антиэрмитовых
$(n+1)\times(n+1)$ матриц $Z$ с нулевым следом, $\tr Z=0$.
\item $\Gs\Gl(n+1,\MR)$ -- некомпактная алгебра Ли всех вещественных
$(n+1)\times(n+1)$ матриц $X$ с нулевым следом, $\tr X=0$.
\item $\Gs\Gu(p,q)$, $p+q=n+1$, $p\ge q$, -- некомпактная алгебра Ли всех матриц
вида
\begin{equation*}
  \begin{pmatrix} Z_1 & Z_2 \\ Z_2^\dagger & Z_3 \end{pmatrix},
\end{equation*}
где $Z_1,Z_3$ -- антиэрмитовы матрицы порядков $p$ и $q$ соответственно,
$\tr Z_1+\tr Z_3=0$, матрица $Z_2$ произвольна.
\item $\Gs\Gu^*\big(2(n+1)\big)$ -- некомпактная алгебра Ли всех
$2(n+1)\times2(n+1)$ матриц вида
\begin{equation*}
  \begin{pmatrix} Z_1 & Z_2 \\ -\overline Z_2 & \overline Z_1 \end{pmatrix},
\end{equation*}
где $Z_1,Z_2$ -- комплексные $(n+1)\times(n+1)$ матрицы,
$\tr Z_1+\tr\overline Z_1=0$.
\end{enumerate}
\subsubsection{Вещественные формы алгебры Ли $\Gb_n\simeq\Gs\Go(2n+1,\MC)$,
$n\ge2$}
\begin{enumerate}
\item $\Gs\Go(2n+1)$ -- компактная алгебра Ли всех вещественных антисимметричных
$(2n+1)\times(2n+1)$ матриц.
\item $\Gs\Go(p,q)$, $p+q=2n+1$, $p\ge q$, -- некомпактная алгебра Ли всех
вещественных $(2n+1)\times(2n+1)$ матриц вида
\begin{equation*}
  \begin{pmatrix} X_1 & X_2 \\ X_2^\St & X_3 \end{pmatrix},
\end{equation*}
где $X_1,X_3$ -- антисимметричные матрицы порядков $p$ и $q$ соответственно,
матрица $X_2$ произвольна.
\end{enumerate}
\subsubsection{Вещественные формы алгебры Ли $\Gc_n\simeq\Gs\Gp(n,\MC)$,
$n\ge3$}
\begin{enumerate}
\item $\Gs\Gp(n)$ -- компактная алгебра Ли всех антиэрмитовых бесследовых
комплексных $2n\times2n$ матриц вида
\begin{equation*}
  \begin{pmatrix} Z_1 & Z_2 \\ Z_3 & -Z_1^\St \end{pmatrix},
\end{equation*}
где все матрицы $Z_{1,2,3}$ имеют порядок $n$, и матрицы $Z_2$ и $Z_3$
симметричные (т.е.\ $\Gs\Gp(n)=\Gs\Gp(n,\MC)\cap \Gs\Gu(2n)$).
\item $\Gs\Gp(n,\MR)$ -- некомпактная алгебра Ли всех вещественных
$2n\times2n$ матриц вида
\begin{equation*}
  \begin{pmatrix} X_1 & X_2 \\ X_3 & -X_1^\St \end{pmatrix},
\end{equation*}
где $X_{1,2,3}$ -- вещественные $n\times n$ матрицы и $X_{2,3}$ -- симметричные
матрицы.
\item $\Gs\Gp(p,q)$, $p+q=n$, $p\ge q$, -- некомпактная алгебра Ли всех
комплексных $2n\times 2n$ матриц вида
\begin{equation*}
  \begin{pmatrix} Z_{11} & Z_{12} & Z_{13} & Z_{14}  \\
  Z_{12}^\dagger & Z_{22} & Z_{14}^\St & Z_{24} \\
  -\overline Z_{13} & \overline Z_{14} & \overline Z_{11} & -\overline Z_{12} \\
  Z_{14}^\dagger & -\overline Z_{24} & -Z_{12}^\St & \overline Z_{22}
  \end{pmatrix},
\end{equation*}
где $Z_{11},Z_{13}$ -- матрицы порядка $p$, $Z_{12}$ и $Z_{14}$ -- $p\times q$
матрицы, $Z_{11}$ и $Z_{22}$ -- антиэрмитовы, $Z_{13}$ и $Z_{24}$ --
симметричные.
\end{enumerate}
\subsubsection{Вещественные формы алгебры Ли $\Gd_n\simeq\Gs\Go(2n,\MC)$,
$n\ge4$}
\begin{enumerate}
\item $\Gs\Go(2n)$ -- компактная алгебра Ли всех вещественных антисимметричных
$2n\times2n$ матриц.
\item $\Gs\Go(p,q)$, $p+q=2n$, $p\ge q$, -- некомпактная алгебра Ли всех
вещественных $2n\times2n$ матриц вида
\begin{equation*}
  \begin{pmatrix} X_1 & X_2 \\ X_2^\St & X_3 \end{pmatrix},
\end{equation*}
где $X_1,X_3$ -- антисимметричные матрицы порядков $p$ и $q$ соответственно,
матрица $X_2$ произвольна.
\item $\Gs\Go^*(2n)$ -- некомпактная алгебра Ли всех комплексных $2n\times2n$
матриц вида
\begin{equation*}
  \begin{pmatrix} Z_1 & Z_2 \\ -\overline Z_2 & \overline Z_1 \end{pmatrix},
\end{equation*}
где $Z_1,Z_2$ -- комплексные $n\times n$ матрицы, $Z_1$ -- антисимметричная,
$Z_2$ -- эрмитова.
\end{enumerate}

Вещественные формы алгебр Ли, приведенные выше, определены при всех $n\ge1$.
Ограничения на номера алгебр $n$ возникают из-за наличия изоморфизмов
между комплексными алгебрами Ли, которые индуцируют изоморфизмы их вещественных
форм. Запишем эти изоморфизмы в виде таблицы. Случай
$\Gd_2\simeq\Ga_1\oplus\Ga_1$ включен для удобства. Отметим также, что
алгебра $\Gd_1$ не является полупростой.
\begin{table}[ht]
\begin{center}
  \begin{tabular}{|l|l|}                                              \hline
  \begin{minipage}[c]{.4\textwidth} Изоморфизмы комплексных \\ алгебр Ли
  \end{minipage}  &
  \begin{minipage}[c]{.5\textwidth} Изоморфизмы вещественных \\ форм алгебр Ли
  \end{minipage}  \\ \hline
  $\Ga_1\simeq\Gb_1\simeq\Gc_1$ & $\Gs\Gu(2)\simeq\Gs\Go(3)\simeq\Gs\Gp(1)$ \\
  &$\Gs\Gl(2,\MR)\simeq\Gs\Gu(1,1)\simeq\Gs\Go(2,1)\simeq\Gs\Gp(1,\MR)$\\ \hline
  $\Gb_2\simeq\Gc_2$ & $\Gs\Go(5)\simeq\Gs\Gp(2)$ \\
  & $\Gs\Go(3,2)\simeq\Gs\Gp(2,\MR)$ \\
  & $\Gs\Go(4,1)\simeq\Gs\Gp(1,1)$ \\ \hline
  $\Gd_2\simeq\Ga_1\oplus\Ga_1$ & $\Gs\Go(4)\simeq\Gs\Go(3)\oplus\Gs\Go(3)$ \\
  & $\Gs\Go(2,2)\simeq\Gs\Gl(2,\MR)\oplus\Gs\Gl(2,\MR)$ \\
  & $\Gs\Gl(2,\MC)\simeq\Gs\Go(3,1)$ \\
  & $\Gs\Go^*(4)\simeq\Gs\Gl(2,\MR)\oplus\Gs\Gu(2)$ \\ \hline
  $\Ga_3\simeq\Gd_3$ & $\Gs\Gu(4)\simeq\Gs\Go(6)$ \\
  & $\Gs\Gl(4,\MR)\simeq\Gs\Go(3,3)$ \\
  & $\Gs\Gu(2,2)\simeq\Gs\Go(4,2)$ \\
  & $\Gs\Gu(3,1)\simeq\Gs\Go^*(6)$ \\
  & $\Gs\Gu^*(4)\simeq\Gs\Go(5,1)$ \\
  & $\Gs\Go^*(8)\simeq\Gs\Go(6,2)$ \\ \hline
  \end{tabular}
  \caption{Изоморфизм комплексных и вещественных алгебр Ли}
  \label{tdegko}
\end{center}
\end{table}

Вещественные формы исключительных простых алгебр Ли приведены, например, в
\cite{HauSch68}.

Как уже отмечалось, каждой комплексной простой алгебре Ли соответствует только
одна вещественная компактная форма.

Приведем также классификацию комплексных простых групп и их вещественных
компактных форм (см., например, \cite{Helgas01R}). В таблице колонка простых
комплексных групп Ли обозначена ${}^\MC\MG$, а их вещественных компактных форм
-- $\MG$. Универсальные накрывающие вещественных компактных групп Ли отмечены
знаком тильды $\widetilde\MG$, а через $\MZ(\widetilde\MG)$ обозначен центр
универсальной накрывающей. То есть $\MG=\widetilde\MG/\MZ(\widetilde\MG)$.
Приведены также размерности вещественных групп Ли.
\begin{table}[ht]
\setlength{\extrarowheight}{4pt}
\begin{center}
  \begin{tabular}{|l|l|l|l|l|}                                            \hline
  $ \Gg$ & ${}^\MC\MG$ & $\MG$ & $\MZ(\widetilde\MU)$ & $\dim\MU$ \\ \hline
  $ \Ga_n\quad (n\ge1)$ & $\MS\ML(n+1,\MC)$ & $\MS\MU(n+1)$ & $\MZ_{n+1}$ &
  $n(n+2)$ \\ \hline
  $ \Gb_n\quad (n\ge2)$ & $\MS\MO(2n+1,\MC)$ & $\MS\MO(2n+1)$ & $\MZ_2$ &
  $n(2n+1)$ \\ \hline
  $ \Gc_n\quad (n\ge3)$ & $\MS\MP(n,\MC)$ & $\MS\MP(n)$ & $\MZ_2$ &
  $n(2n+1)$ \\ \hline
  $ \Gd_n\quad (n\ge4)$ & $\MS\MO(2n,\MC)$ & $\MS\MO(2n)$ &
  \begin{minipage}[l]{0.25\textwidth}$\MZ_4^{\vphantom{2}}$,\quad $n$ нечетно \\
  $\MZ_2\oplus\MZ_2$, \quad $n$ четно \end{minipage} & $n(2n-1)$ \\ \hline
  $ \Gg_2$ & ${}^\MC\MG_2$ & $\MG_2$ & $\MZ_1$ & $14$ \\ \hline
  $ \Gf_4$ & ${}^\MC\MF_4$ & $\MF_4$ & $\MZ_1$ & $52$ \\ \hline
  $ \Ge_6$ & ${}^\MC\ME_6$ & $\ME_6$ & $\MZ_3$ & $78$ \\ \hline
  $ \Ge_7$ & ${}^\MC\ME_7$ & $\ME_7$ & $\MZ_2$ & $133$ \\ \hline
  $ \Ge_8$ & ${}^\MC\ME_8$ & $\ME_8$ & $\MZ_1$ & $248$ \\ \hline
  \end{tabular}
  \caption{Простые группы Ли, соответствующие простым комплексным алгебрам Ли и
  их компактным вещественным формам}
  \label{tdegkl}
\end{center}
\end{table}
\chapter{Группы преобразований                                   \label{stragf}}
Теория групп имеет важные приложения в математической физике, и почти всегда
группы появляются как группы преобразований чего либо. Это может быть группа
преобразований пространства-времени, группа симметрий изотопического
пространства (например, $\MS\MU(3)$ симметрия в физике элементарных частиц),
группа симметрий уравнений движения или что нибудь более экзотическое. В
настоящем разделе мы рассмотрим группы преобразований $\MG$ многообразия $\MM$.
В общем случае группа $\MG$ может быть конечной, счетной или группой Ли. В
основном мы будем рассматривать группы Ли преобразований, хотя многие
определения и утверждения справедливы и для произвольных групп.
\section{Действие групп преобразований                           \label{stragr}}
Пусть на многообразии $\MM$ задано {\em правое} действие группы $\MG$, т.е.\
задано отображение
\index{Правое действие группы (right action of a group)}%
\index{Действие группы справа (right action of a group)}%
\begin{equation}                                                  \label{egract}
  \MM\times \MG\ni\quad x,a\mapsto xa\quad\in\MM,
\end{equation}
которое предполагается достаточно гладким по $x$.
\begin{defn}
Пара $(\MM,\MG)$ называется {\em группой правых преобразований} многообразия
$\MM$, если действие группы $\MG$ на $\MM$ справа удовлетворяет следующим
условиям:\newline
\indent 1) \parbox[t]{.92\linewidth}{при любом $a\in\MG$ отображение
$\MM\ni x\mapsto xa\in\MM$ является диффеоморфизмом,}\newline
\indent 2) \parbox[t]{.92\linewidth}{$(xa)b=x(ab)$ для всех $x\in\MM$ и
$a,b\in \MG$.}\newline
В этом случае многообразие $\MM$ называется {\em $\MG$-многообразием}.
\qed\end{defn}
\index{Группа правых преобразований (group of right transformations)}%
\index{Правых преобразований группа (group of right transformations)}%
\index{$\MG$-многообразие ($\MG$-manifold)}%

Если $\MG$ -- группа Ли, то отображение (\ref{egract}) предполагается
достаточно гладким также по $a\in\MG$ и, следовательно, непрерывным для всех
$a\in\MG$ и для всех $x\in\MM$. Для каждой фиксированной точки $x\in\MM$ мы
имеем достаточно гладкое отображение $\MG\rightarrow\MM$.
\begin{com}
Иногда в качестве $\MM$ мы будем рассматривать многообразия с краем. Под
дифференцируемой функцией $f$ на $\MM$ в таком случае мы понимаем наличие
всех частных производных у $f$ во всех внутренних точках и существование предела
этих производных на крае $\pl\MM$.
\qed\end{com}
Каждому элементу группы $a\in\MG$ соответствует достаточно гладкое отображение
$\MM\rightarrow\MM$, заданное правилом $x\mapsto xa$. Поэтому группа
преобразований определяет отображение $\MG\rightarrow\diff(\MM)$ группы $\MG$ в
группу диффеоморфизмов многообразия $\diff(\MM)$.
\begin{prop}
Пара $(\MM,\MG)$ является группой преобразований многообразия $\MM$ тогда и
только тогда, когда отображение $\MG\rightarrow\diff(\MM)$ является
гомоморфизмом групп и отображение (\ref{egract}) является достаточно гладким.
\end{prop}
\begin{proof}
Условие 1) в определении группы преобразований означает, что отображение
$x\mapsto xa$ принадлежит группе $\diff(\MM)$. Групповая операция сохраняется
при отображении в силу свойства 2). Осталось доказать, что единица $e\in\MG$
отображается в тождественное преобразование многообразия $\MM$. Действительно,
отображение $x\mapsto xa$  является диффеоморфизмом и, следовательно, биекцией
для всех $a\in\MG$. В частности, для каждого $x\in\MM$ найдется такая точка
$y\in\MM$, что $x=ye$. Умножив это уравнение справа на $e$, получаем цепочку
равенств:
\begin{equation*}
  xe=ye^2=ye=x,\qquad\forall x\in\MM.
\end{equation*}
То есть единичный элемент группы не сдвигает ни одну точку многообразия.

Обратно. Каждому элементу $a$ соответствует некоторый диффеоморфизм, т.е.\
условие 1) выполнено. Поскольку отображение $\MG\rightarrow\diff(\MM)$ --
гомоморфизм, то выполнено условие 2).
\end{proof}

Аналогично определяется {\em левое} действие группы $\MG$ на $\MM$, т.е.\
отображение
\begin{equation*}
  \MG\times\MM\ni\quad a,x\mapsto ax\quad\in\MM.
\end{equation*}
\index{Левое действие группы (left action of a group)}%
\index{Действие группы слева (left action of a group)}%
\index{Группа левых преобразований (group of left transformations)}%
\index{Левых преобразований группа (group of left transformations)}%

В дальнейшем, если не оговорено противное, под {\em группой преобразований}
многообразия будем понимать группу правых преобразований и будем говорить,
\index{Группа преобразований (transformation group)}%
\index{Преобразований группа (transformation group)}%
что $\MG$ действует на $\MM$, подразумевая действие группы справа.
\begin{exa}
Если задано представление группы $\MG$, то задана группа линейных преобразований
$(\MV,\MG)$ векторного пространства $\MV$ (автоморфизмов), которое снабжено
естественной структурой многообразия.
\qed\end{exa}
\begin{exa}
Группа Ли является группой преобразований самой себя
$\MG\times\MG\rightarrow\MG$, которую можно рассматривать, как действующую
слева или справа.
\qed\end{exa}

\index{Тривиальное действие группы (trivial action of a group)}%
\index{Действие группы тривиальное (trivial action of a group)}%
\begin{defn}
Группа $\MG$ действует на $\MM$ {\em тривиально}, если равенство $xa=x$
выполнено для всех $x\in\MM$ и всех $a\in \MG$. В этом случае группа
преобразований превращается в тривиальное главное расслоение
$\MM\times\MG\xrightarrow{\pi}\MM$ с естественной проекцией
$\pi:~(x,a)\mapsto x$.

Говорят, что $\MG$ действует {\em эффективно} на $\MM$, если из равенства $xa=x$
для всех $x\in\MM$ следует, что $a=e$. Если для любой точки $x\in\MM$ уравнение
$xa=x$ имеет единственное решение $a=e$, то говорят, что группа преобразований
действует {\em свободно}.
\qed\end{defn}
\index{Эффективное действие группы (effective action of a group)}%
\index{Действие группы эффективное (effective action of a group)}%
\index{Свободное действие группы (free action of a group)}%
\index{Действие группы свободное (free action of a group)}%
Другими словами, действие группы $\MG$ на $\MM$ является эффективным, если
любой элемент группы, отличный от единичного, перемещает хотя бы одну
точку. Действие группы свободно, если любой элемент группы, отличный от
единичного, перемещает все точки многообразия. Группа, действующая свободно,
является эффективной. Обратное утверждение неверно, что показывает следующий
\begin{exa}
Группа вращений $\MS\MO(3)$ трехмерного евклидова пространства $\MR^3$ действует
эффективно, но не свободно, т.к.\ начало координат остается неподвижным.
\qed\end{exa}
Если действие группы преобразований свободно, то это значит, что для любой пары
точек $x_1,x_2\in\MM$ либо не существует преобразования, связывающего эти точки,
$x_2=x_1a$, либо элемент $a\in\MG$ {\em единственен}.
\begin{defn}
{\em Ядром $\MK$ группы преобразований} $(\MM,\MG)$ называется множество
элементов группы, которое действует тривиально на все точки многообразия,
\begin{equation*}                                                    \tag*{\qed}
  \MK:=\lbrace a\in\MG:~xa=x,~\forall x\in\MM\rbrace.
\end{equation*}
\renewcommand{\qed}{}\end{defn}
\index{Ядро группы преобразований (kernel of a transformation group)}%
Действие группы эффективно, если $\MK=\lbrace e\rbrace$.
\begin{prop}
Ядро группы преобразований является нормальной подгруппой в $\MG$. Кроме того,
эта подгруппа замкнута в $\MG$.
\end{prop}
\begin{proof}
Отображение $\MG\rightarrow\diff(\MM)$ непрерывно и ядро $\MK$ отображается в
единицу группы диффеоморфизмов $\diff(\MM)$. Поскольку единица замкнута в
$\diff(\MM)$, то ее прообраз $\MK$ замкнут в $\MG$ по критерию 1) теоремы
\ref{tconma}.
\end{proof}
Ядро группы преобразований представляет собой то множество элементов группы,
которое вообще не действует на многообразии $\MM$. Поэтому действие группы $\MG$
естественным образом сводится к действию фактор группы $\MG/\MK$, которая
действует эффективно.
\begin{exa}
Рассмотрим присоединенное действие группы Ли на себя
$\MG\times\MG\rightarrow\MG$, заданное преобразованием подобия
$a\mapsto bab^{-1}$, для
всех $a,b\in\MG$. Ядром этого действия является центр группы $\MK\subset\MG$.
Присоединенное действие группы Ли не является свободным, так как единичный
элемент группы $e\in\MG$ остается неподвижным, $aea^{-1}=e$, $\forall a\in\MG$.
Присоединенное действие фактор группы $\MG/\MK$ является эффективным.
\qed\end{exa}

\begin{defn}
Пусть точка $x\in\MM$ фиксирована. Множество точек
\begin{equation*}
  x\MG:=\left\lbrace xa\in\MM:~\forall a\in \MG\right\rbrace
\end{equation*}
называется {\em орбитой} точки $x$ относительно действия группы $\MG$.
\qed\end{defn}
\index{Орбита действия группы (orbit of a group action)}%
Используя групповые свойства, нетрудно проверить, что орбиты двух точек
$x_1,x_2\in\MM$ либо совпадают, либо не имеют общих точек. Орбитой произвольного
подмножества $\MU\subset\MM$ будем называть объединение орбит всех точек из
$\MU$. Обозначим его $\MU\MG$.

Если группа $\MG$ конечна, то орбита любой точки также конечна. Если группа
преобразований счетна, то ее орбиты либо счетны, либо конечны.
\begin{exa}
Рассмотрим поворот евклидовой плоскости $\MR^2$ на конечный угол $\theta$.
Начало координат неподвижно относительно вращений. Все остальные точки имеют
нетривиальные орбиты, если угол поворота не кратен $2\pi$.
Пусть $\theta=2\pi a$, где $a\in\MR$. Если число $a$ рационально, $a=m/n$, где
$m$ и $n$ -- взаимно простые натуральные числа, то орбиты состоят из $n$ точек.
Если $a$ -- иррациональное число, то орбита произвольной точки $x\ne0$ счетна и
являются всюду плотным подмножеством на окружности $\MS^1\hookrightarrow\MR^2$,
проходящей через точку $x$.
\qed\end{exa}

Если задана группа Ли преобразований $(\MM,\MG)$, то орбита $x\MG$ произвольной
точки $x\in\MM$ является подмногообразием в $\MM$. Это следует из
дифференцируемости отображения (\ref{egract}) для всех $x\in\MM$ и групповых
свойств $\MG$. Размерность орбиты не превосходит размерности группы,
$\dim(x\MG)\le\dim\MG$.
\begin{exa}
Рассмотри евклидово пространство $\MR^n$ с декартовой системой координат
$x^\al$, $\al=1,\dotsc,n$. Пусть группа вещественных чисел $\MR$ по сложению
действует на $\MR^n$, как трансляции вдоль первой оси,
\begin{equation*}
  \MR^n\times\MR\ni\quad \lbrace x^1,x^2,\dotsc,x^n\rbrace,a
  \mapsto\lbrace x^1+a,x^2,\dotsc,x^n\rbrace\quad \in\MR^n.
\end{equation*}
Тогда орбитой произвольной точки $x\in\MR^n$ является прямая, проходящая через
$x$ и параллельная оси $x^1$.
\qed\end{exa}

\begin{prop}                                                      \label{pclosm}
Орбита $x\MG$ произвольной точки $x\in\MM$ является замкнутым подмногообразием в
$\MM$. Кроме того, если $\MG$ -- связная группа Ли, то орбита является связным
подмногообразием.
\end{prop}
\begin{proof}
Подмножество $\MM\setminus x$ открыто в $\MM$. Поэтому объединение открытых
подмножеств $(\MM\setminus x)\MG$ открыто в $\MM$. Следовательно, орбита
произвольной точки  замкнута, как дополнение открытого подмножества,
$x\MG=\MM\setminus\big((\MM\setminus x)\MG\big)$. Поскольку отображение
(\ref{egract}) является непрерывным для всех $x$, то связность сохраняется.
\end{proof}
\begin{defn}
Обозначим через $\MM/\MG$ множество орбит $\breve x:=x\MG$ относительно действия
группы $\MG$ на $\MM$. То есть $\breve x_1=\breve x_2$ тогда и только тогда,
когда точки $x_1$ и $x_2$ лежат на одной орбите: $x_2=x_1a$ для некоторого
$a\in\MG$. Пусть $\pi:~\MM\rightarrow\MM/\MG$
-- проекция, сопоставляющая каждой точке $x\in\MM$ ее орбиту $\breve x$. Тогда
пространство орбит наделяется фактортопологией, т.е.\ множество $\MV\in\MM/\MG$
является открытым тогда и только тогда, когда множество $\pi^{-1}(\MV)$ открыто
в $\MM$. Если $\MU\subset\MM$ -- окрестность точки $x\in\MM$, то множество орбит
$\MV=\MU\MG$ -- окрестность точки $\breve x\in\MM/\MG$. Множество орбит
$\MM/\MG$ с фактортопологией называется {\em пространством орбит} многообразия
$\MM$ относительно действия группы $\MG$ или {\em факторпространством}.
\qed\end{defn}
\index{Пространство орбит (orbit apace)}%
\index{Факторпространство (coset space)}%

Хотя $\MM$ и $\MG$ -- многообразия, пространство орбит с фактортопологией
может оказаться весьма сложным. Оно, как покажут дальнейшие примеры, может не
иметь структуры многообразия. В общем случае пространство орбит может не быть
даже хаусдорфовым топологическим пространством.

\begin{defn}
Действие группы $\MG$ на многообразии $\MM$ называется {\em транзитивным},
если для любых двух точек $x_1,x_2\in\MM$ найдется такой элемент $a\in \MG$,
что $x_1a=x_2$.
\qed\end{defn}
\index{Транзитивное действие группы (transitive action of a group)}%
\index{Действие группы транзитивное (transitive action of a group)}
Если действие группы $\MG$ на многообразии $\MM$ транзитивно, то орбита
произвольной точки $x\in\MM$ совпадает со всем $\MM$. Если на многообразии
существует одна точка $x_0\in\MM$, которую можно преобразовать во все другие
точки $\MM$, то этого достаточно для того, чтобы действие группы $\MG$ на $\MM$
было транзитивным. Действительно, если $x_1=x_0a$ и $x_2=x_0b$, то
$x_2=x_1a^{-1}b$.
\begin{exa}
Действие группы Ли $\MG$ на себе является транзитивным и свободным.
\qed\end{exa}
\begin{exa}                                                       \label{ehomsp}
Пусть $\MG$ -- группа и $\MH$ -- ее нетривиальная подгруппа. Рассмотрим
пространство правых смежных классов $\MG/\MH$, состоящее из элементов вида
$\MH a$, где $a\in\MG$. Определим действие группы $\MG$ в пространстве смежных
классов $\MG/\MH$ соотношением
\begin{equation*}
  \MG/\MH\times\MG\ni\quad \MH a,b\mapsto \MH ab\quad \in\MG/\MH.
\end{equation*}
Это действие транзитивно для любых $\MG$ и $\MH$. Действие $\MG$ на $\MG/\MH$
эффективно тогда и только тогда, когда $\MH$ не содержит нормальных подгрупп.
Действительно, если $\MH$ -- нормальная подгруппа, то для всех $h\in\MH$ и
$a\in\MG$ выполнены равенства:
\begin{equation*}
  (\MH a)h=a\MH h=a\MH=\MH a.
\end{equation*}
Действие группы никогда не является свободным, т.к.\ $(\MH e)h=\MH e$ для всех
$h\in\MH$.
\qed\end{exa}

В теории групп преобразований важную роль играет
\begin{theorem}                                                   \label{tfacma}
Пусть $\MG$ -- группа Ли и $\MH$ -- ее замкнутая подгруппа. Тогда на фактор
пространстве правых смежных классов $\MG/\MH$ существует единственная структура
вещественно аналитического многообразия такая, что действие $\MG$ на $\MG/\MH$
вещественно аналитично. В частности, проекция $\MG\rightarrow\MG/\MH$
вещественно аналитична.
\end{theorem}
\begin{proof}
См., например, в \cite{Cheval46R}.
\end{proof}
\begin{defn}
Точка $x\in\MM$ называется {\em стационарной} или {\em неподвижной}
относительно действия группы преобразований, если $xa=x$ для всех $a\in \MG$.
\qed\end{defn}
\index{Стационарная точка (fixed point)}%
\index{Точка стационарная (fixed point)}%
\index{Неподвижная точка (fixed point)}%
\index{Точка неподвижная (fixed point)}%
Если группа действует на многообразии свободно или транзитивно, то стационарные
точки отсутствуют.
\begin{exa}
Представление группы $(\MV,\MG)$ не может быть транзитивной или свободной
группой преобразований, т.к.\ нулевой вектор из $\MV$ является стационарной
точкой.
\qed\end{exa}
\begin{defn}
{\em Группой изотропии} $\MG_x$ точки многообразия $x\in\MM$ называется
множество элементов группы $\MG$, оставляющих точку $x$ неподвижной,
\begin{equation*}                                                    \tag*{\qed}
  \MG_x:=\left\lbrace a\in \MG:~xa=x\right\rbrace.
\end{equation*}
\end{defn}
\index{Группа изотропии (isotropy group)}%
\index{Изотропии группа (isotropy group)}%
Нетрудно проверить, что это множество элементов образует группу. Группы
изотропии $\MG_{x_1}$ и $\MG_{x_2}$ двух точек, лежащих на одной орбите,
$x_2 a=x_1$, являются сопряженными подгруппами в $\MG$:
$$
  \MG_{x_2}=a\MG_{x_1}a^{-1}.
$$

Ядро $\MK$ группы преобразований $(\MM,\MG)$ связано с группами изотропии
простым соотношением
\begin{equation*}
  \MK=\bigcap_{x\in\MM}\MG_x,
\end{equation*}
где объединение берется по всем точкам $x\in\MM$.

Из теоремы \ref{tfacma} следует, что пара $(\MG/\MH,\MG)$ является группой
транзитивных преобразований. Представляет большой интерес обратное утверждение:
если пара $(\MM,\MG)$ -- транзитивная группа преобразований, то она имеет вид
$(\MG/\MH,\MG)$ для некоторой подгруппы $\MH\subset\MG$ и при некоторых
дополнительных предположениях, которые мы уточним ниже. Для ясности сначала
будет доказана теорема безотносительно топологии и дифференцируемой структуры,
заданной на $\MM$.
\begin{theorem}                                                   \label{tisogh}
Пусть $\MG$ -- группа, которая действует транзитивно на некотором множестве
$\MM$. Тогда для каждой точки $x\in\MM$ существует биекция правых смежных
классов $\MG/\MG_x$ на множество $\MM$, заданное отображением
\begin{equation}                                                  \label{emapgx}
  \MG/\MG_x\ni\quad\MG_xa\mapsto xa\quad\in\MM,
\end{equation}
где $\MG_x$ -- группа изотропии точки $x$.
\end{theorem}
\begin{proof}
Отображение (\ref{emapgx}) определено для каждого представителя из правого
смежного класса. Покажем сначала, что это отображение не зависит от выбора
представителя. Пусть $a$ и $b$ -- два представителя какого либо одного
смежного класса. Тогда существует элемент $h\in\MG_x$ такой, что $a=hb$ и
поэтому $xa=xhb=xb$. То есть отображение (\ref{emapgx}) не зависит от
выбора представителя смежного класса и, следовательно, определено корректно.

Выберем по представителю из двух смежных классов $\MG_x a$ и $\MG_xb$ и
подействуем на точку $x$. В результате получим точки $x_1=xa$ и $x_2=xb$.
Эти точки совпадают тогда и только тогда, когда группа изотропии содержит
элемент $ba^{-1}\in\MG_x$. Это означает, что отображение (\ref{emapgx})
является взаимно однозначным.

Поскольку группа преобразований $(\MM,\MG)$ является транзитивной, то
отображение (\ref{emapgx}) является взаимно однозначным отображением на, т.е.\
биекцией.
\end{proof}
\begin{cor}
Если группа преобразований $(\MM,\MG)$ транзитивна и свободна, т.е.\ $\MG_x=e$
для всех точек $x\in\MM$, то отображение
\begin{equation*}
  \MG\ni\quad a\mapsto xa\quad\in\MM
\end{equation*}
является биекцией.
\qed\end{cor}

Теперь разберемся с дифференцируемой структурой.
\begin{prop}
Пусть $(\MM,\MG)$ -- группа транзитивных преобразований. Выберем произвольную
точку $x\in\MM$. Тогда ее группа изотропии $\MG_x$ является замкнутой подгруппой
в $\MG$.
\end{prop}
\begin{proof}
Поскольку группа преобразований транзитивна, то отображение
$\MG\ni a\mapsto xa\in\MM$, определенное для каждого $x\in\MM$, является
сюрьективным и непрерывным. Точка $x$ является замкнутым подмножеством в $\MM$,
и, в силу теоремы критерия 1) из теоремы \ref{tconma}, группа изотропии $\MG_x$
замкнута в $\MG$.
\end{proof}
\begin{cor}
Группа изотропии $\MG_x$ для всех $x\in\MM$ является замкнутой подгруппой в
$\MG$ для произвольной группы Ли преобразований $(\MM,\MG)$.
\end{cor}
\begin{proof}
Каждая точка $x\in\MM$ принадлежит орбите $x\MG$, которая является
подмногообразием в $\MM$. На каждой орбите группа $\MG$ действует транзитивно,
и поэтому пара $(x\MG,\MG)$ -- транзитивная группа преобразований.
Следовательно, группа изотропии $\MG_x$ является замкнутой подгруппой в $\MG$.
\end{proof}

Согласно теореме \ref{tfacma} на пространстве смежных классов $\MG/\MG_x$
существует единственная дифференцируемая структура. Поскольку отображение
(\ref{emapgx}) является биекцией, то эту дифференцируемую структуру можно
перенести на $\MM$, превратив его тем самым в многообразие. Тогда отображение
(\ref{emapgx}) становится диффеоморфизмом.

Таким образом, если группа преобразований $(\MM,\MG)$ транзитивна, то
многообразие $\MM$ представляет собой фактор пространство $\MG/\MG_x$. При этом
\begin{equation*}
  \dim\MM=\dim\MG-\dim\MG_x.
\end{equation*}
Отсюда следует, что размерность многообразия $\MM$, на котором действует
транзитивная группа преобразований, не может превышать размерность группы.
\begin{defn}
Произвольное множество $\MM$, на котором группа преобразований $(\MM,\MG)$
действует транзитивно называется {\em однородным пространством}. Если, вдобавок,
действие группы эффективно, то это множество называется {\em главным однородным
пространством} над $\MG$.
\qed\end{defn}
\index{Однородное пространство}%
\index{Однородное пространство}%
\index{Главное однородное пространство}%
\index{Пространство главное однородное}%
\begin{exa}
Пара $(\MG/\MH,\MG)$ (пример \ref{ehomsp}) является однородным пространством.
Если подгруппа $\MH\subset\MG$ не содержит нормальных подгрупп, то пара
$(\MG/\MH,\MG)$ -- главное однородное пространство.
\qed\end{exa}
\section{Инфинитезимальные преобразования                        \label{sinftr}}
Пусть задана группа преобразований $(\MM,\MG)$, где $\MG$ -- группа Ли. Тогда
имеет смысл говорить об инфинитезимальных преобразованиях, которые
взаимно однозначно определяются векторными полями на $\MM$ и $\MG$. Ниже мы
рассмотрим соотношения между алгебрами Ли векторных полей на $\MG$ и $\MM$.

Построим отображение $\mu:~\Gg\rightarrow\CX(\MM)$ алгебры Ли $\Gg$ группы Ли
$\MG$ (множества левоинвариантных векторных полей в $\CX(\MG)$) в
бесконечномерную алгебру Ли векторных полей $\CX(\MM)$ на многообразии $\MM$.
Это можно сделать тремя эквивалентными способами.

1) Каждому элементу алгебры Ли $X\in\Gg$ соответствует однопараметрическая
подгруппа (см.\ раздел \ref{sexpli})
\begin{equation*}
  \exp_X:\quad \MR\ni\quad t\mapsto a(t)\quad\in\MG,
\end{equation*}
где $a(t):=\exp_X(t)$ (экспоненциальное отображение) и $a(0)=e$ -- единица
группы. Тогда через каждую точку $x\in\MM$ проходит единственная кривая $xa(t)$.
Касательные векторы к этим кривым образуют векторное поле $X^*(x)$ на
многообразии $\MM$. Таким образом мы построили отображение $\mu$ алгебры Ли
$\Gg$ в алгебру векторных полей $\CX(\MM)$:
\begin{equation*}
  \mu:\quad \Gg\ni\quad X\mapsto \mu(X):=X^*\quad\in\CX(\MM).
\end{equation*}

2) Для каждой точки $x\in\MM$ определим отображение
$\mu_x:~G\ni a\mapsto xa\in\MM$. Тогда отображение $\mu$ определено соотношением
\begin{equation}                                                  \label{esedef}
  X^*(x)=\mu(X)|_x:=(\mu_x)_*X_e,
\end{equation}
где $(\mu_x)_*$ -- дифференциал отображения $\mu_x$, действующий на касательный
вектор $X_e:=X(e)$ к единице группы, который соответствует элементу алгебры Ли
$X\in\Gg$.

3) Каждое векторное поле взаимно однозначно определяется дифференцированием в
алгебре функций. Поэтому рассмотрим гладкую функцию $f\in\CC^\infty(\MM)$ и
определим векторное поле соотношением
\begin{equation*}
  X^*(f):=\left.\frac d{dt}f\big(xa(t)\big)\right|_{t=0},
\end{equation*}
где $a(t):=\exp_X(t)$.

Поскольку произвольное левоинвариантное векторное поле $X$ на группе Ли $\MG$
полно, то полно также векторное поле $X^*$ на $\MM$. Это следует из определения,
т.к.\ отображение $\mu$ определено для всех $t\in(-\infty,\infty)$.
\begin{prop}                                                      \label{pinftr}
Отображение $\mu$ есть гомоморфизм алгебр Ли. Если $\MG$ действует эффективно на
$\MM$, то $\mu$ есть изоморфизм алгебры Ли $\Gg$ и ее образа
$\mu(\Gg)\subset\CX(\MM)$. Если $\MG$ действует свободно на $\MM$, то для
каждого ненулевого элемента алгебры Ли $X\in\Gg$, его образ $\mu(X)=X^*$ не
обращается в нуль на $\MM$.
\end{prop}
\begin{proof}
Поскольку дифференциал отображения -- это линейная операция, то из определения
(\ref{esedef}) следует, что $\mu$ есть линейное отображение из $\Gg$ в
$\CX(\MM)$. Доказательство того факта, что отображение $\mu$ сохраняет
коммутатор векторных полей:
\begin{equation*}
  [X^*,Y^*]=\mu[X,Y],\qquad \text{где}\quad X^*,Y^*\in\CX(\MM),~X,Y\in\Gg,
\end{equation*}
технично и приведено, например, в \cite{KobNom6369R}.

Докажем второе утверждение. Пусть $\mu(X)=0$ всюду на $\MM$. Это
значит, что однопараметрическая группа действует тривиально на $\MM$, т.е.\
$a(t)=e$ для всех $t$. Верно и обратное утверждение. Пусть теперь $X\in\ker \mu$,
т.е.\ $\mu(X)=0$. Это значит, что $a(t)=e$ для всех $t$. Отсюда следует, что
ядро $\ker\mu$ тривиально, $\mu(X)=0$. Поэтому отображение $\mu$ -- изоморфизм.

Пусть $\mu(X)=0$ в некоторой точке $x\in\MM$. Тогда точка $x$ неподвижна,
$xa(t)=x$ для всех $t$. Обратно. Если точка $x$ неподвижна, то $a(t)=e$ для всех
$t$ и, следовательно, $X=0$.
\end{proof}

Это предложение является основой для рассмотрения главных расслоений со
структурной группой Ли в разделе \ref{sprfib}.

Мы видим, что транзитивные группы преобразований устроены довольно просто.
В этом случае многообразие $\MM$ -- это одна орбита, и фактор пространство
$\MM/\MG$ состоит из одной точки. Если действие группы $\MG$ на многообразии
$\MM$ не является транзитивным, то орбит может быть много. В общем случае
пространство орбит устроено сложно, что показывает следующий
\begin{exa}                                                       \label{exsotr}
Стационарной точкой в евклидовом пространстве $\MR^3$ относительно действия
группы трехмерных вращений $\MS\MO(3)$ является начало координат. Группа
изотропии начала координат совпадает со всей группой, а группой изотропии любой
другой точки является подгруппа двумерных вращений $\MS\MO(2)\subset\MS\MO(3)$
вокруг оси, проходящей через данную точку и начало координат. Пространство орбит
представляет собой множество сфер $\MS^2_r$, которое можно параметризовать их
радиусом $r\in\MR_+$, объединенное с началом координат
$\lbrace 0\rbrace\in\MR^3$,
$$
  \left.\MR^3\right/\MS\MO(3)\approx\MR_+\cup\lbrace 0\rbrace=\overline\MR_+.
$$
То есть факторпространство представляет собой положительную полуось
$\overline{\MR}_+$, включая начало координат, и не является многообразием, т.к.\
не является открытым подмножеством в $\MR$ (оно является многообразием с краем).
\qed\end{exa}

Если группа преобразований $(\MM,\MG)$ не является транзитивной, то многообразие
$\MM$ представляет собой объединение орбит, на каждой из которых группа $\MG$
действует транзитивно. Выберем точку $x_i$ на каждой орбите и обозначим через
$\MG_i$ ее группу изотропии. Тогда существует диффеоморфизм каждой орбиты на
фактор пространство $\MG/\MG_i$. Другими словами, каждая орбита является
однородным пространством и на нем существует естественная дифференцируемая
структура (теорема \ref{tfacma}). Согласно предложению \ref{pclosm} каждая
орбита является замкнутым подмногообразием в $\MM$.

Рассмотрим частный случай. Пусть $L_\Sa$, $\Sa=1,\dotsc,\dim\MG$, -- базис
алгебры Ли. Если векторные поля $\mu(L_\Sa)$ линейно независимы в каждой точке,
то они определяют инволютивное распределение на многообразии $\MM$. Тогда,
согласно теореме Фробениуса, каждая орбита с естественной дифференцируемой
структурой является подмногообразием в $\MM$. В общем случае, несмотря на
изоморфизм отображения $\mu$, векторные поля $\mu(L_\Sa)$ могут быть линейно
зависимы. Это показывает пример \ref{exsotr}, в котором орбиты двумерны (за
исключением $0$-мерного начала координат), а алгебра Ли $\Gs\Go(3)$ --
трехмерна. Вопрос о том можно ли ввести на пространстве орбит (фактор
пространстве $\MM/\MG$) дифференцируемую структуру сложен и выходит за рамки
данной книги.

В заключение настоящего раздела докажем утверждение, которое будет использовано
при построении связностей в главном расслоении.
\begin{prop}                                                      \label{plodpa}
Пусть $a$ есть диффеоморфизм многообразия $\MM$,
\begin{equation*}
  a:\quad \MM\ni\quad x\mapsto a(x)\quad\in\MM.
\end{equation*}
Если векторное поле $X\in\CX(\MM)$ порождает однопараметрическую группу
преобразований $s_t:~\MM\rightarrow\MM$, то векторное поле $a_*X$, где $a_*$ --
дифференциал отображения $a$, порождает однопараметрическую группу
преобразований $a\circ s_t\circ a^{-1}$.
\end{prop}
\begin{proof}
Ясно, что $a\circ s_t\circ a^{-1}$ есть однопараметрическая группа
преобразований. Покажем, что эта группа преобразований индуцирует векторное поле
$a_*X$. Пусть $x\in\MM$ -- произвольная точка и $y=a^{-1}(x)$. Так как
отображение $s_t$ индуцирует векторное поле $X$, то вектор $X_y\in\MT_y(\MM)$
касается кривой $\g_1(t)=s_t(y)$ в точке $y=\g_1(0)$. Из этого следует, что
вектор
\begin{equation*}
  (a_*X)_x=a_*X_y\in\MT_x(\MM)
\end{equation*}
касается кривой $\g_2(t)=a\circ s_t(y)=a\circ s_t\circ a^{-1}(x)$.
\end{proof}
\begin{com}
В данном предложении преобразование $a$ не обязано принадлежать какой либо
группе. Это может быть произвольный фиксированный диффеоморфизм многообразия
$\MM$.
\qed\end{com}
\begin{cor}
Векторное поле $X\in\CX(\MM)$ инвариантно относительно диффеоморфизма
$a:~\MM\rightarrow\MM$, т.е.\ $a_*X=X$, тогда и только тогда, когда
диффеоморфизм $a$ перестановочен с однопараметрической группой преобразований
$s_t$, порожденной векторным полем $X$.
\end{cor}
\section{Инвариантные структуры                                  \label{sinvst}}
Рассмотрим группу преобразований $(\MM,\MG)$. Предположим, что на многообразии
задана некоторая геометрическая структура, например, тензорное поле. Возникает
вопрос при каких условиях эта структура является инвариантной относительно
данной группы преобразований ? Ответ на этот вопрос очень важен для приложений,
поскольку часто мы ищем решения некоторой системы уравнений, обладающие
определенной симметрией. Такие структуры называются {\em $\MG$-инвариантными}.
\index{$\MG$-инвариантная структура ($\MG$-invariant structure)}%
\index{Структура $\MG$-инвариантная ($\MG$-invariant structure)}%

Начнем с простейшего случая. Предположим, что на многообразии $\MM$ задана
функция и действие группы преобразований $(\MM,\MG)$. Определим действие
элемента группы преобразований $a\in\MG$ на функцию формулой
\begin{equation*}
  f(x)\mapsto \hat f(x):=f(xa^{-1}).
\end{equation*}
Отсюда следует равенство $\hat f(xa)=f(x)$. То есть значение новой функции
$\hat f$ в точке $xa$ равно значению исходной функции в точке $x$. Это --
обычное преобразование функции при диффеоморфизмах.
\begin{defn}
Назовем функцию $\MG$-{\em инвариантной}, если она удовлетворяет условию
\begin{equation}                                                  \label{eginsc}
  f(xa)=f(x)
\end{equation}
для всех $a\in\MG$.
\qed\end{defn}
\index{Функция $\MG$-инвариантная ($\MG$-invariant function)}%
\index{$\MG$-инвариантная функция ($\MG$-invariant function)}%
Из определения следует, что $\MG$-инвариантные функции постоянны на орбитах
группы преобразований.
\begin{exa}
Сферически симметричные функции в трехмерном евклидовом пространстве $\MR^3$
постоянны на сферах $\MS^2\hookrightarrow\MR^3$ с центром в начале координат.
Эти функции в сферической системе координат зависят только от радиуса $f=f(r)$.
\qed\end{exa}
\begin{exa}
Если группа преобразований $(\MM,\MG)$ транзитивна, то $\MG$-инвариантные
функции на $\MM$ постоянны.
\qed\end{exa}

Рассмотрим более сложную ситуацию.
\begin{defn}
Пусть на многообразии $\MM$, на котором действует группа преобразований
$(\MM,\MG)$, задано поле $\vf=\lbrace\vf^i(x)\rbrace$, имеющее $N$ компонент,
$i=1,\dotsc,N$, которое преобразуются по некоторому представлению группы $\MG$.
Это значит, что группа преобразований действует не только на точки многообразия,
но одновременно и на компоненты $\vf^i$ по правилу
\begin{align}                                                     \nonumber
  x&\mapsto x'=xa,
\\                                                                \label{etruvf}
  \vf(x)&\mapsto \vf^{\prime i}(x'):=\vf^j(x)S_j{}^i(a),
\end{align}
где $S_j{}^i(a)$ -- матрица представления, соответствующая элементу группы
$a\in\MG$. Поле $\vf(x)$ называется $\MG$-инвариантным, если для его компонент
выполнено условие
\begin{equation}                                                  \label{eginvf}
  \vf^i(xa)=\vf^j(x)S_j{}^i(a). \qed
\end{equation}
\end{defn}
В отличие от правила преобразования поля (\ref{etruvf}) в левой части
этого равенства опущен штрих у поля, и поэтому условие $\MG$-инвариантности
представляет собой уравнение на поле $\vf^i(x)$. В случае, если
представление группы преобразований тривиально, $S_j{}^i(a)=\dl_j^i$ для всех
$a\in\MG$, мы имеем набор $\MG$-инвариантных скалярных полей (\ref{eginsc}).

Условие $\MG$-инвариантности можно переписать в эквивалентном виде
\begin{equation*}
  \vf^i(x)=\vf^j(xa^{-1})S_j{}^i(a).
\end{equation*}
\begin{exa}
Рассмотрим сферически инвариантные векторные поля в трехмерном евклидовом
пространстве $\MR^3$. Нетрудно проверить, что векторное поле с компонентами
\begin{equation*}
  \vf^i(x)=x^if(r),\qquad i=1,2,3,
\end{equation*}
заданными в декартовой системе координат $x^i$, где $f(r)$ -- произвольная
функция от радиуса, является сферически симметричным. Можно также доказать
обратное утверждение, что любое сферически симметричное векторное поле имеет
такой вид для некоторой функции $f(r)$. Таким образом, сферически симметричные
векторные поля на $\MR^3$ параметризуются одной функцией радиуса $f(r)$. В
разделе \ref{spheco} приведены также сферически симметричные тензоры второго
ранга.
\qed\end{exa}
\begin{com}
При исследовании моделей математической физики предположение о симметрии полей
позволяет существенно уменьшить число независимых переменных и число аргументов.
Это позволяет в ряде случаев найти явно точное решение уравнений
Эйлера--Лагранжа.
\qed\end{com}
\begin{com}
При рассмотрении групп преобразований мы предполагаем, что преобразования
многообразия $\MM$ являются глобальными, т.е.\ элемент группы $a\in\MG$ не
зависит от точки многообразия $x\in\MM$. При таких преобразованиях калибровочные
поля преобразуются, как тензоры. Поэтому условие $\MG$-инвариантности
(\ref{eginvf}) можно распространить и на калибровочные поля.
\qed\end{com}

Рассмотрим общую ситуацию. Пусть задана группа преобразований $(\MM,\MG)$.
Каждому элементу группы $a\in\MG$ соответствует диффеоморфизм
$\MM\xrightarrow{a}\MM$, который мы обозначим той же буквой, что и элемент
группы. У этого отображения существует дифференциал $a_*$ и возврат отображения
$a^*$. Возврат отображения обратим, т.к.\ $a$ -- диффеоморфизм. Пусть в точке
$x\in\MM$ задан тензор $T(x)\in\MT^r_{s,x}(\MM)$ типа $(r,s)$. Тогда его
можно перенести из точки $x$ в точку $xa$ с помощью дифференциала
отображения и возврата отображения, которые действуют соответственно на
контравариантные и ковариантные индексы, $T(xa)=(a_*)^r(a^{*-1})^s T(x)$.

В координатах это отображение записывается следующим образом. Пусть
$x=\lbrace x^\al\rbrace$ и $xa=\lbrace y^\al\rbrace$, $\al=1,\dotsc,\dim\MM$, --
координаты точек. Тогда
\begin{equation*}
  T_{\bt_1\dotsc\bt_s}{}^{\al_1\dotsc\al_r}(xa)=
  \frac{\pl x^{\g_1}}{\pl y^{\bt_1}}\dots\frac{\pl x^{\g_s}}{\pl y^{\bt_s}}
  T_{\g_1\dotsc\g_s}{}^{\dl_1\dotsc\dl_r}(x)
  \frac{\pl y^{\al_1}}{\pl x^{\dl_1}}\dots\frac{\pl y^{\al_r}}{\pl x^{\dl_r}},
\end{equation*}
где матрицы Якоби вычисляются в точке $x$.
\begin{defn}
Произвольное тензорное поле $T\in\CT^r_s(\MM)$, заданное на многообразии $\MM$,
называется {\em $\MG$-инвариантным}, если выполнено условие
\begin{equation}                                                  \label{einvte}
  T(xa):=(a_*)^r(a^{*-1})^s T(x)
\end{equation}
для всех $x\in\MM$ и $a\in\MG$.
\qed\end{defn}
\index{$\MG$-инвариантное тензорное поле ($\MG$-invariant tensor field)}%
\index{Тензорное поле $\MG$-инвариантное ($\MG$-invariant tensor field)}%
\begin{com}
Это определение применимо как для счетных групп, так и для групп Ли.
\qed\end{com}

Поскольку действие группы преобразований транзитивно на каждой орбите, то
тензорное поле на всей орбите однозначно задается его значением в некоторой
точке $x$ на данной орбите. Это не значит, что тензор в точке $x$ можно задать
произвольно, а затем разнести по орбите. Если группа изотропии $\MG_x$
нетривиальна, то в соответствии с определением (\ref{einvte}) тензор в точке
$x$ должен быть выбран симметричным относительно преобразований из $\MG_x$.

В случае, когда $\MG$ -- группа Ли, условие (\ref{einvte}) можно записать в
дифференциальной форме. Генераторы группы Ли $L_\Sa$, $\Sa=1,\dotsc,\dim\MG$, --
это базис левоинвариантных векторных полей на $\MG$. Им соответствуют векторные
поля $L^*_\Sa$ на многообразии $\MM$. Из определения производной Ли следует, что
если тензорное поле является $\MG$-инвариантным, то
\begin{equation*}
  \Lie_{L^*_\Sa}T=0
\end{equation*}
для всех $\Sa$ и $x\in\MM$. Это -- система дифференциальных уравнений в частных
производных первого порядка на компоненты $\MG$-инвариантного тензорного поля.
\begin{com}
В математической физике, как правило, ставится одна из двух задач. 1) По
заданному тензорному полю определить его группу симметрии и 2) найти тензорное
поле, обладающее определенной симметрией.
\qed\end{com}

В моделях гравитации важную роль играют однопараметрические группы
преобразований, которые оставляют инвариантной метрику, заданную на многообразии
$\MM$. Этим группам преобразований соответствуют векторные поля, которые
называются векторными полями Киллинга. Им будет посвящен отдельный раздел
\ref{skilve}.
\section{Отображения групп преобразований                        \label{seqyim}}
Пусть задано две группы преобразований $(\MM,\MG)$ и $(\MN,\MG)$ двух
многообразий $\MM$ и $\MN$ с одной и той же группой $\MG$. В частном случае
многообразия могут совпадать, $\MM=\MN$, но действия группы $\MG$ на $\MM$ можно
задавать различными способами.
\begin{exa}
Можно рассматривать группу вращений плоскости $\big(\MR^2,\MS\MO(2)\big)$
относительно произвольной точки $x\in\MR^2$.
\qed\end{exa}
Обозначим действие элемента группы $a\in\MG$ на $\MM$ и $\MN$ соответственно
через $a_\MM$ и $a_\MN$:
\begin{align*}
  a_\MM:&\qquad \MM\times\MG\ni\quad x,a\mapsto xa\quad\in\MM,
\\
  a_\MN:&\qquad \MN\times\MG\ni\quad y,a\mapsto ya\quad\in\MN.
\end{align*}
\begin{defn}
Отображение $f:~\MM\rightarrow\MN$ называется {\em эквивариантным}, если
диаграмма
\begin{equation*}
\begin{diagram}
  \MM\times\MG & \rTo^{f\times\id} & \MN\times\MG \\
  \dTo^{a_\MM} & & \dTo_{a_\MN} \\
  \MM & \rTo^{f} & \MN
\end{diagram},
\end{equation*}
где $\id$ -- тождественное отображение $\MG\rightarrow\MG$, коммутативна. То
есть
\begin{equation*}
  f\circ a_\MM=a_\MN\circ f\quad \text{или}\quad f(xa)=f(x)a
\end{equation*}
для всех $a\in\MG$ и $x\in\MM$.
\qed\end{defn}
\index{Эквивариантное отображение (equivariant map}%
Понятие эквивариантного отображения можно обобщить. Рассмотрим две группы
преобразований $(\MM,\MG)$ и $(\MN,\MH)$. Пусть задан гомоморфизм групп
$\rho:~\MG\rightarrow\MH$ и достаточно гладкое отображение многообразий
$f:~\MM\rightarrow\MN$.
\begin{defn}
Отображение $f\times\rho:~\MM\times\MG\rightarrow\MN\times\MH$ называется
{\em эквивариантным}, если диаграмма
\begin{equation*}
\begin{diagram}
  \MM\times\MG & \rTo^{f\times\rho} & \MN\times\MH \\
  \dTo^{a_\MM} & & \dTo_{\rho(a)_\MN} \\
  \MM & \rTo^{f} & \MN
\end{diagram}
\end{equation*}
коммутативна. То есть
\begin{equation*}
  f\circ a_\MM=\rho(a)_\MN\circ(f\times\rho)\quad \text{или}\quad f(xa)=f(x)\rho(a)
\end{equation*}
для всех $a\in\MG$ и $x\in\MM$.
\qed\end{defn}
\begin{exa}                                                       \label{esuspi}
Важным физическим приложением указанной конструкции является понятие спинора.
Рассмотрим две группы преобразований: группу вращений $\MS\MO(3)$ трехмерного
евклидова пространства $\MR^3$ и группу $\MS\MU(2)$ линейных преобразований
двумерного комплексного пространства $\MC^2$. Согласно теореме \ref{tsusot},
группа $\MS\MU(2)$ два раза накрывает группу вращений $\MS\MO(3)$. Обозначим
соответствующий гомоморфизм $\rho:~\MS\MU(2)\rightarrow\MS\MO(3)$. Тогда для
спиноров имеем коммутативную диаграмму
\begin{equation*}
\begin{diagram}
  \MC^2\times\MS\MU(2) & \rTo^{f\times\rho} & \MR^3\times\MS\MO(3) \\
  \dTo^{a_{\MC^2}} & & \dTo_{\rho(a)_{\MR^3}} \\
  \MC^2 & \rTo^{f} & \MR^3
\end{diagram},
\end{equation*}
где $a_{\MC^2}=U$ -- унитарная $2\times2$ матрица с единичным определителем
линейных преобразований двумерного комплексного пространства $\MC^2$, и
$\rho(a)_{\MR^3}$ -- соответствующая ортогональная $3\times3$ матрица вращений
трехмерного евклидова пространства $\MR^3$. Отображение $f$ задается формулой
$x^i=(\bar\psi\s^i\psi)$, где $\s^i$, $i=1,2,3$, -- матрицы Паули,
$\psi=\begin{pmatrix}\psi^1 \\ \psi^2\end{pmatrix}\in\MC^2$ -- столбец из двух
комплексных чисел (компоненты спинора) и $\bar\psi=(\psi^*_1,\psi^*_2)$ --
строка из двух комплексно сопряженных чисел (компоненты сопряженного спинора).
В физике принято называть матрицы $U$ -- спинорным представлением группы
вращений $\MS\MO(3)$. Строго говоря, соответствие
\begin{equation*}
  \MS\MO(3)\ni\quad \rho(a)\mapsto a\quad\in\MS\MU(2)
\end{equation*}
не является отображением и, следовательно, представлением, так как каждому
вращению $\rho(a)\in\MS\MO(3)$ соответствуют две унитарные матрицы
$\pm a\in\MS\MU(2)$, которые отличаются знаком.
Поэтому спинорное представление называют двузначным представлением.
\qed\end{exa}
\chapter{Гомотопии и фундаментальная группа                      \label{sfugro}}
Все многообразия локально устроены одинаково так же как и евклидовы
пространства. Однако их глобальная структура может существенно отличаться.
Описание глобальной структуры многообразий является сложной задачей, и для ее
решения применяются различные методы. Например, на каждом многообразии можно
задать некоторую алгебраическую структуру, которая является топологическим
инвариантом, т.е.\ сохраняется при диффеоморфизмах. В настоящем разделе мы
опишем один из простейших топологических инвариантов -- фундаментальную группу.
Этот инвариант довольно груб и не различает все топологически различные
многообразия. Однако он полезен в приложениях, т.к.\ если фундаментальная группа
двух многообразий различна, то можно с уверенностью сказать, что данные
многообразия недиффеоморфны.
\section{Пути                                                    \label{spaths}}
Важным методом алгебраического анализа глобальной структуры многообразий
является нахождение фундаментальной группы многообразия, которая строится
на основе понятия {\em пути} или {\em кривой}. Определение кривой в евклидовом
пространстве (\ref{ecurve}) дословно переносится на произвольное многообразие
$\MM$ размерности $n$.
\begin{defn}
{\em Путем} $\g$ называется кусочно дифференцируемое отображение единичного
отрезка $[0,1]\in\MR$ в $\MM$,
\begin{equation}                                                  \label{epathm}
  \g:\quad [0,1]\ni\quad t\mapsto x(t)\quad\in\MM,
\end{equation}
где $t$ -- вещественный параметр вдоль кривой. Говорят, что путь соединяет
{\em начало} пути  $x(0)$ и его {\em конец} $x(1)$. Если начало и конец пути
совпадают, $x(0)=x(1)$, то путь называют {\em замкнутым}. Замкнутый путь
называют также {\em петлей}. Мы будем также писать $\g=x(t)$.
\qed\end{defn}
\index{Замкнутый путь (closed curve)}\index{Путь замкнутый (closed curve)}%
\index{Начало пути (beginning of a curve)}%
\index{Конец пути (end of a curve)}%
\index{Петля (loop)}%
В координатах отображение (\ref{epathm}) задается набором $n$ кусочно
дифференцируемых функций $x(t)=\lbrace x^\al(t)\rbrace$, $\al=1,\dotsc,n$, одной
переменной $t\in[0,1]$.
\begin{com}
В общем случае кривая $x(t)$ может иметь точки самопересечения, и тогда
отображение $\g$ не является взаимно однозначным отображением отрезка $[0,1]$
на его образ $\g([0,1])$.
\qed\end{com}
\begin{com}
В алгебраической топологии, где в роли $\MM$ выступают топологические
пространства, от отображения (\ref{epathm}) требуется только непрерывность.
Для алгебраического анализа многообразий удобно рассматривать кусочно
дифференцируемые пути, т.е.\ функции $x^\al(t)$ считать кусочно
дифференцируемыми. Это удобно при умножении путей, которое будет введено позже.
\qed\end{com}

Замкнутую кривую в $\MM$ можно рассматривать, как непрерывный образ окружности,
т.к.\ при совпадении начала и конца пути можно отождествить концы единичного
интервала $t\in[0,1]$.
\begin{defn}
{\em Простой замкнутой} или {\em жордановой} кривой называется гомеоморфный
образ окружности, т.е.\ замкнутая кривая без точек самопересечения. В общем
случае простая замкнутая кривая может иметь изломы, т.е.\ от нее требуется
только непрерывность. Жорданова кривая на евклидовой плоскости, состоящая из
конечного числа прямолинейных отрезков, называется {\em жордановым}
многоугольником.
\qed\end{defn}
\index{Простая замкнутая кривая}\index{Кривая простая замкнутая}%
\index{Жорданова кривая (Jordan curve)}%
\index{Кривая жорданова (Jordan curve)}%
\index{Жорданов многоугольник (Jordan polygon)}%
\index{Многоугольник жорданов (Jordan polygon)}%
\begin{theorem}[\bf Жордан]
Пусть $\MS$ -- простая замкнутая кривая на евклидовой плоскости $\MR^2$. Тогда
$\MR^2\setminus\MS$ несвязно и состоит из двух компонент, общей границей которых
является $\MS$. В точности одна из этих компонент ограничена.
\end{theorem}
\begin{proof}
Несмотря на интуитивную очевидность утверждения теоремы, ее доказательство
сложно. Оно приведено, например, в \cite{Kosnio80R}.
\end{proof}

В определении пути (\ref{epathm}) область определения параметра вдоль пути
фиксирована единичным отрезком. Этого можно не делать, считая, что параметр
принимает значения на отрезке произвольной длины $[a,b]\subset\MR$.
В этом случае все пути можно разделить на классы эквивалентности. А именно,
два пути $x_1(t)$ и $x_2(s)$ называются {\em эквивалентными},
\index{Эквивалентный путь (equivalent path)}%
\index{Путь эквивалентный (equivalent path)}%
если они связаны гладкой монотонной заменой параметра $s=s(t)$:
$$
  x_1(t)=x_2\big(s(t)\big),\qquad \frac{ds}{dt}>0.
$$
где $s(a)=0$ и $s(b)=1$. В частном случае, возможно преобразование параметра $t$
внутри единичного отрезка, $a=0$, $b=1$. В дальнейшем под путем будем понимать
класс эквивалентных путей, определенных с точностью до гладкой репараметризации.
\begin{com}
Иногда понятие пути и кривой различают, понимая под путем отображение
(\ref{epathm}), а под кривой -- образ пути $\g([0,1])$ в многообразии $\MM$.
Как правило, мы этого не делаем и употребляем термины путь и кривая как
синонимы. При изучении фундаментальных групп отображение (\ref{epathm}) будем
называть путем, как это принято в алгебраической топологии.
\qed\end{com}
\begin{exa}
Простейшим примером пути является {\em тождественный} путь $e_x$ в некоторой
точке $x\in\MM$, определенный равенством
\begin{equation}                                                  \label{eidpat}
  e_x:\quad [0,1]\ni\quad t\mapsto x\quad\in\MM
\end{equation}
для всех $t$, где $x=\lbrace x^\al=\const\rbrace\in\MM$ -- произвольная
фиксированная точка многообразия.
\qed\end{exa}
\index{Тождественный путь}\index{Путь тождественный}%
\begin{exa}
В математической физике под многообразием $\MM$ часто понимается пространство, а
под вещественным параметром $t$ -- время. Тогда путь
$x(t)=\lbrace x^\mu(t)\rbrace$ -- это траектория точечной частицы.
\qed\end{exa}

При рассмотрении евклидова пространства $\MR^n$ как топологического пространства
в разделе \ref{seucto} было введено понятие связности. Введем новое понятие
связности, которое удобно, в частности, при рассмотрении фундаментальных групп.
\begin{defn}
Топологическое пространство $\MM$ называется {\em линейно связным}, если для
любых двух точек $x_1$, $x_2\in\MM$ найдется соединяющий их путь, целиком
лежащий в $\MM$. Топологическое пространство $\MM$ называется {\em локально
линейно связным}, если для любого $x\in\MM$ всякая открытая окрестность точки
$x$ содержит линейно связную окрестность этой точки.
\qed\end{defn}
\index{Топологическое пространство линейно связное %
(arcwise connected, path connected topological space)}%
\index{Линейно связное топологическое пространство %
(arcwise connected, path connected topological space)}%
\index{Топологическое пространство локально линейно связное %
(locally arcwise connected, path connected topological space)}%
\index{Локально линейно связное топологическое пространство %
(arcwise connected, path connected topological space)}%
Если какую либо фиксированную точку $x_0\in\MM$ можно соединить путем с любой
другой точкой $x_1\in\MM$, то этого достаточно для линейной связности
многообразия, потому что тогда две произвольные точки $x_1,x_2\in\MM$ можно соединить
путем, проходящим через $x_0$.

Между просто связностью и линейной связностью топологических прост\-ранств
имеется тесная связь, которая дается следующей теоремой.
\begin{theorem}                                                   \label{tlicon}
Всякое линейно связное топологическое пространство связно.
\end{theorem}
\begin{proof}
Зафиксируем точку $x_0\in\MM$. Тогда ее можно соединить некоторой кривой $\g_x$,
целиком лежащей в $\MM$, с произвольной точкой $x\in\MM$, потому что $\MM$ --
линейно связно. Каждая кривая $\g_x$ является связным подмножеством в $\MM$.
Тогда все пространство $\MM$ является объединением, $\MM=\bigcup_x\g_x$, и
связно, как объединение связных подмножеств, имеющих непустое пересечение
(теорема \ref{tconun}).
\end{proof}
Обратное утверждение теоремы \ref{tlicon} неверно. Приведем пример связного, но
не линейно связного топологического пространства.
\begin{exa}[\bf Блоха и гребенка]
Пусть подмножество комплексной плоскости $\MW\subset\MC$ является
объединением двух множеств $\MW=\MU\bigcup\MV$ (см.\ рис.~\ref{fblokh}), где
$$
\begin{array}{ccll}
  \MU &=& \left\lbrace i\right\rbrace  &\qquad \text{-- блоха} \\
  \MV &=& [0,1]\bigcup\left\lbrace 1/n+iy:\quad n\in\MN,~0\le y\le1\right\rbrace
  &\qquad \text{-- гребенка}.
\end{array}
$$
\begin{figure}[h,b,t]
\hfill\includegraphics[width=.25\textwidth]{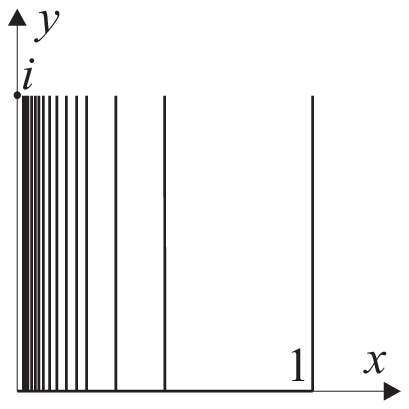}
\hfill {}
\centering\caption{Пример связного, но не линейно связного топологического
 пространства ``блоха и гребенка''.}
\label{fblokh}
\end{figure}
Будем считать, что топология $\MW$ индуцирована вложением в комплексную
плоскость $\MC\approx\MR^2$. Прежде всего заметим, что множество $\MV$ линейно
связно как объединение линейно связных множеств (теорема \ref{tconun}) и потому
связно. Пусть $\MU_\e$ -- $\e$-окрестность точки $i$, т.е.\ пересечение
множества $\left\lbrace z:\quad |z-i|<\e\right\rbrace $ с $\MW$. Подмножество $\MU$,
состоящее их одной точки, принадлежит замыканию $\overline{\MV}$, т.к.\ для
любого $\e>0$, пересечение $\MU_\e\bigcap\MV$ не пусто, поскольку для достаточно
больших $n$ точки $1/n+i\in\MV$ лежат в $\e$-окрестности точки $i$ (лемма
\ref{lpcous}). Отсюда вытекает, что связная компонента $\MV$ содержит
$\MU$ и поэтому совпадает со всем $\MW$.

Покажем теперь, что топологическое пространство $\MW$ не является линейно
связным, т.к.\ точку $i\in\MW$ нельзя соединить путем с любой другой точкой из
$\MW$. Предположим, что существует путь с началом в точке $\g(0)=i$ и концом в
некоторой точке $\g(1)\in\MV$. Не ограничивая общности можно считать, что
$\g(t)\ne i$ при $t>0$. Из непрерывности отображения $\g(t)$ следует, что для
любого положительного $\e$ существует такое положительное число $\dl$, что при
$t<\dl$ выполняется неравенство $|x(t)-i|<\e$, т.е.\ путь лежит в
$\e$-окрестности $\MU_\e$ точки $i$. Выберем конечное число $\e<1$ и некоторое
значение параметра $0<t_0<\dl$. По предположению, $x(t_0)\in\MV$. Окрестность
$\MU_\e$ многосвязна, т.к.\ состоит из самой точки $i$ и объединения
непересекающихся интервалов. Отрезок $[0,t_0]$ связен и, как показано в примере
\ref{eontwm}, его невозможно непрерывно отобразить на многосвязную область таким
образом, чтобы начало и конец отрезка лежали в разных компонентах.
\qed\end{exa}

В рассмотренном примере топологическое пространство $\MW$ не является
многообразием по двум причинам. Во-первых, окрестность точки $i$ нельзя
отобразить на интервал и, во вторых, множество $\MV$ имеет точки
самопересечения.

\begin{figure}[h,b,t]
\hfill\includegraphics[width=.35\textwidth]{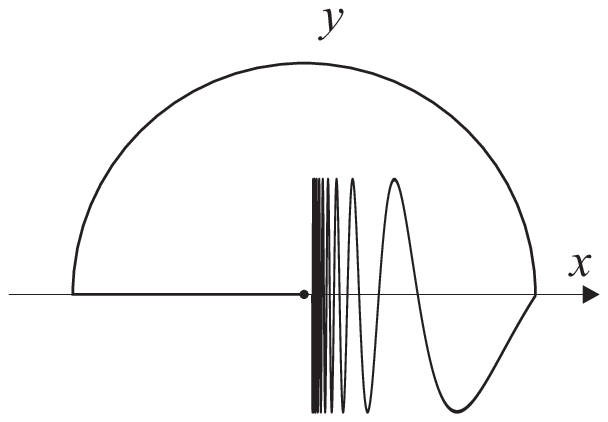}
\hfill {}
\centering\caption{Пример линейно связного, но не локально линейного связного
 топологического пространства.}
\label{fnoloc}
\end{figure}

Следующий пример показывает, что не всякое линейно связное топологическое
пространство является локально линейно связным.
\begin{exa}[\bf Польская окружность]
Рассмотрим подмножество на евклидовой плоскости $\MU\subset\MR^2$, которое
имеет вид $\MU=\MU_1\cup\MU_2\cup\MU_3$, где
\begin{align*}
  \MU_1&=\lbrace(x,y)\in\MR^2:\quad x^2+y^2=1,\quad y\ge0\rbrace,
\\
  \MU_2&=\lbrace(x,y)\in\MR^2:\quad -1\le x\le0,\quad y=0\rbrace,
\\
  \MU_3&=\lbrace(x,y)\in\MR^2:\quad 0<x\le1,\quad
  y=\dfrac{\raise-.7ex\hbox{$1$}}{\raise .3ex\hbox{$2$}}\sin
  \dfrac{\raise-.7ex\hbox{$\pi$}}{\raise .7ex\hbox{$x$}}\rbrace,
\end{align*}
с индуцированной топологией, см.рис.\ref{fnoloc}. Это пространство линейно
связно, но не локально линейно связно, т.к.\ начало координат $(0,0)$ не имеет
окрестности, содержащей линейно связную окрестность. Это топологическое
пространство не является многообразием, поскольку начало координат не имеет
окрестности, гомеоморфной прямой $\MR$.
\qed\end{exa}
Следующие свойства справедливы для произвольных топологических пространств.
\begin{prop}                                                      \label{tliceq}
Образ линейно связного многообразия при непрерывном отображении линейно связен.
\end{prop}
\begin{cor}
Если $\MM_1$ и $\MM_2$ два гомеоморфных топологических пространства, то $\MM_1$
линейно связно тогда и только тогда, когда $\MM_2$ линейно связно.
\qed\end{cor}
\begin{proof}
См., например, в \cite{Kosnio80R}.
\end{proof}
\begin{prop}                                                      \label{tlindp}
Многообразия $\MM_1$ и $\MM_2$ линейно связны тогда и только тогда, когда их
прямое произведение $\MM_1\times\MM_2$ линейно связно.
\end{prop}
\begin{proof}
См., например, в \cite{Kosnio80R}.
\end{proof}

Для многообразий понятие связности и линейной связности эквивалентны.
\begin{theorem}                                                   \label{tmalco}
Многообразие $\MM$ связно тогда и только тогда, когда $\MM$ -- линейно связно.
\end{theorem}
\begin{proof}
Достаточность является утверждением теоремы \ref{tlicon}.

Необходимость докажем от противного. Пусть $\MM$ связное, но не линейно связное
многообразие. Прежде всего заметим, что каждое многообразие можно покрыть
счетным числом карт. Каждая карта является линейно связным подмножеством, потому
что гомеоморфна $\MR^n$. Если две карты пересекаются, то их объединение также
линейно связно. Это значит, что многообразие можно представить в виде
объединения не более, чем счетного числа непересекающихся подмножеств,
$\MM=\bigcup_i\MU_i$, где каждое $\MU_i$ линейно связно и
$\MU_i\cap\MU_j=\emptyset$ при $i\ne j$. Поскольку $\MM$ связно, то среди
подмножеств $\MU_i$ существует по крайней мере одно подмножество, пусть это
будет $\MU_i$, которое содержит предельную точку некоторого другого подмножества
$\MU_j$. Пусть точка $x_0\in\MU_i$ является предельной для подмножества $\MU_j$.
Это значит, что в $\MU_j$ существует
последовательность точек $\lbrace x_k\rbrace\in\MU_j$, которая сходится к $x_0$.
Поэтому, начиная с некоторого $\Sn$, все точки последовательности $x_k$,
$k>\Sn$, лежат в окрестности $\MU_0\ni x_0$. Следовательно, точку $x_0$ можно
соединить кривой с $x_k$ и подмножества $\MU_i$ и $\MU_j$ -- линейно связны, что
противоречит предположению.
\end{proof}
\begin{exa}
Евклидово пространство $\MR^n$ линейно связно, т.к.\ любые две точки
$x_0,x_1\in\MR^n$ можно соединить путем $x(t)=x_0+(x_1-x_0)t$, который
представляет собой прямолинейный отрезок.
\qed\end{exa}
Нетрудно также доказать
\begin{prop}
Каждое многообразие $\MM$ локально линейно связно.
\end{prop}
В общем случае многосвязное многообразие представляет собой набор компонент
связности, каждая из которых является линейно связной.

Для полноты картины приведем достаточное условие линейной связности
произвольного топологического пространства.
\begin{prop}
Если топологическое пространство $\MM$ связно и локально линейно связно, то оно
линейно связно.
\end{prop}
\begin{proof}
См., например, \cite{Kosnio80R}.
\end{proof}
\section{Гомотопия непрерывных отображений                       \label{shomot}}
На множестве непрерывных отображений двух топологических пространств
$f:~\MM_1\rightarrow\MM_2$ и, в частности, многообразий можно ввести отношение
эквивалентности -- гомотопию. Понятие гомотопии является топологическим, потому
что при его определении используется только непрерывность отображений.
\begin{defn}
Два непрерывных отображения
\begin{equation*}
\begin{split}
  f_0:&\qquad \MM_1\ni\quad x\mapsto f_0(x)\quad\in\MM_2,
\\
  f_1:&\qquad \MM_1\ni\quad x\mapsto f_1(x)\quad\in\MM_2,
\end{split}
\end{equation*}
называются {\em гомотопными}, если существует непрерывное отображение
$$
  F:\quad \MM_1\times[0,1]\ni\quad x,t\mapsto F(x,t)\quad\in\MM_2
$$
такое, что $F(x,0)=f_0(x)$ и $F(x,1)=f_1(x)$. При этом само отображение
$F$ называется {\em гомотопией} между $f_0$ и $f_1$:
$$
  F:\quad f_0\sim f_1.
$$
Гомотопия $F(x,t)$ называется также {\em деформацией} отображения $f_0$ в $f_1$.
\qed\end{defn}
\index{Гомотопные отображения (homotopic maps)}%
\index{Отображения гомотопные (homotopic maps)}%
\index{Гомотопия (homotopy)}%
\index{Деформация отображения (deformation of a map)}%
\index{Отображения деформация (deformation of a map)}%

Другими словами, гомотопия позволяет непрерывно деформировать одно отображение в
другое. Нетрудно проверить, что гомотопия является отношением эквивалентности на
множестве всех отображений $\MM_1\rightarrow\MM_2$. Отображения, гомотопные
отображению $f$, образуют гомотопический класс отображений, который будем
обозначать квадратными скобками $[f]$. Множество всех гомотопических классов
отображений $\MM_1\rightarrow\MM_2$ будем обозначать через $[\MM_1;\MM_2]$.

Гомотопию $F(x,t)$ можно представить себе следующим образом. Зафиксируем точку
$x\in\MM_1$ и будем менять параметр $t\in[0,1]$. Тогда мы получим траекторию
точки $x$ в пространстве $\MM_2$. Тогда гомотопия $F(x,t)$ представляет собой
семейство всех траекторий, зависящих от параметра $x\in\MM_1$.
\begin{exa}                                                       \label{ehocos}
Любой путь $\g:~[0,1]\ni s\mapsto x(s)\in\MM$ гомотопен постоянному пути
$e_{x_0}$ с началом и концом в точке $x_0:=x(0)$ посредством гомотопии
$$
  F:\quad [0,1]\times[0,1]\ni\quad s,t\mapsto F(s,t)\quad\in\MM,\qquad
  \text{где}~F(s,t):=x\big((1-t)s\big).
$$
В этой гомотопии конец пути меняется: он скользит вдоль кривой $\g(s)$,
приближаясь к началу пути при $t\rightarrow1$. При данном выше определении
гомотопии, когда отображения $f_0$ и $f_1$ не обязаны совпадать ни в одной
точке, все пути на связном многообразии гомотопны между собой.
\qed\end{exa}

Понятие гомотопных отображений можно использовать для определения отношения
эквивалентности самих топологических пространств.
\begin{defn}
Говорят, что пространства $\MM_1$ и $\MM_2$ имеют один и тот же
{\em гомотопический тип}, если существуют непрерывные отображения
$f:~\MM_1\rightarrow\MM_2$ и $g:~\MM_2\rightarrow\MM_1$ такие, что композиции
этих отображений гомотопны тождественным отображениям:
$$
  g\circ f\sim\id_1:\quad \MM_1\rightarrow\MM_1,\qquad
  f\circ g\sim\id_2:\quad \MM_2\rightarrow\MM_2,
$$
где $\id_{1,2}$ -- тождественные отображения пространств $\MM_1$ и $\MM_2$.
Отображения $f$ и $g$ называются в этом случае {\em гомотопическими
эквивалентностями}. Говорят также, что пространства $\MM_1$ и $\MM_2$ в этом
случае {\em гомотопически эквивалентны}.
\qed\end{defn}
\index{Гомотопический тип (homotopic type)}%
\index{Тип гомотопический (homotopic type)}%
\index{Гомотопическая эквивалентность (homotopic equivalence)}%
\index{Эквивалентность гомотопическая (homotopic equivalence)}%
Это определение отличается от определения гомеоморфизма тем, что вместо знака
равенства стоит знак гомотопической эквивалентности. Поэтому два гомеоморфных
пространства имеют одинаковый гомотопический тип. Обратное утверждение неверно.
\begin{exa}
$n$-мерный шар $\MB^n\in\MR^n$ гомотопически эквивалентен точке $x\in\MB^n$, но
не гомеоморфен ей.
\qed\end{exa}
\begin{defn}
В общем случае топологическое пространство, гомотопически эквивалентное точке,
называется {\em стягиваемым}.
\qed\end{defn}
\index{Стягиваемое пространство (contractible space)}%
\index{Пространство стягиваемое (contractible space)}%
Другими словами, пространство стягиваемо, если его можно непрерывно
деформировать по себе в точку.
\begin{exa}
Нестягиваемыми пространствами являются $n$-мерная сфера $\MS^n$ и тор $\MT^n$.
\qed\end{exa}
\begin{exa}
Евклидово пространство $\MR^n$, а также любая выпуклая область $\MU\subset\MR^n$
гомотопически эквивалентна одной точке $x_0\in\MU$. Напомним, что область
$\MU\subset\MR^n$ называется {\em выпуклой}, если наряду с произвольными точками
$x_1,x_2\in\MU$ она содержит также и отрезок $x_1+(x_2-x_1)t$, их соединяющий.
\qed\end{exa}
\index{Выпуклая область (convex domain)}%
\index{Область выпуклая (convex domain)}%
\begin{exa}
Евклидово пространство с выколотой точкой, $\MR^n\setminus x_0$, гомотопически
эквивалентно сфере $\MS^{n-1}$, содержащей точку $x_0$ внутри. В частности,
плоскость с выколотой точкой гомотопически эквивалентна окружности.
\qed\end{exa}
\begin{exa}
Плоскость с двумя выколотыми точками $x_1,x_2\in\MR^2$ гомотопически
эквивалентна {\em букету окружностей}, изображенному на рис.~\ref{fcirun}.
Букет окружностей представляет собой объединение двух окружностей, имеющих одну
общую точку. Он является топологическим пространством, но не многообразием.
\index{Букет окружностей (union of circles)}%
\begin{figure}[h,b,t]
\hfill\includegraphics[width=.3\textwidth]{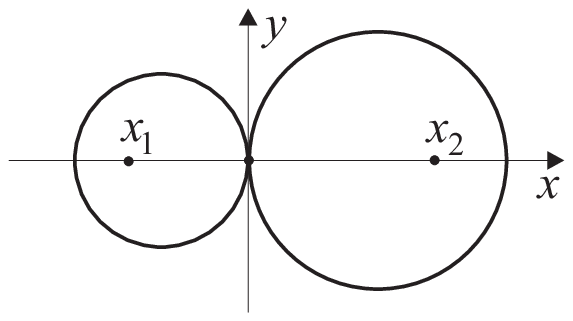}
\hfill {}
\centering\caption{Букет двух окружностей.}
\label{fcirun}
\end{figure}
\qed\end{exa}
\begin{exa}
Гомотопически эквивалентными пространствами является тройка: цилиндр, лист
Мебиуса и окружность.
\qed\end{exa}
Из этих примеров видно, что гомотопически эквивалентные многообразия могут иметь
разную размерность и ориентируемость. Многообразия могут быть также
гомотопически эквивалентны топологическим пространствам, которые не являются
многообразиями, как показывает пример плоскости с двумя выколотыми точками.
Тем не менее некоторые топологические свойства являются общими для гомотопически
эквивалентных пространств.
\begin{prop}
Если пространство $\MM_1$ связно и имеет тот же гомотопический тип, что и
$\MM_2$, то пространство $\MM_2$ также связно.
\end{prop}
Понятие гомотопии может быть использовано для доказательства нетривиальных
утверждений.
\begin{theorem}[\bf О невозможности причесать ежа]                \label{tvecsp}
На четномерной сфере $\MS^n$ нельзя задать непрерывное векторное поле, которое
не обращается в нуль ни в одной точке.
\end{theorem}
\index{Теорема о невозможности причесать ежа}%
\begin{proof}
От противного. Рассмотрим сферу единичного радиуса
\begin{equation*}
  \MS^n:=\lbrace x\in\MR^{n+1}:\quad x^\al x_\al=1,\quad \al=1,\dotsc,n+1\rbrace,
\end{equation*}
вложенную в евклидово пространство $\MR^{n+1}$. Допустим, что на сфере задано
непрерывное векторное поле $X$, не имеющее нулей. Тогда векторное поле
$T:=X/\sqrt{X^2}$ имеет единичную длину и поэтому задает непрерывное отображение
сферы на себя $\MS^n\rightarrow\MS^n$. Касательный вектор к сфере
$T=\lbrace T^\al(x)\rbrace$ ортогонален радиус-вектору $\lbrace x^\al\rbrace$,
т.е.\ $x^\al T_\al=0$. При этом мы рассматриваем координаты точки на сфере и
касательный вектор как векторы евклидова пространства $\MR^{n+1}$. Тогда
отображение
\begin{equation*}
  F(x,t):\qquad x^\al,t\mapsto x^\al\cos(\pi t)+T^\al\sin(\pi t)
\end{equation*}
задает гомотопию
\begin{equation*}
  F:\qquad \MS^n\times[0,1]\rightarrow\MS^n.
\end{equation*}
Эта гомотопия переводит произвольную точку на сфере $\lbrace x^\al\rbrace$ в
ее антиподальную точку с координатами $\lbrace -x^\al\rbrace$.  Выберем
ориентацию на сфере, согласованную с канонической ориентацией евклидова
пространства (\ref{ecaore}). Тогда объем сферы равен
\begin{equation}                                                  \label{evolsp}
  V=\int_{\MS^n}\!\!\!\upsilon\ne 0,
\end{equation}
где форма объема на сфере $\upsilon$ получена сужением канонической формы
объема евклидова пространства $\upsilon_0$ на сферу $\MS^n$:
\begin{equation*}
  \upsilon=\inm_N\upsilon_0,\qquad
  \upsilon_0:=dx^1\wedge dx^2\wedge\dotsc\wedge dx^{n+1}.
\end{equation*}
Здесь $\inm_N$ -- внутреннее умножение формы объема $\upsilon_0$ на внешнюю
нормаль к сфере $N$. Интеграл (\ref{evolsp}) задает непрерывное отображение
$\Lm_{n+1}(\MR^{n+1})\rightarrow\MR$. При отображении $x^\al\rightarrow-x^\al$
объем сферы преобразуется по правилу
\begin{equation*}
  V\mapsto(-1)^{n+1}V.
\end{equation*}
Поскольку изменение знака объема при гомотопии невозможно, т.к.\ она
является непрерывным отображением, то для четных $n$ приходим к противоречию.
\end{proof}
\begin{com}
На нечетномерной сфере непрерывные векторные поля, нигде не обращающиеся в нуль,
существуют. А именно, можно доказать \cite{Adams62}, что максимальное число
линейно независимых нигде не обращающихся в нуль векторных полей на
нечетномерной сфере $\MS^{n-1}$ равно $2^c+8d-1$, где $c$ и $d$ --
неотрицательные целые числа, которые определяются следующим образом. Поскольку
$n$ четно, то его единственным образом можно представить в виде $n=(2a-1)2^b$,
где $a,b$  -- натуральные числа. Тогда $c=b\mod 4$ и $d=(b-c)/4$.
\qed\end{com}

Рассмотренный выше пример \ref{ehocos} гомотопных путей показывает, что
определение гомотопии является грубым, т.к.\ гомотопические классы путей
содержат слишком много элементов. Более тонкое деление на классы дает понятие
отображений, гомотопных относительно некоторого подмножества $\MU\subset\MM_1$.
\begin{defn}
Два непрерывных отображения топологических пространств
$$
  f_0,f_1:\quad \MM_1\rightarrow\MM_2
$$
называются {\em гомотопными относительно подмножества $\MU\subset\MM_1$},
если существует гомотопия
$$
  F:\quad \MM_1\times[0,1]\rightarrow\MM_2,
$$
такая, что $F(x,t)$ не зависит от $t$ при $x\in\MU$. Мы пишем
$f_0\sim f_1(\rel\MU)$.
\qed\end{defn}
\index{Отображения, гомотопные относительно подмножества}%
В частности, отображения $f_0$ и $f_1$ должны совпадать на подмножестве $\MU$.
Другими словами, при непрерывной деформации отображений допускается их изменение
только на множестве $\MM\setminus\MU$. Соответствующая гомотопия называется
{\em относительной гомотопией} и также определяет отношение эквивалентности на
множестве непрерывных отображений топологических пространств, которое является
более тонким.
\index{Гомотопия относительная (relative homotopy)}%
\index{Относительная гомотопия (relative homotopy)}%
\begin{exa}
Можно рассмотреть класс путей, гомотопных относительно начала $x(0)$ и конца
$x(1)$ некоторого пути. Этот класс содержит все пути, которые имеют
фиксированное начало и конец и которые можно непрерывно деформировать друг в
друга по многообразию $\MM$. Более того, как показывает пример тора, не все пути
с одинаковым началом и концом можно непрерывно деформировать друг в друга, не
выходя из многообразия $\MM$. Если начало и конец пути отличаются, то пути
относительно негомотопны тождественному пути. Это говорит о том, что понятие
относительной гомотопии более тонко, чем просто гомотопия.
\qed\end{exa}
Понятие относительной гомотопии для путей будет использовано при определении
фундаментальной группы многообразия.
\section{Фундаментальная группа                                  \label{sfundg}}
Существует два простых, но важных способа получения новых путей из старых.
А именно, можно определить произведение путей, когда сначала проходится путь
$\g_1$, а затем -- путь $\g_2$, если такое возможно. Кроме того, можно
определить {\em обратный} путь $\g^{-1}$, который проходится в обратном
направлении. Сформулируем это в виде следующего утверждения.
\index{Обратный путь (inverse path)}%
\index{Путь обратный (inverse path)}%
\begin{prop}
1) Если начало второго пути $\g_2=x_2(t)$ совпадает с концом первого пути
$\g_1=x_1(t)$, то можно определить произведение путей $\g_2\circ\g_1$ следующей
формулой:
\begin{equation}                                                  \label{epamul}
  (\g_2\circ\g_1)=x(t):=\left\lbrace
  \begin{array}{llc}
  x_1(2t)   & \text{при} & 0<t<\frac12,\\
  x_2(2t-1) & \text{при} & \frac12<t<1,
  \end{array}\right.
\end{equation}
которое является путем в $\MM$.

2)
Для любого пути $\g$ можно определить путь $\g^{-1} $, получаемый как путь
$\g$, проходимый в обратном направлении:
\begin{equation}                                                  \label{epathi}
  \g^{-1}:=x(1-t).
\end{equation}
\end{prop}

Эти операции похожи, но не являются групповыми операциями на множестве всех
путей. Действительно, умножение путей определено только в том случае, если конец
первого пути совпадает с началом второго. Кроме того, произведения
$\g^{-1}\circ\g$ и $\g\circ\g^{-1}$ дают не тождественные, а замкнутые пути в
началом и концом в точках $x(0)$ и $x(1)$ соответственно.

Разобьем все множество путей $\g$ на классы эквивалентных путей $[\g]$, которые
гомотопны относительно начала и конца пути. То есть каждый элемент $[\g]$
содержит все пути, которые можно непрерывно деформировать друг в друга при
фиксированном начале и конце путей: $\g_1\in[\g]$ тогда и только тогда, когда
$\g_1\sim\g(\rel\lbrace 0,1\rbrace)$. В классах эквивалентных путей также можно
ввести умножение путей и обратный элемент, выбрав из каждого класса по
представителю и воспользовавшись формулами (\ref{epamul}) и (\ref{epathi}).
Нетрудно показать, что эти операции на множестве относительно гомотопных путей
корректно определены, т.к.\ не зависят от выбора представителя в каждом классе.
В этих классах произведения $[\g^{-1}]\circ[\g]=[e_{x(0)}]$ и
$[\g]\circ[\g^{-1}]=[e_{x(1)}]$ дают классы путей, гомотопных тождественным
путям. Однако умножение определено не для всех классов путей.

Ситуацию можно исправить и превратить некоторое подмножество всех путей на
$\MM$ в группу, если зафиксировать точку многообразия и ограничиться
замкнутыми путями, имеющими начало и конец в данной точке.

Зафиксируем произвольную точку многообразия $x_0\in\MM$ и рассмотрим
все замкнутые пути, имеющие начало и конец в данной точке. Множество этих путей
обозначим через $\Om(\MM,x_0)$. При этом допускается, чтобы пути имели точки
самопересечения. Поскольку начала и концы всех путей совпадают, то на этом
множестве определена операция умножения (\ref{epamul}). Для каждого пути
определен также обратный путь (\ref{epathi}). Множество $\Om(\MM,x_0)$ с
введенной операцией умножения и обратного элемента группы не образует, потому
что произведения двух различных путей на их обратные не совпадают, и,
следовательно, невозможно ввести понятие единственного единичного элемента. Эту
трудность можно обойти, если разбить множество всех путей $\Om(\MM,x_0)$ на
классы путей, гомотопных относительно начала и конца, которые обозначим через
$\pi(\MM,x_0)$. Справедлива следующая
\begin{theorem}                                                   \label{thomot}
Гомотопические классы путей $\pi(\MM,x_0)$ образуют группу относительно операции
умножения путей, причем обратным элементом является гомотопический класс
обратного пути, а единицей -- гомотопический класс единичного пути $e_{x_0}$.
\end{theorem}
\begin{proof}
Сводится к проверке корректности операций и групповых свойств.
Детали можно найти, например, в \cite{Kosnio80R}.
\end{proof}
Таким образом, для определения группы было сделано два шага. Во-первых, чтобы
определить операцию умножения на всех элементах, множество всех путей было
ограничено только замкнутыми путями, имеющими начало и конец в фиксированной
точке многообразия. Во-вторых, для корректного определения единичного элемента
мы перешли от умножения отдельных путей к умножению классов относительно
гомотопных путей.
\begin{defn}
Группа $\pi(\MM,x_0)$ называется {\em фундаментальной группой} топологического
пространства или многообразия $\MM$ в точке $x_0$.
\qed\end{defn}
\index{Фундаментальная группа (fundamental group)}%
\index{Группа фундаментальная (fundamental group)}%

В этом определении используется фиксированная точка $x_0$. Для
связных многообразий роль фиксированной точки не является существенной.
\begin{theorem}                                                   \label{tfipho}
Пусть $\MM$ -- линейно связное топологическое пространство (многообразие). Тогда
для любых точек $x_0,x_1\in\MM$ фундаментальные группы
$\pi(\MM,x_0)$ и $\pi(\MM,x_1)$ изоморфны.
\end{theorem}
\begin{proof}
Пусть $\dl$ -- путь из $x_0$ в $x_1$ и $\g$ -- произвольный замкнутый путь в
точке $x_0$. Тогда $\dl\circ\g\circ\dl^{-1}$ -- замкнутый путь в точке $x_1$.
Поэтому между множествами замкнутых путей $\Om(\MM,x_0)$ и $\Om(\MM,x_1)$
устанавливается взаимно однозначное соответствие. Отсюда следует, что между
фундаментальными группами существует биекция
$f_\dl:~\pi(\MM,x_0)\rightarrow\pi(\MM,x_1)$. Эта биекция -- гомоморфизм групп,
т.к.
\begin{equation*}
  f_\dl([\g_2]\circ[\g_1])=[\dl]\circ[\g_2]\circ[\g_1]\circ\dl^{-1}
  =[\dl]\circ[\g_2]\circ\dl^{-1}\circ\dl\circ[\g_1]\circ\dl^{-1}
  =f_\dl([\g_2])\circ f_\dl([\g_2]).
\end{equation*}
Обратное отображение $f_\dl^{-1}$ определяется обратным путем $\dl^{-1}$.
\end{proof}
В силу этой теоремы при обозначении фундаментальной группы фиксированную точку
$x_0$ мы часто будем опускать: $\pi(\MM,x_0)=\pi(\MM)$. Изоморфизм между
фундаментальными группами $f_\dl$ определяется гомотопическим классом пути
$[\dl]$ из $x_0$ в $x_1$. Поскольку нет естественного (предпочтительного) выбора
такого класса путей, то естественного изоморфизма между фундаментальными
группами в различных точках не существует.
\begin{exa}
В евклидовом пространстве $\MR^n$ при любом $n$ произвольный замкнутый путь для
любой отмеченной точки гомотопен тождественному пути в этой точке. Поэтому
фундаментальная группа евклидова пространства тривиальна $\pi(\MR^n)=e$.
\qed\end{exa}
\begin{defn}
Топологическое пространство (многообразие) $\MM$ называется {\em односвязным},
если его фундаментальная группа тривиальна, $\pi(\MM)=e$.
\qed\end{defn}
\index{Односвязное многообразие}\index{Многообразие односвязное}%
\begin{prop}
Если $\MM_1$ и $\MM_2$ -- односвязные многообразия, то их прямое
произведение $\MM_1\times\MM_2$ односвязно.
\end{prop}
\begin{proof}
Простая проверка.
\end{proof}
\begin{exa}
Фундаментальная группа окружности с отмеченной точкой изоморфна группе целых
чисел по сложению, $\pi(\MS^1,x_0)\simeq\MZ$. При этом целое число равно
количеству полных обходов окружности, причем положительные и отрицательные числа
соответствуют обходам против и по часовой стрелке соответственно. Хотя
изоморфизм фундаментальной группы окружности группе целых чисел интуитивно
очевиден, доказательство громоздко. Его можно найти, например, в
\cite{Kosnio80R}.
\qed\end{exa}
\begin{exa}
Фундаментальная группа цилиндра совпадает с фундаментальной группой окружности и
изоморфна группе целых чисел по сложению $\MZ$.
\qed\end{exa}
В общем случае вычисление фундаментальной группы является сложной задачей.
\begin{prop}
Фундаментальная группа группы Ли $\MU(n)$ при $N\ge1$ совпадает с группой целых
чисел по сложению $\MZ$. Группа специальных унитарных матриц $\MS\MU(n)$ при
$n\ge1$ односвязна. Фундаментальная группа группы вращений $\MS\MO(n)$ при
$n\ge3$ равна $\MZ_2$. Фундаментальная группа группы вращений плоскости
$\MS\MO(2)$ равна $\MZ$.
\end{prop}
\begin{proof}
См., например, \cite{Postni82R}.
\end{proof}
В общем случае фундаментальная группа не является абелевой.
\begin{exa}
Рассмотрим евклидову плоскость $\MR^2$ с двумя выколотыми точками $x_1$ и $x_2$
и фиксированной точкой $x_0$ (см.\ рис.\ref{fundnoncom}).
\begin{figure}[h,b,t]
\hfill\includegraphics[width=.95\textwidth]{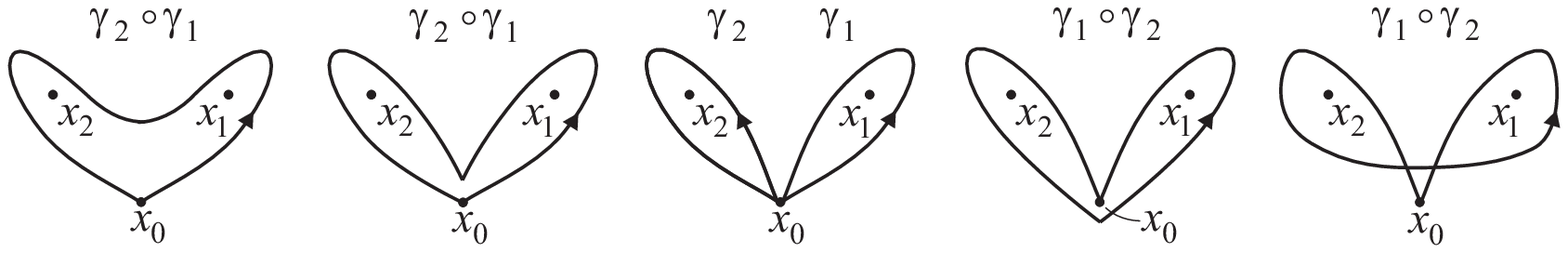}
\hfill {}
\centering\caption{Замкнутые пути $\g_1$ и $\g_2$ вокруг точек $x_1$ и $x_2$ на
евклидовой плоскости (в центре). Непрерывная деформация произведения путей
$[\g_2]\circ[\g_1]$ (влево) и $[\g_1]\circ[\g_2]$ (вправо).}
\label{fundnoncom}
\end{figure}
Пусть $\g_1$ и $\g_2$ -- замкнутые пути с началом и концом в точке $x_0$,
которые охватывают соответственно точки $x_1$ и $x_2$. На
рис.\ref{fundnoncom} показаны представители из классов гомотопных
путей $[\g_2]\circ[\g_1]$ и $[\g_1]\circ[\g_2]$. Ясно, что эти пути не могут
быть непрерывно деформированы друг в друга без пересечения выколотых точек и
поэтому негомотопны друг другу. Следовательно, фундаментальная группа евклидовой
плоскости с двумя выколотыми точками некоммутативна.
\qed\end{exa} В общем случае фундаментальная группа многообразия представляет
собой свободную группу с конечным числом образующих. Напомним
\begin{defn}
Группа $\MG$ называется {\em свободной} группой с системой $E\subset\MG$
порождающих (образующих) элементов, если любое отображение множества $E$ в любую
группу $\MH$ продолжается до гомоморфизма $\MG\to\MH$. Система $E\subset\MG$
называется {\em системой свободных порождающих}. Ее мощность называется
называется {\em рангом свободной группы} $\MG$. Множество $E$ называется также
{\em алфавитом}. Элементы $g\in\MG$ представляют собой {\em слова} в алфавите
$E$, т.е.\ выражения в виде конечного произведения
\begin{equation}                                                  \label{ewordd}
  g=e_{i_1}^{\e_1}e_{i_2}^{\e_2}\dotsc e_{i_k}^{\e_k},\qquad e_{i_j}\in E,\quad
  \e_j=\pm1,\quad \forall j=1,\dotsc,k,
\end{equation}
а также пустое слово. Слово называется {\em несократимым}, если
\begin{equation*}
  e_{i_j}^{\e_j}\ne e_{i_{j+1}}^{-\e_{j+1}},\qquad \forall j=1,\dotsc,k-1.
\end{equation*}
Несократимые слова являются разными элементами свободной группы $\MG$, и каждое
слово равно единственному несократимому слову. Число $k$ называется {\em длиной}
слова $g$, если оно несократимо.
\qed\end{defn}
\index{Свободная группа}%
\index{Порождающие свободной группы}%
\index{Ранг свободной группы}%
\index{Алфавит (alphabet)}%
\index{Слово (word)}%
\index{Несократимое слово (word)}%
\index{Длина слова (word length)}%
\begin{exa}
Рассмотрим расширенную комплексную плоскость $\overline\MC$ (сферу Римана).
\begin{defn}
Область $\MD\subset\overline\MC$ называется {\em $m$-связной}, если ее граница
состоит из $m$ связных компонент.
\qed\end{defn}
\index{$m$-связная область}%
Любая $m$-связная область диффеоморфна комплексной плоскости $\MC$ с $m-1$
дырками.
\begin{prop}
Фундаментальная группа произвольной $m$-связной области в расширенной
комплексной плоскости $\overline\MC$ является свободная группа с $m-1$
образующими.
\end{prop}
\begin{proof}
Каждой связной компоненте границы соответствует ровно одна компонента дополнения
$\overline\MC\setminus\MD$. Обозначим эти компоненты
$\MB_0,\MB_1,\dotsc,\MB_{m-1}$ (см.\ рис.\ref{fobrazuyushchie}).
\begin{figure}[h,b,t]
\hfill\includegraphics[width=.3\textwidth]{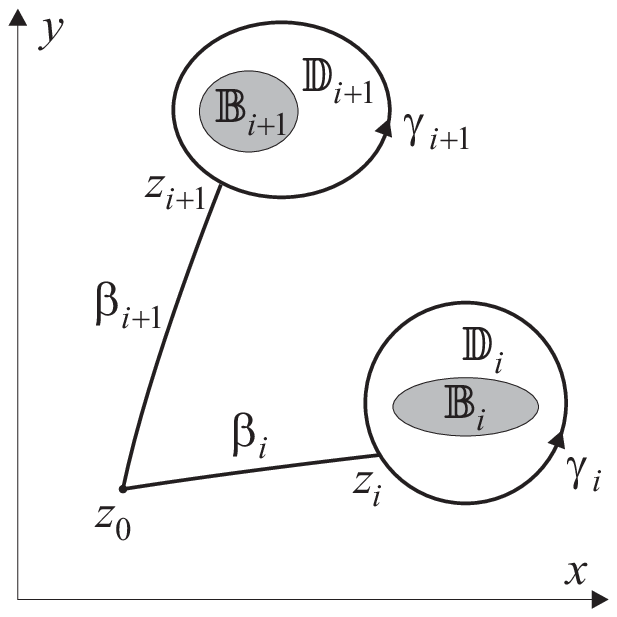}
\hfill {}
\\
\centering \caption{Образующие $m$-связной области в расширенной комплексной
плоскости.\label{fobrazuyushchie}}
\end{figure}
Заключим каждую компоненту $\MB_i$ в некоторую
окрестность $\MD_i$, которая ограничена простой замкнутой кривой $\g_i$, целиком
лежащей в $\MD$. Области $\MD_i$ выберем такими, чтобы кривые $\g_i$ не
пересекались. На каждой кривой выберем некоторую точку $z_i$, которую будем
считать началом и концом кривой $\g_i$. Из фиксированной точки $z_0$ проведем в
каждую точку $z_i$ кривую $\bt_i$, целиком лежащую в области $\MD$. Обозначим
\begin{equation*}
  e_i:=\bt_i^{-1}\circ\g_i\circ\bt_i,\qquad i=0,1,\dotsc,m-1.
\end{equation*}
Заметим, что справедливо тождество
\begin{equation*}
  [e_0]\circ[e_1]\circ\dotsc\circ[e_{m-1}]=1.
\end{equation*}
Поэтому гомотопические классы $[e_1],\dotsc,[e_{m-1}]$ можно выбрать в качестве
образующих фундаментальной группы $\pi(\MD,z_0)$. Можно проверить, что любой
путь $\g$ гомотопен конечному произведению:
\begin{equation*}
  [\g]=[\g_{i_1}^{\e_1}]\circ[\g_{i_2}^{\e_2}]\circ\dotsc\circ[e_{i_k}^{\e_k}],
\end{equation*}
где все $i_j$, $j=1,\dotsc,k$, принадлежат множеству
$\lbrace1,\dotsc,m-1\rbrace$. При этом данное представление единственно, если
оно несократимо. \qed
\end{proof}
\end{exa}

Нетрудно доказать, что при непрерывном отображении $f:~\MM_1\rightarrow\MM_2$
гомотопически эквивалентные пути $\g_1\sim\g_2$ переходят в гомотопически
эквивалентные пути $f\g_1\sim f\g_2$  и что замкнутые пути переходят в
замкнутые. При этом между фундаментальными группами возникает связь.
\begin{theorem}                                                   \label{thomap}
При непрерывном отображении $f:~\MM_1\rightarrow\MM_2$ фундаментальные группы
испытывают гомоморфизм
$$
  f_*:\quad \pi(\MM_1,x)\rightarrow\pi\big(\MM_2,f(x)\big),
$$
где $x$ -- произвольная точка из $\MM_1$, который называется индуцированным и
одинаков для всех отображений, гомотопных относительно точки $x$.
\end{theorem}
\begin{proof}
См., например, \cite{Kosnio80R}.
\end{proof}
\index{Индуцированный гомоморфизм (induced homomorphism)}%
\index{Гомоморфизм индуцированный (induced homomorphism)}%
\begin{com}
Если отображение двух пространств является гомеоморфизмом, то гомоморфизм
фундаментальных групп становится изоморфизмом.
\qed\end{com}

Понятие фундаментальной группы позволяет осуществить переход от топологии
пространств к алгебре и является одним из разделов алгебраической топологии.
Этот переход позволяет использовать алгебраические методы при изучении
топологических пространств и, следовательно, многообразий. При этом 1) каждому
топологическому пространству с отмеченной точкой ставится в соответствие
фундаментальная группа, 2) каждому непрерывному отображению пространств ставится
в соответствие гомоморфизм групп, 3) тождественному отображению отвечает
тождественный гомоморфизм, 4) гомеоморфизму отвечает изоморфизм, 5) композиции
непрерывных отображений сопоставляется композиция гомоморфизмов.
\begin{com}
Отмеченные выше свойства 1) -- 5) дают пример функтора из категории
топологических пространств с отмеченной точкой и непрерывными отображениями,
сохраняющими отмеченную точку, в категорию групп и их гомоморфизмов.
\qed\end{com}

Рассмотренные примеры окружности и цилиндра показывают, что изоморфизм
фундаментальных групп не означает гомеоморфизма пространств. Однако, если
фундаментальные группы не изоморфны, то соответствующие пространства заведомо не
гомеоморфны.

\begin{theorem}                                                   \label{tisoho}
Пусть $f:~\MM_1\rightarrow\MM_2$ -- гомотопическая эквивалентность двух
пространств, тогда индуцированное отображение
$f_*:~\pi(\MM_1,x)\rightarrow\pi\big(\MM_2,f(x)\big)$ является изоморфизмом для
любой точки $x\in\MM_1$.
\end{theorem}
\begin{proof}
См., например, \cite{Kosnio80R}.
\end{proof}
\begin{cor}                                                       \label{tconmf}
Стягиваемое пространство имеет тривиальную фундаментальную группу.
\qed\end{cor}
Мы видим, что любое стягиваемое пространство является односвязным. Обратное
утверждение неверно.
\begin{exa}
Сфера $\MS^n$, $n\ge1$, является односвязным, но не стягиваемым многообразием.
\qed\end{exa}
\begin{exa}
Шар $\MB^n$, $n\ge2$, с выколотой точкой является односвязным, но не стягиваемым
многообразием.
\qed\end{exa}
\begin{theorem}
Пусть $\MM_1$ и $\MM_2$ -- два линейно связных топологических пространства.
Тогда фундаментальная группа произведения $\MM_1\times\MM_2$ изоморфна прямому
произведению фундаментальных групп,
$\pi(\MM_1\times\MM_2)\simeq\pi(\MM_1)\times\pi(\MM_2)$.
\end{theorem}
\begin{proof}
Доказательство заключается в явном построении изоморфизма. См., например,
\cite{Kosnio80R}.
\end{proof}
\begin{cor}
Произведение $\MM_1\times\MM_2$ двух односвязных пространств является
односвязным.
\qed\end{cor}
\begin{exa}
Фундаментальная группа тора $\MT^n\approx\MS^1\times\dotsc\times\MS^1$ изоморфна
прямому произведению групп целых чисел,
$\pi(\MT^n)\simeq\underbrace{\MZ\times\dotsc\times\MZ}_n$.
\qed\end{exa}
\section{Фундаментальная группа и ориентируемость                \label{sfunor}}
Все связные многообразия можно разделить на два класса: ориентируемые и
неориентируемые. Напомним, что многообразие $\MM$ называется ориентируемым (см.\
раздел \ref{sdifma}), если его можно покрыть системой карт,
$\MM=\bigcup_i\MU_i$, причем якобиан преобразования координат (\ref{ejacob}) во
всех пересечениях $\MU_i\cap\MU_j$ положителен.

Рассмотрим класс многообразий, на которых можно задать репер
$\left\lbrace e_a\right\rbrace$, $a=1,\dotsc,n$, т.е.\ $n$ достаточно гладких
векторных полей, которые линейно независимы в каждой точке многообразии $\MM$,
$\dim\MM=n$. Отсюда, в частности, следует, что векторные поля $e_a(x)$ не могут
обращаться в нуль ни в какой точке многообразия $\MM$, т.к.\ в противном случае
они были бы линейно зависимы в этой точке. Репер $\lbrace e_a\rbrace$
образует базис для касательных векторных полей. Локально репер всегда
существует. Однако его глобальное существование является сильным предположением,
и далеко не каждое ориентируемое многообразие допускает существование репера.
\begin{exa}                                                       \label{espori}
Двумерная сфера $\MS^2$ является ориентируемым многообразием, однако она не
допускает существование репера, так как на ней не существует векторных полей,
которые не обращаются в нуль (теорема о невозможности причесать ежа
\ref{tvecsp}).
\qed\end{exa}

В локальной системе координат репер раскладывается по координатному базису,
$e_a=e^\al{}_a\pl_\al$, $\det e^\al{}_a\ne0$.

Введенное ниже определение класса ориентации репера основано на локальном
существовании репера и поэтому применимо ко всем многообразиям.

Если от отдельных реперов перейти к классам ориентирующих реперов, то можно дать
новое эквивалентное определение ориентируемости многообразий. А именно, пусть в
точке $x\in\MM$ задан {\em класс ориентации}, т.е.\ класс реперов
$\left\lbrace e_a\right\rbrace$, связанных друг с другом
линейным преобразованием с положительным определителем. Подробнее, два репера
$\lbrace e^\prime_a\rbrace$ и $\lbrace e_a\rbrace$ принадлежат одному
классу ориентации, если они связаны преобразованием $e^\prime_a=S_a{}^b e_b$,
причем $\det S(x)>0$. Если $\det S<0$, то будем говорить, что в точке $x$ реперы
принадлежат классам противоположной ориентации. Таким образом, в каждой точке
многообразия имеются ровно два класса ориентации. Поскольку репер можно
непрерывно смещать из точки $x$ в близкие точки многообразия, то имеет смысл
говорить о непрерывной зависимости классов ориентации от точки многообразия.
\begin{defn}
Если в каждой точке многообразия $\MM$ существует класс ориентации реперов,
который непрерывно зависит от точки многообразия, то многообразие $\MM$
называется {\em ориентируемым}.
\qed\end{defn}
\index{Класс ориентации (class of orientation)}%
\index{Ориентации класс (class of orientation)}%
\index{Ориентируемое многообразие (orientable manifold)}%
\index{Многообразие ориентируемое (orientable manifold)}%
Класс ориентации принимает только два значения, например, $+1$ или $-1$. Поэтому
непрерывность означает, что во всех точках ориентируемого многообразия можно
выбрать класс ориентации $+1$.

Докажем эквивалентность нового определения и определения ориентируемости
многообразий, которое было дано в разделе \ref{sdifma}. Пусть многообразие
ориентируемо. Тогда на нем существует координатное покрытие $\MM=\bigcup_i\MU_i$
такое, что якобиан преобразования координат положителен,
$\det\pl_\al x^{\al'}>0$, во всех областях пересечения карт. В каждой
координатной окрестности выберем координатный репер $\lbrace\pl_\al\rbrace$.
Множество этих реперов определяет класс ориентации, который непрерывно зависит
от точки многообразия. Действительно, если точка принадлежит различным картам,
то координатные реперы принадлежат одному классу ориентации. Обратно. Пусть
существует класс ориентации реперов, непрерывно зависящий от точки многообразия.
Выберем координатное покрытие многообразия, $\MM=\bigcup_i\MU_i$. Тогда в двух
пересекающихся картах $\MU_i$ и $\MU_j$ можно выбрать координаты $x^\al$ и
$x^{\al'}$ таким образом, что
\begin{equation*}
  e_a=e^\al{}_a\pl_\al=e^{\al'}{}_a\pl_{\al'},\qquad \det e^\al{}_a>0,\quad
  \det e^{\al'}{}_a>0,
\end{equation*}
где $\lbrace e_a\rbrace$ -- некоторый представитель заданного класса ориентации.
Действительно, если в какой то системе координат $\det e^\al{}_a<0$, то,
переставив две произвольные координаты, получим $\det e^\al{}_a>0$.
Отсюда следует положительность якобиана преобразования координат в области
пересечения карт,
\begin{equation}                                                  \label{ejatrc}
  J=\det\pl_\al x^{\al'}=\det(e^a{}_\al e^{\al'}{}_a)>0.
\end{equation}
Это справедливо для любого представителя из класса ориентации.

На произвольном многообразии, в том числе и неориентируемом, ориентацию можно
переносить вдоль путей. Пусть на многообразии $\MM$ задан кусочно
дифференцируемый путь $\g=x(t)$, $t\in[0,1]$, и класс ориентации реперов
$\lbrace e_a\big(x(t)\big)\rbrace$ во всех точках пути, который
непрерывно зависит от точки пути. Тогда класс ориентации $\lbrace e_a(1)\rbrace$
в конечной точке $x(1)$ называется {\em переносом класса ориентации} репера
$\lbrace e_a(0)\rbrace$ из точки $x(0)$ в точку $x(1)$ вдоль пути $\g$.

Операция переноса ориентации вдоль путей обладает следующими свойствами.

1) Из любой точки $x\in\MM$ ориентацию можно однозначно перенести во все
близлежащие точки многообразия вдоль путей, целиком лежащих в пределах
координатной окрестности точки $x$.

2) Для любого кусочно дифференцируемого пути перенос ориентации существует и не
зависит от выбора репера вдоль пути. Существование очевидно, т.к.\ достаточно
задать произвольным образом компоненты $n$ линейно независимых векторов
$\lbrace e_a(t)\rbrace$, как дифференцируемые функции одного переменного.
Независимость от выбора репера доказывается просто. Пусть
$\lbrace e_a(t)\rbrace$ и $\lbrace e^\prime_a(t)\rbrace$ -- два репера вдоль
одной кривой $x(t)$, имеющие одинаковую ориентацию при $t=0$. Матрица перехода
от одного репера к другому, $e^\prime_a=S_a{}^b e_b$, невырождена,
$\det S(t)\ne0$ для всех $t\in[0,1]$. Поэтому, если в начальный момент
$\det S(0)>0$, то из непрерывности следует, что $\det S(t)>0$ для всех
$t\in[0,1]$, поскольку матрица $S$ не может быть вырождена.

3) Если два пути $x_1(t)$ и $x_2(t)$ гомотопны относительно начала и конца,
то переносы класса ориентации вдоль этих путей совпадают. Действительно, для
гомотопных путей существует непрерывное отображение
$F(t,s):~[0,1]\times[0,1]\rightarrow\MM$ такое, что $F(t,0)=x_1(t)$ и
$F(t,1)=x_2(t)$. Пусть $\lbrace e_a(t)\rbrace$ -- репер вдоль $x_1(t)$.
Перенесем каким либо непрерывным образом этот репер из каждой точки кривой
$\g_1$ вдоль кривых $F(t,s)$ по второму аргументу $s$. Перенос ориентации в
конечные точки не зависит от выбора репера $\lbrace e_a(t)\rbrace$ вдоль $\g_1$
в силу свойства 2). В результате получим некоторый репер вдоль второй кривой
$\g_2$. Поскольку в начальной и конечной точке реперы по построению совпадают,
то совпадают и переносы классов ориентации вдоль гомотопных путей.

Приведенные выше определения удобны для доказательства неориентируемости
некоторых многообразий в силу следующей теоремы.
\index{Перенос класса ориентации (transport of a class of orientation)}%
\index{Класс ориентации перенос (transport of a class of orientation)}%
\begin{theorem}                                                   \label{tnonom}
Связное многообразие ориентируемо тогда и только тогда, когда перенос класса
ориентации репера вдоль любого замкнутого пути сохраняет ориентацию.
\end{theorem}
\begin{proof}
Пусть многообразие $\MM$ ориентируемо. Выберем координатное покрытие такое,
чтобы якобиан преобразования координат был положителен,
$\det \pl_\al x^{\al'}>0$,
во всех областях пересечения карт. Выберем произвольную точку $x_0\in\MM$ и
зафиксируем какой либо репер, который принадлежал бы тому же классу ориентации,
что и координатный базис, т.е.\ $\det\big(e^\al{}_a(x_0)\big)>0$. Эту ориентацию
можно перенести в любую точку многообразия, т.к.\ оно линейно связно. Результат
переноса ориентации не зависит от пути, потому что ориентацию можно переносить
внутри произвольной карты, а при переходе от карты к карте принадлежность репера
определенному классу ориентации сохраняется. В частности, принадлежность репера
данному классу ориентации сохраняется при переносе ориентации вдоль
произвольного замкнутого пути.

Обратно. Выберем произвольным образом репер в какой либо точке $x_0\in\MM$ и
разнесем эту ориентацию во все другие точки $x\in\MM$. Результат будет
однозначен, т.к.\ перенос класса ориентации вдоль произвольного замкнутого пути
сохраняет ориентацию. Таким образом в каждой точке $x\in\MM$ определен класс
ориентации $\lbrace e_a(x)\rbrace$. Пусть $\MU_i$ -- координатное покрытие
многообразия $\MM$. На каждой координатной окрестности можно выбрать систему
координат такую, что $\det e^\al{}_a>0$ для некоторого представителя из класса
ориентации. Следовательно, из (\ref{ejatrc}) следует, что якобиан преобразования
координат во всех пересекающихся областях будет положителен.
\end{proof}
Из этой теоремы вытекает, что если можно указать замкнутый путь, вдоль которого
не существует переноса класса ориентации репера, то многообразие является
неориентируемым. В частности, неориентируемое многообразие не может быть
односвязным. Действительно, любой замкнутый путь на односвязном многообразии
можно непрерывно стянуть в точку. Поскольку замкнутые пути в достаточно малой
окрестности произвольной точки сохраняют ориентацию, то из непрерывности
следует, что на односвязном многообразии любой замкнутый путь сохраняет
ориентацию. То есть каждая односвязная компонента многообразия всегда
ориентируема. Неодносвязные многообразия могут быть как ориентируемыми, так и
неориентируемыми.
\begin{exa}
Цилиндр и лист Мебиуса имеют изоморфные фундаментальные группы $\pi\simeq\MZ$,
однако являются соответственно ориентируемой и неориентируемой поверхностями.
\qed\end{exa}

Пусть на многообразии $\MM$ задан ориентирующий репер $\lbrace e_a\rbrace$
глобально. Рассмотрим этот репер в качестве базиса касательных векторных полей
$X=X^ae_a\in\CX(\MM)$. Зададим на $\MM$ тривиальную линейную
$\MG\ML(n,\MR)$-связность: $\om_{\al a}{}^b=0$ в каждой карте. Это значит, что
при параллельном переносе векторов их компоненты относительно выбранного репера
не меняются. Тогда рассматриваемое многообразие будет пространством абсолютного
параллелизма, т.к.\ тензор кривизны для тривиальной связности равен нулю,
$R_{\al\bt a}{}^b=0$. По этой причине, если на многообразии можно задать репер
глобально, то оно называется {\em параллелизуемым}.
\index{Параллелизуемое многообразие (parallelizable manifold)}%
\index{Многообразие параллелизуемое (parallelizable manifold)}%
В то же время тензор кручения для тривиальной связности в общем случае будет
отличен от нуля. При этом тензор кручения на односвязных многообразиях для
заданной нулевой связности однозначно определяется репером. Верно также и
обратное утверждение: в односвязном пространстве абсолютного параллелизма, в
котором тензор кривизны всюду равен нулю, $R_{\al\bt a}{}^b=0$, всегда можно
задать репер глобально. Для этого достаточно выбрать $n$ линейно независимых
векторов в касательном пространстве произвольной точки, а затем разнести их по
всему многообразию, используя то свойство, что результат не зависит от пути
параллельного переноса.

Очевидно, что любое параллелизуемое многообразие является ориентируемым.
\begin{exa}
Любая группа Ли является параллелизуемым многообразием, если групповое умножение
слева (справа) принять за определение параллельного переноса (см.\ раздел
\ref{sligrc}). Поэтому все группы
Ли представляют собой ориентируемые многообразия. При этом левоинвариантные
(правоинвариантные) векторные поля представляют собой глобально определенный
репер.
\qed\end{exa}
Обратное утверждение, что все ориентируемые многообразия являются
параллелизуемыми, неверно, как показывает пример сферы \ref{espori}.

Приведем критерий параллелизуемости многообразий.
\begin{theorem}
Многообразие $\MM$ является параллелизуемым тогда и только тогда, когда
касательное расслоение тривиально, т.е.\ диффеоморфно прямому произведению
$\MT(\MM)\approx\MM\times\MR^n$.
\end{theorem}
\begin{proof}
Проверяется тот факт, что глобально определенный репер устанавливает
диффеоморфизм расслоений.
\end{proof}

Ориентируемость многообразий можно также определить с помощью дифференциальных
форм, рассмотренных в разделе \ref{sdifin}. Рассмотрим связное многообразие
$\MM$, $\dim\MM=n$. Пространство дифференциальных форм максимальной степени
$\Lm_n(\MM)$ в каждой точке многообразия одномерно. Пусть $A,B\in\Lm_n(\MM)$ и
$A\ne0$, $B\ne0$ во всех точках $x\in\MM$.
Отношение $A=f(x) B$, где $f\in\CC(\MM)$ и $f>0$, задает отношение
эквивалентности в пространстве $n$-форм, не обращающихся в нуль ни в одной точке
многообразия. Это отношение определяет ровно два класса эквивалентности.
Назовем {\em ориентацией} многообразия $\MM$ соответствующий класс
эквивалентности $n$-форм, если такой класс существует. Это определение
ориентируемости эквивалентно приведенному выше в силу следующей теоремы.
\index{Ориентация многообразия (orientation of a manifold)}%
\index{Многообразия ориентация (orientation of a manifold)}%
\begin{theorem}                                                   \label{torivo}
Связное многообразие $\MM$, $\dim\MM=n$, ориентируемо тогда и только тогда,
когда оно допускает непрерывную нигде не обращающуюся в нуль $n$-форму.
\end{theorem}
\begin{proof}
Достаточность. Пусть на связном многообразии $\MM$ ориентация задана с помощью
$n$-формы $A$, которая не равна нулю ни в одной точке.
Пусть $\lbrace\MU_i\rbrace$ -- локально конечное покрытие многообразия
$\MM$. Пусть $x^\al$, $\al=1,\dotsc,n$ -- система координат на некоторой
окрестности $\MU_i$. Тогда $n$-форма $dx^1\wedge\dotsc\wedge dx^n$ не равна
нулю ни в какой точке $\MU_i$ и поэтому отличается от $A$ на некоторый
отличный от нуля множитель $f(x)$,
\begin{equation*}
  dx^1\wedge\dotsc\wedge dx^n=fA.
\end{equation*}
Если $f>0$, то мы говорим, что система координат $x^\al$ согласована с
ориентацией на $\MM$. В противном случае, при $f<0$, достаточно переставить две
координаты для изменения знака $n$-формы. Обозначим координаты на области
$\MU_j$, которая пересекается с $\MU_i$, через $x^{\al'}$. Эти координаты всегда
можно выбрать так, чтобы они были согласованы с ориентацией $A$. Тогда $n$-форма
$dx^{1'}\wedge\dotsc\wedge dx^{n'}$ также отличается от формы $A$ на некоторый
положительный множитель $g(x)>0$,
\begin{equation*}
  dx^{1'}\wedge\dotsc\wedge dx^{n'}=gA.
\end{equation*}
Поскольку
\begin{equation*}
  dx^{1'}\wedge\dotsc\wedge dx^{n'}
  =\det\left(\frac{\pl x^{\al'}}{\pl x^\al}\right)dx^1\wedge\dotsc\wedge dx^n,
\end{equation*}
то якобиан соответствующего преобразования
координат $x^\al~\rightarrow~x^{\al'}$ положителен,
\begin{equation*}
  \det\left(\frac{\pl x^{\al'}}{\pl x^\al}\right)=gf^{-1}>0.
\end{equation*}
Полученное выражение для якобиана преобразования координат корректно,
что легко проверяется для пересечения трех карт.

Необходимость. Пусть $f_i$ -- разбиение единицы, подчиненное заданному счетному
локально конечному покрытию $\lbrace\MU_i\rbrace$ (теорема \ref{trazed}).
Выберем координаты на каждой области $\MU_i$ таким образом, чтобы все якобианы
преобразований координат в областях пересечения были положительны. На каждой
области $\MU_i$ можно построить отличную от нуля $n$-форму
$\om_i=dx^1\wedge\dotsc\wedge dx^n$. Тогда $n$-форма $\sum_i f_i\om_i$
определена на всем многообразии и отлична от нуля во всех точках.
\end{proof}
\begin{com}
В доказательстве теоремы использовано разбиение единицы, которое всегда
существует, т.к.\ в определение многообразия мы включили счетность базы
топологии. Если это требование исключено из определения, как это часто делается,
то в условии теоремы необходимо дополнительно потребовать, например,
паракомпактность многообразия.
\qed\end{com}

В роли $n$-формы, которая не обращается в нуль, может выступать форма объема
(\ref{evolel}). Если многообразие не допускает существования непрерывной
невырожденной $n$-формы, то оно будет неориентируемым. Если на связном
многообразии задана невырожденная $n$-форма, то мы говорим, что на многообразии
задана ориентация. Если многообразие несвязно и состоит из нескольких
компонент, то на каждой компоненте ориентацию можно задавать независимо, если
каждая компонента связности ориентируема.

Отметим, что каждый репер $e_a$, заданный глобально, определяет невырожденную
$n$-форму. Действительно, обозначим дуальный базис кокасательного пространства
через $e^a$. Тогда $n$-форма
\begin{equation*}
  \upsilon:=e^1\wedge\dotsc\wedge e^n
\end{equation*}
определена глобально и невырождена. Ее можно выбрать в качестве формы объема.
Обратное утверждение неверно: не каждая форма объема определяет репер.
\begin{exa}
На двумерной сфере $\MS^2$ существует форма объема, однако репер определить
глобально нельзя, так как на ней не существует векторных полей, не обращающихся
в нуль.
\qed\end{exa}

Обсудим более подробно связь неориентируемости многообразия с фундаментальной
группой. Перенос ориентации с помощью репера показывает, что каждый
гомотопический класс замкнутого пути с началом и концом в точке $x_0$ (т.е.\
элемент фундаментальной группы $\pi(\MM,x_0)$) либо сохраняет, либо меняет
ориентацию репера на противоположную. Это означает, что существует гомоморфизм
$\s$ фундаментальной группы в группу $\MZ_2$,
$$
  \s:\quad \pi(\MM,x_0)\rightarrow\MZ_2=\left\lbrace \pm1\right\rbrace .
$$
Другими словами, каждому замкнутому пути ставится в соответствие $+1$ или $-1$,
в зависимости от того, сохраняется ориентация репера при переносе или меняется
на противоположную. Если многообразие ориентируемо, то гомоморфизм $\s$
тривиален. Для неориентируемых многообразий ввиду наличия путей, обращающих
ориентацию, гомоморфизм $\s$ нетривиален. Как следствие получаем, что
фундаментальная группа неориентируемых многообразий не может быть тривиальной.
\begin{exa}
Фундаментальная группа листа Мебиуса совпадает с группой целых чисел,
$\pi(\MM)=\MZ$, т.к.\ он стягивается к центральной окружности. При гомоморфизме
$\s$ все четные и нечетные числа отображаются соответственно в $+1$ и $-1$.
\qed\end{exa}
\begin{exa}
Фундаментальная группа проективного пространства $\MR\MP^n$ нетривиальна:
$\pi(\MR\MP^n)=\MZ_2$, и гомоморфизм $\s$ является изоморфизмом для четных
$n$. Для нечетных $n$ гомоморфизм $\s$ тривиален, т.к.\ проективные пространства
нечетной размерности ориентируемы.
Проективное пространство можно параметризовать точками шара, у которого
отождествлены диаметрально противоположные точки граничной сферы
(см.\ пример \ref{eprspa}). Тогда диаметр шара будет замкнутым путем (образом
окружности), который не стягивается в точку. В то же время, если при
отображении окружности в проективное пространство диаметр проходится два раза,
то такой образ окружности можно стянуть в точку, как показано на
рис.~\ref{fsohom}.
\begin{figure}[hbt]
\hfill\includegraphics[width=.7\textwidth]{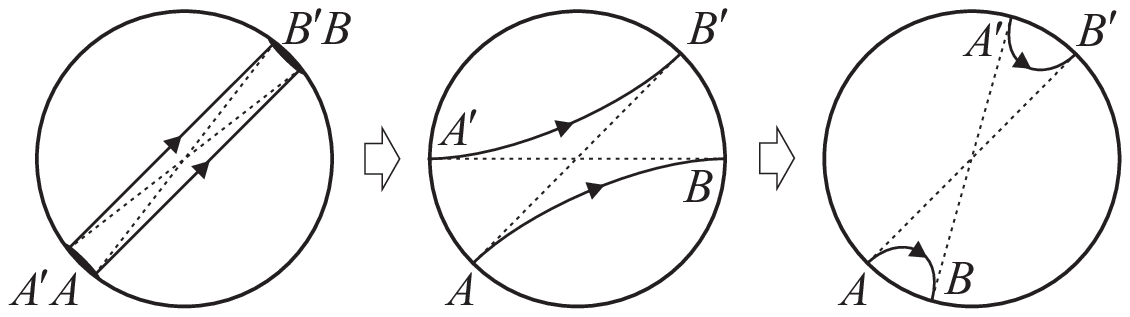}
\hfill {}
\centering\caption{Непрерывная деформация образа окружности в проективном
 пространстве $\MR\MP^2$, который проходит по диаметру $AB$ два раза, в точку.
 Пунктиром показано отождествление противоположных точек граничной окружности.}
\label{fsohom}
\end{figure}
Аналогично стягивается в точку любой образ окружности, который проходит
диаметр произвольное четное число раз. Это поясняет, почему фундаментальная
группа проективного пространства равна $\MZ_2$.
\qed\end{exa}
\chapter{Накрытия                                                \label{scover}}
Важным классом непрерывных отображений топологических пространств являются
накрытия, которые представляют собой локальный гомеоморфизм. Все, сказанное в
настоящей главе о топологических пространствах, относится также к многообразиям.
\section{Определения и примеры}
\begin{defn}
Непрерывное отображение топологических пространств
$p:~\widetilde\MM\rightarrow\MM$ называется {\em накрытием}, если выполнены
следующие условия:

1) \parbox[t]{.92\linewidth}{$p$ сюрьективно;}

2) \parbox[t]{.92\linewidth}{для любого $x\in\MM$ найдется открытая окрестность
$\MU_x$ точки $x$ такая, что $p^{-1}(\MU_x)=\bigcup_{j\in\CJ}\MU_j$
для некоторого семейства открытых подмножеств $\MU_j\subset\widetilde\MM$,
удовлетворяющих условиям $\MU_j\bigcap\MU_k=\emptyset$ при $j\ne k$ и
сужение отображения $p|_{\MU_j}:~\MU_j\rightarrow\MU_x$ -- гомеоморфизм для
всех $j\in\CJ$.}\newline
Мы говорим также, что окрестность $\MU_x$ {\em просто накрыта} отображением $p$.
Топологическое пространство $\MM$ называется {\em базой} накрытия, а
$\widetilde\MM$ -- {\em накрывающим пространством}. Каждое множество $\MU_j$
называется {\em листом} накрытия области $\MU_x$. Если множество $\CJ$ содержит
$n$ элементов, то говорят об $n$-листном накрытии. Если топологическое
пространство $\widetilde\MM$ является односвязным, т.е.\ фундаментальная группа
тривиальна, $\pi\big(\widetilde\MM\big)=e$, то накрытие называется
{\em универсальным}.
\qed\end{defn}
\index{Накрытие (covering)}%
\index{Накрытие универсальное (universal covering)}%
\index{Универсальное накрытие (universal covering)}%
\index{База накрытия (base of a covering)}%
\index{Накрывающее пространство (covering space)}%
\index{Пространство накрывающее (covering space)}%
\index{Лист накрытия (sheet of a covering)}%
\index{Накрытия лист (sheet of a covering)}%
\begin{prop}
Если база $\MM$ накрытия $p:~\widetilde\MM\rightarrow\MM$ линейно связна, то
мощность множества $\CJ$ не зависит от точки $x\in\MM$ при линейно связной базе.
\end{prop}
\begin{proof}
См., например, \cite{Spanie66R}.
\end{proof}
Это утверждение говорит о том, что определение $n$-листного накрытия корректно,
т.к.\ не зависит от точки $x\in\MM$.
\begin{com}
Если $\widetilde\MM$ и $\MM$ являются не просто топологическими пространствами,
а многообразиями, то вместо непрерывности отображения $p$ мы требуем его
дифференцируемость достаточное число раз, а гомеоморфизм заменяем на
диффеоморфизм областей $\MU_j$ и $\MU_x$ для всех $j\in\CJ$. Для многообразий
$\dim\MU_j=\dim\MU_x$ для всех $j\in\CJ$, т.к.\ сужение отображения $p$ на
$\MU_j$ -- диффеоморфизм. Кроме того, каждое подмножество $\MU_j$,
по-определению, открыто в $\widetilde\MM$. Это возможно только если
$\dim\widetilde\MM=\dim\MU_j$. Таким образом, если
$p:~\widetilde\MM\rightarrow\MM$ накрытие многообразий, то необходимо, чтобы
накрывающее многообразие и база имели одинаковую размерность,
$\dim\widetilde\MM=\dim\MM$. Для многообразий отображение $p$ является
погружением.
\qed\end{com}
\begin{exa}
Всякий гомеоморфизм является однолистным накрытием. В этом случае у каждой
области $\MU_x$ имеется только один лист.
\qed\end{exa}
\begin{exa}
Если $\widetilde\MM=\MM\times\MX$ -- топологическое произведение топологического
пространства $\MM$ на счетное множество $\MX$, то естественная проекция
$p:~\MM\times\MX\rightarrow\MM$ является накрытием. Число листов равно числу
элементов множества $\MX$.
\qed\end{exa}
\begin{exa}
Пусть задана аналитическая функция $F(z)$ в некоторой области $\MD\subset\MC$.
Если в каждой точке $z\in\MD$ функция $F(z)$ имеет $\Sn$ значений, то риманова
поверхность функции $F(z)$ является $\Sn$-листным накрытием области $\MD$.
\qed\end{exa}
\begin{exa}
Отображение вещественной прямой в окружность,
\begin{equation*}
  p:\quad \MR\ni\quad t\mapsto \ex^{2\pi i t}\quad\in\MS^1,
\end{equation*}
является накрытием с бесконечным числом листов. Это накрытие является
универсальным.
\qed\end{exa}
\begin{exa}                                                       \label{esircc}
Пусть окружность $\MS^1$ задана на комплексной плоскости $\MC$ уравнением
$|z|=1$. Тогда отображение окружности в окружность,
\begin{equation*}
  p:\quad \MS^1\ni\quad z\mapsto z^n\quad\in\MS^1,\qquad \forall n\in\MN,
\end{equation*}
является накрытием с $n$ листами. Это же отображение является $n$-листным
накрытием комплексной плоскости с выколотым началом координат
$p:~\MC\setminus\lbrace0\rbrace\rightarrow\MC\setminus\lbrace0\rbrace$.
\qed\end{exa}
\begin{exa}
Пусть $\MG$ -- группа Ли и $\MH$ -- ее дискретная подгруппа. Поскольку
подгруппа $\MH$ замкнута, то по теореме \ref{tfacma} на фактор пространстве
$\MG/\MH$ существует дифференцируемая структура многообразия. Тогда проекция
$p:~\MG\rightarrow\MG/\MH$, где $\MG/\MH$ -- пространство левых или правых
смежных классов, является накрытием. Число листов равно числу элементов группы
$\MH$.
\qed\end{exa}
\begin{prop}
Пусть $\ME(\MM,\pi,\MF)$ расслоение, определенное в разделе \ref{sfibun},
типичным слоем $\MF$ которого является $0$-мерное многообразие. Тогда
проекция $\pi:~\ME\rightarrow\MM$ является накрытием. Наоборот, любое накрытие
является расслоением $\widetilde\MM(\MM,p,\CJ)$ с $0$-мерным типичным слоем
$\CJ$.
\qed\end{prop}
\begin{proof}
Определение накрытия и расслоения в случае $0$-мерного типичного слоя совпадают,
если записать
$\pi^{-1}(\MU_x)=\chi^{-1}(\MU_x\times\MF)=\bigcup_{j\in\CJ}\MU_j$, где точки
$0$-мерного многообразия $j\in\CJ=\MF$ нумеруют листы накрытия области $\MU_x$.
Второе условие в определении расслоения при $0$-мерном слое выполняется
автоматически, т.к.\ отображение $\pi\circ\chi^{-1}$ дифференцируемо, а это
возможно только при проекции на первый сомножитель,
$\pi\circ\chi^{-1}:~\MU_x\times\MF\ni(y,f)\mapsto y\in\MM$.
\end{proof}
\begin{com}
Если типичным слоем расслоения $\ME(\MM,\pi,\MF)$ является многообразие более
высокой размерности, $\dim\MF\ge1$, то расслоение не является накрытием, потому
что, например, $\dim\pi^{-1}(\MU_x)=\dim\MM+\dim\MF>\dim\MM$.
\qed\end{com}

\begin{figure}[h,b,t]
\hfill\includegraphics[width=.25\textwidth]{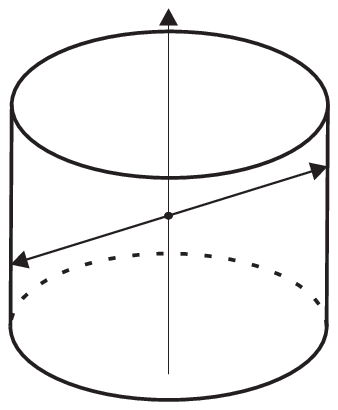}
\hfill {}
\centering\caption{Накрытие листа Мебиуса цилиндром. Отождествив центрально
симметричные точки цилиндра, мы получим лист Мебиуса в виде полосы с
отождествленными краевыми точками.}
\label{fmeban}
\end{figure}

Отображение накрытия позволяет установить связь между ориентируемыми и
неориентируемыми многообразиями.
\begin{theorem}                                                   \label{tcorin}
Для любого неориентируемого многообразия $\MM$ существует двулистное
ориентируемое накрытие $p:~\widetilde\MM\rightarrow\MM$.
\end{theorem}
\begin{proof}
См., например, \cite{Narasi71R}.
\end{proof}
\begin{exa}
Четномерная сфера $\MS^n$, $n=2,4,\dotsc$, ориентируема и является двулистным
универсальным накрывающим пространством для проективной плоскости $\MR\MP^n$,
которая неориентируема. Накрытие $p:~\MS^n\rightarrow\MR\MP^n$ осуществляется
путем отождествления диаметрально противоположных точек сферы.
\end{exa}
\begin{exa}
Лист Мебиуса можно получить из цилиндра путем отождествления центрально
симметричных точек, как показано на рис.~\ref{fmeban}.
Это накрытие двулистно, но не является универсальным. Универсальные накрывающие
цилиндра и листа Мебиуса совпадают -- это плоскость $\MR^2$.
\qed\end{exa}

Следующее утверждение облегчает изучение накрытий за счет ограничения класса
рассматриваемых баз.

\begin{theorem}
Если пространство $\MM$ -- локально связно, то непрерывное отображение
$p:~\widetilde\MM\rightarrow\MM$ тогда и только тогда является накрытием, когда
для каждой компоненты связности $\MU\subset\MM$ сужение отображения
\begin{equation*}
  p|_{p^{-1}(\MU)}:\quad p^{-1}(\MU)\rightarrow\MU
\end{equation*}
является накрытием.
\end{theorem}
\begin{proof}
См., например, \cite{Spanie66R}.
\end{proof}
Поскольку все многообразия являются локально связными, то в дифференциальной
геометрии мы можем ограничиться рассмотрением накрытий над связными базами
$\MM$. Поэтому в дальнейшем мы будем рассматривать только линейно связные и,
следовательно, связные базы накрытий.

Из определения накрытия следует, что прообраз $p^{-1}(\MU_x)$ является
объединением непересекающихся открытых подмножеств $\MU_j$, каждое из которых
гомеоморфно $\MU_x$. Если $\MU_x$ связно, то каждое из подмножеств $\MU_j$ также
связно. Отсюда следует
\begin{prop}
Пусть $\MU$ -- открытое связное подмножество пространства $\MM$, просто накрытое
отображением $p:~\widetilde\MM\rightarrow\MM$. Тогда $p$ гомеоморфно отображает
каждую компоненту связности множества $p^{-1}(\MU)$ на $\MU$.
\end{prop}
\begin{cor}
Рассмотрим два накрытия $p_1:~\widetilde\MM_1\rightarrow\MM$,
$p_2:~\widetilde\MM_2\rightarrow\MM$ с одинаковыми базами и коммутативную
диаграмму
\begin{equation*}
\begin{diagram}
  \widetilde\MM_1 & \rTo^p & \widetilde\MM_2 \\
  & \rdTo_{p_1} &  \dTo_{p_2} \\
  &  & \MM
\end{diagram}
\end{equation*}
где база $\MM$ локально связна. Если $p$ сюрьективно, то оно также является
накрытием.
\qed\end{cor}
\begin{theorem}
Пусть $p:~\widetilde\MM\rightarrow\MM$ -- накрытие. Тогда

1) $p$ -- открытое отображение;

2) $\MM$ имеет фактор топологию относительно $p$.
\end{theorem}
\begin{proof}
См., например, \cite{Kosnio80R}.
\end{proof}

Если на базе накрытия $\MM$ задан путь $\g:~[0,1]\ni t\mapsto x(t)\in\MM$, то
возникает вопрос, что представляет собой прообраз $p^{-1}(\g)$
при накрытии $p:~\widetilde\MM\rightarrow\MM$ ? Для ответа введем новое понятие.

Пусть задано накрытие $p:~\widetilde\MM\rightarrow\MM$ и непрерывное отображение
некоторого пространства $\MN$ в базу накрытия, $f:~\MN\rightarrow\MM$. (Для
путей в роли $\MN$ выступает единичный отрезок $[0,1]$.)
\begin{defn}
{\em Поднятием} непрерывного отображения $f$ называется непрерывное отображение
$\tilde f:~\MN\rightarrow\widetilde\MM$ такое, что $p\circ\tilde f=f$,
т.е.\ диаграмма
\begin{equation*}
\begin{diagram}
  \MN & \rTo^{\tilde f} & \widetilde\MM \\
  & \rdTo_f &  \dTo_p \\
  &  & \MM
\end{diagram}
\end{equation*}
коммутативна.
\qed\end{defn}
\index{Поднятие отображения (lifting of a map)}%
\index{Отображения поднятие (lifting of a map)}%
Следующий результат показывает, что, если поднятие отображения $f$ существует,
то оно, по-существу, единственно.
\begin{prop}
Пусть $p:~\widetilde\MM\rightarrow\MM$ -- накрытие и
$\tilde f,\tilde{\tilde f}:~\MN\rightarrow\widetilde\MM$ -- два поднятия
отображения $f:~\MN\rightarrow\MM$. Если $\MN$ связно и поднятия совпадают
$\tilde f(y)=\tilde{\tilde f}(y)$ хотя бы в одной точке $y\in\MN$, то поднятия
совпадают во всех точках, $\tilde f=\tilde{\tilde f}$.
\end{prop}
\begin{proof}
См., например, \cite{Kosnio80R}.
\end{proof}
\begin{theorem}[\bf О накрывающей гомотопии для путей]             \label{tcoho}
Пусть $p:~\widetilde\MM\rightarrow\MM$ -- накрытие. Тогда \newline
\indent 1) \parbox[t]{.92\linewidth}{для пути $\g:~[0,1]\rightarrow\MM$ и
точки $\tilde x\in\widetilde\MM$ такой, что $p(\tilde x)=\g(0)$ существует
единственный путь $\tilde\g:~[0,1]\rightarrow\widetilde\MM$, который является
поднятием, $p\circ\tilde\g=\g$, пути $\g$ и для которого
$\tilde\g(0)=\tilde x$;} \newline
\indent 2) \parbox[t]{.92\linewidth}{для непрерывного отображения (гомотопии
двух путей в базе) $F:~[0,1]\times[0,1]\rightarrow\MM$ и точки
$\tilde x\in\widetilde\MM$ такой, что $p(\tilde x)=F(0,0)$ существует
единственное непрерывное отображение (гомотопия двух путей в накрывающем
пространстве)  $\widetilde F:~[0,1]\times[0,1]\rightarrow\widetilde\MM$,
которое является поднятием, $p\circ\widetilde F=F$, отображения $F$ и для
которого $\widetilde F(0,0)=\tilde x$.}
\end{theorem}
\begin{proof}
Теорема доказывается путем явного построения отображений. Детали можно найти,
например, в \cite{Kosnio80R}.
\end{proof}
\begin{cor}[\bf Теорема о монодромии]
Пусть $\g_1\sim\g_2(\rel\lbrace0,1\rbrace)$ -- два пути в базе $\MM$, гомотопные
относительно начала и конца, и начальные точки их поднятий совпадают,
$\tilde\g_1(0)=\tilde\g_2(0)$. Тогда конечные точки поднятий также совпадают,
$\tilde\g_1(1)=\tilde\g_2(1)$.
\qed\end{cor}
Напомним, что единица фундаментальной группы $\pi(\MM,x_0)$ -- это класс
замкнутых путей с началом и концом в точке $x_0$, которые гомотопны
тождественному пути $e_{x_0}$. Из теоремы о монодромии следует, что поднятия
всех этих путей также замкнуты в накрывающем пространстве $\widetilde\MM$.
Следующая теорема говорит о том, что поднятие замкнутых путей, которые не
стягиваются в точку $x_0$ будет незамкнутым.
\begin{theorem}                                                   \label{tfrdkk}
Пусть $p:~\widetilde\MM\rightarrow\MM$ -- универсальное накрытие, т.е.\
накрывающее пространство $\widetilde\MM$ -- односвязно. Тогда существует
взаимно однозначное соответствие между элементами фундаментальной группы
$\pi\big(\MM,p(\tilde x)\big)$ и множеством прообразов
$p^{-1}\big(p(\tilde x)\big)$, где $\tilde x$ -- произвольная точка
$\widetilde\MM$.
\end{theorem}
\begin{proof}
Пусть $[\g]\in\pi\big(\MM,p(\tilde x)\big)$ -- класс гомотопных замкнутых путей
на базе и $\tilde\g$ -- поднятие какого либо представителя из данного класса.
Определим отображение
\begin{equation*}
  \vf:\quad \pi\big(\MM,p(\tilde x)\big)\ni\quad[\g]\mapsto\vf([\g])=\tilde\g(1)
  \quad\in p^{-1}\big(p(\tilde x)\big).
\end{equation*}
Это отображение определено корректно согласно теореме о монодромии.

Теперь определим обратное отображение
$\vf^{-1}:~p^{-1}\big(p(\tilde x)\big)\rightarrow\pi\big(\MM,p(\tilde x)\big)$.
Выбираем точку $\tilde x_1\in p^{-1}\big(p(\tilde x)\big)$ и некоторый путь
$\tilde\g$ из $\tilde x$ в $\tilde x_1$. Поскольку $\widetilde\MM$ односвязно,
то любые такие пути гомотопны. Поэтому класс замкнутых путей $[p\circ\tilde\g]$
-- корректно определенный элемент фундаментальной группы
$\pi\big(\MM,p(\tilde x)\big)$. Легко проверить, что
$\vf\circ\vf^{-1}=\vf^{-1}\circ\vf=e$, так что $\vf$ и $\vf^{-1}$
-- биекции.
\end{proof}
Эта теорема дает возможность в некоторых случаях вычислить фундаментальную
группу. Для этого надо найти универсальное накрытие
$p:~\widetilde\MM\rightarrow\MM$ и групповую структуру на прообразе
$p^{-1}(x_0)$, где $x_0\in\MM$, так, чтобы биекция
$\vf:~\pi(\MM,x_0)\rightarrow p^{-1}(x_0)$ стала гомоморфизмом групп. Вообще
говоря, сделать это сложно.
\section{Фундаментальная группа пространства орбит}
В настоящем разделе мы рассмотрим группы преобразований $(\widetilde\MM,\MG)$ и
приведем достаточные условия на группу преобразований, которые обеспечивают то,
что проекция пространства $\widetilde\MM$ в пространство орбит
$\MM:=\widetilde\MM/\MG$ является накрытием. Обсудим также связь группы
преобразований $(\widetilde\MM,\MG)$ с фундаментальными группами базы
$\MM:=\widetilde\MM/\MG$ и накрывающего пространства $\widetilde\MM$. Кроме
того, обсудим достаточные условия существования в пространстве орбит
дифференцируемой структуры.
\begin{com}
Многие определения и теоремы настоящего раздела справедливы не только для
многообразий, но и для произвольных топологических пространств. Поэтому мы
будем употреблять термин топологическое пространство, имея в виду, что
соответствующее утверждение верно и для многообразий. Там, где важна специфика
дифференцируемой структуры, мы будем писать многообразие.
\qed\end{com}
\begin{defn}
Группа $\MG$ действует на топологическом пространстве $\widetilde\MM$
{\em собственно разрывно}, если выполнены следующие условия:\newline
\indent 1) \parbox[t]{.92\linewidth}{для любых двух точек $\tilde x_1$ и
$\tilde x_2$ из $\widetilde\MM$, не лежащих на одной орбите, существуют
окрестности $\widetilde\MU_1\supset \tilde x_1$ и
$\widetilde\MU_2\supset \tilde x_2$ такие, что
$\widetilde\MU_1\MG\bigcap\widetilde\MU_2=\emptyset$,}\newline
\indent 2) \parbox[t]{.92\linewidth}{группа изотропии $\MG_{\tilde x}$ любой
точки $\tilde x\in\widetilde\MM$ конечна,}\newline
\indent 3) \parbox[t]{.92\linewidth}{для каждой точки $\tilde x$ существует
окрестность $\widetilde\MU$, устойчивая относительно группы изотропии
$\MG_{\tilde x}$, т.е.\ $\widetilde\MU\MG_{\tilde x}=\widetilde\MU$, и
такая, что для всех $a\in \MG$, не лежащих в $\MG_{\tilde x}$, пересечение
$\widetilde\MU a\bigcap\widetilde\MU=\emptyset$.}
\end{defn}
\index{Собственно разрывная группа преобразований %
(properly discontinuous transformation group)}%
\index{Группа преобразований собственно разрывная %
(properly discontinuous transformation group)}%

Ниже мы увидим, что первое условие в данном определении эквивалентно
хаусдорфовости фактор пространства $\MM$ с фактортопологией.
\begin{prop}
Если группа $\MG$ действует на многообразии $\widetilde\MM$ собственно разрывно,
то она конечна или счетна.
\end{prop}
\begin{proof}
В силу условия 3) орбита произвольной точки является дискретным подмножеством
многообразия. Поскольку множество таких точек на многообразии счетно, то, вместе
с условием 2) это влечет счетность элементов группы $\MG$. В частности, оно
может быть конечно.
\end{proof}
Для доказательства следующей теоремы понадобится лемма, которая представляет и
самостоятельный интерес.
\begin{lemma}                                                     \label{lhomma}
Пусть $(\widetilde\MM,\MG)$ -- группа преобразований топологического
пространства $\widetilde\MM$ и фактор пространство $\MM=\widetilde\MM/\MG$
снабжено фактор топологией. Тогда каноническая проекция
$p:~\widetilde\MM\rightarrow\MM$ является открытым отображением.
\end{lemma}
\begin{proof}
Пусть $\widetilde\MU$ -- открытое подмножество $\widetilde\MM$. Тогда
подмножество $p(\widetilde\MU)$ открыто в $\widetilde\MM/\MG$, т.к.\ оно
снабжено фактор топологией. Прообраз этого подмножества можно представить в виде
объединения
\begin{equation}                                                  \label{eproob}
  p^{-1}\big(p(\widetilde\MU)\big)=\bigcup_{a\in\MG}\widetilde\MU a.
\end{equation}
Поскольку отображение $\widetilde\MU\rightarrow\widetilde\MU a$ -- гомеоморфизм
для всех $a\in\MG$, то $p^{-1}\big(p(\widetilde\MU)\big)$ открыто в
$\widetilde\MM$ как объединение открытых подмножеств.
\end{proof}

\begin{theorem}
Пусть группа $\MG$ действует в топологическом пространстве $\widetilde\MM$
собственно разрывно и свободно, а пространство орбит $\widetilde\MM/\MG$
снабжено фактор топологией. Тогда каноническая проекция
$p:~\widetilde\MM\rightarrow\widetilde\MM/\MG$ является накрытием.
\end{theorem}
\begin{proof}
По построению отображение $p$ сюрьективно и непрерывно. По лемме \ref{lhomma}
это отображение открыто. Пусть $\widetilde\MU$ -- окрестность точки
$\tilde x\in\widetilde\MM$, удовлетворяющая условию 3) в определении собственно
разрывной группы преобразований. Так как $p$ -- открытое отображение, то
$p(\widetilde\MU)$ -- открытая окрестность орбиты $p(\tilde x)=\tilde x\MG$ и ее
прообраз имеет вид (\ref{eproob}). Поскольку действие группы свободно, то все
области $\widetilde\MU a$, $a\in\MG$, открыты в $\widetilde\MM$ и
не пересекаются. Сужение отображения на каждую подобласть
$p|_{\widetilde\MU a}:~\widetilde\MU a\rightarrow p(\widetilde\MU)$ является
непрерывным открытым биективным отображением и, следовательно, гомеоморфизмом.
\end{proof}

Теперь обсудим связь между группой преобразований $(\widetilde\MM,\MG)$,
действующей собственно разрывно и свободно, и фундаментальной группой
пространства орбит $\MM=\widetilde\MM/\MG$, на котором определена фактор
топология. Пусть $\tilde x\in\widetilde\MM$ и
$x=p(\tilde x)\in\widetilde\MM/\MG$. Заметим, что
\begin{equation*}
  p^{-1}(x)=\lbrace \tilde xa\in\widetilde\MM:\quad a\in\MG\rbrace,
\end{equation*}
т.е.\ точки орбиты находятся во взаимно однозначном соответствии с элементами
группы. Если класс относительно гомотопных путей принадлежит фундаментальной
группе пространства орбит, $[\g]\in\pi(\widetilde\MM/\MG,x)$, то существует
единственное поднятие $\tilde\g$ пути $\g$ с началом в точке
$\tilde x\in\widetilde\MM$ (теорема \ref{tcoho}). Конец поднятого пути
$\tilde x(1)\in p^{-1}(x)$, и, поскольку группа преобразований действует
свободно, то существует единственный элемент $a_{(\g)}\in\MG$ такой, что
$\tilde x(1)=\tilde xa_{(\g)}$. Следовательно, определено отображение
\begin{equation}                                                  \label{emapat}
  \vf:\quad \pi(\widetilde\MM/\MG,x)\ni\quad[\g]\mapsto a_{(\g)}\quad\in\MG.
\end{equation}
\begin{theorem}
Отображение (\ref{emapat}) является гомоморфизмом групп.
\end{theorem}
\begin{proof}
Сводится к проверке равенства
$\vf([\g_1][\g_2])=\vf([\g_1])\vf([\g_2])$. Детали можно найти, например, в
\cite{Kosnio80R}.
\end{proof}
Если фундаментальная группа накрывающего пространства нетривиальна, то ядро
гомоморфизма также нетривиально.
\begin{prop}
Ядро гомоморфизма (\ref{emapat}) совпадает с индуцированной подгруппой
$p_*\pi(\widetilde\MM,\tilde x)\subset\pi(\widetilde\MM/\MG,x)$.
\end{prop}
\begin{proof}
См., например, \cite{Kosnio80R}.
\end{proof}
В частности, $p_*\pi(\widetilde\MM,\tilde x)$ -- нормальная подгруппа в
$\pi(\widetilde\MM/\MG,x)$, и поэтому определена фактор группа
$\pi(\widetilde\MM/\MG,x)/p_*\pi(\widetilde\MM,\tilde x)$.
\begin{theorem}                                                   \label{tisofu}
Группы $\pi(\widetilde\MM/\MG,x)/p_*\pi(\widetilde\MM,\tilde x)$ и $\MG$
изоморфны.
\end{theorem}
\begin{proof}
См., например, \cite{Kosnio80R}.
\end{proof}
\begin{cor}
Если накрывающее топологическое пространство $\widetilde\MM$ односвязно, то
$\pi(\widetilde\MM/\MG,x)\simeq\MG$.
\qed\end{cor}
Это следствие позволяет в ряде случаев найти фундаментальную группу
многообразия.
\begin{exa}
Нетрудно проверить, что окружность $\MS^1$ гомеоморфна пространству орбит
$\MR/\MZ$, где группа $\MZ$ действует на $\MR$ сдвигами на постоянное число.
Действие группы $\MZ$ на $\MR$ свободно и собственно разрывно. Поскольку
вещественная прямая $\MR$ односвязна, то отображение $p:~\MR\rightarrow\MR/\MZ$
-- универсальное накрытие. Следовательно $\pi(\MS^1)\simeq\MZ$.
\qed\end{exa}
\begin{exa}
Тор является факторпространством, $\MT^n\approx\MR^n/\MZ^n$. Поскольку
отображение $p:~\MR^n\rightarrow\MR^n/\MZ^n$ -- универсальное накрытие, то
$\pi(\MT^n)\simeq\MZ^n$.
\qed\end{exa}
Следующая теорема дает достаточное условие существования дифференцируемой
структуры на пространстве орбит $\widetilde\MM/\MG$.
\begin{theorem}                                                   \label{tfsman}
Пусть группа преобразований $(\widetilde\MM,\MG)$ действует на многообразии
$\widetilde\MM$ собственно разрывно и свободно, тогда факторпространство
$\widetilde\MM/\MG$ с фактор топологией имеет структуру дифференцируемого
многообразия такую, что проекция $p:~\widetilde\MM\rightarrow\widetilde\MM/\MG$
дифференцируема.
\end{theorem}
\begin{proof}
Условие 1) в определении собственно разрывной группы преобразований эквивалентно
хаусдорфовости факторпространства $\widetilde\MM/\MG$. Действительно,
окрестностью орбиты $x=\tilde x\MG$ является множество орбит
$\MU=\widetilde\MU\MG$, где $\widetilde\MU$ -- окрестность точки $\tilde x$, а
условие 1) можно переписать в виде
$\widetilde\MU_1\MG\cap\widetilde\MU_2\MG=\emptyset$, т.к.\ орбиты либо не
имеют общих точек, либо совпадают. Это и есть условие хаусдорфовости
пространства орбит с фактортопологией.

В условиях теоремы группа $\MG$ действует свободно и условие 2) в определении
собственно разрывной группы преобразований выполняется автоматически, т.к.\
группа изотропии любой точки состоит только из одного элемента -- единицы.

Пусть $\MU=\widetilde\MU\MG\subset\widetilde\MM/\MG$ -- достаточно малая
окрестность орбиты $x\in\widetilde\MM/\MG$. Условие 3) вместе с условием 2)
значит, что каждая точка $x$ в пространстве орбит имеет окрестность $\MU$ такую,
что прообраз $\widetilde\MU=p^{-1}(\MU)$ состоит не более, чем из счетного
числа компонент, $\widetilde\MU=\bigcup_i\widetilde\MU_i$, и проекция $p$ каждой
связной компоненты $\widetilde\MU_i$ на $\MU$ есть гомеоморфизм, т.к.\ множество
орбит можно параметризовать точками из какой либо окрестности $\MU_i$.
Зафиксируем связную компоненту $\widetilde\MU_i$ в $p^{-1}(\MU)$. Тогда $p$ --
гомеоморфизм $\widetilde\MU_i$ на $\MU$ и существует обратное непрерывное
отображение $p^{-1}:~\MU\rightarrow\widetilde\MU_i$. Выбрав $\MU$ достаточно
малым, можно считать, что имеется допустимая карта $(\widetilde\MU_i,\vf)$, где
$\vf:~\widetilde\MU_i\rightarrow\MR^n$, многообразия $\widetilde\MM$.
Теперь можно ввести дифференцируемую структуру в пространстве орбит,
рассматривая $(\MU,\psi)$, где $\psi=\vf\circ p^{-1}$, как допустимую карту.
\end{proof}
Последняя теорема является достаточным, но не необходимым условием того, что
пространство орбит является многообразием.
\begin{exa}
Рассмотрим вращения евклидовой плоскости $\MR^2$ вокруг начала координат на
фиксированный угол $\al=2\pi/n$, где $n$ -- одно из натуральных чисел
$2,3,\dotsc$. При этом мы отождествляем
поворот на $2\pi$ с единичным элементом группы. Эта группа $\MG$ абелева и
состоит из $n$ элементов. Действие группы является эффективным. Начало координат
является неподвижной точкой и одной из орбит группы. Ее группа изотропии
совпадает с $\MG$. На остальной части плоскости $\MR^2\setminus\lbrace0\rbrace$
группа $\MG$ действует свободно. Нетрудно проверить, что группа $\MG$ действует
на плоскость собственно разрывно. Пространство орбит $\MR^2/\MG$ представляет
собой конус с углом дефицита $2\pi/n-2\pi$ (знак минус означает, что угол
вырезается, а не вставляется). Конус является гладким двумерным многообразием,
т.е.\ пространство орбит допускает гладкую структуру, несмотря на то, что
действие группы не является свободным. Вершина конуса -- это ``дефект'' не
многообразия, а вложения. В этом примере каноническая проекция
$p:~\MR^2\rightarrow\MR^2/\MG$ не является накрытием, так как прообраз
$p^{-1}(\MU_0)$ окрестности $\MU_0\subset\MR^2/\MG$, содержащей начало
координат, состоит из одного связного листа, который не гомеоморфен самой
окрестности (нет взаимной однозначности). Фундаментальные группы плоскости
$\MR^2$ и конуса $\MR^2/\MG$ тривиальны и совпадают.
\qed\end{exa}

Если пара $(\widetilde\MM,\MG)$ -- группа преобразований, причем группа $\MG$
действует на многообразии $\widetilde\MM$ свободно и собственной разрывно, то
одним из способов изучения фактор пространства $\widetilde\MM/\MG$ является
построение фундаментальной области.
\begin{defn}
Подмножество $\MD\subset\widetilde\MM$ называется {\em фундаментальной областью}
многообразия $\widetilde\MM$ для группы преобразований $(\widetilde\MM,\MG)$,
действующей свободно и собственно разрывно, если выполнены следующие условия:

1) \parbox[t]{.92\linewidth}{$\MD$ является замкнутым подмножеством в
$\widetilde\MM$;}

2) \parbox[t]{.92\linewidth}{
орбита $\MD\MG$ совпадает со всем многообразием $\widetilde\MM$;}

3) \parbox[t]{.92\linewidth}{покрытие $\widetilde\MM$ множествами $\MD a$,
$a\in \MG$, таково, что с достаточно малой окрестностью произвольной
точки $\widetilde\MM$ пересекается лишь конечное число множеств вида $\MD a$;}

4) \parbox[t]{.92\linewidth}{образ множества всех внутренних точек
фундаментальной области, $(\Int\MD)a$, при действии любого преобразования
$a\in \MG$, отличного от единичного, не пересекается с множеством внутренних
точек фундаментальной области, $(\Int\MD)a\cap\Int\MD=\emptyset$, $a\ne e$.\qed}
\end{defn}
\index{Фундаментальная область (fundamental domain)}%
\index{Область фундаментальная (fundamental domain)}%
Фундаментальная область $\MD$ всегда является многообразием с краем той же
размерности, что и само $\widetilde\MM$.
\begin{exa}
Группа трансляций $\MG$ евклидовой плоскости $\MR^2$ на всевозможные векторы
с целочисленными компонентами
\begin{equation*}
  \MR^2\ni\quad(x,y)\mapsto(x+m,y+n)\quad\in\MR^2,\qquad \forall m,n\in\MZ,
\end{equation*}
действует гладко, свободно и собственно разрывно. Фактор пространство
$\MR^2/\MG$ представляет собой тор $\MT^2$. В качестве фундаментальной области
можно выбрать единичный квадрат
\begin{equation*}
  \MD=\lbrace (x,y)\in\MR^2:\quad 0\le x\le1,~0\le y\le1\rbrace.
\end{equation*}
Тор можно представить в виде квадрата на плоскости, у которого отождествлены
противоположные стороны.
\qed\end{exa}
Из определения фундаментальной области следует, что отображение
$\widetilde\MM\rightarrow\Int\MD$ всегда представляет собой накрытие. Если
фундаментальная область известна, то пространство орбит $\widetilde\MM/\MG$
получается из фундаментальной области путем склеивания граничных точек.
\section{Группа скольжений и существование накрытий}
В предыдущем разделе было показано, что отображение многообразия в пространство
орбит для группы преобразований, действующей свободно и собственно разрывно,
является накрытием. Теперь мы рассмотрим обратный вопрос о том, в каком случае
заданное накрытие можно представить, в виде отображения накрывающего
пространства в пространство орбит относительно действия некоторой группы
преобразований и какова эта группа.

Введем новое понятие, которое дает возможность описать произвол, существующий
при построении накрывающего пространства, если база задана.
\begin{defn}
{\em Группой скольжений} накрытия $p:~\widetilde\MM\rightarrow\MM$ называется
группа всех гомеоморфизмов $h:~\widetilde\MM\rightarrow\widetilde\MM$, при
которых $p\circ h=p$, т.е.\ диаграмма
\begin{equation*}
\begin{diagram}
  \widetilde\MM & \rTo^h & \widetilde\MM \\
  & \rdTo_p &  \dTo_p \\
  &  & \MM
\end{diagram}
\end{equation*}
коммутативна. Эта группа обозначается $\MG(\widetilde\MM,p,\MM)$.
\qed\end{defn}
\index{Группа скольжений (sliding group)}%
\index{Скольжений группа (sliding group)}%
\begin{theorem}
Если топологическое пространство $\widetilde\MM$ связно и локально линейно
связно, то действие группы скольжений $\MG(\widetilde\MM,p,\MM)$ на
$\widetilde\MM$ свободно и собственно разрывно.
\end{theorem}
\begin{proof}
См., например, \cite{Kosnio80R}.
\end{proof}
В частности, действие группы скольжений на связном многообразии $\widetilde\MM$
является свободным и собственно разрывным. В этом случае группа скольжений
$\MG(\widetilde\MM,p,\MM)$ конечна или счетна. Поэтому пара
$\big(\widetilde\MM,\MG(\widetilde\MM,p,\MM)\big)$ является группой
преобразований.
\begin{theorem}                                                   \label{tbasho}
Пусть $p:~\widetilde\MM\rightarrow\MM$ -- накрытие и накрывающее топологическое
пространство $\widetilde\MM$ связно и локально линейно связно. Если
индуцированная группа $p_*\pi(\widetilde\MM,\tilde x)$ является нормальной
подгруппой $\pi(\MM,x)$, где $x=p(\tilde x)$, то база $\MM$ гомеоморфна
пространству орбит $\widetilde\MM/\MG(\widetilde\MM,p,\MM)$.
\end{theorem}
\begin{proof}
См., например, \cite{Kosnio80R}.
\end{proof}
\begin{cor}
Пусть пространство $\widetilde\MM$ связно и локально линейно связно. Если
$p_*\pi(\widetilde\MM,\tilde x)$ -- нормальная подгруппа в $\pi(\MM,x)$, то
$\pi(\MM,x)/p_*\pi(\widetilde\MM,\tilde x)\simeq\MG(\widetilde\MM,p,\MM)$.
\qed\end{cor}
\begin{proof}
Прямое следствие теорем \ref{tbasho} и \ref{tisofu}.
\end{proof}
\begin{cor}
Если накрывающее пространство $\widetilde\MM$ односвязно и локально линейно
связно, то $\pi(\MM,x)\simeq\MG(\widetilde\MM,p,\MM)$.
\qed\end{cor}

По своей сути две предыдущие теоремы и следствия означают, что произвольное
накрытие можно представить, как отображение многообразия $\widetilde\MM$ в
пространство орбит $\MM=\widetilde\MM/\MG(\widetilde\MM,p,\MM)$. При этом роль
группы преобразований играет группа скольжений $\MG(\widetilde\MM,p,\MM)$. Если
накрывающее пространство $\widetilde\MM$ -- многообразие, то действие группы
скольжений сводится к перестановке листов накрытия.

Теперь обсудим вопрос о существовании универсального накрытия для заданной базы
$\MM$. Универсальное накрытие существует, если на топологическое пространство
$\MM$ наложен ряд условий. Чтобы не вводить новых понятий для их формулировки,
мы ограничимся многообразиями, для которых эти условия выполняются.
\begin{theorem}                                                   \label{texunc}
Произвольное связное $n$-мерное многообразие $\MM$ имеет универсальное накрытие
$p:~\widetilde\MM\rightarrow\MM$, где накрывающее пространство $\widetilde\MM$
-- также $n$-мерное многообразие (связное и односвязное). Универсальное
накрывающее пространство определено с точностью до действия группы скольжений
$\MG(\widetilde\MM,p,\MM)$, которая действует на $\widetilde\MM$ свободно и
собственно разрывно.
\end{theorem}
\begin{proof}
См., например, \cite{Kosnio80R}.
\end{proof}
Эта теорема очень важна, т.к.\ позволяет разделить задачу классификации
многообразий на два этапа: сначала описать все односвязные многообразия, а затем
найти все группы преобразований, действующие на них свободно и собственно
разрывно. Данная задача решена для поверхностей. Для многообразий размерности
три и выше вопрос остается открытым.
\chapter{Главные и ассоциированные расслоения                    \label{sprifb}}
Теория расслоений или расслоенных пространств играет важнейшую роль в
современной математической физике. Достаточно отметить, что в основе общей
теории относительности лежит расслоение реперов, которое является главным
расслоением со структурной группой $\MG\ML(n,\MR)$. В основе калибровочных
моделей элементарных частиц лежит главное расслоение с полупростой компактной
структурной группой, которой обычно является некоторая подгруппа унитарной
группы $\MU(n)$. В настоящем разделе мы дадим определения и рассмотрим основные
свойства главных и ассоциированных расслоений.
\section{Главные расслоения                                      \label{sprfib}}
В дифференциальной геометрии важнейшую роль играют главные расслоения.
Фактически, они лежат в основе многих геометрических конструкций.
\begin{defn}
{\em Главным расслоением} называется четверка $\MP(\MM,\pi,\MG)$, где $\MP$ и
$\MM$ -- многообразия, $\MG$ -- группа Ли, $\pi:~\MP\rightarrow\MM$ --
отображение, которые удовлетворяют следующим условиям:\newline
\indent 1) \parbox[t]{.92\linewidth}{определено свободное дифференцируемое
действие группы $\MG$ на $\MP$ справа:
\begin{equation}                                                  \label{epract}
  \MP\times\MG\ni\quad (p,a)\mapsto pa\quad\in\MP;
\end{equation}
}\newline
\indent 2) \parbox[t]{.92\linewidth}{$\MM$ есть факторпространство для $\MP$ по
отношению эквивалентности, индуцированному действием группы $\MG$, и
каноническая проекция $\pi:~\MP\rightarrow\MM=\MP/\MG$ дифференцируема;}\newline
\indent 3) \parbox[t]{.92\linewidth}{каждая точка $x\in\MM$ имеет окрестность
$\MU_x$ такую, что существует диффеоморфизм
\begin{equation}                                                  \label{eloctr}
  \chi:\quad\pi^{-1}(\MU_x)\ni\quad p\mapsto\chi(p)=\big(\pi(p),\vf(p)\big)
  \quad\in \MU_x\times\MG
\end{equation}
такой, что отображение $\vf:~\pi^{-1}(\MU_x)\rightarrow\MG$ удовлетворяет
условию $\vf(pa)=\big(\vf(p)\big)a$ для всех $p\in\pi^{-1}(\MU_x)$ и $a\in\MG$
(локальная тривиальность).}
\newline
Многообразие $\MP$ называется {\em пространством расслоения}, $\MM$ --
{\em базой расслоения}, $\MG$ -- {\em структурной группой} и $\pi$ --
{\em проекцией}.
\qed\end{defn}
\index{Главное расслоение (principal fiber bundle)}%
\index{Расслоение главное (principal fiber bundle)}%
\index{Пространство расслоения (fiber bundle space)}%
\index{База расслоения (base of a fiber bundle}%
\index{Проекция (projection)}%

Поскольку отображение (\ref{eloctr}) является диффеоморфизмом, то
\begin{equation*}
  \dim\MP=\dim\MM+\dim\MG.
\end{equation*}
Иногда в качестве структурной группы мы будем рассматривать не группу Ли, а
группу, состоящую из конечного или счетного набора элементов. Для удобства такие
группы мы будем считать $0$-мерными группами Ли. В этом случае
$\dim\MP=\dim\MM$.
\begin{com}
Каждое главное расслоение является расслоением в смысле определения, данного в
разделе \ref{sfibun}. В дополнение к общему определению расслоения мы
зафиксировали типичный слой, предположив, что им является группа Ли $\MG$, и
добавили действие этой группы на пространстве расслоения $\MP$ так, чтобы оно
было согласовано с проекцией. Дифференцируемая структура на $\MP$ согласована с
дифференцируемыми структурами на $\MM$ и $\MG$, поскольку отображение $\chi$, по
предположению, является диффеоморфизмом. На самом деле, можно было бы
потребовать только непрерывность отображения $\chi$, а затем с его помощью
перенести дифференцируемую структуру с $\MM$ и $\MG$ на пространство расслоения
$\MP$.
\qed\end{com}
\begin{com}
Пара $(\MP,\MG)$ является группой преобразований, определенной в разделе
\ref{stragf}. Однако, не всякая группа преобразований есть главное расслоение.
Напомним, что в общем случае пространство орбит $\MM/\MG$ группы преобразований
$(\MM,\MG)$ может оказаться нехаусдорфовым топологическим пространством и на нем
нельзя ввести дифференцируемую структуру.
\qed\end{com}

Условие локальной тривиальности главного расслоения можно изобразить в виде
коммутативной диаграммы
\begin{equation*}
\begin{diagram}
  \MP\supset\pi^{-1}(\MU_x) & \rTo^{\chi} & \MU_x\times\MG \\
  & \rdTo_\pi & \dTo_{\pr=\pi\circ\chi^{-1}} \\ & & \MU_x
\end{diagram}
\end{equation*}
где $\pr$ -- естественная проекция прямого произведения
$\MU_x\times\MG\rTo^\pr \MU_x$ на первый сомножитель. Групповое действие на
пространстве расслоения описывается эквивариантной диаграммой
\begin{equation*}
\begin{diagram}
  \pi^{-1}(\MU_x)\times\MG & \rTo^{\chi\circ\id} & (\MU_x\times\MG)\times\MG \\
  \dTo^{a_\MP} & & \dTo_{a_{\MU_x\times\MG}} \\
  \pi^{-1}(\MU_x)& \rTo^\chi & \MU_x\times\MG
\end{diagram},
\end{equation*}
где $a_\MP$ и $a_{\MU_x\times\MG}$ обозначают действие элемента группы $a\in\MG$
на пространстве расслоения и прямом произведении $\MM\times\MG$.
\begin{defn}
Для каждой точки базы $x\in\MM$ множество $\pi^{-1}(x)$ есть замкнутое
подмногообразие в пространстве расслоения $\MP$, которое называется {\em слоем}
над $x$. {\em Сечением} или {\em глобальным сечением} главного расслоения
$\MP(\MM,\pi,\MG)$ называется дифференцируемое отображение
$\s:~\MM\rightarrow\MP$ такое, что $\pi\circ\s=\id_\MM$. Дифференцируемое
отображение $\s:~\MU\rightarrow\MP$, где $\MU\subset\MM$ -- некоторая
окрестность базы, называется {\em локальным сечением} над $\MU$, если
$\pi\circ\s=\id_\MU$.
\index{Сечение расслоения (fiber bundle cross-section)}%
\index{Локальное сечение расслоения (local fiber bundle cross-section)}%
\index{Глобальное сечение расслоения (global fiber bundle cross-section)}%
\qed\end{defn}
\index{слой (fiber)}%
Каждый слой является орбитой $p\MG$ какой либо точки $p\in\pi^{-1}(x)$.
То, что каждый слой представляет собой замкнутое подмногообразие в $\MP$
является следствием предложения \ref{pclosm}. Поскольку действие группы Ли $\MG$
в каждом слое свободно и транзитивно, то между точками слоя и структурной группы
имеется взаимно однозначное соответствие. То есть каждый слой $\pi^{-1}(x)$
диффеоморфен $\MG$. Этот диффеоморфизм осуществляет функция $\vf$ в отображении
(\ref{eloctr}) при фиксированном $x\in\MM$.

При каждом фиксированном $p\in\MP$ отображение (\ref{epract}) дифференцируемо.
Поэтому каждой точке главного расслоения соответствует диффеоморфизм структурной
группы на типичный слой в данной точке
\begin{equation}                                                  \label{epdema}
  p:\quad \MG\ni\quad a\mapsto pa\quad\in\pi^{-1}(x),
\end{equation}
где $x=\pi(p)$. В дальнейшем мы иногда будем рассматривать точку $p$ именно в
этом смысле, как отображение.

Сечение $\s$, если оно существует, не может быть сюрьективным отображением,
т.к.\ размерность базы в общем случае меньше размерности расслоения. Для
накрытий размерность базы совпадает с размерностью главного расслоения. Если
накрытие многолистно, то сечение также не является сюрьективным отображением.
Сечение сюрьективно в одном случае, когда структурная группа состоит из
единственного элемента -- единицы. Сужение проекции $\pi$ на образ базы
$\s(\MM)$ в $\MP$ является дифференцируемым отображением, которое обратно к
сечению $\s$. Поэтому пара $(\s,\MM)$ является подмногообразием в $\MP$. Если
$\dim\MG\ge1$, то это подмногообразие замкнуто в $\MP$ (см.\ раздел
\ref{subman}).
\begin{exa}                                                       \label{extrib}
Пусть $\MG$ -- группа Ли и $\MM$ -- многообразие. Определим действие группы
$\MG$ справа на прямом произведении $\MP=\MM\times\MG$:
\begin{equation*}
  \MG\ni b:\quad \MM\times\MG\ni\quad x,a\mapsto x,ab\quad\in\MM\times\MG.
\end{equation*}
Тогда четверка $\MP(\MM,\pi,\MG)$, где $\pi:~\MM\times\MG\rightarrow\MM$ --
каноническая проекция, является главным расслоением. Это главное расслоение
имеет глобальное сечение
$\s:~\MM\rightarrow\MP$, $\pi\circ\s=\id_\MM$. Например,
\begin{equation*}                                                    \tag*{\qed}
  \s:\quad\MM\ni\quad x\mapsto x,e\quad\in\MM\times\MG
\end{equation*}
\end{exa}
\begin{defn}
Главное расслоение $\MP(\MM,\pi,\MG)$ называется {\em тривиальным}, если оно
изоморфно главному расслоению вида $\MM\times\MG\xrightarrow{\pr_\MM}\MM$.
\qed\end{defn}
\index{Тривиальное главное расслоение (trivial principal fiber bundle)}%
\index{Главное расслоение тривиальное (trivial principal fiber bundle)}%
\begin{exa}
Расслоение реперов $\ML(\MM)=\MP\big(\MM,\pi,\MG\ML(n,\MR)\big)$, рассмотренное
в разделе \ref{scorep}, является главным расслоением со структурной группой
$\MG\ML(n,\MR)$.
\qed\end{exa}
\begin{exa}
Пусть $\MG$ -- группа Ли и $\MH\subset\MG$ -- ее замкнутая подгруппа. Согласно
теореме \ref{tfacma} на пространстве правых смежных классов $\MH a\in\MG/\MH$,
где $a\in\MG$, можно задать дифференцируемую структуру. Тогда
$\MG(\MG/\MH,\pi,\MH)$ -- главное расслоение с базой $\MG/\MH$, структурной
группой $\MH$ и проекцией $\pi:~\MG\rightarrow\MG/\MH$.
\qed\end{exa}
\begin{exa}
Накрытия $\widetilde\MM\big(\MM,\pi,\MG(\widetilde\MM,p,\MM)\big)$, где
$p:~\widetilde\MM\rightarrow\MM$ -- отображение накрытия, рассмотренные в
разделе \ref{scover}, являются главными расслоениями с $0$-мерной структурной
группой Ли, которой является группа скольжений $\MG(\widetilde\MM,p,\MM)$.
\qed\end{exa}
\begin{exa}
Множество вещественных чисел без нуля образует абелеву группу относительно
умножения, которую мы обозначим $\MR_0:=\MR\setminus\lbrace0\rbrace$. Пусть эта
группа действует в евклидовом пространстве без начала координат
$\MR^n\setminus\lbrace0\rbrace$ посредством умножения декартовых координат:
\begin{equation*}
  \MR_0\ni a:\quad (x^1,\dotsc,x^n)\mapsto(ax^1,\dotsc,ax^n).
\end{equation*}
Это действие дифференцируемо, свободно и транзитивно на орбитах. Поэтому
$\MR^n\setminus\lbrace0\rbrace\xrightarrow{\pi}\MR\MP^{n-1}$ -- тривиальное
главное расслоение со структурной группой $\MR_0$, базой которого является
вещественное проективное пространство
$\MR\MP^{n-1}=\big(\MR^n\setminus\lbrace0\rbrace\big)/\MR_0$.
\qed\end{exa}
\begin{exa}
Если в предыдущем примере группу $\MR_0$ заменить на группу положительных
чисел $\MR_+$ по умножению, то получим тривиальное главное расслоение
$\MR^n\setminus\lbrace0\rbrace\xrightarrow{\pi}\MS^{n-1}$ со структурной группой
$\MR_+$, базой которого является сфера
$\MS^{n-1}=\big(\MR^n\setminus\lbrace0\rbrace\big)/\MR_+$.
\qed\end{exa}
\begin{exa}
Пусть $\MC_0:=\MC\setminus\lbrace0\rbrace$ -- группа комплексных чисел по
отношению к умножению, которая действует в $n$-мерном комплексном
пространстве без начала координат $\MC^n\setminus\lbrace0\rbrace$ умножением,
\begin{equation*}
  \MC_0\ni a:\quad (z^1,\dotsc,z^n)\mapsto(az^1,\dotsc,az^n).
\end{equation*}
Тогда $\MC^n\setminus\lbrace0\rbrace\xrightarrow{\pi}\MC\MP^{n-1}$ --
тривиальное главное расслоение со структурной группой $\MC_0$, базой которого
является комплексное проективное пространство
$\MC\MP^{n-1}=\big(\MC^n\setminus\lbrace0\rbrace\big)/\MC_0$.
\qed\end{exa}
В примере \ref{extrib} мы отметили, что произвольное тривиальное расслоение
имеет глобальное сечение. Справедливо и обратное утверждение.
\begin{prop}                                                      \label{ptrpri}
Если главное расслоение $\MP(\MM,\pi,\MG)$ имеет глобальное сечение
$\s:~\MM\rightarrow\MP$, то оно изоморфно тривиальному, т.е.\ существует
диффеоморфизм
\begin{equation}                                                  \label{edidfp}
  f_\MP:\quad \MP\ni\quad p\mapsto\big(\pi(p),\vf(p)\big)\quad\in\MM\times\MG
\end{equation}
такой, что следующая диаграмма коммутативна:
\begin{equation}                                                  \label{etrivp}
\begin{diagram}
  \MP & \rTo^{f_\MP} & \MM\times\MG \\
  & \rdTo_\pi & \dTo_{\pr_\MM} \\
  &  & \MM
\end{diagram}
\end{equation}
где $\pr_\MM$ -- проекция на первый сомножитель, и справедливо равенство
\begin{equation}                                                  \label{equima}
  f_\MP(pa)=\big(\pi(p),\vf(p)a\big)\qquad \forall~p\in\MP,~a\in\MG.
\end{equation}
\end{prop}
\begin{proof}
Пусть
\begin{equation*}
  \s:\quad \MM\ni\quad x\mapsto\s(x)\quad\in\MP,\qquad \pi\circ\s=\id_\MM,
\end{equation*}
-- глобальное сечение главного расслоения. Для каждой точки слоя
$p\in\pi^{-1}(x)$ определим единственный элемент группы $a\in\MG$, для которого
$p=\s(x)a(p)$. Таким образом определено отображение
$p\mapsto f_\MP(p)=\big(\pi(p),a(p)\big)$. Легко проверить, что это отображение
удовлетворяет требуемым свойствам.
\end{proof}
\begin{cor}
Нетривиальные главные расслоения имеют только локальные сечения.
\qed\end{cor}
В общем случае изоморфизм расслоений будет определен позже в разделе
\ref{shofbu}.

Диаграмму (\ref{etrivp}) вместе с условием (\ref{equima}) можно изобразить в
виде коммутативной диаграммы
\begin{equation*}
\begin{diagram}
  \MP\times\MG & \rTo^{f_\MP\circ\id} & (\MM\times\MG)\times\MG \\
  \dTo^{a_\MP} &  & \dTo_{a_{\MM\times\MG}} \\
  \MP & \rTo^{f_\MP} & \MM\times\MG
\end{diagram},
\end{equation*}
где $a_\MP$ и $a_{\MM\times\MG}$ обозначает действие элемента $a\in\MG$
соответственно на $\MP$ и $\MM\times\MG$. Это означает, что отображение
(\ref{edidfp}) является эквивариантным (см.\ раздел \ref{seqyim}).
\begin{theorem}                                                   \label{trigla}
Если база главного расслоения $\MP(\MM,\pi,\MG)$ диффеоморфна евклидову
пространству, $\MM\approx\MR^n$, то главное расслоение изоморфно тривиальному,
$\MP(\MM,\pi,\MG)\simeq\MM\times\MG$.
\end{theorem}
\begin{proof}
Рассмотрим главное расслоение $\MP(\MR^n,\pi,\MG)$, базой которого является
евклидово пространство. Согласно теореме \ref{tsushc} любое главное расслоение
допускает связность. Предположим, что на $\MP(\MR^n,\pi,\MG)$ задана какая либо
связность. Рассмотрим в $\MR^n$ сферическую систему координат. Тогда каждая
отличная от начала координат точка $x\in\MR^n$ параметризуется парой $(r,n)$,
где $r$ -- расстояние от начала координат и $n$ -- единичный вектор,
определяющий направление луча, выходящего из начала координат и на котором лежит
точка $x$. Пусть $p_0\in\pi^{-1}(0)$ -- произвольная точка из слоя над началом
координат. Согласно предложению \ref{pholcu} для каждого луча существует его
единственный горизонтальный лифт в пространство расслоения $\MP$ с началом в
точке $p_0$. Обозначим через $p(x)$ единственную точку на горизонтальном лифте
луча, которая лежит над $x$, т.е.\ $\pi(p)=x$. Таким образом, $p(x)$ -- это
глобальное сечение главного расслоения. Следовательно, по предложению
\ref{ptrpri} главное расслоение тривиально $\MP=\MR^n\times\MG$. Если база
диффеоморфна $\MR^n$, то сферическая система координат просто переносится на
$\MM$ с помощью диффеоморфизма, и построение глобального сечения повторяется.
\end{proof}
При рассмотрении многообразий в разделе \ref{sdefma} были введены функции
склейки, с помощью которых осуществляется преобразование координат в двух
пересекающихся картах. Обобщением этого понятия на случай главных расслоений
являются функции перехода, которые вводятся следующим образом.
\begin{defn}
В силу условия 3) в определении главного расслоения на базе можно выбрать такое
координатное покрытие, $\MM=\bigcup_i\MU_i$, что
\begin{equation*}
  \chi_i:\quad
  \pi^{-1}(\MU_i)\ni\quad p\mapsto\big(\pi(p),\vf_i(p)\big)\quad
  \in\MU_i\times\MG,
\end{equation*}
причем $\vf_i(pa)=\vf_i(p)a$.

Пусть две карты пересекаются, $\MU_i\cap\MU_j\ne\emptyset$. Если
$p\in\pi^{-1}(\MU_i\cap\MU_j)$, то
\begin{equation*}
  \vf_j(pa)\circ\vf_i(pa)^{-1}=\vf_j(p)\circ\vf_i(p)^{-1},
\end{equation*}
где $\vf_i$ рассматривается как отображение фиксированного слоя
$\pi^{-1}\big(\pi(p)\big)$ в группу $\MG$. Следовательно, отображение
$\vf_j(p)\circ\vf_i(p)^{-1}$ зависит только от точки базы $x=\pi(p)$.
Поэтому определено отображение
\begin{equation}                                                  \label{etrasf}
  a_{ji}:\quad \MU_i\cap\MU_j\ni\quad x=\pi(p)\mapsto a_{ji}(x)
  :=\vf_j(p)\circ\vf_i(p)^{-1}\quad\in\MG.
\end{equation}
Эти функции на $\MM$ со значениями в $\MG$ называются {\em функциями перехода}
или {\em функциями склейки} главного расслоения $\MP(\MM,\pi,\MG)$,
соответствующими координатному покрытию $\MM=\bigcup_i\MU_i$. Набор функций
склейки называется {\em склеивающим коциклом} главного расслоения $\MP$.
\qed\end{defn}
\index{Функция перехода (transition function)}%
\index{Перехода функция (transition function)}%
\index{Функция склейки (sewing function)}%
\index{Склейки функция (sewing function)}%
\index{Склеиващий коцикл (sewing cocycle)}%
\index{Коцикл склеиващий (sewing cocycle)}%

Функции перехода показывают насколько в различных картах ``сдвинуты'' образы
слоя в структурной группе над фиксированной точкой $x\in\MM$.

Из определения (\ref{etrasf}) следует, что функции склейки обладают следующим
свойством
\begin{equation}                                                  \label{einglf}
  a_{ij}(x)=\big(a_{ji}(x)\big)^{-1}\qquad \forall x\in\MU_i\cap\MU_j.
\end{equation}
Кроме того, нетрудно проверить, что для трех пересекающихся карт выполнено
равенство
\begin{equation}                                                  \label{etrfco}
  a_{ij}(x)a_{jk}(x)a_{ki}(x)=e,\qquad \forall x\in\MU_i\cap\MU_j\cap\MU_k,
\end{equation}
где $e$ -- единица структурной группы. Поэтому определение функций перехода
корректно.
\begin{exa}
Пусть $\MM$, $\dim\MM=n$, -- многообразие. Рассмотрим расслоение реперов
$\ML(\MM)$ (см.\ раздел \ref{scorep}). Пусть $\MU_i$ и $\MU_j$ -- две
пересекающиеся карты с координатами $x^\al$ и $x^{\al'}$, $\al,\al'=1,\dotsc,n$.
Репер в этих картах имеет компоненты $e^\al{}_a\in\MG\ML(n,\MR)$ и
$e^{\al'}{}_a\in\MG\ML(n,\MR)$, $a=1,\dotsc,n$, которые связаны между собой
преобразованием
\begin{equation*}
  e^{\al'}{}_a=e^\al{}_a\pl_\al x^{\al'}.
\end{equation*}
Таким образом, функциями перехода для расслоения реперов являются матрицы Якоби
преобразования координат. Эти матрицы, как легко проверить, удовлетворяют
условиям (\ref{einglf}) и (\ref{etrfco}).
\qed\end{exa}

Таким образом, для каждого главного расслоения можно однозначно построить
семейство функций перехода, соответствующих заданному координатному покрытию
базы, и эти функции перехода удовлетворяют равенствам (\ref{einglf}),
(\ref{etrfco}). Справедливо также обратное утверждение.
\begin{theorem}                                                   \label{tlofib}
Пусть $\MM$ -- многообразие с координатным покрытием $\MM=\bigcup_i\MU_i$ и
$\MG$ -- группа Ли. Если заданы отображения
$a_{ji}:~\MU_i\cap\MU_j\rightarrow\MG$ для всех непустых пересечений
$\MU_i\cap\MU_j$ такие, что выполнены условия (\ref{einglf}) и (\ref{etrfco}) во
всех областях пересечения карт, то существует единственное с точностью до
изоморфизма главное расслоение $\MP(\MM,\pi,\MG)$ с функциями перехода $a_{ji}$.
\end{theorem}
\begin{proof}
Рассмотрим несвязное объединение $\MQ:=\sqcup_i(\MU_i\times\MG)$. Введем на этом
множестве отношение эквивалентности:
\begin{align*}
  (x,a)&\sim\big(x,aa_{ij}(x)\big),     &\forall\quad  &(x,a)\in\MU_i\times\MG,
  \quad\big(x,aa_{ij}(x)\big)\in\MU_j\times\MG,
\\
  (x,a)&\sim(x,a),      &\forall\quad  &(x,a)\in\MU_i\times\MG.
\end{align*}
Из свойств (\ref{einglf}), (\ref{etrfco}) следует, что это действительно
отношение эквивалентности. Обозначим фактор пространство $\MQ/\sim$ через
$\MP$ и введем на нем естественную дифференцируемую структуру. Пусть
$p:=\langle x,a\rangle$ -- точка $\MP$, т.е.\ класс эквивалентности пары
$(x,a)$. Определим действие группы $\MG$ на $\MP$ формулой
\begin{equation*}
  \MP\times\MG\ni\quad\langle x,a\rangle,b\mapsto\langle x,ab\rangle\quad\in\MP.
\end{equation*}
Определим также проекцию
\begin{equation*}
  \pi:\quad \MP\ni\quad\langle x,a\rangle\mapsto\pi(\langle x,a\rangle):=x
  \quad\in\MM.
\end{equation*}
Нетрудно проверить, что все свойства главного расслоения для четверки
$\MP(\MM,\pi,\MG)$ выполнены.

Теперь докажем единственность построенного главного расслоения с точностью до
изоморфизма. Пусть множество функций склеек $\lbrace a_{ji}(x)\rbrace$ построено
для некоторого главного расслоения $\MP'(\MM,\pi',\MG)$ с фиксированным
покрытием базы $\MM=\cup_i\MU_i$. Построим гомеоморфизм
$f_\MP:~\MP\rightarrow\MP'$, где $f_\MP$ совпадает с $\chi_i^{-1}$ на каждом
$\MU_i\times\MG$. Для корректности этого определения нужно убедиться, что
отображения $\chi_i^{-1}$ и $\chi_j^{-1}$ совпадают на общей области
определения. Действительно, если $x\in\MU_i\cap\MU_j$ и
$(x,a)\in\MU_i\times\MG$, то по построению пары $(x,a)$ и $(x,aa_{ji})$
определяют одну и ту же точку $\langle x,a\rangle\in\MP$. Из определения функций
склейки $a_{ji}(x)$ следуют равенства:
\begin{equation*}
  \chi_i^{-1}(x,a)=\chi_j^{-1}\circ\chi_j\circ\chi_i^{-1}(x,a)
  =\chi_j^{-1}(x,aa_{ji}).
\end{equation*}
Эквивариантность отображения $f_\MP$ и коммутативность диаграммы
\begin{equation}                                                  \label{epeqvi}
\begin{diagram}
  \MP & \rTo^{f_\MP} & \MP' \\
  & \rdTo_{\pi} & \dTo_\pi' \\
  &  & \MM
\end{diagram}
\end{equation}
следуют из построения. Это и означает, что главные расслоения $\MP(\MM,\pi,\MG)$
и $\MP'(\MM,\pi',\MG)$ изоморфны. (Подробнее изоморфизм расслоений
рассматривается в разделе \ref{shofbu}. Там будет показано, что построенный
изоморфизм $f_\MP$ относится к классу вертикальных автоморфизмов.).
\end{proof}

Склеивающий коцикл определяется расслоением не однозначно. Он зависит от выбора
координатного покрытия базы $\MM=\bigcup_i\MU_i$ и тривиализаций
$\lbrace\chi_i\rbrace$.
\begin{defn}
Два коцикла $\lbrace a_{ji}\rbrace$ и $\lbrace a_{ji}^\prime\rbrace$,
соответствующие заданному координатному покрытию базы $\MM=\bigcup_i\MU_i$,
называются {\em эквивалентными}, если существуют отображения
$b_i:~\MU_i\rightarrow\MG$ такие, что выполнены равенства
\begin{equation*}                                                    \tag*{\qed}
  a_{ji}^\prime=b_ja_{ji}b_i^{-1},\qquad \forall x\in\MU_i\cap\MU_j.
\end{equation*}
\end{defn}
\begin{theorem}
Два главных расслоения $\MP(\MM,\pi,\MG)$ и $\MP'(\MM,\pi',\MG)$ изоморфны
тогда и только тогда, когда их склеивающие коциклы $\lbrace a_{ji}\rbrace$ и
$\lbrace a_{ji}^\prime\rbrace$, соответствующие некоторому координатному
покрытию базы $\MM=\bigcup_i\MU_i$, эквивалентны.
\end{theorem}
\begin{proof}
См., например, \cite{Bolibr00R}.
\end{proof}
\begin{exa}[\bf Расслоение Хопфа]
\index{Расслоение Хопфа (Hopf fiber bundle)}%
\index{Хопфа расслоение (Hopf fiber bundle)}%
Реализуем трехмерную сферу $\MS^3$ в двумерном комплексном пространстве $\MC^2$
с помощью вложения
\begin{equation}                                                  \label{esthco}
  |z^1|^2+|z^2|^2=1,\qquad (z^1,z^2)\in\MC^2.
\end{equation}
Каждое уравнение $az^1+bz^2=0$, где $(a,b)\in\MC^2$ и по крайней мере одно из
комплексных чисел $a,b$ отлично от нуля, задает {\em комплексную прямую},
комплексной размерности один,
проходящую через начало координат. Обратно. Комплексная прямая определяет пару
комплексных чисел $(a,b)\in\MC^2$ с точностью до отношения эквивалентности
$(a,b)\sim(a',b')$, если и только если $|a'|=\lm|a|$, $|b'|=\lm|b|$,
$\lm\in\MR_+$, или $a'=a\ex^{it}$, $b'=b\ex^{it}$, $t\in[0,2\pi]$. Множество
всех комплексных прямых, проходящих через начало координат, является одномерным
комплексным проективным пространством $\MC\MP^1$ и называется {\em комплексной
проективной прямой}. Каждая комплексная прямая, проходящая через начало
координат, пересекает трехмерную сферу $\MS^3$ по большой окружности $\MS^1$,
называемой {\em окружностью Хопфа}, которая задается одной из двух систем
уравнений:
\begin{equation*}
\begin{split}
  |z^1|^2&=\frac1{1+|k|^2},\qquad z^2=kz^1,\qquad k:=-\frac ab,\quad b\ne0,
\\
  |z^2|^2&=\frac1{1+|k|^2},\qquad z^1=kz^2,\qquad k:=-\frac ba,\quad a\ne0.
\end{split}
\end{equation*}
\index{Комплексная прямая (complex line)}%
\index{Прямая комплексная (complex line)}%
\index{Комплексная проективная прямая (complex projective line)}%
\index{Прямая комплексная проективная (complex projective line)}%
\index{Окружность Хопфа (Hopf circle)}%
\index{Хопфа окружность (Hopf circle)}%
Поскольку ровно одна окружность Хопфа проходит через каждую точку $\MS^3$,
эти окружности заполняют всю трехмерную сферу. При этом окружности Хопфа
взаимно однозначно соответствуют комплексным прямым в $\MC^2$, проходящим через
начало координат. Две точки $(z^1,z^2)$ и $(z^{\prime1},z^{\prime2})$ лежат на
одной окружности Хопфа тогда и только тогда, когда их координаты отличаются на
фазовый множитель: $z^{\prime1}=z^1\ex^{it}$, $z^{\prime2}=z^2\ex^{it}$. Выбрав
подходящим образом параметр $t$, всегда можно добиться, например, чтобы
$\im z^2=0$. Тогда каждой прямой, проходящей через начало координат, будет
соответствовать одна и только одна точка на окружности Хопфа. Множество таких
точек образует двумерную сферу $\MS^2$, которая задается в $\MC^2$ уравнением
\begin{equation}                                                  \label{ecirho}
  \big(x^1\big)^2+\big(y^1\big)^2+\big(x^2\big)^2=1,\qquad y^2=0,
\end{equation}
где $z^1=x^1+iy^1$, $z^2=x^2+iy^2$. Таким образом, комплексная проективная
прямая диффеоморфна двумерной сфере, $\MC\MP^1\approx\MS^2$. Соответствующая
проекция $\MS^3\xrightarrow{\pi}\MS^2$ определяет главное расслоение с базой
$\MS^2$ и типичным слоем $\MG=\MU(1)\approx\MS^1$, которое называется
{\em расслоением Хопфа}. Структурная группа действует на пространстве расслоения
$\MS^3$ умножением комплексных координат на $\ex^{it}$, $t\in[0,2\pi]$.
\index{Расслоение Хопфа (Hopf fiber bundle)}%
\index{Хопфа расслоение (Hopf fiber bundle)}%

Отображение $g_t:~\MS^3\rightarrow\MS^3$, $t\in\MR$, задаваемое умножением
комплексных координат в $\MC^2$ на $\ex^{it}$, является изометрией для метрики,
индуцированной вложением (\ref{esthco}).  Отсюда следует, что у трехмерной
сферы $\MS^3$ существуют изометрии, не имеющие выделенной оси: в каждой
точке это движение выглядит точно так же, как и в любой другой. Это показывает,
что трехмерная сфера $\MS^3$ ``более круглая'', чем ее двумерный аналог $\MS^2$.
Однопараметрическая группа преобразований $g_t$ называется {\em потоком Хопфа}.
\index{Поток Хопфа (Hopf flow)}\index{Хопфа поток (Hopf flow)}%

Очевидно, что метрика проективной прямой $\MC\MP^1\approx\MS^2$,
индуцированная вложением (\ref{ecirho}), является стандартной метрикой сферы
(см.\ раздел \ref{sphere}).

На расслоение Хопфа можно взглянуть с другой точки зрения. Вспомним, что
трехмерная сфера может быть оснащена групповой структурой,
$\MS^3\approx\MS\MU^2$. (см.\ раздел \ref{sthecs}).
Двумерные унитарные матрицы с единичным определителем могут быть параметризованы
двумя комплексными числами
\begin{equation*}
  U=\begin{pmatrix} z^1 & z^2 \\ -\bar z^2 & \bar z^1 \end{pmatrix}\in\MS\MU(2),
\end{equation*}
где $z^1$ и $z^2$ лежат на сфере (\ref{esthco}). Группа $\MS\MU(2)$ содержит
подгруппу $\MU(1)$, которую можно реализовать в виде диагональных матриц
\begin{equation*}
  \begin{pmatrix} \ex^{it} & 0 \\ 0 & \ex^{-it} \end{pmatrix}\in\MU(1).
\end{equation*}
Эта подгруппа не является нормальной. Рассмотрим множество правых смежных
классов $\MS\MU(2)/\MU(1)$. Два элемента группы $U,U'\in\MS\MU(2)$ принадлежат
одному смежному классу тогда и только тогда, когда $z^{\prime1}=z^1\ex^{it}$ и
$z^{\prime2}=z^2\ex^{it}$. Поэтому расслоение Хопфа можно представить в
эквивалентном виде $\MS\MU(2)\xrightarrow{\pi}\MM:=\MS\MU(2)/\MU(1)$.

Действия группы $\MS\MU(2)$ слева и справа на сферу $\MS^3$, которую мы
отождествим с $\MS\MU(2)$,
\begin{equation*}
  U\mapsto a^{-1}Ub,\qquad a,b\in\MS\MU(2)
\end{equation*}
коммутируют и являются изометриями. Тем самым мы имеем гомоморфизм
$\MS\MU(2)\times\MS\MU(2)\rightarrow\MS\MO(4)$, где $\MS\MO(4)$ является
группой движений евклидова пространства $\MR^4\approx\MC^2$ и вложенной
трехмерной сферы. При этом действие группы $\MS\MU(2)$ слева и справа
является транзитивным, а группой изотропии являются преобразования вида
$\pm(1,1)\in\MS\MU(2)\times\MS\MU(2)$. Тем самым мы имеем сюрьективный
гомоморфизм с ядром $\MZ_2$ или изоморфизм
\begin{equation}                                                  \label{eisofu}
  \MS\MO(4)\simeq\frac{\MS\MU(2)\times\MS\MU(2)}{\MZ_2}.
\end{equation}

Теперь дадим координатное описание расслоения Хопфа. Покроем базу
$\MS^2=\overline\MC$, которую мы отождествим с расширенной комплексной
плоскостью, двумя картами следующим образом. Введем координату ``окрестности
нуля'' $z:=z^2/z^1$ при $z_1\ne0$ и ``окрестности бесконечности'' $w:=z^1/z^2$,
при $z^2\ne0$. Тогда база $\MS^2=\overline\MC$ покрывается двумя картами:
\begin{equation*}
  \MU_0:=\lbrace z:~z\in\MC\rbrace,\quad \text{и}\quad
  \MU_\infty:=\lbrace w:~w\in\MC\rbrace.
\end{equation*}
Тривиализация расслоения Хопфа задается двумя отображениями:
\begin{equation*}
\begin{split}
  \chi_0:\quad &\MS^3\ni\quad(z^1,z^2)\mapsto (z,\ex^{i\arg z^1})\quad
  \in\MC\times\MS^1,
\\
  \chi_\infty:\quad &\MS^3\ni\quad(z^1,z^2)\mapsto (w,\ex^{i\arg z^2})\quad
  \in\MC\times\MS^1.
\end{split}
\end{equation*}
В области пересечения карт $\MU_0\cap\MU_\infty$ координаты связаны
преобразованием $z\mapsto w=1/z$. Следовательно, расслоение Хопфа голоморфно.
В области пересечения карт задано отображение:
\begin{equation*}
  \chi_\infty\circ\chi_0^{-1}:\quad \MC\times\MS^1\ni\quad(z,\ex^{i\arg z^1})
  \mapsto(z,\ex^{i\arg z^2})\quad\in(\MC\setminus\lbrace0\rbrace)\times\MS^1,
\end{equation*}
которое мы записали в координате $z$. Поскольку
\begin{equation*}
  \ex^{i\arg z^2}=\ex^{i\arg zz^1}=\ex^{i\arg z^1}\ex^{i\arg z},
\end{equation*}
то функция склейки имеет вид $a_{\infty0}=\ex^{i\arg z}$. Таким образом,
расслоение Хопфа имеет следующее координатное описание:
\begin{align}                                                          \nonumber
  &\MU_0=\MC,\qquad \MU_\infty=\overline\MC\setminus\lbrace0\rbrace,
\\                                                                   \tag*{\qed}
  a_{\infty0}:\quad &\MC\setminus\lbrace0\rbrace\ni\quad z\mapsto
  \ex^{i\arg z}\quad\in\MU(1)\approx\MS^1.
\end{align}
\renewcommand{\qed}{}\end{exa}

В заключение настоящего раздела поясним соотношение между связностью
(топологической) пространства главного расслоения $\MP$ и связностями базы $\MM$
и структурной группы $\MG$.
\begin{prop}
Если база $\MM$ и структурная группа $\MG$ главного расслоения
$\MP(\MM,\pi,\MG)$ связны, то пространство расслоения $\MP$ также связно.
\end{prop}
\begin{proof}
Пусть $\MM=\bigcup_i\MU_i$ покрытие базы открытыми множествами такое, что для
каждого $i$ существует тривиализация
$\chi_i:~\pi^{-1}(\MU_i)\rightarrow\MU_i\times\MG$. Поскольку $\MU_i$ и $\MG$
связны, то $\pi^{-1}(\MU_i)$ также связно. Поэтому $\MP$ тоже связно, как
объединение пересекающихся связных множеств.
\end{proof}
Это предложение дает достаточное, но не необходимое условие связности
пространства главного расслоения.
\begin{exa}
Рассмотрим главное расслоение $\MS^1(\MS^1,\pi,\MZ^2)$, базой которого является
окружность, а структурной группой -- $\MZ_2=\lbrace\pm1\rbrace$. Структурная
группа не является связной, однако пространство расслоения связно. Это --
двулистное накрытие окружности окружностью, рассмотренное в примере
\ref{esircc}.
\qed\end{exa}
\section{Ассоциированные расслоения                              \label{sassfi}}
Дать определение расслоения, ассоциированного с заданным главным расслоением, в
нескольких предложениях довольно затруднительно. Ассоциированные расслоения,
хотя и просты по своему содержанию, требуют некоторой конструкции, которую мы
сейчас опишем.
\begin{defn}
Пусть $\MP(\MM,\pi,\MG)$ -- главное расслоение, и $(\MF,\MG)$ -- группа
преобразований, т.е.\ задано дифференцируемое отображение
\begin{equation}                                                  \label{etrasc}
  \MF\times\MG\ni\quad(v,a)\mapsto va\quad\in\MF.
\end{equation}
Сейчас мы построим расслоение $\ME(\MM,\pi_\ME,\MF,\MG,\MP)$, которое
ассоциировано с главным расслоением $\MP$, и типичным слоем которого является
многообразие $\MF$. Во-первых, определим действие группы $\MG$ на прямом
произведении многообразий $\MP\times\MF$ по формуле
\begin{equation*}
  \MG\ni a:\quad \MP\times\MF\ni\quad(p,v)\mapsto(pa,va)\quad\in\MP\times\MF.
\end{equation*}
Факторпространство для $\MP\times\MF$ относительно такого действия группы
обозначается $\ME=\MP\times_\MG\MF$. Немного позже мы введем на $\ME$
дифференцируемую структуру, а пока $\ME$ -- всего лишь множество. Проекция $\pi$
в главном расслоении $\MP$ определяет отображение, которое мы обозначим той же
буквой
\begin{equation*}
  \pi:\quad \MP\times\MF\ni\quad(p,v)\mapsto x=\pi(p)\quad\in\MM.
\end{equation*}
Это отображение индуцирует отображение факторпространства,
\begin{equation*}
  \pi_\ME:\quad \ME\ni\quad u\mapsto x=\pi_\ME(u)=\pi(p)\quad\in\MM,
\end{equation*}
где $u=(p\MG,v\MG)$ -- орбита точки $(p,v)\in\MP\times\MF$, которое называется
проекцией $\ME$ на $\MM$. Для каждой точки базы $x\in\MM$ множество
$\pi^{-1}_\ME(x)$ называется слоем в $\ME$ над $x$. Поскольку главное расслоение
$\MP$ локально тривиально, то каждая точка $x\in\MM$ имеет окрестность
$\MU\subset\MM$ такую, что прообраз $\pi^{-1}(\MU)$ диффеоморфен прямому
произведению $\MU\times\MG$. Отождествляя $\pi^{-1}(\MU)$ с $\MU\times\MG$, мы
видим, что действие структурной группы $\MG$ на произведении
$\pi^{-1}(\MU)\times\MF$ задается отображением
\begin{equation*}
  \MG\ni b:\quad\MU\times\MG\times\MF\ni\quad(x,a,v)
  \mapsto(x,ab,vb)\quad\in\MU\times\MG\times\MF.
\end{equation*}
При этом орбита произвольной точки $(p,v)\in\MP\times\MF$ имеет вид
\begin{equation*}
  (x, a\MG,v\MG)=(x,\MG,v\MG),\qquad \forall~a\in\MG,~v\in\MF.
\end{equation*}
На каждой орбите можно выбрать по одному представителю, соответствующему,
например, единице группы $e\in\MG$. Тогда множество орбит прямого произведения
$\MU\times\MG$ будет параметризовано парой элементов $(x,e,v)$, $x\in\MM$,
$v\in\MF$.
Отсюда следует, что диффеоморфизм $\pi^{-1}(\MU)\approx\MU\times\MG$ индуцирует
изоморфизм $\pi^{-1}_\ME(\MU)\approx\MU\times\MF$ (локальная тривиальность).
Поэтому на множестве $\ME$ можно ввести дифференцируемую структуру. А именно, мы
потребуем, чтобы множество $\pi^{-1}_\ME(\MU)$ было открытым подмногообразием в
$\ME$, диффеоморфным прямому произведению $\MU\times\MF$ относительно
изоморфизма $\pi^{-1}_\ME(\MU)\approx\MU\times\MF$. Таким образом построено
отображение $\chi_\ME:~\pi^{-1}_\ME(\MU)\rightarrow\MU\times\MF$, которое
является диффеоморфизмом. При этом проекция $\pi_\ME$ будет дифференцируемым
отображением $\pi_\ME:~\ME\rightarrow\MM$. Назовем объект
$\ME(\MM,\pi_\ME,\MF,\MG,\MP)$ {\em расслоением} с базой $\MM$, проекцией
$\pi_\ME$, типичным слоем $\MF$ и структурной группой $\MG$,
{\em ассоциированным} с главным расслоением $\MP(\MM,\pi,\MG)$.
\qed\end{defn}
\index{Ассоциированное расслоение (associated fiber bundle)}%
\index{Расслоение ассоциированное (associated fiber bundle)}%

Поскольку отображение $\chi_\ME$ -- диффеоморфизм, то
\begin{equation*}
  \dim\ME=\dim\MM+\dim\MF.
\end{equation*}

Легко проверить, что расслоение, ассоциированное с главным расслоением, является
расслоением в смысле определения, данного в разделе \ref{sfibun}. В отличии от
данного ранее определения на ассоциированном расслоении определено действие
структурной группы $\MG$ справа. Действительно, согласно построению у каждой
точки базы $x\in\MM$ существует окрестность $\MU$ такая, что прообраз
$\pi^{-1}_\ME(\MU)$ диффеоморфен прямому произведению $\MU\times\MF$.
Отождествляя $\pi^{-1}_\ME(\MU)$ с $\MU\times\MF$, действие структурной группы
справа задано равенством
\begin{equation*}
  \MG\ni a:\quad \MU\times\MF\ni\quad (x,v)\mapsto (x,va)\quad\in\MU\times\MF,
\end{equation*}
где произведение $va$ было определено в самом начале формулой (\ref{etrasc}).
Таким образом, пара $(\ME,\MG)$ представляет собой группу преобразований.
Структурная группа действует внутри каждого слоя и не перемешивает слои между
собой.

В терминах представителей действие структурной группы на $\ME$ описывается
следующим образом. Если $(p,v)\in\MP\times\MF$ -- представитель точки $u\in\ME$,
то представителем точки $ua\in\ME$ является пара $(p,va)\in\MP\times\MF$.
\begin{com}
В моделях математической физики наиболее распространенным примером
типичного слоя $\MF$ является векторное пространство, в котором задано
представление структурной группы $\MG$. В калибровочных моделях Янга--Миллса
поля материи, которым соответствуют скалярные или спинорные поля в
пространстве-времени, имеют дополнительный изотопический индекс. Этот индекс
нумерует компоненты вектора в некотором векторном (изотопическом) пространстве
(типичном слое $\MF$), в котором задано представление калибровочной группы
(структурной группы $\MG$).
\qed\end{com}
В дальнейшем часть аргументов у ассоциированного расслоения
$\ME(\MM,\pi_\ME,\MF,\MG,\MP)$ мы, для краткости, будем опускать, оставляя лишь
те, которые наиболее существенны в данный момент.
\begin{prop}                                                      \label{poinre}
Пусть $\MP(\MM,\pi,\MG)$ -- главное расслоение и $(\MF,\MG)$ -- группа
преобразований. Пусть $\ME$ -- расслоение, ассоциированное с $\MP$. Для каждого
$p\in\MP$ и каждого $v\in\MF$ обозначим через $p(v)$ образ элемента
$(p,v)\in\MP\times\MF$ при действии естественной проекции
$\MP\times\MF\rightarrow\ME$. Тогда каждая точка $p\in\MP$ определяет
отображение типичного слоя $\MF$ в слой $\MF_x=\pi^{-1}_\ME(x)$,
\begin{equation}                                                  \label{edefpm}
  p:\quad \MF\ni\quad v\mapsto p(v)\quad\in\MF_x,
\end{equation}
где $x=\pi(p)$, и $(pa^{-1})(v)=p(va)$ для всех $p\in\MP$, $a\in\MG$ и
$v\in\MF$, т.е.\ диаграмма
\begin{equation*}
\begin{diagram}
  \MF & \rTo^a & \MF \\
  & \rdTo_{pa^{-1}} & \dTo_p \\
  &  & \MF_x
\end{diagram}
\end{equation*}
коммутативна.
\end{prop}
\begin{proof}
Следует прямо из определения, если учесть, что точки $(pa,va)$ и $(p,v)$ из
произведения $\MP\times\MF$ проектируются в одну и ту же точку $\ME$.
\end{proof}
Под изоморфизмом слоя $\MF_x=\pi^{-1}_\ME(x)$ на другой слой
$\MF_y=\pi^{-1}_\ME(y)$, где $x,y\in\MM$, мы понимаем диффеоморфизм, который
представим в виде $q\circ p^{-1}$, где
\begin{equation*}
\begin{split}
  \MF_x\ni p:&\quad \MF\rightarrow\MF_x,
\\
  \MF_y\ni q:&\quad \MF\rightarrow\MF_y.
\end{split}
\end{equation*}
В частности, автоморфизм слоя $\MF_x$ -- это отображение вида $q\circ p^{-1}$,
где $p,q\in\MF_x$. В этом случае $q=pa$ для некоторого $a\in\MG$, так что
любой автоморфизм слоя $\MF_x$ представим в виде $p\circ a\circ p^{-1}$, где
$p$ -- произвольная фиксированная точка в том же слое $\MF_x$. Поэтому группа
автоморфизмов каждого слоя $\MF_x$ изоморфна структурной группе $\MG$.

Опишем несколько примеров ассоциированных расслоений.
\begin{exa}
Рассмотрим группу преобразований $(\MG,\MG)$, когда $\MG$ действует на себе
справа. Тогда ассоциированное расслоение изоморфно исходному главному
расслоению, $\ME(\MM,\pi_\ME,\MG,\MG,\MP)\simeq\MP(\MM,\pi,\MG)$. В общем случае
они могут не совпадать, т.к.\ существует произвол в параметризации фактор
пространства $\ME=\MP\times_\MG\MF$. Таким образом, произвольное главное
расслоение ассоциировано само с собой. Если пару $(\MG,\MG)$ рассматривать, как
группу преобразований, действующую слева, то ассоциированное расслоение также
будет изоморфно главному расслоению $\MP(\MM,\pi,\MG)$, но структурная группа
будет действовать на $\MP$ слева.
\qed\end{exa}
\begin{exa}[\bf Цилиндр, лист Мебиуса, бутылка Клейна]            \label{ecymek}
Рассмотрим главное расслоение $\MS^1(\MS^1,\pi,\MZ_2)$, где пространство
расслоения реализовано в виде единичной окружности $\MP\approx\MS^1$ на
евклидовой плоскости $\MR^2$ и структурная группа которого состоит из двух
элементов $\MZ_2=\lbrace1,-1\rbrace$ с обычным умножением. Действие структурной
группы $\MZ_2$ на пространстве расслоения определено отражением относительно
центра окружности,
\begin{equation*}
  \MZ_2\ni\pm1:\quad \MS^1\ni\quad x\mapsto \pm x\quad\in\MS^1,
  \qquad x=(x^1,x^2)\in\MR^2.
\end{equation*}
База расслоения $\pi(\MS^1)=\MS^1$ также является окружностью. Это ни
что иное, как двулистное накрытие окружности окружностью, рассмотренное в
примере \ref{esircc}. Построенное главное расслоение $\MS^1(\MS^1,\pi,\MZ_2)$
нетривиально, т.к.\ не диффеоморфно прямому произведению $\MS^1\times\MZ_2$.

Построим для этого главного расслоения три ассоциированных расслоения
(см.\ рис.\ref{fassocfibbun}).
\begin{figure}[h,b,t]
\hfill\includegraphics[width=.95\textwidth]{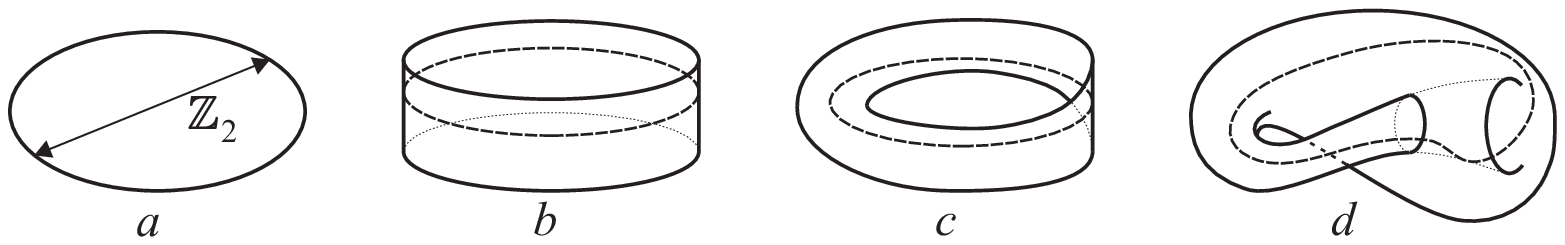}
\hfill {}
\\
\centering \caption{Примеры ассоциированных расслоений. (\textit{a}) Главное
расслоение. (\textit{b}) Цилиндр. (\textit{c}) Лист Мёбиуса. (\textit{d})
Бутылка Клейна. \label{fassocfibbun}}
\end{figure}

1). Пусть группа $\MZ_2$ действует на отрезок $\MF=[-1,1]$ тривиально,
\begin{equation*}
  \MZ_2\ni\pm1:\quad [-1,1]\ni\quad v\mapsto v\quad\in[-1,1].
\end{equation*}
Тогда ассоциированное расслоение представляет собой {\em цилиндр}, который
является прямым произведением $\MS^1\times\MF$.

2). Пусть группа $\MZ_2$ действует на отрезок $\MF=[-1,1]$ нетривиально,
\begin{equation*}
  \MZ_2\ni\pm1:\quad [-1,1]\ni\quad v\mapsto \pm v\quad\in[-1,1].
\end{equation*}
Тогда ассоциированное расслоение представляет собой {\em лист Мёбиуса}. Лист
Мёбиуса является нетривиальным расслоением, т.к.\ не имеет вида прямого
произведения $\MS^1\times\MF$.

3). Пусть группа $\MZ_2$ действует на окружность $\MF=\MS^1$, отображая точки
окружности относительно диаметра,
\begin{equation*}
  \MZ_2\ni\pm1:\quad \MS^1\ni\quad \ex^{iv}\mapsto \ex^{\pm iv}\quad\in\MS^1,
\end{equation*}
где окружность $\MS^1$ вложена в $\MC=\MR^2$. Тогда ассоциированное расслоение
представляет собой {\em бутылку Клейна}.

В первых двух примерах типичным слоем является многообразие с краем. Все три
ассоциированных расслоения имеют глобальные сечения, несмотря но то, что во
втором и третьем случаях ассоциированные расслоения не имеют вида прямого
произведения. На рисунках они показаны пунктирными линиями.
\qed\end{exa}
\index{Цилиндр (cylinder)}%
\index{Лист Мебиуса (M\"obius strip)}\index{Мебиуса лист (M\"obius strip)}%
\index{Бутылка Клейна (Klein bottle)}\index{Клейна бутылка (Klein bottle)}%
Приведенные примеры показывают, что многие хорошо известные примеры многообразий
можно рассматривать, как ассоциированные расслоения. Приведем еще два примера,
которые играют важную роль в дифференциальной геометрии.
\begin{exa}[\bf Касательное расслоение]
\index{Касательное расслоение (tangent fiber bundle)}%
\index{Расслоение касательное (tangent fiber bundle)}%
Рассмотрим расслоение реперов $\ML(\MM)$, $\dim\MM=n$, определенное в разделе
\ref{scorep}. Оно состоит из упорядоченных наборов (реперов)
$p=\lbrace x,e_a\rbrace$, $a=1,\dotsc,n$, где
$x\in\MM$ и $\lbrace e_a\rbrace$ -- упорядоченный набор линейно независимых
векторов касательного пространства, т.е.\ базис в $\MT_x(\MM)$. При этом
проекция определена отображением $\pi(p):=x\in\MM$. Общая линейная
группа $\MG\ML(n,\MR)$ действует в $\ML(\MM)$ справа так. Если
$a=\lbrace S_a{}^b\rbrace\in\MG\ML(n,\MR)$ и $p=\lbrace x,e_a\rbrace$ -- репер,
то преобразованный репер $pa=\lbrace x,e^\prime_a\rbrace$ есть, по определению,
репер, состоящий из векторов $e^\prime_a:=S_a{}^b e_b$. Мы записываем правое
действие структурной группы как матричное умножение слева из-за принятого
нами соглашения суммирования ``с десяти до четырех''. Отметим, что порядок
записи является условным, $S_a{}^b e_b= e_b S_a{}^b$, так как суммирование
проводится по одному нижнему и одному верхнему индексу. Группа $\MG\ML(n,\MR)$
действует на расслоении реперов $\ML(\MM)$ свободно и сохраняет слои, так как
$\pi(p)=\pi(q)$, $p,q\in\ML(\MM)$ тогда и только тогда, когда $p=qa$ для
некоторого $a\in\MG\ML(n,\MR)$.

Построим теперь ассоциированное расслоение. Пусть
$\big(\MR^n,\MG\ML(n,\MR)\big)$ -- группа преобразований. Группа
$\MG\ML(n,\MR)$, по определению, действует в $\MR^n$ следующим образом. Пусть
$b_a$, $a=1,\dots,n$, -- некоторый базис в $\MR^n$. Если
$a=\lbrace S_a{}^b\rbrace\in\MG\ML(n,\MR)$ -- матрица преобразования, то
$b_aa:=S_a{}^b b_b\in\MR^n$. Покажем, что
ассоциированное расслоение $\ME\big(\MM,\pi_\ME,\MR^n,\MG\ML(n,\MR),\ML\big)$
можно отождествить с касательным расслоением $\MT(\MM)$, определенным в разделе
\ref{sglove}. Действительно, базы расслоений совпадают, а слои изоморфны как
векторные пространства одинаковой размерности. Покажем, что слои
ассоциированного расслоения $\pi^{-1}_\ME(x)$ для всех $x\in\MM$ можно
рассматривать, как касательные пространства $\MT_x(\MM)$. Пусть
$p=\lbrace x,e_a\rbrace$ -- репер в точке $x\in\MM$. Он же является точкой в
главном расслоении $p\in\ML(\MM)$. Согласно предложению \ref{poinre}, каждому
реперу ставится в соответствие невырожденное линейное отображение. Выберем такой
репер, что
\begin{equation*}
  p:\quad \MR^n\ni\quad b_a\mapsto e_a\quad\in\pi^{-1}_\ME(x).
\end{equation*}
С другой стороны, по определению, набор векторов $\lbrace e_a\rbrace$ образует
базис касательного пространства $\MT_x(\MM)$. Поэтому
$\pi^{-1}_\ME(x)=\MT_x(\MM)$.

В компонентах это отождествление выглядит следующим образом. Произвольный
касательный вектор $X\in\MT_x(\MM)$ в точке $x\in\MM$ можно разложить по
координатному базису и реперу: $X=X^\al\pl_\al=X^a e_a$, где
$e_a=e^\al{}_a\pl_\al$. Компоненты вектора $X$ связаны линейным преобразованием
$X^a=X^\al{}e_\al{}^a$, где $e_\al{}^a$ -- компоненты обратного репера.
Структурная группа $\MG\ML(n,\MR)$ действует на латинские индексы, не меняя
касательного вектора, $X=X^ae_a=X^{\prime a}e^\prime_a$, где
$X^{\prime a}=X^b S^{-1}_b{}^a$ и $e^\prime_a=S_a{}^b e_b$. При преобразовании
координат преобразуются греческие индексы с помощью взаимно обратных матриц
Якоби, $X=X^\al\pl_\al=X^{\al'}\pl_{\al'}$, где $X^{\al'}=X^\al\pl_\al x^{\al'}$
и $\pl_{\al'}=\pl_{\al'}x^\al\pl_\al$. В каждой точке $x\in\MM$ матрицы Якоби
также принадлежат общей линейной группе $\MG\ML(n,\MR)$.
\qed\end{exa}
\begin{exa}[\bf Тензорные расслоения]
\index{Тензорное расслоение (tensor fiber bundle)}%
\index{Расслоение тензорное (tensor fiber bundle)}%
Пусть $\MT^r_s$ -- пространство тензоров типа $(r,s)$ над векторным
пространством $\MR^n$. Общая группа линейных преобразований $\MG\ML(n,\MR)$
действует в $\MT^r_s$ обычным образом:
\begin{multline*}
  e_{a_1}\otimes\dotsc\otimes e_{a_r}\otimes
  e^{b_1}\otimes\dotsc\otimes e^{b_s}\mapsto
\\
  \mapsto S_{a_1}{}^{c_1}e_{c_1}\otimes\dotsc\otimes S_{a_r}{}^{c_r}e_{c_r}
  \otimes e^{d_1}S^{-1}_{d_1}{}^{b_1}\otimes\dotsc\otimes
  e^{d_s}S^{-1}_{d_s}{}^{b_s},
\end{multline*}
где $e_a$ -- базис в $\MR^n$, $e^a$ -- базис в сопряженное пространстве
$\MR^{n*}$ и $\lbrace S_a{}^b\rbrace\in\MG\ML(n,\MR)$. $a,b,...=1,\dotsc,n$.
Расслоение $\ME\big(\MM,\pi_\ME,\MT^r_s,\MG\ML(n,\MR),\ML\big)$, ассоциированное
с расслоением реперов $\ML(\MM)$, так же как и касательное расслоение
отождествляется с тензорным расслоением $\MT^r_s(\MM)$ типа $(r,s)$,
определенным в разделе \ref{stenfi}.
\qed\end{exa}

В разделе \ref{svecbu} было дано определение векторного расслоения
$\ME\big(\MM,\pi_\ME,\MV,\MG\ML(\Sn,\MR),\MP\big)$, $\dim\MV=\Sn$,
ассоциированного с главным расслоением $\MP\big(\MM,\pi,\MG\ML(\Sn,\MR)\big)$.
При этом размерность векторного пространства $\Sn$ не обязана совпадать с
размерностью базы $n=\dim\MM$. Дадим координатное
описание векторных расслоений. Рассмотрим координатное покрытие базы
$\MM=\bigcup_i\MU_i$. Каждая область предполагается односвязной, и,
следовательно, над каждой областью $\MU_i$ расслоение является тривиальным,
т.е.\ существуют диффеоморфизмы (тривиализации)
\begin{equation*}
  \chi_i:\quad \pi_\ME^{-1}(\MU_i)\rightarrow\MU_i\times\MV.
\end{equation*}
Отсюда следует, что определено отображение
\begin{equation*}
  f_{ji}:=\chi_j\circ\chi_i^{-1}:\quad \MV\rightarrow\MV,
\end{equation*}
которое называется {\em функциями склейки}.
\index{Функции склейки (sewing functions)}%
\index{Склейки функции (sewing functions)}%
Если базис векторного пространства $\MV$, $\dim\MV=\Sn$, фиксирован, то функции
склейки принимают значения в группе невырожденных матриц:
\begin{equation}                                                  \label{esevef}
  f_{ji}:\quad \MU_i\cap\MU_j\rightarrow\MG\ML(\Sn,\MR).
\end{equation}
\begin{theorem}
Пусть задано векторное расслоение
$\ME\big(\MM,\pi_\ME,\MV,\MG\ML(\Sn,\MR),\MP\big)$, координатное покрытие базы
$\MM=\bigcup_i\MU_i$ и соответствующие тривиализации $\chi_i$. Тогда однозначно
определен набор функций склеек (\ref{esevef}), обладающих свойствами:
\begin{align}                                                     \label{esefip}
  f_{ij}&=f_{ji}^{-1}, && \forall\quad x\in\MU_i\cap\MU_j,
\\                                                                \label{esesep}
  f_{ij}f_{jk}f_{ki}&=\id, && \forall\quad x\in\MU_i\cap\MU_j\cap\MU_k.
\end{align}
Обратно. По любому координатному покрытию $\MM=\bigcup_i\MU_i$ и по любому
набору отображений (\ref{esevef}) со свойствами (\ref{esefip}), (\ref{esesep})
Можно построить векторное расслоение
$\ME'\big(\MM,\pi_\ME^{\prime},\MV,\MG\ML(\Sn,\MR),\MP\big)$. Если набор функций
склеек был построен по векторному расслоению $\ME$, то расслоения $\ME'$ и $\ME$
изоморфны.
\end{theorem}
\begin{proof}
Аналогично доказательству теоремы \ref{tlofib} для главных расслоений.
\end{proof}
\begin{cor}
Главное расслоение $\MP\big(\MM,\pi,\MG(\Sn,\MR)\big)$ и ассоциированные с ним
векторные расслоения $\ME\big(\MM,\pi_\ME,\MV,\MG(\Sn,\MR),\MP\big)$ определяют
эквивалентные склеивающие коциклы.
\end{cor}
\begin{proof}
См., например, \cite{Bolibr00R}.
\end{proof}

Следующая конструкция показывает, что коциклы можно использовать для построения
новых ассоциированных расслоений.
\begin{prop}
Пусть дано векторное расслоение $\ME(\MM,\pi_\ME,\MV)$, ассоциированное с
главным расслоением $\MP\big(\MM,\pi,\MG\ML(\Sn,\MR)\big)$, где $\Sn=\dim\MV$, с
коциклом $\lbrace f_{ji}\rbrace$, соответствующим некоторому координатному
покрытию базы $\MM=\bigcup_i\MU_i$. Тогда существует {\em детерминантное
расслоение} $|\ME|(\MM,\pi_{|\ME|},\MR)$ с одномерным слоем $\MR$, определенное
коциклом $\lbrace\det f_{ji}\rbrace$, которое ассоциировано с тем же главным
расслоением $\MP$.
\end{prop}
\begin{proof}
Проверка свойств (\ref{esefip}), (\ref{esesep}). Кроме того, отображение
$f_{ji}\mapsto\det f_{ji}$ определяет одномерное представление группы
$\MG\ML(\Sn,\MR)$ в $\MR$.
\end{proof}
\index{Детерминантное расслоение (determinant fiber bundle)}%
\index{Расслоение детерминантное (determinant fiber bundle)}%
\begin{exa}
Пусть $\ME(\MM,\pi_\ME,\MV)=\MT(\MM)$ -- касательное расслоение к многообразию
$\MM$. Тогда скалярные плотности степени 1, введенные в разделе \ref{scfden},
являются сечениями детерминантного расслоения.
\qed\end{exa}

Вернемся к общему случаю. Пусть $\MP(\MM,\pi,\MG)$ -- главное расслоение, и
$\MH$ -- замкнутая подгруппа в $\MG$. Пара $(\MG/\MH,\MG)$, где $\MG/\MH$ --
пространство правых смежных классов, является группой преобразований. Поэтому
определено ассоциированное расслоение $\ME(\MM,\pi_\ME,\MG/\MH,\MG,\MP)$ со
стандартным слоем $\MG/\MH$. С другой стороны, группа $\MH$, будучи подгруппой
в структурной группе $\MG$, действует на $\MP$ справа. Поэтому определено фактор
пространство $\MP/\MH$. Между построенными ассоциированным расслоением и фактор
пространством существует связь.
\begin{prop}                                                      \label{pfafib}
Расслоение $\ME=\MP\times_\MG(\MG/\MH)$, ассоциированное с $\MP(\MM,\pi,\MG)$,
со стандартным слоем $\MG/\MH$ можно отождествить с $\MP/\MH$ так. Пусть
$(p,\MH a)\in\MP\times(\MG/\MH)$, где $a\in\MG$, -- представитель элемента
ассоциированного расслоения $\ME$. Отождествим его с элементом $pa\in\MP$,
который является представителем некоторого элемента из фактор пространства
$\MP/\MH$.
\end{prop}
\begin{proof}
Каждый представитель однозначно определяет элемент соответствующего
пространства. Поэтому отождествление представителей приводит к отождествлению
точек ассоциированного расслоения $\ME(\MM,\pi_\ME,\MG/\MH,\MG,\MP)$ и фактор
пространства $\MP/\MH$. Легко проверить, что это отождествление не зависит от
выбора представителей.
\end{proof}
\begin{cor}
Четверка $\MP(\ME,\nu,\MH)$, где $\nu:~\MP\rightarrow\MP/\MH$ -- естественная
проекция, является главным расслоением, базой которого является ассоциированное
расслоение $\ME$, а структурной группой $\MH\subset\MG$.
\end{cor}
\begin{proof}
Из предложения \ref{pfafib} следует, что фактор пространство $\MP/\MH$ является
многообразием и поэтому может являться базой некоторого расслоения. Действие
группы $\MH$ на $\MP$ не двигает точки базы $\ME$ по построению. Локальная
тривиальность $\MP(\ME,\nu,\MH)$ следует из локальной тривиальности расслоений
$\ME(\MM,\pi_\ME,\MG/\MH,\MG,\MP)$ и $\MG(\MG/\MH,\MH)$. Действительно, пусть
$\MU$ -- окрестность в $\MM$ такая, что $\pi^{-1}(\MU)\approx\MU\times(\MG/\MH)$
и $\MV$ -- окрестность в $\MG/\MH$, для которой
$\pi^{-1}_\MG(\MV)\approx\MV\times\MH$, где $\pi_\MG:~\MG\rightarrow\MG/\MH$ --
естественная проекция. Пусть $\MW$ -- открытое подмножество в
$\pi^{-1}_\ME(\MU)\subset\ME$, которое соответствует прямому произведению
$\MU\times\MV$ при отождествлении $\pi^{-1}(\MU)\approx\MU\times(\MG/\MH)$.
Тогда $\nu^{-1}(\MW)\approx\MW\times\MH$.
\end{proof}
Выпишем связь размерностей многообразий, которые участвовали в приведенной выше
конструкции:
\begin{equation*}
\begin{split}
  \dim\MP(\MM,\pi,\MG)&=\dim\MM+\dim\MG,
\\
  \dim\MG&=\dim\MG/\MH+\dim\MH,
\\
  \dim\ME&=\dim\MM+\dim\MG/\MH,
\\
  \dim\MP(\ME,\nu,\MH)&=\dim\MM+\dim\MG/\MH+\dim\MH=\dim\MP(\MM,\pi,\MG).
\end{split}
\end{equation*}

Теперь обсудим возможность продолжения сечения ассоциированного расслоения,
которое задано на некотором подмножестве базы $\MN\subset\MM$. Для формулировки
теоремы нам понадобится
\begin{defn}
Отображение $f:~\MN\rightarrow\ME$ подмножества $\MN$ многообразия $\MM$ в
многообразие $\ME$ называется {\em дифференцируемым на $\MN$}, если для каждой
точки $x\in\MN$ существует дифференцируемое отображение
$f_x:~\MU_x\rightarrow\ME$, где $\MU_x\subset\MM$ -- некоторая окрестность точки
$x\in\MM$, такое, что $f_x=f$ на пересечении $\MU_x\cap\MN$. Если задано
ассоциированное расслоение $\ME(\MM,\pi_\ME,\MF,\MG,\MP)$ и некоторое
подмножество базы $\MN\subset\MM$, то {\em сечением на $\MN$} называется
дифференцируемое отображение $\s:~\MN\rightarrow\ME$ такое, что
$\pi_\ME\circ\s=\id_\MM$.
\qed\end{defn}
\index{Отображение дифференцируемое на подмножестве %
(map differentiable on a subset)}%
\begin{com}
В данном определении подмножество $\MN\subset\MM$ может быть произвольным и в
общем случае не является подмногообразием в $\MM$.
\qed\end{com}
\begin{exa}
Пусть $f:~\MU\rightarrow\ME$ -- дифференцируемое отображение некоторой
окрестности $\MU\subset\MM$ в $\ME$. Тогда сужение $f|_\MN$ на любое
подмножество $\MN\subset\MU$ является дифференцируемым на $\MN$.
\qed\end{exa}
\begin{theorem}                                                   \label{tasstr}
Пусть $\ME(\MM,\pi_\ME,\MF,\MG,\MP)$ -- ассоциированное расслоение такое, что
типичный слой диффеоморфен евклидову пространству, $\MF\approx\MR^\Sn$. Пусть
$\MN$ -- замкнутое подмножество (возможно, пустое) в базе $\MM$. Тогда любое
сечение $\s:~\MN\rightarrow\ME$, определенное на $\MN$, может быть продолжено до
глобального сечения, определенного на всем $\MM$. В частности, если подмножество
$\MN$ пусто, то на ассоциированном расслоении $\ME$ с типичным слоем
$\MF\approx\MR^\Sn$ существует глобальное сечение $\s:~\MM\rightarrow\ME$.
\end{theorem}
\begin{proof}
Явно строится сечение, при этом используется паракомпактность многообразия.
Детали можно найти, например, в \cite{Godeme58R}.
\end{proof}
\begin{cor}
На любом многообразии $\MM$ существует векторное поле, например, тождественно
равное нулю. Более того, если векторное поле задано на произвольном замкнутом
подмножестве $\MN\subset\MM$, то его всегда можно продолжить до векторного поля
на всем многообразии $\MM$. Замкнутость подмножества $\MN$ существенна. Пример
\ref{evenoo} показывает, что, если векторное поле задано на открытом
подмножестве, то продолжение может не существовать.
\qed\end{cor}
\begin{com}
Данная теорема показывает, что ассоциированные расслоения существенно отличаются
от главных. Согласно предложению \ref{ptrpri} главное расслоение $\MP$ допускает
глобальное сечение тогда и только тогда, когда оно тривиально. Если
ассоциированное расслоение имеет вид прямого произведения, $\ME=\MM\times\MF$,
то оно, очевидно, также допускает глобальные сечения. Последняя теорема
утверждает, что ассоциированное расслоение $\ME$ имеет глобальные сечения даже
если оно не имеет вида прямого произведения $\MM\times\MF$. Однако, взамен мы
требуем, чтобы типичный слой был диффеоморфен евклидову пространству,
$\MF\approx\MR^\Sn$ некоторой размерности $\Sn$.
\qed\end{com}
\begin{exa}
Заменим единичный отрезок $[-1,1]$ для листа Мёбиуса из примера \ref{ecymek} на
вещественную прямую $\MR$. Тогда мы попадаем в условия теоремы \ref{tasstr}, и,
следовательно лист Мёбиуса имеет глобальное сечение.
\qed\end{exa}
\begin{defn}
Ассоциированное расслоение $\ME(\MM,\pi_\ME,\MF,\MG,\MP)$ называется
{\em тривиальным}, если соответствующее главное расслоение $\MP(\MM,\pi,\MG)$
тривиально.
\qed\end{defn}
\index{Тривиальное ассоциированное расслоение %
(trivial associated fiber bundle)}%
\index{Ассоциированное расслоение тривиальное %
(trivial associated fiber bundle)}%
Из данного определения следует, что тривиальное ассоциированное расслоение
всегда имеет вид прямого произведения, $\ME\approx\MM\times\MF$, т.е.\ оно
тривиально как расслоение. Обратное утверждение неверно.
\begin{exa}
Пусть $\MP(\MM,\pi,\MG)$ -- нетривиальное главное расслоение и $\MF$ --
произвольное многообразие, на котором структурная группа $\MG$ действует
тривиально. Тогда ассоциированное расслоение $\ME(\MM,\pi_\ME,\MF,\MG,\MP)$
тривиально как расслоение, т.е.\ $\ME\approx\MM\times\MF$, однако оно не
является тривиальным ассоциированным расслоением.
\qed\end{exa}
В общем случае справедливо
\begin{prop}
Векторное расслоение $\ME\big(\MM,\pi_\ME,\MV,\MG\ML(\Sn,\MR),\MP\big)$,
$\dim\MV=\Sn$, имеет вид прямого произведения $\ME\simeq\MM\times\MV$ тогда и
только тогда, когда оно имеет $\Sn$ сечений (того же класса гладкости), линейно
независимых над каждой точкой базы.
\end{prop}
\begin{proof}
См., например, \cite{Bolibr00R}.
\end{proof}
\section{Отображение расслоений                                  \label{shofbu}}
Рассмотрим два главных расслоения $\MP_1(\MM_1,\pi_1,\MG_1)$ и
$\MP_2(\MM_2,\pi_2,\MG_2)$. Оба расслоения являются группами преобразований
$(\MP_1,\MG_1)$ и $(\MP_2,\MG_2)$. Поэтому при определении гомоморфизма
расслоений используется понятие эквивариантного отображения, введенного в
разделе \ref{seqyim}.
\begin{defn}
{\em Гомоморфизмом} расслоений
$f:\quad \MP_1(\MM_1,\pi_1,\MG_1)\rightarrow\MP_2(\MM_2,\pi_2,\MG_2)$ называется
эквивариантное отображение $f=f_\MP\times f_\MG$ (пара отображений),
\begin{equation*}
\begin{split}
  f_\MP:&\quad \MP_1\rightarrow\MP_2,
\\
  f_\MG:&\quad \MG_1\rightarrow\MG_2,
\end{split}
\end{equation*}
где $f_\MP$ -- дифференцируемое отображение пространств расслоений и $f_\MG$ --
гомоморфизм структурных групп, таких, что $f_\MP(pa)=f_\MP(p)f_\MG(a)$ для всех
$p\in\MP_1$ и $a\in\MG_1$, т.е.\ диаграмма
\begin{equation*}
\begin{diagram}
  \MP_1\times\MG_1 & \rTo^{f_\MP\times f_\MG} & \MP_2\times\MG_2 \\
  \dTo^{a} & & \dTo_{f_\MG(a)} \\
  \MP_1 & \rTo^{f_\MP} & \MP_2
\end{diagram}
\end{equation*}
коммутативна.

Расслоения $\MP_1(\MM_1,\pi_1,\MG_1)$ и $\MP_2(\MM_2,\pi_2,\MG_2)$
{\em изоморфны}, если $f_\MP$ -- диффеоморфизм пространств расслоений и $f_\MG$
-- изоморфизм структурных групп. Если $\MP_1=\MP_2=\MP$, то изоморфизм
расслоений называется {\em автоморфизмом} расслоения $\MP$.
\qed\end{defn}
\index{Гомоморфизм расслоений (homomorphism of fiber bundles}%
\index{Изоморфизм расслоений (isomorphism of fiber bundles}%
\index{Автоморфизм расслоения (automorphism of a fiber bundle)}%
Из приведенной диаграммы следует, что отображение $f_\MP:~\MP_1\rightarrow\MP_2$
отображает каждый слой $p\MG_1\subset\MP_1$ расслоения $\MP_1$ в слой
$f_\MP(p)\MG_2$ из $\MP_2$ и поэтому индуцирует дифференцируемое отображение баз
\begin{equation*}
  f_\MM:\quad \MM_1\ni\quad\pi_1(p)\mapsto\pi_2\big(f_\MP(p)\big)\quad\in\MM_2.
\end{equation*}
Легко проверить, что это отображение не зависит от выбора точки в слое
$p\in\pi^{-1}_1(x)$.

Если $f:~\MP_1\rightarrow\MP_2$ -- изоморфизм расслоений, то индуцированное
отображение баз $f_\MM$ является диффеоморфизмом и существует обратный
изоморфизм $f^{-1}:~\MP_2\rightarrow\MP_1$.
\begin{exa}[\bf Вертикальный автоморфизм]                         \label{everau}
\index{Автоморфизм вертикальный (vertical automorphism)}%
\index{Вертикальный автоморфизм (vertical automorphism)}%
Пусть $a(x)$ -- произвольная дифференцируемая функция на базе $\MM$ со
значениями в структурной группе $\MG$. Умножим каждую точку главного расслоения
$p\in\MP(\MM,\pi,\MG)$ на $a\big(\pi(p)\big)$, что соответствует повороту
каждого слоя. В результате получим главное расслоение $\MP'(\MM,\pi,\MG)$,
которое изоморфно исходному. Этот автоморфизм задается функциями:
\begin{align*}
  f_\MP:&\quad \MP\ni\quad p\mapsto pa\big(\pi(p)\big)\quad\in\MP'
\\
  f_\MG=\id_\MG:&\quad \MG\ni\quad b\mapsto b\quad\in\MG.
\end{align*}
При этом отображение баз является тождественным преобразованием, $f_\MM=\id$.
Этот автоморфизм называется {\em вертикальным}.
\qed\end{exa}
\begin{defn}
Гомоморфизм расслоений называется {\em вложением} или {\em инъекцией}, если
индуцированное отображение баз $f_\MM:~\MM_1\rightarrow\MM_2$ есть вложение
многообразий и $f_\MG:~\MG_1\rightarrow\MG_2$ -- мономорфизм (инъективный
гомоморфизм). Отождествляя пространство первого расслоения $\MP_1$ с его образом
$f_\MP(\MP_1)$, $\MG_1$ с $f_\MG(\MG_1)$ и $\MM_1$ с $f_\MM(\MM_1)$, мы говорим,
что $\MP_1(\MM_1,\pi_1,\MG_1)$ есть {\em подрасслоение} для
$\MP_2(\MM_2,\pi_2,\MG_2)$. Если, кроме того, $\MM_1=\MM_2=\MM$ и индуцированное
отображение баз $f_\MM=\id_\MM$ есть тождественное отображение, то гомоморфизм
расслоений $f:~\MP_1(\MM,\pi_1,\MG_1)\rightarrow\MP_2(\MM,\pi_2,\MG_2)$
называется {\em редукцией} структурной группы $\MG_2$ главного расслоения
$\MP_2(\MM,\pi_2,\MG_2)$ к подгруппе $\MG_1$. Само расслоение
$\MP_1(\MM,\pi_1,\MG_1)$ называется {\em редуцированным расслоением}.
Если задано главное расслоение $\MP(\MM,\pi,\MG)$ и подгруппа Ли $\MH$ в $\MG$,
то мы говорим, что {\em структурная группа $\MG$ редуцируема к $\MH$}, если
существует редуцированное главное расслоение $\MP'(\MM,\pi',\MH)$.
\qed\end{defn}
\index{Вложение расслоений (embedding of fiber bundles)}%
\index{Инъекция расслоений (injection of fiber bundles)}%
\index{Подрасслоение (sub fiber bundle)}%
\index{Редукция расслоений (reduction of fiber bundles)}%
\index{Редуцированное расслоение (reduced fiber bundle)}%
\index{Расслоение редуцированное (reduced fiber bundle)}%
\index{Автоморфизм главного расслоения %
(automorphism of a principal fiber bundle)}%

\begin{com}
Поскольку при вложении главных расслоений $f_\MM$ -- вложение и $f_\MG$ --
мономорфизм, то
\begin{equation*}
  \dim\MM_1\le\dim\MM_2,\quad \text{и}\quad \dim\MG_1\le\dim\MG_2.
\end{equation*}
Кроме того, мы не требуем, чтобы подгруппа $\MH$ была замкнутой в $\MG$. Эта
общность необходима в теории связностей.
\qed\end{com}
\begin{theorem}
Структурная группа $\MG$ главного расслоения $\MP(\MM,\pi,\MG)$ редуцируема к
подгруппе Ли $\MH$ тогда и только тогда, когда существует координатное покрытие
базы $\MM=\bigcup_i\MU_i$ такое, что все функции перехода $a_{ji}$ принимают
значение в $\MH$.
\end{theorem}
\begin{proof}
Доказательство проводится путем построения редуцированного расслоения. См.,
например, \cite{KobNom6369R}.
\end{proof}
Приведем еще один критерий редуцируемости расслоений.
\begin{theorem}                                                   \label{tredas}
Структурная группа $\MG$ главного расслоения $\MP(\MM,\pi,\MG)$ редуцируема к
подгруппе $\MH$ тогда и только тогда, когда ассоциированное расслоение
$\ME(\MM,\pi_\ME,\MG/\MH,\MG,\MP)=\MP/\MH$ (см.\ предложение \ref{pfafib})
допускает глобальное сечение $\s:~\MM\rightarrow\ME$. Между редуцированными
главными расслоениями $\MQ(\MM,\pi,\MH)$ и сечениями $\s$ существует
естественное взаимно однозначное соответствие.
\end{theorem}
\begin{proof}
См., например, \cite{KobNom6369R}.
\end{proof}
\begin{cor}
Структурная группа $\MG$ главного расслоения $\MP(\MM,\pi,\MG)$ редуцируема к
единичному элементу $e\in\MG$ тогда и только тогда, когда оно имеет глобальное
сечение и, следовательно, тривиально, $\MP\simeq\MM\times\MG$.
\end{cor}
\begin{com}
Если структурная группа состоит только из единичного элемента, то пространство
редуцированного главного расслоения $\MP'(\MM,\pi,e)$ естественным образом
отождествляется с базой, $\MP'=\MM$. Поэтому существование отображения
$f_\MP:~\MP'=\MM\rightarrow\MP$ для редуцированного расслоения эквивалентно
существованию глобального сечения для $\MP(\MM,\pi,\MG)$.
\qed\end{com}
Теорема о существовании глобальных сечений на ассоциированном расслоении
позволяет по новому доказать и взглянуть на существование римановой метрики на
произвольном многообразии. Для формулировки результата нам понадобится
дополнительное утверждение.

Из разложения Ивасавы \cite{Iwasaw49} следует, что любая связная группа Ли
диффеоморфна прямому произведению, $\MG\approx\MH\times\MR^n$, где $\MH$ --
максимальная компактная подгруппа в $\MG$ и $\MR^n$ -- евклидово пространство
размерности $n=\dim\MG-\dim\MH$ \cite{Helgas01R}.
\begin{prop}
Пусть $\MP(\MM,\pi,\MG)$ -- главное расслоение со связной структурной группой
Ли $\MG$. Тогда структурная группа редуцируема к максимальной компактной
подгруппе $\MH\subset\MG$.
\end{prop}
\begin{proof}
Рассмотрим ассоциированное расслоение
$\ME(\MM,\pi_\ME,\MG/\MH,\MG,\MP)=\MP/\MH$, где $\MH$ -- максимальная компактная
подгруппа в $\MG$. Поскольку типичный слой $\MG/\MH$ диффеоморфен евклидову
пространству (разложение Ивасавы), то, согласно теореме \ref{tasstr},
ассоциированное расслоение допускает глобальное сечение. Поэтому из теоремы
\ref{tredas} вытекает, что структурная группа главного расслоения
$\MP(\MM,\pi,\MG)$ редуцируема к максимальной компактной подгруппе $\MH$.
\end{proof}
\begin{exa}[\bf Существование римановой метрики]                  \label{exriex}
Пусть $\ML(\MM)$ -- расслоение линейных реперов над многообразием $\MM$,
$\dim\MM=n$. Известно, что максимальной компактной подгруппой в группе общих
линейных преобразований $\MG\ML(n,\MR)$ является группа вращений $\MO(n)$
размерности $\frac12n(n-1)$. В силу разложения Ивасавы однородное пространство
$\MG\ML(n,\MR)/\MO(n)$ диффеоморфно евклидову пространству $\MR^d$ размерности
\begin{equation*}
  d=n^2-\frac{n(n-1)}2=\frac{n(n+1)}2.
\end{equation*}
Из теоремы \ref{tasstr} следует, что ассоциированное расслоение
$\ME=\ML(\MM)/\MO(n)$ со слоем $\MG\ML(n,\MR)/\MO(n)$ имеет глобальное сечение.
Поэтому структурная группа $\MG\ML(n,\MR)$ может быть редуцирована к группе
вращений $\MO(n)$ как следствие теоремы \ref{tredas}.

Пусть в евклидовом пространстве $\MR^n$ в декартовых координатах задано
скалярное произведение
\begin{equation}                                                  \label{escvef}
  (\widetilde X,\widetilde Y):=\widetilde X^a \widetilde Y^b\dl_{ab},\qquad
  \widetilde X,\widetilde Y\in\MR^n,
\end{equation}
где $\dl_{ab}$ -- евклидова метрика (\ref{eclmet}). Это скалярное произведение
инвариантно относительно вращений евклидова пространства $\MO(n)$.

Покажем, что каждая редукция структурной группы $\MG\ML(n,\MR)$ расслоения
реперов $\ML(\MM)$ порождает риманову метрику $g$ на $\MM$. Пусть
$\MO(\MM):=\MO\big(\MM,\pi,\MO(n)\big)$ -- редуцированное подрасслоение для
расслоения реперов $\ML(\MM)$. Типичный слой редуцированного расслоения состоит
из реперов вида $p=\lbrace x,S_a{}^b e_b\rbrace$, где $e_b$ -- некоторый
фиксированный базис касательного пространства и $S_a{}^b\in\MO(n)$ --
произвольная матрица вращений. Если мы рассматриваем репер $p\in\ML(\MM)$
как линейный изоморфизм из $\MR^n$ на касательное пространство $\MT_x(\MM)$, где
$x=\pi(p)$, то каждый репер из редуцированного расслоения $p\in\MO(\MM)$
определяет скалярное произведение в касательном пространстве $\MT_x(\MM)$ по
формуле
\begin{equation}                                                  \label{edefri}
  g(X,Y)=(p^{-1}X,p^{-1}Y),\qquad X,Y\in\MT_x(\MM).
\end{equation}
Инвариантность скалярного произведения (\ref{escvef}) относительно вращений
$\MO(n)$ влечет за собой независимость $g(X,Y)$ от выбора репера $p\in\MQ$.
Таким образом, мы доказали, что каждое многообразие допускает риманову метрику.

В компонентах определение (\ref{edefri}) имеет хорошо знакомый вид
\begin{equation*}
  g(X,Y)=X^\al Y^\bt g_{\al\bt}=\widetilde X^a \widetilde Y^b\dl_{ab},
  \qquad g_{\al\bt}=e_\al{}^a e_\bt{}^b\dl_{ab},
\end{equation*}
где $\widetilde X^a=X^\al e_\al{}^a$ и $\widetilde Y^a=Y^\al e_\al{}^a$. В таком
виде скалярное произведение уже встречалось в разделе \ref{sunhba} (без знаков
тильды).

Верно также обратное утверждение. Пусть на $\MM$ задана риманова метрика $g$.
Пусть $\MO(\MM)$ -- подмножество в расслоении реперов $\ML(\MM)$, состоящее из
реперов $p=\lbrace x,e_a\rbrace$, которые ортонормальны относительно $g$,
т.е.\
\begin{equation*}
  (e_a,e_b)=e^\al{}_a e^\bt{}_b g_{\al\bt}=\dl_{ab}.
\end{equation*}
Если
репер $p\in\ML(\MM)$ рассматривается как изоморфизм из $\MR^n$ в $\MT_x(\MM)$,
то $p$ принадлежит редуцированному подрасслоению $\MO(\MM)$ тогда и только
тогда, когда $(\widetilde X,\widetilde Y)=g(p\widetilde X,p\widetilde Y)$ для
всех $\widetilde X,\widetilde Y\in\MR^n$. Легко проверить, что $\MO(\MM)$
образует редуцированное подрасслоение в $\ML(\MM)$ над базой $\MM$ со
структурной группой $\MO(n)$. Расслоение $\MO(\MM)$ называется {\em расслоением
ортонормальных реперов} над $\MM$.
\index{Расслоение ортонормальных реперов (fiber bundle of orthonormal frames}%
Каждый элемент из $\MO(\MM)$ есть {\em ортонормальный репер}.
\index{Ортонормальный репер (orthonormal frame)}%
\index{Репер ортонормальный (orthonormal frame)}%

Чтобы подчеркнуть принадлежность вектора евклидову пространству, в настоящем
примере был использован знак тильды. Поскольку репер устанавливает изоморфизм
$\MR^n$ и $\MT_x(\MM)$, то эти пространства мы отождествим и опустим знак
тильды, считая, что $X^\al$ и $X^a$ -- это компоненты одного касательного
вектора $X\in\MT_x(\MM)$ относительно координатного базиса и репера. Чтобы их не
путать, мы используем буквы греческого и латинского алфавитов.
\qed\end{exa}
В рассмотренном примере мы не только доказали существование римановой метрики
на произвольном многообразии, но и следующее утверждение.
\begin{theorem}
Существует взаимно однозначное соответствие между римановыми метриками на
многообразии $\MM$ и редукциями структурной группы $\MG\ML(n,\MR)$ расслоения
реперов $\ML(\MM)$ к группе вращений $\MO(n)$
\end{theorem}

В начале настоящего раздела мы установили, что гомоморфизм расслоений индуцирует
дифференцируемое отображение баз расслоений. Теперь мы рассмотрим обратную
задачу: в какой степени отображение некоторого многообразия в базу заданного
расслоения может быть сопоставлено некоторому гомоморфизму расслоений ? Ответ на
этот вопрос дает
\begin{theorem}                                                   \label{tinfib}
Пусть дано главное расслоение $\MP(\MM,\pi,\MG)$ и дифференцируемое отображение
многообразий $f_\MN:~\MN\rightarrow\MM$. Тогда существует единственное с
точностью до изоморфизма главное расслоение $\MQ(\MN,\pi_\MQ,\MG)$ с
гомоморфизмом $f:~\MQ\rightarrow\MP$, индуцирующим отображение баз
$f_\MN:~\MN\rightarrow\MM$ и соответствующим тождественному автоморфизму
структурной группы $\MG$.
\end{theorem}
\index{Индуцированное расслоение (induced fiber bundle)}%
\index{Расслоение индуцированное (induced fiber bundle)}%
\begin{proof}
В прямом произведении $\MN\times\MP$ рассмотрим подмножество $\MQ$, состоящее из
точек $(y,p)\in\MN\times\MP$ таких, что $f_\MN(y)=\pi(p)$. Определим действие
структурной группы $\MG$ справа на построенном множестве $\MQ$:
\begin{equation*}
  \MG\ni a:\quad \MQ\ni\quad(y,p)\mapsto (y,p)a:=(y,pa)\quad\in\MQ.
\end{equation*}
Это отображение не зависит от точки слоя $\pi^{-1}\big(\pi(p)\big)$.
Нетрудно проверить, что $\MG$ действует свободно на $\MQ$ и что множество $\MQ$
представляет собой главное расслоение $\MQ(\MN,\pi_\MQ,\MG)$ с базой $\MN$,
структурной группой $\MG$ и проекцией $\pi_\MQ:~(y,p)\mapsto y$.

Единственность.
Пусть $\MQ'(\MN,\pi'_\MQ,\MG)$ -- другое главное расслоение с базой $\MN$ и
структурной группой $\MG$ и $f':~\MQ'\rightarrow\MP$ -- гомоморфизм,
индуцирующий заданное отображение баз $f_\MN:~\MN\rightarrow\MM$ и
соответствующий тождественному автоморфизму структурной группы $\MG$. Определим
отображение $\MQ'$ на $\MQ$
\begin{equation*}
  \tilde f:\quad \MQ'\ni\quad p'\mapsto\big(\pi'_\MQ(p'),f'(p')\big)\quad\in\MQ.
\end{equation*}
Тогда отображение $\tilde f:~\MQ'\rightarrow\MQ$ -- изоморфизм расслоений,
индуцирующий тождественное преобразование базы $\MN$ и соответствующий
тождественному автоморфизму структурной группы $\MG$.
\end{proof}
\begin{defn}
Расслоение $\MQ(\MN,\pi_\MQ,\MG)$ в утверждении теоремы \ref{tinfib} называется
{\em расслоением, индуцированным отображением баз $f_\MN:~\MN\rightarrow\MM$ из
главного расслоения $\MP(\MM,\pi,\MG)$}, или просто {\em индуцированным
расслоением}. Оно обозначается $f_\MN^{-1}\MP$.
\qed\end{defn}
\begin{com}
Если отображение $f_\MN$ является вложением, то главное расслоение
$\MQ(\MN,\pi_\MQ,\MG)$ есть подрасслоение для $\MP(\MM,\pi,\MG)$.
\qed\end{com}
\begin{exa}
Рассмотрим главное расслоение $\MP(\MM,\pi,\MG)$ и некоторую окрестность базы
$\MU\subset\MM$ тогда $\MP|_\MU=\pi^{-1}(\MU)$ есть подрасслоение в $\MP$. Оно
же является индуцированным расслоением $f^{-1}_\MU\MP$, где
$f:~\MU\rightarrow\MP$ -- вложение.
\qed\end{exa}
\chapter{Связности на главных и ассоциированных расслоениях      \label{sconne}}
Теория связностей на расслоениях играет исключительно важную роль в моделях
математической физики, т.к.\ позволяет определить ковариантную производную. В
свою очередь ковариантная производная используется для записи ковариантных
уравнений. В настоящем разделе мы определим связность на главном
и ассоциированных расслоениях. Покажем, что калибровочные поля Янга--Миллса
представляют собой компоненты локальной формы связности. Кроме того, будет
рассмотрена группа голономии, которая является одной из важнейших характеристик
связности.
\section{Связность на главном расслоении                         \label{scofib}}
Пусть задано главное расслоение $\MP(\MM,\pi,\MG)$ с базой $\MM$, $\dim\MM=n$, и
структурной группой Ли $\MG$, $\dim\MG=\Sn$. Пара $(\MP,\MG)$ представляет собой
группу преобразований.
При этом группа Ли $\MG$ действует на многообразии $\MP$ свободно. Каждый
слой $\pi^{-1}(x)$, где $x\in\MM$, диффеоморфен структурной группе и ее действие
на нем транзитивно. Из предложения \ref{pinftr} следует, что действие
структурной группы $\MG$ на $\MP$ индуцирует гомоморфизм $\mu$ алгебры Ли $\Gg$
в алгебру Ли векторных полей $\CX(\MP)$. Этот гомоморфизм является
мономорфизмом, т.к.\ действие группы свободно и, следовательно, эффективно.
\begin{defn}
Векторное поле $X^*:=\mu(X)\in\CX(\MP)$, где $X\in\Gg$ -- произвольный элемент
алгебры Ли структурной группы Ли $\MG$, называется {\em фундаментальным
векторным полем, соответствующим} $X\in\Gg$.
\qed\end{defn}
\index{Фундаментальное векторное поле (fundamental vector field)}%
\index{Векторное поле фундаментальное (fundamental vector field)}%
Поскольку действие структурной группы на $\MP$ отображает каждый слой в себя,
то каждое фундаментальное векторное поле $X^*_p$ касается слоя
$\pi^{-1}\big(\pi(p)\big)$ в каждой точке $p\in\MP$. Поскольку действие группы
свободно, то по предложению \ref{pinftr} фундаментальные векторные поля нигде не
обращаются в нуль. Так как размерность каждого слоя равна размерности алгебры Ли
$\Gg$, то отображение
\begin{equation*}
  \Gg\ni\quad X\mapsto X^*_p\quad\in\MT_p(\MP)
\end{equation*}
есть линейный мономорфизм алгебры Ли $\Gg$ в касательное пространство в точке
$p\in\MP$ к слою, проходящему через $p$.
\begin{defn}
Образ алгебры Ли $\Gg$ в касательном пространстве $\MT_p(\MP)$ называется
{\em вертикальным подпространством} в $\MT_p(\MP)$ и обозначается $\MV_p(\MP)$.
\qed\end{defn}
\index{Вертикальное подпространство (vertical subspace)}%
\index{Подпространство вертикальное (vertical subspace)}%

Пусть $\lbrace L_\Sa\rbrace$, $\Sa=1,\dotsc,\Sn=\dim\MG$, -- базис алгебры Ли
$\Gg$ с коммутационными соотношениями $[L_\Sa,L_\Sb]=f_{\Sa\Sb}{}^\Sc L_\Sc$,
где $f_{\Sa\Sb}{}^\Sc$ -- структурные константы группы Ли $\MG$. Тогда
индуцированные фундаментальные векторные поля $L^*_\Sa$ удовлетворяют тем же
коммутационным соотношениям, $[L^*_\Sa,L^*_\Sb]=f_{\Sa\Sb}{}^\Sc L^*_\Sc$.
Множество фундаментальных векторных полей $\lbrace L^*_\Sa\rbrace$ задает
дифференцируемое инволютивное распределение вертикальных подпространств
$p\mapsto\MV_p(\MP)$ на
пространстве главного расслоения $\MP$ (см., раздел \ref{sfrote}). Согласно
теореме Фробениуса \ref{tfrofi} через каждую точку $p\in\MP$ проходит
интегральное подмногообразие этого распределения. Это интегральное
подмногообразие есть ни что иное, как слой $\pi^{-1}\big(\pi(p)\big)$.
\begin{prop}                                                      \label{pfuadj}
Пусть $X^*$ -- фундаментальное векторное поле, соответствующее элементу алгебры
Ли $X=X_0^\Sa L_\Sa\in\Gg$, $X_0^\Sa=\const$ для всех $\Sa$. Тогда для каждого
элемента группы $a\in\MG$ векторное поле $r_{a*}X^*$, где $r_{a*}$ --
дифференциал отображения, индуцированного действием элемента $a$ справа,
является фундаментальным векторным полем, которое соответствует
левоинвариантному векторному полю
$\ad(a^{-1})X=X_0^\Sb S^{-1}_{\quad \Sb}{}^\Sa(a) L_\Sa\in\Gg$, где $\ad$
обозначает присоединенное представление $\MG$ в $\Gg$ и $S_\Sb{}^\Sa(a)$ --
матрица присоединенного представления.
\end{prop}
\begin{proof}
Фундаментальное векторное поле $X^*$ индуцируется однопараметрической группой
преобразований $r_{b(t)}$, где $b(t)=\exp(tX)$. Векторное поле $r_{a*}X^*$
индуцируется  однопараметрической группой преобразований
$r_a\circ r_{b(t)}\circ r_{a^{-1}}=r_{a^{-1}b(t)a}$ по предложению \ref{plodpa}.
Утверждение предложения следует из того, что преобразования вида $a^{-1}b(t)a$
представляют собой однопараметрическую группу преобразований, порожденную
элементом алгебры Ли $\ad(a^{-1})X\in\Gg$.
\end{proof}
\begin{exa}[\bf Локальное рассмотрение]                           \label{elocon}
Чтобы прояснить абстрактное построение, которое было проведено выше, повторим
его в компонентах. В настоящей главе мы будем возвращаться к этому примеру
неоднократно. Пусть $\MU\subset\MM$ -- окрестность базы, для которой
определено отображение (\ref{eloctr}), с координатами $x^\al$, $\al=1,\dotsc,n$.
Используя диффеоморфизм $\chi:~\pi^{-1}(\MU)\rightarrow \MU\times\widetilde\MG$,
мы отождествим подрасслоение $\MQ=\pi^{-1}(\MU)$ с прямым произведением
$\MU\times\widetilde\MG$. Мы отметили структурную группу в прямом произведении
знаком тильды, потому что в дальнейшем нам будет необходимо различать точку
главного расслоения и точку структурной группы. То есть $p=(x,a)$, где
$x\in\MU$, $a\in\widetilde\MG$, для всех $p\in\MQ$. Действие структурной группы
$\MG$ на $\MQ$ имеет вид
\begin{equation*}
  \MG\ni a:\quad \MQ\ni\quad p=(x,b)\mapsto pa=(x,ba)\quad\in\MQ.
\end{equation*}

Фактически, данный пример относится к произвольному тривиальному главному
расслоению, база которого покрывается одной картой, $\MM\approx\MR^n$.

Любой элемент алгебры Ли $\Gg$ (левоинвариантное векторное поле $X$ на группе
Ли $\MG$) имеет постоянные компоненты $X^\Sa_0$ относительно левоинвариантного
базиса, $X=X^\Sa_0L_\Sa$, где $L_\Sa$ -- базис алгебры Ли $\Gg$ (см.\ раздел
\ref{sleacg}). Базису алгебры Ли $L_\Sa$ соответствуют фундаментальные векторные
поля $L_\Sa^*$ на главном расслоении, которые образуют базис вертикальных
подпространств. Выберем базис касательных пространств $\MT_p(\MQ)$ в виде
$\lbrace\pl_\al,L_\Sa^*\rbrace$. Тогда фундаментальное векторное поле,
соответствующее произвольному элементу алгебры Ли $X\in\Gg$, будет иметь только
вертикальные компоненты, $X^*=X^\Sa_0L_\Sa^*$. Утверждение предложения
\ref{pfuadj} сводится к равенству
\begin{equation}                                                  \label{eridif}
  r_{a*}X^*=X_0^\Sa S_\Sa{}^\Sb(a^{-1})L_\Sb^*,
\end{equation}
где $S_\Sa{}^\Sb(a^{-1})=S^{-1}_{\quad \Sa}{}^\Sb(a)$ -- матрица присоединенного
представления для обратного элемента $a^{-1}\in\MG$. Особенно просто это
равенство проверяется вблизи единицы группы, где определена функция композиции.
Пусть $X^*(e)$ -- значение фундаментального векторного поля в единице группы.
Тогда
\begin{equation*}
  r_{a*}\big(X^*(e)\big)=X^\Sa_0R_\Sa{}^\Sb(a)\pl_\Sb|_a
  =X_0^\Sa R_\Sa{}^\Sb(a)L^{-1}_{\quad \Sb}{}^\Sc(a)L_\Sc^*(a)
  =X_0^\Sa S_\Sa{}^\Sb(a^{-1})L_\Sb^*(a),
\end{equation*}
где матрицы $R_\Sa{}^\Sb$, $L_\Sa{}^\Sb$ и $S_\Sa{}^\Sb$ были определены в
разделах \ref{sleacg}--\ref{sadpre}.

Произвольное векторное поле $\widetilde X\in\CX(\MQ)$ имеет вид
\begin{equation*}
  \widetilde X=\widetilde X^\al\pl_\al+\widetilde X^\Sa L^*_\Sa.
\end{equation*}
После правого действия группы $r_a$, оно преобразуется в новое векторное поле
\begin{equation}                                                  \label{endiri}
  r_{a*}\widetilde X=\widetilde X^\al\pl_\al
  +\widetilde X^\Sb S^{-1}_{\quad \Sb}{}^\Sa(a)L^*_\Sa,
\end{equation}
т.к.\ дифференциал отображения $r_{a*}$ действует только на вертикальные
компоненты по правилу (\ref{eridif}).
\qed\end{exa}
Продолжим общее построение.
\begin{defn}
{\em Связностью} $\Gamma$ на главном расслоении $\MP(\MM,\pi,\MG)$ называется
распределение
\begin{equation*}
  \Gamma:\quad \MP\ni\quad p\mapsto\MH_p(\MP)\quad\subset\MT_p(\MP),
\end{equation*}
которое каждой точке $p\in\MP$ ставит в соответствие подпространство
$\MH_p(\MP)$ в касательном пространстве $\MT_p(\MP)$ такое, что\newline
\indent 1) \parbox[t]{.92\linewidth}{в каждой точке $p$ касательное пространство
$\MT_p(\MP)$ разлагается в прямую сумму:
\begin{equation*}
  \MT_p(\MP)=\MV_p(\MP)\oplus\MH_p(\MP);
\end{equation*}
}\newline
\indent 2) \parbox[t]{.92\linewidth}{подпространства $\MH_p(\MP)$ инвариантны
относительно правого действия структурной группы:
\begin{equation}                                                  \label{einvho}
  r_{a*}\MH_p(\MP)=\MH_{pa}(\MP);
\end{equation}
}\newline
\indent 3) \parbox[t]{.92\linewidth}{$\MH_p(\MP)$ зависит дифференцируемо от
$p$.}\newline
Множество касательных векторов $\MH_p(\MP)$ в точке $p\in\MP$ называется {\em
горизонтальным подпространством} в $\MT_p(\MP)$. Вектор
$\widetilde X\in\MT_p(\MP)$ называется {\em вертикальным} или
{\em горизонтальным}, если он лежит соответственно в $\MV_p(\MP)$ или
$\MH_p(\MP)$.
\qed\end{defn}
\index{Связность на главном расслоении (connection on a principal fiber bundle)}%
\index{Горизонтальное подпространство (horizontal subspace)}%
\index{Подпространство горизонтальное (horizontal subspace)}%
\index{Вертикальный вектор (vertical vector)}%
\index{Вектор вертикальный (vertical vector)}%
\index{Горизонтальный вектор (horizontal vector)}%
\index{Вектор горизонтальный (horizontal vector)}%
Из условия 1) следует, что каждый вектор $\widetilde X\in\MT_p(\MP)$ может быть
единственным образом представлен в виде суммы
\begin{equation*}
  \widetilde X=\ver\widetilde X+\hor\widetilde X,\qquad
  \ver\widetilde X\in\MV_p(\MP),~\hor\widetilde X\in\MH_p(\MP),
\end{equation*}
где $\ver$ и $\hor$ -- проекторы на соответствующие подпространства.
Векторы $\ver\widetilde X$ и $\hor\widetilde X$ называются соответственно
{\em вертикальной} и {\em горизонтальной} компонентами касательного вектора
$\widetilde X$. Ясно, что
\begin{equation*}
  \dim\MV_p(\MP)=\dim\MG,\quad \dim\MH_p(\MP)=\dim\MM,\qquad \forall p\in\MM.
\end{equation*}

Если главное расслоение тривиально, то у него существует глобальное сечение. В
этом случае условие 2) означает, что связность достаточно задать на каком либо
сечении, а затем разнести по всему пространству расслоения $\MP$ с помощью
группового действия. В дальнейшем мы сформулируем теорему \ref{troloc},
определяющую связность на главном расслоении общего вида через семейство
локальных форм связности, которые заданы на координатном покрытии базы.

По-определению, условие 3) значит, что, если $\widetilde X$ -- дифференцируемое
векторное поле на $\MP$, то таковы же вертикальная и горизонтальная компоненты
$\ver\widetilde X$ и $\hor\widetilde X$.
\begin{exa}
Накрытие $\widetilde\MM\rightarrow\MM$ (см.\ раздел \ref{scover}) является
главным расслоением с 0-мерной структурной группой. В этом случае вертикальные
подпространства отсутствуют, а горизонтальное подпространство в точке
$p\in\widetilde\MM$ совпадает с касательным пространством
$\MT_p(\widetilde\MM)$. Это означает, что связность для накрытий единственна и
распределение горизонтальных подпространств совпадает с касательны расслоением
$\MT(\widetilde\MM)$.
\qed\end{exa}
Пусть на главном расслоении $\MP(\MM,\pi,\MG)$ задана связность $\Gamma$. Для того,
чтобы конструктивно описать распределение горизонтальных подпространств построим
на главном расслоении $\MP$ 1-форму связности $\om$ со значениями в алгебре Ли
$\Gg$, т.е.\ $\om=\om^\Sa L_\Sa$, где $\om^\Sa\in\Lm_1(\MP)$ для всех $\Sa$. Как
было отмечено в начале настоящего раздела, каждому элементу алгебры Ли $X\in\Gg$
соответствует единственное фундаментальное векторное поле $X^*$. При этом
отображение $X\mapsto X^*_p$ представляет собой линейный мономорфизм $\Gg$ в
$\MT_p(\MP)$ для всех точек главного расслоения $p\in\MP$.
\begin{defn}
Для каждого касательного вектора $\widetilde X_p\in\MT_p(\MP)$ определим
значение 1-формы $\om(\widetilde X_p)$ на векторе $\widetilde X_p$, как
единственный элемент алгебры Ли $X\in\Gg$ такой, что фундаментальное векторное
поле $X^*_p$ в точке $p\in\MP$ совпадает с вертикальной компонентой вектора,
$X^*_p=\ver\widetilde X_p$. 1-форма $\om$ на $\MP$ со значениями в алгебре Ли
$\Gg$ называется {\em формой связности} для заданной связности $\Gamma$ на
$\MP(\MM,\pi,\MG)$.
\qed\end{defn}
\index{Форма связности (connection form)}%
\index{Связности форма (connection form)}%
По построению, $\om(\widetilde X)=0$ тогда и только тогда, когда векторное поле
$\widetilde X$ горизонтально. В силу следующего предложения 1-форма $\om$
взаимно однозначно определяет распределение горизонтальных подпространств и,
следовательно, связность $\Gamma$ на $\MP$.
\begin{prop}                                                      \label{pcofop}
Форма связности $\om$ на $\MP(\MM,\pi,\MG)$ имеет следующие свойства:\newline
\indent 1) \parbox[t]{.92\linewidth}{$\om(X^*)=X$ для всех $X\in\Gg$;}\newline
\indent 2) \parbox[t]{.92\linewidth}{$r_a^*\om=\ad(a^{-1})\om$, т.е.\
\begin{equation*}
  (r_a^*\om)(\widetilde X)=\ad(a^{-1})\om(\widetilde X)
  =\om(\widetilde X)^\Sb S^{-1}_{\quad \Sb}{}^\Sa(a) L_\Sa
\end{equation*}
для всех $a\in\MG$ и каждого векторного поля $\widetilde X$ на $\MP$.} \newline
Обратно. Если на главном расслоении $\MP(\MM,\pi,\MG)$ задана 1-форма $\om$ со
значениями в алгебре Ли $\Gg$, удовлетворяющая условиям 1) и 2), то на $\MP$
существует единственная связность $\Gamma$ такая, что ее форма связности есть $\om$.
\end{prop}
\begin{proof}
Пусть $\om$ -- форма связности на $\MP(\MM,\pi,\MG)$. Условие 1) следует
немедленно из определения $\om$. Поскольку каждое векторное поле
$\widetilde X\in\CX(\MP)$ разлагается на вертикальную и горизонтальную
составляющую, то достаточно проверить условие 2) для этих двух компонент. Если
$\widetilde X$ горизонтально, то $r_{a*}\widetilde X$ также горизонтально для
всех $a\in\MG$, что следует из условия 2) в определении связности. Поэтому
$(r_a^*\om)_p(\widetilde X)=\om_{pa}(r_{a*}\widetilde X)$ и
$\ad(a^{-1})\om_p(\widetilde X)$ одновременно обращаются в нуль. Если
$\widetilde X$ вертикально, то его значение в точке $p\in\MP$ определяется
некоторым фундаментальным векторным полем, $\widetilde X_p=X^*_p$, для некоторого
$X\in\Gg$. Тогда из предложения \ref{pfuadj} следует, что векторное поле
$r_{a*}X^*$ соответствует элементу алгебры Ли $\ad(a^{-1})X\in\Gg$. Поэтому
\begin{equation*}
  (r_a^*\om)_p(X^*)=\om_{pa}(r_{a*}X^*)=\ad(a^{-1})X
  =\ad(a^{-1})\big(\om_p(X^*)\big).
\end{equation*}

Обратно. Пусть задана 1-форма $\om$, удовлетворяющая условиям 1) и 2). Определим
распределение горизонтальных подпространств
\begin{equation*}
  \MH_p(\MP):=\lbrace \widetilde X\in\MT_p(\MP):\quad \om(\widetilde X)=0\rbrace
  \qquad \forall p\in\MP.
\end{equation*}
Теперь нетрудно проверить, что распределение $\Gamma:~p\mapsto\MH_p(\MP)$ является
связностью, для которой $\om$ -- форма связности.
\end{proof}

Пусть на главном расслоении $\MP(\MM,\pi,\MG)$ задано две связности $\Gamma_1$ и
$\Gamma_2$ с формами связности $\om_1$ и $\om_2$ соответственно. Нетрудно проверить,
что 1-форма $(1-t)\om_1+t\om_2$, где $t\in[0,1]$, также задает некоторую
связность на $\MP$. Таким образом, любые две связности, заданные на одном
главном расслоении, гомотопны.

\begin{exa}[\bf Локальное рассмотрение]
Продолжим локальное построение, начатое в примере \ref{elocon}. У нас есть
левоинвариантный базис $\lbrace\pl_\al,L_\Sa^*\rbrace$ касательных
пространств $\MT_p(\MQ)$ к главному подрасслоению
$\MQ(\MU,\pi,\MG)\subset\MP(\MM,\pi,\MG)$. Обозначим дуальный к нему базис
1-форм на $\MQ$ через $\lbrace dx^\al,\om^{*\Sa}\rbrace$. Теперь построим форму
связности $\om$ на $\MQ$. Произвольную 1-форму
на $\MQ$ со значениями в алгебре Ли $\Gg$ можно разложить по этому базису,
\begin{equation*}
  \om(x,a)=\big(dx^\al\om_\al{}^\Sa+\om^{*\Sb}\om_\Sb{}^\Sa\big)L_\Sa,
\end{equation*}
где $\om_\al{}^\Sa(x,a)$ и $\om_\Sb{}^\Sa(x,a)$ -- некоторые компоненты,
зависящие от точки $(x,a)\in\MQ$. Свойство 1) предложения \ref{pcofop} взаимно
однозначно определяет компоненту $\om_\Sb{}^\Sa$:
\begin{equation*}
  \om(X^*)=X\quad \Leftrightarrow\quad \om_\Sb{}^\Sa(x,a)=\dl_\Sb^\Sa.
\end{equation*}
Нетрудно проверить, что 1-форма $\om^{*\Sa}L_\Sa$ удовлетворяет свойству 2)
предложения \ref{pcofop}. Действительно,
\begin{equation*}
  r^*_a(\om^{*\Sa}L_\Sa)(\widetilde X)=(\om^{*\Sa}L_\Sa)(r_{a*}\widetilde X)
  =\widetilde X^\Sb S^{-1}_{\quad \Sb}{}^\Sa(a) L_\Sa,
\end{equation*}
где мы учли действие дифференциала отображения на векторное поле (\ref{endiri}).
Отсюда следует, что форма $\om^{*\Sa}L_\Sa$ действительно удовлетворяет условию
2),
\begin{equation*}
  r^*_a(\om^{*\Sa}L_\Sa)=\om^{*\Sb}S^{-1}_{\quad \Sb}{}^\Sa(a) L_\Sa.
\end{equation*}

Проведя аналогичные вычисления, получаем, что 1-форма со значениями в алгебре
Ли $dx^\al\om_\al{}^\Sa L_\Sa$ удовлетворяет свойству 2) предложения
\ref{pcofop} тогда и только тогда, когда ее компоненты имеют вид
\begin{equation}                                                  \label{ecoomc}
  \om_\al{}^\Sa(x,b)=\overset{\,\,\circ}A_\al{}^\Sb(x)S^{-1}_{\quad \Sb}{}^\Sa(b),
\end{equation}
где $\overset{\,\,\circ}A_\al{}^\Sb(x)$ -- произвольные функции от $x\in\MU$.
Таким образом, форма связности на $\MQ(\MU,\pi,\MG)=\MU\times\widetilde\MG$
имеет вид
\begin{equation}                                                  \label{ecjfor}
  \om(x,a)=\big[dx^\al\om_\al{}^\Sa(x,a)+\om^{*\Sa}(a)\big]L_\Sa,
\end{equation}
где компоненты $\om_\al{}^\Sa$ определены равенством (\ref{ecoomc}). Мы видим,
что форма связности $\om$ на главном расслоении $\MQ=\MU\times\widetilde\MG$
параметризуется $n\times\Sn$ произвольными функциями
$\overset{\,\,\circ}A_\al{}^\Sb(x)$ на координатной окрестности $\MU\subset\MM$.
Функции $\overset{\,\,\circ}A_\al{}^\Sb(x)$ определяют компоненты формы
связности на {\em нулевом локальном сечении}
\index{локальное сечение нулевое (zero local cross-section)}%
\index{нулевое локальное сечение (zero local cross-section)}%
\begin{equation}                                                  \label{ezerse}
  \s_0:\quad \MU\ni\quad x\mapsto(x,e)\quad\in\MU\times\widetilde\MG.
\end{equation}

Вектор $\widetilde X_p=\widetilde X_p^\al\pl_\al
+\widetilde X_p^\Sa L^*_\Sa\in\MT_x(\MQ)$ горизонтален тогда и только тогда,
когда выполнено равенство
\begin{equation*}
  \om(\widetilde X_p)
  =\big(\widetilde X_p^\al\om_\al{}^\Sa+\widetilde X_p^\Sa\big)L_\Sa=0,
\end{equation*}
где $\om_\al{}^\Sa$ имеет вид (\ref{ecoomc}). Поэтому его вертикальная и
горизонтальная составляющие равны
\begin{equation}                                                  \label{ehovec}
\begin{split}
  \ver\widetilde X_p&
  =\left(\widetilde X_p^\Sa+\widetilde X_p^\al\om_\al{}^\Sa\right) L^*_\Sa,
\\
  \hor\widetilde X_p&=\widetilde X^\al_p\pl_\al
  -\widetilde X^\al_p\om_\al{}^\Sa L^*_\Sa.
\end{split}
\end{equation}
Перемешивание компонент $\widetilde X_p^\al$ и $\widetilde X_p^\Sa$ связано с
тем, что векторные поля $\pl_\al$ в общем случае не являются горизонтальными.
Они горизонтальны тогда и только тогда, когда
$\om_\al{}^\Sa=0\Leftrightarrow\overset{\,\,\circ}A_\al{}^\Sa=0$.
\qed\end{exa}
\begin{com}
Из вида формы связности (\ref{ecjfor}) следует, что она совпадает с канонической
1-формой $\theta$ на группе Ли (\ref{exanfo}), если база расслоения состоит из
одной точки. В этом смысле форма связности $\om$ на главном расслоении $\MP$
является обобщением канонической 1-формы $\theta$ на группе Ли $\MG$.
\qed\end{com}

Проекции $\pi:~\MP\rightarrow\MM$ соответствует дифференциал отображения
$\pi_*:~\MT_p(\MP)\rightarrow\MT_x(\MM)$, который линейно отображает касательное
пространство к главному расслоению в каждой точке $p\in\MP$ в касательное
пространство к базе в точке $x=\pi(p)\in\MM$. Ядром дифференциала проекции
$\pi_*$ является вертикальное подпространство $\MV_p(\MP)$. Действительно, любой
вертикальный вектор касается некоторой кривой, целиком лежащей в слое. При
проекции вся кривая отображается в одну точку $x$. Поэтому каждый вертикальный
вектор отображается в нулевой вектор из $\MT_x(\MM)$. Поскольку дифференциал
проекции -- это сюрьективный гомоморфизм, то отсюда следует, что отображение
горизонтального подпространства $\pi_*:~\MH_p(\MP)\rightarrow\MT_x(\MM)$
является изоморфизмом.
\begin{defn}
{\em Горизонтальным лифтом} или {\em подъемом} (или просто {\em лифтом})
векторного поля $X$ на базе $\MM$ называется единственное векторное поле
$\widetilde X$ на $\MP$, которое горизонтально, $\widetilde X_p\in\MH_p(\MP)$,
и проектируется на $X$, т.е.\ $\pi_*(\widetilde X_p)=X_{\pi(p)}$ для всех
$p\in\MP$.
\index{Горизонтальный лифт векторного поля (horizontal lift of a vector field)}%
\index{Лифт векторного поля (lift of a vector field)}%
\qed\end{defn}
\begin{prop}
Если на главном расслоении $\MP(\MM,\pi,\MG)$ задана связность $\Gamma$ и на базе
$\MM$ задано векторное поле $X$, то существует единственный горизонтальный лифт
$\widetilde X$ векторного поля $X$. Лифт $\widetilde X$ инвариантен относительно
действия структурной группы, $r_{a*}\widetilde X_p=\widetilde X_{pa}$, для
всех $a\in\MG$. Обратно. Каждое горизонтальное векторное поле $\widetilde X$ на
$\MP$, инвариантное относительно действия структурной группы, является
горизонтальным лифтом некоторого векторного поля $X$ на базе $\MM$.
\end{prop}
\begin{proof}
См., например, \cite{KobNom6369R}.
\end{proof}
\begin{prop}
Пусть $\widetilde X$ и $\widetilde Y$ -- горизонтальные лифты соответственно
векторных полей $X$ и $Y$, заданных на базе $\MM$. Тогда:\newline
\indent 1) \parbox[t]{.92\linewidth}{$\widetilde X+\widetilde Y$ --
горизонтальный лифт векторного поля $X+Y$;}\newline
\indent 2) \parbox[t]{.92\linewidth}{для каждой функции $f\in\CC^\infty(\MM)$
произведение $\tilde f\widetilde X$ есть горизонтальный лифт для
$fX\in\CX(\MM)$, где функция $\tilde f\in\CC^\infty(\MP)$ постоянна на слоях и
определена равенством $\tilde f=f\circ\pi$;}\newline
\indent 3) \parbox[t]{.92\linewidth}{горизонтальная компонента коммутатора
$[\widetilde X,\widetilde Y]$ есть горизонтальный лифт коммутатора $[X,Y]$.}
\end{prop}
\begin{proof}
Первые два утверждения очевидны. Доказательство третьего просто:
\begin{equation*}                                                    \tag*{\qed}
  \pi_*(\hor[\widetilde X,\widetilde Y])=\pi_*([\widetilde X,\widetilde Y])
  =[X,Y].
\end{equation*}
\renewcommand{\qed}{}\end{proof}
Пусть $x^\al$, $\al=1,\dotsc,\dim\MM$ -- система координат на некоторой
окрестности базы $\MU\subset\MM$. Пусть $D_\al$ -- горизонтальный лифт
векторного поля $\pl_\al$ в $\pi^{-1}(\MU)$ для всех $\al$. Тогда векторные
поля $\lbrace D_\al\rbrace$ образуют локальный базис распределения
$\Gamma:~p\mapsto\MH_p(\MP)$ в окрестности $\pi^{-1}(\MU)\subset\MP$.

\begin{exa}[\bf Локальное рассмотрение]
Продолжим локальное построение, начатое в примере \ref{elocon}. Пусть
$X(x)=X^\al_x\pl_\al$ -- произвольное векторное поле на базе $\MM$. Его
горизонтальный лифт в $\MQ$ имеет вид
\begin{equation*}
  \widetilde X_p=X^\al_x\pl_\al-X^\al_x\om_\al{}^\Sa L^*_\Sa,\qquad x=\pi(p),
\end{equation*}
для всех $p\in\MQ$. Вблизи единицы группы второе слагаемое имеет вид
\begin{equation*}
  X^\al_x\overset{\,\,\circ}A_\al{}^\Sa S^{-1}_{\quad \Sa}{}^\Sb L^*_\Sb
  =X^\al_x\overset{\,\,\circ}A_\al{}^\Sa R^*_\Sa,
\end{equation*}
где $R^*_\Sa$ правоинвариантные векторные поля на группе Ли $\widetilde\MG$ и мы
воспользовались выражением матрицы присоединенного представления через
производные от функции композиции (\ref{eadsma}). Отсюда сразу следует, что
горизонтальное векторное поле $\widetilde X$ инвариантно относительно действия
структурной группы справа. Утверждения предложения \ref{pcofop} становятся
тривиальными.

В частности, горизонтальный лифт координатного базиса $\pl_\al$ на $\MM$ имеет
вид
\begin{equation}                                                  \label{qgjkol}
  D_\al=\pl_\al-\om_\al{}^\Sa(p) L^*_\Sa=\pl_\al
  -\overset{\,\,\circ}A_\al{}^\Sb(x)S^{-1}_{\quad \Sb}{}^\Sa(a) L^*_\Sa
\end{equation}
для всех точек $p=(x,a)\in\MP$.
\qed\end{exa}

Теперь спустимся на базу и определим форму связности на $\MP$ через семейство
1-форм на $\MM$.
Пусть $\MP(\MM,\pi,\MG)$ -- главное расслоение с формой связности $\om$. Пусть
$\MU\subset\MM$ -- некоторая координатная окрестность базы, на которой задано
локальное сечение $\s:~\MU\rightarrow\MP$. Тогда форма связности $\om$
определяет на $\MU$ 1-форму с помощью возврата отображения $\s^*$.
\begin{defn}
1-форма $\om_\s:=\s^*\om$ со значениями в алгебре Ли $\Gg$ структурной группы,
$\om_\s=\om_\s^\Sa L_\Sa$, где $\om_\s^\Sa\in\Lm_1(\MU)$ называется
{\em локальной формой связности} на $\MU\subset\MM$. В компонентах она имеет вид
\begin{equation}                                                  \label{egafde}
  \om_\s=dx^\al A_\al{}^\Sa L_\Sa,
\end{equation}
где $A_\al{}^\Sa(x)$ -- некоторые функции на $\MU$.
\qed\end{defn}
\index{Локальная форма связности (local connection form)}%
\index{Форма связности локальная (local connection form)}%
Напомним, что, по определению, $\om_\s(X)=\om(\s_*X)$ для всех векторных полей
на базе $X\in\CX(\MM)$, где $\s_*$ -- дифференциал локального сечения.

Каждая форма связности $\om$ на $\MP$ однозначно определяет локальную форму
связности $\om_\s$ на $\MU\subset\MM$. Обратное, конечно, неверно, т.к.\
локальная форма связности $\om_\s$ определена только на окрестности $\MU$ и,
вдобавок, зависит от сечения $\s$.

\begin{exa}[\bf Локальное рассмотрение]
Рассмотрим зависимость локальной формы связности от сечения. Пусть на главном
расслоении $\MQ=\MU\times\widetilde\MG$ задано два произвольных сечения $\s$ и
$\s'$. Поскольку две точки одного слоя связаны некоторым групповым
преобразованием, то $\s'$ можно выразить через первое сечение,
$\s'(x)=\s(x)a(x)$, где $a(x):~\MU\rightarrow\MG$ -- некоторая функция. Таким
образом,
\begin{equation}                                                  \label{esecde}
  \s:~x\mapsto\big(x,b(x)\big)\qquad \text{и}\qquad
  \s':~x\mapsto \big(x,b(x)a(x)\big).
\end{equation}
Рассмотрим случай, когда все элементы структурной группы Ли $\MG$ находятся в
окрестности единицы группы, соответствующей локальной группе Ли (см.\ раздел
\ref{slolir}). То есть каждый элемент группы имеет координаты,
$a=\lbrace a^\Sa\rbrace\in\MG$, $\Sa=1,\dotsc,\Sn$, и задана функция композиции
$ba=f=\lbrace f^\Sa(b,a)\rbrace$. Тогда действие дифференциалов сечений на
векторное поле $X=X^\al\pl_\al\in\CX(\MM)$ имеет вид
\begin{equation*}
\begin{split}
  \s_*X&=X^\al\pl_\al+X^\al\pl_\al b^\Sa\pl_\Sa
  =X^\al\pl_\al+X^\al\pl_\al b^\Sa L^{-1}_{\quad \Sa}{}^\Sb(b) L_\Sb^*(b),
\\
  \s'_*X&=X^\al\pl_\al+X^\al\pl_\al f^\Sa\pl_\Sa
  =X^\al\pl_\al+X^\al\pl_\al f^\Sa L^{-1}_{\quad \Sa}{}^\Sb(f) L_\Sb^*(f),
\end{split}
\end{equation*}
где использован явный вид формы связности (\ref{ecjfor}).
Локальные формы связности $\om_\s$ и $\om_{\s'}$ являются 1-формами на $\MU$ со
значениями в алгебре Ли $\Gg$. Поэтому они представимы в виде
\begin{equation}                                                  \label{edisfo}
\begin{split}
  \om_\s&=dx^\al A_\al{}^\Sa(x)L_\Sa,
\\
  \om_{\s'}&=dx^\al A'_\al{}^\Sa(x)L_\Sa,
\end{split}
\end{equation}
где $A_\al{}^\Sa(x)$ и $A'_\al{}^\Sa(x)$ -- компоненты этих форм (некоторые
дифференцируемые функции от $x\in\MU$). С другой стороны, по определению,
\begin{equation}                                                  \label{elocof}
\begin{split}
  \om_{\s}(X)&=\om(\s_* X)
  =X^\al\left[\overset{\,\,\circ}A_\al{}^\Sb S^{-1}_{\quad \Sb}{}^\Sa(b)
  +\pl_\al b^\Sb L^{-1}_{\quad \Sb}{}^\Sa(b)\right]L_\Sa,
\\
  \om_{\s'}(X)&=\om(\s'_* X)
  =X^\al\left[\overset{\,\,\circ}A_\al{}^\Sb S^{-1}_{\quad \Sb}{}^\Sa(f)
  +\pl_\al f^\Sb L^{-1}_{\quad \Sb}{}^\Sa(f)\right]L_\Sa.
\end{split}
\end{equation}

Рассмотрим нулевое сечение $\s_0=(x,e)$, где $e$ -- единица группы. Тогда
\begin{equation*}
  \om_0=dx^\al\overset{\,\,\circ}A_\al{}^\Sa L_\Sa.
\end{equation*}
То есть функции $\overset{\,\,\circ}A_\al{}^\Sa(x)$, параметризующие форму
связности в (\ref{ecjfor}), представляют собой компоненты локальной формы
связности для нулевого сечения. Тогда компоненты локальной формы связности
$\om_\s$ связаны с компонентами локальной формы связности для нулевого сечения
простым соотношением
\begin{equation}                                                  \label{ecozef}
  A_\al{}^\Sa=\overset{\,\,\circ}A_\al{}^\Sb S^{-1}_{\quad \Sb}{}^\Sa(b)
  +\pl_\al b^\Sb L^{-1}_{\quad \Sb}{}^\Sa(b).
\end{equation}
Сравнивая формулы (\ref{elocof}), получим связь между компонентами локальных
форм связности,
\begin{equation}                                                  \label{elotrs}
   A'_\al{}^\Sa=A_\al{}^\Sb S^{-1}_{\quad \Sb}{}^\Sa(a)
  +\pl_\al f^\Sb L^{-1}_{\quad \Sb}{}^\Sa(f)
  -\pl_\al b^\Sb L^{-1}_{\quad \Sb}{}^\Sc(b) S^{-1}_{\quad \Sc}{}^\Sa(a),
\end{equation}
где функция $a(x)$ связывает данные сечения $\s'=\s a=(x,f=ba)$.

Полученная связь компонент двух локальных форм связности (\ref{elotrs}),
заданных на одной окрестности базы $\MU\subset\MM$ неудобна в приложениях, т.к.\
содержит координатные функции $a^\Sa(x)$, $b^\Sa(x)$, $f^\Sa$, которые в явном
виде можно задать только в редких случаях. Чтобы обойти эту трудность
выберем присоединенное представление алгебры Ли $\ad(\Gg)$. То есть вместо
локальных 1-форм $\om_\s$ со значениями в алгебре Ли $\Gg$ будем рассматривать
1-формы связности со значениями в присоединенном представлении алгебры Ли
$\ad(\Gg)$. Этому соответствует переход к матрицам
\begin{equation*}
  A_\al{}^\Sa\mapsto A_{\al\Sb}{}^\Sc:=-A_\al{}^\Sa f_{\Sa\Sb}{}^\Sc,
\end{equation*}
где $f_{\Sa\Sb}{}^\Sc$ -- структурные константы группы Ли $\MG$. Компоненты
локальной формы связности в присоединенном представлении мы будем обозначать
$A_\al=\lbrace A_{\al\Sb}{}^\Sc\rbrace$, опуская, для краткости, матричные
индексы. Продифференцируем матрицу присоединенного представления
$S_\Sa{}^\Sb(b)$,
\begin{equation*}
  \pl_\al S_\Sa{}^\Sb=\pl_\al b^\Sc\pl_\Sc S_\Sa{}^\Sb
  =\pl_\al b^\Sc L^{-1}_{\quad \Sc}{}^\Sd f_{\Sd\Sa}{}^\Se S_\Se{}^\Sb,
\end{equation*}
где мы воспользовались формулой дифференцирования (\ref{eaddil}). Умножив это
равенство справа на $S^{-1}$, получим равенство
\begin{equation*}
  \pl_\al S_\Sa{}^\Sb S^{-1}_{\quad \Sb}{}^\Sc
  =\pl_\al b^\Sb L^{-1}_{\quad \Sb}{}^\Sd f_{\Sd\Sa}{}^\Sc.
\end{equation*}
Тогда для компонент локальной формы связности (\ref{ecozef}) со значениями в
присоединенном представлении алгебры Ли справедлива формула
\begin{equation}                                                  \label{egatra}
  A_\al=S\overset{\,\,\circ}A_\al S^{-1}+\pl_\al S S^{-1},
\end{equation}
где мы опустили матричные индексы. Отсюда следует правило
преобразования компонент локальной формы связности при переходе от одного
сечения $\s(x):~\big(x,b(x)\big)$ к другому $\s'(x):~\big(x,b(x)a(x)\big)$,
\begin{equation}                                                  \label{egatrd}
  A'_\al=SA_\al S^{-1}+\pl_\al S S^{-1},
\end{equation}
где $S=\big(S_\Sa{}^\Sb(a)\big)$ -- матрица присоединенного представления.
Это есть ни что иное, как хорошо известная формула калибровочного преобразования
полей Янга--Миллса. Эта формула имеет явные преимущества по сравнению с
(\ref{elotrs}), т.к.\ позволяет проводить вычисления с матрицами присоединенного
представления, элементы которых зависят от точки базы $x\in\MM$.

Вместо присоединенного можно рассматривать произвольное представление алгебры
Ли $\Gg$. Если $L_{\Sa i}{}^j$ -- матрицы, соответствующие левоинвариантным
векторным полям $L_\Sa$, и $A_{\al i}{}^j:=A_\al{}^\Sa L_{\Sa i}{}^j$ --
компоненты локальной формы  связности в данном представлении, то формулы
преобразования компонент при переходе между сечениями (\ref{egatrd})
сохраняются, если под $S$ понимать матрицу соответствующего преобразования,
$S=\lbrace S_i{}^j(a)\rbrace$.
\qed\end{exa}

Функции $a(x)$ можно рассматривать как функции, задающие вертикальный
автоморфизм главного расслоения $\MQ=\MU\times\widetilde\MG$ (см.\ пример
\ref{everau}). Тогда формула (\ref{egatrd}) задает преобразование компонент
локальной формы связности при вертикальном автоморфизме.

Отображение $\MU\ni x\mapsto S_\Sa{}^\Sb(x)\in\ad(\Gg)$ сопоставляет каждой
точке базы матрицу присоединенного представления структурной группы. В формуле
(\ref{egatrd}) каждое слагаемое определяет 1-форму на $\MU$ и не зависит явно от
функции композиции. Поэтому, несмотря на то, что формулы преобразования
компонент локальной формы связности были получены в окрестности единицы группы,
они справедливы для произвольных локальных сечений на $\MU\subset\MM$.

Чтобы установить связь рассмотренной конструкции с понятиями, которые широко
используются в математической физике, дадим
\begin{defn}
{\em Калибровочным полем} или {\em полем Янга--Миллса} $A_\al{}^\Sa(x)$ на
координатной окрестности $\MU\subset\MM$ называются компоненты локальной формы
связности (\ref{egafde}), после добавления соответствующих уравнений движения.
{\em Калибровочным преобразованием} называется переход между двумя локальными
сечениями $\s$ и $\s'$ на $\MU$ или, что эквивалентно, преобразование компонент
локальной формы связности при вертикальном автоморфизме. Структурная группа
$\MG$ называется {\em калибровочной} группой.
В электродинамике структурной группой является абелева группа $\MU(1)$, а
компоненты локальной формы связности $A_\al(x)$ на $\MU$, после добавления
уравнений Максвелла, называются {\em электромагнитным потенциалом}. Конечно,
когда мы говорим про калибровочные поля, то подразумеваем, что они удовлетворяют
некоторой системе уравнений движения (уравнения Янга--Миллса или уравнения
Максвелла).
\qed\end{defn}
\index{Калибровочное поле (gauge field)}%
\index{Поле калибровочное поле (gauge field)}%
\index{Калибровочное преобразование (gauge transformation)}%
\index{Преобразование калибровочное (gauge transformation)}%
\index{Янга--Миллса поле (Yang--Mills field)}%
\index{Поле Янга--Миллса (Yang--Mills field)}%
\index{Калибровочная группа (gauge group)}%
\index{Группа калибровочная (gauge group)}%
\index{Электромагнитный потенциал (electromagnetic potential)}%
\index{Потенциал электромагнитный (electromagnetic potential)}%
\begin{com}
Форма связности $\om$ на $\MP$ определена инвариантным образом и ее компоненты
$\om_\al{}^\Sa(p)$ в (\ref{ecjfor}) являются тензорными полями на пространстве
главного расслоения $\MP$. Важное неоднородное слагаемое в калибровочном
преобразовании $(\ref{egatrd})$ для компонент локальной формы связности
$A_\al{}^\Sa(x)$ возникает только при переходе к локальным сечениям.
\qed\end{com}

Как уже было отмечено, локальная форма связности $\om_\s$ не определяет форму
связности $\om$ и, следовательно, связность $\Gamma$ на главном расслоении
$\MP(\MM,\pi,\MG)$. Однако, если задано координатное покрытие базы $\MM$ и
семейство локальных форм связности на каждой координатной окрестности, то это
семейство однозначно определяет связность на $\MP$. Опишем соответствующую
конструкцию. Пусть задано координатное покрытие базы, $\MM=\bigcup_i\MU_i$,
семейство изоморфизмов
\begin{equation*}
  \chi_i:\quad \pi^{-1}(\MU_i)\rightarrow\MU_i\times\MG
\end{equation*}
и соответствующие функции перехода
\begin{equation*}
  a_{ji}:\quad \MU_i\cap\MU_j\rightarrow\MG.
\end{equation*}
Это необходимо для однозначного определения главного расслоения в терминах
функций перехода (теорема \ref{tlofib}).
На каждой окрестности $\MU_i$ выберем сечение $\s_i:~\MU_i\rightarrow\MP$,
которое соответствует единичному элементу группы $e\in\MG$, положив
$\s_i(x)=\chi^{-1}_i(x,e)$. Пусть $\theta$ -- (левоинвариантная $\Gg$-значная)
каноническая 1-форма на $\MG$, которая определена в разделе \ref{sleacg}.

Для каждого непустого пересечения $\MU_i\cap\MU_j$ определим $\Gg$-значную
1-форму на $\MU_i\cap\MU_j$ с помощью возврата отображения $a_{ji}$,
\begin{equation}                                                  \label{ethegr}
  \theta_{ji}=a_{ji}^*\theta.
\end{equation}
Для каждой окрестности $\MU_i$ определим локальную форму связности на $\MU_i$ с
помощью возврата сечения $\s_i$,
\begin{equation*}
  \om_i=\s_i^*\om,
\end{equation*}
где $\om$ -- форма связности на $\MP$. Тогда справедлива
\begin{theorem}                                                   \label{troloc}
Формы $\theta_{ji}$ и $\om_i$ удовлетворяют условиям:
\begin{equation}                                                  \label{econff}
  \om_j=\ad(a_{ji}^{-1})\om_i+\theta_{ji},\qquad \text{на}\quad \MU_i\cap\MU_j.
\end{equation}
Обратно. Для каждого семейства локальных форм связности $\lbrace\om_i\rbrace$,
заданных на координатном покрытии $\MM=\bigcup_i\MU_i$ и удовлетворяющих
условиям (\ref{econff}) во всех пересечениях карт, существует единственная форма
связности $\om$ на $\MP$, которая порождает семейство 1-форм
$\lbrace\om_i\rbrace$ вышеописанным образом.
\end{theorem}
\begin{proof}
См., например, \cite{KobNom6369R}.
\end{proof}
\begin{exa}[\bf Локальное рассмотрение]
Продолжим локальное рассмотрение, чтобы прояснить содержание последнего
утверждения. Допустим, что мы находимся в окрестности единицы структурной
группы. Тогда каноническая форма на группе Ли имеет вид
$\theta=da^\Sb L^{-1}_{\quad \Sb}{}^\Sa L_\Sa$, где $L_\Sa$ -- левоинвариантный
базис алгебры Ли $\Gg$. Пусть в пересекающихся картах $\MU_i$ и $\MU_j$ заданы
координаты $x^\al$ и $x^{\al'}$ соответственно. Пусть
\begin{equation*}
  X=X^\al\pl_\al=X^{\al'}\pl_{\al'}\in\CX(\MU_i\cap\MU_j)
\end{equation*}
-- касательный вектор к базе на пересечении окрестностей. Тогда дифференциал
функций перехода отображает этот вектор в касательное пространство группы
\begin{equation*}
  a_{ji*}X=X^\al\pl_\al a_{ji}^\Sa\pl_\Sa
  =X^\al\pl_\al a_{ji}^\Sb L^{-1}_{\quad \Sb}{}^\Sa L_\Sa.
\end{equation*}
Такая же формула имеет место в штрихованной системе координат. Поэтому
$\Gg$-значная 1-форма (\ref{ethegr}) имеет вид
\begin{equation*}
  \theta_{ji}=dx^\al \pl_\al a_{ji}^\Sb L^{-1}_{\quad \Sb}{}^\Sa(a_{ji})L_\Sa
  =dx^{\al'} \pl_{\al'}a_{ji}^\Sb L^{-1}_{\quad \Sb}{}^\Sa(a_{ji})L_\Sa.
\end{equation*}
Пусть
\begin{equation*}
  \om_i=dx^\al A_\al{}^\Sa L_\Sa\qquad \text{и}\qquad
  \om_j=dx^{\al'}A_{\al'}{}^\Sa L_\Sa
\end{equation*}
-- локальные формы связности соответственно на $\MU_i$ и $\MU_j$. Тогда формула
(\ref{econff}) приводит к равенству
\begin{equation}                                                  \label{egatrc}
  A_{\al'}{}^\Sa=\pl_{\al'}x^\al\left[A_\al{}^\Sb S^{-1}_{\quad \Sb}{}^\Sa(a_{ji})
  +\pl_\al a_{ji}^\Sb L^{-1}_{\quad \Sb}{}^\Sa(a_{ji})\right].
\end{equation}
Это преобразование калибровочного поля отличается от описанного ранее
(\ref{ecozef}) только множителем $\pl_{\al'}x^\al$, соответствующим
преобразованию координат в пересечении $\MU_i\cap\MU_j$.
\qed\end{exa}
Таким образом, для того, чтобы описать связность $\Gamma$ на главном расслоении
$\MP(\MM,\pi,\MG)$ в локальных терминах, необходимо задать координатное покрытие
базы $\MM=\bigcup_i\MU_i$, функции перехода $a_{ji}(x)$ в каждом непустом
пересечении $\MU_i\cap\MU_j$, которые удовлетворяют условию (\ref{etrfco}) в
областях пересечения трех карт, и семейство локальных форм связности
(калибровочных полей), заданных на каждой координатной окрестности, такое, что в
областях пересечения карт выполнено условие (\ref{egatrc}).
\begin{com}
В квантовой теории поля обычно рассматривают тривиальные главные расслоения
$\MP=\MR^{1,3}\times\MG$, где $\MR^{1,3}$ -- четырехмерное пространство
Минковского и $\MG$ -- калибровочная группа. В этом частном случае локальная
форма связности $dx^\al A_\al{}^\Sa L_\Sa$, заданная на всем
пространстве-времени $\MR^{1,3}$, взаимно однозначно определяет форму связности
$\om$ и, следовательно, связность $\Gamma$ на $\MP$.
\qed\end{com}
В заключение данного раздела обсудим вопрос о существовании связностей. Пусть
$\MP(\MM,\pi,\MG)$ -- главное расслоение и $\MN\subset\MM$ -- некоторое
подмножество базы. Мы говорим, что связность определена над $\MN$, если в каждой
точке $p\in\pi^{-1}(\MN)\subset\MP$ из прообраза $\MN$ определено горизонтальное
подпространство $\MH_p(\MP)$ таким образом, что выполнены первые два условия
(прямая сумма и инвариантность относительно действия структурной группы) в
определении связности и распределение $\MH_p(\MP)$ дифференцируемо зависит от
точки $p$ в следующем смысле. Для каждой точки $x\in\MN$ существует окрестность
$\MU\ni x$ и связность на $\pi^{-1}(\MU)$ такая, что ее сужение на $\pi^{-1}(x)$
совпадает с $\MH_p(\MP)$ для всех точек слоя $p\in\pi^{-1}(x)$.
\begin{theorem}                                                   \label{tsushc}
Пусть $\MP(\MM,\pi,\MG)$ -- главное расслоение и $\MN$ -- замкнутое подмножество
(возможно, пустое) в базе $\MM$. Любая связность, определенная над $\MN$ может
быть продолжена до связности на всем $\MP$. В частности, любое главное
расслоение $\MP$ допускает связность.
\end{theorem}
\begin{proof}
Доказательство проводится путем явного построения связности. При этом
существенно используется паракомпактность многообразия. См., например,
\cite{KobNom6369R}.
\end{proof}
\begin{exa}
Рассмотрим главное расслоение $\MP(\MM,\pi,e)$, структурная группа которого
состоит из единственного элемента -- единицы. Тогда на нем существует
единственная связность. В этом случае вертикальные касательные пространства
отсутствуют, а распределение горизонтальных пространств совпадает с касательным
расслоением $\MT(\MM)$. Форма связности и форма кривизны, которые будут
определены в следующем разделе, при этом тождественно равны нулю.
\qed\end{exa}
Продолжение связности не является единственным.
\begin{exa}
Рассмотрим расслоение реперов $\ML(\MR^n)$ над евклидовым пространством $\MR^n$.
Поскольку $\MR^n$ покрывается одной картой, то форма связности
$\om=dx^\al\om_{\al a}{}^b \Be_b{}^a$, где $\Be_b{}^a$ -- базис алгебры Ли
$\Gg\Gl(n,\MR)$, взаимно однозначно определятся своими компонентами
$\om_{\al a}{}^b$ на $\ML$. Чтобы их задать, достаточно задать компоненты
локальной формы связности на каком либо сечении, а затем разнести их по всему
пространству расслоения с помощью действия структурной группы. Если компоненты
локальной формы связности $\om_{\al a}{}^b(x)$ заданы на замкнутом подмножестве
$\MN\subset\MR^n$, то их можно продолжить на все евклидово пространство $\MR^n$
многими достаточно гладкими способами. Тем самым мы определим связность на всем
$\ML(\MR^n)$.
\qed\end{exa}
\section{Форма кривизны и структурное уравнение                  \label{scurds}}
Пусть задано главное расслоение $\MP(\MM,\pi,\MG)$ со связностью
$\Gamma:~\MP\ni p\mapsto\MH_p\subset\MT_p(\MP)$. Каждая связность $\Gamma$ на $\MP$
взаимно однозначно определяет $\Gg$-значную 1-форму связности
$\om=\om^\Sa L_\Sa$, где $\om^\Sa\in\Lm_1(\MP)$ для всех $\Sa=1,\dotsc,\Sn$,
которой мы поставим в соответствие единственную 2-форму кривизны
$R=R^\Sa L_\Sa$, где $R^\Sa\in\Lm_2(\MP)$ для всех $\Sa$, со значениями в
алгебре Ли $\Gg$. Форма кривизны $R$ является важнейшей характеристикой заданной
связности $\Gamma$. Для ее определения нам понадобятся новые понятия.

Пусть задано представление $\rho$ структурной группы Ли $\MG$ в конечномерном
векторном пространстве $\MV$,
\begin{equation*}
  \rho:\quad \MG\ni\quad a\mapsto\rho(a)\quad\in\aut(\MV).
\end{equation*}
При этом $\rho(ab)=\rho(a)\circ\rho(b)$ для всех $a,b\in\MG$.
\begin{defn}
$r$-форма $\vf\in\Lm_r(\MP)$ на главном расслоении $\MP$ со значениями в
векторном пространстве $\MV$ называется {\em псевдотензориальной $r$-формой типа
$(\rho,\MV)$}, если выполнено условие
\begin{equation*}
  r_a^*\vf|_p=\rho(a^{-1})\vf|_{pa}\qquad \forall a\in\MG,~\text{и}~
  \forall p\in\MP,
\end{equation*}
где $r_a^*$ -- возврат отображения $r_a:~p\mapsto pa$. Псевдотензориальная
$r$-форма $\vf$ называется {\em тензориальной} $r$-формой типа $(\rho,\MV)$,
если она {\em горизонтальна}, т.е.\
$\vf(\widetilde X_1,\dotsc,\widetilde X_r)=0$, как только один из касательных
векторов $\widetilde X_i\in\CX(\MP)$ вертикален. Под тензориальной 0-формой типа
$(\rho,\MV)$ мы понимаем функцию $\vf:~\MP\rightarrow\MV$, удовлетворяющую
условию $\vf(p)=\rho(a^{-1})\vf(pa)$ или, что эквивалентно,
$\vf(pa)=\rho(a)\vf(p)$.
\qed\end{defn}
\index{Псевдотензориальная $r$-форма (pseudotensorial $r$-form)}%
\index{$r$-форма псевдотензориальная (pseudotensorial $r$-form)}%
\index{Тензориальная $r$-форма (pseudotensorial $r$-form)}%
\index{$r$-форма тензориальная (pseudotensorial $r$-form)}%
\index{Горизонтальная $r$-форма (horizontal $r$-form)}%
\index{$r$-форма горизонтальная (horizontal $r$-form)}%
\begin{exa}                                                       \label{epsome}
В силу свойства 2) предложения \ref{pcofop} форма связности $\om$ является
псевдотензориальной 1-формой типа $(\ad,\Gg)$. Она не является тензориальной,
т.к.\ не обращается в нуль на вертикальных векторных полях. Для краткости,
будем говорить, что форма типа $(\ad,\Gg)$ является формой типа $\ad\MG$.
\qed\end{exa}
\begin{exa}
Пусть $\rho_0$ -- тривиальное представление группы $\MG$ в $\MV$, т.е.\
$\rho_0(a)$ есть тождественное преобразование $\MV$ для всех $a\in\MG$. Тогда
каждая тензориальная $r$-форма $\vf$ типа $(\rho_0,\MV)$ может быть взаимно
однозначно представлена в виде $\vf=\pi^*\vf_\MM$, где $\vf_\MM$ некоторая
$r$-форма на базе $\MM$ со значениями в $\MV$. Для этого достаточно показать,
что тензориальная форма $\vf$ однозначно определяет форму на базе $\vf_\MM$.
Поскольку $r$-форма $\vf$ горизонтальна, то ее значения на произвольном наборе
касательных векторных полей определяются только горизонтальными компонентами,
\begin{equation*}
  \vf(\widetilde X_1,\dotsc,\widetilde X_r)=
  \vf(\hor\widetilde X_1,\dotsc,\hor\widetilde X_r).
\end{equation*}
Так как $r$-форма $\vf$ правоинвариантна, то ее значения на правоинвариантных
векторных полях не зависят от точки слоя $p\in\pi^{-1}(x)$. Поэтому $r$-форма
$\vf$ взаимно однозначно определяет $r$-форму $\vf_\MM$ на базе следующим
равенством
\begin{equation*}
  \vf_\MM(X_1,\dotsc,X_r)=\vf(\widetilde X_1,\dotsc,\widetilde X_r),
\end{equation*}
где $\widetilde X_1,\dotsc,\widetilde X_r$ -- произвольный набор
правоинвариантных векторных полей таких, что $X_i=\pi_*\widetilde X_i$ для всех
$i=1,\dotsc,r$.
\qed\end{exa}
\begin{exa}                                                       \label{easfis}
Пусть $\ME(\MM,\pi_\ME,\MV,\MG,\MP)$ -- ассоциированное с $\MP(\MM,\pi,\MG)$
расслоение, типичным слоем которого является векторное пространство $\MV$, в
котором задано представление $\rho$ структурной группы $\MG$. Тогда
тензориальную $r$-форму $\vf$ типа $(\rho,\MV)$ можно рассматривать как
сопоставление каждой точке $x\in\MM$ мультилинейного антисимметричного
отображения
\begin{equation}                                                  \label{emuanm}
  \tilde\vf_x:\quad \underbrace{\MT_x(\MM)\times\dotsc\times\MT_x}_{r}\rightarrow
  \pi_\ME^{-1}(x)
\end{equation}
с помощью равенства
\begin{equation*}
  \tilde\vf_x(X_1,\dotsc,X_r)=
  p\big(\vf(\widetilde X_1,\dotsc,\widetilde X_r)\big),\qquad X_i\in\MT_x(\MM),
\end{equation*}
где $p$ -- любая точка слоя $\pi^{-1}(x)$ и $\widetilde X_i$ -- произвольный
касательный вектор в $p$, для которого $\pi_*\widetilde X_i=X_i$ для каждого
$i$. Подробнее, $\vf(\widetilde X_1,\dotsc,\widetilde X_r)$ есть тогда элемент
стандартного слоя $\MV$, а $p$ -- линейное отображение из $\MV$ на слой
$\pi_\ME^{-1}(x)$. Поэтому
$p\big(\vf(\widetilde X_1,\dotsc,\widetilde X_r)\big)$ является элементом слоя
$\pi_\ME^{-1}(x)$. Нетрудно проверить, что этот элемент не зависит от выбора
точки $p$ и вектора $\widetilde X_i$ в слое $\pi^{-1}(x)$.

Обратно. Пусть задано мультилинейное антисимметричное отображение (\ref{emuanm})
для всех $x\in\MM$. Тогда можно определить тензориальную $r$-форму типа
$(\rho,\MV)$ на $\MP$ следующим образом
\begin{equation*}
  \vf(\widetilde X_1,\dotsc,\widetilde X_r)=p^{-1}
  \big(\tilde\vf_x(\pi_*\widetilde X_1,\dotsc,\pi_*\widetilde X_r)\big),\qquad
  \widetilde X_i\in\MT_p(\MP),
\end{equation*}
где $x=\pi(p)$. В частности, тензориальная 0-форма типа $(\rho,\MV)$, т.е.\
функция $\vf:~\MP\rightarrow\MV$, удовлетворяющая условию
$\vf(pa)=\rho(a)\vf(p)$, может быть отождествлена с сечением
ассоциированного расслоения $\s_\ME:~\MM\rightarrow\ME$. Несколько специальных
случаев данного примера будет использовано в дальнейшем.
\qed\end{exa}
Продолжим обсуждение связности $\Gamma$ на главном расслоении $\MP(\MM,\pi,\MG)$.
Пусть $\MV_p(\MP)$ и $\MH_p(\MP)$ -- вертикальное и горизонтальное
подпространства в $\MT_p(\MP)$ и $\hor:~\MT_p(\MP)\rightarrow\MH_p(\MP)$ --
проекция на горизонтальное подпространство. Введем новое понятие внешней
ковариантной производной от псевдотензориальной $r$-формы, с помощью которой
будет определена кривизна.
\begin{prop}                                                      \label{ecodee}
Пусть $\vf$ -- псевдотензориальная $r$-форма на $\MP$ типа $(\rho,\MV)$. Тогда:
\newline
\indent 1) \parbox[t]{.92\linewidth}{$r$-форма $\vf\hor$, определенная
равенством
\begin{equation*}
  \vf\hor(\widetilde X_1,\dotsc,\widetilde X_r):=
  \vf(\hor\widetilde X_1,\dotsc,\hor\widetilde X_r),\qquad
  \widetilde X_i\in\MT_p(\MP),\quad i=1,\dotsc,r,
\end{equation*}
есть тензориальная форма типа $(\rho,\MV)$;}\newline
\indent 2) \parbox[t]{.92\linewidth}{$d\vf$ есть псевдотензориальная
$(r+1)$-форма типа $(\rho,\MV)$;}\newline
\indent 3) \parbox[t]{.92\linewidth}{$(r+1)$-форма $D\vf$, определенная как
$D\vf:=(d\vf)\hor$, является тензориальной формой типа $(\rho,\MV)$.}
\end{prop}
\begin{proof}
Распределение горизонтальных векторных полей инвариантно относительно действия
структурной группы, поэтому $r_a\circ\hor=\hor\circ r_a$. Отсюда следует, что
$\vf\hor$ является псевдотензориальной $r$-формой типа $(\rho,\MV)$.
Очевидно, что
\begin{equation*}
  \vf\hor(\widetilde X_1,\dotsc,\widetilde X_r)=0,
\end{equation*}
если один из касательных векторов $\widetilde X_i$ вертикален, и, следовательно,
форма $\vf\hor$ горизонтальна. Второе утверждение следует из равенства
$r_a^*\circ d= d\circ r_a^*$ для всех $a\in\MG$ (\ref{epufor}). Третье
утверждение является прямым следствием двух первых.
\end{proof}
\begin{defn}
Форма $D\vf=(d\vf)\hor$ называется {\em внешней ковариантной производной} от
псевдотензориальной $r$-формы $\vf$ на $\MP$. Оператор $D$ называется
{\em внешним ковариантным дифференцированием}.
\qed\end{defn}
\index{Внешняя ковариантная производная (external covariant derivative)}%
\index{Ковариантная производная внешняя (external covariant derivative)}%
\index{Ковариантное дифференцирование внешнее %
(external covariant differentiation)}%
\index{Внешнее ковариантное дифференцирование %
(external covariant differentiation)}%
\begin{exa}[\bf Локальное рассмотрение]
Выпишем ковариантную производную от тензориальной 0-формы на
$\MQ=\MU\times\widetilde\MG$ типа $(\rho,\MV)$ в компонентах. Пусть $e_i$,
$i=1,\dotsc,\dim\MV$, -- базис векторного пространства $\MV$ и
\begin{equation*}
  \rho:\quad \MG\ni\quad a\mapsto S_i{}^j(a)\quad\in\aut\MV
\end{equation*}
-- представление структурной группы. Пусть $L_{\Sa j}{}^i$ -- представление
генераторов (базиса) $L_\Sa$ структурной группы. Тогда справедливы формулы
(\ref{ediinm}) и (\ref{einvge}). Поскольку $\vf$ -- тензориальная 0-форма типа
$(\rho,\MV)$, то она имеет вид
\begin{equation*}
  \vf(p)=\vf^i(p)e_i=\overset\circ\vf{}^j(x)S^{-1}_{\quad j}{}^i(b)e_i,\qquad
  p=(x,b)\in\MQ,
\end{equation*}
где $\overset\circ\vf(x,0)=\overset\circ\vf{}^i(x)e_i$ -- значение этой
функции на нулевом сечении $\s_0(x)=(x,0)\in\MP$. Внешний дифференциал от
компонент $\vf^i$ равен
\begin{equation*}
\begin{split}
  d\vf^i&=dx^\al\pl_\al\overset\circ\vf{}^jS^{-1}_{\quad j}{}^i
  +db^\Sa\overset\circ\vf{}^j\pl_\Sa S^{-1}_{\quad j}{}^i=
  dx^\al\pl_\al\overset\circ\vf{}^jS^{-1}_{\quad j}{}^i
  +\om^{*\Sa}\overset\circ\vf{}^j L^*_\Sa S^{-1}_{\quad j}{}^i=
\\
  &=dx^\al\pl_\al\overset\circ\vf{}^j S^{-1}_{\quad j}{}^i
  +\om^{*\Sa}\overset\circ\vf{}^j S^{-1}_{\quad j}{}^k L_{\Sa k}{}^i.
\end{split}
\end{equation*}
Поскольку горизонтальная составляющая вектора имеет вид (\ref{ehovec}), то
значение внешней ковариантной производной на произвольном векторном поле
$\widetilde X\in\MX(\MP)$ равно
\begin{equation}                                                  \label{ecotef}
  D\vf^i(\widetilde X)=d\vf^i(\hor\widetilde X)
  =\widetilde X^\al\overset\circ\nb_\al\overset\circ\vf{}^j S^{-1}_{\quad j}{}^i,
\end{equation}
где введено обозначение
\begin{equation*}
  \overset\circ\nb_\al\overset\circ\vf{}^j=\pl_\al\overset\circ\vf{}^j
  +\overset\circ\vf{}^k\overset{\,\,\circ}A_{\al k}{}^j,\qquad
  \overset{\,\,\circ}A_{\al k}{}^j
  =-\overset{\,\,\circ}A_{\al}{}^\Sa L_{\Sa k}{}^j.
\end{equation*}
Теперь спустимся на базу $\MU$. Рассмотрим два произвольных сечения
\begin{equation*}
  \s:\quad x\mapsto\big(x,b(x)\big),\qquad \text{и}\qquad
  \s':\quad x\mapsto\big(x,b(x)a(x)\big),
\end{equation*}
связанных калибровочным преобразованием, которое параметризуется функцией
$a(x)$. Обозначим значение компонент $\vf^i(p)$ на этих сечениях через
\begin{equation}                                                  \label{etwsec}
  \vf^i(x)=\vf^i|_\s,\qquad \text{и}\qquad \vf^{\prime i}(x)=\vf^i|_{\s'}.
\end{equation}
Эти функции на $\MU$ связаны калибровочным преобразованием
\begin{equation}                                                  \label{egavft}
  \vf^{\prime i}=\vf^j S^{-1}_{\quad j}{}^i(a),\qquad a=a(x).
\end{equation}
Внешний ковариантный дифференциал $D\vf$ после проектирования на базу с помощью
возвратов отображений $\s^*$ и $\s^{\prime *}$ принимает вид
\begin{equation}                                                  \label{extcod}
\begin{split}
  \s^*(D\vf^i)&=dx^\al\nb_\al\vf^i,
\\
  \s^{\prime *}(D\vf^i)&=dx^\al\nb'_\al\vf^{\prime i},
\end{split}
\end{equation}
где
\begin{equation}                                                  \label{ecovym}
\begin{split}
  \nb_\al\vf^i&:=\pl_\al\vf^i+\vf^j A_{\al j}{}^i,
\\
  \nb'_\al\vf^{\prime i}&:=\pl_\al\vf^{\prime i}+\vf^{\prime j} A'_{\al j}{}^i.
\end{split}
\end{equation}
Выше мы ввели обозначения: $A_{\al j}{}^i:=-A_\al{}^\Sa L_{\Sa j}{}^i$ и
$A'_{\al j}{}^i:=-A'_\al{}^\Sa L_{\Sa j}{}^i$, где $A_\al{}^\Sa$ и
$A'_\al{}^\Sa$ -- компоненты локальных форм связности (\ref{edisfo}). Нетрудно
проверить, что при калибровочном преобразовании (\ref{egavft}), (\ref{egatra})
ковариантная производная ведет себя ковариантно:
\begin{equation}                                                  \label{ecoder}
  \nb'_\al\vf^{\prime i}=\nb_\al\vf^jS^{-1}_{\quad j}{}^i.
\end{equation}
Этого следовало ожидать, так как определение ковариантной производной было дано
в инвариантном виде. Таким образом, мы видим, что инвариантное определение
внешней ковариантной производной для тензориальной 0-формы типа $(\rho,\MV)$
совпадает с обычным определением ковариантной производной в теории калибровочных
полей. В примере \ref{easfis} было показано, что тензориальную 0-форму типа
$(\rho,\MV)$ можно отождествить с сечением ассоциированного расслоения, типичным
слоем которого является векторное пространство $\MV$. Таким образом, формулы
(\ref{ecovym}) определяют ковариантные производные от сечений
$\s_\ME:~\MM\rightarrow\ME$ ассоциированного расслоения
$\ME(\MM,\pi_\ME,\MV,\MG,\MP)$.
\qed\end{exa}

В примере \ref{epsome} было отмечено, что форма связности $\om$ на $\MP$ есть
псевдотензориальная 1-форма типа $\ad\MG$. Используя предложение \ref{ecodee},
дадим
\begin{defn}
Внешняя ковариантная производная $R:=D\om$ от формы связности $\om$ является
тензориальной 2-формой на $\MP$ типа $\ad\MG$ и называется {\em формой кривизны}
для формы связности $\om$. Если $L_\Sa$ базис алгебры Ли, то $R=R^\Sa L_\Sa$,
где $R^\Sa\in\Lm_2(\MP)$ для всех $\Sa=1,\dotsc,\Sn$.
\qed\end{defn}
\index{Форма кривизны (curvature form)}%
\index{Кривизны форма (curvature form)}%
\begin{theorem}[\bf Структурное уравнение]                        \label{tstreq}
Пусть $\om$ -- форма связности на главном расслоении $\MP(\MM,\pi,\MG)$ и
$R:=D\om$ -- ее форма кривизны. Тогда
\begin{equation}                                                  \label{estreq}
  d\om(\widetilde X,\widetilde Y)
  =-\frac12\left[\om(\widetilde X),\om(\widetilde Y)\right]
  +R(\widetilde X,\widetilde Y)
\end{equation}
для всех $\widetilde X,\widetilde Y\in\MT_p(\MP)$ и $p\in\MP$.
\end{theorem}
\begin{proof}
Каждый вектор в $\MT_p(\MP)$ единственным образом разлагается в сумму
вертикального и горизонтального векторов. Так как каждый член в (\ref{estreq})
билинеен и антисимметричен по $\widetilde X$ и $\widetilde Y$, то достаточно
проверить равенство (\ref{estreq}) в трех случаях.

1) $\widetilde X_p$ и $\widetilde Y_p$ горизонтальны для всех $p\in\MP$. В этом
случае $\om(\widetilde X)=\om(\widetilde Y)=0$ и равенство (\ref{estreq})
сводится к определению формы кривизны $R$, т.к.\ $D\om=d\om$ для горизонтальных
векторных полей.

2) $\widetilde X_p$ и $\widetilde Y_p$ вертикальны для всех $p\in\MP$. В этом
случае их значения в точке $p$ соответствуют некоторым фундаментальным векторным
полям, т.е.\ $\widetilde X_p=X^*_p$ и $\widetilde Y_p=Y^*_p$ для некоторых
$X,Y\in\Gg$. Из формулы (\ref{edacov}) следует равенство
\begin{equation*}
\begin{split}
  2d\om(X^*,Y^*)&=X^*\big(\om(Y^*)\big)-Y^*\big(\om(X^*)\big)
  -\om\big([X^*,Y^*]\big)=
\\
  &=-[X,Y]=-[\om(X^*),\om(Y^*)],
\end{split}
\end{equation*}
т.к.\ $\om(X^*)=X$, $\om(Y^*)=Y$ и $[X^*,Y^*]=[X,Y]^*$. Поскольку для
фундаментальных векторных полей $R(X^*,Y^*)=0$, то формула (\ref{estreq}) в
рассматриваемом случае имеет место.

3) $\widetilde X_p$ горизонтально, $\widetilde Y_p$ вертикально для всех
$p\in\MP$. Продолжим
$\widetilde X_p\in\MH_p(\MP)$ до горизонтального векторного поля $\widetilde X$
на $\MP$. Это всегда возможно в силу следствия теоремы \ref{tasstr}. Пусть
$\widetilde Y_p=Y^*_p$ для некоторого $Y\in\Gg$. В рассматриваемом случае правая
часть равенства (\ref{estreq}) равна нулю, поэтому достаточно доказать равенство
$d\om(\widetilde X,Y^*)=0$. Из формулы (\ref{edacov}) следует, что
\begin{equation*}
  2d\om(\widetilde X,Y^*)=\widetilde X\big(\om(Y^*)\big)
  -Y^*\big(\om(\widetilde X)\big)-\om\big([\widetilde X,Y^*]\big)
  =-\om\big([\widetilde X,Y^*]\big).
\end{equation*}
Теперь достаточно доказать следующее утверждение
\begin{lemma}
Пусть $\widetilde X$ -- горизонтальное векторное поле и $Y^*$ -- фундаментальное
векторное поле, соответствующее элементу алгебры $Y\in\Gg$. Тогда коммутатор
$[\widetilde X,Y^*]$ горизонтален.
\end{lemma}
{\em Доказательство.}
Фундаментальное векторное поле $Y^*$ индуцировано действием $r_{a(t)}$, где
$a(t)$ -- 1-параметрическая подгруппа в $\MG$, порожденная элементом алгебры
$Y\in\Gg$ (экспоненциальное отображение). Поскольку коммутатор векторных полей
совпадает с производной Ли, то из (\ref{elieve}) следует равенство
\begin{equation*}
  [\widetilde X,Y^*]=-\Lie_{Y^*}\widetilde X
  =-\underset{t\to0}\lim\frac{\widetilde X-r_{a(t)*}\widetilde X}t.
\end{equation*}
Если векторное поле $\widetilde X$ горизонтально, то $r_{a(t)*}\widetilde X$
тоже горизонтально. Поэтому коммутатор $[\widetilde X,Y^*]$ горизонтален.
\qed\end{proof}
\begin{cor}
Если $\widetilde X$ и $\widetilde Y$ -- горизонтальные векторные поля, то
\begin{equation}                                                  \label{econho}
  \om([\widetilde X,\widetilde Y])=-2R(\widetilde X,\widetilde Y). \qed
\end{equation}
\end{cor}
\begin{proof}
Для горизонтальных векторных полей $\om(\widetilde X)=\om(\widetilde Y)=0$ и
\begin{equation*}
  2d\om(\widetilde X,\widetilde Y)=-\om([\widetilde X,\widetilde Y])
\end{equation*}
как следствие (\ref{edacov}).
\end{proof}
Структурное уравнение (\ref{estreq}) называют также {\em структурным уравнением
Картана} и часто для простоты записывают в виде
\begin{equation}                                                  \label{ecastr}
  d\om=-\frac12[\om,\om]+R.
\end{equation}
\index{Структурное уравнение Картана (E.~Cartan's structure equation)}%
\index{Картана структурное уравнение (E.~Cartan's structure equation)}%

Приведем еще одну форму записи структурного уравнения. Пусть $L_\Sa$,
$\Sa=1,\dotsc,\Sn$, -- базис алгебры Ли $\Gg$ с коммутационными соотношениями
\begin{equation*}
  [L_\Sa,L_\Sb]=f_{\Sa\Sb}{}^\Sc L_\Sc,
\end{equation*}
где $f_{\Sa\Sb}{}^\Sc$ -- структурные константы алгебры. Тогда формы связности и
кривизны можно разложить по базису, $\om=\om^\Sa L_\Sa$ и $R=R^\Sa L_\Sa$, и
структурные уравнения принимают вид
\begin{equation}                                                  \label{estrba}
  d\om^\Sa=-\frac12\om^\Sb\wedge\om^\Sc f_{\Sb\Sc}{}^\Sa+R^\Sa.
\end{equation}
\begin{com}
Структурное уравнение (\ref{estreq}) отличается от формулы Маурера--Картана для
групп Ли (\ref{emacaf}) только слагаемым с кривизной. Для фундаментальных
векторных полей $R(X^*,Y^*)=0$ и формулы просто совпадают.
\qed\end{com}

Теперь спустимся на базу. Пусть задано локальное сечение
$\s:~\MU\rightarrow\MP$ на некоторой окрестности $\MU\subset\MM$.
\begin{defn}
2-форма на $\MU$ со значениями в алгебре Ли $\Gg$, определенная возвратом
сечения, $R_\s:=\s^*R$, где $R$ -- форма кривизны на $\MP(\MM,\pi,\MG)$,
называется {\em локальной формой кривизны} формы связности $\om$.
\qed\end{defn}
\index{Локальная форма кривизны (local curvature form)}%
\index{Форма кривизны локальная (local curvature form)}%
В компонентах локальная форма кривизны имеет вид
\begin{equation}                                                  \label{elocur}
  R_\s=F^\Sa L_\Sa=\frac12dx^\al\wedge dx^\bt F_{\al\bt}{}^\Sa L_\Sa,
\end{equation}
где $F^\Sa=\frac12dx^\al\wedge dx^\bt F_{\al\bt}{}^\Sa\in\Lm_2(\MU)$ для всех
$\Sa=1,\dotsc,\Sn$.
\begin{exa}[\bf Локальное рассмотрение]
В настоящем примере мы выразим компоненты локальной формы кривизны через
компоненты локальной формы связности. Чтобы это сделать, сначала определим
компоненты формы кривизны на $\MQ=\MU\times\widetilde\MG$ через компоненты формы
связности. Внешняя производная от формы связности (\ref{ecjfor}) имеет вид
\begin{equation*}
\begin{split}
  d\om&=\left[\frac12dx^\al\wedge dx^\bt
  \left(\pl_\al\om_\bt{}^\Sa-\pl_\bt\om_\al{}^\Sa\right)
  -dx^\al\wedge db^\Sb\pl_\Sb\om_\al{}^\Sa+d\om^{*\Sa}\right]L_\Sa=
\\
  &=\left[\frac12dx^\al\wedge dx^\bt
  \left(\pl_\al\om_\bt{}^\Sa-\pl_\bt\om_\al{}^\Sa\right)
  +dx^\al\wedge \om^{*\Sb}\om_\al{}^\Sc f_{\Sb\Sc}{}^\Sa
  -\frac12\om^{*\Sb}\wedge\om^{*\Sc}f_{\Sb\Sc}{}^\Sa\right]L_\Sa,
\end{split}
\end{equation*}
где
$\om_\al{}^\Sa(x,b)=\overset{\,\,\circ}A_\al{}^\Sb(x)S^{-1}_{\quad \Sb}{}^\Sa(b)$.
Кроме того мы воспользовались формулой Маурера--Картана (\ref{emacar}) для
структурной группы и правилом дифференцирования матрицы присоединенного
представления (\ref{ediadm}). Теперь нетрудно вычислить значение формы кривизны
на векторных полях
\begin{equation*}
  D\om(\widetilde X,\widetilde Y)=d\om(\hor\widetilde X,\hor\widetilde Y)
  =\widetilde X^\al\widetilde Y^\bt R_{\al\bt}{}^\Sa L_\Sa,
\end{equation*}
где
\begin{equation}                                                  \label{ecycol}
  R_{\al\bt}{}^\Sa=\pl_\al\om_\bt{}^\Sa-\pl_\bt\om_\al{}^\Sa
  -\om_\al{}^\Sb\om_\bt{}^\Sc f_{\Sb\Sc}{}^\Sa
\end{equation}
-- компоненты формы кривизны. Эти компоненты можно выразить через компоненты,
заданные на нулевом сечении
\begin{equation*}
    R_{\al\bt}{}^\Sa(x,a)=\overset\circ F_{\al\bt}{}^\Sb(x)
    S^{-1}_{\quad \Sb}{}^\Sa(a),
\end{equation*}
где
\begin{equation}                                                  \label{ecurze}
  \overset{\circ}F_{\al\bt}{}^\Sa=\pl_\al\overset{\,\,\circ}A_\bt{}^\Sa
  -\pl_\bt\overset{\,\,\circ}A_\al{}^\Sa
  -\overset{\,\,\circ}A_\al{}^\Sb\overset{\,\,\circ}A_\bt{}^\Sc f_{\Sb\Sc}{}^\Sa
\end{equation}
-- компоненты тензора кривизны для нулевого сечения. Таким образом, форма
кривизны имеет вид
\begin{equation}                                                  \label{ecurvf}
  R=\frac12dx^\al\wedge dx^\bt R_{\al\bt}{}^\Sa L_\Sa,
\end{equation}
где компоненты определены равенствами (\ref{ecycol}).

Теперь спустимся на базу $\MU$. Для сечений (\ref{etwsec}) получаем следующие
выражения для локальных форм кривизны
\begin{equation*}
\begin{split}
  R_\s:=\s^* R&=\frac12dx^\al\wedge dx^\bt F_{\al\bt}{}^\Sa L_\Sa,
\\
  R_{\s'}:=\s^{\prime*} R&=\frac12dx^\al\wedge dx^\bt F'_{\al\bt}{}^\Sa L_\Sa,
\end{split}
\end{equation*}
где
\begin{equation}                                                  \label{eymcur}
  F_{\al\bt}{}^\Sa=\pl_\al A_\bt{}^\Sa-\pl_\bt A_\al{}^\Sa
  -A_\al{}^\Sb A_\bt{}^\Sc f_{\Sb\Sc}{}^\Sa
\end{equation}
и такое же выражение для $F'_{\al\bt}{}^\Sa$ через штрихованные компоненты
$A'_\al{}^\Sa$. Нетрудно проверить, что компоненты локальной формы кривизны
преобразуются ковариантным образом,
\begin{equation}                                                  \label{qcohtd}
  F'_{\al\bt}{}^\Sa=F_{\al\bt}{}^\Sb S^{-1}_{\quad \Sb}{}^\Sa,
\end{equation}
при калибровочном преобразовании (\ref{egatrc}).

Переходя к присоединенному представлению
\begin{equation*}
  F_{\al\bt}{}^\Sa\mapsto F_{\al\bt\Sb}{}^\Sc
  :=-F_{\al\bt}{}^\Sa f_{\Sa\Sb}{}^\Sc,
\end{equation*}
получаем следующее выражение для локальной формы кривизны
\begin{equation}                                                  \label{elocuf}
  F_{\al\bt\Sa}{}^\Sb=\pl_\al A_{\bt\Sa}{}^\Sb-\pl_\bt A_{\al\Sa}{}^\Sb
  -A_{\al\Sa}{}^\Sc A_{\bt\Sc}{}^\Sb+A_{\bt\Sa}{}^\Sc A_{\al\Sc}{}^\Sb.
\end{equation}
Это выражение совпадает с выражением для локальной формы кривизны в аффинной
геометрии (\ref{ecucav}), которое было получено ранее, после замены
$A_{\al\Sa}{}^\Sb\mapsto\om_{\al a}{}^b$. Тем самым мы показали, что аффинная
связность, которая была введена ранее независимым образом, является частным
случаем связности на главном расслоении общего вида. Выражение для компонент
локальной формы кривизны (\ref{elocuf}) можно записать в виде
\begin{equation}                                                  \label{etentr}
  F_{\al\bt}=\pl_\al A_\bt-\pl_\bt A_\al-[A_\al,A_\bt],
\end{equation}
где мы, для краткости, опустили матричные индексы и квадратные скобки обозначают
коммутатор матриц. При калибровочном преобразовании (\ref{egatrd}) компоненты
локальной формы кривизны преобразуются ковариантно:
\begin{equation*}
  F'_{\al\bt}=SF_{\al\bt}S^{-1},
\end{equation*}
как и следовало ожидать.

Форма кривизны играет важную роль в приложениях. Обращение в нуль ее
компонент дает критерий локальной тривиальности связности. Действительно, при
доказательстве локальной тривиальности линейной связности в разделе
\ref{slojnf} конкретный вид структурной группы не был использован. Поэтому
справедлива
\begin{theorem}
Пусть в некоторой односвязной области $\MU\subset\MM$ заданы компоненты
локальной формы связности $A_{\al\Sa}{}^\Sb$. Если соответствующая локальная
форма кривизны равна нулю на $\MU$, то существует такое калибровочное
преобразование, после которого компоненты локальной формы связности обратятся
в нуль, возможно, в меньшей окрестности. Или, существует такая матрица
калибровочного преобразования $S$, что компоненты локальной формы связности
имеют вид чистой калибровки
\begin{equation}                                                  \label{ejhbtr}
  A_\al=\pl_\al SS^{-1},
\end{equation}
где мы, для краткости, опустили матричные индексы.
\end{theorem}

При проведении вычислений на пространстве главного расслоения $\MP$, например,
в моделях типа Калуцы--Клейна, в касательном расслоении $\MT(\MP)$ удобно
использовать базис $\lbrace D_\al,L^*_\Sa\rbrace$, состоящий из горизонтальных
векторных полей
\begin{equation}                                                  \label{ehobaf}
  D_\al:=\pl_\al-\om_\al{}^\Sa L^*_\Sa,
\end{equation}
и фундаментальных векторных полей $L^*_\Sa$. Этот базис неголономен,
\begin{align}                                                     \label{efinob}
  [D_\al,D_\bt]&=-R_{\al\bt}{}^\Sa L^*_\Sa,
\\                                                                \label{esenob}
  [D_\al,L^*_\Sa]&=0,
\\                                                                \label{ethnob}
  [L^*_\Sa,L^*_\Sb]&=f_{\Sa\Sb}{}^\Sc L^*_\Sc,
\end{align}
где $R_{\al\bt}{}^\Sa$ -- компоненты формы кривизны (\ref{ecycol}). Второе
коммутационное соотношение (\ref{esenob}) является следствием инвариантности
распределения горизонтальных подпространств относительно действия группы справа
(напомним, что левоинвариантные векторные поля генерируют действие группы
справа, а правоинвариантные -- слева, раздел \ref{sriact}). Заметим, что
ковариантную производную $D\vf$ от тензориальной 0-формы типа $(\rho,\MV)$ можно
записать в виде
\begin{equation*}
  D\vf=dx^\al(D_\al\vf^i)\Be_i,
\end{equation*}
где векторное поле $D_\al$ действует как дифференцирование. Это следует из
определения (\ref{ecotef}). В приведенной формуле $\vf^i=\vf^i(x,a)$ в отличии
от формул (\ref{ecovym}), где ковариантная производная берется от сечений
$\vf^i=\vf^i\big(x,\s(x)\big)$. Кроме того, справедливо равенство
\begin{equation}                                                  \label{ecohov}
  [D_\al,D_\bt]\vf^i=\vf^j R_{\al\bt j}{}^i,
\end{equation}
где $R_{\al\bt j}{}^i:=R_{\al\bt}{}^\Sa L_{\Sa j}{}^i$. Эта формула является
аналогом формулы (\ref{emeacu}), полученной в аффинной геометрии. В аффинной
геометрии в правой части стоит дополнительное слагаемое с тензором кручения.
\qed\end{exa}
Для связи с моделями математической физики дадим
\begin{defn}
Компоненты локальной формы кривизны $F_{\al\bt}{}^\Sa$ называются
{\em напряженностью калибровочного поля} или {\em напряженностью поля
Янга--Миллса}. В электродинамике калибровочной группой является абелева группа
$\MU(1)$, а компоненты локальной формы кривизны $F_{\al\bt}$ называются {\em
напряженностью электромагнитного поля}.
\qed\end{defn}
\index{Напряженность калибровочного поля (gauge field strength)}%
\index{Напряженность поля Янга--Миллса (Yang--Mills field strength)}%
\index{Напряженность электромагнитного поля (electromagnetic field strength)}%

Продолжим общее рассмотрение.
\begin{theorem}[\bf Тождества Бианки]                             \label{tbiang}
Пусть на главном расслоении $\MP(\MM,\pi,\MG)$ задана форма кривизны $\om$.
Тогда форма кривизны $R=D\om$ удовлетворяет тождествам Бианки:
\begin{equation}                                                  \label{ebiapr}
  DR=0,
\end{equation}
где $D$ -- внешняя ковариантная производная.
\end{theorem}
\begin{proof}
Из определения внешней ковариантной производной следует, что
$DR(\widetilde X,\widetilde Y,\widetilde Z)=0$, если
хотя бы один из векторов $\widetilde X,\widetilde Y,\widetilde Z$ вертикален.
Поэтому достаточно доказать, что $dR(\widetilde X,\widetilde Y,\widetilde Z)=0$,
когда все три вектора горизонтальны. Возьмем внешнюю производную от структурного
уравнения (\ref{estrba}):
\begin{equation*}
  0=dd\om^\Sa=-\frac12d\om^\Sb\wedge\om^\Sc f_{\Sb\Sc}{}^\Sa
  +\frac12\om^\Sb\wedge d\om^\Sc f_{\Sb\Sc}{}^\Sa+dR^\Sa.
\end{equation*}
Поскольку $\om^\Sa(\widetilde X)=0$, если вектор $\widetilde X$ горизонтален, то
\begin{equation*}
  dR^\Sa(\widetilde X,\widetilde Y,\widetilde Z)=0
\end{equation*}
если все три вектора горизонтальны.
\end{proof}
\begin{theorem}
Пусть $\om$ -- форма связности на главном расслоении $\MP(\MM,\pi,\MG)$ и $\vf$
-- произвольная тензориальная 1-форма типа $\ad\MG$. Тогда
\begin{equation*}
  D\vf(\widetilde X,\widetilde Y)=d\vf(\widetilde X,\widetilde Y)
  +\frac12[\vf(\widetilde X),\om(\widetilde Y)]
  +\frac12[\om(\widetilde X),\vf(\widetilde Y)]
\end{equation*}
для всех $\widetilde X,\widetilde Y\in\MT_p(\MP)$ и $p\in\MP$.
\end{theorem}
\begin{proof}
Аналогично доказательству теоремы \ref{tstreq} о структурном уравнении. См.,
например, \cite{KobNom6369R}.
\end{proof}
\begin{exa}[\bf Локальное рассмотрение]
Запишем тождества Бианки в компонентах. Внешняя производная от компонент формы
кривизны (\ref{ecurvf}) имеет вид
\begin{equation*}
\begin{split}
  dR^\Sa&=\frac12dx^\al\wedge dx^\bt\wedge dx^\g\pl_\al R_{\bt\g}{}^\Sa
  +\frac12dx^\al\wedge dx^\bt\wedge\om^{*\Sb}L^*_\Sb R_{\al\bt}{}^\Sa=
\\
  &=\frac12dx^\al\wedge dx^\bt\wedge dx^\g\pl_\al R_{\bt\g}{}^\Sa
  -\frac12dx^\al\wedge dx^\bt\wedge\om^{*\Sb}
  R_{\al\bt}{}^\Sc f_{\Sb\Sc}{}^\Sa.
\end{split}
\end{equation*}
Ее значение на горизонтальных векторных полях равно
\begin{equation*}
  dR^\Sa(\hor\widetilde X,\hor\widetilde Y,\hor\widetilde Z)
  =3\widetilde X^\al\widetilde Y^\bt\widetilde Z^\al D_{[\al}R_{\bt\g]}{}^\Sa,
\end{equation*}
где квадратные скобки обозначают антисимметризацию по трем индексам, и
\begin{equation*}
  D_\al R_{\bt\g}{}^\Sa=\pl_\al R_{\bt\g}{}^\Sa
  +\om_\al{}^\Sb R_{\bt\g}{}^\Sc f_{\Sb\Sc}{}^\Sa.
\end{equation*}
Таким образом, тождества Бианки в компонентах имеют вид
\begin{equation}                                                  \label{ebiaid}
  D_\al R_{\bt\g}{}^\Sa+D_\bt R_{\g\al}{}^\Sa+D_\g R_{\al\bt}{}^\Sa=0.
\end{equation}
Если задано локальное сечение расслоения, то эти тождества можно спустить на
базу, используя возврат отображения. Тогда тождества Бианки для компонент
локальных форм кривизны и связности примут следующий вид
\begin{equation*}                                                 \label{qbiabh}
  \nb_\al F_{\bt\g}{}^\Sa+\nb_\bt F_{\g\al}{}^\Sa+\nb_\g F_{\al\bt}{}^\Sa=0,
\end{equation*}
где
\begin{equation*}
  \nb_\al F_{\bt\g}{}^\Sa=\pl_\al F_{\bt\g}{}^\Sa
  +A_\al{}^\Sb F_{\bt\g}{}^\Sc f_{\Sb\Sc}{}^\Sa
\end{equation*}
и напряженность $F_{\al\bt}{}^\Sa$ имеет вид (\ref{eymcur}). Именно в таком виде
они, как правило, используются в приложениях.
\qed\end{exa}

************************************************************************
\subsection{Связность на $\MP(\MR^2,\pi,\MR)=\MR^3$}
Чтобы лучше представить себе довольно сложное понятие связности на главном
расслоении, в настоящем разделе мы рассмотрим простой и наглядный пример, когда
пространство главного расслоение совпадает с обычным трехмерным евклидовым
пространством.

Рассмотрим трехмерное евклидово пространство $\MR^3$ с декартовой системой
координат $x,y,z$ как главное расслоение. В качестве базы расслоения выберем
плоскость $x,y\in\MR^2$, а типичным слоем будем считать ось $z\in\MR$,
которая рассматривается, как группа трансляций. Выберем также естественную
проекцию
\begin{equation*}
  \pi:~~\MR^3\ni\quad(x,y,z)\mapsto (x,y)\quad\in\MR^2.
\end{equation*}
Таким образом, построено гладкое тривиальное главное расслоение
$\MP(\MR^2,\pi,\MR)=\MR^2\times\MR=\MR^3$.

Вертикальные подпространства $\MV(\MP)$ в касательном расслоении $\MT(\MP)$
одномерны, и в каждой точке $p=(x,y,z)\in\MP$ натянуты на векторы,
параллельные оси $z$.

Зададим на главном расслоении связность, т.е.\ инвариантное распределение
горизонтальных подпространств. Для этого можно задать достаточно гладкую
поверхность $z(x,y)$, где $x,y\in\MR^2$ -- координаты на поверхности. Тем самым
мы выбираем сечение главного расслоения $\MP(\MR^2,\pi,\MR)$. После этого мы
сдвигаем поверхность на все возможные постоянные векторы вдоль оси $z$. Теперь
отождествим распределение горизонтальных подпространств с касательными
пространствами ко всем поверхностям. По построению, такое распределение будет
инвариантно относительно трансляций. При этом мы требуем также выполнение
неравенств
\begin{equation*}
  \frac{\pl z}{\pl x}\ne\infty,\quad\frac{\pl z}{\pl y}\ne\infty,\qquad
  \forall(x,y)\in\MR^2,
\end{equation*}
Это является необходимым и достаточным условием разложения касательного
пространства
\begin{equation*}
  \MT_p(\MP)=\MV_p(\MP)\oplus\MH_p(\MP)
\end{equation*}
в каждой точке $p\in\MP$ в прямую сумму вертикальных $\MV_p(\MP)$ и
горизонтальных $\MH_p(\MP)$ подпространств. Таким образом, мы построили
связность $\Gamma$ на главном расслоении $\MP(\MR^2,\pi,\MR)$.

Рассмотрим касательный вектор к главному расслоению в координатном базисе
\begin{equation*}
  \widetilde X=\widetilde X^x\pl_x+\widetilde X^y\pl_y+\widetilde X^z\pl_z
  \in\MT_p(\MP).
\end{equation*}
Он раскладывается единственным образом на вертикальную и горизонтальные
составляющие
\begin{equation*}
  \widetilde X=\ver \widetilde X+\hor \widetilde X,
\end{equation*}
где
\begin{align}                                                     \label{qfarys}
  \ver \widetilde X&=\widetilde X^z\pl_z-\widetilde X^x\frac{\pl z}{\pl x}\pl_z
  -\widetilde X^y\frac{\pl z}{\pl y}\pl_z,
\\                                                                \label{qkjiof}
  \hor\widetilde X&=\widetilde X^x\pl_x+\widetilde X^y\pl_y+
  \widetilde X^x\frac{\pl z}{\pl x}\pl_z+\widetilde X^y\frac{\pl z}{\pl y}\pl_z.
\end{align}

Форма связности $\om=dx\om_x+dy\om_y+dz$ в рассматриваемом случае имеет только
две неизвестные компоненты: $\om_x$ и $\om_y$, т.к.\ структурная группа
одномерна, и $L_\Sa\mapsto \pl_z$ и $\om^{*\Sa}\mapsto dz$. Сравнение разложения
(\ref{qkjiof}) с общей формулой (\ref{ehovec}) показывает, что компоненты
связности имеют вид
\begin{equation*}
  \om_x=-\frac{\pl z}{\pl x},\quad\om_y=-\frac{\pl z}{\pl y}.
\end{equation*}
В рассматриваемом случае компоненты формы связности не зависят от $z$ в силу
трансляционной инвариантности. Легко видеть, что значение формы связности на
любом горизонтальном векторном поле равно нулю:
\begin{equation}                                                  \label{qjhudt}
  \om(\hor\widetilde X)=0.
\end{equation}
Верно также и обратное утверждение: любое решение уравнения (\ref{qjhudt}) имеет
вид (\ref{qfarys}).

Единственная возможная нетривиальная компонента тензора кривизны равна нулю:
\begin{equation*}
  R_{xy}=\pl_x\om_y-\pl_y\om_x
  =-\frac{\pl^2z}{\pl x\pl y}+\frac{\pl^2z}{\pl y\pl x}=0.
\end{equation*}

Рассмотрим произвольное сечение главного расслоения
\begin{equation}                                                  \label{qhjsio}
  \s:\qquad\MR^2\ni\quad(x,y)\mapsto\big(x,y,w(x,y)\big)\quad\in\MR^3.
\end{equation}
Дифференциал этого отображения отображает касательные векторы к базе в
касательные векторы к пространству главного расслоения
\begin{multline*}
  \MT(\MR^2)\ni\quad X=X^x\pl_x+X^y\pl_y\mapsto
\\
  \mapsto\s_*X=X^z\pl_x+X^y\pl_y
  +X^x\frac{\pl w}{\pl x}\pl_z+X^y\frac{\pl w}{\pl y}\pl_z\quad\in\MT(\MR^2).
\end{multline*}
Из равенства $\om_\s(X)=\om(\s_*X)$ следуют выражения для компонент локальной
формы связности $\om_\s=dxA_x+dyA_y$:
\begin{equation*}
  A_x=\frac{\pl(w-z)}{\pl x},\qquad A_y=\frac{\pl(w-z)}{\pl y}.
\end{equation*}
Если задано другое сечение $w'(x,y)$, то компоненты локальной формы связности
преобразуются по правилу
\begin{equation}                                                  \label{qdhrku}
  A'_x=A_x+\pl_x a,\qquad A'_y=A_y+\pl_y a,
\end{equation}
где $a(x,y):=w'(x,y)-w(x,y)$.

Заметим, что сама форма связности вообще никак не преобразуется при изменении
сечений, т.к.\ задана на пространстве расслоения и определена до рассмотрения
каких либо сечений.

Локальная форма кривизны для введенной выше связности также равна нулю.

Таким образом, построенная связность является плоской. Распределение
горизонтальных векторных полей в данном случае находится в инволюции и, согласно
теореме Фробениуса, через каждую точку проходит интегральное подмногообразие.
Им является поверхность $z(x,y)$, проходящая через эту точку.

Выше мы рассмотрели связность специального вида, которая задается поверхностями
$z(x,y)$. В общем случае распределение горизонтальных векторных полей задается
произвольной формой связности
\begin{equation*}
  \om=dx\om_x+dy\om_y+dz,
\end{equation*}
где $\om_x$ и $\om_y$ -- некоторые достаточно гладкие функции от $x,y$. Теперь
форма кривизны может иметь нетривиальную компоненту
\begin{equation*}
  R_{xy}=\pl_x\om_y-\pl_y\om_x.
\end{equation*}

Если задано сечение (\ref{qhjsio}), то компоненты локальной формы связности
$\om_\s=dxA_x+dyA_y$ будут иметь вид
\begin{equation*}
  A_x=\om_x+\pl_x w,\qquad A_y=\om_y+\pl_yw.
\end{equation*}
Калибровочное преобразование при переходе к другому сечению при этом останется
прежним (\ref{qdhrku}).

Локальная форма кривизны в общем случае имеет одну независимую компоненту
\begin{equation*}
  F_{xy}=\pl_xA_y-\pl_yA_x,
\end{equation*}
которая может быть нетривиальной.
\section{Параллельный перенос                                    \label{spartr}}
\begin{figure}[h,b,t]
\hfill\includegraphics[width=.4\textwidth]{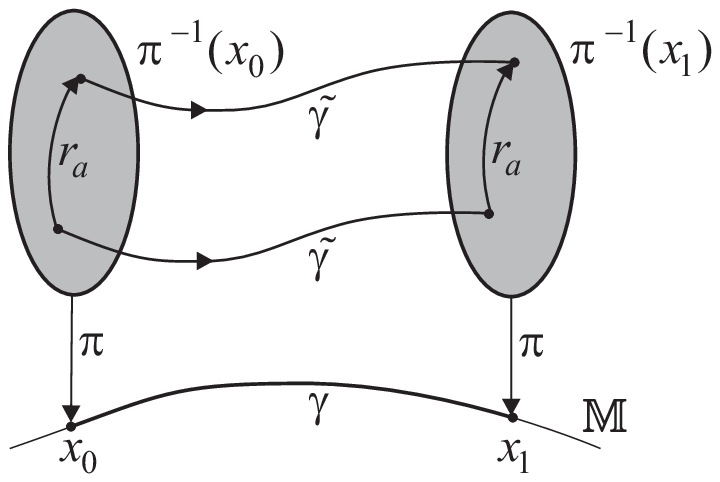}
\hfill {}
\\
\centering \caption{Параллельный перенос слоев вдоль кривой $\g$.\label{fpartr}}
\end{figure}

Пусть на главном расслоении $\MP(\MM,\pi,\MG)$ задана связность
$\Gamma:~\MP\ni p\mapsto\MH_p\subset\MT_p(\MP)$. Определим понятие параллельного
переноса слоя $\pi^{-1}(x_0)$ над точкой базы $x_0\in\MM$ вдоль произвольной
кусочно дифференцируемой кривой
\begin{equation}                                                 \label{earcsm}
  \g:\quad [0,1]\ni\quad t\mapsto x(t)\quad\in\MM
\end{equation}
с началом в точке $x_0$.  Для наших целей достаточно рассматривать кусочно
дифференцируемые кривые класса $\CC^1$.
\begin{defn}
{\em Горизонтальной кривой} в $\MP$ называется кусочно дифференцируемая кривая,
все касательные векторы которой горизонтальны. {\em Горизонтальным лифтом} или
{\em подъемом} (или просто {\em лифтом}) кривой $\g$ (\ref{earcsm}), заданной на
базе $\MM$, называется горизонтальная кривая в $\MP$,
\begin{equation*}
  \tilde\g:\quad [0,1]\ni\quad t\mapsto p(t)\quad\in\MP,
\end{equation*}
такая, что $\pi\circ\tilde\g=\g$.
\qed\end{defn}
\index{Горизонтальный лифт кривой (horizontal lift of a curve)}%
\index{Горизонтальный подъем кривой (horizontal lift of a curve)}%
\index{Лифт кривой (lift of a curve)}%
\index{Горизонтальная кривая (horizontal curve)}%
Понятие горизонтального лифта кривой соответствует понятию лифта векторного
поля. Действительно, если $\widetilde X\in\CX(\MP)$ -- лифт дифференцируемого
векторного поля $X$, заданного на базе $\MM$, то интегральная кривая $\tilde\g$
векторного поля $\widetilde X$, проходящая через точку $p_0\in\MP$, есть
горизонтальный лифт интегральной кривой $\g$ поля $X$, проходящей через точку
$x_0=\pi(p_0)$.
\begin{prop}                                                      \label{pholcu}
Пусть $\g=x(t)$, $t\in[0,1]$, -- кусочно дифференцируемая кривая класса
$\CC^1$ в $\MM$ с началом в точке $x_0\in\MM$. Тогда для произвольной
точки слоя $p_0\in\pi^{-1}(x_0)$ существует единственный горизонтальный лифт
$\tilde\g=p(t)$ кривой $\g$ с началом в точке $p_0$.
\end{prop}
\begin{proof}
Состоит в явном построении лифта $\tilde\g$. См., например,
\cite{KobNom6369R}.
\end{proof}
Используя предложение \ref{pholcu}, определим параллельное перенесение слоев
следующим образом.
\begin{defn}
Пусть $\g=x(t)$ -- кривая в $\MM$ с началом и концом в точках $x_0$ и $x_1$.
Пусть $\tilde\g$ -- единственный горизонтальный лифт кривой $\g$, который
начинается в точке $p_0$, находящейся в слое $\pi^{-1}(x_0)$. Лифт $\tilde\g$
имеет конечную точку $p_1\in\MP$ такую, что $\pi(p_1)=x_1$. Меняя начальную
точку $p_0$ в слое $\pi^{-1}(x_0)$, мы получаем отображение слоя $\pi^{-1}(x_0)$
в слой $\pi^{-1}(x_1)$, которое переводит точку $p_0$ в $p_1$. Это отображение
называется {\em параллельным переносом слоя} из точки $x_0$ в точку $x_1$ вдоль
кривой $\g$. Параллельный перенос слоев обозначается той же буквой, что и
кривая, $\g:~\pi^{-1}(x_0)\rightarrow\pi^{-1}(x_1)$.
\qed\end{defn}
\index{Параллельный перенос слоя (parallel transport of a fiber}%
Параллельный перенос слоев является изоморфизмом, что вытекает из следующего
утверждения.
\begin{prop}                                                      \label{parrit}
Параллельный перенос слоя $\g:~\pi^{-1}(x_0)\rightarrow\pi^{-1}(x_1)$ вдоль
любой  кривой перестановочен с действием структурной группы $\MG$ на $\MP$:
$\g\circ r_a=r_a\circ\g$ для всех $a\in\MG$.
\end{prop}
\begin{proof}
Это следует из того, что правое действие структурной группы $r_a$ отображает
каждую горизонтальную кривую в горизонтальную, см.\ рис.\ref{fpartr}.
\end{proof}
Параллельный перенос слоя вдоль кривой $\g$ не зависит от выбора параметризации
кривой. Кроме того, если слой переносится из точки $x_0$ в точку $x_1$
параллельно вдоль кривой $\g$, то он параллельно переносится вдоль этой же
кривой из точки $x_0$ в любую промежуточную точку $x(t)$, $t\in[0,1]$, на $\g$.
\begin{com}
Если точки $x_0$ и $x_1$ в базе $\MM$ фиксированы, то в общем случае
параллельный перенос слоя $\pi^{-1}(x_0)$ в слой $\pi^{-1}(x_1)$ зависит от
кривой $\g$, соединяющей эти точки. Для односвязных баз эта зависимость
характеризуется формой кривизны $R$ формы связности $\om$ и будет обсуждаться в
следующих разделах.
\qed\end{com}

При рассмотрении фундаментальной группы в разделе \ref{sfundg} мы определили
произведение путей (кривых) $\g_2\circ\g_1$ как последовательный проход вдоль
путей $\g_1$ и $\g_2$ (\ref{epamul}) и обратный путь $\g^{-1}$ как путь $\g$,
проходимый в обратном направлении (\ref{epathi}). Следующее предложение
очевидно.
\begin{prop}                                                      \label{pgrpou}
1) Если $\g_1$ -- путь из $x_0$ в $x_1$ и $\g_2$ -- путь из $x_1$ в $x_2$, то
параллельный перенос слоя $\pi^{-1}(x_0)$ в слой $\pi^{-1}(x_2)$ вдоль
произведения путей $\g_2\circ\g_1$ равен произведению отображений слоев
$\g_2\circ\g_1:~\pi^{-1}(x_0)\rightarrow\pi^{-1}(x_2)$.

2) Если $\g^{-1}$ -- обратный путь для пути $\g$ из точки $x_0$ в точку $x_1$, то
параллельный перенос слоя $\pi^{-1}(x_1)$ в слой $\pi^{-1}(x_0)$ вдоль пути
$\g^{-1}$ является обратным отображением
$\g^{-1}=\g^{-1}:~\pi^{-1}(x_1)\rightarrow\pi^{-1}(x_0)$.
\end{prop}
\section{Группы голономии                                        \label{sholde}}
Пусть задано главное расслоение $\MP(\MM,\pi,\MG)$ со связностью $\Gamma$. Используя
понятие параллельного переноса, определим группу голономии данной связности
$\Gamma$.

Обозначим через $\Om(\MM,x)$ множество замкнутых кусочно дифференцируемых кривых
(петель) на базе $\MM$ с началом и концом в точке $x\in\MM$. Подмножество,
состоящее из путей, гомотопных постоянному пути в точке $x$, обозначим
$\Om_0(\MM,x)\subset\Om(\MM,x)$. Произведение и
обратный путь для всех путей из $\Om(\MM,x)$ были определены в разделе
\ref{sfundg}. В разделе \ref{spartr} было показано, что параллельный перенос
слоя $\pi^{-1}(x)$ вдоль замкнутого пути $\g\in\Om(\MM,x)$ есть изоморфизм слоя
$\pi^{-1}(x)$ на себя. В общем случае этот изоморфизм будет нетривиален, т.к.\
после параллельного переноса вдоль замкнутого пути слой может повернуться.
Множество всех таких изоморфизмов образует группу в силу предложения
\ref{pgrpou}.
\begin{defn}
Группа, состоящая из изоморфизмов слоя $\pi^{-1}(x)$, которые соответствуют
параллельным переносам данного слоя вдоль всех замкнутых кусочно
дифференцируемых путей $\g\in\Om(\MM,x)$, называется {\em группой голономии}
$\Phi(x)$ связности $\Gamma$ в точке $x\in\MM$. Подгруппа $\Phi_0(x)\subset\Phi(x)$,
соответствующая параллельным переносам вдоль замкнутых путей, стягиваемых в
точку, $\g\in\Om_0(\MM,x)$, называется {\em суженной группой голономии}
связности $\Gamma$ в точке $x\in\MM$.
\qed\end{defn}
\index{Группа голономии (holonomy group)}%
\index{Голономии группа (holonomy group)}%
\index{Суженная группа голономии (restricted holonomy group)}%
\index{Группа голономии суженная (restricted holonomy group)}%

Группу голономии $\Phi(x)$ и суженную группу голономии $\Phi_0(x)$ можно
считать подгруппами в структурной группе $\MG$ следующим образом.
\begin{defn}
Зафиксируем некоторую точку слоя $p\in\pi^{-1}(x)$. После параллельного переноса
слоя вдоль пути $\g\in\Om(\MM,x)$ эта точка отобразится в некоторую точку
$\g(p)=pa\in\pi^{-1}(x)$, где $a\in\MG$ -- некоторый элемент структурной группы.
Если задан другой путь $\g'\in\Om(\MM,x)$, которому соответствует элемент
$b\in\MG$, то произведение путей $\g'\circ\g$ определяет элемент $ba\in\MG$,
поскольку
\begin{equation*}
  \g'\circ\g(p)=\g'(pa)=\big(\g'(p)\big)a=pba.
\end{equation*}
По предложению \ref{pgrpou} множество элементов $a\in\MG$, определенных всеми
путями $\g\in\Om(x)$, образует группу, которая называется {\em группой
голономии} $\Phi(p)$ связности $\Gamma$ в точке $p\in\MP$. Замкнутым путям
$\g\in\Om_0(\MM,x)$ соответствует {\em суженная группа голономии} $\Phi_0(p)$
в точке $p\in\MP$.
\qed\end{defn}
\begin{com}
$\Phi(x)$ есть группа изоморфизмов слоя $\pi^{-1}(x)$ на себя, a $\Phi(p)$ есть
подгруппа в $\MG$. Выше мы построили единственный изоморфизм из $\Phi(x)$ на
$\Phi(p)$, который делает коммутативной следующую диаграмму:
\begin{equation*}                                                    \tag*{\qed}
\begin{diagram}
  \Om(\MM,x) &  &  \\
  \dTo & \rdTo &  \\
  \Phi(x) & \rTo & \Phi(p)
\end{diagram}
\renewcommand{\qed}{}\end{equation*}
\qed\end{com}

Группу голономии $\Phi(p)$ можно определить другим образом. Введем на
пространстве главного расслоения отношение эквивалентности $p\sim q$, где
$p,q\in\MP$, если точки $p$ и $q$ можно можно соединить горизонтальной кривой.
При этом точки $p$ и $q$ не обязательно лежат в одном слое. Нетрудно проверить,
что это действительно отношение эквивалентности. Тогда группа голономии
$\Phi(p)$ совпадает с множеством тех элементов $a\in\MG$, для которых
$p\sim pa$. Легко проверить, что это множество элементов образует подгруппу в
$\MG$, т.к.\ $p\sim q$ влечет за собой $pa\sim qa$.
\begin{prop}                                                      \label{pcohog}
Пусть дано главное расслоение $\MP(\MM,\pi,\MG)$ со связностью $\Gamma$.
Тогда:\newline
\indent 1) \parbox[t]{.92\linewidth}{Если $q=pa$, $a\in\MG$, то
$\Phi(q)=\ad(a^{-1})\Phi(p)$, т.е.\ группы голономии точек одного слоя $\Phi(q)$
и $\Phi(p)$ сопряжены в $\MG$. Аналогично, $\Phi_0(q)=\ad(a^{-1})\Phi_0(p)$.}\newline
\indent 2) \parbox[t]{.92\linewidth}{Если точки $p,q\in\MP$ можно соединить
горизонтальной кривой, т.е.\ $p\sim q$, то $\Phi(p)=\Phi(q)$ и
$\Phi_0(p)=\Phi_0(q)$.}
\end{prop}
\begin{proof}
1). Пусть $b\in\Phi(p)$, т.е.\ $p\sim pb$. Тогда $pa\sim pba$ и, следовательно,
$q\sim qa^{-1}ba$. Поэтому $\ad(a^{-1})b=a^{-1}ba\in\Phi(q)$. Отсюда вытекает,
что $\Phi(q)=\ad(a^{-1})\Phi(p)$ и $\Phi_0(q)=\ad(a^{-1})\Phi_0(p)$.

2) Отношение $p\sim q$ влечет за собой $pb\sim qb$. Из транзитивности отношения
эквивалентности $\sim$ следует, что $p\sim pb$ тогда и только тогда, когда
$q\sim qb$, т.е.\ $b\in\Phi(p)$ тогда и только тогда, когда $b\in\Phi(q)$. Тем
самым $\Phi(p)=\Phi(q)$. Чтобы доказать равенство $\Phi_0(p)=\Phi_0(q)$,
допустим, что $\tilde\dl$ -- горизонтальная кривая в $\MP$ из $p$ в $q$. Если
$b\in\Phi_0(p)$, то существует горизонтальная кривая $\tilde\g$ в $\MP$ из $p$ в
$pb$ такая, что кривая в базе $\pi\tilde\g$ является замкнутым путем с началом и
концом в точке $\pi(p)$, которая гомотопна постоянному пути в точке $\pi(p)$.
Тогда композиция $(r_b\tilde\dl)\circ\tilde\g\circ\tilde\dl^{-1}$ есть
горизонтальная кривая в $\MP$ из $q$ в $qb$ и ее проекция на базу $\MM$ есть
замкнутый путь с началом и концом в точке $\pi(q)$, который гомотопен
постоянному пути. Поэтому $b\in\Phi_0(q)$.
\end{proof}
Если база $\MM$ связна, то для любой пары точек $p,q\in\MP$ найдется элемент
$a\in\MG$ такой, что $q\sim pa$. Поэтому из предложения \ref{pcohog} следует, что
группы голономии $\Phi(p)$ для всех точек $p\in\MP$ сопряжены друг другу в $\MG$
и поэтому изоморфны. По тем же причинам все суженные группы голономии
$\Phi_0(p)$ также изоморфны друг другу.

Итак, мы определили группу голономии $\Phi(p)$, суженную группу голономии
$\Phi_0(p)$ и показали, что с точностью до преобразования подобия они не зависят
от точки расслоения $\MP$. Теперь сформулируем несколько общих свойств групп
голономий.
\begin{theorem}                                                   \label{tholie}
Пусть $\MP(\MM,\pi,\MG)$ -- главное расслоение со связной базой $\MM$. Пусть
$\Phi(p)$ и $\Phi_0(p)$ -- группа голономии и суженная группа голономии
связности $\Gamma$ в точке $p\in\MP$. Тогда:\newline
\indent 1) \parbox[t]{.92\linewidth}{$\Phi_0(p)$ есть связная подгруппа Ли в
$\Phi(p)$;}\newline
\indent 2) \parbox[t]{.92\linewidth}{$\Phi_0(p)$ есть нормальная подгруппа в
$\Phi(p)$ и фактор группа $\Phi(p)/\Phi_0(p)$ счетна.}
\end{theorem}
\begin{proof}
Используется паракомпактность базы $\MM$. См., например, \cite{KobNom6369R}.
\end{proof}
\begin{cor}
Группа голономии $\Phi(p)$ является подгруппой Ли в структурной группе $\MG$ с
компонентой единицы $\Phi_0(p)$. В частности, $\dim\Phi(p)=\dim\Phi_0(p)$.
\qed\end{cor}
При определении групп голономий мы не оговорили класс дифференцируемости
рассматриваемых кусочно дифференцируемых кривых $\g\in\Om(\MM,x)$. Чем ниже
класс дифференцируемости, тем больше множество кривых. Поэтому могло бы
оказаться так, что группы голономии зависят от класса дифференцируемости кривых.
Это оказывается не верно. Пусть $\Om^k(\MM,x)$ -- множество замкнутых кусочно
дифференцируемых путей в $\MM$ с началом и концом в точке $x\in\MM$ класса
$\CC^k$. Обозначим соответствующую группу голономии через $\Phi^k(p)$. Очевидно,
что $\Phi^1(p)\supset\Phi^2(p)\supset\dotsc\supset\Phi^\infty(p)$. Верны также и
обратные включения.
\begin{theorem}[\bf Номидзу, Одзеки]
Все группы голономии $\Phi^k(p)$, $1\le k\le\infty$, совпадают.
\end{theorem}
\begin{proof}
См.\ \cite{NomOze62}.
\end{proof}
\begin{cor}
Все суженные группы голономии $\Phi^k_0(p)$, $1\le k\le\infty$, совпадают.
\qed\end{cor}
\begin{proof}
Согласно теореме \ref{tholie} суженная группа голономии $\Phi^k_0(p)$ есть
связная компонента единицы группы $\Phi^k(p)$.
\end{proof}
\begin{com}
В случае, когда $\MP(\MM,\pi,\MG)$ является вещественно аналитическим главным
расслоением с аналитической связностью $\Gamma$, можно определить группу голономии
$\Phi^\om(p)$, используя только кусочно аналитические кривые. Можно доказать,
что $\Phi^\om(p)=\Phi^1(p)$ и $\Phi^\om_0(p)=\Phi^1_0(p)$ \cite{KobNom6369R}.
\qed\end{com}
Поскольку группы голономии не зависят от класса дифференцируемости путей, то в
дальнейшем мы не будем его указывать.
\section{Петля Вильсона                                          \label{swilop}}
Настоящий раздел посвящен одному из способов вычисления группы голономии.
Пусть задано главное расслоение $\MP(\MM,\pi,\MG)$ со связностью $\Gamma$.
Рассмотрим окрестность $\MU\subset\MM$ с координатами $x^\al$, $\al=1,\dotsc,n$,
содержащую точку базы $x_0\in\MM$, и отождествим подрасслоение $\pi^{-1}(\MU)$ с
прямым произведением $\MU\times\MG$. Тогда точка подрасслоения задается парой
элементов $p=(x,a)\in\MU\times\MG$. Пусть $\g=x(t)$ -- произвольная кривая в
$\MU$ с началом в точке $x_0$ и $\tilde\g=\big(x(t),a(t)\big)$ -- ее
единственный горизонтальный лифт с началом в точке $p_0=(x_0,a_0)$. При этом
функция $a(t)$ определяет некоторую кривую в структурной группе $\MG$ с началом
в точке $a_0$. В инвариантном виде мы пишем
\begin{equation}                                                  \label{eqgrho}
  \tilde\g=\overset\circ\g\circ a(t),
\end{equation}
где $\overset\circ\g=\big(x(t),e\big)$ -- опорная кривая в $\pi^{-1}(\MU)$,
лежащая в нулевом сечении, для которой $\pi\big(\overset\circ\g\big)=\g$. При
этом мы рассматриваем равенство (\ref{eqgrho}) как уравнение на $a(t)$ при
заданной кривой $\g$, которая однозначно определяет опорную кривую
$\overset\circ\g$.

Касательный вектор к горизонтальной кривой $\tilde\g$ имеет вид
$\dot{\tilde\g}=(\dot x,\dot a)$. Уравнение для $a(t)$ получается из условия
горизонтальности. Рассмотрим окрестность единицы группы, где форма связности
имеет вид (\ref{ecjfor}). Горизонтальность касательного вектора к кривой
записывается в виде равенства
\begin{equation*}
  \om(\dot{\tilde\g})=
  (\dot x^\al\om_\al{}^\Sa+\dot a^\Sb L^{-1}_{\quad \Sb}{}^\Sa)L_\Sa=0.
\end{equation*}
Отсюда следует система уравнений на $a^\Sa(t)$:
\begin{equation}                                                  \label{egrcuh}
  \dot x^\al\overset{~\circ}A_\al{}^\Sb S^{-1}_{\quad \Sb}{}^\Sa
  +\dot a^\Sb L^{-1}_{\quad \Sb}{}^\Sa=0,
\end{equation}
где $\overset{~\circ}A_\al{}^\Sa(x)$ -- компоненты локальной формы связности
на нулевом сечении (поле Янга--Миллса) и $S_\Sb{}^\Sa(a)$ -- матрица
присоединенного представления, соответствующая элементу $a\in\MG$.

Уравнение (\ref{egrcuh}) неудобно для приложений, т.к.\ содержит матрицу
$L_\Sb{}^\Sa(a)$, которая определена только в окрестности единицы группы. Чтобы
устранить это неудобство, перейдем к какому либо представлению структурной
группы
\begin{equation*}
  \rho:\quad \MG\ni\quad a\mapsto\big(S^{-1}_{\quad j}{}^i(a)\big)\quad
  \in\aut\MV,\qquad i,j=1,\dotsc,\dim\MV.
\end{equation*}
Мы выбрали представление в виде обратных матриц, чтобы не менять общепринятого
определения упорядоченного (хронологического) произведения, которое будет дано
ниже. Теперь заметим, что
\begin{equation*}
  \dot S^{-1}_{\quad i}{}^j=\dot a^\Sa\pl_\Sa S^{-1}_{\quad i}{}^j
  =\dot a^\Sb L^{-1}_{\quad \Sb}{}^\Sa L_\Sa S^{-1}_{\quad i}{}^j
  =\dot a^\Sb L^{-1}_{\quad \Sb}{}^\Sa S^{-1}_{\quad i}{}^k L_{\Sa k}{}^j,
\end{equation*}
где $L_{\Sa i}{}^j$ -- представление генераторов $L_\Sa$ алгебры Ли $\Gg$, и мы
воспользовались правилом дифференцирования матриц представления (\ref{eredim}).
Умножив уравнение (\ref{egrcuh}) на $S^{-1}_{\quad i}{}^k L_{\Sa k}{}^j$ и
воспользовавшись инвариантностью генераторов $L_{\Sa i}{}^j$ (\ref{einvge}),
получаем уравнение на матрицу представления
\begin{equation*}
  \dot S^{-1}_{\quad i}{}^j
  -\dot x^\al\overset{~\circ}A_{\al i}{}^k S^{-1}_{\quad k}{}^j=0,
\end{equation*}
где
$\overset{~\circ}A_{\al i}{}^j:=-\overset{~\circ}A_\al{}^\Sa L_{\Sa i}{}^j$.
Это уравнение можно переписать в почти ковариантном виде
\begin{equation*}
  \dot x^\al\left(\pl_\al S^{-1}_{\quad i}{}^j
  -\overset{~\circ}A_{\al i}{}^k S^{-1}_{\quad k}{}^j \right)=0.
\end{equation*}
Для ковариантности не хватает одного слагаемого с калибровочным полем для
индекса $j$. Опустив, для краткости, матричные индексы, получаем уравнение
\begin{equation}                                                  \label{ehomat}
  \dot S^{-1}=\dot x^\al \overset{~\circ}A_\al S^{-1},
\end{equation}
которое можно записать в интегральном виде:
\begin{equation*}
  S^{-1}(t)=S^{-1}_0+\int_0^t \!\!\!ds\,\dot x^\al\overset{~\circ} A_\al S^{-1}
  =S^{-1}_0+\int_{x(0)}^{x(t)}\!\!\!dx^\al\overset{~\circ}A_\al S^{-1}.
\end{equation*}
где $S^{-1}_0:=S^{-1}(a_0)$, точка обозначает дифференцирование по $s$ и второй
интеграл берется вдоль кривой $\g$ от точки $x(0)$ до точки $x(t)$. Решение
этого уравнения записывается в виде упорядоченной $\P$-экспоненты
\begin{equation}                                                  \label{esohom}
  S^{-1}(t)=\P\exp\left(\int_0^t ds\,\dot x^\al\overset{~\circ}A_\al\right)
  S^{-1}_0,
\end{equation}
которая определена разложением в ряд
\begin{equation}                                                  \label{etexpd}
  \P\exp\left(\int_0^t ds\,\dot x^\al\overset{~\circ}A_\al\right)
  =1+\int_0^tds\,\dot x^\al\overset{~\circ}A_\al
  +\int_0^t ds_1\int_0^{s_1}ds_2\,\dot x^{\al_1}\overset{~\circ}A_{\al_1}
  \dot x^{\al_2}\overset{~\circ}A_{\al_2}+\dotsc,
\end{equation}
где
\begin{equation*}
  \dot x^{\al_k}\overset{~\circ}A_{\al_k}
  =\frac{dx^{\al_k}(s_k)}{ds_k}\overset{~\circ}A_{\al_k}(s_k),\qquad
  \forall k=1,2,\dotsc.
\end{equation*}
Напомним общее определение $\P$-произведения.
\begin{defn}
Пусть задано семейство операторов $A(t)$, непрерывно зависящих от вещественного
параметра $t$, тогда
\begin{equation}                                                  \label{etprod}
  \P[A(t_1)A(t_2)]=\begin{cases} A(t_1)A(t_2),\qquad ~t_1\ge t_2, \\
  A(t_2)A(t_1),\qquad t_1<t_2 \end{cases}
\end{equation}
называется {\em $\P$-произведением} или {\em упорядоченным произведением}
операторов $A(t_1)$ и $A(t_2)$.
\qed\end{defn}
\index{$\P$-произведение ($\P$-product)}%
\index{Упорядоченное произведение (ordered product)}%
\index{Произведение упорядоченное (ordered product)}%
Если операторы для различных точек коммутируют, то $\P$-произведение совпадает с
обычным произведением.
\begin{com}
В квантовой теории поля роль параметра $t$ часто играет время. Поэтому
$\P$-произведение называют также {\em хронологическим произведением}.
\qed\end{com}
\index{Хронологическое произведение (chronological product)}%
\index{Произведение хронологическое (chronological product)}%

Нетрудно проверить равенство
\begin{multline*}
  \int_0^t \!\!\!ds_1\int_0^{s_1}\!\!\!ds_2\dotsc\int_0^{s_{k-1}}\!\!\!ds_k
  A(s_1)A(s_2)\dotsc A(s_k)=
\\
  =\frac1{k!}\int_0^t \!\!\!ds_1\int_0^t\!\!\!ds_2\dotsc
  \int_0^t\!\!\!ds_k\P[A(s_1)A(s_2)\dotsc A(s_k)].
\end{multline*}
Поэтому разложение (\ref{etexpd}) имеет место.

Известно, что ряд (\ref{etexpd}) равномерно сходится в шаре произвольного
радиуса.

Продолжим общее построение. Если представление структурной группы является
точным, то матрица представления $S^{-1}(a)$ однозначно определяет элемент
структурной группы $a=\rho^{-1}(S^{-1})$. В этом случае решение (\ref{esohom})
определяет кривую $a(t)$ с началом в точке $a_0\in\MG$. Меняя точку $a_0$, мы
получаем отображение
\begin{equation}                                                  \label{eparfi}
  \pi^{-1}(x_0)\ni\quad(x_0,a_0)\mapsto\big(x(t),a(t)\big)\quad
  \in\pi^{-1}\big(x(t)\big),\qquad \forall a_0\in\MG.
\end{equation}
Поскольку кривая $\tilde\g$ горизонтальна, то это отображение задает
параллельный перенос слоя $\pi^{-1}(x_0)$ главного расслоения $\MP$ вдоль пути
$\g\in\MM$.

Таким образом, решение (\ref{esohom}) определяет параллельный перенос слоев
(\ref{eparfi}) через компоненты связности $\overset{~\circ}A_\al{}$ на опорном
нулевом сечении. Отображение (\ref{eparfi}) можно построить для произвольного
опорного сечения следующим образом. Пусть задано произвольное сечение
$\s(x)=\big(x,b(x)\big)$. Тогда компоненты локальной формы связности для
рассматриваемого представления на этом сечении имеют вид (\ref{egatra})
\begin{equation*}
  A_\al=S_b\overset{~\circ}A_\al S^{-1}_b+\pl_\al S_b S^{-1}_b,\qquad S_b:=S(b).
\end{equation*}
Для горизонтальной кривой $\tilde\g$ имеем
\begin{equation}                                                  \label{ehorcu}
  \tilde\g=\big(x(t),a(t)\big)=\big(x(t),b(t)c(t)\big),
\end{equation}
где $c(t)$ -- некоторая новая кривая в $\MG$, связывающая кривую
$\big(x(t),b(t)\big)$ на сечении $\s$ с горизонтальной кривой $\tilde\g$.
Поскольку $S^{-1}_a=S^{-1}_bS^{-1}_c$, то простые вычисления приводят
(\ref{ehomat}) к уравнению
\begin{equation*}
  \dot S^{-1}_c=\dot x^\al A_\al S^{-1}_c,
\end{equation*}
определяющему кривую $c(t)$ с начальным условием $c_0=b_0^{-1}a_0$. Решение
этого уравнения также дается упорядоченной экспонентой
\begin{equation*}
  S^{-1}_c(t)=\P\exp\left(\int_0^t ds\,\dot x^\al A_\al\right)S^{-1}_{c(0)}.
\end{equation*}
Эта формула также определяет параллельный перенос (\ref{eparfi}), что следует из
(\ref{ehorcu}).

Поскольку $S^{-1}_0=S^{-1}_{b(0)}S^{-1}_{c(0)}$, то простые вычисления приводят
к следующему правилу преобразования упорядоченной экспоненты при изменении
сечения $\s_0\mapsto\s$
\begin{equation}                                                  \label{etrrex}
  \P\exp\left(\int_0^t ds\,\dot x^\al\overset{~\circ}A_\al\right)
  =S(b_0)\P\exp\left(\int_0^t ds\,\dot x^\al A_\al\right)S^{-1}(b_t).
\end{equation}
Этот закон преобразования похож на тензорный, однако таковым не является, потому
что слева и справа стоят матрицы преобразования, взятые в различных точках: в
начале и конце пути.

Если задано два произвольных сечения (\ref{esecde}) то имеет место аналогичная
формула.

Рассмотрим замкнутый путь в базе $\g\in\Om(\MU,x_0)$. Поскольку мы ограничились
координатной окрестностью $\MU$, то все пути стягиваемы к точке $x_0$ (гомотопны
постоянному пути в $x_0$). Для этих путей $S(1)=S(0)$ и $\P$-экспонента при
калибровочном преобразовании $\s_0\mapsto\s$ преобразуется по тензорному закону
\begin{equation}                                                  \label{etewol}
  \P\exp\left(\oint_\g dx^\al \overset{~\circ}A_\al\right)
  =S(b_0)\P\exp\left(\oint_\g dx^\al A_\al\right)S^{-1}(b_0).
\end{equation}

При параллельном переносе слоя $\pi^{-1}(x_0)$ вдоль замкнутого пути $\g$ точка
$p_0=(x_0,a_0)$ отобразится в точку $p_1=(x_0,a_1)$. Пусть $a_0=e$, тогда
$S_0=\one$. В этом случае
\begin{equation}                                                  \label{ehowil}
  \hat a=\rho^{-1}\left[\P\exp\left(\oint_\g
  dx^\al\overset{~\circ}A_\al\right)\right]\quad \in\Phi_0\big(\s_0(x_0)\big)
\end{equation}
-- элемент суженной группы голономии $\Phi_0\big(\s_0(x_0)\big)$. Здесь мы
предполагаем, что представление $\rho$ является точным. Этот элемент
определяется калибровочным полем, соответствующим нулевому сечению $\s_0$,
которое проходит через начальную точку кривой $(x_0,e)$. Если $a_0\ne e$, то
элемент группы голономии имеет вид $a_0^{-1}a_1\in\Phi_0(p_0)$ и равен
\begin{equation*}
  \rho^{-1}\left[S^{-1}_0\P\exp\left(\oint_\g dx^\al\overset{~\circ}A_\al\right)
  S_0\right]=a_0^{-1}\hat a a_0\quad \in\Phi_0\big(\s(x_0)\big).
\end{equation*}
То есть он сопряжен элементу $\hat a$ в соответствии с утверждением 1)
предложения \ref{pcohog}.

Поскольку $\P$-экспонента для замкнутого пути при изменении сечения
преобразуется по тензорному закону (\ref{etewol}), то для произвольного сечения
$\s$ $\P$-экспонента определяет элемент суженной группы голономии
$\Phi_0\big(\s(x_0)\big)$,
\begin{equation}                                                  \label{eholgr}
  \rho^{-1}\left[\P\exp\left(\oint_\g dx^\al A_\al\right)\right]
  \quad \in\Phi_0\big(\s(x_0)\big),
\end{equation}
где $A_\al{}$ -- калибровочное поле, соответствующее произвольному сечению,
проходящему через точку $\s(x_0)\in\pi^{-1}(\MU)$. Таким образом, рассматривая
все замкнутые пути $\g\in\MM$ с началом в точке $x_0$, можно определить
суженную группу голономии.
\begin{defn}
След от $\P$-экспоненты
\begin{equation}                                                  \label{ewilde}
  W_\g[A]:=\tr\left[\P\exp\left(\oint_\g dx^\al A_\al\right)\right].
\end{equation}
называется {\em петлей Вильсона}.
\qed\end{defn}
\index{Петля Вильсона (Wilson loop)}%
\index{Вильсона петля (Wilson loop)}%
\begin{prop}
Петля Вильсона инвариантна относительно калибровочных преобразований.
\end{prop}
\begin{proof}
Прямое следствие равенства (\ref{etewol}).
\end{proof}
\begin{com}
Петля Вильсона играет важную роль в решеточной формулировке квантовых
калибровочных моделей.
\qed\end{com}
Полученное выражение для упорядоченной экспоненты (\ref{esohom}) не зависит от
выбора координат на $\MU\subset\MM$. Это значит, что понятие упорядоченной
экспоненты, которое было получено в одной карте, без труда переносится на
произвольные пути в $\MM$, которые в общем случае не покрываются одной картой.
Для этого весь путь надо разбить на отрезки, каждый из которых покрывается одной
картой, и взять сумму интегралов вдоль каждого отрезка.
\section{Отображение связностей}
В разделе \ref{shofbu} мы изучили отображение расслоений. В частности, было
определено вложение (инъекция) расслоений, редукция структурной группы, а также
индуцированное расслоение. Ниже мы изучим вопрос о том, как ведут себя связности
и соответствующие им группы голономий при отображении расслоений. В дальнейшем
эти результаты будут использованы при изучении групп голономий.
\begin{prop}                                                      \label{phocon}
Пусть $f:~\MP_1(\MM_1,\pi_1,\MG_1)\rightarrow\MP_2(\MM_2,\pi_2,\MG_2)$ --
гомоморфизм главных расслоений, состоящий из дифференцируемого отображения
пространств расслоений $f_\MP:~\MP_1\rightarrow\MP_2$ и гомоморфизма структурных
групп $f_\MG:~\MG_1\rightarrow\MG_2$, такой, что индуцированное отображение баз
$f_\MM:~\MM_1\rightarrow\MM_2$ есть диффеоморфизм. Пусть $\Gamma_1$ -- связность на
$\MP_1$ с формой связности $\om_1$ и формой кривизны $R_1$. Тогда:\newline
\indent 1) \parbox[t]{.92\linewidth}{Существует единственная связность $\Gamma_2$ на
$\MP_2$ такая, что $f_\MP$ отображает горизонтальные подпространства связности
$\Gamma_1$ в горизонтальные подпространства связности $\Gamma_2$.}\newline
\indent 2) \parbox[t]{.92\linewidth}{Если $\om_2$ и $R_2$ -- формы связности и
кривизны для $\Gamma_2$, то
\begin{equation*}
  f^*_\MP\om_2=f_\Gg\om_1\qquad \text{и}\qquad f^*_\MP R_2=f_\Gg R_1,
\end{equation*}
где правые части $f_\Gg\om_1$ и $f_\Gg R_1$ обозначают $\Gg_2$-значные формы на
$\MP_1$, определенные соотношениями:
\begin{equation*}
  (f_\Gg\om_1)(\widetilde X)=f_\Gg\big(\om_1(\widetilde X)\big)\qquad \text{и}\qquad
  (f_\Gg R_1)(\widetilde X,\widetilde Y)
  =f_\Gg\big(R_1(\widetilde X,\widetilde Y)\big)\qquad
  \forall\widetilde X,\widetilde Y\in\CX(\MP_1),
\end{equation*}
где $f_\Gg=f_{\MG*}$ -- гомоморфизм алгебр Ли $\Gg_1\rightarrow\Gg_2$,
индуцированный отображением структурных групп $f_\MG:~\MG_1\rightarrow\MG_2$
(дифференциал отображения $f_\MG$).}\newline
\indent 3) \parbox[t]{.92\linewidth}{Если $p_2=f_\MP(p_1)\in\MP_2$ -- образ
точки $p_1\in\MP_1$, то $f_\MG$ гомоморфно отображает группу голономии
$\Phi(p_1)$ в точке $p_1$ на $\Phi(p_2)$ и ограниченную группу голономии
$\Phi_0(p_1)$ на $\Phi_0(p_2)$.}
\end{prop}
\begin{proof}
Проводится путем явного построения связности $\Gamma_2$ на $\MP_2$.
См., например, \cite{KobNom6369R}.
\end{proof}
\begin{defn}
В ситуации, описанной в предложении \ref{phocon}, говорят, что $f$ отображает
связность $\Gamma_1$ в связность $\Gamma_2$. В частном случае, если
$\MP_1(\MM_1,\pi_1,\MG_1)$ -- редуцированное подрасслоение в
$\MP_2(\MM_2,\pi_2,\MG_2)$, т.е.\ $f_\MG$ -- мономорфизм, $\MM_1=\MM_2=\MM$ и
$f_\MM=\id_\MM$, то говорят, что связность $\Gamma_2$ на $\MP_2$ {\em редуцируема} к
связности $\Gamma_1$ на $\MP_1$. Автоморфизм $f$ главного расслоения
$\MP(\MM,\pi,\MG)$ называется {\em автоморфизмом связности} $\Gamma$ на $\MP$, если
он отображает $\Gamma$ в $\Gamma$. В этом случае говорят, что связность $\Gamma$
{\em инвариантна} относительно $f$.
\qed\end{defn}
\index{Редуцированная связность (reduced connection)}%
\index{Связность редуцированная (reduced connection)}%
\index{Инвариантная связность (invariant connection)}%
\index{Связность инвариантная (invariant connection)}%
\index{Автоморфизм связности (connection automorphism)}%
\begin{prop}
Любая связность $\Gamma$ на главном расслоении $\MP(\MM,\pi,\MG)$ инвариантна
относительно вертикальных автоморфизмов (пример \ref{everau}).
\qed\end{prop}
\begin{proof}
Прямое следствие свойства 2) в определении связности.
\end{proof}
\begin{cor}
Пусть $\MQ(\MM,\pi,\MH)$ -- подрасслоение в $\MP(\MM,\pi,\MG)$, где $\MH$ --
подгруппа Ли в $\MG$. Пусть $\Gamma$ - связность на $\MP$ с формой связности $\om$.
Тогда связность $\Gamma$ на $\MP$ редуцируема к связности $\Gamma'$ на $\MQ$ тогда и
только тогда, когда сужение формы связности $\om$ на $\MQ$ является
$\Gh$-значным, где $\Gh$ -- подалгебра Ли в $\Gg$, соответствующая подгруппе Ли
$\MH\subset\MG$. Если связность $\Gamma$ редуцируема к $\Gamma'$, то группы голономии
$\Phi$ и суженные группы голономии $\Phi_0$ для $\MQ$ и $\MP$ изоморфны.
\qed\end{cor}
\begin{proof}
Пусть $\Gamma'$ -- связность на $\MQ$. Ее форма связности, по определению,
$\Gh$-значна и по предложению \ref{phocon} продолжается единственным образом
до связности $\Gamma$ на $\MP$. Обратно. Если связность $\Gamma$ на $\MP$ редуцируема к
связности на $\MQ$, то сужение $\om$ на $\MQ$ $\Gh$-значно. Пусть связность
$\Gamma$ редуцируема к $\Gamma'$. Отождествим множество точек $\MQ$ с его образом
$f_\MP(\MQ)$ в $\MP$. Тогда любая горизонтальная кривая в $\MP$ с началом в
произвольной точке $p'\in f_\MP(\MQ)$ будет целиком лежать в $f_\MP(\MQ)$, т.к.\
сужение распределения горизонтальных подпространств в $\MP$ на $\MQ$
совпадает со связностью на $\MQ$. Поскольку группы голономии для всех точек
$p\in\MP$ изоморфны, то изоморфны  также все группы голономии для $\MQ$ и $\MP$.
\end{proof}
\begin{prop}                                                      \label{predco}
Пусть $\MQ(\MM,\pi,\MH)$ -- подрасслоение в главном расслоении
$\MP(\MM,\pi,\MG)$, где $\MH$ -- подгруппа Ли в $\MG$. Допустим, что алгебра Ли
$\Gg$ для $\MG$ допускает подпространство $\Gm$ такое, что $\Gg=\Gh\oplus\Gm$ и
$\ad(\MH)\Gm=\Gm$, где $\Gh$ подалгебра Ли для $\MH$. Тогда для каждой формы
связности $\om$ на $\MP$ $\Gh$-компонента $\om'$ формы связности $\om$,
суженная на $\MQ$ является формой связности на $\MQ$.
\end{prop}
\begin{proof}
См., например, \cite{KobNom6369R}.
\end{proof}
\begin{com}
В силу следствия из предложения \ref{phocon} форма связности $\om'$ совпадает с
$\om$ на $\MQ$.
\qed\end{com}

В предложении \ref{phocon} мы рассматривали отображение связности $\Gamma_1$ на
$\MP_1$ в некоторую связность $\Gamma_2$ на $\MP_2$. При определенных условиях
справедливо также обратное утверждение, и связность с $\MP_2$ можно перенести
на $\MP_1$.
\begin{prop}                                                      \label{pindco}
Пусть $f:~\MP_1(\MM_1,\pi_1,\MG_1)\rightarrow\MP_2(\MM_2,\pi_2,\MG_2)$ --
гомоморфизм главных расслоений такой, что гомоморфизм структурных групп
$f_\MG:~\MG_1\rightarrow\MG_2$ является изоморфизмом. Пусть $\Gamma_2$ -- связность
на $\MP_2$ с формой связности $\om_2$ и формой кривизны $R_2$. Тогда:\newline
\indent 1) \parbox[t]{.92\linewidth}{Существует единственная связность $\Gamma_1$ на
$\MP_1$ такая, что $f_\MP$ отображает горизонтальные подпространства связности
$\Gamma_1$ в горизонтальные подпространства связности $\Gamma_2$.}\newline
\indent 2) \parbox[t]{.92\linewidth}{Если $\om_1$ и $R_1$ -- формы связности и
кривизны для $\Gamma_1$, то
\begin{equation*}
  f^*_\MP\om_2=f_\Gg\om_1\qquad \text{и}\qquad f^*_\MP R_2=f_\Gg R_1,
\end{equation*}
где праве части определены в предложении \ref{phocon}.
}\newline
\indent 3) \parbox[t]{.92\linewidth}{Если $p_2=f_\MP(p_1)\in\MP_2$ -- образ
точки $p_1\in\MP_1$, то изоморфизм $f_\MG$ гомоморфно отображает группу
голономии $\Phi(p_1)$ в точке $p_1$ в $\Phi(p_2)$ и ограниченную группу
голономии $\Phi_0(p_1)$ в $\Phi_0(p_2)$.}
\end{prop}
\begin{proof}
Проводится путем явного построения связности $\Gamma_1$ на $\MP_1$.
См., например, \cite{KobNom6369R}.
\end{proof}
\begin{defn}
В ситуации, описанной в предложении \ref{pindco}, говорят, что связность $\Gamma_1$
{\em индуцирована гомоморфизмом} $f$ из связности $\Gamma_2$.
\qed\end{defn}
\index{Индуцированная связность (induced connection)}%
\index{Связность индуцированная (induced connection)}%
\begin{cor}
Пусть $f:~\MP_1(\MM_1,\pi_1,\MG)\rightarrow\MP_2(\MM_2,\pi_2,\MG)$ --
отображение расслоений с одинаковой структурной группой такое, что
$f_\MG=\id_\MG$ -- тождественный автоморфизм. Если $\om_2$ -- форма связности
на $\MP_2$, то отображение $f_\MP$ индуцирует связность на $\MP_1$:
$\om_1=f^*_\MP\om_2$. В частности, для данного главного расслоения
$\MP(\MM,\pi,\MG)$ и отображения баз $f_\MN:~\MN\rightarrow\MM$ каждая связность
на $\MP$ индуцирует связность на $f^{-1}_\MN(\MP)$. В частном случае, если
$\MU\subset\MM$ -- открытое подмножество, то связность $\Gamma$ на
$\MP(\MM,\pi,\MG)$ индуцирует связность на индуцированном подрасслоении
$\MP|_\MU=\pi^{-1}(\MU)$.
\qed\end{cor}
\begin{com}
В предложении \ref{phocon} связность $\Gamma_2$ на $\MP_2$ строилась таким образом,
что гомоморфное отображение групп голономий является сюрьективным. В предложении
\ref{pindco} утверждается, что отображение групп голономий является только
гомоморфизмом.
\qed\end{com}
\section{Связность на ассоциированном расслоении}
Пусть дано главное расслоение $\MP(\MM,\pi,\MG)$ и ассоциированное с ним
расслоение $\ME(\MM,\pi_\ME,\MF,\MG,\MP)$ с типичным слоем $\MF$ (см.\ раздел
\ref{sassfi}). Если на $\MP$ задана связность, то она определяет связность на
$\ME$ следующим образом.
\begin{defn}
Пусть $u\in\ME$ -- произвольная точка ассоциированного расслоения.
{\em Вертикальным подпространством} $\MV_u(\ME)$ в касательном пространстве
$\MT_u(\ME)$ называется касательное пространство к слою
$\pi^{-1}_\ME\big(\pi_\ME(u)\big)$, которое лежит в $\MT_u(\ME)$.
\qed\end{defn}
Чтобы определить горизонтальное подпространство, вспомним, что ассоциированное
расслоение строилось с помощью естественной проекции на фактор пространство
\begin{equation*}
  \MP\times\MF\rightarrow\ME=\MP\times_\MG\MF.
\end{equation*}
Выберем точку $(p,v)\in\MP\times\MF$, которая проектируется на $u\in\ME$.
Теперь зафиксируем точку типичного слоя $v\in\MF$ и рассмотрим отображение
\begin{equation}                                                  \label{emapsv}
  v:~\MP\ni\quad p\mapsto v(p)=u\quad\in\ME,
\end{equation}
отображающее точку $p\in\MP$ в $v(p)\in\ME$. То есть каждой точке типичного слоя
$v$ ставится в соответствие отображение (\ref{emapsv}), которое мы обозначаем
той же буквой.
\begin{com}
В отличии от отображения $p:~\MF\rightarrow\MF_x$, определенного ранее
(\ref{edefpm}), это отображение в общем случае не является диффеоморфизмом, так
как размерности главного и ассоциированного расслоения могут отличаться. Даже
если размерности совпадают, $\dim\MG=\dim\MF$, то этого недостаточно для того,
чтобы отображение (\ref{emapsv}) было диффеоморфизмом. Действительно, если $v_0$
-- неподвижная точка группы преобразований, то отображение $v_0(p)$ переводит
все точки слоя $\pi^{-1}(x)$ в одну фиксированную точку ассоциированного
расслоения $u_0\in\pi^{-1}_\ME(x)$.
\qed\end{com}
Поскольку точки $(p,v)$ и $(pa,va)$ из $\MP\times\MF$ проектируются в одну и ту
же точку ассоциированного расслоения $u\in\ME$, то $(va^{-1})(p)=v(pa)$, т.е.\
диаграмма
\begin{equation*}
\begin{diagram}
  \MP & \rTo^a & \MP \\
  & \rdTo_{va^{-1}} & \dTo_v \\
  &  & \ME
\end{diagram}
\end{equation*}
коммутативна. Используя построенное отображение, определим горизонтальные
подпространства в ассоциированное расслоении.
\begin{defn}
{\em Горизонтальным подпространством} в точке $u\in\ME$ называется образ
$\MH_u(\ME)=v_*\big(\MH_p(\MP)\big)$, где $v_*$ -- дифференциал отображения
(\ref{emapsv}).
\qed\end{defn}
Легко видеть, что подпространство $\MH_u(\ME)$ не зависит от выбора точек
$(p,v)\in\MP\times\MF$, которые проектируются в точку $u\in\ME$. Действительно,
поскольку $(va)(p)=v(p a^{-1})$, то $(va)_*=v_*\circ r^{-1}_{a*}$. Поэтому для
точки $(pa,va)$, которая проектируется в ту же точку $u\in\ME$, что и $(p,v)$,
справедливо равенство
\begin{equation*}
  \MH_u=(va)_*\MH_{pa}=v_*\circ r^{-1}_{a*}\circ r_{a*}\MH_p=v_*\MH_p,
\end{equation*}
где мы использовали инвариантность (\ref{einvho}) распределения горизонтальных
подпространств на $\MP$. В следующем примере мы докажем, что касательное
пространство к ассоциированному расслоению представляет собой прямую сумму,
\begin{equation}                                                  \label{edirsu}
  \MT_u(\ME)=\MV_u(\ME)\oplus\MH_u(\ME).
\end{equation}
\index{Вертикальное подпространство (vertical subspace)}%
\index{Подпространство вертикальное (vertical subspace)}%
 \index{Горизонтальное подпространство (horizontal subspace)}%
\index{Подпространство горизонтальное (horizontal subspace)}%
Этого достаточно для определения связности на ассоциированном расслоении. Тем не
менее построенная связность на ассоциированном расслоении обладает
дополнительным свойством, которое наследуется из главного расслоения: она
инвариантна относительно действия группы справа. Действительно, по-построению,
отображение (\ref{emapsv}) перестановочно с групповым действием. Поэтому
перестановочны также дифференциалы этих отображений,
\begin{equation*}
  v_*\circ r_{a*}=r_{a*}\circ v_*.
\end{equation*}
Тогда из правой инвариантности связности на главном расслоении (\ref{einvho})
следует инвариантность связности на ассоциированном расслоении:
\begin{equation}                                                  \label{einhas}
  r_{a*}\MH_u(\ME)=\MH_{ua}(\ME).
\end{equation}

\begin{exa}[\bf Локальное рассмотрение]
Рассмотрим достаточно малую координатную окрестность на базе, $\MU\subset\MM$, с
координатами $x^\al$, $\al=1,\dotsc,n$, такую, что выполнены условия локальной
тривиализации расслоений, $\pi^{-1}(\MU)\approx\MU\times\MG$ и
$\pi^{-1}_\ME(\MU)\approx\MU\times\MF$. Отождествим $\pi^{-1}(\MU)$ с
$\MU\times\MG$ и $\pi^{-1}_\ME(\MU)$ с $\MU\times\MF$. Ограничим наше
рассмотрение окрестностью единицы группы Ли, $\MU_\MG\subset\MG$, где определены
координаты $a^\Sa$, $\Sa=1,\dotsc,\Sn$, и функция композиции (см.\ раздел
\ref{slolir}). Выберем также некоторую координатную окрестность в типичном слое,
$\MU_\MF\subset\MF$, где определены координаты $v^i$, $i=1,\dotsc,\dim\MF$, и
задано отображение $v\mapsto va$ в координатной форме. Тогда точки расслоений
будут иметь координаты $p=\lbrace x^\al,a^\Sa\rbrace\in\MP$ и
$u=\lbrace x^\al,v^i\rbrace\in\ME$. По определению, векторы $\pl_i$ касательны к
слою $\pi^{-1}_\ME(x)$ и, следовательно, образуют базис вертикальных
подпространств $\MV_u(\ME)$ для всех $u\in\ME$. Пусть
$v_0=\lbrace v^i_0\rbrace\in\MU_\MF$ -- фиксированная точка типичного слоя.
Тогда соответствующее этой точке отображение (\ref{emapsv}) в координатах имеет
вид
\begin{equation*}
  v_0:\quad \MU\times\MU_\MG\ni\quad\lbrace x^\al,a^\Sa\rbrace\mapsto
  \lbrace x^\al,v^i(v_0,a)\rbrace\quad\in\MU\times\MU_\MF,
\end{equation*}
где $v^i(v_0,a)$ -- некоторая функция координат $v^i_0$ и $a^\Sa$. Пусть
$D_\al=\pl_\al-\om_\al{}^\Sa L^*_\Sa$, где $\om_\al{}^\Sa$ есть компоненты
формы связности (\ref{ecoomc}), -- базис горизонтальных подпространств
$\MH_p(\MP)$ в главном расслоении, который был построен ранее (\ref{ehobaf}).
Тогда он отображается в касательное пространство к ассоциированному расслоению,
\begin{equation*}
  v_{0*}:\quad \MH_p(\MP)\ni\quad D_\al\mapsto\overline D_\al
  =\pl_\al-\om_\al{}^\Sa L^*_\Sa v^i\pl_i\quad\in\MH_u(\ME),
\end{equation*}
где $L^*_\Sa$ действует на $v^i(v_0,a)$ как дифференцирование по $a$. Конкретный
вид функций $v^i(v_0,a)$ зависит от типичного слоя и действия на нем структурной
группы. Независимо от вида функций $v^i(v_0,a)$ векторы $\overline D_\al$
линейно независимы, т.к.\ содержат $\pl_\al$, и поэтому образуют базис
горизонтального подпространства $\MH_u(\ME)$. Таким образом совокупность
векторов $\lbrace\overline D_\al,\pl_i\rbrace$ образует базис касательного
пространства $\MT_u(\ME)$, который соответствует разложению (\ref{edirsu}).
Аналогичное построение можно выполнить в окрестности произвольной точки
ассоциированного расслоения $u\in\ME$. Следовательно, разложение касательного
пространства $\MT_u(\ME)$ в прямую сумму (\ref{edirsu}) имеет место в общем
случае.

Если типичный слой -- это векторное пространство $\MV$, на котором
задано представление $\rho$ (см., раздел \ref{slieho}), то
\begin{equation*}
  v^i(v_0,a)=v^j_0S^{-1}_{\quad j}{}^i(a),
\end{equation*}
где $S_j{}^i(a)$ -- матрица представления элемента $a\in\MG$. В этом случае
\begin{equation*}
  \om_\al{}^\Sa L^*_\Sa v^i=\om_\al{}^\Sa v^j_0 S_j{}^k L_{\Sa k}{}^i
  =-v^j\om_\al{}_j{}^i,
\end{equation*}
где введено обозначение $\om_{\al j}{}^i:=-\om_\al{}^\Sa L_{\Sa j}{}^i$ и мы
воспользовались правилом дифференцирования матриц представления  (\ref{ediinm}).
Тогда базис горизонтальных векторных полей на ассоциированном расслоении имеет
вид
\begin{equation}                                                  \label{ehobaa}
  \overline D_\al=\pl_\al+v^j\om_{\al j}{}^i\pl_i.
\end{equation}
Этот базис инвариантен относительно действия структурной группы $\MG$ на
ассоциированном расслоении $\ME$ справа. Действительно, действие элемента
$b\in\MG$ на базисный вектор (\ref{ehobaa}) имеет вид
\begin{equation*}
  r_{b*}\overline D_\al=\pl_\al+v^i\om_{\al i}{}^j(u)S^{-1}_{\quad j}{}^k(b)\pl_k,
\end{equation*}
т.к.\ $v^i(v_0,ab)=v^j(v_0,a)S^{-1}_{\quad j}{}^i(b)$. Поскольку
$v^i(ub)=v^j(u)S^{-1}_{\quad j}{}^i(b)$ и
$\om_\al{}^\Sa(pb)=\om_\al{}^\Sb S^{-1}_{\quad \Sb}{}^\Sa(b)$, то
\begin{equation*}
  r_{b*}\overline D_\al|_u=\overline D_\al|_{ub},
\end{equation*}
где мы воспользовались инвариантностью (\ref{einvge}) матриц представления
генераторов структурной группы. Это соответствует инвариантности распределения
горизонтальных подпространств на ассоциированном расслоении (\ref{einhas}).
\qed\end{exa}

Определения горизонтального лифта и параллельного переноса для ассоциированных
расслоений дословно повторяют определения, данные для главных расслоений.
\begin{defn}
Кривая $\hat\g$ в ассоциированном расслоении $\ME$ называется
{\em горизонтальной}, если касательный к ней вектор горизонтален в каждой точке.
Если задана кривая $\g$ в базе $\MM$, то {\em горизонтальным лифтом} или просто
{\em лифтом} этой кривой называется такая горизонтальная кривая $\hat\g$ в
$\ME$, что $\pi_\ME(\hat\g)=\g$.
\qed\end{defn}
\index{Горизонтальная кривая (horizontal curve)}%
\index{Кривая горизонтальная (horizontal curve)}%
\index{Горизонтальный лифт кривой (horizontal lift of a curve)}%
\index{Лифт кривой (lift of a curve)}%
\begin{prop}                                                      \label{pholca}
Пусть $\g=x(t)$, $t\in[0,1]$, -- кусочно дифференцируемая кривая класса
$\CC^1$ в $\MM$ с началом в точке $x_0\in\MM$. Тогда для произвольной точки слоя
$u_0\in\pi^{-1}_\ME(x_0)$ существует единственный горизонтальный лифт
$\hat\g=u(t)$ кривой $\g$ с началом в точке $u_0$.
\end{prop}
\begin{proof}
Сначала докажем существование горизонтального лифта. Выберем точку
$(p_0,v_0)\in\MP\times\MF$ такую, что $p_0(v_0)=u_0$, где отображение $p$
определено формулой (\ref{edefpm}). Согласно предложению \ref{pholcu} существует
единственный горизонтальный лифт $\tilde\g$ кривой $\g$ в главное расслоение
$\MP$ с началом в точке $p_0\in\MP$. Тогда кривая $\hat\g=\tilde\g(v_0)$
является горизонтальным лифтом кривой в базе $\g$ в ассоциированное расслоение
$\ME$. Действительно, касательный вектор к кривой $\hat\g$ лежит в
$\MH_{\g}(\ME)$, что сразу следует из определения дифференциала отображения
(\ref{etamap}). Единственность горизонтального лифта следует из единственности
решения системы линейных дифференциальных с заданными начальными условиями.
\end{proof}
Для формулировки следующего утверждения нам понадобится естественное
\begin{defn}
Локальное сечение ассоциированного расслоения $\s:~\MU\rightarrow\ME$,
определенное на открытом подмножестве $\MU\subset\MM$, называется
{\em параллельным или горизонтальным}, если образ $\s_*\MT_x(\MM)$, где $\s_*$
-- дифференциал сечения, горизонтален при всех $x\in\MU$, т.е.\ для любой кривой
$\g$ в $\MU$, соединяющей точки $x_0$ и $x_1$, точка слоя $\s(x_0)$ при
параллельном переносе слоя вдоль кривой $\g$ переходит в точку $\s(x_1)$.
\qed\end{defn}
\index{Параллельное сечение (parallel cross-section)}%
\index{Сечение параллельное (parallel cross-section)}%
\index{Горизонтальное сечение (horizontal cross-section)}%
\index{Сечение горизонтальное (horizontal cross-section)}%
В предложении \ref{pfafib} мы отождествили ассоциированное расслоение
$\ME(\MM,\pi_\ME,\MG/\MH,\MG,\MP)$ с фактор пространством $\MP/\MH$. Затем в
теореме \ref{tredas} привели критерий редуцируемости структурной группы $\MG$
главного расслоения $\MP(\MM,\pi,\MG)$ к подгруппе $\MH$, который заключается в
существовании глобального сечения $\s$ ассоциированного расслоения $\ME$. Кроме
того, была установлена естественная взаимно однозначная связь между сечениями
$\s$ ассоциированного расслоения $\ME$ и редуцированными главными расслоениями
$\MQ(\MM,\pi,\MH)$. В примере \ref{exriex} эта теорема была использована для
доказательства существования римановой метрики на произвольном многообразии.
Возникает вопрос о том, в каком случае связность, заданная на главном расслоении
$\MP$ редуцируема к связности на редуцированном расслоении $\MQ$ ? Ответ дает
следующее утверждение.
\begin{prop}
Пусть $\MP(\MM,\pi,\MG)$ -- главное расслоение и
$\ME(\MM,\pi_\ME,\MG/\MH,\MG,\MP)$ ассоциированное расслоение со стандартным
слоем $\MG/\MH$, где $\MH$ -- замкнутая подгруппа в $\MG$. Пусть
$\s:~\MM\rightarrow\ME$ -- глобальное сечение ассоциированного расслоения и
$\MQ(\MM,\pi,\MH)$ редуцированное подрасслоение в $\MP(\MM,\pi,\MG)$,
соответствующее сечению $\s$. Связность $\Gamma$ на $\MP$ редуцируема к связности
$\Gamma'$ на $\MQ$ тогда и только тогда, когда сечение $\s$ параллельно относительно
$\Gamma$.
\end{prop}
\begin{proof}
См., например, \cite{KobNom6369R}.
\end{proof}
\section{Свойства групп голономий                                \label{sholpr}}
Продолжим изучение свойств групп голономий, которое было начато в разделе
\ref{sholde}.
\begin{theorem}[\bf Теорема редукции]                             \label{treduc}
Пусть $\MP(\MM,\pi,\MG)$ -- главное расслоение со связностью $\Gamma$ и $p$ --
произвольная точка в $\MP$. Обозначим через $\MP(p)$ множество точек в $\MP$,
которые можно соединить с точкой $p$ горизонтальными кусочно дифференцируемыми
кривыми. Тогда:\newline
\indent 1) \parbox[t]{.92\linewidth}{$\MP(p)$ -- редуцированное главное
расслоение со структурной группой $\Phi(p)$;}\newline
\indent 2) \parbox[t]{.92\linewidth}{Связность $\Gamma$ редуцируема к связности на
$\MP(p)$.}
\end{theorem}
\begin{proof}
См., например, \cite{KobNom6369R}.
\end{proof}
\index{Теорема редукции (reduction theorem)}%
\index{Редукции теорема (reduction theorem)}%
Эта теорема оправдывает следующее
\begin{defn}
Главное расслоение $\MP(p)$, с базой $\MM$, проекцией $\pi$ и структурной
группой $\Phi(p)$, состоящее из множества точек в главном расслоении
$\MP(\MM,\pi,\MG)$, которые можно соединить с точкой $p$ горизонтальными
кусочно дифференцируемыми кривыми, называется {\em расслоением голономии} через
$p$.
\qed\end{defn}
\index{Расслоение голономии (holonomy fiber bundle)}%
\index{Голономии расслоение (holonomy fiber bundle)}%
Очевидно, что $\MP(p)=\MP(q)$ тогда и только тогда, когда точки $p$ и $q$ можно
соединить горизонтальной кривой. В разделе \ref{sholde} было введено отношение
эквивалентности: $p\sim q$, если $p$ и $q$ можно соединить горизонтальной
кривой. Поэтому для каждой пары точек из главного расслоения $\MP(\MM,\pi,\MG)$
либо $\MP(p)=\MP(q)$, либо $\MP(p)\cap\MP(q)=\emptyset$. Другими словами,
главное расслоение $\MP(\MM,\pi,\MG)$ разлагается в объединение,
\begin{equation*}
  \MP(\MM,\pi,\MG)=\bigcup_{p\in\MP}\MP(p),
\end{equation*}
попарно непересекающихся расслоений голономии. Так как каждый элемент $a\in\MG$
отображает каждую горизонтальную кривую в горизонтальную, то $r_a\MP(p)=\MP(pa)$
и отображение
\begin{equation*}
   r_a:\quad \MP(p)\rightarrow\MP(pa)
\end{equation*}
индуцирует изоморфизм расслоений $f=(f_\MP,f_\MG)$, где $f_\MP=r_a$, с
соответствующим изоморфизмом структурных групп:
\begin{equation*}
  f_\MG=\ad(a^{-1}):\quad \Phi(p)\ni\quad b\mapsto a^{-1}ba\quad\in\Phi(pa),
\end{equation*}
т.к.\ группы голономии в различных точках сопряжены друг другу (предложение
\ref{pcohog}). Легко видеть, что для двух произвольных точек
$p,q\in\MP(\MM,\pi,\MG)$ существует такой элемент $a\in\MG$, что
$\MP(p)=\MP(qa)$. Поэтому все расслоения голономий $\MP(p)$ изоморфны друг
другу.

\begin{theorem}[\bf Амброз--Зингер]                               \label{tambsi}
Пусть $\MP(\MM,\pi,\MG)$ -- главное расслоение со связной базой $\MM$. Пусть
$\Gamma$ -- связность на $\MP$ с формой кривизны $R$, $\Phi(p_0)$ -- группа голономии
в точке $p_0\in\MP$ и $\MP(p_0)$ -- расслоение голономии через $p_0$. Тогда
алгебра Ли группы голономии $\Phi(p_0)$ совпадает с подпространством в алгебре
Ли $\Gg$ структурной группы $\MG$, порожденной всеми элементами вида
$R_p(\widetilde X,\widetilde Y)$ для всех $p\in\MP(p_0)$ и всех горизонтальных
векторных полей $\widetilde X,\widetilde Y$ в точке $p$.
\end{theorem}
\begin{proof}
Используется паракомпактность $\MM$ \cite{AmbSin53}.
\end{proof}
\begin{theorem}                                                   \label{tcospa}
Пусть $\MP(\MM,\pi,\MG)$ -- главное расслоение со связным пространством
расслоения $\MP$. Если $\dim\MM\ge2$, то существует связность на $\MP$ такая,
что расслоения голономии $\MP(p)$ для всех $p\in\MP(\MM,\pi,\MG)$ совпадают с
главным расслоением $\MP(\MM,\pi,\MG)$.
\end{theorem}
\begin{proof}
Явное построение связности. При этом используется паракомпактность $\MM$
\cite{KobNom6369R}. Для линейных связностей это утверждение было доказано в
\cite{HanOze56}. В общем случае доказательство дано в \cite{Nomizu56}.
\end{proof}
\begin{cor}
Любая связная группа Ли $\MG$ может быть реализована как группа голономии
некоторой связности в тривиальном главном расслоении $\MP=\MM\times\MG$, где
$\MM$ -- произвольное дифференцируемое многообразие размерности $\dim\MM\ge2$.
\qed\end{cor}
\begin{proof}
Выберем связную окрестность $\MU\subset\MM$. Тогда расслоение $\MU\times\MG$
связно и мы попадаем в зону деятельности теоремы \ref{tcospa}. Связность с
$\MU\times\MG$ продолжается на связность на $\MM\times\MG$ согласно предложению
\ref{phocon}.
\end{proof}
\section{Плоские связности                                       \label{sflaco}}
Рассмотрим тривиальное главное расслоение $\MP=\MM\times\MG$. Для каждого
элемента структурной группы $a\in\MG$ множество $\MM\times\lbrace a\rbrace$ есть
подмногообразие в $\MP$. В частности, $\MM\times\lbrace e\rbrace$, где $e$ --
единица группы, есть редуцированное подрасслоение в $\MP$.
\begin{defn}
{\em Канонической плоской связностью} на тривиальном главном расслоении
$\MP=\MM\times\MG$ называется распределение горизонтальных подпространств
$\MH_p(\MP)$, образованное касательными пространствами к
$\MM\times\lbrace a\rbrace$ для всех точек $p=(x,a)\in\MM\times\MG$.
\qed\end{defn}
\index{Каноническая плоская связность (canonical flat connection)}%
\index{Плоская связность каноническая (canonical flat connection)}%
\index{Связность каноническая плоская (canonical flat connection)}%
\begin{prop}                                                      \label{pflcon}
Связность $\Gamma$ на тривиальном главном расслоении $\MP=\MM\times\MG$ является
канонической плоской тогда и только тогда, когда она редуцируема к единственной
связности на $\MM\times\lbrace e\rbrace$.
\end{prop}
\begin{proof}
Выберем нулевое сечение $\s_0:~x\mapsto(x,e)$. Это сечение является главным
расслоением $\MP(\MM,\pi,e)$, на котором существует единственная связность. Эта
связность взаимно однозначно определяет каноническую плоскую связность на
$\MP=\MM\times\MG$.
\end{proof}

Пусть $\theta$ -- каноническая форма на группе Ли $\MG$, определенная в разделе
\ref{sleacg}. Обозначим через $\pr:~\MM\times\MG\rightarrow\MG$ естественную
проекцию и положим
\begin{equation}                                                  \label{eflcof}
  \om:=\pr^*\theta=\om^{*\Sa}L_\Sa.
\end{equation}
Эта 1-форма является частным случаем формы связности (\ref{ecjfor}) и определяет
каноническую плоскую связность на $\MP$. Формула Маурера--Картана для
канонической 1-формы (\ref{emucac}) влечет, что каноническая плоская связность
имеет нулевую кривизну, так как
\begin{equation*}
  d\om=d(\pr^*\theta)=\pr^*(d\theta)=\pr^*\left(-\frac12[\theta,\theta]\right)
  =-\frac12[\pr^*\theta,\pr^*\theta]=-\frac12[\om,\om].
\end{equation*}
Сравнивая полученное равенство со структурным уравнением (\ref{estreq}),
заключаем, что форма кривизны канонической плоской связности тождественно равна
нулю, $R=0$.

Теперь рассмотрим случай произвольного главного расслоения. Дадим общее
\begin{defn}
Связность $\Gamma$ на главном расслоении $\MP(\MM,\pi,\MG)$ называется
{\em плоской}, если каждая точка базы $x\in\MM$ имеет окрестность $\MU$ такую,
что индуцированная связность на $\MP|_\MU=\pi^{-1}(\MU)$ изоморфна канонической
плоской связности на $\MU\times\MG$. Другими словами, существует изоморфизм
$\chi:~\pi^{-1}(\MU)\rightarrow\MU\times\MG$, отображающий горизонтальное
подпространство в каждой точке $p\in\pi^{-1}(\MU)$ в горизонтальное
подпространство канонической плоской связности на $\MU\times\MG$ в точке
$\chi(p)$.
\qed\end{defn}
\index{Плоская связность (flat connection)}%
\index{Связность плоская (flat connection)}%
\begin{prop}
Пусть задано главное расслоение $\MP(\MM,\pi,\MG)$. Тогда плоская связность на
$\MP$ существует и единственна с точностью до изоморфизма.
\end{prop}
\begin{proof}
Если задано главное расслоение, то определен атлас на базе $\MM=\bigcup_i\MU_i$
и семейство функций перехода (\ref{etrasf}), которые по теореме \ref{tlofib} с
точностью до изоморфизма определяют главное расслоение. Выбрав координатное
покрытие базы достаточно малым, можно считать, что все координатные окрестности
$\MU_i$ соответствуют окрестностям, входящим в определение плоской связности.
Согласно теореме \ref{troloc} для однозначного задания связности на $\MP$
достаточно задать семейство локальных форм связности на каком либо атласе базы.
Это означает, что плоская связность на произвольном главном расслоении
$\MP(\MM,\pi,\MG)$ существует и единственна с точностью до изоморфизма.
\end{proof}
\begin{theorem}                                                   \label{tflzec}
Связность $\Gamma$ на главном расслоении $\MP(\MM,\pi,\MG)$ является плоской тогда и
только тогда, когда ее форма кривизны равна нулю, $R=0$.
\end{theorem}
\begin{proof}
Необходимость очевидна. Обратно. Допустим, что форма кривизны равна нулю. Пусть
$\MU$ -- односвязная окрестность точки $x\in\MP$ и рассмотрим индуцированную
связность на $\MP|_\MU=\pi^{-1}(\MU)$. По теоремам \ref{tholie} и
Амброза--Зингера \ref{tambsi} группа голономии индуцированной связности на
$\MP|_\MU$  состоит только из единицы. Применяя теорему редукции \ref{treduc},
мы видим, что индуцированная связность на $\MP|_\MU$ изоморфна канонической
плоской связности на $\MU\times\MG$.
\end{proof}
\begin{cor}
Любая связность $\Gamma$ на главном расслоении $\MP(\MM,\pi,\MG)$ с одномерной
базой $\MM$ является плоской.
\qed\end{cor}
\begin{proof}
Любая 2-форма на одномерном многообразии равна нулю. Отсюда следует, что все
локальные формы кривизны тоже равны нулю. Так как для формы кривизны только
горизонтальные компоненты являются нетривиальными, то она также обращается в
нуль.
\end{proof}
\begin{cor}
Пусть $\Gamma$ -- связность на главном расслоении $\MP(\MM,\pi,\MG)$ такая, что ее
форма кривизны равна нулю, $R=0$. Если база $\MM$ односвязна, то главное
расслоение $\MP$ изоморфно тривиальному расслоению $\MM\times\MG$ и связность
$\Gamma$ изоморфна канонической плоской связности на $\MM\times\MG$.
\qed\end{cor}
\begin{proof}
Группа голономии в рассматриваемом случае состоит из единственного элемента --
единицы. Поэтому расслоение голономии $\MP(p)$ пересекает каждый слой ровно в
одной точке. Следовательно, каждое расслоение голономии задает глобальное
сечение, и поэтому главное расслоение тривиально. При этом горизонтальные
подпространства касательны к расслоению голономии. Пусть $q\in\MP(p)$ --
произвольная точка расслоения голономии через $p$ и $\s_0=(x,e)$ -- нулевое
сечение главного расслоения $\MP=\MM\times\MG$. Тогда для каждой точки базы
$x\in\MM$ существует единственный элемент $a(x)\in\MG$ такой, что $\s_0=qa$,
где $x=\pi(q)$. При этом вертикальный автоморфизм $p\mapsto pa$ переводит
связность $\Gamma$ на $\MP=\MM\times\MG$ в каноническую плоскую связность.
\end{proof}
Если форма кривизны равна нулю, то распределение горизонтальных векторных полей
находится в инволюции. Это сразу следует из (\ref{efinob}), т.к.\ векторы
$D_\al$ образуют базис распределения горизонтальных векторных полей. Согласно
теореме Фробениуса для плоской связности через каждую точку $p$ главного
расслоения $\MP(\MM,\pi,\MG)$ проходит интегральное подмногообразие. Все
касательные векторы к интегральным подмногообразиям горизонтальны и интегральное
подмногообразие, проходящее через точку $p$, совпадает с расслоением голономии
$\MP(p)$ через $p$.
\begin{exa}[\bf Локальное рассмотрение]
Пусть $\MQ=\MU\times\MG$ -- тривиальное главное расслоение, база $\MU$ которого
покрыта одной картой. Общий вид формы связности на $\MQ$ был найден ранее, см.\
(\ref{ecjfor}). Сравнение этого выражения с выражением (\ref{eflcof})
показывает, что связность на $\MQ$ является канонической плоской тогда и только
тогда, когда часть ее компонент, описывающих произвол в выборе связности на
нулевом сечении, обращаются в нуль, $\overset{~\circ}A_\al{}^\Sa(x)=0$. Из
равенства (\ref{ecozef}) следует, что компоненты локальной формы канонической
плоской связности для произвольного сечения $\s:~x\mapsto\big(x,b(x)\big)$ имеют
вид
\begin{equation*}
  A_\al{}^\Sa(x)=\pl_\al b^\Sb L^{-1}_{\quad \Sb}{}^\Sa(b).
\end{equation*}
После перехода к какому либо представлению структурной группы
\begin{equation*}
  \rho:\quad \MG\ni\quad a\mapsto\lbrace S_i{}^j(a)\rbrace\quad\in\aut\MV,
\end{equation*}
для компонент локальной формы связности справедливо равенство (\ref{egatra}).
Поскольку $\overset{~\circ}A_\al{}^\Sa=0$, то компоненты плоской связности в
общем случае имеют вид
\begin{equation}                                                  \label{elofla}
  A_\al=\pl_\al S S^{-1},
\end{equation}
где мы, для простоты, опустили матричные индексы.
\begin{defn}
Калибровочное поле $A_\al$, заданное равенством (\ref{elofla}) на координатной
окрестности $U$, называется {\em чистой калибровкой}.
\qed\end{defn}
\index{Чистая калибровка (pure gauge)}%
\index{Калибровка чистая (pure gauge)}%
\end{exa}

Теперь рассмотрим случай, когда база $\MM$ главного расслоения не является
односвязной. Пусть $\Gamma$ -- связность на главном расслоении $\MP(\MM,\pi,\MG)$
со связной базой $\MM$. Выберем произвольную точку $p_0\in\MP(\MM,\pi,\MG)$ и
обозначим через $\widetilde\MM=\MP(p_0)$ расслоение голономии через $p_0$.
Тогда $\widetilde\MM\big(\MM,\pi,\Phi(p_0)\big)$ есть главное расслоение над
$\MM$ со структурной группой $\Phi(p_0)$. Так как суженная группа голономии
$\Phi_0(p_0)$ для плоской связности всегда тривиальна, то по теоремам
\ref{tholie} и Амброза--Зингера \ref{tambsi} группа голономии $\Phi(p_0)$
при неодносвязной базе дискретна. Поэтому отображение
$\pi:~\widetilde\MM\rightarrow\MM$ является накрытием со связным накрывающим
пространством.

Пусть $x_0=\pi(p_0)\in\MM$. Каждая замкнутая кривая в базе $\MM$, исходящая из
$x_0$, при помощи параллельного переноса слоев вдоль нее определяет некоторый
элемент группы голономии $\Phi(p_0)$. Поскольку суженная группа голономии
тривиальна, то любые две замкнутые и гомотопные относительно начала кривые,
представляющие один и тот же элемент фундаментальной группы $\pi(\MM,x_0)$,
порождают один и тот же элемент из $\Phi(p_0)$. Таким образом мы получаем
сюрьективное отображение фундаментальной группы $\pi(\MM,p_0)$ на группу
голономии $\Phi(p_0)$. Легко видеть, что это отображение является гомоморфизмом
групп. Пусть $\MH$ -- нормальная подгруппа в $\Phi(p_0)$ и положим
$\MM'=\widetilde\MM/\MH$. Тогда
$\MM'$ -- главное расслоение над $\MM$ со структурной группой $\Phi(p_0)/\MH$.
В частности, отображение $\pi:~\MM'\rightarrow\MM$ -- накрытие. Пусть
$\MP'(\MM',\pi,\MG)$ -- главное расслоение, индуцированное из $\MP(\MM,\pi,\MG)$
накрывающей проекцией $f_\MM=\pi:~\MM'\rightarrow\MM$. Пусть
$f:~\MP'(\MM',\pi,\MG)\rightarrow\MP(\MM,\pi,\MG)$ -- естественный гомоморфизм
главных расслоений, см.\ теорему \ref{tinfib}, тогда справедливо
\begin{prop}                                                      \label{phozer}
Существует единственная связность $\Gamma'$ на главном расслоении
$\MP'(\MM',\pi,\MG)$, которая отображается на связность $\Gamma$ на
$\MP(\MM,\pi,\MG)$ гомоморфизмом $f:~\MP'\rightarrow\MP$. Связность $\Gamma'$
плоская. Если точка $p'_0\in\MP'$ такова, что $f_\MM(p'_0)=p_0$, то группа
голономии $\Phi(p'_0)$ для связности $\Gamma'$ изоморфно отображается на $\MH$
гомоморфизмом $f_\MG$.
\end{prop}
\begin{proof}
См., например, \cite{KobNom6369R}.
\end{proof}
В данном утверждении всегда можно выбрать в качестве нормальной подгруппы
единицу группы голономии $e\in\Phi(p_0)$.
\begin{cor}
Пусть $\widetilde\MP(\widetilde\MM,\pi,\MG)$ -- главное расслоение,
индуцированное из $\MP(\MM,\pi,\MG)$ накрывающей проекцией
$f_\MM=\pi:~\widetilde\MM\rightarrow\MM$. Тогда существует единственная
связность $\widetilde\Gamma$ на главном расслоении
$\widetilde\MP(\widetilde\MM,\pi,\MG)$, которая отображается на связность $\Gamma$
на $\MP(\MM,\pi,\MG)$ гомоморфизмом $f:~\widetilde\MP\rightarrow\MP$. Связность
$\widetilde\Gamma$ плоская, и соответствующая ей группа голономии тривиальна.
\end{cor}
\section{Локальные и инфинитезимальные группы голономии}
Группа голономии $\Phi(p)$ в точке $p\in\MP(\MM,\pi,\MG)$, которая является
важнейшей глобальной характеристикой связности, была определена при помощи
множества всех замкнутых путей в базе $\Om\big(\MM,\pi(p)\big)$. Это крайне
неудобно для практических вычислений, так как зачастую структура многообразия
$\MM$ просто неизвестна. Возникает вопрос, можно ли каким либо образом вычислить
группу голономии, исходя из локальных характеристик связности? В некоторых
случаях суженная группа голономии $\Phi_0(p)$ действительно определяется
локальными свойствами связности. В настоящем разделе мы опишем два подхода к
этой проблеме, которые основаны на понятиях локальной и инфинитезимальной групп
голономии.

Начнем рассмотрение с локальной группы голономии.
Пусть на главном расслоении $\MP(\MM,\pi,\MG)$ со связной базой $\MM$ задана
связность $\Gamma$. Для каждого связного и односвязного открытого подмножества базы
$\MU\subset\MM$ обозначим через $\Gamma_\MU$ связность на $\MP|_\MU=\pi^{-1}(\MU)$,
которая индуцирована из связности $\Gamma$. В силу следствия из предложения
\ref{pindco} связность $\Gamma_\MU$ существует и единственна -- это сужение
связности $\Gamma$ на $\pi^{-1}(\MU)$. Для каждой точки
$p\in\pi^{-1}(\MU)$ обозначим через $\Phi_0(p,\MU)$ и $\MP(p,\MU)$ суженную
группу голономии с опорной точкой $p$ и расслоение голономии через точку $p$ для
связности $\Gamma_\MU$ соответственно. Напомним, что расслоение голономии
$\MP(p,\MU)$ состоит из тех точек $q\in\pi^{-1}(\MU)$, которые можно соединить с
точкой $p$ горизонтальной кривой, целиком лежащей в $\pi^{-1}(\MU)$.

Рассмотрим две окрестности $\MU_2\subset\MU_1$, которые содержат точку
$x=\pi(p)$. Тогда всякая замкнутая петля, целиком лежащая в $\MU_2$, будет также
петлей в $\MU_1$. Поэтому справедливо включение
\begin{equation*}
  \Phi_0(p,\MU_2)\subset\Phi_0(p,\MU_1).
\end{equation*}
Как подгруппа суженной группы голономии $\Phi_0(p)$ группа $\Phi_0(p,\MU_1)$
однозначно определяется своей алгеброй Ли. Поэтому из равенства размерностей
$\dim\Phi_0(p,\MU_2)=\dim\Phi(p,\MU_1)$ следует совпадение групп голономии
$\Phi_0(p,\MU_2)=\Phi_0(p,\MU_1)$. Это наблюдение приводит к следующему понятию.

\begin{defn}
{\em Локальной группой голономии} $\Phi_{\rm loc}(p)$ в точке $p$ называется
пересечение
\begin{equation}                                                  \label{elocho}
  \Phi_{\rm loc}(p)=\bigcap_\MU\Phi_0(p,\MU)
\end{equation}
по всем связным и односвязным открытым окрестностям $\MU$ точки $x=\pi(p)$.
\qed\end{defn}
\index{Локальная группа голономии (local holonomy group)}%
\index{Группа голономии локальная (local holonomy group)}%
\begin{com}
В настоящем разделе мы будем рассматривать только связные и односвязные открытые
окрестности точек. Поэтому в дальнейшем, для краткости, мы будем говорить просто
окрестности.
\qed\end{com}
Пусть $\lbrace\MU_k\rbrace$, $k=1,2,\dotsc$, -- последовательность окрестностей,
сходящихся к точке $x$, т.е.\ $\MU_k\supset\overline\MU_{k+1}$ и
$\bigcap_{k=1}^\infty\MU_k=\lbrace x\rbrace$. Тогда, очевидно, имеют место
включения
\begin{equation*}
  \Phi_0(p,\MU_1)\supset\Phi_0(p,\MU_2)\supset\Phi_0(p,\MU_3)\supset\dotsc
\end{equation*}
Поскольку для каждой окрестности $\MU$ точки $x$ существует целое число $k_\MU$
такое, что $\MU_k\subset\MU$ для всех $k>k_\MU$, то локальная группа голономии
представима в виде
\begin{equation*}
  \Phi_{\rm loc}(p)=\bigcap_{k=1}^\infty\Phi_0(p,\MU_k).
\end{equation*}
Так как каждая суженная группа голономии $\Phi_0(p,\MU_k)$ есть связная
подгруппа Ли в структурной группе $\MG$ (теорема \ref{tholie}), то отсюда
следует, что размерность суженной группы голономии $\dim\Phi_0(p,\MU_k)$
постоянна для достаточно больших $k$. Поэтому для больших значений $k$
справедливо равенство $\Phi_{\rm loc}(p)=\Phi_0(p,\MU_k)$.
\begin{com}
Конечно, суженная группа голономии $\Phi_0(p,x)$ для ``окрестности'', состоящей
из одной точки $x$, состоит ровно из одного элемента -- единицы, и ее
размерность равна нулю. Допустим, что параметр $k$ в последовательности
$\lbrace\MU_k\rbrace$ непрерывен. Это может быть, например, радиус шара, если
последовательность $\MU_k$ состоит из шаров. Тогда функция
$\dim\Phi_0(p,\MU_k)$ от $k$ принимает значения в целых числах и не может быть
непрерывной, если суженная группа голономии $\Phi_0(p)$ для всего главного
расслоения нетривиальна. В дальнейшем мы увидим, что при определенных условиях
локальная группа голономии $\Phi_{\rm loc}(p)$ совпадает с суженной группой
$\Phi_0(p)$.
\qed\end{com}
\begin{prop}                                                      \label{prohlo}
Локальные группы голономии имеют следующие свойства:\newline
\indent 1) \parbox[t]{.92\linewidth}{локальная группа голономии
$\Phi_{\rm loc}(p)$ есть связная подгруппа Ли в структурной группе $\MG$,
содержащаяся в суженной группе голономии $\Phi_0(p)$;}\newline
\indent 2) \parbox[t]{.92\linewidth}{каждая точка $x=\pi(p)$ имеет окрестность
$\MU$ такую, что $\Phi_{\rm loc}(p)=\Phi_0(p,\MV)$ для любой окрестности
$\MV\ni x$, содержащейся в $\MU$;}\newline
\indent 3) \parbox[t]{.92\linewidth}{если $\MU$ -- окрестность точки $x=\pi(p)$,
о которой говорится в свойстве 2), то
$\Phi_{\rm loc}(p)\supset\Phi_{\rm loc}(q)$ для всех точек
$q\in\MP(p,\MU)$;}\newline
\indent 4) \parbox[t]{.92\linewidth}{для каждого $a\in\MG$ справедливо
равенство}
\begin{equation*}
  \Phi_{\rm loc}(pa)=\ad(a^{-1})\Phi_{\rm loc}(p);
\end{equation*}
\indent 5) \parbox[t]{.92\linewidth}{для каждого целого $m$ множество точек базы
\begin{equation*}
  \lbrace\pi(p)=x\in\MM:~\dim\Phi_{\rm loc}(p)\le m\rbrace
\end{equation*}
открыто в $\MM$.}
\end{prop}
\begin{proof}
Свойства 1)--4) очевидны. Докажем свойство 5). Из свойства 4) следует, что
функция $\dim\Phi_{\rm loc}(p)$ постоянна на каждом слое и поэтому ее можно
рассматривать как функцию на базе $\MM$, принимающую целые значения. Из свойств
3) и 4) вытекает, что
\begin{equation*}
  \dim\Phi_\loc(q)\le\dim\Phi_\loc(p),
\end{equation*}
для всех точек $q\in\pi^{-1}(\MU)$. Отсюда вытекает свойство 5).
\end{proof}
\begin{theorem}                                                   \label{trehol}
Пусть $\Gg_0(p)$ и $\Gg_{\rm loc}(p)$ -- алгебры Ли для групп голономий
$\Phi_0(p)$ и $\Phi_{\rm loc}(p)$ соответственно. Тогда $\Phi_0(p)$ и $\Gg_0(p)$
порождаются соответственно всеми $\Phi_{\rm loc}(q)$ и $\Gg_{\rm loc}(q)$ для
всех точек $q\in\MP(p)$.
\end{theorem}
\begin{proof}
См., например, \cite{KobNom6369R}.
\end{proof}
\begin{theorem}                                                   \label{tcondi}
Если $\dim\Phi_{\rm loc}(p)$ постоянна на главном расслоении $\MP(\MM,\pi,\MG)$,
то локальная и суженная группы голономии совпадают,
$\Phi_{\rm loc}(p)=\Phi_0(p)$, для всех $p\in\MP$.
\end{theorem}
\begin{proof}
По свойству 3) предложения \ref{prohlo} каждая точка $x=\pi(p)$ имеет
окрестность $\MU$ такую, что $\Phi_{\rm loc}(p)\supset\Phi_{\rm loc}(q)$ для
каждой точки $q$ из расслоения голономии $\MP(p,\MU)$. Так как
$\dim\Phi_\loc(p)=\dim\Phi_\loc(q)$, то сами группы голономии в различных
точках совпадают, $\Phi_\loc(p)=\Phi_\loc(q)$. Отсюда следует, что если
$q\in\MP(p)$, то $\Phi_\loc(p)=\Phi_\loc(q)$ для всех $p\in\MP(\MM,\pi,\MG)$.
Из теоремы \ref{trehol} вытекает равенство $\Phi_0(p)=\Phi_\loc(p)$.
\end{proof}

Теперь перейдем к определению инфинитезимальной группы голономии и изучим ее
связь с локальной группой голономии. Инфинитезимальная группа голономии может
быть определена только для гладких $\CC^\infty$ главных расслоений
$\MP(\MM,\pi,\MG)$ с гладкой $\CC^\infty$ связностью $\Gamma$. В дальнейшем мы будем
считать, что условие гладкости выполнено.

Инфинитезимальная группа голономии в точке главного расслоения
$p\in\MP(\MM,\pi,\MG)$ определяется при помощи формы кривизны $R$ связности
$\Gamma$, заданной на $\MP$. Сначала определим индуктивно семейство подпространств
$\Gm_k(p)$, $k=0,1,2,\dotsc$, в алгебре Ли $\Gg$ структурной группы $\MG$. Пусть
$\Gm_0(p)$ подпространство в $\Gg$, порожденное всеми элементами вида
$R_p(X,Y)$, где $X,Y$ -- произвольные горизонтальные векторы в точке $p\in\MP$.
Рассмотрим $\Gg$-значную функцию на $\MP$ вида
\begin{equation}                                                  \label{edecuh}
  f_k=Z_k\dotsc Z_1\big(R(X,Y)\big),
\end{equation}
где $X,Y,Z_1,\dotsc,Z_k$ -- произвольные горизонтальные векторные поля на $\MP$
и векторы $Z_1,\dotsc,Z_k$ действуют на $\Gg$-значную функцию $R(X,Y)$ как
дифференцирования. Пусть $\Gm_k(p)$ -- подпространство в алгебре Ли $\Gg$,
порожденное подпространством $\Gm_{k-1}$ и значениями в точке $p$ всех функций
$f_k$ вида (\ref{edecuh}). Положим
\begin{equation}                                                  \label{elihoi}
  \Gg_\nf(p)=\bigcup_{k=0}^\infty\Gm_k(p).
\end{equation}
По сути дела, это то подмножество в алгебре Ли структурной группы, которое
порождается формой кривизны и всеми ее производными.
\begin{prop}                                                      \label{psualh}
Подпространство $\Gg_\nf(p)$ в $\Gg$ есть подалгебра Ли в алгебре Ли
$\Gg_\loc(p)$ локальной группы голономии.
\end{prop}
\begin{proof}
См.\ \cite{Ozeki56}.
\end{proof}
Это предложение позволяет ввести новое понятие.
\begin{defn}
Связная подгруппа Ли $\Phi_\nf(p)$ в структурной группе $\MG$, порожденная
подалгеброй $\Gg_\nf(p)$ называется {\em инфинитезимальной группой голономии}
связности $\Gamma$ в точке $p\in\MP(\MM,\pi,\MG)$.
\qed\end{defn}                                                    \label{pinfpr}
\index{Инфинитезимальная группа голономии (infinitesimal holonomy group)}%
\index{Группа голономии инфинитезимальная (infinitesimal holonomy group)}%
\begin{prop}
Инфинитезимальная группа голономии $\Phi_\nf(p)$ имеет следующие свойства:
\newline
\indent 1) \parbox[t]{.92\linewidth}{$\Phi_\nf(p)$ является связной подгруппой
Ли локальной группы голономии $\Phi_\loc(p)$;}\newline
\indent 2) \parbox[t]{.92\linewidth}{$\Phi_\nf(pa)=\ad(a^{-1})\Phi_\nf(p)$ и
$\Gg_\nf(pa)=\ad(a^{-1})\Gg_\nf(p)$;}\newline
\indent 3) \parbox[t]{.92\linewidth}{для каждого целого $m$ множество точек базы
\begin{equation}                                                  \label{eprhom}
  \lbrace\pi(p)=x\in\MM:\quad \dim\Phi_\nf(p)\ge m\rbrace
\end{equation}
открыто в $\MM$;}\newline
\indent 4) \parbox[t]{.92\linewidth}{если $\Phi_\nf(p)=\Phi_\loc(p)$ в точке
$p$, то существует окрестность $\MU$ точки $x=\pi(p)$ такая, что
\begin{equation*}
  \Phi_\nf(q)=\Phi_\loc(q)=\Phi_\nf(p)=\Phi_\loc(p)
\end{equation*}
для всех $q\in\MU$.}
\end{prop}
\begin{proof}
Свойство 1) следует из предложения \ref{psualh}.

Свойство 2) вытекает из интуитивно понятного равенства
\begin{equation*}
  \Gm_k(pa)=\ad(a^{-1})\Gm_k(p),\qquad \forall k.
\end{equation*}
Детали доказательства приведены в \cite{KobNom6369R}.

3). Размерность инфинитезимальной группы голономии $\dim\Phi_\nf(p)$ в силу
свойства 2) можно рассматривать как функцию на $\MM$ со значениями в целых
числах. Если значения конечного числа функций $f_k$ вида (\ref{edecuh}) линейно
независимы в точке $p$, то они таковы и в любой точке из некоторой окрестности
точки $p$. Поэтому, если свойство (\ref{eprhom}) выполнено в точке $x$, то оно
имеет место и в некоторой окрестности $x$.

4). Допустим, что $\Phi_\nf(p)=\Phi_\loc(p)$. Из свойства 3) предложения
\ref{pinfpr} и свойства 5) предложения \ref{prohlo} следует, что точка
$x=\pi(p)$ имеет окрестность $\MU$ такую, что
\begin{equation*}
  \dim\Phi_\nf(q)\ge\dim\Phi_\nf(p)\qquad \text{и}\qquad
  \dim\Phi_\loc(q)\le\dim\Phi_\loc(p),\qquad \forall q\in\pi^{-1}(\MU).
\end{equation*}
С другой стороны, $\Phi_\loc(q)\supset\Phi_\nf(q)$ для каждого
$q\in\pi^{-1}(\MU)$. Отсюда
\begin{equation*}
  \dim\Phi_\loc(q)=\dim\Phi_\nf(q)=\dim\Phi_\loc(p)=\dim\Phi_\nf(p)
\end{equation*}
и, следовательно, $\Phi_\loc(q)=\Phi_\nf(q)$ для каждого $q\in\pi^{-1}(\MU)$.
Применяя теорему \ref{tcondi} к индуцированному расслоению $\MP|_\MU$, видим,
что $\Phi_0(p,\MU)=\Phi_\loc(p)$ и $\Phi_0(q,\MU)=\Phi_\loc(q)$. Если
$q\in\MP(p,\MU)$, то $\Phi_0(p,\MU)=\Phi_0(q,\MU)$. Поэтому
$\Phi_\loc(p)=\Phi_\loc(q)$.
\end{proof}
\begin{theorem}                                                   \label{tinfho}
Если размерность инфинитезимальной группы голономии $\dim\Phi_\nf(q)$ постоянна
в некоторой окрестности точки $p\in\MP(\MM,\pi,\MG)$, то локальная и
инфинитезимальная группы голономий в точке $p$ совпадают,
$\Phi_\loc(p)=\Phi_\nf(p)$.
\end{theorem}
\begin{proof}
См.\ \cite{Ozeki56}
\end{proof}
\begin{cor}
Если размерность инфинитезимальной группы голономии $\dim\Phi_\nf(p)$ постоянна
на всем главном расслоении $\MP(\MM,\pi,\MG)$, то
\begin{equation}                                                  \label{eqhigr}
  \Phi_0(p)=\Phi_\loc(p)=\Phi_\nf(p)
\end{equation}
для всех $p\in\MP$.
\qed\end{cor}
\begin{proof}
Вытекает из теорем \ref{tcondi} и \ref{tinfho}.
\end{proof}
\begin{theorem}
Для вещественно аналитической связности $\Gamma$ на вещественно аналитическом
главном расслоении $\MP(\MM,\pi,\MG)$ следующие группы голономии равны
\begin{equation*}
  \Phi_0(p)=\Phi_\loc(p)=\Phi_\nf(p)
\end{equation*}
для всех $p\in\MP$.
\end{theorem}
\begin{proof}
См.\ \cite{Ozeki56}.
\end{proof}
Равенство групп голономий (\ref{eqhigr}) позволяет вычислить суженную группу
голономии $\Phi_0(p)$. Действительно, алгебра Ли инфинитезимальной группы
голономии (\ref{elihoi}) порождена всеми функциями вида (\ref{edecuh}).
Пусть $x^\al$, $\al=1,\dotsc,n$, -- система координат в
окрестности $\MU$ точки $x=\pi(p)$ и $\s:~\MU\rightarrow\MP$ -- локальное
сечение. Тогда алгебра  инфинитезимальной группы голономии порождается
значениями компонент локальной формы кривизны $F_{\al\bt}{}^\Sa$ и всех ее
ковариантных производных:
\begin{equation*}
  F_{\al\bt}{}^\Sa,\quad ~\nb_{\g_1}F_{\al\bt}{}^\Sa,\quad ~
  \nb_{\g_2}\nb_{\g_1}F_{\al\bt}{}^\Sa,\quad ~\dotsc
\end{equation*}
Ясно, что подпространство $\Gm_0\subset\Gg$ порождено всеми компонентами тензора
кривизны $F_{\al\bt}{}^\Sa$, подпространство $\Gm_1\subset\Gg$ -- всеми
компонентами тензора кривизны $F_{\al\bt}{}^\Sa$ и их первых ковариантных
производных $\nb_{\g_1}F_{\al\bt}{}^\Sa$ и так далее. Таким образом, при
заданной связности на главном расслоении, инфинитезимальная группа голономии
позволяет в принципе вычислить суженную группу голономии. Конечно, после этого
необходимо проверить, что условие следствия выполнено.
\section{Инвариантные связности                                           }
Прежде чем рассматривать инвариантные связности общего вида, мы опишем важный
частный случай.
\begin{theorem}                                                   \label{tinvsu}
Пусть $\MG$ -- связная группа Ли, и $\MH$ -- ее замкнутая подгруппа Ли. Пусть
$\Gg$ и $\Gh$ -- алгебры Ли для $\MG$ и $\MH$ соответственно.\newline
\indent 1) \parbox[t]{.92\linewidth}{Если существует линейное подпространство
$\Gm$ в $\Gg$ такое, что $\Gg=\Gh\oplus\Gm$ и $\ad(\MH)\Gm=\Gm$, то
$\Gh$-компонента $\om$ канонической 1-формы $\theta$ в $\MG$ определяет
связность на главном расслоении $\MG(\MG/\MH,\pi,\MH)$, инвариантную
относительно действия левых сдвигов из $\MG$.}\newline
\indent 2) \parbox[t]{.92\linewidth}{Обратно, любая связность на главном
расслоении $\MG(\MG/\MH,\pi,\MH)$, инвариантная относительно действия левых
сдвигов из $\MG$ (если она существует), определяет разложение $\Gg=\Gh\oplus\Gm$
и может быть получена так, как это описано в пункте 1).}\newline
\indent 3) \parbox[t]{.92\linewidth}{Форма кривизны $R$ инвариантной связности,
определенная формой $\om$ из пункта 1), равна
\begin{equation*}
  R(X,Y)=-\frac12[X,Y]_\Gh,
\end{equation*}
где $X,Y\in\Gm$ -- произвольные левоинвариантные векторные поля на $\MG$ из
$\Gm$ и в правой части равенства взята $\Gh$-компонента коммутатора.}\newline
\indent 4) \parbox[t]{.92\linewidth}{Пусть $\Gg(e)$ -- алгебра Ли группы
голономии $\Phi(e)$ в единице $e$ группы Ли $\MG$ для инвариантной связности,
определенной в пункте 1). Тогда $\Gg(e)$ порождается всеми элементами вида
$[X,Y]_\Gh$, где $X,Y\in\Gm$.}
\end{theorem}
\begin{proof}
1). Пусть $Z^*$ -- фундаментальное векторное поле, соответствующее элементу
подалгебры $Z\in\Gh$. Из определения канонической 1-формы (см.\ раздел
\ref{sleacg}) следует, что $\om(Z^*)=\theta(Z^*)=Z$. Пусть $\theta_\Gm$ есть
$\Gm$-компонента канонической формы $\theta$. Для любого $a\in\MH$ и
$X\in\MT_p(\MG)$ справедливы равенства
\begin{align*}
  \theta(r_{a*}X)&=\om(r_{a*}X)+\theta_\Gm(r_{a*}X),
\\
  \ad(a^{-1})\theta(X)&=\ad(a^{-1})\om(X)+\ad(a^{-1})\theta_\Gm(X).
\end{align*}
Левые части этих равенств совпадают. Поскольку по условию теоремы
$\ad(a^{-1})\Gm=\Gm$, то сравнение $\Gh$-компонент правых частей приводит к
равенству
\begin{equation*}
  \om(r_{a*}X)=\ad(a^{-1})\om(X),
\end{equation*}
т.е.\ 1-форма $\om$ определяет связность на $\MG$. Эта связность инвариантна
относительно действия группы слева по построению.

2). Пусть $\om$ -- форма связности на главном расслоении $\MG(\MG/\MH,\pi,\MH)$,
инвариантная относительно действия левых сдвигов из $\MG$. Пусть $X\in\Gm$ --
множество левоинвариантных векторных полей на $\MG$ таких, что $\om(X)=0$. Тогда
алгебра Ли на $\MG$ разлагается в прямую сумму, $\Gg=\Gh\oplus\Gm$.

3). Левоинвариантное векторное поле горизонтально тогда и только тогда, когда
оно лежит в $\Gm$. Поэтому утверждение 3) следует из равенства (\ref{econho}).

4). Пусть $\Gg_1$ -- подпространство в $\Gg$, порожденное множеством элементов
вида $R_e(X,Y)$, где $X,Y\in\Gm$.  Пусть $\Gg_2$ -- подпространство в $\Gg$,
порожденное множеством элементов $R_a(X,Y)$, где $X,Y\in\Gm$ для всех $a\in\MG$.
Тогда по теореме Амброза--Зингера $\Gg_1\subset\Gg(e)\subset\Gg_2$. С другой
стороны, $\Gg_1=\Gg_2$, т.к.\ $R_a(X,Y)=R_e(X,Y)$ для любых $X,Y\in\Gm$ и
$a\in\MG$. Теперь 4) следует из 3).
\end{proof}
\begin{com}
Линейное подпространство алгебры Ли $\Gm\subset\Gg$ представляет собой
распределение горизонтальных векторных полей на $\MG$, т.е.\ связность $\Gamma$ на
главном расслоении $\MG(\MG/\MH,\pi,\MH)$, инвариантную относительно действия
группы слева. В общем случае $\Gm\subset\Gg$ является только линейным
подпространством, а не подалгеброй.
\qed\end{com}
\begin{com}
Утверждение 1) теоремы \ref{tinvsu} можно рассматривать как частный случай
предложения \ref{predco}. Пусть $\MP=(\MG/\MH)\times\MG$ -- тривиальное главное
расслоение над базой $\MG/\MH$ со структурной группой $\MG$. Вложим расслоение
$\MG(\MG/\MH,\pi,\MH)$ в $\MP$ при помощи отображения
\begin{equation*}
  f(a)=\big(\pi(a),a\big),\qquad a\in\MG,
\end{equation*}
где $\pi:~\MG\rightarrow\MG/\MH$ -- естественная проекция. Пусть $\phi$ --
форма связности, определяющая каноническую плоскую связность на $\MP$. Ее
$\Gh$-компонента, суженная на подрасслоение $\MG(\MG/\MH,\pi,\MH)$, по
предложению \ref{predco} определяет связность и совпадает с формой связности
$\om$ в утверждении 1).
\qed\end{com}

Возвращаясь к общему случаю, сначала докажем предложение, которое является
основой для многих приложений.
\begin{prop}                                                      \label{ponepa}
Пусть $s_t$ -- однопараметрическая группа автоморфизмов главного расслоения
$\MP(\MM,\pi,\MG)$ и $\widetilde X$ -- векторное поле на $\MP$, индуцированное
$s_t$. Пусть $\Gamma$ -- связность на $\MP$, инвариантная относительно действия
$s_t$. Для произвольной точки главного расслоения $p_0\in\MP$ определим четыре
кривые $p_t$, $x_t$, $\tilde x_t$ и $a_t$ следующим образом:
\begin{equation*}
  p_t:=s_t(p_0)\in\MP,\qquad x_t:=\pi(p_t)\in\MM,
\end{equation*}
$\tilde x_t$ есть горизонтальный лифт $x_t$ такой, что $\tilde x_0=p_0$ и
\begin{equation}                                                  \label{eonpgr}
  p_t:=\tilde x_t a_t,\qquad a_t\in\MG.
\end{equation}
Тогда $a_t$ является однопараметрической подгруппой в структурной группе $\MG$,
порожденной элементом $X=\om_{p_0}(\widetilde X)$, где $\om$ -- форма связности
для $\Gamma$.
\end{prop}
\begin{proof}
Поскольку
\begin{equation*}
  \dot p_t=r_{a_t*}\dot{\tilde x}_t+\tilde x_{t*}\dot a_t,
\end{equation*}
то
\begin{equation*}
  \om(\dot p_t)=\ad(a_t^{-1})\om(\dot{\tilde x}_t)+a_{t*}^{-1}\dot a_t,
\end{equation*}
где $\tilde x_{t*}$ и $a_{t*}$ -- дифференциалы отображений
$\tilde x_t:~\MG\rightarrow\pi^{-1}(x_t)$ и $a_t:~\MG\rightarrow\MG$
соответственно. Первое слагаемое в правой части равно нулю, т.к.\ кривая
$\tilde x_t$ горизонтальна. Следовательно, $\om(\dot p_t)=a_{t*}^{-1}\dot a_t$.
С другой стороны, $\dot p_t=s_{t*}\widetilde X_{p_0}$, где $s_{t*}$ --
дифференциал отображения $s_t:~\MP\rightarrow\MP$, и поэтому
$\om(\dot p_t)=\om(\widetilde X_{p_0})=X$, т.к.\ форма связности инвариантна
относительно однопараметрической группы преобразований $s_t$. Отсюда следует
равенство $X=a_{t*}^{-1}\dot a_t$.
\end{proof}
\begin{defn}
Кривая $a_t$ в структурной группе Ли $\MG$ из условия предложения \ref{ponepa}
называется {\em разверткой} кривой $p_t$ в главном расслоении
$\MP(\MM,\pi,\MG)$.
\qed\end{defn}
\index{Развертка кривой}%
\begin{com}
В этом определении кривая $p_t$ может быть произвольной кривой в главном
расслоении $\MP(\MM,\pi,\MG)$, не обязательно связанной с группой симметрии
связности.
\qed\end{com}
\begin{defn}
Пусть $\MK$ -- группа Ли автоморфизмов главного расслоения $\MP(\MM,\pi,\MG)$ с
алгеброй Ли $\Gk$. Выберем в главном расслоении опорную точку $p_0\in\MP$.
Каждый элемент из $\MK$ индуцирует некоторое преобразование базы при помощи
проекции. Множество $\MJ$ всех элементов из $\MK$, которые оставляют неподвижной
точку $x_0=\pi(p_0)$, образуют замкнутую подгруппу в $\MK$, которая называется
{\em подгруппой изотропии} в $\MK$ для точки $x_0\in\MM$.
\index{Подгруппа изотропии (isotropy subgroup)}%
\index{Изотропии подгруппа (isotropy subgroup)}%
Определим гомоморфизм групп Ли
\begin{equation}                                                  \label{elmapa}
  \lm:\quad \MJ\ni\quad j\mapsto \lm(j)=a\quad\in\MG
\end{equation}
следующим образом. Каждый автоморфизм $j\in\MJ$ переводит $p_0$ в точку
$j(p_0)$, которая принадлежит тому же слою $\pi^{-1}\big(\pi(p_0)\big)$, т.к.\
точка $x_0$ неподвижна. Следовательно, $j(p_0)=p_0a$ для некоторого $a\in\MG$.
Положим $\lm(j)=a$. Тогда для двух элементов подгруппы изотропии,
$j_1,j_2\in\MJ$, справедливы равенства
\begin{equation*}
  p_0\lm(j_1j_2)=(j_1j_2)(p_0)=j_1\big(p_0\lm(j_2)\big)
  =\big(j_1(p_0)\big)\lm(j_2)=p_0\lm(j_1)\lm(j_2).
\end{equation*}
Тем самым $\lm(j_1j_2)=\lm(j_1)\lm(j_2)$ и, следовательно, построенное
отображение является гомоморфизмом групп. Нетрудно проверить, что отображение
$\lm:~\MJ\rightarrow\MG$ дифференцируемо. Гомоморфизм групп $\lm$ индуцирует
гомоморфизм алгебр Ли,
\begin{equation*}
  \lm:\quad \Gj\rightarrow\Gg,
\end{equation*}
который мы обозначили той же буквой.
\qed\end{defn}

\begin{com}
Отображение (\ref{elmapa}) зависит от выбора опорной точки $p_0\in\MP$. В
настоящем разделе мы будем считать, что опорная точка $p_0$ выбрана и
зафиксирована.
\qed\end{com}

\begin{prop}                                                      \label{pdeflm}
Пусть $\MK$ -- группа Ли автоморфизмов главного расслоения $\MP(\MM,\pi,\MG)$ с
алгеброй Ли $\Gk$ и $\Gamma$ -- связность на $\MP$ с формой связности $\om$ и
формой кривизны $R$, инвариантная относительно автоморфизмов $\MK$. Пусть
$\MJ\subset\MK$ -- подгруппа изотропии в $\MK$ для точки $x_0=\pi(p_0)$ с
алгеброй Ли $\Gj$. Определим линейное отображение
\begin{equation*}
  \Lm:\quad \Gk\ni\quad X\mapsto\Lm(X)=\om_{p_0}(\widetilde X)\quad\in\Gg,
\end{equation*}
где $\widetilde X$ -- векторное поле на $\MP$, индуцированное полем $X$. Тогда:
\newline
\indent 1) \parbox[t]{.92\linewidth}{$\Lm(X)=\lm(X)$, $\forall X\in\Gj$;}
\newline
\indent 2) \parbox[t]{.92\linewidth}{
$\Lm\big(\ad(j)X\big)=\ad\big(\lm(j)\big)\Lm(X)$, $\forall j\in\MJ$ и
$\forall X\in\Gk$,
где $\ad(j)$ обозначает присоединенное представление подгруппы изотропии $\MJ$ в
$\Gk$ и $\ad\big(\lm(j)\big)$ -- присоединенное представление структурной группы
$\MG$ в $\Gg$;}\newline
\indent 3) \parbox[t]{.92\linewidth}{$2R_{p_0}(\widetilde X,\widetilde Y)
=[\Lm(X),\Lm(Y)]-\Lm\big([X,Y]\big),\qquad \forall X,Y\in\Gk$.}
\end{prop}
\begin{proof}
См., например, \cite{KobNom6369R}.
\end{proof}
\begin{com}
Геометрический смысл отображения $\Lm$ дается предложением \ref{ponepa}.
$\Lm(X)$ -- это тот элемент алгебры Ли $\Gg$, который порождает
однопараметрическую подгруппу $a_t$ в структурной группе $\MG$, определенную
равенством (\ref{eonpgr}).
\qed\end{com}
\begin{com}
Отображение $\Lm$ в предложении \ref{pdeflm} является только линейным. В общем
случае оно не является гомоморфизмом алгебр Ли.
\qed\end{com}
\begin{defn}
Группа автоморфизмов $\MK$ действует на главном расслоении $\MP(\MM,\pi,\MG)$
{\em слой-транзитивно}, если для любых двух слоев из $\MP$ существует элемент в
$\MK$, отображающий один слой в другой, т.е.\ если действие $\MK$ на базе $\MM$
транзитивно.
\qed\end{defn}
\index{Слой-транзитивное действие (cross-section transitive action)}%
\index{Действие слой-транзитивное (cross-section transitive action)}%
\begin{prop}
Если $\MJ$ -- подгруппа изотропии для слой-транзитивной группы автоморфизмов
$\MK$ в точке $x_0=\pi(p_0)$, то база $\MM$ является однородным пространством,
$\MM\approx\MK/\MJ$.
\end{prop}
\begin{proof}
Вытекает из теоремы \ref{tisogh}.
\end{proof}

Описание $\MK$-инвариантных связностей на главном расслоении в случае
слой-транзитивного действия групп автоморфизмов дает
\begin{theorem}                                                   \label{twangt}
Если связная группа Ли $\MK$ является слой-транзитивной группой автоморфизмов
главного расслоения $\MP(\MM,\pi,\MG)$ и если $\MJ$ -- подгруппа изотропии в
$\MK$ для точки $x_0=\pi(p_0)$, то существует взаимно однозначное соответствие
между множеством $\MK$-инвариантных связностей на $\MP$ и множеством линейных
отображений $\Lm:~\Gk\rightarrow\Gg$, которые удовлетворяют условиям 1) и 2)
предложения \ref{pdeflm}. Соответствие задается следующим образом:
\begin{equation*}
  \Lm:\quad \Gk\ni\quad X\mapsto\Lm(X)=\om_{p_0}(\widetilde X)\quad\in\Gg,
\end{equation*}
где $\widetilde X$ -- векторное поле на $\MP$, индуцированное полем $X$.
\end{theorem}
\begin{proof}
См.\ \cite{Wang58}.
\end{proof}
\begin{com}
Из теоремы \ref{troloc} следует, что связность на главном расслоении
$\MP(\MM,\pi,\MG)$ однозначно определяется заданием семейства локальных форм
связности на произвольном покрытии базы $\MM$. Это происходит потому что при
помощи действия структурной группы распределение горизонтальных подпространств с
локальных сечений можно разнести на все пространство главного расслоения $\MP$.
Если известно, что связность инвариантна также относительно слой-транзитивных
автоморфизмов $\MK$, то локальную форму связности достаточно задать в одной
точке базы $\MM$. Действительно, при помощи автоморфизмов $\MK$ она разносится
по всей базе $\MM$, а затем, действуя структурной группой, ее можно разнести по
всему пространству расслоения $\MP$.

Если группа автоморфизмов $\MK$ не является слой-транзитивной, тогда в задании
инвариантной связности появляется значительный произвол. В этом случае форму
связности можно определить на каждой орбите действия группы $\MK$ на $\MP$. Для
этого достаточно задать ее в какой либо одной точке на каждой орбите и при
помощи $\MK$ разнести ее по орбитам и, наконец, по всему пространству главного
расслоения $\MP$, действуя структурной группой $\MG$. После этого необходимо
проверить дифференцируемость полученного распределения горизонтальных
подпространств.
\qed\end{com}
Если группа автоморфизмов $\MK$ действует на $\MP(\MM,\pi,\MG)$
слой-транзитивно, то форма кривизны $R$, которая является тензориальной формой
типа $\ad\MG$, инвариантная относительно $\MK$, полностью определяется своими
значениями в опорной точке $R_{p_0}(\widetilde X,\widetilde Y)$, где
$\widetilde X$ и $\widetilde Y$ -- векторные поля на $\MP$, индуцированные
элементами $X,Y\in\Gk$ алгебры Ли группы автоморфизмов $\MK$. В этом случае
утверждение 3) предложения \ref{pdeflm} выражает форму кривизны
$R_{p_0}(\widetilde X,\widetilde Y)$ в терминах $\Lm$.

Из предложения \ref{pdeflm} и теорем \ref{tflzec} и \ref{twangt} получаем
\begin{cor}
$\MK$-инвариантная связность на главном расслоении $\MP(\MM,\pi,\MG)$,
определенная отображением $\Lm$, является плоской тогда и только тогда, когда
отображение $\Lm:~\Gk\rightarrow\Gg$ есть гомоморфизм алгебр Ли.
\qed\end{cor}
\begin{theorem}                                                   \label{tcainc}
Допустим, что в теореме \ref{twangt} алгебра Ли $\Gk$ содержит линейное
подпространство $\Gm$ такое, что $\Gk=\Gj\oplus\Gm$ и $\ad(\MJ)\Gm=\Gm$, где
$\ad(\MJ)$ -- присоединенное представление $\MJ$ в $\Gk$. Тогда:\newline
\indent 1) \parbox[t]{.92\linewidth}{Существует взаимно однозначное соответствие
между множеством $\MK$-инвариантных связностей на главном расслоении
$\MP(\MM,\pi,\MG)$ и множеством линейных отображений
$\Lm_\Gm:~\Gm\rightarrow\Gg$ таких, что
\begin{equation*}
  \Lm_\Gm\big(\ad(j)X\big)=\ad\big(\lm(j)\big)\Lm_\Gm(X),\qquad \forall X\in\Gm~
  \text{и}\quad \forall j\in\MJ;
\end{equation*}
соответствие задается теоремой \ref{twangt} следующим образом
\begin{equation*}
  \Lm(X)=\begin{cases} \lm(X), & \text{если}\quad  X\in\Gj, \\
          \Lm_\Gm(X), & \text{если}\quad X\in\Gm.\end{cases}
\end{equation*}  }\newline
\indent 2) \parbox[t]{.92\linewidth}{Форма кривизны $R$ для $\MK$-инвариантной
связности, определяемой при помощи отображения $\Lm_\Gm$, удовлетворяет
следующему равенству
\begin{equation*}
  2R_{p_0}(\widetilde X,\widetilde Y)=[\Lm_\Gm(X),\Lm_\Gm(Y)]
  -\Lm_\Gm\big([X,Y]_\Gm\big)-\lm\big([X,Y]_\Gj\big),\qquad \forall X,Y\in\Gm,
\end{equation*}
где $[X,Y]_\Gm$ и $[X,Y]_\Gj$ обозначают соответственно $\Gm$- и
$\Gj$-компоненту коммутатора $[X,Y]\in\Gk$.}
\end{theorem}
\begin{proof}
Пусть $\Lm:~\Gk\rightarrow\Gg$ -- линейное отображение, удовлетворяющее
утверждениям 1) и 2) предложения \ref{pdeflm}. Пусть $\Lm_\Gm$ -- сужение
отображения $\Lm$ на $\Gm$. Нетрудно проверить, что отображение
$\Lm\rightarrow\Lm_\Gm$ является взаимно однозначным и согласно теореме
\ref{twangt} дает желаемое соответствие. Утверждение 2) следует из утверждения
3) предложения \ref{pdeflm}.
\end{proof}
\begin{defn}
В теореме \ref{tcainc} $\MK$-инвариантная связность на главном расслоении
$\MP(\MM,\pi,\MG)$, определяемая условием $\Lm_\Gm=0$ называется
{\em канонической инвариантной связностью} относительно разложения
$\Gg=\Gj\oplus\Gm$.
\qed\end{defn}
\index{Каноническая инвариантная связность (canonical invariant connection}%
\index{Связность каноническая инвариантная (canonical invariant connection}%

Следующая теорема определяет алгебру Ли группы голономии $\MK$-инвариантной
связности.
\begin{theorem}                                                   \label{thoinv}
В предположениях и обозначениях теоремы \ref{twangt} алгебра Ли $\Gg(p_0)$
группы голономии $\Phi(p_0)$ для $\MK$-инвариантной связности, определяемой при
помощи линейного отображения $\Lm:~\Gk\rightarrow\Gg$, задается суммой
\begin{equation*}
  \Gm_0+[\Lm(\Gk),\Gm_0]+\big[\Lm(\Gk),[\Lm(\Gk),\Gm_0]\big]+\dotsc,
\end{equation*}
где $\Gm_0$ -- линейное подпространство в $\Gg$, порожденное множеством
\begin{equation*}
  \lbrace[\Lm(X),\Lm(Y)]-\Lm([X,Y])\rbrace,\qquad \forall X,Y\in\Gk.
\end{equation*}
\end{theorem}
\begin{proof}
См., например, \cite{KobNom6369R}.
\end{proof}
\begin{com}
Утверждения 1) и 3) теоремы \ref{tinvsu} следуют из теоремы \ref{tcainc}, если
в качестве главного расслоения $\MP(\MM,\pi,\MG)$ выбрать $\MG(\MG/\MH,\pi,\MH)$
и положить $\MK=\MG$. Тогда инвариантная связность из теоремы \ref{tinvsu}
является канонической инвариантной связностью. Утверждение 4) теоремы
\ref{tinvsu} следует из теоремы \ref{thoinv}.
\qed\end{com}
\chapter{Приложения в квантовой механике}
В настоящей главе рассмотрены некоторые приложения дифференциальной геометрии в
нерелятивистской квантовой механике. Нетривиальные геометрические структуры, а
речь идет о нетривиальной связности на главном расслоении, часто возникают при
решении уравнений математической физики. В настоящей главе будет показано, как
возникает нетривиальная связность на главном расслоении со структурной группой
$\MU(1)$ или $\MU(\Sn)$ в нерелятивистской квантовой механике при решении
уравнения Шредингера. Удивительно не столько то, что главное расслоение
возникает естественным образом, а то, что предсказанные эффекты были
подтверждены экспериментально.

Сначала мы дадим геометрическую интерпретацию нерелятивистской квантовой
механике в конечномерном случае. Будет показано, что гамильтониан квантовой
системы задает компоненты локальной формы связности на главном расслоении, а
уравнение Шредингера определяет параллельный перенос слоев. При этом базой
является одномерное многообразие, соответствующее времени, а структурной группой
-- унитарная группа $\MU(\Sn)$, где $\Sn$ -- размерность гильбертова
пространства состояний квантовомеханической системы. Решение
квантовомеханической задачи не зависит от выбора базиса в гильбертовом
пространстве, и его выбирают из соображений удобства. Использование базиса,
состоящего из собственных векторов гамильтониана, позволяет упростить
доказательство адиабатической теоремы и сделать его более прозрачным.

В качестве приложения адиабатической теоремы рассмотрена фаза Берри
\cite{Berry84}. В заключительном разделе настоящей главы рассмотрен эффект
Ааронова--Бома \cite{AhaBoh59}, который, хотя и не имеет прямого отношения к
адиабатической теореме, с геометрической точки зрения аналогичен фазе Берри.

Эффект Ааронова--Бома и фаза Берри привлекают
большое внимание теоретиков и экспериментаторов в течении многих лет. Интерес
вызван двумя обстоятельствами. Во-первых, в обоих случаях при решении уравнения
Шредингера естественным образом возникает $\MU(1)$-связность. Во-вторых, в
теории калибровочных полей распространено мнение, что к наблюдаемым эффектам
может приводить только нетривиальная напряженность поля, а не сами потенциалы,
которые не являются калибровочно инвариантными. Вопреки этому мнению Ааронов и
Бом, а также Берри показали, что интеграл от калибровочного поля вдоль замкнутой
кривой может привести к наблюдаемым эффектам. Эти выводы вскоре были
подтверждены экспериментально.

Понятие фазы Берри было обобщено на неабелев случай, соответствующий вырожденным
уровням энергии гамильтониана, Вилчеком и Зи \cite{WilZee84}. В этом случае при
решении уравнения Шредингера естественным образом возникают неабелевы
калибровочные поля.

Во всех перечисленных выше случаях к наблюдаемым эффектам приводят элементы
группы голономии (см.\ раздел \ref{sholde}) соответствующих связностей. Элементы
группы голономии в общем случае являются ковариантными объектами, а для абелевой
группы $\MU(1)$ -- инвариантными. Мы покажем, что главное расслоение
может быть тривиальным, но связность, которая на нем возникает, в общем случае
имеет нетривиальную группу голономии и приводит к наблюдаемым эффектам. Отсюда
следует, что фаза Берри и эффект Ааронова--Бома имеют геометрическую природу.
\section{Адиабатическая теорема}
Адиабатическая теорема \cite{BorFoc28} занимает одно из центральных мест в
нерелятивистской квантовой механике, т.к.\ позволяет находить приближенное
решение уравнения Шредингера при медленном изменении гамильтониана во времени.
Первоначально она была доказана для дискретного (возможно, бесконечного) спектра
гамильтониана при некоторых ограничениях на возможное пересечение уровней
энергии \cite{BorFoc28}. Ниже приведено доказательство адиабатической теоремы в
наиболее простом конечномерном случае.

В нерелятивистской квантовой механике состояние системы описывается вектором
гильбертова пространства ({\em волновой функцией}) $\psi(t)\in\MH$, зависящим
от времени $t$ и некоторого набора других переменных, который
определяется рассматриваемой задачей. Эволюция квантовой системы во времени
описывается {\em уравнением Шредингера} \cite{Schrod26A,Schrod26B}
\begin{equation}                                                  \label{escheq}
  i\hbar\frac{\pl\psi}{\pl t}=H\psi,
\end{equation}
где $H$ -- самосопряженный линейный оператор, действующий в гильбертовом
пространстве $\MH$, который называется {\em гамильтонианом} системы, и $\hbar$
-- постоянная Планка.

Для уравнения Шредингера, как правило, ставится задача Коши с начальным условием
\begin{equation}                                                  \label{echaus}
  \psi(0)=\psi_0,
\end{equation}
где $\psi_0\in\MH$ -- некоторый фиксированный вектор гильбертова пространства.
\index{Уравнение Шредингера (Schr\"odinger equation)}%
\index{Шредингера уравнение (Schr\"odinger equation)}%
\index{Волновая функция (wave function)}%
\index{Функция волновая (wave function)}%
\index{Гамильтониан (Hamiltonian)}%

В дальнейшем, для простоты, положим $\hbar=1$ и обозначим частную производную по
времени, точкой, $\dot\psi:=\pl_t\psi$.

Предположим, для простоты, что гильбертово пространство представляет собой
конечномерное комплексное пространство $\MH=\MC^\Sn$ комплексной размерности
$\dim\MH=\Sn$. В гильбертовом пространстве задано скалярное произведение,
которое обозначим круглыми скобками,
\begin{equation*}
  \MH\times\MH\ni\quad\psi,\phi\mapsto(\psi,\phi)\quad\in\MC.
\end{equation*}
По определению, скалярное произведение линейно по первому аргументу $\psi$ и
выполнено равенство: $(\psi,\phi)^\dagger=(\phi,\psi)$, где символ $\dagger$
обозначает комплексное сопряжение. Квадрат вектора гильбертова пространства
$(\psi,\psi)$ является вещественным числом, при этом мы требуем, чтобы
квадратичная форма $(\psi,\psi)$ была строго положительно определена, т.е.\
$(\psi,\psi)\ge0$, причем $(\psi,\psi)=0$ тогда и только тогда, когда $\psi=0$.
Тогда скалярное произведение определяет норму вектора гильбертова пространства:
\begin{equation*}
  \|\psi\|:=\sqrt{(\psi,\psi)}.
\end{equation*}
Поскольку уравнение Шредингера линейно по $\psi$, а гамильтониан самосопряжен,
то норма произвольного решения уравнений Шредингера сохраняется во времени.
Отсюда следует, что векторы состояния можно нормировать. Обычно предполагается,
что векторы состояния нормированы на единицу,
\begin{equation}                                                  \label{enorco}
  \|\psi\|=1.
\end{equation}
Нормировка вектора состояния не устраняет полностью произвол в выборе вектора
гильбертова пространства, т.к.\ остается произвол в выборе постоянного фазового
множителя.
\begin{com}
Мы используем символ $\dagger$ для обозначения эрмитова сопряжения матриц, т.е.\
транспонирования матрицы и комплексного сопряжения всех элементов. В частном
случае, когда матрица состоит из одного элемента, эрмитово сопряжение совпадает
с комплексным.
\qed\end{com}

Пусть в гильбертовом пространстве $\MH$ выбран некоторый базис $e_k$,
$k=1,\dotsc,\Sn$. Тогда гамильтониан квантовомеханической системы задается
эрмитовой $\Sn\times\Sn$-матрицей, а вектор состояния $\psi=\psi^k(t)e_k$ --
строкой из $\Sn$ компонент,
\begin{equation*}
  \psi=(\psi^1,\dotsc,\psi^\Sn),
\end{equation*}
где $\psi^1(t),\dotsc,\psi^\Sn(t)$ -- комплекснозначные компоненты вектора. Если
базис гильбертова пространства ортонормирован,
\begin{equation*}
  (e_k,e_l)=\dl_{kl},
\end{equation*}
то скалярное произведение задается равенством
\begin{equation*}
  (\psi,\phi)=\psi\phi^\dagger
  =\psi^1\phi^\dagger_1+\dotsc+\psi^\Sn\phi^\dagger_\Sn.
\end{equation*}

Рассмотрим задачу Коши (\ref{escheq}), (\ref{echaus}) в общем случае, когда
гамильтониан системы зависит от времени $H=H(t)$. Для решения этой задачи
необходимо выбрать базис в гильбертовом пространстве $\MH$. Конечно, решение
задачи от выбора базиса не зависит, и его выбирают из соображений удобства.
Рассмотрим два случая.

Пусть базис $e_k\in\MH$, $k=1,\dotsc,\Sn$, ортонормирован и фиксирован,
$\dot e_k=0$. Произвольный вектор можно разложить по этому базису
$\psi=\psi^k e_k$. При этом гамильтониан задается эрмитовой
$\Sn\times\Sn$-матрицей $H_l{}^k$, а задача Коши для уравнения Шредингера в
компонентах примет вид системы обыкновенных дифференциальных уравнений с
некоторыми начальными условиями:
\begin{equation}                                                  \label{echfib}
\begin{split}
  i\dot\psi^k=\psi^lH_l{}^k,
\\
  \psi^k(0)=\psi_0^k.
\end{split}
\end{equation}
\begin{com}
Мы записываем действие гамильтониана в конечномерном случае справа, чтобы
согласовать наши обозначения с обозначениями, принятыми в дифференциальной
геометрии. Напомним, что в дифференциальной геометрии для векторного поля в
координатном базисе принята запись $X=X^\al\pl_\al$. (Суммирование с десяти до
четырех по циферблату часов.) Альтернативная запись $X=\pl_\al X^\al$
используется для обозначения дивергенции векторного поля. Преобразование
координат мы записываем в виде
\begin{equation*}
  X=X^\al\pl_\al=X^\al\frac{\pl x^{\al'}}{\pl x^\al}
  \frac{\pl x^\bt}{\pl x^{\al'}}\pl_\bt.
\end{equation*}
То есть действие матрицы преобразования координат записывается справа. С другой
стороны, действие операторов в квантовой механике общепринято записывать слева.
Поэтому, если мы хотим использовать единообразные обозначения, то необходимо чем
то пожертвовать. В конечномерном случае мы будем записывать матрицу, задающую
линейный оператор, справа. Это вопрос соглашения, и к нему легко привыкнуть.
\qed\end{com}

Рассмотрим теперь другой ортонормированный базис $b_k$, который может зависеть
от времени, $b_k=b_k(t)$. Такой базис может оказаться более удобным для решения
некоторых задач. Вектор гильбертова пространства $\psi$ можно разложить также
по этому базису $\psi=\psi^{\prime k}b_k$. Тогда задача Коши (\ref{echfib})
будет выглядеть по другому:
\begin{equation}                                                  \label{echneb}
\begin{split}
  i\dot\psi^{\prime k}=\psi^{\prime l}H^\prime_l{}^k,
\\
  \psi^{\prime k}(0)=\psi^{\prime k}_0,
\end{split}
\end{equation}
где $H^\prime_l{}^k$ -- компоненты гамильтониана относительно нового базиса,
которые будут определены ниже. Поскольку базисы ортонормированы, то они связаны
между собой некоторым унитарным преобразованием:
\begin{equation}                                                  \label{ebatra}
  b_k=S_k{}^l e_l,\qquad S\in\MU(\Sn),
\end{equation}
которое в общем случае зависит от времени, $S=S(t)$. При этом компоненты вектора
гильбертова пространства преобразуются с помощью обратной матрицы,
\begin{equation*}
  \psi^{\prime k}=\psi^lS^{-1}_{\quad l}{}^k.
\end{equation*}
Отсюда следует выражение для компонент начального вектора гильбертова
пространства $\psi^{\prime k}_0=\psi^l_0S^{-1}_{\quad l}{}^k(0)$. Переписав
уравнение Шредингера (\ref{echfib}) в базисе $b_k$, получим компоненты
гамильтониана относительно нового базиса:
\begin{equation}                                                  \label{enewhs}
  H'=SHS^{-1}+iS\dot S^{-1}=SHS^{-1}-i\dot SS^{-1},
\end{equation}
где мы, для краткости, опустили матричные индексы. Мы видим, что компоненты
гамильтониана преобразуются также, как компоненты локальной формы
$\MU(\Sn)$-связности (\ref{egatrd}).

Теперь можно дать геометрическую интерпретацию нерелятивистской квантовой
механике. Пусть время пробегает всю вещественную прямую, $t\in\MR$. Тогда
мы имеем главное расслоение
$\MP\big(\MR,\pi,\MU(\Sn)\big)\approx\MR\times\MU(\Sn)$ с базой $\MR$,
типичным слоем $\MU(\Sn)$ и проекцией $\pi:~\MP\rightarrow\MR$ (см.\ раздел
\ref{sprfib}). Это расслоение тривиально, т.к.\ базой является вещественная
прямая. Гамильтониан квантовой системы задает компоненты локальной формы
$\MU(\Sn)$-связности (1-форма на $\MR$ со значениями в алгебре Ли):
\begin{equation*}
  A_t=\big(iH_l{}^k\big)\in\Gu(\Sn),
\end{equation*}
где $t$ -- координатный ковариантный индекс, принимающий одно значение, который
раньше обозначался греческой буквой $\al$.
Вектор гильбертова пространства $\psi\in\MH$ -- это сечение ассоциированного
расслоения $\ME\big(\MR,\pi_\ME,\MH,\MU(\Sn),\MP\big)$, типичным слоем
которого является гильбертово пространство $\MH$. Уравнение Шредингера
имеет вид равенства нулю ковариантной производной,
\begin{equation*}
  \nb_t\psi=\dot\psi+\psi A_t=0,
\end{equation*}
т.е.\ задает параллельный перенос вектора гильбертова пространства. При
изменении сечения компоненты связности преобразуются по-правилу
\begin{equation*}
  A'_t=SA_tS^{-1}+\dot SS^{-1},
\end{equation*}
как и положено компонентам локальной формы связности (\ref{egatrd}). Кривизна
этой связности равна нулю, поскольку база одномерна. Поэтому связность является
плоской согласно теореме \ref{tflzec}.

Решение задачи Коши для уравнения Шредингера (\ref{escheq}), (\ref{echaus}) не
зависит от выбора базиса. Поэтому его выбирают из соображений удобства.
\begin{exa}
Пусть в фиксированном базисе $e_k$ компоненты вектора гильбертова пространства
имеют вид $\psi^k=\psi_0^lU_l{}^k$, где унитарная матрица $U_l{}^k(t)$ задает
оператор эволюции квантовой системы, который, по-определению, удовлетворяет
дифференциальному уравнению
\begin{equation*}
  i\dot U=UH,
\end{equation*}
с начальным условием $U_l{}^k(0)=\dl_l^k$. Тогда нетрудно проверить, что
оператор эволюции задает переход к такому базису гильбертова пространства
$b_k:=U^{-1}_{\quad k}{}^le_l$, в котором гамильтониан равен нулю, $H'=0$.
Следовательно, вектор гильбертова пространства, описывающий эволюцию квантовой
системы, в этом базисе имеет постоянные компоненты $\psi_0^k$, которые
определяются начальным состоянием.
\qed\end{exa}
\begin{com}
При преобразовании базиса, которое зависит от времени, эрмитова матрица,
соответствующая гамильтониану, испытывает калибровочное преобразование
(\ref{enewhs}). При этом преобразовании в общем случае собственные значения
матрицы меняются. Рассмотренный выше пример показывает, что если гамильтониан,
заданный в постоянном базисе, имел некоторый спектр, то после перехода к новому
базису, заданному оператором эволюции, гамильтониан $H'$ обращается в нуль, и
имеет только нулевые собственные значения. В квантовой механике уравнение
Шредингера обычно задают, определив гамильтониан в постоянном базисе, исходя из
физических соображений. Затем, если это удобнее, можно перейти к новому базису,
зависящему от времени.
\qed\end{com}

Перейдем к определению адиабатического предела и описанию базиса $b_k(t)$,
который будет использован при доказательстве адиабатической теоремы.
Адиабатическая теорема справедлива для гамильтонианов, которые медленно меняются
со временем. А именно, мы предполагаем, что гамильтониан некоторой квантово
механической системы достаточно гладко зависит от вещественного параметра
$\nu=\e t$, где $\e>0$, который меняется на конечном отрезке, $\nu\in[0,\nu_0]$.
Тогда медленное изменение гамильтониана означает, что параметр $\nu$ меняется на
конечную величину при малых $\e$ и больших временах $t$.
\begin{defn}
Двойной предел в решении задачи Коши для уравнения Шредингера (\ref{escheq}) и
(\ref{echaus}) на отрезке $[0,t]$
\begin{equation}                                                  \label{eadlim}
  \e\to0,\qquad t\to\infty,\qquad \text{при условии}\quad \e t=\nu=\const.
\end{equation}
называется {\em адиабатическим}.
\qed\end{defn}
\index{Адиабатический предел (adiabatic limit)}%
\index{Предел адиабатический (adiabatic limit)}%
При исследовании адиабатического предела время $t$ в уравнении Шредингера удобно
заменить на параметр $\nu$:
\begin{equation}                                                  \label{eschnu}
  i\e \frac{\pl \psi}{\pl\nu}=\psi H(\nu).
\end{equation}
Вектор состояния $\psi(\nu,\e)$ в таком случае зависит также от параметра $\e$,
а адиабатический предел соответствует простому пределу $\e\to0$ при каждом
значении параметра $\nu\in[0,\nu_0]$.

Асимптотическое решение уравнения вида (\ref{eschnu}) в общем случае построено в
\cite{VlaVol84CR,VlaVol85R}.

Для доказательства адиабатической теоремы нам понадобится специальный базис
$b_k(\nu)$, зависящий от $\nu$. Пусть исходный гамильтониан $H(\nu)$ квантовой
системы задан в некотором фиксированном базисе $e_k$. Тогда существует унитарная
матрица $S(\nu)$, которая диагонализирует гамильтониан:
\begin{equation}                                                  \label{ediaog}
  SH(\nu)S^{-1}=H_\Sd(\nu)=\diag\big(E_1(\nu),\dotsc,E_\Sn(\nu)\big),
\end{equation}
где $E_1\le E_2\le\dotsc\le E_\Sn$ -- уровни энергии собственных состояний
гамильтониана $H$, которые будем считать упорядоченными. Пусть $b_k$ --
собственные векторы исходного гамильтониана:
\begin{equation}                                                  \label{eighal}
  b_kH=E_kb_k,
\end{equation}
для всех $\nu$. Как известно, строками матрицы преобразования $S_k{}^l$, где
индекс $k$ фиксирован и $l=1,\dotsc,\Sn$, являются компоненты собственных
векторов $b_k=b_k{}^le_l$ гамильтониана $H$: $S_k{}^l=b_k{}^l$. То есть
гамильтониан $H(\nu)$ в базисе $b_k$ диагонален. Унитарная матрица $S$
определена неоднозначно, и произвол в ее выборе в дальнейшем рассмотрении будет
использован.

Мы допускаем, что часть уровней энергии может быть вырождена. Обозначим
через $\Upsilon_n$ множество индексов, для которых $E_j(\nu)=E_n(\nu)$ при
$j\in\Upsilon_n$. Конечно, в качестве индекса $n$ можно выбрать любой индекс,
принадлежащий $\Upsilon_n$. Если уровень $E_n$ невырожден, то множество индексов
состоит из одного элемента: $\Upsilon_n=\lbrace n\rbrace$. Мы докажем
адиабатическую теорему в случае, когда множество индексов $\Upsilon_n$ для всех
$n$ не меняется со временем, т.е.\ уровни энергии не пересекаются.

Мы также предполагаем, что гамильтониан $H$, уровни энергии $E_1,\dotsc,E_\Sn$ и
матрица преобразования $S$ достаточно гладко зависят от $\nu$ на конечном
отрезке $[0,\nu_0]$.

Для доказательства адиабатической теоремы нам понадобится
\begin{lemma}                                                     \label{ladthe}
Существует унитарная матрица $S$ в (\ref{ediaog}) такая, что выполнено условие
\begin{equation}                                                  \label{eadcom}
  \left(\frac{dS}{d\nu}S^{-1}\right)_k{\vphantom{\frac{dS}{d\nu}}}^j=0,
  \qquad \forall k\in\Upsilon_j.
\end{equation}
\end{lemma}
\begin{proof}
Рассмотрим два случая. Пусть уровень энергии $E_k$ невырожден. Матрица
преобразования $S$ в формуле (\ref{ediaog}) определена с точностью до умножения
каждой строки на фазовый множитель: $S_k{}^j\mapsto S_k{}^j\ex^{i\al_k(\nu)}$
для всех $j=1,\dotsc,\Sn$. Это соответствует произволу в выборе фазового
множителя у собственного вектора состояния (\ref{eighal}). Пусть фазовый
множитель удовлетворяет уравнению
\begin{equation*}
  \frac{d\al_k}{d\nu}=i\sum_{j=1}^\Sn\frac{dS_k{}^j}{d\nu}S_{\quad j}^{-1}{}^k,
\end{equation*}
где суммирование по $k$ в правой части отсутствует. Тогда нетрудно проверить,
что после преобразования для любого решения этого уравнения выполнено равенство
\begin{equation}                                                  \label{ediael}
  \left(\frac{dS}{d\nu}S^{-1}\right)_k{\vphantom{\frac{dS}{d\nu}}}^k=0.
\end{equation}
Это можно проделать для всех невырожденных уровней одновременно, выбрав
подходящим образом фазы $\al_k(\nu)$.

Теперь предположим, что все уровни энергии вырождены, $E_1=\dotsc=E_\Sn$. Тогда
матрица $S$ формуле (\ref{ediaog}) определена с точностью до унитарного
преобразования:
\begin{equation*}
  S\mapsto WS,\qquad W(\nu)\in\MU(\Sn).
\end{equation*}
Пусть матрица $W$ удовлетворяет уравнению
\begin{equation*}
  \frac{dW}{d\nu}+W\frac{dS}{d\nu}S^{-1}=0,
\end{equation*}
которое всегда имеет решение. Тогда после преобразования для любого решения
будет выполнено равенство (\ref{eadcom}) для всех $j,k$.

Если вырождена только часть уровней, то соответствующее унитарное преобразование
необходимо проделать только с этими уровнями. Таким образом, равенство
(\ref{eadcom}) будет выполнено для всех уровней с $E_j=E_k$.
\end{proof}
Доказательство адиабатической теоремы будет проведено в ортонормированном
базисе (\ref{ebatra}), где матрица $S$ выбрана таким образом, как описано в
лемме \ref{ladthe}. Этот базис состоит из собственных векторов исходного
гамильтониана и гамильтониан $H(\nu)$ в нем диагонален (\ref{ediaog}).
Компоненты вектора состояния в базисе $b_k$, как и ранее, пометим штрихом,
$\psi=\psi^{\prime k}b_k$. Поскольку гамильтониан $H$ в этом базисе диагонален,
то квадрат модуля $k$-той компоненты вектора состояния
\begin{equation*}
  |(\psi,b_k)|^2=|\psi^{\prime k}|^2,
\end{equation*}
где круглые скобки обозначают скалярное произведение в $\MH$, равен вероятности
обнаружить квантовую систему в состоянии $E_k$.

Для формулировки теоремы нам понадобится функция
\begin{equation*}
  \triangle E_n(\nu)=\underset{j,\s}{\min}|E_j(\s)-E_n(\s)|,\qquad
  \forall\s\in[0,\nu],
\end{equation*}
где минимум $|E_j-E_n|$ берется по всем $j$, для которых $E_j\ne E_n$, и всем
$\s\in[0,\nu]$. Поскольку уровни энергии не пересекаются, то для каждого
значения параметра $\nu$ функция $\triangle E_n(\nu)$ конечна и равна
минимальному расстоянию от уровня энергии $E_n$ до остальных уровней энергии.
\begin{theorem}[\bf Адиабатическая теорема]
Пусть гамильтониан $H=H(\nu)$, его собственные состояния $b_k(\nu)$ и уровни
энергии $E_k(\nu)$ достаточно гладко зависят от $\nu$ на конечном отрезке
$\nu\in[0,\nu_0]$. Предположим, что число вырожденных собственных состояний
постоянно во времени. Пусть $\psi_{(n)}(\nu,\e)$ -- решение уравнения
Шредингера, которое в начальный момент времени совпадает с собственным
состоянием $b_n(0)$ гамильтониана $H(0)$, соответствующим уровню энергии
$E_n(0)$. Тогда в адиабатическом пределе (\ref{eadlim}) справедлива следующая
оценка нормы
\begin{equation}                                                  \label{eotsen}
   1-\sum_{j\in\Upsilon_n}|(\psi_{(n)},b_j)|^2
   =\frac{\obig(\e^2)}{\triangle E^2_n(\nu)},\qquad \forall\nu\in[0,\nu_0].
\end{equation}
То есть в процессе эволюции квантовая система будет оставаться в собственном
состоянии гамильтониана $H(\nu)$, соответствующим уровню энергии $E_n(\nu)$, с
точностью порядка $\e^2$.
\end{theorem}
\begin{proof}
Будем решать задачу Коши (\ref{echneb}) в базисе (\ref{ebatra}). Гамильтониан,
который входит в уравнение Шредингера, в этом базисе диагонален с точностью до
линейных членов по $\e$,
\begin{equation*}
  H'=H_\Sd-i\e \frac{dS}{d\nu}S^{-1}.
\end{equation*}
Пусть матрица $S$ выбрана таким образом, как описано в лемме \ref{ladthe}.
Предположим, что в начальный момент времени система находится в собственном
состоянии гамильтониана $H_\Sd$ и, следовательно, в собственном состоянии
исходного гамильтониана $H=S^{-1}H_\Sd S$. Это значит, что начальное условие в
базисе $b_k$ имеет вид
\begin{equation*}
  \psi_{(n)}(0,\e)=b_n(0)=(\underbrace{0,\dotsc,0}_{n-1},1,0\dotsc,0).
\end{equation*}
Любое решение уравнения Шредингера представимо в виде
\begin{equation}                                                  \label{esodia}
  \psi_{(n)}(\nu,\e)
  =\phi_{(n)}(\nu,\e)\exp\left(-\frac i\e\int_0^\nu \!\!\! d\s H_\Sd(\s)\right),
\end{equation}
где $\phi_{(n)}$ -- некоторый вектор гильбертова пространства $\MH$. Тогда для
вектора $\phi_{(n)}$ получаем уравнение
\begin{equation*}
  \frac{\pl\phi_{(n)}}{\pl\nu}=
  -\phi_{(n)}\exp\left(-\frac i\e\int_0^\nu \!\!\! d\s H_\Sd\right)
  \frac{dS}{d\nu}S^{-1}
  \exp\left(\frac i\e\int_0^\nu \!\!\! d\s H_\Sd\right),
\end{equation*}
Полученное уравнение вместе с начальным условием перепишем в виде интегрального
уравнения
\begin{equation}                                                  \label{einteq}
  \phi_{(n)}(\nu,\e)=\psi'_{(n)}(0)-\int_0^\nu\!\!\! d\s\phi_{(n)}
  \exp\left(-\frac i\e\int_0^\s \!\!\! d\lm H_\Sd\right)
  \frac{dS}{d\s}S^{-1}\exp\left(\frac i\e\int_0^\s \!\!\! d\lm H_\Sd\right).
\end{equation}
При $\e\to0$ подынтегральное выражение содержит быстро осциллирующий множитель
и его легко оценить. Рассмотрим модуль компоненты решения
$\psi^{\prime j}_{(n)}$, которая соответствует собственному состоянию
гамильтониана $H$ с энергией $E_j$, где $E_j\ne E_n$,
\begin{equation}                                                  \label{ecopgj}
  \left|\psi^{\prime j}_{(n)}\right|=\left|\phi^j_{(n)}\right|
  =\left|\sum_{k=1}^\Sn\int_0^{\nu}\!\!\! d\s
  \exp\left(\frac i\e\int_0^\s\!\!\! d\lm(E_j-E_k)\right)\phi^k_{(n)}
  \left(\frac{dS}{d\nu}S^{-1}\right)_k{\vphantom{\frac{dS}{d\nu}}}^j\right|.
\end{equation}
В сумме справа слагаемые с $E_k=E_j$ вклада не дают в силу равенства
(\ref{eadcom}). При $E_k\ne E_j$ каждое слагаемое проинтегрируем по частям:
\begin{multline}                                                  \label{eotska}
  \left.\frac\e{i(E_j-E_k)}
  \exp\left(\frac i\e\int_0^\s\!\!\! d\lm(E_j-E_k)\right)\phi^k_{(n)}
  \left(\frac{dS}{d\nu}S^{-1}\right)_k{\vphantom{\frac{dS}{d\nu}}}^j
  \right|_0^\nu-
\\
  -\frac\e i\int_0^{\nu}\!\!\! d\s
  \exp\left(\frac i\e\int_0^\s\!\!\! d\lm(E_j-E_k)\right)
  \frac1{E_j-E_k}\frac d{d\s}\left[\phi^k_{(n)}
  \left(\frac{dS}{d\nu}S^{-1}\right)_k{\vphantom{\frac{dS}{d\nu}}}^j\right].
\end{multline}
По предположению подынтегральное выражение во втором слагаемом является
дифференцируемой функцией и его снова можно проинтегрировать по частям. В
результате получим, что оно имеет порядок $\e^2$, и им можно пренебречь. Модуль
первого слагаемого, очевидно, ограничен. Таким образом получаем оценку
\begin{equation}                                                  \label{eotsji}
  \left|\psi^{\prime j}_{(n)}(\nu,\e)\right|=
  \frac{\obig(\e)}{\min\big|E_j(\s)-E_k(\s)\big|},\qquad \forall j\notin\Upsilon_n,
\end{equation}
где минимум берется по всем $k$, для которых $E_k\ne E_j$, и всем
$\s\in[0,\nu]$.

Теперь снова вернемся к выражению (\ref{eotska}). Из оценки (\ref{eotsji})
вытекает, что $|\phi_{(n)}^k|$ для всех $k$ при $E_k\ne E_n$ имеет порядок не
ниже $\e$. Поэтому в сумме (\ref{ecopgj}) все слагаемые с индексом
$k\notin\Upsilon_n$ дают вклад не ниже $\e^2$, и ими можно пренебречь. Поэтому
оценку (\ref{eotsji}) можно улучшить
\begin{equation*}
  \left|\psi^{\prime j}_{(n)}(\nu,\e)\right|=
  \frac{\obig(\e)}{\min\big|E_j(\s)-E_n(\s)\big|},\qquad \forall j\notin\Upsilon_n,
\end{equation*}
где минимум берется только по $\s\in[0,\nu]$.

Норма произвольного решения сохраняется во времени и равна единице.
Следовательно,
\begin{equation*}
  1-\sum_{j\in\Upsilon_n}|\psi^{\prime j}_{(n)}(\nu,\e)|^2
  =\sum_{j\notin\Upsilon_n}|\psi^{\prime j}_{(n)}(\nu,\e)|^2.
\end{equation*}
Поскольку число уровней конечно, то отсюда вытекает оценка (\ref{eotsen}).
\end{proof}
\begin{com}
В теореме функция $\triangle E_n(\nu)$ для каждого $\nu$ равна константе и ее
можно включить в $\obig(\e^2)$. Тем не менее мы выделили множитель
$\triangle E_n$ с тем, чтобы показать, что предположение о том, что уровни
энергии не пересекаются, является существенным. При пересечении уровней энергии
знаменатель в (\ref{eotsen}) обращается в нуль и доказательство не проходит.
В этом случае требуются дополнительные предположения о степени касания уровней
энергии и дополнительное исследование. В своей оригинальной статье
\cite{BorFoc28} Борн и Фок рассмотрели случай, когда спектр гамильтониана
дискретен, но может быть неограничен. Неявно ими было сделано предположение о
невырожденности спектра почти для всех моментов времени. Кроме того, допускалась
возможность определенного пересечения уровней энергии с течением времени. Мы
рассмотрели более простой конечномерный случай, когда уровни энергии не
пересекаются. Это позволило упростить доказательство и выявить наиболее
существенные моменты. Оценка (\ref{eotsen}) согласуется с оценкой, приведенной
в \cite{BorFoc28}.
\qed\end{com}
Адиабатическая теорема утверждает, что, если в начальный момент времени система
находилась в собственном состоянии гамильтониана, соответствующем уровню энергии
$E_n(0)$, и этот уровень невырожден, то в адиабатическом пределе она будет
оставаться в собственном состоянии $E_n(\nu)$ с точностью порядка $\e^2$ при
конечных значениях параметра $\nu$. Если уровень энергии $E_n$ вырожден, то
система будет находиться в одном из собственных состояний $E_j$, где
$j\in\Upsilon_n$, с той же точностью. В следующем разделе мы увидим, что оценка
(\ref{eotsen}) неулучшаема, а в процессе эволюции система может оказаться в
любом из вырожденных состояний $E_j$, $j\in\Upsilon_n$, с вероятностью порядка
единицы. Эти утверждения, естественно, не зависят от выбора базиса, который
использовался при доказательстве адиабатической теоремы.

Рассмотрим теперь, как выглядит в адиабатическом пределе решение задачи Коши
(\ref{echfib}) в фиксированном базисе в невырожденном случае. Пусть $\vf(\nu)$
-- собственная функция гамильтониана $H(\nu)$, отвечающая невырожденному
собственному значению энергии $E(\nu)$,
\begin{equation*}
  \vf H=E\vf,\qquad \forall \nu\in[0,\nu_0].
\end{equation*}
Эти собственные функции определены с точностью до фазового множителя, который
может зависеть от $\nu$. Пусть в начальный момент времени система находилась в
собственном состоянии $\psi_0=\vf(0)$. В адиабатическом приближении она будет
находиться в собственном состоянии, соответствующем энергии $E(\nu)$. Поскольку
собственное состояние невырождено, то решение задачи Коши (\ref{echfib}) может
отличаться от $\vf$ не более, чем на фазовый множитель. Поэтому будем искать
решение в виде $\psi=\ex^{i\Theta}\vf$, где $\Theta(t)$ -- неизвестная функция
времени. Тогда из уравнения Шредингера следует уравнение на фазу
\begin{equation}                                                  \label{eberfa}
  \dot\Theta=i(\dot\vf,\vf)-E(\e t),
\end{equation}
где точка обозначает дифференцирование по времени.
Поскольку в начальный момент времени $\Theta(0)=0$, то фаза имеет вид
\begin{equation}                                                  \label{ephaex}
  \Theta(t)=i\int_0^t\!\!ds\left(\frac{d\vf}{ds},\vf\right)
  -\int_0^t\!\!dsE(\e s)
  =i\int_0^\nu\!\!d\s\left(\frac{d\vf}{d\s},\vf\right)-\int_0^t\!\!dsE(\e s),
\end{equation}
где $\s:=\e s$.

Покажем, что фазу собственной функции $\vf$ всегда можно выбрать таким образом,
что будет выполнено равенство
\begin{equation}                                                  \label{egavfc}
  \left(\frac{d\vf}{d\nu},\vf\right)=0,
\end{equation}
если $\nu\in[0,\infty)$. Действительно, пусть $\vf=\ex^{i\beta}\chi$, где
функция $\beta(\nu)$ удовлетворяет уравнению
\begin{equation}                                                  \label{epgasa}
  i\frac{d\beta}{d\nu}=\left(\frac{d\vf}{d\nu},\vf\right)
\end{equation}
с некоторым начальным условием, например, $\bt(0)=0$.
Тогда нетрудно проверить, что для новых собственных функций выполнено равенство
$(d\chi/d\nu,\chi)=0$. Поскольку уравнение (\ref{epgasa}) всегда имеет решение
на полупрямой, то собственные функции $\vf$ гамильтониана всегда можно
выбрать таким образом, что будет выполнено равенство (\ref{egavfc}).

Однако уравнение (\ref{epgasa}) может не иметь решения на окружности $\MS^1$.
Будем считать, что на окружности $\nu\in[0,2\pi]$. Тогда необходимым условием
существования решения является равенство
\begin{equation*}
  i\int_0^{2\pi}\!\!\!d\nu\left(\frac{d\vf}{d\nu},\vf\right)
  =2\pi m,\qquad \quad m=0,\pm1,\pm2,\dotsc.
\end{equation*}
Ясно, что это условие в общем случае не выполняется. Поэтому уравнение
(\ref{epgasa}) может не иметь решения на окружности. В этом случае первое
слагаемое в выражении для фазы (\ref{ephaex}) устранить нельзя. По сути дела
это и есть фаза Берри.

Решение задачи Коши на окружности $\nu\in\MS^1$ означает наличие машины времени.
Эти решения можно отбросить как нефизические. Однако Берри предложил другую
схему рассуждений, которая будет рассмотрена в разделе \ref{sberph}.
\subsection{Двухуровневая система}
В настоящем разделе в качестве примера мы рассмотрим двухуровневую
квантовомеханическую систему, для которой уравнение Шредингера решается явно.
Будет показано, что оценка, данная в адиабатической теореме, является
неулучшаемой.

Чтобы упростить задачу, поступим следующим образом. Вместо того, чтобы задать
исходный гамильтониан в фиксированном базисе, а затем его диагонализировать, мы
зададим диагональную матрицу $H_\Sd$ и унитарную матрицу $S$, которые определяют
исходный гамильтониан $H=S^{-1}H_\Sd S$. Пусть диагонализированный гамильтониан
имеет вид
\begin{equation*}
  H_\Sd=\begin{pmatrix} E_1(\nu) & 0 \\ 0 & E_2(\nu)\end{pmatrix},
\end{equation*}
где $E_{1,2}(\nu)$ -- некоторые заданные функции. Унитарную матрицу $S$ в
(\ref{ediaog}) выберем в виде
\begin{equation*}
  S=\begin{pmatrix} \cos\frac\al2 & i\sin\frac\al2 \\[2mm]
  i\sin\frac\al2 & \cos\frac\al2\end{pmatrix},
\end{equation*}
где $\al(\nu)\in\MR$ -- также некоторая заданная функция. Следовательно,
исходный гамильтониан задачи имеет вид
\begin{equation*}
  H=S^{-1}H_\Sd S=\begin{pmatrix} E_1\cos^2\frac\al2+E_2\sin^2\frac\al2
  & -\frac i2(E_2-E_1)\sin\al \\[2mm] \frac i2(E_2-E_1)\sin\al &
  E_1\sin^2\frac\al2 +E_2\cos^2\frac\al2 \end{pmatrix}
\end{equation*}
и зависит от трех, пока произвольных, функций $E_1(\nu)$, $E_2(\nu)$ и
$\al(\nu)$ параметра $\nu$, которые предполагаются достаточно гладкими.

Будем решать уравнение Шредингера в базисе (\ref{ebatra}), в котором
гамильтониан имеет вид (\ref{enewha}). Простые вычисления приводят к
гамильтониану
\begin{equation*}
  H'=\begin{pmatrix} E_1(\nu) & \frac{\dot\al}2 \\[2mm]
  \frac{\dot\al}2 & E_2(\nu) \end{pmatrix},
\end{equation*}
где точка обозначает дифференцирование по времени $t$. Ищем решение уравнения
Шредингера (\ref{echneb}) в виде строки
\begin{equation*}
  \psi'=\left(\exp\left(-i\int_0^t\!\!\!dsE_1\right)\phi,
  \exp\left(-i\int_0^t\!\!\!dsE_2\right)\chi\right),
\end{equation*}
где $\phi(t)$ и $\chi(t)$ -- неизвестные функции. Подстановка этого выражения в
уравнение Шредингера приводит к системе уравнений для компонент:
\begin{equation}                                                  \label{esypch}
\begin{split}
  i\dot\phi&=\frac{\dot\al}2\exp\left(-i\int_0^t\!\!\!ds(E_2-E_1)\right)\chi,
\\
  i\dot\chi&=\frac{\dot\al}2\exp\left(i\int_0^t\!\!\!ds(E_2-E_1)\right)\phi.
\end{split}
\end{equation}
При $\dot\al\ne0$ из первого уравнения следует равенство
\begin{equation}                                                  \label{echdph}
  \chi=\frac{2i}{\dot\al}\exp\left(i\int_0^t\!\!\!ds(E_2-E_1)\right)\dot\phi.
\end{equation}
Продифференцируем это равенство по времени и подставим во второе уравнение. В
результате получим уравнение второго порядка для $\phi$:
\begin{equation}                                                  \label{eqphis}
  \ddot\phi+\left(i(E_2-E_1)-\frac{\ddot\al}{\dot\al}\right)\dot\phi
  +\left(\frac{\dot\al}2\right)^2\phi=0.
\end{equation}
Для того, чтобы решить это уравнение в явном виде зафиксируем произвольные
функции, которые входят в задачу:
\begin{equation}                                                  \label{earfud}
\begin{split}
  E_1&=E^{(0)}_1+\e t,\qquad E_{1(0)}=\const,
\\
  E_2&=E^{(0)}_2+\e t,\qquad E_{2(0)}=\const,
\\
  \al&=2\e t.
\end{split}
\end{equation}
Тогда уравнение (\ref{eqphis}) примет простой вид
\begin{equation}                                                  \label{ephedi}
  \ddot\phi+2i\triangle E\dot\phi+\e^2\phi=0,
\end{equation}
где $\triangle E=E^{(0)}_2-E^{(0)}_1$ -- расстояние между уровнями энергии.
Общее решение этого уравнения зависит от двух постоянных интегрирования
$C_{1,2}$ и имеет вид
\begin{equation*}
  \phi=\ex^{-i\triangle Et}\left(C_1\ex^{i\om_\e\,t}
  +C_2\ex^{-i\om_\e\,t}\right),
\end{equation*}
где
\begin{equation*}
  \om_\e:=\sqrt{\triangle E^2+\e^2}.
\end{equation*}

Компонента $\chi$ определяется по формуле (\ref{echdph}).

Предположим, что в начальный момент времени система находилась в состоянии
$E_1$, т.е.\
\begin{equation}                                                  \label{echdat}
  \phi(0)=1,\qquad \chi(0)=0.
\end{equation}
Простые вычисления дают решение задачи Коши для системы уравнений
(\ref{esypch}):
\begin{equation}                                                  \label{esolss}
\begin{split}
  \phi&=\ex^{-i\triangle Et}\left[\cos\left(\om_\e t\right)
  +\frac{i\triangle E}{\om_\e}
  \sin\left(\om_\e t\right)\right],
\\
  \chi&=~\ex^{i\triangle Et}\left[-\frac{i\e}{\om_\e}
  \sin\left(\om_\e t\right)\right].
\end{split}
\end{equation}
Выпишем также компоненты соответствующего вектора состояния
\begin{equation}                                                  \label{esolsv}
\begin{split}
  \psi^{\prime 1}&=\ex^{-i\left(\frac{\nu^2}{2\e}
  +E^{(0)}_1\frac\nu\e-\triangle E\frac\nu\e\right)}
  \left[\cos\frac{\om_\e\nu}\e+\frac{i\triangle E}{\om_\e}
  \sin\frac{\om_\e\nu}\e\right],
\\
  \psi^{\prime 2}&=\ex^{-i\left(\frac{\nu^2}{2\e}
  +E^{(0)}_2\frac\nu\e+\triangle E\frac\nu\e\right)}
  \left[-\frac{i\e}{\om_\e}\sin\frac{\om_\e\nu}\e\right],
\end{split}
\end{equation}
где мы перешли от $t,\e$ к переменным $\nu,\e$.
Отсюда следует, что адиабатический предел для самого вектора состояния
неопределен, т.к.\ его фаза стремится к бесконечности. Однако оценку квадрата
модуля компонент можно дать. Для решения (\ref{esolsv}) следует оценка
\begin{equation*}
  1-|\psi^{\prime1}(\nu,\e)|^2=\frac{\obig(\e^2)}{(\triangle E)^2},
  \qquad |\psi^{\prime2}(\nu,\e)|^2=\frac{\obig(\e^2)}{(\triangle E)^2},
\end{equation*}
которая совпадает с оценкой в адиабатической теореме. Отсюда следует, что
данная оценка неулучшаема.

Теперь рассмотрим случай вырожденных состояний, $E_1=E_2$, при заданных ранее
функциях (\ref{earfud}). В этом случае уравнение (\ref{ephedi}) сводится к
уравнению для свободного осциллятора:
\begin{equation*}
  \ddot\phi+\e^2\phi=0,
\end{equation*}
и легко интегрируется. Выпишем соответствующее решение задачи Коши
(\ref{echdat}) для компонент вектора состояния:
\begin{equation*}
\begin{split}
  \psi^{\prime 1}&=\quad ~\ex^{-i\left(\frac{\nu^2}{2\e}-E^{(0)}_1\frac\nu\e\right)}
  \cos\nu,
\\
  \psi^{\prime 2}&=-i\ex^{-i\left(\frac{\nu^2}{2\e}-E^{(0)}_1\frac\nu\e\right)}
  \sin\nu.
\end{split}
\end{equation*}
Мы снова видим, что адиабатический предел у вектора состояния отсутствует.
Однако квадраты модулей компонент хорошо определены:
\begin{equation*}
  |\psi^{\prime1}|^2=\cos^2\nu,\qquad |\psi^{\prime2}|^2=\sin^2\nu.
\end{equation*}
Отсюда следует, что по мере увеличения параметра $\nu$ вектор состояния $\psi'$
осциллирует между вырожденными состояниями. Это показывает, что, если в
начальный момент времени система находится в одном из вырожденных состояний, то
в процессе эволюции она может оказаться в любом из вырожденных состояний  с
вероятностью порядка единицы.
\section{Фаза Берри}
Перейдем к задаче, которую рассмотрел М.~Берри \cite{Berry84}, в ее простейшем
варианте.

Пусть гильбертово пространство конечномерно и гамильтониан $H=H(\lm)$ достаточно
гладко зависит от точки некоторого многообразия $\lm\in\MM$ размерности
$\dim\MM=n$. Если на $\MM$ выбрать координатную окрестность $\MU\subset\MM$, то
гамильтониан будет зависеть от $n$ параметров $\lm^k$, $k=1,\dotsc,n$,
(координат точки $\lm$). Будем считать, что положение точки $\lm$ на $\MM$
зависит от времени $t$ некоторым наперед заданным образом, т.е.\ гамильтониан
зависит от точки некоторой кривой $\g=\lm(t)$, $t\in[0,t_0]$. Мы будем решать
уравнение Шредингера в адиабатическом приближении, т.е.\ величина $t_0$ должна
быть достаточно велика. Предположим также, что гамильтониан зависит от времени
только через точку $\lm(t)\in\MM$ как сложная функция.
\subsection{Абелев случай: невырожденное состояние               \label{sberph}}
Рассмотрим задачу на собственные значения
\begin{equation*}
  \phi H=E\phi,\qquad E=\const,
\end{equation*}
где $\phi\in\MH$ при всех $\lm\in\MM$. Предположим, что существует невырожденное
собственное значение энергии $E$ и соответствующее собственное состояние $\phi$,
которые достаточно гладко зависят от $\lm\in\MM$. Не ограничивая общности,
предположим, что собственная функция $\phi$ нормирована на единицу,
$(\phi,\phi)=1$. Тогда она единственна с точностью до умножения на фазовый
множитель, который может зависеть от $\lm$. Зафиксируем этот фазовый множитель
каким либо образом.

Теперь будем решать задачу Коши для уравнения Шредингера (\ref{escheq}) с
начальным условием
\begin{equation}                                                  \label{einval}
  \psi|_{t=0}=\phi_0,
\end{equation}
где $\phi_0:=\phi\big(\lm(0)\big)$. В адиабатическом приближении квантовая
система в процессе эволюции будет оставаться в собственном состоянии,
соответствующем уровню энергии $E(\lm)$. Поэтому ищем решение в виде
\begin{equation*}
  \psi=\ex^{i\Theta}\phi,
\end{equation*}
где $\Theta=\Theta(t)$ -- некоторая неизвестная функция от времени. Тогда из
уравнения Шредингера следует уравнение на фазу (\ref{eberfa}) с начальным
условием $\Theta|_{t=0}=0$. Поскольку $\dot\phi=\dot\lm^k\pl_k\phi$, то решение
задачи Коши для уравнения (\ref{eberfa}) имеет вид
\begin{equation}                                                  \label{esothe}
  \Theta=\int_0^t\!\! dt\dot\lm^k A_k-\int_0^t\!\! ds E\big(\lm(s)\big)
  =\int_{\lm(0)}^{\lm(t)}\!\! d\lm^k A_k-\int_0^t\!\! ds E\big(\lm(s)\big),
\end{equation}
где введено обозначение
\begin{equation}                                                  \label{edeloc}
  A_k(\lm):=i(\phi,\pl_k\phi)
\end{equation}
и интеграл по $\lm$ берется вдоль кривой $\lm(t)$.

Таким образом, интеграл (\ref{esothe}) в адиабатическом приближении дает решение
задачи Коши для уравнения Шредингера (\ref{escheq}) с начальным условием
(\ref{einval}). Первое слагаемое в (\ref{esothe}) называется {\em геометрической
фазой} или {\em фазой Берри}, а второе -- {\em динамической фазой}.
\index{Фаза Берри (Berry phase)}\index{Берри фаза (Berry phase)}%
\index{Фаза динамическая (dynamical phase)}%
\index{Динамическая фаза (dynamical phase)}%
\index{Геометрическая фаза (geometric phase)}%
\index{Фаза динамическая (geometric phase)}%

Заметим, что компоненты (\ref{edeloc}) вещественны вследствие нормировки
волновой функции. Действительно, дифференцируя условие нормировки
$(\phi,\phi)=1$, получаем равенство
\begin{equation*}
  (\pl_k\phi,\phi)+(\phi,\pl_k\phi)=(\phi,\pl_k\phi)^\dagger+(\phi,\pl_k\phi)=0.
\end{equation*}
Отсюда вытекает вещественность компонент (\ref{edeloc}) и, следовательно, фазы
Берри.

Теперь рассмотрим множество замкнутых кривых $\g=\lm(t)\in\Om(\MM,\lm_0)$ на
многообразии параметров $\MM$ с началом и концом в некоторой фиксированной точке
$\lm_0\in\MM$. Тогда для полного изменения фазы волновой функции получаем ответ
\begin{equation*}
  \Theta=\Theta_\Sb-\int_0^{t_0}\!\!dtE\big(\lm(t)\big),
\end{equation*}
где
\begin{equation}                                                  \label{ebphas}
  \Theta_\Sb=\oint_{\g}\! d\lm^k A_k.
\end{equation}
Динамическая часть фазы волновой функции расходится при $t_0\to\infty$. Однако в
экспериментах наблюдается разность фаз двух векторов состояний с одинаковой
динамической фазой, которая определяется фазой Берри. Поэтому рассмотрим фазу
Берри подробнее.

Заметим, что выражение для фазы Берри не зависит от параметризации кривой
$\g$. Это значит, что переход к адиабатическому пределу в уравнении
Шредингера влияет на фазу Берри только через компоненты локальной формы
связности (\ref{edeloc}).

В таком виде можно дать геометрическую интерпретацию фазе Берри $\Theta_\Sb$,
которая определяется первым слагаемым в полученном выражении (\ref{esothe}). А
именно, мы имеем главное расслоение $\MP\big(\MM,\pi,\MU(1)\big)$, базой
которого является многообразие параметров $\lm\in\MM$, а структурной группой --
группа $\MU(1)$ (фаза вектора состояния $\ex^{i\Theta}$). Вектор гильбертова
пространства $\phi\in\MH$ представляет собой локальное сечение ассоциированного
расслоения $\ME\big(\MM,\pi_\ME,\MH,\MU(1),\MP\big)$, типичным слоем которого
является гильбертово пространство $\MH$.

Рассмотрим изменение локального сечения ассоциированного расслоения, которое
вызвано умножением вектора гильбертова пространства на фазовый множитель
(вертикальный автоморфизм),
\begin{equation*}
  \phi'=\ex^{ia}\phi,
\end{equation*}
где $a=a(\lm)\in\CC^2(\MM)$ -- произвольная дважды дифференцируемая функция.
Тогда компоненты (\ref{edeloc}) преобразуются по-правилу
\begin{equation*}
  A'_k=A_k-\pl_ka.
\end{equation*}
Сравнивая это правило с преобразованием компонент локальной формы связности,
заключаем, что поля $A_k(\lm)$ можно интерпретировать, как
компоненты локальной формы связности для группы $\MU(1)$. Другими словами,
$A_k(\lm)$ -- это калибровочное поле для одномерной унитарной группы $\MU(1)$.
Если база ассоциированного расслоения $\ME\big(\MM,\pi_\ME,\MH,\MU(1),\MP\big)$
покрыта некоторым семейством карт, $\MM=\cup_j\MU_j$, то множество сечений,
заданных в каждой координатной окрестности $\MU_j$, определяет семейство
локальных форм связности на главном расслоении $\MP\big(\MM,\pi,\MU(1)\big)$.
Семейство локальных форм связности $d\lm^kA_k$ определяет единственную с
точностью до изоморфизма связность на $\MP$ (теорема \ref{troloc}).

Вспомним выражение для элемента группы голономии в виде упорядоченной
$\P$-экспоненты (\ref{ehowil}). В рассматриваемом случае группа $\MU(1)$
абелева и $\P$-экспонента совпадает с обычной экспонентой. Поэтому фаза Берри
(\ref{ebphas}) определяет элемент $\ex^{i\Theta_\Sb}$ группы голономии
$\Phi(\lm_0,e)\subset\MU(1)$ главного расслоения  в точке $(\lm_0,e)\in\MP$,
соответствующей нулевому сечению $\MM\ni\lm\mapsto(\lm,e)\in\MP$, где
$\lm_0:=\lm(0)$ и $e$ -- единица структурной группы $\MU(1)$. Сечение является
нулевым, поскольку в начальный момент времени фаза Берри равна нулю,
$\Theta_\Sb|_{t=0}=0$. Локальная форма связности $d\lm^kA_k$ также соответствует
нулевому сечению.

Если база $\MM$ односвязна, то выражение для фазы Берри (\ref{ebphas}) можно
переписать в виде поверхностного интеграла от компонент локальной формы
кривизны. Используя формулу Стокса, получаем следующее выражение
\begin{equation}                                                  \label{ebersu}
  \Theta_\Sb=\frac12\iint_S d\lm^k\wedge d\lm^l F_{kl},
\end{equation}
где $S$ -- поверхность в $\MM$ с границей $\g\in\Om(\MM,\lm_0)$ и
$F_{kl}=\pl_k A_l-\pl_l A_k$ -- компоненты локальной формы кривизны
(напряженности калибровочного поля). Если база $\MM$ не является односвязной, то
выражение для фазы Берри в виде поверхностного интеграла (\ref{ebersu}) имеет
место только для тех замкнутых путей, которые стягиваются в точку.
\subsection{Частица со спином $1/2$ в магнитном поле}
В качестве примера вычислим фазу Берри для частицы со спином 1/2, находящейся во
внешнем однородном магнитном поле. В нерелятивистской квантовой механике частица
со спином 1/2 описывается двухкомпонентной волновой функцией
\begin{equation*}
  \psi=(\psi_+,\psi_-).
\end{equation*}
Будем считать, что она находится в евклидовом пространстве $\MR^3$ с заданным
однородным магнитным полем. Пусть напряженность магнитного поля $H^k(t)$,
$k=1,2,3$, не зависит от точки пространства, но меняется со временем $t$
некоторым заданным образом. Кроме того, для простоты, пренебрежем кинетической
энергией частицы и будем считать, что другие поля отсутствуют. В этом случае
гильбертово пространство $\MH$ двумерно, и гамильтониан частицы состоит из
одного слагаемого -- взаимодействия магнитного момента частицы с внешним
магнитным полем (см., например, \cite{Messia62R,Fock76R}),
\begin{equation*}
  H=-\mu H^k\s^\St_k,
\end{equation*}
где $\s^\St_k$ -- транспонированные матрицы Паули (поскольку в наших
обозначениях они действуют справа) и $\mu$ -- магнетон (размерная постоянная,
равная отношению магнитного момента частицы к ее спину). Чтобы привести
гамильтониан к виду, который был рассмотрен ранее, ведем новые переменные
$\lm^k:=-\mu H^k$. Тогда гамильтониан примет вид
\begin{equation}                                                  \label{ehaspo}
  H=\lm^k\s^\St_k=\begin{pmatrix}\lm^3 & ~\lm^+ \\ \lm^- & -\lm^3 \end{pmatrix},
\end{equation}
где $\lm^\pm:=\lm^1\pm i\lm^2$.

Собственные значения гамильтониана (\ref{ehaspo}) находятся из уравнения
\begin{equation*}
  \det(H-E\one)=0,
\end{equation*}
которое имеет два вещественных решения
\begin{equation}                                                  \label{ehaigm}
  E_\pm=\pm|\lm|,
\end{equation}
где
\begin{equation*}
  |\lm|=\sqrt{(\lm^1)^2+(\lm^2)^2+(\lm^3)^2}
\end{equation*}
-- длина вектора $\lm=\lbrace\lm^k\rbrace\in\MR^3$. Нетрудно проверить, что
уравнение на собственные функции,
\begin{equation*}
  \phi_\pm H=E_\pm\phi_\pm,
\end{equation*}
имеет два решения
\begin{equation}                                                  \label{eigfuh}
  \phi_\pm=\frac1{\sqrt{2|\lm|}}
  \left(\pm\frac{\lm^-}{\sqrt{|\lm|\mp\lm^3}},~
  \sqrt{|\lm|\mp\lm^3}\right).
\end{equation}
В полученном выражении множитель выбран таким образом, что собственные функции
нормированы на единицу,
\begin{equation*}
  (\phi_\pm,\phi_\pm)=1.
\end{equation*}
Таким образом, гамильтониан (\ref{ehaspo}) частицы со спином 1/2, находящейся во
внешнем однородном магнитном поле, имеет два невырожденных собственных состояния
(\ref{eigfuh}), соответствующих уровням энергии (\ref{ehaigm}).

Для дальнейших вычислений в пространстве параметров $\lm\in\MR^3$ удобно ввести
сферические координаты $|\lm|,\theta,\vf$:
\begin{align*}
  \lm^1&=|\lm|\sin\theta\cos\vf,
\\
  \lm^2&=|\lm|\sin\theta\sin\vf,
\\
  \lm^3&=|\lm|\cos\theta.
\end{align*}
Тогда собственные функции примут вид
\begin{equation*}
  \phi_+=\left(\cos\frac\theta2\ex^{-i\vf},\sin\frac\theta2\right),\qquad
  \phi_-=\left(-\sin\frac\theta2\ex^{-i\vf},\cos\frac\theta2\right).
\end{equation*}

Допустим, что экспериментатор, наблюдающий за частицей, достаточно гладко
меняет однородное магнитное со временем. То есть параметры $\lm^k(t)$, от
которых зависит гамильтониан, достаточно гладко зависят от времени. Предположим
также, что в начальный момент времени $t=0$ частица находилась в состоянии
$\phi_+$. Соответствующее решение уравнения Шредингера (\ref{escheq}) в
адиабатическом приближении имеет вид
\begin{equation*}
  \psi=\ex^{i\Theta}\phi_+,
\end{equation*}
где фаза $\Theta$ удовлетворяет уравнению (\ref{eberfa}). Нетрудно вычислить
компоненты локальной формы связности $A_k=i(\phi_+,\pl_k\phi_+)$ для
собственного состояния $\phi_+$:
\begin{equation*}
  A_{|\lm|}=0,\qquad A_\theta=0,\qquad A_\vf=\cos^2\frac\theta2.
\end{equation*}
Соответствующая локальная форма кривизны имеет только две отличные от нуля
компоненты:
\begin{equation*}
  F_{\theta\vf}=-F_{\vf\theta}=-\frac12\sin\theta.
\end{equation*}

Теперь вычислим фазу Берри для замкнутой кривой в пространстве параметров
$\g=\lbrace\lm^k(t)\rbrace\in\MM$,
\begin{equation}                                                  \label{eberyp}
\begin{split}
  \Theta_\Sb&=\oint_\g d\lm^k A_k
  =\frac12\iint_S d\lm^k\wedge d\lm^l F_{kl}=
\\
  &=\iint_S d\theta\wedge d\vf F_{\theta\vf}
  =-\frac12\iint_S d\theta\wedge d\vf\sin\theta=-\frac12\Om(\g),
\end{split}
\end{equation}
где $S$ -- поверхность в $\MR^3$ с границей $\g$ и $\Om(\g)$ -- телесный
угол, который занимает контур $\g$, если смотреть из начала координат.

Если в начальный момент времени частица находилась в состоянии $\phi_-$, то
вычисления проводятся аналогично. В этом случае
\begin{equation*}
  A_{|\lm|}=0,\qquad A_\theta=0,\qquad A_\vf=\sin^2\frac\theta2,
\end{equation*}
и компоненты локальной формы кривизны отличаются знаком:
\begin{equation*}
  F_{\theta\vf}=-F_{\vf\theta}=\frac12\sin\theta.
\end{equation*}
Следовательно, фаза Берри также отличается только знаком.

Таким образом, если в начальный момент времени частица находилась в одном из
состояний $\phi_\pm$, то после изменения однородного магнитного поля вдоль
замкнутой кривой $\g$, ее волновая функция изменится на фазовый множитель,
геометрическая часть которого равна
\begin{equation}                                                  \label{eberco}
  \Theta_{\Sb\pm}=\mp\frac12\Om(\g),
\end{equation}
где $\Om(\g)$ -- телесный угол, под которым виден замкнутый контур $\g$
из начала координат. Этот результат имеет место в адиабатическом приближении,
когда параметры $\lm(t)$ медленно меняются со временем.

Выражение для фазы Берри (\ref{eberco}) было подтверждено экспериментально
\cite{BitDub87} для рассеяния поляризованных нейтронов в спиральном магнитном
поле.

В рассмотренном примере однородное магнитное поле может иметь произвольное
направление и величину. Следовательно, база $\MM$ главного расслоения
$\MP\big(\MM,\pi,\MU(1)\big)$ совпадает с евклидовым пространством, $\MM=\MR^3$.
Поэтому, согласно теореме \ref{trigla}, главное расслоение $\MP$ тривиально,
$\MP\simeq\MR^3\times\MU(1)$. В случае фазы Берри связность на этом расслоении
определяется сечением ассоциированного расслоения, например, $\phi_+$, которое
находится путем решения уравнения Шредингера. Это сечение (\ref{eigfuh}), как
нетрудно проверить, имеет особенность на положительной полуоси $\lm^3\ge0$.
Компоненты локальной формы связности относительно декартовой
системы координат имеют вид
\begin{equation}                                                  \label{elocco}
\begin{split}
  A_1&=\frac{\pl\vf}{\pl\lm^1}A_\vf
  =-\frac{\sin\vf\cos\frac\theta2}{2|\lm|\sin\frac\theta2},
\\
  A_2&=\frac{\pl\vf}{\pl\lm^2}A_\vf
  =\quad \frac{\cos\vf\cos\frac\theta2}{2|\lm|\sin\frac\theta2},
\\
  A_3&=\frac{\pl\vf}{\pl\lm_3}A_\vf=0.
\end{split}
\end{equation}
Здесь мы вынуждены перейти в декартову систему координат, поскольку сферическая
система координат сингулярна на оси $\lm^3$ и непригодна для исследования
особенностей, которые здесь расположены. Как видим, компоненты локальной формы
связности имеют особенность на положительной полуоси $\lm^3\ge0$ как и вектор
$\phi_+$. Теперь вычислим компоненты локальной формы тензора кривизны. У нее
отличны от нуля все компоненты:
\begin{align*}
  F_{12}&=-F_{21}=-\frac{\cos\theta}{2|\lm|^2},
\\
  F_{13}&=-F_{31}=\quad \frac{\sin\theta\sin\vf}{2|\lm|^2},
\\
  F_{23}&=-F_{32}=-\frac{\sin\theta\cos\vf}{2|\lm|^2}.
\end{align*}
Наконец, вычислим квадрат тензора кривизны, который является инвариантом,
\begin{equation*}
  F^2=2\big(F_{12}^2+F_{13}^2+F_{23}^2\big)=\frac1{2|\lm|^4}.
\end{equation*}
Таким образом, форма кривизны сингулярна только в начале координат.

Вернемся к нашему главному расслоению $\MR^3\times\MU(1)$. Локальная форма
связности (\ref{elocco}) на нем неопределена, т.к.\ имеет особенность на полуоси
$\lm^3\ge0$, которую мы обозначим $\lbrace\lm^3_+\rbrace$. Поэтому, чтобы
построить главное расслоение с заданной связностью, мы вынуждены удалить
прообраз $\pi^{-1}\big(\lbrace \lm^3_+\rbrace)$, где
$\pi:~\MR^3\times\MU(1)\rightarrow\MR^3$ -- естественная проекция. В результате
получаем тривиальное главное расслоение
$\big(\MR^3\setminus\lbrace \lm^3_+\rbrace\big)\times\MU(1)$, которое является
подрасслоением исходного расслоения. На этом главном расслоении локальная форма
(\ref{elocco}) бесконечно дифференцируема.

Можно рассуждать по-другому. Поскольку магнитное поле является внешним, то мы
вправе предположить, что оно меняется, например, в полупространстве $\MR^3_+$,
определяемом условием $\lm_1>0$. Поскольку полупространство $\MR^3_+$
диффеоморфно всему евклидову пространству $\MR^3$, то соответствующее главное
расслоение тривиально: $\MP\approx\MR^3_+\times\MU(1)$. В этом случае никаких
вопросов в определением связности вообще не возникает, т.к.\ локальная форма
связности (\ref{elocco}) гладкая. При этом выражение для фазы Берри
(\ref{eberco}) останется прежним.

Таким образом, фаза Берри является не топологическим понятием, а геометрическим,
т.к.\ топология главного расслоения тривиальна. Она обязана своим происхождением
нетривиальной связности, которая определяется сечениями ассоциированного
расслоения.
\subsection{Неабелев случай: вырожденное состояние               \label{swizee}}
Понятие фазы Берри было обобщено на случай, когда уровни энергии гамильтониана
$H(\lm)$ вырождены \cite{WilZee84}. В этом случае при решении уравнения
Шредингера возникает главное расслоение $\MP\big(\MM,\pi,\MU(r)\big)$ со
структурной группой $\MU(r)$, где $r$ -- количество независимых собственных
функций, соответствующих вырожденному уровню энергии $E$. Опишем эту конструкцию
подробно.

Предположим, что гамильтониан квантовой системы зависит от точки некоторого
многообразия $\lm(t)\in\MM$, как и ранее. Пусть $E$ -- вырожденное  собственное
значение гамильтониана $H$, которому соответствуют $r$ независимых собственных
функций $\phi^a\in\MH$, $a=1,\dotsc,r$,
\begin{equation*}
  \phi^aH=E\phi^a
\end{equation*}
для всех моментов времени. Мы предполагаем, что $E(\lm)$ и $\phi^a(\lm)$
являются достаточно гладкими функциями от точки многообразия $\lm$, и число
собственных функций $r$ не меняется со временем.

Собственные функции можно выбрать ортонормированными,
\begin{equation*}                                                 \label{eortnf}
  (\phi^a,\phi_b)=\dl^a_b,
\end{equation*}
где $\dl_b^a$ -- символ Кронекера. Будем искать решение задачи Коши $\psi^a$
для уравнения Шредингера (\ref{escheq}) с начальным условием
\begin{equation*}
  \psi^a(0)=\psi^a_0:=\phi^a\big(\lm(0)\big).
\end{equation*}
То есть в начальный момент времени система находится в одном из собственных
состояний $\phi^a$. В адиабатическом приближении решение $\psi^a$ для всех
моментов времени является собственной функцией гамильтониана $H(\lm)$,
соответствующей значению энергии $E(\lm)$. Поэтому его можно разложить по
собственным функциям вырожденного состояния
\begin{equation}                                                  \label{esutra}
  \psi^a=\phi^bU^{-1}_{\quad b}{}^a,
\end{equation}
где $U(\lm)\in\MU(r)$ -- некоторая унитарная матрица, которая достаточно гладко
зависит от точки $\lm\in\MM$.

Унитарность матрицы $U$ обусловлена следующим обстоятельством. Рассмотрим
решения $\psi^a$ для всех значений индекса $a=1,\dotsc,r$. Дифференцируя
скалярное произведение $(\psi^a,\psi_b)$ по времени и используя уравнение
Шредингера, получаем уравнение
\begin{equation*}
  \frac\pl{\pl t}(\psi^a,\psi_b)=-i(\psi^aH,\psi_b)+i(\psi^a,H\psi_b)=0.
\end{equation*}
Последнее равенство следует из самосопряженности гамильтониана. Отсюда следует,
что, если в начальный момент времени векторы $\psi_0^a:=\phi^a\big(\lm(0)\big)$
ортонормированы, то соответствующие решения уравнения Шредингера останутся
таковыми и во все последующие моменты времени. Поэтому матрица $U$ в разложении
(\ref{esutra}) унитарна.

Для искомого решения (\ref{esutra}) уравнение Шредингера сводится к уравнению
\begin{equation*}
  i\phi^c\dot U^{-1}_{\quad c}{}^b+i\dot\phi^cU^{-1}_{\quad c}{}^b
  =\phi^cHU^{-1}_{\quad c}{}^b.
\end{equation*}
Возьмем скалярное произведение левой и правой части с $\phi_a$. В результате
получаем уравнение на унитарную матрицу
\begin{equation}                                                  \label{euneqm}
  \dot U^{-1}_{\quad a}{}^b=\dot\lm^k A_{ka}{}^cU^{-1}_{\quad c}{}^b-iEU^{-1}_{\quad a}{}^b,
\end{equation}
где введено обозначение
\begin{equation}                                                  \label{edegau}
  A_{ka}{}^c:=-(\pl_k\phi^c,\phi_a).
\end{equation}

Из условия ортонормированности собственных функций $\phi^a$ следует
антиэрмитовость компонент $A_{ka}{}^b$ для всех $k=1,\dotsc,n$, если индексы
$a,b$ рассматриваются, как матричные. Действительно, дифференцируя условие
ортонормированности $(\phi^b,\phi_a)=\dl_a^b$, получаем равенство
\begin{equation*}
  (\pl_k\phi^b,\phi_a)+(\phi^b,\pl_k\phi_a)=(\phi^a,\pl_k\phi_b)^\dagger
  +(\phi^b,\pl_k\phi_a)=0.
\end{equation*}
То есть матрицы $A_k$ антиэрмитовы и поэтому принадлежат алгебре Ли $\Gu(r)$.
Следовательно, матрицы $A_k$ определяют 1-формы в некоторой окрестности
$\MU\subset\MM$ со значениями в алгебре Ли, как и компоненты локальной формы
связности.

Начальное условие для унитарной матрицы имеет вид
\begin{equation*}
  U^{-1}_{\quad a}{}^b|_{t=0}=\dl_a^b.
\end{equation*}
Решение задачи Коши для уравнения (\ref{euneqm}) можно записать в виде
упорядоченного $\P$-произведения (см.\ раздел \ref{swilop})
\begin{equation}                                                  \label{ebenab}
\begin{split}
  U^{-1}(t)&=\P\exp\left(\int_0^t\!\! ds\dot\lm^k(s)A_k(s)
  -i\int_0^t\!\! dsE\big(\lm(s)\big)\right)=
\\
  &=\P\exp \left(\int_{\lm(0)}^{\lm(t)}\!\!d\lm^k A_k\right)
  \times\exp\left(-i\int_0^t\!\! dsE\big(\lm(s)\big)\right),
\end{split}
\end{equation}
где мы, для краткости, опустили матричные индексы.

Первый сомножитель является обобщением фазы Берри на случай вырожденных
состояний, а второй сомножитель -- это динамическая фаза. Динамическая фаза
имеет тот же вид, что и в случае невырожденного состояния.

Первый сомножитель в решении (\ref{ebenab}) представляет собой унитарную матрицу
Вилчека--Зи
\begin{equation}                                                  \label{ewizem}
  U^{-1}_{\Sw\Sz}:=\P\exp \left(\int_{\lm(0)}^{\lm(t)}\!\!\!d\lm^k A_k\right),
\end{equation}
которой можно дать следующую геометрическую интерпретацию. Мы имеем главное
расслоение $\MP\big(\MM,\pi,\MU(r)\big)$ со структурной группой $\MU(r)$
(преобразование (\ref{esutra})). Набор собственных функций $\phi^a$ представляет
собой сечение ассоциированного расслоения
$\ME\big(\MM,\pi_\ME,\MH^r,\MU(r),\MP\big)$, типичным слоем которого является
тензорное произведение гильбертовых пространств
\begin{equation*}
  \MH^r:=\underbrace{\MH\otimes\dotsc\otimes\MH}_r.
\end{equation*}
При вертикальном автоморфизме, который задан унитарной матрицей
$U(\lm)\in\MU(r)$,
\begin{equation*}
  \phi^{\prime a}=\phi^b U^{-1}_{\quad b}{}^a,\qquad ~\phi'_a=U_a{}^b\phi_b,
\end{equation*}
поля (\ref{edegau}) преобразуются по правилу
\begin{equation}                                                  \label{elotrx}
  A'_k=UA_kU^{-1}+\pl_k U U^{-1},
\end{equation}
где мы опустили матричные индексы. Отсюда следует, что поля $A_k$ можно
интерпретировать, как компоненты локальной формы связности или поля
Янга--Миллса. Совокупность этих компонент, заданная на координатном покрытии
базы $\MM$, однозначно задает связность на главном расслоении
$\MP\big(\MM,\pi,\MU(r)\big)$.

Если путь замкнут, $\g\in\Om(\MM,\lm_0)$, то унитарная матрица Вилчека--Зи
(\ref{ewizem}) представляет собой элемент группы голономии
$U^{-1}_{\Sw\Sz}\in\Phi(\lm_0,e)$, в точке $(\lm_0,e)\in\MP$,
соответствующей нулевому сечению $\MM\ni\lm\mapsto(\lm,e)\in\MP$, где
$\lm_0:=\lm(0)$ и $e$ -- единица структурной группы $\MU(r)$.

Таким образом, в случае вырожденного уровня энергии гамильтониана возникает
главное расслоение $\MP\big(\MM,\pi,\MU(r)\big)$. В рассматриваемом случае базой
$\MM$ является многообразие параметров $\lm\in\MM$, от точки которого зависит
гамильтониан. Мы предполагаем, что это многообразие конечномерно. Структурной
группой является унитарная группа $\MU(r)$, которая также конечномерна.
Связность на главном расслоении определяется сечениями ассоциированного
расслоения $\ME\big(\MM,\pi_\ME,\MH^r,\MU(r),\MP\big)$. В общем случае типичным
слоем ассоциированного расслоения может быть бесконечномерное гильбертово
пространство $\MH^r$. В настоящей монографии мы не рассматриваем
бесконечномерных многообразий и расслоений, чтобы избежать возникающих при этом
трудностей \cite{Zharin08R}. Однако в данном случае все, что нужно, это формула
преобразования компонент локальной формы связности (\ref{elotrx}), которую легко
проверить в каждом конкретном случае. Если база $\MM$ не покрывается одной
картой, то состояние квантовой системы задается семейством локальных сечений на
координатном покрытии базы. Оно определяет семейство локальных форм связности
(\ref{edegau}). В свою очередь семейство локальных форм связности однозначно с
точностью до изоморфизма определяет связность на главном расслоении
$\MP\big(\MM,\pi,\MU(r)\big)$.

Опять мы видим, что главные и ассоциированные расслоения могут быть тривиальными
или нет, это зависит от рассматриваемой задачи. Связность на главном расслоении
$\MP\big(\MM,\pi,\MU(r)\big)$ может быть нетривиальна и приводить к
нетривиальной матрице Вилчека--Зи (\ref{ewizem}), которая описывает параллельный
перенос слоев вдоль пути на базе $\lm(t)\in\MM$, даже для тривиальных
расслоений. Это говорит о ее геометрическом, а не топологическом происхождении.
Для замкнутых путей $\g\in\Om(\MM,\lm_0)$ с началом и концом в точке
$\lm_0\in\MM$ матрица Вилчека--Зи определяет элемент группы голономии
$U_{\Sw\Sz}\in\Phi(\lm_0,e)\subset\MU(r)$.

При рассмотрении фазы Берри и матрицы Вилчека--Зи мы, для простоты,
предположили, что гильбертово пространство квантовой системы конечномерно. Это
предположение можно существенно ослабить. Полученные формулы справедливы для
всех уровней, для которых справедлива адиабатическая теорема. То есть это
должен быть изолированный уровень, энергия которого отделена от остального
спектра.
\section{Эффект Ааронова--Бома                                   \label{sahabo}}
Другой пример возникновения нетривиальной связности на тривиальном главном
расслоении $\MP\big(\MR^4,\pi,\MU(1)\big)\simeq\MR^4\times\MU(1)$ в
нерелятивистской квантовой механике дает эффект Ааронова--Бома \cite{AhaBoh59}.
В этом случае в отличии от фазы Берри в качестве базы $\MM$ главного расслоения
выступает не пространство параметров, а пространство-время $\MR^4$, в котором
частица движется. Эффект Ааронова--Бома не связан с адиабатической теоремой, и
обсуждается только с геометрической точки зрения.

Рассмотрим уравнение Шредингера (\ref{escheq}), в котором гамильтониан описывает
движение точечной частицы массы $m$ в трехмерном евклидовом пространстве $\MR^3$
с декартовыми координатами $x^\mu$, $\mu=1,2,3$,
\begin{equation}                                                  \label{ezehal}
  H_0=-\frac{\eta^{\mu\nu}p_\mu p_\nu}{2m}+U=-\frac{\hbar^2}{2m}\triangle+U,
\end{equation}
где $p_\mu=i\hbar\pl_\mu$ -- оператор импульса частицы,
$\eta_{\mu\nu}=\diag(---)$ -- отрицательно определенная пространственная
метрика, $\triangle:=\pl_1^2+\pl_2^2+\pl_3^2$ -- оператор Лапласа и $U(x)$ --
потенциальная энергия частицы.

Четырехмерный оператор импульса имеет вид $p_\al=i\hbar\pl_\al$, $\al=0,1,2,3$.
При этом нулевая компонента 4-импульса $p_0=i\hbar\pl_0=i\hbar\pl_t$ имеет
физический смысл оператора энергии частицы.

Если частица взаимодействует с электромагнитным полем, то это взаимодействие
описывается с помощью минимальной подстановки для всех четырех компонент
импульса
\begin{equation}                                                  \label{emisus}
  p_\al\quad \mapsto\quad i\hbar\pl_\al-\frac ecA_\al,
\end{equation}
где $e$ -- заряд частицы, $c$ -- скорость света и $A_\al$ -- потенциал
электромагнитного поля (компоненты локальной формы $\MU(1)$-связности). При этом
нулевая компонента, разделенная на скорость
света, $A_0/c$, имеет физический смысл потенциала электрического поля, а
пространственные компоненты $A_\mu$ -- ковекторного потенциала магнитного поля.
Таким образом, точечная частица, находящаяся в электромагнитном поле,
описывается уравнением Шредингера
\begin{equation}                                                  \label{eshrch}
  i\hbar\frac{\pl\psi}{\pl t}=\left[\frac{\hbar^2}{2m}
  \eta^{\mu\nu}\left(\pl_\mu+i\frac e{\hbar c}A_\mu\right)
  \left(\pl_\nu+i\frac e{\hbar c}A_\nu\right)+\frac ecA_0\right]\psi+U\psi.
\end{equation}
Здесь мы вернулись к стандартным обозначениям квантовой механики, когда
гамильтониан действует на вектор гильбертова пространства слева, поскольку он
содержит производные.

В дальнейшем, для простоты, положим $\hbar=1$ и $c=1$.

С геометрической точки зрения минимальная подстановка (\ref{emisus}) с точностью
до постоянных совпадает с заменой частной производной на ковариантную:
\begin{equation*}
  \pl_\al\quad \mapsto\quad \nb_\al:=\pl_\al+ieA_\al.
\end{equation*}

Посмотрим на уравнение Шредингера (\ref{eshrch}) с геометрической точки зрения.
Оно решается во всем пространстве, $\psi=\psi(t,x)$, поэтому базой расслоения
является четырехмерное евклидово пространство, $(t,x)\in\MR\times\MR^3=\MR^4$.
При этом $\MR^4$ рассматривается просто как четырехмерное многообразие без какой
либо четырехмерной метрики. При желании метрику можно ввести, однако ее наличие
никак не влияет на структуру главного расслоения и связности. Метрика
$\eta_{\mu\nu}$ определена только на пространственных сечениях $t=\const$,
поскольку она входит в уравнение Шредингера. Волновая функция $\psi(t,x)$
представляет собой сечение ассоциированного расслоения
$\ME\big(\MR^4,\pi_\ME,\MC,\MU(1),\MP\big)$, типичным слоем которого является
комплексная плоскость $\MC$ и которое ассоциировано с некоторым главным
расслоением $\MP\big(\MR^4,\pi,\MU(1)\big)$. Это главное расслоение всегда
тривиально, $\MP\simeq\MR^4\times\MU(1)$, т.к.\ базой является четырехмерное
евклидово пространство. На этом главном расслоении задана локальная форма
$\MU(1)$-связности, которая определяется электромагнитным потенциалом
$A_\al(t,x)$. В нерелятивистской квантовой механике рассматривается не все
множество сечений ассоциированного расслоения, а лишь подмножество, состоящее из
тех дифференцируемых функций $\psi(t,x)$, которые в каждый момент времени $t$
принадлежат гильбертову пространству квадратично интегрируемых функций
$\MH=\CL_2(\MR^3)$ на пространственных сечениях $\MR^3$.

Рассмотрим два случая.
\subsection{Электрический потенциал}
Предположим, что магнитный потенциал равен нулю, $A_\mu=0$, $\mu=1,2,3$. Запишем
уравнение Шредингера в виде
\begin{equation}                                                  \label{eschre}
  i\dot\psi=(H_0+eA_0)\psi,
\end{equation}
где $H_0$ -- гамильтониан системы в отсутствии электромагнитного поля
(\ref{ezehal}). Предположим также, что электрический потенциал зависит только от
времени, $A_0=A_0(t)$. Будем искать решение уравнения Шредингера (\ref{eschre})
в виде $\psi=\ex^{-i\Theta}\phi$, где $\phi$ -- решение свободного уравнения
Шредингера,
\begin{equation}                                                  \label{esheqf}
  i\dot\phi=H_0\phi,
\end{equation}
и $\Theta=\Theta(t)$ -- некоторая фаза, не зависящая от точки пространства.
Подстановка $\psi=\ex^{-i\Theta}\phi$ в исходное уравнение Шредингера
(\ref{eschre}) приводит к уравнению на фазу
\begin{equation*}
  \dot\Theta=eA_0,
\end{equation*}
где мы сократили общий фазовый множитель $\ex^{-i\Theta}$ и $\phi$. Это можно
сделать, т.к.\ уравнение Шредингера должно выполняться для всех $t$ и $x$.
Решение полученного уравнения имеет вид
\begin{equation}                                                  \label{ewavab}
  \Theta(t)=\Theta_0+e\int_0^t\!\!\!dsA_0(s),
\end{equation}
где $\Theta_0$ -- значение фазы волновой функции в начальный момент времени.

\begin{figure}[h,b,t]
\hfill\includegraphics[width=.6\textwidth]{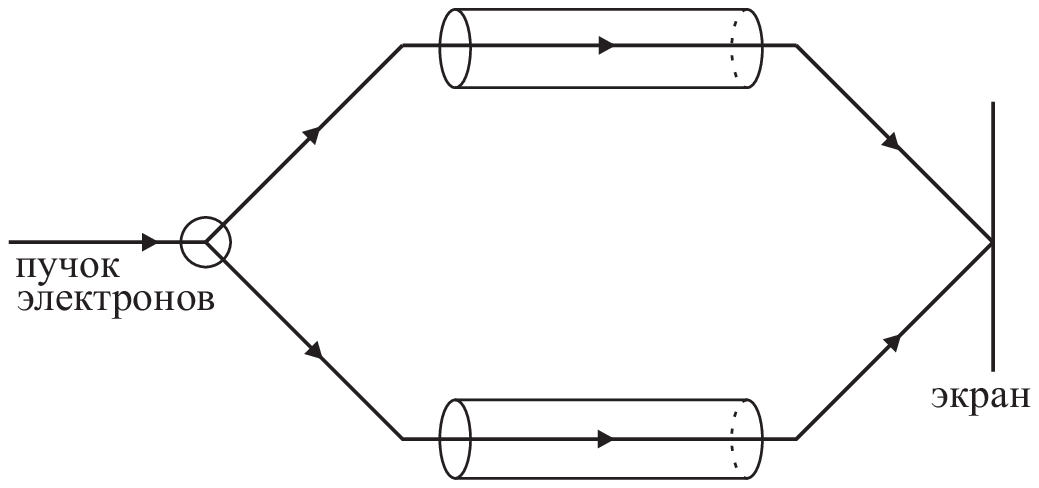}
\hfill {}
\\
\centering \caption{Пучок электронов делится на два пучка, которые пропускаются
через две металлические трубки, имеющие разные потенциалы. Затем пучки вновь
собираются и наблюдается интерференционная картина, которая зависит от разности
фаз электронов в разных пучках.\label{ahabohel}}
\end{figure}

Ааронов и Бом предложили следующий эксперимент, схема которого показана на
рис.\ref{ahabohel}. Пучок электронов делится на два пучка, которые пропускаются
через две металлические трубки, на которые подается различный потенциал. Затем
пучки собираются и на экране наблюдается интерференционная картина.
Электрический потенциал, который подается на трубки, зависит от времени.
Предполагается, что он равен нулю, пока оба пучка не окажутся в своих трубках.
Затем он возрастает до некоторых значений, которые отличаются внутри каждой
трубки, и снова падает до нуля перед выходом пучков из трубок. Таким образом,
пучки находятся в поле $A_0$ только внутри трубок. Интерференционная картина
зависит от разности фаз электронов в пучках, которую можно приближенно оценить
следующим образом.

Предположим, что электрон описывается волновой функцией $\psi(t,x)$, которая
удовлетворяет уравнению Шредингера (\ref{eschre}) во всем пространстве-времени
$\MR^4$. Мы считаем, что в каждый момент времени носитель волновой функции
отличен от нуля в небольшой окрестности пространства вблизи траектории частицы.
Только в этом случае вообще можно говорить о траектории частицы. В частности,
при прохождении электрона сквозь металлическую трубку предполагается, что
носитель волновой функции целиком лежит внутри трубки. Математически это можно
описать, выбрав в уравнении Шредингера (\ref{eschre}) соответствующий потенциал.
Этот гипотетический потенциал не меняет пространство-время, т.е.\ базу главного
расслоения, а только обеспечивает движение электронов по заданной траектории.
Изменение фазы электрона в верхнем пучке оценим следующим образом. Поскольку
потенциал электрического поля однороден внутри трубки и носитель волновой
функции целиком содержится внутри трубки, то можно считать, что фаза волновой
функции определяется интегралом (\ref{ewavab}). Обозначим моменты времени,
соответствующие расщеплению пучка и достижению экрана, соответственно $t_1$ и
$t_2$. Тогда фаза волновой функции электрона в верхнем пучке при достижении
экрана изменится на величину, задаваемую интегралом
\begin{equation*}
  \Theta_1=e\int_{t_1}^{t_2}\!\!\!dtA^{(1)}_0(t),
\end{equation*}
где $A_0^{(1)}(t)$ -- потенциал электрического поля в момент времени $t$, т.е.\
в той точке пространства, где в момент времени $t$ находится электрон из
верхнего пучка. Аналогично, изменение фазы волновой функции электрона из нижнего
пучка равно
\begin{equation*}
  \Theta_2=e\int_{t_1}^{t_2}\!\!\!dtA^{(2)}_0(t),
\end{equation*}
где $A_0^{(2)}$ -- потенциал электрического поля вдоль нижней траектории. Ясно,
что разность фаз электронов в верхнем и нижнем пучке,
$\Theta_{\Sa\Sb}=\Theta_2-\Theta_1$, можно записать в виде интеграла
\begin{equation}                                                  \label{eahboe}
  \Theta_{\Sa\Sb}=e\oint_\g dtA_0(t).
\end{equation}
вдоль замкнутого контура $\g$ в пространстве-времени, когда сначала проходится
нижняя половина контура, изображенного на рис.\ref{ahabohel}, а затем -- верхняя
половина в обратную сторону. На рис.\ref{ahabohel} показана проекция контура
$\g$ на пространственную плоскость.

Вернемся к геометрической интерпретации. Разность фаз электронов дается
интегралом (\ref{eahboe}), который однозначно определяется контуром $\g$ и
заданным на нем потенциалом $A_0$. Электрический потенциал $A_0$ представляет
собой временн\'ую компоненту локальной формы $\MU(1)$-связности на главном
расслоении $\MP\big(\MR^4,\pi,\MU(1)\big)$. Поэтому разность фаз (\ref{eahboe})
определяет элемент группы голономии
$\ex^{i\Theta_{\Sa\Sb}}\in\Phi\big((t_0,x_0),e\big)\subset\MU(1)$ в точке
$(t_0=0,x_0)$, где $x_0$ -- точка пространства, в которой пучок расщепляется, а
$e$ -- единица группы.

В заключение данного раздела рассмотрим преобразование компоненты локальной
формы $\MU(1)$-связности при изменении сечения. Из уравнения Шредингера
(\ref{eschre}) следует, что при вертикальном автоморфизме
\begin{equation*}
  \psi'=\ex^{ia}\psi,
\end{equation*}
где $a=a(t)$ -- дифференцируемая функция времени, компонента локальной формы
$\MU(1)$ связности преобразуется по правилу
\begin{equation*}
  eA'_0=eA_0+\dot a,
\end{equation*}
как и подобает компонентам локальной формы $\MU(1)$-связности.

Таким образом, в основе эффекта Ааронова--Бома, так же как и для фазы Берри,
лежит не топология, а нетривиальная геометрия, т.е.\ связность с нетривиальной
группой голономии. При этом топология пространства может быть как тривиальной,
так и нетривиальной.
\subsection{Магнитный потенциал}
Рассмотрим теперь случай, когда потенциал электрического поля равен нулю,
$A_0=0$. Предположим, что ковекторный потенциал магнитного поля зависит только
от пространственных координат $x^\mu$ и не завит от времени $t$ (статическое
поле). Тогда уравнение Шредингера примет вид
\begin{equation}                                                  \label{eshmaf}
\begin{split}
  i\dot\psi&=\frac1{2m}\eta^{\mu\nu}(\pl_\mu+ieA_\mu)
  (\pl_\nu+ieA_\nu)\psi+U\psi
\\
  &=\frac1{2m}\eta^{\mu\nu}(\pl^2_{\mu\nu}\psi+2ieA_\mu\pl_\nu\psi
  +ie\pl_\mu A_\nu\psi-e^2A_\mu A_\nu\psi)+U\psi.
\end{split}
\end{equation}
Пусть $\phi$ -- решение уравнения Шредингера в отсутствии потенциала магнитного
поля (\ref{esheqf}). Тогда нетрудно проверить, что функция
\begin{equation*}
  \psi=\ex^{-i\Theta}\phi,
\end{equation*}
где фаза $\Theta$ удовлетворяет уравнению
\begin{equation}                                                  \label{emagpa}
  \pl_\mu\Theta=eA_\mu
\end{equation}
является решением исходного уравнения Шредингера (\ref{eshmaf}).

\begin{figure}[h,b,t]
\hfill\includegraphics[width=.60\textwidth]{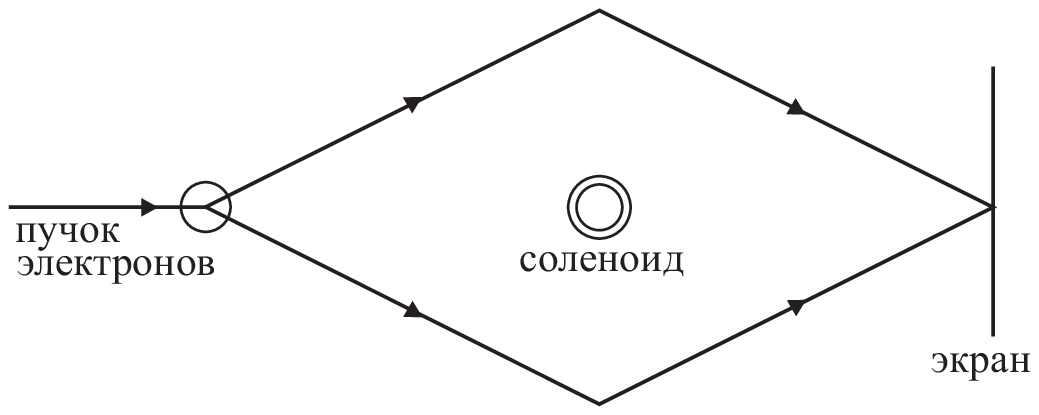}
\hfill {}
\\
\centering \caption{Пучок электронов делится на два пучка, которые огибают
тонкий соленоид с разных сторон. Затем пучки вновь собираются и наблюдается
интерференционная картина, которая зависит от разности
фаз электронов в разных пучках.\label{ahabohma}}
\end{figure}

Ааронов и Бом предложили эксперимент для определения фазы $\Theta$, схема
которого показана на рис.\ref{ahabohma}. В этом эксперименте пучок электронов
делится на два пучка, которые огибают бесконечно длинный соленоид с постоянным
магнитным потоком $\Phi$, который перпендикулярен плоскости рисунка, с разных
сторон. Затем пучки собираются вместе и на экране наблюдается интерференционная
картина, которая зависит от разности фаз электронов в разных пучках.

Для оценки разности фаз электронов сделаем те же предположения, что и в случае
электрического поля. А именно, будем считать, что уравнение Шредингера без
магнитного потенциала имеет решение с носителем, который сосредоточен в малой
окрестности траектории электрона. Мы предполагаем, что это можно осуществить
путем введения в уравнение (\ref{esheqf}) соответствующего потенциала. Этот
потенциал не меняет топологию пространства-времени, а только обеспечивает
движение электронов вдоль заданной траектории. Тогда для фазы решения уравнения
Шредингера с магнитным потенциалом справедливы уравнения (\ref{emagpa}).
Поскольку магнитное поле вне соленоида равно нулю,
$\pl_\mu A_\nu-\pl_\nu A_\mu=0$, то условия интегрируемости для системы
уравнений (\ref{emagpa}) выполнены. Поэтому разность фаз можно представить в
виде контурного интеграла
\begin{equation}                                                  \label{eahbkm}
  \Theta_{\Sa\Sb}=e\oint_\g dx^\mu A_\mu,
\end{equation}
где $\g$ -- замкнутый контур в четырехмерном пространстве времени, который
охватывает соленоид. Отметим, что слагаемое $dx^0A_0$ в подынтегральном
выражении равно нулю, т.к.\ $A_0=0$ по-предположению. Этот интеграл не зависит
от выбора контура, охватывающего соленоид, поскольку магнитное поле вне
соленоида равно нулю, $\pl_\mu A_\nu-\pl_\nu A_\mu=0$.

Фазу Ааронова--Бома, используя формулу Стокса, можно записать в виде
поверхностного интеграла
\begin{equation}                                                  \label{eahbom}
  \Theta_{\Sa\Sb}=\frac12e\iint_Sdx^\mu\wedge dx^\nu F_{\mu\nu}=e\Phi,
\end{equation}
где $F_{\mu\nu}$ -- напряженность магнитного поля (компоненты локальной 2-формы
кривизны) и $\Phi$ -- полный поток магнитного поля через соленоид. Заметим, что
для применения формулы Стокса, необходимо считать, что магнитный потенциал
определен в пространстве $\MR^3$ всюду, включая сам соленоид.

Геометрическая интерпретация рассмотренного эффекта Ааронова--Бома состоит в
следующем. Мы имеем то же самое главное расслоение, что и в случае
электрического потенциала, $\MP\big(\MR^4,\pi,\MU(1)\big)$, базой которого
является четырехмерное евклидово пространство, $(t,x)\in\MR\times\MR^3=\MR^4$, в
котором движутся пучки электронов, а структурной группой -- унитарная группа
$\MU(1)$ (фазовый множитель $\ex^{i\Theta}$ волновой функции). Однако связность
на нем другая: отличны от нуля только пространственные компоненты локальной
формы связности $A_\mu$, $\mu=1,2,3$. Разность фаз Ааронова--Бома (\ref{eahbkm})
однозначно определяется контуром $\g$ (тем же, что и в случае электрического
потенциала) и значениями компонент связности $A_\mu$ на нем. При записи
контурного интеграла в виде поверхностного (\ref{eahbom}) предполагается, что
связность задана на всем пространстве-времени $\MR^4$. Тем самым, мы
рассматриваем соленоид конечного радиуса, чтобы избежать сингулярностей.

Таким образом, главное расслоение тривиально, а фаза Ааронова--Бома
$\Theta_{\Sa\Sb}$, зависящая от связности и контура, однозначно определяет
элемент группы голономии.

Волновая функция электрона, как и в случае электрического потенциала, является
сечением ассоциированного расслоения
$\ME\big(\MR^4,\pi_\ME,\MC,\MU(1),\MP\big)$, базой которого является евклидово
пространство $\MR^4$, а структурной группой -- унитарная группа $\MU(1)$.
При вертикальном автоморфизме
\begin{equation*}
  \psi'=\ex^{ia}\psi,
\end{equation*}
где $a=a(x)$ -- дифференцируемая функция пространственных координат $x^\mu$,
$\mu=1,2,3$, потенциал магнитного поля преобразуется по-правилу
\begin{equation*}
  eA'_\mu=eA_\mu-\pl_\mu a.
\end{equation*}
Это следует из уравнения Шредингера (\ref{eshmaf}). Таким образом, компоненты
потенциала магнитного поля действительно ведут себя, как компоненты локальной
формы связности.

Поскольку разность фаз электронов (\ref{eahbkm}) определяется значениями
компонент локальной формы связности только вблизи контура интегрирования, то
базу тривиального главного расслоения $\MP\big(\MR^4,\pi,\MU(1)\big)$ можно
сузить, не меняя ответа. Например, можно удалить область пространства-времени,
лежащую внутри контура $\g$ и содержащую соленоид. Тогда база расслоения
перестанет быть односвязной. По этой причине эффект Ааронова--Бома часто
называют топологическим. Как было показано выше, это совершенно необязательно.
Достаточно считать, что магнитное поле отлично от нуля в ограниченной области на
плоскости рис.\ref{ahabohma} внутри контура интегрирования. Если считать, что
базой расслоения является евклидово пространство $\MR^4$ с выколотым соленоидом,
то формулу Стокса применить нельзя, т.к.\ она применима только для стягиваемых
контуров. Таким образом, эффект Ааронова--Бома, вызванный магнитным потенциалом,
также как и фазу Берри следует рассматривать как геометрический, а не
топологический.

Эффект Ааронова--Бома как с электрическим, так и с магнитным потенциалом привлек
большое внимание физиков по следующей причине. Согласно современным
представлениям в калибровочных моделях наблюдаемыми величинами являются только
калибровочно инвариантные функции. С этой точки зрения потенциал
электромагнитного поля $A_\al$, $\al=0,1,2,3$, сам по себе ненаблюдаем, т.к.\ не
является калибровочно инвариантным. В рассмотренных примерах электрического и
магнитного полей пучки электронов не подвергаются действию электромагнитных сил,
поскольку напряженности электрического и магнитного поля в областях, через
которые пролетают электроны, равны нулю. Поэтому, казалось бы, разность фаз в
пучках электронов должна быть равна нулю. Однако из уравнения Шредингера
следует, что это не так. Следует отметить, что наблюдаемым является не сам
потенциал электромагнитного поля, а интеграл от него по замкнутому контуру,
который определяет элемент группы голономии $\MU(1)$-связности, который является
инвариантным объектом.

Вскоре после публикации статьи, эффект Ааронова--Бома был подтвержден
экспериментально. Эффект, вызванный магнитным потенциалом наблюдался в
экспериментах \cite{WerBri60,Chambe60,BoHaWoGr60}.

\chapter{Векторные поля Киллинга                                 \label{skilve}}
В разделе \ref{sinvst} были рассмотрены геометрические структуры на
многообразии, которые инвариантны относительно действия некоторой группы
преобразований. Вопрос ставился так. Пусть задана группа преобразований
$(\MM,\MG)$, и требуется найти такие структуры, которые инвариантны относительно
этих преобразований. Обратная задача нахождения группы преобразований, которую
допускает заданная геометрическая структура на многообразии $\MM$ также очень
важна.

Изучение преобразований, которые сохраняют метрику пространства-времени играет
исключительно важную роль в математической физике. Достаточно сказать, что с
такими преобразованиями связаны наиболее важные законы сохранения. В настоящей
главе мы рассмотрим (псевдо-)риманово многообразие $(\MM,g)$ и найдем условия,
при которых метрика инвариантна относительно действия группы преобразований
$(\MM,\MG)$. Дадим определение векторных полей Киллинга, которые являются
генераторами локальных симметрий метрики, а также изучим некоторые из их
свойств.
\section{Изометрии и инфинитезимальные изометрии}
Рассмотрим $n$-мерное (псевдо-)риманово многообразие $(\MM,g)$ с метрикой
$g(x)=$\linebreak$g_{\al\bt}(x)dx^\al\otimes dx^\bt$, $\al,\bt=0,1,\dotsc,n-1$ и
соответствующей связностью Леви-Чивиты $\Gamma$.
\begin{defn}
Диффеоморфизм
\begin{equation*}
  \imath:\quad \MM\ni\quad x\mapsto x'=\imath(x)\quad\in\MM
\end{equation*}
называется {\em изометрией} или {\em движением} многообразия $\MM$, если он
сохраняет метрику,
\begin{equation}                                                  \label{eisoma}
  g(x)=\imath^*g(x'),
\end{equation}
где $\imath^*$ -- возврат отображения $\imath$.
\qed\end{defn}
\index{Изометрия (isometry)}\index{Движение (movement)}%
В настоящей главе, для простоты, мы не будем использовать знак тильды для
обозначения компонент связности Леви--Чивиты, т.к.\ аффинная связность общего
вида с кручением и неметричностью использоваться не будет.

Из условия инвариантности метрики (\ref{eisoma}) следует инвариантность
скалярного произведения векторов. Пусть $X,Y\in\MT_x(\MM)$ -- два произвольных
вектора из касательного пространства в точке $x\in\MM$. Тогда справедливо
равенство
\begin{equation*}
  g(X,Y)|_x=\imath^*g(\imath_*X,\imath_*Y)|_{\imath(x)}
  =g(\imath_*X,\imath_*Y)|_{\imath(x)},
\end{equation*}
которое эквивалентно определению (\ref{eisoma}).

Поскольку изометрия сохраняет метрику, то она сохраняет также связность
Леви-Чивиты, соответствующий тензор кривизны, экстремали и, вообще, все
геометрические объекты, которые определяются только метрикой.

Запишем отображение (\ref{eisoma}) в координатах. Пусть обе точки $x$ и $x'$
принадлежат одной координатной окрестности и имеют, соответственно, координаты
$x^\al$ и $x^{\prime\al}$. Тогда изометрия $\imath$ в координатах запишется в
виде условия
\begin{equation}                                                  \label{eisomt}
  g_{\al\bt}(x)=\frac{\pl x^{\prime\g}}{\pl x^\al}
  \frac{\pl x^{\prime\dl}}{\pl x^\bt}g_{\g\dl}(x'),
\end{equation}
связывающего компоненты метрики в различных точках многообразия. Это условие по
виду совпадает с правилом преобразования компонент метрики при преобразовании
координат (\ref{emetra}). Разница заключается в следующем. При преобразовании
координат мы считаем, что одной и той же точке $x\in\MM$ соответствует два
набора координат $\lbrace x^\al\rbrace$ и
$\lbrace x^{\al'}:=x^{\prime\al}\rbrace$ в двух различных системах координат.
При рассмотрении изометрий $x$ и $x'$ -- это две различные точки одного и того
же многообразия $\MM$.
\begin{prop}
Множество всех изометрий данного (псевдо-)риманова многообразия $(\MM,g)$
является группой, которую обозначим $\imath\in\MI(\MM)$.
\end{prop}
\begin{proof}
Две последовательных изометрии также являются изометрией. Тождественное
отображение многообразия $\MM$ является изометрией и представляет собой единицу
группы. У каждого диффеоморфизма $\imath$ есть обратной диффеоморфизм
$\imath^{-1}$, который является обратной изометрией.
\end{proof}

Если метрика на многообразии задана, т.е.\ определены значения ее компонент во
всех точках $x$, то соотношение (\ref{eisomt}) представляет собой уравнение на
функции $x'(x)$, которые определяют изометрию. В общем случае это уравнение не
имеет решений и у соответствующего (псевдо-)риманова многообразия нет никаких
нетривиальных изометрий. В этом случае группа изометрий состоит из одного
единичного элемента. Чем шире группа изометрий, тем уже класс соответствующих
(псевдо-)римановых многообразий.
\begin{exa}
Евклидово пространство $\MR^n$ с евклидовой метрикой $\dl_{\al\bt}$ допускает
группу изометрий, которая состоит из преобразований неоднородной группы вращений
$\MI\MO(n,\MR)$, $\dim\MI\MO(n,\MR)=\frac12n(n+1)$, состоящей из вращений,
сдвигов и отражений.
\qed\end{exa}

Группа изометрий $\MI(\MM)$ может быть дискретной или группой Ли.
\begin{defn}
Если группа изометрий $\MI(\MM)$ является группой Ли, то имеет смысл говорить об
инфинитезимальных преобразованиях (см.\ раздел \ref{sinftr}). В этом случае мы
говорим об {\em инфинитезимальных изометриях}. Каждая инфинитезимальная
изометрия генерируется некоторым достаточно гладким векторным полем, которое
называется векторным {\em полем Киллинга}.
\qed\end{defn}
\index{Инфинитезимальная изометрия (infinitesimal isometry)}%
\index{Векторное поле Киллинга (Killing vector field)}%
\index{Киллинга векторное поле (Killing vector field)}
\begin{com}
Дискретные изометрии (псевдо-)риманова многообразия, например, отражения, не
генерируются никакими векторными полями.
\qed\end{com}

Запишем условие инвариантности метрики относительно инфинитезимальных
преобразований из группы изометрий в координатах. В разделе \ref{svechs} было
показано, что каждое векторное поле генерирует однопараметрическую группу
преобразований, которая называется экспоненциальным отображением. Формально
условие инвариантности метрики записывается в виде равенства нулю производной Ли
вдоль векторного поля Киллинга $K=K^\al\pl_\al$ от метрики \cite{Killin92}
\begin{equation}                                                  \label{elidmk}
  \Lie_K g=0.
\end{equation}
Используя явное выражение для производной Ли (\ref{elidet}), это уравнение в
локальной системе координат принимает вид
\begin{equation}                                                  \label{ekileq}
   \nb_\al K_\bt+ \nb_\bt K_\al=0,
\end{equation}
где $K_\al:=K^\bt g_{\bt\al}$, а ковариантная производная
\begin{equation*}
  \nb_\al K_\bt=\pl_\al K_\bt-\Gamma_{\al\bt}{}^\g K_\g
\end{equation*}
строится по символам Кристоффеля $\Gamma_{\al\bt}{}^\g$.
\begin{defn}
Уравнение (\ref{ekileq}) называется {\em уравнением Киллинга}, а интегральные
кривые полей Киллинга называются {\em траекториями Киллинга}. Если
$K=K^\al\pl_\al$ -- векторное поле Киллинга, то ему соответствует 1-форма
$K=dx^\al K_\al$, где $K_\al:=K^\bt g_{\bt\al}$, которая называется
{\em формой Киллинга}, и для которой мы сохранили то же обозначение.
\qed\end{defn}
\index{Уравнение Киллинга (Killing equation)}%
\index{Киллинга уравнение (Killing equation)}%
\index{Траектория Киллинга (Killing trajectory)}%
\index{Киллинга траектория (Killing trajectory)}%
\index{Форма Киллинга (Killing form)}%

На любом (псевдо-)римановом многообразии $(\MM,g)$ уравнения Киллинга
(\ref{elidmk}) всегда имеют тривиальное решение $K=0$. Если уравнения Киллинга
имеют только тривиальное решение, то в этом случае непрерывные изометрии
отсутствуют.

Траектории Киллинга $\lbrace x^\al(t)\rbrace\in\MM$, где $t\in\MR$,
определяются системой обыкновенных дифференциальных уравнений
\begin{equation}                                                  \label{qlkodn}
  \dot x^\al=K^\al.
\end{equation}
Если траектория Киллинга при $t=0$ проходит через точку
$p=\lbrace p^\al\rbrace\in\MM$, то при малых $t$ она имеет вид
\begin{equation}                                                  \label{ekiltr}
  x^\al(t)=p^\al+tK^\al(p)+\osmall(t).
\end{equation}
Если в некоторой точке векторное поле Киллинга равно нулю, то эта точка
остается неподвижной, т.е.\ является стационарной точкой группы изометрий.
Поскольку изометрии определены для всего многообразия $\MM$ и образуют группу,
то векторные поля Киллинга обязаны быть полными, т.е.\ параметр $t$ должен
меняться на всей вещественной прямой $\MR$.

Если для (псевдо-)риманова многообразия $(\MM,g)$ известно векторное поле
Киллинга, то оно определяет не только инфинитезимальные изометрии, но и всю
однопараметрическую группу диффеоморфизмов. Для этого нужно найти интегральные
кривые $x(t)$, проходящие, через все точки многообразия $p\in\MM$. Если
$x(0)=p$, то каждому значению $t\in\MR$ соответствует диффеоморфизм
\begin{equation*}
  \imath:\quad \MM\ni\quad p\mapsto x(t)\quad\in\MM.
\end{equation*}

Уравнения для векторных полей Киллинга в ковариантной форме (\ref{ekileq}) можно
переписать в частных производных,
\begin{equation*}
  \pl_\al K_\bt+\pl_\bt K_\al-2 \Gamma_{\al\bt}{}^\g K_\g=0.
\end{equation*}
В моделях математической физики часто ставится задача нахождения векторов
Киллинга для заданной метрики на многообразии. Для решения этой задачи
бывает удобнее использовать контравариантные компоненты векторов Киллинга, для
которых уравнение Киллинга принимает вид
\begin{equation}                                                  \label{ekilco}
  g_{\al\g}\pl_\bt K^\g+g_{\bt\g}\pl_\al K^\g+K^\g\pl_\g g_{\al\bt}=0.
\end{equation}
Полученное уравнение линейно по компонентам метрики и компонентам векторов
Киллинга. Отсюда сразу следует, что две метрики, которые отличаются постоянным
множителем, имеют один и тот же набор векторов Киллинга. Кроме того, векторное
поле Киллинга определено с точностью до умножения на произвольную постоянную,
отличную от нуля. В частности, если $K^\al$ -- векторное поле Киллинга, то и
$-K^\al$ также является полем Киллинга. Также любая линейная комбинация полей
Киллинга является полем Киллинга.
\begin{com}
Векторные поля Киллинга не выдерживает умножения на функцию. Поэтому они не
образуют $\CC^\infty(\MM)$-модуль в отличие от множества всех векторных полей
$\CX(\MM)$.
\qed\end{com}
\begin{prop}
Пусть (псевдо-)риманово многообразие $(\MM,g)$ имеет $k<\dim\MM$ отличных от
нуля коммутирующих между собой и линейно независимых векторных полей Киллинга
$K_i$, $i=1,\dotsc,k$. Тогда существует такая система координат, в которой все
компоненты метрики не зависят от $k$ координат, соответствующих траекториям
Киллинга.
\end{prop}
\begin{proof}
В разделе \ref{svechs} была построена специальная система координат, связанная с
произвольным векторным полем, отличным от нуля. Применительно к векторным полям
Киллинга $K_i$ это означает, что существует такая система координат
$(x^1,\dotsc,x^n)$, в которой каждое поле Киллинга имеет только одну постоянную
компоненту, $K_i=\pl_i$. В этой системе координат уравнение (\ref{ekilco}) для
каждого поля Киллинга принимает особенно простой вид
\begin{equation*}
  \pl_i g_{\al\bt}=0,\qquad i=1,\dotsc,k.
\end{equation*}
Это значит, что все компоненты метрики не зависят от координат $x^i$. В этой
системе координат траектории Киллинга определяются уравнениями
\begin{equation*}
  \dot x^i=1,\quad\dot x^\mu=0,\qquad\mu=1,\dotsc,i-1,i+1,\dotsc,n.
\end{equation*}
Отсюда следует, что координатными линиями $x^i$ являются траектории Киллинга.
\end{proof}
Если риманово многообразие $(\MM,g)$ имеет два или более некоммутирующих
векторных полей Киллинга, то это отнюдь не означает, что существует такая
система координат, в которой компоненты метрики не зависят от двух или более
координат.
\begin{exa}
Рассмотрим двумерную сферу $\MS^2\hookrightarrow\MR^3$. Пусть метрика $g$ на
сфере индуцирована вложением. Риманово пространство $(\MS^2,g)$ имеет три
векторных поля Киллинга, соответствующих $\MS\MO(3)$ вращениям евклидова
пространства $\MR^3$. Легко понять, что на сфере не существует локальной системы
координат, в которой компоненты метрики не зависели бы от двух координат.
Действительно, это означает, что в данной системе координат компоненты метрики
постоянны, и, следовательно, кривизна равна нулю. Но это невозможно, поскольку
кривизна сферы постоянна.
\qed\end{exa}

В общей теории относительности мы предполагаем, что пространство-время является
псевдоримановым многообразием с метрикой лоренцевой сигнатуры. Используя
понятие векторного поля Киллинга, можно дать инвариантное
\begin{defn}
Пространство-время или его область называются {\em статическими}, если на них
существует времениподобное векторное поле Киллинга.
\qed\end{defn}
\index{Статическое пространство-время (static space-time)}%
\index{Пространство-время статическое (static space-time)}%

\begin{figure}[h,b,t]
\hfill\includegraphics[width=.4\textwidth]{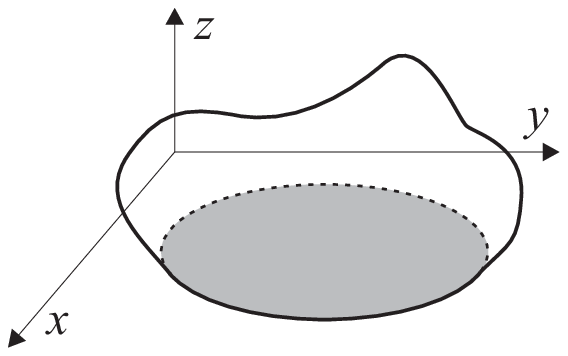}
\hfill {}
\centering\caption{Двумерная поверхность, вложенная в трехмерное евклидово
пространство. Нижняя часть поверхности является плоской.}
\label{fsefla}
\end{figure}

Векторные поля Киллинга определены глобально и удовлетворяют уравнениям
Киллинга на всем $\MM$. В то же время уравнения Киллинга -- это локальный
объект, в том смысле, что они определены в каждой окрестности и могут иметь
нетривиальные решения только на некотором подмногообразии $\MU\subset\MM$.
\begin{exa}
Рассмотрим гладкую замкнутую двумерную поверхность $\MM$, вложенную в трехмерное
евклидово пространство $\MR^3$, как показано на рис.\ref{fsefla}. Отличительной
особенностью этой поверхности является то, что ее нижняя часть является плоской.
Пусть метрика на $\MM$ индуцирована вложением $\MM\hookrightarrow\MR^3$. Тогда
уравнения Киллинга в нижней части поверхности легко интегрируются, как и на
евклидовой плоскости. Однако найденные нетривиальные решения не будут в общем
случае нетривиальными на всем $\MM$. Действительно, верхняя часть поверхности
может быть искривлена так, что уравнения Киллинга на ней имеют только
тривиальное решение. Следовательно, векторные поля Киллинга могут быть
нетривиальны только на части многообразия $\MM$.
\qed\end{exa}

Группа изометрий $\MI(\MM)$ является группой преобразований многообразия $\MM$,
которая действует слева или справа, в зависимости от нашего соглашения. Эта
группа преобразований может иметь подгруппу $\MH\subset\MI(\MM)$, действующую на
$\MM$ свободно и собственно разрывно. Тогда фактор пространство $\MM/\MH$
является многообразием. Более того, на нем можно определить метрику, поскольку
$\MH$ -- изометрия. Таким образом, фактор пространство $\MM/\MH$ превращается в
(псевдо-)риманово многообразие. Поскольку каждое векторное поле Киллинга на
$\MM$ при факторизации переходит в некоторое поле Киллинга на $\MM/\MH$ и
линейная независимость полей Киллинга при этом сохраняется, то группа изометрий
фактор пространства $\MM/\MH$ совпадает с $\MI(\MM)$.
\section{Однородные и изотропные многообразия}
С каждым полем Киллинга как и с произвольным векторным полем связана
однопараметрическая группа преобразований, которая в данном случае
сохраняет метрику. Произвольная линейная комбинация векторов Киллинга ввиду
линейности уравнения Киллинга (\ref{ekileq}) снова дает вектор Киллинга.
То есть поля Киллинга образуют векторное пространство над полем вещественных
чисел. В этом векторном пространстве можно ввести билинейную операцию.
Простые вычисления показывают, что коммутатор двух векторных полей Киллинга
$K_1$ и $K_2$ снова дает поле Киллинга:
\begin{equation*}
  \Lie_{[K_1,K_2]}g=\Lie_{K_1}\circ\Lie_{K_2}g-\Lie_{K_2}\circ\Lie_{K_1}g=0,
\end{equation*}
Отсюда следует, что векторные поля Киллинга образуют алгебру Ли $\Gi(\MM)$ над
полем вещественных чисел, которая является подалгеброй алгебры Ли множества всех
векторных полей, $\Gi(\MM)\subset\CX(\MM)$. Эта алгебра является алгеброй Ли
группы изометрий $\MI(\MM)$.

Уравнения Киллинга (\ref{ekileq}) накладывают сильные ограничения на векторные
поля Киллинга, которые мы сейчас обсудим. Воспользовавшись тождеством для
коммутатора ковариантных производных (\ref{eonfco}), получаем равенство
\begin{equation}                                                  \label{eqcoki}
  \nb_\al \nb_\bt K_\g-\nb_\bt \nb_\al K_\g=-R_{\al\bt\g}{}^\dl K_\dl.
\end{equation}
Теперь воспользуемся тождеством (\ref{eanfti}) для тензора кривизны и
уравнениями Киллинга (\ref{ekileq}). В результате получим тождество для
векторных полей Киллинга:
\begin{equation*}
  \nb_\al\nb_\bt K_\g+\nb_\bt\nb_\g K_\al+\nb_\g\nb_\al K_\bt=0,
\end{equation*}
где слагаемые отличаются циклической перестановкой индексов. Это равенство
позволяет переписать уравнение (\ref{eqcoki}) в виде
\begin{equation}                                                  \label{emaeqk}
  \nb_\g\nb_\al K_\bt= R_{\al\bt\g}{}^\dl K_\dl.
\end{equation}

Полученное равенство является следствием уравнений Киллинга, но не эквивалентно
им. Оно позволяет сделать важные выводы. Предположим, что компоненты вектора
Киллинга бесконечно дифференцируемы в некоторой точке многообразия $p\in\MM$ и
разлагаются в ряд Тейлора, который сходится в некоторой окрестности $\MU_p$.
Допустим, что в точке $p\in\MM$ нам известны компоненты формы Киллинга
$K_\al(p)$ и их первых производных $\pl_\bt K_\al(p)$. Тогда соотношения
(\ref{emaeqk}) позволяют вычислить все вторые производные от компонент формы
Киллинга $\pl^2_{\bt\g}K_\al$. Теперь возьмем ковариантную производную от
равенства (\ref{emaeqk}) и получим некоторое соотношение, линейное по третьим
производным. Из него можно найти все третьи производные от вектора Киллинга и
т.д.\ до бесконечности. Важно отметить, что все соотношения линейны по
компонентам формы Киллинга и их производным. Это значит, что в окрестности
$\MU_p$ компоненты векторного поля Киллинга имеют вид
\begin{equation}                                                  \label{ekilde}
  K_\al(x)=A_\al{}^\bt(x,p)K_\bt(p)
  +B_{\al}{}^{\bt\g}(x,p)\left[\pl_\bt K_\g(p)-\pl_\bt K_\al(p)\right],
\end{equation}
где $A_\al{}^\bt(x,p)$ и $B_{\al}{}^{\bt\g}(x,p)$ -- некоторые функции.
Антисимметрия последнего слагаемого по индексам $\bt,\g$ связана с тем, что
симметризованная частная производная выражается через компоненты формы Киллинга
в силу уравнения Киллинга (\ref{ekileq}). Таким образом, компоненты формы
Киллинга в окрестности $\MU_p$ являются линейными функциями от компонент формы
Киллинга в точке $p$ и ее внешней производной в той же точке.

Пусть векторные поля Киллинга гладкие на $\MM$ и их компоненты разлагаются в
ряды Тейлора в окрестности каждой точки $p\in\MM$. Обозначим через $\MU_p$
окрестность точки $p$, в которой разложение (\ref{ekilde}) справедливо и
обратимо, т.е.\ аргументы $x$ и $p$ можно поменять местами для некоторых новых
матриц $A$ и $B$. Рассмотрим точку $q$, которая лежит вне $\MU_p$. Для этой
точки также справедливо обратимое разложение вида (\ref{ekilde}) в некоторой
окрестности $\MU_q$. Предположим, что точка $q$ лежит достаточно близко к
$\MU_p$ так, что окрестности пересекаются, $\MU_p\cap\MU_q\ne\emptyset$. Тогда
для всех точек из пересечения $x\in\MU_p\cap\MU_q$ справедливо разложение
(\ref{ekilde}) по компонентам форм Киллинга $K(p)$ и $K(q)$ и их внешним
производным. Отсюда следует, что компоненты формы Киллинга и ее внешней
производной в точке $q$ линейно выражаются через компоненты формы Киллинга и ее
внешней производной в точке $p$. Таким образом, разложение (\ref{ekilde})
справедливо также в объединении $\MU_p\cup\MU_q$. Это построение можно
продолжить на все многообразие $\MM$. Поэтому разложение (\ref{ekilde})
справедливо для всех точек $x\in\MM$.

Теперь предположим, что (псевдо-)риманово многообразие $(\MM,g)$ имеет несколько
векторных полей Киллинга $K_i$, $i=1,\dotsc,\Sn$. Тогда для каждого векторного
поля Киллинга справедливо разложение (\ref{ekilde})
\begin{equation}                                                  \label{ekildr}
  K_{i\al}(x)=A_\al{}^\bt(x,p)K_{i\bt}(p)
  +B_{\al}{}^{\bt\g}(x,p)\left[\pl_\bt K_{i\g}(p)-\pl_\g K_{i\bt}(p)\right],
  \qquad i=1,\dotsc,\Sn,
\end{equation}
с теми же функциями $A_\al{}^\bt(x,p)$ и $B_{\al}{}^{\bt\g}(x,p)$. Эти функции
одинаковы для всех полей Киллинга, потому что соотношения (\ref{emaeqk}) линейны
по компонентам векторов Киллинга и их производным. Они полностью определяются
метрикой, тензором кривизны и их ковариантными производными. В полученном
разложении точка $p\in\MM$ произвольна, но фиксирована, а точка $x\in\MM$
пробегает все многообразие.

Соотношение (\ref{emaeqk}) представляет собой систему уравнений в частных
производных на компоненты формы Киллинга, у которой есть нетривиальные условия
разрешимости. Одно из этих условий в ковариантной форме имеет вид
\begin{equation*}
  \left[\nb_\g\nb_\dl\right]\nb_\al K_\bt
  =-R_{\g\dl\al}{}^\e\nb_\e K_\bt-R_{\g\dl\bt}{}^\e\nb_\al K_\e,
\end{equation*}
где квадратные скобки обозначают коммутатор ковариантных производных.
Подстановка в левую часть этого уравнения исходного выражения для вторых
производных от формы Киллинга (\ref{emaeqk}) после несложных алгебраических
преобразований приводит к равенству
\begin{equation}                                                  \label{erelkd}
  \left(R_{\al\bt\g}{}^\e\dl_\dl^\z-R_{\al\bt\dl}{}^\e\dl_\g^\z
  +R_{\g\dl\al}{}^\e\dl_\bt^\z-R_{\g\dl\bt}{}^\e\dl_\al^\z\right)\nb_\z K_\e
  =\left(\nb_\g R_{\al\bt\dl}{}^\e-\nb_\dl R_{\al\bt\g}{}^\e\right)K_\e.
\end{equation}
Если кривизна нетривиальна, то это уравнение дает некоторые линейные соотношения
между компонентами формы Киллинга $K_\al$ и их ковариантными производными
$\nb_\bt K_\al$. Наоборот, если существует некоторая информация в
формах Киллинга, то полученное уравнение может определить структуру тензора
кривизны. В теореме \ref{tkildi}, которая сформулирована ниже, соотношение
(\ref{erelkd}) будет использовано для доказательства того, что однородное и
изотропное многообразие является пространством постоянной кривизны.

Перейдем к определениям.
\begin{defn}
(Псевдо-)риманово многообразие $(\MM,g)$ размерности $\dim\MM=n$ называется
{\em однородным в точке} $p\in\MM$, если существуют инфинитезимальные изометрии,
которые переводят эту точку в любую другую точку из некоторой окрестности
$\MU_p$. Другими словами, метрика должна допускать такие векторные поля
Киллинга, которые в точке $p$ имеют все возможные направления. Поскольку векторы
Киллинга образуют линейное пространство, то в сопряженном пространстве
достаточно существования такого набора из $n$ форм Киллинга
$K^{(\g)}=dx^\al K_\al{}^{(\g)}(x,p)$, где индекс $\g$ в скобках нумерует формы
Киллинга, что выполнены условия:
\begin{equation}                                                  \label{ehocok}
  K_\al{}^{(\g)}(p,p)=\dl_\al^\g.
\end{equation}
Если (псевдо-)риманово многообразие $(\MM,g)$ однородно в каждой своей точке, то
оно называется {\em однородным}. Другими словами, группа преобразований
действует транзитивно.

(Псевдо-)риманово многообразие $(\MM,g)$ называется {\em изотропным в точке}
$p\in\MM$, если существуют такие инфинитезимальные изометрии с формами Киллинга
$K(x)$, которые оставляют эту точку на месте, т.е.\ $K(p)=0$, и для которых
внешняя производная $dK(p)$ в точке $p$ принимает любое значение в пространстве
2-форм $\Lm_2(\MM)|_p$ в точке $p$. Для этого достаточно существования такого
набора из $\frac12n(n-1)$ форм Киллинга
$K^{[\g\dl]}=-K^{[\dl\g]}=dx^\al K_\al{}^{[\g\dl]}(x,p)$, где индексы $\g,\dl$
нумеруют формы Киллинга, что выполнены условия:
\begin{equation}                                                  \label{eprkif}
\begin{split}
  K_\al{}^{[\g\dl]}(p,p)&=0,
\\
  \left.\frac{\pl K_\bt{}^{[\g\dl]}(x,p)}{\pl x^\al}\right|_{x=p}
  &=\dl_{\al\bt}^{\g\dl}-\dl_{\al\bt}^{\dl\g}.
\end{split}
\end{equation}
Если (псевдо-)риманово многообразие $(\MM,g)$ изотропно в каждой своей точке, то
оно называется {\em изотропным}.
\qed\end{defn}
\index{Однородное пространство (homogeneous space)}%
\index{Изотропное пространство (isotropic space)}%

В силу непрерывности, наборы форм $K^{(\g)}$ и $K^{[\g\dl]}$ линейно независимы
в некоторой окрестности точки $p$.
\begin{prop}                                                      \label{pisoho}
Любое изотропное (псевдо-)риманово многообразие $(\MM,g)$ является также
однородным.
\end{prop}
\begin{proof}
Если многообразие изотропно, то формы Киллинга $K^{[\g,\dl]}(x,p)$ и
$K^{[\g,\dl]}(x,p+dp)$ удовлетворяют условиям (\ref{eprkif}) в близких точках
$p$ и $p+dp$ соответственно. Любая их линейная комбинация будет формой Киллинга
и, следовательно, производная
\begin{equation*}
  \frac{\pl K^{[\g\dl]}(x,p)}{\pl p^\al}
\end{equation*}
также будет формой Киллинга. Вычислим производную по $x$ этой формы Киллинга в
точке $p$. Из первого условия в (\ref{eprkif}) следует равенство
\begin{equation*}
  \frac\pl{\pl p^\al}K_\bt{}^{[\g\dl]}(p,p)
  =\left.\frac{\pl K_\bt{}^{[\g\dl]}(x,p)}{\pl x^\al}\right|_{x=p}
  +\left.\frac{\pl K_\bt{}^{[\g\dl]}(x,p)}{\pl p^\al}\right|_{x=p}=0.
\end{equation*}
Откуда, с учетом второго условия в (\ref{eprkif}), получаем равенство
\begin{equation*}
  \left.\frac{\pl K_\bt{}^{[\g\dl]}(x,p)}{\pl p^\al}\right|_{x=p}=
  -\dl_{\al\bt}^{\g\dl}+\dl_{\al\bt}^{\dl\g}.
\end{equation*}
Отсюда следует, что из форм Киллинга $K^{[\g\dl]}$ можно построить форму
Киллинга, которая в точке $p$ принимает любое заданное значение $dx^\al a_\al$,
где $a_\al\in\MR$. Для этого достаточно положить
\begin{equation*}
  K_\al=\frac{a_\g}{n-1}\frac{\pl K_\al{}^{[\g\dl]}(x,p)}{\pl p^\dl}.
\end{equation*}
Выбрав соответствующим образом постоянные $a_\g$, получим набор форм Киллинга,
который удовлетворяет условиям (\ref{ehocok}).
\end{proof}

\begin{theorem}                                                   \label{tkildi}
Алгебра Ли $\Gi(\MM)$ инфинитезимальных изометрий связного (псевдо-)риманова
многообразия $\MM$ имеет размерность не более, чем $\frac12n(n+1)$, где
$n=\dim\MM$. Если размерность максимальна, $\dim\Gi(\MM)=\frac12n(n+1)$, то
$\MM$ есть пространство постоянной кривизны.
\end{theorem}
\begin{proof}
Размерность алгебры Ли $\Gi(\MM)$ равна максимальному числу линейно независимых
векторных полей Киллинга на многообразии $\MM$. Из равенства (\ref{ekildr})
следует, что число независимых векторных полей Киллинга $\Sn$ не может превышать
числа независимых компонент формы $\lbrace K_\al(p)\rbrace$ и ее внешней
производной $\lbrace \pl_\bt K_\al(p)-\pl_\al K_\bt(p)\rbrace$ в фиксированной
точке $p\in\MM$. Число независимых компонент любой 1-формы в фиксированной точке
не превосходит $n$, а число независимых компонент внешней производной не может
превышать $\frac12n(n-1)$. Поэтому справедливо следующее ограничение на
размерность алгебры Ли векторных полей Киллинга:
\begin{equation*}
  \dim\Gi(\MM)\le n+\frac12n(n-1)=\frac12n(n+1).
\end{equation*}
Это доказывает первое утверждение теоремы.

Однородные и изотропные многообразия имеют максимальное число $\frac12n(n+1)$
векторных полей Киллинга и, в силу разложения (\ref{ekildr}), определяют все
возможные векторы Киллинга на многообразии $\MM$. Следовательно, если некоторое
многообразие имеет максимальное число независимых полей Киллинга, то оно с
необходимостью должно быть однородным и изотропным.

Теперь докажем, что любое однородное и изотропное пространство является
пространством постоянной кривизны. Если пространство однородно и изотропно, то
для каждой точки $x\in\MM$ найдутся такие формы Киллинга, для которых
$K_\al(x)=0$, а $\nb_\bt K_\al(x)$ является произвольной антисимметричной
матрицей. Отсюда следует, что антисимметризованный коэффициент при $\nb_\z K_\e$
в уравнении (\ref{erelkd}) должен быть равен нулю, что приводит к равенству
\begin{equation}                                                  \label{erecuk}
   R_{\al\bt\g}{}^\e\dl_\dl^\z-R_{\al\bt\dl}{}^\e\dl_\g^\z
  +R_{\g\dl\al}{}^\e\dl_\bt^\z-R_{\g\dl\bt}{}^\e\dl_\al^\z
  =R_{\al\bt\g}{}^\z\dl_\dl^\e-R_{\al\bt\dl}{}^\z\dl_\g^\e
  +R_{\g\dl\al}{}^\z\dl_\bt^\e- R_{\g\dl\bt}{}^\z\dl_\al^\e.
\end{equation}
Ранее было доказано (доказательство предложения \ref{pisoho}), что для
произвольной точки $x\in\MM$ существуют также такие формы Киллинга, которые
принимают в этой точке произвольные значения. Следовательно, из уравнений
(\ref{erelkd}) и (\ref{erecuk}) вытекает равенство
\begin{equation}                                                  \label{erecud}
  \nb_\g R_{\al\bt\dl}{}^\e=\nb_\dl R_{\al\bt\g}{}^\e.
\end{equation}
Теперь свернем уравнение (\ref{erecuk}) по индексам $\dl,\z$ и опустим верхний
индекс. В результате получим выражение тензора кривизны через тензор Риччи и
метрику:
\begin{equation}                                                  \label{ecurik}
  (n-1)R_{\al\bt\g\dl}=R_{\bt\dl}g_{\al\g}-R_{\al\dl}g_{\bt\g}.
\end{equation}
Правая часть этой формулы должна быть антисимметрична по индексам $\dl,\g$.
Поэтому возникает дополнительное ограничение
\begin{equation*}
  R_{\bt\dl}g_{\al\g}-R_{\al\dl}g_{\bt\g}=
  -R_{\bt\g}g_{\al\dl}+ R_{\al\g}g_{\bt\dl}.
\end{equation*}
Свертка полученного равенства по индексам $\bt,\g$ дает связь между тензором
Риччи и скалярной кривизной:
\begin{equation}                                                  \label{ericsk}
  R_{\al\dl}=\frac1n Rg_{\al\dl}.
\end{equation}
Подстановка этого выражения в (\ref{ecurik}) приводит к следующему выражению для
тензора кривизны
\begin{equation}                                                  \label{ecumek}
  R_{\al\bt\g\dl}=\frac R{n(n-1)}
  \left(g_{\al\g}g_{\bt\dl}-g_{\al\dl}g_{\bt\g}\right).
\end{equation}

Теперь осталось доказать, что скалярная кривизна $ R$ однородного и изотропного
пространства постоянна. Для этой цели используем свернутые тождества Бианки
\begin{equation*}
  2\nb_\bt R_\al{}^\bt-\nb_\al R=0.
\end{equation*}
Подставляя в это тождество выражение для тензора Риччи (\ref{ericsk}), получаем
условие
\begin{equation*}
  \left(\frac2n-1\right)\pl_\al R=0.
\end{equation*}
При $n\ge3$ отсюда следует $R=\const$.

Случай $n=2$ требует особого рассмотрения. Свертка равенства (\ref{erecud}) по
индексам $\bt,\e$ приводит к равенству
\begin{equation*}
  \nb_\g R_{\al\dl}-\nb_\dl R_{\al\g}=0.
\end{equation*}
дальнейшая свертка с $g^{\al\dl}$ с учетом уравнения (\ref{ericsk}) приводит к
условию $\pl_\g R=0$, т.е.\ $R=\const$ и при $n=2$.

Таким образом, скалярная кривизна в выражении для полного тензора кривизны
(\ref{ecumek}) равна константе, $R=\const$, и максимально симметричное
(псевдо-)риманово многообразие является пространством постоянной кривизны.
\end{proof}
\begin{com}
Если тензор кривизны имеет вид (\ref{ecumek}), где $R=\const$, то
соответствующее многообразие является пространством постоянной кривизны,
т.к.\ ковариантная производная от метрики в римановой геометрии равна нулю,
$\nb_\e R_{\al\bt\g\dl}=0$. Обратное, вообще говоря, неверно. У пространства
постоянной кривизны тензор кривизны не обязательно имеет вид (\ref{ecumek}).
Примером является полупростая группа Ли (см.\ раздел \ref{sligri}). Отсюда
следует, что не всякое пространство постоянной кривизны является максимально
симметричным.
\qed\end{com}

\begin{exa}
Рассмотрим евклидово пространство $\MR^n$, на котором задана метрика нулевой
кривизны, т.е.\ $R_{\al\bt\g\dl}=0$. Ясно, что это пространство постоянной
нулевой кривизны. Тогда в $\MR^n$ существует такая система координат $x^\al$,
$\al=1,\dotsc,n$ в которой все компоненты метрики постоянны. В этой системе
координат символы Кристоффеля равны нулю и уравнение для векторов Киллинга
(\ref{emaeqk}) принимает простой вид:
\begin{equation*}
  \pl^2_{\bt\g}K_\al=0.
\end{equation*}
Общее решение этого уравнения имеет вид
\begin{equation*}
  K_\al(x)=a_\al+b_{\al\bt}x^\bt,
\end{equation*}
где $a_\al$ и $b_{\al\bt}$ -- некоторые постоянные. Из уравнения Киллинга
(\ref{ekileq}) следует, что это выражение задает форму Киллинга тогда и только
тогда, когда матрица $b_{\al\bt}$ антисимметрична, т.е.\
$b_{\al\bt}=-b_{\bt\al}$. Следовательно, можно задать $\frac12n(n+1)$ линейно
независимых форм Киллинга:
\begin{align*}
  K_\al{}^{(\g)}(x)&=\dl_\al^\g,
\\
  K_\al{}^{[\g\dl]}(x)&=\dl_\al^\dl x^\g-\dl_\al^\g x^\dl.
\end{align*}
Тогда произвольная форма Киллинга выражается в виде линейной комбинации
\begin{equation*}
  K_\al=a_\g K_\al{}^{(g)}+\frac12b_{\dl\g}K_\al{}^{[\g\dl]}.
\end{equation*}
В полученном выражении $n$ векторов Киллинга $K^{(\g)}$ генерируют трансляции в
$\MR^n$ вдоль осей координат, а $\frac12n(n-1)$ векторов $K^{[\g\dl]}$ --
вращения вокруг начала координат. Таким образом, метрика пространства нулевой
кривизны допускает максимальное число $\frac12n(n+1)$ векторов Киллинга и
поэтому является однородным и изотропным пространством.

Известно, что линейным преобразованием координат $x^\al$ метрику можно
преобразовать к диагональному виду, когда на диагонали будут стоять $\pm1$, в
зависимости от сигнатуры исходной метрики. Если метрика риманова, то после
преобразования координат, она примет стандартный вид $g_{\al\bt}=\dl_{\al\bt}$.
Эта метрика инвариантна относительно неоднородной группы вращений $\MI\MO(n)$.
\qed\end{exa}
\section{Свойства векторных полей Киллинга}
Векторные поля Киллинга обладают рядом замечательных свойств. Начнем с
простейших.
\begin{prop}                                                      \label{poriki}
Длина вектора Киллинга остается постоянной вдоль траектории Киллинга:
\begin{equation}                                                  \label{ekilen}
  K^\al\pl_\al K^2=0.
\end{equation}
\end{prop}
\begin{proof}
Свернем уравнения Киллинга (\ref{ekileq}) с $K^\al K^\bt$:
\begin{equation*}                                                    \tag*{\qed}
  2K^\al K^\bt\nb_\al K_\bt=K^\al\nb_\al K^2=K^\al\pl_\al K^2=0.
\end{equation*}
\renewcommand{\qed}{}\end{proof}

\begin{cor}
Если векторные поля Киллинга существуют на лоренцевом многообразии, то они имеют
определенную ориентацию: времениподобную, светоподобную или
пространственноподобную.
\qed\end{cor}

Сравним траектории Киллинга с экстремалями \cite{Bianch18}.
\begin{prop}                                                      \label{pkiext}
Пусть $(\MM,g)$ -- (псевдо-)риманово многообразие с векторным полем Киллинга
$K$. Траектории Киллинга являются экстремалями тогда и только тогда, когда их
длина постоянна на $\MM$, $K^2=\const$.
\end{prop}
\begin{proof}
Рассмотрим траектории Киллинга $x^\al(t)$, которые определяются системой
уравнений
\begin{equation}                                                  \label{eqkilt}
  \dot x^\al=K^\al.
\end{equation}
Дифференцируя эти уравнения по параметру $t$, получим равенство
\begin{equation*}
  \ddot x^\al=\pl_\bt K^\al\dot x^\bt
  =(\nb_\bt K^\al-\Gamma_{\bt\g}{}^\al K^\g)\dot x^\bt.
\end{equation*}
С учетом уравнения (\ref{eqkilt}) это приводит к уравнению
\begin{equation}                                                  \label{eqslik}
  \ddot x^\al=K^\bt\nb_\bt K^\al-\Gamma_{\bt\g}{}^\al
  \dot x^\bt\dot x^\g.
\end{equation}
Уравнения Киллинга позволяют переписать первое слагаемое в правой части в виде
\begin{equation*}
  K^\bt\nb_\bt K^\al=-\frac12g^{\al\bt}\pl_\bt K^2.
\end{equation*}
Тогда уравнения (\ref{eqslik}) примут вид
\begin{equation*}
  \ddot x^\al=-\frac12g^{\al\bt}\pl_\bt K^2-\Gamma_{\bt\g}{}^\al
  \dot x^\bt\dot x^\g.
\end{equation*}
Это уравнение совпадает с уравнением для экстремалей (\ref{eextre}) тогда и
только тогда, когда $K^2=\const$.
\end{proof}
Доказанное утверждение показывает, что далеко не каждая траектория Киллинга
является экстремалью.
\begin{exa}
Рассмотрим евклидову плоскость $\MR^2$ с евклидовой метрикой. Эта метрика
инвариантна относительно трехпараметрической неоднородной группы вращений
$\MI\MO(2)$. Обозначим декартовы и полярные координаты на плоскости
соответственно $x,y$ и $r,\vf$. Тогда векторные поля Киллинга имеют вид
$K_1=\pl_\vf$ для вращений и $K_2=\pl_x$, $K_3=\pl_y$ для сдвигов. Квадраты
длин векторов Киллинга равны:
\begin{equation*}
  K^2_1=r^2,\qquad K^2_2=K^2_3=1.
\end{equation*}
Векторы Киллинга $K_2$ и $K_3$ имеют постоянную длину, их траекториями Киллинга
являются прямые линии, которые являются экстремалями. Это согласуется с
предложением \ref{pkiext}. Траекториями Киллинга для вращений $K_1$ являются
концентрические окружности с центром в начале координат. Длина вектора Киллинга
$K_1$ постоянна на траекториях в соответствии с предложением \ref{poriki},
однако непостоянна на всей плоскости $\MR^2$. Соответствующие траектории
Киллинга -- окружности -- не являются экстремалями.
\qed\end{exa}
\begin{exa}
Рассмотри полупростую группу Ли $\MG$, как (псевдо-)риманово пространство с
формой Киллинга--Картана в качестве метрики (см.\ раздел \ref{sligri}). Это --
пространство постоянной кривизны. Левоинвариантные векторные поля генерируют
групповые преобразования справа, а правоинвариантные -- слева. Групповые
преобразования слева и справа сохраняют метрику, и, следовательно, лево- и
правоинвариантные векторные поля являются полями Киллинга. Длина этих полей
Киллинга равна $\pm1$. Поэтому соответствующие траектории Киллинга являются
экстремалями.
\qed\end{exa}

Свертывая уравнения Киллинга (\ref{ekileq}) с метрикой, получаем, что
дивергенция поля Киллинга равна нулю:
\begin{equation}                                                  \label{edikiv}
  \nb_\al K^\al=0.
\end{equation}
Ковариантная производная $\nb^\bt$ со связностью Леви--Чивита от
уравнения Киллинга (\ref{ekileq}) с учетом уравнения (\ref{eonfco}) для
перестановки ковариантных производных и уравнения (\ref{edikiv}) приводит к
уравнению
\begin{equation}                                                  \label{eivkil}
  \triangle K_\al=R_{\al\bt} K^\bt,
\end{equation}
$\triangle:=\nb^\bt\nb_\bt$ -- оператор Лапласа--Бельтрами на многообразии
$\MM$. Для пространства постоянной кривизны тензор Риччи выражается через
скалярную кривизну (\ref{ericsk}), и уравнение (\ref{eivkil}) принимает вид
\begin{equation*}
  \triangle K_\al=\frac Rn K_\al,\qquad R=\const.
\end{equation*}
То есть каждая компонента формы Киллинга является собственным вектором оператора
Лапласа--Бельтрами.
\begin{prop}
Пусть $X,Y\in\CX(\MM)$ -- два произвольных векторных векторных поля на
(псевдо-)римановом многообразии $(\MM,g)$ и $K$ -- векторное поле Киллинга.
Тогда справедливо равенство
\begin{equation*}
  g\big((\Lie_K-\nb_K)X,Y\big)+g\big(X,(\Lie_K-\nb_K)Y\big)=0,
\end{equation*}
где $\Lie_KX=[K,X]$ -- производная Ли и
$\nb_K X=K^\al(\pl_\al X^\bt+\Gamma_{\al\g}{}^\bt X^\g)\pl_\bt$
-- ковариантная производная векторного поля $X$ вдоль поля Киллинга $K$.
\end{prop}
\begin{proof}
Прямая проверка с учетом явного выражения для символов Кристоффеля
(\ref{echris}) и уравнения Киллинга (\ref{ekileq}).
\end{proof}
\section{Лоренц-инвариантные метрики                             \label{slorin}}
Рассмотрим (псевдо-)риманово многообразие $(\MM,g)$, которое допускает
векторное поле Киллинга $K$. Пусть на этом многообразии заданы дополнительно
какие либо тензорные поля. Тогда представляет интерес задача отыскания таких
полей, которые также инвариантны относительно преобразований изометрии.
Допустим, например, что на $\MM$ задана функция $f\in\CC^\infty(\MM)$. Эта
функция инвариантна относительно инфинитезимальных преобразований, генерируемых
векторным полем Киллинга $K$, если производная Ли равна нулю:
\begin{equation}                                                  \label{ekiscf}
  \Lie_Kf=K^\al\pl_\al f=0,
\end{equation}
т.е.\ если производная $f$ вдоль векторного поля $K$ равна нулю.

Для произвольного тензорного поля $T$ условие инвариантности имеет имеет вид
\begin{equation}                                                  \label{ekisct}
  \Lie_K T=0
\end{equation}
В качестве примера
рассмотрим пространство Минковского $\MR^{1,n-1}$ произвольной размерности $n$.
В нем задана метрика Лоренца $\eta_{\al\bt}$ в декартовой системе координат
$x^\al$. Эта метрика инвариантна относительно преобразований из группы Пуанкаре,
в частности, относительно преобразований Лоренца (см.\ раздел \ref{sminsp}).
Соответствующие векторные поля Киллинга имеют вид
\begin{equation}                                                  \label{ekilor}
  K_{\e\dl}=\frac12\left(x_\dl\pl_\e-x_\e\pl_\dl\right)
  =\frac12\left(x_\dl\dl_\e^\g-x_\e\dl_\dl^\g\right)\pl_\g,
\end{equation}
где индексы $\e,\dl$ нумеруют $n(n-1)/2$ векторов Киллинга и
$x_\al:=x^\bt\eta_{\bt\al}$. Пусть в пространстве Минковского $\MR^{1,n-1}$
задана вторая метрика $g_{\al\bt}$. Поставим следующую задачу: найти все метрики
$g_{\al\bt}$, инвариантные относительно действия группы Лоренца $\MS\MO(1,n-1)$.

Уравнения (\ref{ekilco}) для векторных полей Киллинга (\ref{ekilor}) принимают
вид
\begin{equation}                                                  \label{ekillo}
  g_{\al\e}\eta_{\bt\dl}-g_{\al\dl}\eta_{\bt\e}+g_{\bt\e}\eta_{\al\dl}
  -g_{\bt\dl}\eta_{\al\e}+x_\dl\pl_\e g_{\al\bt}-x_\e\pl_\dl g_{\al\bt}=0.
\end{equation}

При преобразованиях Лоренца компоненты метрики $g_{\al\bt}$ ведут себя как
компоненты ковариантного тензора второго ранга. Их необходимо построить из
метрики Лоренца $\eta_{\al\bt}$ и векторов $x=\lbrace x^\al\rbrace$.
Единственная возможность -- это метрика вида
\begin{equation*}
  g_{\al\bt}=A\eta_{\al\bt}+Bx^\al x^\bt,
\end{equation*}
где $A$ и $B$ -- некоторые функции на $\MR^{1,n-1}$. Подстановка этой метрики в
уравнение Киллинга (\ref{ekillo}) ограничивает вид функций $A$ и $B$. Можно
доказать, что они могут быть произвольными функциями от одной переменной
$$
  s:=x^\al x^\bt\eta_{\al\bt},
$$
которая инвариантна относительно преобразований Лоренца. Таким образом, метрика,
инвариантная относительно преобразований Лоренца, параметризуется двумя
произвольными функциями $A(s)$ и $B(s)$. Ее удобно записать в несколько другом
виде
\begin{equation}                                                  \label{elocom}
  g_{\al\bt}=f(s)\Pi^\St_{\al\bt}+g(s)\Pi^\Sl_{\al\bt}=
             f\eta_{\al\bt}+(g-f)\frac{x_\al x_\bt}s,
\end{equation}
где $\Pi^\St$ и $\Pi^\Sl$ -- проекционные операторы:
\begin{align*}
  \Pi^\St_{\al\bt}&:=\eta_{\al\bt}-\frac{x_\al x_\bt}s,
\\
  \Pi^\Sl_{\al\bt}&:=\frac{x_\al x_\bt}s.
\end{align*}
На произвольные функции $f(s)$ и $g(s)$ наложено только одно условие:
существование предела
\begin{equation*}
  \underset{s\to 0}\lim\frac{f(s)-g(s)}s,
\end{equation*}
которое необходимо, чтобы метрика была определена при $s=0$.

Здесь и до конца настоящего раздела, если не оговорено противное, для подъема и
опускания индексов будет использоваться метрика Минковского.

Лоренц инвариантная метрика вида (\ref{elocom}) при $f=g$ рассматривалась
В.~А.~Фоком \cite{Fock61R}. Метрике (\ref{elocom}) соответствует инвариантный
интервал
$$
  ds^2=fdx_\al dx^\al+(g-f)\frac{(x_\al dx^\al)^2}s.
$$

Метрический тензор (\ref{elocom}) имеет одинаковый вид во всех системах
координат, связанных между собой преобразованиями Лоренца. Однако его вид
меняется при сдвигах $x^\al\rightarrow x^\al+a^\al$, поскольку метрика явно
зависит от координат, и начало системы отсчета выделено.

Используя представление (\ref{edetab}) для определителя суммы двух матриц и
полагая
$$
  A_{\al\bt}=f\eta_{\al\bt},\qquad B_{\al\bt}=(g-f)\frac{x_\al x_\bt}s,
$$
можно вычислить определитель метрики (\ref{elocom}). В результате получим
равенство
\begin{equation}                                                  \label{edelme}
  \det g_{\al\bt}=(-f)^{n-1}g.
\end{equation}
Таким образом, лоренц инвариантная метрика вырождена тогда и только тогда, когда
$fg=0$. Мы будем предполагать, что функции $f$ и $g$ являются достаточно
гладкими, и $f>0$ и $g\ne0$.

При $g>0$ метрике (\ref{elocom}) можно поставить в соответствие репер
\begin{equation}                                                  \label{eveilo}
  e_\al{}^a=\sqrt f\dl_\al^a+(\sqrt{g\vphantom{f}}-\sqrt f)\frac{x_\al x^a}s.
\end{equation}
Он инвариантен относительно одновременного действия группы Лоренца на греческие
и латинские индексы.

Нетрудно проверить следующие свойства проекционных операторов
\begin{align*}
  \Pi^{\St\al\bt}x_\bt&=0,    &\Pi^\St_\al{}^\al&=n-1,
  &\pl_\al\Pi^\St_{\bt\g}&=-\frac{\Pi^\St_{\al\bt}x_\g+\Pi^\St_{\al\g}x_\bt}s,
\\
  \Pi^{\Sl\al\bt}x_\bt&=x^\al,&\Pi^\Sl_\al{}^\al&=1,
  &\pl_\al\Pi^\Sl_{\bt\g}&=\quad \frac{\Pi^\St_{\al\bt}x_\g+\Pi^\St_{\al\g}x_\bt}s,
\end{align*}
которые будут использоваться при проведении вычислений.

Несложные вычисления приводят к следующему выражению для символов Кристоффеля,
соответствующих метрике (\ref{elocom}),
\begin{align}                                                          \nonumber
  \Gamma_{\al\bt\g}&
  =f'(x_\al\Pi^\St_{\bt\g}+x_\bt\Pi^\St_{\al\g}-x_\g\Pi^\St_{\al\bt})
  +g'(x_\al\Pi^\Sl_{\bt\g}+x_\bt\Pi^\Sl_{\al\g}-x_\g\Pi^\Sl_{\al\bt})
  +\frac{g-f}s x_\g\Pi^\St_{\al\bt},
\\                                                                \label{echcoa}
  \Gamma_{\al\bt}{}^\g&=\frac{f'}f(x_\al\Pi^\St_\bt{}^\g+x_\bt\Pi^\St_\al{}^\g)
  +\frac{g'}g(x_\al\Pi^\Sl_\bt{}^\g+x_\bt\Pi^\Sl_\al{}^\g
  -x^\g\Pi^\Sl_{\al\bt})+\frac{g-f-f's}{sg} x^\g\Pi^\St_{\al\bt},
\end{align}
где штрих обозначает дифференцирование по аргументу $s$. Для подъема индекса в
последнем выражении использовалась обратная метрика
\begin{equation}                                                  \label{einmef}
  g^{\al\bt}=\frac1f\Pi^{\St\al\bt}+\frac1g\Pi^{\Sl\al\bt}.
\end{equation}
Тензор кривизны для метрики (\ref{elocom}) имеет вид
\begin{align}                                                          \nonumber
  R_{\al\bt\g}{}^\dl&=\Pi^\St_{\al\g}\Pi^\St_\bt{}^\dl
  \left[\frac{(f+f's)^2}{sfg}-\frac1s\right]
\\                                                                     \nonumber
  &+\Pi^\Sl_{\al\g}\Pi^\St_\bt{}^\dl
  \left[2\left(\frac{f+f's}f\right)^\prime+\left(\frac{f'}f-\frac{g'}g\right)
  \frac{f+f's}f\right]
\\                                                                \label{eculoc}
  &+\Pi^\Sl_\al{}^\dl\Pi^\St_{\bt\g}
  \left[-2\left(\frac{f+f's}g\right)^\prime+\left(\frac{f'}f-\frac{g'}g\right)
  \frac{f+f's}g\right]-(\al\leftrightarrow\bt).
\end{align}
Свернув это выражение по индексам $\bt$ и $\dl$, получим тензор Риччи
\begin{align}                                                          \nonumber
  R_{\al\bt}&=\Pi^\St_{\al\bt}
  \left[\frac{n-2}s\left(\frac{(f+f's)^2}{fg}-1\right)
  +2\frac{(f+f's)'}g-\left(\frac{f'}f+\frac{g'}g\right)
  \frac{f+f's}g\right]
\\                                                                \label{ericom}
  &+\Pi^\Sl_{\al\bt}(n-1)\left[2\frac{(f+f's)'}f
  -\left(\frac{f'}f+\frac{g'}g\right)\frac{f+f's}f\right].
\end{align}
Дальнейшая свертка с обратной метрикой (\ref{einmef}) дает скалярную кривизну
\begin{equation}                                                  \label{escurf}
  R=(n-1)\left[\frac{n-2}{fs}\left(\frac{(f+f's)^2}{fg}-1\right)
  +4\frac{(f+f's)'}{fg}-2\left(\frac{f'}f+\frac{g'}g\right)
  \frac{f+f's}{fg}\right]
\end{equation}

Как было отмечено ранее, пространства постоянной кривизны, определяемые
уравнением
\begin{equation}                                                  \label{ecolor}
  R_{\al\bt\g\dl}=-\frac{2K}{n(n-1)}(g_{\al\g}g_{\bt\dl}-g_{\al\dl}g_{\bt\g}),
\end{equation}
с некоторой постоянной $K$, автоматически удовлетворяют вакуумным
уравнениям Эйнштейна с космологической постоянной. Решим эти уравнения
для лоренц инвариантной метрики (\ref{elocom}). Для этого опустим
последний индекс у тензора кривизны (\ref{eculoc}) с помощью метрики
(\ref{elocom})
\begin{align}                                           \label{eculoi}
  R_{\al\bt\g\dl}&=\Pi^\St_{\al\g}\Pi^\St_{\bt\dl}\frac1s
  \left[\frac{(f+f's)^2}g-f\right]
\\                                                      \nonumber
  &+(\Pi^\Sl_{\al\g}\Pi^\St_{\bt\dl}-\Pi^\Sl_{\al\dl}\Pi^\St_{\bt\g})
  \left[2(f+f's)'-\left(\frac{f'}f+\frac{g'}g\right)(f+f's)\right]
  -(\al\leftrightarrow\bt)
\end{align}
и подставим в уравнение (\ref{ecolor}). В результате получим систему
дифференциальных уравнений на функции $f$ и $g$
\begin{align}                                                     \label{efgeqo}
  \frac{(f+f's)^2}{sg}-\frac fs&=-\frac{2K}{n(n-1)}f^2,
\\                                                                \label{efgeqt}
  2(f+f's)'-\left(\frac{f'}f+\frac{g'}g\right)(f+f's)&=-\frac{2K}{n(n-1)}fg.
\end{align}
Из первого уравнения получаем решение для функции $g$
\begin{equation}                                                  \label{egfexe}
  g=\frac{(f+f's)^2}{f\left(1-\frac{2K}{n(n-1)}fs\right)}.
\end{equation}
Поскольку для невырожденности метрики необходимо, чтобы $g\ne0$, то
при $s\ne0$ функция $f$ должна удовлетворять неравенству
\begin{equation}                                                  \label{enfufu}
  f\ne\frac{n(n-1)}{2Ks},\qquad s\ne0.
\end{equation}
Подстановка этого выражения во второе уравнение (\ref{efgeqt}) приводит к
тождеству. Таким образом мы доказали следующее утверждение.
\begin{theorem}
Лоренц инвариантная метрика
\begin{equation}                                                  \label{eloinm}
  g_{\al\bt}=f\Pi^\St_{\al\bt}
  +\frac{(f+f's)^2}{f\left(1-\frac{2K}{n(n-1)}fs\right)}\Pi^\Sl_{\al\bt},
\end{equation}
где $f(s)$ -- произвольная положительная функция, удовлетворяющая условию
(\ref{enfufu}), является метрикой пространства постоянной кривизны. Обратно.
Метрику пространства постоянной кривизны можно записать в лоренц инвариантном
виде (\ref{eloinm}) для некоторой функции $f(s)$.
\end{theorem}
\begin{proof}
Нам осталось доказать, что произвольную метрику пространства постоянной
кривизны можно привести к лоренц инвариантному виду (\ref{elocom}). Чтобы
ответить на этот вопрос, запишем метрику (\ref{eloinm}) в более известной форме.
С этой целью зафиксируем функцию $f$, положив $f=g$. Это равенство с учетом
(\ref{egfexe}) сводится к уравнению
$$
  f^{\prime2}s+2f'f+\frac{2K}{n(n-1)}f^3=0,
$$
общее решение которого имеет вид
$$
  f=\frac C{(C+\frac K{2n(n-1)} s)^2},\qquad C=\const.
$$
Постоянная интегрирования убирается растяжкой координат. Поэтому без ограничения
общности положим $C=1$. В результате получим метрику постоянной кривизны
\begin{equation}                                                  \label{ecomeu}
  g_{\al\bt}=\frac{\eta_{\al\bt}}{(1+\frac K{2n(n-1)}s)^2}.
\end{equation}
То, что метрику пространства постоянной кривизны можно привести к такому виду --
хорошо известный факт. Доказательство этого утверждения нетривиально (см.,
например, \cite{Wolf72R}).
\end{proof}
\begin{com}
Проведенные вычисления просто переносятся на случай метрики в евклидовом
пространстве, которая инвариантна относительно $\MS\MO(n)$-вращений. Для этого
во всех формулах метрику Лоренца $\eta_{\al\bt}$ нужно заменить на евклидову
метрику $\dl_{\al\bt}$.
\qed\end{com}

Поскольку предел функции $(g-f)/s$ равен
\begin{equation*}
  \underset{s\to0}\lim\frac{g-f}s=2f'+\frac{2K}{n(n-1)}f^2,
\end{equation*}
то выражение для метрики (\ref{eloinm}) определено и при $s=0$.

Решим вакуумные уравнения Эйнштейна с космологической постоянной
$$
  R_{\al\bt}=\Lm g_{\al\bt}
$$
для лоренц инвариантной метрики. Поскольку число этих уравнений меньше, чем
число уравнений в условии постоянства кривизны (\ref{ecolor}), то можно было бы
ожидать, что они допускают решения не только с постоянной кривизной. Однако для
лоренц инвариантных метрик классы решений совпадают. Действительно, подстановка
тензора Риччи (\ref{ericom}) в уравнения Эйнштейна приводит к следующей системе
уравнений:
\begin{align}                                                     \label{einone}
  \frac{n-2}s\left[\frac{(f+f's)^2}{fg}-1\right]
  +2\frac{(f+f's)'}g-\left(\frac{f'}f+\frac{g'}g\right)\frac{f+f's}g
  &=\Lm f,
\\                                                                \label{eintwo}
  (n-1)\left[2\frac{(f+f's)'}f-\left(\frac{f'}f+\frac{g'}g\right)
  \frac{f+f's}f\right] &=\Lm g.
\end{align}
Второе уравнение при
$$
  \Lm=-\frac{2K}n
$$
совпадает с уравнением (\ref{efgeqt}). Линейная комбинация
$(\ref{einone})/f$$-(\ref{eintwo})/g$ эквивалентна уравнению (\ref{efgeqo}).

Таким образом мы доказали, что все лоренц инвариантные решения вакуумных
уравнений Эйнштейна с космологической постоянной исчерпываются пространствами
постоянной кривизны.

В последнее время многие авторы понятие векторного поля Киллинга понимают в
более широком смысле: как симметрию не только метрики, но и других тензорных
полей.
\begin{exa}
Рассмотрим ``векторное поле Киллинга'' на плоскости, которое в декартовой
системе координат $u,v$ имеет вид
\begin{equation*}
  K=u\pl_u+av\pl_v,\qquad a=\const\ne0.
\end{equation*}
Тогда уравнение Киллинга (\ref{ekiscf}) для функции $f(u,v)$ принимает вид
\begin{equation}                                                  \label{eselsi}
  Kf=0.
\end{equation}
Любое решение этого уравнения имеет вид $f=f(\x)$, где
\begin{equation*}
  \x=\frac{(av)^{1/a}}u.
\end{equation*}
Это значит, что решения уравнения (\ref{eselsi}) по существу являются функциями
одной переменной $\x$. Если функция $f$ одновременно удовлетворяет некоторому
уравнению в частных производных на поверхности, то его решения, зависящие от
переменной $\x$ называются {\em автомодельными}. Для таких решений уравнение в
частных производных сводится к обыкновенному дифференциальному уравнению
относительно $\x$, что существенно упрощает задачу.
\index{Автомодельное решение (self-similar solution)}%
\index{Решение автомодельное (self-similar solution)}%
\qed\end{exa}
\chapter{Геодезические и экстремали                              \label{sgextr}}
Пусть на многообразии $\MM$ задана аффинная геометрия, т.е.\ метрика $g$ и
аффинная связность $\Gamma$. Тогда можно построить два типа выделенных кривых:
геодезические и экстремали. Геодезические линии определяются только связностью
как линии, касательный вектор к которым остается касательным при параллельном
переносе. Экстремали, напротив, определяются только метрикой как линии
экстремальной длины. Поскольку метрика и связность являются независимыми
геометрическими объектами, то в общем случае геодезические линии и экстремали
различны. В частном случае (псевдо-)римановой геометрии, когда связностью
является связность Леви--Чивиты, геодезические и экстремали совпадают.

В настоящей главе мы рассмотрим оба типа кривых, т.к.\ они играют важную роль в
моделях математической физики. Достаточно сказать, что одним из постулатов
общей теории относительности является предположение о том, что свободные
точечные частицы, подверженные действию только гравитационных сил, движутся по
экстремалям. Кроме того, понятие полноты многообразий связано также с
экстремалями.
\section{Геодезические                                           \label{sgeode}}
В аффинной геометрии $(\MM,g,\Gamma)$ существует выделенное семейство линий, которые
называются геодезическими. Рассмотрим произвольную кривую
$\g=x(t)=\lbrace x^\al(t)\rbrace$, где $-\infty\le t_1<t<t_2\le\infty$, на
многообразии $\MM$. Вектор скорости кривой,
$u(t)=\lbrace u^\al(t):=\dot x^\al(t)\rbrace$, как всегда, предполагается
отличным от нуля.
\begin{defn}
{\em Геодезической} линией на многообразии $\MM$ называется кривая $x(t)$ класса
$\CC^2([t_1,t_2])$, касательный вектор к которой остается касательным при
параллельном переносе вдоль нее.
\qed\end{defn}
\index{Геодезическая (geodesic)}%
\begin{com}
В определении геодезической линии присутствует только аффинная связность.
Поэтому понятие геодезической линии никакого отношения к метрике не имеет,
которой может вообще не быть на многообразии.
\qed\end{com}
Получим уравнения, которым должны удовлетворять координатные функции $x^\al(t)$
для того, чтобы кривая $\g$ была геодезической. Выберем произвольный отличный от
нуля вектор $X_0$, который касается кривой $\g$ в некоторой точке $x(t_0)$, и
разнесем его вдоль всей кривой с помощью параллельного переноса. В результате
получим векторное поле $X\big(x(t)\big)$, определенное на кривой $\g$. Множество
векторных полей, определенных на кривой $\g$, будем обозначать $\CX(\MM,\g)$.
Из определения геодезической следует, что это векторное поле всюду касается $\g$
и, следовательно, пропорционально векторному полю скорости:
$X^\al(t)=f(t)u^\al(t)$, где $f$ -- некоторая отличная от нуля функция на $\g$.
Изменим параметризацию кривой $t\mapsto s(t)$. Тогда условие параллельности
примет вид
\begin{equation}                                                  \label{ecapad}
  X^\al=f\frac{ds}{dt}\frac{dx^\al}{ds}.
\end{equation}
Выберем новый параметр вдоль геодезической таким образом, чтобы было выполнено
уравнение
\begin{equation*}
  \frac{ds}{dt}=\frac1f,
\end{equation*}
которое всегда имеет решение, поскольку $f\ne0$. Таким образом, на геодезической
линии существует такая параметризация, что вектор скорости $u$ при параллельном
переносе остается вектором скорости. В дальнейшем мы предполагаем, что параметр
$t$ вдоль геодезической выбран таким образом, что $f=1$. Если вектор скорости
геодезической при параллельном переносе остается касательным, то ковариантная
производная от него вдоль геодезической равна нулю (\ref{epartc}):
\begin{equation}                                                  \label{egeoin}
  \nb_u u=u^\al(\pl_\al u^\bt+\Gamma_{\al\g}{}^\bt u^\g)\pl_\bt=0.
\end{equation}
Поскольку $d/dt=u^\al\pl_\al$, то это уравнение в компонентах принимает вид
\begin{equation}                                                  \label{egeode}
  \ddot x^\al=-\Gamma_{\bt\g}{}^\al\dot x^\bt\dot x^\g.
\end{equation}
Это уравнение не инвариантно относительно перепараметризации кривой. Однако оно
допускает линейную замену параметра
\begin{equation}                                                  \label{qdskhm}
  t\mapsto at+b,\qquad a,b\in\MR,~a\ne0.
\end{equation}
Таким образом, мы получили критерий того, что кривая $x(t)$ является
геодезической.
\begin{prop}
Кривая $x(t)$ на многообразии $\MM$ с заданной аффинной связностью $\Gamma$
является геодезической тогда и только тогда, когда существует такая
параметризация, что ее координатные функции $x^\al(t)$ удовлетворяют системе
уравнений (\ref{egeode}).
\end{prop}
\begin{com}
В уравнение (\ref{egeoin}) входит частная производная $\pl_\al u^\bt$ от вектора
скорости. Эта производная неопределена, т.к.\ векторное поле скорости
$u^\bt\big(x(t)\big)$ задано только вдоль кривой $\g$. Тем не менее в уравнение
входит производная по направлению $u^\al\pl_\al u^\bt=\ddot x^\al$, которая
имеет смысл. Это значит, что, для определения частной производной
$\pl_\al u^\bt$, векторное поле скорости можно продолжить в некоторую
окрестность кривой $\g$ произвольным дифференцируемым образом, а конечный ответ
от такого продолжения не зависит. Подобный трюк будет часто встречаться в
дальнейшем.
\qed\end{com}
\begin{defn}
Векторное поле $a(t)$, определенное вдоль произвольной кривой $x(t)$ на
многообразии $\MM$ с заданной связностью $\Gamma$,
\begin{equation}                                                  \label{eaccla}
  a:=\nb_u u
  =\left(\ddot x^\al+\Gamma_{\bt\g}{}^\al\dot x^\bt\dot x^\g\right)\pl_\al,
\end{equation}
называется {\em ускорением} кривой.
\qed\end{defn}
\index{Ускорение кривой (acceleration of a curve)}%
Сравнивая ускорение кривой с уравнением для геодезической, получаем, что кривая
будет геодезической тогда и только тогда, когда ее ускорение равно нулю.

Ускорение кривой, так же как и скорость, является векторным полем вдоль кривой,
$u,a\in\CX(\MM,\g)$, и зависит от ее параметризации. При выводе уравнений
геодезической линии (\ref{egeode}) была выбрана специальная параметризация
кривой $\g$.
\begin{defn}
Параметр $t$, по которому проводится дифференцирование в системе обыкновенных
дифференциальных уравнений (\ref{egeode}), определяющих геодезическую линию,
называется {\em каноническим} или {\em аффинным}.
\qed\end{defn}
Поскольку уравнение (\ref{ecapad}) имеет тензорную форму, то канонический
параметр не зависит от выбора системы координат на $\MM$.
\index{Канонический параметр (canonical parameter)}%
\index{Параметр канонический (canonical parameter)}%
\index{Аффинный параметр (affine parameter)}%
\index{Параметр аффинный (affine parameter)}%
\begin{prop}
Любые два канонических параметра вдоль геодезической связаны между собой
линейным преобразованием (\ref{qdskhm}).
\end{prop}
\begin{proof}
Рассмотрим вопрос как меняется уравнение для геодезических при произвольной
замене канонического параметра. При другой параметризации геодезической
$x^\al(s)$, где $s=s(t)$, $ds/dt\ne0$, уравнение (\ref{egeode}) изменится:
\begin{equation}                                        \label{ecaptg}
  \left(\frac{ds}{dt}\right)^2\frac{d^2x^\al}{ds^2}
  +\frac{d^2s}{dt^2}\frac{dx^\al}{ds}
  =-\left(\frac{ds}{dt}\right)^2\Gamma_{\bt\g}{}^\al
  \frac{dx^\bt}{ds}\frac{dx^\g}{ds}.
\end{equation}
Отсюда следует, что форма уравнений (\ref{egeode}) не изменится тогда и только
тогда, когда $d^2s/dt^2=0$, т.е.\ замена параметра является аффинной.
\end{proof}
\begin{exa}
В пространстве Минковского $\MR^{1,3}$ точечная частица движется по некоторой
мировой линии $x(t)$. Если в качестве параметра вдоль траектории выбрано время
$t=x^0$, то скорость и ускорение кривой совпадают со скоростью и ускорением
частицы. Если частица свободна, т.е.\ на нее не действуют никакие силы и,
следовательно, ее ускорение равно нулю, то траекторией частицы будет одна из
геодезических линий. Поскольку связность в $\MR^{1,3}$ в декартовой системе
координат имеет равные нулю компоненты, то уравнения (\ref{egeode}) сводятся к
уравнениям
$$
  \ddot x^\al=0,
$$
которые определяют прямые линии. Таким образом, в пространстве Минковского
$\MR^{1,3}$ прямые и только они являются геодезическими. Это значит, что
свободная частица в пространстве Минковского равномерно движется вдоль прямой
линии.
\qed\end{exa}
Геодезическая линия в аффинной геометрии обобщает понятие прямой в
(псевдо-)евклидовом пространстве, сохраняя то свойство, что касательный вектор
остается касательным при параллельном переносе.
\begin{exa}
В пространствах абсолютного параллелизма, для которых тензор кривизны равен
нулю $R_{\al\bt\g}{}^\dl=0$, локально существует система координат, в которой
симметричная часть аффинной связности равна нулю
$\Gamma_{\lbrace\al\bt\rbrace}{}^\g=0$ (см., раздел \ref{sloccu}). Поскольку
геодезические линии (\ref{egeode}) определяются только симметричной частью
аффинной связности, то в этой системе координат геодезические являются прямыми
линиями. В частности, если на группе Ли задана каноническая связность, т.е.\
параллельный перенос отождествлен с групповым действием справа, то такая система
координат локально существует. Заметим, что в правоинвариантном базисе на группе
Ли компоненты связности равны нулю, однако он не является голономным и не
определяет ту систему координат, о которой идет речь.
\qed\end{exa}
Для двух параметризаций $x^\al(t)$ и $x^\al(s)$ одной геодезической справедливо
равенство
\begin{equation*}
  \frac{d^2s}{dt^2}=\frac{ds}{dt}\frac{dx^\bt}{ds}\pl_\bt
  \left(\frac{ds}{dt}\right),
\end{equation*}
и уравнение (\ref{ecaptg}) переписывается в эквивалентной форме
\begin{align}                                                          \nonumber
  \frac{d^2x^\al}{ds^2}
  &=-\Gamma_{\bt\g}{}^\al\frac{dx^\bt}{ds}\frac{dx^\g}{ds}
  -\frac{dx^\al}{ds}\frac{dx^\bt}{ds}
  \pl_\bt\ln\left|\frac{ds}{dt}\right|
\\                                                                \label{ecaptr}
  &=-\Gamma_{\bt\g}{}^\al\frac{dx^\bt}{ds}\frac{dx^\g}{ds}
  -\frac12\left(\dl^\al_\bt\pl_\g\ln\left|\frac{ds}{dt}\right|
  +\dl^\al_\g\pl_\bt\ln\left|\frac{ds}{dt}\right|\right)
  \frac{dx^\bt}{ds}\frac{dx^\g}{ds}.
\end{align}

Будем говорить, что две геодезические линии $x^\al(t)$ и $x^\al(s)$ для
различных связностей $\Gamma_{1\al\bt}{}^\g$ и $\Gamma_{2\al\bt}{}^\g$ на одном и том же
многообразии $\MM$ имеют одинаковую форму, если множества точек, через которые
проходят геодезические линии, совпадают. Тогда из уравнения (\ref{ecaptr})
следует критерий совпадения формы геодезических для двух аффинных связностей.
\begin{theorem}                                           \label{tcrgef}
Для того, чтобы форма геодезических $x^\al(t)$ и $x^\al(s)$, где $t$ и $s$ --
канонические параметры, для двух аффинных
связностей $\Gamma_{1\al\bt}{}^\g$ и $\Gamma_{2\al\bt}{}^\g$ на многообразии $\MM$
совпадала, необходимо и достаточно, чтобы симметричные части этих
связностей были связаны соотношением
\begin{equation}                                                  \label{erecog}
  \Gamma_{2\left\lbrace \al\bt\right\rbrace }{}^\g
  =\Gamma_{1\left\lbrace \al\bt\right\rbrace }{}^\g
  +\frac12\left(\dl^\g_\bt\pl_\al\vf+\dl^\g_\al\pl_\bt\vf\right),
\end{equation}
где $\vf=\vf(x)$ -- некоторая функция, определяющая связь
канонических параметров для каждой геодезической линии
\begin{equation}                                                  \label{erecag}
  \frac{ds}{dt}=e^\vf.
\end{equation}
\end{theorem}
Пусть задан диффеоморфизм многообразий
\begin{equation*}
  f:\quad \MM\rightarrow\MN
\end{equation*}
такой, что аффинная связность на $\MM$ отображается в аффинную связность на
$\MN$. Тогда диффеоморфизм $f$ отображает геодезическую линию в геодезическую,
поскольку определение геодезической линии инвариантно (не зависит от системы
координат).

Уравнения для геодезических (\ref{egeode}) -- это система нелинейных
обыкновенных дифференциальных уравнений второго порядка. Поэтому при достаточно
гладких компонентах связности для однозначного решения задачи Коши необходимо
задать начальную точку $\lbrace x^\al(0)\rbrace$ и вектор скорости
$\lbrace \dot x^\al(0)\rbrace$. Геометрически это означает, что через данную
точку многообразия в данном направлении проходит одна и только одна
геодезическая.
\begin{prop}                                                      \label{pinfsm}
Если аффинная связность $\Gamma$ на многообразии $\MM$ гладкая, то геодезические
линии также являются гладкими $\CC^\infty$ кривыми.
\end{prop}
\begin{proof}
Продифференцируем уравнение геодезических (\ref{egeode}) по каноническому
параметру. Правая часть полученного равенства зависит от компонент связности и
их первых производных, а также от компонент вектора скорости и их первой
производной. Поэтому правая часть определена и непрерывна. Следовательно третья
производная от координатных функций $x^\al(t)$ существует и непрерывна.
Дальнейшее дифференцирование приводит к гладкости геодезических.
\end{proof}

\begin{figure}[h,b,t]
\hfill\includegraphics[width=.4\textwidth]{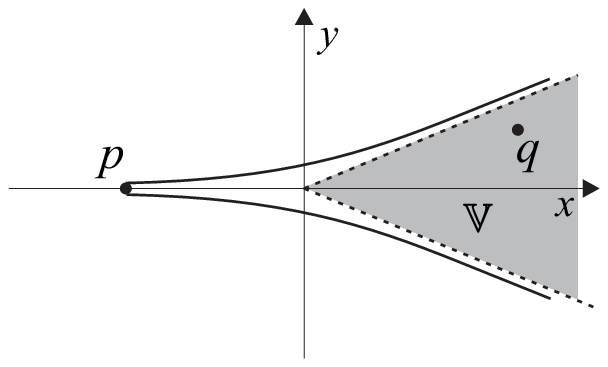}
\hfill {}
\centering\caption{Точки $p$ и $q$ на плоскости с конической сингулярностью и
 положительным углом дефицита нельзя соединить геодезической.}
\label{fconpo}
\end{figure}

Для геодезической линии можно также поставить краевую задачу: найти
геодезическую, соединяющую две фиксированные точки многообразия, которое
предполагается линейно связным. Эта задача разрешима в малом, т.е.\ любые две
достаточно близкие точки можно соединить геодезической и притом только одной.
Для удаленных точек эта задача может не иметь решения или иметь несколько
решений.
\begin{defn}
(Псевдо-)риманово многообразие $\MM$ называется {\em геодезически выпуклым},
если любые две его точки могут быть соединены геодезической.
\end{defn}
\index{Многообразие геодезически выпуклое (geodesically convex manifold)}%
\index{Геодезически выпуклое многообразие (geodesically convex manifold)}%
\begin{exa}
Рассмотрим плоскость с конической сингулярностью и положительным углом дефицита
$\MM=\MU\cup\MV$, где $\MU$ -- область, изометричная евклидовой плоскости
$\MR^2$ с выколотой полупрямой
$\overline\MR_+:=\lbrace(x,y)\in\MR^2:~x\ge0, y=0\rbrace$, и
$\MV$ -- клин евклидовой плоскости, который вставлен (см.\ рис.\ref{fconpo}).
Тогда на ней существуют точки, которые нельзя соединить геодезической.
Действительно, все геодезические, проходящие через точку $p$, соединяют ее со
всеми точками евклидовой плоскости $\MR^2$, до того, как клин был вставлен.
Поскольку при диффеоморфизме ($\MR^2\setminus\overline\MR_+$ на $\MU$)
геодезическая переходит в геодезическую и их число сохраняется, то точку $p$
нельзя соединить геодезической ни с какой точкой $q$, лежащей на клине $\MV$,
который вставлен со стороны, противоположной конической сингулярности.
Это многообразие не является геодезически выпуклым.
\qed\end{exa}
\begin{exa}
Рассмотрим сферу, вложенную в евклидово пространство,
$\MS^2\hookrightarrow\MR^3$. Пусть метрика на сфере индуцирована вложением и
на ней задана связность Леви--Чивиты. Тогда любые две точки на сфере $\MS^2$, не
являющиеся диаметрально противоположными, можно соединить двумя разными
геодезическими (две дуги большой окружности, проходящей через эти точки).
Диаметрально противоположные точки соединяются бесконечным числом геодезических.
Сфера $\MS^2$ является геодезически выпуклым многообразием.
\qed\end{exa}

Введем важное понятие полноты геодезической по каноническому параметру $t$,
В силу однозначного разрешения задачи Коши, через каждую точку многообразия в
заданном направлении проходит одна геодезическая. Это значит, что геодезическую
линию, если она заканчивается в некоторой точке $q\in\MM$, всегда можно
продолжить. Действительно, если при конечном значении канонического параметра
геодезическая попадает в точку $q$, то продолжим ее, склеив с геодезической,
выходящей из точки $q$ в том же направлении.
\begin{defn}
Геодезическая в $\MM$ называется {\em полной}, если ее можно продолжить в обе
стороны до бесконечного значения канонического параметра,
$t\in(-\infty,\infty)$.
\qed\end{defn}
\index{Геодезическая полная (complete geodesic)}%
\index{Полнота геодезической (completeness of geodesic)}%
\begin{com}
Поскольку канонический параметр определен с точностью до аффинного
преобразования и не зависит от выбора системы координат, то данное определение
корректно.
\qed\end{com}
Геодезическую линию можно рассматривать как интегральную кривую векторного
поля скорости $u:=\lbrace\dot x^\al\rbrace$. Тогда полнота геодезической
означает полноту векторного поля $u$.

Пусть на многообразии $\MM$ задана не только аффинная связность $\Gamma$, но и
метрика $g$. Тогда компоненты аффинной связности можно выразить через метрику,
кручение и неметричность по формуле (\ref{elicon}). Хотя в уравнение для
геодезических входит только симметричная часть связности, тем не менее оно
нетривиально зависит от кручения и неметричности. Действительно, из формулы
(\ref{elicon}) следует, что симметричная часть аффинной связности имеет вид
\begin{align}                                                          \nonumber
  \Gamma_{\lbrace \bt\g\rbrace }{}^\al&=\frac12
  (\Gamma_{\bt\g}{}^\al+\Gamma_{\g\bt}{}^\al)=
\\                                                                \label{emecsy}
  &=\widetilde\Gamma_{\bt\g}{}^\al+\frac12(T^\al{}_{\bt\g}+T^\al{}_{\g\bt})
  +\frac12(Q_{\bt\g}{}^\al+Q_{\g\bt}{}^\al-Q^\al{}_{\bt\g}),
\end{align}
где $\widetilde\Gamma_{\bt\g}{}^\al$ -- символы Кристоффеля. Ясно, что две
связности, имеющие одинаковую симметричную часть определяют одно и то же
семейство геодезических.

Рассмотрим два вектора $X$ и $Y$, которые параллельно переносятся вдоль
геодезической $\g$.
\begin{prop}                                                      \label{pscaex}
Зависимость скалярного произведения $(X,Y)$ двух векторов, которые параллельно
переносятся вдоль $\g$, от точки геодезической определяется только тензором
неметричности:
\begin{equation}                                                  \label{eparsc}
  \pl_u(X,Y)=\nb_u(X^\al Y^\bt g_{\al\bt})=-u^\g X^\al Y^\bt Q_{\g\al\bt}.
\end{equation}
Отсюда следует, что в римановой геометрии и геометрии Римана--Картана, где
$Q=0$, скалярное произведение двух векторов при параллельном переносе вдоль
геодезической сохраняется. В частности, квадрат вектора скорости геодезической
постоянен вдоль нее.
\end{prop}
\begin{proof}
Следует из определения тензора неметричности (\ref{enonme}).
\end{proof}
\begin{com}
Это утверждение верно и для произвольной кривой (предложение \ref{pscver}).
Скалярное произведение двух векторных полей, полученных в результате
параллельного переноса двух векторов вдоль произвольной кривой $\g$ в римановой
геометрии и геометрии Римана--Картана, не зависит от точки кривой.
\qed\end{com}
\begin{cor}
Если неметричность лоренцева многообразия $\MM$ равна нулю, то квадрат вектора
скорости постоянен вдоль геодезических, и их можно разделить на три класса:
времениподобные, пространственноподобные и светоподобные (изотропные). При этом
тип геодезической не может меняться от точки к точке.
\qed\end{cor}

В заключение настоящего раздела получим уравнение девиации геодезических. Пусть
на многообразии $\MM$ задана произвольная кривая $\s=y(s)$, $s\in[s_1,s_2]$ и
отличное от нуля векторное поле $X(s)\in\CX(\MM,\s)$, определенное вдоль этой
кривой и которое нигде не касается кривой $\s$. Рассмотрим семейство
геодезических $\g_s=x(t,s)$, проходящих через каждую точку кривой $\s$ в
направлении $X(s)$:
\begin{equation*}
  x(0,s)=y(s),\qquad\dot x(0,s)=X(s),
\end{equation*}
где точка, как и раньше, обозначает дифференцирование по каноническому параметру
$t$. Если кривая $y(s)$ и векторное поле $X(s)$ достаточно гладкие, то множество
точек $x(t,s)$ образует двумерное подмногообразие (поверхность) $\Sigma$ в $\MM$
при достаточно малых $t$. При этом параметры $t,s$ можно выбрать в качестве
координат на $\Sigma$. Этим координатам соответствуют векторные поля
\begin{equation}                                                  \label{qlkhii}
  u:=\pl_t=\frac{\pl x^\al}{\pl t}\pl_\al,\qquad
  Y:=\pl_s=\frac{\pl x^\al}{\pl s}\pl_\al,
\end{equation}
которые коммутируют между собой, $[u,Y]=0$. После замены в коммутаторе частных
производных на ковариантные, получим равенство
\begin{equation}                                                  \label{erecot}
  u^\bt\nb_\bt Y^\al=Y^\bt\nb_\bt u^\al+u^\bt Y^\g T_{\bt\g}{}^\al,
\end{equation}
где $T_{\bt\g}{}^\al:
=\Gamma_{\bt\g}{}^\al-\Gamma_{\g\bt}{}^\al$ -- тензор кручения.
\begin{defn}
Векторное поле $Y\in\CX(\MM,\Sigma)$ определенное равенством (\ref{qlkhii})б
называется {\em вектором девиации} геодезических. Векторные поля с компонентами:
\begin{align}                                                     \label{evedev}
  V^\al&:=u^\bt\nb_\bt Y^\al,
\\                                                                \label{eacdev}
  A^\al&:=u^\bt\nb_\bt V^\al,
\end{align}
называются соответственно {\em скоростью} и {\em ускорением девиации} близких
геодезических.
\qed\end{defn}
\index{Вектор девиации (deviation vector)}%
\index{Девиации девиации (deviation vector)}%
\index{Скорость девиации (deviation velocity)}%
\index{Девиации скорость (deviation velocity)}%
\index{Ускорение девиации (deviation acceleration)}%
\index{Девиации ускорение (deviation acceleration)}%
Векторное поле девиации определяет расположение близких геодезических на
поверхности $\Sigma$, а векторные поля скорости и ускорения девиации
характеризуют, как меняются расположение геодезических при движении вдоль
геодезической.
\begin{prop}
Ускорение девиации геодезических определяется тензором кривизны и кручения:
\begin{equation}                                                  \label{edecto}
  A^\al=u^\g u^\dl Y^\bt R_{\g\bt\dl}{}^\al
  +u^\g u^\dl\nb_\dl\left(Y^\bt T_{\g\bt}{}^\al\right).
\end{equation}
\end{prop}
\begin{proof}
Прямая проверка:
\begin{equation*}
\begin{split}
  &u^\g\nb_\g(u^\bt\nb_\bt Y^\al)
  =u^\g\nb_\g(Y^\bt\nb_\bt u^\al+u^\bt Y^\dl T_{\bt\dl}{}^\al)=
\\
  &=u^\g\nb_\g Y^\bt\nb_\bt u^\al+u^\g Y^\bt\nb_\g\nb_\bt u^\al
  +u^\bt u^\g\nb_\g(Y^\dl T_{\bt\dl}{}^\al)=
\\
  &=u^\g\nb_\g Y^\bt\nb_\bt u^\al+u^\g Y^\bt\nb_\bt\nb_\g u^\al
  +u^\g u^\dl Y^\bt R_{\g\bt\dl}{}^\al-u^\g Y^\bt T_{\g\bt}{}^\dl\nb_\dl u^\al
  +u^\bt u^\g\nb_\g(Y^\dl T_{\bt\dl}{}^\al)=
\\
  &=Y^\g\nb_\g(u^\bt\nb_\bt u^\al)+u^\g u^\dl Y^\bt R_{\g\bt\dl}{}^\al
  +u^\bt u^\g\nb_\g(Y^\dl T_{\bt\dl}{}^\al)=
\\
  &=u^\g u^\dl Y^\bt R_{\g\bt\dl}{}^\al
  +u^\bt u^\g\nb_\g(Y^\dl T_{\bt\dl}{}^\al),
\end{split}
\end{equation*}
где мы учли равенство (\ref{erecot}), уравнения геодезических (\ref{egeoin}) и
выражение коммутатора ковариантных производных через тензоры кривизны и кручения
(\ref{emeacu}).
\end{proof}
В (псевдо-)римановой геометрии уравнение (\ref{edecto}) упрощается
\begin{equation}                                                  \label{edects}
  A^\al=u^\g u^\dl Y^\bt \widetilde R_{\g\bt\dl}{}^\al.
\end{equation}
В литературе по общей теории относительности оно известно, как {\em уравнение
девиации геодезических}.
\index{Уравнение девиации геодезических (geodesic deviation equation)}%
\index{Девиация геодезических (geodesic deviation)}%
Это уравнение можно переписать в виде
\begin{equation}                                                  \label{qbvvgt}
  \nb_u^2Y^\al+u^\dl u^\bt Y^\g \widetilde R_{\g\dl\bt}{}^\al=0.
\end{equation}
Оно совпадает с равенством нулю второй вариации действия для экстремалей
(\ref{exseva}), которое будет получено позже  в разделе \ref{saferk}.
\section{Экстремали                                              \label{sextre}}
Другим выделенным типом кривых в аффинной геометрии $(\MM,g,\Gamma)$ являются
экстремали, которые определяются как линии экстремальной длины. Рассмотрим
произвольную достаточно гладкую кривую $\g=x(t)\in\MM$, $t\in[t_1,t_2]$.
Предположим, что квадрат вектора скорости кривой, $u^\al:=\dot x^\al(t)$, всюду
отличен от нуля, $u^2:=u^\al u^\bt g_{\al\bt}\ne0$.
\begin{com}
Для римановой метрики это условие автоматически выполняется, поскольку вектор
скорости предполагается отличным от нуля. Если метрика не является
знакоопределенной, то это условие нетривиально. Например, для лоренцевой метрики
это условие равносильно тому, что мы рассматриваем либо времени-, либо
пространственноподобные кривые.
\qed\end{com}
Пусть кривая соединяет две точки $p=\lbrace x^\al(t_1)\rbrace $ и
$q=\lbrace x^\al(t_2)\rbrace$. Тогда длина этой кривой задается интегралом
\begin{equation}                                                  \label{eactim}
  S=\int\limits_p^q ds,\qquad
  ds= dt\sqrt{|g_{\al\bt}\dot x^\al\dot x^\bt|}=dt\sqrt{|u^2|}.
\end{equation}
Этот функционал инвариантен относительно общих преобразований координат $x^\al$
и произвольной перепараметризации кривой $t\rightarrow \tau(t)$.
\begin{defn}
{\em Экстремалью}, соединяющей две точки (псевдо-)риманова многообразия
$p,q\in\MM$, если она существует, называется неизотропная кривая $\g$ класса
$\CC^2([t_1,t_2])$, для которой функционал (\ref{eactim}) принимает
экстремальное значение.
\qed\end{defn}
\index{Экстремаль (extremal)}
\begin{com}
Если метрика на многообразии $\MM$ не является знакоопределенной, то существуют
изотропные кривые, для которых функционал длины (\ref{eactim}) равен нулю и
данное выше определение экстремалей не проходит. Определение изотропных
экстремалей будет дано ниже.
\qed\end{com}

\begin{figure}[h,b,t]
\hfill\includegraphics[width=.4\textwidth]{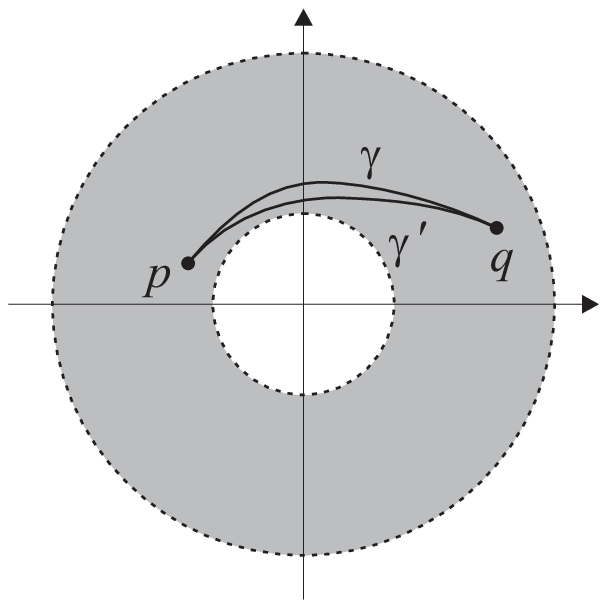}
\hfill {}
\centering\caption{Точки $p$ и $q$ нельзя соединить экстремалью.}
\label{fering}
\end{figure}

Экстремали в римановом пространстве обобщают понятие прямой в евклидовом
пространстве, сохраняя свойство быть линиями минимальной (экстремальной) длины,
соединяющей две точки.
\begin{exa}
Не любые две точки линейно связного многообразия $p,q\in\MM$, на котором задана
риманова метрика, можно соединить кривой наименьшей длины. Рассмотрим кольцо на
евклидовой плоскости $\MR^2$, рис.\ref{fering}. Это -- риманово некомпактное
связное неодносвязное двумерное многообразие. Если отрезок, соединяющий точки
$p$ и $q$, пересекает вырезанную дырку, то эти точки нельзя соединить кривой
наименьшей длины. Действительно, для любой кривой $\g$, соединяющей точки $p$ и
$q$, всегда найдется кривая $\g'$ меньшей длины. Построение ясно из рисунка.
\qed\end{exa}

Найдем уравнения, которым должны удовлетворять координатные функции $x^\al(t)$
для того, чтобы кривая $x(t)$ была экстремалью. Пусть экстремаль соединяет две
точки $p$ и $q$ многообразия. Выберем такую карту на многообразии, которая
целиком содержит данную экстремаль. Для этого достаточно взять объединение всех
достаточно малых шаров, центры которых лежат на экстремали. Пусть в этой карте
экстремаль и ее вариация задаются набором функций $x^\al(t)$ и $\dl x^\al(t)$.
Мы предполагаем, что вариации кривой в конечных точках $p,q\in\MM$ равны нулю,
$\dl x^\al(p)=\dl x^\al(q)=0$, и поэтому вкладом граничных членов при
интегрировании по частям можно пренебречь. Вариация длины кривой (\ref{eactim})
с точностью до знака имеет вид
\begin{equation*}
  \dl S=\int \frac{dt}{2\sqrt{|u^2|}}\left[2\dl(\dot x^\al)\dot x^\bt g_{\al\bt}
  +\dot x^\al\dot x^\bt\dl g_{\al\bt}\right].
\end{equation*}
Проинтегрируем первое слагаемое по частям и воспользуемся равенством
\begin{equation*}
  \dl g_{\bt\g}=\pl_\al g_{\bt\g}\dl x^\al.
\end{equation*}
Тогда вариация длины кривой принимает вид
\begin{equation*}
  \dl S=-\int dt\left[\frac d{dt}\left(\frac{\dot x^\bt g_{\al\bt}}{\sqrt{|u^2|}}
  \right)
  -\frac1{2\sqrt{|u^2|}}\dot x^\bt\dot x^\g\pl_\al g_{\bt\g}\right]\dl x^\al.
\end{equation*}
Поскольку
\begin{equation*}
  ds=\sqrt{|u|^2}dt,\qquad \text{и}\qquad
  \dot x^\al=\sqrt{|u^2|}\frac{dx^\al}{ds},
\end{equation*}
то вариацию длины кривой можно переписать в виде
\begin{equation*}
  \dl S=-\int ds\left(\frac{d^2 x^\bt}{ds^2}+\widetilde\Gamma_{\g\dl}{}^\bt
  \frac{dx^\g}{ds}\frac{dx^\dl}{ds}\right)g_{\al\bt}\dl x^\al,
\end{equation*}
где $\widetilde\Gamma_{\bt\g}{}^\al$ -- символы Кристоффеля (\ref{echris}).
Фактически, на этом этапе мы воспользовались инвариантностью интеграла
(\ref{eactim}) относительно перепараметризации кривой, выбрав длину кривой в
качестве параметра, $t\rightarrow s$. Параметр $s$ вдоль экстремали называется
{\em каноническим}. В дальнейшем канонический параметр мы будем обозначать
буквой $t$. Таким образом, мы доказали следующее утверждение.
\index{Канонический параметр (canonical parameter)}%
\index{Параметр канонический (canonical parameter)}%
\begin{theorem}
Для того, чтобы кривая $x(t)$ была экстремалью, необходимо и достаточно, чтобы
координатные функции $x^\al(t)$ удовлетворяли системе обыкновенных
дифференциальных уравнений
\begin{equation}                                                  \label{eextre}
  \ddot x^\al=-\widetilde\Gamma_{\bt\g}{}^\al\dot x^\bt\dot x^\g,
\end{equation}
где точка обозначает дифференцирование по каноническому параметру $t$.
\end{theorem}
\begin{com}
Параметр $t$, по которому проводится дифференцирование в уравнении
(\ref{eextre}) также, как и для геодезических, называется каноническим или
аффинным. Он определен с точностью до аффинного преобразования. Таким образом,
канонический параметр в общем случае не равен, а пропорционален длине экстремали.
\qed\end{com}
\begin{exa}
Пусть в некоторой области $\MU\subset\MM$ (псевдо-)риманова многообразия
существует такая система координат, в которой метрика имеет постоянные
компоненты. Тогда в этой области символы Кристоффеля равны нулю и,
следовательно, уравнения (\ref{eextre}) принимают вид $\ddot x^\al=0$. Поэтому
прямые и только они являются экстремалями в рассматриваемой области $\MU$.
\qed\end{exa}

Мы получили критерий того, что неизотропная кривая является экстремалью. При
выводе уравнений (\ref{eextre}) из вариационного принципа для действия
(\ref{eactim}) существенно используется условие $u^2\ne0$, которое исключает
изотропные (светоподобные) экстремали. Поэтому изотропные экстремали определим
не с помощью функционала длины, а непосредственно уравнениями (\ref{eextre}).
Для любой изотропной кривой $u^2=0$ и интеграл (\ref{eactim}) равен нулю. В то
же время уравнения (\ref{eextre}) имеют смысл.
\begin{defn}
{\em Изотропной экстремалью} называется изотропная кривая $\g$ класса
$\CC^2([t_1,t_2])$, которая задана функциями $x^\al(t)$, удовлетворяющими
системе обыкновенных дифференциальных уравнений (\ref{eextre}).
\qed\end{defn}
\index{Изотропная экстремаль (isotropic extremal)}%
\index{Экстремаль изотропная (isotropic extremal)}%
Это определение корректно, т.к.\ квадрат вектора скорости экстремали постоянен.
При этом не всякая изотропная кривая является экстремалью.
\begin{exa}
В четырехмерном пространстве Минковского $\MR^{1,3}$ любая изотропная кривая
имеет вид (\ref{enulfm}). В то же время экстремалями являются прямые линии и
только они. Следовательно, не всякая изотропная кривая является экстремалью.
Отметим, что в двумерном пространстве-времени любая изотропная кривая
является экстремалью, что нетрудно доказать в изотермических координатах
(конформной калибровке).
\qed\end{exa}
Определения изотропных и неизотропных экстремалей можно объединить.
\begin{defn}
Кривая $\g$ класса $\CC^2([t_1,t_2])$ называется {\em экстремалью}, если она
задается функциями $x^\al(t)$, где $t$ -- канонический параметр,
удовлетворяющими системе дифференциальных
уравнений (\ref{eextre}). Другими словами, экстремалью на (псевдо-)римановом
многообразии $(\MM,g)$ называется геодезическая для связности Леви-Чивиты.
\qed\end{defn}
\begin{com}
Мы начали с другого определения экстремалей, т.к.\ в последнем определении
теряется основное свойство неизотропных экстремалей реализовывать экстремумы
функционала длины (\ref{eactim}).
\qed\end{com}
Уравнение для экстремалей (\ref{eextre}) можно переписать в явно ковариантном
виде
\begin{equation}                                                  \label{excoeq}
  \dot x^\bt\widetilde\nb_\bt\dot x^\al=0.
\end{equation}
Это просто уравнение геодезических (\ref{egeoin}) для связности Леви--Чивиты.
Опустив в этом уравнении индекс $\al$ и расписав ковариантную производную,
получим альтернативную форму уравнения для экстремалей
\begin{equation}                                                  \label{eexrne}
  \frac\pl{\pl t}(g_{\al\bt}\dot x^\bt)
  -\frac12\pl_\al g_{\bt\g}\dot x^\bt\dot x^\g=0.
\end{equation}

Уравнения (\ref{eextre}) определяются только метрикой, поскольку функционал
длины не зависит от аффинной связности. Тем самым экстремали не зависят от того
заданы ли на многообразии $\MM$ тензоры кручения и неметричности или нет.
Сравнение уравнений (\ref{eextre}) с уравнением для геодезических (\ref{egeode})
показывает, что экстремали являются геодезическими линиями по отношению к
параллельному переносу, определяемому символами Кристоффеля. Это означает, что
все свойства геодезических справедливы также и для экстремалей. В частности,
канонический параметр вдоль экстремали инвариантен относительно преобразования
координат. При произвольной параметризации уравнение для экстремалей имеет вид
(\ref{ecaptg}).

Из предложения \ref{pinfsm} следует, что если метрика $g$ на многообразии
является гладкой функцией, то экстремали являются гладкими кривыми класса
$\CC^\infty$.

Через произвольную точку $x\in\MM$ в направлении $X^\al$ проходит одна и только
одна экстремаль, поскольку она определяется системой обыкновенных
дифференциальных уравнений второго порядка (\ref{eextre}). Это значит что любую
экстремаль, которая заканчивается в некоторой точке $q\in\MM$ можно продолжить.
Действительно, если она заканчивается в точке $q\in\MM$ при значении
канонического параметра $t_2$, то существует единственная экстремаль, проходящая
через $q$ и имеющая тот же касательный вектор.
\begin{defn}
Экстремаль $\g$ в $\MM$ называется {\em полной}, если ее можно продолжить до
бесконечного значения канонического параметра в обе стороны,
$t\in(-\infty,\infty)$.
\qed\end{defn}
\index{Экстремаль полная (complete extremal)}%
\index{Полнота экстремали (completeness of extremal)}%
\begin{com}
При продолжении экстремали возможны два случая. Во-первых, она может оказаться
полной и иметь бесконечную длину. К полным экстремалям мы относим также и
замкнутые экстремали, которые имеют конечную длину. Хотя их длина конечна, но
канонический параметр продолжается до бесконечности, что соответствует
бесконечному числу проходов вдоль экстремали. Во-вторых, при конечном значении
канонического параметра экстремаль может попасть в такую точку многообразия, в
которой один из геометрических инвариантов, например, скалярная кривизна,
неопределен. Эта точка является сингулярной, и продолжение экстремали через нее
не имеет смысла.
\qed\end{com}

Очевидно, что в (псевдо-)римановой геометрии экстремали и геодезические
совпадают, поскольку совпадают уравнения (\ref{egeode}) и (\ref{eextre})
при $\Gamma_{\al\bt}{}^\g=\widetilde\Gamma_{\al\bt}{}^\g$. В общем случае
экстремали и геодезические не совпадают и их поведение должно
исследоваться отдельно. Критерий совпадения экстремалей и геодезических
дает следующая
\begin{theorem}                                                   \label{textge}
В аффинной геометрии $(\MM,g,\Gamma)$ экстремали и геодезические совпадают тогда и
только тогда, когда кручение и неметричность удовлетворяют соотношению
\begin{equation}                                                  \label{etotno}
  \frac12(T_{\al\bt\g}-T_{\g\al\bt})=Q_{\al\bt\g}.
\end{equation}
\end{theorem}
\begin{proof}
Экстремали и геодезические совпадают если и только если симметричная часть
компонент связности равна символам Кристоффеля,
$\Gamma_{\lbrace\bt\g\rbrace}{}^\al=\widetilde\Gamma_{\bt\g}{}^\al$. Тогда из
(\ref{emecsy}) следует соотношение между тензором кручения и неметричностью:
\begin{equation}                                                  \label{eqproe}
  T_{\al\bt\g}-T_{\g\al\bt}=-Q_{\bt\g\al}-Q_{\g\al\bt}+Q_{\al\bt\g}.
\end{equation}
Сделав циклическую перестановку индексов $\al\bt\g\mapsto\bt\g\al$ и сложив
полученное уравнение с (\ref{eqproe}) получим (\ref{etotno}). Обратно.
Подстановка тензора неметричности (\ref{etotno}) в уравнение (\ref{eqproe}), как
легко проверить, приводит к тождеству.
\end{proof}
\begin{cor}
В геометрии Римана--Картана, где $Q=0$, экстремали и геодезические совпадают
тогда и только тогда, когда тензор кручения с опущенным индексом антисимметричен
по всем трем индексам, $T_{\al\bt\g}=T_{[\al\bt\g]}$.
\end{cor}
\section{Интегрирование уравнений для экстремалей и геодезических\label{sgexin}}
Уравнения для экстремалей и геодезических в ряде случаев имеют первые интегралы,
наличие которых существенно упрощает их исследование. Начнем с универсального
закона сохранения. Из сравнения уравнений (\ref{eextre}) и (\ref{egeode})
следует, что экстремаль является геодезической линией для связности,
определяемой символами Кристоффеля. Поскольку при параллельном переносе и в
римановой геометрии, и в геометрии Римана--Картана длины векторов не меняется,
то отсюда сразу следует, что длина вектора скорости $u=\lbrace\dot u^\al\rbrace$
вдоль экстремалей и геодезических постоянна.
\begin{prop}
В геометрии Римана--Картана и (псевдо-)римановой геометрии для уравнений
геодезических (\ref{egeode}) и экстремалей (\ref{eextre}) существует первый
интеграл
\begin{equation}                                                  \label{efirin}
  C_0=u^2:=\dot x^\al\dot x^\bt g_{\al\bt}=\const,
\end{equation}
квадратичный по первым производным (скоростям).
\end{prop}
\begin{proof}
Следствие предложения \ref{pscaex}. Можно доказать и формально,
продифференцировав уравнение (\ref{efirin}) по каноническому параметру и
воспользовавшись уравнением для геодезических или экстремалей.
\end{proof}
Рассмотрим как меняется квадрат длины касательного вектора
$C_0(t)=\dot x^\al\dot x^\bt g_{\al\bt}$ вдоль геодезической линии в аффинной
геометрии общего вида. Дифференцируя это соотношение по каноническому параметру
и используя уравнение для геодезических, получим
\begin{equation}                                                  \label{evalge}
  \frac{du^2}{dt}=\dot x^\al\dot x^\bt\dot x^\g Q_{\al\bt\g}.
\end{equation}
Отсюда следует, что $C_0=\const$ при нулевом тензоре неметричности. В геометрии
Римана--Картана--Вейля уравнение (\ref{evalge}) принимает вид
\begin{equation}                                                  \label{egelwe}
  \frac{du^2}{dt}=\dot x^\al W_\al C_0(t).
\end{equation}

Первый интеграл (\ref{efirin}) имеет кинематический характер и существует для
любой экстремали и геодезической в (псевдо-)римановой геометрии и геометрии
Римана--Картана. Если метрика имеет лоренцеву сигнатуру, то экстремали и
геодезические можно разделить на три класса: времениподобные, $C_0>0$,
изотропные или светоподобные, $C_0=0$, и пространственноподобные, $C_0<0$.
Поскольку канонический параметр определен с точностью до аффинных
преобразований, то для времениподобных и пространственноподобных экстремалей его
всегда можно выбрать таким образом, что $C_0=\pm1$. В этом случае для
времениподобных экстремалей канонический параметр называется {\em собственным
временем}, а для пространственноподобных -- {\em длиной} экстремали.
\index{Собственное время (proper time)}\index{Время собственное (proper time)}%
\index{Длина экстремали (length of extremal)}%
\begin{com}
Отметим, что, если некоторая кривая, не обязательно экстремаль или
геодезическая, имеет определенный тип, то вдоль нее параметр всегда можно
выбрать таким образом, что будет выполнено условие (\ref{efirin}). Для
изотропных кривых равенство (\ref{efirin}), очевидно, выполняется. Допустим, что
кривая имеет определенный тип в некоторой области, т.е.\ $C_0(t)\ne0$. Тогда,
вводя новый параметр $s(t)$, получим
$$
  C_0(t)=\left(\frac{ds}{dt}\right)^2\dot x^\al\dot x^\bt g_{\al\bt},
$$
где точка обозначает дифференцирование по $s$. Уравнение
$$
  \frac{ds}{dt}=\sqrt{|C_0|}
$$
всегда имеет решение. Поэтому условие (\ref{efirin}) будет выполнено
относительно нового параметра $s$.
\qed\end{com}

Существование других первых интегралов связано с инфинитезимальными симметриями
метрики, которые определяются векторными полями Киллинга.
\begin{prop}
Если метрика на многообразии имеет один или несколько векторов Киллинга
$K_i=\lbrace K_i^\al\rbrace$, $i=1,\dots,m$, то для каждого вектора Киллинга
имеется свой интеграл движения и для экстремалей, и для геодезических
\begin{equation}                                                  \label{ekilin}
  C_i=K_i^\al\dot x^\bt g_{\al\bt}=\const,\qquad i=1,\dots,m,
\end{equation}
который линеен по компонентам скорости.
\end{prop}
\begin{proof}
Дифференцируем соотношения (\ref{ekilin}) по каноническому параметру и
используем уравнения (\ref{egeode}) или (\ref{eextre}).
\end{proof}
\section{Вторая вариация уравнений для экстремалей               \label{saferk}}
Допустим, что метрика положительно определена, т.е.\ многообразие риманово.
Экстремали на (псевдо-)римановом многообразии $(\MM,g)$ определяются действием
(\ref{eactim}). Тогда любая экстремаль удовлетворяет уравнениям Эйлера--Лагранжа
(\ref{eextre}). Это свойство является необходимым условием для того, чтобы
экстремаль была линией наименьшей или наибольшей длины. Для того, чтобы найти
достаточное условие реализации минимума или максимума функционалов
(\ref{eactim}) или (\ref{elagex}) необходимо исследовать вторую вариацию
функционала.

Напомним общее определение второй вариации. Пусть задана достаточно гладкая
кривая $\g=\lbrace x^\al(t)\rbrace\in\MM$, где $t\in[t_1,t_2]$, соединяющая
две фиксированные точки многообразия $p=\g(t_1)$ и $q=\g(t_2)$. Рассмотрим
функционал
\begin{equation}                                                  \label{eacext}
  S[\g]=\int_\g dtL(x,\dot x).
\end{equation}
Допустим, что заданы два произвольных векторных поля $X,Y\in\CX(\MM,\g)$,
которые определены на кривой $\g$ и равны нулю в точках $p$ и $q$. Обозначим
через $\lm$ и $\mu$ два вещественных параметра.
\begin{defn}
Назовем {\em первой вариацией} функционала (\ref{eacext}) частную производную
\begin{equation}                                                  \label{efirva}
  \left.\frac\pl{\pl\lm}S[\g+\lm X]\right|_{\lm=0}
  =\int_\g dt\frac{\dl S}{\dl x^\al}X^\al,
\end{equation}
Частная производная
\begin{equation}                                                  \label{esevar}
  G_\g[X,Y]=G_\g[Y,X]
  =\left.\frac{\pl^2}{\pl\lm\pl\mu}S_\g[\g+\lm X+\mu Y]\right|_{\lm=0,~\mu=0}
\end{equation}
называется {\em второй вариацией} функционала (\ref{eacext}).
\qed\end{defn}
\index{Первая вариация (first variation)}%
\index{Вариация первая (first variation)}%
\index{Вторая вариация (second variation)}%
\index{Вариация вторая (second variation)}%
Поскольку векторное поле $X$ произвольно, то равенство нулю первой вариации
эквивалентно уравнениям Эйлера--Лагранжа, которые мы записываем в виде
\begin{equation}                                                  \label{eilaex}
  \frac{\dl S}{\dl x^\al}=\frac{\pl L}{\pl x^\al}
  -\frac\pl{\pl t}\left(\frac{\pl L}{\pl\dot x^\al}\right)=0.
\end{equation}

\begin{prop}
Если кривая $\g=\lbrace x^\al(t)\rbrace$ удовлетворяет уравнениям
Эйлера--Лагранжа (\ref{eilaex}), то вторая вариация функционала (\ref{eacext})
имеет вид
\begin{equation}                                                  \label{esevaf}
  G_\g[X,Y]=-\int_{t_1}^{t_2}\!\!\!dt\,\left(J_{\al\bt}X^\bt\right)Y^\al,
\end{equation}
где
\begin{equation}                                                  \label{edefju}
  J_{\al\bt}X^\bt=\frac d{dt}
  \left(\frac{\pl^2L}{\pl\dot x^\al\pl\dot x^\bt}\dot X^\bt
  +\frac{\pl^2L}{\pl\dot x^\al\pl x^\bt}X^\bt\right)
  -\frac{\pl^2L}{\pl x^\al\pl\dot x^\bt}\dot X^\bt
  -\frac{\pl^2L}{\pl x^\al\pl x^\bt}X^\bt.
\end{equation}
\end{prop}
\begin{proof}
Прямые вычисления с учетом выражения для первой вариации:
\begin{multline*}
  G_\g[X,Y]=\left.\frac\pl{\pl\mu}\int dt\left(\frac{\pl L}{\pl x^\bt}X^\bt
  +\frac{\pl L}{\pl\dot x^\bt}\dot X^\bt\right)\right|_{\mu=0}=
\\
  =\int dt\left(\frac{\pl^2L}{\pl x^\al\pl x^\bt}X^\bt Y^\al
  +\frac{\pl^2L}{\pl\dot x^\al\pl x^\bt}X^\bt\dot Y^\al
  +\frac{\pl^2L}{\pl x^\al\pl\dot x^\bt}\dot X^\bt Y^\al
  +\frac{\pl^2L}{\pl\dot x^\al\pl\dot x^\bt}\dot X^\bt\dot Y^\al\right).
\end{multline*}
После интегрирования по частям двух слагаемых, пропорциональных $\dot Y^\al$,
получаем выражение (\ref{edefju}).
\end{proof}
\begin{defn}
Линейный оператор $J=\lbrace J^\al{}_\bt:=g^{\al\g}J_{\g\bt}\rbrace$,
\begin{equation*}
  J:\quad \CX(\MM,\g)\rightarrow\CX(\MM,\g),
\end{equation*}
который действует на векторные поля $X$, определенные вдоль кривой $\g$, по
правилу (\ref{edefju}), называется {\em оператором Якоби}.
\qed\end{defn}
\index{Оператор Якоби (Jacobi operator)}%
\index{Якоби оператор (Jacobi operator)}%

Теперь применим формулу (\ref{esevaf}) для вычисления второй вариации действия
для экстремалей. Во-первых, отметим, что действие для экстремалей в виде
(\ref{eactim}) не является единственным. Рассмотрим действие
\begin{equation}                                                  \label{extrim}
  S=\int_\g\!dt\frac12g_{\al\bt}\dot x^\al\dot x^\bt,
\end{equation}
\begin{prop}
Действие (\ref{extrim}) имеет тот же набор экстремалей, что и действие
(\ref{eactim}).
\end{prop}
\begin{proof}
Вариация действия (\ref{extrim}) по $x^\al(t)$ приводит к уравнениям
(\ref{eextre}).
\end{proof}
В отличие от действия (\ref{eactim}) функционал (\ref{extrim}) не инвариантен
относительно перепараметризации кривой. Это значит, что параметр $t$ в
действии (\ref{extrim}) является каноническим.

Пусть $\g=\lbrace x^\al(t)\rbrace$, $t\in[t_1,t_2]$, -- экстремаль и
$u:=\lbrace\dot x^\al\rbrace$ -- вектор скорости экстремали.
\begin{theorem}
Вторая вариация действия для экстремалей (\ref{extrim}) имеет вид
\begin{equation}                                                  \label{exseva}
  G_\g(X,Y)=-\int_{t_1}^{t_2}\!\!\!dt\left(\nb^2_u Y^\al
  +u^\dl u^\bt Y^\g \widetilde R_{\g\dl\bt}{}^\al\right)X_\al,
\end{equation}
где
\begin{equation}                                                  \label{ereydd}
\begin{split}
  \nb_u Y^\al&:=u^\bt\left(\pl_\bt Y^\al+\widetilde\Gamma_{\bt\g}{}^\al Y^\g\right)
  =\dot Y^\al +u^\bt\widetilde\Gamma_{\bt\g}{}^\al Y^\g,
\\
  \nb^2_u Y^\al&:=u^\dl\nb_\dl\left(\dot Y^\al
  +u^\bt\widetilde\Gamma_{\bt\g}{}^\al Y^\g\right)=
\\
  &=\ddot Y^\al+2u^\bt\widetilde\Gamma_{\bt\g}{}^\al\dot Y^\g
  +u^\dl u^\e Y^\g\left(\pl_\dl\widetilde\Gamma_{\e\g}{}^\al
  -\widetilde\Gamma_{\dl\e}{}^\bt\widetilde\Gamma_{\g\bt}{}^\al
  +\widetilde\Gamma_{\dl\g}{}^\bt\widetilde\Gamma_{\e\bt}{}^\al\right)
\end{split}
\end{equation}
-- ковариантные производные вдоль вектора скорости и
$\widetilde R_{\g\dl\bt}{}^\al$ -- тензор кривизны.
\end{theorem}
\begin{proof}
Первая вариация действия (\ref{extrim}) имеет вид
\begin{equation*}
  \left.\frac{\pl S[\g+\lm X]}{\pl\lm}\right|_{\lm=0}
  =-\int_{t_1}^{t_2}\!\!\!dt\left(\ddot x^\al
  +\widetilde\Gamma_{\bt\g}{}^\al\dot x^\bt\dot x^\g\right)X_\al.
\end{equation*}
Поэтому
\begin{equation*}
\begin{split}
  &G_\g(X,Y)=
\\
  &=-\frac\pl{\pl\mu}
  \int_{t_1}^{t_2}\!\!\!dt\left(\ddot x^\al+\mu\ddot Y^\al
  +\widetilde\Gamma_{\bt\g}{}^\al\dot x^\bt\dot x^\g
  +2\mu\widetilde\Gamma_{\bt\g}{}^\al\dot x^\bt\dot Y^\g
  +\mu\pl_\e\widetilde\Gamma_{\g\dl}{}^\al\dot x^\bt\dot x^\g Y^\e\right)X_\al=
\\
  &=-\int_{t_1}^{t_2}\!\!\!dt\left(\ddot Y^\al
  +2\widetilde\Gamma_{\bt\g}{}^\al\dot x^\bt\dot Y^\g
  +\pl_\e\widetilde\Gamma_{\g\dl}{}^\al\dot x^\bt\dot x^\g Y^\e\right),
\end{split}
\end{equation*}
где мы оставили только линейные по $\mu$ слагаемые и учли, что вклад производной
$\pl X_\al/\pl\mu$ равен нулю при выполнении уравнений для экстремалей
$\nb_u\dot x^\al=0$. Исключив из полученного выражения производные
$\ddot Y^\al$ и $\dot Y^\al$ с помощью второго из соотношений (\ref{ereydd}),
получим равенство (\ref{exseva}).
\end{proof}
Ясно, что билинейная форма (\ref{exseva}) симметрична, $G_\g(X,Y)=G_\g(Y,X)$,
т.к.\ ковариантное дифференцирование можно перекинуть на вектор $X$, а тензор
кривизны симметричен относительно перестановки пар индексов:
$\widetilde R_{\al\bt\g\dl}=\widetilde R_{\g\dl\al\bt}$ (\ref{esymct}).

Предположим, что метрика $g_{\al\bt}$ риманова. Тогда условие минимальности
экстремали $\g$, соединяющей точки $p$ и $q$, состоит в том, что квадратичная
форма $G_\g(X,X)$ положительно определена для всех векторных полей, определенных
на экстремали $\g$ и обращающихся в нуль на ее концах.
\begin{defn}
Векторное поле $X$, которое определено на экстремали $\g$, соединяющей точки $p$
и $q$, называется {\em якобиевым}, если оно удовлетворяет {\em уравнению Якоби}
\begin{equation}                                                  \label{ejaequ}
  J^\al{}_\bt X^\bt=0,
\end{equation}
где $J$ -- оператор Якоби (\ref{edefju}), и обращается в нуль на концах $p$ и
$q$. Точки $p$ и $q$ называются {\em сопряженными} вдоль экстремали $\g$, если
существует ненулевое якобиево поле $X$ вдоль $\g$.
\qed\end{defn}
\index{Якобиево векторное поле (Jacobi vector field)}%
\index{Векторное поле якобиево (Jacobi vector field)}%
\index{Уравнение Якоби (Jacobi equation)}%
\index{Уравнение Якоби (Jacobi equation)}%
\index{Сопряженные точки (conjugate points)}%
\index{Точки сопряженные (conjugate points)}%

Для действия (\ref{extrim}) уравнение Якоби принимает вид
\begin{equation}                                                  \label{ejaeqe}
  \nb^2_u X^\al+u^\dl u^\bt X^\g\widetilde R_{\g\dl\bt}{}^\al=0.
\end{equation}
Это уравнение совпадает с уравнением девиации геодезических (\ref{qbvvgt}).
\begin{prop}
Билинейная форма $G_\g(X,Y)$ невырождена тогда и только тогда, когда концевые
точки $p$ и $q$ экстремали $\g$ не сопряжены вдоль $\g$.
\end{prop}
\begin{proof}
Пусть $X$ и $Y$ произвольные векторные поля вдоль $\g$, которые обращаются в
нуль в концевых точках. Напомним, что билинейная форма $G_\g(X,Y)$ называется
невырожденной, если не существует такого векторного поля $X$, что $G_\g(X,Y)=0$
для всех $Y$. Если поле $X$ якобиево, то $G_\g(X,Y)=0$ при любом $Y$.

Обратно. Допустим, что для вектора $X$ выполнено уравнение $G_\g(X,Y)=0$ для
всех $Y$. Положим $Y=f(t)JX$, где функция $f(t)$ неотрицательна и обращается в
нуль на концах экстремали. Тогда из выражения для второй вариации (\ref{esevaf})
имеем
\begin{equation*}
  G_\g(X,Y)=-\int_{t_1}^{t_2}\!\!\!dt \,fg(JX,JX)=0.
\end{equation*}
Отсюда следует, что $JX=0$. Поэтому концы сопряжены.
\end{proof}
Теперь сформулируем необходимое условие минимальности экстремали.
\begin{theorem}
Если экстремаль $\g$, соединяющая точки $p$ и $q$, содержит внутри себя пару
сопряженных точек $p'$ и $q'$, то она не является минимальной.
\end{theorem}
\begin{proof}
См., например, \cite{DuNoFo98R}.
\end{proof}
\begin{theorem}
Для достаточно малых отрезков $l=[t_1,t_2]$ экстремали задают минимум
функционала (\ref{extrim}). Поэтому каждая экстремаль является кратчайшей линией
в классе дважды дифференцируемых кривых, соединяющих достаточно близкие точки
$p$ и $q$.
\end{theorem}
\begin{proof}
Экстремаль $\g$ реализует минимум функционала (\ref{extrim}), если форма
$G_\g(X,X)$ положительна для всех ненулевых векторных полей $X\in\CX(\MM,\g)$,
которые обращаются в нуль на концах $p$ и $q$. Из формулы для второй вариации
(\ref{exseva}) следует, что
\begin{align*}
  G_\g(X,X)&=-\int_p^q\!\!\!dt\left[(\nb_u^2X,X)-u^\al X^\bt u^\g X^\dl
  \widetilde R_{\al\bt\g\dl}\right]
\\
  &=\int_p^q\!\!\!dt\left[(\nb_u X,\nb_u X)+u^\al X^\bt u^\g X^\dl
  \widetilde R_{\al\bt\g\dl}\right],
\end{align*}
где мы проинтегрировали первое слагаемое по частям с учетом равенств
$X(p)=X(q)=0$. Можно доказать, что для достаточно коротких экстремалей
справедлива оценка
\begin{equation*}
  \left|\int_p^q\!\!\!dt\,u^\al X^\bt u^\g X^\dl
  \widetilde R_{\al\bt\g\dl}\right|=\obig(l)\int_p^q\!\!\!dt(\nb_u X,\nb_u X)
\end{equation*}
при $l\to0$. Поскольку квадратичная форма $(\nb_u X,\nb_u X)$ положительна, то
для достаточно коротких экстремалей положительна и форма $G_\g(X,X)$.
\end{proof}
\section{Уравнение Гамильтона--Якоби для экстремалей             \label{shajae}}
Уравнения для экстремалей $\g=x(t)$ вытекают из вариационного принципа для
действия (\ref{eactim}). Важным обстоятельством является то, что уравнения для
экстремалей являются уравнениями Эйлера--Лагранжа также и для другого действия
(\ref{extrim}). А именно, рассмотрим действие
\begin{equation}                                                  \label{elagex}
  S_\mathrm{m}=\int_a^b \!\!\!dt L_\mathrm{m},\qquad
  L_\mathrm{m}=-\frac12 mg_{\al\bt}\dot x^\al\dot x^\bt,
\end{equation}
где точка обозначает дифференцирование по каноническому параметру $t$ и
$m=\const$ -- постоянная, имеющая физический смысл массы точечной частицы. Это
действие совпадает с (\ref{extrim}) при $m=-1$. Действие (\ref{elagex}) приводит
к уравнениям для экстремалей, в которых переменная $t$ уже является каноническим
параметром. Это согласуется с тем обстоятельством, что рассмотренное действие
инвариантно относительно общих преобразований координат и сдвигов параметра $t$.
Для сравнения напомним, что исходное действие для экстремалей (\ref{eactim})
инвариантно также относительно произвольных преобразований параметра $t$ вдоль
экстремали.

Действие (\ref{elagex}) имеет простой физический смысл. Предположим, что метрика
имеет лоренцеву сигнатуру и зафиксируем временн\'ую калибровку (\ref{etemga})
\begin{equation}                                                  \label{emebld}
  g_{\al\bt}=\begin{pmatrix} 1 & 0 \\ 0 & g_{\mu\nu} \end{pmatrix},
  \qquad\sign g_{\mu\nu}=(-\dotsc -).
\end{equation}
Символы Кристоффеля для этой метрики имеют вид (\ref{echrti}).
Предположим также, что пространственная часть метрики $g_{\mu\nu}$ не
зависит от времени $x^0$. Тогда уравнения для экстремалей расщепляются:
\begin{align}                                                     \label{extzet}
  \ddot x^0&=0,
\\                                                                \label{extspt}
  \ddot x^\mu&=-\widetilde\Gamma_{\nu\rho}{}^\mu\dot x^\nu\dot x^\rho.
\end{align}
Из первого уравнения следует, что, не ограничивая общности, канонический
параметр $t$ можно отождествить с временем $x^0=ct$, где $c=\const$ -- скорость
света. Тогда лагранжиан (\ref{elagex}) имеет прямой физический смысл -- с
точностью до аддитивной постоянной это кинетическая энергия точечной частицы,
которая движется в римановом пространстве со статической метрикой
$g_{\mu\nu}(x)$. Несмотря на то, что потенциальная энергия частицы равна нулю,
ее траекториями уже не будут прямые линии, если метрика нетривиально зависит от
точки пространства.

Вернемся к исходному действию (\ref{elagex}) до фиксирования временн\'ой
калибровки. Переформулируем эту лагранжеву систему на гамильтоновом языке,
рассматривая канонический параметр $t$ в качестве параметра эволюции. Под
временем мы подразумеваем координату $x^0$ и, соответственно, предполагаем,
что $g_{00}>0$. Кроме этого мы предполагаем, что все сечения постоянного времени
$x^0=\const$ пространственноподобны. Импульс, сопряженный координатам $x^\al$, и
гамильтониан системы равны
\begin{equation}                                                  \label{efrham}
\begin{split}
  p_\al&=\frac{\pl L_\mathrm{m}}{\pl\dot x^\al}=-mg_{\al\bt}\dot x^\bt,
\\
  H_\mathrm{m}&=-\frac1{2m}g^{\al\bt}p_\al p_\bt.
\end{split}
\end{equation}
Для действия (\ref{elagex}) связи на канонические переменные отсутствуют, т.к.\
метрика рассматривается как внешнее поле и варьирование по ней не проводится.
Соответствующие уравнения Гамильтона (уравнения движения) имеют вид
\begin{align}                                                     \label{edtxal}
  \dot x^\al&=[x^\al,H_\mathrm{m}]=-\frac1m g^{\al\bt}p_\bt,
\\                                                                \label{edmoal}
  \dot p_\al&=[p_\al,H_\mathrm{m}]=\frac1{2m}\pl_\al g^{\bt\g}p_\bt p_\g.
\end{align}
Дифференцируя первое из этих уравнений по каноническому параметру и исключая
импульсы $p_\al$ и производные $\dot p_\al$ с помощью уравнений движения,
нетрудно проверить, что система уравнений (\ref{edtxal}), (\ref{edmoal})
эквивалентна системе уравнений для экстремалей (\ref{eextre}). Тем самым мы
переписали уравнения для экстремалей в виде канонической системы уравнений
движения.

Ранее было доказано, что длина касательного вектора к экстремали постоянна
(\ref{efirin}). В гамильтоновом форме это утверждение имеет вид
\begin{equation}                                                  \label{ecolao}
  C_0=\frac1{m^2}g^{\al\bt}p_\al p_\bt=\const.
\end{equation}
По своей физической сути это есть закон сохранения энергии точечной частицы. В
данном случае только кинетической, т.к.\ потенциальная энергия тождественно
равна нулю.

В дальнейшем нам понадобятся гамильтоновы уравнения для нулевых экстремалей,
где в качестве параметра эволюции выбрано время $x^0$, а не канонический
параметр $t$. Они получаются следующим образом. Для нулевых экстремалей
интеграл движения (\ref{ecolao}) принимает вид
\begin{equation*}
  g^{00}p_0^2+2g^{0\mu}p_0p_\mu+g^{\mu\nu}p_\mu p_\nu=0.
\end{equation*}
Это квадратное уравнение решается относительно $p_0$:
\begin{equation}                                                  \label{epzeip}
  p_0=N^\mu p_\mu\pm N\hat p,\qquad\hat p:=\sqrt{-\hat g^{\mu\nu}p_\mu p_\nu},
\end{equation}
где $N=1/\sqrt{g^{00}}\ne0$ и $N^\mu=\hat g^{\mu\nu}g_{0\nu}$ -- функции хода и
сдвига, используемые в АДМ параметризации метрики (\ref{eadmme}), а
$\hat g^{\mu\nu}$ -- метрика, обратная к пространственной метрике $g_{\mu\nu}$.
Если частица движется, то $p_\mu\ne0$ и $\hat p\ne0$. Тогда из уравнения
(\ref{edtxal}) находим производную координаты $x^0$ по каноническому параметру
\begin{equation*}
  \dot x^0=\mp\frac{\hat p}{mN}.
\end{equation*}
Отсюда следует, что выбор знака в (\ref{epzeip}) соответствует выбору взаимной
ориентации канонического параметра $t$ и времени $x^0$. После этого канонические
уравнения (\ref{edtxal}), (\ref{edmoal}) можно записать в виде системы уравнений
движения только для пространственных координат и импульсов:
\begin{equation}                                                  \label{ecoemx}
\begin{split}
  \pl_0x^\mu&=\frac{\dot x^\mu}{\dot x^0}
  =-N^\mu\pm\frac{N\hat g^{\mu\nu}p_\nu}{\hat p},
\\
  \pl_0p_\mu&=\frac{\dot p_\mu}{\dot x^0}=\pl_\mu N^\nu p_\nu
  \pm\pl_\mu N\hat p\mp\frac{N\pl_\mu\hat g^{\nu\rho}p_\nu p_\rho}{2\hat p}.
\end{split}
\end{equation}
Таким образом, из системы канонических уравнений (\ref{edtxal}), (\ref{edmoal})
для нулевых экстремалей мы исключили в явном виде канонический параметр $t$ и
нулевую компоненту импульса $p_0$.
\begin{prop}                                                      \label{enucga}
Система уравнений для пространственных компонент канонически сопряженных
переменных $x^\mu$ и $p_\mu$ (\ref{ecoemx}) является гамильтоновой. При этом
эволюция системы уравнений рассматривается по отношению к времени $x^0$, а
гамильтонианом является выражение для $p_0$ (\ref{epzeip}).
\end{prop}
\begin{proof}
Прямое сравнение уравнений
\begin{align}                                                          \nonumber
  \pl_0x^\mu&=[x^\mu,p_0],
\\                                                                   \tag*{\qed}
  \pl_0p_\mu&=[p_\mu,p_0].
\end{align}
\renewcommand{\qed}{}\end{proof}
Число гамильтоновых уравнений движения, определяющих экстремаль, сократилось
с $2n$ в (\ref{edtxal}), (\ref{edmoal}) до $2n-2$ в (\ref{ecoemx}). Это
достигнуто за счет использования интеграла движения (\ref{ecolao}) и выбора
специального параметра эволюции $x^0$.

Продолжим анализ гамильтоновой формы уравнений для экстремалей. Функция действия
$S_\mathrm{m}(x,t)$ (\ref{efunac}) удовлетворяет уравнению Гамильтона--Якоби
(\ref{eyfjae})
\begin{equation}                                                  \label{ehamja}
  \frac{\pl S_\mathrm{m}}{\pl t}-\frac1{2m} g^{\al\bt}
  \frac{\pl S_\mathrm{m}}{\pl x^\al}\frac{\pl S_\mathrm{m}}{\pl x^\bt}=0.
\end{equation}
Поскольку гамильтониан (\ref{efrham}) не зависит от параметра $t$ явно, то
функция действия имеет вид
$$
  S_\mathrm{m}(x^\al,t)=\frac{mC_0}2t+W_\mathrm{m}(x^\al),\qquad C_0=\const,
$$
где укороченная функция действия $W_\mathrm{m}$ удовлетворяет укороченному
уравнению Гамильтона--Якоби
\begin{equation}                                                  \label{ehajas}
  g^{\al\bt}\frac{\pl W_\mathrm{m}}{\pl x^\al}
  \frac{\pl W_\mathrm{m}}{\pl x^\bt}=m^2C_0.
\end{equation}

Так как $p_\al=\pl W_\mathrm{m}/\pl x^\al$, то постоянная $C_0$ равна
длине касательного вектора к экстремали
\begin{equation*}
  g_{\al\bt}\dot x^\al\dot x^\bt=\frac1{m^2}g^{\al\bt}p_\al p_\bt=C_0.
\end{equation*}
Отметим, что для экстремалей нулевой длины $C_0=0$, и укороченное действие
совпадает с полным $W_\mathrm{m}=S_\mathrm{m}$.
\section{Волновое уравнение                                      \label{swaequ}}
\index{Волновое уравнение (wave equation)}%
\index{Уравнение волновое (wave equation)}%
\index{Принцип Гюйгенса (Huygens's principle)}%
\index{Гюйгенса принцип (Huygens's principle)}%
Пусть задано многообразие $\MM$, $\dim\MM=n$, с метрикой лоренцевой сигнатуры,
$\sign g_{\al\bt}=(+-\dotsc-)$. Рассмотрим волновое уравнение для скалярного
поля $\vf(x)\in\CC^2(\MM)$:
\begin{align}                                                     \label{ewaeqs}
  &g^{\al\bt}\widetilde\nb_\al\widetilde\nb_\bt\vf=g^{\al\bt}\pl_\al\pl_\bt\vf
  -g^{\al\bt}\widetilde\Gamma_{\al\bt}{}^\g\pl_\g\vf=0,
\\ \intertext{или, эквивалентно,}                                      \nonumber
  &\frac1\vol\pl_\al\left(\vol g^{\al\bt}\pl_\bt\vf\right)=0,\qquad
  g:=\det g_{\al\bt},
\end{align}
где мы воспользовались тождеством (\ref{elassr}). Это -- линейное
дифференциальное уравнение в частных производных второго порядка с переменными
коэффициентами гиперболического типа, т.к.\ метрика имеет лоренцеву сигнатуру.
Важным понятием в теории дифференциальных уравнений является характеристика (или
характеристическая поверхность), которая для дифференциальных уравнений второго
порядка определяется квадратичной формой $g^{\al\bt}$.
\begin{defn}
{\em Характеристикой} гиперболического дифференциального уравнения второго
порядка (\ref{ewaeqs}) называется $\CC^1$ гиперповерхность в многообразии $\MM$,
которая задается уравнением
\begin{equation}                                                  \label{ecared}
  W(x)=0,
\end{equation}
где функция $W\in\CC^1(\MM)$ на поверхности $W=0$ удовлетворяет условию
\begin{equation}                                                  \label{echaeq}
  \left.g^{\al\bt}\pl_\al W\pl_\bt W\right|_{W=0}=0.
\end{equation}
При этом требуется, чтобы по крайней мере одна из частных производных
$\pl_\al W$ была отлична от нуля на гиперповерхности (\ref{ecared}).
\qed\end{defn}
\index{Характеристика (characteristic)}%
\begin{com}
Отметим, что в определении характеристики важна гиперболичность, т.к.\ при
положительно или отрицательно определенной метрике уравнение (\ref{echaeq}) не
имеет вещественных решений.
\qed\end{com}
Уравнение (\ref{echaeq}) для характеристики совпадает с
укороченным уравнением Гамильтона--Якоби для экстремалей (\ref{ehajas}) при
$C_0=0$. Это значит, что характеристика соответствует укороченной функции
действия для экстремалей нулевой длины. Напомним, что для экстремалей нулевой
длины укороченная и полная функции действия совпадают. Однако условие
(\ref{echaeq}) является более слабым, т.к.\ мы требуем выполнения (\ref{echaeq})
только на характеристике, а не во всем пространстве-времени.

Из определения характеристик следует, что они являются изотропными
гиперповерхностями, которые будут рассмотрены в разделе \ref{slicga}. Метрика,
индуцированная на таких поверхностях, по-определению, вырождена, и все
нормальные векторы изотропны (теорема \ref{tisohy}).

Характеристики обладают следующим важным свойством. Допустим, что каждая
гиперповерхность $W(x)-y^0=0$, где $-\e<y^0<\e$, $\e>0$, есть характеристика
уравнения (\ref{ewaeqs}). Другими словами, мы требуем, чтобы уравнение
характеристик (\ref{echaeq}) выполнялось не только на самой характеристике, но и
в некоторой ее окрестности. Поскольку на каждой характеристике по крайней мере
одна из частных производных $\pl_\al W$ отлична от нуля, то это семейство
заполняет некоторую достаточно малую область, через каждую точку которой
проходит одна и только одна характеристика. Тогда можно перейти в новую систему
координат $x^\al\rightarrow y^\al(x)$, где $y^0=W$. При этом обратная метрика
преобразуется по тензорному закону
\begin{equation*}
  g^{\al\bt}(y)=\frac{\pl y^\al}{\pl x^\g}
  \frac{\pl y^\bt}{\pl x^\dl}g^{\g\dl}(x).
\end{equation*}
Отсюда следует, что в новой системе координат $g^{00}=0$. Это обстоятельство
имеет важное следствие. Рассмотрим задачу Коши для волнового уравнения
(\ref{ewaeqs}) в новой системе координат. Пусть на характеристике заданы
начальные условия
\begin{equation*}
  \vf|_{y^0=0}=\vf_0,\qquad \pl_0\vf|_{y^0=0}=\vf_1,
\end{equation*}
где $\vf_0$ и $\vf_1$ -- достаточно гладкие функции от ``пространственных''
координат $\lbrace y^\mu\rbrace=(y^1,\dotsc,y^{n-1})$. По начальным данным на
характеристике при $y^0=0$ можно вычислить все частные производные по
``пространственным'' координатам $\pl_\mu\vf,\pl^2_{\mu\nu}\vf,\dotsc$ и все
частные производные с одной производной по ``времени''
$\pl^2_{0\mu}\vf,\pl^3_{0\mu\nu}\vf,\dotsc$. Для определения эволюции скалярного
поля необходимо знать вторые производные по ``времени'' $\pl^2_{00}\vf$. Если
задача Коши корректно поставлена, то вторые производные по ``времени'' находятся
из волнового уравнения. На характеристике это не так, поскольку $g^{00}=0$.
Вместо определения второй производной $\pl^2_{00}\vf$ волновое уравнение
накладывает ограничение (связь) на возможный выбор начальных условий:
\begin{equation*}
  g^{0\mu}\pl_\mu\vf_1-g^{\al\bt}\widetilde\Gamma_{\al\bt}{}^0\vf_1
  -g^{\al\bt}\widetilde\Gamma_{\al\bt}{}^\mu\pl_\mu\vf_0=0.
\end{equation*}
Для нахождения вторых производных $\pl_{00}^2\vf$ волновое уравнение нужно
продифференцировать по $y^0$. Однако после этой процедуры вторые производные не
будут определены однозначно. Таким образом, при постановке задачи Коши на
характеристике, во-первых, начальные данные нельзя задавать произвольно, и,
во-вторых, волновое уравнение не определяет эволюцию поля единственным образом.
Тем самым задача Коши на характеристике не допускает корректную постановку.

Выше мы взяли слова ``пространственный'' и ``время'' в кавычки, т.к.\
характеристика является изотропной поверхностью и не может быть
пространственноподобной, а координата $y^0$ светоподобна и не может играть роль
времени.

Поскольку вторые производные от неизвестной функции по одну сторону
характеристики не определяются уравнением (\ref{ewaeqs}) и значениями функции
$\vf$ по другую сторону от характеристики, то они могут иметь разрывы. Это
значит, что решения уравнения (\ref{ewaeqs}) могут иметь разрывы производных,
которые распространяются в пространстве вдоль характеристик.

Для корректной постановки задачи Коши для волнового уравнения (\ref{ewaeqs}) на
многообразии с метрикой лоренцевой сигнатуры по $x^0$ координаты выбираются
таким образом, что $g^{00}>0$, а сечение $x^0=0$ является
пространственноподобным. Поскольку пространственноподобное сечение не может быть
изотропным, то оно не может быть также характеристикой. Оно не может также
касаться характеристики, потому что в этом случае один из касательных векторов
был бы изотропным, что противоречит отрицательной определенности метрики,
индуцированной на гиперповерхности.
\begin{com}
В разделе \ref{signme} мы определили время $x^0$ как такую координату, что
касательное векторное поле $\pl_0$ является времениподобным. Это соответствует
условию $g_{00}>0$. Отметим, что условия $g_{00}>0$ и $g^{00}>0$ неэквивалентны.
Подробнее эти условия будут обсуждаться в разделе \ref{sadmpa} при АДМ
параметризации метрики.
\qed\end{com}

Уравнение характеристик (\ref{echaeq}) представляет собой нелинейное
дифференциальное уравнение в частных производных первого порядка. В общем случае
оно очень сложно и, как правило, имеет решения с особыми точками.
Характеристика, например, может быть не гладкой поверхностью.

Относительно частной производной по времени $\pl_0W$ уравнение характеристик
является алгебраическим квадратным уравнением, и имеет не более двух
вещественных корней. Допустим, что метрика имеет лоренцеву сигнатуру и
$g_{00}>0$, тогда оно имеет два вещественных корня разных знаков
\begin{equation}                                                  \label{erooch}
  \pl_0W=N^\mu\pl_\mu W\pm N\sqrt{-\hat g^{\mu\nu}\pl_\mu W\pl_\nu W},
\end{equation}
где использована АДМ параметризаций метрики (\ref{eadmme}). Здесь мы
предполагаем, что все сечения $x^0=\const$ являются пространственноподобными
и, следовательно, метрика $g_{\mu\nu}$ и ее обратная $\hat g^{\mu\nu}$
отрицательно определены. Обозначим $p_\mu:=\pl_\mu W$. Введем функцию Гамильтона
в фазовом пространстве $\lbrace x^\mu,p_\mu\rbrace$ (см.\ главу \ref{scanfp})
\begin{equation}                                                  \label{ehabic}
  H(x^0,x^\mu,p_\mu)=-N^\mu p_\mu\mp N\sqrt{-\hat g^{\mu\nu}p_\mu p_\nu},
\end{equation}
где зависимость от координат $x^0$ и $x^\mu$ входит через метрику
$g^{\al\bt}=g^{\al\bt}(x^0,x^\mu)$, и время $x^0$ рассматривается в качестве
параметра эволюции.
\begin{defn}
Траектории $x^\mu(x^0)$ в конфигурационном пространстве, соответствующем
гамильтониану (\ref{ehabic}), называются {\em бихарактеристиками} волнового
уравнения (\ref{ewaeqs}).
\qed\end{defn}
\index{Бихарактеристика (bicharacteristic)}%
\begin{prop}
Бихарактеристики волнового уравнения (\ref{ewaeqs}) совпадают с образами
нулевых экстремалей для метрики $g_{\al\bt}$.
\end{prop}
\begin{proof}
Следствие предложения \ref{enucga}.
\end{proof}
Единственное отличие бихарактеристик и нулевых экстремалей сводится к тому, что
экстремали определяются в параметрическом виде $x^\al(t)$, а бихарактеристики
задаются в виде явной зависимости координат $x^\mu(x^0)$.

Гамильтониан (\ref{ehabic}) совпадает с временн\'ой компонентой импульса
(\ref{epzeip}). Функция $W(x^0,x^\mu)$, определяющая характеристику, является
действием для бихарактеристик.
\begin{com}
Характеристики и бихарактеристики имеют физическую интерпретацию. Из теории
дифференциальных уравнений известно, что если в момент времени $x^0$ в точке
$\Bx$ произошло некоторое возмущение решения $\vf$ волнового уравнения
(\ref{ewaeqs}), то оно будет распространяться в виде волны. При этом сечения
$x^0=\const$ соответствующей характеристической поверхности (характеристического
коноида с вершиной в точке $\lbrace x^0,\Bx\rbrace$) являются фронтом волны в
момент времени $x^0$, а бихарактеристики --  это лучи, вдоль которых
распространяется волна.
\qed\end{com}
\begin{prop}
Если бихарактеристика касается характеристики в некоторой точке, то она
целиком лежит на этой характеристике.
\end{prop}
\begin{proof}
Для экстремалей лагранжиан можно выбрать в виде
\begin{equation*}
  L=\frac12g_{\al\bt}\dot x^\al\dot x^\bt.
\end{equation*}
Для нулевых экстремалей он равен нулю, $L=0$. Следовательно, функция действия
$S(x,t)$ и укороченная функция действия $W=S+Et$ равны нулю, т.к.\ для нулевых
экстремалей $E=0$. Тем самым для бихарактеристик $W=0$ и они целиком лежат на
какой-либо из характеристик.
\end{proof}
\section{Приближение эйконала}
Посмотрим на волновое уравнение (\ref{ewaeqs}) с другой точки зрения. Любое
решение волнового уравнения (\ref{ewaeqs}) можно представить в виде\footnote{В
этом представлении $\hbar\in\MR$ рассматривается, как вещественный параметр.
Обозначение продиктовано аналогией с квантовой механикой, где $\hbar$ -- это
постоянная Планка.}
\begin{equation}                                                  \label{esowae}
  \vf=A\ex^{iW/\hbar},
\end{equation}
где $A(x)\ne0$ и $W(x)$ -- амплитуда и фаза волны. Тогда волновое уравнение
примет вид
\begin{multline*}
  g^{\al\bt}\frac{\pl^2_{\al\bt}A}A
  +\frac1\hbar 2ig^{\al\bt}\frac{\pl_\al A}A\pl_\bt W
  +\frac1\hbar ig^{\al\bt}\pl^2_{\al\bt}W-
\\
  -\frac1{\hbar^2}g^{\al\bt}\pl_\al W\pl_\bt W
  -g^{\al\bt}\Gamma_{\al\bt}{}^\g\left(\frac{\pl_\g A}A
  +\frac1\hbar i\pl_\g W\right)=0.
\end{multline*}
В формальном пределе $\hbar\to0$ волновое уравнение сводится к уравнению на фазу
\begin{equation}                                                  \label{eikoeq}
  g^{\al\bt}\pl_\al W\pl_\bt W=0.
\end{equation}
Оно совпадает с уравнением для характеристик (\ref{echaeq}), однако должно
выполняться во всем пространстве-времени.
\begin{defn}
Для волновых решений предел  $\hbar\to0$ означает, что относительное изменение
амплитуды мало, по сравнению с изменением фазы. Этот предел называют {\em
приближением эйконала}. Фазу $W(x)$ называют {\em эйконалом}, а уравнение
(\ref{eikoeq}) -- {\em уравнением эйконала}. В электродинамике это приближение
называют также приближением {\em геометрической оптики}. Для плоских волн оно
соответствует высоким частотам излучения и маленьким длинам волн.
\qed\end{defn}
\index{Приближение эйконала (eikonal approximation)}%
\index{Эйкональное приближение (eikonal approximation)}%
\index{Эйконал (eikonal)}%
\index{Уравнение эйконала (eikonal equation)}%
\index{Эйконала уравнение (eikonal equation)}%
\index{Приближение геометрической оптики (geometric optic approximation}%
\index{Геометрическая оптика (geometric optic)}%

Параметр $\hbar$ был введен с единственной целью -- определить эйкональное
приближение. Поэтому в дальнейшем, для краткости, мы включим его в определение
фазы $W/\hbar\rightarrow W$.
\begin{exa}                                                       \label{emipch}
Рассмотрим волновое уравнение в трехмерном пространстве-времени Минковского
$\MR^{1,2}$ в декартовой системе координат
$(x^\al)=(x^0,x^1,x^2)$:
\begin{equation*}
  \pl_0^2\vf-\pl_1^2\vf+\pl_2^2\vf=0.
\end{equation*}
В данном примере трехмерное пространство Минковского выбрано для наглядности.
Все следующие формулы естественным образом переносятся на пространство
Минковского произвольной размерности $\MR^{1,n-1}$.

Уравнение характеристик имеет вид
\begin{equation}                                                  \label{echfla}
  \left.\left( (\pl_0W)^2-(\pl_1W)^2-(\pl_2W)^2\right)\right|_{W=0}=0.
\end{equation}
Это уравнение допускает два семейства характеристик (изотропные поверхности
в примере \ref{eishyp}).

{\it Первое семейство характеристик}
\begin{equation*}
  W=(x^0-x_0^0)^2-(x^1-x_0^1)^2-(x^2-x_0^2)^2=0,
\end{equation*}
где $x_0=(x_0^0,x_0^1,x_0^2)$ -- три произвольных вещественных
параметра (координаты точки $x_0$). Нетрудно проверить, что
\begin{equation*}
  (\pl_0W)^2-(\pl_1W)^2-(\pl_2W)^2=4W.
\end{equation*}
Это значит, что уравнение характеристик (\ref{echfla}) выполняется только
на характеристиках, а не во всем пространстве-времени. Для каждой точки $x_0$
характеристики первого семейства состоят из двух конусов (конуса прошлого,
$x^0<x_0^0$, и будущего, $x^0>x_0^0$) с общей вершиной в точке $x_0$.

{\it Второе семейство характеристик}
\begin{equation*}
  W=x^0+k_\mu x^\mu-C,\qquad\mu=1,2
\end{equation*}
параметризуется единичным вектором в пространстве $|\Bk|:=\sqrt{k_1^2+k_2^2}=1$
и постоянной $C$. В этом случае уравнение характеристик выполняется во всем
пространстве, а не только на поверхности $W=0$. Характеристики второго семейства
параметризуются двумя параметрами, и представляют собой плоскости,
перпендикулярные ковектору $(1,k_1,k_2)$ нулевой длины и пересекающие плоскость
$x^0=0$ по прямой $k_\mu x^\mu=C$. Эти характеристики касаются характеристик
первого семейства (конусов).

Напомним, что нулевыми экстремалями в пространстве Минковского являются прямые и
только они с нулевым касательным вектором. Мы видим, что если произвольная
нулевая экстремаль (или бихарактеристика) в некоторой точке касается
характеристики, то она целиком принадлежит этой характеристике. Кроме того,
через каждую регулярную точку характеристики (исключение составляют вершины
конусов первого семейства) проходит одна и только одна нулевая экстремаль.
Поэтому нулевые экстремали полностью заметают характеристические поверхности.
Отметим, что одна и та же нулевая экстремаль может принадлежать разным
характеристическим конусам.

Из уравнения для характеристик (\ref{echfla}) можно найти производную по
времени
\begin{equation*}
  \pl_0W=\pm\sqrt{(\pl_1W)^2+(\pl_2W)^2}.
\end{equation*}
Отсюда следует выражение для гамильтониана, определяющего бихарактеристики
(траектории)
\begin{equation*}
  H=\pm\sqrt{p_1^2+p_2^2}=\pm\hat p.
\end{equation*}
Уравнения движения для бихарактеристик имеют вид
\begin{equation*}
\begin{split}
  \dot x^\mu&=\pm\frac{p^\mu}{\hat p},
\\
  \dot p_\mu&=0,\qquad \qquad \Rightarrow\quad p_\mu=\const,
\end{split}
\end{equation*}
где точка обозначает дифференцирование по времени $t=x^0$. Отсюда следует, что
импульс бихарактеристик постоянен, а траектории -- это прямые
$x^\mu=\pm n^\mu t+x^\mu_0$, проходящие через все точки пространства
$x_0\in\MR^2$ во всех возможных направлениях $n^\mu:=p^\mu/\hat p$. Поскольку
пространственный вектор $n^\mu$ имеет единичную длину, то бихарактеристики в
пространстве-времени совпадают с нулевыми экстремалями.

В пространстве Минковского уравнение (\ref{ewaeqs}) допускает решение в виде
плоской волны
\begin{equation*}
  \vf=A_0\ex^{ik_\al x^\al}=A_0\ex^{i(\om t-\Bk\Bx)},\qquad A_0=\const\ne0,
\end{equation*}
где $\Bk\Bx=-k_\mu x^\mu$. Сравнение этого выражения с (\ref{esowae}) дает
выражения для {\em частоты} $\om$ и {\em волнового вектора} $\Bk$ (в
рассматриваемом примере мы, для краткости, полагаем $\hbar=1$)
\index{Частота (frequency)}%
\index{Волновой вектор (wave vector)}\index{Вектор волновой (wave vector)}%
\begin{equation*}
  \om=k_0=\pl_0 W,\qquad \Bk=\lbrace\pl_1W,\pl_2W\rbrace
\end{equation*}
Тогда волновое уравнение (\ref{echaeq}) сводится к соотношению между
частотой и волновым вектором
\begin{equation*}
  g^{\al\bt}k_\al k_\bt=\om^2-\Bk^2=0,
\end{equation*}
где $\Bk^2:=-g^{\mu\nu}k_\mu k_\nu$. Для плоской волны в пространстве
Минковского эйконал $W(x)=x^\al k_\al$ является линейной функцией от
координат, а частота и волновой вектор постоянны.
\qed\end{exa}
В рассмотренном примере плоских волн зависимость эйконала от координат
была линейной. В общем случае эта зависимость является более сложной.
Тогда частота и волновой вектор определяются соотношениями
\begin{equation}                                                  \label{efrwav}
  \om=\pl_0W,\qquad k_\mu=\pl_\mu W,
\end{equation}
и зависят от точки пространства-времени. В этом случае уравнение эйконала
(\ref{eikoeq}) определяет зависимость частоты от волнового вектора
\begin{equation*}
  g^{00}\om^2+2g^{0\mu}\om k_\mu+g^{\mu\nu}k_\mu k_\nu=0.
\end{equation*}
Эта зависимость называется {\em дисперсией}, а производная
\index{Дисперсия (dispersion)}%
\begin{equation*}
  v_\mathrm{g}^\mu:=\frac{\pl\om}{\pl k_\mu}
\end{equation*}
называется {\em групповой скоростью}. Сравнение уравнения, определяющего
дисперсию, с формулой (\ref{erooch}) показывает, что частота $\om$ и волновой
вектор $k_\mu$ по сути дела совпадают, соответственно, с гамильтонианом и
импульсами бихарактеристик волнового уравнения.
\index{Групповая скорость (group velocity)}%
\index{Скорость групповая (group velocity)}%
\section{Гармонические координаты}
Для исследования волнового уравнения (\ref{ewaeqs}) на многообразии $\MM$,
$\dim\MM=n$, с метрикой лоренцевой сигнатуры $g_{\al\bt}$ удобно использовать
гармонические координаты. Как отмечено в разделе \ref{scooch} функции перехода
к новой системе координат являются скалярными полями на $\MM$.
\begin{defn}
Рассмотрим волновое уравнение (\ref{ewaeqs}) и допустим, что в некоторой области
$\MU\subset\MM$ оно имеет $n$ функционально независимых решений (это так при
достаточно общих предположениях). Пронумеруем эти
решения $\vf^\Sa$, $\Sa,\dotsc=0,1,\dotsc,n-1$. Тогда в области $\MU$
решения волнового уравнения задают систему координат $x^\Sa:=\vf^\Sa$, которая
называется {\em гармонической}.
\qed\end{defn}
\index{Гармонические координаты (harmonic coordinates)}%
\index{Координаты гармонические (harmonic coordinates)}%
\begin{com}
Гармоническая система координат была введена T.~де Дондером \cite{DeDond21} и
К.~Ланцосом \cite{Lanczo22} и получила физическую интерпретацию в работах
В.~А.~Фока \cite{Fock39R,Fock61R}.
\qed\end{com}
Гармоническая система координат обладает следующим важным свойством.
\begin{prop}
Система координат является гармонической тогда и только тогда, когда выполнены
условия
\begin{equation}                                                  \label{efogau}
  \pl_\Sa(\sqrt{|g|}g^{\Sa\Sb})=0\quad \Leftrightarrow\quad
  \Xi^\Sc=g^{\Sa\Sb}\Gamma_{\Sa\Sb}{}^\Sc=0,
\end{equation}
где $g:=\det g_{\Sa\Sb}$.
\end{prop}
\begin{proof}
Справедливо тождество
\begin{equation*}
  \pl_\al\left[\det\left(\frac{\pl x}{\pl \vf}\right)
  \frac{\pl\vf^\Sa}{\pl x^\bt}\right]\frac{\pl x^\al}{\pl \vf^\Sa}=0,
\end{equation*}
где $\pl x/\pl\vf$ -- обратная матрица Якоби преобразования координат
(\ref{ejacma}). Это тождество доказывается прямым дифференцированием с
учетом правила дифференцирования определителя матрицы (\ref{edetde}).
С учетом доказанного равенства и правила преобразования координат справедлива
следующая цепочка равенств:
\begin{multline*}
  \pl_\Sa(\sqrt{|g|}g^{\Sa\Sb})=\frac{\pl x^\g}{\pl\vf^\Sa}\pl_\g
  \left[\sqrt{|\overset\circ g|}\det\left(\frac{\pl x}{\pl\vf}\right)
  \frac{\pl\vf^\Sa}{\pl x^\al}
  \frac{\pl\vf^\Sb}{\pl x^\bt}g^{\al\bt}\right]=
\\
  =\det\left(\frac{\pl x}{\pl\vf}\right)\pl_\al
  \left(\sqrt{|\overset\circ g|}g^{\al\bt}\pl_\bt\vf^\Sb\right)=0,
\end{multline*}
где $\overset\circ g:=\det g_{\al\bt}$. Таким образом, из гармоничности
координат следует первое равенство (\ref{efogau}) и наоборот. Эквивалентность
равенств (\ref{efogau}) между собой проверяется прямой проверкой.
\end{proof}

Условия гармоничности координат можно записать с помощью принципа наименьшего
действия. Пусть задано действие для $n$ скалярных полей $\vf^\Sa$:
\begin{equation*}
  S=\int \!dx\vol\,\frac12g^{\al\bt}\pl_\al\vf^\Sa\pl_\bt\vf^\Sa\eta_{\Sa\Sb},
\end{equation*}
где $\eta_{\Sa\Sb}$ -- произвольная симметричная матрица (метрика в
пространстве-мишени). Тогда соответствующие уравнения Эйлера--Лагранжа примут
вид
\begin{equation*}
  S,_\Sa:=\frac{\dl S}{\dl\vf^\Sa}
  =-\vol\, g^{\al\bt}\widetilde\nb_\al\widetilde\nb_\bt\vf^\Sb
  \eta_{\Sb\Sa}=0.
\end{equation*}
Эти уравнения эквивалентны волновому уравнению (\ref{ewaeqs}.

Уравнения на метрику (\ref{efogau}) называются {\em условиями гармоничности}.
В дальнейшем будем, как обычно, обозначать гармоническую систему координат
снова греческими буквами $x^\al$, предполагая, что условия гармоничности
(\ref{efogau}) выполнены.
\index{Условие гармоничности (harmonic condition)}%
\index{Гармоничности условие (harmonic condition)}%
\begin{prop}
В гармонической системе координат волновой оператор, действующий на
функцию $f$, принимает вид
\begin{equation}                                                  \label{elahar}
  \widetilde\square f:=g^{\al\bt}\widetilde\nb_\al\widetilde\nb_\bt
  =g^{\al\bt}\pl^2_{\al\bt}f
\end{equation}
и не содержит первых частных производных $\pl_\al f$.
\end{prop}
\begin{proof}
Прямая проверка.
\end{proof}
\begin{exa}
В пространстве-времени Минковского $\MR^{1,n-1}$ декартова система координат
является гармонической.
\qed\end{exa}

Допустим, что на произвольном лоренцевом многообразии $\MM$, которое
топологически совпадает с $\MR^{1,n-1}$, выбрана гармоническая система
координат $x^\al$. Тогда произвольная линейная комбинация координат также
удовлетворяет волновому уравнению (\ref{elahar}). Можно доказать, что все
гармонические системы координат на $\MM$, которые являются асимптотически
декартовыми, связаны между собой преобразованиями Лоренца \cite{Fock61R}.
В.~А.~Фок придавал гармоническим системам координат в общей теории
относительности физический смысл, считая, что на произвольном лоренцевом
многообразии $\MM$ они выделены и играют ту же роль, что и декартовы
координаты в пространстве-времени Минковского.
\begin{prop}
В переменных Картана (см.\ раздел \ref{scorep}) условие гармоничности
(\ref{efogau}) имеет вид
\begin{equation*}
  \tilde\om^a=g^{\al\bt}\pl_\al e_\bt{}^a=0.
\end{equation*}
\end{prop}
\begin{proof}
Прямая проверка.
\end{proof}
\section{Нормальные, геодезические или римановы координаты       \label{srinog}}
Нормальные координаты, которые называются также геодезическими или римановыми
играют большую роль при изучении свойств метрики и связности, а также в
приложениях в математической физике. Сначала мы введем эту систему координат с
помощью рядов, что является более наглядным и важным для приложений. Затем
дадим определение с помощью экспоненциального отображения и сформулируем ряд
утверждений, связанных с полнотой римановых многообразий.
\begin{exa}
В дальнейшем мы увидим, что декартовы координаты являются нормальными
координатами в евклидовом пространстве $\MR^n$.
\qed\end{exa}
В определенном смысле нормальные координаты являются обобщением декартовой
системы координат на общий случай многообразия с заданной аффинной геометрией.
\subsection{Нормальные координаты. Локальное рассмотрение.}
Пусть на многообразии $\MM$ задана аффинная геометрия $(\MM,g,\Gamma)$. В настоящем
разделе мы предполагаем, что все геометрические объекты (в частности метрика и
связность) являются вещественно аналитическими, т.е.\ их компоненты в
окрестности каждой точки $x_0\in\MM$ представимы в виде сходящихся степенных
рядов. Мы рассмотрим специальную систему координат в окрестности точки $x_0$,
которая является аналогом декартовой системы координат в евклидовом пространстве
и имеет многочисленные приложения. В литературе эта система координат
встречается под разными названиями: геодезические, римановы или нормальные
координаты, и обычно строится в (псевдо-)римановом пространстве. Мы построим
такую систему координат в более общем случае произвольной аффинной геометрии.

Сначала мы рассмотрим геодезические линии вблизи точки $x_0$. Пусть на
многообразии задана кривая $x(t)$, проходящая через точку $x_0$ в заданном
направлении,
\begin{equation}                                                  \label{epodir}
  x^\al(0)=x_0^\al,\qquad\dot x^\al(0)=\dot x_0^\al.
\end{equation}
Для того, чтобы кривая была геодезической, функции $x^\al(t)$ должны
удовлетворять уравнениям для геодезических (\ref{egeode}) с начальными
условиями (\ref{epodir}). При достаточно малых $t$ будем искать решение
уравнений для геодезических в виде степенного ряда
\begin{equation}                                                  \label{egeota}
  x^\al=x_0^\al+\dot x_0^\al t+\frac12\ddot x_0^\al t^2
  +\frac1{3!}\stackrel{\dots}{x}_0^\al t^3+\dots
\end{equation}
Первые два члена определяются начальными условиями (\ref{epodir}), а все
последующие -- уравнениями (\ref{egeode}). В нулевом порядке по $t$ из уравнений
для геодезических следует равенство
$$
  \ddot x_0^\al=-\overset\circ\Gamma_{\bt\g}{}^\al\dot x_0^\bt\dot x_0^\g,
  \qquad \text{где}\quad \overset\circ\Gamma_{\bt\g}{}^\al=\Gamma_{\bt\g}{}^\al(x_0).
$$
Первый порядок уравнений по $t$ определяет кубический член разложения в
(\ref{egeota}):
\begin{equation*}
  \stackrel{\dots}{x}_0^\al=(-\pl_\dl\overset\circ\Gamma_{\bt\g}{}^\al
  +2\overset\circ\Gamma_{\bt\g}{}^\e\overset\circ\Gamma_{\lbrace\dl\e\rbrace}{}^\al)
  \dot x^\bt_0\dot x^\g_0\dot x^\dl_0,
\end{equation*}
где, для краткости, мы допускаем некоторую вольность в обозначениях:
\begin{equation*}
  \pl_\dl\overset\circ\Gamma_{\bt\g}{}^\al
  :=\left.\pl_\dl\Gamma_{\bt\g}{}^\al\right|_{x=x_0},
\end{equation*}
и фигурные скобки обозначают симметризацию по индексам.

Следовательно, решение задачи Коши для геодезических в третьем порядке по $t$
имеет вид
\begin{equation}                                                  \label{ekosog}
  x^\al=x_0^\al+\dot x_0^\al t
  -\frac12\overset\circ\Gamma_{0\bt\g}{}^\al\dot x_0^\bt\dot x_0^\g t^2
  -\frac16\left(\pl_\dl\overset\circ\Gamma_{\bt\g}{}^\al-2\overset\circ
  \Gamma_{\bt\g}{}^\e\overset\circ\Gamma_{\lbrace\dl\e\rbrace}{}^\al\right)
  \dot x^\bt_0\dot x^\g_0\dot x^\dl_0 t^3+\dots.
\end{equation}
Коэффициенты при более высоких степенях $t^k$, $k\ge4$ пропорциональны
$(\dot x_0)^k t^k$ с коэффициентами, зависящими от аффинной связности и их
частных производных вплоть до $k-2$ порядка, вычисленными в точке $x_0$.
Интервал сходимости ряда (\ref{ekosog}) определяется начальными данными и
компонентами аффинной связности. Мы, конечно, предполагаем, что он больше
нуля.

Перейдем к определению нормальных координат в окрестности $\MU_0\subset\MM$
точки $x_0$. Пусть $\MM$ -- многообразие с заданной аффинной связностью без
кручения, т.е.\ $\Gamma_{\bt\g}{}^\al=\Gamma_{\g\bt}{}^\al$. Перепишем закон
преобразования компонент аффинной связности (\ref{econtr}) в виде
\begin{equation*}
  \Gamma_{\bt'\g'}{}^{\al'}=\pl_{\bt'}x^\bt\pl_{\g'}x^\g
  (\Gamma_{\bt\g}{}^\al\pl_\al x^{\al'}-\pl^2_{\bt\g}x^{\al'}).
\end{equation*}
Совершим преобразование координат, которое задается квадратичным полиномом с
постоянными коэффициентами:
\begin{equation}                                                  \label{etrgeo}
  x^{\al'}=B_\al{}^{\al'}(x^\al-x^\al_0)
  +\frac12\overset\circ\Gamma_{\bt\g}{}^\al
  B_\al{}^{\al'}(x^\bt-x^\bt_0)(x^\g-x^\g_0),
\end{equation}
где $B_\al{}^{\al'}$ -- произвольная невырожденная матрица. Тогда в новой
системе координат компоненты аффинной связности будут равны
\begin{equation*}
  \Gamma_{\bt'\g'}{}^{\al'}=\pl_{\bt'}x^\bt\pl_{\g'}x{}^\g
  \left[\Gamma_{\bt\g}{}^\al\left(B_\al{}^{\al'}
  +\overset\circ\Gamma_{\al\dl}{}^\e B_\e{}^{\al'}(x^\dl-x^\dl_0)\right)
  -\overset\circ\Gamma_{\bt\g}{}^\al B_\al{}^{\al'}\right].
\end{equation*}
Отсюда следует, что в точке $x_0$ компоненты аффинной связности без кручения
обращаются в нуль.
\begin{com}
Если связность обладает кручением, то, поскольку кручение является тензором, его
нельзя обратить в нуль никаким преобразованием координат даже в одной точке.
\qed\end{com}
Выше мы доказали, что для произвольной аффинной связности в произвольной
заданной точке $x_0$ всегда можно обратить в нуль симметричную часть связности.
Для этого достаточно выбрать соответствующим образом только квадратичные члены в
функциях преобразований координат (\ref{etrgeo}). Данная система координат
существует в окрестности произвольной точки $x_0\in\MM$ и определена
неоднозначно. Во-первых, матрица $B_\al{}^{\al'}$ в (\ref{etrgeo}) является
невырожденной, а в остальном произвольна. Во-вторых, к правилу преобразования
координат (\ref{etrgeo}) можно добавить произвольные слагаемые третьей и более
высокой степени по $(x-x_0)$. Этот произвол в выборе системы координат можно
использовать для дальнейшей специализации системы координат.

Геометрический смысл построенной системы координат состоит в том, что в
достаточно малой окрестности точки $x_0$ свойства многообразия близки к
свойствам аффинного пространства, т.к.\ при параллельном переносе компоненты
тензоров в линейном приближении по вектору смещения не меняются. Как и декартовы
координаты в евклидовом пространстве, данные координаты в точке $x_0$
определены, в частности, с точностью до линейных преобразований.

Можно доказать, что симметричную часть компонент аффинной связности можно
обратить в нуль не только в фиксированной точке, но и вдоль произвольной
кривой $\g$ на многообразии $\MM$ \cite{Fermi22}. Соответствующая система
координат называется {\em геодезической вдоль кривой} $\g\in\MM$.

Оставшийся произвол в выборе системы координат можно использовать для дальнейшей
специализации геометрических объектов. Воспользуемся свободой добавления высших
степеней $(x-x_0)^k$, $k\ge3$, в закон преобразования координат (\ref{etrgeo}) и
покажем, что в произвольной точке $x_0\in\MM$ можно обратить в нуль не только
симметричные части самих компонент аффинной связности, но и все их полностью
симметризованные частные производные.
\begin{theorem}
Если компоненты аффинной связности вещественно аналитичны, то в окрестности
произвольной точки $x_0\in\MM$ существует такая система координат, что выполнены
равенства
\begin{equation}                                                  \label{esypac}
  \overset\circ\Gamma_{\lbrace\g_1\g_2\rbrace}{}^\al=0,\qquad
  \pl_{\lbrace\g_3}\overset\circ\Gamma_{\g_1\g_2\rbrace}{}^\al=0,\qquad \dotsc\qquad
  \pl^{k-2}_{\lbrace\g_3\dotsc\g_k}\overset\circ\Gamma_{\g_1\g_2\rbrace}{}^\al=0,
\end{equation}
где фигурные скобки обозначают симметризацию по всем индексам, заключенным
между ними.
\end{theorem}
\begin{proof}
Перепишем закон преобразования компонент аффинной связности (\ref{econts}) в
виде
\begin{equation}                                                  \label{etrafc}
  \Gamma_{\bt'\g'}{}^{\al'}A_{\al'}{}^\al
  =A_{\bt'}{}^\bt A_{\g'}{}^\g\Gamma_{\bt\g}{}^\al+\pl_{\bt'}A_{\g'}{}^\al,
\end{equation}
где матрица $A_{\al'}{}^\al(x)=J^{-1}_{~\al'}{}^\al=\pl_{\al'}x^\al$ является
обратным якобианом преобразования координат. Дифференцирование этого соотношения
по $x^{\dl'}$ приводит к равенству
\begin{multline}                                                  \label{eafrar}
  \pl_{\dl'}\Gamma_{\bt'\g'}{}^{\al'}A_{\al'}{}^\al
  +\Gamma_{\bt'\g'}{}^{\al'}\pl_{\dl'}A_{\al'}{}^\al
  =\pl_{\dl'}A_{\bt'}{}^\bt A_{\g'}{}^\g\Gamma_{\bt\g}{}^\al+
\\
  +A_{\bt'}{}^\bt\pl_{\dl'} A_{\g'}{}^\g\Gamma_{\bt\g}{}^\al
  +A_{\bt'}{}^\bt A_{\g'}{}^\g\pl_{\dl'}\Gamma_{\bt\g}{}^\al
  +\pl^2_{\dl'\bt'}A_{\g'}{}^\al.
\end{multline}
Последовательное дифференцирование этого соотношения приводит к равенствам,
содержащим старшие производные от компонент аффинной связности и обратной
матрицы Якоби $A_{\al'}{}^\al$. Рассматривая эти тождества в точке $x^0$, мы
докажем возможность выбора системы координат, в которой выполнены равенства
(\ref{esypac}), по теории возмущений.

Совершим преобразование координат $x\mapsto y(x)$. Предположим, что вблизи точки
$x_0\in\MM$ обратное преобразование задается степенным рядом
\begin{equation}                                                  \label{etrser}
  x^\al=x^\al_0+\left.\frac{\pl x^\al}{\pl y^\g}\right|_0 y^\g
  +\frac12\left.\frac{\pl^2 x^\al}{\pl y^{\g_1}\pl y^{\g_2}}\right|_0
  y^{\g_1}y^{\g_2}+\dotsc
\end{equation}
Здесь для новой системы координат мы используем букву $y$, чтобы избежать
штрихов у индексов. Выберем первые три члена разложения в виде
\begin{align*}
  \left.\frac{\pl x^\al}{\pl y^\g}\right|_0&=\dl^\al_\g,
\\
  \left.\frac{\pl^2 x^\al}{\pl y^{\g_1}\pl y^{\g_2}}\right|_0
  &=-\overset\circ\Gamma_{\lbrace \g_1\g_2\rbrace}{}^\al,
\\
  \left.\frac{\pl^3 x^\al}{\pl y^{\g_1}\pl y^{\g_2}\pl y^{\g_3}}\right|_0
  &=2\overset\circ\Gamma_{\lbrace \g_1\g_2}{}^\dl
  \overset\circ\Gamma_{\dl\g_3\rbrace}{}^\al
  -\pl_{\lbrace\g_3}\overset\circ\Gamma_{\g_1\g_2\rbrace}{}^\al.
\end{align*}
Тогда из уравнений (\ref{etrafc}), (\ref{eafrar}) следует, что в новой системе
координат
\begin{equation*}
  \overset\circ\Gamma_{\lbrace\g_1\g_2\rbrace}{}^\al=0,\qquad
  \pl_{\lbrace\g_3}\overset\circ\Gamma_{\g_1\g_2\rbrace}{}^\al=0.
\end{equation*}
Выбор квадратичного члена ряда (\ref{etrser}) уже был использован ранее. Выбор
кубического члена разложения позволил обратить в нуль симметризованную первую
частную производную аффинной связности в той же точке.

В дифференциальные тождества, получаемые последовательным дифференцированием
(\ref{etrafc}), максимальная степень производной от обратной матрицы Якоби
$A_{\al'}{}^\al$ всегда входит линейно и на единицу превышает максимальную
степень производной от компонент аффинной связности. Это значит, что
коэффициенты ряда (\ref{etrser}) всегда можно подобрать таким образом, чтобы
обратить в нуль все симметризованные частные производные от коэффициентов
аффинной связности (\ref{esypac}). В этом нет ничего удивительного, т.к.\
члены ряда (\ref{etrser}), начиная с квадратичного, находятся во взаимно
однозначном соответствии с условиями на аффинную связность (\ref{esypac}).
\end{proof}
Нетрудно проверить, что ряд (\ref{etrser}), определяющий преобразование
координат, в точности совпадает с рядом (\ref{ekosog}), определяющим
геодезические линии, если положить $y^\al=\dot x_0^\al t$. Это означает, что в
новой системе координат прямые линии $y^\al=X_0^\al t$, где $t\in\MR$, и
$X_0^\al$ -- произвольные числа, из которых по-крайней мере одно отлично от
нуля, являются геодезическими линиями на $\MM$. В этом месте прослеживается
аналогия с декартовой системой координат в евклидовом пространстве:
геодезические являются прямыми линиями.

Построенная система координат является нормальной системой координат в
окрестности $\MU_0$ точки $x_0\in\MM$, определяется только симметричной частью
компонент аффинной связностью и никак от метрики не зависит. При этом у нас
остался произвол в выборе первых двух членов разложения функций преобразования
координат (\ref{etrser}). Слагаемое нулевого порядка определим так, чтобы точка
$x_0$ отображается в начало координат $y_0=(0,\dotsc,0)$. Слагаемое первого
порядка по $y^\al$ можно использовать для фиксирования значения метрики в начале
координат. Очевидно, что его всегда можно подобрать таким образом, чтобы
метрика в точке $x_0$ была диагональной
\begin{equation}                                                  \label{emezev}
  \overset\circ g_{\al\bt}=\eta_{\al\bt}.
\end{equation}
Таким образом мы однозначно фиксировали все члены разложения функций
преобразования координат (\ref{etrser}). Это доказывает следующее утверждение.
\begin{theorem}                                                   \label{tnorco}
Пусть в окрестности произвольной точки $x_0\in\MM$ многообразия, на котором
задана аффинная геометрия, метрика и связность заданы вещественно аналитическими
функциями. Тогда существует такая система координат $y^\al$, что точка $x_0$
соответствует началу координат, а также выполнены равенства (\ref{esypac}) и
(\ref{emezev}). Такая система координат определена однозначно.
\end{theorem}
\begin{com}
В доказательстве теоремы нигде не использовалась сигнатура метрики. Поэтому
сформулированная теорема справедлива как для римановых, так и для
псевдоримановых метрик.
\qed\end{com}
\begin{defn}
Система координат $y^\al$ в теореме \ref{tnorco} называется {\em нормальной},
{\em геодезической} или {\em римановой}.
\qed\end{defn}
\index{Римановы координаты (Riemannian coordinates)}%
\index{Координаты римановы (Riemannian coordinates)}%
\index{Нормальные координаты (normal coordinates)}%
\index{Координаты нормальные (normal coordinates)}%
\index{Нормальные координаты (geodesic coordinates)}%
\index{Координаты нормальные (geodesic coordinates)}%

\begin{prop}
Система координат $y^\al$ в некоторой окрестности начала координат является
нормальной тогда и только тогда, когда выполнены условия:
\begin{equation}                                                  \label{egeocj}
  \Gamma_{\bt\g}{}^\al(y) y^\bt y^\g=0\qquad \text{и}\qquad
  g_{\al\bt}(0)=\eta_{\al\bt}.
\end{equation}
\end{prop}
\begin{proof}
В нормальной системе координат геодезическая линия, проходящая через точку $x_0$
в направлении $u_0$, является прямой и задается параметрически в виде
\begin{equation}                                                  \label{egeric}
  y^\al(t)=u^\al_0t.
\end{equation}
Подстановка этого выражения в уравнение для геодезических линий (\ref{egeode})
дает
\begin{equation*}
  u^\al_0 u^\bt_0\Gamma_{\al\bt}{}^\g=0.
\end{equation*}
Умножив это уравнение на $t^2$, получим (\ref{egeocj}).

Обратно. Пусть выполнено уравнение (\ref{egeocj}). Разложим компоненты связности
в ряд Тейлора вблизи начала координат. Тогда уравнение (\ref{egeocj})
эквивалентно цепочке равенств (\ref{esypac}). Затем разложим уравнение
геодезических и сами геодезические в ряды Тейлора по $t$. Приравняв нулю
коэффициенты при одинаковых степенях $t$, получаем, что прямые линии и
только они являются геодезическими линиями, проходящими через начало координат.
\end{proof}
\begin{com}
Сравнение первого условия в (\ref{egeocj}) со вторым условием в (\ref{efogau})
показывает, что нормальные координаты в общем случае не являются гармоническими.
\qed\end{com}

Нормальная система координат обладает рядом замечательных свойств. В частности,
в окрестности начала координат компоненты произвольного тензорного поля можно
представить в виде ряда, коэффициенты которого зависят от ковариантных
производных этого поля и тензоров кривизны, кручения и их ковариантных
производных, вычисленных в точке $x_0$. Доказательство этого утверждения
проводится конструктивно, путем явного построения соответствующего разложения.

Для доказательства нам понадобится предварительный результат.
\begin{lemma}
Пусть связность $\Gamma$ на $\MM$ вещественно аналитична. Тогда в окрестности начала
нормальной системы координат все производные
\begin{equation}                                                  \label{epasyg}
  \pl^{k-1}_{\lbrace\g_1\dotsc\g_{k-1}}\overset\circ\Gamma_{\g_k\rbrace\al}{}^\bt,
\end{equation}
где, в отличие от (\ref{esypac}), симметризация проводится только по части
нижних индексов, выражаются через тензоры кривизны, кручения и их ковариантные
производные.
\end{lemma}
\begin{proof}
Лемма доказывается прямыми, но громоздкими вычислениями. Поэтому мы ограничимся
только первыми двумя слагаемыми.

Используя свойства (\ref{esypac}), нетрудно доказать равенства
\begin{equation}                                                  \label{eparco}
\begin{split}
  \pl_{\lbrace\g_1}\overset\circ\Gamma_{\g_2\rbrace\al}{}^\bt
  &=\frac13\left(\pl_{\lbrace\g_1}\overset\circ\Gamma_{\al\g_2\rbrace}{}^\bt
  -\pl_\al\overset\circ\Gamma_{\lbrace\g_1\g_2\rbrace}{}^\bt
  +2\pl_{\lbrace\g_1}\overset\circ T_{\g_2\rbrace\al}{}^\bt\right),
\\
  \pl^2_{\lbrace\g_1\g_2}\overset\circ\Gamma_{\g_3\rbrace\al}{}^\bt
  &=\frac13\left(\pl^2_{\lbrace\g_1\g_2}\overset\circ\Gamma_{\al\g_3\rbrace}{}^\bt
  -\pl^2_{\al\lbrace\g_1}\overset\circ\Gamma_{\g_2\g_3\rbrace}{}^\bt
  +2\pl^2_{\lbrace\g_1\g_2}\overset\circ T_{\g_3\rbrace\al}{}^\bt\right).
\end{split}
\end{equation}
Здесь и в дальнейшем фигурные скобки обозначают симметризацию {\em только}
по индексам $\g_1$, $\g_2,\dotsc$.

В точке $x_0$ аффинная связность полностью определяется тензором кручения
\begin{equation}                                                  \label{etoric}
  \overset\circ\Gamma_{\g\al}{}^\bt=\frac12\overset\circ T_{\g\al}{}^\bt.
\end{equation}
Далее, прямые вычисления приводят к следующим равенствам
\begin{align*}
  \overset\circ R_{\al\lbrace\g_1\g_2\rbrace}{}^\bt
  &=\pl_\al\overset\circ\Gamma_{\lbrace\g_1\g_2\rbrace}{}^\bt
  -\pl_{\lbrace\g_1}\overset\circ\Gamma_{\al\g_2\rbrace}{}^\dl
  -\frac14\overset\circ T_{\al\lbrace\g_1}{}^\dl
  \overset\circ T_{\g_2\rbrace\dl}{}^\bt,
\\
  \nb_{\lbrace\g_1}\overset\circ R_{\al\g_2\g_3\rbrace}{}^\bt
  &=\pl^2_{\al\lbrace\g_1}\overset\circ\Gamma_{\g_2\g_3\rbrace}{}^\bt
  -\pl^2_{\lbrace\g_1\g_2}\overset\circ\Gamma_{\al\g_3\rbrace}{}^\bt
  +\frac23\overset\circ R_{\al\lbrace\g_1\g_2}{}^\dl
  \overset\circ T_{\g_3\rbrace\dl}{}^\bt
  +\frac23\overset\circ T_{\al\lbrace\g_1}{}^\dl
  \overset\circ R_{\dl\g_2\g_3\rbrace}{}^\bt
\\
  &+\frac16\nb_{\lbrace\g_1}\overset\circ T_{\g_2\al}{}^\dl
  \overset\circ T_{\g_3\rbrace\dl}{}^\bt
  -\frac13\overset\circ T_{\al\lbrace\g_1}{}^\dl\nb_{\g_2}
  \overset\circ T_{\g_3\rbrace\dl}{}^\bt,
\\
  \nb_{\lbrace\g_1}\overset\circ T_{\g_2\rbrace\al}{}^\bt
  &=\pl_{\lbrace\g_1}\overset\circ T_{\g_2\rbrace\al}{}^\bt,
\\
  \nb_{\lbrace\g_1}\nb_{\g_2}\overset\circ T_{\g_3\rbrace\al}{}^\bt
  &=\pl^2_{\lbrace\g_1\g_2}\overset\circ T_{\g_3\rbrace\al}{}^\bt
  +\frac16\overset\circ R_{\al\lbrace\g_1\g_2}{}^\dl
  \overset\circ T_{\g_3\rbrace\dl}{}^\bt
  +\frac16\overset\circ T_{\al\lbrace\g_1}{}^\dl
  \overset\circ R_{\dl\g_2\g_3\rbrace}{}^\bt
\\
  &+\frac16\nb_{\lbrace\g_1}\overset\circ T_{\g_2\al}{}^\dl
  \overset\circ T_{\g_3\rbrace\dl}{}^\g
  +\frac16\overset\circ T_{\al\lbrace\g_1}{}^\dl
  \nb_{\g_2}\overset\circ T_{\g_3\rbrace\dl}{}^\bt.
\end{align*}
Используя полученные формулы, правые части (\ref{eparco}) можно записать в
ковариантном виде
\begin{align}                                                     \label{esecor}
  \pl_{\lbrace\g_1}\overset\circ\Gamma_{\g_2\rbrace\al}{}^\bt&=\frac13\left(
  -\overset\circ R_{\al\lbrace\g_1\g_2\rbrace}{}^\bt
  +2\nb_{\lbrace\g_1}\overset\circ T_{\g_2\rbrace\al}{}^\bt
  -\frac14\overset\circ T_{\al\lbrace\g_1}{}^\dl
  \overset\circ T_{\g_2\rbrace\dl}{}^\bt\right),
\\                                                                     \nonumber
  \pl^2_{\lbrace\g_1\g_2}\overset\circ \Gamma_{\g_3\rbrace\al}{}^\bt
  &=\frac13\left(-\nb_{\lbrace\g_1}\overset\circ R_{\al\g_2\g_3\rbrace}{}^\bt
  +\frac13\overset\circ R_{\al\lbrace\g_1\g_2}{}^\dl
  \overset\circ T_{\g_3\rbrace\dl}{}^\bt
  +\frac13\overset\circ T_{\al\lbrace\g_1}{}^\dl
  \overset\circ R_{\dl\g_2\g_3\rbrace}{}^\bt\right.
\\                                                                \label{esydsg}
  &\left.
  +2\nb_{\lbrace\g_1}\nb_{\g_2}\overset\circ T_{\g_3\rbrace\al}{}^\bt
  -\frac16\nb_{\lbrace\g_1}\overset\circ T_{\g_2\al}{}^\dl
  \overset\circ T_{\g_3\rbrace\dl}{}^\bt
  -\frac23\overset\circ T_{\al\lbrace\g_1}{}^\dl\nb_{\g_2}
  \overset\circ T_{\g_3\rbrace\dl}{}^\bt\right),
\end{align}
где, напомним, симметризация проводится только по индексам $\g_1$,
$\g_2,\dotsc$. Аналогичным образом все частные производные от связности вида
(\ref{epasyg}) можно выразить через ковариантные объекты.
\end{proof}

Теперь можно доказать утверждение о разложении компонент произвольного
тензорного поля в окрестности начала нормальной системы координат в ряд Тейлора.
Для определенности рассмотрим произвольный ковариантный тензор второго ранга
$A_{\al\bt}$ в нормальных координатах и разложим его в ряд Тейлора вблизи начала
координат:
\begin{equation}                                                  \label{eteate}
  A_{\al\bt}(y)=\overset\circ A_{\al\bt}+\pl_\g\overset\circ A_{\al\bt}\,y^\g
  +\frac12\pl^2_{\g_1\g_2}\overset\circ A_{\al\bt}\,y^{\g_1}y^{\g_2}+\dotsc
\end{equation}
Эта процедура явно нековариантна, потому что координаты $y^\g$ не являются
компонентами вектора, и коэффициенты этого ряда также нековариантны. Основным
достоинством нормальных координат является то, что все коэффициенты этого ряда
тем не менее можно выразить через ковариантные величины. Для этого необходимо
выразить все частные производные через ковариантные и выразить компоненты
аффинной связности в начале координат через тензор кривизны, кручения и их
ковариантные производные. Начнем с первой производной
\begin{equation*}
  \pl_\g\overset\circ A_{\al\bt}=\nb_\g\overset\circ A_{\al\bt}
  +\overset\circ \Gamma_{\g\al}{}^\dl\overset\circ A_{\dl\bt}
  +\overset\circ \Gamma_{\g\bt}{}^\dl\overset\circ A_{\al\dl}
  =\nb_\g\overset\circ A_{\al\bt}
  +\frac12\overset\circ T_{\g\al}{}^\dl\overset\circ A_{\dl\bt}
  +\frac12\overset\circ T_{\g\bt}{}^\dl\overset\circ A_{\al\dl},
\end{equation*}
где мы воспользовались формулами (\ref{etoric}). Прямые, но более громоздкие
выкладки позволяют представить вторые частные производные также в ковариантном
виде:
\begin{equation*}
\begin{split}
  \pl^2_{\g_1\g_2}\overset\circ A_{\al\bt}
  &=\nb_{\lbrace\g_1}\nb_{\g_2\rbrace}\overset\circ A_{\al\bt}
+\overset\circ T_{\lbrace\g_1\al}{}^\dl\nb_{\g_2\rbrace}\overset\circ A_{\dl\bt}
+\overset\circ T_{\lbrace\g_1\bt}{}^\dl\nb_{\g_2\rbrace}\overset\circ A_{\al\dl}
+\frac12\overset\circ T_{\lbrace\g_1\al}{}^\dl\overset\circ T_{\g_2\rbrace}{}^\e
  \overset\circ A_{\dl\e}+
\\
  &+\frac13\left(-\overset\circ R_{\al\lbrace\g_1\g_2\rbrace}{}^\dl
  +2\nb_{\lbrace\g_1}\overset\circ T_{\g_2\rbrace\al}{}^\dl
  -\overset\circ T_{\al\lbrace\g_1}{}^\e\overset\circ T_{\g_2\rbrace\e}{}^\dl
  \right)\overset\circ A_{\dl\bt}+
\\
  &+\frac13\left(-\overset\circ R_{\bt\lbrace\g_1\g_2\rbrace}{}^\dl
  +2\nb_{\lbrace\g_1}\overset\circ T_{\g_2\rbrace\bt}{}^\dl
  -\overset\circ T_{\bt\lbrace\g_1}{}^\e\overset\circ T_{\g_2\rbrace\e}{}^\dl
  \right)\overset\circ A_{\al\dl}.
\end{split}
\end{equation*}
Эту процедуру можно продолжить вплоть до произвольного порядка. Однако уже в
третьем порядке формулы настолько громоздки, что нет смысла их приводить в явном
виде.

Ясно, что аналогичное представление справедливо для тензорного поля
произвольного ранга и типа. Поэтому справедлива следующая
\begin{theorem}
Компоненты произвольного вещественно аналитического тензорного поля в некоторой
окрестности начала нормальной системы координат представимы в виде ряда Тейлора,
коэффициенты которого определяются ковариантными производными компонент данного
тензорного поля, а также тензором кривизны, тензором кручения и их ковариантными
производными, взятыми в начале координат.
\end{theorem}
В математической физике такое представление часто бывает очень удобным при
проведении вычислений.
\subsection{Нормальные координаты в (псевдо-)римановом пространстве
                                                                 \label{snorco}}
Основное достоинство нормальных координат заключается в том, что вещественно
аналитические тензорные поля в окрестности $\MU_0\subset\MM$ произвольной
точки $x_0\in\MM$ представляются в виде ряда Тейлора, коэффициенты которого
задаются только ковариантными объектами: ковариантными производными данного
тензорного поля, а также тензорами кривизны и кручения и их ковариантными
производными. В предыдущем разделе были явно вычислены первые два члена этого
ряда для ковариантного тензора второго ранга. Эти члены содержат много слагаемых
с тензором кручения. Поэтому в (псевдо-)римановой геометрии, где кручение
тождественно равно нулю, формулы упрощаются, и можно продвинуться значительно
дальше в вычислениях.

Кроме того, если на многообразии $\MM$ задана аффинная геометрия $(\MM,g,\Gamma)$,
то в качестве определяющего пучка кривых, проходящих через точку $x_0\in\MM$
можно выбрать не геодезические линии, а экстремали, которые определяются
исключительно метрикой. Это также дает основание рассмотреть нормальные
координаты в (псевдо-)римановой геометрии более подробно.

Два слова об обозначениях. Мы часто используем знак тильды для геометрических
объектов в (псевдо-)римановой геометрии. Поскольку значок окружности над
символом уже используется для обозначения геометрических объектов,
рассматриваемых в точке $x_0$, то, чтобы не загромождать обозначений, знак
тильды в настоящем разделе мы опустим.

В (псевдо-)римановой геометрии в нормальных координатах
$\overset\circ\Gamma_{\al\bt}{}^\g=0$, и ковариантная производная произвольного
тензора в этой точке совпадает с частной производной. При этом в выражении для
тензора кривизны пропадают квадратичные слагаемые по связности:
\begin{align*}
  \overset\circ R_{\al\bt\g}{}^\dl&=\pl_\al\overset\circ \Gamma_{\bt\g}{}^\dl
  -\pl_\bt\overset\circ \Gamma_{\al\g}{}^\dl
\\ \intertext{или}
  \overset\circ R_{\al\bt\g\dl}&=\frac12
  (\pl_{\al\g}^2\overset\circ g_{\bt\dl}-\pl_{\al\dl}^2\overset\circ g_{\bt\g}
  -\pl_{\bt\g}^2\overset\circ g_{\al\dl}+\pl_{\bt\dl}^2\overset\circ g_{\al\g}).
\end{align*}

Вычисления, аналогичные тем, что привели к формулам (\ref{esecor}),
(\ref{esydsg}), в (псевдо-)римановой геометрии дают равенства
\begin{equation}                                                  \label{erisyg}
\begin{split}
  \pl_{\lbrace\g_1}\overset\circ\Gamma_{\g_2\rbrace\al}{}^\bt
  &=-\frac13\overset\circ R_{\al\lbrace\g_1\g_2\rbrace}{}^\bt,
\\
  \pl^2_{\lbrace\g_1\g_2}\overset\circ\Gamma_{\g_3\rbrace\al}{}^\bt
  &=-\frac12\nb_{\lbrace\g_1}\overset\circ R_{\al\g_2\g_3\rbrace}{}^\bt,
\\
  \pl^3_{\lbrace\g_1\g_2\g_3}\overset\circ\Gamma_{\g_4\rbrace\al}{}^\bt
  &=-\frac35\nb_{\lbrace\g_1}\nb_{\g_2}
  \overset\circ R_{\al\g_3\g_4\rbrace}{}^\bt-\frac2{15}
  \overset\circ R_{\al\lbrace\g_1\g_2}{}^\dl
  \overset\circ R_{\dl\g_3\g_4\rbrace}{}^\bt,
\end{split}
\end{equation}
где симметризация проводится только по индексам $\g_1$, $\g_2,\dotsc$. Здесь мы
дополнительно вычислили третью симметризованную производную от связности.

Разложим ковариантный тензор $A_{\al_1\dotsc\al_s}$ произвольного ранга $s$ в
окрестности точки $x_0$ в ряд по нормальным координатам:
\begin{equation*}
  A_{\al_1\dotsc\al_s}=\overset\circ A_{\al_1\dotsc\al_s}
  +\pl_\g\overset\circ A_{\al_1\dotsc\al_s}\,y^\g
  +\frac12\pl^2_{\g_1\g_2}\overset\circ A_{\al_1\dotsc\al_s}\,y^{\g_1}y^{\g_2}
  +\dotsc,
\end{equation*}
где $x_0=\lbrace y^\al=0\rbrace$.
Преобразовав частные производные в ковариантные и исключив слагаемые со
связностью с помощью формул (\ref{erisyg}), получим следующий ряд
\begin{equation}                                                  \label{emesno}
\begin{split}
  A_{\al_1\dotsc\al_s}&=\overset\circ A_{\al_1\dotsc\al_s}
  +\nb_\g\overset\circ A_{\al_1\dotsc\al_s}y^\g+
\\
  &+\frac1{2!}\left(\nb_{\g_1}\nb_{\g_2}\overset\circ A_{\al_1\dotsc\al_s}
  -\frac13\sum_{k=1}^s\overset\circ R_{\al_k\g_1\g_2}{}^{\bt_k}
  \overset\circ A_{\al_1\dotsc\bt_k\dotsc\al_s}\right)y^{\g_1}y^{\g_2}+
\\
  &+\frac1{3!}\left(\nb_{\g_1}\nb_{\g_2}\nb_{\g_3}
  \overset\circ A_{\al_1\dotsc\al_s}
  -\sum_{k=1}^s\overset\circ R_{\al_k\g_1\g_2}{}^{\bt_k}\nb_{\g_3}
  \overset\circ A_{\al_1\dotsc\bt_k\dotsc\al_s}\right.-
\\
  &\qquad\qquad\qquad\qquad\left.-\frac12\sum_{k=1}^s\nb_{\g_1}
  \overset\circ R_{\al_k\g_2\g_3}{}^{\bt_k}
  \overset\circ A_{\al_1\dotsc\bt_k\dotsc\al_s}\right)y^{\g_1}y^{\g_2}y^{\g_3}
  +\dotsc
\end{split}
\end{equation}
Под знаком суммы подразумевается суммирование по немому индексу $\bt_k$,
который стоит на $k$-том месте в
$\overset\circ A_{\al_1\dotsc\bt_k\dotsc\al_s}$. В полученном выражении
симметризацию по индексам $\g_1$, $\g_2,\dotsc$ можно не указывать, т.к.\
происходит свертка с симметричным произведением $y^{\g_1}\dotsc y^{\g_k}$.

Аналогичные ряды можно построить для тензоров, содержащих произвольные наборы
ковариантных и контравариантных индексов. Во всех случаях коэффициенты ряда
вплоть до любого порядка могут быть выражены только через ковариантные объекты в
точке $x_0$.

Нормальные координаты особенно удобны при анализе (псевдо-)римановой метрики. В
этом случае возникает дополнительное упрощение, поскольку все ковариантные
производные от метрики тождественно равны нулю. Несложные вычисления показывают,
что ряд (\ref{emesno}) для метрики в нормальных координатах принимает вид
\begin{equation}                                                  \label{emenoc}
\begin{split}
  g_{\al\bt}&=\eta_{\al\bt}
  -\frac13\overset\circ R_{\al\g_1\g_2\bt}\,y^{\g_1}y^{\g_2}
 -\frac1{3!}\nb_{\g_1}\overset\circ R_{\al\g_2\g_3\bt}\,y^{\g_1}y^{\g_2}y^{\g_3}
 +
\\
  &\quad+\frac1{5!}\left(-6\nb_{\g_1}\nb_{\g_2}\overset\circ R_{\al\g_3\g_4\bt}
  +\frac{16}3\overset\circ R_{\al\g_1\g_2}{}^\dl
  \overset\circ R_{\dl\g_3\g_4\bt}\right)y^{\g_1}y^{\g_2}y^{\g_3}y^{\g_4}
  +\dotsc,
\end{split}
\end{equation}
где мы также вычислили слагаемое четвертого порядка.
\subsection{(Псевдо-)римановы пространства постоянной кривизны}
Выражение для метрики в нормальных координатах (\ref{emenoc}) принимает особенно
простой вид для пространств постоянной кривизны, которые определяются
равенством $\nb_\e R_{\al\bt\g}{}^\dl=0$.
\begin{prop}
В пространстве постоянной кривизны (псевдо-)риманова метрика в окрестности
произвольной точки $x_0\in\MM$ в нормальных координатах $y^\al$ имеет вид
\begin{equation}                                                  \label{emenos}
  g_{\al\bt}=\eta_{\al\bt}+\frac12\sum_{k=1}^\infty
  \frac{(-1)^k2^{2k+2}}{(2k+2)!}V_\al{}^{\e_1}V_{\e_1}{}^{\e_2}\dotsc
  V_{\e_{k-1}}{}^{\e_k}\eta_{\e_k\bt},
\end{equation}
где
\begin{equation*}
  V_{\e_{k-1}}{}^{\e_k}:=\overset\circ R_{\e_{k-1}\g_1\g_2}{}^{\e_k}\,
  y^{\g_1}y^{\g_2},\qquad\e_0=\al.
\end{equation*}
\end{prop}
\begin{proof}
Сначала проверяем, что первые два члена суммы действительно совпадают с
(\ref{emenoc}) для пространств постоянной кривизны. Далее доказательство
проводится по индукции.
\end{proof}
Ряд для метрики (\ref{emenos}) можно просуммировать для пространств
постоянной кривизны специального вида (\ref{qdfrss}). В этом случае в начале
координат справедливо равенство
\begin{equation}                                                  \label{ecocus}
  \overset\circ R_{\al\bt\g\dl}
  =-\frac{2K}{n(n-1)}(\eta_{\al\g}\eta_{\bt\dl}-\eta_{\al\dl}\eta_{\bt\g}),
  \qquad n=\dim\MM,\quad K=\const.
\end{equation}
Тогда матрица $V_\al{}^\bt$ пропорциональна проекционному оператору:
\begin{equation*}
  V_\al{}^\bt=as\left(\dl_\al^\bt-\frac{y_\al y^\bt}s\right),
\end{equation*}
где
\begin{equation*}
  a:=\frac{2K}{n(n-1)},\qquad s:=y^\al y_\al,\qquad y_\al:=y^\bt\eta_{\al\bt},
\end{equation*}
и ряд для метрики принимает вид
\begin{equation}                                                  \label{emecos}
  g_{\al\bt}=\eta_{\al\bt}
  +\frac12\left(\eta_{\al\bt}-\frac{y_\al y_\bt}s\right)
  \sum_{k=1}^\infty\frac{2^{2k+2}}{(2k+2)!}(-as)^k.
\end{equation}
Это выражение для метрики определено и для $s=0$, т.к.\ ряд начинается с
линейного по $s$ члена. Теперь ряд можно просуммировать, что приводит к
следующему выражению для метрики
\begin{equation}                                                  \label{emeccn}
  g_{\al\bt}=f(s)\Pi^\St_{\al\bt}+\Pi^\Sl_{\al\bt}
  =f\eta_{\al\bt}+(1-f)\frac{y_\al y_\bt}s,
\end{equation}
где метрика представлена в виде суммы проекционных операторов
\begin{equation*}
  \Pi^\St_{\al\bt}:=\eta_{\al\bt}-\frac{y_\al y_\bt}s,\qquad
  \Pi^\Sl_{\al\bt}:=\frac{y_\al y_\bt}s,
\end{equation*}
а функция $f(s)$ определена рядом
\begin{equation}                                                  \label{efunme}
  f(s):=\sum_{k=0}^\infty\frac{2^{2k+2}}{(2k+2)!}(-as)^k,
\end{equation}
который сходится на всей комплексной плоскости $s$. Тем самым доказано следующее
утверждение.
\begin{theorem}
Пусть на многообразии задана (псевдо-)риманова метрика $g$ класса $\CC^2$ такая,
что многообразие является пространством постоянной кривизны специального вида:
\begin{equation}                                                  \label{ecocuc}
  R_{\al\bt\g\dl}
  =-\frac{2K}{n(n-1)}(g_{\al\g}g_{\bt\dl}-g_{\al\dl}g_{\bt\g}).
\end{equation}
Тогда метрика $g$ на $\MM$ вещественно аналитична.
\end{theorem}
\begin{proof}
Прямая проверка показывает, что вещественно аналитическая метрика
(\ref{emeccn}), (\ref{efunme}) описывает многообразие постоянной кривизны
(\ref{ecocuc}). Единственность метрики следует из единственности решения задачи
Коши для уравнений геодезических.
\end{proof}

Ряд (\ref{efunme}) можно просуммировать. Для псевдоримановых пространств
положительной кривизны $K>0,a>0$, и функция $f(s)$ имеет вид
\begin{equation}                                                  \label{efunsm}
  f(s)=\begin{cases}\quad \displaystyle{\frac{\sin^2\sqrt{as}}{as}}, & s>0,
\\
  \quad 1, & s=0,
\\
    -\displaystyle{\frac{\sh^2\sqrt{-as}}{as}}, & s<0.\end{cases}
\end{equation}
Функция $f(s)$, как нетрудно проверить, аналитична. Напомним, что
$\sin ix=i\sh x$.

Для псевдоримановых пространств отрицательной кривизны $K<0,a<0$ имеем равенство
\begin{equation}                                                  \label{efunsn}
  f(s)=\begin{cases}-\displaystyle{\frac{\sh^2\sqrt{-as}}{as}}, & s>0,
\\
  \quad 1, & s=0,
\\
  \quad \displaystyle{\frac{\sin^2\sqrt{as}}{as}}, & s<0.  \end{cases}
\end{equation}
Отметим, что на световом конусе, $s=0$, метрика в обоих случаях совпадает
с метрикой Минковского.

Если кривизна псевдориманова пространства постоянной кривизны равна нулю, $K=0$,
то $f=1$.

Для римановых пространств постоянной кривизны всегда $s>0$, и
\begin{equation}                                                  \label{efunsr}
  f(s)=\begin{cases}\quad \displaystyle{\frac{\sin^2\sqrt{as}}{as}}, & K>0,
\\
  \quad 1, & K=0,
\\
  -\displaystyle{\frac{\sh^2\sqrt{-as}}{as}}, & K<0.  \end{cases}
\end{equation}

Тот факт, что функция $f(s)$ действительно приводит к ряду (\ref{emecos})
проверяется прямой проверкой. Не зная функции $f(s)$, ряд (\ref{emecos}) можно
просуммировать следующим образом. Представление (\ref{emecos}) задает тензорную
структуру метрики. В разделе \ref{slorin} для метрики более общего вида, которая
параметризуется двумя произвольными функциями $f(s)$ и $g(s)$, были вычислены
символы Кристоффеля и тензор кривизны. Рассматриваемому случаю соответствует
условие
\begin{equation*}
  g=\frac{(f+f's)^2}{f-afs}=1,
\end{equation*}
где штрих обозначает дифференцирование по $s$. Последнее равенство задает
обыкновенное дифференциальное уравнение первого порядка на $f(s)$. Это уравнение
легко решается после подстановки
\begin{equation}                                                  \label{esufum}
  U=fs
\end{equation}
Постоянная интегрирования находится из условия $U(0)=0$ (ограниченность
метрики).

Рассмотрим римановы пространства постоянной кривизны, для которых
$g_{\al\bt}(0)=\dl_{\al\bt}$. Метрика (\ref{emeccn}) принимает особенно простой
вид в сферической системе координат евклидова пространства $\MR^n$. В
сферических координатах $s=y^\al y^\bt\dl_{\al\bt}=r^2$, где $r$ -- радиальная
координата, и справедливы тождества
\begin{equation*}
  \Pi^\Sl_{\al\bt}dy^\al dy^\bt=dr^2,\qquad
  \Pi^\St_{\al\bt}dy^\al dy^\bt=r^2d\Om,
\end{equation*}
где $d\Om$ -- элемент телесного угла в евклидовом пространстве $\MR^n$. Таким
образом метрику пространств постоянной кривизны в нормальных координатах можно
записать в виде
\begin{equation}                                                  \label{emeris}
  ds^2=dr^2+r^2fd\Om.
\end{equation}

Проанализируем пространство $\MR^n$ с метрикой (\ref{emeris}) подробнее. Объем
сферы радиуса $r\sqrt f$ в $\MR^n$ с метрикой (\ref{emeris}) равен
\begin{equation*}
  S^{n-1}_r=\frac{2\pi^{n/2}}{\Gamma(\frac n2)}(r^2f)^{(n-1)/2}.
\end{equation*}
Мы видим, что нули функции $f(r^2)=0$ определяют те сферы в $\MR^n$, площадь
которых равна нулю. Поскольку площадь поверхности является инвариантным
объектом, то это означает, что на самом деле эти сферы соответствуют отдельным
точкам пространства постоянной кривизны.

Нормальные координаты для многообразий постоянной кривизны определены для всех
$\lbrace y^\al\rbrace\in\MR^n$. При этом все геодезические, проходящие через
точку $x_0$, полны, т.к.\ канонический параметр пробегает всю вещественную
прямую $t\in(-\infty,\infty)$.

Проведенное рассмотрение доказывает
\begin{theorem}
Нормальные координаты на (псевдо-)римановом многообразии $\MM$ постоянной
кривизны вида (\ref{ecocuc}) в каждой точке $x_0\in\MM$ задают гладкое
сюрьективное отображение
\begin{equation}                                                  \label{emacoc}
  \MR^n\rightarrow\MM.
\end{equation}
Для римановых пространств нулевой кривизны нормальные координаты совпадают с
декартовыми.
\end{theorem}
\begin{com}
В дальнейшем мы увидим, что отображение (\ref{emacoc}) не является накрытием.
\qed\end{com}

В общем случае отображение (\ref{emacoc}) не является взаимно однозначным.
Поэтому в области определения нормальных координат $\MR^n$ можно задать
отношение эквивалентности, отождествив те точки, которые отображаются на одну
и ту же точку из $\MM$. Таким образом пространство постоянной кривизны вида
(\ref{ecocuc}) можно рассматривать как евклидово пространство $\MR^n$, в котором
задано некоторое отношение эквивалентности между точками.
\begin{exa}
Рассмотрим двумерную сферу $\MS^2\hookrightarrow\MR^3$ единичного радиуса в
качестве пространства постоянной положительной кривизны. В этом случае
$K=1,a=1,n=2$. Функция $f(r^2)$ в полярных координатах на плоскости $\MR^2$
имеет вид
\begin{equation*}
  f=\frac{\sin^2r}{r^2}.
\end{equation*}
Это соответствует метрике
\begin{equation*}
  ds^2=dr^2+\sin^2rd\vf^2.
\end{equation*}

Длина окружности на плоскости $\MR^2$ радиуса $r$ с центром в начале
координат равна
\begin{equation*}
  L=\int_0^{2\pi}\!\!\!d\vf\sin r=2\pi\sin r.
\end{equation*}
Отсюда следует, что окружности радиусов $r=\pi k$, $k=1,2,\dotsc$ отображаются в
одну точку сферы. При этом плоскость $\MR^2$ бесконечное число раз ``накрывает''
сферу $\MS^2$. Если условиться, что начало координат соответствует южному полюсу
сферы, то при отображении $\MR^2\rightarrow\MS^2$ все окружности радиуса
$r=2\pi k$, $k=0,1,\dotsc$, соответствуют южному полюсу, а все окружности
радиуса $r=\pi+2\pi k$ -- северному. В рассматриваемом случае между точками
евклидовой плоскости возникает отношение эквивалентности
\begin{equation*}                                                    \tag*{\qed}
  y^\al\sim y^\al+\frac{y^\al}r 2\pi k,\qquad k=0,1,\dotsc.
\end{equation*}
\renewcommand{\qed}{}\end{exa}

Нормальные координаты были определены таким образом, что геодезические линии в
них совпадают с прямыми. В (псевдо-)римановом пространстве экстремали совпадают
с геодезическими и поэтому также являются прямыми. Проверим это для пространств
постоянной кривизны, которые были рассмотрены выше. Уравнения для экстремалей,
определяемых метрикой (\ref{emeccn}) можно проинтегрировать. Из выражения для
символов Кристоффеля (\ref{echcoa}), которые в рассматриваемом случае имеют вид
\begin{equation}                                                  \label{echsco}
  \Gamma_{\al\bt}{}^\g=\frac{f'}f\left(y_\al\Pi^{\St\g}_\bt
  +y_\bt\Pi^{\St\g}_\al\right)+\frac{1-f-f's}sy^\g\Pi^\St_{\al\bt},
\end{equation}
следуют уравнения для экстремалей (\ref{eextre})
\begin{equation}                                                  \label{extcos}
  \ddot y^\al=-2\frac{f'}f\dot y^\al y_\bt\dot y^\bt
  +2\frac{f'}fy^\al\frac{(y_\bt\dot y^\bt)^2}s
  +\frac{1-f-f's}{s^2}y^\al(y_\bt\dot y^\bt)^2
  -\frac{1-f-f's}sy^\al(\dot y_\bt\dot y^\bt).
\end{equation}
Эти уравнения имеют интеграл (\ref{efirin})
\begin{equation*}
  C_0=\dot y^\al\dot y^\bt g_{\al\bt}=\dot y_\al\dot y^\al f
  -\frac{(y_\al\dot y^\al)^2}s(f-1)=\const.
\end{equation*}

Прямая проверка показывает, что все прямые линии, проходящие через начало
координат,
\begin{equation*}
  y^\al=v^\al t,\qquad v^\al=\const,\quad t\in\MR,
\end{equation*}
являются экстремалями. Эти экстремали, очевидно, полны. Отметим, что символы
Кристоффеля (\ref{echsco}) равны нулю только в начале координат. В близких
точках они отличны от нуля, и среди экстремалей прямыми являются только те,
которые проходят через начало координат.

Уравнения для экстремалей (\ref{extcos}) можно проинтегрировать и в общем
случае. То есть найти те экстремали, которые не проходят через начало координат.
Для этой цели рассмотрим зависимость $s:=y^\al y^\bt\eta_{\al\bt}$ от
$t$. Учитывая равенства
\begin{equation*}
\begin{split}
  \dot s&=2y_\al\dot y^\al,
\\
  \ddot s&=2y_\al\ddot y^\al+2\dot y_\al\dot y^\al,
\end{split}
\end{equation*}
из уравнений для экстремалей получаем обыкновенное уравнение на $s(t)$:
\begin{equation*}
  \ddot s=\left(\frac1{2s}-\frac{U'}{2U}\right)\dot s^2+\frac{2C_0U's}U,
\end{equation*}
где функция $U(s)$ определена равенством (\ref{esufum})
\begin{equation*}
  U(s)=\frac{\sin^2\sqrt{as}}a.
\end{equation*}
Рассмотрим случай $C_0a>0$. Введя новую переменную $z^2=as$, при $as>0$, и
растянув канонический параметр $t\rightarrow t/\sqrt{C_0a}$, приходим к
уравнению
\begin{equation*}
  \ddot z=(1-\dot z^2)\ctg z,
\end{equation*}
которое можно явно проинтегрировать. Общее решение этого уравнения имеет вид
\begin{equation*}
  \cos\sqrt{as}=\sqrt{1-\frac1{C_1}}\sin(t+t_0),\qquad
  C_1=\const>1,\quad t_0=\const,
\end{equation*}

Аналогично можно рассмотреть все остальные случаи.

Таким образом, для пространств постоянной кривизны специального вида
(\ref{ecocus}) в нормальных координатах можно найти и проанализировать поведение
всех экстремалей. В дальнейшем мы рассмотрим эту задачу в ряде конкретных
случаев.
\subsection{Нормальные координаты и экспоненциальное отображение}
Пусть задана связность $\Gamma$ на многообразии $\MM$ и $\g=x(t)$ -- геодезическая.
Если $t$ -- канонический параметр вдоль геодезической, то координатные функции
$\lbrace x^\al(t)\rbrace$ удовлетворяют системе уравнений (\ref{egeode}).
Зафиксируем канонический параметр $t$ каким либо образом и начальную точку
$x_0=x(0)$. Тогда касательный вектор $u_0:=\dot x_0\in\MT_0(\MM)$ к
геодезической в точке $x_0$ определен однозначно. Верно также обратное
утверждение. Если задана точка $x_0\in\MM$ и касательный вектор
$u_0\in\MT_0(\MM)$, то существует единственная геодезическая, проходящая
через $x_0$ с начальным вектором $u_0$, при этом канонический параметр определен
с точностью до сдвига. Таким образом, каждая геодезическая однозначно определена
парой $(x_0,u_0)$.

В разделе \ref{svechs} было определено экспоненциальное отображение для
гладких полных векторных полей на многообразии. Геодезическая линия $x(t)$
является интегральной кривой для векторного поля скорости $u(t):=\dot x(t)$.
Тогда для полных геодезических определено экспоненциальное отображение
\begin{equation}                                                  \label{expgeo}
  \exp tu_0:\quad x_0\mapsto x(t),
\end{equation}
которое отображает начальную точку $x_0$ в точку $x(t)$. Если вектор скорости
$u_0$ принимает все возможные направления в касательном пространстве
$\MT_0(\MM)$, то экспоненциальное отображение (\ref{expgeo}) можно
рассматривать, как отображение касательного пространства в многообразие
\begin{equation}                                                  \label{expmag}
  \exp tu_0:\quad \MT_0(\MM)\ni\quad tu_0\mapsto x(t)\quad\in\MM,
\end{equation}
для которого мы сохраним прежнее обозначение.
\begin{com}
В разделе \ref{svechs} экспоненциальное отображение было определено для каждого
полного векторного поля. При этом вопрос о том, каким образом данное векторное
поле задано, не рассматривался. Если векторное поле одно, то экспоненциальное
отображение нельзя рассматривать, как отображение касательного пространства
(\ref{expmag}), т.к.\ в точке $x_0$ имеется только один вектор. В
рассматриваемом случае ситуация другая. На многообразии $\MM$ задана связность
$\Gamma$, которая определяет все множество геодезических. Для фиксированной точки
$x_0$ мы рассматриваем множество геодезических, проходящих через данную точку во
всех возможных направлениях. То есть вектор скорости $u_0$ в (\ref{expmag})
принимает все возможные направления. Следовательно, экспоненциальное отображение
можно рассматривать, как отображение касательного пространства.
\qed\end{com}

По построению, каждая прямая в касательном пространстве $\MT_0(\MM)$, проходящая
через начало координат, отображается в соответствующую геодезическую. Ясно, что
такое отображение можно построить для каждой точки $x_0\in\MM$.

Если геодезическая неполна, то экспоненциальное отображение (\ref{expmag})
определено только для некоторого интервала значений канонического параметра
$-\e_1<t<\e_2$, где $\e_{1,2}>0$. В результате экспоненциальное отображение
будет определено в некоторой окрестности начала координат касательного
пространства. Если связность на $\MM$ класса $\CC^\infty$, то экспоненциальное
отображение будет того же класса гладкости $\CC^\infty$. Поскольку дифференциал
экспоненциального отображения в точке $x_0$ невырожден, то существует
окрестность $\MU_0\ni x_0$, такая, что экспоненциальное отображение
(\ref{expmag}) является диффеоморфизмом
$\MU_0\rightarrow\MV_0\subset\MT_0(\MM)$, где $\MV_0$ -- некоторая окрестность
касательного пространства, содержащая начало координат. Выберем координатный
репер $\lbrace \pl_\al\rbrace$ в точке $x_0$ и предположим, что векторы скорости
принимают значение на единичной сфере
\begin{equation*}
  \sum_{\al=1}^n(u^\al_0)^2=1.
\end{equation*}
Теперь отождествим касательное пространство $\MT_0(\MM)$ с евклидовым
пространством $\MR^n$ естественным образом, отождествив координаты касательного
вектора $\lbrace tu_0^\al\rbrace$ с декартовыми координатами точки
$\lbrace y^\al\rbrace$ в $\MR^n$. В результате получим координатную систему,
определенную на $\MU_0$.
\begin{defn}
Система координат в окрестности $\MU_0\subset\MM$ точки $x_0$, определенная
экспоненциальным отображением (\ref{expmag}),
\begin{equation*}
  \vf:\quad \MM\supset\MU_0\ni\quad\lbrace x^\al(t)\rbrace\mapsto
  \lbrace y^\al:=tu_0^\al\rbrace\quad\in\MV_0\subset\MR^n,
\end{equation*}
называется {\em нормальной}.
\qed\end{defn}
\index{Нормальная система координат (normal coordinate system)}%
\index{Система координат нормальная (normal coordinate system)}%
\begin{com}
Подчеркнем, что нормальная система координат определена исключительно
связностью, а не метрикой, которой вообще может не быть на многообразии.
\qed\end{com}

Если на многообразии $\MM$ помимо связности $\Gamma$ задана также метрика $g$, то
исходную систему координат $x^\al$ в окрестности точки $x_0$ можно всегда
выбрать таким образом, что координатный базис $\pl_\al$ будем ортонормальным в
данной точке $x_0$, т.е.\ $g_{\al\bt}(x_0)=\eta_{\al\bt}$. Мы всегда
предполагаем, что для нормальной системы координат при наличии метрики данное
условие выполнено.
\begin{theorem}[\bf Уайтхед]
Пусть $y^\al$ -- нормальная система координат в окрестности точки $x_0\in\MM$.
Определим окрестность $\MU_0(\rho)$ точки $x_0$ равенством
$\sum_\al(y^\al)^2<\rho^2$. Тогда существует положительное число $r$ такое, что
если $0<\rho<r$, то:\newline
\indent 1) \parbox[t]{.92\linewidth}{окрестность $\MU_0(\rho)$ является
геодезически выпуклой, т.е.\ любые две точки из $\MU_0(\rho)$ можно соединить
геодезической, целиком лежащей в $\MU_0(\rho)$;}\newline
\indent 2) \parbox[t]{.92\linewidth}{каждая точка из $\MU_0(\rho)$ имеет
нормальную координатную окрестность, содержащую $\MU_0(\rho)$.}
\end{theorem}
\begin{proof}
См.\ \cite{Whiteh32}.
\end{proof}
\begin{com}
Данная теорема справедлива для произвольных достаточно гладких связностей,
независимо от того задана ли на многообразии метрика или нет.
\qed\end{com}
\section{Полнота римановых многообразий                          \label{sglstr}}
Пусть задано риманово многообразие $(\MM,g)$. В настоящем разделе положительная
определенность метрики существенна, и мы будем ее предполагать.
\begin{defn}
Рассмотрим две произвольные точки многообразия $p,q\in\MM$. {\em Расстоянием}
между этими точками называется нижняя грань интегралов
\begin{equation}                                                  \label{edispq}
  l(p,q):=\inf\int_p^q\!\!\!ds,\qquad ds:=\sqrt{\dot x^\al\dot x^\bt g_{\al\bt}}
\end{equation}
по всем кусочно дифференцируемым кривым класса $\CC^1$, соединяющим эти точки.
\qed\end{defn}
\index{Расстояние (distance)}%
Функция расстояния определяет отображение
\begin{equation*}
  l:\quad \MM\times\MM\ni\quad(p,q)\mapsto l(p,q)\quad\in\MR_+.
\end{equation*}
Нетрудно проверить, что все свойства топологической метрики, рассмотренной в
разделе (\ref{seucto}) выполнены и, следовательно, функция $l(p,q)$ задает
топологическую метрику на $\MM$.
\begin{defn}
Множество точек
\begin{equation}                                                  \label{edebal}
  \MB_r(p):=\lbrace x\in\MM:\quad l(x,p)<r\rbrace
\end{equation}
называется {\em шаром} радиуса $r$ с центром в точке $p\in\MM$.
\qed\end{defn}
\index{Шар (ball)}%

\begin{prop}
Функция расстояния (\ref{edispq}) непрерывна и семейство метрических шаров для
всех $p\in\MM$ и $r\in\MR_+$ образует базу исходной топологии многообразия.
\end{prop}
\begin{proof}
Следует из непрерывной зависимости интеграла от пределов интегрирования.
\end{proof}

Для связностей Леви--Чивиты на римановых пространствах справедлива
\begin{theorem}[\bf Майерс, Стинрод]
Пусть $(\MM_1,g_1)$ и $(\MM_2,g_2)$ -- римановы многообразия. Пусть $l_1$ и
$l_2$ -- функции расстояния на $\MM_1$ и $\MM_2$ соответственно. Если задано
отображение $f:~\MM_1\rightarrow\MM_2$ (которое не предполагается непрерывным
или дифференцируемым), такое, что $l_1(p,q)=l_2\big(f(p),f(q)\big)$ для всех
$p,q\in\MM_1$, то $f$ есть диффеоморфизм из $\MM_1$ на $\MM_2$, который
отображает метрику $g_1$ в $g_2$.

В частности, каждое отображение $f$ из $\MM$ на себя, сохраняющее функцию
расстояния $l$, есть изометрия, т.е.\ отображение $f$ сохраняет метрику $g$.
\end{theorem}
\begin{proof}
См.\ \cite{MyeSte39}.
\end{proof}

Напомним, что топологическое пространство $\MM$ и, в частности, многообразие,
называется метрически полным, если любая фундаментальная последовательность в
$\MM$ сходится к некоторой точке из $\MM$.

С другой стороны, в разделе \ref{sextre} мы определили экстремали как линии, для
которых первая вариация интеграла (\ref{edispq}) равна нулю. Экстремали являются
одновременно геодезическими линиями для связности Леви--Чивиты. Эта связность
является полной, если любую экстремаль можно продолжить в обе стороны до
бесконечных значений канонического параметра. В этом случае мы говорим, что
риманово многообразие полно.

Таким образом было введено два понятия полноты многообразия: метрическая полнота
и полнота связности Леви--Чивиты. Оба эти понятия определяются одним
геометрическим объектом -- римановой метрикой и поэтому между ними существует
тесная связь, которую мы здесь рассмотрим.
\begin{theorem}[\bf Хопф, Ринов]                                  \label{thopri}
Для связного риманова многообразия следующие условия эквивалентны:\newline
\indent 1) \parbox[t]{.92\linewidth}{$(\MM,g)$ -- полное риманово многообразие;}
\newline
\indent 2) \parbox[t]{.92\linewidth}{$(\MM,l)$ -- полное метрическое
                                      пространство;} \newline
\indent 3) \parbox[t]{.92\linewidth}{каждый замкнутый метрический шар
$\overline\MB_r(p)$ в $\MM$ компактен;}\newline
\indent 4) \parbox[t]{.92\linewidth}{для каждой точки $x_0\in\MM$
экспоненциальное отображение (\ref{expmag}) определено на всем касательном
пространстве $\MT_0(\MM)$.}
\end{theorem}
\index{Теорема Хопфа--Ринова (Hopf--Rinov theorem)}%
\index{Хопфа--Ринова теорема (Hopf--Rinov theorem)}%
\begin{proof}
См.\ \cite{HopRin31}.
\end{proof}
\begin{theorem}
Если $\MM$ -- связное полное риманово многообразие, то $\MM$ геодезически
выпукло, т.е.\ любые две точки $p,q\in\MM$ можно соединить минимизирующей
экстремалью.
\end{theorem}
\begin{proof}
См., например, \cite{KobNom6369R}.
\end{proof}
\begin{cor}
Если все геодезические, исходящие из любой выбранной точки $p$ связного риманова
многообразия полны, то $(\MM,l)$ -- геодезически и метрически полно.
\end{cor}
\begin{proof}
См., например, \cite{KobNom6369R}.
\end{proof}
\begin{cor}                                                       \label{comcom}
Каждое компактное риманово многообразие $(\MM,g)$ метрически полно.
\end{cor}
\begin{proof}
Следствие импликации $3)\rightarrow 1)$ в теореме \ref{thopri}.
\end{proof}
Напомним, что риманово пространство $(\MM,g)$ называется {\em однородным}, если
группа изометрий действует на $\MM$ транзитивно.
\begin{theorem}
Каждое однородное риманово многообразие $(\MM,g)$ полно.
\end{theorem}
\begin{proof}
Пусть $x$ -- точка однородного риманова пространства $(\MM,g)$. Тогда существует
положительное число $r>0$ такое, что для каждого единичного вектора
$u\in\MT_x(\MM)$ геодезическая $\exp tu$ определена для каждого $|t|\le r$.
Пусть $\g=x(t)$, $0\le t\le s$, -- произвольная геодезическая в $\MM$ с
каноническим параметром $t$. Покажем, что эта геодезическая $\g$ может быть
продолжена до геодезической, определенной при $0\le t\le s+r$.
Пусть $\imath$ -- изометрия многообразия $\MM$, которая отображает точку $x$ в
$x(s)$. Тогда дифференциал обратной изометрии $\imath^{-1}_*$ отображает
касательный вектор $\dot x(s)$ в $u$
\begin{equation*}
  u=\imath^{-1}_*\dot x(s).
\end{equation*}
Поскольку $\exp tu$ есть геодезическая, проходящая через точку $x$, то
$\imath(\exp tu)$ -- геодезическая, проходящая через $x(s)$. Положим
\begin{equation*}
  x(s+t):=\imath(\exp tu)\qquad \text{для}\qquad 0\le t\le r.
\end{equation*}
Тогда кривая $\g=x(t)$ при $0\le t\le s+r$ является продолжением геодезической.
\end{proof}
\begin{com}
Доказанная теорема следует также из того общего факта, что каждое локально
компактное метрическое однородное пространство полно.
\qed\end{com}
\begin{theorem}                                                   \label{tcomco}
Пусть $\widetilde\MM$ и $\MM$ -- связные римановы многообразия одинаковой
размерности. Допустим, что существует изометрическое погружение
$p:~\widetilde\MM\rightarrow\MM$. Тогда:\newline
\indent 1) \parbox[t]{.92\linewidth}{Если $\widetilde\MM$ полно, то отображение
$p:~\widetilde\MM\rightarrow\MM$ является накрытием и $\MM$ полно.}\newline
\indent 2) \parbox[t]{.92\linewidth}{Обратно, если
$p:~\widetilde\MM\rightarrow\MM$ -- накрытие и $\MM$ полно, то $\widetilde\MM$
также полно.}
\end{theorem}
\begin{proof}
См., например, \cite{KobNom6369R}.
\end{proof}
\begin{cor}
Пусть $\widetilde\MM$ и $\MM$ -- связные многообразия одинаковой размерности и
$p:~\widetilde\MM\rightarrow\MM$ -- погружение. Тогда, если многообразие
$\widetilde\MM$ компактно, то $\MM$ также компактно, а $p$ -- накрывающее
отображение.
\end{cor}
\begin{proof}
Возьмем любую риманову метрику $g$ на $\MM$. Тогда $p^*g$ -- единственная
метрика на $\widetilde\MM$ такая, что $p$ -- изометрическое погружение.
Поскольку $\widetilde\MM$ компактно, то, по следствию \ref{comcom}, оно полно.
Тогда из теоремы \ref{tcomco} следует, что $p$ -- накрывающее отображение и
отсюда $\MM$ компактно.
\end{proof}
\begin{com}
В предыдущих теореме и следствии требование одинаковой размерности многообразий
$\widetilde\MM$ и $\MM$ является излишним, т.к.\ погружение возможно только для
многообразий одинаковой размерности.
\qed\end{com}
\begin{defn}
Говорят, что связное риманово пространство {\em непродолжаемо}, если его нельзя
изометрически вложить в другое связное риманово пространство как собственное
открытое подмногообразие.
\qed\end{defn}
\index{Непродолжаемое риманово пространство (noncontinuable Riemannian space}%
\index{Риманово пространство непродолжаемое (noncontinuable Riemannian space}%
\index{Пространство риманово непродолжаемое (noncontinuable Riemannian space}%
Теорема \ref{tcomco} показывает, что каждое полное связное риманово многообразие
непродолжаемо. Обратное утверждение неверно.
\begin{exa}
Пусть $\MM$ есть евклидова плоскость с выколотым началом координат, а
$\widetilde\MM$ -- его универсальное накрывающее пространство. Как открытое
подмногообразие евклидовой плоскости $\MM$ имеет естественную евклидову метрику,
которая, очевидно, неполна. На универсальной накрывающей $\widetilde\MM$ также
задана естественная евклидова метрика. Риманово многообразие $\widetilde\MM$
неполно по теореме \ref{tcomco}. Может быть доказано, что $\widetilde\MM$
непродолжаемо. Таким образом, неполное связное риманово многообразие в общем
случае может быть непродолжаемо.
\qed\end{exa}
\begin{cor}
Пусть $\MI(\MM)$ -- группа изометрий связного риманова многообразия $(\MM,g)$.
Если орбита $x\MI$ точки $x\in\MM$ содержит открытое подмножество из $\MM$, то
орбита $x\MI$ совпадает со всем $\MM$. Тем самым риманово многообразие $\MM$
однородно.
\end{cor}
\section{Формулы Френе                                           \label{sfrefj}}
Рассмотрим произвольную гладкую кривую $\g=\lbrace x^\al(t)\rbrace $ в
трехмерном римановом многообразии $\MM$, $\dim\MM=3$, с положительно
определенной метрикой $g$. Тогда длина кривой (\ref{eculei}) отлична от нуля.
Выберем длину кривой в качестве канонического параметра $t$ вдоль кривой.
Предположим также, что на многообразии задана метрическая связность $\Gamma$
(геометрия Римана--Картана). Единичный касательный вектор к кривой определяется
вектором скорости
$$
  u^\al:=\dot x^\al,
$$
где точка обозначает дифференцирование по каноническому параметру $t$.
Дифференцируя тождество $u^2=1$, получим равенство
\begin{equation}                                                  \label{efreno}
  \nb_\bt u^2=2u_\al\nb_\bt u^\al=0.
\end{equation}
Отсюда вытекает, что ковариантная производная (\ref{ecodcu}) от $u^\al$ вдоль
кривой $\g$ (ускорение кривой),
\begin{equation}                                                  \label{eprcul}
  \frac{Du^\al}{dt}=u^\bt\nb_\bt u^\al=\frac1\rho n^\al,
\end{equation}
где $\rho(t)$ -- некоторая функция вдоль кривой, ортогональна вектору скорости.
Здесь мы предполагаем, что $u^\bt\nb_\bt u^\al\ne0$, т.е.\ кривая $\g$ не
является геодезической. Вектор $n^\al$ всегда можно
выбрать единичным:
\begin{equation}                                                  \label{eidtnl}
   (u,n)=0,\qquad n^2=1.
\end{equation}
\begin{defn}
Единичное векторное поле $n^\al(s)\in\CX(\MM,\g)$, определенное вдоль кривой,
называется {\em главной нормалью кривой}. Функция $1/\rho(t)$ называется
{\em кривизной кривой}.

Поскольку пространство трехмерно, то дополним векторы $u^\al$ и $n^\al$ до
ортонормированного базиса в касательном пространстве с помощью вектора
{\em бинормали} к кривой, определяемого следующими соотношениями:
\begin{equation}                                                  \label{enbide}
  (u,b)=0,\qquad (n,b)=0,\qquad b^2=1. \qed
\end{equation}
\end{defn}
\index{Главная нормаль (principle normal)}%
\index{Кривизна кривой (curvature of a curve}%
\index{Бинормаль к кривой (binormal to a curve)}%
Ковариантные производные от $n^\al$ и $b^\al$ можно разложить по этому базису с
некоторыми коэффициентами
\begin{equation}                                                  \label{ecodnl}
\begin{split}
  \frac{Dn^\al}{dt}&=au^\al+bn^\al+cb^\al,
\\
  \frac{Db^\al}{dt}&=a'u^\al+b'n^\al+c'b^\al.
\end{split}
\end{equation}
Дифференцируя тождества (\ref{eidtnl}), (\ref{enbide}) вдоль кривой, получим
условия на коэффициенты разложения:
$$
  \frac1\rho+a=0,\qquad b=0,\qquad a'=0,\qquad c+b'=0,\qquad c'=0.
$$
Тогда из соотношений (\ref{eprcul}) и (\ref{ecodnl}) следуют {\em формулы
 Френе}:
\index{Формулы Френе (Frenet formulae)}%
\index{Френе формулы (Frenet formulae)}%
\begin{equation}                                                  \label{efrefo}
\begin{split}
  \frac{Du^\al}{dt}&=\quad \frac1\rho n^\al,
\\
  \frac{Dn^\al}{dt}&=-\frac1\rho u^\al+\frac1\tau b^\al,
\\
  \frac{Db^\al}{dt}&=-\frac1\tau n^\al,
\end{split}
\end{equation}
где введено обозначение
$$
  c=-b'=\frac1\tau.
$$
Функция $1/\tau(t)$ называется {\em кручением} кривой.
\index{Кручение кривой (torsion of a curve)}%

Если кривая задана, то при желании можно найти явные выражения для векторных
полей $u^\al$, $n^\al$ и $b^\al$, а также вычислить кривизну и кручение кривой.
В трехмерном евклидовом пространстве $\MR^3$ верно также обратное утверждение
(см. \cite{DuNoFo98R}). А именно, если известны кривизна и кручение как функции
канонического параметра вдоль кривой, то можно восстановить кривую в $\MR^3$ с
точностью до движений (сдвигов, вращений и отражений) всего пространства.
Таким образом, кривизна и кручение кривой в трехмерном евклидовом пространстве
представляют собой полный набор геометрических инвариантов кривой.
\begin{exa}
Кривизна и кручение прямой линии в трехмерном евклидовом пространстве $\MR^3$
равны нулю.
\end{exa}
\begin{exa}
Рассмотрим спираль в трехмерном евклидовом пространстве $\MR^3$, ось которой
совпадает с осью $z$:
\begin{equation}                                                  \label{espira}
\begin{split}
  x&=R\cos\frac t{\sqrt{R^2+v^2}},
\\
  y&=R\sin\frac t{\sqrt{R^2+v^2}},
\\
  z&=\frac{vt}{\sqrt{R^2+v^2}},
\end{split}
\end{equation}
где $R>0$ и $v$ -- постоянные и параметр $t\in\MR$ совпадает с длиной спирали.
Касательный вектор к спирали имеет следующие компоненты:
\begin{equation*}
\begin{split}
  u^x&=-\frac R{\sqrt{R^2+v^2}}\sin\frac t{\sqrt{R^2+v^2}},
\\
  u^y&=\quad \frac R{\sqrt{R^2+v^2}}\cos\frac t{\sqrt{R^2+v^2}},
\\
  u^z&=\quad \frac v{\sqrt{R^2+v^2}}.
\end{split}
\end{equation*}
Дифференцирование этих равенств по $t$ приводит к следующему вектору главной
нормали:
\begin{equation*}
\begin{split}
  n^x&=-\cos\frac t{\sqrt{R^2+v^2}},
\\
  n^y&=-\sin\frac t{\sqrt{R^2+v^2}},
\\
  n^z&=0.
\end{split}
\end{equation*}
При этом кривизна спирали равна
\begin{equation*}
  \frac1\rho=\frac R{R^2+v^2}.
\end{equation*}

Дальнейшее дифференцирования полученных равенств по $t$ определяет вектор
бинормали:
\begin{equation*}
\begin{split}
  b^x&=\quad \frac v{\sqrt{R^2+v^2}}\sin\frac t{\sqrt{R^2+v^2}},
\\
  b^y&=-\frac v{\sqrt{R^2+v^2}}\cos\frac t{\sqrt{R^2+v^2}},
\\
  b^z&=\quad \frac R{\sqrt{R^2+v^2}}
\end{split}
\end{equation*}
и кручение спирали
\begin{equation*}
  \frac1\tau=\frac v{R^2+v^2}.
\end{equation*}
Таким образом вычислены все характеристики спирали.

Если $v=0$, то спираль вырождается в окружность. Для окружности кривизна и
кручение равны
\begin{equation*}                                                    \tag*{\qed}
  \frac1\rho=\frac1R,\qquad \frac1\tau=0.
\end{equation*}
\renewcommand{\qed}{}\end{exa}
\begin{com}
Кривизна $1/\rho$ и кручение $1/\tau$ кривой зависят от метрики $g$ и
тензора кручения $T$ трехмерного многообразия $\MM$, что следует из определения
метрической связности. При выводе формул Френе условие метричности связности
важно, т.к.\ при отличной от нуля неметричности формула (\ref{efreno}) неверна и
ковариантная производная (\ref{eprcul}) не будет ортогональна вектору скорости.
Заметим также, что кривизна и кручение кривой являются понятиями, отличными от
кривизны и кручения аффинной связности, введенных ранее.
\qed\end{com}

Посмотрим на геодезические линии с точки зрения введенных выше понятий.
Следующее утверждение очевидно.
\begin{prop}
Кривая $\g$ в трехмерном пространстве Римана--Картана является геодезической
тогда и только тогда, когда ее кривизна $1/\rho$ равна нулю.
\end{prop}
Для геодезической линии на $\MM$ правая часть уравнения (\ref{eprcul}) равна
нулю, и, следовательно, вектор нормали $n$ к геодезической нельзя определить
соотношением (\ref{eprcul}). Кручение геодезической линии также неопределено.

При смещении вдоль кривой, отличной от геодезической, на расстояние $dt$
компоненты касательного вектора и вектора главной нормали получают приращение
$$
  Du^\al=\frac{dt}\rho n^\al,\qquad
  Dn^\al=-\frac{dt}\rho u^\al+\frac{dt}\tau b^\al.
$$
Отсюда следует, что при параллельном переносе вдоль кривой на бесконечно малое
расстояние касательный вектор и вектор главной нормали остаются в плоскостях,
натянутых на векторы $u^\al$ и $n^\al$, тогда и только тогда, когда кручение
кривой равно нулю, $1/\tau=0$. В этом случае векторы $u^\al$ и $n^\al$
поворачиваются на угол $d\vf=-dt/\rho$.

Если рассматривать кривую $\g$ на двумерном многообразии с заданной римановой
метрикой и метрической связностью, то вектор бинормали тождественно равен нулю,
а формулы Френе принимают вид
\begin{equation*}
\begin{split}
  \frac{Du^\al}{dt}&=\quad \frac1\rho n^\al,
\\
  \frac{Dn^\al}{dt}&=-\frac1\rho u^\al.
\end{split}
\end{equation*}
В этом случае кривые нулевой кривизны и только они являются геодезическими.
\chapter{Симплектические и пуассоновы многообразия}
Симплектические и пуассоновы многообразия играют важную роль в дифференциальной
геометрии в связи с применениями, в первую очередь, к гамильтоновой динамике,
рассмотренной в главе \ref{scanfp}.
\section{Симплектические группы}                                  \label{simgru}
Симплектические группы играют в симплектической геометрии ту же роль, что и
группы вращений в евклидовой геометрии. Поскольку эти группы устроены намного
сложнее, чем ортогональные и унитарные матричные группы, то мы посвятим
симплектическим группам и их свойствам целый раздел. Более подробное изложение
содержится в \cite{Fomenk88R}.
\begin{defn}
Рассмотрим антисимметричную $2n\times 2n$ матрицу ({\em каноническую
симплектическую форму}),
\begin{equation}                                                  \label{esymfo}
  \varpi=(\varpi_{\al\bt})=\begin{pmatrix} 0 & -\one \\ \one & 0 \end{pmatrix},
\end{equation}
где $\al,\bt=1,\dotsc,2n$ и $\one$ -- единичная $n\times n$ матрица. Эта матрица
определяет билинейную квадратичную форму в евклидовом пространстве $\MR^{2n}$,
рассматриваемом, как векторное пространство,
\begin{equation*}
  \varpi:\quad \MR^{2n}\times\MR^{2n}\ni\quad X,Y\mapsto\varpi(X,Y)
  :=X^\al Y^\bt\varpi_{\al\bt}\quad\in\MR,
\end{equation*}
где $X^\al$ и $Y^\bt$ -- компоненты векторов в декартовой системе координат.
Квадратные матрицы $A$ размера $2n\times 2n$ с вещественными элементами,
оставляющие каноническую симплектическую форму $\varpi$ инвариантной,
\begin{equation}                                                  \label{edspde}
  A^\St\varpi A=\varpi,
\end{equation}
образуют группу Ли $\MS\MP(n,\MR)$, которая называется {\em вещественной
симплектической группой}.
\qed\end{defn}
\index{Каноническая симплектическая форма (canonical symplectic form)}%
\index{Симплектическая форма каноническая (canonical symplectic form)}%
\index{Вещественная симплектическая группа (real symplectic group)}%
\index{Симплектическая группа вещественная (real symplectic group)}%
\index{Группа симплектическая вещественная (real symplectic group)}%

Взятие определителя от обеих частей определения (\ref{edspde}) приводит к
равенству $\det A=\pm1$, поскольку $\det\varpi=1$. Отсюда следует, что для любой
матрицы $A$ существует обратная. Нетрудно убедиться, что обратная матрица также
является симплектической, а также в том, что произведение двух симплектических
матриц снова дает симплектическую матрицу. Тем самым все групповые аксиомы
выполнены.

Каноническая симплектическая форма универсальна в следующем смысле. Если $\om$
-- произвольная невырожденная антисимметричная матрица, то из курса линейной
алгебры известно, что за счет линейного преобразования базиса в $\MR^{2n}$ ее
всегда можно преобразовать к каноническому виду (\ref{esymfo}).
\begin{prop}                                                      \label{pslalg}
Алгебра Ли $\Gs\Gp(n,\MR)$ группы $\MS\MP(n,\MR)$ состоит из матриц вида
\begin{equation}                                                  \label{espalm}
\begin{pmatrix} B & ~C \\ D & -B^\St \end{pmatrix}\in\Gs\Gp(n,\MR),
\end{equation}
где $B$ -- произвольная вещественная $n\times n$ матрица, а вещественные
$n\times n$ матрицы $C$ и $D$ симметричны.
\end{prop}
\begin{proof}
Вблизи единицы группы симплектическая матрица представима в виде
\begin{equation*}
  A=\ex^{tM}=\one+tM+\dotsc,\qquad t\in\MR,
\end{equation*}
где $M\in\Gs\Gp(n,\MR)$ -- элемент алгебры Ли. Подставляя это разложение в
(\ref{edspde}) в линейном по $t$ порядке получаем равенство
\begin{equation}                                                  \label{epalge}
  M^\St\varpi+\varpi M=0.
\end{equation}
Представим элемент алгебры в блочном виде
\begin{equation*}
  M=\begin{pmatrix} B & C \\ D & E \end{pmatrix}.
\end{equation*}
Тогда из (\ref{epalge}) следуют равенства:
\begin{equation*}                                                    \tag*{\qed}
  C=C^\St,\qquad D=D^\St,\qquad E=-B^\St
\end{equation*}
\renewcommand{\qed}{}\end{proof}
\begin{cor}
Размерность симплектической группы $\MS\MP(n,\MR)$ равна $n(2n+1)$.
\qed\end{cor}

Выше было отмечено, что $\det A=\pm1$. Справедливо более сильное утверждение.
\begin{prop}
Определитель любой симплектической матрицы равен единице
\begin{equation}                                                  \label{edetsy}
  \det A=1,\qquad A\in\MS\MP(n,\MR).
\end{equation}
\end{prop}
\begin{proof}
Любую симплектическую матрицу $A=(A_\al{}^\bt)\in\MS\MP(n,\MR)$,
$\al,\bt=1,\dotsc,2n$, можно рассматривать, как невырожденное линейное
преобразование $2n$-мерного евклидова пространства $\MR^{2n}$, которое оставляет
каноническую 2-форму $\varpi$ инвариантной. Следовательно, симплектическое
преобразование сохраняет инвариантной и любую внешнюю степень формы $\varpi$. В
частности, симплектическое преобразование сохраняет $2n$-форму объема
$\varpi^n$. Известно, что при преобразовании координат форма объема умножается
на якобиан преобразования координат. Таким образом для симплектических
преобразований имеем
\begin{equation*}
  \varpi^n=\varpi^n\det A.
\end{equation*}
Отсюда следует равенство (\ref{edetsy}).
\end{proof}

Симплектические группы $\MS\MP(n,\MR)$ существенно отличаются от групп
вращений евклидова пространства. В частности, они являются некомпактными.
Продемонстрируем это на примере простейшей группы $\MS\MP(1,\MR)$.
\begin{theorem}
Группа $\MS\MP(1,\MR)$ изоморфна группе $2\times 2$ матриц $\MS\ML(2,\MR)$.
С топологической точки зрения эта группа трехмерна некомпактна и гомеоморфна
прямому произведению окружности на двумерную плоскость,
$\MS\MP(1,\MR)\approx\MS^1\times\MR^2$. Она неодносвязна, и ее фундаментальная
группа изоморфна группе целых чисел
\begin{equation*}
  \pi\big(\MS\MP(1,\MR)\big)\approx\MZ.
\end{equation*}
\end{theorem}
\begin{proof}
В двумерном случае каноническая симплектическая форма с точностью до знака
совпадает с полностью антисимметричным тензором $\varpi_{\al\bt}=-\ve_{\al\bt}$
(\ref{easttd}). Рассмотрим матрицу $A\in\MS\MP(1,\MR)$, как линейное
преобразование двумерной евклидовой плоскости $\MR^2$. Тогда уравнение
(\ref{edspde}) запишется в виде
\begin{equation*}
  A_\al{}^\g A_\bt{}^\dl\ve_{\g\dl}=\ve_{\al\bt}.
\end{equation*}
Поскольку левая часть уравнения антисимметрична по индексам $\al$ и $\bt$, то
она равна $\det A\,\ve_{\al\bt}$. Отсюда следует, что уравнение (\ref{edspde})
эквивалентно одному уравнению $\det A=1$. Таким образом группа $\MS\MP(1,\MR)$
изоморфна группе вещественных $2\times2$ матриц с единичным определителем, т.е.\
группе $\MS\ML(2,\MR)$.

Из курса линейной алгебры известно, что любое линейное преобразование плоскости,
для которого $\det A=1$, можно однозначно представить в виде композиции двух
преобразований: ортогонального поворота плоскости (группа $\MU(1)\approx\MS^1$)
и преобразования, задающегося верхнетреугольной матрицей вида
\begin{equation*}
\begin{pmatrix} a & b \\ 0 & 1/a \end{pmatrix},\qquad a>0.
\end{equation*}
Вещественные числа $a$ и $b$ можно рассматривать в качестве координат на
полуплоскости $a>0$, которая гомеоморфна всей евклидовой плоскости. Таким
образом мы получаем топологическое разложение группы $\MS\MP(1,\MR)$ на прямое
произведение окружности и двумерной плоскости. Прямое произведение
$\MS^1\times\MR^2$ гомотопически эквивалентно окружности (стягивается к
окружности), и поэтому фундаментальная группа $\MS\MP(1,\MR)$ изоморфна
фундаментальной группой окружности (теорема \ref{tisoho}).
\end{proof}

Симплектические группы $\MS\MP(n,\MR)$ при $n>1$ имеют более сложную структуру,
и их описание выходит за рамки настоящей монографии. Отметим лишь, что все
группы $\MS\MP(n,\MR)$ некомпактны. В дальнейшем нам понадобятся другой класс
групп Ли, который обозначается $\MS\MP(n)$ и состоит из компактных
симплектических групп. Они существуют и строятся, как подгруппы в комплексных
симплектических группах $\MS\MP(n)\subset\MS\MP(n,\MC)$.

Рассмотрим $2n$-мерное комплексное пространство $\MC^{2n}$ с координатами
$z^\al$, $\al=1,\dotsc,2n$. Каноническая симплектическая форма $\varpi$
задает на $\MC^{2n}$ симплектическую структуру (билинейную квадратичную форму),
т.е.\ двум векторам $X$ и $Y$ ставится в соответствие число
\begin{equation*}
  \varpi:\quad \MC^{2n}\times\MC^{2n}\ni\quad X,Y\mapsto\varpi(X,Y)
  :=X^\al Y^\bt\varpi_{\al\bt}\quad\in\MC.
\end{equation*}
По-определению, симплектическая структура антисимметрична
$\varpi(X,Y)=-\varpi(Y,X)$. Невырожденное комплексное линейное преобразование
координат называется {\em симплектическим}, если оно сохраняет каноническую
симплектическую структуру. Это преобразование задается комплексной $2n\times 2n$
матрицей $A\in\MS\MP(2n,\MC)$, которая удовлетворяет тому же равенству
(\ref{edspde}), что и в вещественном случае. Единственное отличие состоит
в том, что элементами матрицы $A$ теперь являются комплексные числа.
Эти матрицы образуют {\em комплексную симплектическую группу} $\MS\MP(n,\MC)$.
\index{Комплексная симплектическая группа (complex symplectic group)}%
\index{Симплектическая группа комплексная (complex symplectic group)}%
\index{Группа комплексная симплектическая (complex symplectic group)}%
Ясно, что группа $\MS\MP(n,\MR)$ содержится в группе $\MS\MP(n,\MC)$, как
подгруппа вещественных симплектических преобразований.

Представление элементов алгебры Ли $\Gs\Gp(n,\MC)$ в виде (\ref{espalm})
справедливо и для комплексных симплектических групп, только матрицы $B,C$ и $D$
будут комплексными. Поэтому размерность группы $\MS\MP(n,\MC)$ в два раза больше
размерности вещественной группы $\MS\MP(n,\MR)$ и равна $2n(2n+1)$. Так же, как
и в вещественном случае, группа $\MS\MP(n,\MC)$ при всех $n$ является
некомпактной.
\begin{prop}
Симплектические группы $\MS\MP(n,\MR)$ и $\MS\MP(n,\MC)$ являются связными,
т.е.\ состоят из одной компоненты.
\end{prop}
\begin{prop}
Характеристический полином
\begin{equation*}
  f(\lm)=\det (A-\lm\one)=\sum_{k=1}^{2n}a_k\lm^k
\end{equation*}
симплектического вещественного преобразования $A\in\MS\MP(n,\MR)$ обладает
свойством
\begin{equation*}
  f(\lm)=\lm^{2n}f(1/\lm),
\end{equation*}
что означает симметричность его коэффициентов $a_k=a_{2n-k}$. В частности,
если $\lm$ собственное число симплектического преобразования, то $1/\lm$
также собственное число.
\end{prop}
\begin{proof}
Из определения симплектического преобразования (\ref{edspde}) следует, что
$A=-\varpi A^{-1\St}\varpi$, т.к.\ $\varpi^2=-\one$. Отсюда вытекает
цепочка равенств
\begin{align*}
  f(\lm)=\det(-\varpi A^{-1\St}\varpi-\lm\one)&=\det(-A^{-1\St}+\lm\one)=
\\
  &=\det(-\one+\lm A)=\lm^{2n}\det\left(A-\frac1\lm\one\right),
\end{align*}
где мы воспользовались равенством $\det A=\det A^\St=1$.
\end{proof}
Отметим, что у характеристического полинома не может быть нулевого
собственного значения, т.к.\ $\det A=1$. Поскольку характеристический
полином является вещественным, то, если $\lm$ -- комплексное собственное
число, то $\bar\lm$ -- также собственное число. Таким образом, в случае
общего положения собственные числа вещественного симплектического
преобразования разбиваются на четверки $\lm,\bar\lm,1/\lm,1/\bar\lm$,
т.е.\ собственные числа расположены на комплексной плоскости симметрично
относительно вещественной оси и единичной окружности.

По-построению, комплексная симплектическая группа $\MS\MP(n,\MC)$ содержит
некомпактную вещественную подгруппу $\MS\MP(n,\MR)$. Оказывается, что группа
$\MS\MP(n,\MC)$ содержит компактную подгруппу, которая называется компактной
симплектической группой и обозначается через $\MS\MP(n)$. Эту подгруппу удобно
определить с использованием алгебры кватернионов $\MH$ (см.\ приложение
\ref{squate}).
\begin{defn}
Рассмотрим $n$-мерное кватернионное пространство $\MH^n$ с базисом $e_\Sa$,
$\Sa=1,\dotsc,n$. Каждый вектор $q\in\MH^n$ однозначно представим в виде
$q=q^\Sa e_\Sa$, где каждая координата является кватернионом $q^\Sa\in\MH$.
Каждый кватернион разлагается по базису $\lbrace 1,i,j,k\rbrace$:
\begin{equation*}
  q^\Sa=a^\Sa+b^\Sa i+c^\Sa j+d^\Sa k.
\end{equation*}
Вещественная размерность $\MH^n$ равна $4n$. Очевидно, что $\MH^1=\MH$.

Рассмотрим в $\MH^n$ симметричное вещественнозначное скалярное произведение
\begin{equation}                                                  \label{equasc}
  (q_1,q_2):=\re q_1\bar q_2=\sum_{\Sa=1}^n
  (a_1^\Sa a_2^\Sa+b_1^\Sa b_2^\Sa+c_1^\Sa c_2^\Sa+d_1^\Sa d_2^\Sa).
\end{equation}
Множество всех линейных кватернионных преобразований $A\in\MG\ML(n,\MH)$
пространства $\MH^n$, не меняющих начало координат и сохраняющих скалярное
произведение
\begin{equation*}
  (Aq_1,Aq_2)=(q_1,q_2),
\end{equation*}
называется {\em симплектической компактной группой} $\MS\MP(n)$.
\qed\end{defn}
\index{Компактная симплектическая группа (compact symplectic group)}%
\index{Симплектическая группа компактная (compact symplectic group)}%

Кватернионное пространство $\MH^n$ естественным образом отождествляется с
евклидовым пространством $\MR^{4n}$. Тогда скалярное произведение (\ref{equasc})
совпадает с обычным евклидовым скалярным произведением в $\MR^{4n}$. Это значит,
что группа $\MS\MP(n)$ является подгруппой в $\MO(4n,\MR)$. Поскольку группа
вращений $\MO(4n,\MR)$ компактна, то и симплектическая группа $\MS\MP(n)$
также компактна.

В приложении \ref{squate} показано, что алгебра кватернионов $\MH$ естественным
образом отождествляется с двумерным комплексным пространством $\MC^2$. Выполняя
эту операцию вдоль каждой из кватернионных координат, мы получим отождествление
$\MH^n$ с $\MC^{2n}$. При отождествлении
\begin{equation*}
  q=z_1+j\bar z_2,\qquad z_1=a+ib,\quad z_2=c+id,
\end{equation*}
квадратичная форма двух кватернионов $q_1=z_{11}+jz_{12}$ и
$q_2=z_{21}+jz_{22}$ перейдет в сумму комплексных квадратичных форм
\begin{equation*}
  (q_1,q_2)_\MH:=q_1\bar q_2=(q_1,q_2)_\MC+\langle q_1,q_2\rangle_\MC,
\end{equation*}
где
\begin{equation*}
  (q_1,q_2)_\MC:=\sum_\Sa (z_{11}^\Sa\bar z_{21}^\Sa+z_{12}^\Sa\bar z_{22}^\Sa),
  \qquad \langle q_1,q_2\rangle_\MC
  :=\sum_\Sa (z_{12}^\Sa z_{21}^\Sa-z_{11}^\Sa z_{22}^\Sa).
\end{equation*}
Квадратичная форма $(q_1,q_2)_\MC$ эрмитова, т.е.\
$\overline{(q_1,q_2)}_\MC=(q_2,q_1)_\MC$, а форма $\langle q_1,q_2\rangle_\MC$
антисимметрична: $\langle q_1,q_2\rangle_\MC=-\langle q_1,q_2\rangle_\MC$.
Отметим, что квадратичная форма $(q_1,q_2)_\MH$ отличается от квадратичной формы
(\ref{equasc}) отсутствием знака реальной части.
\begin{theorem}
Множество элементов $\MS\MP(n)$ является связной компактной группой Ли
вещественной размерности $n(2n+1)$. При отождествлении $\MH^n$ с $\MC^{2n}$
группа $\MS\MP(n)$
вкладывается как подгруппа в унитарную группу $\MU(2n)$. При этом вложении
алгебра Ли $\Gs\Gp(n)$ группы $\MS\MP(n)$ состоит из комплексных $2n\times 2n$
матриц вида
\begin{equation}                                                  \label{espinu}
\begin{pmatrix}
  \quad Z_1 & Z_2 \\-\bar Z_2 & \bar Z_1,
\end{pmatrix}
\end{equation}
где $Z_1$ -- комплексная антиэрмитова $n\times n$ матрица, а $Z_2$ --
комплексная симметричная $n\times n$ матрица. Если матрицы из унитарной группы
$\MU(2n)$ представить в виде
\begin{equation*}
\begin{pmatrix}
  U_1 & U_2 \\ U_3 & U_4,
\end{pmatrix}
\end{equation*}
где $U_1,U_2,U_3$ и $U_4$ -- комплексные $n\times n$ матрицы, то подгруппа
$\MS\MP(n)$ в $\MU(2n)$ состоит из унитарных матриц вида
\begin{equation}                                                  \label{espunu}
\begin{pmatrix}
  \quad U_1 & U_2 \\-\overline U_2 & \overline U_1.
\end{pmatrix}
\end{equation}
При этом матрица (\ref{espunu}) является унитарной тогда и только тогда, когда
комплексные матрицы $U_1$ и $U_2$ удовлетворяют уравнениям
$U_1U_1^\dagger+U_2U_2^\dagger=\one$ и $U_2U_1^\St=U_1U_2^\St$.
\end{theorem}
\begin{proof}
Проводится прямой проверкой \cite{Fomenk88R}.
\end{proof}

Связь симплектических групп с другими матричными группами дается следующими
двумя теоремами, доказательство которых дано, например, в \cite{Fomenk88R}.
\begin{theorem}
Рассмотрим стандартные вложения групп $\MO(n)\hookrightarrow\MU(n)$,
$\MU(n)\hookrightarrow\MS\MO(2n)$, $\MS\MP(n)\hookrightarrow\MU(2n)$. Тогда
имеют место следующие соотношения:\newline
\indent 1) \parbox[t]{.92\linewidth}{$\MS\MO(2n)\cap\MS\MP(n)=\MU(n)$; здесь
группы $\MS\MO(2n)$ и $\MS\MP(n)$ рассматриваются, как подгруппы в одной группе
$\MU(2n)$;}\newline
\indent 2) \parbox[t]{.92\linewidth}{$\MS\MP(n,\MC)\cap\MU(2n)=\MS\MP(n)$;}
\newline
\indent 3) \parbox[t]{.92\linewidth}{$\MS\MP(n,\MR)\cap\MG\ML(2n,\MC)=\MU(n)$;}
\newline
\indent 4) \parbox[t]{.92\linewidth}{$\MS\MP(n,\MR)\cap\MU(2n)=\MU(n)$.}
\end{theorem}
\begin{theorem}
Для компактных симплектических групп $\MS\MP(1)$ и $\MS\MP(2)$ имеют место
изоморфизмы:
\begin{equation*}
  \MS\MP(1)\simeq\MS\MU(2)\simeq\MS\MP\MI\MN(3),\qquad
  \MS\MP(2)\simeq\MS\MP\MI\MN(5).
\end{equation*}
\end{theorem}
При больших $n>2$ компактные симплектические группы $\MS\MP(n)$ уже не сводятся
к унитарным и ортогональным группам.
\begin{theorem}
Группа $\MS\MP(n)$ является максимальной компактной подгруппой в комплексной
симплектической группе $\MS\MP(n,\MC)$.
\end{theorem}
\begin{proof}
См., например, \cite{Helgas01R}.
\end{proof}
\begin{theorem}
Все компактные симплектические группы $\MS\MP(n)$ односвязны.
\end{theorem}
\begin{proof}
См., например, \cite{Postni82R}.
\end{proof}
\section{Симплектические многообразия                            \label{ssymma}}
\begin{defn}
Многообразие $\MM$ четной размерности, $\dim\MM=2n$, называется
{\em симплектическим}, если на нем задана достаточно гладкая 2-форма
\begin{equation*}
  \om=\frac12dx^\al\wedge dx^\bt\om_{\al\bt}\in\Lm_2(\MM),
\end{equation*}
удовлетворяющая двум условиям:
\begin{align*}
  1)&\quad \det\om_{\al\bt}\ne0,\quad \forall x\in\MM &&
  \text{ -- невырожденность,}
\\
  2)&\quad d\om=0 &&\text{ -- замкнутость}.
\end{align*}
Форма $\om$ называется {\em симплектической}.
\qed\end{defn}
\index{Симплектическое многообразие (symplectic manifold)}%
\index{Многообразие симплектическое (symplectic manifold)}%
\index{Симплектическая форма (symplectic form)}%
\index{Форма симплектическая (symplectic form)}%

Пусть на многообразии $\MM$ задано два векторных поля $X,Y\in\CX(\MM)$. Тогда
форма $\om$ называется невырожденной, если из условия $\om(X,Y)=0$ для всех
$Y\in\CX(\MM)$ следует $X=0$. Легко проверить, что данное инвариантное
определение невырожденности 2-формы эквивалентно условию $\det\om_{\al\bt}\ne0$,
которое должно выполняться во всех картах и для всех точек $x\in\MM$.

По-определению, компоненты симплектической формы антисимметричны относительно
перестановки индексов. Это значит, что ее невырожденность возможна только на
многообразиях четной размерности. Поэтому размерность многообразия включена в
определение.

В координатах замкнутость симплектической формы $\om$ записывается в виде
дифференциального уравнения
\begin{equation}                                                  \label{eclsif}
  \pl_\al\om_{\bt\g}+\pl_\bt\om_{\g\al}+\pl_\g\om_{\al\bt}=0.
\end{equation}

Рассмотрим связь симплектических форм с ориентацией многообразий и формами
объема.
\begin{theorem}
Пусть $\om$ -- замкнутая 2-форма на многообразии $\MM$, $\dim\MM=2n$. Для того,
чтобы эта форма была симплектической необходимо и достаточно, чтобы $2n$-форма
$\om^n$ была формой объема на $\MM$, т.е.\ нигде не обращалась в нуль.
\end{theorem}
\begin{proof}
Следствие предложения \ref{pseome}.
\end{proof}
\begin{cor}
Любое симплектическое многообразие $(\MM,\om)$ ориентируемо.
\qed\end{cor}
\begin{proof}
Отличие от нуля формы $\om^n$ согласно теореме \ref{torivo} достаточно для
ориентируемости.
\end{proof}

Обычно симплектическое многообразие $(\MM,\om)$ ориентируют формой объема
$\upsilon$ с дополнительным множителем:
\begin{equation}                                                  \label{evosim}
  \upsilon:=\frac1{n!}(-1)^{\frac{n(n+1)}2}\om^n,
\end{equation}
чтобы согласовать выбор канонических координат (см.\ ниже) с канонической
ориентацией евклидова пространства.
\begin{exa}
Рассмотрим каноническую 2-форму (\ref{esymfo}) на евклидовом пространстве
$\MR^{2n}$
\begin{equation}                                                  \label{esicac}
  \varpi=\frac12dx^\al\wedge dx^\bt\varpi_{\al\bt}
  =dx^{n+1}\wedge dx^1+dx^{n+2}\wedge dx^2+\dotsc+dx^{2n}\wedge dx^n.
\end{equation}
Она невырождена, т.к.\ $\det\varpi_{\al\bt}=1$, и замкнута, поскольку компоненты
формы постоянны. Тем самым форма $\varpi$ определяет симплектическую форму на
евклидовом пространстве. Поскольку
\begin{equation*}
  \varpi^n=(-1)^{\frac{n(n+1)}2}n!dx^1\wedge\dotsc\wedge dx^{2n},
\end{equation*}
то соответствующая форма объема
\begin{equation}                                                  \label{ecavos}
  \upsilon:=\frac1{n!}(-1)^{\frac{n(n+1)}2}\varpi^n
  =dx^1\wedge\dotsc\wedge dx^{2n}
\end{equation}
является канонической формой объема евклидова пространства (\ref{ecaore}).
\qed\end{exa}

Обратное утверждение о связи между наличием формы объема и симплектической формы
в общем случае неверно. Исключение составляют двумерные многообразия: всякая
ориентируемая поверхность допускает симплектическую структуру. Это просто
полностью антисимметричный тензор второго ранга:
\begin{equation*}
  \om=\upsilon=\frac12dx^\al\wedge dx^\bt\ve_{\al\bt}=dx^1\wedge dx^2\vol.
\end{equation*}
Для ориентируемых многообразий более высокой размерности $\dim\MM=2n$, $n>1$,
симплектическая форма существует не всегда.

Следующая конструкция позволяет строить симплектические структуры на
произвольном кокасательном расслоении $\MT^*(\MM)$, $\dim\MM=n$, которое имеет
размерность $2n$.
\begin{defn}
Пусть $X_r\in\MT_r\big(\MT^*(\MM)\big)$ -- произвольный касательный вектор к
кокасательному расслоению в точке $r\in\MT^*(\MM)$. Определим {\em каноническую
линейную форму} $\Theta$ на $\MT^*(\MM)$, которая называется {\em формой
Лиувилля} или {\em относительным интегральным инвариантом Пуанкаре},
следующим соотношением
\begin{equation}                                                  \label{efunsy}
  \Theta(X_r)=r(\pi_*X_r),
\end{equation}
где $\pi_*$ -- дифференциал проекции кокасательного расслоения,
$\pi:~\MT^*(\MM)\rightarrow\MM$.
\qed\end{defn}
\index{Каноническая линейная форма (canonical linear form)}%
\index{Линейная форма каноническая (canonical linear form)}%
\index{Относительный интегральный инвариант Пуанкаре%
(relative integral Poincar\'e invariant)}%
\index{Интегральный инвариант Пуанкаре относительный%
(relative integral Poincar\'e invariant)}%
\index{Форма Лиувилля (Liouville form)}%
\index{Лиувилля форма (Liouville form)}%
\begin{theorem}
На кокасательном расслоении $\MT^*(\MM)$ к произвольному многообразию $\MM$
существует симплектическая структура.
\end{theorem}
\begin{proof}
Заметим, что $r$ является линейной формой на $\MT_{\pi(r)}(\MM)$ и $\pi_*X_r$
-- касательный вектор в точке $\pi(r)$. Следовательно, форма Лиувилля
(\ref{efunsy}) определена всюду на кокасательном расслоении $\MT^*(\MM)$. Пусть
$q^\al$ -- локальная система координат на $\MU\subset\MM$. Введем на
$\pi^{-1}(\MU)$ координаты $(x^1,\dotsc,x^{2n})=(q^1,\dotsc,q^n,p_1,\dotsc,p_n)$
следующим образом. Пусть $q^\al$, $\al=1,\dotsc,n$ -- координаты на $\MU$. Формы
$dx^1,\dotsc,dx^n$ порождают пространство 1-форм на $\MU$. Поэтому для всякой
1-формы $r\in\pi^{-1}(\MU)$ имеем равенство
\begin{equation*}
  r=dq^\al p_\al.
\end{equation*}
Разложим касательный вектор $X_r$ по координатному базису $(q,p)$:
\begin{equation*}
  X_r=X_r^\al\frac\pl{\pl q^\al}+X_{r\al}\frac\pl{\pl p_\al}.
\end{equation*}
Тогда
\begin{equation*}
  \pi_*X_r=X_r^\al\frac\pl{\pl x^\al},
\end{equation*}
и условие (\ref{efunsy}) принимает вид
\begin{equation*}
  \Theta(X_r)=X_r^\al p_\al.
\end{equation*}
Таким образом получаем выражение для формы Лиувилля в локальных координатах:
\begin{equation}                                                  \label{eliofo}
  \Theta=dx^\al p_\al.
\end{equation}
Внешняя производная от формы Лиувилля имеет вид
\begin{equation*}
  d\Theta=dp_\al\wedge dq^\al=\frac12dx^\Sa\wedge dx^\Sb\varpi_{\Sa\Sb},
  \qquad \Sa,\Sb=1,\dotsc,2n,
\end{equation*}
т.е.\ совпадает с канонической симплектической формой $\varpi$. Таким образом
построена невырожденная замкнутая 2-форма $d\Theta$ на произвольном
кокасательном расслоении.
\end{proof}
Форма Лиувилля естественна в том смысле, что для любого сечения
$\s:~\MM\rightarrow\MT^*(\MM)$ возврат отображения $\s^*$ переводит 1-форму
$\Theta$ в $r$: $\s^*(\Theta)=r$.
\begin{cor}
Любое кокасательное расслоение $\MT^*(\MM)$ ориентируемо.
\qed\end{cor}
\begin{com}
В гамильтоновом подходе к описанию динамики точечных частиц (см.\ главу
\ref{scanfp}) многообразие $\MM$ и его кокасательное расслоение $\MT^*(\MM)$
отождествляются, соответственно, с конфигурационным и фазовым пространствами.
\qed\end{com}

Поскольку ранг симплектической 2-формы максимален и равен $2n$, то ее класс
также равен $2n$. В этом случае теорема Дарбу (\ref{tdartw}) формулируется
следующим образом.
\begin{theorem}[\bf Дарбу]
Пусть задано симплектическое многообразие $(\MM,\om)$. Тогда у каждой точки
$x\in\MM$ существует такая координатная окрестность $\MU_x\subset\MM$, в которой
симплектическая форма принимает канонический вид
\begin{equation*}
  \om=\varpi.
\end{equation*}
\end{theorem}
\begin{defn}
Координаты, в которых симплектическая форма имеет канонический вид, называются
{\em координатами Дарбу}.
\qed\end{defn}
\index{Координаты Дарбу (Darboux coordinates)}%
\index{Дарбу координаты (Darboux coordinates)}%
\begin{com}
В главе \ref{smetri} мы видели, что метрику $g$ на многообразии $\MM$ за счет
выбора системы координат можно привести к каноническому виду только в
фиксированной точке $x\in\MM$. Лучшее, что можно сделать в общем случае, это
привести метрику к каноническому виду вдоль произвольной кривой $\g\in\MM$. В
этом отношении симплектические многообразия проще. Согласно теореме Дарбу
симплектическую форму можно привести к каноническому виду не только в данной
точке многообразия, но и в некоторой окрестности этой точки.
\qed\end{com}

Поскольку симплектическая структура $\om_{\al\bt}(x)$ является невырожденной,
то в каждой точке $x\in\MM$ существует обратная матрица $\om^{-1\,\al\bt}(x)$,
которая также антисимметрична:
\begin{equation*}
  \om^{-1\,\al\bt}\om_{\bt\g}=\om_{\g\bt}\om^{-1\,\bt\al}=\dl^\al_\g,\qquad
  \om^{-1\,\al\bt}=-\om^{-1\,\bt\al}.
\end{equation*}
\begin{com}
Если на симплектическом многообразии $(\MM,\om)$ задана также метрика $g$, то в
общем случае
\begin{equation*}                                                    \tag*{\qed}
  \om^{-1\,\al\bt}\ne\om^{\al\bt}:=g^{\al\g}g^{\bt\dl}\om_{\g\dl}.
\end{equation*}
\end{com}
\begin{exa}
Для канонической симплектической структуры (\ref{esymfo})
\begin{equation*}                                                    \tag*{\qed}
  \varpi^{-1}=(\varpi^{-1\al\bt})=-\varpi=\varpi^\St=
  \begin{pmatrix} 0 & \one \\-\one & 0 \end{pmatrix}.
\end{equation*}
\renewcommand{\qed}{}\end{exa}
Симплектическая структура устанавливает взаимно однозначное соответствие сечений
касательных и кокасательных расслоений. В компонентах данное соответствие
задается простым правилом:
\begin{equation}                                                  \label{efovec}
  A_\al=X^\bt\om_{\bt\al},\quad X^\al=A_\bt\om^{-1\,\bt\al},\qquad
  X\in\CX(\MM),\quad A\in\Lm_1(\MM).
\end{equation}

Замкнутость симплектической формы (\ref{eclsif}) для обратного тензора
$\om^{-1\,\al\bt}(x)$ можно переписать в виде
\begin{equation*}
  \om^{-1\,\al\dl}\pl_\dl\om^{-1\,\bt\g}+\om^{-1\,\bt\dl}\pl_\dl\om^{-1\,\g\al}
  +\om^{-1\,\g\dl}\pl_\dl\om^{-1\,\al\bt}=0.
\end{equation*}
Для этого достаточно свернуть равенство (\ref{eclsif}) с обратными матрицами
$\om^{-1\,\al\bt}$.

Используя внутреннее умножение (\ref{einpro}), формулы (\ref{efovec}) можно
переписать в инвариантном виде:
\begin{equation*}
  A=\inm_{X_A}\om=X_A\rfloor\om,
\end{equation*}
где $X_A$ -- векторное поле, соответствующее 1-форме $A$.

Поскольку симплектическая структура на $\MM$ устанавливает взаимно однозначное
соответствие между множеством векторных полей $\CX(\MM)$ и пространством 1-форм
$\Lm_1(\MM)$, то структуру алгебры Ли на $\CX(\MM)$ можно перенести на
$\Lm_1(\MM)$. Для двух произвольных 1-форм $A,B\in\Lm_1(\MM)$ положим
\begin{equation}                                                  \label{eliefo}
  [A,B]:=[X_A,X_B]\rfloor\om.
\end{equation}
Тем самым линейное пространство $\Lm_1(\MM)$ приобретает структуру алгебры Ли.
Таким образом коммутатору векторных полей ставится в соответствие {\em скобка
Пуассона} 1-форм (\ref{eliefo}).
\index{Скобка Пуассона (Poisson bracket)}%
\index{Пуассона скобка (Poisson bracket)}%

В компонентах $A=dx^\al A_\al$, $B=dx^\al B_\bt$ и
\begin{equation}                                                  \label{epocof}
  [A,B]=dx^\al\left[(A_\bt\pl_\g B_\al+B_\g\pl_\bt A_\al)\om^{-1\bt\g}
  +A_\bt B_\g\pl_\al\om^{-1\bt\g}\right].
\end{equation}
Эта формула получается прямыми вычислениями.

Скобку Пуассона двух 1-форм можно выразить через производную Ли:
\begin{equation*}
\begin{split}
  \Lie_{X_A}B&=\Lie_{X_A}(X_B\rfloor\om)
  =\Lie_{X_A}X_B\rfloor\om+X_B\rfloor(\Lie_{X_A}\om)=
\\
  &=[X_A,X_B]\rfloor\om+X_B\rfloor\left(X_A\rfloor d\om+d(X_A\rfloor\om)\right)
  =[A,B]+X_B\rfloor dA,
\end{split}
\end{equation*}
где мы воспользовались линейностью производной Ли, основной формулой гомотопии
(\ref{eldefo}) и замкнутостью симплектической формы. Отсюда следует выражение
для скобки Пуассона через производную Ли:
\begin{equation}                                                  \label{elipoe}
  [A,B]=L_{X_A}B-X_B\rfloor dA.
\end{equation}

Если 1-формы $A$ и $B$ замкнуты, то полученное выражение упрощается:
\begin{equation*}
  [A,B]=L_{X_A}B=-\Lie_{X_B}A.
\end{equation*}
\begin{prop}
Скобка Пуассона двух замкнутых форм является точной формой.
\end{prop}
\begin{proof}
Из основной формулы гомотопии (\ref{eldefo}) следует, что для любой замкнутой
формы $A$ справедливо равенство $\Lie_X A=d(X\rfloor A)$.
\end{proof}

В компонентах скобка Пуассона двух замкнутых 1-форм (\ref{epocof}) принимает
простой вид
\begin{equation*}
  [A,B]=dx^\al\pl_\al(\om^{-1\bt\g}A_\bt B_\g).
\end{equation*}

Скобку Пуассона можно также определить в алгебре функций на $\MM$. А именно,
каждая симплектическая структура определяет скобку Пуассона двух
дифференцируемых функций $f(x)$ и $g(x)$ класса $\CC^2$:
\begin{equation}                                                  \label{eposis}
  [f,g]:=\om^{-1\,\al\bt}\frac{\pl f}{\pl x^\al}\frac{\pl g}{\pl x^\bt}.
\end{equation}
Дважды непрерывная дифференцируемость функций необходима для выполнения тождеств
Якоби.

Поля $\om^{-1\,\al\bt}(x)$ являются компонентами антисимметричного
контравариантного тензора второго ранга. Они являются частным случаем
пуассоновой структуры, рассмотренной в следующем разделе.

Пусть на симплектическом многообразии $(\MM,\om)$ задана также аффинная
геометрия, т.е.\ метрика $g$ и аффинная связность $\Gamma$. Назовем аффинную
связность {\em согласованной с симплектической структурой}, если ковариантная
производная симплектической формы равна нулю:
\begin{equation}                                                  \label{eacfom}
  \nb_\al\om_{\bt\g}=\pl_\al\om_{\bt\g}-\Gamma_{\al\bt}{}^\dl\om_{\dl\g}
  -\Gamma_{\al\g}{}^\dl\om_{\bt\dl}=0.
\end{equation}
\index{Аффинная связность, согласованная с симплектической структурой%
(affine connection compatible with symplectic structure)}%
Рассмотрим линейную комбинацию ковариантных производных, в которой слагаемые
отличаются циклической перестановкой индексов
\begin{equation*}
  \nb_\al\om_{\bt\g}+\nb_\bt\om_{\g\al}+\nb_\g\om_{\al\bt}.
\end{equation*}
Слагаемые с производными от симплектической формы сокращаются ввиду замкнутости
$\om$, и мы получаем уравнение на тензор кручения
\begin{equation}                                                  \label{etosig}
  T_{\al\bt}{}^\dl\om_{\dl\g}+T_{\bt\g}{}^\dl\om_{\dl\al}
  +T_{\g\al}{}^\dl\om_{\dl\bt}=0.
\end{equation}
Это необходимое условие для того, чтобы аффинная связность была согласована с
симплектической структурой, но не достаточное.
\section{Пуассоновы многообразия                                 \label{spoist}}
Рассмотрим множество скалярных полей $f,g,\dotsc\in\CC^\infty(\MM)$
(алгебру гладких функций) на многообразии $\MM$ произвольной размерности $n$.
\begin{defn}
Билинейное отображение
\begin{equation*}
  J:\quad \CC^\infty(\MM)\times\CC^\infty(\MM)\ni\quad f,g\mapsto
  [f,g]\quad\in\CC^\infty(\MM)
\end{equation*}
называется {\em пуассоновой структурой} или {\em скобкой Пуассона}, если
выполнены следующие четыре условия:
  \begin{align*}
  1)\quad &[af+bg,h]=a[f,h]+b[g,h],
  \quad a,b\in\MR & &\text{-- линейность},\\
  2)\quad &[f,g]=-[g,f]&
  &\text{-- антисимметрия},\\
  3)\quad &[f,gh]=[f,g]h+g[f,h] & &
  \text{-- правило Лейбница},\\
  4)\quad &[f,[g,h]]+[g,[h,f]]+[h,[f,g]]=0 &&\text{-- тождество Якоби}.
  \end{align*}
Многообразие с заданной скобкой Пуассона называется {\em пуассоновым}.
\qed
\end{defn}
\index{Пуассонова структура (Poisson structure)}%
\index{Структура пуассонова (Poisson structure)}%
\index{Скобка Пуассона (Poisson bracket)}%
\index{Пуассона скобка (Poisson bracket)}%
\index{Пуассоново многообразие (Poisson manifold)}%
\index{Многообразие пуассоново (Poisson manifold)}%

Из линейности скобки Пуассона и правила Лейбница следует, что скобка Пуассона
постоянной функции $f=C=\const$ с произвольной функцией равна нулю:
$$
  [C,g]=0.
$$
Используя это свойство, получим явное выражение для скобки Пуассона в локальной
системе координат $x^\al$. С этой целью разложим функции $f$ и $g$ в ряды
Тейлора в окрестности произвольной точки $x_0$:
\begin{align*}
  f(x)&=f_0+(x^\al-x^\al_0)\left.\frac{\pl f}{\pl x^\al}\right|_{x=x_0}+\dotsc,
\\
  g(x)&=g_0+(x^\al-x^\al_0)\left.\frac{\pl g}{\pl x^\al}\right|_{x=x_0}+\dotsc.
\end{align*}
Тогда
\begin{align*}
  [f,g]&=\left.[x^\al-x^\al_0,x^\bt-x^\bt_0]
  \frac{\pl f}{\pl x^\al}\frac{\pl g}{\pl x^\bt}\right|_{x=x_0}+\osmall(x-x_0)=
\\
  &=\left.[x^\al,x^\bt]\frac{\pl f}{\pl x^\al}\frac{\pl g}{\pl x^\bt}
  \right|_{x=x_0}+\osmall(x-x_0).
\end{align*}
Таким образом, в точке $x_0$ скобка определяется первым слагаемым. Поскольку
точка $x_0$ произвольна, то получаем явное выражение для скобки Пуассона в
координатах
\begin{equation}                                                  \label{epobra}
  [f,g]=J^{\al\bt}\frac{\pl f}{\pl x^\al}\frac{\pl g}{\pl x^\bt},
\end{equation}
где введены {\em структурные функции} пуассонова многообразия:
\begin{equation}                                                  \label{epocoo}
  J^{\al\bt}:=[x^\al,x^\bt]
\end{equation}
\index{Структурные функции (structure functions)}%
\index{Функции структурные (structure functions)}%
Напомним, что каждую координату можно рассматривать, как функцию на многообразии
$x^\al:~\MM\ni x\mapsto x^\al(x)\in\MR$, поэтому для координатных функций скобка
Пуассона (\ref{epocoo}) определена.

Скобка Пуассона (\ref{epobra}) совпадает со скобкой Пуассона (\ref{eposis}),
введенной для симплектической структуры, при $J^{\al\bt}=\om^{-1\al\bt}$.

По-построению структурные функции $J^{\al\bt}(x)$ представляют собой компоненты
антисимметричного тензора второго ранга, и выражение в правой части равенства
(\ref{epobra}) является функцией. Из определения скобки Пуассона следует, что
структурные функции антисимметричны,
$$
  J^{\al\bt}=-J^{\bt\al},
$$
и удовлетворяют тождеству Якоби:
\begin{equation}                                                  \label{estjai}
  J^{\al\dl}\pl_\dl J^{\bt\g}+J^{\bt\dl}\pl_\dl J^{\g\al}
  +J^{\g\dl}\pl_\dl J^{\al\bt}=0,
\end{equation}
где слагаемые отличаются циклической перестановкой индексов $\al,\bt,\g$.
Легко проверяется, что произвольный антисимметричный тензор второго
ранга, удовлетворяющий тождеству Якоби, взаимно однозначно определяет
скобку Пуассона (\ref{epobra}).
\begin{exa}
Произвольная постоянная антисимметричная матрица определяет пуассонову структуру
в некоторой системе координат. Действительно, тождества Якоби в этом случае
выполняются.
\qed\end{exa}

Пуассонова структура является билинейным дифференциальным оператором и
записывается также в виде
\begin{equation*}
  J(f,g):=[f,g]=J^{\al\bt}\pl_\al f\pl_\bt g.
\end{equation*}
Поэтому для нее применяют иногда обозначение
\begin{equation*}
  J=\frac12J^{\al\bt}\pl_\al\wedge\pl_\bt,
\end{equation*}
где знак $\wedge$ обозначает внешнее умножение (см.\ раздел \ref{sdifin}).

Если на многообразии помимо пуассоновой структуры $J$ задана также аффинная
связность $\Gamma$, то тождество Якоби (\ref{estjai}) можно переписать в явно
ковариантном виде
\begin{align*}
  J^{\al\dl}\nb_\dl J^{\bt\g}&+J^{\bt\dl}\nb_\dl J^{\g\al}
  +J^{\g\dl}\nb_\dl J^{\al\bt}-
\\
  &-J^{\al\dl}T_{\dl\e}{}^\bt J^{\e\g}-J^{\bt\dl}T_{\dl\e}{}^\g J^{\e\al}
  -J^{\g\dl}T_{\dl\e}{}^\al J^{\e\bt}=0,
\end{align*}
где $T_{\al\bt}{}^\g$ -- тензор кручения аффинной связности. Это доказывает, что
тождества Якоби ковариантны, и их выполнение в одной системе координат влечет за
собой их выполнение во всех других системах. Впрочем, это можно было бы и не
доказывать, т.к.\ определение пуассоновой структуры было дано в инвариантном
виде.

Если матрица структурных функций невырождена, $\det J^{\al\bt}\ne0$, то она
имеет обратную $J^{-1}_{\al\bt}$:
$$
  J^{\al\bt} J^{-1}_{\bt\g}=\dl^\al_\g,
$$
которая также антисимметрична. Это значит, что контравариантный тензор
$J^{\al\bt}$ взаимно однозначно определяет 2-форму
$\frac12dx^\al\wedge dx^\bt J^{-1}_{\al\bt}$ на многообразии. Из тождеств Якоби
(\ref{estjai}) следует, что 2-форма $J^{-1}_{\al\bt}$ замкнута,
$$
  \pl_\al J^{-1}_{\bt\g}+\pl_\bt J^{-1}_{\g\al}+\pl_\g J^{-1}_{\al\bt}=0,
$$
т.е.\ является симплектической (см.\ раздел \ref{ssymma}). Легко проверяется и
обратное утверждение: произвольная симплектическая форма
$\om_{\al\bt}=J^{-1}_{\al\bt}$ определяет пуассонову
структуру на многообразии с невырожденными структурными функциями. При этом
тождества Якоби следуют из условия замкнутости формы. Таким образом пуассоновы
структуры с невырожденными структурными функциями находятся во взаимно
однозначном соответствии с симплектическими формами на многообразии.

В общем случае структурные функции пуассоновой структуры могут быть вырождены.
Это значит, в частности, что, в отличие от симплектической структуры, скобка
Пуассона может быть также определена на многообразии нечетной размерности.
{\em Рангом} пуассоновой структуры называется ранг матрицы $J^{\al\bt}$
\index{Ранг пуассоновой структуры (rank of a Poisson structure)}%
структурных функций. Ввиду антисимметрии $J^{\al\bt}$ ранг пуассоновой структуры
может быть только четным. В общем случае ранг пуассоновой структуры может
меняться от точки к точке.
\begin{defn}
Если пуассонова структура вырождена, $\det J^{\al\bt}=0$, то существуют
{\em функции Казимира} $c\in\CC^\infty(\MU)$, определенные, возможно, только в
некоторой области $\MU\subset\MM$. Они определяются следующим равенством
\begin{equation}                                                  \label{eqcafy}
  [c,f]=0,\quad \forall f\in\CC^\infty(\MU). \qed
\end{equation}
\end{defn}
\index{Функция Казимира (Casimir function)}%
\index{Казимира функция (Casimir function)}%
В координатах это равенство принимает вид
\begin{equation*}
  J^{\al\bt}\pl_\bt c=0.
\end{equation*}
Очевидно, что любое решение этих уравнений определено с точностью до аддитивной
постоянной. Если пуассонова структура невырождена, то константы $c=\const$
являются единственными функциями Казимира.

Пусть ранг матрицы $J^{\al\bt}$ постоянен на $\MM$ и равен
$\rank J^{\al\bt}=2m<n$. Количество функционально независимых функций Казимира
$c^\Sa$, $\Sa=1,\dotsc,n-2m$, равно числу функционально независимых решений
уравнения (\ref{eqcafy}), т.е.\ разности между размерностью многообразия $n$ и
рангом пуассоновой структуры $2m$. Если все функции Казимира известны и
определены на всем $\MM$, то условия $c^\Sa=\const$ выделяет в $\MM$
подмногообразие $\MU$ размерности $2m$. В разделе \ref{spoima} мы увидим, что
сужение $J$ на это подмногообразие приводит к невырожденной пуассоновой
структуре на $\MU$.

Из определения функций Казимира следует, в частности, что скобка Пуассона
двух функций Казимира равна нулю:
\begin{equation}                                                  \label{epocaf}
  [c^\Sa,c^\Sb]=J^{\al\bt}\pl_\al c^\Sa\pl_\bt c^\Sb=0.
\end{equation}

Пусть задано пуассоново многообразие $(\MM,J)$. Наличие пуассоновой структуры
задает билинейное отображение на алгебре функций $\CC^\infty(\MM)$, которое
превращает множество функций в алгебру Ли. Эта алгебра Ли гомоморфно
отображается в алгебру Ли векторных полей на $\MM$ следующим образом. Скобка
Пуассона координатных функций $x^\al$ с произвольной функцией $f$ определяет
компоненты векторного поля $X_f=X_f^\al\pl_\al$, где
$$
  X^\al_f:=[f,x^\al]=-J^{\al\bt}\pl_\bt f.
$$
Таким образом, мы имеем отображение множества гладких функций в алгебру Ли
векторных полей
\begin{equation}                                                  \label{emafve}
  \CC^\infty(\MM)\ni\quad f\mapsto X_f\quad\in\CX(\MM).
\end{equation}
\begin{prop}                                                      \label{poisth}
Скобка Пуассона двух функций $f,g\in\CC^\infty(\MM)$ связана с коммутатором
соответствующих векторных полей соотношением
\begin{equation}                                                  \label{ehopov}
  [X_f,X_g]=X_{[f,g]}.
\end{equation}
То есть отображение (\ref{emafve}) является гомоморфизмом.
\end{prop}
\begin{proof}
Простое следствие тождеств Якоби.
\end{proof}
Если пуассонова структура невырождена, то гомоморфизм (\ref{emafve}) имеет
нетривиальное ядро, состоящее из функций, постоянных на $\MM$. Если пуассонова
структура вырождена, то ядро отображения (\ref{ehopov}) включает также линейную
оболочку функций Казимира.
\begin{prop}                                                      \label{plipob}
Векторное поле $X_f$, порожденное произвольной функцией $f$, сохраняет скобку
Пуассона. А именно, производная Ли вдоль векторного поля $X_f$ от структурных
функций равна нулю:
$$
  \Lie_{X_f}J=0.
$$
\end{prop}
\begin{proof}
Простая проверка.
\end{proof}

Обсудим связь пуассоновых структур с гамильтоновой динамикой точечных частиц.
Рассмотрим произвольную функцию $H\in\CC^\infty(\MM)$ и соответствующее
векторное поле $X_H$. Тогда экспоненциальное отображение, определяемое системой
уравнений
$$
  \dot x^\al=-X^\al_H=[x^\al,H]=J^{\al\bt}\pl_\bt H,
$$
является ни чем иным, как уравнениями движения механической системы точечных
частиц в канонической (гамильтоновой) форме. При этом точка обозначает
дифференцирование по времени, и $H=H(x)$ -- гамильтониан системы. Векторное поле
\begin{equation}                                                  \label{ehavef}
  X_H=-J^{\al\bt}\pl_\bt H\pl_\al
\end{equation}
называется {\em гамильтоновым}.
\index{Гамильтоново векторное поле (Hamiltonian vector field)}%
\index{Векторное поле гамильтоново (Hamiltonian vector field)}%
Другими словами, каждую функцию на пуассоновом многообразии можно рассматривать,
как гамильтониан некоторой механической системы. При этом каждому гамильтониану
соответствует свое гамильтоново векторное поле.

Легко проверить, что для каждого гамильтонова векторного поля справедливо
равенство
\begin{equation}                                                  \label{ehacoe}
  X_H f=[f,H].
\end{equation}
\begin{exa}
Механическая система, состоящая из $\Sn$ точечных частиц, в фазовом пространстве
описывается координатами $q^i$ и импульсами $p_i$, $i=1,\cdots,\Sn$, со скобками
Пуассона
$$
  [q^i,p_j]=\dl^i_j,\qquad [q^i,q^j]=0,\qquad [p_i,p_j]=0.
$$
Фазовое пространство представляет собой $2\Sn$-мерное пуассоново многообразие.
В координатах
$\lbrace x^\al\rbrace=\lbrace q^1,\dotsc,q^\Sn,$$p_1,\dotsc,p_\Sn\rbrace$
структурные функции имеют вид
\begin{equation}                                                  \label{ecapos}
  J^{\al\bt}=\varpi^{-1\al\bt}=\begin{pmatrix}0&\one\\-\one&0\end{pmatrix}.
\end{equation}
Эта пуассонова структура невырождена и называется {\em канонической}.
\index{Каноническая пуассонова структура (canonical Poisson structure)}%
\index{Пуассонова структура каноническая (canonical Poisson structure)}%
Как матрица каноническая пуассонова структура с точностью до знака совпадает с
канонической симплектической формой (\ref{esymfo}), однако ее индексы
расположены сверху, что соответствует контравариантному тензору второго ранга.

Координаты $q^i$ и сопряженные импульсы $p_i$ являются системой локальных
координат фазового пространства $\MT^*(\MM)$, которое является кокасательным
расслоением к конфигурационному пространству $\MM$ с координатами $q^i$.

Канонические уравнения движения механической системы
\begin{equation*}
  \dot q^i=\frac{\pl H}{\pl p_i},\qquad \dot p_i=-\frac{\pl H}{\pl q^i}
\end{equation*}
записываются в виде
$$
  \dot x^\al=[x^\al,H],
$$
где $H=H(q,p)$ -- гамильтониан системы.
Гамильтоново векторное поле на фазовом пространстве $\MT^*(\MM)$ имеет вид
\begin{equation*}                                                    \tag*{\qed}
  X_H=\frac{\pl H}{\pl p_i}\frac\pl{\pl q^i}
  -\frac{\pl H}{\pl q^i}\frac\pl{\pl p_i}.
\end{equation*}
\end{exa}
\begin{com}
С точки зрения динамики функции Казимира представляют собой первые интегралы
движения механической системы, которые существуют для широкого класса функций
Гамильтона. В этом смысле они носят кинематический характер.
\qed\end{com}
\section{Структура Ли--Пуассона}
Теперь рассмотрим важный класс пуассоновых структур, связанных с алгебрами Ли.
\begin{defn}
Пусть $\Gg$ -- алгебра Ли размерности $\Sn$. Это -- линейное пространство,
каждая точка которого $y=y^\Sa L_\Sa\in\Gg$, $\Sa=1,\dotsc,\Sn$, разлагается по
базису $L_\Sa$ с коммутационными соотношениями
\begin{equation*}
  [L_\Sa,L_\Sb]=f_{\Sa\Sb}{}^\Sc L_\Sc,
\end{equation*}
где $f_{\Sa\Sb}{}^\Sc$ -- структурные константы алгебры Ли $\Gg$. Обозначим
дуальное пространство к алгебре Ли через $\Gg^*$ (множество линейных
функционалов на $\Gg$). Пусть $\om^\Sa$ -- базис в $\Gg^*$, который дуален к
$L_\Sa$, т.е.\ $\om^\Sa(L_\Sb)=\dl^\Sa_\Sb$. Тогда точка дуального пространства
представима в виде $x=\om^\Sa x_\Sa\in\Gg^*$. Значение функционала $x$ на
векторе $y$ равно
\begin{equation*}
  x(y)=y^\Sa x_\Sa.
\end{equation*}
Рассмотрим функцию $f(x)\in\CC^\infty(\Gg^*)$ на дуальном пространстве $\Gg^*$.
Ее дифференциал равен
\begin{equation*}
  df=dx_\Sa\pl^\Sa f.
\end{equation*}
Мы пишем индекс у частной производной $\pl^\Sa$ вверху, т.к.\ координаты
дуального пространства $\Gg^*$ нумеруются нижним индексом. Каждый дифференциал
естественным образом отождествляется с элементом алгебры Ли
$\nb f:=\pl^\Sa fL_\Sa\in\Gg$. Определим скобку Пуассона для функций на дуальном
пространстве $f,g\in\CC^\infty(\Gg^*)$ в произвольной точке $x\in\Gg^*$:
\begin{equation}                                                  \label{epolie}
  [f,g](x):=x\big([\nb f,\nb g]\big),
\end{equation}
где $[\nb f,\nb g]$ -- коммутатор в алгебре Ли $\Gg$. В компонентах данная
скобка Пуассона записывается в виде
\begin{equation}                                                  \label{elipox}
  [f,g]=J_{\Sa\Sb}\pl^\Sa f\pl^\Sb g,
\end{equation}
где структурные функции
\begin{equation}                                                  \label{elipsr}
  J_{\Sa\Sb}=f_{\Sa\Sb}{}^\Sc x_\Sc
\end{equation}
линейны по координатам $x_\Sc$. Нетрудно проверить, что тождества Якоби для
структурных функций (\ref{elipsr}) совпадают с тождествами Якоби для структурных
констант алгебры Ли $\Gg$:
\begin{equation*}
  f_{\Sa\Sb}{}^\Sd f_{\Sc\Sd}{}^\Se+f_{\Sb\Sc}{}^\Sd f_{\Sa\Sd}{}^\Se
  +f_{\Sc\Sa}{}^\Sd f_{\Sb\Sd}{}^\Se=0.
\end{equation*}
Скобка Пуассона (\ref{epolie}) называется скобкой {\em Ли--Пуассона}.
\qed\end{defn}
\index{Скобка Ли--Пуассона (Lie--Poisson bracket)}%
\index{Ли--Пуассона скобка (Lie--Poisson bracket)}%
Таким образом, скобка Ли--Пуассона является частным случаем скобки Пуассона с
линейными структурными функциями (\ref{elipsr}). Ранг этой скобки в нуле всегда
равен нулю, $J_{\Sa\Sb}|_{x=0}=0$.

\begin{exa}{\bf (Вращение твердого тела.)}                        \label{elipos}
В трехмерном евклидовом пространстве с декартовыми координатами $x_i\in\MR^3$,
$i=1,2,3$, которое мы отождествим с дуальной алгеброй Ли трехмерных вращений
$\Gs\Go(3)^*$, скобка Ли--Пуассона равна
\begin{equation}                                                  \label{expoth}
  [x_i,x_j]=-\ve_{ijk}x^k,
\end{equation}
где $\ve_{ijk}$ -- полностью антисимметричный тензор третьего ранга, а подъем и
опускание индексов производится с помощью евклидовой метрики $\dl_{ij}$.
Эта структура вырождена, ее ранг равен двум всюду, кроме начала координат, где
он равен нулю. Для пуассоновой структуры (\ref{expoth}) существует единственная
функция Казимира
$$
  c=x^i x_i.
$$
Скобку Ли--Пуассона (\ref{expoth}) можно сузить на сферу произвольного
радиуса, соответствующую постоянному значению функции Казимира $c=\const$.
Соответствующая пуассонова структура на сфере $x^i x_i=\const>0$ невырождена. В
качестве  координат Дарбу, которые будут определены в следующем разделе, на
сфере можно выбрать ось $z=x^3$ и полярный угол $\vf=\arctg(x^2/x^1)$. Используя
определение (\ref{expoth}), нетрудно проверить, что
\begin{equation*}
  [z,\vf]=1,\qquad [\vf,\vf]=0,\qquad [z,z]=0.
\end{equation*}
Таким образом, симплектическими слоями, соответствующими постоянной функции
Казимира, являются сферы, а координатами Дарбу -- цилиндрические координаты.

Скобке Ли--Пуассона (\ref{expoth}) соответствует хорошо известный пример из
механики твердого тела. Рассмотрим вращающееся твердое тело с покоящимся центром
масс в декартовой системе координат, оси которой направлены по главным осям
инерции. Гамильтониан системы в этом случае имеет вид
\begin{equation*}
  H(x)=\frac{x_1^2}{2I_1}+\frac{x_2^2}{2I_2}+\frac{x_3^2}{2I_3},
\end{equation*}
где $x_i$, $i=1,2,3$ -- моменты количества движения и $I_{1,2,3}$ -- моменты
инерции твердого тела. Если для координат $x_i$ определить скобку Пуассона
(\ref{expoth}), то гамильтоновы уравнения движения примут вид
\begin{align*}
  \dot x_1&=\frac{I_2-I_3}{I_2I_3}x_2x_3,
\\
  \dot x_2&=\frac{I_3-I_1}{I_1I_3}x_1x_3,
\\
  \dot x_3&=\frac{I_1-I_2}{I_1I_2}x_1x_2.
\end{align*}
Это есть {\em уравнения Эйлера} вращения твердого тела (см., например,
\cite{LanLif88RA}).
\qed\end{exa}
\index{Уравнения Эйлера вращения твердого тела%
(Euler's equations of rigid body rotations)}%
\index{Эйлера уравнения вращения твердого тела%
(Euler's equations of rigid body rotations)}%
\section{Отображения пуассоновых многообразий                    \label{spoima}}
\begin{defn}
Пусть $\vf:~\MM\rightarrow\MN$ -- гладкое отображение двух пуассоновых
многообразий. Это отображение называется {\em пуассоновым}, если оно сохраняет
скобку Пуассона:
\begin{equation}                                                  \label{epusum}
  [f\circ\vf,g\circ\vf]_\MM=[f,g]_\MN\circ\vf,
\end{equation}
где $f$ и $g$ -- произвольные функции на многообразии $\MN$ и, следовательно,
$f\circ\vf$ и $g\circ\vf$ -- функции на $\MM$.
Соответственно, подмногообразие $\MM\subset\MN$ называется {\em пуассоновым},
если вложение $\MM\hookrightarrow\MN$ является пуассоновым отображением.
В этом случае мы говорим, что пуассонова структура на $\MN$ {\em сужена}
до пуассоновой структуры на подмногообразии $\MM\subset\MN$.

С другой стороны, если пуассонова структура задана на многообразии $\MM$, то ее
всегда можно отобразить на образ $\vf(\MM)\subseteq\MN$ с помощью дифференциала
отображения. Эту пуассонову структуру на образе $\vf(\MM)\subset\MN$ будем
называть {\em индуцированной}.

В классической механике пуассоново отображение фазового пространства на себя
называется {\em каноническим преобразованием}.
\qed\end{defn}
\index{Пуассоново отображение (Poisson map)}%
\index{Отображение пуассоново (Poisson map)}%
\index{Пуассоново подмногообразие (Poisson submanifold)}%
\index{Подмногообразие пуассоново (Poisson submanifold)}%
\index{Ограничение пуассоновой структуры (restriction of a Poisson structure)}%
\index{Индуцированная пуассонова структура (induced Poisson structure)}%
\index{Пуассонова структура индуцированная (induced Poisson structure)}%
\index{Каноническое преобразование (canonical transformation)}%
\index{Преобразование каноническое (canonical transformation)}%

Выпишем структурные функции для индуцированной пуассоновой структуры.
Пусть отображение $\vf:~\MM\rightarrow\MN$ в координатах задается функциями
$y^\Sa(x)$, где $y^\Sa$, $\Sa=1,\dotsc,\dim\MN$ и $x^\al$,
$\al=1,\dotsc,\dim\MM$, -- координаты соответственно на $\MN$ и $\MM$. Тогда
индуцированная пуассонова структура на образе $\vf(\MM)\subseteq\MN$ всегда
определена и имеет следующие структурные функции:
\begin{equation*}
  J^{\Sa\Sb}=J^{\al\bt}\pl_\al y^\Sa\pl_\bt y^\Sb.
\end{equation*}
Из свойств умножения матриц следует, что если отображение $\vf$ -- вложение, то
ранг индуцированной пуассоновой структуры $J^{\Sa\Sb}$ равен рангу пуассоновой
структуры на подмногообразии $\MM$.

Рассмотренный ниже пример показывает, что пуассонову структуру на $\MN$ в общем
случае нельзя сузить на произвольное подмногообразие $\MM\subset\MN$.
\begin{exa}
Приведем простой пример непуассонова вложения, которое, на первый взгляд, должно
быть таковым. Пусть многообразие $\MN$ является прямым произведением двух
многообразий, $\MN:=\MM\times\MK$. Обозначим координаты и пуассонову структуру
на $\MM$ через $x^\al$, $\al=1,\dotsc,\dim\MM$, и $J^{\al\bt}$. Пусть на $\MK$
также заданы координаты $y^\mu$, $\mu=1,\dotsc,\dim\MK$, и пуассонова структура
$J^{\mu\nu}$. Тогда матрица
\begin{equation*}
  J=\begin{pmatrix} J^{\al\bt} & 0 \\ 0 & J^{\mu\nu} \end{pmatrix}
\end{equation*}
определяет пуассонову структуру на $\MN$ в координатах $\lbrace x,y\rbrace$.
Это пуассонова структура индуцирована двумя естественными вложениями
$\MM\hookrightarrow\MM\times\MK$ и $\MK\hookrightarrow\MM\times\MK$.
Рассмотрим вложение
\begin{equation*}
  \vf:\quad \MM\ni\quad x\mapsto(x,y_0)\quad\in\MN=\MM\times\MK,
\end{equation*}
где $y_0\in\MK$ -- произвольная фиксированная точка.
Пусть на $\MN$ заданы две произвольные функции $f(x,y)$ и $g(x,y)$. Вычислим
левую и правую части равенства (\ref{epusum}):
\begin{align*}
  [f\circ\vf,g\circ\vf]_\MM&=J^{\al\bt}\pl_\al f\pl_\bt g\big|_{y=y_0},
\\
  [f,g]_\MN\circ\vf&=\left(J^{\al\bt}\pl_\al f\pl_\bt g
  +J^{\mu\nu}\pl_\mu f\pl_\nu g\right)\big|_{y=y_0}.
\end{align*}
Мы видим, что в общем случае равенство (\ref{epusum}) не выполняется. Отсюда
следует, что вложение $\vf$ не является пуассоновым.
\qed\end{exa}
\begin{prop}
Пусть $(\MM,J)$ -- пуассоново многообразие и $X_H$ -- гамильтоново векторное
поле. Тогда поток векторного поля
\begin{equation*}
  s_t:=\exp(tX_H):\quad \MM\ni\quad x(0)\mapsto x(t)\quad\in\MM,
\end{equation*}
где $x(t)$ -- интегральная кривая для $X_H$, определяет пуассоново отображение.
\end{prop}
\begin{proof}
Пусть $f,g\in\CC^\infty(\MM)$. При малых $t$ поток векторного поля действует
на $x$ следующим образом $s_t:~x^\al\mapsto x^\al+tX^\al_H$. Поэтому
\begin{equation*}
  f\circ s_t=f(x^\al+tX^\al_H)\simeq f(x)+tX_Hf.
\end{equation*}
Продифференцируем по $t$ условие пуассоновости (\ref{epusum}) и положим $t=0$.
Тогда условие пуассоновости примет вид
\begin{equation*}
  [X_H f,g]+[f,X_H g]=X_H[f,g].
\end{equation*}
Из формулы (\ref{ehacoe}) следует, что полученное равенство совпадает с
тождествами Якоби. При $t=0$ экспоненциальное отображение является
тождественным и, следовательно, пуассоново. Поэтому интегрирование условия
(\ref{epusum}) вдоль $X_H$ доказывает его выполнение при всех $t$, для которых
определены интегральные кривые.
\end{proof}
\begin{exa}
Пусть на фазовой плоскости $q,p\in\MR^2$ задана каноническая пуассонова
структура (\ref{ecapos}). Рассмотрим гармонический осциллятор с гамильтонианом
\begin{equation*}
  H=\frac12(q^2+p^2).
\end{equation*}
Соответствующее гамильтоново векторное поле имеет вид
\begin{equation*}
  X_H=q\pl_p-p\pl_q
\end{equation*}
и задает группу вращений плоскости. Таким образом, каждое вращение фазовой
плоскости задает пуассоново преобразование для гармонического осциллятора. Оно
является каноническим, т.к.\ сохраняет скобку Пуассона.
\qed\end{exa}

Поскольку любой гамильтонов поток сохраняет скобку Пуассона, то он, в частности
сохраняет ее ранг. Поэтому справедливо
\begin{cor}
Для любого гамильтонова векторного поля $X_H$ на пуассоновом многообразии
$(\MM,J)$ ранг пуассоновой структуры $J$ постоянен вдоль произвольной
интегральной кривой для $X_H$.
\qed\end{cor}
Это следствие является, по существу, переформулировкой предложения \ref{plipob}.
\begin{prop}
Если ранг пуассоновой структуры в какой либо точке $x\in\MM$ пуассонова
многообразия $(\MM,J)$ равен нулю, то эта точка является неподвижной для любой
гамильтоновой системы $H$ на $\MM$.
\end{prop}
\begin{proof}
Из курса линейной алгебры известно, что если матрица антисимметрична и ее ранг
равен нулю, то она сама равна нулю. Если в точке $x\in\MM$ ранг пуассоновой
структуры равен нулю, то гамильтоново векторное поле $X_H$ в этой точке для
произвольного гамильтониана $H$ обращается в нуль. Следовательно, точка $x$
является неподвижной.
\end{proof}
\begin{exa}
Пуассонова структура в начале координат в примере \ref{elipos} имеет нулевой
ранг. Она остается неподвижной для любого гамильтонова потока.
\qed\end{exa}

Пусть $(\MM,J)$ -- пуассоново многообразие. Тогда для каждой точки $x\in\MM$
существует единственное линейное отображение кокасательного пространства
в соответствующее касательное
\begin{equation}                                                  \label{epomap}
  J:\quad \MT^*_x(\MM)\rightarrow\MT_x(\MM),
\end{equation}
такое, что для любой функции $f(x)$ справедливо равенство
\begin{equation*}
  J(df)=[f,x^\al]\pl_\al\in\MT_x(\MM).
\end{equation*}
Это есть рассмотренное ранее отображение (\ref{emafve}). Для произвольной
1-формы $A=dx^\al A_\al$ отображение $J$ в компонентах задается матрицей
структурных функций:
\begin{equation*}
  X^\al=-J^{\al\bt}A_\bt.
\end{equation*}
Для симплектических многообразий, для которых матрица $J^{\al\bt}$ невырождена,
отображение $J$ является взаимно однозначным.

Ясно, что ядром отображения (\ref{epomap}) является линейная оболочка
дифференциалов функций Казимира $dc^\Sa$.

Обозначим образ отображения (\ref{epomap}) в точке $x\in\MM$ через
\begin{equation*}
  \MJ_x(\MM):=\lbrace J(A)\in\MT_x(\MM):\quad \forall A\in\MT^*_x(\MM)\rbrace.
\end{equation*}
Размерность векторного подпространства $\MJ_x(\MM)\subset\MT_x(\MM)$ равна
рангу пуассоновой структуры $\dim\MJ_x(\MM)=\rank J^{\al\bt}(x)$. Если ранг
пуассоновой структуры на $\MM$ является постоянным и равен $2m$, то совокупность
подпространств $\MJ_x(\MM)$ для всех точек $x$ задает на $\MM$
распределение векторных полей $\CJ_{2m}(\MM)$ размерности $2m$, которое было
определено в разделе \ref{sfrote}. При этом, если пуассонова структура
дифференцируема, то соответствующее распределение векторных полей также
дифференцируемо.

В общем случае ранг пуассоновой структуры может меняться от точки к точке. Из
определения отображения $J$ следует, что образ $\MJ_x(\MM)$ является линейной
оболочкой всех гамильтоновых векторных полей на $\MM$ в точке $x$.

Введенные понятия позволяют сформулировать следующее утверждение.
\begin{theorem}                                                   \label{tposub}
Подмногообразие $\MM$ пуассонова многообразия $(\MN,J)$ является пуассоновым
тогда и только тогда, когда $\MJ_x(\MM):=\MJ_x(\MN)|_\MM\subset\MT_x(\MM)$ для
всех $x\in\MM$, т.е.\ каждое гамильтоново векторное поле на $\MN$ всюду касается
$\MM$. В частности, если $\MJ_x(\MM)=\MT_x(\MM)$ для всех $x\in\MM$, то $\MM$
является симплектическим подмногообразием в $\MN$.
\end{theorem}
\begin{proof}
См., например, \cite{Olver86R}.
\end{proof}

Из этой теоремы следует, что, поскольку размерность пространства $\MJ_x(\MM)$
совпадает с рангом пуассоновой структуры в данной точке, то размерность
пуассонова подмногообразия не может быть меньше ранга, $\dim\MM\ge\rank J$.

Пусть $\MM\subset\MN$ -- пуассоново подмногообразие. Как было отмечено в теореме
\ref{tposub}, любое гамильтоново векторное поле $X_H$ на $\MN$ касается
подмногообразия $\MM$. Это означает, что его сужение на $\MM$ может быть
получено из сужения гамильтониана на $\MM$:
\begin{equation*}
  \left.X_H\right|_\MM=X_{\widetilde H},\quad \text{где}\quad
  \widetilde H=H|_\MM.
\end{equation*}
Допустим, что нас интересуют траектории движения гамильтоновой системы с
некоторым гамильтонианом $H$, которые начинаются в некоторой точке пуассонова
подмногообразия $x\in\MM\subset\MN$. В этом случае можно ограничиться движением
точки в подмногообразии $\MM$, которое порождается суженным гамильтонианом
$\widetilde H$, тем самым понизив порядок гамильтоновой системы. Можно поставить
вопрос, каково минимальное пуассоново подмногообразие для данных начальных
условий. Ответ дает следующая
\begin{theorem}
Пусть $(\MN,J)$ -- пуассоново многообразие. Тогда соответствующее распределение
гамильтоновых векторных полей $\CJ(\MN)$ на $\MN$ интегрируемо, т.е.\ через
каждую точку $x\in\MN$ проходит интегральное подмногообразие $\MM$ распределения
$\CJ(\MN)$, для которого $\MT_y(\MM)=\MJ_y(\MM)$ в любой точке $y\in\MM$.
Всякое интегральное подмногообразие является симплектическим подмногообразием в
$\MN$, и в совокупности эти подмногообразия определяют симплектическое слоение
пуассонова многообразия $\MN$. Кроме того, если $H:~\MN\rightarrow\MR$ --
произвольный гамильтониан и $x(t)$ -- соответствующая траектория системы,
проходящая через точку $x_0\in\MN$, то $x(t)$ остается в одном и том же
интегральном подмногообразии $\MM$ при всех $t$.
\end{theorem}
\begin{proof}
Как уже отмечалось, распределение $\CJ(\MM)$ является линейной оболочкой всех
гамильтоновых векторных полей на $\MM$. Поскольку скобка Пуассона гамильтоновых
векторных полей является гамильтоновым векторным полем (\ref{ehopov}), то
распределение $\CJ(\MM)$ находится в инволюции. Отсюда по теореме Фробениуса
следует существование интегрального подмногообразия. Остальные утверждения
теоремы следуют из теоремы \ref{tposub} и инвариантности ранга пуассоновой
структуры вдоль гамильтоновых векторных полей.
\end{proof}

Проиллюстрируем данную теорему для пуассонова многообразия $(\MN,J)$,
$\dim\MN=n$, с пуассоновой структурой постоянного ранга $\rank J^{\al\bt}=2m<n$
путем построения специальной системы координат. В некоторой окрестности
$\MU\subset\MN$ существует $(n-2m)$ функционально независимых функций Казимира
$c^\Sa(x)$, $\Sa=1,\dotsc,n-2m$. Поверхности уровня функций Казимира
$c^\Sa=\const$ определяют $2m$-мерное подмногообразие $\MM\subset\MU$. Дополним
функции Казимира $2m$ функционально независимыми скалярными полями $g^\Sm$,
$\Sm=1,\dotsc,2m$, (координаты на подмногообразии $\MM$) таким образом, чтобы
выполнялись условия
\begin{equation*}
  \rank[g^\Sm,g^\Sn]_\MM=2m.
\end{equation*}
Это всегда можно сделать, т.к.\ ранг пуассоновой структуры равен $2m$.
Совокупность функций $\lbrace c^\Sa,g^\Sm\rbrace$ выберем в качестве новой
системы координат на $\MU$. Пуассонова структура в этой системе координат примет
вид
\begin{equation*}
  J=\begin{pmatrix} J^{\Sa\Sb}=[c^\Sa,c^\Sb] & J^{\Sa\Sn}=[c^\Sa,g^\Sn] \\
    J^{\Sm\Sb}=[g^\Sm,c^\Sb] & J^{\Sm\Sn}=[g^\Sm,g^\Sn]   \end{pmatrix}
    =\begin{pmatrix} 0 & 0 \\ 0 & J^{\Sm\Sn} \end{pmatrix},
\end{equation*}
т.к.\ скобка Пуассона функции Казимира с любой функцией на $\MN$ равна нулю.
Таким образом, сужение Пуассоновой структуры на подмногообразие $\MM$ определено
и невырождено. То есть поверхности уровней функций Казимира представляют собой
симплектические многообразия. Пусть на $\MN$ задан гамильтониан. Соответствующие
гамильтоновы уравнения
\begin{align}                                                     \label{ehacas}
  \dot c^\Sa&=J^{\Sa\Sb}\pl_\Sb H+J^{\Sa\Sn}\pl_\Sn H=0,
\\                                                                \label{ehaadi}
  \dot g^\Sm&=J^{\Sm\Sb}\pl_\Sb H+J^{\Sm\Sn}\pl_\Sn H=J^{\Sm\Sn}\pl_\Sn H
\end{align}
определяют траектории системы. Поскольку уравнение (\ref{ehacas}) имеет решение
$c^\Sa=\const$, то траектория гамильтоновой системы, проходящей через точку
$x_0\in\MN$, принадлежит соответствующему симплектическому подмногообразию
$\MM$:
\begin{equation*}
  \dot g^\Sm=\widetilde J^{\Sm\Sn}\pl_\Sn\widetilde H,
\end{equation*}
где
\begin{equation*}
  \widetilde J^{\Sm\Sn}=J^{\Sm\Sn}|_\MM\qquad {\text и}\qquad
  \widetilde H=H|_\MM.
\end{equation*}

Таким образом, всякое пуассоново многообразие $(\MN,J)$ расщепляется на
симплектические подмногообразия -- слои симплектического слоения. Размерность
любого такого слоя $\MM$ равна рангу пуассоновой структуры в произвольной точке
$x\in\MM$. Это означает, что, если ранг пуассоновой структуры на $\MN$
непостоянен, то размерность симплектических слоев будет различной.
\begin{theorem}[\bf Дарбу]                                        \label{tdrpoi}
Пусть $(\MM,J)$ -- пуассоново многообразие размерности $n$. Если пуассонова
структура на многообразии имеет постоянный ранг, $\rank J=2m\le n$, то в
некоторой окрестности $\MU$ произвольной точки $x\in\MM$ существует такая
система координат
\begin{equation*}
  \lbrace x^\al\rbrace=\lbrace q^\Sm,p_\Sm,c^\Sa\rbrace
  =(q^1,\dotsc,q^m,p_1,\dotsc,p_m,c^1,\dotsc,c^{n-2m}),
\end{equation*}
в которой скобка Пуассона двух произвольных функций $f,g\in\CC^\infty(\MU)$
имеет вид
\begin{equation}                                                  \label{edarpo}
  [f,g]=\frac{\pl f}{\pl q^\Sm}\frac{\pl g}{\pl p_\Sm}
  -\frac{\pl f}{\pl p_\Sm}\frac{\pl g}{\pl q^\Sm}.
\end{equation}
Координаты $c^\Sa$ являются функциями Казимира пуассоновой структуры $J$, и их
постоянные значения, $c_\Sa=\const$, определяет симплектическое слоение
пуассонова многообразия $(\MU,J)$.
\end{theorem}
\begin{proof}
Если ранг пуассоновой структуры равен нулю, то $J^{\al\bt}=0$ и доказывать
нечего. Любая система координат на $\MU$ удовлетворяет утверждению теоремы.

Допустим, что ранг отличен от нуля. Зафиксируем точку $x_0\in\MM$ и выберем в
ее окрестности $\MU$ две такие функции $Q,f\in\CC^\infty(\MU)$, что
\begin{equation*}
  [Q,f]=X_Q f\ne0.
\end{equation*}
В частности $X_Q|_{x_0}\ne0$. Согласно теореме \ref{tvecfi}, возможно, в меньшей
окрестности существует такая система координат $Q,x^2,\dotsc,x^n$, что
$X_Q=\pl_Q$. Тогда существует такая функция $P\in\CC^\infty(\MU)$, что выполнено
условие
\begin{equation*}
  X_QP=[Q,P]=1.
\end{equation*}
Из предложения \ref{poisth} следует равенство
\begin{equation*}
  [X_Q,X_P]=X_{[Q,P]}=0.
\end{equation*}
Кроме того в силу антисимметрии $[X_Q,X_Q]=0$ и $[X_P,X_P]=0$. Таким образом,
$X_Q,X_P$ образуют пару коммутирующих векторных полей. Теорема Фробениуса
позволяет дополнить пару функций $q^1=Q$, $p_1=P$ до локальной системы координат
$q^1,p_1,y^3,\dotsc,y^n$ в некоторой, возможно, меньшей окрестности точки $x_0$.
Поскольку $J^{\al\bt}=[x^\al,x^\bt]$ и выполнены равенства $[q^1,y^i]=0$,
$[p_1,y^i]=0$, $i=3,\dotsc,n$, то структурные функции в данной системе координат
принимают вид
\begin{equation*}
  J=\begin{pmatrix}\quad 0 & 1 & 0 \\ -1 & 0 & 0 \\ \quad 0 & 0 & J^{ij}
    \end{pmatrix},
\end{equation*}
где $J^{ij}=[y^i,y^j]$. Покажем, что матрица $J^{ij}$ не зависит от координат
$q^1$ и $p_1$. Действительно,
\begin{equation*}
  \frac{\pl J^{ij}}{\pl q^1}=[q^1,J^{ij}]=\big[q^1,[y^i,y^j]\big]=0,
\end{equation*}
где мы воспользовались тождествами Якоби. Аналогично доказывается, что матрица
$J^{ij}$ не зависит от $p_1$. Таким образом, матрица $J^{ij}$ задает пуассонову
структуру на подмногообразии $q^1=\const$, $p_1=\const$. Ранг этой структуры на
два меньше исходного, $\rank J^{ij}=2m-2$. Поэтому, если $m>1$, то этот процесс
можно продолжить.
\end{proof}
\chapter{Принцип наименьшего действия                            \label{sacpri}}
Можно с уверенностью сказать, что в основе построения моделей современной
математической физики лежит принцип наименьшего действия. Этот принцип требует
стационарности некоторого функционала -- действия -- относительно вариаций
полей, описывающих данную модель. В результате мы получаем систему уравнений
Эйлера--Лагранжа, которая принимается в качестве уравнений движения, уравнений
равновесия и т.д.\ для данной модели. При этом инвариантность действия
относительно некоторой группы преобразований приводит к ковариантным уравнениям
движения и к законам сохранения, которые играют важнейшую роль в физике.
\section{Постановка вариационных задач                          \label{svarpr}}
Начнем с постановки вариационной задачи в евклидовом пространстве. Предположим,
для простоты, что $\MM\subset\MR^n$ -- ограниченная область евклидова
пространства с кусочно гладкой границей $\pl\MM$. Пусть в этой области задан
некоторый набор дважды непрерывно дифференцируемых функций вплоть до границы,
$\vf=\lbrace\vf^a\rbrace\in\CC^2(\overline\MM)$, $a=1,\dotsc,\Sn$. Это значит,
что все функции и их производные до второго порядка непрерывны и ограничены в
$\MM$ и имеют конечный предел на границе. Если $x^\al$, $\al=1,\dotsc,n$, --
система координат на $\MM$, то обозначим, для краткости, все первые производные
полей через $\pl\vf=\lbrace\pl_\al\vf^a\rbrace$. Предположим, что на $\MM$
определен {\em функционал действия} или, просто, {\em действие}
\index{Функционал действия (action functional)}%
\index{Действия функционал (action functional)}\index{Действие (action)}%
\begin{equation}                                                  \label{eacgen}
  S[\vf]=\int_\MM\!dx\,L(x,\vf,\pl\vf),
\end{equation}
где $L$ -- некоторая функция от $n$ переменных $x^\al$, $\Sn$ переменных $\vf^a$
и $n\Sn$ переменных $\pl_\al\vf^a$. Она предполагается дважды непрерывно
дифференцируемой функцией переменных $x\in\MM$ и всех остальных переменных для
всех конечных значений $\vf$ и $\pl\vf$. Назовем функцию $L(x,\vf,\pl\vf)$
{\em лагранжевой плотностью} или {\em лагранжианом} данной модели, которая
описывается набором полей $\vf$.
\index{Лагранжева плотность (Lagrangian density)}%
\index{Плотность лагранжева (Lagrangian density)}%
\index{Лагранжиан (Lagrangian)}%

Для определения функционала действия мы ограничили себя классом дважды
непрерывно дифференцируемых функций $\CC^2(\overline\MM)$. Тогда функционал
действия задает отображение
\begin{equation}                                                  \label{eacmap}
  S:\quad \big[\CC^2(\overline\MM)\big]^\Sn\ni\quad\vf\mapsto S[\vf]\quad\in\MR
\end{equation}
бесконечномерного функционального пространства в поле вещественных чисел. Для
постановки вариационной задачи нам важно, что функциональное пространство
$\big[\CC^2(\overline\MM)\big]^\Sn$ снабжено структурой линейного пространства с
обычным поточечным сложением и умножением на вещественные числа. Для того, чтобы
говорить о непрерывности и вариационных производных отображения $S$, введем на
$\big[\CC^2(\overline\MM)\big]^\Sn$ норму:
\begin{equation}                                                  \label{enorfu}
  \|\vf\|:=\sum_{a=1}^\Sn\left(\underset{x\in\MM}{\sup}|\vf^a|
  +\sum\limits_{\al=1}^n\underset{x\in\MM}{\sup}|\pl_\al\vf^a|\right).
\end{equation}
Тем самым класс рассматриваемых функций $\big[\CC^2(\overline\MM)\big]^\Sn$
становится нормированным линейным функциональным пространством. По данной норме
строится метрика и определяется топология, относительно которой отображение
(\ref{eacmap}) непрерывно.

Назовем {\em вариацией} функции $\vf^a$ разность двух представителей из класса
рассматриваемых функций: $\dl\vf^a:=\vf^{\prime a}-\vf^a$, где $\vf^{\prime a}$
-- произвольная функция из $\CC^2(\overline\MM)$. Ясно, что вариация функции
принадлежит тому же классу, что и сама функция. Для малых вариаций функций
$\e\dl\vf^a$, где $\e>0$ -- малая величина, вариация (главная линейная часть)
функционала действия, если она существует, равна
\begin{equation}                                                  \label{evarfn}
\begin{split}
  \dl S&=S[\vf+\e\dl\vf]-S[\vf]=
\\
  &=\int_\MM\!dx\left[\frac{\pl L}{\pl\vf^a}
  -\pl_\al\left(\frac{\pl L}{\pl(\pl_\al\vf^a)}\right)\right]\e\dl\vf^a
  +\int_{\pl\MM}\!ds_\al\frac{\pl L}{\pl(\pl_\al\vf^a)}\e\dl\vf^a+\osmall(\e),
\end{split}
\end{equation}
где второе слагаемое возникло при интегрировании по частям, и $ds_\al$
обозначает ориентированный элемент объема границы $\pl\MM$. Отметим, что при
вычислении вариации действия (\ref{eacgen}) область интегрирования $\MM$
считалась неизменной.

В рассматриваемом классе функций $\vf$ и лагранжианов $L$ вариация функционала
всегда существует.

Назовем набор функций $\vf$ {\em стационарной}, или {\em критической} точкой,
или {\em экстремалью} функционала $S[\vf]$, если в этой точке линейная часть
вариации действия равна нулю, $\dl S=\osmall(\e)$. Такие точки соответствуют
либо локальному минимуму, либо локальному максимуму, либо седловой точке
функционала $S$. Это можно проверить после нахождения экстремали функционала,
рассмотрев члены более высокого порядка по $\e$.
\index{Стационарная точка (stationary point)}%
\index{Точка стационарная (stationary point)}%
\index{Критическая точка (critical point)}%
\index{Точка критическая (critical point)}%
\index{Экстремаль функционала (extremal of a functional)}%

Для действия (\ref{eacgen}) можно поставить различные вариационные задачи.
Рассмотрим задачи, которые наиболее часто встречаются в физике.
\subsection{Задача с заданными граничными условиями}
Вариационная задача с заданными граничными условиями является наиболее
распространенной и самой простой с точки зрения постановки. Рассмотрим класс
функций в $\CC^2(\overline\MM)$ с заданными граничными условиями
\begin{equation}                                                  \label{ebocfi}
  \left.\vf\vphantom{\sum}\right|_{\pl\MM}=\vf_0.
\end{equation}
Поскольку граничные условия для всех функций при фиксированном индексе $a$ одни
и те же, то вариации полей обращаются в нуль на границе:
\begin{equation}                                                  \label{ebodlv}
  \left.\dl\vf\vphantom{\sum}\right|_{\pl\MM}=0.
\end{equation}
Тогда интеграл по границе области в вариации действия (\ref{evarfn}) обращается
в нуль в силу граничных условий (\ref{ebodlv}). В рассматриваемом случае
существует предел
\begin{equation}                                                  \label{elimfd}
  \underset{\e\rightarrow0}\lim\frac{S[\vf+\e\dl\vf]-S[\vf]}\e
  =\int_\MM\!dx\,\frac{\dl S}{\dl\vf^a}\dl\vf^a.
\end{equation}
Функция $\dl S/\dl\vf^a$, стоящая под знаком интеграла, называется {\em
вариационной производной} функционала $S$ по полю $\vf^a$ и обозначается также
запятой:
\index{Вариационная производная (variational derivative)}%
\index{Производная вариационная (variational derivative)}%
\begin{equation}                                                  \label{evadno}
  S,{}_a:=\frac{\dl S}{\dl\vf^a}.
\end{equation}
Из вида вариации (\ref{evarfn}) получаем явное выражение для вариационной
производной
\begin{equation}                                                  \label{evadel}
  S,{}_a=\frac{\pl L}{\pl\vf^a}-\pl_\al\frac{\pl L}{\pl(\pl_\al\vf^a)}.
\end{equation}

Вариационная производная $S,{}_a$ является ядром линейного оператора, который
есть производная по Фреше и, следовательно, по Гато отображения (\ref{eacmap}).

Формула (\ref{elimfd}) позволяет дать другое определение вариационной
производной. А именно, вариационной производной функционала $S$ в точке
$\vf\in[\CC^2(\overline\MM)]^\Sn$ назовем производную
\begin{equation}                                                  \label{evards}
  \frac{\dl S}{\dl\vf^a}=\left.\frac{\pl S(\vf+\lm\psi)}{\pl\lm}\right|_{\lm=0},
\end{equation}
где $\psi\in[\CC^2(\overline\MM)]^\Sn$. Ясно, что если главная линейная часть
приращения функционала существует, то она равна производной (\ref{evards}).
Обратное утверждение в общем случае неверно. Существуют примеры функционалов
(менее гладких, чем мы рассматриваем), для которых производная (\ref{evards})
определена, однако из их приращения нельзя выделить главную линейную часть.
Поэтому определение вариационной производной (\ref{evards}) является более
общим. В рассматриваемом нами классе функций и лагранжианов данные выше
определения эквивалентны.

Определение вариационной производной (\ref{evards}) просто обобщается на
случай вариационных производных более высокого порядка. Вторые вариационные
производные (\ref{esevar}) были рассмотрены для экстремалей функционала длины
кривой.

Из условия стационарности действия $\dl S=\osmall(\e)$ в силу произвольности
вариации $\dl\vf^a$ и основной леммы вариационного исчисления следует
\begin{theorem}                                                   \label{teilat}
Набор функций $\vf$ при заданных граничных условиях является стационарной точкой
действия $S[\vf]$ тогда и только тогда, когда он удовлетворяет системе уравнений
Эйлера--Лагранжа
\begin{equation}                                                  \label{eulage}
  S,{}_a=\frac{\pl L}{\pl\vf^a}-\pl_\al\frac{\pl L}{\pl(\pl_\al\vf^a)}=0.
\end{equation}
\end{theorem}
\index{Уравнения Эйлера--Лагранжа (Euler--Lagrange equations)}%
\index{Эйлера--Лагранжа уравнения (Euler--Lagrange equations)}%

В общем случае уравнения Эйлера--Лагранжа представляют собой систему нелинейных
дифференциальных уравнений в частных производных второго порядка, число которых
$\Sn$ равно числу функций, от которых зависит функционал действия. Их надо
решать при заданных граничных условиях (\ref{ebocfi}). Решение поставленной
вариационной задачи может не существовать, а если оно существует, то может быть
неединственно. Это зависит от вида лагранжиана  и области $\MM$.
\begin{com}
В общем случае уравнения Эйлера--Лагранжа для заданного лагранжиана могут
приводить к противоречию. Например, пусть лагранжиан зависит от одной функции
$\vf$ и имеет вид $L=\vf$. Тогда уравнения Эйлера--Лагранжа приводят к
противоречию $1=0$. Таким образом, для того, чтобы уравнения Эйлера--Лагранжа
имели решение, лагранжиан не может быть произвольной функцией полей и их
производных. В дальнейшем мы предполагаем, что лагранжиан выбран таким образом,
что соответствующие уравнения Эйлера--Лагранжа непротиворечивы.
\qed\end{com}

Действие (\ref{eacgen}) для заданных уравнений Эйлера--Лагранжа при постановке
задачи с фиксированными граничными условиями определено неоднозначно.
Действительно, рассмотрим новый лагранжиан, $\widetilde L_\al=L+\pl_\al F$,
который отличается от исходного на частную производную от некоторой достаточно
гладкой функции $F(x,\vf,\pl\vf)$. Тогда действие получит дополнительный вклад,
сводящийся к интегралу по границе. Вариация дополнительного слагаемого равна
нулю, т.к.\ вариации всех полей на границе равны нулю. Отсюда следует, что
уравнения Эйлера--Лагранжа не изменятся при добавлении к лагранжиану частных
производных $\pl_\al F$ от произвольной функции.

Рассмотренная вариационная задача наиболее часто рассматривается в моделях
математической физики. При этом уравнения Эйлера--Лагранжа приводят к уравнениям
движения, равновесия и т.д.
\begin{com}
Поверхностный интеграл в вариации действия (\ref{evarfn}) в задаче с заданными
граничными условиями равен нулю, поскольку вариации полей обращаются в нуль на
границе (\ref{ebodlv}). Этого достаточно, если область $\MM$ ограничена. Однако
для действия (\ref{eacgen}), рассматриваемого во всем евклидовом пространстве
$\MR^n$, ситуация усложняется. В этом случае интеграл по границе является
несобственным, т.к.\ площадь бесконечно удаленной границы стремится к
бесконечности. Тогда важно не только граничное условие на вариации, но и их
асимптотическое поведение. Так, в общей теории относительности в асимптотически
плоском пространстве-времени поверхностный интеграл отличен от нуля, т.к.\
метрика недостаточно быстро стремится к метрике Минковского на пространственной
бесконечности. Анализ граничного поведения полей важен и в общем случае сложен,
поскольку зависит от рассматриваемой задачи. На данном этапе мы пренебрежем
граничными слагаемыми.
\qed\end{com}
\begin{com}
Во многих моделях математической физики, например, в электродинамике, в качестве
области $\MM$ выбирается все пространство Минковского $\MR^{1,3}$. При этом
функционал действия не ограничен ни снизу, ни сверху. Кроме того, для многих
решений уравнений Эйлера--Лагранжа он расходится. Поэтому говорить о принципе
наименьшего действия в строгом смысле не приходится. Тем не менее уравнения
Эйлера--Лагранжа имеют смысл, поскольку являются локальным объектом. Поэтому для
действия часто пишут формальное выражение, не заботясь о сходимости интеграла.
Этот способ очень удобен для получения уравнений с заданными свойствами
симметрии.
\qed\end{com}
\subsection{Задача со свободными граничными условиями}
Для действия (\ref{eacgen}) можно поставить другую вариационную задачу, расширив
класс рассматриваемых функций. Пусть, по-прежнему, все функции дважды непрерывно
дифференцируемы $\vf^a\in\CC^2(\overline\MM)$ для всех $a=1,\dotsc,\Sn$, но
теперь снимем ограничения, накладываемые граничными условиями (\ref{ebocfi}). В
этом случае вариации полей $\dl\vf^a$ также ничем не ограничены на границе
$\pl\MM$, и из явного вида вариации действия (\ref{evarfn}) следует
\begin{theorem}                                                   \label{teilst}
Набор функций $\vf$ является стационарной точкой действия $S[\vf]$ тогда и
только тогда, когда он удовлетворяет системе уравнений Эйлера--Лагранжа
(\ref{eulage}) и граничным условиям
\begin{equation}                                                  \label{eboele}
  \left.\frac{\pl L}{\pl(\pl_\al\vf^a)}\right|_{\pl\MM}=0.
\end{equation}
\end{theorem}
Таким образом, в задаче со свободными граничными условиями граничные условия
все таки возникают из условия стационарности действия.
\begin{exa}
Вариационная задача со свободными граничными условиями рассматривается в теории
открытых бозонных струн, для которых из вариационного принципа вытекают
граничные условия Неймана.
\qed\end{exa}

Постановка вариационной задачи со свободными граничными условиями, зависит от
добавления к действию граничных слагаемых. Иногда появления граничных условий
(\ref{eboele}) можно избежать, если из исходного действия вычесть все граничные
вклады, которые возникают при интегрировании по частям. В этом случае условие
(\ref{eboele}) тождественно выполняется.
\begin{com}
В общем случае можно рассматривать смешанные вариационные задачи, когда
граничные условия ставятся только для части полей. Именно такого типа задача
естественным образом возникает в теории гравитации, где можно считать заданными
на границе только физические поля. Нефизические поля находятся, как решение
уравнений связей и калибровочных условий, и для них граничные значения не могут
быть фиксированы произвольным образом.
\qed\end{com}
\subsection{Задача с подвижной границей}
Возможна также более общая постановка вариационной задачи для функционала
(\ref{eacgen}), когда рассматриваются не только вариации полей, но и вариации
самой области $\MM$. Сначала уточним постановку задачи. Предположим, что
координаты и поля преобразуются следующим образом:
\begin{align}                                                     \label{etracn}
  x^\al&\mapsto x^{\prime\al}(x,\vf,\pl\vf,\e),
\\                                                                \label{etrafi}
  \vf^a&\mapsto\vf^{\prime a}(x,\vf,\pl\vf,\e),
\end{align}
где $\e$ -- параметр преобразования. Если поля $\vf$ являются заданными
функциями от координат, $\vf=\vf(x)$, то из уравнения (\ref{etracn}) можно
выразить старые координаты через новые: $x=x(x',\e)$. Тогда подстановка
найденных функций в формулу (\ref{etrafi}) позволяет рассматривать новые поля,
как функции от новых координат, $\vf'=\vf'(x',\e)$.

Мы считаем, что до и после преобразования координаты определены соответственно
на ограниченных областях $\MM$ и $\MM'$ евклидова пространства $\MR^n$, т.е.\
$x\in\MM$ и $x'\in\MM'$. Тогда под вариацией функционала подразумевается
разность
\begin{equation*}
  \dl S:=S[\vf'(x')]-S[\vf(x)]
  =\int_{\MM'}\!\!\!dx'L(x',\vf',\pl'\vf')-\int_\MM\!\!\!dx L(x,\vf,\pl\vf).
\end{equation*}
Нам требуется найти эту вариацию в линейном по $\e$ приближении. При этом мы
считаем, что при $\e=0$ преобразование координат и полей является тождественным.
Разлагая формулы преобразования в ряд Тейлора при малых $\e$, получаем
\begin{equation}                                                  \label{elotrg}
\begin{split}
  x^\al&\mapsto x^{\prime\al}=x^\al+\dl x^\al+\osmall(\e),
\\
  \vf^a(x)&\mapsto\vf^{\prime a}(x')=\vf^a(x)+\bar\dl\vf^a(x)+\osmall(\e),
\end{split}
\end{equation}
где независимые вариации
\begin{align}                                                     \label{evarxg}
  \dl x^\al=\e R^\al(x,\vf,\pl\vf),
\\                                                                \label{evavfu}
  \bar\dl\vf^a=\vf^{\prime a}(x')-\vf^a(x)=\e R^a(x,\vf,\pl\vf)
\end{align}
определяются некоторыми функциями $R^\al(x,\vf,\pl\vf)$ и $R^a(x,\vf,\pl\vf)$.
Отметим, что вариация полей (\ref{evavfu}) определена, как разность значений
полей в различных точках и не совпадает с вариацией формы поля, рассмотренной в
разделе \ref{sinfct}. Напомним, что под вариацией формы функции мы понимаем
разность значений этой функции после и до преобразования в одной и той же точке:
\begin{equation*}
  \dl\vf^a(x):=\vf^{\prime a}(x)-\vf^a(x).
\end{equation*}
При этом преобразования (\ref{elotrg}) рассматриваются, как активные (см.\
раздел \ref{scooch}). Тогда для вариации формы функции справедливо равенство
\begin{equation}                                                  \label{evarfg}
  \dl\vf^a(x)=\bar\dl\vf^a-\dl x^\al\pl_\al\vf^a=\e(R^a-R^\al\pl_\al\vf^a).
\end{equation}
По-построению, для постоянного параметра преобразования вариация формы функции
$\dl$ перестановочна с операцией частного дифференцирования $\pl_\al$. Вариация
действия относительно преобразований (\ref{elotrg}) имеет вид
$$
  \dl S=\int_{\MM'}\!\!\!dx'\,L(x')-\int_\MM\!\!\!dx\,L(x)
  =\int_{\MM'}\!\!\!dx'\,\left(L(x)+\bar\dl L(x)\right)-\int_\MM\!\!\!dx\,L(x),
$$
где $L(x)=L\big(x,\vf(x),\pl_\al\vf(x)\big)$. Здесь и в дальнейшем мы будем
рассматривать вариацию действия только с точностью до слагаемых, линейных по
$\e$, и не будем это указывать. Вариации лагранжиана и элемента объема равны,
соответственно,
\begin{align}                                                     \label{evarlx}
  \bar\dl L(x)&=\frac{\pl L}{\pl x^\al}\dl x^\al
  +\frac{\pl L}{\pl\vf^a}\bar\dl\vf^a
  +\frac{\pl L}{\pl(\pl_\al\vf^a)}\bar\dl(\pl_\al\vf^a)
  =\dl L+\frac{dL}{dx^\al}\dl x^\al,
\\                                                                \label{evarvg}
  dx'&\simeq dx+dx\frac{\pl(\dl x^\al)}{\pl x^\al},
\end{align}
где
\begin{equation*}
  \frac{dL}{dx^\al}=\frac{\pl L}{\pl x^\al}+\frac{\pl L}{\pl\vf^a}\pl_\al\vf^a
  +\frac{\pl L}{\pl(\pl_\bt\vf^a)}\pl^2_{\al\bt}\vf^a.
\end{equation*}
Используя преобразование (\ref{evarvg}), перейдем от интегрирования по $\MM'$ к
интегрированию по исходной области $\MM$ и перепишем вариацию действия в виде
\begin{equation}                                                  \label{evarag}
  \dl S=\int_\MM\!\!\!dx\,\left(\dl L+\frac{dL}{dx^\al}\dl x^\al
  +L\frac{\pl(\dl x^\al)}{\pl x^\al}\right)
  =\int_\MM\!\!\!dx\,\big(\dl L+\pl_\al(L\dl x^\al)\big).
\end{equation}
Теперь перепишем вариацию формы лагранжиана:
\begin{equation}                                                  \label{evarlg}
\begin{split}
  \dl L&=\frac{\pl L}{\pl\vf^a}\dl\vf^a
  +\frac{\pl L}{\pl(\pl_\al\vf^a)}\dl(\pl_\al\vf^a)=
\\
  &=\left(\frac{\pl L}{\pl\vf^a}-\pl_\al\frac{\pl L}{\pl(\pl_\al\vf^a)}\right)
  \dl\vf^a+\pl_\al\left(\frac{\pl L}{\pl(\pl_\al\vf^a)}\dl\vf^a\right).
\end{split}
\end{equation}
При этом существенно, что параметр преобразования постоянен, $\e=\const$, и,
следовательно,
$$
  \dl(\pl_\al\vf^a)=\pl_\al(\dl\vf^a).
$$
Воспользовавшись формулой Стокса, дивергентные слагаемые можно переписать в виде
поверхностного интеграла. Таким образом для вариации действия получаем
окончательное выражение
\begin{equation}                                                  \label{efivaa}
  \dl S=\int_\MM\! dx \left(\frac{\pl L}{\pl\vf^a}
  -\pl_\al\frac{\pl L}{\pl(\pl_\al\vf^a)}\right)\dl\vf^a
  +\int_{\pl\MM}\!\!\!ds_\al\left(\frac{\pl L}{\pl(\pl_\al\vf^a)}\dl\vf^a
  +L\dl x^\al\right).
\end{equation}
Поверхностный интеграл перепишем с учетом выражения для вариации формы функции
(\ref{evarfg}):
\begin{equation*}
  \int_{\pl\MM}\!\!\!ds_\al\left(\frac{\pl L}{\pl(\pl_\al\vf^a)}\bar\dl\vf^a
  +\left(L\dl_\bt^\al-\frac{\pl L}{\pl(\pl_\al\vf^a)}\pl_\bt\vf^a\right)
  \dl x^\bt\right).
\end{equation*}
поскольку вариации координат $\dl x^\al$ и функций $\bar\dl\vf^a$ произвольны и
независимы, то из полученного выражения для вариации действия вытекает
\begin{theorem}
Набор функций $\vf$ является стационарной точкой действия $S[\vf]$ в
вариационной задаче с подвижной границей тогда и только тогда, когда он
удовлетворяет системе уравнений Эйлера--Лагранжа (\ref{eulage}) и граничным
условиям:
\begin{equation}                                                  \label{eboels}
\begin{split}
  \left.\frac{\pl L}{\pl(\pl_\al\vf^a)}\right|_{\pl\MM}&=0,
\\
  \left. L\dl_\bt^\al
  -\frac{\pl L}{\pl(\pl_\al\vf^a)}\pl_\bt\vf^a\right|_{\pl\MM}&=0.
\end{split}
\end{equation}
\end{theorem}
В последней вариационной задаче с подвижной границей мы имеем $n(n+\Sn)$
граничных условий, число которых быстро растет с увеличением размерности
пространства. Соответствующая вариационная задача далеко не всегда имеет
решение. Чтобы уменьшить число независимых граничных условий, предположим, что в
окрестности границы $\pl\MM$ поля принимают наперед заданные значения:
\begin{equation*}
  \vf^a(x)=\Phi^a(x),
\end{equation*}
где достаточно гладкие функции $\Phi^a$ заданы в некоторой окрестности $\pl\MM$.
Тогда вариации функций и координат связаны соотношением
\begin{equation*}
  \left.\bar\dl\vf^a\right|_{\pl\MM}
  =\left.\dl x^\al\pl_\al\Phi^a\right|_{\pl\MM}.
\end{equation*}
В этом случае вместо $n(n+\Sn)$ граничных условий (\ref{eboels}) остаются $n^2$
условий:
\begin{equation}                                                  \label{etraco}
  \left[L\dl_\bt^\al+\frac{\pl L}{\pl(\pl_\al\vf^a)}
  (\pl_\bt\Phi^a-\pl_\bt\vf^a)\right]_{\pl\MM}=0.
\end{equation}
Эти граничные условия называются {\em условиями трансверсальности}.
\index{Условия трансверсальности (transversality conditions)}%
\index{Трансверсальности условия (transversality conditions)}%

Стационарные точки действия для задачи с подвижной границей являются также
стационарными точками для двух задач с фиксированной областью $\MM$,
рассмотренных ранее. Выполнение уравнений Эйлера--Лагранжа является необходимым
условием во всех трех рассмотренных вариационных задачах. В задачах со
свободными граничными условиями и с подвижной границей к уравнениям
Эйлера--Лагранжа добавляются граничные условия.
\subsection{Задача на условную стационарную точку                \label{solusp}}
В настоящем разделе мы ограничимся рассмотрением вариационных задач на конечном
отрезке $[x_1,x_2]\subset\MR$ с заданными граничными условиями. Пусть требуется
найти стационарную точку действия (\ref{eacgen}) в классе функций
$\vf^a\in\CC^2\big([x_1,x_2]\big)$, $a=1,\dotsc,\Sn$ при наличии $\Sm<\Sn$
независимых дополнительных условий, которые называются {\em связями}
\index{Связь (constraint)}%
\begin{equation}                                                  \label{eextco}
  G_\Sa(x,\vf,\pl\vf)=0,\qquad \Sa=1,\dots,\Sm<\Sn,
\end{equation}
где $G_\Sa$ -- достаточно гладкие функции своих аргументов. Мы предполагаем, что
связи $G_\Sa$ не противоречат граничным условиям и функционально независимы,
т.е.\ ни одна из связей не является следствием остальных. В частности, ни одна
из связей не выполняется тождественно для всех функций $\vf$. Функциональная
независимость связей означает, что матрица производных
\begin{equation*}
  \frac{\pl G_\Sa}{\pl\big(\vf^a,\pl_x\vf^b\big)}
\end{equation*}
имеет постоянный ранг $\Sm$. Отсюда следует, что локально связи можно разрешить
относительно $\Sm$ функций или их первых производных, рассматривая остальные
$2(\Sn-\Sm)$ функции и их производные как независимые.

В общем случае связи являются дифференциальными уравнениями, и их решения
содержат произвольные постоянные. Мы предполагаем, что этот произвол устранен,
например, наложением граничных условий, либо каким либо иным образом.

В частном случае связи могут быть алгебраическими уравнениями на неизвестные
функции $G_\Sa(x,\vf)=0$. В этом случае они называются {\em голономными}. В
общем случае связи (\ref{eextco}) называются {\em неголономными}.
\index{Голономная связь (holonomic constraint)}%
\index{Связь голономная (holonomic constraint)}%
\index{Неголономная связь (holonomic constraint)}%
\index{Связь неголономная (holonomic constraint)}%

При наличии связей вариации функций не являются независимыми, и выполнение
уравнений Эйлера--Лагранжа для исходного действия (\ref{eacgen}) не является
необходимым условием. Прямым способом решения задачи на условный экстремум
является явное разрешение связей относительно $\Sm$ функций, подстановка
полученного решения в исходное действие и исследование нового действия от
$\Sn-\Sm$ функций на безусловный экстремум.

Задачи на условную стационарную точку часто встречаются в математической
физике. В частности, к ним приводят все калибровочные модели, инвариантные
относительно локальных преобразований полей. В связи с этим введем удобную
терминологию, которая часто используется в физике. А именно, назовем поля {\em
нефизическими}, если связи разрешаются относительно этих полей. Остальные поля,
относительно которых после исключения нефизических полей возникает задача на
безусловную стационарную точку, называются {\em физическими}. Деление полей на
физические и нефизические условно, т.к.\ связи можно разрешать относительно
различных переменных. В то же время число физических ($\Sn-\Sm$) и нефизических
($\Sm$) полей, по предположению, постоянно.
\index{Нефизическое поле (unphysical field)}%
\index{Поле нефизическое (unphysical field)}%
\index{Физическое поле (physical field)}%
\index{Поле физическое (physical field)}%

Прямой способ исключения нефизических полей неприменим, если связи не решаются
явно. Кроме этого исключение части полей может нарушить симметрию задачи,
например, лоренц-инвариантность, что часто приводит к существенному усложнению
вычислений. Поэтому используют метод неопределенных множителей Лагранжа. А
именно, строят полное (total) действие
\begin{equation}                                                  \label{eneact}
  S_\St=\int_{x_1}^{x_2}\!\!\!dx(L-\lm^\Sa G_\Sa),
\end{equation}
где $\lm(x)\in\CC^1\big([x_1,x_2]\big)$ -- новые функции, которые называются
{\em множителями Лагранжа.} Это действие исследуется на безусловный экстремум.
Вариация действия (\ref{eneact}) по полям $\vf^a$ и множителям Лагранжа
$\lm^\Sa$ приводит к $\Sn+\Sm$ уравнениям Эйлера--Лагранжа, $\Sm$ из которых,
возникших при вариации по множителям Лагранжа, совпадают с уравнениями связей
(\ref{eextco}). При этом вариации множителей Лагранжа на границе не обязаны быть
равными нулю, т.к.\ они входят в действие без производных и никаких
дополнительных граничных условий не возникает. Решение новой задачи на
безусловный экстремум дает решение исходной задачи на условный экстремум, что
является содержанием следующего утверждения.
\index{Множитель Лагранжа (Lagrange multiplier)}%
\index{Лагранжа множитель (Lagrange multiplier)}%
\begin{theorem}
Для функций $\vf$, на которых функционал (\ref{eacgen}) имеет стационарное
значение при выполнении уравнений связей (\ref{eextco}), существует такой
набор множителей Лагранжа, что они вместе с полями $\vf$ удовлетворяют
уравнениям Эйлера--Лагранжа для действия (\ref{eneact}):
$$
  \frac{\dl S_\St}{\dl\vf^a}=0,\qquad \frac{\dl S_\St}{\dl\lm^\Sa}=G_\Sa=0.
$$
\end{theorem}
\begin{proof}
См., например, \cite{Elsgol98R}.
\end{proof}

Сформулированная теорема позволяет свести вариационную задачу на условный
экстремум к вариационной задаче на безусловный экстремум, но для действия,
зависящего от большего числа функций. В теории поля вариационные задачи
рассматриваются не на прямой $\MR$, а в евклидовом пространстве $\MR^n$. Тогда
связи представляют собой в общем случае дифференциальные уравнения в частных
производных. Для того чтобы доказать аналог теоремы о множителях Лагранжа
необходимо зафиксировать каким либо образом класс рассматриваемых связей, что
является сложной задачей. На практике метод неопределенных множителей Лагранжа
часто используют, не заботясь о его применимости. В таком случае применимость
метода необходимо доказывать в каждом конкретном случае.
\subsection{Другие задачи и терминология}
В общем случае функционал действия может зависеть от частных производных полей
$\vf^a$ любого порядка, вплоть до бесконечного:
\begin{equation}                                                  \label{eacind}
  S[\vf]=\int_\MM\!dx\,L(x,\vf,\pl\vf,\pl^2\vf,\dots).
\end{equation}
Тогда уравнения Эйлера--Лагранжа примут вид
\begin{equation}                                                  \label{eilaid}
  S,{}_a=\frac{\pl L}{\pl(\vf^a)}-\pl_\al\frac{\pl L}{\pl(\pl_\al\vf^a)}
         +\pl^2_{\al\bt}\frac{\pl L}{\pl(\pl^2_{\al\bt}\vf^a)}-\dotsc=0.
\end{equation}
Для простоты в настоящем разделе мы не будем обсуждать возможные граничные
слагаемые.
\begin{defn}
Модели теории поля, лагранжиан которых зависит от производных бесконечного
порядка, называются {\em нелокальными}. Для {\em локальных} моделей, порядок
производных ограничен и ряд (\ref{eilaid}) обрывается. Модели, для
которых уравнения Эйлера--Лагранжа содержат производные третьего или более
высокого, но конечного порядка называются {\em моделями с высшими производными}.
\qed\end{defn}
\index{Нелокальная модель (nonlocal model)}%
\index{Модели нелокальные (nonlocal model)}%
\index{Локальная модель (local model)}%
\index{Модель локальная (local model)}%
\index{Модель с высшими производными (model with higher derivatives)}%

Хорошо известно, что порядок уравнений можно понизить, рассматривая частные
производные в качестве новых независимых переменных. В этом смысле любую теорию
с высшими производными можно свести к модели без высших производных.
\begin{com}
С физической точки зрения теории с высшими производными представляют
определенные трудности, т.к.\ наличие векторных лоренцевых индексов у полей, как
правило, приводит к каноническому гамильтониану, неограниченному снизу за счет
вклада временн\'ых компонент. После квантования такие теории приводят к
гильбертову пространству с индефинитной метрикой, которая на допускает
вероятностной интерпретации квантовой теории. Эти трудности можно избежать за
счет выбора лагранжиана специального вида или налагая условие калибровочной
инвариантности, которое позволяет исключить вклад временн\'ых компонент в
канонический гамильтониан для физических степеней свободы. В квантовой теории
поля модели с высшими производными принято считать неудовлетворительными до тех
пор, пока не доказана положительная определенность канонического гамильтониана
для физических степеней свободы.

Нелокальные теории поля представляют собой еще б\'ольшие трудности для
физической интерпретации, т.к.\ помимо проблем с индефинитной метрикой
гильбертова пространства, в общем случае они нарушают причинность. Это следует
из того, что значение функции, разложимой в ряд Тейлора в точке $x$, в точке
$y\ne x$ выражаются через ее значения и значения ее производных в точке $x$ в
виде бесконечного ряда. Следовательно, значение функции в некоторой точке может
зависеть от ее значений в конусе будущего.
\qed\end{com}

Допустим, что задана система уравнений Эйлера--Лагранжа. Функционал действия,
приводящий к этим или эквивалентным уравнениям Эйлера--Лагранжа, определен
неоднозначно. Во-первых, как уже отмечалось при рассмотрении вариационной задачи
с заданными граничными условиями, уравнения не изменятся, если к лагранжиану
добавить частную производную от функции, зависящей произвольным образом от полей
и их частных производных. В частности, можно добавлять члены, имеющие вид
дивергенции. Отметим, что если сам лагранжиан равен полной дивергенции, то он не
приводит ни к каким уравнениям Эйлера--Лагранжа (получается тождество $0=0$).
Во-вторых, вместо одного набора полей $\vf^a$ можно выбрать другой:
$\vf^{\prime a}=\vf^{\prime a}(\vf)$. Если это преобразование полей невырождено,
то новая система уравнений Эйлера--Лагранжа будет эквивалентна старой. Пример
дают канонические преобразования в гамильтоновом формализме. Иногда можно
изменить даже число новых переменных в лагранжиане, и доказать биективность
пространств возникающих решений уравнений Эйлера--Лагранжа. В частности,
лагранжев и гамильтонов способы описания динамики точечных частиц, рассмотренные
в следующей главе, приводят к эквивалентным системам уравнений, но для разного
числа переменных.
\begin{com}
Выбор независимых переменных в действии, по которым проводится варьирование,
чрезвычайно важен, поскольку может привести к существенному упрощению
возникающих уравнений движения, особенно в нелинейных теориях. С другой стороны,
квантования моделей теории поля, основанные на различном выборе динамических
переменных, может привести к различным квантовым теориям. В последнем случае
теоретическим критерием выбора способа квантования является простота и
самосогласованность конечной квантовой теории. Этот вопрос актуален для
построения самосогласованной квантовой теории гравитации, которая в настоящее
время отсутствует.
\qed\end{com}
В релятивистских моделях математической физики, т.е.\ в моделях, инвариантных
относительно преобразований из группы Пуанкаре, действие записывается в виде
интеграла по всему пространству Минковского $\MR^{1,3}$. В этом случае уравнения
Эйлера--Лагранжа (\ref{eulage}) называются также {\em уравнениями движения},
поскольку описывают эволюцию системы во времени. При этом для уравнений движения
часто ставится не краевая задача, а задача Коши.
\index{Уравнения движения (equations of motion)}%
\index{Движения уравнения (equations of motion)}%

При рассмотрении моделей теории поля в пространстве Минковского $\MR^{1,3}$
действие, как правило, расходится. Например, действие в электродинамике для
электромагнитных волн расходится. Это связано с бесконечным объемом
интегрирования. Тем не менее с действием проводятся формальные выкладки, которые
приводят к уравнениям Эйлера--Лагранжа, которые локальны и хорошо определены.
При рассмотрении законов сохранения, связанных с первой теоремой Нетер, мы
рассматриваем либо поля в конечном объеме, либо достаточно быстро убывающие на
бесконечности.

Для корректной постановки вариационной задачи необходим глубокий анализ
уравнений Эйлера--Лагранжа. Во многих важных случаях эти уравнения настолько
сложны, что корректность постановки вариационной задачи доказать не удается.
Поэтому в теоретической физике выбор действия означает, как правило, просто
удобный способ задания уравнений движения для модели с заданными свойствами
симметрии, что, конечно, чрезвычайно важно.
\section{Первая теорема Нетер                                    \label{sfinet}}
\index{Первая теорема Нетер (first Nether's theorem)}%
\index{Теорема Нетер первая (first Nether's theorem)}%
В наиболее содержательных моделях математической физики функционал действия
инвариантен относительно глобальных или локальных преобразований симметрии. С
каждым преобразованием симметрии связан закон сохранения, что было установлено
Эмми Нетер \cite{Nether18R} в первой и второй теореме, соответственно, для
глобальных и локальных преобразований.

Пусть функционал действия (\ref{eacgen}) инвариантен относительно бесконечно
малых преобразований
\begin{align}                                                     \label{elorhu}
  x^\al&\mapsto x^{\prime\al}=x^\al+\dl x^\al,
\\                                                                \label{elotrf}
  \vf^a&\mapsto\vf^{\prime a}(x')=\vf^a(x)+\bar\dl\vf^a(x).
\end{align}
Рассмотрим независимые вариации координат и полей:
\begin{align}                                                     \label{evarxe}
  \dl x^\al&=\e^\Sa R_\Sa{}^\al(x,\vf,\pl\vf),
\\                                                                \label{evarfe}
  \bar\dl\vf^a&=\vf^{\prime a}(x')-\vf^a(x)=\e^\Sa R_\Sa{}^a(x,\vf,\pl\vf),
\end{align}
где $R_\Sa{}^\al(x,\vf,\pl\vf)$ и $R_\Sa{}^a(x,\vf,\pl\vf)$ -- некоторые
достаточно гладкие функции своих аргументов, которые называются
{\em генераторами} преобразований симметрии, а $\e^\Sa(x)$, $\Sa=1,2,\dots,\Sk$,
-- постоянные или локальные параметры преобразований, число которых зависит от
рассматриваемой модели. Мы говорим, что каждому значению индекса $\Sa$
соответствует одно преобразование симметрии.
\index{Генератор преобразования симметрии %
(generator of symmetry transformation)}%
\index{Преобразование симметрии генератор %
(generator of symmetry transformation)}%

Преобразования (\ref{elorhu}) и (\ref{elotrf}) уже рассматривались нами при
обсуждении вариационной задачи с подвижной границей. Разница заключается в том,
что сейчас у нас не один, а $\Sk$ параметров преобразования, которые, вдобавок,
могут зависеть от точки $x\in\MM$.

Начнем с доказательства первой теоремы Нетер, т.е.\ будем считать параметры
преобразований постоянными, $\e^\Sa=\const$. Под инвариантностью функционала
действия мы понимаем следующее равенство
\begin{equation}                                                  \label{einacf}
  S=\int_\MM\!\!\!dx\,L(x,\vf,\pl\vf)=\int_{\MM'}\!\!\!dx'\,L(x',\vf',\pl'\vf'),
\end{equation}
где интегрирование производится по ограниченной области $\MM\subset\MR^n$,
которая отображается в $\MM'$ при преобразовании (\ref{elorhu}).

Преобразования (\ref{evarxe}), (\ref{evarfe}) нетривиально действуют как на
поля, так и на координаты. В дальнейшем нам понадобится вариация формы функции
в данной точке $x\in\MM$ (см., раздел \ref{sinfct})
\begin{equation*}
  \dl\vf^a(x):=\vf^{\prime a}(x)-\vf^a(x),
\end{equation*}
которая определяется разностью значений полей после и до преобразования в точке
$x$. Она связана с вариацией (\ref{elotrf}) следующим соотношением
\begin{equation}                                                  \label{evarfo}
  \dl\vf^a(x)=\bar\dl\vf^a-\dl x^\al\pl_\al\vf^a
  =\e^\Sa(R_\Sa{}^a-R_\Sa{}^\al\pl_\al\vf^a).
\end{equation}
По-построению, для постоянных параметров преобразований вариация формы функции
$\dl$ перестановочна с операцией частного дифференцирования $\pl_\al$. Вариация
действия относительно преобразований (\ref{elorhu}), (\ref{elotrf}) была
вычислена ранее (\ref{efivaa}). Если выполнены уравнения Эйлера--Лагранжа, то
вариацию действия при постоянных параметрах преобразований симметрии запишем в
виде
\begin{equation}                                                  \label{eacvaf}
  \dl S=-\int_\MM\!\!\!dx\,\e^\Sa\pl_\al J_\Sa{}^\al,
\end{equation}
где
\begin{equation}                                                  \label{ethgen}
  J_\Sa{}^\al:=-\frac{\pl L}{\pl(\pl_\al\vf^a)}
  (R_\Sa{}^a-R_\Sa{}^\bt\pl_\bt\vf^a)-LR_\Sa{}^\al.
\end{equation}
Совокупность величин $J_\Sa{}^\al$, $\al=1,\dotsc,n$ можно рассматривать, как
компоненты некоторого вектора (точнее, векторной плотности) $J_\Sa$, который
называется {\em сохраняющимся током} для каждого преобразования симметрии с
параметром $\e^\Sa$. Из полученного выражения следует
\index{Сохраняющийся ток (conserved current)}%
\index{Ток сохраняющийся (conserved current)}%

\begin{theorem}[\bf Первая теорема Нетер]
Если действие (\ref{eacgen}) инвариантно относительно преобразований
(\ref{elorhu})--(\ref{evarfe}) с постоянными параметрами $\e^\Sa$, то для
каждого преобразования симметрии и любого решения уравнений Эйлера--Лагранжа
токи сохраняются:
\begin{equation}                                                  \label{ecudic}
  \pl_\al J_\Sa{}^\al=0,\qquad \Sa=1,\dotsc,\Sk.
\end{equation}
\end{theorem}
Заметим, что для сохранения тока достаточно глобальной инвариантности, когда
параметр преобразования не зависит от точек пространства-времени.

Поскольку лагранжиан не содержит производных выше первого порядка, то компоненты
токов в общем случае зависят только от координат, полей и их первых производных.

Закон сохранения (\ref{ecudic}) не нарушится, если к току (\ref{ethgen})
добавить слагаемое
\begin{equation}                                                  \label{earbcu}
  J^\prime_\Sa{}^\al=J_\Sa{}^\al+\pl_\bt f_\Sa{}^{\bt\al},
\end{equation}
где $f_\Sa{}^{\bt\al}=-f_\Sa{}^{\al\bt}$ -- произвольная антисимметричная по
индексам $\al$, $\bt$ функция. Чтобы не менять структуры тока (\ref{ethgen}),
будем считать, что она зависит только от координат $x^\al$, полей $\vf$ и их
первых производных $\pl\vf$. Это преобразование часто используется, чтобы
упростить выражения для токов.

Перепишем закон сохранения (\ref{ecudic}) в интегральной форме и используем
формулу Стокса
\begin{equation*}
  \int_\MM\!\!\!dx\,\pl_\al J_\Sa{}^\al
  =\int_{\pl\MM}\!\!\!ds_\al\, J_\Sa{}^\al=0,
\end{equation*}
где интегрирование ведется по многообразию $\MM$ и его краю $\pl\MM$. Пусть
на $\MM$ задана (псевдо-)риманова геометрия, т.е.\ метрика $g_{\al\bt}$ и
связность Леви--Чивиты $\widetilde\Gamma_{\al\bt}{}^\g$. Тогда, если индекс $\Sa$ не
преобразуется при преобразовании координат, то первый интеграл по
(псевдо-)риманову многообразию, можно переписать в ковариантной форме:
\begin{equation*}
  \int_\MM\!\!\!dx\,\pl_\al J_\Sa{}^\al
  =\int_\MM\!\!\!dx\,\vol\widetilde\nb_\al\left(\frac1\vol J_\Sa{}^\al\right),
\end{equation*}
где
\begin{equation*}
  \widetilde\nb_\al\left(\frac1\vol J_\Sa{}^\al\right)
  =\pl_\al\left(\frac1\vol J_\Sa{}^\al\right)
  +\widetilde\Gamma_{\al\bt}{}^\al\frac1\vol J_\Sa{}^\bt
\end{equation*}
-- ковариантная производная от вектора тока, и мы воспользовались формулой для
дивергенции (\ref{edivrf}).

Рассмотрим действие (\ref{eacgen}) в пространстве Минковского $\MR^{1,n-1}$.
Пусть $\lbrace x^\al\rbrace=\lbrace x^0,\Bx\rbrace $ -- декартова система
координат, и все поля достаточно быстро убывают на пространственной
бесконечности:
$$
  \underset{|\Bx|\rightarrow\infty}\lim\vf^a=0
$$
для всех моментов времени. Тогда, интегрируя уравнение (\ref{ecudic}) по области
пространства Минковского, которая ограниченна двумя пространственноподобными
сечениями $x^0_1=\const$ и $x^0_2=\const$, получим закон сохранения
\begin{equation}                                                  \label{econsl}
  Q_\Sa=\int_\MS\!d\Bx\,J_\Sa{}^0=\const,
\end{equation}
где $\MS$ -- произвольное сечение $x^0=\const$. Это означает, что каждому
преобразованию симметрии соответствует закон сохранения: для любого решения
уравнений движения, достаточно быстро убывающего на пространственной
бесконечности, интеграл (\ref{econsl}) не зависит от времени. Интеграл
(\ref{econsl}) называется {\em сохраняющимся зарядом}, соответствующим току
$J_\Sa{}^\al$. Если для уравнений движения поставлена задача Коши, то значение
заряда $Q_\Sa$ однозначно определяется начальными условиями.
\index{Заряд сохраняющийся (conserved charge)}%
\index{Сохраняющийся заряд (conserved charge)}%
\subsection{Тензор энергии-импульса                              \label{senmot}}
Предположим, что некоторая модель описывается набором полей $\vf^a$ в
пространстве Минковского $\MR^{1,n-1}$ с декартовыми координатами $x^\al$. При
этом метрика $\eta_{\al\bt}=\diag(+-\dotsc-)$ является заданной функцией в
действии, по которой варьирование не проводится. Пусть действие инвариантно
относительно трансляций
\begin{align}                                                     \label{etracz}
  &\dl x^\al=\e^\al=\const,
\\                                                                \label{etrafu}
  &\bar\dl\vf^a=0.
\end{align}
Для этого достаточно, чтобы лагранжиан модели $L(\vf,\pl\vf)$ не зависел явно от
координат. Для трансляций индекс $\Sa$ в (\ref{evarxe}) пробегает те же
значения, что и $\al$, генератор трансляций совпадает с символом Кронекера,
$R_\Sa{}^\al\mapsto\dl_\bt^\al$, и $R_\Sa{}^a=0$. В этом случае выражение
для тока (\ref{ethgen}) имеет вид
\begin{equation}                                                  \label{etenma}
  T_\al{}^\bt=\pl_\al\vf^a\frac{\pl L}{\pl(\pl_\bt\vf^a)}-\dl_\al^\bt L.
\end{equation}
Это выражение называется {\em тензором энергии-импульса} полей $\vf^a$. В силу
первой теоремы Нетер, он сохраняется:
\begin{equation}                                                  \label{enmocm}
  \pl_\bt T_\al{}^\bt=0.
\end{equation}
Тензор энергии-импульса (\ref{etenma}) будем называть {\em каноническим}.
\index{Тензор энергии-импульса (energy-momentum tensor)}%
\index{Энергии-импульса тензор (energy-momentum tensor)}%
\index{Канонический тензор энергии-импульса (canonical energy-momentum tensor)}%
\index{Тензор энергии-импульса канонический (canonical energy-momentum tensor)}%

Если лагранжиан модели является скалярным полем (функцией) относительно
глобальных преобразований Лоренца $\MO(1,n-1)$, то выражение (\ref{etenma})
представляет собой тензор второго ранга типа $(1,1)$. Ясно, что выражение для
$T_0{}^0$ всегда совпадает с плотностью гамильтониана для полей $\vf^a$, и это
оправдывает название ``канонический''.
\begin{com}
В общей теории относительности, основанной на псевдоримановой геометрии,
постулируется, что тензор Эйнштейна пропорционален тензору энергии-импульса
материи. При этом тензор энергии-импульса материи (\ref{edenmo}) определяется,
как вариационная производная действия для полей материи по метрике. При таком
определении тензор энергии-импульса всегда симметричен. Для скалярного поля
вариационная производная действия по метрике является ковариантным обобщением
тензора (\ref{etenma}). В других случаях связь двух определений сложнее и будет
обсуждаться в каждом конкретном случае.
\qed\end{com}
Вообще говоря, тензор энергии-импульса с опущенным верхним индексом $T_{\al\bt}$
не является симметричным. Если это так, то в ряде случаев можно провести
симметризацию, добавив соответствующую дивергенцию (\ref{earbcu}). Однако, это
не всегда возможно. Действительно, после добавления дивергенции получим новый
тензор энергии-импульса
\begin{equation*}
  T^\prime_{\al\bt}=T_{\al\bt}+\pl^\g f_{\al\g\bt}.
\end{equation*}
Из условия симметрии $T^\prime_{\al\bt}-T^\prime_{\bt\al}=0$ следуют уравнения
на неизвестную функцию $f_{[\al\g\bt]}=0$, в которые входят только полностью
антисимметричные компоненты,
\begin{equation*}
  \pl^\g f_{[\al\g\bt]}=-\frac12(T_{\al\bt}-T_{\bt\al}).
\end{equation*}
Таким образом мы имеем $n(n-1)/2$ дифференциальных уравнений на $n(n-1)(n-2)/3!$
неизвестных компонент. При $n=4$ возникает 6 уравнений на 4 неизвестные функции,
которые не всегда имеют решения.

Введем стандартные $3$-формы на координатных трехмерных гиперповерхностях в
четырехмерном пространстве-времени $\MR^{1,3}$:
\begin{equation}                                                  \label{efoths}
  ds_\al=\frac16\ve_{\al\bt\g\dl}dx^\bt\wedge dx^\g\wedge dx^\dl.
\end{equation}
Определим сохраняющийся во времени {\em ковектор энергии-импульса} с помощью
интеграла
\index{Ковектор энергии-импульса (energy-momentum covector)}%
\index{Энергии-импульса ковектор (energy-momentum covector)}%
\begin{equation}                                                  \label{enmove}
  P_\al=\int\limits_{x^0=\const}\!\!\!ds_\bt\,T_\al{}^\bt,
\end{equation}
где по индексу $\bt$ производится суммирование. Полученное выражение
(\ref{enmove}), по-построению, является ковектором относительно глобальных
лоренцевых вращений. В предположении, что все поля достаточно быстро убывают на
пространственной бесконечности ковектор энергии-импульса определяется одним
интегралом по пространству,
\begin{equation}                                                  \label{enmovs}
  P_\al=\int\limits_{x^0=\const}\!\!\!d\Bx\,T_\al{}^0.
\end{equation}
Выражение для нулевой компоненты $P_0$ совпадает с гамильтонианом системы полей
$\vf^a$, т.е.\ равно сохраняющейся полной энергии. Это оправдывает название
ковектора энергии-импульса. Пространственные компоненты тензора энергии-импульса
\begin{equation}                                                  \label{emoden}
  T_i{}^0=\frac{\pl L}{\pl(\pl_0\vf^a)}\pl_i\vf^a,\qquad i=1,\dotsc,n-1,
\end{equation}
определяют сохраняющийся полный импульс системы полей $\vf^a$
\begin{equation*}
  P_i=\int\limits_{x^0=\const}\!\!\!d\Bx\,T_i{}^0.
\end{equation*}
Полная энергия системы $P_0$ и каждая компонента полного импульса $P_i$
относительно данной декартовой системы сохраняются во времени. В другой
декартовой системе координат они тоже сохраняются, но имеют другие численные
значения.
\subsection{Тензор момента количества движения                   \label{sanmot}}
Пусть действие $S[\vf]$ в пространстве Минковского $\MR^{1,n-1}$ инвариантно
относительно лоренцевых вращений. Мы предполагаем, что набор полей $\vf^a$
преобразуется по некоторому, возможно, приводимому представлению группы Лоренца
$\MS\MO_\uparrow(1,n-1)$. Обозначим представление генераторов группы для полей
через $L_{\g\dl b}{}^a=-L_{\dl\g b}{}^a$. Тогда в инфинитезимальной форме
лоренцевы вращения примут вид
\begin{align}                                                     \label{elorox}
  \dl x^\al&=-x^\bt\om_\bt{}^\al
  =\sum_{\g<\dl}\om^{\g\dl}(x_\dl\dl_\g^\al-x_\g\dl_\dl^\al),
\\                                                                \label{eloroz}
  \bar\dl\vf^a&=\sum_{\g<\dl}\om^{\g\dl}L_{\g\dl b}{}^a\vf^b,
\end{align}
$\om^{\g\dl}=-\om^{\dl\g}$ -- параметры преобразований, которые предполагаются
постоянными. Для инвариантности действия достаточно, чтобы лагранжиан был
скалярным полем (функцией) от координат, полей и их производных. Для лоренцевых
вращений индекс $\Sa\mapsto(\al\bt)=-(\bt\al)$ представляет собой пару
антисимметричных векторных индексов.

Выражение для тока (\ref{ethgen}) приводит к следующему тензору
{\em момента количества движения}, который мы представим в виде суммы двух
слагаемых:
\index{Тензор момента количества движения (angular-momentum tensor)}%
\index{Момента количества движения тензор (angular-momentum tensor)}%
\begin{align}                                                     \label{eanmot}
  J_{\g\dl}{}^\al&=-\frac{\pl L}{\pl(\pl_\al\vf^a)}
  \left(L_{\g\dl b}{}^a\vf^b-(x_\dl\dl_\g^\bt
  -x_\g\dl_\dl^\bt)\pl_\bt\vf^a\right)-L(x_\dl\dl_\g^\al-x_\g\dl_\dl^\al)
\\                                                                     \nonumber
  &=M_{\g\dl}{}^\al+S_{\g\dl}{}^\al,
\end{align}
где введен {\em орбитальный} и {\em спиновый} моменты, соответственно,
\index{Орбитальный момент (orbital momentum)}%
\index{Момент орбитальный (orbital momentum)}%
\index{Спиновый момент (spin momentum)}%
\index{Момент спиновый (spin momentum)}%
\begin{align}                                                     \label{eorbmo}
  M_{\al\bt}{}^\g&:=x_\bt T_\al{}^\g-x_\al T_\bt{}^\g,
\\                                                                \label{espimo}
  S_{\al\bt}{}^\g&:=-\frac{\pl L}{\pl(\pl_\g\vf^a)}L_{\al\bt b}{}^a\vf^b.
\end{align}
Здесь $T_\al{}^\bt$ -- канонический тензор энергии-импульса (\ref{etenma}).
Оба объекта являются тензорами третьего ранга относительно преобразований
Лоренца. Обратим внимание, что орбитальный момент (\ref{eorbmo}) не инвариантен
относительно трансляций, т.к.\ явно зависит от координат. В противоположность
этому спиновый момент инвариантен относительно трансляций.

Если все поля $\vf^a$ являются скалярами относительно лоренцевых вращений, то
$L_{\al\bt b}{}^a=0$ и спиновый момент равен нулю, $S_{\al\bt}{}^\g=0$.

Допустим, что действие для некоторой системы полей инвариантно относительно
трансляций и лоренцевых вращений (группы Пуанкаре), и спиновый момент равен нулю
$S_{\al\bt}{}^\g=0$, как для скалярных полей. Тогда закон сохранения момента
количества движения принимает вид
$$
  \pl_\g M_{\al\bt}{}^\g=T_{\al\bt}-T_{\bt\al}
  +x_\bt\pl_\g T_{\al}{}^\g-x_\al\pl_\g T_\bt{}^\g.
$$
С учетом закона сохранения тензора энергии-импульса (\ref{enmocm}) отсюда
вытекает, что для такой системы ковариантный тензор энергии-импульса
симметричен:
\begin{equation}                                                  \label{esyenm}
  T_{\al\bt}=T_{\bt\al}.
\end{equation}

Так же, как и для канонического тензора энергии-импульса, для тензора момента
количества движения можно ввести полный момент системы. Для полей, достаточно
быстро убывающих на пространственной бесконечности, он равен интегралу по
пространству:
\begin{equation}                                                  \label{emoant}
  J_{\al\bt}=\int\limits_{x^0=\const}\!\!\!d\Bx\,J_{\al\bt}{}^0.
\end{equation}
Полный момент количества движения является антисимметричным тензором второго
ранга относительно преобразований Лоренца.
\begin{com}
Требование инвариантности моделей математической физики относительно
преобразований группы Пуанкаре имеет глубокий физический смысл и составляет
основное содержание специальной теории относительности. Инвариантность действия
относительно трансляций означает однородность пространства-времени. То есть все
точки пространства-времени равноправны, и законы природы имеют одинаковый вид в
декартовых координатах с произвольно выбранным началом. Инвариантность
относительно преобразований Лоренца означает изотропность пространства-времени.
То есть равноправие всех направлений и одинаковый вид законов природы в
декартовых координатах с произвольной ориентацией осей. Законы сохранения
энергии-импульса и момента количества движения к настоящему времени нашли
многочисленные экспериментальные подтверждения в различных областях физики.
Поэтому инвариантность фундаментальных моделей математической физики
относительно действия группы Пуанкаре следует считать экспериментально
установленным фактом.

Помимо этого требование инвариантности функционала действия относительно
преобразований группы Пуанкаре в квантовой теории поля означает, что все
элементарные частицы должны описываться полями, принадлежащими одному из
неприводимых представлений группы Пуанкаре, которые характеризуются массой и
спином. Использование этих понятий в экспериментальной физике элементарных
частиц также чрезвычайно плодотворно. Это также можно рассматривать, как
экспериментальное подтверждение инвариантности законов природы относительно
преобразований группы Пуанкаре.
\qed\end{com}
\section{Вторая теорема Нетер                                    \label{sfinez}}
\index{Вторая теорема Нетер (second Nether's theorem)}%
\index{Теорема Нетер вторая (second Nether's theorem)}%
Рассмотрим действие (\ref{eacgen}), которое инвариантно относительно
преобразований (\ref{elorhu})--(\ref{evarfe}) с локальными параметрами
$\e^\Sa(x)$, зависящим от точек пространства-времени. Мы допускаем, что эти
преобразования могут зависеть от частных производных $\pl_\al\e^\Sa$ первого и
более высокого порядка. Чтобы упростить формулы будем использовать обозначения
Девитта \cite{DeWitt65R}, т.е.\ суммирование по индексу $\Sa$ в формулах
(\ref{elorhu}), (\ref{elotrf}) подразумевает интегрирование, а генераторы
локальных преобразований рассматриваются, как двухточечные функции, содержащие
$\dl$-функции и (или) их производные.
\begin{exa}
Калибровочное преобразование в электродинамике
$$
  \dl A_\al=\pl_\al\e
$$
будем записывать в виде
\begin{equation}                                                  \label{egtred}
  \dl A_\al=\e R_\al=\pl_\al\int\!\!dx'\,\e(x')\dl(x'-x),
\end{equation}
где
\begin{equation}                                                  \label{egegte}
  R_\al(x',x):=\frac\pl{\pl x^\al}\dl(x'-x). \qed
\end{equation}
\end{exa}
\begin{exa}
Бесконечно малые общие преобразования координат для электромагнитного
поля (\ref{eintof}) можно записать в виде
\begin{align}                                                     \label{einfcx}
  \dl x^\al&=\e^\al=\e^\bt R_\bt{}^\al,
\\                                                                \label{einfca}
  \dl A_\al&=-\pl_\al\e^\bt A_\bt-\e^\bt\pl_\bt A_\al=\e^\bt N_{\bt\al},
\end{align}
где
\begin{align}                                                     \label{egectx}
  R_{\bt}{}^\al&:=\dl_\bt^\al\dl(x'-x),
\\                                                                \label{egecta}
  N_{\bt\al}&:=F_{\al\bt}(x')\dl(x'-x)
  -A_\bt(x')\frac\pl{\pl x^\al}\dl(x'-x). \qed
\end{align}
\end{exa}
\begin{defn}
Преобразования полей (\ref{evarfe}) с локальными параметрами $\e^\Sa(x)$
называются {\em калибровочными}.
\qed\end{defn}
\index{Калибровочное преобразование (gauge transformation)}%
\index{Преобразования калибровочные (gauge transformation)}%

Рассмотрим одну из вариационных задач. Будем считать, что параметры $\e^\Sa$ и
их производные равны нулю на границе области. Тогда инвариантность действия
относительно калибровочных преобразований можно записать в виде
\begin{equation}                                                  \label{eacvar}
  \dl S=\int\!\!dx\,\dl\vf^aS,{}_a
  =\int\!\!dx\,\e^\Sa(R_\Sa{}^a-R_\Sa{}^\al\pl_\al\vf^a)S,{}_a=0.
\end{equation}
При этом были отброшены все граничные слагаемые. Отсюда следует
\begin{theorem}[\bf Вторая теорема Нетер]                         \label{tdepeq}
Если функционал действия (\ref{eacgen}) инвариантен относительно калибровочных
преобразований, которые параметризуются $\Sk$ произвольными функциями
$\e^\Sa(x)$, $\Sa=1,\dotsc,\Sk$, то уравнения Эйлера--Лагранжа удовлетворяют
$\Sk$ тождествам:
\begin{equation}                                                  \label{edepel}
  (R_\Sa{}^a-R_\Sa{}^\al\pl_\al\vf^a)S,{}_a=0.
\end{equation}
\end{theorem}
\begin{com}
В формулировке теоремы мы отбросили предположение о том, что параметры
преобразований и их производные равны нулю на границе. Если это не так, то
зависимость уравнений Эйлера--Лагранжа все равно сохранится. В этом случае из
требования инвариантности действия появятся дополнительные следствия для
граничных условий, которые мы не рассматриваем.
\qed\end{com}
Напомним, что в линейном соотношении между уравнениями движения (\ref{edepel})
суммирование по индексу $a$ предполагает интегрирование. Отсюда следует, что
если калибровочные преобразования зависят от частных производных $l$-того
порядка от параметра преобразования, то соотношения (\ref{edepel}) представляют
собой систему $K$ линейных дифференциальных уравнений в частных производных
$l$-того порядка относительно вариационных производных $S,{}_a$.

Вторая теорема Нетер утверждает, что в калибровочных моделях, а также моделях,
инвариантных относительно общих преобразований координат, не все уравнения
движения являются линейно независимыми. Это указывает на то, что в решениях
задачи Коши будет содержаться функциональный произвол, т.к.\ количества
уравнений недостаточно для однозначного определения решений по начальным данным.

Для доказательства теоремы существенно, что параметры преобразований $\e^\Sa(x)$
является произвольными функциями, т.к.\ только в этом случае подынтегральное
выражение в (\ref{eacvar}) согласно основной лемме вариационного исчисления
должно обращаться в нуль.
\begin{exa}
Проведем аналогию с теорией функций многих переменных. Пусть $f=f(x)$ -- функция
$n$ переменных $x=\lbrace x^\al\rbrace$. Аналогом вариационной производной
действия в таком случае является обычная частная производная $\pl_\al f$.
Допустим, что $f$ инвариантна относительно калибровочных преобразований
$\dl x^\al=\e X^\al$, где $\e=\e(x)$ -- параметр преобразования и $X^\al$ --
векторное поле (генератор калибровочного преобразования), которое предполагается
отличным от нуля. Тогда ``зависимость уравнений движения'' сводится к линейной
зависимости частных производных $X^\al\pl_\al f=0$. Поэтому функция $f$
постоянна вдоль интегральной кривой $x(t)$ векторного поля $X^\al$:
\begin{equation*}
  f\big(x(t)\big)=\const,\qquad \dot x^\al=X^\al.
\end{equation*}
Это значит, что локальный экстремум $\pl_\al f=0$ достигается не в точке, а на
интегральной кривой $x(t)$.
\qed\end{exa}
Из второй теоремы Нетер следует, что функционал действия для калибровочных
моделей достигает экстремального значения не на отдельных функциях, а на классах
функций, которые связанных между собой калибровочными преобразованиями.

Если некоторая модель инвариантна относительно локальных преобразований, то она,
в частности, инвариантна относительно тех же преобразований с постоянными
параметрами. Это значит, что токи (\ref{ethgen}) приводят к законам сохранения
(\ref{econsl}) и для локальных преобразований. Поэтому в моделях, инвариантных
относительно локальных преобразований можно применить обе теоремы Нетер. При
этом первая теорема дает выражения для сохраняющихся токов, а вторая --
зависимость уравнений движения. В общем случае это не одно и то же.
\begin{exa}
Рассмотрим модели математической физики, инвариантные относительно общих
преобразований координат. Пусть действие $S=S(g,\Gamma)$ зависит только от метрики
$g_{\al\bt}$ и аффинной связности $\Gamma_{\al\bt}{}^\g$. Обозначим вариационные
производные действия следующим образом:
\begin{equation}                                                  \label{eqmmco}
  \vol S,{}^{\al\bt}:=\frac{\dl S}{\dl g_{\al\bt}},\qquad
  \vol S,{}^{\al\bt}{}_\g:=\frac{\dl S}{\dl \Gamma_{\al\bt}{}^\g}.
\end{equation}
Здесь мы явно ввели в качестве множителя определитель репера
$\vol=\det e_\al{}^a$, поскольку вариационные производные так же, как и
лагранжиан, являются тензорными плотностями степени $-1$. Инвариантность
действия относительно общих преобразований координат означает равенство нулю
вариации
$$
  \dl S=\int dx\vol(S,{}^{\al\bt}\dl g_{\al\bt}+S,{}^{\al\bt}{}_\g
  \dl\Gamma_{\al\bt}{}^\g)=0.
$$
Подставляя сюда вариации метрики (\ref{eitcms}) и связности (\ref{ecoitc}) и
интегрируя по частям (\ref{eintpa}), получим тождества
\begin{multline}                                                  \label{edepgc}
  2\widetilde\nb_\al S,{}^\al{}_\g-\nb_\bt\nb_\al S,{}^{\al\bt}{}_\g+
\\
  +\nb_\bt(T_\al+{\textstyle\frac12}Q_\al)S,{}^{\al\bt}{}_\g
 +(T_\al+{\textstyle\frac12}Q_\al)\nb_\bt(S,{}^{\al\bt}{}_\g+S,{}^{\bt\al}{}_\g)
  -\nb_\al S,{}^{\al\bt}{}_\dl T_{\bt\g}{}^\dl-\qquad
\\
  -(T_\al+{\textstyle\frac12}Q_\al)
  (T_\bt+{\textstyle\frac12}Q_\bt)S,{}^{\al\bt}{}_\g
  +(T_\al+{\textstyle\frac12}Q_\al)T_{\bt\g}{}^\dl S,{}^{\al\bt}{}_\dl
  +S,{}^{\al\bt}{}_\dl R_{\al\g\bt}{}^\dl=0,
\end{multline}
где $\widetilde\nb_\al$ и $\nb_\al$ -- ковариантные производные соответственно
со связностью Леви--Чивиты $\widetilde\Gamma_{\al\bt}{}^\g$ и аффинной связностью
$\Gamma_{\al\bt}{}^\g$, а подъем и опускание индексов производится с помощью метрики
$g_{\al\bt}$. Таким образом, в моделях, инвариантных относительно общих
преобразований координат, уравнения движения удовлетворяют $n$ линейным
дифференциальным тождествам.

В (псевдо-)римановой геометрии, когда гравитационная часть действия зависит
только от метрики, эти тождества значительно упрощаются:
\begin{equation}                                                  \label{erdeeq}
  \widetilde\nb_\al S,{}^\al{}_\bt=0.
\end{equation}
В общей теории относительности для действия Гильберта--Эйнштейна справедливо
равенство
\begin{equation}                                                  \label{eeihid}
  \widetilde\nb_\al\left(\widetilde R^\al{}_\bt
  -\frac12\dl^\al_\bt\widetilde R\right)=0.
\end{equation}
Это тождество совпадает со свернутыми тождествами Бианки (\ref{ebieit}).
\qed\end{exa}
\begin{exa}
Модели гравитации, построенные в рамках геометрии Римана--Картана, в переменных
Картана инвариантны относительно общих преобразований координат и локальных
преобразований Лоренца. Как следствие второй теоремы Нетер не все уравнения
движения являются независимыми, поскольку удовлетворяют тождествам. Пусть
действие $S=S(e,\om)$ зависит только от репера $e_\al{}^a$ и лоренцевой
связности $\om_\al{}^{ab}$. Обозначим вариационные производные следующим
образом:
\begin{equation}                                                  \label{eqreld}
  \vol S,{}^\al{}_a:=\frac{\dl S}{\dl e_\al{}^a},\qquad
  \vol S,{}^\al{}_{ab}:=\frac{\dl S}{\dl\om_\al{}^{ab}}.
\end{equation}
Бесконечно малые преобразования Лоренца для репера и лоренцевой связности
при локальных лоренцевых вращениях имеют вид
\begin{align}                                                     \label{elrrep}
  \dl e_\al{}^a&=-e_\al{}^b\om_b{}^a,
\\                                                                \label{elrloc}
  \dl\om_{\al a}{}^b&=~\om_a{}^c\om_{\al c}{}^b
  -\om_{\al a}{}^c\om_c{}^b+\pl_\al\om_a{}^b.
\end{align}
Отсюда вытекает следующая зависимость уравнений движения
\begin{equation}                                                  \label{eqdelo}
  \widetilde\nb_\al S,{}^\al{}_{ab}+\frac12(S,{}_{ab}-S,{}_{ba})=0,
\end{equation}
где переход от греческих индексов к латинским осуществляется с помощью репера.
Полученная зависимость соответствует инвариантности действия относительно
локальных лоренцевых вращений.

Вариации полей $e$ и $\om$ при общих преобразованиях координат имеют вид
(\ref{eintre}) и (\ref{eintrl}). Поэтому инвариантность действия относительно
общих преобразований координат приводит к тождествам:
\begin{equation}                                                  \label{elodel}
  \nb_\al S,{}^\al{}_\bt+T_\al S,{}^\al{}_\bt+S,{}^\al{}_a T_{\al\bt}{}^a
  +S,{}^\al{}_{ab}R_{\al\bt}{}^{ab}=0,
\end{equation}
где мы учли полученное ранее тождество (\ref{eqdelo}).

При добавлении к гравитационному действию слагаемых, зависящих от других полей
(полей материи), тождества, которым удовлетворяют уравнения движения меняются,
т.к.\ необходимо учитывать вариации всех полей.
\qed\end{exa}
\section{Эффективное действие                                    \label{sredac}}
При исследовании моделей математической физики, действие которых зависит от
нескольких полей, иногда удается решить часть уравнений Эйлера--Лагранжа явно.
В этом случае вариационную задачу можно свести к новому эффективному действию,
зависящему от меньшего числа переменных. В настоящем разделе мы докажем простую
теорему, позволяющую строить эффективное действие в случае вариационной задачи с
фиксированными граничными условиями. То есть будем пренебрегать всеми граничными
слагаемыми. Обобщение на более сложные случаи будет ясно из дальнейшего
рассмотрения.

Начнем с простейшего случая.
Пусть на ограниченной области $\MM\subset\MR^n$ заданы два скалярных поля
$\vf$ и $\psi$. Предположим, что функция $\psi=\psi(x,\vf,\pl\vf,\pl^2\vf)$ в
каждой точке $x\in\MM$ задана как функция $\vf$, ее первых и вторых частных
производных: $\pl_\al\vf$ и $\pl_\al\pl_\bt\vf$. Представим значение функции
$\psi(x):=\psi\big[x,\vf(x),\pl\vf(x),\pl^2\vf(x)\big]$ в точке $x\in\MM$ в виде
функционала, использую $\dl$-функцию,
$$
  \psi(x)=\int_\MM\!\!\!dy\,\psi(y)\dl(y-x).
$$
Вариация функционала $\psi(x)$, вызванная вариацией $\dl\vf$, имеет вид
$$
  \dl\psi(x)=\int_\MM\!\!\!dy\left(
  \frac{\pl\psi}{\pl\vf}\dl\vf+\frac{\pl\psi}{\pl(\pl_\al\vf)}\dl(\pl_\al\vf)
  +\frac{\pl\psi}{\pl(\pl_\al\pl_\bt\vf)}\dl(\pl_\al\pl_\bt\vf)\right)\dl(y-x).
$$
Проинтегрировав второе и третье слагаемые по частям, получим выражение для
вариационной производной
\begin{equation}                                                  \label{evarde}
  \frac{\dl\psi(x)}{\dl\vf(y)}=\frac{\pl\psi}{\pl\vf}\dl(y-x)
  -\frac\pl{\pl y^\al}
  \left(\frac{\pl\psi}{\pl(\pl_\al\vf)}\dl(y-x)\right)
  +\pl_\al\pl_\bt\left(\frac{\pl\psi}{\pl(\pl_\al\pl_\bt\psi)}\dl(y-x)\right),
\end{equation}
где в правой части $\psi=\psi(y)$ и $\vf=\vf(y)$.

Теперь обсудим вариационную задачу. Пусть действие $S[\vf,\psi]$ зависит от двух
функций $\vf$ и $\psi$. Тогда из принципа наименьшего действия следуют два
уравнения Эйлера--Лагранжа:
\begin{align}                                                     \label{eillaf}
  \frac{\dl S}{\dl\vf}&=0,
\\                                                                \label{eillap}
  \frac{\dl S}{\dl\psi}&=0.
\end{align}
Допустим, что второе уравнение Эйлера--Лагранжа допускает общее решение для
$\psi$, как функции от $\vf$ и ее производных:
\begin{equation}                                                  \label{egesoe}
  \psi=\psi(x,\vf,\pl\vf,\pl^2\vf).
\end{equation}
При этом мы предполагаем, что общее решение не имеет особенностей. Если действие
зависит только от самих функций и их первых производных, то в общее решение
будут входить производные от $\vf$ не выше второго порядка. Поскольку уравнение
Эйлера--Лагранжа (\ref{eillap}) является дифференциальным уравнением в частных
производных, то общее решение зависит также от некоторого набора произвольных
функций и постоянных. Часть этих произвольных функций и постоянных фиксируется,
если это возможно, граничными условиями
$\psi|_{\pl\MM}=\psi_0$ и $\vf|_{\pl\MM}=\vf_0$. Используем полученное решение
для построения нового {\em эффективного} действия
\index{Эффективное действие (effective action)}%
\index{Действие эффективное (effective action)}%
\begin{equation}                                                  \label{effacf}
  S_{\text{eff}}[\vf]:=S[\vf,\psi(\vf)],
\end{equation}
которое зависит только от одной функции $\vf$. Тогда уравнение Эйлера--Лагранжа
для $\vf$ связано со старыми уравнениями (\ref{eillaf}), (\ref{eillap}) простым
соотношением
\begin{equation}                                                  \label{eilaff}
  \frac{\dl S_{\text{eff}}}{\dl\vf(x)}=\frac{\dl S}{\dl\vf(x)}
  +\left.\frac{\dl S}{\dl\psi(y)}\frac{\dl\psi(y)}{\dl\vf(x)}
  \right|_{\psi=\psi(\vf)}=0,
\end{equation}
где во втором слагаемом подразумевается интегрирование по аргументу поля
$\psi(y)$, которое снимается $\dl$-функцией в вариационной производной. Ясно,
что второе слагаемое равно нулю, если выполнено уравнение Эйлера--Лагранжа для
$\psi$ (\ref{eillap}).

Проведенные вычисления остаются в силе и в том случае, когда мы рассматриваем
наборы полей $\vf=\lbrace\vf^a\rbrace$, $a=1,\dotsc,\Sn$ и $\psi^\Sa$,
$\Sa=1,\dotsc,m$. Отсюда следует
\begin{theorem}                                                   \label{teffac}
Пусть дано действие $S[\vf,\psi]$, зависящее от двух наборов полей $\vf$ и
$\psi$. Тогда множество множество решений уравнений Эйлера--Лагранжа для
вариационной задачи с заданными граничными условиями совпадает с множеством
решений уравнений Эйлера--Лагранжа для эффективного действия (\ref{effacf}),
дополненным выражением $\psi$ через $\vf$ (\ref{egesoe}).
\end{theorem}
\begin{com}
Если рассматриваются другие вариационные задачи, то слова ``множество
экстремалей'' нужно заменить на ``''.
\qed\end{com}

Эта теорема важна, поскольку позволяет строить эффективное действие, которое
зависит от меньшего числа полей, подставляя решение части системы уравнений
Эйлера--Лагранжа непосредственно в исходное действие.

При доказательстве теоремы \ref{teffac} мы предположили, что всеми граничными
слагаемыми можно пренебречь. В полевых моделях математической физики это не
всегда так. В разделе \ref{sexava} будет построен пример, где подстановка
решения части уравнений Эйлера--Лагранжа в действие не воспроизводит оставшиеся
уравнения движения. Это связано с нетривиальной ролью граничных слагаемых в
действие для полевых моделей.
\section{Редуцированное действие}
В настоящем разделе мы рассмотрим еще один способ сведения сложной вариационной
задачи к более простой.
Пусть задано действие $S[\vf]$. Рассмотрим для него вариационную задачу с
фиксированными граничными условиями. Как правило, уравнения Эйлера--Лагранжа
(уравнения движения) настолько сложны, что не позволяют найти все решения. В
таких случаях для нахождения частных решений делают упрощающие предположения:
решение уравнений движения ищется в определенном классе функций. Например,
ищутся статические или сферически симметричные решения. Чтобы найти стационарную
точку действия, упрощающую подстановку, которую часто называют {\em анзац}
(от немецкого ansatz $\simeq$ исходное математическое выражение), следует
производить в уравнения Эйлера--Лагранжа (\ref{eulage}), а не в действие.
В этом случае найденное точное решение уравнений движения действительно
будет стационарной точкой исходного действия.
\index{Анзац (ansatz)}%

Однако в ряде случаев подстановки можно производить непосредственно в действие.
Это означает следующее. Пусть в результате некоторых упрощающих предположений
исходный набор функций $\vf^a$, $a=1,\dotsc,\Sn$, будет выражен через меньшее
число независимых функций $\psi^\Sa$, $\Sa=1,\dotsc,m$, и координаты:
\begin{equation}                                                  \label{eoldne}
  \vf^a=\vf^a(x,\psi).
\end{equation}
При этом функции $\psi$ могут зависеть от меньшего числа координат. В результате
подстановки будет получено новое {\em редуцированное действие}
\index{Редуцированное действие (reduced action)}%
\index{Действие редуцированное (reduced action)}%
\begin{equation}                                                  \label{eredac}
  S_\text{red}[\psi]:=S[\vf(\psi)],
\end{equation}
зависящее от меньшего числа независимых полей.

Допустим, что найдено решение уравнений движения для редуцированного действия
\begin{equation*}
  \frac{\dl S_\text{red}}{\dl\psi^\Sa}=0.
\end{equation*}
Тогда функции (\ref{eoldne}), как правило, не будут удовлетворять исходным
уравнениям (\ref{eulage}). Тем не менее в ряде случаев исходные уравнения все же
будут удовлетворены. Это замечательные случаи, которые позволяют существенно
упростить вычисления. Вместе с этим наличие редуцированного действия помогает в
анализе свойств рассматриваемой модели.

Опишем достаточные условия для возможности подстановки (\ref{eoldne})
непосредственно в действие. Пусть на многообразии $\MM$ действует группа Ли
преобразований $\MG$ справа:
\begin{equation*}
  \MM\times\MG\ni\quad x,g\mapsto xg\qquad\in\MM.
\end{equation*}
Предположим, что набор полей $\vf^a(x)$ при этом преобразуются по некоторому
представлению $T(g)_a{}^b$ группы Ли $\MG$:
\begin{equation}                                                  \label{egrtrv}
  \vf^{\prime a}(xg)=\vf^b(x)T(g)_b{}^a,
\end{equation}
где штрихом обозначены новые поля на $\MM$, полученные в результате действия
группы преобразований $\MG$. Поля $\vf^a(x)$ называются
$\MG$-{\em инвариантными}, если в равенстве (\ref{egrtrv}) можно убрать штрих в
левой части. То есть выполнено условие
\index{$\MG$-инвариантные поля ($\MG$-invariant fields)}%
\begin{align}
  \vf^a(xg)&=\vf^b(x)T(g)_b{}^a,                                       \nonumber
\\ \intertext{или}
  \vf^a(x)&=\vf^b(xg^{-1})T(g)_b{}^a.                             \label{einvfi}
\end{align}
Эти условия представляют собой уравнения, определяющие $\MG$-инвариантные
поля на многообразии $\MM$.

\begin{theorem}[\bf Принцип Коулмена]
Пусть исходный функционал действия $S[\vf]$ инвариантен относительно действия
группы Ли преобразований (\ref{egrtrv}),
\begin{equation*}
  S[\vf']=S[\vf].
\end{equation*}
Допустим, что множество всех $\MG$-инвариантных полей на $\MM$ параметризуется
некоторым набором функций $\psi^\Sa$, которые могут зависеть от меньшего числа
координат и не определяются из уравнений (\ref{einvfi}). В результате
$\MG$-инвариантные функции будут представлены в виде (\ref{eoldne}). Тогда поля
(\ref{eoldne}), построенные для стационарных точек редуцированного действия
(\ref{eredac}), будут удовлетворять уравнениям Эйлера--Лагранжа исходного
действия.
\end{theorem}
Это утверждение известно, как {\em принцип Коулмена}.
Оно было высказано Коулменом и проиллюстрировано на нескольких примерах
\cite{Colema75} (см.\ также \cite{Faddee77R}). Строгое доказательство вместе с
ограничениями на его применимость было дано в статье \cite{Palais79}. Этот
результат затем был обобщен в работах \cite{LadKap83R,SchSim89}. Другая его
формулировка, на языке теории стратов была дана еще до Коулмена \cite{MicRad68}.
В настоящее время доказано, что принцип Коулмана справедлив для всех компактных
групп преобразований, полупростых групп, а также для унитарных представлений
некомпактных групп.
\index{Принцип Коулмана (Coleman principle)}%
\index{Коулмана принцип (Coleman principle)}%
\chapter{Канонический формализм                                  \label{scanfp}}
Трудно переоценить роль канонического (гамильтонова) формализма в классической и
квантовой механике, а также в теории поля. Он предоставляет наиболее мощные
методы интегрирования уравнений движения и является основой для канонического
квантования различных моделей математической физики. К его недостаткам относится
явное нарушение Лоренц-инвариантности моделей теории поля, поскольку время в
гамильтоновом формализме играет выделенную роль. Это усложняет вычисления,
проводимые в рамках теории возмущений. Однако принципиальные вопросы, связанные
с физической интерпретацией математических моделей, невозможно решить без
обращения к гамильтоновой формулировке. В настоящей главе рассматривается
канонический формализм для системы точечных частиц, формализм Дирака для систем
со связями и его обобщение на теорию поля.
\section{Канонический формализм в механике точечных частиц}
В настоящем разделе мы во многом следуем \cite{Arnold89R}
\subsection{Преобразование Лежандра                              \label{sletra}}
\begin{figure}[h,b,t]
\hfill\includegraphics[width=.3\textwidth]{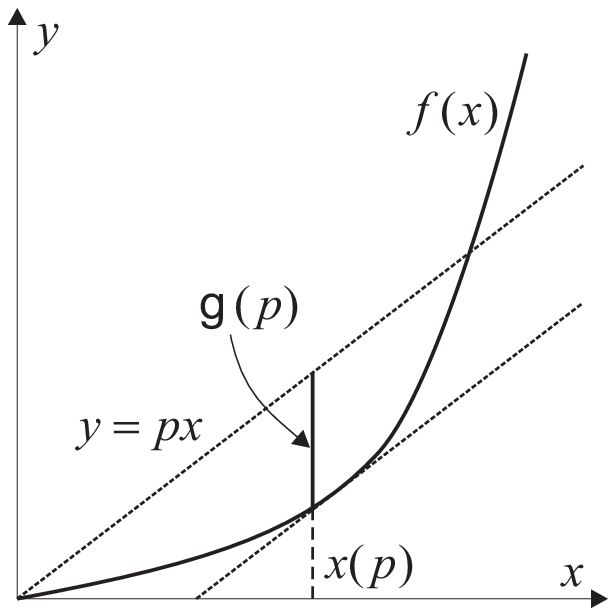}
\hfill {}
\centering\caption{Преобразование Лежандра $g(p)$ выпуклой функции $f(x)$.}
\label{fletra}
\end{figure}

\index{Преобразование Лежандра (Legendre's transformation)}%
\index{Лежандра преобразование (Legendre's transformation)}%
\index{Выпуклая функция (convex function)}%
\index{Функция выпуклая (convex function)}%
Рассмотрим {\em выпуклую функцию} $y=f(x)$ на интервале
$-\infty\le a<x<b\le\infty$, т.е.\ функцию, у которой $f''(x)>0$ при всех
$x\in(a,b)$. Преобразованием Лежандра функции $f$ на интервале $(a,b)$
называется новая функция $g(p)$ нового переменного $p$, которая строится
следующим образом (см.\ рис.~\ref{fletra}). Нарисуем на плоскости $x,y$ график
функции $f$.
Рассмотрим прямую $y=px$, где $p$ -- фиксированное число. Найдем точку $x(p)$, в
которой кривая дальше всего от прямой по вертикали, т.е.\ функция
$F(x,p):=px-f(x)$ имеет максимум по $x$ при фиксированном $p$. Тогда,
по-определению, $g(p):=F\big(x(p),p\big)$. Функция $g(p)$ определена на
некотором интервале $-\infty\le c<p<d\le\infty$.

Точка $x(p)$ определяется из условия экстремума $\pl F/\pl x=0$ или
$$
  p=f'(x)
$$
Ввиду выпуклости $f$, если это уравнение имеет решение, то оно единственно.

\begin{exa}
Нетрудно проверить, что функция $f(x)=\frac{mx^2}2$, $m=\const>0$ выпукла на
всей вещественной оси $x\in\MR$ и $p=mx$. Ее преобразование Лежандра определено
для всех $p\in\MR$ и имеет вид
\begin{equation*}                                                    \tag*{\qed}
  g(p)=\left.(xp-f)\right|_{x=p/m}=\frac{p^2}{2m}
\end{equation*}
\renewcommand{\qed}{}\end{exa}

Пусть функция $f$ достаточно гладкая. Тогда преобразование Лежандра переводит
выпуклые функции в выпуклые. Это значит, что преобразование Лежандра можно
применить дважды. Можно доказать \cite{Arnold89R}, что преобразование Лежандра
{\em инволютивно}, т.е.\ его квадрат равен тождественному преобразованию.
\index{Инволютивное преобразование (involute transformation)}%
\index{Преобразование инволютивное (involute transformation)}%

По-определению, $F(x,p):=px-f(x)\le g(p)$ для всех $x$ и $p$. Отсюда вытекает
{\em неравенство Юнга}
\index{Неравенство Юнга (Young inequality)}%
\index{Юнга неравенство (Young inequality)}%
$$
  px\le f(x)+g(p)
$$

Преобразование Лежандра без труда обобщается на функции нескольких переменных.
Пусть $f(\Bx)$ -- выпуклая функция нескольких переменных
$\Bx=(x^1,\dotsc,x^n)\in\MR^n$, т.е.\ квадратичная форма
$dx^\al dx^\bt\pl^2_{\al\bt}f$, $\al=1,\dotsc,n$, положительно определена в
некоторой окрестности $\MU\subset\MR^n$. Тогда преобразованием Лежандра
называется функция $g(\Bp)$ того же числа переменных $\Bp=(p_1,\dotsc,p_n)$,
которая строится аналогично случаю одного переменного:
$$
  g(\Bp)=F\big(\Bx(\Bp),\Bp\big)=\underset\Bx\max F(\Bx,\Bp),
$$
где
$$
  F(\Bx,\Bp)=x^\al p_\al-f(\Bx),\qquad \text{и}\qquad
  p_\al=\frac{\pl f}{\pl x^\al}.
$$
Преобразование Лежандра $g(\Bp)$ определено в некоторой окрестности
$\Bp\in\MV\subset\MR^n$, которая определяется исходной функцией $f(\Bx)$.

В этом определении мы различаем верхние и нижние индексы по следующей причине.
Поскольку $\Bx$ -- точка многообразия $\MR^n$, то индексы ее координат мы пишем
сверху. По-определению, $f$ -- функция на $\MR^n$. Поэтому набор переменных
$p_\al$ определяет компоненты некоторого ковектора (1-формы). Отметим также, что
в этом определении все координаты $x^\al$ равноправны.
\begin{exa}
Преобразованием Лежандра положительно определенной квадратичной формы
\begin{equation*}
  f(\Bx)=\frac12g_{\al\bt}x^\al x^\bt,
\end{equation*}
где $g_{\al\bt}=g_{\bt\al}$ -- постоянная матрица, снова является положительно
определенная квадратичная форма
\begin{equation*}
  g=\left.\big(x^\al p_\al-f(\Bx)\big)\right|_{x^\al=g^{\al\bt}p_\bt}
  =\frac12g^{\al\bt}p_\al p_\bt,
\end{equation*}
где $g^{\al\bt}$ -- матрица, обратная к $g_{\al\bt}$. При этом значения обеих
форм в соответствующих точках совпадают:
\begin{equation*}
  f\big(\Bx(\Bp)\big)=g(\Bp),\qquad g\big(\Bp(\Bx)\big)=f(\Bx).
\end{equation*}
Обе функции $f(\Bx)$ и $g(\Bp)$ определены на всем евклидовом пространстве
$\MR^n$.
\qed\end{exa}
\subsection{Гамильтонова динамика точечных частиц                \label{shadyp}}
Механика точечных частиц может быть описана на двух эквивалентных языках:
лагранжевом и гамильтоновом. Каждый подход имеет свои преимущества и недостатки.
В лагранжевом подходе совокупность $\Sn$ частиц описывается обобщенными
координатами $q^i(t)$, $i=1,\dotsc,\Sn$, зависящих от времени $t$. Если каждая
частица движется в трехмерном пространстве, то $q^i$ представляет собой
трехмерный вектор в евклидовом пространстве $\MR^3$ для каждого значения индекса
$i$. Для определенности будем считать, что каждая частица движется в одномерном
пространстве. Тогда совокупность всех координат $q^i$ можно также рассматривать,
как координаты одной частицы в {\em конфигурационном} пространстве $\MR^\Sn$.
Размерность конфигурационного пространства называется {\em числом степеней
свободы} механической системы. Говорят, что механическая система имеет $\Sn$
степеней свободы.
\index{Конфигурационное пространство (configuration space)}%
\index{Пространство конфигурационное (configuration space)}%
\index{Число степеней свободы (the number of degrees of freedom)}%

Для простоты обозначений условимся считать, что символ с индексом $q^i$
обозначает $i$-ю координату частицы, а по повторяющимся индексам производится
суммирование. Если индекс отсутствует, то символ $q$ обозначает весь набор
координат $q:=(q^1,\dotsc,q^\Sn)$. В лагранжевом подходе уравнения движения
имеют второй порядок, поэтому будем рассматривать дважды непрерывно
дифференцируемые функции $q^i\in\CC^2\big([t_1,t_2]\big)$ для всех $i$ на
конечном отрезке $[t_1,t_2]$.

Предположим, что механическая система описывается некоторым действием
\begin{equation}                                                  \label{eacpoi}
  S[q]=\int_{t_1}^{t_2}\!\!\!dt\,L(q,\dot q,t),
\end{equation}
где функция Лагранжа или лагранжиан $L(q,\dot q,t)$ зависит только от обобщенных
координат $q$ и их первых производных по времени $\dot q$, которые называются
обобщенными {\em скоростями}. Рассмотрим вариационную задачу с фиксированными
граничными условиями
\begin{equation}                                                  \label{ebocco}
  q(t_1)=q_1,\qquad q(t_2)=q_2,
\end{equation}
т.е.\ траектория механической системы представляет собой кривую в
конфигурационном пространстве, которая соединяет две фиксированные точки $q_1$ и
$q_2$. Тем самым вариации координат на концах интервала обращаются в нуль.
\index{Обобщенная скорость (generalized velocity)}%
\index{Скорость обобщенная (generalized velocity)}%

Обычно предполагают, что конфигурационное пространство представляет собой
евклидово пространство $\MR^\Sn$ с заданной метрикой $\dl_{ij}$, а обобщенные
координаты -- это декартовы координаты в $\MR^\Sn$. Кроме того, мы предполагаем,
что лагранжиан в действии (\ref{eacpoi}) представляет собой функцию (скалярное
поле) от своих аргументов. Поэтому метрика необходима для построения инвариантов
из координат $q^i$ и скоростей $\dot q^i$, и она входит в действие. В этом
действии евклидова метрика рассматривается как внешнее поле, и по ней
варьирование не проводится. Конечно, после того, как задача поставлена, действие
можно переписать в произвольной криволинейной системе координат в $\MR^\Sn$.
Тогда обобщенные координаты $q^i$ станут криволинейными координатами в
$\MR^\Sn$, а компоненты евклидовой метрики $g_{ij}(x)$ будут нетривиальными
функциями от $x$. Действие по этим компонентам не варьируется.
\begin{com}
При рассмотрении движения механической системы, естественная линейная структура
в $\MR^\Sn$ не играет никакой роли и нигде не используется. Важно только наличие
метрики. В общем случае можно считать, что конфигурационное пространство -- это
произвольное риманово многообразие $(\MM,g)$ с заданной метрикой $g$, которая
необходима для построения инвариантов. Действие в этом случае будет зависеть от
метрики, которая рассматривается как внешнее поле, и по ней варьирование не
проводится. Заметим также, что задание лагранжиана ничего не говорит о топологии
конфигурационного пространства. Для того, чтобы задать топологию $\MM$
необходимо сделать какие либо дополнительные предположения. В дальнейшем мы
будем рассматривать в основном топологически тривиальные конфигурационные
пространства, $\MM\approx\MR^\Sn$ с естественной топологией.
\qed\end{com}

Уравнения Эйлера--Лагранжа (уравнения движения) для действия (\ref{eacpoi}),
\begin{equation}                                                  \label{eeilae}
  \frac{\dl S}{\dl q^i}=\frac{\pl L}{\pl q^i}
  -\frac{d}{dt}\frac{\pl L}{\pl\dot q^i}=0,
\end{equation}
представляют собой систему уравнений не выше второго порядка. Функции $q^i(t)$,
удовлетворяющие уравнениям Эйлера--Лагранжа, соответствуют стационарным точкам
действия (\ref{eacpoi}). Они определяют {\em траекторию} механической системы.
Траектории частиц не зависят от выбора координат в конфигурационном пространстве
$\MR^\Sn$, которые выбираются из соображений удобства.
\index{Траектория частицы (particle trajectory)}%

Как правило, решение системы уравнений движения (\ref{eeilae}) содержит $2\Sn$
произвольных постоянных, которые находятся из граничных условий (\ref{ebocco}).
У краевой задачи решение может не существовать, а если оно существует, то может
не быть единственным. Вместо краевой задачи можно также поставить задачу Коши,
которая имеет единственное решение при корректной постановке. В этом случае
произвольные постоянные, возникающие в решении уравнений движения, находятся из
начальных данных для $q$ и $\dot q$ в начальный момент времени $t=0$. Именно эта
задача наиболее часто ставится в физических приложениях для уравнений движения
(\ref{eeilae}).

Перейдем к гамильтонову или каноническому формализму. В этом случае механическая
система, состоящая из $\Sn$ частиц, описывается $\Sn$ обобщенными координатами
$q^i$ и $\Sn$ обобщенными импульсами
\begin{equation}                                                  \label{emogen}
  p_i:=\frac{\pl L}{\pl\dot q^i},
\end{equation}
которые рассматриваются, как независимые переменные. Обобщенные координаты и
импульсы являются координатами механической системы в $2\Sn$-мерном
{\em фазовом} пространстве. Мы говорим, что координаты и импульсы являются
канонически сопряженными переменными.
\index{Фазовое пространство (phase space)}%
\index{Пространство фазовое (phase space)}%

Предположим, что конфигурационное пространство с координатами $q^i$ является
$\Sn$-мерным многообразием $\MM$. Тогда производные $\dot q^i$ являются
компонентами касательного вектора вдоль траектории частицы. Это значит, что
лагранжиан $L(q,\dot q)$ является функцией (скалярным полем) на касательном
расслоении $\MT(\MM)$. Тогда формула (\ref{emogen}) определяет компоненты
ковариантного вектора (1-формы). Это значит, что фазовое пространство есть ни
что иное, как кокасательное расслоение $\MT^*(\MM)$ к конфигурационному
пространству $\MM$.

Поскольку $q^i$ и $p_i$ являются координатами в базе и в кокасательном слое, то
мы пишем координатные индексы, соответственно, вверху и внизу.

Предположим, что функция Лагранжа $L(q,\dot q)$ выпукла по скоростям, т.е.\
квадратичная форма
\begin{equation}                                                  \label{edehes}
  \pl^2L/\pl\dot q^i\pl\dot q^j,
\end{equation}
которая называется {\em гессианом}, положительно определена. Это условие не
является ковариантным относительно произвольной замены координат на касательном
расслоении $\MT(\MM)$. Однако оно ковариантно относительно общих преобразований
координат в конфигурационном пространстве $\MM$. Действительно, при замене
координат $q\mapsto q'(q)$ скорости преобразуются как векторы:
\begin{equation*}
  \dot q^i\mapsto \dot q^{\prime i}=\frac{\pl q^{\prime i}}{\pl q^j}\dot q^j.
\end{equation*}
\index{Гессиан (Hessian)}%
Поскольку якобиан преобразования $\pl q^{\prime i}/\pl q^j$ не зависит от
скоростей, то гессиан преобразуется как ковариантный тензор второго ранга:
\begin{equation*}
  \frac{\pl^2L}{\pl\dot q^i\pl\dot q^j}\mapsto
  \frac{\pl^2L}{\pl\dot q^{\prime i}\pl\dot q^{\prime j}}=
  \frac{\pl q^k}{\pl q^{\prime i}}\frac\pl{\pl\dot q^k}
  \left(\frac{\pl q^l}{\pl q^{\prime j}}\frac{\pl L}{\pl\dot q^l}\right)=
  \frac{\pl q^k}{\pl q^{\prime i}}\frac{\pl q^l}{\pl q^{\prime j}}
  \frac{\pl^2 L}{\pl\dot q^k\pl\dot q^l}.
\end{equation*}
Таким образом, понятие выпуклости лагранжиана по скоростям не зависит от выбора
координат в конфигурационном пространстве $\MM$.
\begin{com}
Рассмотрим координаты $q$ и время $t$ в определении обобщенных импульсов
(\ref{emogen}) как параметры. Тогда формулы (\ref{emogen}) задают скорости
$\dot q$ как неявные функции обобщенных импульсов $p$. Из курса математического
анализа известно, что уравнения (\ref{emogen}) локально разрешимы относительно
скоростей тогда и только тогда, когда гессиан (\ref{edehes}), который в данном
случае совпадает с якобианом преобразования координат $\dot q\mapsto p$,
является невырожденной матрицей.
\qed\end{com}
Если функция Лагранжа выпукла по скоростям, то можно построить ее преобразование
Лежандра (по скоростям):
\begin{equation}                                                  \label{ehadef}
  H(q,p):=p_i\dot q^i-L(q,\dot q),
\end{equation}
которое называется {\em функцией Гамильтона} или {\em гамильтонианом} системы.
\index{Функция Гамильтона (Hamilton's function)}%
\index{Гамильтона функция (Hamilton's function)}%
\index{Гамильтониан (Hamiltonian)}%
В силу определения преобразования Лежандра гамильтониан системы зависит только
от обобщенных координат и импульсов. Это можно проверить и непосредственно
\begin{equation*}
  \frac{\pl H}{\pl\dot q^i}=p_i-\frac{\pl L}{\pl\dot q^i}=0,
\end{equation*}
что следует из определения обобщенных импульсов (\ref{emogen}).

Используя связь между гамильтонианом и лагранжианом (\ref{ehadef}), рассмотрим
действие
\begin{equation}                                                  \label{ehaact}
  S[q,p]=\int_{t_1}^{t_2}\!\!\!dt\,\big(p_i\dot q^i-H(q,p)\big)
\end{equation}
как функционал от канонически сопряженных координат и импульсов. Соответствующая
система уравнений Эйлера--Лагранжа имеет вид
\begin{align}                                                     \label{eilhae}
  \frac{\dl S}{\dl q^i}&=-\dot p_i-\frac{\pl H}{\pl q^i}=0,
\\                                                                \label{einhax}
  \frac{\dl S}{\dl p_i}&=\quad \dot q^i-\frac{\pl H}{\pl p_i}=0.
\end{align}
Нетрудно проверить, что $\Sn$ уравнений Эйлера--Лагранжа второго порядка
(\ref{eeilae}) эквивалентны $2\Sn$ уравнениям первого порядка (\ref{eilhae}).
Понижение порядка уравнений движения произошло за
счет введения новых независимых переменных -- импульсов.

Поскольку конфигурационное и фазовое пространства имеют разную размерность, то
уточним понятие эквивалентности. Для однозначного определения траектории частицы
$q(t)$ в конфигурационном пространстве необходимо решить уравнения
Эйлера--Лагранжа второго порядка (\ref{eeilae}), например, с начальными
условиями $q(0)=q_0$, $\dot q(0)=q_1$. Траектория движения в фазовом
пространстве $q(t),p(t)$ однозначно находится решением уравнений движения
первого порядка (\ref{eilhae}), (\ref{einhax}) с начальными условиями
$q(0)=q_0$, $p(0)=p_0$. Каждая траектория в фазовом пространстве естественным
образом проектируется на траекторию в конфигурационном пространстве
$\lbrace q(t),p(t)\rbrace\mapsto\lbrace q(t)\rbrace$, при этом начальное
условие для скорости $q_1$ определяется из уравнения (\ref{einhax}). Обратно.
Для любой траектории в конфигурационном пространстве $q(t)$ уравнение
(\ref{einhax}) определяет импульсы $p(t)$ и начальные условия $p_0$ так, что
пара $q(t),p(t)$ является траекторией в фазовом пространстве. Здесь мы
предполагаем, что уравнение (\ref{einhax}) имеет единственное решение для $p$ (в
противном случае преобразование Лежандра не определено).

При постановке граничной задачи для действия (\ref{ehaact}) мы не можем
рассматривать траектории, соединяющие две произвольные точки $(q_1,p_1)$ и
$(q_2,p_2)$ фазового пространства, т.к.\ тогда возникнет $4\Sn$ условий для
$2\Sn$ обыкновенных дифференциальных уравнений первого порядка (\ref{eilhae}),
(\ref{einhax}). В этом случае количество граничных условий превышает количество
уравнений, и задача может не иметь решения. Поэтому можно рассматривать,
например, те траектории, которые имеют начало и конец на $\Sn$-мерных
подмногообразиях фазового пространства, определяемых условиями (\ref{ebocco}),
как и в лагранжевом подходе. При этом никаких граничных условий на импульсы не
возникает, поскольку они входят в подынтегральное выражение без производных, и
при вариации импульсов интегрирования по частям не происходит.
\begin{com}
Как правило для уравнений Эйлера--Лагранжа в фазовом пространстве ставится
задача Коши, которая имеет единственное решение. При этом начальной точкой
траектории может быть произвольная точка фазового пространства
$\MT^*(\MR^\Sn)=\MR^{2\Sn}$. Тем самым мы предполагаем, что фазовое пространство
так же как и конфигурационное пространство топологически тривиально. Обсуждение
области определения различных функций в конфигурационном и фазовом пространствах
требует знания явного вида лагранжиана, гамильтониана и анализа уравнений
движения. Поскольку в общем случае учесть все возможности нельзя, то в
дальнейшем при обсуждении общей схемы все рассматриваемые функции будут
считаться определенными на всем фазовом пространстве и достаточно гладкими. В
каждом конкретном случае области определения должны быть проанализированы, а
постановка задачи уточнена.
\qed\end{com}
Посмотрим на уравнения (\ref{eilhae}), (\ref{einhax}) с другой точки зрения.
Пусть заданы две функции от канонических переменных $f(q,p)$ и $g(q,p)$, т.е.\
два скалярных поля на фазовом пространстве. Определим для них {\em скобку
Пуассона}
\index{Скобка Пуассона (Poisson bracket)}%
\index{Пуассона скобка (Poisson bracket)}%
\begin{equation}                                                  \label{epobar}
  [f,g]:=\frac{\pl f}{\pl q^i}\frac{\pl g}{\pl p_i}
  -\frac{\pl f}{\pl p_i}\frac{\pl g}{\pl q^i}.
\end{equation}
Легко проверить, что она обладает следующими свойствами:
  \begin{align*}
  1)\quad &[ af+bg,h] =a[ f,g]+b[ g,h],
  \quad a,b\in\MR & &\text{-- линейность},\\
  2)\quad &[ f,g] =-[ g,f] &
  &\text{-- антисимметрия},\\
  3)\quad &[ f,gh] =[ f,g] h+g[ f,h] & &
  \text{-- правило Лейбница},\\
  4)\quad &[ f,[ g,h]]
  +[ g,[ h,f]]
  +[ h,[ f,g]]=0 &&\text{-- тождество Якоби},
  \end{align*}
и, следовательно, определяет пуассонову структуру на фазовом пространстве (см.\
раздел \ref{spoist}.

Скобку Пуассона (\ref{epobar}) можно переписать в эквивалентном виде. Обозначим
координаты фазового пространства через
$\lbrace x^\al\rbrace:=(q^1\dotsc q^\Sn,p_1\dotsc p_\Sn$,
$\al=1,\dotsc,2\Sn$, или, короче, $x=(q,p)$. Тогда
\begin{equation}                                                  \label{epoica}
  [f,g]=\varpi^{-1\al\bt}\frac{\pl f}{\pl x^\al}\frac{\pl g}{\pl x^\bt},
\end{equation}
где
\begin{equation*}
  \varpi^{-1\al\bt}=[x^\al,x^\bt]
  =\begin{pmatrix}0&\one \\ -\one&0\end{pmatrix}
\end{equation*}
-- каноническая пуассонова структура. Каноническая форма $\varpi$ невырождена и
точна, $d\varpi=0$, поскольку ее компоненты постоянны. Поэтому фазовое
пространство представляет собой также симплектическое многообразие (см.\ раздел
\ref{ssymma}).

Рассмотрим простейшие свойства скобки Пуассона. Очевидно, скобки Пуассона
произвольной функции $f$ с самой собой и константой $c$ равны нулю:
$$
  [f,f]=0,\qquad [ f,c]=0.
$$

Из определения (\ref{epobar}) следует, что скобка Пуассона обобщенной координаты
с импульсом равна символу Кронекера:
\begin{equation}                                                  \label{eqtipo}
  [q^i,p_j]=\dl^i_j.
\end{equation}

Рассмотрим траекторию в фазовом пространстве $q(t)$, $p(t)$, где $t\in\MR$.
Тогда, используя скобку Пуассона, уравнения движения (\ref{eilhae}),
(\ref{einhax}) можно переписать в виде
\begin{equation}                                                  \label{eqgenq}
\begin{split}
  \dot q^i&=[ q^i,H]=\quad \frac{\pl H}{\pl p_i},
\\
  \dot p_i&=[ p_i,H]=-\frac{\pl H}{\pl q^i}.
\end{split}
\end{equation}
При этом скобка Пуассона для координат фазового пространства (\ref{eqtipo})
рассматривается как одновременная:
\begin{equation}                                                  \label{eeqtoc}
  [q^i(t),p_j(t)]=\dl^i_j,\qquad \forall t\in\MR.
\end{equation}
Скобка Пуассона для различных моментов времени $[q^i(t_1),p_j(t_2)]$ при
$t_1\ne t_2$ не определена. Вообще, эволюция во времени любой функции
$f(t,q,p)$, зависящей от времени и канонических переменных определяется
уравнением
\begin{equation}                                                  \label{edotfh}
  \dot f=\frac{df}{dt}=\frac{\pl f}{\pl t}+[f,H].
\end{equation}
В частности, если функция Гамильтона не зависит от времени явно, то
$$
  \frac{dH}{dt}=[H,H]=0,
$$
ввиду антисимметрии скобки Пуассона. Это значит, что для заданной траектории
механической системы в фазовом пространстве гамильтониан является интегралом
движения и его численное значение сохраняется. Это значение называется {\em
энергией} механической системы и определяется начальными данными. Системы, у
которых гамильтониан не зависит от времени, называются {\em консервативными}.
\index{Энергия механической системы (energy of mechanical system)}%
\index{Консервативная система (conservative system}%
\index{Система консервативная (conservative system}%
\begin{exa}[\bf Гармонический осциллятор с затуханием]
Рассмотрим функцию Лагранжа для одной точечной частицы (осциллятора), которая
явно зависит от времени
\begin{equation*}
  L=\frac12\ex^{2\mu t}(\dot q^2-\om^2q^2),\qquad \mu,\om=\const.
\end{equation*}
Постоянные $\om$ и $\mu$ называются соответственно собственной частотой и
коэффициентом затухания осциллятора. Обобщенный импульс частицы равен
\begin{equation*}
  p:=\frac{\pl L}{\pl\dot q}=\ex^{2\mu t}\dot q.
\end{equation*}
Гамильтониан осциллятора также зависит от времени явно:
\begin{equation*}
  H=\frac12\ex^{-2\mu t}p^2+\frac12\ex^{2\mu t}\om^2q^2.
\end{equation*}
Гамильтоновы и лагранжевы уравнения движения для осциллятора с затуханием имеют
вид
\begin{equation*}
\left.
\begin{aligned}
  \dot q&=\quad \ex^{-2\mu t}p,
\\
  \dot p&=-\ex^{2\mu t}\om^2 q,
\end{aligned}
\right\rbrace\qquad \Leftrightarrow\qquad
\ex^{2\mu t}(\ddot q+2\mu\dot q+\om^2q)=0.
\end{equation*}
Общее решение этих уравнений параметризуется двумя постоянными: амплитудой
$A_0$ и фазой $\vf$,
\begin{equation*}
  q=A_0\ex^{-\mu t}\cos(\tilde\om t+\vf),\qquad \tilde\om^2:=\om^2-\mu^2.
\end{equation*}
Произвольная фаза $\vf$ соответствует произволу в выборе начала отсчета времени,
и в дальнейшем мы положим $\vf=0$. При малых коэффициентах затухания
$\mu^2<\om^2$ амплитуда колебаний экспоненциально затухает, если $\mu>0$.
Затухающие колебания происходят с частотой, меньшей собственной частоты
осциллятора $\tilde\om<\om$. Для произвольной траектории численное значение
гамильтониана зависит от времени явно
\begin{equation*}
  H=\frac12A_0^2\om^2\left[1-\frac12\cos 2\tilde\om t
  -\frac12\cos(2\tilde\om t+2\psi)\right],
\end{equation*}
где $\cos\psi:=\mu/\om$.
\qed\end{exa}

Функция канонических переменных $f(q,p)$ называется {\em интегралом движения},
если $f(q,p)=\const$ для любого решения канонических уравнений движения, т.е.\
является интегралом уравнений движения.
\index{Интеграл движения (integral of motion)}%
\begin{theorem}
Если известны два интеграла движения $f$ и $g$, то их скобка Пуассона $[f,g]$
также является интегралом движения.
\end{theorem}
\begin{proof}
Прямая проверка равенства $d[f,g]/dt=0$.
\end{proof}
\begin{com}
Теорема не гарантирует того, что вычисление скобки Пуассона двух интегралов
движения дает новый интеграл движения. Часто она равна нулю, или интегралы
движения $f$, $g$ и $[f,g]$ -- функционально зависимы.
\qed\end{com}

Если гамильтониан не зависит от какой-либо из координат, например, от $q^1$,
т.е.\ $\pl H/\pl q^1=0$, то эта координата называется {\em циклической}, а
соответствующий обобщенный импульс сохраняется в силу второго уравнения
(\ref{eqgenq})
\index{Циклическая координата (cyclic coordinate)}%
\index{Координата циклическая (cyclic coordinate)}%
$$
  p_1=c=\const.
$$
При этом изменение остальных координат и импульсов во времени такое же,
как в системе с координатами $q^2,\dotsc,q^\Sn$, импульсами $p_2,\dotsc,p_\Sn$
и функцией Гамильтона $H(q^2,\dots,q^\Sn,c,p_2,\dots,p_\Sn)$.

В большинстве физических приложений для системы уравнений Гамильтона
(\ref{eqgenq}) решается задача Коши, т.е.\ ищется решение системы уравнений
(\ref{eqgenq}) с заданными начальными условиями
\begin{equation}                                                  \label{einche}
  q(0)=q_0,\qquad p(0)=p_0.
\end{equation}

Канонические уравнения движения (\ref{eqgenq}) можно записать в виде
\begin{equation*}
  \dot x=X,
\end{equation*}
где $x=(q,p)$, а правая часть уравнений определяются векторным полем
$$
  X=\frac{\pl H}{\pl p_i}\frac\pl{\pl q^i}
  -\frac{\pl H}{\pl q^i}\frac\pl{\pl p_i}
$$
на фазовом пространстве. С геометрической точки зрения решение уравнений
Гамильтона с заданными начальными условиями задает интегральную кривую этого
векторного поля, проходящую через точку (\ref{einche}). Как было показано в
разделе \ref{svechs}, интегральные кривые векторного поля задают абелеву
однопараметрическую группу преобразований многообразия, которым в данном случае
является фазовое пространство. Эта группа преобразований в гамильтоновой
динамике имеет специальное название.
\begin{defn}
{\em Фазовым потоком} называется однопараметрическая группа преобразований
фазового пространства
$$
  s_t:\quad \MT^*(\MM)\ni\quad q(0),p(0)\mapsto q(t),p(t)\quad\in\MT^*(\MM),
$$
где $q(t),p(t)$ -- решение системы уравнений Гамильтона (\ref{eqgenq}).
\qed\end{defn}
\index{Фазовый поток (phase flow)}\index{Поток фазовый (phase flow)}%
\subsection{Потенциальное движение точечной частицы              \label{spotmo}}
Потенциальное движение точечной частицы массы $m$ в евклидовом пространстве
$\MR^\Sn$ с декартовыми координатами $q^i$, $i=1,\dotsc,\Sn$ описывается
следующей функцией Лагранжа
\begin{equation}                                                  \label{epodpo}
  L=m\frac{\dot q^i\dot q_i}2-U(q),
\end{equation}
где $U(q)\ge0$ -- некоторая положительно определенная функция координат. В
настоящем разделе подъем и опускание индексов производится с помощью евклидовой
метрики $\dl_{ij}=\diag(+\dotsc+)$. Уравнения движения ({\em уравнения Ньютона})
при этом имеют вид
\index{Уравнения Ньютона (Newton's equations)}%
\index{Ньютона уравнения (Newton's equations)}%
\begin{equation}                                                  \label{enewte}
  m\ddot q^i=-\frac{\pl U}{\pl q_i}.
\end{equation}
Это -- система обыкновенных дифференциальных уравнений второго порядка и их
решение зависит от $2\Sn$ произвольных постоянных. Для их определения решают,
как правило, задачу Коши. То есть ищется решение уравнений (\ref{enewte}) с
заданными начальными условиями:
$$
  q(0)=q_0,\qquad \dot q(0)=v_0.
$$
В физических приложениях функции $U(q)$ обычно таковы, что решение этой задачи
существует, единственно и определено при всех $t\in(-\infty,\infty)$. Заметим,
что потенциальная энергия определена с точностью до постоянной, которая не
влияет на уравнения движения.

Если потенциальная энергия имеет локальный экстремум в точке $q_0$, то
\begin{equation*}
  \left.\frac{\pl U}{\pl q^i}\right|_{q=q_0}=0.
\end{equation*}
Следовательно, постоянная траектория $q=q_0$ удовлетворяет системе уравнений
(\ref{enewte}). Таким образом, локальные экстремумы потенциальной энергии
определяют положения равновесия частицы. Эти положения могут быть устойчивы или
не устойчивы по отношению к малым возмущениям в зависимости от того является ли
локальный экстремум минимумом или максимумом потенциальной энергии. Для
положительно определенной потенциальной энергии существует по крайней мере одна
точка равновесия.

Посмотрим на эту задачу с точки зрения принципа наименьшего действия. Во-первых,
действие
$$
  S[q]=\int\!dt\left(m\frac{\dot q^i\dot q_i}2-U(q)\right)
$$
не является положительно определенным. Поэтому нужно говорить не о минимуме
действия, а только о стационарных точках. Во-вторых, для свободного движения,
$U=0$, траектории представляют собой прямые линии
$$
  q=v_0t+q_0,
$$
по которым частица движется с постоянной скоростью $v_0$. Соответствующее
действие при интегрировании по бесконечному интервалу расходится и говорить даже
о стационарности действия в этом случае не имеет смысла. Подытожить сделанные
замечания можно следующим образом. Потенциальное движение точечной частицы
происходит таким образом, что для каждого конечного интервала времени
$(t_1,t_2)$ действие стационарно среди всех возможных траекторий, соединяющих
точки $q_1$ и $q_2$. При этом первую граничную точку можно выбрать произвольным
образом, а вторая должна быть такой, чтобы задача Коши, поставленная в первой
точке, имела решение для некоторой скорости $\dot q_1=v_0$.

Переформулируем потенциальное движение точечной частицы на гамильтоновом языке.
Компоненты импульса и гамильтониан точечной частицы равны
\begin{align}                                                          \nonumber
  p_i&=m\dot q_i,
\\                                                                \label{ehapoi}
  H&=\frac{p^ip_i}{2m}+U(q).
\end{align}
Гамильтоновы уравнения движения принимают вид
\begin{equation}
\begin{split}
  \dot q^i&=\quad \frac{p^i}m,
\\
  \dot p_i&=-\frac{\pl U}{\pl q^i},
\end{split}
\end{equation}
которые эквивалентны уравнениям Ньютона (\ref{enewte}). Заметим, что в данном
случае первое уравнение совпадает с определением импульса частицы.

Гамильтониан точечной частицы (\ref{ehapoi}) положительно определен и для
заданной траектории его значение (энергия) постоянно. Первое и второе слагаемые
в (\ref{ehapoi}) называются соответственно {\em кинетической} и
{\em потенциально} энергией точечной частицы.
\index{Кинетическая энергия (kinetic energy)}%
\index{Энергия кинетическая (kinetic energy)}%
\index{Потенциальная энергия (potential energy)}%
\index{Энергия потенциальная (potential energy)}%
Следующий пример показывает, что функция Гамильтона зависит от выбора системы
координат в фазовом пространстве, а каноническая пуассонова структура -- нет.
\begin{exa}                                                       \label{epocax}
Рассмотрим частицу массы $m$, которая движется в трехмерном конфигурационном
пространстве $q\in\MR^3$ с заданным потенциалом $U$. Гамильтониан является
функцией на фазовом пространстве и в произвольной криволинейной системе
координат имеет вид
\begin{equation*}
  H=\frac1{2m}g^{ij}p_ip_j+U(q),
\end{equation*}
где $g^{ij}=g^{ij}(q)$ -- обратная метрика в выбранной системе отсчета.
Например, в декартовых координатах $x,y,z$
\begin{align}                                                          \nonumber
  H&=\frac1{2m}\left(p_x^2+p_y^2+p_z^2\right)+U(x,y,z).
\\ \intertext{В цилиндрических координатах $r,\vf,z$}                  \nonumber
  H&=\frac1{2m}\left(p_r^2+\frac{p_\vf^2}{r^2}+p_z^2\right)+U(r,\vf,z).
\\ \intertext{В сферических координатах $r,\theta,\vf$}              \tag*{\qed}
  H&=\frac1{2m}\left(p_r^2+\frac{p_\theta^2}{r^2}
  +\frac{p_\vf^2}{r^2\sin^2\theta}\right)+U(r,\theta,\vf).
\end{align}

Каноническая пуассонова структура была определена в произвольной, в общем случае
криволинейной, системе координат. Проверим корректность этого определения
относительно преобразования координат, поскольку при переходе от одной системы
координат к другой структурные функции могли бы измениться. Допустим, что мы
определили каноническую пуассонову структуру в декартовой системе координат. В
рассматриваемом случае преобразование координат имеет вид
$q,p\mapsto Q(q),P(q,p)$, где большие буквы обозначают криволинейные
(сферические или цилиндрические) координаты в конфигурационном пространстве.
Скобки Пуассона новых координат и импульсов равны:
\begin{align*}
  [Q^i,Q^j]&=0,
\\
  [Q^i,P_j]&=\frac{\pl Q^i}{\pl q^k}\frac{\pl P_j}{\pl p_k},
\\
  [P_i,P_j]&=\frac{\pl P_i}{\pl q^k}\frac{\pl P_j}{\pl p_k}
             -\frac{\pl P_i}{\pl p_k}\frac{\pl P_j}{\pl q^k}.
\end{align*}
Поскольку компоненты обобщенных импульсов являются компонентами ковектора, то
они преобразуются по правилу
\begin{equation*}
  P_i=\frac{\pl q^k}{\pl Q^i}p_k.
\end{equation*}
Теперь нетрудно проверить выполнение скобок Пуассона:
\begin{equation*}
  [Q^i,P_j]=\dl^i_j,\qquad [Q^i,Q^j]=0,\qquad [P_i,P_j]=0.
\end{equation*}
Таким образом, структурные функции имеют канонический вид и не зависят от выбора
системы координат в конфигурационном пространстве $\MR^3$. Рассматриваемые
преобразования координат $q,p\mapsto Q,P$ в фазовом пространстве относятся к
классу канонических преобразований, которые будут рассмотрены позже в разделе
\ref{scantr}.
\end{exa}
\begin{defn}
Преобразование координат $q=q(Q)$ конфигурационного пространства $\MR^\Sn$
называется {\em точечным}.
\qed\end{defn}
\index{Точечное преобразование (pointwise transformation)}%
\index{Преобразование точечное (pointwise transformation)}%
\subsection{Лемма Стокса}
В этом и следующих разделах механика частиц будет рассмотрена с более общей
точки зрения. Начнем с геометрического рассмотрения, которое затем применим к
гамильтоновой динамике точечных частиц. Пусть на многообразии $\MM$ нечетной
размерности $\dim\MM=2\Sn+1$ задана $2$-форма
$$
  B=\frac12dx^\al\wedge dx^\bt B_{\al\bt}.
$$
Поскольку матрица, задающая $2$-форму в локальной системе координат
антисимметрична, $B_{\al\bt}=-B_{\bt\al}$, а многообразие нечетномерно, то ее
определитель равен нулю. Это значит, что в каждой точке $x\in\MM$ у нее
существует по крайней мере один нетривиальный собственный вектор
$X=X^\al\pl_\al$ с нулевым собственным значением. Отсюда вытекает, что для
$2$-формы $B$ существует {\em нулевое} векторное поле $X$ со свойством
\index{Нулевое векторное поле (null vector field)}%
\index{Векторное поле нулевое (null vector field)}%
$$
  B(X,Y)=X^\al Y^\bt B_{\al\bt}=0,\qquad \forall\,Y=Y^\al\pl_\al.
$$
Пространство нулевых векторных полей линейно.
\begin{defn}
$2$-форма называется {\em неособой}, если размерность пространства нулевых
векторов минимальна, т.е.\ равна нулю или единице на многообразиях четной и
нечетной размерности соответственно.
\qed\end{defn}
\index{Неособая $2$-форма (nonsingular $2$-form)}%
\index{$2$-форма неособая (nonsingular $2$-form)}%
\begin{exa}
Рассмотрим $2$-форму $\varpi$ в фазовом пространстве $\MR^{2\Sn}$ с координатами
$x=(x^1\dotsc x^{2\Sn})=(q^1\dotsc q^\Sn,p_1\dotsc p_\Sn)$,
\begin{equation}                                                  \label{eomfao}
  \varpi=dp_i\wedge dq^i
  =dx^{\Sn+1}\wedge dx^1+dx^{\Sn+2}\wedge dx^2+\dotsc+dx^{2\Sn}\wedge dx^\Sn
  =\frac12dx^\al\wedge dx^\bt\varpi_{\al\bt},
\end{equation}
где $\varpi$ -- каноническая симплектическая форма (\ref{esymfo}). Эта форма
неособа, т.к.\ ее определитель равен единице $\det\varpi_{\al\bt}=1$ и
размерность пространства нулевых векторов равна нулю.
\qed\end{exa}

Пусть на многообразии $\MM$, $\dim\MM=2\Sn+1$, задана $1$-форма
$A=dx^\al A_\al$. Предположим, что внешний дифференциал этой формы является
неособым. Это значит, что в каждой точке многообразия $x\in\MM$ существует
единственный, с точностью до умножения на постоянную, нулевой вектор
$X$ такой, что
\begin{equation}                                                  \label{enuvec}
  dA(X,Y)=0,\qquad \forall\,Y.
\end{equation}
Тем самым $1$-форма $A$ определяет нулевое векторное поле $X$ на $\MM$ с
точностью до умножения на отличную от нуля функцию. Назовем интегральные кривые
$x^\al(t)$,
$$
  \frac{dx^\al}{dt}=X^\al,
$$
нулевого векторного поля {\em характеристиками} формы $A$. Далее, пусть $\g_1$
-- замкнутая кривая на $\MM$, которая ни в одной своей точке не касается
характеристик. Тогда множество характеристик, выходящих из точек кривой $\g_1$,
образует {\em трубку характеристик}. Трубка характеристик определена по крайней
мере в некоторой окрестности кривой $\g_1$.
\index{Характеристика $1$-формы (characteristic of $1$-form)}%
\index{Трубка характеристик (characteristic tube)}%
\begin{lemma}[\bf Стокс]
Интеграл от $1$-формы $A$ с неособым внешним дифференциалом $dA$ по любой из
двух замкнутых кривых $\g_1$ и $\g_2$, охватывающих одну и ту же трубку
характеристик, одинаков:
\begin{equation*}
  \oint_{\g_1}\!\!\!A=\oint_{\g_2}\!\!\!A,
\end{equation*}
если $\g_1-\g_2=\pl S$, где $S$ -- часть трубки характеристик.
\end{lemma}
\index{Лемма Стокса (Stokes lemma)}\index{Стокса лемма (Stokes lemma)}%
\begin{proof}
По формуле Стокса справедливы равенства
$$
  \oint_{\g_1}\!\!\!A-\oint_{\g_2}\!\!\!A=\oint_{\pl S}\!\!\!A=\int_S\!dA.
$$
Этот интеграл обращается в нуль, т.к.\ значение формы $dA$ на любой паре
векторов, касательных к трубке характеристик, равен нулю в силу (\ref{enuvec}).
\end{proof}
\subsection{Канонические уравнения Гамильтона                    \label{scaham}}
Из безобидной, на первый взгляд, леммы Стокса непосредственно вытекают все
основные положения гамильтоновой динамики. Предположим, что механической системе
соответствует топологически тривиальное фазовое пространство
$\MT^*(\MR^\Sn)\approx\MR^{2\Sn}$. Рассмотрим {\em расширенное фазовое
пространство} $\MR^{2\Sn+1}$ с координатами
$q^1,\dotsc,q^\Sn,$\linebreak[3]$p_1,\dotsc,p_\Sn,t$, в котором время $t$
рассматривается, как независимая дополнительная координата.
\begin{com}
Предположение о том, что частица движется в тривиальном конфигурационном
пространстве $\MR^\Sn$ и, следовательно, в тривиальном фазовом пространстве
$\MT^*(\MR^\Sn)\approx\MR^{2\Sn}$ не означает, что приведенные ниже формулы
справедливы только для движения в топологически тривиальных пространствах.
Данное предположение просто означает, что мы рассматриваем механическую систему
в определенной карте. При движении частицы по нетривиальному многообразию
необходимо дополнительно проследить за склейкой карт. В дальнейшем мы не будем
обсуждать этот вопрос.
\qed\end{com}

Пусть на расширенном фазовом пространстве задана некоторая функция Гамильтона
$H(q,p,t)$. Определим $1$-форму на расширенном фазовом пространстве
\index{Расширенное фазовое пространство (extended phase space)}%
\index{Фазовое пространство расширенное (extended phase space)}%
\begin{equation}                                                  \label{eaform}
  A:=dq^ip_i-dtH.
\end{equation}
Эта форма называется {\em интегральным инвариантом Пуанкаре--Картана}. Точнее,
интегральный инвариант получается после интегрирования формы $A$ по замкнутому
контуру в расширенном фазовом пространстве. Смысл названия будет ясен из
дальнейшего рассмотрения. Внешний дифференциал 1-формы (\ref{eaform}) является
неособым, т.к.\ каноническая форма $\varpi=dp_i\wedge dq^i$ неособа.
\index{Интегральный инвариант Пуанкаре--Картана%
(Poincar\'e--Cartan integral invariant}%
\index{Пуанкаре--Картана интегральный инвариант%
(Poincar\'e--Cartan integral invariant}%
\begin{theorem}                                                   \label{tcharp}
Характеристики формы $A$ в расширенном фазовом пространстве $\MR^{2\Sn+1}$
однозначно проектируются на фазовое пространство $q,p$, т.е.\ задаются
функциями $q(t)$, $p(t)$. Эти функции удовлетворяют уравнениям Гамильтона:
\begin{equation}                                                  \label{ehameq}
\begin{split}
  \frac{dq^i}{dt}&=\quad \frac{\pl H}{\pl p_i},
\\
  \frac{dp_i}{dt}&=-\frac{\pl H}{\pl q^i},
\end{split}
\end{equation}
Другими словами, характеристики формы $A$ представляют собой траектории фазового
потока в расширенном фазовом пространстве, т.е.\ интегральные кривые
канонических уравнений (\ref{ehameq}).
\end{theorem}
\begin{proof}
Дифференциал формы (\ref{eaform}) равен
$$
  dA=dp_i\wedge dq^i-\frac{\pl H}{\pl q^i}dq^i\wedge dt
  -\frac{\pl H}{\pl p_i} dp_i\wedge dt
$$
Эта $2$-форма неособа, т.к.\ ранг матрицы, составленной из ее координат, равен
$2\Sn$. Прямая подстановка показывает, что вектор
\begin{equation}                                                  \label{evecha}
  X=\frac{\pl H}{\pl p_i}\frac\pl{\pl q^i}
  -\frac{\pl H}{\pl q^i}\frac\pl{\pl p_i}+\frac\pl{\pl t}
\end{equation}
является нулевым вектором формы $A$
$$
  dA(X,Y)=-\frac{\pl H}{\pl q^i}Y^i-Y_i\frac{\pl H}{\pl p_i}
  -\frac{\pl H}{\pl p_i}Y^0\frac{\pl H}{\pl q^i}
  +Y^i\frac{\pl H}{\pl q^i}+\frac{\pl H}{\pl q^i}Y^0\frac{\pl H}{\pl p_i}
  +Y_i\frac{\pl H}{\pl p_i}=0.
$$
Это значит, что векторное поле (\ref{evecha}) задает направление характеристик
$1$-формы (\ref{eaform}). С другой стороны, вектор (\ref{evecha}) является
вектором скорости фазового потока (\ref{ehameq}). Действительно, интегральные
кривые $\lbrace q(\tau),p(\tau),t(\tau)\rbrace$ в расширенном фазовом
пространстве задаются уравнениями
\begin{equation}                                                  \label{eincep}
  \frac{d q^i}{d\tau}=\frac{\pl H}{\pl p_i},\qquad
  \frac{d p_i}{d\tau}=-\frac{\pl H}{\pl q^i},\qquad \frac{dt}{d\tau}=1.
\end{equation}
В силу последнего уравнения они однозначно проектируются на фазовое
пространство. Таким образом, интегральные кривые (\ref{ehameq}) представляют
собой проекции характеристик формы (\ref{eaform}) на фазовое пространство.
\end{proof}
Применим теперь лемму Стокса и интегральному инварианту Пуанкаре--Картана.
\begin{theorem}                                                   \label{tinpot}
Пусть две замкнутые кривые $\g_1$ и $\g_2$ охватывают одну и ту же трубку
характеристик формы (\ref{eaform}). Тогда интегралы по ним от интегрального
инварианта Пуанкаре--Картана одинаковы:
$$
  \oint_{\g_1}\!\!\!\big(dq\,p-dtH\big)=\oint_{\g_2}\!\!\!\big(dq\,p-dtH\big).
$$
\end{theorem}
Именно поэтому 1-форма $dqp-dtH$ называется интегральным инвариантом.

В частном случае, когда замкнутые кривые лежат в фазовом подпространстве,
соответствующем постоянному значению времени, $t=\const\Leftrightarrow dt=0$,
получаем
\begin{cor}
Фазовый поток сохраняет интеграл
$$
  \oint_\g\!dq\,p
$$
вдоль произвольной замкнутой кривой в фазовом пространстве.
\end{cor}
Форма $dq\,p:=dq^ip_i$ в фазовом пространстве называется {\em относительным
интегральным инвариантом Пуанкаре} или {\em формой Лиувилля}. Точнее,
интегральный инвариант получается после интегрирования формы Лиувилля по
замкнутому контуру $\g$.
\index{Форма Лиувилля (Liouville's form)}%
\index{Лиувилля форма (Liouville's form)}%
\index{Относительный интегральный инвариант Пуанкаре%
(relative integral Poincer\'e's invariant)}%
\index{Пуанкаре относительный интегральный инвариант%
(relative integral Poincer\'e's invariant)}%
Он имеет простой геометрический смысл. Пусть $S$ -- двумерная ограниченная
ориентированная поверхность с кусочно гладкой границей такая, что $\pl S=\g$,
тогда по формуле Стокса
$$
  \oint_\g\!dqp=\int_S\!dp\wedge dq,
$$
где $dp\wedge dq:=dp_i\wedge dq^i$. Отсюда вытекает
\begin{cor}
Фазовый поток $s_t:=\exp(tX_H)$ сохраняет сумму ориентированных площадей
проекций поверхности на $\Sn$ координатных плоскостей $q^i,p_i$:
$$
  \int_S\!dp\wedge dq=\int_{s_t S}\!dp\wedge dq.
$$
\end{cor}
В этом смысле каноническая $2$-форма $\varpi=dp\wedge dq$ является абсолютным
интегральным инвариантом фазового потока $s_t$.
\begin{com}
Эпитеты ``относительный'' и ``абсолютный'' интегральный инвариант не несут
глубокого смысла. Исторически сложилась так, что относительным называют
инвариант, полученный после интегрирования по замкнутому (компактному и без
края) многообразию. Абсолютным называют инвариант, возникающий после
интегрирования некоторой формы по компактному многообразию с краем.
\qed\end{com}
\subsection{Принцип Мопертюи}
В приложениях часто встречаются гамильтонианы, не зависящие от времени явно
$H=H(q,p)$. В этом случае численное значение гамильтониана (энергия) на фазовой
траектории сохраняется
$$
  \frac{dH}{dt}=[ H,H]=0.
$$
Покажем, что наличие интеграла энергии позволяет понизить размерность
расширенного фазового пространства $\MR^{2\Sn+1}$ на две единицы и свести задачу
к интегрированию некоторой системы канонических уравнений в $(2\Sn-1)$-мерном
пространстве.

Предположим, что в некоторой области фазового пространства уравнение
$$
  H(q,p)=E
$$
можно решить относительно $p_1$:
$$
  p_1=K(Q,P,T;E),
$$
где $Q:=(q^2,\dotsc,q^\Sn)$, $P:=(p_2,\dotsc,p_\Sn)$ и $T=-q^1$. Тогда
\begin{equation}                                                  \label{eregdi}
  pdq-Hdt=PdQ-KdT-d(Ht)+tdH.
\end{equation}
Пусть $\g$ -- интегральная кривая канонических уравнений (\ref{ehameq}). Она
лежит на $2\Sn$-мерном подмногообразии $H(q,p)=E$ в расширенном фазовом
пространстве $\MR^{2\Sn+1}$. Спроектируем расширенное фазовое пространство на
фазовое пространство $\MR^{2\Sn}$. При этом поверхность $H=E$ проектируется на
$(2\Sn-1)$-мерное подмногообразие $\MM^{2\Sn-1}$ фазового пространства, которое
определяется тем же уравнением $H=E$, а кривая $\g$ -- на кривую $\bar\g$,
лежащую на этом подмногообразии. Переменные $Q,P,T$ образуют локальную систему
координат на $\MM^{2\Sn-1}$. Но характеристики формы $PdQ-KdT$ удовлетворяют
уравнениям Гамильтона, поскольку $dH=0$ на $\MM^{2\Sn-1}$, а полный дифференциал
$d(Ht)$ в (\ref{eregdi}) не влияет на уравнения движения. Это доказывает
следующее утверждение.
\begin{theorem}
Если гамильтониан не зависит от времени явно, то фазовые траектории канонических
уравнений (\ref{ehameq}) на подмногообразии $\MM^{2\Sn-1}$ фазового
пространства, определяемом уравнением $H(q,p)=E$, удовлетворяют каноническим
уравнениям
\begin{equation}                                                  \label{eshcae}
  \frac{dq^i}{dq^1}=-\frac{\pl K}{\pl p_i},\qquad
  \frac{dp_i}{dq^1}=\frac{\pl K}{\pl q^i},\qquad i=2,\dotsc,\Sn,
\end{equation}
где функция $K(q^1,\dotsc,q^\Sn,p_2,\dotsc,p_\Sn,E)$ определяется уравнением
$$
  H(q^1,\dotsc,q^\Sn,K,p_2,\dotsc,p_\Sn)=E.
$$
\end{theorem}
\begin{com}
В этой теореме роль времени играет первая координата $q^1$, и канонические
уравнения определяют не динамику системы, а форму траектории.
\qed\end{com}

Покажем, каким образом канонические уравнения Гамильтона связаны с принципом
наименьшего действия. Рассмотрим интегральную кривую уравнений (\ref{eincep}) в
расширенном фазовом пространстве $(q,p,t)$, соединяющую две точки
$(q_0,p_0,t_0)$ и $(q_1,p_1,t_1)$.
\begin{theorem}
Кривая $\g$ является стационарной точкой интеграла
\begin{equation}                                                  \label{eacexp}
  \int_\g\!(dq\,p-dtH)
\end{equation}
при таких вариациях $\g$, когда концы кривой остаются на $\Sn$-мерных
подмногообразиях $(q=q_0,t=t_0)$ и $(q=q_1,t=t_1)$.
\end{theorem}
\begin{proof}
В силу третьего уравнения (\ref{eincep}) вариацию интеграла (\ref{eacexp})
можно записать как вариацию функционала действия
$$
  \dl\int_\g\!dt(p\dot q-H)=\left.p\dl q\vphantom{\sum}\right|_{q_0}^{q_1}
  +\int_\g\!dt\left[\left(\dot q^i-\frac{\pl H}{\pl p_i}\right)\dl p_i
  -\left(\dot p_i+\frac{\pl H}{\pl q^i}\right)\dl q^i\right].
$$
Отсюда вытекает, что для исчезновения граничных вкладов достаточно зафиксировать
только значения обобщенных координат при $t=t_{0,1}$.
\end{proof}

Понижение размерности расширенного фазового пространства при гамильтониане, не
зависящем от времени, позволяет по новому взглянуть и в определенной степени
оправдать употребление термина ``принцип наименьшего действия'' в механике.
Фазовые траектории исходных канонических уравнений (\ref{ehameq}) целиком лежат
на подмногообразии $\MM\subset\MR^{2\Sn+1}$, $\dim\MM=2\Sn-1$, соответствующем
фиксированному значению энергии, и являются характеристиками формы
$dq\,p=dQP-dTK$. Отсюда следует
\begin{theorem}
Если функция Гамильтона не зависит от времени явно, то фазовые траектории
канонических уравнений (\ref{ehameq}), лежащие на подмногообразии $\MM^{2n-1}$,
соответствующем фиксированному значению энергии $H(q,p)=E$, являются
стационарными точками интеграла
$$
  \int\!dq\,p
$$
в классе кривых, лежащих на $\MM^{2\Sn-1}$ и соединяющих подпространства $q=q_0$
и $q=q_1$.
\end{theorem}

Рассмотрим теперь проекцию экстремали, лежащую на подмногообразии постоянной
энергии $\MM$ на конфигурационное пространство. Эта кривая соединяет точки с
координатами $q_0$ и $q_1$. Пусть $q=q(\tau)$, $\tau\in[a,b]$ -- некоторая
кривая, соединяющая те же точки $q(a)=q_0$ и $q(b)=q_1$ в конфигурационном
пространстве. Она является проекцией некоторой кривой $\g$ на подмногообразии
$\MM$. Эту кривую нетрудно построить. Для этого достаточно найти обобщенный
импульс $p=\pl L/\pl\dot q$, где $\dot q:=dq/d\tau$ -- вектор скорости кривой в
конфигурационном пространстве. Если параметр $\tau$ подобран так, что
$H(q,p)=E$, то мы получаем кривую $q=q(\tau)$, $p=\pl L/\pl\dot q$ на
поверхности $\MM$. Применяя предыдущую теорему, получаем
\begin{cor}
Среди всех кривых $q=q(\tau)$, $\tau\in[a,b]$ соединяющих точки $q_0$ и $q_1$ в
конфигурационном пространстве и параметризованных так, что функция Гамильтона
имеет фиксированное значение
$$
  H(q,\pl L/\pl\dot q)=E
$$
траекторией движения механической системы является экстремаль {\em укороченного
действия}
\begin{equation}                                                  \label{eshact}
  \int_a^b\!\!\!d\tau\, p_i\dot q^i
  =\int_a^b\!\!\!d\tau\frac{\pl L}{\pl\dot q^i}\dot q^i.
\end{equation}
\end{cor}
\index{Укороченное действие (truncated action)}%
\index{Действие укороченное (truncated action)}%
Это следствие называется {\em принципом наименьшего действия Мопертюи}. Важно
отметить, что отрезок $\tau\in[a,b]$ не фиксирован и может быть разным у
сравниваемых кривых. Зато одинаковой должна быть энергия. Принцип Мопертюи
определяет только форму траектории, а для определения зависимости координат от
времени необходимо воспользоваться условием постоянства энергии. Заметим также,
что во многих задачах подынтегральное выражение в (\ref{eshact}) положительно
определено. В таких случаях движение происходит действительно по экстремалям
укороченного действия, и можно говорить о принципе наименьшего действия.
\index{Принцип наименьшего действия Мопертюи%
 (Maupertius's principle of least action)}%
\index{Мопертюи принцип наименьшего действия%
 (Maupertius's principle of least action)}%
\begin{exa}
Покажем, что если материальная точка движется по риманову многообразию $(\MM,g)$
только под действием сил инерции, то движение происходит по экстремалям. Пусть
$q(t)=\lbrace q^i(\tau)\rbrace$ -- траектория частицы и $ds^2=dq^i dq^j g_{ij}$
-- интервал риманова многообразия. Тогда функция Лагранжа для инерциального
движения частицы определяется только кинетической энергией:
$$
  L=T=H=\frac12\left(\frac{ds}{d\tau}\right)^2
  =\frac12g_{ij}\frac{dq^i}{d\tau}\frac{dq^j}{d\tau},\qquad
  \frac{\pl L}{\pl\dot q^i}\dot q^i=2T=\left(\frac{ds}{d\tau}\right)^2.
$$
Чтобы обеспечить фиксированное значение энергии вдоль траектории, параметр
$\tau$ необходимо выбрать пропорциональным длине траектории
$d\tau=ds/\sqrt{2E}$ (канонический параметр). Тогда укороченное действие
принимает вид
$$
  \int_a^b\!\!\!d\tau\frac{\pl L}{\pl\dot q^i}\dot q^i=\sqrt{2E}\int_a^b\!\!ds.
$$
Это означает, что инерциальное движение происходит по экстремалям римановой
метрики $g_{ij}$.
\qed\end{exa}
\begin{exa}
Пусть движение материальной точки происходит по риманову многообразию $(\MM,g)$
с интервалом $ds^2=dq^i dq^j g_{ij}$ в потенциальном поле $U(q)\in\CC^k(\MM)$.
Функция Лагранжа и гамильтониан имеют вид
$$
  L=T-U,\qquad H=T+U,\qquad T=\frac12\left(\frac{ds}{d\tau}\right)^2.
$$
Чтобы обеспечить фиксированное значение энергии $H=E$, параметр $\tau$ вдоль
траектории необходимо выбрать пропорциональным длине:
$$
  d\tau=\frac{ds}{\sqrt{2(E-U)}}.
$$
Тогда укороченное действие примет вид
$$
  \int_\g\!d\tau\frac{\pl L}{\pl\dot q^i}\dot q^i=\int_\g\!ds\sqrt{2(E-U)}.
$$
Это значит, что движение материальной частицы по риманову многообразию $(\MM,g)$
в потенциальном поле $U$ происходит вдоль экстремалей метрики
$\rho=g\,2(E-U)$, которая связана с исходной метрикой множителем $2(E-U)$. Эта
метрика имеет особенность на границе $U(q)=E$.
\qed\end{exa}

Если начальная и конечная точки экстремали в рассмотренных примерах достаточно
близки, то экстремум длины является минимумом. Это оправдывает название
``принцип наименьшего действия''.

\begin{exa}[\bf Метод факторизации]
\index{Метод факторизации (factorization method)}%
Для нахождения траектории частицы иногда удобно выделить из гамильтониана
$H(q,p)$ отличный от нуля множитель. Пусть $f(q,p)>0$ -- некоторая положительная
функция на фазовом пространстве. Рассмотрим новый гамильтониан
\begin{equation*}
  \widetilde H:=f(q,p)\big(H(q,p)-E\big),\qquad E=\const.
\end{equation*}
Соответствующие уравнения движения имеют вид
\begin{equation}                                                  \label{edfach}
\begin{split}
  \frac{dq}{ds}&=\quad f\frac{\pl H}{\pl p}+\frac{\pl f}{\pl p}(H-E),
\\
  \frac{dp}{ds}&=-f\frac{\pl H}{\pl q}-\frac{\pl f}{\pl q}(H-E).
\end{split}
\end{equation}
Эти уравнения на инвариантной поверхности фазового пространства, определяемой
уравнением $\widetilde H=0$ или $H=E$, эквивалентны гамильтоновым уравнениям
(\ref{eqgenq}) для исходного гамильтониана $H$. Уравнения (\ref{edfach})
позволяют определить только форму траектории. Для определения эволюции частицы
во времени достаточно проинтегрировать уравнение
\begin{equation*}
  \frac{dt}{ds}=f\big(q(s),p(s)\big),
\end{equation*}
где $q(s),p(s)$ -- соответствующая траектория.
\qed\end{exa}
\subsection{Уравнение Гамильтона--Якоби                          \label{shaiae}}
Рассмотрим {\em расширенное конфигурационное} пространство $\MR^{\Sn+1}$ с
координатами $\lbrace q,t\rbrace=(q^1,\dotsc,q^\Sn,t)$, которое
получается из конфигурационного пространства $\MR^\Sn$ добавлением еще одного
измерения -- времени.
\index{Расширенное конфигурационное пространство (extended configuration space)}%
\index{Конфигурационное пространство расширенное (extended configuration space)}%
\begin{defn}
{\em Функцией действия} $S(q,t)$ называется интеграл
\begin{equation}                                                  \label{efunac}
  S(q,t)=\int_\g\!\!d\tau L,
\end{equation}
где $L(q,\dot q,\tau)$ -- функция Лагранжа, и интегрирование ведется вдоль
экстремали $\g:=\lbrace q(\tau),t(\tau)\rbrace$, с фиксированным началом
$q_0,t_0$ и переменным концом $q,t$ расширенного конфигурационного пространства.
\qed\end{defn}
\index{Функция действия (action function)}%
\index{Действия функция (action function)}%
Это определение корректно по крайней мере в малой окрестности начальной точки
$q_0,t_0$. Точнее, функция действия (\ref{efunac}) определена в некоторой
окрестности $\MU$ точки $q_0,t_0$, если любую точку этой окрестности можно
соединить с точкой $q_0,t_0$ экстремалью, целиком лежащей в $\MU$, и эта
экстремаль единственна. Если точка $q,t$ лежит далеко от $q_0,t_0$, то интеграл
(\ref{efunac}) может не определить функцию действия $S(q,t)$ по двум причинам.
Во-первых, может существовать несколько экстремалей, соединяющих точки $q_0,t_0$
и $q,t$. Во-вторых, возможно, что точки $q_0,t_0$ и $q,t$ вообще нельзя
соединить экстремалью. С другой стороны, для любой точки $q,t$ расширенного
конфигурационного пространства можно так подобрать начальную точку $q_0,t_0$,
что функция действия будет определена в некоторой окрестности точки $q,t$. В
дальнейшем мы будем считать, что точка $q_0,t_0$ фиксирована, и рассматривать те
области расширенного конфигурационного пространства, где функция действия
определена.

Аргументом функции действия является верхний предел интеграла (\ref{efunac}),
поэтому ее дифференциал равен
\begin{equation}                                                  \label{ediacf}
  dS(q,t)=dq^ip_i-dtH,
\end{equation}
где $p_i:=\pl L/\pl\dot q^i$ и $H:=\dot q^ip_i-L$ определяются в конечной точке
кривой $\g$. Таким образом, дифференциал функции действия $dS$ является
интегральным инвариантом Пуанкаре--Картана (\ref{eaform}).
\begin{theorem}
Функция действия удовлетворяет уравнению Гамильтона--Якоби
\begin{equation}                                                  \label{eyfjae}
  \frac{\pl S}{\pl t}+H\left(q,\frac{\pl S}{\pl q},t\right)=0.
\end{equation}
\end{theorem}
\begin{proof}
Достаточно заметить, что из выражения (\ref{ediacf}) следуют равенства
\begin{equation*}
  \frac{\pl S}{\pl t}=-H(q,p,t),\qquad p_i=\frac{\pl S}{\pl q^i}.\tag*{\qed}
\end{equation*}
\renewcommand{\qed}{}\end{proof}
\index{Уравнение Гамильтона--Якоби (Hamilton--Jacobi equation)}%
\index{Гамильтона--Якоби уравнение (Hamilton--Jacobi equation)}%
Очевидно, что любое решение уравнения Гамильтона--Якоби определено с точностью
до аддитивной постоянной, которая соответствует произволу в выборе начальной
точки $q_0,t_0$ экстремали $\g$ в определении функции действия (\ref{efunac}).

Уравнение Гамильтона--Якоби представляет собой уравнение в частных производных
первого порядка. Покажем, что его интегрирование сводится к интегрированию
обыкновенных дифференциальных уравнений Гамильтона. Поставим для уравнения
Гамильтона--Якоби задачу Коши, т.е.\ будем искать решение уравнения
(\ref{eyfjae}) с начальным условием
\begin{equation}                                                  \label{einchj}
  S(q,t_0)=S_0(q).
\end{equation}
Чтобы построить решение этой задачи, рассмотрим задачу Коши для уравнений
Гамильтона (\ref{ehameq}) с начальными условиями:
$$
  q^i(t_0)=q^i_0,\qquad p_i(t_0)=\left.\frac{\pl S_0}{\pl q^i}\right|_{q_0}.
$$
Соответствующее этим начальным условиям решение изображается в расширенном
конфигурационном пространстве кривой $q(t)$, которая является стационарной
точкой действия $\int\!dt L$, где $L(q,\dot q,t)$ есть преобразование Лежандра по
импульсу $p$ от гамильтониана $H(q,p,t)$. Эта траектория называется
{\em характеристикой} задачи Коши для уравнения Гамильтона--Якоби, выходящей из
точки $q_0,t_0$. При временах, достаточно близких к $t_0$, значения $q(t),t$
можно принять за координаты в окрестности точки $q_0,t_0$ расширенного
конфигурационного пространства. Построим теперь функцию действия
\index{Характеристика уравнения Гамильтона--Якоби%
(characteristic of the Hamilton--Jacobi equation)}%
$$
  S(q,t)=S_0(q_0)+\int_{q_0,t_0}^{q,t}\!\!\!d\tau L(q,\dot q,\tau),
$$
где интегрирование ведется вдоль экстремали, соединяющей точки $q_0,t_0$ и
$q,t$. После этого проверяется, что эта функция действия удовлетворяет уравнению
Гамильтона--Якоби (\ref{eyfjae}) и начальному условию (\ref{einchj}). Можно
также доказать, что это решение задачи Коши для уравнения Гамильтона--Якоби
единственно.

Если гамильтониан не зависит от времени явно, то зависимость функции действия от
времени легко находится:
\begin{equation}                                                  \label{eacfce}
  S(q,t)=-Et+W(q),
\end{equation}
где $W(q)$ -- {\em укороченная функция действия}, зависящая только от координат,
и $E=\const$. Из уравнения Гамильтона--Якоби (\ref{eyfjae}) следует, что
укороченная функция действия должна удовлетворять уравнению в частных
производных
\index{Укороченная функция действия (truncated action function)}%
\index{Функция действия укороченная (truncated action function)}%
\begin{equation}                                                  \label{eshaja}
 H\left(q,\frac{\pl W}{\pl q}\right)=E.
\end{equation}
Отсюда следует, что постоянная $E$ равна энергии системы. Это уравнение
называется {\em укороченным} уравнением Гамильтона--Якоби.
\index{Укороченное уравнение Гамильтона--Якоби%
(truncated Hamilton--Jacobi equation)}%
\index{Уравнение Гамильтона--Якоби укороченное%
(truncated Hamilton--Jacobi equation)}%
\begin{exa}
Поясним определение функции действия на простом примере свободной точечной
частицы единичной массы, движущейся в евклидовом пространстве $\MR^\Sn$.
В этом случае лагранжиан и гамильтониан частицы имеют вид
\begin{equation*}
  L=\frac12\dot q^i\dot q_i,\qquad H=\frac12p^ip_i,
\end{equation*}
где подъем и опускание индексов осуществляется с помощью символов Кронекера.
Траекториями свободной частицы являются прямые линии и только они. Экстремаль
$q^i(\tau)$, соединяющая точки $q_0^i,t_0$ и $q^i,t$, задается линейной функцией
\begin{equation*}
  q^i(\tau)=q_0^i+\frac{q^i-q_0^i}{t-t_0}(\tau-t_0),
\end{equation*}
описывающей равномерное прямолинейное движение частицы со скоростью
$\dot q^i=(q^i-q_0^i)/(t-t_0)$.
Нетрудно вычислить соответствующую функцию действия
\begin{equation*}
  S(q,t)=\int_{t_0}^t\!\!\!d\tau\frac12\left(\frac{q-q_0}{t-t_0}\right)^2
  =\frac12\frac{(q-q_0)^2}{t-t_0},
\end{equation*}
где
\begin{equation*}
  (q-q_0)^2:=(q^i-q^i_0)(q_i-q_{0i}).
\end{equation*}
Функция действия определена для всех $q\in\MR^\Sn$, $t>t_0$ и удовлетворяет
уравнению Гамильтона--Якоби:
\begin{equation*}
  \frac{\pl S}{\pl t}+\frac12\frac{\pl S}{\pl q^i}\frac{\pl S}{\pl q_i}=0.
\end{equation*}

Поскольку гамильтониан частицы не зависит от времени явно, то для любой
траектории частицы энергия сохраняется:
\begin{equation*}
  E=\frac12\left(\frac{q-q_0}{t-t_0}\right)^2=\const.
\end{equation*}
Отсюда можно выразить время через координаты
\begin{equation*}
  t-t_0=\sqrt{\frac{(q-q_0)^2}{2E}}.
\end{equation*}
Поскольку $S=E(t-t_0)$, то укороченное действие равно
\begin{equation*}
  W=Et+S=2E(t-t_0)+Et_0=\sqrt{2E(q-q_0)^2}+Et_0.
\end{equation*}
Нетрудно проверить, что укороченное действие удовлетворяет укороченному
уравнению Гамильтона--Якоби
\begin{equation*}
  E=\frac12\frac{\pl W}{\pl q^i}\frac{\pl W}{\pl q_i}.               \tag*{\qed}
\end{equation*}
\renewcommand{\qed}{}\end{exa}

Уравнение Гамильтона--Якоби предоставляет мощный метод решения задач
механики точечных частиц. При обобщении этого метода на теорию поля
возникают существенные усложнения. В настоящее время, насколько известно
автору, это интересное обобщение не развито.
\subsection{Принцип Гюйгенса}
\index{Принцип Гюйгенса (Huygens principle)}%
\index{Гюйгенса принцип (Huygens principle)}%
Многие понятия гамильтоновой механики возникли при перенесении  на общие
вариационные принципы весьма простых и наглядных понятий геометрической оптики.
В настоящем разделе мы обсудим некоторые аспекты геометрической оптики.

Рассмотрим свет, распространяющийся в среде (конфигурационном пространстве),
которую мы отождествим с евклидовым пространством $\MR^3$ с декартовыми
координатами $q^i$, $i=1,2,3$. В общем случае скорость света зависит от точки
$q\in\MR^3$ (неоднородная среда) и направления луча света (неизотропная среда).
Неизотропность среды можно описать, задав в каждой точке $q$ поверхность в
касательном пространстве $\MT_q(\MR^3)$. Для этого отложим в начале координат
касательного пространства $\MT_q(\MR^3)$ вектор скорости распространения света в
точке $q$. Эта поверхность называется {\em индикатрисой}. Согласно {\em принципу
Ферма}, свет распространяется в среде из точки $q_0$ в точку $q$ за кратчайшее
время.
\index{Индикатриса (indicatrix)}%
\index{Принцип Ферма (Fermat principle)}%
\index{Ферма принцип (Fermat principle)}%

Если среда изотропна, то принцип Ферма приобретает простую математическую форму.
Пусть $q(t)$ -- траектория луча. Тогда для скорости света $v^i:=dq^i/dt$
справедливо равенство
\begin{equation*}
  v^2dt^2=ds^2,
\end{equation*}
где $v^2:=v^iv^j\dl_{ij}$ и $ds^2:=dq^idq^j\dl_{ij}$. Отсюда следует, что время
прохождения луча между точками $q_0$ и $q$ вдоль траектории $q(t)$ равно
\begin{equation*}
  t=\int_{q_0}^q\!\frac{ds}{v},
\end{equation*}
где $v:=\sqrt{v^2}$ и $ds:=\sqrt{ds^2}$. Для изотропной среды показатель
преломления $n$ равен отношению скорости света $c$ в вакууме к скорости света
$v$ в среде, $n=c/v$. Поэтому время распространения луча света дается интегралом
\begin{equation}                                                  \label{eferpr}
  t=\frac1c\int_{q_0}^q\!\!\!ds\,n.
\end{equation}
Таким образом, принцип Ферма для распространения лучей света в изотропной среде
сводится к вариационной задаче для действия (\ref{eferpr}), в котором показатель
преломления $n(q)$ является заданной функцией на конфигурационном пространстве.

Распространение света в среде можно также описать на языке волновых фронтов.
Допустим, что в момент времени $t=0$ в точке $q_0$ произошла вспышка света.
Рассмотрим множество точек, до которых свет дойдет за время, меньшее или равное
$t>0$. Граница этого множества $\Phi_0(t)$, которая представляет собой некоторую
поверхность в $\MR^3$, называется {\em волновым фронтом} точки $q_0$ через время
$t$ и состоит из точек, до которых свет дойдет за время $t$ и не может дойти
быстрее. При этом мы предполагаем, что среда такова, что волновые фронты
представляют собой гладкие поверхности в $\MR^3$. Тогда между волновыми фронтами
для разных моментов времени $t$ имеется замечательное соотношение.
\index{Волновой фронт (wave front)}\index{Фронт волновой (wave front)}%
\begin{theorem}[\bf Принцип Гюйгенса]
Рассмотрим волновой фронт $\Phi_0(t)$ точки $q_0$ в момент времени $t$. Для
каждой точки этого фронта $q\in\Phi_0(t)$ построим волновой фронт $\Phi_q(s)$
через время $s>0$. Тогда волновой фронт $\Phi_0(t+s)$ точки $q_0$ через время
$t+s$ будет огибающей поверхностью всех фронтов $\Phi_q(s)$ для $q\in\Phi_0(t)$.
\end{theorem}
\begin{proof}
Пусть $q_{t+s}\in\Phi_0(t+s)$. Тогда существует путь из начальной точки $q_0$ в
точку $q_{t+s}$, по которому свет распространяется за время $t+s$ и нет более
короткого. Рассмотрим точку $q_t$ на этом пути, до которой свет идет время $t$.
Никакого более короткого пути из $q_0$ в $q_t$ не существует, иначе путь не был
бы кратчайшим. Поэтому точка $q_t$ лежит на фронте $\Phi_0(t)$. Точно так же
\begin{figure}[h,b,t]
\hfill\includegraphics[width=.35\textwidth]{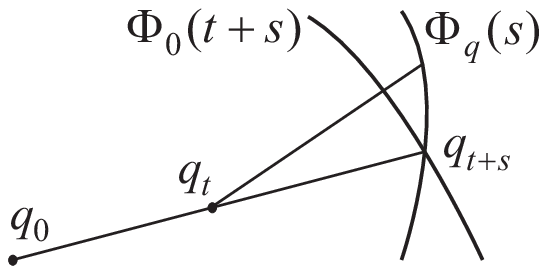}
\hfill {}
\centering\caption{Невозможность пересечения фронтов}
\label{feldef}
\end{figure}
путь из $q_t$ в $q_{t+s}$ свет проходит за время $s$ и нет более короткого пути
между этими точками. Поэтому точка $q_{t+s}$ лежит на фронте $\Phi_q(s)$ точки
$q(t)$ через время $s$. Осталось показать, что фронты $\Phi_q(s)$ и
$\Phi_0(t+s)$ в точке $q(t+s)$ касаются. Действительно, предположим, что фронты
пересекаются, как показано на рис.\ref{feldef}. Тогда в некоторые точки
фронта $\Phi_0(t+s)$ из точки $q_t$ можно было бы добраться за время, меньшее
$s$. В свою очередь это значит, что в эти точки из $q_0$ можно было бы добраться
за время, меньшее $t+s$, что противоречит определению фронта $\Phi_0(t+s)$.
\end{proof}
Разумеется, точку $q_0$ в принципе Гюйгенса можно заменить на кривую,
поверхность или вообще на произвольное замкнутое множество, а трехмерное
евклидово пространство на произвольное дифференцируемое многообразие $\MM$. При
этом свет можно заменить на распространение любого возмущения, передающегося
локально, т.е.\ вдоль линии $\g\in\MM$, на которой некоторый функционал
принимает наименьшее значение.

Принцип Гюйгенса приводит к двум способам описания распространения возмущений.
Во-первых, можно следить за {\em лучами}, т.е.\ кратчайшими путями, вдоль
которых распространяется свет (корпускулярная точка зрения). В этом случае
распространение света задается вектором скорости $\dot q$ в каждой точке
конфигурационного пространства, т.е.\ некоторой точкой на индикатрисе.
Во-вторых, мы можем следить за распространением волнового фронта (волновая точка
зрения).
\index{Луч (ray)}%

Определим скорость движения $p$ волнового фронта следующим образом. Для каждой
точки $q_0$ определим функцию $S_0(q)$, как наименьшее время распространения
света из точки $q_0$ в точку $q$. По-построению, поверхность уровня функции
$S_0(q)$ является волновым фронтом:
\begin{equation*}
  \Phi_0(t):=\lbrace q\in\MR^3:\quad S_0(q)=t\rbrace.
\end{equation*}
\begin{defn}
Ковектор с компонентами
\begin{equation*}
  p_i:=\frac{\pl S_0(q)}{\pl q^i}
\end{equation*}
называется {\em ковектором нормальной медлительности фронта}.
\qed\end{defn}
\index{Ковектор нормальной медлительности фронта}%
Такое название связано с тем, что чем больше градиент функции $S_0(q)$, тем
медленнее движется фронт. Действительно, в линейном приближении справедливо
равенство
\begin{equation*}
  S_0(q+dq)=S_0(q)+p_idq^i=t+dt.
\end{equation*}
Отсюда следует, что чем больше расстояние $dq$, пройденное светом за время $dt$,
тем меньше ковектор нормальной медлительности фронта $p$. Векторы $X$,
касательные к волновому фронту, определяются, как ортогональное дополнение к
ковектору нормальной медлительности фронта:
\begin{equation*}
  X^ip_i=0.
\end{equation*}
Если на конфигурационном пространстве определена риманова метрика $g_{ij}$, то
можно определить вектор нормальной медлительности фронта с компонентами
$p^i:=g^{ij}p_j$. По-построению, этот вектор ортогонален волновому фронту.

Если среда анизотропна, то направление лучей света $\dot q$ и направление
движения фронта $p$ не совпадают. Однако они связаны между собой простым
соотношением, которое легко выводится из принципа Гюйгенса. Напомним, что
оптические свойства среды в точке $q$ характеризуются поверхностью в касательном
пространстве $\MT_q(\MR^3)$ -- индикатрисой.
\begin{defn}
Направление гиперплоскости (т.е.\ вектор, перпендикулярный к данной
гиперплоскости), касающейся индикатрисы в точке $\dot q$, называется
{\em сопряженным} к направлению $\dot q$.
\qed\end{defn}
\index{Сопряженное направление}%
\begin{prop}
Направление $p$ волнового фронта $\Phi_0(t)$ в точке $q_t$ сопряжено направлению
луча $\dot q$ в данной точке.
\end{prop}
\begin{proof}
Рассмотрим точки $q_{t-\e}$, лежащие на луче с началом в точке $q_0$ и концом в
$q_t$, при малых $\e>0$. Фронт волны $\Phi_{q_{t-\e}}(\e)$ в точке $q_{t-\e}$ в
момент времени $\e$ можно представить в виде
\begin{equation*}
  \Phi_{q_{t-\e}}(\e)=\lbrace q^i_{t-\e}+\dot q^i_t\e\rbrace,
\end{equation*}
где вектор скорости $\dot q$ пробегает всю индикатрису, с точностью порядка
$\osmall(\e)$. Из принципа Гюйгенса следует, что фронт $\Phi_{q_{t-\e}}(\e)$
касается фронта $\Phi_0(t)$ в точке $q_t$. В пределе $\e\to0$ получаем
утверждение предложения.
\end{proof}
Теперь сравним геометрическую оптику с построениями предыдущего раздела.
Лучом является траектория частицы $q(t)$ в конфигурационном пространстве
$\MR^3$. Ковектор нормальной медлительности фронта -- это импульс частицы $p$.
Принцип Ферма соответствует принципу наименьшего действия. Функция
$S_0(q)$ есть ни что иное, как функция действия (\ref{efunac}), в которой время
$t$ соответствует концу траектории $q=q(t)$. Поверхности уровня функции действия
$S(q,t)$ соответствуют волновым фронтам.
\subsection{Переменные действие-угол}
Знание интегралов движения позволяет упросить задачу. При этом интерес
представляют функционально независимые интегралы движения. Из определения
функциональной независимости следует, что на фазовом пространстве размерности
$2\Sn$ может существовать не более $2\Sn$ функционально независимых интегралов
движения.

Чтобы проинтегрировать систему из $2\Sn$ обыкновенных дифференциальных
уравнений первого порядка, достаточно знать $2\Sn$ независимых первых
интегралов. Оказывается, что если задана каноническая система уравнений
движения, то ситуация существенно проще: достаточно знать только $\Sn$
независимых первых интегралов. Это происходит потому что каждый интеграл
движения позволяет понизить порядок системы уравнений не на одну, а на две
единицы.
\begin{defn}
Функция $F\in\CC^1(\MM)$ на пуассоновом многообразии $\MM$ называется {\em
первым интегралом} механической системы с гамильтонианом $H$, если ее скобка
Пуассона с гамильтонианом равна нулю, $[F,H]=0$. Две функции
$F_1,F_2\in\CC^1(\MM)$ находятся в {\em инволюции}, если их скобка Пуассона
равна нулю, $[F_1,F_2]=0$.
\qed\end{defn}
\index{Первый интеграл (first integral)}%
\index{Интеграл первый (first integral)}%
\index{Инволюция (involution)}%
Заметим, что условие $[F,H]=0$ эквивалентно условию $\dot f=0$, т.к.\
$\dot f=[F,H]$
\begin{theorem}[\bf Лиувилль]                                     \label{tliovs}
Предположим, что на симплектическом многообразии $\MM$ размерности $2\Sn$ заданы
$\Sn$ функционально независимых дифференцируемых функций $\lbrace F_i\rbrace$,
$i=1,\dotsc\Sn$, которые находятся в инволюции, $[F_i,F_j]=0$. Рассмотрим
$\Sn$-мерное подмногообразие $\MM_f\hookrightarrow\MM$, которое является
множеством уровня функций $F_i$:
\begin{equation*}
  \MM_f:=\lbrace x\in\MM:\quad F_i=f_i=\const,~i=1,\dotsc,\Sn\rbrace.
\end{equation*}
Тогда справедливы следующие утверждения:\newline
\indent 1) \parbox[t]{.93\linewidth}{Подмногообразие $\MM_f$ инвариантно
относительно фазового потока с функцией Гамильтона $H=F_i$ при любом
фиксированном $i$.}\newline
\indent 2) \parbox[t]{.93\linewidth}{Если подмногообразие $\MM_f$ компактно и
связно, то оно диффеоморфно $\Sn$-мерному тору $\MT^\Sn$.}\newline
\indent 3) \parbox[t]{.93\linewidth}{Фазовый поток с функцией Гамильтона $H$
определяет на $\MM_f$ условно-периодическое движение, т.е.\ в угловых
координатах $\lbrace \vf^i,~\mod 2\pi\rbrace$ на торе уравнения движения имеют
вид
\begin{equation}                                                  \label{eangci}
  \frac{d\vf^i}{dt}=\om^i,\qquad \om^i=\om^i(f)=\const,\quad \forall i.
\end{equation}  }
\indent 4) \parbox[t]{.93\linewidth}{Канонические уравнения движения с функцией
Гамильтона $H$ интегрируются в квадратурах.}
\end{theorem}
\begin{proof}
См., например, \cite{Arnold89R}.

Прокомментируем первые два утверждения теоремы.

Поскольку фазовое пространство $\MM$ является симплектическим и, следовательно,
пуассоновым многообразием, то каждой функции $F_i$ ставится в соответствие
векторное поле $X_i$ по правилу (\ref{emafve}). При этом инволютивность первых
интегралов движения влечет за собой инволютивность распределения векторных
полей $\lbrace X_i\rbrace$. Согласно теореме Фробениуса у этого распределения
существует интегральное подмногообразие, которым является
$\MM_f\hookrightarrow\MM$. Отсюда следует инвариантность $\MM_f$ относительно
фазового потока для любой функции $F_i$.

Тор возникает из-за того, что векторные поля $X_i$, которые являются
генераторами группы однопараметрических преобразований $\MM_f$, не просто
находятся в инволюции, а коммутируют между собой.
\end{proof}
\begin{com}
Если гамильтониан механической системы $H$ не зависит от времени, то его можно
выбрать в качестве одного из первых интегралов $F_i$.
\qed\end{com}
\begin{defn}
Гамильтонова механическая система, называется {\em интегрируемой}, если она
имеет $\Sn$ или более функционально независимых интегралов движения.
\qed\end{defn}
\index{Интегрируемая гамильтонова система (integrable Hamiltonian system)}%
\index{Гамильтонова система интегрируемая (integrable Hamiltonian system)}%

Если фазовое пространство механической системы таково, что выполнены условия
теоремы Лиувилля, то на фазовом пространстве $\MM$ существует выделенная система
координат (действие-угол), в которой уравнения движения выглядят особенно
просто. Для определенности будем считать, что гамильтониан системы совпадает с
первым интегралом движения: $H=F_1$. В теореме Лиувилля утверждается, что
подмногообразие $\MM_f\hookrightarrow\MM$ является $\Sn$-мерным тором,
инвариантно относительно фазового потока и на нем существуют угловые координаты
$\vf^i$, для которых уравнения движения имеют вид (\ref{eangci}). Общее решение
этих уравнений имеет простой вид
\begin{equation*}
  \vf=\vf_0+\om t,\qquad \vf_0=\const.
\end{equation*}
Поскольку интегралы движения функционально независимы, то в некоторой
окрестности подмногообразия $\MM_f$ в качестве координат можно выбрать
совокупность функций $\lbrace F,\vf\rbrace$. В этой системе координат уравнения
движения принимают простой вид
\begin{equation*}
  \frac{dF}{dt}=0,\qquad \frac{d\vf}{dt}=\om(F),
\end{equation*}
и легко интегрируются:
\begin{equation*}
  F(t)=F(0),\qquad \vf(t)=\vf_0+\om\big(F(0)\big)t.
\end{equation*}

В общем случае в координатах $\lbrace F,\vf\rbrace$ симплектическая форма не
будет иметь канонического вида. Однако существует такой набор функционально
независимых функций $\lbrace I_i=I_i(F)\rbrace$, $i=1,\dotsc,\Sn$, что
переменные $\lbrace I,\vf\rbrace$ образуют такую систему координат в окрестности
$\MM_f$, в которой симплектическая форма имеет канонический вид
\begin{equation}                                                  \label{ecasyp}
  \varpi=dI_i\wedge d\vf^i.
\end{equation}
\begin{defn}
Координаты $\lbrace \vf^1,\dotsc,\vf^\Sn,I_1,\dotsc,I_\Sn\rbrace$ в некоторой
окрестности подмногообразия фазового пространства $\MM_f\subset\MR^{2\Sn}$, в
которых канонические уравнения движения имеют вид
\begin{equation*}
  \frac{dI}{dt}=0,\qquad \frac{d\vf}{dt}=\om(I),
\end{equation*}
и симплектическая форма является канонической (\ref{ecasyp}) называются
{\em переменными действие-угол}.
\qed\end{defn}
\index{Переменные действие-угол (action-angle variables)}%
\index{Действие-угол переменные (action-angle variables)}%

Переменные действие-угол являются координатами на фазовом пространстве. При этом
угловые переменные удобно рассматривать как обобщенные координаты, а переменные
действия -- как сопряженные импульсы. Из определения сразу следует, что
преобразование координат $q,p\mapsto\vf,I$, если оно существует, является
каноническим.

Переменные $I$ так же как и функции $F$ являются первыми интегралами движения.
В переменных действие-угол гамильтониан механической системы имеет вид
$H=F_1=H(I)$ и не зависит от угловых переменных $\vf$. То есть каждая из
угловых координат $\vf$ является циклической.
\begin{theorem}
Координаты действие-угол $\lbrace\vf,I\rbrace$ существуют в некоторой
окрестности подмногообразия $\MM_f\hookrightarrow\MR^{2\Sn}$.
\end{theorem}
\begin{proof}
Рассмотрим множество торов $\MM_f$, соответствующих различным значениям
интегралов движения $F_i$. Пусть $\g_i$, $i=1,\dotsc,\Sn$, -- базисные
одномерные циклы (окружности) на торах $\MM_f$, т.е.\ приращение координаты
$\vf^i$ на цикле $\g_j$ равно $2\pi$, если $i=j$ и 0, если $i\ne j$. Каждое
подмногообразие $\MM_f\hookrightarrow\MR^{2\Sn}$ является нулевым.
Действительно, векторные поля $X_i$, соответствующие интегралам движения $F_i$,
образуют базис касательных пространств $\MT_x(\MM_F)$ для всех $x\in\MM_f$ и
попарно коммутируют. Поэтому значение канонической симплектической формы
$\varpi=dp_i\wedge dq^i$ на двух произвольных базисных полях равно нулю:
\begin{equation*}
  \varpi(X_i,X_j)=\varpi_{kl}X^k_i X^l_j=\varpi^{-1kl}\pl_k F_i\pl_l F_j
  =[F_i,F_j]=0.
\end{equation*}
Следовательно, подмногообразие $\MM_f$ является нулевым. Отсюда вытекает, что
1-форма $dq\,p$ замкнута на $\MM_f$, т.е.\ ее внешняя производная $dp\wedge dq$
обращается в нуль на подмногообразии $\MM_f$. Положим
\begin{equation*}
  I_i(F):=\oint_{\g_i}\!\!\! dq\,p.
\end{equation*}
Эти функции зависят от значений интегралов движения $F_i$, определяющих тор
$\MM_f$, но не зависят от выбора базисных циклов $\g_i$, т.к.\ форма $dq\,p$
замкнута. Это означает, что функции $I_i(F)$ не зависят от координат $\vf$ на
торе.

Теперь совершим два канонических преобразования. Поскольку скобка Пуассона
интегралов движения $F_i$ между собой равна нулю, то их можно выбрать в качестве
новых импульсов: $P_i=F_i$. Совершим первое каноническое преобразование
$q,p\mapsto Q,F$ с производящей функцией $S_4(p,F)$ (см.\ следующий раздел),
зависящей от новых и старых импульсов $F$ и $p$. Тогда старые и новые координаты
определены соотношениями (\ref{ethctf}):
\begin{equation*}
  q=-\frac{\pl S_4}{\pl p},\qquad Q=\frac{\pl S_4}{\pl F}.
\end{equation*}
Подмногообразие $\MM_f$ в новой системе координат задано соотношениями $P=0$.
Следовательно, новые координаты $Q$ образуют некоторую систему координат на
$\MM_f$.

Зафиксируем точку $Q_0\in\MM_f$. В некоторой ее окрестности можно выбрать
систему координат $Q,I$, т.к.\ функции $I=I(P)$ зависят только от импульсов
$P=F$. При этом импульсы $P$ можно выразить, как функции от $Q$ и $I$, т.е.\
$P=P(Q,I)$. Поскольку преобразование $q,p\mapsto Q,P$ каноническое, то 1-форма
$dQP$ замкнута. Поэтому в односвязной окрестности $\MU$ точки
$Q_0\in\MU\subset\MM_f$ определена функция
\begin{equation*}
  S_2(Q,I)=\int_{Q_0}^Q\!\!\!dQ'P(Q',I).
\end{equation*}
Этот интеграл не зависит от кривой, целиком лежащей в $\MU$ и соединяющей точки
$Q_0$ и $Q$. Следовательно функцию $S_2(Q,I)$ можно выбрать в качестве
производящей функции второго канонического преобразования $Q,P\mapsto \vf,I$,
зависящей от старых координат $Q$ и новых импульсов $I$. Формулы преобразования
имеют вид (\ref{esecat}):
\begin{equation*}
  P=\frac{\pl S_2}{\pl Q},\qquad \vf=\frac{\pl S_2}{\pl I}.
\end{equation*}

Таким образом, преобразование координат $q,p\mapsto \vf,I$ является
произведением двух канонических преобразований $q,p\mapsto Q,P$ и
$Q,P\mapsto\vf,I$ и, следовательно, само является каноническим. По-построению,
оно определено в окрестности тора $\MM_f$.
\end{proof}
\begin{com}
В доказательстве теоремы переход от координат $q,p$ к переменным действие-угол
$\vf,I$ содержит только алгебраические операции и интегрирование. Это доказывает
утверждение 4 теоремы Лиувилля \ref{tliovs}.
\qed\end{com}

Переменные действие-угол определены неоднозначно. Очевидно, что координаты можно
сдвигать:
\begin{equation*}
  I'=I+\const,\qquad \vf'=\vf+\const(I).
\end{equation*}
\subsection{Канонические преобразования                          \label{scantr}}
Наиболее мощный и гибкий метод нахождения точных решений уравнений движения дают
канонические преобразования. По-прежнему, обозначим конфигурационное и фазовое
пространства соответственно через $\MR^\Sn$, и
$\MT^*(\MR^\Sn)\approx\MR^{2\Sn}$. Будем считать, что на фазовом пространстве
заданы координаты  $\lbrace x^\al\rbrace=\lbrace q^i,p_i\rbrace
:=(q^1,\dotsc,q^\Sn,p_1,\dotsc,p_\Sn)$ и каноническая пуассонова структура
(\ref{ecapos}).
\begin{defn}
Достаточно гладкое биективное отображение $g$ фазового пространства
(диффеоморфизм) $\MT^*(\MR^\Sn)$ на себя называется каноническим, если оно
сохраняет каноническую симплектическую форму:
\begin{equation*}
  g^*\varpi=\varpi,
\end{equation*}
где $\varpi:=dp\wedge dq:=dp_i\wedge dq^i$.
\qed\end{defn}
В координатах это определение записывается следующим образом. Отображение $g$
фазового пространства задается $2\Sn$ функциями, которые в общем случае могут
зависеть от времени,
\begin{equation}                                                  \label{ecantc}
  g:\quad \MT^*(\MR^\Sn)\ni\quad(q^i,p_i)\mapsto\big(Q^i(q,p,t),P_i(q,p,t)\big)
  \quad\in\MT^*(\MR^\Sn),
\end{equation}
где время $t$ входит в качестве параметра. Поскольку ранг 2-формы $\varpi$ равен
$2\Sn$, то канонические преобразования задают преобразование координат
(диффеоморфизм) фазового пространства с отличным от нуля якобианом:
\begin{equation*}
  \frac{\pl(Q,P)}{\pl(q,p)}\ne0.
\end{equation*}
Условие сохранения канонической симплектической формы $\varpi$ принимает вид
\begin{equation}                                                  \label{ecanco}
  dp_i\wedge dq^i=dP_i\wedge dQ^i.
\end{equation}
Это равенство приводит к дифференциальным уравнениям на функции $Q^i(q,p,t)$,
$P_i(q,p,t)$, которые мы рассмотрим несколько позже.

Определение канонических преобразований можно записать в эквивалентном
интегральном виде
\begin{equation}                                                  \label{eincat}
  \int_\MS\!dp\wedge dq=\int_{g\MS}\!\!\!dp\wedge dq,
\end{equation}
где интегрирование ведется по произвольной двумерной поверхности
$\MS\subset\MT^*(\MR^\Sn)$ с кусочно гладкой границей. Поскольку кокасательное
расслоение $\MT^*(\MR^\Sn)$ односвязно, то к интегралу (\ref{eincat}) можно
применить формулу Стокса:
\begin{equation*}
  \int_\MS\!dp\wedge dq=\oint_{\pl\MS}\!\!\!dq\,p,
\end{equation*}
где мы предположили, что граница $\pl\MS$ такова, что $dq$ на ней может
обратиться в нуль только в изолированных точках.
Поэтому для односвязных фазовых пространств интегральное определение
канонического преобразования (\ref{eincat}) можно переписать в виде
\begin{equation}                                                  \label{eincag}
  \oint_\g\!dq\,p=\oint_{g\g}\!\!\!dq\,p,
\end{equation}
где $\g$ -- произвольная кусочно гладкая замкнутая кривая в фазовом
пространстве.

В общем случае неодносвязных многообразий из равенства (\ref{eincag}) следует
(\ref{eincat}), но не наоборот. То есть сохранение 1-формы $dq\,p$
является достаточным условием того, что преобразование будет каноническим.
\begin{exa}
Каноническая пуассонова структура определена на произвольном кокасательном
расслоении $\MT^*(\MM)$. При этом вид структурных функций не зависит от выбора
координат на конфигурационном пространстве $\MM$. Поэтому преобразование
$q,p\mapsto Q(q),P(q,P)$, где $Q$ зависит только от $q$, является каноническим
(см., пример \ref{epocax}).
\qed\end{exa}
\begin{exa}
Пусть фазовое пространство топологически тривиально,
$\MT^*(\MM)\approx\MR^{2\Sn}$. Поскольку линейное преобразование координат
фазового пространства, порождаемое симплектической группой $\MS\MP(\Sn,\MR)$
(см.\ раздел \ref{simgru}), сохраняет каноническую симплектическую форму, то оно
является каноническим.
\qed\end{exa}
На практике наиболее полезными и содержательными являются нелинейные
канонические преобразования, связанные со спецификой задачи: симметриями и
наличием интегралов движения.
\begin{exa}
Фазовый поток является каноническим отображением в силу следствия из теоремы
\ref{tinpot}.
\qed\end{exa}
\begin{exa}[\bf Теория возмущений]
Допустим, что гамильтониан механической системы состоит из двух слагаемых,
\begin{equation}                                                  \label{efuham}
  H(q,p)=H_0(q,p)+H_1(q,p).
\end{equation}
При этом гамильтониан $H_0$ настолько прост, что для него известно общее решение
\begin{equation}                                                  \label{egesis}
  q=q(t;q_0,p_0),\qquad p=p(t;q_0,p_0),
\end{equation}
задачи Коши с начальными условиями:
\begin{equation*}
  q(0;q_0,p_0)=q_0,\qquad p(0;q_0,p_0)=p_0.
\end{equation*}
Как было отмечено, фазовый поток определяет каноническое преобразование, которое
задается функциями (\ref{egesis}). Обратное преобразование $q,p\mapsto q_0,p_0$
также является каноническим. При этом преобразовании гамильтониан $H_0$
обращается в константу и уравнения движения исходной гамильтоновой системы
упрощаются:
\begin{equation*}
  \dot q_0=\frac{\pl\tilde H_1}{\pl p_0},\qquad
  \dot p_0=-\frac{\pl\tilde H_1}{\pl q_0},
\end{equation*}
где
\begin{equation}                                                  \label{epehat}
  \tilde H_1(q_0,p_0,t)=H_1\big(q(t;q_0,p_0),p(t;q_0,p_0)\big).
\end{equation}
Обозначим через $\tilde q_0(t;q_0.p_0)$ и $\tilde p_0(t;q_0.p_0)$ решение задачи
Коши для полученной системы уравнений движения с начальными данными:
\begin{equation*}
  \tilde q_0(0;q_0,p_0)=q_0,\qquad \tilde p_0(0;q_0,p_0)=p_0.
\end{equation*}
Это решение обладает замечательным свойством: для невозмущенной системы оно
постоянно и совпадает с начальными данными. Таким образом, задача Коши для
гамильтоновой системы (\ref{efuham}) сведена к задаче Коши для возмущения,
которое определяется гамильтонианом (\ref{epehat}). Формулы для решения исходной
задачи в старых координатах $q,p$ получаются после подстановки решения
$\tilde q_0,\tilde p_0$ в общее решение (\ref{egesis}) невозмущенной задачи.
\qed\end{exa}
Из определения канонических преобразований фазового пространства следует, что
фазовое пространство вместе с каноническими отображениями образуют группу
преобразований. Эта группа является подгруппой группы общих преобразований
координат фазового пространства $\diff\MT^*(\MM)$. В общем случае канонические
преобразования $q^i,p_i\mapsto Q^i,P_i$ не сохраняют структуру кокасательного
расслоения на фазовом пространстве, соответствующую координатам $q^i,p_i$, т.к.\
могут перемешивать канонические координаты и импульсы.

По-определению, каноническое преобразование сохраняет $2$-форму $\varpi$
(\ref{eomfao}) и, следовательно, все ее внешние степени. Отсюда вытекает
\begin{theorem}
Канонические преобразования сохраняют интегральные инварианты
$\varpi^2,\dotsc,\varpi^{\Sn}$.
\end{theorem}
Поскольку $2\Sn$-форма $\varpi^{\Sn}$ существует и невырождена, то на фазовом
пространстве существует также форма объема $\upsilon$, которая ей
пропорциональна (\ref{ecavos}). Отсюда вытекает
\begin{cor}
Канонические преобразования сохраняют форму объема $\upsilon$ фазового
пространства. Другими словами, якобиан канонического преобразования координат
равен единице,
\begin{equation*}
  \det\frac{\pl(Q,P)}{\pl(q,p)}=1.
\end{equation*}
Это значит, что объем
\begin{equation*}
  V=\int_\MU\!\upsilon
\end{equation*}
произвольной области $\MU$ фазового пространства сохраняется при каноническом
преобразовании: $gV=V$.
\end{cor}
Поскольку фазовый поток для произвольной гамильтоновой системы является
каноническим преобразованием, то он также имеет интегральные инварианты
$\varpi,\varpi^2,\dotsc,\varpi^{\Sn}$. Последний из этих инвариантов есть
фазовый объем. Отсюда следует
\begin{theorem}[\bf Лиувилль]                                     \label{tliovf}
Фазовый поток сохраняет объем произвольной области фазового пространства.
\end{theorem}
\index{Теорема Лиувилля (Liouville's theorem)}%
\index{Лиувилля теорема (Liouville's theorem)}%
Верно также обратное утверждение. А именно, пусть задан функционал
\begin{equation*}
  I_k=\int_\MU\!d^{2k}xF(t,x),\qquad k=1,\dotsc,\Sn,
\end{equation*}
где $\MU\subset\MR^{2\Sn}$ -- произвольное четномерное подмногообразие фазового
пространства размерности $\dim\MU=2k$ с кусочно гладкой границей и $F(t,x)$ --
некоторая функция времени и канонических переменных. Если этот функционал
инвариантен относительно фазового потока с произвольным гамильтонианом, то он
называется {\em интегральным инвариантом}.
\index{Интегральный инвариант (integral invariant}%
\index{Инвариант интегральный (integral invariant}%
\begin{theorem}                                                   \label{tintin}
Любой интегральный инвариант $I_k$ фазового потока с произвольным гамильтонианом
имеет вид
\begin{equation*}
  I_k=c_k\int_\MU\!\varpi^k,
\end{equation*}
где $c_k$ -- некоторые постоянные, для всех $k=1,\dotsc,\Sn$.
\end{theorem}
\begin{proof}
Дано сравнительно недавно \cite{HwaChu47}.
\end{proof}

Важность канонических преобразований заключается в том, что они позволяют менять
координаты в фазовом пространстве не меняя вида канонических уравнений движения.
Чтобы показать это, рассмотрим преобразование координат в расширенном фазовом
пространстве $\MR^{2\Sn+1}$, с координатами $q,p,t$. Как было показано в разделе
\ref{scaham}, фазовый поток взаимно однозначно связан с характеристиками
интегрального инварианта Пуанкаре--Картана в расширенном фазовом пространстве.
Поскольку характеристики $1$-формы являются инвариантным объектом, то замену
координат $q,p,t\mapsto Q,P,T$ удобно провести в расширенном фазовом
пространстве.
\begin{theorem}
Пусть $Q,P,T$ -- новая система координат в расширенном фазовом пространстве
$\MR^{2\Sn+1}$ и $K(Q,P,T)$, $S(Q,P,T)$ -- такие функции, что выполнено условие
\begin{equation}                                                  \label{ecatfo}
  dq\,p-dtH=dQP-dTK+dS.
\end{equation}
Тогда траектории фазового потока (\ref{ehameq}) изображаются в координатах
$Q,P,T$ интегральными кривыми канонических уравнений
\begin{equation}                                                  \label{etrcae}
  \frac{dQ^i}{dT}=\frac{\pl K}{\pl P_i},\qquad
  \frac{dP_i}{dT}=-\frac{\pl K}{\pl Q^i}.
\end{equation}
\end{theorem}
\begin{proof}
Поскольку $ddS=0$, то $1$-форма $dS$ не влияет на характеристики формы
(\ref{ecatfo}). Это значит, что фазовый поток определяется также
характеристиками формы $dQP-dTK$. Уравнения (\ref{etrcae}) следуют из теоремы
\ref{tcharp}.
\end{proof}
Каноническим преобразованиям координат в фазовом пространстве $q,p\mapsto Q,P$
соответствует преобразование координат $q,p,t\mapsto Q,P,t$ в расширенном
фазовом пространстве, которое не меняет времени.
\begin{theorem}
В новых координатах $Q,P$ канонические уравнения движения (\ref{ehameq}) имеют
канонический вид (\ref{etrcae}) со старой функцией Гамильтона, выраженной через
новые координаты $K(Q,P,t)=H\big(q(Q,P),p(Q,P),t\big)$.
\end{theorem}
\begin{proof}
Рассмотрим $1$-форму $dqp-dQP$ в фазовом пространстве. Тогда для любой замкнутой
кривой $\g$ справедливо равенство
$$
  \oint_\g\!(dq\,p-dQP)=0,
$$
поскольку преобразование является каноническим. Это равенство можно записать в
дифференциальной форме $dq\,p-dQP=dS$, где $S(Q,P)$ -- некоторая функция на
фазовом пространстве. Следовательно, в расширенном фазовом пространстве имеем
равенство
\begin{equation}                                                  \label{qffrsg}
  dq\,p-dtH=dQP-dtH+dS,
\end{equation}
и справедлива предыдущая теорема.
\end{proof}

Таким образом, мы показали, что при канонических преобразованиях вид
гамильтоновых уравнений движения не меняется. Часто это свойство принимается за
определение канонических преобразований. Следующий пример показывает, что такое
определение является более общим, чем принятое в настоящей монографии.
\begin{exa}
Преобразование координат, которое заключается в растяжке импульсов:
\begin{equation}                                                  \label{edeffc}
  \vf_c:\quad \MR^{2\Sn}\ni\quad(q,p)\mapsto(Q=q,P=cp)\quad\in\MR^{2\Sn},
\end{equation}
где $c\ne0$ -- некоторая постоянная, сохраняет гамильтонову форму уравнений
движения с гамильтонианом $K=cH$. Тем не менее оно не является каноническим,
т.к.\ каноническая форма
\begin{equation*}
  \varpi=dp\wedge dq=\frac1c dP\wedge dQ
\end{equation*}
меняет свой вид.

В более общем случае преобразование $Q=bq,P=cp$, где $b$ и $c$ -- произвольные
отличные от нуля постоянные, также сохраняет вид гамильтоновых уравнений
движения. При этом новый гамильтониан равен $K=bcH$.
\qed\end{exa}
По сути дела этот пример описывает все отличия в разных определениях
канонического преобразования.
\begin{prop}
Пусть при преобразовании координат $\vf:~q,p\mapsto Q,P$ фазового пространства
$\MR^{2\Sn}$ гамильтоновы уравнения движения (\ref{eqgenq}) сохраняют свой вид:
\begin{equation*}
  \dot Q^i=\frac{\pl K}{\pl P_i},\qquad \dot P_i=-\frac{\pl K}{\pl Q^i},
\end{equation*}
где $K(Q,P)$ -- некоторый новый гамильтониан. Тогда
\begin{equation}                                                  \label{epocat}
  \oint_{\vf(\g)}\!\!\!dQP=c\oint_\g\!dq\,p,
\end{equation}
где $c\ne0$ -- некоторая постоянная.
\end{prop}
\begin{proof}
Рассмотрим расширенное фазовое пространство $\MR^{2\Sn+1}$. Пусть $\g$ --
замкнутый контур в фазовом пространстве, соответствующий фиксированному моменту
времени $t=\const$. Тогда в силу следствия из теоремы \ref{tinpot} интегралы
\begin{equation*}
  \oint_\g\!dq\,p\qquad\text{и}\qquad\oint_{\vf(\g)}\!\!\!dQP
\end{equation*}
являются интегральными инвариантами фазовых потоков. По теореме Стокса интеграл
по контуру $\g$ преобразуется в интеграл по поверхности, натянутой на данный
контур. Используя теорему \ref{tintin}, заключаем, что
\begin{equation*}
  \oint_{\vf(\g)}\!\!\!dQP=c\oint_\g\!dq\,p,
\end{equation*}
где $c\ne0$ -- некоторая постоянная.
\end{proof}
Ясно, что множество преобразований, сохраняющих вид гамильтоновых уравнений,
образует группу преобразований фазового пространства.
\begin{cor}
Любое преобразование координат $q,p\mapsto Q,P$ фазового пространства, которое
сохраняет форму гамильтоновых уравнений движения, является композицией
преобразования $\vf_c$ (\ref{edeffc}) и некоторого канонического преобразования.
\qed\end{cor}
\begin{proof}
При преобразовании координат $q,p\mapsto Q,P$ относительный интегральный
инвариант Пуанкаре--Картана умножается на некоторую постоянную $c$
(\ref{epocat}). Представим это преобразование в виде двух последовательных
преобразования координат:
\begin{equation*}
  \vf_c:~q,p\mapsto\tilde q,\tilde p\qquad \text{и}\qquad
  g:~\tilde q,\tilde p\mapsto Q,P.
\end{equation*}
Поскольку при первом преобразовании инвариант умножается на ту же постоянную
$c$, то при втором преобразовании координат он сохраняется,
\begin{equation*}
  g:\quad \oint_{\vf_c(\g)}\!\!\!d\tilde q\,\tilde p
  =\oint_{g\circ\vf_c(\g)}\!\!\!dQP.
\end{equation*}
Ввиду произвольности контура $\g$ это значит, что преобразование $g$ является
каноническим.
\end{proof}
Поскольку преобразование $\vf_c$ (\ref{edeffc}) является простой растяжкой
координат, то канонические преобразования составляют нетривиальную и наиболее
содержательную часть преобразований координат, сохраняющих вид гамильтоновых
уравнений движения.

Сформулируем три критерия для канонических преобразований. С этой целью введем
новое понятие. Обозначим координаты фазового пространства через
$\lbrace x^\al\rbrace=\lbrace q^i,p_i\rbrace$, $\al=1,\dotsc,2\Sn$.
\begin{defn}
Пусть на фазовом пространстве $\MR^{2\Sn}$ с координатами
$\lbrace x^\al\rbrace=\lbrace q^i,p_i\rbrace$ и канонической симплектической
формой $\varpi$ задано преобразование координат
$\lbrace x^\al\rbrace\mapsto\lbrace y^\al\rbrace=\lbrace Q^i(x),P_i(x)\rbrace$.
Назовем {\em скобкой Лагранжа} двух новых координат $y^\al$ и $y^\bt$ для
заданного набора функций $q^i(y^\al,y^\bt)$, $p_i(y^\al,y^\bt)$, определяющих
преобразование координат, следующее
выражение
\begin{equation}                                                  \label{elegbr}
  [y^\al,y^\bt]_\Sl:=
  \varpi_{\dl\g}\frac{\pl x^\g}{\pl y^\al}\frac{\pl x^\dl}{\pl y^\bt}
  =\frac{\pl q^i}{\pl y^\al}\frac{\pl p_i}{\pl y^\bt}
  -\frac{\pl q^i}{\pl y^\bt}\frac{\pl p_i}{\pl y^\al}.\qquad \qed
\end{equation}
\end{defn}
\index{Скобка Лагранжа (Lagrange bracket)}%
\index{Лагранжа скобка (Lagrange bracket)}%
Скобка Лагранжа антисимметрична и определена для любой пары новых координат
$y^\al$ и $y^\bt$. При этом мы не требуем, чтобы преобразование координат
$x\mapsto y(x)$ было каноническим.
\begin{com}
Если на фазовом пространстве просто заданы две дифференцируемые функции
$F,G\in\CC^1(\MR^{2\Sn})$, то скобка Лагранжа для них неопределена, т.к.\
не определены частные производные $\pl x^\al/\pl F$ и $\pl x^\al/\pl G$.
\qed\end{com}

\begin{theorem}
Для того, чтобы преобразование координат фазового пространства
$q,p\mapsto Q,P$ было каноническим, необходимо и достаточно выполнения
следующих условий для скобок Лагранжа новых координат:
\begin{equation}                                                  \label{elacri}
  [Q^i,Q^j]_\Sl=0,\qquad [P_i,P_j]_\Sl=0,\qquad [Q^i,P_j]_\Sl=\dl^i_j.
\end{equation}
\end{theorem}
\begin{proof}
Подставляя в определение канонических преобразований (\ref{ecanco}) функции
$q^i(Q,P)$, $p_i(Q,P)$, определяющие канонические преобразования, получим
равенство
\begin{equation}                                                  \label{eqcafu}
\begin{split}
  dP_i\wedge dQ^i&=\frac12dQ^l\wedge dQ^k
  \left(\frac{\pl q^i}{\pl Q^k}\frac{\pl p_i}{\pl Q^l}
  -\frac{\pl q^i}{\pl Q^l}\frac{\pl p_i}{\pl Q^k}\right)+
\\
  &+dP_l\wedge dQ^k\left(\frac{\pl q^i}{\pl Q^k}\frac{\pl p_i}{\pl P_l}
  -\frac{\pl q^i}{\pl P_l}\frac{\pl p_i}{\pl Q^k}\right)+
\\
  &+\frac12dP_l\wedge dP_k
  \left(\frac{\pl q^i}{\pl P_k}\frac{\pl p_i}{\pl P_l}
  -\frac{\pl q^i}{\pl P_l}\frac{\pl p_i}{\pl P_k}\right).
\end{split}
\end{equation}
Отсюда вытекает первый критерий каноничности преобразований координат.
\end{proof}
Таким образом, при каноническом преобразовании скобки Лагранжа для новых
координат обязаны совпадать со скобкой Пуассона.

Матрицы Якоби для канонических преобразований обладают замечательным свойством,
которое можно сформулировать в виде второго критерия канонических
преобразований.
\begin{theorem}
Преобразование координат фазового пространства является каноническим тогда и
только тогда, когда матрица Якоби этого преобразования
\begin{equation}                                                  \label{ecasja}
  J_\al{}^\bt:=\frac{\pl y^\bt}{\pl x^\al}
  =\begin{pmatrix}
  \dfrac{\raise-.7ex\hbox{$\pl Q$}}{\pl q} &
  \dfrac{\raise-.3ex\hbox{$\pl P$}}{\pl q} \\[4mm]
  \dfrac{\raise-.7ex\hbox{$\pl Q$}}{\pl p} &
  \dfrac{\raise-.3ex\hbox{$\pl P$}}{\pl p}
  \end{pmatrix}
\end{equation}
является симплектической, $J_\al{}^\bt\in\MS\MP(n,\MR)$.
\end{theorem}
\begin{proof}
Доказательство проводится прямой проверкой равенства
\begin{equation}                                                  \label{edespf}
  \varpi_{\al\bt}=J_\al{}^\g J_\bt{}^\dl\varpi_{\g\dl}
\end{equation}
с учетом свойств (\ref{elacri}). Вычисления проще проводить в блочно
диагональном виде (\ref{ecasja}).
\end{proof}
Поскольку определитель симплектической матрицы равен единице (\ref{edetsy}), то
форма объема фазового пространства сохраняется при каноническом преобразовании.
Отсюда также следует теорема Лиувилля \ref{tliovf}.

Третий критерий канонических преобразований формулируется в терминах скобок
Пуассона.
\begin{theorem}
Преобразование координат фазового пространства является каноническим тогда и
только тогда, когда оно сохраняет скобки Пуассона
\begin{equation}                                                  \label{ethcrp}
  [F,G]_{q,p}=[F,G]_{Q,P}.
\end{equation}
где $F$ и $G$ -- две произвольные дифференцируемые функции на фазовом
пространстве, и скобки Пуассона вычислены, соответственно, в координатах $q,p$ и
$Q,P$ c одинаковыми каноническими структурными функциями $\varpi^{-1\al\bt}$.
\end{theorem}
\begin{proof}
Канонические преобразования и только они сохраняют каноническую симплектическую
форму. Сохранение этой формы эквивалентно сохранению матрицы
$\varpi^{-1\al\bt}$, т.е.\ канонической пуассоновой структуры.
\end{proof}
\begin{exa}                                                       \label{ecatra}
Преобразование
\begin{equation}                                                  \label{ecatrt}
  Q^i=q^i,\qquad P_i=p_i+\frac{\pl F(q)}{\pl q^i},
\end{equation}
где $F(q)\in\CC^2(\MR^\Sn)$ -- произвольная функция от обобщенных координат $q$,
является каноническим. Действительно, скобки Пуассона
\begin{equation*}
  [Q^i,Q^j]=0,\qquad [Q^i,P_j]=\dl^i_j
\end{equation*}
остаются прежними. Нетрудно также проверить, что
\begin{equation*}
  [P_i,P_j]=\left[P_i,\frac{\pl F}{\pl q^j}\right]
  +\left[\frac{\pl F}{\pl q^i},P_j\right]=0.
\end{equation*}
Аналогично, преобразование
\begin{equation*}
  Q^i=q^i+\frac{\pl G(p)}{\pl p_i},\qquad P_i=p_i,
\end{equation*}
где $G(p)\in\CC^2(\MR^\Sn)$ -- произвольная функция импульсов, также является
каноническим.
\qed\end{exa}
Важность канонических преобразований сводится к следующему. Если удается найти
такое каноническое преобразование, при котором гамильтониан упрощается
настолько, что уравнения Гамильтона явно интегрируются, то тогда можно построить
и решение исходной задачи. Соответствующие канонические преобразования зависят
от конкретной задачи, и общего метода их нахождения не существует.
\subsection{Производящие функции канонических преобразований     \label{sgefca}}
Широкий класс канонических преобразований (но не все) можно описать на языке
производящих функций, которые строятся следующим образом. Рассмотрим фазовое
пространство $\MT^*(\MR^\Sn)\approx\MR^{2\Sn}$ с координатами $q,p$ и
канонической пуассоновой структурой. Пусть $2\Sn$ функций $Q^i(q,p)$ и
$P_i(q,p)$ от $2\Sn$ переменных задают каноническое преобразование.
Тогда $1$-форма
\begin{equation}                                                  \label{etrusa}
  dq\,p-dQP=dS(q,p)
\end{equation}
есть полный дифференциал.

Предположим, что в окрестности некоторой точки старые и новые координаты $(q,Q)$
можно выбрать в качестве координат фазового пространства. Это значит, что
\begin{equation}                                                  \label{edeqqu}
  \det\frac{\pl(q,Q)}{\pl(q,p)}=\det\begin{pmatrix}
  \dfrac{\raise-.7ex\hbox{$\pl q$}}{\raise .5ex\hbox{$\pl q$}} &
  \dfrac{\raise-.7ex\hbox{$\pl Q$}}{\raise .5ex\hbox{$\pl q$}} \\[3mm] 0 &
  \dfrac{\raise-.3ex\hbox{$\pl Q$}}{\raise .5ex\hbox{$\pl p$}}  \end{pmatrix}
  =\det\frac{\pl Q}{\pl p}\ne0.
\end{equation}
Тогда функцию $S$ можно локально выразить через эти координаты:
$$
  S(q,p)=S_1(q,Q).
$$
Функция $S_1(q,Q)$ называется {\em производящей функцией} канонического
преобразования.
\index{Производящая функция канонического преобразования%
(generating function of canonical transformation)}%

Из определения производящей функции и уравнения (\ref{etrusa}) следует, что
\begin{equation}                                                  \label{eficat}
  p=\frac{\pl S_1}{\pl q},\qquad P=-\frac{\pl S_1}{\pl Q}.
\end{equation}
Конечно, эти формулы являются сокращенной записью равенств
\begin{equation*}
  p_i=\frac{\pl S_1}{\pl q^i},\qquad P_i=-\frac{\pl S_1}{\pl Q^i}.
\end{equation*}
Полученные уравнения служат для нахождения канонического преобразования
$q,p\mapsto$\linebreak[3]$Q(q,p)$, $P(q,p)$. При этом выполнено равенство
\begin{equation*}
  \frac{\pl p_i}{\pl Q^j}=\frac{\pl^2 S_1}{\pl q^i\pl Q^j}.
\end{equation*}
Таким образом, каждая производящая функция $S_1(q,Q)$ такая, что
$$
  \det\frac{\pl^2 S_1}{\pl q^i\pl Q^j}\ne0,
$$
определяет некоторое каноническое преобразование по формулам (\ref{eficat}).
\ref{edeqqu}).

Для того, чтобы получить гамильтониан $K(Q,P)$ для новых канонических
переменных, необходимо просто подставить в старое выражение $H(q,p)$ функции
$q=q(Q,P)$ и $p=p(Q,P)$:
\begin{equation}                                                  \label{enewha}
  K(Q,P):=H\big(q(Q,P),p(Q,P)\big).
\end{equation}
\begin{exa}                                                       \label{eidcat}
Пусть $S_1=q^iQ_i$. Тогда $Q_i=p_i$ и $P^i=-q^i$. Тем самым координаты и
импульсы меняются местами. Этот пример показывает, что в каноническом формализме
координаты и импульсы играют совершенно равноправную роль.
\qed\end{exa}

Пусть гамильтониан не зависит от времени явно. Покажем, что если производящая
функция $S_1(q,Q)$ сама удовлетворяет уравнению Гамильтона--Якоби по переменным
$q$, то уравнения движения интегрируются в квадратурах. Для этого заметим, что
если гамильтониан системы зависит только от новых координат $H=K(Q)$, то
канонические уравнения
$$
  \dot Q=0,\qquad \dot P=-\frac{\pl K}{\pl Q}
$$
просто интегрируются
$$
  Q=Q_0,\qquad P=P_0-t\left.\frac{\pl K}{\pl Q}\right|_{Q_0}.
$$
Это, конечно, соответствует переходу к переменным действие-угол.
Теперь будем искать производящую функцию $S_1(q,Q)$ этого канонического
преобразования. Из первого условия (\ref{eficat}) следует, что такая
производящая функция должна удовлетворять уравнению
\begin{equation}                                                  \label{ehajag}
  H\left(q,\frac{\pl S_1}{\pl q}\right)=K(Q).
\end{equation}
То есть при каждом фиксированном значении $Q$ производящая функция $S_1(q,Q)$
должна удовлетворять укороченному уравнению Гамильтона--Якоби (\ref{eshaja}).
Верно также и обратное утверждение.
\begin{theorem}[\bf Якоби]
Если найдено решение $S_1(q,Q)$ укороченного уравнения Гамильтона--Якоби
(\ref{ehajag}), зависящее от $\Sn$ параметров $Q^i$ и такое, что
$$
  \det\frac{\pl^2 S_1}{\pl q^i\pl Q^j}\ne0,
$$
то канонические уравнения (\ref{ehameq}) решаются в квадратурах, причем функции
$Q(q,p)$, определяемые уравнениями $\pl S_1/\pl q^i=p_i$, являются $\Sn$ первыми
интегралами уравнений Гамильтона.
\end{theorem}
\begin{proof}
См., например, \cite{Arnold89R}.
\end{proof}
Теорема Якоби сводит решение канонических уравнений к нахождению полного
интеграла укороченного уравнения Гамильтона--Якоби. Может показаться
удивительным, что сведение более простого к более сложному доставляет
эффективный метод решения конкретных задач. Между тем оказывается, что это --
наиболее мощный из существующих в настоящее время методов интегрирования
уравнений движения, и многие задачи, решенные методом Якоби, вообще не поддаются
решению другими способами.

В качестве дополнительного аргумента производящей функции можно рассматривать
время, $S_1=S_1(q,Q,t)$. При этом все предыдущие формулы остаются в силе за
исключением выражения (\ref{enewha}) для нового гамильтониана. Оно получает
дополнительное слагаемое
\begin{equation}                                                  \label{efiham}
  K=H+\frac{\pl S_1}{\pl t},
\end{equation}
что следует из выражения для действия (\ref{qffrsg}) в расширенном
конфигурационном пространстве.

Если после канонического преобразования новый гамильтониан тождественно равен
нулю, то уравнения движения, очевидно, интегрируются. Для этого необходимо
выполнение уравнения
$$
  \frac{\pl S_1}{\pl t}+H(q,p,t)=0.
$$
Учитывая выражение для импульсов через производящую функцию (\ref{eficat})
получаем, что производящая функция в этом случае должна удовлетворять уравнению
Гамильтона--Якоби для всех значений $Q$. Тем самым доказана
\begin{theorem}[\bf Якоби]                                        \label{tjacth}
Если найдено решение $S_1(q,Q,t)$ уравнения Гамильтона--Якоби (\ref{eyfjae}),
зависящее от $\Sn$ параметров $Q^i$ и такое, что
$$
  \det\frac{\pl^2 S_1}{\pl q^i\pl Q^j}\ne0,
$$
то канонические уравнения (\ref{ehameq}) решаются в квадратурах, причем
координаты $Q^i=\const$ и импульсы $P_i:=-\pl S_1/\pl Q^i=\const$ дают $2\Sn$
первых интегралов уравнений Гамильтона.
\end{theorem}

Частное решение уравнения Гамильтона--Якоби, зависящее от $\Sn+1$ параметра
(по числу независимых переменных), называется {\em полным интегралом}. Функция
от $\Sn+1$ параметра $S_1(q,Q,t)+\const$, где $S_1$ -- решение уравнения
Гамильтона--Якоби из предыдущей теоремы является полным интегралом уравнения
Гамильтона--Якоби. В том случае, когда для уравнения Гамильтона--Якоби можно
определить общее решение, оно будет зависеть от произвольных функций. Это
значит, что полный интеграл уравнения Гамильтона--Якоби дает лишь незначительную
часть всех решений.
\index{Полный интеграл уравнения Гамильтона--Якоби%
(complete integral of Hamilton--Jacobi equation)}%

Если известен полный интеграл уравнения Гамильтона--Якоби $S_1(q,Q,t)+\const$,
то для нахождения траектории частицы, проходящей через заданную точку фазового
пространства в начальный момент времени, необходимо решить дополнительные
уравнения:
\begin{equation}                                                  \label{eaddhj}
  \frac{\pl S}{\pl Q^i}=c_i,
\end{equation}
где $c_i$ -- некоторые постоянные, определяемые начальным положением частицы.

Продолжим изучение производящих функций. Выше был рассмотрен случай, когда
производящая функция зависит от старых и новых координат. Аналогичным образом
можно показать, что канонические преобразования генерируются производящими
функциями $S_2(q,P,t)$, зависящими от старых координат и новых импульсов. Если
$$
  \det\frac{\pl^2 S_2}{\pl q^i\pl P_j}\ne0,
$$
то каноническое преобразование задается формулами
\begin{equation}                                                  \label{esecat}
  p:=\frac{\pl S_2}{\pl q},\qquad Q:=\frac{\pl S_2}{\pl P},\qquad
  K:=H+\frac{\pl S_2}{\pl t}.
\end{equation}
Переход от производящей функции $S_1(q,Q)$ к $S_2(q,P)$ является преобразованием
Лежандра по переменной $Q$ при этом
$$
  S_2(q,P,t)=S_1(q,Q,t)+PQ.
$$
\begin{exa}
Каноническое преобразование, генерируемое производящей функцией
$S_2=qP:=q^i P_i$ является тождественным.
\qed\end{exa}
\begin{exa}
Пусть $S_2=q^iS_i{}^jP_j$, где $S_i{}^j$ -- произвольная невырожденная матрица.
Тогда $P_i=S^{-1}_i{}^j p_j$ и $Q^i=q^jS_j{}^i$. То есть это преобразование
генерирует линейное однородное преобразование фазового пространства.
\qed\end{exa}
\begin{exa}                                                       \label{ehamge}
Выберем производящую функцию бесконечно малых канонических преобразований в виде
$$
  S_2=qP+\e S(q,P,\e),\qquad \e\ll1,
$$
где $S(q,P,\e)$ -- произвольная функция, зависящая от параметра $\e$. Тогда
$$
  Q=q+\e\frac{\pl S}{\pl P},\qquad p=P+\e\frac{\pl S}{\pl q}.
$$
Отсюда следует, что бесконечно малые канонические преобразования удовлетворяют
каноническим уравнениям
$$
  \left.\frac{dQ}{d\e}\right|_{\e=0}=\frac{\pl H}{\pl p},\qquad
  \left.\frac{dP}{d\e}\right|_{\e=0}=-\frac{\pl H}{\pl q}
$$
с функцией Гамильтона $H(q,p)=S(q,p,0)$, т.к.\ в главном порядке $p=P$. Тем
самым гамильтониан произвольной механической системы можно рассматривать, как
генератор бесконечно малых канонических преобразований по времени $t=\e$,
которые происходят по мере эволюции механической системы.
\qed\end{exa}
На самом деле движение частицы в фазовом пространстве $q(t)$, $p(t)$ можно
рассматривать как последовательность канонических преобразований,
параметризующихся одним параметром $t$. Допустим, что решение канонических
уравнений движения определено для всех $t\in\MR$. Тогда канонические уравнения
движения (\ref{eqgenq}) определяют однопараметрическую группу преобразований
фазового пространства. Для каждого момента времени, по-определению, выполнены
канонические коммутационные соотношения (\ref{eeqtoc}). Следовательно, для
каждого момента времени $t$ отображение фазового пространства
\begin{equation*}
  \MR^{2\Sn}\ni\quad\big(q(0),p(0)\big)\mapsto\big(q(t),p(t)\big)\quad
  \in\MR^{2\Sn}
\end{equation*}
представляет собой каноническое преобразование.

В заключение рассмотрим еще два вида производящих функций, часто встречающихся в
приложениях. Если производящая функция $S_3(p,Q,t)$ зависит от старых импульсов
и новых координат и при этом
\begin{equation*}
  \det\frac{\pl^2 S_3}{\pl p_i\pl Q^j}\ne0,
\end{equation*}
то каноническое преобразование задается формулами:
\begin{equation}                                                  \label{ethctr}
  q:=-\frac{\pl S_3}{\pl p},\qquad P:=-\frac{\pl S_3}{\pl Q},\qquad
  K:=H+\frac{\pl S_3}{\pl t}.
\end{equation}
При этом производящая функция $S_3$ является преобразованием Лежандра функции
$S_1$ по переменной $q$:
$$
  S_3(p,Q,t)=S_1(q,Q,t)-pq.
$$
\begin{exa}
Каноническое преобразование, генерируемое производящей функцией $S_3=-p_iQ^i$,
является тождественным.
\qed\end{exa}
Производящая функция $S_3(p,Q,t)$ полезна в следующем случае. Допустим, что нам
заранее известно, как удобно выбрать координаты в конфигурационном пространстве,
т.е.\ заданы функции $q=q(Q)$. Например, если конфигурационное пространство
инвариантно относительно действия некоторой группы преобразований, то новые
координаты $Q$ удобно выбрать в соответствии с симметрией задачи. Тогда
производящая функция $S_3=-p_iq^i(Q)$ дает выражения для новых обобщенных
импульсов $P_i$, сопряженных к координатам $Q^i$.

Точечные преобразования, рассмотренные в разделе \ref{spotmo}, относятся именно
к этому типу. Точечные преобразования -- это все преобразования координат,
которые допустимы в лагранжевой механике. Рассмотренный выше пример показывает,
что точечные преобразования составляют лишь небольшой класс канонических
преобразований. Все множество канонических преобразований намного шире, и это
является причиной большей гибкости канонического формализма по сравнению с
лагранжевым.

Продолжим изучение производящих функций. Если производящая функция $S_4(p,P,t)$
зависит от старых и новых импульсов и
\begin{equation*}
  \det\frac{\pl^2 S_4}{\pl p_i\pl P_j}\ne0,
\end{equation*}
то каноническое преобразование задается формулами:
\begin{equation}                                                  \label{ethctf}
  q:=-\frac{\pl S_4}{\pl p},\qquad Q:=\frac{\pl S_4}{\pl P},\qquad
  K:=H+\frac{\pl S_4}{\pl t}.
\end{equation}
При этом производящая функция $S_4$ является двойным преобразованием Лежандра
функции $S_1(q,Q)$ по переменным $q$ и $Q$:
$$
  S_4(p,P,t)=S_1(q,Q,t)-pq+PQ.
$$
\begin{exa}                                                       \label{exchqp}
Каноническое преобразование, генерируемое производящей функцией $S_4=p_iP^i$,
меняет местами координаты и импульсы: $Q_i=p_i$, $P^i=-q_i$. Это преобразование
совпадает с каноническим преобразованием из примера \ref{eidcat}.
\qed\end{exa}
\begin{exa}[\bf Гармонический осциллятор]
Продемонстрируем сложные построения последних разделов на примере гармонического
осциллятора. Пусть задана функция Лагранжа
\begin{equation*}
  L=\frac12\dot q^2-\frac12\om^2q^2,
\end{equation*}
где $\om=\const$ -- собственная частота гармонического осциллятора.
Конфигурационное пространство представляет собой прямую $q\in\MR$. Импульс,
сопряженный координате $q$, равен скорости
\begin{equation*}
  p=\frac{\pl L}{\pl\dot q}=\dot q.
\end{equation*}
Преобразование Лежандра приводит к положительно определенному гамильтониану:
\begin{equation*}
  H=\frac12p^2+\frac12\om^2q^2.
\end{equation*}
Фазовым пространством для гармонического осциллятора является евклидова
плоскость $(q,p)\in\MT^*(\MR)\approx\MR^2$ с канонической пуассоновой
структурой. Соответствующие уравнения движения имеют вид
\begin{equation}                                                  \label{eharmo}
\left.
\begin{aligned}
  \dot q&=\quad p,
\\
  \dot p&=-\om^2q,
\end{aligned}\right\rbrace\quad \Leftrightarrow\quad
  \ddot q+\om^2q=0.
\end{equation}
Общее решение этих уравнений параметризуется двумя произвольными постоянными:
амплитудой $A_0$ и фазой $\vf$,
\begin{equation}                                                  \label{ehages}
\begin{split}
  q&=\quad A_0\cos(\om t+\vf),
\\
  p&=-A_0\om\sin(\om t+\vf).
\end{split}
\end{equation}
Фаза $\vf$ соответствует выбору начала отсчета времени. Полученное решение
соответствует решению задачи Коши с начальными данными
\begin{equation*}
\begin{split}
  q_0&=\quad A_0\cos\vf,
\\
  p_0&=-A_0\om\sin\vf.
\end{split}
\end{equation*}
В терминах начальных условий общее решение (\ref{ehages}) перепишется в виде
\begin{equation*}
\begin{split}
  q&=\quad q_0\cos\om t+\frac{p_0}\om\sin\om t,
\\
  p&=-\om q_0\sin\om t+p_0\cos\om t.
\end{split}
\end{equation*}
Траекториями гармонического осциллятора в фазовом пространстве являются эллипсы.
Как видим, эволюция во времени представляет собой эллиптическое вращение
фазового пространства $(q_0,p_0)\mapsto\big(q(t),p(t)\big)$. Если
$\MV_0\subset\MR^2$ -- произвольная ограниченная область фазового пространства,
то в соответствии с теоремой Лиувилля (\ref{tliovf}) объем этой области
сохраняется во времени:
\begin{equation*}
  \int_{\MV_0}\!\!\!dq_0dp_0=\int_{\MV(t)}\!\!\!\!\!dqdp
  =\int_{\MV(t)}\!\!\!\!\!dq_0dp_0,
\end{equation*}
т.к.\ якобиан преобразования координат равен единице,
\begin{equation*}
  \det\frac{\pl(q,p)}{\pl(q_0,p_0)}=1.
\end{equation*}
Поскольку гамильтониан не зависит от времени явно, то энергия на каждой
траектории постоянна $E=H=\const$.

Найдем функцию действия $S(q,t)$. Чтобы вычислить интеграл (\ref{efunac})
необходимо исключить скорости $\dot q$ из лагранжиана. Поскольку
интегрирование ведется вдоль траектории частицы, то в лагранжиан необходимо
подставить
\begin{equation*}
  \dot q=p=\sqrt{2E-\om^2q^2}.
\end{equation*}
В этом выражении $E$ рассматривается, как некоторая постоянная.
После несложных вычислений получаем функцию действия
\begin{equation*}
\begin{split}
  S(q,t)&=-\int dtE+\int dq\sqrt{2E-\om^2q^2}=
\\
  &=-Et+\frac E\om\arcsin\frac{\om q}{\sqrt{2E}}+\frac q2\sqrt{2E-\om^2q^2}
\end{split}
\end{equation*}
с точностью до несущественной постоянной, определяемой начальными данными.
Нетрудно проверить, что
\begin{equation*}
  \frac{\pl S}{\pl q}=\sqrt{2E-\om^2q^2}=p.
\end{equation*}
Поэтому уравнение Гамильтона--Якоби (\ref{eyfjae}) выполняется.
Укороченное действие имеет вид
\begin{equation*}
  W(q,E)=\frac E\om\arcsin\frac{\om q}{\sqrt{2E}}+\frac q2\sqrt{2E-\om^2q^2}.
\end{equation*}
Таким образом, нам известно укороченное действие, зависящее от параметра
$E$, причем
\begin{equation*}
  \frac{\pl^2 W}{\pl q\pl E}\ne0.
\end{equation*}
Поэтому его можно выбрать в качестве производящей функции канонического
преобразования $q,p\rightarrow Q,P$
\begin{equation*}
  S_2(q,P)=\frac P\om\arcsin\frac{\om q}{\sqrt{2P}}
  +\frac q2\sqrt{2P-\om^2q^2}.
\end{equation*}
Воспользовавшись формулами (\ref{esecat}), получаем явный вид канонических
преобразований
\begin{equation*}
\begin{split}
  q&=\frac{\sqrt{2P}}\om\sin(\om Q),
\\
  p&=\sqrt{2P}\cos(\om Q).
\end{split}
\end{equation*}
Отсюда вытекает связь дифференциалов:
\begin{equation*}
\begin{split}
  dq&=\quad \sqrt{2P}\cos(\om Q)dQ+\frac1{\om\sqrt{2P}}\sin(\om Q)dP,
\\
  dp&=-\om\sqrt{2P}\sin(\om Q)dQ+\frac1{\sqrt{2P}}\cos(\om Q)dP.
\end{split}
\end{equation*}
Теперь нетрудно убедиться в том, что построенное преобразование
действительно является каноническим
\begin{equation*}
  dp\wedge dq=dP\wedge dQ.
\end{equation*}
В новых переменных гамильтониан имеет вид $H=P$ и уравнения движения
\begin{equation*}
  \dot Q=P,\qquad \dot P=0
\end{equation*}
тривиально интегрируются. Таким образом укороченная функция действия
для гармонического осциллятора является производящей функцией к
переменным ``действие-угол''. Фазовым пространством в переменных
``действие-угол'' является полуплоскость $Q\in\MR$, $P\ge0$, а траекториями
-- прямые линии $P=\const$.
\qed\end{exa}

В настоящем разделе мы рассмотрели широкий класс канонических преобразований,
которые генерируются четырьмя видами производящих функций: $S_1(q,Q,t)$,
$S_2(q,P,t)$, $S_3(p,Q,t)$ и $S_4(p,P,t)$. Существуют также производящие
функции более общего вида, которые мы сейчас опишем.
\begin{prop}
Пусть заданы две системы координат $q,p$ и $Q,P$ в фазовом пространстве
пространстве $\MR^{2\Sn}$ с канонической пуассоновой структурой. Тогда среди
$4\Sn$ координатных функций $q,p,Q,P$ всегда можно выбрать $2\Sn$ независимых
функций так, чтобы среди них не было ни одной пары канонически сопряженных
функций $q^i,p_i$ или $Q^i,P_i$.
\end{prop}
\begin{proof}
См., например, \cite{Gantma01R}.
\end{proof}
Из данного утверждения следует, что в качестве координат на фазовом пространстве
$\MR^{2\Sn}$ всегда можно выбрать набор функций
\begin{equation*}
  q^1,\dotsc,q^L,~p_{L+1},\dotsc,p_\Sn,~Q^1,\dotsc,Q^M,~P_{M+1},\dotsc,P^\Sn,
  \qquad 0\le L,M\le\Sn,
\end{equation*}
состоящий из $\Sn$ старых и $\Sn$ новых координат. Введем обозначения:
\begin{align*}
  \lbrace z^i\rbrace&:=\lbrace q^1,\dotsc,q^L,p_{L+1},\dotsc,p_\Sn\rbrace,\qquad
\\
  \lbrace Z^i\rbrace&:=\lbrace Q^1,\dotsc,Q^M,P_{M+1},\dotsc,P^\Sn\rbrace.
\end{align*}
Допустим, что на фазовом пространстве задана функция $S(z,Z)$ такая, что
\begin{equation}                                                  \label{edegef}
  \det\frac{\pl^2 S}{\pl z^i\pl Z^j}\ne0.
\end{equation}
Тогда нетрудно проверить, что функция $S(z,Z)$ определяет каноническое
преобразование:
\begin{align*}
  p_i&:=~~\frac{\pl S}{\pl q^i}, && i=1,\dotsc,L,
\\
  q^i&:=-\frac{\pl S}{\pl p_i}, && i=L+1,\dotsc,\Sn,
\\
  P_i&:=-\frac{\pl S}{\pl Q^i}, && i=1,\dotsc,M,
\\
  Q^i&:=~~\frac{\pl S}{\pl P_i}, && i=M+1,\dotsc,\Sn.
\end{align*}

Рассмотренное каноническое преобразование включает в себя четыре предыдущих при
$L=0,\Sn$ и $M=0,\Sn$. В общем случае производящая функция $S(z,Z)$ зависит от
старых координат и импульсов и новых координат и импульсов. При этом среди них
нет ни одной пары канонически сопряженных переменных. Примеры \ref{eidcat} и
\ref{exchqp} показывают, что различные производящие функции могут приводить к
одинаковым каноническим преобразованиям. Во всех случаях от производящей функции
требуется отличие от нуля определителя (\ref{edegef}). Поэтому производящие
функции описывают широкий класс канонических преобразований, но не все.
\begin{exa}
В примере \ref{ecatra} из предыдущего раздела было рассмотрено каноническое
преобразование, которое не охватывается производящими функциями, рассмотренными
в настоящем разделе. Действительно, для преобразования (\ref{ecatrt}) формула
(\ref{etrusa}) принимает вид
\begin{equation*}
  -dq\frac{\pl F(q)}{\pl q}=dS(q,p)=dq\frac{\pl S}{\pl q}+dp\frac{\pl S}{\pl p}.
\end{equation*}
Откуда следуют равенства:
\begin{equation*}
  \frac{\pl S}{\pl q}=-\frac{\pl F}{\pl q},\qquad \frac{\pl S}{\pl p}=0.
\end{equation*}
Таким образом, для канонических преобразований вида (\ref{ecatrt}) функция
$S(q,p)$ имеет вид
\begin{equation*}
  S(q,p)=-F(q)+\const
\end{equation*}
и не относится ни к одному типу производящих функций настоящего раздела.
\qed\end{exa}
\subsection{Разделение переменных в уравнении Гамильтона--Якоби}
Теорема Якоби \ref{tjacth} утверждает, что если найден полный интеграл уравнения
Гамильтона--Якоби, то канонические уравнения решаются в квадратурах.
По-видимому, все известные в настоящее время полные интегралы могут быть найдены
путем {\em разделения переменных} в уравнении Гамильтона--Якоби. Суть метода
состоит в следующем.
\index{Разделение переменных (separation of variables)}%

Допустим, что одна из координат -- обозначим ее $q^\Sn$ -- и соответствующая ей
производная $\pl S/\pl q^\Sn$ входят в уравнение Гамильтона--Якоби только в виде
комбинации $f(q^\Sn,\pl S/\pl q^\Sn)$, где $f$ -- некоторая функция двух
переменных, не зависящая от других координат, соответствующих производных
действия и времени. Тогда будем искать решение уравнения Гамильтона--Якоби
(\ref{eyfjae}) в виде
\begin{equation*}
  S(q,t)=\check S(\check q,t)+S_\Sn(q^\Sn),
\end{equation*}
где $\check S(\check q,t)$ -- функция оставшихся переменных
$\check q=(q^1,\dotsc,q^{\Sn-1})$ и времени, а $S_\Sn(q^\Sn)$ зависит только от
отделяемой координаты $q^\Sn$. Тогда уравнение Гамильтона--Якоби принимает вид
\begin{equation}                                                  \label{ehajad}
  \frac{\pl\check S}{\pl t}+H\left(\check q,\frac{\pl\check S}{\pl\check q},
  f\left(q^\Sn,\frac{dS_\Sn}{dq^\Sn}\right),t\right)=0.
\end{equation}
Допустим, что решение этого уравнения найдено. Тогда, после подстановки
найденного решения в уравнение (\ref{ehajad}) мы получаем тождество,
справедливое для всех значений координат. При изменении координаты $q^\Sn$
меняется только функция $f$. Поэтому для выполнения уравнения (\ref{ehajad})
необходимо, чтобы функция $f$ сама была равна постоянной, которую мы обозначим
$Q^\Sn$, для произвольного решения. Таким образом, уравнение Гамильтона--Якоби
(\ref{ehajad}) распадается на два уравнения:
\begin{align}                                                     \label{esepeq}
  f\left(q^\Sn,\frac{dS_\Sn}{dq^\Sn}\right)&=Q^\Sn,
\\                                                                \label{esepla}
  \frac{\pl\check S}{\pl t}+H\left(\check q,\frac{\pl\check S}{\pl\check q},
  Q^\Sn,t\right)&=0.
\end{align}
Первое уравнение представляет собой обыкновенное дифференциальное уравнение
первого порядка, из которого можно определить функцию
$S_\Sn=S_\Sn(q^\Sn;Q^\Sn)$. Если функция $f$ достаточно проста, то это уравнение
иногда удается проинтегрировать явно. Постоянная $Q^\Sn$ является первым
интегралом исходного уравнения Гамильтона--Якоби. После этого остается решить
оставшееся уравнение (\ref{esepla}), которое также имеет вид уравнения
Гамильтона--Якоби, но для системы с меньшим числом степеней свободы. Эта
система описывается гамильтонианом, который получается из исходного путем замены
функции $f$ для отделенной степени свободы на некоторую константу. Единственное
ограничение на постоянную $Q^\Sn$ состоит в том, чтобы уравнение (\ref{esepeq})
имело решение.

Иногда этот процесс отделения переменных можно продолжить. Если таким образом
можно разделить все $\Sn$ координат, то тогда полный интеграл уравнения
Гамильтона--Якоби находится в квадратурах. В этом случае мы говорим, что
уравнение Гамильтона--Якоби допускает {\em разделение переменных}.
\begin{com}
Разделение переменных в уравнении Гамильтона--Якоби для одной и той же
механической системы возможно не во всякой системе координат конфигурационного
пространства. Как правило системы координат, в которых переменные делятся
полностью или частично связаны с симметрией системы. Процедура нахождения такой
системы координат, в которой переменные для интегрируемой системы разделяются,
представляет собой отдельную задачу.
\qed\end{com}
Допустим, что механическая система консервативна и переменные делятся. Тогда
полный интеграл уравнения Гамильтона--Якоби имеет вид
\begin{equation}                                                  \label{ecoija}
  S(q,t)=-E(Q^1,\dotsc,Q^\Sn)t+\sum_{k=1}^\Sn S_k(q^k;Q^k,\dotsc,Q^\Sn),
\end{equation}
где каждая функция $S_k$ зависит от одной координаты $q^k$, а энергия $E$ как
функция первых интегралов $Q^1,\dotsc,Q^\Sn$ получается подстановкой
укороченного действия
\begin{equation*}
  W=\sum_{k=1}^\Sn S_k(q^k,Q^k,\dotsc,Q^\Sn)
\end{equation*}
в укороченное уравнение Гамильтона--Якоби (\ref{eshaja}).
\begin{exa}[\bf Циклические координаты]
Пусть $q^\Sn$ -- циклическая координата, т.е.\ гамильтониан от нее вообще не
зависит. В этом случае функция $f(q^\Sn,\pl S/\pl q^\Sn)$ сводится просто к
$\pl S/\pl q^\Sn$. Тогда из уравнения (\ref{esepeq}) следует, что
$S_\Sn=q^\Sn Q_\Sn$ (никакого суммирования). Тогда функция действия принимает
вид
\begin{equation*}
  S(q,t)=\check S(\check q,t)+q^\Sn Q_\Sn.
\end{equation*}
В этом случае первый интеграл $Q_\Sn=\pl S/\pl q^\Sn=p_\Sn$ имеет смысл
обобщенного импульса, сопряженного циклической координате $q^\Sn$.

Отметим, что если время $t$ рассматривать, как обобщенную координату, то
отделение времени в виде слагаемого $-Et$ в функции действия (\ref{eacfce}) для
консервативной системы соответствует методу разделения переменных для
циклической координаты $t$. При этом энергия $E$ сохраняется и соответствует
импульсу для обобщенной координаты $t$.
\qed\end{exa}

Разделение переменных для консервативной системы возможно также в следующем
более общем случае. Пусть гамильтониан механической системы имеет вид
\begin{equation*}
  H(q,p)=\frac{f_1(q^1,p_1)+\dotsc+f_\Sn(q^\Sn,p_\Sn)}
  {g_1(q^1,p_1)+\dotsc+g_\Sn(q^\Sn,p_\Sn)},
\end{equation*}
где $f_i(q^i,p_i)$ и $g_i(q^i,p_i)$ -- некоторые функции только от указанных
аргументов. Предыдущий случай возникает, например, при $g_i=\const$,
$i=1,\dotsc,\Sn$, $\sum_ig_i\ne0$. Будем искать решение уравнения
Гамильтона--Якоби в виде
\begin{equation*}
  S(q,t)=-Et+\sum_{i=1}^\Sn W_i(q^i).
\end{equation*}
Тогда укороченное уравнение Гамильтона--Якоби (\ref{eshaja}) примет вид
\begin{equation*}
  \sum_{i=1}^\Sn\left[f_i\left(q^i,\frac{dW_i}{dq_i}\right)
  -Eg_i\left(q^i,\frac{dW_i}{dq_i}\right)\right]=0.
\end{equation*}
Поскольку координаты $q_i$ меняются независимо, то каждое слагаемое в данной
сумме должно быть равно некоторой постоянной,
\begin{equation}                                                  \label{ecohja}
  f_i\left(q^i,\frac{dW_i}{dq_i}\right)
  -Eg_i\left(q^i,\frac{dW_i}{dq_i}\right)=c_i.
\end{equation}
При этом не все постоянные $c_i$ являются независимыми, т.к. должно быть
выполнено равенство
\begin{equation*}
  c_1+\dotsc+c_\Sn=0.
\end{equation*}

Допустим, что все уравнения (\ref{ecohja}) можно разрешить относительно
производных:
\begin{equation*}
  \frac{dW_i}{dq^i}=F_i(q^i,c_i).
\end{equation*}
Тогда полный интеграл уравнения Гамильтона--Якоби равен сумме
\begin{equation*}
  S(q,t)=-Et+\sum_{i=1}^\Sn\int\! dq^iF_i(q^i,c_i,E).
\end{equation*}
\begin{exa}[\bf Задача Кеплера]
Рассмотрим частицу массы $m$, которая движется по плоскости с полярными
координатами $r,\vf$ под действием центральной силы с потенциалом $U(r)$.
Гамильтониан такой системы имеет вид
\begin{equation}                                                  \label{eflsef}
  H(r,\vf)=\frac1{2m}\left(p_r^2+\frac{p_\vf^2}{r^2}\right)+U(r).
\end{equation}
К этой задаче сводится задача движения двух тел, находящихся под действием
гравитационного взаимодействия c потенциалом $U(r)=-m_1m_2/r$ после отделения
движения центра масс. Поскольку гамильтониан не зависит от угла, то координата
$\vf$ является циклической, и переменные делятся. Поэтому функция действия
имеет вид
\begin{equation}                                                  \label{etwboi}
  S=-Et+W(r,E,M)+\vf M,
\end{equation}
где $p_\vf=M=\const$ -- сохраняющийся угловой импульс частицы (момент количества
движения). В этом случае уравнение Гамильтона--Якоби сводится к одному
обыкновенному дифференциальному уравнению первого порядка
\begin{equation}                                                  \label{ewtwob}
  \left(\frac{dW}{dr}\right)^2=2m(E-U)-\frac{M^2}{r^2}.
\end{equation}
Произвол в решении данного уравнения фиксируется положением частицы в начальный
момент времени. В соответствии с общими утверждениями, функция действия
(\ref{etwboi}), где $W$ -- решение уравнения (\ref{ewtwob}), зависит от двух
постоянных: энергии $E$ и момента количества движения $M$, и представляет собой
полный интеграл уравнения Гамильтона--Якоби. Таким образом, задача двух тел,
взаимодействующих посредством центральных сил с произвольным потенциалом $U(r)$,
решается в квадратурах. Чтобы найти траекторию частицы, проходящую через
заданную точку фазового пространства в начальный момент времени, в соответствии
с теоремой Якоби (\ref{tjacth}) необходимо решить дополнительные уравнения:
\begin{equation*}
  \frac{\pl S}{\pl M}=\frac{\pl W}{\pl M}+\vf=c_1,\qquad
  \frac{\pl S}{\pl E}=-t+\frac{\pl W}{\pl E}=c_2,
\end{equation*}
где $c_{1,2}$ -- некоторые постоянные (\ref{esecat}), определяемые начальным
положением частицы.

Заметим, что переменные в задаче Кеплера на плоскости делятся в полярных
координатах (в  трехмерном пространстве -- в сферических), которые связаны со
сферической симметрией потенциала взаимодействия. В декартовой системе координат
переменные в уравнении Гамильтона--Якоби не делятся.
\qed\end{exa}
\section{Гамильтонова динамика частиц со связями                 \label{scohal}}
В современной математической физике большую роль играют модели, инвариантные
относительно действия локальных групп преобразований, которые принято называть
калибровочными. С точки зрения гамильтонова формализма такие модели
соответствуют системам со связями. В калибровочных моделях связи возникают не
как дополнительные условия на канонические переменные, наложенные извне, а при
переходе от вырожденного инвариантного лагранжиана к гамильтониану. Основы
гамильтонова формализма для калибровочных моделей были заложены П.~Дираком на
примере точечных частиц \cite{Dirac58AR,Dirac58BR,Dirac64R}. Обобщение этого
формализма на теорию поля представляет значительные трудности. Несмотря на важность, эта задача до сих пор не
решена. В теории поля до настоящего времени, как правило, используют методы,
развитые для точечных частиц. В простейших случаях это приводит к разумным
результатам.

В последующих разделах мы обсудим вопрос о гамильтоновой динамике со связями для
систем точечных частиц, оставив в стороне важную задачу перехода от вырожденных
калибровочно инвариантных лагранжианов к гамильтонианам, которая подробно
рассматривается в монографиях \cite{GitTyu90,HenTei92}. Квантование
калибровочных моделей с помощью функционального интеграла в фазовом пространстве
было развито в \cite{Faddee69R,FraVil77}.
\subsection{Связи в гамильтоновом формализме}
Рассмотрим систему $\Sn$ точечных частиц с действием (\ref{ehaact}).
Предположим, что фазовое пространство представляет собой дифференцируемое
многообразие $\MN$ класса $\CC^k$, $\dim\MN=2\Sn$. Для простоты предположим, что
фазовое пространство топологически тривиально $\MN\approx\MR^{2\Sn}$ и
покрывается одной картой. Координаты на фазовом пространстве $\MN$, как и
раньше, обозначим $(q,p)=(q^1,\dotsc,q^\Sn,p_1,\dotsc,p_\Sn)$. Мы предполагаем,
что на фазовом пространстве задана каноническая пуассонова структура
$[q^i,p_j]=\dl^i_j$.

Пусть на канонические переменные наложены {\em связи} в виде $2\Sm$
алгебраических условий:
\index{Связь (constraint)}%
\begin{equation}                                                  \label{econtw}
  \Phi^\mu(q,p)=0,\qquad \Phi^\mu\in\CC^k(\MN),\quad \mu=1,\dotsc,2\Sm<2\Sn.
\end{equation}
Мы пишем индекс связи вверху, поскольку в дальнейшем связи будут частью новых
координатных функций. Условия (\ref{econtw}) представляют собой систему
уравнений на канонические переменные. Будем считать, что все $\Phi^\mu$
функционально независимы, т.е.\ ранг матрицы $\pl\Phi^\mu/\pl x^\al$, где
$x=(q,p)$, максимален и равен $2\Sm$ в окрестности всех точек, координаты
которых удовлетворяют системе уравнений (\ref{econtw}). Если все связи
принадлежат классу $\CC^k$, то их совокупность, согласно теореме \ref{tsuman},
определяет подмногообразие
\begin{equation*}
  \MM=\lbrace x\in\MN:\quad \Phi=0\rbrace
\end{equation*}
того же класса гладкости и размерности $\dim\MM=2(\Sn-\Sm)$. Это подмногообразие
называется {\em поверхностью связей}, которую будем обозначать равенством
$\Phi=0$ без индекса.
\index{Поверхность связей (constraint surface)}%
\index{Связей поверхность (constraint surface)}%
\begin{com}
Вообще говоря, $\Phi^\mu(q,p)$ не являются функциями на фазовом пространстве,
т.к.\ могут иметь тензорные индексы. Например, в электродинамике первичная связь
имеет вид $p_0=0$.
\qed\end{com}

Условие функциональной независимости связей накладывает жесткие ограничения на
вид функций $\Phi^\mu$.
\begin{exa}
Пусть на фазовом пространстве $\MN$ задана одна связь $\Phi(q,p)=0$. Условие
функциональной независимости означает, что ранг матрицы $\pl_\al\Phi$, которая в
рассматриваемом случае представляет собой строку, равен единице, т.е.\ по
крайней мере одна из частных производных в каждой точке поверхности связей $\MM$
отлична от нуля. Если возвести связь в некоторую степень, то уравнение
$\Phi^n=0$ при $0<n<1$ и $n>1$ определяет ту же поверхность связей
$\MM\subset\MN$ (как множество точек). Однако связь $\Phi^n=0$ не является
функционально независимой. Действительно,
\begin{equation*}
  \pl_\al\Phi^n|_{\Phi=0}=n\Phi^{n-1}\pl_\al\Phi|_{\Phi=0}=
  \begin{cases}
    0, & n>1, \\ \infty, & 0<n<1.
\end{cases}
\end{equation*}
Поэтому ранг матрицы $\pl_\al\Phi^n$ либо равен нулю, либо неопределен на
поверхности связей, и связь $\Phi^n=0$ не является функционально независимой на
$\MM\subset\MN$, хотя и определяет ту же поверхность связей.
\qed\end{exa}
Рассмотренный пример показывает, что условие функциональной независимости связей
означает не только определение поверхности связей $\MM\subset\MN$, но и
возможность выбора функций $\Phi^\mu$ в качестве трансверсальных координат к
$\MM$ в некоторой окрестности поверхности связей.

Уравнения, задающие поверхность связей, определены неоднозначно. Например,
невырожденные линейные комбинации исходных связей
\begin{equation}                                                  \label{elinco}
  \Psi^\mu(q,p):=\Phi^\nu(q,p)S_\nu{}^\mu(q,p),
\end{equation}
где матрица $S_\nu{}^\mu$ невырождена на всем фазовом пространстве $\MN$,
определяют ту же поверхность связей $\MM\subset\MN$ и функционально независимы.
\begin{defn}
Системы связей $\Phi=0$ и $\Psi=0$ называются {\em эквивалентными}, если они
связаны между собой невырожденным линейным преобразованием (\ref{elinco}) во
всем фазовом пространстве.
\qed\end{defn}
\index{Эквивалентные связи (equivalent constraints)}%
\index{Связи эквивалентные (equivalent constraints)}%
\begin{com}
Если потребовать выполнения равенства (\ref{elinco}) не во всем фазовом
пространстве, а только на поверхности связей, то этого недостаточно для
эквивалентности. Действительно, пусть поверхность связей $\MM\subset\MN$
определена уравнениями $\Phi=0$. Определим новые связи $\Psi=0$ на поверхности
связей с помощью некоторой невырожденной матрицы $S$, заданной на $\MM$. После
этого продолжим функции $\Psi$ на все фазовое пространство таким образом, чтобы
уравнения $\Psi=0$ определяли некоторое подмногообразие вида
$\MM\cup\MU\subset\MN$, где $\MU$ -- некоторое собственное подмногообразие
$\MN$ такое, что $\MU\cap\MM=\emptyset$. Никаких препятствий для этого нет.
Таким образом, для новых связей условие (\ref{elinco}) будет выполнено на $\MM$,
однако условия $\Psi=0$ выделяют в фазовом пространстве $\MN$ другое
подмногообразие. Это означает, что матрица $S$ будет вырождена на $\MU$.
\qed\end{com}

\begin{prop}                                                      \label{panyco}
Любая функция $f(q,p)\in\CC^k(\MN)$, обращающаяся в нуль на поверхности связей
$\MM\subset\MN$, в некоторой окрестности $\MU_0\subset\MN$ произвольной
точки $(q_0,p_0)\in\MM$ представима в виде
\begin{equation}                                                  \label{efucon}
  f(q,p)=\Phi^\mu(q,p)f_\mu(q,p),\qquad (q,p)\in\MU_0,
\end{equation}
с достаточно гладкими коэффициентами $f_\mu(q,p)\in\CC^k(\MU_0)$.
\end{prop}
\begin{proof}
Поскольку связи $\Phi^\mu$ функционально независимы, то в некоторой окрестности
произвольной точки на поверхности связей $(q_0,p_0)\in\MM$ их можно дополнить до
системы координат $\lbrace x^\al\rbrace\rightarrow\lbrace y^a,\Phi^\mu\rbrace$,
где $y^a$, $a=1,\dotsc,2(\Sn-\Sm)$, -- координаты на поверхности связей. Тогда
согласно предложению \ref{lrecfu} произвольная достаточно гладкая функция
$f\in\CC^k(\MU_0)$ представима в виде
\begin{equation*}
  f(q,p)=f(q_0,p_0)+y^a(q,p)f_a(q,p)+\Phi^\mu(q,p)f_\mu(q,p)
\end{equation*}
с достаточно гладкими коэффициентами $f_\mu$, $f_a$. Так как функция $f$ равна
нулю на поверхности связей $\MM$, то $f(q_0,p_0)=0$ и $f_a(q,p)=0$ для всех $a$
и $(q,p)\in\MU_0$. Поэтому справедливо представление (\ref{efucon}).
\end{proof}
\subsection{Гамильтонова динамика частиц со связями II рода      \label{seccog}}
Начнем рассмотрение гамильтоновой динамики частиц со связями с обсуждения связей
второго рода, т.к.\ она значительно проще. Связи первого рода будут рассмотрены
в следующем разделе. Там будет показано, что динамика частиц со связями I рода
после наложения канонической калибровки сводится к рассмотрению частиц со
связями II рода.

\begin{defn}
Связи $\Phi^\mu$, $\mu=1,\dotsc,2\Sm$, называются {\em связями II рода}, если
определитель матрицы, составленной из скобок Пуассона связей между собой,
отличен от нуля на поверхности связей:
\begin{equation}                                                  \label{eseclc}
  \det[\Phi^\mu,\Phi^\nu]_{\Phi=0}\ne0. \qed
\end{equation}
\end{defn}
\index{Связи II рода (second class constraints)}%
Отметим, что элементы матрицы, составленной из скобок Пуассона
$[\Phi^\mu,\Phi^\nu]$, в общем случае зависят от точки фазового пространства.
Условие (\ref{eseclc}) означает отличие определителя от нуля в каждой точке
подмногообразия $\MM\subset\MN$, определяемого связями (\ref{econtw}). Из
непрерывности функций $\Phi^\mu$ следует, что определитель (\ref{eseclc})
отличен от нуля не только на $\MM$, но и в некоторой окрестности подмногообразия
$\MM$.

Поскольку матрица, составленная из скобок Пуассона,
\begin{equation}                                                  \label{epoima}
  J^{\mu\nu}=-J^{\nu\mu}:=[\Phi^\mu,\Phi^\nu],
\end{equation}
антисимметрична, то ее определитель может быть отличен от нуля только при четном
числе связей. Поэтому мы с самого начала предположили наличие четного числа
связей $2\Sm$. Это значит, что поверхность связей также имеет четную размерность
$2(\Sn-\Sm)$.

С точки зрения вариационного принципа при наличии связей на фазовом пространстве
мы имеем задачу на условный экстремум, которую можно решать методом
неопределенных множителей Лагранжа. С этой целью рассмотрим {\em обобщенный
(extended) гамильтониан}, который получается после добавления к исходному
гамильтониану всех связей второго рода
\index{Обобщенный гамильтониан (extended Hamiltonian)}%
\index{Гамильтониан обобщенный (extended Hamiltonian)}%
\begin{equation}                                                  \label{ettham}
  H_\Se=H+\lm_\mu\Phi^\mu,
\end{equation}
где $\lm_\mu=\lm_\mu(q,p,t)$ -- неопределенные множители Лагранжа. Этому
гамильтониану соответствует {\em обобщенное действие}
\index{Обобщенное действие (extended action)}%
\index{Действие обобщенное (extended action)}%
\begin{equation}                                                  \label{etacpr}
  S_\Se=\int_{t_1}^{t_2}\!\!\!dt(p\dot q-H-\lm_\mu \Phi^\mu).
\end{equation}
При варьировании этого действия мы считаем вариации координат на границе
нулевыми, $\dl q(t_{1,2})=0$, а вариации импульсов и множителей Лагранжа могут
быть произвольны, т.к.\ они входят в действие без производных. Канонические
уравнения движения для обобщенного гамильтониана имеют вид
\begin{equation}                                                  \label{ecoint}
\begin{split}
  \dot q^i&=[q^i,H_\Se]=[q^i,H]+\lm_\mu[q^i,\Phi^\mu]+[q^i,\lm_\mu]\Phi^\mu,
\\
  \dot p_i&=[p_i,H_\Se]=[p_i,H]+\lm_\mu[p_i,\Phi^\mu]+[p_i,\lm_\mu]\Phi^\mu.
\end{split}
\end{equation}
Эти уравнения необходимо дополнить уравнениями связей (\ref{econtw}),
возникающими при варьировании обобщенного действия по множителям Лагранжа.
В уравнениях движения (\ref{ecoint}) последние слагаемые можно отбросить, т.к.\
они равны нулю на поверхности связей (\ref{econtw}). Таким образом, мы имеем
$2(\Sn+\Sm)$ уравнений (\ref{ecoint}), (\ref{econtw}) на $2(\Sn+\Sm)$ переменных
$q,p,\lm$, из которых $2\Sn$ уравнений являются дифференциальными. Решение этой
задачи можно провести следующим образом.
\begin{prop}                                                      \label{pconsu}
Для того, чтобы фазовая траектория гамильтоновой системы (\ref{etacpr}),
проходящая через произвольную точку поверхности связей, целиком лежала на этой
поверхности необходимо и достаточно, чтобы производная по времени от всех
связей обращалась в нуль на поверхности связей,
\begin{equation}                                                  \label{ecoinc}
  \dot \Phi^\mu=[\Phi^\mu,H_\Se]_{\Phi=0}
  =[\Phi^\mu,H]_{\Phi=0}+\lm_\nu[\Phi^\mu,\Phi^\nu]_{\Phi=0}=0.
\end{equation}
\end{prop}
\begin{proof}
Достаточность. Допустим, что в начальный момент времени траектория находилась на
поверхности связей, т.е.\ $\Phi^\mu(q_0,p_0)=0$, где $q_0:=q(0)$ и $p_0:=p(0)$.
Тогда с течением времени уравнения $\Phi^\mu\big(q(t),p(t)\big)=0$ будут
выполнены на решении задачи Коши для системы уравнений (\ref{ecoinc}).

Необходимость. Допустим, что каждая траектория, имеющая хотя бы одну точку на
поверхности связей, целиком принадлежит этой поверхности. Тогда производная
вдоль траектории от каждой связи должна быть равна нулю на поверхности связей,
поскольку мы можем произвольно менять исходную точку на поверхности. Это
означает выполнение условий (\ref{ecoinc}) для всех значений индекса $\mu$.
\end{proof}
Уравнения (\ref{ecoinc}) можно рассматривать как уравнения на множители
Лагранжа на поверхности связей. При этом условие, определяющее связи второго
рода (\ref{eseclc}), является необходимым и достаточным для однозначного
определения множителей Лагранжа $\lm_\mu(x)$ на поверхности связей $\MM$. Вне
поверхности связей множители Лагранжа можно продолжить любым достаточно гладким
образом, поскольку это не влияет на динамику частицы на поверхности связей. Для
определенности, положим
\begin{equation}                                                  \label{elasec}
  \lm_\mu=-J^{-1}_{\mu\nu}[\Phi^\nu,H]
\end{equation}
в некоторой окрестности поверхности связей. Если продолжить множители Лагранжа
вне поверхность связей каким либо иным образом, то это изменит только
траектории, лежащие вне поверхности связей, которые нас не интересуют. Множители
Лагранжа (\ref{elasec}) подставляем в уравнения (\ref{ecoint}) и решаем задачу
Коши для координат и импульсов. Тогда, если в начальный момент времени точка
фазового пространства находилась на поверхности связей, то она там и останется
при эволюции системы. Это означает, что метод неопределенных множителей Лагранжа
позволяет из вариационного принципа для обобщенного действия (\ref{etacpr})
определить множители Лагранжа и эволюцию динамических переменных.

Решение (\ref{elasec}) для множителей Лагранжа означает, что, если связи
выполняются в начальный момент времени для некоторой траектории, то множители
Лагранжа всегда можно подобрать таким образом, что связи будут выполнены и во
все последующие моменты времени.

Исключим множители Лагранжа (\ref{elasec}) из уравнений движения (\ref{ecoint}):
\begin{equation*}                                        
\begin{split}
  \dot q^i&=[q^i,H]-[q^i,\Phi^\mu]J^{-1}_{\mu\nu}[\Phi^\nu,H],
\\
  \dot p_i&=[p_i,H]-[p_i,\Phi^\mu]J^{-1}_{\mu\nu}[\Phi^\nu,H].
\end{split}
\end{equation*}
Эти уравнения можно записать в компактном виде с помощью нового важного понятия,
которое было введено в \cite{Dirac64R}.
\begin{defn}
Для любых функций $f,g\in\CC^1(\MN)$ билинейная операция
\begin{equation}                                                  \label{edibrd}
  [f,g]_\Sd:=[f,g]-[f,\Phi^\mu]J^{-1}_{\mu\nu}[\Phi^\nu,g],
\end{equation}
называется {\em скобкой Дирака}.
\qed\end{defn}
\index{Скобка Дирака (Dirac's bracket)}%
\index{Дирака скобка (Dirac's bracket)}%
Скобка Дирака, очевидно, антисимметрична и билинейна. Кроме того, нетрудно
проверить, что для нее справедливо правило Лейбница и тождества Якоби.
Следовательно, скобка Дирака определяет на фазовом пространстве $\MN$ новую
пуассонову структуру (см.\ раздел \ref{spoist}). Напомним, что на фазовом
пространстве уже существует каноническая пуассонова структура для координат и
сопряженных импульсов $[f,g]$. Скобка Дирака определяет на $\MN$ вторую
{\em диракову пуассонову структуру}, которая обладает рядом замечательных
свойств для гамильтоновых систем со связями II рода.
\index{Диракова пуассонова структура (Dirac's Poisson structure)}%
\index{Пуассонова структура Дирака (Dirac's Poisson structure)}%

Сначала заметим, что уравнения движения (\ref{ecoint}) можно записать в
компактном виде
\begin{equation}                                                  \label{ecoind}
  \dot q^i=[q^i,H]_\Sd,\qquad \dot p_i=[p_i,H]_\Sd.
\end{equation}
В эти уравнения не входят явно множители Лагранжа, потому что достаточная
информация о связях содержится в определении скобки Дирака.

Из определения (\ref{edibrd}) следует, что скобка Дирака каждой связи
(\ref{econtw}) с произвольной функцией $F\in\CC^1(\MN)$ равна нулю:
\begin{equation*}
  [F,\Phi^\mu]_\Sd=0.
\end{equation*}
Отсюда вытекает, что диракова пуассонова структура вырождена и все связи второго
рода являются функциями Казимира для скобки Дирака. Покажем, что других
функционально независимых функций Казимира не существует.
\begin{prop}
На поверхности связей $\Phi=0$ существует такая система координат $y^{a}$,
$a=1,\dotsc,2(\Sn-\Sm)$, что выполнены следующие условия:
\begin{equation*}
  \det[y^a,y^b]_{\Phi=0}\ne0,\qquad [y^a,\Phi^\mu]_{\Phi=0}=0.
\end{equation*}
\end{prop}
\begin{proof}
Поскольку связи функционально независимы, то их можно выбрать в качестве части
координатных функций новой системы координат в некоторой окрестности поверхности
связей. Выберем какие либо координаты $\tilde y^a$, $a=1,\dotsc,2(\Sn-\Sm)$, на
поверхности связей. Тогда совокупность функций
$\lbrace\tilde y^a,\Phi^\mu\rbrace$ образует систему координат в окрестности
поверхности связей. В общем случае $[\tilde y^a,\Phi^\mu]\ne0$. Тогда введем
новые координаты на поверхности
связей $y^a=\tilde y^a+\lm^a_\mu\Phi^\mu$, где $\lm^a_\mu$ -- некоторые функции.
На поверхности связей $\Phi=0$ эти координаты совпадают со старыми. Выберем
неизвестные функции $\lm^a_\mu$ таким образом, чтобы на поверхности связей
выполнялись уравнения:
\begin{equation*}
  [y^a,\Phi^\mu]_{\Phi=0}=[\tilde y^a,\Phi^\mu]_{\Phi=0}
  +\lm^a_\nu[\Phi^\nu,\Phi^\mu]_{\Phi=0}=0.
\end{equation*}
Это всегда можно сделать, поскольку связи $\Phi^\mu$ второго рода. В новой
системе координат структурные функции канонической пуассоновой структуры на
поверхности связей примут вид
\begin{equation*}
J=\begin{pmatrix}[y^a,y^b]_{\Phi=0} & 0 \\ 0 & [\Phi^\mu,\Phi^\nu]_{\Phi=0}
\end{pmatrix}.
\end{equation*}
Поскольку пуассонова структура не может быть вырожденной, то
$\det[y^a,y^b]_{\Phi=0}\ne0$.
\end{proof}
В системе координат $\lbrace y^a,\Phi^\mu\rbrace$ из только что доказанного
предложения структурные функции дираковой пуассоновой структуры примут вид
\begin{equation*}
  J_\Sd=\begin{pmatrix}[y^a,y^b]_{\Phi=0} & 0 \\ 0 & 0 \end{pmatrix},
\end{equation*}
поскольку $[y^a,y^b]_\Sd{}_{\Phi=0}=[y^a,y^b]_{\Phi=0}$. Отсюда следует, что в
окрестности связей ранг $J_\Sd$ равен $2(\Sn-\Sm)$. Таким образом мы доказали
следующее утверждение.
\begin{theorem}
Связи второго рода $\Phi^\mu$ и только они вместе со всеми их линейными
комбинациями являются функциями Казимира для пуассоновой структуры Дирака
(\ref{edibrd}).
\end{theorem}

Используя скобку Дирака, для каждой связи $\Phi^\mu$ можно построить векторное
поле  с компонентами
\begin{equation*}
  X^\al_{\Phi^\mu}=[\Phi^\mu,x^\al]_\Sd,
\end{equation*}
где $x=(q,p)$ -- координаты на фазовом пространстве $\MN$.
Согласно предложению \ref{poisth} коммутаторы этих векторных полей равны нулю:
\begin{equation*}
  [X_{\Phi^\mu},X_{\Phi^\nu}]=X_{[\Phi^\mu,\Phi^\nu]_\Sd}=0.
\end{equation*}
Следовательно, распределение векторных полей $\lbrace X_{\Phi^\mu}\rbrace$,
$\mu=1,\dotsc,2\Sm$, находится в инволюции и определяет интегральные
подмногообразия в $\MN$. Таким образом, постоянные значения функций Казимира
слоят фазовое пространство $\MN$ на симплектические сечения, одним из которых
является поверхность связей $\MM\subset\MN$.

Каноническая пуассонова структура на $\MN$ не приспособлена для описания систем
со связями, потому что ограничение скобки Пуассона двух функций на поверхность
связей нельзя проводить до вычисления самой скобки:
\begin{equation*}
  \big[f,g\big]_{\Phi=0}\ne\big[f_{\Phi=0},g_{\Phi=0}\big].
\end{equation*}
Например, мы сразу приходим к противоречию, если $f=\Phi^\mu$ и $g=\Phi^\nu$ для
некоторых $\mu$ и $\nu$. В то же время ограничение скобки Дирака на поверхность
связей можно проводить до вычисления самой скобки:
\begin{equation}                                                  \label{eduunv}
  \big[f,g\big]_\Sd\big|_{\Phi=0}=\big[f_{\Phi=0},g_{\Phi=0}\big]_\Sd.
\end{equation}
Так как поверхность связей является подмногообразием $\MM\subset\MN$, то ее
можно рассматривать, как вложение $\MM\hookrightarrow\MN$. Тогда равенство
(\ref{eduunv}) означает, что вложение $\MM\hookrightarrow\MN$ является
пуассоновым отображением для скобки Дирака, но не для канонической пуассоновой
структуры на $\MN$.

Поскольку скобки Дирака достаточно для описания эволюции всех переменных, то ее
использование приносит существенные упрощения в описание динамики систем со
связями II рода.

Скобка Дирака (\ref{edibrd}) определена с помощью канонической пуассоновой
структуры на фазовом пространстве $\MN$. В разделе \ref{scantr} было показано,
что канонические преобразования сохраняют вид канонической скобки Пуассона.
Поэтому скобка Дирака двух функций также инвариантна относительно канонических
преобразований, т.е.\ она имеет одинаковый вид в любых координатах на $\MN$,
связанных между собой каноническим преобразованием.

В физических приложениях важную роль играет специальная система координат на
фазовом пространстве, которая строится с учетом вложения
$\MM\hookrightarrow\MN$. Ее существование обеспечивается следующим важным
утверждением.
\begin{theorem}                                                   \label{tphyco}
Пусть задано фазовое пространство $\MN$ и набор функционально независимых связей
второго рода (\ref{econtw}). Тогда существует такое каноническое преобразование
\begin{equation*}
  \lbrace q^i,p_i\rbrace\mapsto\lbrace q^{*a},p^*_a,Q^\Sa,P_\Sa\rbrace,
  \quad a=1,\dotsc,\Sn-\Sm,\quad \Sa=1,\dotsc,\Sm,
\end{equation*}
что набор связей $\Phi=0$ эквивалентен связям
\begin{equation*}
  Q=0,\qquad P=0.
\end{equation*}
При этом координаты $q^*,p^*$ являются канонически сопряженными координатами и
импульсами на поверхности связей $\MM\subset\MN$.
\end{theorem}
\begin{proof}
Поскольку связи $\Phi^\mu$ являются связями второго рода, то согласно теореме
Дарбу существуют такие координаты $Q^\Sa,P_\Sa$, что $[Q^\Sa,P_\Sb]=\dl^\Sa_\Sb$
и уравнения $\Phi=0$ эквивалентны уравнениям $Q=0$, $P=0$. После этого из
теоремы Дарбу \ref{tdrpoi} следует существование системы координат
$q^{*a},p^*_a,Q^\Sa,P_\Sa$. Поскольку скобка Пуассона в новых координатах имеет
канонический вид, то преобразование координат является каноническим.
\end{proof}
Эта теорема локальна. Если найдены координаты $q^*,p^*,Q,P$ в явном виде, то
ограничение функций на поверхность связей особенно просто: нужно просто положить
$Q=0$ и $P=0$. В частности,
\begin{equation*}
  H_\Se|_{\Phi=0}=H|_{Q=0,\,P=0}.
\end{equation*}

Поскольку
\begin{equation*}
  [q^*,Q]=0,\quad [q^*,P]=0\qquad \text{и}\qquad [p^*,Q]=0,\quad [p^*,P]=0,
\end{equation*}
то уравнения движения (\ref{ecoind}) в новой системе координат примут вид
\begin{equation*}
\begin{split}
  \dot q^*=[q^*,H_\mathrm{ph}],\qquad &\dot p^*=[p^*,H_\mathrm{ph}]
\\
  \dot Q^\Sa=0,\qquad &\dot P_\Sa=0,
\end{split}
\end{equation*}
где
\begin{equation*}
  H_\mathrm{ph}(q^*,p^*)=H_\Se|_{\Phi=0}=H(q^*,p^*,Q,P)|_{Q=0,\,P=0}.
\end{equation*}
Мы видим, что динамика системы $\Sn$ частиц, на которую наложено $2\Sm$ связей
второго рода свелась к обычной динамике системы из $\Sn-\Sm$ частиц, на которую
уже не наложено никаких связей. По этой причине мы говорим, что система имеет
$\Sn-\Sm$ {\em физических степеней свободы} $q^*,p^*$. Координаты $Q,P$
описывают {\em нефизические степени свободы}, поскольку определяют связи.
Динамика физических степеней свободы задается {\em эффективным (физическим)
гамильтонианом} $H_\mathrm{ph}$, зависящим только от физических координат и
импульсов. Координаты на поверхности связей определены неоднозначно. Например,
на $\MM$ всегда можно совершить каноническое преобразование. Однако размерность
поверхности связей (удвоенное число физических степеней свободы) фиксирована и
всегда равна $2(\Sn-\Sm)$. Мы говорим, что система $\Sn$ частиц с $2\Sm$ связями
II рода описывает $\Sn-\Sm$ физических степеней свободы.
\index{Физическая степень свободы}\index{Степень свободы физическая}%
\index{Нефизическая степень свободы}\index{Степень свободы нефизическая}%
\index{Эффективный гамильтониан}\index{Гамильтониан эффективный}%
\index{Физический гамильтониан}\index{Гамильтониан физический}%
\begin{com}
На практике найти в явном виде координаты $q^*,p^*$ для физических степеней
свободы удается только в простейших случаях. Кроме того, в теории поля связи
часто представляют собой дифференциальные уравнения по пространственным
координатам. В этих случаях переход к координатам $q^*,p^*$ задается
нелокальными выражениями, как, например, в свободной электродинамике (см.\
раздел \ref{semfhf}). Поэтому, вычисления обычно проводят в исходных координатах
$q^i,p_i$, используя теорему \ref{tphyco} для доказательства общих утверждений.
\qed\end{com}
Таким образом, если на фазовом пространстве $\MN$ с канонической пуассоновой
структурой задана система связей $\lbrace\Phi^\mu \rbrace$ второго рода, то она
определяет некоторое подмногообразие $\MM\subset\MN$, на котором естественным
образом определена каноническая пуассонова структура. Посмотрим на эту задачу с
обратной точки зрения. Пусть задано вложение $\MM\hookrightarrow\MN$, и мы
знаем, что подмногообразие $\MM$ является фазовым пространством некоторой
механической системы с канонической пуассоновой структурой. Возникает
естественный вопрос может ли пространство-мишень также быть фазовым
пространством ? Ниже мы покажем, что ответ на этот вопрос положительный.

Пусть задано вложение
\begin{equation*}
  \vf:\quad \MM~\hookrightarrow~\MN,\qquad
  \dim\MM=2(\Sn-\Sm),\quad \dim\MN=2\Sn,~\Sm<\Sn,
\end{equation*}
фазового пространства $\MM$ с канонической пуассоновой структурой. Тогда в
окрестности каждой точки $\MM$ существует такая система координат $y^a$,
$a=1,\dotsc,2(\Sn-\Sm)$, в которой пуассонова структура имеет канонический вид
\begin{equation*}
  [y^a,y^b]=\varpi^{-1ab}.
\end{equation*}
Эта пуассонова структура определяет пуассонову структуру на образе $\vf(\MM)$ с
помощью дифференциала отображения $\vf_*$. В дальнейшем образ отображения
вложения мы отождествим с самим многообразием, т.е.\ положим
$\MM=\vf(\MM)\subset\MN$. Если $x^\al$, $\al=1,\dotsc,2\Sn$, -- система
координат на $\MN$, то индуцированная пуассонова структура на $\MM$ в
координатах $x^\al\in\MN$ задается антисимметричной матрицей
\begin{equation}                                                  \label{epostn}
  J^{\al\bt}=\varpi^{-1ab}\pl_a x^\al\pl_b x^\bt,
\end{equation}
где функции $x^\al(y)$ описывают вложение. Из свойств произведения матриц
следует, что ранг этой пуассоновой структуры равен $2(\Sn-\Sm)$, и поэтому она
всегда вырождена. Следовательно, для индуцированной на $\MN$ пуассоновой
структуры существует $2\Sm$ независимых функций Казимира $c^\mu$,
$\mu=1,\dotsc2\Sm$. Выберем координаты в окрестности $\MM\subset\MN$ следующим
образом
\begin{equation*}
  \lbrace x^\al\rbrace=\lbrace y^a,c^\mu\rbrace.
\end{equation*}
Тогда подмногообразие $\MM$ задается постоянными значениями функций Казимира
$c^\mu=\const$. Отсюда следует, что уравнения $c^\mu=\const$ эквивалентны
исходной системе связей (\ref{econtw}).

Проведенное построение показывает, что на фазовом пространстве $\MN$ существует
такая система координат, в которой индуцированная пуассонова структура имеет вид
\begin{equation*}
  J^{\al\bt}=\begin{pmatrix}\varpi^{-1ab} & 0 \\ 0 & 0 \end{pmatrix}.
\end{equation*}
Тем самым можно отождествить индуцированную пуассонову структуру с дираковой
пуассоновой структурой, а функции Казимира $c^\mu$ со связями $\Phi^\mu$.

Ранее скобка Дирака была определена через каноническую пуассонову структуру на
исходном многообразии $\MN$ с помощью связей. В обратную сторону однозначного
рецепта определения невырожденной пуассоновой структуры на $\MN$ не существует,
поскольку ранг индуцированной пуассоновой структуры $2(\Sn-\Sm)$ меньше
размерности многообразия $\dim\MN=2\Sn$. Здесь существует много возможностей.
Например, можно положить $J^{\al\bt}=\varpi^{-1\al\bt}$ в координатах
$\lbrace y^a,c^\mu\rbrace$. Тогда получим каноническую пуассонову структуру на
$\MN$.
\subsection{Гамильтонова динамика частиц со связями I рода       \label{sficon}}
Пусть действие для $\Sn$ точечных частиц в фазовом пространстве $\MN$ имеет
обычный вид (\ref{ehaact}). Будем искать решение канонических уравнений
движения (\ref{ehameq}) при наличии $\Sm<\Sn$ связей на канонические переменные:
\begin{equation}                                                  \label{ecopop}
  G_\Sa(q,p)=0,\qquad \Sa=1,\dotsc,\Sm<\Sn.
\end{equation}
Другими словами, будем считать, что частицы не могут покинуть
$(2\Sn-\Sm)$-мерное подмногообразие фазового пространства $\MU\subset\MN$
({\em поверхность связей}), определенного уравнениями (\ref{ecopop}). Из
дальнейшего рассмотрения будет ясно, почему мы выбрали число связей первого рода
в два раза меньшим числа связей второго рода и почему индекс у связей первого
рода пишется внизу.

Предположим, что связи являются достаточно гладкими функциями и функционально
независимы на поверхности связей (\ref{ecopop}), т.е.\ ранг матрицы
$\pl G_\Sa/\pl x^\al$, где $x=(q^1,\dotsc,q^\Sn,p_1,\dotsc,p_\Sn)$ -- координаты
фазового пространства, максимален и равен $\Sm$. Будем считать, что
рассматриваемая гамильтонова система со связями находится в {\em инволюции}:
\index{Инволюция}%
\begin{align}                                                     \label{einvco}
  [G_\Sa,G_\Sb]&=f_{\Sa\Sb}{}^\Sc G_\Sc\approx0,
\\                                                                \label{einvha}
  [G_\Sa,H]&=v_\Sa{}^\Sb G_\Sb\approx0,
\end{align}
где $f_{\Sa\Sb}{}^\Sc(q,p)=-f_{\Sb\Sa}{}^\Sc(q,p)$ и $v_\Sa{}^\Sb(q,p)$ --
некоторые функции от точки фазового пространства $(q,p)\in\MN$. Волнистый знак
равенства $\approx$ обозначает, что данная функция обращается в нуль на
поверхности связей:
\begin{equation*}
  f\approx0\quad \Leftrightarrow\quad f|_{G=0}=0.
\end{equation*}
При этом говорят, что скобки Пуассона связей между собой и с гамильтонианом
{\em слабо равны нулю}.

Для выполнения равенств (\ref{einvco}) необходимо, чтобы количество связей не
превосходило половины размерности фазового пространства $\Sm\le\Sn$, что мы
потребовали с самого начала (\ref{ecopop}). Действительно, при преобразованиях
координат фазового пространства каноническая пуассонова структура не может
вырождаться. В силу функциональной независимости функций $G_\Sa$ их можно
выбрать в качестве части новых координат. Если $\Sm>\Sn$, то пуассонова
структура была бы вырожденной, что невозможно для фазового пространства.
\begin{defn}
Связи (\ref{ecopop}), наложенные на канонические переменные механической системы
с гамильтонианом $H(q,p)$, которые удовлетворяют условиям (\ref{einvco}),
(\ref{einvha}), называются {\em связями I рода}.
\qed\end{defn}
\index{Связи первого рода (first class constraints)}%
Мы предполагаем, что все связи I рода учтены в системе уравнений (\ref{ecopop}),
т.е.\ для заданной гамильтоновой системы не существует большего числа
функционально независимых соотношений между каноническими переменными, для
которых выполнены условия (\ref{einvco}) и (\ref{einvha}).

Скобки Пуассона (\ref{einvco}) по виду напоминают коммутатор базисных векторных
полей в алгебре Ли. Однако в рассматриваемом случае допускается нетривиальная
зависимость {\em структурных функций} от точки фазового пространства:
$f_{\Sa\Sb}{}^\Sc=f_{\Sa\Sb}{}^\Sc(q,p)$.
\index{Структурные функции (structure functions)}%
\index{Функции структурные (structure functions)}%

Напомним, что связи определены неоднозначно. А именно, невырожденные линейные
комбинации связей определяют ту же поверхность связей. Это приведет к изменению
структурных функций $f_{\Sa\Sb}{}^\Sc$ и $v_\Sa{}^\Sb$, что важно при решении
уравнений движения.

Функции $f_{\Sa\Sb}{}^\Sc$ и $v_\Sa{}^\Sb$ не могут быть произвольными.
По-определению, скобки Пуассона удовлетворяют тождеству Якоби. Рассмотрев скобки
Пуассона $\big[[G_\Sa,G_\Sb],G_\Sc\big]$ и $\big[[G_\Sa,G_\Sb],H\big]$ и их
циклические перестановки, получим ограничения на структурные функции:
\begin{align}                                                     \label{ealgco}
  f_{\Sa\Sb}{}^\Sd f_{\Sd\Sc}{}^\Se+f_{\Sb\Sc}{}^\Sd f_{\Sd\Sa}{}^\Se
  +f_{\Sc\Sa}{}^\Sd f_{\Sd\Sb}{}^\Se+[f_{\Sa\Sb}{}^\Se,G_\Sc]
  +[f_{\Sb\Sc}{}^\Se,G_\Sa]+[f_{\Sc\Sa}{}^\Se,G_\Sb]&=0,
\\                                                                     \nonumber
  f_{\Sa\Sb}{}^\Sd v_\Sd{}^\Sc+v_\Sb{}^\Sd f_{\Sd\Sa}{}^\Sc
  -v_\Sa{}^\Sd f_{\Sd\Sb}{}^\Sc+[f_{\Sa\Sb}{}^\Sc,H]
  +[v_\Sb{}^\Sc,G_\Sa]-[v_\Sa{}^\Sc,G_\Sb]&=0.
\end{align}
Тождество Якоби для двойной скобки Пуассона $\big[[G_\Sa,H],H\big]$
удовлетворяются автоматически в силу уравнения (\ref{einvha}).

Если структурные функции постоянны, $f_{\Sa\Sb}{}^\Sc=\const$, то множество всех
линейных комбинаций $a^\Sa G_\Sa$, $a^\Sa\in\MR$, образует алгебру Ли с базисом
$G_\Sa$. Тогда уравнение (\ref{einvco}) задает коммутатор базисных векторов, а
соотношение (\ref{ealgco}) сводится к тождеству Якоби для структурных констант.
\begin{prop}                                                      \label{pfiers}
Для того, чтобы фазовые траектории для гамильтониана $H$, проходящие через
произвольную точку на поверхности связей, целиком лежали на этой поверхности
необходимо и достаточно, чтобы были выполнены условия (\ref{einvha}).
\end{prop}
\begin{proof}
Совпадает с доказательством предложения \ref{pconsu}. Достаточно только
заметить, что
\begin{equation}                                                  \label{eficce}
  \dot G_\Sa=[G_\Sa,H]\approx0,
\end{equation}
и воспользоваться предложением \ref{panyco}.
\end{proof}

С точки зрения вариационного принципа решение канонических уравнений
(\ref{ehameq}) при наличии связей (\ref{ecopop}) является задачей на условный
экстремум. Применим к решению этой задачи метод неопределенных множителей
Лагранжа. С этой целью построим {\em полный (total) гамильтониан}, добавив к
исходному гамильтониану линейную комбинацию связей первого рода,
\index{Полный гамильтониан (total Hamiltonian)}%
\index{Гамильтониан полный (total Hamiltonian)}%
\begin{equation}                                                  \label{etoham}
  H_\St=H+\lm^\Sa G_\Sa,
\end{equation}
где $\lm^\Sa=\lm^\Sa(q,p,t)$ -- неопределенные множители Лагранжа.
Соответствующее действие, которое мы назовем {\em полным}, имеет вид
\index{Полное действие (total action)}%
\index{Действие полное (total action)}%
$$
  S_\St=\int_{t_1}^{t_2}\!\!\!dt(p\dot q-H-\lm^\Sa G_\Sa).
$$
Из него вытекают уравнения Эйлера--Лагранжа:
\begin{align}                                                     \label{eplmmo}
\begin{split}
  \frac{\dl S_\St}{\dl p_i}&=\phantom{-}\dot q^i-\frac{\pl H}{\pl p_i}
  -\lm^\Sa\frac{\pl G_\Sa}{\pl p_i}=0,
\\
  \frac{\dl S_\St}{\dl q^i}&=-\dot p_i-\frac{\pl H}{\pl q^i}
  -\lm^\Sa\frac{\pl G_\Sa}{\pl q^i}=0,
\end{split}&
\\                                                                \label{elagfi}
  &\frac{\dl S_\St}{\dl\lm^\Sa}=-G_\Sa=0,
\end{align}
где в правых частях уравнений движения (\ref{eplmmo}) мы отбросили слагаемые,
пропорциональные связям. При варьировании действия мы считаем вариации координат
на границе равными нулю, $\dl q(t_{1,2})=0$, а вариации импульсов и множителей
Лагранжа могут быть произвольны.
\begin{prop}
Для того, чтобы фазовые траектории для полного действия $S_\St$, проходящие
через произвольную точку на поверхности связей, целиком лежали на этой
поверхности достаточно, чтобы были выполнены условия
(\ref{einvco}) и (\ref{einvha}).
\end{prop}
\begin{proof}
Повторяет доказательство предложения \ref{pconsu}. Достаточно заметить, что
\begin{equation*}                                                    \tag*{\qed}
  \dot G_\Sa=[G_\Sa,H]+\lm^\Sb[G_\Sa,G_\Sb]\approx0.
\end{equation*}
\renewcommand{\qed}{}\end{proof}
Уравнения движения для систем со связями первого рода (\ref{eplmmo}) существенно
отличаются от уравнений для систем со связями второго рода. Дело в том, что они
не позволяют определить ни одного множителя Лагранжа. Это связано с тем, что
уравнения (\ref{ecoinc}), из которых находились множители Лагранжа для связей
второго рода в рассматриваемом случае (\ref{eficce}) на поверхности связей
вовсе не содержат множителей Лагранжа. Тем самым любое решение уравнений
движения для систем со связями первого рода содержит произвольные функции
$\lm^\Sa(q,p,t)$, число которых совпадает с числом связей. Причиной
функционального произвола в решении уравнений движения является калибровочная
инвариантность.
\begin{theorem}                                                   \label{tlochs}
Полное действие $S_\St$ инвариантно относительно локальных инфинитезимальных
преобразований, генерируемых каждой связью первого рода:
\begin{equation}                                                  \label{elotrl}
\begin{split}
  \dl q^i&=\e^\Sa[ q^i,G_\Sa]=\phantom{-}\e^\Sa\frac{\pl G_\Sa}{\pl p_i},
\\
  \dl p_i&=\e^\Sa[ p_i,G_\Sa]=-\e^\Sa\frac{\pl G_\Sa}{\pl q^i},
\\
  \dl\lm^\Sa&=\dot\e^\Sa+\e^\Sb v_\Sb{}^\Sa+\e^\Sb\lm^\Sc f_{\Sb\Sc}{}^\Sa,
\end{split}
\end{equation}
где $\e^\Sa=\e^\Sa(q,p,t)$ -- малый параметр преобразований и
$\dot\e=\pl\e/\pl t$. Параметр преобразований может быть произвольной функцией
координат, импульсов и времени с нулевыми граничными условиями
$\e^\Sa(t_{1,2})=0$ для всех $q$ и $p$.
\end{theorem}
\begin{proof}
Вариация действия имеет вид
\begin{equation*}
  \dl S_\St=\int dt\left(-\e^\Sa\frac{\pl G_\Sa}{\pl q^i}\dot q^i
  -\dot p_i\e^\Sa\frac{\pl G_\Sa}{\pl p_i}
  +\e^\Sa[ G_\Sa,H]+\e^\Sa\lm^\Sb[ G_\Sa,G_\Sb]-\dl\lm^\Sa G_\Sa\right).
\end{equation*}
Подставляя сюда вариацию множителей Лагранжа и интегрируя слагаемое
$\dot\e^\Sa G_\Sa$ по частям с учетом уравнений (\ref{einvco}) и (\ref{einvha}),
получаем $\dl S_\St=0$. Условия на параметры калибровочных преобразований
$\e^\Sa(t_{1,2})=0$ достаточны для того, чтобы интегрирование слагаемого
$\dot\e^\Sa G_\Sa$ по частям было возможно.
\end{proof}

Согласно второй теореме Нетер локальная инвариантность приводит к линейной
зависимости уравнений движения:
\begin{equation}                                                  \label{elodeo}
  \frac{\dl S_\St}{\dl q}\frac{\pl G_\Sa}{\pl p}
  -\frac{\dl S_\St}{\dl p}\frac{\pl G_\Sa}{\pl q}
  -\frac\pl{\pl t}\left(\frac{\dl S_\St}{\dl\lm^\Sa}\right)
  +\frac{\dl S_\St}{\dl\lm^\Sb}(v_\Sa{}^\Sb+\lm^\Sc f_{\Sa\Sc}{}^\Sb)=0,
\end{equation}
в чем нетрудно убедиться и прямой проверкой. В этом случае для любого решения
системы уравнений (\ref{eplmmo}) будет автоматически выполнено уравнение
\begin{equation*}
  \dot G_\Sa=(v_\Sa{}^\Sb+\lm^\Sc f_{\Sa\Sc}{}^\Sb)G_\Sb.
\end{equation*}
Для этой системы уравнений точка $G=0$ является неподвижной. Поэтому, если в
начальный момент времени фазовая траектория находится на поверхности связей, то
для любого решения системы уравнений (\ref{eplmmo}) при любых множителях
Лагранжа $\lm^\Sa(q,p,t)$ связи $G_\Sa=0$ будут автоматически удовлетворены.
Это и является причиной возникновения функционального произвола в решениях
уравнений движения.
\begin{com}
Преобразования координат фазового пространства $q,p$ (\ref{elotrl}) с малым
постоянным параметром $\e=\const$ являются инфинитезимальной формой
канонического преобразования, описанного в примере \ref{ehamge},
\begin{equation*}
  q^i(t)~\mapsto~q^i(t,\e),\qquad p_i(t)~\mapsto~p_i(t,\e),
\end{equation*}
определяемого уравнениями
\begin{equation*}                                                    \tag*{\qed}
  \frac{\pl q^i}{\pl\e^\Sa}=[q^i,G_\Sa],\qquad
  \frac{\pl p_i}{\pl\e^\Sa}=[p_i,G_\Sa].
\end{equation*}
\renewcommand{\qed}{}\end{com}
\begin{defn}
Модели, функционал действия которых инвариантен относительно локальных
преобразований, называют {\em калибровочными}, а сами преобразования -- {\em
калибровочными}.
\qed\end{defn}
\index{Калибровочная модель (gauge model)}%
\index{Модель калибровочная (gauge model)}%
\index{Калибровочное преобразование (gauge transformation)}%
\index{Преобразование калибровочное (gauge transformation)}%
\begin{com}
Это название пришло из физики, где все калибровочные модели, в частности,
электродинамика и поля Янга--Миллса обладают этим свойством.
\qed\end{com}
Тем самым мы доказали, что каждой связи первого рода соответствует локальная
инвариантность полного действия, а сама связь является генератором калибровочных
преобразований. По аналогии с генераторами групп Ли мы пишем индексы у связей
первого рода внизу.

Проанализируем уравнения движения подробнее. Поскольку среди $2\Sn+\Sm$
уравнений Эйлера--Лагранжа (\ref{eplmmo}) и (\ref{elagfi}) только $2\Sn$
являются независимыми, то в отличии от задачи на условный экстремум,
рассмотренной в разделе \ref{svarpr}, этих уравнений недостаточно для
определения всех неизвестных функций $q(t)$, $p(t)$ и $\lm(t)$. Допустим, что
систему уравнений (\ref{eplmmo}) можно решить относительно канонических
переменных, для которых поставлена задача Коши: $q(t_1)=q_1$, $p(t_1)=p_1$.
Тогда решение этой задачи будет зависеть от $\Sm$ произвольных функций времени,
которыми являются множители Лагранжа, и $2\Sn$ постоянных интегрирования. Все
постоянные интегрирования определяются начальными данными. В этом случае через
одну точку фазового пространства проходит множество фазовых траекторий, которые
параметризуется множителями Лагранжа.

С физической точки зрения это означает следующее. Пусть некоторое физическое
явление описывается калибровочной моделью. Тогда начальное состояние системы не
определяет однозначно последующую эволюцию, что противоречит экспериментальным
данным, если не принимать во внимание квантовомеханическую неопределенность. Тем
не менее калибровочные модели в настоящее время широко используются в
теоретической физике: классическим примером служит электродинамика.

Выход из этого противоречия прост. Для калибровочных моделей вводится постулат:
все физические наблюдаемые калибровочно инвариантны, т.е.\ не зависят от
произвольных функций, которые могут содержаться в решении уравнений движения.
Отсюда следует, что физические наблюдаемые описываются функциями
$f\in\CC^k(\MN)$ на фазовом пространстве $\MN$, которые являются калибровочно
инвариантными. А именно, мы требуем, чтобы скобка Пуассона каждой наблюдаемой
функции со связями обращалась в нуль на поверхности связей:
\begin{equation}                                                  \label{egainc}
  [f,G_\Sa]=d_\Sa{}^\Sb G_\Sb\approx0,
\end{equation}
где $d_\Sa{}^\Sb(q,p)$ -- некоторые достаточно гладкие функции канонических
переменных. Тогда в уравнении движения для калибровочно инвариантной функции
\begin{equation*}
  \dot f=[f,H]+\lm^\Sa[f,G_\Sa]\approx[f,H],
\end{equation*}
все слагаемые с множителями Лагранжа обратятся в нуль на поверхности связей и,
следовательно, на поверхности связей никакого произвола в эволюции калибровочно
инвариантной функции нет.

Проведенное рассмотрение требует комментария, потому что каждой физической
наблюдаемой соответствует не одна, а целый класс калибровочно инвариантных
функций. Поскольку мы рассматриваем динамику частиц на поверхности связей
(\ref{ecopop}), то физические наблюдаемые определяются значениями калибровочно
инвариантных функций на поверхности связей. Пусть две калибровочно инвариантные
функции $f_1$ и $f_2$ совпадают на поверхности связей $\MU$. Тогда они могут
отличаться только на линейную комбинацию связей:
\begin{equation}                                                  \label{etwfud}
  f_2=f_1+\mu^\Sa G_\Sa,
\end{equation}
где $\mu^\Sa(q,p)$ -- некоторые достаточно гладкие функции. Следовательно, все
множество калибровочно инвариантных функций разбивается на классы
эквивалентности (\ref{etwfud}). При этом каждый класс эквивалентности
соответствует одной физической наблюдаемой.

Другими словами, калибровочно инвариантная функция задается на поверхности
связей $\MU\subset\MN$, а затем продолжается на все фазовое пространство в
значительной степени произвольным образом. Степень произвола описывается
произвольными функциями $\mu^\Sa(q,p)$. При этом физические наблюдаемые не
зависят от способа продолжения.

Оценим произвол, с которым калибровочно инвариантная функция может быть задана
на поверхности связей $\MU\subset\MN$. Условие калибровочной инвариантности
(\ref{egainc}) представляет собой систему $\Sm$ дифференциальных уравнений в
частных производных первого порядка, которую можно переписать в виде
\begin{equation*}
  X_{G_\Sa}f=0,
\end{equation*}
где $X_{G_\Sa}$ -- векторное поле, соответствующее связи $G_\Sa$ (\ref{emafve}).
Для этой системы уравнений условиями совместности являются уравнения
(\ref{einvco}). Действительно,
\begin{equation*}
  [X_{G_\Sa},X_{G_\Sb}]f=X_{[G_\Sa,G_\Sb]}f=f_{\Sa\Sb}{}^\Sc X_{G_\Sc}f=0,
\end{equation*}
где мы воспользовались равенством (\ref{ehopov}). Поэтому функция $f$ однозначно
определяется начальными данными на некотором собственном подмногообразии
$\MM\subset\MU\subset\MN$ размерности $2\Sn-\Sm-\Sm=2(\Sn-\Sm)$ и существенно
зависит только от части координат на $\MU$. Подмногообразие $\MM$ можно задать с
помощью $\Sm$ дополнительных функционально независимых соотношений между
координатами и импульсами:
\begin{equation}                                                  \label{egacof}
  F^\Sa(q,p)=0,
\end{equation}
которые называются {\em калибровочными условиями}. Эти условия должны
удовлетворять неравенству
\index{Калибровочное условие (gauge condition)}%
\index{Условие калибровочное (gauge condition)}%
\begin{equation}                                                  \label{edegfa}
  \det[F^\Sa,G_\Sb]\left.\vphantom{\sum}\right|_{F=0,\,G=0}\ne0,
\end{equation}
т.к.\ только в этом случае на $\MM$ можно задать начальные данные для системы
уравнений (\ref{egainc}). Условия (\ref{egacof}) при выполнении (\ref{edegfa})
называются {\em канонической калибровкой}. Для точечных частиц отличие от нуля
определителя (\ref{edegfa}) является необходимым и достаточным условием
однозначного определения множителей Лагранжа в полном гамильтониане
(\ref{etoham}). В общем случае функции $F^\Sa$, определяющие калибровочные
условия, могут зависеть также от времени и множителей Лагранжа $\lm^\Sa$.
\index{Каноническая калибровка (canonical gauge)}%
\index{Калибровка каноническая (canonical gauge)}%

После фиксирования канонической калибровки на рассматриваемую гамильтонову
систему будет наложено $2\Sm$ связей (\ref{ecopop}) и (\ref{egacof}). Введем
для полной совокупности связей и калибровочных условий следующее обозначение
\begin{equation}                                                  \label{etocon}
  \lbrace\Phi^\mu\rbrace=(F^1,\dotsc,F^\Sm,G_1,\dotsc,G_\Sm),
  \qquad \mu=1,\dotsc,2\Sm.
\end{equation}
Очевидно, что
\begin{equation}                                                  \label{edetfc}
  \det[\Phi^\mu,\Phi^\nu]_{\Phi=0}
  =\det\begin{pmatrix}[F^\Sa,F^\Sb]&[F^\Sa,G_\Sb]
  \\ [G_\Sa,F^\Sb]&[G_\Sa,G_\Sb]\end{pmatrix}_{\Phi=0}
  =\det\!{}^2[F^\Sa,G_\Sb]_{\Phi=0},
\end{equation}
т.к.\ $[G_\Sa,G_\Sb]\approx0$. Поскольку определитель скобок Пуассона для
канонических калибровочных условий со связями первого рода, по-построению,
отличен от нуля (\ref{edegfa}), то полная совокупность связей $\Phi^\mu$
представляет собой систему связей второго рода, рассмотренную в предыдущем
разделе. Таким образом, калибровочные модели в канонической калибровке сведены
к гамильтоновым системам со связями второго рода, для которых метод множителей
Лагранжа применим в полном объеме. Заметим, что значение скобок Пуассона для
канонических калибровочных условий между собой $[F^\Sa,F^\Sb]$ несущественно.

После наложения калибровочных условий возникает обобщенное действие
\begin{equation*}
  S_\Se=\int\! dt(p_i\dot q^i-H-\lm^\Sa G_\Sa-\pi_\Sa F^\Sa)
  =\int\! dt(p_i\dot q^i-H-\lm_\mu\Phi^\mu),
\end{equation*}
которое совпадает с выражением (\ref{etacpr}) для систем со связями II рода.

Каноническая калибровка выделяет в фазовом пространстве $\MN$ единственную
траекторию, проходящую через данную точку физического подпространства $\MM$. Это
происходит благодаря тому, что каноническая калибровка однозначно определяет
множители Лагранжа на поверхности связей $\Phi=0$. Верно также обратное
утверждение: произвольный выбор множителей Лагранжа эквивалентен некоторой
канонической калибровке. Действительно, при фиксированных множителях Лагранжа
граничная задача для уравнений (\ref{eplmmo}) имеет единственное решение
$x(t)=\lbrace q(t),p(t)\rbrace$. Перепишем для данного решения уравнения
(\ref{eplmmo}), но теперь уже с неопределенными множителями Лагранжа. В
результате получим переопределенную, но совместную систему $2\Sn$ линейных
алгебраических уравнений на $\Sm$ множителей Лагранжа. Тогда соответствующие
соотношения $x^\Sa=x^\Sa(t)$ можно принять в качестве канонических калибровочных
условий. Это следует из того, что, поскольку условия $F^\Sa=x^\Sa-x^\Sa(t)$
позволяют однозначно определить множители Лагранжа, то условие (\ref{edegfa})
выполнено. Напомним, что это условие является необходимым и достаточным для
однозначного определения множителей Лагранжа $\lm^\Sa$ на поверхности связей,
как следует из анализа, проведенного в предыдущем разделе, и (\ref{edetfc}).
\begin{com}
В приложениях часто рассматриваются неканонические калибровки. Например,
лоренцева или временн\'ая калибровки в электродинамике не являются
каноническими. Они не фиксируют калибровочную свободу полностью, и это создает
определенные трудности, например, при квантовании. В электродинамике
неканонические калибровочные условия можно использовать, т.к.\ уравнения
движения линейны и их можно проанализировать. В общем случае существенно
нелинейных моделей формализм для неканонических калибровок не развит и мы на них
останавливаться не будем.
\qed\end{com}
Таким образом, динамика частиц со связями I рода в канонических калибровках
сводится к гамильтоновым моделям со связями II рода. Это сведение не является
однозначным, т.к.\ от калибровочных условий требуется выполнение только
неравенств (\ref{edegfa}). Выбор той или иной системы калибровочных условий
диктуется рассматриваемой задачей и соображениями простоты. Как правило,
исследование калибровочных моделей проводится в различных калибровках, каждая из
которых имеет свои преимущества и недостатки.
\subsection{Калибровочная модель нерелятивистской частицы        \label{sgauno}}
В настоящем разделе мы рассмотрим динамику точечной частицы с точки зрения
калибровочных моделей и покажем трудности в определении энергии, которые при
этом возникают.

Рассмотрим движение точечной частицы в фазовом пространстве $\MT^*(\MR^\Sn)$ с
координатами $q^*=\lbrace q^{*a}\rbrace$ и $p^*=\lbrace p^*_a\rbrace$,
где $a=1,\dotsc,\Sn$. Пусть динамика частицы задана гамильтонианом
$H^*(q^*,p^*)$, не зависящим явно от времени. Действие такой частицы имеет
обычный вид
\begin{equation}                                                  \label{eorapo}
  S=\int_{t_1}^{t_2}\!\!\!dt(\dot q^{*a}p^*_a-H^*),
\end{equation}
где точка обозначает дифференцирование по времени $t$. Будем считать, что все
координаты имеют фиксированные значения на границе $t=t_{1,2}$. Для каждой
траектории частицы $q^*(t),p^*(t)$, которая определяется каноническими
уравнениями движения
\begin{equation*}
  \dot q^*=\frac{\pl H^*}{\pl p^*},\qquad \dot p^*=-\frac{\pl H^*}{\pl q^*},
\end{equation*}
выполнено равенство $\dot H^*=0$. Это значит, что на каждой траектории
гамильтониан имеет постоянное численное значение, которое называется энергией
точечной частицы.

Переформулируем модель точечной частицы, как калибровочную. С этой целью
расширим фазовое пространство $\MT^*(\MR^\Sn)\rightarrow\MT^*(\MR^{\Sn+1})$
путем введения дополнительной пары сопряженных канонических переменных $Q,P$, и
рассмотрим новое действие
\begin{equation}                                                  \label{epogaa}
  S_\St=\int_{\tau_1}^{\tau_2}\!\!\!d\tau(\dot QP+\dot q^*p_*-\lm G),\qquad
  G=-Q+H^*,
\end{equation}
где $\lm$ -- множитель Лагранжа, а точка обозначает дифференцирование по
некоторому параметру $\tau$, играющему роль времени. Уравнения движения для этой
модели имеют вид
\begin{align}                                                     \label{epoeqo}
  \dot Q&=0,
\\                                                                \label{epoeqt}
  \dot P&=\lm,
\\                                                                \label{epoeqh}
  \dot q^*&=\lm\frac{\pl H^*}{\pl p_*},
\\                                                                \label{epoeqf}
  \dot p_*&=-\lm\frac{\pl H^*}{\pl q^*},
\\                                                                \label{epoeqi}
  G&=0.
\end{align}
Последнее уравнение представляет собой уравнение связи. Эта связь, очевидно,
является связью первого рода, которая определяет гамильтониан системы.

Согласно общей теории действие (\ref{epogaa}) инвариантно относительно
инфинитезимальных калибровочных преобразований с параметром $\e(\tau)$,
генерируемых связью $G$,
\begin{equation}                                                  \label{egatrp}
\begin{split}
  \dl Q&=\e[Q,G]=0,
\\
  \dl P&=\e[P,G]=\e,
\\
  \dl q^*&=\e[q^*,G]=\e\frac{\pl H^*}{\pl p_*},
\\
  \dl p_*&=\e[p_*,G]=-\e\frac{\pl H^*}{\pl q^*}.
\end{split}
\end{equation}
При этом множитель Лагранжа преобразуется по правилу
\begin{equation}                                                  \label{elmtrp}
  \dl\lm=\dot\e.
\end{equation}
Для того, чтобы вариации $\dl q^*$ на границе были равны нулю, необходимо
предположить, что $\e(\tau_{1,2})=0$. Вариация множителя Лагранжа на границе
несущественна, т.к.\ он входит в действие без производных. Отметим, что эволюцию
во времени канонических переменных можно рассматривать, как калибровочное
преобразование с параметром $\e=\lm d\tau$.

Приведенные выше калибровочные преобразования представляют собой бесконечно
малые преобразования, соответствующие инвариантности действия (\ref{epogaa})
относительно перепараметризации времени. При произвольном преобразовании
временн\'ого параметра $\tau'=\tau'(\tau)$ мы постулируем, что координаты
фазового пространства преобразуются, как скаляры:
\begin{equation*}
  Q'(\tau')=Q(\tau),\qquad P'(\tau')=P(\tau)
\end{equation*}
(такие же формулы преобразования постулируются для $q^*$ и $p_*$). При этом
множитель Лагранжа преобразуется, как 1-форма:
\begin{equation*}
  \lm'(\tau')=\frac{d\tau}{d\tau'}\lm(\tau).
\end{equation*}
Нетрудно проверить, что действие (\ref{epogaa}) инвариантно относительно
произвольной перепараметризации времени $\tau$. Это и есть калибровочная
инвариантность. При бесконечно малом преобразовании $\tau'=\tau+u(\tau)$ для
вариаций формы функций имеем следующие формулы:
\begin{equation}                                                  \label{efofuq}
\begin{split}
  \dl Q&=-u\dot Q=0,
\\
  \dl P&=-u\dot P=-u\lm,
\\
  \dl q^*&=-u\dot q^*=-u\lm\frac{\pl H^*}{\pl p_*},
\\
  \dl p_*&=-u\dot p_*=u\lm\frac{\pl H^*}{\pl q^*},
\\
  \dl\lm&=-u\dot\lm-\dot u\lm=-\frac d{d\tau}(u\lm),
\end{split}
\end{equation}
где в первых четырех уравнениях были использованы уравнения движения. Полученные
преобразования совпадают с инфинитезимальными калибровочными преобразованиями
(\ref{egatrp}), (\ref{elmtrp}) при $\e=-u\lm$.

Таким образом, мы показали, что действие (\ref{epogaa}) инвариантно относительно
произвольной перепараметризации времени. В модели имеется одна связь первого
рода, и для фиксирования соответствующего произвола в решениях уравнений
движения необходимо наложить одно калибровочное условие. Поскольку множитель
Лагранжа $\lm$ является произвольной функцией времени, то из уравнения движения
(\ref{epoeqt}) следует, что импульс $P$ также произволен. Чтобы устранить
этот произвол, зафиксируем калибровку
\begin{equation*}
  F_1=P-\const=0.
\end{equation*}
Тогда из уравнений движения определяется множитель Лагранжа $\lm=0$. При этом
для действия получаем следующее выражение
\begin{equation*}
  \left.S_\St\right|_{F_1=0,\,G=0}=\int\!d\tau \dot q^*p^*.
\end{equation*}
Это значит, что в выбранной калибровке мы получили ``замороженную'' теорию,
в которой не происходит никакой эволюции. При этом вся эволюция заменяется
калибровочным преобразованием.

Можно рассмотреть класс калибровок, явно зависящих от времени. Пусть
калибровочное условие имеет вид
\begin{equation*}
  F_2=P-\tau=0.
\end{equation*}
В этом случае из уравнений движения следует $\lm=1$, и эффективный гамильтониан
для физических степеней свободы становится нетривиальным. Поскольку
\begin{equation*}
  \dot QP=\frac d{d\tau}(QP)-Q\dot P,
\end{equation*}
и
\begin{equation*}
  Q\dot P|_{F_2=0,G=0}=H^*(q^*,p^*),
\end{equation*}
то эффективное действие для физических степеней свободы равно
\begin{equation*}
  \left.S_\St\right|_{F_2=0,G=0}=\int\! d\tau(\dot q^*p^*-H^*),
\end{equation*}
что совпадает с исходным действием (\ref{eorapo}) для точечной частицы. Мы
видим, что в калибровке $F_2=0$ нетривиальный эффективный гамильтониан возникает
из кинетического слагаемого $\dot QP$ для нефизической степени свободы.

Можно рассмотреть более общий класс калибровок
\begin{equation*}
  F_3=P-f(\tau)=0,
\end{equation*}
где $f(\tau)$ -- произвольная функция времени с положительной производной,
$\dot f>0$. Для этой калибровки $\lm=\dot f$, и полное действие принимает вид
\begin{equation*}
  \left.S_\St\right|_{F_3=0,G=0}=\int\! d\tau(\dot q^*p_*-\dot fH^*).
\end{equation*}
После перепараметризации траектории $dt:=d\tau\dot f$, мы возвращаемся к
исходному действию для точечной частицы (\ref{eorapo}).

Таким образом, действие (\ref{epogaa}) калибровочно инвариантно и после
наложения калибровочного условия (из достаточно широкого класса калибровок)
эквивалентно обычному действию для точечной частицы. В рассмотренном примере
нефизическую степень свободы удалось в явном виде исключить из теории после
решения связи и калибровочного условия. В подавляющем большинстве моделей
математической физики это сделать не удается. Даже в электродинамике связи и
калибровочные условия нельзя решить в явном виде. Ситуация в моделях
Янга--Миллса и гравитации еще более сложная. Поэтому для проведения вычислений в
калибровочных моделях используют методы, учитывающие как физические, так и
нефизические степени свободы.

А теперь обратимся к вопросу об определении энергии в калибровочно инвариантных
теориях. Во многих моделях математической физики исходное действие в
гамильтоновой форме имеет вид (\ref{epogaa}). В таком виде гамильтониан системы
при выполнении уравнений движения тождественно равен нулю, и принимать его за
энергию системы не имеет никакого смысла. В рассмотренной модели за энергию
частицы естественно принять численное значение гамильтониана $H^*$ для
физических степеней свободы. Для того, чтобы его построить, исходя из действия
(\ref{epogaa}), необходимо сначала зафиксировать калибровку, зависящую явно от
времени, а затем решить уравнения движения для нефизических степеней свободы и
связь. При этом, выбирая различные функции времени в калибровочном условии
$F_3=0$, можно получить, что множитель Лагранжа и, следовательно, эффективный
гамильтониан будут явно зависеть от времени. Для простоты картины, следует
выбрать такую функцию времени, чтобы эта зависимость исчезла. Если это возможно,
то построенный таким образом гамильтониан следует принять за определение
энергии, а соответствующий ему временн\'ой параметр назвать временем $\tau=t$.
\subsection{Частица в псевдоримановом пространстве               \label{schaps}}
Рассмотрим точечную частицу постоянной массы $m>0$, которая движется в
произвольном псевдоримановом пространстве $(\MM,g)$ размерности $n$, на
котором задана достаточно гладкая метрика лоренцевой сигнатуры. Пусть
$x^\al$, $\al=0,1,\dotsc,n-1$, -- локальная система
координат в некоторой окрестности $\MU\subset\MM$. Будем считать, что координаты
выбраны таким образом, что $x^0$ является временн\'ой координатой, т.е.\
$g_{00}>0$, и все сечения $x^0=\const$ пространственноподобны, т.е.\
пространственная часть метрики $g_{\mu\nu}$, $\mu,\nu=1,\dotsc,n-1$ отрицательно
определена или $N^2>0$, где $N$ -- функция хода в АДМ параметризации метрики
(см.\ раздел \ref{sadmpa}).

Поскольку на многообразии задана метрика лоренцевой сигнатуры, то в каждой точке
заданы световые конусы прошлого и будущего. Будем считать, что на $\MM$ выбрана
ориентация во времени, т.е.\ световые конусы будущего непрерывно зависят от
точки многообразия.

Рассмотрим времениподобную кривую $\g=q(t)$, $t\in[t_0,t_1]\subset\MR$,
соединяющую две причинно связанные точки выбранной координатной окрестности
$q(t_0),q(t_1)\in\MU$ и целиком лежащую в $\MU$. Действие для точечной
частицы, по-определению, пропорционально длине траектории (\ref{eactim}) и имеет
вид
\begin{equation}                                                  \label{epopac}
  S=\int_\g\!dtL(q,\dot q):=-m\int_\g\!dt\sqrt{g_{\al\bt}\dot q^\al\dot q^\bt}.
\end{equation}
Поскольку траектория частицы предполагается времениподобной, т.е.\
\begin{equation*}
  \dot q^2:=g_{\al\bt}\dot q^\al\dot q^\bt>0,
\end{equation*}
то подынтегральное выражение определено. В рассматриваемом действии метрика
$g_{\al\bt}\big(q(t)\big)$ является внешним заданным полем и по ней варьирование
не производится пока не включено взаимодействие с гравитационным полем, т.е.\ не
добавлено, например, действие Гильберта--Эйнштейна.

Обозначения выбраны таким образом, чтобы производная, например, от метрики вдоль
траектории частицы записывалась в виде
\begin{equation*}
  \dot g_{\al\bt}:=\frac{dg_{\al\bt}}{dt}
  =\left.\dot q^\al\frac{\pl g_{\al\bt}}{\pl x^\al}\right|_{x=q},
\end{equation*}
что следует из правила дифференцирования сложной функции.

Мы рассмотрим случай как положительных функций хода $N>0$, так и отрицательных
$N<0$. Если в некоторой области функция хода меняет знак, то из непрерывности
следует, что она где то обращается в нуль. В таких точках метрика становится
вырожденной, и этот вопрос требует отдельного рассмотрения. Пока же предположим,
что в области $\MU$ функция хода либо положительна, либо отрицательна.

Мы рассматриваем оба возможных случая знака производных $\dot q^0>0$ и
$\dot q^0<0$. В дальнейшем мы увидим, что в случае $\dot q^0<0$ действие
(\ref{epopac}) описывает античастицу, т.е.\ частицу той же массы $m$, но
противоположного электрического заряда.

Следуя общим правилам, построим гамильтонов формализм для точечной массивной
частицы, описываемой действием (\ref{epopac}). Обобщенные импульсы,
сопряженные координатам точечной частицы $q^\al$ имеют вид
\begin{equation}                                                  \label{edempo}
  p_\al:=\frac{\pl L}{\pl\dot q^\al}
  =-m\frac{g_{\al\bt}\dot q^\bt}{\sqrt{\dot q^2}}.
\end{equation}
Отсюда следует, что импульсы удовлетворяют соотношению
\begin{equation}                                                  \label{efipoc}
  p^2:=g^{\al\bt}(q)p_\al p_\bt=m^2,
\end{equation}
которое должно быть выполнено для всех траекторий, вдоль которых частица может
двигаться. Поэтому соотношение
\begin{equation}                                                  \label{ecopap}
  \widetilde G:=p^2-m^2=0
\end{equation}
является первичной связью для точечной частицы.

\begin{exa}
В пространстве Минковского связь (\ref{ecopap}) зависит только от импульсов:
\begin{equation*}
  p_0^2+\eta^{\mu\nu}p_\mu p_\nu-m^2=0,
\end{equation*}
где $\eta^{\mu\nu}=\diag(-\dotsc-)$, и выделяет в фазовом пространстве
$(q,p)\in\MT^*(\MR^{1,n-1})$ двуполостный гиперболоид в кокасательном
пространстве, умноженный на пространство Минковского $\MR^{1,n-1}$, которое
соответствует координатам $q$. Поверхность связей является несвязным
подмногообразием в фазовом пространстве $\MT^*(\MR^{1,n-1})$ и состоит из двух
компонент связности, определяемых неравенствами $p_0>0$ и $p_0<0$. Топологически
поверхность связей в произвольном псевдоримановом многообразии устроена также.
\qed\end{exa}

Используя АДМ параметризацию метрики (см.\ раздел \ref{sadmpa}), связь
(\ref{ecopap}) перепишем в виде произведения двух сомножителей:
\begin{equation}                                                  \label{eproco}
  p^2-m^2=\left[\frac1N(p_0-N^\mu p_\mu)+\sqrt{\hat p^2+m^2}\right]
  \left[\frac1N(p_0-N^\mu p_\mu)-\sqrt{\hat p^2+m^2}\right]=0,
\end{equation}
где введено обозначение $\hat p^2:=-\hat g^{\mu\nu}p_\mu p_\nu>0$ для
положительно определенного квадрата пространственных компонент импульса частицы.
Напомним, что индексы из середины греческого алфавита пробегают только
пространственные значения: $\mu,\nu,\dotsc=1,\dotsc,n-1$ и $\hat g^{\mu\nu}$ --
матрица, обратная к $g_{\mu\nu}$. В дальнейшем для подъема пространственных
индексов всегда используется обратная метрика $\hat g^{\mu\nu}$.

Равенство нулю одного из сомножителей выделяет в фазовом пространстве одну полу
``гиперболоида''. То, на какой именно пол\'е ``гиперболоида'' находится частица,
определяется начальными данными. Если в начальный момент времени частица
находилась, скажем, на пол\'е, определяемой первым сомножителем в
(\ref{eproco}), то из непрерывности следует, что она на ней и останется в
процессе эволюции.

Из определения импульсов (\ref{edempo}) следует равенство
\begin{equation}                                                  \label{epropd}
  p_0-N^\mu p_\mu=-\frac{mN^2}{\sqrt{\dot q^2}}\dot q^0.
\end{equation}
Мы видим, что знак производной $\dot q^0$ всегда противоположен знаку функции
$p_0-N^\mu p_\mu$. Поэтому, если $\dot q^0N>0$, то в нуль обращается первый
сомножитель. В противном случае, $\dot q^0N<0$, равен нулю второй сомножитель.
Следовательно, связь (\ref{ecopap}) эквивалентна связи
\begin{equation}                                                  \label{efipra}
  G:=\sqrt{\hat p^2+m^2}-\left|\frac{p_0-N^\mu p_\mu}N\right|=0,
\end{equation}
где мы использовали знак модуля, чтобы объединить оба случая. В таком виде
первичная связь будет удобна для дальнейших вычислений.

Мы предполагаем, что если в начальный момент времени частица находилась на
какой то одной поле гиперболоида, то в процессе движения она на ней и останется.
В противном случае траектория в фазовом пространстве не будет непрерывной. Это
означает, что если в некоторой области пространства-времени функция хода меняет
знак, то с точки зрения внешнего наблюдателя частица будет восприниматься как
частица в области $N>0$ и античастица в области $N<0$.

Перейдем к вычислению гамильтониана $H:=p_\al\dot q^\al-L$. Поскольку в теории
есть первичная связь, то гессиан модели вырожден и из определения обобщенных
импульсов (\ref{edempo}) нельзя определить все скорости, как функции импульсов и
координат. Ранг гессиана в рассматриваемом случае равен $n-1$, что позволяет
определить пространственные компоненты скорости. Чтобы их найти, заметим, что
\begin{equation}                                                  \label{esqdoq}
  \dot q^2=N^2(\dot q^0)^2
  +g_{\mu\nu}(\dot q^\mu+N^\mu\dot q^0)(\dot q^\nu+N^\nu\dot q^0).
\end{equation}
Далее, из определения импульсов (\ref{edempo}) следует равенство
\begin{equation*}
  p_\mu\sqrt{\dot q^2}=-m(N_\mu\dot q^0+g_{\mu\nu}\dot q^\nu).
\end{equation*}
Возведение этого равенства в квадрат с помощью метрики $\hat g^{\mu\nu}$
позволяет найти квадрат пространственных компонент скорости:
\begin{equation*}
  g_{\mu\nu}(\dot q^\mu+N^\mu\dot q^0)(\dot q^\nu+N^\nu\dot q^0)
  =-\frac{N^2\hat p^2}{\hat p^2+m^2}(\dot q^0)^2.
\end{equation*}
Подстановка полученного выражения в формулу (\ref{esqdoq}) приводит к равенству
\begin{equation*}
  \sqrt{\dot q^2}=\frac{m|\dot q^0N|}{\sqrt{\hat p^2+m^2}},
\end{equation*}
которое позволяет решить выражение для импульсов (\ref{edempo}) относительно
компонент скорости:
\begin{align}                                                     \label{qlkugg}
  \frac{\dot q^0}{|\dot q^0|}&=-\frac{p_0-N^\nu p_\nu}{|N|\sqrt{\hat p^2+m^2}},
\\                                                                \label{qnfyrt}
  \frac{\dot q^\mu}{|\dot q^0|}&=-\frac{N^2\hat g^{\mu\nu}p_\nu
  -N^\mu(p_0-N^\nu p_\nu)}{|N|\sqrt{\hat p^2+m^2}},
\end{align}
Мы видим, что уравнения (\ref{edempo}) определяют только знак производной
$\dot q^0$ и пространственные компоненты скорости $\dot q^\mu$. При этом модуль
временн\'ой компоненты скорости $|\dot q^0|$ является произвольной функцией.
Позже мы увидим, что она соответствует перепараметризации мировой линии частицы.

Поскольку мы нашли все $n-1$ пространственные компоненты скоростей, то это
доказывает, что гессиан модели имеет ранг $n-1$, и других первичных связей в
теории нет.

Теперь нетрудно вычислить гамильтониан
\begin{multline}                                                  \label{einhan}
  H:=p\dot q-L=p_0\dot q^0+p_\mu\dot q^\mu+m\sqrt{\dot q^2}=
\\
  =\frac{|\dot q^0|}{|N|\sqrt{\hat p^2+m^2}}\left[p_0|N|\sqrt{\hat p^2+m^2}
  +N^\mu p_\mu(p_0-N^\nu p_\nu)+N^2(\hat p^2+m^2)\right]=
\\
  =\frac{|\dot q^0|}{|N|\sqrt{\hat p^2+m^2}}\left[-(p_0-N^\mu p_\mu)^2
  +N^2(\hat p^2+m^2)\right]
  =-\frac{|\dot q^0N|}{\sqrt{\hat p^2+m^2}}\widetilde G,
\end{multline}
где мы использовали связь (\ref{efipra}) во второй строке. Таким образом,
гамильтониан определен и пропорционален первичной связи $\widetilde G$. Он
определен неоднозначно, т.к.\ в теории есть связь.

Перепишем гамильтониан в более удобной форме. С этой целью в предпоследнем
выражении (\ref{einhan}) используем связь (\ref{efipra}):
\begin{equation}                                                  \label{qinham}
  H=\frac{|\dot q^0|}{|N|\sqrt{\hat p^2+m^2}}
  \left[\big|(p_0-N^\mu p_\mu)N\big|\sqrt{\hat p^2+m^2}+N^2(\hat p^2+m^2)\right]
  =|\dot q^0N|G.
\end{equation}
Теперь гамильтониан пропорционален связи в форме (\ref{efipra}).

Действие (\ref{epopac}) в гамильтоновой форме принимает вид
\begin{equation}                                                  \label{epohas}
  S=\int\!dt(p_\al\dot q^\al-H)=\int\!dt\left(p_\mu\dot q^\mu
  -|\dot q^0N|\sqrt{\hat p^2+m^2}+\dot q^0N^\mu p_\mu\right).
\end{equation}
Отметим сокращение кинетических слагаемых $p_0\dot q^0$. Как и исходное действие
полученное выражение не зависит от знака функции хода и параметризационно
инвариантно, если перепараметризация не меняет ориентацию кривой.

Действие (\ref{epohas}) приводит к следующим уравнениям движения для
пространственных координат и импульсов:
\begin{align}                                                     \label{qkjjuh}
  \dot q^\mu&=-\left.\frac{|\dot q_0N|}{\sqrt{\hat p^2+m^2}}\right|_{x=q}
  p^\mu-\dot q^0N^\mu\big|_{x=q},
\\                                                                \label{qlkiij}
  \dot p_\mu&=-\pl_\mu\left[|\dot q^0N|\sqrt{\hat p^2+m^2}
  -\dot q^0N^\nu p_\nu\right]_{x=q}=
\\
  &=-\left.|\dot q^0|\sqrt{\hat p^2+m^2}\,\pl_\mu |N|\right|_{x=q}
  -\left.\frac{|\dot q^0N|p^\nu p^\rho\hat\Gamma_{\mu\nu\rho}}
  {\sqrt{\hat p^2+m^2}}
  \right|_{x=q}+\dot q^0\pl_\mu N^\nu p_\nu\big|_{x=q},
\end{align}
где $\hat\Gamma_{\mu\nu\rho}$ -- символы Кристоффеля для пространственной
метрики $g_{\mu\nu}$ и $p^\mu:=\hat g^{\mu\nu}p_\nu$.
В таком виде эквивалентность гамильтоновых и лагранжевых уравнений движения
совсем не очевидна. В дальнейшем эквивалентность гамильтоновых и лагранжевых
уравнений движения будет доказана для уравнений, записанных в другой форме.

Поскольку исходное действие инвариантно относительно произвольной
перепараметризации мировой линии частицы (локальные преобразования), то
рассматриваемая модель является калибровочной. Этому обстоятельству
соответствует наличие одной связи первого рода (\ref{ecopap}). Поэтому частица
имеет физические и нефизические степени свободы. В качестве физических степеней
свободы, для которых можно поставить задачу Коши, выберем пространственные
компоненты координат и импульсов $q^\mu,p_\mu$, а нефизических степеней свободы
-- $q^0,p_0$. Посмотрим с этой точки зрения на действие (\ref{epohas}). Для
определенности предположим, что $\dot q_0N>0$. Тогда действие можно переписать
в виде
\begin{equation*}
  S=\int dt\left(p_\mu\dot g^\mu-\dot q^0H_{\rm eff}\right),
\end{equation*}
где
\begin{equation*}
  H_{\rm eff}:=N\sqrt{\hat p^2+m^2}-N^\mu p_\mu
\end{equation*}
-- эффективный гамильтониан для физических степеней свободы. При этом уравнения
движения примут вид
\begin{align*}
  \dot q^\mu&=~~\dot q_0\frac{\pl H_{\rm eff}}{\pl p_\mu},
\\
  \dot p_\mu&=-\dot q^0\frac{\pl H_{\rm eff}}{\pl q^\mu}.
\end{align*}
Легко проверить, что если выполнены уравнения движения, то энергия сохранятся во
времени,
\begin{equation*}
  E:=H_{\rm eff}=\const,
\end{equation*}
для произвольной функции $q^0(t)$. Мы видим, что произвольная функция $q^0(t)$
не определяется уравнениями движения и соответствует свободе в выборе параметра
вдоль мировой линии. От нее всегда можно избавиться, переопределив параметр, что
соответствует выбору калибровки. Таким образом нефизические степени свободы
убираются из модели путем решения связи (\ref{efipra}) относительно временн\'ой
компоненты импульса $p_0$ и наложения калибровочного условия на произвольную
функцию $q^0(t)$.

Заметим, что если проварьировать действие (\ref{epohas}) по $q_0$, то получим
закон сохранения энергии $\dot E=0$, т.е.\ никакой дополнительной информации.

Если предположить, что допускаются только те перепараметризации, которые
сохраняют ориентацию мировой линии, т.е.\ $\dot q^0(t)>0$, то отрицательной
функции хода будет соответствовать эффективный гамильтониан
\begin{equation*}
  H_{\rm eff}:=-N\sqrt{\hat p^2+m^2}-N^\mu p_\mu.
\end{equation*}
В любом случае эффективный гамильтониан для физических степеней свободы
массивной точечной частицы положительно определен при $N^\mu=0$.

В соответствии с общим методом, полный гамильтониан на первом этапе получается
путем добавления первичной связи. Так как исходный гамильтониан уже
пропорционален связи, то полный гамильтониан равен
\begin{equation}                                                  \label{etopop}
  H_\St=\lm G,
\end{equation}
где $\lm=\lm(t)$ -- неопределенный множитель Лагранжа.

Поскольку скобка Пуассона связи с собой равна нулю,
\begin{equation*}
  [G,G]=0
\end{equation*}
(она должна быть антисимметрична, что невозможно для одной связи),
то вторичных связей не возникает и связь $G=0$ является связью первого рода.
Поэтому система находится в инволюции, т.е.\ выполнены условия (\ref{einvco}),
(\ref{einvha}). Отсюда вытекает, что полный гамильтониан системы (\ref{etoham})
определяется единственной связью первого рода (\ref{etopop}).

Гамильтониан (\ref{etopop}) приводит к уравнениям движения, которые эквивалентны
уравнениям движения для гамильтониана
\begin{equation}                                                  \label{eesefo}
  \widetilde H_\St=\mu\widetilde G,
\end{equation}
где $\mu=\mu(t)$ -- множитель Лагранжа, на поверхности связей, т.к.\ связи
$\widetilde G$ и $G$ эквивалентны. Действительно, связи $G$ и $\widetilde G$
отличаются на отличный от нуля множитель (для каждой полы гиперболоида):
\begin{equation*}
  \widetilde G=fG,\qquad f(q,p)\ne0.
\end{equation*}
Поэтому
\begin{equation*}
  \dot q^\al=\mu[q^\al,\widetilde G]=\mu[q^\al,f]G+\mu f[q^\al,G].
\end{equation*}
На поверхности связи $G=0$, и поэтому первое слагаемое в правой части исчезает.
Такой же вид имеет уравнение для импульсов. Следовательно, замена связи в полном
гамильтониане $H_\St$ приводит к переопределению множителя Лагранжа:
$\lm=\mu f$.

В дальнейшем, из соображений удобства, мы будем выбирать тот или иной вид
полного гамильтониана.

Рассмотрим гамильтоновы уравнения движения (\ref{eplmmo}) для гамильтониана
(\ref{eesefo})
\begin{equation}                                                  \label{eqmopa}
\begin{split}
  \dot q^\al&=2\mu g^{\al\bt}p_\bt,
\\
  \dot p_\al&=\mu p^\bt p^\g\pl_\al g_{\bt\g}\big|_{x=q}.
\end{split}
\end{equation}
Для сравнения полученных уравнений движения с уравнениями для экстремалей
(\ref{eextre}), продифференцируем первое уравнение по времени и воспользуемся
вторым уравнением для исключения $\dot p$. В результате получим уравнение для
координат частицы
\begin{equation}                                                  \label{ecoorp}
  \ddot q^\al
  =\frac{\dot\mu}{\mu}\dot q^\al-\Gamma_{\bt\g}{}^\al\dot q^\bt\dot q^\g.
\end{equation}
С точностью до первого слагаемого в правой части оно совпадает с уравнением для
экстремалей (\ref{eextre}). Это слагаемое связано с произволом в выборе
параметризации мировой линии частицы. Действительно, после преобразования
$t\mapsto t'(t)$, где функция $t'(t)$ удовлетворяет дифференциальному уравнению
\begin{equation*}
  \frac{dt'}{dt}=\mu(t),\qquad \mu>0,
\end{equation*}
гамильтоновы уравнения движения будут иметь вид (\ref{ecoorp}), но с $\mu=1$. В
этом случае уравнения движения совпадут с уравнениями для экстремалей, и,
значит, параметр $t'$ является каноническим параметром вдоль экстремали.

Обратное утверждение также верно. Если выполнены уравнения для экстремалей
\begin{equation*}
  \ddot q^\al=-\Gamma_{\bt\g}{}^\al\dot q^\bt\dot q^\g,
\end{equation*}
то из них вытекают гамильтоновы уравнения (\ref{eqmopa}) при $\mu=1$. Таким
образом мы доказали эквивалентность гамильтоновых и лагранжевых уравнений
движения.

Поскольку модель содержит связь первого рода, то ей соответствует калибровочная
инвариантность полного действия
\begin{equation}                                                  \label{etoapo}
  S_\St=\int_\g\!dt(p_\al\dot q^\al-\mu\widetilde G).
\end{equation}
Чтобы найти соответствующие преобразования симметрии, рассмотрим
инфинитезимальные преобразования, которые генерируются связью (\ref{elotrl}):
\begin{align*}
  \dl q^\al&=\e[q^\al,\widetilde G]=2\e g^{\al\bt}p_\bt,
\\
  \dl p_\al&=\e[p_\al,\widetilde G]=\e p^\bt p^\g\pl_\al g_{\bt\g},
\\
  \dl\mu&=\dot\e.
\end{align*}
Сравнение этих преобразований с уравнениями движения (\ref{eqmopa}) показывает,
что эволюция во времени канонических переменных представляет собой
последовательность калибровочных преобразований, где $\e=\mu dt$. Эта
калибровочная симметрия описывает произвол в выборе параметра $t$ вдоль
траектории частицы. Уравнения движения (\ref{eqmopa}) вместе со связью
(\ref{efipra}) не определяют множитель Лагранжа $\mu$. Для его определения
необходимо зафиксировать калибровку.

Если учесть выражение для импульсов (\ref{edempo}), то кинетический член в
действии (\ref{etoapo}) примет вид
\begin{equation*}
  p_\al\dot q^\al=-m\sqrt{\dot q^2}.
\end{equation*}
После интегрирования кинетического члена получается исходное действие
(\ref{epopac}). Таким образом, на поверхности связей гамильтониан равен нулю, а
исходное действие определяется только кинетическим слагаемым.
\subsubsection{Временн\'ая калибровка}
Продолжим исследование модели в соответствии с общей схемой. Зафиксируем
{\em временн\'ую калибровку}
\index{Временн\'ая калибровка (time gauge)}%
\index{Калибровка временн\'ая (time gauge)}%
\begin{equation}                                                  \label{egapop}
  F_\St:=q^0-bt=0,\qquad b=\const\ne0.
\end{equation}
Модуль свободного параметра $b$ связан с выбором единицы измерения времени, что
не существенно. Поэтому, без ограничения общности, положим $|b|=1$, т.е.\
$b=1$, если время $q^0$ для внешнего наблюдателя увеличивается при увеличении
параметра $t$ вдоль мировой линии частицы, и $b=-1$, если при увеличении $t$
время $q^0$ уменьшается.

Скобка Пуассона этого калибровочного условия со связью равна
\begin{equation}                                                  \label{epospo}
  [F_\St,G]=-\frac1{|N|}\left[q^0,|p_0-N^\mu p_\mu|\right]=
  \begin{cases} \quad 1/|N|, & b>0, \\ -1/|N|, &b<0, \end{cases}
\end{equation}
где мы использовали равенство (\ref{epropd}) для определения знака выражения,
стоящего под модулем. Поскольку функция хода отлична от нуля, $N\ne0$, то скобка
Пуассона (\ref{epospo}) отлична от нуля во всем фазовом пространстве и, в
частности, на поверхности связи $G=0$. Следовательно, условие $q^0=bt$
определяет каноническую калибровку.

Расширенный гамильтониан с учетом связи и калибровочного условия имеет вид
\begin{equation*}
  H_\Se=\lm G+\pi F_\St,
\end{equation*}
где $\lm$ и $\pi$ -- множители Лагранжа. Из условий сохранения связи и
калибровочного условия во времени,
\begin{align*}
  \dot G&=\lm[G,G]+\pi[G,F_\St]=\pi[G,F_\St]\approx0,
\\
  \dot F_\St&=\frac{\pl F_\St}{\pl t}+\lm[F_\St,G]+\pi[F_\St,F_\St]
  =-b+\lm[F_\St,G]\approx0,
\end{align*}
находим множители Лагранжа:
\begin{equation*}
  \lm=|N|,\qquad \pi=0,
\end{equation*}
где мы объединили оба случая, $b>0$ и $b<0$, и учли равенство $|b|=1$. Таким
образом в данной канонической калибровке полный гамильтониан (\ref{etopop})
совпадает с исходным гамильтонианом (\ref{qinham}), т.к.\ $|\dot q^0|=|b|=1$.

Связь (\ref{efipra}) и калибровочное условие (\ref{egapop}) можно решить
относительно нефизических переменных $q^0$ и $p_0$:
\begin{align*}
  q^0&=bt,
\\
  p_0&=-b|N|\sqrt{\hat p^2+m^2}+N^\mu p_\mu.
\end{align*}

Так как решения уравнений Эйлера--Лагранжа можно подставлять в действие (см.\
раздел \ref{sredac}), то эффективное действие для физических переменных
$q^\mu,p_\mu$ принимает вид
\begin{equation*}
  S_\text{eff}=\int\!dt\left(p_0\dot q^0+p_\mu\dot q^\mu\right)
  =\int\!dt\left(p_\mu\dot q^\mu-H_\text{eff}\right),
\end{equation*}
где эффективный гамильтониан для физических степеней свободы имеет вид
\begin{equation}                                                  \label{efhapa}
  H_\text{eff}=|N|\sqrt{\hat p^2+m^2}-bN^\mu p_\mu.
\end{equation}
Как и в случае нерелятивистской точечной частицы эффективный гамильтониан для
физических степеней свободы полностью определяется кинетическим слагаемым
$p_0\dot q^0$ для нефизических степеней свободы. Эффективный гамильтониан
зависит только от физических степеней свободы, которыми являются
пространственные координаты точечной частицы и соответствующие импульсы,
$\lbrace q^\mu,p_\mu\rbrace$, $\mu=1,\dotsc,n-1$. Компоненты метрики $N,N^\mu$ и
$g_{\mu\nu}$ входят в гамильтониан в качестве внешних полей. Уравнения движения
для физических степеней свободы во временн\'ой калибровке (\ref{egapop}) имеют
вид
\begin{align}                                                     \label{eqcopa}
  \dot q^\mu&=\quad \frac{|N|p^\mu}{\sqrt{\hat p^2+m^2}}-bN^\mu,
\\                                                                \label{epcopa}
  \dot p_\mu&=-\sqrt{\hat p^2+m^2}\pl_\mu |N|
  -\frac{|N|\pl_\mu\hat g^{\nu\rho}p_\nu p_\rho}{2\sqrt{\hat p^2+m^2}}
  +b\pl_\mu N^\nu p_\nu,
\end{align}
где $p^\mu:=\hat g^{\mu\nu}p_\nu$.

Чтобы дать физическую интерпретацию двух возможных ориентаций мировой линии
$b=\pm1$ рассмотрим следующий
\begin{exa}
В пространстве Минковского $\MR^{1,n-1}$ компоненты метрики имеют вид
\begin{equation*}
  N=1,\qquad N^\mu=0,\qquad g_{\mu\nu}=\eta_{\mu\nu}=\diag(-\dotsc-),
\end{equation*}
и уравнения движения (\ref{eqcopa}), (\ref{epcopa}) существенно упрощаются:
\begin{align*}
  \dot q^\mu&=\frac{p^\mu}{\sqrt{\hat p^2+m^2}},
\\
  \dot p_\mu&=0.
\end{align*}
Эти уравнения имеют хорошо известный в специальной теории относительности вид.
Возводя первое уравнение в квадрат с помощью пространственной метрики
$\eta_{\mu\nu}$, получим выражение для пространственного импульса частицы через
ее скорость:
\begin{equation}                                                  \label{emopoi}
  p^\mu=\frac{m\dot q^\mu}{\sqrt{1-\Bu^2}},
\end{equation}
где мы ввели квадрат пространственной скорости частицы
$\Bu^2:=-\dot q^\mu\dot q^\nu \eta_{\mu\nu}>0$. Тогда уравнения движения для
свободной точечной частицы в пространстве Минковского во временн\'ой калибровке
сводятся просто к условию сохранения импульса: $p_\mu=\const$.

Для того, чтобы дать физическую интерпретацию двум возможным ориентациям
$\dot q^0=\pm1$ мировой линии частицы относительно временн\'ой координаты $q^0$
внешнего наблюдателя, рассмотрим взаимодействие частицы с внешним магнитным
полем в четырехмерном пространстве Минковского $\MR^{1,3}$. Такое взаимодействие
описывается дополнительным слагаемым в лагранжиане:
\begin{equation*}
  L\mapsto L+eA_\al\dot q^\al,
\end{equation*}
где $A_\al(x)$ -- четырехмерный потенциал электромагнитного поля и $e=\const$ --
заряд частицы. С точки зрения внешнего наблюдателя, для которого временем
является координата $q^0$, добавочное слагаемое в лагранжиане имеет вид
\begin{equation*}
  gA_\al\frac{dq^\al}{dq^0},
\end{equation*}
где $g:=be=\pm e$ -- наблюдаемый заряд частицы. Таким образом, для внешнего
наблюдателя частица имеет либо положительный, либо отрицательный заряд. В
остальном частицы совпадают. Такие частицы в физике принято называть частицей и
античастицей. Например, электрон и позитрон (они, правда, имеют спин 1/2, в то
время как в рассматриваемом случае спин частиц равен нулю).

Таким образом, свобода в выборе ориентации мировой линии частицы относительно
временн\'ой координаты внешнего наблюдателя соответствует двум возможным зарядам
частицы $\pm e$. Это означает, что исходное действие (\ref{epopac}) описывает
частицу и античастицу, у которых массы совпадают, а заряды при включении
внешнего электромагнитного поля имеют противоположный знак. Если заряд равен
нулю, $e=0$, то действие (\ref{epopac}) описывает одну нейтральную частицу.

Заметим, что и частица, и античастица движутся вперед по времени $x^0$ с точки
зрения внешнего наблюдателя. Однако собственное время античастицы $t$ движется
в обратную сторону.

Гамильтоновы уравнения движения для точечной заряженной частицы принимают
особенно простой вид при движении в постоянном магнитном поле. В этом случае
временн\'ая компонента потенциала равна нулю, а пространственные компоненты
зависят только от пространственных координат:
\begin{equation*}
  \lbrace A_\al\rbrace=\lbrace A_0=0,A_\mu(\Bx)\rbrace.
\end{equation*}
Тогда все изменения в гамильтоновом формализме сводятся к переопределению
пространственных компонент импульсов:
\begin{equation*}
  p_\mu\mapsto \tilde p_\mu:=\frac{\pl(L+eA_\nu\dot q^\nu)}{\pl\dot q^\mu}
  =p_\mu+eA_\mu.
\end{equation*}
В частности, гамильтониан для физических степеней свободы и уравнения движения
принимают вид
\begin{align*}
  H_\text{eff}&=\sqrt{\tilde p^2+m^2},
\\
  \dot q^\mu&=\frac{p^\mu+eA^\mu}{\sqrt{\tilde p^2+m^2}},
\\
  \dot p_\mu&=-e\frac{\pl_\mu A_\nu(p^\nu+eA^\nu)}{\sqrt{\tilde p^2+m^2}},
\end{align*}
где введено обозначение
\begin{equation*}
  \tilde p^2:=-g^{\mu\nu}(p_\mu+eA_\mu)(p_\nu+eA_\nu).
\end{equation*}
Как видим, при взаимодействии с внешним полем импульс больше не сохраняется, и
траектория частицы в общем случае отличается от экстремали.
\qed\end{exa}
Точечные частицы под действием гравитационного поля двигаются в
пространстве-времени вдоль экстремалей. В разделе \ref{sextre} было показано,
что через данную точку в данном направлении проходит
одна и только одна экстремаль. Это значит, что при постановке задачи Коши для
экстремали достаточно задать точку $x\in\MM$ и вектор $X\in\MT_x(\MM)$. Анализ
настоящего раздела показывает, что эта информация является избыточной.
Действительно, длина касательного вектора к экстремали постоянна вдоль
экстремали. Это значит, что для однозначного восстановления экстремали,
проходящей через данную точку, достаточно задать не сам вектор, а его
направление, которое определяется $n-1$ параметром. Кроме того, остается еще
произвол в выборе параметризации. Во временн\'ой калибровке в качестве параметра
выбирается наблюдаемое время $q^0=\pm t$. Следовательно, для задания траектории
частицы достаточно задать пространственные координаты $q^\mu$ и пространственные
компоненты импульсов $p_\mu$ в начальный момент времени $q^0$. Поэтому точечная
частица на псевдоримановом многообразии описывает $n-1$ степень свободы.
\begin{exa}
В четырехмерном пространстве-времени Минковского $\MR^{1,3}$ точечная массивная
частица имеет три степени свободы.
\qed\end{exa}
\subsubsection{Калибровка светового конуса}
Для многих приложений, например, в суперсимметричных моделях, удобно
использовать калибровку светового конуса. Эта калибровка упрощает многие
формулы, если частица движется в пространстве Минковского $\MR^{1,n-1}$, что мы
и предположим.

Чтобы определить калибровку светового конуса, вместо двух первых координат
частицы $q^0$ и $q^1$ введем новые конусные переменные
\begin{equation*}
  q^\pm:=\frac1{\sqrt2}(q^0\pm q^1),
\end{equation*}
оставив остальные координаты без изменения. Поскольку известно выражение новых
координат через старые, то совершим каноническое преобразование в фазовом
пространстве с производящей функцией
\begin{equation*}
  S_3=-\frac1{\sqrt2}(p_0+p_1)q^+-\frac1{\sqrt2}(p_0-p_1)q^--p_2q^2-\dotsc
  -p_{n-1}q^{n-1},
\end{equation*}
зависящей от новых координат и старых импульсов (см.\ раздел \ref{sgefca}).
Отсюда следуют выражения для новых импульсов, сопряженных $q^\pm$:
\begin{equation*}
  p_\pm=-\frac{\pl S_3}{\pl q^\pm}=\frac1{\sqrt2}(p_0\pm p_1).
\end{equation*}
Остальные импульсы при этом не меняются.

В новых канонических переменных первичная связь (\ref{ecopap}) принимает вид
\begin{equation}                                                  \label{elicap}
  \widetilde G=2p_+p_-+p^\Sa p_\Sa-m^2=0,
\end{equation}
где суммирование ведется только по $n-2$ значениям индексов: $\Sa=2,\dotsc,n-1$.
Поскольку квадратичная форма $p^\Sa p_\Sa-m^2$ отрицательно определена, то
$p_+p_->0$. Полы гиперболоида, соответствующего поверхности связей, определяются
неравенствами ($p_+>0$, $p_->0$) и ($p_+<0$, $p_-<0$). И в любом случае
$p_+p_-\ne0$. Связь (\ref{elicap}) просто решается
\begin{align}                                                     \label{esolcg}
  p_+&=\frac1{2p_-}(-p^\Sa p_\Sa+m^2),
\\ \intertext{или}                                                     \nonumber
  p_-&=\frac1{2p_+}(-p^\Sa p_\Sa+m^2).
\end{align}

Зафиксируем калибровку {\em светового конуса}
\begin{equation}                                                  \label{elicog}
  F_{\Sl\Sc}:=q^+-at=0,\qquad a=\pm1.
\end{equation}
\index{Калибровка светового конуса (light cone gauge)}%
\index{Светового конуса калибровка (light cone gauge)}%
Легко видеть, что
\begin{equation*}
  [\widetilde G,F_{\Sl\Sc}]=-2p_-.
\end{equation*}
Поскольку на поверхности связей $p_-\ne0$, то условие (\ref{elicog}) определяет
каноническую калибровку.

Так как связи $\widetilde G$ и $G$ пропорциональны, то обобщенный гамильтониан
можно записать в виде
\begin{equation*}
  H_\Se=\lm\widetilde G+\pi F_{\Sl\Sc},
\end{equation*}
где $\lm$ и $\pi$ -- множители Лагранжа. Из условий сохранения связи и
калибровочного условия во времени,
\begin{align*}
  \dot{\widetilde G}&=\lm[\widetilde G,\widetilde G]
  +\pi[\widetilde G,F_{\Sl\Sc}]=\pi[\widetilde G,F_{\Sl\Sc}]\approx0,
\\
  \dot F_{\Sl\Sc}&=\frac{\pl F_{\Sl\Sc}}{\pl t}+\lm[F_{\Sl\Sc},\widetilde G]
  +\pi[F_{\Sl\Sc},F_{\Sl\Sc}]=-a+\lm[F_{\Sl\Sc},\widetilde G]\approx0,
\end{align*}
находим множители Лагранжа:
\begin{equation*}
  \lm=\frac a{2p_-},\qquad \pi=0.
\end{equation*}

В калибровке светового конуса физическими переменными являются
$\lbrace q^-,q^\Sa,p_-,p_\Sa\rbrace$, $\Sa=2,\dotsc,n-1$. После подстановки
решения уравнения связи (\ref{esolcg}) и калибровочного условия (\ref{elicog}) в
действие,
\begin{equation*}
  S_\text{eff}=\int dt(p_+\dot q^++p_-\dot q^-+p_\Sa\dot q^\Sa)
  =\int dt(p_-\dot q^-+p_\Sa\dot q^\Sa-H_\text{eff}),
\end{equation*}
получаем выражение для эффективного гамильтониана
\begin{equation}                                                  \label{eefhli}
  H_\text{eff}=a\frac{p^\Sa p_\Sa-m^2}{2p_-}.
\end{equation}
Как и раньше, нетривиальный эффективный гамильтониан для физических степеней
свободы возникает из кинетического слагаемого $p_+\dot q^+$ для нефизической
степени свободы.

Уравнения движения для физических степеней свободы имеют вид
\begin{align*}
  \dot q^-&=-a\frac{p^\Sa p_\Sa-m^2}{2p_-^2},&
  \dot q^\Sa&=a\frac{p^\Sa}{p_-},
\\
  \dot p_-&=0,&
  \dot p_\Sa&=0.
\end{align*}
Мы видим, что как и во временн\'ой калибровке уравнения движения в калибровке
светового конуса сводятся к сохранению обобщенных импульсов свободной точечной
частицы.
\begin{com}
В калибровке светового конуса параметр эволюции $t\in\MR$ целиком лежит на
световом конусе в пространстве Минковского, который является характеристикой
волнового уравнения (см.\ раздел \ref{swaequ}).
\qed\end{com}
Эффективный гамильтониан (\ref{eefhli}) в калибровке светового конуса мало чем
напоминает эффективный гамильтониан (\ref{efhapa}) во временн\'ой калибровке.
Тем на менее оба гамильтониана описывают одну и ту же массивную частицу. Ее
траектории в конфигурационном пространстве -- это экстремали. Выбор той или иной
калибровки является существенным для анализа уравнений движения и квантования.
Часто калибровка светового конуса упрощает вычисления, особенно в квантовой
теории поля.
\subsection{Граничные слагаемые в калибровочных моделях          \label{sexava}}
В настоящем разделе мы рассмотрим простой пример вариационной задачи, который
проанализируем с различных точек зрения. Этот пример позволяет
продемонстрировать тонкости вариационной задачи в теории поля, важную роль
граничных слагаемых в действии и связь вариационной задачи на условную
стационарную точку с фиксированием калибровки в калибровочных моделях.
\subsubsection{Лагранжева формулировка}
Обозначим декартовы координаты на евклидовой плоскости $\MR^2$ через $x^0=\tau$,
$x^1=\s$. Будем называть координату $\tau$ временем, а $\s$ -- пространством,
хотя мы не предполагаем наличие на $\MR^2$ какой либо метрики. Пусть в конечном
прямоугольнике задано действие
\begin{equation}                                                  \label{efiact}
  S=\int_{\tau_1}^{\tau_2}\!\!\!d\tau\int_{\s_1}^{\s_2}\!\!\!d\s
  (p\dot q+p^*\dot q^*),
\end{equation}
зависящее от четырех полей $q,p,q^*,p^*\in\CC^1(\MR^2)$. В действии точка
обозначает дифференцирование по $\tau$. В дальнейшем пределы интегрирования, для
краткости, будем опускать. Рассмотрим задачу на условный экстремум для действия
(\ref{efiact}). Пусть на поля наложены две связи:
\begin{align}                                                     \label{econso}
  G&:=-\pl_1q+H^*(q^*,p^*)=0,
\\                                                                \label{econst}
  F&:=p=0.
\end{align}
Будем считать, что функция $H^*(q^*,p^*)\ge0$ зависит только от полей и не
зависит от их производных. Предположим также, что поля $q$ и $q^*$ имеют
определенные граничные условия при $\tau=\tau_{1,2}$. Этого достаточно для того,
чтобы избежать граничных вкладов в вариацию действия (\ref{efiact}), возникающих
при интегрировании по частям. Тогда для действия (\ref{efiact}) определена
задача на условный экстремум (см.\ раздел \ref{solusp}).

Решим эту задачу прямым методом и методом множителей Лагранжа. В первом
случае исключим из действия переменные $q$ и $p$ с помощью уравнений связей.
Поскольку на поверхности связей $p=0$, то, независимо от вида функции
$q(\tau,\s)$, первое слагаемое в (\ref{efiact}) обращается в нуль, и мы получаем
эффективное действие
\begin{equation}                                                  \label{eacfgz}
  \left.S\vphantom{\sum}\right|_{G=0,F=0}=\int\!d\tau d\s p^*\dot q^*,
\end{equation}
в котором переменные $q^*$ и $p^*$ уже рассматриваются, как независимые
переменные. Поскольку вариация $\dl q^*$ равна нулю на границе
$\tau=\tau_{1,2}$, то из вариационного принципа следуют только уравнения
Эйлера--Лагранжа, которые просто интегрируются:
\begin{align}                                                     \label{efirqe}
  \frac{\dl S}{\dl p^*}&=\quad \dot q^*=0,& &\Rightarrow & q^*&=q^*(\s),
\\                                                                \label{efirpe}
  \frac{\dl S}{\dl q^*}&=-\dot p^*=0,& &\Rightarrow & p^*&=p^*(\s).
\end{align}
Мы видим, что решением уравнений Эйлера--Лагранжа являются произвольные функции
от $\s$. Вид произвольной функции $q^*$ находится из граничных условий, которые
должны быть заданы одинаковыми при $\tau=\tau_{1,2}$. Затем можно определить
$q$ из уравнения связи (\ref{econso}). Таким образом, задача на условный
экстремум имеет решение, хотя и не для очень широкого класса граничных условий.

Поскольку с помощью уравнений связей мы исключили переменные $q$ и $p$, то будем
называть их нефизическими, а переменные $q^*$ и $p^*$ -- физическими.

Теперь воспользуемся методом неопределенных множителей Лагранжа. Построим
расширенное действие
\begin{equation}                                                  \label{estiln}
  S_\Se=\int\!d\tau d\s(p\dot q+p^*\dot q^*-\lm G-\mu F),
\end{equation}
где $\lm,\mu\in\CC^1(\MR^2)$ -- множители Лагранжа. Для этого действия
нефизическими полями являются $q,p,\lm$ и $\mu$. Полная система уравнений
Эйлера--Лагранжа имеет вид
\begin{align}                                                     \label{exelon}
  \frac{\dl S_\Se}{\dl p}&=\quad \dot q-\mu=0,
\\                                                                \label{exeltw}
  \frac{\dl S_\Se}{\dl q}&=-\dot p-\pl_1\lm=0,
\\                                                                \label{exelth}
  \frac{\dl S_\Se}{\dl p^*}&=\quad \dot q^*-\lm\frac{\pl H^*}{\pl p^*}=0,
\\                                                                \label{exelfo}
  \frac{\dl S_\Se}{\dl q^*}&=-\dot p^*-\lm\frac{\pl H^*}{\pl q^*}=0,
\\                                                                \label{exelfi}
  \frac{\dl S_\Se}{\dl\lm}&=-G=0,
\\                                                                \label{exelsi}
  \frac{\dl S_\Se}{\dl\mu}&=-F=0.
\end{align}
При вариации слагаемого $-\lm\pl_1q$ по $q$ возникает также граничное условие на
множитель Лагранжа
\begin{equation}                                                  \label{ebolam}
  \lm|_{\s_{1,2}}=0,
\end{equation}
поскольку значения переменной $q$ фиксированы только на пространственноподобной
границе $\tau=\tau_{1,2}$ и, следовательно, вариации $\dl q$ на времениподобной
границе $\s=\s_{1,2}$ произвольны.

Перейдем к анализу уравнений Эйлера--Лагранжа. Решение последней связи
(\ref{exelsi}) тривиально. Решение связи (\ref{exelfi}) имеет вид
\begin{equation}                                                  \label{ecjsoq}
  q=\int_{\s_1}^\s\!\!\!d\s'H^*+q_0(\tau),
\end{equation}
где $q_0$ -- произвольная функция $\tau$. Отсюда следует, что задание граничных
условий $q|_{\s_2}$, будет противоречить уравнению связи (\ref{exelfi}), т.к.\
значение поля $q$ при $\s=\s_2$ определяется значениями физических полей $q^*$ и
$p^*$ во внутренних точках области. Дифференцируя решение (\ref{ecjsoq}) по
$\tau$ и используя уравнения (\ref{exelth}) и (\ref{exelfo}), получим
$\dot q=\dot q_0$. Затем решаем уравнения (\ref{exelon}) и (\ref{exeltw})
относительно множителей Лагранжа:
\begin{equation}                                                  \label{elamue}
  \mu=\dot q_0,\qquad \lm=\lm_0(\tau),
\end{equation}
где $\lm_0(\tau)$ -- произвольная функция. Таким образом, мы решили уравнения
движения для нефизических переменных $q,p$ и множителей Лагранжа $\lm,\mu$, и
это решение зависит от двух произвольных функций $q_0(\tau)$ и $\lm_0(\tau)$.
Для физических переменных $q^*$ и $p^*$ остаются уравнения (\ref{exelth}),
(\ref{exelfo}), которые имеют вид обычной гамильтоновой системы. Эта система
уравнений действительно воспроизводит уравнения Эйлера--Лагранжа на условный
экстремум (\ref{efirqe}), (\ref{efirpe}) при $\lm_0=0$. Заметим, что только это
значение согласуется с граничным условием (\ref{ebolam}).

Таким образом, мы решили вариационную задачу для действия (\ref{efiact}) с
заданными граничными условиями для полей $q$ и $q^*$ на границе
$\tau=\tau_{1,2}$ прямым способом и методом множителей Лагранжа. Как и следовало
ожидать, результат одинаков, а класс решений очень беден.

Однако для расширенного действия (\ref{estiln}) можно поставить более
содержательную вариационную задачу. Предположим, что нефизическая переменная
задана на всей границе $q|_{\s_{1,2}}$ и $q|_{\tau_{1,2}}$, а физическое поле --
только на пространственноподобной границе $q^*|_{\tau_{1,2}}$. В этом случае
вариации $\dl q$ равны нулю на границе и граничного условия на множитель
Лагранжа (\ref{ebolam}) не возникнет. Тогда, при $\lm_0\ne0$, вместо $\tau$
можно ввести новую переменную $t$, определяемую дифференциальным уравнением
$$
  \frac{dt}{d\tau}=\lm_0(\tau).
$$
Это уравнение определяет координату $t$ с точностью до сдвига на постоянную
величину, что несущественно. Тогда уравнения для физических полей
(\ref{exelth}), (\ref{exelfo}) примут вид
\begin{equation}                                                  \label{eqphva}
\begin{split}
  \frac{dq^*}{dt}&=\quad \frac{\pl H^*}{\pl p^*},
\\
  \frac{dp^*}{dt}&=-\frac{\pl H^*}{\pl q^*}.
\end{split}
\end{equation}
Таким образом, мы получили систему гамильтоновых уравнений движения для
физических полей, динамика которых определяется гамильтонианом $H^*(q^*,p^*)$.
Это показывает, что модель, основанная на действии $S_\Se$ с множителями
Лагранжа допускает постановку более широкого класса вариационных задач, чем
исходная задача на условную стационарную точку, и является более содержательной.

Здесь выявляется специфика полевых моделей, поскольку важно, что поля зависят не
только от времени $\tau$, но и от пространственной координаты $\s$.
Действительно, если бы связь имела вид $-q+H^*=0$, то уравнение (\ref{exeltw})
приняло бы вид $-\dot p+\lm_0=0$. Откуда следовало бы единственное решение
$\lm_0=0$ при $p=0$. (Мы употребили термины время и пространство, исходя из
аналогии с теорией относительности, не смотря на то, что на плоскости $\MR^2$
никакой метрики не задано.)

Заметим, что подстановка решения связей в расширенное действие $S_\Se$ снова
приводит к тривиальному действию (\ref{eacfgz}), которое не воспроизводит
уравнения Эйлера--Лагранжа (\ref{eqphva}). Это показывает, что подстановки
решения части уравнений Эйлера--Лагранжа в действие и в оставшиеся уравнения в
общем случае не эквивалентны. Это связано с тем, что в общем случае мы не можем
накладывать граничные условия на нефизические поля произвольным образом и
рассматривать исчезающие на границе вариации. Действительно, вариация $\dl q$
определяется физическими полями и их вариациями во внутренних точках области:
\begin{equation}                                                  \label{evaunq}
  \dl q=\int_{\s_1}^\s\!\!\!d\s'\left(\frac{\pl H^*}{\pl q^*}\dl q^*
  +\frac{\pl H^*}{\pl p^*}\dl p^*\right),
\end{equation}
и в общем случае не равна нулю на границе $\s=\s_2$. То есть задание граничного
условия $q|_{\s_2}$ противоречит уравнению связи (\ref{exelfi}).

Изменим постановку вариационной задачи таким образом, чтобы уравнения
Эйлера--Лагранжа остались прежними, а исключение нефизических полей в действии
приводило бы к новому действию, воспроизводящему уравнения (\ref{exelth}),
(\ref{exelfo}). Введем новое действие
\begin{equation}                                                  \label{ephace}
  S_{\rm ph}
  :=S_\Se-\int_{\tau_1}^{\tau_2}\!\!\!d\tau(\lm q)\big|_{\s=\s_1}^{\s_2},
\end{equation}
которое отличается от расширенного действия с множителями Лагранжа граничным
слагаемым. Как и раньше, мы считаем, что на границе $\tau_{1,2}$ заданы значения
переменных $q$ и $q^*$. Добавление граничного члена не меняет уравнений
Эйлера--Лагранжа, но меняет граничные условия. Этот граничный член подобран
таким образом, чтобы компенсировать граничный вклад в вариацию действия,
обусловленный вариацией $\dl q$. Его необходимо добавить, если мы не хотим
получить граничное условие $\lm|_{\s_2}=0$ при произвольной вариации $\dl q$ на
границе. Теперь нетрудно проверить, что на решениях уравнений (\ref{exeltw}) и
(\ref{exelfi}) действие принимает вид
\begin{equation}                                                  \label{ephacf}
  \left.S_{\rm ph}\vphantom{\sum}\right|_{G=0,F=0}
  =\int_{\tau_1}^{\tau_2}\!\!\!d\tau\int_{\s_1}^{\s_2}\!\!\!d\s p^*\dot q^*
  -\int_{\tau_1}^{\tau_2}\!\!\!d\tau\lm(\tau,\s_2) q(\s_2)
  =\int_{\tau_1}^{\tau_2}\!\!\!d\tau\int_{\s_1}^{\s_2}\!\!\!d\s
  (p^*\dot q^*-\lm_0H^*),
\end{equation}
где мы положили $\lm_0(\tau):=\lm(\tau,\s_2)$. Это действие воспроизводит
гамильтоновы уравнения движения для физических полей. При этом мы отбросили
слагаемое с $q_0(\tau)$, которое не влияет на уравнения Эйлера--Лагранжа, т.к.\
в этом действии функция $\lm_0(\tau)$ рассматривается как заданная и не
варьируется. Таким образом, добавление граничного слагаемого не меняет уравнений
Эйлера--Лагранжа, позволяет избежать граничных условий на множители Лагранжа и,
что самое важное, разрешает подстановку решений уравнений Эйлера--Лагранжа
непосредственно в действие. При этом возникает нетривиальный гамильтониан для
физических полей.

В предыдущем построении была выделена роль точки $\s_2$ в определении функции
$\lm_0(\tau)$. Это не существенно и связано с выбором начальной точки в решении
уравнения связи (\ref{ecjsoq}). Замена $\s_1\mapsto\s_2$ в нижнем пределе
этого интеграла приведет к переопределению $\lm_0(\tau)=\lm(\tau,\s_1)$.

В принципе, можно было бы ограничиться случаем $\lm=0$ и не добавлять граничный
член. Однако исходное действие в калибровочных моделях и в моделях, инвариантных
относительно общих преобразований координат, в канонической формулировке имеют
вид расширенного действия $S_\Se$, содержащего множители Лагранжа. При этом
многие решения, важные с физической точки зрения, соответствуют $\lm\ne0$.
Например, решение Шварцшильда соответствует нетривиальному множителю Лагранжа,
роль которого играет функция хода $N$.
\subsubsection{Гамильтонова формулировка}
Обозначения в предыдущем разделе были выбраны не случайно. По сути дела модель
(\ref{estiln}) уже записана в гамильтоновой форме, при этом переменные $p$ и
$p^*$ являются импульсами, сопряженными координатам $q$ и $q^*$. Нашей исходной
точкой будет полное действие
\begin{equation}                                                  \label{eacexa}
  S_\St=\int\!\!d\tau d\s
  (p\dot q+p^*\dot q^*-\lm G),
\end{equation}
которое получится из действия (\ref{estiln}), если положить $\mu=0$. Координаты
$\tau$ и $\s$ будем считать временн\'ой и пространственной соответственно.
В рассматриваемом случае гамильтониан системы задан единственной связью
\begin{equation}                                                  \label{einzha}
  H=\int\!\!d\s \lm G,\qquad G:=-\pl_1q+H^*(q^*,p^*).
\end{equation}
Будем считать, что интегрирование в (\ref{eacexa}) проводится по всей плоскости
$\tau,\s$. При этом все возникающие интегралы предполагаются сходящимися.

Таким образом, модель описывается двумя парами канонически сопряженных
переменных $q,p$ и $q^*,p^*$, на которые наложена одна связь $G=0$. Уравнения
движения имеют прежний вид (\ref{exelon})--(\ref{exelfi}) (при $\mu=0$), где
точка обозначает дифференцирование по времени. Нетрудно проверить, что связь $G$
является связью первого рода:
$$
  [ G,G']=0,
$$
где штрих обозначает, что соответствующие полевые переменные рассматриваются в
точке $\s'$. Поэтому на поверхности связей
$$
  \dot G=[ G,H]\approx0,
$$
и никаких дополнительных связей в модели не возникает. Наличие связи первого
рода означает, что действие (\ref{eacexa}) калибровочно инвариантно. Генератором
калибровочных преобразований для канонических переменных является функционал
$$
  T=\int\!d\s \e G,
$$
где $\e(\tau,\s)$ -- малый параметр локальных преобразований. Бесконечно малые
преобразования имеют вид
\begin{equation}                                                  \label{eingex}
\begin{split}
  \dl q&=[ q,T]=0,
\\
  \dl p&=[ p,T]=-\pl_1\e,
\\
  \dl q^*&=[ q^*,T]=\quad \e\frac{\pl H^*}{\pl p^*},
\\
  \dl p^*&=[ p^*,T]=-\e\frac{\pl H^*}{\pl q^*}.
\end{split}
\end{equation}
Если дополнить эти преобразования преобразованием множителя Лагранжа
$$
  \dl\lm=\dot\e,
$$
то, как нетрудно убедиться с помощью прямой подстановки, действие (\ref{eacexa})
является калибровочно инвариантным (см.\ раздел \ref{sficon}). Отметим, что
эволюция во времени канонических переменных (\ref{exelon})--(\ref{exelfo})
является в рассматриваемом случае калибровочным преобразованием с параметром
$\e=\lm d\tau$.

Согласно второй теореме Нетер калибровочная инвариантность приводит к
зависимости уравнений движения:
$$
  \pl_1\left(\frac{\dl S_\St}{\dl p}\right)
  +\frac{\dl S_\St}{\dl q^*}\frac{\pl H^*}{\pl p^*}
  -\frac{\dl S_\St}{\dl p^*}\frac{\pl H^*}{\pl q^*}
  -\pl_\tau\frac{\dl S_\St}{\dl\lm}=0.
$$

С математической точки зрения наличие калибровочной инвариантности отражается в
том, что решение уравнений движения зависит от произвольной функции $\lm$,
которая ничем не фиксирована. Основным предположением в моделях с калибровочной
симметрией является утверждение о том, что все физические наблюдаемые
калибровочно инвариантны. В данном случае это означает, что наблюдаемые функции
от канонических переменных не зависят от $\lm$. Чтобы исключить произвол в
решениях уравнений движения и исключить нефизические переменные необходимо
наложить калибровочное условие. Согласно канонической процедуре фиксирования
калибровки мы должны наложить одно калибровочное условие по числу связей первого
рода. Выберем его в виде
\begin{equation}                                                  \label{eqacex}
  F=p-p_0(\s)=0,
\end{equation}
где $p_0(\s)$ -- произвольная функция только от $\s$. Скобка Пуассона связи с
калибровочным условием имеет вид
$$
  [G,F']=\dl'(\s'-\s),
$$
где $\dl'$ обозначает производную от $\dl$-функции. Поскольку скобка Пуассона
связи с калибровочным условием не обращается в нуль при выполнении связи, то
вместе они не представляют собой систему связей первого рода. В то же время
пара функций $G,F$ не представляет собой также и систему связей второго рода,
поскольку $\det[ G,F']=0$, т.к.\ у $\dl'$ нетривиально ядро, состоящее из
констант.

Канонически сопряженные переменные $q,p$ являются нефизическими переменными, и
могут быть исключены из рассмотрения. С этой целью решим их уравнения движения и
связь, как это было сделано в предыдущем разделе. В результате получим, что
модель описывает одну физическую пару канонически сопряженных полей $q^*,p^*$ с
эффективным гамильтонианом $\lm_0 H^*(q^*,p^*)$. При этом подстановка связи и
калибровочного условия в действие (\ref{eacexa}) приводит к неверному
результату, который не воспроизводит уравнения движения для физических полей
$q^*$ и $p^*$. Причина этого и решение проблемы то же, что и в предыдущем
разделе -- к действию необходимо добавить граничный член (\ref{ephace}).
Заметим, что при выполнении уравнения связи граничный член в исходном действии
превращается в интеграл по пространству от некоторой гамильтоновой плотности для
физических переменных.

Исключение нефизических полей из уравнений движения не зависело от глобальной
структуры пространства-времени, т.к.\ при этом решаются только уравнения
движения, связи и калибровочные условия, которые локальны. Нетривиальный
эффективный гамильтониан из ``нулевого'' исходного гамильтониана (\ref{einzha})
для замкнутых многообразий можно получить следующим образом. Предположим, что
пространство-время является прямым произведением $\MR\times\MS^1$, где первый
сомножитель соответствует времени, а второй -- пространству. Поскольку
окружность $\MS^1$ -- компактное многообразие без края, то, казалось бы,
никаких граничных вкладов не возникает, и можно свободно интегрировать по
частям. Однако при внимательном рассмотрении оказывается, что при таком подходе
можно потерять много решений уравнений движения, представляющих физический
интерес. Опишем это подробнее.

Пусть пространственная координата $\s\in[0,2\pi]$ параметризует окружность
$\MS^1$. Поскольку физические поля $q^*$ и $p^*$ ничем не ограничены, то их
можно считать достаточно гладкими функциями на окружности. В то же время
нефизические поля должны удовлетворять уравнению связи (\ref{econso}), которое в
общем случае не имеет непрерывных решений (\ref{ecjsoq}) на окружности:
\begin{equation*}
  q(0)=0,\qquad q(2\pi)=\int_0^{2\pi}\!\!\!d\s H^*\ne0.
\end{equation*}
Кроме того, вариация нефизического поля (\ref{evaunq}) в общем случае не может
быть определена как непрерывная функция на окружности. Это значит, что при
постановке вариационной задачи необходимо сделать разрез и добавить к действию
(\ref{eacexa}) граничный член, который и приведет к нетривиальному эффективному
действию для физических переменных. Поэтому, если пространство представляет
собой окружность, то в физическом действии и эффективном гамильтониане
достаточно просто изменить пределы интегрирования по $\s$.

Покажем, что нетривиальный эффективный гамильтониан возникает также в более
общих калибровках, зависящих от времени явно. Пусть калибровочное условие имеет
вид
$$
  F=p-p_0(\tau,\s)=0,
$$
где функция $p_0(\tau,\s)$ задана. В этом случае действие на поверхности связей
получит дополнительный вклад за счет слагаемого $p\dot q$. Нетрудно проверить,
что дополнительный вклад не меняет окончательного ответа. Действительно,
$$
  \triangle S\Big|_{F=0,G=0}=\int\!\!d\tau d\s\,p\dot q
  =\int\!\!d\tau d\s\,(-\dot p q),
$$
поскольку интегрирование по частям по времени $\tau$ допустимо. Используя
уравнение (\ref{exeltw}), определяющее $\lm$, и интегрируя по частям, получим
равенство
$$
  \triangle S\Big|_{F=0,G=0}=\int\!\!d\tau d\s(-\lm\pl_1q)
  +\int\!\!d\tau\,(\lm q)\Big|_{\s=\s_1}^{\s_2}.
$$
Теперь первое слагаемое воспроизводит эффективный гамильтониан $\lm H^*$, где
$\lm=\lm(\tau,\s)$, а второе слагаемое сокращается с граничным членом,
добавленным в физическое действие (\ref{ephace}).

Таким образом, при подстановке связей и калибровочных условий в действие,
необходимо проявлять осторожность. Рассмотренный пример показывает, что
эффективный гамильтониан и действие для физических полей может полностью
определяться граничным членом в исходном действии. Причина этого кроется в том,
что связи могут не иметь решений, убывающих в бесконечности, и предположение о
финитности вариации неправомерно. К сожалению, для определения явного вида
граничных членов необходим глубокий анализ уравнений связей, что не всегда
возможно, из-за их сложности.
\chapter{Основы общей теории относительности}
\index{Общая теория относительности (general theory of relativity)}%
\index{Теория относительности общая (general theory of relativity)}%
В настоящей главе мы приступим к изложению основ общей теории относительности,
которая в настоящее время рассматривается в качестве основной модели
гравитационных взаимодействий. После вступительного раздела, будут написаны
основные уравнения и поставлена задача, которая решается в теории гравитации.
\section{Пространство-время, метрика и гравитация}
В основе общей теории относительности лежит ряд постулатов. Выделим среди них
четыре, на наш взгляд, основных.
\begin{enumerate}
\item Пространство-время $\MM$, в котором мы живем, является четырехмерным
многообразием.
\item Гравитационное взаимодействие между материальными телами описывается
метрикой $g$ лоренцевой сигнатуры, $\sign g=(+---)$, заданной на $\MM$.
\item Метрика пространства-времени удовлетворяет уравнениям Эйнштейна.
\item Пробная точечная частица, собственным гравитационным полем которой в
данной задаче можно пренебречь, под действием только гравитационного поля
движется по экстремалям (геодезическим) пространства-времени $(\MM,g)$.
\end{enumerate}

Первые два постулата являются ``кинематическими''. Из них следует, что в общей
теории относительности все законы природы формулируются на четырехмерном
псевдоримановом многообразии (пространстве-времени) $(\MM,g)$. Если выбрана
некоторая система координат $x^\al$, $\al=0,1,2,3$, то метрика имеет вид
$g=dx^\al\otimes dx^\bt g_{\al\bt}$, $\sign g_{\al\bt}=\diag(+---)$. В общем
случае система координат может быть выбрана только локально.

В общей теории относительности не предполагается, что пространство-время
снабжено какой либо линейной структурой, как это было в механике Ньютона и
специальной теории относительности.

В общем случае глобальная структура (топология) пространства-времени $\MM$ может
быть нетривиальной и отличаться от пространства Минковского. Поскольку
глобально структура $\MM$ не фиксирована, то в моделях гравитации вводится новое
требование. Пространство-время, по-определению, должно быть максимально
продолжено вдоль геодезических (экстремалей). Это значит, что любая
геодезическая в
пространстве-времени может быть продолжена либо до бесконечного значения
канонического параметра в обе стороны, либо при конечном значении канонического
параметра она попадет в сингулярную точку, где какой либо из геометрических
инвариантов обращается в бесконечность. Поскольку канонический параметр вдоль
экстремалей определен с точностью до линейных преобразований (см., главу
\ref{sgextr}), то данное требование инвариантно, т.е.\ не зависит от выбора
системы координат.
\begin{com}
Требование максимального продолжения пространства-времени вдоль геодезических
нельзя заменить на более жесткое требование геодезической полноты, т.к.\ многие
важные точные решения уравнений Эйнштейна не являются геодезически полными.
Например, для решений, описывающих черные дыры, времениподобные геодезические
линии достигают сингулярного края, в которой квадрат тензора кривизны обращается
в бесконечность, при конечном значении собственного времени.
\qed\end{com}

Первые две аксиомы важны, поскольку позволяют описывать окружающий нас мир с
помощью некоторого набора полей и формулировать законы природы в виде системы
дифференциальных уравнений на $\MM$. Этот подход оказался самым плодотворным в
последние три столетия.

Третья и четвертая аксиома являются ``динамическими''. В общей теории
относительности постулируется, что метрика на $\MM$ должна удовлетворять
уравнениям Эйнштейна (\ref{einequ}). Тем самым компоненты метрики
пространства-времени удовлетворяют некоторой  системе уравнений движения так же,
как и все другие поля. Это -- очень важное отличие общей теории относительности
от специальной, где метрика Лоренца $\eta_{\al\bt}:=\diag(+---)$ в пространстве
Минковского $\MR^{1,3}$ была постулирована.

В правой части уравнений Эйнштейна стоит тензор энергии-импульса полей материи.
Выбор полей материи зависит от рассматриваемой модели. Это может быть, например,
сплошная среда, точечные массивные частицы, электромагнитное поле или что то
еще. Возможны также произвольные комбинации полей материи.

Сама по себе система уравнений Эйнштейна неполна. Если мы выбрали какой либо
набор полей материи, то уравнения Эйнштейна необходимо дополнить уравнениями
движения полей материи. Вид дополнительных уравнений зависит от рассматриваемой
задачи.

Последний выделенный постулат говорит о следующем. Допустим, что мы выбрали
некоторый набор полей материи, записали и решили полную систему уравнений для
метрики и полей материи. В результате мы получим псевдориманово многообразие
$(\MM,g)$, на котором заданы также поля материи. Теперь допустим, что к нашей
системе добавлена точечная массивная частица, масса которой настолько мала,
что она не влияет на решение уравнений Эйнштейна. То есть мы пренебрегаем
собственным гравитационным полем частицы. Такую частицу назовем {\em пробной}.
Тогда возникает вопрос, по какой траектории будет двигаться пробная частица под
действием только гравитационных сил? Ответ на этот вопрос дает четвертый
постулат: пробная частица будет двигаться по $\MM$ вдоль экстремалей
(геодезических), определяемых метрикой $g$. Мы также предполагаем, что
безмассовые частицы (например, фотоны) также распространяются вдоль экстремалей.
\index{Пробная частица (test particle)}\index{Частица пробная (test particle)}%

На четвертой аксиоме основано экспериментальное подтверждение общей теории
относительности. Два классических теста: смещение перигелия Меркурия и
отклонение лучей света в поле тяготения основаны на анализе геодезических
для решения Шварцшильда, о котором речь пойдет позже. Третий классический тест
-- красное смещение частоты излучения -- это следствие второй аксиомы.

Приведенные аксиомы выделены, потому что лежат в основе любой модели,
построенной в рамках общей теории относительности. Их недостаточно для
построения конкретной модели гравитации, т.к.\ необходимо выбрать поля материи и
дополнить уравнения Эйнштейна. При этом используются дополнительные аксиомы,
которые мы не выделяем, поскольку их столько же, сколько и моделей.

Остановимся на некоторых общих свойствах и терминологии.

Все поля, кроме метрики, мы делим на поле излучения, описывающее безмассовое
электромагнитное поле, и поля материи, описывающие массивные частицы. Многие
современные модели математической физики содержат дополнительные безмассовые
поля, например, глюоны, существование которых в природе пока экспериментально не
подтверждено. Для определенности, все безмассовые поля мы также относим к полям
излучения. Таким образом, все поля, за исключением метрики, мы делим на два
класса: {\em излучение} (безмассовые поля или частицы) и {\em поля материи}
(массивные поля или частицы).
\index{излучение (radiation)}%
\index{поля материи (matter fields)}\index{Материи поля (matter fields)}%

Теперь скажем несколько слов о гравитационном взаимодействии.
Для описания движения планет в солнечной системе с хорошей точностью
используется механика Ньютона и закон всемирного тяготения. Мы говорим, что
между планетами действуют гравитационные силы, которые определяют их движение.
При этом движение происходит в плоском трехмерном евклидовом пространстве
$\MR^3$, а время играет роль параметра. Основное свойство гравитационного взаимодействия
заключается в том, что движение пробной частицы, при заданных начальных
условиях, не зависит от ее массы.
\begin{exa}
Ускорение свободного падения на Земле не зависит от массы падающего тела. Это
утверждение в настоящее время экспериментально проверено с высокой степенью
точности.
\qed\end{exa}
Независимость ускорения от массы частицы означает, что при одних и тех же
начальных условиях траектории и мировые линии пробных частиц разной массы
совпадают.

Рассмотрим движение пробной частицы в специальной теории относительности в
инерциальной системе отсчета. По-определению, если гравитационное поле
отсутствует и на частицу не действуют никакие другие силы, то она движется
равномерно и прямолинейно. Теперь рассмотрим движение той же частицы, но в
неинерциальной системе отсчета, которая движется, например, с постоянным
ускорением относительно инерциальной системы отсчета. В этой системе координат
свободная частица движется с ускорением и наблюдатель в этой системе отсчета
может сказать (если не наблюдает за другими телами), что его система
инерциальна, а частица движется в постоянном и однородном гравитационном поле,
которое и вызывает ускорение. При этом ускорение не зависит от массы частицы и
определяется только неинерциальной системой координат. Поэтому часто формулируют
\newline
{\bf Принцип эквивалентности}. Движение свободной пробной частицы в
неинерциальной системе отсчета такое же, как движение частицы в инерциальной
системе отсчета, но находящейся под действием некоторого гравитационного поля.
\qed
\index{Принцип эквивалентности (equivalence principle)}%
\index{Эквивалентности принцип (equivalence principle)}%

В дальнейшем мы увидим, что никакой эквивалентности нет. А именно, обратное
утверждение неверно. Если пробная частица движется в гравитационном поле, то,
строго говоря, в общем случае не существует такой системы координат, в которой
она бы двигалась равномерно и прямолинейно.

В инерциальной системе координат в пространстве Минковского метрика диагональна
и имеет постоянные компоненты, $\eta_{\al\bt}=\diag(+---)$. В такой системе
координат все экстремали и только они являются прямыми линиями. Поэтому можно
сказать, что свободная пробная частица движется вдоль одной из экстремалей
пространства Минковского. Если перейти в неинерциальную (криволинейную) систему
координат, то в общем случае метрика перестанет быть диагональной, и ее
компоненты станут зависеть от координат точки пространства-времени. В этой
системе координат траектория свободной частицы уже не будет выглядеть
прямолинейной, а движение -- равномерным. Тем не менее траектория, конечно,
будет оставаться экстремалью пространства Минковского, т.к.\ понятие экстремали
инвариантно и не зависит от выбора системы координат.

В дифференциальной геометрии метрика является инвариантным объектом и
определяется независимо от выбора системы координат. Метрика Лоренца не зависит
от точки пространства Минковского. Однако ее компоненты могут быть
непостоянными функциями от координат точки: это зависит от системы координат.

Таким образом, утверждение о том, что свободная пробная частица движется в
пространстве Минковского вдоль одной из экстремалей инвариантно относительно
выбора системы координат и лежит в основе перехода от механики Ньютона к общей
теории относительности. Как уже было сказано, в общей теории относительности мы
предполагаем, что пространство-время представляет собой четырехмерное
многообразие $\MM$, на котором задана метрика лоренцевой сигнатуры. Мы
постулируем, что любая пробная частица движется вдоль одной из экстремалей
пространства-времени. Этот постулат согласуется с упомянутыми выше свойствами
гравитационного взаимодействия: мировая линия пробной частицы не зависит от ее
массы. При этом принцип эквивалентности является лишь наводящим соображением о
том, что метрика с нетривиальными компонентами описывает гравитационное
взаимодействие.

Тем самым метрика пространства-времени в общей теории относительности играет
выделенную роль. Мы считаем, что метрика описывает гравитационные
взаимодействия материальных тел и излучения. А именно, если частица движется в
плоском пространстве-времени, которое изометрично пространству Минковского с
лоренцевой метрикой, то на нее не действуют гравитационные силы. В этом случае
частица в инерциальной системе координат движется равномерно и прямолинейно.
Если гравитационное поле нетривиально, то частицы (массовые и безмассовые)
движутся по экстремалям в искривленном пространстве-времени, т.е.\ по
многообразию $\MM$ с метрикой $g$ и связностью Леви--Чивиты $\widetilde\Gamma$, для
которой тензор кривизны отличен от нуля. В этом случае отсутствует понятие
инерциальной системы отсчета, а экстремали отличаются от прямых линий.

Поскольку в пространстве-времени $\MM$ задана метрика, то она однозначно
определяет связность Леви--Чивиты или символы Кристоффеля. Это позволяет
использовать аппарат ковариантного дифференцирования для построения инвариантов
и записи ковариантных уравнений движения. Введение связности Леви--Чивиты на
многообразии $\MM$ является постулатом общей теории относительности. То есть
в теории тяготения Эйнштейна мы постулируем, что кручение и неметричность
аффинной связности тождественно равны нулю.

В настоящее время теория тяготения Эйнштейна имеет много обобщений. Большой
класс таких обобщений представляют собой модели, в которых на многообразии $\MM$
помимо метрики задается также независимая аффинная связность $\Gamma$ с
нетривиальным кручением $T$ и неметричностью $Q$. Эти обобщения естественны с
геометрической точки зрения, т.к.\ метрика и аффинная связность являются
совершенно независимыми геометрическими объектами. В общем случае, даже если
ограничиться инвариантными лагранжианами, приводящими к уравнениям движения
второго порядка, существует очень много возможностей для построения моделей,
которые в настоящее время не исследованы в полной мере.

Считается, что общая теория относительности согласуется со всеми
экспериментальными данными. Однако, поскольку мы не знаем экспериментальных
следствий упомянутых выше геометрических обобщений теории тяготения, говорить о
том, что они противоречат экспериментальным данным нельзя. Различные
геометрические обобщения теории тяготения Эйнштейна представляют самостоятельный
математический интерес и могут быть полезны при построении квантовой теории
гравитации и единых моделей. В современной математической физике такие модели
привлекают исследователей постоянно со времен создания общей теории
относительности.
\section{Действие Гильберта--Эйнштейна                           \label{shieia}}
В общей теории относительности постулируется, что пространство-время является
псевдоримановым многообразием $\MM$, $\dim\MM=n$, с метрикой лоренцевой
сигнатуры $g_{\al\bt}$. При этом считается, что метрика описывает гравитационные
взаимодействия. Мы намеренно не ограничиваемся наиболее интересным случаем
$n=4$, потому что модели гравитации в большем и меньшем числе измерений также
важны для приложений.

Мы рассматриваем метрику пространства-времени в качестве одной из полевых
переменных и постулируем для нее {\em уравнения Эйнштейна:}
\index{Уравнения Эйнштейна (Einstein's equations)}%
\index{Эйнштейна уравнения (Einstein's equations)}%
\begin{equation}                                                  \label{einequ}
  \kappa\left(\widetilde R_{\al\bt}-\frac12g_{\al\bt}\widetilde R\right)
  +g_{\al\bt}\frac{n-2}2\Lm=-\frac12T_{\Sm\al\bt}.
\end{equation}
В левой части этой системы уравнений для метрики стоит {\em тензор Эйнштейна}
\index{Тензор Эйнштейна (Einstein tensor)}%
\index{Эйнштейна тензор (Einstein tensor)}%
\begin{equation}                                                  \label{eitens}
  G_{\al\bt}:=\widetilde R_{\al\bt}-\frac12g_{\al\bt}\widetilde R,
\end{equation}
умноженный на {\em гравитационную постоянную} $\kappa$, и {\em космологическая
постоянная} $\Lm$. В правой части уравнений Эйнштейнв стоит {\em тензор
энергии-импульса материи} $T_{\Sm\al\bt}$. Эти уравнения при $\Lm=0$ и $n=4$
были впервые предложены в статье \cite{Einste15}.
\index{Гравитационная постоянная (gravitational constant)}%
\index{Постоянная гравитационная (gravitational constant)}%
\index{Космологическая постоянная (cosmological constant)}%
\index{Космологическая постоянная (cosmological constant)}%
\index{Тензор энергии-импульса материи (energy-momentum tensor of matter)}%
\index{Энергии-импульса тензор материи (energy-momentum tensor of matter)}%

Тензор энергии-импульса материи зависит от рассматриваемой модели, и в общем
случае уравнения Эйнштейна необходимо дополнить уравнениями для полей материи.
То есть сама по себе система уравнений Эйнштейна неполна.
\begin{com}
В уравнении (\ref{einequ}) мы оставили знак тильды, чтобы подчеркнуть, что
тензор кривизны строится только по метрике при нулевом кручении и неметричности.
То есть метрика $g$ на $\MM$ определяет связность Леви--Чивиты (символы
Кристоффеля), которые в свою очередь задают тензор кривизны.
\qed\end{com}
\begin{com}
Вклад космологической постоянной в уравнения Эйнштейна (\ref{einequ}) можно
перенести в правую часть
\begin{equation*}
  \kappa\left(\widetilde R_{\al\bt}-\frac12g_{\al\bt}\widetilde R\right)
  =-\frac12\big(T_{\Sm\al\bt}+T_{\Lm\al\bt}\big),
\end{equation*}
где
\begin{equation*}
  T_{\Lm\al\bt}:=(n-2)\Lm g_{\al\bt}
\end{equation*}
и рассматривать его как дополнение к тензору энергии-импульса материи
$T_{\Sm\al\bt}$. Сравнивая это выражение с тензором энергии-импульса непрерывной
среды (\ref{enmotl}), который будет рассмотрен позже, его можно
интерпретировать, как вклад среды с давлением $\CP=-(n-2)\Lm$ и плотностью
энергии противоположного знака $\CE=-\CP=(n-2)\Lm$. Разность знаков давления и
плотности энергии не позволяет интерпретировать космологическую постоянную, как
постоянное распределение некоторой обычной материи.
\qed\end{com}

Обсудим некоторые общие свойства уравнений Эйнштейна и введем терминологию.

Тензор Эйнштейна (\ref{eitens}) инвариантен относительно вейлевского
преобразования метрики с постоянным параметром
$$
  g_{\al\bt}\mapsto kg_{\al\bt},\qquad k=\const\ne0.
$$
Эти преобразования меняют длины векторов, но сохраняют углы между ними.

Уравнения Эйнштейна при заданном тензоре энергии-импульса представляют собой
систему из $n(n+1)/2$, где $n$ -- размерность пространства-времени, нелинейных
уравнений в частных производных второго порядка для метрики. В частности, в
четырехмерном пространстве-времени мы имеем десять уравнений. Уравнения
Эйнштейна чрезвычайно сложны, и в настоящее время известны лишь отдельные классы
решений, часть из которых будет обсуждаться в дальнейшем.

Уравнения Эйнштейна можно переписать в другом виде. След (\ref{einequ})
эквивалентен уравнению
$$
  \kappa\widetilde R=n\Lm+\frac1{n-2}T_\Sm,
$$
где $T_\Sm:=T_{\Sm\al}{}^\al$ -- след тензора энергии-импульса материи. Исключив
скалярную кривизну из (\ref{einequ}) с помощью этого равенства, получим
эквивалентную систему уравнений
\begin{equation}                                                  \label{eineqe}
  \kappa\widetilde R_{\al\bt}-\Lm g_{\al\bt}=-\frac12\rho_{\al\bt}.
\end{equation}
где
\begin{equation*}
  \rho_{\al\bt}:=T_{\Sm\al\bt}-\frac1{n-2}g_{\al\bt}T_\Sm.
\end{equation*}

Пространство-время называется {\em пустым}, если тензор энергии-импульса
материи всюду равен нулю. В этом случае уравнения Эйнштейна (\ref{eineqe})
принимают вид
\index{Пустое пространство-время (empty space-time)}%
\index{Пространство-время пустое (empty space-time)}%
\begin{equation}                                                  \label{eqevac}
  \kappa\widetilde R_{\al\bt}=\Lm g_{\al\bt}.
\end{equation}
Это -- {\em вакуумные уравнения Эйнштейна} с космологической постоянной.
\index{Вакуумные уравнения Эйнштейна (vacuum Einstein's equations)}%
\index{Уравнения Эйнштейна вакуумные (vacuum Einstein's equations)}%
Отсюда следует, что скалярная кривизна пустого пространства постоянна:
$$
  \widetilde R=\frac{n\Lm}\kappa.
$$
\begin{com}
Коэффициент перед космологической постоянной в уравнениях Эйнштейна
(\ref{einequ}) подобран таким образом, чтобы вакуумные уравнения Эйнштейна имели
вид (\ref{eqevac}) и не зависели от размерности пространства-времени.
\qed\end{com}

При ненулевой космологической постоянной уравнения (\ref{eqevac}) означают, что
тензор Риччи пропорционален метрике. Частным случаем таких пространств являются
пространства постоянной кривизны. При нулевой космологической постоянной,
$\Lm=0$, пустое пространство является {\em Риччи плоским}:
\index{Риччи плоское пространство (Ricci flat space)}%
\index{Пространство Риччи плоское (Ricci flat space)}%
\begin{equation}                                                  \label{ericfl}
  \widetilde R_{\al\bt}=0.
\end{equation}
Следовательно, в этом случае скалярная кривизна также равна нулю,
$\widetilde R=0$.
\begin{exa}
Для метрики Лоренца тензор кривизны равен нулю. Следовательно, пространство
Минковского является пространством постоянной -- нулевой -- кривизны. В
частности, оно является Риччи плоским. Ясно, что метрика Лоренца удовлетворяет
вакуумным уравнениям Эйнштейна с нулевой космологической постоянной.
\qed\end{exa}

В двумерном пространстве-времени полный тензор кривизны однозначно
восстанавливается по скалярной кривизне, а в трехмерном -- по тензору Риччи и
скалярной кривизне. Следовательно, полный тензор кривизны равен нулю, если
выполнено условие (\ref{ericfl}). Это значит, что двумерное и трехмерное Риччи
плоское пространство локально является пространством Минковского. То есть может
быть либо пространством Минковского, либо цилиндром или тором. В четырех
измерениях и выше равенства нулю тензора Риччи недостаточно для обращения в нуль
полного тензора кривизны.
\begin{exa}
Решение Шварцшильда, описывающее черную и белую дыру, является Риччи плоским, но
полный тензор кривизны отличен от нуля. А именно, отличен от нуля тензор Вейля.
\qed\end{exa}
Физическая интерпретация уравнений Эйнштейна при нулевой космологической
постоянной следующая. В общей теории относительности постулируется, что метрика
пространства-времени не является метрикой Лоренца, а находится, как решение
уравнений Эйнштейна. Таким образом, пространство-время представляет собой
псевдориманово многообразие с метрикой специального вида, удовлетворяющей
уравнениям (\ref{einequ}). Эти пространства называются {\em пространствами
Эйнштейна}. Следующий постулат состоит в том, что пробные частицы под действием
гравитационных сил двигаются по экстремалям в пространстве Эйнштейна. При этом в
правой части уравнений Эйнштейна подразумевается тензор энергии-импульса всей
остальной материи. При этом мы говорим следующее. Пустое пространство при
нулевой космологической постоянной и отсутствии гравитационных волн является
пространством Минковского, и точечные частицы двигаются по прямым линиям. Это
соответствует отсутствию сил тяготения. При наличии полей материи в уравнениях
Эйнштейна появляется нетривиальная правая часть, что приводит к тому, что
пространство-время становится нетривиальным псевдоримановым многообразием. В
этом пространстве-времени экстремали уже не являются прямыми линиями, что
интерпретируется, как наличие сил тяготения. Мы говорим, что пробная частица
движется в поле тяготения, созданном остальной материей. При этом закон
всемирного тяготения является следствием уравнений Эйнштейна в определенном
приближении, которое рассмотрено в разделе \ref{snelim}.
\index{Пространство Эйнштейна (Einstein space)}%
\index{Эйнштейна пространство (Einstein space)}%

Существующие экспериментальные данные свидетельствуют о том, что в отсутствие
сил тяготения пространство-время в масштабах солнечной системы близко к
пространству Минковского. Это значит, что если космологическая постоянная
существует, то является в определенном смысле малой величиной. Отметим, что
равенство или неравенство космологической постоянной нулю имеет принципиальное
значение. Действительно, наличие даже малой космологической постоянной приводит
к тому, что метрика Лоренца уже не будет удовлетворять вакуумным уравнениям
Эйнштейна.

Теперь обсудим принцип наименьшего действия для уравнений Эйнштейна. Левую часть
уравнений (\ref{einequ}) можно получить из {\em действия Гильберта--Эйнштейна}
\cite{Hilber15R,Einste16R}
\index{Действие Гильберта--Эйнштейна (Hilbert--Einstein actio)}%
\index{Гильберта--Эйнштейна действие (Hilbert--Einstein actio)}%
\begin{equation}                                                  \label{ehieia}
  S_{\Sh\Se}=\int_\MM\!\!\!dx\,\vol\left(\kappa\widetilde R-(n-2)\Lm\right),
\end{equation}
где интегрирование ведется по всему пространству-времени $\MM$. Конечно, мы
предполагаем, что интеграл сходится. Это действие было впервые предложено
Д.~Гильбертом в 1915 году в четырехмерном пространстве-времени. Он предложил
действие в более общем виде, включающем также электромагнитное поле
\cite{Hilber15R}. Несколько позже А.~Эйнштейн тоже рассмотрел это действие для
вывода уравнений общей теории относительности в такой системе координат, где
$\det g_{\al\bt}=1$ \cite{Einste16R}.

В следующем разделе мы покажем, что вариационная производная действия
Гильберта--Эйнштейна по метрике имеет вид
\begin{equation}                                                  \label{evahie}
  \frac{\dl S_{\Sh\Se}}{\dl g_{\al\bt}}
  =-\vol\kappa\left(\widetilde R^{\al\bt}-\frac12g^{\al\bt}\widetilde R\right)
  -\vol\frac{n-2}2\Lm g^{\al\bt}.
\end{equation}
При доказательстве этого равенства были отброшены все граничные вклады,
возникающие при интегрировании по частям.

При наличии полей материи чаще удобнее варьировать по обратной метрике, что
приводит к изменению знака вариационной производной:
\begin{equation}                                                  \label{ehieiv}
  \frac{\dl S_{\Sh\Se}}{\dl g^{\al\bt}}
  =\vol\kappa\left(\widetilde R_{\al\bt}-\frac12g_{\al\bt}\widetilde R\right)
  +\vol\frac{n-2}2\Lm g_{\al\bt}.
\end{equation}

Полное действие для гравитационного поля и полей материи имеет вид суммы
\begin{equation}                                                  \label{ehieit}
  S=S_{\Sh\Se}+S_\Sm,
\end{equation}
где $S_\Sm$ -- действие для полей материи. Обычно действие для полей материи в
теории гравитации получают путем {\em минимальной подстановки}: выбирают
лоренц-инвариантное действие в пространстве Минковского, заменяют лоренцеву
метрику на псевдориманову $\eta_{\al\bt}\mapsto g_{\al\bt}(x)$, обычные
производные -- на ковариантные $\pl_\al\mapsto\widetilde\nb_\al$, и умножают
лагранжиан на определитель репера $\vol$, чтобы получить инвариантную меру
интегрирования. В результате получим действие для полей материи, инвариантное
относительно общих преобразований координат. Сравнивая правую часть уравнений
Эйнштейна (\ref{einequ}) с вариационной производной (\ref{ehieiv}), получаем
выражение для тензора энергии-импульса материи
\index{Минимальная подстановка (minimal substitution)}%
\index{Подстановка минимальная (minimal substitution)}%
\begin{equation}                                                  \label{edenmo}
  T_{\Sm\al\bt}:=\frac2\vol\frac{\dl S_\Sm}{\dl g^{\al\bt}}.
\end{equation}
Эту вариационную производную часто принимают за определение {\em тензора
энергии-импульса} полей материи в общей теории относительности. При таком
определении тензор энергии-импульса всегда симметричен. В ряде случаев,
например, для скалярного и калибровочного полей, это определение совпадает с
ковариантным обобщением канонического тензора энергии-импульса, т.е.\ получается
из выражения (\ref{etenma}) путем минимальной подстановки. Однако в общем
случае это не так, потому что действие для полей материи (например, спинорных
полей) не всегда может быть выражено через метрику.
\subsubsection{Размерности постоянных и полей}
В теории поля важную роль играет анализ размерностей. Он помогает контролировать
проведение вычислений путем сравнения размерностей различных слагаемых и в
некоторых случаях делать общие выводы. В квантовой теории поля, например,
размерность констант связи позволяет судить о перенормируемости моделей. Здесь
принят упрощенный вариант подсчета размерностей, где все измеряется в единицах
длины. Однако, для сравнения общей теории относительности с теорией гравитации
Ньютона нам этого будет недостаточно. Поэтому мы опишем оба подхода к
определению размерностей.

В квантовой теории поля, по-определению, действие и метрика являются
безразмерными величинами, а координаты имеют размерность длины (скорость света
мы принимаем за единицу):
\begin{equation*}
  [S_{\Sh\Se}]:=[g_{\al\bt}]:=1,\qquad [x^\al]:=l.
\end{equation*}
Тогда компоненты тензора кривизны имеют следующую размерность:
\begin{equation*}
  [\widetilde R_{\al\bt\g\dl}]=[\widetilde R_{\al\bt}]=[\widetilde R]=l^{-2},
\end{equation*}
т.к.\ содержит две производные. Отсюда следует, что гравитационная и
космологическая постоянные являются размерными величинами:
\begin{equation*}
  [\kappa]=l^{2-n},\qquad [\Lm]=l^{-n}.
\end{equation*}
Для контроля вычислений такого описания размерностей, как правило, достаточно.

Теперь опишем размерности величин, как это принято в системе СГС, которые нам
понадобятся в дальнейшем. При этом мы считаем, что размерность
пространства-времени равна четырем, $n=4$. Исходными являются размерности массы
(грамм), расстояния (сантиметр) и времени (секунда):
\begin{equation*}
  [m]:=\text{г},\qquad [x^\mu]:=\text{см},\qquad [t]:=\text{сек},
\end{equation*}
где индекс $\mu=1,2,3$ пробегает только пространственные значения.
Координата $x^0:=ct$, где $c$ -- скорость света, также измеряется в сантиметрах.
По-определению, компоненты метрики безразмерны:
\begin{equation*}
  [g_{\al\bt}]=[g^{\al\bt}]:=1.
\end{equation*}
Поскольку кривизна содержит две производные по координатам, то
\begin{equation*}
  [\widetilde R_{\al\bt\g\dl}]=[\widetilde R_{\al\bt}]
  =[\widetilde R]=\frac1{\text{см}^2}.
\end{equation*}
Действие, как это принято в механике, имеет ту же размерность, что и
произведение импульса на скорость $p\,dq$ или энергии на время $Edt$:
\begin{equation*}
  [S]:=\frac{\text{г}\cdot\text{см}^2}{\text{сек}}.
\end{equation*}
Учитывая, что в действии $[dt\,d^3x]=\text{сек}\cdot\text{см}^3$, определяем
размерность гравитационной постоянной:
\begin{equation}                                                  \label{egrcod}
  [\kappa]=\frac{\text{г}\cdot\text{см}}{\text{сек}^2}.
\end{equation}
В разделе \ref{snelim} мы сравним эту гравитационную постоянную с той, которая
входит во всемирный закон тяготения. Наконец, размерность космологической
постоянной равна
\begin{equation*}
  [\Lm]=\frac{\text{г}}{\text{см}\cdot\text{сек}^2}.
\end{equation*}

В дальнейшем для упрощения формул мы часто будем полагать $c=1$ и $\kappa=1$.
Там, где это необходимо, степени скорости света легко восстановить, исходя из
соображений размерности.
\section{Вариация действия Гильберта--Эйнштейна                  \label{svarhe}}
Докажем равенство (\ref{evahie}) в более общем виде, который полезен при
рассмотрении моделей, основанных на геометрии Римана--Картана или аффинной
геометрии. А именно, рассмотрим инвариантное действие
\begin{equation}                                                  \label{ehieph}
  S=\int_\MM\!\!\! dx\,L=\int_\MM\!\!\! dx\,\vol\,\vf R,
\end{equation}
зависящее от скалярного поля $\vf(x)\in\CC^2(\MM)$ и скалярной кривизны
$R(g,\Gamma)$, построенной по метрике $g_{\al\bt}$ и аффинной связности общего вида
$\Gamma_{\al\bt}{}^\g$. Мы предполагаем, что компоненты метрики и связности являются
достаточно гладкими функциями, и интеграл (\ref{ehieph}) сходится. Кроме этого
мы предположим, что всеми граничными слагаемыми, возникающими при интегрировании
по частям, можно пренебречь.

Подстановка в действие (\ref{ehieph}) римановой кривизны $\widetilde R$,
зависящей только от метрики, приводит к чрезвычайно трудоемкой вариационной
задаче. Это связано с тем, что при дифференцировании по частям необходимо
дифференцировать также и скалярное поле. Поскольку скалярная кривизна
$\widetilde R$ содержит вторые производные от метрики, то интегрировать по
частям необходимо два раза, и это приводит к большому числу слагаемых.
Значительное упрощение вносят последовательные действия. Сначала варьируем по
метрике $g_{\al\bt}$ и связности $\Gamma_{\al\bt}{}^\g$, рассматривая их, как
независимые переменные, а затем подставляем вариацию связности, выраженную через
вариацию метрики, тензора кручения и неметричности. В общей теории
относительности такой подход называется {\em формализмом первого порядка}.
\index{Формализм первого порядка (first order formalism)}%
\index{Первого порядка формализм (first order formalism)}%

Начнем с нескольких вспомогательных формул, необходимых в дальнейшем. Варьируя
определение обратной метрики,
$$
  g^{\al\bt}g_{\bt\g}=\dl^\al_\g,
$$
получаем тождество
$$
  \dl g^{\al\bt}g_{\bt\g}+g^{\al\bt}\dl g_{\bt\g}=0.
$$
Отсюда следует связь между вариацией самой метрики и ее обратной:
\begin{equation}                                                  \label{evarin}
  \dl g^{\al\bt}=-g^{\al\g}g^{\bt\dl}\dl g_{\g\dl}.
\end{equation}
Из теории матриц известно, что для произвольной квадратной обратимой матрицы
$A=(A_{\al\bt})$ справедливо тождество
$$
  \dl\det A=\det A\, A^{-1\al\bt}\dl A_{\al\bt}.
$$
Отсюда следует, что вариация определителя метрики $g:=\det g_{\al\bt}$ равна
\begin{equation}                                                  \label{evamde}
  \dl g=gg^{\al\bt}\dl g_{\al\bt}=-gg_{\al\bt}\dl g^{\al\bt}.
\end{equation}
Эту вариацию мы записали в двух видах, так как в приложениях часто бывает
удобнее варьировать действие не по самой метрики, а по ее обратной. Наличие
квадратного корня в мере объема $\vol:=\sqrt{|\det g_{\al\bt}|}$ приводит к
появлению множителя $1/2$. Поэтому для ее вариации справедливы равенства
\begin{equation}                                                  \label{evavof}
  \dl\vol=\frac12\vol g^{\al\bt}\dl g_{\al\bt}
  =-\frac12\vol\dl g^{\al\bt}g_{\al\bt}.
\end{equation}

Приступим к вариации действия (\ref{ehieph}). Вариационная производная по
скалярному полю $\vf$ очевидна
\begin{equation}
  S,_\vf:=\frac{\dl S}{\dl\vf}=\vol R.
\end{equation}
Метрика входит в действие (\ref{ehieph}) дважды: в форму объема и в определение
скалярной кривизны (\ref{escurv}), причем без производных. Поэтому нетрудно
проверить, что
\begin{equation}                                                  \label{evahem}
  S,^{\al\bt}:=\frac{\dl S}{\dl g_{\al\bt}}
  =-\vol\vf\left(R^{\al\bt}-\frac12g^{\al\bt}R\right).
\end{equation}
Вариация действия по аффинной связности $\Gamma_{\al\bt}{}^\g$, все компоненты
которой рассматриваются, как независимые переменные, более трудоемка. Это
связано с тем, что приходится интегрировать по частям, так как тензор кривизны
(\ref{ecurva}) зависит от производных аффинной связности. Прямые вычисления
приводят к следующему выражению для вариации лагранжиана после интегрирования по
частям
\begin{equation}                                                  \label{evacoa}
\begin{split}
  \dl L=&-\vol\pl_\al\vf g^{\al\g}\dl\Gamma_{\bt\g}{}^\bt
  +\vol\pl_\g\vf g^{\al\bt}\dl\Gamma_{\al\bt}{}^\g
  -\vol\vf\left(Q_\al{}^{\al\g}-\frac12Q^\g\right)\dl\Gamma_{\bt\g}{}^\bt
\\
  &+\vol\vf\left[-\left(T_{\dl\g}{}^\dl+\frac12Q_\g\right)g^{\al\bt}
  -T_\g{}^{\bt\al}-T_\g{}^{\al\bt}+Q_\g{}^{\al\bt}\right]\dl\Gamma_{\al\bt}{}^\g,
\end{split}
\end{equation}
где $T_{\al\bt}{}^\g$ и $Q_{\al\bt\g}$ -- тензоры кручения и неметричности.
Для облегчения вычислений следует помнить, что выражение, стоящее перед
вариацией связности, должно быть тензорным полем. Это поможет правильно
сгруппировать слагаемые. Заметим также, что выражение, стоящее в квадратных
скобках, симметрично по индексам $\al$ и $\bt$.

Теперь вычислим очень важную для приложений вариационную производную действия
(\ref{ehieph}) по метрике $g_{\al\bt}$ в римановой геометрии. Обозначим
соответствующее действие, зависящее только от скалярного поля и метрики, через
\begin{equation}                                                  \label{ehisct}
  \widetilde S =\int_\MM\!\!\! dx\,\widetilde L
  =\int_\MM\!\!\! dx\,\vol\,\vf\widetilde R.
\end{equation}
Поскольку кручение и неметричность в римановой геометрии равны нулю, то из
(\ref{evahem}) и (\ref{evacoa}) следует, что вариация подынтегрального выражения
в действии по метрике равна
$$
  \dl\widetilde L
  =-\vol\vf\left(\widetilde R^{\al\bt}-\frac12g^{\al\bt}\widetilde R\right)
  \dl g_{\al\bt}-\vol\pl_\g\vf g^{\g\bt}\dl\widetilde\Gamma_{\al\bt}{}^\al
  +\vol\pl_\g\vf g^{\al\bt}\dl\widetilde\Gamma_{\al\bt}{}^\g.
$$
Выразив вариацию связности $\dl\widetilde\Gamma_{\al\bt}{}^\g$ через вариацию
метрики, проинтегрировав по частям и приведя подобные члены, получим
окончательное выражение для вариационной производной
\begin{equation}                                                  \label{evastm}
  \widetilde S,^{\al\bt}:=\frac{\dl\widetilde S}{\dl g_{\al\bt}}
  =-\vol\vf\left(\widetilde R^{\al\bt}-\frac12g^{\al\bt}\widetilde R\right)
  +\vol\big(\widetilde\square\vf g^{\al\bt}
  -\widetilde\nb^\al\widetilde\nb^\bt\vf\big).
\end{equation}
Напомним, что в римановой геометрии ввиду симметрии символов Кристоффеля по
нижним индексам вторая ковариантная производная от скалярного поля симметрична:
$\widetilde\nb^\al\widetilde\nb^\bt\vf=\widetilde\nb^\bt\widetilde\nb^\al\vf$
(сравните с равенством (\ref{ecocds})).

Если скалярное поле равно единице, $\vf=1$, то действие (\ref{ehisct}) совпадает
с действием Гильберта--Эйнштейна (\ref{ehieia}) без космологической постоянной,
и мы получаем выражение для вариационной производной (\ref{evahie}).

Формула для вариационной производной (\ref{evastm}) со скалярным полем важна,
например, при вычислении скобок Пуассона в канонической формулировке общей
теории относительности.

Вариация действия (\ref{ehieph}) была проведена в пространстве произвольной
размерности. Отметим, что в двумерном случае тензор Эйнштейна (\ref{eitens})
тождественно равен нулю, и первое слагаемое в (\ref{evastm}) отсутствует
(см.\ раздел \ref{sricas}).
\section{Зависимость уравнений Эйнштейна                         \label{seigud}}
В настоящем разделе мы считаем, что кручение и неметричность равны нулю, а
связностью является связность Леви--Чивиты, построенная по заданной метрике.

Важным обстоятельством в общей теории относительности является линейная
зависимость уравнений Эйнштейна (\ref{einequ}). Предположим, что эти уравнения
получены вариацией по метрике действия (\ref{ehieit}), которое инвариантно
относительно общих преобразований координат. Если действие инвариантно
относительно локальных преобразований, то согласно второй теореме Нетер между
уравнениями движения существует линейная зависимость (\ref{edepel}). Рассмотрим
эту зависимость в случае общих преобразований координат. Для простоты
предположим, что действие полей материи зависит только от некоторого набора
скалярных полей $\vf^a(x)$, $a=1,\dotsc,N$. При бесконечно малых преобразованиях
координат с параметром $\e^\al(x)$ метрика и скалярные поля преобразуются по
правилам (\ref{eitcms}) и (\ref{einftr}):
\begin{equation*}
\begin{split}
  \dl g_{\al\bt}&=-\widetilde\nb_\al\e_\bt-\widetilde\nb_\bt\e_\al,
\\
  \dl\vf^a&=-\e^\al\pl_\al\vf^a,
\end{split}
\end{equation*}
где $\e_\al:=g_{\al\bt}\e^\bt$. Следовательно инвариантность действия
записывается в виде
\begin{equation*}
  \dl S=\int dx\left(\frac{\dl S}{\dl g_{\al\bt}}
  (-2\widetilde\nb_\bt\e_\al\big)
  +\frac{\dl S}{\dl\vf^a}(-\e^\al\pl_\al\vf^a)\right)=0.
\end{equation*}
После интегрирования по частям первого слагаемого получаем искомую зависимость
уравнений движения
\begin{equation*}
  2\widetilde\nb_\bt\left(\frac{\dl(S_{\Sh\Se}+S_\Sm)}{\dl g_{\al\bt}}\right)
  -\pl_\al\vf^a\frac{\dl S_\Sm}{\dl\vf^a}=0,
\end{equation*}
т.к.\ действие Гильберта--Эйнштейна не зависит от полей материи. Это --
тождества, которые выполняются независимо от того удовлетворяют поля уравнениям
движения или нет. Поскольку каждое слагаемое в действии инвариантно само по
себе, то выполняются два тождества:
\begin{align}                                                     \label{ecobia}
  \widetilde\nb_\bt G^{\bt\al}&=0,
\\                                                                \label{ecjenc}
  \vol\widetilde\nb_\bt T_{\Sm}^\bt{}_\al
  -\frac{\dl S_\Sm}{\dl\vf^a}\pl_\al\vf^a&=0,
\end{align}
где мы воспользовались определением тензора энергии-импульса материи
(\ref{edenmo}) в общей теории относительности. Первое из этих уравнений
представляет собой свернутые тождества Бианки (\ref{ebieit}), а второе --
ковариантный ``закон сохранения'' тензора энергии-импульса материи.
Действительно, если выполнены уравнения для полей материи,
\begin{equation*}
  \frac{\dl S_\Sm}{\dl\vf^a}=0,
\end{equation*}
то дивергенция тензора энергии-импульса материи обращается в нуль
\begin{equation}                                                  \label{ecoems}
  \widetilde\nb_\bt T_{\Sm}^\bt{}_\al=0.
\end{equation}

Нетрудно видеть, что аналогичные выкладки можно проделать для любого набора
полей материи. При этом второе слагаемое в (\ref{ecjenc}) может усложниться, но
оно всегда будет пропорционально уравнениям движения для полей материи.
Единственное условие -- это инвариантность действия. Таким образом получаем
следующее утверждение.
\begin{prop}
Если действие полей материи инвариантно относительно общих преобразований
координат и поля материи удовлетворяют своим уравнениям Эйлера--Лагранжа, то
дивергенция тензора энергии-импульса (\ref{ecoems}) равна нулю.
\end{prop}

На формулу (\ref{ecoems}) можно взглянуть с другой точки зрения. Допустим, что
нам заданы уравнения Эйнштейна (\ref{einequ}), а про инвариантное действие,
приводящее к этим уравнениям, ничего не известно. Уравнения Эйнштейна -- это
система дифференциальных уравнений на метрику, и у них есть условия
интегрируемости. Чтобы их получить возьмем ковариантную производную от обеих
частей уравнений Эйнштейна. Дивергенция тензора Эйнштейна равна нулю
(\ref{ecobia}) как следствие тождеств Бианки (\ref{ebieit}). Дивергенция метрики
тоже равна нулю, т.к.\ связность Леви--Чивиты является метрической.
Следовательно, ковариантный ``закон сохранения'' тензора энергии-импульса
материи (\ref{ecoems}) является условием интегрируемости системы
дифференциальных уравнений Эйнштейна для метрики (\ref{einequ}). Это важно
учитывать в тех случаях, когда тензор энергии-импульса материи получен не из
принципа наименьшего действия, а из каких либо других соображений.
\begin{exa}
Если в качестве материи рассматривать жидкость или газ, для которой уравнения
движения не следуют из принципа наименьшего действия, то условие (\ref{ecoems})
является независимым уравнением.
\qed\end{exa}
\section{Действие для полей материи в обобщенных моделях гравитации}
В настоящем разделе мы покажем, что при минимальной подстановке ковариантное
обобщение канонического тензора энергии-импульса материи является источником в
уравнении для репера $e_\al{}^a$, а ковариантное обобщение спинового момента
полей материи -- источником в уравнении для лоренцевой связности
$\om_{\al a}{}^b$.

В различных обобщенных моделях гравитации мы обычно предполагаем, что
инвариантное действие состоит из двух слагаемых
\begin{equation}                                                  \label{egeact}
  S=S_\Sg+S_\Sm,
\end{equation}
где $S_\Sg$ -- гравитационная часть действия и $S_\Sm$ -- действие для полей
материи, в которое для простоты мы включили также калибровочные поля
(электромагнитное поле и поле Янга--Миллса). В общей теории относительности
$S_\Sg=S_{\Sh\Se}$ -- это действие Гильберта--Эйнштейна (\ref{ehieia}), равное
интегралу от скалярной кривизны с возможным добавлением космологической
постоянной. Это действие зависит только от метрики или репера. В более общих
моделях гравитационная часть действия может включать также инварианты более
высокого порядка по кривизне, кручению, тензору неметричности и их ковариантных
производных. В таких случаях мы рассматриваем в качестве независимых переменных,
по которым производится варьирование, переменные Картана: репер $e_\al{}^a$ и
линейную $\MG\ML(n,\MR)$-связность $\om_{\al a}{}^b$ (см.\ раздел \ref{scorep}).

Если действие для полей материи может быть выражено через метрику и аффинную
связность, то в качестве независимых переменных можно рассматривать также
метрику, кручение и неметричность. Однако это не всегда имеет место. Например,
для спинорного поля в геометрии Римана--Картана лоренцева связность не может
быть выражена через метрику и аффинную связность. В этом случае введение репера
необходимо. Таким образом, в общем случае использование переменных Картана
предпочтительнее.

В настоящем разделе мы обсудим общие свойства уравнений движения для полей
материи, не используя конкретный вид гравитационной части действия $S_\Sg$.

Остановимся более подробно на действии $S_\Sm$ для полей материи. К полям
материи в настоящем разделе мы относим скалярные, спинорные поля,
электромагнитное поле, поля Янга--Миллса и все другие поля, кроме репера и
линейной связности. Обозначим всю совокупность полей материи через
$\vf=\lbrace\vf^\Sa\rbrace$, $\Sa=1,2,\dotsc$. Пусть действие для
полей материи в плоском пространстве-времени Минковского $\MR^{1,n-1}$ с
координатами $x^a$, $a=0,1,\dotsc,n-1$, имеет вид
\begin{equation*}
  S_\Sm=\int_{\MR^{1,n-1}}\!\!\!dxL_{\Sm}(\eta,\hat\eta,\vf,\pl\vf),
\end{equation*}
где лагранжиан полей материи $L_\Sm$ зависит от метрики Минковского
$\eta=\lbrace \eta_{ab}\rbrace$, инвариантной метрики в пространстве-мишени для
полей материи $\hat\eta=\lbrace \eta_{\Sa\Sb}\rbrace$ (во многих случаях это
просто символ Кронекера, $\eta_{\Sa\Sb}=\dl_{\Sa\Sb}$), полей материи
$\vf=\lbrace\vf^\Sa\rbrace$ и их частных производных первого порядка
$\pl\vf=\lbrace\pl_a\vf^\Sa\rbrace$.

В моделях гравитации мы предполагаем, что пространство-время $\MM$ является
многообразием той же размерности, что и исходное пространство Минковского
$\MR^{1,n-1}$. Обозначим локальные координаты на $\MM$ через $x^\al$,
$\al=0,1,\dotsc,n-1$. В простейшем случае инвариантное действие для полей
материи получается из действия в пространстве Минковского с помощью
{\em минимальной подстановки}:
\index{Минимальная подстановка (minimal substitution)}%
\index{Подстановка минимальная (minimal substitution)}%
\begin{align*}
  \MR^{1,n-1}&\rightarrow\MM,\qquad \dim\MM=n,
\\
  L_\Sm&\mapsto\vol L_\Sm,
\\
  \eta_{ab}&\mapsto g_{\al\bt}=e_\al{}^a e_\bt{}^b g_{ab},
\\
  \eta_{\Sa\Sb}&\mapsto g_{\Sa\Sb}(x),
\\
  \vf^\Sa&\mapsto\vf^\Sa,
\\
  \pl_a\vf^\Sa&\mapsto\nb_\al\vf^\Sa,\qquad
  \nb_\al\vf^\Sa:=\pl_\al\vf^\Sa+\om_{\al\Sb}{}^\Sa\vf^\Sb,
\end{align*}
где $\om_{\al\Sa}{}^\Sb$ -- компоненты линейной связности в том представлении, в
котором преобразуются поля материи.

В случае общей линейной группы преобразований $\MG\ML(n,\MR)$ инвариантной
метрики в пространстве-мишени не существует (если поля материи преобразуются по
точному представлению $\MG\ML(n,\MR)$), и мы вынуждены рассматривать
нетривиальную метрику в пространстве-мишени $g_{\Sa\Sb}(x)$ как дополнительное
поле. Метрика $g_{ab}$ -- это произвольная симметричная невырожденная матрица
лоренцевой сигнатуры, с помощью которой определяется репер.
\begin{exa} Рассмотрим набор массивных скалярных полей одинаковой массы,
лагранжиан которых в пространстве Минковского имеет вид
(см.\ раздел \ref{srescf})
\begin{equation}                                                  \label{emscaf}
  L_\Sm=\frac12\eta^{ab}\pl_a\vf^\Sa\pl_b\vf^\Sb\eta_{\Sa\Sb}
  -\frac12 m^2\vf^\Sa\vf^\Sb\eta_{\Sa\Sb}.
\end{equation}
Мы считаем, что скалярные поля преобразуются по некоторому представлению группы
Лоренца (индекс $\Sa$), и $\eta_{\Sa\Sb}$ -- инвариантная метрика. Этот
лагранжиан инвариантен относительно глобальных преобразований Лоренца,
действующих на координаты $x^a$ и поля $\vf^\Sa$. После минимальной подстановки
он примет вид
\begin{equation*}
  L_\Sm=\vol\left(\frac12g^{\al\bt}\nb_\al\vf^\Sa\nb_\bt\vf^\Sb g_{\Sa\Sb}
  -\frac12m^2\vf^\Sa\vf^\Sb g_{\Sa\Sb}\right),
\end{equation*}
где ковариантная производная определяется линейной связностью (см.\ раздел
\ref{scovec})
\begin{equation}                                                  \label{ecplid}
  \nb_\al\vf^\Sa:=\pl_\al\vf^\Sa+\om_{\al\Sb}{}^\Sa\vf^\Sb,\qquad
  \om_{\al\Sb}{}^\Sa:=\om_\al{}^{ab}L_{ab\Sb}{}^\Sa,
\end{equation}
$g_{\Sa\Sb}$ -- метрика в пространстве-мишени и
$g^{\al\bt}=e^\al{}_a e^\bt{}_b g^{ab}$. В приведенной формуле $L_{ab\Sb}{}^\Sa$
-- представление генераторов линейной группы для выбранного набора скалярных
полей. Соответствующее действие инвариантно относительно общих преобразований
координат и локальных $\MG\ML(n,\MR)$ вращений в пространстве-мишени:
\begin{align*}
  e_\al{}^b&\mapsto e_\al{}^b S^{-1}_{\quad b}{}^a,
\\
  g_{ab}&\mapsto S_a{}^c S_b{}^d g_{cd},
\\
  \om_{\al a}{}^b&\mapsto S_a{}^c\om_{\al c}{}^d S^{-1}_{\quad d}{}^b
  +\pl_\al S_a{}^c S^{-1}_{\quad c}{}^b,
\\
  \vf^\Sa&\mapsto\vf^\Sb S^{-1}_{~~\Sb}{}^\Sa,
\\
  g_{\Sa\Sb}&\mapsto S_\Sa{}^\Sc S_\Sb{}^\Sd g_{\Sc\Sd},
\end{align*}
где $S_a{}^b(x)\in\MG\ML(n,\MR)$ -- матрица локальных вращений и $S_\Sa{}^\Sb$
-- ее представление, которое соответствует выбранному набору скалярных полей.
Если ограничить группу $\MG\ML(n,\MR)$ до группы Лоренца $\MO(1,n-1)$, то в
качестве метрики $g_{ab}(x)$ можно выбрать инвариантную метрику Лоренца
$\eta_{ab}$, которая уже не будет зависеть от точки многообразия. Ей
соответствует некоторая инвариантная метрика $\eta_{\Sa\Sb}$ в
пространстве-мишени.
\qed\end{exa}

Ограничимся моделями, основанными на геометрии Римана--Картана. Тогда
$g_{ab}=\eta_{ab}$ и $g_{\Sa\Sb}=\eta_{\Sa\Sb}$.

Из общего выражения для действия (\ref{egeact}) следуют уравнения
Эйлера--Лагранжа для геометрических переменных $e_\al{}^a$, $\om_{\al a}{}^b$
и полей материи $\vf^\Sa$:
\begin{align}                                                     \label{eveieq}
  \frac{\dl S}{\dl e_\al{}^a}&=0,
\\                                                                \label{econeq}
  \frac{\dl S}{\dl\om_{\al a}{}^b}&=0,
\\
  \frac{\dl S}{\dl\vf^\Sa}=\frac{\dl S_\Sm}{\dl\vf^\Sa}&=0.
\end{align}
Поскольку при минимальной подстановке репер входит в действие полей материи
только в качестве общего множителя $\vol:=\det e_\al{}^a$ и в ковариантную
производную с латинским индексом $\nb_a\vf^\Sa=e^\al{}_a\nb_\al\vf^\Sa$,
то первое уравнение можно переписать в виде
\begin{equation}                                        \label{erepva}
  \frac{\dl S_\Sg}{\dl e_\al{}^a}
  =\vol\frac{\pl L_\Sm}{\pl(\nb_\al\vf^\Sa)}e^\bt{}_a\nb_\bt\vf^\Sa
  -\vol e^\al{}_aL_\Sm=\vol e^\bt{}_aT_{\Sm\bt}{}^\al,
\end{equation}
где
\begin{equation*}
  T_{\Sm\bt}{}^\al:=-\frac1\vol\frac{\dl S_\Sm}{\dl e_\al{}^a} e_\bt{}^a
  =\nb_\bt\vf^\Sa\frac{\pl L_\Sm}{\pl(\nb_\al\vf^\Sa)}
  -\dl^\al_\bt L_\Sm.
\end{equation*}
Сравнивая последнее выражение с каноническим тензором энергии-импульса
(\ref{etenma}), полученным из теоремы Нетер, мы видим, что в правой части
гравитационных уравнений (\ref{erepva}), полученных после варьирования по
реперу, стоит его ковариантное обобщение. Поэтому говорят, что тензор
энергии-импульса материи является источником для репера.

Линейная связность $\om_\al{}^{ab}$ входит в действие полей материи только
через ковариантную производную. Поэтому уравнение (\ref{econeq}) можно
переписать в виде
\begin{equation*}
  \frac{\dl S_\Sg}{\dl\om_\al{}^{ab}}=-\vol\frac{\pl L_\Sm}{\pl(\nb_\al\vf^\Sa)}
  L_{ab\Sb}{}^\Sa\vf^\Sb=\vol S_{\Sm ab}{}^\al,
\end{equation*}
где
\begin{equation}                                                  \label{espite}
  S_{\Sm ab}{}^\al
  :=-\frac{\pl L_\Sm}{\pl(\nb_\al\vf^\Sa)}L_{ab\Sb}{}^\Sa\vf^\Sb.
\end{equation}
В геометрии Римана--Картана общая линейная группа $\MG\ML(n,\MR)$ сужается
до группы Лоренца $\MO(1,n-1)$, линейная связность $\om_\al{}^{ab}$ становится
лоренцевой связностью, $L_{ab\Sb}{}^\Sa$ -- генераторы группы Лоренца,
$g_{ab}=\eta_{ab}$, $g_{\Sa\Sb}$ -- инвариантная метрика в пространстве-мишени,
которая всегда существует, т.к.\ группа Лоренца проста. В этом случае выражение
(\ref{espite}) является ковариантным обобщением спинового момента полей материи
(\ref{espimo}). Поэтому мы говорим, что спиновый момент полей материи является
источником для лоренцевой связности.

Выбор переменных Картана для моделей гравитации позволил дать физическую
интерпретацию источников гравитационного поля: тензор энергии-импульса
материи является источником для репера и спиновый момент -- источником
для лоренцевой связности.
\begin{com}
При выборе в качестве независимых переменных в геометрии Римана--Картана метрики
и кручения, в правой части соответствующих уравнений будут стоять выражения,
которые не имеют столь простой и привлекательной интерпретации.
\qed\end{com}
\section{Скалярно-тензорные модели                               \label{sctemo}}
В качестве одного из возможных обобщений эйнштейновской теории гравитации,
основанного на римановой геометрии, рассматривают скалярно-тензорные модели, в
которых гравитационное взаимодействие описывается метрикой $g_{\al\bt}$ и
скалярным полем $\phi$. Лагранжиан скалярно-тензорных моделей обычно записывают
в виде
\begin{equation}                                                  \label{esctel}
  L=\kappa\vol \phi R+\vol\frac{\om(\phi)}\phi\pl\phi^2-\vol V(\phi)+L_\Sm,
\end{equation}
где $R(g)$ -- псевдориманова кривизна (в настоящем разделе знаки тильды
опущены),
$$
  \pl\phi^2:=g^{\al\bt}\pl_\al\phi\pl_\bt\phi
$$
-- кинетический член для скалярного поля, $\kappa$ -- гравитационная постоянная,
$\om(\phi)$ и $V(\phi)$ -- некоторые дифференцируемые функции от скалярного
поля, характеризующие неминимальность взаимодействия и потенциал самодействия.
Все остальные поля включены в лагранжиан полей материи $L_\Sm$.

С точки зрения общей теории относительности можно сказать следующее. Лагранжиан
(\ref{esctel}) описывает гравитационное взаимодействие полей материи, причем
гравитационная ``постоянная'' $\kappa\phi(x)$ зависит от точки
пространства-времени, то есть является скалярным полем, которое удовлетворяет
своим уравнениям движения. При этом связь этого скалярного поля с метрикой
неминимальна и включается некоторое самодействие, описываемое потенциалом $V$.

При $n=2$ лагранжиан вида (\ref{esctel}) описывает двумерную дилатонную
гравитацию общего вида, а скалярное поле $\phi$ называется полем дилатона.
Этот класс моделей характеризуется двумя произвольными функциями
$U(\phi):=2\om(\phi)/\phi$ и $V(\phi)$ и приводит к интегрируемым уравнениям
движения.

Впервые действие вида (\ref{esctel}) было рассмотрено М.~Фирцем в 1956 году
\cite{Fierz56}. Скалярно-тензорные модели гравитации привлекли значительное
внимание после работ П.~Йордана \cite{Jordan59} и С.~Бранса и Р.~Дике
\cite{BraDic61}. Основная идея этих исследований восходит к работе П.~Дирака,
который предположил, что гравитационная постоянная может меняться со временем
\cite{Dirac38}. М.~Фирц пошел дальше, выдвинув гипотезу о том, что
гравитационная постоянная описывается независимым скалярным полем,
удовлетворяющим некоторому нелинейному уравнению движения.

Получим уравнения движения для скалярно-тензорных моделей (\ref{esctel}).
Используя вид вариационной производной (\ref{evastm}), нетрудно получить
вариационные производные действия (\ref{esctel}) по обратной метрике и
скалярному полю
\begin{align}                                                          \nonumber
  \frac1\vol\frac{\dl S}{\dl g^{\al\bt}}:&&&
  \kappa\phi\left(R_{\al\bt}-\frac12g_{\al\bt}R\right)
  -\kappa(\square\phi g_{\al\bt}-\nb_\al\nb_\bt\phi)+
\\                                                                \label{esctem}
  &&&\qquad
  +\frac\om\phi\left(\pl_\al\phi\pl_\bt\phi-\frac12g_{\al\bt}\pl\phi^2\right)
  +\frac12g_{\al\bt}V+\frac12T_{\Sm\al\bt}=0.
\\                                                                \label{esctsf}
  \frac1\vol\frac{\dl S}{\dl\phi}:&&&
  \kappa R-2\frac\om\phi\square\phi-\frac{\om'}\phi\pl\phi^2
  +\frac\om{\phi^2}\pl\phi^2-V'=0,
\end{align}
где $T_{\Sm\al\bt}$ -- тензор энергии-импульса материи (\ref{edenmo}), а
штрих обозначает дифференцирование функции по аргументу. След уравнения
(\ref{esctem}) имеет вид
$$
  \phi({\textstyle\frac n2}-1)\left(\kappa R+\frac\om{\phi^2}\pl\phi^2\right)
  +\kappa(n-1)\square\phi-\frac n2V-T_\Sm=0,
$$
где $T_\Sm:=T_{\Sm\al}{}^\al$ -- след тензора энергии-импульса материи. Это
уравнение при $n\ne2$ позволяет исключить сумму
$$
  \kappa R+\frac\om{\phi^2}\pl\phi^2
$$
из уравнения для скалярного поля (\ref{esctsf}). В результате получим
эквивалентное уравнение
\begin{equation}                                                  \label{esctes}
  2\left(\frac{n-1}{n-2}\kappa+\om\right)\square\phi
  +\frac1{n-2}(nV+2T_\Sm)-\om'\pl\phi^2-\phi V'=0.
\end{equation}
Таким образом систему уравнений движения (\ref{esctem}), (\ref{esctsf}) можно
переписать в эквивалентном виде, заменив уравнение для скалярного поля
(\ref{esctsf}) на уравнение (\ref{esctes}). В четырехмерном пространстве-времени
уравнения движения обычно записывают в виде
\begin{align}                                                          \nonumber
  \kappa\phi\left(R_{\al\bt}-\frac12g_{\al\bt}R\right)=&-\frac12T_{\Sm\al\bt}
  +\kappa(\square\phi g_{\al\bt}-\nb_\al\nb_\bt\phi)-
\\                                                                \label{escton}
  &-\frac\om\phi\left(\pl_\al\phi\pl_\bt\phi-\frac12g_{\al\bt}\pl\phi^2\right)
  -\frac12g_{\al\bt}V,
\\                                                                \label{esctto}
  \square\phi=&\frac1{2\om+3\kappa}(2V+T_\Sm-\om'\pl\phi^2-\phi V').
\end{align}

С математической точки зрения уравнения движения скалярно-тензорных моделей
значительно сложнее уравнений общей теории относительности и изучены
недостаточно полно. Поэтому практически ничего нельзя сказать о виде функций
$\om(\phi)$ и $V(\phi)$, которые приводят к удовлетворительным результатам с
теоретической точки зрения и не противоречат существующим экспериментальным
данным.
\section{Полиномиальная форма действия \\ Гильберта--Эйнштейна   \label{spolaf}}
Рассмотрим пространство-время $\MM$ произвольной размерности $\dim\MM=n\ge3$.
Обозначим локальные координаты пространства-времени через $x^\al$,
$\al=0,1,\dotsc,n-1$. Уравнения движения для метрики $g_{\al\bt}(x)$ в общей
теории относительности без полей материи следуют из вариационного принципа для
действия Гильберта--Эйнштейна (\ref{ehieia}).

Действие Гильберта--Эйнштейна неполиномиально по компонентам метрики
по двум причинам. Во-первых, оно содержит неполиномиальный элемент объема
$\vol$. Во-вторых, выражение для скалярной кривизны содержит обратную
метрику $g^{\al\bt}$, компоненты которой также неполиномиальны по $g_{\al\bt}$.
Поэтому действие Гильберта--Эйнштейна в теории возмущений представляет собой
очень сложный бесконечный ряд, что является основной технической трудностью
при анализе уравнений движения и квантовании. По этим же причинам действие
Гильберта--Эйнштейна неполиномиально по компонентам обратной метрики.

Покажем, что конфигурационное пространство общей теории относительности можно
расширить, включив определитель метрики в качестве дополнительной независимой
переменной таким образом, что действие примет полиномиальный вид.
Эквивалентность полиномиального действия исходному действию достигается за счет
наложения в расширенном конфигурационном пространстве связи, которая также
полиномиальна по полям. Изложение настоящего раздела следует
\cite{Peres63,Katana06A}.

Координатами конфигурационного пространства $\CM$ общей теории относительности
являются компоненты метрики $g_{\al\bt}(x)$. Размерность этого пространства
равна
\begin{equation*}
  \dim\CM=\frac{n(n+1)}2\times \infty^{(n-1)},
\end{equation*}
где символический множитель $\infty^{(n-1)}$ соответствует точкам пространства
в пространстве-времени $\MM$.

Рассмотрим другое конфигурационное пространство $\CN$ с координатами
$\varrho(x),k_{\al\bt}(x)$. При этом будем считать, что $\varrho>0$ и матрица
$k_{\al\bt}$ симметрична и невырождена в каждой точке пространства-времени:
\begin{equation*}
  k_{\al\bt}=k_{\bt\al},\qquad \det k_{\al\bt}\ne0.
\end{equation*}
Мы предполагаем также, что матрицы $g_{\al\bt}$ и $k_{\al\bt}$ имеют одинаковую
лоренцевы сигнатуру
\begin{equation*}
  \sign g_{\al\bt}=\sign k_{\al\bt}=(+\underbrace{-\dotsc-}_{n-1}).
\end{equation*}
Размерность нового конфигурационного пространства равна
\begin{equation*}
  \dim\CN=\left(\frac{n(n+1)}2+1\right)\times\infty^{(n-1)}.
\end{equation*}

Выделим в $\CN$ подпространство $\CM'$ с помощью дополнительного условия
\begin{equation}                                                  \label{emehed}
  \det k_{\al\bt}=\begin{cases}\quad 1, & \det g_{\al\bt}>0,\quad
  \text{нечетные}~n,
  \\-1, &\det g_{\al\bt}<0,\quad \text{четные}~n. \end{cases}
\end{equation}
Тогда между точками подпространства $\CM'\subset\CN$ и исходного
конфигурационного пространства $\CM$ можно установить взаимно однозначное
соответствие
\begin{equation}                                                  \label{eghrel}
  g_{\al\bt}:=\varrho^2k_{\al\bt}.
\end{equation}
Обратное преобразование имеет вид
\begin{equation}                                                  \label{erhokg}
  \varrho=|\det g_{\al\bt}|^{1/2n},\qquad k_{\al\bt}=\varrho^{-2}g_{\al\bt}.
\end{equation}
Поэтому мы отождествим $\CM=\CM'$. Представление для обратной метрики
$g^{\al\bt}=\varrho^{-2}k^{\al\bt}$, где $k^{\al\bt}k_{\bt\g}=\dl^\al_\g$
следует из уравнения (\ref{eghrel}).

Сделаем два важных замечания. Во-первых, компоненты обратной матрицы
$k^{\al\bt}$ полиномиальны по $k_{\al\bt}$. В общем случае из условия
(\ref{emehed}) следует, что они являются полиномами степени $n-1$ по
компонентам $k_{\al\bt}$. Во-вторых, компоненты $k_{\al\bt}$ являются
компонентами не тензора, а тензорной плотности второго ранга. Действительно,
потребуем, чтобы равенство (\ref{eghrel}) было выполнено в произвольной системе
координат. Поскольку определитель метрики $g$ представляет собой скалярную
плотность степени $\deg g=-2$, то матрица $k_{\al\bt}$ является симметричной
тензорной плотностью второго ранга и степени $\deg k_{\al\bt}=2/n$, а поле
$\varrho$ -- скалярной плотностью степени $\deg\varrho=-1/n$. То есть при
преобразовании координат $x^\al\rightarrow x^{\al'}(x)$ новые переменные
преобразуется по-правилу
\begin{equation*}
  k_{\al'\bt'}=\pl_{\al'}x^\al\pl_{\bt'}x^\bt k_{\al\bt} J^{2/n},
  \qquad \varrho'=\varrho J^{-1/n},
\end{equation*}
где $J:=\det\pl_\al x^{\al'}$ -- якобиан преобразования координат. Это значит,
что дополнительное условие (\ref{emehed}) инвариантно относительно
преобразования координат. В связи с этим будем называть поле $k_{\al\bt}(x)$
{\em плотностью метрики}.
\index{Плотность метрики (metric density)}%

Явная формула для компонент обратной плотности метрики имеет вид
\begin{equation*}
  k^{\al\bt}=\frac1{(n-1)!}\hat\ve^{\al\g_2\dotsc\g_n}
  \hat\ve^{\bt\dl_2\dotsc\dl_n}k_{\g_2\dl_2}\dotsc k_{\g_n\dl_n},
\end{equation*}
где $\hat\ve^{\al_1\dotsc\al_n}$ -- полностью антисимметричная тензорная
плотность степени $-1$ с единичными компонентами
$|\hat\ve^{\al_1\dotsc\al_n}|=1$. Данная формула показывает, что компоненты
$k^{\al\bt}$ -- это полиномы степени $n-1$ от компонент $k_{\al\bt}$ и наоборот.

Перепишем действие Гильберта--Эйнштейна в новых переменных $\varrho,k_{\al\bt}$.
Для вычислений нам понадобятся следующие тождества, которые следуют после
дифференцирования равенства (\ref{emehed})
\begin{align*}
  k^{\g\dl}\pl_\al k_{\g\dl}&=0,
\\
  k^{\g\dl}\pl^2_{\al\bt}k_{\g\dl}
  -k^{\g\dl}k^{\e\z}\pl_\al k_{\g\e}\pl_\bt k_{\dl\z}&=0.
\end{align*}
Прямые вычисления приводят к равенствам
\begin{equation}                                                  \label{eiqugh}
\begin{split}
  \pl_\al g_{\bt\g}&=\varrho^2\left(\pl_\al k_{\bt\g}
  +2 \pl_\al\ln\varrho\,k_{\bt\g}\right),
\\
  \pl^2_{\al\bt}g_{\g\dl}&=\varrho^2\left(\pl^2_{\al\bt}k_{\g\dl}
  +2\pl^2_{\al\bt}\ln\varrho\, k_{\g\dl}+\right.
\\
  &\qquad \qquad \qquad \left.+2\pl_\al\ln\varrho\,\pl_\bt k_{\g\dl}
  +2\pl_\bt\ln\varrho\,\pl_\al k_{\g\dl}
  +4\pl_\al\ln\varrho\,\pl_\bt\ln\varrho\, k_{\g\dl}\right),
\\
  g^{\bt\dl}\pl_\al g_{\dl\g}&=k^{\bt\dl}\pl_\al k_{\dl\g}
  +2\pl_\al\ln\varrho\dl^\bt_\g,
\\
  g^{\e\z}\pl_\al g_{\g\e}\pl_\bt g_{\dl\z}
  &=\varrho^2\left(k^{\e\z}\pl_\al k_{\bt\e}\pl_\bt k_{\dl\z}
  +2\pl_\al\ln\varrho\,\pl_\bt k_{\g\dl}+2\pl_\bt\ln\varrho\,\pl_\al k_{\g\dl}+
  \right.
\\
  &\qquad\qquad\qquad\qquad\qquad\qquad\qquad\qquad\qquad\qquad\qquad\qquad
  \left.+4\pl_\al\ln\varrho\,\pl_\bt\ln\varrho\, k_{\g\dl}\right),
\\
  g^{\g\dl}g^{\e\z}\pl_\al g_{\g\e}\pl_\bt g_{\dl\z}
  &=k^{\g\dl}k^{\e\z}\pl_\al k_{\g\e}\pl_\bt k_{\dl\z}
  +4n\pl_\al\ln\varrho\,\pl_\bt\ln\varrho,
\\
  g^{\g\dl}\pl^2_{\al\bt}g_{\g\dl}&=k^{\g\dl}\pl^2_{\al\bt}k_{\g\dl}
  +2n\pl^2_{\al\bt}\ln\varrho+4n\pl_\al\ln\varrho\,\pl_\bt\ln\varrho,
\\
  \pl^2_{\al\bt}g_{\g\dl}-\frac12g^{\e\z}\pl_\al g_{\g\e}&\pl_\bt g_{\dl\z}
  -\frac12g^{\e\z}\pl_\bt g_{\g\e}\pl_\al g_{\dl\z}=
\\
  =\varrho^2&\left(\pl^2_{\al\bt}k_{\g\dl}
  -\frac12k^{\e\z}\pl_\al k_{\g\e}\pl_\bt k_{\dl\z}
  -\frac12k^{\e\z}\pl_\bt k_{\g\e}\pl_\al k_{\dl\z}
  +2\pl^2_{\al\bt}\ln\varrho\,k_{\g\dl}\right).
\end{split}
\end{equation}
Теперь нетрудно найти выражение для скалярной кривизны
\begin{equation}                                                  \label{escurh}
  R=\vr^{-4}\left[\vr^2R^{(k)}+2(n-1)\vr\pl_\al(k^{\al\bt}\pl_\bt\vr)
  +(n-1)(n-4)\pl\vr^2\right],
\end{equation}
где
\begin{equation*}
  \pl\varrho^2:=k^{\al\bt}\pl_\al\varrho\pl_\bt\varrho,
\end{equation*}
и ``скалярная'' кривизна $R^{(k)}$ для плотности метрики $k_{\al\bt}$
принимает удивительно простой вид
\begin{equation}                                                  \label{escrha}
  R^{(k)}=\pl_{\al\bt}k^{\al\bt}
  +\frac12k^{\al\bt}\pl_\al k^{\g\dl}\pl_\g k_{\bt\dl}
  -\frac14k^{\al\bt}\pl_\al k^{\g\dl}\pl_\bt k_{\g\dl}.
\end{equation}
Отметим, что это выражение полиномиально и по плотности метрики $k_{\al\bt}$,
и по ее обратной $k^{\al\bt}$.

Заметим, что для заданной плотности метрики $k_{\al\bt}$ мы можем формально
построить символы Кристоффеля, тензоры кривизны и Риччи, а также скалярную
кривизну. Соответствующие ``символы Кристоффеля'' не определяют связность на
$\MM$, а ``кривизна'' не является тензором, потому что новая переменная
$k_{\al\bt}$ является не тензором, а тензорной плотностью. Например, скаляром в
выражении (\ref{escurh}) является не просто $R^{(k)}$, а вся правая часть
выражения. Вместе с этим, группа общих преобразований координат содержит
подгруппу, состоящую из преобразований с единичным якобианом. Относительно этой
подгруппы ``символы Кристоффеля'' $\Gamma^{(k)}_{\al\bt}{}^\g$ для $k_{\al\bt}$
преобразуются, как компоненты связности, а ``кривизна''
$R^{(k)}_{\al\bt\g}{}^{\dl}$ является тензором.

Вторые производные $\pl^2_{\al\bt}k^{\al\bt}$ и $\pl_\al(k^{\al\bt}\pl_\bt\vr)$,
как и в общей теории относительности, можно исключить из действия
Гильберта--Эйнштейна (\ref{ehieia}), вычтя граничный член
\begin{equation*}
  \pl_\al\big(\vr^{n-2}\pl_\bt k^{\al\bt}
  +2(n-1)\vr^{n-3}k^{\al\bt}\pl_\bt\vr\big).
\end{equation*}
Здесь, для краткости, мы положили $\kappa=0$. В результате действие примет вид
\begin{equation}                                                  \label{ehirid}
  S_{\Sh\Se}\overset{\div}{=}\int dxL_{\Sh\Se},
\end{equation}
где
\begin{equation}                                                  \label{ehieil}
\begin{split}
  L_{\Sh\Se}
  =\vr^{n-4}&\left[\frac12\vr^2k^{\al\bt}\pl_\al k^{\g\dl}\pl_\g k_{\bt\dl}
  -\frac14k^{\al\bt}\pl_\al k^{\g\dl}\pl_\bt k_{\g\dl}-\right.
\\
  &\qquad \left.-(n-2)\vr\pl_\al k^{\al\bt}\pl_\bt\vr
  -(n-1)(n-2)\pl\vr^2-(n-2)\Lm\vr^4\phantom{\frac12}\right].
\end{split}
\end{equation}
При размерности пространства-времени $n\ge4$ этот лагранжиан полиномиален по
полям $\vr,k_{\al\bt}$ со степенями $n$ и $n+1$ соответственно. Общая степень
полинома равна $2n-1$. По-построению, с точностью до граничных слагаемых это
действие инвариантно относительно общих преобразований координат. Необходимо
только помнить, что поля $\vr$ и $k_{\al\bt}$ являются не тензорами, а
тензорными плотностями.

Лагранжиан (\ref{ehieil}) имеет вид лагранжиана дилатонной гравитации, где роль
дилатона играет определитель метрики. От обычных моделей он отличается тем, что
содержит перекрестный член с производными $\pl_\al k^{\al\bt}\pl_\bt\vr$. Кроме
того, он содержит меньшее число независимых полей, так как на плотность метрики
наложено условие (\ref{emehed}).

Дополнительное условие на плотность метрики (\ref{emehed}) можно учесть, добавив
к лагранжиану связь,
\begin{equation*}
  L_{\Sh\Se}\mapsto L_{\Sh\Se}+\lm(\det k_{\al\bt}\pm1),
\end{equation*}
где $\lm$ -- множитель Лагранжа.

Введение множителей Лагранжа необязательно. Из условия (\ref{emehed}) следует
условие на вариации плотности метрики
\begin{equation}                                                  \label{ecomev}
  \dl\det k_{\al\bt}=\pm k^{\al\bt}\dl k_{\al\bt}
  =\mp\dl k^{\al\bt}k_{\al\bt}=0.
\end{equation}
Поэтому действие (\ref{ehirid}) можно варьировать по $k_{\al\bt}$ или
$k^{\al\bt}$, рассматривая все компоненты как независимые, а затем взять
бесследовую часть получившихся уравнений.

Уравнения Эйлера--Лагранжа для действия (\ref{ehirid}) эквивалентны вакуумным
уравнениям Эйнштейна (\ref{eqevac}). Вариация действия по $\varrho$ приводит к
уравнению
\begin{equation}                                                  \label{etreie}
  2(n-1)\vr\square\vr+(n-1)(n-4)\pl\vr^2+R^{(k)}\vr^2=n\Lm\vr^4,
\end{equation}
которое эквивалентно следу уравнений Эйнштейна (\ref{eqevac}). Вариация действия
(\ref{ehirid}) по плотности метрики дает $n(n+1)/2-1$ уравнение
\begin{multline}                                                  \label{etrles}
  \vr^2\left(R^{(k)}_{\al\bt}-\frac1nk_{\al\bt}R^{(k)}\right)+
\\
  +(n-2)(\vr\nb_\al\nb_\bt\vr-\nb_\al\vr\nb_\bt\vr)
  -\frac{n-2}nk_{\al\bt}(\vr\square\vr-\pl\vr^2)=0,
\end{multline}
где
\begin{equation}                                                  \label{edefno}
\begin{split}
  \nb_\al\vr&=\pl_\al\vr,
\\
  \nb_\al\nb_\bt\vr&=\pl^2_{\al\bt}\vr-\Gamma^{(k)}_{\al\bt}{}^\g\pl_\g\vr,
\\
  \square\vr&=\pl_\al(k^{\al\bt}\pl_\bt\vr),
\end{split}
\end{equation}
и символы Кристоффеля $\Gamma^{(k)}_{\al\bt}{}^\g$ построены по плотности метрики
$k_{\al\bt}$. Эти уравнения эквивалентны бесследовой части уравнений Эйнштейна,
потому что при варьировании необходимо учесть условие на вариации компонент
плотности метрики (\ref{ecomev}). Заметим, что ``ковариантные'' производные
тензоров и тензорных плотностей при условии (\ref{emehed}) совпадают, так как
$\Gamma^{(k)}_{\al\bt}{}^\bt=0$. ``Тензор Риччи'' в (\ref{etrles}) имеет вид
\begin{equation*}
\begin{split}
  R^{(k)}_{\al\bt}&=\frac12k^{\g\dl}(\pl^2_{\g\dl}k_{\al\bt}
  -\pl^2_{\al\g}k_{\bt\dl}-\pl^2_{\bt\g}k_{\al\dl})-\frac12\pl_\g k^{\g\dl}
  (\pl_\al k_{\bt\dl}+\pl_\bt k_{\al\dl}-\pl_\dl k_{\al\bt})
\\
  &-\frac14\pl_\al k_{\g\dl}\pl_\bt k^{\g\dl}-\frac12k^{\g\dl}k^{\e\z}
  (\pl_\g k_{\al\e}\pl_\dl k_{\bt\z}-\pl_\g k_{\al\e}\pl_\z k_{\bt\dl}).
\end{split}
\end{equation*}
Уравнения (\ref{etreie}) и (\ref{etrles}) представляют собой соответственно след
и бесследовую часть уравнений Эйнштейна (\ref{eqevac}). Для любого решения
уравнений (\ref{etreie}) и (\ref{etrles}) существует единственная метрика
(\ref{eghrel}), которая удовлетворяет уравнениям Эйнштейна (\ref{eqevac}).
Наоборот, для любого решения уравнений Эйнштейна (\ref{eqevac}) можно построить
единственные плотность метрики и поле $\varrho$ (\ref{erhokg}), которые будут
удовлетворять уравнениям (\ref{etreie}) и (\ref{etrles}).

Таким образом, при $n\ge4$, действие (\ref{ehirid}), уравнения Эйлера--Лагранжа
(\ref{etreie}), (\ref{etrles}) и связь (\ref{emehed}) полиномиальны по полям
$\vr,k_{\al\bt}$. Это существенное упрощение теории достигнуто за счет
расширения конфигурационного пространства путем введения дополнительной полевой
переменной $\vr$. Если связь (\ref{emehed}) явно решить относительно одной из
компонент плотности метрики $k_{\al\bt}$ и подставить найденное решение в
действие (\ref{ehirid}), то полиномиальность теории будет утеряна. Отметим, что
введение нового поля и связи не является чем-то, из ряда вон выходящим: исходная
метрика $g_{\al\bt}$ и так содержит нефизические степени свободы, которые
исключаются из теории путем решения калибровочных условий и связей, содержащихся
в общей теории относительности. Мы лишь увеличили число переменных и связей,
оставив физические степени свободы прежними.

На гамильтоновом языке проведенная процедура означает следующее. Фазовое
пространство, соответствующее переменным $\vr,k_{\al\bt}$, также расширится,
при этом возникнет дополнительная связь на импульсы (равенство нулю следа
импульсов, соответствующих $k_{\al\bt}$), которая вместе со связью
(\ref{emehed}) образует пару связей второго рода. Однако полное фазовое
пространство уже будет не симплектическим, а только пуассоновым, так как
матрица скобок Пуассона координат этого пространства будет вырождена.
Подробнее этот вопрос рассмотрен далее в разделе \ref{scapol}.

Добавление скалярного $\vf(x)$ и электромагнитного $A_\al(x)$ полей сохраняет
действие и уравнения Эйлера--Лагранжа полиномиальными. При минимальной
подстановке получаем следующие лагранжианы:
\begin{align*}
  L_{\Ss}&=\frac12\sqrt g \big[g^{\al\bt}\pl_\al\vf\pl_\bt\vf-V(\vf)\big]
  =\frac12\vr^{n-2}\big[k^{\al\bt}\pl_\al\vf\pl_\bt\vf-\vr^2V(\vf)\big],
\\
  L_{\Se\Sm}&=-\frac14\sqrt gg^{\al\bt}g^{\g\dl}F_{\al\g}F_{\bt\dl}
  =-\frac14\vr^{n-4}k^{\al\bt}k^{\g\dl}F_{\al\g}F_{\bt\dl},
\end{align*}
где $V(\vf)$ -- потенциал скалярного поля, включая массовый член, и
$F_{\al\bt}:=\pl_\al A_\bt-\pl_\bt A_\al$ -- напряженность электромагнитного
поля.

В четырехмерном пространстве-времени уравнение (\ref{etreie}) упрощается
\begin{equation}                                                  \label{efodtr}
  \square\vr+\frac16R^{(k)}\vr=\frac23\Lm\vr^3.
\end{equation}
При этом мы разделили его на $\vr$, что возможно в силу предположения $\vr>0$.

Отбросим на время условие на плотность метрики (\ref{efodtr}) и будем считать
$k_{\al\bt}$ метрикой, а $\vr$ -- скалярным полем. Тогда уравнение
(\ref{efodtr}) ковариантно относительно конформных преобразований
\begin{equation}                                                  \label{econmt}
  \bar k_{\al\bt}=\Om^2k_{\al\bt},\qquad \bar\vr=\Om^{-1}\vr,
\end{equation}
где $\Om(x)>0$ -- дважды дифференцируемая функция. Оно рассматривалось в
\cite{Penros65,CheTag68,Ibragi68R} при $\Lm=0$ и в \cite{CaCoJa70} при
$\Lm\ne0$. Рассматриваемый подход существенно отличается, поскольку на плотность
метрики $k_{\al\bt}$ наложено дополнительно условие (\ref{emehed}), которое явно
нарушает конформную инвариантность. Однако появление множителя $1/6$ в
уравнении (\ref{efodtr}) не случайно,
так как параметризация (\ref{eghrel}) по виду совпадает с конформным
преобразованием метрики (\ref{econmt}).

Модели гравитации для метрики с единичным определителем рассматривались в физике
неоднократно. В общековариантных теориях, которой является общая теория
относительности, существует произвол в выборе системы координат, которым можно
воспользоваться. Нетрудно доказать, что в окрестности произвольной точки
пространства-времени существует такая система координат, что определитель
метрики равен по модулю единице $|\det g_{\al\bt}|=1$. Такие системы координат
определены с точностью до преобразований координат с единичным якобианом.
Условие $|\det g_{\al\bt}|=1$ существенно упрощает многие формулы, в частности,
выражение для скалярной кривизны. Этим обстоятельством пользовался Эйнштейн при
создании общей теории относительности \cite{Einste16R}.
\section{Точечные частицы в теории гравитации}
Пусть задано пространство-время, т.е.\ многообразие $\MM$ с метрикой $g$
лоренцевой сигнатуры. Будем считать, что кручение и неметричность на $\MM$
равны нулю, и, для простоты, не будем помечать это обстоятельство знаком
тильды. Размерность пространства-времени пока не фиксируем, $\dim\MM=n\ge2$.
Точечная частица движется в пространстве-времени $\MM$ по некоторой
дифференцируемой времениподобной кривой $\lbrace q^\al(\tau)\rbrace\in\MM$,
где $\tau\in\MR$ -- произвольный параметр вдоль этой кривой. Напомним, что
кривая называется времениподобной, если вектор скорости кривой, $u^\al:=\dot
q^\al:=dq^\al/d\tau$, времениподобен, т.е.\ $u^2:=u^\al u^\bt g_{\al\bt}>0$.
Мы считаем, что частица движется в будущее, т.е.\ $u^0>0$. Форма кривой
определяется рассматриваемой задачей и силами, которые действуют на частицу.
В общем случае параметр вдоль кривой произволен, и его выбирают из
соображений удобства. Наиболее часто в качестве параметра вдоль траектории
частицы выбирают ее длину $s$ (канонический параметр). Это всегда возможно,
т.к.\ обыкновенное дифференциальное уравнение $ds=\sqrt{\dot q^\al\dot q^\bt
g_{\al\bt}}\,d\tau$ разрешимо.
\begin{defn}
Времениподобная дифференцируемая кривая $\lbrace q^\al(\tau)\rbrace\in\MM$,
вдоль которой движется точечная частица, называется {\em траекторией} или
{\em мировой линией} частицы. Если параметр вдоль траектории частицы
совпадает с ее длиной, $\tau=s$, то он называется  {\em собственным
временем}. Векторное поле с компонентами
\begin{equation}                                                  \label{evelos}
  u^\al:=\frac{dq^\al}{ds},
\end{equation}
определенное на траектории частицы, называется {\em собственной скоростью}
частицы. Ковариантная производная от скорости частицы вдоль ее траектории
\begin{equation}                                                  \label{eacdep}
  w^\al:=\frac{dq^\bt}{ds}\nb_\bt u^\al=u^\bt\nb_\bt u^\al
\end{equation}
называется {\em ускорением} частицы. Частица называется {\em свободной}, если
ее ускорение равно нулю. Производная вдоль траектории частицы
\begin{equation}                                                  \label{evelob}
  v^\al:=\frac{dq^\al}{dq^0}=\frac{\dot q^\al}{\dot q^0}
\end{equation}
называется {\em наблюдаемой скоростью} частицы в системе координат $x^\al$.
\qed\end{defn}
\index{Траектория частицы (particle trajectory)}%
\index{Мировая линия частицы (particle world line)}%
\index{Линия мировая частицы (particle world line)}%
\index{Собственное время (proper time)}\index{Время собственное (proper time)}%
\index{Собственная скорость (proper velocity)}%
\index{Скорость собственная (proper velocity)}%
\index{Ускорение (acceleration)}%
\index{Свободная частица (free particle)}%
\index{Частица свободная (free particle)}%
\index{Наблюдаемая скорость}%
Собственная скорость и ускорение частицы являются $n$-мерными векторами,
определенными вдоль траектории частицы. Собственное время -- это то время,
которое показывают часы наблюдателя, движущегося вместе с частицей. Когда
наблюдатель движется вместе с частицей, то он может измерить свою скорость
относительно системы координат $x^\al$, это и будут компоненты собственной
скорости. Наблюдаемая скорость, как следует из определения, не является
векторным полем. Это та скорость, которую измеряет внешний наблюдатель в
выбранной системе отсчета.

Равенство нулю ускорения частицы
\begin{equation*}
  u^\bt\nb_\bt u^\al=0,
\end{equation*}
определяет экстремали (\ref{excoeq}). Это значит, что свободные частицы в
теории тяготения движутся вдоль экстремалей пространства-времени. Если на
частицу действуют негравитационные силы, например, электромагнитные, то в
уравнении движения $mw^\al=f^\al$, где $m=\const$ -- масса частицы, появится
внешняя сила с компонентами $f^\al$. В этом случае ее траектория будет
отличаться от экстремали.
\begin{prop}
Для любой точечной частицы квадрат собственной скорости равен единице,
\begin{equation}                                                  \label{evesqu}
  u^2:=u^\al u^\bt g_{\al\bt}=1.
\end{equation}
При этом ускорение всегда ортогонально скорости
\begin{equation*}
  u^\al w_\al=0.
\end{equation*}
\end{prop}
\begin{proof} Первое утверждение следует из определения:
\begin{equation*}
  u^2=\frac{dx^\al}{ds}\frac{dx^\bt}{ds}g_{\al\bt}=\frac{ds^2}{ds^2}=1.
\end{equation*}
Продифференцируем это равенство вдоль траектории
\begin{equation*}
  u^\bt\nb_\bt(u^2)=2u^\bt u^\al\nb_\bt u_\al=2u^\al w_\al=0,
\end{equation*}
где мы воспользовались тем, что ковариантная производная от метрики для
связности Леви--Чивиты равна нулю. Отсюда вытекает второе утверждение
предложения.
\end{proof}

Наблюдаемая скорость является нековариантным объектом. Из определения следует
соотношение между компонентами собственной и наблюдаемой скоростями
\begin{equation}                                                  \label{eobved}
  v^\al=\frac{u^\al}{\dot q^0}.
\end{equation}
Более подробно,
\begin{equation*}
  v^0=1,\qquad v^\mu=\frac{dq^\mu}{dq^0}.
\end{equation*}
Возведем равенство $v^\al\dot q^0=u^\al$ в квадрат и учтем, что $u^2=1$.
Тогда получим, что компоненты собственной скорости частицы можно записать в
следующем виде
\begin{equation}                                                  \label{eobser}
  u^0:=\dot q^0=\frac1{\sqrt{v^2}},\qquad\qquad
  u^\mu:=\dot q^\mu=\frac{v^\mu}{\sqrt{v^2}},
\end{equation}
где $v^2:=v^\al v^\bt g_{\al\bt}$ -- квадрат наблюдаемой скорости. Отсюда
также следует, что $v^2>0$. Понятие наблюдаемой скорости частицы будет
использовано в дальнейшем при определении ультрарелятивистского предела для
точечной частицы.

\begin{com}
Определения даны для точечной частицы, находящейся под действием произвольных
сил, не только гравитационных. Мы также предполагаем, что, если наблюдатель
находится в определенной точке пространства в определенный момент времени, то
ему известны значения всех координат $x^\al$, соответствующих данной точке
пространства-времени. Мы также предполагаем, что ему известны координаты всех
близлежащих точек, что необходимо для вычисления производных. Эти упрощающие
предположения сделаны для того, чтобы обойти важный и интересный, но сложный
вопрос измерений. \qed\end{com} В общем случае, когда на частицу действуют
произвольные силы, она может двигаться по любой времениподобной кривой. В
моделях гравитации мы предполагаем, что точечная частица, на которую
действуют только гравитационные силы, описывается инвариантным действием
\begin{equation}                                                  \label{epoacp}
  S_\text{m}:=-mc\int_\Sp^\Sq\!\!\!ds
  =-mc\int_{\tau_1}^{\tau_2}\!\!\!d\tau\sqrt{\dot q^\al\dot q^\bt g_{\al\bt}},
\end{equation}
где $m=\const>0$ -- масса частицы, $c$ -- скорость света и интегрирование
проводится вдоль времениподобной кривой $q(\tau)$, $\tau\in\MR$, соединяющей
точки $\Sp=q(\tau_1)$ и $\Sq=q(\tau_2)$. Действие (\ref{epoacp}) отличается
от длины мировой линии частицы (\ref{eactim}) постоянным множителем $-mc$ и,
если метрика задана, варьируется только по траектории частицы $\dl
q^\al(\tau)$. В дальнейшем положим $c=1$.

Допустим, что точки $\Sp,\Sq\in\MM$ можно соединить времениподобной
экстремалью и притом только одной. Ясно, что эти точки всегда можно соединить
также ломанными светоподобными кривыми $\g$. Длина этих кривых равна нулю,
потому что каждый сегмент ломанных светоподобен. Если некоторая
времениподобная кривая аппроксимирует одну из этих кривых $\g$, то ее длина
будет близка к нулю. Остальные времениподобные кривые, соединяющие точки
$\Sp$ и $\Sq$ будут иметь некоторую положительную длину. Поэтому, если точки
$\Sp$ и $\Sq$ соединяются только одной экстремалью, то это будет
времениподобная кривая наибольшей длины. Благодаря знаку минус перед
действием (\ref{epoacp}), экстремаль соответствует минимуму функционала
(\ref{epoacp}) среди всех времениподобных кривых, соединяющих точки $\Sp$ и
$\Sq$. Отсюда следует, что точечные частицы движутся вдоль экстремалей
пространства-времени, если на них действуют только внешние гравитационные
силы.

Если метрика пространства-времени считается заданной, то это означает, что мы
не учитываем гравитационное поле, создаваемое самой частицей. Во многих
случаях такая постановка задачи вполне допустима, например, если масса
частицы мала. Как уже говорилось, такие частицы называются {\em пробными}.

\begin{com}
Мы предполагаем, что точечная частица описывается действием (\ref{epoacp})
независимо от того, задана ли на $\MM$ аффинная геометрия общего вида с
кручением и неметричностью или задана только метрика. Тогда в общем случае
точечная частица будет двигаться по экстремали, а не по геодезической. Если
кручение и неметричность пространства-времени равны нулю, то экстремали
совпадают с геодезическими. Поэтому в общей теории относительности можно
также сказать, что точечные частицы под действием гравитационных сил движутся
по геодезическим. Если, помимо гравитационных сил, присутствуют также другие,
например, электромагнитные взаимодействия, то вид действия (\ref{epoacp})
изменится. Поэтому при наличии сил негравитационного происхождения траектории
частиц, вообще говоря, отличаются от экстремалей. \qed\end{com}

Выше было введено понятие собственного времени для произвольной
времениподобной линии. Для экстремали собственное время по существу совпадает
с каноническим параметром. Действительно, для любой экстремали квадрат
вектора скорости сохраняется, $u^2=\const$, (\ref{efirin}). Поскольку
канонический параметр определен с точностью до умножения на отличную от нуля
постоянную, то его всегда можно выбрать таким образом, чтобы квадрат вектора
скорости частицы был равен единице (\ref{evesqu}). Другими словами, в
качестве канонического параметра выбирается длина экстремали, $\tau=s$.

Предположим, что в пространстве-времени находится $\Sn$ частиц с массами
$m_\Si$, $\Si=1,\dotsc,\Sn$, которые взаимодействуют между собой только
посредством гравитационных сил. В общей теории относительности суммарное
действие гравитационного поля и совокупности точечных частиц равно сумме
действия Гильберта--Эйнштейна (\ref{ehieia}) и действий для каждой частицы,
\begin{equation*}
\begin{split}
  S&=S_{\Sh\Se}+\sum_{\Si=1}^\Sn S_\Si=
\\
  &=\kappa\int_\MM\!\! dx\vol R
  -\sum_\Si m_\Si\int_{\tau_{\Si_1}}^{\tau_{\Si_2}}
  \!\! d\tau_\Si\sqrt{\dot q^\al_\Si\dot q^\bt_\Si g_{\al\bt}},
\end{split}
\end{equation*}
где мы, для простоты, опустили космологическую постоянную и знак тильды у
скалярной кривизны. Во втором слагаемом метрика рассматривается как сложная
функция $g_{\al\bt}(\tau_\Si):=g_{\al\bt}\big(q(\tau_\Si)\big)$, и параметры
$\tau_\Si$ могут быть выбраны произвольно для каждой частицы. Первый интеграл
берется по всему пространству-времени, а последующие -- в пределах
$\tau_{\Si_{1,2}}$ (возможно, бесконечных), которые соответствуют пересечению
мировых линий частиц с краем $\pl\MM$. Для простоты, предположим, что мировые
линии частиц нигде не пересекаются, т.е.\ частицы не сталкиваются между
собой.

В настоящем разделе нас не будут интересовать граничные эффекты. Поэтому
пределы интегрирования, для простоты, мы в дальнейшем опустим.

Действие (\ref{epohep}) инвариантно относительно общих преобразований
координат и независимой перепараметризации параметров $\tau_\Si$ вдоль каждой
траектории. Более того, поскольку в действие входит сумма интегралов вдоль
траекторий, то индекс $\Si$ у параметров $\tau_\Si$ можно опустить:
\begin{equation}                                                  \label{epohep}
    S=\int\!\! dx\vol\kappa R
  -\int\!\! d\tau\sum_\Si m_\Si\sqrt{\dot q^\al_\Si\dot q^\bt_\Si g_{\al\bt}}.
\end{equation}
В общем случае пределы интегрирования для различных частиц могут отличаться.
Однако, поскольку нас не интересуют граничные эффекты, мы этого указывать не
будем.

В разделе \ref{sextre} действие для экстремали было проварьировано в
предположении, что вдоль нее выбран канонический параметр. Сейчас мы получим
уравнения без этого предположения. Рассмотрим одну точечную частицу. Простые
вычисления приводят к следующим уравнениям движения
\begin{equation}                                                  \label{qspjum}
  S_{\rm m},_\al:=\frac{\dl S_{\rm m}}{\dl q^\al}=m\frac{g_{\al\bt}}{u^2}
  \left(\ddot q^\bt+\Gamma_{\g\dl}{}^\bt\dot q^\g\dot q^\dl-\frac{\dot x^\bt}{2u^2}
  \frac{d u^2}{d\tau}\right)=0.
\end{equation}
Поскольку исходное действие инвариантно относительно произвольной замены
параметра вдоль мировой линии, то согласно второй теореме Нетер между
уравнениями существует линейная зависимость. Чтобы ее найти рассмотрим
бесконечно малое изменение параметра
\begin{equation*}
  \tau\mapsto\tau+\e(\tau).
\end{equation*}
Соответствующая вариация формы функций $q^\al(\tau)$ (см., раздел
\ref{sinfct}) имеет вид
\begin{equation*}
  \dl q^\al(\tau)=-\e\dot q^\al.
\end{equation*}
Следовательно, вариация действия равна
\begin{equation*}
  \dl S_{\rm m}=-\int d\tau S_{\rm m},_\al\e\dot q^\al.
\end{equation*}
Поскольку функции $\e(\tau)$ произвольны, то из инвариантности действия, $\dl
S_{\rm m}=0$, следует зависимость уравнений движения:
\begin{equation}                                                  \label{qggftl}
  S_{\rm m},_\al\dot q^\al=0.
\end{equation}
В этом тождестве нетрудно убедиться прямой проверкой.

При произвольной параметризации мировой линии квадрат вектора скорости не
является постоянным, $u^2\ne\const$. Если выбран канонический параметр вдоль
экстремали, то $d u^2/d\tau=0$ и последнее слагаемое в уравнении
(\ref{qspjum}) равно нулю.

Действие для совокупности частиц (\ref{epohep}) инвариантно относительно
независимой перепараметризации каждой мировой линии. Поэтому вдоль каждой
кривой можно выбрать свой канонический параметр. В дальнейшем мы это
предположим.

Для вывода полной системы уравнений движения, действие (\ref{epohep})
необходимо проварьировать по метрике $g_{\al\bt}(x)$ и траекториям частиц
$q^\al_\Si(\tau)$. По траекториям частиц мы уже проварьировали. Поэтому
рассмотрим вариацию действия по метрике. Ниже мы проведем формальные
выкладки, считая компоненты метрики достаточно гладкими функциями всюду,
включая мировые линии частиц. На самом деле это не так. Детальный анализ
уравнений движения показывает, что компоненты метрики расходятся на мировых
линиях.

Вариация действия для частиц (\ref{epohep}) по компонентам метрики $\dl
g_{\al\bt}(x)$ неопределена, т.к.\ оно записано только вдоль траекторий.
Поэтому мы преобразуем интегралы вдоль траекторий в интегралы по всему
пространству-времени. Для этого вставим в подынтегральное выражение единицу,
\begin{equation*}
  1=\int\!\! dx\,\dl(x-q_\Si):=\int\!\!dx\,\dl(x^0-q_\Si^0)\dl(x^1-q_\Si^1)
  \dots\dl(x^{n-1}-q_\Si^{n-1}),
\end{equation*}
где
\begin{equation*}
  \dl(x):=\dl(x^0)\dl(x^1)\dots\dl(x^{n-1})
\end{equation*}
-- $n$-мерная $\dl$-функция, и изменим порядок интегрирования, предположив,
что это возможно. Тогда действие примет вид
\begin{equation}                                                  \label{epopgr}
  S=\int\!\! dx\left[\kappa\vol R-\int\!\!d\tau\sum_\Si m_\Si
  \sqrt{\dot q^\al_\Si\dot q^\bt_\Si g_{\al\bt}}\,\dl(x-q_\Si)\right].
\end{equation}
Теперь метрику во втором слагаемом можно рассматривать, как функцию от точки
пространства-времени, $g_{\al\bt}=g_{\al\bt}(x)$. Вариационные производные
этого действия по траекториям частиц и метрике равны
\begin{align}                                                     \label{epoint}
  \frac{\dl S}{\dl q_\Si^\al}
  &=m_\Si\left(\ddot q_\Si^\bt+\Gamma_{\g\dl}{}^\bt\dot q_\Si^\g\dot
  q_\Si^\dl\right)g_{\bt\al},
\\                                                                \label{epopem}
  \frac{\dl S}{\dl g_{\al\bt}}
  &=-\vol\kappa\left(R^{\al\bt}-\frac12g^{\al\bt}R\right)
  -\int\!\!d\tau\sum_\Si\frac{m_\Si}{2\sqrt{\dot q_\Si^\g\dot
  q_\Si^\dl g_{\g\dl}}}\dot q_\Si^\al\dot q_\Si^\bt\,\dl(x-q_\Si).
\end{align}
Вариационная производная действия Гильберта--Эйнштейна также была получена
ранее в разделе (\ref{svarhe}). Для любого решения уравнения (\ref{epoint})
параметр $\tau$ можно выбрать так, что $\sqrt{\dot q_\Si^\g\dot q_\Si^\dl
g_{\g\dl}}=1$ для каждой частицы. Поэтому, не ограничивая общности,
знаменатель во втором слагаемом (\ref{epopem}) можно упростить, отбросив
квадратный корень. Таким образом, полная система уравнений движения
гравитационного поля и системы точечных частиц примет вид
\begin{align}                                                     \label{eincou}
 \kappa\vol\left(R^{\al\bt}-\frac12g^{\al\bt}R\right)&
 =-\frac12\vol\, T_{\text{m}}^{\al\bt},
\\                                                                \label{eingeo}
  g_{\al\bt}\left(\ddot q_\Si^\bt+\Gamma_{\g\dl}{}^\bt\dot q_\Si^\g
  \dot q_\Si^\dl\right)&=0,
\end{align}
где
\begin{equation}                                                  \label{enmopo}
  T_{\text{m}}{}^{\al\bt}
  =\frac1\vol\int\!\!d\tau\sum_\Si m_\Si\dot q_\Si^\al\dot q_\Si^\bt\dl(x-q_\Si)
\end{equation}
-- тензор энергии-импульса точечных частиц. Интегрирование по каноническому
параметру $\tau$ в тензоре энергии-импульса можно снять, использовав одну
$\dl$-функцию, а именно $\dl\big(x^0-q_\Si^0(\tau)\big)$. Поскольку $\dot
q_\Si^0>0$ (все частицы движутся в будущее), то для тензора энергии-импульса
точечных частиц получаем следующее выражение
\begin{equation}                                                  \label{enmopi}
  T_\text{m}{}^{\al\bt}=\frac1\vol\sum_\Si
  \frac{m_\Si\dot q_\Si^\al\dot q_\Si{}^\bt}{\dot q^0_\Si}\dl(\Bx-\Bq_\Si),
\end{equation}
где
\begin{equation*}
  \dl(\Bx-\Bq_\Si):=\dl(x^1-q^1_\Si)\dotsc\dl(x^{n-1}-q^{n-1}_\Si)
\end{equation*}
-- пространственная $(n-1)$-мерная $\dl$-функция, и параметр $\tau$ является
неявной функцией $x^0$, заданной уравнением $x^0=q^0(\tau)$.
\begin{com}
Появление множителя $1/\vol$ в выражении для тензора энергии-импульса
точечных частиц неслучайно. Напомним, что $\dl$-функция является не функцией
на многообразии, а скалярной плотностью степени $-1$, как и элемент объема
$\vol$. \qed\end{com} Таким образом, для точечных частиц, на которые
действуют только гравитационные силы, мы имеем связанную систему уравнений
(\ref{eincou}), (\ref{eingeo}). Каждая частица движется по экстремали
пространства-времени в соответствии с уравнением (\ref{eingeo}), где метрика
определяется уравнениями Эйнштейна (\ref{eincou}). В свою очередь, метрика
зависит от распределения частиц, так как в правой части уравнений Эйнштейна
стоит нетривиальный тензор энергии-импульса.

Отметим трудности, которые возникают при решении уравнений (\ref{eincou}) и
(\ref{eingeo}) в связи с наличием $\dl$-функций.

Уравнение для экстремалей (траекторий точечных частиц) (\ref{enmopo}) хорошо
определено, если компоненты метрики -- дифференцируемые функции,
$g_{\al\bt}\in\CC^1(\MU)$, $\MU\subset\MM$. Однако детальный анализ системы
уравнений движения показывает, что это условие не выполняется, т.к.\
компоненты метрики имеют особенности на мировых линиях частиц. Именно по этой
причине мы не опустили индексы в уравнениях (\ref{eincou}). В рассматриваемом
случае уравнения Эйнштейна с контравариантными и ковариантными индексами не
эквивалентны. По этой же причине мы не сократили множитель $\vol$ в
уравнениях Эйнштейна (\ref{eincou}), т.к.\ он обращается в нуль на мировых
линиях частиц. В уравнениях для экстремалей (\ref{eingeo}) мы также не
произвели каких либо манипуляций с индексами.

Наличие $\dl$-функций в полной системе уравнений приводит к серьезным
математическим трудностям. Поскольку есть $\dl$-функции, то решения системы
уравнений Эйнштейна надо понимать в обобщенном смысле после интегрирования с
пробными функциями. Если в качестве пробных функций выбрать пространство
гладких функций с финитными носителями $\CD(\MR^{(n-1)})$ (см., например,
\cite{Vladim88R}), то компоненты метрики должны лежать в сопряженном
пространстве $\CD'(\MR^{(n-1)})$. Однако уравнения Эйнштейна нелинейны, а
умножение в $\CD'(\MR^{(n-1)})$ в общем случае определить нельзя. Насколько
известно автору, решение этой проблемы в настоящее время отсутствует. Поэтому
вычисления данного раздела следует рассматривать, как ориентир, с которым
необходимо будет сравнивать более строгие выкладки.

В заключение раздела проведем еще одно обобщение. Действие (\ref{epohep})
описывает совокупность точечных частиц, взаимодействующих с гравитационным
полем. Если частицы находятся дополнительно под действием некоторых
потенциальных сил негравитационного происхождения, то действие можно
обобщить, вставив соответствующий потенциал $V(x)$,
\begin{equation}                                                  \label{epkhep}
    S=\int\!\! dx\vol\kappa R
  -\int\!\! d\tau \sum_\Si m_\Si V\sqrt{\dot q^\al_\Si\dot q^\bt_\Si
  g_{\al\bt}},
\end{equation}
где потенциал рассматривается на мировых линиях,
$V(\tau):=V\big(q_\Si(\tau)\big)$. При этом мы не нарушаем инвариантность
действия относительно независимого преобразования параметров мировых линий
частиц. Если вдоль каждой мировой линии выбрать канонический параметр, то
уравнения движения для действия (\ref{epkhep}) примут вид
\begin{align}                                                     \label{eincoj}
 \kappa\vol\left(R^{\al\bt}-\frac12g^{\al\bt}R\right)
 =-\frac12\vol\, T_{\text{m}}^{\al\bt} V,
\\                                                                \label{eingel}
  Vg_{\al\bt}\left(\ddot q_\Si^\bt+\Gamma_{\g\dl}{}^\bt\dot q_\Si^\g
  \dot q_\Si^\dl\right)+\dot x^\g\pl_\g Vg_{\al\bt}\dot q_\Si^\bt-\pl_\al V=0.
\end{align}
При $V\ne0$ второе уравнение можно разделить на $V$ и переписать в следующем
виде
\begin{equation}                                                  \label{efscju}
  g_{\al\bt}\left(\ddot q_\Si^\bt+\Gamma_{\g\dl}{}^\bt\dot q_\Si^\g
  \dot q_\Si^\dl\right)
  =\Pi^\St_{~\al}{}^\bt\pl_\bt \ln|V|,
\end{equation}
где
\begin{equation*}
  \Pi^\St_{~\al}{}^\bt:=\dl_\al^\bt-\dot q_{\Si\al}\dot q_\Si^\bt
\end{equation*}
-- проекционный оператор на направление, перпендикулярное к мировой линии
$\Si$-той частицы.
\subsection{Нерелятивистский предел для точечной частицы         \label{sextrp}}
В настоящем разделе будет показана связь между уравнениями движения для
точечных частиц (\ref{eingeo}) и хорошо знакомыми уравнениями движения частиц
под действием гравитационного поля в механике Ньютона.

Для простоты, рассмотрим движение одной частицы. В пространстве-времени $\MM$
с нетривиальной метрикой $g_{\al\bt}(x)$ функции $\lbrace
q^{\al}(\tau)\rbrace$ задают мировую линию точечной частицы. Пусть $\tau=t$
-- канонический параметр (собственное время). Используя инвариантность
действия (\ref{epoacp}) относительно общих преобразований координат, выберем
такие координаты, чтобы координата $x^0$ на траектории частицы совпадала с
собственным временем, $c\tau=ct=x^0$. Здесь мы ввели явно скорость света $c$
для того, чтобы в дальнейшем строить разложение по малому параметру
$\Bu^2/c^2\ll1$, где $\Bu$ -- пространственная часть собственной скорости
частицы. Условимся нумеровать, как обычно, пространственные координаты
буквами из середины греческого алфавита:
\begin{equation*}
  \lbrace x^\al\rbrace=\lbrace x^0,x^\mu\rbrace,\qquad\mu=1,\dotsc,n-1.
\end{equation*}
Поскольку исходное действие (\ref{epoacp}) инвариантно относительно общих
преобразований координат, то у нас имеется возможность дополнительно
фиксировать $n-1$ компонент метрики. Положим $g_{0\mu}=0$. Тогда метрика
примет блочно диагональный вид
\begin{equation}                                                  \label{etiblm}
  g_{\al\bt}=\begin{pmatrix}g_{00} & 0 \\0&g_{\mu\nu}\end{pmatrix}
\end{equation}
Другими словами, система координат выбрана таким образом, чтобы
времениподобный вектор $\pl_0$ был ортогонален всем касательным векторам к
пространственным сечениям $x^0=\const$.

Введем два параметра разложения. Во-первых, нерелятивистский предел
соответствует скоростям, малым по сравнению со скоростью света,
\begin{equation*}
  \frac{\Bu^2}{c^2}\sim\e\ll1,
  \qquad\Bu^2:=-\eta_{\mu\nu}\dot q^\mu \dot q^\nu=\dl_{\mu\nu}u^\mu u^\nu\ge0.
\end{equation*}
Во-вторых, слабому гравитационному полю соответствует метрика, которая мало
отличается от метрики Минковского:
\begin{equation*}
  g_{00}=1+h_{00},\qquad g_{\mu\nu}=\eta_{\mu\nu}+h_{\mu\nu},\qquad
  h_{00},h_{\mu\nu}\sim\e\ll1.
\end{equation*}
Поправки к метрике $h_{00}(x)$ и $h_{\mu\nu}(x)$ в общем случае зависят от всех
координат.

Поскольку в выбранной системе координат $\dot q^0=c$, то наблюдаемая и
собственная скорости частицы совпадают, $v^\al=u^\al$. Это следует из
определения (\ref{evelob}), если в нем восстановить скорость света. Поэтому
пределы $\Bu^2\to0$ и $\Bv^2\to0$, где $\Bv^2:=-\eta_{\mu\nu}v^\mu
v^\nu\to0$, эквивалентны.

Нерелятивистской частице в слабом гравитационном поле соответствует интервал
\begin{equation*}
  ds^2=(c^2+c^2h_{00}+\eta_{\mu\nu}u^\mu u^\nu+h_{\mu\nu}u^\mu u^\nu)dt^2,
\end{equation*}
где мы ограничились первыми неисчезающими поправками к метрике.
В силу сделанных предположений о малости гравитационного поля и скоростей
последним слагаемым в этом представлении можно пренебречь. Тогда в
нерелятивистском пределе с учетом только первой поправки интервал для
точечной частицы примет вид
\begin{equation}                                                  \label{eintpp}
  ds^2\simeq\left(c^2+\frac{2U}m-\Bu^2\right)dt^2,
\end{equation}
где введено обозначение
\begin{equation}                                                  \label{ehzzde}
  h_{00}:=\frac{2U}{mc^2}.
\end{equation}
Подставим приближенное выражение для интервала (\ref{eintpp}) в действие для
точечной частицы (\ref{epoacp}), умноженное на скорость света, и разложим по
степеням $\e$. Тогда в первом порядке по $\e$ получим приближенное выражение
\begin{equation}                                                  \label{eponor}
\begin{split}
  S_\mathrm{m}&\simeq-mc\int d\tau\sqrt{c^2+\frac{2U}m-\Bu^2}\approx
\\
  &\simeq\int d\tau\left(-mc^2+\frac{m\Bu^2}2-U\right).
\end{split}
\end{equation}
С точностью до энергии покоя точечной частицы с обратным знаком $-mc^2$
подынтегральное выражение совпадает с хорошо известным выражением для
лагранжиана точечной частицы в нерелятивистской механике (\ref{epodpo}). Тем
самым мы показали, что в нерелятивистском пределе поправка к временн\'ой
компоненте метрики, умноженной на $mc^2$, следует интерпретировать, как
потенциальную энергию $U:=mc^2h_{00}/2$ точечной частицы, находящейся во
внешнем гравитационном поле.

Отметим, что разумный нерелятивистский предел обусловливает также выбор
общего знака минус в исходном действии для точечной частицы (\ref{epoacp}).

Мы проанализировали нерелятивистский предел для действия точечной частицы.
Однако на этом уровне остается вопрос об уравнениях движения. Дело в том, что
исходное действие для релятивистской частицы (\ref{epoacp}) варьируется по
$n$ компонентам траектории $q^\al$. В то же время действие для
нерелятивистской частицы (\ref{eponor}) варьируется только по
пространственным компонентам $q^\mu$. Следовательно, одно уравнение потеряно.
Поэтому необходимо проследить, что происходит в нерелятивистском пределе не
только с действием, но и с уравнениями движения.

Интервалу (\ref{eintpp}) соответствует метрика
\begin{equation}                                                  \label{exppau}
  g_{\al\bt}=\diag(1+2U/m,-1,\dots,-1).
\end{equation}
Символы Кристоффеля в первом порядке малости имеют только три нетривиальные
компоненты:
\begin{equation*}
  \Gamma_{00}{}^\mu=-\frac{\eta^{\mu\nu}\pl_\nu U}m,\qquad
  \Gamma_{0\mu}{}^0=\Gamma_{\mu0}{}^0=\frac{\pl_\mu U}m.
\end{equation*}
Соответствующие уравнения для экстремалей (\ref{eextre}) принимают вид
\begin{align}                                                     \label{extpoz}
  \ddot q^0&=-2\frac{\pl_\mu U}m\dot q^\mu,
\\                                                                \label{extpos}
  \ddot q^\mu&=-\frac{\dl^{\mu\nu}\pl_\nu U}m.
\end{align}
Поскольку в выбранной системе координат $\ddot q^0=0$, то первое уравнение
удовлетворяется с точностью $\e^2$. Второе уравнение совпадает с {\em вторым
законом Ньютона} для движения точечной частицы в гравитационном поле
\index{Второй закон Ньютона}\index{Ньютона второй закон}%
\begin{equation}                                                  \label{esenil}
  m\ddot q^\mu=-\dl^{\mu\nu}\pl_\nu U.
\end{equation}
Таким образом, потерянное уравнение выполняется с точностью $\e^2$.

Тот факт, что уравнение для $q^0$ удовлетворяется с рассматриваемой степенью
точности не является удивительным. Действительно, среди $n$ исходных уравнений
для точечной частицы имеется одна линейная зависимость (\ref{qggftl}). Поэтому
только $n-1$ уравнений являются независимыми, которые в нерелятивистском пределе
сводятся к уравнениям Ньютона (\ref{esenil}).

В общей теории относительности $(n=4)$ метрика вдали от точечной массы $M$
дается решением Шварцшильда
\begin{equation*}
  ds^2=\left(1-\frac {2M}r\right)dt^2-\frac{dr^2}{1-\frac {2M}r}
  -r^2\left(d\theta^2+\sin^2\theta d\vf^2\right).
\end{equation*}
Соответствующее нерелятивистское выражение для потенциальной энергии имеет вид
\begin{equation}                                                  \label{egralo}
  U=-G\frac{mM}r,
\end{equation}
где $G$ -- гравитационная постоянная, что совпадает с {\em законом всемирного
тяготения}. Таким образом мы показали, что закон всемирного тяготения
вытекает из общей теории относительности в нерелятивистском пределе.

В конце предыдущего раздела было рассмотрено действие точечных частиц,
взаимодействующих не только с гравитационным полем, но и с другими
потенциальными полями (\ref{epkhep}). Нерелятивистский предел в этом случае
определяется также, как и раньше. Дополнительно мы требуем, чтобы потенциал
$V$ мало отличался от единицы:
\begin{equation*}
  V=1+\frac W{mc^2},\qquad W\sim\e\ll1.
\end{equation*}
Тогда в нерелятивистском пределе действие точечной частицы (\ref{eponor})
примет вид
\begin{equation}                                                  \label{eponol}
  S_\mathrm{m}\simeq\int d\tau\left(-mc^2+\frac{m\Bu^2}2-U-W\right).
\end{equation}
В этом случае движение частицы определяется суммой гравитационного потенциала
$U$, который возник из $g_{00}$ компоненты метрики, и негравитационного
потенциала $W$, который возник из множителя $V$.
\subsection{Теория гравитации Ньютона}
В настоящем разделе мы опишем гравитационное взаимодействие точечных частиц в
механике Ньютона. При этом, по возможности, мы будем следовать общей схеме,
принятой в общей теории относительности.

Пусть пространство-время $\MM$ -- это тривиальное четырехмерное многообразие,
$\MM\approx\MR^4$, с декартовой системой координат $x=\lbrace
x^\al\rbrace=\lbrace x^0,x^\mu\rbrace=\lbrace t,\Bx\rbrace$, где
$\al=0,1,2,3$ и $\mu=1,2,3$. Мы отождествим нулевую координату с временем,
$x^0=t$. Пусть в $\MM$ движутся $\Sn$ точечных частиц по траекториям
$q_\Si(t)=\lbrace q_\Si^\mu(t)\rbrace$, $\Si=1,\dotsc,\Sn$. В механике
Ньютона время имеет абсолютный характер в том смысле, что оно одинаково для
всех частиц и играет роль параметра вдоль траектории каждой частицы.
Предположим, что между частицами действуют только гравитационные силы. Это
значит, что каждая частица движется в гравитационном поле, которое создается
другими частицами. В свою очередь каждая частица создает гравитационное поле,
которое влияет на движение других частиц. Обозначим суммарный потенциал
гравитационного поля через $\vf(x)$, который, по определению, является
функцией (скалярным полем) на $\MR^4$.

В механике Ньютона мы считаем, что на каждом пространственноподобном сечении
$t=\const$ задана евклидова метрика. В наших обозначениях она отрицательно
определена $\eta_{\mu\nu}=-\dl_{\mu\nu}$.

Действие для точечных частиц, взаимодействующих посредством гравитационного
поля, является суммой действия для гравитационного поля
\begin{equation}                                                  \label{eacgrn}
  S_{\rm g}
  =\frac1{4\pi G}\int_{\MR^4}
  \!\!\!dx\,\frac12\eta^{\mu\nu}\pl_\mu\vf \pl_\nu\vf,
\end{equation}
где $G$ -- гравитационная постоянная, и действия для точечных частиц
\begin{equation}                                                  \label{eacgpo}
  S_{\rm m}=-\sum_{\Si=1}^\Sn\int_{-\infty}^\infty\!\!\!dt\,
  \frac12m_\Si\dot q_\Si^\mu\dot q_\Si^\nu\eta_{\mu\nu}
  -\sum_{\Si=1}^\Sn\int_{\MR^4}\!\!\!dx\,m_\Si\dl(\Bx-\Bq_\Si)\vf.
\end{equation}
Действие для гравитационного поля (\ref{eacgrn}) отрицательно определено и
равно потенциальной энергии гравитационного поля, взятой с обратным знаком.
Действие для точечных частиц (\ref{eacgpo}), как обычно, представляет собой
разность кинетической и потенциальной энергии. Отметим, что потенциальная
энергия взаимодействия точечных частиц с гравитационным полем содержит только
трехмерную дельта-функцию.

Варьирование суммарного действия, $S=S_{\rm g}+S_{\rm m}$, по гравитационному
полю и траекториям частиц дает уравнения движения
\begin{align}                                                     \label{eqmogr}
  \frac{\dl S}{\dl\vf}&=\frac1{4\pi G}\triangle\vf
  -\sum_\Si m_\Si\dl(\Bx-\Bq_\Si)=0,
\\                                                                \label{eqmpon}
  \frac{\dl S}{\dl q_\Si}&=m_\Si\left(\ddot q_{\Si\mu}-\pl_\mu\vf\right)=0,
\end{align}
где $\triangle:=\pl_1^2+\pl_2^2+\pl_3^2$ -- трехмерный лапласиан, $\ddot
q_{\Si\mu}:=\ddot q_\Si^\nu\eta_{\mu\nu}$ и градиент потенциала
гравитационного поля $\pl_\mu\vf$ во втором уравнении берется в той точке
$q_\Si=\lbrace t,q_\Si^\mu\rbrace\in\MM$, где в данный момент времени
расположена частица.

Уравнение для гравитационного поля (\ref{eqmogr}) представляет собой
уравнение Пуассона
\begin{equation}                                                  \label{epoise}
  \triangle\vf=4\pi G\sum_\Si m_\Si\dl(\Bx-\Bq_\Si).
\end{equation}
Мы рассматриваем решения этого уравнения в слабом смысле, т.е.\ равенство
достигается после свертки левой и правой части с основными функциями. Если
ограничить класс рассматриваемых решений только теми решениями, которые равны
нулю на бесконечности, то решение единственно и имеет вид (см., например,
\cite{Vladim88R})
\begin{equation}                                                  \label{esolpo}
  \vf(t,\Bx)=-G\sum_\Si\frac{m_\Si}{|\Bx-\Bq_\Si(t)|},
\end{equation}
где
\begin{equation*}
  |\Bx-\Bq_\Si|:=\sqrt{(x^1-q_\Si^1)^2+(x^2-q_\Si^2)^2+(x^3-q_\Si^3)^2}
  =\sqrt{-\eta_{\mu\nu}(x^\mu-q_\Si^\mu)(x^\nu-q_\Si^\nu)}
\end{equation*}
-- расстояние от точки пространства $\Bx=\lbrace x^\mu\rbrace\in\MR^3$ до
$\Si$-той частицы в момент времени $t$.

Таким образом, мы нашли потенциал гравитационного поля для произвольного
движения частиц, которое описывается функциями $\Bq_\Si(t)$. Принято
говорить, что гравитационное взаимодействие в механике Ньютона является
дальнодействующим и распространяется с бесконечной скоростью. Это отражает
тот факт, что решение (\ref{esolpo}) в произвольной точке пространства в
произвольный момент времени однозначно определяется только массами частиц и
их расположением в тот же момент времени. Можно сказать по другому: изменение
положения частицы мгновенно приводит к изменению гравитационного поля во всем
пространстве.

Теперь подставим решение для потенциала гравитационного поля (\ref{esolpo}) в
уравнения движения точечных частиц (\ref{eqmpon}),
\begin{equation*}
  \ddot q_{\Si\mu}=-G\frac\pl{\pl q_\Si^\mu}
  \sum_\Sj\frac{m_\Sj}{|\Bq_\Si-\Bq_\Sj|},
\end{equation*}
чтобы полностью исключить потенциал гравитационного поля. Однако на этом
этапе возникает серьезная трудность, т.к.\ правая часть уравнения расходится
в точке $\Bq_\Si=\Bq_\Sj$ и поэтому неопределена. Чтобы устранить эту
трудность мы отбросим в сумме слагаемое с $\Si=\Sj$. Физически это означает,
что частица не движется под действием собственного гравитационного поля.
Таким образом, получаем систему, состоящую из $3\Sn$ обыкновенных
дифференциальных уравнений:
\begin{equation}                                                  \label{eeqpon}
  \ddot q_\Si^\mu=-G\sum_{\Si\ne \Sj}m_\Sj\frac{q_\Si^\mu-q_\Sj^\mu}
  {|\Bq_\Si-\Bq_\Sj|^3}.
\end{equation}

В механике Ньютона уравнение (\ref{eeqpon}) интерпретируется следующим
образом. Если имеется всего две частицы с массами $m_\Si$ и $m_\Sj$, то между
ними возникает притяжение, обусловленное гравитационным взаимодействием. При
этом сила $\BF=\lbrace F^\mu\rbrace$, действующая на частицу $m_\Si$ со
стороны частицы $m_\Sj$, равна
\begin{equation}                                                  \label{enewlo}
  F^\mu=-Gm_\Si m_\Sj\frac{q_\Si^\mu-q_\Sj^\mu}{|\Bq_\Si-\Bq_\Sj|^3}.
\end{equation}
Это -- {\em закон всемирного тяготения} Ньютона.
\index{Закон всемирного тяготения (Newton's gravitational law)}%
\index{Всемирного тяготения закон (Newton's gravitational law)}%
\begin{com}
В настоящее время закон всемирного тяготения Ньютона подтвержден
экспериментально с высокой степенью точности в лабораторных условиях и в
небесной механике. С его помощью в первом приближении рассчитывают движение
планет в солнечной системе и звезд в галактиках. Численное значение
гравитационной постоянной в системе СГС равно
\begin{equation*}
  G=(6,673\pm0,003)\cdot 10^{-8} \frac{\text{см}^3}{\text{г}\cdot\text{сек}^2}.
\end{equation*}
Общая теория относительности в первом приближении приводит к результатам,
которые совпадают с результатами, полученными в рамках механики Ньютона.
Кроме этого общая теория относительности приводит к поправкам, которые
называются {\em постньютоновскими}. \qed\end{com}
\index{Постньтоновские поправки (post Newtonian corrections)}%
\index{Поправки постньтоновские (post Newtonian corrections)}%
Систему уравнений движения для точечных частиц (\ref{eeqpon}) можно получить
из эффективного действия
\begin{equation}                                                  \label{eacpog}
  S_\text{eff}=\int\!\!dt\left(-\sum_{\Si=1}^\Sn\frac12
  \dot q_\Si^\mu\dot q_\Si^\nu\eta_{\mu\nu}
  +\frac12G\sum_{\Si=1}^\Sn\sum_{\Sj\ne \Si}\frac{m_\Si m_\Sj}
  {|\Bq_\Si-\Bq_\Sj|}\right),
\end{equation}
которое варьируется только по траекториям частиц. Это действие получается из
исходного действия для точечных частиц (\ref{eacgpo}) подстановкой в него
общего решения уравнения Эйлера--Лагранжа для потенциала (\ref{esolpo}). Эта
процедура подстановки решения части уравнений Эйлера--Лагранжа в исходное
действие с целью исключения некоторых динамических переменных была описана в
разделе эффективное действие (\ref{sredac}) в общем виде. Множитель $1/2$ во
втором слагаемом в действии (\ref{eacpog}) возникает из-за двойной суммы, где
каждое слагаемое встречается дважды.

Таким образом, движение точечных частиц, между которыми действуют только
гравитационные силы, сводится к системе обыкновенных дифференциальных
уравнений (\ref{eeqpon}). Это полное описание, и в таком виде оно обычно
встречается в курсах классической механики. Как видим, введение
гравитационного поля $\vf(x)$ в механике Ньютона совсем необязательно. Мы
проделали более длинный путь с тем, чтобы показать аналогию с общей теорией
относительности. К сожалению, решить уравнения Эйнштейна (\ref{eincou}) для
компонент метрики при произвольном движении частиц, как в случае механики
Ньютона (\ref{esolpo}), не удается. Поэтому эффективное действие для точечных
частиц в общей теории относительности в настоящее время неизвестно. Более
того, его просто не существует. Дело в том, что при постановке задачи Коши в
общей теории относительности необходимо задать не только начальные данные для
точечных частиц, но и для части компонент метрики (гравитационные волны).
Поэтому полного описания на языке эффективного действия для частиц не может
существовать.
\subsection{Свойства тензора энергии-импульса точечных частиц    \label{senmop}}
В настоящем разделе мы обсудим некоторые свойства тензора энергии-импульса
точечных частиц: ковариантное сохранение тензора энергии-импульса, аналогию с
тензором энергии-импульса сплошной среды, неотрицательность следа тензора
энергии-импульса и ультрарелятивистский предел. Для краткости, мы не будем
писать знак тильды над геометрическими объектами, построенные при нулевом
кручении и неметричности.
\subsubsection{Ковариантное сохранение тензора энергии-импульса}
Ниже мы проведем формальные выкладки, предполагая, что компоненты метрики
являются достаточно гладкими функциями на мировых линиях частиц. Как уже
отмечалось, это предположение не верно: компоненты метрики расходятся на
мировых линиях. Тем не менее мы приведем эти выкладки по нескольким причинам.
Во-первых, рассматриваемый вопрос очень важен, во-вторых, приведенные
выкладки часто можно встретить в научной литературе и, в-третьих, данный
пример покажет насколько легко получить неправильный результат в случае
работы с обобщенными функциями.

Поскольку тензор энергии-импульса (\ref{enmopo}) получен из вариации
действия, инвариантного относительно общих преобразований координат, то, в
силу второй теоремы Нетер (см.\ раздел \ref{sfinez}), между уравнениями
движения существует зависимость,
\begin{equation*}
  2\vol\nb_\g\left(\frac1\vol\frac{\dl S}{\dl g_{\bt\g}}\right)g_{\al\bt}
  +\sum_\Si\frac{\dl S}{\dl q_i^\al}=0.
\end{equation*}
На языке уравнений движения (\ref{eincou}), (\ref{eingeo}) это означает
следующее. Возьмем ковариантную дивергенцию от уравнения (\ref{eincou}).
Поскольку тензор Эйнштейна удовлетворяет свернутым тождествам Бианки
(\ref{eeihid}),
\begin{equation*}
  \nb_\g G^{\bt\g}=0,
\end{equation*}
то тензор энергии-импульса точечных частиц ковариантно сохраняется,
\begin{equation}                                                  \label{ecopoi}
  \nb_\al T_\text{m}{}^{\al\bt}=0,
\end{equation}
для любого решения системы уравнений движения для точечных частиц
(\ref{eingeo}). То есть условия интегрируемости уравнений Эйнштейна
выполняются автоматически, если выполнены уравнения движения точечных частиц.

Явная проверка ковариантного сохранения тензора энергии-импульса точечных
частиц (\ref{ecopoi}) требует осторожности. Поэтому проведем соответствующие
вычисления, для простоты, для одной частицы,
\begin{equation*}
\begin{split}
  \nb_\al T_{\rm m}{}^{\al\bt}&=\pl_\al T_{\rm m}{}^{\al\bt}
  +\Gamma_\al T_{\rm m}{}^{\al\bt}+\Gamma_{\al\g}{}^\bt T_{\rm m}{}^{\al\g}=
\\
  &=\frac m\vol\left[\pl_\al\int\!\! d\tau\dot q^\al\dot q^\bt\dl(x-q)
  +\Gamma_{\al\g}{}^\bt\int\!\! d\tau\dot q^\al\dot q^\g\dl(x-q)\right],
\end{split}
\end{equation*}
где мы воспользовались определением тензора энергии-импульса (\ref{enmopo}) и
тождеством (\ref{etrchs})
\begin{equation*}
  \pl_\al\vol=\vol\Gamma_\al.
\end{equation*}
Во втором слагаемом символы Кристоффеля можно внести под знак интеграла и
рассматривать их, как функции от $q$ ввиду наличия $\dl$-функции. Для первого
слагаемого в слабом смысле справедливо равенство
\begin{equation}                                                  \label{eweaki}
  \pl_\al\int\!\! d\tau\dot q^\al\dot q^\bt\dl(x-q)
  =\int\!\! d\tau\ddot q^\bt\dl(x-q).
\end{equation}
Чтобы доказать это равенство, напомним определения \cite{Vladim88R}.
\begin{defn}
Назовем функцию $\vf$ на многообразии $\MM$, $\dim\MM=n$, {\em финитной},
если ее носитель (замыкание множества точек, в которых функция отлична от
нуля) компактен. Пространство $\CD(\MM)$, состоящее из всех финитных
бесконечно дифференцируемых функций на $\MM$ называется пространством {\em
основных} ({\em пробных}) функций. {\em Обобщенной функцией} или {\em
распределением} $f$ на многообразии $\MM$ называется линейный непрерывный
функционал на пространстве основных функций. Равенство двух обобщенных
функций $f=g$ называется {\em слабым}, если для любой пробной функции
справедливо равенство
\begin{equation*}
  \int_\MM\!\!\!dx\,\vf f=\int_\MM\!\!\!dx\,\vf g,
\end{equation*}
для всех $\vf\in\CD(\MM)$. \qed\end{defn}
\index{Финитная функция (compactly supported function)}%
\index{Функция финитная (compactly supported function)}%
\index{Основная функция (test function)}%
\index{Функция основная (test function)}%
\index{Пробная функция (test function)}%
\index{Функция пробная (test function)}%
\index{Обобщенная функция (generalized function, distribution)}%
\index{Функция обобщенная (generalized function, distribution)}%
\index{Равенство слабое (weak equality)}%
\index{Слабое равенство (weak equality)}%
Мы не будем останавливаться на определении топологии в пространствах основных
и обобщенных функций, отсылая читателя к \cite{Vladim88R}.

Вернемся к равенству (\ref{eweaki}). Умножим левую часть на пробную функцию
$\vf\in\CD(\MM)$ и проинтегрируем по $\MM$:
\begin{multline*}
  \int_\MM\!\!\!dx\,\vf\pl_\al\left(\int\!\!d\tau\dot q^\al\dot q^\bt
  \dl(x-q)\right)
  =-\int_\MM\!\!\!dx\,\pl_\al\vf\int\!\!d\tau\dot q^\al\dot q^\bt
  \dl(x-q)=
\\
  =-\int_\MM\!\!\!dx\int\!\!d\tau\,\frac{d\vf}{d\tau}\dot q^\bt\dl(x-q)
  =-\int\!\!d\tau\,\frac{d\vf}{d\tau}\dot q^\bt
  =\int\!\!d\tau\,\vf\ddot q^\bt,
\end{multline*}
где мы два раза проинтегрировали по частям, воспользовались равенством
$d\vf/d\tau=\dot q^\al\pl\vf/\pl q^\al$ и взяли интеграл по $\MM$, используя
$\dl$-функцию. Интеграл по $\MM$ от правой части (\ref{eweaki}) с пробной
функцией приводит, очевидно, к тому же результату. Таким образом, слабое
равенство (\ref{eweaki}) доказано.

Теперь равенство (\ref{ecopoi}) можно переписать в виде
\begin{equation}                                                  \label{ecocom}
  \nb_\al T_\text{m}{}^{\al\bt}=\frac1\vol\int\!\!d\tau\sum_\Si m_\Si
  \left(\ddot q^\bt_\Si+\Gamma_{\al\g}{}^\bt\dot q^\al_\Si\dot q^\g_\Si\right)
  \dl(x-q),
\end{equation}
где мы вернулись к общему случаю $\Sn$ частиц. Отсюда следует, что тензор
энергии-импульса точечных частиц ковариантно сохраняется, если выполнены
уравнения движения (\ref{eingeo}). Таким образом мы привели прямое
доказательство следующего утверждения.
\begin{prop}                                                      \label{ptepoi}
Тензор энергии-импульса точечных частиц ковариантно сохраняется
(\ref{ecopoi}), если выполнены уравнения движения для точечных частиц.
\end{prop}

Зададимся вопросом: ``Верно ли обратное утверждение ?''. Для этого умножим
равенство (\ref{ecocom}) на пробную функцию и проинтегрируем по $\MM$,
используя $\dl$-функцию. Если тензор энергии-импульса ковариантно
сохраняется, то должно быть выполнено равенство
\begin{equation*}
  \int\!\!d\tau\vf\frac1\vol\sum_\Si m_\Si
  \left(\ddot q^\bt_\Si+\Gamma_{\al\g}{}^\bt\dot q^\al_\Si\dot q^\g_\Si\right)=0.
\end{equation*}
Поскольку пробная функция $\vf$ произвольна, и ее носитель может быть отличен
от нуля для одной произвольной частицы, то отсюда следуют уравнения движения
для точечных частиц (\ref{eingeo}).

Вывод о том, что уравнения геодезических следуют из ковариантного закона
сохранения тензора энергии-импульса точечных частиц, что, в свою очередь,
является необходимым условием совместности уравнений Эйнштейна, широко
распространен, см., например, \cite{InfPle60R}. Однако этот вывод не верен. А
именно, если найдено точное решение уравнений Эйнштейна с источниками в виде
точечных частиц, то отсюда не следует, что уравнения геодезических будут
автоматически удовлетворены. Это связано с расходимостями компонент метрики
на мировых линиях частиц.
\subsubsection{Аналогия с тензором энергии-импульса сплошной среды}
Введем новую временн\'ую координату $x^0\mapsto\tau=\tau(x^0,x^\mu)$ в
пространстве-времени $\MM$ таким образом, чтобы вдоль каждой мировой линии
частицы она совпадала с собственным временем $\tau(x^\al=q_\Si^\al)=\tau$.
Это всегда можно сделать, причем не единственным образом, так как траектории
всех частиц не пересекаются, времениподобны, а канонический параметр
определен с точностью до сдвигов. Тогда в новой системе координат
$\tau,x^\mu$ производные $\dot q_\Si^\al$ в выражении (\ref{enmopo}) можно
заменить на частные производные $\dot x^\al:=\pl x^\al/\pl\tau$ ввиду наличия
$\dl$-функций, и вынести за знак интегрирования:
\begin{equation}                                                  \label{epocon}
  T_\text{m}{}^{\al\bt}=\frac{\dot x^\al\dot x^\bt}\vol\int\!\! d\tau
  \sum_\Si m_\Si\dl(x-q_\Si)
  =\frac{\dot x^\al\dot x^\bt}\vol\sum_\Si\frac{m_\Si}{\dot q^0_\Si}
  \dl(\Bx-\Bq_i).
\end{equation}
Полученное выражение для тензора энергии-импульса точечных частиц имеет такой
же вид, как и для сплошной среды (\ref{enmotl}), которая будет рассмотрена
позже. Для точечных частиц давление равно нулю, $\CP=0$, а плотность энергии
принимает вид
\begin{equation*}
  \CE=\frac1\vol\sum_\Si\frac{m_\Si}{\dot q^0_\Si}\dl(\Bx-\Bq_\Si).
\end{equation*}
Поскольку временн\'ая координата $x^0$ на траекториях частиц совпадает с
собственным временем $\tau$, то $\dot q_\Si^0=1$ и выражение для плотности
энергии приобретает интуитивно ясную форму,
\begin{equation*}
  \CE=\frac1\vol\sum_\Si m_\Si\dl(\Bx-\Bq_\Si).
\end{equation*}
То есть энергия сосредоточена в точках расположения частиц, и каждая частица
несет энергию, которая равна ее массе. Тензор энергии-импульса точечных
частиц соответствует пылевидной материи, поскольку давление равно нулю.
\subsubsection{След тензора энергии-импульса}
Вернемся в произвольную систему координат. Из формулы (\ref{enmopi}) следует
выражение для следа тензора энергии-импульса произвольного распределения
точечных частиц
\begin{equation}                                                  \label{enmutr}
  T_\text{m}{}^\al{}_\al
  =\frac1\vol\sum_\Si\frac{m_\Si}{\dot q^0_\Si}\dl(\Bx-\Bq_\Si).
\end{equation}
Поскольку $m_\Si>0$ и $\dot q^0_\Si>0$, то след тензора энергии-импульса
положителен (при этом мы рассматриваем $\dl$-функцию, как положительную).

Поскольку след тензора энергии-импульса положителен для {\em произвольного}
распределения точечных частиц, то в моделях математической физики делается
предположение о том, что след тензора энергии-импульса для любой обычной
(наблюдаемой) материи всегда неотрицателен, $T^\al{}_\al\ge0$. При этом
равенство следа тензора энергии-импульса нулю достигается только для частиц,
движущихся со скоростью света, или излучения.

След тензора энергии-импульса электромагнитного поля \ref{enmoem}, который
соответствует электромагнитному излучению, равен нулю. Это согласуется с
утверждением о том, что след тензора энергии-импульса произвольного
распределения ультрарелятивистских частиц равен нулю. Напомним, что в
квантовой электродинамике электромагнитное поле описывает безмассовые частицы
-- фотоны, которые распространяются со скоростью света.

Мы выделили рассмотрение следа тензора энергии-импульса точечных частиц в
отдельный пункт именно в свете последнего замечания, т.к.\ положительность
следа тензора энергии-импульса для произвольной материи ниоткуда больше не
следует, и в то же время ведет к важным следствиям.
\subsubsection{Ультрарелятивистский предел}
Рассмотрим ультрарелятивистский предел для тензора энергии-импульса точечных
частиц (\ref{epocon}). Этот предел, прежде всего, требует определения, потому
что, глядя на определение собственной скорости частицы (\ref{evelos}),
непонятно что и куда стремить. Тензор энергии-импульса точечных частиц и его
след можно выразить через наблюдаемую скорость
\begin{align}                                                     \label{enmoob}
  T_\text{m}{}^{\al\bt}
  &=\frac{v^\al v^\bt}{\sqrt{|g|v^2}}\sum_\Si m_\Si\dl(\Bx-\Bq_\Si),
\\                                                                \label{enmatr}
  T_\text{m}{}^\al{}_\al
  &=\frac{\sqrt{v^2}}\vol\sum_\Si m_\Si\dl(\Bx-\Bq_\Si),
\end{align}
где мы воспользовались равенствами (\ref{eobved}) и (\ref{eobser}).
\begin{defn}
Предел, когда квадрат наблюдаемой скорости частицы (\ref{evelob}) стремится к
нулю,
\begin{equation}                                                  \label{eultra}
  v^2:=v^\al v^\bt g_{\al\bt}\rightarrow0,
\end{equation}
называется {\em ультрарелятивистским пределом} для точечной частицы.
\qed\end{defn}
\index{Ультрарелятивистский предел (Ultrarelativistic limit}%
\begin{com}
Если метрика имеет блочно диагональный вид (\ref{etiblm}), то
\begin{equation*}
  v^2=g_{00}-\Bv^2,\qquad \text{и}\qquad \lbrace u^\al\rbrace
  =\left\lbrace\frac1{\sqrt{g_{00}-\Bv^2}},
  \frac{v^\mu}{\sqrt{g_{00}-\Bv^2}}\right\rbrace
\end{equation*}
где $\Bv^2:=-v^\mu v^\nu g_{\mu\nu}\ge0$ -- квадрат пространственной
наблюдаемой скорости и собственная скорость частицы $u^\al$ определена
формулой (\ref{evelos}). В этом случае ультрарелятивистский предел
соответствует пределу $\Bv^2\to g_{00}$. Для пространства Минковского это
означает, что наблюдаемая скорость частицы стремится к скорости света.
\qed\end{com} В ультрарелятивистском пределе компоненты самого тензора
энергии-импульса точечных частиц (\ref{enmoob}) не определены, однако след
тензора энергии-импульса (\ref{enmatr}) стремится к нулю
\begin{equation*}
  \underset{v^2\to0}{\lim}T_\text{m}{}^\al{}_\al=0.
\end{equation*}
\begin{com}
Это утверждение справедливо для произвольного количества и распределения
точечных частиц, находящихся только под действием гравитационных сил. Если
присутствуют другие взаимодействия, то действие для точечных частиц
(\ref{epohep}) может измениться и, следовательно, изменится выражение для
тензора энергии-импульса. В таком случае требуется дополнительное
исследование. \qed\end{com}
\section{Ньютонов предел                                         \label{snelim}}
Для того, чтобы сказать, что общая теория относительности не противоречит
экспериментальным данным, желательно показать, что теория тяготения Ньютона в
каком то смысле (приближении) следует из уравнений Эйнштейна. Поскольку
гравитация Ньютона находится в хорошем согласии с экспериментом, то в этом
случае можно утверждать, что общая теория относительности описывает
гравитационные взаимодействия по крайней мере не хуже, чем законы Ньютона.
Такое приближение существует, и будет описано в настоящем разделе.

Сначала сделаем общее замечание. Уравнения Эйнштейна существенно нелинейны, в
то время как гравитация Ньютона линейна: гравитационные потенциалы различных
массивных тел просто складываются. Поэтому естественно ожидать, что закон
всемирного тяготения вытекает из уравнений Эйнштейна в линейном приближении.

Рассмотрим вакуумные уравнения Эйнштейна без космологической постоянной в
четырехмерном пространстве-времени
\begin{equation}                                                  \label{ericwa}
  \kappa\left(R_{\al\bt}-\frac12g_{\al\bt}R\right)=-\frac12T_{\Sm\al\bt}.
\end{equation}
Будем считать, что пространство-время топологически тривиально, и существует
глобальная система координат, в которой метрика, удовлетворяющая уравнениям
(\ref{ericwa}), мало отличается от метрики Лоренца в пространстве Минковского
$\MR^{1,3}$:
\begin{equation}                                                  \label{esmmeh}
  g_{\al\bt}=\eta_{\al\bt}+\e h_{\al\bt},\qquad\e\ll1,
\end{equation}
где $h_{\al\bt}(x)$ -- некоторые достаточно гладкие функции. При этом мы
считаем малыми также все частные производные: $\pl_\al g_{\bt\g}\sim\e$.
Символы Кристоффеля пропорциональны производным $\pl_\al g_{\bt\g}$, и
поэтому их квадраты дают вклад в тензор кривизны порядка $\e^2$. В
(псевдо-)римановой геометрии тензор кривизны имеет вид (\ref{ecutrl}). Тем самым
вкладом квадратичных слагаемых по символам Кристоффеля в тензор кривизны
можно пренебречь по сравнению со вторыми производными от
компонент метрики, которые дают вклад прядка $\e$. Таким образом, в линейном
приближении по $\e$ тензор Риччи имеет вид
\begin{equation*}
  \frac1\e R_{\al\bt}=\frac12\eta^{\g\dl}(\pl^2_{\al\bt}h_{\g\dl}
  -\pl^2_{\al\g}h_{\bt\dl}-\pl^2_{\bt\g}h_{\al\dl}+\pl^2_{\g\dl}h_{\al\bt}).
\end{equation*}
В правой части равенства свертка проводится с помощью метрики Лоренца, т.к.\
выражение в скобках имеет первый порядок. Скалярная кривизна имеет вид
\begin{equation*}
  \frac1\e R=\eta^{\al\bt}\eta^{\g\dl}(\pl^2_{\al\bt}h_{\g\dl}
  -\pl^2_{\al\g}h_{\bt\dl}).
\end{equation*}
Введем новые переменные
\begin{equation}                                                  \label{ehbard}
  \bar h_{\al\bt}:=h_{\al\bt}-\frac12\eta_{\al\bt}h,
\end{equation}
где $h:=\eta^{\al\bt}h_{\al\bt}$ -- след возмущения метрики. Обратное
преобразование имеет вид
\begin{equation}                                                  \label{emebam}
  h_{\al\bt}=\bar h_{\al\bt}-\frac12\eta_{\al\bt}\bar h,\qquad
  \bar h:=\eta^{\al\bt}\bar h_{\al\bt}.
\end{equation}
Тогда тензор Эйнштейна в линейном приближении примет вид
\begin{equation*}
  \frac1\e\left(R_{\al\bt}-\frac12g_{\al\bt}R\right)=
  \frac12\eta^{\g\dl}\big(\pl^2_{\g\dl}\bar h_{\al\bt}
  -\pl^2_{\al\g}\bar h_{\bt\dl}-\pl^2_{\bt\g}\bar h_{\al\dl}
  +\eta_{\al\bt}\eta^{\e\z}\pl^2_{\g\e}\bar h_{\dl\z}\big).
\end{equation*}

Теперь воспользуемся инвариантностью действия Гильберта--Эйнштейна
относительно общих преобразований координат. Рассмотрим бесконечно малые
преобразования координат, которые генерируются некоторым векторным полем
(\ref{einfct}),
\begin{equation}                                                  \label{elicoc}
  x^\al\mapsto x^\al+\e u^\al(x).
\end{equation}
При этом компоненты метрики получат приращение (\ref{eitcms}), или
\begin{equation}                                                  \label{ecoiht}
  h_{\al\bt}\mapsto h_{\al\bt}+\pl_\al u_\bt+\pl_\bt u_\al.
\end{equation}
Возьмем в качестве векторного поля $u^\al$ произвольное решение уравнения
\begin{equation*}
  \square u^\al=-\pl_\bt\bar h_\al{}^\bt,
\end{equation*}
где $\square:=\eta^{\al\bt}\pl_\al\pl_\bt$ -- оператор Даламбера. Тогда в
новой системе координат возмущение компонент метрики будет удовлетворять
уравнению
\begin{equation}                                                  \label{egastw}
  \pl_\bt\bar h_\al{}^\bt=0.
\end{equation}
Это есть ни что иное как условие гармоничности координат (\ref{efogau}) в
линейном приближении.
\begin{com}
Гармонические координаты в общей теории относительности являются аналогом
лоренцевой калибровки в электродинамике (см.\ раздел \ref{selema}).
\qed\end{com}

С учетом условия гармоничности уравнения Эйнштейна (\ref{ericwa}) принимают
вид
\begin{equation}                                                  \label{eilitw}
  \e\square\bar h_{\al\bt}=-\frac1\kappa T_{\Sm\al\bt}.
\end{equation}
\begin{com}
Если поля материи отсутствуют, $T_{\Sm\al\bt}=0$, то система уравнений
(\ref{egastw}), (\ref{eilitw}) совпадает с уравнениями для безмассового поля
со спином 2 в плоском пространстве-времени Минковского \cite{FiePau39}.
Поэтому общую теорию относительности в целом можно рассматривать как теорию
безмассового поля со спином 2 и с некоторым самодействием, которое
соответствует отброшенным нелинейным членам. Следует однако заметить, что
понятие массы и спина требует наличия метрики Лоренца, которая является
фоновой метрикой для линейного приближения. В общем случае, без обращения к
линейному приближению, утверждению о том, что метрика описывает безмассовое
поле спина 2 придать точный смысл нельзя. \qed\end{com} Рассмотрим в качестве
источника в уравнениях Эйнштейна одну частицу массы $M$. Поскольку мы
рассматриваем слабые гравитационные поля, то будем считать, что $M\sim\e$.
Этой частице соответствует тензор энергии-импульса (\ref{enmopi})
\begin{equation*}
  T_{\Sm\al\bt}=\frac1\vol M\frac{\dot q_\al\dot q_\bt}{\dot q^0}
  \dl(\Bx-\Bq).
\end{equation*}
В линейном приближении по $\e$ можно сделать замену $\vol\mapsto1$.

Предположим, что частица покоится в начале координат, т.е.\ $\lbrace\dot
q^\al\rbrace=(1,0,0,0)$ и $\lbrace q^\al\rbrace=(\tau,0,0,0)$. Предположим
также, что компоненты метрики не зависят от времени (статическое решение).
Тогда полная система уравнений Эйнштейна примет вид
\begin{align}                                                     \label{enewla}
  \kappa\e\triangle\bar h_{00}=M\dl(\Bx),
\\                                                                \label{eneade}
  \triangle\bar h_{0\mu}=\triangle\bar h_{\mu\nu}=0,
\end{align}
где $\triangle:=\pl^2_1+\pl^2_2+\pl^2_3$ -- лапласиан в трехмерном евклидовом
пространстве. Если предположить, что компоненты возмущений метрики $\bar
h_{\al\bt}$ стремятся к нулю на бесконечности, то уравнения (\ref{eneade})
имеют единственное решение
\begin{equation*}
  \bar h_{0\mu}=0,\qquad \bar h_{\mu\nu}=0.
\end{equation*}
Для сравнения уравнения (\ref{enewla}) с законом всемирного тяготения,
необходимо восстановить размерные постоянные. Во-первых, положим
\begin{equation*}
  \e\bar h_{00}=\frac{4\vf}{c^2},
\end{equation*}
где $\vf$ -- потенциал гравитационного поля. Это следует из нерелятивистского
предела для точечной частицы (\ref{ehzzde}). Кроме того, в правую часть
уравнения (\ref{enewla}) надо вставить множитель $c^2$: один множитель $c$
следует из опущенного множителя в действии для точечной частицы
(\ref{epoacp}), а второй -- из равенства $\dot q^0=c$. Если после этого
положить
\begin{equation}                                                  \label{egravc}
  \kappa:=\frac{c^4}{16\pi G},
\end{equation}
где $G$ -- гравитационная постоянная в законе тяготения Ньютона
(\ref{enewlo}), то уравнение (\ref{enewla}) совпадет с уравнением Пуассона
для гравитационного поля (\ref{epoise}). В этом случае решение уравнения
Пуассона (\ref{enewla}) примет вид
\begin{equation*}
  \vf=-G\frac Mr,
\end{equation*}
где $r:=|\Bx|$.

Ясно, что для найденных компонент метрики калибровочное условие
(\ref{egastw}) выполнено, и, следовательно, найдено самосогласованное решение
задачи.

Поскольку след $\e\bar h=4\vf/c^2$, то из формулы для обратного
преобразования возмущения компонент метрики(\ref{emebam}), получаем, что в
ньтоновом приближении метрика имеет вид
\begin{equation}                                                  \label{eneame}
  g=\begin{pmatrix}
    1-G\displaystyle\frac{2M}r & 0 & 0 & 0 \\
    0 & -1-G\displaystyle\frac{2M}r & 0 & 0 \\
    0 & 0 & -1-G\displaystyle\frac{2M}r & 0 \\
    0 & 0 & 0 & -1-G\displaystyle\frac{2M}r \end{pmatrix}.
\end{equation}
Напомним, что метрика имеет такой вид в линейном приближении в гармонической
системе координат.

Таким образом, мы показали, что теория тяготения Ньютона согласуется с общей
теорией относительности. Она возникает в статическом случае для слабых
гравитационных полей в линейном приближении. При этом константа связи перед
действием Гильберта--Эйнштейна имеет вид (\ref{egravc}). По этой причине многие
авторы записывают действие Гильберта--Эйнштейна именно с такой константой связи,
часто полагая $c=1$.
\begin{com}
Выбор константы связи перед скалярной кривизной в действии
Гильберта--Эйнштейна зависит от определения кривизны и ее сверток через
метрику. Наш выбор выражения для тензора кривизны, тензора Риччи и скалярной
кривизны через метрику и аффинную связность согласован с обычными
обозначениями в теории калибровочных полей и приспособлен к анализу общей
теории относительности. В то же время он отличается от стандартных
определений, принятых в математической литературе. Например, скалярная
кривизна сферы радиуса $r$, вложенной в трехмерное евклидово пространство,
$\MS_r^2\hookrightarrow\MR^3$, равна $-2/r^2$. В этом нет ничего страшного.
Просто об этом следует помнить. \qed\end{com} Несмотря на то, что общая
теория относительности содержит в себе теорию тяготения Ньютона в качестве
предельного случая, отметим принципиальное отличие. В механике Ньютона
свободная частица движется по прямой линии. Если она находится в поле другой
массивной частицы, то на нее действует сила гравитационного притяжения.
Теперь она уже не является свободной и ее траектория отличается от прямой
линии в соответствии с законом всемирного тяготения. В общей теории
относительности ситуация совершенно иная. Массивная частица искривляет
пространство-время в соответствии с уравнениями Эйнштейна. Пробная частица в
гравитационном поле остается свободной и движется вдоль экстремали. Однако
теперь экстремаль не является прямой линией для внешнего наблюдателя,
поскольку пространство-время перестает быть плоским из-за наличия массивной
частицы. Это искривление траекторий воспринимается наблюдателем, как
проявление гравитационного воздействия.
\section{Гравитационные волны}
В механике Ньютона гравитационных волн нет, что вытекает из системы уравнений
(\ref{eqmogr}), (\ref{eqmpon}). При этом изменение положения одного из массивных
тел мгновенно приводит к изменению гравитационного поля во всем пространстве.
Кроме этого, если массивные тела отсутствуют, то потенциал гравитационного
поля равен нулю. В общей теории относительности ситуация другая. Во-первых,
гравитационные взаимодействия распространяются с конечной постоянной скоростью
света $c$ в локально инерциальной системе отсчета. Во-вторых, даже если
материальные тела отсутствуют, уравнения Эйнштейна допускают нетривиальные
решения в виде гравитационных волн. То есть гравитационное поле может быть
отлично от нуля даже если материальные тела отсутствуют. В настоящем разделе
мы изучим решения уравнений Эйнштейна, описывающие гравитационные волны.

Рассмотрим вакуумные уравнения Эйнштейна без космологической постоянной
\begin{equation}                                                  \label{egraeq}
  R_{\al\bt}=0.
\end{equation}
Как и в предыдущем разделе будем считать, что пространство-время топологически
тривиально, $\MM\approx\MR^4$, и существует глобальная система координат, в
которой метрика мало отличается от метрики Лоренца (\ref{esmmeh}). В нулевом
порядке по $\e$ вакуумные уравнения Эйнштейна, очевидно, удовлетворяются, т.к.\
кривизна пространства Минковского равна нулю. Найдем решение уравнений
(\ref{egraeq}) в первом порядке по $\e$.

Используя инвариантность действия Гильберта--Эйнштейна относительно общих
преобразований координат, выберем систему отсчета таким образом, чтобы
выполнялось калибровочное условие (\ref{egastw}) (гармонические координаты).
Тогда все компоненты метрики в первом порядке будут удовлетворять волновому
уравнению
\begin{equation*}
  \square\bar h_{\al\bt}=0\quad \Leftrightarrow\quad \square h_{\al\bt}=0,
\end{equation*}
где компоненты $\bar h_{\al\bt}$ определены уравнением (\ref{ehbard}) и
$\square$ -- оператор Даламбера в пространстве Минковского. Однако
далеко не все компоненты метрики являются независимыми и описывают физические
степени свободы, от которых нельзя избавиться путем выбора соответствующей
системы координат.

Покажем это. Во-первых, выберем гармоническую систему координат. Тогда в
линейном приближении выполнено уравнение (\ref{egastw}), которое рассматривается
как калибровочное условие. Это условие не фиксирует систему координат
однозначно. Действительно, допустим, что в некоторой системе координат это
условие выполнено. Совершим преобразование координат
$x^\al\mapsto x^\al+\e u^\al$, где все компоненты векторного поля $u$
удовлетворяют волновому уравнению
\begin{equation}                                                  \label{esquau}
  \square u^\al=0.
\end{equation}
Нетрудно проверить, что в новой системе координат калибровочное условие
(\ref{egastw}) будет также выполнено. Следовательно, оставшуюся свободу в
выборе системы координат можно использовать для того, чтобы зафиксировать
дополнительные компоненты метрики.

Чтобы найти подходящие дополнительные калибровочные условия, необходимо решить
уравнения (\ref{esquau}) c некоторыми начальными условиями. Посмотрим как
преобразуется след возмущений метрики $h:=h_\al{}^\al$ и компоненты $h_{0\mu}$,
$\mu=1,2.3$, при бесконечно малых преобразованиях координат (\ref{elicoc}):
\begin{align*}
  h&\mapsto h+2\pl_\al u^\al,
\\
  h_{0\mu}&\mapsto h_{0\mu}+\pl_0 u_\mu+\pl_\mu u_0.
\end{align*}
Рассмотрим пространственное сечение $x^0:=t=\const$ в качестве поверхности Коши
для уравнения (\ref{esquau}).
На этой поверхности найдем какое либо решение системы линейных уравнений
\begin{align}                                                     \label{egaufi}
  2\left(\dot u_0+\pl_\mu u^\mu\right)&=-h,
\\                                                                \label{egause}
  2\left(\triangle u_0+\pl_\mu\dot u^\mu\right)&=-\dot h,
\\                                                                \label{egauth}
  \dot u_\mu+\pl_\mu u_0&=-h_{0\mu},
\\                                                                \label{egaufo}
  \triangle u_\mu+\pl_\mu\dot u_0&=-\dot h_{0\mu},
\end{align}
где точка обозначает дифференцирование по времени $t$. Из этой системы уравнений
определяем компоненты $u^\al$ и их производные по времени $\dot u^\al$ на
поверхности Коши. После этого решаем задачу Коши для уравнений (\ref{esquau}) с
найденными начальными условиями в обе стороны по времени и определяем векторное
поле $u$ во всем пространстве-времени. Поскольку выполнены уравнения
(\ref{egaufi})--(\ref{egaufo}), то на поверхности Коши справедливы равенства
\begin{equation}                                                  \label{egarac}
\begin{aligned}
  h&=0, & \dot h&=0,
\\
  h_{0\mu}&=0, & \dot h_{0\mu}&=0.
\end{aligned}
\end{equation}
Это очевидно, если заметить, что уравнения (\ref{egause}) и (\ref{egaufo})
возникают после дифференцирования уравнений (\ref{egaufi}) и (\ref{egauth}) по
времени и использования уравнений движения.
Поскольку компоненты $h$ и $h_{0\mu}$ удовлетворяют волновому уравнению, то
условия (\ref{egarac}) выполнены также во всем пространстве-времени.
\begin{defn}
Система координат, в которой в линейном приближении выполнены условия
\begin{equation}                                                  \label{eradga}
  \pl_\bt h_\al{}^\bt=0,\qquad h=0,\qquad h_{0\mu}=0
\end{equation}
называется {\em радиационной калибровкой}.
\qed\end{defn}
\index{Радиационная калибровка (radiative gauge)}%
\index{Калибровка радиационная (radiative gauge)}%
\begin{prop}
При отсутствии полей материи и космологической постоянной радиационная
калибровка существует в линейном приближении к метрике Лоренца.
\end{prop}
\begin{proof}
Было приведено выше.
\end{proof}
Несмотря на то, что в четырехмерном пространстве-времени общие преобразования
координат параметризуются четырьмя произвольными функциями, мы сумели наложить
восемь калибровочных условий (\ref{eradga}).
\begin{com}
В электродинамике аналогичная калибровка называется кулоновской.
\qed\end{com}

Из первого условия (\ref{eradga}) с учетом того, что $h_{0\mu}=0$, получаем
уравнение $\dot h_{00}=0$, которое должно быть выполнено во всем
пространстве-времени. Тогда уравнение движения для временн\'ой компоненты
сводится к уравнению Лапласа $\triangle h_{00}=0$. С учетом нулевых граничных
условий на бесконечности получаем дополнительное условие на компоненты метрики:
$h_{00}=0$.

Рассмотрим плоскую волну, которая распространяется в направлении волнового
вектора $k=\lbrace k^\al\rbrace$:
\begin{equation}                                                  \label{eflaea}
  h_{\al\bt}=H_{\al\bt}\ex^{ik_\g x^\g},
\end{equation}
где $H_{\al\bt}$ -- некоторая постоянная матрица и $k^2:=k^\al k_\al=0$.
Радиационная калибровка (\ref{eradga}) для этого решения уравнений движения
задается следующими восемью условиями:
\begin{equation*}
  k^\bt H_{\al\bt}=0,\qquad H_\al{}^\al=0,\qquad H_{0\mu}=0.
\end{equation*}
Из первого и третьего условия вытекает равенство $k^0 H_{00}$. Для
нетривиального решения $k^0\ne0$, и поэтому $H_{00}=0$. Ввиду симметрии по
индексам матрица $H_{\al\bt}$ имеет 10 независимых элементов. Радиационная
калибровка накладывает на них 8 независимых условий. Отсюда вытекает, что в
выбранной системе координат только 2 компоненты возмущения метрики являются
независимыми.

Допустим, что гравитационная волна распространяется вдоль оси $x^1$, т.е.\
нормированный волновой вектор имеет только две отличные от нуля компоненты
$k=(1,1,0,0)$. Тогда матрица $H_{\al\bt}$ в радиационной калибровке имеет вид
\begin{equation*}
  H=\begin{pmatrix} 0 & 0 & 0 & 0 \\ 0 & 0 & 0 & 0 \\
  0 & 0 & A & \quad B \\ 0 & 0 & B & -A \end{pmatrix},
\end{equation*}
где $H_{22}=-H_{33}=A$ и $H_{23}=H_{32}=B$ -- два произвольных числа (амплитуды
волн).

Введем новое понятие спиральности плоской волны. Для этого рассмотрим вращение
пространства $\MR^3\subset\MR^{1,3}$ на угол $\om$ вокруг оси $x^1$. Матрица
вращения задана в явном виде формулой (\ref{eormtr}), и ясно, что такое вращение
не меняет волнового вектора $k$. При преобразовании координат компоненты метрики
преобразуются по тензорному закону
\begin{equation*}
  g_{\al\bt}\mapsto g'_{\al\bt}=S_\al{}^\g S_\bt{}^\dl g_{\g\dl}
\end{equation*}
С соответствующей матрицей вращений $S$. Простые вычисления приводят к
равенствам для новых амплитуд:
\begin{align*}
  A'&=\quad \cos2\om A+\sin2\om B,
\\
  B'&=-\sin2\om A+\cos2\om B.
\end{align*}
Это означает, что при повороте системы координат на угол $\om$ амплитуда волны
поворачивается на удвоенный угол $2\om$. В физике часто рассматривают
комплексные амплитуды
\begin{equation*}
  H_\pm:=H_{22}\mp iH_{23}=A\mp iB.
\end{equation*}
При вращении они преобразуются по правилу
\begin{equation*}
  H'_\pm=\ex^{\pm2i\om}H_\pm.
\end{equation*}
\begin{defn}
Если амплитуда плоской поперечной волны при повороте на угол $\om$ вокруг
направления распространения волны поворачивается на угол $h\om$, то говорят,
что волна имеет {\em спиральность} $h$.
\qed\end{defn}
\index{Спиральность (helicity)}%

Таким образом, плоские гравитационные волны описывают поперечные волны
спиральности два.

Тензор Риччи и скалярная кривизна для данного решения вакуумных уравнений
Эйнштейна в линейном приближении равны, конечно, нулю. Это следует из того, что
мы решаем вакуумные уравнения Эйнштейна без космологической постоянной
(\ref{ericfl}). Тем не менее полный тензор кривизны отличен от нуля. В линейном
приближении тензор кривизны имеет вид (см.\ выражение (\ref{ecutrl}))
\begin{equation}                                                  \label{elicuy}
  \frac1\e R_{\al\bt\g\dl}=\frac12(\pl^2_{\al\g} h_{\bt\dl}
  -\pl^2_{\al\dl}h_{\bt\g}-\pl^2_{\bt\g}h_{\al\dl}+\pl^2_{\bt\dl}h_{\al\g}).
\end{equation}
Простые вычисления показывают, что среди 20 независимых компонент тензора
кривизны только 9 отличны от нуля:
\begin{align*}
  R_{0303}&=-R_{0202}=R_{1313}=-R_{1212}=R_{0212}=-R_{0313}
  =\frac12\e A\ex^{i(t-x^1)},
\\
  R_{0203}&=R_{1213}=-R_{0213}=-\frac12\e B\ex^{i(t-x^1)}.
\end{align*}

При преобразовании координат компоненты тензора кривизны ведут себя ковариантным
образом -- на то он и тензор. Однако в линейном приближении они не просто
ковариантны, а инвариантны. Нетрудно проверить, что выражение (\ref{elicuy})
действительно инвариантно относительно преобразований (\ref{ecoiht}) с
произвольным вектором $u$.

Из явного выражения для нетривиальных компонент тензора кривизны вытекает, что
амплитуды волн $A$ и $B$ нельзя обратить в нуль никаким преобразованием
координат. Следовательно, они описывают физические распространяющиеся степени
свободы.

Поскольку вакуумные уравнения Эйнштейна в линейном приближении линейны, то им
будет удовлетворять произвольная суперпозиция плоских волн. В частности,
поправки к метрике вида
\begin{equation*}
  h_{\al\bt}=H_{\al\bt}\big[f(x^1-t)+g(x^1+t)\big],
\end{equation*}
где $f$ и $g$ -- произвольные функции, описывающие распространение волн вдоль
оси $x^1$ в положительную и отрицательную стороны, также удовлетворяют
линеаризованным уравнениям Эйнштейна в радиационной калибровке. Отсюда, в
частности, следует, что для однозначного задания волнового решения уравнений
Эйнштейна необходимо задать четыре функции на поверхности Коши: по две для
каждой волны. При этом функции можно задать произвольным образом. Таким образом,
вакуумные уравнения Эйнштейна без космологической постоянной описывают
распространение двух физических степеней свободы, которые порождают
нетривиальную кривизну пространства-времени и не устраняются никаким
преобразованием координат. Данный подсчет степеней свободы приводит к тому же
результату, что и общий подход, основанный на гамильтоновом формализме, который
будет рассмотрен позже в главе \ref{shamgr}.
\section{Сплошная среда в общей теории относительности           \label{shhfyj}}
В правой части уравнений Эйнштейна (\ref{eincos}) находится тензор
энергии-импульса материи $T_{\Sm\al\bt}$. В случае скалярного, электромагнитного
и других полей, уравнения движения которых следуют из вариационного принципа,
правая часть уравнений Эйнштейна определяется вариацией соответствующего
действия по метрике. В этом случае вопросов с определением тензора
энергии-импульса материи не возникает. Некоторые из этих тензоров будут
рассмотрены в дальнейшем.

В то же время в общей теории относительности существует ряд важных моделей
(особенно в космологии), для которых тензор энергии-импульса материи не следует
из вариационного принципа. В настоящем разделе мы определим тензор
энергии-импульса материи $T_\Sm^{\al\bt}$, которая рассматривается, как сплошная
среда, и изучим некоторые из его свойств. При этом мы не будем опираться на
вариационный принцип.

Пусть пространство-время $(\MM,g)$ топологически тривиально
$\MM\approx\MR^{1,3}$ и покрыто одной картой. Мы предполагаем, что координаты
$\lbrace x^\al\rbrace=\lbrace x^0,x^\mu\rbrace$ выбраны таким образом, что
координата $x^0$ является временем, т.е.\ $g_{00}>0$. Кроме того, мы считаем,
что все сечения $x^0=\const$ -- пространственноподобны.

Можно привести ряд физических аргументов \cite{LanLif88R} в пользу того, что
тензор энергии-импульса материи, которая рассматривается как сплошная среда,
имеет вид
\begin{equation}                                                  \label{enmotl}
  T_\Sm^{\al\bt}:=(\CE+\CP)u^\al u^\bt-\CP g^{\al\bt},
\end{equation}
где $\CE(x)$ и $\CP(x)$ -- плотность энергии и давление материи в точке
$x\in\MM$, и
\begin{equation*}
  u^\al:=dx^\al/ds,\qquad ds:=\sqrt{|ds^2|},\qquad
  ds^2:=g_{\al\bt}dx^\al dx^\bt,
\end{equation*}
-- четырехмерная скорость материи в точке $x\in\MM$, которая удовлетворяет
тождеству $u^\al u_\al=1$. Здесь мы предполагаем, что каждая точка материи
движется вдоль времениподобной мировой линии $x^\al(s)$ в будущее, т.е.\
$u^0>0$. Ясно, что мировые линии точек материи -- это интегральные кривые
векторного поля скорости $u$.

Поскольку $u^\al$ и $g^{\al\bt}$ являются соответственно компонентами вектора и
тензора относительно преобразований координат, то мы считаем, что плотность
энергии $\CE$ и давление материи $\CP$ являются достаточно гладкими скалярными
полями на пространстве-времени $\MM$. В этом случае правая часть равенства
(\ref{enmotl}) представляет собой ковариантный симметричный тензор второго
ранга. Для обычной (наблюдаемой) материи плотность энергии предполагается
положительной, $\CE>0$, а давление -- неотрицательным, $\CP\ge0$\footnote{
Давление, в принципе, может быть отрицательным. Примером является резина. Для
нее увеличение объема по сравнению с состоянием равновесия приводит к увеличению
давления. Поскольку выбор точки отсчета давления в классической механике
сплошных сред является условным, то нельзя утверждать, что ограничение $\CP\ge0$
обосновано с физической точки зрения.}.

Мы не обсуждаем физических аргументов, приводящих к тензору энергии-импульса
(\ref{enmotl}), отсылая читателя к монографии \cite{LanLif88R}. В настоящем
разделе мы рассматриваем, в основном, математические свойства данного
определения.
\begin{com}
В гидро- и газодинамике все уравнения записываются таким образом, что в них
входит не сама энергия и давление, а только их градиенты. Это означает, что
энергия и давление определены с точностью до добавления произвольной постоянной.
В общей теории относительности ситуация отличается принципиально, т.к.\
уравнения меняются, если к $\CE$ или $\CP$ добавить постоянную. В частности,
наличие космологической постоянной можно интерпретировать как среду с постоянной
плотностью энергии $\CE=\const$ и постоянным давлением $\CP=-\CE$. Если $\CE>0$,
то давление отрицательно, $\CP<0$. Поэтому космологическую постоянную можно
интерпретировать, как некоторую среду, заполняющую все пространство-время со
свойствами обыкновенной резины.
\qed\end{com}

Из общих физических представлений следует, что след тензора энергии-импульса для
обычной материи должен быть неотрицательным \cite{LanLif88R}
\begin{equation}                                                  \label{enmetr}
  T_\Sm^\al{}_\al=\CE-3\CP\ge0.
\end{equation}
Этим свойством обладает, в частности, тензор энергии-импульса для произвольного
распределения точечных частиц (\ref{enmutr}). Отсюда вытекает ограничение на
давление
\begin{equation}                                                  \label{eprere}
  \CP\le\frac\CE3.
\end{equation}

Поскольку давление, по предположению, неотрицательно, то с учетом равенства
(\ref{eprere}) существует два крайних случая. Если материя, которой заполнена
вселенная, настолько разрежена, что давление можно считать равным нулю, то
говорят, что материя пылевидна. Максимальное возможное давление, $\CP=\CE/3$,
соответствует газу ультрарелятивистских частиц, скорости которых близки к
скорости света (см.\ раздел \ref{senmop}). В этом случае говорят, что вселенная
заполнена газом излучения или, просто, излучением.
\begin{align*}
  \CP&=0, && \text{-- \em пыль},
\\
  \CP&=\frac\CE3, && \text{-- \em излучение}.
\end{align*}
\index{Пыль (dust)}\index{Пылевидная материя (dust matter)}%
\index{Материя пылевидная (dust matter)}%
\index{Излучение (radiation)}\index{Газ излучения (gas of radiation)}%
\begin{exa}[\bf Нерелятивистская гидродинамика]
Рассмотрим пространство Минковского $\MR^{1,3}$ в декартовой системе координат
с метрикой Лоренца $\eta_{\al\bt}=\diag(+---)$. Пусть пространство-время
заполнено идеальной (без вязкости) жидкостью. Течение жидкости описывается
плотностью $\rho$, давлением $\CP$ и трехмерной скоростью $u^\mu$, $\mu=1,2,3$.
Тензор энергии-импульса идеальной жидкости, по-определению, имеет вид
\begin{equation}                                                  \label{enmono}
  T_\Sm^{\al\bt}=(\CE+\CP)u^\al u^\bt-\CP\eta^{\al\bt}.
\end{equation}
Покажем, что уравнения движения нерелятивистской идеальной жидкости (если не
считать уравнения состояния) следуют из закона сохранения четырехмерного тензора
энергии-импульса, $\pl_\bt T_\Sm^{\al\bt}=0$.

В нерелятивистском приближении мы считаем, что пространственные компоненты
скорости малы: $u^0\simeq1$, $u^\mu\ll1$, где
$\lbrace u^\al\rbrace=\lbrace u^0,u^\mu\rbrace$, $\al=0,1,2,3$. Кроме того,
давление мало, $\CP\ll\CE$, и в нулевом приближении плотность энергии совпадает
с плотностью жидкости, $\CE\simeq\rho$. Тогда в низшем приближении компоненты
тензора энергии-импульса равны:
\begin{equation}                                                  \label{enhyfl}
\begin{aligned}
  T_\Sm^{00}&=(\CE+\CP)u^0u^0-\CP&&\simeq\rho,
\\
  T_\Sm^{0\mu}=T_\Sm^{\mu0}&=(\CE+\CP)u^0u^\mu&&\simeq\rho u^\mu,
\\
  T_\Sm^{\mu\nu}&=(\CE+\CP)u^\mu u^\nu-\CP\eta^{\mu\nu}&&\simeq\rho u^\mu u^\nu
  -\CP\eta^{\mu\nu}.
\end{aligned}
\end{equation}
Рассмотрим закон сохранения энергии-импульса $\pl_\bt T_\Sm^{\al\bt}=0$. Нулевая
компонента этого равенства в главном приближении имеет вид
\begin{equation}                                                  \label{eilequ}
  \pl_\bt T_\Sm^{0\bt}=\pl_0T_\Sm^{00}+\pl_\mu T_\Sm^{0\mu}
  \simeq\pl_0\rho+\pl_\mu(\rho u^\mu)=0.
\end{equation}
Полученное уравнение совпадает с {\em уравнением непрерывности}.
\index{Уравнение непрерывности (continuity equation)}%
\index{Непрерывности уравнение (continuity equation)}%
Пространственные компоненты закона сохранения энергии-импульса в главном
приближении приводят к равенству
\begin{equation*}
  \pl_\bt T_\Sm^{\mu\bt}=\pl_0T_\Sm^{\mu0}+\pl_\nu T_\Sm^{\mu\nu}\simeq
  \rho\pl_0u^\mu+u^\mu\left[\pl_0\rho+\pl_\nu(\rho u^\nu)\right]
  +\rho u^\nu\pl_\nu v^\mu-\pl^\mu\CP=0,
\end{equation*}
что, с учетом уравнения непрерывности (\ref{eilequ}), дает {\em уравнение
Эйлера}
\begin{equation}                                                  \label{eilera}
  \pl_0u^\mu+u^\nu\pl_\nu u^\mu=\frac1\rho\eta^{\mu\nu}\pl_\nu\CP.
\end{equation}
\index{Уравнение Эйлера (the Euler equation)}%
\index{Эйлера уравнение (the Euler equation)}%
Если дополнить уравнение непрерывности и уравнение Эйлера уравнением
состояния идеальной жидкости $\CP=\CP(\rho)$, связывающим давление и плотность,
то получим полную систему уравнений для идеальной жидкости. Таким образом,
уравнения движения нерелятивистской идеальной жидкости следуют из закона
сохранения четырехмерного тензора энергии-импульса (\ref{enmono}), дополненного
уравнениям состояния.

Эта же система уравнений (\ref{eilequ}), (\ref{eilera}) описывает движение
идеального газа. Разница заключается только в уравнении состояния. Для
идеального газа уравнение состояния имеет вид
\begin{equation}                                                  \label{eidgas}
  \CP=\frac\rho\mu RT,
\end{equation}
где $\mu,R$ и $T$ есть, соответственно, молекулярный вес, универсальная газовая
постоянная и абсолютная температура. При постоянной температуре $T=\const$
давление идеального газа прямо пропорционально плотности.
\qed\end{exa}

Наличие в пространстве-времени метрики и времениподобного векторного поля $u$,
$u^2=1$, позволяет определить проекционные операторы (\ref{epropv}):
\begin{equation*}
  \Pi^\Sl_\al{}^\bt:=u_\al u^\bt,\qquad
  \Pi^\St_\al{}^\bt:=\dl_\al^\bt-u_\al u^\bt.
\end{equation*}
В каждой точке $x\in\MM$ эти операторы проектируют тензорные поля,
соответственно, на направление вектора скорости $u$ и перпендикулярную
гиперплоскость в касательном пространстве $\MT_x(\MM)$. Например, проекция
метрики имеет вид
\begin{align*}
  g^\Sl_{\al\bt}&:=\Pi^\Sl_\al{}^\g\Pi^\Sl_\bt{}^\dl g_{\g\dl}=u_\al u_\bt,
\\
  g^\St_{\al\bt}&:=\Pi^\St_\al{}^\g\Pi^\St_\bt{}^\dl g_{\g\dl}
  =g_{\al\bt}-u_\al u_\bt.
\end{align*}
Ясно также, что
\begin{equation*}
  u^{\Sl\al}:=u^\bt\Pi^\Sl_\bt{}^\al=u^\al,\qquad
  u^\St:=u^\bt\Pi^\St_\bt{}^\al=0.
\end{equation*}
Поэтому тензор энергии-импульса (\ref{enmotl}) можно переписать с помощью
проекционных операторов
\begin{equation}                                                  \label{qmmjng}
  T_\Sm^{\al\bt}=\CE g^{\Sl\al\bt}-\CP g^{\St\al\bt}.
\end{equation}

Поскольку тензор энергии-импульса сплошной среды (\ref{enmotl}) не был получен
из вариационного принципа, то на него необходимо наложить дополнительное условие
\begin{equation}                                                  \label{ecosem}
  \nb_\bt T_\Sm^{\bt\al}=0,
\end{equation}
которое является условием совместности уравнений Эйнштейна. Более подробно
\begin{equation*}
  (\CE+\CP)u^\bt\nb_\bt u^\al+u^\al\nb_\bt\big[(\CE+\CP)u^\bt\big]
  -g^{\al\bt}\pl_\bt\CP=0,
\end{equation*}
где мы воспользовались условием метричности связности Леви-Чивиты
$\nb_\bt g^{\g\al}=0$. Проекции этого уравнения на вектор $u$ и перпендикулярную
гиперплоскость имеют следующий вид
\begin{align}                                                     \label{eupogi}
  (\CE+\CP)\nb_\al u^\al+u^\al\pl_\al\CE&=0,
\\                                                                \label{epeugi}
  (\CE+\CP)u^\bt\nb_\bt u^\al-(g^{\al\bt}-u^\al u^\bt)\pl_\bt\CP&=0,
\end{align}
где мы воспользовались уравнением $u_\al\nb_\bt u^\al=0$, которое следует из
условия $u^2=0$ после дифференцирования. Легко проверить, что свертка уравнений
(\ref{epeugi}) с ковектором $u_\al$ тождественно обращается в нуль.
Следовательно, только четыре уравнения из (\ref{eupogi}), (\ref{epeugi})
являются независимыми, и они эквивалентны условию ковариантного сохранения
тензора энергии-импульса $\nb_\bt T_\Sm^{\bt\al}=0$.

Уравнение (\ref{eupogi}) является ковариантным обобщением уравнения
непрерывности для нерелятивистской жидкости (\ref{eilequ}), а уравнение
(\ref{epeugi}) -- ковариантным обобщением уравнения Эйлера (\ref{eilera}). Эти
уравнения представляют собой систему уравнений {\em релятивистской
гидродинамики}.
\index{Релятивистская гидродинамика (relativistic hydrodynamics)}%
\index{Гидродинамика релятивистская (relativistic hydrodynamics)}%

Система уравнений (\ref{eupogi}), (\ref{epeugi}) вместе с уравнениями Эйнштейна
не образует полной системы уравнений релятивистской гидродинамики. Ее необходимо
дополнить уравнением состояния. Широкий класс моделей описывается уравнением
состояния $\CP=\CP(\CE)$, связывающим давление с плотностью энергии в каждой
точке пространства-времени. Такие жидкости называются {\em баротропными}.
\index{Баротропная жидкость (borotropic fluid)}%
\index{Жидкость баротропная (borotropic fluid)}%

Второе слагаемое в уравнении Эйлера (\ref{epeugi}) после опускания индекса имеет
вид
\begin{equation*}
  \Pi^\St_\al{}^\bt\pl_\bt\CP=(\dl_\al^\bt-u_\al u^\bt)\pl_\bt\CP.
\end{equation*}
Если оно равно нулю, т.е.\ градиент давления параллелен вектору скорости, то
уравнение Эйлера упрощается $u^\bt\nb_\bt u^\al=0$. Это есть уравнение
экстремалей. В этом случае точки жидкости движутся так же, как и точечные
частицы.

Для пылевидной материи давление равно нулю и система уравнений релятивистской
гидродинамики существенно упрощается:
\begin{equation}                                                  \label{eduseq}
  \nb_\al(\CE u^\al)=0,\qquad u^\bt\nb_\bt u^\al=0.
\end{equation}
Мы видим, что пылевидная материя движется вдоль экстремалей, как множество
точечных частиц.

При анализе общих свойств решений уравнений Эйнштейна на тензор энергии-импульса
материи накладываются определенные {\em энергетические условия.} Перечислим три
наиболее распространенных.
\begin{defn}
Говорят, что источники (поля материи) удовлетворяют: \newline
\indent 1) \parbox[t]{.93\linewidth}{{\em слабому энергетическому условию}, если
для любого времениподобного векторного поля $X$ выполнено неравенство
\begin{equation}                                                  \label{ewenco}
  T_{\Sm\al\bt}X^\al X^\bt\ge0;
\end{equation}  }
\indent 2) \parbox[t]{.93\linewidth}{{\em сильному энергетическому условию},
если для любого времениподобного векторного поля $X$ выполнено неравенство
\begin{equation}                                                  \label{estenc}
  \rho_{\al\bt}X^\al X^\bt\ge0,\qquad\text{где}\quad
  \rho_{\al\bt}:=T_{\Sm\al\bt}-\frac1{n-2} g_{\al\bt}T_{\Sm\g}{}^\g,
\end{equation}
где $n$ -- размерность пространства-времени. Это условие называется также
{\em условием положительности Риччи} (см.\ уравнение (\ref{eineqe}); }
\indent 3) \parbox[t]{.93\linewidth}{{\em доминантному энергетическому условию},
если для произвольного времениподобного векторного поля $X$, направленного в
будущее, векторное поле с компонентами $T_{\Sm\bt}{}^\al X^\bt$ также
времениподобно и направлено в будущее. }
\qed\end{defn}
\index{Энергетическое условие (energy condition)}%
\index{Условие энергетическое (energy condition)}%
\index{Слабое энергетическое условие (weak energy condition)}%
\index{Условие слабое энергетическое (weak energy condition)}%
\index{Сильное энергетическое условие (strong energy condition)}%
\index{Условие сильное энергетическое (strong energy condition)}%
\index{Условие положительности Риччи (Ricci positivity condition}%
\index{Доминантное энергетическое условие (dominant energy condition)}%
\index{Условие доминантное энергетическое (dominant energy condition)}%
В данном определении векторное поле $X$ времениподобно, т.е.\
$X^2:=X^\al X_\al>0$. Поэтому, не ограничивая общности, можно считать, что оно
нормировано на единицу, $X^2=1$. Кроме того, всегда можно выбрать такую систему
координат, в которой произвольное времениподобное векторное поле имеет только
одну нетривиальную компоненту $X=(1,0,\dotsc,0)$. Поэтому слабое и сильное
энергетические условия можно переформулировать. В произвольной системе
координат, в которой координата $x^0$ является временем, должны выполняться
равенства: $T_{\Sm00}\ge0$ или $\rho_{00}\ge0$, соответственно.

Ясно, что если выполнено доминантное энергетическое условие, то
$T_{\Sm\al\bt}X^\al X^\bt>0$. Следовательно из 3) следует 1). Кроме того, если
след тензора энергии-импульса материи положителен, как, например, для точечных
частиц (см.\ раздел \ref{senmop}), то из сильного энергетического условия
вытекает слабое, причем получается строгое неравенство
$T_{\Sm\al\bt}X^\al X^\bt>0$. Если же след тензора энергии-импульса материи
отрицателен, то, наоборот, слабое энергетическое условие сильнее, чем сильное
энергетическое условие.

\begin{prop}
Если выполнены условия $\CE>0$, $\CP\ge0$, то тензор энергии-импульса сплошной
среды удовлетворяет слабому и сильному энергетическим условиям.
\end{prop}
\begin{proof}
Зафиксируем систему координат, в которой времениподобное векторное поле имеет
вид $X=(1,0,0,0)$. Тогда слабое энергетическое условие эквивалентно неравенству
\begin{equation*}
  T_{\Sm00}=\CE u_0^2+\CP(u_0^2-g_{00})\ge0.
\end{equation*}
Первое слагаемое в этом выражении, очевидно, неотрицательно (в принципе оно
может обратиться в нуль, т.к.\ условия $u^0>0$ недостаточно для положительности
нулевой ковариантной компоненты скорости $u_0$). Поскольку
\begin{equation*}
  u_0:=g_{00}u^0+g_{0\mu}u^\mu,
\end{equation*}
то второе слагаемое принимает вид
\begin{equation*}
  \CP\left(g_{00}^2(u^0)^2+2g_{00}g_{0\mu}u^0u^\mu+(g_{0\mu}u^\mu)^2
  -g_{00}\right)
  =-\CP g_{00}\left(g_{\mu\nu}-\frac{g_{0\mu}g_{0\nu}}{g_{00}}\right)
  u^\mu u^\nu,
\end{equation*}
где мы воспользовались тождеством $u^2=1$. Согласно теореме \ref{tlosim} на
лоренцевом многообразии матрица в скобках в правой части равенства отрицательно
определена, и поэтому слабое энергетическое условие выполнено.

Для тензора энергии-импульса сплошной среды справедливо равенство
\begin{equation*}
  \rho_{\al\bt}=(\CE+\CP)u_\al u_\bt-\CP g_{\al\bt}-\frac12(\CE-3\CP)g_{\al\bt}.
\end{equation*}
Отсюда следует выражение для нулевой компоненты
\begin{equation*}
  \rho_{00}=\CE(u_0^2-g_{00})+\CP(u_0^2-g_{00})
  +\frac12(\CE+3\CP)g_{00}.
\end{equation*}
Первые два слагаемых неотрицательны в силу предыдущих рассуждений. Последнее
слагаемое положительно, т.к.\ координата $x^0$ является временем и,
следовательно, $g_{00}>0$.
\end{proof}
\begin{com}
В доказательстве предложения мы не использовали неравенство $\CP\le\CE/3$,
обеспечивающее неотрицательность следа тензора энергии-импульса.
\qed\end{com}
\begin{prop}
Если выполнено условие $|\CP|<|\CE|$, то тензор энергии-импульса сплошной среды
удовлетворяет условию энергодоминантности.
\end{prop}
\begin{proof}
В системе координат, где $X=(1,0,0,0)$, условие энергодоминантности имеет вид
\begin{equation*}
  T_{\Sm0}{}^\al T_{\Sm0\al}>0.
\end{equation*}
Подставляя в это неравенство выражение для тензора энергии-импульса
(\ref{enmotl}), получим соотношение
\begin{equation*}
  (\CE^2-\CP^2)u_0^2+\CP^2 g_{00}>0.
\end{equation*}
Очевидно, что неравенство $|\CP|<|\CE|$ достаточно для выполнения условия
энергодоминантности.
\end{proof}
\subsection{Акустические фононы в нерелятивистской гидродинамике \label{seldef}}
В физике всегда большой интерес вызывали аналогии между явлениями из разных
областей. В настоящем разделе мы покажем, что некоторые явления в
нерелятивистской гидродинамике и общей теории относительности описываются
уравнениями, которые обладают рядом одинаковых свойств.

Уравнения движения для акустических фононов следуют из классических
нерелятивистских уравнений гидродинамики следующим образом. Рассмотрим некоторое
точное решение уравнений гидродинамики. Тогда уравнение для малых возбуждений
(фононов) вблизи этого решения сводится к уравнению Даламбера с нетривиальной
эффективной четырехмерной метрикой лоренцевой сигнатуры, в которой роль скорости
света играет скорость звука в жидкости. Отличие от общей теории относительности
сводится к тому, что эффективная метрика для фононов определяется уравнениями
гидродинамики, а не уравнениями Эйнштейна. Тем не менее уравнение для фононов
задается нетривиальной четырехмерной метрикой, для которой тензор кривизны
отличен от нуля. Другими словами, фононы двигаются на многообразии с
нетривиальной геометрией. При этом возможно возникновение горизонтов, когда
скорость течения жидкости превышает скорость звука, и, следовательно,
образование акустических аналогов черных дыр.

В настоящем разделе мы следуем выводу уравнения для фононов, предложенному в
\cite{Unruh81,Unruh95} (см.\ также \cite{Visser98}).

Рассмотрим $4$-мерное галилеево пространство-время с декартовой системой
координат $\lbrace x^\al\rbrace$, $\al=0,1,2,3$, которые мы будем обозначать
индексами из начала греческого алфавита $\al,\bt,\dotsc$. Координату $x^0\in\MR$
мы отождествляем с временем, $x^0=t$. Пространственные координаты
$\lbrace x^\mu\rbrace\in\MR^3$ мы будем обозначать индексами из середины
греческого алфавита $\mu,\nu,\dotsc$. Жидкость без вязкости называется
{\em идеальной} и описывается плотностью $\rho(x)$, давлением $\CP(x)$ и
вектором скорости $\Bu=\lbrace u^\mu(x)\rbrace$, где $x=\lbrace x^\al\rbrace$.
Движение идеальной жидкости или идеального газа в пространстве определяется
следующей замкнутой системой из пяти нелинейных уравнений для пяти переменных
(см., например, \cite{LanLif86R})
\index{Идеальная жидкость (ideal fluid)}%
\index{Жидкость идеальная (ideal fluid)}%
\begin{align}                                                     \label{eulere}
  \rho\dot{\Bu}+\rho(\Bu\nb)\Bu &=-\nb\CP+\Bf,
\\                                                                \label{ecoequ}
  \dot\rho+\div(\rho\Bu)&=0,
\\                                                                \label{eqstat}
  \CP&=\CP(\rho),
\end{align}
где точка обозначает дифференцирование по времени и $\nb$ -- градиент.
Уравнение (\ref{eulere}) называется уравнением Эйлера (\ref{eilera}) и
представляет собой второй закон Ньютона для элемента объема жидкости. Здесь
$\Bf(x)$ -- плотность внешних сил. Например, в гравитационном поле
$\Bf=-\rho\nb \vf$, где $\vf(x)$ -- потенциал гравитационного поля. В дальнейшем
мы ограничимся только этим случаем. Уравнение (\ref{ecoequ}) является уравнением
непрерывности (\ref{eilequ}). Последнее уравнение (\ref{eqstat}) является
уравнением состояния жидкости, которое характеризует саму жидкость и считается
заданным. Здесь мы предполагаем, что давление жидкости зависит только от ее
плотности, т.е.\ она является баротропной.

Для несжимаемой жидкости $\rho=\const$ и уравнение непрерывности принимает
вид $\div\Bu=0$.

Уравнения (\ref{eulere})--(\ref{eqstat}) записаны в стандартном для
гидродинамики виде, где не делается различие между верхними и нижними индексами.

Если уравнение состояния задано, то давление $\CP$ (или плотность $\rho$) можно
исключить из системы уравнений движения. Для этого заметим, что
\begin{equation*}
  \nb\CP=\frac{d\CP}{d\rho}\nb\rho=c^2\nb\rho,
\end{equation*}
где введена скорость звука $c(\rho)$
\begin{equation}                                                  \label{esouve}
  c^2:=\frac{d\CP}{d\rho}.
\end{equation}
Для обычных жидкостей с увеличением плотности давление увеличивается, и поэтому
$c^2>0$. Тогда уравнение Эйлера можно переписать в виде
\begin{equation}                                                  \label{eeulse}
  \dot{\Bu}+(\Bu\nb)\Bu=-c^2\frac{\nb\rho}\rho-\nb\vf,
\end{equation}
где скорость звука $c=c(\rho)$ рассматривается, как заданная функция от
плотности жидкости. Уравнение Эйлера (\ref{eeulse}) вместе с уравнением
непрерывности (\ref{ecoequ}) однозначно описывают движение такой жидкости при
заданных начальных и граничных условиях.

Воспользовавшись тождеством
\begin{equation*}
  \frac12\nb\Bu^2=[\Bu,\rot\Bu]+(\Bu\nb)\Bu,
\end{equation*}
где квадратные скобки обозначают векторное произведение, уравнение Эйлера
(\ref{eeulse}) можно переписать в виде
\begin{equation*}
  \dot{\Bu}-[\Bu,\rot\Bu]=-\nb\left(h+\frac{\Bu^2}2+\vf\right),
\end{equation*}
где введена {\em энтальпия} жидкости
\index{Энтальпия (enthalpy)}%
\begin{equation*}
  h(\CP):=\int_0^\CP\frac{d\CP'}{\rho(\CP')}.
\end{equation*}

Предположим, что движение жидкости является безвихревым:
\begin{equation}                                                  \label{eirrol}
  \rot\Bu=0\quad \Leftrightarrow\quad \pl_\mu u_\nu-\pl_\nu u_\mu=0.
\end{equation}
На геометрическом языке это означает, что 1-форма $dx^\mu u_\mu$ является
замкнутой на пространственном сечении $x^0=\const$. Тогда локально (во всем
$\MR^3$) существует потенциальное поле $\psi(x)$ (потенциал) такое, что
\begin{equation}                                                  \label{eirrve}
  u_\mu=-\pl_\mu\psi.
\end{equation}
Для безвихревой жидкости уравнение Эйлера эквивалентно {\em уравнению Бернулли}
\index{Уравнение Бернулли (Bernoulli equation)}%
\index{Бернулли уравнение (Bernoulli equation)}%
\begin{equation}                                                  \label{eberne}
  -\dot\psi+h+\frac{(\nb\psi)^2}2+\vf=F(t),
\end{equation}
где $F(t)$ -- произвольная функция времени. Поскольку потенциал $\psi$ определен
с точностью до добавления произвольной функции времени, то, без ограничения
общности, положим $F(t)=0$.

Теперь можно приступить к изучению фононов в нерелятивистской гидродинамике.
Допустим, что нам известно точное решение уравнений гидродинамики $\rho_0(x)$,
$\CP_0(x)$ и $\Bu_0(x)=-\nb\psi_0$. Получим уравнение, описывающее
распространение акустических возбуждений (фононов) вблизи этого решения. Пусть
\begin{equation}                                                  \label{eapphy}
\begin{split}
  \rho&\simeq\rho_0+\e\rho_1,
\\
  \CP&\simeq\CP_0+\e\CP_1,
\\
  u^\mu&\simeq u_0^\mu+\e u_1^\mu,
\\
  \psi&\simeq\psi_0+\e\psi_1.
\end{split}
\end{equation}
где $\e\ll 1$ -- малый параметр разложения. При этом мы считаем внешние силы
заданными $\vf=\vf_0$.

В дальнейшем мы будем использовать обозначения, принятые в дифференциальной
геометрии, и различать верхние и нижние индексы. Пространственные индексы в
декартовой системе координат поднимаются и опускаются с помощью метрики
$\eta_{\mu\nu}=-\dl_{\mu\nu}$ и ее обратной, которая отличается от евклидовой
метрики знаком. В наших обозначениях $u_\mu=\pl_\mu\psi$,
$u^\mu=\eta^{\mu\nu}\pl_\nu\psi=-\pl_\mu\psi$.

Уравнение Бернулли (\ref{eberne}) в нулевом и первом порядке по $\e$ имеет вид
\begin{align}                                                     \label{eberze}
  \e^0:\qquad &-\pl_0\psi_0+h_0
  -\frac12\eta^{\mu\nu}\pl_\mu\psi_0\pl_\nu\psi_0+\vf_0=0,
\\                                                                \label{eberfi}
  \e^1:\qquad &-\pl_0\psi_1+\frac{\CP_1}{\rho_0}-u_0^\mu\pl_\mu\psi_1=0,
\end{align}
где учтено разложение для энтальпии
\begin{equation*}
  h(\CP)\simeq h(\CP_0)+\e\frac{\CP_1}{\rho_0}.
\end{equation*}
Учтем, что
\begin{equation*}
  \rho_1=\frac{d\rho}{d\CP}\CP_1=\frac{\CP_1}{c^2},
\end{equation*}
и найдем поправку к плотности $\rho_1$ из уравнения (\ref{eberfi})
\begin{equation}                                                  \label{erhfio}
  \rho_1=\frac{\rho_0}{c^2}(\pl_0\psi_1+u_0^\mu\pl_\mu\psi_1).
\end{equation}

Уравнение непрерывности в нулевом и первом порядке по $\e$ имеет вид
\begin{align}                                                     \label{ecoeze}
  \e^0:\qquad &\pl_0\rho_0+\pl_\mu(\rho_0u_0^\mu)=0,
\\                                                                \label{ecoefi}
  \e^1:\qquad &\pl_1\rho_1+\pl_\mu(\rho_0u_1^\mu+\rho_1u_0^\mu)=0.
\end{align}
Подставим во второе уравнение решение для поправки к плотности (\ref{erhfio}).
В результате получим уравнение для поправки к потенциалу скорости:
\begin{equation}                                                  \label{evepot}
  \pl_0\left[\frac{\rho_0}{c^2}(\pl_0\psi_1+u_0^\mu\pl_\mu\psi_1)\right]
  +\pl_\mu\left[\rho_0\pl^\mu\psi_1+\frac{\rho_0}{c^2}u_0^\mu
  (\pl_0\psi_1+u_0^\nu\pl_\nu\psi_1)\right]=0.
\end{equation}
Это волновое уравнение для $\psi_1(x)$ полностью определяет распространение
акустических колебаний в движущейся жидкости, описываемой плотностью $\rho_0(x)$
и полем скоростей $u^\mu_0(x)$, с заданным уравнением состояния $c=c(\rho)$.
В случае, когда скорость жидкости равна нулю, $u_0=0$, а плотность $\rho_0$ и
величина $c$ постоянны, уравнение (\ref{evepot}) сводится к уравнению
Даламбера, которое описывает распространение акустических возбуждений со
скоростью $c$. Это оправдывает введенное выше понятие скорости звука
(\ref{esouve}).

Если решение для $\psi_1(x)$ известно, то поправка к плотности $\rho_1$
однозначно определяется формулой (\ref{erhfio}).

Уравнение (\ref{evepot}), как легко проверить, можно записать в матричных
обозначениях
\begin{equation*}
  \pl_\al(f^{\al\bt}\pl_\bt\psi_1)=0,
\end{equation*}
где
\begin{equation*}
    f^{\al\bt}:=\frac{\rho_0}{c^2}
  \begin{pmatrix}
    1 & u_0^\nu \\[1mm] u_0^\mu & c^2\eta^{\mu\nu}+u_0^\mu u_0^\nu.
  \end{pmatrix}
\end{equation*}

Введем метрику в галилеевом пространстве
\begin{align}                                                     \label{emetef}
  g_{\al\bt}&:=\frac{\rho_0}{c}
  \begin{pmatrix}
    c^2+u_0^\mu u_{0\mu} & -u_{0\nu} \\[1mm] -u_{0\mu} & \eta_{\mu\nu},
  \end{pmatrix}
\\  \intertext{и ее обратную}                                     \label{emeten}
  g^{\al\bt}&=\frac1{\rho_0c}
  \begin{pmatrix}
    1 & u_0^\nu \\[1mm] u_0^\mu & c^2\eta^{\mu\nu}+u_0^\mu u_0^\nu.
  \end{pmatrix}
\end{align}
Интервал, соответствующий метрике (\ref{emetef}), можно записать в виде
\begin{equation}                                                  \label{effinh}
  ds^2=\frac{\rho_0}c\left[c^2dt^2
  +\eta_{\mu\nu}(dx^\mu-u_0^\mu dt)(dx^\nu-u_0^\nu dt)\right].
\end{equation}

``Эффективная'' метрика (\ref{emetef}) имеет лоренцеву сигнатуру $(+---)$, и ее
определитель равен
\begin{equation*}
  g:=\det g_{\al\bt}=-\frac{\rho_0^4}{c^2}.
\end{equation*}

Сравнение метрики (\ref{emetef}) с АДМ параметризацией произвольной
псевдоримановой метрики (\ref{eadmme}) дает следующее выражение для функций хода
и сдвига:
\begin{equation*}
  N=\sqrt{\rho_0c},\qquad N^\mu=-\frac{\rho_0u_0^\mu}c.
\end{equation*}

Вернемся к уравнению для фононов. Обратная метрика $g^{\al\bt}$ отличается от
матрицы $f^{\al\bt}$ простым множителем
\begin{equation*}
  g^{\al\bt}=\frac c{\rho_0^2}f^{\al\bt}.
\end{equation*}
Теперь уравнение для акустических фононов можно переписать в инвариантном
относительно общих преобразований координат виде
\begin{equation}                                                  \label{egepho}
  \frac1{\sqrt{|g|}}\pl_\al\big(\sqrt{|g|}g^{\al\bt}\pl_\bt\psi\big)=0,
\end{equation}
где мы, для простоты обозначений, отбросили индекс у поправки к потенциалу
скорости.

Таким образом, распространение фононов в движущейся жидкости описывается
инвариантным волновым уравнением в четырехмерном пространстве-времени с
нетривиальной метрикой лоренцевой сигнатуры (\ref{emetef}). Эта метрика
определяется плотностью $\rho_0$, скоростью звука $c$ и полем скоростей
$\Bu_0$, которые удовлетворяют исходным уравнениям (\ref{eberze}),
(\ref{ecoefi}). Подчеркнем, что движение самой жидкости происходит в плоском
галилеевом пространстве-времени, а распространение акустических возбуждений в
этой движущейся жидкости описывается волновым уравнением на псевдоримановом
пространстве-времени с нетривиальной ``эффективной'' метрикой.

Мы считаем, что уравнение состояния жидкости задано, и, следовательно, задана
скорость звука в жидкости, как функция плотности. Тогда эффективная метрика
определяется четырьмя функциями $\rho_0(x)$ и $\Bu_0(x)$, которые
удовлетворяют уравнениям гидродинамики. При постановке задачи Коши для
однозначного определения этих функций необходимо задать четыре произвольные
функции в качестве начальных условий. В общей теории относительности метрика
удовлетворяет уравнениям Эйнштейна и имеет две распространяющихся степени
свободы. При постановке задачи Коши для уравнений Эйнштейна также необходимо
задать четыре произвольные функции на пространственноподобном сечении: по две
на каждую степень свободы, так как уравнения движения второго порядка.

Нулевая компонента метрики $g_{00}$ в (\ref{emetef}) меняет знак в тех точках
пространства-времени, где течение жидкости становится сверхзвуковым:
$c^2=\Bu^2:=-u^\mu u_\mu$. Эти поверхности в пространстве соответствуют
горизонтам черных дыр. Действительно, поскольку скорость фононов ограничена
скоростью звука в жидкости, то они не могут покинуть область сверхзвукового
течения. Следовательно, в быстро текущей жидкости для фононов могут
образовываться аналоги черных дыр в общей теории относительности.
\begin{figure}[h,b,t]
\hfill\includegraphics[width=.35\textwidth]{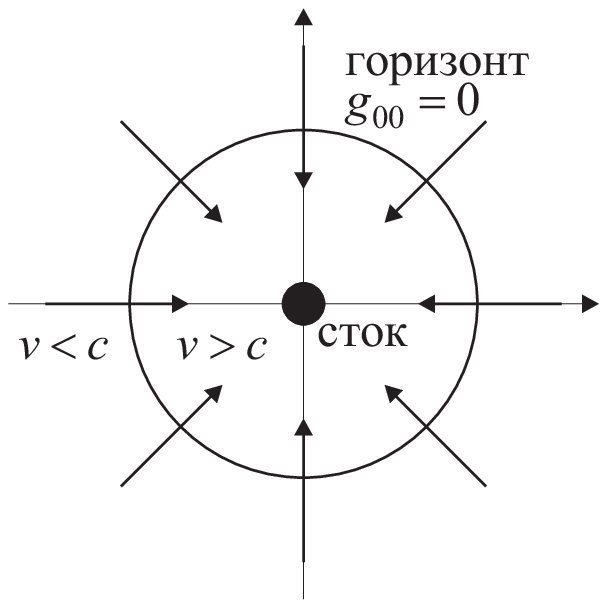}
\hfill {}
\centering\caption{Черная дыра для акустических фононов. Горизонт определяется
равенством модуля скорости жидкости $v:=\sqrt{-u_0^\mu u_{0\mu}}$ скорости
звука $c$.}
\label{facblho}
\end{figure}
На рис.\ref{facblho} показана качественная картинка образования черной дыры в
жидкости. Представим, что все трехмерное пространство $\MR^3$ заполнено
несжимаемой жидкостью, и в центре декартовой системы координат находится
точечный сток, в который жидкость засасывается. Предположим, что жидкость на
бесконечности покоится, и ее движение радиально. Поскольку скорость жидкости
стремится к бесконечности при приближении к стоку, то существует сфера
некоторого радиуса, на которой скорость жидкости равна скорости звука. Эта сфера
называется горизонтом. Если фонон испущен в некоторой точке, лежащей вне
горизонта, то он может либо попасть в сток, либо уйти на бесконечность. Если же
фонон испущен из точки, лежащей внутри горизонта, то он неминуемо попадет в
сток, т.к.\ его скорости недостаточно для прохождения через горизонт. Описанная
ситуация является аналогом черных дыр в общей теории относительности. Надо
только скорость звука заменить на скорость света. Аналогом фононов является
электромагнитное излучение -- фотоны.
\section{Выбор системы координат                                 \label{sdeawh}}
Уравнения общей теории относительности ковариантны относительно общих
преобразований координат. Эту свободу можно использовать для выбора подходящей
системы отсчета, которая может упростить уравнения Эйнштейна. В настоящем
разделе будут описаны несколько широко распространенных способа фиксирования
системы координат.
\subsection{Сопутствующая система координат}
Рассмотрим уравнения Эйнштейна
\begin{equation}                                                  \label{einmat}
  \kappa\left(R_{\al\bt}-\frac12g_{\al\bt}R\right)=-\frac12T_{\Sm\al\bt}
\end{equation}
для сплошной среды с тензором энергии-импульса (см.\ раздел \ref{shhfyj})
\begin{equation}                                                  \label{enmosf}
  T_\Sm^{\al\bt}=(\CE+\CP)u^\al u^\bt-\CP g^{\al\bt}.
\end{equation}
Для получения замкнутой системы уравнений уравнения Эйнштейна необходимо
дополнить законом сохранения (уравнениями релятивистской гидродинамики)
\begin{equation}                                                  \label{ecoema}
  \nb_\bt T_\Sm^{\bt\al}=0
\end{equation}
и уравнением состояния среды
\begin{equation}                                                  \label{eqsteq}
  \CP=\CP(\CE),
\end{equation}
предполагая среду баротропной. Система уравнений (\ref{einmat}), (\ref{ecoema})
и (\ref{eqsteq}) образуют полную систему для неизвестных функций: $g_{\al\bt}$,
$u^\al$, $\CP$ и $\CE$. Нетрудно проверить, что число уравнений равно числу
неизвестных (напомним, что на вектор скорости $u$ наложено условие $u^2=1$ и
число его независимых компонент на единицу меньше размерности
пространства-времени).

По-построению, все уравнения ковариантны. Поэтому преобразования координат
можно использовать для упрощения системы уравнений. Обычно преобразования
координат используют для фиксирования части компонент метрики. Однако для
системы уравнений (\ref{einmat}), (\ref{ecoema}) и (\ref{eqsteq}) существует
другая естественная возможность. Если размерность пространства-времени равна
$n$, то в нашем распоряжение имеется $n$ функций, которых достаточно для
фиксирования векторного поля скорости. Тем самым число неизвестных функций
уменьшится, и задача упростится. Такой подход часто используется в космологии.

Опишем этот способ задания системы координат.
Рассмотрим псевдориманово многообразие $\MM$, $\dim\MM=n$, с метрикой $g$
лоренцевой сигнатуры. Пусть на нем задано произвольное достаточно гладкое
времениподобное векторное поле $u=u^\al(x)\pl_\al$, всюду отличное от нуля,
$u^2\ne0$. Не ограничивая общности, будем считать, что
\begin{equation*}
  u^2:=g_{\al\bt}u^\al u^\bt=1.
\end{equation*}
В противном случае можно просто заменить $u^\al\mapsto u^\al/u^2$.
\begin{exa}
Пусть пространство-время $\MM$ заполнено сплошной средой. Тогда задано векторное
поле скорости каждой точки среды $u$, для которого $u^2=1$.
\qed\end{exa}

Если на $\MM$ задано единичное времениподобное векторное поле, то определены
продольный и поперечный проекционные операторы:
\begin{equation}                                                  \label{eprtiv}
  \Pi^\Sl_\al{}^\bt:=u_\al u^\bt,\qquad
  \Pi^\St_\al{}^\bt:=\dl_\al^\bt-u_\al u^\bt.
\end{equation}
Произвольное векторное поле $X^\al\pl_\al$ можно спроектировать на векторное
поле $u$ и ортогональное дополнение
\begin{equation*}
  X^{\Sl\al}=(Xu)u^\al,\qquad X^{\St\al}=X^\al-(Xu)u^\al.
\end{equation*}
Аналогично проектируются тензоры произвольного ранга.

Согласно общему рассмотрению в разделе \ref{svechs} в окрестности произвольной
точки существует такая система координат, в которой все компоненты векторного
поля, кроме одной, равны нулю. Для единичного времениподобного векторного поля
нетривиальная компонента равна единице
\begin{equation}                                                  \label{ecomof}
  \lbrace u^\al\rbrace=\left\lbrace 1,0,\dotsc,0\right\rbrace.
\end{equation}
\begin{defn}
Система координат, в которой для единичного времениподобного векторного поля
выполнено условие (\ref{ecomof}), называется {\em сопутствующей векторному полю}
$u$.
\qed\end{defn}
\index{Сопутствующая система координат (comoving coordinate frame)}%
\index{Система координат сопутствующая (comoving coordinate frame)}%

В этой системе координат ковариантная производная векторного поля $u$ равна
символам Кристоффеля
\begin{equation*}
  \nb_\al u^\bt=\pl_\al u^\bt+\Gamma_{\al\g}{}^\bt u^g=\Gamma_{\al0}{}^\bt.
\end{equation*}

В сопутствующей системе координат $\pl_\al u^\bt=0$, поэтому при бесконечно
малых преобразованиях координат вариация формы компонент вектора (\ref{eintvf})
и ковектора (\ref{eintof}) содержит только одно слагаемое. Аналогично
преобразуются компоненты произвольного тензора. Отсюда вытекает, что производная
Ли от любого тензора $Y$ типа $(r,s)$ вдоль векторного поля $u$ совпадает с
производной по времени $x^0$:
\begin{equation*}
  \Lie_u Y^{\al_1\dotsc\al_r}{}_{\bt_1\dotsc\bt_s}
  =\pl_0 Y^{\al_1\dotsc\al_r}{}_{\bt_1\dotsc\bt_s}
\end{equation*}
и не зависит от связности, как, впрочем, любая производная Ли.

С физической точки зрения сопутствующую систему координат можно представить
следующим образом. Допустим, что некоторая среда заполняет все
пространство-время. Тогда с каждой точкой среды связана мировая линия
$x(s)$ (линия тока). Мы предполагаем, что касательные векторы к мировым линиям
образуют достаточно гладкое времениподобное векторное поле (вектор скорости)
\begin{equation*}
  u=u^\al\pl_\al,\qquad u^\al:=\frac{dx^\al}{ds},\qquad
  ds:=\sqrt{g_{\al\bt}dx^\al dx^\bt},
\end{equation*}
на многообразии $\MM$. Выберем произвольное сечение $\MS$, которое пересекает
все линии тока один раз, и зададим произвольную систему координат $x^\mu$,
$\mu=1,\dotsc,n-1$, на $\MS$. Это сечение совсем не обязано быть
пространственноподобным. Тогда сопутствующими координатами произвольной
точки $y\in\MM$ является набор чисел $\lbrace x^0:=s,x^\mu\rbrace$, где $x^\mu$
-- координаты точки пересечения поверхности $\MS$ с кривой $x(s)$, проходящей
через точку $y$.
\begin{com}
В предыдущем разделе мы установили, что пылевидная материя движется вдоль
экстремалей (\ref{eduseq}). Это значит, что в общем случае при наличии давления
или других негравитационных сил линии тока среды отличаются от экстремалей.
\qed\end{com}

Если производная Ли от некоторого тензора вдоль векторного поля скорости $u$
равна нулю, то в сопутствующей системе координат компоненты этого тензора могут
зависеть только от координат $x^\mu$, $\mu=1,\dotsc,n-1$ на сечении $\MS$. Это
значит, что соответствующий тензор жестко связан с движущейся средой и движется
вместе с ней. В дальнейшем мы будем предполагать, что все сечения $s=const$ и, в
частности, сечение $\MS$ пространственноподобны.

Сопутствующая векторному полю система координат определена неоднозначно.
Действительно, совершим преобразование координат $x^\al\mapsto x^{\al'}(x)$.
Тогда компоненты скорости преобразуются по тензорному закону:
\begin{equation*}
  u^\al\mapsto u^{\al'}:=\frac{\pl x^{\al'}}{\pl x^\al}u^\al.
\end{equation*}
Если до и после преобразования координат система координат является
сопутствующей, то функции преобразования координат должны удовлетворять
следующей системе уравнений
\begin{equation*}
  1=\frac{\pl x^{0'}}{\pl x^0},\qquad 0=\frac{\pl x^{\mu'}}{\pl x^0}.
\end{equation*}
Общее решение данной системы уравнений имеет вид
\begin{equation}                                                  \label{ereinv}
  x^0\mapsto x^0+f(\Bx),\qquad x^\mu\mapsto x^{\mu'}=x^\mu+f^\mu(\Bx),
\end{equation}
где $f,f^\mu$ -- $n$ произвольных функций координат на сечении $\MS\subset\MM$ и
$\Bx=\lbrace x^\mu\rbrace$. Функция $f$ соответствует произволу в выборе сечения
$x^0=\const$, и функции $f^\mu$ -- свободе в выборе координат $x^\mu$ на данных
сечениях.

Таким образом мы устранили $n-1$ неизвестную функцию в полной системе уравнений
(\ref{einmat}), (\ref{ecoema}) и (\ref{eqsteq}). В этой системе координат
тензор энергии-импульса принимает вид
\begin{equation*}
  T_\Sm^{00}=(\CE+\CP)-\CP g^{00},\qquad T_{\Sm}^{0\mu}=-\CP g^{0\mu},\qquad
  T_\Sm^{\mu\nu}=-\CP g^{\mu\nu}.
\end{equation*}
В общем случае ни он, ни тензор энергии-импульса с одним опущенным индексом
индексом $T_{\Sm\bt}^\al$ не будут диагональны.

Если задано единичное времениподобное векторное поле $u$, то в каждой точке
пространства-времени $x\in\MM$ в касательном пространстве $\MT_x(\MM)$ его можно
дополнить $n-1$ линейно независимыми векторами $e_\mu$, $\mu=1,\dotsc,n-1$,
которые перпендикулярны вектору $u$. Тогда совокупность векторов
$\lbrace u,e_\mu\rbrace$ образует в каждой точке репер. Ясно, что векторы
$e_\mu$ пространственноподобны, и их можно выбрать достаточно гладкими. Тогда
они задают $n-1$ мерное распределение векторных полей на $\MM$ (см.\ раздел
\ref{sfrote}). Согласно теореме Фробениуса для этого распределения существуют
интегральные подмногообразия тогда и только тогда, когда векторные поля $e_\mu$
находятся в инволюции. В общем случае это не так (это зависит от метрики).
Отсюда следует, что остаточного произвола в выборе сопутствующей системы
координат (\ref{ereinv}) недостаточно для того, чтобы выбрать секущую
поверхность $\MS$ таким образом, чтобы вектор $u$ был к ней всюду ортогонален.

При определении системы координат, сопутствующей векторному полю, за основу
взяты мировые линии точек среды. Теперь мы введем другое понятие сопутствующей
системы координат, где за основу определения будет взята система
гиперповерхностей, а не векторное поле.

Рассмотрим систему локальных координат $x^\al$ на $\MM$, где координата $x^0$
является временем, и все сечения $x^0=\const$ пространственноподобны.
По-предположению, на $\MM$ задана метрика лоренцевой сигнатуры, для которой мы
будем использовать АДМ параметризацию  (\ref{eadmme}). Выберем базис $n,e_\mu$ в
касательном пространстве, состоящий из пространственных векторов
$e_\mu=e_\mu{}^\al\pl_\al=\pl_\mu$ (символ $e_\mu{}^\al:=\dl_\mu^\al$ введен для
удобства последующих выкладок), касательных к гиперповерхностям $x^0=\const$, и
векторное поле
\begin{equation}                                                  \label{enovee}
  n:=\frac1N(\pl_0-N^\mu\pl_\mu),
\end{equation}
где $N$ -- функция хода и $N^\mu$ -- функции сдвига. Это векторное поле
перпендикулярно пространственным гиперповерхностям:
\begin{equation*}
  (n,e_\mu):=n^\al e_\mu{}^\bt g_{\al\bt}
  =n^0 e_\mu{}^\nu g_{0\nu}+n^\rho e_\mu{}^\nu g_{\rho\nu}=0,
\end{equation*}
и имеет единичную длину, $n^2=1$.

В общем случае коммутатор векторных полей $[n,e_\mu]$ отличен от нуля. Поэтому
пространственные координаты $x^\mu$ нельзя дополнить времениподобной координатой
$x^{0'}$ так, чтобы вектор $n$ был касателен к соответствующей координатной
линии: $n=\pl_{0'}$.

Нормальному вектору соответствует ортонормальная 1-форма
\begin{equation}                                                  \label{eotanf}
  n=dx^\al n_\al=dx^0N,
\end{equation}
где $n_\al:=n^\bt g_{\bt\al}$, для которой мы будем использовать то же
обозначение.

Для векторов и 1-форм справедливы разложения на перпендикулярную и касательные
составляющие:
\begin{equation*}
  X^\al=X^\bot n^\al+\tilde X^\mu e_\mu{}^\al,\qquad
  X_\al=X_\bot n_\al+\tilde X_\mu e^\mu{}_\al,
\end{equation*}
где
\begin{align*}
  X^\bot&=X^0N, & \tilde X^\mu&=X^0 N^\mu+X^\mu,
\\
  X_\bot&=\frac1N(X_0-N^\mu X_\mu), & \tilde X_\mu&=X_\mu.
\end{align*}
Поскольку $X_0:=X^\al g_{\al 0}$ и $X_\mu:=X^\al g_{\al\mu}$, то нетрудно
проверить, что $X_\bot=X^\bot$ и $\tilde X_\mu=\tilde X^\nu g_{\nu\mu}$.

Аналогично раскладываются тензоры произвольного ранга.

По-построению, векторное поле $n$ времениподобно и имеет единичную длину.
Поэтому его можно (как мы увидим, не всегда) отождествить с полем скоростей
материи $u$.
\begin{defn}
Система координат называется {\em сопутствующей}, если в этой системе координат
скорость $\lbrace u^\al\rbrace$ каждой точки сплошной среды имеет вид
\begin{equation}                                                  \label{ecomfr}
  \lbrace u^\al\rbrace=\left\lbrace\sqrt{g^{00}},\frac{g^{0\mu}}{\sqrt{g^{00}}}
  \right\rbrace=\left\lbrace\frac1N,-\frac{N^\mu}N\right\rbrace. \qed
\end{equation}
\end{defn}
\index{Сопутствующая система координат (comoving coordinate frame)}%
\index{Система координат сопутствующая (comoving coordinate frame)}%
Это определение корректно, т.к.\ условие $u^2=1$, как легко проверить,
выполнено.

В правой части равенства (\ref{ecomfr}), определяющего сопутствующую систему
координат, стоят определенные компоненты метрики, которые не образуют компонент
вектора. Следовательно, равенство (\ref{ecomfr}) нековариантно и действительно
фиксирует систему координат.

Название сопутствующая система координат оправдано следующим образом. Поскольку
вектор нормали $n$ к пространственному сечению $x^0=\const$ совпадает с вектором
скорости $u$, то репер $n,e_\mu$ привязан к среде и движется вместе с ней.

Если метрика имеет блочно диагональный вид \ref{etigam}, т.е.\ $N^\mu=0$ и
$N=1$, и, следовательно, выбрана временн\'ая калибровка (см.\ следующий раздел),
то вектор скорости материи в каждой точке пространства-времени имеет только
временн\'ую составляющую $\lbrace u^\al\rbrace=(1,0,0,0)$. В этом
случае сопутствующая система координат совпадает с системой координат,
сопутствующей векторному полю скорости $u$, которая была введена в начале
раздела. Это значит, что в сопутствующей системе координат каждая точка материи
покоится. Другими словами, система координат движется вместе с материей. При
этом каждая точка среды движется по времениподобной экстремали (геодезической).

При наличии сил негравитационного происхождения, например, давления, точки
материи могут двигаться не по экстремалям, и метрика в общем случае не будет
блочно-диагональна. Математически это означает, что в общем случае нельзя
одновременно удовлетворить условию (\ref{ecomfr}) и условию блочной
диагональности метрики (\ref{etigam}). Поэтому данные выше два определения
сопутствующей системы координат неэквивалентны.

Обозначим тензорные индексы по отношению к базису
$\lbrace e_a\rbrace:=\lbrace n,e_\mu\rbrace$ латинскими буквами $a,b,\dotsc$.
Тогда они принимают значения
$\lbrace a\rbrace=\lbrace\bot,\mu\rbrace=(\bot,1,\dotsc,n-1)$.
Отличие этого базиса от координатного базиса в касательном пространстве
заключается в том, что он неголономен, т.е.\ в общем случае не существует такой
системы координат $y^a(x)$, в которой были бы выполнены условия:
\begin{equation*}
  \frac{\pl x^\al}{\pl y^0}=n^\al,\qquad \frac{\pl x^\nu}{\pl y^\mu}=\dl_\mu^\nu
\end{equation*}
(см.\ раздел \ref{sunhba}). В этом базисе метрика имеет блочно диагональный вид
\begin{equation}                                                  \label{emetts}
  g_{ab}=\begin{pmatrix} 1 & 0 \\ 0 & g_{\mu\nu} \end{pmatrix},
\end{equation}
поскольку базисные векторы $n$ были выбраны единичными и перпендикулярными к
гиперповерхностям. В каждой точке пространства-времени векторы четырехмерной
скорости (\ref{ecomfr}) в базисе $n,\pl_\mu$ имеют только одну отличную от нуля
компоненту,
\begin{equation*}
  \lbrace u^a\rbrace=(1,0,0,0).
\end{equation*}
Это значит, что в рассматриваемом базисе материя покоится, что оправдывает
название сопутствующая.

Понятие сопутствующей системы координат полезно, т.к.\ позволяет исключить из
тензора энергии-импульса сплошной среды компоненты скорости $u^\al$, заменив их
на компоненты метрики следующим образом. Мы предполагаем, что тензор
энергии-импульса материи с одним контравариантным и одним ковариантным индексом
в сопутствующей системе координат (\ref{ecomfr}), и, следовательно, относительно
базиса $n,\pl_\mu$ является диагональным
\begin{equation}                                                  \label{engmom}
  T_\Sm^a{}_b=\begin{pmatrix} {\cal E} & 0 & 0 & 0 \\ 0 & -{\cal P} & 0 & 0 \\
  0 & 0 & -{\cal P} & 0 \\ 0 & 0 & 0 & -{\cal P} \end{pmatrix}.
\end{equation}
Тогда в координатном базисе $\lbrace \pl_\al\rbrace$ его контравариантные
компоненты имеют вид
\begin{equation}                                                  \label{etenmo}
  T_\Sm^{00}=\CE\frac1{N^2},\qquad
  T_\Sm^{0\mu}=T_\Sm^{\mu0}=-\CE\frac{N^\mu}{N^2},
  \qquad T_\Sm^{\mu\nu}=\CE\frac{N^\mu N^\nu}{N^2}-\CP\hat g^{\mu\nu}.
\end{equation}
Тем самым компоненты контравариантного тензора энергии-импульса не зависят от
векторного поля скоростей сплошной среды.
\begin{prop}
Функции $\CE$ и $\CP$ инвариантны относительно масштабного преобразования
(растяжки, гомотетии) пространственных координат:
\begin{equation}                                                  \label{erasco}
  x^0\mapsto x^0,\qquad x^\mu\mapsto x^{\prime\mu}=k(t)x^\mu,
\end{equation}
где $k(t)\ne0$ -- достаточно гладкая функция времени $x^0:=t$.
\end{prop}
\begin{proof}
Координатный базис преобразуется по-правилам:
\begin{equation*}
\begin{aligned}
  \pl_0&=\pl'_0+\dot kx^\mu\pl'_\mu, &\qquad dx^0&=d x^{\prime0},
\\
  \pl_\mu&=k\pl'_\mu, &\qquad
  dx^\mu&=\frac{dx^{\prime\mu}}k-\frac{x^{\prime\mu}\dot kdt}{k^2},
\end{aligned}
\end{equation*}
где $\dot k:=dk/dt$. Тогда в новой системе координат компоненты метрики примут
вид
\begin{equation*}
\begin{aligned}
  N'&=N, &&
\\
  N'_\mu&=\frac{N_\mu}k-\frac{\dot k x_\mu}{k^2}, &\qquad
  N^{\prime\mu}&=kN^\mu-\dot k x^\mu,
\\
  g'_{\mu\nu}&=\frac{g_{\mu\nu}}{k^2}, &\qquad
  \hat g{}^{\prime\mu\nu}&=k^2\hat g^{\mu\nu}.
\end{aligned}
\end{equation*}
Теперь из формул (\ref{etenmo}) следует, что плотность энергии и давление в
сопутствующей системе координат не меняются при масштабном преобразовании
(\ref{erasco})
\begin{equation*}
  \CE'=\CE,\qquad \CP'=\CP.
\end{equation*}
При этом компоненты вектора скорости преобразуются по-правилам
\begin{equation*}                                                    \tag*{\qed}
  u^{\prime0}=u^0,\qquad u^{\prime\mu}=ku^\mu+\dot ku^0x^\mu.
\end{equation*}
\renewcommand{\qed}{}\end{proof}
\begin{com}
Инвариантность плотности энергии и давления относительно масштабного
преобразования следовало ожидать, т.к.\ плотность энергии и давление являются
скалярными полями и инвариантны относительно любых преобразований координат
пространства-времени $x^\al$ и, в частности, растяжений (\ref{erasco}).
Нетривиальность проведенного рассмотрения заключается в том, что в правой части
равенства (\ref{ecomfr}), определяющего сопутствующую систему координат, стоят
определенные компоненты метрики, которые в общем случае не совпадают с
компонентами никакого вектора. Другими словами, сопутствующая система координат
(\ref{ecomfr}) определена по крайней мере с точностью до масштабных
преобразований (\ref{erasco}).
\qed\end{com}

Уравнения релятивистской гидродинамики (\ref{ecoema}), которые были выписаны в
начале раздела, можно записать в сопутствующей системе координат (\ref{ecomfr}).
Прямые вычисления приводят к следующей системе уравнений:
\begin{equation}                                                  \label{edepep}
\begin{split}
  \dot\CE-N^\mu\pl_\mu\CE-NK(\CE+\CP)&=0,
\\
  \pl_\mu\CP+\frac{\pl_\mu N}N(\CE+\CP)&=0,
\end{split}
\end{equation}
где точка обозначает дифференцирование по времени
$\dot{\cal E}=\pl{\cal E}/\pl t$ и
\begin{equation}                                                  \label{excusc}
  K=\frac1{2N}(2\hat\nb^\mu N_\mu-\hat g^{\mu\nu}\dot g_{\mu\nu})
\end{equation}
-- внешняя скалярная кривизна пространственной гиперповерхности.
\subsection{Временн\'{а}я калибровка}
Рассмотрим многообразие $\MM$, $\dim\MM=n$, на котором задана метрика лоренцевой
сигнатуры $g_{\al\bt}(x)$, $\sign g_{\al\bt}=(+-\dotsc-)$.
\begin{defn}
Система координат, в которой метрика имеет блочно диагональный вид
(\ref{etigam})
\begin{equation}                                                  \label{etemga}
  g_{\al\bt}=\begin{pmatrix} 1 & 0 \\ 0 & g_{\mu\nu} \end{pmatrix},
\end{equation}
где $g_{\mu\nu}$ -- отрицательно определенная риманова метрика на
пространственноподобных сечениях $x^0$$=\const$, называется {\em временн\'ой
калибровкой}. Эту систему координат называют также {\em синхронной, гауссовой
или полугеодезической}.
\qed\end{defn}
\index{Временн\'ая калибровка (temporal gauge)}%
\index{Калибровка временн\'ая (temporal gauge)}%
\index{Синхронная система координат (synchronous coordinate system)}%
\index{Система координат синхронная (synchronous coordinate system)}%
\index{Полугеодезическая система координат (semigeodesic coordinate system)}%
\index{Система координат полугеодезическая (semigeodesic coordinate system)}%
\index{Гауссова система координат (Gaussian coordinate system)}%
\index{Система координат гауссова (Gaussian coordinate system)}%
В синхронной системе отсчета координата $x^0$ является временем и явно выделена.
Напомним, что греческие буквы из начала алфавита пробегают все значения
индексов, $\al,\bt,\dotsc=0,1,\dotsc,n-1$, а из середины -- только
пространственные, $\mu,\nu,\dotsc=1,2,\dotsc,n-1$.

При переходе в синхронную систему отсчета $n$ произвольных функций,
параметризующих диффеоморфизмы, используются для фиксирования $n$ компонент
метрики:
\begin{equation*}
  g_{00}=1,\qquad  g_{0\mu}=0.
\end{equation*}

В АДМ параметризации метрики (см.\ раздел \ref{sadmpa}) временн\'ая калибровка
соответствует условиям $N=1$, $N_\mu=0$.
\begin{com}
Названия гауссова или полугеодезическая система координат распространены в
математической литературе, когда рассматриваются римановы пространства с
положительно определенной метрикой. В физической литературе, где преимущественно
рассматриваются многообразия с метрикой лоренцевой сигнатуры, чаще употребляют
термины временн\'ая калибровка или синхронная система координат, потому что в
этой системе отсчета координата $x^0$ действительно играет роль наблюдаемого
времени.
\qed\end{com}

Название синхронная система координат для метрики (\ref{etemga}) оправдана
следующим обстоятельством.
\subsubsection{Синхронизация часов}
В общем случае интервал между двумя близкими событиями $\lbrace x^\al\rbrace$ и
$\lbrace x^\al+dx^\al\rbrace$ имеет вид
\begin{equation*}
  ds^2=g_{\al\bt}dx^\al dx^\bt.
\end{equation*}
Предположим, что координата $x^0$ является (наблюдаемым) временем, т.е.\
$g_{00}>0$, и все сечения $x^0=\const$ пространственноподобны. Если два события
$C$ и $D$ произошли в данной системе координат в одной и той же точке
пространства, то они имеют координаты $C=\lbrace x_C^0,x^\mu\rbrace$ и
$D=\lbrace x_D^0,x^\mu\rbrace$. При этом данные события разделены интервалом
собственного времени
\begin{equation}                                                  \label{eprotd}
  \triangle s=\int_{x_C^0}^{x_D^0}\!\!\!dx^0\sqrt{g_{00}}.
\end{equation}
Этот интеграл равен длине времениподобной кривой
\begin{equation*}
  x^0=x^0_C+(x^0_D-x^0_C)t,\qquad x^\mu=\const,\qquad t\in[0,1],
\end{equation*}
соединяющей события $C$ и $D$. Конечно, в другой системе координат события могут
произойти не только в разное время, но и в разных точках пространства.

Таким образом, если два события, произошедшие в одной точке пространства в
данной системе координат, разделены наблюдаемым временем $x^0_D-x^0_C$, то они
разделены интервалом собственного времени (\ref{eprotd}). При этом нулевая
компонента метрики $g_{00}$ определяет различие собственного и наблюдаемого
времени для событий, произошедших в одной точке.

\begin{figure}[h,b,t]
\hfill\includegraphics[width=.3\textwidth]{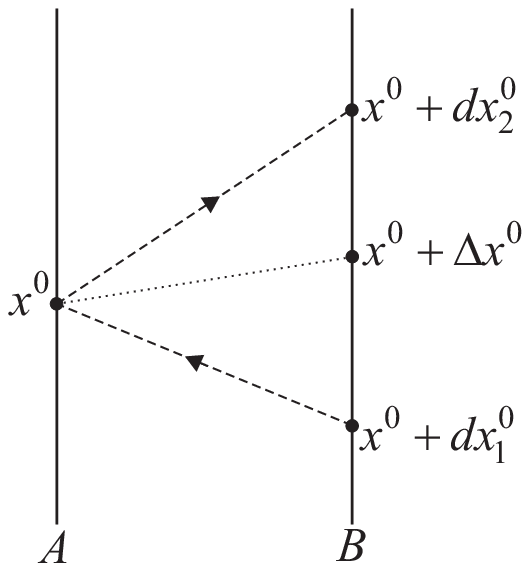}
\hfill {}
\centering\caption{Одновременность близких событий $A$ и $B$.}
\label{ftempogauge}
\end{figure}

Теперь определим понятие одновременности для событий, которые произошли в двух
разных, но близких точках пространства в данной фиксированной системе координат.
Пусть событие $A$ имеет пространственные координаты $x^\mu$, а событие $B$ --
близкие координаты $x^\mu+dx^\mu$. На рис.\ref{ftempogauge} показаны временн\'ые
оси, проходящие через точки $A$ и $B$. Возникает следующий вопрос
одновременности. Допустим, что событие $A$ имеет координаты
$\lbrace x^0,x^\mu\rbrace$. Какова временн\'ая координата  $x^0+\triangle x^0$
события, произошедшего в точке $B$, которое можно назвать одновременным с
событием $A$ ?

Чтобы определить одновременность, испустим свет в точке $B$ в некоторый момент
времени $x^0+dx^0_1$ (величина $dx_1^0$ отрицательна). Как только свет попадет в
точку $A$, сразу отразим его. Допустим, что свет вернулся в точку $B$ в момент
времени $x^0+dx^0_2$. Поскольку для света $ds^2=0$, то изменение наблюдаемого
времени в обоих случаях должно удовлетворять уравнению
\begin{equation*}
  g_{00}\big(dx^0_{1,2}\big)^2+2g_{0\mu}dx^0_{1,2}dx^\mu
  +g_{\mu\nu}dx^\mu dx^\nu=0.
\end{equation*}
Это квадратное уравнение имеет два решения:
\begin{align*}
  dx^0_1&=\frac1{g_{00}}\left[-g_{0\mu}dx^\mu
  -\sqrt{(g_{0\mu}g_{0\nu}-g_{\al\bt}g_{00})dx^\mu dx^\nu}\right],
\\
  dx^0_2&=\frac1{g_{00}}\left[-g_{0\mu}dx^\mu
  +\sqrt{(g_{0\mu}g_{0\nu}-g_{\al\bt}g_{00})dx^\mu dx^\nu}\right].
\end{align*}
Поскольку мы предположили, что $g_{00}>0$ и метрика $g_{\mu\nu}$ отрицательно
определена, то отсюда вытекает, что $dx^0_2>0$, а $dx^0_1<0$.

Назовем событие в точке $B$ {\em одновременным} событию
$A=\lbrace x^0,x^\mu\rbrace$, если его временная координата равна
$x^0+\triangle x^0$, где
\begin{equation*}
  \triangle x^0:=\frac{dx^0_1+dx^0_2}2=-\frac{g_{0\mu}dx^\mu}{g_{00}},
\end{equation*}
т.е.\ лежит посередине между $x^0+dx^0_2$ и $x^0+dx^0_1$.

Таким образом можно синхронизировать часы, расположенные в различных, но близко
расположенных точек в пространстве. Этот процесс можно продолжить вдоль
произвольной кривой в пространстве. Конечно, данная процедура синхронизации
часов зависит от выбора системы координат (нековариантна) и зависит также от
выбора кривой, соединяющей две точки в пространстве.

Рассмотрим замкнутую кривую $\g$ в пространстве с началом и концом в точке $A$.
Произведем синхронизацию часов вдоль кривой $\g$ описанным выше способом. Тогда
после возвращения в точку $A$ временн\'ая координата получит приращение
\begin{equation*}
  \triangle x^0:=-\oint_\g \frac{g_{0\mu}dx^\mu}{g_{00}}.
\end{equation*}
Отсюда следует, что синхронизация часов в общем случае невозможна, т.к.\
приращение $\triangle x^0$ в точке $A$ может быть отлично от нуля. Кроме того,
если мы хотим синхронизировать часы во всей области пространства-времени
$\MU\subset\MM$, которая покрывается данной системой координат, то равенство
$\triangle x^0=0$ должно также выполняться для любой замкнутой кривой $\g$,
целиком лежащей в $\MU$. Отсюда вытекает
\begin{prop}
Для того, чтобы в выбранной системе координат $x^\al$, где $x^0$ -- время и все
сечения $x^0=\const$ пространственноподобны, покрывающей некоторую область
$\MU\subset\MM$, можно было синхронизировать часы во всей области $\MU$
необходимо и достаточно чтобы $g_{0\mu}=0$. \qed
\end{prop}

Вернемся к рассмотрению синхронной системы координат и докажем теорему
существования.
\begin{prop}                                                      \label{ptemga}
Пусть задано пространство-время $\MM$ с метрикой лоренцевой сигнатуры. Тогда в
некоторой окрестности каждой точки $x\in\MM$ существует система координат, в
которой метрика имеет блочно диагональный вид (\ref{etemga}).
\end{prop}
\begin{proof}
Выберем в многообразии $\MM$ произвольную достаточно гладкую
пространственноподобную гиперповерхность, содержащую точку
$x\in\MN\hookrightarrow\MM$. Пусть $y^\mu$ -- некоторая система координат на
гиперповерхности в окрестности точки $x$. Построим на $\MN$ векторное поле $n$,
перпендикулярное к гиперповерхности. Через каждую точку $y\in\MN$ в направлении
$n$ проведем экстремаль в обоих направлениях. Мы уже знаем, что такая экстремаль
существует и единственна (см.\ раздел \ref{sgextr}). Поскольку гиперповерхность
пространственноподобна, то векторное поле и экстремали времениподобны. Выберем в
качестве канонического параметра вдоль каждой экстремали ее длину $t$ таким
образом, чтобы гиперповерхность $\MN$ задавалась уравнением $t=0$. Тогда в
некоторой окрестности $\MU$ гиперповерхности, $\MU\subset\MM$, будет определена
система координат $y=\lbrace y^0:=t,y^\mu\rbrace\in\MU$. Это и есть искомая
синхронная система координат.

Покажем это. По-построению, координатная кривая
$\lbrace y^0=t,y^\mu=\const\rbrace$, $t\in\MR$, является экстремалью. Ее вектор
скорости в построенной системе координат имеет одну отличную от нуля компоненту
$\dot y^\al=\dl_0^\al$. Поскольку экстремаль удовлетворяет уравнению
\begin{equation*}
  \ddot y^\al=-\Gamma_{\bt\g}{}^\al\dot y^\bt\dot y^\g,
\end{equation*}
то в построенной системе координат на метрику наложены условия
$\Gamma_{00}{}^\al=0$. Опустив индекс $\al$, получим уравнения на компоненты
метрики:
\begin{equation}                                                  \label{eqcote}
  \pl_0 g_{0\al}-\frac12\pl_\al g_{00}=0.
\end{equation}
Поскольку в качестве параметра вдоль экстремали выбрана ее длина, то касательный
вектор $\pl_0$ имеет единичную длину. Следовательно, в построенной системе
координат $g_{00}=1$. Тогда уравнения (\ref{eqcote}) примут вид
$\pl_0 g_{0\al}=0$, т.е.\ компоненты $g_{0\mu}$ не зависят от времени. Кроме
того, вектор скорости, по-построению, перпендикулярен гиперповерхности на $\MN$.
Это значит, что в начальный момент времени $t$ пространственно-временн\'ые
компоненты метрики равны нулю, $g_{0\mu}=0$. Поскольку они не зависят от
времени, то это равенство выполнено всюду в $\MU$. Тем самым построенная система
координат является синхронной.
\end{proof}

Ниже мы докажем обратное утверждение: если метрика имеет блочно диагональный
вид (\ref{etemga}), то координатные линии, соответствующие времени, являются
экстремалями. Это значит, что единственный произвол при построении синхронной
системы отсчета -- это выбор пространственного сечения $\MN$, которое может быть
произвольно, и выбор пространственных координат на $\MN$.
\begin{com}
Если в качестве исходной гиперповерхности $\MN$ для построения координат выбрать
времениподобную гиперповерхность, то метрика примет вид
\begin{equation*}
  g_{\al\bt}=\begin{pmatrix} -1 & 0 \\ \quad 0 & g_{\mu\nu} \end{pmatrix},
\end{equation*}
где $g_{\mu\nu}$ -- метрика лоренцевой сигнатуры на сечениях $x^0=\const$.
Вообще говоря, построенная система координат не зависит от сигнатуры метрики.
Аналогичные системы координат рассматривались еще Гауссом в римановой геометрии.
\qed\end{com}
\begin{com}
Построение синхронной системы координат начиналось с выбора гиперповерхности
$\MN\hookrightarrow\MM$, которая имеет размерность $n-1$. Аналогичные системы
координат можно строить, стартуя с подмногообразия $\MN$ произвольной
размерности, $0\le\dim\MN<n$. Детали построения таких систем координат, которые
названы полугеодезическими, можно найти в \cite{Petrov66R}. В частном случае,
когда все экстремали стартуют из одной точки, $\dim\MN=0$, получаем нормальные
координаты, рассмотренные в разделе \ref{srinog}.
\qed\end{com}
\begin{defn}
Сечения пространства-времени в синхронной системе координат $x^0=t=\const$
образуют семейство гиперповерхностей, которые называются {\em параллельными}
друг другу.
\qed\end{defn}
\index{Параллельные гиперповерхности (parallel hypersurfaces)}%
\index{Гиперповерхности параллельные (parallel hypersurfaces)}%
Для римановых многообразий понятие экстремали и геодезической совпадают.
Поэтому построенная система координат иногда называется {\em полугеодезической}.
Мы будем использовать принятый в физике термин временн\'ая калибровка для
метрики.

Обратная метрика во временн\'ой калибровке имеет вид
\begin{equation}                                                  \label{eintem}
  g^{\al\bt}=\begin{pmatrix} 1 & 0 \\ 0 & g^{\mu\nu} \end{pmatrix},
\end{equation}
где $g^{\mu\nu}$ -- метрика, обратная к $g_{\mu\nu}$. Ясно, что требование
блочной диагональности метрики (\ref{etemga}) эквивалентно требованию блочной
диагональности обратной метрики.

Чтобы найти явный вид преобразований координат к синхронной системе отсчета
$x^\al\mapsto y^\al(x)$, необходимо проинтегрировать уравнения экстремалей с
заданными начальными условиями на гиперповерхности $\MN\hookrightarrow\MM$. Это
-- обычная механическая задача, которую можно решить, интегрируя уравнение
Гамильтона--Якоби (см.\ раздел \ref{shaiae}). Обрисуем кратко этот подход.

При преобразовании к синхронной системе координат $x^\al\mapsto y^\al(x)$
временн\'ая компонента обратной метрики преобразуется по тензорному закону.
Следовательно, функция перехода $y^0(x)$ должна удовлетворять уравнению
\begin{equation}                                                  \label{extiga}
  1=g^{\al\bt}\frac{\pl y^0}{\pl x^\al}\frac{\pl y^0}{\pl x^\bt}.
\end{equation}
Это уравнение на функцию $y^0(x)$ совпадает с укороченным уравнением
Гамильтона--Якоби (\ref{ehajas}). Поэтому временную координату $y^0$ можно
отождествить с функцией действия точечной частицы. Общий интеграл укороченного
уравнения Гамильтона--Якоби зависит от $n$ постоянных. Поскольку функция
действия определена с точностью до добавления постоянной, то одна из постоянных
интегрирования аддитивна. Поэтому общий интеграл уравнения Гамильтона--Якоби
можно представить в виде
\begin{equation*}
  y^0=f(x,c)+A(c),
\end{equation*}
где $f$ -- некоторая функция от точки многообразия и $n-1$ постоянных
интегрирования $c=\lbrace c^\mu\rbrace=(c^1,\dotsc,c^{n-1})$, а $A(c)$ --
аддитивная постоянная, которую можно рассматривать как некоторую функцию
постоянных $c$. Уравнения $y^0=\const$ определяют гиперповерхность
$\MN\hookrightarrow\MM$. Локально решение этого уравнения можно представить в
виде $x^\al=x^\al(c)$. То есть постоянные $c$ образуют систему координат на
$\MN$. Временн\'ые координатные линии $y^0$ определяются уравнениями
$\pl y^0/\pl c^\mu=0$ или
\begin{equation*}
  \frac{\pl f}{\pl c^\mu}=-\frac{\pl A}{\pl c^\mu}.
\end{equation*}
Если зафиксировать точку на гиперповерхности $\MN$ (постоянные $c$), то в правой
части данной системы уравнений будут стоять некоторые константы. Соответствующее
решение системы $x^\mu=x^\mu(x^0)$ определит кривую, где роль параметра вдоль
кривой играет координата $x^0$, проходящую через данную точку $c\in\MN$. Эти
кривые являются траекториями частиц и, следовательно, экстремалями.

Из вида обратной метрики (\ref{eintem}) следует уравнение
\begin{equation*}
  \dl^{0\g}=g^{\al\bt}\frac{\pl y^0}{\pl x^\al}\frac{\pl y^\g}{\pl x^\bt},
  \qquad \g=0,\dotsc,n-1,
\end{equation*}
которое обобщает уравнение (\ref{extiga}). Умножив его на матрицу Якоби
преобразования координат, получим равенство
\begin{equation}                                                  \label{enocve}
  g^{\al\bt}\frac{\pl y^0}{\pl x^\al}=\frac{\pl x^\bt}{\pl y^0}.
\end{equation}
Вектор нормали к гиперповерхностям имеет вид
\begin{equation*}
  n=g^{\al\bt}\frac{\pl y^0}{\pl x^\bt}\pl_\al.
\end{equation*}
С учетом равенства (\ref{enocve}) отсюда следует, что вектор нормали касателен
к координатным линиям $y^0$:
\begin{equation*}
  \frac{\pl}{\pl y^0}=\frac{\pl x^\al}{\pl y^0}\pl_\al
  =g^{\al\bt}\frac{\pl y^0}{\pl x^\bt}\pl_\al.
\end{equation*}
Таким образом, координатные линии $y^0$ являются экстремалями, перпендикулярными
к гиперповерхностям $y^0=\const$.

Перейдем к вычислению явного вида основных геометрических объектов в синхронной
системе координат. Прямые вычисления приводят к следующим выражениям для
символов Кристоффеля:
(\ref{echris})
\begin{equation}                                                  \label{echrti}
\begin{split}
  \Gamma_{00}{}^0&=\Gamma_{00}{}^\mu=\Gamma_{0\mu}{}^0=\Gamma_{\mu0}{}^0=0,
\\
  \Gamma_{0\mu}{}^\nu&=\Gamma_{\mu0}{}^\nu=\frac12g^{\nu\rho}\pl_0 g_{\mu\rho},
\\
  \Gamma_{\mu\nu}{}^0&=-\frac12\pl_0 g_{\mu\nu},
\\
  \Gamma_{\mu\nu}{}^\rho&=\hat\Gamma_{\mu\nu}{}^\rho,
\end{split}
\end{equation}
где $\hat\Gamma_{\mu\nu}{}^\rho$ -- символы Кристоффеля на пространственноподобном
сечении $x^0=\const$, построенные только по метрике $g_{\mu\nu}$. В настоящем
разделе знак тильды, который мы используем для обозначения геометрических
объектов римановой геометрии при нулевых тензорах кручения и неметричности, для
простоты, опущен.

Несложные вычисления приводят к следующим выражениям для компонент тензора
кривизны со всеми опущенными индексами (\ref{ecurlo})
\begin{equation}                                                  \label{eritct}
\begin{split}
  R_{0\mu0\nu}&=\frac12\pl^2_{00} g_{\mu\nu}
  -\frac14g^{\rho\s}\pl_0 g_{\mu\rho}\pl_0 g_{\nu\s},
\\
  R_{0\mu\nu\rho}&=-R_{\mu0\nu\rho}=\frac12(\hat\nb_\nu\pl_0 g_{\mu\rho}
  -\hat\nb_\rho\pl_0 g_{\mu\nu}),
\\
  R_{\mu\nu\rho\s}&=\hat R_{\mu\nu\rho\s}
  +\frac14(\pl_0 g_{\mu\rho}\pl_0 g_{\nu\s}-\pl_0 g_{\nu\rho}\pl_0 g_{\mu\s}),
\end{split}
\end{equation}
где $\hat\nb_\nu$ обозначает ковариантную производную на пространственноподобном
сечении
\begin{equation*}
  \hat\nb_\nu\pl_0 g_{\mu\rho}:=\pl_\nu\pl_0 g_{\mu\rho}
  -\hat\Gamma_{\nu\mu}{}^\s\pl_0 g_{\s\rho}-\hat\Gamma_{\nu\rho}{}^\s\pl_0 g_{\mu\s},
\end{equation*}
и $\hat R_{\mu\nu\rho\s}$ -- тензор кривизны пространственноподобного сечения
$t=\const$, построенный только по метрике $g_{\mu\nu}$. Свертка с обратной
метрикой дает соответствующие тензор Риччи и скалярную кривизну:
\begin{equation}                                                  \label{eriscc}
\begin{split}
  R_{00}&=\frac12g^{\mu\nu}\pl^2_{00} g_{\mu\nu}
  -\frac14g^{\mu\nu}g^{\rho\s}\pl_0 g_{\mu\rho}\pl_0 g_{\nu\s},
\\
  R_{0\mu}&=R_{\mu0}=\frac12g^{\nu\rho}(\hat\nb_\mu\pl_0 g_{\nu\rho}
  -\hat\nb_\rho\pl_0 g_{\nu\mu}),
\\
  R_{\mu\nu}&=\hat R_{\mu\nu}+\frac12\pl^2_{00} g_{\mu\nu}
  -\frac12g^{\rho\s}\pl_0 g_{\mu\rho}\pl_0 g_{\nu\s}
  +\frac14\pl_0 g_{\mu\nu}g^{\rho\s}\pl_0 g_{\rho\s},
\\
  R&=\hat R+g^{\mu\nu}\pl^2_{00} g_{\mu\nu}
  -\frac34g^{\mu\nu}g^{\rho\s}\pl_0 g_{\mu\rho}\pl_0 g_{\nu\s}
  +\frac14(g^{\mu\nu}\pl_0 g_{\mu\nu})^2.
\end{split}
\end{equation}

Уравнения для экстремалей $x^\al(t)$ во временн\'ой калибровке имеют вид
\begin{align}                                                     \label{extsef}
  \ddot x^0&=\quad \frac12\pl_0 g_{\mu\nu}\dot x^\mu\dot x^\nu,
\\                                                                \label{exrsap}
  \ddot x^\mu&=-g^{\mu\nu}\pl_0 g_{\nu\rho}\dot x^0\dot x^\rho
  -\hat\Gamma_{\nu\rho}{}^\mu\dot x^\nu\dot x^\rho,
\end{align}
где точка обозначает дифференцирование по каноническому параметру $t$.
Из вида уравнений сразу следует
\begin{prop}
Если выбрана синхронная система координат, то временн\'ые координатные линии
$\lbrace x^0=t,x^\mu=\const\rbrace$ являются экстремалями.
\end{prop}
Это утверждение уже было доказано другим способом при рассмотрении укороченного
уравнения Гамильтона--Якоби (\ref{extiga}).

Из уравнений для экстремалей можно исключить временн\'ую компоненту скорости.
Для этого воспользуемся законом сохранения (\ref{efirin})
\begin{equation}                                                  \label{econti}
  (\dot x^0)^2-\dot\Bx^2=C_0=\const,
\end{equation}
где $\dot\Bx^2:=-g_{\mu\nu}\dot x^\mu\dot x^\nu$, и исключим производную
$\dot x^0$ из уравнения (\ref{exrsap}). В результате получим замкнутую систему
уравнений только для пространственных координат экстремали:
\begin{equation*}
  \ddot x^\mu=-g^{\mu\nu}\pl_0 g_{\nu\rho}\dot x^\rho\sqrt{|C_0+\dot\Bx^2|}
  -\hat\Gamma_{\nu\rho}{}^\mu\dot x^\nu\dot x^\rho.
\end{equation*}
Неоднозначность при извлечении корня несущественна, так как соответствует замене
$x^0\mapsto-x^0$. Отсюда следует, что во временн\'ой калибровке пространственные
компоненты экстремали $\lbrace x^\mu(t)\rbrace$ в общем случае не являются
экстремалями для пространственной части метрики $g_{\mu\nu}$. В частном случае,
когда пространственная метрика $g_{\mu\nu}$ не зависит от времени $x^0$,
проекция экстремали
$\lbrace x^\al(t)\rbrace\mapsto\lbrace c^0,x^\mu(t)\rbrace$ на
пространственное сечение $x^0=c^0=\const$ является экстремалью для метрики
$g_{\mu\nu}$ на этом сечении.

Рассмотрим еще один способ записи уравнений для экстремалей.
Пусть $\dot x^0\ne0$. Тогда уравнения для экстремалей во временн\'ой калибровке
(\ref{extsef}), (\ref{exrsap}) можно переписать в эквивалентном виде, заменив
канонический параметр $t$ на наблюдаемое время $x^0$. Другими словами,
рассмотрим траектории $x^\mu(x^0)$. Поскольку
\begin{equation*}
\begin{split}
  \pl_0 x^\mu&=\frac{\dot x^\mu}{\dot x^0},
\\
  \pl^2_{00}x^\mu&=\frac{\ddot x^\mu\dot x^0-\dot x^\mu\ddot x^0}{(\dot x^0)^3},
\end{split}
\end{equation*}
то, воспользовавшись уравнениями для экстремалей, получим равенство
\begin{equation}                                                  \label{extioz}
  \pl^2_{00}x^\mu=
  -\hat\Gamma_{\nu\rho}{}^\mu\pl_0x^\nu\pl_0x^\rho
  -g^{\mu\nu}\pl_0g_{\nu\rho}\pl_0x^\rho
  -\frac12\pl_0x^\mu\pl_0g_{\nu\rho}\pl_0x^\nu\pl_0x^\rho.
\end{equation}
Зависимость канонического параметра $t$ от наблюдаемого времени $x^0$ можно
определить из закона сохранения (\ref{econti}), который можно переписать в виде
\begin{equation*}
  (\dot x^0)^2(1+g_{\mu\nu}\pl_0x^\mu\pl_0x^\nu)=C_0.
\end{equation*}
Отсюда для ненулевых экстремалей, $C_0\ne0$, получаем выражение для производной
\begin{equation*}
  \dot x^0=\pm\sqrt{\frac{C_0}{1+g_{\mu\nu}\pl_0x^\mu\pl_0x^\nu}}.
\end{equation*}
Это уравнение позволяет определить зависимость $x^0(t)$.

Для нулевых экстремалей $C_0=0$ и $1+g_{\mu\nu}\pl_0x^\mu\pl_0x^\nu=0$.
Последнее условие согласуется с уравнениями (\ref{extioz}). Действительно,
прямые вычисления приводят к равенству
\begin{equation*}
  \pl_0(1+g_{\mu\nu}\pl_0x^\mu\pl_0x^\nu)=
  -(1+g_{\mu\nu}\pl_0x^\mu\pl_0x^\nu)\pl_0g_{\rho\s}\pl_0x^\rho\pl_0x^\s.
\end{equation*}
То есть, если условие $1+g_{\mu\nu}\pl_0x^\mu\pl_0x^\nu=0$ выполнено в начальный
момент времени, то оно будет также выполнено во все последующие моменты. Это
значит, что для нулевых экстремалей зависимость $x^0(t)$ не определяется
уравнениями (\ref{extioz}) и может быть произвольна. В этом случае канонический
параметр можно просто отождествить с временем: $x^0=t$.

Теперь покажем, что синхронная система координат в общей теории относительности
при наличии материи не может быть статична в том смысле, что ее пространственные
компоненты $g_{\mu\nu}$ обязательно зависят от $x^0$. Для определенности
рассмотрим четырехмерное пространство-время. С этой целью введем обозначение для
производных по времени от компонент метрики
\begin{equation*}
  p_{\mu\nu}:=\pl_0 g_{\mu\nu}.
\end{equation*}
Это нечто, напоминающее импульсы, сопряженные к компонентам метрики. След новых
переменных равен производной от определителя метрики
\begin{equation*}
  p:=g^{\mu\nu}p_{\mu\nu}=\pl_0[\ln(-g)],\qquad g:=\det g_{\al\bt},
\end{equation*}
т.к.\ во временн\'ой калибровке $\det g_{\al\bt}=\det g_{\mu\nu}$. В новых
переменных компоненты тензора Риччи (\ref{eriscc}) примут вид:
\begin{equation}                                                  \label{erimov}
\begin{split}
  R_{00}&=\frac12\pl_0p+\frac14p_{\mu\nu}p^{\mu\nu},
\\
  R_{0\mu}&=\frac12\hat\nb_\mu p-\frac12\hat\nb_\nu p_\mu{}^\nu,
\\
  R_{\mu\nu}&=\hat R_{\mu\nu}+\frac12\pl_0 p_{\mu\nu}
  -\frac12p_\mu{}^\rho p_{\nu\rho}+\frac14p_{\mu\nu}p.
\end{split}
\end{equation}
Запишем уравнения Эйнштейна (\ref{eineqe}) без космологической постоянной
\begin{align}                                                     \label{eielzl}
  R_{00}&=-\frac1{2\kappa}\left(T_{\Sm00}-\frac12T_\Sm\right),
\\                                                                \label{eielsz}
  R_{0\mu}&=-\frac1{2\kappa}T_{\Sm0\mu},
\\                                                                \label{eielsl}
  R_{\mu\nu}&
  =-\frac1{2\kappa}\left(T_{\Sm\mu\nu}-\frac12g_{\mu\nu}T_{\Sm}\right).
\end{align}
Эти уравнения для непрерывной среды с тензором энергии-импульса (\ref{enmotl})
принимают вид
\begin{align}                                                     \label{eielzz}
  R_{00}&=-\frac1{2\kappa}\left((\CE+\CP)(u_0)^2-\frac12\CE+\frac12\CP\right),
\\                                                                \label{eielzs}
  R_{0\mu}&=-\frac1{2\kappa}(\CE+\CP)u_0 u_\mu,
\\                                                                \label{eielss}
  R_{\mu\nu}&
  =-\frac1{2\kappa}\left((\CE+\CP)u_\mu u_\nu-\frac12g_{\mu\nu}\CE
  +\frac12g_{\mu\nu}\CP\right).
\end{align}
\begin{prop}
Пусть пространство-время заполнено непрерывной средой с положительной
плотностью энергии и давлением, $\CE>0$ и $\CP>0$. Тогда уравнения Эйнштейна
(\ref{eielzz})--(\ref{eielss}) без космологической постоянной в синхронной
системе координат не имеют статических решений.
\end{prop}
\begin{proof}
В синхронной системе координат $u^2=(u_0)^2+g^{\mu\nu}u_\mu u_\nu=1$, и поэтому
$u_0\ge1$. Статичность метрики означает, что $p_{\mu\nu}=0$. В этом случае
$R_{0\mu}=0$, и из уравнения (\ref{eielzs}) следует равенство $u_\mu=0$.
Следовательно, $u_0=1$. Тогда уравнение (\ref{eielzz}) примет вид
\begin{equation*}
  0=\frac12\CE+\frac32\CP,
\end{equation*}
что противоречит сделанным предположениям.
\end{proof}
\begin{prop}
Вакуумные уравнения Эйнштейна без космологической постоянной в синхронной
системе координат имеют только плоские статические решения, для которых тензор
кривизны равен нулю.
\end{prop}
\begin{proof}
При нулевом тензоре энергии-импульса материи, $T_{\Sm\al\bt}=0$, уравнения
(\ref{eielzl}) и (\ref{eielsz}) для $p_{\mu\nu}=0$ удовлетворяются
автоматически. Уравнение (\ref{eielss}) сводится к уравнению
\begin{equation*}
  \hat R_{\mu\nu}=0.
\end{equation*}
Если пространство трехмерно, то полный тензор кривизны взаимно однозначно
определяется тензором Риччи и, следовательно, полный тензор кривизны
пространства равен нулю, $\hat R_{\mu\nu\rho\s}=0$. Теперь из выражения для
компонент тензора кривизны (\ref{eritct}) следует, что полный тензор кривизны
пространства-времени также обращается в нуль, $R_{\al\bt\g\dl}=0$. В свою
очередь, это значит, что локально существует такая система координат, в которой
метрика четырехмерного пространства-времени является лоренцевой,
$g_{\al\bt}=\eta_{\al\bt}$. Это и означает тривиальность решений уравнений
Эйнштейна. Глобально соответствующее пространство-время представляет либо
пространство Минковского, либо все возможные торы или цилиндры, получающиеся из
пространства Минковского путем отображения всех или части декартовых координат
на окружность.
\end{proof}
Теперь предположим, что выполнено сильное энергетическое условие (\ref{estenc}).
Тогда из уравнения (\ref{eielzz}) следует неравенство
\begin{equation*}
  \frac12\pl_0p+\frac14p_\mu{}^\nu p_\nu{}^\mu\le0.
\end{equation*}
Справедливо алгебраическое неравенство
\begin{equation*}
  p_\mu{}^\nu p_\nu{}^\mu\ge\frac13p^2,
\end{equation*}
которое нетрудно доказать путем диагонализации матрицы $p_\mu{}^\nu$. Поэтому
должно быть выполнено неравенство
\begin{equation*}
  \pl_0p+\frac16p^2\le0.
\end{equation*}
Перепишем его в виде
\begin{equation}                                                  \label{enmede}
  \pl_0\frac1p\ge\frac16.
\end{equation}

В заключение раздела рассмотрим еще одно свойство решений уравнений Эйнштейна в
синхронной системе отсчета. Предположим, что в некоторый момент времени $x^0:=t$
величина $1/p$ положительна. Тогда при уменьшении времени $t$ за конечное время
функция $1/p$ обратится в нуль, поскольку производная больше нуля. Это означает,
что модуль определитель метрики $|g|$ обращается в нуль. Допустим, что это
происходит в момент времени $t_0$. Вблизи этой точки положим
\begin{equation*}
  -g=C(t-t_0)^k,\qquad C=\const,
\end{equation*}
с некоторым показателем степени $k\in\MR_+$. Тогда неравенство (\ref{enmede})
ограничивает показатель степени $k\le6$. Отсюда следует, что модуль определителя
метрики обращается в нуль не быстрее, чем $(t-t_0)^6$.

Если в начальный момент времени $t=0$ величина $1/p$ отрицательна, то то же
самое происходит при положительных временах.

Обращение в нуль определителя метрики отнюдь не означает, что возникает
сингулярность в пространстве-времени. Например, определитель евклидовой метрики
в сферической системе координат равен нулю в начале координат. Как правило,
обращение в нуль определителя метрики связано с выбором системы отсчета. Выше
было показано, что синхронная система координат возникает при построении
семейства экстремалей, которые ортогональны некоторой пространственноподобной
гиперповерхности. В общем случае на конечном расстоянии эти экстремали начинают
пересекаться, образуя каустические поверхности. В точках пересечения экстремалей
система координат вырождается, что может приводить к обращению в нуль
определителя метрики.
\subsection{Калибровка светового конуса                          \label{slicga}}
Рассмотрим многообразие $\MM$, $\dim\MM=n$, на котором задана метрика
$g_{\al\bt}$ лоренцевой сигнатуры.
\begin{defn}
Система координат, в которой метрика имеет вид
\begin{equation}                                                  \label{elicga}
  g_{\al\bt}=\begin{pmatrix}
    0 & 1 & 0 & \dotsc & 0 \\ 1 &&&& \\ 0 &&&& \\
    \vdots &&& g_{\mu\nu} & \\ 0 &&&&  \end{pmatrix},\qquad
  \begin{aligned}
    \al,\bt&=0,1,\dotsc,n-1, \\ \mu,\nu&=1,2,\dotsc,n-1,
\end{aligned}
\end{equation}
называется калибровкой {\em светового конуса}.
\qed\end{defn}
\index{Калибровка светового конуса (light cone gauge)}%
\index{Светового конуса калибровка (light cone gauge)}%
Ниже мы докажем, что локально калибровка светового конуса существует. При этом
$n$ произвольных функций, параметризующих преобразования координат, используются
для фиксирования $n$ компонент метрики:
\begin{equation*}
  g_{00}=g_{02}=g_{03}=\dotsc=g_{0\,n-1}=0,\qquad g_{01}=1.
\end{equation*}

Для построения, системы координат, соответствующей калибровке светового конуса,
введем несколько определений. Допустим, что гиперповерхность
$\MS\hookrightarrow\MM$ задана параметрически $x^\al=x^\al(y^\mu)$, где $y^\mu$
-- координаты на гиперповерхности. Вложение $\MS\hookrightarrow\MM$ индуцирует
на гиперповерхности метрику
\begin{equation}                                                  \label{einlom}
  ds^2=h_{\mu\nu}dy^\mu dy^\nu=
  g_{\al\bt}\frac{\pl x^\al}{\pl y^\mu}\frac{\pl x^\bt}{\pl y^\nu}dy^\mu dy^\nu.
\end{equation}
В этом выражении индуцированная метрика $h_{\mu\nu}$ в общем случае не совпадает
с блоком $g_{\mu\nu}$, входящим в выражение (\ref{elicga}).
\begin{defn}
Если индуцированная метрика $h_{\mu\nu}$ на гиперповерхности невырождена, то
гиперповерхность называется {\em неизотропной}. В противном случае, если
$\det h_{\mu\nu}=0$, гиперповерхность называется {\em изотропной}.
\qed\end{defn}
\index{Изотропная гиперповерхность (isotropic hypersurface)}%
\index{Гиперповерхность изотропная (isotropic hypersurface)}%
\index{Неизотропная гиперповерхность (non isotropic hypersurface)}%
\index{Гиперповерхность неизотропная (non isotropic hypersurface)}%
В дальнейшем мы увидим, что индуцированная метрика может быть вырождена только
тогда, когда метрика на исходном многообразии $\MM$ не является положительно или
отрицательно определенной. Например, она может иметь лоренцеву сигнатуру.

Набор $n-1$ векторных полей на $\MS$, которые нумеруются индексом $\mu$,
\begin{equation*}
  e_\mu:=\pl_\mu x^\al\pl_\al:=\frac{\pl x^\al}{\pl y^\mu}\pl_\al
\end{equation*}
образуют базис касательного пространства к гиперповерхности
$\MS\hookrightarrow\MM$. Вектор $m=m^\al\pl_\al$ ортогонален к гиперповерхности,
если его скалярное произведение со всеми касательными векторами равно нулю:
\begin{equation}                                                  \label{edenof}
  (m,e_\mu):=g_{\al\bt}m^\al\pl_\mu x^\bt=m_\al\pl_\mu x^\al=0,
  \qquad \forall\mu=1,2,\dotsc,n-1,
\end{equation}
где скобки означают скалярное произведение.
Поскольку ранги матриц $g_{\al\bt}$ и $\pl_\mu x^\al$ равны, соответственно, $n$
и $n-1$, то ранг матрицы $g_{\al\bt}\pl_\mu x^\bt$ равен $n-1$. Это означает,
что система линейных однородных уравнений на компоненты нормального вектора
$m^\al$ имеет единственное нетривиальное решение с точностью до умножения на
произвольную отличную от нуля постоянную.

Важную роль в дальнейшем будет играть следующее
\begin{prop}                                                   \label{tisohy}
Нормаль к гиперповерхности изотропна тогда и только тогда, когда поверхность
изотропна.
\end{prop}
\begin{proof}
Индуцированную метрику $h_{\mu\nu}$ (\ref{einlom}) можно рассматривать, как
произведение двух прямоугольных матриц $\pl_\mu x^\al$ и
$g_{\al\bt}\pl_\mu x^\bt$ размеров, соответственно, $(n-1)\times n$ и
$n\times(n-1)$. Тогда из формулы Бине--Коши (\ref{ebicho}) следует, что
\begin{equation*}
  \det h_{\mu\nu}=\sum_{\g=1}^n\det_\g(\pl_\mu x^\al)
  \det_\g(g_{\al\bt}\pl_\mu x^\bt),
\end{equation*}
где $\det_\g$ обозначает определитель квадратной матрицы, полученной из матрицы
размера $(n-1)\times n$ вычеркиванием $\g$-того столбца. Поскольку
$m_\al\pl_\mu x^\al=0$, то между столбцами матрицы $\pl_\mu x^\al=0$ имеется
линейная зависимость. Предположим, что некоторая компонента ковектора
$\lbrace m_\al\rbrace$ отлична от нуля, например, $m_{\dl_1}\ne0$. Тогда, в силу
свойств определителя и линейной зависимости столбцов, справедливо равенство
\begin{equation*}
  \det_\g(\pl_\mu x^\al)
  =m_\g\frac{(-1)^{\dl_1-\g}\det_{\dl_1}(\pl_\mu x^\al)}{m_{\dl_1}}.
\end{equation*}
Аналогичная формула имеет место и для второго определителя под знаком суммы
\begin{equation*}
  \det_\g(g_{\al\bt}\pl_\mu x^\bt)
  =m^\g\frac{(-1)^{\dl_2-\g}\det_{\dl_2}(g_{\al\bt}\pl_\mu x^\bt)}{m^{\dl_2}}
\end{equation*}
для некоторого индекса $\dl_2$. Таким образом, определитель индуцированной
метрики пропорционален длине нормального вектора к гиперповерхности
\begin{equation*}
  \det h_{\mu\nu}=(-1)^{\dl_1+\dl_2}\frac{\det_{\dl_1}(\pl_\mu x^\al)
  \det_{\dl_2}(g_{\al\bt}\pl_\mu x^\bt)}{m_{\dl_1}m^{\dl_2}}
  m_\g m^\g,
\end{equation*}
где в правой части предполагается суммирование только по индексу $\g$. Индексы
$\dl_1$ и $\dl_2$ фиксированы и выбраны таким образом, что $m_{\dl_1}\ne0$ и
$m^{\dl_2}\ne0$. Поскольку коэффициент пропорциональности между
$\det h_{\mu\nu}$ и $m_\g m^\g$ отличен от нуля, то из этого равенства следует
утверждение предложения.
\end{proof}
Рассмотрим некоторые свойства изотропных гиперповерхностей.

Допустим, что изотропная гиперповерхность $\MS$ задана уравнением
\begin{equation}                                                  \label{eisueq}
  W(x)=\const,
\end{equation}
где $W(x)$ -- некоторая достаточно гладкая функция координат.
При этом мы считаем, что различным значениям постоянной соответствуют различные
непересекающиеся изотропные гиперповерхности. Обозначим, как и ранее, координаты
на $\MS$ через $y^\mu$. То есть функции $x^\al(y)$ удовлетворяют уравнению
(\ref{eisueq}) для всех $y$. Продифференцируем это уравнение:
\begin{equation*}
  \frac{\pl W}{\pl x^\al}\frac{\pl x^\al}{\pl y^\mu}
  =\frac{\pl W}{\pl x^\al}e_\mu{}^\al=0.
\end{equation*}
Сравнение полученного равенства с определением нормали (\ref{edenof})
показывает, что вектор нормали к гиперповерхности имеет вид
\begin{equation}                                                  \label{enocov}
  m=m^\al\pl_\al=g^{\al\bt}\frac{\pl W}{\pl x^\bt}\pl_\al.
\end{equation}
Поскольку вектор нормали к изотропной гиперповерхности имеет нулевую длину, то
выполнено равенство
\begin{equation}                                                  \label{edeiss}
  m^2=g^{\al\bt}\frac{\pl W}{\pl x^\al}\frac{\pl W}{\pl x^\bt}=0.
\end{equation}
Таким образом доказано
\begin{prop}
Поверхность $\MS$, определяемая уравнением (\ref{eisueq}), изотропна тогда и
только тогда, когда выполнено равенство (\ref{edeiss}).
\end{prop}
Следующее утверждение не имеет аналога в римановой геометрии.
\begin{prop}
Нормальный вектор к изотропной поверхности $\MS$ является также и касательным к
ней.
\end{prop}
\begin{proof}
Дополним координаты на гиперповерхности до системы координат в $\MM$ еще одной
координатой $y^0:=W$. В новой системе координат гиперповерхности определяются
условиями $y^0=\const$, и уравнение (\ref{edeiss}) примет вид $g^{00}=0$. Тогда
нулевая компонента нормального вектора (\ref{enocov}) обратится в нуль, $m^0=0$,
т.к.\ $\pl W/\pl y^\mu=0$. Это значит, что нормальный вектор является также и
касательным.
\end{proof}
На самом деле сделанное утверждение очевидно. Действительно, вектор $k$ будет
касательным к гиперповерхности $\MS$ тогда и только тогда, когда его скалярное
произведение с вектором нормали равно нулю, $(k,m)=0$. Это и означает, что
вектор нормали $m$ является касательным к $\MS$, т.к.\ $m^2=0$.

Рассмотрим интегральные кривые $x(u)$ векторного поля $m$:
\begin{equation}                                                  \label{edemcu}
  \dot x^\al=m^\al,\qquad m^\al:=g^{\al\bt}\frac{\pl W}{\pl x^\bt},
\end{equation}
где точка обозначает дифференцирование по параметру $u\in\MR$. Эти кривые
изотропны, т.к.\ выполнено равенство (\ref{edeiss}). Кроме того, они
перпендикулярны изотропной поверхности $\MS$. С другой стороны, вдоль каждой
кривой выполнено равенство
\begin{equation*}
  \frac{dW}{du}=\frac{\pl W}{\pl x^\al}\frac{d x^\al}{du}
  =g^{\al\bt}\frac{\pl W}{\pl x^\al}\frac{\pl W}{\pl x^\bt}=0.
\end{equation*}
Отсюда следует, что, если интегральная кривая $x(u)$ начинается на изотропной
поверхности $\MS$, то она целиком лежит на этой поверхности.

Можно сказать по-другому. Поскольку векторное поле $m$ касательно к $\MS$, то
его интегральные кривые лежат на $\MS$.

Вычислим теперь ускорение данной кривой:
\begin{equation*}
  \dot x^\al\nb_\al \dot x^\bt
  =\dot x^\al g^{\bt\g}\nb_\al\frac{\pl W}{\pl x^\g}
  =g^{\al\dl}\frac{\pl W}{\pl x^\dl}g^{\bt\g}\nb_\al\frac{\pl W}{\pl x^\g}.
\end{equation*}
С другой стороны, продифференцируем определяющее уравнение (\ref{edeiss}):
\begin{equation*}
  \nb_\al\left(g^{\bt\g}\frac{\pl W}{\pl x^\bt}\frac{\pl W}{\pl x^\g}\right)
  =2g^{\bt\g}\frac{\pl W}{\pl x^\bt}\nb_\al\frac{\pl W}{\pl x^\g}=0.
\end{equation*}
Поскольку $\nb_\al\nb_\g W=\nb_\g\nb_\al W$, то отсюда вытекает, что ускорение
кривой (\ref{edemcu}) равно нулю, $\dot x^\al\nb_\al \dot x^\bt=0$. Таким
образом, интегральные кривые векторного поля $m$ являются экстремалями. Отсюда
следует
\begin{prop}
Все изотропные (нулевые) экстремали, которые касаются изотропной поверхности
$\MS$ хотя бы в одной точке, целиком лежат на этой поверхности.
\end{prop}
Теперь докажем существование калибровки светового конуса.
\begin{prop}                                                      \label{pxlico}
Пусть задано пространство-время $\MM$ с метрикой лоренцевой сигнатуры. Тогда в
некоторой окрестности каждой точки $x\in\MM$ существует система координат, в
которой метрика имеет вид (\ref{elicga}).
\end{prop}
\begin{proof}
Аналогично доказательству предложения \ref{ptemga}. Выберем произвольную
пространственноподобную поверхность $\MN\hookrightarrow\MM$ с координатами
$y^\mu$. В каждой точке $y\in\MN$ строим нормальный вектор $n$. Он
времениподобен. Выберем координаты $y^\mu$ на $\MN$ таким образом, чтобы вектор
\begin{equation*}
  e_1:=\frac{\pl x^\al}{\pl y^1}\pl_\al
\end{equation*}
имел единичную длину, $e_1^2=-1$, и был перпендикулярен всем остальным
касательным векторам:
\begin{equation*}
  (e_1,e_\Sa)=0,\qquad \Sa=2,\dotsc,n-1.
\end{equation*}
Это всегда возможно в силу предложения \ref{ptemga}. То есть мы фиксируем
временн\'ую калибровку на поверхности $\MN$. Затем строим вектор
\begin{equation*}
  m=n-e_1.
\end{equation*}
По построению, этот вектор определен на поверхности $\MN$, имеет нулевую длину,
$m^2=0$, перпендикулярен всем касательным векторам $e_\Sa$, и выполнено
равенство $(m,e_1)=1$. Другими словами, в каждом световом конусе в точке
$y\in\MN$ мы выбираем единственный вектор $m$, скалярное произведение которого с
касательным вектором $e_1$ равно единице, а с остальными касательными
векторами равно нулю.

Затем через каждую точку $y\in\MN$ в направлении $m$ проводим экстремаль.
Выбираем канонический параметр $y^0$ вдоль экстремалей таким образом, чтобы
гиперповерхность $\MN$ определялась уравнением $y^0=0$. Тогда в некоторой
окрестности гиперповерхности $\MN$ определена система координат
$\lbrace y^\al\rbrace=\lbrace y^0,y^\mu\rbrace$.

По-построению, координатные линии $y^0$ являются экстремалями. Поэтому на
компоненты метрики в новой системе координат имеются ограничения (\ref{eqcote}).
В рассматриваемом случае касательный вектор к координатной линии $y^0$
изотропен, и, следовательно, $g_{00}=0$. Поэтому на компоненты метрики возникает
уравнение (\ref{eqcote}), которое сводится к условию
\begin{equation}                                                  \label{elicos}
  \pl_0 g_{0\mu}=0.
\end{equation}
По-построению, координаты $y^\mu$ на $\MN$ и изотропный вектор $m$ выбраны таким
образом, что на поверхности $\MN$ выполнены условия:
\begin{equation*}
  g_{00}=0,\qquad g_{01}=1,\qquad g_{02}=g_{03}=\dotsc=g_{0\,n-1}=0.
\end{equation*}
Тогда из уравнений (\ref{elicos}) следует, что эти условия будут выполнены во
все последующие моменты времени. Следовательно, система координат, в которой
выполнена калибровка светового конуса построена.

Аналогичное построение можно провести, если в качестве исходной поверхности
$\MN$ выбрать поверхность с метрикой лоренцевой сигнатуры.
\end{proof}

Посмотрим теперь на построение калибровки светового конуса с точки зрения
уравнения Гамильтона--Якоби. Предположим, что изотропная гиперповерхность $\MS$
задана уравнением $W(x)=0$. Тогда условие изотропности поверхности примет
вид (\ref{edeiss}), которое должно быть выполнено на изотропной гиперповерхности
$\MS$. Изотропная гиперповерхность, задаваемая уравнением $W=0$, является
ни чем иным, как характеристикой волнового уравнения (\ref{ewaeqs}). Уравнение
(\ref{edeiss}) представляет собой укороченное уравнение Гамильтона--Якоби для
нулевых экстремалей, где $W(x)$ -- укороченная функция действия.

Рассмотрим систему линейных дифференциальных уравнений первого порядка
\begin{equation*}
  g^{\al\bt}\pl_\al W\pl_\bt\vf=0
\end{equation*}
на функцию $\vf(x)$. Известно, что эта система имеет $n-1$ функционально
независимых решений, которые обозначим через $\vf^\mu(x)$, $\mu=1,\dotsc,n-1$.
Среди этих решений содержится, очевидно, и $W(x)$. Не ограничивая общности,
положим $\vf^1(x):=W(x)$. Введем новую систему координат
\begin{equation*}
  y^0:=\psi(x),\quad y^\mu:=\vf^\mu(x),
\end{equation*}
где $\psi(x)$ -- некоторая функция, функционально независимая от $\vf^\mu(x)$,
т.е.
\begin{equation*}
  \frac{\pl y^\al}{\pl x^\bt}\ne0
\end{equation*}
где $\lbrace y^\al\rbrace=\lbrace y^0,y^\mu\rbrace$. В новой системе координат
компоненты обратной метрики примут вид
\begin{equation*}
  g^{\prime\al\bt}
  =g^{\g\dl}\frac{\pl y^\al}{\pl x^\g}\frac{\pl y^\bt}{\pl x^\dl}.
\end{equation*}
По-построению,
\begin{equation*}
  g^{\prime01}=g^{\prime10}=g^{\al\bt}\pl_\al W\pl_\bt\psi\ne0,\qquad
  g^{\prime1\mu}=g^{\prime \mu 1}=0,
\end{equation*}
причем $g^{\prime01}\ne0$, так как $\psi$ функционально независима от $\vf^\mu$.
Эту функцию всегда можно подобрать таким образом, что $g^{\prime01}=1$. Таким
образом, в новой системе координат обратная метрика примет вид
\begin{equation*}
  g^{\prime\al\bt}=\begin{pmatrix} g^{\prime00} & 1 & g^{\prime0\Sb} \\
  1 & 0 & 0 \\ g^{\prime\Sa0} & 0 & g^{\prime\Sa\Sb}  \end{pmatrix},\qquad
  \Sa,\Sb=2,3,\dotsc,n-1.
\end{equation*}
Нетрудно проверить, что метрика $g'_{\al\bt}$, которая обратна к матрице
$g^{\prime\al\bt}$, в новой системе координат имеет вид (\ref{elicga}). Тем
самым локальное существование калибровки светового конуса доказано еще раз. В
этой калибровке изотропные гиперповерхности $\MS$ задаются условиями
$y^1=\const$. Векторы $\pl_0$, касательные к координатным линиям $y^0$ и,
следовательно, касательные также к изотропной гиперповерхности, имеют нулевую
длину, $(\pl_0,\pl_0)=g_{00}=0$. Метрика, индуцированная на гиперповерхностях
$y^1=\const$, имеет вид
\begin{equation*}
  h_{\mu\nu}=\begin{pmatrix}
    0 & 0 & \dotsc & 0 \\ 0 &&& \\ \vdots && g'_{\Sa\Sb} & \\ 0 &&&
  \end{pmatrix},\qquad
  \Sa,\Sb=2,3,\dotsc,n-1,
\end{equation*}
и является, очевидно, вырожденной.

Вернемся к прежним обозначениям координат через $x^\al$ вместо $y^\al$. То есть
будем считать, что в координатах $x^\al$ зафиксирована калибровка светового
конуса, и метрика имеет вид (\ref{elicga}), или
\begin{equation}                                                  \label{emelia}
  ds^2=2dx^0dx^1+g_{\mu\nu}dx^\mu dx^\nu.
\end{equation}

Наша интуиция основана на таких системах координат, в которых метрика
диагональна. Поскольку в калибровке светового конуса метрика недиагональна, то
полезно рассмотреть простой пример плоского пространства Минковского.
\begin{exa}                                                       \label{eishyp}
Рассмотрим трехмерное пространство Минковского $\MR^{1,2}$. В инерциальной
системе отсчета $t,x,y$ метрика Лоренца имеет вид
\begin{equation*}
  ds^2=dt^2-dx^2-dy^2.
\end{equation*}
В плоскости $t,x$ можно ввести координаты светового конуса (\ref{econco})
\begin{equation*}
  u:=\frac1{\sqrt2}(t+x),\qquad v:=\frac1{\sqrt2}(t-x).
\end{equation*}
Рассмотрим более общее преобразование координат
\begin{equation}                                                  \label{eligec}
  \xi=\frac1{\sqrt2}\big((2-a)t+ax\big),\qquad \eta=\frac1{\sqrt2}(t-x),\qquad
  a\in\MR,
\end{equation}
которое параметризуется одним вещественным параметром $a$. Эти координаты в
плоскости $t,x$ показаны на рис.~\ref{falcoc}.
\begin{figure}[h,b,t]
\hfill\includegraphics[width=.35\textwidth]{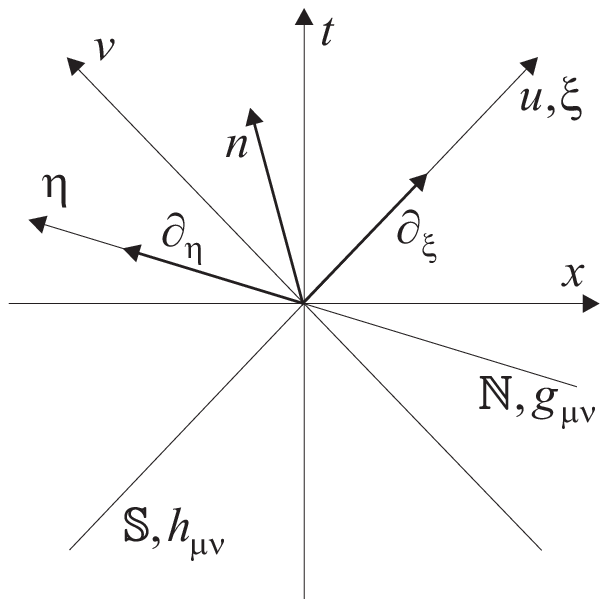}
\hfill {}
\centering\caption{Координаты $\xi,\eta$ в плоскости $t,x$.\label{falcoc}}
\end{figure}
На рисунке показаны конусные оси $v$ и $u$, которые определяются уравнениями,
соответственно, $u=0$ и $v=0$. Ось $\xi$ совпадает с осью $u$. Показана также
ось $\eta$ при $0<a<1$. В рассматриваемой системе координат, координата
$\xi$ по-прежнему является конусной, а координата $\eta$ совпадает с конусной
координатой $v$ только при $a=1$. Легко найти обратное преобразование координат
\begin{equation*}
  t=\frac1{\sqrt2}(\xi+a\eta),\qquad x=\frac1{\sqrt2}\big(\xi+(a-2)\eta\big).
\end{equation*}
Метрика Лоренца в координатах $(x^0,x^1,x^2):=(\xi,\eta,y)$ имеет вид
\begin{equation*}
  ds^2=2d\xi d\eta+2(a-1)d\eta^2-dy^2.
\end{equation*}
Компоненты метрики и ее обратной в координатах $\xi,\eta,y$ равны
\begin{equation*}
  g_{\al\bt}=\begin{pmatrix}0&1&\quad0\\1&2(a-1)&\quad0\\ 0&0&-1 \end{pmatrix},
  \qquad
  g^{\al\bt}=\begin{pmatrix} -2(a-1)&1&\quad0\\1&0&\quad0\\0&0&-1 \end{pmatrix}.
\end{equation*}
Следовательно, координаты $(x^0,x^1,x^2)=(\xi,\eta,y)$ задают калибровку
светового конуса.

Отметим, что метрика $g_{\mu\nu}$ на плоскостях $\xi=\const$ является
отрицательно определенной при $a<1$ и имеет лоренцеву сигнатуру при $a>1$.

В трехмерном пространстве Минковского существует два типа изотропных
гиперповерхностей: плоскости и конусы (характеристики в примере \ref{emipch}).
Изотропные плоскости -- это все плоскости, имеющие угол $\pi/4$ с осью $t$.

Сравним координаты светового конуса данного примера при $0<a<1$ с системой
координат, построенной в предложении (\ref{pxlico}) (см.\ рис.\ref{falcoc}).
Пространственноподобная поверхность $\MN$ натянута на векторы $\pl_\eta,\pl_y$ и
определяется уравнением $x^0=\xi=0$. Изотропные поверхности $\MS$ натянуты на
векторы $\pl_\xi,\pl_y$ и определяются уравнениями $x^1=\eta=\const$. Вектор
\begin{equation*}
  \pl_\xi=\frac1{\sqrt2}\pl_t+\frac1{\sqrt2}\pl_x
\end{equation*}
касается изотропной поверхности $\MS$. Вектор
\begin{equation*}
  \pl_\eta=\frac a{\sqrt2}\pl_t-\frac{2-a}{\sqrt2}\pl_x
\end{equation*}
касается поверхности $\MN$. Нормированный вектор $e_1$ пропорционален
$\pl_\eta$:
\begin{equation*}
  e_1=\frac1{\sqrt{2(1-a)}}\pl_\eta=\frac a{2\sqrt{1-a}}\pl_t
  -\frac{2-a}{2\sqrt{1-a}}\pl_x.
\end{equation*}
Нормированный вектор нормали $n$ к пространственноподобной поверхности $\MN$
имеет вид
\begin{equation*}
  n=\frac{2-a}{2\sqrt{1-a}}\pl_t-\frac a{2\sqrt{1-a}}\pl_x.
\end{equation*}
Вычислим изотропный вектор $m$:
\begin{equation*}
  m:=n-e_1=\sqrt{1-a}\,\pl_t+\sqrt{1-a}\,\pl_x=\sqrt{2(1-a)}\,\pl_\xi.
\end{equation*}
Как и следовало ожидать, он пропорционален вектору $\pl_\xi$.
\qed\end{exa}

Рассмотренный пример показывает, что калибровка светового конуса (\ref{emelia})
определена неоднозначно и может быть зафиксирована таким образом, что метрика
$g_{\mu\nu}$ на гиперповерхностях $x^0=\const$ будет либо отрицательно
определена, либо иметь лоренцеву сигнатуру в произвольной конечной области
$\MU$.
\section{О постановке задач в теории гравитации                  \label{sprobl}}
Отметим специфику задач, возникающих при рассмотрении произвольного функционала
действия, инвариантного относительно общих преобразований координат и
содержащего метрику. Уравнения Эйлера--Лагранжа записываются и решаются в
произвольной, но фиксированной системе координат, т.е.\ локальны. Допустим, что
мы нашли какое-то решение уравнений Эйлера--Лагранжа в системе координат,
совпадающей со всем $\MR^n$. Поскольку исходное действие инвариантно
относительно общих преобразований координат, то без потери общности можно
отобразить все $\MR^n$, и тем самым найденное решение, на ограниченную область,
например, в шар конечного радиуса. В связи с этим возникает вопрос нельзя ли
найденное решение продолжить, т.е.\ существует ли решение в большей области,
сужение которого совпадает с уже найденным решением в шаре. Для ответа на этот
вопрос необходимо инвариантное определение глобальности решения, которое дается
через полноту геодезических и экстремалей.

В теории гравитации, основанной на аффинной геометрии, также, как и в общей
теории относительности, понятие геодезической и экстремали является чрезвычайно
важным, поскольку позволяет установить связь между локальными решениями
уравнений движения и глобальными свойствами многообразий.
\begin{defn}
Назовем многообразие $\MM$ с заданной аффинной геометрией, т.е.\ тройку
$(\MM,g,\Gamma)$, {\em полным}, если любую геодезическую и экстремаль в $\MM$ можно
продолжить до бесконечного значения канонического параметра в обе стороны. На
практике часто локальное решение уравнений движения нельзя продолжить до полного
решения, поскольку возможно появление сингулярностей. Назовем точку многообразия
$\MM$ {\em сингулярной}, если в этой точке по крайней мере одно скалярное поле,
построенное из метрики и (или) аффинной связности обращается в бесконечность.
\qed\end{defn}
\index{Полное многообразие (complete manifold)}%
\index{Многообразие полное (complete manifold)}%
\index{Сингулярная точка (singular point)}%
\index{Точка сингулярная (singular point)}%
В этом определении важно, чтобы в сингулярной точке именно скалярная комбинация
геометрических объектов обращалась в бесконечность, поскольку наряду с истинными
сингулярностями могут существовать координатные сингулярности, связанные с
неудачным выбором системы координат. Например, хорошо известна координатная
особенность метрики на горизонте событий для решения Шварцшильда, от которой
можно избавиться, перейдя, например, к системе координат
Эддингтона--Финкельстейна или Крускала--Секереша. Простейшими функциями,
определяющими положение сингулярностей являются скалярная кривизна и квадрат
тензора кручения, имеющие одинаковую размерность. Примером истинной
сингулярности может служить черная дыра в решении Шварцшильда, в которой квадрат
тензора кривизны обращается в бесконечность.

В связи с возможным существованием сингулярностей оказывается полезным понятие
максимально продолженного многообразия.
\begin{defn}
Многообразие с заданной аффинной геометрией назовем {\em максимально
продолженным}, если любую геодезическую и экстремаль можно либо продолжить до
бесконечного значения канонического параметра, либо они продолжаются до
сингулярной точки при конечном значении канонического параметра. Соответствующую
тройку $(\MM,g,\Gamma)$ назовем {\em глобальным решением} в теории гравитации.
\qed\end{defn}
\index{Многообразие максимально продолженное (maximally extended manifold)}%
\index{Максимально продолженное многообразие (maximally extended manifold)}%
\index{Глобальное решение (global solution)}%
\index{Решение глобальное (global solution)}%
Слово ``продолжение'' в этом определении связано с тем, что в моделях
математической физики, как правило, метрика и кручение сначала находятся только
локально в какой либо области, как решение уравнений движения в заданной системе
координат, а затем, если необходимо, это решение гладко продолжается в соседние
области до тех пор, пока дальнейшее продолжение становится невозможным либо
вследствие сингулярностей, либо в связи с тем, что полученное решение станет
полным.

Важно отметить, что определение максимально продолженного многообразия является
инвариантным и не зависит от выбора локальной системы координат. Это связано с
тем, что канонический параметр определяется однозначно с точностью до линейного
преобразования и не зависит от системы координат.
\begin{exa}
Рассмотрим евклидову плоскость $\MR^2$ в полярных координатах $r,\vf$. Будем
считать, что полярный угол меняется в бесконечном интервале, $0<\vf<\infty$.
\begin{defn}
{\em Гиперболической спиралью} называется кривая на евклидовой плоскости,
которая в полярных координатах задана уравнением
\begin{equation}                                                  \label{ehypsp}
  r=\frac1\vf,\qquad 0<\vf<\infty. \qed
\end{equation}
\end{defn}
\index{Гиперболическая спираль (hyperbolic spiral)}%
\index{Спираль гиперболическая (hyperbolic spiral)}%
Гиперболическая спираль соединяет бесконечно удаленную точку с декартовыми
координатами $(\infty,0)$ и начало координат $(0,0)$, вокруг которого она
наматывается бесконечное число раз. Если полярный угол меняется в интервале
$\vf\in(\vf_1,\vf_2)\subset(0,\infty)$, то ее длина определяется интегралом
\begin{equation*}
  l(\vf_1,\vf_2)=\int_{\vf_1}^{\vf_2}\!\!\!d\vf\sqrt{dr^2+r^2d\vf^2}=
  \int_{\vf_1}^{\vf_2}\!\!\!d\vf \frac{\sqrt{1+\vf^2}}{\vf^2}.
\end{equation*}
Этот интеграл расходится на обоих концах гиперболической спирали, как при
$\vf\to0$, так и при $\vf\to\infty$. Расходимость интеграла при $\vf\to0$
естественна, т.к.\ любая кривая, уходящая в бесконечность, имеет бесконечную
длину. Начало координат является конечной точкой, т.к.\ любая экстремаль,
проходящая через эту точку, проходит ее при конечном значении канонического
параметра. В то же время не любая кривая, подходящая к этой точке, имеет
конечную длину. Это связано с тем, что гиперболическая спираль достаточно быстро
наматывается вокруг начала координат.
\qed\end{exa}
Данный пример показывает, что в определении полноты многообразий требование
полноты экстремалей, а не произвольных кривых является существенным.

С физической точки зрения требование полноты или максимального продолжения
пространства-времени $\MM$ является естественным. Действительно, если
рассмотреть движение точечной частицы в сопутствующей системе отсчета, в которой
время является каноническим параметром, то естественно предположить, что
эволюция продолжается либо бесконечно долго, либо обрывается в сингулярной
точке.

Отметим, что полное многообразие не может иметь края. При этом существуют две
возможности: либо пространство-время $\MM$ некомпактно, либо компактно и без
края (замкнутая вселенная). В первом случае геометрические инварианты в
бесконечности могут стремиться как к конечным, так и бесконечным значениям.
В космологии принята следующая терминология. Если все пространственные сечения
пространства-времени $\MM$ некомпактны, то вселенная {\em бесконечна}. Если все
пространственные сечения компактны и без края, то говорят, что вселенная
{\em замкнута}.
\index{Открытая вселенная (open universe)}%
\index{Вселенная открытая (open universe)}%
\index{Замкнутая вселенная (closed universe)}%
\index{Вселенная замкнутая (closed universe)}%
При наличии сингулярностей, соответствующих конечному значению канонического
параметра, сингулярные точки образуют край пространства-времени, находящийся на
конечном расстоянии (при конечных значениях канонического параметра).

Чтобы найти максимально продолженное решение необходимо пройти несколько этапов:
1) решить уравнения Эйлера--Лагранжа в некоторой области, 2) найти и
проанализировать полноту всех геодезических и экстремалей для соответствующей
метрики и связности, 3) если область, где найдено решение, оказалась неполной,
то продолжить решение. Первые два этапа чрезвычайно сложны, поскольку
предполагают решение нелинейных систем дифференциальных уравнений. Последний
этап также сложен. Его можно осуществить по крайней мере двумя способами. Либо
перейти в новую систему координат, охватывающую б\'ольшую область, либо найти
решение в соседней области, а затем доказать гладкость склейки.

В общей теории относительности существуют только отдельные примеры максимально
продолженных многообразий. Например, расширение Крускала--Секереша решения
Шварцшильда, которое будет обсуждаться в дальнейшем.
\chapter{Гамильтонова формулировка общей теории относительности  \label{shamgr}}
В настоящей главе мы перепишем вакуумные (без полей материи) уравнения Эйнштейна
в гамильтоновом виде. Эта задача важна как для исследования классических
уравнений движения, в частности, при рассмотрении задачи Коши, так и для
построения квантовой теории гравитации.

Каноническая формулировка общей теории относительности была впервые дана Дираком
в формализме второго порядка \cite{Dirac58BR}. Он показал, что гамильтониан
гравитационного поля равен линейной комбинации связей и нашел его явное
выражение. Позже Арновитт, Дезер и Мизнер в серии статей, завершившихся обзором
\cite{ArDeMi62} (без ссылки на Дирака), существенно упростили вычисления и
прояснили геометрический смысл канонических импульсов, выразив их через внешнюю
кривизну пространственной гиперповерхности, вложенной в четырехмерное
пространство-время. Выражение для гамильтониана было найдено ими в формализме
первого порядка, когда метрика $g_{\al\bt}$ и симметричная аффинная связность
$\Gamma_{\lbrace\al\bt\rbrace}{}^\g$ рассматриваются в качестве независимых
переменных. По сути дела этот подход и упростил вычисления.
\section{Лагранжиан Гильберта--Эйнштейна                         \label{shieil}}
Рассмотрим пространство-время $\MM$ произвольной размерности, $\dim\MM=n\ge3$.
Локальные координаты, как и ранее, мы нумеруем греческими буквами из начала
алфавита $x^\al$, $\al=0,1,\dotsc,n-1$. Мы предполагаем, что на $\MM$ задана
метрика $g_{\al\bt}$ лоренцевой сигнатуры, а кручение и тензор неметричности
тождественно равны нулю (псевдориманова геометрия). Как и ранее мы предполагаем,
что координата $x^0$ является временем, и все сечения $x^0=\const$
пространственноподобны. Уравнения движения для метрики в общей теории
относительности при отсутствии полей материи следуют из вариационного принципа
для действия Гильберта--Эйнштейна (\ref{ehieia})
\begin{equation}                                                  \label{ehilea}
  S_{\Sh\Se}=\int dx \vol R,\qquad
  \vol:=\det e_\al{}^a=\sqrt{|\det g_{\al\bt}|},
\end{equation}
где $R$ -- псевдориманова скалярная кривизна, построенная по метрике
$g_{\al\bt}$. В настоящей главе мы опускаем знак тильды для геометрических
объектов псевдоримановой геометрии, чтобы упростить обозначения. Гравитационную
постоянную перед действием Гильберта--Эйнштейна мы положили равной единице,
$\kappa=1$, поскольку уравнения не меняются при умножении полного действия на
постоянную. Космологическую постоянную мы пока положим равной нулю, $\Lm=0$, так
как ее учет приводит только к изменению потенциала для метрики и не вызывает
трудностей при канонической формулировке общей теории относительности.

Действие Гильберта--Эйнштейна в виде (\ref{ehilea}) содержит вторые производные
от метрики, что создает определенные трудности для перехода к каноническому
формализму. Однако, из явного вида тензора кривизны (\ref{ecutrl}) следует, что
вторые производные входят линейно. Поэтому от них можно избавиться, добавив к
подынтегральному выражению полную производную. При этом мы потеряем явную
ковариантность, зато действие будет зависеть только от первых производных
метрики. С этой целью представим скалярную кривизну в виде двух слагаемых, что
проверяется прямой проверкой,
\begin{equation}                                                  \label{escder}
  R=\pl\Gamma-L_{\Sh\Se},
\end{equation}
где
\begin{align}                                                     \label{elaghe}
  L_{\Sh\Se}&:=g^{\al\bt}(\Gamma_{\al\bt}{}^\g\Gamma_\g
  -\Gamma_{\al\dl}{}^\g\Gamma_{\bt\g}{}^\dl),
\\                                                                \label{edertl}
  \pl\Gamma&:=g^{\al\bt}(\pl_\al\Gamma_\bt-\pl_\g\Gamma_{\al\bt}{}^\g)
  =\frac1\vol\pl_\al\left(\vol g^{\al\bt}\Gamma_\bt-\vol g^{\bt\g}\Gamma_{\bt\g}{}^\al
  \right)
  +2L_{\Sh\Se}.
\end{align}
Отсюда следует, что лагранжиан Гильберта--Эйнштейна $\vol L_{\Sh\Se}$ отличается
от скалярной кривизны, умноженной на определитель репера, на полную производную:
\begin{equation*}                                                 \label{elahed}
  \vol R-\pl_\al\left(\vol g^{\al\bt}\Gamma_\bt-\vol g^{\bt\g}\Gamma_{\bt\g}{}^\al
  \right)=\vol L_{\Sh\Se},
\end{equation*}
и, следовательно, приводит к тем же уравнениям Эйлера--Лагранжа. Он квадратичен
по символам Кристоффеля и поэтому зависит только от первых производных метрики.
\section{АДМ параметризация метрики и репера                     \label{sadmpa}}
\index{АДМ параметризация метрики (ADM parametrization of metric)}%
\index{Параметризация метрики АДМ (ADM parametrization of metric)}%
При анализе гамильтоновой структуры уравнений общей теории относительности
Арновитт, Дезер и Мизнер (АДМ) \cite{ArDeMi62} использовали специальную
параметризацию метрики, которая существенно упростила вычисления. В настоящем
разделе будет описана АДМ параметризация метрики и соответствующая
параметризация репера, которая упрощает вычисления в переменных Картана.

Рассмотрим многообразие $\MM$, $\dim\MM=n$, с метрикой лоренцевой сигнатуры
$(+-\dotsc-)$. Пусть $\lbrace x^\al\rbrace$, $\al=0,1,\dotsc,n-1$, -- система
локальных координат. Выделим среди координат время $t=x^0$, тогда
$\lbrace x^\al\rbrace=\lbrace x^0,x^\mu\rbrace$, $\mu=1,\dotsc,n-1$. В
дальнейшем буквы из начала греческого алфавита $\al,\bt,\dotsc$ будут пробегать
все значения индексов, а из середины $\mu,\nu,\dotsc$ -- только пространственные
значения. Это правило легко запомнить по следующим включениям:
$\lbrace\mu,\nu,\dotsc\rbrace\subset\lbrace\al,\bt,\dotsc\rbrace$ и
$\lbrace1,2,\dotsc\rbrace\subset\lbrace0,1,2,\dotsc\rbrace$.
АДМ параметризация метрики имеет следующий вид
\begin{equation}                                                  \label{eadmme}
g_{\al\bt}=\begin{pmatrix} N^2+N^\rho N_\rho & N_\nu \\
           N_\mu & g_{\mu\nu}\end{pmatrix},
\end{equation}
где $g_{\mu\nu}$ -- метрика на $(n-1)$-мерных сечениях многообразия
$x^0=\const$, которые предполагаются пространственноподобными. В выбранной
параметризации вместо $n$ компонент метрики, содержащих хотя бы один временн\'ой
индекс, $g_{00}$ и $g_{0\mu}$, введены $n$ функций $N$ и $N_\mu$. Выше
$N^\rho:=\hat g^{\rho\mu}N_\mu$, где $\hat g^{\rho\mu}$ -- $(n-1)\times(n-1)$
матрица, обратная к $g_{\mu\nu}$:
\begin{equation*}
  \hat g^{\rho\mu}g_{\mu\nu}=\dl^\rho_\nu,
\end{equation*}
которую мы называем обратной пространственной метрикой. В настоящей главе подъем
пространственных индексов будет всегда осуществляться с помощью обратной метрики
$\hat g^{\rho\mu}$, помеченной шляпкой, которая в общем случае не совпадает с
``пространственной'' частью метрики $g^{\al\bt}$, обратной к $g_{\al\bt}$:
$\hat g^{\rho\mu}\ne g^{\rho\mu}$. Функция $N=N(x)$ называется {\em функцией
хода}, а функции $N_\mu=N_\mu(x)$ -- {\em функциями сдвига}. Не ограничивая
общности, можно считать, что функция хода положительна $N>0$. В этом случае АДМ
параметризация метрики (\ref{eadmme}) является взаимно однозначной.
\index{Функция хода (lapse function)}\index{Хода функция (lapse function)}%
\index{Функция сдвига (shift function)}\index{Сдвига функция (shift function)}%

Интервал, соответствующий параметризации (\ref{eadmme}), можно записать в виде
\begin{equation*}
  ds^2=N^2dt^2+g_{\mu\nu}(dx^\mu+N^\mu dt)(dx^\nu+N^\nu dt).
\end{equation*}
Вообще, квадрат произвольного вектора $X=X^\al\pl_\al$ в АДМ параметризации
метрики (\ref{eadmme}) имеет вид суммы двух квадратов:
\begin{equation*}
  X^2:=g_{\al\bt}X^\al X^\bt
  =N^2(X^0)^2+g_{\mu\nu}(X^\mu+N^\mu X^0)(X^\nu+N^\nu X^0),
\end{equation*}
где выделен квадрат временн\'ой компоненты. Первое слагаемое в этом выражении
положительно, а второе -- отрицательно, т.к.\ метрика на пространственноподобных
сечениях отрицательно определена.

Мы предполагаем, что координата $x^0=t$ является временем, т.е. касательный
вектор $\pl_0$ к координатной линии $x^0$ времениподобен. Формально это условие
записывается в виде
\begin{equation}                                                  \label{etimec}
  (\pl_0,\pl_0)=g_{00}=N^2+N^\rho N_\rho>0.
\end{equation}
В этом случае метрика $g_{\al\bt}$ имеет лоренцеву сигнатуру тогда и только
тогда, когда матрица (теорема \ref{tlosim})
\begin{equation}                                                  \label{espadm}
  g_{\mu\nu}-\frac{N_\mu N_\nu}{N^2+N^\rho N_\rho}
\end{equation}
отрицательно определена. Отметим, что сама метрика $g_{\mu\nu}$, индуцированная
на сечении $x^0=\const$, может и не быть отрицательно определенной. Это значит,
что сечение $x^0=\const$ в общем случае не является пространственноподобным. В
дальнейшем мы будем дополнительно предполагать, что координаты выбраны таким
образом, чтобы все сечения $x^0=\const$ были пространственноподобны, т.е.\
метрика $g_{\mu\nu}$ также отрицательно определена. Это удобно для постановки
задачи Коши, когда начальные данные задаются на пространственноподобной
поверхности, и рассматривается их эволюция во времени.
\begin{com}
Аналогичным образом можно параметризовать метрику и на римановом многообразии с
положительно определенной метрикой, для этого вместо времени достаточно явно
выделить произвольную координату.
\qed\end{com}

Метрика, обратная к (\ref{eadmme}), имеет вид
\begin{equation}                                                  \label{eadmmi}
  g^{\al\bt}=\left(\begin{aligned}  &\dfrac{\raise-.7ex\hbox{1}}{N^2}
  & &\!-\!\dfrac{\raise-.7ex\hbox{$N^\nu$}}{N^2} \\
  -&\dfrac{\raise-.7ex\hbox{$N^\mu$}}{N^2}
  & ~\hat g^{\mu\nu}\!\!\!&+\!\dfrac{\raise-.7ex\hbox{$N^\mu N^\nu$}}{N^2}
  \end{aligned}\right).
\end{equation}
\begin{prop}                                                      \label{potrit}
Пространственная матрица в правом нижнем блоке
\begin{equation}                                                  \label{eincpm}
  g^{\mu\nu}=\hat g^{\mu\nu}+\frac{N^\mu N^\nu}{N^2},
\end{equation}
отрицательно определена.
\end{prop}
\begin{proof}
Легко проверить, матрица (\ref{eincpm}) является обратной к метрике
(\ref{espadm}). Это значит, что отрицательная определенность метрики
(\ref{espadm}) эквивалентна отрицательной определенности матрицы $g^{\mu\nu}$.
\end{proof}

Заметим, что, если метрика на многообразии $\MM$ имеет лоренцеву сигнатуру, то
условие пространственноподобности всех сечений $x^0=\const$ эквивалентно
условию $N^2>0$. Действительно, из отрицательной определенности $g_{\mu\nu}$
следует отрицательная определенность обратной матрицы $\hat g^{\mu\nu}$.
Тогда из уравнения (\ref{eincpm}) вытекает отрицательная определенность матрицы
\begin{equation*}
  g^{\mu\nu}-\frac{N^\mu N^\nu}{N^2}.
\end{equation*}
Это, в свою очередь, эквивалентно условию $g^{00}>0$ или $N^2>0$.

Чтобы продемонстрировать тонкости, которые могут возникнуть при АДМ
параметризации метрики, приведем простой
\begin{exa}
Рассмотрим двумерное пространство-время Минковского $\MR^{1,1}$ с декартовыми
координатами $t,x$. Введем новую систему координат $\xi,\eta$, зависящую от двух
вещественных параметров $a$ и $b$ (см.\ рис.\ref{fminpp})
\begin{figure}[h,b,t]
\hfill\includegraphics[width=.35\textwidth]{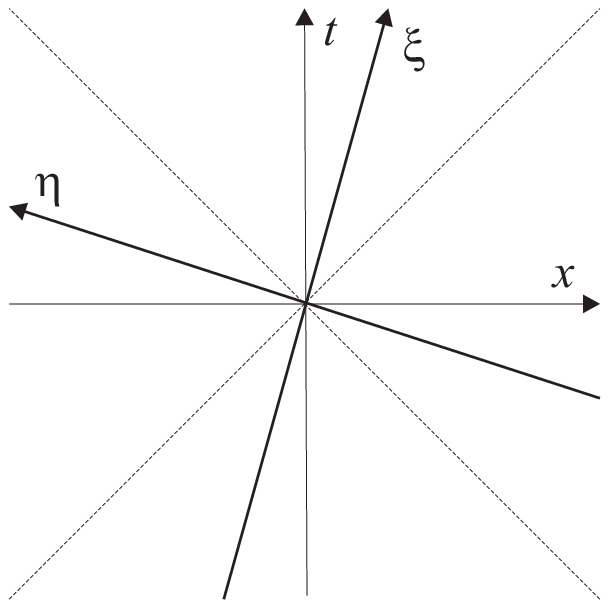}
\hfill {}
\centering\caption{Выбор координат на плоскости Минковского. Рисунок
соответствует следующим значениям параметров: $0<a<1$ и $b>1$.}
\label{fminpp}
\end{figure}
\begin{equation*}
  \xi=t+ax,\quad \eta=t-bx,\qquad |a|\ne1,\quad |b|\ne1,\quad a+b\ne0.
\end{equation*}
Легко получить формулы обратного преобразования
\begin{equation*}
  t=\frac{b\xi+a\eta}{a+b},\qquad x=\frac{\xi-\eta}{a+b}.
\end{equation*}
В новых координатах лоренцева метрика имеет вид
\begin{equation*}
  ds^2=dt^2-dx^2=\frac1{(a+b)^2}\left[(b^2-1)d\xi^2+2(ab+1)d\xi d\eta
  +(a^2-1)d\eta^2\right].
\end{equation*}
Векторы, касательные к координатным линиям $\xi$ и $\eta$ имеют следующие
компоненты:
\begin{equation*}
  \pl_\xi=\frac b{a+b}\pl_t+\frac1{a+b}\pl_x,\qquad
  \pl_\eta=\frac a{a+b}\pl_t-\frac1{a+b}\pl_x.
\end{equation*}

Проанализируем АДМ параметризацию метрики в координатах $x^0=\xi,~x^1=\eta$.
Метрика имеет следующие компоненты:
\begin{equation*}
  g_{00}=\frac{b^2-1}{(a+b)^2},\qquad g_{01}=\frac{ab+1}{(a+b)^2},\qquad
  g_{11}=\frac{a^2-1}{(a+b)^2}.
\end{equation*}
Функции хода и сдвига имеют вид
\begin{equation*}
  N^2=-\frac1{a^2-1},\qquad N_1=\frac{ab+1}{(a+b)^2}.
\end{equation*}
Из условий $g_{00}>0$ и $g_{11}<0$ следует, соответственно, что $|b|>1$ и
$|a|<1$. Мы видим, что эти условия являются необходимыми и достаточными для
того, чтобы координатная линия $\xi$ была времениподобной, а $\eta$ --
пространственноподобной. Нетрудно также проверить эквивалентность условий:
\begin{equation*}
  g_{00}>0\qquad \Leftrightarrow\qquad
  g_{11}-\frac{N_1N_1}{N^2+N^1N_1}=-\frac1{b^2-1}<0,
\end{equation*}
а также условий:
\begin{equation*}                                                    \tag*{\qed}
  g^{00}>0\qquad \Leftrightarrow\qquad
  \hat g^{11}=g^{11}-\frac{N^1 N^1}{N^2}=\frac{(a+b)^2}{a^2-1}<0.
\end{equation*}
\end{exa}

Обратная метрика (\ref{eadmmi}) удобна в приложениях тем, что квадрат
произвольного ковектора (1-формы) $A=dx^\al A_\al$ имеет вид суммы двух
квадратов:
\begin{equation*}
  A^2:=g^{\al\bt}A_\al A_\bt
  =\frac1{N^2}(A_0-N^\mu A_\mu)^2+\hat g^{\mu\nu}A_\mu A_\nu,
\end{equation*}
где выделен квадрат пространственных компонент. Здесь первое слагаемое
положительно, а второе отрицательно.

Используя формулу (\ref{edemnm}) для определителя блочных матриц, получаем
выражение для определителя метрики (\ref{eadmme})
\begin{equation}                                                  \label{edeadm}
  \det g_{\al\bt}=N^2\det g_{\mu\nu}.
\end{equation}
Отсюда следует выражение для элемента объема
\begin{equation}                                                  \label{evolex}
  \vol=N\sqrt{|\hat g|},\qquad
  \vol:=\sqrt{|\det g_{\al\bt}|},\qquad
  \sqrt{|\hat g|}:=\sqrt{|\det g_{\mu\nu}|}.
\end{equation}
Последняя формула является обобщением хорошо известной из школьного курса
геометрии правила: объем призмы равен произведению площади основания на высоту.
В рассматриваемом случае площадью основания является $\sqrt{|\hat g|}$, а
высотой -- функция хода $N$.

При проведении вычислений оказываются полезными следующие формулы, которые
проверяются прямой проверкой,
\begin{equation}                                                  \label{eadmef}
\begin{split}
  g^{00}g^{\mu\nu}-g^{0\mu}g^{0\nu}&=\frac{\hat g^{\mu\nu}}{N^2},
\\
  g^{\s\mu}g^{0\nu}-g^{\s\nu}g^{0\mu}&
  =\frac{N^\mu\hat g^{\s\nu}-N^\nu\hat g^{\s\mu}}{N^2},
\\
  g^{\mu\nu}g_{\nu\s}&=\dl^\mu_\s+\frac{N^\mu N_\s}{N^2},
\\
  g^{\mu\nu}g_{\mu\nu}&=n-1+\frac{N^\mu N_\mu}{N^2}.
\end{split}
\end{equation}

В аффинной геометрии вместо метрики $g_{\al\bt}$ и аффинной связности
$\Gamma_{\al\bt}{}^\g$ удобно использовать переменные Картана: репер $e_\al{}^a$
и $\MG\ML(n,\MR)$-связность $\om_{\al a}{}^b$, где
$a,b,\dotsc=\zero,1,\dotsc,n-1$ (см.\ раздел \ref{scorep}). Здесь и в дальнейшем
мы примем следующее обозначение. Если латинский индекс $a$ принимает значение
нуль, то мы будем писать его курсивом: $a=\zero$. Это правило принято потому что
в дальнейшем значение нуль будут принимать как греческие, так и латинские
индексы, что необходимо различать. Выделим в ортонормальном базисе касательного
пространства временн\'ой вектор $e_\zero$,
$\lbrace e_a\rbrace=\lbrace e_\zero,e_i\rbrace$, тогда метрика Минковского
примет вид
\begin{equation*}
  \eta_{ab}=\begin{pmatrix} 1 & 0 \\ 0 & \eta_{ij}\end{pmatrix},
  \qquad i,j=1,\dotsc,n-1,
\end{equation*}
где $\eta_{ij}=-\dl_{ij}$ -- евклидова метрика с обратным знаком для
пространственных компонент тензоров. Как и раньше, мы принимаем следующее
соглашение. Латинские индексы из начала алфавита $a,b,\dotsc$ пробегают все
значения, а индексы из середины алфавита $i,j,\dotsc$ -- только
пространственные. Поскольку репер содержит большее число независимых компонент,
чем метрика, то его параметризация, соответствующая выделенной временн\'ой
координате, является более громоздкой:
\begin{equation}                                                  \label{eviela}
  e_\al{}^a=\begin{pmatrix}
  N\sqrt{1-M^2}+(MN) & \qquad N^i-NM^i
  +\dfrac{\raise-.3ex\hbox{$\sqrt{1-M^2}-1$}}{M^2}(MN)M^i \\[4mm]
  M_\mu &
  \hat e_\mu{}^i+\dfrac{\raise-.3ex\hbox{$\sqrt{1-M^2}-1$}}{M^2}M_\mu M^i
  \end{pmatrix},
\end{equation}
где $M_\mu$ -- дополнительное $(n-1)$-компонентное пространственноподобное
векторное поле (относительно трехмерных диффеоморфизмов). Ниже будет показано,
что это поле пропорционально трехмерному вектору скорости, параметризующему
преобразования Лоренца. Репер со шляпкой в (\ref{eviela}) $\hat e_\mu{}^i$ на
сечении $x^0=\const$ определяется трехмерной метрикой:
\begin{equation}                                                  \label{ethrem}
  g_{\mu\nu}=\hat e_\mu{}^i \hat e_\nu{}^j\eta_{ij},
\end{equation}
В параметризации репера (\ref{eviela}) использованы также обозначения
$(MN):=M^\mu N_\mu=M^i N_i$ и $M^2:=M^\mu M_\mu=M^i M_i<0$. Переход от
пространственных латинских индексов к пространственным греческим индексам
осуществляется с помощью репера $\hat e_\mu{}^i$ и его обратного
$\hat e^\mu{}_i$, которые мы пометили шляпкой,
\begin{equation*}
  \hat g^{\mu\nu}=\hat e^\mu{}_i\hat e^\nu{}_j\eta^{ij}.
\end{equation*}
Например, $M^i:=M_\mu \hat e^{\mu i}$. Как и для метрики, реперы со шляпкой
и без нее необходимо различать: $\hat e_\mu{}^i\ne e_\mu{}^i$,
$\hat e^\mu{}_i\ne e^\mu{}_i$. Из определения репера со шляпкой (\ref{ethrem})
следует равенство для определителей
\begin{equation*}
  \det \hat e_\mu{}^i=\sqrt{|\det g_{\mu\nu}|},
\end{equation*}
что согласуется с обозначением в (\ref{evolex}).

С помощью прямых вычислений нетрудно проверить, что приведенная параметризация
репера соответствует АДМ параметризации метрики (\ref{eadmme})
\begin{equation*}
  N^2+N^\rho N_\rho=e_0{}^a e_{0a},\qquad N_\mu=e_0{}^a e_{\mu a},\qquad
  g_{\mu\nu}=e_\mu{}^ae_{\nu a}.
\end{equation*}

Обратный репер имеет вид
\begin{equation}                                                  \label{evinad}
  e^\al{}_a=\begin{pmatrix}
  \dfrac{\raise-.3ex\hbox{$\sqrt{1-M^2}$}}N &
  -\dfrac{\raise-.3ex\hbox{$M_i$}}N \\[4mm]
  M^\mu-\dfrac{\raise-.3ex\hbox{$\sqrt{1-M^2}$}}N N^\mu &
  \qquad \hat e^\mu{}_i+\dfrac{\raise-.3ex\hbox{$N^\mu M_i$}}N
  +\dfrac{\raise-.3ex\hbox{$\sqrt{1-M^2}-1$}}{M^2}M^\mu M_i.
  \end{pmatrix}
\end{equation}

Чтобы понять геометрический смысл параметров $M_i$, рассмотрим преобразования
касательного пространства. Связная компонента единицы группы Лоренца
$\MS\MO_\uparrow(1,n-1)$ имеет подгруппу вращений $\MS\MO(n-1)$, действующую на
пространственные компоненты репера $\hat e_\mu{}^i$. Эта подгруппа
параметризуется $(n-1)(n-2)/2$ параметрами. Остальные $n-1$ параметров
соответствуют лоренцевым бустам, которые параметризуются функциями $M_i$. В этом
нетрудно убедиться. Представим репер (\ref{eviela}) в виде
\begin{equation*}
  e_\al{}^a=\overset{\circ}e_\al{}^b S_b{}^a,
\end{equation*}
где
\begin{equation}                                                  \label{erefiz}
  \overset{\circ}e_\al{}^a=\begin{pmatrix} N & N^i
  \\ 0 & \hat e_\mu{}^i \end{pmatrix}
\end{equation}
-- фиксированный репер, который преобразуется с помощью лоренцевых бустов
\begin{equation}                                                  \label{eloram}
  S_b{}^a=\begin{pmatrix}\sqrt{1-M^2} & -M^j \\ M_i & \quad S_i{}^j\end{pmatrix}
  \quad \in\MS\MO_\uparrow(1,n-1),
\end{equation}
где введено обозначение для пространственной части матрицы лоренцевых вращений
\begin{equation}                                                  \label{eijmat}
  S_i{}^j=\dl_i^j+\dfrac{\raise-.3ex\hbox{$\sqrt{1-M^2}-1$}}{M^2}M_i M^j
\end{equation}
Можно проверить, что матрица (\ref{eloram}) действительно задает преобразования
Лоренца. Репер, обратный к (\ref{erefiz}), имеет вид
\begin{equation*}
  \overset{\circ}e{}^\al{}_a=\begin{pmatrix}\quad\dfrac{\raise-.3ex\hbox{1}}N& 0
  \\[4mm]
  -\dfrac{\raise-.3ex\hbox{$N^\mu$}}N & \hat e^\mu{}_i \end{pmatrix}.
\end{equation*}
Формулы для фиксированного репера $\overset{\circ}e_\al{}^a$ и его обратного
$\overset{\circ}e{}^\al{}_a$ получаются из (\ref{eviela}) и (\ref{evinad}) при
$M_i=0$.

Матрица пространственных вращений (\ref{eijmat}) часто используется
в вычислениях, поэтому приведем для нее несколько полезных соотношений, которые
проверяются прямой проверкой,
\begin{align*}
  S^{-1}_{\quad i}{}^j&=\dl_i^j-\dfrac{\raise-.3ex\hbox{$\sqrt{1-M^2}-1$}}
  {M^2\sqrt{1-M^2}}M_i M^j
\\
  S_i{}^j(M)&=S_i{}^j(-M),\qquad \qquad S_{ij}=S_{ji},
\\
  S_i{}^jM_j&=\sqrt{1-M^2}M_i,
\\
  S_i{}^kS_{jk}&=\eta_{ij}-M_iM_j,
\\
  S^{-1}{}_i{}^kS^{-1}{}_{jk}&=\eta_{ij}+\frac{M_iM_j}{\sqrt{1-M^2}}.
\end{align*}

Таким образом, вместо $n^2$ независимых компонент репера мы ввели функцию сдвига
$N$ (1 компонента), функции хода $N^i$ ($n-1$ компонент), лоренцевы бусты $M_i$
($n-1$ компонент) и пространственный репер $\hat e_\mu{}^i$ ($(n-1)^2$
компонент). Эта параметризация при $N>0$ является взаимно однозначной.

Отметим, что единичный нормальный вектор к сечению $x^0=\const$ (\ref{enovee})
относительно неголономного базиса $e_a=e^\mu{}_a\pl_\mu$ имеет вид
\begin{equation*}
  n=\sqrt{1-M^2}e_\zero-M^ie_i.
\end{equation*}
То есть его компоненты относительно неголономного базиса полностью
определяются лоренцевыми бустами.

Приведем более привычную параметризацию лоренцевых вращений $S_a{}^b$ с помощью
вектора скорости. Вектор $M^i$ взаимно однозначно параметризуется вектором
скорости $V^i$:
\begin{equation*}
  M^i=\frac{V^i}{\sqrt{1-\BV^2}}\qquad \Leftrightarrow\qquad
  V^i=\frac{M^i}{\sqrt{1-M^2}},
\end{equation*}
где $\BV^2:=-V^i V_i$. По-определению, $M^2<0$ и, следовательно, $0<\BV^2<1$.
Тогда
\begin{equation*}
  S_a{}^b=\begin{pmatrix}
  \displaystyle{\frac1{\sqrt{1-\BV^2}}}
  & -\displaystyle{\frac{V^j}{\sqrt{1-\BV^2}}}
  \\[4mm]
  \displaystyle{\frac{V_i}{\sqrt{1-V^2}}} &\quad
  \dl_i^j-\displaystyle{\frac{V_iV^j}{\BV^2}\left(\frac1{\sqrt{1-V^2}}-1\right)}
  \end{pmatrix}.
\end{equation*}
Эта матрица для лоренцевых бустов была получена ранее (\ref{elobug}).
\begin{com}
В римановом пространстве с положительно определенной метрикой репер можно
параметризовать похожим образом, заменив матрицу лоренцевых вращений на
ортогональную матрицу.
\qed\end{com}
\section{Геометрия гиперповерхностей                             \label{sgesum}}
При гамильтоновой формулировке моделей гравитации мы рассматриваем
пространство-время как семейство пространственноподобных гиперповерхностей,
которое параметризуется временем. Другими словами, для каждого момента времени
пространство представляет собой гиперповерхность, вложенную в
пространство-время. Поскольку уравнения моделей гравитации определяют геометрию
всего пространства-времени, то полезно знать, какая геометрия возникает при этом
на пространственноподобных сечениях. В настоящем разделе мы подойдем к этому
вопросу с общей точки зрения, предполагая, что на объемлющем многообразии задана
аффинная геометрия общего вида, т.е.\ независимые метрика и связность, не
предполагая лоренцевой сигнатуры метрики.

Рассмотрим $(n-1)$-мерную гиперповерхность $\MU$, вложенную в $n$-мерное
многообразие $\MM$:
\begin{equation}                                                  \label{eimmum}
  f:\qquad \MU\hookrightarrow \MM.
\end{equation}
Обозначим координаты на $\MM$ и $\MU$ соответственно через $x^\al$,
$\al=1,\dotsc,n$, и $u^i$, $i=1,\dotsc,n-1$. Тогда вложение $\MU$ в $\MM$
локально задается $n$ функциями $x^\al(u)$, которые предполагаются достаточно
гладкими. Произвольное векторное поле $X=\lbrace X^i\rbrace\in\CX(\MU)$, на
гиперповерхности, отображается в векторное поле
$Y=\lbrace Y^\al\rbrace\in\CX(\MM)$ на $\MM$ с помощью дифференциала отображения
\begin{equation*}
  f_*:\quad \CX(\MU)\ni\quad X=X^i\pl_i\mapsto Y=Y^\al\pl_\al\quad\in\CX(\MM),
\end{equation*}
где
\begin{equation*}
  Y^\al=e^\al{}_i X^i,\qquad e^\al{}_i:=\pl_i x^\al.
\end{equation*}
Матрица Якоби $e^\al{}_i$ отображения $f$ прямоугольна, имеет размер
$n\times(n-1)$, ранг $n-1$ и, естественно, необратима. Возврат отображения $f$
каждому ковекторному полю (1-форме) на $\MM$ ставит в соответствие ковекторное
поле на $\MU$:
\begin{equation*}
  f^*:\quad \Lm_1(\MM)\ni\quad A=dx^\al A_\al\mapsto B=du^i B_i\quad
  \in\Lm_1(\MU),
\end{equation*}
где
\begin{equation*}
  B_i=A_\al e^\al{}_i.
\end{equation*}

В дальнейшем мы будем отождествлять гиперповерхность $\MU$ с ее образом:
$\MU=f(\MU)\subset\MM$.

1-форма $n=dx^\al n_\al$, определяемая системой алгебраических уравнений
\begin{equation}                                                  \label{eortre}
  n_\al e^\al{}_i=0,\qquad i=1,\dotsc,n-1,
\end{equation}
ортогональна гиперповерхности $\MU$ и задает на многообразии $\MM$ распределение
$(n-1)$-мерных подпространств, касательных к $\MU$. Уравнения (\ref{eortre})
имеют единственное нетривиальное решение с точностью до умножения на
произвольную отличную от нуля функцию, поскольку из определения вложения
следует, что ранг матрицы Якоби равен $n-1$.

Матрица Якоби $e^\al{}_i$ задает в касательном пространстве $\MT_x(\MM)$, где
$x\in\MU\subset\MM$, набор из $n-1$ векторов $e_i=e^\al{}_i\pl_\al$, которые
образуют базис касательного пространства к гиперповерхности $\MU\subset\MM$.

Это все, что можно сказать о гиперповерхности $\MU$, если задано только вложение
(\ref{eimmum}). Теория становится намного более содержательной, если на
многообразии $\MM$ заданы дополнительные структуры. Остановимся на этом вопросе
подробно.

Пусть на $\MM$ задана аффинная геометрия, т.е.\ метрика $g_{\al\bt}$ и
аффинная связность $\Gamma_{\al\bt}{}^\g$. Рассмотрим, какая геометрия возникает на
гиперповерхности $\MU\hookrightarrow\MM$. Возврат отображения $f^*$ индуцирует
на гиперповерхности единственную метрику:
\begin{equation}                                                  \label{eindme}
  f^*:\quad g_{\al\bt}\mapsto g_{ij}=g_{\al\bt}e^\al{}_i e^\bt{}_j.
\end{equation}
Для многообразий с метрикой лоренцевой сигнатуры мы предполагаем, что
гиперповерхность $\MU$ пространственноподобна и, следовательно, индуцированная
метрика невырождена и отрицательно определена. Наличие двух метрик $g_{\al\bt}$
и $g_{ij}$ соответственно на $\MM$ и $\MU$ позволяет опускать и поднимать
индексы у матрицы Якоби:
\begin{equation*}
  e_\al{}^i:=g_{\al\bt}e^\bt{}_j g^{ij}.
\end{equation*}
Эта матрица проектирует произвольный вектор из $\MT_{f(u)}(\MM)$ в касательное
пространство к гиперповерхности $\MT_u(\MU)$
\begin{equation}                                                  \label{eceprj}
  \MT_{f(u)}(\MM)\ni\quad\lbrace X^\al\rbrace\mapsto
  \lbrace X^i:=X^\al e_\al{}^i\rbrace\quad\in\MT_u(\MU).
\end{equation}

Теперь определим связность на гиперповерхности $\MU\hookrightarrow\MM$,
спроектировав ковариантную производную в $\MM$ на гиперповерхность $\MU$:
\begin{equation}                                                  \label{edefco}
  \hat\nb_i X^k:=(\nb_\al X^\bt)e^\al{}_i e_\bt{}^k,
\end{equation}
где компоненты $X^i$ определены отображением (\ref{eceprj}).
\begin{com}
Из приведенного определения нельзя выразить ковариантную производную
$\nb_\al X^\bt$ в $\MM$ через ковариантную производную $\hat\nb_i X^k$ на $\MU$,
т.к.\ оригинал нельзя восстановить по его проекции.
\qed\end{com}
Раскрытие равенства (\ref{edefco}) приводит к выражению для компонент
индуцированной связности на гиперповерхности $\MU$:
\begin{equation}                                                  \label{ecocou}
  \hat\Gamma_{ij}{}^k
  =(\pl^2_{ij} x^\g+\Gamma_{\al\bt}{}^\g e^\al{}_i e^\bt{}_j)e_\g{}^k.
\end{equation}
Эта связность единственна. Отметим, что если исходная связность
$\Gamma_{\al\bt}{}^\g$ симметрична, то индуцированная связность $\hat\Gamma_{ij}{}^k$
также симметрична. Кроме того, связность на $\MU$ определяется единственным
образом только в том случае, если на $\MM$ помимо связности задана также
метрика.

Из уравнения (\ref{ecocou}) следует, что тензор кручения на $\MM$ индуцирует
кручение на гиперповерхности:
\begin{equation}                                                  \label{eindto}
  T_{ij}{}^k=T_{\al\bt}{}^\g e^\al{}_i e^\bt{}_j e_\g{}^k.
\end{equation}

Прямые вычисления дают следующее выражение для ковариантной производной от
индуцированной метрики на гиперповерхности
\begin{equation*}
  \hat\nb_i g_{jk}=\pl_i g_{ij}-\hat\Gamma_{ij}{}^l g_{lk}-\hat\Gamma_{ik}{}^l g_{jl}
  =(\nb_\al g_{\bt\g})e^\al{}_i e^\bt{}_j e^\g{}_k.
\end{equation*}
Отсюда вытекает выражение для тензора неметричности на гиперповерхности
\begin{equation*}
  Q_{ijk}=Q_{\al\bt\g}e^\al{}_i e^\bt{}_j e^\g{}_k.
\end{equation*}
В частности, если связность $\Gamma_{\al\bt}{}^\g$ на $\MM$ является метрической,
то и индуцированная связность $\hat\Gamma_{ij}{}^k$ на $\MU$ также метрическая.

Таким образом, метрика $g_{\al\bt}$ и связность $\Gamma_{\al\bt}{}^\g$ на $\MM$
индуцируют единственные метрику $g_{ij}$ и связность $\hat\Gamma_{ij}{}^k$ на
гиперповерхности $\MU\subset\MM$. Обратное утверждение, конечно, неверно. Если
метрика и связность заданы на гиперповерхности $\MU$, то они не индуцируют на
$\MM$ геометрию единственным образом. Это понятно, поскольку размерность
гиперповерхности меньше размерности самого многообразия.

В дальнейшем все геометрические объекты, относящиеся только к гиперповерхности
и построенные по индуцированной метрике $g_{ij}$ и связности $\hat\Gamma_{ij}{}^k$
мы будем отмечать шляпкой.

Наличие метрики $g_{\al\bt}$ позволяет построить единичное векторное поле
$n=n^\al\pl_\al$, ортогональное к гиперповерхности. Как уже отмечалось, система
уравнений $n_\al e^\al{}_i=0$ определяет 1-форму $dx^\al n_\al$ с точностью до
умножения на произвольную функцию. Воспользуемся этим произволом для того, чтобы
в каждой точке вектор с компонентами $n^\al:=g^{\al\bt}n_\bt$ имел единичную
длину $(n,n)=n^\al n^\bt g_{\al\bt}=1$. По-построению, этот вектор ортогонален
всем векторам, касательным к гиперповерхности:
\begin{equation*}
  (n,e_i)=n^\al e^\bt{}_i g_{\al\bt}=n_\al e^\al{}_i=0.
\end{equation*}

Если на многообразии задана гиперповерхность, то естественно рассматривать базис
$\lbrace n,e_i\rbrace$ в касательном пространстве $\MT_x(\MM)$, $x\in\MU$,
определяемый этой гиперповерхностью. Ему соответствует сопряженный базис
$\lbrace n:=dx^\al n_\al,e^i:=dx^\al e_\al{}^i\rbrace$ в кокасательном
пространстве $\MT^*_x(\MM)$. Тогда произвольный вектор $X$ и 1-форма $A$
раскладываются по этим базисам:
\begin{equation*}
\begin{split}
  X^\al&=X^\bot n^\al+X^i e^\al{}_i,\qquad
  X^\bot=X^\al n_\al,\quad X^i=X^\al e_\al{}^i,
\\
  A_\al&=A_\bot n_\al+A_i e_\al{}^i,\qquad
  A_\bot=A_\al n^\al,\quad A_i=A_\al e^\al{}_i.
\end{split}
\end{equation*}
Аналогично можно разложить тензор произвольного ранга. В частности, разложение
ковариантного тензора второго ранга имеет вид
\begin{equation*}
  A_{\al\bt}=A_{\bot\bot}n_\al n_\bt +A_{\bot i}n_\al e^i{}_\bt
  +A_{i\bot}e^i{}_\al n_\bt+A_{ij}e^i{}_\al e^j{}_\bt,
\end{equation*}
где
\begin{equation*}
  A_{\bot\bot}=A_{\al\bt}n^\al n^\bt,\quad A_{\bot i}=A_{\al\bt}n^\al e^\bt{}_i,
  \quad A_{i\bot}=A_{\al\bt}e^\al{}_i n^\bt,\quad A_{ij}
  =A_{\al\bt}e^\al{}_i e^\bt{}_j.
\end{equation*}

Нетрудно проверить, что разложение для метрики гораздо проще:
\begin{equation}                                                  \label{emetde}
  g_{\al\bt}=n_\al n_\bt+e_\al{}^i e_\bt{}^j g_{ij}.
\end{equation}
Разложение для обратной метрики имеет аналогичный вид:
\begin{equation}                                                  \label{einmer}
  g^{\al\bt}=n^\al n^\bt+e^\al{}_i e^\bt{}_j g^{ij}.
\end{equation}
Из определения обратной метрики $g^{\al\bt}g_{\bt\g}=\dl^\al_\g$ следует правило
суммирования матрицы Якоби по латинским индексам:
\begin{equation}                                                  \label{esulae}
  e^\al{}_i e_\bt{}^i=\dl^\al_\bt-n^\al n_\bt.
\end{equation}

Нетрудно также проверить, что из определения обратной индуцированной метрики
$g^{ij}g_{jk}=\dl^i_k$ следует равенство
\begin{equation}                                                  \label{esugre}
  e^\al{}_i e_\al{}^j=\dl_i^j,
\end{equation}
где суммирование проводится по греческим индексам. С учетом этого правила
из равенства (\ref{einmer}) следует представление для обратной индуцированной
метрики
\begin{equation*}
  g^{ij}=g^{\al\bt}e_\al{}^i e_\bt{}^j.
\end{equation*}

Индуцированные метрика (\ref{eindme}) и связность (\ref{ecocou}) задают
внутреннюю геометрию гиперповерхности $\MU\hookrightarrow\MM$. В частности,
индуцированная связность задает тензор внутренней кривизны гиперповерхности
\begin{equation*}
  \hat R_{ijk}{}^l(\hat\Gamma)=\pl_i\hat\Gamma_{jk}{}^l
  -\hat\Gamma_{ik}{}^m\hat\Gamma_{jm}{}^l-(i\leftrightarrow j).
\end{equation*}

Вложение гиперповерхности $f:~\MU\hookrightarrow\MM$ позволяет определить еще
один важный объект, характеризующий то, как гиперповерхность $\MU$ вложена в
$\MM$.
\begin{defn}
Тензор второго ранга с компонентами
\begin{equation}                                                  \label{excuhy}
  K_{ij}:=-(\nb_\al n_\bt)e^\al{}_i e^\bt{}_j,
\end{equation}
определен на $\MU\subset\MM$ и называется {\em внешней кривизной}
гиперповерхности. Он равен ковариантной производной нормали, спроектированной на
касательное пространство к гиперповерхности и взятой с обратным знаком.
\qed\end{defn}
\index{Внешняя кривизна гиперповерхности (external curvature of hypersurface)}%
\index{Кривизна гиперповерхности внешняя (external curvature of hypersurface)}%
В отличии от тензора внутренней кривизны внешняя кривизна является тензором
второго, а не четвертого ранга и в общем случае никакой симметрии по индексам не
имеет. Этот тензор характеризует изменение нормали при ее параллельном переносе
вдоль кривой на гиперповерхности.

Раскрывая определение тензора внешней кривизны, с учетом равенства
(\ref{eortre}) получим, что
\begin{equation*}
  K_{ij}=n_\al(\pl^2_{ij} x^\al+\Gamma_{\bt\g}{}^\al e^\bt{}_i e^\g{}_j).
\end{equation*}
Отсюда следует, что антисимметричная часть тензора внешней кривизны определяется
тензором кручения
\begin{equation}                                                  \label{etonoc}
  K_{[ij]}=\frac12n_\al T_{\bt\g}{}^\al e^\bt{}_i e^\g{}_j=\frac12T_{ij}{}^\bot.
\end{equation}
Это доказывает следующее
\begin{prop}
Внешняя кривизна гиперповерхности $\MU\hookrightarrow\MM$ симметрична тогда и
только тогда, когда сужение кручения связности $\Gamma_{\bt\g}{}^\al$ на
$\MU\subset\MM$ удовлетворяет условию $T_{ij}{}^\bot=0$.
\end{prop}

Вычислим ковариантную производную от матрицы Якоби
\begin{equation}                                                  \label{ecojam}
  \nb_i e^\al{}_j=e^\bt{}_i(\pl_\bt e^\al{}_j+\Gamma_{\bt\g}{}^\al e^\g{}_j)
  -\hat\Gamma_{ij}{}^k e^\al{}_k.
\end{equation}
Разложение этого равенства (индекс $\al$) по базису $n,e_i$ показывает, что эта
ковариантная производная имеет только нормальную составляющую и пропорциональна
внешней кривизне:
\begin{equation}                                                  \label{enexcu}
  \nb_i e^\al{}_j=n^\al K_{ij}.
\end{equation}

Введем специальное обозначение для ``половины'' ковариантной производной
(\ref{ecojam}), которая содержит связность только для греческих индексов,
\begin{equation*}
  \check\nb_i e^\al{}_j:=e^\bt{}_i(\pl_\bt e^\al{}_j
  +\Gamma_{\bt\g}{}^\al e^\g{}_j).
\end{equation*}
Она ковариантна относительно преобразований координат $x^\al$ в $\MM$, но не на
$\MU$. Разлагая ее по базису $n,e_i$, получим равенство
\begin{equation}                                                  \label{egavei}
  \check\nb_i e^\al{}_j=n^\al K_{ij}+\hat\Gamma_{ij}{}^k e^\al{}_k.
\end{equation}
Это соотношение известно как {\em формула Гаусса--Вейнгартена}.
\index{Формула Гаусса--Вейнгартена (Gauss--Weingarten' formulae)}%
\index{Гаусса--Вейнгартена формула (Gauss--Weingarten' formulae)}%

Из равенств (\ref{enexcu}) и (\ref{egavei}) следует ещё одно представление
для тензора внешней кривизны
\begin{equation}                                                  \label{excuco}
  K_{ij}=n_\al\nb_i e^\al{}_j=n_\al\check\nb_i e^\al{}_j.
\end{equation}

Полный тензор кривизны $R_{\al\bt\g\dl}$, спроектированный на гиперповерхность,
можно выразить через тензор внутренней кривизны $\hat R_{ijkl}$, построенный
только по индуцированной метрике (\ref{eindme}) и связности (\ref{ecocou}),
и тензор внешней кривизны. Для этого рассмотрим коммутатор ковариантных
производных ковекторного поля, который определяется тензором кривизны и кручения
(\ref{eonfco})
\begin{equation}                                                  \label{ecoman}
\begin{split}
  [\nb_\al,\nb_\bt]X_\g&
  =-R_{\al\bt\g}{}^\dl X_\dl-T_{\al\bt}{}^\dl\nb_\dl X_\dl=
\\
  &=-R_{\al\bt\g i}X^i-R_{\al\bt\g\bot}X^\bot-T_{\al\bt}{}^i\nb_i X_\g
  -T_{\al\bt}{}^\bot\nb_\bot X_\g,
\end{split}
\end{equation}
где в правой части сначала вычисляются ковариантные производные, а затем
проектируются на гиперповерхность и ортогональное направления:
\begin{equation*}
  \nb_i X_\g:=e^\al{}_i\nb_\al X_\g,\qquad \nb_\bot X_\g:=n^\al\nb_\al X_\g.
\end{equation*}
Для того, чтобы спроектировать равенство (\ref{ecoman}) на гиперповерхность,
спроектируем сначала ковариантную производную:
\begin{equation*}
  \nb_i X_j:=e^\al{}_i(\nb_\al X_\bt)e^\bt{}_j=e^\al{}_i\nb_\al(X_\bt e^\bt{}_j)
  -e^\al{}_i X_\bt\nb_\al e^\bt{}_j=\hat\nb_i X_j-X_\bot K_{ij},
\end{equation*}
где мы воспользовались равенством (\ref{enexcu}) и
\begin{equation*}
  \hat\nb_i X_j:=\pl_i X_j-\hat\Gamma_{ij}{}^k X_k
\end{equation*}
-- $(n-1)$-мерная ковариантная производная на гиперповерхности. Аналогично
проектируется вторая ковариантная производная:
\begin{equation*}
\begin{split}
  \nb_i\nb_j X_k&=\hat\nb_i(\nb_j X_k)-\nb_\bot X_k K_{ij}-\nb_j X_\bot K_{ik}=
\\
  &=\hat\nb_i\hat\nb_j X_k-\hat\nb_iX_\bot K_{jk}-X_\bot\hat\nb_i K_{jk}
  -\nb_\bot X_k K_{ij}-\hat\nb_jX_\bot K_{ik}-X^lK_{lj}K_{ik},
\end{split}
\end{equation*}
где
\begin{equation*}
  \nb_j X_\bot:=e^\al{}_j(\nb_\al X_\bt)n^\bt
  =\hat\nb_j X_\bot+X^lK_{lj}.
\end{equation*}
Антисимметризация полученного выражения по индексам $i,j$ дает проекцию
коммутатора (\ref{ecoman}) на гиперповерхность:
\begin{equation*}
\begin{split}
  [\nb_i,\nb_j]X_k&=-R_{ijkl}X^l-R_{ijk\bot}X^\bot
  -T_{ij}{}^l\nb_lX_k-T_{ij}{}^\bot\nb_\bot X_k=
\\
  &=-R_{ijkl}X^l-R_{ijk\bot}X^\bot
  -T_{ij}{}^l\hat\nb_lX_k+T_{ij}{}^lX_\bot K_{lk}-T_{ij}{}^\bot\nb_\bot X_k.
\end{split}
\end{equation*}
Учитывая независимость компонент $X^l$ и $X^\bot$ и выражения для компонент
тензора кручения (\ref{eindto}), (\ref{etonoc}), получаем выражение для
проекций полного тензора кривизны на гиперповерхность:
\begin{align}                                                     \label{ecuhyc}
  R_{ijkl}&=\hat R_{ijkl}+K_{ik}K_{lj}-K_{jk}K_{li},
\\                                                                \label{ecunhy}
  R_{ijk\bot}&=\hat\nb_i K_{jk}-\hat\nb_j K_{ik}+T_{ij}{}^lK_{lk}.
\end{align}
Полученные соотношения являются обобщением
{\em уравнений Гаусса--Петерсона--Кодацци} на случай, когда на объемлющем
многообразии $\MM$ задана не риманова геометрия, а произвольная аффинная
геометрия с ненулевым кручением и неметричностью.
\index{Уравнения Гаусса--Петерсона--Кодацци %
(Gauss--Peterson--Codazzi' equations)}%
\index{Гаусса--Петерсона--Кодацци уравнения %
(Gauss--Peterson--Codazzi' equations)}%

В заключение настоящего раздела вычислим нормальные компоненты $G_{\bot\bot}$
и $G_{\bot i}$ тензора Эйнштейна
\begin{equation*}
  G_{\al\bt}:=\widetilde R_{\al\bt}-\frac12g_{\al\bt}\widetilde R,
\end{equation*}
которые играют важную роль при анализе уравнений движения общей теории
относительности. Знак тильды в правой части, как и ранее, означает, что кручение
и неметричность связности, заданной на $\MM$, равны нулю. В этом случае тензор
кривизны обладает дополнительной симметрией относительно перестановки индексов
(см.\ раздел \ref{scurpr}) и тензор внешней кривизны симметричен:
$K_{ij}=K_{ji}$.

Ниже, для простоты, мы опустим знак тильды.

Сначала вычислим скалярную кривизну
\begin{equation*}
  R=g^{\al\g}g^{\bt\dl}R_{\al\bt\g\dl}=2R_{\bot\bot}+g^{ik}g^{jl}R_{ijkl},
\end{equation*}
где $R_{\bot\bot}=g^{ij}R_{i\bot j\bot}$ -- нормальная составляющая тензора
Риччи и учтено представление для обратной метрики (\ref{einmer}). С учетом
уравнений Гаусса--Петерсона--Кодацци (\ref{ecuhyc}) получаем равенство
\begin{equation*}
  g^{ik}g^{jl}R_{ijkl}=\hat R+K^2-K^{ij}K_{ij},
\end{equation*}
где $\hat R$ -- скалярная внутренняя кривизна гиперповерхности,
$K:=g^{ij}K_{ij}$ -- скалярная внешняя кривизна гиперповерхности.
Отсюда следуют выражения для нормальных компонент тензора Эйнштейна:
\begin{equation}                                                  \label{einbom}
\begin{split}
  G_{\bot\bot}&=-\frac12(\hat R+K^2-K^{ij}K_{ij}),
\\
  G_{\bot i}&=\hat\nb_jK_i{}^j-\hat\nb_iK.
\end{split}
\end{equation}
Важным обстоятельством является то, что эти компоненты тензора Эйнштейна
вообще не содержат нормальных производных к гиперповерхности $\nb_\bot$ от
индуцированной метрики и тензора внешней кривизны. На гамильтоновом языке
это означает отсутствие производных по времени, и что вакуумные уравнения
Эйнштейна $G_{\bot\bot}=0$ и $G_{\bot i}=0$ представляют собой связи, поскольку
компоненты тензора внешней кривизны $K^{ij}$, как будет показано в следующем
разделе, пропорциональны импульсам, канонически сопряженным компонентам
индуцированной метрики $g_{ij}$.
\section{Кривизна в АДМ параметризации метрики}
АДМ параметризация метрики (\ref{eadmme}) удобна для канонической формулировки
общей теории относительности, в которой исходными независимыми переменными
являются компоненты метрики $g_{\al\bt}$ (обобщенные координаты) и канонически
сопряженные импульсы $p^{\al\bt}$. Однако, чтобы перейти от лагранжиана к
гамильтониану, необходимо произвести довольно громоздкие вычисления, чем мы и
займемся в настоящем разделе.

Для существенного упрощения вычислений следует воспользоваться результатами
раздела \ref{sgesum}, в котором описана геометрия гиперповерхностей. Будем
считать, что пространство-время является топологическим произведением
$\MM\approx\MR\times\MU$, где первый сомножитель соответствует времени $x^0$.
Тогда сечения $x^0=\const$ пространства-времени $\MM$ задают семейство
гиперповерхностей $\MU\subset\MM$, которые, по-предположению, являются
пространственноподобными. В качестве координат на гиперповерхностях выберем
пространственные координаты
\begin{equation*}
  \lbrace u^i\rbrace \mapsto\lbrace x^\mu\rbrace.
\end{equation*}
При этом мы теряем возможность независимого преобразования координат в
пространстве-времени $\MM$ и пространственноподобной гиперповерхности $\MU$,
зато многие формулы упрощаются.

Матрица Якоби вложения гиперповерхности в рассматриваемом случае имеет вид
\begin{equation*}
  \lbrace e^\al{}_i\rbrace\mapsto\lbrace 0_\nu,\dl^\mu_\nu\rbrace,\qquad
  \lbrace e_\al{}^i\rbrace\mapsto\lbrace N^\mu,\dl^\mu_\nu\rbrace,
\end{equation*}
где $0_\nu$ обозначает строку, состоящую из $n-1$ нулей. Вложение индуцирует на
гиперповерхностях метрику $g_{\mu\nu}$ согласно формуле (\ref{eindme}).

Построим векторное поле $n=n^\al\pl_\al$, ортогональное семейству
пространственноподобных гиперповерхностей $x^0=\const$. Из условия
ортогональности $(n,X)=0$, где $X=X^\mu\pl_\mu$ -- произвольный вектор,
касательный к сечению $x^0=0$, следует, что
\begin{equation*}
  n=n^0(\pl_0-N^\mu\pl_\mu),
\end{equation*}
где $n^0$ -- произвольная отличная от нуля функция. Если, кроме того, положить
$n^0=1/N$, то длина нормального вектора будет равна единице, $n^2=1$. Таким
образом, единичный вектор, нормальный к сечениям $x^0=\const$ имеет вид
\begin{equation}                                                  \label{enovei}
  n=\frac1N(\pl_0-N^\mu\pl_\mu)
\end{equation}
и всегда является времениподобным. Соответствующая ортонормальная 1-форма имеет
вид
\begin{equation}                                                  \label{eoronf}
  n=dx^0N.
\end{equation}

Произвольный тензор на $\MM$ можно разложить по базису $\lbrace n,e_\mu\rbrace$.
В частности, для векторов и 1-форм справедливы разложения
\begin{equation*}
  X^\al=X^\bot n^\al+\tilde X^\mu e_\mu{}^\al,\qquad
  X_\al=X_\bot n_\al+\tilde X_\mu e^\mu{}_\al,
\end{equation*}
где
\begin{align*}
  X^\bot&=X^0N, & \tilde X^\mu&=X^0 N^\mu+X^\mu,
\\
  X_\bot&=\frac1N(X_0-N^\mu X_\mu), & \tilde X_\mu&=X_\mu.
\end{align*}

Представления для метрики (\ref{emetde}) всего пространства-времени и ее
обратной (\ref{einmer}) принимает вид
\begin{equation}                                                  \label{emettr}
\begin{split}
  g_{\al\bt}&=n_\al n_\bt+g_{\mu\nu}e_\al{}^\mu e_\bt{}^\nu,
\\
  g^{\al\bt}&=n^\al n^\bt+\hat g^{\mu\nu}e_\mu{}^\al e_\nu{}^\bt.
\end{split}
\end{equation}

Связность (\ref{ecocou}), индуцированная на гиперповерхностях -- это символы
Кристоффеля $\hat\Gamma_{\mu\nu}{}^\rho$, построенные по пространственной метрике
$g_{\mu\nu}$.

Тензор внешней кривизны (\ref{excuhy}) гиперповерхности $x^0=\const$ в АДМ
параметризации метрики имеет вид
\begin{equation}                                                  \label{excurv}
  K_{\mu\nu}=\Gamma_{\mu\nu}{}^0N=\frac1{2N}(\hat\nb_\mu N_\nu
  +\hat\nb_\nu N_\mu-\dot g_{\mu\nu}),
\end{equation}
где точка обозначает дифференцирование по времени,
$\dot g_{\mu\nu}:=\pl_0 g_{\mu\nu}$, и
$\hat\nb_\mu N_\nu:=\pl_\mu N_\nu-\hat\Gamma_{\mu\nu}{}^\rho N_\rho$ --
пространственная ковариантная производная. Тензор внешней кривизны симметричен,
$K_{\mu\nu}=K_{\nu\mu}$, поскольку кручение в общей теории относительности равно
нулю. В дальнейшем нам понадобится также след тензора внешней кривизны
\begin{equation*}
  K:=K_\mu{}^\mu=\hat g^{\mu\nu}K_{\mu\nu}
  =\frac1{2N}(2\hat\nb_\mu N^\mu-\hat g^{\mu\nu}\dot g_{\mu\nu}).
\end{equation*}
При вычислении тензора кривизны $R_{\al\bt\g\dl}$ пространства-времени $\MM$ все
производные по времени от пространственной части метрики $\dot g_{\mu\nu}$
удобно выражать через $K_{\mu\nu}$. Кроме этого, для исключения вторых
производных по времени $\ddot g_{\mu\nu}$ нам понадобится производная по времени
от тензора внешней кривизны,
\begin{equation*}
  \dot K_{\mu\nu}=\frac1{2N}\left[\hat\nb_\mu\dot N_\nu
  +\hat\nb_\nu\dot N_\mu-\ddot g_{\mu\nu}-N^\rho(\hat\nb_\mu\dot g_{\nu\rho}
  +\hat\nb_\nu\dot g_{\mu\rho}-\hat\nb_\rho\dot g_{\mu\nu})\right]
  -\frac{\dot N}NK_{\mu\nu},
\end{equation*}
где
\begin{equation*}
  \hat\nb_\mu\dot N_\nu=\pl_\mu\dot N_\nu-\hat\Gamma_{\mu\nu}{}^\rho\dot N_\rho
\end{equation*}
и куда при вычислениях мы должны подставить выражение для $\dot g_{\mu\nu}$
через $K_{\mu\nu}$.

Приступим к вычислению тензора кривизны $R_{\al\bt\g\dl}$ для метрики,
записанной в виде (\ref{eadmme}). Первый шаг состоит в вычислении символов
Кристоффеля (\ref{echris}). Прямые выкладки приводят к следующим выражениям для
линейно независимых символов Кристоффеля
\begin{equation}                                                  \label{eadmch}
\begin{split}
  \Gamma_{00}{}^0&=\frac1N\left(\dot N+N^\rho\pl_\rho N
  +N^\rho N^\s K_{\rho\s}\right),
\\
  \Gamma_{00}{}^\mu&=\hat g^{\mu\nu}(\dot N_\nu-N\pl_\nu N
  -N^\rho\hat\nb_\nu N_\rho)
  -\frac{N^\mu}N\left(\dot N+N^\rho\pl_\rho N+N^\rho N^\s K_{\rho\s}\right),
\\
  \Gamma_{0\mu}{}^0&=\frac1N\left(\pl_\mu N+N^\nu K_{\mu\nu}\right),
\\
  \Gamma_{0\mu}{}^\nu&=\hat\nb_\mu N^\nu-NK_\mu{}^\nu-\frac{N^\nu}N
  (\pl_\mu N+N^\rho K_{\mu\rho}),
\\
  \Gamma_{\mu\nu}{}^0&=\frac1NK_{\mu\nu},
\\
  \Gamma_{\mu\nu}{}^\rho&=\hat\Gamma_{\mu\nu}{}^\rho
  -\frac{N^\rho}N K_{\mu\nu}.
\end{split}
\end{equation}
В дальнейшем нам понадобятся два независимых следа символов Кристоффеля:
\begin{equation*}
  \Gamma_\al:=\Gamma_{\al\bt}{}^\bt,\qquad \Xi^\al:=g^{\bt\g}\Gamma_{\bt\g}{}^\al.
\end{equation*}
Несложные вычисления приводят к формулам
\begin{equation}                                                  \label{eadmlc}
\begin{split}
  \Gamma_0&=\frac{\dot N}N+\hat\nb_\mu N^\mu-NK,
\\
  \Gamma_\mu&=\hat\Gamma_\mu+\frac{\pl_\mu N}N,
\\
  \Xi^0&=\frac1NK+\frac1{N^3}(\dot N-N^\mu\pl_\mu N),
\\
  \Xi^\mu&=\left(\hat g^{\rho\s}
  +\frac{N^\rho N^\s}{N^2}\right)\hat\Gamma_{\rho\s}{}^\mu-\frac{N^\mu}N K
  -\frac{N^\mu}{N^3}(\dot N-N^\rho\pl_\rho N)+
\\
  &+\frac1{N^2}\hat g^{\mu\rho}(\dot N_\rho-N\pl_\rho N
  -N^\s\hat\nb_\rho N_\s-2N^\s\hat\nb_\s N_\rho+2NN^\s K_{\rho\s}).
\end{split}
\end{equation}
Выпишем также формулы дифференцирования по времени для символов Кристоффеля:
\begin{equation}                                                  \label{echtid}
\begin{split}
  \pl_0\hat\Gamma_{\mu\nu\rho}&=
  \frac12(\hat\nb_\mu\dot g_{\nu\rho}+\hat\nb_\nu\dot g_{\mu\rho}
  -\hat\nb_\rho\dot g_{\mu\nu})+\hat\Gamma_{\mu\nu}{}^\s\dot g_{\rho\s},
\\
  \pl_0\hat\Gamma_{\mu\nu}{}^\s&=
  \frac12\hat g^{\s\rho}(\hat\nb_\mu\dot g_{\nu\rho}
  +\hat\nb_\nu\dot g_{\mu\rho}-\hat\nb_\rho\dot g_{\mu\nu}),
\\
  \pl_0\hat\Gamma_\mu&=\frac12\hat g^{\nu\rho}\hat\nb_\mu\dot g_{\nu\rho}.
\end{split}
\end{equation}
В этих выражениях производные по времени $\dot g_{\mu\nu}$ также исключаются
с помощью соотношения (\ref{excurv}).

Теперь вычислим линейно независимые компоненты тензора кривизны:
\begin{equation}                                                  \label{eadmcu}
\begin{split}
  R_{0\mu0\nu}=&-N\dot K_{\mu\nu}+\hat R_{\mu\rho\nu\s}N^\rho N^\s
  +NN^\rho(\hat\nb_\mu K_{\nu\rho}+\hat\nb_\nu K_{\mu\rho}
  -\hat\nb_\rho K_{\mu\nu})+
\\
  &+N\hat\nb_\mu\hat\nb_\nu N
  +K_{\mu\nu}N^\rho N^\s K_{\rho\s}+N(K_\mu{}^\rho\hat\nb_\nu N_\rho
  +K_\nu{}^\rho\hat\nb_\mu N_\rho)-
\\
  &-N^2K_\mu{}^\rho K_{\nu\rho}-N^\rho N^\s K_{\mu\rho}K_{\nu\s},
\\
  R_{\mu\nu\rho0}=&\hat R_{\mu\nu\rho\s}N^\s
  +N(\hat\nb_\mu K_{\nu\rho}-\hat\nb_\nu K_{\mu\rho})
  +(K_{\mu\rho}K_{\nu\s}-K_{\nu\rho}K_{\mu\s})N^\s,
\\
  R_{\mu\nu\rho\s}=&\hat R_{\mu\nu\rho\s}
  +K_{\mu\rho}K_{\nu\s}-K_{\mu\s}K_{\nu\rho},
\end{split}
\end{equation}
где мы воспользовались формулой для коммутатора ковариантных производных:
\begin{equation*}
  (\hat\nb_\mu\hat\nb_\nu-\hat\nb_\nu\hat\nb_\mu)N_\rho
  =-\hat R_{\mu\nu\rho\s}N^\s.
\end{equation*}

Компоненты тензора кривизны, имеющие по крайней мере один временн\'ой индекс,
относительно ортонормального базиса $n,e_\mu$ выглядят проще:
\begin{equation}                                                  \label{ecuorn}
\begin{split}
  R_{\bot\mu\bot\nu}&=\frac1N\left(-\dot K_{\mu\nu}+\hat\nb_\mu\hat\nb_\nu N
  +K_{\mu\rho}\hat\nb_\nu N^\rho+K_{\nu\rho}\hat\nb_\mu N^\rho
  -NK_{\mu\rho}K_\nu{}^\rho+N^\rho\hat\nb_\rho K_{\mu\nu}\right),
\\
  R_{\mu\nu\rho\bot}&=\hat\nb_\mu K_{\nu\rho}-\hat\nb_\nu K_{\mu\rho}.
\end{split}
\end{equation}

Фактически, компоненты тензора кривизны $R_{\mu\nu\rho\s}$ и
$R_{\mu\nu\rho\bot}$ уже были получены в предыдущем разделе, см.\ формулы
(\ref{ecuhyc}), (\ref{ecunhy}), без прямых вычислений.

Тензор Риччи имеет следующие линейно независимые компоненты:
\begin{equation}                                                  \label{eadmri}
\begin{split}
  R_{00}=&-N\hat g^{\mu\nu}\dot K_{\mu\nu}+\hat R_{\mu\nu}N^\mu N^\nu
  +NN^\mu(2\hat\nb_\nu K^\nu{}_\mu-\pl_\mu K)+N\hat\nb^\mu\hat\nb_\mu N+
\\
  &+N^\mu N^\nu K_{\mu\nu}K+2NK^{\mu\nu}\hat\nb_\mu N_\nu
  -N^2K^{\mu\nu}K_{\mu\nu}-2N^\mu N^\nu K_\mu{}^\rho K_{\nu\rho}+
\\
  &+\frac{N^\mu N^\nu}N\left(-\dot K_{\mu\nu}+N^\rho\hat\nb_\mu K_{\nu\rho}
  +\hat\nb_\mu\hat\nb_\nu N+2K_\mu{}^\rho\hat\nb_\nu N_\rho\right),
\\
  R_{0\mu}=&\frac{N^\nu}N\left(-\dot K_{\mu\nu}+N^\s\hat\nb_\nu K_{\mu\s}
  +\hat\nb_\mu\hat\nb_\nu N+K_\mu{}^\rho\hat\nb_\nu N_\rho
  +K_\nu{}^\rho\hat\nb_\mu N_\rho\right)
\\
  &+\hat R_{\mu\nu}N^\nu+N\left(\hat\nb_\nu K^\nu{}_\mu-\pl_\mu K\right)
  +K_{\mu\nu}N^\nu K-2K_\mu{}^\rho K_{\nu\rho}N^\nu,
\\
  R_{\mu\nu}=&\hat R_{\mu\nu}+\frac1N\left(-\dot K_{\mu\nu}
  +\hat\nb_\mu\hat\nb_\nu N
  +K_\mu{}^\rho\hat\nb_\nu N_\rho+K_\nu{}^\rho\hat\nb_\mu N_\rho
  +N^\rho\hat\nb_\rho K_{\mu\nu}\right)+
\\
  &+K_{\mu\nu}K-2K_\mu{}^\rho K_{\nu\rho}.
\end{split}
\end{equation}
Опять же, компоненты тензора Риччи относительно базиса $n,e_\mu$ существенно
проще:
\begin{equation}                                                  \label{ericor}
\begin{split}
  R_{\bot\bot}=&-\frac1N\hat g^{\mu\nu}\dot K_{\mu\nu}
  +\frac1N\hat\nb^\mu\hat\nb_\mu N+\frac2N K^{\mu\nu}\hat\nb_\mu N_\nu
  -K^{\mu\nu}K_{\mu\nu}+\frac{N^\mu}N\pl_\mu K,
\\
  R_{\bot\mu}=&\hat\nb_\nu K^\nu{}_\mu-\pl_\mu K.
\end{split}
\end{equation}

Далее, вычисляем скалярную кривизну
\begin{equation}                                                  \label{eadmsc}
\begin{split}
  R&=\hat R
  +\frac2N \left(-\hat g^{\mu\nu}\dot K_{\mu\nu}+\hat\nb^\mu\hat\nb_\mu N
  +2K^{\mu\nu}\hat\nb_\mu N_\nu+N^\mu\pl_\mu K\right)-3K^{\mu\nu}K_{\mu\nu}+K^2.
\end{split}
\end{equation}

Вакуумные уравнения Эйнштейна
\begin{equation*}
  G_{\al\bt}:=R_{\al\bt}-\frac12g_{\al\bt}R=0
\end{equation*}
также можно переписать в базисе $n,e^\mu$. Выражения для $G_{\bot\bot}$ и
$G_{0\bot}$ уже были получены в предыдущем разделе менее трудоемким образом
(\ref{einbom}). Выражение для $G_{\mu\nu}$ в АДМ параметризации метрики следует
из выражения для тензора Риччи (\ref{eadmri}) и скалярной кривизны
(\ref{eadmsc}). Оно громоздко и мы не будем его выписывать, т.к.\ в дальнейшем
оно нам не понадобится.

Как уже отмечалось, тензор Эйнштейна удовлетворяет свернутым тождествам
Бианки (\ref{ebieit}). Эти тождества полезно переписать в АДМ параметризации
метрики. Четыре тождества (\ref{ebieit}) перепишем в следующем виде:
\begin{equation*}
  \nb_\bt G^{\bt\al}=0\quad \Leftrightarrow\quad
  (\nb_\bt G^{\bt\al})n_\al=0,\quad (\nb_\bt G^{\bt\al})e_\al{}^\mu=0.
\end{equation*}
После довольно утомительных вычислений получим равенства:
\begin{align}                                                     \label{ecobio}
 (\nb_\bt G^{\bt\al})n_\al&=\frac1N\dot G^{\bot\bot}
 -\frac{N^\mu}N\pl_\mu G^{\bot\bot}-G^{\bot\bot}K+\hat\nb_\mu G^{\mu\bot}
 +2\frac{\pl_\mu N}NG^{\mu\bot}+G^{\mu\nu}K_{\mu\nu},
\\                                                                     \nonumber
  (\nb_\bt G^{\bt\al})e_\al{}^\mu&=\frac1N\dot G^{\bot\mu}
  -\frac{N^\nu}N\hat\nb_\nu G^{\bot\mu}-G^{\bot\mu}K
  +\hat\nb_\nu G^{\nu\mu}+\frac{\pl_\nu N}NG^{\nu\mu}-
\\                                                                \label{ecobis}
  &-\frac1N G^{\bot\bot}\hat g^{\mu\nu}\pl_\nu N
  +\frac1N G^{\bot\nu}\hat\nb_\nu N^\mu-2G^{\bot\nu}K_\nu{}^\mu.
\end{align}
Эти тождества будут полезны при анализе вторичных связей в общей теории
относительности.
\section{Гамильтониан}
Скалярная кривизна содержит вторые производные от компонент метрики как по
времени, так и по пространственным координатам и поэтому неудобна для
канонической формулировки общей теории относительности. Чтобы исключить из
лагранжиана все вторые производные, в разделе \ref{shieil} к действию был
добавлен граничный член. Однако для канонической формулировки достаточно
исключить из лагранжиана только вторые производные по времени. Наиболее
простой вид лагранжиан принимает после добавления следующего граничного члена
\begin{equation}                                                  \label{ehiela}
  L_{\Sa\Sd\Sm}=N\hat eR
  +2\pl_0\left(\hat eK\right)
  -2\pl_\mu\left(\hat e\hat g^{\mu\nu}\pl_\nu N\right),
\end{equation}
где
\begin{equation*}
  \hat e:=\det\hat e_\mu{}^i=\sqrt{|\hat g|}=\sqrt{|\det g_{\mu\nu}|}.
\end{equation*}
Прямые вычисления приводят к следующему простому выражению
\begin{equation}                                                  \label{eadmla}
  L_{\Sa\Sd\Sm}=N\hat e\left(K^{\mu\nu}K_{\mu\nu}-K^2+\hat R\right).
\end{equation}

Теперь нетрудно перейти к гамильтонову формализму. Во-первых, АДМ-лагранжиан не
содержит производных по времени от функции хода $N$ и функций сдвига $N_\mu$.
Это значит, что теория содержит $n$ первичных связей:
\begin{equation}                                                  \label{eprcge}
  p^0:=\frac{\pl L_{\Sa\Sd\Sm}}{\pl\dot N}=0,\qquad
  p^\mu:=\frac{\pl L_{\Sa\Sd\Sm}}{\pl\dot N_\mu}=0,
\end{equation}
число которых совпадает с числом независимых функций, параметризующих
диффеоморфизмы. Импульсы, канонически сопряженные к пространственной метрике
$g_{\mu\nu}$, имеют вид
\begin{equation}                                                  \label{emomet}
  p^{\mu\nu}=\frac{\pl L_{\Sa\Sd\Sm}}{\pl\dot g_{\mu\nu}}
  =-\frac1{2N}\frac{\pl L_{\Sa\Sd\Sm}}{\pl K_{\mu\nu}}
  =-\hat e\left(K^{\mu\nu}-g^{\mu\nu}K\right),
\end{equation}
где мы воспользовались выражением для внешней кривизны (\ref{excurv}). Отсюда
следует, что обобщенные импульсы пропорциональны тензору внешней кривизны.
Отметим, что импульсы являются не тензорами относительно преобразований
координат $x^\mu$, а тензорными плотностями степени $-1$, как и определитель
репера, степень которого, по-определению, равна
\begin{equation*}
  \deg \hat e=-1.
\end{equation*}
Поэтому ковариантная производная от импульсов в соответствии с правилом
ковариантного дифференцирования (\ref{ecodtd}) имеет вид
\begin{equation*}
  \hat\nb_\mu p^{\nu\rho}=\pl_\mu p^{\nu\rho}+\hat\Gamma_{\mu\s}{}^\nu p^{\s\rho}
  +\hat\Gamma_{\mu\s}{}^\rho p^{\nu\s}-\hat\Gamma_\mu p^{\nu\rho}.
\end{equation*}

Чтобы исключить из АДМ-лагранжиана скорости $\dot g_{\mu\nu}$, разложим импульсы
на неприводимые компоненты, выделив из $p^{\mu\nu}$ след,
\begin{equation}                                                  \label{eirmom}
  p^{\mu\nu}=\tilde p^{\mu\nu}+\frac1{n-1}p\hat g^{\mu\nu},
\end{equation}
где мы ввели след импульсов
\begin{equation}                                                  \label{exptrm}
  p:=p^{\mu\nu}g_{\mu\nu}=\hat e(n-2)K
\end{equation}
и симметричную бесследовую часть
\begin{equation*}
  \tilde p^{\mu\nu}:=\tilde p^{\nu\mu}
  =-\hat e\left(K^{\mu\nu}-\frac1{n-1}\hat g^{\mu\nu}K\right),
  \qquad \tilde p^{\mu\nu}g_{\mu\nu}=0.
\end{equation*}

Импульсы $p^{\mu\nu}$ и внешняя кривизна $K^{\mu\nu}$ находятся во взаимно
однозначном соответствии. Из формулы (\ref{emomet}) следует выражение внешней
кривизны через импульсы
\begin{equation}                                                  \label{extmom}
  K^{\mu\nu}=-\frac1{\hat e}\left(p^{\mu\nu}-\frac1{n-2}g^{\mu\nu}p\right).
\end{equation}

Теперь можно решить уравнение (\ref{emomet}) относительно скоростей,
воспользовавшись соотношением (\ref{excurv}),
\begin{equation*}
  \dot g_{\mu\nu}=\frac{2N}{\hat e}\left(p_{\mu\nu}
  -\frac1{(n-2)}pg_{\mu\nu}\right)+\hat\nb_\mu N_\nu+\hat\nb_\nu N_\mu.
\end{equation*}
Несложные вычисления приводят к гамильтоновой плотности
\begin{equation}                                                  \label{ehawid}
  H=p^{\mu\nu}\dot g_{\mu\nu}-L_{\Sa\Sd\Sm}
  =NH_\bot+N^\mu H_\mu+2\pl_\mu(p^{\mu\nu}N_\nu),
\end{equation}
где
\begin{equation}                                                  \label{ecoadm}
\begin{split}
  H_\bot&:=\frac1{\hat e}\left(p^{\mu\nu}p_{\mu\nu}-\frac1{(n-2)}p^2\right)
  -\hat e\hat R,
\\
  H_\mu&:=-2\hat\nb_\nu p^\nu{}_\mu=-2\pl_\nu p^\nu{}_\mu
  +\pl_\mu g_{\nu\rho}p^{\nu\rho},
\end{split}
\end{equation}
и $p_{\mu\nu}:=g_{\mu\rho}g_{\nu\s}p^{\rho\s}$.
\begin{com}
Ковариантная производная от импульсов в последнем выражении содержит только одно
слагаемое с символами Кристоффеля, так как импульсы являются тензорными
плотностями.
\qed\end{com}

Отбрасывая в выражении для гамильтониана (\ref{ehawid}) дивергенцию, приходим к
окончательному выражению для гамильтониана
\begin{equation}                                                  \label{eadmha}
  \CH_{\Sa\Sd\Sm}=\int d\Bx H_{\Sa\Sd\Sm}=\int d\Bx(NH_\bot+N^\mu H_\mu).
\end{equation}
Для функционалов, полученных интегрированием по пространственным сечениям
$x^0=\const$, будем использовать каллиграфический шрифт.

Перепишем выражение для $H_\bot$ в терминах неприводимых компонент импульсов
(\ref{eirmom}):
\begin{equation*}
  H_\bot=\frac1{\hat e}\left[\tilde p^{\mu\nu}\tilde p_{\mu\nu}
  -\frac1{(n-1)(n-2)}p^2\right]-\hat e\hat R.
\end{equation*}
Отсюда следует, что квадратичная форма импульсов в $H_\bot$ при $n\ge3$ не
является положительно определенной. Отсутствие положительной определенности
квадратичной формы импульсов в гамильтониане представляет серьезные трудности
для физической интерпретации моделей. В рассматриваемом случае не все компоненты
импульсов являются физическими, т.к.\ общая теория относительности инвариантна
относительно общих преобразований координат. После исключения нефизических
степеней свободы с помощью связей и калибровочных условий квадратичная форма
импульсов для физических степеней свободы будет положительно определена.
\section{Вторичные связи                                         \label{salcog}}
Для завершения построения гамильтонова формализма необходимо исследовать
согласованность первичных связей (\ref{eprcge}) с уравнениями движения. Фазовое
пространство общей теории относительности бесконечномерно и описывается $n(n+1)$
сопряженными координатами и импульсами: $(N,p^0)$, $(N_\mu,p^\mu)$,
$(g_{\mu\nu},p^{\mu\nu})$, в каждой точке пространства $\Bx\in\MU$. На этом
фазовом пространстве задана каноническая пуассонова структура с помощью
одновременн\'ых скобок Пуассона:
\begin{equation}                                                  \label{epoisp}
  [N,p^{\prime0}]=\dl(\Bx-\Bx'),\qquad
  [N_\mu,p^{\prime\nu}]=\dl_\mu^\nu\dl(\Bx-\Bx'),\qquad
  [g_{\mu\nu},p^{\prime\rho\s}]=\dl_{\mu\nu}^{\rho\s}\dl(\Bx-\Bx'),
\end{equation}
где штрих у полевой переменной означает, что она рассматривается в точке
$\Bx'=(x^{\prime1},\dotsc,x^{\prime n-1})$. Все поля рассматриваются в момент
времени $t=x^0$. В правых частях скобок Пуассона для краткости использованы
следующие обозначения для пространственной $(n-1)$-мерной $\dl$-функции и
симметризованной комбинации символов Кронекера:
\begin{equation}                                                  \label{epoino}
\begin{split}
  \dl(\Bx-\Bx')&
  :=\dl(x^1-x^{\prime1})\dotsc\dl(x^{n-1}-x^{\prime n-1}),
\\
  \dl_{\mu\nu}^{\rho\s}
  &:=\frac12(\dl_\mu^\rho\dl_\nu^\s+\dl_\mu^\s\dl_\nu^\rho).
\end{split}
\end{equation}

Если заданы два функционала
\begin{equation*}
  \CF=\int d\Bx\, F,\qquad \CG=\int d\Bx\, G,
\end{equation*}
где $F$ и $G$ -- некоторые функции от пространственной метрики $g_{\mu\nu}$,
сопряженных импульсов $p^{\mu\nu}$ и их производных, то скобка Пуассона этих
функционалов определяется следующим интегралом
\begin{equation}                                                  \label{epoibf}
  [\CF,\CG]:=\int d\Bx\left(
  \frac{\dl\CF}{\dl g_{\mu\nu}}\frac{\dl G}{\dl p^{\mu\nu}}
  -\frac{\dl\CF}{\dl p^{\mu\nu}}\frac{\dl G}{\dl g_{\mu\nu}}\right).
\end{equation}
Напомним, что сами компоненты метрики и импульсов также можно рассматривать
в виде функционалов. Например,
\begin{equation*}
  g_{\mu\nu}(\Bx)=\int d\Bx' g_{\mu\nu}(\Bx')\dl(\Bx-\Bx').
\end{equation*}
Тогда последняя скобка Пуассона в (\ref{epoisp}) является следствием общего
определения канонической пуассоновой структуры на множестве функционалов
(\ref{epoibf}).

Аналогично определяется скобка Пуассона для функционалов, зависящих от полного
набора канонических переменных: $(N,p^0)$, $(N_\mu,p^\mu)$,
$(g_{\mu\nu},p^{\mu\nu})$.

Рассмотрим теперь гамильтоновы уравнения движения для первичных связей
(\ref{eprcge}):
\begin{equation*}
\begin{split}
  \dot p^0&=[p^0,\CH_{\Sa\Sd\Sm}]=-\frac{\dl\CH_{\Sa\Sd\Sm}}{\dl N}=-H_\bot,
\\
  \dot p^\mu&=[p^\mu,\CH_{\Sa\Sd\Sm}]
  =-\frac{\dl\CH_{\Sa\Sd\Sm}}{\dl N^\mu}=-H^\mu.
\end{split}
\end{equation*}
Из условия согласованности первичных связей с уравнениями движения $\dot p^0=0$,
$\dot p^\mu=0$ следуют вторичные связи:
\begin{equation}                                                  \label{esecco}
  H_\bot=0,\qquad H_\mu=0.
\end{equation}
Отметим, что вторичные связи являются не тензорами, а тензорными плотностями
степени $-1$. Кроме того, вместо связей $H^\mu$ удобнее рассматривать
эквивалентную систему связей с опущенным индексом $H_\mu:=g_{\mu\nu}H^\nu$. В
следующем разделе будет показано, что эти связи определяют генераторы
преобразований координат на сечениях $x^0=\const$ и удовлетворяют более простой
алгебре.

Связи $H_\mu$ линейны по импульсам и метрике. Связь $H_\bot$ квадратична по
импульсам и неполиномиальна по метрике $g_{\mu\nu}$, поскольку зависит от корня
из определителя метрики $\hat e$ и обратной метрики $\hat g^{\mu\nu}$. Последнее
обстоятельство является существенной технической трудностью при построении
теории возмущений.

Вторичные связи не зависят от  канонических переменных $(N,p^0)$,
$(N_\mu,p^\mu)$, и их можно исключить, рассматривая
$n(n-1)\times\infty^{n-1}$-мерное фазовое пространство переменных $g_{\mu\nu}$ и
$p^{\mu\nu}$, на которые наложены связи (\ref{esecco}). В этом случае функция
хода $N$ и функции сдвига $N_\mu$ рассматриваются, как множители Лагранжа в
задаче на условный экстремум для действия
\begin{equation*}
  S=\int dx(p^{\mu\nu}\dot g_{\mu\nu}-H_{\Sa\Sd\Sm}).
\end{equation*}

Поскольку гамильтониан общей теории относительности (\ref{eadmha}) представляет
собой линейную комбинацию связей, то для исследования согласованности вторичных
связей (\ref{esecco}) с уравнениями движения необходимо вычислить скобки
Пуассона связей между собой. Алгебра связей в общей теории относительности
хорошо известна:
\begin{align}                                                     \label{econoo}
  [H_\bot,H'_\bot]&=-(H_\mu\hat g^{\mu\nu}+H'_\mu\hat g^{\prime\mu\nu})\dl_\nu,
\\                                                                \label{econom}
  [H_\bot,H'_\mu]&=-H'_\bot\dl_\mu,
\\                                                                \label{econmn}
  [H_\mu,H'_\nu]&=-H_\nu\dl_\mu-H'_\mu\dl_\nu,
\end{align}
где введено сокращенное обозначение для производной $\dl$-функции:
\begin{equation*}
  \dl_\mu:=\frac\pl{\pl x^{\prime\mu}}\dl(\Bx'-\Bx).
\end{equation*}

Вид двух скобок Пуассона (\ref{econom}) и (\ref{econmn}) можно
найти, не прибегая к прямым вычислениям. С этой целью рассмотрим функционал
\begin{equation*}
  T_u:=-\int d\Bx~u^\mu H_\mu,
\end{equation*}
где $u^\mu$ малое дифференцируемое векторное поле, которое мы рассматриваем как
инфинитезимальный параметр преобразования. Вычисление скобок Пуассона координат
фазового пространства $g_{\mu\nu}$, $p^{\mu\nu}$ с $T_u$ приводит к равенствам:
\begin{equation*}
\begin{split}
  \dl_u g_{\mu\nu}&:=[g_{\mu\nu},T_u]=-\pl_\mu u^\rho g_{\rho\nu}
  -\pl_\nu u^\rho g_{\mu\rho}-u^\rho\pl_\rho g_{\mu\nu},
\\
  \dl_u p^{\mu\nu}&:=[p^{\mu\nu},T_u]=\pl_\rho u^\mu p^{\rho\nu}
  +\pl_\rho u^\nu p^{\mu\rho}-\pl_\rho(u^\rho p^{\mu\nu}).
\end{split}
\end{equation*}
Это значит, что функционал $T_u$, который определяется связями $H_\mu$, является
генератором общих преобразований координат на гиперповерхностях $x^0=\const$.
Напомним, что импульсы $p^{\mu\nu}$ являются тензорными плотностями степени
$-1$. Алгебра преобразований координат хорошо известна и задается скобкой
Пуассона (\ref{econmn}). Скобку Пуассона (\ref{econom}) также можно не вычислять
явно. Ее вид следует из того, что связь $H_\bot$ является скалярной плотностью
степени $-1$. Таким образом, необходимо вычислить только скобку Пуассона
(\ref{econoo}). Эти вычисления очень громоздки, и впервые, по-видимому, были
проделаны намного позже ДеВиттом \cite{DeWitt67A}. В следующем разделе мы
вычислим эту скобку после канонического преобразования, приводящего связи к
полиномиальному виду, что существенно упрощает вычисления.

Ввиду того, что связи $H_\mu$ задают только пространственные диффеоморфизмы, мы
будем называть их кинематическими. Они также не зависят от констант связи в
действии, если таковые имеются. Связь $H_\bot$ называется динамической, так как
она определяет развитие начальных данных во времени и существенно зависит от
исходного действия, в частности, от констант связи.

Для сравнения приведем скобку Пуассона связей $H^\mu=0$ с контравариантным
индексом, которые эквивалентны связям $H_\mu=0$,
\begin{equation*}
  [H^\mu,H^{\prime \nu}]=\left(\hat g^{\mu\nu}H^\rho
  +\hat g^{\prime\mu\nu}H^{\prime\rho}\right)\dl_\rho
  +\left(\hat g^{\mu\rho}\pl_\rho g_{\s\lm}\hat g^{\nu\s}
  -\hat g^{\nu\rho}\pl_\rho g_{\s\lm}\hat g^{\mu\s}\right)H^\lm\dl.
\end{equation*}
Как видим, данная скобка Пуассона выглядит сложнее (\ref{econmn}). Это говорит о
том, что выбор вида связей из эквивалентных наборов является очень важным
и может привести к существенному упрощению вычислений. К сожалению, общего
рецепта как поступать в таких случаях не существует, и действует метод
``пристального всматривания''.

Теперь посмотрим на вторичные связи с другой точки зрения. Сравнение выражения
для связей (\ref{ecoadm}) с тензором Эйнштейна (\ref{einbom}) приводит к
равенствам
\begin{equation*}
\begin{split}
  H_\bot&=2\hat eG_{\bot\bot}=\frac{2\hat e}{N^2}(G_{00}-2N^\mu G_{0\mu}
  +N^\mu N^\nu G_{\mu\nu}),
\\
  H_\mu&=2\hat eG_{\bot\mu}=\frac{2\hat e}N(G_{0\mu}-N^\nu G_{\nu\mu}).
\end{split}
\end{equation*}
Мы видим, что вторичные связи (\ref{esecco}) эквивалентны четырем уравнениям
Эйнштейна
\begin{equation*}
  G_{\bot\bot}=0,\qquad G_{\bot\mu}=0,
\end{equation*}
которые представляют собой определенные комбинации всех вакуумных уравнений
Эйнштейна $G_{\al\bt}=0$. Если выполнены все уравнения Эйнштейна, а также
вторичные связи в начальный момент времени, то вторичные связи будут выполнены
также во все последующие моменты времени. Это следует из того, что
$\dot G_{\bot\bot}=0$ и $\dot G_{\bot\mu}=0$ как следствие свернутых тождеств
Бианки (\ref{ecobio}), (\ref{ecobis}). Таким образом мы доказали, что вторичные
связи в общей теории относительности согласованы с уравнениями движения, т.е.\
представляют собой связи первого рода. Для этого необязательно вычислять в явном
виде алгебру связей (\ref{econoo})--(\ref{econmn}). Однако знание алгебры связей
полезно.

В заключение данного раздела выпишем канонические уравнения движения для
пространственных компонент метрики и импульсов
\begin{align}                                                     \label{ecamet}
  \dot g_{\mu\nu}=&\frac{2N}{\hat e}p_{\mu\nu}
  -\frac{2N}{\hat e(n-2)}g_{\mu\nu}p+\hat\nb_\mu N_\nu+\hat\nb_\nu N_\mu,
\\                                                                     \nonumber
  \dot p^{\mu\nu}=&\frac N{2\hat e}\hat g^{\mu\nu}\left(p^{\rho\s}p_{\rho\s}
  -\frac1{n-2}p^2\right)-\frac{2N}{\hat e}\left(p^{\mu\rho}p^\nu{}_\rho
  -\frac1{n-2}p^{\mu\nu}p\right)-
\\                                                                     \nonumber
  &-\hat eN\left(\hat R^{\mu\nu}-\frac12\hat g^{\mu\nu}\hat R\right)
  +\hat e(\triangle N \hat g^{\mu\nu}-\hat\nb^\mu\hat\nb^\nu N)-
\\                                                                \label{ecamot}
  &-p^{\mu\rho}\hat\nb_\rho N^\nu-p^{\nu\rho}\hat\nb_\rho N^\mu
  +\hat\nb_\rho(N^\rho p^{\mu\nu}),
\end{align}
где $\triangle:=\hat\nb^\mu\hat\nb_\mu$ -- оператор Лапласа--Бельтрами.

Уравнение (\ref{ecamet}), конечно, совпадает с определением импульсов
(\ref{emomet}). Уравнения для импульсов (\ref{ecamot}) эквивалентны уравнениям
Эйнштейна $G_{\mu\nu}=0$. Чтобы доказать это, запишем уравнения (\ref{ecamot}) в
виде
\begin{equation*}
  \dot p^{\mu\nu}-(\dotsc)^{\mu\nu}=0,
\end{equation*}
где точки обозначают все слагаемые в правой части (\ref{ecamot}).
Тогда можно проверить, что выполнено равенство
\begin{equation*}
  g_{\mu\rho}g_{\nu\s}[\dot p^{\rho\s}-(\dotsc)^{\rho\s}]=N\hat eG_{\mu\nu}.
\end{equation*}
Таким образом, канонические уравнения движения (\ref{ecamet}), (\ref{ecamot}) с
учетом уравнений связей (\ref{esecco}) эквивалентны полной системе вакуумных
уравнений Эйнштейна $G_{\al\bt}=0$.
\section{Полиномиальная гамильтонова форма                       \label{scapol}}
Гамильтониан (\ref{eadmha}), построенный ранее, полиномиален по импульсам
$p^{\mu\nu}$, но неполиномиален по компонентам метрики $g_{\mu\nu}$, т.к.\
содержит скалярную кривизну $\hat R$ и определитель репера, которые входят в
динамическую связь $H_\bot$. В настоящем разделе мы опишем каноническое
преобразование, которое приводит к полиномиальной форме гамильтониана и,
следовательно, к полиномиальной форме уравнений движения. Это преобразование
является аналогом полиномиальной лагранжевой формы действия
Гильберта--Эйнштейна, рассмотренной в разделе \ref{spolaf}.

Идея канонического преобразования состоит в следующем. Импульсы $p^{\mu\nu}$
приводимы и разлагаются на бесследовую часть $\tilde p^{\mu\nu}$ и след $p$ в
соответствии с формулой (\ref{eirmom}). При вычислениях, как правило, удобнее
работать с неприводимыми компонентами, поскольку много слагаемых автоматически
сокращаются. Поставим вопрос: ``Нельзя ли совершить такое каноническое
преобразование, после которого новыми импульсами будут неприводимые компоненты
$\tilde p^{\mu\nu}$ и $p$ ?''. Этот вопрос нетривиален, потому что разложение
импульсов включает метрику, компоненты которой сами являются координатами
фазового пространства. Ответ на поставленный вопрос отрицательный, потому что
скобка Пуассона импульсов между собой отлична от нуля. Например,
$[\tilde p^{\mu\nu},p']\ne0$, где $p':=p(t,\Bx')$. Однако существует такое
каноническое преобразование, что новые импульсы будут пропорциональны
неприводимым компонентам $\tilde p^{\mu\nu}$ и $p$. Построением этого
канонического преобразования мы и займемся в настоящем разделе.

Рассмотрим каноническое преобразование
\begin{equation}                                                  \label{ecatrn}
  (g_{\mu\nu},p^{\rho\s})\quad \mapsto\quad(k_{\mu\nu},P^{\rho\s}), (\varrho,P),
\end{equation}
к новым парам канонически сопряженных координат $k_{\mu\nu},\varrho$ и импульсов
$P^{\mu\nu},P$, где на координаты $k_{\mu\nu}$ и сопряженные им импульсы
$P^{\mu\nu}$ наложены дополнительные условия
\begin{equation}                                                  \label{econev}
  |\det k_{\mu\nu}|=1,\qquad P^{\mu\nu}k_{\mu\nu}=0.
\end{equation}
В качестве производящего функционала канонического преобразования выберем
функционал
\begin{equation}                                                  \label{egefun}
  \CF=-\int\!d\Bx\,\varrho^s k_{\mu\nu}p^{\mu\nu},\qquad s\in\MR,\quad s\ne0,
\end{equation}
зависящий от новых координат $\varrho,k_{\mu\nu}$ и старых импульсов
$p^{\mu\nu}$, а также от вещественного параметра $s$, который будет определен
позже. Тогда старые координаты и новые импульсы определяются вариационными
производными (см. раздел \ref{sgefca})
\begin{align}                                                     \label{eoldme}
  g_{\mu\nu}&:=-\frac{\dl\CF}{\dl p^{\mu\nu}}=\varrho^s k_{\mu\nu},
\\                                                                \label{eoldmo}
  P^{\mu\nu}&:=-\frac{\dl\CF}{\dl k_{\mu\nu}}=\varrho^s\tilde p^{\mu\nu},
\\                                                                \label{eoldmt}
  P&:=-\frac{\dl\CF}{\dl\varrho}=\frac s\varrho p.
\end{align}
При вычислении вариационной производной по $k_{\mu\nu}$ учтено условие
$|\det k_{\mu\nu}|=1$, из которого вытекает ограничение на вариации
$k^{\mu\nu}\dl k_{\mu\nu}=0$, где $k^{\mu\nu}$ -- тензорная плотность, обратная
к $k_{\mu\nu}$: $k^{\mu\nu}k_{\nu\s}=\dl^\mu_\s$. Тем самым равенство нулю следа
импульсов (\ref{econev}) автоматически следует из условия единичности
определителя плотности $k_{\mu\nu}$ для производящего функционала
(\ref{egefun}). В выражении (\ref{eoldmt}) учтено соотношение (\ref{eoldme}).

По сути дела, в качестве новой канонической переменной из метрики выделен ее
определитель в некоторой степени, как следует из (\ref{eoldme}):
\begin{equation*}
  |\det g_{\mu\nu}|=\hat e^2=\varrho^{s(n-1)}.
\end{equation*}
Поскольку метрика $g_{\mu\nu}$ невырождена, то отсюда, в частности, следует
равенство $\varrho>0$.

В дальнейшем симметричную тензорную плотность с единичным определителем
$k_{\mu\nu}$ мы для краткости также будем называть метрикой.

Из формул (\ref{eoldmo}), (\ref{eoldmt}) следует, что новые импульсы
$P^{\mu\nu}$ и $P$ пропорциональны неприводимым компонентам старых импульсов
$\tilde p^{\mu\nu}$ и $p$.

Отметим, что все новые канонические переменные являются тензорными плотностями
следующих степеней:
\begin{alignat*}{2}
  \deg k_{\mu\nu}&=\frac2{n-1}, &\qquad\deg\varrho&=-\frac2{s(n-1)},
\\
  \deg P^{\mu\nu}&=-\frac2{n-1}-1,  &\deg P&=\frac2{s(n-1)}-1.
\end{alignat*}

Чтобы вычислить скалярную кривизну $\hat R$, входящую в связь (\ref{ecoadm}),
заметим, что каноническое преобразование (\ref{eoldme}) имеет вид преобразования
Вейля $g_{\mu\nu}=\ex^{2\phi}k_{\mu\nu}$, где $\phi=s\ln\varrho/2$.
Прямые вычисления приводят к следующему выражению для скалярной кривизны
сечения $x^0=\const$ в новых координатах:
\begin{equation}                                                  \label{escpka}
  \hat R=\varrho^{-s-2}\left[\varrho^2R^{(k)}
  +s(n-2)\varrho\pl_\mu(k^{\mu\nu}\pl_\nu\varrho)+s(n-2)
  \left(s\frac{n-3}4-1\right)(\pl\varrho)^2\right],
\end{equation}
где $(\pl\varrho)^2:=k^{\mu\nu}\pl_\mu\varrho\pl_\nu\varrho$. Скалярная
кривизна, построенная по метрике $k_{\mu\nu}$, принимает удивительно простой вид
\begin{equation}                                                  \label{escuka}
  R^{(k)}=\pl^2_{\mu\nu}k^{\mu\nu}
  +\frac12k^{\mu\nu}\pl_\rho k_{\mu\s}\pl_\nu k^{\rho\s}
  -\frac14k^{\mu\nu}\pl_\mu k_{\rho\s}\pl_\nu k^{\rho\s}.
\end{equation}
Из единичности определителя метрики следует, что компоненты обратной метрики
$k^{\mu\nu}$ являются полиномами степени $n-2$ по компонентам $k_{\mu\nu}$:
\begin{equation*}
  k^{\mu\nu}=\frac1{(n-2)!}\hat\ve^{\mu\rho_1\dotsc\rho_{n-2}}
  \hat\ve^{\nu\s_1\dotsc\s_{n-2}}k_{\rho_1\s_1}\dotsc k_{\rho_{n-2}\s_{n-2}},
\end{equation*}
где $\hat\ve^{\mu_1\dotsc\mu_{n-1}}$ -- полностью антисимметричная тензорная
плотность ранга $n-1$. Поэтому скалярная кривизна $R^{(k)}$ полиномиальна как по
метрике $k_{\mu\nu}$, так и по обратной метрике $k^{\mu\nu}$.

Динамическая связь в новых переменных принимает вид
\begin{multline*}
  H_\bot=\varrho^{-\frac{s(n-1)}2}\left[P^{\mu\nu}P_{\mu\nu}
  -\frac{\varrho^2}{s^2(n-1)(n-2)}P^2\right]-
\\
  -\varrho^{\frac{s(n-1)}2-s-2}\left[\varrho^2R^{(k)}
  +s(n-2)\varrho\pl_\mu(k^{\mu\nu}\pl_\nu\varrho)
  +s(n-2)\left(s\frac{n-3}4-1\right)(\pl\varrho)^2\right],
\end{multline*}
где опускание индексов у импульсов производится с помощью новой метрики,
$P_{\mu\nu}:=k_{\mu\rho}k_{\nu\s}P^{\rho\s}$.

Проанализируем возможность такого выбора постоянной $s$, чтобы динамическая
связь имела полиномиальный вид. Оба выражения в квадратных скобках полиномиальны
по всем динамическим переменным. Поскольку $n\ge3$, то для положительности
степени плотности $\varrho$ перед первой квадратной скобкой необходимо
выполнение неравенства $s<0$. В этом случае степень $\varrho$ перед второй
квадратной скобкой будет отрицательна. Таким образом, за счет выбора параметра
$s$ добиться полиномиальности самой связи $H_\bot$ нельзя. Однако, связь можно
умножить целиком на произвольный множитель, отличный от нуля. При этом
поверхность в фазовом пространстве, определяемая данной связью не изменится.
Минимальная степень $\varrho$, на которую необходимо умножить $H_\bot$ будет
тогда, когда степени $\varrho$ перед квадратными скобками будут равны. Отсюда
следует равенство
\begin{equation*}
  s=\frac2{n-2}.
\end{equation*}
Тогда, умножив динамическую связь на степень $\rho$,
\begin{equation}                                                  \label{edynec}
  K_\bot:=\varrho^{\frac{n-1}{n-2}}H_\bot,
\end{equation}
получим эквивалентную полиномиальную связь
\begin{equation}                                                  \label{enecan}
  K_\bot=P^{\mu\nu}P_{\mu\nu}-\frac{n-2}{4(n-1)}\varrho^2P^2
  -\varrho^2R^{(k)}-2\varrho\pl_\mu(k^{\mu\nu}\pl_\nu\rho)
  +\frac{n-1}{n-2}(\pl\varrho)^2=0.
\end{equation}

Кинематические связи в новых динамических переменных сохраняют свою
полиномиальность:
\begin{equation}                                                  \label{ekicon}
  H_\mu=-2\pl_\nu(P^{\nu\s}k_{\s\mu})+P^{\nu\s}\pl_\mu k_{\nu\s}
  -\frac{n-2}{n-1}\pl_\mu(P\varrho)+P\pl_\mu\varrho=0.
\end{equation}

Исходя из явного выражения для новых канонических переменных
(\ref{eoldme})--(\ref{eoldmt}), вычислим основные скобки Пуассона. Отличными от
нуля являются только три скобки:
\begin{align}                                                     \label{eponek}
  [k_{\mu\nu},P^{\prime\varrho\s}]&=\left(\dl_{\mu\nu}^{\rho\s}
  -\frac1{n-1}k_{\mu\nu}k^{\rho\s}\right)\dl(\Bx'-\Bx),
\\                                                                \label{epobpp}
  [P^{\mu\nu},P^{\prime\rho\s}]&=\frac1{n-1}
  \big(P^{\mu\nu}k^{\rho\s}-P^{\rho\s}k^{\mu\nu}\big)\dl(\Bx'-\Bx),
\\                                                                \label{eponer}
  [\varrho,P']&=\dl(\Bx'-\Bx),
\end{align}
где
\begin{equation*}
  \dl_{\mu\nu}^{\rho\s}
  :=\frac12\big(\dl_\mu^\rho\dl_\nu^\s+\dl_\mu^\s\dl_\nu^\rho\big).
\end{equation*}
Скобка Пуассона (\ref{eponek}) не имеет канонического вида для фазовых
переменных. Появление дополнительного слагаемого в (\ref{eponek}) связано с тем,
что на поля $k_{\mu\nu}$ и $P^{\mu\nu}$ наложены дополнительные условия
(\ref{econev}). По этой же причине отлична от нуля скобка Пуассона
(\ref{epobpp}).

Нетрудно проверить, что пуассонова структура, определяемая скобками Пуассона
(\ref{eponek})--(\ref{eponer}) является вырожденной. Это значит, что все
многообразие $\MN$, задаваемое координатами $(k_{\mu\nu},P^{\mu\nu})$ и
$(\varrho,P)$ является не симплектическим, а только пуассоновым многообразием
(см.\ раздел \ref{spoist}). На пуассоновом многообразии $\MN$ существуют две
функции Казимира:
\begin{equation*}
  c^1:=\det k_{\mu\nu},\qquad c^2:=P^{\mu\nu}k_{\mu\nu}.
\end{equation*}
Действительно, из определения пуассоновой структуры
(\ref{eponek})--(\ref{eponer}) следует, что следующие скобки Пуассона равны
нулю:
\begin{equation*}
  [c^{1,2},k'_{\mu\nu}]=[c^{1,2},P^{\prime\mu\nu}]=[c^{1,2},\varrho']
  =[c^{1,2},P']=0.
\end{equation*}
Отсюда вытекает, что скобка Пуассона $[c^{1,2},f']=0$, где $f\in\CC^1(\MN)$ --
произвольная дифференцируемая функция на $\MN$. Пуассонова структура,
ограниченная на сечения $\MV\subset\MN$, которые определяются уравнениями
$c^1=\const$ и $c^2=\const$, невырождена. Следовательно, эти сечения являются
симплектическими. Новым фазовым пространством общей теории относительности в
рассматриваемом случае является подмногообразие $\MV\subset\MN$, определяемое
двумя фиксированными значениями функций Казимира (\ref{econev}). Строго говоря,
каноническое преобразование (\ref{ecatrn}) является каноническим, т.е.\
сохраняющим вид скобок Пуассона, только между исходным фазовым пространством
и подмногообразием $\MV$. Полиномиальность связей достигнута за счет расширения
исходного фазового пространства до пуассонова многообразия $\MN$. Если
решить дополнительные связи (\ref{econev}) явно, то полиномиальность будет
нарушена. В этом нет ничего необычного. Например, электродинамика содержит
связи, явное решение которых приводит даже к нелокальному действию для
физических степеней свободы (см., например, \cite{GitTyu86R}).

Рассмотрим алгебру связей. Поскольку вместо динамической связи $H_\bot$ мы
ввели новую связь $K_\bot$, то алгебра связей изменится. Несложные вычисления
приводят к равенствам
\begin{align}                                                     \label{enewcz}
  [K_\bot,K'_\bot]&=-(\varrho^2H^\mu+\varrho^{\prime 2}H^{\prime\mu})
  \dl_\mu(\Bx'-\Bx),
\\                                                                \label{eneseo}
  [K_\bot,H'_\mu]&=-(K_\bot+K'_\bot)\dl_\mu(\Bx'-\Bx),
\\                                                                \label{enekip}
  [H_\mu,H'_\nu]&=-H_\nu\dl_\mu(\Bx'-\Bx)-H'_\mu\dl_\nu(\Bx'-\Bx),
\end{align}
где
\begin{equation*}
  H_\mu:=k_{\mu\nu}H^\nu,\quad\text{и}\quad
  \dl_\mu(\Bx'-\Bx):=\frac{\pl}{\pl x^{\prime\mu}}\dl(\Bx'-\Bx).
\end{equation*}
По сравнению с исходной алгеброй (\ref{econoo})--(\ref{econom}), изменения
касаются скобок Пуассона (\ref{enewcz}) и (\ref{eneseo}). Скобка Пуассона
(\ref{enewcz}) является результатом прямого счета. Вторая скобка Пуассона
(\ref{eneseo}) носит кинематический характер и определяется тем, что новая связь
является не функцией, а скалярной плотностью.

Получим явные выражения для геометрических объектов в новых переменных.
Сначала вычислим символы Кристоффеля:
\begin{equation*}
  \hat\Gamma_{\mu\nu}{}^\rho=\frac1{(n-2)\varrho}(\pl_\mu\varrho\dl_\nu^\rho
  +\pl_\nu\varrho\dl_\mu^\rho-k_{\mu\nu}k^{\rho\s}\pl_\s\varrho)
  +\Gamma^{(k)}_{\mu\nu}{}^\rho,
\end{equation*}
где ``символы Кристоффеля'' $\Gamma^{(k)}_{\mu\nu}{}^\rho$ выражаются через
тензорную плотность $k_{\mu\nu}$ по тем же формулам, что и символы Кристоффеля
через метрику. Конечно, ``символы Кристоффеля'' $\Gamma^{(k)}_{\mu\nu}{}^\rho$
никакой связности не определяют. Отсюда следует выражение для следа символов
Кристоффеля, который определяет дополнительные слагаемые в ковариантных
производных тензорных плотностей,
\begin{equation*}
  \hat\Gamma_\mu=\frac{n-1}{n-2}\frac{\pl_\mu\varrho}{\varrho},
\end{equation*}
так как
\begin{equation*}
  \Gamma^{(k)}_\mu:=\Gamma^{(k)}_{\nu\mu}{}^\nu=\frac12k^{\nu\rho}\pl_\mu k_{\nu\rho}=0
\end{equation*}
в силу условия $|\det k_{\mu\nu}|=1$. Нетрудно проверить ковариантное
постоянство новых переменных:
\begin{equation*}
  \hat\nb_\mu k_{\nu\rho}=0,\qquad\hat\nb_\mu\varrho=0.
\end{equation*}

Прямые вычисления приводят к следующему тензору Риччи
\begin{multline}                                                  \label{eriner}
  \hat R_{\mu\nu}=R^{(k)}_{\mu\nu}
  +\frac{n-3}{n-2}\frac{\pl^2_{\mu\nu}\varrho}\varrho
  -\frac{(n-1)(n-3)}{(n-2)^2}\frac{\pl_\mu\varrho\pl_\nu\varrho}{\varrho^2}
  -\frac{n-3}{n-2}\Gamma^{(k)}_{\mu\nu}{}^\rho\frac{\pl_\rho\varrho}\varrho-
\\[1mm]
  -\frac1{(n-2)^2}k_{\mu\nu}\frac{\pl\varrho^2}{\varrho^2}
  +\frac1{n-2}k_{\mu\nu}\frac{\pl_\rho(k^{\rho\s}\pl_\s\varrho)}\varrho,
\end{multline}
где введен ``тензор Риччи'' $R^{(k)}_{\mu\nu}$, построенный по ``метрике''
$k_{\mu\nu}$. Отсюда следует выражение для тензорной плотности скалярной
кривизны, которая возникает после свертки тензора Риччи с $k^{\mu\nu}$:
\begin{equation*}
  \check R:=k^{\mu\nu}\hat R_{\mu\nu}=\varrho^{\frac2{n-2}}\hat R=R^{(k)}
  +2\frac{\pl_\mu(k^{\mu\nu}\pl_\nu\varrho)}\varrho
  -\frac{n-1}{n-2}\frac{(\pl\varrho)^2}{\varrho^2}.
\end{equation*}

После переопределения динамической связи (\ref{edynec}) гамильтониан также равен
линейной комбинации связей:
\begin{equation*}
  \CH=\int\! d\Bx(\tilde N K_\bot+N^\mu H_\mu),
\end{equation*}
только с новым множителем Лагранжа $\tilde N:=\varrho^{-\frac{n-1}{n-2}}N$.
Соответствующее действие содержит дополнительные слагаемые:
\begin{equation*}
  S_{\Sh\Se}=\int dx\big(P^{\mu\nu}\dot k_{\mu\nu}+P\dot\varrho-\tilde NK_\bot
  -N^\mu H_\mu-\lm(|\det k_{\mu\nu}|-1)-\mu P^{\mu\nu}k_{\mu\nu}\big),
\end{equation*}
где мы учли связи (\ref{econev}) с помощью множителей Лагранжа $\lm$ и $\mu$.
Уравнения движения для новых канонических переменных примут вид:
\begin{align}                                           \label{eqrhod}
  \dot\varrho&=-\frac{n-2}{2(n-1)}\tilde N\varrho^2P
  +\frac{n-2}{n-1}\hat\nb_\mu N^\mu\varrho,
\\                                                      \label{eqcapm}
  \dot P&=\frac{n-2}{2(n-1)}\tilde N\varrho P^2+2\tilde N\varrho\check R
  +2k^{\mu\nu}\hat\nb_\mu\hat\nb_\nu\tilde N\varrho
  +\frac1{n-1}\hat\nb_\mu N^\mu P+N^\mu\hat\nb_\mu P,
\\                                                      \label{ekamnu}
  \dot k_{\mu\nu}&=2\tilde NP_{\mu\nu}+\hat\nb_\mu N^\rho k_{\nu\rho}
  +\hat\nb_\nu N^\rho k_{\mu\rho}-\frac2{n-1}\hat\nb_\rho N^\rho k_{\mu\nu},
\\                                                        \nonumber
  \dot P^{\mu\nu}&=-2\tilde NP^{\mu\rho}P^\nu{}_\rho
  +\hat\nb_\rho(N^\rho P^{\mu\nu})+\frac2{n-1}\hat\nb_\rho N^\rho P^{\mu\nu}
  -P^{\mu\rho}\hat\nb_\rho N^\nu-P^{\nu\rho}\hat\nb_\rho N^\mu-
\\                                                      \label{ecapmu}
  &\qquad -\varrho^2k^{\mu\rho}k^{\nu\s}\left(\tilde N\hat R_{\rho\s}
  +\hat\nb_\rho\hat\nb_\s\tilde N\right)
  +\frac{\varrho^2k^{\mu\nu}}{n-1}\left(\tilde N\check R
  +k^{\rho\s}\hat\nb_\rho\hat\nb_\s\tilde N\right).
\end{align}
Отметим, что множители Лагранжа $\lm$ и $\mu$ в уравнения движения вообще не
входят, т.к.\ связи являются функциями Казимира. Эта система уравнений движения
записана для тензорных плотностей:
\begin{equation*}
\begin{split}
  \deg k_{\mu\nu}&=\quad \frac2{n-1},
\\
  \deg P^{\mu\nu}&=-\frac{n+1}{n-1},
\end{split}
\qquad
\begin{split}
  \deg\varrho&=-\frac{n-2}{n-1},
\\
  \deg P&=-\frac1{n-1}.
\end{split}
\end{equation*}
Новый множитель Лагранжа также является тензорной плотностью:
\begin{equation*}
  \deg\tilde N=1,\qquad \deg N^\mu=0.
\end{equation*}
Поскольку степени тензорных плотностей при умножении складываются, то
\begin{equation*}
  \deg(\tilde N\varrho)=\frac1{n-1},\qquad
  \deg(\tilde N\varrho k^{\mu\nu})=-\frac1{n-1}.
\end{equation*}
Ковариантные производные от тензорных плотностей в системе уравнений
(\ref{eqrhod})--(\ref{ecapmu}) имеют следующий вид:
\begin{equation*}
\begin{split}
  \hat\nb_\mu\tilde N&=\pl_\mu\tilde N
  +\frac{n-1}{n-2}\frac{\pl_\mu\varrho}\varrho\tilde N,
\\
  \hat\nb_\mu P&=\pl_\mu P-\frac1{n-2}\frac{\pl_\mu\varrho}\varrho P.
\end{split}
\end{equation*}
Полученная система уравнений движения (\ref{eqrhod})--(\ref{ecapmu}) является
полиномиальной.
\section{Проблема энергии в теории гравитации                    \label{senepr}}
Определение энергии гравитационного поля является одной из главных
нерешенных проблем, которая привлекает большое внимание с момента создания
общей теории относительности. Взгляд на эту проблему существенно меняется
с течением времени, поэтому мы опишем несколько подходов.

Сначала сформулируем в чем именно заключается проблема. В механике точечных
частиц, а также в теории поля в пространстве Минковского под энергией понимают
численное значение гамильтониана системы. Если гамильтониан не зависит от
времени явно, то энергия сохраняется. Закон сохранения энергии можно получить
также из теоремы Нетер. Если действие инвариантно относительно трансляций во
времени $t\mapsto t+\const$, то из первой теоремы Нетер следует закон сохранения
энергии (см.\ раздел \ref{senmot}), при этом компонента $T_0{}^0$
энергии-импульса в теории поля совпадает с определением гамильтоновой плотности
системы полей. Этот подход к определению энергии в теории гравитации приводит к
ответу, мало удовлетворительному с физической точки зрения. В теории гравитации
действие инвариантно относительно общих преобразований координат и, тем более,
относительно трансляций во времени. С другой стороны, мы уже показали в разделе
\ref{sgauno}, что канонический гамильтониан для любой модели, инвариантной
относительно произвольного невырожденного преобразования временн\'ой координаты
пропорционален связи и равен нулю на уравнениях движения. Это значит, что
формальный подход к определению энергии в теории гравитации дает нуль для любого
решения уравнений движения. Этот результат малосодержателен и плохо согласуется
с нашим интуитивным представлением об энергии, т.к.\ наличие материи
ассоциируется с наличием энергии. Заметим также, что в пространстве-времени
Минковского действие инвариантно относительно трансляций только в декартовой
системе координат. В общековариантных моделях гравитации действие инвариантно
относительно трансляций в произвольной криволинейной системе координат.
\begin{com}
В настоящем разделе мы рассмотрим определение энергии в общей теории
относительности, то есть при нулевых тензорах кручения и неметричности.
Обобщение определения на более общий случай аффинной геометрии проблемы не
составляет, так как основная трудность, связанная с общей ковариантностью
уравнений движения, присутствует во всех моделях.
\qed\end{com}
\subsection{Тензор энергии-импульса полей материи}
Первые попытки определить энергию, как сохраняющуюся величину для системы полей
материи и гравитационного поля, были предприняты на заре исследований по общей
теории относительности. В этой модели действие представляет собой сумму действия
Гильберта--Эйнштейна (\ref{ehieia}) и действия для полей
материи $S_{\Sm}$:
\begin{equation}                                                  \label{etoacm}
  S=S_{\Sh\Se}+S_{\Sm}.
\end{equation}
Предположим, что действие полей материи зависит только от полей материи и
метрики. Тогда вариация действия по метрике приводит к уравнениям Эйнштейна
(\ref{einequ}), где
\begin{equation}                                                  \label{enmote}
  T_{{\Sm}\al\bt}:=\frac2\vol\frac{\dl S_{\Sm}}{\dl g^{\al\bt}}
\end{equation}
-- тензор энергии-импульса полей материи.
В общем случае мы предполагаем, что действие для полей материи в моделях
гравитации получено из действия, записанного в плоском пространстве-времени
Минковского, путем минимальной подстановки, т.е.\ замены обычных производных на
ковариантные $\pl_\al\mapsto\nb_\al$ и метрики Минковского на нетривиальную
метрику пространства-времени
$\eta_{ab}\mapsto g_{\al\bt}=e_\al{}^a e_\bt{}^b\eta_{ab}$.
Возможно также введение инвариантных слагаемых неминимального взаимодействия,
которые обращаются в нуль в пространстве Минковского. В таких случаях действие
зависит только от самих полей материи и репера. Соответствующее выражение для
тензора энергии-импульса можно записать через вариационную производную по
реперу:
\begin{equation}                                                  \label{enmotr}
  T_{{\Sm}\al\bt}=\frac2\vol\frac{\dl S_{\Sm}}{\dl e^\al{}_a}e_{\bt a}.
\end{equation}
Если действие для полей материи может быть записано через метрику, то данное
определение совпадает с (\ref{enmote}) и приводит к симметричному тензору
энергии-импульса. В общем случае это не так. Например, лагранжиан спинорного
поля не может быть записан через метрику, так как лоренцева связность выражается
через репер, и ее нельзя выразить через компоненты метрики. Определение тензора
энергии-импульса через репер (\ref{enmotr}) не всегда приводит к симметричному
тензору энергии-импульса.

В разделе (\ref{seigud}) было показано, что инвариантное действие приводит к
ковариантному закону сохранения тензора энергии-импульса:
\begin{equation}                                                  \label{etabco}
  \nb_\bt T_{\Sm\al}{}^\bt=0.
\end{equation}
Однако это ковариантное равенство не является законом сохранения. Покажем это.
Прямые вычисления приводят к равенству
\begin{equation}                                                  \label{ecovde}
  \nb_\bt T_{\Sm\al}{}^\bt=\frac1\vol\pl_\bt\left(\vol T_{\Sm\al}{}^\bt\right)
  -\frac12\pl_\al g_{\bt\g}T_{\Sm}{}^{\bt\g}=0.
\end{equation}
При интегрировании этого равенства по объему первое слагаемое для каждого
значения индекса $\al$ преобразуется с помощью формулы Стокса (\ref{estsec}) в
интеграл по граничной поверхности:
\begin{equation*}
  \int_\MM\!\!\! dx\pl_\bt\left(\vol T_{\Sm\al}{}^\bt\right)
  =\int_{\pl\MM}\!\!\!ds_\bt \vol T_{\Sm\al}{}^\bt,
\end{equation*}
где $ds_\al$ -- ориентированный элемент граничной гиперповерхности $\pl\MM$.
Этот интеграл привел бы к закону сохранения в выбранной системе координат, если
бы не наличие второго слагаемого в равенстве (\ref{ecovde}).
\subsection{Псевдотензор энергии-импульса для гравитации}
Для решения проблемы сохранения энергии к тензору энергии-импульса материи,
который мы будем записывать с одним верхним и одним нижним индексом
$T_{\Sm\al}{}^\bt$, было предложено добавить некоторый объект с компонентами
$t_\al{}^\bt$, зависящими только от метрики и ее первых производных таким
образом, чтобы было выполнено равенство
\begin{equation}                                                  \label{enmoco}
  \pl_\bt\left[\vol(T_{\Sm\al}{}^\bt+t_\al{}^\bt)\right]=0.
\end{equation}
Объект с компонентами $t_\al{}^\bt$ называется {\em псевдотензором
энергии-импульса} гравитационного поля. При этом равенство (\ref{enmoco})
принято рассматривать, как закон сохранения энергии-импульса полей материи и
гравитационного поля. Приставка ``псевдо'' в данном случае означает, что для
выполнения равенства (\ref{enmoco}) компоненты $t_\al{}^\bt$ не могут
образовывать тензор.
\index{Псевдотензор энергии-импульса (energy-momentum pseudotensor)}%
\index{Энергии-импульса псевдотензор (energy-momentum pseudotensor)}%

Задача о нахождении явного вида псевдотензора энергии-импульса гравитационного
поля $t_\al{}^\bt$ может быть решена следующим образом. Закон сохранения
(\ref{enmoco}) имеет тот же вид, что и закон сохранения энергии-импульса,
вытекающий из первой теоремы Нетер (\ref{enmocm}). Из инвариантности полного
действия (\ref{etoacm}) относительно трансляций в пространстве-времени следует
закон сохранения
\begin{equation}                                                  \label{econlo}
  \pl_\bt\left(\vol T_\al{}^\bt\right)=0,
\end{equation}
где суммарный тензор энергии-импульса материи и гравитационного поля,
умноженный на определитель репера, имеет вид
\begin{equation}                                                  \label{enmocn}
\begin{split}
  \vol T_\al{}^\bt&=\pl_\al g^{\g\dl}
  \frac{\pl L_{\Sh\Se}}{\pl(\pl_\bt g^{\g\dl})}
  +\pl_\al\vf^a\frac{\pl L_\Sm}{\pl(\pl_\bt\vf^a)}
  -\dl_\al^\bt(L_{\Sh\Se}+L_\Sm)
\\
  &=\vol T_{\Sm\al}^{(\mathrm{c})}{}^\bt
  +\pl_\al g^{\g\dl}\frac{\pl L_{\Sh\Se}}{\pl(\pl_\bt g^{\g\dl})}
  -\dl_\al^\bt L_{\Sh\Se},
\end{split}
\end{equation}
где $\vf^a$, $a=1,2,\dotsc$, -- совокупность всех полей материи.
В этом выражении $T_{\Sm\al}^{(\mathrm{c})}{}^\bt$ -- канонический тензор
энергии-импульса полей материи, определенный соотношением
\begin{equation*}
  \vol T_{\Sm\al}^{(\mathrm{c})}{}^\bt:=\pl_\al\vf^a
  \frac{\pl L_{\Sm}}{\pl(\pl_\bt\vf^a)}-\dl_\al^\bt L_{\Sm}.
\end{equation*}
В том случае, когда определение тензора энергии-импульса (\ref{enmote})
является ковариантным обобщением канонического тензора энергии-импульса
материи в пространстве Минковского, имеет место формула
\begin{equation*}
  T_{\Sm\al}{}^\bt=T_{\Sm\al}^{(\mathrm{c})}{}^\bt.
\end{equation*}
Тогда последние два слагаемых в выражении (\ref{enmocn}), зависящие только от
метрики и ее первых производных, можно принять за определение псевдотензора
энергии-импульса гравитационного поля
\begin{equation}                                                  \label{epsenm}
  \vol t_\al{}^\bt:=
  \pl_\al g^{\g\dl}\frac{\pl L_{\Sh\Se}}{\pl(\pl_\bt g^{\g\dl})}
  -\dl_\al^\bt L_{\Sh\Se}.
\end{equation}
Напомним, что под лагранжианом $L_{\Sh\Se}$ понимается функция, полученная из
$\kappa\vol R(g)$ добавлением полной производной, которая приводит к
сокращению вторых производных от метрики (см.\ раздел \ref{shieil}). Полученное
выражение для псевдотензора энергии-импульса широко использовалось, в том числе
классиками науки: Г.~Вейлем \cite{Weyl18BR}, П.~Дираком \cite{Dirac75R},
В.~Паули \cite{Pauli58R}, Э.~Шредингером \cite{Schrod50R} и др. После несложных
вычислений можно получить явное выражение для псевдотензора энергии-импульса
\begin{equation}                                                  \label{explte}
  \vol t_\al{}^\bt=\kappa(\Gamma_{\g\dl}{}^\bt-\dl_\g^\bt\Gamma_{\dl\e}{}^\e)
  \pl_\al\big(\vol g^{\g\dl}\big)-\dl_\al^\bt L_{\Sh\Se}.
\end{equation}

Отметим недостатки такого определения псевдотензора энергии-импульса
гравитационного поля. Из полученного выражения следует, что в нормальной системе
координат, где все символы Кристоффеля обращаются в нуль в некоторой точке, все
компоненты псевдотензора энергии-импульса гравитационного поля равны нулю. То
есть, независимо от кривизны пространства, в любой заданной точке
пространства-времени можно обратить в нуль все компоненты псевдотензора
энергии-импульса. В то же время в плоском пространстве-времени (отсутствие
гравитационного поля) в криволинейной системе координат символы Кристоффеля и,
следовательно, компоненты псевдотензора энергии-импульса в общем случае отличны
от нуля. Отметим также, что компонента тензора энергии-импульса $T_0{}^0$ в
законе сохранения (\ref{econlo}) совпадает с гамильтонианом системы и обращается
в нуль на уравнениях движения. Эти замечания ставят под сомнение возможность
физической интерпретации псевдотензора энергии-импульса гравитационного поля.

Если равенство (\ref{enmoco}) принято за определение псевдотензора
энергии-импульса гравитационного поля, то последний определен неоднозначно.
Очевидно, что псевдотензор
\begin{equation}                                                  \label{etodia}
  t^\prime_\al{}^\bt:=t_\al{}^\bt+\pl_\g B_\al{}^{\g\bt},
\end{equation}
где $B_\al{}^{\g\bt}=-B_\al{}^{\bt\g}$ -- произвольный ``тензор'' третьего
ранга, антисимметричный по верхним индексам, также удовлетворяет закону
сохранения. В процессе исследования общей теории относительности было предложено
несколько явных выражений для псевдотензора энергии-импульса, которые обладают
своими достоинствами и недостатками и отличаются между собой на некоторый
``тензор'' $B_\al{}^{\g\bt}$. Мы не будем останавливаться на обсуждении
различных подходов, а отметим лишь общий недостаток. Псевдотензор
энергии-импульса $t_\al{}^\bt$ не является тензором. Это значит, что энергия в
различных системах координат может быть положительна, равна нулю или
отрицательна, что является неудовлетворительным с физической точки зрения.

Аналогичное построение псевдотензора энергии-импульса гравитационного поля можно
провести в реперном формализме. При этом все недостатки псевдотензора
сохраняются.

Проведенное обсуждение показывает, что в теории гравитации невозможно дать
локальное определение энергии-импульса, удовлетворив одновременно двум условиям:
1) компоненты энергии-импульса должны образовывать тензорное поле второго ранга
и 2) должен быть выполнен закон сохранения (\ref{enmoco}).
\subsection{Законы сохранения и векторы Киллинга}
Другой подход к законам сохранения связан с симметриями пространства-времени.
Если метрика пространства-времени допускает группу движений, то появляется
возможность определить законы сохранения для тензора энергии-импульса полей
материи без привлечения понятия псевдотензора энергии-импульса гравитационного
поля. Пусть тензор энергии-импульса полей материи -- это симметричный тензор
второго ранга $T_\Sm^{\al\bt}=T_\Sm^{\bt\al}$, удовлетворяющий условию
(\ref{etabco}). Предположим, что метрика $g_{\al\bt}$ имеет $\Sn$ векторов
Киллинга $K_\Sa=K_\Sa{}^\al\pl_\al$, $\Sa=1,\dotsc,\Sn$. Рассмотрим $\Sn$
векторов
$$
  P_\Sa^\al:=K_\Sa{}^\bt T_{\Sm\bt}{}^\al.
$$
Тогда справедливо равенство
$$
  \nb_\al P_\Sa^\al=\nb_\al K_\Sa{}^\bt T_{\Sm\bt}{}^\al
  +K_\Sa{}^\bt\nb_\al T_{\Sm\bt}{}^\al=0.
$$
Здесь первое слагаемое равно нулю как следствие симметричности тензора
энергии-импульса и уравнения Киллинга (\ref{ekileq}). Второе слагаемое
обращается в нуль в силу уравнения (\ref{etabco}). С другой стороны, справедливо
тождество (\ref{edivrf})
\begin{equation*}
  \nb_\al P_\Sa^\al=\frac1\vol \pl_\al(\vol P_\Sa^\al).
\end{equation*}
Поэтому, если $\MM$ -- компактная ориентируемая область пространства-времени
с краем $\pl\MM$, то объемный интеграл можно преобразовать в поверхностный по
формуле Стокса:
\begin{equation}                                                  \label{eencon}
  \int_\MM\!\!\!dx\,\vol\nb_\al P_\Sa^\al=\int_{\pl\MM}\!\!\!ds_\al P_\Sa^\al=0,
\end{equation}
где $ds_\al$ -- ориентированный элемент площади края. Таким образом каждому
вектору Киллинга соответствует закон сохранения. В частности, если существует
вектор Киллинга $K_0=\pl_0$ и поля материи исчезают на пространственной
бесконечности, то ему соответствует закон сохранения энергии $\pl_0 E=0$, где
\begin{equation*}
  E=\int\!d\Bx\, P^0=\int\! d\Bx\, T_{\Sm0}{}^0.
\end{equation*}
Отметим, что в этом выражении присутствует гамильтонова плотность только для
полей материи.

В четырехмерном пространстве-времени Минковского метрика имеет десять векторов
Киллинга: четыре вектора соответствуют трансляциям и шесть --
(псевдо-)вращениям. Этим векторам Киллинга соответствуют законы сохранения
энергии-импульса и момента количества движения.
\subsection{Полная гравитационная энергия асимптотически плоского
            пространства-времени}
Дальнейшее развитие общей теории относительности изменило подход к определению
полной энергии гравитационного поля и материи. Это определение основано на
сферически симметричном решении Шварцшильда и учете граничных вкладов в действие
Гильберта--Эйнштейна.

Запишем решение Шварцшильда в асимптотически декартовой системе
координат, где координаты $x,y,z$ связаны с координатами $r,\theta,\vf$ обычными
формулами трехмерного евклидова пространства. С этой целью воспользуемся
формулами (\ref{erelqw}) для перехода от сферических координат к декартовым:
\begin{equation}                                                  \label{eshcar}
\begin{split}
  ds^2&=\left(1-\frac{2M}r\right)dt^2-\frac{dr^2}{1-\frac{2M}r}
  -r^2d(d\theta^2+\sin^2\theta d\vf^2)=
\\
  &=\left(1-\frac{2M}r\right)dt^2-(dx^2+dy^2+dz^2)-\frac{2M}{r^2(r-2M)}
  (xdx+ydy+xdz)^2.
\end{split}
\end{equation}
Метрика Шварцшильда на больших расстояниях в первом порядке по $M/r\rightarrow0$
принимает вид
\begin{equation}                                                  \label{eshasc}
\begin{split}
  ds^2\simeq\left(1-\frac{2M}r\right)dt^2
  -\left(1+\frac{2M}r\right)(dx^2+dy^2+dz^2)-&
\\
  -\frac{4M}r(xydxdy+xzdxdz+yzdydz)&.
\end{split}
\end{equation}
Очевидно, что в нулевом порядке эта метрика совпадает с метрикой Лоренца.
Последнее выражение используется для определения асимптотически плоского
\linebreak[4] пространства-времени.

\begin{defn}
Топологически тривиальное пространство-время $\MM\approx\MR^{1,3}$ называется
{\em асимптотически плоским}, если существует такая система координат $t,x,y,z$,
что его метрика при $r\to\infty$, где $r:=\sqrt{x^2+y^2+z^2}$ и всех моментов
времени $t\in\MR$, достаточно быстро стремится к метрике Шварцшильда
(\ref{eshasc}), записанной в декартовой системе координат.
\end{defn}
\index{Асимптотически плоское пространство-время %
(asymptotically flat space-time)}%
\index{Пространство-время асимптотически плоское %
(asymptotically flat space-time)}%

Максимально продолженное пространство-время Шварцшильда описывает черные дыры и
не является топологически тривиальным. Понятие асимптотически плоского
пространства-времени просто обобщается на многообразия с нетривиальной
топологией.
\begin{defn}
Пусть пространство-время имеет вид прямого произведения $\MM=\MR\times\MU$, где
явно выделено время $x^0=t\in\MR$. Допустим, что существует компактное
подмножество $\MK\subset\MU$ такое, что разность $\MU\setminus\MK$ представляет
собой несвязное объединение открытых множеств $\MS_\Sa$, $\Sa=1,\dotsc,\Sn$.
Если в каждом произведении $\MR\times\MS_\Sa$ существует такая система
координат $t,x,y,z$, что метрика при $r\to\infty$ и всех $t\in\MR$ достаточно
быстро стремится к метрике Шварцшильда (\ref{eshasc}) в декартовой системе
координат, то пространство-время $\MM$ называется {\em асимптотически плоским}.
\qed\end{defn}

Рассмотрим действие Гильберта--Эйнштейна в гамильтоновой форме, которое вытекает
из выражения для гамильтониана (\ref{eadmha}),
\begin{equation*}
  S_{\Sh\Se}
  =\int dx\left(p^{\mu\nu}\dot g_{\mu\nu}-NH_\bot-N^\mu H_\mu\right),
\end{equation*}
где функции хода и сдвига рассматриваются в качестве множителей Лагранжа.
Динамическая связь $H_\bot$ (\ref{ecoadm}) содержит трехмерную скалярную
кривизну и, следовательно, вторые производные от пространственных компонент
метрики $g_{\mu\nu}$. При вариации действия $L_{\Sh\Se}$ по компонентам метрики
приходится интегрировать по частям, отбрасывая граничные члены. Ниже будет
показано, что в асимптотически плоском пространстве-времени (для решения
Шварцшильда) граничный вклад, возникающий при интегрировании по частям
слагаемого со вторыми производными, отличен от нуля. В настоящее время это
граничное слагаемое принято в качестве определения полной энергии асимптотически
плоского распределения масс. Рассмотрим данный вопрос подробно.

Воспользуемся формулами (\ref{escder})--(\ref{edertl}) и запишем трехмерную
скалярную кривизну, умноженную на определитель репера, в виде
\begin{equation*}
  \hat e\hat R=\pl_\mu\left[\hat e(\hat g^{\mu\nu}\hat\Gamma_{\rho\nu}{}^\rho
  -\hat g^{\nu\rho}\hat\Gamma_{\nu\rho}{}^\mu)\right]+\hat e\hat L_{\Sh\Se},
\end{equation*}
где последнее слагаемое $\hat L_{\Sh\Se}$ квадратично по символам Кристоффеля
(его вид в настоящий момент не важен). Первое слагаемое в этом представлении
имеет вид дивергенции, которую можно выразить через компоненты метрики:
\begin{equation*}
  \pl_\mu\left[\hat e\hat g^{\mu\nu}\hat g^{\rho\s}
  (\pl_\nu g_{\rho\s}-\pl_\rho g_{\nu\s})\right],
\end{equation*}
где мы воспользовались выражением символов Кристоффеля через метрику
(\ref{echris}). При интегрировании этого слагаемого по частям возникает
граничное слагаемое, которое, как мы увидим ниже, отлично от нуля. Поэтому для
того, чтобы компенсировать его вклад, добавим к исходному действию граничный
член с самого начала:
\begin{equation*}
  S_{\Sh\Se}\mapsto S_{\Sh\Se}+\kappa\int\! dx\,\pl_\mu B^\mu,
\end{equation*}
где
\begin{equation}                                                  \label{ebouve}
  B^\mu:=N\hat e\hat g^{\mu\nu}\hat g^{\rho\s}
  (\pl_\nu g_{\rho\s}-\pl_\rho g_{\nu\s})
\end{equation}
и мы восстановили гравитационную константу связи $\kappa$.
\begin{defn}
{\em Полной энергией} асимптотически плоского распределения масс называется
интеграл
\begin{equation}                                                  \label{enetot}
  E:=\kappa\sum_\Sa\int_\Sa\!\!\! d\Bx\,\pl_\mu B^\mu,
\end{equation}
где компоненты $B^\mu$ определены в (\ref{ebouve}), и интегрирование проводится
по пространственноподобному сечению $x^0=\const$.
\qed\end{defn}
\index{Полная энергия (total energy)}%
\index{Энергия полная (total energy)}%
Компоненты $B^\mu$ образуют вектор относительно глобальных преобразований
координат в пространстве $\MR^3$. В целом, определение полной энергии
неинвариантно и зависит от выбора системы координат.

Для трехмерного евклидова пространства $\MR^3$ в декартовой системе координат
метрика имеет вид $g_{\mu\nu}=-\dl_{\mu\nu}$. Поэтому $B^\mu=0$ и,
следовательно, полная энергия плоского пространства равна нулю, $E=0$. Уже на
этом этапе видна важность выбора системы координат в определении энергии
(\ref{enetot}). Действительно, если система координат не является асимптотически
декартовой, то полная энергия трехмерного евклидова пространства может быть
отлична от нуля.

Вычислим полную энергию для решения Шварцшильда и, тем самым, для произвольного
асимптотически плоского пространства-времени в первом порядке по $M/r\to\infty$.
Для этого запишем решение Шварцшильда в декартовой системе координат
(\ref{eshcar}). Пространственная часть метрики отличается от евклидовой метрики:
$g_{\mu\nu}=\eta_{\mu\nu}+h_{\mu\nu}$, где
\begin{equation*}
  h_{\mu\nu}=-\frac{2M}{r^2(r-2M)}\begin{pmatrix}
    x^2 & xy & xz \\ xy & y^2 & yz \\ xz & yz & z^2 \end{pmatrix}
\end{equation*}
На больших расстояниях, $r\gg M$, эта поправка к метрике стремится к нулю. В
определении вектора $B^\mu$ (\ref{ebouve}) выражение в скобках имеет первый
порядок малости. Поэтому остальные сомножители достаточно участь в нулевом
порядке:
\begin{equation*}
  N=1,\qquad \hat e=1,\qquad \hat g^{\mu\nu}=\eta^{\mu\nu}.
\end{equation*}
Простые вычисления дают следующие выражения для компонент вектора $B^\mu$ в
первом порядке по $M/r$:
\begin{equation*}
  B^x\simeq\frac{4Mx}{r^3},\qquad B^y\simeq\frac{4My}{r^3},\qquad
  B^z\simeq\frac{4Mz}{r^3}.
\end{equation*}
Нормальный единичный ковектор к сфере с центром в начале координат евклидова
пространства имеет вид
\begin{equation*}
  \lbrace n_\mu\rbrace=\left(\frac xr,\frac yr, \frac zr\right),
  \qquad n^2=-1.
\end{equation*}
Поэтому
\begin{equation*}
  B^\mu n_\mu\simeq\frac{4M}{r^2}.
\end{equation*}
Теперь можно вычислить полную энергию. Для каждой компоненты связности $\MS_\Sa$
в первом порядке получаем равенство
\begin{equation}                                                  \label{etoens}
  E=\kappa\int\! d\Bx\, \pl_\mu B^\mu=\int_0^\pi\!\!\! d\theta
  \int_0^{2\pi}\!\!\!d\vf\,r^2\sin\theta(B^\mu n_\mu)\simeq 16\pi\kappa M.
\end{equation}

Чтобы придать полученному выражению более привлекательный вид, восстановим
размерные константы. Это полезно делать хотя бы изредка. Компоненты метрики,
по-определению, являются безразмерными $[g_{\al\bt}]=1$. Поскольку
\begin{equation*}
  \left[\frac Mr\right]=\frac{\text{г}}{\text{см}},
\end{equation*}
то это отношение необходимо обезразмерить, умножив на некоторую комбинацию
гравитационной постоянной и скорости света, которые имеют следующие размерности:
\begin{equation*}
  [G]=\frac{\text{см}^3}{\text{г}\cdot\text{сек}^2},\qquad
  [c]=\frac{\text{см}}{\text{сек}}.
\end{equation*}
Это можно сделать единственным образом путем замены
\begin{equation}                                                  \label{edishs}
  M\mapsto\frac{GM}{c^2}.
\end{equation}

Если воспользоваться выражением (\ref{egravc}) для константы связи $\kappa$
через гравитационную постоянную $G$, которое было получено при рассмотрении
ньютонова предела в общей теории относительности, то выражение для полной
энергии примет вид
\begin{equation}                                                  \label{eshent}
  E=Mc^2.
\end{equation}
То, что гравитационная энергия асимптотически плоского пространства-времени
совпадает с энергией покоя для массы в решении Шварцшильда, привлекательно с
физической точки зрения и оправдывает определение полной энергии через
поверхностный интеграл (\ref{enetot}). Численное значение полной гравитационной
энергии зависит от выбора системы координат, т.к.\ данное определение не
инвариантно относительно преобразований координат.

Поскольку $M$ -- постоянная интегрирования уравнений Эйнштейна, то полная
энергия в асимптотически плоском пространстве-времени сохраняется.

\chapter{Скалярные и калибровочные поля}
Скалярные и электромагнитное поля, вместе со спинорными полями, являются
важнейшими объектами квантовой теории поля. В настоящей главе мы напомним
основные свойства скалярных и калибровочных полей в пространстве Минковского,
рассмотрим эти поля на произвольных многообразиях с заданной аффинной геометрией
и обсудим их уравнения в общей теории относительности.
\section{Действительное скалярное поле                           \label{srescf}}
\subsection{Скалярное поле в пространстве Минковского            \label{srescx}}
Рассмотрим действительное скалярное поле (т.е.\ функцию)
$\vf(x)\in\CC^2(\MR^{1,n-1})$ в пространстве Минковского $\MR^{1,n-1}$
произвольного числа измерений. Обозначим декартовы координаты через $x^\al$,
$\al=0,1,\dotsc,n-1$. Свободное скалярное поле описывается квадратичным
лагранжианом
\begin{equation}                                                  \label{elascf}
  L=\frac12\et^{\al\bt}\pl_\al\vf\pl_\bt\vf-\frac12m^2\vf^2,
\end{equation}
где $\eta^{\al\bt}:=\diag(+-\dotsc-)$ -- обратная метрика Лоренца и $m=\const>0$
-- масса скалярного поля. Соответствующее действие имеет вид
$$
  S=\int\!dx L.
$$
Знаки слагаемых в лагранжиане выбран таким образом, чтобы канонический
гамильтониан, рассмотренный ниже, был положительно определен.

Посчитаем размерности. Действие и компоненты метрики, по-определению,
безразмерны. Координаты имеют размерность длины, $[x^\al]=l$,
$\al=0,1,\dotsc,n-1$ (скорость света мы положили равной единице). Поэтому
скалярное поле и масса имеют следующие размерности:
\begin{equation}                                                  \label{edisca}
  [\vf]=l^{\frac{2-n}2},\qquad [m]=l^{-1}.
\end{equation}

Уравнение движения для свободного скалярного поля линейно:
\begin{equation}                                                  \label{eqmrsc}
  S,{}_\vf:=\frac{\dl S}{\dl\vf}=-(\square+m^2)\vf=0,
\end{equation}
где
\begin{equation*}
  \square:=\eta^{\al\bt}\pl_\al\pl_\bt=\pl_t^2-\pl_1^2-\dotsc-\pl_{n-1}^2
\end{equation*}
-- волновой оператор (оператор Даламбера). Это уравнение называется
{\em уравнением Клейна--Гордона--Фока}, которое при $m=0$ сводится к уравнению
Даламбера. Уравнение (\ref{eqmrsc}) было предложено независимо в работах
\cite{Gordon26,Fock26A,Fock26B,Klein27}.
\index{Уравнение Клейна--Гордона--Фока (Klein--Gordon--Fock equation)}%
\index{Клейна--Гордона--Фока уравнение (Klein--Gordon--Fock equation)}%

Скалярное поле называется свободным, потому что описывается линейным уравнением
движения.

Поскольку свободное скалярной поле удовлетворяет волновому гиперболическому
уравнению, то для постановки задачи Коши в полупространстве $x^0\ge0$ для
однозначного определения решения необходимо задать начальные данные
на гиперповерхности $x^0=0$ для поля $\vf$ и его производной по времени
$\dot\vf:=\pl_0\vf$. Говорят, что скалярное поле описывает одну динамическую
степень свободы. В квантовой теории поля оно описывает нейтральные скалярные
частицы.

Для того, чтобы описать скалярное поле с самодействием, к лагранжиану
(\ref{elascf}) добавляется потенциал взаимодействия $V(\vf)$,
\begin{equation*}
  L\mapsto L=\frac12\et^{\al\bt}\pl_\al\vf\pl_\bt\vf-V(\vf),
\end{equation*}
где $V(\vf)$ -- некоторая достаточно гладкая функция одного аргумента. Для
упрощения обозначений мы включили массовый член в определение потенциала
$V(\vf)$. В общем случае функция $V(\vf)$ зависит от некоторого набора размерных
или безразмерных констант связи. Для самодействующего скалярного поля уравнение
движения становится нелинейным:
\begin{equation}                                                  \label{eqmrsn}
  S,{}_\vf:=\frac{\dl S}{\dl\vf}=-\square\vf-V'(\vf)=0,
\end{equation}
где $V':=dV/d\vf$.

Действие для скалярного поля инвариантно относительно глобального действия
группы Пуанкаре $\MI\MO(1,n-1)$. Согласно первой теореме Нетер эта
инвариантность приводит к законам сохранения энергии-импульса и момента
количества движения (см.\ разделы \ref{senmot}, \ref{sanmot}). Выражение для
сохраняющегося канонического тензора энергии-импульса (\ref{etenmo}) скалярного
поля имеет вид
\begin{equation}                                                  \label{enmots}
  T^{(c)}_{\al\bt}=\pl_\al\vf\pl_\bt\vf-
  \eta_{\al\bt}\left(\frac12\pl\vf^2-V\right),
\end{equation}
где использовано сокращенное обозначение для кинетического слагаемого
$$
  \pl\vf^2:=\eta^{\al\bt}\pl_\al\vf\pl_\bt\vf.
$$
Тензор энергии-импульса, очевидно, симметричен по своим индексам.

Спиновый момент скалярного поля (\ref{espimo}) равен нулю, $S_{\al\bt}{}^\g=0$,
т.к.\ при вращениях равна нулю вариация формы скалярного поля. Отсюда вытекает,
что в квантовой теории поля скалярное поле описывает частицы с нулевым спином.
Момент количества движения скалярного поля полностью определяется орбитальным
моментом по формуле (\ref{eanmot})
\begin{equation}                                                  \label{emokod}
  J_{\al\bt}{}^\g=M_{\al\bt}{}^\g:=x_\bt T^{(c)}_{~\al}{}^\g
  -x_\al T^{(c)}_{~\bt}{}^\g.
\end{equation}

\begin{figure}[h,b,t]
\hfill\includegraphics[width=.8\textwidth]{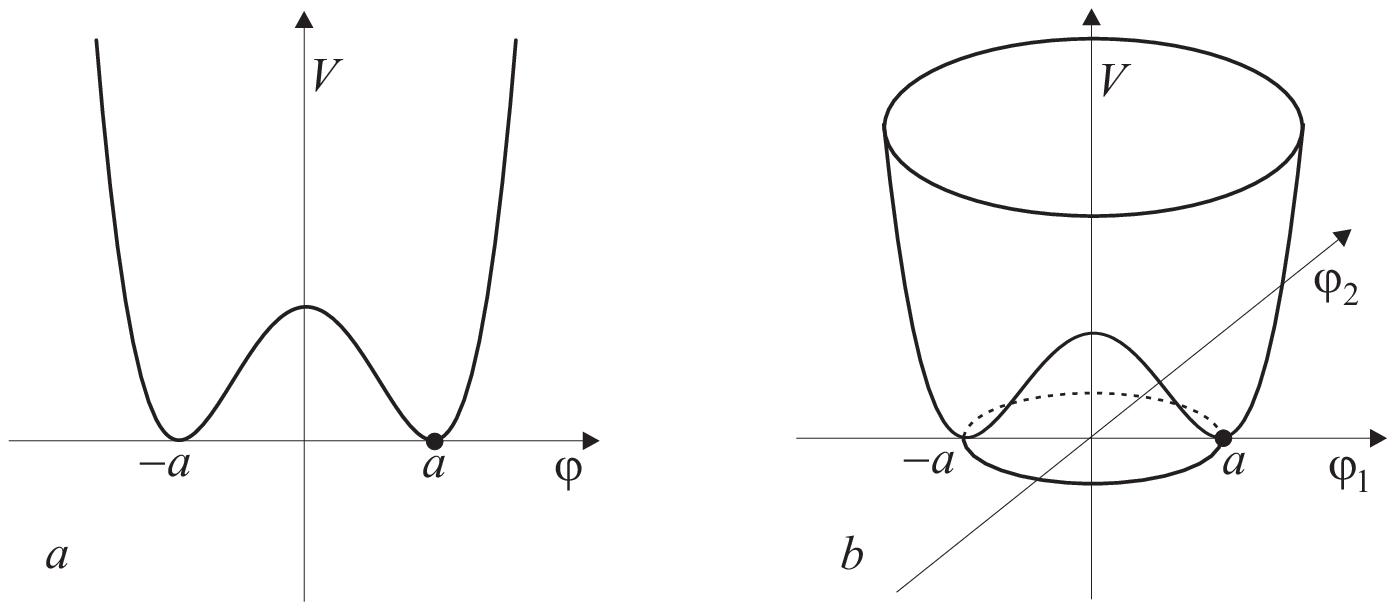}
\hfill {}
\centering\caption{Минимум потенциала действительного $(a)$ и комплексного $(b)$
скалярного поля.}
\label{fminscapot}
\end{figure}

Функция $V(\vf)$ может быть положительно определена и при ``неправильном'' знаке
квадрата массы, $m^2<0$. Например, положим (см.\ рис.\ref{fminscapot},$a$)
\begin{equation}                                                  \label{epolch}
  V(\vf)=\frac14\lm(\vf^2-a^2)^2=\frac14\lm(\vf^4-2\vf^2a^2+a^4),\qquad a>0.
\end{equation}
При $\lm>0$ потенциал положительно определен, однако `` квадрат массы'',
$m^2=-\lm a^2$, отрицателен. Здесь возникает вопрос: ``Что называется массой
частицы, соответствующей полю $\vf$?'' В квантовой теории поля масса частиц
определяется квадратичным приближением лагранжиана, вокруг которого строится
теория возмущений. Для этого необходимо выбрать вакуумное решение уравнений
движения, вблизи которого будет происходить разложение полей. Простейшее решение
уравнения движения (\ref{eqmrsn}) -- это постоянное поле $\vf_0=\const$, которое
удовлетворяет условию
\begin{equation*}
  V'(\vf_0)=0.
\end{equation*}
То есть мы предполагаем, что вакуум однороден и статичен. Для потенциала
(\ref{epolch}) уравнение равновесия принимает вид
\begin{equation*}
  \vf(\vf^2-a^2)=0.
\end{equation*}
Оно имеет три решения: $\vf_0=0$, $\vf_0=\pm a$. Решение $\vf_0=0$ соответствует
неустойчивому положению равновесия. Именно поэтому вблизи данного решения
квадрат массы имеет неправильный знак. Оба решения $\vf_0=\pm a$ соответствуют
устойчивым положениям равновесия и имеют одинаковую нулевую плотность энергии.
Любое из этих решений можно выбрать в качестве вакуума, вокруг которого строится
теория возмущений. Положим
\begin{equation*}
  \vf=a+\tilde\vf.
\end{equation*}
Тогда потенциал примет вид
\begin{equation*}
  V=\lm\left(a^2\tilde\vf^2+a\tilde\vf^3+\frac14\tilde\vf^4\right).
\end{equation*}
Из квадратичного приближения следует, что масса скалярного поля положительна,
$m^2=2\lm a^2$.
То есть масса частиц положительна, что соответствует устойчивому положению
равновесия.

Перейдем к канонической формулировке. Импульс, канонически сопряженный
скалярному полю, равен производной по времени от скалярного поля
\begin{equation}                                                  \label{escfmo}
  p:=\frac{\pl L}{\pl(\dot\vf)}=\dot\vf.
\end{equation}
Одновременн\'ые скобки Пуассона для канонически сопряженных переменных имеют вид
$$
  [\vf,p']:=[\vf(t,\Bx),p(t,\Bx')]=\dl(\Bx-\Bx'),
$$
где введена пространственная $\dl$-функция
\begin{equation*}
  \dl(\Bx-\Bx'):=\dl(x^1-x^{\prime1})\dotsc\dl(x^{n-1}-x^{\prime\, n-1}).
\end{equation*}
Соответствующая плотность гамильтониана равна
\begin{equation}                                                  \label{escfha}
  H=T_{00}^{(c)}=\frac12p^2-\frac12\eta^{\mu\nu}\pl_\mu\vf\pl_\nu\vf+V(\vf),
\end{equation}
где греческие буквы из середины алфавита пробегают, как и раньше, только
пространственные значения: $\mu,\nu=1,\dotsc,n-1$. Если функция $V(\vf)$
положительно определена, то гамильтонова плотность $H$ также положительно
определена, т.к.\ $\eta^{\mu\nu}:=-\dl^{\mu\nu}$.

Гамильтониан скалярного поля получается интегрированием гамильтоновой плотности
по всему пространству:
\begin{equation}                                                  \label{ehascf}
  \CH=\int\! d\Bx H.
\end{equation}
Если этот интеграл сходится, то полная энергия скалярного поля сохраняется.

Уравнение движения второго порядка (\ref{eqmrsc}) в гамильтоновой форме
эквивалентно системе двух уравнений движения первого порядка относительно
производных по времени:
\begin{equation}                                                  \label{ehasco}
\begin{split}
  \dot\vf&=[\vf,\CH]=p,
\\
  \dot p&=[p,\CH]=\triangle\vf-m^2\vf-V',
\end{split}
\end{equation}
где $\triangle$ -- оператор Лапласа.

Компонента $T_{00}^{(c)}$ тензора энергии-импульса скалярного поля
(\ref{enmots}) совпадает с плотностью гамильтониана (\ref{escfha}). Остальные
компоненты канонического тензора энергии-импульса также могут быть выражены
через канонические переменные:
\begin{align*}
  H_\mu&:=T_{0\mu}^{(c)}=p\pl_\mu\vf,
\\
  T_{\mu\nu}&=\pl_\mu\vf\pl_\nu\vf-\left[
  \frac12(p^2+\pl^\rho\vf\pl_\rho\vf-V(\vf)\right]\eta_{\mu\nu}.
\end{align*}

Ковектор энергии-импульса $\lbrace P_\al\rbrace=\lbrace P_0:=\CH,P_\mu\rbrace$
для скалярного поля получается из компонент канонического тензора
энергии-импульса интегрированием по пространству:
\begin{equation}                                                  \label{enmovz}
  P_0:=\int\!d\Bx\,H,\qquad P_\mu:=\int\!d\Bx\,H_\mu,
\end{equation}
где интегрирование проводится по сечениям $x^0=\const$. Численное значение $P_0$
и $P_\mu$ дают полную энергию и полный импульс скалярного поля. Если поле
достаточно быстро убывает на бесконечности, что соответствует, в частности,
отсутствию излучения, то полные энергия и импульс сохраняются:
\begin{equation*}
  \dot P_0=0,\qquad \dot P_\mu=0.
\end{equation*}
Отсюда следует, что компоненты канонического тензора энергии-импульса
$T_{0\mu}^{(c)}$ имеют физический смысл плотности импульса скалярного поля.

Полный момент импульса скалярного поля получается интегрированием нулевых
компонент тензора момента количества движения (\ref{emokod}):
\begin{equation*}
  \CJ_{\al\bt}:=\int\!d\Bx J_{\al\bt}{}^0.
\end{equation*}
Если скалярное поле достаточно быстро убывает на пространственной бесконечности,
то он также сохраняется,
\begin{equation*}
  \dot\CJ_{\al\bt}=0.
\end{equation*}
Строго говоря, полным моментом импульса являются только пространственные
компоненты $\CJ_{\mu\nu}=-\CJ_{\nu\mu}$.
\subsection{Скалярное поле в аффинной геометрии}
Пусть на многообразии (пространстве-времени) $\MM$ произвольной размерности $n$
задана аффинная геометрия, т.е.\ задана метрика $g$ лоренцевой сигнатуры и
связность $\Gamma$. Лагранжиан скалярного поля, минимальным образом
взаимодействующего с гравитацией, выбирается в виде
\begin{align}                                                     \label{escfag}
  L&=\vol\left(\frac12\pl\vf^2-V(\vf)\right),
\\ \intertext{где введено сокращенное обозначение,}                    \nonumber
  \pl\vf^2&:=g^{\al\bt}\pl_\al\vf\pl_\bt\vf,
\end{align}
для кинетической части лагранжиана. Лагранжиан (\ref{escfag}) зависит только от
метрики, а аффинная связность в него не входит. Это означает, что лагранжиан
скалярного поля при минимальной подстановке имеет один и тот же вид как в
римановой, так и в аффинной геометрии.

Вычислим вариационные производные действия:
\begin{align}                                                     \label{eqmsag}
  S,{}_\vf&:=\frac{\dl S}{\dl\vf}=-\vol\left(\widetilde\square\vf+V'(\vf)
  \right)=0,
\\                                                                \label{eqmmag}
  S,{}_{\al\bt}&:=\frac{\dl S}{\dl g^{\al\bt}}=\frac12\vol T_{\al\bt},
\\ \intertext{где}                                                     \nonumber
  \widetilde\square&:=g^{\al\bt}\widetilde\nb_\al\widetilde\nb_\bt.
\end{align}
-- инвариантный волновой оператор, построенный по псевдоримановой метрике
$g_{\al\bt}$ и
\begin{equation}                                                  \label{enmois}
  T_{\al\bt}=\pl_\al\vf\pl_\bt\vf-g_{\al\bt}\left(\frac12\pl\vf^2-V\right)
\end{equation}
-- ковариантное обобщение тензора энергии-импульса (\ref{enmots}) для
пространства Минковского.
Уравнение (\ref{eqmsag}) является инвариантным уравнением движения для
скалярного поля в аффинной геометрии. Однако в него входит не аффинная
связность, которая может быть задана на многообразии, а только символы
Кристоффеля. Это происходит потому что скалярное поле не чувствует кручения и
неметричности при минимальной подстановке. Взаимодействие скалярного поля с
кручением и неметричностью можно ввести неминимальным образом, добавив к
лагранжиану соответствующие слагаемые. При этом существует много возможностей,
поэтому мы их обсуждать не будем. Вариационная производная (\ref{enmois})
определяет тензор энергии-импульса скалярного поля, который служит источником
для гравитационного поля (компонент метрики) в уравнениях Эйнштейна.
\begin{com}
То, что вариационная производная действия по метрике (\ref{enmois})
пропорциональна ковариантному обобщению канонического тензора энергии-импульса
не является общим свойством. Дальнейшие примеры покажут, что вариация действия
по метрике в общем случае не всегда пропорциональна ковариантному обобщению
тензора энергии-импульса.
\qed\end{com}
\begin{prop}                                                      \label{pzezsc}
Если метрика на многообразии $\MM$ имеет лоренцеву сигнатуру и координата $x^0$
является временем, то временн\'ая компонента тензора энергии-импульса
действительного скалярного поля,
\begin{equation}                                                  \label{ettsca}
  T_{00}=\pl_0\vf\pl_0\vf-g_{00}\left(\frac12g^{\al\bt}\pl_\al\vf\pl_\bt\vf
  -V\right),
\end{equation}
при $V\ge0$ положительно определена и, следовательно удовлетворяет слабому
энергетическому условию (\ref{ewenco}).
\end{prop}
\begin{proof}
Обозначим $u^\al:=g^{\al\bt}\pl_\bt\vf$. Тогда
\begin{equation*}
  \pl_0\vf=g_{00}u^0+g_{0\mu}u^\mu.
\end{equation*}
Легко проверить равенство
\begin{equation}                                                  \label{qhhyfr}
  (\pl_0\vf)^2-\frac12g_{00}g^{\al\bt}\pl_\al\vf\pl_\bt\vf
  =\frac12(g_{00}u^0-g_{0\mu}u^\mu)^2-\frac12g_{00}
  \left(g_{\mu\nu}-\frac{g_{0\mu}g_{0\nu}}{g_{00}}\right)u^\mu u^\nu.
\end{equation}

Если метрика имеет лоренцеву сигнатуру и координаты выбраны так, что $x^0$ --
это время, то из теоремы \ref{tlosim} следует, что $g_{00}>0$ и матрица
\begin{equation*}
  g_{\mu\nu}-\frac{g_{0\mu}g_{0\nu}}{g_{00}}
\end{equation*}
отрицательно определена. Это означает, что временн\'ая компонента (\ref{ettsca})
является положительно определенной как сумма положительно определенных
слагаемых в выражении (\ref{qhhyfr}).
\end{proof}

Если действие инвариантно относительно общих преобразований координат, которые
параметризуются $n$ функциями, то согласно второй теореме Нетер уравнения
движения удовлетворяют $n$ тождествам. Допустим, что действие зависит только от
метрики и скалярного поля. Тогда инвариантность действия означает равенство нулю
вариации
$$
  \dl S=\int dx \vol(S,{}^{\al\bt}\dl g_{\al\bt}+S,{}_\vf\dl\vf)=0.
$$
Отсюда с учетом явного вида вариации компонент метрики (\ref{eitcms}) и
скалярного поля (\ref{einfts}) получаем, что уравнения движения (\ref{eqmsag}) и
(\ref{eqmmag}) удовлетворяют $n$ тождествам:
\begin{equation}                                                  \label{edescg}
  2\widetilde\nb_\al S,{}^\al{}_\bt-S,{}_\vf\pl_\bt\vf=0,
\end{equation}
где $\widetilde\nb_\al$ -- ковариантная производная с символами Кристоффеля.

С формальной точки зрения сдвиги на постоянный вектор,
$x^\al\mapsto x^\al+a^\al$, $a^\al=\const$, образуют подгруппу группы общих
преобразований координат. Поэтому, так же, как и в пространстве Минковского,
можно построить полную ``энергию'' и ``импульс'' скалярного поля. Эти величины
будут сохраняться на уравнениях движения, однако им не всегда можно придать
физический смысл, т.к.\ понятие декартовой системы координат в общем случае
отсутствует. Это построение имеет смысл в асимптотически плоском
пространстве-времени, когда на больших расстояниях пространство-время
приближается к пространству Минковского. Это же верно и для момента количества
движения.
\subsection{Скалярное поле в общей теории относительности        \label{scfger}}
Рассмотрим действительное скалярное поле, которое минимальным образом
взаимодействует с гравитацией в общей теории относительности. В этом случае
действие имеет вид
\begin{equation}                                                  \label{escari}
  S=\kappa S_{\Sh\Se}+S_\vf,
\end{equation}
где $S_{\Sh\Se}$ -- действие Гильберта--Эйнштейна (\ref{ehieia}) и $S_\vf$ --
действие для скалярного поля с лагранжианом (\ref{escfag}). Добавление к
действию скалярного поля действия Гильберта--Эйнштейна дает кинетический член
для метрики. Поэтому вариация этого действия по обратной метрике приводит к
уравнениям движения Эйнштейна
\begin{equation}                                                  \label{evarsg}
  \frac1\vol\frac{\dl S}{\dl g^{\al\bt}}:\qquad
  \kappa\left(\widetilde R_{\al\bt}-\frac12g_{\al\bt}\widetilde R\right)
  +g_{\al\bt}\Lm+\frac12 T_{\al\bt}=0,
\end{equation}
где $T_{\al\bt}$ -- тензор энергии-импульса скалярного поля (\ref{enmois}).
Поскольку действие Гильберта--Эйнштейна не зависит от скалярного поля, то
уравнение движения для скалярного поля остается прежним (\ref{eqmsag}). Таким
образом, полная система уравнений движения для скалярного поля в общей теории
относительности состоит из уравнений (\ref{eqmsag}) и (\ref{evarsg}).

Действие (\ref{escari}) инвариантно относительно общих преобразований координат.
Поэтому, согласно второй теореме Нетер, между уравнениями движения существует
линейная зависимость (\ref{edescg}).

Перепишем уравнения Эйнштейна в другом виде. След уравнений Эйнштейна
(\ref{evarsg}),
$$
  \kappa\left(1-\frac n2\right)\widetilde R+n\Lm+\frac12\left(1-\frac n2\right)
  \pl\vf^2+\frac n2V=0,
$$
позволяет исключить скалярную кривизну. В результате уравнения движения
(\ref{evarsg}) можно записать в эквивалентной форме
\begin{equation}                                                  \label{einsca}
  \kappa\widetilde R_{\al\bt}=
  -\frac12\pl_\al\vf\pl_\bt\vf+\frac1{n-2}g_{\al\bt}(2\Lm+V).
\end{equation}

При нулевой космологической постоянной $\Lm=0$ и потенциале $V=0$ уравнения
(\ref{einsca}) существенно упрощаются,
\begin{equation}                                                  \label{esctez}
  \kappa\widetilde R_{\al\bt}=-\frac12\pl_\al\vf\pl_\bt\vf.
\end{equation}
Эта модель безмассового скалярного поля в общей теории относительности
привлекает в последнее время большое внимание из-за своей относительной
простоты.

Покажем, что уравнение движения для скалярного поля (\ref{eqmsag}) являются
следствием уравнений Эйнштейна (\ref{evarsg}).
\begin{prop}                                                      \label{peqsce}
Если в некоторой области пространства-времени $\MU\subset\MM$ градиент
скалярного поля отличен от нуля, $\lbrace\pl_\al\vf\ne\rbrace0$, то в этой
области уравнение для скалярного поля (\ref{eqmsag}) является следствием
уравнений Эйнштейна (\ref{evarsg}).
\end{prop}
\begin{proof}
Подействуем оператором ковариантного дифференцирования $\widetilde\nb^\bt$ на
уравнение (\ref{evarsg}). В силу свернутых тождеств Бианки (\ref{ebieit}),
\begin{equation*}
  \widetilde\nb^\bt\left(
  \widetilde R_{\al\bt}-\frac12g_{\al\bt}\widetilde R\right)=0,
\end{equation*}
получим равенство
\begin{equation*}
  \widetilde\nb^\bt T_{\al\bt}=\pl_\al\vf(\widetilde\square\vf+V')=0.
\end{equation*}
Отсюда вытекает сделанное утверждение.
\end{proof}
Доказанное предложение позволяет вместо решения полной системы уравнений для
скалярного поля и метрики ограничится решением только уравнений Эйнштейна.

В рассматриваемом случае ситуация аналогична точечным частицам в общей теории
относительности, предложение \ref{ptepoi}.

До сих пор мы рассматривали минимальное взаимодействие скалярного с метрикой.
Представляет также интерес неминимальное взаимодействие, поскольку в этом случае
возможно появление дополнительной важной локальной инвариантности.
\begin{theorem}
Действие
\begin{equation}                                                  \label{eacwey}
  S=\int_\MM\!\!\!dx\vol \left(\frac12\pl\vf^2
  -\frac{n-2}{8(n-1)}\vf^2\widetilde R-\lm\vf^{\frac{2n}{n-2}}\right),
  \qquad \lm=\const,
\end{equation}
с точностью до граничных слагаемых инвариантно относительно преобразований
Вейля:
\begin{equation}                                                  \label{eweyte}
\begin{split}
  g_{\al\bt}&\mapsto \bar g_{\al\bt}=\ex^{2\phi}g_{\al\bt},
\\
  \vf&\mapsto\bar\vf=\ex^{-\frac{n-2}2\phi}\vf,
\end{split}
\end{equation}
где $\phi(x)\in\CC^2(\MM)$ -- произвольная функция.
\end{theorem}
\begin{proof}
Прямая проверка.
\end{proof}

Учитывая размерность скалярного поля (\ref{edisca}), получаем, что константа
связи $\lm$ самодействия скалярного поля в действии (\ref{eacwey}) безразмерна.

Коэффициенты перед первыми двумя слагаемыми в действии (\ref{eacwey})
фиксированы требованием инвариантности относительно преобразований Вейля. Их
можно умножить только на общую отличную от нуля постоянную. Фиксирован также
показатель степени в преобразовании Вейля для скалярного поля (\ref{eweyte}).
Нетрудно также проверить, что добавление к действию (\ref{eacwey}) действия
Гильберта--Эйнштейна и космологической постоянной нарушает вейлевскую
инвариантность.

Уравнения движения для действия (\ref{eacwey}) имеют следующий вид
\begin{align}                                                          \nonumber
  \frac1\vol S,^{\al\bt}:\qquad &
  \frac{n-2}{8(n-1)}\left[\vf^2\left(\widetilde R^{\al\bt}
  -\frac12g^{\al\bt}\widetilde R\right)-g^{\al\bt}\square\vf^2
  +\widetilde\nb^\al\widetilde\nb^\bt\vf^2\right]+
\\                                                                \label{einwey}
  &\qquad\qquad\qquad+\frac14g^{\al\bt}\pl\vf^2-\frac12\pl^\al\vf\pl^\bt\vf
  -\frac12\lm g^{\al\bt}\vf^{\frac{2n}{n-2}}=0,
\\                                                                \label{escawe}
  \frac1\vol S,_\vf:\qquad &
  -\widetilde\square\vf-\frac{n-2}{4(n-1)}\widetilde R\vf
  -\lm\frac{2n}{n-2}\vf^{\frac{n+2}{n-2}}=0.
\end{align}
Эти уравнения движения не инвариантны, а ковариантны относительно преобразований
Вейля, т.к.\ при преобразовании умножаются на некоторую отличную от нуля
функцию.

Поскольку действие (\ref{eacwey}) инвариантно относительно преобразований Вейля,
то в силу второй теоремы Нетер между уравнениями движения существует линейная
зависимость. Чтобы ее найти, выпишем вариации формы для метрики и скалярного
поля при вейлевских преобразованиях (\ref{eweyte}):
\begin{equation*}
  \dl g_{\al\bt}=2\e g_{\al\bt},\qquad \dl\vf=-\frac{n-2}2\e\vf,
\end{equation*}
где $\e(x)$ -- бесконечно малый параметр преобразования. Поскольку действие
инвариантно, то выполнено условие
\begin{equation*}
  \dl S=\int dx\left(\frac{\dl S}{\dl g_{\al\bt}}\dl g_{\al\bt}
  +\frac{\dl S}{\dl\vf}\dl\vf\right)=0.
\end{equation*}
Подставив сюда выражения для вариации полей, получим зависимость уравнений
движения:
\begin{equation}                                                  \label{ezaequ}
  2S,^{\al\bt}g_{\al\bt}-\frac{n-2}2\vf S,_\vf=0.
\end{equation}
Нетрудно проверить, что уравнения движения (\ref{einwey}), (\ref{escawe})
действительно удовлетворяют этому тождеству.

В важном случае четырехмерного пространства-времени, действие принимает хорошо
известную форму
\begin{equation}                                                  \label{eacwyy}
  S=\int_\MM\!\!\!dx\vol \left(\frac12\pl\vf^2
  -\frac1{12}\vf^2\widetilde R-\lm\vf^4\right).
\end{equation}
Оно инвариантно относительно преобразований Вейля
\begin{equation}                                                  \label{eweyfo}
  g_{\al\bt}\mapsto\bar g_{\al\bt}=\ex^{2\phi}g_{\al\bt},
  \qquad \vf\mapsto\bar\vf=\ex^{-\phi}\vf.
\end{equation}
Соответствующие уравнения движения,
\begin{align}                                                          \nonumber
  \frac1\vol S,^{\al\bt}:\qquad &
  \frac1{12}\left[\vf^2\left(\widetilde R^{\al\bt}
  -\frac12g^{\al\bt}\widetilde R\right)-g^{\al\bt}\square\vf^2
  +\widetilde\nb^\al\widetilde\nb^\bt\vf^2\right]-
\\                                                                \label{egnwey}
  &\qquad\qquad\qquad-\frac12\pl^\al\vf\pl^\bt\vf+\frac14g^{\al\bt}\pl\vf^2
  -\frac12\lm g^{\al\bt}\vf^4=0,
\\                                                                \label{egcawe}
  \frac1\vol S,_\vf:\qquad &
  -\widetilde\square\vf-\frac16\widetilde R\vf-4\lm\vf^3=0,
\end{align}
ковариантны.

Действие (\ref{eacwyy}) является выделенным случаем скалярно-тензорных моделей
гравитации, рассмотренных в разделе \ref{sctemo}. Чтобы привести лагранжиан
(\ref{esctel}) к рассматриваемому виду, необходимо положить
\begin{equation*}
  \phi=-\frac{\vf^2}{12},\qquad \om(\phi)=-\frac32,\qquad V(\phi)=\lm\vf^4,
  \qquad \kappa=1.
\end{equation*}
Построение действия для других полей, которое было бы инвариантно относительно
преобразований Вейля, представляет собой отдельную задачу, на которой мы
останавливаться не будем.
\section{Комплексное скалярное поле                              \label{scfcom}}
\subsection{Комплексное скалярное поле в пространстве Минковского}
Пусть $\vf(x)=\vf_1(x)+i\vf_2(x)$ -- комплексное скалярное поле (функция) в
пространстве Минковского $\MR^{1,n-1}$, где $\vf_{1,2}(x)\in\CC^2(\MR^{1,n-1})$
-- два действительных скалярных поля (действительная и мнимая части). Комплексно
сопряженное поле будем обозначать крестом $\vf^\dagger:=\vf_1-i\vf_2$, как это
принято в квантовой теории поля. Лагранжиан свободного поля
\begin{equation}                                                  \label{elacsf}
\begin{split}
  L&=\et^{\al\bt}\pl_\al\vf^\dagger\pl_\bt\vf-m^2\vf^\dagger\vf=
\\
   &=\et^{\al\bt}(\pl_\al\vf_1\pl_\bt\vf_1+\pl_\al\vf_2\pl_\bt\vf_2)
   -m^2(\vf_1^2+\vf_2^2)
\end{split}
\end{equation}
с точностью до множителя $1/2$ является суммой лагранжианов (\ref{elascf}) для
действительной и мнимой части. Он является действительной функцией.

Размерность комплексного скалярного поля такая же как и действительного
(\ref{edisca}).

Комплексное скалярное поле задается двумя действительными скалярными полями
$\vf_1$ и $\vf_2$. Для получения уравнений движения действие можно варьировать
либо по действительной и мнимой частям $\vf_1$ и $\vf_2$, либо по полям $\vf$ и
$\vf^\dagger$, рассматривая их в качестве независимых переменных. Мы остановимся
на второй возможности как более удобной. Тогда вариация действия по
$\vf^\dagger$ приводит к волновому уравнению для $\vf$ (уравнению
Клейна--Гордона--Фока)
\begin{equation}                                                  \label{eqmcsf}
  \frac{\dl S}{\dl\vf^\dagger}=-(\square+m^2)\vf=0,
\end{equation}
которое эквивалентно двум волновым уравнениям для действительной и мнимой части.
Вариация действия по $\vf$ дает уравнение, сопряженное к (\ref{eqmcsf}).

При рассмотрении задачи Коши для волнового уравнения (\ref{eqmcsf}) необходимо
задать начальные условия независимо для полей $\vf_1$ и $\vf_2$. Поэтому
комплексное скалярное поле описывает две степени свободы и в квантовой теории
поля описывает заряженные скалярные частицы.

Для описания самодействия комплексного скалярного поля вводится потенциал
$V(\vf^\dagger,\vf)$, который в общем случае является достаточно гладкой
вещественнозначной функцией двух аргументов $\vf^\dagger$ и $\vf$. Мы
предположим, что потенциал зависит только от одного аргумента -- произведения
$\vf^\dagger\vf$ -- и обозначим $V':=dV/d(\vf^\dagger\vf)$. Тогда лагранжиан
примет вид
\begin{equation}                                                  \label{elagsc}
  L=\eta^{\al\bt}\pl_\al\vf^\dagger\pl_\bt\vf-V(\vf^\dagger\vf),
\end{equation}
где, как и в случае действительного скалярного поля, мы включили массовый член в
определение потенциала $V$. Уравнение движения для скалярного поля с
самодействием становится нелинейным:
\begin{equation}                                                  \label{esceco}
  \frac{\dl S}{\dl\vf^\dagger}=-(\square+V')\vf=0,
\end{equation}

Как и в случае действительного скалярного поля, действие для комплексного
скалярного поля инвариантно относительно группы Пуанкаре $\MI\MO(1,n-1)$. При
этом трансляциям соответствует сохраняющийся канонический тензор
энергии-импульса
\begin{equation}                                                  \label{enmotc}
  T^{(c)}_{\al\bt}=\pl_\al\vf^\dagger\pl_\bt\vf+\pl_\bt\vf^\dagger\pl_\al\vf
  -\et_{\al\bt}L.
\end{equation}
Его компоненты, очевидно, вещественны. Спиновый момент комплексного скалярного
поля равен нулю и момент количества движения равен орбитальному моменту
(\ref{emokod}).

Действие для комплексного скалярного поля инвариантно также относительно
глобальных $\MU(1)$ преобразований, меняющих фазу скалярного поля сразу во всем
пространстве
\begin{equation}                                                  \label{eglptr}
  \vf'=e^{i\e}\vf,\qquad \vf^{\prime\dagger}=e^{-i\e}\vf^\dagger,\qquad
  \e=\const.
\end{equation}
Согласно первой теореме Нетер, этой инвариантности соответствует сохраняющийся
ток (\ref{ethgen})
\begin{equation*}
  \pl_\al J^\al=0.
\end{equation*}
где
\begin{equation}                                                  \label{ejcosf}
  J_\al=i(\vf^\dagger\pl_\al\vf-\pl_\al\vf^\dagger\vf).
\end{equation}
Позже мы увидим, что $\MU(1)$ инвариантность модели комплексного скалярного поля
соответствует закону сохранения заряда.

Рассмотрим гамильтонову формулировку теории. Из лагранжиана (\ref{elagsc})
следуют выражения для канонических импульсов, сопряженных полям $\vf$ и
$\vf^\dagger$,
\begin{align}                                                     \label{emoscw}
  p&:=\frac{\pl L}{\pl(\pl_0\vf)}=\pl_0\vf^\dagger,
\\                                                                \label{emoscs}
  p^\dagger&:=\frac{\pl L}{\pl(\pl_0\vf^\dagger)}=\pl_0\vf.
\end{align}
Ненулевые одновременн\'ые скобки Пуассона для координат и импульсов имеют вид
$$
  [\vf(t,\Bx),p(t,\Bx')]=[\vf^\dagger(t,\Bx),p^\dagger(t,\Bx')]=\dl(\Bx-\Bx').
$$

Для положительно определенного потенциала $V$ плотность гамильтониана
комплексного скалярного поля,
\begin{equation}                                                  \label{ecsfha}
  H=p^\dagger p-\eta^{\mu\nu}\pl_\mu\vf^\dagger\pl_\nu\vf+V(\vf^\dagger\vf),
\end{equation}
положительно определена. Канонические уравнения движения принимают вид
\begin{equation}                                                  \label{ecacso}
\begin{split}
  \dot\vf&=p^\dagger,
\\
  \dot\vf^\dagger&=p,
\\
  \dot p&=\triangle\vf^\dagger-V'\vf^\dagger,
\\
  \dot p^\dagger&=\triangle\vf-V'\vf.
\end{split}
\end{equation}

Теперь рассмотрим обобщение модели (\ref{epolch}) для комплексного скалярного
поля. Потенциал самодействия выберем в виде
\begin{equation}                                                  \label{evslmf}
  V(\vf^\dagger\vf)=\frac12\lm\big[(\vf^\dagger\vf)^2-a^2\big]^2
  =\frac12\lm\big[(\vf^\dagger\vf)^2-2\vf^\dagger\vf a^2+a^4\big],\qquad a>0.
\end{equation}
При $\lm>0$ этот потенциал положительно определен. Однородное и статичное
решение $\vf_0$ для вакуума должно удовлетворять уравнению
\begin{equation*}
  \vf_0(\vf^\dagger_0\vf_0-a^2)=0.
\end{equation*}
Это уравнение имеет бесконечно много решений:
\begin{equation*}
  \vf_0=0\qquad \text{и}\qquad \vf_0=a\ex^{i\al},\qquad\al=\const,
\end{equation*}
которые зависят от параметра $\al$. Это связано с наличием унитарной $\MU(1)$
глобальной симметрии (\ref{eglptr}). Решение $\vf_0=0$ соответствует
неустойчивому положению равновесия. Поэтому выберем вакуумное решение в виде
$\vf_0=a$ и разложим поля вблизи данного решения:
\begin{equation}                                                  \label{qvacas}
  \vf_1=a+\tilde\vf_1,\qquad \vf_2=\vf_2.
\end{equation}
В новых переменных потенциал примет вид
\begin{equation*}
  V=\frac12\lm[4a^2\tilde\vf_1^2+4a\tilde\vf_1^3+4a\tilde\vf_1\vf_2^2
  +\tilde\vf_1^4+\vf_2^4+2\tilde\vf_1^2\vf_2^2].
\end{equation*}
Его квадратичная часть равна
\begin{equation*}
  V\simeq2\lm a^2\tilde\vf_1^2.
\end{equation*}
Таким образом, вещественная часть скалярного поля имеет положительную массу, в
то время как мнимая часть вблизи вакуумного решения является безмассовой. Это
ясно из рис.\ref{fminscapot},$b$. Точка $\vf=a$ соответствует минимуму
потенциальной энергии, однако в направлении $\vf_2$ этот минимум является
безразличным.

Описанная выше ситуация возникновения безмассовых полей в моделях с симметриями
является типичной. Допустим, что модель зависит от некоторого набора скалярных
полей $\vf=\lbrace\vf^a\rbrace$, $a=1,\dotsc,\Sn$, которые можно рассматривать,
как компоненты некоторого вектора. Предположим также, что потенциал
взаимодействия инвариантен относительно некоторой группы Ли преобразований
$\MG$. Если группа Ли $\Sk$-мерна и вблизи единицы параметризуется набором
параметров $\e^\Sa$, $\Sa=1,\dotsc,\Sk$, то инфинитезимальные преобразования
полей имеют вид
\begin{equation*}
  \vf(x)\mapsto\vf(x)+\dl\vf(x)=\vf(x)+\e^\Sa R_\Sa{}^a(\vf),
\end{equation*}
где $R_\Sa{}^a(\vf)$ -- некоторые функции, возможно, нелинейные, полей $\vf^a$.
Из инвариантности потенциала вытекает равенство нулю его вариации:
\begin{equation}                                                  \label{epoing}
  \dl V(\vf)=\e^\Sa R_\Sa{}^a\frac{\pl V}{\pl\vf^a}=0.
\end{equation}
Это условие должно выполняться при всех значениях $\vf$. Допустим, что
существует вакуумное решение $\vf_0=\const$, которое соответствует минимуму
потенциала $V(\vf)$. Тогда выполнено уравнение
\begin{equation*}
  \left.\frac{\pl V}{\pl\vf^a}\right|_{\vf_0}=0.
\end{equation*}
Само вакуумное решение может нарушать симметрию полностью или частично. Если все
векторы $R_\Sa(\vf_0)\ne0$ линейно независимы, то симметрия полностью нарушена.
Если существуют некоторые наборы параметров $\e^\Sa$, для которых
$\e^\Sa R_\Sa(\vf_0)=0$, то при соответствующих преобразованиях вакуум не
меняется, и соответствующая симметрия является симметрией вакуумного решения.
Очевидно, что множество преобразований, сохраняющих вакуум, образуют некоторую
подгруппу $\MH\subset\MG$. Описанное явление называется {\em спонтанным
нарушением симметрии}, и было описано в \cite{Goldst61}.
\index{Спонтанное нарушение симметрии (spontaneous symmetry breaking)}%
\index{Нарушение симметрии спонтанное (spontaneous symmetry breaking)}%
Слово ``спонтанный'' в данном случае означает, что симметрия нарушена не руками
в исходном действии, а выбором вакуумного решения уравнений движения.

\begin{theorem}[\bf Голдстоун]
Если потенциал самодействия скалярных полей инвариантен относительно некоторой
группы Ли $\MG$ преобразований (\ref{epoing}) и вакуумное решение
нарушает симметрию до некоторой собственной подгруппы $\MH\subset\MG$, то
соответствующая модель содержит безмассовые скалярные поля, соответствующие
нарушенным симметриям. Число безмассовых полей больше или равно разности
$\dim\MG-\dim\MH$.
\end{theorem}
\index{Теорема Голдстоуна (Goldstone theorem)}%
\index{Голдстоуна теорема (Goldstone theorem)}%
\begin{proof}
Продифференцируем равенство (\ref{epoing}) по $\vf^b$:
\begin{equation*}
  \e^\Sa\left(\frac{\pl R_\Sa{}^a}{\pl\vf^b}\frac{\pl V}{\pl\vf^a}
  +R_\Sa{}^a\frac{\pl^2 V}{\pl\vf^a\pl\vf^b}\right)=0.
\end{equation*}
Если $\vf_0$ -- вакуумное решение уравнений движения, то первое слагаемое
обращается в нуль и, следовательно, должно быть выполнено равенство
\begin{equation}                                                  \label{enegol}
  \left.\e^\Sa R_\Sa{}^a\frac{\pl^2 V}{\pl\vf^a\pl\vf^b}\right|_{\vf_0}=0.
\end{equation}
Массы полей определяются квадратичным приближением потенциала вблизи вакуумного
решения. Поэтому массовая матрица имеет вид
\begin{equation*}
  m^2_{ab}:=\left.\frac{\pl^2 V}{\pl\vf^a\pl\vf^b}\right|_{\vf_0}.
\end{equation*}
Собственные значения этой матрицы не могут быть отрицательны, т.к.\ $\vf_0$
соответствует минимуму потенциала. Она симметрична и может быть приведена к
диагональному виду с помощью ортогональной матрицы (теорема \ref{tdiama}).
Система линейных уравнений (\ref{enegol}) для $\e^\Sa R_\Sa$ имеет нетривиальные
решения тогда и только тогда, когда массовая матрица вырождена. Каждой
нарушенной симметрии соответствует некоторый нетривиальный вектор
$\e^\Sa R_\Sa$. Таких линейно независимых векторов ровно $\dim\MG-\dim\MH$. Это
означает, что массовая матрица должна иметь по крайней мере $\dim\MG-\dim\MH$
нулевых собственных значений. В общем случае возможно появление дополнительных
нулевых собственных значений, вызванных не нарушением симметрии, а спецификой
модели. Нулевые собственные числа массовой матрицы означают наличие безмассовых
скалярных полей.
\end{proof}
\begin{com}
В доказанной теореме возможная зависимость параметров преобразования $\e^\Sa$ от
точки пространства-времени никак не использовалась. Поэтому теорема Голдстоуна
справедлива как для глобальных, так и для локальных (калибровочных)
преобразований.
\qed\end{com}
\begin{exa}
Рассмотрим комплексное скалярное поле с положительно определенным потенциалом
\begin{equation*}
  V=\frac14\nu\big(\vf^\dagger\vf-a^2\big)^4,\qquad \nu,a>0.
\end{equation*}
Вакуумное решение должно удовлетворять уравнению
\begin{equation*}
  \vf_0(\vf^\dagger_0\vf_0-a^2)^3=0.
\end{equation*}
Как и для потенциала (\ref{evslmf}) минимуму энергии соответствует бесконечно
много решений. Выберем вакуумное решение в виде $\vf_0=a$ и произведем вблизи
него разложение (\ref{qvacas}). Тогда потенциал примет вид
\begin{equation*}
  V=\frac14\nu\big(2a\tilde\vf_1+\tilde\vf_1^2+\vf_2^2\big)^4.
\end{equation*}
Этот потенциал вообще не имеет квадратичных слагаемых, и, значит, оба поля
$\tilde\vf_1$ и $\vf_2$ являются безмассовыми. В то же время, вакуумное решение
$\vf_0=a$ нарушает $\MU(1)$-симметрию. Таким образом, число безмассовых полей
превышает разность $\dim\MG-\dim\MH=1$.
\qed\end{exa}
В рассмотренном примере появление дополнительного безмассового поля связано не
с нарушением симметрии, а со спецификой модели. В квантовой теории поля
безмассовые частицы, соответствующие безмассовым скалярным полям в спонтанно
нарушенных моделях теории поля, которые связаны именно с нарушением симметрии,
называются {\em голдстоуновскими бозонами}. Их число равно разности
$\dim\MG-\dim\MH$.
\index{Голдстоуновский бозон (Goldstone boson)}%
\index{Бозон голдстоуновский (Goldstone boson)}%
\subsection{Комплексное скалярное поле в аффинной геометрии}
В аффинной геометрии минимальная подстановка для комплексного скалярного
поля приводит к следующему лагранжиану
\begin{equation}                                                  \label{ecsfrg}
  L=\vol\big[g^{\al\bt}\pl_\al\vf^\dagger\pl_\bt\vf-V(\vf^\dagger\vf)\big],
\end{equation}
и аффинная связность в него не входит. Вариация соответствующего действия по
полям приводит к следующим уравнениям движения и тензору энергии-импульса
\begin{align}                                                     \label{ecsfvs}
  S,{}_{\vf^\dagger}:=\frac{\dl S}{\dl\vf^\dagger}&
  =-\vol(\widetilde\square+V')\vf=0,
\\                                                                \label{ecsfvt}
  S,{}_\vf:=\frac{\dl S}{\dl\vf}&
  =-\vol(\widetilde\square+V')\vf^\dagger=0,
\\                                                                \label{ecsfvm}
  \frac{\dl S}{\dl g^{\al\bt}}&=\quad \frac12\vol T_{\al\bt},
\end{align}
где
\begin{equation}                                                  \label{ecenmc}
  T_{\al\bt}=\pl_\al\vf^\dagger\pl_\bt\vf+\pl_\bt\vf^\dagger\pl_\al\vf
  -g_{\al\bt}(g^{\g\dl}\pl_\g\vf^\dagger\pl_\dl\vf-V)
\end{equation}
-- ковариантное обобщение канонического тензора энергии-импульса (\ref{enmotc}).

Перепишем тензор энергии-импульса через действительную и мнимую части,
\begin{equation*}
  T_{\al\bt}=2\pl_\al\vf_1\pl_\bt\vf_1+2\pl_\al\vf_2\pl_\bt\vf_2-
  g_{\al\bt}\big(\pl\vf_1^2+\pl\vf_2^2-V\big).
\end{equation*}
Слагаемые с производными имеют тот же вид, что и в случае двух действительных
скалярных полей. Поэтому из доказательства предложения \ref{pzezsc} вытекает
положительная определенность их временн\'ых компонент. Следовательно,
справедливо
\begin{prop}
Если метрика на многообразии $\MM$ имеет лоренцеву сигнатуру и координата $x^0$
является временем, то временн\'ая компонента тензора энергии-импульса
комплексного скалярного поля,
\begin{equation}                                                  \label{qzezsc}
  T_{00}=\pl_0\vf^\dagger\pl_0\vf+\pl_0\vf^\dagger\pl_0\vf
  -g_{00}(g^{\g\dl}\pl_\g\vf^\dagger\pl_\dl\vf-V),
\end{equation}
при $V\ge0$ положительно определена и, следовательно, удовлетворяет слабому
энергетическому условию.
\end{prop}

Если инвариантное действие зависит только от метрики и скалярного поля, то из
инвариантности действия относительно общих преобразований координат следует
зависимость уравнений движения:
\begin{equation}                                                  \label{ecscde}
  2\widetilde\nb_\al S,{}^\al{}_\bt-S,{}_\vf\pl_\bt\vf
  -S,{}_{\vf^\dagger}\pl_\bt\vf^\dagger=0.
\end{equation}
Эта зависимость уравнений движения будет иметь место, например, в общей теории
относительности, когда к действию Гильберта--Эйнштейна добавляется действие для
комплексного скалярного поля.

Действие для комплексного скалярного поля инвариантно также относительно
$\MU(1)$ преобразований в пространстве-мишени (\ref{eglptr}). Эта инвариантность
сохраняется и в аффинной геометрии, если компоненты метрики и связности не
преобразуются. Таким образом, закон сохранения заряда для комплексного
скалярного поля
\begin{equation*}
  \pl_\al J^\al=0,
\end{equation*}
где
\begin{equation}                                                  \label{qtoksc}
  J^\al=i\vol g^{\al\bt}(\vf^\dagger\pl_\bt\vf-\vf\pl_\bt\vf^\dagger),
\end{equation}
имеет место также на многообразиях с нетривиальной геометрией.
\subsection{Комплексное скалярное поле в общей теории относительности}
В общей теории относительности действие для комплексного скалярного поля при
минимальной подстановке имеет вид
\begin{equation*}
  S=S_{\Sh\Se}+S_\vf,
\end{equation*}
где
\begin{equation*}
  S_\vf=\int\!dx\vol\big[g^{\al\bt}\pl_\al\vf^\dagger\pl_\bt\vf
  -V(\vf^\dagger\vf)\big].
\end{equation*}
Уравнения Эйнштейна имеют прежний вид (\ref{evarsg}), где тензор
энергии-импульса определен равенством (\ref{ecenmc}). Их необходимо дополнить
двумя уравнениями для скалярного поля (\ref{ecsfvs}) и (\ref{ecsfvt}).

Как и в случае действительного скалярного поля действие оператора ковариантного
дифференцирования на уравнения Эйнштейна приводит к ковариантному закону
сохранения тензора энергии-импульса
\begin{equation*}
  \widetilde\nb^\bt T_{\al\bt}=\pl_\al\vf^\dagger(\widetilde\square\vf+V'\vf)
  +\pl_\al\vf(\widetilde\square\vf^\dagger+V'\vf^\dagger)=0.
\end{equation*}
Ясно, что если выполнены уравнения движения (\ref{ecsfvs}), (\ref{ecsfvt}), то
тензор энергии-импульса сохраняется. Однако обратное утверждение теперь неверно,
и аналога предложения \ref{peqsce} в комплексном случае нет.

Как и в случае действительного скалярного поля (\ref{einsca}) из уравнений
Эйнштейна можно исключить скалярную кривизну и переписать их в эквивалентном
виде
\begin{equation}                                                  \label{qcosce}
  \kappa\widetilde R_{\al\bt}
  =-\pl_\al\vf^\dagger\pl_\bt\vf+\frac2{n-2}g_{\al\bt}(\Lm+V).
\end{equation}

При неминимальной подстановке также возможно появление инвариантности
относительно преобразований Вейля.
\begin{prop}
Действие
\begin{equation}                                                  \label{qwecos}
  S=\int_\MM\!\!\!dx\vol\left(g^{\al\bt}\pl_\al\vf^\dagger\pl_\bt\vf
  -\frac{n-2}{4(n-1)}\vf^\dagger\vf\widetilde R
  -\lm(\vf^\dagger\vf)^{\frac n{n-2}}\right),\qquad \lm=\const,
\end{equation}
с точностью до граничных слагаемых инвариантно относительно преобразований
Вейля
\begin{equation}                                                  \label{qwecom}
\begin{split}
  g_{\al\bt}&\mapsto\bar g_{\al\bt}=\ex^{2\phi}g_{\al\bt},
\\
  \vf&\mapsto\bar\vf=\ex^{-\frac{n-2}2\phi}\vf,
\\
  \vf^\dagger&\mapsto\bar\vf^\dagger=\ex^{-\frac{n-2}2\phi}\vf^\dagger,
\end{split}
\end{equation}
где $\phi(x)\in\CC^2(\MM)$ -- произвольная вещественнозначная функция.
\end{prop}
\begin{proof}
Перепишем действие (\ref{qwecos}) через вещественную и мнимую части:
\begin{equation*}
  S=\int_\MM\!\!\!dx\vol\left(\pl\vf_1^2+\pl\vf_2^2
  -\frac{n-2}{4(n-1)}(\vf_1^2+\vf_2^2)\widetilde R
  -\lm(\vf_1^2+\vf_2^2)^{\frac n{n-2}}\right).
\end{equation*}
Первые три слагаемых инвариантны как сумма двух инвариантных слагаемых для
$\vf_1$ и $\vf_2$. Инвариантность последнего слагаемого просто проверяется.
\end{proof}
\section{Электромагнитное поле                                   \label{selema}}
Электромагнитное поле является важнейшим геометрическим объектом в
математической физике. Оно наблюдается в экспериментах и широко используется на
протяжении многих лет.

С геометрической точки зрения в природе существует следующая конструкция.
Пространство-время, в котором мы живем, представляет собой четырехмерное
многообразие $\MM$. Строится главное расслоение
$\MP\big(\MM,\pi,\MU(1)\big)$ (см.\ главу \ref{sprifb}), базой которого является
пространство-время, а структурной группой -- мультипликативная группа Ли
комплексных чисел $\MU(1)$, равных по модулю единице, которая изоморфна группе
двумерных вращений $\MU(1)\simeq\MS\MO(2)$, и, как многообразие,
диффеоморфна окружности, $\MU(1)\approx\MS^1$. На этом главном расслоении
задается связность (см.\ главу \ref{sconne}). Связность, в свою очередь,
определяет форму связности и, после проектирования на базу с какого либо
сечения, локальную форму связности:
\begin{equation*}
  \om=dx^\al A_\al.
\end{equation*}
Группа $\MU(1)$ одномерна, и поэтому форма $\om$ принимает значения в $\MR$,
которое рассматривается как одномерное векторное пространство.
Компоненты локальной формы связности $A_\al(x)$, после добавления уравнений
Максвелла, называются в физике {\em потенциалом электромагнитного поля}, а
соответствующие компоненты
\begin{equation}                                                  \label{qstrem}
  F_{\al\bt}:=\pl_\al A_\bt-\pl_\bt A_\al,
\end{equation}
локальной формы кривизны,
\begin{equation}                                                  \label{qlocuo}
  F=\frac12 dx^\al\wedge dx^\bt F_{\al\bt},
\end{equation}
-- называются {\em напряженностью электромагнитного поля} или просто
{\em электромагнитным полем}.
\index{Потенциал электромагнитного поля (electromagnetic field potential)}%
\index{Электромагнитное поле потенциал  (electromagnetic field potential)}%
\index{Электромагнитное поле (electromagnetic field)}%
\index{Поле электромагнитное (electromagnetic field)}%
\index{Напряженность электромагнитного поля (electromagnetic field strength)}%
\index{Электромагнитное поле напряженность (electromagnetic field strength)}%

Как и любая другая форма кривизны, напряженность электромагнитного поля
удовлетворяет тождествам Бианки (\ref{ebiaid}), которые в рассматриваемом случае
имеют вид
\begin{equation}                                                  \label{ebidem}
  \pl_\al F_{\bt\g}+\pl_\bt F_{\g\al}+\pl_\g F_{\al\bt}=0.
\end{equation}

Обозначим два локальных сечения через
\begin{equation*}
  \ex^{ia(x)},\ex^{ia'(x)}\in\MU(1),\qquad a,a'\in[0,2\pi].
\end{equation*}
Тогда существует некоторая функция $\phi(x)$, связывающая эти сечения:
\begin{equation*}
  \ex^{ia'(x)}=\ex^{i\phi(x)}\ex^{ia(x)}.
\end{equation*}
При переходе от одного локального сечения к другому компоненты локальной формы
связности меняются по-правилу (\ref{egatrd}), которое в данном случае принимает
простой вид
\begin{equation}                                                  \label{qgaaco}
  A'_\al:=A_\al+\pl_\al\phi.
\end{equation}
Нетрудно видеть, что при калибровочном преобразовании (\ref{qgaaco}) компоненты
напряженности электромагнитного поля (\ref{qstrem}) не меняются.

Подчеркнем, что преобразование компонент локальной формы связности возникает в
дифференциальной геометрии до того, как вводится понятие ковариантной
производной (см.\ раздел \ref{scofib}).

Для построения полевых моделей математической физики одного электромагнитного
поля недостаточно и поэтому выбирается некоторый набор дополнительных полей
$\psi:=\lbrace\psi^\Sa(x)\rbrace$, $\Sa=1,\dotsc,\Sn$. Если поля
вещественнозначные, то группа $\MU(1)$ действует на них тривиально, и они
описывают в квантовой теории поля нейтральные частицы. Поэтому рассмотрим
комплекснозначные поля. Мы предполагаем, что эти поля являются локальными
сечениями некоторого векторного расслоения
$\ME\big(\MM,\pi_\ME,\MV,\MU(1),\MP\big)$, которое ассоциировано с главным
расслоением $\MP\big(\MM,\pi,\MU(1)\big)$. Оно имеет ту же базу $\MM$ и
структурную группу $\MU(1)$. Типичным слоем ассоциированного расслоения является
комплексное векторное пространство $\MV$, комплексной размерности
$\dim_\MC\MV=\Sn$, в котором принимает значения поле $\psi\in\MV$. Структурная
группа $\MU(1)$ действует в $\MV$ в соответствии с некоторым представлением. Все
неприводимые комплексные представления группы $\MU(1)$ хорошо известны (см.,
например, \cite{Wigner59R}). Они одномерны и параметризуются произвольным целым
числом $k$:
\begin{equation}                                                  \label{qtsepa}
  \MU(1)\ni\quad\ex^{i\phi}\mapsto\ex^{ik\phi}\quad\in\MC,\qquad k\in\MZ.
\end{equation}
Число $k$ обязано быть целым, поскольку для группы $\MU(1)$ мы отождествляем
точки $\phi=0$ и $\phi=2\pi$.

Далее, мы предполагаем, что каждая компонента векторного поля $\psi^\Sa$ при
действии структурной группы преобразуется по некоторому неприводимому
представлению:
\begin{equation}                                                  \label{qeasca}
  \MU(1)\ni\ex^{i\phi}:\qquad \psi^\Sa\mapsto\ex^{ik_\Sa\phi}\psi^\Sa,\qquad
  k_\Sa\in\MZ.
\end{equation}
При переходе от одного локального сечения к другому, функция $\phi=\phi(x)$
является некоторой достаточно гладкой функцией на пространстве-времени,
$x\in\MM$, которая принимает значения в интервале $[0,2\pi]$, концы которого
отождествлены.

Преобразование (\ref{qgaaco}), (\ref{qeasca}) является калибровочным, т.к.\
параметр преобразования $\phi(x)$ зависит от точки пространства-времени.
Это преобразование называется также {\em градиентным}.
\index{Градиентное преобразование (gradient transformation)}%
\index{Преобразование градиентное (gradient transformation)}%

Связность в главном расслоении $\MP$ определяет связности во всех
ассоциированных расслоениях $\ME$ и, следовательно, понятие ковариантной
производной (\ref{ecovym}) для локальных сечений. Для определения ковариантной
производной, необходимо знать представление генератора группы $\MU(1)$ в $\MC$.
Разлагая ``матрицу'' представления (\ref{qeasca}) вблизи единицы, получим
\begin{equation*}
  \ex^{ik_\Sa\phi}=1+ik_\Sa\phi+\osmall(\phi).
\end{equation*}
Отсюда вытекает, что числа $ik_\Sa$ задают представления генератора группы
$\MU(1)$, и, следовательно, ковариантная производная (\ref{ecovym}) каждой
компоненты поля $\psi^\Sa$ имеет вид
\begin{equation}                                                  \label{qcobps}
  \nb_\al\psi^\Sa:=(\pl_\al-ik_\Sa A_\al)\psi^\Sa.
\end{equation}

Замечательным свойством ковариантной производной является, как легко проверить,
ее тензорный характер при калибровочных преобразованиях (\ref{qgaaco}) и
(\ref{qeasca}):
\begin{equation*}
  \nb_\al\psi^\Sa\mapsto\nb'_\al\psi^{\prime\Sa}
  =\ex^{ik_\Sa\phi(x)}\nb_\al\psi^\Sa,
\end{equation*}
где
\begin{equation}                                                  \label{qgacjo}
  \nb'_\al\psi^{\prime\Sa}:=(\pl_\al-ik_\Sa A'_\al)\psi^{\prime\Sa}.
\end{equation}

Все приведенные формулы уже встречались нам в главе \ref{sconne} в общем виде.
Калибровочное преобразование (\ref{qeasca}), (\ref{qgacjo}) можно
интерпретировать либо как переход между двумя сечениями главного и
ассоциированного расслоений, которые определяют локальные формы связности и
кривизны, либо как вертикальный автоморфизм (пример \ref{everau}). В первом
случае калибровочное преобразование интерпретируется как пассивное, когда точки
расслоения остаются на месте, а меняется только локальное сечение. Во втором
случае калибровочное преобразование рассматривается как активное, когда точки
расслоения сдвигаются под действием структурной группы.
\begin{com}
Описанная выше конструкция не зависит от того, задана ли или нет на многообразии
$\MM$ аффинная геометрия, т.е.\ метрика и аффинная связность. Ни одна из
приведенных выше формул, включая тождества Бианки (\ref{ebidem}), не содержит
метрики и аффинной связности.
\qed\end{com}

При построении моделей математической физики для полей $\psi$ необходимо
написать некоторую систему уравнений. Мы предполагаем, что уравнения движения
для полей, не взаимодействующих с электромагнитным полем, следуют из
принципа наименьшего действия для некоторого инвариантного интеграла
\begin{equation*}
  S=\int_\MM\!\!\!dx\,L(\psi,\pl\psi),
\end{equation*}
где лагранжиан $L$ зависит от полей $\psi=\lbrace\psi^\Sa\rbrace$ и их
производных $\pl\psi=\lbrace\pl_\al\psi^\Sa\rbrace$. Как правило, лагранжиан
также зависит от метрики и аффинной связности, которые используется для
построения инвариантов. Тогда взаимодействие с электромагнитным полем вводится
путем {\em минимальной подстановки}, которая заключается в замене всех частных
производных на ковариантные:
\begin{equation}                                                  \label{qovsur}
  \pl_\al\psi^\Sa\mapsto\nb_\al\psi^\Sa:=(\pl_\al-ik_\Sa A_\al)\psi^\Sa,
\end{equation}
и добавлению к исходному лагранжиану лагранжиана для свободного
электромагнитного поля:
\begin{equation*}
  L(\psi,\pl\psi)\mapsto -\frac1{4e^2}F^{\al\bt}F_{\al\bt}+L(\psi,\nb\psi),
\end{equation*}
где подъем индексов напряженности электромагнитного поля осуществляется с
помощью пространственно-временн\'ой метрики и $e$ -- универсальная постоянная
электромагнитного взаимодействия.
\index{Минимальная подстановка (minimal substitution)}%
\index{Подстановка минимальная (minimal substitution)}%

Теперь сделаем переобозначение $A_\al\mapsto eA_\al$ для того, чтобы уравнения
движения свободного электромагнитного поля не содержали константу связи $e$ и
можно было бы переходить к пределу $e\to0$. Тогда лагранжиан примет вид
\begin{equation*}
  -\frac14F^{\al\bt}F_{\al\bt}+L(\psi,\nb\psi),
\end{equation*}
где изменится выражение для ковариантной производной
\begin{equation}                                                  \label{qcodza}
  \nb_\al\psi^\Sa:=\pl_\al\psi^\Sa-iek_\Sa A_\al\psi^\Sa.
\end{equation}
Коэффициент $q_\Sa:=ek_\Sa$ описывает ``силу'' взаимодействия компоненты поля
$\psi^\Sa$ с электромагнитным полем и называется {\em электрическим зарядом}
данного поля. Для слабых электромагнитных полей именно величина $q_\Sa$ входит в
закон Кулона (см., например, \cite{LanLif88R}).
\index{Электрический заряд (electric charge)}%
\index{Заряд электрический (electric charge)}%

Мы видим, что при минимальной подстановке заряды всех полей квантуются: они
должны быть пропорциональны некоторому универсальному заряду $e$. Причина этого
в том, что все неприводимые представления группы $\MU(1)$ параметризуются целыми
числами (\ref{qtsepa}). И это соответствует действительности, т.к.\ заряды всех
известных в настоящее время элементарных частиц кратны заряду электрона. Поэтому
естественно считать, что $e$ -- это заряд электрона. Таким образом,
дифференциальная геометрия диктует нам {\em квантование электрического заряда}.
\index{Квантование электрического заряда (electric charge quantization)}%
\index{Электрический заряд квантование (electric charge quantization)}%
\begin{com}
В моделях, объединяющих электрослабые и сильные взаимодействия, существенную
роль играют кварки, электрический заряд которых пропорционален одной трети
заряда электрона. В этом нет ничего страшного. Если кварки будут обнаружены
экспериментально, то под $e$ надо понимать $1/3$ заряда электрона.
\qed\end{com}
\begin{com}
Электроны в квантовой электродинамике описываются спинорным полем, которое имеет
четыре комплексные компоненты. В этом случае для каждой из этих компонент
следует положить $k_\Sa=-1$.
\qed\end{com}

В релятивистских теориях поля пространство-время отождествляется с пространством
Минковского $\MR^{1,n-1}$ некоторого числа измерений. Поскольку пространство
Минковского $\MR^{1,n-1}$ как многообразие диффеоморфно евклидову пространству
$\MR^n$, то согласно теореме \ref{trigla} все расслоения тривиальны:
\begin{equation*}
  \MP\simeq\MR^{1,n-1}\times\MU(1),\qquad \ME\simeq\MR^{1,n-1}\times\MV.
\end{equation*}
Это означает, что топология пространства-времени в релятивистских моделях при
наличии электромагнитного поля всегда тривиальна, и все поля определены
глобально в соответствующей системе координат, покрывающей все пространство
Минковского $\MR^{1,n-1}$. Нетривиальной может быть только геометрия, если
кривизна $\MU(1)$-связности отлична от нуля.

Однако, в моделях математической физики, как правило, компоненты связности
ищутся путем решения некоторых уравнений. Если окажется так, что соответствующие
решения определены не во всем пространстве Минковского, а лишь на некотором
подмногообразии $\MU\subset\MR^{1,n-1}$, то возникает главное расслоение
$\MP\big(\MU,\pi,\MU(1)\big)$ с другой базой $\MU$. В этом случае топология
$\MU$ может быть нетривиальна, и можно говорить о топологических эффектах.
Такова, например, ситуация с монополем Дирака.

Точечные частицы в (псевдо-)римановом пространстве движутся вдоль мировых линий
$q_\Si(t)=\lbrace q^\al_\Si(t)\rbrace\in\MM$, $\Si=1,2,\dotsc$. Если
каждой частице приписать заряд $e_\Si$, то минимальная подстановка состоит в
добавлении к лагранжиану свободной частицы (см.\ \ref{schaps}) лагранжиана
взаимодействия:
\begin{equation*}
  -m\sqrt{g_{\al\bt}\dot q_\Si^\al\dot q_\Si^\bt}\mapsto
  -m\sqrt{g_{\al\bt}\dot q_\Si^\al\dot q_\Si^\bt}+e_\Si A_\al \dot q_\Si^\al,
\end{equation*}
где $\dot q_\Si:=dq_\Si/dt$ -- скорость частицы.

Мы считаем, что при калибровочном $\MU(1)$ преобразовании (\ref{qgaaco})
координаты частиц не меняются. Тогда лагранжиан минимального взаимодействия с
частицей сам по себе не инвариантен:
\begin{equation*}
  e_\Si A_\al\dot q_\Si^\al\mapsto e_\Si A_\al\dot q_\Si^\al
  +e_\Si\pl_\al\phi\dot q_\Si^\al
  =e_\Si A_\al\dot q_\Si^\al+e_\Si\frac{d\phi}{dt}.
\end{equation*}
Однако при интегрировании вдоль мировой линии частицы,
\begin{equation*}
  e_\Si\int_{q_1}^{q_2}\!\!\!dt\,\pl_\al\phi \dot q_\Si^\al
  =e_\Si\big[\phi(q_2)-\phi(q_1)\big],
\end{equation*}
результат зависит только от положения начальной и конечной точек $q_1,q_2$, но
не от траектории частицы. Отсюда вытекает инвариантность действия при
фиксированных граничных условиях и, следовательно, ковариантность уравнений
движения для точечной частицы, минимальным образом взаимодействующей с
электромагнитным полем.
\begin{com}
Если рассматривать только замкнутые траектории с началом и концом в точке $q_1$,
то интеграл
\begin{equation}                                                  \label{qolgre}
  \oint\!dt\,\dot q^\al A_\al,
\end{equation}
определяет элемент группы голономии $\MU(1)$-связности в главном расслоении
$\MP\big(\MM,\pi,\MU(1)\big)$ в точке $q_1\in\MM$ (см.\ разделы \ref{sholde}
и \ref{swilop}).
\qed\end{com}

В моделях математической физики считается, что в экспериментах можно наблюдать
только калибровочно инвариантные объекты. Таковыми являются компоненты
напряженности электромагнитного поля, которые, как мы увидим ниже,
отождествляются с напряженностью электрического и магнитного полей. При этом сам
потенциал электромагнитного поля $A_\al$ считается ненаблюдаемым, т.к.\ зависит
от выбора калибровки. Однако, наблюдаемым также является элемент группы
голономии (\ref{qolgre}), который инвариантен относительно калибровочных
преобразований. Это свойство называется эффектом Ааронова--Бома (см.\ раздел
\ref{sahabo}).

Для точеных частиц заряд $e_\Si$ может быть произволен, и его квантование не
вытекает из описанного выше подхода.

В настоящей главе мы не будем рассматривать взаимодействие точечных частиц с
электромагнитным полем.

Теперь перейдем к описанию свободного электромагнитного поля в пространстве
Минковского.
\subsection{Лагранжева формулировка                              \label{slaemf}}
Пусть задано пространство Минковского $\MR^{1,n-1}$ произвольного числа
измерений и декартова система координат $x^\al$, $\al=0,1,\dotsc,n-1$.
{\em Электромагнитное поле}
\index{Электромагнитное поле (electromagnetic field)}%
\index{Поле электромагнитное (electromagnetic field)}%
описывается компонентами $A_\al(x)$ локальной формы $U(1)$-связности на
пространстве Минковского. Оно является ковекторным полем по отношению к
преобразованиям координат -- в данном случае -- по отношению к преобразованиям
Лоренца. Временн\'ую и пространственные компоненты электромагнитного поля будем
обозначать следующим образом:
\begin{equation}                                                  \label{etispc}
  A_\al=\lbrace A_0,A_\mu\rbrace ,\qquad \mu=1,\dots,n-1.
\end{equation}
То есть греческие буквы из середины алфавита будут пробегать только
пространственные значения.

Действие для свободного электромагнитного поля, по определению, имеет вид
\begin{equation}                                                  \label{elmafl}
  S_{\Se\Sm}=\int\! dxL_{\Se\Sm},\qquad
  L_{\Se\Sm}=-\frac14F^{\al\bt}F_{\al\bt},
\end{equation}
где подъем и опускание индексов производится с помощью метрики
Минковского $\et_{\al\bt}=\diag(+-\dotsc-)$ и $F_{\al\bt}$ -- напряженность
электромагнитного поля (\ref{qstrem}). Поле $F_{\al\bt}=-F_{\bt\al}$ является
антисимметричным ковариантным тензором второго ранга типа $(0,2)$ и представляет
собой компоненты локальной 2-формы кривизны (\ref{qlocuo}) $U(1)$-связности.

Уравнения движения для действия (\ref{elmafl}) линейны:
\begin{equation}                                                  \label{elmaem}
  S_{\Se\Sm},{}^\al:=\frac{\dl S_{\Se\Sm}}{\dl A_\al}=\pl_\bt F^{\bt\al}=0.
\end{equation}
Поэтому мы говорим, что действие (\ref{elmafl}) описывает свободное
электромагнитное поле.

Лагранжиан (\ref{elmafl}), напряженность (\ref{qstrem}) и уравнения движения
(\ref{elmaem}) инвариантны относительно калибровочных преобразований
(\ref{qgaaco}) действующих только на поля и не затрагивающие координаты.
Из калибровочной инвариантности, согласно второй теореме Нетер, следует, что
между уравнениями движения имеется линейная зависимость:
\begin{equation}                                                  \label{emfdem}
  \pl_\al S_{\Se\Sm},{}^\al=0.
\end{equation}
\begin{com}
Легко видеть, что добавление массового члена $-\frac12m^2A^\al A_\al$ к
лагранжиану электромагнитного поля (\ref{elmafl}) нарушает калибровочную
инвариантность электродинамики.
\qed\end{com}
\begin{exa}
Напомним, что в {\em четырехмерном} пространстве Минковского $\MR^{1,3}$
напряженности электрического и магнитного полей связаны с потенциалом
$A_\al=\lbrace A_0,\BA\rbrace$ известными соотношениями (см., например,
\cite{LanLif88R}):
\begin{equation}                                                  \label{elmagf}
  \BE=\grad A_0-\frac{\pl\BA}{dt},\qquad \BH=\rot\BA,
\end{equation}
где жирным шрифтом обозначены трехмерные векторы. Это значит, что временн\'ая
компонента электромагнитного поля $A_0$ является электростатическим потенциалом,
а трехмерный ковектор $\BA$ -- векторным потенциалом магнитного поля. Отсюда
вытекает, что все компоненты напряженности электромагнитного поля определяются
напряженностями электрического и магнитного полей:
\begin{equation}                                                  \label{emfsma}
  F_{\al\bt}=\left(\begin{array}{rrrr}
  0   & -E_1 & -E_2 & -E_3 \\
  E_1 &  0   &  H_3 & -H_2 \\
  E_2 & -H_3 &  0   &  H_1 \\
  E_3 &  H_2 & -H_1 &  0     \end{array}\right)
\end{equation}
Лагранжиан электромагнитного поля (\ref{elmafl}) также можно выразить через
напряженности электрического и магнитного полей:
\begin{equation*}                                                    \tag*{\qed}
  L_{\Se\Sm}=\frac12(\BE^2-\BH^2).
\end{equation*}
\end{exa}

Проанализируем уравнения Эйлера--Лагранжа (\ref{elmaem}) в пространстве
Минковского произвольного числа измерений подробнее. Для временн\'ой и
пространственных компонент электромагнитного поля уравнения движения имеют вид
\begin{align}                                                     \label{eqmotc}
  -\triangle A_0-\pl_0\pl_\mu A^\mu&=0,
\\                                                                \label{eqmsco}
  \square A_\mu-\pl_\mu\pl_\nu A^\nu-\pl_\mu\pl_0A^0&=0,
\end{align}
где введены операторы Лапласа и Даламбера:
$$
  \triangle:=-\pl^\mu\pl_\mu=\pl_1^2+\pl_2^2+\dotsc+\pl_{n-1}^2,\qquad
  \square:=\pl^\al\pl_\al=\pl_0^2-\triangle.
$$
Уравнение для временн\'ой составляющей (\ref{eqmotc}) для каждого момента
времени представляет собой уравнение Пуассона в евклидовом пространстве
$\MR^{n-1}$ с правой частью, определяемой пространственными компонентами.
Сформулируем две теоремы о существовании и единственности решения этого
уравнения во всем пространстве и в ограниченной области, доказательство которых
приведено, например, в \cite{Vladim88R}.
\begin{theorem}                                                   \label{tpuaeo}
Если правая часть уравнения Пуассона дифференцируема и убывает на бесконечности,
то решение уравнения Пуассона во всем пространстве $\MR^{n-1}$, $n\ge2$,
существует и единственно в классе обобщенных функций, обращающихся в нуль на
бесконечности.
При этом решением уравнения Пуассона является дважды непрерывно дифференцируемая
функция во всем пространстве и убывающая на бесконечности.
\end{theorem}
\begin{theorem}                                                   \label{tpuaes}
Если правая часть уравнения Пуассона дифференцируема в ограниченной области
$\MU\subset\MR^{n-1}$, $n\ge2$, с кусочно гладкой границей $\pl\MU$ и непрерывна
в замыкании $\overline\MU$, то решение внутренней задачи Дирихле для уравнения
Пуассона существует, единственно и непрерывно зависит от граничных условий на
$\pl\MU$.
\end{theorem}
Если для внешней задачи Дирихле предположить, что правая часть уравнения
Пуассона и решение стремятся к нулю на бесконечности, то справедлива аналогичная
теорема для внешней задачи Дирихле. Аналогичные теоремы справедливы также
для задачи Неймана, когда на границе задается значение нормальной производной
искомой функции. В последнем случае решение определяется с точностью до
аддитивной постоянной.

Приведенные теоремы показывают, что при достаточно общих предположениях оператор
Лапласа в евклидовом пространстве, а также в ограниченной области с заданными
граничными условиями обратим, и решение уравнения Пуассона (\ref{eqmotc}) можно
формально записать в виде
\begin{equation}                                                  \label{ezecos}
  A_0=-\triangle^{-1}\pl_0\pl_\mu A^\mu,
\end{equation}
где $\triangle^{-1}$ -- ``обратный'' оператор Лапласа. Конечно, в строгом смысле
оператор Лапласа необратим, т.к.\ решение уравнения Пуассона в общем случае
определяется с точностью до произвольной гармонической функции. Тем не менее
такая запись удобна и часто употребляется в физической литературе, предполагая,
что решение соответствующего уравнения Пуассона однозначно выделено, например, с
помощью граничных условий. Запись (\ref{ezecos}) означает, что в каждый момент
времени временн\'ая компонента электромагнитного поля однозначно определяется
пространственными компонентами.

Теперь рассмотрим уравнение для пространственных компонент (\ref{eqmsco}).
Пространственные компоненты электромагнитного поля $A_\mu$ образуют ковектор в
евклидовом пространстве $\MR^{n-1}$. Если поле $A_\mu$ непрерывно
дифференцируемо, то при достаточно общих граничных условиях его можно взаимно
однозначно разложить на {\em поперечную (соленоидальную)} и {\em продольную
(потенциальную)} составляющие:
\index{Продольная составляющая векторного поля %
(longitudinal component of a vector field)}%
\index{Поперечная составляющая векторного поля %
(transverse component of a vector field)}%
\index{Потенциальная составляющая векторного поля %
(potential component of a vector field)}%
\index{Соленоидальная составляющая векторного поля %
(solenoid component of a vector field)}%
\begin{equation}                                                  \label{eemptl}
  A_\mu=A^\St_\mu+A^\Sl_\mu,
\end{equation}
которые удовлетворяют следующим условиям:
\begin{align}                                                     \label{edivtr}
  \pl_\mu A^{\St\mu}&=0,
\\                                                                \label{erotle}
  \pl_\mu A^\Sl_\nu-\pl_\nu A^\Sl_\mu&=0.
\end{align}
Формула (\ref{eemptl}) однозначно определяет ковекторное поле $A_\mu$ при
заданных полях $A^\St_\mu$ и $A^\Sl_\mu$. Покажем, что пространственная часть
электромагнитного поля $A_\mu$ также однозначно определяет свою продольную и
поперечную составляющую. В силу леммы Пуанкаре (\ref{tlempo}) из условия
(\ref{erotle}) следует, что продольная составляющая, по крайней мере локально,
является градиентом от некоторой функции $\vf(x)$, определенной с точностью до
постоянной, которая фиксируется граничными условиями. Для односвязных
многообразий, как в нашем случае, это утверждение является глобальным. Запишем
данное свойство в виде равенства
\begin{equation}                                                  \label{elotrc}
  A_\mu=A^\St_\mu+\pl_\mu\vf.
\end{equation}
Дифференцирование данного соотношения с учетом условия (\ref{edivtr}) приводит
к тому, что скалярное поле, соответствующее продольной составляющей, является
решением уравнения Пуассона:
\begin{equation}                                                  \label{epueff}
  \triangle\vf=-\pl_\mu A^\mu,\qquad \triangle:=-\pl^\mu\pl_\mu,
\end{equation}
которое существует и единственно при достаточно общих граничных условиях
(см.\ теоремы \ref{tpuaeo}, \ref{tpuaes}). После этого поперечная составляющая
однозначно определена:
\begin{equation}                                                  \label{qtraco}
  A^\St_\mu=A_\mu-\pl_\mu\vf.
\end{equation}
Нетрудно проверить, что ее дивергенция действительно равна нулю. Таким образом,
вместо $n-1$ компоненты ковекторного поля $A_\mu$ можно рассматривать скалярное
поле $\vf$, определяемое уравнением (\ref{epueff}), и соленоидальное ковекторное
поле $A^\St_\mu$ (\ref{qtraco}). При таком разложении число независимых
компонент сохранено, т.к.\ поперечное ковекторное поле имеет $n-2$ независимые
компоненты в силу дополнительного условия (\ref{edivtr}).
\begin{com}
При $n=2$ поперечная составляющая у электромагнитного поля отсутствует.
\qed\end{com}

Перепишем уравнения движения (\ref{eqmotc}), (\ref{eqmsco}) для продольной и
поперечной части:
\begin{align}                                                     \label{qtiemc}
  -\triangle A_0+\triangle\dot\vf&=0,
\\                                                                \label{qseloc}
  \pl_\mu\square\vf+\pl_\mu\triangle\vf-\pl_\mu\ddot\vf&=0,
\\                                                                \label{eqtrco}
  \square A^\St_\mu&=0,
\end{align}
где уравнение (\ref{eqmsco}) также расщеплено на продольную и поперечную части.
Для полей, убывающих на бесконечности, уравнение (\ref{qtiemc}) для временн\'ой
компоненты электромагнитного поля легко решается: $A_0=\dot\vf$, поскольку
решение уравнения Лапласа в данном случае единственно. Уравнение (\ref{qseloc})
тождественно удовлетворяется. Поскольку все проделанные операции обратимы, то мы
доказали следующее
\begin{prop}
Для полей, убывающих на бесконечности, лагранжевы уравнения движения для
электромагнитного поля (\ref{elmaem}) эквивалентны следующей системе уравнений:
\begin{align}                                                     \label{qtiems}
  A_0&=\dot\vf,
\\                                                                \label{qsetrd}
  \square A^\St_\mu&=0,
\end{align}
дополненной определением поля $\vf$ (\ref{elotrc}), (\ref{epueff}).
\end{prop}
Тот факт, что $n$ лагранжевых уравнений оказались эквивалентными $n-1$ уравнению
(\ref{qtiems}), (\ref{qsetrd}) не является удивительным. Как уже отмечалось,
из-за калибровочной симметрии, в силу второй теоремы Нетер, между исходными
уравнениями движения имеется одно линейное соотношение (\ref{emfdem}).

Если рассматривать задачу Коши для электромагнитного поля, то мы видим, что она
может быть поставлена только для поперечных компонент электромагнитного
потенциала. Следовательно, электромагнитное поле описывает только $n-2$
физических (распространяющихся) степени свободы. При этом скалярное поле $\vf$
является произвольной функцией (непрерывно дифференцируемой и удовлетворяющей
граничным условиям для рассматриваемой физической задачи), т.к.\ для нее нет
ни одного уравнения. Временн\'ая компонента электромагнитного потенциала
выражается через $\vf$. Конечно, можно считать наоборот, что компонента $A_0$
произвольна, а функция $\vf$ находится из уравнения (\ref{qtiems}).

Поперечные степени свободы являются распространяющимися и соответствуют
электромагнитным волнам. В двух измерениях электромагнитное поле не описывает
ни одной степени свободы и электромагнитное излучение отсутствует. В
пространстве Минковского трех и четырех измерений электромагнитное поле имеет
соответственно одну и две степени свободы.

То обстоятельство, что мы не получили ни одного уравнения на продольную
составляющую является следствием калибровочной инвариантности электродинамики
относительно преобразований (\ref{qgaaco}). Нетрудно проверить, что
напряженность электромагнитного поля при выполнении уравнения движения
(\ref{qtiems}) зависит только от поперечных компонент:
\begin{equation}                                                  \label{efstem}
  F_{0\mu}=\pl_0 A^\St_\mu,\qquad F_{\mu\nu}=\pl_\mu A^\St_\nu-\pl_\nu A^\St_\mu.
\end{equation}
Калибровочные преобразования (\ref{qgaaco}) для компонент электромагнитного
поля выглядят следующим образом:
\begin{align}                                                     \label{egataz}
  A^\prime_0&=A_0+\pl_0\phi,
\\                                                                \label{egatal}
  \vf^\prime&=\vf+\phi,
\\                                                                \label{egatat}
  A^{\prime\St}_\mu&=A^\St_\mu.
\end{align}
Используя произвольную функцию $\phi(x)$ можно зафиксировать одну из компонент
электромагнитного поля $A_0$ или $\vf$, то есть придать им определенные наперед
заданные значения, которые, конечно, должны быть согласованы с уравнением
(\ref{qtiems}). Это называется {\em фиксированием калибровки} и упрощает решение
уравнений Эйлера--Лагранжа, т.к.\ устраняет функциональный произвол в общем
\index{Фиксирование калибровки (gauge fixing)}%
решении и уменьшает число неизвестных функций. Наложение калибровочных условий
необходимо также для построения квантовой теории. Например, можно положить
$A_0=0$ и $\vf=0$. Для полей, убывающих на бесконечности, последнее условие
эквивалентно {\em кулоновской} калибровке $\pl_\mu A^\mu=0$.
\index{Кулоновская калибровка (Coulomb gauge)}%
\index{Калибровка кулоновская (Coulomb gauge)}%
Среди калибровок, наиболее часто использующихся в электродинамике, отметим
следующие\footnote{Лоренцеву калибровку ввел датский физик Ludwig Lorenz в 1867
году, а не голландский физик H.~A.~Lorentz (на русский язык обе фамилии в
настоящее время принято переводить одинаково, хотя ранее фамилия Lorentz
переводилась, как Лорентц), в честь которого названы преобразования Лоренца. По
этому поводу см.~\cite{vanBla91A} }:
\begin{equation}                                                  \label{qdifga}
\begin{aligned}
  \pl_\mu A^\mu&=0 & &\text{-- кулоновская калибровка}, \\
  \pl_\al A^\al&=0 & &\text{-- лоренцева калибровка}, \\
  A_0&=0 & &\text{-- временн\'ая калибровка}, \\
  A_3&=0 & &\text{-- аксиальная калибровка}.
\end{aligned}
\end{equation}
\index{Лоренцева калибровка (Lorenz gauge)}%
\index{Калибровка Лоренца (Lorenz gauge)}%
\index{Временн\'ая калибровка (time gauge)}%
\index{Калибровка временн\'ая (time gauge)}%
\index{Аксиальная калибровка (axial gauge)}%
\index{Калибровка аксиальная (axial gauge)}%

Из этих калибровок только кулоновская и аксиальная являются каноническими (см.\
раздел \ref{sficon}). Лоренцева калибровка содержит производную по времени:
$\pl_0A^0+\pl_\mu A^\mu=0$. Временн\'ая калибровка с гамильтоновой точки зрения
фиксирует множитель Лагранжа и оставляет произвол в определении пространственных
компонент электромагнитного потенциала $A_\mu$.

Мы называем поля $A_0$ и $A_\mu^\Sl$ нефизическими только потому, что для них
нет задачи Коши. Однако это не означает, что они всегда равны нулю и не
наблюдаются. Напротив, электростатический потенциал $A_0$ очень даже
наблюдается, например, при взаимодействии точечных зарядов в законе Кулона.
\subsection{Законы сохранения                                    \label{semfcl}}
Действие для свободного электромагнитного поля в пространстве Минковского
инвариантно относительно глобального действия группы Пуанкаре и локальных
калибровочных преобразований (\ref{qgaaco}). Из теоремы Нетер отсюда следуют
законы сохранения тензора энергии-импульса, момента количества движения и
вектора электромагнитного тока. Поскольку калибровочные преобразования локальны,
то вторая теорема Нетер приводит также к зависимости уравнений движения
(\ref{emfdem}), которая была отмечена раньше.

Явное выражение для тензора энергии-импульса (\ref{etenmo}) в случае
электромагнитного поля принимает вид
\begin{equation}                                                  \label{enmoea}
  \widehat T_{\al\bt}:=\widehat T_\al{}^\g\et_{\g\bt}
  =-\pl_\al A_\g F_\bt{}^\g+\frac14\et_{\al\bt}F^2.
\end{equation}
Первое слагаемое в этой формуле явно несимметрично относительно перестановки
индексов. Воспользуемся произволом (\ref{earbcu}) в определении тензора
энергии-импульса и определим новый симметричный тензор энергии-импульса
\begin{equation}                                                  \label{esyemt}
  T^\Ss_{\al\bt}=-F_{\al\g}F_\bt{}^\g+\frac14\et_{\al\bt}F^2
  =\widehat T_{\al\bt}+\pl_\g A_\al F_\bt{}^\g.
\end{equation}
С учетом уравнений движения для электромагнитного поля (\ref{elmaem})
разность между новым и старым выражением принимает вид
\begin{equation}                                                  \label{editee}
  T^\Ss_{~\al}{}^\bt-\widehat T_\al{}^\bt=\pl_\g(A_\al F^{\bt\g}),
\end{equation}
что согласуется с видом добавочного члена (\ref{earbcu}). Важным обстоятельством
является то, что симметричный тензор энергии-импульса (\ref{esyemt})
калибровочно инвариантен. Закон сохранения энергии-импульса имеет вид
(\ref{enmocm}),
\begin{equation*}
  \pl_\bt T^\Ss_{~\al}{}^\bt=\pl_\bt\widehat T_\al{}^\bt=0,
\end{equation*}
для любого решения уравнений движения.

Лоренцевы вращения для электромагнитного поля записываются в виде
\begin{equation}                                                  \label{emflro}
  \dl x^\al=-x^\bt\om_\bt{}^\al,\qquad \dl A_\al=\om_\al{}^\bt A_\bt.
\end{equation}
Поэтому для спинового момента (\ref{espimo}) получаем следующее выражение
\begin{equation}                                                  \label{espmem}
  S_{\al\bt}{}^\g=A_\al F_\bt{}^\g-A_\bt F_\al{}^\g.
\end{equation}

Полный тензор момента количества движения (\ref{eanmot}) состоит из орбитального
и спинового моментов:
\begin{equation}                                                  \label{etoanm}
  \widehat J_{\al\bt}{}^\g
  =x_\bt\widehat T_\al{}^\g-x_\al\widehat T_\bt{}^\g
  +S_{\al\bt}{}^\g.
\end{equation}
Используя выражения для симметричного тензора энергии-импульса (\ref{esyemt}) и
спинового момента (\ref{espmem}) его можно переписать в виде
\begin{equation}                                                  \label{qstomo}
  \widehat J_{\al\bt}{}^\g=M_{\al\bt}{}^\g
  +\pl_\dl(x_\al A_\bt F^{\g\dl}-x_\bt A_\al F^{\g\dl}),
\end{equation}
где
\begin{equation*}
  M_{\al\bt}{}^\g:=x_\bt T^\Ss_{~\al}{}^\g-x_\al T^\Ss_{~\bt}{}^\g
\end{equation*}
-- орбитальный момент, построенный по симметричному тензору энергии-импульса
(\ref{editee}). Поскольку второе слагаемое в выражении (\ref{qstomo}) имеет вид
(\ref{earbcu}), то тензор момента количества движения можно переопределить,
положив
\begin{equation}                                                  \label{eanmne}
  J_{\al\bt}{}^\g=M_{\al\bt}{}^\g
  =x_\bt T^\Ss_{~\al}{}^\g-x_\al T^\Ss_{~\bt}{}^\g,
\end{equation}
который также сохраняется
\begin{equation*}
  \pl_\g J_{\al\bt}{}^\g=0.
\end{equation*}
Отметим полное отсутствие спинового момента в этом выражении.

С калибровочной инвариантностью электродинамики связан закон сохранения заряда
\begin{equation}                                                  \label{qsohza}
  \pl_\al J^\al=0,
\end{equation}
где
\begin{equation}                                                  \label{qcuele}
  J^\al=\pl_\bt F^{\bt\al}
\end{equation}
-- ток, соответствующий первой теореме Нетер, то есть, когда параметр
калибровочных преобразований не зависит от точки пространства-времени. В данном
случае закон сохранения заряда (\ref{qsohza}) совпадает с зависимостью уравнений
движения, вытекающей из второй теоремы Нетер (\ref{emfdem}). При отсутствии
полей материи этот закон тривиален, т.к.\ ток равен нулю в силу уравнений
движения (\ref{elmaem}).
\subsection{Гамильтонова формулировка                            \label{semfhf}}
Свободное электромагнитное поле является относительно простым и важным примером
калибровочной модели. Ниже мы используем общий гамильтонов анализ для систем со
связями, описанный в разделе \ref{sficon}.

Электромагнитному потенциалу $A_\al$ соответствуют канонически сопряженные
импульсы
\begin{equation}                                                  \label{eemfcm}
  P^\al:=\frac{\pl L_{\Se\Sm}}{\pl(\pl_0A_\al)}=F^{\al0}.
\end{equation}
По определению, скобки Пуассона между координатами и импульсами в каждый момент
времени имеют вид
\begin{equation}                                                  \label{epoibe}
  \big[A_\al,P^{\prime\bt}\big]:=\big[A_\al(t,\Bx),P^\bt(t,\Bx')\big]
  =\dl_\al^\bt\dl(\Bx-\Bx'),
\end{equation}
где $(n-1)$-мерная $\dl$-функция зависит только от пространственных координат.
Ввиду антисимметрии напряженности электромагнитного поля в теории возникает
первичная связь
\begin{equation}                                                  \label{eemffc}
  G_1:=P^0\approx0,
\end{equation}
где волнистый знак равенства $\approx$ обозначает, что равенство должно быть
выполнено на поверхности всех связей. Других первичных связей электродинамика
без полей материи не содержит. Гамильтониан получается простым вычислением
\begin{equation}                                                  \label{emfham}
  \CH=\int\!\!d\Bx\,\left(-\frac12P^\mu P_\mu+\frac14F^{\mu\nu}F_{\mu\nu}
  -A_0\pl_\mu P^\mu+\lm P^0\right),
\end{equation}
где мы добавили первичную связь (\ref{eemffc}) с множителем Лагранжа $\lm$.
Гамильтоновы уравнения движения для построенного гамильтониана имеют
следующий вид:
\begin{align}                                                     \label{qemtuh}
  \dot A_0&=\lm,
\\                                                                \label{qemsvf}
  \dot A_\mu&=-P_\mu+\pl_\mu A_0,
\\                                                                \label{qemzmy}
  \dot P^0&=\pl_\mu P^\mu,
\\                                                                \label{qemlju}
  \dot P^\mu&=\pl_\nu F^{\nu\mu},
\end{align}
которые необходимо дополнить уравнением первичной связи (\ref{eemffc}),
возникающей при варьировании соответствующего действия по множителю Лагранжа.

В уравнении (\ref{qemtuh}) множитель Лагранжа действительно является
произвольной функцией. Чтобы убедиться в этом, необходимо проверить
самосогласованность уравнений движения при произвольных $\lm$. Поскольку в
теории есть связь, то она должна сохраняться во времени. То есть правая часть
уравнения (\ref{qemzmy}) должна быть равна нулю для любого решения. Вычислим
производную по времени:
\begin{equation*}
  \pl_0(\pl_\mu P^\mu)=[\pl_\mu P^\mu,\CH]=\pl^2_{\mu\nu}F^{\nu\mu}=0,
\end{equation*}
где мы использовали антисимметрию компонент напряженности. Поэтому, если
связь (\ref{eemffc}) выполнена в начальный момент времени, то она будет
удовлетворяться и в последующие моменты времени. Таким образом, у нас имеется
только четыре уравнения (\ref{qemtuh})--(\ref{qemlju}) на пять функций. Поэтому
функцию $\lm$ можно рассматривать как произвольную.

\begin{prop}
Для полей, убывающих на бесконечности, гамильтоновы уравнения движения для
электромагнитного поля (\ref{eemffc}), (\ref{qemtuh})--(\ref{qemlju})
эквивалентны лагранжевым уравнениям (\ref{elmaem}).
\end{prop}
\begin{proof}
Пусть выполнены гамильтоновы уравнения. Из уравнения (\ref{qemsvf}) находим
импульсы
\begin{equation*}
  P_\mu=-\dot A_\mu+\pl_\mu A_0=F_{\mu0}.
\end{equation*}
Тогда уравнение (\ref{qemlju}) примет вид лагранжева уравнения
\begin{equation*}
  \pl_\al F^{\al\mu}=0.
\end{equation*}
Поскольку $P^0=0$, то уравнение (\ref{qemzmy}) примет вид $\pl_\mu P^\mu$ или, с
учетом выражения для импульсов
\begin{equation*}
  \pl_\mu F^{\mu0}=0.
\end{equation*}
Таким образом, лагранжевы уравнения являются следствием гамильтоновых. При этом
мы никак не использовали уравнение для временн\'ой компоненты (\ref{qemtuh}).

Обратно. Пусть выполнены лагранжевы уравнения. Обозначим $P^{\al}:=F^{\al0}$.
Тогда выполнены уравнения (\ref{qemsvf}) и связь $P^0=0$. Лагранжевы уравнения
в новых переменных примут вид
\begin{equation*}
  \pl_\mu P^\mu=0,\qquad -\dot P^\mu+\pl_\nu F^{\nu\mu}=0,
\end{equation*}
что эквивалентно гамильтоновым уравнениям (\ref{qemzmy}), (\ref{qemlju}) с
учетом условия $P^0=0$. Осталось показать, что уравнение (\ref{qemtuh}) является
следствием лагранжевых уравнений. Ранее мы доказали, что лагранжевы уравнения
эквивалентны системе уравнений (\ref{qtiems}), (\ref{qsetrd}). В уравнении
(\ref{qtiems}) функция $\vf$, а в уравнении (\ref{qemtuh}) функция $\lm$
являются совершенно произвольными. Поэтому можно положить $\lm=\ddot\vf$, и
тогда эти уравнения совпадут.
\end{proof}
\begin{com}
В гамильтониан (\ref{emfham}) необходимо добавить первичную связь, т.к.\ в
противном случае эквивалентности между гамильтоновыми и лагранжевыми уравнениями
не будет.
\qed\end{com}
Продолжим построение гамильтонова формализма согласно общей схеме анализа систем
со связями. Производная по времени от первичной связи (\ref{eemffc}) равна
\begin{equation}                                                  \label{esecoe}
  \dot G_1=\big[P^0,\CH\big]=\pl_\mu P^\mu.
\end{equation}
Это приводит к вторичной связи
\begin{equation}                                                  \label{esscom}
  G_2:=\pl_\mu P^\mu\approx0.
\end{equation}
Нетрудно проверить, что других связей в теории нет. Связи $G_1$ и $G_2$ являются
функционально независимыми связями первого рода, т.к.\ их скобка Пуассона равна
нулю:
\begin{equation*}
  \big[G_1,G_2^\prime\big]=0.
\end{equation*}
К полному гамильтониану теории мы добавляем все связи:
\begin{equation}                                                  \label{etohae}
  \CH_\St=\int\!d\Bx\,\left(-\frac12P^\mu P_\mu+\frac14F^{\mu\nu}F_{\mu\nu}
      +\lm P^0+\mu\pl_\mu P^\nu\right),
\end{equation}
где $\lm(x)$ и $\mu(x)$ -- множители Лагранжа. На поверхности связей $G_1=0$ и
$G_2=0$ гамильтонова плотность, очевидно, положительно определена.

В лагранжевом формализме действие инвариантно относительно калибровочных
преобразований (\ref{qgaaco}), которые параметризуются одной произвольной
функцией $\phi(x)$. Согласно второй теореме Нетер это приводит к зависимости
уравнений движения (\ref{emfdem}). Покажем, что в гамильтоновом формализме
действие инвариантно относительно калибровочных преобразований,
параметризующихся не одной, а двумя произвольными функциями, по числу связей
первого рода, и уравнения движения удовлетворяют двум тождествам. Удвоение числа
локальных симметрий произошло потому что в гамильтоновом формализме
калибровочный параметр $\phi(x)$ и его временн\'ая производная $\dot\phi(x)$
рассматриваются, как независимые функции и параметризуют, соответственно,
различные калибровочные преобразования.

Выразим полное действие для свободного электромагнитного поля через координаты и
импульсы:
\begin{equation}                                                  \label{emacha}
  S_\St=\int\!dx\,(P^\al\dot A_\al+\frac12P^\mu P_\mu
  -\frac14F^{\mu\nu}F_{\mu\nu}-\lm P^0-\mu\pl_\mu P^\nu).
\end{equation}
Полное действие совпадает с исходным действием (\ref{elmafl}) в лагранжевой
формулировке только при $\lm=0$ и $\mu=A_0$ и поэтому является более общим.
Уравнения Эйлера--Лагранжа для действия (\ref{emacha}) принимают вид
\begin{align}                                                     \label{emfhao}
  \frac{\dl S_\St}{\dl P^0}&=\dot A_0-\lm=0,
\\                                                                \label{emfhas}
  \frac{\dl S_\St}{\dl P^\mu}&=\dot A_\mu+P_\mu+\pl_\mu\mu=0,
\\                                                                \label{emfhat}
  \frac{\dl S_\St}{\dl A_0}&=-\dot P^0=0,
\\                                                                \label{emfhaf}
  \frac{\dl S_\St}{\dl A_\mu}&=-\dot P^\mu-\pl_\nu F^{\mu\nu}=0,
\\                                                                \label{emfcon}
  \frac{\dl S_\St}{\dl\lm}&=-P^0=0,
\\                                                                \label{eleqes}
  \frac{\dl S_\St}{\dl\mu}&=-\pl_\mu P^\mu=0.
\end{align}

Согласно теореме \ref{tlochs} каждой связи первого рода соответствует
калибровочное преобразование, относительно которого полное действие инвариантно.
Первичной связи (\ref{eemffc}) соответствует генератор калибровочных
преобразований
$$
  T_1=\int\!d\Bx\,\e_1 P^0
$$
с некоторым малым параметром $\e_1(x)$. При этом преобразуется только
временн\'ая компонента электромагнитного поля $A_0$ и множитель Лагранжа $\lm$:
\begin{equation}                                                  \label{emgtrp}
  \dl_1A_0=[ A_0,T_1]=\e_1,\qquad \dl_1\lm=\dot\e_1.
\end{equation}
Все остальные поля остаются без изменения. Согласно второй теореме Нетер из
инвариантности действия относительно калибровочного преобразования
(\ref{emgtrp}) следует зависимость уравнений движения
$$
  \frac{\dl S_\St}{\dl A_0}-\pl_t\frac{\dl S_\St}{\dl\lm}=0.
$$
В справедливости этого тождества нетрудно убедиться прямой проверкой.

Вторичной связи первого рода (\ref{esscom}) также соответствует генератор
преобразований
$$
  T_2=\int\!d\Bx\,\e_2\pl_\mu P^\mu,
$$
где $\e_2(x)$ -- малый параметр второго калибровочного преобразования. В этом
случае преобразуются только пространственные компоненты электромагнитного
потенциала $A_\mu$ и множитель Лагранжа $\mu$:
$$
  \dl_2 A_\mu=-\pl_\mu\e_2,\qquad \dl_2\mu=\dot\e_2.
$$
Из инвариантности действия следует зависимость уравнений Эйлера--Лагранжа
$$
  \pl_\mu\frac{\dl S_\St}{\dl A_\mu}-\pl_t\frac{\dl S_\St}{\dl\mu}=0,
$$
что также легко проверить. Таким образом, число параметров калибровочных
преобразований полного действия $S_\St$ в гамильтоновом формализме по сравнению
с лоренцевой формулировкой удвоилось. Эта ситуация типична для калибровочных
моделей, когда в лагранжевом формализме локальные преобразования зависят как от
самого параметра, так и от его производной по времени.

Генератором калибровочных преобразований в лагранжевом формализме (\ref{qgaaco})
является линейная комбинация связей первого рода:
$$
  T=\int\!dx\,(\dot\e P^0-\e\pl_\mu P^\mu).
$$
Действительно,
\begin{align*}
  \dl A_0&=[A_0,T]=\dot\e,
\\
  \dl A_\mu&=[A_\mu,T]=\pl_\mu\e.
\end{align*}

Теперь проанализируем гамильтоновы уравнения движения
(\ref{emfhao})--(\ref{eleqes}) и связи (\ref{eemffc}), (\ref{esscom}) без
фиксирования какой-либо калибровки. Для определенности и простоты будем
рассматривать задачу в евклидовом пространстве $\MR^{n-1}$ для гладких функций
достаточно быстро убывающих на бесконечности. По аналогии с разложением
электромагнитного поля на поперечную и продольную части (\ref{elotrc}) разложим
также импульсы:
\begin{equation*}
  P_\mu=-P^\mu=P^\St_\mu+P^\Sl_\mu,
\end{equation*}
где
\begin{equation*}
  \pl_\mu P^{\St \mu}=0,\qquad P^\Sl_\mu=\pl_\mu \chi.
\end{equation*}
В этом разложении скалярное поле $\chi(x)$ для каждого момента времени является
решением уравнения Пуассона
\begin{equation*}
  \triangle\chi=-\pl_\mu P^\mu=-\pl_\mu P^{\Sl \mu}.
\end{equation*}
Для широкого класса граничных условий в односвязных областях разложение
импульсов на продольную и поперечную части является взаимно однозначным. Тогда
связь (\ref{esscom}) примет вид
\begin{equation*}
  \pl_\mu P^\mu=\pl_\mu P^{\Sl \mu}=0\quad \Leftrightarrow\quad \triangle\chi=0.
\end{equation*}
В рассматриваемом случае мы имеем единственное решение $\chi=0$, что влечет
за собой отсутствие продольной составляющей импульса, $P^{\Sl\mu}=0$.

Запишем уравнение движения для электромагнитного потенциала (\ref{emfhas})
отдельно для продольной и поперечной компоненты:
\begin{equation}                                                  \label{qlocom}
  \dot A^\St_\mu=-P^\St_\mu,\qquad
  \dot A^\Sl_\mu=-P^\Sl_\mu-\pl_\mu\mu.
\end{equation}
Последнее уравнение можно переписать в терминах потенциала
$A^\Sl_\mu=\pl_\mu\vf$ и один раз проинтегрировать:
\begin{equation*}
    \dot\vf=-\chi-\mu=\triangle^{-1}\pl_\mu P^\mu-\mu.
\end{equation*}
Отсюда следует, что эволюция продольной $A^\Sl_\mu$ составляющей
электромагнитного поля произвольна, так как определяется множителем Лагранжа.

С учетом уравнения движения для временн\'ой компоненты $A_0$, мы видим, что
эволюция компонент $A_0$ и $A^\Sl_\mu$ совершенно произвольна, т.к.\
определяется множителями Лагранжа. При этом уравнения связей позволяют найти
сопряженные импульсы $P^0=0$ и $P^{\Sl\mu}=0$. В то же время множители Лагранжа
не входят в уравнения движения для поперечных компонент $A^\St_\mu$ и
$P^{\St\mu}$. Мы видим, что временн\'ая и продольные компоненты соответствуют
калибровочным степеням свободы и являются нефизическими.

В теории свободного электромагнитного поля связи можно решить и выписать в
явном виде эффективное действие для физических степеней свободы (см.\
раздел~\ref{sficon}). Ввиду важности этой процедуры, остановимся на ней
подробно.

Электродинамика без полей материи содержит две связи первого рода
\begin{equation*}
  G_\Sa=\lbrace G_1,G_2\rbrace=0,\qquad \Sa=1,2.
\end{equation*}
В соответствии с общим правилом (см.\ раздел \ref{sficon}) наложим два
калибровочных условия:
\begin{equation*}
  F^\Sa=\lbrace F^1,F^2\rbrace=\lbrace A_0,\pl_\mu A^\mu\rbrace=0.
\end{equation*}
Второе калибровочное условие $\pl_\mu A^\mu=0$ является уравнением только на
продольную составляющую электромагнитного поля:
\begin{equation*}
  \pl_\mu A^\mu=\pl_\mu A^{\Sl\mu}=0\quad \Leftrightarrow\quad \triangle\vf=0.
\end{equation*}
При нулевых граничных условиях на бесконечности это уравнение в евклидовом
пространстве $\MR^{n-1}$ имеет единственное решение $\vf=0$, т.е.\ у
пространственных компонент ковектора $A_\mu$ отсутствует продольная
составляющая. Эта калибровка называется кулоновской.

Обобщенный гамильтониан электродинамики, который получается после добавления
всех связей и калибровочных условий, примет вид
\begin{equation*}
  \CH_\Se=\int d\Bx\left(-\frac12P^\mu P_\mu+\frac14F^{\mu\nu}F_{\mu\nu}
  +\lm P^0+\mu\pl_\mu P^\mu+\nu A_0+\xi\pl_\mu A^\nu\right),
\end{equation*}
где $\lm,\mu,\nu,\xi$ -- множители Лагранжа. Из условия сохранения
калибровочных условий и связей во времени получаем уравнения на
множители Лагранжа:
\begin{equation*}
  \lm=0,\quad \triangle\mu=0,\quad \nu=0,\quad \triangle\xi=0.
\end{equation*}
При сделанных предположениях уравнения на $\mu$ и $\xi$ имеют единственное
решение $\mu=0$, $\xi=0$. Таким образом, кулоновская калибровка является
канонической, поскольку однозначно фиксирует множители Лагранжа.

Для точечных частиц отличие от нуля определителя (\ref{edegfa}) является
необходимым и достаточным условием однозначного определения множителей Лагранжа
в расширенном гамильтониане $S_\Se$. Вопрос о критерии однозначного определения
множителей Лагранжа в теории поля сложен. Действительно, введем обозначение
для полной системы связей и калибровочных условий:
\begin{equation*}
  \Phi^\Sm=\lbrace G_1,F^1,G_2,F^2\rbrace,\qquad \Sm=1,2,3,4.
\end{equation*}
Тогда матрица, составленная из скобок Пуассона, примет вид
\begin{equation*}
  [\Phi^\Sm,\Phi^{\prime\Sn}]=\begin{pmatrix}
  0 & -\dl(\Bx'-\Bx) & 0 & 0 \\ \dl(\Bx'-\Bx) & 0 & 0 & 0 \\
  0 & 0 & 0 & \triangle\dl(\Bx'-\Bx) \\
  0 & 0 & -\triangle\dl(\Bx'-\Bx) & 0 \end{pmatrix}.
\end{equation*}
Вопрос о том, что такое определитель этой матрицы сложен, и мы на нем
останавливаться не будем.

Фазовое пространство электродинамики $\MN$ бесконечномерно,
$\dim\MN=2n\times\infty$, и задается координатами $A_\al(\Bx)$ и сопряженными
импульсами $P^\al(\Bx)$. Физическое подпространство (поверхность связей и
калибровочных условий $\Phi=0$) $\MM\subset\MN$ размерности
$\dim\MM=2(n-2)\times\infty$ определяется уравнениями $\Phi^\Sm=0$. Построим на
$\MN$ такую систему координат, которая фигурирует в теореме \ref{tphyco}.
Сначала заметим, что на поверхности связей и калибровочных условий $\Phi^\Sm=0$
продольные компоненты координат и импульсов отсутствуют:
\begin{equation*}
  A_\mu|_{\Phi=0}=A^\St_\mu,\qquad P^\mu|_{\Phi=0}=P^{\St\mu}.
\end{equation*}

Исходной системой координат на $\MN$ являются координаты $A_\al$ и сопряженные
импульсы $P^\al$ с канонической скобкой Пуассона (\ref{epoibe}). Введем вместо
пространственных компонент $A_\mu$ и $P^\mu$ их разложение на
поперечную и продольную составляющую:
\begin{equation*}
  \lbrace A_\mu,P^\mu\rbrace\quad \mapsto\quad
  \lbrace A^\St_\mu,A^\Sl_\mu,P^{\St\mu},P^{\Sl\mu}\rbrace.
\end{equation*}
Явные формулы для поперечных и продольных компонент имеют вид:
\begin{equation*}
\begin{split}
  A^\St_\mu&=A_\mu+\frac{\pl_\mu\pl_\nu}\triangle A^\nu,\qquad
  A^\Sl_\mu=-\frac{\pl_\mu\pl_\nu}\triangle A^\nu,
\\
  P^{\St\mu}&=P^\mu+\frac{\pl^\mu\pl_\nu}\triangle P^\nu,\qquad
  P^{\Sl\mu}=-\frac{\pl^\mu\pl_\nu}\triangle P^\nu,
\end{split}
\end{equation*}
где $\pl^\mu=-\pl_\mu$. Пуассонова структура в новых координатах уже не будет
иметь канонического вида, т.к.\ на них имеются дополнительные соотношения
(\ref{edivtr}), (\ref{erotle}) и такие же соотношения на импульсы. Простые
вычисления показывают, что для пространственных компонент отличны от нуля только
две скобки Пуассона:
\begin{equation}                                                  \label{epobpe}
  [A^\St_\mu,P^{\prime\St\nu}]=\left(\dl_\mu^\nu+\frac{\pl_\mu\pl^\nu}
  \triangle\right)\dl(\Bx-\Bx'),
  \qquad [A^\Sl_\mu,P^{\prime\Sl\nu}]=-\frac{\pl_\mu\pl^\nu}\triangle\dl(\Bx-\Bx').
\end{equation}

В качестве независимых координат на поверхности связей и калибровочных условий
$\MM$ выберем первые $n-2$ поперечные компоненты электромагнитного поля
\begin{equation*}
  q^*=\lbrace A^\St_a\rbrace,\qquad a=1,\dotsc,n-2.
\end{equation*}
Тогда последняя компонента $A_{n-1}$ находится из уравнения поперечности
(\ref{edivtr})
\begin{equation*}
  A^\St_{n-1}=-\int_{-\infty}^{x^{n-1}}\!\!\!dx'\pl_a
  A^{\St a}(x^0,x^1,\dotsc,x^{n-2},x'):=\frac1{\pl_{n-1}}\pl_a A^{\St a},
\end{equation*}
где мы ввели обозначение $1/\pl_{n-1}$ для оператора, обратного к частной
производной $\pl_{n-1}$. При этом мы предполагаем, что все компоненты полей
достаточно быстро убывают при $x^{n-1}\to-\infty$ и, следовательно, оператор
частной производной $\pl_{n-1}$ обратим. Импульсы, сопряженные к координатам
$q^*$, должны иметь с ними каноническую скобку Пуассона. Учитывая формулу
(\ref{epobpe}), нетрудно проверить, что импульсы, сопряженные к $q^*$, имеют
вид
\begin{equation*}
  p^*=\left\lbrace P^{\St a}
  -\frac1{\pl_{n-1}}\pl^aP^{\St,n-1}\right\rbrace,\qquad a=1,\dotsc,n-2.
\end{equation*}
Координаты $q^*$ и сопряженные им импульсы $p^*$ образуют систему координат на
физическом подмногообразии $\MM\subset\MN$ фазового пространства. Заметим, что
при $x^{n-1}\to\infty$ компоненты импульсов $p^*$ в общем случае не обязательно
убывают, поскольку содержат обратный оператор $1/\pl_{n-1}$. Эта ситуация
типична в теории поля, где связи представляют собой дифференциальные уравнения.

Таким образом, мы построили канонически сопряженные координаты на поверхности
связей. Это -- не единственный выбор. В качестве независимых координат можно
было бы выбрать непосредственно компоненты
$\lbrace q^*,p^*\rbrace=\lbrace A_a,P^a\rbrace$, $a=1,\dotsc,n-2$. В
этом случае преобразования координат фазового пространства были бы проще, однако
выражение для эффективного гамильтониана для физических степеней свободы --
сложнее.

Поверхность связей и калибровочных условий $\Phi^\Sm=0$ можно задать канонически
сопряженными координатами и импульсами:
\begin{equation*}
  Q_\Sa=\lbrace A_0,\pl_\mu A^\mu\rbrace,\qquad
  P^\Sa=\lbrace P^0,\triangle^{-1}\pl_\mu P^\mu\rbrace,\qquad \Sa=1,2.
\end{equation*}
Действительно, нетрудно проверить, что скобки Пуассона имеют канонический вид:
\begin{equation*}
  [Q_\Sa,P^{\prime\Sb}]=\dl_\Sa^\Sb\dl(\Bx-\Bx'),\qquad [Q_\Sa,Q'_\Sb]=0,\quad
  [P^\Sa,P^{\prime\Sb}]=0.
\end{equation*}
Эти координаты описывают нефизические (калибровочные) степени свободы.

Таким образом, в соответствии с теоремой \ref{tphyco} мы построили систему
координат на фазовом пространстве $\MN$
\begin{equation*}
  \lbrace q^*,p^*,Q_\Sa,P^\Sa\rbrace
\end{equation*}
такую, что канонически сопряженные координаты и импульсы $q^*,p^*$ являются
координатами на физическом подпространстве $\MM\subset\MN$, а поверхность
связей и калибровочных условий задается особенно просто:
\begin{equation*}
  \Phi^\Sm=0\quad \Leftrightarrow\quad Q_\Sa=0,~P^\Sa=0.
\end{equation*}

\begin{exa}
В четырехмерном пространстве-времени на поверхности связей и калибровочных
условий физический (эффективный) гамильтониан принимает простой вид
\begin{equation*}
  \CH_\mathrm{ph}=\int d\Bx\left(-\frac12P^{\St\mu}P^\St_\mu
  +\frac14F^{\mu\nu}F_{\mu\nu}\right)=\frac12\int d\Bx(\BE^{\St 2}+\BH^2),
\end{equation*}
где мы учли выражения для напряженностей электрического и магнитного полей
(\ref{elmagf}). Напомним, что из определения напряженности магнитного поля
следует, что $\BH=\BH^\St$. Таким образом, мы получили стандартное выражение
для плотности энергии свободного электромагнитного поля, которое, очевидно,
положительно определено.
\qed\end{exa}
Построенная система координат нелокальна в том смысле, что содержит обратные
дифференциальные операторы. Эта ситуация типична для калибровочных теорий поля,
где связи являются дифференциальными уравнениями по пространственным
координатам, поскольку их решения задаются интегралами по пространству.

Решения дифференциальных уравнений в частных производных существенно зависят
от области, в которой ищется решение, и граничных условий. В частности,
оператор Лапласа может быть необратим, и это может приводить к физическим
следствиям.

Таким образом, электродинамика в пространстве Минковского $\MR^{1,n-1}$
описывает $n-2$ физические (распространяющиеся) степени свободы. В четырехмерном
пространстве-времени мы говорим, что фотон имеет две поляризации. При этом
временн\'ая и продольная компоненты электромагнитного поля являются
нефизическими (нераспространяющимися). Это не значит, что они равны нулю.
Например, в электростатике временн\'ая компонента $A_0$ нетривиальна и
наблюдаема, т.к.\ представляет собой электрический кулонов потенциал
взаимодействия зарядов. Нефизичность в рассматриваемом контексте означает
только то, что для соответствующих полей нельзя поставить задачу Коши.

В двумерном пространстве-времени, $n=2$, пространственная часть
электромагнитного поля $A_\mu$ имеет только одну продольную составляющую. Все
координаты и импульсы находятся из решения связей и калибровочных условий.
Поэтому электромагнитное поле в двух измерениях не имеет физических
(распространяющихся) степеней свободы. В этом смысле модель тривиальна. Тем не
менее она представляет интерес, особенно при включении взаимодействия с другими
полями.

Гамильтонова формулировка свободной электродинамики на произвольном многообразии
$\MM$ с метрикой $g$ проводится, по существу, так же как и в пространстве
Минковского. Необходимо произвести замену $\eta_{\al\bt}\mapsto g_{\al\bt}$ и
рассматривать компоненты метрики $g_{\al\bt}$ как внешнее поле. Почти все
формулы останутся при этом без изменения, только в ковариантных производных
появятся символы Кристоффеля, возникающие при интегрировании по частям.
\subsection{Скалярная электродинамика                            \label{scaedy}}
Рассмотрим простейший пример нетривиального взаимодействия полей материи с
электромагнитным полем в пространстве Минковского $\MR^{1,n-1}$ произвольного
числа измерений. Эта модель -- {\em скалярная электродинамика}
\index{Скалярная электродинамика (scalar electrodynamics)}%
\index{Электродинамика скалярная (scalar electrodynamics)}%
-- описывает комплексное скалярное поле $\vf$, минимально взаимодействующее
с электромагнитным полем $A_\al$. Соответствующий лагранжиан имеет вид
\begin{equation}                                                  \label{elasce}
  L=-\frac1{4}F^{\al\bt}F_{\al\bt}
  +\et^{\al\bt}\nb_\al\vf^\dagger\nb_\bt\vf-V(\vf^\dagger\vf),
\end{equation}
где ковариантные производные определены следующими соотношениями:
\begin{equation}                                                  \label{ecodss}
\begin{split}
  \nb_\al\vf&:=\pl_\al\vf-iqA_\al\vf,
\\
  \nb_\al\vf^\dagger&:=\pl_\al\vf^\dagger+iqA_\al\vf^\dagger,
\end{split}
\end{equation}
и $q=\const$ -- заряд поля (как мы увидим позже).

Уравнения движения для скалярной электродинамики имеют вид
\begin{align}                                                     \label{eqmoaa}
  \frac{\dl S}{\dl A_\al}&=\pl_\bt F^{\bt\al}
  +iq\vf^\dagger\nb^\al\vf-iq\vf\nb^{\al}\vf^\dagger=0,
\\                                                                \label{eqmosw}
  \frac{\dl S}{\dl\vf^\dagger}&=-\nb^\al\nb_\al\vf-V'\vf=0,
\\                                                                \label{eqmoss}
  \frac{\dl S}{\dl\vf}&=-\nb^\al\nb_\al\vf^\dagger-V'\vf^\dagger=0.
\end{align}
Сравнение этих уравнений с уравнениями Максвелла в той форме, как они
обычно записываются в физической литературе (см., например, \cite{LanLif88R}),
показывает, что константа связи $q$ имеет физическую интерпретацию заряда поля.

Лагранжиан (\ref{elasce}) инвариантен относительно калибровочных преобразований
\begin{equation}                                                  \label{egatsw}
\begin{split}
  \vf&\mapsto\vf'=e^{iq\phi}\vf,
\\
  \vf^\dagger&\mapsto\vf^{\prime\dagger}=e^{-iq\phi}\vf^\dagger,
\\
  A_\al&\mapsto A^\prime_\al=A_\al+\pl_\al\phi,
\end{split}
\end{equation}
где $\phi(x)$ -- локальный параметр преобразования. Согласно второй теореме
Нетер отсюда вытекает линейная зависимость уравнений движения:
\begin{equation}                                                  \label{eldeem}
  -\pl_\al\frac{\dl S}{\dl A_\al}+iq\vf\frac{\dl S}{\dl\vf}
  -iq\vf^\dagger\frac{\dl S}{\dl\vf^\dagger}=0.
\end{equation}
В случае свободного электромагнитного поля зависимость уравнений движения
(\ref{emfdem}) выглядит тривиально. Для скалярной электродинамики доказать
зависимость уравнений (\ref{eqmoaa})--(\ref{eqmoss}) без обращения к
калибровочной инвариантности намного сложнее.

В пространстве Минковского лагранжиан инвариантен относительно действия группы
Пуанкаре $\MI\MS\MO(1,n-1)$. Этой инвариантности, согласно первой теореме Нетер,
соответствует закон сохранения энергии импульса и момента количества движения.
Канонический тензор энергии-импульса равен сумме тензора энергии-импульса
электромагнитного (\ref{enmoea}) и скалярного полей (\ref{enmotc}):
\begin{equation}                                                  \label{enmose}
  \widehat T^{(c)}_{\al\bt}=-\pl_\al A_\g F_\bt{}^\g+\frac14\et_{\al\bt}F^2
  +\pl_\al\vf^\dagger\nb_\bt\vf+\nb_\bt\vf^\dagger\pl_\al\vf
  -\et_{\al\bt}\big[\nb^\g\vf^\dagger\nb_\g\vf-V(\vf^\dagger\vf)\big],
\end{equation}
который, очевидно, несимметричен. Как и в случае свободного электромагнитного
поля, добавление дивергенции (\ref{editee}) с учетом уравнений движения
позволяет определить симметричный и калибровочно инвариантный тензор
энергии-импульса
\begin{equation}                                                  \label{enmoss}
  T^{(s)}_{\al\bt}=-F_{\al\g}F_\bt{}^\g+\frac14\et_{\al\bt}F^2
  +\nb_\al\vf^\dagger\nb_\bt\vf+\nb_\bt\vf^\dagger\nb_\al\vf
  -\et_{\al\bt}\big[\nb^\g\vf^\dagger\nb_\g\vf-V(\vf^\dagger\vf)\big].
\end{equation}

Выражение для тензора спина в скалярной электродинамике такое же, как и для
свободного электромагнитного поля (\ref{espmem}). Связь полного момента
количества движения с орбитальным моментом, определяемым симметричным тензором
энергии-импульса также остается прежней (\ref{eanmne}).

Сохраняющийся вектор электромагнитного тока, соответствующий глобальным
преобразованиям (\ref{egatsw}), имеет вид
\begin{equation}                                                  \label{ecoced}
  J^\al=-\pl_\bt F^{\al\bt}+iq(\vf^\dagger\nb^\al\vf-\vf\nb^\al\vf^\dagger).
\end{equation}
Заметим, что зависимость уравнений движения (\ref{eldeem}), вытекающая
из второй теоремы Нетер, и сохранение тока $\pl_\al J^\al$ -- не одно и то же.

Спектр модели, т.е.\ массы, спины и заряды частиц после вторичного квантования,
определяется квадратичным приближением лагранжиана, которое, в свою очередь,
зависит от вида потенциала самодействия скалярного поля. Если скалярное поле
свободно, т.е.\
\begin{equation*}
  V(\vf^\dagger\vf)=m^2\vf^\dagger\vf,\qquad m>0,
\end{equation*}
то модель (\ref{elasce}) в низкоэнергетическом пределе описывает заряженные
скалярные частицы массы $m$, взаимодействующие с безмассовым электромагнитным
полем.

Если потенциал скалярного поля описывает самодействие, то ситуация может
сильно измениться. Продемонстрируем это на примере потенциала ``$\lm\vf^4$''
(\ref{evslmf}). Сначала перепишем лагранжиан (\ref{elasce}) через действительную
и мнимую части:
\begin{equation*}
  L=-\frac1{4}F^{\al\bt}F_{\al\bt}
  +\et^{\al\bt}\big(\nb_\al\vf_1\nb_\bt\vf_1+\nb_\al\vf_2\nb_\bt\vf_2\big)
  -V(\vf_1^2+\vf_2^2),
\end{equation*}
где
\begin{align*}
  \nb_\al\vf_1&:=\pl_\al\vf_1+qA_\al\vf_2,
\\
  \nb
  _\al\vf_2&:=\pl_\al\vf_2-qA_\al\vf_1.
\end{align*}
В таком виде скалярные поля $\vf_1$ и $\vf_2$ преобразуются по неприводимому
двумерному вещественному представлению унитарной группы $\MU(1)$ (векторному
представлению группы вращений $\MS\MO(2)$) Вакуумное решение выберем однородным
и изотропным:
\begin{equation*}
  \vf_0=a,\qquad A_\al=0
\end{equation*}
и произведем разложение скалярного поля (\ref{qvacas}) вблизи вакуума.
В новых переменных лагранжиан (\ref{elasce}) примет вид
\begin{multline*}
  L=-\frac1{4}F^{\al\bt}F_{\al\bt}+\et^{\al\bt}
  (\pl_\al\tilde\vf_1\pl_\bt\tilde\vf_1+\pl_\al\vf_2\pl_\bt\vf_2)+
\\
  +2qA^\al(\vf_2\pl_\al\tilde\vf_1-\tilde\vf_1\pl_\al\vf_2-a\pl_\al\vf_2)
  +q^2A^\al A_\al(a^2+2a\tilde\vf_1+\tilde\vf_1^2+\vf_2^2)+
\\
  -\frac12\lm[4a^2\tilde\vf_1^2+4a\tilde\vf_1^3+4a\tilde\vf_1\vf_2^2
  +\tilde\vf_1^4+\vf_2^4+2\tilde\vf_1^2\vf_2^2].
\end{multline*}
Проанализируем квадратичное приближение. Мы видим, что вблизи вакуума векторное
поле приобрело массовый член $q^2a^2A^\al A_\al$, и появилось слагаемое с
перемешиванием $-2qaA^\al\pl_\al\vf_2$. Заметим, что квадратичные слагаемые по
$A_\al$ и $\pl_\al\vf_2$ можно записать в виде полного квадрата:
\begin{equation*}
  \eta^{\al\bt}\pl_\al\vf_2\pl_\bt\vf_2-2qaA^\al\pl_\al\vf_2+q^2a^2A^\al A_\al
  =q^2a^2\eta^{\al\bt}\left(A_\al-\frac1{qa}\pl_\al\vf_2\right)
  \left(A_\bt-\frac1{qa}\pl_\bt\vf_2\right).
\end{equation*}
Поскольку замена компонент векторного поля
\begin{equation*}
  A_\al\mapsto B_\al:=A_\al-\frac1{qa}\pl_\al\vf_2
\end{equation*}
не изменяет кинетического члена, то это преобразование диагонализирует
квадратичную часть лагранжиана. В результате возникает лагранжиан с квадратичной
частью, которая вообще не содержит голдстоуновское поле $\vf_2$:
\begin{equation}                                                  \label{quadla}
  -\frac14 F^{\al\bt}_\Sb F_{\Sb\al\bt}+q^2a^2B^\al B_\al
  +\pl\tilde\vf_1^2 -2\lm a^2\tilde\vf_1^2,
\end{equation}
где
\begin{equation*}
  F_{\Sb\al\bt}:=\pl_\al B_\bt-\pl_\bt B_\al.
\end{equation*}
Теперь можно описать спектр теории. Модель описывает массивное векторное поле,
взаимодействующее с нейтральным массивным скалярным полем. Таким образом,
безмассовое голдстоуновское поле исчезло, а векторное поле приобрело массу. При
этом общее число степеней свободы не изменилось, т.к.\ массивное векторное поле,
как мы увидим позже в разделе \ref{saadew}, имеет на одну степень свободы
больше, чем безмассовое. То есть одна степень свободы комплексного скалярного
поля перешла в продольную компоненту массивного векторного поля.

Обратим внимание, что массы полей имеют правильные знаки.

Поле $\vf_2$ при описанном преобразовании остается в высших порядках
лагранжиана, однако это не влияет на спектр теории.

Появление массивного векторного поля в физическом секторе модели можно пояснить
по-другому. Параметризуем комплексное векторное поле его амплитудой и фазой:
\begin{equation*}
  \vf:=\rho\ex^{iq\theta},\qquad \rho=\rho(x),\quad \theta=\theta(x).
\end{equation*}
Поскольку исходная теория калибровочно инвариантна, то физические степени
свободы остаются после фиксирования калибровки. Совершим калибровочное
преобразование
\begin{align*}
  \vf&\mapsto\vf'=\ex^{-iq\theta},
\\
  A_\al&\mapsto A'_\al=A_\al-\pl_\al\theta.
\end{align*}
Другими словами, зафиксируем калибровку $\im\vf=0$. Тогда лагранжиан примет
вид (опустив штрихи)
\begin{align*}
  L&=-\frac14F^{\al\bt}F_{\al\bt}+\eta^{\al\bt}(\pl_\al\rho+iqA_\al\rho)
  (\pl_\bt\rho-iqA_\bt\rho)-V(\rho^2)=
\\
  &=-\frac14F^{\al\bt}F_{\al\bt}+\pl\rho^2+q^2A^\al A_\al\rho^2-V(\rho^2).
\end{align*}
Теперь разложим поля вблизи вакуумного решения
\begin{equation*}
  \rho=a+\tilde\rho,\qquad A_\al=A_\al.
\end{equation*}
Тогда квадратичная часть лагранжиана примет вид
\begin{equation*}
  -\frac14F^{\al\bt}F_{\al\bt}+\pl\tilde\rho^2+q^2a^2A^\al A_\al
  -2\lm a^2\tilde\rho^2.
\end{equation*}
Это выражение с точностью до обозначений совпадает с полученным ранее выражением
(\ref{quadla}). То есть мы получили прежний спектр модели. При таком подходе
голдстоуновское нефизическое поле, каковым является фаза скалярного поля, вообще
исключается из лагранжиана.

Описанная выше конструкция обобщается на теорию полей Янга--Миллса,
взаимодействующих со скалярными полями. При этом безмассовые голдстоуновские
бозоны исчезают, а калибровочные поля приобретают массу. Это явление
приобретения масс первоначально безмассовыми векторными полями за счет
спонтанного нарушения симметрии называется {\em механизмом Хиггса}.
\index{Механизм Хиггса (Higgs mechanism)}%
\index{Хиггса механизм (Higgs mechanism)}%
Оно было независимо открыто Энглертом, Броутом и Хиггсом
\cite{EngBro64,Higgs64}.

Как видим, спектр теории, который определяется низкоэнергетическим приближением,
существенно меняется при спонтанном нарушении симметрии. Современные модели,
объединяющие электромагнитные, слабые и сильные взаимодействия, основываются на
калибровочной симметрии и соответствующих полях Янга--Миллса. Поэтому теория
изначально содержит большое число безмассовых векторных полей. С другой стороны,
нам экспериментально известно только одно безмассовое векторное поле --
электромагнитное. Чтобы устранить возникающее противоречие теории и эксперимента
вводится взаимодействие со скалярными полями (хиггсовскими бозонами) таким
образом, чтобы вакуумное решение уравнений движения нарушало калибровочную
симметрию. В результате калибровочные поля приобретают массу за счет механизма
Хиггса, и можно сказать, что мы эти поля не наблюдаем, потому что они слишком
массивны.

Важно отметить, что перенормируемость квантовополевой модели определяется не
низко-, а высокоэнергетическим приближением, которое у спонтанно нарушенной
теории такое же, как и до нарушения симметрии. Поэтому свойство
перенормируемости сохраняется при спонтанном нарушении симметрии.
\subsection{Электромагнитное поле в аффинной геометрии           \label{selmag}}
Лагранжиан электромагнитного поля на многообразии $\MM$ с заданной аффинной
геометрией при минимальной подстановке принимает вид
\begin{equation}                                                  \label{emfage}
  L_{\Se\Sm}=-\frac14\vol g^{\al\bt}g^{\g\dl}F_{\al\g}F_{\bt\dl}
  :=-\frac14\vol F^2,
\end{equation}
где $F_{\al\bt}$ -- напряженность электромагнитного поля (\ref{qstrem}). Ему
соответствует действие
\begin{equation}                                                  \label{qacemf}
  S_{\Se\Sm}=\int_\MM\!\!\!dxL_{\Se\Sm}.
\end{equation}
\begin{com}
Как и в случае скалярных полей, лагранжиан (\ref{emfage}) не зависит от аффинной
связности. Это связано с тем, что $F_{\al\bt}$ -- это компоненты локальной
формы кривизны для $U(1)$-связности и являются компонентами тензора независимо
от того задана или нет на многообразии $\MM$ аффинная геометрия. С
геометрической точки зрения в выражении для $F_{\al\bt}$ при минимальной
подстановке нет никакой необходимости заменять частные производные на
ковариантные, т.к.\ локальная форма кривизны для $U(1)$-связности уже является
тензором относительно общих преобразований координат.

Если все же в определении компонент напряженности электромагнитного поля
заменить частные производные на ковариантные, определяемые аффинной связностью
$\Gamma$, то получим добавочное слагаемое с тензором кручения:
\begin{equation*}
  \nb_\al A_\bt-\nb_\bt A_\al=\pl_\al A_\bt-\pl_\bt A_\al-T_{\al\bt}{}^\g A_\g.
\end{equation*}
Полученное выражение ковариантно относительно общих преобразований координат,
однако добавочный член явно нарушает калибровочную инвариантность. Такая замена
соответствует неминимальной подстановке.
\qed\end{com}
\begin{com}
В настоящее время считается, что спектр модели: массы, спины и заряды частиц,
описываемых данным полем, определяется линейным приближением в уравнениях
движения. При этом метрика, как правило, раскладывается вблизи метрики
Минковского. Это означает, что электромагнитное поле в аффинной геометрии
описывает безмассовые нейтральные частицы со спином единица. Такой подход
вызывает много вопросов. Например, если пространство-время имеет нетривиальную
топологию, то говорить про метрику Лоренца не имеет никакого смысла. Мы не будем
обсуждать эти важные вопросы, потому что лучших предложений пока не поступало.
\qed\end{com}

Уравнения движения для электромагнитного поля и тензор энергии-импульса
получаются варьированием соответствующего действия
\begin{align}                                                     \label{ememag}
  \vol S_{\Se\Sm},{}^\al:=\frac{\dl S_{\Se\Sm}}{\dl A_\al}&
  =\pl_\bt(\vol F^{\bt\al})=\vol \widetilde\nb_\bt F^{\bt\al}=0,
\\                                                                \label{emmeag}
  \vol S_{\Se\Sm},{}_{\al\bt}:=\frac{\dl S_{\Se\Sm}}{\dl g^{\al\bt}}&
  =\frac12\vol T_{\Se\Sm\al\bt}
  =-\frac12\vol F_{\al\g}F_\bt{}^\g+\frac18\vol g_{\al\bt}F^2.
\end{align}
Отсюда следует выражение для тензора энергии-импульса электромагнитного поля
\begin{equation}                                                  \label{enmoem}
  T_{\Se\Sm\al\bt}=-F_{\al\g}F_\bt{}^\g+\frac14g_{\al\bt}F^2.
\end{equation}
\begin{com}
След тензора энергии-импульса электромагнитного поля в четырехмерном
пространстве-времени равен нулю, $T_{\Se\Sm}{}^\al{}_\al=0$. Здесь важна
четырехмерность пространства-времени, т.к.\, если размерность не равна четырем,
то след тензора энергии-импульса отличен от нуля.
\qed\end{com}
\begin{com}
Вариация действия по метрике (\ref{emmeag}) приводит к тензору энергии-импульса
электромагнитного поля (\ref{enmoem}), стоящему в правой части уравнений
Эйнштейна (\ref{einequ}). Он является ковариантным обобщением симметричного
канонического тензора энергии-импульса (\ref{esyemt}) в пространстве
Минковского. В этом отношении ситуация с электромагнитным полем такая же, как и
для скалярного поля. В общем случае вариация действия полей материи по метрике
совсем необязательно приводит к ковариантному обобщению канонического тензора
энергии-импульса, следующему из теоремы Нетер.
\qed\end{com}
Если присутствуют источники электромагнитного поля, то в правой части уравнения
(\ref{ememag}) появляется электрический ток:
\begin{equation}                                                  \label{qqcurt}
  \widetilde\nb_\bt F^{\bt\al}=J^\al,
\end{equation}
где $J=\lbrace J^\al\rbrace$ -- вектор тока. Происхождение электрического тока в
настоящий момент не имеет значения.
\begin{prop}
Если пространство-время топологически тривиально (диффеоморфно $\MR^n$), то
система уравнений второго порядка для потенциала $A_\al$ (\ref{qqcurt})
эквивалентна системе уравнений первого порядка для компонент напряженности
$F_{\al\bt}=-F_{\bt\al}$ электромагнитного поля:
\begin{equation}                                                  \label{qmaxeq}
\begin{split}
  \widetilde\nb_\bt F^{\bt\al}&=J^\al,
\\
  \pl_\al F_{\bt\g}+\pl_\bt F_{\g\al}+\pl_\g F_{\al\bt}&=0
\end{split}
\end{equation}
\end{prop}
\begin{proof}
Пусть $F_{\al\bt}=\pl_\al A_\bt-\pl_\bt A_\al$. Тогда из уравнения
(\ref{qqcurt}) следует система уравнений (\ref{qmaxeq}). Обратно. В силу леммы
Пуанкаре \ref{tlempo} из второго уравнения (\ref{qmaxeq}) для односвязных
многообразий следует существование потенциала $A_\al$. Подстановка
соответствующего выражения (\ref{qstrem}) в первое уравнение (\ref{qmaxeq})
приводит к (\ref{qqcurt}).
\end{proof}
\begin{defn}
Система уравнений первого порядка (\ref{qmaxeq}) для напряженности
электромагнитного поля $F_{\al\bt}=-F_{\bt\al}$ называется {\em уравнениями
Максвелла}.
\qed\end{defn}
\index{Уравнения Максвелла (Maxwell's equations)}%
\index{Максвелла уравнения (Maxwell's equations)}%
Запишем действие для электромагнитного поля в еще одной форме.
\begin{prop}                                                      \label{pfidre}
Уравнения движения для действия (\ref{qacemf}) эквивалентны уравнениям
Эйлера--Лагранжа для действия
\begin{equation}                                                  \label{qemfio}
  S=-\int_\MM\!\!\!dx\vol\left[\frac12F^{\al\bt}(\pl_\al A_\bt-\pl_\bt A_\al)
  -\frac14F^{\al\bt}F_{\al\bt}\right],
\end{equation}
в котором компоненты $F^{\al\bt}$ и $A_\al$ рассматриваются в качестве
независимых переменных, по которым проводится варьирование.
\end{prop}
\begin{proof}
Уравнения движения для компонент напряженности являются алгебраическими:
\begin{equation*}
  \frac{\dl S}{\dl F^{\al\bt}}=\frac12\big(\pl_\al A_\bt-\pl_\bt A_\al
  -F_{\al\bt}\big)=0.
\end{equation*}
Их решение приводит к выражению компонент напряженности электромагнитного поля
через потенциал (\ref{qstrem}). Согласно теореме \ref{teffac} решение уравнений
Эйлера--Лагранжа можно подставлять в действие. После подстановки действие
(\ref{qemfio}) совпадет с (\ref{qacemf}).
\end{proof}
Рассмотрение в качестве независимых переменных и компонент потенциала
электромагнитного поля $A_\al$, и компонент напряженности $F_{\al\bt}$
называется {\em формализмом первого порядка}.
\index{Формализм первого порядка (first-order formalism)}%

Действие для лагранжиана (\ref{emfage}) инвариантно относительно общих
преобразований координат. Соответствующие вариации полей имеют вид (см.\ раздел
\ref{sinfct})
\begin{align*}
  \dl A_\al&=-\pl_\al\e^\bt A_\bt-\e^\bt\pl_\bt A_\al,
\\
  \dl g^{\al\bt}&=g^{\al\g}\pl_\g\e^\bt+g^{\bt\g}\pl_\g\e^\al
  -\e^\g\pl_\g g^{\al\bt}.
\end{align*}
Отсюда, согласно второй теореме Нетер (\ref{edepel}), следует зависимость
уравнений движения
\begin{equation}                                                  \label{emdegc}
  \pl_\bt S,{}^\bt A_\al+S,{}^\bt F_{\bt\al}-\pl_\bt S,{}^\bt{}_\al
  -\pl_\bt S,{}_\al{}^\bt-S,{}_{\bt\g}\pl_\al g^{\bt\g}=0.
\end{equation}
Или, в ковариантном виде,
\begin{equation}                                                  \label{emcold}
  \widetilde\nb_\bt S,{}^\bt A_\al+S,{}^\bt F_{\bt\al}
  -2\widetilde\nb^\bt S,{}_{\bt\al}=0.
\end{equation}

Если выполнены уравнения движения для электромагнитного поля (\ref{ememag}), то
с учетом свернутых тождеств Бианки соотношение (\ref{emcold}) принимает вид
\begin{equation}                                                  \label{ecocem}
  \widetilde\nb_\bt T_{\Se\Sm\al}{}^\bt=0.
\end{equation}
Полученное равенство, выполненное для всех решений уравнений Эйлера--Лагранжа,
можно интерпретировать, как ковариантное обобщение закона сохранения тензора
энергии-импульса.

Действие для электромагнитного поля инвариантно также относительно калибровочных
преобразований (\ref{qgaaco}), при которых метрика не меняется. Из калибровочной
инвариантности в силу второй теоремы Нетер следует зависимость уравнений
движения:
\begin{equation}                                                  \label{ecogai}
  \widetilde\nb_\al S,{}^\al=0.
\end{equation}
С учетом уравнения Максвелла (\ref{qqcurt}) это равенство приводит к закону
сохранения электрического тока: $\widetilde\nb_\al J^\al=0$. Если ток возникает
при варьировании некоторого калибровочно инвариантного действия для полей
заряженной материи, то сохранение тока будет выполняться автоматически в силу
второй теоремы Нетер. Если же электрический ток вводится в уравнения Максвелла
``руками'', то условие $\widetilde\nb_\al J^\al=0$ необходимо для
самосогласованности уравнений.
\subsection{Электромагнитное поле в общей теории относительности \label{semoto}}
В настоящем разделе мы ограничимся четырехмерным пространством-временем $\MM$, в
котором формулы электродинамики имеют свою специфику.

В свободной электродинамике многие формулы принимают особенно простой вид, если
использовать язык дифференциальных форм, описанных в главе \ref{sdifin}. Как
уже было отмечено, электромагнитный потенциал -- это локальная 1-форма
$\MU(1)$-связности,
\begin{equation}                                                  \label{qlocul}
  A:=dx^\al A_\al.
\end{equation}
Слово ``локальная'' в данном случае означает, что форма связности задана на
некоторой координатной окрестности базы $\MU\subset\MM$ главного расслоения
$\MP\big(\MM,\pi,\MU(1)\big)$, а не на самом (пятимерном) пространстве
расслоения $\MP$. Локальная 2-форма кривизны -- это внешняя производная от
локальной формы связности,
\begin{equation}                                                  \label{qcotfo}
  F:=dA=\frac12dx^\al\wedge dx^\bt(\pl_\al A_\bt-\pl_\bt A_\al).
\end{equation}

В дальнейшем слово ``локальная'' мы, для краткости, будем опускать.

Поскольку форма кривизны является точной, то ее внешняя производная равна нулю
(\ref{eptdth}),
\begin{equation}                                                  \label{qbiank}
  dF=\frac12dx^\al\wedge dx^\bt\wedge dx^\g\,\pl_\al F_{\bt\g}=0.
\end{equation}
Это равенство, очевидно, эквивалентно тождествам Бианки (\ref{ebidem}).

Приведенные выше формулы справедливы независимо от того задана ли на $\MM$
аффинная геометрия или нет.

Если на пространстве-времени задана метрика, то можно определить 2-форму,
дуальную к форме кривизны:
\begin{equation}                                                  \label{qducuu}
  *F:=\frac12dx^\al\wedge dx^\bt(*F)_{\al\bt},
\end{equation}
где компоненты дуальной формы кривизны заданы равенством
\begin{equation}                                                  \label{qdyull}
  *F_{\al\bt}:=\frac12\ve_{\al\bt\g\dl}F^{\g\dl}.
\end{equation}
В приведенной формуле $\ve_{\al\bt\g\dl}(x)$ -- полностью антисимметричный
тензор четвертого ранга. В определении дуальной формы кривизны метрика
встречается дважды. Во-первых, с ее помощью производится подъем и опускание
индексов:
\begin{equation}                                                  \label{qdualt}
  F^{\g\dl}:=g^{\g\al}g^{\dl\bt}F_{\al\bt}.
\end{equation}
Во-вторых, метрика входит в определение полностью антисимметричного тензора:
\begin{equation*}
  \ve_{\al\bt\g\dl}:=\vol\hat\ve_{\al\bt\g\dl},
\end{equation*}
где $\hat\ve_{\al\bt\g\dl}$ -- полностью антисимметричная тензорная плотность
четвертого ранга, чьи компоненты равны по модулю единице (см.\ раздел
\ref{santst}).

Преобразование, обратное к преобразованию дуальности (\ref{qdyull}), имеет вид
\begin{equation}                                                  \label{qindur}
  F_{\al\bt}=-\frac12\ve_{\al\bt\g\dl}(*F)^{\g\dl},
\end{equation}
где мы использовали формулы (\ref{etotfm}) для сверток антисимметричных
тензоров. Полученное равенство эквивалентно следующему свойству оператора Ходжа:
\begin{equation*}
  F=**F.
\end{equation*}
\begin{exa}
В четырехмерном пространстве Минковского $\MR^{1,3}$ компоненты напряженности
выражаются через компоненты электрического и магнитного полей (\ref{emfsma}). В
этом случае нетрудно получить выражения для компоненты дуальной напряженности:
\begin{equation}                                                  \label{qduana}
  (*F)_{\al\bt}=\begin{pmatrix} 0 & H_1 & H_2 & H_3 \\ -H_1 & 0 & -E_3 & E_2 \\
  -H_2 & E_3 & 0 & -E_1 \\ -H_3 & -E_2 & E_1 & 0 \end{pmatrix}.
\end{equation}
Это означает, что преобразование дуальности меняет местами электрическое и
магнитное поля:
\begin{equation*}
  \BE\leftrightarrow-\BH.                                            \tag*{\qed}
\end{equation*}
\end{exa}

Используя дуальную форму кривизны, действие для свободного электромагнитного
поля можно переписать в виде интеграла от 4-формы:
\begin{equation}                                                  \label{qacems}
  S_{\Se\Sm}:=-\frac14\int dx\vol F^{\al\bt}F_{\al\bt}
  =-\frac12\int *F\wedge F,
\end{equation}
поскольку, согласно общей формуле для внешнего произведения (\ref{ewpprd}),
\begin{equation*}
  *F\wedge F=\frac14dx^\al\wedge dx^\bt\wedge dx^\g\wedge dx^\dl\,
  *F_{\al\bt}F_{\g\dl}=\frac18dx^\al\wedge dx^\bt\wedge dx^\g\wedge dx^\dl\,
  \ve_{\al\bt\e\z}F^{\e\z}F_{\g\dl}.
\end{equation*}
Кроме этого мы воспользовались формулой (\ref{eintin}) для интегрирования форм
и сверткой полностью антисимметричных тензоров четвертого ранга (\ref{ecfmif}).

Свернув уравнения Максвелла (\ref{qmaxeq}) с антисимметричным тензором, получим
уравнение Максвелла в терминах дуальной формы кривизны:
\begin{equation}                                                  \label{qmazdu}
\begin{split}
  \pl_\al(*F)_{\bt\g}+\pl_\bt(*F)_{\g\al}+\pl_\g(*F)_{\al\bt}&
  =J^\dl\ve_{\dl\al\bt\g},
\\
  \nb_\bt(*F)^{\bt\g}=0.
\end{split}
\end{equation}
Поскольку преобразование дуальности меняет местами электрическое и магнитное
поле, то второе из этих уравнений означает отсутствие в природе магнитных
зарядов.

Первую пару уравнений Максвелла можно записать, используя оператор кограницы
(\ref{edefdl}), который в рассматриваемом случае имеет вид
\begin{equation*}
  \dl:=*d*.
\end{equation*}
Несложные вычисления позволяют переписать первую пару уравнений Максвелла
(\ref{qmazdu}) в виде равенства
\begin{equation*}
  \dl F=\frac12J,
\end{equation*}
где $J:=dx^\al J_\al$ -- 1-форма электрического тока.

Остановимся на уравнениях электромагнитного поля без источников. Тогда первую
пару уравнений Максвелла (\ref{qmazdu}) можно записать в виде
\begin{equation}                                                  \label{qfinax}
  d*F=0.
\end{equation}
Это уравнение можно получить путем варьирования действия (\ref{qacems})
непосредственно в терминах форм. Поскольку $*F$ -- это 2-форма, то из свойства
(\ref{eptdse}) следует формула
\begin{equation*}
  d(*F\wedge A)=d(*F)\wedge A+*F\wedge dA.
\end{equation*}
Если вариации всех полей исчезают на границе, то уравнения движения в форме
(\ref{qfinax}) легко получить, варьируя непосредственно действие (\ref{qacems}):
\begin{equation*}
  \bar\dl S_{\Se\Sm}=-\int *F\wedge\bar\dl(dA)=\int d(*F)\wedge\bar\dl A,
\end{equation*}
где мы пометили вариацию формы чертой, чтобы отличить ее от оператора кограницы.

В четырехмерном пространстве-времени можно построить еще одну 4-форму
\begin{equation*}
  F\wedge F.
\end{equation*}
Эта форма является точной, потому что из тождеств Бианки следует равенство
\begin{equation*}
  F\wedge dA=d(F\wedge A)-dF\wedge A=d(F\wedge A).
\end{equation*}

Рассмотрим {\em уравнение самодуальности}
\index{Уравнение самодуальности (selfdual equation)}%
\begin{equation}                                                  \label{qsefdu}
  F=*F\qquad \Leftrightarrow\qquad F_{\al\bt}=*F_{\al\bt}.
\end{equation}
Если в правой части этих уравнений стоит знак минус, то уравнение называется
{\em антисамодуальным}.
\index{Уравнение антисамодуальности (antiselfdual equation)}%
\begin{prop}
Если потенциал электромагнитного поля удовлетворяет уравнению
(анти-)самодуальности, то он также удовлетворяет уравнениям Эйлера--Лагранжа
(\ref{ememag}).
\end{prop}
\begin{proof}
Возьмем внешнюю производную от уравнения (\ref{qsefdu}):
\begin{equation*}
  dF=d*F.
\end{equation*}
Левая часть полученного соотношения тождественно равна нулю в силу тождеств
Бианки (\ref{qbiank}). Поэтому возникают уравнения Эйлера--Лагранжа в форме
(\ref{qfinax}).
\end{proof}
В отличие от уравнений Эйлера--Лагранжа уравнения (анти-)самодуальности имеют
первый порядок и проще.

Теперь изучим свойства тензора энергии-импульса в четырехмерном
пространстве-времени. Из определения дуального тензора напряженности
(\ref{qducuu}) вытекает равенство
\begin{equation*}
  *F_{\al\g}*F_\bt{}^\g=-\frac12g_{\al\bt}F^{\g\dl}F_{\g\dl}
  +F_{\al\g}F_\bt{}^\g,
\end{equation*}
где мы воспользовались формулами свертки антисимметричных тензоров
(\ref{ecfmif}). Теперь выражение для тензора энергии-импульса можно переписать
в эквивалентном виде
\begin{equation}                                                  \label{qenmoe}
  T_{\Se\Sm\al\bt}=-\frac12\left[F_{\al\g}F_\bt{}^\g
  +*F_{\al\g}*F_\bt{}^\g\right].
\end{equation}

Из приведенного выражения следует, что тензор энергии-импульса электромагнитного
поля инвариантен относительно {\em дуальных $\MS\MO(2)$ вращений}:
\index{Дуальное вращение (dual rotation)}%
\index{Вращение дуальное (dual rotation)}%
\begin{equation}                                                  \label{qdurot}
\begin{split}
  F_{\al\bt}&\mapsto F'_{\al\bt}=F_{\al\bt}\cos\om-*F_{\al\bt}\sin\om,
\\
  *F_{\al\bt}&\mapsto *F'_{\al\bt}=F_{\al\bt}\sin\om+*F_{\al\bt}\cos\om,
\end{split}
\end{equation}
где $\om\in[0,2\pi]$ -- угол вращения. Конечно, второе из этих уравнений
является следствием первого.
\begin{exa}
В пространстве Минковского $\MR^{1,3}$ компоненты тензора напряженности и его
дуального имеют вид (\ref{emfsma}), (\ref{qduana}). Поэтому для временн\'ой
компоненты тензора энергии-импульса (\ref{qenmoe}) получаем хорошо известное
выражение
\begin{equation}                                                  \label{qmiene}
  T_{\Se\Sm00}=\frac12(\BE^2+\BH^2).
\end{equation}

Смешанная компонента контравариантного тензора энергии-импульса
электромагнитного поля имеет вид
\begin{equation*}
  T_{\Se\Sm}^{\mu0}=F^{\mu\nu}F^0{}_\nu.
\end{equation*}
Используя выражение напряженности электромагнитного поля через электрическое и
магнитное поля  (\ref{emfsma}) нетрудно получить соотношение
\begin{equation*}
  T_{\Se\Sm}^{\mu0}=\ve^{\mu\nu\rho}E_\nu H_\rho,
\end{equation*}
где $\ve^{\mu\nu\rho}$ -- полностью антисимметричный тензор третьего ранга
(\ref{etttat}). Полученное равенство можно записать в виде векторного
произведения
\begin{equation}                                                  \label{qvecpo}
  \BT=\BE\times\BH.
\end{equation}
Этот вектор называется {\em вектором Пойнтинга} и имеет смысл плотности
потока энергии электромагнитного поля.
\index{Вектор Пойнтинга (Pointing vector)}%
\index{Пойнтинга вектор (Pointing vector)}%
При этом закон сохранения электромагнитной энергии примет вид
\begin{equation*}
  \pl_0T_{\Se\Sm}^{00}+\pl_\mu T_{\Se\Sm}^{\mu0}=0.                  \tag*{\qed}
\end{equation*}
\end{exa}
\begin{prop}
Если метрика на многообразии $\MM$ имеет лоренцеву сигнатуру и координата $x^0$
является временем, то временн\'ая компонента тензора энергии-импульса
электромагнитного поля
\begin{equation}                                                  \label{qzzcos}
  T_{\Se\Sm00}=-\frac12\left[F_{0\g}F_0{}^\g+*F_{0\g}*F_0{}^\g\right]
\end{equation}
положительно определена и, следовательно, удовлетворяет слабому энергетическому
условию.
\end{prop}
\begin{proof}
Ввиду антисимметрии по индексам первое слагаемое в выражении (\ref{qzzcos})
имеет вид
\begin{equation*}
  -\frac12F_{0\mu}F_{0\nu}g^{\mu\nu}.
\end{equation*}
При сделанных предположениях матрица $g^{\mu\nu}$ согласно предложению
\ref{potrit} отрицательно определена. То же верно и для второго слагаемого.
\end{proof}
\begin{cor}
Временн\'ая компонента тензора энергии-импульса электромагнитного поля
удовлетворяет сильному энергетическому условию (\ref{estenc}).
\end{cor}
\begin{proof}
Поскольку след тензора энергии-импульса равен нулю, то понятие слабого и
сильного энергетического условия эквивалентны.
\end{proof}
\begin{prop}
Действие электромагнитного поля в аффинной геометрии (\ref{qacemf}) инвариантно
относительно преобразований Вейля:
\begin{equation}                                                  \label{qweyam}
  g_{\al\bt}\mapsto\bar g_{\al\bt}=\ex^{2\phi}g_{\al\bt},\qquad
  A_\al\mapsto \bar A_\al=A_\al,
\end{equation}
где $\phi(x)\in\CC^2(\MM)$ -- произвольная вещественнозначная функция.
\end{prop}
\begin{proof}
Утверждение следует из равенства
\begin{equation*}
  \vol g^{\al\bt}g^{\g\dl}=\sqrt{|\bar g|}\bar g^{\al\bt}\bar g^{\g\dl}.
  \tag*{\qed}
\end{equation*}
\renewcommand{\qed}{}\end{proof}
\begin{com}
Вейлевская инвариантность действия электромагнитного поля является спецификой
четырехмерности пространства-времени. Если размерность пространства-времени
отлична от четырех, то след тензора энергии-импульса электромагнитного поля
отличен от нуля.
\qed\end{com}

Равенство нулю следа тензора энергии-импульса электромагнитного поля связано с
наличием вейлевской калибровочной инвариантности. Этот факт является общим.
\begin{prop}                                                      \label{pweyrf}
Пусть действие зависит от метрики $g_{\al\bt}$ и некоторого набора полей материи
$\vf^\Sa$
\begin{equation*}
  S_\Sm=\int\!dxL_\Sm(g,\vf).
\end{equation*}
Если действие инвариантно относительно преобразований Вейля, не затрагивающих
поля материи,
\begin{equation*}
  g_{\al\bt}\mapsto\bar g_{\al\bt}=\ex^{2\phi}g_{\al\bt},\qquad
  \vf^\Sa\mapsto \bar\vf^\Sa=\vf^\Sa,
\end{equation*}
то след тензора энергии-импульса полей материи равен нулю, $T_{\Sm\al}{}^\al=0$.
\end{prop}
\begin{proof}
Поскольку параметр преобразования Вейля зависит от точки пространства-времени,
то справедлива вторая теорема Нетер. В рассматриваемом случае это означает
следующее. Инвариантность действия имеет вид
\begin{equation*}
  \dl S_\Sm=\int dx\frac{\dl S_\Sm}{\dl g^{\al\bt}}\dl g^{\al\bt}=0.
\end{equation*}
Учтем определение тензора энергии-импульса в общей теории относительности
(\ref{edenmo}) и выражение для бесконечно малых преобразований Вейля
\begin{equation*}
  \dl g^{\al\bt}=-2\phi g^{\al\bt}.
\end{equation*}
Тогда равенство нулю вариации действия полей материи равносильно равенству
$g^{\al\bt}T_{\Sm\al\bt}=0$.
\end{proof}
\begin{exa}
Бозонная струна описывается набором скалярных безмассовых полей $X^a(x)$,
$a=0,1,\dotsc,\Sd-1$, заданных на двумерном пространстве-времени (поверхности)
$\MM$ и принимающих значения в $\Sd$-мерном пространстве Минковского
$\MR^{1,\Sd-1}$:
\begin{equation*}
  X:\qquad
  \MM\ni\quad x=\lbrace x^\al\rbrace\mapsto
  X(x)=\lbrace X^a(x)\rbrace\quad\in\MR^{1,\Sd-1}.
\end{equation*}
Действие для бозонной струны можно записать в виде
\begin{equation*}
  S=-\frac12\int_\MM\!\!\!dx\vol g^{\al\bt}\pl_\al X^a\pl_\bt X^b\eta_{ab}.
\end{equation*}
Знак этого действия выбран таким образом, чтобы пространственные компоненты
$X^i$, $i=1,\dotsc,\Sd-1$, описывались обычным действием (\ref{escfag}) и давали
положительный вклад в гамильтониан системы. Это действие инвариантно
относительно преобразований Вейля:
\begin{equation*}
  g_{\al\bt}\mapsto\bar g_{\al\bt}=\ex^{2\phi}g_{\al\bt},\qquad
  X^a\mapsto X^a.
\end{equation*}
Тензор энергии-импульса бозонной струны имеет вид
\begin{equation*}
  T_{\al\bt}:=\frac2\vol\frac{\dl S}{\dl g^{\al\bt}}=-\pl_\al X^a\pl_\bt X_a
  +\frac12 g_{\al\bt}g^{\g\dl}\pl_\g X^a\pl_\dl X_a,
\end{equation*}
где $X_a:=X^b\eta_{ba}$. Его след, очевидно, равен нулю. Вейлевская
инвариантность действия для безмассового скалярного поля является спецификой
двумерия. В пространстве-времени б\'ольшего числа измерений она отсутствует.
\qed\end{exa}
\section{Поле Прок\'а                                            \label{saadew}}
Поле Прок\'а $A=\lbrace A_\al\rbrace$, $\al=0,1,\dotsc,n-1$, описывает массивные
векторные частицы со спином единица \cite{Proca36A,Proca36B}. В пространстве
Минковского $\MR^{1,n-1}$ ему соответствует лагранжиан
\begin{equation}                                                  \label{qprogi}
  L=-\frac14 F^{\al\bt}F_{\al\bt}+\frac12m^2A^\al A_\al,
\end{equation}
где, как и для электромагнитного поля,
\begin{equation*}
  F_{\al\bt}:=\pl_\al A_\bt-\pl_\bt A_\al
\end{equation*}
и $m>0$ -- масса поля. Он отличается от лагранжиана электромагнитного поля
только массовым членом. Знак перед массовым членом выбран так, чтобы вклад
пространственных компонент $A_\mu$, $\mu=1,\dotsc,n-1$, в плотность энергии был
положителен.

Уравнения движения для лагранжиана (\ref{qprogi}) имеют вид
\begin{equation}                                                  \label{qeqpro}
  S,^\al:=\frac{\dl S}{\dl A_\al}=\pl_\bt F^{\bt\al}+m^2A^\al=0.
\end{equation}
Ввиду антисимметрии относительно перестановки индексов,
$F^{\al\bt}=-F^{\bt\al}$, дивергенция этой системы уравнений приводит к
дифференциальному уравнению первого порядка на компоненты поля:
\begin{equation}                                                  \label{qaddlo}
  \pl_\al A^\al=0,
\end{equation}
которое должно быть выполнено для любого решения системы уравнений
(\ref{qeqpro}). Полученное условие имеет вид калибровки Лоренца (\ref{qdifga})
для электромагнитного поля. С учетом условия (\ref{qaddlo}) уравнения движения
(\ref{qeqpro}) принимают вид
\begin{equation}                                                  \label{qproeq}
  \square A^\al+m^2 A^\al=0,\qquad \square:=\pl^\al\pl_\al.
\end{equation}
Таким образом для каждой компоненты векторного поля возникает уравнение
Клейна--Гордона--Фока. Однако поле Прок\'а описывает не $n$ распространяющихся
степеней свободы, а только $n-1$, поскольку на компоненты поля наложено
дополнительное условие (\ref{qaddlo}). Чтобы убедиться в этом проведем более
детальный анализ уравнений движения.

Запишем систему уравнений (\ref{qeqpro}) отдельно для временн\'ой и
пространственных компонент:
\begin{align}                                                     \label{qtilag}
  \pl_\mu F^{\mu0}+m^2A^0&=0,
\\                                                                \label{qsplae}
  \pl_0 F^{0\mu}+\pl_\nu F^{\nu\mu}+m^2A^\mu&=0.
\end{align}

Предположим, что все поля убывают на бесконечности. Тогда согласно теореме
\ref{tpuaeo} оператор Лапласа обратим и пространственные компоненты поля
Прок\'а, как и в случае электромагнитного поля (\ref{edivtr}), можно взаимно
однозначно разложить на поперечную и продольную составляющие:
\begin{equation}                                                  \label{qdevef}
  A_\mu=A^\St_\mu+\pl_\mu\vf.
\end{equation}
Тогда лагранжевы уравнения движения (\ref{qtilag}), (\ref{qsplae}) эквивалентны
следующей системе уравнений:
\begin{align}                                                     \label{qfizec}
  -\triangle A_0+\triangle\dot\vf+m^2A_0&=0,
\\                                                                \label{qsezec}
  \pl_\mu\ddot\vf-\pl_\mu\dot A_0+m^2\pl_\mu\vf&=0,
\\                                                                \label{qtheze}
  \square A^\St_\mu+m^2A^\St_\mu&=0,
\end{align}
где мы расщепили уравнение (\ref{qsplae}) на продольную и поперечную части, а
также ввели обозначения для производной по времени и операторов Лапласа и
Даламбера:
\begin{equation*}
  \dot\vf:=\pl_0\vf,\qquad \triangle:=-\pl^\mu\pl_\mu,\qquad
  \square:=\pl^\al\pl_\al=\pl_0^2-\triangle.
\end{equation*}
Уравнения для поперечных компонент (\ref{qtheze}) -- это уравнения
Клейна--Гордона--Фока, и они полностью отщепились. Рассмотрим уравнение
(\ref{qfizec}) как уравнение на временн\'ую компоненту:
\begin{equation}                                                  \label{qhelti}
  (\triangle-m^2)A_0=\triangle\dot\vf.
\end{equation}
Это -- уравнение Гельмгольца, решение которого существует и единственно для
достаточно широкого класса граничных условий, в частности, для полей, убывающих
на бесконечности \cite{TihSam77R}. Запишем его решение в виде
\begin{equation}                                                  \label{qsoaze}
  A_0=\frac1{\triangle-m^2}\triangle\dot\vf.
\end{equation}
Для полей, убывающих на бесконечности, уравнение (\ref{qsezec}) можно один раз
проинтегрировать:
\begin{equation*}
  \ddot\vf-\dot A_0+m^2\vf=0.
\end{equation*}
Подставив сюда решение для $A_0$ (\ref{qsoaze}) и умножив на оператор
$\triangle-m^2$, получим уравнение Клейна--Гордона--Фока на функцию $\vf$,
\begin{equation}                                                  \label{qklter}
  \square\vf+m^2\vf=0.
\end{equation}
Поскольку все проделанные операции обратимы, то мы доказали
\begin{prop}
Для полей, убывающих на бесконечности, лагранжевы уравнения движения
(\ref{qeqpro}) для поля Прок\'а эквивалентны $n-1$ уравнениям
Клейна--Гордона--Фока (\ref{qtheze}), (\ref{qklter}), дополненных уравнениями
для продольной составляющей (\ref{epueff}) и выражением для временн\'ой
компоненты (\ref{qsoaze}).
\end{prop}

Для тех, у кого остались сомнения по поводу следствия уравнений движения
(\ref{qaddlo}), скажем, что подстановка решения (\ref{qsoaze}) в это условие,
что нетрудно проверить, также приводит к уравнению Клейна--Гордона--Фока
(\ref{qklter}).

Таким образом, чтобы поставить задачу Коши для поля Прок\'а, необходимо и
достаточно поставить задачу Коши для $n-2$ поперечных компонент $A^\St_\mu$ и
поля $\vf$, определяющего продольную составляющую. Поэтому поле Прок\'а
действительно описывает $n-1$ физическую (распространяющуюся) степень свободы.
\subsection{Гамильтонова формулировка}
Векторному полю $A_\al$ соответствуют  сопряженные импульсы
\begin{equation}                                                  \label{qmocoa}
  P^\al:=\frac{\pl L}{\pl(\pl_0A_\al)}=F^{\al0}.
\end{equation}
Скобки Пуассона между координатами и импульсами имеют тот же вид, что и для
электромагнитного поля (\ref{epoibe}). Как и для электромагнитного поля, из
выражения для импульсов (\ref{qmocoa}) следует первичная связь
\begin{equation}                                                  \label{qfipro}
  G_1=P^0\approx0.
\end{equation}
Других первичных связей рассматриваемая модель не содержит. Несложные вычисления
приводят к следующему выражению для гамильтониана
\begin{equation}                                                  \label{qhapoy}
  \CH=\int d\Bx\left(-\frac12P^\mu P_\mu+\frac14F^{\mu\nu}F_{\mu\nu}
  -A_0\pl_\mu P^\mu-\frac12m^2A^\al A_\al+\lm P^0\right),
\end{equation}
где $\lm$ -- множитель Лагранжа.

Гамильтоновым уравнениям движения имеют следующий вид:
\begin{align}                                                     \label{qzeaeq}
  \dot A_0&=\lm,
\\                                                                \label{qspaco}
  \dot A_\mu&=-P_\mu+\pl_\mu A_0,
\\                                                                \label{qzemon}
  \dot P^0&=\pl_\mu P^\mu+m^2A_0,
\\                                                                \label{qspmon}
  \dot P^\mu&=\pl_\nu F^{\nu\mu}+m^2A^\mu.
\end{align}
Кроме этого, вариация действия по множителю Лагранжа приводит к связи
(\ref{qfipro}). Таким образом, полная система гамильтоновых уравнений движения
состоит из уравнений (\ref{qzeaeq})--(\ref{qspmon}) и связи (\ref{qfipro}). На
первый взгляд решения гамильтоновых уравнений содержат функциональный произвол,
т.к.\ множитель Лагранжа $\lm$ произволен. Однако этот вывод не верен.
Действительно, поскольку связь $P^0=0$ для любого решения гамильтоновых
уравнений должна выполняться в любой момент времени, то производная по времени
от уравнения (\ref{qzemon}) должна быть равна нулю. Это приводит к уравнению
\begin{equation}                                                  \label{qsecoe}
  m^2\dot A_0+\pl_\mu\dot P^\mu=m^2(\dot A_0+\pl_\mu A^\mu)=0,
\end{equation}
где мы воспользовались уравнением (\ref{qspmon}). Сравнивая полученное равенство
с уравнением (\ref{qzeaeq}), заключаем, что для любого решения гамильтоновых
уравнений множитель Лагранжа имеет вид
\begin{equation}                                                  \label{qlageq}
  \lm=-\pl_\mu A^\mu.
\end{equation}
Теперь можно доказать эквивалентность лагранжевой и гамильтоновой формулировок.
\begin{prop}                                                      \label{peqhal}
Полная система гамильтоновых уравнений (\ref{qzeaeq})--(\ref{qspmon}) и
(\ref{qfipro}) эквивалентна лагранжевым уравнениям движения (\ref{qeqpro}).
\end{prop}
\begin{proof}
Пусть выполнены гамильтоновы уравнения (\ref{qzeaeq})--(\ref{qspmon}) и связь
(\ref{qfipro}). Из уравнения (\ref{qspaco}) находим импульсы
\begin{equation*}
  P_\mu=\pl_\mu A_0-\dot A_\mu=F_{\mu0}.
\end{equation*}
Подстановка этого выражения в формулу (\ref{qspmon}) приводит к уравнению
(\ref{qsplae}). Уравнение (\ref{qzemon}) с учетом связи (\ref{qfipro}) дает
уравнение (\ref{qtilag}).

Обратно. Пусть выполнены лагранжевы уравнения движения (\ref{qtilag}),
(\ref{qsplae}). Введем обозначение $P^\al:=F^{\al0}$. Отсюда немедленно следует
связь (\ref{qfipro}). А для пространственных значений индекса определение
$P^\mu=F^{\mu0}$ эквивалентно уравнению (\ref{qspaco}). Используя новую
переменную $P^\mu$, лагранжевы уравнения (\ref{qsplae}) переписываются в виде
гамильтонова уравнения (\ref{qspmon}). Соотношение (\ref{qaddlo}) является
следствием лагранжевых уравнений движения, и оно дает уравнение (\ref{qzeaeq})
при выполнении условия (\ref{qlageq}). Наконец, лагранжево уравнение
(\ref{qtilag}) сводится к уравнению (\ref{qzemon}) при выполнении связи
(\ref{qfipro}).
\end{proof}
\begin{com}
Отметим, что для эквивалентности лагранжевой и гамильтоновой формулировок к
гамильтониану необходимо добавить первичную связь (\ref{qfipro}). В противном
случае никакой эквивалентности нет. Действительно, равенство $\lm=0$ явно
противоречит условию (\ref{qlageq}).
\qed\end{com}

Продолжим построение гамильтонова формализма, согласно общей схеме,
рассмотренной в разделе \ref{scohal}.

Производная по времени от первичной связи (\ref{qfipro}) равна
\begin{equation*}
  \dot G_1=[G_1,\CH]=\pl_\mu P^\mu+m^2A^0.
\end{equation*}
Поскольку первичная связь должна сохраняться во времени, то возникает вторичная
связь
\begin{equation}                                                  \label{qsepoc}
  G_2=\pl_\mu P^\mu+m^2A^0\approx0.
\end{equation}
Нетрудно вычислить производную по времени от вторичной связи:
\begin{equation*}
  \dot G_2=[G_2,\CH]=m^2(\pl_\mu A^\mu+\lm).
\end{equation*}
Сохранение вторичной связи во времени определяет множитель Лагранжа:
\begin{equation}                                                  \label{qfilag}
  \lm=-\pl_\mu A^\mu.
\end{equation}
Поэтому в теории не возникает других связей.

Равенство (\ref{qfilag}) уже было получено ранее (\ref{qlageq}) при анализе
гамильтоновых уравнений движения.

Скобка связей между собой имеет вид
\begin{equation*}
  [G_1,G'_2]=-m^2\dl(\Bx-\Bx').
\end{equation*}
Введем обозначение для полного набора связей:
\begin{equation*}
  \lbrace\Phi^\Sm\rbrace=\lbrace G_1,G_2\rbrace,\qquad \Sm=1,2.
\end{equation*}
Тогда матрица скобок Пуассона для связей примет вид
\begin{equation*}
  [\Phi^\Sm,\Phi^{\prime\Sn}]=\begin{pmatrix} 0 & -m^2 \\ m^2 & 0 \end{pmatrix}
  \dl(\Bx-\Bx').
\end{equation*}
Она, очевидно, невырождена, и, следовательно, связи $G_1$ и $G_2$ являются
связями второго рода.

Полный гамильтониан получается после добавления всех связей:
\begin{equation}                                                  \label{qtopty}
  \CH_\St=\int d\Bx\left(-\frac12P^\mu P_\mu+\frac14F^{\mu\nu}F_{\mu\nu}
  -A_0\pl_\mu P^\mu-\frac12m^2A^\al A_\al+\lm P^0
  +\mu(\pl_\mu P^\mu+m^2A^0)\right).
\end{equation}
Поскольку поле Прок\'а содержит только связи второго рода, то полный
гамильтониан совпадает с расширенным, $\CH_\St=\CH_\Se$.

Один из множителей Лагранжа уже найден (\ref{qfilag}). Второй множитель Лагранжа
определяется из условия сохранения первичной связи во времени,
\begin{equation*}
  \dot G_1=[G_1,\CH_\St]=\pl_\mu P^\mu+m^2A^0-m^2\mu=0.
\end{equation*}
Отсюда вытекает равенство
\begin{equation}                                                  \label{qselam}
  \mu=A^0+\frac1{m^2}\pl_\mu P^\mu.
\end{equation}
\begin{prop}
Гамильтоновы уравнения движения для полного гамильтониана (\ref{qtopty})
эквивалентны лагранжевым уравнениям (\ref{qeqpro}).
\end{prop}
\begin{proof}
Добавление к исходному гамильтониану (\ref{qhapoy}) вторичной связи приводит к
возникновению самой связи из вариационного принципа. Кроме того меняются
два гамильтоновых уравнения:
\begin{align*}
  \dot A_\mu&=-P_\mu+\pl_\mu A_0-\pl_\mu\mu,
\\
  \dot P^0&=\pl_\mu P^\mu+m^2A_0-\mu m^2.
\end{align*}
Если выполнена вторичная связь, то из соотношения (\ref{qselam}) следует
равенство $\mu=0$. Поэтому гамильтоновы уравнения для полного гамильтониана
эквивалентны исходным гамильтоновым уравнениям и, в силу предложения
\ref{peqhal}, лагранжевым уравнениям (\ref{qeqpro}).
\end{proof}

Согласно общей схеме, описанной в разделе \ref{seccog}, решение связей второго
рода можно подставлять в действие. В рассматриваемом случае связи легко
решаются:
\begin{equation*}
  A_0=-\frac1{m^2}\pl_\mu P^\mu,\qquad P_0=0.
\end{equation*}
После подстановки этого решения в формулу (\ref{qtopty}) получим гамильтониан
для физических степеней свободы
\begin{equation}                                                  \label{qhysha}
  \CH_\text{ph}=\int d\Bx\left(-\frac12P^\mu P_\mu+\frac1{2m^2}(\pl_\mu P^\mu)^2
  +\frac14F^{\mu\nu}F_{\mu\nu}-\frac12m^2A^\mu A_\mu\right).
\end{equation}
Таким образом, временн\'ая компонента поля Прок\'а является нефизической и
может быть полностью исключена из теории путем решения пары связей второго рода.
Пространственные компоненты $A_\mu$ и сопряженные импульсы $P^\mu$ являются
физическими (т.е.\ распространяющимися). Их динамика полностью определяется
гамильтонианом для физических степеней свободы (\ref{qhysha}). Мы видим, что
поле Прок\'а описывает $n-1$ степень свободы, что на единицу превышает
количество степеней свободы электромагнитного поля.

Гамильтонова плотность для физических степеней свободы поля Прок\'а
(\ref{qhysha}), очевидно, положительно определена. В то же время гамильтоновы
плотности исходного (\ref{qhapoy}) и полного (\ref{qtopty}) гамильтонианов не
являются положительно определенными, т.к.\ содержат множители Лагранжа.

Гамильтоновы уравнения движения для физических степеней свободы, следующие из
гамильтониана (\ref{qhysha}) имеют вид
\begin{align}                                                     \label{qhapse}
  \dot A_\mu&=-P_\mu-\frac1{m^2}\pl_\mu\pl_\nu P^\nu,
\\                                                                \label{qhapht}
  \dot P^\mu&=\pl_\nu F^{\nu\mu}+m^2A^\mu.
\end{align}
Здесь проявляется важная особенность теории поля: в отличие от гамильтонового
описания динамики точечных частиц, уравнение (\ref{qhapse}) для импульсов
является не алгебраическим, а дифференциальным. Поэтому выразить импульсы через
скорости, чтобы вернуться к лагранжеву описанию, не удается.

Чтобы сравнить гамильтонову и лагранжеву динамику физических степеней свободы,
разложим координаты и импульсы на поперечную и продольную составляющие.
Разложение пространственных компонент векторного поля уже было введено
формулой (\ref{qdevef}). Для импульсов мы пишем
\begin{equation*}
  P_\mu=P^\St_\mu+\pl_\mu\chi.
\end{equation*}
Тогда гамильтоновы уравнения (\ref{qhapse}), (\ref{qhapht}) также раскладываются
на поперечные и продольные части:
\begin{align}                                                     \label{qhades}
  \dot A_\mu^\St&=-P^\St_\mu,
\\                                                                \label{qhaloj}
  \pl_\mu\dot\vf&=-\pl_\mu\chi+\frac\triangle{m^2}\pl_\mu\chi,
\\                                                                \label{qhaimf}
  \dot P^\St_\mu&=-\triangle A^\St_\mu+m^2A^\St_\mu,
\\                                                                \label{qthfer}
  \pl_\mu\dot\chi&=m^2\pl_\mu\vf.
\end{align}
Гамильтоновы уравнения для поперечных компонент (\ref{qhades}), (\ref{qhaimf})
отщепились и, как легко видеть, эквивалентны уравнениям Клейна--Гордона--Фока
(\ref{qtheze}).

Для полей, убывающих на бесконечности, уравнения (\ref{qhaloj}) и (\ref{qthfer})
можно один раз проинтегрировать:
\begin{align*}
  \dot\vf&=-\chi+\frac\triangle{m^2}\chi,
\\
  \dot\chi&=m^2\vf.
\end{align*}
Первое уравнение можно решить относительно $\chi$,
\begin{equation*}
  \chi=\frac{m^2}{\triangle-m^2}\dot\vf,
\end{equation*}
и подставить данное решение во второе. В результате получим уравнение
Клейна--Гордона--Фока (\ref{qklter}). Поскольку все проделанные операции были
обратимы, то мы доказали
\begin{prop}
Для полей, убывающих на бесконечности, гамильтоновы уравнения движения для
физических степеней свободы (\ref{qhapse}), (\ref{qhapht}) эквивалентны
лагранжевым уравнениям для физических степеней свободы (\ref{qtheze}),
(\ref{qklter}).
\end{prop}

Сейчас проявилось очень важное преимущество гамильтонова подхода, как более
гибкого. А именно, выделение физических степеней свободы на гамильтоновом языке
было проведено явно, и получено локальное выражение для соответствующего
гамильтониана. В то же время, выделение физических степеней свободы в
лагранжевом формализме нелокально, т.к.\ для нахождения поля $\vf$ необходимо
решить дифференциальное уравнение (\ref{epueff}).
\section{Поля Янга--Миллса}
{\em Поля Янга--Миллса (калибровочные поля)} играют важную роль в математической
физике, потому что лежат в основе современных моделей, объединяющих
электромагнитные, слабые и сильные взаимодействия. Экспериментальное их
обнаружение пока остается под вопросом.
\index{Поле Янга--Миллса (Yang--Mills field)}%
\index{Янга--Миллса поле (Yang--Mills field)}%
\index{Поле калибровочное (gauge field)}%
\index{Калибровочное поле (gauge field)}%

С геометрической точки зрения у нас есть следующая конструкция. Во-первых, мы
предполагаем, что пространство-время представляет собой многообразие $\MM$.
Для общности будем рассматривать пространство-время произвольной размерности
$n$. Затем строится главное расслоение $\MP(\MM,\pi,\MG)$ с некоторой
структурной группой Ли $\MG$ размерности $\dim\MG=\Sn$, о которой мы
поговорим несколько позже, и проекцией $\pi$. Мы
предполагаем, что на главном расслоении задана связность, т.е.\ инвариантное
распределение горизонтальных подпространств (см.\ раздел \ref{scofib}). Каждая
связность взаимно однозначно определяется формой связности, заданной на
пространстве главного расслоения $\MP$. Если задано координатное покрытие базы
$\MM=\cup_i\MU_i$, и семейство локальных сечений $\s_i:~\MU_i\rightarrow\MP$, то
форма связности взаимно однозначно определяет семейство локальных форм
связности, заданных на каждой координатной окрестности $\MU_i$, которые
определяются формой связности с помощью возвратов отображений $\s_i$.
Локальная форма связности -- это 1-форма на координатной окрестности $\MU_i$ со
значениями в алгебре Ли $\Gg$ группы ли $\MG$:
\begin{equation}                                                  \label{qloccj}
  \om=dx^\al A_\al{}^\Sa L_\Sa,
\end{equation}
где $L_\Sa$, $\Sa=1,\dotsc,\Sn$, -- базис алгебры Ли (см.\ раздел \ref{slolir}).
Компоненты локальной формы связности $A_\al{}^\Sa(x)$ это и есть поля
Янга--Миллса (после добавления соответствующих уравнений движения, что будет
сделано позже). Базис алгебры Ли удовлетворяет некоторым коммутационным
соотношениям
\begin{equation*}
  [L_\Sa,L_\Sb]=f_{\Sa\Sb}{}^\Sc L_\Sc,
\end{equation*}
где $f_{\Sa\Sb}{}^\Sc$ -- структурные константы группы Ли $\MG$.

При изменении локального сечения $\s\mapsto\s'$ (или при вертикальном
автоморфизме) компоненты локальной формы связности меняются
$A_\al{}^\Sa\mapsto A'_\al{}^\Sa$. Правило преобразования компонент связности
известно (\ref{elotrs}), однако оно явно содержит функцию композиции для группы
Ли $\MG$ и поэтому неудобно. Инфинитезимальные калибровочные преобразования с
параметрами $\e^\Sa(x)$ имеют вид
\begin{equation}                                                  \label{qinfga}
  A'_\al{}^\Sa=A_{\al}{}^\Sa+\nb_\al\e^\Sa,
\end{equation}
где
\begin{equation*}
  \nb_\al\e^\Sa:=\pl_\al\e^\Sa-A_\al{}^\Sb\e^\Sc f_{\Sb\Sc}{}^\Sa.
\end{equation*}

Чтобы записать конечные калибровочные преобразования в удобном виде, рассмотрим
произвольное точное представление группы Ли $\MG$. Для определенности рассмотрим
присоединенное представление. В этом представлении каждому элементу базиса
алгебры Ли $\Gg$ соответствует матрица: $L_\Sa\mapsto f_{\Sa\Sb}{}^\Sc$, где
первый индекс $\Sa$ нумерует базисные векторы, а второй и третий
рассматриваются, как матричные. Тогда каждой локальной форме связности можно
поставить в соответствие матричнозначную 1-форму с компонентами
\begin{equation}                                                  \label{qmayam}
  A_{\al\Sb}{}^\Sc:=-A_\al{}^\Sa f_{\Sa\Sb}{}^\Sc.
\end{equation}
(Знак минус не имеет принципиального значения.) Тогда поле Янга--Миллса при
изменении локального сечения преобразуется по-правилу (\ref{egatrd})
\begin{equation}                                                  \label{qymtrf}
  A_\al\mapsto A'_\al=SA_\al S^{-1}+\pl_\al S S^{-1},
\end{equation}
где $S(x)=\lbrace S_\Sa{}^\Sb(x)\rbrace$ -- матрица присоединенного
представления, и мы, для краткости, опустили матричные индексы. Это -- хорошо
известное калибровочное преобразование полей Янга--Миллса.

Каждой локальной форме связности соответствует локальная форма кривизны (см.\
раздел \ref{scurds}). Она представляет собой 2-форму на координатной окрестности
$\MU_i$ со значениями в алгебре Ли:
\begin{equation}                                                  \label{qcurfo}
  R=dx^\al\wedge dx^\bt F_{\al\bt}{}^\Sa L_\Sa,
\end{equation}
где компоненты имеют вид (\ref{eymcur})
\begin{equation}                                                  \label{qcohyf}
  F_{\al\bt}{}^\Sa=\pl_\al A_\bt{}^\Sa-\pl_\bt A_\al{}^\Sa
  -A_\al{}^\Sb A_\bt{}^\Sc f_{\Sb\Sc}{}^\Sa.
\end{equation}
Это есть напряженность поля Янга--Миллса. Ранее было показано, что при
калибровочном преобразовании (\ref{qymtrf}) компоненты напряженности
преобразуются ковариантным образом (\ref{qcohtd})
\begin{equation*}
  F_{\al\bt}{}^\Sa\mapsto F'_{\al\bt}{}^\Sa
  =F_{\al\bt}{}^\Sb S^{-1}_{\quad \Sb}{}^\Sa.
\end{equation*}
Если ввести матричнозначную форму кривизны
\begin{equation}
  F_{\al\bt\Sb}{}^\Sc:=-F_{\al\bt}{}^\Sa f_{\Sa\Sb}{}^\Sc,
\end{equation}
то ее закон преобразования будет выглядеть немного иначе:
\begin{equation}                                                  \label{qtrcyt}
  F_{\al\bt}\mapsto F'_{\al\bt}=SF_{\al\bt}S^{-1},
\end{equation}
где мы, для краткости, опустили матричные индексы.

Матричнозначные компоненты напряженности можно выразить через матричнозначные
поля Янга--Миллса:
\begin{equation}                                                  \label{qmayan}
  F_{\al\bt}=\pl_\al A_\bt-\pl_\bt A_\al-[A_\al,A_\bt],
\end{equation}
где $[A_\al,A_\bt]$ -- коммутатор матриц, и мы опустили матричные индексы. Чтобы
получить данное выражение для компонент напряженности из формулы (\ref{qcohyf}),
использованы тождества Якоби для структурных констант (\ref{ejatoj}).

Согласно теореме \ref{tbiang} каждая форма кривизны удовлетворяет тождествам
Бианки. Тождества Бианки для локальной формы кривизны были получены ранее
(\ref{qbiabh}):
\begin{equation}                                                  \label{qgyutr}
  \nb_\al F_{\bt\g}{}^\Sa+\nb_\bt F_{\g\al}{}^\Sa+\nb_\g F_{\al\bt}{}^\Sa=0,
\end{equation}
где
\begin{equation}                                                  \label{econgr}
  \nb_\al F_{\bt\g}{}^\Sa=\pl_\al F_{\bt\g}{}^\Sa
  -A_\al{}^\Sb F_{\bt\g}{}^\Sc f_{\Sb\Sc}{}^\Sa
\end{equation}
-- ковариантная производная (по отношению к калибровочным преобразованиям) от
компонент напряженности поля Янга--Миллса.

\begin{prop}
Для существования такого калибровочного преобразования, чтобы преобразованное
поле Янга--Миллса обратилось бы в нуль в некоторой окрестности $\MU\subset\MM$
необходимо и достаточно, чтобы локальная форма кривизны связности была равна
нулю в $\MU$.
\end{prop}
\begin{proof}
Если локальная форма связности равна нулю, то локальная форма кривизны
(\ref{qcohyf}) также обращается в нуль.

Обратно. Рассмотрим правило преобразования полей Янга--Миллса (\ref{qymtrf}) как
дифференциальное уравнение на матрицу $S$. Если преобразованное поле
Янга--Миллса равно нулю, $A'_\al=0$, то матрица преобразования должна
удовлетворять системе дифференциальных уравнений:
\begin{equation}                                                  \label{qedrfg}
  \pl_\al S=-SA_\al.
\end{equation}
Эта система локально разрешима тогда и только тогда, когда выполнены условия
разрешимости:
\begin{equation*}
  \pl_{[\bt}\pl_{\al]}S=0,
\end{equation*}
где квадратные скобки обозначают антисимметризацию по индексам. Вторая частная
производная от матрицы преобразования имеет вид
\begin{equation*}
  \pl_\bt\pl_\al S=-\pl_\bt S A_\al-S\pl_\bt A_\al=S(A_\bt A_\al-\pl_\bt A_\al),
\end{equation*}
где в первом слагаемом мы воспользовались исходным уравнением (\ref{qedrfg}).
Антисимметризация полученного выражения по индексам $\al$ и $\bt$ приводит к
равенству
\begin{equation*}
  F_{\al\bt}=0.
\end{equation*}
Таким образом, равенство нулю локальной формы кривизны является необходимым и
достаточным условием разрешимости системы уравнений (\ref{qedrfg}) в некоторой
окрестности $\MU\subset\MM$.
\end{proof}
\begin{defn}
Из доказанного утверждения следует, что если локальная форма кривизны равна
нулю, то в некоторой окрестности $\MU\subset\MM$ компоненты соответствующей
локальной формы связности имеют вид
\begin{equation}                                                  \label{qputyh}
  A_\al=-S^{-1}\pl_\al S=\pl_\al S^{-1}S.
\end{equation}
Такое поле Янга--Миллса называется {\em чистой калибровкой}.
\qed\end{defn}
\index{Чистая калибровка (pure gauge}%
\index{Калибровка чистая (pure gauge}%
Легко проверить, что напряженность поля Янга--Миллса для поля (\ref{qputyh})
действительно обращается в нуль.

Описанная выше геометрическая конструкция не зависит от того задана ли на базе
$\MM$ какая либо геометрия, т.е.\ метрика и аффинная связность, или нет. Главное
расслоение, связность и кривизна определяются в дифференциальной геометрии
самостоятельно. Закон преобразования компонент локальных форм связности и
кривизны при изменении сечения следует из определения. Для этого не надо вводить
какие либо дополнительные поля и ковариантные производные. Не смотря на
отсутствие аффинной связности в частных производных при определении локальной
формы кривизны (\ref{qcohyf}) и в тождествах Бианки (\ref{econgr}), все
выписанные соотношения ковариантны относительно общих преобразований координат.

Следующий шаг построения моделей состоит в построении действия. Мы считаем, что
действие должно быть инвариантно относительно общих преобразований координат и
калибровочных преобразований (\ref{qymtrf}). Чтобы построить соответствующий
инвариант нам понадобятся две метрики: метрика $g_{\al\bt}$ на
пространственно-временн\'ом многообразии и двусторонне инвариантная метрика
$\eta_{\Sa\Sb}$ на групповом многообразии. Соответствующее действие имеет вид
\begin{equation}                                                  \label{qyanmi}
  S_{\Sy\Sm}=\int_\MM\!\!\!dx\,L_{\Sy\Sm},\qquad L_{\Sy\Sm}:=-\frac14\vol
  g^{\al\g}g^{\bt\dl}F_{\al\bt}{}^\Sa F_{\g\dl}{}^\Sb\eta_{\Sa\Sb}.
\end{equation}

В квантовой теории поля в качестве пространства-времени выбирается пространство
Минковского $\MR^{1,n-1}$, группа общих преобразования координат сужается до
группы Пуанкаре $\MI\MO(1,n-1)$, и в качестве пространственно-временн\'ой
метрики выбирается метрика Лоренца, $g_{\al\bt}\mapsto\eta_{\al\bt}$.

Если калибровочная группа $\MG$ является абелевой, как, например, в
электродинамике, то в качестве двусторонне инвариантной метрики подойдет любая
постоянная матрица. Действительно, при калибровочных преобразованиях
(\ref{qymtrf}) компоненты напряженности поля Янга--Миллса для абелевой группы
вообще не преобразуются.

Для неабелевых групп в качестве двустронне инвариантной метрики выберем форму
Киллинга--Картана (см.\ раздел \ref{ekilcf}). Согласно теореме \ref{tcarki}
форма Киллинга--Картана невырождена тогда и только тогда, когда группа Ли $\MG$
полупроста. Поэтому мы предполагаем, что структурная группа Ли $\MG$ полупроста.

Согласно теореме \ref{tsemsi} любая полупростая группа Ли единственным образом
представляется в виде прямого произведения простых групп. В этом случае для
каждого сомножителя можно написать отдельный инвариант вида (\ref{qyanmi}) со
своей константой связи. Поэтому в дальнейшем, для простоты, мы ограничимся
рассмотрением простых групп Ли. Если мы научимся работать с простыми группами
Ли, то построить модель, соответствующую их прямому произведению, не составит
труда.

В дальнейшем мы покажем, что канонический гамильтониан для физических степеней
свободы, которые описываются действием (\ref{qyanmi}), положительно определен
тогда и только тогда, когда метрика $\eta_{\Sa\Sb}$ положительно определена.
Согласно теореме \ref{tcomnh} связная полупростая группа Ли компактна тогда и
только тогда, когда ее форма Киллинга--Картана
\begin{equation}                                                  \label{qkilca}
  \eta_{\Sa\Sb}:=-f_{\Sa\Sc}{}^\Sd f_{\Sb\Sd}{}^\Sc
\end{equation}
положительно определена. Поэтому мы предполагаем, что структурная группа Ли
$\MG$ связна, проста и компактна.

Базис алгебры Ли $L_\Sa$ определен с точностью до невырожденных линейных
преобразований
\begin{equation*}
  L_\Sa\mapsto S_\Sa{}^\Sb L_\Sb,\qquad S\in\MG\ML(\Sn,\MR).
\end{equation*}
При этом форма Киллинга--Картана преобразуется по тензорному закону
\begin{equation*}
  \eta_{\Sa\Sb}\mapsto S_\Sa{}^\Sc S_\Sb{}^\Sd\eta_{\Sc\Sd}.
\end{equation*}
Согласно теореме \ref{tdiama} произвольную симметричную матрицу можно привести
к диагональному виду с помощью ортогональной матрицы. Дальнейшей растяжкой
координат можно добиться того, что на диагонали формы Киллинга--Картана будут
стоять единицы. Таким образом, не теряя общности, можно считать, что для связной
компактной простой группы Ли форма Киллинга--Картана совпадает с единичной
матрицей
\begin{equation}                                                  \label{qkicae}
  \eta_{\Sa\Sb}=\dl_{\Sa\Sb}.
\end{equation}
\begin{exa}
В единых моделях теории поля используются, как правило, унитарные $\MS\MU(\Sn)$
и ортогональные $\MS\MO(\Sn)$ группы. Эти группы связны, компактны и просты.
\qed\end{exa}

Если вместо полей Янга--Миллса $A_\al{}^\Sa$ рассматривать их точное матричное
представление, то лагранжиан (\ref{qyanmi}) можно записать, используя след
матриц,
\begin{equation*}
  L_{\Sy\Sm}:=\frac14\vol g^{\al\g}g^{\bt\dl}
  \tr\big(F_{\al\bt}F_{\g\dl}\big).
\end{equation*}
Обратим внимание на изменение общего знака, что связано с определением формы
Киллинга--Картана (\ref{qkilca}).

Для построения физически содержательных моделей математической физики,
необходимо введение дополнительных полей, например, скаляров и спиноров. Мы
предполагаем, что эти поля являются локальными сечениями векторного
ассоциированного расслоения $\ME(\MM,\pi_\ME,\MV,\MG,\MP)$, типичным слоем
которого является векторное пространство $\MV$, в котором задано представление
структурной группы $\MG$. Отображение $\pi_\ME:~\ME\rightarrow\MM$ -- это
проекция. Каждое сечение ассоциированного расслоения задается набором компонент
$\vf=\lbrace\vf^i\rbrace$, $i=1,\dotsc,\dim\MV$. При калибровочных
преобразованиях поля преобразуются по некоторому представлению структурной
группы:
\begin{equation}                                                  \label{qaumji}
  \vf\mapsto\vf'=\vf S^{-1}=\lbrace\vf^j S^{-1}_{\quad j}{}^i\rbrace,
\end{equation}
где $S_j{}^i$ -- матрица представления. Ковариантная производная имеет вид
(\ref{ecovym})
\begin{equation}                                                  \label{qcovfr}
  \nb_\al\vf^i=\pl_\al\vf^i+\vf^j A_{\al j}{}^i,
\end{equation}
в котором
\begin{equation*}
  A_{\al j}{}^i:=-A_\al{}^\Sa L_{\Sa j}{}^i,
\end{equation*}
где $L_{\Sa j}{}^i$ -- представление базиса алгебры Ли $L_\Sa$ в векторном
пространстве $\MV$. Нетрудно убедиться в том, что при калибровочных
преобразованиях (\ref{qymtrf}), (\ref{qaumji}) ковариантная производная ведет
себя ковариантно:
\begin{equation*}
  \nb_\al\vf^i\mapsto\nb'_\al\vf'^i=\nb_\al\vf^j S^{-1}_{\quad j}{}^i,
\end{equation*}
где
\begin{equation*}
  \nb'_\al\vf'^i=\pl_\al\vf'^i+\vf'^j A'_{\al j}{}^i,
\end{equation*}
\begin{defn}
В физике векторное пространство $\MV$ называется {\em изотопическим}, а
калибровочные преобразования (\ref{qaumji}) -- {\em изотопическими вращениями}.
\qed\end{defn}
\index{Изотопическое пространство (isotopic space)}%
\index{Пространство изотопическое (isotopic space)}%
\index{Изотопическое вращение (isotopic rotation)}%
\index{Вращение изотопическое (isotopic rotation)}%

При построении моделей математической физики мы предполагаем, что уравнения
движения для полей $\vf$, не взаимодействующих с полями Янга--Миллса, следуют
из некоторого действия
\begin{equation*}
  S=\int_\MM\!\!\!dx\,L(\vf,\pl\vf),
\end{equation*}
где лагранжиан $L$ зависит от полей $\vf=\lbrace\vf^i\rbrace$ и их частных
производных $\pl\vf=\lbrace\pl_\al\vf^i\rbrace$. Как и для электромагнитного
поля, взаимодействие с полями Янга--Миллса вводится путем {\em минимальной
подстановки}, которая заключается в замене всех частных производных на
ковариантные:
\begin{equation*}
  \pl_\al\vf^i\mapsto\nb_\al\vf^i:=\pl_\al\vf^i+\vf^j A_{\al j}{}^i.
\end{equation*}
\index{Минимальная подстановка (minimal substitution)}%
\index{Подстановка минимальная (minimal substitution)}%
Кроме этого, к лагранжиану добавляется лагранжиан поля Янга--Миллса
\begin{equation*}
  L(\vf,\pl\vf)\mapsto-\frac1{4e^2}F^{\al\bt\Sa}F_{\al\bt\Sa}+L(\vf,\nb\vf),
\end{equation*}
который приводит к динамическим уравнениям движения для самого поля
Янга--Миллса. Выше $e$ -- константа связи и
\begin{equation*}
  F^{\al\bt\Sa}:=g^{\al\g}g^{\bt\dl}\eta^{\Sa\Sb}F_{\g\dl\Sb}.
\end{equation*}

Для построения теории возмущений удобно перейти к новым переменным
$A_\al{}^\Sa\mapsto eA_\al{}^\Sa$. Тогда лагранжиан примет вид
\begin{equation*}
  -\frac14F^{\al\bt\Sa}F_{\al\bt\Sa}+L(\vf,\nb\vf),
\end{equation*}
где
\begin{align*}
  F_{\al\bt}{}^\Sa&=\pl_\al A_\bt{}^\Sa-\pl_\bt A_\al{}^\Sa
  -eA_\al{}^\Sb A_{\bt}{}^\Sc f_{\Sb\Sc}{}^\Sa,
\\
  \nb_\al\vf^i&=\pl_\al\vf^i+e\vf^j A_{\al j}{}^i.
\end{align*}
В таком виде удобно строить теорию возмущений при малой константе связи $e$. В
дальнейшем, для упрощения обозначений, мы положим $e=1$.
\subsection{Лагранжева формулировка}
В квантовой теории поля мы предполагаем, что пространством-временем является
пространство Минковского, $\MM\approx\MR^{1,n-1}$, которое топологически
тривиально, т.е.\ диффеоморфно $\MR^n$. В этом случае  согласно теореме
\ref{trigla} расслоение тривиально
\begin{equation*}
  \MP(\MR^{1,n-1},\pi,\MG)\approx\MR^{1,n-1}\times\MG.
\end{equation*}
Если выбрать систему координат, например, декартову, покрывающую все
пространство-время, то связность на главном расслоении будет находиться во
взаимно однозначном соответствии с локальной формой связности. По этой причине
в физике часто говорят, что поля Янга--Миллса это и есть связность.

Действие для полей Янга--Миллса в пространстве Минковского $\MR^{1,n-1}$ имеет
вид
\begin{equation}                                                  \label{qlagym}
  S_{\Sy\Sm}=\int dx L_{\Sy\Sm},\qquad
  L_{\Sy\Sm}:=-\frac14F^{\al\bt\Sa}F_{\al\bt\Sa}.
\end{equation}
где подъем и опускание греческих индексов производится с помощью метрики
Минковского $\eta_{\al\bt}$, а латинских -- с помощью формы Киллинга--Картана
$\eta_{\Sa\Sb}$. Из этого действия следуют нелинейные уравнения движения
\begin{equation}                                                  \label{qeamij}
  \nb_\bt F^{\bt\al\Sa}:=\pl_\bt F^{\bt\al\Sa}
  -A_\bt{}^\Sb F^{\bt\al\Sc}f_{\Sb\Sc}{}^\Sa=0.
\end{equation}

Добавление к лагранжиану Янга--Миллса массового члена
$\frac12m^2 A^{\al\Sa}A_{\al\Sa}$, как легко видеть, нарушает калибровочную
инвариантность.

В отличие от электродинамики в пространстве Минковского для неабелевых
калибровочных групп $\MG$ уравнения (\ref{qeamij}) нелинейны. Поэтому говорить о
``свободных'' полях Янга--Миллса не имеет смысла. Взаимодействие полей
Янга--Миллса есть всегда. В линейном приближении уравнения (\ref{qeamij}) имеют
тот же вид, что и набор из $\Sn$ электромагнитных полей. При этом
инфинитезимальные калибровочные преобразования (\ref{qymtrf}) в линейном
приближении по полям принимают вид $\MU(1)$ преобразований:
\begin{equation*}
  A'_\al{}^\Sa=A_{\al}{}^\Sa+\pl_\al\e^\Sa+\osmall_\al{}^\Sa(\e,A),
\end{equation*}
где $\e^\Sa(x)$, $\Sa=1,\dotsc,\Sn$, -- параметры преобразований. Таким
образом, в линейном приближении свободное поле Янга--Миллса описывает $\Sn$
безмассовых векторных полей.

По построению, действие калибровочно инвариантно. Согласно второй теореме Нетер
в таком случае между уравнениями движения существует $\Sn$ линейных
зависимостей по числу параметров преобразования. Для нахождения этих
зависимостей достаточно знать вид инфинитезимальных калибровочных преобразований
(\ref{qinfga}). Из определения инвариантности следует равенство
\begin{equation*}
  \dl S_{\Sy\Sm}=\int \!dx\frac{\dl S_{\Sy\Sm}}{\dl A_\al{}^\Sa}
  \nb_\al\e^\Sa=0.
\end{equation*}
После интегрирования по частям, получим зависимость уравнений движения:
\begin{equation}                                                  \label{qlider}
  \nb_\al\nb_\bt F^{\bt\al}{}_\Sa=0,\qquad \Sa=1,\dotsc,\Sn.
\end{equation}

В справедливости полученного равенства можно убедиться с помощью прямых
вычислений. Действительно, в силу антисимметрии напряженности,
$F^{\al\bt}{}_\Sa=-F^{\bt\al}{}_\Sa$, равенство (\ref{qlider}) можно переписать
в виде
\begin{equation*}
  \frac12(\nb_\al\nb_\bt-\nb_\bt\nb_\al)F^{\bt\al}{}_\Sa
  =\frac12 F_{\al\bt}{}^\Sc F^{\bt\al\Sb} f_{\Sc\Sa\Sb}=0,
\end{equation*}
где мы воспользовались свойством коммутатора ковариантных производных
(\ref{ecohov}), симметрией тензора $F_{\al\bt}{}^\Sc F^{\bt\al\Sb}$ по
индексам $\Sc$ и $\Sb$ и антисимметрией структурных констант (предложение
\ref{panstr}).

Если в правой части уравнений движения (\ref{qeamij}) стоит ток,
\begin{equation}                                                  \label{qyemcu}
  \nb_\bt F^{\bt\al}{}_\Sa=J^\al{}_\Sa,
\end{equation}
То из тождества (\ref{qlider}) следует закон ковариантного сохранения тока:
\begin{equation}                                                  \label{qcucon}
  \nb_\al J^\al{}_\Sa=0,
\end{equation}
выполнение которого необходимо для самосогласованности уравнений движения. В
свою очередь, если ток получен в результате варьирования некоторого калибровочно
инвариантного действия для других полей, то равенство (\ref{qcucon}) будет
автоматически выполнено на уравнениях движения, опять же, в силу второй теоремы
Нетер. Если же ток $J^\al{}_\Sa$ вводится в уравнения движения ``руками'', то на
него необходимо наложить условие сохранения (\ref{qcucon}).

Эффективным инструментом исследования свободных уравнений движения в
электродинамике явилось разложение пространственных компонент электромагнитного
потенциала на поперечную и продольную составляющие. Для каждого значения индекса
$\Sa$ поле Янга--Миллса также можно разложить на поперечную и продольные части.
Однако это не будет столь же эффективно, т.к.\ уравнения ``свободного'' поля
Янга--Миллса (\ref{qeamij}) нелинейны. Такое разложение эффективно в квантовой
теории поля при построении теории возмущений.

Теперь рассмотрим законы сохранения, вытекающие из первой теоремы Нетер.
По-построению, действие Янга--Миллса инвариантно относительно глобального
действия группы Пуанкаре $\MI\MO(1,n-1)$. Из инвариантности относительно
трансляций, согласно первой теореме Нетер, следует закон сохранения
энергии-импульса (см.\ раздел \ref{senmot}),
\begin{equation*}
  \pl_\bt \widehat T_\al{}^\bt=0,
\end{equation*}
где
\begin{equation*}
  \widehat T_{\al\bt}=-\pl_\al A_{\g\Sa}F_\bt{}^{\g\Sa}+\frac14\eta_{\al\bt}F^2
\end{equation*}
-- канонический тензор энергии-импульса поля Янга--Миллса. Это выражение
страдает двумя недостатками. Во-первых, оно не инвариантно относительно
калибровочных преобразований. Во-вторых, отсутствует симметрия относительно
перестановки индексов. Для их устранения используем произвол в выборе
канонического тензора энергии-импульса. Заметим, что выполнено равенство
\begin{equation*}
  \pl_\g(A_{\al\Sa}F_\bt{}^{\g\Sa})=\pl_\g A_{\al\Sa}F_\bt{}^{\g\Sa}
  +A_{\al\Sa}\nb_\g F_\bt{}^{\g\Sa}
  +A_{\al\Sa}A_\g{}^\Sc f_{\Sc\Sb}{}^\Sa F_\bt{}^{\g\Sb}.
\end{equation*}
Если выполнены уравнения движения, что предполагается в первой теореме Нетер, то
второе слагаемое обращается в нуль, и можно переопределить тензор
энергии-импульса
\begin{equation}                                                  \label{qsyenm}
  T^\Ss_{\al\bt}:=\widehat T_{\al\bt}+\pl_\g(A_{\al\Sa}F_\bt{}^{\g\Sa})
  =-F_{\al\g\Sa}F_\bt{}^{\g\Sa}+\frac14\eta_{\al\bt}F^2.
\end{equation}
Этот тензор энергии-импульса, очевидно, калибровочно инвариантен и симметричен.
На уравнениях движения он сохраняется:
\begin{equation}                                                  \label{qsohtr}
  \pl_\bt T^\Ss_{~\al}{}^\bt=0.
\end{equation}

При лоренцевых вращениях поле Янга--Миллса преобразуется, как и электромагнитное
поле (\ref{emflro}). Поэтому для спинового момента получаем следующее выражение
\begin{equation}                                                  \label{qspmem}
  S_{\al\bt}{}^\g=A_{\al\Sa} F_\bt{}^{\g\Sa}-A_{\bt\Sa} F_\al{}^{\g\Sa}.
\end{equation}

Полный тензор момента количества движения (\ref{eanmot}) состоит из орбитального
и спинового моментов:
\begin{equation}                                                  \label{qtoanm}
  \widehat J_{\al\bt}{}^\g
  =x_\bt\widehat T_\al{}^\g-x_\al\widehat T_\bt{}^\g
  +S_{\al\bt}{}^\g.
\end{equation}
Используя выражения для симметричного тензора энергии-импульса (\ref{qsyenm}) и
спинового момента (\ref{qspmem}) его можно переписать в виде
\begin{equation}                                                  \label{qstoms}
  \widehat J_{\al\bt}{}^\g=M_{\al\bt}{}^\g
  +\pl_\dl(x_\al A_{\bt\Sa} F^{\g\dl\Sa}-x_\bt A_{\al\Sa} F^{\g\dl\Sa}),
\end{equation}
где
\begin{equation*}
  M_{\al\bt}{}^\g:=x_\bt T^\Ss_{~\al}{}^\g-x_\al T^\Ss_{~\bt}{}^\g
\end{equation*}
-- орбитальный момент, построенный по симметричному тензору энергии-импульса
(\ref{qsyenm}). Поскольку второе слагаемое в выражении (\ref{qstoms}) имеет вид
(\ref{earbcu}), то тензор момента количества движения можно переопределить,
положив
\begin{equation}                                                  \label{qanmne}
  J_{\al\bt}{}^\g=M_{\al\bt}{}^\g
  =x_\bt T^\Ss_{~\al}{}^\g-x_\al T^\Ss_{~\bt}{}^\g,
\end{equation}
который также сохраняется
\begin{equation*}
  \pl_\g J_{\al\bt}{}^\g=0.
\end{equation*}
Отметим полное отсутствие спинового момента в этом выражении.

Поскольку действие для поля Янга--Миллса инвариантно относительно калибровочных
преобразований, то оно также инвариантно относительно глобальных преобразований
из калибровочной группы. Поэтому можно воспользоваться первой теоремой Нетер.
С калибровочной инвариантностью действия для поля Янга--Миллса связан закон
сохранения заряда
\begin{equation}                                                  \label{qsohzg}
  \pl_\al J^\al{}_\Sa=0,
\end{equation}
где
\begin{equation}                                                  \label{qcuell}
  J^\al{}_\Sa=\nb_\bt F^{\bt\al}{}_\Sa
\end{equation}
-- ток, соответствующий первой теореме Нетер, то есть, когда параметр
калибровочных преобразований не зависит от точки пространства-времени. При
отсутствии полей материи этот закон тривиален, т.к.\ ток равен нулю в силу
уравнений движения (\ref{qeamij}).
\subsection{Гамильтонова формулировка}
Рассмотрим гамильтонову формулировку поля Янга--Миллса в пространстве
Минковского $\MR^{1,n-1}$ с действием (\ref{qlagym}). По-определению, импульсы,
сопряженные компонентам $A_\al{}^\Sa$, имеют вид
\begin{equation}                                                  \label{qmonau}
  P^\al{}_\Sa:=\frac{\pl L_{\Sy\Sm}}{\pl\dot A_\al{}^\Sa}=F^{0\al}{}_\Sa,
\end{equation}
где $\dot A_\al{}^\Sa:=\pl_0 A_\al{}^\Sa$. Выпишем отличные от нуля
одновременн\'ые скобки Пуассона:
\begin{equation*}
  [A_\al{}^\Sa,P'^\bt{}_\Sb]:=[A_\al{}^\Sa(x^0,\Bx),P^\bt{}_\Sb(x^0,\Bx')]
  =\dl^\bt_\al\dl^\Sa_\Sb\dl(\Bx-\Bx').
\end{equation*}

Как и в случае электромагнитного поля в теории возникают первичные связи. Из
определения импульсов (\ref{qlagym}) следует $\Sn$ первичных связей:
\begin{equation}                                                  \label{qprico}
  G_{1\Sa}=P^0{}_\Sa=0,\qquad \Sa=1,\dotsc,\Sn.
\end{equation}
Других первичных связей в модели нет. Скобки Пуассона первичных связей между
собой, очевидно, равны нулю:
\begin{equation*}
  [G_{1\Sa},G'_{1\Sb}]=0.
\end{equation*}

Несложные вычисления приводят к гамильтониану
\begin{equation}                                                  \label{qhayub}
  \CH=\int d\Bx\left(-\frac12P^{\mu\Sa}P_{\mu\Sa}
  +\frac14F^{\mu\nu\Sa}F_{\mu\nu\Sa}-A_0{}^\Sa\nb_\mu P^\mu{}_\Sa
  +\lm^\Sa P^0{}_\Sa\right),
\end{equation}
где мы добавили первичные связи с множителями Лагранжа $\lm^\Sa$ и ковариантная
производная от импульсов равна
\begin{equation*}
  \nb_\mu P^\nu{}_\Sa:=\pl_\mu P^\nu{}_\Sa-A_{\mu\Sa}{}^\Sb P^\nu{}_\Sb.
\end{equation*}

Гамильтоновы уравнения движения для построенного гамильтониана имеют вид
\begin{align}                                                     \label{qhaaze}
  \dot A_0{}^\Sa&=\lm^\Sa,
\\                                                                \label{qhaasp}
  \dot A_\mu{}^\Sa&=-P_\mu{}^\Sa+\nb_\mu A_0{}^\Sa,
\\                                                                \label{qhamoz}
  \dot P^0{}_\Sa&=\nb_\mu P^\mu{}_\Sa,
\\                                                                \label{qhamol}
  \dot P^\mu{}_\Sa&=\nb_\nu F^{\nu\mu}{}_\Sa+A_{0\Sa}{}^\Sb P^\mu{}_\Sb,
\end{align}
где
\begin{align*}
  \nb_\mu A_0{}^\Sa&:=\pl_\mu A_0{}^\Sa+A_{\mu\Sb}{}^\Sa A_0{}^\Sb,
\\
  \nb_\nu F^{\rho\mu}{}_\Sa&:=\pl_\nu F^{\rho\mu}{}_\Sa
  -A_{\nu\Sa}{}^\Sb F^{\rho\mu}{}_\Sb.
\end{align*}
Конечно, эти уравнения необходимо дополнить уравнениями первичных связей
(\ref{qprico}), которые возникают при варьировании соответствующего действия по
множителям Лагранжа.

Гамильтоновы уравнения движения (\ref{qhaaze})--(\ref{qhamol}), содержат
множители Лагранжа $\lm^\Sa$, которые в исходном действии рассматриваются в
качестве произвольных функций. Однако, уравнения движения, в принципе, могут
наложить на них некоторые ограничения. Чтобы проверить совместность
гамильтоновых уравнений, рассмотрим эволюцию первичных связей во времени.
Поскольку $P^0{}_\Sa=0$, то из уравнения (\ref{qhamoz}) следуют равенства
\begin{equation}                                                  \label{qserfd}
  \nb_\mu P^\mu{}_\Sa=0,
\end{equation}
которые должны быть выполнены для любого решения гамильтоновых уравнений.
С учетом уравнения (\ref{qhamol}) производная по времени от полученных равенств
имеет вид
\begin{equation}                                                  \label{qsevtu}
  \pl_0(\nb_\mu P^\mu{}_\Sa)=\nb_\mu\nb_\nu F^{\nu\mu}{}_\Sa
  +\nb_\mu A_{0\Sa}{}^\Sb P^\mu{}_\Sb+A_{0\Sa}{}^\Sb\nb_\mu P^\mu{}_\Sb
  -\dot A_{\mu\Sa}{}^\Sb P^\mu{}_\Sb.
\end{equation}
Первое слагаемое обращается в нуль:
\begin{equation*}
  \frac12(\nb_\mu\nb_\nu-\nb_\nu\nb_\mu)F^{\nu\mu}{}_\Sa
  =\frac12 F_{\mu\nu}{}^\Sc F^{\nu\mu\Sb} f_{\Sc\Sa\Sb}=0,
\end{equation*}
где мы воспользовались свойством коммутатора ковариантных производных
(\ref{ecohov}), симметрией тензора $F_{\mu\nu}{}^\Sc F^{\nu\mu\Sb}$ по
индексам $\Sc$ и $\Sb$ и антисимметрией структурных констант (предложение
\ref{panstr}). Второе и четвертое слагаемые сокращаются с учетом уравнения
(\ref{qhaasp}). В результате возникает равенство
\begin{equation}                                                  \label{qdosed}
  \pl_0(\nb_\mu P^\mu{}_\Sa)=A_{0\Sa}{}^\Sb\nb_\mu P^\mu{}_\Sb\approx0,
\end{equation}
т.к.\ правая часть равна нулю на уравнениях движения (\ref{qserfd}). На этом
анализ заканчивается, и мы видим, что никаких ограничений на множители Лагранжа
гамильтоновы уравнения движения не накладывают.

Продолжим общий гамильтонов анализ. Производная по времени от первичных связей
равна
\begin{equation*}
  \dot G_1=[G_1,\CH]=\nb_\mu P^\mu{}_\Sa.
\end{equation*}
Это приводит к вторичным связям:
\begin{equation}                                                  \label{qsedco}
  G_{2\Sa}=\nb_\mu P^\mu{}_\Sa\approx0.
\end{equation}
Скобка Пуассона первичных и вторичных связей между собой равна нулю,
\begin{equation*}
  [G_{1\Sa},G'_{2\Sb}]=0.
\end{equation*}
Несложно вычислить скобки Пуассона вторичных связей между собой:
\begin{equation}                                                  \label{qansec}
  [G_{2\Sa},G'_{2\Sb}]=-f_{\Sa\Sb}{}^\Sc G_{2\Sc}.
\end{equation}
Правая часть, очевидно, обращается в нуль на поверхности связей.

Выше мы уже показали, что вторичные связи сохраняются во времени (\ref{qdosed}).

Таким образом, полная система связей
\begin{equation}                                                  \label{qtocas}
  \lbrace G_a\rbrace:=\lbrace G_{1\Sa},G_{2\Sa}\rbrace,\qquad a=1,\dotsc,2\Sn,
\end{equation}
находится в инволюции, и модель содержит $2\Sn$ связей первого рода.

Полный гамильтониан получается после добавления всех связей:
\begin{equation*}
  \CH_\St=\int d\Bx\left(-\frac12P^{\mu\Sa}P_{\mu\Sa}
  +\frac14F^{\mu\nu\Sa}F_{\mu\nu\Sa}
  +\lm^\Sa P^0{}_\Sa+\mu^\Sa\nb_\mu P^\mu{}_\Sa\right),
\end{equation*}
где слагаемые $A_0{}^\Sa\nb_\mu P^\mu{}_\Sa$ мы абсорбировали, переопределив
множители Лагранжа $\mu^\Sa$. Ему соответствует полное действие
\begin{equation}                                                  \label{qtoasd}
  S_\St=\int dx\left(P^\al{}_\Sa\dot A_\al{}^\Sa+\frac12P^{\mu\Sa}P_{\mu\Sa}
  -\frac14F^{\mu\nu\Sa}F_{\mu\nu\Sa}
  -\lm^\Sa P^0{}_\Sa-\mu^\Sa\nb_\mu P^\mu{}_\Sa\right).
\end{equation}
Это действие является более общим, чем исходное (\ref{qlagym}), т.к.\ совпадает
с ним только при $\lm^\Sa=0$ и $\mu^\Sa=-A_0{}^\Sa$.

Гамильтоновы уравнения движения для действия (\ref{qtoasd}) имеют вид:
\begin{align}                                                     \label{qhatoe}
  \frac{\dl S_\St}{\dl P^0{}_\Sa}&=\dot A_0{}^\Sa-\lm^\Sa=0,
\\                                                                \label{qhasza}
  \frac{\dl S_\St}{\dl P^\mu{}_\Sa}&=\dot A_\mu{}^\Sa+P_\mu{}^\Sa
  +\nb_\mu\mu^\Sa=0,
\\                                                                \label{qhamoi}
  \frac{\dl S_\St}{\dl A_0{}^\Sa}&=-\dot P^0{}_\Sa=0,
\\                                                                \label{qhamos}
  \frac{\dl S_\St}{\dl A_\mu{}^\Sa}&=-\dot P^\mu{}_\Sa+\nb_\nu F^{\nu\mu}{}_\Sa
  -\mu^\Sb f_{\Sa\Sb}{}^\Sc P^\mu{}_\Sc=0,
\\                                                                \label{qhalmt}
  \frac{\dl S_\St}{\dl\lm^\Sa}&=-P^0{}_\Sa=0,
\\                                                                \label{qhamuj}
  \frac{\dl S_\St}{\dl\mu^\Sa}&=-\nb_\mu P^\mu{}_\Sa=0.
\end{align}

Согласно теореме \ref{tlochs} каждой связи первого рода соответствует
калибровочное преобразование, относительно которого полное действие инвариантно.
Первичным связям (\ref{qprico}) соответствует генератор калибровочных
преобразований
$$
  T_1:=\int\!d\Bx\,\e_1^\Sa P^0{}_\Sa
$$
с некоторыми малыми параметрами $\e_1^\Sa(x)$. При этом преобразуются только
временн\'ые компоненты поля Янга--Миллса $A_0{}^\Sa$ и множители Лагранжа
$\lm^\Sa$:
\begin{equation}                                                  \label{emgtro}
  \dl_1A_0{}^\Sa=[ A_0{}^\Sa,T_1]=\e_1^\Sa,\qquad \dl_1\lm^\Sa=\dot\e_1^\Sa.
\end{equation}
Все остальные поля остаются без изменения. Согласно второй теореме Нетер из
инвариантности действия относительно калибровочного преобразования
(\ref{emgtro}) следует зависимость уравнений движения:
$$
  \frac{\dl S_\St}{\dl A_0{}^\Sa}-\pl_t\frac{\dl S_\St}{\dl\lm^\Sa}=0.
$$
В справедливости этого тождества нетрудно убедиться прямой проверкой.

Вторичным связям первого рода (\ref{qsedco}) также соответствует генератор
преобразований
$$
  T_2:=\int\!d\Bx\,\e_2^\Sa\nb_\mu P^\mu{}_\Sa,
$$
где $\e_2^\Sa(x)$ -- малые параметры второго калибровочного преобразования. В
этом случае преобразуются только пространственные компоненты поля Янга--Миллса
$A_\mu{}^\Sa$, сопряженные импульсы $P^\mu{}_\Sa$ и множители Лагранжа
$\mu^\Sa$:
\begin{align*}
  \dl_2 A_\mu{}^\Sa&=[A_\mu{}^\Sa,T_2]=-\nb_\mu\e_2^\Sa,
\\
  \dl_2 P^\mu{}_\Sa&=[P^\mu{}_\Sa,T_2]=-\e_2^\Sb f_{\Sa\Sb}{}^\Sc P^\mu{}_\Sc,
\\
  \dl_2\mu^\Sa&=\dot\e_2^\Sa-\e_2^\Sb\mu^\Sc f_{\Sb\Sc}{}^\Sa.
\end{align*}
Согласно второй теореме Нетер, из локальной инвариантности действия следует
зависимость уравнений Эйлера--Лагранжа:
$$
  \nb_\mu\frac{\dl S_\St}{\dl A_\mu{}^\Sa}
  -\frac{\dl S_\St}{\dl P^\mu{}_\Sb}f_{\Sa\Sb}{}^\Sc P^\mu{}_\Sc
  -\pl_t\frac{\dl S_\St}{\dl\mu^\Sa}
  -\mu^\Sb f_{\Sa\Sb}{}^\Sc\frac{\dl S_\St}{\dl\mu^\Sc}=0,
$$
что также легко проверить. Таким образом, как и в электродинамике, число
параметров калибровочных преобразований полного действия $S_\St$ в гамильтоновом
формализме по сравнению с лоренцевой формулировкой удвоилось.

Генератором калибровочных преобразований в лагранжевом формализме (\ref{qymtrf})
является линейная комбинация связей первого рода:
$$
  T=\int\!dx\,\left[(\dot\e^\Sa-A_0{}^\Sb\e^\Sc f_{\Sb\Sc}{}^\Sa)P^0{}_\Sa
  -\e^\Sa\nb_\mu P^\mu{}_\Sa)\right].
$$
Действительно,
\begin{align*}
  \dl A_0{}^\Sa&=[A_0^\Sa,T]=\dot\e^\Sa-A_0{}^\Sb\e^\Sc f_{\Sb\Sc}{}^\Sa
  =\nb_0\e^\Sa,
\\
  \dl A_\mu{}^{\Sa}&=[A_\mu{}^\Sa,T]=\nb_\mu\e^\Sa,
\\
  \dl P^0{}_\Sa&=\e^\Sb f_{\Sa\Sb}{}^\Sc P^0{}_\Sc,
\\
  \dl P^\mu{}_\Sa&=\e^\Sb f_{\Sa\Sb}{}^\Sc P^\mu{}_\Sc.
\end{align*}
По сравнению с электродинамикой возникли нетривиальные преобразования импульсов.
Это связано с тем, что в электродинамике компоненты напряженности инвариантны, а
для полей Янга--Миллса -- ковариантны.

Решения гамильтоновых уравнений движения содержат произвол, соответствующий
множителям Лагранжа $\lm^\Sa$ и $\mu^\Sa$. Для того, чтобы устранить этот
произвол необходимо наложить $2\Sn$ калибровочных условий. В квантовой теории
поля обычно используют такие же калибровочные условия (\ref{qdifga}) как и в
электродинамике для каждого значения индекса $\Sa$. Если зафиксировать,
например, кулоновскую калибровку $\pl_\mu A^\mu=0$, то это даст $\Sn$
калибровочных условий. Недостающие $\Sn$ калибровочных условий в гамильтоновом
формализме можно получить, рассматривая эволюцию этих калибровочных условий во
времени. Из уравнения (\ref{qhaasp}) следует
\begin{equation*}
  \pl_0(\pl_\mu A^\mu{}_\Sa)=-\pl_\mu P^\mu{}_\Sa+\pl^\mu\nb_\mu A_{0\Sa}.
\end{equation*}
Тогда кулоновскую калибровку для потенциала можно дополнить $\Sn$ калибровочными
условиями
\begin{equation*}
  -\pl_\mu P^\mu{}_\Sa+\pl^\mu\nb_\mu A_{0\Sa}=0.
\end{equation*}
В этом случае калибровочные условия будут согласованы с гамильтоновыми
уравнениями для исходного гамильтониана (\ref{qhayub}). Вместо этих
калибровочных можно выбрать другие, например, $A_{0\Sa}=0$.

Гамильтонова формулировка поля Янга--Миллса на произвольном многообразии $\MM$
с метрикой $g$ проводится, по существу, так же как и в пространстве Минковского.
Необходимо произвести замену $\eta_{\al\bt}\mapsto g_{\al\bt}$ и рассматривать
компоненты $g_{\al\bt}$ как внешнее поле. Почти все формулы останутся при этом
без изменения, только в ковариантных производных появятся символы Кристоффеля,
возникающие при интегрировании по частям.
\subsection{Поля Янга--Миллса в аффинной геометрии}
Как уже было отмечено, для построения моделей математической физики на
многообразии $\MM$ (пространстве-времени) кроме поля Янга--Миллса желательно
иметь метрику и аффинную связность, т.е.\ аффинную геометрию. Поскольку поле
Янга--Миллса -- это компоненты локальной формы связности, то действие для
``свободного'' поля Янга--Миллса (\ref{qyanmi}) зависит только от метрики, но не
от связности.

Уравнения движения и тензор энергии-импульса поля Янга--Миллса получаются
простым варьированием действия (\ref{qyanmi}):
\begin{align}                                                     \label{qfryre}
  \vol S_{\Sy\Sm},^\al{}_\Sa:=\frac{\dl S_{\Sy\Sm}}{\dl A_\al{}^\Sa}
  &=\pl_\bt(\vol F^{\bt\al}{}_\Sa)
  =\vol\widetilde\nb_\bt F^{\bt\al}{}_\Sa=0,
\\                                                                \label{qemtym}
  \vol S_{\Sy\Sm},_{\al\bt}:=\frac{\dl S_{\Sy\Sm}}{\dl g^{\al\bt}}
  &=\frac12\vol T_{\Sy\Sm\al\bt}=-\frac12\vol F_{\al\g\Sa}F_\bt{}^{\g\Sa}
  +\frac18\vol g_{\al\bt}F^2,
\end{align}
где
\begin{equation*}
  \widetilde\nb_\bt F^{\bt\al}{}_\Sa:=\pl_\bt F^{\bt\al}{}_\Sa+
  F^{\g\al}{}_\Sa\widetilde\Gamma_{\bt\g}{}^\bt-A_{\bt\Sa}{}^\Sb F^{\bt\al}{}_\Sb.
\end{equation*}
Слагаемое с символами Кристоффеля $\widetilde\Gamma$ возникло при интегрировании
по частям (см.\ раздел \ref{susefg}). Отсюда вытекает выражение для тензора
энергии-импульса поля Янга--Миллса
\begin{equation}                                                  \label{qemygt}
  T_{\Sy\Sm\al\bt}=-F_{\al\g\Sa}F_\bt{}^{\g\Sa}+\frac14g_{\al\bt}F^2.
\end{equation}
Полученное выражение является ковариантным обобщением симметричного
канонического тензора энергии-импульса (\ref{qsyenm}), полученного в
пространстве Минковского из первой теоремы Нетер.

Если поле Янга--Миллса взаимодействует с другими полями, то в правой части
уравнений движения (\ref{qfryre}) возникают токи:
\begin{equation}                                                  \label{qymtou}
  \widetilde\nb_\bt F^{\bt\al}{}_\Sa=J^\al{}_\Sa.
\end{equation}

Как и в случае электродинамики, для поля Янга--Миллса можно воспользоваться
формализмом первого порядка.
\begin{prop}
Уравнения движения для действия (\ref{qyanmi}) эквивалентны уравнениям
Эйлера--Лагранжа для действия
\begin{equation}                                                  \label{qemfko}
  S=-\int_\MM\!\!\!dx\vol\left[\frac12F^{\al\bt\Sa}(\pl_\al A_{\bt\Sa}
  -\pl_\bt A_\al)-\frac14F^{\al\bt\Sa}F_{\al\bt\Sa}\right],
\end{equation}
в котором компоненты $F^{\al\bt\Sa}$ и $A_{\al\Sa}$ рассматриваются в качестве
независимых переменных.
\end{prop}
\begin{proof}
Повторяет доказательство предложения \ref{pfidre} для электродинамики.
\end{proof}

Действие для поля Янга--Миллса (\ref{qyanmi}) инвариантно относительно общих
преобразований координат. Соответствующие вариации полей имеют вид (см.\ раздел
\ref{sinfct})
\begin{align*}
  \dl A_\al{}^\Sa&=-\pl_\al\e^\bt A_\bt{}^\Sa-\e^\bt\pl_\bt A_\al{}^\Sa,
\\
  \dl g^{\al\bt}&=g^{\al\g}\pl_\g\e^\bt+g^{\bt\g}\pl_\g\e^\al
  -\e^\g\pl_\g g^{\al\bt}.
\end{align*}
Отсюда, согласно второй теореме Нетер (\ref{edepel}), следует зависимость
уравнений движения
\begin{equation}                                                  \label{qmdegc}
  \pl_\bt S,{}^{\bt\Sa} A_{\al\Sa}+S,{}^{\bt\Sa} F_{\bt\al\Sa}
  -\pl_\bt S,{}^\bt{}_\al
  -\pl_\bt S,{}_\al{}^\bt-S,{}_{\bt\g}\pl_\al g^{\bt\g}=0.
\end{equation}
Или, в ковариантном виде,
\begin{equation}                                                  \label{qmcold}
  \widetilde\nb_\bt S,{}^{\bt\Sa} A_{\al\Sa}+S,{}^{\bt\Sa} F_{\bt\al\Sa}
  -2\widetilde\nb^\bt S,{}_{\bt\al}=0.
\end{equation}

Если выполнены уравнения движения для поля Янга--Миллса (\ref{qfryre}), то
с учетом свернутых тождеств Бианки соотношение (\ref{qmcold}) принимает вид
\begin{equation}                                                  \label{qcocem}
  \widetilde\nb_\bt T_{\Sy\Sm\al}{}^\bt=0.
\end{equation}
Полученное равенство, выполненное для всех решений уравнений Эйлера--Лагранжа,
можно интерпретировать, как ковариантное обобщение закона сохранения тензора
энергии-импульса (\ref{qsohtr}).

Действие для поля Янга--Миллса инвариантно также относительно калибровочных
преобразований (\ref{qinfga}). Этой инвариантности, согласно второй теореме
Нетер, соответствует зависимость уравнений движения:
\begin{equation}                                                  \label{qcogai}
  \widetilde\nb_\al S,{}^\al{}_\Sa=0.
\end{equation}
С учетом уравнения (\ref{qymtou}) это равенство приводит к закону сохранения
тока: $\widetilde\nb_\al J^\al{}_\Sa=0$. Если ток возникает при варьировании
некоторого калибровочно инвариантного действия для полей материи, то сохранение
тока будет выполняться автоматически в силу второй теоремы Нетер. Если же ток
вводится в уравнения для поля Янга--Миллса ``руками'', то условие
$\widetilde\nb_\al J^\al{}_\Sa=0$ необходимо для самосогласованности уравнений.
\subsection{Поле Янга--Миллса в общей теории относительности}
Если ограничиться рассмотрением четырехмерного пространства-времени, то в
теории поля Янга--Миллса возникают определенные специфические свойства. Как и в
электродинамике поле Янга--Миллса определяет локальные формы связности
\begin{equation*}
  A^\Sa:=dx^\al A_\al{}^\Sa,\qquad \Sa=1,\dotsc,\Sn.
\end{equation*}
Теперь локальная форма кривизны (\ref{qcurfo}) не является внешней производной
от локальной формы связности. Однако ее можно записать с помощью внешней
ковариантной производной, введенной в разделе \ref{scurds},
\begin{equation}                                                  \label{qymcuy}
  R^\Sa=DA^\Sa:=dA^\Sa-\frac12A^\Sb\wedge A^\Sc f_{\Sb\Sc}{}^\Sa.
\end{equation}
Тождества Бианки (\ref{qgyutr}) также можно записать в компактном виде с помощью
оператора внешнего ковариантного дифференцирования (\ref{ebiapr}):
\begin{equation*}
  DR^\Sa=0.
\end{equation*}
На этом, собственно, преимущество использования обозначений дифференциальных
форм в теории полей Янга--Миллса заканчивается.

В общей теории относительности для каждого значения изотопического индекса
$\Sa$ можно ввести 2-формы, дуальные к форме локальной кривизны:
\begin{equation}                                                  \label{qducul}
  *F^\Sa:=\frac12dx^\al\wedge dx^\bt(*F^\Sa)_{\al\bt},
\end{equation}
где компоненты дуальной формы кривизны заданы равенством
\begin{equation}                                                  \label{qdyual}
  *F_{\al\bt}{}^\Sa:=\frac12\ve_{\al\bt\g\dl}F^{\g\dl}{}^\Sa.
\end{equation}
В приведенной формуле $\ve_{\al\bt\g\dl}(x)$ -- полностью антисимметричный
тензор четвертого ранга.

Дальнейшая перезапись основных уравнений в точности следует анализу,
проведенному для электромагнитного поля в разделе \ref{semoto}, надо лишь
добавить изотопический индекс $\Sa$ там, где это необходимо. В частности,
{\em уравнение (анти-) самодуальности} имеет вид
\begin{equation}                                                  \label{qsefdl}
  F_{\al\bt}{}^\Sa=\pm*F_{\al\bt}{}^\Sa.
\end{equation}
\index{Уравнение самодуальности (selfdual equation)}%
\index{Уравнение антисамодуальности (antiselfdual equation)}%
Справедливо также
\begin{prop}
Если поле Янга--Миллса удовлетворяет уравнению (анти-) самодуальности, то оно
также удовлетворяет уравнениям Эйлера--Лагранжа (\ref{qfryre}).
\end{prop}
Решения уравнений (анти-) самодуальности важны в теории полей Янга--Миллса. Для
евклидовой сигнатуры метрики соответствующие решения в евклидовом пространстве
$\MR^4$ называются {\em (анти-)инстантонами}.
\index{Инстантон (instanton)}%
\index{Антиинстантон (antiinstanton)}%

Используя дуальную форму кривизны, тензор энергии-импульса поля Янга--Миллса
(\ref{qemygt}) можно переписать в виде
\begin{equation}                                                  \label{qknmoe}
  T_{\Sy\Sm\al\bt}=-\frac12\left[F_{\al\g\Sa}F_\bt{}^{\g\Sa}
  +*F_{\al\g\Sa}*F_\bt{}^{\g\Sa}\right].
\end{equation}
Он, очевидно, инвариантен относительно дуальных вращений (\ref{qdurot}) для
каждого значения индекса $\Sa$. Если базис алгебры Ли выбран таким образом, что
форма Киллинга--Картана совпадает с символом Кронекера, то эту инвариантность
можно расширить до группы $\MS\MO(2\Sn)$.

Так же как и в электродинамике, доказывается
\begin{prop}
Если метрика на многообразии $\MM$ имеет лоренцеву сигнатуру и координата $x^0$
является временем, то временн\'ая компонента тензора энергии-импульса поля
Янга--Миллса
\begin{equation}                                                  \label{qzzcoe}
  T_{\Se\Sm00}=-\frac12\left[F_{0\g\Sa}F_0{}^{\g\Sa}+*F_{0\g\Sa}*F_0{}^{\g\Sa}
  \right]
\end{equation}
положительно определена и, следовательно, удовлетворяет слабому энергетическому
условию. Она также удовлетворяет сильному энергетическому условию.
\end{prop}

Ситуация с полем Янга--Миллса во многих отношениях аналогична ситуации в
электродинамике.
\begin{prop}
Действие для полей Янга--Миллса в аффинной геометрии (\ref{qyanmi}) инвариантно
относительно преобразований Вейля, не затрагивающих само поле Янга--Миллса:
\begin{equation}                                                  \label{qweyal}
  g_{\al\bt}\mapsto\bar g_{\al\bt}=\ex^{2\phi}g_{\al\bt},\qquad
  A_\al{}^\Sa\mapsto \bar A_\al{}^\Sa=A_\al{}^\Sa,
\end{equation}
где $\phi(x)\in\CC^2(\MM)$ -- произвольная вещественнозначная функция.
\end{prop}
\begin{proof}
Утверждение следует из равенства
\begin{equation*}
  \vol g^{\al\bt}g^{\g\dl}=\sqrt{|\bar g|}\bar g^{\al\bt}\bar g^{\g\dl}.
  \tag*{\qed}
\end{equation*}
\renewcommand{\qed}{}\end{proof}

Как следствие общего утверждения \ref{pweyrf}, получаем, что след тензора
энергии-импульса поля Янга--Миллса (\ref{qemygt}) для четырехмерного
пространства-времени равен нулю, $T_{\Sy\Sm\al}{}^\al=0$. Это, конечно, легко
проверить.
\chapter{Геометрия поверхностей                                  \label{sagtwo}}
Геометрия поверхностей является классическим разделом дифференциальной
геометрии. Она относительно проста, и многие вопросы можно решить аналитически,
что позволяет глубже понять проблемы, возникающие в более высоких размерностях.
Кроме этого геометрия поверхностей в последние годы приобрела особую
актуальность в связи с большим числом физических приложений и, в первую очередь,
в связи с интересом к теории струн, мировая поверхность которых представляет
собой двумерное многообразие с метрикой лоренцевой сигнатуры.
\section{Геометрия Римана--Картана на поверхности                \label{sricas}}
Минимальная размерность многообразия, на котором возможно существование
нетривиального кручения и кривизны, равна двум. Настоящая глава посвящена именно
этому случаю.

Пусть на поверхности задана геометрия Римана--Картана, т.е.\ задана метрика и
кручение, а тензор неметричности тождественно равен нулю, $Q_{\al\bt\g}=0$. Мы
предполагаем, что на поверхности задана метрика $g$ либо лоренцевой, либо
евклидовой сигнатуры.

В двумерном случае многие формулы дифференциальной геометрии значительно
упрощаются, благодаря наличию антисимметричного (псевдо)тензора второго ранга
\begin{equation}                                                  \label{eastsr}
  \ve_{ab}=-\ve_{ba},\qquad \ve_{01}=\ve_{12}=1.
\end{equation}
Приставка ``псевдо'' означает, что при пространственном отражении или обращении
времени $\ve_{ab}$ меняет знак. В евклидовом случае, конечно, есть только
пространственное отражение одной из координат. Напомним, что в псевдоевклидовом
и евклидовом случаях латинские индексы принимают значения $\lbrace 0,1\rbrace $
и $\lbrace 1,2\rbrace $, соответственно. Контравариантные компоненты этого
тензора зависят от сигнатуры метрики и отличаются знаком
\begin{equation}                                                  \label{eansec}
  \ve^{01}=-1,\qquad \ve^{12}=1.
\end{equation}
Свойства полностью антисимметричного тензора второго ранга приведены в
приложении \ref{scomat}. Существование антисимметричного тензора второго ранга
означает, что произвольное антисимметричное тензорное поле второго ранга,
$X_{ab}=-X_{ba}$, взаимно однозначно определяется псевдоскалярным полем $X^*$:
\begin{equation}                                                  \label{eseaps}
  X_{ab}=\opm\frac12\ve_{ab}X^*,\qquad \text{где}\quad X^*=\ve^{ab}X_{ab}.
\end{equation}
Здесь и далее верхний знак внутри окружности $\opm$ или $\omp$ относится к
римановой поверхности, а нижний -- к лоренцевой.

Благодаря антисимметрии, тензор кривизны в геометрии Римана--Картана всегда
можно представить в виде
\begin{equation}                                                  \label{ecusct}
  R_{abcd}=\opm\frac12\ve_{ab}\ve_{cd}R
  =\frac12(\et_{ac}\et_{bd}-\et_{ad}\et_{bc})R,
\end{equation}
где $R$ -- скалярная кривизна. Здесь и далее в этом разделе римановы и
лоренцевы поверхности будут рассматриваться параллельно. Поэтому символ
$\et_{ab}$ будет обозначать и символ Кронекера, и метрику Минковского. Из
формулы (\ref{ecusct}) следует, что у тензора кривизны только одна неприводимая
компонента -- скалярная кривизна -- отлична от нуля. Соотношение (\ref{ecusct})
справедливо как в римановой геометрии, так и в геометрии Римана--Картана при
нетривиальном кручении. В частности, тензор Риччи всегда симметричен
\begin{equation}                                                  \label{qeieqt}
  R_{ab}=\frac12\et_{ab}R.
\end{equation}
Отсюда следует, что вакуумные уравнения Эйнштейна (\ref{einequ}) при нулевой
космологической постоянной в двумерном пространстве-времени удовлетворяются
тождественно.

Все квадратичные инварианты тензора кривизны выражаются через квадрат скалярной
кривизны. Например,
\begin{align}                                                     \label{ecurin}
  R_{abcd}R^{abcd}&=R^2,
\\                                                                \label{ericin}
  R_{ab}R^{ab}&=\frac12R^2.
\end{align}

В двумерном случае упрощается также общий вид тензора кручения. Нетрудно
проверить, что он взаимно однозначно представляется либо своим следом:
\begin{equation}                                                  \label{etortr}
    T_{abc}=\et_{ac}T_b-\et_{bc}T_a,\qquad T_b:=T_{ab}{}^a,
\end{equation}
либо псевдоследом:
\begin{equation}                                                  \label{etorpt}
    T_{abc}=\opm\frac12\ve_{ab}T_c^*,\qquad T_c^*:=T_{abc}\ve^{ab}.
\end{equation}
Псевдовектор $T_c^*$ и след $T_a$ тензора кручения взаимно однозначно связаны
между собой простыми соотношениями:
$$
    T_a^*=2\ve_{ab}T^b,\qquad T_a=\frac12\ve_{ab}T^{*b}.
$$
Квадратичный инвариант тензора кручения имеет вид
\begin{equation}                                                  \label{etorin}
  T_{abc}T^{abc}=2T_aT^a=\opm\frac12T_a^*T^{*a}.
\end{equation}

При проведении вычислений часто бывает удобно переходить от ковариантной
производной $\widetilde\nb_\al$, определяемой символами Кристоффеля и
возникающей при интегрировании по частям, к ковариантной производной $\nb_\al$,
определяемой полной метрической связностью с кручением. Эта связь осуществляется
с помощью соотношения
$$
  \Gamma_{\al\bt\g}=\widetilde\Gamma_{\al\bt\g}+g_{\al\g}T_\bt-g_{\al\bt}T_\g,
$$
которое следует из (\ref{emecon}), (\ref{etortr}).

Поскольку тождества Бианки антисимметричны по трем индексам, то они
автоматически удовлетворяются в двух измерениях и не накладывают никаких
ограничений на тензор кривизны и кручения.

Теперь обсудим переменные Картана: репер $e_\al{}^a$ и лоренцеву связность
$\om_\al{}^{ab}$. Наличие антисимметричного (псевдо)тензора второго ранга
позволяет воспользоваться следующей параметризацией лоренцевой связности в
геометрии Римана--Картана:
\begin{equation}                                                  \label{elcopa}
    \om_\al{}^{ab}=\om_\al\ve^{ab}.
\end{equation}
$1$-форму $\om_\al$ также будем называть лоренцевой связностью. В евклидовом
случае вместо термина ``лоренцева связность'' следует говорить
``$\MS\MO(2)$-связность''. Для определенности, мы будем говорить лоренцева
связность, имея в виду оба случая. При пространственных отражениях или обращении
времени эта форма, также как и антисимметричный тензор $\e^{ab}$, меняет знак.

Из определения (\ref{elcopa}) следует, что
\begin{equation}                                                  \label{elocsu}
  \om_a=\opm\frac12\om_{abc}\ve^{bc}=\omp\ve_a{}^b(c_b+T_b),
\end{equation}
где $c_b$ -- след коэффициентов неголономности (\ref{etranc}).

Закон преобразования связности $\om_\al$ при локальных лоренцевых вращениях
неоднороден. Прямые вычисления показывают, что при
вращении на угол $\om(x)$, лоренцева связность преобразуется так же, как и
электромагнитное поле
\begin{equation}                                                  \label{qlotrf}
  \om_\al\mapsto\om'_\al=\om_\al+\pl_\al\om.
\end{equation}

Тензор кривизны поверхности в переменных Картана представляется в виде
\begin{equation}                                                  \label{ecurtw}
    R_{\al\bt}{}^{ab}=F_{\al\bt}\ve^{ab},\qquad
    F_{\al\bt}:=\pl_\al\om_\bt-\pl_\bt\om_\al.
\end{equation}
Отсутствие квадратичных слагаемых в тензоре кривизны связано с тем, что в
двумерном случае группы Лоренца $\MS\MO_\uparrow(1,1)$ и вращений $\MS\MO(2)$
являются абелевыми. При этом скалярная кривизна имеет вид
\begin{equation}                                                  \label{esctdi}
  R=2\ve^{\al\bt}\pl_\al\om_\bt=\ve^{\al\bt}F_{\al\bt},\qquad
  F_{\al\bt}=\opm\frac12\ve_{\al\bt}R.
\end{equation}
где введен антисимметричный тензор
\begin{equation}                                                  \label{eatwoi}
  \ve^{\al\bt}=e^\al{}_ae^\bt{}_b\ve^{ab},\qquad
  \nb_\al\ve^{\bt\g}=0.
\end{equation}
Выше введен репер $e_\al{}^a$, который в двумерном случае называется
{\em диадой}, и определяется соотношением
\begin{equation*}
  g_{\al\bt}=e_\al{}^a e_\bt{}^b\eta_{ab}
\end{equation*}
с точностью до локальных лоренцевых вращений.
\index{Диада (diad)}%

Приведем также явное выражение для псевдоследа тензора кручения в картановых
переменных, которое часто используется в приложениях,
\begin{equation}                                                  \label{eptrto}
  T^{*a}=2\ve^{\al\bt}(\pl_\al e_\bt{}^a+\om_\al e_\bt{}^b\ve_b{}^a).
\end{equation}
В римановом пространстве кручение равно нулю, и из уравнения (\ref{eptrto})
следует выражение лоренцевой связности через производные от диады:
\begin{equation}                                                  \label{etdlct}
  \widetilde\om_\al=-e_\al{}^a\ve^{\bt\g}\pl_\bt e_{\g a}.
\end{equation}

В приложениях часто используется также полностью антисимметричная тензорная
плотность единичного веса
\begin{equation}                                                  \label{eteade}
  \hat\ve_{\al\bt}=\frac1\vol\ve_{\al\bt},\qquad
  \hat\ve^{\al\bt}=\vol\ve^{\al\bt}.
\end{equation}
Ее компоненты во всех системах координат равны по модулю единице, и частная
производная обращается в нуль:
\begin{equation}                                                  \label{etendc}
  \pl_\al\hat\ve_{\bt\g}=\pl_\al\hat\ve^{\bt\g}=0.
\end{equation}
Определитель диады можно записать в различных видах:
\begin{equation}                                                  \label{edetdi}
  \vol=\det e_\al{}^a=e_0{}^ae_1{}^b\ve_{ab}
  =\omp\hat\ve^{\al\bt}e_\al{}^0e_\bt{}^1
  =\omp\frac12\hat\ve^{\al\bt}e_\al{}^ae_\bt{}^b\ve_{ab}.
\end{equation}
Для вычислений, проводимых с мономами третьего и более высокого порядка по
лоренцевым векторам, полезны следующие тождества:
\begin{align}                                                     \label{etoant}
  \ve_{ab}\ve_{cd}+\ve_{bc}\ve_{ad}+\ve_{ca}\ve_{bd}&=0,
\\                                                                \label{efirve}
  \eta_{ab}\ve_{cd}+\eta_{da}\ve_{bc}+\eta_{cd}\ve_{ab}+\eta_{bc}\ve_{da}&=0.
\end{align}
Тождество (\ref{etoant}), в котором слагаемые отличаются циклической
перестановкой первых трех индексов, следует из того, что в двумерном
пространстве не существует тензора, антисимметричного по трем индексам.
Тождество (\ref{efirve}) аналогично тождествам Фирца для $\g$ матриц Дирака и
проверяется прямой подстановкой.

Тождество, связывающее скалярную кривизну в (псевдо-)римановой геометрии со
скалярной кривизной в геометрии Римана--Картана (\ref{eident}), в двумерном
случае существенно упрощается:
\begin{equation}                                                  \label{qffgtd}
  R-\frac1\vol\pl_\al\left(\vol\ve^\al{}_\bt T^{*\bt}\right)=\widetilde R.
\end{equation}
То есть скалярные кривизны отличаются на дивергенцию.
\section{Неголономный базис в двух измерениях                   \label{sanholb}}
В двумерном пространстве-времени Римана--Картана удобно использовать
неголономный базис в касательном пространстве, определяемый диадой:
\begin{equation}                                                  \label{eanhol}
  e_a:=\pl_a:=e^\al{}_a \pl_\al.
\end{equation}
Многие геометрические объекты выражаются через коэффициенты неголономности
$c_{ab}{}^c$, которые определяются коммутатором базисных векторных полей,
\begin{equation}
  \left[e_a,e_b\right]:=c_{ab}{}^c e_c.
\end{equation}
Из выражения (\ref{eanhol}) следует, что коэффициенты неголономности имеют вид
\begin{equation}                                                   \label{eanho}
\begin{aligned}
  c_{ab}{}^c &=\left(e^\al{}_a\pl_\al e^\bt{}_b-e^\al{}_b\pl_\al e^\bt{}_a
  \right) e_\bt{}^c \\
  &=-e^\al{}_a e^\bt{}_b\left(\pl_\al e_\bt{}^c-\pl_\bt e_\al{}^c\right).
\end{aligned}
\end{equation}
В двух измерениях они взаимно однозначно определяются следом
\begin{equation}                                                 \label{eanhotr}
  c_{ab}{}^c=\dl_a^c c_b-\dl_b^c c_a,\qquad c_b:=c_{ab}{}^a.
\end{equation}
Из уравнения (\ref{eanho}) вытекает, что коэффициенты неголономности инвариантны
относительно общих преобразований координат. При локальном лоренцевом повороте
на угол $\om(x)$ они преобразуются, как лоренцева связность:
\begin{equation}
  \dl c_b=-\om \ve_b{}^c c_c-\ve_b{}^c\pl_c\om.
\end{equation}
Приведем два полезных тождества:
\begin{equation}
  \ve^{ab}\pl_a e^\al{}_b=\ve^{\al b}c_b,\quad\quad
  \ve^{\al\bt}\pl_\al e_\bt{}^a=-\ve^{ab}c_b.
\end{equation}

Лоренцева связность в неголономном базисе определяется соотношением
\begin{equation}
  \om_a:=e_a{}^\al\om_\al .
\end{equation}
Скалярная кривизна $R$ и след тензора кручения $T_b:=T_{ab}{}^a$ в неголономном
базисе имеют вид
\begin{align}                                                      \label{escur}
  R&=2\ve^{ab}(\pl_a\om_b+c_a\om_b),
\\                                                                \label{etetru}
  T_b&=-c_b+\ve_b{}^c\om_c.
\end{align}

В геометрии Римана--Картана диада и лоренцева связность являются независимыми
переменными. В то же время по заданной диаде всегда можно построить вторую
лоренцеву связность и другие геометрические объекты, которые соответствуют
нулевому кручению. В этом случае, т.е.\ в (псевдо)римановой геометрии, лоренцева
связность взаимно однозначно определяется коэффициентами неголономности:
\begin{equation}
  \tilde\om_a = \ve_a{}^b c_b.
\end{equation}
Разность между двумя лоренцевыми связностями определяется тензором кручения:
\begin{equation}                                                  \label{erelco}
  \om_a = \tilde\om_a + \e_a{}^b T_b.
\end{equation}
Из уравнения (\ref{escur}) следует, что скалярная кривизна в римановой геометрии
определяется коэффициентами неголономности следующей простой формулой
\begin{equation}
  \widetilde R=2\pl_a c^a+2 c_a c^a
\end{equation}
\section{Выбор системы координат                                 \label{sccudg}}
В двумерном пространстве метрика содержит три независимые компоненты, при этом
две из них можно фиксировать за счет выбора системы координат (фиксирования
калибровки). Таким образом после фиксирования координат многие геометрические
характеристики выражаются через одну функцию, что существенно упрощает анализ
двумерных моделей. В настоящем разделе приведен ряд формул в различных системах
координат, которые часто встречаются в приложениях.
\subsection{Диагональная калибровка                              \label{sdioga}}
В {\em диагональной калибровке}, которую используют в вычислениях, метрика и ее
обратная имеют диагональный вид:
\index{Диагональная калибровка (diagonal gauge)}%
\index{Калибровка диагональная (diagonal gauge)}%
\begin{equation}                                                  \label{ediame}
  g_{\al\bt}=\begin{pmatrix} g_{00} & 0 \\ 0 & g_{11} \end{pmatrix},\qquad
  g^{\al\bt}=\begin{pmatrix} 1/g_{00} & 0 \\ 0 & 1/g_{11} \end{pmatrix}.
\end{equation}
Эта калибровка фиксирует только одну из двух произвольных функций,
параметризующих преобразования координат. Оставшийся функциональный произвол
можно использовать для фиксирования других переменных, входящих в модель.
Символы Кристоффеля (\ref{echris}) для метрики (\ref{ediame}) имеют следующий
вид:
\begin{equation}                                                  \label{echsdi}
\begin{split}
  \widetilde\Gamma_{00}{}^0 &=\quad \frac12\frac{\dot g_{00}}{g_{00}}, \\
  \widetilde\Gamma_{00}{}^1 &=-\frac12\frac{g_{00}^\prime}{g_{11}},
\end{split}\qquad
\begin{split}
  \widetilde\Gamma_{01}{}^0 &=\widetilde\Gamma_{10}{}^0
  =\frac12\frac{g_{00}^\prime}{g_{00}}, \\
  \widetilde\Gamma_{01}{}^1 &=\widetilde\Gamma_{10}{}^1
  =\frac12\frac{\dot g_{11}}{g_{11}},
\end{split}\qquad
\begin{split}
  \widetilde\Gamma_{11}{}^0 &=-\frac12\frac{\dot g_{11}}{g_{00}}, \\
  \widetilde\Gamma_{11}{}^1 &=\quad \frac12\frac{g_{11}^\prime}{g_{11}},
\end{split}
\end{equation}
Где точка и штрих обозначают дифференцирование по $x^0$ и $x^1$ соответственно.
У тензора кривизны со всеми опущенными индексами имеется только одна
нетривиальная линейно независимая компонента
\begin{equation}                                                  \label{ecutdg}
  R_{0101}=\frac12(\ddot g_{11}+g''_{00})
  -\frac14\left(\frac{\dot g_{11}^2}{g_{11}}
  +\frac{g_{00}^{\prime2}}{g_{00}}+\frac{\dot g_{00}\dot g_{11}}{g_{00}}
  +\frac{g'_{00}g'_{11}}{g_{11}}\right).
\end{equation}
Остальные отличные от нуля компоненты получаются простой перестановкой индексов.
Тензор Риччи в рассматриваемой калибровке диагонален:
\begin{equation}                                                  \label{eritdg}
  R_{00}=\frac1{g_{11}}R_{0101},\qquad R_{11}=\frac1{g_{00}}R_{0101}.
\end{equation}
Скалярная кривизна равна следующему выражению
\begin{equation}                                                  \label{escudg}
  R=\frac{\ddot g_{11}+g''_{00}}{g_{00}g_{11}}-\frac1{2g_{00}g_{11}}\left(
  \frac{\dot g_{11}^2}{g_{11}}+\frac{g_{00}^{\prime2}}{g_{00}}
  +\frac{\dot g_{00}\dot g_{11}}{g_{00}}
  +\frac{g_{00}^\prime g_{11}^\prime}{g_{11}}\right).
\end{equation}
\subsection{Конформная калибровка для римановых поверхностей     \label{sconga}}
Для двумерной римановой метрики часто используют {\em конформную калибровку},
т.е.\ такую систему координат, в которой метрика является вейлевски евклидовой:
\index{Конформная калибровка (conformal gauge)}%
\index{Калибровка конформная (conformal gauge)}%
\begin{equation}                                                  \label{econga}
  ds^2=\Phi(dx^2+dy^2),
\end{equation}
где $\Phi=\Phi(x,y)$ -- некоторая функция от координат $x=x^1$ и $y=x^2$.
{\em Конформный множитель} $\Phi$ предполагается всюду отличным от нуля, иначе
метрика будет вырождена, а соответствующая скалярная кривизна может обратиться в
бесконечность (это зависит от того, как функция $\Phi$ стремится к нулю).
Координаты в которых метрика имеет вид (\ref{econga}) называются
{\em изотермическими}. Для положительно определенной метрики известна локальная
теорема существования изотермических координат.
\begin{theorem}                                                   \label{qisoco}
Если метрика принадлежит классу ${\cal C}^3(\MM)$, тогда для любой точки
$x\in\MM$ существует окрестность, в которой можно выбрать изотермические
координаты.
\end{theorem}
\begin{proof}
См., например, \cite{Wolf72R}.
\end{proof}
Заметим, что конформный множитель не является скалярным полем, т.к.\ в других
системах координат метрика в общем случае не будет иметь вейлевски евклидова
вида.
\index{Конформный множитель (conformal factor)}%
\index{Множитель конформный (conformal factor)}%
\index{Изотермические координаты (isothermal coordinates)}%
\index{Координаты изотермические (isothermal coordinates)}%

Если метрика имеет вид (\ref{econga}), то удобно ввести  комплексные координаты
(\ref{epofcn}). Тогда интервал запишется в виде
$$
  ds^2=\Phi dzd\bar z.
$$
После конформного преобразования $z\rightarrow\z(z)$ интервал сохраняет
вейлевски евклидову форму
$$
  ds^2=\Phi\frac{dz}{d\z}\frac{d\bar z}{d\bar\z}d\z d\bar\z.
$$
То есть изотермические координаты определены с точностью до конформных
преобразований. По этой причине калибровка (\ref{econga}) называется конформной.
В изотермических координатах отличны от нуля только два символа Кристоффеля:
$$
  \widetilde\Gamma_{zz}{}^z=\frac{\pl_z \Phi}\Phi,\qquad
  \widetilde\Gamma_{\bar z\bar z}{}^{\bar z}=\frac{\pl_{\bar z} \Phi}\Phi.
$$
Отличные от нуля линейно независимые компоненты тензора кривизны имеют вид
\begin{equation}                                                  \label{ecotwc}
  \widetilde R_{z\bar z\bar z}{}^{\bar z}=\widetilde R_{\bar z zz}{}^z
  =\frac{\Phi\pl^2_{z\bar z}\Phi-\pl_z \Phi\pl_{\bar z}\Phi}{\Phi^2}
\end{equation}
У тензора Риччи отличны от нуля только две компоненты:
\begin{equation}                                                  \label{erictw}
  \widetilde R_{z\bar z}=\widetilde R_{\bar z z}
  =\frac{\Phi\pl^2_{z\bar z}\Phi-\pl_z \Phi\pl_{\bar z}\Phi}{\Phi^2}.
\end{equation}
Это приводит к следующему выражению для скалярной кривизны
\begin{equation}                                                  \label{escisc}
  \widetilde R=4\frac{\Phi\pl^2_{z\bar z}\Phi
  -\pl_z \Phi\pl_{\bar z}\Phi}{\Phi^3}=\frac4\Phi\pl^2_{z\bar z}\ln|\Phi|
\end{equation}

Часто для конформного множителя используют следующую параметризацию
\begin{equation}                                                  \label{econpa}
  \Phi=\ex^{2\phi}.
\end{equation}
Тогда
\begin{equation}                                                  \label{escalo}
  \widetilde R=2\ex^{-2\phi}\triangle\phi,
\end{equation}
где $\triangle\phi:=4\pl^2_{z\bar z}\phi$ -- обычный лапласиан.

Для поверхностей постоянной кривизны (см. раздел \ref{sconcu}) скалярная
кривизна постоянна,
\begin{equation}                                                  \label{ecoscp}
  \widetilde R=-2K=\const,
\end{equation}
где $K$ -- гауссова кривизна. Это уравнение в изотермических координатах
сводится к {\em уравнению Лиувилля}
\index{Уравнение Лиувилля (Liouville equation)}%
\index{Лиувилля уравнение (Liouville equation)}
\begin{equation}                                                  \label{elioem}
  \triangle\phi+Ke^{2\phi}=0.
\end{equation}
Уравнение Лиувилля является интегрируемым, и его общее решение приведено в
разделе \ref{slioeq}.
\subsection{Конформная калибровка для лоренцевых поверхностей    \label{scongl}}
На псевдоримановой поверхности $\MM$ интервал в конформной калибровке имеет вид
\begin{equation}                                                  \label{emetcc}
  ds^2=\Phi(dt^2-dx^2)=\Phi dudv
\end{equation}
где $\Phi=\Phi(u,v)$ -- конформный множитель и $u$ и $v$ -- координаты светового
конуса (\ref{econco}). Соответствующая метрика и ее обратная в конусных
координатах имеют вид
\begin{equation}                                                  \label{qincib}
  g_{\al\bt}=\frac \Phi2\begin{pmatrix} 0 & 1 \\ 1 & 0 \end{pmatrix},\qquad
  g^{\al\bt}=\frac2\Phi\begin{pmatrix} 0 & 1 \\ 1 & 0 \end{pmatrix}.
\end{equation}

\begin{theorem}                                                   \label{tlokju}
Если метрика дважды дифференцируема $g_{\al\bt}\in\CC^2(\MM)$, то каждая точка
$\MM$ имеет координатную окрестность, в которой метрика является вейлевски
лоренцевой (\ref{emetcc}).
\end{theorem}
\begin{proof}
Докажем, что преобразованием координат произвольную обратную метрику можно
привести к виду (\ref{qincib}). Рассмотрим преобразование координат
$x^\al\mapsto y^\al(x)$ в произвольной координатной окрестности. Тогда
компоненты обратной метрики преобразуются по правилу
\begin{equation*}
  g^{\al\bt}\mapsto\tilde g^{\al\bt}=g^{\g\dl}\frac{\pl y^\al}{\pl x^\g}
  \frac{\pl y^\bt}{\pl x^\dl}.
\end{equation*}
Для упрощения формул введем обозначение
\begin{equation*}
  g^{\al\bt}=\begin{pmatrix} a & b \\ b & c \end{pmatrix},
\end{equation*}
где $a,b,c,d\in\CC^2(\MM)$.
Не ограничивая общности, можно считать, что исходная система координат выбрана
таким образом, что $a>0$. Потребуем, чтобы преобразованные компоненты метрики
имели вид (\ref{qincib}). Тогда функции перехода должны удовлетворять двум
уравнениям:
\begin{equation}                                                  \label{qfyper}
\begin{split}
  a\left(\frac{\pl y^0}{\pl x^0}\right)^2+2b\frac{\pl y^0}{\pl x^0}
  \frac{\pl y^0}{\pl x^1}+c\left(\frac{\pl y^0}{\pl x^1}\right)^2&=0,
\\
  a\left(\frac{\pl y^1}{\pl x^0}\right)^2+2b\frac{\pl y^1}{\pl x^0}
  \frac{\pl y^1}{\pl x^1}+c\left(\frac{\pl y^1}{\pl x^1}\right)^2&=0.
\end{split}
\end{equation}
Эти квадратные уравнения относительно $\pl y^0/\pl x^0$ и $\pl y^1/\pl x^0$
легко решаются. В результате получаем два эквивалентных линейных уравнения:
\begin{equation}                                                  \label{qtwely}
  \frac{\pl y^0}{\pl x^0}+\lm_0\frac{\pl y^0}{\pl x^1}=0,\qquad
  \frac{\pl y^1}{\pl x^0}+\lm_1\frac{\pl y^1}{\pl x^1}=0,
\end{equation}
где
\begin{equation*}
  \lm_0:=\frac{b-\sqrt{b^2-ac}}a,\qquad \lm_1:=\frac{b+\sqrt{b^2-ac}}a.
\end{equation*}
Поскольку метрика лоренцева, то $b^2-ac>0$.

Предположим, что существуют решения  уравнений (\ref{qtwely}) такие, что
$\lbrace\pl_\al y^0\rbrace\ne0$ и $\lbrace\pl_\al y^1\rbrace\ne0$. Тогда кривые
\begin{equation}                                                  \label{qchers}
  y^0(x)=C^0,\qquad y^1(x)=C^1,\qquad C^{0,1}=\const,
\end{equation}
определяют два семейства характеристик (\ref{echaeq}) для волнового уравнения
на лоренцевой поверхности. Действительно, уравнения (\ref{qfyper}) это и есть
уравнения характеристик.

Для дальнейшего нам понадобится
\begin{lemma}
Пусть функция $W(x)\in\CC^2$ такова, что $\pl W/\pl x^1\ne0$. Для того, чтобы
семейство кривых $W=C=\const$ давало семейство характеристик волнового уравнения
(\ref{ewaeqs}), необходимо и достаточно, чтобы выражение $W=C$ было общим
интегралом одного из обыкновенных дифференциальных уравнений
\begin{equation}                                                  \label{qfrtyi}
  \frac{dx^1}{dx^0}=\lm_0(x^0,x^1),\qquad \frac{dx^1}{dx^0}=\lm_1(x^0,x^1).
\end{equation}
\end{lemma}
\begin{proof}
Из теоремы о неявной функции следует, что уравнение $W=C$ при $\pl_1W\ne0$
определяет неявную функцию $x^1=x^1(x^0,C)$, при этом
\begin{equation*}
  \frac{dx^1}{dx^0}=-\frac{\pl_0W}{\pl_1W}.
\end{equation*}
Если функция $W-C$ является характеристикой волнового уравнения, то она
удовлетворяет квадратному уравнению
\begin{equation*}
  a\pl_0W^2+2b\pl_0W\pl_1W+c\pl_1W^2=0,
\end{equation*}
которое приводит к уравнению для неявной функции
\begin{equation*}
  a\left(\frac{dx^1}{dx^0}\right)^2-2b\frac{dx^1}{dx^0}+c=0,
\end{equation*}
которое сводится к одному из двух уравнений (\ref{qfrtyi}).

Легко проверить также обратное утверждение.
\end{proof}

Согласно общей теории дифференциальных уравнений решения обыкновенных
дифференциальных уравнений (\ref{qfrtyi}) существуют в, возможно, меньшей
окрестности. Согласно доказанной лемме, это означает существование двух семейств
характеристик $y^0(x)=C_0$ и $y^1(x)=C_1$ класса $\CC^1$. Поскольку
$\lm_{0,1}\in\CC^2$, то и $y^{0,1}\in\CC^2$.

Далее, из уравнений (\ref{qtwely}) следует неравенство
\begin{equation*}
  \det\frac{\pl y}{\pl x}:=\pl_0y^0\pl_1y^1-\pl_1y^0\pl_0y^1
  =(\lm_1-\lm_0)\pl_1y^0\pl_1y^1=\frac{2\sqrt{b^2-ac}}a\pl_1y^0\pl_1y^1\ne0.
\end{equation*}
Это означает, что преобразование координат $x\mapsto y$ невырождено.
\end{proof}
\begin{com}
Приведенное доказательство существования системы координат, в которой метрика
имеет вейлевски лоренцев вид (\ref{emetcc}), можно обобщить на евклидов
случай (см., например, \cite{VlaZha00R}). Однако при этом нужно потребовать
аналитичность компонент метрики.
\qed\end{com}

Система координат, в которой метрика является вейлевски лоренцевой, определена
неоднозначно. Преобразования светоподобных координат
\begin{equation}                                                  \label{econtl}
  u,v\quad \mapsto\quad u'(u),v'(v),
\end{equation}
которые приводят только к изменению конформного множителя, будем, как и в
евклидовом случае, называть конформными.

Выражения для символов Кристоффеля и тензора кривизны аналогичны соответствующим
формулам в римановом случае. В конусных координатах отличны от нуля только два
символа Кристоффеля:
\begin{equation}                                                  \label{echsyc}
  \widetilde\Gamma_{uu}{}^u=\frac{\pl_u \Phi}\Phi,\qquad
  \widetilde\Gamma_{vv}{}^v=\frac{\pl_v \Phi}\Phi.
\end{equation}
Линейно независимые отличные от нуля компоненты тензора кривизны имеют вид
\begin{equation}                                                  \label{ecurwc}
  \widetilde R_{uvv}{}^v=\widetilde R_{vuu}{}^u
  =\frac{\Phi\pl^2_{uv}\Phi-\pl_u \Phi\pl_v \Phi}{\Phi^2}.
\end{equation}
У тензора Риччи отличны от нуля только недиагональные компоненты:
\begin{equation}                                                  \label{erictc}
  \widetilde R_{uv}=\widetilde R_{vu}
  =\frac{\Phi\pl^2_{uv}\Phi-\pl_u \Phi\pl_v \Phi}{\Phi^2}.
\end{equation}
Это приводит к следующему выражению для скалярной кривизны
\begin{equation}                                                  \label{escicc}
  \widetilde R=4\frac{\Phi\pl^2_{uv}\Phi-\pl_u \Phi\pl_v \Phi}{\Phi^3}
  =\frac4\Phi\pl^2_{uv}\ln|\Phi|.
\end{equation}

Если конформный множитель $\Phi$ положителен, то его можно параметризовать с
помощью экспоненты
\begin{equation}                                                  \label{ecfpar}
  \Phi:=e^{2\phi},
\end{equation}
где $\phi=\phi(u,v)$ -- некоторая функция координат. Тогда скалярная кривизна
(\ref{escicc}) принимает вид
\begin{equation}                                                  \label{esconf}
  \widetilde R=2e^{-2\phi}\square\phi,
\end{equation}
где $\square$ -- оператор Даламбера на плоскости Минковского (\ref{edalot}).

Для поверхностей постоянной кривизны (см. раздел \ref{sconcu}) скалярная
кривизна постоянна,
\begin{equation}                                                  \label{ecoscu}
  \widetilde R=-2K=\const,
\end{equation}
где $K$ -- гауссова кривизна. Это уравнение в конформной калибровке сводится к
{\em уравнению Лиувилля}
\index{Уравнение Лиувилля (Liouville equation)}%
\index{Лиувилля уравнение (Liouville equation)}
\begin{equation}                                                  \label{elioeq}
  \square\phi+Ke^{2\phi}=0.
\end{equation}
Уравнение Лиувилля является интегрируемым, и его общее решение приведено в
разделе \ref{slioeq}.
\section{Координаты светового конуса                             \label{slicoo}}
В трех предыдущих разделах мы использовали преобразования координат для
уменьшения числа неизвестных компонент метрики. В настоящем разделе мы
рассмотрим удобный способ введения базиса в касательном пространстве для {\em
лоренцевой} поверхности, оставляя полную свободу в выборе координат на
многообразии.

Если на поверхности задана метрика лоренцевой сигнатуры, то при проведении
вычислений бывает удобно использовать координаты светового конуса для векторов
касательного пространства, которые совпадают с координатами светового конуса
(\ref{econco}) на плоскости Минковского. Преимущество этих координат связано с
тем, что локальное лоренцево вращение не перемешивает компоненты векторного
поля, т.к.\ каждая из них преобразуется по неприводимому одномерному
представлению связной компоненты единицы группы $\MS\MO_\uparrow(1,1)$.

Преобразование компонент вектора $X=X^ae_a$ относительно ортонормального базиса
в касательном пространстве при переходе от декартовых координат к {\em
координатам светового конуса} определим следующим образом:
\index{Координаты светового конуса (light cone coordinates)}%
\index{Светового конуса координаты (light cone coordinates)}%
\begin{equation}                                                  \label{elicop}
  X^\pm:=\frac1{\sqrt2}(X^0\pm X^1),
\end{equation}
Якобиан преобразования $X^0,X^1\mapsto X^+,X^-$ равен единице. Обратное
преобразование имеет вид
\begin{equation}                                                  \label{elicoi}
  X^0=\frac1{\sqrt2}(X^++X^-),\qquad X^1=\frac1{\sqrt2}(X^+-X^-).
\end{equation}
В частности, конусные координаты репера (диады) равны
\begin{equation}                                                  \label{ezwelc}
  e_\al{}^\pm=\frac1{\sqrt2}(e_\al{}^0\pm e_\al{}^1).
\end{equation}
Нетрудно проверить, что определитель репера можно записать в виде
\begin{equation}                                                  \label{edelcc}
  \vol=\det e_\al{}^a=e_0{}^-e_1{}^+-e_0{}^+e_1{}^-
  =\hat\ve^{\al\bt}e_\al{}^+e_\bt{}^-,
\end{equation}
что эквивалентно тождеству
\begin{equation}                                                  \label{eidrep}
  \ve^{\al\bt}e_\al{}^+e_\bt{}^-=1.
\end{equation}
При проведении расчетов часто приходится обращать диаду. Выпишем соответствующие
формулы:
\begin{equation}                                                  \label{eindia}
\begin{aligned}
  e^0{}_+&=-\frac1\vol e_1{}^-,
\\
  e^0{}_-&=\quad \frac1\vol e_1{}^+,
\end{aligned}\qquad
\begin{aligned}
   e^1{}_+&=\quad \frac1\vol e_0{}^-
\\
   e^1{}_-&=-\frac1\vol e_0{}^+.
\end{aligned}
\end{equation}

Скалярное произведение векторов в координатах светового конуса имеет вид
\begin{equation}                                                  \label{escprl}
  X^aY^b\eta_{ab}=X^-Y^++X^+Y^-.
\end{equation}
\begin{com}
Важным является то обстоятельство, что лоренцев квадрат вектора,
\begin{equation}                                                  \label{esclcc}
  X^2=2X^+X^-,
\end{equation}
линеен по компонентам, а не квадратичен, как в декартовой системе координат. Это
помогает в ряде случаев записать явно решение некоторых уравнений, не извлекая
квадратного корня.
\qed\end{com}
Скалярное произведение можно записывать в виде $X^\pm Y^\pm\eta_{\pm\pm}$, где
индекс $\pm=\lbrace +,-\rbrace$ пробегает два значения, а лоренцева метрика и ее
обратная в координатах светового конуса имеют одинаковый вид,
\begin{equation}                                                  \label{elomlc}
  \eta_{\pm\pm}=\eta^{\pm\pm}=\begin{pmatrix} 0 & 1 \\ 1 & 0 \end{pmatrix}.
\end{equation}
Ко- и контравариантные компоненты векторов связаны простыми соотношениями:
\begin{equation}                                                  \label{eraloi}
  X^+=X_-,\qquad X^-=X_+,
\end{equation}
при этом соотношения между ковариантными компонентами такие же, как и для
контравариантных векторов (\ref{elicop}):
$$
  X_\pm=\frac1{\sqrt2}(X_0\pm X_1).
$$

Псевдоскаляр, построенный из двух векторов, в координатах светового конуса
принимает вид
\begin{equation}                                                  \label{epslcc}
  X^aY^b\ve_{ab}=X^\pm Y^\pm\ve_{\pm\pm}=X^-Y^+-X^+Y^-,
\end{equation}
где
\begin{align}                                                     \label{eantlc}
  \ve_{\pm\pm}=-\ve^{\pm\pm}&=\begin{pmatrix} 0 & -1 \\ 1 & ~0 \end{pmatrix},
\\                                                                \label{etntlc}
  \ve^\pm{}_\pm=-\ve_\pm{}^\pm&=\begin{pmatrix} 1 & ~0 \\ 0 & -1 \end{pmatrix}.
\end{align}

Используя выражение для псевдоследа тензора кручения (\ref{eptrto}), получаем
его компоненты в координатах светового конуса
\begin{equation}                                                  \label{etptlc}
  T^{*\pm}=2\ve^{\al\bt}(\pl_\al\mp\om_\al)e_\bt{}^\pm.
\end{equation}
Отсюда следует, что каждая компонента репера $e_\al{}^\pm$ определяет одну
компоненту псевдоследа. Тождество (\ref{qffgtd}) в координатах светового конуса
принимает вид
\begin{equation}                                                  \label{eidscc}
  R=\widetilde R+2\nb_aT^a-4T^+T^-,
\end{equation}
где
$$
  \nb_\al T^{*\pm}=\pl_\al T^{*\pm}\mp\om_\al T^{*\pm}.
$$
Координаты светового конуса играют большую роль в общем решении уравнений
движения двумерной гравитации.
\chapter{Поверхности постоянной кривизны                         \label{sconcu}}
Поверхности постоянной кривизны являются относительно простыми объектами,
которые хорошо изучены и в некоторых случаях классифицированы. Универсальными
накрывающими для любой полной римановой поверхности постоянной кривизны являются
либо сфера $\MS^2$, либо евклидова плоскость $\MR^2$, либо одна компонента
связности двуполостного гиперболоида $\MH^2$, соответственно для положительной,
нулевой и отрицательной гауссовой кривизны. Все полные римановы поверхности
постоянной кривизны получаются из этих трех как фактор пространства по действию
свободной и собственно разрывной группы преобразований (теорема \ref{tfsman}).

В случае псевдоримановых поверхностей, метрика которых имеет лоренцеву
сигнатуру, можно не различать поверхности постоянной положительной и
отрицательной кривизны, т.к.\ они связаны между собой простым изменением знака
метрики: $g_{\al\bt}\mapsto-g_{\al\bt}$, не меняющим сигнатуру. Для нулевой
кривизны, универсальной накрывающей является плоскость Минковского $\MR^{1,1}$.
Если кривизна отлична от нуля, то универсальной накрывающей поверхности
постоянной кривизны является универсальная накрывающая однополостного
гиперболоида $\ML^2$.

В настоящей главе мы подробно опишем все универсальные накрывающие поверхности
постоянной кривизны, за исключением евклидовой плоскости $\MR^2$ и двумерного
пространства Минковского $\MR^{1,1}$, которые достаточно полно были рассмотрены
во Введении.
\section{Сфера $\MS^2$                                           \label{sphere}}
\begin{figure}[h,b,t]
\hfill\includegraphics[width=.8\textwidth]{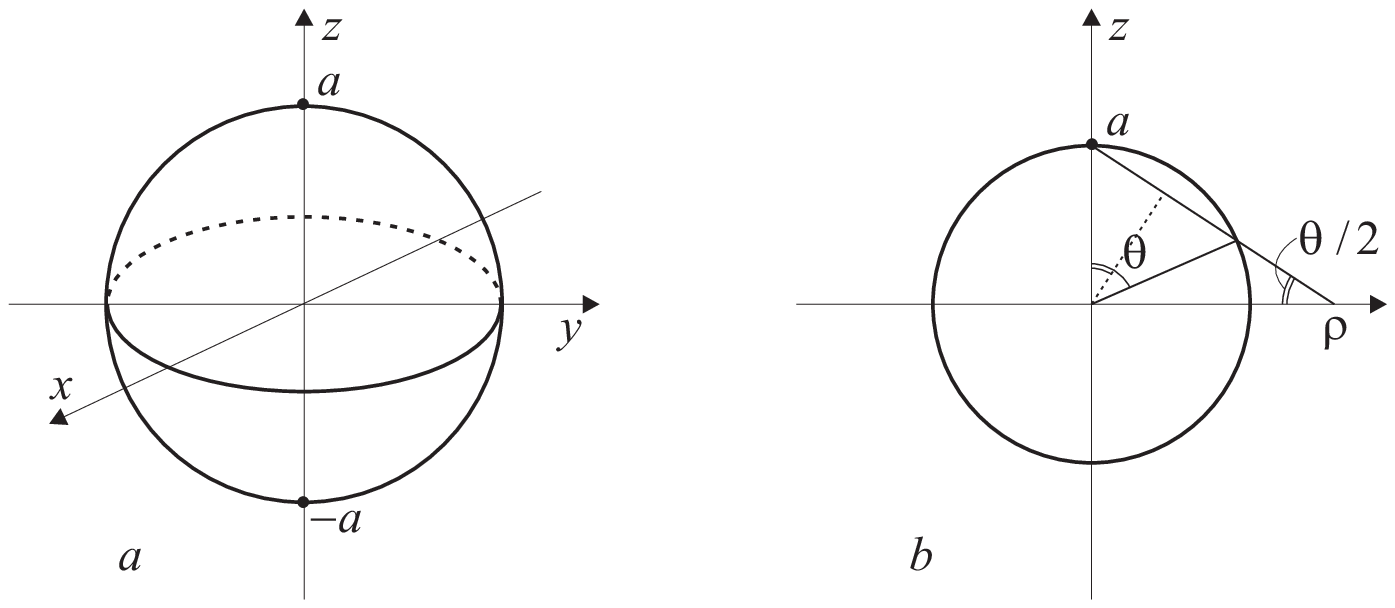}
\hfill {}
\centering\caption{Сфера радиуса $a$ с центром в начале координат {(\it a)}.
 Стереографическая проекция точки сферы на плоскость $x,y$ {(\it b)}.}
\label{fspher}
\end{figure}

Рассмотрим трехмерное евклидово пространство $\MR^3$ с декартовыми координатами
$x,y,z$. Двумерная сфера $\MS^2_a$ радиуса $a>0$ с центром в начале системы
координат задается уравнением (см.\ рис.~\ref{fspher},{\it a})
\begin{equation}                                                  \label{espher}
  x^2+y^2+z^2=a^2.
\end{equation}
Сфера, очевидно, является связным и односвязным многообразием.

Вложение сферы в трехмерное евклидово пространство
$\MS^2_a\hookrightarrow\MR^3$, заданное уравнением (\ref{espher}), индуцирует на
ней положительно определенную метрику с помощью возврата отображения (см.\
раздел \ref{smapma}). Явный вид индуцированной метрики проще всего выглядит в
сферической системе координат $r,\theta,\vf$ (см.\ раздел \ref{spheco}). Чтобы
получить его, в выражение для евклидовой метрики в $\MR^3$, записанной в
сферической системе координат,
\begin{equation*}
  ds^2=dr^2+r^2(d\theta^2+\sin^2\theta d\vf^2)
\end{equation*}
достаточно подставить уравнение сферы $r=a$. В результате метрика,
индуцированная на сфере, в сферической системе координат примет вид
\begin{equation}                                                  \label{einmes}
  d\Om:=a^2(d\theta^2+\sin^2\theta d\vf^2),
\end{equation}
где координаты на сфере $\theta$ и $\vf$ называются, соответственно,
{\em азимутальным} и {\em полярным} углами.
\index{Азимутальный угол (azimuth angle)}%
\index{Угол азимутальный (azimuth angle)}%
\index{Полярный угол (polar angle)}\index{Угол полярный (polar angle)}%
Координаты $\theta$, $\vf$ покрывают всю сферу, но не определяют глобальную
систему координат на сфере, т.к.\ необходимо провести отождествление, например,
$\vf\sim\vf+2\pi$. Соответствующую метрику и ее обратную можно записать в
матричных обозначениях:
$$
  g_{\al\bt}=a^2\begin{pmatrix} 1 & 0\\0 & \sin^2\theta \end{pmatrix},\qquad
  g^{\al\bt}=\frac1{a^2}
  \begin{pmatrix} 1 & 0\\ 0 & \sin^{-2}\theta\end{pmatrix}.
$$
Из символов Кристоффеля (\ref{echris}) отличны от нуля только три:
$$
  \Gamma_{\theta\vf}{}^\vf=\Gamma_{\vf\theta}{}^\vf=\frac{\cos\theta}{\sin\theta},\qquad
  \Gamma_{\vf\vf}{}^\theta=-\sin\theta\cos\theta.
$$
Тензор кривизны имеет только две нетривиальные линейно независимые компоненты:
\begin{equation}                                                  \label{ecursp}
  R_{\theta\vf\theta}{}^\vf=-1,\qquad R_{\theta\vf\vf}{}^\theta=\sin^2\theta.
\end{equation}
Тензор кривизны со всеми опущенными индексами имеет только одну независимую
нетривиальную компоненту
$$
  R_{\theta\vf\theta\vf}=-a^2\sin^2\theta.
$$
Тензор Риччи диагонален:
$$
  R_{\theta\theta}=-1,\qquad R_{\vf\vf}=-\sin^2\theta.
$$
Соответствующая скалярная кривизна (\ref{escurv}) равна константе
$$
  R=-2K=-\frac2{a^2}.
$$
Отсюда следует, что согласно нашему определению тензора кривизны, тензора Риччи
и скалярной кривизны, скалярная кривизна сферы единичного радиуса равна $-2$.
Поэтому мы также используем {\em гауссову кривизну} $K$, которая отличается от
скалярной кривизны множителем $-2$.
\index{Гауссова кривизна (Gaussian curvature)}%
\index{Кривизна гауссова (Gaussian curvature)}%
Для сферы единичного радиуса гауссова кривизна равна единице, $K=1$.

Экстремали (линии наименьшей длины) на сфере определяются следующей теоремой.
\begin{theorem}                                                   \label{texsph}
Сечения сферы плоскостями, проходящими через ее центр и только они являются
экстремалями. Все экстремали полны.
\end{theorem}
\begin{proof}
См., например, \cite{DuNoFo98R}.
\end{proof}
Таким образом все экстремали на сфере являются большими окружностями и замкнуты.
Ясно, что они полны, т.к.\ их можно проходить бесконечное число раз. Любые две
точки, которые не являются диаметрально противоположными, можно соединить двумя
экстремалями. Одна из них будет кратчайшей, а другая -- длиннейшей.
Диаметрально противоположные точки соединяются бесконечно большим числом
экстремалей, которые имеют одинаковую длину.

Метрика евклидова пространства $\MR^3$ инвариантна относительно действия группы
трехмерных вращений $\MS\MO(3)$. Кроме того уравнение сферы (\ref{espher})
также инвариантно относительно вращений. Отсюда следует, что метрика,
индуцированная на сфере, допускает по крайней мере три вектора Киллинга по числу
параметров группы Ли преобразований. С другой стороны, поверхность постоянной
кривизны допускает существование не более трех линейно независимых векторов
Киллинга (теорема \ref{tkildi}). Поэтому сфера представляет собой максимально
симметричную поверхность постоянной положительной кривизны. Любой вектор
Киллинга на сфере представляет собой линейную комбинацию векторов Киллинга,
соответствующих вращениям. В декартовой системе координат
$\lbrace x^i\rbrace=(x,y,z)$, $i=1,2,3$, векторы Киллинга евклидова пространства
$\MR^3$ имеют следующие компоненты:
\begin{equation}                                                  \label{ekilca}
  K_i=\ve_{ij}{}^k x^j\pl_k.
\end{equation}
где $\ve_{ijk}$ -- полностью симметричный тензор третьего ранга. Они
удовлетворяют коммутационным соотношениям (\ref{ealatr}) алгебры $\Gs\Go(3)$:
\begin{equation}                                                  \label{qsoark}
  [K_i,K_j]=-\ve_{ij}{}^kK_k.
\end{equation}
Используя формулы перехода от декартовых координат к сферическим (\ref{esphct}),
получаем явный вид векторных полей Киллинга на сфере $\MS^2_a$ в сферических
координатах:
\begin{equation}                                                  \label{ekilsp}
\begin{split}
  K_1&=-\sin\vf\pl_\theta-\frac{\cos\theta\cos\vf}{\sin\theta}\pl_\vf,
\\
  K_2&=\quad \cos\vf\pl_\theta-\frac{\cos\theta\sin\vf}{\sin\theta}\pl_\vf,
\\
  K_3&=\quad \pl_\vf.
\end{split}
\end{equation}
Можно проверить, что они удовлетворяют той же алгебре (\ref{qsoark}), что и
векторы Киллинга евклидова пространства.

В приложениях часто используются {\em стереографические координаты} $\rho,\vf$
\index{Стереографические координаты (stereographic coordinates)}%
\index{Координаты стереографические (stereographic coordinates)}%
на сфере $\MS^2_a$, в которых индуцированная метрика (\ref{einmes}) является
вейлевски евклидовой. Они строятся следующим образом. Спроектируем точку
$(\theta,\vf)$, не находящуюся на северном полюсе, на плоскость $x,y$ из
северного полюса, как показано на рис.~\ref{fspher},{\it b}. При этом точки
нижней и верхней полусферы отображаются соответственно на внутренность и
дополнение к кругу радиуса $a$ на плоскости $x,y$. Тогда полярный радиус $\rho$
на плоскости $x,y$ связан с азимутальным углом $\theta$ следующим соотношением
\begin{equation*}
  \frac a\rho=\tg\frac\theta2=\frac{1-\cos\theta}{\sin\theta}.
\end{equation*}
Отсюда находим радиус
\begin{equation}                                                  \label{eazpoa}
  \rho=a\frac{\sin\theta}{1-\cos\theta}.
\end{equation}
Подставляя соотношение между дифференциалами
\begin{equation*}
  d\rho=-a\frac1{1-\cos\theta}d\theta\end{equation*}
в выражение для индуцированной метрики (\ref{einmes}), получаем выражение для
метрики на сфере в стереографических координатах
\begin{equation}                                                  \label{espstr}
  ds^2=\frac4{(1+\rho^2/a^2)^2}(d\rho^2+\rho^2d\vf^2),\qquad 0\le\rho<\infty,
  \quad 0\le\vf<2\pi.
\end{equation}
Отсюда следует, что в стереографических координатах метрика на сфере является
вейлевски евклидовой.

Векторы Киллинга в стереографических координатах имеют вид:
\begin{equation}                                                  \label{ekilsc}
\begin{split}
  K_1&=\quad \frac{\rho^2+a^2}{2a}\sin\vf\pl_\rho
  -\frac{\rho^2-a^2}{2a\rho}\cos\vf\pl_\vf,
\\
  K_2&=-\frac{\rho^2+a^2}{2a}\cos\vf\pl_\rho
  -\frac{\rho^2-a^2}{2a\rho}\sin\vf\pl_\vf,
\\
  K_3&=\quad \pl_\vf.
\end{split}
\end{equation}

Стереографические координаты покрывают все точки сферы, за исключением северного
полюса, который отображается в бесконечность на плоскости $x,y$. Аналогично
можно построить проекцию точек сферы из южного полюса. Тогда сфера будет покрыта
двумя картами.

Рассмотрим плоскость $x,y$ как комплексную плоскость $z=x+iy\in\MC$. В
комплексных координатах метрика (\ref{espstr}) примет вид
\begin{equation}                                                  \label{espstc}
  ds^2=\frac4{(1+z\bar z/a^2)^2}dzd\bar z.
\end{equation}
При стереографической проекции северный полюс сферы отображается в бесконечно
удаленную точку на комплексной плоскости. Комплексная плоскость, объединенная с
бесконечно удаленной точкой, называется {\em расширенной комплексной плоскостью}
и обозначается $\overline\MC$. Комплексная плоскость $\MC\approx\MR^2$ и
расширенная комплексная плоскость $\overline\MC\approx\MS^2$, как было отмечено
в примере \ref{ecofie}, имеют разные топологии. Сфера $\MS^2$, на которую
взаимно однозначно отображается расширенная комплексная плоскость с помощью
стереографической проекции называется {\em сферой Римана}.
\index{Расширенная комплексная плоскость (extended complex plane)}%
\index{Комплексная плоскость расширенная (extended complex plane)}%
\index{Сфера Римана (Riemann sphere)}\index{Римана сфера (Riemann sphere)}%

Отображение расширенной комплексной плоскости $\overline\MC$ на сферу $\MS^2$
можно записать в комплексном виде. Обозначим декартовы координаты точки,
расположенной на сфере, через $\xi,\eta,t$. По-определению, они удовлетворяют
уравнению
\begin{equation*}
  \xi^2+\eta^2+t^2=a^2,
\end{equation*}
т.е.\ в качестве координат на сфере можно выбрать координаты $\xi$ и $\eta$ или
комплексную переменную $\z:=\xi+i\eta$. Они связаны с комплексными координатами
на плоскости $z=x+iy$ соотношением
\begin{equation*}
  \frac\z z=\frac{a-t}a.
\end{equation*}
Исключив из этих соотношений $t$, получим равенство
\begin{equation*}
  \z=\frac{2z}{1+z\bar z/a^2}.
\end{equation*}
Выпишем также формулы, для координаты $t$ и обратного преобразования
$\xi,\eta,t\rightarrow z$
\begin{equation*}
  t=\frac{z\bar z-a^2}{z\bar z+a^2}a,\qquad z=\frac a{a-t}\z.
\end{equation*}
Эти формулы часто используются в приложениях.
\section{Двуполостный гиперболоид $\MH^2$                        \label{sutwse}}
\begin{figure}[h,b,t]
\hfill\includegraphics[width=.8\textwidth]{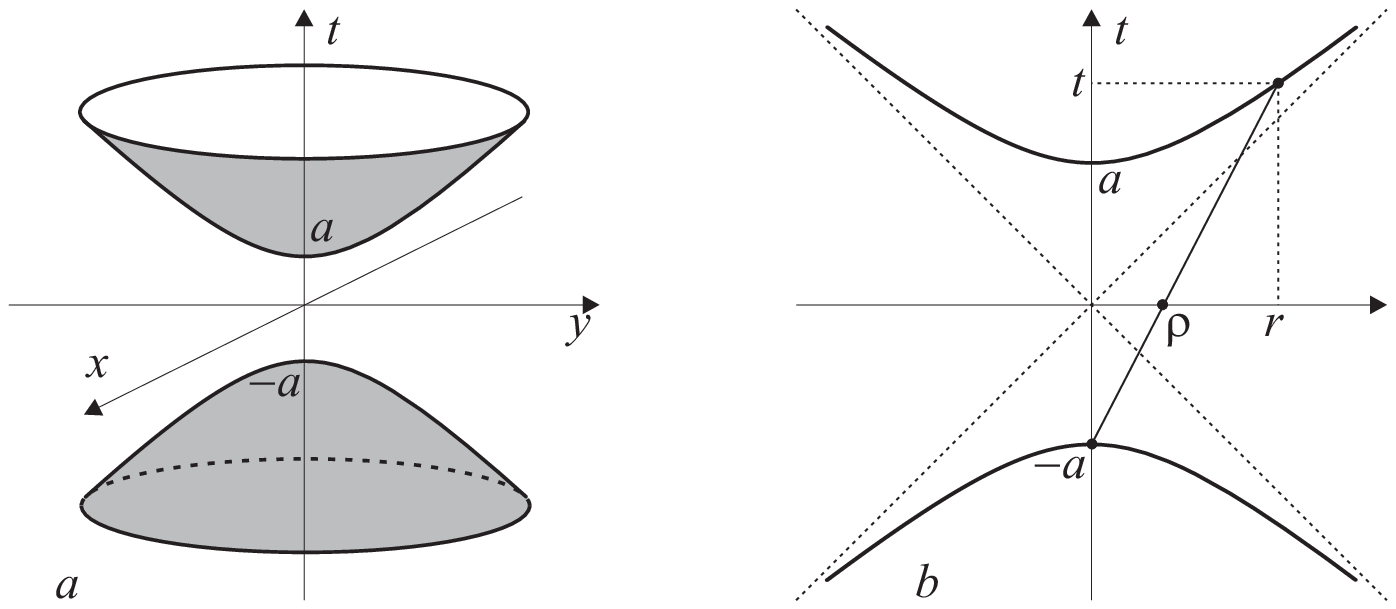}
\hfill {}
\centering\caption{Двуполостный гиперболоид $\MH^2$, вложенный в трехмерное
пространство Минковского $\MR^{1,2}$ ({\it a}). Стереографические координаты
$\rho,\vf$ на двуполостном гиперболоиде ({\it b}).}
\label{ftwhyp}
\end{figure}

Рассмотрим трехмерное пространство Минковского $\MR^{1,2}$ с декартовыми
координатами $t,x,y$. Метрика Лоренца на этом многообразии определяется
квадратичной формой
\begin{equation}                                                  \label{qlotyr}
  ds^2=dt^2-dx^2-dy^2.
\end{equation}
{\em Двуполостный гиперболоид}  ``радиуса'' $a>0$, вложенный в пространство
Минковского $\MR^{1,2}$, задается уравнением
\index{Двуполостный гиперболоид (two sheeted hyperboloid)}%
\index{Гиперболоид двуполостный (two sheeted hyperboloid)}%
\begin{equation}                                                  \label{etwhyp}
  t^2-x^2-y^2=a^2,\qquad a>0,
\end{equation}
и показан на рис.~\ref{ftwhyp},{\it a}. Это уравнение имеет решения только при
$|t|\ge a$. Верхняя и нижняя полы гиперболоида соответствуют значениям $t\ge a$
и $t\le-a$. Как многообразие двуполостный гиперболоид состоит из двух компонент
связности: верхней и нижней полы. Каждая пола является связным и односвязным
многообразием. Для определенности выберем верхнюю полу гиперболоида и обозначим
ее $\MH^2$. В дальнейшем, говоря про двуполостный гиперболоид, будем
подразумевать именно его верхнюю полу $\MH^2$.
Верхняя (или нижняя) пола двуполостного гиперболоида называется также
{\em плоскостью Лобачевского} или {\em гиперболической плоскостью}.
\index{Плоскость Лобачевского (Lobachevsky plane)}%
\index{Лобачевского Плоскость (Lobachevsky plane)}%
\index{Гиперболическая плоскость (hyperbolic plane)}%
\index{Плоскость гиперболическая (hyperbolic plane)}%

Вложение $\MH^2\hookrightarrow\MR^{1,2}$ двуполостного гиперболоида в
пространство Минковского (\ref{etwhyp}) индуцирует на нем метрику евклидовой
сигнатуры (риманову метрику). Это очевидно, поскольку все касательные векторы к
поверхности (\ref{etwhyp}) пространственноподобны.

Вычисление геометрических характеристик двуполостного гиперболоида наиболее
просто проводится в гиперболических координатах:
\begin{equation}                                                  \label{qhybco}
\begin{split}
  x&=r\sh\theta\cos\vf,
\\
  y&=r\sh\theta\sin\vf,
\\
  t&=\pm r\ch\theta,
\end{split}
\end{equation}
которые мы обозначим теми же буквами $r,\theta,\vf$, что и сферические
координаты. Преобразования координат (\ref{qhybco}) определены при всех
значениях координат $r,\theta,\vf$. В приведенных формулах знак $\pm$ в формуле
для $t$ относится к гиперболическим координатам, соответственно, в верхнем $t>0$
и нижнем $t<0$ полупространствах.

Якобиан преобразования легко вычислить
\begin{equation*}
  J=\frac{\pl(t,x,y)}{\pl(r,\theta,\vf)}=\pm r^2\sh\theta.
\end{equation*}
Таким образом, преобразование координат вырождено при $r=0$ и $\theta=0$, т.е.\
на оси времени $t$.

Обратные преобразования запишем в следующем виде
\begin{equation}                                                  \label{qinhyt}
\begin{split}
  r&=\sqrt{t^2-x^2-y^2},
\\
  \theta&=\pm \arcth\frac{\sqrt{x^2+y^2}}t,
\\
  \vf&=\arctg\frac yx.
\end{split}
\end{equation}
Они определены при $t^2\ge x^2+y^2$, и знак $\pm$ относится соответственно к
верхней и нижней поле гиперболоида. Мы видим, что внутренность светового конуса
будущего, $t>0$, с вершиной в начале координат и удаленной полуплоскостью $y=0$,
$x\ge0$ отображается на область
\begin{equation}                                                  \label{qpolyt}
  0<r<\infty,\qquad 0<\theta<\infty,\qquad 0<\vf<2\pi.
\end{equation}
При этом внутренность светового конуса прошлого, $t<0$, с удаленной
полуплоскостью также отображается на область (\ref{qpolyt}). Заметим, что
полуплоскость необходимо удалить из-за полярного угла $0<\vf<2\pi$.

В гиперболических координатах двуполостный гиперболоид (\ref{etwhyp}) задается
особенно просто: $r=a=\const>0$. Поэтому углы $\theta$ и $\vf$ можно
рассматривать в качестве координат на двуполостном гиперболоиде $\MH^2$.

Лоренцева метрика (\ref{qlotyr}) в гиперболических координатах (\ref{qhybco})
принимает вид
\begin{equation*}
  ds^2=dr^2-r^2(d\theta^2+\sh^2\theta d\vf^2).
\end{equation*}
Поэтому индуцированная метрика на двуполостном гиперболоиде $\MH^2$,
$r=a=\const$, задается квадратичной формой
\begin{equation}                                                  \label{eintwh}
  dl^2=a^2(d\theta^2+\sh^2\theta d\vf^2).
\end{equation}
При этом мы изменили общий знак двумерной индуцированной метрики для того, чтобы
она стала положительно определенной. Соответствующая метрика и ее обратная в
матричных обозначениях имеют вид
\begin{equation}                                                  \label{emeinm}
  g_{\al\bt}=a^2\begin{pmatrix} 1&0\\ 0&\sh^2\theta\end{pmatrix}\qquad
  g^{\al\bt}=a^{-2}\begin{pmatrix} 1&0\\ 0&\sh^{-2}\theta\end{pmatrix}.
\end{equation}
\begin{com}
Если бы мы рассматривали вложение (\ref{etwhyp}) двуполостного гиперболоида в
трехмерное евклидово пространство $\MR^3$, то индуцированная метрика имела бы
другой вид. Поэтому важно, что рассматривается вложение именно в пространство
Минковского.
\qed\end{com}

Прямые вычисления показывают, что только три символа Кристоффеля (\ref{echris})
отличны от нуля:
\begin{equation*}
  \Gamma_{\theta\vf}{}^\vf=\Gamma_{\vf\theta}{}^\vf =\frac{\ch\theta}{\sh\theta},
  \qquad \Gamma_{\vf\vf}{}^\theta=-\sh\theta\ch\theta.
\end{equation*}
Полный тензор кривизны имеет две отличные от нуля независимые компоненты:
\begin{equation}                                                  \label{ecursh}
  R_{\theta\vf\theta}{}^\vf=1,\qquad R_{\theta\vf\vf}{}^\theta=-\sh^2\theta
\end{equation}
При этом тензор кривизны со всеми опущенными индексами имеет только одну
отличную от нуля независимую компоненту
\begin{equation*}
  R_{\theta\vf\theta\vf}=a^2\sh^2\theta.
\end{equation*}
Соответствующий тензор Риччи диагонален:
\begin{equation*}
  R_{\theta\theta}=1,\qquad R_{\vf\vf}=\sh^2\theta,
\end{equation*}
и скалярная кривизна постоянна
\begin{equation*}
  \widetilde R=-2K=\frac2{a^2}.
\end{equation*}
Для гиперболоида единичного радиуса, $a=1$, гауссова кривизна $K=-1$. Таким
образом, двуполостный гиперболоид является поверхностью постоянной отрицательной
гауссовой кривизны.

Очевидно, что группа Лоренца $\MO(1,2)$, действующая глобально в пространстве
Минковского $\MR^{1,2}$, является группой симметрии и лоренцевой метрики
(\ref{qlotyr}), и  двуполостного гиперболоида (\ref{etwhyp}). Поэтому группой
симметрии индуцированной метрики на двуполостном гиперболоиде (\ref{eintwh})
является группа Ли псевдоортогональных вращений (группа Лоренца) $\MS\MO(1,2)$.
Соответствующая алгебра $\Gs\Go(1,2)$ в пространстве Минковского $\MR^{1,2}$
задается тремя векторами Киллинга:
\begin{equation}                                                  \label{ekisod}
\begin{split}
  K_0&=-x\pl_y+y\pl_x=-\pl_\vf,
\\
  K_1&=\quad y\pl_t+t\pl_y
  =\quad \sin\vf\pl_\theta+\frac{\ch\theta\cos\vf}{\sh\theta}\pl_\vf,
\\
  K_2&=-t\pl_x-x\pl_t
  =-\cos\vf\pl_\theta+\frac{\ch\theta\sin\vf}{\sh\theta}\pl_\vf,
\end{split}
\end{equation}
которые мы записали в декартовых и гиперболических координатах. Векторы Киллинга
в гиперболических координатах не содержат компоненты вдоль $\pl_r$. Это
означает, что они касательны к двуполостному гиперболоиду. Коммутационные
соотношения для векторных полей Киллинга имеют вид
\begin{equation}                                                  \label{ecjrso}
  [K_0,K_1]=K_2,\qquad [K_2,K_0]=K_1,\qquad [K_1,K_2]=-K_0.
\end{equation}
Отметим, что инверсия времени $t\rightarrow-t$ преобразует
$K_{1,2}\rightarrow-K_{1,2}$ и не меняет $K_0$. При этом алгебра Ли полей
Киллинга остается неизменной.

Введем обозначение $\lbrace K_i\rbrace=\lbrace K_0,K_1,K_2\rbrace$, $i=0,1,2$.
Тогда алгебру $\Gs\Go(1,2)$ можно записать в компактном виде:
\begin{equation*}
  [K_i,K_j]=-\ve_{ij}{}^k K_k,
\end{equation*}
где $\ve_{ijk}$ -- полностью антисимметричный тензор третьего ранга,
$\ve_{012}=1$, а подъем индексов, в отличие от алгебры вращений $\Gs\Go(3)$,
осуществляется с помощью обратной метрики Минковского $\eta^{ij}=\diag(+--)$.

Поскольку в общем случае поверхность допускает не более трех векторов Киллинга,
то двуполостный гиперболоид представляет собой максимально симметричную
поверхность.
\begin{com}
Внимательный читатель, конечно, заметил, что формулы для двуполостного
гиперболоида и сферы, рассмотренной в разделе \ref{sphere}, очень похожи.
Соответствующие выражения отличаются частью знаков и заменой тригонометрических
функций от $\theta$ на гиперболические.
\qed\end{com}

Группа $\MS\MO_\uparrow(1,2)$ некомпактна и ее алгебра Ли $\Gs\Go(1,2)$ имеет
нетривиальную неабелеву подалгебру, натянутую на векторные поля $K_1$ и
$K_0+K_2$. Действительно, из коммутационных соотношений (\ref{ecjrso}) следует
равенство
\begin{equation}                                                  \label{qojkuh}
  [K_1,K_0+K_2]=-(K_0+K_2).
\end{equation}
Соответствующая двумерная неабелева группа Ли подробно рассмотрена в разделе
\ref{stwnog}, где $L_x=-\al K_1$, $L_y=K_0+K_2$. Отметим, что алгебра вращений
$\Gs\Go(3)$ не имеет неабелевых подгрупп.

Легко найти интегральные кривые полей Киллинга в пространстве Минковского
$\MR^{1,2}$. Например, интегральными кривыми поля $K_2$ в плоскости $t,x$
являются гиперболы
\begin{equation*}
  t^2-x^2=\pm c^2,\qquad c\in\MR.
\end{equation*}
Нетрудно проверить, что двуполостные гиперболоиды, параметризуемые ``радиусом''
$a$, являются интегральными подмногообразиями полей Киллинга (\ref{ekisod}).

Нетрудно проверить, что векторы Киллинга линейно зависимы:
\begin{equation}                                                  \label{qlokij}
  tK_0+xK_1+yK_2=0.
\end{equation}

С точки зрения теоремы Фробениуса \ref{tfrofi} векторные поля $K_1$ и $K_0+K_2$
линейно независимы и определяют распределение в пространстве Минковского
$\MR^{1,2}$. Они находятся в инволюции и, следовательно, определяют семейство
интегральных подмногообразий, которыми являются двуполостные гиперболоиды.

Двуполостный гиперболоид является римановым многообразием постоянной
отрицательной гауссовой кривизны. Выше было показано, что эта поверхность
естественным образом вкладывается в трехмерное пространство Минковского
$\MR^{1,2}$, что связано с $\MS\MO(1,2)$ симметрией метрики (\ref{eintwh}).
Можно доказать, что не существует $\CC^2$ вложения полной поверхности постоянной
отрицательной гауссовой кривизны в трехмерное евклидово пространство
\cite{Milnor72}. В то же время в классе $\CC^1$ вложение полной поверхности
постоянной отрицательной гауссовой кривизны уже возможно \cite{Kuiper55}.

Теперь обсудим экстремали.
\begin{theorem}                                                   \label{texthy}
Сечения двуполостного гиперболоида $\MH^2$ плоскостями, проходящими через начало
координат, и только они является экстремалями. Все экстремали на двуполостном
гиперболоиде полны.
\end{theorem}
\begin{proof}
Уравнения для экстремалей интегрируются в явном виде.
\end{proof}

Из рис.~\ref{ftwhyp},{\it a} ясно, что все экстремали незамкнуты. Через две
произвольные точки на гиперболоиде $\MH^2$ и начало координат проходит одна и
только одна плоскость. Это означает, что две произвольные точки на гиперболоиде
можно соединить единственной экстремалью, и это будет линия наименьшей длины.

При преобразовании Лоренца плоскость переходит в плоскость. Поэтому экстремаль
переходит в экстремаль. Это следовало ожидать, поскольку метрика двуполостного
гиперболоида (\ref{eintwh}) инвариантна относительно преобразований Лоренца.

Для того, чтобы записать метрику двуполого гиперболоида (\ref{eintwh}) в
вейлевски евклидовом виде, введем стереографические координаты $\rho,\vf$. Это
делается аналогично введению стереографических координат на сфере. Сначала
введем обычные полярные координаты $r,\vf$ на плоскости $x,y$. Выберем точку
$(t,r,\vf)\in\MH^2$ на верхней пол\'е гиперболоида и спроектируем ее из вершины
нижней полы $x=y=0$, $t=-a$ на плоскость $x,y$. На плоскости $x,y$ ей будет
соответствовать точка с координатами $\rho,\vf$ (см.\ рис.~\ref{ftwhyp},$b$).
Из рисунка ясно, что справедливо соотношение
\begin{equation*}
  \frac r\rho=\frac{t+a}a.
\end{equation*}
Подставив сюда время $t=\sqrt{a^2+r^2}$ из уравнения гиперболоида
(\ref{etwhyp}), получим связь радиусов:
\begin{equation*}
  r=\frac{2a^2\rho}{a^2-\rho^2}.
\end{equation*}
Это приводит к соотношению между дифференциалами
\begin{equation*}
  dr=2a^2\frac{a^2+\rho^2}{(a^2-\rho^2)^2}d\rho.
\end{equation*}

Чтобы записать метрику двуполостного гиперболоида в стереографических
координатах, запишем ее сначала в полярных координатах (после изменения общего
знака):
\begin{equation}                                                  \label{qtwers}
  ds^2=\frac{a^2}{a^2+r^2}dr^2+r^2d\vf^2.
\end{equation}
Для получения этой формулы надо в евклидовой метрике (\ref{qlotyr}) произвести
замену координат $x,y\mapsto r,\vf$ и подставить выражение для времени
$t=\sqrt{a^2+r^2}$. Замена $r\mapsto\rho$ приводит к метрике в стереографических
координатах
\begin{equation}                                                  \label{qtrfca}
  ds^2=\frac4{(1-\rho^2/a^2)^2}(d\rho^2+\rho^2d\vf^2).
\end{equation}
Эта метрика является вейлевски евклидовой и определена при
\begin{equation*}
  0\le\rho<a,\qquad 0\le\vf<2\pi.
\end{equation*}
Она отличается от метрики сферы (\ref{espstr}) только знаком перед $\rho^2$ в
знаменателе, и этим определяется различие в области определения
стереографических координат для сферы $\MS^2$ и двуполостного гиперболоида
$\MH^2$.
\section{Однополостный гиперболоид $\ML^2$                       \label{shyper}}
\begin{figure}[b,t]
\hfill\includegraphics[width=.35\textwidth]{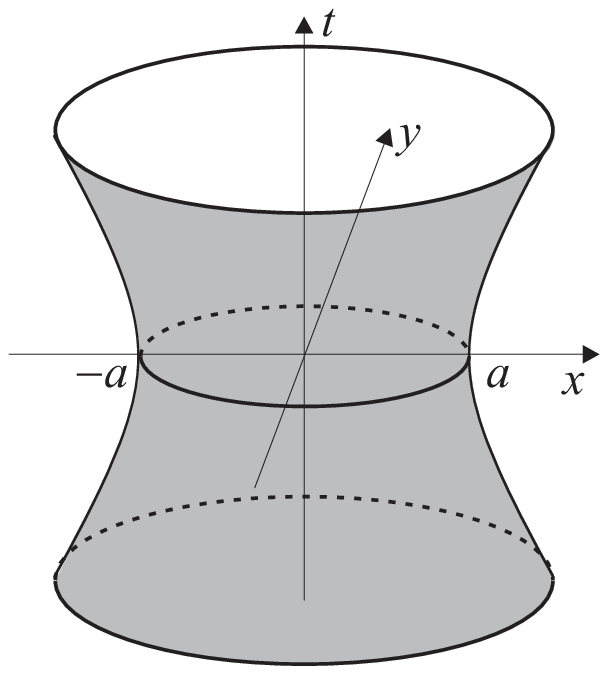}
\hfill {}
\centering\caption{Однополостный гиперболоид $\ML^2$, вложенный в трехмерное
пространство Минковского $\MR^{1,2}$.}
\label{fonhyp}
\end{figure}

Как и раньше, будем считать, что в трехмерном пространстве Минковского
$\MR^{1,2}$ заданы декартовы координаты $t,x,y$ и метрика Лоренца
(\ref{qlotyr}). Рассмотрим {\em однополостный гиперболоид} $\ML^2$, вложенный в
пространство-время Минковского $\MR^{1,2}$. Он задается уравнением (см.\
рис.~\ref{fonhyp}).
\index{Однополостный гиперболоид (one sheet hyperboloid)}%
\index{Гиперболоид однополостный (one sheet hyperboloid)}%
\begin{equation}                                                  \label{eoshyp}
  t^2-x^2-y^2=-a^2,\qquad a>0,
\end{equation}
которое отличается от уравнения для двуполостного гиперболоида (\ref{etwhyp})
знаком правой части. Это уравнение имеет решения только при $x^2+y^2\ge a^2$.

С топологической точки зрения однополостный гиперболоид является связным, но не
односвязным многообразием. Его фундаментальная группа совпадает с
фундаментальной группой окружности, $\pi(\ML^2)=\MZ$, а универсальная
накрывающая, очевидно, диффеоморфна евклидовой плоскости $\MR^2$.

Из рисунка ясно, что в любой точке гиперболоида касательное пространство
содержит как времениподобные, так и пространственноподобные векторы по отношению
к метрике Лоренца (\ref{qlotyr}). Это значит, что вложение
$\ML^2\hookrightarrow\MR^{1,2}$ индуцирует на поверхности $\ML^2$ метрику
лоренцевой сигнатуры.

Для проведения вычислений удобно воспользоваться гиперболическими координатами
$r,\theta,\vf$:
\begin{equation}                                                  \label{qhylok}
\begin{split}
  x&=r\ch\theta\cos\vf,
\\
  y&=r\ch\theta\sin\vf,
\\
  t&=r\sh\theta.
\end{split}
\end{equation}
Эти формулы определены при всех значениях $r,\theta,\vf$. Якобиан преобразования
координат (\ref{qhylok}) равен
\begin{equation*}
  J=\frac{\pl(t,x,y)}{\pl(r,\theta,\vf)}=r^2\ch\theta.
\end{equation*}
Таким образом, преобразование координат вырождено только в начале координат
$r=0$. Поскольку якобиан преобразования координат неотрицателен, то переход к
новым координатам сохраняет ориентацию.

Обратные преобразования запишем в виде
\begin{equation}                                                  \label{qinhyg}
\begin{split}
  r&=\sqrt{x^2+y^2-t^2},
\\
  \theta&=\arccth\frac{\sqrt{x^2+y^2}}t,
\\
  \vf&=\arctg\frac yx.
\end{split}
\end{equation}
Они определены при $x^2+y^2\ge t^2$. Мы видим, что внешность светового конуса с
вершиной в начале координат и удаленной полуплоскостью $y=0$, $x\ge0$
отображается на область
\begin{equation*}
  0<r<\infty,\qquad -\infty<\theta<\infty,\qquad 0<\vf<2\pi.
\end{equation*}

В гиперболических координатах однополостный гиперболоид определяется уравнением
$r=a=\const$. Поэтому углы $\theta$ и $\vf$ можно выбрать в качестве координат
на однополостном гиперболоиде $\ML^2$.

Метрика Лоренца (\ref{qlotyr}) в гиперболических координатах (\ref{qinhyg})
принимает вид
\begin{equation*}
  ds^2=-dr^2+r^2(d\theta^2-\ch^2\theta d\vf^2).
\end{equation*}
Следовательно, индуцированная метрика на однополостном гиперболоиде $r=a=\const$
имеет вид
\begin{equation}                                                  \label{qingtf}
  ds^2=a^2(d\theta^2-\ch^2\theta d\vf^2).
\end{equation}
Она имеет, очевидно, лоренцеву сигнатуру.

В матричных обозначениях метрика и ее обратная имеют вид
\begin{equation}                                                  \label{qmemas}
  g_{\al\bt}=a^2\begin{pmatrix} 1 & 0 \\ 0 & -\ch^2\theta \end{pmatrix}\qquad
  g^{\al\bt}=a^{-2}\begin{pmatrix} 1 & 0 \\ 0 & -\ch^{-2}\theta \end{pmatrix}.
\end{equation}

Из символов Кристоффеля только три отличны от нуля:
\begin{equation*}
  \Gamma_{\theta\vf}{}^\vf=\Gamma_{\vf\theta}{}^\vf=\frac{\sh\theta}{\ch\theta},\qquad
  \Gamma_{\vf\vf}{}^\theta\sh\theta\ch\theta.
\end{equation*}
У полного тензора кривизны только две линейно независимые компоненты отличны от
нуля:
\begin{equation*}
  R_{\theta\vf\theta}{}^\vf=1,\qquad
  R_{\theta\vf\vf}{}^\theta=\ch^2\theta.
\end{equation*}
Тензор кривизны со всеми опущенными индексами имеет только одну линейно
независимую компоненту
\begin{equation*}
  R_{\theta\vf\theta\vf}=-a^2\ch^2\theta.
\end{equation*}
Тензор Риччи диагонален:
\begin{equation*}
  R_{\theta\theta}=1,\qquad R_{\vf\vf}=-\ch^2\theta.
\end{equation*}
Наконец, скалярная кривизна равна
\begin{equation*}
  R=-2K=\frac2{a^2}.
\end{equation*}

Таким образом, однополостный гиперболоид $\ML^2$ является поверхностью
постоянной отрицательной гауссовой кривизны, $K=-1/a^2$. Чтобы получить
поверхность постоянной положительной гауссовой кривизны с метрикой лоренцевой
сигнатуры, можно рассмотреть тот же однополостный гиперболоид, но вложенный в
трехмерное пространство с метрикой
$$
  ds^2=-dt^2+dx^2+dy^2,
$$
что приведет к изменению знака метрики на поверхности, а, значит, и знака
скалярной кривизны.

Метрика трехмерного пространства Минковского (\ref{qlotyr}) и уравнение
гиперболоида (\ref{eoshyp}) инвариантны относительно трехмерной группы
лоренцевых вращений $\MS\MO(1,2)$. Это означает, что преобразования из этой
группы являются изометриями. Группа Лоренца $\MS\MO(1,2)$ трехмерна, и ее
действие в пространстве Минковского $\MR^{1,2}$ задается тремя векторами
Киллинга, которые в декартовых координатах имеют тот же вид, что и в случае
двуполостного гиперболоида (\ref{ekisod}):
\begin{equation}                                                  \label{qkisod}
\begin{split}
  K_0&=-x\pl_y+y\pl_x,
\\
  K_1&=\quad y\pl_t+t\pl_y,
\\
  K_2&=-t\pl_x-x\pl_t.
\end{split}
\end{equation}
Конечно, они удовлетворяют той же алгебре Ли $\Gs\Go(1,2)$ (\ref{ecjrso}),
поскольку пространство Минковского $\MR^{1,2}$ не изменилось. Изменится только
их вид в гиперболических координатах:
\begin{equation}                                                  \label{qkisoh}
\begin{split}
  K_0&=-\pl_\vf,
\\
  K_1&=\quad \sin\vf\pl_\theta+\frac{\sh\theta\cos\vf}{\ch\theta}\pl_\vf,
\\
  K_2&=-\cos\vf\pl_\theta+\frac{\sh\theta\sin\vf}{\ch\theta}\pl_\vf.
\end{split}
\end{equation}
Поскольку максимальное число полей Киллинга для поверхности равно трем, то для
однополостного гиперболоида векторы Киллинга (\ref{qkisoh}) определяют
максимальный набор изометрий.

Как и в предыдущем разделе, векторы Киллинга (\ref{qkisoh}) на однополостном
гиперболоиде линейно зависимы (\ref{qlokij}). Векторы $K_1$ и $K_0+K_2$
образуют подалгебру в $\Gs\Go(1,2)$ и определяют двумерное инволютивное
распределение в пространстве Минковского $\MR^{1,2}$. Согласно теореме
Фробениуса \ref{tfrofi} это распределение определяет семейство интегральных
подмногообразий, которыми являются однополостные гиперболоиды.

Теперь исследуем экстремали (\ref{eextre}) на однополостном гиперболоиде
$\ML^2$.
\begin{theorem}                                                   \label{texthl}
Сечения однополостного гиперболоида $\ML^2$ плоскостями, проходящими через
начало координат, и только они является экстремалями. Все экстремали на
однополостном гиперболоиде полны.
\end{theorem}
\begin{proof}
См., например, в \cite{Katana93A}.
\end{proof}
Обозначим угол между секущей плоскостью, проходящей через начало координат, и
плоскостью $x,y$ через $\psi$. В зависимости от величины этого угла все экстремали
делятся на три класса:
$$
\begin{array}{rl}
  \psi \in [0,\pi/4) &\qquad -\quad \text{пространственноподобные},
\\
  \psi =\pi/4        &\qquad -\quad \text{светоподобные},
\\
  \psi \in (\pi/4,\pi/2] &\qquad -\quad \text{времениподобные}.
\end{array}
$$
Пространственноподобные экстремали замкнуты, а остальные -- нет.

Через любые две различные точки на гиперболоиде $\ML^2$ и начало координат можно
провести одну и только одну плоскость. Это значит, что две произвольные точки
однополостного гиперболоида можно соединить по крайней мере одной экстремалью.
Если экстремали, соединяющие две точки пространственноподобны, то таких
экстремалей две, причем по крайней мере одна из них имеет минимальную длину. В
остальных случаях экстремаль, соединяющая две точки, единственна.

При преобразовании Лоренца плоскость переходит в плоскость. Следовательно,
экстремаль отображается в экстремаль.

Вычислим длину пространственноподобной замкнутой экстремали. Рассмотрим
плоскость
\begin{equation*}
  t=kx,\qquad 0<k<1,
\end{equation*}
которая параллельна оси $y$ и определяет пространственноподобную экстремаль.
В гиперболических координатах это уравнение принимает вид
\begin{equation*}
  \tanh\theta=k\cos\vf.
\end{equation*}
Отсюда следует равенство для дифференциалов
\begin{equation*}
  d\theta=-\frac{k\sin\vf}{1-k^2\cos^2\vf}d\vf.
\end{equation*}
Подставив это выражение в метрику (\ref{qingtf}), получим квадрат элемента длины
пространственноподобной экстремали
\begin{equation*}
  ds^2=-a^2\frac{1-k^2}{(1-k^2\cos^2\theta)^2}d\vf^2.
\end{equation*}
Чтобы вычислить длину экстремали, надо извлечь квадрат и проинтегрировать:
\begin{equation*}
  L:=a\sqrt{1-k^2}\int_0^{2\pi}\frac{d\vf}{1-k^2\cos^2\vf}=2\pi a.
\end{equation*}
Поскольку задача симметрична относительно вращений вокруг оси $t$, то отсюда
следует, что длина произвольной пространственноподобной экстремали не зависит от
угла, под которым плоскость пересекает двуполостный гиперболоид, и равна длине
окружности, по которой гиперболоид пересекает плоскость $x,y$.

Теперь обсудим {\em стереографические координаты}, для однополостного
гиперболоида, в которых она является конформно плоской. Их введение в данном
случае не так наглядно, как для сферы, и сложнее. Определим стереографические
координаты $\tau,\s$ для гиперболоида, как координаты точки на плоскости $t,x$,
которая является центральной проекцией точки гиперболоида из точки $t=x=0$,
$y=-a$.
\index{Стереографические координаты (stereographic coordinates)}%
\index{Координаты стереографические (stereographic coordinates)}%
\begin{figure}[h,b,t]
\hfill\includegraphics[width=.8\textwidth]{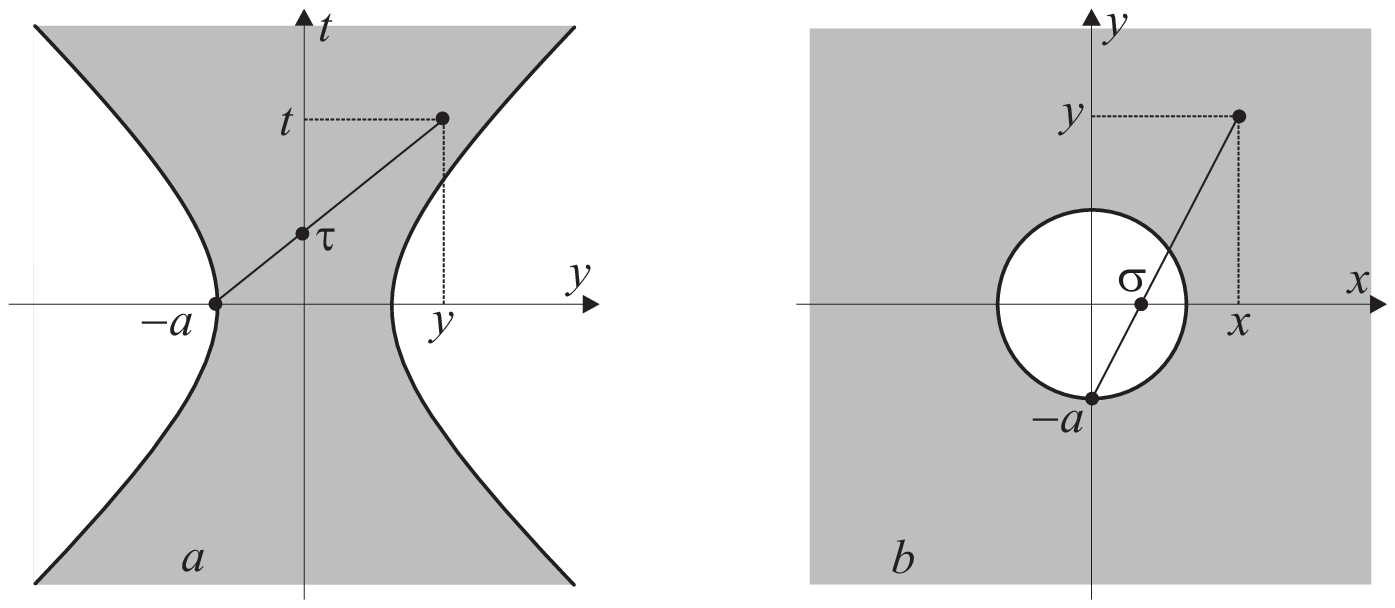}
\hfill {}
\centering\caption{Стереографические координаты $\tau,\s$ на однополостном
гиперболоиде. Проекции на плоскости $t,y$ $(a)$ и $x,y$ $(b)$.}
\label{fprojonhyper}
\end{figure}

Чтобы явно построить функции перехода требуется некоторое геометрическое
построение. Пусть точка на гиперболоиде имеет координаты $t,x,y$.
По-определению, они удовлетворяют равенству (\ref{eoshyp}), т.е.\ в качестве
координат на гиперболоиде можно выбрать, например, декартовы координаты $t,x$,
которые покрывают половину гиперболоида. Соединим эту точку с точкой $t=x=0$,
$y=-a$ прямой. Соответствующие проекции на плоскости $t,y$ и $x,y$ показаны на
рис.\ref{fprojonhyper} $a$ и $b$. Из рисунков следуют равенства:
\begin{equation}                                                  \label{qfrets}
  \frac t\tau=\frac{y+a}a,\qquad\frac x\s=\frac{y+a}a.
\end{equation}
Из этих уравнений выразим $t$ и $x$ и подставим в уравнение гиперболоида
(\ref{eoshyp}). В результате возникнет квадратное уравнение на $y$, которое
можно решить:
\begin{equation*}
  y=a\frac{a^2+\tau^2-\s^2}{a^2-\tau^2+\s^2},
\end{equation*}
где мы выбрали больший корень. Тогда
\begin{equation*}
  \frac{y+a}a=\frac{2a^2}{a^2-\tau^2+\s^2}.
\end{equation*}
Теперь уравнения (\ref{qfrets}) дают функции перехода:
\begin{equation}                                                  \label{ehystf}
  t=\ta\frac{2a^2}{a^2-\ta^2+\s^2},\qquad x=\s\frac{2a^2}{a^2-\ta^2+\s^2}.
\end{equation}

Якобиан преобразования координат $\tau,\s\mapsto t,x$ легко вычислить
$$
  J=\frac{\pl(t,x)}{\pl(\ta,\s)}
  =4a^4\frac{a^2+\tau^2-\s^2}{(a^2-\tau^2+\s^2)^3}.
$$
Он равен нулю и обращается в бесконечность, соответственно, на гиперболах
$\tau^2-\s^2=-a^2$ и $\tau^2-\s^2=a^2$ и имеет разный знак по разные стороны
гипербол. Таким образом, стереографические координаты пробегают всю плоскость
$-\infty<\ta,\s<\infty$ за исключением четырех ветвей гипербол.

На гиперболах $\tau^2-\s^2=-a^2$ из формул преобразования координат
(\ref{ehystf}) следуют равенства $t=\tau$ и $x=\s$, и, следовательно,
$t^2-x^2=-a^2$. То есть гиперболы $\tau^2-\s^2=-a^2$ соответствует сечениям
гиперболоида плоскостью $y=0$.

\begin{figure}[b,t]
\hfill\includegraphics[width=.8\textwidth]{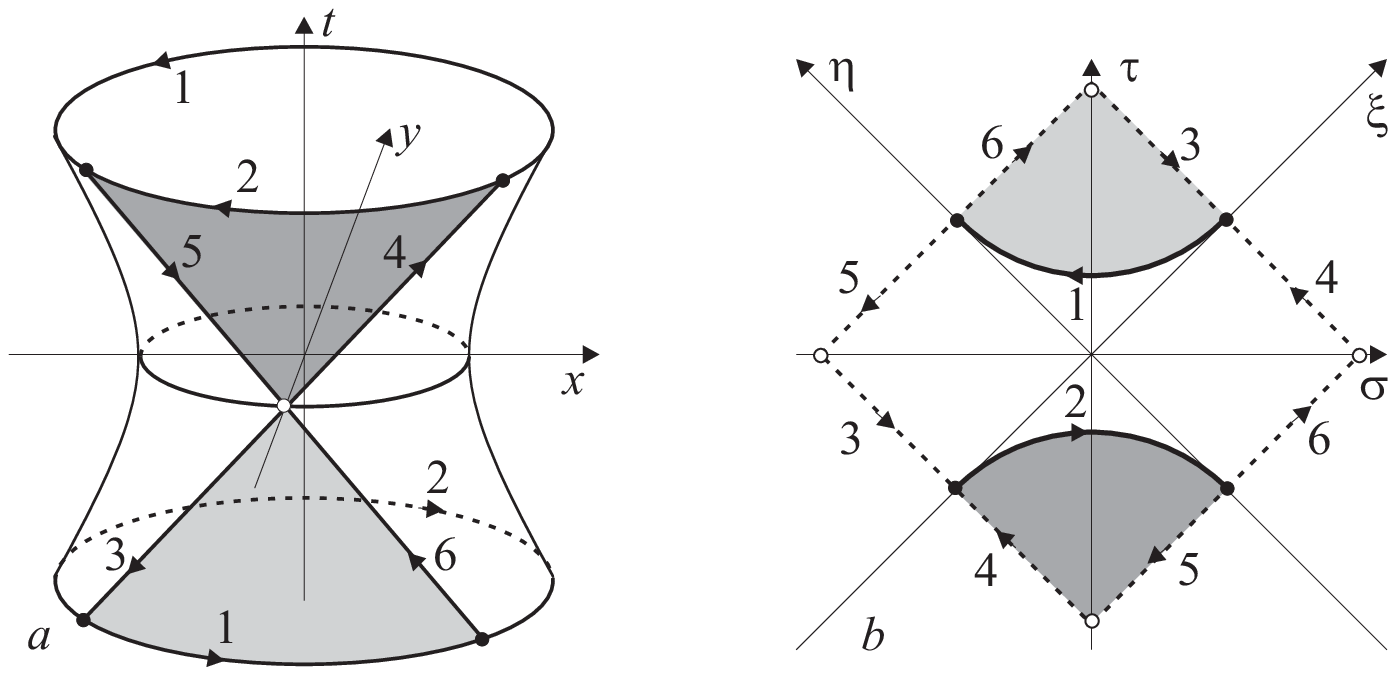}
\hfill {}
\centering\caption{Стереографические координаты на однополостном гиперболоиде.
Центральная проекция гиперболоида с координатами $t,x$ $(a)$. Стереографические
координаты $\tau,\s$ $(b)$.}
\label{fonhyf}
\end{figure}

Отображение однополостного гиперболоида на плоскость $\tau,\s$ показано на
рис.\ref{fonhyf}. Для наглядности изображен  гиперболоид конечной высоты. Мы
предполагаем, что его верхняя и нижняя граничные окружности бесконечно удалены.
Также для наглядности мы сжали плоскость $\tau,\s$ вдоль светоподобных
направлений
\begin{equation}                                                  \label{qlikoj}
  \xi=\tau+\s,\qquad\eta=\tau-\s,
\end{equation}
и изобразили ее в виде квадрата. Этого можно добиться с помощью невырожденной
замены координат, например,
\begin{equation*}
  \xi\mapsto\arctg\xi,\qquad\eta\mapsto\arctg\eta.
\end{equation*}
Мы также приняли следующие обозначения. Закрашенные черным кружки обозначают
точки, находящиеся на бесконечном расстоянии, а полые -- на конечном. Точке
$t=x=0$, $y=-a$, из которой ведется проектирование, соответствуют четыре
незакрашенные вершины квадрата. Кривые между выделенными точками на гиперболоиде
и их образы обозначены одинаковыми цифрами. Стрелки показывают выбранную
ориентацию кривых.

При отображении на плоскость $\tau,\s$ весь гиперболоид разбивается на три
области, ограниченные кривыми (1,5,3,2,6,4), (2,5,4) и (1,6,3). То есть
гиперболоид разрезается вдоль двух прямых $t=\pm x$, $y=-a$. Центральной точке
$t=x=0$, $y=-a$ при стереографической проекции соответствуют четыре точки --
вершины квадрата -- и их необходимо отождествить. Проще всего проследить
отображение областей, рассмотрев отображение их границ. Область, ограниченная
кривыми (1,5,3,2,6,4) отображается на среднюю часть плоскости $\tau,\s$ между
гиперболами $\tau^2-\s^2=a^2$. Области, ограниченные кривыми (2,5,4) и (1,6,3),
отображаются, соответственно, в области на плоскости $\tau,\s$, расположенные
ниже гиперболы $\tau=-\sqrt{a^2+\s^2}$ и выше гиперболы $\tau=\sqrt{a^2+\s^2}$.

Каждая точка гиперболоида из областей (1,5,3,2,6,4), (2,5,4) и (1,6,3) имеет
единственные стереографические координаты. При этом две светоподобные прямые
$t^2=x^2,~y=-a$ (линии 3,4 и 5,6 на рис.~\ref{fonhyf}) отображаются в
бесконечность на плоскости $\ta,\s$. Точки прямых $t^2=x^2,~y=-a$ лежат на
конечном расстоянии, и поэтому стороны квадрата на плоскости $\tau,\s$
нарисованы пунктиром. Напротив, бесконечно удаленная граница гиперболоида
отображается на гиперболы $\ta^2-\s^2=a^2$, и поэтому нарисованы жирной сплошной
линией.

Противоположные стороны квадрата на $\tau,\s$ плоскости лежат на конечном
расстоянии и отображаются в одну прямую на гиперболоиде. Это означает, что их
необходимо отождествить (за исключением бесконечно удаленных точек
$\xi=0$, $\eta=\pm\infty$ и $\xi=\pm\infty$, $\eta=0$).

Чтобы получить метрику однополостного гиперболоида в стереографических
координатах, сначала запишем ее в координатах $t,x$. Это легко сделать путем
исключения координаты $y$ из метрики Лоренца (\ref{qlotyr}) с помощью уравнения
(\ref{eoshyp}). В результате метрика примет вид
\begin{equation}                                                  \label{ehymes}
  ds^2=\frac{(a^2-x^2)dt^2+2txdtdx-(a^2+t^2)dx^2}{a^2+t^2-x^2}.
\end{equation}
Подставив сюда выражения для дифференциалов из формул преобразования координат
(\ref{ehystf}), мы получим, что в стереографических координатах метрика
однополостного гиперболоида является вейлевски лоренцевой:
\begin{equation}                                                  \label{ehystc}
  ds^2=\frac4{\left(1-\frac{\tau^2-\s^2}{a^2}\right)^2}(d\tau^2-d\s^2).
\end{equation}
В светоподобных координатах (\ref{qlikoj}) метрика примет вид
\begin{equation}                                                  \label{ehysts}
  ds^2=\frac4{\big(1-\xi\eta/a^2\big)^2}d\xi d\eta.
\end{equation}

Произведем еще одну замену координат так, чтобы интервал зависел только от
времениподобной координаты. В этой системе координат двумерное
пространство-время является однородным и анализ экстремалей значительно
упрощается. Сначала с помощью дробно линейного преобразования преобразуем
светоподобные координаты $\xi,\eta\mapsto u,v$:
\begin{equation}                                                  \label{qtfetr}
  \xi:=a\frac{u+a}{u-a},\qquad \eta:=a\frac{v+a}{v-a}.
\end{equation}
\begin{figure}[b,t]
\hfill\includegraphics[width=.35\textwidth]{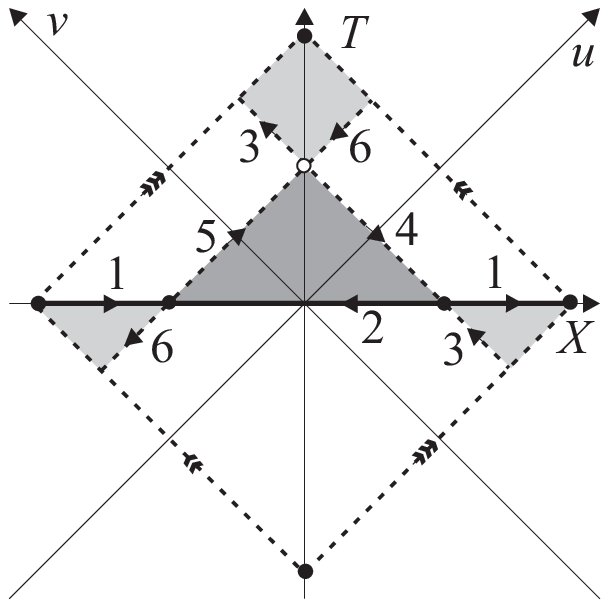}
\hfill {}
\centering\caption{Координаты на однополостном гиперболоиде, в которых
          метрика зависит только от времени.}
\label{fcocoo}
\end{figure}

На рис.~\ref{fcocoo} показан однополостный гиперболоид в новых координатах.
Цифрами обозначены образы соответствующих кривых на рис.~\ref{fonhyp}. Ось
абсцисс нарисована жирной линией, что означает ее полноту (бесконечную
удаленность). Верхняя и нижняя вершины квадрата также лежат на бесконечном
расстоянии и показаны закрашенными кружками. Противоположные стороны квадрата
(за исключением вершин) лежат на конечном расстоянии и показаны пунктиром. Их
необходимо отождествить, т.к.\ они соответствуют одним и тем же точкам
гиперболоида. Все четыре вершины квадрата также отождествляются. Центральная
точка стереографической проекции $t=x=0$, $y=-a$ (четыре вершины квадрата на
рис.~\ref{fonhyp},$b$) отображается в точку $u=v=a$. Следуя стрелкам, легко
проследить отображение областей (1,5,3,2,6,4), (2,5,4) и (1,6,3).

Из формул преобразования координат (\ref{qtfetr}) вытекают формулы
преобразования дифференциалов:
\begin{align*}
  d\xi&=-2a^2\frac{du}{(u-a)^2},
\\
  d\eta&=-2a^2\frac{dv}{(v-a)^2}.
\end{align*}
Якобиан преобразования координат сингулярен на линиях $u=a$ и $v=a$:
$$
  \frac{\pl(\xi,\eta)}{\pl(u,v)}=\frac{4a^2}{(u-a)^2(v-a)^2}.
$$
В новых светоподобных координатах интервал принимает вид
\begin{equation}                                                  \label{eliele}
  ds^2=\frac{4a^2}{(u+v)^2}dudv.
\end{equation}
Теперь снова введем временн\'ую $T$ и пространственную $X$ координаты с помощью
соотношений:
\begin{equation*}
  u:=T+X,\qquad v:=T-X.
\end{equation*}
Тогда метрика однополостного гиперболоида примет вид
\begin{equation}                                                  \label{qmetrt}
  ds^2=\frac{a^2}{T^2}(dT^2-dX^2).
\end{equation}
Она является вейлевски лоренцевой, а соответствующее пространство -- однородным.

При решении уравнений движения в различных моделях гравитации мы получаем для
метрики некоторое выражение, например, формулу (\ref{ehystc}) или
(\ref{qmetrt}). То есть мы имеем плоскость $\tau,\s$ или $T,X$ с заданной на ней
метрикой. Восстановить по этой метрике глобальную структуру пространства-времени
(однополостный гиперболоид) довольно сложно, но можно. В разделе \ref{sglosu}
будет описан общий подход к этой проблеме, основанный на анализе экстремалей.
Если окажется, что экстремали для найденного решения уравнений движения неполны,
то решение необходимо продолжить и восстановить глобальную структуру. Это можно
сделать с помощью метода конформных блоков. Процедура восстановления поверхности
по виду метрики, допускающей один вектор Киллинга, конструктивна, и в результате
возникает диаграмма Картера--Пенроуза, представляющая глобальную структуру
пространства-времени.
\section{Уравнение Лиувилля                                   \label{slioeq}}
Рассмотрим евклидову плоскость $\MR^2$ с декартовыми координатами $x,y$ или, что
эквивалентно, комплексную плоскость $\MC$ с координатами $z:=x+iy$ и
$\bar z:=x-iy$. Пусть задана вещественнозначная достаточно гладкая функция
$\phi(z,\bar z)$. Тогда {\em уравнением Лиувилля} для функции
$\phi$ называется нелинейное уравнение второго порядка в частных производных
\index{Уравнение Лиувилля (Liouville equation)}%
\index{Лиувилля уравнение (Liouville equation)}
\begin{equation}                                                  \label{qliouh}
  \triangle\phi+K\ex^{2\phi}=0,\qquad K=\const,
\end{equation}
где $\triangle:=4\pl^2_{z\bar z}$ -- оператор Лапласа.

Уравнение Лиувилля часто встречается в моделях математической физики и является
одним из немногих нелинейных явно интегрируемых уравнений.
\begin{theorem}
Общее решение уравнения Лиувилля в односвязной области имеет вид
\begin{equation}                                                  \label{qliojs}
  \ex^{2\phi}=\frac{4F'\bar F'}{(aF\bar F+bF+\bar b\bar F+d)^2},
\end{equation}
где $F=F(z)$ -- произвольная голоморфная функция, введены обозначения:
\begin{equation*}
  F':=\pl_z F,\quad \bar F:=\bar F(\bar z),\quad \bar F':=\pl_{\bar z}\bar F,
\end{equation*}
и произвольные постоянные $a,d\in\MR$ и $b\in\MC$ удовлетворяют одному условию
\begin{equation}                                                  \label{qposcg}
  ad-b\bar b=K.
\end{equation}
При этом, если $K=0$, то по крайней мере одна из постоянных $a,b,$ или $d$
должна быть отлична от нуля. Мы также предполагаем, что $F'\ne0$.
\end{theorem}
\begin{proof}
См.\ \cite{Liouvi53}.
\end{proof}

Таким образом, произвольное решение уравнения Лиувилля параметризуется одной
голоморфной функцией $F(z)$, двумя вещественными $(a,d)$ и одной комплексной
$(b)$ постоянными, на которые наложено одно условие (\ref{qliojs}).

Если задана поверхность $\MM$ с римановой метрикой $g\in\CC^3(\MM)$, то согласно
теореме \ref{qisoco} локально существует изотермическая система координат, в
которой метрика является вейлевски евклидовой:
\begin{equation*}
  ds^2=\ex^{2\phi}dz d\bar z.
\end{equation*}
Если поверхность имеет постоянную кривизну, то ее гауссова кривизна также
постоянна, $K=\const$. Это условие в изотермических координатах сводится к
уравнению Лиувилля (\ref{qliouh}) (см.\ раздел \ref{scongl}). Отсюда вытекает
следующая
\begin{theorem}
Любая поверхность с римановой метрикой $g\in\CC^3(\MM)$ постоянной кривизны
локально изометрична сфере $\MS^2$ $(K>0)$, евклидовой плоскости $\MR^2$ $(K=0)$
или двуполостному гиперболоиду $\MH^2$ $(K<0)$.
\end{theorem}
\begin{proof}
При умножением метрики на положительную постоянную $k$,
$g_{\al\bt}\mapsto kg_{\al\bt}$, гауссова кривизна делится на ту же постоянную,
$K\mapsto K/k$. Поэтому достаточно рассмотреть три случая $K=1,0,-1$.
Выберем произвольную точку на поверхности. Тогда в некоторой окрестности этой
точки существуют изотермические координаты. Следовательно, для конформного
множителя возникнет уравнение Лиувилля. Рассмотрим три случая.

$K=1$. Положим в общем решении уравнения Лиувилля
\begin{equation*}
  d=1,\quad a=1,\quad b=0,\quad F=z.
\end{equation*}
Тогда решение (\ref{qliojs}) примет вид
\begin{equation*}
  ds^2=\frac4{(1+z\bar z)^2}dzd\bar z.
\end{equation*}
Это -- метрика сферы единичного радиуса (\ref{espstc}).

$K=0$. Положим
\begin{equation*}
  d=1,\quad a=0,\quad b=0,\quad F=z/2.
\end{equation*}
Тогда получим евклидову метрику $ds^2=dzd\bar z$.

$K=-1$. Положим
\begin{equation*}
  d=1,\quad a=-1,\quad b=0,\quad F=z.
\end{equation*}
Тогда метрика примет вид
\begin{equation*}
  ds^2=\frac4{(1-z\bar z)^2}dzd\bar z.
\end{equation*}
Это -- метрика (\ref{qtrfca}) двуполостного гиперболоида единичного ``радиуса''
в комплексных координатах.
\end{proof}
\begin{cor}
Риманова метрика поверхности постоянной кривизны бесконечно дифференцируема.
\qed\end{cor}
\begin{proof}
Поскольку функция $F(z)$ голоморфна, то функция (\ref{qliojs}) бесконечно
дифференцируема.
\end{proof}
Доказанная теорема локальна. Согласно теореме \ref{texunc} каждая поверхность
имеет единственную с точностью до диффеоморфизма универсальную накрывающую
поверхность. Поскольку поверхности $\MS^2$, $\MR^2$ и $\MH^2$ связны и
односвязны, то они являются универсальными накрывающими поверхностями для всех
поверхностей, соответственно, положительной, нулевой и отрицательной кривизны.
Кроме того, они полны, т.е.\ любую экстремаль можно продолжить до бесконечного
значения канонического параметра в обе стороны. Чтобы найти все полные
поверхности постоянной кривизны, для поверхностей $\MS^2$, $\MR^2$ и $\MH^2$
необходимо найти все группы преобразований $\MG$, действующие собственно
разрывно и свободно. Тогда все полные поверхности постоянной кривизны будут
являться фактор пространствами $\widetilde\MM/\MG$, где
$\widetilde\MM=\MS^2,\MR^2,\MH^2$ (см.\ главу \ref{scover}).

Теперь рассмотрим уравнение Лиувилля на плоскости Минковского $\MR^{1,1}$.
Обозначим декартовы координаты через $t,x$. Введем также конусные координаты
\begin{equation*}
  u:=t+x,\qquad v:=t-x.
\end{equation*}
Тогда аналогом уравнения Лиувилля (\ref{qliouh}) будет являться нелинейное
уравнение в частных производных
\begin{equation}                                                  \label{elioea}
  \square\phi+Ke^{2\phi}=0,\qquad K=\const.
\end{equation}
где $\square:=4\pl^2_{uv}$ -- оператор Даламбера на плоскости Минковского
(\ref{edalot}). Это уравнение также будем называть {\em уравнением Лиувилля}.

Уравнение Лиувилля на плоскости Минковского также явно интегрируется. Его общее
решение имеет вид
\begin{equation}                                                  \label{egesle}
  e^{2\phi}=\frac{4F'G'}{(aFG+bF+cG+d)^2},
\end{equation}
где $F=F(u)$ и $G=G(v)$ -- произвольные дифференцируемые монотонные функции от
светоподобных координат $u$ и $v$, соответственно, $F'G'\ne0$. Четыре
вещественные постоянные $a$, $b$, $c$ и $d$, должны удовлетворять одному условию
\begin{equation}                                                  \label{econsc}
  ad-bc=K.
\end{equation}
Если $K=0$, то хотя бы одна из постоянных $a,b,c$ или $d$ должна быть отлична от
нуля.

Рассмотрим псевдориманову поверхность. Пусть метрика дважды непрерывно
дифференцируема. Тогда согласно теореме \ref{tlokju} в некоторой окрестности
произвольной точки поверхности существует система координат, в которой метрика
является вейлевски лоренцевой (\ref{emetcc}). Если поверхность имеет постоянную
кривизну, то для конформного множителя возникает уравнение Лиувилля
(\ref{elioeq}).
\begin{theorem}
Любая псевдориманова поверхность с метрикой $g\in\CC^2(\MM)$ постоянной кривизны
локально изометрична однополостному гиперболоиду $\ML^2$ $(K\ne0)$ или плоскости
Минковского $\MR^{1,1}$ $(K=0)$.
\end{theorem}
\begin{proof}
Для лоренцевых поверхностей изменение знака метрики меняет знак гауссовой
кривизны. С другой стороны, это эквивалентно перестановке местами времени и
пространства. Поэтому можно не различать псевдоримановы поверхности
положительной и отрицательной кривизны. Так же, как и для римановых поверхностей
метрику можно умножить на положительную постоянную. Следовательно, достаточно
рассмотреть два случая $K=1$ и $K=0$.

$K=1$. Положим в решении уравнения Лиувилля
\begin{equation*}
  d=1,\quad a=1,\quad b=c=0,\quad F=u,\quad G=v.
\end{equation*}
Тогда метрика примет вид
\begin{equation*}
  ds^2=\frac4{(1+uv)^2}dudv.
\end{equation*}
Это -- метрика однополостного гиперболоида (\ref{ehysts}) единичного ``радиуса''
(с точностью до замены общего знака метрики).

$K=0$. Положим
\begin{equation*}
  d=1,\quad a=0,\quad b=c=0,\quad F=u/2,\quad G=v/2.
\end{equation*}
Тогда получим метрику Лоренца $ds^2=dudv$.
\end{proof}
Доказанная теорема локальна. Как и в случае римановых поверхностей любая полная
псевдориманова поверхность является фактор пространством $\widetilde\MM/\MG$,
где $\widetilde\MM=\ML^2,\MR^{1,1}$ для поверхностей, соответственно, ненулевой
и нулевой гауссовой кривизны, а $\MG$ -- группа преобразований, действующая
собственно разрывно и свободно.

Произвольные функции $F(u)$ и $G(v)$, входящие в решение (\ref{egesle})
уравнения Лиувилля, соответствуют конформным преобразованиям светоподобных
координат (\ref{econtl}), в которых метрика является вейлевски лоренцевой.
Поскольку $F'\ne0$ и $G'\ne0$, то всегда можно совершить конформное
преобразование координат $u,v\mapsto F,G$, в которых метрика примет вид
\begin{equation}                                                  \label{elimet}
  ds^2=e^{2\phi}dudv=\frac{4dFdG}{(aFG+bF+cG+d)^2}.
\end{equation}

Область определения решения уравнения Лиувилля (\ref{egesle}) зависит от выбора
функций $F$ и $G$, а также от значения постоянных $a$, $b$, $c$ и $d$.

При $K>0$, положим в решении (\ref{egesle})
\begin{equation*}
  F=u,\quad G=v,\quad a=d=0,\quad b=-c=\sqrt{K}.
\end{equation*}
Тогда метрика примет вид
\begin{equation}                                                  \label{emccpo}
  ds^2=\frac{4dudv}{K(u-v)^2}=\frac1{K}\frac{dt^2-dx^2}{x^2}.
\end{equation}
Эта метрика определена либо в правой, $x>0$, либо в левой, $x<0$, полуплоскости.

Аналогично, при $K<0$ положим
\begin{equation*}
  F=u,\quad G=v,\quad a=d=0,\quad b=c=\sqrt{|K|}.
\end{equation*}
Тогда метрика примет вид
\begin{equation}                                                  \label{eccpxf}
  ds^2=\frac{4dudv}{|K|(u+v)^2}=\frac1{|K|}\frac{dt^2-dx^2}{t^2}.
\end{equation}
В таком виде метрика определена в верхней, $t>0$, или нижней, $t<0$,
полуплоскости. Она уже была получена ранее (\ref{qmetrt}).

Два последних выражения для метрики связаны между собой перестановкой координат
$t\leftrightarrow x$ и изменением знака метрики
$g_{\al\bt}\rightarrow-g_{\al\bt}$. Это означает, что поверхности постоянной
положительной и отрицательной кривизны, как уже отмечалось, в случае
псевдоримановой метрики эквивалентны.
\begin{com}
Модели математической физики, как правило, помимо метрики включают также другие
поля. Поэтому может случиться так, что суммарное действие для всех полей не
будет инвариантно относительно изменения знака метрики. В этом случае знак
гауссовой кривизны является существенным.
\qed\end{com}

При любом значении гауссовой кривизны $K$ мы можем положить
\begin{equation*}
  F=\frac u2,\quad G=\frac v2,\quad  a=K,\quad d=1,\quad b=c=0.
\end{equation*}
Тогда метрика примет вид
$$
  ds^2=\frac{dudv}{(1+\frac K4uv)^2}.
$$
Эта метрика определена на всей плоскости за исключением двух ветвей гиперболы
$uv=-4/K$.

Положим $F=u$ и $G=v$ в общем решении уравнения Лиувилля (\ref{egesle}). Тогда
метрика примет вид
\begin{equation}                                                  \label{qgftre}
  ds^2=\frac{4dudv}{(auv+bu+cv+d)^2}.
\end{equation}
\begin{prop}
При дробно линейном преобразовании светоподобных координат
\begin{equation}                                                  \label{qfrads}
  u\mapsto u':=\frac{\al_1u+\bt_1}{\g_1u+\dl_1},\qquad
  v\mapsto v':=\frac{\al_2v+\bt_2}{\g_2v+\dl_2},
\end{equation}
где $\al_1,\dots,\dl_1,\al_2,\dots,\dl_2$ -- некоторые постоянные, такие что
$\det_1:=\al_1\dl_1-\bt_1\g_1\ne0$ и $\det_2:=\al_2\dl_2-\bt_2\g_2\ne0$, метрика
(\ref{qgftre}) принимает вид
\begin{equation}                                                  \label{qnerda}
  ds^2=\frac{4\det_1\det_2\,\, du'dv'}{(a'u'v'+b'u'+c'v'+d')^2},
\end{equation}
где
\begin{align*}
  a'&=a\al_1\al_2+b\al_1\g_2+c\g_1\al_2+d\g_1\g_2,
\\
  b'&=a\al_1\bt_2+b\al_1\dl_2+c\g_1\bt_2+d\g_1\dl_2,
\\
  c'&=a\bt_1\al_2+b\bt_1\g_2+c\dl_1\al_2+d\dl_1\g_2,
\\
  d'&=a\bt_1\bt_2+b\bt_1\dl_2+c\dl_1\bt_2+d\dl_1\dl_2.
\end{align*}
\end{prop}
\begin{proof}
Прямые вычисления.
\end{proof}
Дробно линейные преобразования (\ref{qfrads}) образуют подгруппу группы
конформных преобразований координат (\ref{econtl}). Если $\det_1=\det_2=1$, то
соответствующие дробно линейные преобразования образуют подгруппу (модулярная
группа), которая сохраняет вид метрики (\ref{qgftre}). Эти преобразования можно
использовать для фиксирования некоторых постоянных из $a$, $b$, $c$ и $d$.
\chapter{Лоренцевы поверхности с одним вектором Киллинга         \label{sglosu}}
(Псевдо-)римановы поверхности с одним вектором Киллинга часто встречаются при
построении глобальных решений в различных моделях гравитации. Дадим
\begin{defn}
Пара $(\MM,g)$, где $\MM$ -- многообразие и $g$ -- заданная на нем метрика,
называется {\em глобальным решением} в данной модели гравитации, если метрика
удовлетворяет соответствующим уравнениям движения на $\MM$, а само многообразие
$\MM$ является максимально продолженным.
\qed\end{defn}
Напомним, что максимальное продолжение означает, что любую экстремаль можно либо
продолжить в обе стороны до бесконечного значения канонического параметра, либо
при конечном значении канонического параметра она попадает в сингулярную точку,
где один из геометрических инвариантов становится сингулярным. Это определение
инвариантно и не зависит от выбора системы отсчета, т.к.\ канонический параметр
инвариантен и определен с точностью до линейного преобразования.
\begin{exa}
Нетривиальным глобальным решением в общей теории относительности является
расширение Крускала--Секереша \cite{Kruska60,Szeker60} решения Шварцшильда
\cite{Schwar16R} (независимо найденного также Дросте
\cite{Droste15,Droste16,Droste17}). В этом
случае четырехмерное пространство-время представляет собой топологическое
произведение сферы $\MS^2$ на псевдориманову поверхность с одним вектором
Киллинга, которая обычно изображается в виде диаграммы Картера--Пенроуза и
будет описана в настоящем разделе.
\qed\end{exa}
\index{Глобальное решение (global solution)}%
\index{Решение глобальное (global solution)}%

В настоящей главе развит конструктивный метод конформных блоков построения
глобальных решений $(\MM,g)$ для двумерных метрик лоренцевой и евклидовой
сигнатуры специального вида, обладающих одним вектором Киллинга.

Необходимость построения глобальных решений, т.е.\ одновременного построения и
самих многообразий и заданных на них метрик, связана с инвариантностью моделей
гравитации относительно общих преобразований координат. Действительно, для
решения уравнений движения, как правило, фиксируется некоторая система
координат, что уменьшает число искомых функций. Но, с другой стороны,
фиксирование системы отсчета означает, что, за исключением тривиальных случаев,
найденное решение является только локальным и представляет собой, как правило,
только часть некоторого б\'ольшего пространства-времени. Отметим также, что
только глобальная структура многообразия позволяет дать физическую интерпретацию
решений уравнений движения для метрики.

Построение глобальных решений в гравитации является трудной задачей, поскольку,
помимо решения уравнений движения, что само по себе сложно, предполагает решение
уравнений для экстремалей, анализ их полноты и продолжения многообразия, если
оно оказалось неполным. В общей теории относительности из-за сложности уравнений
движения известно лишь небольшое число глобальных решений. Некоторые из них
будут описаны в главе \ref{swarso}). В двумерных моделях гравитации ситуация
проще и удалось построить все глобальные решения \cite{Katana93A} в двумерной
гравитации с кручением \cite{KatVol86}, а также в широком классе моделей
дилатонной гравитации \cite{KloStr96C,KaKuLi96,KaKuLi97}. Конструктивный метод
конформных блоков построения глобальных решений для псевдоримановых поверхностей
был предложен в \cite{Katana93A} для двумерной гравитации с кручением в
конформной калибровке. Для поверхностей с римановой метрикой метод конформных
блоков был развит в статье \cite{Katana09}.

\begin{com}
В настоящей главе мы рассматриваем продолжение решений только вдоль экстремалей.
В аффинной геометрии с кручением и неметричностью геодезические и экстремали в
общем случае различны. Поэтому имеет смысл рассматривать максимальное
продолжение многообразий как вдоль экстремалей, так и вдоль геодезических. Если
геодезические и экстремали совпадают как, например, в общей теории
относительности, то продолжение решения вдоль экстремалей автоматически влечет
за собой продолжение вдоль геодезических.
\qed\end{com}
\section{Локальный вид лоренцевой метрики                        \label{sglocm}}
Рассмотрим плоскость $\MR^2$ с декартовыми координатами
$\lbrace\z^\al\rbrace=(\tau,\s)$, $\al=0,1$. Пусть в некоторой области на
плоскости задана метрика лоренцевой сигнатуры, которая имеет вейлевски плоский
вид
\begin{equation}                                                  \label{emetok}
  ds^2=|\Phi(q)|(d\tau^2-d\s^2).
\end{equation}
Будем считать, что конформный множитель $\Phi(q)\in {\cal C}^l$, $l\ge2$,
является $l$ раз непрерывно дифференцируемой функцией одного аргумента $q\in\MR$
за исключением конечного числа степенных особенностей. Этого достаточно для
того, чтобы компоненты метрики (\ref{emetok}) могли удовлетворять некоторой
системе уравнений второго порядка вне особых точек. Пусть аргумент $q$ зависит
только от одной из координат, $\tau$ или $\s$, и эта связь задается обыкновенным
дифференциальным уравнением
\begin{equation}                                                  \label{eshiff}
  \left|\frac{dq}{d\z}\right|=\pm \Phi(q),
\end{equation}
где выполняется следующее правило знаков:
\begin{equation}                                                  \label{esignr}
\begin{array}{ccl}
\Phi>0: & \quad \z=\s,   &\quad \text{знак $+$ (статическое локальное решение)},
\\
\Phi<0: & \quad \z=\tau, &\quad \text{знак $-$ (однородное локальное решение)}.
\end{array}
\end{equation}

На первый взгляд двумерная метрика (\ref{emetok}) имеет очень специфический вид.
Однако, это -- метрика достаточно общего вида. В дальнейшем будет показано, что
некоторые модели гравитации при решении полной системы уравнений движения
приводят к двумерной метрике вида (\ref{emetok}). Например, при отыскании
сферически симметричных вакуумных решений уравнений Эйнштейна в четырехмерном
пространстве-времени (см.\ следующую главу), метрика (\ref{emetok}) возникает на
двумерных поверхностях, точки которых параметризуются временем и радиусом.
Метрика вида (\ref{emetok}) возникает также при решении уравнений движения
двумерной гравитации. По этой причине в настоящей главе мы будем, для краткости,
называть метрику вида (\ref{emetok}) просто локальным решением.

Мы допускаем, что конформный множитель метрики (\ref{emetok}) может обращаться в
нуль и иметь особенности в конечном числе точек, которые мы пронумеруем в
порядке возрастания: $q_i$, $i=1,\dots,k$. В эту последовательность включены
также бесконечно удаленные точки $q_1=-\infty$ и $q_k=\infty$. Будем считать,
что вблизи каждой точки $q_i$ конформный множитель ведет себя степенным образом:
\begin{align}                                                     \label{ecfapo}
  |q_i|<\infty:&\qquad \Phi(q)\sim(q-q_i)^m,
\\                                                                \label{ecfasi}
  |q_i|=\infty:&\qquad \Phi(q)\sim q^m.
\end{align}
При конечных значениях $q_i$ показатель степени не равен нулю, $m\ne0$, т.к.\
конформный множитель в этой точке, по-предположению, либо равен нулю, либо
сингулярен.

Строго говоря, показатель $m$ различен для различных точек, и мы должны были бы
писать $m_i$, а не просто $m$. Однако эта упрощенная запись не приводит к
недоразумениям, поскольку при анализе полноты экстремалей будет рассматриваться
одна конкретная точка $q_i$. Показатель степени $m$ может быть любым
вещественным числом, для которого правые части (\ref{ecfapo}), (\ref{ecfasi})
определены.

В промежутках между нулями и особенностями, где функция $\Phi$ или положительна,
или отрицательна, локальное решение, соответственно, статично с вектором
Киллинга $K=\pl_\tau$ или однородно с вектором Киллинга $K=\pl_\s$. Квадрат
длины вектора Киллинга в обоих случаях равен конформному множителю, $K^2=\Phi$.
На горизонтах, которые, как будет показано ниже, соответствуют нулям конформного
множителя, $\Phi(q)=0$, вектор Киллинга является светоподобным. При этом он
отличен от нуля, что будет продемонстрировано в координатах
Эддингтона--Финкельстейна (см.\ раздел \ref{sedfic}).

При конформных преобразованиях вид метрики (\ref{emetok}), (\ref{eshiff})
меняется, т.к.\ переменная $q$ становится зависящей от обеих координат на
плоскости. При растяжке координат $\z\mapsto\tilde\z=k\z$, $k=\const\ne0$,
которые образуют подгруппу группы конформных преобразований, метрика сохраняет
свой вид c $\tilde q=k^{-1}q$, $\tilde \Phi=k^{-2}\Phi$.

По сути дела, формулы (\ref{emetok}), (\ref{eshiff}) задают четыре различные
метрики: из-за наличия знака модуля у производной в (\ref{eshiff}) есть две
области со статической метрикой и две области с однородной метрикой,
отличающиеся знаком производной $dq/d\z$. Будем обозначать эти области римскими
цифрами:
\begin{equation}                                                  \label{emetdo}
\begin{array}{rcl}
  \text{I}:  \quad & \Phi>0,\quad & dq/d\s>0,\\
  \text{II}: \quad & \Phi<0,\quad & dq/d\tau<0,\\
  \text{III}:\quad & \Phi>0,\quad & dq/d\s<0,\\
  \text{IV}: \quad & \Phi<0,\quad & dq/d\tau>0.
\end{array}
\end{equation}
Порядок следования областей будет ясен из дальнейшего построения.

Статическое локальное решение в области {\rm III} получается из локального
решения в области {\rm I} пространственным отражением $\s\rightarrow-\s$, а
однородное локальное решение в области {\rm IV} связано с локальным решением в
области {\rm II} обращением времени $\tau\rightarrow-\tau$. Поскольку изменение
знака конформного множителя у метрики можно компенсировать перестановкой
пространственной и временн\'ой координаты $\tau\leftrightarrow\s$, то однородные
локальные решения в областях {\rm II} и {\rm IV} можно получить из стационарных
локальных решений для конформного множителя $-\Phi$ в областях {\rm III} и
{\rm I} с последующим поворотом плоскости $\tau,\s$ на угол $\pi/2$ по часовой
стрелке.
\begin{exa}
Рассмотрим $t,r$ компоненты метрики Шварцшильда вне горизонта при $r>2M>0$
\begin{equation*}
  ds^2=\left(1-\frac{2M}r\right)dt^2-\frac{dr^2}{1-\displaystyle\frac{2M}r}.
\end{equation*}
Эту двумерную метрику можно записать в вейлевски плоском виде
\begin{equation*}
  ds^2=\left(1-\frac{2M}r\right)(dt^2-d\s^2),
\end{equation*}
преобразовав радиальную координату $r=r(\s)$, где функция $r(\s)$ удовлетворяет
уравнению
\begin{equation*}
  \left|\frac{dr}{d\s}\right|=1-\frac{2M}r.
\end{equation*}
Это -- статические области I,III. Здесь роль параметра $q$ играет радиус $r$.

Если координата $r$ лежит в интервале $0<r<2M$, то ее следует назвать временем,
т.е.\ произвести замену $r\leftrightarrow t$. Следовательно, при $0<t<2M$ (под
горизонтом) метрика Шварцшильда имеет вид
\begin{equation*}
  ds^2=-\frac{dt^2}{1-\displaystyle\frac{2M}t}+\left(1-\frac{2M}t\right)dr^2.
\end{equation*}
Ее также можно записать в вейлевски плоском виде
\begin{equation*}
  ds^2=\left|1-\frac{2M}t\right|(d\tau^2-dr^2),
\end{equation*}
где
\begin{equation*}
  \left|\frac{dt}{d\tau}\right|=-\left(1-\frac{2M}t\right).
\end{equation*}
Таким образом мы получили однородную метрику в областях II, IV. В этом случае
роль параметра $q$ играет параметр $t$.

Как видим, двумерная $t,r$ часть метрики Шварцшильда в точности имеет вид
(\ref{emetok}) с конформным множителем
\begin{equation}                                                  \label{qsgcre}
  \Phi(q)=1-\frac{2M}q,
\end{equation}
где $q=r,t$. Конформный множитель $\Phi(q)$ для метрики Шварцшильда имеет
простой полюс при $q=0$ и простой нуль (горизонт) в точке $q=2M$.
\qed\end{exa}

Для метрики (\ref{emetok}) в областях I и III можно перейти к координатам
$\tau,q$, в которых метрика примет вид
\begin{equation*}
  ds^2=\Phi(q)d\tau^2-\frac{dq^2}{\Phi(q)}.
\end{equation*}
В областях II и IV можно перейти к координатам $q,\s$:
\begin{equation*}
  ds^2=-\frac{dq^2}{\Phi(q)}+\Phi(q)d\s^2.
\end{equation*}
Координаты $\tau,q$ в областях I, III и $q,\s$ в областях II, IV называются
{\em координатами Шварцшильда}. Метрика в этих координатах имеет простой вид,
причем все компоненты метрики заданы явно. Недостатком координат Шварцшильда
является то, что они не различают между собой области I и III, а также II и IV.
Это существенно, потому что для построения глобальных решений необходимо
использовать все области.
\index{Координаты Шварцшильда (Schwarzschild coordinates)}%
\index{Шварцшильда координаты (Schwarzschild coordinates)}%

Перейдем к вычислению геометрических объектов для метрики (\ref{emetok}).
Символы Кристоффеля имеют различный вид в различных областях. Ниже приведены
явные выражения только для ненулевых компонент:
\begin{align}                                                     \label{echrso}
\text{I}: &\quad
  \Gamma_{\tau\tau}{}^\s=\Gamma_{\tau\s}{}^\tau
  =\Gamma_{\s\tau}{}^\tau=\Gamma_{\s\s}{}^\s=\frac12\Phi',\\
\text{II}: &\quad                                                     \label{echrss}
  \Gamma_{\tau\tau}{}^\tau=\Gamma_{\tau\s}{}^\s
  =\Gamma_{\s\tau}{}^\s=\Gamma_{\s\s}{}^\tau=\frac12\Phi',\\
\text{III}: &\quad                                                    \label{echrst}
  \Gamma_{\tau\tau}{}^\s=\Gamma_{\tau\s}{}^\tau
  =\Gamma_{\s\tau}{}^\tau=\Gamma_{\s\s}{}^\s=-\frac12\Phi',\\
\text{IV}: &\quad                                                     \label{echrsf}
  \Gamma_{\tau\tau}{}^\tau=\Gamma_{\tau\s}{}^\s
  =\Gamma_{\s\tau}{}^\s=\Gamma_{\s\s}{}^\tau=-\frac12\Phi',
\end{align}
где штрих обозначает производную по $q$. В статичных областях отличны от нуля
символы Кристоффеля с нечетным числом пространственных индексов, а в однородных
областях -- с нечетным числом временн\'ых индексов. В статичных и однородных
областях символы Кристоффеля отличаются знаком. Это является следствием того,
что однотипные области связаны преобразованием $\z\rightarrow-\z$, а символы
Кристоффеля линейны по производным. У тензора кривизны в каждой области отличны
от нуля только четыре компоненты:
\begin{align}                                                     \label{ecurtc}
  \text{I,~III}: &\quad
  R_{\tau\s\tau}{}^\s=-R_{\s\tau\tau}{}^\s
  =R_{\tau\s\s}{}^\tau=-R_{\s\tau\s}{}^\tau=-\frac12\Phi''\Phi,
\\                                                                \label{ecurtt}
  \text{II,~IV}: &\quad
  R_{\tau\s\tau}{}^\s=-R_{\s\tau\tau}{}^\s
  =R_{\tau\s\s}{}^\tau=-R_{\s\tau\s}{}^\tau=\quad \frac12\Phi''\Phi.
\end{align}
Тензор Риччи диагонален:
\begin{align}                                                     \label{erictz}
  \text{I,~III}: &\quad
  R_{\tau\tau}=-R_{\s\s}=-\frac12\Phi''\Phi,
\\                                                                \label{erictt}
  \text{II,~IV}: &\quad
  R_{\tau\tau}=-R_{\s\s}=\frac12\Phi''\Phi.
\end{align}
При этом во всех четырех областях скалярная кривизна одинакова,
\begin{equation}                                                  \label{esctmk}
  \text{I,~II,~III,~IV}:\quad R=-\Phi''.
\end{equation}
\begin{com}
При заданном конформном множителе $\Phi$ это алгебраическое уравнение связывает
$q$ с $R$. Отсюда вытекает, что функция $q=q(\tau,\s)$, также как и скалярная
кривизна, может рассматриваться, как функция на плоскости. В той области, где
уравнение (\ref{esctmk}) разрешимо относительно $q$ в качестве одной из
координат можно выбрать скалярную кривизну $R$.
\qed\end{com}

Максимальное продолжение многообразия означает, что, если у поверхности
(пространства-времени) с метрикой (\ref{emetok}) есть край, лежащий на конечном
расстоянии, т.е.\ соответствующий конечному значению канонического параметра для
экстремалей, то он может быть только сингулярным. В противном случае
многообразие можно и необходимо продолжить.

Поскольку в двумерном случае все компоненты тензора кривизны определяются
скалярной кривизной, то проанализируем ее поведение подробнее.

Сначала рассмотрим
\begin{exa}
Из уравнения (\ref{esctmk}) вытекает, что для поверхностей постоянной кривизны,
которые рассмотрены в разделе \ref{scocus}, конформный множитель представляет
собой квадратичный полином:
\begin{equation}                                                  \label{ecocsp}
  \Phi(q)=-(aq^2+bq+c),\qquad a,b,c=\const.
\end{equation}
При этом $R=2a$.
\qed\end{exa}
Из уравнения (\ref{esctmk}) следует, что скалярная кривизна сингулярна вблизи
точки $q_i$ при следующих показателях степени асимптотического поведения
(\ref{ecfapo}):
\begin{align}                                                     \label{esfpsc}
  |q_i|<\infty:&\qquad m<0,\quad 0<m<1,\quad 1<m<2,
\\                                                                \label{esfpcc}
  |q_i|=\infty:&\qquad ~m>2.
\end{align}
Отметим, что при конечных значениях $q_i$ и при $m=1$ скалярная кривизна может
иметь отличные от нуля значения за счет поправок второго порядка.
При бесконечных значениях параметра, $q\rightarrow\pm\infty$, скалярная кривизна
стремится к отличной от нуля постоянной при $m=2$ и к нулю при $m<2$. Отметим,
что ненулевое значение кривизны в конечной точке $|q_i|<\infty$ может возникнуть
и при $m=1$ за счет поправок следующего порядка в разложении (\ref{ecfapo}).
\section{Конформные блоки                                        \label{scoblo}}
При построении глобальных решений будет использовано понятие конформного блока,
который ставится в соответствие каждому интервалу $(q_i,q_{i+1})$. Внутри
каждого интервала конформный множитель $\Phi$ либо строго положителен, либо
строго отрицателен.

Сначала найдем область определения соответствующей метрики (\ref{emetok}) на
плоскости $\tau,\s$. Для определенности, рассмотрим статическое локальное
решение {\rm I} на интервале $(q_i,q_{i+1})$. Тогда временн\'ая координата
пробегает всю числовую ось, $\tau\in\MR$. Уравнение (\ref{eshiff}) для
пространственной координаты $\s$ принимает вид
\begin{equation}                                                  \label{estsos}
  \frac{dq}{d\s}=\Phi(q).
\end{equation}
Постоянная интегрирования этого уравнения соответствует сдвигу пространственной
координаты, $\s\mapsto\s+\const$, т.е.\ выбору начала отсчета $\s$. Из
уравнения (\ref{estsos}) следует, что область изменения координаты $\s$ зависит
от сходимости интеграла
\begin{equation}                                                  \label{eforin}
  \s_{i,i+1}\sim\int^{q_i,q_{i+1}}\!\!\!\frac{dq}{\Phi(q)}
\end{equation}
в граничных точках. Интеграл (\ref{eforin}) сходится или расходится в
зависимости от показателя степени $m$:
\begin{equation}                                                  \label{eboucb}
\begin{array}{rl}
  |q_i|<\infty:\qquad &\left\lbrace
  \begin{array}{rll}
  m<1  \quad &\text{-- сходится,}  &\text{\qquad прямая,}\\
  m\ge1\quad &\text{-- расходится,}&\text{\qquad угол,}
  \end{array}\right. \\
  |q_i|=\infty:\qquad &\left\lbrace
  \begin{array}{rll}
  m\le1\quad &\text{-- расходится,}&\text{\qquad угол,}\\
  m>1  \quad &\text{-- сходится,}  &\text{\qquad прямая.}
  \end{array}\right.
\end{array}
\end{equation}
В этой таблице справа указана форма границы соответствующих конформных блоков,
которые будут введены ниже. Если на обоих концах интервала $(q_i,q_{i+1})$
интеграл расходится, то $\s\in(-\infty,\infty)$, и метрика определена на всей
плоскости $\tau,\s$. Если на одном из концов $q_{i+1}$ или $q_i$ интеграл
сходится, то метрика определена на полуплоскости $\s\in(-\infty,\s_{i+1})$ или
$\s\in(\s_i,\infty)$, соответственно. При этом выбор граничных точек $\s_{i+1}$
и $\s_i$ произволен, и, не ограничивая общности, можно положить $\s_{i,i+1}=0$.
Если на обоих концах интервала $(q_i,q_{i+1})$ интеграл сходится, то локальное
решение определено в полосе $\s\in(\s_i,\s_{i+1})$, при этом совместить с
началом координат можно только один из концов интервала.

\begin{figure}[h,b,t]
\hfill\includegraphics[width=.95\textwidth]{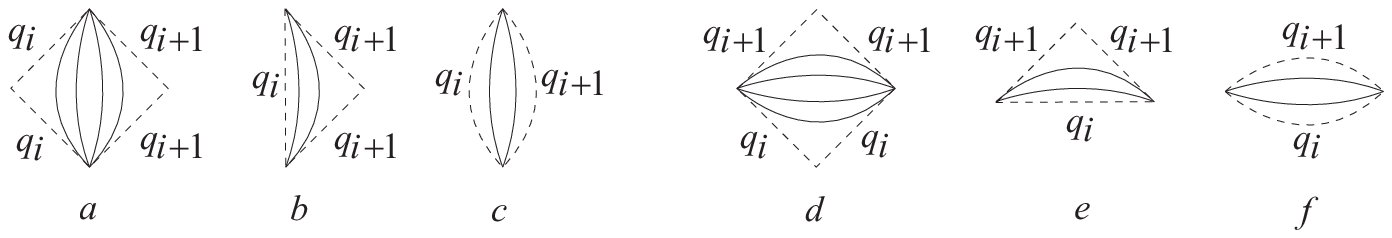}
\hfill {}
\centering\caption{Конформные блоки для статических ({\it a,b,c}) и однородных
({\it d,e,f}) локальных решений. Тонкие сплошные линии обозначают траектории
Киллинга.}
\label{fcoblo}
\end{figure}

Для наглядного изображения максимально продолженных решений введем понятие
конформного блока. С этой целью отобразим плоскость $\tau,\s$ на квадрат вдоль
светоподобных направлений:
\begin{equation}                                                  \label{elicco}
  \x=\tau+\s,\qquad \eta=\tau-\s.
\end{equation}
Для этого произведем конформную замену переменных:
\begin{equation}                                                  \label{etrfun}
  u=u(\x),\qquad v=v(\eta),\qquad u,v\in{\cal C}^{l+1},
\end{equation}
где функции $u$ и $v$ ограничены на всей оси, а их первые производные
положительны. Например, $u=\arctg\xi$, $v=\arctg\eta$. Предположение о классе
гладкости функций перехода сохраняет класс гладкости метрики после
преобразования координат. Тогда статическому локальному решению, определенному
на всей плоскости $\tau,\s$, будет поставлен в соответствие квадратный
{\em конформный блок}, изображенный на рис.~\ref{fcoblo},{\it a}.
\index{Конформный блок (conformal block)}%
\index{Блок конформный (conformal block)}%

В общем случае под конформным блоком понимается конечная часть плоскости $u,v$ и
заданная на ней метрика (\ref{emetok}) в координатах (\ref{elicco}),
(\ref{etrfun}). Если потребовать, чтобы экстремали подходили к границе
конформного блока под всеми возможными углами, то это однозначно фиксирует
асимптотическое поведение функций перехода (\ref{etrfun}) \cite{Katana93A}. В
остальном функции перехода остаются произвольными. Мы не будем останавливаться
на этом вопросе подробно, т.к.\ гладкость метрики на горизонтах будет доказана
путем перехода к координатам Эддингтона--Финкельстейна в разделе \ref{sedfic}.

Значение переменной $q$ и, значит, скалярной кривизны постоянно вдоль
времениподобных траекторий Киллинга, которые внутри блока показаны тонкими
сплошными линиями. Переменная $q$ непрерывно меняется слева направо: монотонно
возрастает в области {\rm I} и монотонно убывает в области {\rm III}. При этом
на двух левых сторонах квадрата она принимает значение $q_i$, а на двух правых
-- значение $q_{i+1}$. Нижняя и верхняя вершины квадрата являются существенно
особыми точками, т.к.\ в этих точках предел $q$ зависит от пути, по которому
произвольно выбранная последовательность подходит к вершине. Будем говорить, что
у статического конформного блока есть две границы: левая и правая, на которых
переменная $q$ принимает значения $q_i$ и $q_{i+1}$, соответственно.

Если решение уравнения (\ref{estsos}) определено на полуинтервале
$(\s_i=0,\s_{i+1}=\infty)$, то статическому локальному решению ставится в
соответствие треугольный конформный блок, пример которого показан на
рис.~\ref{fcoblo},{\it b}. При этом конечной граничной точке $\s_i=0$ ставится в
соответствие вертикальная прямая. Этого всегда можно добиться путем выбора
функций $u(\x)$ и $v(\eta)$.

В случае, когда решение уравнения (\ref{estsos}) определено на конечном
интервале $(\s_i,\s_{i+1})$, конформный блок изображается в виде линзы,
изображенной на рис.~\ref{fcoblo},{\it c}. За счет выбора функций $u(\x)$ и
$v(\eta)$ левую или правую границу этого конформного блока можно сделать
вертикальной, но не обе границы одновременно.

Напомним, что каждому статическому интервалу $(q_i,q_{i+1})$ ставится в
соответствие два конформных блока, т.к.\, благодаря знаку модуля, уравнение
(\ref{eshiff}) инвариантно относительно инверсии пространственной координаты
$\s\leftrightarrow-\s$.

Аналогично каждому однородному локальному решению ставится в соответствие один
из конформных блоков, изображенных на рис.~\ref{fcoblo},{\it d,e,f}, или
перевернутый блок, полученный обращением времени $\tau\leftrightarrow-\tau$. Для
этих конформных блоков переменная $q$ постоянна вдоль пространственноподобных
траекторий Киллинга и монотонно возрастает или убывает от значения $q_i$ на
нижней границе, до значения $q_{i+1}$ на верхней границе. Левая и правая
вершины однородных конформных блоков являются существенно особыми точками.

Конформные блоки для статических и однородных локальных решений будем называть,
соответственно, статическими и однородными.

Итак, мы показали, что лоренцева поверхность, на которой задана конформно
плоская метрика (\ref{emetok}) на каждом интервале $(q_i,q_{i+1})$ диффеоморфна
одному из конформных блоков. Для построения максимально продолженных решений
необходимо найти и проанализировать экстремали на этих поверхностях, при
необходимости продолжить многообразие и метрику, а также доказать
дифференцируемость метрики при склеивании конформных блоков. Использование
конформных блоков для построения глобальных решений удобно и наглядно, т.к.\
все светоподобные экстремали изображаются в виде двух семейств прямых линий,
проходящих под углом $\pm\pi/4$ к оси времени. Поэтому при продолжении
поверхности конформные блоки можно склеивать вдоль ребер, сохраняя гладкость
светоподобных экстремалей.
\section{Экстремали                                              \label{sextsm}}
\subsection{Форма экстремалей                                    \label{sextfo}}
Для того, чтобы понять, как устроено глобальное решение для метрики
(\ref{emetok}), необходимо проанализировать поведение экстремалей
$\lbrace\z^\al(t)\rbrace=\left\lbrace \tau(t),\s(t)\right\rbrace$, $t\in\MR$,
которые подчиняются системе обыкновенных дифференциальных уравнений второго
порядка
\begin{equation}                                                  \label{qexsda}
  \ddot\z^\al=-\Gamma_{\bt\g}{}^\al\dot\z^\bt\dot\z^\g,
\end{equation}
где точка обозначает дифференцирование по каноническому параметру $t$ и
$\Gamma_{\bt\g}{}^\al$ -- символы Кристоффеля.

Исследуем подробно поведение экстремалей для статического локального решения в
области
\begin{equation}                                                  \label{emetfd}
  \text{I}:\quad ds^2=\Phi(q)(d\tau^2-d\s^2),\qquad \frac{dq}{d\s}=\Phi(q)>0.
\end{equation}
Используя выражение для символов Кристоффеля (\ref{echrso}), получаем уравнения
для экстремалей
\begin{align}                                                     \label{eqexts}
  \ddot\tau &=-\Phi'\dot\tau\dot\s,
\\                                                                \label{eqextt}
  \ddot\s   &=-\frac12\Phi'(\dot\tau^2+\dot\s^2).
\end{align}
Эта система уравнений имеет два первых интеграла (см.\ раздел \ref{sgexin}):
\begin{align}                                                     \label{efisco}
  \Phi(\dot\tau^2-\dot\s^2)&=C_0=\const,
\\                                                                \label{efisct}
  \Phi\dot\tau&=C_1=\const.
\end{align}
Интеграл движения (\ref{efisco}) существует для любой метрики, поскольку для
параметризации экстремалей используется канонический параметр.
Значение постоянной $C_0$ определяет тип экстремали:
\begin{align*}
  C_0>0 &\text{ -- времениподобные экстремали},
\\
  C_0=0 &\text{ -- светоподобные экстремали},
\\
  C_0<0 &\text{ -- пространственноподобные экстремали}.
\end{align*}
Отметим, что тип экстремали не может меняться от точки к точке. Второй из
интегралов (\ref{efisct}) связан с наличием вектора Киллинга, т.е.\ с симметрией
задачи, и для произвольной метрики он отсутствует.

В системе уравнений для экстремалей (\ref{eqexts}), (\ref{eqextt}) конформный
множитель $\Phi$ рассматривается, как сложная функция от $\s$:
$\Phi=\Phi\big(q(\s)\big)$, где $q(\s)$ -- решение уравнения (\ref{emetfd}).
Если форма экстремали $\s(\tau)$ найдена, то конформный множитель можно
рассматривать также как сложную функцию от $\tau$ вдоль каждой экстремали:
$\Phi=\Phi\big(\s(\tau)\big)$. В дальнейшем анализе это будет подразумеваться.
\begin{theorem}                                                   \label{textrf}
Любая экстремаль в статическом пространстве-времени типа {\rm I, III}
принадлежит одному из следующих четырех классов.\newline
1) Светоподобные экстремали
\begin{equation}                                                  \label{extlif}
  \tau=\pm\s+\const,
\end{equation}
для которых канонический параметр определен уравнением
\begin{equation}                                                  \label{extlip}
  \dot\tau=\frac1\Phi.
\end{equation}
2) Экстремали общего вида, форма которых определена уравнением
\begin{equation}                                                  \label{extgef}
  \frac{d\tau}{d\s}=\pm\frac1{\sqrt{1-C_2 \Phi}},
\end{equation}
где $C_2$ -- произвольная отличная от нуля постоянная, причем значения $C_2<0$ и
$C_2>0$ описывают соответственно пространственно- и времениподобные экстремали.
Канонический параметр определяется любым из двух уравнений:
\begin{align}                                                     \label{extgpf}
  \dot\tau&=\frac1\Phi,
\\                                                                \label{extgps}
  \dot\s  &=\pm\frac{\sqrt{1-C_2\Phi}}\Phi.
\end{align}
При этом в уравнениях (\ref{extgef}) и (\ref{extgps}) знаки плюс или минус
выбираются одновременно.\newline
3) Прямые пространственноподобные экстремали, параллельные оси $\s$ и проходящие
через каждую точку $\tau=\const$. Канонический параметр определен уравнением
\begin{equation}                                                  \label{extstp}
  \dot\s=\frac1{\sqrt \Phi}.
\end{equation}
4) Прямые вырожденные времениподобные экстремали, параллельные оси $\tau$ и
проходящие через критические точки $\s_0=\const$, в которых
\begin{equation}                                                  \label{extdcp}
  \Phi'(\s_0)=0.
\end{equation}
Канонический параметр для них совпадает с временн\'ой координатой,
\begin{equation}                                                  \label{extdgp}
  t=\tau.
\end{equation}
\end{theorem}
\begin{proof}
Доказательство теоремы сводится к интегрированию системы уравнений
(\ref{eqexts}), (\ref{eqextt}). При получении интеграла движения (\ref{efisct})
необходимо делить на $\dot\tau$ и $\dot\s$, поэтому эти вырожденные случаи
рассматриваются отдельно.

Начнем с вырожденных случаев. Для $\tau=\const$ уравнение (\ref{eqexts})
выполняется автоматически, а уравнение (\ref{eqextt}) после интегрирования
приводит к условию
$$
  -\Phi\dot\s^2=C_0<0.
$$
Это уравнение сводится к (\ref{extstp}) растяжкой канонического параметра.

Второй вырожденный случай соответствует $\dot\s=0$. При этом уравнение
(\ref{eqextt}) выполняется только в тех точках $\s=\s_0$, в которых
$\Phi'(\s_0)=0$, а уравнение (\ref{eqexts}) сводится к уравнению $\ddot\tau=0$.
Отсюда вытекает, что координату $\tau$ можно выбрать в качестве канонического
параметра. Здесь можно лишь заметить справедливость равенства
$$
  \frac{d\Phi}{d\s}=\Phi\frac{d\Phi}{dq}=0,
$$
и, поскольку $\Phi>0$, то в уравнении (\ref{extdcp}) дифференцирование по $q$
можно заменить дифференцированием по $\s$. Таким образом исчерпываются оба
специальных случая.

Остальные экстремали можно получить из анализа интегралов (\ref{efisco}),
(\ref{efisct}), причем константа $C_0$ может быть произвольной, а $C_1\ne0$,
т.к.\ случай $C_1=0$ относится к прямым экстремалям третьего класса. Уравнение
(\ref{efisco}) с учетом (\ref{efisct}) приводит к равенству
\begin{equation}                                                  \label{extsgt}
  \dot\s=\pm\frac{C_1}\Phi\sqrt{1-C_2\Phi},
\end{equation}
где
\begin{equation}                                                  \label{econct}
  C_2:=\frac{C_0}{C_1^2}.
\end{equation}
Уравнение для формы экстремалей общего вида (\ref{extgef}) является следствием
(\ref{efisct}) и (\ref{extsgt}), причем значение постоянной $C_2$ определяет тип
экстремали. Уравнения для канонического параметра получаются из (\ref{efisct}) и
(\ref{extsgt}) после растяжки. В частном случае $C_2=0$ получаем светоподобные
экстремали (\ref{extlif}).
\end{proof}
\begin{com}
Из уравнения (\ref{extgef}) следует, что постоянная $C_2$ параметризует угол,
под которым экстремаль общего вида проходит через заданную точку. Можно
проверить, что через произвольную точку в каждом направлении проходит одна и
только одна экстремаль. Это подтверждает тот факт, что найдены все экстремали.
\qed\end{com}

Доказанная теорема позволяет качественно понять поведение экстремалей для
произвольного конформного множителя.

\begin{figure}[h,b,t]
\hfill\includegraphics[width=.95\textwidth]{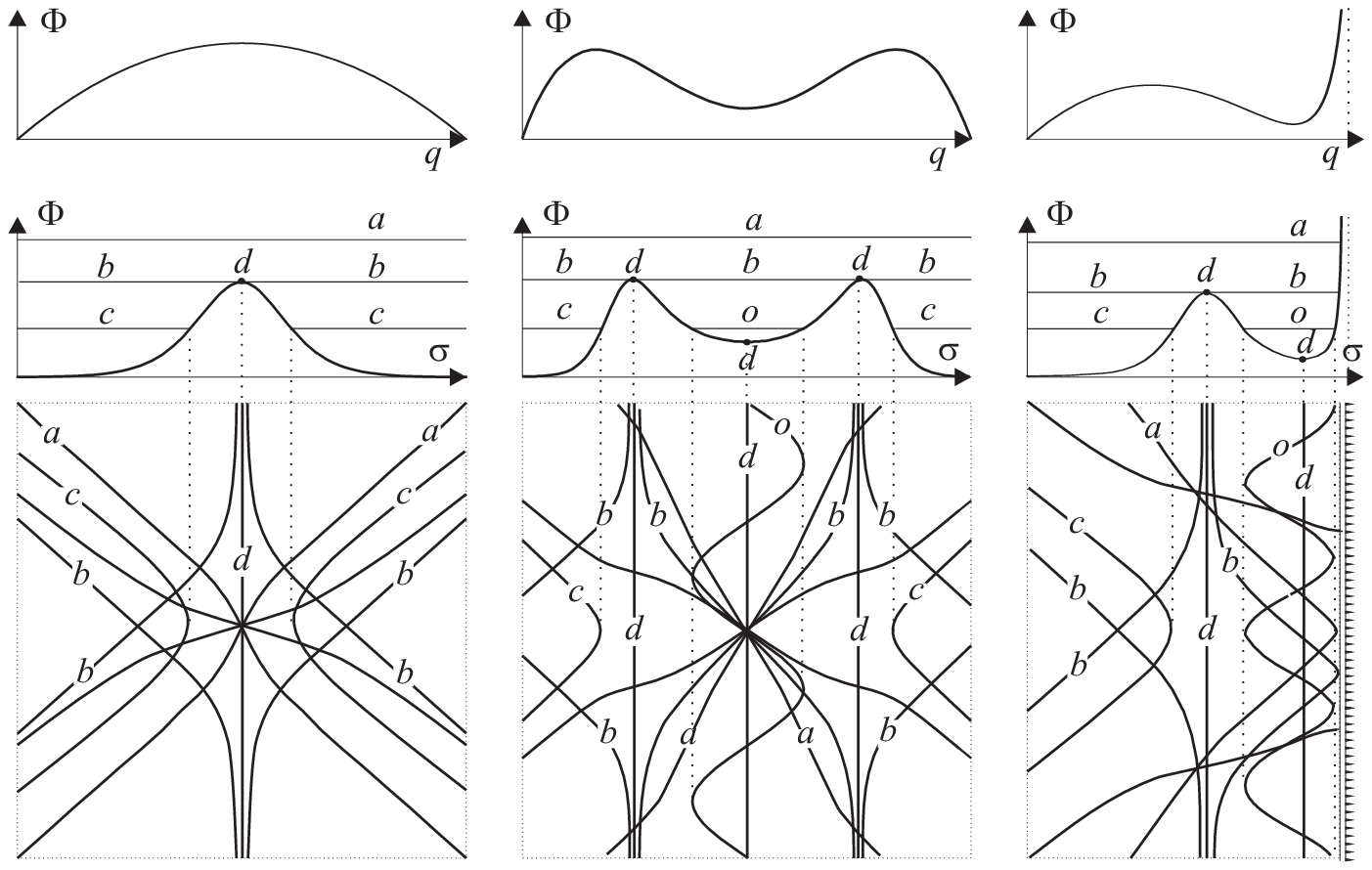}
\hfill {}
\centering\caption{Типичное поведение конформного множителя $\Phi(q)$ между
двумя нулями и нулем и сингулярностью для статического локального решения
(верхний ряд). Зависимость конформного множителя $\Phi(\s)$ от пространственной
координаты (средний ряд). Типичные времени- и пространственно подобные
экстремали (нижний ряд). Буквами отмечены времениподобные экстремали, имеющие
качественно различное поведение при разных значениях постоянной $C_2$. Через
каждый локальный экстремум проходит вырожденная экстремаль {\it d}. Во втором и
третьем случае имеются осциллирующие экстремали {\it o}.}
\label{fextre}
\end{figure}

На рис.~\ref{fextre} в первом ряду показано типичное поведение конформного
множителя $\Phi(q)$ с одним и тремя локальными экстремумами между двумя нулями и
с двумя локальными экстремумами между нулем и особенностью для статических
конформных блоков. Во втором ряду показана зависимость этих конформных
множителей $\Phi(\s)$ от пространственной координаты. В третьем ряду показано
качественное поведение времени- и пространственноподобных экстремалей. Чтобы не
загромождать рисунок, мы опустили светоподобные экстремали, проходящие через
каждую точку пространства-времени, а также экстремали, параллельные оси $\s$.
Все экстремали можно сдвигать вдоль времениподобной координаты $\tau$.

Если обе граничные точки интервала $q_i$ и $q_{i+1}$ являются нулями, то из
непрерывности конформного множителя $\Phi$ следует, что у него есть по крайней
мере один максимум, через который проходит вырожденная экстремаль {\it d}.
В общем случае через каждый локальный экстремум функции $\Phi(\s)$ проходит
вертикальная экстремаль. На рисунках они помечены буквой {\it d}.

Пространственноподобные экстремали общего вида соответствуют отрицательным
значениям постоянной $C_2$ в уравнении (\ref{extgef}), и, поскольку $\Phi>0$,
подынтегральное выражение всюду определено. Они начинаются на левой границе
конформного блока и заканчиваются на правой. Пространственноподобные экстремали
общего вида показаны на рисунке без каких либо пометок.

Времениподобные экстремали общего вида имеют качественно различное поведение при
различных значениях постоянной $C_2>0$. Они помечены буквами {\it a,b,c} и
{\it o}.

Если постоянная $C_2$ достаточно мала,
\begin{equation*}
  C_2<\frac1{\max\Phi(\s)},
\end{equation*}
то времениподобная экстремаль общего вида определена при всех значениях
$\s\in(\s_i,\s_{i+1})$. Такие экстремали соединяют левую и правую границу
конформного блока, пересекая все вырожденные экстремали, и помечены на рисунке
буквой $a$.

Во втором ряду рис.\ref{fextre} горизонтальными линиями показаны области
определения координаты $\s$,
\begin{equation}                                                  \label{qdesrf}
  C_2\Phi(\s)\le1,
\end{equation}
при различных значениях постоянной $C_2$.

Если у функции $\Phi$ имеется локальный минимум, которому также соответствует
вырожденная экстремаль, то среди экстремалей общего вида существуют
времениподобные экстремали, осциллирующие вблизи локального минимума, которые
помечены буквой {\it o}. Это следует из уравнения (\ref{extgef}), т.к.\ при
достаточно больших положительных значениях $C_2$ область изменения координаты
$\s$ определяется неравенством (\ref{qdesrf}). Критические точки $\s_*$, в
которых $\Phi(\s_*)=1/C_2$, являются точками поворота при конечном значении
$\tau=\tau_*$ тогда и только тогда, когда интеграл
\begin{equation}                                                  \label{eintpo}
  \int^{\tau_*}\!\!\! d\tau=\pm\int^{\s_*}\!\!\!\frac{d\s}{\sqrt{1-C_2\Phi}}
\end{equation}
сходится. Этот интеграл сходится, если точка $\s_*$ не совпадает с локальным
максимумом. То есть поворот экстремали происходит при конечном значении $\tau_*$
и канонического параметра, что вытекает из уравнения (\ref{extgps}).

При том же значении постоянной $C_2$ существуют времениподобные неосциллирующие
экстремали общего вида, которые приходят и уходят на одну и ту же (левую или
правую) границу. У них есть точка поворота при конечных значениях $\tau$. Такие
экстремали на рисунке помечены буквой {\it c}.

Если критическая точка $\s_*$ совпадает с локальным максимумом, то
\begin{equation*}
  \Phi\simeq \frac1{C_2}-C_3(\s-\s_*)^2+\dotsc,\qquad C_3=\const>0,
\end{equation*}
и интеграл (\ref{eintpo}) расходится. Это означает, что критическая точка
достигается при бесконечном значении $\tau_*$ и канонического параметра. То есть
соответствующая экстремаль в этом направлении полна. На рисунке такие экстремали
помечены буквой {\it b}.

Приведенная теорема описывает все экстремали для статических локальных решений.
Поведение экстремалей для однородных локальных решений аналогично. Они
получаются заменой $\Phi\mapsto-\Phi$, перестановкой временн\'ой и
пространственной координат $\tau\leftrightarrow\s$ и поворотом на угол $\pi/2$
всей плоскости.
\subsection{Асимптотика экстремалей                              \label{sextac}}
Для решения вопроса о том является ли метрика, определенная на $\tau,\s$
плоскости, полной по экстремалям или локальное решение следует продолжить,
необходимо исследовать асимптотику и полноту экстремалей вблизи границы
конформного блока. Забегая вперед, заметим, что все экстремали, за исключением
осциллирующих, имеют асимптотику при приближении к границе конформного блока.

Рассмотрим квадратный конформный блок. В верхнюю и нижнюю вершины конформного
блока попадают все экстремали, для которых
$$
  \underset{\tau\to\pm\infty}{\lim}\left|\frac{d\tau}{d\s}\right|>1,
$$
если предел существует. В левую и правую вершины конформного блока попадают
экстремали, для которых
\begin{equation}                                                  \label{qspasy}
  \underset{\s\to\pm\infty}\lim\left|\frac{d\tau}{d\s}\right|<1.
\end{equation}
На стороны квадратного конформного блока могут попасть только экстремали,
имеющие светоподобную асимптотику:
\begin{equation}                                                  \label{qliasx}
  \underset{\tau,\s\to\pm\infty}\lim\left|\frac{d\tau}{d\s}\right|=1.
\end{equation}
В принципе, в вершину конформного блока могут попасть также экстремали, имеющие
светоподобную асимптотику. Однако, как будет показано ниже, этого не происходит.

Если одна из границ $q_i$, например, статического конформного блока прямая, то
на нее попадают экстремали со светоподобной (\ref{qliasx}) и
пространственноподобной (\ref{qspasy}) асимптотикой.

Осциллирующие экстремали всегда зажаты прямыми вырожденными экстремалями.
Поэтому они попадают в ту же вершину конформного блока, что и прямые вырожденные
экстремали 4) из теоремы \ref{textrf}.

Рассмотрим асимптотику экстремалей вблизи границы статических конформных блоков.
Любая светоподобная экстремаль на квадратном конформном блоке начинается и
заканчивается на противоположных сторонах. В случае треугольного конформного
блока или блока типа линзы светоподобные экстремали начинаются или заканчиваются
на времениподобных границах.

Для статических локальных решений типа I, III прямые экстремали, параллельные
пространственной оси $\s$ начинаются в левой и заканчиваются в правой вершине
квадратного конформного блока. В случае треугольного блока и блока типа линзы
эти экстремали начинаются или (и) заканчиваются на времениподобной границе.
Вырожденные экстремали начинаются в нижней и заканчиваются в верхней вершине
статического конформного блока. Осциллирующие экстремали зажаты вырожденными
экстремалями и поэтому также начинаются и заканчиваются в нижней и верхней
вершине.

Таким образом, мы поняли в какие граничные точки конформного блока попадают все
прямые экстремали, а также осциллирующие экстремали общего вида.

Осталось рассмотреть поведение неосциллирующих экстремалей общего вида вблизи
границы. Сначала определим точки границы конформного блока, в которые они
попадают. Допустим, что граница $q_i$ конформного блока является нулем,
$\Phi(q_i)=0$, $|q_i|<\infty$. Этому соответствует положительный показатель
$m>0$ (\ref{ecfapo}). Соответствующая граница конформного блока является углом
при $m\ge1$ и прямой при $0<m<1$ (\ref{eboucb}). В этом случае из уравнения для
экстремалей (\ref{extgef}) следует, что векторы, касательные к ним, имеют
светоподобную асимптотику
$$
  \underset{q\to q_i}\lim\left|\frac{d\tau}{d\s}\right|=1.
$$
Если граница конформного блока является углом, то экстремали общего вида
попадают на стороны угла, а не в вершину. Действительно, из уравнений
(\ref{extgpf}), (\ref{extgps}) для экстремалей, идущих в правую верхнюю сторону
квадратного конформного блока, следует равенство
$$
  \frac{d\eta}{d\x}=\frac{1-\sqrt{1-C_2\Phi}}{1+\sqrt{1+C_2\Phi}}.
$$
Для малых $\Phi\ll1$ разложим правую часть этого равенства в ряд. В первом
порядке по $\Phi$ получим равенство
\begin{equation*}
  \frac{d\eta}{d\xi}\simeq\frac12\left[1-\left(1-\frac{C_2\Phi}2\right)\right]
  =\frac{C_2}4\Phi.
\end{equation*}
Откуда вытекает, что интеграл
$$
  \int d\eta\sim\int d\x \Phi\sim\int d\s\Phi\sim\int^{q_i}\!\!\!dq=q_i
$$
сходится. Это означает, что точка $\xi\to\infty$ достигается при конечных
значениях $\eta$, и это соответствует некоторой точке на стороне угла, а не
вершина. Если граница $q_i$ конформного блока является времениподобной, то
экстремали общего вида попадают на нее.

Пусть теперь на границе $\Phi(q_i)=\infty$, $|q_i|<\infty$. В этом случае
$m<0$, граница времениподобна и достигается при конечных $\s_i$. Из равенства
(\ref{extgef}) следует, что касательный вектор к пространственноподобной
экстремали при $\Phi\to\infty$ имеет асимптотику
$$
  \frac{d\tau}{d\s}\sim\frac1{\sqrt \Phi}\to0.
$$
То есть, они попадают на границу $|\s_i|<\infty$ под прямым углом.

Времениподобные экстремали общего вида не могут попасть на времениподобную
границу, потому что правая часть в (\ref{extgef}) при $C_2>0$ вблизи особенности
$\Phi(q_i)=\infty$ становится отрицательной. Эти экстремали вблизи границы
$|\s_i|<\infty$ имеют точку поворота при конечных значениях $\tau$, по той же
причине, что и осциллирующие экстремали.
\subsection{Полнота экстремалей                                  \label{sextco}}
Теорема \ref{textrf} позволяет проанализировать полноту экстремалей при подходе
к границе конформного блока. Для определенности рассмотрим стационарное
локальное решение. Прежде всего заметим, что вырожденные экстремали всегда
полны, т.к.\ канонический параметр для них совпадает с временем (\ref{extdgp}).
Осциллирующие экстремали также полны поскольку делают бесконечное число
осцилляций, каждая из которых соответствует конечному изменению канонического
параметра. Если же экстремаль общего вида при $\tau\to\pm\infty$ приближается к
вырожденной экстремали, то она полна, поскольку из уравнения для канонического
параметра (\ref{extgpf}) следует, что
$$
  \underset{\s\to\s_0}\lim t\to\int^{\infty}\!\!\!d\tau \Phi(\s_0)\to\infty.
$$
Для стационарного локального решения в нижнюю и верхнюю вершины попадают только
вырожденные и осциллирующие экстремали, и поэтому они всегда полны. Возможна
ситуация, когда эти экстремали отсутствуют. В этом случае будем считать эти
вершины полными, потому что любая времениподобная кривая имеет бесконечную длину
при $\tau\to\pm\infty$. Это означает, что нижняя и верхняя вершины стационарного
блока всегда полны, т.е.\ представляют собой временн\'ую бесконечность прошлого
и будущего, соответственно.

Полнота светоподобных экстремалей определяется уравнением (\ref{extlip}),
которое позволяет вычислить предел
\begin{equation}                                                  \label{elieco}
  \underset{\s\to\pm\infty}\lim t\to\int^{\s_i}\!\!\!d\s \Phi
  =\int^{q_i}\!\!\!dq=q_i.
\end{equation}
Это значит, что при приближении к границе конформного блока они неполны при
конечных $q_i$ и полны при $|q_i|=\infty$.

Полнота неосциллирующих экстремалей общего вида, попадающих на границу
конформного блока, соответствующую значению $q_i$, вытекает из уравнения
(\ref{extgps}):
\begin{equation}                                                  \label{extgtc}
  \underset{q\to q_i}\lim t\to\int^{\s_i}\!\!\!d\s\frac \Phi{\sqrt{1-C_2\Phi}}
  \sim\int^{q_i}\!\!\!\frac{dq}{\sqrt{1-C_2\Phi}}.
\end{equation}
Вблизи нулей $\Phi\rightarrow0$ поведение экстремалей общего вида такое же, как
и у светоподобных экстремалей (\ref{elieco}). Особенность
$\Phi\rightarrow\infty$ достигается только пространственноподобными
экстремалями, $C_2<0$, полнота которых определяется интегралом
\begin{equation}                                                  \label{extgpc}
  \underset{q\to q_i}\lim t\to\int^{q_i}\!\!\!\frac{dq}{\sqrt{\Phi}}.
\end{equation}
Эти экстремали неполны в конечных точках $|q_i|<\infty$ при $m<0$. В бесконечно
удаленных точках $|q_i|=\infty$ пространственноподобные экстремали полны при
$0<m\le2$ и неполны при $m>2$.

Полнота прямых экстремалей, параллельных оси $\s$, определяется уравнением
(\ref{extstp}) и при приближении к границе $q_i$ дается интегралом
(\ref{extgpc}). Это значит, что их полнота вблизи особенностей такая же, как и
у экстремалей общего вида. Вблизи нулей они всегда полны за исключением случая
простого нуля в конечной точке, где они неполны.

\begin{figure}[h,b,t]
\hfill\includegraphics[width=.95\textwidth]{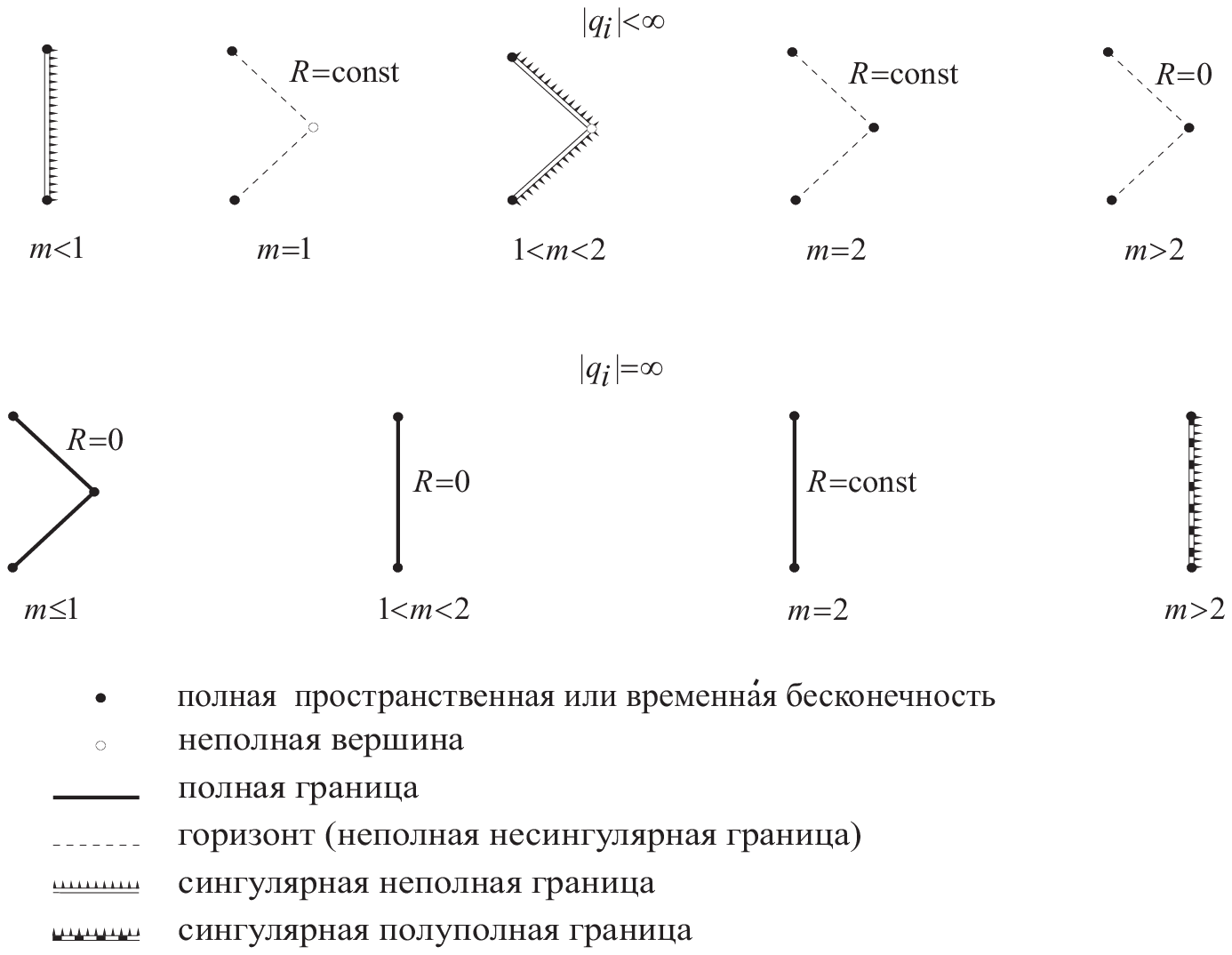}
\hfill {}
\centering\caption{Форма правой границы статических конформных блоков в
зависимости от показателя степени $m$. В верхнем и нижнем рядах показаны границы
для конечных и бесконечных значений $q_i$, соответственно.}
\label{fboprr}
\end{figure}

Таким образом проанализирована полнота всех элементов границы. Итог анализа
приведен на рис.~\ref{fboprr}.
Здесь в зависимости от показателя степени $m$ (\ref{ecfapo}) для конечных и
бесконечных значений $q_i$ показаны соответствующие элементы границы конформных
блоков. Для определенности, показана правая граница статических конформных
блоков типа {\rm I}. На рисунке времениподобная граница показана вертикальной
линией, а светоподобная -- углом. Если на границе скалярная кривизна сингулярна
(\ref{esfpsc}), (\ref{esfpcc}), то на ней присутствуют выступы. Неполные и
полные границы показаны соответственно пунктиром и жирной сплошной линией.
Исключение составляет полуполная граница, соответствующая $|q_i|=\infty$ при
$m>2$, рис.~\ref{fboprr}. Светоподобные экстремали, попадающие на эту
сингулярную границу полны, а пространственноподобные экстремали -- неполны.
Нижние и верхние вершины всех статических конформных блоков (времениподобные
бесконечности прошлого и будущего) являются существенно особыми точками и всегда
полны, что отмечено закрашенными окружностями. Полнота или неполнота правых
вершин границы в форме угла отмечена закрашенной или незакрашенной окружностью,
соответственно. Левые границы статических конформных блоков имеют такое же
строение, только углы нужно отразить относительно вертикальной прямой.

Качественное поведение всех экстремалей в однородном пространстве-времени типа
{\rm II, IV} определяется нулями и особенностями конформного множителя так же,
как и для стационарного локального решения. При этом соответствующие конформные
блоки должны быть повернуты на угол $\pi/2$.
\section{Построение глобальных решений                           \label{sglrul}}
В разделе \ref{scoblo} каждому локальному решению типа {\rm I-IV} был сопоставлен
конформный блок, определенный на интервале $(q_i,q_{i+1})$. Затем в разделе
\ref{sextco} из анализа экстремалей была доказана полнота или неполнота границ
этих блоков. В результате оказалось, что нули конформного множителя, которые
соответствуют {\em горизонтам} пространства-времени, и только они определяют
неполную по экстремалям границу конформного блока, на которой кривизна конечна.
Во всех остальных случаях граница либо полна, либо на ней кривизна сингулярна.
Поэтому решения вида (\ref{emetok}) необходимо продолжить только через нули
функции $\Phi(q)$. Напомним, что нулям конформного множителя (горизонтам)
соответствуют границы в виде угла (см.\ рис.\ \ref{fboprr}).
\index{Горизонт (horizon)}%

Сформулируем правила, по которым осуществляется максимальное продолжение
поверхности с метрикой (\ref{emetok}) при заданной функции $\Phi(q)$.
Дифференцируемость максимально продолженной по экстремалям псевдоримановой
поверхности, построенной по этим правилам дается приведенной ниже теоремой
\ref{tunisp}.
\begin{enumerate}
\item Каждое глобальное решение для метрики (\ref{emetok}) соответствует
некоторому интервалу значений переменной $q\in(q_-,q_+)$, где $q_\pm$ либо
бесконечно удаленные точки $q_\pm=\pm\infty$, либо сингулярности кривизны,
определяемые условием (\ref{esfpsc}). Внутри интервала сингулярности должны
отсутствовать.
\item Если внутри интервала $(q_-,q_+)$ конформный блок не имеет нулей, то
соответствующий конформный блок представляет собой глобальное решение.
\item Если внутри интервала $(q_-,q_+)$ нули существуют, то пронумеруем их в
порядке возрастания, $\Phi(q_j)=0$, $j=1,\dots,n$, и поставим
в соответствие каждому интервалу $(q_-,q_1),\dots,(q_n,q_+)$ пару статических
или однородных конформных блоков для $\Phi>0$ и $\Phi<0$, соответственно.
\item Склеим конформные блоки вдоль границ, соответствующих нулям $q_j$, причем
склеиваются только блоки, соответствующие соседним интервалам $(q_{j-1},q_j)$ и
$(q_j,q_{j+1})$.
\item Диаграмма Картера--Пенроуза получается склеиванием всех соседних
конформных блоков. Она представляет собой связную фундаментальную область, если
внутри интервала $(q_-,q_+)$ конформный множитель меняет знак. Если $\Phi\ge0$
или $\Phi\le0$ всюду внутри интервала $(q_-,q_+)$, то получим две
фундаментальные области, которые получаются друг из друга отражением
пространства или времени.
\item При наличии в интервале $(q_-,q_+)$ только одного нуля нечетного порядка
граница фундаментальной области состоит из границ конформных блоков,
соответствующих точкам $q_-$ и $q_+$, и диаграмма Картера--Пенроуза представляет
глобальное решение.
\item При наличии одного нуля четного порядка или двух и более нулей
произвольного порядка граница фундаментальной области включает нули конформного
множителя, и ее необходимо либо продолжить периодически в пространстве и (или)
во времени, либо отождествить противоположные стороны вдоль горизонтов.
\item Если у диаграммы Картера--Пенроуза фундаментальная группа тривиальна,
то она представляет собой универсальную накрывающую глобального решения.
\item Если у диаграммы Картера--Пенроуза фундаментальная группа нетривиальна, то
следует построить соответствующую универсальную накрывающую.
\end{enumerate}
\begin{defn}
{\em Диаграммой Картера--Пенроуза} называется образ на плоскости максимально
продолженной поверхности, соответствующей интервалу $(q_-,q_+)$ и полученный в
результате склеивания всех конформных блоков для интервала $(q_-,q_+)$ по
правилам 1)--9).
\qed\end{defn}
\index{Диаграмма Картера--Пенроуза (Carter--Penrose diagram)}%
\index{Картера--Пенроуза диаграмма (Carter--Penrose diagram)}%
Диаграммы Картера--Пенроуза являются наглядным изображением максимально
продолженной поверхности, т.к.\ оба класса непересекающихся между собой
светоподобных экстремалей изображаются двумя классами перпендикулярных прямых,
как на плоскости Минковского.

Поясним приведенные правила.
Утверждение 1) является следствием того, что продолжение решения через точки
$q_\pm$ невозможно, т.к.\ эти точки либо полны, либо сингулярны. Решения
продолжаются только через горизонты $|q_j|<\infty$ с $m=1$ или $m\ge2$. В этих
случаях граница конформного блока является углом. Если точка $q_j$ -- нуль
нечетного порядка, то все четыре конформных блока, соответствующих интервалам
$(q_{j-1},q_j)$ и $(q_j,q_{j+1})$ склеиваются согласно правилам 3) и 4) вокруг
вершины $q_j$, причем эта точка является седловой для конформного множителя
$\Phi(q)$, рассматриваемого как функция на диаграмме Картера--Пенроуза. Если
$q_j$ -- нуль четного порядка, то вдоль соответствующего горизонта склеиваются
между собой только области одного типа, например, I-I или III-III. Если внутри
интервала $(q_-,q_+)$ функция $\Phi$ не меняет знака, то возникают две несвязные
между собой фундаментальные области, каждую из которых можно периодически
продолжить. При изменении знака $\Phi$ возникает седловая точка, и области
различных типов образуют связную фундаментальную область, которую можно либо
периодически продолжить, либо нет, в зависимости от значений $q$ на границе. Это
составляет содержание правил 5) и 6).

Правила 5), 6) и 7) являются утверждениями, которые будут доказаны дальнейшим
построением.

Сформулируем основную теорему, оправдывающую приведенные выше правила построения
глобальных решений.
\begin{theorem}                                                   \label{tunisp}
Универсальное накрывающее пространство, построенное по правилам 1)--9),
представляет собой максимально продолженную вдоль экстремалей дифференцируемую
псевдориманову поверхность класса $\CC^{l+1}$ с метрикой класса ${\cal C}^l$,
$l\ge2$, такую, что внутренность каждого конформного блока изометрична
поверхности с метрикой (\ref{emetok}).
\end{theorem}
\begin{proof}
По построению, внутренность каждого конформного блока покрывается одной картой и
является многообразием класса ${\cal C}^{l+1}$, что следует из (\ref{etrfun}).
Поэтому внутри конформного блока метрика принадлежит тому же классу гладкости,
что и конформный множитель. При этом функции перехода (\ref{etrfun}) изометрично
отображают плоскость $\tau,\s$ или ее часть, на которой определена метрика, на
внутренность конформного блока. Поэтому необходимо доказать дифференцируемость
многообразия и метрики только на горизонтах и в седловых точках. Это сделано
путем перехода к новым системам координат в разделах \ref{sedfic} и
\ref{sedpom}, соответственно.
\end{proof}
Перед завершением доказательства теоремы рассмотрим несколько примеров, чтобы
пояснить, как пользоваться правилами построения глобальных решений в конкретных
случаях, и какие глобальные решения могут возникнуть.
\section{Примеры                                                 \label{sexamz}}
Если в результате решения уравнений движения какой либо модели гравитации
возникает двумерная метрика вида (\ref{emetok}), то, следуя правилам 1)--9) из
предыдущего раздела можно построить глобальное решение, не заботясь о переходе к
новым координатным системам, покрывающим б\'ольшие области. Достоинством
рассматриваемого метода является его конструктивность, т.к.\ для построения
глобальных решений достаточно элементарного анализа конформного множителя
$\Phi(q)$. Начнем с трех известных примеров из общей теории относительности.
\subsection{Решение Шварцшильда                                  \label{schwas}}
\begin{figure}[h,b,t]
\hfill\includegraphics[width=.95\textwidth]{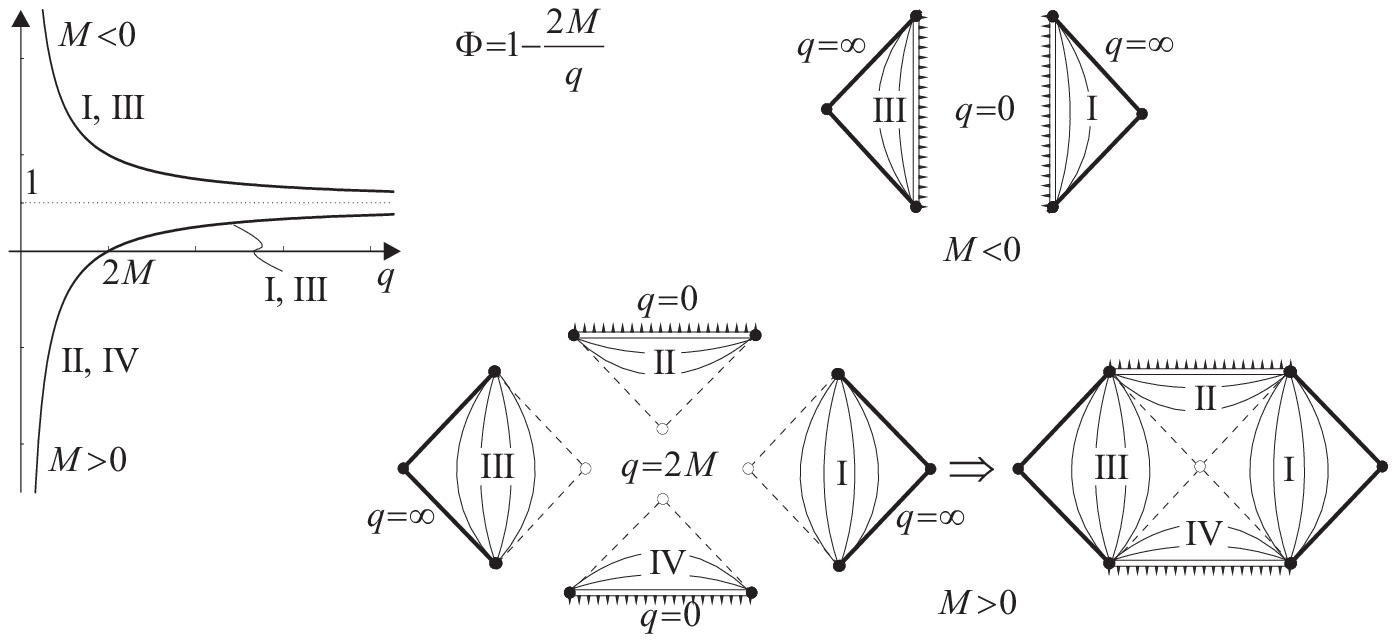}
\hfill {}
\centering\caption{Поведение конформного множителя для решения Шварцшильда для
положительной и отрицательной массы. При $M<0$ существует два глобальных
решения, представленных треугольными диаграммами Картера--Пенроуза. При $M>0$
четыре конформных блока, соответствующих двум интервалам $(0,2M)$ и
$(2M,\infty)$, склеиваются в одно глобальное решение.}
\label{fschws}
\end{figure}

Конформный множитель для решения Шварцшильда имеет вид (\ref{qsgcre}). Он имеет
простой полюс в точке $q=0$ $(m=-1)$, который соответствует сингулярности
кривизны (\ref{esfpsc}). Точка $q_1=2M$ $(m=1)$ является простым нулем и
соответствует горизонту. При $M>0$ нуль лежит на положительной полуоси. Значения
$q=\pm\infty$ $(m=0)$ соответствуют асимптотически плоской пространственной
бесконечности $(R=0)$. Поведение конформного множителя показано на
рис.~\ref{fschws} слева.

Бесконечному интервалу $q\in(-\infty,\infty)$ соответствует два глобальных
решения для положительных и отрицательных $q$. Рассмотрим случай $M>0$. Для
положительных значений $q\in(0,\infty)$ имеем $q_-=0$, $q_1=2M$ и $q_+=\infty$.
Каждому интервалу $(q_-,q_1)=(0,2M)$ и $(q_1,q_+)=(2M,\infty)$ соответствует по
два однородных (II, IV) и статических (I, III) конформных блока, показанных на
рис.~\ref{fschws} внизу в центре. При этом элементы границы определяются по
рис.~\ref{fboprr}. Из этих четырех конформных блоков по правилу 4) единственным
образом склеивается глобальное решение, показанное на рис.~\ref{fschws}. Эта
диаграмма Картера--Пенроуза представляет собой расширение Крускала--Секереша
решения Шварцшильда \cite{Kruska60,Szeker60}. Напомним, что расширение
Крускала--Секереша представляет собой запись метрики Шварцшильда в такой системе
координат, которая покрывает сразу все области I--IV. Картер впервые изобразил
это расширение в виде конечной области на плоскости \cite{Carter73}.

Поскольку фундаментальная группа для полученной диаграммы тривиальна, то
построенное глобальное решение является универсальной накрывающей для других
глобальных решений. Оба статических конформных блока I или III диффеоморфны
внешнему решению Шварцшильда при $r>2M$. Теорема \ref{tunisp} обеспечивает, в
данном случае, гладкость соответствующего глобального решения, и нет
необходимости в явном построении глобальной системы координат.

Аналогичные глобальные решения будут возникать для широкого класса метрик, у
которых конформный множитель ведет себя качественно так же, как и нижняя ветвь
на рис.~\ref{fschws}. То есть определен на полуинтервале $(q_-,\infty)$, имеет
особенность в $q_-$, один нуль и стремится к постоянной на бесконечности.

Отметим, что поверхность, представленная диаграммой Картера--Пенроуза имеет
непостоянную скалярную кривизну,
\begin{equation}                                                  \label{escshs}
  R=\frac{4M}{q^3}.
\end{equation}
Это -- двумерная скалярная кривизна на лоренцевой поверхности. Скалярная
кривизна, соответствующая четырехмерной метрике Шварцшильда с учетом угловой
зависимости тождественно равна нулю в силу уравнений Эйнштейна. Заметим, что
двумерная скалярная кривизна (\ref{escshs}) совпадает с инвариантным собственным
значением четырехмерного тензора Вейля \cite{LanLif88R}. Центр диаграммы
Картера--Пенроуза является седловой точкой для переменной $q$ и, следовательно,
для скалярной кривизны (\ref{escshs}). Эта точка неполна и отмечена
незакрашенной окружностью.

Физическая интерпретация диаграммы Картера--Пенроуза для решения Шварцшильда
состоит в следующем. В общей теории относительности пространство-время
четырехмерно. Поэтому максимально продолженное сферически симметричное решение
вакуумных уравнений Эйнштейна представляет собой топологическое произведение
сферы $\MS^2$ на диаграмму Картера--Пенроуза, изображенную на рис.~\ref{fschws}
и представляющую собой ту часть пространства-времени, которая описывается
координатами $t$ и $r$. Нижняя и верхняя сингулярные границы диаграммы
Картера--Пенроуза, умноженные на сферу, называются, соответственно, {\em белой}
и {\em черной дырой}. Сфера $r=2M$ называется {\em горизонтом} черной дыры.
Конформный блок II, лежащий под
\index{Белая дыра (white hole)}%
\index{Дыра белая (white hole)}%
\index{Черная дыра (black hole)}%
\index{Дыра черная (black hole)}%
\index{Горизонт (horizon)}%
горизонтом черной дыры, соответствует внутренности черной дыры. Если выбрать
произвольную точку в области II, то любая времениподобная или светоподобная
кривая неизбежно попадет на сингулярную пространственноподобную границу (черную
дыру) за конечное собственное время. При этом нет никакой возможности пересечь
горизонт. Внешние области I и III соответствуют двум разным вселенным для одной
и той же черной дыры. Эти вселенные причинно не связаны между собой, поскольку
их точки можно соединить только пространственноподобными кривыми. Если
наблюдатель находится, например, во вселенной I, то у него есть две возможности:
либо за конечное время упасть на черную дыру, либо жить бесконечно долго,
поскольку бесконечность будущего (верхняя вершина конформного блока I) полна.

Для отрицательных значений $q$ горизонты отсутствуют. Поэтому треугольные
конформные блоки, показанные на рис.~\ref{fschws} вверху справа, представляют
собой максимально продолженные решения. В этом пространстве-времени
сингулярность кривизны располагается вдоль времениподобной границы, находящейся
на конечном расстоянии и не окруженной горизонтом. Такие сингулярности в общей
теории относительности называются {\em голыми}. Заметим, однако, что на голые
сингулярности попадают только свето- и пространственноподобные экстремали при
конечном значении канонического параметра. Времениподобные экстремали отражаются
вблизи сингулярности (см.\ рис.~\ref{fextre}).

Глобальное решение для отрицательных $q$ можно рассматривать как решение для
положительных $q$ (что соответствует интерпретации координаты $q$, как радиуса),
но с отрицательной массой, $M<0$. Эти решения рассматриваются в общей теории
относительности, как нефизические.
\index{Голая сингулярность (naked singularity)}%
\index{Сингулярность голая (naked singularity)}%
\subsection{Решение Рейснера--Нордстрема                         \label{srenos}}
\begin{figure}[h,b,t]
\hfill\includegraphics[width=.95\textwidth]{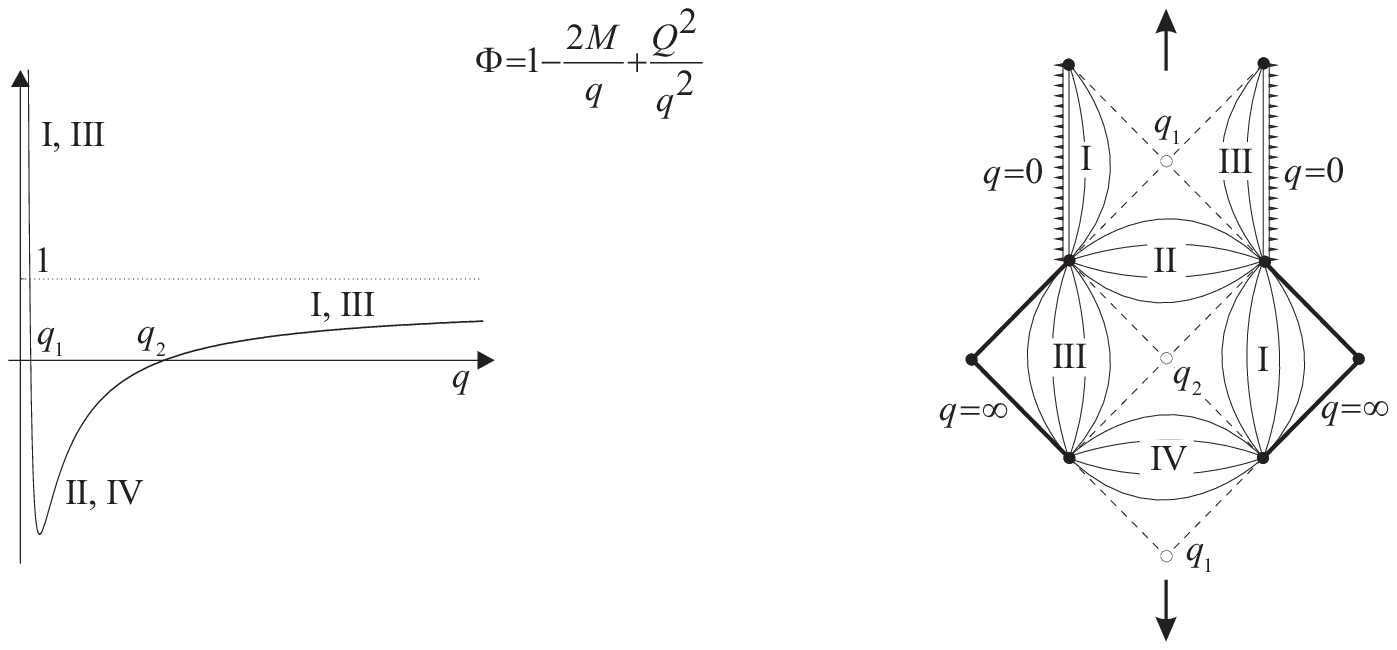}
\hfill {}
\centering\caption{Конформный множитель и фундаментальная область для решения
Рейснера--Нордстрема. Стрелки показывают возможное продолжение или
отождествление решения вдоль горизонтов.}
\label{frenpd}
\end{figure}

Конформный множитель для решения Рейснера--Нордстрема \cite{Reissn16,Nordst18},
описывающего заряженную черную дыру, имеет вид
\begin{equation}                                                  \label{erenom}
  \Phi=1-\frac{2M}q+\frac{Q^2}{q^2},\qquad 0<Q<M,
\end{equation}
где $M$ и $Q$ -- масса и заряд черной дыры. Функцию $\Phi$ при заданных
соотношениях между константами можно записать в виде
\begin{equation}                                                  \label{erenoq}
  \Phi=\frac{(q-q_1)(q-q_2)}{q^2},
\end{equation}
где
$$
  q_{1,2}=M\pm\sqrt{M^2-Q^2}.
$$
Отсюда следует, что конформный множитель имеет полюс второго порядка при $q=0$ и
два положительных простых нуля (горизонта) $q_{1,2}$. Его поведение для
положительных $q$ показано на рис.~\ref{frenpd} слева.

Глобальная структура решения для положительных $q$, по сравнению с решением
Шварцшильда, меняется качественно из-за наличия двух горизонтов. Каждому из
интервалов $(0,q_1)$, $(q_1,q_2)$ и $(q_2,\infty)$ соответствует по два
конформных блока. Поскольку внутри интервала $(0,\infty)$ конформный множитель
меняет знак, то по правилу 6) диаграмма Картера--Пенроуза, склеенная из шести
конформных блоков, рис.~\ref{frenpd}, представляет собой фундаментальную область
для решения Рейснера--Нордстрема. Ее граница состоит не только из сингулярных и
бесконечно удаленных точек, но также включает отрезки, соответствующие
горизонтам. В соответствии с правилом 7) ее можно периодически продолжить,
склеивая одинаковые фундаментальные области вдоль горизонтов в направлениях,
показанных стрелками. В этом случае мы получим универсальную накрывающую для
решения Рейснера--Нордстрема. Другая возможность заключается в том, что после
склеивания произвольного конечного числа фундаментальных областей, граничные
точки на горизонтах снизу и сверху можно отождествить. Тогда глобальное решение
с топологической точки зрения будет представлять собой цилиндр.

Сингулярности в решении Рейснера--Нордстрема являются голыми.

Скалярная кривизна (\ref{esctmk}) для решения Рейснера--Нордстрема непостоянна:
\begin{equation}                                                  \label{escren}
  R=\frac{4M}{q^3}-\frac{6Q^2}{q^4}.
\end{equation}
Это -- двумерная скалярная кривизна поверхности, а не четырехмерного
пространства-времени.

Левая ветвь конформного множителя $-\infty<q<0$ для решения Рейснера--Нордстрема
соответствует голой сингулярности и рассматриваться не будет, т.к.\ изменение
порядка полюса не влияет на структуру глобального решения.
\subsection{Экстремальная черная дыра                            \label{sextbh}}
\begin{figure}[h,b,t]
\hfill\includegraphics[width=.95\textwidth]{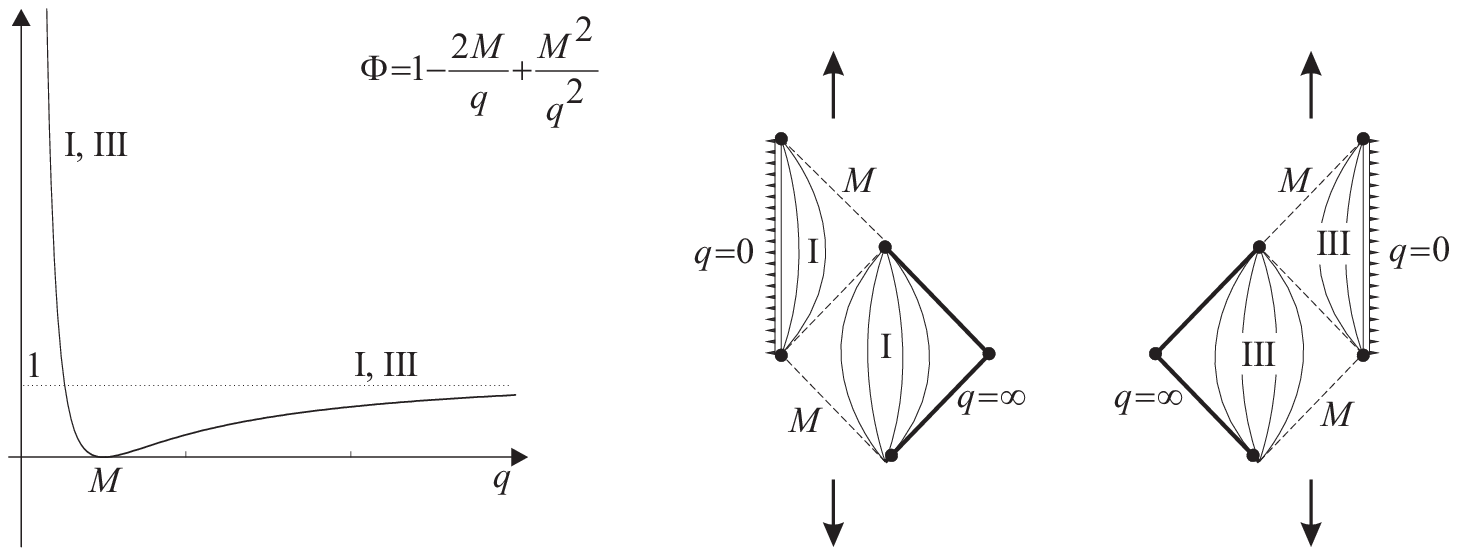}
\hfill {}
\centering\caption{Конформный множитель и две фундаментальные области для
экстремальной черной дыры. Стрелками показаны направления, вдоль которых решения
периодически продолжаются.}
\label{fextbh}
\end{figure}

Экстремальная черная дыра возникает из решения Рейснера--Нордстрема
(\ref{erenom}) в том случае, когда заряд равен массе, $Q=M$. Соответствующий
конформный множитель,
\begin{equation}                                                  \label{eexblf}
  \Phi=\frac{(q-M)^2}{q^2},
\end{equation}
имеет полюс второго порядка при $q=0$ и положительный нуль $q=M$ также второго
порядка. Его поведение показано на рис.~\ref{fextbh} слева.
Левая ветвь $-\infty<q<0$, как и в предыдущих случаях, описывает голую
сингулярность.

Положительные значения $q$ разбиваются на два интервала $(q_-,q_1)=(0,M)$ и
$(q_1,q_+)=(M,\infty)$. Этим интервалам ставятся в соответствие два конформных
блока и их пространственное отражение. Поскольку конформный множитель не меняет
знак, то в соответствии с правилом 6) имеются две несвязные фундаментальные
области, показанные на рис.~\ref{fextbh}. Вторая фундаментальная область,
построенная из областей типа {\rm III}, получается пространственным отражением
из области типа {\rm I}. Как и в случае решения Рейснера--Нордстрема границы
фундаментальных областей включают горизонты и их можно либо склеивать до
бесконечности, либо отождествить, что приводит к универсальной накрывающей и
цилиндрам, соответственно.
\subsection{Плоскость Минковского                                \label{sminpl}}
Предыдущие примеры демонстрируют правила построения глобальных лоренцевых
поверхностей непостоянной кривизны с одним вектором Киллинга. Следующие два
примера показывают, как эти правила работают в случае поверхностей постоянной
кривизны, которые имеют по три вектора Киллинга. Пример плоскости Минковского
важен также потому, что будет использован в дальнейшем для доказательства
дифференцируемости глобальных решений в седловых точках.

Для плоскости Минковского скалярная кривизна равна нулю, а конформный множитель
является линейной функцией
\begin{equation}                                                  \label{ecfmis}
  \Phi=bq+c,\qquad b,c=\const.
\end{equation}
При этом возможны два качественно отличных случая: $b=0$ и $b\ne0$.

\begin{figure}[h,b,t]
\hfill\includegraphics[width=.8\textwidth]{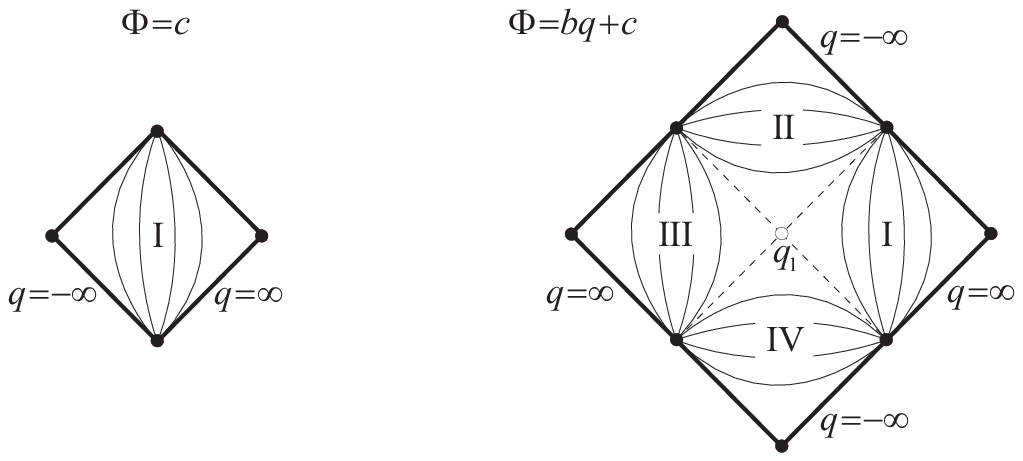}
\hfill {}
\centering\caption{Два представления плоскости Минковского при отсутствии и
 наличии одного горизонта. В последнем случае плоскость Минковского
 склеивается из четырех конформных блоков.}
\label{fminpl}
\end{figure}

При $b=0$ метрика (\ref{emetok}) является метрикой Минковского
\begin{equation}                                                  \label{eminmo}
  ds^2=c(d\tau^2-d\s^2),
\end{equation}
где, для определенности, будем считать, что $c>0$. Конформный множитель $\Phi=c$
не имеет ни особенностей, ни нулей. При этом уравнение (\ref{eshiff}) имеет
общее решение
$$
  q=\pm c\s,
$$
где мы отбросили несущественную постоянную интегрирования, соответствующую
сдвигу $\s$. Знак $\pm$ соответствует различной ориентации переменной $q$
относительно пространственной координаты. Поскольку эта переменная не входит в
метрику, то ее можно исключить из рассмотрения. Поэтому интервалу
$q\in(-\infty,\infty)$ ставится в соответствие один квадратный конформный блок,
показанный на рис.~\ref{fminpl} слева.

Теперь рассмотрим случай $b\ne0$. Не ограничивая общности, положим $b>0$. Тогда
конформный множитель (\ref{ecfmis}) имеет один простой нуль в точке $q_1=-c/b$.
Интервалам $(-\infty,q_1)$ и $(q_1,\infty)$ соответствует по два квадратных
конформных блока, которые склеиваются вдоль горизонтов, как показано на
рис.~\ref{fminpl} справа.

Покажем, как связано это глобальное решение с предыдущим представлением
плоскости Минковского в виде одного конформного блока. Не ограничивая общности,
положим $c=0$, т.е.\
\begin{equation}                                                  \label{ecomph}
  ds^2=b|q|(d\tau^2-d\s^2),\qquad b>0.
\end{equation}
Этого всегда можно добиться путем сдвига $q\rightarrow q+\const$, который не
влияет на глобальную структуру решения. Рассмотрим по очереди все четыре
области.

{\bf Область {\rm I}.} Уравнение (\ref{eshiff}) для $\Phi=bq$ имеем следующее
решение
\begin{equation}                                                  \label{eqsfir}
  q=\ex^{b\s}
\end{equation}
с точностью до сдвига $\s$. Соответствующая метрика статична,
\begin{equation}                                                  \label{emifir}
  ds^2=b\ex^{b\s}(d\tau^2-d\s^2),\qquad \tau,\s\in\MR^2,
\end{equation}
и определена на всей плоскости. Переход к конусным координатам (\ref{elicco}) и
конформное преобразование
\begin{equation}                                                  \label{ectrfi}
  u=\frac2b\ex^{\frac{b\x}2}>0,\qquad v=-\frac2b\ex^{-\frac{b\eta}2}<0
\end{equation}
приводит к метрике Минковского
\begin{equation}                                                  \label{emimec}
  ds^2=b\,dudv,
\end{equation}
определенной в первом квадранте.

{\bf Область {\rm III}}. Для переменной $q$ и метрики справедливы равенства:
\begin{align}                                                     \label{eqsthi}
  q&=\ex^{-b\s},
\\                                                                \label{emithi}
  ds^2&=b\ex^{-b\s}(d\tau^2-d\s^2),\qquad \tau,\s\in\MR^2.
\end{align}
Конформное преобразование, приводящее к метрике Минковского (\ref{emimec}),
имеет вид
\begin{equation}                                                  \label{ectrth}
  u=-\frac2b\ex^{-\frac{b\x}2}<0,\qquad v=\frac2b\ex^{\frac{b\eta}2}>0,
\end{equation}
что соответствует третьему квадранту плоскости Минковского.

{\bf Область {\rm II}}. Эта область однородна:
\begin{align}                                                     \label{eqssec}
  q&=-\ex^{b\tau},
\\                                                                \label{emisec}
  ds^2&=b\ex^{b\tau}(d\tau^2-d\s^2),\qquad \tau,\s\in\MR^2.
\end{align}
Конформное преобразование, приводящее к метрике Минковского (\ref{emimec}),
имеет вид
\begin{equation}                                                  \label{ectrse}
  u=\frac2b\ex^{\frac{b\x}2}>0,\qquad v=\frac2b\ex^{\frac{b\eta}2}>0,
\end{equation}
что соответствует второму квадранту.

{\bf Область {\rm IV}}. Аналогично, для переменной $q$ и метрики справедливы
равенства
\begin{align}                                                     \label{eqsfor}
  q&=-\ex^{-b\tau},
\\                                                                \label{emifor}
  ds^2&=b\ex^{-b\tau}(d\tau^2-d\s^2),\qquad \tau,\s\in\MR^2.
\end{align}
Конформное преобразование, приводящее к метрике Минковского имеет вид
\begin{equation}                                                  \label{ectfor}
  u=-\frac2b\ex^{-\frac{b\x}2}<0,\qquad v=-\frac2b\ex^{-\frac{b\eta}2}<0,
\end{equation}
что соответствует четвертому квадранту.

Для всех четырех областей горизонт $q_1=0$ соответствует координатным прямым
$u=0$ и $v=0$. Простые вычисления показывают, что во всех областях
\begin{equation}                                                  \label{emisqc}
  q=-\frac{b^2}4(t^2-x^2),
\end{equation}
где
$$
  u=t+x,\qquad v=t-x.
$$
Эти координаты дают простейший пример координат Крускала--Секереша для метрики
(\ref{ecomph}). То есть с точностью до знака переменная $q$ представляет собой
квадрат радиуса гиперболической полярной системы координат (\ref{epcmip}) на
плоскости Минковского. Для гиперболического полярного угла в статичных областях
$$
  \vf=\arcth\left(\frac tx\right),
$$
справедливы равенства:
$$
  {\rm I}:\quad \vf=\frac b2\tau,\qquad {\rm III}:\quad \vf=-\frac b2\tau.
$$
В однородных областях
$$
  \vf=\arcth\left(-\frac xt\right),
$$
и
$$
  {\rm II}:\quad \vf=-\frac b2\s,\qquad {\rm IV}:\quad \vf=\frac b2\s.
$$

Поскольку конформные преобразования (\ref{ectrfi}), (\ref{ectrth}),
(\ref{ectrse}) и (\ref{ectfor}) во всех четырех областях дают одну и ту же
метрику Минковского (\ref{emimec}), но в разных квадрантах, то это доказывает
гладкость метрики на горизонтах и в седловой точке для переменной $q$
(\ref{emisqc}).

Таким образом, линейный конформный множитель (\ref{ecfmis}) при $b\ne0$
описывает плоскость Минковского, рис.\ \ref{fminpl} справа, в
гиперболической полярной системе координат.
\subsection{Поверхности постоянной кривизны                      \label{scocus}}
Рассмотрим поверхности постоянной ненулевой кривизны, для которых конформный
множитель является квадратичным полиномом (\ref{ecocsp}). Для определенности,
рассмотрим поверхности положительной скалярной (отрицательной гауссовой)
кривизны, $a>0$. Поверхности отрицательной скалярной кривизны получаются из
поверхностей положительной кривизны преобразованием $\tau\leftrightarrow\s$,
т.е.\ поворотом всех диаграмм на угол $\pi/2$.

В зависимости от значений постоянных $a,b,c$ уравнение $\Phi(q)=0$ может
не иметь решений или иметь один или два корня. Рассмотрим последовательно все
три случая, которые приводят к разным метрикам и различным диаграммам
Картера--Пенроуза для поверхностей постоянной кривизны. Не ограничивая общности,
положим $b=0$, чего всегда можно добиться сдвигом переменной $q$.

\begin{figure}[h,b,t]
\hfill\includegraphics[width=.95\textwidth]{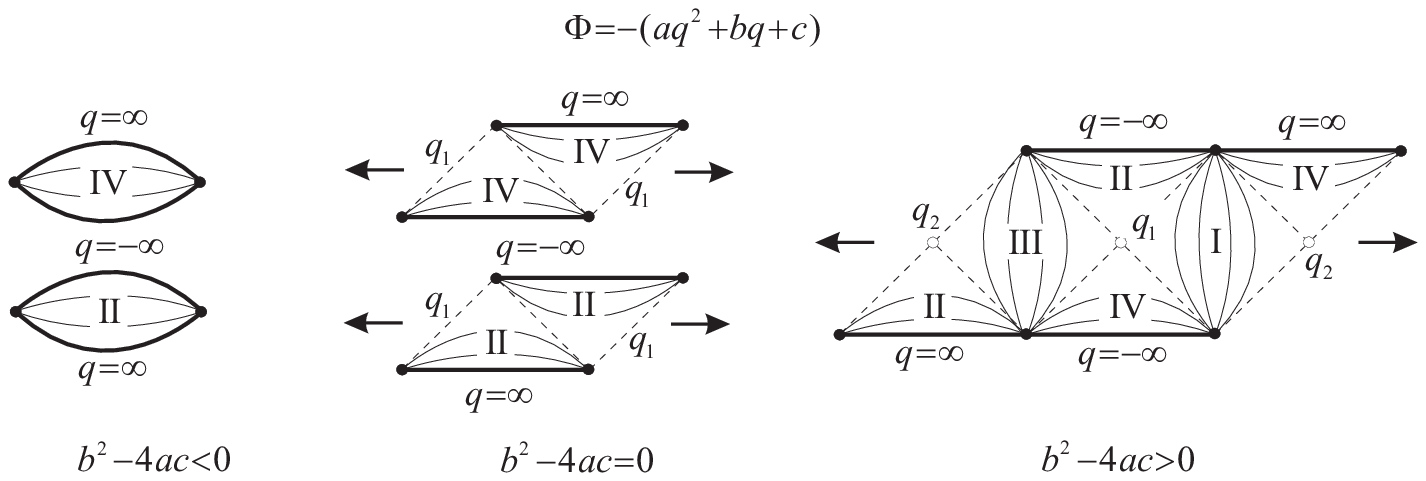}
\hfill {}
\centering\caption{Диаграммы Картера--Пенроуза для поверхностей постоянной
 положительной скалярной кривизны. Возможно три случая: отсутствие, один или два
 горизонта.}
\label{fcocsu}
\end{figure}

{\bf Отсутствие горизонта.} Рассмотрим конформный множитель вида
\begin{equation}                                                  \label{ecocsw}
  \Phi=-(aq^2+c),\qquad a>0,~c>0.
\end{equation}
Поскольку $\Phi<0$, а сингулярности и нули отсутствуют, то глобальное решение
соответствует бесконечному интервалу $q\in(-\infty,\infty)$. Оно представляется
однородным конформным блоком типа линзы II или IV, показанными на
рис.~\ref{fcocsu} слева.
В области IV уравнение (\ref{eshiff}) для конформного множителя (\ref{ecocsw})
приводит к равенству
\begin{equation}                                                  \label{ecjcso}
  q=\sqrt{\frac ca}\,\tg(\sqrt{ac}\,\tau)
\end{equation}
с точностью до сдвига $\tau$. Временн\'ая координата меняется в конечном
интервале
$$
  -\frac\pi2<\sqrt{ac}\,\tau<\frac\pi2,
$$
что соответствует диаграмме типа линзы. Метрика в плоскости $\tau,\s$ имеет вид
\begin{equation}                                                  \label{eccmef}
  ds^2=\frac c{\cos^2(\sqrt{ac}\,\tau)}(d\tau^2-d\s^2).
\end{equation}
Эта метрика является глобальной и определена на всей универсальной накрывающей
постоянной кривизны.

В области II метрика имеет такой же вид, меняется только ориентация переменной
$q$ по отношению к $\tau$.

{\bf Один горизонт.} Пусть конформный множитель имеет вид
\begin{equation}                                                  \label{ecohsw}
  \Phi=-aq^2,\qquad a>0.
\end{equation}
В точке $q_1=0$ он имеет нуль второго порядка, соответствующий горизонту. В
области IV для обоих интервалов $q\in(-\infty,0)$ и $q\in(0,\infty)$ получаем
следующую метрику
\begin{align}                                                     \label{eqfccs}
  q&=-\frac1{a\tau},
\\                                                                \label{eincct}
  ds^2&=\frac1{a\tau^2}(d\tau^2-d\s^2).
\end{align}
Фундаментальная область в этом случае состоит из двух треугольных конформных
блоков, показанных на рис.~\ref{fcocsu} в центре. Чтобы получить универсальную
накрывающую поверхность, диффеоморфную диаграмме типа линзы из предыдущего
случая, фундаментальную область необходимо продолжить в обе стороны. Если
отождествить левую и правую границы фундаментальной области, то получим
поверхность, диффеоморфную однополостному гиперболоиду, рассмотренному в разделе
\ref{shyper}.

{\bf Два горизонта.} Пусть конформный множитель имеет вид
\begin{equation}                                                  \label{ecthsw}
  \Phi=-(aq^2+c),\qquad a>0,~c<0.
\end{equation}
Он имеет два простых нуля в точках $q_{1,2}=\mp\sqrt{-c/a}$, которые
соответствуют горизонтам. Поэтому трем интервалам $(-\infty,q_1)$, $(q_1,q_2)$ и
$(q_2,\infty)$ соответствует шесть конформных блоков. Из симметрии конформного
множителя $q\mapsto-q$ следует, что однородные конформные блоки II, IV и
стационарные конформные блоки I, III связаны преобразованием $q\mapsto-q$.
Для однородной области типа {\rm IV} получаем метрику в виде
\begin{align}                                                     \label{eqthcs}
  q&=-\sqrt{-\frac ca}\,\cth(\sqrt{-ac}\,\tau),
\\                                                                \label{einfth}
  ds^2&=\frac{-c}{\sh^2(\sqrt{-ac}\,\tau)}(d\tau^2-d\s^2).
\end{align}
Для стационарной области {\rm I} метрика имеет вид
\begin{align}                                                     \label{eqshcs}
  q&=\sqrt{-\frac ca}\,\tanh(\sqrt{-ac}\,\s),
\\                                                                \label{eisfth}
  ds^2&=\frac{-c}{\ch^2(\sqrt{-ac}\,\s)}(d\tau^2-d\s^2).
\end{align}

На рис.\ \ref{fcocsu} справа изображены конформные блоки и их склейка
(диаграмма Картера--Пенроуза) для поверхностей постоянной кривизны.

Различные виды метрики при отсутствии или наличии горизонтов обусловлены
различным выбором систем координат. Выпишем формулы преобразования координат,
которые связывают между собой метрики в отсутствие горизонта, а также с одним и
двумя горизонтами. Поскольку все метрики являются вейлевски плоскими, то связь
между ними осуществляется конформным преобразованием координат. Запишем метрику
с одним горизонтом (\ref{eincct}) в конусных координатах (\ref{elicco}) и
совершим конформное преобразование $\x\mapsto u(\x)$, $\eta\mapsto v(\eta)$,
тогда метрика примет вид
\begin{equation}                                                  \label{ecotrm}
  ds^2=\frac1{a\tau^2}(d\tau^2-d\s^2)
  =\frac4{a(\x+\eta)^2}\frac{d\x}{du}\frac{d\eta}{dv}dudv.
\end{equation}
Преобразование конусных координат
\begin{equation}                                                  \label{efcotr}
  \x=\frac2{\sqrt{ac}}\tg\left(\frac{\sqrt{ac}}2u\right),\qquad
  \eta=-\frac2{\sqrt{ac}}\ctg\left(\frac{\sqrt{ac}}2v\right)
\end{equation}
дает метрику (\ref{eccmef}), которая соответствует отсутствию горизонта.
Преобразования конусных координат
\begin{align}                                                     \label{esectx}
  \x&=\frac2{\sqrt{-ac}}\tanh\left(\frac{\sqrt{-ac}}2u\right), &
  \eta&=\frac2{\sqrt{-ac}}\tanh\left(\frac{\sqrt{-ac}}2v\right)
\\ \intertext{и}                                                  \label{etectx}
  \x&=\frac2{\sqrt{-ac}}\tanh\left(\frac{\sqrt{-ac}}2u\right), &
  \eta&=-\frac2{\sqrt{-ac}}\cth\left(\frac{\sqrt{-ac}}2v\right)
\end{align}
приводят к однородной (\ref{einfth}) и статической (\ref{eisfth}) метрикам при
наличии двух горизонтов.

Рассмотренные примеры показывают, что одной и той же поверхности могут
соответствовать различные вейлевски плоские метрики, связанные конформным
преобразованиям координат. Однако независимо от выбора исходной метрики метод
конформных блоков позволяет восстановить всю поверхность единственным образом.
\section{Координаты Эддингтона--Финкельстейна                    \label{sedfic}}
В теореме \ref{tunisp} был описан конструктивный метод конформных блоков
построения максимально продолженных поверхностей с одним вектором Киллинга.
Дифференцируемость метрики внутри каждого конформного блока следует из
построения. Конформные блоки склеиваются вдоль горизонтов $q_j$ (нулей
конформного множителя $\Phi(q_j)=0$) с показателями асимптотики $m=1$ (простой
нуль) и $m\ge2$. В настоящем разделе мы докажем дифференцируемость метрики на
горизонтах после склейки, записав ее в новой системе координат, которая
покрывает горизонты.

Пусть внутри интервала $(q_-,q_+)$ есть нуль $q_j$ нечетного порядка. Для
определенности, будем считать, что в соседних интервалах $(q_{j-1},q_j)$ и
$(q_j,q_{j+1})$ конформный множитель положителен и отрицателен соответственно.
В соответствии с правилом 4) стационарный конформный блок типа I должен быть
склеен с двумя однородными блоками типа II и IV. Соответствующие горизонты можно
накрыть {\em координатами Эддингтона--Финкельстейна} \cite{Edding24A,Finkel58},
которые вводятся следующим образом.
\index{Координаты Эддингтона--Финкельстейна
(Eddington--Finkelstein coordinates)}%
\index{Эддингтона--Финкельстейна координаты
(Eddington--Finkelstein coordinates)}%

{\bf Склейка I--II.} Для доказательства дифференцируемости метрики в склеенных
областях I и II, включая горизонт, введем координаты Эддингтона--Финкельстейна
$\tau,\s\mapsto\x,q$. Эти координаты полностью покрывают указанные области, при
этом функции перехода внутри областей I и II принадлежат классу
${\cal C}^{l+1}$, а метрика в новых координатах -- классу ${\cal C}^l$.

В области I метрика для $q\in(q_{j-1},q_j)$ имеет вид
\begin{equation}                                                  \label{qmutyi}
  {\rm I}:\qquad ds^2=\Phi(q)(d\tau^2-d\s^2),\qquad \frac{dq}{d\s}=\Phi(q)>0.
\end{equation}
перейдем от декартовых координат $\tau,\s$ к координатам
Эддингтона--Финкельстейна $\x,q$:
\begin{equation}                                                  \label{eedfit}
  \tau: = \x-\int^q\!\!\!\frac{dr}{\Phi(r)},\qquad
  \s:=\int^q\!\!\!\frac{dr}{\Phi(r)}.
\end{equation}
Интегралы в этих формулах на горизонте $q\rightarrow q_j$ расходятся, однако это
не важно, т.к.\ преобразование координат (\ref{eedfit}) рассматривается только
во внутренних точках области I. Координата $\x=\tau+\s$ -- это обычная
``светоподобная'' координата на плоскости $\tau,\s$, а в качестве второй
координаты выбрана сама переменная $q$. Эпитет светоподобная взят в кавычки,
т.к.\ вскоре мы увидим, что координатные линии $\x$ в действительности не
являются светоподобными.

Очевидно, что во внутренних точках функции перехода принадлежат классу
${\cal C}^{l+1}$.

Подставляя выражения для дифференциалов
\begin{equation}                                                  \label{edfidt}
  d\tau = d\x-\frac{dq}{\Phi},\qquad d\s=\frac{dq}{\Phi},
\end{equation}
в выражение для метрики (\ref{qmutyi}), получим квадратичную форму
\begin{equation}                                                  \label{edfint}
  ds^2=\Phi d\x^2-2dqd\x.
\end{equation}
Определитель этой метрики равен $\det g_{\al\bt}=-1$, и поэтому метрика
(\ref{edfint}) определена для всех $\x\in(-\infty,\infty)$ и, что важно, при
всех $q\in(q_-,q_+)$, а не только при $q\in(q_{j-1},q_j)$. Вектор $\pl_q$,
касательный к координатным линиям $q$, имеет нулевую длину, $(\pl_q,\pl_q)=0$,
и, следовательно, координатные линии $q$ светоподобны. Вектор $\pl_\x$,
касательный к координатным линиям $\x$, не является светоподобным. Его квадрат
равен $(\pl_\x,\pl_\x)=\Phi$. Он положителен в статичных областях и стремится к
нулю при приближении к горизонту.

\begin{figure}[h,b,t]
\hfill\includegraphics[width=.95\textwidth]{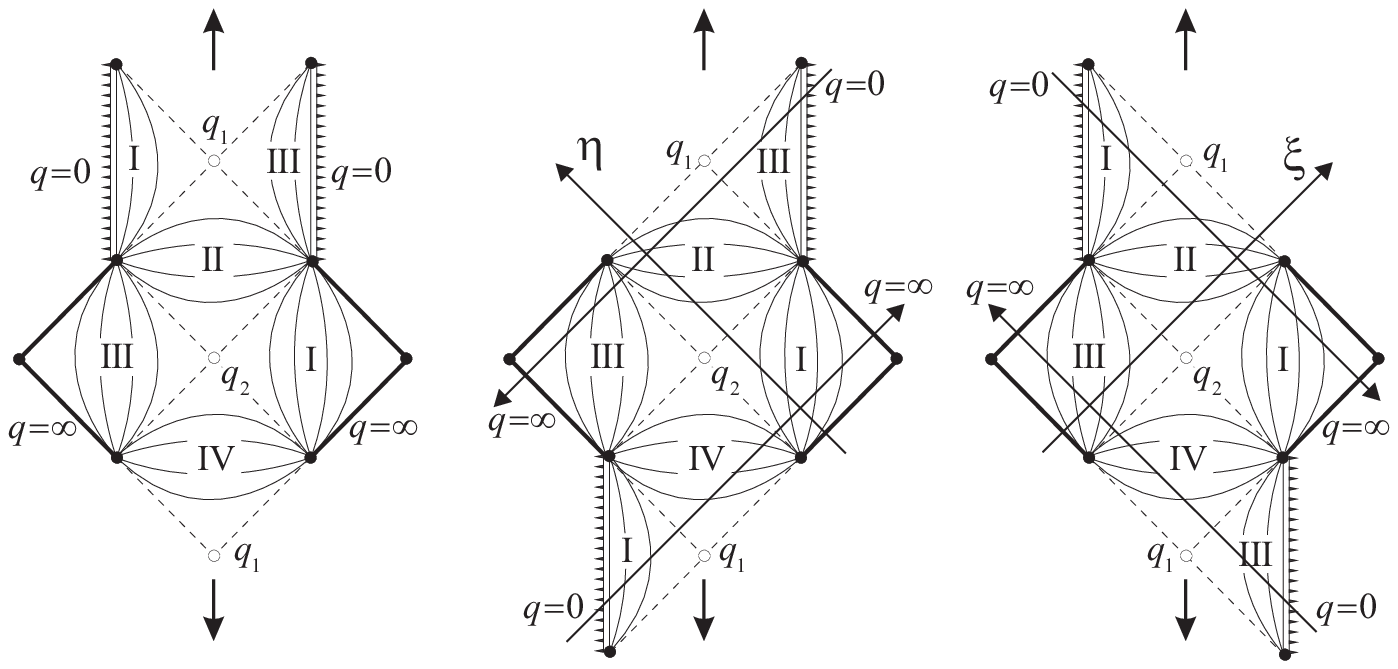}
\hfill {}
\centering\caption{Эквивалентные диаграммы Картера--Пенроуза для решения
Рейснера--Нордстрема. Координаты Эддингтона--Финкельстейна покрывают
диагональные цепочки конформных блоков, идущие снизу слева наверх направо (в
центре) или снизу справа наверх налево (справа).}
\label{feddfink}
\end{figure}

Метрика (\ref{edfint}) определена в большей области, чем исходная метрика
(\ref{emetok}). Эта область -- диагональная цепочка конформных блоков, идущих
сверху слева вниз направо и склеенных вдоль координаты $q$
(см.\ рис.~\ref{feddfink} справа). При этом гладкость метрики (\ref{edfint})
совпадает с гладкостью конформного множителя всюду, включая горизонты.

Отметим, что световые конусы в исходных координатах $\tau,\s$ были такими же,
как и на плоскости Минковского, т.е.\ их образующие -- это прямые линии, идущие
под углом $\pm 45^\circ$. В координатах $\x,q$ так, как они показаны на рисунке,
образующие световых конусов будут отличаться. Одна из образующих будет
параллельна оси $q$, а вторая будет иметь наклон
\begin{equation*}
  \frac{d\x}{dq}=\frac2{1-\frac{2M}q}.
\end{equation*}

В однородной области для $q\in(q_j,q_{j+1})$ типа
\begin{equation}                                                  \label{emetht}
  \text{II}:\quad ds^2=-\Phi(q)(d\tau^2-d\s^2),\qquad \frac{dq}{d\tau}=\Phi(q)<0,
\end{equation}
перейдем к координатам Эддингтона--Финкельстейна $\xi,q$:
\begin{equation}                                                  \label{edfict}
  \tau:=\int^q\!\!\!\frac{dr}{\Phi(r)},\qquad
  \s:=\x-\int^q\!\!\!\frac{dr}{\Phi(r)}.
\end{equation}
Преобразование координат в области II отличается от преобразования
(\ref{eedfit}). Тем не менее метрика (\ref{emetht}) снова принимает вид
(\ref{edfint}). Это означает, что координаты Эддингтона--Финкельстейна и
заданная в них метрика (\ref{edfint}) изометрично покрывают области I и II и,
кроме того, метрика определена на их объединении и горизонте. Это доказывает,
что на горизонте дифференцируемость метрики на продолженной поверхности
совпадает с дифференцируемостью конформного множителя.

Можно также проверить, что горизонт $q_j$ в координатах
Эддингтона--Финкельстейна сам является светоподобной экстремалью.

Метрика (\ref{edfint}) в координатах Эддингтона--Финкельстейна имеет вектор
Киллинга $\pl_\x$. Квадрат его длины равен конформному множителю
$(\pl_\x,\pl_\x)=\Phi(q)$. Поэтому на горизонте длина вектора Киллинга равна
нулю.

{\bf Склейка III--IV.} Области {\rm III}--{\rm IV} склеиваются аналогично
областям I и II. Приведем явные формулы перехода к координатам
Эддингтона--Финкельстейна:
\begin{equation}                                                  \label{eedfth}
{\rm III}:\qquad
  \tau:=\x+\int^q\!\!\!\frac{dr}{\Phi(r)},\qquad
  \s:=-\int^q\!\!\!\frac{dr}{\Phi(r)}.
\end{equation}
и
\begin{equation}                                                  \label{edfifo}
{\rm IV}:\qquad
  \tau:=-\int^q\!\!\!\frac{dr}{\Phi(r)},\qquad
  \s:=\x+\int^q\!\!\!\frac{dr}{\Phi(r)}.
\end{equation}
При этом в обеих областях метрика имеет одинаковый вид
\begin{equation}                                                  \label{edfitf}
  ds^2=\Phi d\x^2+2dqd\x.
\end{equation}
Эта метрика определена на всей диагональной цепочке конформных блоков
$q\in(q_-,q_+)$, идущей снизу справа наверх налево (см.\ рис.~\ref{feddfink}
справа).

Между собой метрики (\ref{edfint}) и (\ref{edfitf}) связаны преобразованием
$\x\mapsto-\x$.

{\bf Склейка I--IV.} В области I перейдем к координатам
Эддингтона--Финкельстейна $\eta,q$:
\begin{equation}                                                  \label{edfiff}
  {\rm I}:\qquad \tau:= \eta+\int^q\!\!\!\frac{dr}{\Phi(r)},\qquad
  \s:=\int^q\!\!\!\frac{dr}{\Phi(r)},
\end{equation}
где $\eta=\tau-\s$ -- светоподобная координата. В новых координатах метрика
принимает вид
\begin{equation}                                                  \label{eintff}
  ds^2= \Phi d\eta^2+2dqd\eta.
\end{equation}
Эта метрика определена на всей диагональной цепочке конформных блоков
$q\in(q_-,q_+)$, идущей снизу слева наверх направо (см.\ рис.~\ref{feddfink} в
центре).

К этой же метрике в области IV приводят координаты
\begin{equation}                                                  \label{edffou}
  {\rm IV}:\qquad \tau:= -\int^q\!\!\!\frac{dr}{\Phi(r)},\qquad
  \s:=-\eta-\int^q\!\!\!\frac{dr}{\Phi(r)}.
\end{equation}

Аналогичные координаты вводятся при склейке областей II и III, при этом
соответствующая метрика связана с метрикой (\ref{eintff}) преобразованием
$\eta\mapsto-\eta$.

Если горизонт $q_j$ имеет четный порядок, и в обоих интервалах $(q_{j-1},q_j)$ и
$(q_j,q_{j+1})$ конформный множитель, например, положителен, то, согласно
правилу 4), склеиваются только однотипные блоки. Для доказательства
дифференцируемости склейки на блоках типа I и III вводятся координаты
(\ref{eedfit}) и (\ref{eedfth}), соответственно.

Таким образом доказывается дифференцируемость метрики и, следовательно, символов
Кристоффеля и тензора кривизны на всех горизонтах. Метрика в координатах
Эддингтона--Финкельстейна (\ref{edfint}) при $q\in(q_-,q_+)$ покрывает всю
диагональную цепочку конформных блоков типа I и II для фундаментальной области.
Для этой цепочки блоков координата $q$ убывает с возрастанием времени $\tau$.
Цепочка блоков представляет собой многообразие класса $\CC^{l+1}$, т.к.\ функции
перехода (\ref{eedfit}), (\ref{edfict}) принадлежат этому классу. Метрика
(\ref{edfitf}) покрывает параллельную цепочку блоков типа III и IV. Метрика в
виде (\ref{eintff}) покрывает перпендикулярную диагональную цепочку конформных
блоков типа I и IV. Внутри каждой цепочки метрика в координатах
Эддингтона--Финкельстейна невырождена и класса $\CC^l$.

Если внутри интервала $(q_-,q_+)$, который соответствует глобальному решению,
имеется $n$ нулей, то для получения фундаментальной области необходимо склеить
$2(n+1)$ конформных блоков (удвоенное число интервалов $(q_j,q_{j+1})$). Их
можно склеить в две диагональных цепочки блоков, например, идущих снизу слева
наверх направо, как показано на рис.~\ref{feddfink} в центре. В каждой цепочке
склеиваются между собой либо только блоки I, IV, либо только блоки II, III.
По-построению, каждая цепочка является связным многообразием. Если внутри
интервала есть хотя бы один нуль нечетного порядка, то вокруг него происходит
склейка цепочек, как для решения Шварцшильда. При этом достаточно одного нуля
нечетного порядка, чтобы превратить фундаментальную область в связное
многообразие. Если же конформный множитель внутри интервала $(q_-,q_+)$ знак не
меняет, то диагональные цепочки блоков представляет собой две компоненты
связности фундаментальной области. Это доказывает утверждение в правиле 5).

Нетрудно проверить, что переход от декартовых координат $\tau,\s$ к координатам
Эддингтона--Финкельстейна вырожден на всех горизонтах.

Координаты Эддингтона--Финкельстейна являются естественными в следующем смысле.
Прежде всего отметим, что продолжать локальное решение надо по переменной $q$, т.к.\
именно она определяет полноту многообразия и сингулярности кривизны. Поскольку
для стационарного решения переменная $q$ зависит от $\s$, то вместо $\s$ можно
выбрать само $q$ (\ref{eedfit}), (\ref{eedfth}), (\ref{edfiff}), при этом связь
координат определяется уравнением (\ref{eshiff}) взаимно однозначно для каждого
типа области с точностью до несущественного сдвига $\s$. После этого вместо
временн\'ой координаты вводится светоподобная координата (\ref{elicco}). При
этом существует две возможности, и обе они реализуются, задавая координаты на
взаимно перпендикулярных диагональных цепочках конформных блоков. Аналогично
вводятся координаты Эддингтона--Финкельстейна на однородных блоках.
\section{Дифференцируемость метрики в седловой точке             \label{sedpom}}
Координаты Эддингтона--Финкельстейна не покрывают седловых точек, расположенных
в местах пересечения горизонтов. Седловые точки соответствуют нулям конформного
множителя $\Phi(q_j)=0$, расположенным в конечных точках, $|q_j|<\infty$ с
показателем степени $m=1$ или $m\ge2$ (см.\ рис.~\ref{fboprr}). Если точка $q_j$
является нулем кратности два или больше, то седловая точка полна, и доказывать
ничего не надо. Если точка $q_j$ -- простой нуль, то можно ввести координаты,
покрывающие окрестность седловой точки. Поскольку поведение конформного
множителя вблизи простого нуля такое же, как и на плоскости Минковского, то
перейдем к светоподобным {\em координатам Крускала--Секереша}
$\tau,\s\mapsto u,v$. Явные формулы преобразования в областях I--IV имеют вид
(\ref{ectrfi}), (\ref{ectrse}), (\ref{ectrth}) и (\ref{ectfor}), соответственно.
При этом мы положим $b:=\Phi'(q_j)$ (см.\ раздел \ref{sminpl}).

Рассмотрим область I более подробно. Метрика имеет вид
\begin{equation*}
  ds^2=\Phi(d\tau^2-d\s^2)=\Phi d\xi d\eta,
\end{equation*}
где $\xi,\eta$ -- светоподобные координаты (\ref{elicco}). Преобразуем
светоподобные координаты $\xi,\eta\mapsto u,v$ по формулам (\ref{ectrfi}). Тогда
метрика примет вид
\begin{equation*}
  ds^2=\Phi\ex^{-b\s}dudv=\Phi\exp\left(-b\int^q\frac{dp}{\Phi(p)}\right)dudv,
\end{equation*}
где мы учли связь $\s$ с $q$, которая дается уравнением
\begin{equation*}
  \frac{dq}{d\s}=\Phi.
\end{equation*}

Проведем аналогичное преобразование координат во всех четырех квадрантах и
везде положим $b:=\Phi'(q_j)$. Тогда во всех четырех квадрантах метрика
(\ref{emetok}) примет одинаковый вид
\index{Координаты Крускала--Секереша (Cruskal--Sekeres coordinates)}%
\index{Крускала--Секереша координаты (Cruskal--Sekeres coordinates)}%
\begin{equation}                                                  \label{emesap}
  ds^2=|\Phi|\exp\left(-\int^q\!\!\! dp\frac{\Phi'(q_j)}{\Phi(p)}\right)\,dudv
  =\exp\left(\int^q\!\!\! dp\frac{\Phi'(p)-\Phi'(q_j)}{\Phi(p)}\right)\,dudv.
\end{equation}

Как и в случае плоскости Минковского координаты $u,v$ покрывают окрестность
седловой точки, состоящую из областей всех четырех типов, склеенных вдоль
горизонтов.

Вблизи седловой точки при $m=1$ справедливо разложение
\begin{equation*}
  \Phi(p)=\Phi_1(p-q_j)+\frac12\Phi_2(p-q_j)^2+\osmall\big((p-q_j)^2\big),
  \qquad\Phi_{1,2}=\const,\quad \Phi_1\ne0.
\end{equation*}
Поэтому
\begin{equation*}
  \underset{p\to q_j}\lim\frac{\Phi'(p)-\Phi'(q_j)}{\Phi(p)}
  =\frac{\Phi_2}{\Phi_1}.
\end{equation*}
Поскольку в седловой точке предел подынтегрального выражения конечен, то
интеграл является достаточно гладкой функцией. Это доказывает невырожденность и
дифференцируемость метрики в седловой точке, которая совпадает с
дифференцируемостью конформного множителя.

Очевидно, что вблизи каждой седловой точки первого порядка можно ввести
координаты Крускала--Секереша. Поскольку функции перехода к координатам
Крускала--Секереша класса $\CC^\infty$, то класс гладкости универсальной
накрывающей поверхности определяется гладкостью функций перехода к координатам
Эддингтона--Финкельстейна, $\CC^{l+1}$. Тем самым доказательство основной
теоремы \ref{tunisp} завершено.

Таким образом, мы показали, что локальное решение вида (\ref{emetok}) имеет
дифференцируемое максимальное продолжение. Описанная процедура является
однозначной. Однако, это не доказывает единственность продолжения. Дело в том,
что преобразование координат к координатам Эддингтона--Финкельстейна и
Крускала--Секереша вырождено на горизонтах. Поэтому нельзя утверждать, что
указанное продолжение решения вдоль экстремалей является единственным.

Метод построения глобальных решений для двумерных метрик лоренцевой сигнатуры,
рассмотренный в настоящей главе, является простым и конструктивным. Если
уравнения теории гравитации приводят к метрике (\ref{emetok}) с некоторым
конформным множителем, то для построения глобального решения достаточно
проанализировать его нули и особенности. При этом восстанавливается
универсальная накрывающая поверхность. Остальные глобальные решения являются
факторпространствами универсальной накрывающей по группе преобразований,
действующей собственно разрывно и свободно. Выше мы отметили возможность
построения решений с нетривиальной фундаментальной группой путем отождествления
элементов границы (горизонтов) фундаментальной области. В общем случае
нахождение групп преобразований, действующих собственно разрывно и свободно,
зависит от конкретного вида конформного множителя и является предметом
самостоятельного исследования.
\chapter{Римановы поверхности с одним вектором Киллинга}
Римановы поверхности с одним вектором Киллинга включают в себя как частный
случай римановы поверхности постоянной кривизны, которые имеют по три вектора
Киллинга. Этот более широкий класс поверхностей имеет важное значение для
общей теории относительности, т.к.\ позволяет дать физическую интерпретацию
многим точным решениям уравнений Эйнштейна.
\section{Локальный вид римановой метрики                         \label{slovri}}
Чтобы преобразовать лоренцеву метрику (\ref{emetok}) к метрике евклидовой
сигнатуры, совершим поворот в комплексной плоскости той из координат $\tau$ или
$\s$, от которой конформный множитель не зависит. Поскольку метрика вида
(\ref{emetok}), по-предположению, возникает в результате решения системы
уравнений движения для некоторой модели гравитации, то соответствующая риманова
метрика также будет являться локальным решением той же системы уравнений после
поворота в комплексной плоскости.

В явном виде комплексный поворот задается следующим образом. Введем новую
координату $\tau:=i\rho$ в статичных областях I, III и $\s:=i\rho$ -- в
однородных областях II, IV. В результате получим метрику
\begin{equation}                                                  \label{eemtvi}
  ds^2=-\Phi(q)(d\s^2+d\rho^2),
\end{equation}
где в областях II и IV мы переобозначили координату $\tau\mapsto\s$. Знак
конформного множителя $\Phi(q)$ не фиксирован, т.е.\ рассматриваются как
положительно, так и отрицательно определенные метрики. После комплексного
поворота аргумент конформного множителя $q$ зависит только от $\s$, и эта связь
задается обыкновенным дифференциальным уравнением
\begin{equation}                                                  \label{emdmee}
  \left|\frac{dq}{d\s}\right|=\left|\Phi(q)\right|.
\end{equation}
Знаки модуля в этом уравнении раскрываются следующим образом:
\begin{equation}                                                  \label{edomva}
\begin{aligned}
  {\rm I:}   && \Phi>0,\qquad dq/d\s&>0, & \quad\sign g_{\al\bt}&=(--), \\
  {\rm II:}  && \Phi<0,\qquad dq/d\s&<0, & \sign g_{\al\bt}&=(++), \\
  {\rm III:} && \Phi>0,\qquad dq/d\s&<0, & \sign g_{\al\bt}&=(--), \\
  {\rm IV:}  && \Phi<0,\qquad dq/d\s&>0, & \sign g_{\al\bt}&=(++),
\end{aligned}
\end{equation}
где в последней колонке показана сигнатура метрики (\ref{eemtvi}).

Как и случае метрики лоренцевой сигнатуры, у нас есть четыре области: две
области с положительно определенной и две области с отрицательно определенной
метрикой. Метрики в областях I и III, а также II и IV по существу одинаковы,
т.к.\ связаны между собой простым преобразованием $\s\mapsto-\s$.

Метрика вида (\ref{eemtvi}) обладает по-крайней мере одним вектором Киллинга
$K=\pl_\rho$, квадрат длины которого равен $(\pl_\rho,\pl_\rho)=-\Phi(q)$.

Риманова метрика, заданная уравнениями (\ref{eemtvi}), (\ref{emdmee}), и будет
предметом исследования в настоящем разделе. Мы допускаем, что конформный
множитель может обращаться в нуль и иметь особенности в конечном числе точек
$q_i$, $i=1,\dots,k$. В эту последовательность включены также бесконечно
удаленные точки $q_1=-\infty$ и $q_k=\infty$. Таким образом, вещественная ось
$q$ разбивается точками $q_i$ на интервалы, внутри которых конформный множитель
либо строго больше, либо строго меньше нуля. Будем считать, что вблизи граничных
точек $q_i$ конформный множитель ведет себя, как и в лоренцевом случае,
степенным образом (\ref{ecfapo}), (\ref{ecfasi}). При конечных значениях $q_i$
показатель степени не равен нулю, $m\ne0$, так как конформный множитель в этой
точке, по-предположению, либо равен нулю, либо сингулярен.

Символы Кристоффеля имеют следующие нетривиальные компоненты:
\begin{align}                                                     \label{ecrson}
  {\rm I, II}:&\qquad \Gamma_{\s\s}{}^\s=\Gamma_{\s\rho}{}^\rho=\Gamma_{\rho\s}{}^\rho
  =-\Gamma_{\rho\rho}{}^\s=\frac12\Phi',
\\                                                                \label{ecrstw}
  {\rm III, IV}:&\qquad \Gamma_{\s\s}{}^\s=\Gamma_{\s\rho}{}^\rho=\Gamma_{\rho\s}{}^\rho
  =-\Gamma_{\rho\rho}{}^\s=-\frac12\Phi',
\end{align}
где штрих обозначает дифференцирование по аргументу, $\Phi':=d\Phi/dq$, и
никакого суммирования по индексам $\rho$ и $\s$ не проводится. Тензор кривизны
одинаков для всех областей и имеет только одну независимую компоненту
\begin{equation}                                                  \label{eriete}
  R_{\s\rho\s}{}^\rho=\frac12\Phi\Phi''
\end{equation}
Тензор Риччи диагонален. Его ненулевые компоненты и скалярная кривизна равны
\begin{align}                                                     \label{eriecl}
  R_{\s\s}&=R_{\rho\rho}=\frac12\Phi\Phi'',
\\                                                                \label{ecusce}
  R&=-\Phi''.
\end{align}
Отсюда следует, что для поверхностей постоянной кривизны, $R=\const$, как и в
лоренцевом случае, конформный множитель $\Phi$ является квадратичным полиномом.

Из уравнения (\ref{ecusce}) следует, что скалярная кривизна сингулярна вблизи
граничной точки $q_i$ при тех же показателях (\ref{esfpsc}), (\ref{esfpcc}), что
и в лоренцевом случае. В граничных точках $q_i=\pm\infty$ скалярная кривизна
стремится к отличной от нуля постоянной при $m=2$ и к нулю при $m<2$. Отметим,
что ненулевое значение кривизны в конечной точке $|q_i|<\infty$ может возникнуть
и при $m=1$ за счет поправок следующего порядка в разложении (\ref{ecfapo}).

Значение переменной $q$ и, значит, скалярной кривизны постоянно вдоль траекторий
Киллинга $\s=\const$. В соответствии с определением областей (\ref{edomva}) $q$
монотонно возрастает по $\s$ в областях I, IV и монотонно убывает в областях II,
III.

Область определения метрики (\ref{eemtvi}) на плоскости $\s,\rho$ зависит от
вида конформного множителя. Координата $\rho\in\MR$ пробегает всю вещественную
прямую, поскольку от нее ничего не зависит, а область изменения координаты $\s$
определяется уравнением (\ref{emdmee}). Каждому интервалу переменной
$q\in(q_i,q_{i+1})$ соответствует конечный, полубесконечный или бесконечный
интервал координаты $\s$ в зависимости от сходимости или расходимости интеграла
\begin{equation}                                                  \label{eferin}
  \s_{i,i+1}\sim\int^{q_i,q_{i+1}}\!\!\!\frac{dq}{\Phi(q)}
\end{equation}
в граничных точках. Сходимость интеграла определяется показателем степени $m$:
\begin{equation}                                                  \label{eboucl}
\begin{array}{rl}
  |q_i|<\infty:\qquad &\left\lbrace
  \begin{array}{rll}
  m<1,  \quad &\text{сходится,}  \\
  m\ge1,\quad &\text{расходится,}
  \end{array}\right. \\
  |q_i|=\infty:\qquad &\left\lbrace
  \begin{array}{rll}
  m\le1,\quad &\text{расходится,}\\
  m>1,  \quad &\text{сходится,}
  \end{array}\right.
\end{array}
\end{equation}
Если на обоих концах интервала $(q_i,q_{i+1})$ интеграл расходится, то
координата $\s$ пробегает всю вещественную ось, $\s\in(-\infty,\infty)$, и
метрика определена на всей плоскости $\s,\rho\in\MR^2$. Если на одном из концов
$q_{i+1}$ или $q_i$ интеграл сходится, то метрика определена на полуплоскости
$\s\in(-\infty,\s_{i+1})$ или $\s\in(\s_i,\infty)$, соответственно. При этом
выбор граничных точек $\s_{i+1}$ и $\s_i$ произволен, и, не ограничивая
общности, можно положить $\s_{i,i+1}=0$. Если на обоих концах интервала интеграл
сходится, то локальное решение определено в полосе $\s\in(\s_i,\s_{i+1})$, при
этом можно приравнять нулю только один из концов интервала.

В лоренцевом случае для построения максимально продолженных поверхностей каждому
интервалу $(q_i,q_{i+1})$ ставятся в соответствие конформные блоки определенной
формы, которые затем склеиваются между собой. При евклидовой сигнатуре метрики в
этом нет необходимости, поскольку ``светоподобные'' экстремали с асимптотикой
$\rho=\pm\s+\const$, как будет показано ниже, в общем случае отсутствуют.
\section{Экстремали                                              \label{sexesm}}
Для того чтобы описать максимально продолженную поверхность для римановой
метрики (\ref{eemtvi}) с одним вектором Киллинга, необходимо проанализировать
поведение экстремалей $\lbrace \z^\al(t)\rbrace=\lbrace\s(t),\rho(t)\rbrace$,
$t\in\MR$, которые задаются системой нелинейных обыкновенных дифференциальных
уравнений (\ref{qexsda}), где точка обозначает дифференцирование по
каноническому параметру $t$.

Для определенности, рассмотрим область I. Используя выражение для символов
Кристоффеля (\ref{ecrson}), получаем систему уравнений для экстремалей:
\begin{align}                                                     \label{eqexte}
  \ddot\s  &=\frac12\Phi'(\dot\rho^2-\dot\s^2),
\\                                                                \label{eqexty}
  \ddot\rho&= -\Phi'\dot\s\dot\rho.
\end{align}
Эта система уравнений имеет два первых интеграла:
\begin{align}                                                     \label{efiice}
  -\Phi(\dot\s^2+\dot\rho^2)&=C_0=\const,
\\                                                                \label{efisce}
  -\Phi\dot\rho&=C_1=\const.
\end{align}
Существование интеграла движения (\ref{efiice}) позволяет выбрать длину
экстремали в качестве канонического параметра. Второй интеграл (\ref{efisce})
связан с наличием вектора Киллинга. На этом этапе проявляется важность наличия
у метрики одного вектора Киллинга, т.к.\ существование двух интегралов позволяет
проанализировать экстремали в общем случае, не конкретизируя вид конформного
множителя $\Phi(q)$.

В системе уравнений для экстремалей (\ref{eqexte}), (\ref{eqexty}) конформный
множитель $\Phi$ рассматривается, как сложная функция от $\s$:
$\Phi=\Phi\big(q(\s)\big)$, где $q(\s)$ -- решение уравнения (\ref{emdmee}).
Если форма экстремали $\s(\rho)$ найдена, то конформный множитель можно
рассматривать также как сложную функцию от $\rho$ вдоль каждой экстремали:
$\Phi=\Phi\big(\s(\rho)\big)$. В дальнейшем анализе это будет подразумеваться.
\begin{theorem}                                                   \label{textre}
Любая экстремаль в области I принадлежит одному из следующих четырех классов.
\newline
1. Прямые экстремали вида (аналог светоподобных экстремалей)
\begin{equation}                                                  \label{extlie}
  \rho=\pm\s+\const,
\end{equation}
существуют только для евклидовой метрики $\Phi=\const$, при этом канонический
параметр можно выбрать в виде $\s=t$.\\
2. Экстремали общего вида, форма которых определена уравнением
\begin{equation}                                                  \label{extgee}
  \frac{d\rho}{d\s}=\pm\frac1{\sqrt{-1-C_2 \Phi}},
\end{equation}
где $C_2<0$ -- отрицательная постоянная. Соответствующий канонический параметр
определяется любым из двух уравнений:
\begin{align}                                                     \label{exttpe}
  \dot\s  &=\pm\frac{\sqrt{-1-C_2\Phi}}\Phi,
\\                                                                \label{extgpe}
  \dot\rho&=\frac1\Phi.
\end{align}
При этом в уравнениях (\ref{extgee}) и (\ref{exttpe}) знаки плюс или минус
выбираются одновременно.\newline
3. Прямые экстремали, параллельные оси $\s$ и проходящие через каждую точку
$\rho=\const$. Канонический параметр определен уравнением
\begin{equation}                                                  \label{extste}
  \dot\s=\frac1{\sqrt \Phi}.
\end{equation}
4. Прямые вырожденные экстремали, параллельные оси $\rho$ и проходящие через
критические точки $\s_0=\const$, в которых
\begin{equation}                                                  \label{extdce}
  \Phi'(\s_0)=0.
\end{equation}
Канонический параметр для них можно выбрать в виде
\begin{equation}                                                  \label{extdge}
  t=\rho.
\end{equation}
\end{theorem}
\begin{proof}
Почти дословное повторение доказательства теоремы \ref{textrf} в лоренцевом
случае.
\end{proof}
Постоянная $C_2$, как и в лоренцевом случае, определяется интегралами
(\ref{efiice}) и (\ref{efisce}):
\begin{equation}                                                  \label{econcl}
  C_2:=\frac{C_0}{C_1^2}.
\end{equation}
Для области I $\Phi>0$ и, следовательно, $C_0<0$ и $C_2<0$. Из уравнения
(\ref{extgee}) следует, что постоянная $C_2$ параметризует угол, под которым
экстремаль общего вида проходит через заданную точку.

Подчеркнем качественное отличие в поведении экстремалей для метрик евклидовой и
лоренцевой сигнатуры. Во-первых, для римановой метрики в общем случае
отсутствует аналог светоподобных экстремалей. Это важно, так как в лоренцевом
случае светоподобные экстремали были неполны на горизонтах и их необходимо было
продолжить. Для римановых метрик эта проблема снимается. Во-вторых, уравнения
для экстремалей общего вида (\ref{extgee}) и (\ref{exttpe}) отличаются от
соответствующих уравнений в лоренцевом случае знаком перед единицей, стоящей под
знаком корня. Это небольшое на первый взгляд отличие приводит к тому, что у
экстремалей общего вида уже нет светоподобной асимптотики при $q\rightarrow q_i$
вблизи нулей конформного множителя $\Phi(q_i)=0$, которые определяют горизонты.

\begin{figure}[p]
\hfill\includegraphics[width=.95\textwidth]{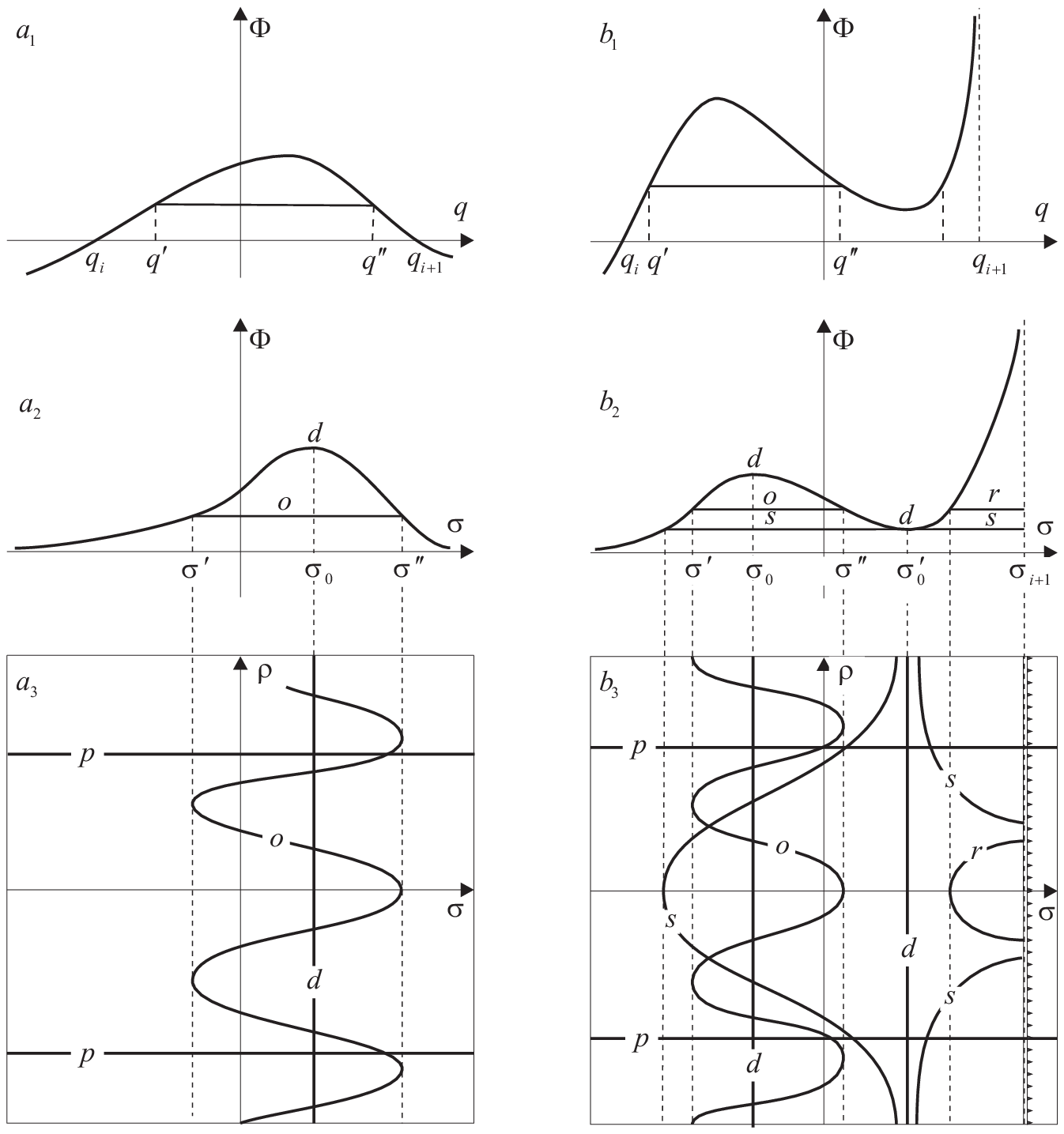}
\hfill {}
\centering\caption{Верхний ряд ($a_1$, $b_1$): типичное поведение конформного
множителя $\Phi(q)$ между двумя нулями и нулем и сингулярностью. Средний ряд
($a_2$, $b_2$): зависимость конформного множителя $\Phi(\s)$ от координаты $\s$.
Точка $\s_{i+1}$ может находиться как на конечном, так и на бесконечном
расстоянии от начала координат. Нижний ряд ($a_3$, $b_3$): типичные экстремали
общего вида при разных значениях постоянной $C_2$. Вблизи локального максимума
экстремали ($o$) осциллируют между $\s'$ и $\s''$. Через каждый локальный
экстремум проходит вырожденная экстремаль ($d$). Существуют экстремали ($s$),
которые асимптотически приближаются к вырожденной экстремали в локальном
минимуме $\s'_0$ и могут заканчиваться на сингулярной границе $\s_{i+1}$.
Экстремали ($r$) конечной длины начинаются и заканчиваются на сингулярной
границе. Все экстремали можно произвольно сдвигать вдоль оси $\rho$. Через
каждую точку $\rho$ проходит также экстремаль ($p$), параллельная оси $\s$.}
\label{fcoext}
\end{figure}

Качественное поведение экстремалей общего вида анализируется в общем случае, для
чего достаточно знать поведение конформного множителя $\Phi(q)$ вблизи граничных
точек $q_i$. Поскольку конформный множитель в области I положителен, то
экстремали общего вида существуют только при отрицательных значениях постоянной
$C_2$, т.к.\ в противном случае правая часть уравнения (\ref{extgpe}) становится
мнимой. При этом должно быть выполнено неравенство
$$
  \Phi(q)\ge-1/C_2.
$$
При достаточно больших значениях модуля $C_2$ это неравенство определяет
интервал изменения переменной $q\in(q',q'')$, где граничные точки $q'$ и $q''$
определяются уравнением $\Phi(q)=-1/C_2$. Точкам $q',q''$ соответствуют
некоторые точки $\s'$ и $\s''$. Экстремаль общего вида не может выйти за пределы
полосы $\s\in(\s',\s'')$, $\rho\in(-\infty,\infty)$. Несложный анализ уравнения
(\ref{extgpe}) показывает, что экстремали общего вида осциллируют между
значениями $\s'$ и $\s''$, как показано на рис.\ref{fcoext} $a_3$, $b_3$.
Осциллирующие экстремали всегда полны, поскольку правая часть уравнения
(\ref{extgpe}) ограничена сверху значением $1/\Phi(q')$ или $1/\Phi(q'')$ и
снизу $1/\max \Phi(q)$.

Если конформный множитель $\Phi(q)$ равен бесконечности в точке $q_{i+1}$, как
показано на рис.\ref{fcoext} $b$, то экстремаль общего вида может начинаться и
заканчиваться на границе, где $\Phi(q_{i+1})=\infty$. Эта граница соответствует
конечному значению $\s_{i+1}$ при $|q_{i+1}|<\infty$, $m<1$ и
$|q_{i+1}|=\infty$, $m>1$. При этом все экстремали подходят к границе
$\Phi=\infty$ под прямым углом, т.к.\ правая часть уравнения (\ref{extgee})
стремится к нулю. Полнота экстремалей общего вида, подходящих к границе
$\Phi=\infty$, определяется интегралом
\begin{equation}                                                  \label{extgpt}
  \underset{q\to q_{i+1}}\lim t\to\int^{q_{i+1}}\!\!\!\frac{dq}{\sqrt{\Phi}},
\end{equation}
что следует из уравнений (\ref{exttpe}) и (\ref{emdmee}). Тем самым полнота
экстремалей общего вида, подходящих к сингулярной границе, такая же, как и
прямых экстремалей, параллельных оси $\s$ (\ref{extste}). Они полны только при
$|q_{i+1}|=\infty$ и $1<m\le2$.

В лоренцевом случае в области I к границе $\Phi=\infty$ подходят только прямые и
пространственноподобные экстремали (см.\ раздел \ref{sextsm}). Их полнота
связана с показателем степени $m$ так же, как и для римановой метрики.

На рис.\ref{fcoext} в верхнем ряду ($a_1$, $b_1$) показано типичное поведение
конформного множителя с одним экстремумом между двумя нулями и с двумя
локальными экстремумами между нулем и особенностью. В последнем случае мы
предполагаем, что $|q_{i+1}|<\infty$ и $m<0$, что соответствует сингулярной
кривизне. В среднем ряду ($a_2$, $b_2$) представлена зависимость этих конформных
множителей от координаты $\s$. На рис.\ref{fcoext}, $b_2$ значение координаты
$\s_{i+1}$ конечно, что следует из сходимости интеграла (\ref{eferin}). В
нижнем ряду ($a_3$, $b_3$) показано качественное поведение экстремалей на
$\s,\rho$ плоскости. Через каждый локальный экстремум конформного множителя
проходит вырожденная экстремаль $(d)$. Экстремали общего вида $(o)$ осциллируют
вблизи локального максимума $\Phi(\s_0)$ между значениями $\s'$ и $\s''$,
которые определяются постоянной $C_2$. Экстремали общего вида $(r)$ могут
также начинаться и заканчиваться на границе $\Phi(\s_{i+1})=\infty$,
имея при этом конечную длину. Существуют также экстремали общего вида $(s)$,
которые асимптотически приближаются при $\rho\to\pm\infty$ к вырожденной
экстремали, проходящей через локальный минимум $\s'_0$. Часть из этих
экстремалей заканчивается на границе $\Phi(\s_{i+1})=\infty$ при конечном
значении канонического параметра.

Все экстремали можно произвольно сдвигать вдоль оси $\rho$. Через каждую точку
$\rho$ проходит также экстремаль $(p)$, параллельная оси $\s$.

Если обе граничные точки интервала $q_i$ и $q_{i+1}$ являются нулями, то из
непрерывности конформного множителя $\Phi$ следует, что у него есть по крайней
мере один экстремум, через который проходит вырожденная экстремаль ($d$).
Вырожденные экстремали всегда полны, т.е.\ имеют бесконечную длину, т.к.\
канонический параметр совпадает с координатой $\rho$ (\ref{extdge}).

Проведенный анализ экстремалей показывает, что неполнота экстремалей в полосе
$\s\in(\s_i,\s_{i+1})$, $\rho\in(-\infty,\infty)$ полностью определяется
поведением экстремалей, параллельных оси $\s$, при подходе к граничным точкам
$\s_i,\s_{i+1}$. Что, в свою очередь, определяется сходимостью интеграла
(\ref{extgpt}). Эти экстремали неполны в конечных точках $|q_i|<\infty$ при
$m<2$. В бесконечно удаленных точках $|q_i|=\infty$ они неполны при $m>2$ и
полны во всех остальных случаях. Учитывая, что продолжать поверхность необходимо
только при конечных значениях кривизны, приходим к выводу, что продолжение
поверхности необходимо только в случае простого нуля, $m=1$, у конформного
множителя в конечной точке $|q_i|<\infty$. Напомним, что простой нуль
конформного множителя в лоренцевом случае соответствует горизонту.
\section{Построение глобальных решений                           \label{sgloeu}}
Чтобы построить максимальное продолжение поверхностей с римановой метрикой
(\ref{eemtvi}), имеющей по крайней мере один вектор Киллинга, произведем
следующее построение, которое будет также полезно для наглядного изображения
поверхностей. Поскольку конформный множитель не зависит от $\rho$, то можно
отождествить прямые $\rho$ и $\rho+L$, где $L>0$ -- произвольная положительная
постоянная. При таком отождествлении плоскость $\s,\rho$ сворачивается в
цилиндр. Длина направляющей окружности $P$ определяется конформным множителем и
на границе $q\to\pm\infty$ равна либо постоянной, либо бесконечности:
\begin{equation}                                                  \label{elenci}
  P^2=-L^2\Phi(q)\overset{q\to\pm\infty}\longrightarrow
  \begin{cases} 0, & \Phi(\pm\infty)=0,
  \\ \const\ne0, & \Phi(\pm\infty)=\const\ne0,
  \\ \mp\infty & \Phi(\pm\infty)=\pm\infty. \end{cases}
\end{equation}
В конечных точках $q\to q_i\ne\pm\infty$ квадрат длины направляющей окружности
может принимать только два значения:
\begin{equation}                                                  \label{elencu}
  P^2=-L^2\Phi(q)\overset{q\to q_i}\longrightarrow
  \begin{cases} 0, & \Phi(q_i)=0,
  \\ \mp\infty & \Phi(q_i)=\pm\infty. \end{cases}
\end{equation}
При этом плоскость $\s,\rho$ является универсальным накрывающим пространством
цилиндра.

Подытожим в таблице \ref{tprobo} свойства граничных точек $q_i$.
\begin{table}
\begin{center}
\begin{tabular}{|c|c|c|c|c|c|c|}
     \multicolumn{7}{c}{$|q_i|<\infty$}\\ \hline
     & $m<0$ & $0<m<1$ & $m=1$ & $1<m<2$ & $m=2$ & $m>2$ \\ \hline
$R$    & $\infty$ & $\infty$ & $\const$ & $\infty$ & $\const$ & $0$ \\
$\s_i$ & $\const$ & $\const$ & $\infty$ & $\infty$ & $\infty$ & $\infty$ \\
$P^2$    & $\infty$ & $0$ & $0$ & $0$ & $0$ & $0$ \\
Полнота& $-$ & $-$ & $-$ & $-$ & $+$ & $+$ \\  \hline
     \multicolumn{7}{c}{$|q_i|=\infty$}\\ \hline
     & $m<0$ & $m=0$ & $0<m\le 1$ & $1<m<2$ & $m=2$ & $m>2$ \\ \hline
$R$    & $0$ & $0$ & $0$ & $0$ & $\const$ & $\infty$ \\
$\s_i$ & $\infty$ & $\infty$ & $\infty$ & $\const$ & $\const$ & $\const$ \\
$P^2$    & $0$ & $\const$ & $\infty$ & $\infty$ & $\infty$ & $\infty$ \\
Полнота& $+$ & $+$ & $+$ & $+$ & $+$ & $-$ \\ \hline
\end{tabular}
\end{center}
 \caption{Свойства граничных точек в зависимости от показателя степени
 $m$. Символ $\const$ в строках для скалярной кривизны и длины направляющей
 окружности обозначает отличную от нуля константу.
 \label{tprobo}}
\end{table}
В зависимости от показателя степени $m$ здесь указаны значения скалярной
кривизны $R$ на соответствующей границе, конечность значения координаты $\s_i$
в точке $q_i$, квадраты длин направляющих окружностей цилиндров $P^2$ (с
точностью до знака) и полнота экстремалей, параллельных оси $\s$.

\begin{figure}[t,b,p]
\hfill\includegraphics[width=.95\textwidth]{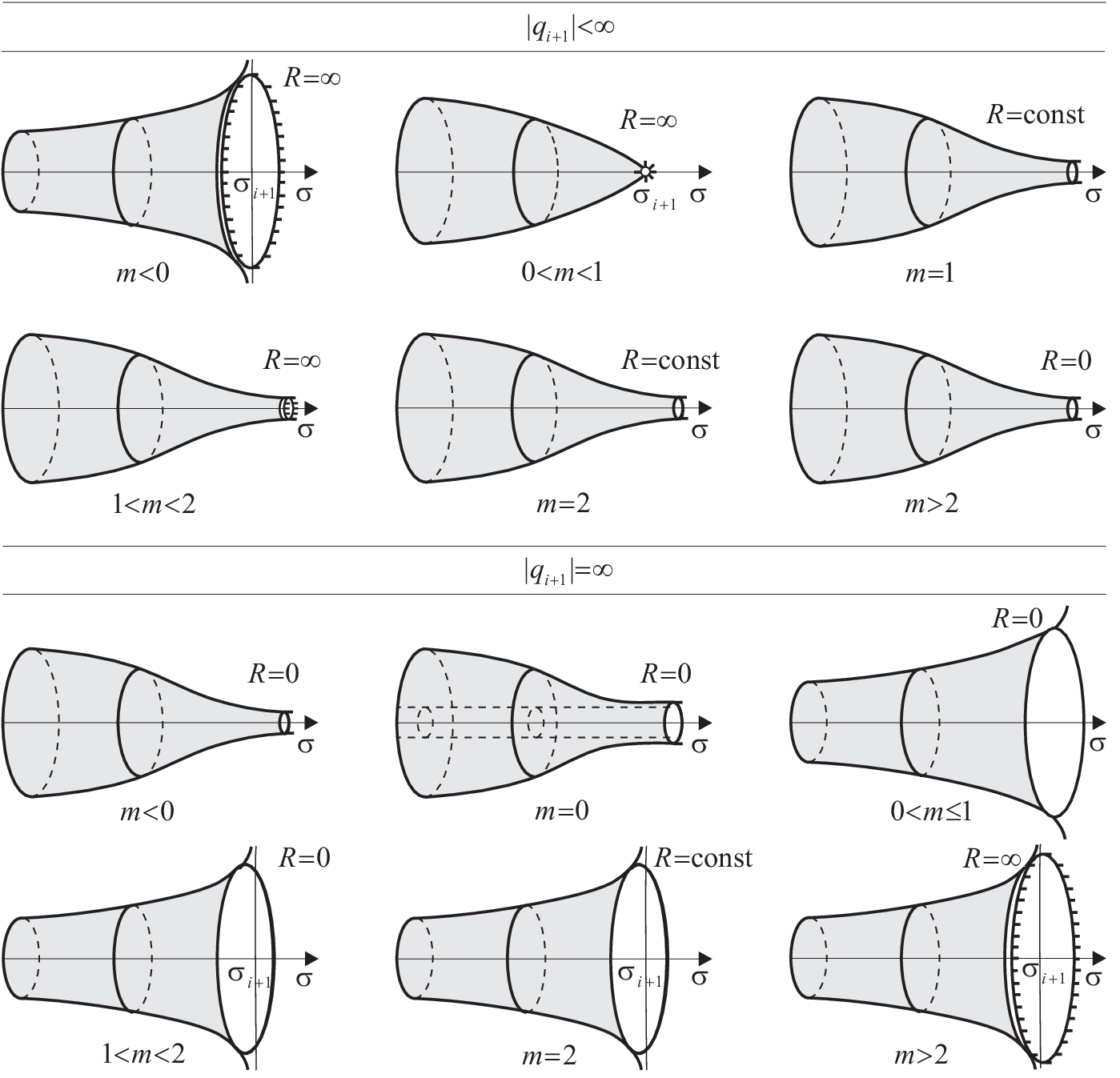}
\hfill {}
\\
\centering
 \caption{Вид поверхности вблизи граничной точки $q_{i+1}$ после отождествления
 $\rho\sim\rho+L$. Мы предполагаем, что координата $\s$ возрастает слева направо
 и $\s_{i+1}>\s_i$. Поверхности вблизи точки $q_i$ имеют такой же вид, но
 повернуты в другую сторону. Поверхность для интервала $q_i,q_{i+1}$ получается
 после склейки двух таких поверхностей для точек $q_i$ и $q_{i+1}$.
 \label{fsurec}}
\end{figure}

На рис.~\ref{fsurec} показан вид поверхностей вблизи граничной точки $q_{i+1}$
после отождествления $\rho\sim\rho+L$. Поверхности вблизи точки $q_i$ имеют
такой же вид, но повернуты в другую сторону. Поверхность, соответствующая
всему интервалу $(q_i,q_{i+1})$, получается после склейки двух таких
поверхностей для граничных точек $q_i$ и $q_{i+1}$.
Из таблицы \ref{tprobo} следует, что точка $q_i$ является неполной по
экстремалям, и при этом кривизна несингулярна в единственном случае, когда
$|q_i|<\infty$ при $m=1$. Таким образом, мы видим, что необходимо продолжить
экстремали только вблизи точки $|q_i|<\infty$ при $m=1$. Поскольку эта точка при
лоренцевой сигнатуре метрики соответствует горизонту, то и в евклидовом случае
оставим это же название.

Продолжение поверхности через горизонт проводится следующим образом. Прежде
всего отметим, что горизонт в евклидовом случае представляет собой точку, так
как длина направляющей окружности стремится к нулю. Кроме того, эта ``бесконечно
удаленная'' точка $\s=\infty$ в плоскости $\rho,\s$ на самом деле находится на
конечном расстоянии, т.к.\ все экстремали достигают этой точки при конечном
значении канонического параметра. В окрестности точки $q_i$ с точностью до
членов более высокого порядка конформный множитель имеет вид
\begin{equation}                                                  \label{eclcod}
  \Phi(q)=\Phi'_i(q-q_i),
\end{equation}
где $\Phi'_i:=\Phi'(q_i)=\const\ne0$. В области I при $\Phi'_i>0$ и $q>q_i$
уравнение (\ref{emdmee}) легко интегрируется
$$
  q-q_i=e^{\Phi'_i\s},
$$
где отброшена несущественная постоянная интегрирования, соответствующая сдвигу
$\s$. Таким образом, граничная точка $|q_i|<\infty$ достигается при
$\s\rightarrow-\infty$. В координатах $q,\rho$, которые играют роль
шварцшильдовских координат в лоренцевом случае, метрика (\ref{eemtvi})
принимает вид
\begin{equation}                                                  \label{eclmes}
  -ds^2=\frac{dq^2}{\Phi'_i(q-q_i)}+\Phi'_i(q-q_i)d\rho^2.
\end{equation}
В полярных координатах $r,\vf$, определенных формулами
\begin{equation}                                                  \label{epocoe}
  q-q_i:=\frac{\Phi'_i}4r^2,\qquad \rho:=\frac2{\Phi'_i}\vf,
\end{equation}
метрика становится евклидовой
$$
  -ds^2=dr^2+r^2d\vf^2.
$$
При этом полярный угол, что важно, меняется в интервале $\vf\in(0,L\Phi'_i/2)$
при $\rho\in[0,L]$, а радиус $r$ определен в окрестности граничной точки $q_i$
соотношением (\ref{epocoe}). Рассмотренное преобразование координат отображает
``бесконечно удаленную'' точку $\s_i=-\infty$ в начало координат евклидовой
плоскости. Поскольку полярный угол меняется в интервале, который в общем случае
отличен от интервала $(0,2\pi)$, то в начале координат возможно наличие
конической сингулярности. Соответствующий угол дефицита равен
\begin{equation}                                                  \label{edefae}
  2\pi\theta:=\frac{L\Phi'_i}2-2\pi.
\end{equation}
Таким образом, мы получили евклидову метрику на плоскости $\MR^2$ с конической
сингулярностью. При
\begin{equation}                                                  \label{qvbttf}
  L=4\pi/\Phi_i
\end{equation}
угол дефицита равен нулю, коническая
сингулярность отсутствует, и мы получаем плоскую евклидову метрику, гладкость
которой в начале координат очевидна. В общем случае конформный множитель
(\ref{eclcod}) вблизи граничной точки $q_i$ содержит поправки более высокого
порядка, и переход к полярным координатам при нулевом угле дефицита дает метрику
того же класса гладкости, что и конформный множитель.

\begin{figure}[h,t,b]
\hfill\includegraphics[width=.35\textwidth]{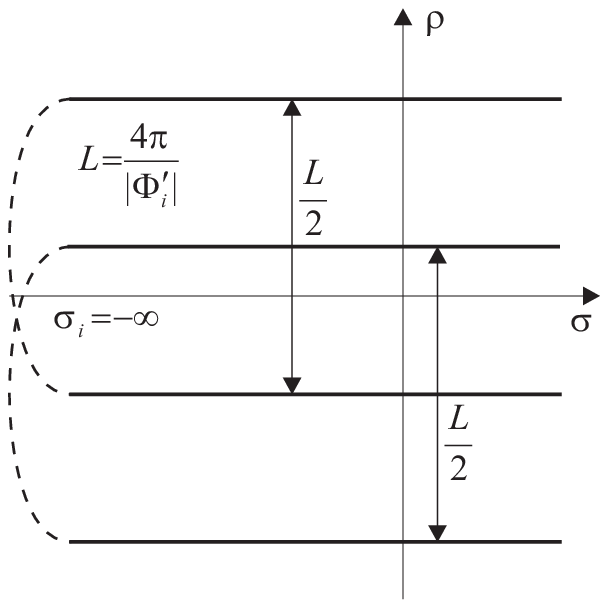}
\hfill {}
\caption{Продолжение прямых экстремалей, проходящих через точки $\rho$ и
 $\rho+L/2$ в отсутствие конической сингулярности $L=4\pi/|\Phi'_i|$.
 Отождествление происходит в точке $\s_i=-\infty$.}
 \label{fcorec}
\end{figure}

Таким образом, в общем случае продолжение решения через точку $|q_i|<\infty$ при
$m=1$ не имеет смысла, так как эта точка соответствует конической сингулярности.
Мы считаем, что точка, в которой сосредоточена коническая сингулярность, как и
носитель всякой другой сингулярности, не принадлежит многообразию. Поэтому
универсальной накрывающей для такого решения будет плоскость $\s,\rho$ или ее
часть с метрикой (\ref{eemtvi}). В исключительном случае отсутствия конической
сингулярности при $L=4\pi/\Phi'_i$ продолжение необходимо. При этом прямые
экстремали, параллельные оси $\s$ и проходящие через точки $\rho$ и $\rho+L/2$,
представляют собой две половины одной экстремали, как показано на
рис.\ref{fcorec}.
При отсутствии конической сингулярности фундаментальная группа поверхности
тривиальна, и поэтому соответствующая поверхность сама представляет собой
универсальную накрывающую.

\begin{figure}[htb]
\hfill\includegraphics[width=.95\textwidth]{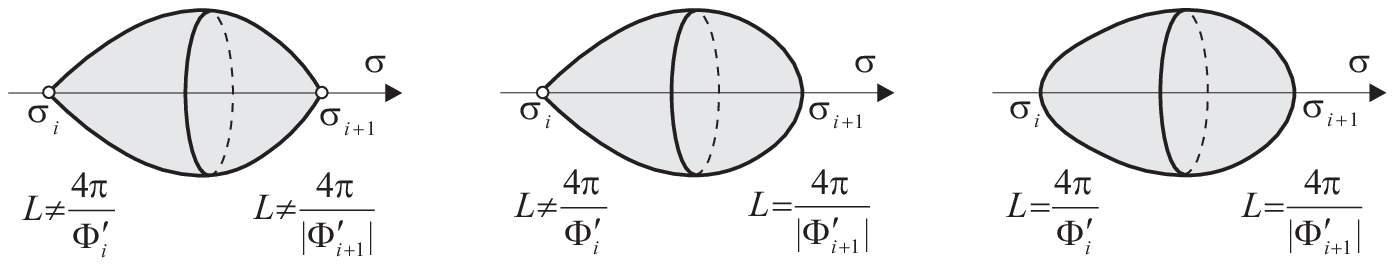}
\hfill {}
\\
\centering
 \caption{Три возможных вида поверхностей, соответствующих интервалу
 $(q_i,q_{i+1})$, когда конформный множитель имеет асимптотику
 $\Phi\sim q-q_{i,i+1}$ в граничных точках. При отождествлении $\rho\sim\rho+L$
 в общем случае возникают две конические сингулярности. При
 $L|\Phi'_{i,i+1}|=4\pi$ конические сингулярности отсутствуют.
 \label{fglmon}}
\end{figure}

Если на обоих концах интервала $(q_i,q_{i+1})$ конформный множитель имеет
асимптотики $\Phi\sim(q-q_i)$ и $\Phi\sim(q_{i+1}-q)$, то поверхность необходимо
продолжить в обеих точках $\s=\pm\infty$. После отождествления $\rho\sim\rho+L$
в общем случае в граничных точках возникают конические сингулярности. Если
$L\Phi'_i=4\pi$ и $L\Phi'_{i+1}=4\pi$, то конические сингулярности отсутствуют.
На рис.\ref{fglmon} показаны три возможных вида глобальных поверхностей,
соответствующих наличию или отсутствию конических сингулярностей. Эти
поверхности имеют топологию цилиндра, плоскости и сферы, соответственно.

Сформулируем правила построения максимально продолженных решений с
метрикой вида (\ref{eemtvi})
\begin{enumerate}
\item После отождествления $\rho\sim\rho+L$ каждому интервалу $(q_i,q_{i+1})$
соответствует максимально продолженное решение, которое получается склеиванием
двух поверхностей, изображенных на рис.\ref{fsurec} и соответствующих граничным
точкам $q_i$ и $q_{i+1}$.
\item Во всех случаях, кроме отсутствия конической сингулярности $|q_i|<\infty$,
$L\Phi'_i\ne4\pi$ или $|q_{i+1}|<\infty$, $L|\Phi'_{i+1}|\ne4\pi$, полоса
$\s\in(\s_i,\s_{i+1})$, $\rho\in\MR$ с метрикой (\ref{eemtvi}) является
универсальным накрывающим пространством для соответствующего максимально
продолженного решения.
\item При отсутствии одной из конических сингулярностей, $|q_i|<\infty$,
$L\Phi'_i=4\pi$, или $|q_{i+1}|<\infty$, $L|\Phi'_{i+1}|=4\pi$, поверхность,
полученная из плоскости $\s,\rho$ путем отождествления $\rho\sim\rho+L$,
представляет собой максимально продолженное решение с тривиальной
фундаментальной группой.
\end{enumerate}

Отметим, что в общем случае при построении максимально продолженных решений
нам не приходилось склеивать карты. Это значит, что соответствующие
поверхности принадлежат классу $\CC^\infty$. При отсутствии конической
сингулярности переход к полярным координатам (\ref{epocoe}) не использует
явный вид конформного множителя, и возникающая поверхность так же,
как и евклидова плоскость является бесконечно дифференцируемым
многообразием. Тем самым мы доказали теорему, оправдывающую приведенные
выше правила построения глобальных решений.
\begin{theorem}                                                   \label{tunisl}
Универсальное накрывающее пространство, построенное по правилам 1--3, является
максимально продолженным гладким многообразием класса $\CC^\infty$ с римановой
метрикой класса ${\cal C}^l$, $l\ge2$, таким, что область вне горизонта
изометричная (части) плоскости $\rho,\s$ с метрикой (\ref{eemtvi}).
\end{theorem}

Таким образом построены глобальные решения для широкого класса двумерных метрик
евклидовой сигнатуры, которые имеют один вектор Киллинга. Доказана гладкость
возникающих двумерных поверхностей и дифференцируемость метрики. Этот метод
конструктивен и является аналогом метода конформных блоков для двумерных метрик
лоренцевой сигнатуры. Почти всегда каждый конформный блок представляет собой
универсальное накрывающее пространство для максимально продолженной римановой
поверхности. Исключение составляют поверхности с горизонтом, который в
евклидовом случае представляет собой точку.
\section{Решение Шварцшильда}
\begin{figure}[h,b,t]
\hfill\includegraphics[width=.95\textwidth]{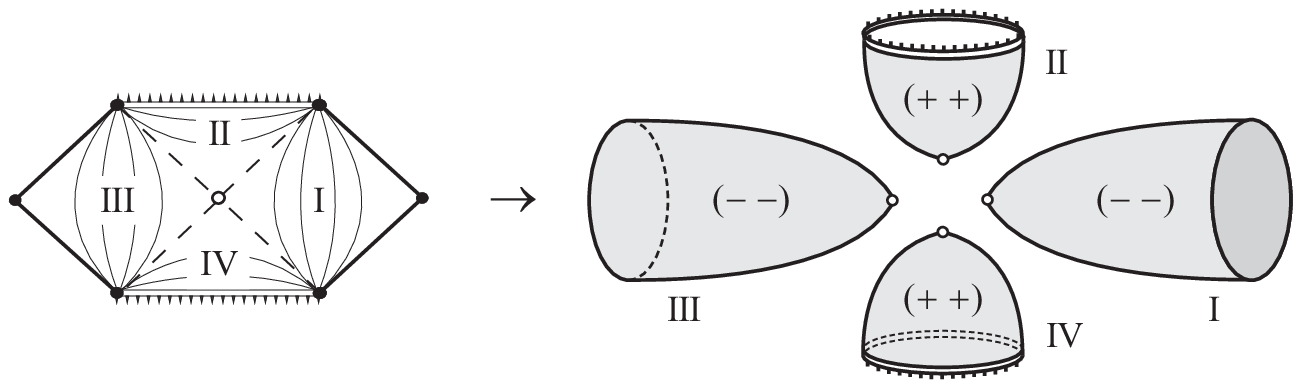}
\hfill {}
\\
\centering \caption{Диаграмма Картера--Пенроуза для решения Шварцшильда при
евклидовой сигнатуре метрики распадается на четыре несвязных между собой
поверхности. Областям вне горизонта соответствуют две поверхности с отрицательно
определенной метрикой I, III, а областям внутри горизонта -- две поверхности с
положительно определенной метрикой. \label{feushw}}
\end{figure}

В качестве примера рассмотрим решение Шварцшильда. Для этого решения $t,r$
часть метрики имеет вид
\begin{equation}                                                  \label{eschco}
  ds^2=|\Phi(q)|(d\tau^2-d\s^2),
\end{equation}
где
\begin{equation*}
  \Phi(q)=1-\frac{2M}q,\qquad q=\begin{cases} r, & \Phi(q)>0 \\ t, & \Phi(q)<0.
\end{cases}
\end{equation*}
При этом связь переменной $q$ с одной из координат $\tau$ или $\s$ задается
дифференциальным уравнением (\ref{eshiff}). Максимально продолженное решение
Шварцшильда имеет вид топологического произведения двух поверхностей
$\MU\times\MS^2$, где $\MS^2$ -- сфера и $\MU$ -- двумерная лоренцева
поверхность, которая наглядно изображается в виде диаграммы Картера--Пенроуза,
рис.\ref{feushw} слева.

Диаграмма Картера--Пенроуза после изменения сигнатуры метрики распадается на
четыре несвязные между собой поверхности, которые показаны на рис.\ref{feushw}
справа. Областям вне горизонта черной дыры соответствуют две римановы
поверхности I и III с отрицательно определенной метрикой. Четырехмерная метрика
является отрицательно определенной, $\sign g_{\mu\nu}=(----)$. Отрицательная
определенность евклидовой метрики связана с выбором сигнатуры для метрики
Шварцшильда (\ref{eschco}) и может быть изменена. Это решение обычно
рассматривается, как евклидова версия решения Шварцшильда. Области под
горизонтами черной и белой дыр соответствуют поверхностям II и IV
соответственно. При этом сигнатура полной метрики равна
$\sign g_{\mu\nu}=(++--)$.

Ситуация с евклидовой версией решения Шварцшильда типична. Связная диаграмма
Картера--Пенроуза в лоренцевом случае разрезается вдоль всех горизонтов и
переходит в несвязные римановы поверхности с отрицательно или положительно
определенной метрикой. При этом переход через горизонт с нечетным $m$
соответствует изменению знака римановой метрики.
\chapter{Сплетенные решения в общей теории относительности       \label{swarso}}
В настоящей главе построены глобальные решения вакуумных уравнений общей теории
относительности (\ref{eqevac}) с космологической постоянной в предположении, что
четырехмерное пространство-время является {\em сплетенным произведением} двух
поверхностей. При этом не делается никаких предположений о симметрии решений.
Как следствие уравнений движения по крайней мере одна из двух поверхностей
должна быть поверхностью постоянной кривизны. Отсюда вытекает, что метрика имеет
по крайней мере три вектора Киллинга. Другими словами, свойства симметрии
решений при таком подходе являются следствием самих уравнений движения.
Построенные решения включают, в частности, сферически симметричные решения,
которые соответствуют произведению некоторой лоренцевой поверхности на сферу.
Многие глобальные решения имеют интересную физическую интерпретацию. В
частности, построены решения, описывающие кротовые норы, доменные стенки
сингулярностей кривизны, космические струны, комические струны, окруженные
доменными стенками, решения с замкнутыми времениподобными кривыми и др.
\cite{KaKlKu99}
\section{Сплетенное произведение}
Дадим
\begin{defn}
Пусть задано два многообразия $\MM_1$ и $\MM_2$ с метриками $g$ и $h$
соответственно. Касательное пространство в каждой точке топологического
произведения $(x_1,x_2)\in\MM_1\times\MM_2$ разлагается в прямую сумму:
\begin{equation*}
  \MT_{(x_1,x_2)}(\MM_1\times\MM_2)=\MT_{x_1}(\MM_1)\oplus\MT_{x_2}(\MM_2).
\end{equation*}
{\em Сплетенным произведением} (warped product) двух многообразий называется
их топологическое произведение $\MM_1\times\MM_2$ с метрикой $\hat g$, которая
определена следующим соотношением
\begin{equation}                                                  \label{qwarfd}
  \hat g(X,Y):=k(x_2)g(X_1,Y_1)+m(x_1)h(X_2,Y_2),
\end{equation}
где $k(x_2)\in\CC^{n_2}(\MM_2)$ и $m(x_1)\in\CC^{n_1}(\MM_1)$ -- достаточно
гладкие отличные от нуля функции на многообразиях $\MM_2$ и $\MM_1$, и
\begin{align*}
  \MT(\MM_1\times\MM_2)\ni\quad X&=X_1\oplus X_2\quad
  \in\MT(\MM_1)\oplus\MT(\MM_2),
\\
  \MT(\MM_1\times\MM_2)\ni\quad Y&=Y_1\oplus Y_2\quad
  \in\MT(\MM_1)\oplus\MT(\MM_2),
\end{align*}
-- разложение векторных полей $X,Y$, касательных к $\MM_1\times\MM_2$, в прямую
сумму.
\qed\end{defn}
\index{Сплетенное произведение (warped product)}%
\index{Произведение сплетенное (warped product)}%

Предположим, что четырехмерное пространство-время является сплетенным
произведением двух поверхностей: $\MM=\MU\times \MV$, где $\MU$ -- поверхность
с лоренцевой метрикой $g$ и $\MV$ -- поверхность с римановой метрикой $h$.
Обозначим локальные координаты на $\MU$ и $\MV$ соответственно через $x^\al$
($\al,\bt,\dotsc =0,1$) и $y^\mu$ ($\mu,\nu,\dotsc=2,3$). Тогда топологическому
произведению $\MU\times\MV$ соответствуют координаты
$\lbrace x^i\rbrace:=\lbrace x^\al,y^\mu\rbrace$ ($i,j,\dotsc =0,1,2,3$).
В этой системе координат четырехмерная метрика имеет блочно-диагональный вид:
\begin{equation}                                                  \label{emetbl}
  \widehat g_{ij}
  =\begin{pmatrix} k(y)g_{\al\bt}(x) & 0 \\ 0 & m(x)h_{\mu\nu}(y) \end{pmatrix},
\end{equation}
где $k(y)$ и $m(x)$ -- достаточно гладкие отличные от нуля функции на $\MV$ и
$\MU$ соответственно.

В настоящей главе шляпка над символом означает, что соответствующий
геометрический объект относится ко всему четырехмерному пространству-времени
$\MM$, а символы без шляпки относятся к двумерным поверхностям $\MU$ или $\MV$.
Соответственно $g_{\al\bt}$ и $h_{\mu\nu}$ являются метриками на $\MU$ и $\MV$.
Греческие буквы из начала ($\al,\bt,\dots$) и середины ($\mu,\nu,\dots$)
алфавита всегда относятся к координатам на первой и второй поверхностям
соответственно.

В физике функции $k(y)$ и $m(x)$ часто называют ({\em дилатонными}) полями на
поверхностях $\MV$ и $\MU$.
\index{Дилатонное поле (dilaton field)}\index{Поле дилатона (dilaton field)}%

Для определенности будем считать поверхность $\MU$ псевдоримановым многообразием
с метрикой лоренцевой сигнатуры, а поверхность $\MV$ -- римановым многообразием
с положительно определенной метрикой. Тогда с точностью до перестановки первых
двух координат сигнатура метрики на $\MM$ будет либо $(+---)$, либо $(-+++)$ в
зависимости от знака $m$. Эти метрики связаны между собой инверсией
$\widehat g_{ij}\mapsto-\widehat g_{ij}$, относительно которой уравнения
Эйнштейна при отсутствии полей материи и космологической постоянной инвариантны.
Предположим также, что обе поверхности являются ориентируемыми.

Отметим, что относительно вида метрики (\ref{emetbl}) не делается никаких
дополнительных предположений, связанных с симметрией. Однако в дальнейшем мы
увидим, что уравнения Эйнштейна и требование полноты многообразий приводят к
тому, что по крайней мере одна из поверхностей $\MU$ или $\MV$ должна быть
поверхностью постоянной кривизны. То есть любое максимально продолженное
решение уравнений Эйнштейна вида (\ref{emetbl}) допускает по крайней мере три
вектора Киллинга. Следовательно, в рассматриваемом случае симметрия решений
является следствием уравнений движения. В частном случае будут получены
сферически симметричные решения, когда поверхность $\MV$ является сферой
$\MS^2$.

Все решения, рассмотренные в настоящей главе относятся к решениям типа $D$ по
классификации Петрова \cite{Petrov69}. Физическая интерпретация решений
опирается на глобальную структуру пространства-времени.
\section{Двумерная редукция                                      \label{stpodr}}
Вид метрики (\ref{emetbl}) позволяет решить явно четырехмерные вакуумные
уравнения Эйнштейна с космологической постоянной $\Lm$
\begin{equation}                                                  \label{einseq}
  \widehat R_{ij}=\Lm\widehat g_{ij},
\end{equation}
и построить глобальные (максимально продолженные) решения.

Мы увидим, что уравнения Эйнштейна существенно ограничивают дилатонные поля: по
крайней мере одно дилатонное поле должно быть постоянно. Поэтому все решения
делятся на три основных класса: оба дилатонных поля постоянны (случай {\sf A}),
только $k = \const$ (случай {\sf B}) или $m=\const$ (случай {\sf C}). В первом
случае решение уравнений Эйнштейна представляет собой топологическое
произведение двух поверхностей постоянной кривизны. В случае {\sf B} риманова
поверхность $\MV$ должна быть поверхностью постоянной кривизны. Сюда входят
сферически симметричные решения, а также другие решения, когда поверхность $\MV$
представляет собой евклидову плоскость или плоскость Лобачевского (двуполостный
гиперболоид). Последние решения соответствуют кротовым норам. В случае {\sf C}
поверхность $\MU$ должна быть поверхностью постоянной кривизны. Эти решения
описывают космические струны и доменные стенки сингулярности кривизны.

Приступим к решению уравнений Эйнштейна (\ref{einseq}). Метрика, обратная к
(\ref{emetbl}), имеет вид
\begin{equation}                                                  \label{ennvme}
  \widehat g^{ij}=\begin{pmatrix} {\displaystyle\frac1k} g^{\al\bt} & 0 \\
  0 & {\displaystyle\frac1m} h^{\mu\nu} \end{pmatrix},
\end{equation}
где $g^{\al\bt}$ и $h^{\mu\nu}$ матрицы, обратные соответственно к $g_{\al\bt}$
и $h_{\mu\nu}$. Символы Кристоффеля (\ref{echris}) равны
\begin{equation}                                                 \label{echris1}
\begin{split}
  \widehat\Gamma_{\al\bt}{}^\g&=\Gamma_{\al\bt}{}^\g,
\\
  \widehat\Gamma_{\al\bt}{}^\mu&=-\frac12g_{\al\bt}\frac{h^{\mu\nu}\pl_\nu k}m,
\\
  \widehat\Gamma_{\al\mu}{}^\g&=\widehat\Gamma_{\mu\al}{}^\g
  =\frac12\dl_\al^\g\frac{\pl_\mu k}k,
\\
  \widehat\Gamma_{\al\mu}{}^\nu&=\widehat\Gamma_{\mu\al}{}^\nu
  =\frac12\dl_\mu^\nu\frac{\pl_\al m}m,
\\
  \widehat\Gamma_{\mu\nu}{}^\al&=-\frac12h_{\mu\nu}\frac{g^{\al\bt}\pl_\bt m}k,
\\
  \widehat\Gamma_{\mu\nu}{}^\rho&=\Gamma_{\mu\nu}{}^\rho.
\end{split}
\end{equation}

Компоненты тензора Риччи (\ref{ericte}) принимают вид
\begin{equation}                                                  \label{eritab}
\begin{split}
  \widehat R_{\al\bt}&=R_{\al\bt}+\frac{\nabla_\al\nabla_\bt m}m
  -\frac{\nabla_\al m\nabla_\bt m}{2m^2}+\frac{g_{\al\bt}\nabla^2 k}{2m}
\\
  \widehat R_{\al\mu}&=\widehat R_{\mu\al}=-\frac{\nabla_\al m\nabla_\mu k}{2mk}
\\
 \widehat R_{\mu\nu}&=R_{\mu\nu}+\frac{\nabla_\mu\nabla_\nu k}k
  -\frac{\nabla_\mu k\nabla_\nu k}{2k^2}+\frac{h_{\mu\nu}\nabla^2 m}{2k},
\end{split}
\end{equation}
где, для краткости, введены обозначения
\begin{equation}                                                  \label{edalan}
  \nabla^2 m:=g^{\al\bt}\nabla_\al\nabla_\bt m,\qquad
  \nabla^2 k:=h^{\mu\nu}\nabla_\mu\nabla_\nu k.
\end{equation}
Здесь и далее в этой главе символ $\nb$ обозначает ковариантную производную с
соответствующими символами Кристоффеля. Четырехмерная скалярная кривизна равна
\begin{equation}                                                  \label{escacu}
  \widehat R=\frac1kR^g+2\frac{\nabla^2 m}{km}-\frac{(\nabla m)^2}{2km^2}
  +\frac1mR^h+2\frac{\nabla^2 k}{km}-\frac{(\nabla k)^2}{2k^2m},
\end{equation}
где введены обозначения
\begin{equation}                                                  \label{egrsqn}
  (\nabla m)^2:=g^{\al\bt}\pl_\al m\pl_\bt m,\qquad
  (\nabla k)^2:=h^{\mu\nu}\pl_\mu k\pl_\nu k.
\end{equation}
Скалярные кривизны поверхностей $\MU$ и $\MV$ обозначены через $R^g$ и $R^h$
соответственно.

Таким образом, уравнений Эйнштейна (\ref{einseq}) для метрики (\ref{emetbl})
принимают вид
\begin{align}                                                     \label{eineaa}
  R_{\al\bt}+\frac{\nabla_\al\nabla_\bt m}m
  -\frac{\nabla_\al m\nabla_\bt m}{2m^2}+\frac12g_{\al\bt}
  \left(\frac{\nabla^2 k}m-2k\Lm\right)&=0,
\\                                                                \label{einemm}
  R_{\mu\nu}+\frac{\nabla_\mu\nabla_\nu k}k
  -\frac{\nabla_\mu k\nabla_\nu k}{2k^2}+\frac12h_{\mu\nu}
  \left(\frac{\nabla^2 m}m-2m\Lm\right)&=0,
\\                                                                \label{eineam}
  \frac{\nabla_\al m\nabla_\mu k}{mk}&=0.
\end{align}
Перепишем уравнения (\ref{eineaa}) и (\ref{einemm}) в более удобном виде,
выделив из них след, который определяет скалярные кривизны поверхностей:
\begin{align}                                                     \label{escuru}
  R^g+\frac{\nabla^2 m}m-\frac{(\nabla m)^2}{2m^2}
  +\frac{\nabla^2 k}m-2k\Lm&=0,
\\                                                                \label{esccur}
 R^h+\frac{\nabla^2 k}k-\frac{(\nabla k)^2}{2k^2}
  +\frac{\nabla^2 m}k-2m\Lm&=0.
\end{align}
Бесследовые части уравнений (\ref{eineaa}) и (\ref{einemm}), умноженные на $m$ и
$k$, принимают простой вид
\begin{align}                                                     \label{einnaa}
  \nabla_\al\nabla_\bt m
  -\frac{\nabla_\al m\nabla_\bt m}{2m}-\frac12g_{\al\bt}
  \left[\nabla^2 m-\frac{(\nabla m)^2}{2m}\right]&=0,
\\                                                                \label{einnmm}
  \nabla_\mu\nabla_\nu k
  -\frac{\nabla_\mu k\nabla_\nu k}{2k}-\frac12h_{\mu\nu}
  \left[\nabla^2 k-\frac{(\nabla k)^2}{2k}\right]&=0.
\end{align}
Они не содержат слагаемые с кривизной вовсе, потому что в двух измерениях тензор
Риччи полностью определяется скалярной кривизной (\ref{ericte}) и не имеет
бесследовой части.

Отметим, что наличие сингулярности у двумерной скалярной кривизны на поверхности
означает ее наличие в полном тензоре кривизны в соответствии сформулой
(\ref{escacu}).

Таким образом, четырехмерные уравнения Эйнштейна (\ref{einseq}) для метрики вида
(\ref{emetbl}) эквивалентны системе уравнений (\ref{eineam})--(\ref{einnmm}).
Уравнения (\ref{einnaa}) и (\ref{einnmm}) содержат функции, зависящие только от
координат $x$ и $y$ соответственно. В то же время координаты различных
поверхностей в уравнениях (\ref{eineam}), (\ref{escuru}) и (\ref{esccur})
перемешаны.

Уравнение (\ref{eineam}) накладывает жесткие ограничения. Как следствие имеем,
что либо поле дилатона $k$, либо поле дилатона $m$, либо $k$ и $m$ одновременно
должны быть постоянны. Соответственно, возможны три случая:
\begin{equation}                                                  \label{ecases}
\begin{array}{lrr}
  {\sf A}:  & \qquad k=\const\ne0,      & \qquad m=\const\ne0, \\
  {\sf B}:  & k=\const\ne0,      & \nabla_\al m\ne0, \\
  {\sf C}:  & \nabla_\mu k\ne0,  & m=\const\ne0.
\end{array}
\end{equation}
Рассмотрим эти случаи подробно.
\section{Произведение поверхностей постоянной кривизны           \label{soocus}}
Наиболее симметричные решения вакуумных уравнений Эйнштейна (\ref{einseq}) в
виде топологического произведения двух поверхностей постоянной кривизны
возникают, когда оба дилатонных поля $k$ и $m$ постоянны (случай {\sf A} в
(\ref{ecases})). Если $k$ и $m$ постоянны, то уравнения (\ref{eineam}),
(\ref{einnaa}), и (\ref{einnmm}) удовлетворяются, и скалярные кривизны обеих
поверхностей $\MU$ и $\MV$ должны быть постоянны как следствие уравнений
(\ref{escuru}), (\ref{esccur}), которые принимают следующий вид
\begin{align}                                                     \label{ecscrg}
  R^g=2k\Lm,\qquad  R^h=2m\Lm.
\end{align}

Если $\Lm=0$, то обе поверхности $\MU$ и $\MV$ имеют нулевую кривизну, и все
пространство-время $\MM$ представляет пространство Минковского или его
фактор-пространство по группе преобразований, действующей свободно и собственно
разрывно, (цилиндр или тор) с метрикой Лоренца
\begin{equation}                                                  \label{eminme}
  \widehat g_{ij}=\diag (+---)\qquad \text{или}\qquad
  \widehat g_{ij}=\diag (-+++).
\end{equation}

При ненулевой космологической постоянной $\Lm\ne0$ обе поверхности $\MU$ и $\MV$
имеют постоянную ненулевую кривизну. Поверхности постоянной кривизны были
рассмотрены в главе \ref{sconcu}, и мы приведем лишь выражения для
соответствующих метрик. Если $\MU$ является полной псевдоримановой поверхностью
ненулевой кривизны $R^g=-2K=\const$, то она представляет собой однополостный
гиперболоид $\ML^2$, вложенный в трехмерное пространство Минковского, с
индуцированной метрикой или его универсальной накрывающей (раздел \ref{shyper}).
Его группой симметрии является группа Лоренца $\MS\MO(1,2)$. В
стереографических координатах метрика однополостного гиперболоида $\ML^2$ имеет
хорошо известный вид (\ref{ehystc})
\begin{equation}                                                  \label{ecoclm}
  ds^2_\ML=g_{\al\bt}dx^\al dx^\bt=\frac{dt^2-dx^2}
  {\left[1+\frac K4(t^2-x^2)\right]^2},
\end{equation}
где введены обозначения $t:=x^0$ и $x:=x^1$. В отличие от риманова случая
псевдориманова поверхность постоянной кривизны одна и та же как для
положительной, так и для отрицательной гауссовой кривизны $K$, при этом меняется
только общий знак метрики (\ref{ecoclm}), что соответствует перестановке
координат $t\leftrightarrow x$.

При $K=0$ метрика (\ref{ecoclm}) совпадает с обычной двумерной метрикой
Минковского, и соответствующая поверхность представляет собой плоскость
Минковского $\MR^{1,1}$ с группой Пуанкаре $\MI\MO(1,1)$ в качестве группы
симметрии.

Положительно определенная метрика на двумерной римановой поверхности постоянной
кривизны $R^h=-2K\ne0$ в стереографических координатах имеет вид (\ref{qtrfca})
\begin{equation}                                                  \label{ecocrm}
  ds^2_\MH=h_{\mu\nu}dy^\mu dy^\nu=\frac{dy^2+dz^2}
  {\left[1+\frac K4(y^2+z^2)\right]^2},
\end{equation}
где $y:=y^1$ и $z:=y^2$. Эта метрика отличается от (\ref{ecoclm}) только
знаками.

Для положительных $K>0$ она соответствует сфере $\MS^2$, рассмотренной в разделе
\ref{sphere}. При $K=0$ метрика (\ref{ecocrm}) соответствует евклидовой
плоскости $\MR^2$, или цилиндру, или тору. При отрицательных $K<0$ мы имеем
плоскость Лобачевского (гиперболическую плоскость) $\MH^2$, рассмотренную в
разделе \ref{sutwse}, или компактную риманову поверхность рода два или выше.
Группами симметрии сферы $\MS^2$, евклидовой плоскости $\MR^2$, и плоскости
Лобачевского $\MH^2$ являются, соответственно, группы $\MO(3)$, $\MI\MO(2)$ и
$\MO(1,2)$.

При ненулевой постоянной кривизне всегда можно произвести растяжку координат
таким образом, чтобы $K=\pm 1$.

Если скалярные кривизны постоянны (\ref{ecscrg}), то решение для ненулевой
космологической постоянной $\Lm\ne0$ является топологическим произведением двух
поверхностей постоянной кривизны с метрикой
\begin{equation}                                                  \label{esymef}
  ds^2=k\frac{dt^2-dx^2}{\left[1-\frac{k\Lm}4(t^2-x^2)\right]^2}
  +m\frac{dy^2+dz^2}{\left[1-\frac{m\Lm}4(y^2+z^2)\right]^2}.
\end{equation}
В данном случае можно не говорить о сплетенном произведении поверхностей, т.к.\
дилатонные поля постоянны. Растягивая координаты, всегда можно добиться
выполнения равенств $k=\pm1$, $m=\pm1$. Выберем $k=1$ и $m=-1$ с тем, чтобы
метрика имела сигнатуру $(+---)$. Тогда возможны три качественно отличных
случая, соответствующих положительной, нулевой и отрицательной космологической
постоянной:
\begin{equation}                                                  \label{econcs}
\begin{array}{llll}
  \Lm<0:\quad & K^g=+|\Lm|,\quad & K^h=-|\Lm|,\quad & \MM=\ML^2\times \MH^2, \\
  \Lm=0:\quad& K^g=0,\quad& K^h=0,\quad & \MM=\MR^{1,1}\times \MR^2=\MR^{1,3},\\
  \Lm>0:\quad & K^g=-|\Lm|,\quad & K^h=+|\Lm|,\quad & \MM=\ML^2\times \MS^2,
\end{array}
\end{equation}
где $K^g$ и $K^h$ -- гауссовы кривизны соответственно поверхностей $\MU$ и
$\MV$.

Напомним, что вакуумные уравнения Эйнштейна (\ref{einseq}) допускают решение в
виде пространства-времени постоянной кривизны, которое называется
пространством-временем (анти-) де Ситтера. Это пространство-время имеет
максимальное число -- десять -- векторов Киллинга. Хотя полная (четырехмерная)
скалярная кривизна для решения в виде произведения двух поверхностей постоянной
кривизны с метрикой (\ref{esymef}), впрочем как и для всех других решений
вакуумных уравнений Эйнштейна (\ref{einseq}), постоянна, $\widehat R=4\Lm$,
решения (\ref{econcs}) при $\Lm\ne0$ не совпадают с решением (анти-) де
Ситтера. Действительно, каждая из поверхностей $\ML^2$, $\MH^2$ и $\MS^2$ имеет
по три вектора Киллинга, и можно показать (см.\ например \cite{KrStMaHe80}), что
четырехмерное пространство-время имеет всего шесть векторов Киллинга. Поэтому
решения в виде произведения двух поверхностей не совпадают с решением (анти-) де
Ситтера.

Все решения в виде произведения двух поверхностей постоянной кривизны относятся
к типу $D$ по классификации Петрова \cite{Petrov69}.
\section{Пространственно симметричные решения                    \label{solcog}}
Во втором случае {\sf B} (\ref{ecases}) дилатонное поле $k$ постоянно. Не
ограничивая общности, положим $k=1$. Тогда вся система уравнений Эйнштейна
(\ref{eineam})--(\ref{einnmm}) сводится к следующей системе:
\begin{align}                                                     \label{einmaa}
  \nabla_\al\nabla_\bt m
  -\frac{\nabla_\al m\nabla_\bt m}{2m}-\frac12g_{\al\bt}
  \left[\nabla^2 m-\frac{(\nabla m)^2}{2m}\right]&=0,
\\                                                                \label{einmmk}
  R^h+\nabla^2 m -2m\Lm&=0,
\\                                                                \label{einscf}
  R^g+\frac{\nabla^2 m}m-\frac{(\nabla m)^2}{2m^2}-2\Lm&=0.
\end{align}

Уравнение (\ref{einmmk}) представляет собой сумму двух слагаемых, зависящих от
координат на разных поверхностях, $x\in\MU$ и $y\in\MV$, которая должна быть
равна нулю. Это значит, что каждое слагаемое равно некоторой постоянной.
Зафиксируем эту постоянную следующим уравнением $R^h=-2K=\const$. Таким образом,
в случае {\sf B} поверхность $\MV$ -- это поверхность постоянной кривизны. При
этом возможны три случая, когда гауссова кривизна римановой поверхности $\MV$
положительна, $\MV=\MS^2$, равна нулю, $\MV=\MR^2$, или отрицательна,
$\MV=\MH^2$. Тогда соответствующие решения вакуумных уравнений Эйнштейна
инвариантны относительно групп преобразований $\MO(3)$, $\MI\MO(2)$ или
$\MO(1,2)$, которые являются группами изометрий поверхностей $\MS^2$, $\MR^2$ и
$\MH^2$. Соответствующее четырехмерное пространство-время представляет собой
сплетенное произведение поверхности $\MU$ с одной из поверхностей $\MS^2$,
$\MR^2$ или $\MH^2$, где $\MU$ представляется диаграммой Картера--Пенроуза. В
частном случае при $K=1$, возникают сферически симметричные решения. Таким
образом, в рассматриваемом случае группа симметрии пространства-времени
возникает как следствие уравнений движения.

При $K=\const$ уравнение (\ref{einmmk}) принимает вид
\begin{equation}                                                  \label{einrhk}
  \nabla^2 m -2(m\Lm+K)=0.
\end{equation}
Исключая случай {\sf A}, рассмотренный в предыдущем разделе, двинемся дальше,
считая, что $\nabla_\al m\ne0$.
\begin{prop}
Уравнение (\ref{einrhk}) является первым интегралом уравнений (\ref{einmaa}) и
(\ref{einscf}).
\end{prop}
\begin{proof}
Продифференцируем уравнение (\ref{einrhk}), используем тождество
$$
  \left[\nb_\al,\nb_\bt\right]A_\g=-R^g_{\al\bt\g}{}^\dl A_\dl\,,
$$
где $A_\al$ -- компоненты произвольного ковекторного поля, для изменения порядка
ковариантных производных и используем уравнение (\ref{einmaa}) три раза для
исключения вторых производных от $m$. После небольших алгебраических выкладок мы
получим уравнение (\ref{einscf}).
\end{proof}
Из доказательства предложения следует, что достаточно решить только уравнения
(\ref{einmaa}) и (\ref{einrhk}), при этом уравнение (\ref{einscf}) будет
удовлетворено автоматически.
\begin{com}
Исходное действие Гильберта--Эйнштейна инвариантно относительно общих
преобразований координат, и, согласно второй теореме Нетер, между уравнениями
движения существует линейная зависимость. Поэтому зависимость уравнений
(\ref{einmaa})--(\ref{einscf}) не является чем то удивительным и связана с
инвариантностью исходного действия.
\qed\end{com}
Для явного решения уравнений движения (\ref{einmaa}) и (\ref{einrhk})
зафиксируем конформную калибровку для метрики $g_{\al\bt}$ на лоренцевой
поверхности $\MU$:
\begin{equation}                                                  \label{ecogag}
  g_{\al\bt}dx^\al dx^\bt=\Phi d\xi d\eta,
\end{equation}
где $\Phi(\xi,\eta)$ -- конформный множитель, который зависит от координат
светового конуса $\xi,\eta$ на $\MU$. Соответствующая четырехмерная метрика
примет вид
\begin{equation}                                                  \label{emetko}
  ds^2=\Phi d\xi d\eta+md\Om,
\end{equation}
где $d\Om$ -- метрика на римановой поверхности постоянной кривизны $\MV=\MS^2$,
$\MR^2$ или $\MH^2$. Знак конформного множителя $\Phi$ пока не фиксируем.

При $\Phi>0$ и $m<0$ сигнатура метрики (\ref{emetko}) равна $(+---)$, если
пространственная и временн\'ая координаты определены так же, как и раньше
(\ref{elicco}). Если
изменить знак $m>0$, то сигнатура метрики станет $(+-++)$. Такое же
преобразование сигнатуры можно получить, изменив общий знак метрики,
$\hat g_{ij}\mapsto-\hat g_{ij}$, и переставив первые две координаты. Вакуумные
уравнения Эйнштейна (\ref{einseq}) инвариантны относительно одновременного
изменения знаков метрики и космологической постоянной. Поскольку в дальнейшем мы
построим глобальные решения для всех возможных значений космологической
постоянной, $\Lm\in\MR$, то, не ограничивая общности, достаточно рассмотреть
случай $m<0$. При отрицательных $m$ удобно ввести параметризацию
\begin{equation}                                                  \label{qparam}
  m=-q^2,\qquad q(\xi,\eta)>0.
\end{equation}

Символы Кристоффеля для метрики (\ref{ecogag}) в конформной калибровке имеют
только две ненулевые компоненты:
\begin{equation}                                                  \label{echsyu}
  \Gamma_{\xi\xi}{}^\xi=\frac{\pl_\xi\Phi}\Phi,\qquad
  \Gamma_{\eta\eta}{}^\eta=\frac{\pl_\eta\Phi}\Phi,
\end{equation}
и уравнения (\ref{einmaa}), (\ref{einrhk}) принимают простой вид
\begin{align}                                                     \label{egcoga}
  -\pl^2_{\xi\xi}q+\frac{\pl_\xi\Phi\pl_\xi q}\Phi&=0,
\\                                                                \label{egcogb}
  -\pl^2_{\eta\eta}q+\frac{\pl_\eta\Phi\pl_\eta q}\Phi&=0,
\\                                                                \label{egcogc}
  -2\frac{\pl^2_{\xi\eta}q^2}\Phi-(K-\Lm q^2)&=0.
\end{align}
Таким образом, полная система уравнений (\ref{einmaa})--(\ref{einscf}) в
конформной калибровке (\ref{ecogag}) сводится к трем уравнениям на две
неизвестные функции $q$ и $\Phi$. Первые два уравнения являются обыкновенными
дифференциальными уравнениями, и определяют функции $q$ и $\Phi$ с точностью до
умножения на произвольную постоянную. Система уравнений
(\ref{egcoga})--(\ref{egcogc}) переопределена и может быть проинтегрирована
явно.
\begin{prop}
Условия $\pl_\xi q=0$ и $\pl_\eta q=0$ эквивалентны.
\end{prop}
\begin{proof}
Если $\pl_\xi q=0$, то из уравнения (\ref{egcogc}) следует $q^2=K/\Lm=\const$ и,
следовательно, $\pl_\eta q=0$. Обратное утверждение верно по той же причине.
\end{proof}
Поскольку $q=\const$ соответствует уже рассмотренному случаю {\sf A}, то
предположим, что $\pl_\xi q\ne0$ и $\pl_\eta q\ne0$. Тогда, разделив уравнения
(\ref{egcoga}) и (\ref{egcogb}) соответственно на $\pl_\xi q$ и $\pl_\eta q$,
они легко интегрируются:
\begin{align}                                                     \label{qeqwsa}
  -\ln|\pl_\xi q|+\ln|\Phi|&=\tilde G(\eta),
\\                                                                \label{qeaswq}
  -\ln|\pl_\eta q|+\ln|\Phi|&=\tilde F(\xi).
\end{align}
При этом возникают две произвольные функции $\tilde F(\xi)$ и $\tilde G(\eta)$.
Введем монотонные функции $F(\eta)$ и $G(\xi)$ при помощи дифференциальных
уравнений
\begin{equation*}
  F':=\frac{dF}{d\xi}=C\ex^{\tilde F}>0,\qquad
  G':=\frac{dG}{d\eta}=C\ex^{\tilde G}>0,
\end{equation*}
где $C>0$ -- некоторая положительная постоянная, которую мы зафиксируем чуть
позже. Тогда разность уравнений (\ref{qeqwsa}) и (\ref{qeaswq}) примет вид
\begin{equation}                                                   \label{qfrtd}
  \frac{|\pl_\xi q|}{F'}=\frac{|\pl_\eta q|}{G'}.
\end{equation}

Конформная калибровка для двумерной метрики (\ref{ecogag}) определена с
точностью до конформных преобразований. Воспользуемся этой свободой и перейдем
к новым координатам $\xi,\eta\mapsto F,G$. Это всегда можно сделать, т.к.\
якобиан преобразования координат отличен от нуля, $F'G'\ne0$. При конформном
преобразовании координат конформный множитель преобразуется по-правилу
$\Phi\mapsto\Phi/(F'G')$, что следует из вида метрики в конформной калибровке
(\ref{ecogag}).
\begin{prop}
Уравнения (\ref{egcoga})--(\ref{egcogc}) ковариантны относительно конформных
преобразований
\begin{equation}                                                  \label{qcased}
  \xi,\eta\mapsto F,G,\qquad \Phi\mapsto\tilde\Phi=\frac\Phi{F'G'}.
\end{equation}
\end{prop}
\begin{proof}
Прямая проверка.
\end{proof}
Таким образом, произвольные функции $\tilde F(\xi)$ и $\tilde G(\eta)$,
возникшие в первых интегралах (\ref{qeqwsa}), (\ref{qeaswq}), соответствуют
конформным преобразованиям.

Перейдем к новым координатам $\xi,\eta\mapsto F,G$.
\begin{prop}
Если $\pl_\xi q\,\pl_\eta q>0$, то в новых координатах функция $q(\tau)$ зависит
только от временн\'ой координаты $\tau:=\frac12(F+G)$. Если
$\pl_\xi q\,\pl_\eta q<0$, то функция $q(\s)$ зависит только от пространственной
координаты $\s:=\frac12(F-G)$.
\end{prop}
\begin{proof}
Из-за знаков модулей в уравнении (\ref{qfrtd}), возможны два случая.

Если $\pl_\xi q\,\pl_\eta q>0$, то справедливо равенство
\begin{equation}                                                  \label{qfitye}
  \frac{\pl q}{\pl(F-G)}=\pl_\xi q\frac{\pl\xi}{\pl(F-G)}
  +\pl_\eta q\frac{\pl\eta}{\pl(F-G)}=\frac{\pl_\xi q}{F'}
  -\frac{\pl_\eta q}{G'}=0.
\end{equation}
Последнее равенство вытекает из уравнения (\ref{qfrtd}). Поэтому, переходя к
координатам $F,G\mapsto\tau,\s$, получаем сделанное утверждение.

Аналогично, если $\pl_\xi q\,\pl_\eta q<0$, то выполнено равенство
\begin{equation}                                                  \label{qtersw}
  \frac{\pl q}{\pl(F+G)}=\pl_\xi q\frac{\pl\xi}{\pl(F+G)}
  +\pl_\eta q\frac{\pl\eta}{\pl(F+G)}=\frac{\pl_\xi q}{F'}
  +\frac{\pl_\eta q}{G'}=0. \qed
\end{equation}
\renewcommand{\qed}{}\end{proof}
Теперь из каждого из двух уравнений (\ref{qeqwsa}) или (\ref{qeaswq}) следует
одно и то же равенство
\begin{equation*}
  |\Phi|=\frac1{2C}F'G'|q'|,
\end{equation*}
где $q'$ обозначает производную функции $q$ либо по $\tau:=\frac12(F+G)$, либо
по $\s:=\frac12(F-G)$. Постоянная $C$ соответствует растяжке новых координат
$F,G$, и, для упрощения последующих формул, положим $C=1/2$. При конформном
преобразовании конформный множитель преобразуется по правилу (\ref{qcased}).
Поэтому после конформного преобразования (\ref{qcased}) будет выполнено
равенство
\begin{equation}                                                  \label{qsoiht}
  |\tilde\Phi|=|q'|.
\end{equation}
В дальнейшем знак тильды у конформного множителя мы, для краткости, опустим.

Таким образом, координаты всегда можно выбрать таким образом, чтобы функции
$q$ и $\Phi$ зависели одновременно только от времениподобной или
пространственноподобной координаты
\begin{equation}                                                  \label{eindva}
  \z=\frac12(F\pm G):=\begin{cases} \tau,\qquad \pl_\xi q\,\pl_\eta q>0,
  \\ \s,\qquad \pl_\xi q\,\pl_\eta q<0. \end{cases}
\end{equation}
Это значит, что двумерная метрика (\ref{ecogag}) обладает вектором Киллинга,
$\pl_\s$ или $\pl_\tau$, как следствие уравнений (\ref{egcoga}) и
(\ref{egcogb}). Назовем эти решения соответственно однородными и статическими,
хотя это и относится только к определенной системе координат. Существование
вектора Киллинга является обобщением {\em теоремы Бирхгоффа} \cite{Birkho23},
утверждающей, что произвольное сферически симметричное решение вакуумных
уравнений Эйнштейна должно быть статическим. (Это утверждение было опубликовано
ранее в статье \cite{Jebsen21}.) Обобщение заключается в том, что наличие
вектора Киллинга доказано не только для сферически симметричных решений $(K=1)$,
но и для решений, инвариантных относительно групп преобразований $\MI\MO(2)$
$(K=0)$ и $\MS\MO(1,2)$ $(K=-1)$.
\index{Теорема Бирхгоффа (Birkhoff's theorem)}%
\index{Бирхгоффа теорема (Birkhoff's theorem)}%

Окончательно, решением уравнений (\ref{egcoga}), (\ref{egcogb}) в фиксированной
системе координат является равенство (\ref{qsoiht}) и утверждение о том, что
функции $q$ и $\Phi$ зависят только от одной переменной $\z$ (\ref{eindva}).
Осталось решить только одно уравнение (\ref{egcogc}).

В статическом, $q=q(\s)$, и однородном, $q=q(\tau)$, случаях уравнение
(\ref{egcogc}) принимает вид
\begin{align}                                                     \label{qfijhg}
  (q^2)''&=\quad 2(K-\Lm q^2)\Phi, & q&=q(\s),
\\                                                                \label{qsedrf}
  (q^2)''&=-2(K-\Lm q^2)\Phi, & q&=q(\tau).
\end{align}
Чтобы проинтегрировать полученные уравнения, необходимо выразить $\Phi$ через
$q$ с помощью уравнения (\ref{qsoiht}), а для этого необходимо раскрыть знаки
модулей.

Рассмотрим подробно статический случай $q=q(\s)$, $\Phi>0$ и $q'>0$. Тогда
уравнение (\ref{qfijhg}) с учетом (\ref{qsoiht}) примет вид
\begin{equation*}
  (q^2)''=2(K-\Lm q^2)q'.
\end{equation*}
Его легко проинтегрировать
\begin{equation*}
  (q^2)'=2\left(Kq-\frac{\Lm q^3}3-2M\right),
\end{equation*}
где $M=\const$ -- постоянная интегрирования. В дальнейшем мы увидим, что она
совпадает с массой в решении Шварцшильда. Выполнив дифференцирование в левой
части и поделив на $2q>0$, получим уравнение
\begin{equation*}
  q'=K-\frac{2M}q-\frac{\Lm q^2}3.
\end{equation*}
Поскольку в рассматриваемом случае $q'=\Phi$, то отсюда следует выражение для
конформного множителя через переменную $q$:
\begin{equation}                                                  \label{qcohjy}
  \Phi(q)=K-\frac{2M}q-\frac{\Lm q^2}3.
\end{equation}

Если $q=q(\s)$, $\Phi>0$ и $q'<0$, то аналогичное интегрирование приводит к
уравнению
\begin{equation*}
  q'=-\Phi(q),
\end{equation*}
где в правой части стоит тот же самый конформный множитель (\ref{qcohjy}). Этот
случай можно объединить с предыдущим, записав уравнение для $q$ в виде
\begin{equation}                                                  \label{qeqsra}
  |q'|=\Phi(q),\qquad q=q(\s),\quad \Phi>0.
\end{equation}

Аналогично интегрируется статический случай при $\Phi<0$:
\begin{equation}                                                  \label{qftrye}
  |q'|=-\Phi(q),\qquad q=q(\s),\quad \Phi<0.
\end{equation}

Если решение однородно, $q=q(\tau)$ и $\Phi>0$, $q'>0$, то интегрирование
уравнения (\ref{qsedrf}) приводит к равенству
\begin{equation*}
  q'=-\left(K-\frac{2M}q-\frac{\Lm q^2}3\right).
\end{equation*}
То есть в этом случае конформный множитель надо отождествить с правой частью
\begin{equation}                                                  \label{qnewsd}
  \hat\Phi=-\left(K-\frac{2M}q-\frac{\Lm q^2}3\right).
\end{equation}
Поскольку выражение конформного множителя в однородном случае через $q$
отличается знаком, то мы пометили его шляпкой. Таким образом, однородные
решения уравнений Эйнштейна можно записать в виде
\begin{align}                                                     \label{qfredl}
  |q'|&=\quad \hat\Phi(q),\qquad q=q(\tau),\quad \hat\Phi>0.
\\                                                                \label{qsdeax}
  |q'|&=-\hat\Phi(q),\qquad q=q(\tau),\quad \hat\Phi<0.
\end{align}

Если конформный множитель отрицателен, то сигнатура метрики равна $(-+--)$. В
этом случае, сделав замену $\tau\leftrightarrow\s$, мы вернемся к прежней
сигнатуре метрики $(+---)$. Это преобразование позволяет объединить стационарные
и однородные решения, написав знак модуля у конформного множителя в выражении
для метрики (\ref{emetko}). Тогда общее решение вакуумных уравнений Эйнштейна
(\ref{einseq}) в соответствующей системе координат примет вид
\begin{equation}                                                  \label{qnhtsd}
 ds^2=|\Phi|(d\tau^2-d\s^2)-q^2d\Om,
\end{equation}
где конформный множитель $\Phi$ имеет вид (\ref{qcohjy}). При этом переменная
$q$ зависит либо от $\s$ (статическое локальное решение), либо от $\tau$
(однородное локальное решение) через дифференциальное уравнение
\begin{equation}                                                  \label{qdtres}
  \left|\frac{dq}{d\z}\right|=\pm\Phi(q),
\end{equation}
где выполнено правило знаков:
\begin{equation}                                                  \label{esignk}
\begin{array}{ccl}
\Phi>0: & \quad \z=\s, &\quad \text{знак $+$ (статическое локальное решение)},\\
\Phi<0: & \quad \z=\tau, &\quad \text{знак $-$ (однородное локальное решение)}.
\end{array}
\end{equation}
Таким образом, из четырехмерных уравнений Эйнштейна вытекает, что на поверхности
$\MU$ возникает метрика с одним вектором Киллинга, которая была подробно
рассмотрена в главе \ref{sglosu}. Теперь с помощью метода конформных блоков
можно построить глобальные (максимально продолженные вдоль экстремалей) решения
вакуумных уравнений Эйнштейна. Число особенностей и нулей конформного множителя
(\ref{qcohjy}) зависит от соотношения между постоянными $K$, $M$ и $\Lm$.
Поэтому возможно существование многих существенно различных глобальных решений,
которые мы рассмотрим в следующих разделах.

Конформный множитель (\ref{qcohjy}) имеет одну сингулярность: простой полюс при
$q=0$. Поэтому, согласно правилам построения глобальных решений из раздела
\ref{sglrul} каждое глобальное решение соответствует одному из интервалов
$(-\infty,0)$ или $(0,\infty)$. Из вида конформного множителя (\ref{edefus})
следует, что эти глобальные решения получаются друг из друга преобразованием
$M\mapsto-M$. Поэтому, не ограничивая общности, мы опишем только глобальные
решения, соответствующие положительному интервалу $(q_-,q_+)=(0,\infty)$.
Впрочем, это предположение уже было сделано при параметризации (\ref{qparam}).

Поскольку конформный множитель $\Phi(q)$ является гладкой функцией при $q>0$, то
все возникающие лоренцевы поверхности $\MU$ и метрика на них, являются гладкими.

Используя уравнения (\ref{einscf}) и (\ref{esecom}), нетрудно
вычислить скалярную кривизну поверхности $\MU$:
\begin{equation}                                                  \label{escsok}
  R^g=\frac23\Lm+\frac{4M}{q^3}.
\end{equation}
Она не зависит от гауссовой кривизны $K$ римановой поверхности $\MV$ и
сингулярна при $q=0$, если $M\ne0$. Отметим, что сингулярная часть двумерной
скалярной кривизны (\ref{escsok}) пропорциональна собственному значению
четырехмерного тензора Вейля (см., например, \cite{LanLif88R}):
\begin{equation}                                                  \label{ectsqu}
  \frac1{48}\widehat C_{ijkl}\widehat C^{ijkl}=\left(-\frac M{q^3}\right)^2.
\end{equation}

Теперь перейдем к описанию всех пространственно симметричных глобальных решений
вакуумных уравнений Эйнштейна.
\subsection{Сферически симметричные решения $K=1$                \label{sphers}}
При $K=1$ риманова поверхность $\MV$ представляет собой сферу $\MS^2$, и все
решения сферически симметричны. Для сферы единичного радиуса метрику
(\ref{ecocrm}) запишем в сферических координатах
\begin{equation}                                                  \label{eliesp}
  d\Om=d\theta^2+\sin^2\theta d\vf^2.
\end{equation}
Сферически симметричную метрику пространства-времени, которая удовлетворяет
уравнениям Эйнштейна, можно записать в виде
\begin{equation}                                                  \label{esphsi}
  ds^2=|\Phi(q)|(d\tau^2-d\s^2)-q^2(d\theta^2+\sin^2\theta d\vf^2),
\end{equation}
где
\begin{equation}                                                  \label{edefus}
  \Phi(q)=1-\frac{2M}q-\frac{\Lm q^2}3.
\end{equation}
Переменная $q$ связана с $\s$ или $\tau$ дифференциальным уравнением
(\ref{qdtres}), где выполнено правило знаков (\ref{esignk}).
\begin{defn}
Координаты $\tau,\s$, в которых записана сферически симметричная метрика
(\ref{esphsi}), называются {\em черепашьими}. Это название, по-видимому,
произошло потому что диаграммы Картера--Пенроуза чем то напоминают рисунок
панциря черепахи.
\qed\end{defn}
\index{Черепашьи координаты (tortoise coordinates)}%
\index{Координаты черепашьи (tortoise coordinates)}%

Обобщение решения Шварцшильда на случай ненулевой космологической постоянной
(\ref{esphsi}) было получено Коттлером \cite{Kottle18}.

В рассматриваемом случае все решения параметризуются двумя постоянными:
космологической постоянной $\Lm$ и массой $M$. Вторую постоянную мы будем
называть массой, хотя она и не имеет физического смысла массы для большинства
решений, отличных от решения Шварцшильда.

Как было отмечено в предыдущем разделе, мы рассматриваем глобальные решения
вакуумных уравнений Эйнштейна, соответствующих положительному интервалу
$(q_-,q_+)=(0,\infty)$. Структура глобальных решений определяется количеством и
типом положительных корней уравнения $\Phi(q)=0$ или эквивалентного кубического
уравнения
\begin{equation}                                                  \label{erootn}
  \frac\Lm3q^3-q+2M=0.
\end{equation}
При ненулевой космологической постоянной это уравнение может иметь до трех
нулей. Элементарный анализ показывает, что по крайней мере один из корней
отрицателен. Это значит, что при положительных $q$ возможно существование не
более двух горизонтов.

\begin{figure}[h,b,t]
\hfill\includegraphics[width=.95\textwidth]{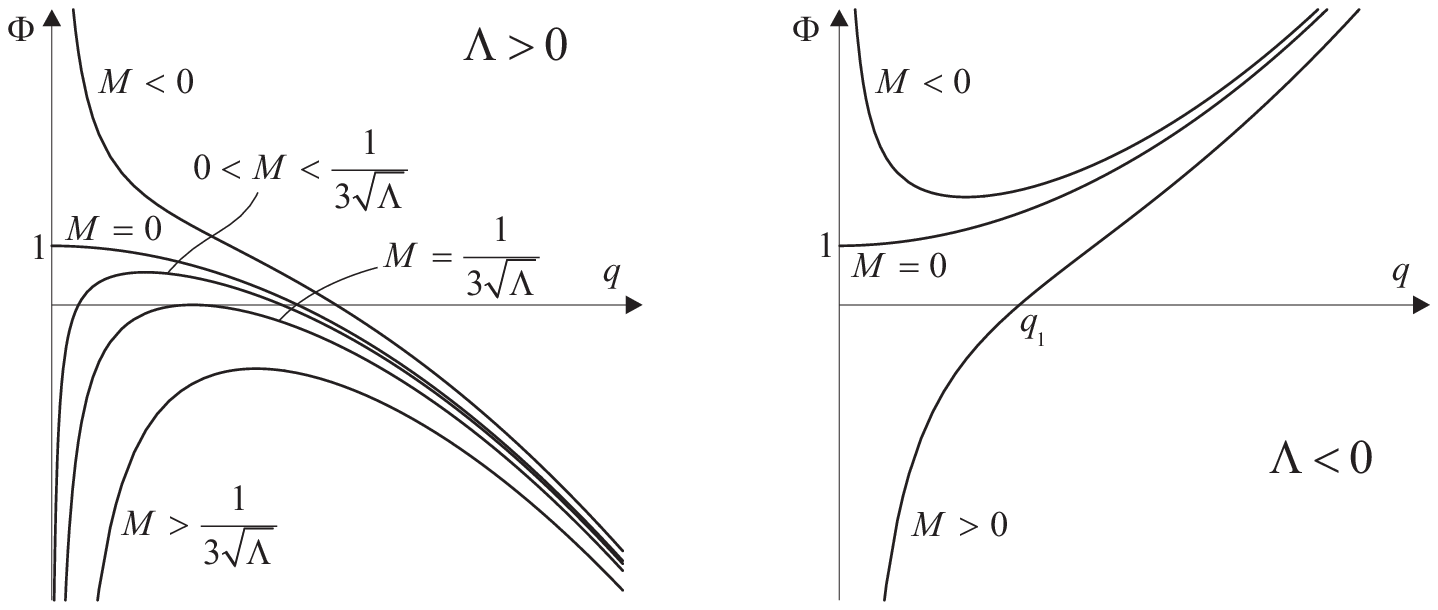}
\hfill {}
\centering\caption{Поведение конформного множителя
 $\Phi=1-\frac{2M}q-\frac{\Lm q^2}3$ в зависимости от величины постоянной $M$
 (массы) при положительной (слева) и отрицательной (справа) космологической
 постоянной $\Lm$.}
\label{fspfercofa}
\end{figure}

При положительной космологической постоянной $\Lm>0$ существуют следующие
возможности в зависимости от величины массы $M$ (см.\
рис.\ref{fspfercofa} слева). Если $M>\frac1{3\sqrt\Lm}$, то уравнение
(\ref{erootn}) не имеет положительных корней. При $M=\frac1{3\sqrt\Lm}$
возникает один положительный корень второго порядка. В интервале значений
$0<M<\frac1{3\sqrt\Lm}$ существуют два положительных корня. При неположительных
значениях $M\le0$ имеется один простой положительный нуль.

При отрицательных значениях космологической постоянной $\Lm<0$ мы имеем один
простой положительный нуль при $M>0$ и ни одного нуля при $M\le0$ (см.\
рис.\ref{fspfercofa} справа).

Перейдем к классификации решений.
\subsubsection{Пространство-время Минковского $\Lm=0$, $M=0$}
Наиболее простое сферически симметричное решение получается при $\Lm=0$ и $M=0$
(см.\ рис.\ref{fsphersym}). В этом случае $\Phi=1$, и метрика принимает вид
\begin{equation}                                                  \label{emilis}
  ds^2=d\tau^2-dr^2-r^2(d\theta^2+\sin^2\theta d\vf^2),\qquad r\in(0,\infty),
\end{equation}
где мы переобозначили $\s\mapsto r$. Точка $r=0$ является координатной
сингулярностью. Переходя к декартовым координатам в четырехмерном
пространстве-времени и добавляя мировую линию начала сферической системы
координат $r=0$, мы получим пространство-время Минковского $\MR^{1,3}$. При
этом пространственная координата $r$ естественным образом отождествляется
с радиусом сферической системы координат. В этом случае пространство-время
нельзя представить в виде топологического произведения $\MU\times\MS^2$, и
поэтому диаграмма Картера--Пенроуза на рис.\ref{fsphersym} не продолжается.
\subsubsection{Черная дыра Шварцшильда $\Lm=0$, $M>0$}
\begin{figure}[p,b,t]
\hfill\includegraphics[width=.95\textwidth]{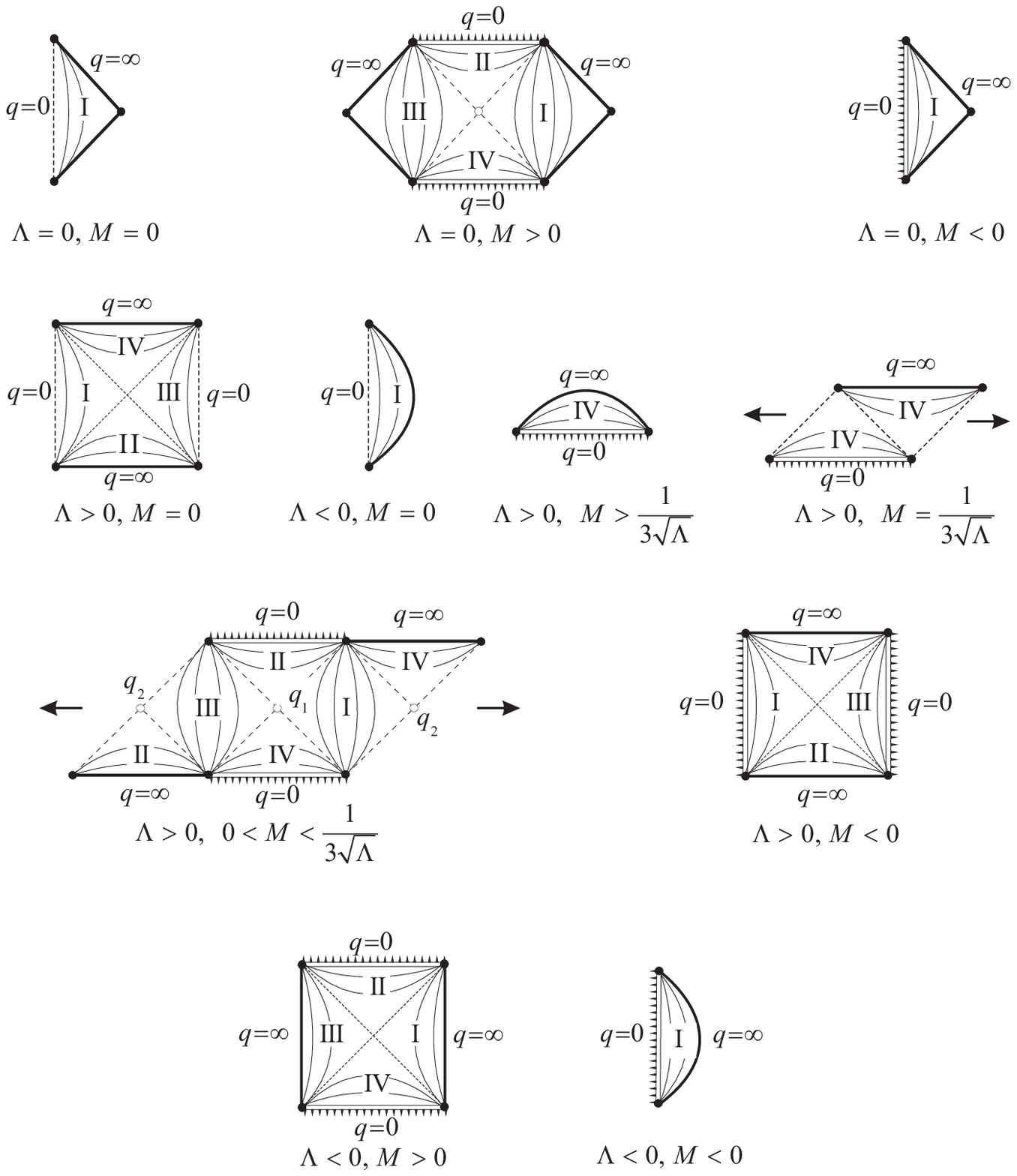}
\hfill {}
\centering\caption{Сферически симметричные решения для конформного множителя
$\Phi=1-\frac{2M}q-\frac{\Lm q^2}3$. Условные обозначения
 те же, что и на рис.\ref{fboprr}. Стрелки показывают возможное
 периодическое продолжение решений.}
\label{fsphersym}
\end{figure}

Решение Шварцшильда соответствует нулевой космологической постоянной $\Lm=0$ и
положительной массе $M>0$. Оно уже обсуждалось в разделе \ref{schwas}. В этом
случае конформный множитель (\ref{edefus}) имеет один простой нуль в точке
$q_1=2M$ и, следовательно, один горизонт. Соответствующая диаграмма
Картера--Пенроуза также изображена на рис.\ref{fsphersym}.
\subsubsection{Голая сингулярность $\Lm=0$, $M<0$}
При нулевой космологической постоянной $\Lm=0$ и отрицательной массе $M<0$
уравнение (\ref{erootn}) не имеет положительных корней. Этот случай также был
рассмотрен в разделе \ref{schwas}. При указанных значениях постоянных существуют
две диаграммы Картера--Пенроуза, каждая из которых состоит из одного
треугольного конформного блока, изображенного на рис.\ref{fsphersym}, и его
пространственного отражения. Каждый из конформных блоков представляет собой
максимально продолженную поверхность $\MU$. Сингулярная граница времениподобна и
называется {\em голой сингулярностью}, т.к.\ не окружена горизонтом.
\index{Голая сингулярность (naked singularity)}%
\index{Сингулярность голая (naked singularity)}%
\subsubsection{Решение де Ситтера $\Lm>0$, $M=0$}
Решение де Ситтера соответствует положительной космологической постоянной и
нулевой массе. При этом пространство-время представляет собой многообразие
постоянной (в наших обозначениях положительной) скалярной кривизны и может быть
представлено, как четырехмерный гиперболоид, вложенный в пятимерное пространство
Минковского $\MR^{1,4}$, с индуцированной метрикой. Его группой симметрии
является группа Лоренца $\MO(1,4)$, а метрика имеет максимальное число -- десять
-- векторов Киллинга, что совпадает с размерностью группы симметрии. Конформный
множитель (\ref{edefus}) имеет один простой положительный нуль, соответствующий
горизонту.

Статические и однородные решения в координатах Шварцшильда имеют вид
\begin{align}                                                     \label{edesss}
  ds^2=\left(1-{\frac\Lm3}r^2\right)d\tau^2
  -\frac{dr^2}{1-\frac\Lm3r^2}-r^2d\Om^2,
\\                                                                     \nonumber
  0<r<{\sqrt{\textstyle\frac3\Lm}}\,,\quad -\infty<\tau<\infty\,,
\\                                                                \label{edeshs}
  ds^2=-\frac{dt^2}{1-\frac\Lm3t^2}
  +\left(1-{\frac\Lm3}t^2\right)d\s^2-t^2d\Om^2,
\\                                                                     \nonumber
  {\sqrt{\textstyle\frac3\Lm}}<t<\infty\,,\quad -\infty<\s<\infty \,.
\end{align}
Поскольку уравнение (\ref{qdtres}) в этом случае интегрируется явно, то метрику
можно записать также в конформно плоском виде
\begin{align}                                                     \label{edescs}
  ds^2&=\frac1{\ch^2\left(\sqrt{\frac\Lm3}\s\right)}(d\tau^2-d\s^2)
  -\frac3\Lm\tanh^2\left({\textstyle\sqrt{\frac\Lm3}}\s\right)d\Om^2,
\\                                                                \label{edesch}
  ds^2&=\frac1{\sh^2\left(\sqrt{\frac\Lm3}\tau\right)}(d\tau^2-d\s^2)
  -\frac3\Lm\cth^2\left({\textstyle\sqrt{\frac\Lm3}}\tau\right)d\Om^2,
\end{align}
соответственно для статического и однородного случая. Область определения
$r\in\left(0,\sqrt{3/\Lm}\right)$ переходит в $\s\in(0,\infty)$, а
$t\in\left(\sqrt{3/\Lm},\infty\right)$ -- в $\tau\in(0,\infty)$.

Поведение конформного множителя показано на рис.\ref{fspfercofa} слева. Он имеет
один простой нуль в некоторой точке, которую мы обозначим $q_1$. В интервале
$(0,q_1)$ конформный множитель положителен и ему ставится в соответствие два
треугольных статических конформных блока I и III с метрикой (\ref{edescs}). В
интервале $(q_1,\infty)$ конформный множитель отрицателен, и ему соответствуют
два треугольных однородных конформных блока II и IV с метрикой (\ref{edesch}).
Эти четыре конформных блока склеиваются единственным образом по правилам,
сформулированным в разделе \ref{sglrul}. Соответствующая диаграмма
Картера--Пенроуза показана на рис.\ref{fsphersym}. Решение де Ситтера получится
после добавления времениподобной границы $r=0$ к сплетенному произведению
$\MU\times\MS^2$.

Заметим, что конформный множитель для решения де Ситтера является квадратичным
полиномом, $\Phi=1-\frac{\Lm q^2}3$. Как показано в разделе \ref{scocus}, при
максимальном продолжении поверхности с такой метрикой возникает однополостный
гиперболоид. В рассматриваемом случае мы этого не делаем не потому что возникают
какие то проблемы с поверхностью $\MU$, а потому что при $q=0$ коэффициент при
угловой части метрики (\ref{esphsi}) становится равным нулю и метрика
четырехмерного пространства-времени вырождается. Здесь возникает такая же
ситуация, как в пространстве Минковского $\MR^{1,3}$ в сферической системе
координат.
\subsubsection{Решение анти-де Ситтера $\Lm<0$, $M=0$}
Изменение знака космологической постоянной в уравнениях Эйнштейна приводит к
качественному изменению решений. При нулевой массе мы получаем решение анти-де
Ситтера. Соответствующее пространство-время представляет собой многообразие
постоянной отрицательной кривизны, которое можно представить в виде
четырехмерного гиперболоида, вложенного в плоское пространство $\MR^{2,3}$ с
метрикой $\eta_{\al\bt}=\diag(1,1,-1,-1,-1)$. Метрика анти де Ситтера
симметрична относительно действия группы вращений $\MO(2,3)$ и имеет также
максимальное число -- десять -- векторов Киллинга. Конформный множитель
(\ref{edefus}) не имеет нулей и всегда положителен. Поэтому решение статично и
не имеет горизонтов. В координатах Шварцшильда метрика имеет тот же вид
(\ref{edesss}), что и для решения де Ситтера, однако из-за отрицательного знака
$\Lm$ область изменения $r$ совпадает со всем положительным интервалом
$(0,\infty)$. В конформно плоском виде метрика принимает вид
\begin{equation}                                                  \label{eadscg}
  ds^2=\frac1{\cos^2\left(\sqrt{\frac{|\Lm|}3}\s\right)}(d\tau^2-d\s^2)
  -\frac3{|\Lm|}\tanh^2\left({\textstyle\sqrt{\frac{|\Lm|}3}}\s\right)d\Om.
\end{equation}
При этом координата $\s$ меняется в конечном интервале
$\s\in\left(0,\frac\pi2\sqrt{\frac3{|\Lm|}}\right)$. Соответствующая диаграмма
Картера--Пенроуза имеет вид линзы и изображена на рис.\ref{fsphersym}. Это
решение неполно на границе $r=0$ и не может быть продолжено, что является
следствием выбора сферической системы координат. Для получения полного решения
анти-де Ситтера мировую линию $r=0$ необходимо добавить к многообразию, что
можно сделать путем перехода к другой системе координат.
\subsubsection{Однородная пространственная сингулярность
                                       $\Lm>0$, $M>\frac1{3\protect\sqrt{\Lm}}$}
При $\Lm>0$, $M>\frac1{3\protect\sqrt{\Lm}}$ конформный множитель $\Phi$ для
положительных $q>0$ не имеет нулей и всегда отрицателен. Решение однородно и не
имеет горизонтов. Поверхность $\MU$ в этом случае представляется диаграммой
Картера--Пенроуза в виде линзы, изображенной на рис.\ref{fsphersym}, и ее
отражения во времени. При конечном значении времени в прошлом имеется истинная
сингулярность и у четырехмерной кривизны, и у тензора кривизны поверхности
$\MU$. Многообразие не может быть продолжено через эту сингулярность.
Космологическая интерпретация этого решения простая. Вселенная рождается в
конечном прошлом из сингулярности и развивается вечно. В пределе
$t\rightarrow\infty$ метрика на поверхности $\MU$ стремится к метрике де
Ситтера. Это же верно для трех следующих глобальных решений.
\subsubsection{Двойной горизонт        $\Lm>0$, $M=\frac1{3\protect\sqrt{\Lm}}$}
Если $\Lm>0$, $M=\frac1{3\protect\sqrt{\Lm}}$ то конформный множитель $\Phi$
имеет один нуль второго порядка, соответствующий горизонту $q_1=1/\sqrt\Lm$. При
$q>0$ функция $\Phi$ неотрицательна и, следовательно, все конформные блоки
однородны. Два однородных конформных блока соответствуют интервалу $(0,q_1)$ и
два -- интервалу $(q_1,\infty)$. Они склеиваются в фундаментальную область,
показанную на рис.\ref{fsphersym}. Эту фундаментальную область можно либо
продолжить влево и вправо, что приводит к универсальному накрывающему
пространству, либо отождествить противоположные неполные края. В последнем
случае возникающая поверхность диффеоморфна цилиндру. В случае одного двойного
горизонта имеются также поверхности, возникающие из построенных выше отражением
во времени.
\subsubsection{Два горизонта $\Lm>0$, $0<M<\frac1{3\protect\sqrt{\Lm}}$}
При $\Lm>0$, $0<M<\frac1{3\protect\sqrt{\Lm}}$ конформный множитель $\Phi$ имеет
два нуля и, значит, два горизонта в точках $q_1$ и $q_2$. Двум интервалам
$(0,q_1)$ и $(q_2,0)$, в которых конформный множитель отрицателен, ставится в
соответствие по два однородных треугольных конформных блока. В интервале
$(q_1,q_2)$ конформный множитель положителен, и ему ставится в соответствие
два статических квадратных конформных блока. Фундаментальная область возникает
после склейки всех шести конформных блоков по правилам, описанным в разделе
\ref{sglrul}. Как и в предыдущем примере, фундаментальную область можно либо
продолжить влево и вправо, и получить универсальное накрывающее пространство,
как показано стрелками на рис.\ref{fsphersym}, либо отождествить неполные края.
\subsubsection{Две статические сингулярности $\Lm>0$, $M<0$}
При положительной космологической постоянной $\Lm>0$ и отрицательной массе
$M<0$ конформный множитель $\Phi$ имеет один простой нуль в точке $q_1$.
Интервалу $(0,q_1)$, в котором конформный множитель положителен, ставится в
соответствие два треугольных статических конформных блока. Интервалу
$(q_1,\infty)$, где конформный множитель отрицателен, ставится в соответствие
два треугольных однородных конформных блока. После склейки этих блоков
возникает диаграмма Картера--Пенроуза, показанная на рис.\ref{fsphersym}. Мы
имеем две сингулярности в статических областях I, III, разделенных горизонтами.
Из области I вдоль времениподобной кривой можно достичь только правую
сингулярность. Это же верно для области III и левой сингулярности. Из области II
имеется возможность попасть в обе сингулярности. Из области IV никакой
сингулярности достичь невозможно, и жизнь там продолжается вечно, если вы
родились в этой области.
\subsubsection{Черная дыра анти-де Ситтера $\Lm<0$, $M>0$}
Если $\Lm<0$ и $M>0$, то конформный множитель $\Phi$ имеет один нуль и,
соответственно, один горизонт в точке $q_1$, как показано на
рис.\ref{fspfercofa} справа. Интервалу $(0,q_1)$, где конформный множитель
отрицателен, ставится в соответствие два треугольных однородных конформных
блока. В интервале $(q_1,\infty)$ конформный множитель положителен, и ему
соответствует два треугольных статических конформных блока. Эти четыре
конформных блока склеиваются в диаграмму Картера--Пенроуза, показанную на
рис.\ref{fsphersym}. Она аналогична диаграмме для шварцшильдовской черной дыры.
Отличие заключается в том, что полные левая и правая границы в рассматриваемом
случае времениподобны. Сингулярности кривизны расположены в конечном прошлом
(белая дыра) и в конечном будущем (черная дыра). Они имеют тот же характер, что
и в решении Шварцшильда. Метрика пространства-времени не является асимптотически
плоской и стремится к метрике анти-де Ситтера при $q=r\rightarrow\infty$.
\subsubsection{Голая сингулярность $\Lm<0$, $M<0$}
При $\Lm<0$ и $M<0$ имеется голая сингулярность без горизонтов. Соответствующая
диаграмма Картера--Пенроуза совпадает с диаграммой Картера--Пенроуза для
$\Lm>0$, $M>\frac1{3\protect\sqrt{\Lm}}$ изображенной на рис.\ref{fsphersym}, но
повернутой на угол $\pi/2$, т.к.\ конформный множитель положителен $\Phi>0$, и,
следовательно, метрика статична. Метрика при $q=r\rightarrow\infty$ также
стремится к метрике анти-де Ситтера.
\subsection{Планарные решения $K=0$                              \label{spherp}}
В случае $K=0$ метрика на римановой поверхности $\MV$ (\ref{ecocrm}) становится
евклидовой
\begin{equation}                                                  \label{evplso}
  d\Om_\Sp:=dy^2+dz^2.
\end{equation}
Это значит, что соответствующая максимально продолженная поверхность $\MV$
является либо евклидовой плоскостью $\MR^2$ с группой симметрии Пуанкаре
$\MI\MO(2)$, либо ее компактификацией (цилиндр, тор). Будем называть
соответствующие четырехмерные глобальные решения вакуумных уравнений Эйнштейна
{\em планарными}.
\index{Планарное решение (planar solution)}%
\index{Решение планарное (planar solution)}%
Для решений этого типа метрика в координатах Шварцшильда имеет вид
\begin{equation}                                                  \label{eplsoi}
  ds^2=\Phi(q)d\z^2-\frac{dq^2}{\Phi(q)}-q^2\Om_\Sp,
\end{equation}
где
\begin{equation}                                                  \label{edefut}
  \Phi (q)=-\frac{2M}q-\frac{\Lm q^2}3.
\end{equation}
Координаты $q$ и $\z$ определены в (\ref{qdtres}). Множество планарных решений
так же как и в сферически симметричном случае параметризуется двумя постоянными:
космологической постоянной $\Lm$ и массой $M$.
\begin{com}
Для планарных решений координата $q$ не может быть интерпретирована, как радиус
пространства. Поэтому мы с самого начала отказались от ее обозначения через $r$,
т.к.\ решение Шварцшильда является лишь частным случаем описываемого общего
подхода.
\qed\end{com}

Физическая интерпретация планарных решения весьма интересна. Например, при
$\MV = \MT^2$ трехмерное пространство представляет собой прямое произведение
тора $\MT^2$ и прямой $\MR$. Это пространство содержит нестягиваемые замкнутые
пространственноподобные кривые, т.е.\ имеет нетривиальную фундаментальную
группу. С физической точки зрения такие пространства описывают кротовые норы.
В этом случае все горизонты также представляют собой торы.

\begin{figure}[h,b,t]
\hfill\includegraphics[width=.95\textwidth]{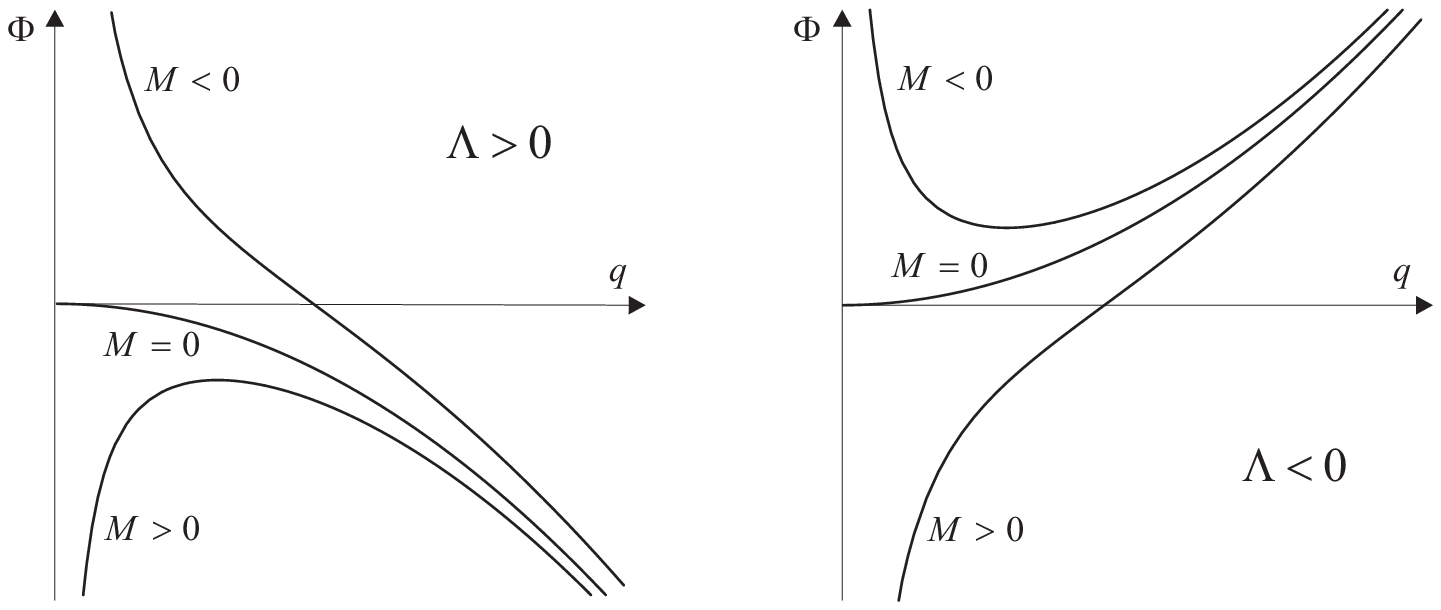}
\hfill {}
\centering\caption{Поведение конформного множителя
 $\Phi=-\frac{2M}q-\frac{\Lm q^2}3$ в зависимости от величины массы $M$ при
 положительной (слева) и отрицательной (справа) космологической постоянной
 $\Lm$.}
\label{fconfacpl}
\end{figure}

Для построения глобальных планарных решений вакуумных уравнений Эйнштейна
необходимо знать поведение конформного множителя (\ref{edefut}) при различных
значениях $\Lm$ и $M$. При $\Lm\ne0$ оно ясно из рис.\ref{fspfercofa} и
получается сдвигом оси абсцисс на единицу вверх, как показано на
рис.\ref{fconfacpl}.

При $M=0$ конформный множитель является квадратичным полиномом. Этот случай
был рассмотрен в разделе \ref{scocus} и соответствует поверхностям $\MU$
постоянной кривизны.

Если космологическая постоянная положительна, $\Lm>0$, то конформный множитель
при $M>0$ не имеет нулей. При $M<0$ существует один положительный нуль.

Для отрицательной космологической постоянной, $\Lm<0$, ситуация противоположна.
При $M>0$ есть один простой нуль, а при $M<0$ нули отсутствуют.

Перечислим все возможные максимально продолженные поверхности $\MU$. Как и
ранее, достаточно рассмотреть глобальные решения, соответствующие интервалу
$q\in(0,\infty)$.
\subsubsection{Однородные и голые сингулярности $\Lm=0$, $M\ne0$}
\begin{figure}[h,b,t]
\hfill\includegraphics[width=.95\textwidth]{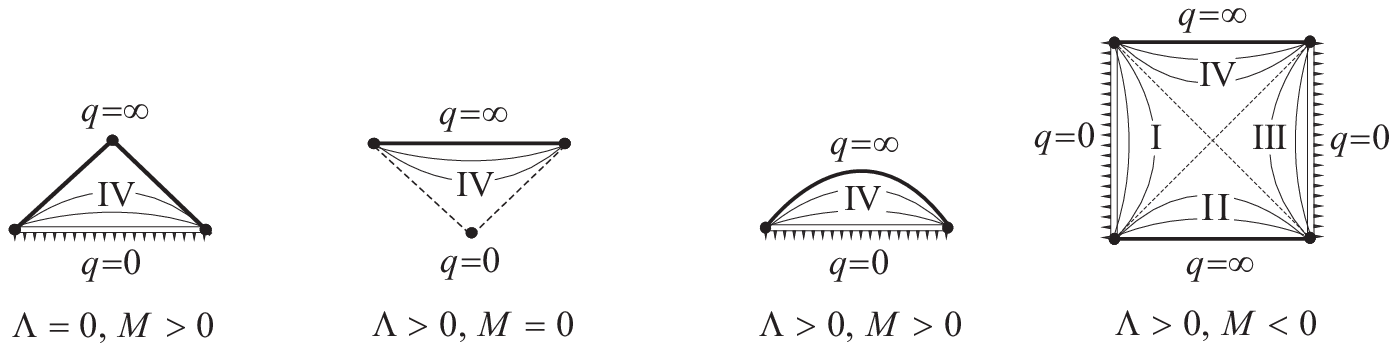}
\hfill {}
\centering\caption{Планарные решения $K=0$. Условные обозначения
 те же, что и на рис.\ref{fboprr}.}
\label{fplansol}
\end{figure}

При нулевой космологической постоянной и положительной массе $M>0$ конформный
множитель при $q>0$ отрицателен и не имеет нулей (горизонтов). Поэтому
глобальное решение состоит из одного треугольного конформного блока типа IV,
однородно и имеет пространственноподобную сингулярность при $q=0$. В
шварцшильдовских и черепашьих координатах метрика имеет вид
\begin{align}                                                     \label{eplmas}
  ds^2&=\frac t{2M}dt^2-\frac{2M}td\s^2-t^2d\Om_\Sp,
  &&t\in(0,\infty),
\\
  ds^2&=\sqrt{\textstyle\frac M{|\tau|}}(d\tau^2-d\s^2)-4M|\tau|d\Om_\Sp,
  &&\mbox{$\tau<0$\quad или\quad $\tau>0$}.
\end{align}
Соответствующая диаграмма Картера--Пенроуза изображена на рис.\ref{fplansol}.

Как и ранее, существует также глобальное решение типа II, полученное из
показанного на рисунке обращением времени.

Если масса отрицательна, $M<0$, то оба глобальных решения имеют такой же вид как
и при $M>0$, но являются статическими. Поэтому соответствующие диаграммы
Картера--Пенроуза необходимо просто повернуть на угол $\pi/2$. В таком
случае возникают голые сингулярности.
\subsubsection{Решение (анти-) де Ситтера            $\Lm\ne0$, $M=0$}
При положительной космологической постоянной $\Lm>0$ и нулевой массе $M=0$
конформный множитель
$$
  \Phi =-\frac{\Lm t^2}3,\qquad q:=t\in(0,\infty),
$$
всюду отрицателен и не имеет нулей. Соответствующее глобальное решение однородно
и не имеет горизонтов. Оно представляется треугольным конформным блоком типа IV,
изображенном на рис.\ref{fplansol}. Есть также глобальное решение типа II,
полученное из решения типа IV обращением времени.

В шварцшильдовских и черепашьих координатах метрика имеет вид
\begin{align}                                                     \label{enlzms}
  ds^2&=\frac3{\Lm t^2}dt^2-\frac{\Lm t^2}3d\s^2-t^2d\Om_\Sp,
  &&t\in(0,\infty),
\\                                                                \label{enlzmc}
  ds^2&=\frac3{\Lm\tau^2}(d\tau^2-d\s^2)-\frac9{\Lm^2\tau^2}d\Om_\Sp,
  &&\tau\in(0,\infty).
\end{align}

Вдоль границы $q=0$ (т.е.\ $t=0$) диаграммы Картера--Пенроуза $4$-х мерная
метрика вырождается и ее нельзя продолжать. Эти светоподобные края показаны на
рис.\ref{fplansol} пунктиром.

Можно проверить, что метрики (\ref{enlzms}) и (\ref{enlzmc}) описывают
пространства постоянной четырехмерной кривизны. Это означает, что эти метрики
представляют собой решение де Ситтера.

В случае отрицательной космологической постоянной $\Lm<0$ зависимость компонент
метрики от времени в выражениях (\ref{enlzms}) и (\ref{enlzmc}) необходимо
заменить на зависимость от пространственной координаты, и соответствующие
диаграммы Картера--Пенроуза повернуть на угол $\pi/2$. В этом случае
пространство-время снова является пространством постоянной кривизны, и возникает
новое представление решения анти-де Ситтера.
\subsubsection{Однородные и голые сингулярности $\Lm M>0$}
При положительной космологической постоянной, $\Lm>0$, и положительной массе,
$M>0$, конформный множитель
\begin{equation}                                                  \label{enfuzk}
  \Phi =-\frac{\Lm t^3+6M}{3t}<0,\qquad q:=t\in(0,\infty),
\end{equation}
отрицателен и не имеет нулей (горизонтов). В шварцшильдовских и черепашьих
координатах метрика имеет вид
\begin{align}                                                     \label{epslms}
  ds^2&=\frac{3t}{\Lm t^3+6M}dt^2-\frac{\Lm t^3+6M}{3t}d\s^2-t^2d\Om_\Sp,
  &&t\in(0,\infty),
\\
  ds^2&=|\Phi|(d\tau^2-d\s^2)-t^2d\Om_\Sp,
  &-\frac\pi{12\sqrt3M}&<\tau<\frac\pi{4\sqrt3M},
\end{align}
где функция $t=t(\tau)$ определена уравнением (\ref{qdtres}), которое
интегрируется и дает неявную связь $t=t(\tau)$ (при $q'>0$):
\begin{equation}                                                  \label{erects}
  \frac\Lm3(\tau+\const)=\frac1{6a}\ln\frac{a^2-at+t^2}{(a+t)^2}
  +\frac1{a\sqrt3}\arcth\left(\frac{2t-a}{a\sqrt3}\right),
  \qquad a:=6M/\Lm.
\end{equation}
Соответствующее максимально продолженное решение представляется диаграммой
Картера--Пенроуза в виде линзы типа IV, которая изображена на
рис.\ref{fplansol}, и ее отражением во времени (диаграмма типа II). Оно
описывает однородную пространственную сингулярность при $t=0$ и является
асимптотически пространством-временем де Ситтера при $t\rightarrow\infty$.

Глобальные решения, возникающие при $\Lm<0$, $M<0$, получается из рассмотренного
выше решения преобразованием $\tau\leftrightarrow\s$. Соответствующие решения
статичны. Диаграммы Картера--Пенроуза также имеют вид линзы и получаются из
предыдущего случая поворотом на угол $\pi/2$.
\subsubsection{Черное кольцо $\Lm M<0$}
Если $\Lm>0$ и $M<0$, то конформный множитель (\ref{enfuzk}) имеет один простой
нуль, и пространство-время содержит один горизонт. В этом случае имеется два
статических и два однородных конформных блока, которые вместе склеиваются вдоль
горизонта в диаграмму Картера--Пенроуза, изображенную на рис.\ref{fplansol}
справа. Она имеет тот же вид, что и в сферически симметричном случае $K=1$,
$\Lm>0$ и $M<0$ (рис.\ref{fsphersym}). Метрика для однородного решения имеет вид
(\ref{epslms}), но из-за условия (\ref{erects}) области определения координат
$t$ и $\tau$ меняются. Для статического решения метрика задается уравнением
(\ref{eplsoi}) с соответствующей областью определения координат $r$ и $\s$,
связанных тем же уравнением, что и $t$ и $\tau$.

Решения, возникающие при $\Lm<0$, $M>0$, получаются из решений, соответствующих
$\Lm>0$, $M<0$, путем перестановки пространственной и временной координаты
$\tau\leftrightarrow\s$. Соответствующую диаграмму Картера--Пенроуза необходимо
повернуть на угол $\pi/2$. По виду она совпадает с черной дырой анти де Ситтера.
В этом случае после компактификации $\MV=\MT^2$ горизонт черной дыры имеет вид
тора $\MT^2$, а пространство-время является асимптотически пространством
анти-де Ситтера. Следовательно, в пространстве возникает кольцевая черная дыра
(черное кольцо). Напомним, что в этом случае все пространство имеет вид
топологического произведения прямой $\MR$ и тора $\MT^2$.
\subsection{Гиперболические глобальные решения $K=-1$            \label{sootss}}
При $K=-1$ поверхность $\MV$ представляет собой двуполостный гиперболоид
$\MH^2$ (плоскость Лобачевского). Точнее, верхнюю полу двуполостного
гиперболоида (см.\ раздел \ref{sutwse}). После компактификации $\MH^2$ в
качестве поверхности $\MV$ получится компактная риманова поверхность рода два и
выше. Отметим, что группой изометрий однополостного гиперболоида $\MH^2$
является группа преобразований Лоренца $\MO(1,2)$. То есть возникают решения
вакуумных уравнений Эйнштейна симметричные относительно действия группы Лоренца
$\MS\MO(1,2)$ не в пространстве-времени, а на пространственных сечениях
$t=\const$.

Метрика двуполостного гиперболоида (\ref{ecocrm}) единичного радиуса в
гиперболической системе координат имеет вид (\ref{eintwh})
\begin{equation}                                                  \label{eilpac}
  d\Om_\Sh=d\theta^2+\sh^2\theta d\vf^2.
\end{equation}
Соответствующее вакуумное решение уравнений Эйнштейна можно записать в
координатах Шварцшильда
\begin{equation}                                                  \label{elosoi}
  ds^2=\Phi (q)d\z^2-\frac{dq^2}{\Phi (q)}-q^2d\Om_\Sh,
\end{equation}
где конформный множитель имеет вид
\begin{equation}                                                  \label{edelos}
  \Phi(q)=-1-\frac{2M}q-\frac{\Lm q^2}3.
\end{equation}

\begin{figure}[h,b,t]
\hfill\includegraphics[width=.35\textwidth]{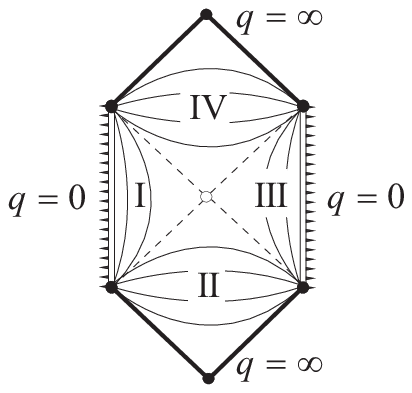}
\hfill {}
\centering\caption{Статические сингулярности в гиперболическом случае.}
\label{fhypsol}
\end{figure}

Чтобы описать глобальные гиперболические решения, заметим, что конформный
множитель $\Phi$ общего вида (\ref{qcohjy}), меняет знак на противоположный при
преобразовании всех постоянных:
\begin{equation*}
  K\mapsto-K,\qquad \Lm\mapsto-\Lm, \qquad M\mapsto-M.
\end{equation*}
Это означает, что все глобальные решения в случае $K=-1$, можно получить из
сферически симметричных решений, если поменять пространственную и временн\'ую
координату $\tau\leftrightarrow\s$ на поверхности $\MU$, а также изменить знак
космологической постоянной и массы.
\begin{exa}
Аналогом решения Шварцшильда является решение с нулевой космологической
постоянной $\Lm=0$ и отрицательной массой $M<0$. В этом случае метрика для
однородного и статичного конформного блока принимает вид
\begin{align}
  ds^2&=\frac{dt^2}{1+\frac{2M}t}-\left(1+{\textstyle\frac{2M}t}\right)d\s^2
       -t^2d\Om_\Sh,\qquad-2M<t<\infty,
\\                                                                \label{eschse}
  ds^2&=\left(1+{\textstyle\frac{2M}r}\right)d\tau^2
       -\frac{dr^2}{1+\frac{2M}r}-r^2d\Om_\Sh,\qquad 0<r<-2M.
\end{align}
При этом диаграмму Картера--Пенроуза для решения Шварцшильда необходимо
повернуть на угол $\pi/2$, рис.\ref{fhypsol}. Соответствующее глобальное решение
имеет левую и правую статические сингулярности при $q=r=0$. Свойства этого
глобального решения похожи на свойства двух статических сферически симметричных
сингулярностей, рассмотренных в разделе \ref{sphers}, но с другим
асимптотическим поведением. Пространство-время является асимптотически плоским в
бесконечно удаленном прошлом и будущем (оба случая соответствуют пределу
$q=t\rightarrow\infty$).
\qed\end{exa}

Остальные глобальные решения для $K=-1$ анализируются аналогично, и мы оставляем
это любознательному читателю.

Не так легко представить себе пространство (трехмерное сечение четырехмерного
пространства-времени, задаваемое уравнением $t=\const$) при $K=-1$. В
рассматриваемом случае оно представляет собой произведение интервала (конечного
или бесконечного в зависимости от значений $\Lm$ и $M$) и однополостного
гиперболоида $\MH^2$. После компактификации, когда поверхность $\MV$
представляет собой замкнутую риманову поверхность рода два и выше, глобальное
решение  можно интерпретировать, как множество кротовых нор, число которых
совпадает с родом поверхности (числом ручек). Топология горизонта совпадает с
топологией поверхности $\MV$.
\section{Лоренц-инвариантные решения                             \label{solcos}}
В настоящем разделе мы рассмотрим случай {\sf C} (\ref{ecases}), когда второе
дилатонное поле в сплетенном произведении (\ref{emetbl}) постоянно, $m=\const$.
В этом случае, как будет показано ниже, псевдориманова поверхность $\MU$ должна
быть поверхностью постоянной кривизны. Следовательно, она представляет собой
однополостный гиперболоид, $\MU=\ML^2$ или его универсальную накрывающую,
которые были подробно описаны в разделе \ref{shyper}. В этом случае глобальные
решения вакуумных уравнений Эйнштейна имеют вид топологического произведения
$\MM=\ML^2\times\MV$. Второй сомножитель $\MV$ представляет собой максимально
продолженную риманову поверхность с одним вектором Киллинга, которые были
рассмотрены в разделе \ref{sgloeu}. Как было показано, поверхность $\MV$ может
иметь конические сингулярности или сингулярности кривизны вдоль края поверхности
$\MV$. С физической точки зрения этим сингулярностям соответствуют космические
струны или сингулярные доменные стенки, которые эволюционируют во времени.

Случай {\sf C} похож на пространственно симметричные решения, рассмотренные в
случае {\sf B}, однако имеет также несколько существенно новых черт. Во-первых,
мы не можем ограничить себя только положительно определенными метриками
$h_{\mu\nu}$ на $\MV$, потому что уравнения Эйнштейна (\ref{esccur})
неинвариантны относительно преобразования $h_{\mu\nu}\rightarrow-h_{\mu\nu}$ при
заданном $m$. Отметим, что при $k=\const$ преобразование
$g_{\al\bt}\rightarrow-g_{\al\bt}$ всегда можно дополнить перестановкой
пространственной и временн\'ой координаты на $\MU$, $\tau\leftrightarrow\s$, что
вместе оставляют уравнение (\ref{escuru}) инвариантным. В случае евклидовой
метрики на $\MV$ это невозможно. Поэтому, не ограничивая общности, мы
зафиксируем $m=1$, но допустим, что метрика $h_{\mu\nu}$ может быть как
положительно, так и отрицательно определена. В обоих случаях сигнатура
четырехмерной метрики будет лоренцевой: либо $(+---)$, либо $(+-++)$.

Решение уравнений (\ref{escuru})--(\ref{einnmm}) проводится так же, как и для
метрики лоренцевой сигнатуры, при этом необходимо функцию $m$ заменить на $k$ и
метрику $g_{\al\bt}$ на $h_{\mu\nu}$. Поэтому мы только кратко обозначим
основные этапы вычислений, подчеркнув те моменты, которые специфичны для
евклидовой сигнатуры.

При $m=1$ полная система вакуумных уравнений Эйнштейна
(\ref{escuru})--(\ref{einnmm}) принимает вид
\begin{align}                                                     \label{qhygtr}
  \nb_\mu\nb_\nu k-\frac{\nb_\mu k\nb_\nu k}{2k}-\frac12h_{\mu\nu}
  \left[\nb^2k-\frac{(\nb k)^2}{2k}\right]&=0,
\\                                                                \label{qfreds}
  R^g+\nb^2 k-2k\Lm&=0,
\\                                                                \label{qgeras}
  R^h+\frac{\nb^2 k}k-\frac{(\nb k)^2}{2k^2}-2\Lm&=0.
\end{align}
Как и в случае {\sf B}, в уравнение (\ref{qfreds}) входит сумма функций от
разных аргументов: $R^g=R^g(x)$ и $k=k(y)$. Поэтому скалярная кривизна
поверхности $\MU$ должна быть постоянна, $R^g=-2K=\const$. Отсюда вытекает, что
поверхность $\MU$ является однополостным гиперболоидом $\ML^2$ или его
универсальной накрывающей.

Это -- очень важное следствие вакуумных уравнений Эйнштейна, т.к.\ в
рассматриваемом случае {\sf C} все решения должны быть $\MS\MO(1,2)$
инвариантны, где группа преобразований Лоренца $\MS\MO(1,2)$ действует на
однополостном гиперболоиде. Поэтому глобальные решения класса ${\sf C}$ названы
лоренц-инвариантными.

Тогда уравнение (\ref{qfreds}) принимает вид
\begin{equation}                                                  \label{qneasw}
  \nb^2 k-2(k\Lm+K)=0.
\end{equation}

Как и в случае {\sf B} уравнение (\ref{qgeras}) является следствием уравнений
(\ref{qhygtr}) и (\ref{qneasw}). Поэтому для нахождения решений вакуумных
уравнений Эйнштейна достаточно решить уравнения (\ref{qhygtr}) и (\ref{qneasw}).

Следующий шаг состоит в фиксировании координат на поверхности $\MV$. Конформно
евклидова метрика на поверхности $\MV$ имеет вид
\begin{equation}                                                  \label{ecogae}
  h_{\mu\nu}dy^\mu dy^\nu=\Phi dzd\bar z=\Phi(d\s^2+d\rho^2).
\end{equation}
Здесь $\Phi(z,\bar z)$ является функцией комплексных координат
\begin{equation}                                                  \label{ecocov}
  z:=\s+i\rho,\qquad \bar z=\s-i\rho,
\end{equation}
где $\s=y^2$, $\rho=y^3$. При этом метрика всего четырехмерного
пространства-времени равна
\begin{equation}                                                  \label{eincoe}
  ds^2=kd\Om_\Sl+\Phi dzd\bar z,
\end{equation}
где $d\Om_\Sl$ -- метрика постоянной кривизны на однополостном гиперболоиде
$\ML^2$, заданная, например, уравнением (\ref{ecoclm}).

Не ограничивая общности, рассмотрим положительные $k>0$. В противном случае
можно просто переставить первые две координаты. Тогда удобно ввести
параметризацию
\begin{equation*}
  k=q^2,\qquad q>0.
\end{equation*}

Для двух неизвестных функций $q$ и $\Phi$ вместо уравнений
(\ref{egcoga})--(\ref{egcogc}) возникает следующая система уравнений
\begin{align}                                                     \label{egcoha}
  \pl^2_{zz}q-\frac{\pl_z\Phi\pl_zq}\Phi&=0,
\\                                                                \label{egcohb}
  \pl^2_{\bar z\bar z}q-\frac{\pl_{\bar z}\Phi\pl_{\bar z}k}\Phi&=0,
\\                                                                \label{egcohc}
  2\frac{\pl_z\pl_{\bar z}q^2}\Phi-(K+\Lm q^2)&=0.
\end{align}
Аналогично случаю {\sf B}, решением уравнений (\ref{egcoha}) и (\ref{egcohb})
являются функции одного аргумента: $q=q(z\pm\bar z)$ и $\Phi=\Phi(z\pm\bar z)$,
при этом функция $\Phi$ определяется уравнением
\begin{equation}                                                  \label{ehintk}
  |\Phi|=|q'|,
\end{equation}
где штрих обозначает дифференцирование по соответствующему аргументу. В
полученной формуле нижний и верхний знаки соответствуют положительно и
отрицательно определенной римановой метрике на $\MV$. Таким образом, функции $q$
и $\Phi$ зависят либо от координаты $\s$, либо от $i\rho$. Поскольку, благодаря
вращательной $\MS\MO(2)$ симметрии конформно евклидовой метрики (\ref{ecogae}),
оба выбора равнозначны, то для определенности мы предположим, что функции
$q(\s)$ и $\Phi(\s)$ зависят от $\s$.

После этого уравнение (\ref{egcohc}) упростится:
\begin{equation}                                                  \label{esecok}
  \frac12(q^2)''=(K+\Lm q^2)\Phi,
\end{equation}
где штрих обозначает дифференцирование по $\s$. Чтобы его проинтегрировать, в
уравнении (\ref{ehintk}) необходимо раскрыть знаки модулей.

Рассмотрим случай $\Phi q'>0$. Тогда уравнение (\ref{esecok}) с учетом
(\ref{ehintk}) примет вид
\begin{equation*}
  \frac12(q^2)''=(K+\Lm q^2)q',
\end{equation*}
и его легко проинтегрировать:
\begin{equation}                                                  \label{efiokw}
  q'=K-\frac{2M}q+\frac{\Lm q^2}3,
\end{equation}
где $M$ -- произвольная постоянная интегрирования. Хотя в рассматриваемом случае
ее нельзя интерпретировать как массу, мы будем использовать старые обозначения,
чтобы облегчить сравнение. Учитывая уравнение (\ref{ehintk}) получаем выражение
для конформного множителя
\begin{equation}                                                  \label{efiokp}
  \Phi(q)=K-\frac{2M}q+\frac{\Lm q^2}3
\end{equation}

Случай $\Phi q'<0$ интегрируется аналогично.

Окончательно, общее решение вакуумных уравнений Эйнштейна в случае {\sf C} имеет
вид
\begin{equation}                                                  \label{egshty}
  ds^2=q^2d\Om_\Sl+\Phi (q)(d\s^2+d\rho^2),
\end{equation}
где конформный множитель задан уравнением (\ref{efiokp}) и функция $q=q(\s)$
определяется уравнением (\ref{ehintk}). Таким образом, метрика на поверхности
$\MV$ имеет один вектор Киллинга $\pl_\rho$ и имеет тот же вид, что и в разделе
\ref{slovri}. Поэтому мы можем использовать развитую там технику для построения
глобальных решений вакуумных уравнений Эйнштейна.

Выбирая функцию $q(\s)$ в качестве одной из координат, метрику (\ref{egshty})
можно записать в виде, напоминающем метрику Шварцшильда,
\begin{equation}                                                  \label{egssty}
  ds^2=q^2d\Om_\Sl+\frac{dq^2}{\Phi(q)}+\Phi (q)d\s^2.
\end{equation}

Результирующая метрика имеет три вектора Киллинга, соответствующих группе
симметрии $\MS\MO(1,2)$ однополостного гиперболоида постоянной кривизны $\ML^2$,
и один дополнительный вектор Киллинга $\pl_\rho$ на поверхности $\MV$.

Вычисления, аналогичные случаю $k=1$, приводят к следующему выражению для
скалярной кривизны поверхности $\MV$
$$
  R^h=\frac23\Lm+\frac{4M}{q^3}.
$$
При этом для инвариантного собственного значения тензора Вейля мы получаем то же
выражение (\ref{ectsqu}), что и в случае {\sf B}.
\subsection{Лоренц-инвариантные решения $K=1$                    \label{solhyp}}
Прежде всего отметим, что случаи $K=1$ и $K=-1$ связаны между собой
перестановкой первых двух координат $\tau\leftrightarrow\s$. Мы выберем значение
$K=1$, чтобы выражение для конформного множителя $\Phi$ имело, с точностью до
изменения знака космологической постоянной, тот же вид, что и для сферически
симметричного случая. Метрика для однополостного гиперболоида в
гиперболической полярной системе координат имеет вид (\ref{qingtf}). Поэтому
четырехмерная метрика пространства-времени в координатах Шварцшильда запишется
следующим образом
\begin{equation}                                                 \label{egshts}
  ds^2=q^2(d\theta^2-\ch^2\theta d\vf^2)+\frac{dq^2}{\Phi(q)}+\Phi (q)d\rho^2,
\end{equation}
где конформный множитель,
\begin{equation}                                                  \label{qcoklj}
  \Phi=1-\frac{2M}q+\frac{\Lm q^2}3,
\end{equation}
имеет тот же вид, что и в решении Коттлера \cite{Kottle18}, но в рассматриваемом
случае этот нетривиальный конформный множитель входит в евклидову часть метрики.

На поверхности $\MV$ метрика может быть как отрицательно ($\Phi<0$), так и
положительно ($\Phi>0$) определена. Для отрицательно определенной метрики
сигнатура метрики пространства-времени равна $(+---)$, и роль времени играет
координата $\theta$. Поэтому времениподобная координата принимает значения
на всей вещественной оси $\theta\in\MR$, и трехмерное пространство
представляет собой произведение окружности $\vf\in[0,2\pi)$ и поверхности $\MV$,
которая будет построена ниже. Если в качестве $\MU$ выбрать универсальную
накрывающую однополостного гиперболоида $\ML^2$, то трехмерное пространство
будет представлять произведение $\MR\times\MV$. Эволюция этих пространств во
времени длится вечно, и если поверхность $\MV$ имеет сингулярность, то ей будет
соответствовать времениподобная кривая.

Для положительно определенной метрики на $\MV$ сигнатура четырехмерной метрики
равна $(+-++)$, и времениподобной координатой является угол $\vf$. При
$\MU=\ML^2$ он принимает значения на окружности $\vf\in[0,2\pi)$, и трехмерное
пространство представляет собой произведение прямой $\theta\in\MR$
и поверхности $\MV$. Соответствующее пространство-время содержит замкнутые
времениподобные кривые (включая экстремали), если только в качестве поверхности
$\MU$ не выбрано универсальное накрывающее пространство для $\ML^2$.

Поскольку метрика на поверхности $\MV$ имеет тот же локальный вид, что и в
разделе \ref{slovri}, то мы может построить максимально продолженные поверхности
$\MV$. Вид римановой поверхности определяется конформным множителем, показанным
на рис.\ref{fspfercofa} при $\Lm\ne0$. Надо только помнить, что в выражение для
конформного множителя космологической постоянная входит с противоположным
знаком. Кроме того, при отождествлении $\rho\sim\rho+L$ возможно появление
конических сингулярностей, которые соответствуют космическим струнам.

Перейдем к классификации глобальных решений вакуумных уравнений Эйнштейна
в рассматриваемом случае {\sf C}.
\subsubsection{Пространство-время Минковского                    $\Lm=0$, $M=0$}
При нулевых значениях постоянных поверхность $\MV$ представляет собой
полуплоскость $q\in(0,\infty)$, $\rho\in(-\infty,\infty)$. Она неполна при $q=0$
из-за координатной сингулярности, аналогичной сферически симметричному случаю.
\subsubsection{Космическая струна                                $\Lm=0$, $M>0$}
\begin{figure}[h,b,t]
\hfill\includegraphics[width=.7\textwidth]{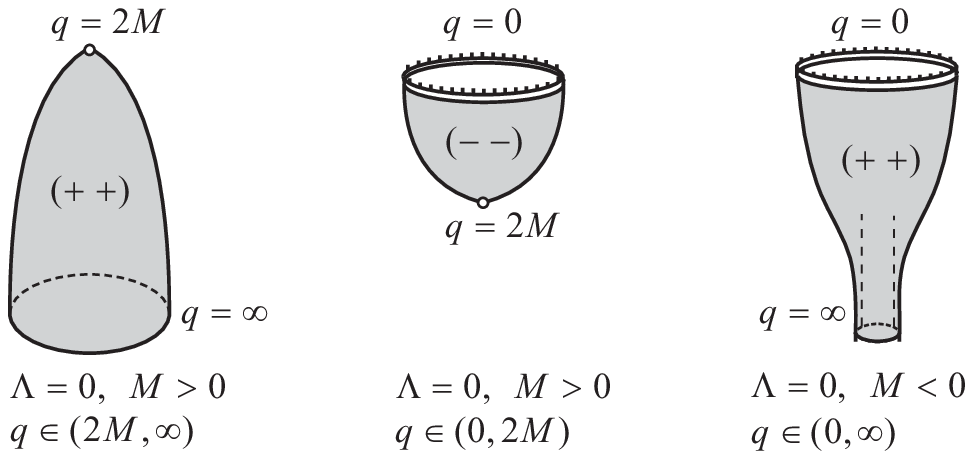}
\hfill {}
\centering\caption{Максимально продолженная риманова поверхность $\MV$ при
нулевой космологической постоянной $\Lm=0$. Условные обозначения те же, что и на
рис.~\ref{fboprr}. Незакрашенные окружности обозначают возможные конические
сингулярности.}
\label{flmzero}
\end{figure}

При нулевой космологической постоянной $\Lm=0$ и положительных значениях $M>0$
возникает решение, соответствующее решению Шварцшильда. В этом случае существуют
две несвязные между собой максимально продолженные римановы поверхности,
изображенные на рис.\ref{flmzero} слева и в центре с положительно и отрицательно
определенной метрикой. Для наглядности мы отождествили точки с координатами
$\rho$ и $\rho+L$.

Внешнему решению Шварцшильда соответствует интервал
$q\in(2M,\infty)$. В этой области конформный множитель положителен и
четырехмерная метрика имеет сигнатуру $(+-++)$. Соответствующая поверхность
$\MV$ является гладким многообразием. При этом возможно появление конической
сингулярности на горизонте $q=2M$. С топологической точки зрения коническая
сингулярность представляет собой плоскость (мировую поверхность космической
струны), а все пространство-время описывает бесконечную эволюцию бесконечной
космической струны. Для этого решения кривизна не имеет больше никаких
сингулярностей, и если выполнено уравнение (\ref{qvbttf}), то пространство-время
вообще не имеет сингулярностей. Специфическим свойством этого решения является
то, что периметр цилиндра, изображенного на рис.\ref{flmzero} слева, стремится
к постоянной при $q\rightarrow\infty$. Это свойство соответствует тому, что
решение Шварцшильда является асимптотически плоским.

Для внутреннего решения Шварцшильда $q\in(0,2M)$, конформный множитель
отрицателен, и метрика пространства-времени имеет сигнатуру $(+---)$.
Максимально продолженная поверхность $\MV$ представляет собой в этом случае диск
конечного радиуса с сингулярным краем $q=0$, который изображен на
рис.\ref{flmzero} в центре. Если выполнено уравнение (\ref{qvbttf}), то
поверхность $\MV$ гладкая и имеет особенность только на крае. В противном случае
в центре диска $q=2M$ возникает коническая сингулярность, соответствующая
космической струне. Поэтому трехмерное пространство представляет собой
космическую струну, окруженную цилиндрической доменной стенкой, на которой
кривизна сингулярна. Сингулярный край невозможно адекватно изобразить на
рисунке, поскольку он находится на конечном расстоянии, но в то же время имеет
бесконечный периметр.
\subsubsection{Доменная стенка                                   $\Lm=0$, $M<0$}
Если космологическая постоянная равна нулю, а масса отрицательна, то решение
Шварцшильда описывает голую сингулярность. В этом случае конформный множитель
при  $q\in(0,\infty)$ положителен и сигнатура метрики равна $(+-++)$.
Максимально продолженная риманова поверхность $\MV$ представляет
собой полуплоскость, $q\in(0,\infty)$, $\rho\in\MR$. Сингулярность кривизны
расположена вдоль края $q=0$, что можно интерпретировать, как доменную стенку,
на которой кривизна сингулярна. С четырехмерной точки зрения это решение
описывает эволюцию доменной стенки, которая расположена на конечном расстоянии.

Благодаря трансляционной симметрии, точки с координатами $\rho$ и $\rho+L$
можно отождествить. Возникающая при этом поверхность изображена на
рис.\ref{flmzero} справа. Соответствующее трехмерное пространство представляет
собой цилиндрическую доменную стенку.
\subsubsection{Решение де Ситтера                                $\Lm>0$, $M=0$}
\begin{figure}[h,b,t]
\hfill\includegraphics[width=.7\textwidth]{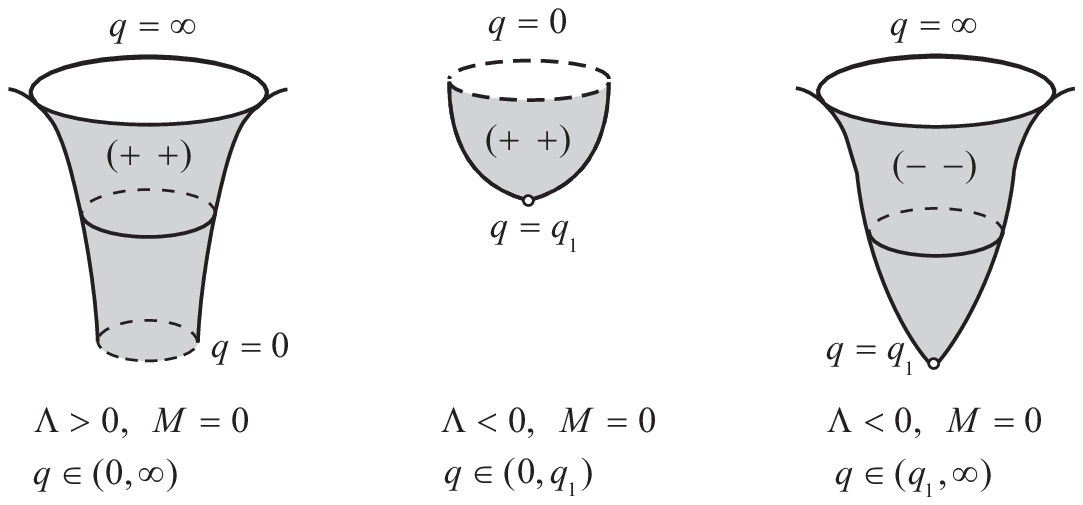}
\hfill {}
\centering\caption{Максимально продолженные римановы поверхности $\MV$ при
нулевой массе $M=0$. Условные обозначения те же, что и на рис.~\ref{fboprr}.
Незакрашенные окружности обозначают возможные конические сингулярности.}
\label{fmazero}
\end{figure}

При нулевой массе и положительной космологической постоянной конформный
множитель (\ref{efiokp}) положителен для $q\in(0,\infty)$. Поверхность
$\MV$ имеет метрику с сигнатурой $(+-++)$ и показана на рис.~\ref{fmazero}
слева. Для наглядности она изображена после отождествления точек с координатами
$\rho$ и $\rho+L$. Пространство де Ситтера соответствует универсальной
накрывающей поверхности $\MV$. Поверхность $\MV$ является гладкой и не имеет
сингулярностей. Однако она неполна при $q=0$. Ее продолжение через край $q=0$
невозможно, т.к.\ четырехмерная метрика пространства-времени вырождается.
\subsubsection{Решение анти-де Ситтера                           $\Lm<0$, $M=0$}
При отрицательной космологической постоянной $\Lm<0$ и нулевой константе $M=0$
возникает решение анти-де Ситтера. Конформный множитель при этом имеет один
простой нуль (горизонт) в точке $q_1=\sqrt{-3/\Lm}$. Следовательно, существуют
две несвязные между собой поверхности $\MV$: одна с положительно, а другая с
отрицательно определенной метрикой.

При $q\in(0,q_1)$ конформный множитель положителен, метрика пространства-времени
имеет сигнатуру $(+-++)$, и поверхность $\MV$ показана на рис.\ref{fmazero} в
центре. С топологической точки зрения эта поверхность представляет собой диск
конечного радиуса с положительно определенной метрикой. В центре диска $q=q_1$
возможно появление конической сингулярности. Ее продолжение через край $q=0$
невозможно, т.к.\ четырехмерная метрика вырождается.

Если $q\in(q_1,\infty)$, то конформный множитель отрицателен и метрика
пространства-времени имеет сигнатуру $(+---)$. Соответствующая поверхность
$\MV$ показана на рис.\ref{fmazero} справа. С топологической точки зрения она
представляет собой плоскость, в центре которой возможно появление конической
сингулярности.
\subsubsection{Доменная стенка        $\Lm<0$, $M>\frac1{3\protect\sqrt{-\Lm}}$}
При отрицательной космологической постоянной $\Lm<0$ и
$M>\frac1{3\protect\sqrt{-\Lm}}$ конформный множитель отрицателен при
$q\in(0,\infty)$, не имеет нулей, и сигнатура метрики пространства-времени равна
$(+---)$. С топологической точки зрения поверхность $\MV$ представляет собой
полуплоскость $q>0$, $\rho\in\MR$ с отрицательно определенной метрикой и
сингулярной кривизной на границе $q=0$. Глобальное решение описывает бесконечную
эволюцию бесконечной плоской доменной стенки сингулярной кривизны.

Компактификация по координате $\rho\sim\rho+L$ дает внутренность цилиндрической
доменной стенки, показанной на рис.\ref{fgerah}. Длина направляющей окружности
цилиндра стремится к бесконечности при $q\to0$ и $q\to\infty$. Сингулярный край
$q=0$ находится на конечном расстоянии.
\subsubsection{Двойной горизонт       $\Lm<0$, $M=\frac1{3\protect\sqrt{-\Lm}}$}
\begin{figure}[h,b,t]
\hfill\includegraphics[width=.9\textwidth]{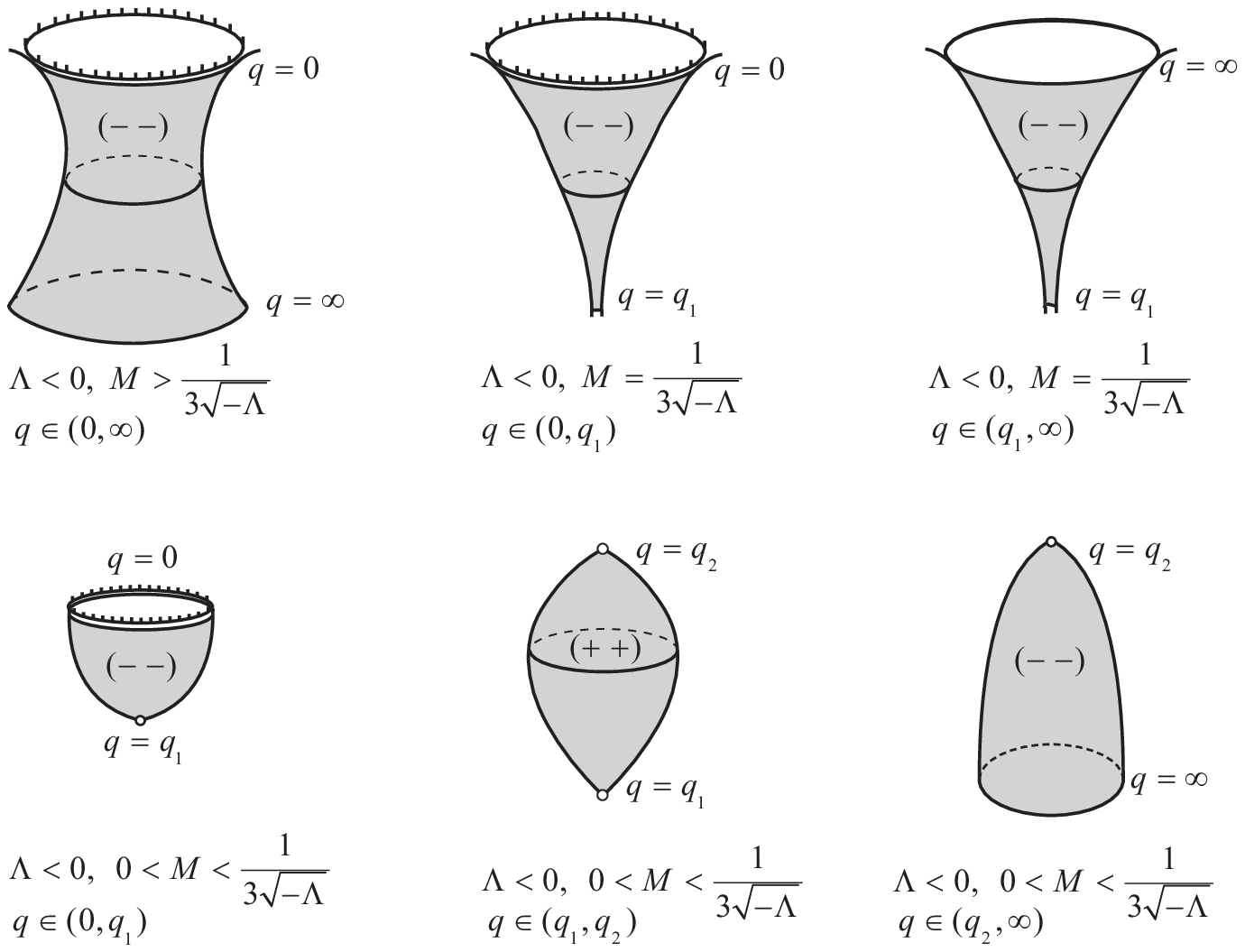}
\hfill {}
\centering\caption{Максимально продолженные римановы поверхности $\MV$ при
отрицательной космологической постоянной $\Lm<0$ и $M\ne0$. Незакрашенные
окружности обозначают возможные конические сингулярности.}
\label{fgerah}
\end{figure}

При указанных значениях постоянных конформный множитель отрицателен и имеет
двойной нуль в точке $q_1$. Метрика пространства-времени в обоих интервалах
$q\in(0,q_1)$ и $q\in(q_1,\infty)$ имеет сигнатуру $(+---)$.
В этом случае имеем две поверхности, изображенные на рис.\ref{fgerah}.
Поверхность $\MV$, соответствующая интервалу $q\in(q_1,\infty)$, не является
односвязной. Ее можно представить себе в виде плоскости, центр которой $q=q_1$
удален в бесконечность, т.к.\ он лежит на бесконечном расстоянии.

Поверхность для интервала $(0,q_1)$  представляет собой цилиндрическую доменную
стенку сингулярностей кривизны. Сингулярный край $q=0$ лежит на конечном
расстоянии.
\subsubsection{Два горизонта        $\Lm<0$, $0<M<\frac1{3\protect\sqrt{-\Lm}}$}
Если космологическая постоянная отрицательна $\Lm<0$ и масса лежит в интервале
$0<M<\frac1{3\protect\sqrt{-\Lm}}$, то конформный множитель имеет максимальное
число нулей: два простых нуля, расположенных в точках $q_{1,2}$
(см.\ рис.\ref{fspfercofa} слева). В интервалах $(0,q_1)$ и $(q_2,\infty)$
конформный множитель отрицателен, и сигнатура метрики пространства-времени
равна $(+---)$. В интервале $(q_1,q_2)$ конформный множитель положителен, и
сигнатура метрики имеет вид $(+-++)$. Максимально продолженная поверхность $\MV$
является одной из трех поверхностей, показанных на рис.~\ref{fgerah} в нижнем
ряду.

При $q\in(0,q_1)$ поверхность $\MV$ представляет собой диск конечного радиуса,
на крае которого кривизна сингулярна. В центре диска, возможно, расположена
коническая сингулярность. Это зависит от отождествления $\rho\sim\rho+L$.

Если $q\in(q_1,q_2)$, то поверхность $\MV$ топологически представляет собой
сферу, в общем случае имеющую две конические сингулярности. Алгебраический
анализ системы уравнений $\Phi(q_1)=\Phi (q_2)=0$ и $\Phi'(q_1)=-\Phi'(q_2)$
при $q_1\ne q_2$ показывает, что таких решений не существует. Поэтому, подгоняя
период компактификации $L$, можно устранить одну из конических сингулярностей,
но не обе одновременно. Это значит, что в рассматриваемом случае всегда
должна существовать космическая струна. С топологической точки зрения
трехмерное пространство представляет собой произведение сферы, содержащей
одну или две конические сингулярности, и прямой.

При $q\in(q_2,\infty)$ поверхность $\MV$ -- это плоскость, в центре которой,
возможно, расположена коническая сингулярность. Экстремали при $q\to\infty$
полны, а кривизна стремится к постоянной.
\subsubsection{Космическая струна                                $\Lm<0$, $M<0$}
Если и космологическая постоянная $\Lm<0$, и постоянная $M<0$, то конформный
множитель $\Phi$ имеет один простой нуль в точке $q_1$. В этом случае
возможно существование двух поверхностей $\MV$.

При $q\in(0,q_1)$ конформный множитель положителен, и сигнатура метрики
пространства-времени равна $(+-++)$. Топологически поверхность $\MV$ имеет то же
вид, что и рассмотренная в предыдущем случае поверхность для $\Lm<0$,
$0<M<\frac1{3\protect\sqrt{-\Lm}}$ и $q\in(0,q_1)$, но с отрицательно
определенной метрикой.

Вторая поверхность $\MV$ соответствует интервалу $q\in(q_1,\infty)$.
Конформный множитель отрицателен, и сигнатура метрики пространства-времени
равна $(+---)$. Поверхность $\MV$ имеет тот же вид, что и в случае $\Lm<0$,
$M=0$ и $q\in(q_1,\infty)$ на рис.\ref{fmazero}. Глобальное решение вакуумных
уравнений Эйнштейна в общем случае описывает бесконечную космическую струну,
причем других сингулярностей у кривизны нет.
\subsubsection{Космическая струна                                $\Lm>0$, $M>0$}
При положительной космологической постоянной $\Lm>0$ и массе $M>0$. Конформный
множитель имеет один простой нуль в точке $q_1$, как видно из
рис.\ref{fspfercofa}, справа. Он отрицателен в интервале $q\in(0,q_1)$ и
положителен при $q\in(q_1,\infty)$. Этот случай аналогичен предыдущему. Мы имеем
две максимально продолженные поверхности $\MV$, на которых необходимо просто
изменить сигнатуру метрики.
\subsubsection{Доменная стенка                                   $\Lm>0$, $M<0$}
Если космологическая постоянная положительна, $\Lm>0$, и постоянная $M$
отрицательна, то конформный множитель положителен при $q\in(0,\infty)$.
Сигнатура метрики пространства-времени равна $(+-++)$. В этом случае имеется
одна поверхность $\MV$, т.к.\ нули у конформного множителя отсутствуют.
Топологически поверхность $\MV$ такая же, как и в случае $\Lm<0$,
$M>1/3\sqrt{-\Lm}$, но с положительно определенной метрикой.
\subsection{Решения с плоскостью Минковского $K=0$               \label{spmprp}}
При $K=0$ поверхность $\MU$ представляет собой или плоскость Минковского
$\MR^{1,1}$, или цилиндр, или тор. При этом возникают новые решения, интересные
с топологической точки зрения. Соответствующая метрика в координатах Шварцшильда
имеет вид
\begin{equation}                                                  \label{einmps}
  ds^2=q^2(dt^2-dx^2)+\frac{dq^2}{\Phi(q)}+\Phi(q)d\rho^2,
\end{equation}
где
$$
  \Phi(q)=-\frac{2M}q+\frac{\Lm q^2}3.
$$
В зависимости от значения постоянных $\Lm$ и $M$, входящих в конформный
множитель, возможны четыре качественно отличных случая.
\subsubsection{Доменная стенка                                 $\Lm=0$, $M\ne0$}
\begin{figure}[h,b,t]
\hfill\includegraphics[width=.5\textwidth]{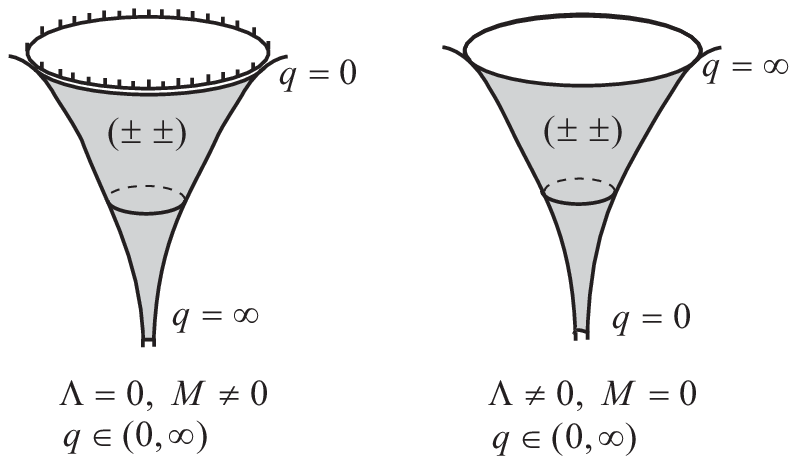}
\hfill {}
\centering\caption{Глобальные решения с плоскостью Минковского, $K=0$.}
\label{fgeram}
\end{figure}

При нулевой космологической постоянной $\Lm=0$, но отличной от нуля постоянной
$M\ne0$ максимально продолженная поверхность $\MV$ изображена на
рис.~\ref{fgeram}. Метрика на этой поверхности положительно и отрицательно
определена соответственно при $M<0$ и $M>0$.
\subsubsection{Решение (анти-)де Ситтера             $\Lm\ne0$, $M=0$}
Если масса равна нулю, то возникает еще одно представление решения
(анти-)де Ситтера для положительной и отрицательной космологической постоянной.
Соответствующая поверхность $\MV$ показана на рис.~\ref{fgeram}.

Пусть $\Phi q'>0$ или, что эквивалентно, $\Lm q'>0$. Тогда уравнение
(\ref{emdmee}) принимает вид
\begin{equation*}
  q'=\frac{\Lm q^2}3
\end{equation*}
и легко интегрируется:
\begin{equation*}
  q=-\frac3{\Lm\s},
\end{equation*}
где мы опустили несущественную постоянную интегрирования, соответствующую сдвигу
$\s$. Тогда метрика в черепашьих координатах равна
\begin{equation}                                                  \label{edescm}
  ds^2=\frac9{\Lm^2\s^2}(dt^2-dx^2)+\frac3{\Lm\s^2}(d\s^2+d\rho^2).
\end{equation}
Растяжением координат $t$ и $x$ она приводится к конформно плоскому виду.
\subsubsection{Космическая струна                                $\Lm>0$, $M>0$}
Конформный множитель $\Phi$ при $\Lm>0$ и $M>0$ имеет один простой нуль в точке
$q_1$. При $q\in(0,q_1)$ и $q\in(q_1,\infty)$, конформный множитель,
соответственно, отрицателен и положителен. Поэтому метрики пространства-времени
имеют сигнатуры $(+---)$ и $(+-++)$. Максимально продолженные поверхности $\MV$
топологически такие же как и для конформного множителя
$\Phi=1-\frac{2M}q+\frac{\Lm q^2}3$, который был рассмотрен ранее. Меняется
только величина $q_1$. Глобальные решения описывают космическую струну и
космическую струну, окруженную доменной стенкой сингулярностей кривизны.

Изменение знака обеих постоянных $\Lm<0$, $M>0$ приводит к изменению сигнатуры
метрики. Качественные свойства решений остаются прежними.
\subsubsection{Доменная стенка                                   $\Lm>0$, $M<0$}
В рассматриваемом случае конформный множитель $\Phi$ положителен и не имеет
нулей. Сигнатура метрики пространства-времени равна $(+-++)$. Существует всего
одна максимально продолженная поверхность $\MV$. Топологически она такая же, как
и для конформного множителя $\Phi=1-\frac{2M}q+\frac{\Lm q^2}3$, рассмотренного
ранее.

При $\Lm<0$ и $M>0$ необходимо просто изменить сигнатуру всей метрики.
\section{Итоги главы}
В двух последних главах были найдены и классифицированы все глобальные вакуумные
решения уравнений Эйнштейна с космологической постоянной, которые имеют вид
сплетенного произведения двух поверхностей. Явное построение и классификация
решений проведена в зависимости от значений постоянной скалярной кривизны одной
из поверхностей, значения космологической постоянной $\Lm$ и единственной
постоянной интегрирования $M$, которая для решения Шварцшильда имеет физический
смысл массы черной дыры. Мы видим, что требование максимального продолжения
решений практически однозначно определяет глобальную структуру
пространства-времени. Важно отметить, что при решении уравнений движения мы не
ставим никаких граничных условий. Подчеркнем, что решение уравнений Эйнштейна в
какой то фиксированной системе координат само по себе дает не так уж много. Для
того, чтобы дать физическую интерпретацию решений необходимо знать глобальную
структуру пространства-времени. Эта задача сложна, но обойти ее, по-видимому,
невозможно.

Предположение о том, что метрика пространства-времени имеет вид сплетенного
произведения метрик двух поверхностей влечет за собой симметрию метрики, если
потребовать выполнения вакуумных уравнений Эйнштейна. Например, для решения
Шварцшильда мы не требовали сферической симметрии метрики -- она возникла в
процессе решения уравнений Эйнштейна.

Построенные решения представляют значительный физический интерес. Мы показали,
что вакуумные решения уравнений Эйнштейна описывают черные дыры, космические
струны, кротовые норы, доменные стенки сингулярностей кривизны и другие. В
настоящей монографии мы лишь кратко обсудили свойства построенных глобальных
решений.
\chapter{Дополнение                                              \label{smatre}}
\section{Матрицы                                                 \label{smatri}}
Мы рассматриваем вещественные или комплексные матрицы. Б\'ольшую часть
утверждений и формул настоящего дополнения можно найти, например, в
монографиях \cite{Gantma88R,HorJoh86R}.

Для любого натурального $n\in\MN$ справедливо разложение
\begin{align}                                                          \nonumber
  (a+b)^n &=\sum_{k=0}^n C_n^ka^{n-k}b^k=
\\                                                                \label{ebinew}
  &=a^n+na^{n-1}b+\frac{n(n-1)}2a^{n-2}b^2+\dotsc+nab^{n-1}+b^n,
\end{align}
которое называется {\em биномом Ньютона}, где
\index{Бином Ньютона (binomial formula)}%
\index{Ньютона бином (binomial formula)}%
\begin{equation}                                                  \label{ebinco}
  C_n^k:=\frac{n!}{k!(n-k)!}
\end{equation}
-- {\em биномиальные коэффициенты}.
\index{Биномиальный коэффициент (binomial coefficient)}%
\index{Коэффициент биномиальный (binomial coefficient)}%
В частности, при $a=b=1$ и $a=-b=1$ справедливы равенства
\begin{equation}                                                  \label{ebicpr}
  2^n=\sum_{k=0}^nC_n^k,\qquad 0=\sum_{k=0}^n(-1)^kC_n^k.
\end{equation}

Пусть дана $2\times2$ матрица
$$
  A=\begin{pmatrix} a & b \\ c & d \end{pmatrix}.
$$
Если ее определитель отличен от нуля, $\det A:=ad-bc\ne0$, то у нее существует
обратная матрица, которая имеет вид
$$
  A^{-1}=\frac1{\det A}\begin{pmatrix} d & -b \\ -c & a \end{pmatrix}.
$$
Для двух $2\times2$ матриц $A$ и $B$ справедлива формула
\begin{equation}                                                  \label{edesum}
  \det(A+B)=\det A+\tr A\tr B-\tr(AB)+\det B.
\end{equation}

Пусть $A$ -- произвольная $3\times 3$ матрица. Тогда справедлива формула
(Bengtsson gr-qc/0703114)
\begin{equation}                                                  \label{ethtth}
  A^3-A^2\tr A-\frac12A\left(\tr A^2-(\tr A)^2\right)-\det A=0.
\end{equation}
След этого выражения имеет вид
\begin{equation*}
  \tr A^3-\frac32\tr A^2\tr A+\frac12(\tr A)^3-3\det A=0.
\end{equation*}

Пусть $A=(A_{ab})$ -- произвольная квадратная невырожденная $n\times n$ матрица,
$\det A\ne0$. Тогда элементы обратной матрица $A^{-1ab}$ равны
\begin{equation*}
  A^{-1ab}=\frac{A^{ba}}{\det A},
\end{equation*}
где $A^{ba}$ -- алгебраическое дополнение элемента $A_{ba}$. Если матрица $A$
симметрична, то обратная матрица $A^{-1}$ также симметрична. Для произведения
двух невырожденных квадратных матриц справедлива формула
$(AB)^{-1}=B^{-1}A^{-1}$. Кроме того $\det A^{-1}\det A=1$.

Определитель квадратной $n\times n$ матрицы $A$ является полиномом от $n^2$
элементов матрицы. Поэтому его можно дифференцировать по каждому элементу. Если
матрица $A$ невырождена, то справедлива формула
\begin{equation}                                                  \label{ematdi}
  A^{-1ba}=\frac1{\det A}\frac{\pl \det A}{\pl A_{ab}}.
\end{equation}

Пусть $A$ и $B$ -- две квадратные обратимые матрицы размеров $m\times m$ и
$n\times n$, соответственно, и пусть $C$ -- произвольная матрица размера
$n\times m$. Тогда для квадратных $(m+n)\times(m+n)$ матриц справедливо
равенство
\begin{equation}                                                  \label{edemnm}
\begin{split}
  \det\begin{pmatrix} A & -C^\mathrm{T} \\ C & B \end{pmatrix}
  &=\det A\,\det(B+CA^{-1}C^\mathrm{T})=
\\
  &=\det B\,\det(A+C^\mathrm{T}B^{-1}C).
\end{split}
\end{equation}
Определитель нижне треугольной матрицы равен произведению определителей
диагональных блоков:
\begin{equation}                                                  \label{edettr}
  \det\begin{pmatrix}A & 0 \\ C & B\end{pmatrix}
  =\det A\det B.
\end{equation}

Рассмотрим квадратную $m\times m$ матрицу $C$, которая равна произведению
двух прямоугольных матриц
\begin{equation*}
  C=AB.
\end{equation*}
Размеры матриц $A$ и $B$ равны, соответственно, $m\times n$ и $n\times m$.
Тогда справедлива формула Бине--Коши \cite{Gantma88R}
\index{Формула Бине--Коши (Binet--Cauchy formula)}%
\index{Бине--Коши формула (Binet--Cauchy formula)}%
\begin{equation}                                                  \label{ebicho}
  \det C=\sum_{1\le\bt_1<\bt_2<\dotsc<\bt_m\le n}
  \det \begin{pmatrix} A_{1\bt_1} & \dotsc & A_{1\bt_m} \\
    \vdots & \vdots & \vdots \\ A_{m\bt_1} & \dotsc & A_{m\bt_m} \end{pmatrix}
  \det \begin{pmatrix} B_{\bt_11} & \dotsc & B_{\bt_m1} \\
    \vdots & \vdots & \vdots \\ B_{\bt_1m} & \dotsc & B_{\bt_mm} \end{pmatrix},
\end{equation}
где сумма берется по всем упорядоченным выборкам $m$ индексов
$\bt_1,\dotsc,\bt_m$ из $1,\dotsc,n$. Если $m>n$, то сумма считается равной
нулю. Согласно формуле Бине--Коши определитель матрицы $C$ равен сумме
произведений всех возможных миноров максимального $m$-того порядка матриц $A$ и
$B$. При $m>n$ $\det C=0$.

Если матрица $A$ невырождена, то справедливо тождество
\begin{align}                                                          \nonumber
  \det(A+B)&=\det A\exp\left(\tr\ln(1+A^{-1}B)\right)=
\\                                                                \label{edetab}
  &=\det\,A\,\exp
  \left[\sum_{n=1}^\infty\frac{(-1)^{n-1}}n\tr(A^{-1}B)^n\right].
\end{align}
В частном случае, если $A=1$, то
\begin{equation}                                                  \label{edeopa}
  \det (1+B)=1+\tr B+\frac12(\tr^2B-\tr B^2)+\dotsc+\det B,
\end{equation}
где точки обозначают сумму определителей матриц, получающихся из $B$ всеми
возможными вычеркиваниями одинаковых строк и столбцов.

Если норма матрицы $A$ меньше единицы, то матрица $1-A$ обратима, и
справедлива формула
\begin{equation}                                                  \label{einvma}
  \sum^\infty_{k=1}A^k=(1-A)^{-1},
\end{equation}
которая является аналогом формулы для суммы членов геометрической прогрессии.

Для произвольных квадратных некоммутирующих матриц $A$ и $B$ справедливо
равенство
\begin{equation}                                                  \label{enocme}
  \ex^{-B}A\,\ex^B=A+[A,B]+\frac12\big[[A,B],B\big]+\dotsc
  =\sum_{k=0}^\infty\frac1{k!}[A,B]_{(k)},
\end{equation}
где члены ряда определены рекуррентными соотношениями
\begin{equation*}
  [A,B]_{(0)}=A,\qquad [A,B]_{(k+1)}=\big[[A,B]_{(k)},B\big].
\end{equation*}
В частности, если $[B,A]=\lm A$, то равенство (\ref{enocme}) принимает вид
\begin{equation}                                                  \label{emaeql}
  \ex^BA\,\ex^{-B}=\ex^\lm A.
\end{equation}
Если коммутатор двух матриц пропорционален единичной матрице,
$$
  [A,B]=c\one,
$$
то ряд (\ref{enocme}) обрывается и справедливы следующие формулы:
\begin{align}                                                     \label{eexpmu}
  \ex^A\ex^B&=\ex^{[A,B]}\ex^B\ex^A=\ex^{\frac12[A,B]}\ex^{A+B},
\\                                                                \label{ecomem}
  [A,\ex^B]&=[A,B]\ex^B.
\end{align}
Последнее равенство можно переписать в эквивалентном виде
\begin{equation}                                                  \label{eqvfco}
  \ex^{-B}A\,\ex^B=A+[A,B].
\end{equation}
Равенство (\ref{eexpmu}) известно, как {\em формула Хаусдорфа}.
\index{Формула Хаусдорфа (Hausdorff formula)}%
\index{Хаусдорфа формула (Hausdorff formula)}%
\begin{theorem}[\bf Троттер]
Для двух произвольных вещественных или комплексных квадратных матриц $A$ и $B$
справедлива формула
\begin{equation}                                                  \label{etrott}
  \ex^{t(A+B)}=\underset{n\to\infty}\lim\left(\ex^{\frac tn A}\ex^{\frac tn B}
  \right)^n.
\end{equation}
\end{theorem}
\begin{proof}
См.\ \cite{Trotte59}.
\end{proof}

Пусть заданы две произвольные симметричные матрицы $A$ и $B$. Тогда, если
равенство
\begin{equation}                                                  \label{etwmai}
  A_{ab}B_{cd}-A_{ad}B_{cb}-A_{cd}B_{ab}+A_{cb}B_{ad}=0
\end{equation}
выполнено для всех значений индексов, то матрицы $A$ и $B$ пропорциональны.

Если элементы квадратной матрицы $A_{ab}(x)$ являются дифференцируемыми
функциями, то производная от определителя равна
\begin{equation}                                                  \label{edetde}
  \pl_\al\det A=\det A\,A^{-1ab}\pl_\al A_{ba}
  =-\det A\,\pl_\al A^{-1ab}A_{ba}.
\end{equation}

Приведем без доказательства несколько теорем о свойствах матриц с вещественными
элементами, которые часто используются в приложениях. Напомним, что
{\em собственными значениями} $\lm_i$, $i=1,\dotsc,n$ квадратной $n\times n$
матрицы $A$ называются корни характеристического (векового) уравнения
\index{Собственное значение (eigenvalue)}%
\index{Значение собственное (eigenvalue)}%
$$
  \det(A-\lm\one)=0.
$$
Справедливы формулы
\begin{equation}                                                  \label{edetrm}
  \det A=\lm_1\dotsc\lm_n,\qquad \tr A^k=\sum_{i=1}^n\lm_i^k,
\end{equation}
для любого натурального $k\in\MN$. У симметричной матрицы все собственные
значения вещественны. При этом симметричная матрица положительно определена
тогда и только тогда, когда все собственные значения вещественны и
положительны $\lm_i>0$.

Если симметричная $2\times2$ матрица $A=A^\St$ имеет положительные собственные
значения, то из нее можно извлечь квадратный корень, т.е.\ представить в виде
$A=B^2$. Этот корень не является единственным. Однако, с точностью до знака
существует единственный симметричный корень $B=B^\St$. Поскольку собственные
значения матрицы $A$ положительны, то на диагонали матрицы должны стоять
положительные числа. Пусть
\begin{equation*}
  A=\begin{pmatrix} a & b \\ b & d \end{pmatrix},\qquad \det A=ad-b^2>0,
  \qquad a,d>0.
\end{equation*}
тогда ``положительный'' квадратный корень имеет вид
\begin{equation*}
  B=\begin{pmatrix} \sqrt a\cos\theta & \sqrt a\sin\theta \\
  \sqrt a\sin\theta & \sqrt d\cos\psi \end{pmatrix},
\end{equation*}
где
\begin{equation*}
\begin{split}
  \sin^2\theta&=\frac{b^2}{a(a+d+2\sqrt{ad-b^2})},
\\
  \sin^2\psi&=\frac{b^2}{d(a+d+2\sqrt{ad-b^2})},
\end{split}
\end{equation*}
и для углов выбираются положительные корни: $0\le\theta,\psi\le\pi/2$.
Отметим, что справедливо тождество
\begin{equation*}
  \sqrt a\sin\theta=\sqrt d\sin\psi.
\end{equation*}

\begin{theorem}
Пусть $A$ -- произвольная квадратная невырожденная матрица. Тогда матрица $A^TA$
симметрична и положительно определена.
\end{theorem}
Пусть $\lm_i(A^\St A)\ge0$ -- собственные значения симметричной матрицы
$A^\St A$. Тогда {\em сингулярными числами} $\nu_i(A)$ матрицы $A$ называется
набор чисел
\index{Сингулярные числа (singular number)}%
\index{Числа сингулярные (singular number)}%
$$
  \nu_i(A):=\sqrt{\lm_i(A^\St A)}.
$$
\begin{theorem}{\bf О сингулярном разложении матриц.}
Пусть $A$ произвольная квадратная матрица и $\nu_i(A)$ -- ее сингулярные
числа, занумерованные произвольным образом, тогда найдутся ортогональные
матрицы $R$ и $S$, такие что
$$
  A=R^\St\diag\lbrace\nu_1(A),\dotsc,\nu_n(A)\rbrace S,
$$
где $\diag\lbrace\nu_1(A),\dotsc,\nu_n(A)\rbrace$ -- диагональная матрица, у
которой на диагонали стоят сингулярные числа.
\end{theorem}
\begin{theorem}
Пусть $A$ и $B$ произвольные квадратные матрицы с сингулярными числами
$\nu_i(A)$ и $\nu_i(B)$, занумерованными в порядке возрастания, тогда
справедлива оценка
$$
  |\tr(AB)|\le\sum_{i=1}^n\nu_i(A)\nu_i(B).
$$
\end{theorem}
\begin{theorem}
Пусть $B$ -- произвольная симметричная положительно определенная матрица. Тогда
существует единственная симметричная положительно определенная матрица $A$,
такая что $A^2=B$. При этом мы пишем $A=\sqrt B$.
\end{theorem}
\begin{theorem}{\bf Полярное разложение вещественных матриц.}     \label{tpomad}
Любая вещественная матрица $A$ может быть представлена в виде произведения
\begin{equation}                                                  \label{eporem}
  A=RS,
 \end{equation}
где $R$ -- неотрицательно определенная симметричная матрица и $S$ --
ортогональная матрица. Матрица $R$ всегда однозначно определена соотношением
$$
  R:=\sqrt{AA^\St}.
$$
Если матрица $A$ невырождена, то симметричная матрица $R$ является положительно
определенной, а ортогональная матрица $S$ также определена однозначно,
$S=R^{-1}A$.
\end{theorem}
\begin{proof}
См., например, \cite{HorJoh86R}.
\end{proof}
\begin{exa}
Вещественное число $a\in\MR$ является $1\times1$ матрицей и его можно
представить в виде $a=|a|\sign a$, где
\begin{equation*}
  \sign a:=\begin{cases} \quad 1 & a>0, \\ -1, &a<0, \end{cases}
\end{equation*}
-- ортогональная $1\times1$ матрица. Если $a\ne0$, то представление существует
и единственно. При $a=0$ модуль числа также равен нулю, $|a|=0$, а матрица
$S=\sign a$ неопределена.
\qed\end{exa}
Полярное разложение вещественных матриц имеет простую геометрическую
интерпретацию. Рассмотрим $n\times n$ матрицу $A$, как линейный оператор,
действующий в евклидовом пространстве $\MR^n$. Тогда ортогональной матрице
соответствует некоторое вращение евклидова пространства вокруг начала координат.
Симметричная положительно определенная матрица осуществляет дилатацию евклидова
пространства вдоль $n$ взаимно перпендикулярных направлений с различными в общем
случае коэффициентами растяжения. Следовательно, полярное разложение матриц
означает последовательное выполнение некоторого вращения и некоторой дилатации.
\begin{theorem}{\bf Полярное разложение комплексных матриц.}      \label{tpomac}
Любая комплексная матрица $A$ может быть представлена в виде произведения
\begin{equation}                                                  \label{epocor}
  A=RU,
\end{equation}
где $R$ -- неотрицательно определенная эрмитова матрица и $U$ -- унитарная
матрица. Матрица $R$ всегда однозначно определена соотношением
$$
  R=\sqrt{AA^\dagger}.
$$
Если матрица $A$ невырождена, то эрмитова матрица $R$ является положительно
определенной, а унитарная матрица $U$ также определена однозначно,
$U=R^{-1}A$.
\end{theorem}
\begin{proof}
См., например, \cite{HorJoh86R}.
\end{proof}
\begin{exa}
Комплексные числа $z=x+iy$ являются комплексными $1\times1$ матрицами. Для них
полярное разложение имеет хорошо известный вид $z=r\ex^{i\vf}$. Если $z\ne0$, то
модуль $r$ и аргумент $\vf$ комплексного числа определены однозначно. При $z=0$
модуль комплексного числа также равен нулю, $r=0$, а аргумент неопределен.
\qed\end{exa}
\begin{com}
Полярные разложения матриц справедливы также для другого порядка матриц в
формулах (\ref{eporem}) и (\ref{epocor}): $A=S'R'$ и $A=U'R'$. Вообще говоря,
новые матрицы будут отличаться от старых. Сомножители $R,S$ и $R,U$ в полярном
разложении матриц перестановочны между собой тогда и только тогда, когда матрицы
$A,A^\St$ и $A,A^\dagger$ коммутируют, соответственно, для вещественного и
комплексного случая.
\qed\end{com}

\begin{defn}
Две квадратные $n\times n$ матрицы $A$ и $B$ называются {\em подобными}, если
существует такая невырожденная матрица $S$, что
\begin{equation}                                                  \label{esimat}
  B=SAS^{-1}.
\end{equation}
Отображение $A\mapsto SAS^{-1}$ называется {\em преобразованием подобия}.
\qed\end{defn}
\index{Подобная матрица (similar matrix)}%
\index{Матрица подобная (similar matrix)}%
\index{Преобразование подобия (similarity transformation)}%
\index{Подобия преобразование (similarity transformation)}%
Легко проверить, что преобразование подобия является отношением эквивалентности
в множестве квадратных $n\times n$ матриц. Если матрицы $A$ и $B$ подобны, то
матрица $S$, с помощью которой проводится преобразование, определена
неоднозначно. Ее можно умножить слева на произвольную невырожденную матрицу,
коммутирующую с $A$.

С помощью преобразования подобия любую матрицу можно преобразовать к некоторому
каноническому виду. Одним из таких видов является жорданова форма матрицы.
\begin{defn}
{\em Жордановым блоком} или {\em жордановой клеткой} $J_k(\lm)$ называется
верхняя треугольная $k\times k$ матрица вида
\begin{equation}                                                  \label{ejorno}
  J_k(\lm):=\begin{pmatrix}
    \lm & 1 &&&&& 0 \\ & \lm & 1 &&&&& \\ && \cdot & \cdot &&& \\
    &&& \cdot & \cdot && \\ &&&& \cdot & \cdot & \\ &&&&& \cdot & 1 \\
    0 &&&&&& \lm \end{pmatrix},
\end{equation}
у которой на главной диагонали стоят одинаковые числа $\lm$, а над ними --
единицы. Все остальные элементы матрицы равны нулю. По определению
$J_1(\lm)=\lm$. {\em Жордановой $n\times n$ матрицей} называется любая прямая
сумма жордановых клеток:
\begin{equation}                                                  \label{ejorma}
  J:=\begin{pmatrix} J_{n_1}(\lm_1) & & & 0 \\ & J_{n_2}(\lm_2) && \\
  && \ddots & \\ 0 &&& J_{n_k}(\lm_k) \end{pmatrix},\qquad n_1+n_2+\dotsc+n_k=n,
\end{equation}
где порядки $n_i$ каких то клеток могут совпадать и числа $\lm_i$ не обязательно
различны.
\qed\end{defn}
\index{Жорданов блок (Jordan box)}%
\index{Блок жорданов (Jordan box)}%
\index{Жорданова клетка (Jordan box)}%
\index{Клетка жорданова (Jordan box)}%
\index{Жорданова матрица (Jordan matrix)}%
\index{Матрица жорданова (Jordan matrix)}%
Можно рассматривать, например, вещественные или комплексные жордановы матрицы.

Все собственные числа жордановой клетки одинаковы и равны $\lm$, т.е.\ $\lm$
является собственным числом жордановой клетки (\ref{ejorno}) кратности $k$.
$\lm_i$, $i=1,\dotsc,n$ -- собственные числа матрицы $A$. Если $\lm_i=\lm_j$ и
отличается от всех остальных собственных чисел, то кратность собственного числа
$\lm_i$ равна $n_i+n_j$.

\begin{theorem}
Пусть задана комплексная $n\times n$ матрица $A$. Тогда существует невырожденная
$n\times n$ матрица $S$ такая, что
\begin{equation}                                                  \label{ematjo}
  A=SJS^{-1},
\end{equation}
где $J$ -- жорданова матрица (\ref{ejorma}). Матрица $A$ определяет подобную ей
жорданову матрицу однозначно с точностью до перестановки жордановых клеток на ее
главной диагонали. Собственные значения $\lm_i$ не обязательно различны. Если
матрица $A$ вещественна и обладает только вещественными собственными значениями,
то подобие может быть реализовано с помощью вещественной матрицы $S$.
\end{theorem}
Число $k$ жордановых клеток матрицы $A$ (с учетом возможных повторов) равно
максимальному числу ее линейно независимых векторов. При этом для каждой
жордановой клетки существует только один независимый собственный вектор. Матрица
$A$ диагонализируема тогда и только тогда, когда $k=n$. В частности, если все
собственные числа различны, то матрицу $A$ можно привести к диагональному виду с
помощью преобразования подобия.

Если порядок всех жордановых клеток равен единице, $n_i=1$, $i=1,\dotsc,k$, то
жорданова матрица является диагональной. Если один из жордановых блоков имеет
б\'ольший порядок, $n_i>1$, то жорданова матрица не может быть приведена к
диагональному виду преобразованием подобия.
\begin{prop}
Любая комплексная матрица $A$ подобна своей транспонированной $A^\St$.
\end{prop}

Любая жорданова клетка записывается в виде суммы единичной и нильпотентной
матриц:
\begin{equation}                                                  \label{eyilpe}
  J_k(\lm)=\lm\one+N_k,\qquad (N_k)^k=0.
\end{equation}
Любая жорданова матрица записывается в виде суммы диагональной и нильпотентной
матриц:
\begin{equation}                                                  \label{esunid}
  J=D+N,\qquad N^k=0,
\end{equation}
где $k$ -- порядок наибольшей жордановой клетки. Если задана произвольная
матрица и ее жорданова форма, то
\begin{equation*}
  A=SJS^{-1}=SDS^{-1}+SNS^{-1}:=A_D+A_N,
\end{equation*}
где $A_D$ диагонализируемая и $A_N$ нильпотентная матрицы. При этом
$[A_D,A_N]=0$.

Приведем некоторые результаты для эрмитовых и симметричных матриц.
\begin{defn}
Комплексная $n\times n$ матрица $A$ называется {\em эрмитовой}, если выполнено
равенство $A^\dagger=A$, где символ $\dagger$ обозначает эрмитово сопряжение:
$A^\dagger:=(\bar A)^\St$. Если выполнено равенство $A^\dagger=-A$, то матрица
$A$ называется {\em антиэрмитовой}.
\qed\end{defn}
\index{Эрмитова матрица (Hermitian matrix)}%
\index{Матрица эрмитова (Hermitian matrix)}%
\index{Антиэрмитова матрица (anti-Hermitian matrix)}%
\index{Матрица антиэрмитова (anti-Hermitian matrix)}%
Если матрица $A$ вещественна, то эрмитовы и антиэрмитовы матрицы соответствуют
симметричным и антисимметричным матрицам.
\begin{theorem}                                                   \label{tdiama}
Все собственные значения эрмитовой $n\times n$ матрицы $A$ вещественны. Матрица
$A$ эрмитова тогда и только тогда, когда существует унитарная матрица $U$ такая,
что выполнено равенство
\begin{equation}                                                  \label{eundia}
  A=U\Lm U^{-1},\qquad U\in\MU(n),
\end{equation}
где $\Lm$ -- вещественная диагональная матрица. Матрица $\Lm$ составлена из
собственных значений матрицы $A$. Унитарную матрицу $U$ можно выбрать из
односвязной подгруппы $\MS\MU(n)\subset\MU(n)$.

Все собственные значения вещественной симметричной $n\times n$ матрицы $B$
вещественны. Матрица $B$ симметрична тогда и только тогда, когда существует
ортогональная матрица $S$ такая, что выполнено равенство
\begin{equation}                                                  \label{edeobe}
  B=S\Lm S^{-1},\qquad S\in\MO(n),
\end{equation}
где $\Lm$ -- вещественная диагональная матрица. Матрица $\Lm$ составлена из
собственных значений матрицы $B$. Ортогональную матрицу $S$ можно выбрать из
связной компоненты единицы $\MS\MO(n)$.
\end{theorem}

С помощью одной унитарной матрицы иногда удается диагонализировать сразу
несколько эрмитовых матриц.
\begin{theorem}
Пусть $\CF$ -- семейство эрмитовых матриц. Унитарная матрица $U$ такая, что
матрица $UAU^{-1}$ -- диагональна для всех $A\in\CF$, существует тогда и только
тогда, когда $[A,B]=0$ для всех $A,B\in\CF$.
\end{theorem}

В заключение рассмотрим некоторые свойства суперматриц, элементами которых
являются коммутирующие и антикоммутирующие объекты. Пусть задана квадратная
суперматрица
\begin{equation}                                                  \label{esupma}
  M=\begin{pmatrix} A_m{}^n & B_m{}^\nu \\ C_\mu{}^n & D_\mu{}^\nu
  \end{pmatrix},
\end{equation}
где $A$ и $D$ -- квадратные невырожденные матрицы с коммутирующими элементами, а
$B$ и $C$ -- матрицы с антикоммутирующими элементами, которые могут быть
прямоугольными. Суперслед и суперопределитель матрицы $M$ определены
соотношениями:
\begin{align}                                                     \label{edtrac}
  \str M &:=\tr A-\tr D,
\\                                                                \label{esdete}
  \sdet M &:=\frac{\det A}{\det(D-CA^{-1}B)}
  =\frac{\det(A-BD^{-1}C)}{\det D}.
\end{align}

Из определения вытекает, что ряд свойств обычного следа и определителя
наследуется суперматрицами:
\begin{align}                                                          \nonumber
  \str(M+N)&=\str M+\str N,
\\                                                                     \nonumber
  \str(MN)&=\str(NM),
\\                                                                \label{esdeto}
  \sdet M^{-1}&=\frac1{\sdet M},
\\                                                                     \nonumber
  \sdet(MN)&=\sdet M\:\sdet N,
\\                                                                     \nonumber
  \sdet M&=\exp\str\ln M.
\end{align}
\section{Матрицы Паули                                           \label{spamat}}
\begin{defn}
Три эрмитовых $2\times2$ матрицы $\boldsymbol{\s}=\lbrace \s_i\rbrace$,
$i=1,2,3$:
\begin{equation}                                                  \label{epamat}
  \s_1:=\begin{pmatrix} 0 & 1 \\ 1 & 0 \end{pmatrix},\qquad
  \s_2:=\begin{pmatrix} 0 & -i \\ i & 0 \end{pmatrix},\qquad
  \s_3:=\begin{pmatrix} 1 & 0 \\ 0 & -1 \end{pmatrix}
\end{equation}
называются {\em матрицами Паули}.
\qed\end{defn}
\index{Матрицы Паули (Pauli matrices)}%
\index{Паули матрицы (Pauli matrices)}%
Они обладают следующими свойствами:
\begin{equation}                                                  \label{eprpam}
\begin{split}
  \s_1\s_2&=-\s_2\s_1=i\s_3, \\
  \s_2\s_3&=-\s_3\s_2=i\s_1, \\
  \s_3\s_1&=-\s_1\s_3=i\s_2, \\
  \s_i^2&=1\qquad (\text{суммирования по $i$ нет}), \\
  \s_1\s_2\s_3&=i.
\end{split}
\end{equation}
Свойства (\ref{eprpam}) можно записать в виде коммутатора и антикоммутатора
матриц Паули:
\begin{equation}                                                  \label{epapro}
  [\s_i,\s_j]=2i\ve_{ij}{}^k\s_k,\qquad \lbrace \s_i,\s_j\rbrace =2\dl_{ij}
\end{equation}
или
\begin{equation}                                                  \label{epamro}
  \s_i\s_j=\dl_{ij}+i\ve_{ij}{}^k\s_k,
\end{equation}
где $\e_{ijk}$ -- полностью антисимметричный тензор третьего ранга и подъем
индексов осуществляется с помощью евклидовой метрики $\dl_{ij}$.

Матрицы Паули встречаются в различных контекстах. Во-первых, они являются
генераторами группы двумерных унитарных матриц $\MS\MU(2)$. Во-вторых, второе из
соотношений (\ref{epapro}) означает, что матрицы Паули являются образующими
алгебры Клиффорда $\MC\ML(3)$. В-третьих, матрицы Паули, умноженные на $-i$,
можно рассматривать, как представление образующих алгебры кватернионов
(\ref{equpam}).

Для следов произведения матриц Паули справедливы тождества:
\begin{equation}                                                  \label{etrpam}
  \tr\s_i=0,\qquad \tr(\s_i\s_j)=2\dl_{ij},\qquad
  \tr(\s_i\s_j\s_k)=2i\ve_{ijk}.
\end{equation}

Рассмотрим вектор $\Bx=\lbrace x^i\rbrace $ в трехмерном евклидовом пространстве
с обычным скалярным произведением. Тогда справедлива формула
\begin{equation}                                                  \label{exputw}
  e^{i(\Bx,\boldsymbol{\s})}
  =\cos|\Bx|+i\frac{(\Bx,\boldsymbol{\s})}{|\Bx|}\sin|\Bx|,
\end{equation}
где
$$
  (\Bx,\boldsymbol{\s}):=x^1\s_1+x^2\s_2+x^3\s_3,\qquad
  |\Bx|=\sqrt{(x^1)^2+(x^2)^2+(x^3)^2}.
$$
Свертка равенства (\ref{epamro}) с произвольными векторами $\Bx$ и $\By$
приводит к тождеству
\begin{equation}                                                  \label{evetop}
  (\Bx,\boldsymbol{\s})(\By,\boldsymbol{\s})
  =(\Bx,\By)+i([\Bx,\By],\boldsymbol{\s}),
\end{equation}
где $[\Bx,\By]^i:=x^jy^k\ve_{ik}{}^i$ -- векторное произведение векторов. Из
свойства (\ref{epamro}) следует, что всякая функция от матриц Паули сводится к
линейной функции.

С групповой точки зрения матрицы Паули являются генераторами группы $\MS\MU(2)$.
``Скалярное произведение'' $(\Bx,\boldsymbol{\s})$ и экспонента
$e^{i(\Bx,\boldsymbol{\s})}$ являются соответственно элементами алгебры Ли
$\Gs\Gu(2)$ и группы Ли $\MS\MO(2)$. Отображение (\ref{exputw}) представляет
собой экспоненциальное отображение алгебры Ли в группу Ли.
\section{Кватернионы                                             \label{squate}}
\begin{defn}
Рассмотрим четырехмерное векторное пространство $\MH$ над полем вещественных
чисел, которое содержит поле вещественных чисел как подпространство. Обозначим
базис этого пространства символами $1,i,j,k$, где 1 -- единица из поля
вещественных чисел. Тогда произвольный элемент $q\in\MH$ представим в виде
\begin{equation}                                                  \label{equatb}
  q=a+bi+cj+dk,
\end{equation}
где $a,b,c,d$ -- вещественные числа. Введем билинейное умножение в $\MH$,
положив следующее правило умножения базисных векторов
\begin{equation}                                                  \label{equmul}
\begin{split}
  ij&=-ji=k,\\ jk&=-kj=i,\\ ki&=-ik=j,\\ i^2&=j^2=k^2=-1.
\end{split}
\end{equation}
Элементы вида (\ref{equatb}) с операциями сложения, умножения на действительные
числа и умножения (\ref{equmul}) называются {\em кватернионами}.
\qed\end{defn}
\index{Кватернион (quaternion)}%
С помощью прямых вычислений можно проверить, что кватернионы с так определенным
умножением образуют ассоциативную, но не коммутативную алгебру с единицей $\MH$
над полем вещественных чисел. Эту алгебру можно представить в виде комплексных
матриц $A(q)\in\MG\ML(2,\MC)$
\begin{equation}                                                  \label{equare}
  A(q)=\left(\begin{array}{rr} a+bi & c+di\\ -c+di & a-bi \end{array}\right),
\end{equation}
где $i$ -- мнимая единица. Прямые вычисления показывают, что отображение
$q\mapsto A(q)$ является гомоморфизмом, то есть
$$
  A(q_1q_2)=A(q_1)A(q_2).
$$
Для доказательства этого утверждения достаточно проверить его для
$q=i,j,k$. В частности, при $q=i,j,k$ имеем равенства:
\begin{equation}                                                  \label{equijk}
  A(i)=\left(\begin{array}{rr} i & 0 \\ 0 & -i \end{array} \right),\qquad
  A(j)=\left(\begin{array}{rr} 0 & 1 \\ -1 & 0 \end{array} \right),\qquad
  A(k)=\left(\begin{array}{rr} 0 & i \\ i & 0 \end{array} \right).
\end{equation}
Эти матрицы с точностью до множителя $-i$ совпадают с матрицами Паули
(\ref{epamat})
\begin{equation}                                                  \label{equpam}
  \s_1=-iA(k),\qquad \s_2=-iA(j),\qquad \s_3=-iA(i).
\end{equation}

Введем операцию сопряжения в $\MH$, то есть поставим каждому кватерниону вида
(\ref{equatb}) его сопряженный элемент
\begin{equation}                                                  \label{equcon}
  \bar q:=a-bi-cj-dk.
\end{equation}
Это отображение $q\rightarrow\bar q$ является антиавтоморфизмом алгебры
кватернионов, то есть
$$
  \overline{q_1+q_2}=\bar q_1+\bar q_2,\qquad \overline{q_1q_2}=\bar q_2\bar q_1.
$$

Число $a$ называется вещественной частью $\re q:=a$, а остаток $\im q:=bi+cj+dk$
-- мнимой частью кватерниона. В алгебре кватернионов можно ввести симметричное
скалярное произведение
\begin{equation}                                                  \label{escqut}
  (q_1,q_2):=\re q_1\bar q_2=a_1a_2+b_1b_2+c_1c_2+d_1d_2,
\end{equation}
которое совпадает с обычным евклидовым произведением в $\MR^4$. Поскольку каждый
кватернион представим в виде $q=\re q+\im q$, то алгебра кватернионов
распадается в прямую сумму ортогональных подпространств $\MH=\MR\oplus\MR^3$,
где $\MR$ -- прямая, состоящая из вещественных кватернионов, а $\MR^3$ --
ортогональное трехмерное евклидово пространство мнимых кватернионов.

Определим норму кватерниона
\begin{equation}                                                  \label{equnor}
  \parallel q\parallel^2:=q\bar q=\bar q q=a^2+b^2+c^2+d^2.
\end{equation}
Нетрудно проверить, что
$$
  \parallel q\parallel^2=\det A(q).
$$
Отсюда следует, что
$\parallel q_1q_2\parallel^2=\parallel q_1\parallel^2\parallel q_2\parallel^2$.

Каждый отличный от нуля кватернион обладает единственным обратным элементом
$$
  qq^{-1}=q^{-1}q=1,
$$
где
$$
  q^{-1}=\frac{\bar q}{\parallel q\parallel^2}.
$$
Это означает, что множество кватернионов относительно операции сложения и
умножения (\ref{equmul}) (без умножения на действительные числа) образует тело,
так как умножение некоммутативно.

Алгебра кватернионов $\MH$ допускает естественное отождествление с двумерным
комплексным пространством $\MC^2$. Используя определение умножения в алгебре
кватернионов $k=-ji$ (\ref{equmul}), запишем кватернион в виде
\begin{equation*}
  q=(a+ib)+j(c-id)=z_1+j\bar z_2,
\end{equation*}
где линейные комбинации $z_1:=a+ib$ и $z_2:=c+id$ можно отождествить с парой
комплексных чисел. При этом $1$ и $j$ рассматриваются, как базис в $\MC^2$.

Рассмотрим мнимый кватернион, не содержащий вещественной части,
\begin{equation*}
  r=bi+cj+dk,\qquad \parallel r\parallel=b^2+c^2+d^2.
\end{equation*}
Его можно представить, как вектор в трехмерном евклидовом пространстве
$r\in\MR^3$. Пусть $q\in\MH$ произвольный обратимый (в смысле кватернионного
умножения) кватернион. Тогда кватернион $qrq^{-1}$ снова является мнимым
кватернионом с той же нормой
\begin{equation*}
  qrq^{-1}\in\MR^3,\qquad \parallel qrq^{-1}\parallel=\parallel r\parallel.
\end{equation*}
Это значит, что каждому обратимому кватерниону, т.е.\ отличному от нуля,
ставится в соответствие некоторое
вращение. При этом двум кватернионам $q$ и $-q$, отличающимся знаком,
соответствует одно и то же вращение. Можно проверить, что любое вращение можно
реализовать с помощью кватерниона единичной нормы. Поскольку кватернионы
единичной нормы параметризуют трехмерную сферу единичного радиуса, то существует
гомеоморфизм
\begin{equation*}
  \MS\MO(3)\approx\frac{\MS^3}{\MZ_2}.
\end{equation*}

Рассмотрим отображение кватернионов $q\in\MH$, задаваемое двумя кватернионами
$a,b\in\MH$ единичной нормы
\begin{equation*}
  q\mapsto aqb^{-1},\qquad \parallel a\parallel=\parallel b\parallel=1.
\end{equation*}
Это отображение задает вращение четырехмерного евклидова пространства
$\MH=\MR^4$, поскольку сохраняет норму $q$. Можно проверить, что любое вращение
определяет пару $a$ и $b$ с точностью до знака, так как кватернионы $a,b$ и
$-a,-b$ определяют одно и то же вращение. Это значит, что существует
гомеоморфизм
\begin{equation*}
  \MS\MO(4)\approx\frac{\MS^3\times\MS^3}{\MZ_2}.
\end{equation*}
\section{Полностью антисимметричные тензоры                      \label{scomat}}
Полностью антисимметричный тензор ранга $n$ в $n$-мерном евклидовом пространстве
единственен с точностью до умножения на произвольную ненулевую постоянную и
имеет всего одну независимую компоненту. Его всегда можно представить в виде
константы, умноженной на тензор, все ненулевые компоненты которого равны по
модулю единице. Этот тензор часто используется в приложениях и имеет специальное
обозначение
\begin{equation}                                                  \label{etotan}
\begin{aligned}
  \ve^{a_1\dots a_n}&=\ve^{[a_1\dots a_n]}, \\
  \ve_{a_1\dots a_n}&=\ve^{a_1\dots a_n},
\end{aligned}\qquad
\begin{aligned}
  \ve^{12\dots n}&=1, \\  \ve_{12\dots n}&=1.
\end{aligned}
\end{equation}
Здесь подъем индексов произведен с помощью евклидовой метрики. Произведения и
свертки двух таких тензоров выражаются через символы Кронекера:
\begin{equation}                                                  \label{etoprz}
\begin{aligned}
  \ve^{a_1\dots a_n}\ve_{b_1\dots b_n}&=
  \vphantom{ \begin{pmatrix} \dl^{a_1}_{b_1} \\ \dl^{a_1}_{b_2} \\
    \vdots \\ \dl^{a_1}_{b_n} \end{pmatrix}}
\\
  \ve^{b_1a_2\dots a_n}\ve_{b_1\dots b_n}&=
  \vphantom{\displaystyle\frac{n!}{C_n^k}}
\\
  \ve^{b_1b_2a_3\dots a_n}\ve_{b_1\dots b_n}&=
  \vphantom{\displaystyle\frac{n!}{C_n^k}}
\\
  \ve^{b_1\dots b_ka_{k+1}\dots a_n}\ve_{b_1\dots b_n}&=
  \vphantom{\displaystyle\frac{n!}{C_n^k}} \\
  \ve^{b_1\dots b_{n-1}a_n}\ve_{b_1\dots b_n}&=
  \vphantom{\displaystyle\frac{n!}{C_n^k}} \\
  \ve^{b_1\dots b_n}\ve_{b_1\dots b_n}&=
  \vphantom{\displaystyle\frac{n!}{C_n^k}}
\end{aligned}
\begin{aligned}
  n!&~\dl_{~b_1\dots b_n}^{[a_1\dots a_n]}
 \vphantom{ \begin{pmatrix} \dl^{a_1}_{b_1} \\ \dl^{a_1}_{b_2} \\
    \vdots \\ \dl^{a_1}_{b_n} \end{pmatrix}} \\
  (n-1)!&~\dl_{~b_2\dots b_n}^{[a_2\dots a_n]}
  \vphantom{\displaystyle\frac{n!}{C_n^k}} \\
  2(n-2)!&~\dl_{~b_3\dots b_n}^{[a_3\dots a_n]}
  \vphantom{\displaystyle\frac{n!}{C_n^k}} \\
  k!(n-k)!&~\dl_{~b_{k+1}\dots b_n}^{[a_{k+1}\dots a_n]}
  \vphantom{\displaystyle\frac{n!}{C_n^k}} \\
  (n-1)!&~\dl_{b_n}^{a_n}
  \vphantom{\displaystyle\frac{n!}{C_n^k}} \\
  n!&
  \vphantom{\displaystyle\frac{n!}{C_n^k}}
\end{aligned}
\begin{aligned}
  {}&=\det
  \begin{pmatrix}
    \dl^{a_1}_{b_1} & \dl^{a_2}_{b_1} & \cdots & \dl^{a_n}_{b_1} \\
    \dl^{a_1}_{b_2} & \dl^{a_2}_{b_2} & \cdots & \dl^{a_n}_{b_2} \\
    \vdots & \vdots & \ddots & \vdots \\
    \dl^{a_1}_{b_n} & \dl^{a_2}_{b_n} & \cdots & \dl^{a_n}_{b_n}
\end{pmatrix} \\
  {}&={\displaystyle\frac{n!}{C_n^1}}\dl_{~b_2\dots b_n}^{[a_2\dots a_n]}\\
  {}&={\displaystyle\frac{n!}{C_n^2}}\dl_{~b_3\dots b_n}^{[a_3\dots a_n]}\\
  {}&={\displaystyle\frac{n!}{C_n^k}}
  \dl_{~b_{k+1}\dots b_n}^{[a_{k+1}\dots a_n]} \\
  {}&={\displaystyle\frac{n!}{C_n^{n-1}}}\dl_{b_n}^{a_n}\\
  {}&
  \vphantom{\displaystyle\frac{n!}{C_n^k}}
\end{aligned}
\end{equation}
Полностью антисимметричный тензор максимального ранга в пространстве с метрикой
произвольной сигнатуры определяется соотношениями
\begin{equation}                                                  \label{elovae}
  \ve_{12\dotsc n}=1,\qquad \ve^{12\dotsc n}\sgn,
\end{equation}
где множитель $\sgn:=\sign(\det g_{\al\bt})$ равен знаку определителя метрики.
Этот множитель возникает при подъеме индексов. Для таких пространств правые
части (\ref{etoprz}) необходимо умножить на множитель $\sgn$.

Определитель произвольной квадратной матрицы $A$ в евклидовом пространстве можно
записать в виде
\begin{equation}                                                  \label{edetms}
\begin{split}
  \det(A_a{}^b)&=\ve_{b_1b_2\dots b_n}A_1{}^{b_1}A_2{}^{b_2}\dots A_n{}^{b_n}
  =\ve^{a_1a_2\dots a_n}A_{a_1}{}^1A_{a_2}{}^2\dots A_{a_n}{}^n
\\
  &=\frac1{n!}\ve^{a_1a_2\dots a_n}\ve_{b_1b_2\dots b_n}
  A_{a_1}{}^{b_1}A_{a_2}{}^{b_2}\dots A_{a_n}{}^{b_n}.
\end{split}
\end{equation}
Полностью антисимметричный тензор в $n$-мерном евклидовом пространстве
инвариантен относительно $\MS\MO(n)$ вращений с матрицей
$S_a{}^{a'}\in \MS\MO(n)$:
\begin{equation}                                                  \label{eaninv}
  \ve^{a'_1\dots a'_n}=\ve^{a_1\dots a_n}S_{a_1}{}^{a'_1}\dots
  S_{a_n}{}^{a'_n}
  =\det(S_a{}^{a'})\ve^{a'_1\dots a'_n}.
\end{equation}

Полностью антисимметричный тензор в голономном базисе на многообразии имеет
следующие компоненты:
\begin{equation}                                                  \label{ehoast}
  \ve^{\al_1\dots\al_n}=e^{\al_1}{}_{a_1}\dots e^{\al_n}{}_{a_n}
  \ve^{a_1\dots a_n},
\end{equation}
и в геометрии Римана--Картана является ковариантно постоянным:
\begin{equation}                                                  \label{ecpans}
  \nb_\bt\ve^{\al_1\dots\al_n}=0.
\end{equation}

По-определению, полностью антисимметричный тензор в пространстве Минковского
четного числа измерений меняет знак при пространственном отражении
$(x^0,x^i)\mapsto(x^0,-x^i)$, то есть является псевдотензором. Приставку
псевдо- мы, как правило, опускаем.

Полностью антисимметричная тензорная плотность $\hat e^{\al_1\dots\al_n}$ веса
$-1$ определяется соотношениями
\begin{equation}                                                  \label{eastde}
\begin{split}
  \hat\ve^{\al_1\dots\al_n}&:=\vol\ve^{\al_1\dots\al_n},
\\
  \hat\ve_{\al_1\dots\al_n}&:=\frac1\vol\ve_{\al_1\dots\al_n}.
\end{split}
\end{equation}
Ее ненулевые компоненты равны по модулю единице в любой системе координат,
\begin{equation}                                                  \label{ecoate}
  \hat\ve^{\al_1\dots\al_n}=
\begin{cases}
  +1&\text{-- четная перестановка индексов},\\
  -1&\text{-- нечетная перестановка индексов},\\
  \quad 0&\text{-- совпадение любой пары индексов}.
\end{cases}
\end{equation}
Отсюда следует, что все частные производные антисимметричной тензорной плотности
равны нулю:
\begin{equation}                                                  \label{epdatd}
  \pl_\bt\hat\ve^{\al_1\dots\al_n}=0.
\end{equation}

Приведем явные формулы для трехмерного евклидова пространства $\MR^3$, которые
часто используются в физических приложениях,
\begin{equation}                                                  \label{etttat}
  \ve_{123}=1,\qquad \ve^{123}=1.
\end{equation}
Свертки:
\begin{equation}                                                  \label{ecotof}
\begin{split}
  \ve^{lmn}\ve_{ijk}&=
  \dl_{ijk}^{lmn}+\dl_{ijk}^{mnl}+\dl_{ijk}^{nlm}
  -\dl_{ijk}^{lnm}-\dl_{ijk}^{mln}-\dl_{ijk}^{nml}
  =\det\begin{pmatrix} \dl_i^l & \dl_i^m & \dl_i^n \\
  \dl_j^l & \dl_j^m & \dl_j^n \\ \dl_k^l & \dl_k^m & \dl_k^n \end{pmatrix}
\\
  \ve^{klm}\ve_{ijm}&=\dl_{ij}^{kl}-\dl_{ij}^{lk},
\\
  \ve^{jkl}\ve_{ikl}&=2\dl_{i}^{j},
\\
  \ve^{ijk}\ve_{ijk}&=6.
\end{split}
\end{equation}

Полностью антисимметричный тензор второго ранга в евклидовом пространстве имеет
две ненулевые компоненты
\begin{equation}                                                  \label{etoasr}
  \ve_{12}=1,\qquad \ve^{12}=1
\end{equation}
или
\begin{equation}                                                  \label{easttd}
  \ve_{ab}=\ve^{ab}=\ve^a{}_b=\ve_a{}^b=\begin{pmatrix}\quad 0&1
  \\-1&0\end{pmatrix}.
\end{equation}
Свертки:
\begin{equation}                                                  \label{ecotsf}
\begin{split}
  \ve^{ab}\ve_{cd}&=\dl_{cd}^{ab}-\dl_{cd}^{ba}
  =\det\begin{pmatrix} \dl_c^a & \dl_c^b \\ \dl_d^a & \dl_d^b \end{pmatrix}
\\
  \ve^{ab}\ve_{ad}&=\dl_{d}^{b},
\\
  \ve^{ab}\ve_{ab}&=2.
\end{split}
\end{equation}

В пространстве Минковского нечетного числа измерений формулы для сверток
полностью антисимметричных тензоров те же, что и в евклидовом пространстве. В
пространстве Минковского четного числа измерений полностью антисимметричные
тензоры с верхними и нижними индексами отличаются знаком. Определим
антисимметричный тензор четвертого ранга в пространстве Минковского следующим
образом
\begin{equation}                                                  \label{etotfm}
  \ve_{0123}=1,\qquad \ve^{0123}=-1.
\end{equation}
Свертки:
\begin{equation}                                                  \label{ecfmif}
\begin{split}
  \ve^{abcd}\ve_{efgh}=
  &-\dl_{efgh}^{abcd}-\dl_{efgh}^{acdb}-\dl_{efgh}^{adbc}
  +\dl_{efgh}^{acbd}+\dl_{efgh}^{abdc}+\dl_{efgh}^{adcb}
\\
  &+\dl_{efgh}^{bcda}+\dl_{efgh}^{bdac}+\dl_{efgh}^{bacd}
  -\dl_{efgh}^{bdca}-\dl_{efgh}^{bcad}-\dl_{efgh}^{badc}
\\
  &-\dl_{efgh}^{cdab}-\dl_{efgh}^{cabd}-\dl_{efgh}^{cbda}
  +\dl_{efgh}^{cadb}+\dl_{efgh}^{cdba}+\dl_{efgh}^{cbad}
\\
  &+\dl_{efgh}^{dabc}+\dl_{efgh}^{dbca}+\dl_{efgh}^{dcab}
  -\dl_{efgh}^{dbac}-\dl_{efgh}^{dacb}-\dl_{efgh}^{dcba}
\\
  \ve^{abcd}\ve_{afgh}=&
  -\dl_{fgh}^{bcd}-\dl_{fgh}^{cdb}-\dl_{fgh}^{dbc}
  +\dl_{fgh}^{cbd}+\dl_{fgh}^{bdc}+\dl_{fgh}^{dcb}
\\
  \ve^{abcd}\ve_{abgh}=&-2(\dl_{gh}^{cd}-\dl_{gh}^{dc}),
\\
  \ve^{abcd}\ve_{abch}=&-6\dl_{h}^{d},
\\
  \ve^{abcd}\ve_{abcd}=&-24.
\end{split}
\end{equation}
Полностью антисимметричные тензоры в трехмерном евклидовом пространстве и
пространстве Минковского связаны следующими соотношениями
\begin{equation}                                                  \label{etottf}
  \ve_{ijk}=\ve_{0ijk},\qquad \ve^{ijk}=-\ve^{0ijk}.
\end{equation}

Полностью антисимметричный тензор второго ранга в пространстве Минковского
$\MR^{1,1}$ имеет две ненулевые компоненты:
\begin{equation}                                                  \label{etoasm}
  \ve_{01}=1,\qquad \ve^{01}=-1
\end{equation}
или
\begin{equation}                                                  \label{eateet}
  \ve_{ab}=-\ve^{ab}=\begin{pmatrix}\quad 0&1\\-1&0\end{pmatrix},\qquad
  \ve^a{}_b=-\ve_a{}^b
  =\begin{pmatrix}0&1\\1&0\end{pmatrix},
\end{equation}
Свертки:
\begin{equation}                                                  \label{ecotsm}
\begin{split}
  \ve^{ab}\ve_{cd}&=-\dl_{cd}^{ab}+\dl_{cd}^{ba},
\\
  \ve^{ab}\ve_{ad}&=-\dl_{d}^{b},
\\
  \ve^{ab}\ve_{ab}&=-2.
\end{split}
\end{equation}

В двумерном пространстве Минковского $\MR^{1,1}$ можно ввести проекционные
операторы
\begin{equation*}
  \Pi^\pm_a{}^b:=\frac12(\dl_a^b\pm \ve_a{}^b).
\end{equation*}
Отметим, что аналогичная конструкция в евклидовом пространстве не приводит
к проекционным операторам.

Пусть в римановом пространстве $(\MM,g)$, $\dim\MM=n$, задано тензорное поле с
компонентами $K_{\al\bt}$ типа $(0,2)$ такое, что $\det K_{\al\bt}\ne0$. Тогда
существует обратное тензорное поле $K^{\al\bt}$ типа $(2,0)$,
$$
  K^{\al\bt}K_{\bt\g}=\dl^\al_\g,
$$
и его можно представить в виде
\begin{equation}                                                  \label{eintes}
  K^{\al\bt}=\frac1{|\det K|(n-1)!}\hat\ve^{\al\g_2\dots\g_n}
  \hat\ve^{\bt\dl_2\dots\dl_n}K_{\dl_2\g_2}\dots K_{\dl_n\g_n},
\end{equation}
Если метрика имеет лоренцеву сигнатуру, то в пространстве четного
числа измерений в правой части появляется знак минус.

\newpage
\small
\printindex

\addcontentsline{toc}{chapter}{Библиография}
\begin{thebibliography}{100}

\bibitem{Arnold75R}
В.~И. Арнольд.
\newblock {\em Обыкновенные диференциальные уравнения}.
\newblock Наука, Москва, 1975.
\newblock Второе изд. 240 с.

\bibitem{ArKoNe02R}
В.~И. Арнольд, В.~В. Козлов, and А.~И. Нейштадт.
\newblock {\em Математические аспекты классической и небесной механики}.
\newblock Эдиториал УРСС, Москва, 2002.
\newblock Второе изд. 414 с.

\bibitem{Vladim88R}
В.~С. Владимиров.
\newblock {\em Уравнения математической физики}.
\newblock Наука, Москва, издание пятое, 1988.

\bibitem{Gelfan98R}
И.~М. Гельфанд.
\newblock {\em Лекции по линейной алгебре}.
\newblock МЦНМО, Москва, 1998.
\newblock Пятое изд.

\bibitem{GeMiSh58R}
И.~М. Гельфанд, Р.~А. Минлос, and З.~Я. Шапиро.
\newblock {\em Представления группы вращений и группы Лоренца, их применения}.
\newblock Физ.-мат.\ лит., Москва, 1958.

\bibitem{DroZav06R}
Ю.~Н. Дрожжинов and Б.~Н. Завьялов.
\newblock {\em Введение в теорию обобщенных функций}.
\newblock МИАН, Москва, 2006.

\bibitem{DuNoFo98R}
Б.~А. Дубровин, С.~П. Новиков, and А.~Т. Фоменко.
\newblock {\em Современная геометрия. Методы и приложения.}
\newblock Наука, Москва, издание четвертое, 1998.

\bibitem{Zharin08R}
В.~В. Жаринов.
\newblock {\em Алгебро-геометриеские основы математической физики}.
\newblock Изд-во МИАН им.~В.~А.~Стеклова, Москва, 2008.

\bibitem{Kelley57R}
J.~L. Kelley.
\newblock {\em General Topology}.
\newblock D. Van Nostrand Company, Inc., Toronto -- London, 1957.
\newblock Перевод: Келли Дж. Л. {\it Общая топология.} М.: Наука, 1968.

\bibitem{Kirill78R}
А.~А. Кирилов.
\newblock {\em Элементы теории представлений}.
\newblock Наука, Москва, Второе edition, 1978.

\bibitem{Kostri00R}
А.~И. Кострикин.
\newblock {\em Введение в алгебру. Часть I. {\it Основы алгебры.} Часть II.
  {\it Линейная алгебра.} Часть III {\it Основные структуры алгебры.}}
\newblock Наука, Москва, 2000.

\bibitem{KosMan86R}
А.~И. Кострикин and Ю.~И. Манин.
\newblock {\em Линейная алгебра и геометрия}.
\newblock Наука, Москва, 1986.

\bibitem{NovTai05R}
С.~П. Новиков and И.~А. Тайманов.
\newblock {\em Современные геометрические структуры и поля}.
\newblock МЦНМО, Москва, 2005.

\bibitem{Postni79R}
М.~М. Постников.
\newblock {\em Аналитическая геометрия}.
\newblock Наука, Москва, 1979.

\bibitem{Rashev67R}
П.~К. Рашевский.
\newblock {\em Риманова геометрия и тензорный анализ}.
\newblock Наука, Москва, 1967.
\newblock Третье изд.

\bibitem{RohFuk77R}
В.~А. Рохлин and Д.~Б. Фукс.
\newblock {\em Начальный курс топологии. Геометрические главы.}
\newblock Наука, Москва, 1977.

\bibitem{Gantma88R}
Ф.~Р. Гантмахер.
\newblock {\em Теория матриц}.
\newblock Наука, Москва, 4-е изд. edition, 1988.

\bibitem{Evgraf91R}
М.~А. Евграфов.
\newblock {\em Аналитические функции}.
\newblock Наука, Москва, Третье edition, 1991.

\bibitem{Engelk85R}
R.~Engelking.
\newblock {\em General Topology}.
\newblock Pa\'nstwowe Wydawnictwo Naukowe, Warszawa, second edition, 1977.
\newblock Перевод: Энгелькинг,~Р. {\it Общая топология}. М.: Мир, 1986. 751 с.

\bibitem{Tychon35}
A.~N. Tychonoff.
\newblock Ein {F}ixpunktsatz.
\newblock {\em Math.\ Ann.}, 11:767--776, 1935.

\bibitem{HorJoh86R}
R.~A. Horn and C.~R. Johnson.
\newblock {\em Matrix Analysis}.
\newblock Cambridge U.P., Cambridge, 1986.
\newblock Перевод: Хорн Р., Джонсон Ч. {\it Матричный анализ} М.: Мир. 1989.

\bibitem{Kosnio80R}
Czes Kosniowski.
\newblock {\em A First Course In Algebraic Topology}.
\newblock Cambridge University Press, Cambridge -- London -- New York, 1980.
\newblock Перевод: Коснёвски Ч. {\it Начальный курс алгебраической топологии}.
  М.: Мир, 1983.

\bibitem{Nimark58R}
М.~А. Наймарк.
\newblock {\em Линейные представления группы {Л}оренца}.
\newblock Физматгиз., Москва, 1958.

\bibitem{Einste05R}
A.~Einstein.
\newblock Zur {E}lektrodynamik der bewegter {K}\"orper.
\newblock {\em Ann.\ Phys.}, 17(2):891--921, 1905.

\bibitem{Cunnin09}
E.~Cunningham.
\newblock The principle of relativity in electrodynamics and an extension
  thereof.
\newblock {\em Proc.\ London Math.\ Soc.}, 8, Series 2:77--98, 1909.

\bibitem{Batema09}
H.~Bateman.
\newblock The transformation of the electrodynamical equations.
\newblock {\em Proc.\ London Math.\ Soc.}, 8, Series 2:223--264, 1909.

\bibitem{Dirac36R}
P.~A.~M. Dirac.
\newblock Wave equations in conformal space.
\newblock {\em Ann.\ Math.}, 37(2):429--442, 1936.
\newblock Перевод в сб. П.~А.~М.~Дирак ``Собрание научных трудов. Том 2.'' М.:
  Наука, 2003, сс. 485--497.

\bibitem{FraPal96}
E.~S. Fradkin and M.~Ya. Palchik.
\newblock {\em Conformal Quantum Field Theory in $D$-dimensions}.
\newblock Kluwer Academic Publishers, Dordrecht, 1996.

\bibitem{Naimar58R}
М.~А. Наймарк.
\newblock {\em Линейные представления группы Лоренца}.
\newblock Физматгиз, Москва, 1958.

\bibitem{BarRac77R}
A.~O. Barut and R.~R\c{a}czka.
\newblock {\em Theory of Group Representations and Applications}.
\newblock PWN -- Polish Scientific Publishers, Warszawa, 1977.
\newblock Перевод: А.~Барут, Р.~Рончка. {\em Теория представлений групп и ее
  приложения.} Т.1,2. М.: Мир, 1980.

\bibitem{MicMor87}
A.~A. Michelson and E.~W. Morley.
\newblock On the relative motion of the earth and the luminiferous ether.
\newblock {\em Am.\ J.\ Sci}, 34:333--345, 1887.
\newblock Воспроизведена в кн.: Relativity Theory: Its Origins and Impact on
  Modern Thought. Ed.~L.~Piarce Williams, John Wiley and Sons, 1968.

\bibitem{JaJaMuTo64}
T.~S. Jaseja, A.~Javan, J.~Murray, and C.H. Townes.
\newblock Test of special relativity or of the isotropy of space by use of
  infrared masers.
\newblock {\em Phys.\ Rev.}, 133:A1221--A1225, 1964.

\bibitem{Herglo11}
G.~Herglotz.
\newblock {\"Uber die Mechanik des deformierbaren K\"orpers vom Standpunkte der
  Relativit\"atstheorie}.
\newblock {\em Ann.\ der Physik}, 36(3):493--533, 1911.

\bibitem{Whitne36}
H.~Whitney.
\newblock Differentiable manifolds.
\newblock {\em Ann.\ of Math. (2)}, 37(5):645--680, 1936.

\bibitem{Munkre60}
J.~Munkres.
\newblock Obstructions to the smoothing of piecewise differentiable
  homeomorphisms.
\newblock {\em Ann.\ of Math. (2)}, 72:521--554, 1960.

\bibitem{Milnor56R}
J.~Milnor.
\newblock On manifolds homeomorphic to the 7-sphere.
\newblock {\em Ann.\ Math.}, 64:394--405, 1956.
\newblock Перевод: Дж.~Милнор. {Сб.\ переводов \em "Математика".} 1990, т.\ 1,
  \No.\ 3, сс.\ 394--405.

\bibitem{DonKro91}
S.~K. Donaldson and P.~B. Kronheimer.
\newblock {\em The Geometry of Four-Manifolds}.
\newblock Clarendon Press, Oxford, 1991.

\bibitem{Kervai60}
M.~Kervaire.
\newblock A manifold which does not admit any differentiable structure.
\newblock {\em Comment.\ Math.\ Helv.}, 34:257--270, 1960.

\bibitem{Aubin01}
T.~Aubin.
\newblock {\em A Course in Differential Geometry}.
\newblock AMS, Providence, Rhode Island, 2001.

\bibitem{Warner83R}
F.~W. Warner.
\newblock {\em Foundations of Differentiable Manifolds and Lie Groups}.
\newblock Springer--Verlag, Berlin -- Heidelberg, 1983.
\newblock Перевод: Ф.~Уорнер. {\it Основы теории гладких многообразий и групп
  Ли.} М.: Мир, 1987.

\bibitem{ChChLa00}
S.~S. Chern, W.~H. Chen, and K.~S. Lam.
\newblock {\em Lectures on Differential Geometry}.
\newblock World Scientific, Singapore, 2000.

\bibitem{MilSta74R}
J.~W. Milnor and J.~D. Stasheff.
\newblock {\em Characteristic Classes}.
\newblock Princeton U.\ Press, Princeton, New Jersey, 1974.
\newblock Перевод: Дж. Милнор, Дж. Сташеф {\it Характеристические классы}. М.:
  Мир, 1979.

\bibitem{Chern67A}
S.~S. Chern.
\newblock Curves and surfaces in euclidean space.
\newblock In S.~S. Chern, editor, {\em "Global Geometry and Analysis". {MAA}
  studies in Mathematics. Vol.~4.}, pages 16--56, Inglewood Cliffs, 1967. The
  Mathematical Association of America, Prentice Hall, Inc.

\bibitem{Isham99}
C.~J. Isham.
\newblock {\em Modern Differential Geometry}.
\newblock World Scientific, Singapore, 1999.

\bibitem{KobNom6369R}
S.~Kobayashi and K.~Nomizu.
\newblock {\em Foundations of differential geometry}, volume 1, 2.
\newblock Interscience publishers, New York -- London, 1963.
\newblock Перевод: Кобаяси Ш., Номидзу К. {\it Основы дифференциальной
  геометрии}. Том 1, 2. -- М.: Наука, 1981.

\bibitem{Schwar67R}
L.~Schwartz.
\newblock {\em Analyse Math\'ematique. Vols. I, II}.
\newblock Hermann, Paris, 1967.
\newblock Перевод: Л.~Шварц. {\it Анализ.} Том 1,2. М.: Мир,1972.

\bibitem{Cheval46R}
C.~Chevalley.
\newblock {\em Theory of Lie Groups}.
\newblock Princeton University Press, Princeton, 1946.
\newblock К.~Шевалле. {\it Теория групп Ли. Т. 1. М.: ИЛ, 1948.}

\bibitem{Godbil69R}
C.~Godbillon.
\newblock {\em G\'eom\'etrie Diff\'eretielle et M\'ecanique Analytique}.
\newblock Hermann, Paris, 1969.
\newblock Перевод: Годбийон~К. {\it Дифференциальная геометрия и аналитическая
  механика}. М.: Мир, 1973. 188 с.

\bibitem{deRham46}
G.~de~Rham.
\newblock Sur la th\'eorie des formes diff\'erentielles harmoniques.
\newblock {\em Ann.\ Univ.\ Grenoble}, 22:132--152, 1946.

\bibitem{deRham55R}
G.~de~Rham.
\newblock {\em Vari'et\'es Diff\'erentiables}.
\newblock Hermann, Paris, 1955.
\newblock Перевод: де Рам~Ж. {\it Дифференцируемые многообразия}. М.: URSS,
  2006. 2-е изд.

\bibitem{Wolf72R}
J.~A. Wolf.
\newblock {\em Spaces of constant curvature}.
\newblock University of California, Berkley, California, 1972.
\newblock Перевод: Вольф~Дж. {\it Пространства постоянной кривизны}. М.: Наука,
  1982. 480 с.

\bibitem{Steenr51}
N.~E. Steenrod.
\newblock {\em The Topology of Fiber Bundles}.
\newblock Princeton University Press, Princeton, New Jersey, 1951.

\bibitem{Weyl18AR}
H.~Weyl.
\newblock Gravitation und {E}lektrizit\"at.
\newblock {\em Sitz.\ Preuss.\ Akad.\ Wiss.}, page S.\ 465, 1918.
\newblock Перевод в сб.\ {``Альберт Эйнштейн и теория гравитации''. М.: Мир,
  1979, с.~513.}

\bibitem{LeviCi17A}
T.~Levi-Civita.
\newblock Nozione di parallelismo una variet\'a qualungue e consequente
  specificazione geometrica della curvatura {R}iemanniana.
\newblock {\em Rendicinti di Palermo}, 42:173--205, 1917.

\bibitem{Katana05R}
М.~О. Катанаев.
\newblock Геометрическая теория дефектов.
\newblock {\em УФН}, 175(7):705--733, 2005.

\bibitem{Gleaso52}
A.~M. Gleason.
\newblock Groups without small subgroups.
\newblock {\em Ann.\ Math.}, 56(2):193--212, 1952.

\bibitem{MonZip52}
D.~Montgomery and L.~Zippin.
\newblock Small subgroups of finite-dimensional groups.
\newblock {\em Ann.\ Math.}, 56(2):213--241, 1952.

\bibitem{Pontry84R}
Понтрягин~Л. С.
\newblock {\em Непрерывные группы}.
\newblock Наука, Москва, 1984.
\newblock Четвертое изд.

\bibitem{GotGro78R}
M.~Goto and Grosshans F.
\newblock {\em Semisimple Lie Algebras}.
\newblock Marcel Dekker, New York, 1978.

\bibitem{Helgas62}
S.~Helgason.
\newblock {\em Differential Geometry and Symmetric Spaces}.
\newblock Academic Press, New York, 1962.

\bibitem{Helgas62R}
S.~Helgason.
\newblock {\em Differential Geometry and Symmetric Spaces}.
\newblock Academic Press, New York, 1962.
\newblock Перевод: Хелгасон С. {\it Дифференциальная геометрия и симметрические
  пространства}. М.: Мир, 1983.

\bibitem{Helgas01R}
S.~Helgason.
\newblock {\em Differential Geometry, Lie Groups, and Symmetric Spaces}.
\newblock American Mathematical Society, Rhode Island, 2001.
\newblock Перевод: Хелгасон С. {\it Дифференциальная геометрия, группы {Ли} и
  симметрические пространства}. М.: Факториал Пресс, 2005.

\bibitem{Postni82R}
М.~М. Постников.
\newblock {\em Группы и алгебры Ли}.
\newblock Наука, Москва, 1982.

\bibitem{Ado47R}
И.~Д. Адо.
\newblock Представление алгебр {Л}и матрицами.
\newblock {\em Усп.\ Матем.\ наук}, 2(6):159, 1947.

\bibitem{Cartan29}
E.~Cartan.
\newblock Groupes simples clos et ouverts et g'eometrie riemanniene.
\newblock {\em J.\ Math.\ pure appl.}, 8:1--33, 1929.

\bibitem{HauSch68}
M.~Hausner and J.~Schwartz.
\newblock {\em Lie Groups; Lie algebras}.
\newblock Gordon \& Breach, New York--London--Paris, 1968.

\bibitem{Adams62}
J.~F. Adams.
\newblock Vector fields on spheres.
\newblock {\em Ann.\ Math.}, 75:603--632, 1962.

\bibitem{Spanie66R}
E.~H. Spanier.
\newblock {\em Algebraic Topology}.
\newblock McGraw--Hill Book Company, New York -- London, 1966.
\newblock Перевод: Э.~Спеньер. {\it Алгебраическая топология.} М.: Мир, 1971.

\bibitem{Narasi71R}
R.~Narasimhan.
\newblock {\em Analysis on Real and Complex Manifolds}.
\newblock Masson and North--Holland, Paris and Amsterdam, 1971.
\newblock Перевод: Р. Нарасимхан. {\it Анализ на действительных и комплексных
  многообразиях}. М.: Мир, 19??.

\bibitem{Bolibr00R}
А.~А. Болибрух.
\newblock {\em Фуксовы дифференциальные уравнения и голоморфные расслоения}.
\newblock МЦНМО, Москва, 2000.

\bibitem{Godeme58R}
R.~Godement.
\newblock {\em Topologie Algebrique et Theorie des Faisceaux}.
\newblock Hermann, Paris, 1958.
\newblock Перевод: Годеман~Р. {\it Алгебраическая топология и теория пучков}.
  М.: ИЛ, 1961.

\bibitem{Iwasaw49}
K.~Iwasawa.
\newblock On some types of topological groups.
\newblock {\em Ann.\ Math.}, 50:507--558, 1949.

\bibitem{NomOze62}
K.~Nomizu and H.~Ozeki.
\newblock On the degree of differentiability of curves used in the definition
  of the holonomy groups.
\newblock {\em Bull.\ Amer.\ Math.\ Soc.}, 68:74--75, 1962.

\bibitem{AmbSin53}
W.~Ambrose and I.~M. Singer.
\newblock A theorem on holonomy.
\newblock {\em Trans.\ Amer.\ Math.\ Soc.}, 75:428--443, 1953.

\bibitem{HanOze56}
J.~Hano and H.~Ozeki.
\newblock On the holonomy groups of linear connexions.
\newblock {\em Nagoya Math.\ J.}, 10:97--100, 1956.

\bibitem{Nomizu56}
K.~Nomizu.
\newblock Un th\'eor\`eme sur les groupes d'holonomie.
\newblock {\em Nagoya Math.\ J.}, 10:101--103, 1956.

\bibitem{Ozeki56}
H.~Ozeki.
\newblock Infinitesimal holonomy groups of bundle connections.
\newblock {\em Nagoya Math.\ J.}, 10:105--123, 1956.

\bibitem{Wang58}
H.~C. Wang.
\newblock On invariant connections over a principal fibre bundle.
\newblock {\em Nagoya Math.\ J.}, 13:1--19, 1958.

\bibitem{Berry84}
M.~V. Berry.
\newblock Quantal phase factors accompanying adiabatic changes.
\newblock {\em Proc.\ Roy.\ Soc.\ London}, A392(1802):45--57, 1984.

\bibitem{AhaBoh59}
Y.~Aharonov and D.~Bohm.
\newblock Significance of electromagnetic potentials in the quantum theory.
\newblock {\em Phys.\ Rev.}, 115(3):485--491, 1959.

\bibitem{WilZee84}
F.~Wilczek and A.~Zee.
\newblock Appearance of gauge structure in simple dynamical systems.
\newblock {\em Phys.\ Rev.\ Lett.}, 52:2111, 1984.

\bibitem{BorFoc28}
M.~Born and V.~Fock.
\newblock Beweis des {A}diabatensatzes.
\newblock {\em Z.\ Phys.}, 51:165--180, 1928.
\newblock English translation in {\it ``V.A. Fock -- Selected Works: Quantum
  Mechanics and Quantum Field Theory''} ed.\ by L.D.~Faddeev, L.A.~Khalfin,
  I.V.~Komarov. Chapman \& Hall/CRC, Boca Raton, 2004.

\bibitem{Schrod26A}
E.~Schr\"odinger.
\newblock {Quantizierung als Eigenwertproblem (Erste Mitteilung)}.
\newblock {\em Ann.\ Phys.\ Leipzig}, 79(4):361--376, 1926.

\bibitem{Schrod26B}
E.~Schr\"odinger.
\newblock {Quantizierung als Eigenwertproblem (Zweite Mitteilung)}.
\newblock {\em Ann.\ Phys.\ Leipzig}, 79(6):489--527, 1926.

\bibitem{VlaVol84CR}
В.~С. Владимиров and И.~В. Волович.
\newblock Локальные и нелокальные токи для нелинейных уравнений.
\newblock {\em ТМФ}, 61(3):3--29, 1984.

\bibitem{VlaVol85R}
В.~С. Владимиров and И.~В. Волович.
\newblock Законы сохранения для нелинейных уравнений.
\newblock {\em УМН}, 40(4(244)):17--26, 1985.

\bibitem{Messia62R}
A.~Messiah.
\newblock {\em Quantum Mechanics}, volume~2.
\newblock North Holland, Amsterdam, 1962.
\newblock Перевод: А. Мессиа. {\it Квантовая механика}. Т.2. М.: Наука, 1979.

\bibitem{Fock76R}
В.~А. Фок.
\newblock {\em Начала квантовой механики.}
\newblock Наука, Москва, 2-е изд. edition, 1976.

\bibitem{BitDub87}
T.~Bitter and D.~Dubbers.
\newblock Manifestation of berry's topological phase in neutron spin rotation.
\newblock {\em Phys.\ Rev.\ Lett.}, 59:251--254, 1987.

\bibitem{WerBri60}
F.~G. Werner and D.~R. Brill.
\newblock Significance of electromagnetic potentials on the quantum theory in
  the interpretation of electron interferometer fringe observations.
\newblock {\em Phys.\ Rev.\ Lett.}, 4(7):344--347, 1960.

\bibitem{Chambe60}
R.~G. Chambers.
\newblock Shift of an electron interference pattern by enclosed magnetic flux.
\newblock {\em Phys.\ Rev.\ Lett.}, 5(1):3--5, 1960.

\bibitem{BoHaWoGr60}
H.~Boersch, H.~Hamisch, D.~Wohlleben, and K.~Grohmann.
\newblock Weissche bereiche als bi-prisme f\"ur elektroneninterferenzen.
\newblock {\em Z.\ Phys.}, 159:397--404, 1960.

\bibitem{Killin92}
W.~Killing.
\newblock {\"Uber die Grunglagen der Geometrie}.
\newblock {\em J.\ Reine Angew.\ Math.}, 109:121--186, 1892.

\bibitem{Bianch18}
L.~Bianchi.
\newblock {\em Lezioni sulla teoria dei gruppi continui finiti di
  transformazioni}.
\newblock Spoerri, Pisa, 1918.

\bibitem{Fock61R}
В.~А. Фок.
\newblock {\em Теория пространства, времени и тяготения. Издание второе.}
\newblock Физматгиз, Москва, 1961.

\bibitem{DeDond21}
Th. De~Donder.
\newblock {\em La Gravifique Einsteinienne}.
\newblock Gauthier-Villars \& cie, Paris, 1921.

\bibitem{Lanczo22}
K.~Lanczos.
\newblock Ein vereinfachendes {K}oordinatensystem f\"ur die {E}insteinschen
  {G}ravitationsgleichungen.
\newblock {\em Phys.\ Zs.}, 23:537--539, 1922.

\bibitem{Fock39R}
В.~А. Фок.
\newblock О движении конечных масс в общей теории относительности.
\newblock {\em ЖЭТФ}, 9(4):375--410, 1939.

\bibitem{Fermi22}
E.~Fermi.
\newblock Sopra i fenomeni che avvengono in vicinanza di una linea oraria.
\newblock {\em Atti Acad.\ Naz.\ Lincei Rend.\ Cl.\ Sci.\ Fiz.\ Mat.\ Nat.},
  31:Three parts, pp. 21--23, 51--52, 101--103, 1922.

\bibitem{Whiteh32}
J.~H.~C. Whitehead.
\newblock Convex regions in the geometry of paths.
\newblock {\em Quart.\ J.\ Math.\ Oxford Ser.}, 3:33--42, 1932.

\bibitem{MyeSte39}
S.~Myers and N.~Steenrod.
\newblock The group of isometries of a {R}iemannian manifold.
\newblock {\em Ann.\ Math.}, 40:400--416, 1939.

\bibitem{HopRin31}
H.~Hopf and W.~Rinow.
\newblock {Uber den Begriff des vollst\"andigen differentialgeometrischen
  Fl\"ache}.
\newblock {\em Comment.\ Math.\ Helv.}, 3(2):209--225, 1931.

\bibitem{Fomenk88R}
А.~Т. Фоменко.
\newblock {\em Симплектическая геометрия}.
\newblock Издательство МГУ, Москва, 1988.

\bibitem{LanLif88RA}
Л.~Д. Ландау and Е.~М. Лифшиц.
\newblock {\em Механика}.
\newblock Наука, Москва, четвертое edition, 1988.

\bibitem{Olver86R}
P.~J. Olver.
\newblock {\em Application of Lie Groups to Differential Equations}.
\newblock Springer--Verlag, Berlin -- Heidelberg, 1986.
\newblock Перевод: Олвер П. {\it Приложения групп Ли к дифференциальным
  уравнениям.} М.: Мир, 1989. 637 с.

\bibitem{Elsgol98R}
Л.~Э. Эльсгольц.
\newblock {\em Дифференциальные уравнения и вариационное исчисление}.
\newblock УРСС, Москва, 1998.

\bibitem{Nether18R}
E.~Noether.
\newblock Invariante variationsprobleme.
\newblock {\em Nachr. D. K\"onig. Gesellsch. D. Wiss. Zu G\"ottingen,
  Math-phys. Klasse}, pages 235--257, 1918.
\newblock Перевод в кн.\ {\it Вариационные принципы в механике}. М.: Физматгиз,
  1959, с. 611.

\bibitem{DeWitt65R}
B.~DeWitt.
\newblock {\em Dynamical Theory of Groups and Fields}.
\newblock Gordon and Breach, New York -- London, 1965.
\newblock Перевод: ДеВитт~Б.~С. {\it Динамическая теория групп и полей}. М.:
  Наука, 1987. 287 с.

\bibitem{Colema75}
S.~Coleman.
\newblock Classical lumps and their quantum descendants. lectures at the 1975
  international school of subnuclear physics "ettore majorana" (erice).
\newblock In A.~Zichichi, editor, {\em New Phenomena in Subnuclear Physics},
  pages 297--421, New York, 1977. Plenum Press.

\bibitem{Faddee77R}
Л.~Д. Фаддеев.
\newblock В поисках многомерных солитонов.
\newblock In Д2-9788, editor, {\em Нелокальные, нелинейные и неренормируемые
  теории поля}, pages 207--223, Дубна, 1977. ОИЯИ.

\bibitem{Palais79}
R.~Palais.
\newblock The principle of symmetric criticality.
\newblock {\em Comm.\ Math.\ Phys.}, 69(1):19--30, 1979.

\bibitem{LadKap83R}
О.~А. Ладыженская and Л.~В. Капитанский.
\newblock О принципе Коулмена нахождения стационарных точек инвариантных
  функционалов.
\newblock {\em Зап. науч. сем. ЛОМИ}, 127(15):84--102, 1983.

\bibitem{SchSim89}
R.~Schmid and L.~Simoni.
\newblock On infinite-dimensional variational principles with constraints.
\newblock {\em J.\ Math.\ Phys.}, 30(5):1171--1176, 1989.

\bibitem{MicRad68}
L.~Michel and L.~Radicati.
\newblock On the dynamical breaking of $su(3)$.
\newblock In B.~Krsunoglu A.~Pearlmutter, C. A.~Hurst, editor, {\em Proc.\ 5th
  Coral Gables Conf. Symmetry Principle at High Energy}, New York, 1968.
  Benjamin.

\bibitem{Arnold89R}
В.~И. Арнольд.
\newblock {\em Математические методы классической механики}.
\newblock Наука, Москва, 1989.
\newblock Третье изд. 472 с.

\bibitem{HwaChu47}
Lee Hwa-Chung.
\newblock Invariants of hamilton systems and applications to to the theory of
  canonical transformations.
\newblock {\em Proc.\ Roy.\ Soc.\ Edinburg}, A62:237--247, 1947.

\bibitem{Gantma01R}
Ф.~Р. Гантмахер.
\newblock {\em Лекции по аналитической механике}.
\newblock Физматлит, Москва, 3-е изд. edition, 2001.

\bibitem{Dirac58AR}
P.~A.~M. Dirac.
\newblock Generalized {H}amiltonian dynamics.
\newblock {\em Proc.\ Roy.\ Soc.\ London}, A246:326--332, 1958.
\newblock Перевод в сб. "Новейшие проблемы гравитации" под редакцией
  Д.~Иваненко. М.: ИЛ, 1961.

\bibitem{Dirac58BR}
P.~A.~M. Dirac.
\newblock The theory of gravitation in hamiltonian form.
\newblock {\em Proc.\ Roy.\ Soc.\ London}, A246(1246):333--343, 1958.
\newblock Перевод в сб. ``Новейшие проблемы гравитации'' под редакцией
  Д.~Иваненко. М.: ИЛ, 1961.

\bibitem{Dirac64R}
P.~A.~M. Dirac.
\newblock {\em Lectures on Quantum Mechanics}.
\newblock Belfer, Yeshiva University, New York, 1964.
\newblock Перевод в кн.: Дирак~П.~А.~М. {\it Принципы квантовой механики.} М.:
  Наука, 1979. С.~408--475.

\bibitem{GitTyu90}
D.~M. Gitman and I.~V. Tyutin.
\newblock {\em Quantization of Fields with Constraints}.
\newblock Springer--Verlag, Berlin -- Heidelberg, 1990.

\bibitem{HenTei92}
M.~Henneaux and C.~Teitelboim.
\newblock {\em Quantization of Gauge Systems}.
\newblock Princeton University Press, Princeton, 1992.

\bibitem{Faddee69R}
Л.~Д. Фаддеев.
\newblock Интеграл Фейнмана для сингулярных лагранжианов.
\newblock {\em ТМФ}, 1(1):3--18, 1969.

\bibitem{FraVil77}
E.~S. Fradkin and G.~A. Vilkovisky.
\newblock Quantization of relativistic systems with constraints -- equivalence
  of canonical and covariant formalism in quantum theory of gravitational
  field.
\newblock {\em CERN Preprint TH 2332}, 1977.

\bibitem{Einste15}
A.~Einstein.
\newblock Die {F}eldgleichungen der {G}ravitation.
\newblock {\em Sitzungsber.\ preuss.\ Akad.\ Wiss.}, 48(2):844--847, 1915.

\bibitem{Hilber15R}
D.~Hilbert.
\newblock Die grundlagen der physik.
\newblock {\em Nachrichten K. Gesellschaft Wiss. G\"ottingen, Math.-phys.},
  Heft 3:395, Klasse 1915.
\newblock Перевод в сб.: Альберт Эйнштейн и теория гравитации. М.: Мир, 1979.
  С.~133--145.

\bibitem{Einste16R}
A.~Einstein.
\newblock Die-{G}rundlage der allgemeinen {R}elativit\"tstheorie.
\newblock {\em Ann.\ d.\ Phys.}, 49:769--822, 1916.
\newblock Перевод в сб.: Альберт Эйнштейн и теория гравитации. М.: Мир, 1979.
  С.~146--196.

\bibitem{Fierz56}
von~M. Fierz.
\newblock \"uber die physikalische deutung der erweiterten gravitationstheorie
  {P.~J}ordans.
\newblock {\em Helv.\ Phys.\ Acta}, 29:128--134, 1956.

\bibitem{Jordan59}
P.~Jordan.
\newblock Zum gegenw\"artigen stand der diracschen kosmologischen hypothesen.
\newblock {\em Z.\ Phys.}, 157:112--121, 1959.

\bibitem{BraDic61}
C.~Brans and R.~H. Dicke.
\newblock Mach's prinsiple and a relativistic theory of gravitation.
\newblock {\em Phys.\ Rev.}, 124(3):925--935, 1961.

\bibitem{Dirac38}
P.~A.~M. Dirac.
\newblock A new basis for cosmology.
\newblock {\em Proc.\ Roy.\ Soc.\ London}, A165(921):199--208, 1938.

\bibitem{Peres63}
A.~Peres.
\newblock Polynomial expansion of gravitational lagrangian.
\newblock {\em Nuovo Cimento}, 28(4):865--867, 1963.

\bibitem{Katana06A}
M.~O. Katanaev.
\newblock Polynomial form of the {H}ilbert--{E}instein action.
\newblock {\em Gen.\ Rel.\ Grav.}, 38:1233--1240, 2006.
\newblock gr-qc/0507026.

\bibitem{Penros65}
R.~Penrose.
\newblock A remarkable property of plane waves in general relativity.
\newblock {\em Rev.\ Mod.\ Phys.}, 37(1):215--220, 1965.

\bibitem{CheTag68}
N.~A. Chernikov and E.~A. Tagirov.
\newblock Quantum theory of scalar field in de {S}itter space-time.
\newblock {\em Ann.\ Inst.\ Henri Poincar\'e}, A9(2):109--141, 1968.

\bibitem{Ibragi68R}
Н.~Х. Ибрагимов.
\newblock К групповой классификации дифференциальных уравнений второго порядка.
\newblock {\em ДАН}, 183(2):274--277, 1968.

\bibitem{CaCoJa70}
C.~G. Callan, S.~Coleman, and R.~Jackiw.
\newblock A new improved energy--momentum tensor.
\newblock {\em Ann.\ Phys.}, 59:42--73, 1970.

\bibitem{InfPle60R}
L.~Infeld and J.~Pleba\'nski.
\newblock {\em Motion and Relativity}.
\newblock Pergamon Press, Oxford -- London, second edition, 1962.
\newblock Перевод: Л.~Инфельд, Е.~Плебаньский. {\it Движение и релятивизм.} М.:
  ИЛ. 1962.

\bibitem{FiePau39}
M.~Fierz and W.~Pauli.
\newblock Relativistic wave equations for particles of arbitrary spin in an
  electromagnetic field.
\newblock {\em Proc.\ Roy.\ Soc.\ London}, A173:211--232, 1939.

\bibitem{LanLif88R}
Л.~Д. Ландау and Е.~М. Лифшиц.
\newblock {\em Теория поля}.
\newblock Наука, Москва, седьмое edition, 1988.

\bibitem{Unruh81}
W.~G. Unruh.
\newblock Experimental black-hole evaporation?
\newblock {\em Phys.\ Rev.\ Lett.}, 46:1351--1353, 1981.

\bibitem{Unruh95}
W.~G. Unruh.
\newblock Sonic analogue of black holes and the effect of high frequencies on
  black holes evaporation.
\newblock {\em Phys.\ Rev.}, D51(6):2827--2838, 1995.

\bibitem{Visser98}
M.~Visser.
\newblock Acoustic black holes: Horizons, ergospheres, and {H}awking radiation.
\newblock {\em Class.\ Quantum Grav.}, 15:1767--1791, 1998.

\bibitem{LanLif86R}
Л.~Д. Ландау and Е.~М. Лифшиц.
\newblock {\em Гидродинамика}.
\newblock Наука, Москва, третье edition, 1986.

\bibitem{Petrov66R}
А.~З. Петров.
\newblock {\em Новые методы в общей теории относительности}.
\newblock Наука, Москва, 1966.

\bibitem{ArDeMi62}
R.~Arnowitt, S.~Deser, and S.~W. Misner.
\newblock The dynamics of general general relativity.
\newblock In L.~Witten, editor, {\em Gravitation: an introduction to current
  research}, New York -- London, 1962. John Wiley \& Sons, Inc.
\newblock gr-qc/0405109.

\bibitem{DeWitt67A}
B.~S. DeWitt.
\newblock Quantum theory of gravity. {I}. {T}he canonical theory.
\newblock {\em Phys.\ Rev.}, 160(5):1113--1148, 1967.

\bibitem{GitTyu86R}
Д.~М. Гитман and И.~В. Тютин.
\newblock {\em Каноническое квантование полей со связями}.
\newblock Наука, Москва, 1986.

\bibitem{Weyl18BR}
H.~Weyl.
\newblock {\em Raum -- Zeit -- Materie}.
\newblock Springer, Berlin, 1918.
\newblock Перевод: Г.~Вейль. {\it Пространство, время, материя}. М.: Янус,
  1996.

\bibitem{Dirac75R}
P.~A.~M. Dirac.
\newblock {\em General Theory of Relativity}.
\newblock John Wiley \& Sons, Inc., New York -- London, 1975.
\newblock Перевод: {\it Общая теория относительности}. М.: Атомиздат, 1978.

\bibitem{Pauli58R}
W.~Pauli.
\newblock {\em Theory of Relativity}.
\newblock Pergamon Press, New York, 1958.
\newblock Перевод: Паули В. {\it Теория относительности}. М.: Наука, 1991.

\bibitem{Schrod50R}
E.~Schr\"odinger.
\newblock {\em Space-time Structure}.
\newblock Cambridge U.P., Cambridge, 1950.
\newblock Перевод: Шредингер Э. {\it Пространственно-временн\'ая структура
  Вселенной}. М.: Наука, 1986.

\bibitem{Gordon26}
W.~Gordon.
\newblock Der {C}omptoneffekt nach der {S}chr\"odingerschen {T}heorie.
\newblock {\em Zs.\ f.\ Phys.}, 40(1,2):117--133, 1926.

\bibitem{Fock26A}
V.~A. Fock.
\newblock Zur {S}chr\"odingerschen {W}ellenmechanik.
\newblock {\em Zs.\ f.\ Phys.}, 38(3):242--250, 1926.

\bibitem{Fock26B}
V.~A. Fock.
\newblock \"uber die invariante {F}orm der {W}ellen- und der {B}ewegungs-
  gleichungen f\"ur einen geladenen {M}assenpunkt.
\newblock {\em Zs.\ f.\ Phys.}, 39(2,3):226--232, 1926.

\bibitem{Klein27}
O.~Klein.
\newblock Elektrodynamik und {W}ellenmechanik vom {S}tandpunkt des
  {K}orrespondenzprinzips.
\newblock {\em Zs.\ f.\ Phys.}, 41(10):407--442, 1927.

\bibitem{Goldst61}
J.~Goldstone.
\newblock Field theories with ``superconductor'' solutions.
\newblock {\em Nuovo Cim.}, 19(1):154--164, 1961.

\bibitem{Wigner59R}
E.~P. Wigner.
\newblock {\em Group Theory and its application to the quantum mechanics of
  atomic spectra}.
\newblock Academic Press, New York -- London, 1959.
\newblock Перевод: Вейль Е. {\it Теория групп и ее приложения к
  квантовомеханической теории атомных спектров}. М.: Иностраннная литература,
  1961.

\bibitem{vanBla91A}
J.~van Bladel.
\newblock Lorenz or {L}orentz ?
\newblock {\em IEEE Antennas and Propagation Magazin}, 33(2):69, 1991.

\bibitem{EngBro64}
F.~Englert and R~Brout.
\newblock Broken symmetry and the mass of gauge vector mesons.
\newblock {\em Phys.\ Rev.\ Lett.}, 13(9):321--323, 1964.

\bibitem{Higgs64}
P.~W. Higgs.
\newblock Broken symmetries, massless particles and gauge fields.
\newblock {\em Phys.\ Lett.}, 12(2):132--133, 1964.

\bibitem{Proca36A}
A.~Proca.
\newblock Sur la th\'eorie ondulatoire des \'electrons positifs et n\'egatifs.
\newblock {\em J.\ Physique}, 8:347--353, 1936.

\bibitem{Proca36B}
A.~Proca.
\newblock Sur la th\'eorie du positon.
\newblock {\em C.\ R.\ Acad.\ Sci.\ Paris}, 202:1366, 1936.

\bibitem{TihSam77R}
А.~Н. Тихонов and А.~А. Самарский.
\newblock {\em Уравнения математической физики}.
\newblock Наука, Москва, пятое edition, 1977.

\bibitem{VlaZha00R}
В.~С. Владимиров and В.~В. Жаринов.
\newblock {\em Уравнения математической физики}.
\newblock Физ-мат литература, Лаборатория базовых знаний, Москва, 2000.

\bibitem{Milnor72}
T.~K. Milnor.
\newblock Efimov's theorem about complete immersed surfaces of negative
  curvature.
\newblock {\em Advances in Math.}, 8:474--543, 1972.

\bibitem{Kuiper55}
N.~H. Kuiper.
\newblock On {$\CC^1$} isometric embeddings. {II}.
\newblock {\em Nederl.\ Akad.\ Wetensch.\ Proc.\ Ser.\ A}, 58:683--689, 1955.

\bibitem{Katana93A}
M.~O. Katanaev.
\newblock All universal coverings of two-dimensional gravity with torsion.
\newblock {\em J.\ Math.\ Phys.}, 34(2):700--736, 1993.

\bibitem{Liouvi53}
L.~Liouville.
\newblock Sur l'\'equation aux diff\'erences partielles
  $\frac{\pl^2\ln\lambda}{\pl u\pl v}\pm\lambda a^2=0$.
\newblock {\em J.\ Math.\ Pures Appl.}, 18:71--72, 1853.

\bibitem{Kruska60}
M.~D. Kruskal.
\newblock Maximal extension of {S}chwarzschild metric.
\newblock {\em Phys.\ Rev.}, 119(5):1743--1745, 1960.

\bibitem{Szeker60}
G.~Szekeres.
\newblock On the singularities of a riemannian manifold.
\newblock {\em Publ.\ Mat.\ Debrecen}, 7(1--4):285--301, 1960.

\bibitem{Schwar16R}
K.~Schwarzschild.
\newblock {\"Uber das Gravitationsfeld eines Massenpunktes nach der
  Einsteinschen Theorie}.
\newblock {\em Sitzungsber.\ Akad.\ Wiss.\ Berlin}, pages 189--196, 1916.
\newblock Перевод в сб.: Альберт Эйнштейн и теория гравитации. М.: Мир, 1979.
  С.~199--207.

\bibitem{Droste15}
J.~Droste.
\newblock Over het veld van een enkel centrum in einstein's theorie der
  zwaarte-kracht.
\newblock {\em Proc.\ Kon.\ Ned.\ Akad.\ Wetensch.\ Amsterdam}, 23:968--981,
  1914--1915.
\newblock English translation: ``On the field of a single centre in Einstein's
  theory of gravitation''. Proc.\ Acad.\ Sci.\ Amsterdam 17(1915)998-- 1011.

\bibitem{Droste16}
J.~Droste.
\newblock Het veld van twee bolvormige restunde centra in einstein's theorie
  der zwaartekracht.
\newblock {\em Proc.\ Kon.\ Ned.\ Akad.\ Wetensch.\ Amsterdam}, 24:749--757,
  1915--1916.
\newblock English translation: ``On the field of two Spherical Fixed Centres in
  Einstein's Theory of Gravitation''. Proc.\ Acad.\ Sci.\ Amsterdam
  18(1916)760--769.

\bibitem{Droste17}
J.~Droste.
\newblock Het veld van een enkel centrum in einstein's theorie der
  zwaartekracht, en de beweging van een stoffelijk punt in dat veld.
\newblock {\em Proc.\ Kon.\ Ned.\ Akad.\ Wet.\ Amsterdam}, 25:163--180,
  1916--1917.
\newblock English translation: ``The Field of a Single Centre in Einstein's
  Theory of Gravitation, and the Motion of a Particle in That Field''. Proc.\
  Acad.\ Sci.\ Amsterdam 19(1917)197--215. Reprinted, with historical comments,
  in {\it Gen.\ Rel.\ Grav.} 34(2002)1545.

\bibitem{KatVol86}
M.~O. Katanaev and I.~V. Volovich.
\newblock String model with dynamical geometry and torsion.
\newblock {\em Phys.\ Lett.}, 175B(4):413--416, 1986.

\bibitem{KloStr96C}
T.~Kl\"osch and T.~Strobl.
\newblock Classical and quantum gravity in $1+1$ dimensions: {II}. {T}he
  universal coverings.
\newblock {\em Class.\ Quantum Grav.}, 13:2395--2421, 1996.

\bibitem{KaKuLi96}
M.~O. Katanaev, W.~Kummer, and H.~Liebl.
\newblock Geometric interpretation and classification of global solutions in
  generalized dilaton gravity.
\newblock {\em Phys.\ Rev.\ D}, 53(10):5609--5618, 1996.

\bibitem{KaKuLi97}
M.~O. Katanaev, W.~Kummer, and H.~Liebl.
\newblock On the completeness of the black hole singularity in 2d dilaton
  theories.
\newblock {\em Nucl.\ Phys.}, B486:353--370, 1997.

\bibitem{Katana09}
M.~O. Katanaev.
\newblock Global solutions in gravity: Euclidean signature.
\newblock In D.~Vassilevich D.~Grumiller, A.~Rebhan, editor, {\em In
  ``Fundumental Interactions. A Memorial Volume for Wolfgang Kummer’’}, pages
  249--266, Singapore, 2010. World Scientific.
\newblock gr-qc/0808.1559.

\bibitem{Carter73}
B.~Carter.
\newblock Black hole equilibrium states.
\newblock In C.~DeWitt and B.~C. DeWitt, editors, {\em Black Holes}, pages
  58--214, New York, 1973. Gordon \& Breach.

\bibitem{Reissn16}
H.~Reissner.
\newblock {\"Uber die Eigengravitation des elektrischen Feldes nach der
  Einsteinschen Theorie}.
\newblock {\em Ann.\ Physik (Leipzig)}, 50:106--120, 1916.

\bibitem{Nordst18}
G~Nordstr\"om.
\newblock On the energy of the gravitational field in einstein's theory.
\newblock {\em Proc.\ Kon.\ Ned.\ Akad.\ Wet.}, 20:1238--1245, 1918.

\bibitem{Edding24A}
A.~S. Eddington.
\newblock A comparison of {W}hitehead's and {E}instein's formulae.
\newblock {\em Nature}, 113:192, 1924.

\bibitem{Finkel58}
D.~Finkelstein.
\newblock Past-future asymmetry of the gravitational field of a point particle.
\newblock {\em Phys.\ Rev.}, 110(4):965--967, 1958.

\bibitem{KaKlKu99}
M.~O. Katanaev, T.~Kl\"osch, and W.~Kummer.
\newblock Global properties of warped solutions in general relativity.
\newblock {\em Ann.\ Phys.}, 276:191--222, 1999.

\bibitem{Petrov69}
A.~Z. Petrov.
\newblock {\em Einstein Spaces}.
\newblock Pergamon Press, Oxford -- London, 1969.

\bibitem{KrStMaHe80}
D.~Kramer, H.~Stephani, M.~MacCallum, and E.~Herlt.
\newblock {\em Exact Solutions of the Einsteins Field Equations}.
\newblock Deutscher Verlag der Wissenschaften, Berlin, 1980.

\bibitem{Birkho23}
G.~D. Birkhoff.
\newblock {\em Relativity and modern physics}.
\newblock Cambridge, Harvard University Press, Cambridge, 1923.

\bibitem{Jebsen21}
J.~T. Jebsen.
\newblock Uber die allgemeinen kugelsymmetrischen losungen der einsteinschen
  gravitationsgleichungen im vakuum.
\newblock {\em Ark.\ Mat.\ Ast.\ Fys.}, 15(18):1--9, 1921.

\bibitem{Kottle18}
F.~Kottler.
\newblock {\"Uber die physikalischen Grundlagen der Einsteinschen
  Gravitationstheorie}.
\newblock {\em Ann.\ Physik (Leipzig), ser.~4}, 56(14):401--462, 1918.

\bibitem{Trotte59}
H.~F. Trotter.
\newblock On the product of semi-groups of operators.
\newblock {\em Proc.\ Amer.\ Math.\ Soc.}, 10(4):545--551, 1959.

\end{thebibliography}
\end{document}